\documentclass{Book_KW}
\title{Principles of\\Quantum Communication Theory:\\A Modern Approach}
\author{Sumeet Khatri, Ludovico Lami, and Mark M. Wilde}
\date{\today}

\numberwithin{equation}{section}

\usepackage[round]{natbib}

\usepackage{anyfontsize}  
\usepackage{microtype} 

\begin{document}

\hypersetup{pageanchor=false}
\maketitle 
\thispagestyle{plain}
\hypersetup{pageanchor=true}

\begin{frontmatter} 
\addtocounter{page}{1}
\pagestyle{plain}

\chapter*{Preface}

	[IN PROGRESS]

\newpage

\chapter*{Acknowledgements}

	[IN PROGRESS]

We dedicate this book to the memory of Jonathan P. Dowling. Jon was generous and kind-hearted, and he always gave all of his students his full, unwavering support. His tremendous impact on the lives of everyone who met him will ensure that his memory lives on and that he will not be forgotten. We will especially remember Jon's humor and his sharp wit. We are sure that, as he had promised, this book would have made the perfect doorstop for his office.

Sumeet Khatri acknowledges support from the National Science Foundation under Grant No.~1714215 and the Natural Sciences and Engineering Research Council of Canada postgraduate scholarship. Mark M.~Wilde acknowledges support from the National Science Foundation over the past decade (specifically from Grant Nos.~1350397, 1714215, 1907615, 2014010), and is indebted and grateful to Patrick Hayden for hosting him for a sabbatical at Stanford University during calendar year 2020, with support from Stanford QFARM and AFOSR (FA9550-19-1-0369).

\newpage





\tableofcontents 
	\cleardoublepage\phantomsection 
	
\newpage

%


\end{frontmatter}

\cleardoublepage 

\begin{mainmatter}

\chapter{Introduction}\label{chap-introduction}

	[IN PROGRESS]

\part{Preliminaries}[

	Before starting our journey though quantum communication protocols, it is necessary for us to learn about and understand the various mathematical and physical concepts involved in their construction and analysis. To this end, we begin in Chapter~\ref{chap-math_tools} by providing an overview of the mathematics required for understanding quantum communication protocols, and quantum information more broadly. Then, in Chapters~\ref{chap-QM_states_meas}--\ref{chap-QM_dist_meas}, we study the basic axioms of quantum mechanics, including quantum states and measurements (Chapter~\ref{chap-QM_states_meas}); followed by quantum channels, with the general theory and many examples (Chapter~\ref{chap-QM_channels}); followed by fundamental quantum information processing tasks, such as teleportation, super-dense coding, and hypothesis testing (Chapter~\ref{chap-QM_protocols}); and then distinguishability measures for states and channels, such as fidelity, trace distance, and diamond distance (Chapter~\ref{chap-QM_dist_meas}). Entropies and entanglement measures are crucial in quantifying the performance of quantum communication protocols, but they are also interesting in their own right, and they have applications in other areas of physics. In Chapters~\ref{chap-entropies}--\ref{chap-ent_measures}, we study these quantities in detail.
]\label{part-prelims}

\chapter{Mathematical Tools}\label{chap-math_tools}

	In this chapter, we learn about the various mathematical concepts required for the analysis of quantum communication protocols. We mostly provide a summary of the main definitions and results needed in later chapters, and we omit some proofs. For further details on the concepts presented here, as well as for proofs not explicitly given here, please consult the Bibliographic Notes (Section~\ref{math-tools:sec:bib-notes}) at the end of the chapter.
	
	Linear algebra forms the core mathematical foundation of quantum information theory for finite-dimensional quantum systems, and thus it is worthwhile for us to start by reviewing the basics of linear algebra, with an emphasis on linear operators. We then proceed to give a summary of several relevant definitions and results in real and convex analysis, probability theory, and semi-definite programming. Concepts from real analysis play an important role in quantum information theory. Indeed, as we discover later, the capacity of a quantum channel is defined as a limit, which is a core notion in real analysis. Convexity plays a prominent role as well. Not only is the set of quantum states a convex set, but also the operator Jensen inequality, a foundational statement about operator convex functions, is a fundamental inequality that leads to various quantum data-processing inequalities. The latter data-processing principle is one of the central tenets of quantum information that allows for placing limitations on the communication capacities of quantum channels. Probability theory is essential as well, due to the probabilistic nature of quantum mechanics and the inevitable and unpredictable errors that occur when communicating information over quantum channels. Finally, semi-definite programming is a remarkably useful tool, not only for analytical manipulations but also for numerically calculating relevant quantities of interest. Semi-definite programming has also played a pivotal role in many of the substantive advances that have taken place in quantum information theory during the past several decades, and so it has become one of the standard tools in the quantum information theorist's toolkit.

\section{Finite-Dimensional Hilbert Spaces}

	The primary mathematical object in quantum theory is the Hilbert space. We consider only finite-dimensional Hilbert spaces, denoted by $\mathcal{H}$, throughout this book, and we use $\dim(\mathcal{H})$ to denote the dimension of $\mathcal{H}$. Although we consider finite-dimensional spaces exclusively in this book, we note here that many of the statements and claims extend directly to the case of separable, infinite-dimensional Hilbert spaces, especially for operationally-defined tasks and information quantities. However, we do not delve into these details.
	
	A $d$-dimensional Hilbert space ($1 \leq d<\infty$) is defined to be a complex vector space equipped with an inner product\footnote{This definition suffices in the finite-dimensional case. More generally, a Hilbert space is a complete inner product space; please consult the Bibliographic Notes (Section~\ref{math-tools:sec:bib-notes}).}. We use the notation $\ket{\psi}$ to denote a vector in $\mathcal{H}$. An inner product is a function $\braket{\cdot}{\cdot}:\mathcal{H}\times\mathcal{H}\to\mathbb{C}$ that satisfies the following properties:
	\begin{itemize}
		\item\textit{Non-negativity}: $\braket{\psi}{\psi}\geq 0$ for all $\ket{\psi}\in\mathcal{H}$, and $\braket{\psi}{\psi}=0$ if and only if $\ket{\psi}=0$.
		\item\textit{Conjugate bilinearity}: For all $\ket{\psi_1},\ket{\psi_2},\ket{\phi_1},\ket{\phi_2}\in\mathcal{H}$ and  $\alpha_1,\beta_1,\alpha_2,\beta_2\in\mathbb{C}$,
			\begin{align}
				\braket{\alpha_1\psi_1+\beta_1\phi_1}{\alpha_2\psi_2+\beta_2\phi_2}&=\conj{\alpha_1}\alpha_2\braket{\psi_1}{\psi_2}+\conj{\alpha_1}\beta_2\braket{\psi_1}{\phi_2}\nonumber\\
				&\qquad +\conj{\beta_1}\alpha_2\braket{\phi_1}{\psi_2}+\conj{\beta_1}\beta_2\braket{\phi_1}{\phi_2}.
			\end{align}
		\item\textit{Conjugate symmetry}: $\braket{\psi}{\phi}=\conj{\braket{\phi}{\psi}}$ for all $\ket{\psi},\ket{\phi}\in\mathcal{H}$.
	\end{itemize}
	In the above, $\conj{\alpha}$ denotes the complex conjugate of $\alpha \in \mathbb{C}$. Throughout this book, the term ``Hilbert space'' always refers to a finite-dimensional Hilbert space.
	
	All $d$-dimensional Hilbert spaces are isomorphic to the vector space $\mathbb{C}^d$ equip\-ped with the Euclidean inner product. By two Hilbert spaces $\mathcal{H}$ and $\mathcal{H}'$ being isomorphic, we mean that there is a bijective linear mapping $U:\mathcal{H} \to \mathcal{H}'$ such that
	\begin{equation}
	\langle U \varphi | U \psi \rangle = \langle  \varphi |  \psi \rangle,
	\end{equation}
	for all $|\varphi\rangle,|\psi\rangle \in \mathcal{H}$, and $U$ is called an isomorphism. For the finite-dimensional case of interest for us, $U$ is a unitary operator (discussed in more detail in Section~\ref{sec-MT:notable-lin-ops}).
	Note that $\mathbb{C}^d$ is the vector space of $d$-dimensional column vectors with elements in $\mathbb{C}$. We let $\{\ket{i}\}_{i=0}^{d-1}$ denote an orthonormal basis, called the \textit{standard basis} or \textit{computational basis}, for the Hilbert space with respect to the Euclidean inner product. The vector~$\ket{i}$ is defined to be a column vector with its $(i+1)^{\text{th}}$ entry equal to one and all others equal to zero, so that
	\begin{equation}\label{eq-standard_basis_vectors}
		\ket{0}=\begin{pmatrix}1\\0\\0\\\vdots\\0\end{pmatrix},\quad\ket{1}=\begin{pmatrix} 0\\1\\0\\\vdots\\ 0\end{pmatrix},\quad\dotsc\quad,\quad\ket{d-1}=\begin{pmatrix}0\\0\\0\\\vdots\\ 1\end{pmatrix}.
	\end{equation}
	The inner product $\braket{i}{j}$ evaluates to $\braket{i}{j}=\delta_{i,j}$ for all $i,j \in \left\{0, \ldots,  d-1\right\}$, where the Kronecker delta function is defined as
	\begin{equation}
	\delta_{i,j} \coloneqq
	\begin{cases}
	0 & \text{ if } i \neq j\\
	1 & \text{ if } i = j.
	\end{cases}
	\end{equation}
	More generally, for two vectors $\ket{\psi}=\sum_{i=0}^{d-1} \alpha_i \ket{i}$ and $\ket{\phi}=\sum_{i=0}^{d-1} \beta_i\ket{i}$, with $\alpha_i=\braket{i}{\psi}\in\mathbb{C}$ and $\beta_i=\braket{i}{\phi}\in\mathbb{C}$ being the respective components of $\ket{\psi}$ and $\ket{\phi}$ in the standard basis, the inner product $\braket{\psi}{\phi}$ is defined as
	\begin{equation}
		\braket{\psi}{\phi} \coloneqq \sum_{i=0}^{d-1} \conj{\alpha_i} \beta_i .
	\end{equation} 
	The Euclidean norm of a vector $\ket{\psi}\in\mathcal{H}$, denoted by $\norm{\ket{\psi}}_2$, is the norm induced by the inner product, i.e.,
	\begin{equation}\label{eq:math-tools:euclid-norm}
		\norm{\ket{\psi}}_2\coloneqq\sqrt{\braket{\psi}{\psi}}.
	\end{equation}
	
	The \textit{Cauchy--Schwarz inequality} establishes an upper bound on the modulus of the inner product of two vectors $\ket{\psi},\ket{\phi}\in\mathcal{H}$ in terms of the product of their norms:
	\begin{equation}\label{eq-Cauchy_Schwarz}
		\abs{\braket{\psi}{\phi}}^2\leq\braket{\psi}{\psi}\cdot\braket{\phi}{\phi}=\norm{\ket{\psi}}_2^2\cdot\norm{\ket{\phi}}_2^2,
	\end{equation}
	with equality if and only if $\ket{\phi}=\alpha\ket{\psi}$ for some $\alpha\in\mathbb{C}$.
	
	Given a vector $\ket{\psi}\in\mathcal{H}$, its \textit{dual vector}, denoted by $\bra{\psi}$, is defined to be a linear functional from $\mathcal{H}$ to $\mathbb{C}$ such that $\bra{\psi}(\ket{\phi})=\braket{\psi}{\phi}$ for all $\ket{\phi}\in\mathcal{H}$. If $\ket{\psi}=\sum_{i=0}^{d-1}\alpha_i\ket{i}$, then $\bra{\psi}=\sum_{i=0}^{d-1}\conj{\alpha_i}\bra{i}$, where $\bra{i}$ can be interpreted, based on \eqref{eq-standard_basis_vectors}, as a row vector with its $(i+1)^{\text{th}}$ entry equal to one and all other entries equal to zero; i.e., $\bra{i}=(\ket{i})^{\t}$, where $(\cdot)^{\t}$ denotes the matrix transpose.

	The tensor product of vectors, operators, and Hilbert spaces plays an important role in quantum theory. For example, it is used to describe the state of multiple quantum systems. For two Hilbert spaces $\mathcal{H}_A$ and $\mathcal{H}_B$ with dimensions~$d_A$ and~$d_B$, respectively, along with associated orthonormal bases $\{\ket{i}_A\}_{i=0}^{d_A-1}$ and $\{\ket{j}_B\}_{j=0}^{d_B-1}$, the tensor product vector $\ket{i}_A \otimes \ket{j}_B$ is a vector in a $(d_Ad_B)$-dimensional Hilbert space with a one in its $(i\cdot d_B + j + 1)^{\text{th}}$ entry and zeros elsewhere. Notice here that we have employed the labels $A$ and $B$ in order to keep track of the Hilbert spaces of the vectors in the tensor product. Later on, when we move to the study of quantum information, we will see that the label $A$ can be associated to a quantum system in possession of ``Alice'' and the label $B$ can be associated to a quantum system in possession of ``Bob.'' As an example of the tensor-product vector $\ket{i}_A \otimes \ket{j}_B$, if $d_A = 2$, $d_B = 3$, $i=0$, and $j=2$, then
	\begin{equation}
	\ket{i}\otimes \ket{j} = \ket{0}\otimes \ket{2} = \begin{pmatrix}1\\0\end{pmatrix} \otimes \begin{pmatrix}0\\0\\1\end{pmatrix} = 
	\begin{pmatrix}1\cdot  \begin{pmatrix}0\\0\\1\end{pmatrix}\\0\cdot \begin{pmatrix}0\\0\\1\end{pmatrix}\end{pmatrix} =
	\begin{pmatrix} 0 \\ 0 \\ 1 \\ 0 \\ 0 \\ 0\end{pmatrix}. \label{eq:math-tools:baby-TP}
	\end{equation}
	More generally, for vectors $\ket{\psi}_A = \sum_{i=0}^{d_A-1} \alpha_i \ket{i}_A$ and $\ket{\phi}_B = \sum_{j=0}^{d_B -1} \beta_j \ket{j}_B$, the tensor-product vector $\ket{\psi}_A \otimes \ket{\phi}_B$ is given by
	\begin{align}
		\ket{\psi}_A \otimes \ket{\phi}_B &= \sum_{i=0}^{d_A-1}\alpha_i\ket{i}_A\otimes\ket{\phi}_B\label{eq-math_tools_vector_TP1}\\
		&=\sum_{i=0}^{d_A-1}\sum_{j=0}^{d_B -1} \alpha_i \beta_j \ket{i}_A \otimes\ket{j}_B.\label{eq-math_tools_vector_TP2}
	\end{align}
	As an example with $d_A =2$ and $d_B = 3$, we find that $\ket{\psi}_A \otimes \ket{\phi}_B$ can be calculated by a generalization of the ``stack-and-multiply'' procedure used in~\eqref{eq:math-tools:baby-TP}:
	\begin{equation}\label{eq:math-tools:baby-TP-plusplus}
	\ket{\psi}_A \otimes \ket{\phi}_B  = \begin{pmatrix}\alpha_0\\\alpha_1\end{pmatrix} \otimes \begin{pmatrix}\beta_0 \\ \beta_1 \\ \beta_2 \end{pmatrix} = 
	\begin{pmatrix}\alpha_0 \cdot  \begin{pmatrix}\beta_0 \\ \beta_1 \\ \beta_2\end{pmatrix}\\\alpha_1 \cdot \begin{pmatrix}\beta_0 \\ \beta_1 \\ \beta_2 \end{pmatrix}\end{pmatrix} =
	\begin{pmatrix} \alpha_0 \beta_0 \\ \alpha_0 \beta_1 \\ \alpha_0 \beta_2 \\ \alpha_1 \beta_0 \\ \alpha_1 \beta_1 \\ \alpha_1 \beta_2  \end{pmatrix}. 
	\end{equation}
	
	The tensor-product Hilbert space $\mathcal{H}_A\otimes\mathcal{H}_B$ is defined to be the Hilbert space spanned by the vectors $\ket{i}_A \otimes \ket{j}_B$ defined above:
	\begin{equation}\label{eq-math_tools_tensor_prod_def}
		\mathcal{H}_A\otimes\mathcal{H}_B \coloneqq \text{span}\{\ket{i}_A\otimes\ket{j}_B:0\leq i\leq d_A-1,0\leq j\leq d_B-1\}.
	\end{equation}
	The inner product on $\mathcal{H}_A\otimes\mathcal{H}_B$ is given by
	\begin{equation}
		(\bra{i}_A\otimes\bra{j}_B)(\ket{i'}_A\otimes\ket{j'}_B)=\langle i |i' \rangle \langle j |j' \rangle=\delta_{i,i'}\delta_{j,j'}
	\end{equation}
	for all $i,i',j,j'$ satisfying $0\leq i,i'\leq d_A-1$ and $0\leq j,j'\leq d_B-1$. The Hilbert space $\mathcal{H}_A\otimes\mathcal{H}_B$ consequently has dimension $d_Ad_B$. We often use the notation $\mathcal{H}_{AB}\equiv\mathcal{H}_A\otimes\mathcal{H}_B$, as well as the abbreviation $\ket{i,j}_{AB}\equiv\ket{i}_A\otimes\ket{j}_B$. We often also use the notation $\mathcal{H}_{A^n}\equiv\mathcal{H}_A^{\otimes n}$ to refer to the $n$-fold tensor product of $\mathcal{H}_A$.
	
	The \textit{direct sum} of $\mathcal{H}_A$ and $\mathcal{H}_B$, denoted by $\mathcal{H}_A\oplus\mathcal{H}_B$, is defined to be the Hilbert space of vectors of the form $\ket{\psi}_A\oplus\ket{\phi}_B$, with $\ket{\psi}_A\in\mathcal{H}_A$ and $\ket{\phi}_B\in\mathcal{H}_B$, where
	\begin{equation}
		\ket{\psi}_A\oplus\ket{\phi}_B\coloneqq  \begin{pmatrix} \ket{\psi}_A \\ \ket{\phi}_B \end{pmatrix}.
	\end{equation}
	In other words, $\mathcal{H}_A\oplus\mathcal{H}_B$ can be viewed as the Hilbert space of column vectors formed by stacking elements of the constituent Hilbert spaces. Observe that if $\mathcal{H}_A$ has the same dimension as $\mathcal{H}_B$, then we can write
	\begin{equation}\label{eq:math-tools:direct-sum-to-TP}
		\ket{\psi}_A\oplus\ket{\phi}_B = \ket{0} \otimes \ket{\psi}_A + \ket{1}\otimes \ket{\phi}_B, 
	\end{equation}
	where $\{ \ket{0}, \ket{1} \}$ is the standard basis for a two-dimensional Hilbert space.
	
	\begin{exercise}{exer-direct_sum_to_tensor_prod}
		Verify \eqref{eq:math-tools:direct-sum-to-TP}.
	\end{exercise}
	
	If $\{\ket{i}_A\}_{i=0}^{d_A-1}$ and $\{\ket{j}_B\}_{j=0}^{d_B-1}$ are orthonormal bases for $\mathcal{H}_A$ and $\mathcal{H}_B$, respectively, then
	\begin{equation}
		\left\{\begin{pmatrix} \ket{i}_A\\ 0\end{pmatrix}:0\leq i\leq d_A-1\right\}\cup\left\{\begin{pmatrix} 0\\\ket{j}_B\end{pmatrix}:0\leq j\leq d_B-1\right\}
	\end{equation}
	is an orthonormal basis for $\mathcal{H}_A\oplus\mathcal{H}_B$ under the inner product
	\begin{equation}
		(\bra{i}_A\oplus \bra{j}_B)(\ket{i'}_A\oplus\ket{j'}_B)=\braket{i}{i'}+\braket{j}{j'}=\delta_{i,i'}+\delta_{j,j'}.
	\end{equation}
	for all $0\leq i,i'\leq d_A-1$ and $0\leq j,j'\leq d_B-1$. Consequently, $\mathcal{H}_A\oplus\mathcal{H}_B$ has dimension $d_A+d_B$. One of the simplest examples of a direct-sum Hilbert space is $\mathbb{C}\oplus\mathbb{C}=\mathbb{C}^2$. More generally, the $d$-fold direct sum $\mathbb{C}^{\oplus d}$ is equal to $\mathbb{C}^d$.
	
	If $\mathcal{H}$ is a $d$-dimensional Hilbert space, then the $k$-fold direct sum $\mathcal{H}^{\oplus k}$ is a $kd$-dimensional Hilbert space. Consequently, it is isomorphic to $\mathbb{C}^k\otimes\mathcal{H}$, and the isomorphism is a generalization of the simple example presented in \eqref{eq:math-tools:direct-sum-to-TP}. Indeed, let $\mathcal{H}_X\equiv\mathbb{C}^k$, with orthonormal basis $\{\ket{i}_X\}_{i=0}^{k-1}$, and let $\mathcal{H}_A\equiv\mathcal{H}$, with orthonormal basis $\{\ket{j}_A\}_{j=0}^{d-1}$. We then have the correspondence 
	\begin{equation}\label{eq-block_diag_iso}
		\ket{i}_X\otimes \ket{j}_A \leftrightarrow \begin{pmatrix} 0 \\ \vdots \\ 0 \\ \ket{j}_A \\ 0 \\ \vdots \\ 0 \end{pmatrix},
	\end{equation}
	holding for all $0\leq i\leq k-1$ and all $0\leq j\leq d-1$, where on the right-hand side there is a one in the $(i\cdot d + j+1)^{\text{th}}$ entry of the column vector and zeros elsewhere. Then, for an element $\ket{\psi_0}_A\oplus\ket{\psi_1}_A\oplus\dotsb\oplus\ket{\psi_{k-1}}_A\in\mathcal{H}_A^{\oplus k}$, we have 
	\begin{equation}\label{eq-block_diag_iso_2}
		\begin{pmatrix} \ket{\psi_0}_A\\ \ket{\psi_1}_A \\ \vdots \\ \ket{\psi_{k-1}}_A \end{pmatrix} \leftrightarrow \sum_{i=0}^{k-1} \ket{i}_X\otimes\ket{\psi_i}_A.
	\end{equation}
	The isomorphism between $\mathcal{H}^{\oplus k}$ and $\mathbb{C}^k\otimes\mathcal{H}$ given by \eqref{eq-block_diag_iso} and \eqref{eq-block_diag_iso_2} is relevant in the context of superpositions of quantum states and entanglement.

\section{Linear Operators}\label{sec-math_tools-lin_ops}

	Linear operators are relevant in quantum theory for describing states of quantum systems, as well as physical evolutions of the states, including measurements and unitary evolutions as special cases of general physical evolutions. Given a Hilbert space $\mathcal{H}_A$ with dimension $d_A$ and a Hilbert space $\mathcal{H}_B$ with dimension $d_B$, a linear operator $X:\mathcal{H}_A\to \mathcal{H}_B$ is defined to be a function such that
	\begin{equation}
		X(\alpha\ket{\psi}_A+\beta\ket{\phi}_A)=\alpha X\ket{\psi}_A+\beta X\ket{\phi}_A
	\end{equation}
	for all $\alpha,\beta\in\mathbb{C}$ and $\ket{\psi}_A,\ket{\phi}_A\in\mathcal{H}_A$. For clarity, we sometimes write $X_{A\to B}$ to explicitly indicate the input and output Hilbert spaces of the linear operator $X$.
 
	Given two vectors $\ket{\psi}_A\in\mathcal{H}_A$ and $\ket{\phi}_B\in\mathcal{H}_B$, one can construct an associated linear operator from $A$ to $B$ given by $\ket{\phi}_B\bra{\psi}_A$, which acts on an arbitrary input $\ket{\chi}_A\in\mathcal{H}_A$ as
        \begin{equation}
            \left(\ket{\phi}_B\bra{\psi}_A\right)\ket{\chi}_A = \ket{\phi}_B \,\braket{\psi}{\chi}.
        \end{equation}

	We use $\mathbbm{1}$ to denote the identity operator, which is defined as the unique linear operator such that $\mathbbm{1}\ket{\psi}=\ket{\psi}$ for every vector $\ket{\psi}$. For clarity, when needed, we write $\mathbbm{1}_d$ to indicate the identity operator acting on a $d$-dimensional Hilbert space.

	\begin{exercise}{exer-identity_operator}
		Given an orthonormal basis $\{\ket{e_k}\}_{k=1}^d$ for a $d$-dimensional Hilbert space, prove that
		\begin{equation}\label{eq-math_tools_identity_ONB}
			\mathbbm{1}_d=\sum_{k=1}^d \ketbra{e_k}{e_k}.
		\end{equation}
	\end{exercise}
	
	We denote the set of all linear operators from $\mathcal{H}_A$ to $\mathcal{H}_B$  by $\Lin(\mathcal{H}_A,\mathcal{H}_B)$. If $\mathcal{H}_A=\mathcal{H}_B$, then $\Lin(\mathcal{H}_A)\coloneqq\Lin(\mathcal{H}_A,\mathcal{H}_A)$, and we sometimes indicate that $\mathcal{H}_A$ is the input and output Hilbert space of $X\in\Lin(\mathcal{H}_A)$ by writing $X_A$. In particular, we often write $X_{AB}$ when referring to linear operators in $\Lin(\mathcal{H}_A\otimes\mathcal{H}_B)$, i.e., when referring to linear operators acting on a tensor-product Hilbert space.
	
	The set $\Lin(\mathcal{H}_A,\mathcal{H}_B)$ is itself a $d_Ad_B$-dimensional vector space. The standard basis for $\Lin(\mathcal{H}_A,\mathcal{H}_B)$ is defined to be
	\begin{equation}\label{eq-standard_linear_op_basis}
		\{\ket{i}_B\bra{j}_A:0\leq i\leq d_B-1,\, 0\leq j\leq d_A-1\}.
	\end{equation}
	By applying \eqref{eq-standard_basis_vectors}, we see that the operator $\ket{i}_B\bra{j}_A$ has a matrix representation as a $d_B\times d_A$ matrix with the $\left(i+1,j+1\right)^{\text{th}}$ entry equal to one and all other entries equal to zero, i.e.,
	\begin{equation}
		\begin{aligned}
		\ket{0}_B\bra{0}_A=\begin{pmatrix} 1 & 0 & \dotsb & 0 \\ 0 & 0 & \dotsb & 0 \\ \vdots & \vdots & \ddots & 0 \\ 0 & 0 & \dotsb & 0\end{pmatrix},&\quad \ket{0}_B\bra{1}_A=\begin{pmatrix} 0 & 1 & \dotsb & 0 \\ 0 & 0 & \dotsb & 0 \\ \vdots & \vdots & \ddots & 0 \\ 0 & 0 & \dotsb & 0\end{pmatrix},\dotsc,\\
		\ket{d_B-1}_B\bra{d_A-1}_A&=\begin{pmatrix} 0 & 0 & \dotsb & 0 \\ 0 & 0 & \dotsb & 0 \\ \vdots & \vdots & \ddots & 0 \\ 0 & 0 & \dotsb & 1\end{pmatrix}.
		\end{aligned}
	\end{equation}
	Using this basis, we can write a linear operator $X\in\Lin(\mathcal{H}_A,\mathcal{H}_B)$ as
	\begin{equation}\label{eq-linear_op_basis_expansion}
		X_{A\to B}=\sum_{i=0}^{d_B-1}\sum_{j=0}^{d_A-1}X_{i,j}\ket{i}_B\bra{j}_A,
	\end{equation}
	where $X_{i,j}\coloneqq\bra{i}_BX\ket{j}_A$. This follows because
	\begin{align}
	X_{A\to B} & = \mathbbm{1}_B X_{A\to B} \mathbbm{1}_A \\
	& = \left(\sum_{i=0}^{d_B-1} |i\rangle\!\langle i|_B\right) X_{A\to B} \left(\sum_{j=0}^{d_A-1} |j\rangle\!\langle j|_A \right)\\
    & = \sum_{i=0}^{d_B-1}\sum_{j=0}^{d_A-1} |i\rangle\!\langle i|_B X_{A\to B}  |j\rangle\!\langle j|_A \\
	&  = \sum_{i=0}^{d_B-1} \sum_{j=0}^{d_A-1} \langle i|_B X_{A\to B}  |j\rangle_A |i\rangle_B \langle j|_A \\
	& = \sum_{i=0}^{d_B-1}\sum_{j=0}^{d_A-1}X_{i,j}\ket{i}_B\bra{j}_A.
	\end{align}
	We can thus interpret a linear operator $X\in\Lin(\mathcal{H}_A,\allowbreak\mathcal{H}_B)$ as a $d_B\times d_A$ matrix with the $(i+1,j+1)^{\text{th}}$ element equal to $X_{i,j}=\bra{i}_BX\ket{j}_A$, where $0\leq i\leq d_B-1$ and $0\leq j\leq d_A-1$. For example, if $d_A=2$ and $d_B=3$, then
	\begin{equation}
		X=\begin{pmatrix} X_{0,0} & X_{0,1} \\ X_{1,0} & X_{1,1} \\ X_{2,0} & X_{2,1} \end{pmatrix}.
	\end{equation}
	
	\begin{exercise}{exer-lin_op_rows_columns}
		Show that every linear operator $X\in\Lin(\mathcal{H}_A,\mathcal{H}_B)$, expressed as in \eqref{eq-linear_op_basis_expansion}, can be written as
		\begin{equation}\label{eq-linear_op_row_column_abstract}
			X_{A\to B}=\sum_{i=0}^{d_B-1}\ket{i}_B\bra{\psi_i}_A=\sum_{j=0}^{d_A-1}\ket{\phi_j}_B\bra{j}_A,
		\end{equation}
		where $\{\bra{\psi_i}_A\}_{i=0}^{d_B-1}$ and $\{\ket{\phi_j}_B\}_{j=0}^{d_A-1}$ are the rows and columns, respectively, of~$X$.
	\end{exercise}
	
	\subsection{Tensor Product}
	
	Given two linear operators $X\in\Lin(\mathcal{H}_A,\mathcal{H}_B)$ and $Y\in\Lin(\mathcal{H}_{A'},\mathcal{H}_{B'})$, their tensor product $X\otimes Y$ is the unique linear operator in $\Lin(\mathcal{H}_{A}\otimes\mathcal{H}_{A'},\mathcal{H}_B\otimes\mathcal{H}_{B'})$ such that 
	\begin{equation}
		(X\otimes Y)(\ket{\psi}_A\otimes\ket{\phi}_{A'})=X\ket{\psi}_{A}\otimes Y\ket{\phi}_{A'}
	\end{equation}
	for all $\ket{\psi}_A\in\mathcal{H}_A$ and $\ket{\phi}_{A'}\in\mathcal{H}_{A'}$. The matrix representation of $X\otimes Y$ is the Kronecker product of the matrix representations of $X$ and $Y$, which is a matrix generalization of the ``stack-and-multiply'' procedure from \eqref{eq:math-tools:baby-TP-plusplus}. For example, if $d_A=d_B=2$ and $d_{A'} = d_{B'} =3$, then
	\begin{align}
	X \otimes Y & =
	\begin{pmatrix}
	X_{0,0} & X_{0,1} \\
	X_{1,0} & X_{1,1}
	\end{pmatrix} \otimes 
	\begin{pmatrix}
	Y_{0,0} & Y_{0,1} & Y_{0,2} \\
	Y_{1,0} & Y_{1,1} & Y_{1,2} \\
	Y_{2,0} & Y_{2,1} & Y_{2,2} 
	\end{pmatrix} \\
	& = 	\begin{pmatrix}
	X_{0,0} \cdot	\begin{pmatrix}
	Y_{0,0} & Y_{0,1} & Y_{0,2} \\
	Y_{1,0} & Y_{1,1} & Y_{1,2} \\
	Y_{2,0} & Y_{2,1} & Y_{2,2} 
	\end{pmatrix} & X_{0,1}\cdot	\begin{pmatrix}
	Y_{0,0} & Y_{0,1} & Y_{0,2} \\
	Y_{1,0} & Y_{1,1} & Y_{1,2} \\
	Y_{2,0} & Y_{2,1} & Y_{2,2} 
	\end{pmatrix} \\
	X_{1,0} \cdot 	\begin{pmatrix}
	Y_{0,0} & Y_{0,1} & Y_{0,2} \\
	Y_{1,0} & Y_{1,1} & Y_{1,2} \\
	Y_{2,0} & Y_{2,1} & Y_{2,2} 
	\end{pmatrix}& X_{1,1} \cdot	\begin{pmatrix}
	Y_{0,0} & Y_{0,1} & Y_{0,2} \\
	Y_{1,0} & Y_{1,1} & Y_{1,2} \\
	Y_{2,0} & Y_{2,1} & Y_{2,2} 
	\end{pmatrix}
	\end{pmatrix} \\
	& =\begin{pmatrix}
	 	X_{0,0} Y_{0,0} & X_{0,0} Y_{0,1} & X_{0,0}Y_{0,2} & 	X_{0,1} Y_{0,0} & X_{0,1} Y_{0,1} & X_{0,1} Y_{0,2} \\
	X_{0,0}Y_{1,0} & X_{0,0}Y_{1,1} & X_{0,0}Y_{1,2} &	X_{0,1} Y_{1,0} & X_{0,1} Y_{1,1} & X_{0,1} Y_{1,2} \\
	X_{0,0} Y_{2,0} & X_{0,0} Y_{2,1} & X_{0,0} Y_{2,2} 
 &
	X_{0,1} Y_{2,0} & X_{0,1} Y_{2,1} & X_{0,1} Y_{2,2} 
 \\
	 	X_{1,0} Y_{0,0} & X_{1,0} Y_{0,1} & X_{1,0}Y_{0,2} &	X_{1,1} Y_{0,0} & X_{1,1} Y_{0,1} & X_{1,1} Y_{0,2} \\
	X_{1,0}Y_{1,0} & X_{1,0}Y_{1,1} & X_{1,0}Y_{1,2} &	X_{1,1} Y_{1,0} & X_{1,1} Y_{1,1} & X_{1,1} Y_{1,2} \\
	X_{1,0} Y_{2,0} & X_{1,0} Y_{2,1} & X_{1,0} Y_{2,2} 
 &
	X_{1,1} Y_{2,0} & X_{1,1} Y_{2,1} & X_{1,1} Y_{2,2} 
	\end{pmatrix}.
	\end{align}
	
	\subsection{Image, Kernel, and Support}
	
	The \textit{image} of a linear operator $X\in\Lin(\mathcal{H}_A,\mathcal{H}_B)$, denoted by $\text{im}(X)$, is the set defined as
	\begin{equation}\label{eq-image}
		\text{im}(X)\coloneqq\{\ket{\phi}_B \in \mathcal{H}_B : \ket{\phi}_B = X\ket{\psi}_A, \ \ket{\psi}_A\in\mathcal{H}_A\}.
	\end{equation}
	It is also known as the column space or range of $X$.
	The image of $X$ is a subspace of~$\mathcal{H}_B$. The \textit{rank} of $X$, denoted by $\text{rank}(X)$, is defined\footnote{The rank of a linear operator can also be equivalently defined as the number of its singular values; see Theorem~\ref{thm-SVD}.} to be the dimension of $\text{im}(X)$. Note that $\text{rank}(X)\leq\min\{d_A,d_B\}$ for all $X\in\Lin(\mathcal{H}_A,\mathcal{H}_B)$.
	
	The \textit{kernel} of a linear operator 
$X\in\Lin(\mathcal{H}_A,\mathcal{H}_B)$, denoted by $\ker(X)$, is defined to be the set of vectors in the input space $\mathcal{H}_A$ of $X$ for which the output is the zero vector; i.e.,
	\begin{equation}\label{eq-kernel-operator}
		\ker(X)\coloneqq\{\ket{\psi}_A \in\mathcal{H}_A: X \ket{\psi}_A = 0\}.
	\end{equation}
	It is also known as the null space of $X$.
	The following dimension formula holds:
	\begin{equation}\label{eq-math_tools_rank_nullity}
		d_A = \text{rank}(X) + \dim(\ker(X)),
	\end{equation}
	and it is known as the \textit{rank-nullity theorem} (the quantity $\dim(\ker(X))$ is called the nullity of $X$).
	
	The \textit{support} of a linear operator $X\in\Lin(\mathcal{H}_A,\mathcal{H}_B)$, denoted by $\supp(X)$, is defined to be the orthogonal complement of its kernel:
	\begin{equation}\label{eq-support}
		\supp(X)\coloneqq \ker(X)^\perp \coloneqq \{\ket{\psi}\in\mathcal{H}_A:\braket{\psi}{\phi}=0 \ \ \forall \ket{\phi} \in \ker(X) \}.
	\end{equation}
	It is also known as the row space or coimage of $X$.
	
	\begin{figure}
		\centering
		\includegraphics[scale=1]{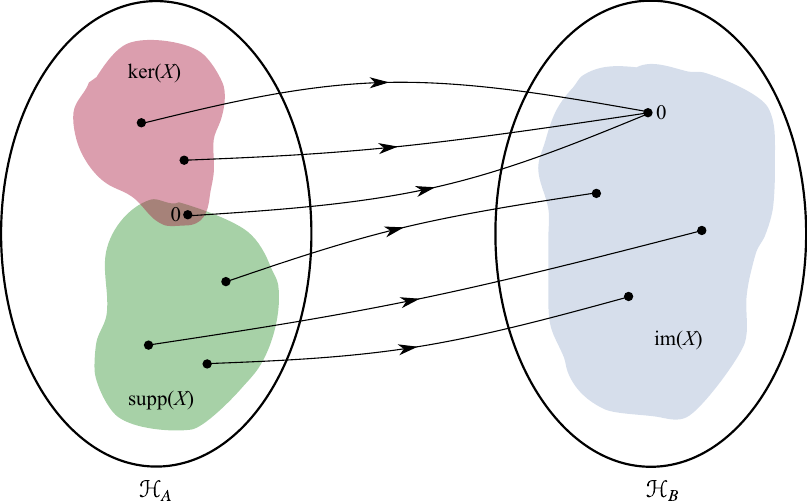}
		\caption{Visual representation of the subspaces $\text{im}(X)$, $\ker(X)$, and $\supp(X)$ corresponding to a linear operator $X\in\Lin(\mathcal{H}_A,\mathcal{H}_B)$. Note that only the zero vector is contained in both $\ker(X)$ and $\supp(X)$.}\label{fig-lin_operator}
	\end{figure}
	
	See Figure~\ref{fig-lin_operator} for a visual representation of the subspaces $\text{im}(X)$, $\ker(X)$, and $\supp(X)$ corresponding to a linear operator $X\in\Lin(\mathcal{H}_A,\mathcal{H}_B)$. We use the notions of support and kernel extensively in Chapter~\ref{chap-entropies}, when proving properties of quantum relative entropy and its variants, which are core distinguishability measures in quantum information.

	A linear operator $X\in\Lin(\mathcal{H}_A,\mathcal{H}_B)$ is called \textit{injective} (or \textit{one-to-one}) if, for all $\ket{\psi},\ket{\phi}\in\mathcal{H}_A$, $X\ket{\psi}=X\ket{\phi}$ implies $\ket{\psi}=\ket{\phi}$. A necessary and sufficient condition for $X$ to be injective is that the kernel of $X$ contains only the zero vector (i.e., the column vector for which all of the elements are equal to zero), which implies that $\dim(\ker(X))=0$.
	
	A linear operator $X\in\Lin(\mathcal{H}_A,\mathcal{H}_B)$ is called \textit{surjective} (or \textit{onto}) if, for all $\ket{\phi}\in\mathcal{H}_B$, there exists $\ket{\psi}\in\mathcal{H}_A$ such that $X\ket{\psi}=\ket{\phi}$. A necessary and sufficient condition for $X$ to be surjective is that $\rank(X)=d_B$.
	
	\begin{exercise}{exer-injective_surjective}
		Prove that a linear operator $X\in\Lin(\mathcal{H})$ with the same, finite-dimensional input and output Hilbert space $\mathcal{H}$ is injective if and only if it is surjective. (\textit{Hint}: use the rank-nullity theorem in \eqref{eq-math_tools_rank_nullity}.)
	\end{exercise}
	
	A linear operator $X\in\Lin(\mathcal{H})$ that is both injective and surjective is known as a \textit{bijection}. By definition, every bijection is \textit{invertible}, meaning that there exists a unique linear operator, denoted by $X^{-1}$, such that $XX^{-1}=X^{-1}X=\mathbbm{1}$.
	
	\subsection{Trace}
	
	The \textit{trace} of a linear operator $X\in\Lin(\mathcal{H})$ acting on a $d$-dimensional Hilbert space~$\mathcal{H}$ is defined as
	\begin{equation}\label{eq-trace}
		\Tr[X]\coloneqq \sum_{i=0}^{d-1}\bra{i}X\ket{i},
	\end{equation}
	which can be interpreted as the sum of the diagonal elements of the matrix corresponding to $X$ in the standard basis.
	
	\begin{exercise}{exer-trace}
		Prove that the trace of a linear operator is independent of the choice of the basis used in \eqref{eq-trace}. In other words, prove that $\sum_{i=0}^{d-1} \bra{i}X\ket{i}=\sum_{k=1}^d\bra{e_k}X\ket{e_k}$ for every orthonormal basis $\{\ket{e_k}\}_{k=1}^d$. (\textit{Hint}: use \eqref{eq-math_tools_identity_ONB}.)
	\end{exercise}
	
	The trace satisfies the \textit{cyclicity} property: for all pairs of operators $X,Y\in\Lin(\HH)$, it holds that
        \begin{equation}
            \Tr[XY]=\Tr[YX].
        \end{equation}
    Similarly, for $X,Y,Z\in\Lin(\mathcal{H})$,
	\begin{equation}\label{eq-trace_cyclic}
		\Tr[XYZ]=\Tr[YZX]=\Tr[ZXY].
	\end{equation}
	More generally, the cyclicity property holds for linear operators with different input and output Hilbert spaces: for $Z_{A\to B} \in \Lin(\mathcal{H}_A,\mathcal{H}_{B})$, $Y_{B\to C} \in \Lin(\mathcal{H}_B,\mathcal{H}_{C})$, and $X_{C\to A} \in \Lin(\mathcal{H}_C,\mathcal{H}_{A})$,
	\begin{align}
	\Tr[X_{C\to A} Y_{B\to C} Z_{A\to B}] & =\Tr[Y_{B\to C}Z_{A\to B}X_{C\to A}]\label{eq-trace_cyclic1}\\
	& =\Tr[Z_{A\to B} X_{C\to A}Y_{B\to C}].\label{eq-trace_cyclic2}
	\end{align}
	
	\begin{exercise}{exer-trace_basic}
		\begin{enumerate}
			\item Prove the equalities in \eqref{eq-trace_cyclic1} and \eqref{eq-trace_cyclic2}.
			\item Prove that $\Tr[X\otimes Y]=\Tr[X]\Tr[Y]$ for all $X\in\Lin(\mathcal{H}_A)$ and $Y\in\Lin(\mathcal{H}_B)$.
		\end{enumerate}
	\end{exercise}
	
	\subsection{Transpose and Conjugate Transpose}
	
	Consider $X\in\Lin(\mathcal{H}_A,\mathcal{H}_B)$ as written in \eqref{eq-linear_op_basis_expansion}. The \textit{transpose} of $X$ is denoted by~$X^{\t}$ or alternatively by $\T(X)$, and it is defined as
	\begin{equation}\label{eq:math-tools:transpose-op}
		X^{\t}\equiv\T(X)\coloneqq\sum_{i=0}^{d_B-1}\sum_{j=0}^{d_A-1}X_{i,j}\ket{j}_A\bra{i}_B. 
	\end{equation}
	Note that the transpose is basis dependent, in the sense that it is defined with respect to a particular basis (in the case above, we have defined it with respect to the standard bases of $\mathcal{H}_A$ and $\mathcal{H}_B$). Furthermore, taking the transpose with respect to one orthonormal basis can lead to an  operator different from that found by taking the transpose with respect to a different orthonormal basis. In this sense, we could more precisely refer to the operation in \eqref{eq:math-tools:transpose-op} as the ``standard transpose.'' The standard transpose can also be understood as a linear superoperator (an operator on operators) with the following representation:
	\begin{equation}
	\T(X) = \sum_{i=0}^{d_B-1} \sum_{j=0}^{d_A - 1} \left(\ket{j}_A \bra{i}_B \right) X \left(\ket{j}_A \bra{i}_B\right) = \sum_{i=0}^{d_B-1}\sum_{j=0}^{d_A-1}X_{i,j}\ket{j}_A\bra{i}_B.
	\label{eq:math-tools:transpose-superop}
	\end{equation}
	Superoperators are discussed in more detail in Section~\ref{sec:math-tools:super-ops}.
	
	The \textit{conjugate transpose} of $X\in\Lin(\mathcal{H}_A,\mathcal{H}_B)$, also known as the \textit{Hermitian conjugate} or the \textit{adjoint} of $X$, is the linear operator $X^\dagger\in\Lin(\mathcal{H}_B,\mathcal{H}_A)$ defined as
	\begin{equation}
		X^{\dagger}\coloneqq \sum_{i=0}^{d_B-1}\sum_{j=0}^{d_A-1}\conj{X_{i,j}}\ket{j}_A\bra{i}_B.
	\end{equation}
	The adjoint of $X$ is the unique linear operator that satisfies
	\begin{equation}\label{eq:math-tools:herm-conj-basis-ind}
		\braket{\phi}{X\psi}=\braket{X^\dagger\phi}{\psi}
	\end{equation}
	for all $\ket{\psi}\in\mathcal{H}_A$ and  $\ket{\phi}\in\mathcal{H}_B$.
	 
	\begin{exercise}{exer-dagger}
		Prove that the conjugate transpose is a basis-independent operation, i.e., that one does not need to specify a basis in order to take the conjugate transpose of a linear operator.
	\end{exercise}
	
	\subsection{Hilbert--Schmidt Inner Product, Vectorization, and Transpose Trick}
	
	On the vector space $\Lin(\mathcal{H}_A,\mathcal{H}_B)$, we define the \textit{Hilbert--Schmidt inner product} as follows:
	\begin{equation}\label{eq-Hilbert_Schmidt_inner_prod}
		\inner{X}{Y}\coloneqq \Tr[X^\dagger Y],\quad X,Y\in\Lin(\mathcal{H}_A,\mathcal{H}_B).
	\end{equation}
	A key application of the Hilbert--Schmidt inner product in quantum mechanics is the Born rule: In Section~\ref{subsec-meas}, we will learn that the probability of obtaining an outcome in an experiment is equal to the Hilbert--Schmidt inner product of the state in which the system is prepared and a measurement operator corresponding to the measurement outcome.
	With this inner product, the vector space $\Lin(\mathcal{H}_A,\mathcal{H}_B)$ becomes an inner product space, and thus a Hilbert space. The basis defined in \eqref{eq-standard_linear_op_basis} is an orthonormal basis for $\Lin(\mathcal{H}_A,\mathcal{H}_B)$ under this inner product. The \textit{Hilbert--Schmidt norm} (or \textit{Schatten 2-norm}) of an operator $X \in \Lin(\mathcal{H}_A,\mathcal{H}_B)$ is defined from the Hilbert--Schmidt inner product as follows:
	\begin{equation}
		\norm{X}_2 \coloneqq \sqrt{\inner{X}{X}},
	\end{equation}
	which is a generalization of the Euclidean norm in \eqref{eq:math-tools:euclid-norm}. The Cauchy--Schwarz inequality for the Hilbert--Schmidt inner product is as follows:
	\begin{equation}\label{eq-Cauchy_Schwarz_HS}
		\abs{\inner{X}{Y}}^2\leq\inner{X}{X}\cdot \inner{Y}{Y} = \norm{X}^2_2 \cdot \norm{Y}^2_2,
	\end{equation}
	for all $X,Y\in\Lin(\mathcal{H}_A,\mathcal{H}_B)$. Observe that the inequality in \eqref{eq-Cauchy_Schwarz_HS}  is a generalization of that in \eqref{eq-Cauchy_Schwarz}.
	
	The Hilbert space $\Lin(\mathcal{H}_A,\mathcal{H}_B)$ is isomorphic to the tensor-product Hilbert space $\mathcal{H}_A\otimes\mathcal{H}_B$, where the isomorphism is defined by
	\begin{equation}
	\label{eq-vec_operation}
		\ket{i}_B\bra{j}_A\leftrightarrow \ket{j}_A\otimes\ket{i}_B\eqqcolon\text{vec}(\ket{i}_B\bra{j}_A)
	\end{equation}
	for all $i \in \{0, \ldots, d_B-1\}$ and  $j \in \{0,\ldots, d_A-1\}$. The operation vec on the right in \eqref{eq-vec_operation} is the ``vectorize'' operation, which transposes the rows of a $d_B \times d_A$ matrix with respect to the standard basis and then stacks the resulting columns in order one on top of the next in order to construct a $d_A d_B$-dimensional column vector. Specifically, for a linear operator $X\in\Lin(\mathcal{H}_A,\mathcal{H}_B)$ written as in \eqref{eq-linear_op_basis_expansion},
	\begin{equation}\label{eq-math_tools:lin_op_vec}
		\text{vec}(X)=\sum_{i=0}^{d_B-1}\sum_{j=0}^{d_A-1}X_{i,j}\ket{j}_A\otimes\ket{i}_B.
	\end{equation}
	
	The following are useful identities involving the vec operation that we call upon repeatedly throughout this book.
	\begin{enumerate}
		\item For every linear operator $X\in\Lin(\mathcal{H}_A,\mathcal{H}_B)$,
			\begin{equation}\label{eq-math_tools:vec}
				\text{vec}(X)=(\mathbbm{1}_A\otimes X_{A\to B})\ket{\Gamma}_{AA},
			\end{equation}
			where
			\begin{equation}\label{eq-max_ent_vector}
				\ket{\Gamma}_{AA}\coloneqq\sum_{i=0}^{d_A-1} \ket{i,i}_{AA}.
			\end{equation}
			For reasons that will become clear later, we refer to $\ket{\Gamma}_{AA}$ as the ``maximally entangled vector.'' Note that $\ket{\Gamma}_{AA}=\text{vec}(\mathbbm{1}_A)$. For clarity, when needed, we write $\ket{\Gamma_d}$ to refer to the vector defined in \eqref{eq-max_ent_vector} when each Hilbert space has dimension $d$.
			
			We also note that, for two vectors $\ket{\psi}_B\in\mathcal{H}_B$ and $\ket{\phi}_A\in\mathcal{H}_A$,
			\begin{equation}\label{eq-vec_operation_vectors}
				\text{vec}(\ket{\psi}_B\bra{\phi}_A)=\conj{\ket{\phi}}_A\otimes\ket{\psi}_B.
			\end{equation}
			
			\begin{exercise}{exer-vec_basic}
				\begin{enumerate}
					\item[1.] Prove \eqref{eq-math_tools:vec}.
					\item[2.] Prove the equality in \eqref{eq-vec_operation_vectors} by writing both $\ket{\psi}_B$ and $\ket{\phi}_A$ in terms of the orthonormal bases $\{\ket{i}_B\}_{i=0}^{d_B-1}$ and $\{\ket{j}_A\}_{j=0}^{d_A-1}$, respectively, and using~\eqref{eq-vec_operation}.
				\end{enumerate}
			\end{exercise}
		
		\item For every vector $\ket{\psi}_{AB}\in\mathcal{H}_A\otimes\mathcal{H}_{B}$, there exists a unique linear operator $X_{A\to B}^\psi\in\Lin(\mathcal{H}_A,\mathcal{H}_B)$ such that
			\begin{equation}\label{eq-pure_state_vec2}
				\ket{\psi}_{AB}=(\mathbbm{1}_A\otimes X_{A\to B}^\psi)\ket{\Gamma}_{AA}=\text{vec}(X_{A\to B}^{\psi}).
			\end{equation}
			In particular, if $\ket{\psi}_{AB}=\sum_{i=0}^{d_A-1}\sum_{j=0}^{d_B-1}\alpha_{i,j}\ket{i,j}_{AB}$, then
			\begin{equation}
				X_{A\to B}^{\psi}=\sum_{j=0}^{d_B-1}\sum_{i=0}^{d_A-1}\alpha_{i,j}\ket{j}_B\bra{i}_A. \label{eq-MT:bipartite-vec-to-X}
			\end{equation}
			Alternatively, there exists a linear operator $Y_{B\to A}^\psi\allowbreak\in\Lin(\mathcal{H}_B,\mathcal{H}_A)$ such that
			\begin{equation}\label{eq-pure_state_vec}
				\ket{\psi}_{AB}=(Y_{B\to A}^{\psi}\otimes\mathbbm{1}_B)\ket{\Gamma}_{BB}.
			\end{equation}
                This operator is given by
			\begin{equation}
				Y_{B\to A}^{\psi}=\sum_{j=0}^{d_B-1}\sum_{i=0}^{d_A-1}\alpha_{i,j}\ket{i}_A\bra{j}_B,\label{eq-MT:bipartite-vec-to-Y}
			\end{equation}
			and one finds by inspection of \eqref{eq-MT:bipartite-vec-to-X}--\eqref{eq-MT:bipartite-vec-to-Y} that $X_{A\to B}^{\psi} = \T(Y_{B\to A}^{\psi})$.
	
		\item\textit{Transpose trick}: For every linear operator $X\in\Lin(\mathcal{H}_A,\mathcal{H}_B)$, the following equality holds:
			\begin{equation}\label{eq-transpose_trick}
				(\mathbbm{1}_A\otimes X_{A\to B})\ket{\Gamma}_{AA}=((X^{\t})_{B\to A}\otimes\mathbbm{1}_B)\ket{\Gamma}_{BB}.
			\end{equation}
		
		\item For every linear operator $X\in\Lin(\mathbb{C}^d)$,
			\begin{equation}\label{eq-trace_identity}
				\Tr[X]=\bra{\Gamma}(X\otimes\mathbbm{1})\ket{\Gamma}=\bra{\Gamma}(\mathbbm{1}\otimes X)\ket{\Gamma}=\bra{\Gamma}\text{vec}(X).
			\end{equation}
	\end{enumerate}
	
	\begin{exercise}{exer-vec_identities}
		\begin{enumerate}
			\item Prove \eqref{eq-transpose_trick}.
			\item Prove \eqref{eq-trace_identity}.
			\item Let $X\in\Lin(\mathcal{H}_A,\mathcal{H}_B)$, $Y\in\Lin(\mathcal{H}_C,\mathcal{H}_A)$, and $Z\in\Lin(\mathcal{H}_C,\mathcal{H}_D)$. Prove that
				\begin{equation}\label{eq-vec_expand}
					\text{vec}(XYZ^{\dagger})=(\conj{Z}\otimes X)\text{vec}(Y).
				\end{equation}
			\item Prove that $\norm{X}_2=\norm{\text{vec}(X)}_2$ for every linear operator $X\in\Lin(\mathcal{H}_A,\mathcal{H}_B)$.
		\end{enumerate}
	\end{exercise}
	
	\subsection{Notable Classes of Linear Operators}
	
	\label{sec-MT:notable-lin-ops}
	
	The following classes of linear operators are  particularly notable.
	\begin{itemize}
		\item \textit{Hermitian operators}: Also known as \textit{self-adjoint} operators, they are linear operators that are equal to their conjugate transpose. That is, $X\in\Lin(\mathcal{H})$ is Hermitian if $X^{\dagger}=X$. The set of all Hermitian operators is a real vector space with dimension $d^2$, where $d=\text{dim}(\mathcal{H})$. This means that every Hermitian operator $X$ can be expanded in terms of an orthonormal basis $\{B_{k}\}_{k=1}^{d^2}$ of Hermitian operators such that
			\begin{equation}
				X=\sum_{k=1}^{d^2}x_{k}B_k,
			\end{equation}
			where $x_{k}$ are \textit{real numbers}. Orthonormality is with respect to the Hilbert--Schmidt inner product, which means that $\inner{B_k}{B_{\ell}}=\Tr[B_kB_{\ell}]=\delta_{k,l}$ and $x_k=\inner{B_k}{X}=\Tr[B_kX]$ for all $k,\ell\in\{1,2,\dotsc,d^2\}$.

	An example of an orthonormal basis of $d^2$ Hermitian operators is the following:		
		\begin{align}
				\sigma_{0,0}^{(d)} & \coloneqq \frac{\mathbbm{1}_d}{\sqrt{d}},\label{eq-su_generators_0}\\
				\sigma_{k,\ell}^{(d;+)} & \coloneqq \frac{1}{\sqrt{2}} \left(\ket{k}\!\bra{\ell}+\ket{\ell}\!\bra{k}\right),\quad 0\leq k<\ell\leq d-1, \label{eq-su_generators_1}\\
				\sigma_{k,\ell}^{(d;\I)}& \coloneqq \frac{1}{\sqrt{2}}\left(-\I\ket{k}\!\bra{\ell}+\I\ket{\ell}\!\bra{k}\right), \quad 0\leq k<\ell\leq d-1,\label{eq-su_generators_2}\\
				\sigma_{k,k}^{(d)}& \coloneqq \frac{1}{\sqrt{k(k+1)}} \left(\left(\sum_{j=0}^{k-1} \ket{j}\!\bra{j}\right)-k\ket{k}\!\bra{k}\right),\quad 1\leq k\leq d-1, \label{eq:math-tools:gell-mann-scaled-mats}
			\end{align}
			Observe that 
                all of the above operators, with the exception of $\sigma_{0,0}^{(d)}$, are traceless. If we scale each of them by $\sqrt{d}$, then they are called the \textit{generalized Gell-Mann matrices}.
			
			When $d=2$, the generalized Gell-Mann matrices reduce to the \textit{Pauli matrices}:
			\begin{align}
				\mathbbm{1} & \coloneqq \begin{pmatrix}
				1 & 0 \\
				0 & 1
				\end{pmatrix} = \sqrt{2}\sigma_{0,0}^{(2)}, \quad
				X \coloneqq \begin{pmatrix}
				0 & 1 \\
				1 & 0
				\end{pmatrix} = \sqrt{2}\sigma_{0,1}^{(2;+)}, \label{eq-Pauli_mat_1}\\
				Y & \coloneqq \begin{pmatrix}
				0 & -\I \\
				\I & 0
				\end{pmatrix} = \sqrt{2}\sigma_{0,1}^{(2;\I)},\quad
				Z \coloneqq \begin{pmatrix}
				1 & 0 \\
				0 & -1
				\end{pmatrix} = \sqrt{2}\sigma_{1,1}^{(2)}. \label{eq-Pauli_mat_2}
			\end{align}
			The Pauli matrices are important in the context of quantum mechanics, and quantum information more generally, as they can be used to describe the quantum states of two-dimensional quantum systems, as well as their evolutions. They are also involved in fundamental quantum information processing protocols such as quantum teleportation. We elaborate upon these points in Chapters~\ref{chap-QM_states_meas}--\ref{chap-QM_protocols}.
			
			\begin{exercise}{exer-su_generators}
				Prove that the operators in \eqref{eq-su_generators_0}--\eqref{eq:math-tools:gell-mann-scaled-mats} form an orthonormal basis for the vector space of Hermitian operators. More generally, prove that they form an orthonormal basis for the vector space $\Lin(\mathbb{C}^d)$ of \textit{all} linear operators.
			\end{exercise}
		
		\item \textit{Positive semi-definite operators}: A Hermitian operator $X\in\Lin(\mathcal{H})$ is positive semi-definite if $\bra{\psi}X\ket{\psi}\geq 0$ for all $\ket{\psi}\in\mathcal{H}$. For every positive semi-definite operator $X$, there exists a linear operator $Y$ such that $X=Y^\dagger Y$. $X$ is called \textit{positive definite} if $\bra{\psi}X\ket{\psi}>0$ for all $\ket{\psi}\in\mathcal{H}$ such that $\ket{\psi} \neq 0$. We write $X\geq 0$ if $X$ is positive semi-definite, and $X>0$ if $X$ is positive definite. If $X-Z$ is positive semi-definite for Hermitian operators $X$ and $Z$, then we write $X\geq Z$, and if $X-Z$ is positive definite, then we write $X>Z$. The ordering $X\geq Z$ on Hermitian operators is a partial order called the L\"owner order and is discussed more in Definition~\ref{def-Loewner}.

        \begin{exercise}{exer-PSD_basic}
			Let $X\in\Lin(\mathcal{H}_A,\mathcal{H}_B)$ be a linear operator, with $\mathcal{H}_A$ and $\mathcal{H}_B$ arbitrary. Prove that $X^{\dagger}X$ is positive semi-definite.
		\end{exercise} 

        \begin{exercise}{exer-pos_semi_to_pos_def}
    		Let $X$ be a positive semi-definite operator. Prove that the operator $X+\varepsilon \mathbbm{1}$ is positive definite for all $\varepsilon > 0$.
    	\end{exercise}
		
		\item \textit{Density operators}: These are Hermitian operators that are positive semi-definite and have unit trace.
		Density operators are generalizations of probability distributions from classical information theory and describe the states of a quantum system, as detailed in Chapter~\ref{chap-QM_states_meas}.
		
		\item \textit{Unitary operators}: These are linear operators whose inverses are equal to their adjoints. That is, $U\in\Lin(\mathcal{H})$ is unitary if $U^\dagger U=UU^\dagger=\mathbbm{1}$. Unitary operators generalize invertible maps or permutations from classical information theory and describe the noiseless evolution of the state of a quantum system, as detailed in Chapters~\ref{chap-QM_states_meas} and \ref{chap-QM_channels}.
		
			\begin{exercise}{exer-unitaries_max_ent}
				Let $U\in\Lin(\mathcal{H})$ be a unitary operator acting on a $d$-dimensional Hilbert space $\mathcal{H}$.
				\begin{enumerate}[topsep=0.3cm]
					\item Given an orthonormal basis $\{\ket{e_k}\}_{k=1}^d$ for $\mathcal{H}$, prove that the set $\{\ket{f_k}\}_{k=1}^d$, with $\ket{f_k}\coloneqq U\ket{e_k}$ for all $1\leq k\leq d$, is another orthonormal basis for~$\mathcal{H}$.
					
					\item Using the transpose trick identity in \eqref{eq-transpose_trick}, prove that
						\begin{equation}\label{eq-transpose_trick_unitary}
							\ket{\Gamma}=(\conj{U}\otimes U)\ket{\Gamma}.
						\end{equation}
						
					\item Using 1. and 2., conclude that the following identity holds for every orthonormal basis $\{\ket{e_k}\}_{k=1}^d$ for $\mathcal{H}$:
						\begin{equation}\label{eq:math-tools:Gamma-any-ortho}
							\ket{\Gamma}= \sum_{k=1}^d \conj{\ket{e_k}} \otimes \ket{e_k}.
						\end{equation}
				\end{enumerate}
			\end{exercise}
		
		\item \textit{Isometric operators} or \textit{isometries}: A linear operator $V\in\Lin(\mathcal{H}_A,\mathcal{H}_B)$ is isometric if $V^\dagger V=\mathbbm{1}_{A}$ (we also say that $V$ is an isometry). Isometries also describe the noiseless evolution of the state of a quantum system, as detailed in Chapters~\ref{chap-QM_states_meas} and \ref{chap-QM_channels}.
		
			\begin{exercise}{exer-isometries}
				Let $V\in\Lin(\mathcal{H}_A,\mathcal{H}_B)$ be an isometry.
				\begin{enumerate}[topsep=0.3cm]
					\item Prove that $\braket{V\psi}{V\phi}=\braket{\psi}{\phi}$ for all $\ket{\psi},\ket{\phi}\in\mathcal{H}_A$.
					\item Using 1., prove that $V$ is injective.
					\item Using 2., prove that $d_B\geq d_A$. (\textit{Hint}: use the rank-nullity theorem in \eqref{eq-math_tools_rank_nullity}.)
					\item Prove that, if $d_A=d_B$, then $V$ is a unitary. (\textit{Hint}: use the result of Exercise~\ref{exer-injective_surjective}.)
				\end{enumerate}
			\end{exercise}
	
		\item \textit{Projection operators}: A linear operator $P\in\Lin(\mathcal{H})$ is a projection if it is Hermitian and satisfies $P^2=P$. Such operators are also sometimes called \textit{orthogonal projection operators}.
		
	\end{itemize}
	
	Note that every linear operator $X$ can be decomposed as
	\begin{equation}\label{eq-operator_Herm_decomp}
		X=\operatorname{Re}[X] +\I \operatorname{Im}[X],
	\end{equation}
	where $\operatorname{Re}[X] \coloneqq \frac{1}{2}(X+X^\dagger)$ and $\operatorname{Im}[X] \coloneqq \frac{1}{2\I}(X-X^\dagger)$ are both Hermitian operators, generalizing how complex numbers can be decomposed into real and imaginary parts.

\subsection{Singular Value, Schmidt, and Polar Decompositions}

	An important fact that we make use of throughout this book is the \textit{singular value decomposition theorem}.
	
	\begin{theorem*}{Singular Value Decomposition}{thm-SVD}
		For every linear operator $X\in\Lin(\mathcal{H}_A,\mathcal{H}_B)$ with $r \coloneqq \rank(X)$, there exist strictly positive real numbers $\{s_k>0:1\leq k\leq r\}$, called the \textit{singular values of $X$}, and orthonormal vectors $\{\ket{e_k}_B:1\leq k\leq r\}$ and $\{\ket{f_k}_A:1\leq k\leq r\}$, such that
		\begin{equation}\label{eq-SVD}
			X=\sum_{k=1}^r s_k\ket{e_k}_B\bra{f_k}_A.
		\end{equation}
		This is called the \textit{singular value decomposition of $X$}.
	\end{theorem*}
	
	\begin{remark}
		From \eqref{eq-SVD}, we see that the rank of a linear operator $X$, which we defined earlier as the dimension of the image of $X$, is equal to the number of singular values of $X$.
	\end{remark}
	
	The singular value decomposition theorem can be written in the following matrix form that is familiar from linear algebra. We first extend the orthonormal vectors $\{\ket{e_k}_B:1\leq k\leq r\}$ in $\mathcal{H}_B$ to an orthonormal basis $\{\ket{e_k}_B:1\leq k\leq d_B\}$ for $\mathcal{H}_B$, and similarly, we extend the set $\{\ket{f_k}_A:1\leq k\leq r\}$ of orthonormal vectors in $\mathcal{H}_A$ to an orthonormal basis $\{\ket{f_k}_A:1\leq k\leq d_A\}$ for $\mathcal{H}_A$. Then, by defining the unitaries
	\begin{equation}
	W\coloneqq \sum_{k=1}^{d_B} \ket{e_k}_B\bra{k-1}_B, \qquad V\coloneqq \sum_{k=1}^{d_A} \ket{f_k}_A\bra{k-1}_A,
	\end{equation}
	we can write \eqref{eq-SVD} as 
	\begin{equation}
		X=WS V^\dagger,
	\end{equation}
	where
	\begin{equation}\label{eq:math-tools:svd-diag}
		S\coloneqq \sum_{k=1}^r s_k\ket{k-1}_B\bra{k-1}_A
	\end{equation}
	is the matrix of singular values. Note that the matrix $S$ is a $d_B \times d_A$ rectangular matrix and \eqref{eq:math-tools:svd-diag} is only specifying the non-zero entries of $S$ along the diagonal.
	
	The singular value decomposition can be used to prove the following useful theorem for vectors in a tensor-product Hilbert space.
	
	\begin{theorem*}{Schmidt Decomposition}{thm-Schmidt}
		Let $\ket{\psi}_{AB}$ be a vector in the tensor-product Hilbert space $\mathcal{H}_{AB}$. Let $X_{A\to B}$ be the linear operator with matrix elements $\bra{j}_B X \ket{i}_A=\braket{i,j}{\psi}_{AB}$, and let $r=\rank(X)$. Then, there exist strictly positive \textit{Schmidt coefficients} $\{\lambda_k\}_{k=1}^r$, and orthonormal vectors $\{\ket{e_k}_A\}_{k=1}^r$ and $\{\ket{f_k}_B\}_{k=1}^r$, such that
		\begin{equation}\label{eq-Schmidt_decomp}
			\ket{\psi}_{AB}=\sum_{k=1}^r \sqrt{\lambda_k}\, \ket{e_k}_A\otimes\ket{f_k}_B.
		\end{equation}
		The quantity $r$ is called the \textit{Schmidt rank}, satisfying $r\leq\min\{d_A,d_B\}$.
	\end{theorem*}
	
	\begin{Proof}
		This is a direct application of the singular value decomposition  (Theorem~\ref{thm-SVD}). Consider the operator $X$ defined in the statement of the theorem, having matrix elements $\bra{j}_B X \ket{i}_A =\braket{i,j}{\psi}_{AB}$. Observe then that $\text{vec}(X)=\ket{\psi}_{AB}$ (see \eqref{eq-vec_operation}).
		
		Now, by the singular-value decomposition (Theorem~\ref{thm-SVD}), we can write $X$ as $X=\sum_{k=1}^r s_k \ket{f_k}_B \bra{g_k}_A$, where $s_k$ are strictly positive numbers, $r=\rank(X)\leq \min\{d_A,d_B\}$, and $\{\ket{f_k}_B:1\leq k\leq r\}$ and $\{\ket{g_k}_A:1\leq k\leq r\}$ are sets of orthonormal vectors in $\mathcal{H}_B$ and $\mathcal{H}_A$, respectively. Letting $\lambda_k=s_k^2$, and upon vectorizing $X$ as written in this form (see \eqref{eq-vec_operation}, \eqref{eq-math_tools:vec}, and \eqref{eq-vec_operation_vectors}), we find that
		\begin{equation}
			\ket{\psi}_{AB}=\text{vec}(X)=\sum_{k=1}^r \sqrt{\lambda_k}\,\,\conj{\ket{g_k}}_A \otimes\ket{f_k}_B.
		\end{equation}
		Letting $\ket{e_k}_A \coloneqq \conj{\ket{g_k}}_A$, the result follows.
	\end{Proof}

	A simple but important consequence of the Schmidt decomposition theorem is that every vector $\ket{\psi}_{AB}$ in a tensor-product Hilbert space $\mathcal{H}_{AB}$ can be written as
	\begin{equation}\label{eq-Schmidt_decomp_gen}
		\ket{\psi}_{AB}=\sum_{k=1}^{\min\{d_A,d_B\}}\sqrt{p_k}\ket{u_k}_A\otimes\ket{v_k}_B,
	\end{equation}
	where $p_k=s_k^2$ for $1\leq k\leq r$ and $p_k=0$ for $r<k\leq\min\{d_A,d_B\}$. The vectors $\ket{u_k}_A$ and $\ket{v_k}_B$ are such that $\ket{u_k}_A=\ket{e_k}_A$ and $\ket{v_k}_B=\ket{f_k}_B$ for $1\leq k\leq r$, and the remaining vectors combine to form orthonormal bases for a $\min\{d_A,d_B\}$-dimensional subspace of $\mathcal{H}_A \otimes \mathcal{H}_B$.
	
	\begin{exercise}{exer-op_Schmidt_decomp}
		Using arguments similar to the proof of Theorem~\ref{thm-Schmidt}, prove that every linear operator $X_{AB}\in\Lin(\mathcal{H}_A\otimes\mathcal{H}_B)$ can be written as
		\begin{equation}
			X_{AB}=\sum_{k=1}^r \sqrt{\smash[b]{\lambda_k}}\, E_A^k\otimes F_B^k,
		\end{equation}
		where the set $\{\lambda_k\}_{k=1}^r$ consists of strictly positive reals, $\{E_A^k\}_{k=1}^r$ and $\{F_B^k\}_{k=1}^r$ are orthonormal sets of linear operators acting on $\mathcal{H}_A$ and $\mathcal{H}_B$, respectively, and $r=\rank(M)$, where $M\in\Lin(\mathcal{H}_A\otimes\mathcal{H}_A,\mathcal{H}_B\otimes\mathcal{H}_B)$ is defined by $\bra{j,\ell}_{BB}M\ket{i,k}_{AA}=\bra{i,j}_{AB}X\ket{k,\ell}_{AB}$ for all $0\leq i,k\leq d_A-1$ and $0\leq j,\ell\leq d_B-1$.
	\end{exercise}

	Another important decomposition of a linear operator is the \textit{polar decomposition}. 
	
	\begin{theorem*}{Polar Decomposition}{thm-polar_decomposition}
		Every linear operator $X\in\Lin(\mathcal{H})$ can be written as $X=UP$, where $U$ is a unitary operator and $P$ is a positive semi-definite operator. In particular, $P=\abs{X}\coloneqq\sqrt{X^\dagger X}$.
	\end{theorem*}
	\begin{Proof}
	If $X=WSV^\dagger$ is the singular value decomposition of $X$, then we can take $P=VSV^\dagger$ and $U=WV^\dagger$.
	\end{Proof}
	
	The polar decomposition generalizes the polar form $z = r \e^{\I \theta}$ of a complex number $z$. Indeed, the fact that $r \geq 0$ is in correspondence with $P$ being a positive semi-definite operator, and the phase $\e^{\I \theta}$ is in correspondence with the unitary $U$, the latter having eigenvalues on the unit circle (i.e., of the form $\e^{\I \theta}$).

\subsection{Spectral Theorem}\label{sec:math_tools-spec_theorem}

	Given a linear operator $X\in\Lin(\mathcal{H})$ acting on some Hilbert space $\mathcal{H}$, if there exists a vector $\ket{\psi}\in\mathcal{H}$ such that
	\begin{equation}
		X\ket{\psi}=\lambda\ket{\psi},
	\end{equation}
	then $\ket{\psi}$ is said to be an \textit{eigenvector of $X$} with associated \textit{eigenvalue} $\lambda$. The set of all eigenvectors associated with an eigenvalue $\lambda$ is a subspace of $\mathcal{H}$ called the \textit{eigenspace of $X$ associated with $\lambda$}, and the \textit{multiplicity} of $\lambda$ is the number of linearly independent eigenvectors of $X$ that are associated with $\lambda$ (in other words, it is the dimension of the eigenspace of $X$ associated with $\lambda$). The eigenspace of $X$ associated with $\lambda$ is equal to $\ker(X-\lambda I)$.
	
	The spectral theorem, which we state below, allows us to decompose every \textit{normal operator} $X$, i.e., an operator that commutes with its adjoint, so that $XX^\dagger=X^\dagger X$, in terms of its eigenvalues and projections onto its corresponding eigenspaces. We employ it most often when analyzing quantum states and observables.
	
	\begin{theorem*}{Spectral Theorem}{thm-spectral_theorem}
		For every normal operator $X\in\Lin(\mathcal{H})$, there exists $n\in\mathbb{N}$ such that
		\begin{equation}\label{eq-spectral_decomp}
			X=\sum_{j=1}^n \lambda_j\Pi_j,
		\end{equation}
		where $\lambda_1,\lambda_2,\dotsc,\lambda_n\in\mathbb{C}$ are the distinct eigenvalues of $X$ and $\Pi_1,\Pi_2,\dotsc,\Pi_n$ are the \textit{spectral projections} onto the corresponding eigenspaces, which satisfy $\Pi_1+\Pi_2+\dotsb+\Pi_n=\mathbbm{1}_{\mathcal{H}}$ and $\Pi_i\Pi_j=\delta_{i,j}\Pi_i$. 	The decomposition in \eqref{eq-spectral_decomp} is unique and is called the \textit{spectral decomposition} of $X$.
		The spectrum of $X$ is denoted by $\operatorname{spec}(X) \coloneqq \{ \lambda_1,\lambda_2,\dotsc,\lambda_n\}$.
	\end{theorem*}
	
	\begin{remark}
		The multiplicity of an eigenvalue $\lambda_i$ is equal to the rank of the corresponding projection $\Pi_i$. 
		
		The property $\Pi_i\Pi_j=\delta_{i,j}\Pi_i$ of the spectral projections indicates that the eigenspaces of a linear operator are orthogonal to each other. In other words, for every eigenvector $\ket{\psi_{\lambda_1}}$ associated with the eigenvalue $\lambda_1$ and every eigenvector $\ket{\psi_{\lambda_2}}$ associated with the eigenvector $\lambda_2$, where $\lambda_1 \neq \lambda_2$, we have that $\langle \psi_{\lambda_1} |\psi_{\lambda_2} \rangle =0$.
	\end{remark}
	
	If we take the spectral decomposition of a normal operator $X\in\Lin(\mathcal{H})$ in \eqref{eq-spectral_decomp} and find orthonormal bases for the corresponding eigenspaces, then $X$ can be written as
	\begin{equation}\label{eq-spectral_decomp_alt}
		X=\sum_{k=1}^{d}\lambda_k \ket{\psi_k}\!\bra{\psi_k},
	\end{equation}
	where $d=\dim(\mathcal{H})$ and $\{\ket{\psi_k}\}_{k=1}^{d}$ is a set of orthonormal vectors such that
	\begin{equation}
		\ket{\psi_1}\!\bra{\psi_1}+\ket{\psi_2}\!\bra{\psi_2}+\dotsb+\ket{\psi_d}\!\bra{\psi_d}=\mathbbm{1}_{\mathcal{H}}.
	\end{equation}
	From this decomposition, we see clearly that $\ket{\psi_k}$ is an eigenvector of $X$ with associated eigenvalue $\lambda_k$.
	
	The spectral theorem is a statement of a fact familiar from linear algebra, that every normal operator can be diagonalized by a unitary. Indeed, if we consider the spectral decomposition in \eqref{eq-spectral_decomp_alt}, and we define the unitary operator $U\coloneqq \sum_{k=1}^d \ket{\psi_k}\!\bra{k-1}$, then \eqref{eq-spectral_decomp_alt} can be written as $X=UDU^\dagger$, where $D\coloneqq\sum_{k=1}^d \lambda_k\ket{k-1}\!\bra{k-1}$ is a diagonal matrix.
	
	Note that for the decomposition in \eqref{eq-spectral_decomp_alt} the numbers labeled by $\lambda_k\in\mathbb{C}$ are not all distinct because the eigenspace associated with each eigenvalue can have dimension greater than one. Also, note that the decomposition in \eqref{eq-spectral_decomp_alt} is generally not unique because the decomposition of the spectral projections into orthonormal vectors is not unique.
	
	From \eqref{eq-spectral_decomp_alt}, it is evident that the rank of a normal operator $X$ (recall the discussion around \eqref{eq-image}) is equal to the number of non-zero eigenvalues of $X$ (including their multiplicities). Furthermore, the support of $X$ (recall \eqref{eq-support}) is equal to the span of all eigenvectors of $X$ associated with the non-zero eigenvalues of $X$. In particular,
	\begin{equation}\label{eq-proj_support}
		\Pi_X\coloneqq\sum_{k:\lambda_k\neq 0}\ket{\psi_k}\!\bra{\psi_k}
	\end{equation}
	is the projection onto the support of $X$. It is also evident that the trace of $X$ is equal to the sum of its eigenvalues, i.e., $\Tr[X]=\sum_{k=1}^{d}\lambda_k=\sum_{k:\lambda_k\neq 0}\lambda_k$.
	
	\begin{exercise}{exer-projections}
		Let $P$ be a projection operator.
		\begin{enumerate}[topsep=0.3cm]
			\item Prove that the eigenvalues of $P$ are either 0 or 1. Prove that $\Tr[P]=\rank(P)$.
			
			\item Using 1., conclude that $\rank(X)=\Tr[\Pi_X]$ for every linear operator $X$, where $\Pi_X$ is the projection onto the support of $X$, as defined in \eqref{eq-proj_support}.
		\end{enumerate}
	\end{exercise}
	
	The singular values of a linear operator $X$ (not necessarily normal) are related to the eigenvalues of $X^\dagger X$ and $XX^\dagger$ in the following way. Let $\{s_k\}_{k=1}^{\rank(X)}$ be the set of singular values of $X$, and let $\{\lambda_k\}_{k=1}^{\rank(X)}$ be the non-zero eigenvalues of $X^\dagger X$, which are the same as the non-zero eigenvalues of $XX^\dagger$. (Note that both $X^\dagger X$ and $XX^\dagger$ are normal operators, so that the spectral theorem applies to them.) Then, $s_k=\sqrt{\lambda_k}$ for all $1\leq k\leq\rank(X)$. In particular, if $X$ is a Hermitian operator with non-zero eigenvalues $\{\omega_k\}_{k=1}^{\rank(X)}$, then $s_k=\sqrt{\smash[b]{\omega_k^2}}=\abs{\omega_k}$ for all $1\leq k\leq\rank(X)$.
	
	\begin{exercise}{exer-Herm_op_real_evals}
		\begin{enumerate}
			\item Prove that every Hermitian operator has real eigenvalues.
			\item Prove that every unitary operator has eigenvalues with unit modulus; i.e., if $\lambda$ is an eigenvalue of a given unitary operator, then $\abs{\lambda}=1$.
		\end{enumerate}
	\end{exercise}
	
	If $X$ is a Hermitian operator, then we can split it into a \textit{positive part}, denoted by~$X_+$, and a \textit{negative part}, denoted by $X_-$, so that
	\begin{equation}\label{eq-operator_pos_neg_decomp}
		X=X_+-X_- \, .
	\end{equation}
	This follows because $X$ has real eigenvalues (see Exercise~\ref{exer-Herm_op_real_evals}).
	In particular, if $X=\sum_{k=1}^d\lambda_k\ket{\psi_k}\!\bra{\psi_k}$ is a spectral decomposition of $X$, then
	\begin{equation}\label{eq-MT:Jordan-Hahn}
		X_+\coloneqq\sum_{k:\lambda_k\geq 0}\lambda_k\ket{\psi_k}\!\bra{\psi_k},\quad X_-\coloneqq \sum_{k:\lambda_k<0}\abs{\lambda_k}\ket{\psi_k}\!\bra{\psi_k}.
	\end{equation}
	Note that both $X_+$ and $X_-$ are positive semi-definite operators, and they are supported on orthogonal subspaces, meaning that $X_+X_-=0$. Such a decomposition of a Hermitian operator $X$ into positive and negative parts is called the \textit{Jordan--Hahn decomposition} of $X$.
	
	\begin{exercise}{exer-lin_op_span_PSD}
		Using the Jordan--Hahn decomposition, prove that there exists a basis for $\Lin(\mathbb{C}^d)$ consisting entirely of positive semi-definite operators, for all $d\geq 2$.
	\end{exercise}
	
	Another important fact is that a Hermitian operator $X$ is positive semi-definite if and only if all of its eigenvalues are non-negative, and positive definite if and only if all of its eigenvalues are strictly positive. The latter implies that every positive definite operator $X\in\Lin(\mathcal{H})$ has full support, i.e., $\supp(X)=\mathcal{H}$ and $\rank(X)=\dim(\mathcal{H})$. Thus, positive definite operators are invertible.

\subsubsection{Functions of Hermitian Operators}\label{sec:math-tools:functions-herm-ops}
	
	Using the spectral decomposition as in \eqref{eq-spectral_decomp_alt}, we can define a function of a Hermitian operator. Basic functions of density operators that we employ extensively are $x \to -x \log_2 x$ for  the quantum entropy and $x \to x^\alpha$ for the R\'enyi entropy.
	
	Let $f\colon \mathbb{R}\to\mathbb{R}$ be a real-valued function with domain $\text{dom}(f)$. For every Hermitian operator $X\in\Lin(\mathcal{H})$ with spectrum $\operatorname{spec}(X) \subseteq \text{dom}(f)$ and spectral decomposition $X = \sum_{j=1}^n \lambda_j \Pi_j$ as in~\eqref{eq-spectral_decomp}, 
    we define $f(X)$ as the 
    operator
	\begin{equation}\label{eq-function_Hermitian_ext}
		f(X) \coloneqq \sum_{j=1}^n f(\lambda_j)\Pi_j.
        \end{equation}
        Note that using the alternative form of the spectral decomposition given in~\eqref{eq-spectral_decomp_alt} we could also write
        \begin{equation}
        f(X)\coloneqq\sum_{k=1}^d f(\lambda_k) \ket{\psi_k}\!\bra{\psi_k},
        \label{eq-function_Hermitian_ext_alt}
        \end{equation}
	where $d=\dim(\mathcal{H})$. An immediate consequence of these definitions is that if $X$ is a Hermitian operator and $U$ is a unitary operator, then
	\begin{equation}\label{eq-function_Hermitian_unitary}
		f(UXU^{\dagger})=Uf(X)U^{\dagger}.
	\end{equation}
	This is due to the fact that $X$ and $UXU^{\dagger}$ have the same eigenvalues and multiplicities.
	
	Functions that arise frequently throughout this book are as follows:
	\begin{itemize}
	\item\textit{Power functions}: for some real number $\alpha$, the function $f_\alpha(x)=x^{\alpha}$ extends to Hermitian operators via~\eqref{eq-function_Hermitian_ext_alt} as
        \begin{equation} \label{eq-power-Hermitian-operator}
		 X^{\alpha} \coloneqq \sum_{k=1}^d \lambda_k^{\alpha} \ket{\psi_k}\!\bra{\psi_k}.
 	\end{equation}
        However, in order to apply $f_\alpha$ we need to make sure that the spectrum of $X$ is contained in the domain of $f_\alpha$. We distinguish four cases:
        \begin{itemize}
            \item If $\alpha \in \mathbb{N}$ is a non-negative integer, then the domain of $f_\alpha$ is the entire real line $\mathbb{R}$, and hence~\eqref{eq-power-Hermitian-operator} is well defined for every Hermitian operator $X$. Here we follow the convention that $0^0=0$.
            
            \item If $\alpha\in \mathbb{Z}\setminus \mathbb{N}$ is a negative integer, then the domain of $f_\alpha$ is $\mathbb{R}\setminus\{0\}$, and therefore~\eqref{eq-power-Hermitian-operator} applies to every full-rank Hermitian operator $X$, i.e., to all $X$ such that $\rank(X)=d$.
            
            \item If $\alpha>0$ but $\alpha\notin \mathbb{N}$, then the domain of $f_\alpha$ is $[0,\infty)$, and therefore~\eqref{eq-power-Hermitian-operator} applies to every positive semi-definite operator $X\geq 0$.
            
            \item If $\alpha<0$ but $\alpha\notin \mathbb{Z}$, then the domain of $f_\alpha$ is $(0,\infty)$, and therefore~\eqref{eq-power-Hermitian-operator} applies to every positive definite operator $X>0$.
        \end{itemize}
        
        When $\alpha=\frac{1}{2}$ and $X\geq 0$, we typically use the notation $\sqrt{X}$ to refer to $X^{\frac{1}{2}}$. In particular, $\sqrt{X}$ is the unique positive semi-definite operator such that $\sqrt{X}\sqrt{X}=X$.

        For $\alpha=0$ and full-rank $X$, the following equality holds:
		\begin{equation}
			X^0 = \sum_{k:\lambda_k\neq 0}\ket{\psi_k}\!\bra{\psi_k} = \Pi_X,
		\end{equation}
	where $\Pi_X$ is the projection onto the support of $X$; recall~\eqref{eq-proj_support}.

		\item\textit{Logarithm functions}: For the function $\log_b:(0,\infty)\rightarrow\mathbb{R}$ with base $b>0$, we define
			\begin{equation}
				\log_b(X)\coloneqq \sum_{k:\lambda_k> 0}\log_b(\lambda_k)\ket{\psi_k}\!\bra{\psi_k}.
			\end{equation}
			We deal throughout this book exclusively with the base-2 logarithm $\log_2$ and the base-$\e$ logarithm $\log_{\e}\equiv\ln$.
		
	\end{itemize}
	
		
	We end this section with an intriguing lemma that finds application several times in Chapter~\ref{chap-entropies}.
	
	\begin{Lemma}{lem:sing-val-lemma-pseudo-commute}
		Let $X\in\Lin(\mathcal{H})$, and let $f$ be a function such that the squares of the singular values of $X$ are in the domain of $f$. Then,
		\begin{equation}
			Xf\big(X^{\dag}X\big)=f\big(XX^{\dag}\big)X.
		\end{equation}
	\end{Lemma}

	\begin{Proof}
		This is a direct consequence of the singular value decomposition theorem (Theorem~\ref{thm-SVD}). Let $X=WSV^{\dagger}$ be a singular value decomposition of $L$, where $W$ and $V$ are unitary operators and $S$ is a diagonal, positive semi-definite operator. Then,
		\begin{align}
			Xf(X^{\dag}X)  &  =WSV^{\dagger}f\left(\left(WSV^{\dagger}\right)^{\dag}WSV^{\dagger}\right)\\
			&  =WSV^{\dagger}f\big(VSW^{\dag}WSV^{\dagger}\big)\\
			&  =WSV^{\dagger}f\big(VS^{2}V^{\dagger}\big).
		\end{align}
		Now, we use the fact that $f\big(VS^2V^{\dagger}\big)=Vf\big(S^2\big)V^{\dagger}$, which holds because 
            $V$ is unitary. Using as well the fact that $Sf\big(S^2\big)=f\big(S^2\big)S$, we obtain
		\begin{align}
			Xf\big(X^{\dagger}X\big)&=WSf\big(S^2\big)V^{\dagger}\\
			&=Wf\big(S^2\big)SV^{\dagger}\\
			&=Wf\big(SVV^{\dagger}S\big)W^{\dagger}WSV^{\dagger}\\
			&=f\big(WSV^{\dagger}VSW^{\dagger}\big)WSV^{\dagger}\\
			&=f\big(XX^{\dagger}\big)X.
		\end{align}
		This concludes the proof.
	\end{Proof}

        \begin{exercise}{exer_Gram_matrices}
            A \textit{Gram matrix} is a $d\times d$ matrix $X$ for which there exists a Hilbert space $\mathcal{H}$ and a collection of vectors $\{\ket{\psi_i}\}_{i=1}^d$, with $\ket{\psi_i}\in \mathcal{H}$, such that $X_{ij} = \braket{\psi_i}{\psi_j}$ for all $1\leq i,j\leq d$.
            \begin{enumerate}
                \item Prove that every Gram matrix is positive semi-definite.
                \item Vice versa, prove that every positive semi-definite matrix is a Gram matrix. (\emph{Hint:} write $X = \sqrt{X}\sqrt{X}$.)
            \end{enumerate}
        \end{exercise}
	
\subsection{Norms}\label{sec:math-tools:norms}

	A \textit{norm} on a Hilbert space $\mathcal{H}$ (more generally a vector space) is a function $\norm{\cdot}:\mathcal{H}\to\mathbb{R}$ that satisfies the following properties:
	\begin{itemize}
		\item\textit{Non-negativity}: $\norm{\ket{\psi}}\geq 0$ for all $\ket{\psi}\in \mathcal{H}$, and $\norm{\ket{\psi}}=0$ if and only if $\ket{\psi}=0$.
		\item\textit{Scaling} (also called \textit{homogeneity}): For every $\alpha\in\mathbb{C}$ and $\ket{\psi}\in\mathcal{H}$, $\norm{\alpha\ket{\psi}}=\abs{\alpha}\norm{\ket{\psi}}$.
		\item\textit{Triangle inequality}: For all $\ket{\psi},\ket{\phi}\in\mathcal{H}$, $\norm{\ket{\psi}+\ket{\phi}}\leq\norm{\ket{\psi}}+\norm{\ket{\phi}}$.
	\end{itemize}
An immediate consequence of the scaling property and the triangle inequality is that a norm is convex:
\begin{equation}
\norm{\lambda\ket{\psi}+(1-\lambda)\ket{\phi}}\leq\lambda\norm{\ket{\psi}}+(1-\lambda)\norm{\ket{\phi}}
\end{equation}
 for all vectors $\ket{\psi}$ and $\ket{\phi}$ and all $\lambda \in [0,1]$.

	In this section, we are primarily interested in the Hilbert space $\Lin(\mathcal{H})$ of linear operators $X:\mathcal{H}\to \mathcal{H}$ for some Hilbert space $\mathcal{H}$. The following norm for linear operators is used extensively in this book.

	\begin{definition}{Schatten Norm}{def-Schatten_norm}
		For every linear operator $X\in\Lin(\mathcal{H})$ acting on a Hilbert space $\mathcal{H}$, we define its \textit{Schatten $\alpha$-norm} as
		\begin{equation}\label{eq-Schatten-norm}
			\norm{X}_\alpha\coloneqq (\Tr[\abs{X}^\alpha])^{\frac{1}{\alpha}},
		\end{equation}
		for all $\alpha\in[1,\infty)$, where $\abs{X}\coloneqq\sqrt{X^\dagger X}$. We let
		\begin{equation}
		\norm{X}_{\infty}\coloneqq\lim_{\alpha\to\infty}\norm{X}_{\alpha}.
		\label{eq-MT:orig-def-inf-norm-Schatten}
		\end{equation}
	\end{definition}
	
	Throughout this book, we extend the function $\norm{\cdot}_{\alpha}$ to include $\alpha\in(0,1)$ (with the definition exactly as in \eqref{eq-Schatten-norm}), although in this case it is not a norm because it does not satisfy the triangle inequality.
	
	Norms are typically employed in pure mathematics to measure the lengths of vectors or operators, and different norms give different ways of measuring length. In quantum information, we employ norms to measure entropy  and information of quantum states and channels (see Chapter~\ref{chap-entropies}). The parameter $\alpha$ for the Schatten norm then becomes the R\'enyi parameter for the R\'enyi entropy.
	
	\begin{exercise}{exer-Schatten_norm_SVD}
		Let $X$ be a linear operator, and let $\{s_k\}_{k=1}^{r}$ be the set of singular values of $X$, where $r \coloneqq \text{rank}(X)$. Prove that
		\begin{equation}\label{eq:math-tools:schatten-norm-from-sing-vals}
			\norm{X}_{\alpha}=\left(\sum_{k=1}^{r}s_k^{\alpha}\right)^{\frac{1}{\alpha}}
		\end{equation}
		for all $\alpha\in(0,\infty)$.
	\end{exercise}
	
	If $X$ is a Hermitian operator with non-zero eigenvalues $\{\lambda_k\}_{k=1}^{r}$, then its singular values are $s_k=\abs{\lambda_k}$, where $k \in\{1,\ldots,  r\}$ (see Section~\ref{sec:math_tools-spec_theorem}). Therefore, for all~$\alpha\in(0,\infty)$, 
	\begin{equation}
		\norm{X}_{\alpha}=\left(\sum_{k=1}^{r}\abs{\lambda_k}^{\alpha}\right)^{\frac{1}{\alpha}}\quad\text{($X$ Hermitian).}
	\end{equation}
	 
	\begin{exercise}{exer-Schatten_norm_alt}
		Let $X$ be a linear operator and $\alpha\in(0,\infty)$. Prove that
		\begin{equation}
			\norm{X^\dagger X}_{\alpha}=\norm{XX^\dagger}_{\alpha}=\norm{X}_{2\alpha}^2.
		\end{equation}
	\end{exercise}
	
	We now state several important properties of the Schatten norm. 
	
	\begin{proposition*}{Properties of Schatten Norm}{prop-Schatten_norm_prop}
		For all $\alpha\in[1,\infty]$, the Schatten norm $\norm{\cdot}_{\alpha}$ has the following properties.
		
		\begin{enumerate}
			\item\textit{Monotonicity}: For every linear operator $X$, the function $\alpha \mapsto \norm{X}_\alpha$ is monotonically non-increasing with $\alpha$; i.e., for $\alpha\geq\beta>0$, the following holds:
	\begin{equation}
	\norm{X}_{\alpha}\leq\norm{X}_{\beta}.
	\label{eq:math-tools:monotonicity-alpha-norms}
	\end{equation}
	In particular, we then have that $\norm{X}_{\infty}\leq \norm{X}_{\alpha}\leq\norm{X}_{1}$ for all $\alpha \in [1,\infty]$ and every  linear operator $X$.
			
			\item\textit{Isometric invariance}: For all isometries $U$ and $V$,
				\begin{equation}\label{eq-Schatten_norm_iso_invar}
					\norm{X}_{\alpha}=\norm{UXV^\dagger}_{\alpha}.
				\end{equation}
			
			\item\textit{Submultiplicativity}: For all linear operators $X$, $Y$, and $Z$,
				\begin{equation}\label{eq-math_tools_submult_alpha_strong}
					\norm{XYZ}_{\alpha}\leq \norm{X}_{\infty}\norm{Y}_{\alpha}\norm{Z}_{\infty}.
				\end{equation}
				Consequently, for all linear operators $X$ and $Y$,
				\begin{equation}\label{eq:math-tools:submult-alpha-norms}
					\norm{XY}_{\alpha}\leq\norm{X}_{\alpha}\norm{Y}_{\alpha}.
				\end{equation}
				
			\item\textit{Multiplicativity with respect to tensor product}: For all linear operators $X$ and $Y$,
				\begin{equation}\label{eq-Schatten_norm_mult}
					\norm{X\otimes Y}_{\alpha}=\norm{X}_{\alpha}\norm{Y}_{\alpha}.
				\end{equation}
				
			\item\textit{Direct-sum property}: Given a collection $\{X_x\}_{x\in\mathcal{X}}$ of linear operators indexed by a finite alphabet $\mathcal{X}$, the following equality holds for every orthonormal basis $\{\ket{x}\}_{x\in\mathcal{X}}$:
				\begin{equation}\label{eq-trace_norm_blkdiag}
					\norm{\sum_{x\in\mathcal{X}}\ket{x}\!\bra{x}\otimes X_x}^{\alpha}_{\alpha}=\sum_{x\in\mathcal{X}}\norm{X_x}^{\alpha}_{\alpha}.
				\end{equation}

			\item\textit{Duality}: For every linear operator $X$,
				\begin{equation}\label{eq-Schatten_norm_var}
					\norm{X}_{\alpha}=\sup_Y \left\{\abs{\Tr[Y^\dagger X]}:\norm{Y}_{\beta}\leq 1, \ \frac{1}{\alpha}+\frac{1}{\beta}=1\right\}.
				\end{equation}
				The equality above implies \textit{H\"{o}lder's inequality}:
				\begin{equation}\label{eq-Schatten_norm_duality}
					\abs{\Tr[Z^\dagger X]}\leq\norm{X}_{\alpha}\norm{Z}_{\beta}
				\end{equation}
				which holds for all linear operators $X$ and $Z$, where $\alpha, \beta \in[1,\infty]$ satisfy $\frac{1}{\alpha}+\frac{1}{\beta}=1$. In this sense, the norms $\norm{\cdot}_{\alpha}$ and $\norm{\cdot}_{\beta}$, with $\frac{1}{\alpha}+\frac{1}{\beta}=1$, are said to be \textit{dual} to each other.
		\end{enumerate}
	\end{proposition*}
	
	\begin{Proof}
		\hfill\begin{enumerate}
		\item[1.] By the scaling property, up to dividing by $\norm{X}_\beta$, it suffices to prove~\eqref{eq:math-tools:monotonicity-alpha-norms} for the case where $\norm{X}_\beta = 1$. Let $\{s_{k}\}_{k=1}^{r}$ denote the singular values of $X$, where $r\coloneqq \rank(X)$. From
        \begin{equation}
            1 = \norm{X}_\beta^\beta = \sum_{k=1}^r s_k^\beta,
        \end{equation}
        we deduce that $s_k\leq 1$ for all $k$. Since $\beta\leq \alpha$, this implies that $s_k^\beta \geq s_k^\alpha$ for all $k$. Summing over $k$ yields precisely
        \begin{equation}
            1 = \sum_{k=1}^r s_k^\beta \geq \sum_{k=1}^r s_k^\alpha = \norm{X}_\alpha^\alpha,
        \end{equation}
        i.e., $\norm{X}_\alpha \leq 1 = \norm{X}_\beta$, as claimed. 
In fact, 
this proof shows that $\norm{X}_\alpha$ is monotone non-increasing for all $\alpha > 0$, not only for $\alpha\geq 1$.
		
		\item[2.] Isometric invariance holds because the singular values of a linear operator $X$ do not change under the action $X\mapsto UXV^\dagger$, for all isometries $U$ and~$V$.
		
		\item[3.] For a proof of \eqref{eq-math_tools_submult_alpha_strong}, see the Bibliographic Notes (Section~\ref{math-tools:sec:bib-notes}). Submultiplicativity in \eqref{eq:math-tools:submult-alpha-norms} follows immediately from \eqref{eq-math_tools_submult_alpha_strong} by taking $Z=\mathbbm{1}$, using the fact that $\norm{\mathbbm{1}}_{\infty}=1$ (see Section~\ref{sec-Schatten_infty_norm} below), and using monotonicity, which implies that $\norm{X}_{\infty}\leq\norm{X}_{\alpha}$.

		\item[4.-5.] Multiplicativity with respect to the tensor product and the direct-sum property follow immediately from the definition of $\norm{\cdot}_{\alpha}$.
		
		\item[6.] We provide a proof of \eqref{eq-Schatten_norm_var} in the special case $\alpha=1,\beta=\infty$ in Proposition~\ref{prop:math-tools:var-char-t-norm} below. For all other values of $\alpha$ and $\beta$, please consult the Bibliographic Notes (Section~\ref{math-tools:sec:bib-notes}). Given \eqref{eq-Schatten_norm_var}, for all linear operators $X$ and $Z$, let $Y=\frac{Z}{\norm{Z}_{\beta}}$. Then, $\norm{Y}_{\beta}\leq 1$, which means that
		\begin{equation}
			\frac{1}{\norm{Z}_{\beta}}\abs{\Tr[Z^\dagger X]}=\abs{\Tr[Y^\dagger X]}\leq \norm{X}_{\alpha}\Rightarrow \abs{\Tr[Z^\dagger X]}\leq\norm{X}_{\alpha}\norm{Z}_{\beta},
		\end{equation}
		which is the inequality in \eqref{eq-Schatten_norm_duality}. \qedhere
		\end{enumerate}
	\end{Proof}
	
	In addition to the variational characterization of the Schatten norm $\norm{\cdot}_{\alpha}$ given in \eqref{eq-Schatten_norm_var}, we have the following variational characterization, which extends to $\alpha\in(0,1)$.
	
	\begin{proposition}{prop-Schatten_pos_var}
		Let $\alpha\in (0,1)\cup (1,\infty]$. Then, for every $\beta$ such that $\frac{1}{\alpha}+\frac{1}{\beta}=1$, and every positive semi-definite operator $X$,
		\begin{equation}
			\norm{X}_{\alpha}=\left\{\begin{array}{l l} \inf\{\Tr\big[XY^{\frac{1}{\beta}}\big]:Y> 0,~\Tr[Y]= 1\} & \text{if }\alpha\in[0,1),\\
			\sup\{\Tr\big[XY^{\frac{1}{\beta}}\big]:Y \geq  0,~\Tr[Y]= 1\} & \text{if }\alpha\in [1,\infty).\end{array}\right.
		\end{equation}
	\end{proposition}
	
	\begin{Proof}
		Please consult the Bibliographic Notes (Section~\ref{math-tools:sec:bib-notes}).
	\end{Proof}

\subsubsection{Schatten \texorpdfstring{$\infty$}{infinity}-Norm (Spectral/Operator Norm)}\label{sec-Schatten_infty_norm}
	
	An important case of the Schatten norms is the Schatten $\infty$-norm, which we recall from \eqref{eq-MT:orig-def-inf-norm-Schatten} is defined as
	\begin{equation}\label{eq:math-tools:infty-norm-as-limit}
		\norm{X}_{\infty}\coloneqq\lim_{\alpha\to\infty}\norm{X}_{\alpha},
	\end{equation}
	
	\begin{proposition*}{Schatten $\infty$-Norm is Largest Singular Value}{prop:math-tools:infty-norm-from-alpha}
		For every linear operator $X\in \Lin(\mathcal{H}_A,\mathcal{H}_B)$, $\norm{X}_{\infty}$ is equal to the largest singular value of $X$, which we denote by $s_{\max}$, i.e.,
		\begin{equation}
			\norm{X}_{\infty} = s_{\max}.
		\end{equation}
	\end{proposition*}
	
	\begin{Proof}
        Let $\vec{s}\coloneqq (s_k)_{k=1}^r$ denote the vector of singular values of $X$, where $r \coloneqq \rank(X)$. Then for all $\alpha \geq 1$ and $k \in \{1,\ldots, r\}$, the inequalities $s_{\max}^{\alpha} \leq \sum_{k=1}^r s_k^{\alpha} \leq r s_{\max}^\alpha$ hold. Raising everything to the power of $1/\alpha$ and taking the limit $\alpha\to\infty$ concludes the proof, because $\lim_{\alpha\to\infty} r^{1/\alpha} = 1$, given that $r$ is fixed.
	\end{Proof}
	
	Due to Proposition~\ref{prop:math-tools:infty-norm-from-alpha}, the term \textit{spectral norm} is often used to refer to the Schatten $\infty$-norm. It is also referred to as the \textit{operator norm}, because it is the norm induced by the Euclidean norm on the underlying Hilbert space on which the operator $X$ acts, i.e.,
	\begin{equation}\label{eq-inf_spectral_norm}
		\norm{X}_{\infty}=\sup_{\ket{\psi}\neq 0}\frac{\norm{X\ket{\psi}}_2}{\norm{\ket{\psi}}_2}=\sup_{\ket{\psi}:\norm{\ket{\psi}}_2=1}\norm{X\ket{\psi}}_2.
	\end{equation}
	In the equation above, we have employed the shorthand $\sup$, which stands for supremum. We also often employ $\inf$ for infimum. These concepts are reviewed in Section~\ref{sec:math-tools:cont-inf-sup}.
	
	\begin{exercise}{exer-inf_spec_norm}
		Using the fact that $X$ has a singular value decomposition of the form $X=\sum_{k=1}^{\rank(X)}s_k\ketbra{e_k}{f_k}$ (see Theorem~\ref{thm-SVD}), prove \eqref{eq-inf_spectral_norm}. Similarly, prove that
		\begin{equation}\label{eq-inf_norm_matrix_sing_decomp}
			\norm{X}_{\infty}=\sup\{\abs{\bra{\psi}X\ket{\phi}}:\norm{\ket{\psi}}_2=\norm{\ket{\phi}}_2=1\}.
		\end{equation}
	\end{exercise}
	
	If $X$ is Hermitian and positive semi-definite, then $\norm{X}_{\infty}$ is equal to the largest eigenvalue of $X$, and we can write
	\begin{align}
		\norm{X}_{\infty}&=\sup_{\ket{\psi}:\norm{\ket{\psi}}_2=1}\bra{\psi}X\ket{\psi}\\
		&=\sup_{\rho\geq 0}\{ \Tr[X\rho]: \Tr[\rho]=1\} \quad \text{($X$ positive semi-definite)}.\label{eq-X_PSD_largest_eig}
	\end{align}
	More generally, if $X$ is Hermitian and $\{\lambda_k\}_{k=1}^{\rank(X)}$ is the set of its eigenvalues, then
	\begin{align}\label{eq-inf_norm_Hermitian}
		\norm{X}_{\infty}&=\sup_{\ket{\psi}:\norm{\ket{\psi}}_2=1}\abs{\bra{\psi}X\ket{\psi}}\\
		&=\max_{1\leq k\leq\rank(X)}\abs{\lambda_k}\quad \text{($X$ Hermitian)}.
	\end{align}
	
	\begin{exercise}{exer-inf_norm_unitaries}
		Let $U$ be a unitary operator. Prove that $\norm{U}_{\infty}=1$. More generally, prove that $\norm{V}_{\infty}=1$ for every isometry $V$.
	\end{exercise}

\subsubsection{Schatten 1-Norm (Trace Norm)}\label{sec-math_tools_trace_norm}
	
	Another important special case of the Schatten $\alpha$-norm is $\alpha=1$. In this case, we refer to it as the \textit{trace norm}, and by applying \eqref{eq:math-tools:schatten-norm-from-sing-vals}, it is equal to the sum of the singular values of $X$:
	\begin{equation}
	\norm{X}_1 = \sum_{k=1}^{\rank(X)} s_k.
	\end{equation}
	If $X$ is Hermitian and positive semi-definite, then $\norm{X}_1$ is equal to the sum of the eigenvalues of $X$, i.e., to the trace of $X$:
	\begin{equation} \label{eq-trace_norm_psd}
		\norm{X}_1=\Tr[X]\quad\text{(X positive semi-definite).}
	\end{equation}
	More generally, if $X$ is Hermitian and $\{\lambda_k\}_{k=1}^{\rank(X)}$ is the set of its eigenvalues, then
	\begin{equation}\label{eq-trace_norm_Herm}
		\norm{X}_1=\sum_{k=1}^{\rank(X)}\abs{\lambda_k}\quad\text{(X Hermitian)}.
	\end{equation}
	
	\begin{exercise}{exer-trace_distance}
		Consider two vectors $\ket{\psi},\ket{\phi}\in\mathbb{C}^d$, with $d\geq 2$. Show that
		\begin{equation}
			\norm{\ketbra{\psi}{\psi}-\ketbra{\phi}{\phi}}_1^2=\left(\braket{\psi}{\psi}+\braket{\phi}{\phi}\right)^2-4\abs{\braket{\psi}{\phi}}^2.
		\end{equation}
	\end{exercise}
	
	We now provide a proof of the variational characterization of the Schatten norm in \eqref{eq-Schatten_norm_var} for the special case of $\alpha=1$ and $\beta=\infty$.
	
	\begin{proposition*}{Variational Characterization of Trace Norm}{prop:math-tools:var-char-t-norm}
		For all $X \in \Lin(\mathcal{H}_A,\mathcal{H}_B)$, the trace norm of $X$ has the following variational characterization:
		\begin{equation}\label{eq-trace_norm_variational}
			\norm{X}_1=\sup_{Y:\norm{Y}_{\infty}\leq 1}\abs{\Tr[Y^{\dagger}X]},
		\end{equation}
		where the optimization is with respect to all  $Y\in\Lin(\mathcal{H}_A,\mathcal{H}_B)$ with spectral norm bounded from above by one.
	\end{proposition*}
	
	\begin{Proof}
		Let $X=\sum_{k=1}^r s_k \ket{e_k}_B \bra{f_k}_A $ be the singular value decomposition of $X$, where $r \coloneqq \rank(X)$. Let $Y\in\Lin(\mathcal{H}_A,\mathcal{H}_B)$ be such that $\norm{Y}_{\infty}\leq 1$. Then,
		\begin{align}
			\abs{\Tr[Y^{\dagger}X]} & = \abs{\Tr\!\left[Y^{\dagger}\left(\sum_{k=1}^r s_k \ket{e_k}_B \bra{f_k}_A\right)\right]}\\
		 	&= \abs{\sum_{k=1}^r s_k \bra{e_k}_B Y \ket{f_k}_A}\\
			& \leq \sum_{k=1}^r s_k \abs{\bra{e_k}_B Y \ket{f_k}_A},
		\end{align}
		where the last line is due to the triangle inequality. Now, using \eqref{eq-inf_norm_matrix_sing_decomp}, we have
		\begin{equation}
			\abs{\bra{e_k}_BY\ket{f_k}_A}\leq\norm{Y}_{\infty}\leq 1,
		\end{equation}
		for every $k\in\{1,2,\dotsc,r\}$. Therefore,
		\begin{equation}
			\abs{\Tr[Y^{\dagger}X]} \leq \sum_{k=1}^r s_k = \norm{X}_1,
		\end{equation}
		which holds for every  $Y\in\Lin(\mathcal{H}_A,\mathcal{H}_B)$ satisfying $\norm{Y}_{\infty}\leq 1$, so that the inequality
		\begin{equation}
			\sup_{Y:\norm{Y}_{\infty}\leq 1}\abs{\Tr[Y^{\dagger}X]}\leq\norm{X}_1 
		\end{equation}
		holds. The opposite inequality holds by making a particular choice for $Y$. We pick $Y$ to be the following linear operator defined from the singular value decomposition of $X$: $Y=\sum_{k=1}^{r} \ket{e_k}_B \bra{f_k}_A $. Observe that $\norm{Y}_{\infty}=1$. Thus,
		\begin{align}
			\sup_{Y:\norm{Y}_{\infty}\leq 1}\abs{\Tr[Y^{\dagger}X]}&\geq\abs{\Tr\left[\left(\sum_{k'=1}^{r} \ket{f_{k'}}_A \bra{e_{k'}}_B\right) \left(\sum_{k=1}^r s_k \ket{e_k}_B \bra{f_k}_A\right) \right]} \\
			& =\abs{\sum_{k=1}^r s_k} \\
			& =\norm{X}_1.
		\end{align}
		This completes the proof.
	\end{Proof}
	
	\begin{remark}
		Observe that Proposition~\ref{prop:math-tools:var-char-t-norm} can be generalized as follows for every linear operator $X_{A\to B}\in\Lin(\mathcal{H}_A,\mathcal{H}_B)$:
		\begin{equation}\label{eq-var_char_trace_norm_alt}
			\norm{X}_1=\sup_{Y:\norm{Y}_{\infty}\leq 1}\text{Re}\left(\Tr[Y^{\dagger}X]\right),
		\end{equation}
		where, as before, the optimization is with respect to every  operator $Y\in\Lin(\mathcal{H}_A,\mathcal{H}_B)$ with spectral norm bounded from above by one. Indeed, for every complex number $z\in\mathbb{C}$, the inequality $\text{Re}(z)\leq\abs{\text{Re}(z)}\leq\abs{z}$ holds, which means that
		\begin{equation}
			\sup_{Y:\norm{Y}_{\infty}\leq 1}\text{Re}\left(\Tr[Y^{\dagger}X]\right)\leq\sup_{Y:\norm{Y}_{\infty}\leq 1}\abs{\Tr[Y^{\dagger}X]}=\norm{X}_1.
		\end{equation}
		Then, to obtain the opposite inequality, the same choice for $Y$ as in the the proof of Proposition~\ref{prop:math-tools:var-char-t-norm} can be made, because for that choice of $Y$ we have $\Tr[Y^{\dagger}X]=\norm{X}_1$, which is real, so that $\text{Re}(\Tr[Y^{\dagger}X])=\norm{X}_1$. We can thus conclude \eqref{eq-var_char_trace_norm_alt}.
		
		We also remark that in both \eqref{eq-trace_norm_variational} and \eqref{eq-var_char_trace_norm_alt}, it suffices to optimize with respect to isometries. In particular, because $\norm{U}_{\infty}=1$ for every isometry $U$ (see Exercise~\ref{exer-inf_norm_unitaries}), using similar techniques as in the proof of Proposition~\ref{prop:math-tools:var-char-t-norm}, it is straightforward to prove that for all $X\in\Lin(\mathcal{H}_A,\mathcal{H}_B)$,
		\begin{align}
			\norm{X}_1&=\sup_{\substack{U_{B\to A}\\\text{isometry}}} \abs{\Tr[U_{B\to A}X_{A\to B}]}=\sup_{\substack{U_{B\to A}\\\text{isometry}}}\text{Re}\left(\Tr[U_{B\to A}X_{A\to B}]\right),\quad d_A\geq d_B,\label{eq-trace_norm_var_iso1}\\
			\norm{X}_1&=\sup_{\substack{V_{A\to B}\\\text{isometry}}} \abs{\Tr[V_{A\to B}(X_{A\to B})^{\dagger}]}=\sup_{\substack{V_{A\to B}\\\text{isometry}}}\text{Re}\left(\Tr[V_{A\to B}(X_{A\to B})^{\dagger}]\right),\quad d_A\leq d_B.\label{eq-trace_norm_var_iso2}
		\end{align}
		In particular, if $d_A=d_B=d$, then the optimization in \eqref{eq-trace_norm_var_iso1} and \eqref{eq-trace_norm_var_iso2} is with respect to unitary operators, so that for all $X\in\Lin(\mathbb{C}^d)$,
		\begin{equation}\label{eq-var_trace_norm_unitary}
			\norm{X}_1=\sup_{\substack{U\in\Lin(\mathbb{C}^d)\\\text{unitary}}}\abs{\Tr[UX]}=\sup_{\substack{U\in\Lin(\mathbb{C}^d)\\\text{unitary}}}\text{Re}\left(\Tr[UX]\right).
		\end{equation}
	\end{remark}
	
	The monotonicity result in Proposition~\ref{prop-Schatten_norm_prop} implies that
	\begin{equation}\label{eq:math-tools:infty-norm-to-trace-norm}
		\norm{X}_\infty \leq \norm{X}_1. 
	\end{equation}
	for every linear operator $X$. The following proposition gives a tighter bound than the one in \eqref{eq:math-tools:infty-norm-to-trace-norm} for the special case when $X$ is a traceless Hermitian operator.
	
	\begin{Lemma}{lem-QCAP:traceless-Hermitian}
		Let $X$ be a Hermitian operator satisfying $\Tr[X]=0$. Then,
		\begin{equation}
			\norm{X}_{\infty}\leq\frac{1}{2}\norm{X}_{1}.
		\end{equation}
	\end{Lemma}

	\begin{Proof}
		Let the Jordan--Hahn decomposition of $X$ be given by
		\begin{equation}
			X=X_{+}-X_{-},
		\end{equation}
		where $X_{+},X_{-}\geq 0$ and $X_{+}X_{-}=0$. Then,
		\begin{equation}
			\norm{X} _{1}=\Tr[X_{+}]+\Tr[X_{-}].
		\end{equation}
		Since $\Tr[X]=0$, we have that $\Tr[X_{+}]=\Tr[X_{-}]$, which means that
		\begin{equation}
			\norm{X} _{1}=2\Tr[X_{+}].
		\end{equation}
		We also have that
		\begin{equation}
			\norm{X} _{\infty}=\max\{\norm{ X_{+}}_{\infty},\norm{X_{-}} _{\infty}\},
		\end{equation}
		because $X_{+}X_{-}=0$. Then, since $\|X_\pm\|_\infty \leq \|X_\pm\|_1 = \Tr X_\pm$ by \eqref{eq:math-tools:infty-norm-to-trace-norm} and \eqref{eq-trace_norm_psd}, we deduce that
		\begin{equation}
			\norm{ X} _{\infty}=\max\left\{  \norm{ X_{+}}_{\infty},\norm{ X_{-}} _{\infty}\right\} \leq \Tr[X_{+}]=\frac{1}{2}\norm{ X} _{1},
		\end{equation}
        thus concluding the proof.
	\end{Proof}
	
	We remark that the monotonicity inequality $\norm{X}_{\infty}\leq\norm{X}_1$ in \eqref{eq:math-tools:infty-norm-to-trace-norm} can be reversed to give
	\begin{equation}
		\norm{X}_1\leq d\norm{X}_{\infty}
	\end{equation}
	for every linear operator $X$ acting on a $d$-dimensional Hilbert space. This holds because
	\begin{equation}
	\norm{X}_1 = \sum_{k=1}^r s_k
	\leq \sum_{k=1}^r \max_{k \in \{1,\ldots,r\}}s_k = r \max_{k \in \{1,\ldots,r\}}s_k = r \norm{X}_{\infty} \leq d \norm{X}_{\infty},
	\end{equation}
	where $\{s_k\}_{k=1}^r$ is the set of singular values of $X$ and $r\coloneqq \rank(X)$. (We used this very same reasoning in the proof of Proposition~\ref{prop:math-tools:infty-norm-from-alpha}.)
	
	Using Proposition~\ref{prop:math-tools:var-char-t-norm}, we can establish the following slight strengthening of the H\"older inequality in \eqref{eq-Schatten_norm_duality}:
	\begin{equation}\label{eq:math-tools:holder++}
		\norm{Z^\dag X}_1 \leq \norm{X}_\alpha \norm{Z}_\beta,
	\end{equation}
	which holds for all linear operators $X$ and $Z$ and $\alpha,\beta \in [1,\infty]$ satisfying $\frac{1}{\alpha} + \frac{1}{\beta}=1$. This actually follows by a direct application of the H\"older inequality itself:
	\begin{equation}
		\abs{\Tr[UZ^{\dag}X]}\leq\norm{X}_{\alpha}\norm{ZU^{\dagger}}_{\beta}=\norm{X}_{\alpha}\norm{Z}_{\beta},
	\end{equation}
	which holds for every isometry $U$. Therefore, it follows from \eqref{eq-trace_norm_variational} that
	\begin{equation}
		\norm{Z^{\dag}X}_{1}=\sup_{U}\abs{\Tr[UZ^{\dagger}X]}\leq\norm{X}_{\alpha}\norm{Z}_{\beta}.
	\end{equation}

	

\subsection{Operator Inequalities} \label{subsec:operator_inequalities}

Throughout this book, we make use of the \textit{L\"{o}wner partial order} for Hermitian operators. It is useful as a way of comparing two Hermitian operators in $\Lin(\mathcal{H})$, generalizing the way in which we compare two real numbers.

	\begin{definition}{L\"{o}wner Partial Order for Hermitian Operators}{def-Loewner}
		For two Hermitian operators $X$ and $Y$, the expression $X\geq Y$ is an \textit{operator inequality} and means that $X-Y\geq 0$, i.e., that $X-Y$ is positive semi-definite. We also write $X\leq Y$ to mean $Y-X\geq 0$. The expressions $X > Y$ and $Y <X$ mean that $X -Y$ is positive definite.
	\end{definition}

	The relations ``$\geq$'' and ``$\leq$'' satisfy the following expected properties: $X\leq Y$ and $X\geq Y$ imply that $X=Y$, and $X\leq Y$ and $Y\leq Z$ imply that $X\leq Z$. The term ``partial order'' is used because not every pair $(X,Y)$ of Hermitian operators satisfies either $X\geq Y$ or $X\leq Y$.
	
	\begin{definition}{Operator Convex, Concave, Monotone Functions}{def-op_conv_conc_mono}
		Let $X$ and $Y$ be Hermitian operators, and let $f\colon \mathbb{R}\to\mathbb{R}$ be a function extended to Hermitian operators as in \eqref{eq-function_Hermitian_ext}.
		\begin{enumerate}
			\item The function $f$ is called \textit{operator convex} if for all $\lambda\in[0,1]$ and Hermitian operators $X$ and $Y$, the following inequality holds:
				\begin{equation}
					f(\lambda X+(1-\lambda) Y)\leq \lambda f(X)+(1-\lambda) f(Y). 
				\end{equation}
				We call $f$ \textit{operator concave} if $-f$ is operator convex.
		
			\item The function $f$ is called \textit{operator monotone} if $X\leq Y$ implies $f(X)\leq f(Y)$ for all Hermitian operators $X$ and $Y$. We call $f$ \textit{operator anti-monotone} if $-f$ is operator monotone.
		\end{enumerate}
	\end{definition}
	
	The functions considered in Section~\ref{sec:math-tools:functions-herm-ops} have the following properties with respect to Definition~\ref{def-op_conv_conc_mono}:
	\begin{itemize}
		\item The function $x\mapsto x^{\alpha}$ is operator monotone for $\alpha\in[0,1]$ and $x\in[0,\infty)$, operator anti-monotone for $\alpha\in[-1,0)$ and $x\in(0,\infty)$, operator convex for $\alpha\in[-1,0)$ and $x\in(0,\infty)$, operator convex for $\alpha\in[1,2]$ and $x\in[0,\infty)$, and operator concave for $\alpha\in(0,1]$ and $x\in[0,\infty)$. Note that the function $x\mapsto x^{\alpha}$ is neither operator monotone, operator convex, nor operator concave for $\alpha<-1$ and $\alpha>2$.
		\item The function $x\mapsto\log_b(x)$, for every base $b>0$ and $x\in(0,\infty)$, is operator monotone and operator concave.
		\item The function $x\mapsto x\log_b(x)$, for every base $b>0$ and $x\in[0,\infty)$, is operator convex\footnote{Note that, because $\lim_{x\to 0}x\log_b(x)=0$, we take the convention that $0\log_b(0)=0$ throughout this book.}.
	\end{itemize}
		For proofs of these properties, please see the Bibliographic Notes (Section~\ref{math-tools:sec:bib-notes}). We note here that these properties are critical for understanding quantum entropies, as detailed in Chapter~\ref{chap-entropies}. Especially, the data-processing inequality for quantum relative entropy, which is at the heart of understanding quantum communication limits, is intimately related to operator convexity.

	We now state some basic operator inequalities that we use repeatedly throughout the book.
	
	\begin{Lemma*}{Basic Operator Inequalities}{prop-operator_ineqs}
		Let $X,Y\in\Lin(\mathcal{H})$ be Hermitian operators acting on a Hilbert space $\mathcal{H}$.
		
		\begin{enumerate}
			\item $X\geq 0\Rightarrow Z X Z^\dagger \geq 0$ for all $Z\in\Lin(\mathcal{H},\mathcal{H}')$. In particular, $X\geq Y\Rightarrow  ZXZ^\dagger \geq Z YZ^\dagger$ for all $Z\in\Lin(\mathcal{H},\mathcal{H}')$.
			
			\item $X\geq Y\Rightarrow \Tr[X]\geq \Tr[Y]$. More generally, $X\geq Y\Rightarrow \Tr[WX]\geq \Tr[WY]$ for all $W\in\Lin(\mathcal{H})$ satisfying $W\geq 0$.

			\item For every Hermitian operator $X$ with maximum and minimum eigenvalues $\lambda_{\max}$ and $\lambda_{\min}$, respectively, $\lambda_{\min}\mathbbm{1}\leq X\leq \lambda_{\max}\mathbbm{1}$.
			
			\item Let $X$ and $Y$ have their spectrum in some interval $I\subset\mathbb{R}$, and let $f\colon I\to\mathbb{R}$ be a monotone increasing function. If $X\leq Y$, then $\Tr[f(X)]\leq\Tr[f(Y)]$. In particular, if $X$ and $Y$ are positive semi-definite, then
				\begin{equation}\label{eq-power_func_monotone_PSD}
					0\leq X\leq Y\Rightarrow \Tr[X^\alpha]\leq \Tr[Y^\alpha]\quad\forall~\alpha>0.
				\end{equation}
		(Note that we are \textit{not} requiring $f$ to be \textit{operator} monotone.)
            \end{enumerate}
	\end{Lemma*}
	
	\begin{Proof}
		\hfill\begin{enumerate}
			\item $X\geq 0$ implies that $\bra{\psi}X\ket{\psi}\geq 0$ for all $\ket{\psi}\in\mathcal{H}$. Then, for every vector $\ket{\phi}\in\mathcal{H}'$, we have $\bra{\phi}Z  XZ^\dag \ket{\phi}\geq 0$ because $Z^\dag \ket{\phi}\equiv\ket{\psi}$ is a vector in $\mathcal{H}$. Therefore, $Z XZ^\dag \geq 0$.
			
				Now, $X\geq Y$ is equivalent to  $X-Y\geq 0$. Let $W=X-Y$. Then, from the arguments in the previous paragraph, we have $Z WZ^\dagger\geq 0$ for all $Z$, which implies that $Z XZ^\dagger-Z YZ^\dagger\geq 0$, i.e., $Z XZ^\dagger\geq Z YZ^\dagger$, as required.
			
			\item $X\geq Y$ implies that $X-Y\geq 0$. The trace of a positive semi-definite operator is non-negative, since positive semi-definite operators have non-negative eigenvalues and the trace of every normal operator is equal to the sum of its eigenvalues. Thus, $X-Y\geq 0$ implies $\Tr[X-Y]\geq 0$, which implies that $\Tr[X]\geq \Tr[Y]$, as required.
			
			Next, let $W$ be a positive semi-definite operator. Using 1.~above, $X\geq Y$ implies that $\sqrt{W}X\sqrt{W}\geq \sqrt{W}Y\sqrt{W}$. Then, using the result of the previous paragraph, we obtain $\Tr[\sqrt{W}X\sqrt{W}]\geq \Tr[\sqrt{W}Y\sqrt{W}]$. Finally, by cyclicity of the trace (recall \eqref{eq-trace_cyclic}), we find that $\Tr[WX]\geq \Tr[WY]$, as required.
			
			\item This result follows from the fact that, for every Hermitian operator $X\in\Lin(\mathcal{H})$ with eigenvalues $\{\lambda_k\}_{k=1}^{\dim(\mathcal{H})}$, the eigenvalues of $X+t\mathbbm{1}$ are equal to $\{\lambda_k+t\}_{k=1}^{\dim(\mathcal{H})}$ for every $t\in\mathbb{R}$. In particular, then, by definition of the minimum eigenvalue, $X-\lambda_{\min}\mathbbm{1}\geq 0$, because all of the eigenvalues of $X-\lambda_{\min}\mathbbm{1}$ are non-negative. Similarly, by definition of the maximum eigenvalue, $X-\lambda_{\max}\mathbbm{1}\leq 0$, because all of the eigenvalues of $X-\lambda_{\max}\mathbbm{1}$ are non-positive.
			
			\item Let $\lambda^{\downarrow}_i(X)$ denote the sequence of decreasingly ordered eigenvalues of $X$. Then the inequalities $\lambda^{\downarrow}_i(X) \leq \lambda^{\downarrow}_i(Y)$ hold for all $i\in\{1,\ldots, \dim(\mathcal{H})\}$. These inequalities are a consequence of the Courant--Fischer--Weyl minimax principle (please consult the Bibliographic Notes in Section~\ref{math-tools:sec:bib-notes} for a reference to this principle). Then, the desired inequality follows directly from the fact that $\Tr[f(X)] = \sum_{i=1}^{\dim(\mathcal{H})} f(\lambda^{\downarrow}_i(X))$, as well as the monotonicity of $f$. The inequality in \eqref{eq-power_func_monotone_PSD} follows because the function $x^\alpha$ with domain $x\geq 0$ is monotone for all $\alpha > 0$. \qedhere
			\end{enumerate}
	\end{Proof}
	\bigskip
    
	\begin{Lemma*}{Araki--Lieb--Thirring Inequality}{lem-ALT_ineq}
		Let $X$ and $Y$ be positive semi-definite operators acting on a finite-dimens\-ional Hilbert space. For all $q\geq 0$:
			\begin{enumerate}
				\item $\Tr\!\left[\left(Y^{\frac{1}{2}}XY^{\frac{1}{2}}\right)^{rq}\right]\geq \Tr\!\left[\left(Y^{\frac{r}{2}}X^r Y^{\frac{r}{2}}\right)^q\right]$ for all $0\leq r< 1$.
				\item $\Tr\!\left[\left(Y^{\frac{1}{2}}X Y^{\frac{1}{2}}\right)^{rq}\right]\leq \Tr\!\left[\left(Y^{\frac{r}{2}}X^rY^{\frac{r}{2}}\right)^q\right]$ for all $r\geq 1$.
			\end{enumerate}
	\end{Lemma*}
	
	\begin{Proof}
		Please consult the Bibliographic Notes (Section~\ref{math-tools:sec:bib-notes}).
	\end{Proof}

    The above inequalities relate certain trace expressions that are obtained from one another by redistributing the powers of non-commuting matrices in a product. Intuitively, the general idea is that the trace of a product of positive semidefinite operators is largest when the operators are as little intertwined as possible: for example, taking $q=1$ and $r=2$ in Lemma~\ref{lem-ALT_ineq}, we obtain that
    \begin{equation}
        \Tr\!\left[\left(Y^{\frac{1}{2}}X Y^{\frac{1}{2}}\right)^{2}\right] = \Tr\!\left[ Y X Y X\right] \leq \Tr\!\left[Y^2X^2\right] ,
    \end{equation}
    which can be intuitively understood to hold because the product on the left-hand side is `more intertwined' than the one on the right-hand side. (This particular inequality can also be shown directly with Cauchy--Schwarz.)
    
	The operator Jensen inequality below is the linchpin of several quantum data-process\-ing inequalities presented later on in Chapter~\ref{chap-entropies}. These in turn are repeatedly used in Parts~\ref{part:q-comm-prots} and \ref{part-feedback} to place fundamental limits on quantum communication protocols. As such, the operator Jensen inequality is a significant bridge that connects convexity to information processing.
		
	\begin{theorem*}{Operator Jensen Inequality}{thm-Jensen}
		Let $f\colon \mathbb{R}\to\mathbb{R}$ be a continuous function with $\text{dom}(f)=I\subset\mathbb{R}$ (where $I$ is an interval). Then the following are equivalent:
		
		\begin{enumerate}
			\item $f$ is operator convex.
			
			\item For all $n\in\mathbb{N}$, the inequality
				\begin{equation}
					f\!\left(\sum_{k=1}^n A_k^\dagger X_k A_k^{\vphantom{\dag}}\right)\leq \sum_{k=1}^n A_k^\dagger f(X_k) A_k^{\vphantom{\dag}}
					\label{eq:math-tools:op-Jen-with-Aks}
				\end{equation}
				holds for every collection $\{X_k\}_{k=1}^n$ of Hermitian operators acting on a Hilbert space $\mathcal{H}$ with spectrum contained in $I$ and every collection $\{A_k\}_{k=1}^n$ of linear operators in $\Lin(\mathcal{H}',\mathcal{H})$  satisfying $\sum_{k=1}^n A_k^\dagger A_k^{\vphantom{\dag}} =\mathbbm{1}_{\mathcal{H}'}$.
				
			\item For every Hermitian operator $X \in\Lin(\mathcal{H})$ with spectrum in $I$ and every isometry $V\in\Lin(\mathcal{H}',\mathcal{H})$, the following inequality holds:
	\begin{equation}
	 f\left(V^\dagger XV\right)\leq V^\dagger f(X) V.
	 \label{eq:math-tools:op-Jen-with-V}
	 \end{equation}
		\end{enumerate}
	\end{theorem*}
	
	\begin{Proof}
We first prove that 2.~$\Rightarrow$ 1. Let $X$ and $Y$ be
Hermitian operators with their eigenvalues in $I$. Let $\lambda\in\left[
0,1\right]  $. We can take $n=2$, $A_{1}=\sqrt{\lambda}\mathbbm{1}$, $X_{1}=X$,
$A_{2}=\sqrt{1-\lambda}\mathbbm{1}$, $X_{2}=Y$, and the following operator inequality is an
immediate consequence of \eqref{eq:math-tools:op-Jen-with-Aks}:
\begin{equation}
f(\lambda X+\left(  1-\lambda\right)  Y)\leq\lambda f(X)+\left(
1-\lambda\right)  f(Y).
\end{equation}
Since $X$, $Y$, and $\lambda$ are arbitrary, it follows that $f$ is operator convex.

We now prove that 2.~$\Rightarrow$ 3. In fact, 3.~is actually a special case of 2., found by setting $n=1$ and taking
$A_{1}=V$ and $X_{k}=X$, with $V$ an isometry and $X$ Hermitian with eigenvalues in $I$.

Now we prove that 3.~$\Rightarrow$ 2. Fix $n\in\mathbb{N}$ and the sets
$\left\{  A_{k}\right\}  _{k=1}^{n}$ and $\left\{  X_{k}\right\}  _{k=1}^{n}$ of operators such that they
satisfy the conditions specified in 2. Define the following Hermitian
operator:
\begin{equation}
X\coloneqq\sum_{k=1}^{n}X_{k}\otimes|k\rangle\!\langle k|,
\end{equation}
as well as the isometry
\begin{equation}
V\coloneqq\sum_{k=1}^{n}A_{k}\otimes|k\rangle,
\end{equation}
where $\{\ket{k}\}_{k=1}^n$ is an orthonormal basis.
The condition $\sum_{k=1}^{n}A_{k}^{\dag}A_{k}=\mathbbm{1}$ and a 
calculation imply that $V$ is an isometry (satisfying $V^\dag V = \mathbbm{1}$). Another calculation implies that
\begin{equation}
V^{\dag}XV=\sum_{k=1}^{n}A_{k}^{\dag}X_{k}A_{k}. \label{eq:math-tools:op-Jens-V-to-Aks1}
\end{equation}
Since
\begin{equation}
f(X)=f\!\left(  \sum_{k=1}^{n}X_{k}\otimes|k\rangle\!\langle k|\right)
=\sum_{k=1}^{n}f(X_{k})\otimes|k\rangle\!\langle k|,
\label{eq:math-tools:f-on-a-block-1}
\end{equation}
which follows as a consequence of \eqref{eq-function_Hermitian_ext}, a similar calculation implies that
\begin{equation}
V^{\dag}f(X)V=\sum_{k=1}^{n}A_{k}^{\dag}f(X_{k})A_{k}.
\label{eq:math-tools:op-Jens-V-to-Aks2}
\end{equation}
Then the desired inequality in \eqref{eq:math-tools:op-Jen-with-Aks} follows from \eqref{eq:math-tools:op-Jens-V-to-Aks1}, \eqref{eq:math-tools:op-Jens-V-to-Aks2}, and \eqref{eq:math-tools:op-Jen-with-V}.

We finally prove that 1.~$\Rightarrow$ 3. Fix the operator $X$ and isometry
$V$, as specified in~3. Let $M$ be a Hermitian operator in $\Lin(\mathcal{H}')$ with spectrum in $I$.
Let $P\coloneqq \mathbbm{1}_{\mathcal{H}}-VV^{\dag}$, and observe that $P$ is a projection (i.e., $P^2=P$), $V^{\dag}P=0$, and $PV=0$. Set
\begin{equation}
Z\coloneqq
\begin{pmatrix}
X & 0\\
0 & M
\end{pmatrix}
,\qquad U\coloneqq
\begin{pmatrix}
V & P\\
0 & -V^{\dag}
\end{pmatrix}
,\qquad W\coloneqq
\begin{pmatrix}
V & -P\\
0 & V^{\dag}
\end{pmatrix}
.
\label{eq:math-tools:op-Jensen-unitary-from-iso}
\end{equation}
Observe that $U$ and $W$ are unitary operators (these are called unitary dilations of the isometry $V$). By direct calculation, we then
find that
\begin{align}
U^{\dag}ZU & =
\begin{pmatrix}
V^{\dag}XV & V^{\dag}XP\\
PXV & PXP+VMV^{\dag}
\end{pmatrix}
,\\
W^{\dag}ZW & =
\begin{pmatrix}
V^{\dag}XV & -V^{\dag}XP\\
-PXV & PXP+VMV^{\dag}
\end{pmatrix}
,
\end{align}
so that
\begin{equation}
\frac{1}{2}\left(  U^{\dag}ZU+W^{\dag}ZW\right)  =
\begin{pmatrix}
V^{\dag}XV & 0\\
0 & PXP+VMV^{\dag}
\end{pmatrix}
.\label{eq:math-tools:op-Jensen-key-step}
\end{equation}
From the same reasoning that leads to \eqref{eq:math-tools:f-on-a-block-1}, and using \eqref{eq:math-tools:op-Jensen-key-step}, we find that
\begin{align}
 & \!\!\!\!\!\!\!\!\!\!\!\!\begin{pmatrix}
f\!\left(  V^{\dag}XV\right)   & 0\\
0 & f\!\left(  PXP+VMV^{\dag}\right)
\end{pmatrix}\notag \\
& =f\!\left(
\begin{pmatrix}
V^{\dag}XV & 0\\
0 & PXP+VMV^{\dag}
\end{pmatrix}
\right)  \\
& =f\!\left(  \frac{1}{2}\left(  U^{\dag}ZU+W^{\dag}ZW\right)  \right)  \\
& \leq\frac{1}{2}f\!\left(  U^{\dag}ZU\right)  +\frac{1}{2}f\!\left(  W^{\dag
}ZW\right)  \\
& =\frac{1}{2}U^{\dag}f\!\left(  Z\right)  U+\frac{1}{2}W^{\dag}f\!\left(
Z\right)  W\\
& =
\begin{pmatrix}
V^{\dag}f(X)V & 0\\
0 & Pf(X)P+Vf(M)V^{\dag}
\end{pmatrix}
.
\end{align}
The inequality follows from the assumption that $f$ is operator convex. The third
equality follows from \eqref{eq-function_Hermitian_ext}. The final equality follows because
\begin{equation}
f(Z)=
\begin{pmatrix}
f(X) & 0\\
0 & f(M)
\end{pmatrix}
,
\end{equation}
and by applying \eqref{eq:math-tools:op-Jensen-key-step} again, with the substitutions $Z\rightarrow f(Z)$, $X\rightarrow f(X)$, and
$M\rightarrow f(M)$. It follows that
\begin{equation}
\begin{pmatrix}
f\left(  V^{\dag}XV\right)   & 0\\
0 & f\left(  PXP+VMV^{\dag}\right)
\end{pmatrix}
\leq
\begin{pmatrix}
V^{\dag}f(X)V & 0\\
0 & Pf(X)P+Vf(M)V^{\dag}
\end{pmatrix}
,
\label{eq:math-tools:op-Jens-final-step-op-conv-to-V}
\end{equation}
and we finally conclude that $f(V^{\dag}XV)\leq V^{\dag}f(X)V$ by examining
the upper left blocks in the operator inequality in \eqref{eq:math-tools:op-Jens-final-step-op-conv-to-V}.
	\end{Proof}

\subsection{Superoperators}\label{sec:math-tools:super-ops}

	Just as we have been considering linear operators of the form $X:\mathcal{H}_A\to\mathcal{H}_B$, with input and output Hilbert spaces $\mathcal{H}_A$ and $\mathcal{H}_B$, we can consider linear operators with input Hilbert space $\Lin(\mathcal{H}_A)$ and output Hilbert space $\Lin(\mathcal{H}_{B})$. We use the term \textit{superoperator} to refer to a linear operator acting on the Hilbert space of linear operators. Specifically, a superoperator is a function $\mathcal{N}:\Lin(\mathcal{H}_A)\to \Lin(\mathcal{H}_{B})$ such that 
	\begin{equation}
		\mathcal{N}(\alpha X+\beta Y)=\alpha\mathcal{N}(X)+\beta\mathcal{N}(Y)
	\end{equation}
	for all $\alpha,\beta\in\mathbb{C}$ and $X,Y\in\Lin(\mathcal{H}_A)$. It is often helpful to indicate explicitly the input and output Hilbert spaces of a superoperator $\mathcal{N}:\Lin(\mathcal{H}_A)\to\Lin(\mathcal{H}_{B})$ by writing $\mathcal{N}_{A\to B}$. We make use of this notation throughout the book.
	
	For every superoperator $\mathcal{N}_{A\to B}$, there exists $n \in \mathbb{N}$, and sets $\{K_i\}_{i=1}^n$ and $\{L_i\}_{i=1}^n$ of operators in $\Lin(\mathcal{H}_A,\mathcal{H}_B)$ such that
	\begin{equation}
    \label{eq-math_tools-superator_op_sum}
	\mathcal{N}_{A\to B}(X_A) = \sum_{i=1}^n K_i X_A L_i^{\dagger},
	\end{equation}
	for all $X_A \in \Lin(\mathcal{H}_A)$.
	This follows as a consequence of the requirement that $\mathcal{N}_{A\to B}$ has a linear action on $X_A$ and the isomorphism in \eqref{eq-vec_operation}. Eq.~\eqref{eq-math_tools-superator_op_sum} is known as the operator-sum representation of the superoperator $\mathcal{N}_{A\to B}$. The transpose operation discussed previously in \eqref{eq:math-tools:transpose-superop} is an example of a superoperator. In Chapter~\ref{chap-QM_channels}, we see that quantum physical evolutions of quantum states, known as quantum channels, are other examples of superoperators with additional constraints on the sets $\{K_i\}_{i=1}^n$ and $\{L_i\}_{i=1}^n$.
	
	We denote the identity superoperator by $\id:\Lin(\mathcal{H})\to\Lin(\mathcal{H})$, and by definition it satisfies $\id(X)=X$ for all $X\in\Lin(\mathcal{H})$.
	
	Given two superoperators $\mathcal{N}_{A\to A'}$ and $\mathcal{M}_{B\to B'}$, their tensor product $\mathcal{N}_{A\to A'}\allowbreak\otimes\mathcal{M}_{B\to B'}$ is the unique superoperator with input Hilbert space $\Lin(\mathcal{H}_A\otimes\mathcal{H}_{B})$ and output Hilbert space $\Lin(\mathcal{H}_{A'}\otimes\mathcal{H}_{B'})$, such that
	\begin{equation}
		(\mathcal{N}_{A\to A'}\otimes\mathcal{M}_{B\to B'})(X_A\otimes Y_B)=\mathcal{N}_{A\to A'}(X_A)\otimes\mathcal{M}_{B\to B'}(Y_B)
	\end{equation}
	for all $X_A\in\Lin(\mathcal{H}_A)$ and $Y_B\in\Lin(\mathcal{H}_B)$. We use the abbreviation
	\begin{equation}
		\mathcal{N}_{A\to A'}\otimes \id_{B\to B'}\equiv\mathcal{N}_{A\to A'},\quad \id_{A\to A'}\otimes\mathcal{M}_{B\to B'}\equiv\mathcal{M}_{B\to B'}
	\end{equation}
	throughout this book whenever a superoperator acts only on one of the tensor factors of the underlying Hilbert space of linear operators.
	
	\begin{definition}{Hermiticity-Preserving Superoperator}{def-Herm_pres_superop}
		A superoperator $\mathcal{N}$ is called \textit{Hermiticity preserving} if $\mathcal{N}(X)$ is Hermitian for every Hermitian input $X$.
	\end{definition}
	
	\begin{exercise}{exer-herm_pres_superoperator}
		Using \eqref{eq-operator_Herm_decomp}, prove that a superoperator $\mathcal{N}$ is Hermiticity preserving if and only if $\mathcal{N}(X^\dagger)=\mathcal{N}(X)^\dagger$ for every linear operator $X$.
	\end{exercise}
	
	\begin{definition}{Adjoint of a Superoperator}{def-adjoint_superop}
		The \textit{adjoint} of a superoperator $\mathcal{N}:\Lin(\mathcal{H}_A)\to\Lin(\mathcal{H}_{A'})$ is the unique superoperator $\mathcal{N}^\dagger:\Lin(\mathcal{H}_{A'})\to\Lin(\mathcal{H}_A)$ that satisfies
		\begin{equation}\label{eq:math-tools:adjoint-unique-def}
			\inner{Y}{\mathcal{N}(X)}=\langle \mathcal{N}^\dagger(Y),X\rangle
		\end{equation}
		for all $X\in\Lin(\mathcal{H}_A)$ and $Y\in\Lin(\mathcal{H}_{A'})$, where $\inner{\cdot}{\cdot}$ is the Hilbert--Schmidt inner product defined in \eqref{eq-Hilbert_Schmidt_inner_prod}.
	\end{definition}
	
	\begin{exercise}{exer-superoperator_adjoint}
		Let $\mathcal{N}_{A\to B}$ be a superoperator represented as in \eqref{eq-math_tools-superator_op_sum}.
		\begin{enumerate}[topsep=0.3cm]
			\item Prove that the adjoint $\mathcal{N}^\dagger$ is given by $\mathcal{N}^{\dagger}(Y)=\sum_{i=1}^n K_i^{\dagger}YL_i$ for every linear operator $Y$.
			
			\item If $\mathcal{N}$ is Hermiticity preserving, then prove that an alternate operator-sum representation of $\mathcal{N}$ is $\mathcal{N}(X)=\sum_{i=1}^n L_i X K_i^{\dagger}$ for all $X\in\Lin(\mathcal{H}_A)$.
			
			\item Using 1. and 2., prove that if $\mathcal{N}$ is Hermiticity preserving, then so is its adjoint $\mathcal{N}^{\dagger}$.
		\end{enumerate}
	\end{exercise}
	
	\begin{definition}{Trace-Preserving and Unital Superoperator}{def-TP_superop}
		Let $\mathcal{N}_{A\to B}$ be a superoperator. 
		\begin{enumerate}[topsep=0.3cm]
			\item $\mathcal{N}$ is called \textit{trace preserving} if $\Tr[\mathcal{N}(X)]=\Tr[X]$ for all $X\in\Lin(\mathcal{H}_A)$.
			
			\item $\mathcal{N}$ is called \textit{unital} if $\mathcal{N}(\mathbbm{1}_A)=\mathbbm{1}_B$.
		\end{enumerate}
	\end{definition}
	
	\begin{remark}
		Observe that if $\mathcal{N}$ is trace preserving and unital, and if $\mathcal{H}_A$ and $\mathcal{H}_B$ have finite dimensions, then we find that $d_A=d_B$, by taking the trace on both sides of $\mathcal{N}(\mathbbm{1}_A)=\mathbbm{1}_B$. This means that, in the finite-dimensional case, it is necessary for trace-preserving and unital superoperators to have the same input and output dimensions.
	\end{remark}
	
	\begin{exercise}{exer-TP_unital_superop}
		Let $\mathcal{N}_{A\to B}$ be a trace-preserving superoperator represented as in \eqref{eq-math_tools-superator_op_sum}.
		\begin{enumerate}[topsep=0.3cm]
			\item Prove that $\sum_{i=1}^n K_i^{\dagger}L_i=\mathbbm{1}_A$.
			\item Using 1., show that the adjoint $\mathcal{N}^{\dagger}$ is unital. Thus, the adjoint of every trace-preserving superoperator is unital.
		\end{enumerate}
	\end{exercise}
	
	For every superoperator $\mathcal{N}_{A\to B}:\Lin(\mathcal{H}_A)\to\Lin(\mathcal{H}_B)$, we define its \textit{induced trace norm} $\norm{\mathcal{N}}_1$ as
	\begin{align}
		\norm{\mathcal{N}}_1&\coloneqq \sup\left\{\frac{\norm{\mathcal{N}(X)}_1}{\norm{X}_1}:X\in\Lin(\mathcal{H}_A),X\neq 0\right\}\\
		&=\sup\{\norm{\mathcal{N}(X)}_1:X\in\Lin(\mathcal{H}_A),\norm{X}_1\leq 1\} . \label{eq-induced_trace_norm}
	\end{align}
	Then, for all $X\in\Lin(\mathcal{H}_A)$, it immediately follows that
	\begin{equation}\label{eq-induced_trace_norm_2}
		\norm{\mathcal{N}(X)}_1\leq\norm{\mathcal{N}}_1\norm{X}_1.
	\end{equation}
	
	\begin{exercise}{exer-induced_Tr_norm_adjoint}
		Prove that
		\begin{equation}\label{eq-induced_Tr_norm_adjoint}
			\norm{\mathcal{N}}_1=\sup_{\substack{U\in\Lin(\mathcal{H}_B)\\\text{unitary}}}\norm{\mathcal{N}^{\dagger}(U)}_{\infty}
		\end{equation}
		for every superoperator $\mathcal{N}_{A\to B}$, where the optimization is with respect to every unitary operator $U$ acting on $\mathcal{H}_B$.
	\end{exercise}
	
	\begin{definition}{Diamond Norm}{def-diamond_norm_superop}
		Let $\mathcal{N}_{A\to B}$ be a superoperator. The quantity
		\begin{equation}\label{eq-diamond_norm_general}
			\norm{\mathcal{N}}_{\diamond}\coloneqq\sup\{\norm{(\id_R\otimes\mathcal{N}_{A\to B})(X_{RA})}_1:X_{RA}\in\Lin(\mathcal{H}_{RA}),\,\norm{X_{RA}}_1\leq 1\}
		\end{equation}
		is known as the \textit{diamond norm} of $\mathcal{N}$, where the optimization is with respect to every linear operator $X_{RA}$, and there is an implicit supremum over the dimension $d_R\geq 1$ of the reference system Hilbert space $\mathcal{H}_R$.
	\end{definition}
	
	\begin{theorem}{thm-diamond_norm}
		For every superoperator $\mathcal{N}_{A\to B}$,
		\begin{align}
			\!\!\!  \norm{\mathcal{N}}_{\diamond}&=\norm{\id_A\otimes\mathcal{N}_{A\to B}}_1\label{eq-diamond_norm_dim}\\
			&=\sup\{\norm{(\id_{A}\otimes\mathcal{N}_{A\to B})(\ketbra{\psi}{\phi}_{AA})}_1:  \norm{\ket{\psi}}_2=\norm{\ket{\phi}}_2=1\}.
		\end{align}
		Furthermore, if $\mathcal{N}$ is Hermiticity preserving, then
		\begin{equation}\label{eq-diamond_norm_Herm_pres}
			\norm{\mathcal{N}}_{\diamond}=\sup\{\norm{(\id_{A}\otimes\mathcal{N}_{A\to B})(\ketbra{\psi}{\psi}_{AA})}_1:\norm{\ket{\psi}_{AA}}_2=1\}.
		\end{equation}
	\end{theorem}
	
	\begin{Proof}
		Please see the Bibliographic Notes in Section~\ref{math-tools:sec:bib-notes}.
	\end{Proof}
	
	We study the diamond norm in detail in Chapter~\ref{chap-QM_dist_meas} in the context of quantum channels.

\section{Analysis and Probability}\label{sec-analysis_probability}

	In this section, we briefly review some essential concepts from mathematical analysis, in particular the concepts of the limit, supremum, and infimum, as well as the continuity of real-valued functions. We also discuss compact sets, convex sets, and functions, as well as the basic notions of probability distributions.

\subsection{Limits, Infimum, Supremum, and Continuity}\label{sec:math-tools:cont-inf-sup}
	
\subsubsection*{Limit of a sequence}
	
	We start with the definition of the limit of a sequence of real numbers. A sequence $\{s_n\}_{n\in\mathbb{N}}\subset\mathbb{R}$ of real numbers is said to have the \textit{limit} $\ell$, written $\lim_{n\to\infty} s_n=\ell$, if for all $\varepsilon>0$ there exists $n_{\varepsilon}\in\mathbb{N}$ such that, for all $n\geq n_{\varepsilon}$, the inequality  $|s_n-\ell|<\varepsilon$ holds.
	
	One can think of the concept of a limit intuitively as a game between two players, an antagonist and a protagonist. The antagonist goes first, and gets to pick an arbitrary $\varepsilon > 0$. The protagonist wins if he reports back an entry in the sequence $\{ s_n\}_n$ such that $|s_n - \ell | < \varepsilon $. If the protagonist reports back an entry $s_n$ such that $|s_n - \ell | \geq \varepsilon$, then the antagonist wins. If the limit exists and is equal to  $\ell$, then the protagonist always wins by taking $n$ sufficiently large (i.e., larger than $n_\varepsilon$) and then reporting back $s_n$. If the limit does not exist or if the limit is not equal to $\ell$, then the protagonist cannot necessarily win with the strategy of taking $n$ sufficiently large; in this case, there exists a choice of $\varepsilon>0$, such that for all $n_\varepsilon \in \mathbb{N}$, there exists $n\geq n_\varepsilon$ such that $|s_n - \ell | \geq \varepsilon$. In this latter case, the choice of $\varepsilon>0$ can again be understood as a strategy of the antagonist.

\subsubsection*{Infimum and supremum}
	
	Let us now recall the concepts of the infimum and supremum of subsets of the real numbers. Roughly speaking, they are generalizations of the concepts of the minimum and maximum, respectively, of a set. Formally, let $E\subset\mathbb{R}$.
	\begin{itemize}
		\item A point $x\in \mathbb{R}$ is a \textit{lower bound} of $E$ if $y\geq x$ for all $y\in E$. If $x$ is the greatest such lower bound, then $x$ is called the \textit{infimum} of $E$, and we write $x=\inf E$.
		\item A point $x\in \mathbb{R}$  is an \textit{upper bound} of $E$ if $y\leq x$ for all $y\in E$. If $x$ is the least such upper bound, then $x$ is called the \textit{supremum} of $E$, and we write $x=\sup E$.
	\end{itemize}
	The supremum and infinum may or may not be contained in the subset $E$. For example, let $E=\{\sfrac{1}{n}\}_{n\in\mathbb{N}}$. Then, $\sup E=1\in E$, but $\inf E=0\notin E$. As another example, let $E = [0,1)$. Then $\sup E = 1 \notin E$ and $\inf E = 0 \in E$. If the supremum is contained in $E$, then it is equal to the maximum element of $E$. Similarly, if the infimum is contained in $E$, then it is equal to the minimum element of $E$.
	
	When considering a function $F\colon S\to\mathbb{R}$ defined on a subset $S$ of $\Lin(\mathcal{H})$, its infimum and supremum are defined for the set $E=\{F(X):X\in S\}$. Specifically,
	\begin{align}
		\inf_{X\in S} F(X) & \coloneqq\inf\{F(X):X\in S\}, \text{ and }\\
		\sup_{X\in S} F(X) & \coloneqq\sup\{F(X):X\in S\}.
	\end{align}

\subsubsection*{Limit inferior and limit superior}
	
	Turning back to limits, the limit of a sequence need not always exist. A particularly illuminating example is the sequence $\{r^n\}_{n \in \mathbb{N}}$ for $r\in \mathbb{R}$. If $-1<r<1$, then the limit exists and is equal to zero. If $r>1$, then the sequence never converges to a finite value and so the limit does not exist. We say that the sequence diverges to $+\infty$ in this case. If $r < -1$, then the sequence oscillates and diverges (but it does not specifically diverge to either $+\infty$ or $-\infty$). If $r=-1$, then the sequence oscillates back and forth between $-1$ and $+1$ and so the limit does not exist.
	
	Given that the limit of a sequence need not always exist, it can be helpful to have a reasonable substitute for this asymptotic concept that does always exist. Such a substitute is provided by two quantities: the limit inferior and limit superior of a sequence. We now define the limit inferior and limit superior, noting that they can be understood as asymptotic versions of the infimum and supremum just discussed. 
	\begin{itemize}
		\item We say that $s$ is an \textit{asymptotic lower bound} on the sequence $\{s_n\}_{n \in \mathbb{N}}$ if for all $\varepsilon >0$ there exists $n_{\varepsilon}\in \mathbb{N}$ such that, for all $n \geq n_{\varepsilon}$, the inequality $s_n > s - \varepsilon$ holds. The \textit{limit inferior} is the greatest asymptotic lower bound and is denoted by
		\begin{equation}
		\liminf_{n \to \infty} s_n.
		\end{equation}

		\item The definition of the limit superior is essentially opposite to that of the limit inferior. We say that $s$ is an \textit{asymptotic upper bound} on the sequence $\{s_n\}_{n \in \mathbb{N}}$ if for all $\varepsilon >0$, there exists $n_{\varepsilon}\in \mathbb{N}$, such that for all $n \geq n_{\varepsilon}$, the inequality $s_n < s + \varepsilon$ holds. The \textit{limit superior} is the least asymptotic upper bound and is denoted by
		\begin{equation}
		\limsup_{n \to \infty} s_n.
		\end{equation}
	
	\end{itemize}
	
	The limit inferior and limit superior always exist by  extending the real line $\mathbb{R}$ to include $-\infty$ and $+\infty$. Furthermore,  every asymptotic lower bound on the sequence cannot exceed an asymptotic upper bound, implying that the following inequality holds for every sequence $\{ s_n\}_{n \in \mathbb{N}}$:
	\begin{equation}\label{eq-liminf_limsup_ineq}
		\liminf_{n \to \infty} s_n \leq \limsup_{n \to \infty} s_n.
	\end{equation}
	If the opposite inequality holds for a sequence $\{ s_n\}_{n \in \mathbb{N}}$,  then the limit of the sequence exists and  we can write
	\begin{equation}
		\liminf_{n \to \infty} s_n = \limsup_{n \to \infty} s_n = \lim_{n \to \infty} s_n .
	\end{equation}
	This collapse is a direct consequence of the definitions of limit, limit inferior, and limit superior.

\subsubsection*{Limits and continuity} 
	
	We now consider the limit and continuity of a function. Specifically, we consider a real-valued function $F\colon \Lin(\mathcal{H})\to\mathbb{R}$ defined on the space of linear operators acting on a Hilbert space $\mathcal{H}$. We view this space as a normed vector space with either the trace norm $\norm{\cdot}_1$ or the spectral norm $\norm{\cdot}_{\infty}$. In the definitions that follow, we use $\norm{\cdot}$ to denote either one of these norms.
	\begin{itemize}
		\item \textit{Limit}: We write $\lim_{X\to X_0}F(X)=y$ to the mean the following: for all $\varepsilon>0$, there exists $\delta_{\varepsilon}>0$ such that $|F(X)-y|<\varepsilon$ for all $X\in\Lin(\mathcal{H})$ for which $\norm{X-X_0}<\delta_{\varepsilon}$. 
		
		\item\textit{Continuity at a point}: We say that $F$ is \textit{continuous at $X_0$} if for all $\varepsilon>0$, there exists $\delta_{\varepsilon}>0$ such that $|F(X)-F(X_0)|<\varepsilon$ for every point $X$ for which $\norm{X-X_0}<\delta_{\varepsilon}$. $F$ is said to be \textit{continuous} if $F$ is continuous at $X_0$ for all $X_0\in\Lin(\mathcal{H})$.
		
		\item\textit{Uniform continuity}: We say that $f$ is \textit{uniformly continuous} if for all $\varepsilon>0$, there exists $\delta_{\varepsilon}>0$ such that $|F(X)-F(X')|<\varepsilon$ for all $X,X'\in\Lin(\mathcal{H})$ for which $\norm{X-X'}<\delta_{\varepsilon}$.
		
		\item\textit{Upper and lower semi-continuity}: We say that $F$ is \textit{upper semi-continuous at $X_0$} if for all $\varepsilon>0$, there exists a neighborhood $N_{X_0,\varepsilon}$ of $X_0$ such that: if $F(X_0)>-\infty$, then $F(X)\leq F(X_0)+\varepsilon$ for all $X\in N_{X_0,\varepsilon}$; if $F(X_0)=-\infty$, then $\lim_{X\to X_0} F(X)=-\infty$. $F$ is called \textit{upper semi-continuous} if it is upper semi-continuous at $X_0$ for all $X_0\in\Lin(\mathcal{H})$. 
		
		We say that $f$ is \textit{lower semi-continuous at $X_0$} if for all $\varepsilon>0$, there exists a neighborhood $N_{X_0,\varepsilon}$ of $X_0$ such that: if $F(X_0)<+\infty$, then $F(X)\geq F(X_0)-\varepsilon$ for all $X\in N_{X_0,\varepsilon}$; if $F(X_0)=+\infty$, then $\lim_{X\to X_0} F(X)=+\infty$.
		
	\end{itemize}

\subsection{Compact Sets}

	A subset $S$ of a finite-dimensional topological vector space is called \textit{compact} if 
 every sequence of elements in $S$ has a subsequence that converges to an element in~$S$. For finite-dimensional vector spaces, a subset $S$ is compact if and only 
 if it is closed and bounded. An important fact about compact sets is that the infimum and supremum of continuous functions defined on compact sets are always achieved; in other words, the minimum and maximum exist and are thus 
 attained by points in the set. In practice, for optimization problems over compact sets, the infimum can be replaced by a minimum and the supremum can be replaced by a maximum. 
 More formally, if $F\colon S\to\mathbb{R}$ is a continuous function defined on a compact subset $S$ of $\Lin(\mathcal{H})$, then
	\begin{equation}
		\inf_{X\in S}F(X)=\min_{X\in S} F(X)\quad\text{and}\quad \sup_{X\in S}F(X)=\max_{X\in S}F(X).
	\end{equation}
	
	An important example of a compact set is the set $\{X\in\Lin(\mathcal{H}):X\geq 0,\Tr[X]\leq 1\}$ of positive semi-definite operators with trace bounded from above by one. This set contains the set of density operators acting on $\mathcal{H}$.

\subsection{Convex Sets and Functions}\label{sec-math_tools_convexity}
	
	A subset $C$ of a vector space is called \textit{convex} if, for all elements $u,v\in C$ and for all $\lambda\in[0,1]$, we have $\lambda u+(1-\lambda)v\in C$. We often call $\lambda u+(1-\lambda)v$ a \textit{convex combination} of $u$ and $v$. More generally, for every set $S=\{v_x\}_{x\in\mathcal{X}}$ of elements of a real vector space indexed by an alphabet $\mathcal{X}$, and every function $p:\mathcal{X}\to[0,1]$ with $p(x)\geq 0$ for all $x\in\mathcal{X}$ and $\sum_{x\in\mathcal{X}}p(x)=1$, the sum $\sum_{x\in\mathcal{X}}p(x) v_x$ is called a convex combination of the vectors in $S$. The \textit{convex hull} of $S$ is the convex set of all possible convex combinations of the vectors in $S$.
	
	Throughout this book, in the context of convex sets and functions, we consider the real vector space of Hermitian operators acting on some Hilbert space. Then, an important example of a convex subset is the set of all positive semi-definite operators. Indeed, if $X$ and $Y$ are positive semi-definite operators, then $\lambda X+(1-\lambda)Y$ is a positive semi-definite operator for all $\lambda\in[0,1]$. From now on, we assume $C$ to be a convex subset of the set of Hermitian operators, and we use $X$, $Y$, and $Z$ to denote arbitrary elements of $C$.
	
	An element $Z\in C$ is called an \textit{extreme point of $C$} if $Z$ cannot be written as a non-trivial convex combination of other vectors in $C$. Formally, $Z$ is an extreme point if  every decomposition of   $Z$ as the convex combination $Z=\lambda X+(1-\lambda)Y$, such that $\lambda\in(0,1)$ (so that the decomposition is non-trivial), implies that $X=Y=Z$.  An important fact is that every convex set is equal to the convex hull of its extreme points.
	
	We now define convex and concave functions.
	
	\begin{definition}{Convex and Concave Functions}{def-convex_concave_func}
		A function $F\colon C\to\mathbb{R}$ defined on a convex subset $C\subseteq\Lin(\mathcal{H})$ is  a \textit{convex function} if, for all $X,Y\in C$ and $\lambda\in[0,1]$, the following inequality holds:
		\begin{equation}
			F(\lambda X+(1-\lambda)Y)\leq \lambda F(X)+(1-\lambda)F(Y).
		\end{equation}
		A function $F\colon C\to\mathbb{R}$ is a \textit{concave function} if $-F$ is a convex function.
	\end{definition}
	
	It follows from an inductive argument using Definition~\ref{def-convex_concave_func} that a convex function $F\colon C\to \mathbb{R}$ on a convex subset $C\subset\Lin(\mathcal{H})$ satisfies
	\begin{equation}
		F\!\left(\sum_{x\in\mathcal{X}} p(x)X_x\right)\leq \sum_{x\in\mathcal{X}}p(x)F(X_x)
	\end{equation}
	for every set $\{X_x\}_{x\in\mathcal{X}}$ of elements in $C$ and every probability distribution $p$ (i.e., a function $p:\mathcal{X}\to[0,1]$ defined on $\mathcal{X}$, such that $p(x)\geq 0$ for all $x\in\mathcal{X}$ and $\sum_{x\in\mathcal{X}}p(x)=1$).
	
	\begin{itemize}
		\item A function $F\colon C\to\mathbb{R}$ defined on a convex subset $C\subseteq\Lin(\mathcal{H})$ is called \textit{quasi-convex} if for all $X,Y\in C$ and all $\lambda\in[0,1]$ the following inequality holds:
			\begin{equation}
				F(\lambda X+(1-\lambda) Y)\leq\max\{F(X),F(Y)\}.
			\end{equation}
		
		\item A function $F\colon C\to\mathbb{R}$ defined on a convex subset $C\subseteq\Lin(\mathcal{H})$ is called \textit{quasi-concave} if $-F$ is quasi-convex. Specifically, $F$ is called quasi-concave if for all $X,Y\in C$ and all $\lambda\in[0,1]$, the following inequality holds:
			\begin{equation}
				F(\lambda X+(1-\lambda) Y)\geq\min\{F(X),F(Y)\}.
			\end{equation}
	\end{itemize}

\subsection{Fenchel--Eggleston--Carath\'{e}odory Theorem}

	We mentioned above that the convex hull of a subset $S$ of a real vector space is the convex set consisting of all convex combinations of the vectors in $S$. A fundamental result is that if the underlying vector space has dimension~$d$, then, in order to obtain an element in the convex hull of $S$, it suffices to take a convex combination of no more than $d+1$ elements of $S$. If $S$ is connected and compact, then no more than $d$ elements are required. We state this formally as follows.
	
	\begin{theorem*}{Fenchel--Eggleston--Carath\'{e}odory Theorem}{thm-Caratheodory}
		Let $S$ be a set of vectors in a real $d$-dimensional vector space ($d<\infty$), and let $\text{conv}(S)$ denote the convex hull of $S$. Then, an arbitrary element $v\in\text{conv}(S)$ can be expressed as a convex combination of at most $m\leq d+1$ elements in $S$. If $S$ is connected and compact, then the same statement holds with $m \leq d$.
	\end{theorem*}
	
	\begin{Proof}
		Please consult the Bibliographic Notes (Section~\ref{math-tools:sec:bib-notes}).
	\end{Proof}

\subsection{Minimax Theorems}

	We often encounter expressions of the following form in quantum information theory:
	\begin{equation}\label{eq-minimax_opt}
		\inf_{X\in S}\sup_{Y\in S'}F(X,Y).
	\end{equation}
	The expression above contains both an infimum and a supremum over subsets $S,S'\subseteq\Lin(\mathcal{H})$ of some real-valued function $F\colon S\times S'\to\mathbb{R}$. 
	
	Expressions such as the one in \eqref{eq-minimax_opt} arise in the context of two-player zero-sum games. In such a game, the function $F(X,Y)$ represents the reward of a protagonist, who chooses elements $Y\in S'$ in order to maximize $F$. The antagonist chooses elements $X\in S$ in order to minimize $F$, i.e., to minimize the reward to the protagonist\footnote{This is due the fact that the reward of the antagonist is equal to $-F(X,Y)$ as a consequence of the zero-sum property of the game.}. The worst-case scenario for the antagonist is that, no matter what element $X\in S$ it chooses, the protagonist chooses the ``best'' possible element in $S'$ for their benefit, so that the reward is $G(X)\coloneqq\sup_{Y\in S'}F(X,Y)$. The optimal value of the antagonist in this scenario is thus given by $\inf_{X\in S}G(X)$, which is the quantity in~\eqref{eq-minimax_opt}.

	On the other hand, the worst-case scenario for the protagonist is that, no matter what element $Y\in S'$ they choose, the antagonist chooses the ``best'' possible element in $S$ for their benefit, so that the reward to the protagonist is $\widetilde{G}(Y)\coloneqq\inf_{X\in S}F(X,Y)$. The optimal reward of the protagonist in this scenario is thus given by $\sup_{Y\in S'}\widetilde{G}(Y)$, i.e.,
	\begin{equation}\label{eq-maximin}
		\sup_{Y\in S'}\inf_{X\in S}F(X,Y).
	\end{equation}
	
	Intuitively, it is advantageous for the protagonist to achieve a higher reward when playing second (in reaction to the antagonist's choice).  This intuition is captured by the following ``max-min inequality'':
	\begin{equation}\label{eq-minimax_gen_inequality}
		\sup_{Y\in S'}\inf_{X\in S}F(X,Y)\leq \inf_{X\in S}\sup_{Y\in S'} F(X,Y),
	\end{equation}
	which always holds. A mnemonic trick to remember it is that the internal optimization ($\inf$ on the left-hand side and $\sup$ on the right-hand side) is the one that determines the inequality, because it represents the player who plays second. We now prove this formally. Observe that for all $X\in S, Y \in S'$, we have that $\widetilde{G}(Y) \leq F(X,Y)$. It then follows that $\sup_{Y \in S'}\widetilde{G}(Y) \leq  \sup_{Y \in S'} F(X,Y)$ by applying the definition of supremum. Since this latter inequality holds for all $X \in S$, the definition of infimum implies that $\sup_{Y \in S'}\widetilde{G}(Y) \leq  \inf_{X\in S}\sup_{Y \in S'} F(X,Y)$, which is precisely the inequality in \eqref{eq-minimax_gen_inequality}.

	Many proofs that we present in this book require determining when the inequality opposite to the one in \eqref{eq-minimax_gen_inequality} holds, i.e., 
	\begin{equation}\label{eq:math-tools:minimax_inequality_rev}
\inf_{X\in S}\sup_{Y\in S'} F(X,Y) \overset{?}{\leq }
		\sup_{Y\in S'}\inf_{X\in S}F(X,Y),
	\end{equation}
		which is known as the ``min-max inequality.'' If it holds, then the inequality in \eqref{eq-minimax_gen_inequality} is saturated and becomes an equality.
		The game-theoretic interpretation of the situation when the reverse inequality holds is that, for the sets $S$ and $S'$ and the reward function $F$, there is no advantage to playing first; i.e., the reward is the same regardless of who goes first, as long as the protagonist and antagonist play optimal strategies. It is thus important to know under what conditions this reverse inequality holds.
		
	We now present theorems for various classes of functions that tell us when the inequality (and thus the equality) in \eqref{eq:math-tools:minimax_inequality_rev} holds.
	
	\begin{theorem*}{Sion Minimax for Extended Reals}{thm-Sion_minimax}
		Let $S$ be a compact and convex subset of a normed vector space, and let~$S'$ be a convex subset of a normed vector space. Let $F\colon S\times S'\to\mathbb{R}\cup\{+\infty,-\infty\}$ be an extended-real-valued function such that
		\begin{enumerate}			
			\item The function $F(\cdot,Y):S\to\mathbb{R}$ is lower semi-continuous and quasi-convex on $S$ for every $Y\in S'$.
            \item The function $F(X,\cdot):S'\to\mathbb{R}$ is upper semi-continuous and quasi-concave on $S'$ for every $X\in S$.
		\end{enumerate}
		Then,
		\begin{equation}
			\inf_{X\in S}\sup_{Y\in S'}F(X,Y)=\sup_{Y\in S'}\inf_{X\in S}F(X,Y).
		\end{equation}
		Furthermore, on both sides of the above equation, the infimum can be replaced with a minimum.
	\end{theorem*}
	
	\begin{Proof}
		Please consult the Bibliographic Notes (Section~\ref{math-tools:sec:bib-notes}).
	\end{Proof}

    \begin{theorem*}{Farkas--R\'ev\'esz Minimax}{thm-Farkas_Revesz_minimax}
		Let $S$ be a compact and convex subset of a normed vector space and let~$S'$ be a convex subset of a normed vector space. Let $F\colon S\times S'\to\mathbb{R}\cup\{+\infty\}$ be an extended-real-valued function such that
		\begin{enumerate}			
			\item The function $F(\cdot,Y):S\to\mathbb{R}$ is lower semi-continuous and convex on $S$ for every $Y\in S'$.
            \item The function $F(X,\cdot):S'\to\mathbb{R}$ is concave on $S'$ for every $X\in S$.
		\end{enumerate}
		Then,
		\begin{equation}
			\inf_{X\in S}\sup_{Y\in S'}F(X,Y)=\sup_{Y\in S'}\inf_{X\in S}F(X,Y).
		\end{equation}
		Furthermore, on both sides of the above equation, the infimum can be replaced with a minimum.
	\end{theorem*}
	
	\begin{Proof}
		Please consult the Bibliographic Notes (Section~\ref{math-tools:sec:bib-notes}).
	\end{Proof}
	
	\begin{theorem*}{Mosonyi--Hiai Minimax}{thm-Mosonyi_minimax}
		Let $S$ be a compact set in a normed vector space, let $S'\subseteq\mathbb{R}$, and let $F\colon S\times S'\to\mathbb{R}\cup\{+\infty,-\infty\}$ an extended-real-valued function such that
		\begin{enumerate}
			\item The function $F(\cdot,y):S\to\mathbb{R}\cup\{+\infty,-\infty\}$ is lower semi-continuous for every $y\in S'$.
			\item The function $F(X,\cdot):S'\to\mathbb{R}\cup\{+\infty,-\infty\}$ is either monotonically increasing or monotonically decreasing for every $X\in S$.
		\end{enumerate}
		Then,
		\begin{equation}
			\inf_{X\in S}\sup_{y\in S'}F(X,y)=\sup_{y\in S'}\inf_{X\in S}F(X,y).
		\end{equation}
		Furthermore, the infimum can be replaced with a minimum.
	\end{theorem*}
	
	\begin{Proof}
		Please consult the Bibliographic Notes (Section~\ref{math-tools:sec:bib-notes}).
	\end{Proof}

    With the three minimax theorems presented above, a natural inclination is to examine their distinctions and to determine the situations in which one might want to apply them. The distinctions have to do with the various assumptions on the objective function $F$. In the Sion Minimax theorem (Theorem~\ref{thm-Sion_minimax}), the function is required to be lower semi-continuous and quasi-convex in $X$ (for fixed $Y$) and upper semi-continuous and quasi-concave in $Y$ (for fixed $X$), and its range is the extended real set $\mathbb{R}\cup\{+\infty,-\infty\}$. Although these conditions are quite broad, they are in fact limiting for our needs in later chapters, where we want to apply minimax theorems to R\'enyi relative entropies, which do not generally satisfy the needed assumption of upper semi-continuity. As such, we do not make much use of Theorem~\ref{thm-Sion_minimax} in later chapters. On the other hand, the Farkas--R\'ev\'esz minimax theorem (Theorem~\ref{thm-Farkas_Revesz_minimax}) has no such stringent condition of upper semi-continuity. Furthermore, the more restrictive conditions (the range being $\mathbb{R}\cup\{+\infty\}$ and convexity and concavity instead of quasi-convexity and quasi-concavity) are not problematic because the theorem is still applicable to a wide range of scenarios involving R\'enyi relative entropies, which are lower semi-continuous functions of their matrix arguments. Finally, the Mosonyi--Hiai minimax theorem (Theorem~\ref{thm-Mosonyi_minimax}) is applicable to a quite different setup, and the conditions needed are not stringent in comparison to the previous minimax theorems. This theorem finds most applicability when performing a minimax exchange involving the R\'enyi parameter $\alpha$ and one of its matrix arguments.

\subsection{Probability Distributions}

	Throughout this book, we are concerned for the most part with discrete probability distributions, and the following definitions suffice for our needs. A \textit{discrete probability distribution} is a function $p:\mathcal{X}\to[0,1]$ defined on an finite alphabet $\mathcal{X}$ such that $p(x)\geq 0$ for all $x\in\mathcal{X}$ and $\sum_{x\in\mathcal{X}}p(x)=1$. Formally, we can consider the alphabet $\mathcal{X}$ to be the set of realizations of a discrete random variable $X:\Omega\to\mathcal{X}$ from the space $\Omega$ of experimental outcomes, called the sample space, to the set $\mathcal{X}$. We then write $p_X$ to denote the probability distribution of the random variable $X$, i.e., $p_X(x)\equiv\Pr[X=x]$.
	
	The \textit{expected value} or \textit{mean} $\mathbb{E}[X]$ of a random variable $X$ taking values in $\mathcal{X}\subset\mathbb{R}$ is defined as
	\begin{equation}
		\mathbb{E}[X]=\sum_{x\in\mathcal{X}} p_X(x)\, x.
	\end{equation}
	For every function $g:\mathcal{X}\to \mathbb{R}$, we define $g(X)$ to be the random variable $g\circ X:\Omega\to\mathbb{R}$ with image $\{g(x):x\in\mathcal{X}\}$. Then,
	\begin{equation}
		\mathbb{E}[g(X)]=\sum_{x\in\mathcal{X}} p_X(x) g(x).
	\end{equation}

	A useful fact is \textit{Markov's inequality}: if $X$ is a non-negative random variable, then, for all $a>0$, we have
	\begin{equation}\label{eq-MT:Markov-ineq}
		\Pr[X\geq a]\leq\frac{\mathbb{E}[X]}{a}.
	\end{equation}
        Intuitively, Markov's inequality tells us that there is a small probability for $X$ to take on values that are much larger than its mean.
        
	\textit{Jensen's inequality} is the following statement: if $X$ is a random variable with finite mean, and $f$ is a real-valued convex function such that $\mathcal{X} \subseteq \text{dom}(f)$, then
	\begin{equation}\label{eq-MT:Jensen_ineq}
		f(\mathbb{E}[X])\leq\mathbb{E}[f(X)].
	\end{equation}
	This is a very special case of the more elaborate operator Jensen inequality presented previously in Theorem~\ref{thm-Jensen}.
	
	As we explain in Chapter~\ref{chap-QM_states_meas}, an observable $O$ in quantum mechanics (which is merely a Hermitian operator) generalizes a random variable, such that its expectation is given by $\mathbb{E}[O]\equiv\left<O\right>_{\rho}\coloneqq\Tr[O\rho]$ for a density operator~$\rho$. In this case, an application of the Jensen inequality in \eqref{eq-MT:Jensen_ineq} leads to the following: for a Hermitian operator $O$, a density operator~$\rho$, and a  convex function $f$,
	\begin{equation} \label{eq-op_Jensen_alt_2}
		f(\Tr[O\rho])\leq\Tr[f(O) \rho ],
	\end{equation}
	which we can alternatively write as
	\begin{equation}\label{eq-op_Jensen_alt}
		f(\left\langle O\right\rangle_{\rho})\leq \left\langle f(O)\right\rangle_{\rho}.
	\end{equation}
	
\begin{exercise}{exer_Jensen_observable_densop}
            Prove the inequality in \eqref{eq-op_Jensen_alt} by applying the Jensen inequality in \eqref{eq-MT:Jensen_ineq} and using the fact that a density operator is a positive semi-definite operator with trace equal to one.
        \end{exercise}

\section{Semi-Definite Programming}
 
\label{sec-SDPs}
	
	\textit{Semi-definite programs} (SDPs) constitute an important class of optimization problems that arise frequently in quantum information theory. An SDP is a constrained optimization problem in which the optimization variable is a positive semi-definite operator $X$, the objective function is linear in $X$, and the constraint is an operator inequality featuring a linear function of $X$. Not only are SDPs useful as an analytical tool, but there also exist a number of numerical solvers that can be used for evaluating these optimization problems (one can use the \texttt{CVX} package for \textsc{Matlab} or the \texttt{CVXPY} package for Python).
	
	Semi-definite programs play an important role in quantum information because, for a number of operational tasks of interest, we are trying to maximize a linear objective function over the sets of quantum states or measurements, which are specified by semi-definite constraints. Furthermore, many of the different communication capacities of quantum channels are difficult to characterize or compute, and it can be helpful to find semi-definite relaxations of them that are efficiently computable. These are two common ways in which semi-definite programs arise in this book.
	
	\begin{definition}{Semi-Definite Program}{def-SDPs}
		Given a Hermiticity-preserving superoperator $\Phi$ and Hermitian operators $A$ and~$B$, a \textit{semi-definite program} (SDP) corresponds to two optimization problems. The first is the \textit{primal SDP}, which is defined as
		\begin{equation}\label{eq-SDP_primal}
			\begin{array}{l l} \text{maximize} & \Tr[A X] \\ \text{subject to} & \Phi(X)\leq B, \\ & X\geq 0. \end{array}
		\end{equation}
		The second optimization problem is the \textit{dual SDP}, which is defined as
		\begin{equation}\label{eq-SDP_dual}
			\begin{array}{l l} \text{minimize} & \Tr[B Y] \\ \text{subject to} & \Phi^\dagger(Y)\geq A, \\ & Y\geq 0. \end{array}
		\end{equation}
		We let 
		\begin{align}
			S(\Phi,A,B) & \coloneqq\sup_{X\geq 0} \{ \Tr[AX] : \Phi(X) \leq B\}, \label{eq-primal_SDP_def}\\
			\widehat{S}(\Phi,A,B) &\coloneqq\inf_{Y\geq 0} \{ \Tr[BY] : \Phi^\dag(Y) \geq A\}\label{eq-dual_SDP_def}
		\end{align}
		denote the optimal values of the primal and dual SDPs, respectively.
	\end{definition}

\begin{remark} \label{rem-equality_constraints}
    Equality constraints of the form $\Tr [CX] = c$, where $C$ is a Hermitian matrix and $c\in \mathbb{R}$ is a real number, can be incorporated into~\eqref{eq-SDP_primal} by rephrasing them as pairs of inequalities $\Tr [CX] \leq c$ and $\Tr [CX] \geq c$.
\end{remark}

	A variable $X$ for the primal SDP in \eqref{eq-primal_SDP_def}  is called a \textit{feasible point} if it is positive semi-definite ($X\geq 0$) and satisfies the constraint $\Phi(X)\leq B$, and it is called a \textit{strictly feasible point} if $X$ is positive definite ($X>0$) and the constraint is satisfied with a strict inequality, i.e., if $\Phi(X)<B$. The same definitions apply to the dual SDP in \eqref{eq-dual_SDP_def}: a variable $Y$ is a feasible point if $Y\geq 0$ and $\Phi^\dagger(Y)\geq A$, and it is strictly feasible if $Y>0$ and $\Phi^\dagger(Y)>A$. By convention, if there is no primal feasible operator $X$, then $S(\Phi,A,B) = -\infty$, and if there is no dual feasible operator $Y$, then $\widehat{S}(\Phi,A,B) = +\infty$. It is also possible for $S(\Phi,A,B) = +\infty$ or $\widehat{S}(\Phi,A,B) = -\infty$. A simple example of $S(\Phi,A,B)=+\infty$ is when $A=1$ is a scalar, $\Phi(X)=0$, and $B=1$ (so that the constraint in \eqref{eq-primal_SDP_def} is trivially satisfied), and a simple example of $\widehat{S}(\Phi,A,B)=-\infty$ is when $B=-1$, $\Phi(Y)=0$, and $A=-1$.

\begin{remark}
As discussed in the remark after Definition~\ref{def-SDPs}, an SDP can incorporate equality constraints by turning each one into a pair of inequality constraints. However, when determining whether a given variable $X$ is strictly feasible, one should group back all such inequality pairs into equality constraints and consider strict feasibility only with respect to the remaining inequality constraints.
\end{remark}
 
	\begin{proposition*}{Weak Duality}{prop:math-tools:weak-duality-SDPs}
		For every SDP corresponding to $\Phi$, $A$, and $B$, the following weak duality inequality holds:
		\begin{equation}\label{eq:math-tools:weak-duality-SDPs}
			S(\Phi,A,B)\leq \widehat{S}(\Phi,A,B).
		\end{equation}
	\end{proposition*}

\begin{Proof}
 Let $X\geq 0$ be primal feasible, and let $Y\geq 0$ be dual feasible. Then the following holds:
	\begin{equation}\label{eq:math-tools:step-4-weak-duality-SDPs}
		\Tr[AX] \leq \Tr[\Phi^\dag(Y)X] = \Tr[Y \Phi(X)] \leq \Tr[YB].
	\end{equation}
	The first inequality follows from the assumption that $Y$ is dual feasible, so that we have $A\leq\Phi^\dagger(Y)$, and by applying 2.~of Lemma~\ref{prop-operator_ineqs}. The equality holds by definition of the adjoint map $\Phi^{\dagger}$; see \eqref{eq:math-tools:adjoint-unique-def}. The last inequality follows from the assumption that $X$ is primal feasible, so that we have $\Phi(X)\leq B$, and by applying 2.~of Lemma~\ref{prop-operator_ineqs}. Since the inequality holds for all primal feasible $X$ and for all dual feasible $Y$, we can take a supremum over the left-hand side of \eqref{eq:math-tools:step-4-weak-duality-SDPs} and an infimum over the right-hand side of \eqref{eq:math-tools:step-4-weak-duality-SDPs}, and we thus arrive at the weak duality inequality in \eqref{eq:math-tools:weak-duality-SDPs}.
	\end{Proof}
	
	There is a deep connection between the weak duality inequality in Proposition~\ref{prop:math-tools:weak-duality-SDPs} and the max-min inequality in \eqref{eq-minimax_gen_inequality}. This is realized by introducing the Lagrangian $\mathcal{L}(\Phi, A, B, X, Y)$ for the SDP as follows:
	\begin{equation}
	\mathcal{L}(\Phi, A, B, X, Y) \coloneqq \Tr[AX] + \Tr[BY] - \Tr[\Phi(X)Y].
	\end{equation}
	Note that the following equalities hold, which are helpful in the discussion below:
	\begin{align}
	\mathcal{L}(\Phi, A, B, X, Y) & = \Tr[AX] + \Tr[(B - \Phi(X))Y] \\
	& = \Tr[BY] + \Tr[(A-\Phi^\dag(Y))X].
	\end{align}
	
	By first taking an infimum over $Y\geq 0$ and then a supremum over $X \geq 0$, we find that 
	\begin{equation}\label{eq:math-tools:connect-weak-dual-maxmin-1}
		\sup_{X\geq 0}\inf_{Y \geq 0} \mathcal{L}(\Phi, A, B, X, Y) = S(\Phi,A,B). 
	\end{equation}
	This equality follows because
	\begin{equation}
	\sup_{X\geq 0}\inf_{Y \geq 0} \mathcal{L}(\Phi, A, B, X, Y) = \sup_{X\geq 0}\left\{\Tr[AX] + \inf_{Y \geq 0}\Tr[(B - \Phi(X))Y]\right\}.
	\end{equation}
	The inner infimum with respect to $Y\geq 0$ forces the outer optimization to be with respect to every feasible point $X$ for the primal SDP in \eqref{eq-primal_SDP_def}. In this sense, the variable $Y$ can be thought of as a ``Lagrange multiplier'', analogous to Lagrange multipliers that are used in constrained optimization problems in calculus. Indeed, suppose that an infeasible $X\geq 0$ is chosen, meaning that the constraint $\Phi(X) \leq B$ is violated. This means that there exists a non-trivial negative eigenspace of $B - \Phi(X)$. Let $\ket{\varphi}$ be a unit vector in this negative eigenspace. We can then pick $Y = c \ket{\varphi}\!\bra{\varphi}$ for $c>0$ and take the limit $c\to\infty$, so that $\inf_{Y \geq 0}\Tr[(B - \Phi(X))Y]=-\infty$. So a violation of the constraint $\Phi(X) \leq B$ incurs an infinite cost for the outer optimization with respect to $X\geq 0$, 
    meaning that the corresponding point $X$ can be effectively discarded. The constraint $\Phi(X)\leq B$ is therefore forced to be satisfied, leading to the equality in \eqref{eq:math-tools:connect-weak-dual-maxmin-1}.
	
	If we instead take a supremum over $X\geq 0$  first and then take an infimum over $Y\geq0$, it follows that
	\begin{equation}
	\inf_{Y \geq 0} \sup_{X\geq 0} \mathcal{L}(\Phi, A, B, X, Y) = \widehat{S}(\Phi,A,B).
	\label{eq:math-tools:connect-weak-dual-maxmin-2}
	\end{equation}
	This time, the equality follows because
	\begin{equation}
	\inf_{Y \geq 0} \sup_{X\geq 0} \mathcal{L}(\Phi, A, B, X, Y)
	= \inf_{Y \geq 0} \left\{\Tr[BY] + \sup_{X\geq 0} \Tr[(A-\Phi^\dag(Y))X]\right\}
	\end{equation}
	Similar to what was argued previously, the inner optimization variable $X$ is a Lagrange multiplier that forces the outer optimization to be 
    restricted to dual feasible points only; the constraint $\Phi^{\dagger}(Y)\geq A$ is thus forced to be satisfied, leading to the equality in \eqref{eq:math-tools:connect-weak-dual-maxmin-2}.
	
	Now, by examining  \eqref{eq-minimax_gen_inequality}, \eqref{eq:math-tools:connect-weak-dual-maxmin-1}, and \eqref{eq:math-tools:connect-weak-dual-maxmin-2}, we see that the weak duality inequality in Proposition~\ref{prop:math-tools:weak-duality-SDPs} can be understood as  a consequence of the max-min inequality in \eqref{eq-minimax_gen_inequality}.
	
	The inequality opposite to the one in \eqref{eq:math-tools:weak-duality-SDPs} does not hold in general; if it does, it implies that $S(\Phi,A,B)=\widehat{S}(\Phi,A,B)$. We then say that the SDP corresponding to $\Phi$, $A$, and $B$ has the \textit{strong duality} property, or that it satisfies strong duality. Considering the  discussion above in terms of the Lagrangian of the SDP, we also can understand strong duality as being equivalent to a minimax theorem holding.

	\begin{theorem*}{Slater's Condition}{thm:math-tools:slater-cond}
	\textit{Slater's condition} is a sufficient condition for strong duality to hold, and it is given as follows:
\begin{enumerate}
\item If there exists $X\geq0$ such that $\Phi(X)\leq B$ and there exists
$Y>0$ such that $\Phi^{\dag}(Y)>A$, then $S(\Phi,A,B)=\widehat{S}(\Phi,A,B)$. Furthermore, there exists a
primal feasible operator $X$ for which $\operatorname{Tr}[AX]=S(\Phi,A,B)$.

\item If there exists $Y\geq0$ such that $\Phi^{\dag}(Y)\geq A$ and there
exists $X>0$ such that $\Phi(X)<B$, then $S(\Phi,A,B)=\widehat{S}(\Phi,A,B)$. Furthermore, there exists a
dual feasible operator $Y$ for which $\operatorname{Tr}[BY]=\widehat{S}(\Phi,A,B)$.
\end{enumerate}
\end{theorem*}

\begin{remark}
The nomenclature Slater's ``condition'' (rather than ``conditions'') is commonly used, but note that one can check either one of the  two sufficient conditions above to determine if strong duality holds.
\end{remark}

For many SDPs of interest, it is straightforward to determine if Slater's condition holds. We provide an example in Section~\ref{sec:math-tools:example-SDP-simple}.

Complementary slackness for SDPs is useful for understanding further constraints on an optimal primal 
operator $X$ and an optimal dual operator~$Y$. 

\begin{proposition*}{Complementary Slackness of SDPs}{prop:math-tools:comp-slack}
Consider an arbitrary SDP corresponding to $\Phi$, $A$, and $B$, and
suppose that strong duality holds. Then the following
complementary slackness conditions hold for feasible $X$ and $Y$ if and only
if they are optimal:
\begin{align}
YB &  =Y\Phi(X),\label{eq:math-tools:comp-slack-1}\\
\Phi^{\dag}(Y)X &  =AX.\label{eq:math-tools:comp-slack-2}
\end{align}
\end{proposition*}

\begin{Proof}
On the one hand, suppose that $X$ is primal feasible, that $Y$ is dual feasible, and that they satisfy \eqref{eq:math-tools:comp-slack-1}--\eqref{eq:math-tools:comp-slack-2}. Then it is clear by inspecting \eqref{eq:math-tools:step-4-weak-duality-SDPs} that the inequalities are saturated, thus implying that $X$ is primal optimal and $Y$ is dual optimal. 

On the other hand, suppose that $X$ is primal optimal and that $Y$ is dual optimal. Then, by this assumption, it follows that $\Tr[AX] = \Tr[BY]$ so that the inequalities in \eqref{eq:math-tools:step-4-weak-duality-SDPs} are saturated. This means that
\begin{align}
\Tr[(\Phi^\dag(Y) -A)X] & = 0 \label{eq:math-tools:comp-slack-one-the-way-1}\\
\Tr[Y(B -\Phi(X))] & = 0.
\label{eq:math-tools:comp-slack-one-the-way-2}
\end{align}
Since $\Phi^\dag(Y) -A$ and $X$ are positive semi-definite, the equality in \eqref{eq:math-tools:comp-slack-one-the-way-1} implies that $(\Phi^\dag(Y) -A)X = 0$, which is equivalent to \eqref{eq:math-tools:comp-slack-1}. Similarly, since $B -\Phi(X)$ and $Y$ are positive semi-definite, the equality in \eqref{eq:math-tools:comp-slack-one-the-way-2} implies that $Y(B -\Phi(X)) = 0$, which is equivalent to \eqref{eq:math-tools:comp-slack-2}. 
\end{Proof}
	
	If the matrices $A$ and $B$ and the map $\Phi$ involved in an SDP are of reasonable size, then the SDP can be computed efficiently using numerical solvers (specifically, the time required is polynomial in the size of these objects and polynomial in the logarithm of the inverse of the numerical accuracy desired). As mentioned earlier, SDPs arise frequently in quantum information, with some examples appearing in Chapter~\ref{chap-QM_dist_meas}. Furthermore, SDPs appear in some of the upper bounds for rates of quantum communication protocols that we consider in Parts \ref{part:q-comm-prots} and \ref{part-feedback}.
	
	\begin{exercise}{exer-SDP_alt_forms}
		Consider the following pair of primal and dual optimization problems:
		\begin{equation}\label{eq-SDP_alt_forms}
			\begin{array}{l l} \text{maximize} & \Tr[CZ] \\ \text{subject to} & \Psi(Z)=D,\\ & Z\geq 0 \end{array} \qquad \begin{array}{l l} \text{minimize} & \Tr[DW] \\ \text{subject to} & \Psi^{\dagger}(W)\geq C,\\ & W \text{ Hermitian}, \end{array}
		\end{equation}
		where $C$ and $D$ are Hermitian operators and $\Psi$ is a Hermiticity-preserving superoperator. The goal of this exercise is to show that these problems are SDPs, i.e., that they are equivalent to the optimization problems in \eqref{eq-SDP_primal} and \eqref{eq-SDP_dual}.
		\begin{enumerate}[topsep=0.3cm]
			\item Given $C$, $D$, and $\Psi$ for the optimization problems in \eqref{eq-SDP_alt_forms}, find $A$, $B$, and $\Phi$ such that these optimization problems can be expressed in the forms presented in \eqref{eq-SDP_primal} and \eqref{eq-SDP_dual}. (\textit{Hint}: Start by using the fact that $\Psi(Z)=D$ if and only if $\Psi(Z)\leq D$ and $-\Psi(Z)\leq -D$.)
			
			\item Conversely, given $A$, $B$, and $\Phi$ for the SDPs in \eqref{eq-SDP_primal} and \eqref{eq-SDP_dual}, find $C$, $D$, and $\Psi$ such that those SDPs can be expressed in the forms in \eqref{eq-SDP_alt_forms}. (\textit{Hint}: Start by using the fact that $\Phi(X)\leq B$ if and only if there exists $S\geq 0$ such that $\Phi(X)+S=B$.)
		\end{enumerate}
	\end{exercise}
	
	Reasoning analogous to that in Exercise~\ref{exer-SDP_alt_forms} can be used to show that the following pair of optimization problems are also SDPs, equivalent to the ones in \eqref{eq-SDP_primal} and \eqref{eq-SDP_dual}:
	\begin{equation}
		\begin{array}{l l} \text{minimize} & \Tr[CZ] \\ \text{subject to} & \Psi(Z)=D, \\ & Z\geq 0 \end{array} \qquad \begin{array}{l l} \text{maximize} & \Tr[DW] \\ \text{subject to} & \Psi^{\dagger}(W)\leq C, \\ & W \text{ Hermitian} \end{array}
	\end{equation}
	
	\begin{exercise}{exer-SDP_basic}
		\begin{enumerate}
			\item Consider the following SDP in primal form:
				\begin{equation}
					\sup_{X\geq 0}\left\{\Tr[AX]: \Phi_1(X)\leq B_1,\,\Phi_2(X)=B_2\right\},
				\end{equation}
				where $A,B_1,B_2$ are Hermitian operators and $\Phi_1,\Phi_2$ are Hermiticity-preserving superoperators. Show that the dual SDP is given by
				\begin{equation}
					\inf_{\substack{Y_1\geq 0,\\Y_2\text{ Hermitian}}}\!\!\!\!\left\{\Tr[B_1Y_1]+\Tr[B_2Y_2]:\Phi_1^{\dagger}(Y_1)+\Phi_2^{\dagger}(Y_2)\geq A\right\}.
				\end{equation}
				Furthermore, determine Slater's conditions and the complementary slackness conditions for this SDP.
				
			\item Now suppose that the primal SDP has the form
				\begin{equation}
					\inf_{Y\geq 0}\left\{\Tr[BY]:\Phi_1(Y)\geq A_1,\,\Phi_2(Y)=A_2\right\},
				\end{equation}
				where $A_1,A_2,B$ are Hermitian operators and $\Phi_1,\Phi_2$ are Hermiticity-preserving superoperators. Show that the dual SDP is given by
				\begin{equation}
					\sup_{\substack{X_1\geq 0,\\X_2\text{ Hermitian}}}\!\!\!\!\left\{\Tr[A_1X_1]+\Tr[A_2X_2]:\Phi_1^{\dagger}(X_1)+\Phi_2^{\dagger}(X_2)\leq B\right\}.
				\end{equation}
				Furthermore, determine Slater's conditions and the complementary slackness conditions for this SDP.
		\end{enumerate}
	\end{exercise}

\subsection{SDPs for Spectral and Trace Norms, Maximum and Minimum Eigenvalues}\label{sec:math-tools:example-SDP-simple}

	In this section, we provide semi-definite programs for calculating the spectral and trace norms of Hermitian operators, as well as their largest and smallest eigenvalues.

	\begin{theorem*}{SDPs for the Spectral Norm of a Hermitian Operator}{thm-spect_norm_Herm}
		Let $H$ be a Hermitian operator, and consider the following functions:
		\begin{align}
			f(H)  & \coloneqq\sup_{X_{1},X_{2}\geq0}\left\{  \operatorname{Tr}[H(X_{1}
			-X_{2})]:\operatorname{Tr}[X_{1}+X_{2}]\leq1\right\},\label{eq:math-tools:example-SDP-primal}\\
			\widehat{f}(H)  & \coloneqq\inf_{t\geq0}\left\{  t:-t\mathbbm{1} \leq H\leq t\mathbbm{1} \right\}.\label{eq:math-tools:example-SDP-dual}
		\end{align}
		The quantities above can be computed via SDPs, and in fact, the following equality holds:
		\begin{equation}\label{eq:math-tools:example-SDP-sd-eq}
			f(H)=\widehat{f}(H)=\left\Vert H\right\Vert _{\infty}.
		\end{equation}
		That is, $f(H)$ is equal to the largest singular value of the Hermitian operator~$H$.
	\end{theorem*}


	\begin{Proof}
		Given that the optimization in \eqref{eq:math-tools:example-SDP-primal} is a maximization, let us first show that \eqref{eq:math-tools:example-SDP-primal} can be written in the form of $S(\Phi,A,B)$ in \eqref{eq-primal_SDP_def}. Indeed if we let
\begin{align}
X  & =
\begin{pmatrix}
X_{1} & Z^{\dag}\\
Z & X_{2}
\end{pmatrix}
,\qquad A=
\begin{pmatrix}
H & 0\\
0 & -H
\end{pmatrix}
,\\
\Phi(X)  & =\operatorname{Tr}[X_{1}+X_{2}],\qquad B=1,
\end{align}
then we have that
\begin{equation}
f(H)=\sup_{X\geq0}\left\{  \operatorname{Tr}[AX]:\Phi(X)\leq B\right\}  .
\end{equation}
The constraint $X\geq0$ implies that $X_{1},X_{2}\geq0$. Furthermore, notice that the operator $Z$ appears neither in the
objective function $\operatorname{Tr}
[H(X_{1}-X_{2})]$ nor in the constraint $\operatorname{Tr}[X_{1}+X_{2}]\leq1$. Thus, the operator $Z$ plays no role in the optimization, and so we can simply set $Z=0$, so
that
\begin{equation}
X=
\begin{pmatrix}
X_{1} & 0\\
0 & X_{2}
\end{pmatrix}
.
\end{equation}
Thus, \eqref{eq:math-tools:example-SDP-primal} is indeed an SDP\ in primal form.

Now, recall from \eqref{eq-inf_norm_Hermitian} that the spectral norm of $H$ is given by the maximum of
the absolute values of the eigenvalues of $H$. In particular, we can write
\begin{equation}
\left\Vert H\right\Vert _{\infty}=\max\left\{  \left\vert \lambda_{\max
}\right\vert ,\left\vert \lambda_{\min}\right\vert \right\}  ,
\end{equation}
where $\lambda_{\max}$ and $\lambda_{\min}$ are the maximum and minimum
eigenvalues, respectively, of $H$. Note that we always have $\lambda_{\max
}\geq\lambda_{\min}$. Let $|\phi_{\max}\rangle$ be an eigenvector of $H$
satisfying $H|\phi_{\max}\rangle=\lambda_{\max}|\phi_{\max}\rangle$, and let
$|\phi_{\min}\rangle$ be an eigenvector of $H$ satisfying $H|\phi_{\min
}\rangle=\lambda_{\min}|\phi_{\min}\rangle$. Let us suppose at first that
$\lambda_{\max}\geq0$. Then one feasible choice of $X_{1}$ and $X_{2}$ in
\eqref{eq:math-tools:example-SDP-primal} is $X_{1}=|\phi_{\max}\rangle
\langle\phi_{\max}|$ and $X_{2}=0$, and for this choice, we find that
$f(H)\geq\lambda_{\max}=\abs{\lambda_{\max}}$. If $\lambda_{\max}\leq0$,
then another feasible choice of $X_{1}$ and $X_{2}$ in
\eqref{eq:math-tools:example-SDP-primal} is $X_{1}=0$ and $X_{2}=|\phi_{\max
}\rangle\!\langle\phi_{\max}|$, and for this choice, we find that $f(H)\geq -\lambda_{\max}=\abs{\lambda_{\max}}$. Therefore, we conclude that
\begin{equation}
f(H)\geq\left\vert \lambda_{\max}\right\vert .
\end{equation}
Now, suppose that $\lambda_{\min}\geq0$. Then a feasible choice of $X_{1}$ and
$X_{2}$ in \eqref{eq:math-tools:example-SDP-primal} is $X_{1}=|\phi_{\min
}\rangle\!\langle\phi_{\min}|$ and $X_{2}=0$, and for this choice, we find that
$f(H)\geq\lambda_{\min}=\abs{\lambda_{\min}}$. If $\lambda_{\min}\leq0$,
then another feasible choice of $X_{1}$ and $X_{2}$ in
\eqref{eq:math-tools:example-SDP-primal} is $X_{1}=0$ and $X_{2}=|\phi_{\min
}\rangle\!\langle\phi_{\min}|$, and for this choice, we find that $f(H)\geq -\lambda_{\min}=\abs{\lambda_{\min}}$. Therefore, we conclude that
\begin{equation}
f(H)\geq\max\left\{  \left\vert \lambda_{\max}\right\vert ,\left\vert
\lambda_{\min}\right\vert \right\}  =\left\Vert H\right\Vert _{\infty
}.\label{eq:math-tools:SDP-example-inf-norm-eigs}
\end{equation}

It now remains to prove the reverse inequality, namely, the inequality $f(H)\leq
\left\Vert H\right\Vert _{\infty}$. To prove this, let us show that
$\widehat{f}(H)$, as defined in \eqref{eq:math-tools:example-SDP-dual}, is given by the SDP dual to the one that defines $f(H)$. In order to do this, we should determine the map $\Phi^{\dag}$, which is the adjoint of
$\Phi$. Since $B=1$ and $\Phi(X)=\operatorname{Tr}[X_{1}+X_{2}]$ are scalars, we take $Y=t$ to be a scalar
also. Then, we find that
\begin{align}
\operatorname{Tr}[Y\Phi(X)]  & =t\operatorname{Tr}[X_{1}+X_{2}]\\
& =\operatorname{Tr}\left[
\begin{pmatrix}
t\mathbbm{1}  & 0\\
0 & t\mathbbm{1} 
\end{pmatrix}
\begin{pmatrix}
X_{1} & 0\\
0 & X_{2}
\end{pmatrix}
\right]  \\
& =\operatorname{Tr}[\Phi^{\dag}(Y)X],
\end{align}
from which we conclude that
\begin{equation}
\Phi^{\dag}(Y)=\Phi^{\dag}(t)=
\begin{pmatrix}
t\mathbbm{1}  & 0\\
0 & t\mathbbm{1} 
\end{pmatrix}
.
\end{equation}
Plugging this into the standard form of the dual in \eqref{eq-dual_SDP_def}, we find that
\begin{align}
\inf_{Y\geq0}\left\{  \operatorname{Tr}[BY]:\Phi^{\dag}(Y)\geq A\right\}  & =\inf_{t\geq0}\left\{  t:
\begin{pmatrix}
t\mathbbm{1}  & 0\\
0 & t\mathbbm{1} 
\end{pmatrix}
\geq
\begin{pmatrix}
H & 0\\
0 & -H
\end{pmatrix}
\right\}  \\
& =\inf_{t\geq0}\left\{  t:t\mathbbm{1} \geq H,\ t\mathbbm{1} \geq-H\right\}  \\
& =\inf_{t\geq0}\left\{  t:-t\mathbbm{1} \leq H\leq t\mathbbm{1} \right\}\label{eq:math-tools:example-SDP-dual2}\\
&=\widehat{f}(H). 
\end{align}

Let us now recall property 3.~of Lemma~\ref{prop-operator_ineqs}, which states that $\lambda_{\min}\mathbbm{1} \leq
H\leq\lambda_{\max}\mathbbm{1} $. By combining with
\eqref{eq:math-tools:SDP-example-inf-norm-eigs}, we find that $\lambda_{\max
}\mathbbm{1} \leq\left\Vert H\right\Vert _{\infty}\mathbbm{1}$ and $\lambda_{\min}\mathbbm{1} \geq-\left\Vert
H\right\Vert _{\infty}\mathbbm{1} $, which implies that
\begin{equation}
-\left\Vert H\right\Vert _{\infty}\mathbbm{1} \leq H\leq\left\Vert H\right\Vert _{\infty
}\mathbbm{1} .
\end{equation}
Thus, we see that $\left\Vert H\right\Vert _{\infty}$ is a feasible choice for
$t$ in \eqref{eq:math-tools:example-SDP-dual2}, which implies that
\begin{equation}
\widehat{f}(H)\leq\left\Vert H\right\Vert _{\infty}
.\label{eq:math-tools:SDP-example-inf-norm-dual}
\end{equation}

Now, combining the inequalities in
\eqref{eq:math-tools:SDP-example-inf-norm-eigs} and
\eqref{eq:math-tools:SDP-example-inf-norm-dual} gives us $f(H)\geq\norm{H}_{\infty}\geq\widehat{f}(H)$. Then, using the weak duality inequality from Proposition~\ref{prop:math-tools:weak-duality-SDPs}, which for our case implies that $f(H)\leq \widehat{f}(H)$, we conclude that the primal
and dual optimal values are equal to each other and equal to the spectral norm
of $H$: $f(H)=\widehat{f}(H)=\left\Vert H\right\Vert _{\infty}$.
\end{Proof}

We proved \eqref{eq:math-tools:example-SDP-sd-eq} by employing clever guesses for
primal feasible and dual feasible points. Doing so is possible in this case because
the problem is simple enough to begin with, and we could apply knowledge from
linear algebra to make these clever guesses. Although it is sometimes possible
to make clever guesses and arrive at analytical solutions like we did above, in many
cases it is not possible. In such cases, it can be helpful to check Slater's
condition in Theorem~\ref{thm:math-tools:slater-cond} explicitly in order to see if strong duality holds. So let us do so for the SDPs corresponding to $f(H)$ and $\widehat{f}(H)$. For the primal SDP\ in
\eqref{eq:math-tools:example-SDP-primal}, a strictly feasible point consists
of the choice $X_{1}=\alpha\frac{\mathbbm{1} }{d}$ and $X_{2}=\beta\frac{\mathbbm{1} }{d}$ such that
$\alpha,\beta>0$ and $\alpha+\beta<1$, where $d$ is the dimension of $\mathbbm{1}$. Then
we clearly have $X_{1}>0$, $X_{2}>0$, and $\operatorname{Tr}[X_{1}+X_{2}]<1$,
so that $X_{1}$ and $X_{2}$ are strictly feasible, as claimed. A feasible point
for the dual consists of the choice $\gamma\geq\left\Vert H\right\Vert
_{\infty}$. Thus, strong duality holds, further confirming that $f(H)=\widehat
{f}(H)$, as shown above.

We now remark about the complementary slackness conditions from Proposition~\ref{prop:math-tools:comp-slack} for the SDPs corresponding to $f(H)$ and $\widehat{f}(H)$,
which apply to optimal primal $X$ and optimal dual $Y$. In this case, the
conditions reduce to
\begin{align}
t  & =t\operatorname{Tr}[X_{1}+X_{2}
],\label{eq:math-tools:comp-slack-example-1}\\
\begin{pmatrix}
t\mathbbm{1} & 0\\
0 & t\mathbbm{1}
\end{pmatrix}
\begin{pmatrix}
X_{1} & 0\\
0 & X_{2}
\end{pmatrix}
& =
\begin{pmatrix}
H & 0\\
0 & -H
\end{pmatrix}
\begin{pmatrix}
X_{1} & 0\\
0 & X_{2}
\end{pmatrix}
,
\end{align}
and the latter is the same as the following two separate conditions:
\begin{equation}
tX_{1}=HX_{1},\qquad-tX_{2}=HX_{2}.\label{eq:math-tools:comp-slack-example-2}
\end{equation}
If we have prior knowledge about the operator $H$ (e.g., that 
it is non-zero), then we conclude that the optimal $t\neq0$ and the
condition in \eqref{eq:math-tools:comp-slack-example-1} implies that
$\operatorname{Tr}[X_{1}+X_{2}]=1$. In this case, we can conclude that the
inequality constraint in \eqref{eq:math-tools:example-SDP-primal} is loose and it suffices to optimize over $X_{1}$ and
$X_{2}$ satisfying the constraint with equality. The conditions in
\eqref{eq:math-tools:comp-slack-example-2} indicate that the image of the optimal $X_{1}$
should be in the eigenspace of $H$ with optimal eigenvalue $t$, and the
image of the optimal $X_{2}$ should be in the eigenspace of $H$ with optimal eigenvalue
$-t$. Observe that these complementary slackness conditions are consistent with the choices that we made above.

As a final remark, if $H$ is actually positive semi-definite, then the lower
bound constraint in \eqref{eq:math-tools:example-SDP-dual} is unnecessary.
Letting $P$ be a positive semi-definite operator, we thus find that
\begin{equation}
f(P)=\left\Vert P\right\Vert _{\infty}=\inf_{t\geq 0}\left\{  t:P\leq t\mathbbm{1} \right\}  .
\label{eq-PSD_largest_eig_SDP}
\end{equation}
Note that, in this case, $\left\Vert P\right\Vert _{\infty}$ is the largest
eigenvalue of $P$.

	\begin{exercise*}{SDPs for the Trace Norm of a Hermitian Operator}{exer-trace_norm_Herm_SDP}
		\begin{enumerate}
			\item Let $H$ be a Hermitian operator. Like the spectral norm of $H$, as shown above, prove that the trace norm of $H$ can also be computed using an SDP. Specifically, prove that
				\begin{equation}\label{eq:MT:trace-norm-herm-op_SDP}
					\norm{H}_1=\sup_{M_1,M_2\geq 0}\{\Tr[H(M_1-M_2)]:M_1+M_2=\mathbbm{1}\}.
				\end{equation}
				(\textit{Hint}: Use \eqref{eq-MT:Jordan-Hahn} and \eqref{eq-trace_norm_Herm}.)
				
			\item Show that an alternate SDP formulation for $\norm{H}_1$ is
				\begin{equation}\label{eq:MT:trace-norm-herm-op-SDP_dual}
					\norm{H}_1=\inf_{Y\geq 0}\{\Tr[Y]:-Y\leq H\leq Y\}.
				\end{equation}
				(\textit{Hint}: Show that the SDP in \eqref{eq:MT:trace-norm-herm-op-SDP_dual} is dual to the one in \eqref{eq:MT:trace-norm-herm-op_SDP}, and then prove strong duality.)
		\end{enumerate}
	\end{exercise*}
	
	\begin{exercise*}{SDPs for the Maximum and Minimum Eigenvalues of a Hermitian Operator}{exer-Herm_eigenvals}
		Let $H$ be a Hermitian operator. Prove that the maximum and minimum eigenvalues of $H$, denoted by $\lambda_{\max}(H)$ and $\lambda_{\min}(H)$, respectively, have the following SDP characterizations:
		\begin{align}
			\lambda_{\min}(H)&=\inf_{\rho\geq 0}\{\Tr[H\rho]:\Tr[\rho]=1\}\\
			&=\sup_{t\in\mathbb{R}}\{t:H\geq t\mathbbm{1}\} \label{eq-math_tools_min_eigenval_SDP_2}
		\end{align}
		and
		\begin{align}
			\lambda_{\max}(H)&=\sup_{\rho\geq 0}\{\Tr[H\rho]:\Tr[\rho]=1\}\\
			&=\inf_{t\in\mathbb{R}}\{t:t\mathbbm{1}\geq H\}.
		\end{align}
		(\textit{Hint}: Use the spectral theorem (Theorem~\ref{thm-spectral_theorem}) and the duality of SDPs.)
	\end{exercise*}

\section{Symmetric Subspace}\label{sec-symm_subspace}

	Given a $d$-dimensional Hilbert space $\mathcal{H}$ and an $n$-fold tensor product $\mathcal{H}^{\otimes n}$ of~$\mathcal{H}$, for $n\geq 2$, it is often important to consider a permutation of the individual Hilbert spaces in the $n$-fold tensor product. This is especially the case in quantum information theory, because we often assume or it is often the case that the resources involved have permutation symmetry.  These permutations can be implemented using a unitary representation of the symmetric group on $n$ elements.
	
	The symmetric group on $n$ elements, denoted by $\mathcal{S}_n$, is defined to be the set of permutations of the set $\{1,2,\dotsc,n\}$. A permutation in $\mathcal{S}_n$ is an invertible function $\pi:\{1,2,\dotsc,n\}\to\{1,2,\dotsc,n\}$ that describes how each element in the set $\{1,2,\dotsc,n\}$ should be rearranged, or permuted. An example of a permutation in $\mathcal{S}_3$ is the function $\pi$ such that $\pi(1)=3$, $\pi(2)=1$, and $\pi(3)=2$. Since there are $n!$ ways to permute $n$ distinct elements, it follows that the set $\mathcal{S}_n$ contains $n!$ elements.  
	
	Given a permutation $\pi\in\mathcal{S}_n$ and an orthonormal basis $\{\ket{i}\}_{i=0}^{d-1}$ for $\mathcal{H}$, we define the unitary permutation operators $W^{\pi}$ acting on $\mathcal{H}^{\otimes n}$ by
	\begin{equation}\label{eq-permutation_rep}
		W^{\pi}\ket{i_1,i_2,\dotsc,i_n}=\ket{i_{\pi^{-1}(1)},i_{\pi^{-1}(2)},\dotsc,i_{\pi^{-1}(n)}},\quad 0\leq i_1,i_2,\dotsc,i_n\leq d-1,
	\end{equation}
        so that an alternative expression for $W^\pi$ is
        \begin{align}
            W^\pi & = \sum_{i_1,i_2,\dotsc,i_n} \ketbra{i_{\pi^{-1}(1)},i_{\pi^{-1}(2)},\dotsc,i_{\pi^{-1}(n)}}{i_1,i_2,\dotsc,i_n} \\
            & = \sum_{i_1,i_2,\dotsc,i_n} \ketbra{i_1,i_2,\dotsc,i_n}{i_{\pi(1)},i_{\pi(2)},\dotsc,i_{\pi(n)}}\, .
            \label{eq:math-tools:permutation-op-rewrite}
        \end{align}
	Since the set $\{\ket{i_1,i_2,\dotsc,i_n}:0\leq i_1,i_2,\dotsc,i_n\leq d-1\}$ is an orthonormal basis for $\mathcal{H}^{\otimes n}$, the definition in \eqref{eq-permutation_rep} extends to every vector in $\mathcal{H}^{\otimes n}$ by linearity. The operators in the set $\{W^{\pi}\}_{\pi\in\mathcal{S}_n}$ constitute a unitary representation of $\mathcal{S}_n$, in the sense that
	\begin{equation}
		(W^{\pi})^\dagger=W^{\pi^{-1}},\quad W^{\pi_1}W^{\pi_2}=W^{\pi_1\circ\pi_2}
  \label{eq:math-tools:permutation-props}
	\end{equation}
	for all $\pi,\pi_1,\pi_2\in\mathcal{S}_n$.
 \begin{exercise}{}
 Prove the equality in \eqref{eq:math-tools:permutation-op-rewrite}, and prove both equalities in \eqref{eq:math-tools:permutation-props}. 
 \end{exercise}
	
	Given a $d$-dimensional Hilbert space $\mathcal{H}$ and the unitary representation $\{W^{\pi}\}_{\pi\in\mathcal{S}_n}$ of $\mathcal{S}_n$ defined in \eqref{eq-permutation_rep}, we are interested in the subspace of vectors $\ket{\psi}\in\mathcal{H}^{\otimes n}$ that are \textit{invariant} under permutations, i.e., $W^{\pi}\ket{\psi}=\ket{\psi}$ for all $\pi\in\mathcal{S}_n$. We call this subspace the \textit{symmetric subspace of $\mathcal{H}^{\otimes n}$}, and it is formally defined as
	\begin{equation}
		\text{Sym}_n(\mathcal{H})\coloneqq\text{span}\{\ket{\psi}\in\mathcal{H}^{\otimes n}:W^{\pi}\ket{\psi}=\ket{\psi}\text{ for all }\pi\in\mathcal{S}_n\}.
		\label{eq:math-tools:def-sym-subspace}
	\end{equation}
	A vector $\ket{\psi}\in\text{Sym}_n(\mathcal{H})$ is sometimes called \textit{symmetric}. The subspace
	\begin{equation}
		\text{ASym}_n(\mathcal{H})\coloneqq\text{span}\{\ket{\psi}\in\mathcal{H}^{\otimes n}:W^{\pi}\ket{\psi}=\text{sgn}(\pi)\ket{\psi}\text{ for all }\pi\in\mathcal{S}_n\}
	\end{equation}
	is called the \textit{anti-symmetric subspace of $\mathcal{H}^{\otimes n}$}, where $\text{sgn}(\pi)$ is the \textit{sign} of the permutation $\pi$, defined as $\text{sgn}(\pi)=(-1)^{T(\pi)}$ where $T(\pi)$ is the number of transpositions into which $\pi$ can be decomposed\footnote{A transposition is a permutation that permutes only two elements of the set $\{1,2,\dotsc,n\}$. Any permutation $\pi\in\mathcal{S}_n$ can be decomposed into a product of transpositions. Although this decomposition is in general not unique, the parity of the number $T(\pi)$ of transpositions into which $\pi$ can be decomposed is unique, so that $\text{sgn}(\pi)$ is well defined.}.
	
	The operator
	\begin{equation}\label{eq-proj_symm_subspace}
		\Pi_{\text{Sym}_n(\mathcal{H})}\coloneqq\frac{1}{n!}\sum_{\pi\in\mathcal{S}_n}W^{\pi}
	\end{equation}
	is the orthogonal projection onto the symmetric subspace of $\mathcal{H}^{\otimes n}$, while
	\begin{equation}\label{eq-proj_asymm_subspace}
		\Pi_{\text{ASym}_n(\mathcal{H})}\coloneqq\frac{1}{n!}\sum_{\pi\in\mathcal{S}_n}\text{sgn}(\pi)W^{\pi}
	\end{equation}
	is the orthogonal projection onto the anti-symmetric subspace of $\mathcal{H}^{\otimes n}$. 
	
	\begin{exercise}{exer-symm_asymm_subspaces}
		\begin{enumerate}
		\item Prove that $\Pi_{\text{Sym}_n(\mathcal{H})}$ and $\Pi_{\text{ASym}_n(\mathcal{H})}$ are projections, as claimed above.
		\item Prove that $\text{Sym}_n(\mathcal{H})$ and $\text{ASym}_n(\mathcal{H})$ are orthogonal subspaces of $\mathcal{H}^{\otimes n}$ by showing that
		\begin{equation}
			\Pi_{\text{Sym}_n(\mathcal{H})}\Pi_{\text{ASym}_n(\mathcal{H})}=0.
		\end{equation}
		This implies that $\braket{\psi_s}{\psi_a}=0$ for all $\ket{\psi_s}\in\text{Sym}_n(\mathcal{H})$ and $\ket{\psi_a}\in\text{ASym}_n(\mathcal{H})$.
		\end{enumerate}
	\end{exercise}
	
	\begin{exercise}{exer-symm_two}
		Let $\mathcal{H}$ be a $d$-dimensional Hilbert space, $d\geq 2$. Show that, for $n=2$,
		\begin{align}
			\Pi_{\text{Sym}_2(\mathcal{H})}&=\frac{1}{2}(\mathbbm{1}_d\otimes\mathbbm{1}_d+F),\label{eq-symm_proj_n2}\\
			\Pi_{\text{ASym}_2(\mathcal{H})}&=\frac{1}{2}(\mathbbm{1}_d\otimes\mathbbm{1}_d-F),\label{eq-asymm_proj_n2}
		\end{align}
		where $F\coloneqq W^{\pi}$ is the unitary representation of the permutation $\pi$ that swaps systems 1 and 2 (i.e., $\pi(1)=2$, $\pi(2)=1$), defined as
		\begin{equation}\label{eq-swap_op_standard_0}
			F=\sum_{k,k'=0}^{d-1}\ketbra{k}{k'}\otimes\ketbra{k'}{k}.
		\end{equation}
		In quantum information theory, $F$ is referred to as the \textit{swap operator}.
	\end{exercise}	
	
	We focus primarily on the symmetric subspace of $\mathcal{H}^{\otimes n}$ in this book, and so we now provide some additional facts about it.
	
	The following set of vectors constitutes an orthonormal basis for the symmetric subspace $\text{Sym}_n(\mathcal{H})$ corresponding to the $d$-dimensional Hilbert space~$\mathcal{H}$:
	\begin{multline}\label{eq:math-tools:occupation-num-basis}
		\ket{n_1,n_2,\dotsc,n_d}\\\coloneqq\frac{1}{\sqrt{n!\left(\prod_{j=1}^d n_j!\right)}}
		\sum_{\pi\in\mathcal{S}_n}W^{\pi}(\ket{0}^{\otimes n_1}\otimes\ket{1}^{\otimes n_2}\otimes\dotsb\otimes \ket{d-1}^{\otimes n_d}),
	\end{multline}
	where $n_1,n_2,\dotsc,n_d\geq 0$ are such that $\sum_{j=1}^d n_j=n$. We often call this the \textit{occupation number basis} for $\text{Sym}_n(\mathcal{H})$. The reason for this name is that, physically, each of the $n$ Hilbert spaces $\mathcal{H}$ corresponds to a quantum system, and each $n_j$ tells us how many of the $n$ quantum systems are in the state given by $\ket{j-1}$. (We formally draw the correspondence between Hilbert spaces and quantum systems in Chapter~\ref{chap-QM_states_meas}.) The number of elements in this basis is equal to the number of ways of selecting $n$ elements, with repetition, from a set of $d$ distinct elements. This number  is equal to $\binom{d+n-1}{n}$. Consequently, the dimension of $\text{Sym}_n(\mathcal{H})$ is
	\begin{equation}
		\dim(\text{Sym}_n(\mathcal{H}))=\binom{d+n-1}{n}=\binom{d+n-1}{d-1}.
	\end{equation}
	
	\begin{exercise}{exer-Sym_n2}
		Let $d\geq 2$ and $n=2$. Show that the basis elements $\ket{n_1,n_2,\dotsc,n_d}$ of $\text{Sym}_2(\mathbb{C}^d)$ are given as follows:
		\begin{equation}
			\ket{n_1,n_2,\dotsc,n_d}=\ket{j-1,j-1},
		\end{equation}
		if $n_j=2,\,n_{\ell}=0\,\,\forall\ell\neq j$, and
		\begin{equation}
			\ket{n_1,n_2,\dotsc,n_d}=\frac{1}{\sqrt{2}}(\ket{j-1,k-1}+\ket{k-1,j-1}),
		\end{equation}
		if $n_j=n_k=1,\,k\neq j\text{ and }n_{\ell}=0\,\,\forall\ell\neq j,k$.
	\end{exercise}
	
	\begin{remark}
		The direct-sum vector space
		\begin{equation}
			\mathcal{F}_B(\mathcal{H})\coloneqq \bigoplus_{n=0}^{\infty} \text{Sym}_n(\mathcal{H})
		\end{equation}
		is called the \textit{bosonic Fock space}. (Note that $\text{Sym}_0(\mathcal{H})$ is the set of complex scalars, i.e., $\text{Sym}_0(\mathcal{H})=\mathbb{C}$.) It is an infinite-dimensional Hilbert space that is relevant for the study of quantum optical and other continuous-variable quantum systems.
	\end{remark}	

	An important fact that we state without proof (please consult the Bibliographic Notes in Section~\ref{math-tools:sec:bib-notes}) is that for every $d$-dimensional Hilbert space~$\mathcal{H}$,
	\begin{equation}\label{eq-sym_proj_integral}
		\Pi_{\text{Sym}_n(\mathcal{H})}=\binom{d+n-1}{n}\int \psi^{\otimes n}~\D\psi,
	\end{equation}
	where the integral on the right-hand side is taken with respect to the Haar measure over all unit vectors.
	
	\begin{remark}
		The measure $\text{d}\psi$ is also called the \textit{Fubini-Study measure}. A concrete coordinate representation of the measure can be obtained by using the following parameterization of every unit vector $\ket{\psi}$ in a $d$-dimensional Hilbert space $\mathcal{H}$:
		\begin{equation}\label{eq-pure_state_dDim_arb}
			\ket{\psi}=\sum_{k=0}^{d-1}r_k \e^{\I\varphi_k} \ket{k},
		\end{equation}
		where $0\leq \varphi_k\leq 2\pi$ and $r_k\geq 0$ for all $0\leq k\leq d-1$. Furthermore, since $\ket{\psi}$ is a unit vector, we require that $\sum_{k=0}^{d-1} r_k^2=1$. The conditions on the coefficients $r_k$ imply that they parameterize the positive octant of a sphere in $d$ dimensions. As such, each $r_k$ can be written as
		\begin{align}
			r_0&=\prod_{k=1}^{d-1}\sin\frac{\theta_k}{2},\\
			r_m&=\cos\frac{\theta_m}{2}\prod_{k=m+1}^{d-1}\sin\frac{\theta_k}{2},\quad 1\leq m\leq d-2,\\
			r_{d-1}&=\cos\frac{\theta_{d-1}}{2},
		\end{align}
		where $0\leq\theta_i\leq \pi$. Similarly, the angles $\varphi_k$ parameterize a torus in $d$ dimensions. The Fubini-Study measure $\text{d}\psi$ is then the volume element of the coordinate system formed from the $r_k$ and the coordinate system formed by the $\varphi_k$:
		\begin{equation}
			\text{d}\psi=\frac{(d-1)!}{(2\pi)^{d-1}}\prod_{i=1}^{d-1} \cos\frac{\theta_i}{2}\sin^{2i-1}\frac{\theta_i}{2}~\text{d}\theta_i~\text{d}\varphi_i.
		\end{equation}
		(Please consult the Bibliographic Notes in Section~\ref{math-tools:sec:bib-notes} for details.) In the case $d=2$, we have that
		\begin{equation}\label{eq-fubini_study_d2}
			\text{d}\psi=\frac{1}{2\pi}\cos\frac{\theta_1}{2}\sin\frac{\theta_1}{2}~\text{d}\theta_1~\text{d}\varphi_1 = \frac{1}{4\pi} \sin(\theta_1)\, \dd\theta_1 \dd\varphi_1\quad (d=2),
		\end{equation}
        recovering the familiar surface element of a $2$-sphere (i.e., the surface of a $3$-dimensional ball).
	\end{remark}

	We often consider the case that the Hilbert space $\mathcal{H}$ is a tensor product of two Hilbert spaces, i.e., $\mathcal{H}=\mathcal{H}_A\otimes\mathcal{H}_B\equiv\mathcal{H}_{AB}$, with $\mathcal{H}_A$ a $d_A$-dimensional Hilbert space and $\mathcal{H}_B$ a $d_B$-dimensional Hilbert space. As we have seen above, if $\{\ket{i}_A\}_{i=0}^{d_A-1}$ is an orthonormal basis for $\mathcal{H}_A$ and $\{\ket{j}_B\}_{j=0}^{d_B-1}$ is an orthonormal basis for $\mathcal{H}_B$, then $\{\ket{i,j}_{AB}\equiv\ket{i}_A\otimes\ket{j}_B:0\leq i\leq d_A-1,0\leq j\leq d_B-1\}$ is an orthonormal basis for $\mathcal{H}_{AB}$. In this case, if we consider the $n$-fold tensor product $\mathcal{H}_{AB}^{\otimes n}$, then the unitary representation $\{W_{(AB)^n}^\pi\}_{\pi\in\mathcal{S}_n}$ defined in \eqref{eq-permutation_rep} acts as follows:
	\begin{multline}
		W_{(AB)^n}^{\pi}(\ket{i_1,j_1}_{A_1B_1}\otimes\ket{i_2,j_2}_{A_2B_2}\otimes\dotsb\otimes\ket{i_n,j_n}_{A_nB_n})\\ = \ket{i_{\pi^{-1}(1)},j_{\pi^{-1}(1)}}_{A_1B_1}\otimes\ket{i_{\pi^{-1}(2)},j_{\pi^{-1}(2)}}_{A_2B_2}\otimes\dotsb\otimes\ket{i_{\pi^{-1}(n)},j_{\pi^{-1}(n)}}_{A_nB_n},
		\label{eq:math-tools:joint-perms}
	\end{multline}
	for all $0\leq i_1,i_2,\dotsc,i_n\leq d_A-1$ and all $0\leq j_1,j_2,\dotsc,j_n\leq d_B-1$. However, by rearranging the tensor factors, we find that the right-hand side of the above equation can be written as
	\begin{equation}
		W_{A^n}^{\pi}\ket{i_1,i_2,\dotsc,i_n}_{A_1\dotsb A_n}\otimes W_{B^n}^{\pi}\ket{j_1,j_2,\dotsc,j_n}_{B_1B_2\dotsb B_n},
	\end{equation}
	where $\{W_{A^n}^{\pi}\}_{\pi\in\mathcal{S}_n}$ and $\{W_{B^n}^{\pi}\}_{\pi\in\mathcal{S}_n}$ are the unitary representations of $\mathcal{S}_n$ acting on $\mathcal{H}_A^{\otimes n}$ and $\mathcal{H}_B^{\otimes n}$, respectively. We can thus write the projection onto $\text{Sym}_n(\mathcal{H}_A\otimes\mathcal{H}_B)$ as
	\begin{equation}\label{eq-proj_symm_bipartite}
		\Pi_{\text{Sym}_n(\mathcal{H}_A\otimes\mathcal{H}_B)}=\frac{1}{n!}\sum_{\pi\in\mathcal{S}_n}W_{(AB)^n}^{\pi}\equiv\frac{1}{n!}\sum_{\pi\in\mathcal{S}_n}W_{A^n}^{\pi}\otimes W_{B^n}^{\pi}.
	\end{equation}

\section{Bibliographic Notes}\label{math-tools:sec:bib-notes}

	The study of inner product spaces, including Hilbert spaces, is the primary focus of functional analysis, for which we refer to the following books: \citep{Reed_Simon_book,kreyszig_book,hall_book}. In the case of finite-dimension\-al Hilbert spaces, which is what we consider throughout this book, many of the concepts studied in functional analysis reduce to those studied in linear algebra and matrix analysis. For these topics, we refer to \citet{Bha97,horn2013_book,Strang16book}.
	
	The generalized Gell-Mann matrices discussed after \eqref{eq:math-tools:gell-mann-scaled-mats} were presented by \citet{HE81,Bertlmann_2008}.
	
	A review of operator monotone, operator concave, and operator convex functions is given by \citet{Bha97}. The short course of \citet{Carlen09} is also helpful. For proofs of the properties listed immediately after Definition \ref{def-op_conv_conc_mono}, see \citep[Chapter~V]{Bha97}. 
	
	The proof of \eqref{eq-math_tools_submult_alpha_strong} follows immediately from \citep[Problem~III.6.2]{Bha97}. A proof of \eqref{eq-Schatten_norm_var}, and therefore, of the H\"{o}lder inequality in \eqref{eq-Schatten_norm_duality}, can be found in \citep[Section IV \& Exercise IV.2.12]{Bha97}.
	
	Lemma~\ref{lem-QCAP:traceless-Hermitian} can be found in \citep[Lemma~4]{AE05}. A proof of Proposition \ref{prop-Schatten_pos_var} can be found in \citet[Lemma 12]{MDSFT13}. Lemma~\ref{lem-ALT_ineq} was proved by \citet{LT76,Araki1990}. The Courant--Fischer--Weyl minimax principle, which is invoked in the proof of property~4 of Lemma~\ref{prop-operator_ineqs}, is presented in \citep[Corollary~III.1.2]{Bha97}.
	
	A proof of the operator Jensen inequality (Theorem \ref{thm-Jensen}) was given by \citet{HP03}. In presenting the implication 1.~$\Rightarrow$~3.~of Theorem~\ref{thm-Jensen}, we followed the proof given by \citet[Theorem~3]{FFN04}.
	
	The notation $\norm{\cdot}_{\diamond}$ for the quantity on the right-hand side of \eqref{eq-diamond_norm_general} was introduced by \citet{Kit97}, and it is known as a \textit{completely bounded trace norm} in the mathematics literature; see, for example, \citep{paulsen_2003}. The result in \eqref{eq-diamond_norm_dim} is due to \citet{Smith83} (see Theorem~2.10 therein), but it can also be found in \citep{Kit97,AKN98}. For a proof of \eqref{eq-diamond_norm_Herm_pres}, see Theorem~3.51 in \citep{Wat18}, which also contains several more properties of the diamond norm.

	For an introduction to real analysis, see \citep{rudin_book}.
	
	For an introduction to convex analysis, see \citep{rockafellar_book,BV04}, and for a proof of the Fenchel--Eggleston--Carath\'{e}odory theorem, see \citep{E58book,rockafellar_book}. 

	Sion's minimax theorem for extended reals (Theorem \ref{thm-Sion_minimax}) is due to \citet[Theorem~2.11]{BB2023}, and it is a generalization of Sion's minimax theorem \citet{Sion1958}, which is in turn a generalization of a minimax theorem of \citet{vonNeumann1928}. A short proof of Sion's minimax theorem can be found in \citep{komiya1988}. The minimax theorem in Theorem \ref{thm-Farkas_Revesz_minimax} was presented by \citet[Theorem~5.2]{Farkas2006}, and that in Theorem \ref{thm-Mosonyi_minimax} was presented by \citet[Corollary~A.2]{MH11}.
	
	For an introduction to probability theory, see \citep{F68book,Ross_2019}. Proofs of Markov's inequality \eqref{eq-MT:Markov-ineq} and Jensen's inequality \eqref{eq-MT:Jensen_ineq} can be found in, e.g., \citep{FG_book_97}. 
	
	For further details on semi-definite programming, see \citet{BV96,Wat18}. Various polynomial-time algorithms for solving semi-definite programs were developed by \citet{Khachiyan1980,AHK05,AK07,AHK12, LSW15}. A proof of Slater's Theorem (Theorem~\ref{thm:math-tools:slater-cond}) can be found in \citep[Section~5.3.2]{BV04}.

	For further details about the symmetric subspace of a tensor product of finite-dimensional Hilbert spaces, as well as for a proof of \eqref{eq-sym_proj_integral}, see \citep{Har13} (see also \citet[Section 12.7]{BZ17_book}). Further details about the Fubini-Study measure $\text{d}\psi$ introduced in \eqref{eq-sym_proj_integral} and elaborated upon in the remark immediately below it may be found in \citep[Chapter 4]{BZ17_book}.

\section{Problems}\label{sec-problems_math_tools}

	\begin{enumerate}[left=0cm,itemsep=1cm]
			
	    \item\label{prob-math_tools_PSD} Let $X\in\Lin(\mathcal{H})$ be a positive semi-definite linear operator. Prove that there exists a Hilbert space $\mathcal{H}'$ and a linear operator $Y\in\Lin(\mathcal{H},\mathcal{H}')$ such that $X=Y^{\dagger}Y$. 

		\item\label{prob-math_tools_isometries} Prove that the columns of every isometry form an orthonormal set of vectors. Similarly, prove that the rows and columns of every unitary operator form orthonormal sets of vectors. (\textit{Hint}: Consider using the expressions in \eqref{eq-linear_op_row_column_abstract}.)

		\item\label{prob-math_tools_kronecker_sum} Let $X\in\Lin(\mathcal{H}_A)$ and $Y\in\Lin(\mathcal{H}_B)$ be normal operators, and consider their so-called \textit{Kronecker sum}:
			\begin{equation}
				X\oplus_{\text{K}} Y\coloneqq X\otimes\mathbbm{1}_B+\mathbbm{1}_A\otimes Y.
			\end{equation}
			Prove that $\text{spec}(X\oplus_{\text{K}} Y)=\{\lambda+\mu:\lambda\in\text{spec}(X),\,\mu\in\text{spec}(Y)\}$. Also prove that the associated eigenvectors are of the form $\ket{\psi}\otimes\ket{\phi}$, where $\ket{\psi}$ is an eigenvector of $X$ and $\ket{\phi}$ is an eigenvector of $Y$.

		\item\label{prob-math_tools_hadamard_prod} The \textit{Hadamard product}, also known as the Schur product, of two linear operators $X,Y\in\Lin(\mathbb{C}^d)$, with $d\geq 2$, is defined to be the element-wise product of $X$ and $Y$: if $X=\sum_{i,j=0}^{d-1} X_{i,j}\ketbra{i}{j}$ and $Y=\sum_{i,j=0}^{d-1}Y_{i,j}\ketbra{i}{j}$, then
			\begin{equation}
				X\ast Y\coloneqq \sum_{i,j=0}^{d-1} X_{i,j}Y_{i,j}\ketbra{i}{j}.
			\end{equation}
			\begin{enumerate}[itemsep=0.3cm,ref=\theenumi.(\alph*)]
				\item Verify that for all $\ket{\psi},\ket{\phi}\in\mathbb{C}^d$,
					\begin{equation}
						\bra{\psi}X\ast Y\ket{\phi}=\Tr\left[X^{\t}\text{diag}(\bra{\psi})Y\text{diag}(\ket{\phi})\right],
					\end{equation}
					where for $\ket{\psi}=\sum_{i=0}^{d-1}\alpha_i\ket{i}$ and $\ket{\phi}=\sum_{j=0}^{d-1}\beta_j\ket{j}$,
					\begin{equation}
						\text{diag}(\bra{\psi})\coloneqq\sum_{i=0}^{d-1}\conj{\alpha_i}\ketbra{i}{i},\quad\text{diag}(\ket{\phi})\coloneqq\sum_{j=0}^{d-1}\beta_j\ketbra{j}{j}.
					\end{equation}
				
				\item\label{prob-math_tools_hadamard_prod_b} Prove that for all $\ket{\psi},\ket{\phi}\in\mathbb{C}^d$,
					\begin{equation}
						\ketbra{\psi}{\psi}\ast\ketbra{\phi}{\phi}=(\ket{\psi}\ast\ket{\phi})(\bra{\psi}\ast\bra{\phi}).
					\end{equation}
				
				\item Prove that the Hadamard product of two positive semi-definite operators is positive semi-definite.
				
			\end{enumerate}

		\item\label{prob-math_tools_ONB} Let $\{\ket{\psi_j}\}_{j=1}^d$ be a set of $d$ linearly independent vectors in $\mathbb{C}^d$, with $d\geq 2$. By definition, this means that, for all $c_1,c_2,\dotsc,c_d\in\mathbb{C}$, the equation $c_1\ket{\psi_1}+c_2\ket{\psi_2}+\dotsb+c_d\ket{\psi_d}=0$ implies $c_1=c_2=\dotsb=c_d=0$.
			\begin{enumerate}[itemsep=0.3cm,topsep=0.3cm,ref=\theenumi.(\alph*)]
				\item\label{prob-math_tools_ONB_a} Let
					\begin{equation}\label{eq-math_tools-pre_frame_operator}
						T\coloneqq\sum_{j=1}^d \ketbra{\psi_j}{j-1}.
					\end{equation}
					The operator $T$ can be thought of as a $d\times d$ matrix whose columns are given by the vectors $\ket{\psi_j}$. Prove that $T$ is invertible. (\textit{Hint}: First prove that $T$ is injective, by showing that its kernel contains only the zero vector. Then use the result of Exercise~\ref{exer-injective_surjective}.)
					
				\item\label{prob-math_tools_ONB_b} Using (a), prove that $\{\ket{\psi_j}\}_{j=1}^d$ is a basis for $\mathbb{C}^d$. In other words, prove that every vector $\ket{\phi}\in\mathbb{C}^d$ can be written as a unique linear combination of the vectors $\ket{\psi_j}$. We thus have that every set of $d$ linearly independent vectors in $\mathbb{C}^d$ is a basis for~$\mathbb{C}^d$.
				
				\item\label{prob-math_tools_ONB_c} Prove that if $\sum_{j=1}^d\ketbra{\psi_j}{\psi_j}=\mathbbm{1}_d$, then $\{\ket{\psi}_j\}_{j=1}^d$ is an orthonormal basis for~$\mathbb{C}^d$.
				
					By combining this result with the result of Exercise~\ref{exer-identity_operator}, we have that a linearly independent set $\{\ket{\psi_j}\}_{j=1}^d$ of vectors in $\mathbb{C}^d$ is an orthonormal basis if and only if $\sum_{j=1}^d\ketbra{\psi_j}{\psi_j}=\mathbbm{1}_d$.
			\end{enumerate}

		\item\label{prob-math_tools_op_ONB} Let $\{B_j\}_{j=1}^{d^2}$ be an orthonormal basis for $\Lin(\mathbb{C}^d)$, with $d\geq 2$.
			\begin{enumerate}[itemsep=0.3cm,topsep=0.3cm,ref=\theenumi.(\alph*)]
				\item Prove that
					\begin{equation}\label{eq-op_ONB_to_Gamma}
						\sum_{j=1}^{d^2} \conj{B_j}\otimes B_j=\Gamma_d,
					\end{equation}
					where we recall that $\Gamma_d=\ketbra{\Gamma_d}{\Gamma_d}=\sum_{i,j=0}^{d-1}\ketbra{i,i}{j,j}$; see \eqref{eq-max_ent_vector}. Similarly, prove that
					\begin{equation}
						\sum_{j=1}^{d^2} B_j^{\dagger}\otimes B_j=F,
					\end{equation}
					where we recall that $F=\sum_{i,j=0}^{d-1}\ketbra{i,j}{j,i}$; see \eqref{eq-swap_op_standard_0}.
					
					(\textit{Hint}: Start by verifying that $\{\conj{B_j}\}_{j=1}^{d^2}$ is an orthonormal basis for $\Lin(\mathbb{C}^d)$. Then, use the fact that every linear operator $Z\in\Lin(\mathbb{C}^d\otimes\mathbb{C}^d)$ can be written as $Z=\sum_{j,k=1}^{d^2}c_{j,k} \conj{B_j}\otimes B_k$ for some coefficients $c_{j,k}\in\mathbb{C}$.)
					
				\item Prove that for all $X\in\Lin(\mathbb{C}^d)$,
					\begin{equation}
						\sum_{j=1}^{d^2} B_jXB_j^{\dagger}=\Tr[X]\mathbbm{1}_d.
					\end{equation}
					(\textit{Hint}: Use \eqref{eq-op_ONB_to_Gamma}, along with the identities in \eqref{eq-transpose_trick}--\eqref{eq-vec_expand}.)
					
				\item\label{prob-math_tools_op_ONB_c} Prove that $\{\text{vec}(B_j)\}_{j=1}^{d^2}$ and $\{(B_j\otimes\mathbbm{1}_d)\ket{\Gamma_d}\}_{j=1}^{d^2}$ are orthonormal bases for $\mathbb{C}^d\otimes\mathbb{C}^d$.
				
			\end{enumerate}

		\item For all $d\geq 2$, construct a basis for $\Lin(\mathbb{C}^d)$ that consists entirely of density operators. (\textit{Hint}: Consider using the eigenvectors of the orthonormal basis of Hermitian operators defined in \eqref{eq-su_generators_1}--\eqref{eq:math-tools:gell-mann-scaled-mats}.)

		\item Let $\{\ket{\psi_j}\}_{j=1}^d$ be a set of linearly independent, normalized, but non-orthogonal vectors in $\mathbb{C}^d$, with $d\geq 2$. We would like to transform these vectors into a new set $\{\ket{\phi_j}\}_{j=1}^d$ of orthonormal vectors via an invertible linear operator $X$, such that $\ket{\phi_j}=X\ket{\psi_j}$ for all $j\in\{1,2,\dotsc, d\}$.
			\begin{enumerate}[itemsep=0.3cm,topsep=0.3cm]
				\item Prove that the operator $S$ defined as
					\begin{equation}
						S\coloneqq\sum_{j=1}^d \ketbra{\psi_j}{\psi_j}
					\end{equation}
					is invertible and positive definite. (\textit{Hint}: Write $S$ in terms of the operator $T$ defined in \eqref{eq-math_tools-pre_frame_operator}.)
					
				\item Let
					\begin{equation}\label{eq-math_tools-symm_ortho}
						\ket{\phi_j}\coloneqq S^{-\frac{1}{2}}\ket{\psi_j}
					\end{equation}
					for all $j\in\{1,2,\dotsc,d\}$. Prove that $\{\ket{\phi_j}\}_{j=1}^d$ is an orthonormal basis for $\mathbb{C}^d$. (\textit{Hint}: See problem~\ref{prob-math_tools_ONB_c}.) Also, prove that $\braket{\phi_i}{\psi_j}=\bra{i-1}G^{\frac{1}{2}}\ket{j-1}$ for all $i,j\in\{1,2,\dotsc,d\}$, where $G\coloneqq T^{\dagger} T$ and $T\coloneqq\sum_{j=1}^d\ketbra{\psi_j}{j-1}$.
				
				\item Let us now show that the vectors defined in \eqref{eq-math_tools-symm_ortho} are optimal with respect to the Euclidean norm, in the following sense:
					\begin{align}
						&\inf_{X}\left\{\sum_{j=1}^d\norm{\ket{\psi_j}-\ket{\phi_j}}_2^2:\ket{\phi_j}=X\ket{\psi_j},\,\,\braket{\phi_i}{\phi_j}=\delta_{i,j}\,\,\forall\,1\leq j\leq d\right\}\label{eq-symm_orth_opt_problem}\\
						&=\sum_{j=1}^d \norm{\ket{\psi_j}-S^{-\frac{1}{2}}\ket{\psi_j}}_2^2,
					\end{align}
					where the optimization in \eqref{eq-symm_orth_opt_problem} is with respect to every invertible linear operator~$X$.
					
					\begin{enumerate}[itemsep=0.3cm,topsep=0.3cm]
						\item Prove that solving the optimization problem given by \eqref{eq-symm_orth_opt_problem} can be reduced to solving the optimization problem given by
							\begin{equation}\label{eq-symm_orth_opt_problem_2}
								\sup_X\left\{\Tr[(X+X^{\dagger})S]:XSX^{\dagger}=\mathbbm{1}_d\right\}.
							\end{equation}
							
						\item  Prove that the constraint $XSX^{\dagger}=\mathbbm{1}_d$ in \eqref{eq-symm_orth_opt_problem_2} implies $X=US^{-\frac{1}{2}}$, where $U$ is a unitary operator. (\textit{Hint}: Consider a polar decomposition of $X$; see Theorem~\ref{thm-polar_decomposition}.)
						
							Hence, show that the optimization problem given by \eqref{eq-symm_orth_opt_problem_2} is equivalent to
							\begin{equation}\label{eq-symm_orth_opt_problem_3}
								\sup_U \text{Re}\left(\Tr[US^{\frac{1}{2}}]\right),
							\end{equation}
							where the optimization is with respect to every unitary operator $U$ acting on~$\mathbb{C}^d$.
						
						\item Prove that the solution to the optimization problem given by \eqref{eq-symm_orth_opt_problem_3} is $U=\mathbbm{1}_d$, implying that the optimal $X$ in \eqref{eq-symm_orth_opt_problem} is indeed $S^{-\frac{1}{2}}$. (\textit{Hint}: Use Proposition~\ref{prop:math-tools:var-char-t-norm}.)
					\end{enumerate}
			\end{enumerate}
			(\textit{Bibliographic Note}: The vectors $\ket{\phi_j}$ defined in \eqref{eq-math_tools-symm_ortho} are known as the \textit{symmetric orthogonalization} of the original vectors $\ket{\psi_j}$, and this construction is attributed to \citet{Lowdin50}; see also \citep{Lowdin70}. An alternate proof of the optimality of this construction, as worked out in part (c) of this problem, can be found in \citep{Mayer2002}.)

		\item For the case $d=2$ and $n=2$, verify the equalities given by \eqref{eq-proj_symm_subspace} and \eqref{eq-sym_proj_integral} by making use of \eqref{eq-fubini_study_d2}.

		\item Prove that the right-hand side of \eqref{eq-proj_symm_subspace} is indeed the projection onto $\text{Sym}_n(\mathcal{H})$ by showing that
			\begin{equation}\label{eq-proj_symm_subspace2}
				\sum_{\substack{n_1,n_2,\dotsc,n_d\geq 0,\\\sum_{j=1}^d n_j=n}} \ketbra{n_1,n_2,\dotsc,n_d}{n_1,n_2,\dotsc,n_d}=\Pi_{\text{Sym}_n(\mathcal{H})}.
			\end{equation}

	\end{enumerate}

\chapter{Quantum States and Measurements}\label{chap-QM_states_meas}

	In the previous chapter, we studied several important topics in mathematics that collectively form one foundational piece for the study of quantum information processing. Another foundational piece is quantum mechanics, and in this and the following chapter, we provide an overview of it, placing particular emphasis on those aspects of it that are useful for the communication protocols that we discuss in later chapters. Many aspects of quantum mechanics cannot be explained by classical reasoning. For example, there is no strong classical analogue for pure quantum states or entanglement, and this leads to stark differences between what is possible in the classical and quantum worlds. However, at the same time, it is important to emphasize that all of classical information theory is subsumed by quantum information theory, so that whatever is possible with classical information processing is also possible with quantum information processing. Interestingly as well, quantum information processing allows for richer possibilities, with protocols such as quantum teleportation and super-dense coding.

\section{Axioms of Quantum Mechanics}\label{sec-QM_axioms}

	The mathematical description of quantum systems can be summarized by the following axioms. Each of these axioms is elaborated upon in the section indicated below.
	\begin{enumerate}
	
		\item \textit{Quantum systems}: A quantum system $A$ is associated with a Hilbert space~$\mathcal{H}_A$. The state of the system $A$ is described by a \textit{density operator}, which is a unit-trace, positive semi-definite operator acting on $\mathcal{H}_A$. (See Section~\ref{sec-QM_states}.)
		
		\item \textit{Bipartite quantum systems}: For distinct quantum systems $A$ and $B$ with associated Hilbert spaces $\mathcal{H}_A$ and $\mathcal{H}_B$, the composite system $AB$ is associated with the Hilbert space $\mathcal{H}_A\otimes\mathcal{H}_B$. (See Section~\ref{subsec-bipartite_state}.)
		
		\item \textit{Measurement}: The measurement of a quantum system $A$ is described by a \textit{positive operator-valued measure (POVM)} $\{M_x\}_{x\in\mathcal{X}}$, which is defined to be a collection of positive semi-definite operators  satisfying $\sum_{x\in\mathcal{X}} M_x=\mathbbm{1}_{\mathcal{H}_A}$, where $\mathcal{X}$ is a finite alphabet\footnote{POVMs need not contain a finite number of elements, but we consider POVMs with a finite number of elements exclusively throughout this book.}. If the system is in the state $\rho$ and the measurement outcome is described by a random variable $X$, then the probability $\Pr[X=x]$ of obtaining the outcome $x$ is given by the \textit{Born rule} as
			\begin{equation}
				\Pr[X=x]=\Tr[M_x\rho].
			\end{equation}
			Furthermore, a physical \textit{observable} $O$ corresponds to a Hermitian operator acting on the underlying Hilbert space. Recall from the spectral theorem (Theorem~\ref{thm-spectral_theorem}) that $O$ has a spectral decomposition as follows:
			\begin{equation}	
			O = \sum_{\lambda\in\text{spec}(O)} \lambda \ \Pi_{\lambda},
			\end{equation}	
			where $\text{spec}(O)$ is the set of distinct eigenvalues of $O$ and $\Pi_\lambda$ is a spectral projection.  
			A measurement of the observable $O$ is described by the POVM $\{\Pi_\lambda\}_{\lambda}$,  which is indexed by the distinct eigenvalues $\lambda$ of $O$.
			The expected value $\left<O\right>_{\rho}$ of the observable $O$ when the state is $\rho$ is given by
			\begin{equation}	
				\left<O\right>_\rho\coloneqq\Tr[O\rho] = \sum_{\lambda\in\text{spec}(O)} \lambda \, \Tr[\Pi_\lambda\rho].
			\end{equation}
			(See Section~\ref{subsec-meas}.)
			
		\item \textit{Evolution}: The evolution of the state of a quantum system is described by a \textit{quantum channel}, which is a linear, completely positive, and trace-preserving map acting on the state of the system. (See Chapter~\ref{chap-QM_channels}.)
	\end{enumerate}
	
	Note that the second axiom for the description of bipartite quantum systems is sufficient to conclude that the multipartite quantum system $A_1A_2\dotsb\allowbreak A_k$, comprising $k$ distinct quantum systems $A_1,A_2,\dotsc, A_k$, is associated with the Hilbert space $\mathcal{H}_{A_1}\otimes\mathcal{H}_{A_2}\otimes\dotsb\otimes\mathcal{H}_{A_k}$.

\section{Quantum Systems and States}\label{sec-QM_states}

	Each quantum system is associated with a Hilbert space. In this book, we consider only finite-dimensional quantum systems, that is, quantum systems described by finite-dimensional Hilbert spaces. In the following, we provide a mathematical description of several finite-dimensional quantum systems, along with examples of how these systems can be physically realized. 
	
	\begin{enumerate}
		\item \textit{Qubit systems}: The qubit is perhaps the most fundamental quantum system and is the quantum analogue of the (classical) bit. Every physical system obeying the laws of quantum mechanics and with two distinct states can be considered a qubit system. The Hilbert space associated with a qubit system is $\mathbb{C}^2$, whose standard orthonormal basis is  denoted by $\{\ket{0},\ket{1}\}$. Three common ways of physically realizing qubit systems are as follows:
			\begin{enumerate}
				\item The two spin states of a spin-$\frac{1}{2}$ particle.
				\item Two distinct energy levels of an atom, such as the ground state and one of the excited states.
				\item Clockwise and counter-clockwise directions of current flow in a superconducting electronic circuit. 
			\end{enumerate}
		
		\item \textit{Qutrit systems}: A qutrit system is a quantum system consisting of three distinct physical degrees of freedom. The Hilbert space of a qutrit is $\mathbb{C}^3$, with the standard orthonormal basis denoted by $\{\ket{0},\ket{1},\ket{2}\}$. Qutrit systems are less commonly considered than qubit systems for implementations, although one important example of an implementation of a qutrit system occurs in quantum optical systems, which we discuss below. Like qubit systems, qutrit systems can also be physically realized using, for example, the spin states of a spin-1 atom or three distinct energy levels of an atom.
		
		\item \textit{Qudit systems}: A qudit system is a quantum system with $d$ distinct degrees of freedom and is described by the Hilbert space $\mathbb{C}^d$, with the standard orthonormal basis  denoted  by $\{\ket{0},\ket{1},\dotsc,\ket{d-1}\}$. The spin states of every spin-$j$ atom can be used to realize a qudit system with $d=2j+1$. Another physical realization of a qudit system is with the $d$ distinct energy levels of an atom.
		
		\item \textit{Quantum optical systems}: An important quantum system, particularly for the implementation of many quantum communication protocols, is a qu\-antum optical system. By a quantum optical system, we mean a physical system, such as an optical cavity or a fiber-optic cable, in which modes of light, with photons as information carriers, propagate. A mode of light has a well defined  momentum, frequency, polarization, and spatial direction. 
			
			Formally, a quantum optical system with $d$ distinct modes is described by the Fock space $\mathcal{F}_{\text{B}}(\mathbb{C}^d)$, which is a Hilbert space equipped with the orthonormal occupation number basis $\{\ket{n_1,\allowbreak\dotsc, n_d}:n_1,\dotsc,n_d\allowbreak \geq 0\}$, where $n_j$, for $j \in \{1, \ldots, d\}$, indicates the number of photons occupied in mode~$j$. See \eqref{eq:math-tools:occupation-num-basis} and the surrounding discussion for a brief review of the occupation number basis and Fock space.
			
			The Fock space is infinite dimensional, but by restricting to particular subspaces, it is possible use photons to physically realize finite-dimensional qua\-ntum systems. The following two realizations of a qubit system are particularly important:
			\begin{enumerate}
				\item A single-mode optical system, with Hilbert space $\mathcal{F}_{\text{B}}(\mathbb{C})$, restricted to the subspace spanned by the orthonormal vectors $\{\ket{0},\ket{1}\}$, interpreted as either zero or one photon occupied in the mode. The vector $\ket{0}$ corresponding to no photons is commonly called the \textit{vacuum state vector} of the mode.
				\item A two-mode optical system, with Hilbert space $\mathcal{F}_{\text{B}}(\mathbb{C}^2)$, restricted to the subspace span\-ned by the orthonormal vectors $\{\ket{0,1},\allowbreak\ket{1,0}\}$, consisting of only one photon in total occupying either one of the two modes. This realization of a qubit system is commonly called the \textit{dual-rail encoding} because it makes use of two modes of light. Two distinct polarization degrees of freedom of photons, such as horizontal and vertical polarizations, are commonly used as the two modes in dual-rail encodings of a qubit. One then usually lets $\ket{H}\equiv\ket{0,1}$ and $\ket{V}\equiv\ket{1,0}$ denote a horizontally- and vertically-polarized photon, respectively.
				
					By considering the three-dimensional subspace spanned by $\{\ket{0,0},\ket{0,1},\allowbreak\ket{1,0}\}$, that is, the dual-rail qubit system with the additional orthogonal vacuum state vector $\ket{0,0}$ of the two modes, we obtain a physical realization of a qutrit system. This particular realization of a qutrit system is relevant for communication protocols in the context of the erasure channel, which is discussed in Section~\ref{sec:QM-over:qudit-erasure}.
			\end{enumerate}
	\end{enumerate}
	
	Having discussed how a quantum system is mathematically described, let us now move on to the mathematical description of the state of a quantum system.
	
	\begin{definition}{Quantum State}{def-quantum_state}
		The state of a quantum system is described by a density operator acting on the underlying Hilbert space of the quantum system. A density operator is a unit-trace, positive semi-definite operator. Throughout the book, we identify a state with its corresponding density operator. We denote the set of density operators on a Hilbert space $\mathcal{H}$ by $\Density(\mathcal{H})$.	
	\end{definition}
	
	We typically use the Greek letters $\rho$, $\sigma$, $\tau$, or $\omega$ to denote quantum states.
	
	\begin{exercise}{exer-QM_states_convex}
		Prove that the set of quantum states is a \textit{convex set}. (Recall the definition of a convex set from Section~\ref{sec-math_tools_convexity}.) In other words, prove that for every alphabet $\mathcal{X}$ and set $\{\rho^x\}_{x\in\mathcal{X}}$ of quantum states, along with every probability distribution $p:\mathcal{X}\rightarrow[0,1]$, the following convex combination is a quantum state:
		\begin{equation}
			\rho=\sum_{x\in\mathcal{X}}p(x)\rho^x.
		\end{equation}
	\end{exercise}
	
	The extremal points in the convex set of quantum states are called \textit{pure states}. A pure state is a rank-one projection onto a unit vector in the Hilbert space. Concretely, pure states are of the form $\ket{\psi}\!\bra{\psi}$ where $\ket{\psi}\in\mathcal{H}$ is a normalized vector. For convenience, we sometimes denote $\ket{\psi}\!\bra{\psi}$ simply as $\psi$, and refer to the unit vector $\ket{\psi}$ as a \textit{state vector}. Since every element of a convex set can be written as a convex combination of the extremal points in the set, every quantum state $\rho$ that is not a pure state can be written as
	\begin{equation}\label{eq-mixed_state}
		\rho=\sum_{x\in\mathcal{X}} p(x)\ketbra{\psi_x}{\psi_x}
	\end{equation}
	for some set $\{\ket{\psi_x}\}_{x\in\mathcal{X}}$ of state vectors defined with respect to a finite alphabet $\mathcal{X}$, where $p:\mathcal{X}\to[0,1]$ is a probability distribution.

	\begin{exercise}{exer-pure_states}
		Prove that a quantum state $\rho$ is pure if and only if $\rho^2=\rho$. More generally, prove that $\rho$ is pure if and only if $\Tr[\rho^2]=1$. The quantity $\Tr[\rho^2]$ is known as the \textit{purity} of $\rho$.
	\end{exercise}

    \begin{exercise}{exer-global_phase_QM_states}
		Prove that the state vectors $\ket{\psi}$ and $e^{\I \theta} \ket{\psi}$, where $\theta \in \mathbb{R}$, lead to the same density operator. (This implies that the global phase $e^{\I \theta}$ is physically unobservable.)
	\end{exercise}

	A state $\rho$ that is not pure is called a \textit{mixed state}, because it can be thought of as arising from the lack of knowledge of which pure state from the set $\{\ket{\psi_x}\}_{x\in\mathcal{X}}$ in \eqref{eq-mixed_state} the system has been prepared. Note that the decomposition in \eqref{eq-mixed_state}, of a quantum state into pure states, is generally not unique.
	
	A state $\rho$ is called \textit{maximally mixed} if the set $\{\ket{\psi_x}\}_{x\in\mathcal{X}}$ in \eqref{eq-mixed_state} consists of $d$ orthonormal state vectors and the probability distribution $\{p(x)\}_{x \in \mathcal{X}}$ is uniform (i.e., $p(x)=\frac{1}{d}$ for all $x\in\mathcal{X}$). In this case, it follows that
	\begin{equation}\label{eq-max_mixed}
		\rho=\frac{\mathbbm{1}_d}{d}\eqqcolon\pi_d,
	\end{equation}
	as a consequence of Exercise~\ref{exer-identity_operator}.
The state $\pi_d$ is called maximally mixed because it corresponds to having the most uncertainty about which state from the set $\{\ket{\psi_k}\}_{k=1}^d$ the system is in. This uncertainty can be quantified by using quantum entropy, and in Chapter~\ref{chap-entropies}, we find that the maximally mixed state $\pi_d$ has the largest entropy among all states of a finite-dimensional system of dimension~$d$, thus justifying the term ``maximally mixed.''
	
	Now, let us recall the orthonormal basis of Hermitian operators defined in \eqref{eq-su_generators_0}--\eqref{eq:math-tools:gell-mann-scaled-mats}. In quantum information, it is common to scale these operators by $\sqrt{d}$, where $d$ is the dimension, so that we have an orthogonal basis $\{S_k^{(d)}\}_{k=0}^{d^2-1}$ of Hermitian operators, with $S_0^{(d)}=\mathbbm{1}_d$ and $S_k^{(d)}$, $k\in\{1,2,\dotsc,d^2-1\}$, equal to the traceless operators in \eqref{eq-su_generators_1}--\eqref{eq:math-tools:gell-mann-scaled-mats} multiplied by $\sqrt{d}$. Note here that we have also relabeled the indices of the set of operators defined in \eqref{eq-su_generators_0}--\eqref{eq:math-tools:gell-mann-scaled-mats}. These operators satisfy $\Tr[(S_k^{(d)})^2]=d$ and $\Tr[S_k^{(d)}S_{\ell}^{(d)}]=d\delta_{k,\ell}$ for all $k,\ell\in\{0,1,\dotsc,d^2-1\}$. We often suppress the dimension and write $S_k\equiv S_k^{(d)}$ if the dimension is unimportant or clear from the context. Using these operators, we can write every density operator $\rho\in\Density(\mathbb{C}^d)$ in the following form:
	\begin{equation}\label{eq-q_state_coh_vec_repr}
		\rho=\frac{1}{d}\left(\mathbbm{1}+\sum_{k=1}^{d^2-1} r_kS_k\right),
	\end{equation}
	where $r_k=\inner{S_k}{\rho}=\Tr[S_k\rho]\in\mathbb{R}$ for all $k\in\{1,2,\dotsc,d^2-1\}$. The vector $\vec{r}_{\rho}\coloneqq(r_1,r_2,\dotsc,r_{d^2-1})\in\mathbb{R}^{d^2-1}$ is sometimes called the \textit{Bloch vector}, or the \textit{coherence vector}, of $\rho$; please see the Bibliographic Notes (Section~\ref{sec:qm:bib-notes}) for more information on this terminology.
	
	\begin{exercise}{exer-coh_vec_pure_state}
		Let $\rho$ be the quantum state represented as in \eqref{eq-q_state_coh_vec_repr}.
		\begin{enumerate}[topsep=0.3cm]
			\item Verify that $\Tr[\rho]=1$.
			\item Prove that $\rho$ is pure if and only if $\sum_{k=1}^{d^2-1}r_k^2=d-1$.
		\end{enumerate}
	\end{exercise}
	
	At this point, it is instructive to look at an example. Let us consider quantum systems with $d=2$, i.e., qubits. The representation of an arbitrary quantum state $\rho$ in \eqref{eq-q_state_coh_vec_repr} becomes
	\begin{equation}\label{eq-q_state_coh_vec_repr_qubits}
		\rho=\frac{1}{2}(\mathbbm{1}+r_1X+r_2Y+r_3Z)\quad\text{(qubit state)},
	\end{equation}
	where $X$, $Y$, and $Z$ are the Pauli operators, which we defined in~\eqref{eq-Pauli_mat_1} and~\eqref{eq-Pauli_mat_2}:
	\begin{equation}\label{eq-QM-Pauli_mat}
		X=\begin{pmatrix} 0 & 1 \\ 1 & 0 \end{pmatrix}, \quad Y=\begin{pmatrix} 0 & -\I \\ \I & 0 \end{pmatrix},\quad Z=\begin{pmatrix} 1 & 0 \\ 0 & -1 \end{pmatrix}.
	\end{equation}
	The fact that $\Tr[\rho]=1$ follows from the fact that $X$, $Y$, and $Z$ are traceless operators, while $\Tr[\mathbbm{1}]=2$. The condition for $\rho$ to be positive semi-definite is left to the following exercise.
	
	\begin{figure}
		\centering
		\includegraphics[scale=1]{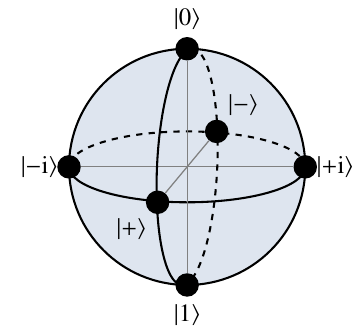}
		\caption{The quantum states in $\Density(\mathbb{C}^2)$ of every qubit system can be represented as a point in the so-called \textit{Bloch ball}. All pure states lie on the surface of the Bloch ball, which is known as the \textit{Bloch sphere}. Shown are the basis state vectors $\ket{0}$ and $\ket{1}$, corresponding to the Bloch vectors $(0,0,1)$ and $(0,0,-1)$, respectively. The superposition state vectors $\ket{\pm}\coloneqq\frac{1}{\sqrt{2}}(\ket{0}\pm\ket{1})$ correspond to the Bloch vectors $(\pm 1,0,0)$, and the superposition state vectors $\ket{\pm\I}\coloneqq\frac{1}{\sqrt{2}}(\ket{0}\pm\I\ket{1})$ correspond to the Bloch vectors $(0,\pm 1,0)$.}\label{fig-Bloch_ball}
	\end{figure}
	
	\begin{exercise}{exer-q_state_coh_vec_repr_qubits}
		Show that the positive semi-definiteness of every qubit state $\rho$, as represented in \eqref{eq-q_state_coh_vec_repr_qubits}, is equivalent to $r_1^2+r_2^2+r_3^2\leq 1$.
	\end{exercise}
	
	The condition $r_1^2+r_2^2+r_3^2\leq 1$ for every qubit state $\rho$ represented as in \eqref{eq-q_state_coh_vec_repr_qubits}, along with the fact that $r_1,r_2,r_3\in\mathbb{R}$, implies that the vector $\vec{r}_{\rho}=(r_1,r_2,r_3)$ lies on or inside the unit sphere in three dimensions, for every qubit state $\rho$. Furthermore, the condition in Exercise~\ref{exer-coh_vec_pure_state} for $\rho$ to be pure implies that a qubit state is pure if and only if the vector $\vec{r}_{\rho}$ lies on the surface of the unit sphere. In quantum mechanics, this unit sphere is known as the \textit{Bloch sphere}, and if we include all mixed states corresponding to the interior of the sphere, then we use the term \textit{Bloch ball} to refer to the set of all qubit states. See Figure~\ref{fig-Bloch_ball} for a visual representation of the Bloch ball.


\subsection{Bipartite States and Schmidt Decomposition}\label{subsec-bipartite_state}

	The joint state of two distinct quantum systems $A$ and $B$ is described by a bipartite quantum state $\rho_{AB}\in\Density(\mathcal{H}_A\otimes\mathcal{H}_B)$. For brevity, the joint Hilbert space $\mathcal{H}_A\otimes\mathcal{H}_B$ of the composite system $AB$ is denoted by $\mathcal{H}_{AB}$.
	
	Let $\{\ket{i}_A\}_{i=0}^{d_A-1}$ and $\{\ket{j}_B\}_{j=0}^{d_B-1}$ be orthonormal bases for $\mathcal{H}_A$ and $\mathcal{H}_B$, respectively. Then, 
	\begin{equation}
	\{\ket{i}_A\otimes\ket{j}_B:0\leq i\leq d_A-1,~0\leq j\leq d_B-1\}
	\end{equation}
	 is an orthonormal basis for $\mathcal{H}_{AB}$. For brevity, we often write $\ket{i,j}_{AB}$ instead of $\ket{i}_A\otimes\ket{j}_B$. Every state vector $\ket{\psi}_{AB}\in\mathcal{H}_{AB}$ can thus be written as
	\begin{equation}\label{eq-bipartite_pure_state}
		\ket{\psi}_{AB}=\sum_{i=0}^{d_A-1}\sum_{j=0}^{d_B-1}\alpha_{i,j}\ket{i,j}_{AB},
	\end{equation}
	where $\alpha_{i,j}=\braket{i,j}{\psi}\in\mathbb{C}$ and $\sum_{i=0}^{d_A-1}\sum_{j=0}^{d_B-1}\abs{\alpha_{i,j}}^2=1$. By the Schmidt decomposition theorem (Theorem~\ref{thm-Schmidt}), we can alternatively write $\ket{\psi}_{AB}$ as
	\begin{equation}\label{eq-Schmidt_decomp_state_vector}
		\ket{\psi}_{AB}=\sum_{k=1}^r \sqrt{\lambda_k}\ket{e_k}_A\otimes\ket{f_k}_B,
	\end{equation}
	where each Schmidt coefficient $\lambda_k$ is strictly positive and they all satisfy $\sum_{k=1}^r\lambda_k=1$, $\{\ket{e_k}_A\}_{k=1}^r$ and $\{\ket{f_k}_B\}_{k=1}^r$ are orthonormal sets of vectors in $\mathcal{H}_A$ and $\mathcal{H}_B$, respectively, and $r=\rank(X)$, where $X\in\Lin(\mathcal{H}_A,\mathcal{H}_B)$ is defined as $\bra{j}_BX\ket{i}_A=\braket{i,j}{\psi}_{AB}$ for all $0\leq i\leq d_A-1$ and $0\leq j\leq d_B-1$.
	
	More generally, recall from Chapter~\ref{chap-math_tools} that we can define the orthonormal bases
	\begin{equation}
		\{\ket{i}\!\bra{i'}_A:0\leq i,i'\leq d_A - 1\},\quad \{\ket{j}\!\bra{j'}_B:0\leq j,j'\leq d_B - 1\},
	\end{equation}
	for $\Lin(\mathcal{H}_A)$ and $\Lin(\mathcal{H}_B)$, respectively. Then, the set
	\begin{equation}\label{eq-ONB_lin_ops_AB}
		\{\ket{i,j}\!\bra{i',j'}_{AB}\equiv\ket{i}\!\bra{i'}_A\otimes\ket{j}\!\bra{j'}_B:0\leq i,i'\leq d_A - 1,~0\leq j,j'\leq d_B - 1\}
	\end{equation}
	is an orthonormal basis for $\Lin(\mathcal{H}_{AB})$. It follows that every mixed state $\rho_{AB}\in\Density(\mathcal{H}_{AB})$ can be written as
	\begin{equation}\label{eq-bipartite_mixed_state}
		\rho_{AB}=\sum_{i,i'=0}^{d_A-1}\sum_{j,j'=0}^{d_B-1}\beta_{i,j;i',j'}\ketbra{i,j}{i',j'}_{AB},
	\end{equation}
	where $\beta_{i,j;i',j'}=\inner{\ketbra{i,j}{i',j'}}{\rho_{AB}}=\bra{i,j}\rho_{AB}\ket{i',j'}\in\mathbb{C}$. Similarly, consider orthogonal bases $\{S_A^k\}_{k=0}^{d_A^2-1}$ and $\{S^{\ell}_B\}_{\ell=0}^{d_B^2-1}$ for $\Lin(\mathcal{H}_A)$ and $\Lin(\mathcal{H}_B)$, respectively, as defined in the paragraph above \eqref{eq-q_state_coh_vec_repr}. Then, $\{S^k_A\otimes S^{\ell}_B:0\leq k\leq d_A^2-1,\,0\leq\ell\leq d_B^2-1\}$ is an orthogonal basis for $\Lin(\mathcal{H}_{AB})$, so that every quantum state $\rho_{AB}\in\Density(\mathcal{H}_{AB})$ can be written as 
	\begin{equation}\label{eq-bipartite_q_state_coh_vec_repr}
		\rho_{AB}=\frac{1}{d_Ad_B}\sum_{k=0}^{d_A^2-1}\sum_{\ell=0}^{d_B^2-1} r_{k,\ell} S_A^k\otimes S^{\ell}_B,
	\end{equation}
	where
	\begin{equation}
	r_{k,\ell}=\inner{S_A^k\otimes S^{\ell}_B}{\rho_{AB}}=\Tr[(S^k_A\otimes S^{\ell}_B)\rho_{AB}]
	\end{equation}
	for all $k\in\{0,1,\dotsc,d_A^2-1\}$ and $\ell\in\{0,1,\dotsc,d_B^2-1\}$. Also, as with the Schmidt decomposition of state vectors in \eqref{eq-Schmidt_decomp_state_vector}, from Exercise~\ref{exer-op_Schmidt_decomp} we have that every mixed state $\rho_{AB}$ can be written as 
	\begin{equation}\label{eq-Schmidt_decomp_mixed_state}
		\rho_{AB}=\sum_{k=1}^r\sqrt{\smash[b]{\lambda_k}}\, E_A^k\otimes F_B^k,
	\end{equation}
	where each coefficient $\lambda_k$ is strictly positive, $\{E_A^k\}_{k=1}^r$ and $\{F_B^k\}_{k=1}^r$ are orthonormal sets of linear operators acting on $\mathcal{H}_A$ and $\mathcal{H}_B$, respectively, and $r=\rank(M)$, where $M\in\Lin(\mathcal{H}_A\otimes\mathcal{H}_A,\mathcal{H}_B\otimes\mathcal{H}_B)$ is defined by $\bra{j,\ell}_{BB}M\ket{i,k}_{AA}=\bra{i,j}\rho_{AB}\ket{k,\ell}$ for all $0\leq i,j\leq d_A-1$ and $0\leq j,\ell\leq d_B-1$.

\subsection{Partial Trace}\label{sec-partial_trace_QM}

	Recall from Section~\ref{sec-math_tools-lin_ops} that the trace of a linear operator $X$ acting on a $d$-dimensional Hilbert space can be written as
	\begin{equation}
		\Tr[X]=\sum_{i=0}^{d-1} \bra{i}X\ket{i},
	\end{equation}
	where $\{\ket{i}\}_{i=0}^{d-1}$ is the standard orthonormal basis. We can interpret the trace as the sum of the diagonal elements of the matrix corresponding to $X$ written in the standard basis. From Exercise~\ref{exer-trace}, however, we know that the trace is independent of the choice of basis used to evaluate it.
	
	\begin{figure}
		\centering
		\includegraphics[scale=1]{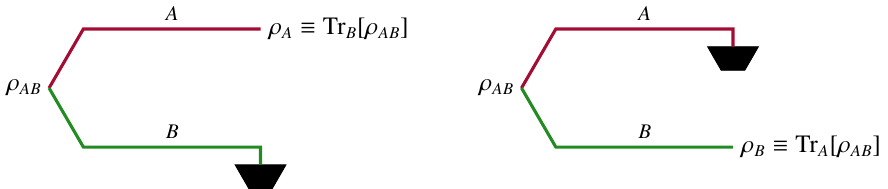}
		\caption{The partial trace superoperator (see Definition~\ref{def-partial_trace}) is the mathematical representation of physically discarding a subsystem of a composite quantum system. In other words, given a bipartite state $\rho_{AB}$ for the two quantum systems $A$ and $B$, the partial trace $\Tr_B$ allows us to determine the quantum state of system $A$ when we do not have access to system $B$ (left), and $\Tr_A$ allows us to determine the quantum state of system $B$ when we do not have access to system~$A$ (right).}\label{fig-partial_trace}
	\end{figure}
	
	The trace is physically relevant, especially when it acts on one part of a bipartite quantum state, in which case we call it the \textit{partial trace}. To be specific, given a state $\rho_{AB}$ for the bipartite system $AB$, we are often interested in determining the state of only one of its subsystems. The partial trace $\Tr_B$, which we define formally below, takes a state $\rho_{AB}$ acting on the space $\mathcal{H}_{AB}$ and returns a state $\rho_A \equiv \Tr_B[\rho_{AB}]$ acting on the space $\mathcal{H}_A$. The partial trace is therefore the mathematical operation used to determine the state of one of the subsystems given the state of a composite system comprising two or more subsystems, and it can be thought of as the action of ``discarding'' one of the subsystems; see Figure~\ref{fig-partial_trace}. The partial trace generalizes the notion of marginalizing a joint probability distribution. Later, in Chapter~\ref{chap-QM_channels}, we see that the partial trace is a particular kind of quantum channel corresponding to this discarding action.
	
	\begin{definition}{Partial Trace}{def-partial_trace}
		Given quantum systems $A$ and $B$, the \textit{partial trace over $B$} is denoted by $\Tr_B\equiv \id_A\otimes \Tr_B$, and it is defined as
		\begin{equation}\label{eq-partial_trace_B}
			\begin{aligned}
			\Tr_B[X_{AB}]&=(\id_A\otimes \Tr)(X_{AB})\\
			&=\sum_{j=0}^{d_B-1}(\mathbbm{1}_A\otimes\bra{j}_B)X_{AB}(\mathbbm{1}_A\otimes\ket{j}_B)
			\end{aligned}
		\end{equation}
		for every linear operator $X_{AB}\in\Lin(\mathcal{H}_A\otimes\mathcal{H}_B)$. The partial trace over $A$ is defined similarly. 
	\end{definition}
	
	\begin{remark}
		For every linear operator $X_{AB}$ acting on $\mathcal{H}_{AB}$, we can define the partial trace $\Tr_B[X_{AB}]$ more abstractly as the unique linear operator $X_A$ acting on $\mathcal{H}_A$ such that
		\begin{equation}
			\Tr[(M_A\otimes\mathbbm{1}_B)X_{AB}]=\Tr[M_A X_A]
		\end{equation}
		for every operator $M_A \in\Lin(\mathcal{H}_A)$. If we let $X_{AB}$ be the state $\rho_{AB}$ and $M_A$ be a Hermitian operator, then we can interpret this equation physically in the following way: in order to determine the expectation value of an observable $M_A$ acting on only one of the subsystems (in this case, the $A$ subsystem), it suffices to know the reduced state $\rho_A$ of the subsystem $A$ rather than the  joint state $\rho_{AB}$ of the total system. 
	\end{remark}
	
	It is clear from Definition~\ref{def-partial_trace} that the partial trace is a linear superoperator. In particular, the expression in \eqref{eq-partial_trace_B} defines the partial trace in precisely the operator-sum form for superoperators shown in \eqref{eq-math_tools-superator_op_sum}.
	
	Now,  in order to explicitly determine the partial trace of a given linear operator $X_{AB}\in\Lin(\mathcal{H}_{AB})$, it suffices to know the action of the partial trace on basis elements of $\Lin(\mathcal{H}_{AB})$ because the action of every linear superoperator is completely defined by its action on basis elements. Using the orthonormal basis for $\Lin(\mathcal{H}_{AB})$ given in \eqref{eq-ONB_lin_ops_AB}, it is straightforward to see that the action of the partial trace $\Tr_B$ on this basis is
	 \begin{equation}\label{eq-partial_trace_B_basis}
		\Tr_B[\ket{i}\!\bra{i'}_A\otimes\ket{j}\!\bra{j'}_B]=\ket{i}\!\bra{i'}_A\delta_{j,j'}
	\end{equation}
	for all $0\leq i,i'\leq d_A - 1,~0\leq j,j'\leq d_B - 1$. Similarly, for $\Tr_A$, we obtain
	\begin{equation}\label{eq-partial_trace_A_basis}
		\Tr_A[\ket{i}\!\bra{i'}_A\otimes\ket{j}\!\bra{j'}_B]=\delta_{i,i'}\ket{j}\!\bra{j'}_B
	\end{equation}
	for all $0\leq i,i'\leq d_A - 1,~0\leq j,j'\leq d_B - 1$. Then, by decomposing every linear operator $X_{AB}$ as
	\begin{equation}
		\begin{aligned}
		X_{AB}&=\sum_{i,i'=0}^{d_A-1}\sum_{j,j'=0}^{d_B-1}X_{i,j;i',j'}\ket{i}\!\bra{i'}_A\otimes\ket{j}\!\bra{j'}_B\\
		&=\sum_{i,i'=0}^{d_A-1}\sum_{j,j'=0}^{d_B-1}X_{i,j;i',j'}\ket{i,j}\!\bra{i',j'}_{AB},
		\end{aligned}
	\end{equation}
	where $X_{i,j;i',j'}\coloneqq \bra{i,j}X_{AB}\ket{i',j'}$, we find that
	\begin{align}
		\Tr_B[X_{AB}]&=\sum_{i,i'=0}^{d_A-1}\left(\sum_{j=0}^{d_B-1}X_{i,j;i',j}\right)\ket{i}\!\bra{i'}_A,\\
		\Tr_A[X_{AB}]&=\sum_{j,j'=0}^{d_B-1}\left(\sum_{i=0}^{d_A-1}X_{i,j;i,j'}\right)\ket{j}\!\bra{j'}_B.
	\end{align}
	
	For every bipartite linear operator $X_{AB}$, we let
	\begin{equation}
		X_A\equiv\Tr_B[X_{AB}]\text{ and }X_B\equiv\Tr_A[X_{AB}]
	\end{equation}
	denote its partial traces. For states, we also use the terms \textit{marginal states} or \textit{reduced states} to refer to their partial traces.
	
	An immediate consequence of the Schmidt decomposition theorem is that the marginal states $\rho_A\coloneqq\Tr_B[\ket{\psi}\!\bra{\psi}_{AB}]$ and $\rho_B\coloneqq \Tr_A[\ket{\psi}\!\bra{\psi}_{AB}]$ of every pure state $\ket{\psi}\!\bra{\psi}_{AB}$ have the same non-zero eigenvalues. Indeed, using \eqref{eq-Schmidt_decomp_state_vector}, we find that
	 \begin{align}
	 	\rho_A&=\sum_{k,k'=1}^r \sqrt{\lambda_k\lambda_{k'}}\Tr_B[\ket{e_k}\!\bra{e_{k'}}_A\otimes \ket{f_k}\!\bra{f_{k'}}_B]\\
	 		&=\sum_{k,k'=1}^r \sqrt{\lambda_k\lambda_{k'}}\ket{e_k}\!\bra{e_{k'}}_A\delta_{k,k'}\\
	 		&=\sum_{k=1}^r \lambda_k\ket{e_k}\!\bra{e_k}_A, \label{eq-partial-trace-eigenvalues-pure-bistates-1}\\
	 	\text{ and \quad} \rho_B&=\sum_{k,k'=1}^r \sqrt{\lambda_k\lambda_{k'}}\Tr_A[\ket{e_k}\!\bra{e_{k'}}_A\otimes\ket{f_k}\!\bra{f_{k'}}_B]\\
	 		&=\sum_{k,k'=1}^r \sqrt{\lambda_k\lambda_{k'}}\delta_{k,k'}\ket{f_k}\!\bra{f_{k'}}_B\\
	 		&=\sum_{k=1}^r \lambda_k\ket{f_k}\!\bra{f_k}_B,
			\label{eq-partial-trace-eigenvalues-pure-bistates-2}
	 \end{align}
	 in which the equalities in \eqref{eq-partial-trace-eigenvalues-pure-bistates-1} and \eqref{eq-partial-trace-eigenvalues-pure-bistates-2} contain spectral decompositions of $\rho_A$ and $\rho_B$.
	 
	 \begin{exercise}{exer-partial_trace}
	 	Consider two quantum systems $A$ and $B$, with $d_A=d_B=d$.
	 	\begin{enumerate}[topsep=0.3cm]
	 		\item Calculate $\Tr_A[\ketbra{\Gamma}{\Gamma}_{AB}]$ and $\Tr_B[\ketbra{\Gamma}{\Gamma}_{AB}]$, where we recall from \eqref{eq-max_ent_vector} that $\ket{\Gamma}_{AB}=\sum_{j=0}^{d-1}\ket{j,j}_{AB}$.
	 		
	 		\item Calculate $\Tr_A[F_{AB}]$ and $\Tr_B[F_{AB}]$, where we recall from \eqref{eq-swap_op_standard_0} that $F_{AB}=\sum_{k,k'=0}^{d-1}\ketbra{k,k'}{k',k}_{AB}$.
	 	\end{enumerate}
	 \end{exercise}
	 
	 Below are two useful lemmas about how the support of a bipartite linear operator (recall the definition of support from Section~\ref{sec-math_tools-lin_ops}) relates to the support of its partial traces. Their proofs are somewhat technical, and so we provide them in Appendices~\ref{app-Q-SM-proof-tech-lem-1} and \ref{app-Q-SM-proof-tech-lem-2}.
	 
	 \begin{Lemma}{lem-app:support-1}
	 	Let $X_{AB}\in\Lin(\mathcal{H}_{A}\otimes\mathcal{H}_{B})$ be positive semi-definite, and let $X_{A}\coloneqq \Tr_{B}[X_{AB}]$ and $X_{B}\coloneqq \Tr_{A}[X_{AB}]$. Then $\supp(X_{AB})\subseteq\supp(X_{A})\otimes\supp(X_{B})$.
	\end{Lemma}

	\begin{Lemma}{lem-app:support-2}
		Let $X_{AB},Y_{AB}\in\Lin(\mathcal{H}_{A}\otimes\mathcal{H}_{B})$ be positive semi-definite, and suppose that $\supp(X_{AB})\subseteq\supp(Y_{AB})$. Then $\supp(X_{A})\subseteq\supp(Y_{A})$, where $X_{A}\coloneqq \Tr_{B}[X_{AB}]$ and $Y_{A}\coloneqq \Tr_{B}[Y_{AB}]$.
	\end{Lemma}

\subsection{Separable and Entangled States}\label{subsubsec-entanglement}

	The concepts of separable and entangled states are at the heart of virtually all of the communication protocols that we consider in this book. More generally, entanglement represents a key distinction between the classical and quantum theories of information; it simply is not present in classical information theory and therefore does not play a role there. Entanglement, in particular, is a key element of private communication and secure key distillation, and the successful distribution of entangled states among several spatially separated parties is a crucial ingredient in the implementation of such protocols over the future quantum internet. If the parties share only separable, unentangled states, then it is not possible for them to distill a key that is secure against a general quantum adversary.
	
	We begin this section by defining separable and entangled states.
	
	\begin{definition}{Separable and Entangled States}{def-sep_ent_state}
		A bipartite state $\sigma_{AB}$ is called \textit{separable} if there exists a finite alphabet $\mathcal{X}$, a probability distribution $p:\mathcal{X}\rightarrow[0,1]$ on $\mathcal{X}$, and sets $\{\omega_A^x\}_{x\in\mathcal{X}}$ and $\{\tau_B^x\}_{x\in\mathcal{X}}$ of states for $A$ and $B$, respectively, such that
		\begin{equation}\label{eq-sep_state}
			\sigma_{AB}=\sum_{x\in\mathcal{X}}p(x)\omega_A^x\otimes\tau_B^x.
		\end{equation}
		In other words, a state is called separable if it can be written as a convex combination of \textit{product states}, each of which has the form $\omega_A\otimes\tau_B$. The set of separable states on $\mathcal{H}_{AB}$ is denoted by $\SEP(A\!:\!B)$.\\[0.1in]
		A state that is not separable is called \textit{entangled}.
	\end{definition}
	
	\begin{remark}
		Note that a separable state can always be written in the form
		\begin{equation}
			\sigma_{AB}=\sum_{x'\in\mathcal{X}'}q(x')\ket{\psi_{x'}}\!\bra{\psi_{x'}}_A\otimes\ket{\phi_{x'}}\!\bra{\phi_{x'}}_B
	\end{equation}
		for some probability distribution $q:\mathcal{X}'\to[0,1]$ on a finite alphabet $\mathcal{X}'$ and sets of pure states $\{\ket{\psi_{x'}}\!\bra{\psi_{x'}}_A:x'\in\mathcal{X}'\}$, $\{\ket{\phi_{x'}}\!\bra{\phi_{x'}}_B:x'\in\mathcal{X}'\}$. In other words, separable states can always be written as a convex combination of pure product states. Indeed, from \eqref{eq-sep_state}, we can take spectral decompositions of $\omega_A^x$ and $\tau_B^x$,
		\begin{equation}
			\omega_A^x=\sum_{k=1}^{r_A^x} t_k^x\ket{e_k^x}\!\bra{e_k^x}_A,\quad \tau_B^x=\sum_{\ell=1}^{r_B^x}s_\ell^x\ket{f_\ell^x}\!\bra{f_\ell^x}_B,
		\end{equation}
		where $r_A^x=\rank(\omega_A^x)$ and $r_B^x=\rank(\tau_B^x)$, so that
		\begin{equation}\label{eq-sep_state_pure_0}
			\rho_{AB}=\sum_{x\in\mathcal{X}}\sum_{k=1}^{r_A^x}\sum_{\ell=1}^{r_B^x}p(x)t_k^x s_\ell^x \ket{e_k^x}\!\bra{e_k^x}_A\otimes\ket{f_\ell^x}\!\bra{f_\ell^x}_B.
		\end{equation}
		Then, define the alphabet $\mathcal{X}'=\{x'\coloneqq (x,k,\ell):x\in\mathcal{X},~1\leq k\leq r_A^x,~1\leq\ell\leq r_B^x\}$, so that $x'$ is a superindex, and the unit vectors
		\begin{equation}
			\ket{\psi_{x'}}_A\coloneqq \ket{e_k^x}_A,\quad \ket{\phi_{x'}}_B\coloneqq \ket{f_\ell^x}_B.
		\end{equation}
		Also, define the probability distribution $q:\mathcal{X}'\to[0,1]$ by
		\begin{equation}
			q(x,k,\ell)=p(x)t_k^x s_\ell^x.
		\end{equation}
		Therefore, \eqref{eq-sep_state_pure_0} can be written as
		\begin{equation}\label{eq-sep_state_pure}
			\sigma_{AB}=\sum_{x'\in\mathcal{X}'} q(x')\ket{\psi_{x'}}\!\bra{\psi_{x'}}_A\otimes\ket{\phi_{x'}}\!\bra{\phi_{x'}}_B.
		\end{equation}
		From the development above, it follows that the set of separable states is the convex hull of the set of pure product states. By an application of the Fenchel--Eggleston--Carath\'{e}odory Theorem (Theorem~\ref{thm-Caratheodory}), it follows that $\sigma_{AB}$ can be written as a convex combination of no more than $\rank(\sigma_{AB})^2$ pure product states. Indeed, in Theorem~\ref{thm-Caratheodory}, it suffices to take as $S$ the set of pure product states of the form $\ketbra{\psi}{\psi}_A \otimes \ketbra{\phi}{\phi}_B$, where $\ket{\psi}_A \otimes \ket{\phi}_B\in \supp(\sigma_{AB})$. The relevant real (affine) space is the space of bipartite Hermitian operators with support contained in $\supp(\sigma_{AB})$ and with trace one, which has real dimension $\rank(\sigma_{AB})^2-1$.
	\end{remark}
	
	In the sense that follows, bipartite separable states can be thought of as exhibiting purely classical correlations between the two parties, Alice and Bob. Suppose that Alice draws $x$ with  probability $p(x)$, prepares her system in the state $\omega_A^x$, sends $x$ to Bob over a classical channel, who then prepares his system in the state $\tau_B^x$, where $x\in\mathcal{X}$ and $\mathcal{X}$ is a finite alphabet. This procedure corresponds to preparing the ensemble $\{(p(x),\omega_A^x\otimes\tau_B^x)\}_{x\in\mathcal{X}}$, and if Alice and Bob discard the label $x$, then their shared joint state is the separable state $\sigma_{AB}=\sum_{x\in\mathcal{X}}p(x)\omega_A^x\otimes\tau_B^x$.
	
	On the other hand, no such procedure consisting of only local operations by Alice and Bob, supplemented by classical communication between them, can ever be used to generate an entangled state between them (without them already sharing some entanglement beforehand). In Section~\ref{subsec-LOCC_channels}, we introduce local operations and classical communication (LOCC) channels and explain this point in more detail. Essentially, two entangled quantum systems are intrinsically linked in such a way that it is insufficient to  describe each one individually. 
	
	Observe that a pure state $\psi_{AB}$ is separable if and only if it is a product state, i.e., if and only if there exist pure states $\phi_A$ and $\varphi_B$ such that $\psi_{AB}=\phi_A\otimes\varphi_B$. Recalling that every pure state has a Schmidt decomposition (see Theorem~\ref{thm-Schmidt}), we obtain the following result:
	\begin{specbox}{0.7\textwidth}
		A pure state is entangled if and only if its Schmidt rank is strictly greater than one.
	\end{specbox}
	
	An important example of an entangled pure state is the state $\Phi_{AB}$ on two $d$-dimensional systems $A$ and $B$, defined as
	$\Phi_{AB} \coloneqq \ket{\Phi}\!\bra{\Phi}_{AB}$, where
	\begin{equation}\label{eq-max_ent_state}
		\ket{\Phi}_{AB}\coloneqq\frac{1}{\sqrt{d}}\sum_{i=0}^{d-1}\ket{i}_A\otimes\ket{i}_B=\frac{1}{\sqrt{d}}\ket{\Gamma}_{AB},
	\end{equation}
	and $\ket{\Gamma}_{AB}=\sum_{i=0}^{d-1}\ket{i}_A\otimes\ket{i}_B$ is the vector defined in \eqref{eq-max_ent_vector}. 
	
	\begin{exercise}{exer-max_ent_marginals}
		\begin{enumerate}
		    \item Show that $\Tr_A[\Phi_{AB}]=\Tr_B[\Phi_{AB}]=\frac{\mathbbm{1}_d}{d}$.
		    \item Define $\Phi_{AB}^U \coloneqq (U_A \otimes \mathbbm{1}_B)\Phi_{AB}(U_A \otimes \mathbbm{1}_B)^\dag$, for $U_A$ a unitary acting on system~$A$. Show that $\Tr_A[\Phi^U_{AB}]=\Tr_B[\Phi^U_{AB}]=\frac{\mathbbm{1}_d}{d}$.
		\end{enumerate}
	\end{exercise}
	
	The state $\Phi_{AB}$ is an example of a maximally entangled state.
	
	\begin{definition}{Maximally Entangled Pure State}{def-max_ent_pure_state}
		A pure state $\psi_{AB}=\ketbra{\psi}{\psi}_{AB}$, for two systems $A$ and $B$ of the same dimension $d$, is called \textit{maximally entangled} if the Schmidt coefficients of $\ket{\psi}_{AB}$ are all equal to $\frac{1}{\sqrt{d}}$, with $d$ being the Schmidt rank of $\ket{\psi}_{AB}$.
	\end{definition}
		
	In other words, $\psi_{AB}$ is called maximally entangled if $\ket{\psi}_{AB}$ has the Schmidt decomposition
	\begin{equation}
		\ket{\psi}_{AB}=\frac{1}{\sqrt{d}}\sum_{k=1}^d \ket{e_k}_A\otimes\ket{f_k}_B
        \label{eq-QM-states-meas:max-ent-schmidt}
	\end{equation}
	for some orthonormal sets $\{\ket{e_k}_A:1\leq k\leq d\}$ and $\{\ket{f_k}_B:1\leq k\leq d\}$. Observe then that
	\begin{equation}
		\Tr_A[\ket{\psi}\!\bra{\psi}_{AB}]=\frac{\mathbbm{1}_d}{d}=\Tr_B[\ket{\psi}\!\bra{\psi}_{AB}].
	\end{equation}
	In other words, like the state $\Phi_{AB}$, the marginal states of every maximally entangled state are maximally mixed for $A$ and $B$. There is a precise sense in which the state $\ket{\psi}_{AB}$ in \eqref{eq-QM-states-meas:max-ent-schmidt} is maximally entangled (i.e., the most entangled state of the $d$-dimensional systems $A$ and $B$), and this point is discussed further in Chapter~\ref{chap-ent_measures}.
	
	Maximally entangled states provide a good example of why entangled quantum systems are, in a sense, greater than the sum of their parts. Since maximally entangled states have maximally mixed marginal states, the individual quantum systems  
    can be viewed as being in a completely random state; indeed, 
    as we have seen, maximally mixed states can be written as the expected state of every ensemble of orthonormal pure states with uniform probability distribution. However, intriguingly, the overall composite system is in a pure, definite state.
	
	\begin{exercise}{exer-maximally_ent_states}
		Prove that every state vector of the form $(\mathbbm{1}_d\otimes U)\ket{\Phi}=\frac{1}{\sqrt{d}}\text{vec}(U)$ and $(U\otimes\mathbbm{1}_d)\ket{\Phi}$, with $d\geq 2$ and $U$ a unitary operator, corresponds to a maximally entangled pure state. Conversely, given a maximally entangled pure state $\ketbra{\psi}{\psi}_{AB}$, prove that there exists a unitary $U_A$ such that $\ket{\psi}_{AB}=(U_A\otimes\mathbbm{1}_B)\ket{\Phi}_{AB}$.
	\end{exercise}

\subsection{Bell States}

	The conclusion of Exercise~\ref{exer-maximally_ent_states} is that every state vector of the form $(\mathbbm{1}\otimes U)\ket{\Phi}$ and $(U\otimes\mathbbm{1})\ket{\Phi}$, with $U$ a unitary operator, is a maximally entangled state. We now provide an important example of a class of maximally entangled states, known as Bell states, for every dimension $d\geq 2$. These states are defined by particular choices for the unitary $U$. The Bell states are essential in many quantum information processing tasks, most prominently quantum teleportation and super-dense coding, which we discuss in Chapter~\ref{chap-QM_protocols}.
	
	We start with dimension $d=2$. Recall the Pauli operators $X$ and $Z$ from \eqref{eq-QM-Pauli_mat}:
	\begin{equation}\label{eq-QM-Pauli_mat_XZ}
		X=\begin{pmatrix} 0 & 1 \\ 1 & 0 \end{pmatrix},\quad Z=\begin{pmatrix} 1 & 0 \\ 0 & 1 \end{pmatrix}.
	\end{equation}
	Observe that, in addition to being Hermitian, these operators are unitary, which is due to the fact that $X^2=Z^2=\mathbbm{1}$. The operator $Y$ defined in \eqref{eq-QM-Pauli_mat} is also unitary, since $Y^2=\mathbbm{1}$, from which it follows that the operator $ZX=\I Y$ is also unitary. Using the operators $X$, $Z$, and $ZX$, we define the following set of four entangled, two-qubit state vectors:
	\begin{equation}\label{eq-two_qubit_Bell_states}
		\ket{\Phi_{z,x}}\coloneqq(Z^z X^x\otimes\mathbbm{1})\ket{\Phi},
	\end{equation}
	for $z,x\in\{0,1\}$, where we recall from \eqref{eq-max_ent_state} that $\ket{\Phi}\coloneqq\frac{1}{\sqrt{2}}(\ket{0,0}+\ket{1,1})$. The corresponding density operators $\Phi_{z,x}$ are known as the \textit{two-qubit Bell states}. The following notation is commonly used:
	\begin{align}
		\ket{\Phi^+}&\equiv\ket{\Phi_{0,0}}=\frac{1}{\sqrt{2}}(\ket{0,0}+\ket{1,1}),\label{eq-two_qubit_Bell_00}\\
		\ket{\Phi^-}&\equiv\ket{\Phi_{1,0}}=\frac{1}{\sqrt{2}}(\ket{0,0}-\ket{1,1}),\label{eq-two_qubit_Bell_10}\\
		\ket{\Psi^+}&\equiv\ket{\Phi_{0,1}}=\frac{1}{\sqrt{2}}(\ket{0,1}+\ket{1,0}),\label{eq-two_qubit_Bell_01}\\
		\ket{\Psi^-}&\equiv\ket{\Phi_{1,1}}=\frac{1}{\sqrt{2}}(\ket{0,1}-\ket{1,0}).\label{eq-two_qubit_Bell_11}
	\end{align}
	
	\begin{exercise}{exer-two_qubit_Bell_ONB}
		\begin{enumerate}
			\item Prove that the two-qubit Bell state vectors defined in \eqref{eq-two_qubit_Bell_states} form an orthonormal basis for $\mathbb{C}^2\otimes\mathbb{C}^2$.
			
			\item Prove that the state vectors $\ket{\Phi^+}$, $\ket{\Phi^-}$, and $\ket{\Psi^+}$ form an orthonormal basis for $\text{Sym}_2(\mathbb{C}^2)$. (\textit{Hint}: See \eqref{eq:math-tools:occupation-num-basis} and Exercise~\ref{exer-Sym_n2}.) For this reason, the subspace $\text{Sym}_2(\mathbb{C}^2)$ is called the triplet subspace. 
			
			\item Prove that $\text{ASym}_2(\mathbb{C}^2) =  \Span\{\ket{\Psi^-}\}$. For this reason, the subspace $\text{ASym}_2(\mathbb{C}^2)$ is called the singlet subspace and the state $\ket{\Psi^-}$ is called the singlet state vector.
		\end{enumerate}
	\end{exercise}
	
	We can generalize the Bell state vectors in \eqref{eq-two_qubit_Bell_states} to systems with dimension $d>2$. Doing so requires a generalization of the qubit Pauli operators $X$ and $Z$ to unitary operators for qudits\footnote{The qudit operators defined in \eqref{eq-su_generators_0}--\eqref{eq:math-tools:gell-mann-scaled-mats} represent one generalization of the qubit Pauli operators. Although they are Hermitian, they are not generally unitary. What we require here is a generalization to qudit operators that are unitary.}.
	
	\begin{definition}{Heisenberg--Weyl Operators}{def-heisenberg-weyl}
		Let $d\geq 2$. The \textit{Heisenberg--Weyl operators} make up a set $\{W_{z,x}:0\leq z,x\leq d-1\}$ of $d^2$ unitary operators acting on $\mathbb{C}^d$. They are defined as follows:
		\begin{align}
			W_{z,x} & =Z(z)X(x), \label{eq-Heisenberg_Weyl_operators}\\
			Z(z) & \coloneqq\sum_{k=0}^{d-1}\e^{\frac{2\pi\I kz}{d}}\ket{k}\!\bra{k},\label{eq-gen_Z_Pauli}\\
			X(x) & \coloneqq\sum_{k=0}^{d-1}\ket{k+x}\!\bra{k},\label{eq-gen_X_pauli}
		\end{align}
		where the addition operation in the definition of $X(x)$ is performed modulo $d$.
	\end{definition}
	
	\begin{exercise}{exer-Heisenberg_Weyl_basic_1}
		\begin{enumerate}
			\item Verify that when $d=2$, the Heisenberg--Weyl operators reduce to the qubit Pauli operators $Z$, $X$, and $ZX$.
			\item Prove that the operators $Z(z)$ and $X(x)$ defined in \eqref{eq-gen_Z_Pauli} and \eqref{eq-gen_X_pauli} satisfy the commutation relation
				\begin{equation}\label{eq-Heisenberg_Weyl_prop1}
					Z(z)X(x)=\e^{\frac{2\pi\I xz}{d}}X(x)Z(z),
				\end{equation}
				for all $z,x\in\{0,1,\dotsc,d-1\}$.
		\end{enumerate}
	\end{exercise}
	
	The Heisenberg--Weyl operators are unitary, just like the Pauli operators; however, unlike the Pauli operators, they are \textit{not} Hermitian. In particular,
	\begin{equation}\label{eq-Heisenberg_Weyl_prop2}
		W_{z,x}^\dagger =\e^{-\frac{2\pi\I xz}{d}}W_{-z,-x}.
	\end{equation}
	It is also straightforward to show that
	\begin{equation}\label{eq-Heisenberg_Weyl_prop3}
		W_{z_1,x_1}W_{z_2,x_2}=\e^{-\frac{2\pi\I x_1z_2}{d}}W_{z_1+z_2,x_1+x_2}.
	\end{equation}
	Furthermore, the Heisenberg--Weyl operators are orthogonal with respect to the Hilbert--Schmidt inner product, meaning that
	\begin{equation}\label{eq:QM-over:HW-orthogonal}
		\inner{W_{z_1,x_1}}{W_{z_2,x_2}}=\Tr[W_{z_1,x_1}^\dagger W_{z_2,x_2}]=d\delta_{z_1,z_2}\delta_{x_1,x_2}
	\end{equation}
	for all $0\leq z_1,z_2,x_1,x_2\leq d-1$. This implies that the scaled Heisenberg--Weyl operators $\left\{\frac{1}{\sqrt{d}}W_{z,x}:0\leq z,x\leq d-1\right\}$ form an orthonormal basis for $\Lin(\mathbb{C}^d)$ for all $d\geq 2$.
	
	\begin{exercise}{exer-Heisenberg_Weyl_properties_basic}
		Prove \eqref{eq-Heisenberg_Weyl_prop2}, \eqref{eq-Heisenberg_Weyl_prop3}, and \eqref{eq:QM-over:HW-orthogonal}.
	\end{exercise}
	
	\begin{exercise}{exer-QFT}
		Let $d\geq 2$, and consider the operator $Q_d$ defined as
		\begin{equation}
			Q_d\coloneqq\frac{1}{\sqrt{d}}\sum_{k,\ell=0}^{d-1} \e^{\frac{2\pi\I k\ell}{d}}\ketbra{k}{\ell}.
		\end{equation}
		\begin{enumerate}[topsep=0.3cm]
			\item Show that $Q_d$ is a unitary operator.
			
			\item Prove that
				\begin{align}
					Q_dX(x)Q_d^{\dagger}&=Z(x),\\
					Q_dZ(z)Q_d^{\dagger}&=X(z)^{\dagger},
				\end{align}
				for all $z,x\in\{0,1,\dotsc,d-1\}$.
		\end{enumerate}
		The unitary operator $Q_d$ is known as the \textit{(discrete) quantum Fourier transform} operator.
	\end{exercise}
	
	Using the Heisenberg--Weyl operators, we now define the set of qudit Bell states in a manner analogous to \eqref{eq-two_qubit_Bell_states}.
	
	\begin{definition}{Qudit Bell States}{def-qudit_Bell_states}
		Let $d\geq 2$. The \textit{qudit Bell states} are $d^2$ pure quantum states $\Phi_{z,x}\coloneqq\ketbra{\Phi_{z,x}}{\Phi_{z,x}}$ in $\Density(\mathbb{C}^d\otimes\mathbb{C}^d)$, where
		\begin{equation}\label{eq-qudit_Bell}
			\ket{\Phi_{z,x}}\coloneqq (W_{z,x}\otimes \mathbbm{1}_d)\ket{\Phi_d}
		\end{equation}
		for all $z,x\in\{0,1,\dotsc,d-1\}$, and $\ket{\Phi_d}=\frac{1}{\sqrt{d}}\sum_{j=0}^{d-1}\ket{j,j}$.
	\end{definition}
	
	\begin{exercise}{exer-two_qudit_Bell_ONB}
		Prove that the two-qudit Bell state vectors defined in \eqref{eq-qudit_Bell} form an orthonormal basis for $\mathbb{C}^d\otimes\mathbb{C}^d$ for all $d\geq 2$.
	\end{exercise}
	
	The fact that the two-qubit Bell state vectors form an orthonormal basis for $\mathbb{C}^d\otimes\mathbb{C}^d$ implies, from Exercise~\ref{exer-identity_operator}, that
	\begin{equation}\label{eq-qudit_Bell_ONB_identity}
		\sum_{z,x=0}^{d-1}\ketbra{\Phi_{z,x}}{\Phi_{z,x}}=\mathbbm{1}_d\otimes\mathbbm{1}_d.
	\end{equation}
	
	A two-qudit state $\rho_{AB}$ is known as a \textit{Bell-diagonal state} if it is diagonal in the two-qudit Bell basis, so that it is of the form
	\begin{equation}\label{eq-Bell_diag_state}
		\sum_{z,x=0}^{d-1} p(z,x) \ketbra{\Phi_{z,x}}{\Phi_{z,x}}
	\end{equation}
	for some probability distribution $p:\{0,1,\dotsc,d-1\}^2\to [0,1]$, meaning that $0\leq p(z,x)\leq 1$ for all $z,x\in\{0,1,\dotsc,d-1\}$ and $\sum_{z,x=0}^{d-1}p(z,x)=1$.

\subsection{Purifications and Extensions}\label{sec:qm:purification-def}

	One of the most useful concepts in quantum information is the notion of purification. There is no strong classical analogue of this concept, and thus this notion represents another distinction between the classical and quantum theories of information. 
	
	\begin{definition}{Purification}{def-purification}
		Let $\rho_A$ be a state of a system $A$. A \textit{purification} of $\rho_A$ is a pure state $\ket{\psi}\!\bra{\psi}_{RA}$ for a bipartite system $RA$ such that
		\begin{equation}
			\Tr_R[\ket{\psi}\!\bra{\psi}_{RA}]=\rho_A.
		\end{equation}
	\end{definition}
	
	We often call the reference system $R$ the ``purifying system.'' 
	
	The following simple theorem establishes that every state $\rho_A$ has a purification.
	
	\begin{theorem*}{State Purification}{thm-purification}
		For every state $\rho_A$ and every reference system $R$ with $d_R\geq \rank(\rho_A)$, there exists a purification $\ket{\psi}\!\bra{\psi}_{RA}$ of $\rho_A$ on $RA$. 
	\end{theorem*}
	
	\begin{Proof}
		Consider a spectral decomposition of $\rho_A$,
		\begin{equation}
			\rho_A=\sum_{k=1}^r \lambda_k\ket{\phi_k}\!\bra{\phi_k},
		\end{equation}
		where $r=\rank(\rho_A)$. Consider a reference system $R$ with $d_R\geq r$ and an arbitrary set $\{\ket{\varphi_k}_R:1\leq k\leq r\}$ of orthonormal states. The unit vector
		\begin{equation}\label{eq-state_purification}
			\ket{\psi}_{RA}\coloneqq \sum_{k=1}^r\sqrt{\lambda_k}\ket{\varphi_k}_R\otimes\ket{\phi_k}_A
		\end{equation}
		then satisfies
		\begin{align}
			\Tr_R[\ket{\psi}\!\bra{\psi}_{RA}]&=\sum_{k,k'=1}^r \sqrt{\lambda_k\lambda_{k'}}\underbrace{\Tr[\ket{\varphi_k}\!\bra{\varphi_{k'}}_R]}_{\delta_{k,k'}}\ket{\phi_k}\!\bra{\phi_{k'}}_A\\
			&=\sum_{k=1}^r\lambda_k\ket{\phi_k}\!\bra{\phi_k}_A\\
			&=\rho_A,
		\end{align}
		so that $\ket{\psi}\!\bra{\psi}_{RA}$ is a purification of $\rho_A$, as required.
	\end{Proof}
	
	\begin{remark}
		The theorem above states that the condition $d_R\geq\rank(\rho_{A})$ on the dimension of the purifying system $R$ is sufficient to guarantee the existence of a purification. This condition is also necessary, meaning that it is not possible to have a purifying system whose dimension is less than the rank of $\rho_{A}$. 
	\end{remark}
	
	The proof of the theorem above not only tells us that every state has a purification, but it also tells us explicitly how to construct one such purification. We can also construct a purification of every state $\rho_A$ as follows:
	\begin{equation}\label{eq-canonical_purification}
		\ket{\psi}_{RA}=(\mathbbm{1}_R\otimes\sqrt{\rho_A})\ket{\Gamma}_{RA}=\mathrm{vec}(\!\sqrt{\rho_A}),
	\end{equation}
	where $\ket{\Gamma}_{RA}=\sum_{i=0}^{d_A-1}\ket{i,i}_{RA}$ and the operation $\text{vec}$ is defined in \eqref{eq-vec_operation}. We often call the state $\ket{\psi}\!\bra{\psi}_{RA}$ the \textit{canonical purification} of $\rho_A$. Note that the canonical purification is very closely related to the purification used in the proof of Theorem~\ref{thm-purification}. Indeed, if
	\begin{equation}
		\rho_A=\sum_{k=1}^r \lambda_k\ket{\phi_k}\!\bra{\phi_k}
	\end{equation}
	is a spectral decomposition of $\rho_A$, with $r=\rank(\rho_A)$, then
	\begin{equation}
		\text{vec}(\!\sqrt{\rho_A})=\sum_{k=1}^r \sqrt{\lambda_k}~\conj{\ket{\phi_k}}_R\otimes \ket{\phi_k}_A,
	\end{equation}
	where we have made use of \eqref{eq-vec_operation_vectors}.
	
	Physically, the fact that every state $\rho_A$ has a purification means that every quantum system $A$ in a mixed state can be viewed as being entangled with \textit{some} system $R$ to which we do not have access, such that the global state is a pure state $\ket{\psi}\!\bra{\psi}_{RA}$. Since we do not have access to $R$, our description of the state of system $A$ has to be as the partial trace of $\ket{\psi}\!\bra{\psi}_{RA}$ over $R$, i.e., by $\rho_A$. 
	
	Observe that if the state $\rho_A$ is pure, i.e., if $\rho_A=\ket{\phi}\!\bra{\phi}_A$, then the only possible purification of it is of the form
	\begin{equation}
	\ket{\psi}\!\bra{\psi}_{RA}=\ket{\varphi}\!\bra{\varphi}_R\otimes\ket{\phi}\!\bra{\phi}_A,
	\label{eq-Q-SM:purification-pure-states}
	\end{equation}
	with $\ket{\varphi}\!\bra{\varphi}_R$ a pure state of the system $R$. In other words, purifications of pure states can only be \textit{pure product states}. Somewhat technically, according to Theorem~\ref{thm-purification}, the dimension of system $R$ need only satisfy $d_R \geq \rank(\rho_A)$. In the case of a pure state, the rank is equal to one, so that the reference system can be a trivial system of dimension one. Thus, in this technical sense, pure states already purify themselves. 
	
	Purifications of states are not unique. In fact, given a state $\rho_A$ and a purification $\ket{\psi}\!\bra{\psi}_{RA}$ of $\rho_A$ as in \eqref{eq-state_purification}, let $V_{R\to R'}$ be an isometry (i.e., a linear operator satisfying $V^\dagger V=\mathbbm{1}_R$) acting on the $R$ system. Defining 
	\begin{equation}\label{eq-purif_iso_equiv_1}
		\ket{\psi'}_{R'A}=(V_{R\to R'}\otimes\mathbbm{1}_A)\ket{\psi}_{RA},
	\end{equation}
	we find that
	\begin{align}
		\Tr_{R'}[\ket{\psi'}\!\bra{\psi'}_{R'A}]&=\sum_{k,k'=1}^r \sqrt{\lambda_k\lambda_{k'}}\Tr[V\ket{\varphi_k}\!\bra{\varphi_{k'}}_R V^\dagger]\ket{\phi_k}\!\bra{\phi_k}_A\label{eq-purif_iso_equiv_2}\\
		&=\sum_{k=1}^r\lambda_k\ket{\phi_k}\!\bra{\phi_k}_A\label{eq-purif_iso_equiv_3}\\
		&=\rho_A,\label{eq-purif_iso_equiv_4}
	\end{align}
	where we conclude that $\Tr[V\ket{\varphi_k}\!\bra{\varphi_{k'}}V^\dagger]=\delta_{k,k'}$ from cyclicity of the trace and $V^\dagger V=\mathbbm{1}_R$. So $\ket{\psi'}\!\bra{\psi'}_{R'A}$ is also a purification of $\rho_A$.
	
	A converse statement holds as well by employing the Schmidt decomposition (Theorem~\ref{thm-Schmidt}): if $\ket{\psi}\!\bra{\psi}_{RA}$ and $\ket{\psi'}\!\bra{\psi'}_{R'A}$ are two purifications of the state~$\rho_A$, then they are related by an isometry taking one reference system to the other. (Either the isometry takes $R$ to $R'$, if the dimension of $R'$ is larger than that of $R$, or vice versa; if the dimensions are equal then the isometry will be a unitary.) By combining this statement and the previous one, we can thus say that purifications are unique ``up to isometries acting on the reference system.''

	A purification is an example of an ``extension'' of a quantum state.

	\begin{definition}{Extension}{def:qm:extension-of-state}
		An extension of a quantum state $\rho_A$ is a state $\omega_{RA}$ satisfying $\Tr_R[\omega_{RA}]=\rho_A$, where $R$ is a reference system. 
	\end{definition}
	
	\begin{remark}
		In order for a purification to exist, it is necessary that $d_R \geq \rank(\rho_A)$. However, there is no such requirement for an extension.
	\end{remark}
	
	Note that if the state $\rho_A$ is pure, i.e., if $\rho_A=\ket{\phi}\!\bra{\phi}_A$, then every extension $\omega_{RA}$ of $\rho_A$ must be a product state, i.e., we must have $\omega_{RA}=\sigma_R\otimes\ket{\phi}\!\bra{\phi}_A$ for some state~$\sigma_R$.

\subsection{Multipartite States and Permutations}\label{sec-multipartite_permutations}

	A multipartite quantum state is a quantum state of more than two quantum systems. Let $A_1,\dotsc,A_n$ denote $n\geq 2$ quantum systems. Then, every quantum state $\rho_{A_1\dotsb A_n}$ can be represented as 
	\begin{equation}
		\rho_{A_1\dotsb A_n}=\sum_{i_1,i_1'=0}^{d_{A_1}-1}\dotsb\sum_{i_n,i_n'=0}^{d_{A_n}-1}\beta_{i_1,\dotsc,i_n;i_1',\dotsc,i_n'} \ketbra{i_1,\dotsc,i_n}{i_1',\dotsc,i_n'}_{A_1\dotsb A_n},
	\end{equation}
	where $\beta_{i_1,\dotsc,i_n;i_1',\dotsc,i_n'}=\bra{i_1,\dotsc,i_n}\rho_{A_1\dotsb A_n}\ket{i_1',\dotsc,i_n'}$. This representation is simply the generalization of the representation in \eqref{eq-bipartite_mixed_state} to $n\geq 2$ quantum systems. Similarly, the generalization of the representation in \eqref{eq-bipartite_q_state_coh_vec_repr} to $n\geq 2$ quantum systems is
	\begin{equation}
		\rho_{A_1\dotsb A_n}=\frac{1}{d_{A_1}\dotsb d_{A_n}}\sum_{k_1=0}^{d_{A_1}^2-1}\dotsb\sum_{k_n=0}^{d_{A_n}^2-1} r_{k_1,\dotsc,k_n} S^{k_1}_{A_1}\otimes\dotsb\otimes S^{k_n}_{A_n}.
	\end{equation}
	
	If the quantum systems $A_1,A_2,\dotsc,A_n$ are identical, meaning that the Hilbert spaces $\mathcal{H}_{A_1},\dotsc,\mathcal{H}_{A_n}$ are isomorphic to each other, so that $d_{A_1}=d_{A_2}=\dotsb=d_{A_n}$, then we can identify them with a single quantum system $A$ of dimension~$d_A$. In this case, for brevity, we often write $\rho_{A^n}\equiv \rho_{A_1\dotsb A_n}$ to denote a quantum state for the $n$ identical systems. For the remainder of this section, we assume that the systems $A_1,A_2,\dotsc,A_n$ are identical.
	
	Unlike for bipartite systems, the entanglement of multipartite systems is less straightforward to define, because there are different notions of separability that one can define. A discussion of these different notions of separability for multipartite quantum systems is beyond of the scope of this book. Please see the Bibliographic Notes (Section~\ref{sec:qm:bib-notes}) for references.
	
	An important consideration for a collection of identical quantum systems is permutations. For example, many quantum systems, such as bosons and fermions, have quantum states that are symmetric and anti-symmetric, respectively, under permutation of the individual systems. This is due to the fact that bosons and fermions are identical particles, and as such they are fundamentally indistinguishable. (See the Bibliographic Notes in Section~\ref{sec:qm:bib-notes} for more information about the quantum theory of bosons and fermions.) For our purposes, in quantum information, permutations are a useful tool for establishing certain information quantities as upper bounds on the rates of some quantum communication tasks.
	
	Recall that we discussed the notion of permutations in Section~\ref{sec-symm_subspace}. Specifically, in \eqref{eq-permutation_rep}, we defined a unitary operator $W_{A^n}^{\pi}$ acting on $\mathcal{H}_A^{\otimes n}$, for every permutation $\pi\in\mathcal{S}_n$, as follows:
	\begin{equation}\label{eq-permutation_rep_QM}
        W_{A^n}^{\pi}\ket{i_1,i_2,\dotsc,i_n}_{A^n}=\ket{i_{\pi^{-1}(1)},i_{\pi^{-1}(2)},\dotsc,i_{\pi^{-1}(n)}}_{A^n},
	\end{equation}
	for all $0\leq i_1,i_2,\dotsc,i_n\leq d-1$. Physically, the operators $W_{A^n}^{\pi}$ correspond to permuting the states of the (identical) systems $A_1,A_2,\dotsc,A_n$. As an example, consider $n$ quantum states $\rho^1,\rho^2,\dotsc,\rho^n\in\Density(\mathcal{H}_A)$. Then, for every permutation $\pi\in\mathcal{S}_n$,
	\begin{equation}\label{eq-permutation_product_state}
		W_{A^n}^{\pi}\left(\rho_{A_1}^1\otimes\rho_{A_2}^2\dotsb\otimes\rho_{A_n}^n\right)W_{A^n}^{\pi\dagger}=\rho_{A_1}^{\pi^{-1}(1)}\otimes\rho_{A_2}^{\pi^{-1}(2)}\dotsb\otimes\rho_{A_n}^{\pi^{-1}(n)},
	\end{equation}
	which follows straightforwardly from the definition in \eqref{eq-permutation_rep_QM}. 
	
	
	\begin{exercise}{exer-permutation_product_state}
		\begin{enumerate}
			\item Verify \eqref{eq-permutation_product_state}.
			
			\item Let $n\geq 2$, and consider the cyclic permutation 
            $\pi$ such that $\pi(i)=i+1$ for all $i\in\{1,2,\dotsc,n-1\}$, and $\pi(n)=1$. Prove that for all quantum states $\rho_{A_1}^1,\,\rho_{A_2}^2,\dotsc,\,\rho_{A_n}^n$, with $A_1,\,A_2,\dotsc,\,A_n$ being identical quantum systems,
				\begin{equation}\label{eq-trace_cyclic_perm}
					\Tr[W_{A^n}^{\pi}(\rho_{A_1}^1\otimes\rho_{A_2}^2\otimes\dotsb\otimes\rho_{A_{n-1}}^{n-1} \otimes\rho_{A_n}^n)]=\Tr[\rho_{A_n}^n\rho_{A_{n-1}}^{n-1} \dotsb \rho_{A_2}^2 \rho_{A_1}^1].
				\end{equation}			
		\end{enumerate}
	\end{exercise}

	If, in \eqref{eq-permutation_product_state}, we have that $\rho^1=\rho^2=\dotsb=\rho^n=\rho$, then the state $\rho^{\otimes n}$ is invariant under every permutation, i.e.,
	\begin{equation}
		W_{A^n}^{\pi}\rho_A^{\otimes n}W_{A^n}^{\pi\dagger}=\rho_A^{\otimes n}
	\end{equation}
	for all $\pi\in\mathcal{S}_n$.

	\begin{definition}{Permutation-Invariant State}{def-perm_invar}
		A state $\rho\in\Density(\mathcal{H}^{\otimes n})$ is called \textit{permutation invariant} if
		\begin{equation}\label{eq-state_perm_invar}
			\rho = W^{\pi}\rho W^{\pi\dagger}
		\end{equation}
		for every permutation $\pi\in\mathcal{S}_n$, where the unitary permutation operator $W^{\pi}$ is defined in \eqref{eq-permutation_rep_QM}.
	\end{definition}
	
	\begin{remark}
		Note that the permutation-invariance condition in \eqref{eq-state_perm_invar} does \textit{not} imply that the state $\rho$ is supported on the symmetric subspace $\text{Sym}_n(\mathcal{H})$ of $\mathcal{H}^{\otimes n}$. In other words, the condition in \eqref{eq-state_perm_invar} does not imply that
		\begin{equation}
			\Pi_{\text{Sym}_n(\mathcal{H})}\rho\Pi_{\text{Sym}_n(\mathcal{H})}=\rho.
		\end{equation}
		As a simple example, suppose that $\mathcal{H}=\mathbb{C}^2$, and let $\rho=\ket{\Psi^-}\!\bra{\Psi^-}$, where $\ket{\Psi^-}=\frac{1}{\sqrt{2}}(\ket{0,1}-\ket{1,0})$ is the two-qubit Bell state defined in \eqref{eq-two_qubit_Bell_11}. Then, it is easy to see that $W^{\pi}\rho W^{\pi\dagger}=\rho$ for all $\pi\in\mathcal{S}_2$, while $\Pi_{\text{Sym}_2(\mathcal{H})}\rho\Pi_{\text{Sym}_2(\mathcal{H})}=0$. The latter is true because $\ket{\Psi^-}$ is an \textit{anti-symmetric state}, i.e., $\ket{\Psi^-}\in\text{ASym}_2(\mathcal{H})$. The state $\rho$ is thus supported on the anti-symmetric subspace, even though it is permutation invariant.
	\end{remark}
	
	\begin{exercise}{exer-k_ext_states}
		Let $\rho_{AB}=\sum_{x\in\mathcal{X}}^n p(x)\sigma_A^x\otimes\tau_B^x$, where $\mathcal{X}$ is a finite alphabet, $p:\mathcal{X}\to[0,1]$ is a probability distribution, and $\{\sigma_A^x\}_{x\in\mathcal{X}}$, $\{\tau_B^x\}_{x\in\mathcal{X}}$ are sets of quantum states.
		\begin{enumerate}[topsep=0.3cm]
			\item  Prove that $\widetilde{\rho}_{ABB'}\coloneqq\sum_{x\in\mathcal{X}}p(x)\sigma_A^x\otimes\tau_B^x\otimes\tau_{B'}^x$ is an extension of $\rho_{AB}$, with $B'$ being the reference system, in accordance with Definition~\ref{def:qm:extension-of-state}, such that $d_{B'}=d_B$. Prove also that $\widetilde{\rho}_{AB'}\coloneqq\Tr_B[\widetilde{\rho}_{ABB'}]=\rho_{AB}$.
			
			\item Now, let $\widetilde{\rho}_{AB_1B_2\dotsb B_k}\coloneqq\sum_{x\in\mathcal{X}}p(x)\sigma_A^x\otimes\tau_{B_1}^x\otimes\tau_{B_2}^x\otimes\dotsb\otimes\tau_{B_k}^x$, where $k\in\mathbb{N}$. Prove that $\widetilde{\rho}_{AB_{\ell}}\coloneqq\Tr_{B_j:j\neq\ell}[\widetilde{\rho}_{AB_1B_2\dotsb B_k}]=\rho_{AB}$ for all $\ell\in\{1,2,\dotsc,k\}$, and that $W_{B_1\dotsb B_k}^{\pi}\widetilde{\rho}_{AB_1\dotsb B_k}W_{B_1\dotsb B_k}^{\pi\dagger}=\widetilde{\rho}_{AB_1\dotsb B_k}$ for all $\pi\in\mathcal{S}_k$.	The notation $\Tr_{B_j:j\neq\ell}$ indicates to take the partial trace over all of the $B$ systems except for $B_{\ell}$.
		\end{enumerate}
	\end{exercise}
	
	In the case $n=2$, meaning that there are only two quantum systems under consideration, there is only one non-trivial permutation, $\pi=(1~2)$, which swaps the two elements of the set $\{1,2\}$. Recall from Exercise~\ref{exer-symm_two} that 
	\begin{equation}\label{eq-swap_operator}
		W^{(1~2)}=F\coloneqq\sum_{k,k'=0}^{d-1}\ketbra{k,k'}{k',k}.
	\end{equation}
	We call $F$ the \textit{swap operator}, because $F(\rho\otimes\sigma)F^{\dagger}=\sigma\otimes\rho$ for all quantum states $\rho$ and $\sigma$, which is a simple consequence of \eqref{eq-permutation_product_state}. In other words, the two states $\rho$ and $\sigma$ become ``swapped'' with respect to the quantum systems after the action of the operator $F$. The swap operator is Hermitian and satisfies $F^2=\mathbbm{1}$, implying that it is also unitary and self-inverse. Also, as a consequence of \eqref{eq-trace_cyclic_perm}, we have
	\begin{equation}
		\Tr[F(\rho\otimes\sigma)]=\Tr[\rho\sigma]
	\end{equation}
	for all quantum states $\rho$ and $\sigma$.
	
	\begin{exercise}{exer-Bell_state_perms}
		\begin{enumerate}
			\item Verify that $F\ket{\Phi_d}=\ket{\Phi_d}$ for all $d\geq 2$.
			
			\item For the two-qubit Bell state vectors defined in \eqref{eq-two_qubit_Bell_states}, prove that $F\ket{\Phi_{z,x}}=(-1)^{zx}\ket{\Phi_{z,x}}$ for all $z,x\in\{0,1\}$.
			
			\item More generally, for the two-qudit Bell state vectors defined in \eqref{eq-qudit_Bell}, prove that 
				\begin{equation}
					F\ket{\Phi_{z,x}}=\e^{\frac{2\pi\I zx}{d}}\ket{\Phi_{z,-x}}
				\end{equation}
				for all $z,x\in\{0,1,\dotsc,d-1\}$.
		\end{enumerate}
	\end{exercise}
	
	\begin{exercise}{exer-symmetrization}
		Let $\rho_{A^n}\in\Density(\mathcal{H}^{\otimes n})$ be an arbitrary quantum state, and consider the state
		\begin{equation}\label{eq-permutation_twirl}
			\sigma_{A^n}\coloneqq\frac{1}{n!}\sum_{\pi\in\mathcal{S}_n} W_{A^n}^{\pi}\rho_{A^n} W_{A^n}^{\pi\dagger}.
		\end{equation}
		\begin{enumerate}[topsep=0.3cm]
			\item Prove that $\sigma_{A^n}$ is a permutation-invariant state.
			\item Let $\ket{\phi}_{RA^n}$ be a purification of $\rho_{A^n}$. Verify that 
				\begin{equation}\label{eq-purif_perm_invar_state}
					\ket{\psi}_{XRA^n}\coloneqq\frac{1}{\sqrt{n!}}\sum_{\pi\in\mathcal{S}_n}\ket{\pi}_X\otimes(\mathbbm{1}_R\otimes W_{A^n}^{\pi})\ket{\phi}_{RA^n}
				\end{equation}
				is a purification of $\sigma_{A^n}$, where $\{\ket{\pi}\}_{\pi\in\mathcal{S}_n}$ is an orthonormal basis indexed by the elements of $\mathcal{S}_n$, such that $\braket{\pi}{\pi'}=\delta_{\pi,\pi'}$ for all $\pi,\pi'\in\mathcal{S}_n$.
		\end{enumerate} 
	\end{exercise}
	
	It turns out that for permutation-invariant states, we can construct a purification that is itself permutation invariant.
	
	\begin{Lemma*}{Purification of Permutation-Invariant States}{lem-symm_purif}
		Let $\rho_{A^n}\in\Density(\mathcal{H}_A^{\otimes n})$ be a permutation-invariant state, i.e.,
		\begin{equation}
			\rho_{A^n} = W_{A^n}^{\pi}\rho_{A^n}W_{A^n}^{\pi\dagger}
		\end{equation}
		for all $\pi\in\mathcal{S}_n$, where the unitary operators in the set $\{W_{A^n}^{\pi}\}_{\pi\in\mathcal{S}_n}$ are defined in \eqref{eq-permutation_rep}. Then, there exists a permutation-invariant purification $\ket{\psi^{\rho}}\!\bra{\psi^{\rho}}_{\hat{A}^{n}A^n}$ of $\rho_{A^n}$, such that $\ket{\psi^{\rho}}_{\hat{A}^{n}A^n}\in\text{Sym}_n(\mathcal{H}_{\hat{A}A})$. This means that
		\begin{equation}
			\ket{\psi^{\rho}}_{\hat{A}^{n}A^n} = W_{\hat{A}^{n}}^{\pi}\otimes W_{A^n}^{\pi}\ket{\psi^{\rho}}_{\hat{A}^{n}A^n}
		\end{equation}
		for all $\pi\in\mathcal{S}_n$, where the dimension of $\hat{A}$ is equal to the dimension of $A$.
	\end{Lemma*}
	
	\begin{Proof}
				Consider the canonical purification of $\rho_{A^n}$ as defined in \eqref{eq-canonical_purification}; i.e., let
				\begin{equation}
				\ket{\psi^{\rho}}_{\hat{A}^nA^n}\coloneqq \left(\mathbbm{1}_{\hat{A}^n}\otimes \sqrt{\rho_{A^n}}\right)\ket{\Gamma}_{\hat{A}^n A^n}.
				\end{equation}
				 Then, because the operators $W_{A^n}^{\pi}$ are real in the standard basis, meaning that $\conj{W_{\hat{A}^n}^{\pi}}=W_{\hat{A}^n}^{\pi}$ for all $\pi\in\mathcal{S}_n$, and using the transpose trick in \eqref{eq-vec_expand}, we obtain
		\begin{align}
			W_{\hat{A}^n}^{\pi}\otimes W_{A^n}^{\pi}\ket{\psi^{\rho}}_{\hat{A}^nA^n}&=\left(\conj{W_{\hat{A}^n}^{\pi}}\otimes W_{A^n}^{\pi}\right)\ket{\psi^{\rho}}_{\hat{A}^nA^n}\\
            & = \left(\conj{W_{\hat{A}^n}^{\pi}}\otimes W_{A^n}^{\pi}\right)\left(\mathbbm{1}_{\hat{A}^n}\otimes \sqrt{\rho_{A^n}}\right)\ket{\Gamma}_{\hat{A}^n A^n}\\
			&=\left(\mathbbm{1}_{\hat{A}^n}\otimes W_{A^n}^{\pi}\sqrt{\rho_{A^n}}W_{A^n}^{\pi\dagger}\right)\ket{\Gamma}_{\hat{A}^n A^n}\\
			&=
			\left(\mathbbm{1}_{\hat{A}^n}\otimes \sqrt{W_{A^n}^{\pi}\rho_{A^n}W_{A^n}^{\pi\dagger}}\right)\ket{\Gamma}_{\hat{A}^n A^n}\\
			&=\left(\mathbbm{1}_{\hat{A}^n}\otimes \sqrt{\rho_{A^n}}\right)\ket{\Gamma}_{\hat{A}^n A^n}\\
			&=\ket{\psi^{\rho}}_{\hat{A}^nA^n}
		\end{align} 
		for all $\pi\in\mathcal{S}_n$, where the fourth equality follows from \eqref{eq-function_Hermitian_unitary} and the penultimate equality follows from the permutation invariance of $\rho_{A^n}$. 
	\end{Proof}

\subsection{Group-Invariant States}\label{sec-QM_group_inv_states}

	So far, we have seen two special types of unitary operators: the Heisenberg--Weyl operators $\{W_{z,x}\}_{z,x=0}^{d-1}$, introduced in Definition~\ref{def-heisenberg-weyl}, and the permutation operators $\{W^{\pi}\}_{\pi\in\mathcal{S}_n}$ defined in \eqref{eq-permutation_rep_QM}. Both sets of operators are examples of projective unitary group representations. Specifically, the Heisenberg--Weyl operators form a projective unitary representation of the group $\mathbb{Z}_d\times\mathbb{Z}_d$, and the permutation operators form a unitary representation of the symmetric group $\mathcal{S}_n$.
	
	Let us now formally define the concepts of a group and group representations.
	A \textit{group} $G$ is a tuple $(\mathcal{G},\ast)$ consisting of a set $\mathcal{G}$ of objects and an associative operation $\ast$ used to combine them. We write $g\in G$ to mean that the object $g$ belongs to the set $\mathcal{G}$. Then, the operation $\ast$ is such that $g\ast g'\in G$ for all $g,g'\in G$. Furthermore, there is an identity element $\id$ such that $g\ast\id=g=\id\ast g$ for all $g\in G$, and corresponding to every element $g\in G$ is an inverse $g^{-1}\in G$ such that $g\ast g^{-1}=g^{-1}\ast g=\id$. We mostly consider finite groups in this book, and we use $|G|$ to denote the number of elements in the associated set $\mathcal{G}$. Please see the Bibliographic Notes (Section~\ref{sec:qm:bib-notes}) for more information about groups.
	
	A \textit{unitary representation} of a group $G$ is a map $g \mapsto U^g$, where $g\in G$ and each $U^g$ is unitary. Furthermore, the unitary operators should respect the group operation $\ast$, in the sense that $U^g U^{g'}=U^{g\ast g'}$ for all $g\in G$. In particular, this implies that $U^{\id}=\mathbbm{1}$ and $U^{g^{-1}}=(U^{g})^{\dagger}$ for all $g\in G$. A unitary representation of a group $G$ is called \textit{projective} if the unitaries respect the group operation up to a phase factor, i.e., if $U^g U^{g'}=\omega(g,g')U^{g\ast g'}$ for all $g,g'\in G$, where $\omega(g,g')\in\mathbb{C}$ satisfies $\abs{\omega(g,g')}=1$ for all $g,g'\in G$. Please see the Bibliographic Notes (Section~\ref{sec:qm:bib-notes}) for more information about group representations.
	
	The \textit{action} of the group representation on a quantum system is defined by the quantum channels $\rho\mapsto U^g\rho U^{g\dagger}$ for all $g\in G$ (see Chapter~\ref{chap-QM_channels} for details of quantum channels), where $\rho$ is a state of the quantum system and $g\mapsto U^g$ is a (projective) unitary representation of the group $G$. Physically, groups are used to model certain types of operations on a system (such as permutations, translations, rotations, etc.). Mathematically, systems that have symmetries are such that the states of the system are invariant under the action of the corresponding group.
	
	\begin{definition}{Group-Invariant State}{def-group_invariant_state}
		Let $G$ be a (finite) group, and let $g \mapsto U^g$ be a $d$-dimensional unitary representation of $G$, with $d\geq 2$. A quantum state $\rho\in\Density(\mathbb{C}^d)$ is called \textit{group invariant}, or \textit{$G$-invariant}, if $\rho=U^g\rho U^{g\dagger}$ for all $g\in G$.
	\end{definition}

	\begin{exercise}{exer-group_inv_twirl}
		Let $\rho_A$ be a quantum state for a $d$-dimensional quantum system $A$, let $G$ be a group, and let $g\mapsto U_A^g$ be a $d$-dimensional unitary representation of $G$.
		\begin{enumerate}[topsep=0.3cm]
			\item Prove that the state
				\begin{equation}\label{eq-QM_twirl_map_states}
					\mathcal{T}^G(\rho_A)\coloneqq\frac{1}{|G|}\sum_{g\in G} U_A^g\rho_A U_A^{g\dagger}
				\end{equation}
				is group invariant. The quantum channel $\mathcal{T}^G$ is known as the \textit{twirl channel} with respect to the unitary representation $g \mapsto U^g$ of the group $G$.
		
			\item Let $\ket{\phi}_{RA}$ be a purification of $\rho_A$. Verify that
				\begin{equation}
					\frac{1}{\sqrt{|G|}}\sum_{g\in G}\ket{g}_X\otimes(\mathbbm{1}_R\otimes U_A^g)\ket{\phi}_{RA}
				\end{equation}
				is a purification of $\mathcal{T}^G(\rho_A)$.
		\end{enumerate}
	\end{exercise}

	The twirl map in \eqref{eq-QM_twirl_map_states} corresponding to the Heisenberg--Weyl unitaries has the following special form.
	
	\begin{Lemma*}{Heisenberg--Weyl Twirl}{lem:QM-over:HW-twirl}
		For every linear operator $M\in\Lin(\mathbb{C}^d)$,
		\begin{equation}\label{eq-HW_twirl}
			\frac{1}{d^2}\sum_{z,x=0}^{d-1}W_{z,x}M W_{z,x}^\dagger=\Tr[M]\frac{\mathbbm{1}_d}{d}. 
		\end{equation}
	\end{Lemma*}
	
	\begin{Proof}
		The Heisenberg--Weyl operators form an irreducible projective unitary representation of the group $\mathbb{Z}_d\times\mathbb{Z}_d$. This fact can be used to prove \eqref{eq-HW_twirl} via Schur's lemma (see Bibliographic Notes in Section~\ref{sec:qm:bib-notes}). Alternatively, the result is immediate using orthonormality of the set $\{\frac{1}{\sqrt{d}}W_{z,x}\}_{z,x=0}^{d-1}$ (see \eqref{eq:QM-over:HW-orthogonal}) along with Problem~\ref{prob-math_tools_op_ONB} in Section~\ref{sec-problems_math_tools}. For an alternative approach, see Exercise~\ref{exer-QM_HW_twirl}.
	\end{Proof}
	
	\begin{exercise}{exer-QM_HW_twirl}
		Provide an alternative proof of Lemma~\ref{lem:QM-over:HW-twirl}. Do this by first showing that $\frac{1}{d^2}\sum_
{z,x=0}^{d-1}W_{z,x}MW_{z,x}^{\dagger}=(\mathcal{D}_X\circ\mathcal{D}_Z)(M)$, where
		\begin{align}
			\mathcal{D}_X(M)&\coloneqq\frac{1}{d}\sum_{x=0}^{d-1}X(x)MX(x)^{\dagger},\\
			\mathcal{D}_Z(M)&\coloneqq\frac{1}{d}\sum_{z=0}^{d-1}Z(z)MZ(z)^{\dagger}.
		\end{align}
	\end{exercise}
	
	Recall that the Heisenberg--Weyl operators reduce to the Pauli operators for $d=2$. In this case, we obtain
	\begin{equation}\label{eq-Pauli_twirl_0}
		\frac{1}{4}M+\frac{1}{4}XMX+\frac{1}{4}YMY+\frac{1}{4}ZMZ=\Tr[M]\frac{\mathbbm{1}_2}{2},
	\end{equation}
	for every $M\in\Lin(\mathbb{C}^2)$.

	\begin{exercise}{exer-group_inv_state_purif}
		Let $G$ be a group, and let $g \mapsto U^g$ be a $d$-dimensional unitary representation of $G$, with $d\geq 2$. If $\rho\in\Density(\mathbb{C}^d)$ is a group-invariant state, then prove that there exists a purification $\ket{\psi^{\rho}}$ of $\rho$ such that $\conj{U^g}\otimes U^g\ket{\psi^{\rho}}=\ket{\psi^{\rho}}$ for all $g\in G$. (\textit{Hint}: Consider an approach similar to the proof of Lemma~\ref{lem-symm_purif}.)
	\end{exercise}

\subsection{Ensembles and Classical--Quantum States}

	For a finite alphabet $\mathcal{X}$, an \textit{ensemble} is a collection $\{(p(x),\rho^x)\}_{x\in\mathcal{X}}$ consisting of a probability distribution $p:\mathcal{X}\to[0,1]$ such that each probability $p(x)$ is paired with a quantum state $\rho^x$. Ensembles are used to describe quantum systems that are known to be in one of a given set of states with some probability.
	
	Suppose that Alice is in possession of a quantum system and that she prepares the system in the state $\rho^x$ with probability $p(x)$. The state of the system can thus be described by the ensemble $\{(p(x),\rho^x)\}_{x\in\mathcal{X}}$. She sends the system to Bob  without telling him in which of the states the system has been prepared, but Bob knows the ensemble $\{(p(x),\rho^x)\}_{x\in\mathcal{X}}$ describing the system. Then, from Bob's perspective, the state of the system is given by the expected state $\rho$ of the ensemble, which is specified as
	\begin{equation}
		\rho=\sum_{x\in\mathcal{X}}p(x)\rho^x.
		\label{eq-QM-SM:ensemble-Bob-state}
	\end{equation}
	
	On the other hand, if Alice sends Bob classical information about which state she has prepared, then from Bob's perspective the state of the system can be described by the following \textit{classical--quantum state}:
	\begin{equation}\label{eq-cq_state_QM}
		\rho_{XB}=\sum_{x\in\mathcal{X}}p(x)\ket{x}\!\bra{x}_X\otimes \rho_B^x,
	\end{equation}
	where $X$ is the $|\mathcal{X}|$-dimensional quantum system corresponding to the register holding the information about which state was sent and $\{\ket{x}\}_{x\in\mathcal{X}}$ is an orthonormal basis for $\mathcal{H}_X$. Furthermore, the reduced state of Bob's system $B$, after discarding~$X$, is
	\begin{equation}
		\Tr_X[\rho_{XB}] = \sum_{x\in \mathcal{X}} p(x) \rho^x = \rho,
	\end{equation}
	consistent with \eqref{eq-QM-SM:ensemble-Bob-state}.
	
	Classical--quantum states have a block-diagonal structure, in the sense that they can equivalently be written as the following block-diagonal matrix:
	\begin{equation}\label{eq-cq_state_block_diag}
		\rho_{XB}=\begin{pmatrix} p(x_1)\rho^{x_1} & & & \\ & p(x_2)\rho^{x_2} & & \\ & & \ddots & \\ & & & p(x_{|\mathcal{X}|})\rho^{x_{|\mathcal{X}|}} \end{pmatrix}.
	\end{equation}
	We can represent \eqref{eq-cq_state_block_diag} more compactly as
	\begin{equation}
		\rho_{XB} = \bigoplus_{x \in \mathcal{X}} p(x)\rho^x.
	\end{equation}

	\begin{exercise}{exer-purif_cq_states}
		Construct a purification of the classical--quantum state $\rho_{XB}$ in \eqref{eq-cq_state_QM}. (\textit{Hint}: Consider a purification analogous to the one in \eqref{eq-purif_perm_invar_state}.)
	\end{exercise}

\subsection{Partial Transpose and PPT States}\label{subsec-PPT}

	In Section~\ref{sec-partial_trace_QM}, we defined the partial trace superoperator as a generalization of the usual trace to the case in which it acts only on one part of a composite quantum system. In a similar manner, in this section, we define the partial transpose superoperator. The partial transpose plays an important role in quantum information theory due to its connection with entanglement. In fact, as we show in this section, it leads to a sufficient condition for a bipartite state to be entangled.

	Recall from \eqref{eq:math-tools:transpose-superop} that the action of the transpose map $\T$ on a linear operator~$X_{A\to A'}$ can be written as
	\begin{equation}
		\T(X)=\sum_{i=0}^{d_A-1}\sum_{j=0}^{d_{A'}-1}\ket{i}_A\bra{j}_{A'} X \ket{i}_A\bra{j}_{A'},
	\end{equation}
	where we have defined the transpose with respect to the orthonormal bases $\{\ket{i}_A:0\leq i\leq d_{A}-1\}$ and $\{\ket{j}_{A'}:0\leq j\leq d_{A'} -1\}$. This is consistent with the familiar definition of the transpose of a matrix $X$ as being the matrix $X^{\t}$ with its rows and columns flipped relative to $X$. Indeed, if $X$ has the matrix representation $X=\sum_{j=0}^{d_{A'}-1}\sum_{i=0}^{d_A-1} X_{j,i}\ket{j}_{A'}\bra{i}_A$, then it follows that the transpose of $X$ is
	\begin{equation}
		X^{\t}=\sum_{j=0}^{d_{A'}-1}\sum_{i=0}^{d_A-1}X_{j,i}\ket{i}_A\bra{j}_{A'}=\T(X).
	\end{equation}
	Note that, unlike the trace or the conjugate transpose, the transpose depends on the choice of orthonormal bases used to evaluate it. Throughout the rest of this book, we use both $\T(X)$ and $X^{\t}$ to refer to the transpose of $X$ with respect to the standard orthonormal basis.
	
	\begin{exercise}{exer-transpose_HW}
		For all $d\geq 2$, prove that the transpose map can be realized as follows, in terms of the Heisenberg--Weyl operators from Definition~\ref{def-heisenberg-weyl}: 
		\begin{equation}\label{eq-transpose_HW}
			\T(X)=\frac{1}{d}\sum_{z,x=0}^{d-1}\e^{\frac{2\pi\I zx}{d}}W_{z,x}^{\dagger} X W_{z,-x},
		\end{equation}
		where the equality holds for every linear operator $X\in\Lin(\mathbb{C}^d)$.
	\end{exercise}
	
	The transpose map is known as the \textit{partial transpose} when it acts on one subsystem of a bipartite operator $X_{AB}$. 

	\begin{definition}{Partial Transpose}{def-partial_transpose}
		Given quantum systems $A$ and $B$, the \textit{partial transpose on $B$} is denoted by $\T_B\equiv \id_A\otimes \T_B$, and it is defined as
		\begin{equation}
			\T_B(X_{AB})\coloneqq\sum_{j,j'=0}^{d_B-1}(\mathbbm{1}_{A}\otimes \ket{j}\!\bra{j'}_{B})X_{AB}(\mathbbm{1}_{A}\otimes\ket{j}\!\bra{j'}_{B})
		\end{equation}
		for every linear operator $X_{AB}\in\Lin(\mathcal{H}_A\otimes\mathcal{H}_B)$. Similarly, the \textit{partial transpose on~$A$} is denoted by $\T_A\equiv \T_A\otimes\id_B$, and it is defined as
		\begin{equation}
			\T_A(X_{AB})\coloneqq\sum_{i,i'=0}^{d_A-1}(\ket{i}\!\bra{i'}_A\otimes\mathbbm{1}_B)X_{AB}(\ket{i}\!\bra{i'}_A\otimes\mathbbm{1}_B)
		\end{equation}
		for all $X_{AB}\in\Lin(\mathcal{H}_A\otimes\mathcal{H}_B)$.
	\end{definition}
	
	Given an expansion of $X_{AB}$ as
	\begin{equation}
		X_{AB}=\sum_{i,j=0}^{d_B-1} X_{A}^{i,j}\otimes\ket{i}\!\bra{j}_{B},
	\end{equation}
	where each $X_{A}^{i,j}\coloneqq\bra{i}_BX_{AB}\ket{j}_B$ is a linear operator acting on system $A$, the partial transpose map $\T_{B}$ has the action
	\begin{equation}
		\T_{B}(X_{AB})=
		\sum_{i,j=0}^{d_B-1} X_{A}^{i,j}\otimes\ket{j}\!\bra{i}_{B}
		=\sum_{i,j=0}^{d_B-1} X_{A}^{j,i}\otimes\ket{i}\!\bra{j}_{B}.
	\end{equation}
	
	The partial transpose map is self-inverse, i.e.,
	\begin{equation}\label{eq:partial-transpose-self-inverse}
		\left(\id_{A}\otimes \T_B\right)\circ\left(\id_{A}\otimes \T_B\right)=\id_{AB},
	\end{equation}
	and it is self-adjoint with respect to the Hilbert--Schmidt inner product, in the sense that
	\begin{equation}\label{eq:partial-transpose-self-adjoint}
		\inner{X_{AB}}{\T_{B}(Y_{AB})}=\inner{\T_{B}(X_{AB})}{Y_{AB}},
	\end{equation}
	for all operators $X_{AB}$ and $Y_{AB}$.
	
	We also have the following generalization of the transpose trick from \eqref{eq-transpose_trick}:
	\begin{equation}
	(X_{RA} \otimes \mathbbm{1}_B) (\mathbbm{1}_R \otimes \ket{\Gamma}_{AB}) = (\mathbbm{1}_A \otimes \T_B(X_{RB})) (\mathbbm{1}_R \otimes \ket{\Gamma}_{AB}),
	\label{eq:extended-trans-trick}
	\end{equation}
	where the Hilbert spaces corresponding to the systems $A$ and $B$ are isomorphic and $X_{RA} \in \Lin(\mathcal{H}_R \otimes \mathcal{H}_A)$.
	
	\begin{exercise}{exer-partial_transpose_basic}
		Verify \eqref{eq:partial-transpose-self-inverse}, \eqref{eq:partial-transpose-self-adjoint}, and \eqref{eq:extended-trans-trick}.
	\end{exercise}
	
	\begin{definition}{PPT State}{def-PPT}
		A bipartite state $\rho_{AB}$ is called a \textit{positive partial transpose (PPT) state} if the partial transpose $\T_B(\rho_{AB})$ is positive semi-definite.\\[0.1in]
		The set of PPT states is denoted by $\PPT(A\!:\!B)$, so that
		\begin{equation}\label{eq:PPT-states-set}
			\PPT(A\!:\!B)\coloneqq \left\{\sigma_{AB}:\sigma_{AB}\geq 0,\,\T_{B}
(\sigma_{AB})\geq 0,\,\Tr[\sigma_{AB}]=1\right\}.
		\end{equation}
	\end{definition}
	
	\begin{Lemma}{lem-PPT_invar}
		Given quantum systems $A$ and $B$, the set $\PPT(A\!:\!B)$ does not depend on which system the transpose is taken, nor does it depend on which orthonormal basis is used to define the transpose map.
	\end{Lemma}
	
	\begin{Proof}
		To see the first statement, suppose that $\rho_{AB}\in\PPT(A\!:\!B)$. This means that $\T_B(\rho_{AB})\geq 0$. But since the eigenvalues are invariant under a full transpose $\T_A\otimes\T_B$, this means that $(\T_A\otimes\T_B)(\T_B(\rho_{AB}))\geq 0$, the latter being the same as $\T_A(\rho_{AB})\geq 0$ due to the self-inverse property of the partial transpose. So $\T_B(\rho_{AB})\geq 0$ implies $\T_A(\rho_{AB})\geq 0$, and vice versa. 
	
		To see the second statement, let $\T_B(\rho_{AB})\geq 0$, and let $\{ \ket{\phi_\ell}_B \}_{\ell=0}^{d_B-1}$ be some other orthonormal basis for $B$. The partial transpose with respect to this basis is given by
		\begin{equation}
			\sum_{\ell,\ell'=0}^{d_B-1}(\mathbbm{1}_A\otimes\ket{\phi_\ell}\!\bra{\phi_{\ell'}}_B)\rho_{AB}(\mathbbm{1}_A\otimes\ket{\phi_{\ell}}\!\bra{\phi_{\ell'}}_{B}).
			\label{eq-QSM:partial-trans-different-basis-1}
		\end{equation}
		Now, consider that $\sum_{\ell=0}^{d_B-1}\ket{\phi_{\ell}}\!\bra{\phi_{\ell}}=\mathbbm{1}_B$, so that
		\begin{align}
			& \T_B(\rho_{AB}) \notag \\&  =
			\sum_{j,j'=0}^{d_B-1}(\mathbbm{1}_{A}\otimes \ket{j}\!\bra{j'}_{B})\rho_{AB}(\mathbbm{1}_{A}\otimes\ket{j}\!\bra{j'}_{B}) \notag\\
			&  =
			\sum_{j,j',\ell, \ell'=0}^{d_B-1}(\mathbbm{1}_{A}\otimes \ket{j}\braket{j'}{\phi_{\ell}}
			\bra{\phi_{\ell}}_{B})\rho_{AB}(\mathbbm{1}_{A}\otimes\ket{\phi_{\ell'}}\braket{\phi_{\ell'}}{j}\!\bra{j'}_{B}) \notag\\
			&  =
			\sum_{j,j',\ell, \ell'=0}^{d_B-1}(\mathbbm{1}_{A}\otimes \braket{\phi_{\ell'}}{j} \ket{j}
			\bra{\phi_{\ell}}_{B})\rho_{AB}(\mathbbm{1}_{A}\otimes\ket{\phi_{\ell'}}\!\bra{j'}_{B}\braket{j'}{\phi_{\ell}}) \notag\\
				&  =
			\sum_{\ell, \ell'=0}^{d_B-1}\left(\mathbbm{1}_{A}\otimes \left(\sum_{j=0}^{d_B-1}\braket{\phi_{\ell'}}{j}\ket{j}\right)
			\bra{\phi_{\ell}}_{B}\right)\rho_{AB}\left(\mathbbm{1}_{A}\otimes\ket{\phi_{\ell'}}
			\left(\sum_{j'=0}^{d_B-1}\bra{j'}_{B}\braket{j'}{\phi_{\ell}}\right)\right) \notag\\
					&  =
			\sum_{\ell, \ell'=0}^{d_B-1}\left(\mathbbm{1}_{A}\otimes \ket{\conj{\phi_{\ell'}}}
			\bra{\phi_{\ell}}_{B}\right)\rho_{AB}\left(\mathbbm{1}_{A}\otimes\ket{\phi_{\ell'}}
			\bra{\conj{\phi_{\ell}}}\right) ,
			\label{eq-QSM:partial-trans-different-basis-2}
		\end{align}
		where in the last line we defined $\ket{\conj{\phi_{\ell}}} \coloneqq  \sum_{j=0}^{d_B-1}\braket{\phi_{\ell}}{j}\ket{j}$ for $0 \leq \ell \leq d_B-1$. Note that the set $\{\ket{\conj{\phi_{\ell}}}\}_{\ell=0}^{d_B-1}$ is orthonormal, so that $U_B \coloneqq  \sum_{\ell=0}^{d_B-1} \ket{\phi_{\ell}}\!\bra{\conj{\phi_{\ell}}}_B$ is a unitary operator. Then we find that
		\begin{equation}
			\sum_{\ell, \ell'=0}^{d_B-1}\left(\mathbbm{1}_{A}\otimes \ket{\phi_{\ell'}}
		\bra{\phi_{\ell}}_{B}\right)\rho_{AB}\left(\mathbbm{1}_{A}\otimes\ket{\phi_{\ell'}}
		\bra{\phi_{\ell}}_B\right) = U_B\T_B(\rho_{AB})U_B^\dag \geq 0, 
		\label{eq-QSM:partial-trans-different-basis-3}
		\end{equation}
		where the last inequality follows from the condition $\T_B(\rho_{AB}) \geq 0$ and property~1 of Lemma~\ref{prop-operator_ineqs}. We thus conclude that the PPT property  does not depend on which orthonormal basis is used to define the transpose map.
	\end{Proof}
		
	We mentioned at the beginning of this section that the partial transpose is useful in quantum information because it leads to a sufficient condition for a bipartite state to be entangled. We now derive the sufficient condition.
	
	Suppose that a state $\sigma_{AB}$ is a separable, unentangled state of the following form:
	\begin{equation}
		\sigma_{AB}=\sum_{x\in\mathcal{X}}p(x)\omega_{A}^{x}\otimes\tau_{B}^{x},
	\end{equation}
	where $p:\mathcal{X}\to[0,1]$ is a probability distribution on a finite alphabet $\mathcal{X}$ and $\{\omega_{A}^{x}\}_{x\in\mathcal{X}}$ and $\{\tau_{B}^{x}\}_{x\in\mathcal{X}}$ are sets of quantum states. Then the action of the partial transpose map $\T_{B}$ on $\sigma_{AB}$ is as follows:
	\begin{equation}
		\T_{B}(\sigma_{AB})=\sum_{x}p(x)\omega_{A}^{x}\otimes \T(\tau_{B}^{x}),
	\end{equation}
	which is a separable quantum state, being the expected state of the ensemble $\{(p(x),\omega_{A}^{x}\otimes\T(\tau_{B}^{x}))\}_{x\in\mathcal{X}}$. Each element of the ensemble is indeed a quantum state because the transpose is a positive map, i.e., $\T(\tau_{B}^{x})\geq 0$ if $\tau_{B}^{x}\geq 0$. Due to this fact, we conclude that $\T_{B}(\sigma_{AB})\geq 0$, so that $\sigma_{AB}$ is a PPT state. Thus, we conclude the following:
	\begin{specbox}{0.6\textwidth}
		If a state is separable, then it is PPT.
	\end{specbox}
	This is called the \textit{PPT criterion}.
	Equivalently, by taking the contrapositive of this statement, we obtain the following:
	\begin{specbox}{0.6\textwidth}
		If a state is not PPT, then it is entangled.
	\end{specbox}
	So the condition of a state being NPT (non-positive partial transpose) is sufficient for detecting entanglement.
	
	As an example, let us consider applying the partial transpose map $\T_{B}$ to the maximally entangled state $\Phi_{AB}$, as defined in \eqref{eq-max_ent_state}. In the case that the bases of the partial transpose and the maximally entangled state are the same, 
	we find that
	\begin{align}
		\T_B(\Phi_{AB})&=\frac{1}{d} \,\T_B\left(\sum_{i,j=0}^{d-1}\ket{i}\!\bra{j}_A\otimes\ket{i}\!\bra{j}_B\right)\label{eq-partial_transpose_gamma_1} \\
		&=\sum_{i,j=0}^{d-1} \ket{i}\!\bra{j}_A\otimes\ket{j}\!\bra{i}_B\label{eq-partial_transpose_gamma_2}\\
		&=\frac{1}{d}F_{AB},\label{eq-partial_transpose_gamma_3}
	\end{align}
	where $F_{AB}$ is the swap operator defined in \eqref{eq-swap_operator}. If the bases are not the same, then we find, by applying the same development from \eqref{eq-QSM:partial-trans-different-basis-1}--\eqref{eq-QSM:partial-trans-different-basis-3}, that there exists a unitary $U_B$ such that
	\begin{equation}
	\T_B(\Phi_{AB}) = \frac{1}{d}U_B F_{AB} U_B^\dag.
	\label{eq-partial_transpose_gamma_different_bases}
	\end{equation}
	
	From \eqref{eq-symm_proj_n2} and \eqref{eq-asymm_proj_n2}, the swap operator has the following spectral decomposition:
	\begin{equation}\label{eq:qm:spectral-decomp-swap-op}
		F_{AB} = \Pi^{\text{Sym}}_{AB} - \Pi^{\text{ASym}}_{AB},
	\end{equation}
	where $\Pi^{\text{Sym}}_{AB}\equiv\Pi_{\text{Sym}_2(\mathbb{C}^d)}$ is the projection onto the symmetric subspace of $\mathbb{C}^d\otimes\mathbb{C}^d$ and $\Pi_{AB}^{\text{ASym}}\equiv\Pi_{\text{ASym}_2(\mathbb{C}^d)}$ is the projection onto the anti-symmetric subspace of $\mathbb{C}^d\otimes\mathbb{C}^d$. Indeed, we have that $\Pi^{\text{Sym}}_{AB} + \Pi^{\text{ASym}}_{AB} = \mathbbm{1}_{AB}$ and $\Pi^{\text{Sym}}_{AB} \Pi^{\text{ASym}}_{AB} = 0$. Thus, the swap operator has negative eigenvalues, which by the PPT criterion means that $\Phi_{AB}$ is an entangled state, as expected.
	
	Although the PPT criterion is generally only a necessary condition for separability of a bipartite state, it is known 
        to be also sufficient for every quantum state $\rho_{AB}$ for which both $A$ and $B$ are qubits or $A$ is a qubit and $B$ is a qutrit; please consult the Bibliographic Notes in Section~\ref{sec:qm:bib-notes}. 
        In higher dimensions, however, there are PPT states that are entangled. These PPT entangled states turn out to be useless for the task of entanglement distillation (see Chapter~\ref{chap-ent_distill}), and thus they are called \textit{bound entangled} (although they are entangled, they cannot be used to extract pure maximally entangled states by means of local operations and classical communication).
	
	\begin{exercise}{exer-SWAP_unitary_invar}
		Prove that the swap operator $F_{AB}$ possesses the following symmetry:
		\begin{equation}
			(U_A\otimes V_B)F_{AB} = F_{AB} (V_A\otimes U_B)
		\end{equation}
		for all unitaries $U$ and $V$. 
	\end{exercise}

\subsection{Isotropic and Werner States}

	In Section~\ref{sec-multipartite_permutations}, we defined permutation-invariant states, which are states that are invariant under the action of the unitary operator $W^{\pi}$, for every permutation $\pi\in\mathcal{S}_n$. Another important class of quantum states in quantum information theory consists of bipartite states that are invariant under certain kinds of unitaries. There are two distinct such classes of states that we define in this section.
	
	\begin{definition}{Isotropic States}{def-isotropic_state}
		Consider two quantum systems $A$ and $B$, with $d_A=d_B=d\geq 2$. A quantum state $\rho_{AB}$ is called an \textit{isotropic state} if it is invariant under the action of $U\otimes\conj{U}$ for every unitary $U$, i.e., if
		\begin{equation}\label{eq-isotropic_state_def}
			\rho_{AB}=(U\otimes\conj{U})\rho_{AB}(U\otimes\conj{U})^{\dagger}
		\end{equation}
		for every unitary $U$. For every isotropic state $\rho_{AB}$, there exists $p\in[0,1]$ such that $\rho_{AB}=\rho_{AB}^{\text{iso};p}$, where
		\begin{equation}\label{eq-isotropic_state}
			\rho_{AB}^{\text{iso};p}\coloneqq p\Phi_{AB}+\frac{1-p}{d^2-1}\left(\mathbbm{1}_{AB}-\Phi_{AB}\right),
		\end{equation}
		where $\Phi_{AB} \equiv \ketbra{\Phi}{\Phi}_{AB}$ and $\ket{\Phi}_{AB}\coloneqq \frac{1}{\sqrt{d}}\sum_{j=0}^{d-1}\ket{j,j}_{AB}$.
	\end{definition}
	
	Using \eqref{eq-qudit_Bell_ONB_identity}, we can write every isotropic state as follows:
	\begin{equation}\label{eq-isotropic_state2}
		\rho_{AB}^{\text{iso};p}=p{\Phi}_{AB}+\frac{1-p}{d^2-1}\sum_{\substack{0\leq z,x\leq d-1\\(z,x)\neq (0,0)}}{\Phi_{z,x}}.
	\end{equation}
	In other words, an isotropic state can be viewed as a probabilistic mixture of the qudit Bell states defined in \eqref{eq-qudit_Bell}, such that the state ${\Phi}_{AB}$ is prepared with probability $p$, and the states ${\Phi_{z,x}}$, such that $(z,x)\neq (0,0)$, are prepared with probability $\frac{1-p}{d^2-1}$. This implies that every isotropic state is a Bell-diagonal state (see \eqref{eq-Bell_diag_state}), that it has full rank for $p\in(0,1)$, and that its eigenvalues are $p$ and $\frac{1-p}{d^2-1}$ (the latter with multiplicity $d^2-1$).
	
	\begin{exercise}{exer-isotropic_state}
		\begin{enumerate}
			\item Verify that the isotropic state in \eqref{eq-isotropic_state} is invariant under $U\otimes\conj{U}$ for every unitary $U$; i.e., verify that
				\begin{equation}\label{eq-isotropic_state_def2}
					(U_A\otimes\conj{U}_B)\rho_{AB}^{\text{iso};p}(U_A\otimes\conj{U}_B)^{\dagger}=\rho_{AB}^{\text{iso};p}
				\end{equation}
				for every unitary $U$ and all $p\in[0,1]$.
		
			\item Show that, for all $p\in[0,1]$, the isotropic state $\rho_{AB}^{\text{iso};p}$ can be represented as $a{\Phi}_{AB}+(1-a)\frac{\mathbbm{1}_{AB}}{d^2}$, where $a=\frac{pd^2-1}{d^2-1}$. Conclude that $a\in\left[\frac{-1}{d^2-1},1\right]$.
		\end{enumerate}
	\end{exercise}
	
	Just as every multipartite quantum state can be made permutation invariant via the construction in \eqref{eq-permutation_twirl}, every bipartite quantum state $\rho_{AB}$, with $d_A=d_B$, can be made invariant under $U\otimes\conj{U}$, i.e., isotropic, via the following construction:
	 \begin{equation}\label{eq-isotropic_twirl}
	 	\int_U (U\otimes\conj{U})\rho_{AB}(U\otimes\conj{U})^{\dagger}~\text{d}U.
	 \end{equation}
	This construction can be thought of as a uniform average over unitaries, analogous to the uniform average over permutations in \eqref{eq-permutation_twirl}. The object ``$\text{d}U$'' is known as the \textit{Haar measure}. Intuitively, an integral is used due to the fact that the set of unitaries is continuous; however, the integral can be evaluated using a uniform average over a discrete set of unitaries known as a \textit{unitary two-design}. In particular, for every state $\rho_{AB}$,
	\begin{equation}\label{eq-isotropic_integral}
		\int_U (U\otimes\conj{U})\rho_{AB}(U\otimes\conj{U})^{\dagger}~\text{d}U = \rho_{AB}^{\text{iso};p},\quad p=\bra{\Phi}\rho_{AB}\ket{\Phi}.
	\end{equation}
	Please see the Bibliographic Notes (Section~\ref{sec:qm:bib-notes}) for more information about this result, as well as about integrals of functions of unitaries with respect to the Haar measure.
	
	The isotropic states constitute one class of bipartite quantum states in which every state is invariant under the action of a unitary acting on the individual subsystems. We now define a second class of such states.
	
	\begin{definition}{Werner States}{def-Werner_state}
		Consider two quantum systems $A$ and $B$, with $d_A=d_B=d\geq 2$. A quantum state $\rho_{AB}$ is called a \textit{Werner state} if it is invariant under the action of $U\otimes U$ for every unitary $U$, i.e., if
		\begin{equation}
			\rho_{AB}=(U\otimes U)\rho_{AB}(U\otimes U)^{\dagger}
		\end{equation}
		for every unitary $U$. For every Werner state $\rho_{AB}$, there exists $p\in[0,1]$ such that $\rho_{AB}=\rho_{AB}^{\text{W};p}$, where
		\begin{equation}\label{eq-Werner_state}
			\rho_{AB}^{\text{W};p}\coloneqq p\zeta_{AB}+(1-p)\zeta_{AB}^{\perp}.
		\end{equation}
		Here, $\zeta_{AB}$ and $\zeta_{AB}^{\perp}$ are quantum states defined as
		\begin{align}
			\zeta_{AB}&\coloneqq \frac{1}{d(d-1)}\left(\mathbbm{1}_{AB}-F_{AB}\right),\label{eq-Werner_zeta}\\
			\zeta_{AB}^{\perp}&\coloneqq\frac{1}{d(d+1)}\left(\mathbbm{1}_{AB}+F_{AB}\right),\label{eq-Werner_zeta_perp}
		\end{align}
		and $F_{AB}\coloneqq \sum_{i,j=0}^{d-1}\ketbra{i,j}{j,i}_{AB}$ is the swap operator. Sometimes $\zeta_{AB}$ is also called the \emph{antisymmetric state}, and $\zeta_{AB}^{\perp}$ the \emph{symmetric state}.
	\end{definition}
	
	To justify these names, observe that the states $\zeta_{AB}$ and $\zeta_{AB}^{\perp}$ in Definition~\ref{def-Werner_state} are proportional to the projections $\Pi_{AB}^{\text{ASym}}\equiv\Pi_{\text{ASym}_2(\mathbb{C}^d)}$ and $\Pi_{AB}^{\text{Sym}}\equiv\Pi_{\text{Sym}_2(\mathbb{C}^d)}$ onto the anti-symmetric and symmetric subspaces, respectively, of $\mathbb{C}^d\otimes\mathbb{C}^d$ (recall \eqref{eq-symm_proj_n2} and \eqref{eq-asymm_proj_n2}). In particular,
	\begin{equation}
		\zeta_{AB} = \frac{2}{d(d-1)}\,\Pi_{AB}^{\text{ASym}},\quad \zeta_{AB}^{\perp} = \frac{2}{d(d+1)}\,\Pi_{AB}^{\text{Sym}}.
	\end{equation}
	
	\begin{exercise}{exer-Werner_state}
		\begin{enumerate}
			\item Verify that the Werner state in \eqref{eq-Werner_state} is invariant under $U\otimes U$ for every unitary $U$; i.e., verify that
				\begin{equation}
					(U_A\otimes U_B)\rho_{AB}^{\text{W};p}(U_A\otimes U_B)^{\dagger}=\rho_{AB}^{\text{W};p}
				\end{equation}
				for every unitary $U$ and all $p\in[0,1]$.
		
			\item Show that, for all $p\in[0,1]$, the Werner state $\rho_{AB}^{\text{W};p}$ can be represented as $\frac{1}{d^2-da}\left(\mathbbm{1}_{AB}-aF_{AB}\right)$, where $a=\frac{d(2p-1)+1}{2p-1+d}$. Conclude that $a\in[-1,1]$.
		
			\item Prove that for $d=2$, $\zeta_{AB}=\Pi_{AB}^{\text{ASym}}=\ketbra{\Psi^-}{\Psi^-}$, where $\ket{\Psi^-}$ is defined in \eqref{eq-two_qubit_Bell_11}.
		
		\end{enumerate}
	\end{exercise}
	
	As with the isotropic states, every bipartite quantum state $\rho_{AB}$, with $d_A=d_B$, can be made into a Werner state via the following construction:
	\begin{equation}
		\int_U (U\otimes U)\rho_{AB}(U\otimes U)^{\dagger}~\text{d}U.
	\end{equation}
	As before, the integral represents the uniform average over all unitaries, which can be evaluated using a unitary two-design. In particular, for every state $\rho_{AB}$,
	\begin{equation}\label{eq-Werner_integral}
		\int_U (U\otimes U)\rho_{AB}(U\otimes U)^{\dagger}~\text{d}U=\rho_{AB}^{\text{W};p},\quad p=\Tr\!\left[\Pi_{AB}^{\text{ASym}} \rho_{AB}\right].
	\end{equation}
	Please see the Bibliographic Notes (Section~\ref{sec:qm:bib-notes}) for more information.

\section{Measurements}\label{subsec-meas}

	Measurements in quantum mechanics are described by \textit{positive operator-valued measures (POVMs)}.
	\begin{definition}{Positive Operator-Valued Measure (POVM)}{def:qm:povm}
	A POVM is a set $\{M_x\}_{x\in\mathcal{X}}$ of operators satisfying
	\begin{equation}
		M_x\geq 0\quad\forall~x\in\mathcal{X},\qquad \sum_{x\in\mathcal{X}}M_x=\mathbbm{1}.
	\end{equation}
	For our purposes, it suffices to consider finite sets of such operators. The elements of the finite alphabet $\mathcal{X}$ are used to label the outcomes of the measurement.
	\end{definition}
	
	The measurement of a quantum system in the state $\rho$ according to the POVM $\{M_x\}_{x\in\mathcal{X}}$ induces a probability distribution $p_X:\mathcal{X}\to[0,1]$. This distribution corresponds to a random variable $X$ that takes values in the alphabet $\mathcal{X}$ and is defined by the \textit{Born rule}:
	\begin{equation}\label{eq-Born_rule}
		p_X(x)=\Pr[X=x]=\Tr[M_x\rho].
	\end{equation}   
	
	If every element $M_x$ of the POVM is a projection, then the corresponding measurement is called a \textit{projective measurement}. If the POVM of a projective measurement consists solely of rank-one projections, then the measurement is also called a \textit{von Neumann measurement}. Observe using \eqref{eq-math_tools_identity_ONB} that every orthonormal basis $\{\ket{e_k}\}_{k=1}^d$ for a $d$-dimensional Hilbert space $\mathcal{H}$ defines a von Neumann measurement via the POVM $\{\ketbra{e_k}{e_k}\}_{k=1}^d$. Furthermore, as stated at the beginning of this chapter, every Hermitian operator defines a projective measurement, with the corresponding POVM given by its spectral projections.
	
	Projective measurements, in particular von Neumann measurements, are viewed as the simplest type of measurement that can be performed on a quantum system. As it turns out, we can actually implement every measurement as a von Neumann measurement if we have access to an auxiliary quantum system, and this is the content of \textit{Naimark's theorem}.
	
	\begin{theorem*}{Naimark's Theorem}{thm-naimark}
		For every POVM $\{M_x\}_{x\in\mathcal{X}}$, there exists an isometry $V$ such that
		\begin{equation}\label{eq-naimark-ext}
			M_x=V^\dagger (\mathbbm{1}\otimes\ket{x}\!\bra{x})V\quad\forall~x\in\mathcal{X},
		\end{equation}
		where $\{\ket{x}\}_{x\in\mathcal{X}}$ is an orthonormal set.
	\end{theorem*}
	
	\begin{Proof}
		This follows immediately by defining the isometry $V$ as
		\begin{equation}
			V=\sum_{x\in\mathcal{X}}\sqrt{M_x}\otimes\ket{x}.
			\label{eq-naimark-ext-choice}
		\end{equation}
		This is indeed an isometry because
		\begingroup
		\allowdisplaybreaks[0]
		\begin{align}
			V^\dagger V&=\left(\sum_{x\in\mathcal{X}}\sqrt{M_x}\otimes\bra{x}\right)\left(\sum_{x'\in\mathcal{X}}\sqrt{M_{x'}}\otimes\ket{x'}\right)\\
			&=\sum_{x,x'\in\mathcal{X}}\sqrt{M_x}\sqrt{M_{x'}}\braket{x}{x'}\\
			&=\sum_{x\in\mathcal{X}}M_x\\
			&=\mathbbm{1},
		\end{align}
		\endgroup
		where the last equality holds because $\{M_x\}_{x\in\mathcal{X}}$ is a POVM. It is straightforward to check that \eqref{eq-naimark-ext} is satisfied for the choice of $V$ in \eqref{eq-naimark-ext-choice}.
	\end{Proof}
	
	\begin{figure}
		\centering
		\includegraphics[scale=0.8]{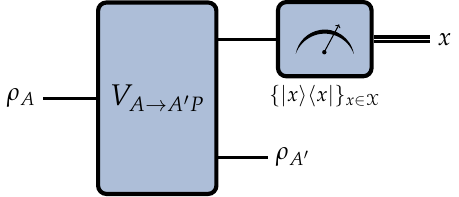}
		\caption{The system-probe model of measurement. In order to measure the system $A$ of interest, we allow it to first interact with a probe system $P$ via an isometry $V_{A\to A'P}$. We then measure the probe with a projective measurement consisting of rank-one elements.}\label{fig-naimark}
	\end{figure}
	
	The physical relevance of Naimark's theorem is illustrated in Figure \ref{fig-naimark}. We model the measurement of the system $A$ of interest as an interaction of the system with a \textit{probe} system $P$ followed by a projective measurement of the probe system described by the POVM $\{\ket{x}\!\bra{x}:x\in\mathcal{X}\}$. The interaction is described by an isometry $V_{A\to A'P}$, and if $\rho_{AP}=V\rho_A V^\dagger$ is the joint state of the system and the probe after the interaction, then the measurement outcome probabilities are
	\begin{align}
		p_X(x)&=\Tr[(\mathbbm{1}_{A'}\otimes\ket{x}\!\bra{x})\rho_{A'P}]\\
		&=\Tr[(\mathbbm{1}_{A'}\otimes\ket{x}\!\bra{x})V\rho_AV^\dagger]\\
		&=\Tr[V^\dagger(\mathbbm{1}_{A'}\otimes\ket{x}\!\bra{x})V\rho_A]\\
		&=\Tr[M_x\rho_A],
	\end{align}
	where we let $M_x\coloneqq V^\dagger(\mathbbm{1}_{A'}\otimes\ket{x}\!\bra{x})V$. This shows us that the system-probe model of measurement, in which the probe is measured after it interacts with the system of interest, can effectively be described by a POVM. Naimark's theorem, on the other hand, tells us the converse: for every measurement described by a POVM, there exists an isometry such that the measurement can be described in the system-probe model, as depicted in Figure \ref{fig-naimark}.
	
	\begin{exercise}{exer-chrysler_POVM}
		Consider the qubit state vectors
		\begin{equation}
			\ket{\psi_k}\coloneqq\cos\!\left(\frac{2\pi k}{5}\right)\ket{0}+\sin\!\left(\frac{2\pi k}{5}\right)\ket{1},\quad k\in\{0,1,2,3,4\}.
		\end{equation}
		Verify that the set $\left\{\frac{2}{5}\ketbra{\psi_k}{\psi_k}\right\}_{k=0}^4$ is a POVM. (This POVM is an example of a non-projective measurement.)
	\end{exercise}
	
	When performing measurements in quantum mechanics, we are typically interested not only in the measurement outcomes and their probabilities but also the \textit{post-measurement} states of the system being measured. That is, we are interested in knowing the state of the system after we have measured it and observed the outcome.
	
	Since every POVM element $M_x$ is positive semi-definite, there exists an operator~$K_x$ such that $M_x=K_x^\dagger K_x$ for all $x\in\mathcal{X}$. For example, we could let $K_x$ be the square root of $M_x$, so that $K_x=\sqrt{M_x}$. Then the Born rule in \eqref{eq-Born_rule} for the probability of the measurement outcome $x\in\mathcal{X}$ can be written as $p_X(x)=\Tr[K_x\rho K_x^\dagger]$. In this case, the post-measurement state corresponding to the outcome $x\in\mathcal{X}$ is as follows:
	\begin{equation}\label{eq-post_meas}
		\rho^x\coloneqq \frac{K_x\rho K_x^\dagger}{\Tr[K_x\rho K_x^\dagger]}.
	\end{equation}
	The state $\rho^x$ can be understood to capture the experimenter's description of the state of the system given that the measurement outcome was observed to be $x$. 
	
		The post-measurement states, labeled by $\rho^x$, give rise to the ensemble $\{(p_X(x),\rho^x)\}_{x\in\mathcal{X}}$. The expected density operator of the ensemble is
	\begin{equation}\label{eq-post_meas_avg}
		\rho_{\mathcal{M}}\coloneqq\sum_{x\in\mathcal{X}}p_X(x)\rho^x=\sum_{x\in\mathcal{X}}K_x\rho K_x^\dagger.
	\end{equation}
	This expected density operator is the state of the system after measurement if the measurement outcome is not available. It can be interpreted as the state of the system after measurement if the experimenter does not have access to the measurement outcome.
	
	Due to the unitary freedom in the decomposition $M_x=K_x^\dagger K_x$, other choices of $K_x$ are given by $K_x=U_x \sqrt{M_x}$ for some unitary $U_x$, so that there is not a unique way to determine the post-measurement state when starting from a POVM. In Section~\ref{sec-quantum_instruments}, we describe an even more general model of measurements, known as the quantum instruments formalism, which includes both the classical outcome and the post-measurement state, while allowing for more general forms of the post-measurement state beyond those discussed above.
	
	 Suppose now that we perform a measurement on a subsystem of a composite system. Specifically, consider measuring a system $A$ that is in a joint state $\rho_{RA}$ with a reference system $R$, and let the measurement be described by the POVM $\{M_A^x\}_{x\in\mathcal{X}}$ for some finite alphabet $\mathcal{X}$. If we let $M_A^x=K_A^{x\dagger}K_A^x$, then according to \eqref{eq-post_meas}, the measurement probabilities are given by $p_X(x)=\Tr[(\mathbbm{1}_R\otimes K_A^x)\rho_{RA}(\mathbbm{1}_R\otimes K_A^{x\dagger})]$ and the post-measurement states are as follows:
	\begin{align}
		\rho_{RA}^x&=\frac{(\mathbbm{1}_R\otimes K_A^x)\rho_{RA}(\mathbbm{1}_R\otimes K_A^{x\dagger})}{\Tr[(\mathbbm{1}_R\otimes K_A^x)\rho_{RA}(\mathbbm{1}_R\otimes K_A^{x\dagger})]}\\
		&=\frac{1}{p_X(x)}(\mathbbm{1}_R\otimes K_A^x)\rho_{RA}(\mathbbm{1}_R\otimes K_A^{x\dagger})
	\end{align}
	for all $x\in\mathcal{X}$. The state of the system $R$ conditioned on the measurement outcome $x$ is then
	\begin{align}
		\rho_R^x&\coloneqq\Tr_A[\rho_{RA}^x]\\
		&=\frac{1}{p_X(x)}\Tr_A[(\mathbbm{1}_R\otimes K_A^x)\rho_{RA}(\mathbbm{1}_R\otimes K_A^{x\dagger})]\\
		&=\frac{1}{p_X(x)}\Tr_A[(\mathbbm{1}_R\otimes K_A^{x\dagger}K_A)\rho_{RA}]\\
		&=\frac{1}{p_X(x)}\Tr_A[(\mathbbm{1}_R\otimes M_A^x)\rho_{RA}].\label{eq-post_meas_state_half}
	\end{align}
	We thus see that, although the post-measurement state on the system $A$ being measured is not uniquely defined due to the unitary freedom in the decomposition $M_A^x=K_A^{x\dagger}K_A^x$, as described earlier, the post-measurement state on the reference system $R$ not being measured is uniquely defined because it depends directly on each POVM element $M_A^x$. If the system $A$ undergoes a measurement for which $M_A^x=\ket{\psi^x}\!\bra{\psi^x}_A$, then \eqref{eq-post_meas_state_half} can be written as
	\begin{equation}\label{eq-post_meas_state_half_proj}
		\rho_R^x=\frac{1}{p_X(x)}\bra{\psi^x}_A\rho_{RA}\ket{\psi^x}_A.
	\end{equation}
	
	\begin{exercise}{exer-PGM}
		Consider an ensemble $\{(p(x),\rho^x)\}_{x\in\mathcal{X}}$ of quantum states, and let $\rho\coloneqq\sum_{x\in\mathcal{X}} p(x) \rho^x$.
		\begin{enumerate}[topsep=0.3cm]
			\item If $\rho$ is invertible, let
				\begin{equation}
					M_x\coloneqq \rho^{-\frac{1}{2}}p(x) \rho^x\rho^{-\frac{1}{2}}\quad\forall~x\in\mathcal{X}.
				\end{equation}
				Prove that $\{M_x\}_{x\in\mathcal{X}}$ is a POVM.
			
			\item If $\rho$ is not invertible, we can still define $M_x \coloneqq \rho^{-\frac{1}{2}}p(x) \rho^x \rho^{-\frac{1}{2}}$ by taking the inverse of $\rho$ on its support. However, in this way the operators $M_x$ no longer add up to the identity. To complete them, we can define
				\begin{equation}
					M_\perp \coloneqq \mathbbm{1}-\Pi_{\rho},
				\end{equation}
				where $\Pi_{\rho}$ is the projection onto the support of $\rho$. Prove that $\{M_x\}_{x\in\mathcal{X} \cup \{\perp\}}$ is a POVM. (\textit{Hint}: Recall the definitions from Section~\ref{sec:math-tools:functions-herm-ops}.)
		\end{enumerate}
	\end{exercise}

\section{Bibliographic Notes}\label{sec:qm:bib-notes}

	More details on many of the concepts that have been presented in this chapter can be found in the  books of \citet{NC00}, \citet{H13book}, \citet{H17}, \citet{Wbook17}, and \citet{Wat18}. For a treatment of these concepts in infinite-dimensional Hilbert spaces, see the books by \citet{H13book} and \citet{HZ12}.
	
	The mathematical theory of quantum mechanics was developed by \citet{Neu1927,Neu32} and \citet{Lan27}. The book by \citet{Hel76} also contains early developments in the theory of quantum measurements. 
	
	We have used quantum-optical modes as a concrete example of qubit encodings throughout this chapter. For a detailed reference on this topic, see the review by \citet{KMNRDM07}.
	
	Recall the vector $\vec{r}_{\rho}$ in \eqref{eq-q_state_coh_vec_repr} of coefficients of a quantum state in terms of the orthogonal basis defined via the operators in \eqref{eq-su_generators_0}--\eqref{eq:math-tools:gell-mann-scaled-mats}. This vector is known as both the ``coherence vector'' and the ``Bloch vector'' of $\rho$. Perhaps the earliest use of the term ``coherence vector'' is in the work of \citet{HE81}, in which the orthogonality convention for the operators in \eqref{eq-su_generators_0}--\eqref{eq:math-tools:gell-mann-scaled-mats} differs from the convention used in this book. Positivity of linear operators in terms of the coherence vector was presented by \citet{BK03,Kim03}. The latter work by \citet{Kim03} uses the term ``Bloch vector'', which has also been used by \citet{Bertlmann_2008}.
	
	Lemma~\ref{lem-symm_purif}, regarding purifications of permutation-invariant states, is due to \citet{RennerThesis}. Separable and entangled states were defined by \citet{Wer89}. For an in-depth review of the properties of entanglement and its various applications in quantum information theory, including a discussion of multipartite entanglement, we refer to the article by \citet{HHHH09}. For an in-depth discussion of multipartite entanglement, we refer to \citep{WGE16}. The relevance of the partial transpose for characterizing entanglement in quantum information was pointed out by \citet{Per96,HHH96}, and the set of positive-partial-transpose states was discussed by \citet{PH97,HHH98}. In particular, the existence of PPT entangled states was found by \citet{PH97}.
	
	Isotropic states were defined by \citet{PhysRevA.59.4206}, and Werner states were defined by \citet{Wer89}. Proofs of formulas for the integration of unitary operators with respect to the Haar measure, including proofs of \eqref{eq-isotropic_integral} and \eqref{eq-Werner_integral}, can be found in \citep{Collins03,CS06,RS09}.

\begin{subappendices}

\section{Proof of Lemma~\ref{lem-app:support-1}}

\label{app-Q-SM-proof-tech-lem-1}

\begin{Proof}
		First suppose that $X_{AB}$ is rank one, so that $X_{AB}=\ketbra{\Psi}{\Psi}_{AB}$ for some vector $|\Psi\rangle_{AB}\in\mathcal{H}_{A}\otimes\mathcal{H}_{B}$. Due to the Schmidt decomposition theorem (Theorem~\ref{thm-Schmidt}), we have that
		\begin{equation}
			|\Psi\rangle_{AB}=\sum_{z\in\mathcal{Z}}\gamma_{z}|\theta_{z}\rangle_{A}\otimes|\xi_{z}\rangle_{B},
		\end{equation}
		where $\left\vert \mathcal{Z}\right\vert \leq\min\{\dim(\mathcal{H}_{A}),\dim(\mathcal{H}_{B})\}$, the set $\{\gamma_{z}\}_z$ is a set of strictly positive numbers, and $\{|\theta_{z}\rangle_{A}\}_z$ and $\{|\xi_{z}\rangle_{B}\}_z$ are orthonormal bases. Then
		\begin{align}
			\supp(X_{AB})  &  =\Span\{|\Psi\rangle_{AB}\}\\
&  \subseteq\Span\{|\theta_{z}\rangle_{A}:z\in\mathcal{Z}
\}\otimes\Span\{|\xi_{z}\rangle_{B}:z\in\mathcal{Z}\}.
		\end{align}
		The statement then follows for this case because $\supp(X_{A})=\Span\{|\theta_{z}\rangle_{A}:z\in\mathcal{Z}\}$ and $\supp(X_{B})=\Span\{|\xi_{z}\rangle_{B}:z\in\mathcal{Z}\}$.

		Now suppose that $X_{AB}$ is not rank one. It admits a decomposition into rank-one vectors of the following form:
		\begin{equation}
			X_{AB}=\sum_{x\in\mathcal{X}}\ketbra{\Psi^{x}}{\Psi^{x}}_{AB},
		\end{equation}
		where $|\Psi^{x}\rangle_{AB}\in\mathcal{H}_{A}\otimes\mathcal{H}_{B}$ for all $x\in\mathcal{X}$. Set $\Psi_{AB}^{x}=|\Psi^{x}\rangle\!\langle\Psi^{x}|_{AB}$, and let $\Psi_{A}^{x}\coloneqq \Tr_{B}[\Psi_{AB}^{x}]$ and $\Psi_{B}^{x}\coloneqq \Tr_{A}[\Psi_{AB}^{x}]$. Then
		\begin{align}
			\supp(X_{AB})  &  =\Span\{|\Psi^{x}\rangle_{AB}:x\in\mathcal{X}\}\\
			&  \subseteq\Span\left[  \bigcup\limits_{x\in\mathcal{X}}\left[\supp(\Psi_{A}^{x})\otimes\supp(\Psi_{B}^{x})\right]  \right] \\
			&  \subseteq\Span\left[  \bigcup\limits_{x\in\mathcal{X}}\supp(\Psi_{A}^{x})\right]  \otimes\Span\left[\bigcup\limits_{x\in\mathcal{X}}\supp(\Psi_{B}^{x})\right] \\
			&  =\supp(X_{A})\otimes\supp(X_{B}),
		\end{align}
		concluding the proof.
	\end{Proof}

\section{Proof of Lemma~\ref{lem-app:support-2}}

\label{app-Q-SM-proof-tech-lem-2}

\begin{Proof}
		First suppose that $X_{AB}$ is rank one, as in the first part of the proof of Lemma~\ref{lem-app:support-1}, and let us use the same notation as given there. Applying the same lemma gives that
		\begin{equation}
			\supp(X_{AB})\subseteq\supp(Y_{AB})\subseteq\supp(Y_{A})\otimes\supp(Y_{B}),
		\end{equation}
		which in turn implies that $\supp(X_{AB})=\Span\{|\Psi\rangle_{AB}\}\subseteq\supp(Y_{A})\otimes\supp(Y_{B})$. This implies that $|\theta_{z}\rangle_{A}\in\supp(Y_{A})$ for all $z\in\mathcal{Z}$, and thus that $\Span\{|\theta_{z}\rangle_{A}\}\subseteq\supp(Y_{A})$. We can then conclude the statement in this case because $\Span\{|\theta_{z}\rangle_{A}\}=\supp(X_{A})$.

		Now suppose that $X_{AB}$ is not rank one. Then it admits a decomposition as given in the proof of the previous lemma. Using the same notation, we have that $\supp(\Psi_{AB}^{x})\subseteq\supp(Y_{AB})$ holds for all $x\in\mathcal{X}$. Since we have proven the lemma for rank-one operators, we can conclude that $\supp(\Psi_{A}^{x})\subseteq\supp(Y_{A})$ holds for all $x\in\mathcal{X}$. As a consequence, we find that
		\begin{equation}
			\supp(X_{A})=\Span\left[  \bigcup\limits_{x\in\mathcal{X}}\supp(\Psi_{A}^{x})\right]  \subseteq \supp(Y_{A}),
		\end{equation}
		concluding the proof.
	\end{Proof}

\end{subappendices}

\chapter{Quantum Channels}\label{chap-QM_channels}

	In the previous chapter, we studied quantum states and measurements, which are regulated by the first three axioms of quantum mechanics, as presented in Section~\ref{sec-QM_axioms}. The fourth and final axiom is about the evolution of quantum systems, which is the subject of this chapter. Mathematically, the evolution is described by a \textit{quantum channel}. As quantum communication necessarily involves the evolution of quantum systems (such as the evolution of photons when travelling through an optical fiber), quantum channels are the primary objects of study in this book. This chapter is devoted to a detailed study of quantum channels, including their properties, representations, and various examples that are relevant for quantum communication and quantum information more broadly.
	
	The fourth axiom in Section~\ref{sec-QM_axioms} states that a quantum channel is a ``linear, completely positive, and trace-preserving map acting on the state of the system.'' At first glance, this appears to be a purely mathematical statement (which we elaborate upon in Section~\ref{sec-QM_channels_basic}), with seemingly little connection to physics. However, we can connect this statement to the axiom of evolution of quantum systems as taught in a basic quantum physics course. There, one learns that the evolution of a (non-relativistic) quantum system is governed by the \textit{Schr\"{o}dinger equation}:
	\begin{equation}\label{eq-Schrodinger}
		\I\hbar\,\frac{\partial}{\partial t}\ket{\psi(t)}=H(t)\ket{\psi(t)},
	\end{equation}
	where $\ket{\psi(t)}$ is the state vector of the system at time $t\geq 0$ and $H(t)$ is the Hamiltonian operator of the system at time $t$. The Hamiltonian operator is a Hermitian operator that describes the energy of the system. Now, we know from Chapter~\ref{chap-QM_states_meas} that the state of a quantum system is described more generally by a density operator. The analogue of \eqref{eq-Schrodinger} for density operators is known as the \textit{von Neumann equation}:
	\begin{equation}\label{eq-von_Neumann}
		\I\hbar\,\frac{\partial\rho(t)}{\partial t}=[H(t),\rho(t)],
	\end{equation}
	where $\rho(t)$ is the density operator describing the state of the system at time $t\geq 0$, and $[H(t),\rho(t)]=H(t)\rho(t)-\rho(t)H(t)$ is the commutator of the Hamiltonian $H(t)$ and the state $\rho(t)$.
	
	Both \eqref{eq-Schrodinger} and \eqref{eq-von_Neumann} describe the evolution of so-called \textit{closed} quantum systems, and this evolution is given by unitary maps. In other words, the solution to \eqref{eq-Schrodinger} is $\ket{\psi(t)}=U(t)\ket{\psi_0}$ for all $t\geq 0$, where $\ket{\psi_0}$ is an initial state vector of the system (at time $t=0$) and $U(t)$ is a unitary operator. Similarly, the solution to \eqref{eq-von_Neumann} is $\rho(t)=U(t)\rho_0U(t)^{\dagger}$ for all $t\geq 0$, where $\rho_0$ is an initial quantum state of the system (at time $t=0$) and $U(t)$ is a unitary operator. We refer to the Bibliographic Notes in Section~\ref{sec:qm_channels:bib-notes} for references on explicit forms for the unitary $U(t)$. We show in this chapter that unitary maps are quantum channels. This fact provides a connection between the mathematical statement of the evolution axiom in Section~\ref{sec-QM_axioms} and the statement of the evolution axiom typically taught in quantum physics courses.
	
	More generally, we are interested in the evolution of \textit{open} quantum systems, i.e., quantum systems that interact with an external environment that is out of our control. For such systems, the same connection as before holds. In fact, the evolution is given by a joint unitary evolution of the system and environment followed by discarding the state of the environment, and as we show in Section~\ref{sec-QM_channel_characterization}, every completely positive trace-preserving map (i.e., every quantum channel) can be viewed in terms of a joint unitary evolution with an environment followed by discarding the state of the environment. (Please see the Bibliographic Notes in Section~\ref{sec:qm_channels:bib-notes} for references on open quantum systems.) Thus, from an abstract, information-theoretic perspective, the evolution of a quantum system is given simply by a quantum channel, and the details of the actual physical system of interest (which would be given by the Hamiltonian operator) are unimportant. This viewpoint is powerful: with it, we realize that virtually every operation on quantum states, including measurements, is a quantum channel.

\section{Definition}\label{sec-QM_channels_basic}

	We can motivate the definition of a quantum channel by using the following basic mathematical facts that should be satisfied by a map $\mathcal{N}:\Lin(\mathcal{H})\to\Lin(\mathcal{H}')$ that represents the evolution of a quantum system:
	
	\begin{enumerate}
		\item If $\mathcal{N}$ acts on a mixture of quantum states, then the output state should be equal to the mixture of the individual outputs. That is,
		\begin{equation}
		\mathcal{N}(\lambda \rho + (1-\lambda)\sigma) = 
		\lambda \mathcal{N}(\rho) + (1-\lambda)\mathcal{N}(\sigma)
		\end{equation}
		for all states $\rho$ and $\sigma$ and $\lambda\in[0,1]$. This requirement is called convex linearity on density operators. Importantly, for each convex linear map acting on the convex set of density operators it is possible to define a unique linear map acting on the space of all linear operators that is an extension of the original map, i.e., that reproduces its action on density operators. The latter is the mathematical object that we employ, and so we require that $\mathcal{N}$ be a linear map, i.e., a superoperator. (Recall the definition of a superoperator from Section~\ref{sec:math-tools:super-ops}.)
		
		\item The map $\mathcal{N}$ should accept a quantum state (or a mixture of quantum states) as input and output a legitimate quantum state. This means that $\mathcal{N}$ should be \textit{trace preserving} and \textit{positive}. However, it is furthermore reasonable to demand that if the channel acts on one share $A$ of a bipartite quantum state $\rho_{RA}$, then the output should be a legitimate bipartite quantum state. So we demand additionally that a quantum channel should be not just positive, but additionally \textit{completely positive}. Let us now define these terms.
			\begin{enumerate}
				\item $\mathcal{N}$ is called \textit{trace preserving} if $\Tr[\mathcal{N}(X)]=\Tr[X]$ for every linear operator~$X$. More generally, $\mathcal{N}$ is called \textit{trace non-increasing} if $\Tr[\mathcal{N}(X)]\leq\Tr[X]$ for every positive semi-definite operator $X$.
				
				\item $\mathcal{N}$ is called \textit{positive} if it maps positive semi-definite operators to positive semi-definite operators, i.e., $\mathcal{N}(X)\geq 0$ for all $X\geq 0$. It is called \textit{$k$-positive}, with $k\geq 1$, if the map $\id_k\otimes\mathcal{N}$ is positive. Note that if $\mathcal{N}$ is a map acting on linear operators in $\Lin(\mathbb{C}^{d})$, then the map $\id_k\otimes\mathcal{N}$ acts on linear operators in $\Lin(\mathbb{C}^{kd})$. In other words, for every linear operator $X$ acting on a $kd$-dimensional Hilbert space, which we can decompose as the block matrix
					\begin{equation}
						X=\begin{pmatrix} X_{0,0} & \dotsb & X_{0,k-1} \\ \vdots & \ddots & \vdots \\ X_{k-1,0} & \dotsb & X_{k-1,k-1} \end{pmatrix},
					\end{equation}
					such that $X_{i,j}$ is a $d\times d$ matrix for all $0\leq i,j\leq k-1$, the action of the map $\id_k\otimes\mathcal{N}$ is defined as
					\begin{equation}
						(\id_k\otimes\mathcal{N})(X)=\begin{pmatrix}\mathcal{N}(X_{0,0}) & \dotsb & \mathcal{N}(X_{0,k-1}) \\ \vdots & \ddots & \vdots \\ \mathcal{N}(X_{k-1,0}) & \hdots & \mathcal{N}(X_{k-1,k-1}) \end{pmatrix}.
					\end{equation}
					We can write this more compactly as follows. Noting that $\Lin(\mathbb{C}^{kd}) $ is isomorphic to $\Lin(\mathbb{C}^{k})\otimes \Lin(\mathbb{C}^{d}) $, we can write $X \in \Lin(\mathbb{C}^{kd})$ as
					\begin{equation}
						X = \sum_{i,j=0}^{k-1} \ket{i}\!\bra{j} \otimes X_{i,j}.
					\end{equation}
					Then the action of $\id_k \otimes \mathcal{N}$ is defined as
					\begin{equation}
						(\id_k \otimes \mathcal{N})(X) = \sum_{i,j=0}^{k-1} \ket{i}\!\bra{j} \otimes \mathcal{N}(X_{i,j}).
					\end{equation}
					The superoperator $\mathcal{N}$ is called \textit{completely positive} if $\id_k\otimes\mathcal{N}$ is positive for every integer $k\geq 1$.
					
					Physically, the complete positivity of $\mathcal{N}$ takes into account the fact that the system of interest might be entangled with another system that is outside of our control, so that simply letting $\mathcal{N}$ be positive is not sufficient to ensure that all positive semi-definite operators get mapped to positive semi-definite operators. Letting $\mathcal{N}$ be completely positive means that positive semi-definite operators are mapped to positive semi-definite operators even in this more general setting.
					
			\end{enumerate}
		
	\end{enumerate}
	
	The defining properties of linearity, complete positivity, and trace preservation for quantum channels together ensure that quantum states for the systems of interest get mapped to quantum states, even if they happen to be entangled with other external systems outside of our control. These properties are consistent with what is observed in real physical systems.
	
	\begin{figure}
		\centering
		\includegraphics[scale=0.8]{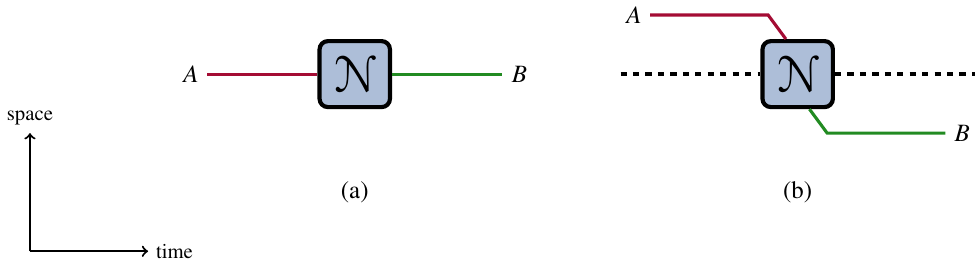}
		\caption{Our convention for drawing quantum channels throughout this book, with time increasing horizontally towards the right, and spatial separations indicated vertically. In (a), the input and output systems $A$ and $B$, respectively, of the quantum channel $\mathcal{N}$ are temporally separated but not spatially separated. In (b), $A$ and $B$ are both spatially and temporally separated. We often draw a dashed line to indicate the spatial separation explicitly.}\label{fig-q_channel_space_time}
	\end{figure}
	
	Throughout this book, we write $\mathcal{N}_{A\to B}$ to denote a map $\mathcal{N}:\Lin(\mathcal{H}_A)\to\Lin(\mathcal{H}_B)$ taking a quantum system $A$ to a quantum system $B$. We sometimes write $\mathcal{N}_A$ if the input and output systems of the channel have the same dimension. We drop the subscript indicating the input and output systems if they are not important in the context being considered. Physically, a quantum channel, being a description of the time evolution of a quantum system, describes the transition of a quantum system between two distinct points in time. The systems $A$ and $B$ at the input and output of the channel, respectively, thus represent quantum systems at two distinct points in time, as shown in Figure~\ref{fig-q_channel_space_time}(a). However, in the context of communication, we regard the systems $A$ and $B$ as being separated both in time as well as in space, with $A$ belonging to an individual ``Alice'' and $B$ belonging to ``Bob.'' We show this physical separation explicitly throughout this book according to the convention shown in Figure~\ref{fig-q_channel_space_time}(b).
	
	\begin{exercise}{exer-channel_Herm_pres}
		Let $\mathcal{N}$ be a $k$-positive superoperator, for an integer $k\geq 1$. Prove that $\mathcal{N}$ is Hermiticity preserving (recall Definition~\ref{def-Herm_pres_superop}). (\textit{Hint}: See Exercise~\ref{exer-lin_op_span_PSD} and use the Jordan--Hahn decomposition.)
	\end{exercise}
	
	\begin{exercise}{exer-TNI_subunital}
		Prove that a  Hermiticity-preserving superoperator $\mathcal{N}_{A\to B}$ is trace-non-increasing if and only if its adjoint is \textit{subunital}, meaning that $\mathcal{N}^{\dagger}(\mathbbm{1}_B)\leq\mathbbm{1}_A$. Prove that the inequality is saturated, i.e., that $\mathcal{N}^{\dagger}(\mathbbm{1}_B)=\mathbbm{1}_A$ ($\mathcal{N}^{\dagger}$ is unital), if and only if $\mathcal{N}$ is trace preserving.
	\end{exercise}
	
	\begin{exercise}{exer-channel_diamond_norm}
		\begin{enumerate}
			\item Let $\mathcal{N}_{A\to B}$ be a positive superoperator. Starting with \eqref{eq-induced_Tr_norm_adjoint}, prove that
				\begin{equation}
					\norm{\mathcal{N}}_1=\norm{\mathcal{N}^{\dagger}(\mathbbm{1}_B)}_{\infty}.
				\end{equation}
			
			\item Using 1., conclude the following: \begin{enumerate}
			\item If $\mathcal{N}_{A\to B}$ is a $k$-positive, trace-non-increasing superoperator, then $\norm{\id_k\otimes\mathcal{N}}_1\leq 1$;
			\item  If $\mathcal{N}_{A\to B}$ is a $k$-positive, trace-preserving superoperator, then $\norm{\id_k\otimes\mathcal{N}}_1=1$.
			\end{enumerate}
			
			\item Using 1. and 2., conclude the following:
			\begin{enumerate}
			\item If $\mathcal{N}_{A\to B}$ is a completely positive, trace-non-increasing superoperator, then $\norm{\mathcal{N}}_{\diamond}\leq 1$ (the diamond norm $\norm{\cdot}_{\diamond}$ is introduced in Definition~\ref{def-diamond_norm_superop});
			\item If $\mathcal{N}_{A\to B}$ is a quantum channel, then $\norm{\mathcal{N}}_{\diamond}=1$.
			\end{enumerate}
		\end{enumerate}
	\end{exercise}
	
	Combining the result of Exercise~\ref{exer-channel_diamond_norm} and \eqref{eq-induced_trace_norm_2}, we conclude that for every positive, trace-non-increasing superoperator $\mathcal{N}_{A\to B}$, and every linear operator $X\in\Lin(\mathcal{H}_A)$,
	\begin{equation}\label{eq-data_proc_trace_norm_0}
		\norm{\mathcal{N}(X)}_1\leq\norm{X}_1.
	\end{equation}
	An inequality of this type is called a \textit{data-processing inequality}, for which we provide an interpretation later on in Section~\ref{sec-QM-trace-distance}. We encounter numerous such inequalities with respect to various different quantities throughout the rest of this book, and they turn out to be of central importance in the analysis of quantum communication protocols, and in quantum information theory more generally.

\section{Choi Representation}
	
	The Choi representation gives a way to represent a quantum channel as a bipartite operator and is an essential concept in quantum information theory.
	
	\begin{definition}{Choi Representation}{def-Choi_rep}
		For every superoperator $\mathcal{N}_{A\to B}$, its \textit{Choi representation}, or \textit{Choi operator}, is defined as
		\begin{equation}\label{eq-Choi_rep}
			\Gamma^{\mathcal{N}}_{AB}\coloneqq\mathcal{N}_{A'\to B}(\ket{\Gamma}\!\bra{\Gamma}_{AA'})=\sum_{i,j=0}^{d_A-1}\ket{i}\!\bra{j}_A\otimes\mathcal{N}(\ket{i}\!\bra{j}_{A'}),
		\end{equation}
		where $\mathcal{H}_{A'}$ is isomorphic to the Hilbert space $\mathcal{H}_{A}$ corresponding to the channel input system $A$. We also define the operator
		\begin{equation}\label{eq-Choi_state}
			\Phi_{AB}^{\mathcal{N}}\coloneqq\frac{1}{d_A}\Gamma_{AB}^{\mathcal{N}}=\mathcal{N}_{A'\to B}(\Phi_{AA'}),
		\end{equation}
		where $\Phi_{AA'}=\ket{\Phi}\!\bra{\Phi}_{AA'}$ is the maximally entangled state defined in~\eqref{eq-max_ent_state}.
	\end{definition}
	
	\begin{figure}
		\centering
		\includegraphics[scale=0.8]{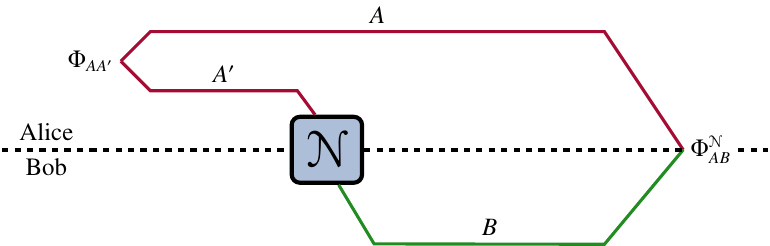}
		\caption{The normalized Choi representation $\Phi_{AB}^{\mathcal{N}}$ of a superoperator $\mathcal{N}_{A\to B}$ is the bipartite operator resulting from sending one share of the maximally entangled state $\Phi_{AA'}$, defined in \eqref{eq-max_ent_state}, through $\mathcal{N}$.}\label{fig-Choi_state}
	\end{figure}
	
	The Choi representation $\Gamma_{AB}^{\mathcal{N}}$ of the superoperator $\mathcal{N}_{A\to B}$ is an operator acting on $\mathcal{H}_{AB}$ and uniquely characterizes the map because it specifies the action of $\mathcal{N}$ on the basis $\{\ket{i}\!\bra{j}_A:0\leq i,j\leq d_A - 1\}$ of linear operators acting on~$\mathcal{H}_A$. As shown in Figure~\ref{fig-Choi_state}, the operator $\Phi_{AB}^{\mathcal{N}}$ is simply the Choi representation normalized by the dimension $d_A$ of the input system $A$ of the superoperator $\mathcal{N}$, and it is the linear operator resulting from sending one share of a maximally entangled state through~$\mathcal{N}$. When $\mathcal{N}$ is a quantum channel, we refer to $\Phi_{AB}^{\mathcal{N}}$ as the \textit{Choi state} of $\mathcal{N}$, because $\Phi_{AB}^{\mathcal{N}}$ is positive semi-definite and has unit trace; see Theorem~\ref{thm-q_channels} below.
	
	\begin{exercise}{exer-Choi_rep_0}
		Let $\mathcal{N}_{A\to B}$ be a superoperator.
		\begin{enumerate}[topsep=0.3cm]
			\item Prove that $\Tr_A[\Gamma_{AB}^{\mathcal{N}}]=\mathcal{N}_{A\to B}(\mathbbm{1}_A)$. 
			
			\item Prove that $\mathcal{N}_{A\to B}$ is trace preserving if and only if $\Tr_B[\Gamma_{AB}^{\mathcal{N}}]=\mathbbm{1}_A$. Conclude that $\Tr[\Phi_{AB}^{\mathcal{N}}]=1$.
			
			\item Prove that $\Inner{\conj{X_A}\otimes Y_B}{\Gamma_{AB}^{\mathcal{N}}}=\inner{Y_B}{\mathcal{N}_{A\to B}(X_A)}$ for all $X_A\in\Lin(\mathcal{H}_A)$ and $Y_B\in\Lin(\mathcal{H}_B)$.
			
			\item Using 3., prove that the Choi representation of $\mathcal{N}$ can be expressed using the adjoint $\mathcal{N}^{\dagger}$ as follows:
				\begin{equation}
					\Gamma_{AB}^{\mathcal{N}}=\sum_{k,\ell=0}^{d_B-1}\conj{\mathcal{N}^{\dagger}(\ket{k}\bra{\ell}_B)}\otimes\ket{k}\bra{\ell}_B.
				\end{equation}
				Conclude that $\Tr_B[\Gamma_{AB}^{\mathcal{N}}]=\conj{\mathcal{N}^{\dagger}(\mathbbm{1}_B)}$.
				
			\item Prove that, for every unitary operator $U_A$,
				\begin{equation}
					\Gamma_{AB}^{\mathcal{N}}=\big(\mathcal{U}_A\otimes\big(\mathcal{N}_{A'\to B}\circ\conj{\mathcal{U}}_{A'}\big)\big)(\Gamma_{AA'}),
				\end{equation}
				where $\mathcal{U}_A(\cdot)\coloneqq U_A(\cdot)U_A^{\dagger}$ and $\conj{\mathcal{U}}_A(\cdot)\coloneqq\conj{U}_A(\cdot)\conj{U}_A^{\dagger}$.
		\end{enumerate}
	\end{exercise}
	
	\begin{proposition*}{Quantum Dynamics from Choi Operator}{prop:qm:dyn-from-choi}
	Let $\mathcal{N}_{A\to B}$ be a superoperator, and let $X_{RA}$ be a bipartite operator, with $R$ an arbitrary reference system. Then the action of the superoperator $\id_R\otimes \mathcal{N}_{A\to B}$ on $X_{RA}$ can be expressed in terms of the Choi operator $\Gamma_{AB}^{\mathcal{N}}$ as follows:
		\begin{align}
	(\id_R\otimes\mathcal{N}_{A\to B})(X_{RA}) & =
	\Tr_A[(\T_A(X_{RA}) \otimes\mathbbm{1}_B)(\mathbbm{1}_R \otimes \Gamma_{AB}^{\mathcal{N}})], 	\label{eq-Choi_rep_action}\\
		 & =\bra{\Gamma}_{A'A}(X_{RA'}\otimes\Gamma_{AB}^{\mathcal{N}})\ket{\Gamma}_{A'A},
		\label{eq-QM:post-selected-TP-Choi-op}
	\end{align}
	where $\T_A$ denotes the partial transpose from Definition~\ref{def-partial_transpose}.
	\end{proposition*}
	
	\begin{Proof}
		Observe that \eqref{eq-Choi_rep} implies that
		\begin{equation}
			\mathcal{N}_{A\to B}(\ket{i}\!\bra{j}_A)=(\bra{i}_A\otimes\mathbbm{1}_B)\Gamma_{AB}^{\mathcal{N}}(\ket{j}_A\otimes\mathbbm{1}_B)
		\end{equation}
		for all $0\leq i,j\leq d_A - 1$. We extend this by linearity to apply to every input operator~$X_A$ by the following reasoning. Expanding $X_A$ as $X_A = \sum_{i,j=0}^{d_A-1} X_{i,j} \ket{i}\!\bra{j}_A$, we find that
		\begin{align}
		\mathcal{N}_{A\to B}(X_A) & = \sum_{i,j=0}^{d_A-1} X_{i,j}	\mathcal{N}_{A\to B}(\ket{i}\!\bra{j}_A)\\
		& =
		\sum_{i,j=0}^{d_A-1} X_{i,j}
		(\bra{i}_A\otimes\mathbbm{1}_B)\Gamma_{AB}^{\mathcal{N}}(\ket{j}_A\otimes\mathbbm{1}_B) \\
	 &  = \sum_{i,j=0}^{d_A-1} X_{i,j}
		\Tr_A[ (\ket{j}\!\bra{i}_A\otimes\mathbbm{1}_B)\Gamma_{AB}^{\mathcal{N}}]\\
		& = \Tr_A[(X_A^{\t}\otimes\mathbbm{1}_B)\Gamma_{AB}^{\mathcal{N}}].
		\end{align}
		So we conclude that the action of $\mathcal{N}_{A\to B}$ on every linear operator $X_A$ can be expressed using the Choi representation as
		\begin{equation}
			\mathcal{N}_{A\to B}(X_A)=\Tr_A[(X_A^{\t}\otimes\mathbbm{1}_B)\Gamma_{AB}^{\mathcal{N}}].
		\end{equation}
		Now employing \eqref{eq-trace_identity} and \eqref{eq-transpose_trick}, we conclude that the action of $\mathcal{N}_{A\to B}$ on every linear operator $X_A$ can be expressed alternatively as
		\begin{equation}
			\mathcal{N}_{A\to B}(X_{A})=\bra{\Gamma}_{A'A}(X_{A'}\otimes\Gamma_{AB}^{\mathcal{N}})\ket{\Gamma}_{A'A}.
			\label{eq-QM:post-selected-TP-Choi-op-simple-1}
		\end{equation}
	The identities in \eqref{eq-Choi_rep_action} and \eqref{eq-QM:post-selected-TP-Choi-op-simple-1} extend more generally to the case of the superoperator $\id_R \otimes \mathcal{N}_{A\to B}$ acting on a bipartite operator $X_{RA}$ by expanding $X_{RA}$ as $X_{RA} = \sum_{i,j=0}^{d_R-1} \ket{i}\!\bra{j}_R \otimes X^{i,j}_A$ and using linearity. We thus conclude \eqref{eq-Choi_rep_action} and \eqref{eq-QM:post-selected-TP-Choi-op}.
	\end{Proof}
	
	\begin{exercise}{exer-Choi_rep_Herm_pres}
		Prove that a superoperator $\mathcal{N}_{A\to B}$ is Hermiticity preserving (recall the definition in Section~\ref{sec:math-tools:super-ops}) if and only if its Choi representation $\Gamma_{AB}^{\mathcal{N}}$ is Hermitian.
	\end{exercise}
	
	Using the definition of the Choi state and the maximally entangled state~$\Phi_{A'A}$, we can write \eqref{eq-QM:post-selected-TP-Choi-op} as
	\begin{equation}\label{eq-Choi_state_post_selected_teleportation}
		\bra{\Phi}_{A'A}(X_{RA'}\otimes\Phi_{AB}^{\mathcal{N}})\ket{\Phi}_{A'A}=\frac{1}{d_A^2}\,\mathcal{N}_{A'\to B}(X_{RA'}).
	\end{equation}
	Comparing this equation with \eqref{eq-post_meas_state_half_proj}, we see that it has the following physical interpretation: if we start with the systems $R$, $A'$, $A$ and $B$ in the state $\rho_{RA'}\otimes\Phi_{AB}^{\mathcal{N}}$ and we measure $A'$ and $A$ according to the POVM $\{\ket{\Phi}\!\bra{\Phi}_{A'A},\mathbbm{1}_{A'A}\allowbreak-\ket{\Phi}\!\bra{\Phi}_{A'A}\}$, then the outcome corresponding to $\ket{\Phi}\!\bra{\Phi}_{A'A}$ occurs with probability $\frac{1}{d_A^2}$ and the post-measurement state on systems $R$ and $B$ is $\mathcal{N}_{A'\to B}(\rho_{RA'})$. The Choi state $\Phi_{AB}^{\mathcal{N}}$ can thus be viewed as a \textit{resource state} for the probabilistic implementation of the channel $\mathcal{N}$. We return to this point in Section~\ref{sec-teleportation} when we discuss post-selected quantum teleportation.
	
	The concept of Choi state allows us to associate to each quantum channel $\mathcal{N}_{A\to B}$ a bipartite quantum state. Conversely, given a bipartite state $\rho_{AB}$, we can associate a map given by
	\begin{equation}
		X_A\mapsto d_A\Tr_A[(X_A^{\t}\otimes\mathbbm{1}_B)\rho_{AB}].
	\end{equation}
	It is straightforward to see that this map is completely positive; however, it is trace preserving if and only if $\Tr_B[\rho_{AB}]=\pi_A$. On the other hand, the map $\mathcal{N}_{A\to B}^{\rho}$ defined as
	\begin{equation}\label{eq-state_to_channel}
		\mathcal{N}_{A\to B}^{\rho}(X_A)\coloneqq\Tr_A\!\left[(X_A^{\t}\otimes\mathbbm{1}_B)\rho_A^{-\frac{1}{2}}\rho_{AB}\rho_A^{-\frac{1}{2}}\right]
	\end{equation}
	is a quantum channel whenever $\rho_A$ is positive definite, where $\rho_A=\Tr_B[\rho_{AB}]$. The operator $\rho_A^{-\frac{1}{2}}\rho_{AB}\rho_A^{-\frac{1}{2}}$ is sometimes called a ``conditional state,'' motivated by the fact that it reduces to a conditional probability distribution when $\rho_{AB}$ is a fully classical state, so that it can be written as $\rho_{AB} = \sum_{x,y} p(x,y) \ket{x}\!\bra{x}_A \otimes \ket{y}\!\bra{y}_B$ where $p(x,y)$ is a probability distribution. Note that if $\rho_A$ is not invertible, then the inverse in \eqref{eq-state_to_channel} should be taken on the support of $\rho_A$, 
 in which case the channel $\mathcal{N}_{A\to B}^{\rho}$ is defined as in \eqref{eq-state_to_channel} only on the support of $\rho_A$.

	\begin{exercise}{exer-Choi_rep}
		\begin{enumerate}
			\item Given two superoperators $(\mathcal{N}_1)_{A_1\to B_1}$ and $(\mathcal{N}_2)_{A_2\to B_2}$, prove that the Choi representation of the tensor-product superoperator $\mathcal{N}_1\otimes\mathcal{N}_2$ is given by
				\begin{equation}\label{eq-Choi_rep_tensor_prod}
					\Gamma_{A_1A_2B_1B_2}^{\mathcal{N}_1\otimes\mathcal{N}_2}=\Gamma_{A_1B_1}^{\mathcal{N}_1}\otimes\Gamma_{A_2B_2}^{\mathcal{N}_2}.
				\end{equation}
				
			\item Given two superoperators $\mathcal{N}_{A\to B}$ and $\mathcal{M}_{B\to C}$, prove that the Choi representation of the composition $(\mathcal{M}\circ\mathcal{N})_{A\to C}$ is given by
				\begin{align}
					\Gamma_{AC}^{\mathcal{M}\circ\mathcal{N}} & = \mathcal{M}_{B\to C}(\Gamma_{AB}^{\mathcal{N}}) \\
					& = \Tr_B[ \T_B(\Gamma_{AB}^{\mathcal{N}}) \Gamma_{BC}^{\mathcal{M}}] \\
					& = \bra{\Gamma}_{BB'}
					\Gamma_{AB}^{\mathcal{N}} \otimes 
					\Gamma_{B'C}^{\mathcal{M}}
					\ket{\Gamma}_{BB'}.	
					\label{eq-QM:serial-comp-choi}	
				\end{align}
		\end{enumerate}
	\end{exercise}
	
	\begin{exercise}{exer-diamond_norm_Choi_bounds}
		Let $\mathcal{N}_{A\to B}$ be a superoperator. Prove that
		\begin{equation}
			\frac{1}{d_A}\norm{\Gamma_{AB}^{\mathcal{N}}}_1\leq\norm{\mathcal{N}}_{\diamond}\leq\norm{\Gamma_{AB}^{\mathcal{N}}}_1.
		\end{equation}
		(\textit{Hint}: Start with Theorem~\ref{thm-diamond_norm}. Then, for the right-most inequality, start with the discussion around \eqref{eq-pure_state_vec} and then use \eqref{eq-math_tools_submult_alpha_strong}.)
	\end{exercise}

\section{Characterizations of Channels: Choi, Kraus, Stinespring}\label{sec-QM_channel_characterization}

	The following theorem provides three useful ways to characterize quantum channels, and as such, it is one of the most important theorems in quantum information theory.
	
	\begin{theorem*}{Characterizations of Quantum Channels}{thm-q_channels}
		Let $\mathcal{N}_{A\to B}$ be a linear map from $\Lin(\mathcal{H}_A)$ to $\Lin(\mathcal{H}_B)$. Then the following are equivalent:
		\begin{enumerate}
			\item $\mathcal{N}$ is a quantum channel.
			\item \textit{Choi}: The Choi representation $\Gamma_{AB}^{\mathcal{N}}$ is positive semi-definite and satisfies $\Tr_B[\Gamma_{AB}^{\mathcal{N}}]=\mathbbm{1}_A$.
			
			\item \textit{Kraus}: There exists a set $\{K_i\}_{i=1}^r$ of finitely many operators, called \textit{Kraus operators}, such that
				\begin{equation}\label{eq-q_channel_Kraus}
					\mathcal{N}(X_A)=\sum_{i=1}^r K_iX_AK_i^\dagger
				\end{equation}
				for every linear operator $X_A$, where $K_i\in\Lin(\mathcal{H}_A,\mathcal{H}_B)$ for all $i\in \{1,\dotsc, r\}$, and $\sum_{i=1}^r K_i^\dagger K_i=\mathbbm{1}_A$.
			
			\item \textit{Stinespring}: There exists an isometry $V_{A\to BE}$, called an \textit{isometric extension}, such that
				\begin{equation}\label{eq-q_channel_Stinespring}
					\mathcal{N}(X_A)=\Tr_E[VX_AV^\dagger]
				\end{equation}
				for every linear operator $X_A$.
		\end{enumerate}
	\end{theorem*}
	
	Please consult the Bibliographic Notes in  Section~\ref{sec:qm:bib-notes} for references containing a proof of this theorem.

        \begin{remark}
            The number $r$ of Kraus operators in part~3 of Theorem~\ref{thm-q_channels} can be chosen to be equal to the rank of the Choi matrix, i.e., $r = \rank(\Gamma_{AB}^{\mathcal{N}})$. Note that this is the minimal number of Kraus operators needed in a Kraus decomposition of the channel.
            Similarly, the dimension $d_E$ of the environment $E$ appearing in the Stinespring extension in part~4 can be chosen to be equal to $\rank(\Gamma_{AB}^{\mathcal{N}})$, and this is the minimal dimension of an environment $E$ needed in a Stinespring representation of the channel.
        \end{remark}
        
	\begin{remark}
		Theorem~\ref{thm-q_channels} holds for quantum channels, i.e., completely positive trace-preserving maps. More generally, if $\mathcal{N}_{A\to B}$ is completely positive and trace non-increasing, then the trace-preserving condition for the Choi, Kraus, and Stinespring representations of $\mathcal{N}_{A\to B}$ changes as follows.
		\begin{itemize}
			\item The trace-preserving condition $\Tr_B[\Gamma_{AB}^{\mathcal{N}}]=\mathbbm{1}_A$ on the Choi representation of $\mathcal{N}_{A\to B}$ changes to $\Tr_B[\Gamma_{AB}^{\mathcal{N}}]\leq\mathbbm{1}_A$ when $\mathcal{N}_{A\to B}$ is trace non-increasing.
			
			\item The trace-preserving condition $\sum_{i=1}^r K_i^{\dagger}K_i=\mathbbm{1}_A$ on a set of Kraus operators changes to $\sum_{i=1}^r K_i^{\dagger}K_i\leq\mathbbm{1}_A$ when $\mathcal{N}_{A\to B}$ is trace non-increasing.
			
			\item The isometric property of the operator $V$ in \eqref{eq-q_channel_Stinespring}, which corresponds to the trace-preserving property of every quantum channel, changes to $V^{\dagger}V\leq\mathbbm{1}_A$ when $\mathcal{N}_{A\to B}$ is trace non-increasing.
		\end{itemize}
		Completely positive trace-non-increasing maps arise in the context of quantum instruments, which we discuss in Section~\ref{sec-quantum_instruments}.
	\end{remark}

	The Kraus operators $K_i$ in \eqref{eq-q_channel_Kraus} have an interpretation in quantum error correction as ``error operators'' that characterize various errors that a quantum system undergoes. Kraus operators for a given quantum channel are, however, not unique in general. If $\{K_i\}_{i=1}^r$ is a set of Kraus operators for the channel $\mathcal{N}$, then, given an $s\times r$ isometric matrix $V$ with elements $\{V_{i,j}: 1\leq i\leq s,~1\leq j\leq r\}$, the operators $\{K_i'\}_{i=1}^s$ defined as
	\begin{equation}
		K_i'=\sum_{j=1}^r V_{i,j}K_j
		\label{eq:qm:isometry-between-kraus-ops}
	\end{equation}
	are also Kraus operators for $\mathcal{N}$. Indeed, for every linear operator $X$, the following equality holds
	\begin{align}
		\sum_{i=1}^s K_i' X (K_i')^\dagger&=\sum_{i=1}^s\sum_{j,j'=1}^r V_{i,j}\conj{V_{i,j'}}K_j X K_{j'}^\dagger\\
		&=\sum_{j,j'=1}^r \underbrace{\left(\sum_{i=1}^s (V^\dagger)_{j',i} V_{i,j}\right)}_{(V^\dagger V)_{j',j}=\delta_{j',j}} K_j X K_{j'}^\dagger\\
		&=\sum_{j=1}^r K_j X K_j^\dagger\\
		&=\mathcal{N}(X), \label{eq-QM-channels:kraus-rel-by-unitaries}
	\end{align}
	where  the second equality follows because $\conj{V_{i,j'}}=(V^\dagger)_{j',i}$.
	
	We note also that a converse statement holds: if $\{K_i\}_{i=1}^r$ and $\{K_i'\}_{i=1}^s$ are two sets of Kraus operators that realize the same quantum channel, then they are related by an isometry as in \eqref{eq:qm:isometry-between-kraus-ops}. This is a dynamical version of the statement made earlier in Section~\ref{sec:qm:purification-def}, the statement there being that all purifications of a state are related by an isometry acting on the purifying system.
	
	\begin{exercise}{exer-Choi_Kraus}
		Show that the Choi representation of a quantum channel $\mathcal{N}_{A\to B}$ can be expressed using a set $\{K_i\}_{i=1}^r$ of its Kraus operators, with $r\geq\rank(\Gamma_{AB}^{\mathcal{N}})$, as
		\begin{equation}
			\Gamma_{AB}^{\mathcal{N}}=\sum_{i=1}^r\text{vec}(K_i)\text{vec}(K_i)^{\dagger}.
		\end{equation}
	\end{exercise}
	
	\begin{figure}
		\centering
		\includegraphics[scale=0.8]{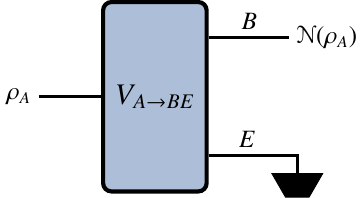}\qquad
		\includegraphics[scale=0.8]{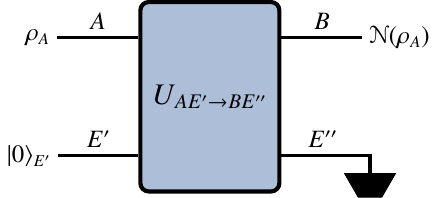}
		\caption{(Left) According to Stinespring's theorem, the evolution of every quantum system $A$ via a quantum channel $\mathcal{N}_{A\to B}$ can be described as an interaction of the system $A$ with its environment $E$ via an isometry $V_{A\to BE}$, followed by discarding the environment. (Right) The isometry $V_{A\to BE}$ can be extended to a unitary $U_{AE'\to BE''}$ using, e.g., the construction in \eqref{eq-chan_iso_ext_to_unitary}, such that $A$ and $E'$ are initially in a product state, with $E'$ starting in a pure state. The two systems then interact accoring to $U$, and after discarding the environment $E''$, the resulting state is the output of the channel.}\label{fig-stinespring}
	\end{figure}

	Physically, the isometric extension $V_{A\to BE}$ in \eqref{eq-q_channel_Stinespring} can be thought of as modelling the interaction of the quantum system of interest with its \textit{environment}, i.e., anything external to the quantum system that is not under our control. Stinespring's theorem then tells us that the evolution of every quantum system can be represented as an interaction of the system with its environment, which is then later discarded; see Figure \ref{fig-stinespring}. Given a set $\{K_i\}_{i=1}^{r}$ of Kraus operators for $\mathcal{N}$, we can let the environment $E$ correspond to a space of dimension $r$ and define the isometry $V_{A\to BE}$ as
	\begin{equation}\label{eq-Kraus_to_Stinespring}
		V_{A\to BE}=\sum_{j=1}^{r} K_j\otimes\ket{j-1}_E.
	\end{equation}
	It is straightforward to show that this is indeed an isometric extension of $\mathcal{N}$ since $\Tr_E[VXV^\dagger]=\sum_{j=1}^{r} K_j X K_j^\dagger=\mathcal{N}(X)$.
	
	\begin{exercise}{exer-Choi_Stinespring}
		Show that the Choi representation of a quantum channel $\mathcal{N}_{A\to B}$ can be expressed using an isometric extension $V_{A\to BE}$ of $\mathcal{N}$, with $d_E\geq\rank(\Gamma_{AB}^{\mathcal{N}})$, as
		\begin{equation}
			\Gamma_{AB}^{\mathcal{N}}=\Tr_E[\text{vec}(V)\text{vec}(V)^{\dagger}].
		\end{equation}
		Hence, conclude that $\frac{1}{\sqrt{d_A}}\,\text{vec}(V)=(\mathbbm{1}_A\otimes V_{A'\to BE})\ket{\Phi}_{AA'}$ is a purification of the Choi state $\Phi_{AB}^{\mathcal{N}}$ of $\mathcal{N}$.
	\end{exercise}
	
	\begin{exercise}{exer-adjoint_unital}
		Let $\mathcal{N}_{A\to B}$ be a quantum channel with the following Kraus and Stinespring representations:
		\begin{equation}
			\mathcal{N}(X)=\sum_{i=1}^r K_i X K_i^\dagger=\Tr_E[VXV^\dagger].
		\end{equation}
		\begin{enumerate}[topsep=0.3cm]
			\item Verify using \eqref{eq:math-tools:adjoint-unique-def} that the adjoint map $\mathcal{N}^\dagger$ can be represented in the following two ways:
				\begin{equation}
					\mathcal{N}^\dagger(Y)=\sum_{i=1}^r K_i^\dagger Y K_i= V^\dagger(Y\otimes\mathbbm{1}_E)V.
				\end{equation}
			
			\item Using 1., verify the following facts:
				\begin{enumerate}
					\item The adjoint of a completely positive map is completely positive.
					\item The adjoint of a trace preserving map is a unital map (recall Definition~\ref{def-TP_superop}). More generally, the adjoint of a trace non-increasing, Hermiticity-preserving map is subunital, meaning that $\mathcal{N}^{\dagger}(\mathbbm{1}_B)\leq\mathbbm{1}_A$.
					\item The adjoint of a unital map is trace preserving, and the adjoint of a subunital Hermiticity-preserving map is trace non-increasing.
				\end{enumerate}
		\end{enumerate}
	\end{exercise}
	
	\begin{exercise}{exer-POVM_adjoint_Naimark}
		\begin{enumerate}
			\item Let $\mathcal{N}_{A\to B}$ be a positive trace-preserving map. Prove that the set $\{\mathcal{N}^{\dagger}(\ket{i}\bra{i})\}_{i=0}^{d_B-1}$ is a POVM. More generally, prove that the set $\{\mathcal{N}^{\dagger}(E_j\rho E_j^{\dagger})\}_{j=1}^{d_B^2}$ is a POVM for every orthonormal basis $\{E_j\}_{j=1}^{d_B^2}$ for $\Lin(\mathcal{H}_B)$ and every quantum state $\rho\in\Density(\mathcal{H}_B)$.
			
			\item Conversely, let $\{M_x\}_{x\in\mathcal{X}}$ be a POVM, where $\mathcal{X}$ is a finite set. Prove that there exists a quantum channel $\mathcal{N}$ such that $M_x=\mathcal{N}^{\dagger}(\ket{x}\bra{x})$ for all $x\in\mathcal{X}$, where $\{\ket{x}\}_{x\in\mathcal{X}}$ is an orthonormal set. (\textit{Hint}: Recall Naimark's Theorem (Theorem~\ref{thm-naimark}).)
		\end{enumerate}
	\end{exercise}

\subsection{Relating Quantum State Extensions and Purifications}

	We can now establish a fundamental relationship between a purification $\psi_{RA}$ of a state $\rho_A$ and an extension $\omega_{R'A}$ of $\rho_A$.

	\begin{proposition}{prop-extension_purif}
		Let $\rho_A$ be a quantum state with purification $\psi_{RA}$. For every extension $\omega_{R'A}$ of~$\rho_A$, there exists a quantum channel $\mathcal{N}_{R\to R'}$ such that
		\begin{equation}
			\mathcal{N}_{R\to R'}(\psi_{RA})=\omega_{R'A}.
		\end{equation}
	\end{proposition}
	
	\begin{Proof}
		Consider a purification $\phi_{R''R'A}$ of $\omega_{R'A}$, with purifying system $R''$ satisfying $d_{R''}\geq\rank(\omega_{R'A})$. Since $\omega_{R'A}$ is an extension of $\rho_A$, we have that
		\begin{equation}
			\Tr_{R''R'}[\phi_{R''R'A}]=\rho_A,
		\end{equation}
		which means that $\phi_{R''R'A}$ is a purification of $\rho_A$. The state $\psi_{RA}$ is also a purification of $\rho_A$, which means that, by the isometric equivalence of purifications (see \eqref{eq-purif_iso_equiv_1}--\eqref{eq-purif_iso_equiv_4} and the paragraph thereafter), there exists an isometry $V_{R\to R''R'}$ such that
		\begin{equation}
			V_{R\to R''R'}\ket{\psi}_{RA}=\ket{\phi}_{R''R'A}.
		\end{equation}
		Now, let us use this isometry to define the channel $\mathcal{N}_{R\to R'}$:
		\begin{equation}
			\mathcal{N}_{R\to R'}(\cdot)=\Tr_{R''}[V_{R\to R''R'}(\cdot)V_{R\to R''R'}^\dagger].
		\end{equation}
		It then follows that
		\begin{equation}
			\mathcal{N}_{R\to R'}(\psi_{RA})=\Tr_{R''}[V_{R\to R''R'}\psi_{RA}V_{R\to R''R'}^\dagger]=\Tr_{R''}[\phi_{R''R'A}]=\omega_{R'A},
		\end{equation}
		as required.
	\end{Proof}
	
	Proposition~\ref{prop-extension_purif} tells us that every extension of a quantum state can be ``reached'' via a quantum channel acting on a purification of the state. In this sense, a purification can be viewed as the ``strongest'' extension of a state.

\subsection{Complementary Channels}\label{sec-QM:complementary-chs}
	
	As stated earlier, the Stinespring representation $\mathcal{N}_{A\to B}(X_A)=\Tr_E[VXV^\dagger]$ of a quantum channel $\mathcal{N}_{A\to B}$, where $V_{A\to BE}$ is an isometry, can be interpreted as an interaction of the quantum system $A$ of interest with its environment~$E$ followed by discarding~$E$. If instead we discard the system $B$ of the isometric channel $\mathcal{V}_{A\to BE}(X_A) = VXV^\dagger$, then we obtain the state of the environment after the interaction with $A$. This defines a channel complementary to $\mathcal{N}_{A \to B}$.
	
	\begin{definition}{Complementary Channel}{def-complementary_chan}
		Let $V_{A\to BE}$ be an isometry and  $\mathcal{N}_{A\to B}$ a quantum channel defined as
		\begin{equation}
			\mathcal{N}_{A\to B}(X_A)=\Tr_E[VXV^\dagger].
		\end{equation}
		The \textit{complementary channel} for $\mathcal{N}_{A\to B}$ associated with the isometric extension $V_{A\to BE}$ is denoted by $\mathcal{N}^c_{A\to E}$ and is defined as
		\begin{equation}
			\mathcal{N}^c_{A\to E}(X_A)\coloneqq\Tr_B[VXV^\dagger].
		\end{equation}
		Related to the above, the channel $\mathcal{M}^c_{A\to E}$ is a complementary channel for the channel $\mathcal{M}_{A\to B}$ if there exists an isometric channel $\mathcal{W}_{A \to BE}$ such that
		\begin{align}
		\mathcal{M}_{A\to B} & = \Tr_E \circ \mathcal{W}_{A \to BE} \\
		\mathcal{M}^c_{A\to E} & = \Tr_B \circ 
		\mathcal{W}_{A \to BE}.
		\end{align}
	\end{definition}

        A quantum channel does not have, in general, a uniquely defined complementary channel, just as it does not have a unique Kraus representation; analogously, a quantum state does not have a unique purification. 
    Similar to the latter scenarios, however, it is possible to show that all complementary channels for $\mathcal{N}_{A\to B}$ are related by an isometric channel acting on their output. That is, let us suppose that $(\mathcal{N}_1)^c_{A\to E}$ and $(\mathcal{N}_2)^c_{A\to E'}$ are complementary channels for $\mathcal{N}_{A\to B}$. Then there exists an isometric channel $\mathcal{S}_{E \to E'}$ such that
	\begin{equation}
	(\mathcal{N}_2)^c_{A\to E'} = \mathcal{S}_{E \to E'} \circ (\mathcal{N}_1)^c_{A\to E}.
	\end{equation}	
	
	\begin{exercise}{exer-complementary_channel}
		Let $\mathcal{N}_{A\to B}$ be a quantum channel with a set $\{K_i\}_{i=1}^r$ of Kraus operators, where $r\geq\rank(\Gamma_{AB}^{\mathcal{N}})$.
		\begin{enumerate}[topsep=0.3cm]
			\item Using \eqref{eq-Kraus_to_Stinespring}, show that a channel complementary to $\mathcal{N}$ is given by
				\begin{equation}\label{eq-QM_channels_complement_standard}
					\mathcal{N}^c(X)=\sum_{i,i'=1}^r \Tr[K_iXK_{i'}^{\dagger}]\ket{i-1}\bra{i'-1}_E
				\end{equation}
				for all $X\in\Lin(\mathcal{H}_A)$.
				
			\item Let $W\coloneqq\sum_{i=1}^r K_i^{\dagger}\otimes\ket{i-1}$. Show that the Choi representation of the complementary channel in \eqref{eq-QM_channels_complement_standard} is
				\begin{equation}
					\Gamma_{AE}^{\mathcal{N}^c}=(WW^{\dagger})^{\t}.
				\end{equation}
			
		\end{enumerate}
	\end{exercise}
	
	The notion of complementary channel arises in the scenario in which two parties, Alice and Bob, wish to communicate to each other using a channel~$\mathcal{N}_{A \to B}$ in the presence of an eavesdropper Eve. In this scenario, we can naturally identify Alice and Bob with the quantum systems $A$ and $B$ and Eve with the system $E$, where $E$ is as given in Definition~\ref{def-complementary_chan}. Any signals sent through the quantum channel by Alice are received by Bob via the action of $\mathcal{N}_{A \to B}$, while Eve receives a signal via the action of $\mathcal{N}^c_{A \to E}$. These concepts are important for private communication over quantum channels, which is the topic of Chapter~\ref{chap-private_capacity}. Two important classes of channels are relevant in this context.
	
	\begin{definition}{Degradable and Anti-Degradable Channels}{def-deg_antideg_chan}
		A channel $\mathcal{N}_{A \to B}$ is called \textit{degradable} if there exists a channel $\mathcal{D}_{B \to E}$, called a \textit{degrading channel}, such that
		\begin{equation}
			\mathcal{D}_{B \to E}\circ\mathcal{N}_{A \to B}=\mathcal{N}^c_{A \to E},
		\end{equation}
		where $\mathcal{N}^c_{A \to E}$ is a complementary channel of $\mathcal{N}_{A \to B}$.
		The channel $\mathcal{N}_{A \to B}$ is called \textit{anti-degradable} if a  complementary channel $\mathcal{N}^c_{A \to E}$ of it is degradable, i.e., if there exists a channel $\mathcal{A}_{E \to B}$, called an \textit{anti-degrading channel}, such that
		\begin{equation}
			\mathcal{A}_{E \to B}\circ\mathcal{N}^c_{A\to E}=\mathcal{N}_{A\to B}.
		\end{equation}
	\end{definition}
	
\begin{figure}
		\centering
		\begin{subfigure}{0.49\textwidth}
			\centering\includegraphics[scale=1]{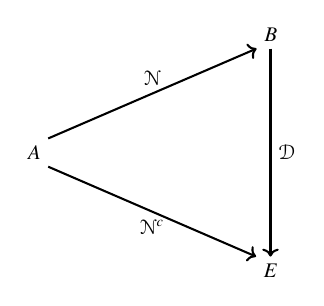}\caption{$\mathcal{N}$ degradable.}\label{subfig-degradable}
		\end{subfigure}~
		\begin{subfigure}{0.49\textwidth}
			\centering\includegraphics[scale=1]{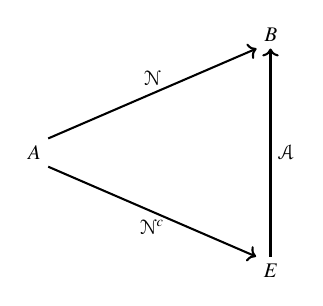}\caption{$\mathcal{N}$ anti-degradable.}\label{subfig-antidegradable}
		\end{subfigure}
		\caption{Degradable and anti-degradable channels. In \subref{subfig-degradable}, the channel $\mathcal{N}$ is degradable, meaning that there exists a channel $\mathcal{D}$ that Bob can apply to the output he receives via $\mathcal{N}$ that can be used to simulate what Eve receives via $\mathcal{N}^c$. In \subref{subfig-antidegradable} on the other hand, $\mathcal{N}$ is anti-degradable since Eve can simulate, using the channel $\mathcal{E}$, what Bob receives.}\label{fig-degradable_antidegradable}
	\end{figure}

	See Figure \ref{fig-degradable_antidegradable} for a schematic depiction of degradable and anti-degradable channels. A degradable channel is one whose complement can be simulated (via~$\mathcal{D}$) using the output of $\mathcal{N}$. This means that Bob can simulate Eve's received signal. On the other hand, an anti-degradable channel is such that $\mathcal{N}$ can be simulated (via $\mathcal{A}$) using the output of $\mathcal{N}^c$, which means that Eve can simulate Bob's received signal. In the above precise sense, therefore, the output of a degradable channel is ``less noisy'' than the environment with which it has interacted; that is, it contains more quantum information. For an anti-degradable channel, instead, the opposite is true. 
	
	\begin{exercise}{exer-QM_antideg_2ext}
		Prove that a quantum channel $\mathcal{N}_{A\to B}$ is anti-degradable if and only if its Choi state $\Phi_{AB}^{\mathcal{N}}$ is \textit{two-extendible}, meaning that there exists a state $\sigma_{ABB'}$, with $\mathcal{H}_B$ and $\mathcal{H}_{B'}$ isomorphic (and thus $d_{B'}=d_B$), such that $\Tr_{B'}[\sigma_{ABB'}]=\Tr_B[\sigma_{ABB'}]=\Phi_{AB}^{\mathcal{N}}$. (\textit{Hint}: Use Proposition~\ref{prop-extension_purif}.)
	\end{exercise}

\subsection{Unitary Extensions of Quantum Channels from Isometric Extensions}
	
	We can always extend every isometric extension $V_{A\to BE}$ of a channel $\mathcal{N}_{A\to B}$ to a unitary $U_{AE'\to BE''}$ in a way similar to what we used in \eqref{eq:math-tools:op-Jensen-unitary-from-iso} in the proof of the operator Jensen inequality (Theorem~\ref{thm-Jensen}). Let the unitary $U_{AE'\to BE''}$ be defined as the following block matrix:
	\begin{equation}\label{eq-chan_iso_ext_to_unitary}
		U=\begin{pmatrix}
		V & \mathbbm{1}-VV^\dagger & 0_{d_B d_E \times d'} \\
		0_{d_A \times d_A} & V^\dagger & 0_{d_A \times d'} \\
		0_{d' \times d_A} & 0_{d' \times d_B d_E} & \mathbbm{1}_{d'}\end{pmatrix},
	\end{equation}
	where we set $d'\coloneqq (d_B-1)d_A$.  Without the various dimensions indicated, $U$ is more simply expressed as
		\begin{equation}
		U=\begin{pmatrix}
		V & \mathbbm{1}-VV^\dagger & 0 \\
		0 & V^\dagger & 0 \\
		0 & 0 & \mathbbm{1}\end{pmatrix}.
	\end{equation}
	Note that $V$ is a $d_Bd_E\times d_A$ matrix and $VV^\dagger$ is a $d_Bd_E\times d_Bd_E$ matrix. Furthermore, let us suppose that $d_E=d_Ad_B$. Note that, by the Stinespring theorem, it is always possible to pick this as the dimension of $\mathcal{H}_E$ since $0<\rank(\Gamma_{AB}^{\mathcal{N}})\leq d_Ad_B$. Then 
	\begin{equation}
	d_Bd_E+d_A+d' = d_A d_B (d_B+1),
	\end{equation}
	and we conclude that $U$ is a $(d_A d_B (d_B+1))\times (d_A d_B (d_B+1))$ matrix. It is also indeed a unitary because
	\begin{align}
		UU^\dagger&=
		\begin{pmatrix}
		V & \mathbbm{1}-VV^\dagger & 0  \\
		0 & V^\dagger & 0 \\
		0 & 0 & \mathbbm{1}
		\end{pmatrix}
		\begin{pmatrix}
		V^\dagger & 0 & 0 \\
		\mathbbm{1}-VV^\dagger & V & 0 \\
		0 & 0 & \mathbbm{1}
		\end{pmatrix}\\
		&=\begin{pmatrix}
		VV^\dagger + (\mathbbm{1}-VV^\dagger)(\mathbbm{1}-VV^\dagger) & (\mathbbm{1}-VV^\dagger)V & 0 \\
		V^\dagger(\mathbbm{1}-VV^\dagger) & V^\dagger V & 0 \\
		0 & 0 & \mathbbm{1}
		\end{pmatrix}\\
		&=\begin{pmatrix}
		\mathbbm{1} & 0 & 0 \\
		0 & \mathbbm{1} & 0 \\
		0 & 0 & \mathbbm{1}
		\end{pmatrix}.
	\end{align}
	Since $U$ is a square matrix, this already implies that $U^{-1} = U^\dag$; i.e., $U$ is unitary. 
	By defining the system $E'$ with dimension $d_{E'}=d_B(d_B+1)$, we can think of $U$ as acting on the input tensor-product space $\mathcal{H}_{E'}\otimes\mathcal{H}_{A}$. Then, we can embed the state $\rho_A$ into this larger space as
	\begin{equation}
		\ket{0}\!\bra{0}_{E'}\otimes\rho_{A}=
		\begin{pmatrix}
		\rho_A & 0 & 0 \\
		0 & 0 & 0\\
		0 & 0 & 0
		\end{pmatrix},
	\end{equation}
	so that
	\begin{equation}
		U\begin{pmatrix}
		\rho_A & 0 & 0 \\
		0&0 & 0\\
		0&0 & 0
		\end{pmatrix}U^\dagger=
		\begin{pmatrix}
		V\rho V^\dagger & 0 & 0 \\
		0 & 0 & 0 \\
		0 & 0 & 0 
		\end{pmatrix}.
	\end{equation}
	By defining the system $E''$ with dimension $d_{E''}=d_A(d_B+1)$, we can think of the output space of $U$ as the tensor-product space $\mathcal{H}_B\otimes\mathcal{H}_{E''}$, so that 
	\begin{equation}
		\mathcal{N}(\rho_A)=\Tr_E[V\rho V]=\Tr_{E''}[U(\ket{0}\!\bra{0}_{E'}\otimes\rho_A)U^\dagger].
	\end{equation}
	
	The above construction is not necessarily an efficient construction (using as few extra degrees of freedom as possible), but it illustrates the principle that every quantum channel can be thought of as arising from
	\begin{enumerate}
	\item adjoining an environment state $\ket{0}\!\bra{0}_{E'}$ to the input system $A$,
	\item performing a unitary interaction $U$, and then
	\item  tracing over an output environment system $E''$,
	\end{enumerate}
	thus assigning a strong physical meaning to the notion of isometric extension of a quantum channel; see Figure \ref{fig-stinespring}.
	
	Another construction of a unitary operator that is less explicit but more efficient is as follows. Let $V_{A\to BE}$ be the isometric extension from \eqref{eq-Kraus_to_Stinespring}, and let $r=\rank(\Gamma_{AB}^{\mathcal{N}})$. Since $r \leq d_A d_B$, without loss of generality, we can take the output environment system $E$ to have dimension $d_E = d_A d_B$. Let $\{\ket{k}_A \otimes \ket{\ell}_{E'}\}_{k,\ell}$ be an orthonormal basis for the input system $A$ and an input environment system $E'$, with $d_{E'}=d_B^2$, where $0 \leq k \leq d_A-1$ and $0 \leq \ell \leq d_{E'}-1$. Then define the orthonormal vectors $\ket{\phi_{k,0}}_{BE}$ for $0 \leq k \leq d_A-1$ as follows:
	\begin{align}
		\ket{\phi_{k,0}}_{BE} & \coloneqq \left(\sum_{j=1}^{d_E} K_j\otimes\ket{j-1}_E\bra{0}_{E'} \right) \ket{k}_A \otimes \ket{0}_{E'} \\
	 & = V_{A\to BE} \ket{k}_A.
	\end{align}
	The fact that these vectors form an orthonormal set is a consequence of the facts that $V^\dag V = \mathbbm{1}_A$ and $\{\ket{k}_A \}_{k=0}^{d_A-1}$ is an orthonormal set. We then define the action of the unitary $U_{AE'\to BE}$ on these vectors $\ket{k}_A \otimes \ket{0}_{E'}$ as follows:
	\begin{equation}
	U_{AE'\to BE} \ket{k}_A \otimes \ket{0}_{E'} = \ket{\phi_{k,0}}_{BE},
	\end{equation}
	for $0 \leq k \leq d_A-1$. 	
    The set of orthonormal vectors $\{\ket{\phi_{k,0}}_{BE}\}_{k=0}^{d_A-1}$ can be completed to an orthonormal basis of the output space by adding suitably chosen $d_B d_E - d_A = (d_B^2-1)d_A$ vectors. Constructing these vectors, denoted as $\{\ket{\phi_{k,\ell}}_{BE}\}_{k,\ell}$, where $0\leq k\leq d_A-1$ and $1\leq \ell\leq d_B^2-1$, is equivalent to choosing a spectral decomposition
    \begin{equation}
    \mathbbm{1}_{BE} - VV^\dag = \sum_{k=0}^{d_A-1}\sum_{\ell = 1}^{d_B^2-1} \ket{\phi_{k,\ell}}\!\bra{\phi_{k,\ell}}_{BE}
    \end{equation}
    of the projection $\mathbbm{1}_{BE} - VV^\dag$. 
    By employing these vectors, 
    we can now extend the action of the unitary $U_{AE'\to BE}$ on the remaining input space vectors as follows:
	\begin{equation}
	U_{AE'\to BE}\ket{k}_A \otimes \ket{\ell}_{E'} = \ket{\phi_{k,\ell}}_{BE},
	\end{equation}
    for $0 \leq k \leq d_A-1$ and $1 \leq \ell \leq d_B^2-1$. Thus, the full unitary $U_{AE'\to BE}$ is specified as
	\begin{equation}
	U_{AE'\to BE} \coloneqq \sum_{k=0}^{d_A-1}\sum_{\ell = 0}^{d_B^2-1} \ket{\phi_{k,\ell}}_{BE}\bra{k}_A \otimes \bra{\ell}_{E'}.
	\end{equation}
	By construction, the following identity holds for every input state $\rho_A$:
	\begin{equation}
	U_{AE'\to BE} \left(\rho_A \otimes \ket{0}\!\bra{0}_{E'}\right) (U_{AE'\to BE})^\dag = V \rho_A V^\dag,
	\end{equation}
	so that we can realize the isometric channel $\rho_A \mapsto V \rho_A V^\dag$ by tensoring in an environment state $\ket{0}\!\bra{0}_{E'}$ and applying the unitary  $U_{AE'\to BE}$. We then realize the original channel $\mathcal{N}_{A\to B}$ by applying a final partial trace over the environment system~$E$:
	\begin{equation}
	\mathcal{N}_{A\to B}(\rho_A) = \Tr_E[U_{AE'\to BE} \left(\rho_A \otimes \ket{0}\!\bra{0}_{E'}\right) (U_{AE'\to BE})^\dag].
	\end{equation}

\section{General Types of Channels}
	
\subsection{Preparation, Appending, and Replacement Channels}

	The preparation of a quantum system in a given (fixed) state, as well as taking the tensor product of a state with a given (fixed) state, can both be viewed as quantum channels.
	
	\begin{definition}{Preparation and Appending Channels}{def-prep_app_channel}
		For a quantum system $A$ and a state $\rho_A$, the \textit{preparation channel}  $\mathcal{P}_{\rho_A}$ is defined for $\alpha\in\mathbb{C}$ as
		\begin{equation}
			\mathcal{P}_{\rho_A}(\alpha)\coloneqq \alpha\rho_A.
		\end{equation}
		When acting in parallel with the identity channel on a linear operator~$Y_B$ of another quantum system $B$, the preparation channel $\mathcal{P}_{\rho_A}$ is called the \textit{appending channel} $\mathcal{P}_{\rho_A}$ and is defined as
		\begin{equation}
			\mathcal{P}_{\rho_A}(Y_B)\equiv (\mathcal{P}_{\rho_A}\otimes\id_B)(Y_B)=\rho_A\otimes Y_B.
		\end{equation}
		In other words, the appending channel $\mathcal{P}_{\rho_A} \otimes\id_B$ takes the tensor product of its argument with $\rho_A$.
	\end{definition}
	
	One way to determine whether a map $\mathcal{N}$ on quantum states is completely positive is to find a Kraus representation for it, i.e., a set $\{K_i\}_i$ of operators such that $\mathcal{N}(X)=\sum_i K_i XK_i^\dagger$ for every operator $X$; if that is possible, then $\mathcal{N}$ is completely positive by Theorem \ref{thm-q_channels}. If in addition the Kraus operators satisfy $\sum_i K_i^\dagger K_i=\mathbbm{1}$, then $\mathcal{N}$ is also trace preserving and therefore a quantum channel.
	
	Supposing that $\rho_A$ has a spectral decomposition of the form
	\begin{equation}
		\rho_A=\sum_k \lambda_k \ket{\phi_k}\!\bra{\phi_k}_A,
	\end{equation}
	one set of Kraus operators for the preparation channel $\mathcal{P}_{\rho_A}$ is $\{\sqrt{\lambda_k}\ket{\phi_k}_A\}_k$. A set of Kraus operators for the appending channel $\mathcal{P}_{\rho_A}\otimes\id_B$ is then $\{\sqrt{\lambda_k}\ket{\phi_k}\otimes\mathbbm{1}_B\}_k$.
	
	\begin{exercise}{exer-prep_channel}
		Determine the Choi representation and a Stinespring representation of the appending channel $\mathcal{P}_{\rho_A} \otimes \operatorname{id}_B$ corresponding to the quantum state $\rho_A$.
	\end{exercise}
	
	\begin{definition}{Replacement Channel}{def-replace_channel}
		For a state $\sigma_B$, the \textit{replacement channel} $\mathcal{R}_{A\to B}^{\sigma_B}$ is defined as the channel that traces out its input and replaces it with the state $\sigma_B$; i.e.,
		\begin{equation}\label{eq-replacement_channel}
			\mathcal{R}_{A\to B}^{\sigma_B}(X_A)=\Tr[X_A]\sigma_B
		\end{equation}
		for every linear operator $X_A$. When acting on one share of a bipartite state $\rho_{RA}$, the replacement channel $\mathcal{R}_{A\to B}^{\sigma_B}$ has the following action:
		\begin{equation}
			\mathcal{R}_{A\to B}^{\sigma_B}(\rho_{RA})=\Tr_A[\rho_{RA}]\otimes\sigma_B=\rho_R\otimes\sigma_B.
		\end{equation}
	\end{definition}
	
	Observe that we can write the replacement channel $\mathcal{R}_{A\to B}^{\sigma_B}$  as the composition of the partial trace over $A$ followed by the preparation/append\-ing channel $\mathcal{P}_{\sigma_B}$:
	\begin{equation}
		\mathcal{R}_{A\to B}^{\sigma_B}=\mathcal{P}_{\sigma_B}\circ\Tr_A.
	\end{equation}
	We often omit the superscript on $\mathcal{R}_{A\to B}^{\sigma_B}$  when it is clear from the context that the replacement state is  $\sigma_B$.
	
	\begin{exercise}{exer-replace_channel}
		Given a quantum state $\sigma_B$, determine the Choi representation, as well as Kraus and Stinespring representations, of the replacement channel $\mathcal{R}_{A\to B}^{\sigma_B}$.
	\end{exercise}

\subsection{Trace and Partial-Trace Channels}

	Recall the trace and partial trace of a linear operator from Definition \ref{def-partial_trace}. As a map acting on linear operators, one can ask whether the partial trace is a channel. The answer, perhaps not surprisingly, is ``yes.'' In fact, observe that the definition in \eqref{eq-partial_trace_B} of the partial trace $\Tr_B$ over $B$ is already in Kraus form, with Kraus operators $K_j=\mathbbm{1}_A\otimes\bra{j}_B$. This means that $\Tr_B$ is completely positive. It is also trace preserving because
	\begin{equation}
		\sum_{j=0}^{d_B-1}K_j^\dagger K_j=\sum_{j=0}^{d_B-1}(\mathbbm{1}_A\otimes\ket{j}_B)(\mathbbm{1}_A\otimes\bra{j}_B)=\mathbbm{1}_A\otimes\sum_{j=0}^{d_B-1}\ket{j}\!\bra{j}_B=\mathbbm{1}_{AB},
	\end{equation}
	where we used the fact that $\sum_{j=0}^{d_B-1}\ket{j}\!\bra{j}_B=\mathbbm{1}_B$.
	
	\begin{exercise}{exer-partial_trace_Choi_Stinespring_adjoint}
		\begin{enumerate}
			\item Determine the Choi representation, as well as a Stinespring representation, of the partial trace channel $\Tr_B$.
			
			\item Prove that the adjoint of the partial trace channel $\Tr_B$ is
				\begin{equation}
					\Tr_B^{\dagger}(X_A)=X_A\otimes\mathbbm{1}_B.
				\end{equation}
		\end{enumerate}
	\end{exercise}
	
	Unlike the trace and partial trace, the transpose and partial transpose are trace-preserving maps but \textit{not} completely positive. Indeed, for the latter, recall from \eqref{eq-swap_operator} that $\T_B(\Phi_{AB})=\frac{1}{d}F_{AB}$, so that its Choi representation is $\Gamma_{AB}^{\T_B}=F_{AB}$, which we know has negative eigenvalues, as shown in \eqref{eq:qm:spectral-decomp-swap-op}. So, by Theorem \ref{thm-q_channels}, the transpose map $\T_B$ is not completely positive.

\subsection{Isometric and Unitary Channels}\label{sec:OM-over:iso-unitary-chs}

	Two more simple examples of quantum channels are \textit{isometric} and \textit{unitary channels}. An isometric channel conjugates  the channel input by an isometry, and a unitary channel conjugates the channel input by a unitary. Specifically, the isometric channel $\mathcal{V}$ corresponding to an isometry $V$ is
	\begin{equation}
		\mathcal{V}(X)\coloneqq VXV^\dagger.
	\end{equation}
	Similarly, the unitary channel $\mathcal{U}$ corresponding to a unitary $U$ is
	\begin{equation}
		\mathcal{U}(X)\coloneqq UXU^\dagger.
	\end{equation}
	Since every unitary is also an isometry, it follows that every unitary channel is an isometric channel.
	Isometric channels are completely positive because they can be described using only one Kraus operator, the isometry $V$. In fact, a quantum channel is isometric if and only if it has a single Kraus operator.
	
	Observe that by the unitarity of $U$, the map
	\begin{equation}
		\mathcal{U}^\dagger(Y)\coloneqq U^\dagger Y U,
	\end{equation}
	i.e., the conjugation by $U^\dagger$, is also a channel. In particular,
	\begin{equation}
		\mathcal{U}^\dagger\circ\mathcal{U}=\mathcal{U}\circ\mathcal{U}^\dagger=\id,
	\end{equation}
	so that $\mathcal{U}^\dagger$ is the inverse channel of $\mathcal{U}$. 
	
	On the other hand, for an isometry $V$, conjugation by $V^\dagger$ is \textit{not} necessarily a channel: although the map $\mathcal{V}^\dagger(Y)\coloneqq V^\dagger YV$ is completely positive, it is not necessarily trace preserving because $VV^\dagger\neq\mathbbm{1}$ in general. However, by defining the reversal channel $\mathcal{R}_V$ as
	\begin{equation}\label{eq-isometry_recover}
		\mathcal{R}_V(Y)\coloneqq \mathcal{V}^\dagger(Y)+\Tr[(\mathbbm{1}-VV^\dagger)Y]\sigma,
	\end{equation}
	where $\sigma$ is an arbitrary (but fixed) state, we find that $\mathcal{R}_V$ is trace preserving:
	\begin{align}
		\Tr[\mathcal{R}_V(Y)]&=\Tr\!\left[\mathcal{V}^\dagger(Y)+\Tr[(\mathbbm{1}-VV^\dagger)Y]\sigma\right]\\
		&=\Tr[V^\dagger Y V]+\Tr[Y]-\Tr[VV^\dagger Y]\\
		&=\Tr[Y].
	\end{align}
	Since it is also completely positive, being the sum of completely positive maps, the map $\mathcal{R}_V$ is indeed a quantum channel. Like $\mathcal{U}^\dagger$, the reversal channel~$\mathcal{R}_V$ reverses the action of $\mathcal{V}$:
	\begin{align}
		(\mathcal{R}_V\circ\mathcal{V})(X)&=\mathcal{V}^\dagger(\mathcal{V}(X))+\Tr[(\mathbbm{1}-VV^\dagger)\mathcal{V}(X)]\sigma \label{eq-QM-ch:reversal-prop-1}\\
		&=V^\dagger V XV^\dagger V+\Tr[(\mathbbm{1}-VV^\dagger)VXV^\dagger]\sigma\\
		&=X+(\Tr[VXV^\dagger]-\Tr[VV^\dagger VXV^\dagger])\sigma\\
		&=X.\label{eq-QM-ch:reversal-prop-last}
	\end{align}
	In the above sense, $\mathcal{R}_V$ is a left inverse of $\mathcal{V}$. Unlike~$\mathcal{U}^\dagger$, however, $\mathcal{R}_V$ is not the right inverse of $\mathcal{V}$ because the equality $(\mathcal{V}\circ\mathcal{R}_V)(Y)\allowbreak=Y$ need not hold.
	
	\begin{exercise}{exer-reversal_channel}
		Determine the Choi representation, as well as a Kraus and Stinespring representation, of the reversal channel $\mathcal{R}_V$ corresponding to an isometry $V$.
	\end{exercise}

\subsection{Classical--Quantum and Quantum--Classical Channels}

	Any classical probability distribution $p:\mathcal{X}\to[0,1]$ over a finite alphabet $\mathcal{X}$ can be represented as a quantum state of a $|\mathcal{X}|$-dimensional system $X$ that is diagonal in an orthonormal basis $\{\ket{x}\}_{x\in\mathcal{X}}$ of $X$. Specifically, the probability distribution can be represented as the state
	\begin{equation}
		\rho_X=\sum_{x\in\mathcal{X}}p(x)\ket{x}\!\bra{x}_X.
	\end{equation}
	States that are diagonal in a preferred basis, as in the equation above, are typically called \textit{classical states}. In addition to being in one-to-one correspondence with classical probability distributions, classical states do not exhibit the quantum properties of coherence and entanglement.
	
	In Chapter~\ref{chap-classical_capacity}, however, we are interested in classical communication over quantum channels. We are then interested in so-called \textit{classical--quantum channels}, which we discuss in this section, that take a classical state as input and output a quantum state. 
	
	\begin{definition}{Classical--Quantum Channel}{def-cq_channel}
		A \textit{classical--quantum channel} is a map from a system $X$ with alphabet $\mathcal{X}$ and orthonormal basis $\{\ket{x}:x\in\mathcal{X}\}$ to a quantum system $A$ with a specified set $\{\sigma_A^x:x\in\mathcal{X}\}$ of states such that
		\begin{equation}
			\ket{x}\!\bra{x'}\mapsto \delta_{x,x'}\sigma_A^x\quad\forall~ x,x'\in\mathcal{X}.
		\end{equation}
	\end{definition}
	
	If $\mathcal{N}^{\text{cq}}$ is a classical--quantum channel, then for every classical state $\rho_X\allowbreak=\sum_{x\in\mathcal{X}}p(x)\ket{x}\!\bra{x}$, we have that
	\begin{equation}
		\mathcal{N}^{\text{cq}}(\rho_X)=\sum_{x\in\mathcal{X}}p(x)\sigma_A^x.
	\end{equation}
	More generally, for every $|\mathcal{X}|$-dimensional system $A$ and every state $\rho_{A}$ that is not necessarily classical and is expressed in the computational basis as $\rho_{A}=\sum_{x,x'\in\mathcal{X}}\bra{x}\rho_A\ket{x'}\ket{x}\!\bra{x'}$, we find that
	\begin{equation}
		\mathcal{N}^{\text{cq}}(\rho_A)=\sum_{x,x'\in\mathcal{X}}\bra{x}\rho_A\ket{x'}\delta_{x,x'}\sigma_A^x=\sum_{x\in\mathcal{X}}\bra{x}\rho_A\ket{x}\sigma_A^x.
	\end{equation}
	Therefore, for every quantum state, the classical--quantum channel $\mathcal{N}^{\text{cq}}$ takes the input state, measures it in the computational basis $\{\ket{x}\}_{x\in\mathcal{X}}$, and outputs the state $\sigma_A^x$ corresponding to the outcome $x\in \mathcal{X}$. 
	
	\begin{exercise}{exer-cq_channel}
		Show that the Choi state of a classical--quantum channel $\mathcal{N}^{\text{cq}}$ is
		\begin{equation}
			\Phi_{XA}^{\mathcal{N}^{\text{cq}}}
			=\frac{1}{|\mathcal{X}|}\sum_{x\in\mathcal{X}}\ket{x}\!\bra{x}_X\otimes\sigma_A^x.
		\end{equation}
		In other words, the Choi state of a classical--quantum channel is a classical--quantum state.
	\end{exercise}
	
	If the states $\sigma_A^x$ have spectral decomposition
	\begin{equation}
		\sigma_A^x=\sum_{j=1}^{r_x}\lambda_j^x\ket{\varphi_j^x}\!\bra{\varphi_j^x},
	\end{equation}
	where $r_x=\rank(\sigma_A^x)$, then $\{K_j^x:x\in\mathcal{X},~1\leq j\leq r_x\}$ is a set of Kraus operators for $\mathcal{N}^{\text{cq}}$, where $K_j^x=\sqrt{\smash[b]{\lambda_j^x}}\ket{\varphi_j^x}_A\bra{x}_X$. Indeed, for all $x,x'\in\mathcal{X}$,
	\begin{align}
		&\sum_{x''\in\mathcal{X}}\sum_{j=1}^{r_{x''}}K_j^{x''}\ket{x}\!\bra{x'}(K_j^{x''})^\dagger \notag\\
		&=\sum_{x''\in\mathcal{X}}\sum_{j=1}^{r_{x''}}\sqrt{\smash[b]{\lambda_j^{x''}}}\ket{\varphi_j^{x''}}\braket{x''}{x}\braket{x'}{x''}\!\bra{\varphi_j^{x''}}\sqrt{\smash[b]{\lambda_j^{x''}}}\\
		&=\delta_{x,x'}\sum_{j=1}^{r_x}\lambda_k^x\ket{\varphi_j^x}\!\bra{\varphi_j^x}\\
		&=\delta_{x,x'}\sigma_A^x\\
		&=\mathcal{N}^{\text{cq}}(\ket{x}\!\bra{x'}),
	\end{align}
	and
	\begin{align}
		\sum_{x\in\mathcal{X}}\sum_{j=1}^{r_x}(K_j^x)^\dagger K_j^x&=\sum_{x\in\mathcal{X}}\sum_{j=1}^{r_x}\lambda_j^x\ket{x}\underbrace{\braket{\smash[b]{\varphi_j^x}}{\smash[b]{\varphi_j^x}}}_{=1}\bra{x}\\
		&=\sum_{x\in\mathcal{X}}\underbrace{\sum_{j=1}^{r_x}\lambda_j^x}_{=1~\forall~x}\ket{x}\!\bra{x}\\
		&=\mathbbm{1}_X.
	\end{align}
	Also, observe from the  construction above that every classical--quantum channel has a Kraus representation with unit-rank Kraus operators.
	
	Having described classical--quantum channels, let us now describe channels for which the situation is opposite, such that they accept quantum inputs and provide classical outputs.

	\begin{definition}{Quantum--Classical Channel}{def-qc_channel}
		Given a quantum system $A$ and a measurement on $A$ with corresponding POVM $\{M_x\}_{x\in\mathcal{X}}$ indexed by elements of a finite alphabet $\mathcal{X}$, a \textit{quantum--classical channel}, or \textit{measurement channel}, is a map $\mathcal{M}_{A\to X}$ from the quantum system $A$ to a classical system $X$ with alphabet $\mathcal{X}$ such that
		\begin{equation}\label{eq-measurement_channel}
			\mathcal{M}_{A\to X}(\rho_A)=\sum_{x\in\mathcal{X}}\Tr[M_x\rho_A]\ket{x}\!\bra{x}_X
		\end{equation}
		for every state $\rho_A$ on $A$, where $\{\ket{x}:x\in\mathcal{X}\}$ is an orthonormal basis for $X$.
	\end{definition}
	
	A measurement channel thus takes the measurement outcome probabilities $\Tr[M_x\rho_A]$ and arranges them into a classical state.
	
	\begin{exercise}{exer-meas_channel}
		Prove that the Choi state of a measurement channel $\mathcal{M}_{A\to X}$, as defined in \eqref{eq-measurement_channel}, is
		\begin{equation}
			\Phi_{AX}^{\mathcal{M}}=\frac{1}{d_A} \sum_{x\in \mathcal{X}}  (M_x^{\t})_A  \otimes   \ket{x}\!\bra{x}_X.
		\end{equation}
	\end{exercise}
	
	By writing a spectral decomposition of $M_x$ as
	\begin{equation}
		M_x=\sum_{j=1}^{r_x}\mu_{j}^x\ket{\phi_j^{x}}\!\bra{\phi_j^{x}},
	\end{equation}
	where $r_x=\rank(M_x)$, we can write
	\begin{equation}
		\Tr[M_x\rho]=\sum_{j=1}^{r_x}\mu_{j}^x\bra{\phi_{j}^x}\rho\ket{\phi_{j}^x}.
	\end{equation}
	Therefore, the action of a quantum--classical channel $\mathcal{N}^{\text{qc}}$ can be written as
	\begin{align}
		\mathcal{N}^{\text{qc}}(\rho)&=\sum_{x\in\mathcal{X}}\Tr[M_x\rho]\ket{x}\!\bra{x}\\
		&=\sum_{x\in\mathcal{X}}\sum_{j=1}^{r_x}\mu_j^x\bra{\phi_{j}^x}\rho\ket{\phi_{j}^x}\ket{x}\!\bra{x}\\
		&=\sum_{x\in\mathcal{X}}\sum_{j=1}^{r_x}\sqrt{\smash[b]{\mu_j^x}}\ket{x}\!\bra{\phi_{j}^x}\rho\ket{\phi_{j}^x}\!\bra{x}\sqrt{\smash[b]{\mu_j^x}}\\
		&=\sum_{x\in\mathcal{X}}\sum_{j=1}^{r_x}K_j^x\rho(K_j^x)^\dagger,
	\end{align}
	where
	\begin{equation}
		K_j^x\coloneqq\sqrt{\smash[b]{\mu_j^x}}\ket{x}\!\bra{\phi_{j}^x}\quad\forall ~x\in\mathcal{X},~1\leq j\leq r_x.
	\end{equation}
	Since
	\begin{align}
		\sum_{x\in\mathcal{X}}\sum_{j=1}^{r_x}(K_j^x)^\dagger K_j^x&=\sum_{x\in\mathcal{X}}\sum_{j=1}^{r_x}\mu_j^x\ket{\phi_{j}^x}\braket{x}{x}\!\bra{\phi_{j}^x}\\
		&=\sum_{x\in\mathcal{X}}\underbrace{\sum_{j=1}^{r_x}\mu_j^x\ket{\phi_{j}^x}\!\bra{\phi_{j}^x}}_{M_x}\\
		&=\sum_{x\in\mathcal{X}}M_x\\
		&=\mathbbm{1}_A,
	\end{align}
	the operators $K_j^x$, where $x\in\mathcal{X}$ and $1\leq j\leq r_x$, form a set of Kraus operators for $\mathcal{N}^{\text{qc}}$. This means that all quantum--classical channels have a Kraus representation with unit-rank Kraus operators.

\subsection{Quantum Instruments}\label{sec-quantum_instruments}
	
	Recall from Section~\ref{subsec-meas} that, for an arbitrary POVM $\{M_x\}_{x\in\mathcal{X}}$, a set of post-measure\-ment states corresponding to an initial state $\rho$ can be given by \eqref{eq-post_meas},
	\begin{equation}
		\rho^x=\frac{K_x\rho K_x^\dagger}{\Tr[K_x\rho K_x^\dagger]}\quad\forall~x\in\mathcal{X},
	\end{equation}
	where $\{K_x\}_{x\in\mathcal{X}}$ is a set of operators such that $M_x=K_x^\dagger K_x$ for all $x\in\mathcal{X}$. Also recall that the expected density operator of the measurement is given by~\eqref{eq-post_meas_avg},
	\begin{equation}
		\sum_{x\in\mathcal{X}}K_x\rho K_x^\dagger.
	\end{equation}
	This expected state can be seen as arising from a quantum channel with Kraus operators $\{K_x\}_{x\in\mathcal{X}}$. Note that this map is indeed a channel since $\sum_{x\in\mathcal{X}} K_x^\dagger K_x=\sum_{x\in\mathcal{X}} M_x=\mathbbm{1}$. We can view the channel as being the sum of the completely positive and trace-non-increasing maps $\mathcal{M}_x$ defined as
	\begin{equation}
		\mathcal{M}_x(\rho)=K_x \rho K_x^\dagger.
	\end{equation}
	
	A quantum instrument generalizes this notion to maps $\mathcal{M}_x$ that are completely positive and trace non-increasing with an arbitrary number of Kraus operators, not just one.
	
	\begin{definition}{Quantum Instrument}{def-q_instrument}
		A \textit{quantum instrument} is a collection $\{\mathcal{M}_x\}_{x\in\mathcal{X}}$ of completely positive maps indexed by elements of a finite alphabet $\mathcal{X}$, such that the sum map $\sum_{x\in\mathcal{X}} \mathcal{M}_x$ is a quantum channel.
	\end{definition}

 \begin{remark}
    In order for the sum map $\sum_{x\in\mathcal{X}} \mathcal{M}_x$ to be a quantum channel, and hence trace preserving, each $\mathcal{M}_x$ needs to be a trace-non-increasing map.
 \end{remark}
	
	A quantum instrument corresponds to a more general notion of a measurement in which, as before, the elements of the alphabet $\mathcal{X}$ label the measurement outcomes. If the initial state of the system is $\rho$, then the corresponding measurement outcome probabilities $p(x)$ are given by
	\begin{equation}
		p(x)=\Tr[\mathcal{M}_x(\rho)],
	\end{equation}
	and the corresponding post-measurement state is
	\begin{equation}
		\rho^x=\frac{\mathcal{M}_x(\rho)}{\Tr[\mathcal{M}_x(\rho)]}.
	\end{equation}
	The expected state of the ensemble $\{(p(x),\rho^x)\}_{x\in\mathcal{X}}$ is then
	\begin{equation}\label{eq-post_meas_avg2}
		\sum_{x\in\mathcal{X}}\mathcal{M}_x(\rho).
	\end{equation}
	
	It is customary to define the output of a quantum instrument $\{\mathcal{M}_x\}_{x\in\mathcal{X}}$ as the channel $\mathcal{M}$ defined by
	\begin{equation}\label{eq-instrument_output}
		\mathcal{M}(\rho)=\sum_{x\in\mathcal{X}}
		\ket{x}\!\bra{x}_X \otimes
		\mathcal{M}_x(\rho).
	\end{equation}
	That is, the output of a quantum instrument is a classical--quantum state in which the classical register $X$ stores the outcome of the measurement. This is unlike the expected state in \eqref{eq-post_meas_avg2}, which represents a lack of knowledge of which measurement outcome occurred.
	
	Note that the channel in \eqref{eq-instrument_output} corresponding to a quantum instrument reduces to the measurement channel defined in \eqref{eq-measurement_channel} if we consider a measurement with POVM $\{M_x\}_{x\in\mathcal{X}}$ and we define the maps $\mathcal{M}_x$ as $\mathcal{M}_x(\rho)=\Tr[M_x\rho]$ for all $x\in\mathcal{X}$. In this case, the channel in \eqref{eq-instrument_output} becomes
	\begin{equation}
		\mathcal{M}(\rho)=\sum_{x\in\mathcal{X}}\Tr[M_x\rho]\ket{x}\!\bra{x}_X,
	\end{equation}
	which is precisely the measurement channel in \eqref{eq-measurement_channel}.

\subsection{Entanglement-Breaking Channels}\label{subsec-ent_break_channels}

	An important class of channels in quantum communication consists of those that, when acting on one share of a bipartite state, eliminate any entanglement between the two systems, such that the resulting state is separable (recall Definition~\ref{def-sep_ent_state}). Such channels are called \textit{entanglement breaking}, and we define them as follows.
	
	\begin{definition}{Entanglement-Breaking Channel}{def-ent_break_chan}
		A channel $\mathcal{N}_{A\to B}$ is called \textit{entanglement breaking} if $(\id_R\otimes\mathcal{N}_{A\to B})(\rho_{RA})$ is a separable state for every state $\rho_{RA}$, where $R$ is an arbitrary reference system.
	\end{definition}

\begin{exercise}{exer-EB_closed_composition}
Show that the set of entanglement-breaking channels is closed under composition with arbitrary channels, both from the left and from the right. Namely, prove that if $\mathcal{N}_{A\to B}$ is entanglement breaking and if $\mathcal{M}_{A'\to A}$ and $\mathcal{L}_{B\to B'}$ are arbitrary channels, then the channel $\big(\mathcal{L} \circ \mathcal{N}\circ \mathcal{M}\big)_{A'\to B'}$ is also entanglement breaking.
\end{exercise}
 
	Although Definition~\ref{def-ent_break_chan} suggests that it is necessary to check an infinite number of input states to determine whether a channel is entanglement breaking, the following proposition states that it is only necessary to check the output of the channel on the maximally entangled state.
	
		\begin{proposition}{prop:qm-over:EB-Choi-sep}
		A channel $\mathcal{N}$ is entanglement breaking if and only if  its Choi state $\Phi_{AB}^{\mathcal{N}}$ is separable.
	\end{proposition}
	
	\begin{Proof}
	Observe that if $\mathcal{N}_{A\to B}$ is entanglement breaking, then its Choi state $\Phi_{AB}^{\mathcal{N}}$ is separable. On the other hand, if the Choi state $\Phi_{AB}^{\mathcal{N}}$ of a given channel $\mathcal{N}$ is separable, then it is of the form
	\begin{equation}
		\Phi_{AB}^{\mathcal{N}}=\sum_{x\in\mathcal{X}}p(x)\sigma_A^x\otimes\tau_B^x
	\end{equation}
	for some probability distribution $p:\mathcal{X}\to[0,1]$ on a finite alphabet $\mathcal{X}$ and sets $\{\sigma_A^x:x\in\mathcal{X}\}$ and $\{\tau_B^x:x\in\mathcal{X}\}$ of states. We note that the property $\Tr_B[\Phi_{AB}^{\mathcal{N}}]=\pi_A$ of the Choi state translates to
	\begin{equation}\label{eq-choi_state_sep_partrace}
		\pi_A = \frac{\mathbbm{1}_A}{d_A}=\sum_{x\in\mathcal{X}} p(x)\sigma_A^x.
	\end{equation}
	Then, for every reference system~$R$ and state $\xi_{RA}$ acting on $\mathcal{H}_{RA}$, we find, by using \eqref{eq-Choi_rep_action}, that
	\begin{align}
		(\id_R\otimes\mathcal{N})(\xi_{RA})&=d_A\Tr_A[(\T_A(\xi_{RA})\otimes\mathbbm{1}_B)(\mathbbm{1}_R\otimes\Phi_{AB}^{\mathcal{N}})]\\
		&=\sum_{x\in\mathcal{X}}q(x)\omega_R^x\otimes\tau_B^x,
	\end{align}
	where
	\begin{align}
		\omega_R^x & \coloneqq\frac{\Tr_A[\T_A(\xi_{RA})(\mathbbm{1}_R\otimes d_Ap(x)\sigma_A^x)]}{q(x)},\\
		 q(x) & \coloneqq p(x)d_A\Tr[\xi_A^{\t}\sigma_A^x].
	\end{align}
	Now, the map $x\mapsto q(x)$ is a probability distribution on $\mathcal{X}$ since $q(x)\geq 0$ for all $x\in\mathcal{X}$ and
	\begin{align}
		\sum_{x\in\mathcal{X}}q(x)&=d_A\Tr\!\left[\xi_A^{\t}\left(\sum_{x\in\mathcal{X}} p(x)\sigma_A^x\right)\right]\\
		&=d_A\Tr[\xi_A^{\t}\pi_A]\\
		&=\Tr[\xi_A]\\
		&=1,
	\end{align}
	where we have made use of \eqref{eq-choi_state_sep_partrace}. So $(\id_R\otimes\mathcal{N})(\xi_{RA})$ is a separable state.
	\end{Proof}

	Another useful characterization of entanglement breaking channels is thr\-ough their Kraus representations, as shown in the following proposition:
	
	\begin{proposition}{prop-ent_break_Kraus}
		A channel $\mathcal{N}$ is entanglement breaking if and only if there exists a set of Kraus operators for $\mathcal{N}$, with each Kraus operator having unit rank.
	\end{proposition}
	
	\begin{Proof}
		First suppose that the Kraus operators of $\mathcal{N}$ have unit rank. They are therefore of the form $\ket{\phi_j}_B\bra{\psi_j}_A\eqqcolon K_j$ for $1\leq j\leq r$. Without loss of generality, we can let each vector in the set $\{\ket{\phi_j}\}_{j=1}^r$ be normalized. Then, since $\mathcal{N}$ is trace preserving, it holds that
		\begin{equation}\label{eq-ent_break_proof}
			\mathbbm{1}_A=\sum_{j=1}^r K_j^\dagger K_j=\sum_{j=1}^r\ket{\psi_j}_A\braket{\phi_j}{\phi_j}\!\bra{\psi_j}_A=\sum_{j=1}^r\ket{\psi_j}\!\bra{\psi_j}_A.
		\end{equation}
		
		Now, for every reference system $R$ of arbitrary dimension and every state $\rho_{RA}$, we find that
		\begin{align}
			(\id_R\otimes\mathcal{N})(\rho_{RA})&=\sum_{j=1}^r (\mathbbm{1}_R\otimes K_j)\rho_{RA}(\mathbbm{1}_R\otimes K_j^\dagger)\\
			&=\sum_{j=1}^r (\mathbbm{1}_R\otimes \ket{\phi_j}_B\bra{\psi_j}_A)\rho_{RA}(\mathbbm{1}_R\otimes \ket{\psi_j}_A\bra{\phi_j}_B)\\
			&=\sum_{j=1}^r (\mathbbm{1}_R\otimes\bra{\psi_j}_A)(\rho_{RA})(\mathbbm{1}_R\otimes\ket{\psi_j}_A)\otimes\ket{\phi_j}\!\bra{\phi_j}_B\\
			&=\sum_{j=1}^r p(j)\sigma_R^j\otimes\ket{\phi_j}\!\bra{\phi_j}_B,
		\end{align}
		where
		\begin{equation}
			\sigma_R^j\coloneqq \frac{(\mathbbm{1}_R\otimes\bra{\psi_j}_A)(\rho_{RA})(\mathbbm{1}_R\otimes\ket{\psi_j}_A)}{p(j)},\quad p(j)\coloneqq\bra{\psi_j}_A\,\rho_A\ket{\psi_j}_A.
		\end{equation}
		Note that $j\mapsto p(j)$ is a probability distribution since $p(j)\geq 0$ for all $j$, and
		\begin{align}
			\sum_{j=1}^r p(j)&=\sum_{j=1}^r\bra{\psi_j}_A\,\rho_A\ket{\psi_j}_A\\
			&=\sum_{j=1}^r\Tr[\ket{\psi_j}\!\bra{\psi_j}_A\,\rho_A]\\
			&=\Tr\!\left[\left(\sum_{j=1}^r \ket{\psi_j}\!\bra{\psi_j}_A\right)\rho_A\right]\\
			&=\Tr[\rho_A]\\
			&=1,
		\end{align}
		where we have made use of \eqref{eq-ent_break_proof}. Therefore, $(\id_R\otimes\mathcal{N})(\rho_{RA})$ is a separable state, so that $\mathcal{N}$ is entanglement breaking.
		
		Now, suppose that $\mathcal{N}$ is entanglement breaking. This means that its Choi state $\Phi_{AB}^{\mathcal{N}}$ is a separable state, which means that it can be written as
		\begin{equation}
			\Phi_{AB}^{\mathcal{N}}=\sum_{x\in\mathcal{X}}p(x)\ket{\psi_x}\!\bra{\psi_x}_A\otimes\ket{\phi_x}\!\bra{\phi_x}_B
		\end{equation}
		for some probability distribution $p:\mathcal{X}\to[0,1]$ on a finite alphabet $\mathcal{X}$ and sets of pure states $\{\ket{\psi_x}_A:x\in\mathcal{X}\}$, $\{\ket{\phi_x}_B:x\in\mathcal{X}\}$. Define the unit-rank operators
		\begin{equation}
			K_x\coloneqq \sqrt{\smash[b]{d_Ap(x)}}\ket{\phi_x}_B\bra{\conj{\psi_x}}_A,\quad x\in\mathcal{X}.
		\end{equation}
		Then, for every orthonormal basis state $\ket{i}_A$ on $A$, we have
		\begin{align}
			K_x\ket{i}_A&=\sqrt{\smash[b]{d_Ap(x)}}\, \ket{\phi_x}_B\braket{\conj{\psi_x}}{i}\\
			&=\sqrt{\smash[b]{d_Ap(x)}}\, \braket{i}{\psi_x}\ket{\phi_x}_B\\
			&=\sqrt{\smash[b]{d_Ap(x)}}\, (\bra{i}_A\otimes\mathbbm{1}_B)(\ket{\psi_x}_A\otimes\ket{\phi_x}_B).
		\end{align}
		Using this, we find that
		\begin{align}
			\mathcal{N}(\ket{i}\!\bra{i'}_A)&=d_A(\bra{i}_A\otimes\mathbbm{1}_B)\Phi_{AB}^{\mathcal{N}}(\ket{i'}_A\otimes\mathbbm{1}_B) \\
			&=\sum_{x\in\mathcal{X}}\sqrt{\smash[b]{d_Ap(x)}}(\bra{i}_A\otimes\mathbbm{1}_B)(\ket{\psi_x}_A\otimes\ket{\phi_x}_B)\nonumber\\[-4pt]
			&\qquad\qquad\times(\bra{\psi_x}_A\otimes\bra{\phi_x}_B)(\ket{i'}_A\otimes\mathbbm{1}_B)\sqrt{\smash[b]{d_Ap(x)}}\\
			&=\sum_{x\in\mathcal{X}}K_x\ket{i}\!\bra{i'}_A K_x^\dagger.
		\end{align}
		This holds for all $0\leq i,i'\leq d_A - 1$, which means that $\{K_x\}_{x\in\mathcal{X}}$ is a set of Kraus operators for $\mathcal{N}$, each of which has unit rank.
	\end{Proof}

	From this proposition, we immediately see that both quantum--class\-ical and classical--quantum channels are entanglement breaking, because each one has a Kraus representation with unit-rank Kraus operators. This implies that a composition of a quantum--classical channel followed by a classical-quan\-tum channel is also entanglement-breaking, and every such map can be written as
	\begin{equation}\label{eq-cq_qc_comp}
		\rho\mapsto \sum_{x\in\mathcal{X}}\Tr[M_x\rho] \sigma^x,
	\end{equation}
	where $\mathcal{X}$ is a finite alphabet, $\{\sigma^x\}_{x\in\mathcal{X}}$ is a set of quantum states, and $\{M_x\}_{x\in\mathcal{X}}$ is a POVM. Indeed, if each POVM element $M_x$ has a spectral decomposition of the form
	\begin{equation}
		M_x=\sum_{k=1}^{r_x}\lambda_k^x\ket{\psi_k^x}\!\bra{\psi_k^x},
	\end{equation}
	where $r_x=\rank(M_x)$, and each state $\sigma^x$ has a spectral decomposition of the form
	\begin{equation}
		\sigma^x=\sum_{\ell=1}^{s_x}\alpha_\ell^x\ket{\phi_\ell^x}\!\bra{\phi_\ell^x},
	\end{equation}
	where $s_x=\rank(\sigma^x)$, then \eqref{eq-cq_qc_comp} can be written as
	\begin{align}        \rho&\mapsto\sum_{x\in\mathcal{X}}\sum_{k=1}^{r_x}\sum_{\ell=1}^{s_x}\lambda_k^x\alpha_\ell^x\ket{\phi_\ell^x}\!\bra{\phi_\ell^x}\!\bra{\psi_k^x}\rho\ket{\psi_k^x}\\
		&=\sum_{x\in\mathcal{X}}\sum_{k=1}^{r_x}\sum_{\ell=1}^{s_x}K_{k,\ell}^x \rho (K_{k,\ell}^x)^\dagger,
	\end{align}
	where $K_{k,\ell}^x\coloneqq\sqrt{\lambda_k^x\alpha_\ell^x}\,\ket{\phi_\ell^x}\!\bra{\psi_k^x}$. Since $\{K_{k,\ell}^x:x\in\mathcal{X},~1\leq k\leq r_x,~1\leq\ell\leq s_x\}$ is a set of Kraus operators for the map, with each Kraus operator having unit rank, by Proposition \ref{prop-ent_break_Kraus} the map in \eqref{eq-cq_qc_comp} is entanglement breaking.
	
	A channel of the form \eqref{eq-cq_qc_comp} is sometimes called a ``measure-and-prepare channel'' or an ``inter\-cept-resend channel'' since the input to the channel is first measured and then replaced by a new state conditioned on the measurement outcome. Remarkably, \textit{any} entanglement breaking channel can be written in the form \eqref{eq-cq_qc_comp}, as we now show.
	
	\begin{theorem}{thm-ent_break_meas_reprepare}
		For every entanglement-breaking channel $\mathcal{N}$, there exists a finite alphabet $\mathcal{X}$, a set $\{\sigma^x\}_{x\in\mathcal{X}}$ of states, and a POVM $\{M_x\}_{x\in\mathcal{X}}$ such that the action of $\mathcal{N}$ can be written as
		\begin{equation}\label{eq-meas_reprep_map_2}
			\mathcal{N}(\rho)=\sum_{x\in\mathcal{X}}\Tr[M_x\rho]\sigma^x
		\end{equation}
		for every state $\rho$.
	\end{theorem}
	
	\begin{Proof}
		By Proposition \ref{prop-ent_break_Kraus}, the action of $\mathcal{N}$ can be written as
		\begin{equation}
			\mathcal{N}(\rho)=\sum_{j=1}^r K_j\rho K_j^\dagger,
		\end{equation}
		where $r=\rank(\Gamma^{\mathcal{N}})$ and $\{K_j\}_{j=1}^r$ is a set of Kraus operators for $\mathcal{N}$, with each Kraus operator having unit rank. 
        Thus, we can write that $K_j=\ket{\phi_j}\!\bra{\psi_j}$ for all $1\leq j\leq r$, where $\{\ket{\phi_j}\}_{j=1}^r$ and $\{\ket{\psi_j}\}_{j=1}^r$ are sets of vectors (without loss of generality, we can take $\{\ket{\phi_j}\}_{j=1}^r$ to be a set of pure states). Since $\mathcal{N}$ is trace preserving, by~\eqref{eq-ent_break_proof} it holds that $\sum_{j=1}^r\ket{\psi_j}\!\bra{\psi_j}=\mathbbm{1}$, implying that $\{\ket{\psi_j}\!\bra{\psi_j}\}_{j=1}^r$ is a POVM. 
        Therefore, defining the alphabet $\mathcal{X}=\{1,2,\dotsc,r\}$, the POVM elements $M_x\coloneqq \ket{\psi_x}\!\bra{\psi_x}$ and states $\sigma^x\coloneqq\ket{\phi_x}\!\bra{\phi_x}$, we have that
		\begin{equation}
			\mathcal{N}(\rho)=\sum_{x\in\mathcal{X}}\Tr[M_x\rho] \sigma^x,
		\end{equation}
		as required.
	\end{Proof}
	
	An extreme example of a measure-and-prepare channel, as in \eqref{eq-meas_reprep_map_2}, is one for which the output is a fixed state $\sigma$ for every outcome of the measurement described by the POVM $\{M_x\}_{x\in\mathcal{X}}$. In this case, the channel in \eqref{eq-meas_reprep_map_2} takes the form
	\begin{equation}
		\mathcal{N}(\rho)=\sum_{x\in\mathcal{X}}\Tr[M_x\rho]\sigma=\Tr[\rho]\sigma=\mathcal{R}^{\sigma}(\rho),
	\end{equation}
	where  the second equality follows because $\sum_{x\in\mathcal{X}}M_x=\mathbbm{1}$, and the last equality from the definition of the replacement channel for $\sigma$ in Definition~\ref{def-replace_channel}. This means that every replacement channel is a measure-and-prepare channel, and in particular an entanglement-breaking channel.
	
	The development above Proposition~\ref{prop-ent_break_Kraus} tells us that to every entanglement breaking channel there is associated a separable state, namely, the Choi state. The converse statement also holds; see Exercise~\ref{exer-ent_breaking_chan}.
	
	\begin{exercise}{exer-ent_breaking_chan}
		Given a separable state $\rho_{AB}=\sum_{x\in\mathcal{X}}p(x)\omega_A^x\otimes\tau_B^x$, show that the channel $\mathcal{N}_{A\to B}^{\rho}$ defined in \eqref{eq-state_to_channel} has the form
		\begin{equation}
			\mathcal{N}_{A\to B}^{\rho}(X_A)=\sum_{x\in\mathcal{X}}\Tr[X_AM_A^x]\tau_B^x,
		\end{equation}
		where
		\begin{equation}
		M_A^x=p(x)\left(\rho_A^{-\frac{1}{2}}\omega_A^x\rho_A^{-\frac{1}{2}}\right)^{\t}
		\end{equation}
		for all $x\in\mathcal{X}$. In other words, by the discussion after \eqref{eq-cq_qc_comp}, every separable state can be associated with an entanglement-breaking channel.
	\end{exercise}

\subsection{Hadamard Channels}\label{subsec-Hadamard_chan}

	It turns out that every entanglement-breaking channel can be regarded as the complement of what is called a \textit{Hadamard channel}, which we now define.
	
	\begin{definition}{Hadamard Channel}{def-Hadamard_chan}
		A channel $\mathcal{N}$ is called a \textit{Hadamard channel} or a \textit{Schur channel} if there exists a positive semi-definite operator $N$ with unit diagonal elements (in the standard basis) and an isometry $V$ such that
		\begin{equation}\label{eq-Hadamard_channel}
			\mathcal{N}(X)=N\ast VXV^\dagger
		\end{equation}
		for every linear operator $X$, where $N\ast VXV^\dagger$ is the \textit{Hadamard product}, also called the \textit{Schur product}, which is defined as the element-wise product of the operators $N$ and $VXV^\dagger$ when represented as matrices with respect to the standard basis.
	\end{definition}
	
	Given linear operators $X$ and $Y$ acting on a $d$-dimensional Hilbert space, with matrix representations in the standard basis as
	\begin{equation}
		X=\sum_{i,j=0}^{d-1} X_{i,j}\ket{i}\!\bra{j},\quad Y=\sum_{i,j=0}^{d-1} Y_{i,j}\ket{i}\!\bra{j},
	\end{equation}
	the Hadamard product $X\ast Y$ is the operator with matrix representation in the standard basis given by the product of the matrix elements of $X$ and $Y$:
	\begin{equation}\label{eq-Hadamard_product}
		X\ast Y=\sum_{i,j=0}^{d-1} X_{i,j}Y_{i,j}\ket{i}\!\bra{j}.
	\end{equation}
	Note that the Hadamard product can be defined as the element-wise product with respect to an arbitrary orthonormal basis; however, in this book, we only consider the Hadamard product with respect to the standard basis.
	
	The positive semi-definiteness of $N$ in the definition of a Hadamard channel is necessary and sufficient for the map $\mathcal{N}$ defined in \eqref{eq-Hadamard_channel} to be completely positive, while the fact that $N$ has unit diagonal elements in the standard basis and that $V$ is an isometry ensures that $\mathcal{N}$ is trace preserving.
	
	\begin{exercise}{exer-Hadamard_chan}
		\begin{enumerate}
			\item Show that a dephasing channel, as defined in \eqref{eq:QM-over:d-dim-dephase}, is a Hadamard channel.
			\item Let $\{K_j\}_{j=1}^r$ be a set of Kraus operators. Prove that the channel defined by the set $\{K_j \otimes \ket{j}\}_{j=1}^r$ of Kraus operators, is a Hadamard channel.
			\end{enumerate}
	\end{exercise}
	
	
	The following fact about complements of Hadamard channels provides an important characterization of Hadamard channels:
		
	\begin{proposition}{prop-Hadamard_chan_comp}
		Any complement of a Hadamard channel is entanglement breaking.
	\end{proposition}
	
	\begin{Proof}
		Suppose $\mathcal{N}_{A\to B}$ is a Hadamard channel between systems $A$ and $B$ with associated positive semi-definite operator $N$ having unit diagonal elements in the standard basis and with associated isometry $V$. Since $N$ is positive semi-definite and it has unit diagonal elements, it can be expressed as the Gram matrix of some set $\{\ket{\psi_i}:1\leq i\leq d\}$ of normalized vectors (Exercise~\ref{exer_Gram_matrices}), so that
		\begin{equation}
			N=\sum_{i,j=0}^{d-1} \braket{\psi_i}{\psi_j}\ket{i}\!\bra{j}.
		\end{equation}
		Therefore, using \eqref{eq-Hadamard_product} and Definition~\ref{def-Hadamard_chan}, the action of $\mathcal{N}$ can be written as
		\begin{equation}
			\mathcal{N}(X)=\sum_{i,j=0}^{d-1} \braket{\psi_i}{\psi_j}\!\bra{i}VXV^\dagger\ket{j}\ket{i}\!\bra{j}_B.
		\end{equation}
		Now, set
		\begin{equation}
			\ket{\phi_i}\coloneqq V^\dagger\ket{i}\Rightarrow \bra{\phi_i}=\bra{i}V,
		\end{equation}
		so that
		\begin{equation}
			\mathcal{N}(X)=\sum_{i,j=0}^{d-1}\braket{\psi_i}{\psi_j}\!\bra{\phi_i}X\ket{\phi_j}\ket{i}\!\bra{j}_B.
		\end{equation}
		Consider the operator $(U^{\mathcal{N}})_{A\to BE}$ defined as
		\begin{equation}\label{eq-Hadamard_chan_iso_ext}
			U^{\mathcal{N}}\coloneqq \sum_{i=0}^{d-1} \ket{i}_B\bra{\phi_i}_A\otimes\ket{\psi_i}_E.
		\end{equation}
		Since $V$ is an isometry, and the vectors $\{\ket{\psi_i}:1\leq i\leq d\}$ are normalized, it follows that
		\begin{equation}
			(U^{\mathcal{N}})^\dagger U^{\mathcal{N}}=\sum_{i=0}^{d-1} \ket{\phi_i}\!\bra{\phi_i}=\sum_{i=0}^{d-1} V^\dagger\ket{i}\!\bra{i}V=V^\dagger V=\mathbbm{1}.
		\end{equation}
		This means that $U^{\mathcal{N}}$ is an isometry. Furthermore,
		\begin{equation}\label{eq-Hadamard_comp_pf}
			U^{\mathcal{N}} X (U^{\mathcal{N}})^\dagger =\sum_{i,j=0}^{d-1} \bra{\phi_i}X\ket{\phi_j}\ket{\psi_i}\!\bra{\psi_j}_E\otimes\ket{i}\!\bra{j}_B,
		\end{equation}
		so that $\Tr_E[U^{\mathcal{N}}X(U^{\mathcal{N}})^\dagger]=\mathcal{N}(X)$. The operator $U^{\mathcal{N}}$ is therefore an isometric extension of $\mathcal{N}$. A complementary channel then results from tracing out $B$ in \eqref{eq-Hadamard_comp_pf}, i.e.,
		\begin{align}
			\mathcal{N}^c(X)&=\Tr_B[U^{\mathcal{N}}X(U^{\mathcal{N}})^\dagger]\\
			&=\sum_{i=0}^{d-1} \bra{\phi_i}X\ket{\phi_i}\ket{\psi_i}\!\bra{\psi_i}_E\\
			&=\sum_{i=0}^{d-1} \ket{\psi_i}\!\bra{\phi_i}X\ket{\phi_i}\!\bra{\psi_i}\\
			&=\sum_{i=0}^{d-1} K_i X K_i^\dagger,
		\end{align}
		where $K_i\coloneqq \ket{\psi_i}\!\bra{\phi_i}$. So $\mathcal{N}^c$ has a Kraus representation with unit-rank Kraus operators, which means, by Proposition \ref{prop-ent_break_Kraus}, that $\mathcal{N}^c$ is entanglement breaking. Every complement of a channel is related to another complement by an isometric channel acting on the output of the complement, and this does not change the entanglement-breaking property. 
	\end{Proof}
	
	By following the proof above  backwards, we find that every entangle\-ment-breaking channel is the complement of some Hadamard channel.

\subsection{Covariant Channels}\label{sec-covariant_channels}

	In Section~\ref{sec-QM_group_inv_states}, we defined states that are invariant under the action of a unitary representation of a group. We now define an analogous notion of invariance for quantum channels.
	
	\begin{definition}{Group-Covariant Channel}{def-group_cov_chan}
		Let $G$ be a group. A channel $\mathcal{N}_{A\to B}$ is called \textit{covariant with respect to $G$}, \textit{group-covariant}, \textit{$G$-covariant}, or simply \textit{covariant}, if there exist unitary representations 
        $G\ni g\mapsto U_A^g$ on $\mathcal{H}_A$ and $G\ni g\mapsto V_B^g$ on $\mathcal{H}_B$ such that for every state $\rho_A$,
		\begin{equation}\label{eq-tele_covariant}
			\mathcal{N}_{A\to B}(U_A^g\rho_A U_A^{g\dagger})=V_B^g\mathcal{N}_{A\to B}(\rho_A)V_B^{g\dagger}
		\end{equation}
		for all $g\in G$.
	\end{definition}
	
	\begin{exercise}{exer-group_cov_chan_reps}
		Let $\mathcal{N}_{A\to B}$ be a group-covariant channel, as per Definition~\ref{def-group_cov_chan}.
		\begin{enumerate}[topsep=0.3cm]
			\item Show that the condition in \eqref{eq-tele_covariant} can be written more compactly as follows:
				\begin{equation}\label{eq-chan_cov_compact}
					\mathcal{N}_{A\to B}=\mathcal{V}_B^{g\dagger}\circ\mathcal{N}_{A\to B}\circ\mathcal{U}_A^g
				\end{equation}
				for all $g\in G$, where $\mathcal{V}_B^{g\dagger}(\cdot)\coloneqq V_B^{g\dagger}(\cdot)V_B^g$ and $\mathcal{U}_A^g(\cdot)\coloneqq U_A^g(\cdot)U_A^{g\dagger}$.
			
			\item Show that the Choi representation of $\mathcal{N}_{A\to B}$ is invariant under the action of $U_A^{g\t}\otimes V_B^{g\dagger}$ for all $g\in G$; i.e., show that
				\begin{equation}
					\Gamma_{AB}^{\mathcal{N}}=(U_A^{g\t}\otimes V_B^{g\dagger})\Gamma_{AB}^{\mathcal{N}}(U_A^{g\t}\otimes V_B^{g\dagger})^{\dagger}
				\end{equation}
				for all $g\in G$.
				
			\item For every set $\{K_i\}_{i=1}^r$ of Kraus operators for $\mathcal{N}$ and for all $g\in G$, 
                show that $\{K_i^g\}_{i=1}^r$, with $K_i^g\coloneqq V_B^{g\dagger}K_i U_A^g$, is another set of Kraus operators for $\mathcal{N}$.
			
			\item For every isometric extension $W_{A\to BE}$ of $\mathcal{N}$ and for all $g\in G$, 
                show that $W_{A\to BE}^g\coloneqq V_B^{g\dagger} WU_A^g$ is another isometric extension of $\mathcal{N}$.
		\end{enumerate}
	\end{exercise}

Group covariant channels have group covariant isometric extensions, as the following lemma states.

\begin{Lemma*}{Isometric Extensions of Group-Covariant Channels}{thm-QM-channels:cov-stinespring}
Suppose that a channel $\mathcal{N}_{A\rightarrow B}$ is covariant  with respect to a group $G$. 
For an isometric extension $U_{A\rightarrow BE}^{\mathcal{N}}$ of
$\mathcal{N}_{A\rightarrow B}$, there is a set of unitaries $\{W_{E}^{g}\}_{g\in G}$ such
that the following covariance holds for all $g \in G$:
\begin{equation}
U_{A\rightarrow BE}^{\mathcal{N}}U_{A}^{g}=\left(  V_{B}^{g}\otimes W_{E}
^{g}\right)  U_{A\rightarrow BE}^{\mathcal{N}}.
\end{equation}
\end{Lemma*}

\begin{Proof}
Let a Kraus representation of $\mathcal{N}_{A\rightarrow B}$ be given as 
\begin{equation}
\mathcal{N}_{A\rightarrow B}(\rho_{A})=\sum_{j}L^{j}\rho_{A}L^{j\dag}.
\end{equation}
By~\eqref{eq-chan_cov_compact}, we can rewrite the group covariance condition as
\begin{equation}
\sum_{j}L^{j}\rho_{A}L^{j\dag}=\sum_{j}V_{B}^{g\dag}L^{j}U_{A}^{g}\rho
_{A}\left(V_{B}^{g\dag}L^{j}U_{A}^{g}\right)  ^{\dag}.
\end{equation}
Thus, the channel has two different Kraus representations $\{L^{j}\}_{j}$ and $\big\{V_{B}^{g\dag}L^{j}U_{A}^{g}\big\}_{j}$, and these are necessarily related by a
unitary with matrix elements $w_{jk}^{g}$ (see the discussion after \eqref{eq-QM-channels:kraus-rel-by-unitaries}):
\begin{equation}
V_{B}^{g\dag}L^{j}U_{A}^{g}=\sum_{k}w_{jk}^{g}L^{k}.
\end{equation}
A canonical isometric extension $U_{A\rightarrow BE}^{\mathcal{N}}$ of
$\mathcal{N}_{A\rightarrow B}$ is given as
\begin{equation}
U_{A\rightarrow BE}^{\mathcal{N}}=\sum_{j}L^{j}\otimes|j\rangle_{E},
\end{equation}
where $\{|j\rangle_{E}\}_j$ is an orthonormal basis. Defining $W_{E}^{g}$ as the unitary
\begin{equation}
W_{E}^{g}|k\rangle_{E}=\sum_{j}w_{jk}^{g}|j\rangle_{E},
\end{equation}
where the states $|k\rangle_{E}$ are chosen from $\{|j\rangle_{E}\}_j$, we have that
\begin{align}
U_{A\rightarrow BE}^{\mathcal{N}}U_{A}^{g}  & =\sum_{j}L^{j}U_{A}^{g}
\otimes|j\rangle_{E}\\
& =\sum_{j}V_{B}^{g}V_{B}^{g\dag}L^{j}U_{A}^{g}\otimes|j\rangle_{E}\\
& =\sum_{j}V_{B}^{g}\left[  \sum_{k}w_{jk}^{g}L^{k}\right]  \otimes
|j\rangle_{E}\\
& =V_{B}^{g}\sum_{k}L^{k}\otimes\sum_{j}w_{jk}^{g}|j\rangle_{E}\\
& =V_{B}^{g}\sum_{k}L^{k}\otimes W_{E}^{g}|k\rangle_{E}\\
& =\left(  V_{B}^{g}\otimes W_{E}^{g}\right)  U_{A\rightarrow BE}
^{\mathcal{N}}.
\end{align}
This concludes the proof.
\end{Proof}

	Recall the definition of the twirling map in Exercise~\ref{exer-group_inv_twirl}:
	\begin{equation}
		\mathcal{T}^G(\rho)\coloneqq\frac{1}{|G|}\sum_{g\in G}U^g\rho U^{g\dagger}.
	\end{equation}
	This map, which is evidently a quantum channel, takes an arbitrary state $\rho$ and makes it invariant under the action of the group $G$ given by the unitary representation $\{U^g\}_{g\in G}$. Similarly, we can define the \textit{twirl of a quantum channel}, which takes an arbitrary quantum channel $\mathcal{N}_{A\to B}$ and makes it group covariant, as per Definition~\ref{def-group_cov_chan}:
	\begin{equation}\label{eq-channel_twirl}
		\mathcal{N}_{A\to B}^G\coloneqq \frac{1}{|G|}\sum_{g\in G} \mathcal{V}_B^g\circ\mathcal{N}_{A\to B}\circ\mathcal{U}_A^{g\dagger}.
	\end{equation}
	
	\begin{exercise}{exer-channel_twirl}
		Given a quantum channel $\mathcal{N}_{A\to B}$, prove that the twirled channel $\mathcal{N}_{A\to B}^G$, as defined in \eqref{eq-channel_twirl}, is group covariant.
	\end{exercise}
	
	\begin{proposition}{prop-diamond_norm_group_cov}
		Let $\mathcal{N}_{A\to B}$ be a Hermiticity-preserving superoperator that is covariant with respect to a group $G$, as defined in Definition~\ref{def-group_cov_chan}. 
		\begin{enumerate}
			\item For every pure state $\psi_{A'A}$, with $d_{A'}=d_A$,
				\begin{equation}\label{eq-diamond_norm_group_cov_1}
					\norm{\mathcal{N}_{A\to B}(\psi_{A'A})}_1\leq\Norm{\mathcal{N}_{A\to B}(\phi_{A'A}^{\overline{\rho}})}_1,
				\end{equation}
				where $\rho_A\coloneqq\psi_A=\Tr_{A'}[\psi_{A'A}]$, $\overline{\rho}_A\coloneqq\mathcal{T}_A^G(\rho_A)=\frac{1}{|G|}\sum_{g\in G}U_A^g\rho_AU_A^{g\dagger}$, and $\phi_{A'A}^{\overline{\rho}}$ is a purification of $\overline{\rho}$.
				
			\item The diamond norm of $\mathcal{N}$ is given by
				\begin{equation}\label{eq-diamond_norm_group_cov_2}
					\norm{\mathcal{N}}_{\diamond}=\sup_{\psi_{A'A}}\left\{\norm{(\id_{A'}\otimes\mathcal{N}_{A\to B})(\psi_{A'A})}_1:\psi_A=\mathcal{T}_A^G(\psi_A)\right\},
				\end{equation}
				where the optimization is with respect to pure states $\psi_{A'A}$, with $d_{A'}=d_A$, such that the reduced state $\psi_A$ is invariant under the twirling channel $\mathcal{T}_A^G(\cdot)\coloneqq\frac{1}{|G|}\sum_{g\in G}U_A^g(\cdot)U_A^{g\dagger}$ defined by the representation $\{U_A^g\}_{g\in G}$.
			
			\item If the representation $\{U_A^g\}_{g\in G}$ is such that $\mathcal{T}_A^G(\cdot)=\Tr[\cdot]\frac{\mathbbm{1}_A}{d_A}$, then the supremum in~\eqref{eq-diamond_norm_group_cov_2} is achieved when $\psi_{A'A}$ is the maximally entangled state $\Phi_{A'A}$, so that
			\begin{equation}
			\norm{\mathcal{N}}_{\diamond}=\norm{\Phi_{AB}^{\mathcal{N}}}_1.
			\end{equation}
		\end{enumerate}
	\end{proposition}
	
	\begin{Proof}
		\hfill\begin{enumerate}
			\item Let $\psi_{A'A}$ be an arbitrary pure state, $\rho_A=\psi_A$, $\overline{\rho}_A=\mathcal{T}_A^G(\rho_A)$, and let $\phi_{A'A}^{\overline{\rho}}$ be a purification of $\overline{\rho}_A$. Let us also consider the following purification of $\overline{\rho}_A$:
				\begin{equation}
					\ket{\psi^{\overline{\rho}}}_{RA'A}\coloneqq\frac{1}{\sqrt{|G|}}\sum_{g\in G}\ket{g}_{R}\otimes U_A^g\ket{\psi}_{A'A},
				\end{equation}
				where $\{\ket{g}\}_{g\in G}$ is an orthonormal set. (Recall Exercise~\ref{exer-group_inv_twirl}.) Now, because all purifications of a state can be mapped to each other via isometries acting on the purifying system (see Section~\ref{sec:qm:purification-def}), there exists an isometry $W_{A'\to RA'}$ such that $\ket{\psi^{\overline{\rho}}}_{RA'A}=W_{A'\to RA'}\ket{\phi^{\overline{\rho}}}_{A'A}$. Therefore,
				\begin{align}
					\norm{\mathcal{N}_{A\to B}(\psi_{RA'A}^{\overline{\rho}})}_1&=\norm{W_{A'\to RA'}\mathcal{N}_{A\to B}(\phi_{A'A}^{\overline{\rho}})(W_{A'\to RA'})^{\dagger}}_1\\
					&=\norm{\mathcal{N}_{A\to B}(\phi_{A'A}^{\overline{\rho}})}_1,\label{eq-diamond_norm_group_cov_pf1}
				\end{align}
				where the last line follows from isometric invariance of the trace norm (see \eqref{eq-Schatten_norm_iso_invar}).
				
				Now, let us apply the quantum channel $X\mapsto\sum_{g\in G}\ket{g}\!\bra{g}X\ket{g}\!\bra{g}$ to the system~$R$. By the data-processing inequality in \eqref{eq-data_proc_trace_norm_0}, we find that
				\begin{align}
					\norm{\mathcal{N}_{A\to B}(\psi_{RA'A}^{\overline{\rho}})}_1&\geq\norm{\frac{1}{|G|}\sum_{g\in G}\ket{g}\!\bra{g}_R\otimes(\mathcal{N}_{A\to B}\circ\mathcal{U}_A^g)(\psi_{A'A})}_1\\
					&=\norm{\frac{1}{|G|}\sum_{g\in G}\ket{g}\!\bra{g}_R\otimes(\mathcal{V}_B^{g\dagger}\circ\mathcal{N}_{A\to B}\circ\mathcal{U}_A^g)(\psi_{A'A})}_1,
				\end{align}
				where the last line follows from applying the unitary channel defined by the unitary $\sum_{g\in G}\ket{g}\!\bra{g}_R\otimes V_B^{g\dagger}$ and from unitary invariance of the trace norm. Finally, using the covariance of $\mathcal{N}$, in particular \eqref{eq-chan_cov_compact}, along with \eqref{eq-diamond_norm_group_cov_pf1}, we obtain
				\begin{align}
					\norm{\mathcal{N}_{A\to B}(\phi_{A'A}^{\overline{\rho}})}_1&=\norm{\mathcal{N}_{A\to B}(\psi_{RA'A}^{\overline{\rho}})}_1\\
					&\geq\norm{\frac{1}{|G|}\sum_{g\in G}\ket{g}\!\bra{g}_R\otimes\mathcal{N}_{A\to B}(\psi_{A'A})}_1\\
					&=\norm{\mathcal{N}_{A\to B}(\psi_{A'A})}_1,
				\end{align}
				 where the last line follows from \eqref{eq-Schatten_norm_mult}. The derived inequality is precisely \eqref{eq-diamond_norm_group_cov_1},
				
			\item Note that, by definition, every purification $\phi_{A'A}^{\overline{\rho}}$ of $\overline{\rho}_A$ is such that its reduced state on $A$ is invariant under the channel $\mathcal{T}_A^G$. Therefore, using \eqref{eq-diamond_norm_group_cov_1}, for every pure state $\psi_{A'A}$, we obtain
				\begin{align}
					\norm{\mathcal{N}_{A\to B}(\psi_{A'A})}_1&\leq\norm{\mathcal{N}_{A\to B}(\phi_{A'A}^{\overline{\rho}})}_1\\
					&\leq\sup_{\phi_{A'A}}\left\{\norm{\mathcal{N}_{A\to B}(\phi_{A'A})}_1:\phi_A=\mathcal{T}_A^G(\phi_A)\right\}.
				\end{align}
				Since this inequality holds for every pure state $\psi_{A'A}$, and because $\mathcal{N}$ is Hermiticity preserving, we can use \eqref{eq-diamond_norm_Herm_pres} to conclude that
				\begin{align}
					\norm{\mathcal{N}}_{\diamond}&=\sup_{\psi_{A'A}}\norm{\mathcal{N}_{A\to B}(\psi_{A'A})}_1\\
					&\leq \sup_{\psi_{A'A}}\left\{\norm{\mathcal{N}_{A\to B}(\psi_{A'A})}_1:\psi_A=\mathcal{T}_A^G(\psi_A)\right\}.
				\end{align}
				The opposite inequality is immediate, because the set $\{\psi_{A'A}:\psi_A=\mathcal{T}_A^G(\psi_A)\}$ is a subset of all pure states. This concludes the proof of \eqref{eq-diamond_norm_group_cov_2}.
			
			\item If $\mathcal{T}_A^G(\cdot)=\Tr[\cdot]\frac{\mathbbm{1}_A}{d_A}$, then the optimization in \eqref{eq-diamond_norm_group_cov_1} is with respect to pure states $\psi_{A'A}$ whose reduced state on $A$ is maximally mixed. All such pure states are maximally entangled (see Definition~\ref{def-max_ent_pure_state} and the discussion below it), which means that there exists a unitary $U$ such that $\psi_{A'A}=\mathcal{U}_{A'}(\Phi_{A'A})$, where $\Phi_{A'A}=\ket{\Phi}\bra{\Phi}_{A'A}$ is the maximally entangled state defined by $\ket{\Phi}_{A'A}=\frac{1}{\sqrt{d_A}}\sum_{i=0}^{d_A-1}\ket{i,i}_{A'A}$ (see Exercise~\ref{exer-maximally_ent_states}). By unitary invariance of the trace norm, we thus immediately obtain $\norm{\mathcal{N}}_{\diamond}=\norm{\mathcal{N}_{A\to B}(\Phi_{A'A})}_1=\norm{\Phi_{A'B}^{\mathcal{N}}}_1$, as required.~\qedhere
		\end{enumerate}
	\end{Proof}

\subsection{Bipartite and Multipartite Channels}\label{sec-bipartite_channels}
	
	
	
	

\section{Examples of Communication Channels}

\label{subsec-qubit_channel}

\subsection{(Generalized) Amplitude Damping Channel}\label{sec:QM-over:amp-damp-ch}

	The \textit{amplitude damping channel with decay parameter $\gamma\in[0,1]$} is the channel $\mathcal{A}_\gamma$ given by $\mathcal{A}_{\gamma}(\rho)=A_1\rho A_1^\dagger+A_2\rho A_2^\dagger$, with the two Kraus operators $A_1$ and $A_2$ defined as
	\begin{equation}\label{eq-amplitude_damping_channel}
		A_1=\sqrt{\gamma}\ket{0}\!\bra{1},\quad A_2=\ket{0}\!\bra{0}+\sqrt{1-\gamma}\ket{1}\!\bra{1}.
	\end{equation}
	It is straightforward to verify that $A_1^\dagger A_1^{\vphantom{\dag}}+A_2^\dagger A_2^{\vphantom{\dag}}=\mathbbm{1}$, so that $\mathcal{A}_\gamma$ is indeed trace preserving. 
	
	Let $\rho$ be a state with a matrix representation in the standard basis $\{\ket{0},\ket{1}\}$ as
	\begin{equation}\label{eq-qubit_state}
		\rho=\begin{pmatrix} 1-\lambda & \alpha \\ \conj{\alpha} & \lambda \end{pmatrix}.
	\end{equation}
	In order for $\rho$ to be a state (positive semi-definite with unit trace),
the conditions $0\leq \lambda\leq 1$ and  $\lambda(1-\lambda)\geq\abs{\alpha}^2$ should hold, where $\alpha\in\mathbb{C}$. The output state $\mathcal{A}_\gamma(\rho)$ has the matrix representation
	\begin{equation}
		\mathcal{A}_\gamma(\rho)=\begin{pmatrix} 1-(1-\gamma)\lambda & \sqrt{1-\gamma}\alpha \\ \sqrt{1-\gamma}\conj{\alpha} & (1-\gamma)\lambda  \end{pmatrix}.
	\end{equation}
	
	To obtain a physical interpretation of the amplitude damping channel, consider that it can written in the Stinespring form as
	\begin{equation}\label{eq-amp_damp_Stinespring}
	\mathcal{A}_\gamma(\rho)=\Tr_E[U^{1-\gamma}(\rho\otimes\ket{0}\!\bra{0}_E)(U^{1-\gamma})^\dagger],
	\end{equation}
	where $E$ is a qubit environment system, 
 and
	\begin{equation}\label{eq-amp_damp_BS}
		U^{\eta}=\begin{pmatrix}1&0&0&0\\0&\sqrt{\eta}&\I\sqrt{1-\eta}&0\\0&\I\sqrt{1-\eta}&\sqrt{\eta}&0\\0&0&0&1\end{pmatrix}
	\end{equation}
	is unitary. Note that the action of $\mathcal{A}_\gamma$ on the pure states $\ket{0}$ and $\ket{1}$ is, respectively,
	\begin{equation}\label{eq-amp_damp_receiver}
		\begin{aligned}
		\mathcal{A}_{\gamma}(\ket{0}\!\bra{0}_A) &= \ket{0}\!\bra{0}_B, \\
		\mathcal{A}_{\gamma}(\ket{1}\!\bra{1}_A) &= \gamma \ket{0}\!\bra{0}_B + (1-\gamma)\ket{1}\!\bra{1}_B.
		\end{aligned}
	\end{equation}
	
	A complementary channel $\mathcal{A}_{\gamma}^c$ for $\mathcal{A}_\gamma$ is defined from
	\begin{equation}
	\mathcal{A}_{\gamma}^c(\rho) \coloneqq \Tr_B[U^{1-\gamma}(\rho\otimes\ket{0}\!\bra{0}_E)(U^{1-\gamma})^\dagger].
	\end{equation}
	
	\begin{exercise}{exer-AD_complement}
		Prove that $\mathcal{A}_{\gamma}^c=\mathcal{A}_{1-\gamma}$ for all $\gamma\in[0,1]$.
	\end{exercise}
	
\begin{figure}
		\centering
		\includegraphics[scale=0.8]{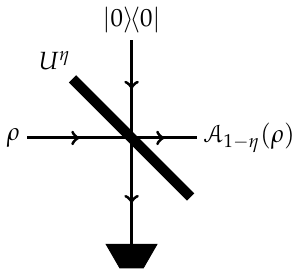}
		\caption{The amplitude damping channel $\mathcal{A}_{1-\eta}$ can be interpreted, using \eqref{eq-amp_damp_Stinespring}, as an interaction with a qubit analogue of a bosonic beamsplitter unitary~$U^\eta$, followed by discarding the output state of the environment. The channel from the sender to the receiver is from the left to the right, while the input and output environment systems are at the top and the bottom, respectively. In the bosonic case, the state $\ket{0}\!\bra{0}$ of the environmental input arm of the beamsplitter corresponds to the vacuum state, which contains no photons, and the channel to the receiver's end is called the pure-loss bosonic channel.}\label{fig-amp_damp}
	\end{figure}

	Using the result of Exercise~\ref{exer-AD_complement}, we have that the action of the complementary channel $\mathcal{A}_{\gamma}^c$ on these states is
	\begin{equation}\label{eq-amp_damp_env}
		\begin{aligned}
		\mathcal{A}_{\gamma}^c(\ket{0}\!\bra{0}_A)&=\ket{0}\!\bra{0}_E,\\
		\mathcal{A}_{\gamma}^c(\ket{1}\!\bra{1}_A)&=(1-\gamma)\ket{0}\!\bra{0}_E + \gamma\ket{1}\!\bra{1}_E.
		\end{aligned}
	\end{equation}
	We see that whenever the state $\ket{0}\!\bra{0}$ is input to the channel, the output systems $B$ and $E$ are both in the state $\ket{0}\!\bra{0}$. On the other hand, if the input state is $\ket{1}\!\bra{1}$, then $B$ receives a mixed state: with probability $\gamma$, the state is $\ket{0}\!\bra{0}$, and with probability $1-\gamma$, the state is $\ket{1}\!\bra{1}$. The situation for $E$ is reversed, receiving $\ket{0}\!\bra{0}$ with probability $1-\gamma$ and $\ket{1}\!\bra{1}$ with probability $\gamma$. The unitary $U^\eta$ can thus be viewed as a qubit analogue of a \textit{beamsplitter} with transmissivity $\eta$, and, accordingly, the amplitude damping channel $\mathcal{A}_{\gamma}$ can be viewed as a qubit analogue of the \textit{pure-loss bosonic channel} with loss probability $\gamma$; see Figure \ref{fig-amp_damp}. 
	
	A beamsplitter is an optical device that takes two beams of light as input and mixes and recombines them into two separate output beams, with the first output beam containing a fraction $\eta$ of the intensity of the first input beam and a fraction $1-\eta$ of the intensity of the second input beam, and vice versa for the second output beam. When one of the input ports of the beamsplitter is empty, i.e., in the vacuum state, and one of the output beams is discarded, the relation between the single remaining input and the single remaining output, interpreted as the receiver, is described by the pure-loss bosonic channel. In the case of a single incoming photon, the pure-loss channel either transmits the photon with probability $\eta$ (allowing it to go to the receiver) or reflects it with probability $1-\eta$ (sending it to the environment). 
	
	To draw the correspondence between the qubit amplitude damping channel and the pure-loss bosonic channel described above, we can think of the qubit state $\ket{0}$ as the vacuum state and the qubit state $\ket{1}$ as the state of a single photon. It is possible to show that the output states of the amplitude damping channel in \eqref{eq-amp_damp_receiver} and \eqref{eq-amp_damp_env} for the receiver and environment, respectively, then exactly match the action of the bosonic pure-loss channel on the subspace spanned by $\ket{0}$ and $\ket{1}$ (please consult the Bibliographic Notes in Section~\ref{sec:qm:bib-notes} for further references on this connection). The amplitude damping channel can indeed, therefore, be viewed as the qubit analogue of the bosonic pure-loss channel.
	
	By replacing the initial state $\ket{0}\!\bra{0}$ of the environment in \eqref{eq-amp_damp_Stinespring} with the state
	\begin{equation}
		\theta_{N_B}\coloneqq (1-N_B)\ket{0}\!\bra{0}+N_B\ket{1}\!\bra{1},\quad N_B\in[0,1],
	\end{equation}
	we define the \textit{generalized amplitude damping channel} $\mathcal{A}_{\gamma,N_B}$ as
	\begin{equation}\label{eq-gen_amp_damp}
		\mathcal{A}_{\gamma,N_B}(\rho) 
        \coloneqq \Tr_E[U^{1-\gamma}(\rho\otimes\theta_{N_B})(U^{1-\gamma})^\dagger].
	\end{equation}
	This channel has the following four Kraus operators:
	\begin{align}
		A_1=\sqrt{1-N_B}\begin{pmatrix}1&0\\0&\sqrt{1-\gamma}\end{pmatrix},&\quad A_2=\sqrt{1-N_B}\begin{pmatrix}0&\sqrt{\gamma}\\0&0\end{pmatrix},\label{eq-gen_amp_damp_Kraus1}\\
		A_3=\sqrt{N_B}\begin{pmatrix}\sqrt{1-\gamma}&0\\0&1\end{pmatrix},&\quad A_4=\sqrt{N_B}\begin{pmatrix}0&0\\\sqrt{\gamma}&0\end{pmatrix}.\label{eq-gen_amp_damp_Kraus2}
	\end{align}
 Note that the amplitude damping channel is a special case of the generalized amplitude damping channel in which the thermal noise $N_B = 0$, so that $\mathcal{A}_{\gamma}=\mathcal{A}_{\gamma,0}$.
 
	\begin{exercise}{exer-GADC}
		\begin{enumerate}
			\item Prove that $\mathcal{A}_{\gamma,N}=(1-N)\mathcal{A}_{\gamma,0}+N\mathcal{A}_{\gamma,1}$ for all $\gamma,N\in[0,1]$.
			
			\item Prove that $\mathcal{A}_{\gamma,N}=\mathcal{A}_{\gamma_2,N_2}\circ\mathcal{A}_{\gamma_1,N_1}$, where $\gamma=\gamma_1+\gamma_2-\gamma_1\gamma_2$ and $N=\frac{\gamma_1(1-\gamma_2)N_1+\gamma_2N_2}{\gamma_1+\gamma_2-\gamma_1\gamma_2}$.
			
			\item Using 2., along with the result of Exercise~\ref{exer-AD_complement}, prove that the amplitude damping channel $\mathcal{A}_{\gamma,0}$ is degradable for all $\gamma\in\left[0,\frac{1}{2}\right)$, with degrading channel $\mathcal{A}_{\frac{1-2\gamma}{1-\gamma},0}$.
		\end{enumerate}
	\end{exercise}
	
	The state $\theta_{N_B}$ in \eqref{eq-gen_amp_damp} is a qubit thermal state and can be regarded as a qubit analogue of the bosonic thermal state. The latter is a state corresponding to a heat bath with an average number of photons equal to $N_B$. Accordingly, the generalized amplitude damping channel can then be seen as a qubit analogue of the bosonic thermal noise channel.
	
	\begin{exercise}{exer-GADC_Pauli}
		Recall the Pauli operators $X$, $Y$, $Z$ from \eqref{eq-QM-Pauli_mat}, and consider the generalized amplitude-damping channel $\mathcal{A}_{\gamma,N}$, with $\gamma,N\in[0,1]$. Show that
		\begin{align}
			\mathcal{A}_{\gamma,N}(X)&=\sqrt{1-\gamma}X,\\
			\mathcal{A}_{\gamma,N}(Y)&=\sqrt{1-\gamma}Y,\\
			\mathcal{A}_{\gamma,N}(Z)&=(1-\gamma)Z,\\
			\mathcal{A}_{\gamma,N}(\mathbbm{1})&=\mathbbm{1}+\gamma(1-2N)Z.
		\end{align}
		From this, conclude that the generalized amplitude-damping channel is covariant with respect to the group defined by the operators $\{\mathbbm{1},Z\}$, so that for all $\gamma,N\in[0,1]$,
		\begin{equation}
		\mathcal{A}_{\gamma,N}(Z\rho Z)=Z\mathcal{A}_{\gamma,N}(\rho)Z
  		\end{equation}
		for every state $\rho$.
	\end{exercise}

\subsection{Erasure Channel}\label{sec:QM-over:qudit-erasure}

	The qudit \textit{erasure channel} $\mathcal{E}_{p}$ with \textit{erasure probability} $p\in[0,1]$ acts on the input by transmitting it unaltered with probability $1-p$, and by replacing it with an ``error flag'' state $\ket{e}\!\bra{e}$ with probability $p$. Mathematically, it is defined as follows for every $\rho\in\Lin(\mathbb{C}^d)$, with $d\geq 2$:
	\begin{equation}\label{eq-erasure_channel}
		\mathcal{E}_{p}(\rho)=(1-p)\rho+p\Tr[\rho]\ket{e}\!\bra{e},
	\end{equation}
	where $\ket{e}$ is some unit vector that is orthogonal to all states in the input qudit Hilbert space and $\ket{e}\!\bra{e}$ is called the \textit{erasure state}. For example, if the input qudit Hilbert space is spanned by the standard basis $\{\ket{0},\ket{1},\dotsc,\ket{d-1}\}$, then we can set~$\ket{e}=\ket{d}$. Thus, the erasure channel takes a linear operator acting on $\mathbb{C}^d$ and outputs a linear operator acting on $\mathbb{C}^{d+1}$. Physically, this ensures that the erasure state is perfectly distinguishable from the input, meaning that \emph{errors can be detected}.
	
	\begin{exercise}{exer-erasure_channel}
		Verify that a set of Kraus operators for the erasure channel $\mathcal{E}_p$ is 
		\begin{equation}
			\left\{\sqrt{1-p}(\ket{0}\!\bra{0}+\cdots + \ket{d-1}\!\bra{d-1}),\sqrt{p}\ket{e}\!\bra{0},\ldots, \sqrt{p}\ket{e}\!\bra{d-1}\right\}.
		\end{equation}
		Also, show that the Choi representation of $\mathcal{E}_p$ is
		\begin{equation}
			\Gamma^{\mathcal{E}_p}=(1-p)\Gamma_d+p\mathbbm{1}_d\otimes\ket{e}\bra{e}.
		\end{equation}
	\end{exercise}
	
\begin{figure}
		\centering
		\includegraphics[scale=0.8]{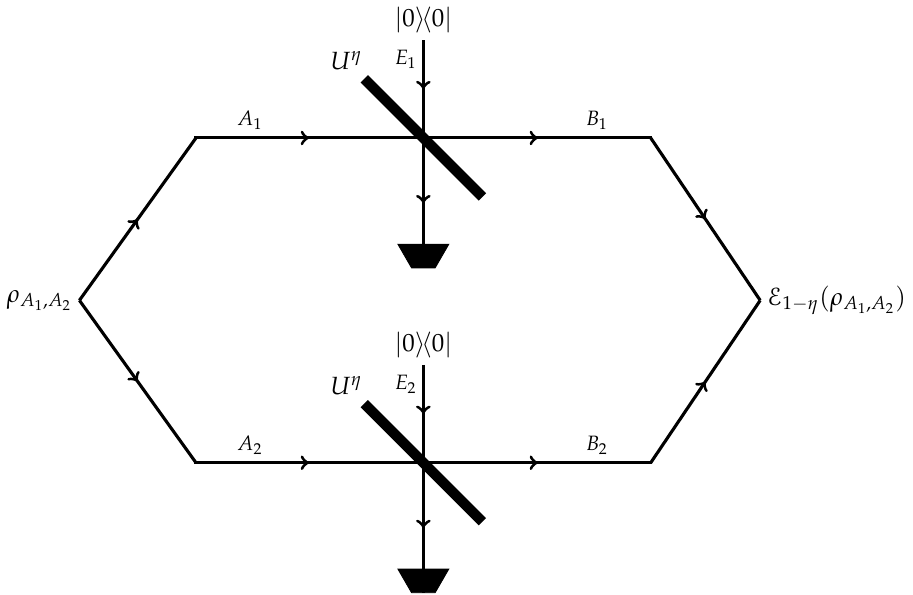}
		\caption{The qubit erasure channel $\mathcal{E}_{1-\eta}$, for $\eta\in[0,1]$, can be physically realized by using a photonic dual-rail qubit system and passing each of the two modes through a beamsplitter, modeled by the unitary $U^\eta$ in \eqref{eq-pure_loss_BS}, such that the input from the environment is the vacuum state. }\label{fig-pure_loss_DR}
	\end{figure}

	We now discuss how the pure-loss bosonic channel restricted to acting on dual-rail single-photon inputs 
    corresponds to a qubit erasure channel. The correspondence is illustrated in Figure \ref{fig-pure_loss_DR}. 
	
	Consider a dual-rail optical system, i.e., a quantum optical system with two distinct optical modes $A_1$ and $A_2$ representing the input to the channel. As described at the beginning of Section~\ref{sec-QM_states}, the two-dimensional subspace spanned by the states $\{\ket{0,1}_{A_1,A_2},\ket{1,0}_{A_1,A_2}\}$, consisting of a total of one photon in either one of the two modes, constitutes a qubit system. We also let~$E_1$ and $E_2$ be two distinct optical modes constituting a dual-rail qubit system representing the environment of the channel. Finally, let $B_1$ and $B_2$ be two distinct optical modes spanned by the states $\{\ket{0,0}_{B_1,B_2},\ket{0,1}_{B_1,B_2},\ket{1,0}_{B_1,B_2}\}$, so that together $B_1$ and $B_2$ constitute a qutrit system.
	
	When the unitary corresponding to a quantum-optical beamsplitter acts on the space spanned by $\{\ket{0,0},\ket{0,1},\ket{1,0}\}$, it is equivalent to the upper left $3\times 3$ corner of the matrix in \eqref{eq-amp_damp_BS}. We can make use of this fact because the environment state for the bosonic pure-loss channel is prepared in the state $\ket{0}\!\bra{0}$. Let $U_{A_iE_i\to B_iE_i}^\eta$ then denote the beamsplitter unitary, for $i\in\{1,2\}$, with the following action on the basis $\{\ket{0,0},\ket{0,1},\ket{1,0}\}$ of the input and output modes (in that order):
	\begin{equation}\label{eq-pure_loss_BS}
		U_{A_iE_i\to B_iE_i}^\eta=\begin{pmatrix} 1&0&0
		\\
		0&\sqrt{\eta}&\I\sqrt{1-\eta}
		\\
		0&\I\sqrt{1-\eta}&\sqrt{\eta}
\end{pmatrix}.
	\end{equation}
	Letting $\rho_{A_1A_2}$ be an arbitrary qubit state on the two modes $A_1$ and $A_2$ defined by
	\begin{equation}\label{eq-pure_loss_input}
		\begin{aligned}
		\rho_{A_1,A_2}&=(1-\lambda)\ket{0,1}\!\bra{0,1}_{A_1,A_2}+\alpha\ket{0,1}\!\bra{1,0}_{A_1,A_2}+\conj{\alpha}\ket{1,0}\!\bra{0,1}_{A_1,A_2}\\
		&\qquad\qquad+\lambda\ket{1,0}\!\bra{1,0}_{A_1,A_2},
		\end{aligned}
	\end{equation}
	the pure-loss bosonic channel on the dual-rail qubit system $A_1A_2$ is given by
	\begin{equation}
		\begin{aligned}
		&\Tr_{E_1,E_2}\!\left[(U_{A_1E_1\to B_1E_1}^\eta\otimes U_{A_2E_2\to B_2E_2}^\eta)(\rho_{A_1,A_2}\otimes\ket{0,0}\!\bra{0,0}_{E_1,E_2})\right.\\
		&\qquad\qquad \left.\times (U_{A_1E_1\to B_1E_1}^\eta\otimes U_{A_2E_2\to B_2E_2}^\eta)^\dagger\right].
		\end{aligned}
	\end{equation}
	Although this has a similar form to \eqref{eq-amp_damp_Stinespring}, which defines the amplitude damping channel, our particular realization of the qubit system in terms of dual-rail single photons results in a completely different output from that of the amplitude damping channel. In particular, using \eqref{eq-pure_loss_input} along with \eqref{eq-pure_loss_BS}, it is straightforward to show that
	\begin{align}
		&\Tr_{E_1,E_2}\!\left[(U_{A_1E_1\to B_1E_1}^\eta\otimes U_{A_2E_2\to B_2E_2}^\eta)(\rho_{A_1,A_2}\otimes\ket{0,0}\!\bra{0,0}_{E_1,E_2})\right. \nonumber \\
		&\qquad\qquad \left. \times(U_{A_1E_1\to B_1E_1}^\eta\otimes U_{A_2E_2\to B_2E_2}^\eta)^\dagger\right]\nonumber\\
		&=\eta\rho_{B_1,B_2}+(1-\eta)\ket{0,0}\!\bra{0,0}_{B_1,B_2} \\
		&=\mathcal{E}_{1-\eta}(\rho_{A_1,A_2}).
	\end{align}
	In other words, the pure-loss bosonic channel on a dual-rail qubit system is simply the qubit erasure channel $\mathcal{E}_{1-\eta}$ with erasure probability $1-\eta$ and erasure state $\ket{0,0}\!\bra{0,0}_{B_1,B_2}$. This means that a dual-rail single-photonic qubit sent through a pure-loss bosonic channel is transmitted to the receiver unchanged with probability~$\eta$, or it is lost, and replaced by the vacuum state $\ket{0,0}\!\bra{0,0}$, with probability $1-\eta$.

\subsection{Pauli Channels}

	The \textit{Pauli channels} constitute an important class of channels on qubit  systems. They are based on the qubit \textit{Pauli operators} $X$, $Y$, $Z$, which we first introduced in \eqref{eq-Pauli_mat_1} and \eqref{eq-Pauli_mat_2} and have the following matrix representation in the standard basis $\{\ket{0},\ket{1}\}$:
	\begin{equation}\label{eq-Pauli_operators}
		X=\begin{pmatrix} 0&1\\1&0\end{pmatrix},\quad Y=\begin{pmatrix} 0&-\I\\\I &0\end{pmatrix},\quad Z=\begin{pmatrix} 1&0\\0&-1\end{pmatrix}.
	\end{equation}
	Recall that the Pauli operators are Hermitian, satisfy $X^2=Y^2=Z^2=\mathbbm{1}$, and mutually anti-commute.
	
	A general Pauli channel is one whose Kraus operators are proportional to the Pauli operators, i.e.,
	\begin{equation}
		\rho\mapsto p_I\rho+p_X X\rho X +p_Y Y\rho Y +p_Z Z\rho Z,
	\end{equation}
	where the four numbers $p_I,p_X,p_Y,p_Z$ form a probability distribution; i.e., they are non-negative and add up to $1$.
	
	Here we highlight two particular Pauli channels of interest.
	\begin{enumerate}
		\item \textit{Dephasing channel}: We let $p_I=1-p$ and $p_Z=p$ for $0\leq p\leq 1,$ and $p_X=p_Y=0$. The action of the dephasing channel is thus
			\begin{equation}
				\rho\mapsto (1-p)\rho+pZ\rho Z.
			\end{equation}
			To see why this is called the dephasing channel, consider again a general state $\rho$ of the form \eqref{eq-qubit_state} and let $p=\frac{1}{2}$. In this case, we call the channel the \textit{completely dephasing channel}, and it is straightforward to see that
			\begin{equation}\label{eq-completely_dephasing_channel}
				\rho=\begin{pmatrix} 1-\lambda & \alpha \\ \conj{\alpha} & \lambda \end{pmatrix}\mapsto \frac{1}{2}\rho+\frac{1}{2}Z\rho Z=\begin{pmatrix} 1-\lambda & 0\\ 0 & \lambda \end{pmatrix}.
			\end{equation}
			In other words, the completely dephasing channel eliminates the off-diagonal elements of the input state (when represented in the standard basis, which is the same basis in which $Z$ is diagonal), so that the relative phase between the $\ket{0}$ state and the $\ket{1}$ state vanishes and the state becomes effectively classical.
			
			In the more general case when $p\neq\frac{1}{2}$, the effect of the dephasing channel is to reduce the magnitude of the off-diagonal elements:
			\begin{align}
				\rho  =\begin{pmatrix} 1-\lambda & \alpha \\ \conj{\alpha} & \lambda \end{pmatrix}
				& \mapsto (1-p)\rho+p Z\rho Z = \begin{pmatrix}
				1-\lambda & (1-2p)\alpha  
				\\ (1-2p)\conj{\alpha} & \lambda \end{pmatrix}.
			\end{align}
		
		\item \textit{Depolarizing Channel}: For $p\in[0,1]$, the depolarizing channel is defined by letting $p_I=1-p$ and $p_X=p_Y=p_Z=\frac{p}{3}$, so that
			\begin{equation}\label{eq-qubit_depolarizing_channel}
				\rho\mapsto (1-p)\rho +\frac{p}{3}(X\rho X+Y\rho Y+Z\rho Z).
			\end{equation}
			By using the identity
			\begin{equation}\label{eq-Pauli_twirl}
				\frac{1}{4}\rho+\frac{1}{4}X\rho X+\frac{1}{4}Y\rho Y+\frac{1}{4}Z\rho Z=\Tr[\rho]\pi,
			\end{equation}
			where $\pi \coloneqq \id/2$ is the maximally mixed state (see Lemma~\ref{lem:QM-over:HW-twirl} and \eqref{eq-Pauli_twirl_0}), we can equivalently write the action of the depolarizing channel as
			\begin{equation}
				\rho\mapsto \left(1-\frac{4p}{3}\right)\rho+\frac{4p}{3}\Tr[\rho]\pi,
			\end{equation}
			which has the interpretation that the state of the system is replaced by the maximally mixed state $\pi$ with probability $\frac{4p}{3}$. Observe, however, that $\frac{4p}{3}$ can be interpreted as a probability only for $0\leq p\leq\frac{3}{4}$.	
	\end{enumerate}

\subsection{Generalized Pauli Channels}\label{sec:QM-over:gen-pauli-chs}
	
	Using the Heisenberg--Weyl operators defined in \eqref{eq-Heisenberg_Weyl_operators}, we can generalize Pauli channels to the qudit case. For all $d\geq 2$, a generalized Pauli channel is defined as
	\begin{equation}\label{eq-Pauli_channel_qudit}
		\rho \mapsto \sum_{z,x=0}^{d-1} p(z,x) W_{z,x} \rho W_{z,x}^{\dagger},
	\end{equation}
	where $p:\{0,1,\dotsc,d-1\}^2\to[0,1]$ is a probability distribution, so that $0\leq p(z,x)\leq 1$ for all $z,x\in\{0,1,\dotsc,d-1\}$ and $\sum_{z,x=0}^{d-1} p(z,x)=1$. In other words, a generalized Pauli channel randomly applies one of the Heisenberg--Weyl operators to the input. The Kraus operators of a generalized Pauli channel are therefore $\left\{\sqrt{p(z,x)}W_{z,x}\right\}_{z,x=0}^{d-1}$.
	
	\begin{exercise}{exer-gen_Pauli_chan_Choi}
		Show that the Choi state of a generalized Pauli channel is a Bell-diagonal state. (Recall the definition of a Bell-diagonal state in \eqref{eq-Bell_diag_state}.)
	\end{exercise}
	
	A special case of a generalized Pauli channel is a \textit{$d$-dimensional dephasing channel}, which is obtained by letting $p(z,x)=0$ for all $x\in\{1,2,\dotsc,d-1\}$:
	\begin{equation}\label{eq:QM-over:d-dim-dephase}
		\rho \mapsto \sum_{z=0}^{d-1} p(z,0) Z(z) \rho Z(z)^{\dagger}.
	\end{equation}
	For this special case, only the phase operators $Z(z)$ are applied randomly to the input $\rho$.
	
	\begin{exercise}{exer-d_dim_dephase}
		For the $d$-dimensional dephasing channel defined in \eqref{eq:QM-over:d-dim-dephase}, prove that, in the standard basis, only the off-diagonal elements of the input state $\rho$ are affected by the channel.
	\end{exercise} 
	
	The \textit{$d$-dimensional depolarizing channel} is defined analogously to the qubit case as follows:
	\begin{equation}\label{eq:QM-over:qudit-depolarizing}
		\mathcal{D}_{p}(\rho)\coloneqq (1-p)\rho+\frac{p}{d^2-1}\sum_{(z,x)\neq (0,0)}W_{z,x}\rho W_{z,x}^\dagger,
	\end{equation}
	for all $p\in[0,1]$.
	
	\begin{exercise}{exer-d_dim_depolar_chan}
		Using \eqref{eq-HW_twirl}, prove that for all $p\in[0,1]$, the action of the depolarizing channel $\mathcal{D}_p$ can be written as
		\begin{equation}\label{eq:QM-over:qudit-depol-reparam}
			\mathcal{D}_{p}(\rho)=\left(  1-q\right)  \rho+q\Tr[\rho]\frac{\mathbbm{1}_d}{d}
		\end{equation}		
		for every linear operator $\rho$, where $q=\frac{pd^{2}}{d^{2}-1}$.
	\end{exercise}

	\begin{exercise}{exer-QM_depol_chan_Choi}
		Prove that the Choi state of the depolarizing channel $\mathcal{D}_p$ is the isotropic state $\rho^{\text{iso};1-p}$ (recall \eqref{eq-isotropic_state2}). Conversely, using \eqref{eq-state_to_channel}, prove that every isotropic state is the Choi state of a depolarizing channel. In other words, prove that for all $p\in[0,1]$,
		\begin{equation}
			\mathcal{D}_p(X)=d\,\Tr[(X^{\t}\otimes\mathbbm{1})\rho^{\text{iso};1-p}]
		\end{equation}
		for all $X\in\Lin(\mathbb{C}^d)$.
	\end{exercise}

	\begin{exercise}{exer-QM_depol_chan_covariance}
		\begin{enumerate}
			\item Using the result of Exercise~\ref{exer-QM_depol_chan_Choi}, along with \eqref{eq-isotropic_state_def2}, show that the depolarizing channel has the following covariance property:
				\begin{equation}
					\mathcal{D}_p=\mathcal{U}^{\dagger}\circ\mathcal{D}_p\circ\mathcal{U},
				\end{equation}
				for all $p\in[0,1]$ and every unitary $U$.
				
			\item Using the result of Exercise~\ref{exer-QM_depol_chan_Choi}, along with \eqref{eq-isotropic_integral}, prove that for every quantum channel $\mathcal{N}_{A\to B}$, with $d_A=d_B=d$,
				\begin{equation}
					\int_U \mathcal{U}\circ\mathcal{N}\circ\mathcal{U}^{\dagger}~\D U=\mathcal{D}_{1-p},
				\end{equation} 
				where $p=\bra{\Phi}\Phi^{\mathcal{N}}\ket{\Phi}$ is the \textit{entanglement fidelity} of $\mathcal{N}$, which we define formally in Section~\ref{sec-QM-DM:fid-meas-chan}.
		\end{enumerate}
	\end{exercise}

\section{Special Types of Channels}

\subsection{Petz Recovery Map}\label{subsec-Petz_channel}

	The following channel plays an important role in variations of the data-processing inequality for the quantum relative entropy and other entropic quantities that we define in the next chapter.
	
	\begin{definition}{Petz Recovery Map}{def-Petz_recovery}
		Let $\sigma\in\Lin(\mathcal{H})$ be a positive semi-definite operator and let $\mathcal{N}:\Lin(\mathcal{H})\to\Lin(\mathcal{H}')$ be a quantum channel. The \textit{Petz recovery map for $\sigma$ and $\mathcal{N}$} is the completely positive and trace-non-increasing map $\mathcal{P}_{\sigma,\mathcal{N}}:\Lin(\mathcal{H}')\to\Lin(\mathcal{H'})$ defined as
		\begin{equation}\label{eq-Petz_chan_def}
			\mathcal{P}_{\sigma,\mathcal{N}}(X)\coloneqq\sigma^{\frac{1}{2}}\mathcal{N}^\dagger\!\left(\mathcal{N}(\sigma)^{-\frac{1}{2}}X\mathcal{N}(\sigma)^{-\frac{1}{2}}\right)\sigma^{\frac{1}{2}}
		\end{equation}
		for all $X\in\Lin(\mathcal{H}')$.
	\end{definition}
	
	\begin{remark}
		If the operator $\mathcal{N}(\sigma)$ is invertible, then the Petz recovery map $\mathcal{P}_{\sigma,\mathcal{N}}$ is a channel. If the operator $\mathcal{N}(\sigma)$ is not invertible, then  the inverse $\mathcal{N}(\sigma)^{-\frac{1}{2}}$ is taken on the support on $\mathcal{N}(\sigma)$, following the convention from Section~\ref{sec:math-tools:functions-herm-ops}. In this latter case, the Petz recovery map $\mathcal{P}_{\sigma,\mathcal{N}}$ is a trace non-increasing map. 
	\end{remark}
	
	The Petz recovery map is indeed completely positive because it is the composition of the following completely positive maps:
	\begin{enumerate}
		\item Sandwiching by the positive semi-definite operator $\mathcal{N}(\sigma)^{-\frac{1}{2}}$.
		\item The adjoint $\mathcal{N}^\dagger$ of $\mathcal{N}$.
		\item Sandwiching by the positive semi-definite operator $\sigma^{\frac{1}{2}}$. 
	\end{enumerate}
	The Petz recovery map is also trace non-increasing, as we can readily verify. For every positive semi-definite operator $X$, the following holds:
	\begin{align}
		\Tr[\mathcal{P}_{\sigma,\mathcal{N}}(X)]&=\Tr\!\left[\sigma^{\frac{1}{2}}\mathcal{N}^\dagger\!\left(\mathcal{N}(\sigma)^{-\frac{1}{2}}X\mathcal{N}(\sigma)^{-\frac{1}{2}}\right)\sigma^{\frac{1}{2}}\right] \label{eq-trace-Petz-map-1} \\
		&=\Tr\!\left[\sigma\mathcal{N}^\dagger\!\left(\mathcal{N}(\sigma)^{-\frac{1}{2}}X\mathcal{N}(\sigma)^{-\frac{1}{2}}\right)\right]\\
		&=\Tr\!\left[\mathcal{N}(\sigma)\mathcal{N}(\sigma)^{-\frac{1}{2}}X\mathcal{N}(\sigma)^{-\frac{1}{2}}\right]\\
		&=\Tr\!\left[\mathcal{N}(\sigma)^{-\frac{1}{2}}\mathcal{N}(\sigma)\mathcal{N}(\sigma)^{-\frac{1}{2}}X\right]\\
		&=\Tr[\Pi_{\mathcal{N}(\sigma)}X] \label{eq-trace-Petz-map-5} \\
		&\leq\Tr[X],
	\end{align}
	where $\Pi_{\mathcal{N}(\sigma)}$ is the projection onto the support of $\mathcal{N}(\sigma)$, which arises because $\mathcal{N}(\sigma)$ need not be invertible. If $X$ is contained in the support of $\mathcal{N}(\sigma)$, then $\Tr[\Pi_{\mathcal{N}(\sigma)}X]=\Tr[X]$, which means that the Petz recovery channel is trace-preserving for all inputs with support contained in the support of $\mathcal{N}(\sigma)$.
	
	One of the important properties of the Petz recovery channel $\mathcal{P}_{\sigma,\mathcal{N}}$ is that it reverses the action of $\mathcal{N}$ on $\sigma$ whenever $\mathcal{N}(\sigma)$ is invertible. In particular,
	\begin{align}
		\mathcal{P}_{\sigma,\mathcal{N}}(\mathcal{N}(\sigma))&=\sigma^{\frac{1}{2}}\mathcal{N}^\dagger\!\left(\mathcal{N}(\sigma)^{-\frac{1}{2}}\mathcal{N}(\sigma)\mathcal{N}(\sigma)^{-\frac{1}{2}}\right)\sigma^{\frac{1}{2}}\\
		&=\sigma^{\frac{1}{2}}\mathcal{N}^\dagger(\mathbbm{1}_{\mathcal{H}'})\sigma^{\frac{1}{2}}\\
		&=\sigma,\label{eq-Petz_recovery_sigma}
	\end{align}
	where we have used the fact that $\mathcal{N}(\sigma)$ is invertible, so that
	\begin{equation}
		\mathcal{N}(\sigma)^{-\frac{1}{2}}\mathcal{N}(\sigma)\mathcal{N}(\sigma)^{-\frac{1}{2}}=\mathbbm{1}_{\mathcal{H}'}.
	\end{equation}
	Then, since $\mathcal{N}$ is a channel, its adjoint is unital, which leads to the final equality.
	
	\begin{remark}
		The equality in \eqref{eq-Petz_recovery_sigma} holds more generally; i.e., it holds even when $\mathcal{N}(\sigma)$ is not invertible. To see this, we use the fact that the projection $\Pi_{\mathcal{N}(\sigma)}$ onto the support of $\mathcal{N}(\sigma)$ satisfies $\Pi_{\mathcal{N}(\sigma)}\leq\mathbbm{1}_{\mathcal{H}'}$. Then, we find that
		\begin{align}
			\mathcal{P}_{\sigma,\mathcal{N}}(\mathcal{N}(\sigma))&=\sigma^{\frac{1}{2}}\mathcal{N}^\dagger\!\left(\mathcal{N}(\sigma)^{-\frac{1}{2}}\mathcal{N}(\sigma)\mathcal{N}(\sigma)^{-\frac{1}{2}}\right)\sigma^{\frac{1}{2}} \label{eq-Petz_recovery-NN_sigma_1} \\
			&=\sigma^{\frac{1}{2}}\mathcal{N}^\dagger(\Pi_{\mathcal{N}(\sigma)})\sigma^{\frac{1}{2}}\\
			&\leq\sigma^{\frac{1}{2}}\mathcal{N}^\dagger(\mathbbm{1}_{\mathcal{H}'})\sigma^{\frac{1}{2}}\\
			&=\sigma. \label{eq-Petz_recovery-NN_sigma_4}
		\end{align}
        On the other hand, substituting $X=\mathcal{N}(\sigma)$ in the chain of equalities \eqref{eq-trace-Petz-map-1}--\eqref{eq-trace-Petz-map-5}, we obtain that
        \begin{align}
            \Tr \left[ \mathcal{P}_{\sigma,\mathcal{N}}(\mathcal{N}(\sigma)) \right] = \Tr \left[\Pi_{\mathcal{N}(\sigma)} \mathcal{N}(\sigma)\right] = 1,
        \end{align}
        implying that
        \begin{equation}
            \left\|\sigma - \mathcal{P}_{\sigma,\mathcal{N}}(\mathcal{N}(\sigma))\right\|_1 = \Tr \left[\sigma - \mathcal{P}_{\sigma,\mathcal{N}}(\mathcal{N}(\sigma))\right] = 0,
        \end{equation}
        where the first equality holds because the operator inside the trace norm is positive semi-definite, as we showed in \eqref{eq-Petz_recovery-NN_sigma_1}--\eqref{eq-Petz_recovery-NN_sigma_4}. Therefore,
        \begin{equation}
		\mathcal{P}_{\sigma,\mathcal{N}}(\mathcal{N}(\sigma))=\sigma
	\end{equation}
	holds even when $\mathcal{N}(\sigma)$ is not invertible.
	\end{remark}

\subsubsection{Petz Recovery Channel for Partial Trace}\label{sec:petz-recovery-partial-trace}

	Let the input Hilbert space to the channel $\mathcal{N}$ in the definition of the Petz recovery map be $\mathcal{H}=\mathcal{H}_{AB}$. Then, let $\mathcal{N}=\Tr_B$ be the partial trace over $B$, and note that (see Exercise~\ref{exer-partial_trace_Choi_Stinespring_adjoint})
	\begin{equation} \label{eq-adjoint_partial_trace}
		\mathcal{N}^\dagger(\rho_A)=\rho_A\otimes\mathbbm{1}_B.
	\end{equation}
	Indeed, using Definition~\ref{def-adjoint_superop} for the adjoint of a superoperator, we have
	\begin{align}
		\langle\mathcal{N}^\dagger(\rho_A),\sigma_{AB}\rangle&=\inner{\rho_A\otimes\mathbbm{1}_B}{\sigma_{AB}}\\
		&=\Tr[(\rho_A\otimes\mathbbm{1}_B)^\dagger\sigma_{AB}]\\
		&=\Tr[\rho_A^\dagger\Tr_B(\sigma_{AB})]\\
		&=\inner{\rho_A}{\mathcal{N}(\sigma_{AB})},
	\end{align}
        for all states $\rho_A$ and $\sigma_{AB}$, implying \eqref{eq-adjoint_partial_trace}. 
	Therefore, for a fixed state $\sigma_{AB}$, the Petz recovery map corresponding to the partial trace over $B$ is
	\begin{equation}\label{eq-Petz_channel_partrace}
		\mathcal{P}_{\sigma_{AB},\Tr_B}(X_A)=\sigma_{AB}^{\frac{1}{2}}\left(\sigma_A^{-\frac{1}{2}}X_A\sigma_A^{-\frac{1}{2}}\otimes\mathbbm{1}_B\right)\sigma_{AB}^{\frac{1}{2}}.
	\end{equation}
	By writing the identity on $B$ as $\mathbbm{1}_B=\sum_{j=0}^{d_B-1}\ket{j}\!\bra{j}_B$, we can write the action $\mathcal{P}_{\sigma_{AB},\Tr_B}$ as
	\begin{align}
		\mathcal{P}_{\sigma_{AB},\Tr_B}(X_A)&=\sigma_{AB}^{\frac{1}{2}}\left(\sigma_A^{-\frac{1}{2}}X_A\sigma_A^{-\frac{1}{2}}\otimes\mathbbm{1}_B\right)\sigma_{AB}^{\frac{1}{2}}\\
		&=\sum_{j=0}^{d_B-1}\sigma_{AB}^{\frac{1}{2}}\left(\sigma_A^{-\frac{1}{2}}X_A\sigma_A^{-\frac{1}{2}}\otimes\ket{j}\!\bra{j}_B\right)\sigma_{AB}^{\frac{1}{2}}\\
		&=\sum_{j=0}^{d_B-1}\sigma_{AB}^{\frac{1}{2}}\left(\sigma_A^{-\frac{1}{2}}\otimes\ket{j}_B\right)X_A\left(\sigma_A^{-\frac{1}{2}}\otimes\bra{j}_B\right)\sigma_{AB}^{\frac{1}{2}}\\
		&=\sum_{j=0}^{d_B-1} K_jX_A K_j^\dagger,
	\end{align}
	where
	\begin{equation}
		K_j\coloneqq \sigma_{AB}^{\frac{1}{2}}\left(\sigma_A^{-\frac{1}{2}}\otimes\ket{j}_B\right).
	\end{equation}
	The operators $K_j$, for $0\leq j\leq d_B-1$, are thus Kraus operators for the Petz recovery map for the partial trace. Using \eqref{eq-Kraus_to_Stinespring}, and letting the environment $E$ be denoted by $\hat{B}$ (since the dimension of the environment in the construction \eqref{eq-Kraus_to_Stinespring} is the same as the number of Kraus operators, which in this case is equal to the dimension of $B$), we find that
	\begin{align}
		V_{A\to B\hat{B}} &= \sum_{j=0}^{d_B-1}\sigma_{AB}^{\frac{1}{2}}(\sigma_A^{-\frac{1}{2}}\otimes\ket{j}_B)\otimes\ket{j}_{\hat{B}}\\
		&=\sum_{j=0}^{d_B-1}\sigma_{AB}^{\frac{1}{2}}(\sigma_A^{-\frac{1}{2}}\otimes\ket{j}_B\otimes\ket{j}_{\hat{B}})\\
		&=(\sigma_{AB}^{\frac{1}{2}}\otimes\mathbbm{1}_{\hat{B}})(\sigma_A^{-\frac{1}{2}}\otimes\mathbbm{1}_B\otimes\mathbbm{1}_{\hat{B}})\left(\mathbbm{1}_A\otimes\sum_{j=0}^{d_B-1}\ket{j,j}_{B\hat{B}}\right)\\
		&= (\sigma_{AB}^{\frac{1}{2}}\otimes\mathbbm{1}_{\hat{B}})(\sigma_A^{-\frac{1}{2}}\otimes\mathbbm{1}_{B\hat{B}})(\mathbbm{1}_A\otimes\ket{\Gamma}_{B\hat{B}}),\label{eq-Petz_PT_iso_ext}
	\end{align}
	which is an isometric extension of the Petz recovery map $\mathcal{P}_{\sigma_{AB},\Tr_B}$. Omitting identity operators, this can be written more simply as follows:
	\begin{equation}
	V_{A\to B\hat{B}} = \sigma_{AB}^{\frac{1}{2}}\sigma_A^{-\frac{1}{2}}\ket{\Gamma}_{B\hat{B}}.
	\end{equation}

	\begin{exercise}{exer-Petz_map}
		Recall the Bayes theorem from probability theory:
\begin{equation}
p_{X|Y}(x|y)p_{Y}(y)=p_{Y|X}(y|x)p_{X}(x),\label{eq-QM-chan:Bayes-thm}
\end{equation}
where $X$ and $Y$ are random variables with joint probability distribution
$p_{XY}(x,y)$ over the alphabets $\mathcal{X}$ and $\mathcal{Y}$, and the
distributions given above are derived from this joint distribution as
\begin{align}
p_{X}(x)  & =\sum_{y\in\mathcal{Y}}p_{XY}(x,y),\qquad p_{Y}(y)=\sum
_{x\in\mathcal{X}}p_{XY}(x,y),\\
p_{Y|X}(y|x)  & =\frac{p_{XY}(x,y)}{p_{X}(x)},\qquad p_{X|Y}(x|y)=\frac
{p_{XY}(x,y)}{p_{Y}(y)}.
\end{align}

We now develop a connection between Bayes theorem and the Petz recovery map.
Let $\{\rho_{x}\}_{x\in\mathcal{X}}$ be a set of states, let $\mathcal{N}$ be
a channel, and let $\{M_{y}\}_{y\in\mathcal{Y}}$ be a POVM. Set
\begin{equation}
p_{Y|X}(y|x)=\operatorname{Tr}[M_{y}\mathcal{N}(\rho_{x})].
\end{equation}

\begin{enumerate}
\item Show that \eqref{eq-QM-chan:Bayes-thm} is satisfied with
\begin{equation}
p_{X|Y}(x|y)=\operatorname{Tr}[L_{x}\mathcal{P}(\sigma_{y})],
\end{equation}
for the set $\{\sigma_{y}\}_{y\in\mathcal{Y}}$ of states, channel $\mathcal{P}$, and POVM\ $\{L_{x}\}_{x\in\mathcal{X}}$ chosen as
\begin{align}
\sigma_{y}  & =\frac{1}{p_{Y}(y)}\left[  \mathcal{N}(\overline{\rho})\right]
^{\frac{1}{2}}M_{y}\left[  \mathcal{N}(\overline{\rho})\right]  ^{\frac{1}{2}
},\\
\mathcal{P}(\cdot)  & =\overline{\rho}^{\frac{1}{2}}\mathcal{N}^{\dag}\left(
\left[  \mathcal{N}(\overline{\rho})\right]  ^{-\frac{1}{2}}(\cdot)\left[
\mathcal{N}(\overline{\rho})\right]  ^{-\frac{1}{2}}\right)  \overline{\rho
}^{\frac{1}{2}},\\
L_{x}  & =p_{X}(x)\left[\overline{\rho}\right]^{-\frac{1}{2}}\rho_{x}\left[\overline{\rho
}\right]^{-\frac{1}{2}},
\end{align}
where
\begin{equation}
\overline{\rho}=\sum_{x\in\mathcal{X}}p_{X}(x)\rho_{x}.
\end{equation}

\item Verify that $\{\sigma_{y}\}_{y\in\mathcal{Y}}$ is a set of states,
$\mathcal{P} = \mathcal{P}_{\overline{\rho},\mathcal{N}}$ is the Petz recovery channel associated with $\overline{\rho}$ and $\mathcal{N}$, and $\{L_{x}\}_{x\in\mathcal{X}}$ is a POVM. For simplicity, suppose that the states $\overline{\rho}$ and $\mathcal{N}(\overline{\rho})$ are positive definite.
\end{enumerate}
	\end{exercise}

\subsection{LOCC Channels}\label{subsec-LOCC_channels}
	
	A very common physical scenario encountered in quantum information theory is one in which two parties, Alice and Bob, who are distantly separated, perform local quantum operations (consisting of channels and/or measurements) at their respective locations and communicate classically with each other in order to transform some initially shared state to some final desired state. This sequence of \textit{local operations and classical communication} is abbreviated \textit{LOCC} and is an important element of many quantum communication protocols, such as entanglement distillation and secret key distillation.
	
	As with every other transformation in quantum theory, an LOCC operation corresponds mathematically to a channel, which we call an \textit{LOCC channel}. Formally, an LOCC channel is defined as follows:
	
	\begin{definition}{LOCC Channel}{def-LOCC}
		Let $\mathcal{X}$ be a finite alphabet, let $\{\mathcal{M}^x\}_{x\in\mathcal{X}}$ be a quantum instrument (a set of completely positive trace-non-increasing maps such that $\sum_{x\in\mathcal{X}}\mathcal{M}^x$ is a channel), and let $\{\mathcal{N}^x\}_{x\in\mathcal{X}}$ be a set of quantum channels. Then, a \textit{one-way LOCC channel from Alice to Bob} is the channel $\mathcal{L}_{AB\to A'B'}^{\rightarrow}$ from Alice's initial and final systems $A$ and $A'$ to Bob's initial and final systems $B$ and $B'$, defined as
		\begin{equation}\label{eq-LOCC_A_to_B}
			\mathcal{L}_{AB\to A'B'}^{\rightarrow}=\sum_{x\in\mathcal{X}}\mathcal{M}_{A\to A'}^x\otimes\mathcal{N}_{B\to B'}^x.
		\end{equation}
		
		A \textit{one-way LOCC channel from Bob to Alice} is the channel $\mathcal{L}_{AB\to A'B'}^{\leftarrow}$ defined as
		\begin{equation}
			\mathcal{L}_{AB\to A'B'}^{\leftarrow}=\sum_{x\in\mathcal{X}}\mathcal{N}_{A\to A'}^x\otimes\mathcal{M}_{B\to B'}^x.
			\label{eq-QM:one-way-LOCC-from-B-to-A}
		\end{equation}
		
		An \textit{LOCC channel} is a composition of a finite, but arbitrarily large number of one-way LOCC channels from Alice to Bob and from Bob to Alice and can be written as
		\begin{equation}
		\label{eq-LOCC_chan_gen}
			\mathcal{L}_{AB\to A'B'}^{\leftrightarrow}=\sum_{y\in\mathcal{Y}}\mathcal{S}_A^y\otimes\mathcal{W}_B^y
		\end{equation}
		for some finite alphabet $\mathcal{Y}$ and sets $\{\mathcal{S}^y\}_{y\in\mathcal{Y}}$, $\{\mathcal{W}^y\}_{y\in\mathcal{Y}}$ of completely positive trace-non-increasing maps such that $\sum_{y\in\mathcal{Y}}\mathcal{S}^y\otimes\mathcal{W}^y$ is trace preserving.
	\end{definition}
	
	Consider the one-way LOCC channel $\mathcal{L}_{AB\to A'B'}^{\rightarrow}$ from Alice to Bob defined as in \eqref{eq-LOCC_A_to_B}. The values in the set $\mathcal{X}$ form the possible messages that can be communicated from Alice to Bob and constitute the ``classical communication'' part of LOCC. The set $\{\mathcal{M}^x\}_{x\in\mathcal{X}}$ of completely positive trace-non-increasing maps and the set $\{\mathcal{N}^x\}_{x\in\mathcal{X}}$ of quantum channels specify the actions of Alice and Bob for each value $x\in\mathcal{X}$ and constitute the ``local operations'' part of LOCC. The set $\{\mathcal{M}^x\}_{x\in\mathcal{X}}$ of completely positive trace-non-increasing maps that sum to a channel essentially specifies a quantum instrument. The operations corresponding to these maps can thus be considered probabilisitic since the maps are not trace preserving. In general, the party that performs the quantum instrument determines the direction of classical communication and thus the direction of the one-way LOCC channel. In this case, Alice performs the classical communication since she is the one applying the quantum instrument. The values in the set $\mathcal{X}$ specify the outcomes of the instrument, and Alice communicates the outcome to Bob, who performs the corresponding channel selected from his set $\{\mathcal{N}^x\}_{x\in\mathcal{X}}$ of channels.
	
	In more detail, let $\rho_{AB}$ be the initial state shared by Alice and Bob. Using the definition in \eqref{eq-instrument_output} for the channel corresponding to the quantum instrument $\{\mathcal{M}^x\}_{x\in\mathcal{X}}$, the state after applying the quantum instrument is
	\begin{equation}\label{eq-1wayLOCC_step1}
		\sum_{x\in\mathcal{X}}\mathcal{M}_{A\to A'}^x(\rho_{AB})\otimes\ket{x}\!\bra{x}_{X_A},
	\end{equation}
	where the system $X_A$ stores the classical information corresponding to the outcome of the instrument. Alice then communicates the outcome of the instrument to Bob. This classical communication can be understood via the noiseless classical channel from $X_A$ to $X_B$ defined by
	\begin{equation}\label{eq-classical_channel_action}
		\theta_{X_A}\mapsto \sum_{x\in\mathcal{X}}\!\bra{x}_{X_A}\theta_{X_A} \ket{x}_{X_A} \ket{x}\!\bra{x}_{X_B}.
	\end{equation}
	The state in \eqref{eq-1wayLOCC_step1} thus gets transformed to
	\begin{equation}
		\sum_{x\in\mathcal{X}}\mathcal{M}_{A\to A'}^x(\rho_{AB})\otimes\ket{x}\!\bra{x}_{X_B}.
	\end{equation}
	Finally, Bob applies the channel specified by
	\begin{equation}\label{eq-1wayLOCC_last_step}
		\tau_B\otimes\omega_{X_B}\mapsto \sum_{x\in\mathcal{X}}\mathcal{N}_{B\to B'}^x(\tau_B)\otimes \bra{x}_{X_B}\omega_{X_B}\ket{x}_{X_B},
	\end{equation}
	which corresponds to Bob measuring his system $X_B$ in the basis $\{\ket{x}_{X_B}\}_{x\in\mathcal{X}}$ and applying a quantum channel from the set $\{\mathcal{N}_{B\to B'}^x\}_{x\in\mathcal{X}}$ to the system $B$ based on the outcome. The final state is then
	\begin{equation}
		\sum_{x\in\mathcal{X}}(\mathcal{M}_{A\to A'}^x\otimes\mathcal{N}_{B\to B'}^x)(\rho_{AB}),
	\end{equation}
	which is precisely the output of the one-way LOCC channel $\mathcal{L}_{AB\to A'B'}^{\rightarrow}$ defined in \eqref{eq-LOCC_A_to_B}. We can succinctly write the steps in \eqref{eq-1wayLOCC_step1}--\eqref{eq-1wayLOCC_last_step} as
	\begin{equation}\label{eq-1wayLOCC_decomp}
		\mathcal{L}_{AB\to A'B'}(\rho_{AB})=(\mathcal{N}_{X_B B\to B'}\circ\mathcal{C}_{X_A \to X_B}\circ\mathcal{M}_{A\to A'X_A})(\rho_{AB}),
	\end{equation}
	where the channel $\mathcal{M}_{A\to A'X_A}$ is defined as
	\begin{equation}
		\mathcal{M}_{A\to A'X_A}(\xi_A)\coloneqq\sum_{x\in\mathcal{X}}\mathcal{M}_{A\to A'}^x(\xi_A)\otimes\ket{x}\!\bra{x}_{X_A},
	\end{equation}
	and the channel $\mathcal{N}_{B X_B\to B'}$ is defined as
	\begin{equation}
		\mathcal{N}_{B X_B\to B'}(\tau_B\otimes\ket{x}\!\bra{x}_{X_B})\coloneqq \mathcal{N}_{B\to B'}^x(\tau_B)
	\end{equation}
	for all $x\in\mathcal{X}$. As indicated above, the channel $\mathcal{C}_{X_A\to X_B}$, defined in \eqref{eq-classical_channel_action}, is a noiseless classical channel that transforms the classical register $X_A$, held by Alice, to the classical register $X_B$ (which is simply a copy of $X$), held by Bob.
	
	\begin{figure}
		\centering
		\includegraphics[scale=0.65]{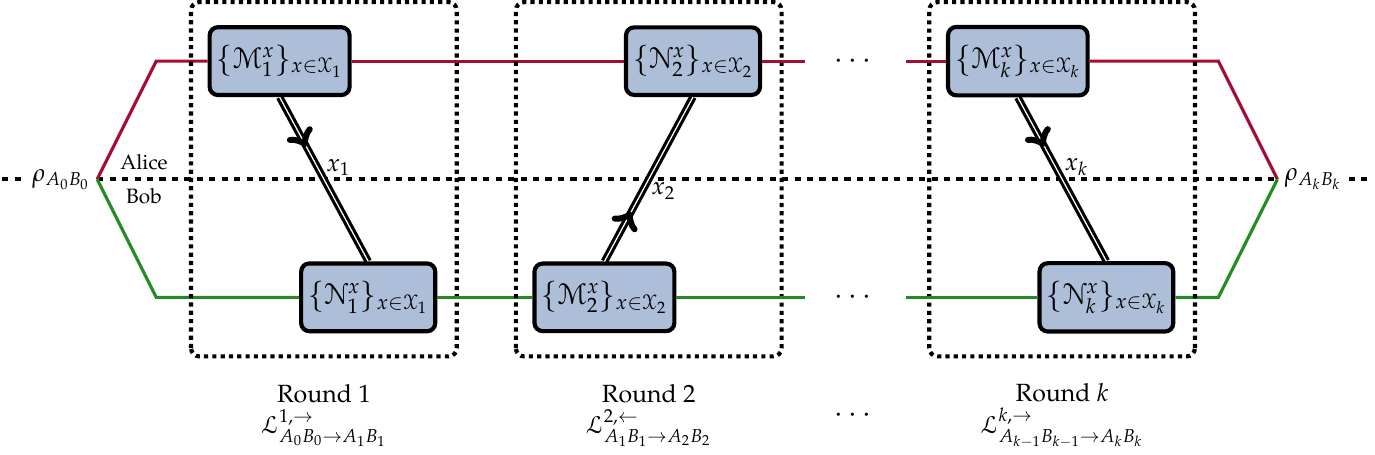}
		\caption{Illustration of an LOCC channel with $k$ rounds of alternating one-way LOCC channels from Alice to Bob and from Bob to Alice. In each round $i$, there is a finite alphabet $\mathcal{X}_i$, a set $\{\mathcal{M}_i^x\}_{x\in\mathcal{X}_i}$ of completely positive trace-non-increasing maps that sum to a channel, and a set $\{\mathcal{N}_i^x\}_{x\in\mathcal{X}_i}$ of quantum channels.}\label{fig-LOCC}
	\end{figure}
	
	An example of an LOCC channel is illustrated in Figure \ref{fig-LOCC}. This is an LOCC channel consisting of $k$ rounds of alternating Alice-to-Bob and Bob-to-Alice one-way LOCC channels and is of the form
	\begin{equation}
		\begin{aligned}
		\mathcal{L}_{A_0B_0\to A_kB_k}^{\leftrightarrow}=\mathcal{L}_{A_{k-1}B_{k-1}\to A_kB_k}^{k,\rightarrow}&\circ\mathcal{L}_{A_{k-2}B_{k-2}\to A_{k-1}B_{k-1}}^{k-1,\leftarrow}\circ\dotsb\\
		&\qquad\circ\mathcal{L}_{A_1B_1\to A_2B_2}^{2,\leftarrow}\circ\mathcal{L}_{A_0B_0\to A_1B_1}^{1,\rightarrow}.
		\end{aligned}
	\end{equation}
	For each round $i$, there is a finite alphabet $\mathcal{X}_i$ consisting of the messages communicated in the round, along with a set $\{\mathcal{M}_i^x\}_{x\in\mathcal{X}_i}$ of completely positive trace-non-increasing maps that sum to a quantum channel and a set $\{\mathcal{N}_i^x\}_{x\in\mathcal{X}_i}$ of quantum channels. Note that, in a multi-round LOCC channel such as this one, it possible for the operation sets $\{\mathcal{M}_i^x\}_{x\in\mathcal{X}_i}$ and $\{\mathcal{N}_i^x\}_{x\in\mathcal{X}_i}$ on the $i$th round to depend on the outcomes and actions taken in previous rounds. Although we sum over the index~$x$ after each one-way LOCC round, seemingly erasing memory of it, a copy of~$x$ can still be retained, if necessary, within the output of either Alice's or Bob's operations, where it can be accessed at any later time. The quantum teleportation protocol in Section~\ref{sec-teleportation}  provides a concrete example of a one-way LOCC protocol from Alice to Bob.

	 \begin{definition}{Separable Channel}{def:QM-over:separable-channel}
	 A \textit{separable channel}  is a quantum channel $\mathcal{S}_{AB\to A'B'}$ such that
	\begin{equation}
	\label{eq-separable_map}
		\mathcal{S}_{AB\to A'B'}=\sum_{x\in\mathcal{X}}\mathcal{C}_{A\to A'}^x\otimes\mathcal{D}_{B\to B'}^x
	\end{equation}
	for some finite alphabet $\mathcal{X}$ and sets of completely positive and trace-non-increasing maps $\{\mathcal{C}^x\}_{x\in\mathcal{X}}$ and $\{\mathcal{D}^x\}_{x\in\mathcal{X}}$ such that $\mathcal{S}$ is trace preserving.
	\end{definition}
	
	Every separable channel has a set of Kraus operators in product form; i.e., for every separable channel $\mathcal{S}_{AB\to A'B'}$ as in \eqref{eq-separable_map} there exists a finite alphabet $\mathcal{Y}$ and sets $\{C^y_{A\to A'}\}_{y\in\mathcal{Y}}$ and $\{D^y_{B\to B'}\}_{y\in\mathcal{Y}}$ such that
	\begin{equation}
		\mathcal{S}_{AB\to A'B'}(\rho_{AB})=\sum_{y\in\mathcal{Y}} (C^y_{A\to A'}\otimes D^y_{B\to B'})\rho_{AB}(C^y_{A\to A'}\otimes D^y_{B\to B'})^\dagger
		\label{eq:QM-over:kraus-form-sep-ch}
	\end{equation}
	for all $\rho_{AB}$.

	A key property of a separable channel is that it outputs a separable state if the input state is separable. To see this, consider the separable state $\sigma_{AB} = \sum_{z\in \mathcal{Z}} p(z) \tau^z_A \otimes \omega^z_B$. Then the output state $\mathcal{S}_{AB\to A'B'}(\sigma_{AB})$ is given by
	\begin{align}
	& \mathcal{S}_{AB\to A'B'}(\sigma_{AB}) \notag \\& = \sum_{y\in\mathcal{Y}} (C^y_{A\to A'}\otimes D^y_{B\to B'})
	\left(\sum_{z\in \mathcal{Z}} p(z) \tau^z_A \otimes \omega^z_B\right)(C^y_{A\to A'}\otimes D^y_{B\to B'})^\dagger
		 \label{eq:QM-over:sep-preserves-sep-1}
		 \\
	& = \sum_{y\in\mathcal{Y},z\in \mathcal{Z}} p(z) C^y_{A\to A'} 
	 \tau^z_A (C^y_{A\to A'})^\dagger \otimes D^y_{B\to B'} \omega^z_B (D^y_{B\to B'})^\dagger,
	 \label{eq:QM-over:sep-preserves-sep-2}
	\end{align}
	and is manifestly separable.
	
	\begin{proposition*}{LOCC Channels are Separable Channels}{prop:QM-over:sep-ch-contains-LOCC}
	From \eqref{eq-LOCC_chan_gen} and \eqref{eq-separable_map}, we conclude that every LOCC channel is a separable channel.
	\end{proposition*}
	
	 The converse of Proposition~\ref{prop:QM-over:sep-ch-contains-LOCC} is not true. For example, let us define the following operators:
	\begin{align}
		K_1&\coloneqq\frac{1}{\sqrt{2}}\ket{0}\!\bra{0}+\ket{1}\!\bra{1},\quad K_2\coloneqq\ket{0}\!\bra{0},\quad K_3\coloneqq\ket{1}\!\bra{1}.
	\end{align}
	Then, following the notation in \eqref{eq-separable_map}, let
	\begin{align}
		\mathcal{C}^1_{A\to A'}(\cdot)&=K_1(\cdot)K_1^\dagger,\\
		\mathcal{C}^2_{A\to A'}(\cdot)&=K_2(\cdot)K_2^\dagger,\\
		\mathcal{C}^3_{A\to A'}(\cdot)&=K_3(\cdot)K_3^\dagger,
	\end{align}
	and
	\begin{align}
		\mathcal{D}^1_{B\to B'}(\cdot)&=K_2(\cdot)K_2^\dagger,\\
		\mathcal{D}^2_{B\to B'}(\cdot)&=K_1(\cdot)K_1^\dagger,\\
		\mathcal{D}^3_{B\to B'}(\cdot)&=K_3(\cdot)K_3^\dagger.
	\end{align}
	Then, the map
	\begin{align}
		\mathcal{S}_{AB\to A'B'}(\cdot)&\coloneqq\sum_{x=1}^3 (\mathcal{C}_{A\to A'}^x\otimes\mathcal{D}_{B\to B'}^x)(\cdot)\\
		&=(K_1\otimes K_2)(\cdot)(K_1\otimes K_2)^\dagger+(K_2\otimes K_1)(\cdot)(K_2\otimes K_1)^\dagger\\
		&\qquad +(K_3\otimes K_3)(\cdot)(K_3\otimes K_3)^\dagger
	\end{align}
	is a separable channel, but it can be shown that it is not an LOCC channel; please consult the Bibliographic Notes in Section~\ref{sec:qm_channels:bib-notes}.

\subsubsection{LOCC and Separable Simulation of Channels}

\begin{figure}
		\centering
		\includegraphics[scale=0.8]{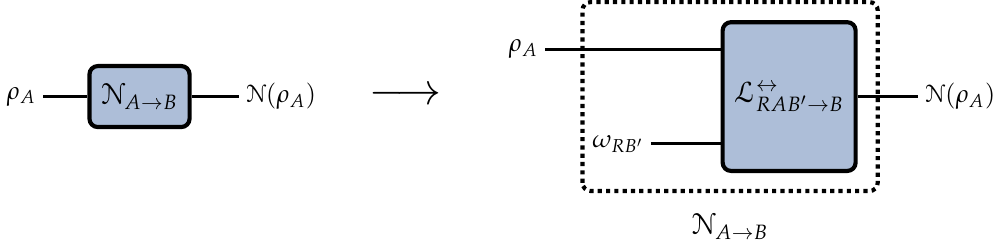}
		\caption{Depiction of an LOCC-simulable channel $\mathcal{N}_{A\to B}$ with associated resource state $\omega_{RB'}$. The channel $\mathcal{N}_{A\to B}$ can be realized via the LOCC channel $\mathcal{L}_{RAB'\to B}^{\leftrightarrow}$ and the resource state $\omega_{RB'}$, for some auxiliary system $R$. The LOCC channel $\mathcal{L}_{RAB'\to B}^{\leftrightarrow}$ could in principle consist of a sequence of several rounds of one-way LOCC channels, as depicted in Figure \ref{fig-LOCC}.}\label{fig-LOCC_sim}
	\end{figure}

	Given a bipartite quantum channel $\mathcal{N}_{AB\to A'B'}$, an important question related to the physical realization of the channel is whether the channel is an LOCC channel. This amounts to determining whether the channel can be decomposed as in \eqref{eq-LOCC_chan_gen}. More generally, one can ask whether a given channel $\mathcal{N}_{A\to B}$ can be \textit{simulated} by an LOCC channel acting on the input and a resource state, a notion that is illustrated in Figure \ref{fig-LOCC_sim} and defined below.
	
	\begin{definition}{LOCC-Simulable Channel}{def-LOCC_sim_chan}
		A channel $\mathcal{N}_{A\to B}$ is called \textit{LOCC-simulable} with associated resource state $\omega_{RB'}$ if there exists an auxiliary system $R$ and an LOCC channel $\mathcal{L}^{\leftrightarrow}_{RAB'\to B}$ such that, for every state $\rho_A$,
		\begin{equation}
			\mathcal{N}(\rho_A)=\mathcal{L}^{\leftrightarrow}_{RAB'\to B}(\rho_A\otimes\omega_{RB'}).
		\end{equation}
		For the LOCC channel $\mathcal{L}^{\leftrightarrow}_{RAB'\to B}$, the input systems $RA$ are Alice's, the input system $B'$ is Bob's, Alice's output system is trivial, and Bob's output system is~$B$.
	\end{definition}
	
	If a channel $\mathcal{N}_{A\to B}$ is LOCC-simulable with associated resource state $\omega_{RB'}$, it means that Alice and Bob, the sender and receiver of the channel, respectively, can execute an LOCC channel of the form as depicted in Figure \ref{fig-LOCC}, with the assistance of the auxiliary system $R$, such that the output on Bob's system at the end is $\mathcal{N}(\rho_A)$. The resource state $\omega_{RB'}$ is fixed, being such that the same resource state can be used for every input state $\rho_A$ on Alice's system. A concrete example of an LOCC simulation of a channel is shown in the context of quantum teleportation in Section~\ref{sec-teleportation} below. 

Due to the fact that separable channels strictly contain LOCC channels, it is sensible to generalize the notion of teleportation simulation even further to this case:
	\begin{definition}{Separable-Simulable Channel}{def-SEP_sim_chan}
		A channel $\mathcal{N}_{A\to B}$ is called \textit{separable-simulable} with associated resource state $\omega_{RB'}$ if there exists an auxiliary system $R$ and a separable channel $\mathcal{S}_{RAB'\to B}$ such that, for every state $\rho_A$,
		\begin{equation}
			\mathcal{N}(\rho_A)=\mathcal{S}_{RAB'\to B}(\rho_A\otimes\omega_{RB'}).
		\end{equation}
		For the separable channel $\mathcal{S}^{\leftrightarrow}_{RAB'\to B}$, the input systems $RA$ are Alice's, the input system $B'$ is Bob's, Alice's output system is trivial, and Bob's output system is~$B$.
	\end{definition}

\subsection{Completely PPT-Preserving Channels}\label{sec-QM:C-PPT-P-chs}

	In Definition \ref{def-PPT}, we introduced PPT states as bipartite states $\rho_{AB}$ such that the partial transpose $\T_B(\rho_{AB})$  is positive semi-definite. A class of channels that preserve PPT states are called \textit{completely PPT-preserving channels}, abbreviated as C-PPT-P channels, and we define them formally as follows.

	\begin{definition}{Completely PPT-Preserving Channel}{def-PPT_pres_chan}
		A channel $\mathcal{P}_{AB\to A'B'}$ is called \textit{completely PPT-preserving} if the following map is a channel:
		\begin{equation}
		\T_{B'}\circ\mathcal{P}_{AB\to A'B'}\circ\T_B.
\end{equation}
	\end{definition}
	
	\begin{proposition}{prop:QM:C-PPT-P-preserves-PPT}
	Completely PPT-preserving channels preserve the set of PPT states.
	\end{proposition}
	
	\begin{Proof}
		Suppose that $\rho_{AB}$ is a PPT state and that $\mathcal{P}_{AB\to A'B'}$ is a completely PPT-preserving channel. If we take the partial transpose $\T_{B'}$ on the output state $\sigma_{A'B'}=\mathcal{P}_{AB\to A'B'}(\rho_{AB})$, then we find that
		\begin{align}
			\T_{B'}(\sigma_{A'B'})&=(\T_{B'}\circ\mathcal{P}_{AB\to A'B'})(\rho_{AB})\\
			&=(\T_B\circ\mathcal{P}_{AB\to A'B'}\circ\T_{B})(\T_B(\rho_{AB})).
		\end{align}
		Since $\mathcal{P}_{AB\to A'B'}$ is completely PPT-preserving, by definition $\T_{B'}\circ\mathcal{P}_{AB\to A'B'}\circ\T_B$ is completely positive. Since $\T_B(\rho_{AB})$ is positive, this implies that $\T_{B'}(\sigma_{A'B'})$ is positive, which means that the output state is a PPT state.
	\end{Proof}

	\begin{proposition}{prop-LOCC_PPT_preserving}
		Every separable channel is a completely PPT-preserving channel.
	\end{proposition}
	
	\begin{Proof}
		Let $\mathcal{S}_{AB\to A'B'}$ be a separable channel. By definition, it has the form
		\begin{equation}
			\mathcal{S}_{AB\to A'B'}=\sum_{x\in\mathcal{X}}\mathcal{R}^x_{A\to A'}\otimes\mathcal{W}_{B\to B'}^x,
		\end{equation}
		where $\mathcal{X}$ is a finite alphabet and $\{\mathcal{R}^x\}_{x\in\mathcal{X}}$ and $\{\mathcal{W}^x\}_{x\in\mathcal{X}}$ are sets of completely positive trace non-increasing maps such that $\sum_{x\in\mathcal{X}}\mathcal{R}^x \otimes \mathcal{W}^x$ is trace preserving. Then,
		\begin{equation}
			\T_{B'}\circ\mathcal{S}_{AB\to A'B'}\circ\T_B=\sum_{x\in\mathcal{X}}\mathcal{R}_{A\to A'}^x\otimes (\T_{B'}\circ\mathcal{W}^x_{B\to B'}\circ \T_B).
		\end{equation}
		By applying Lemma~\ref{lemma-transpose-CPTP} below, we conclude that the maps $\T_{B'}\circ\mathcal{W}_{B\to B'}^x\circ\T_B$ are completely positive for all $x\in\mathcal{X}$, which means that $\T_{B'}\circ\mathcal{S}_{AB\to A'B'}\circ\T_B$ is completely positive. Therefore,  $\mathcal{S}_{AB\to A'B'}$ is completely PPT-preserving.
	\end{Proof}
	
	As a consequence of Proposition~\ref{prop-LOCC_PPT_preserving}, it follows that every LOCC channel is a completely PPT-preserving channel, because every LOCC channel is a separable channel.

The next lemma applies to channels that do not have a bipartite structure, i.e., there is no input or output system for Alice.
	\begin{Lemma}{lemma-transpose-CPTP}
		Let $\mathcal{N}_{B\to B'}$ be a completely positive map. Then the map $\T_{B'} \circ \mathcal{N}_{B\to B'} \circ \T_B$ is completely positive, and its Choi operator is given by the full transpose of the Choi operator for $\mathcal{N}_{B\to B'}$, i.e.,
		\begin{equation}
			\Gamma_{BB'}^{\T_{B'}\circ\mathcal{N}_{B\to B'}\circ\T_{B}}=\T(\Gamma_{BB'}^{\mathcal{N}}).
		\end{equation}
        The above equation holds provided that the basis with respect to which the maximally entangled vector $\ket{\Gamma}_{\hat{B}B}$ is constructed is the same as the basis chosen for the transpose map $\T_B$.
	\end{Lemma}
	
	\begin{Proof}
		Let $\Gamma_{\hat{B}B}=\ket{\Gamma}\!\bra{\Gamma}_{\hat{B}B}$, where $\ket{\Gamma}$ is defined in \eqref{eq-max_ent_vector} and $d_{\hat{B}}=d_B$. Observe that
        \begin{equation} \label{eq:QM-over:transpose-on-ME}
            \T_B(\Gamma_{\hat{B}B}) = \T_{\hat{B}}(\Gamma_{\hat{B}B}),
        \end{equation}
        which follows by taking the partial transpose on $B$ of both sides of the identity $\T_{\hat{B}B}(\Gamma_{\hat{B}B}) = \Gamma_{\hat{B}B}$, which expresses the fact that $\Gamma_{\hat{B}B}$ is a real symmetric matrix. 
		Then the following holds for the Choi representation of $\T_{B'}\circ\mathcal{N}_{B\to B'}\circ \T_B$:
		\begin{align}
			\Gamma_{BB'}^{\T_{B'}\circ\mathcal{N}_{B\to B'}\circ\T_{B}}&=(\T_{B'}\circ\mathcal{N}_{B\to B'}\circ\T_{B})(\Gamma_{\hat{B}B})\\
			& = ((\T_{\hat{B}} \otimes \T_{B'})\circ\mathcal{N}_{B\to B'})(\Gamma_{\hat{B}B}) \\
			&  = \T(\Gamma_{BB'}^{\mathcal{N}}).
			\label{eq:QM-over:transpose-on-ME-last}
		\end{align}
		Since the map $\mathcal{N}_{B\to B'}$ is completely positive, its Choi representation $\Gamma_{BB'}^{\mathcal{N}}$ is positive semi-definite. Since positive semi-definiteness is preserved under transposition, we find that $\T(\Gamma_{BB'}^{\mathcal{N}})$ is positive semi-definite, which means that the map $\T_{B'}\circ\mathcal{N}_{B\to B'}\circ\T_B$ is completely positive (by applying Theorem~\ref{thm-q_channels}). 
	\end{Proof}

As a generalization of the lemma above, we have the following:

\begin{proposition*}{Choi States of C-PPT-P Channels}{prop:QM-over:Choi-C-PPT-P-chs}
Let $\mathcal{N}_{AB \to A'B'}$ be a bipartite channel. The channel $\mathcal{N}_{AB \to A'B'}$ is completely PPT-preserving if and only if its Choi state $\Phi^{\mathcal{N}}_{AB A'B'}$ is a PPT state with respect to the bipartite cut $AA'|BB'$. 
\end{proposition*}

\begin{Proof}
By definition, $\mathcal{N}_{AB \to A'B'}$ is completely PPT-preserving if and only if $\T_{B'} \circ \mathcal{N}_{AB \to A'B'} \circ \T_B$ is completely positive, which is equivalent to the positivity of the corresponding Choi state. To construct the latter, consider two maximally entangled states $\Phi_{A\bar{A}}$ and $\Phi_{B\bar{B}}$ (see~\eqref{eq-max_ent_state} for an explicit definition); here, the Hilbert spaces corresponding to $\bar{A}$ and $\bar{B}$ are isomorphic to the Hilbert spaces corresponding to $A$ and $B$, respectively. A maximally entangled state between $AB$ and $\bar{A}\bar{B}$ can be defined as $\Phi_{AB\bar{A}\bar{B}} \coloneqq \Phi_{A\bar{A}} \otimes \Phi_{B\bar{B}}$. The Choi state of $\T_{B'} \circ \mathcal{N}_{\bar{A}\bar{B} \to A'B'} \circ \T_{\bar{B}}$ then becomes
\begin{align}
& (\T_{B'} \circ \mathcal{N}_{\bar{A}\bar{B} \to A'B'} \circ \T_{\bar{B}})(\Phi_{A\bar{A}} \otimes \Phi_{B\bar{B}}) \notag \\
&\quad = ((\T_{B} \otimes \T_{B'}  )\circ \mathcal{N}_{\bar{A}\bar{B} \to A'B'} )(\Phi_{A\bar{A}} \otimes \Phi_{B\bar{B}}) \\
&\quad = T_{BB'}(\Phi^{\mathcal{N}}_{AB A'B'}) ,
\end{align}
where the first equality is due to~\eqref{eq:QM-over:transpose-on-ME} and~\eqref{eq-max_ent_state}. The left-hand side is positive semi-definite if and only if $\mathcal{N}_{AB \to A'B'}$ is completely PPT-preserving, and the right-hand side is positive semi-definite if and only if $\Phi^{\mathcal{N}}_{AB A'B'}$ is a PPT state with respect to the bipartite cut $AA'|BB'$.
\end{Proof}

\subsubsection{PPT Simulation of Channels}

	Just as we asked whether a given channel $\mathcal{N}_{A\to B}$ can be simulated by an LOCC channel, we can ask whether the channel $\mathcal{N}_{A\to B}$ can be simulated by a completely PPT-preserving channel. This leads to the following definition:
	
	\begin{definition}{PPT-Simulable Channel}{def-PPT_sim_chan}
		A channel $\mathcal{N}_{A\to B}$ is called \textit{PPT-simulable} with associated resource state $\omega_{RB'}$ if there exists an auxiliary system $R$ and a completely PPT-preserving channel $\mathcal{P}_{RAB'\to B}$ such that for every state $\rho_A$
		\begin{equation}
			\mathcal{N}(\rho_A)=\mathcal{P}_{RAB'\to B}(\rho_A\otimes\omega_{RB'}).
		\end{equation}
		For the C-PPT-P channel $\mathcal{P}_{RAB'\to B}$, the input systems $RA$ are Alice's, the input system $B'$ is Bob's, Alice's output system is trivial, and Bob's output system is~$B$.
	\end{definition}
	
\begin{figure}
		\centering
		\includegraphics[scale=0.8]{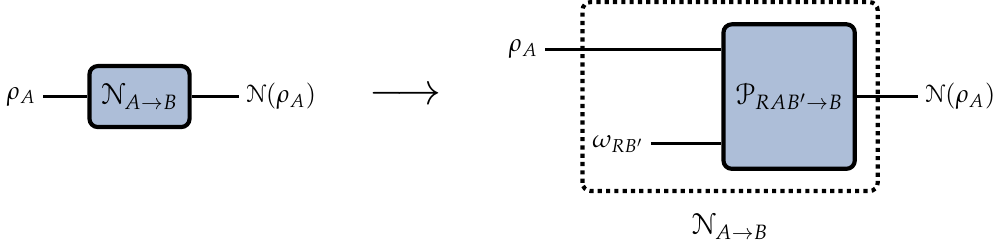}
		\caption{Depiction of a PPT-simulable channel $\mathcal{N}_{A\to B}$ with associated resource state $\omega_{RB'}$. The channel $\mathcal{N}_{A\to B}$ can be realized via the completely PPT-preserving channel $\mathcal{P}_{RAB'\to B}$ and the resource state $\omega_{RB'}$, for some auxiliary system $R$.}\label{fig-PPT_sim}
	\end{figure}

	The definition of a PPT-simulable channel is illustrated in Figure \ref{fig-PPT_sim}.
	
\subsection{Non-Signaling Channels}\label{sec-QM:non-sig-chs}

One of the main applications considered in this book is communication and, more specifically, when communication is possible or impossible. To this end, suppose that Alice and Bob are connected by means of a bipartite channel $\mathcal{N}_{AB\rightarrow A^{\prime}B^{\prime}}$. Such a channel is said to be non-signaling from Alice to Bob if it is impossible for Alice and Bob to make use of it for the purpose of Alice to communicate a message to Bob. We give a precise definition as follows: 

\begin{definition}{Non-Signaling Channel}{def-QM:non-sig-channel}
A bipartite channel $\mathcal{N}_{AB\rightarrow A^{\prime}B^{\prime}}$ is
non-signaling from Alice to Bob  if the following condition
holds 
\begin{equation}
\operatorname{Tr}_{A^{\prime}}\circ\mathcal{N}_{AB\rightarrow A^{\prime
}B^{\prime}}=\operatorname{Tr}_{A^{\prime}}\circ\mathcal{N}_{AB\rightarrow
A^{\prime}B^{\prime}}\circ\mathcal{R}_{A}^{\pi},\label{eq:NS-A-to-B-cond}
\end{equation}
where $\mathcal{R}_{A}^{\pi}$ is a replacer channel, defined as $\mathcal{R}_{A}^{\pi}(\cdot)\coloneqq \operatorname{Tr}_{A}[\cdot]\pi_{A}$, with $\pi_{A}\coloneqq\mathbbm{1}_{A}/d_{A}$ the maximally mixed state on system $A$.
\end{definition}

To interpret the
condition in \eqref{eq:NS-A-to-B-cond}, consider the following. For Bob, the reduced state of his output
system $B^{\prime}$ is obtained by tracing out Alice's output system
$A^{\prime}$. Note that the reduced state on $B^{\prime}$ is all that Bob can
access at the output in this scenario. If the condition in
\eqref{eq:NS-A-to-B-cond} holds, then the reduced state on Bob's output system
$B^{\prime}$ has no dependence on Alice's input system. Thus, if
\eqref{eq:NS-A-to-B-cond} holds, then Alice cannot use $\mathcal{N}
_{AB\rightarrow A^{\prime}B^{\prime}}$ to send a signal to Bob.

We should also remark that the maximally mixed state does not play any special role in Definition~\ref{def-QM:non-sig-channel}, in the sense that it can be substituted by any other fixed state $\sigma_A$. In fact, consider the alternative replacer channel $\mathcal{R}_{A}^{\sigma}(\cdot)\coloneqq \operatorname{Tr}_{A}[\cdot]\sigma_{A}$. If~\eqref{eq:NS-A-to-B-cond} holds, then an analogous identity featuring $\mathcal{R}_{A}^{\sigma}$ instead of $\mathcal{R}_{A}^{\pi}$ also holds, because
\begin{align}
\operatorname{Tr}_{A^{\prime}}\circ\mathcal{N}_{AB\rightarrow A^{\prime}B^{\prime}}\circ\mathcal{R}_{A}^{\sigma} &= \operatorname{Tr}_{A^{\prime}}\circ\mathcal{N}_{AB\rightarrow A^{\prime}B^{\prime}}\circ\mathcal{R}_{A}^{\pi} \circ \mathcal{R}_{A}^{\sigma} \\
&= \operatorname{Tr}_{A^{\prime}}\circ\mathcal{N}_{AB\rightarrow A^{\prime}B^{\prime}}\circ\mathcal{R}_{A}^{\pi} \\
&= \operatorname{Tr}_{A^{\prime}}\circ\mathcal{N}_{AB\rightarrow A^{\prime}B^{\prime}}.
\end{align}
Here, we used~\eqref{eq:NS-A-to-B-cond} twice, in the first and in the third line.

\begin{proposition*}{Choi Operator of a Non-Signaling Channel}{prop:QM-over:Choi-non-sig-chs}
Let $\mathcal{N}_{AB\rightarrow A^{\prime}B^{\prime}}$ be a bipartite channel. Let
\begin{equation}
\Gamma^{\mathcal{N}}_{AB A^{\prime}B^{\prime}} \coloneqq
\mathcal{N}_{\bar{A}\bar{B}\rightarrow A^{\prime}B^{\prime}}(\Gamma_{A\bar{A}} \otimes \Gamma_{B\bar{B}})
\end{equation}
be the Choi operator of $\mathcal{N}_{AB\rightarrow A^{\prime}B^{\prime}}$, where $\bar{A}$ is isomorphic to $A$ and $\bar{B}$ is isomorphic to $B$.
The channel $\mathcal{N}_{AB\rightarrow A^{\prime}B^{\prime}}$ is non-signaling from Alice to Bob if and only if its Choi operator $\Gamma^{\mathcal{N}}_{AB A^{\prime}B^{\prime}}$
 satisfies the following condition:
\begin{equation}
\Tr_{A'}[\Gamma^{\mathcal{N}}_{AB A^{\prime}B^{\prime}}] = 
\pi_A \otimes \Tr_{A'A}[\Gamma^{\mathcal{N}}_{AB A^{\prime}B^{\prime}}].
\label{eq-QM:non-sig-Choi}
\end{equation}
\end{proposition*}
\begin{Proof}
Observe that
\begin{equation}
\operatorname{Tr}_{A^{\prime}}\circ\mathcal{N}_{\bar{A}\bar{B}\rightarrow A^{\prime
}B^{\prime}}(\Gamma_{A\bar{A}} \otimes \Gamma_{B\bar{B}}) = 
\operatorname{Tr}_{A^{\prime}}[\Gamma^{\mathcal{N}}_{AB A^{\prime}B^{\prime}}].
\end{equation}
Also, consider that
\begin{align}
& (\operatorname{Tr}_{A^{\prime}}\circ\mathcal{N}_{\bar{A}\bar{B}\rightarrow
A^{\prime}B^{\prime}}\circ\mathcal{R}_{\bar{A}}^{\pi})(\Gamma_{A\bar{A}} \otimes \Gamma_{B\bar{B}}) \notag \\
& = 
(\operatorname{Tr}_{A^{\prime}}\circ\mathcal{N}_{\bar{A}\bar{B}\rightarrow
A^{\prime}B^{\prime}})(\mathbbm{1}_{A} \otimes \pi_{\bar{A}}\otimes \Gamma_{B\bar{B}}) \\
& = 
(\operatorname{Tr}_{A^{\prime}}\circ\mathcal{N}_{\bar{A}\bar{B}\rightarrow
A^{\prime}B^{\prime}})(\pi_{A} \otimes \mathbbm{1}_{\bar{A}}\otimes \Gamma_{B\bar{B}}) \\
& = 
\pi_{A} \otimes (\operatorname{Tr}_{A^{\prime}}\circ\mathcal{N}_{\bar{A}\bar{B}\rightarrow
A^{\prime}B^{\prime}})( \mathbbm{1}_{\bar{A}}\otimes \Gamma_{B\bar{B}}) \\
& = 
\pi_{A} \otimes (\operatorname{Tr}_{AA^{\prime}}\circ\mathcal{N}_{\bar{A}\bar{B}\rightarrow
A^{\prime}B^{\prime}})( \Gamma_{A\bar{A}}\otimes \Gamma_{B\bar{B}}) \\
& = \pi_{A} \otimes  \Tr_{A'A}[\Gamma^{\mathcal{N}}_{AB A^{\prime}B^{\prime}}].
\end{align}
Since two channels are equal if and only if their Choi operators are equal (see Theorem~\ref{thm-q_channels}), we see that~\eqref{eq:NS-A-to-B-cond} is equivalent to~\eqref{eq-QM:non-sig-Choi}. 
\end{Proof}

A one-way LOCC channel from Bob to Alice is an interesting example of a bipartite channel that is non-signaling from Alice to Bob. Indeed, consider a bipartite channel of the form in \eqref{eq-QM:one-way-LOCC-from-B-to-A}, and let us check that the condition in \eqref{eq:NS-A-to-B-cond} holds for such a channel. By tracing over the output system $A'$, we find that
\begin{align}
 \operatorname{Tr}_{A'}[\mathcal{L}_{AB\to A'B'}^{\leftarrow}(\rho_{AB})] & = 
\sum_{x\in\mathcal{X}} \operatorname{Tr}_{A'}[(\mathcal{N}_{A\to A'}^x\otimes\mathcal{M}_{B\to B'}^x)(\rho_{AB})] \\
& = \sum_{x\in\mathcal{X}} \mathcal{M}_{B\to B'}^x(\operatorname{Tr}_{A}[\rho_{AB}]) \\
& = \sum_{x\in\mathcal{X}} \mathcal{M}_{B\to B'}^x(\rho_{B}).
\end{align}
The second equality follows because $\mathcal{N}_{A\to A'}^x$ is trace preserving for all $x$.
Also,
\begin{align}
&(\operatorname{Tr}_{A'} \circ\mathcal{L}_{AB\to A'B'}^{\leftarrow}\circ \mathcal{R}_{A}^{\pi})(\rho_{AB}) \notag \\
&\quad = (\operatorname{Tr}_{A'} \circ\mathcal{L}_{AB\to A'B'}^{\leftarrow})(\pi_A \otimes \rho_{B}) \\
&\quad = \sum_{x\in\mathcal{X}} \operatorname{Tr}_{A'}[(\mathcal{N}_{A\to A'}^x\otimes\mathcal{M}_{B\to B'}^x)(\pi_A \otimes \rho_{B})] \\
&\quad = \sum_{x\in\mathcal{X}} \mathcal{M}_{B\to B'}^x(\operatorname{Tr}_{A}[\pi_A \otimes \rho_{B}]) \\
&\quad = \sum_{x\in\mathcal{X}} \mathcal{M}_{B\to B'}^x (\rho_{B}).
\end{align}
Thus, the condition in \eqref{eq:NS-A-to-B-cond} holds, and, as expected, $\mathcal{L}_{AB\to A'B'}^{\leftarrow}$ is non-signaling from Alice to Bob.

\section{Summary}

\section{Bibliographic Notes}\label{sec:qm_channels:bib-notes}

	For an introduction to quantum dynamics, and in particular to the Schr\"{o}dinger equation and explicit forms for the unitaries $U(t)$ describing the evolution of closed quantum systems, please see the book of \citet{Sakurai94_book}. The book of \citet{BP02_book} provides a general introduction to the evolution of open quantum systems as a quantum generalization of classical stochastic processes, and shows how completely positive trace-preserving maps (quantum channels) arise from extensions of the Schr\"{o}dinger and von Neumann equations to ``master equations''. We also refer to the tutorial article of \citet{MM21} for a similar exposition on quantum channels from the perspective of stochastic processes and master equations.
	
	The Choi representation of a linear map is named after \citet{C75}, with his paper establishing the equivalence between complete positivity of a map and positivity of its Choi representation. Naimark's dilation theorem for positive operator-valued measures was proved by \citet{N40} (see also \citet{GN43}). Stinespring's dilation theorem for completely positive maps was established by \citet{Sti55}. An early exposition of the theory of quantum channels was presented by \citet{K83}. \citet{Leifer07} proposed the concept of conditional states (see also \citep{LS13}). Proposition~\ref{prop-extension_purif} was found by \citet{CW04}. The channel in \eqref{eq-isometry_recover} for reversing the action of an isometric channel was presented by \citet{Wbook17}. The notions of a complementary channel and a degradable channel were defined by \citet{DS05}. Anti-degradable channels were defined by \citet{CG06}. The quantum erasure channel was defined by \citet{GBP97}. The connection of the amplitude damping channel to the bosonic pure-loss channel was realized by \citet{GF05}. The quantum instrument formalism was developed by \citet{DL70} and further developed by \citet{Ozawa1984}. Entanglement-breaking channels were defined by \citet{HSR03} and several of their properties were established therein. Hadamard channels were defined by \citet{KMNR07}. 

	The Petz recovery map was constructed by \citet{Petz1986} and \citet{Petz1988} in the context of proving the conditions under which equality holds in the data-processing inequality for the quantum relative entropy. The results therein are stated for von Neumann algebras. A more accessible exposition that considers operators acting only on finite-dimensional Hilbert spaces can be found in \citet{Petz2001} and \citet{MP04} (see also \citet{HJPW04}). 
	
	The paradigm of local operations and classical communication was defined by \citet{BDSW96}, and its mathematical properties were explored in more detail by \citet{CLM+14}. See Section~4.3 of \citet{CLM+14} for a justification of the example provided in Section \ref{subsec-LOCC_channels} of a separable channel that is not an LOCC channel. Separable channels were defined by \citet{VPRK97,BNS98}. The existence of a separable channel that is not LOCC was found by \citet{PhysRevA.59.1070}. Completely PPT-preserving channels were defined by \citet{R99} and further developed by \citet{CVGG17}. Non-signaling channels were introduced by \citet{BGNP01} and further considered by \citet{ESW02,PHHH06}.

\section{Problems}

	\begin{enumerate}[left=0cm,itemsep=1cm]
		\item Let $\mathcal{N}$ be a quantum channel, and let $\{K_i\}_{i=1}^r$ and $\{K_i'\}_{i=1}^s$ be two sets of Kraus operators for $\mathcal{N}$. Prove that these sets of Kraus operators are related by an isometry as in \eqref{eq:qm:isometry-between-kraus-ops}.
		
		\item Let $F_{AA'}=\sum_{i,j=0}^{d_A-1}\ket{j,i}\bra{i,j}_{AA'}$ be the swap operator, as defined in \eqref{eq-swap_operator}, and let $\mathcal{N}_{A\to B}$ be a superoperator. Consider the operator
			\begin{equation}\label{eq-dePillis_Jam_rep}
				F_{AB}^{\mathcal{N}}\coloneqq\mathcal{N}_{A'\to B}(F_{AA'})=\sum_{i,j=0}^{d_A-1}\ket{j}\bra{i}_A\otimes\mathcal{N}(\ket{i}\bra{j}_{A'}).
			\end{equation}
			\begin{enumerate}[itemsep=0.3cm,topsep=0.3cm]
				\item Prove that $F_{AB}^{\mathcal{N}}=\T_A(\Gamma_{AB}^{\mathcal{N}})$.
				
				\item Prove that $\Tr_A[F_{AB}^{\mathcal{N}}]=\mathcal{N}_{A\to B}(\mathbbm{1}_A)$.
				
				\item If $\mathcal{N}_{A\to B}$ is trace preserving, then prove that $\Tr_B[F_{AB}^{\mathcal{N}}]=\mathbbm{1}_A$.
				
				\item Prove that $\Inner{X_A^{\dagger}\otimes Y_B}{F_{AB}^{\mathcal{N}}}=\inner{Y_B}{\mathcal{N}_{A\to B}(X_A)}$ for all $X_A\in\Lin(\mathcal{H}_A)$ and $Y_B\in\Lin(\mathcal{H}_B)$.
				
				\item Prove that $F_{AB}^{\mathcal{N}}$ uniquely characterizes $\mathcal{N}$, just as the Choi representation, by showing that 
					\begin{equation}
						\mathcal{N}_{A\to B}(X_A)=\Tr_A[(X_A\otimes\mathbbm{1}_B)F_{AB}^{\mathcal{N}}]
					\end{equation}
					for every linear operator $X_A$.
				
				\item Prove that $F_{AB}^{\mathcal{N}}$ can be expressed in terms of the adjoint $\mathcal{N}^{\dagger}$ as follows:
					\begin{equation}
						F_{AB}^{\mathcal{N}}=\sum_{k,\ell=0}^{d_B-1}\mathcal{N}^{\dagger}(\ket{k}\bra{\ell}_B)^{\dagger}\otimes\ket{k}\bra{\ell}_B.
					\end{equation}
					Conclude that $\Tr_B[F_{AB}^{\mathcal{N}}]=\mathcal{N}^{\dagger}(\mathbbm{1}_B)^{\dagger}$. Furthermore, if $\mathcal{N}$ is Hermiticity preserving, then conclude that $\Tr_B[F_{AB}^{\mathcal{N}}]=\mathcal{N}^{\dagger}(\mathbbm{1}_B)$ and that $F_{AB}^{\mathcal{N}}=(\mathcal{N}^{\dagger})_{B'\to A}(F_{B'B})$, where $F_{B'B}=\sum_{k,\ell=0}^{d_B-1}\ket{\ell,k}\bra{k,\ell}_{B'B}$ is the swap operator for system $B$.
					
				\item Prove that, for every unitary operator $U_A$,
					\begin{equation}
						F_{AB}^{\mathcal{N}}=(\mathcal{U}_A\otimes\mathcal{N}_{A'\to B}\circ\mathcal{U}_{A'})(F_{AA'}).
					\end{equation}
					
				\item Prove that $\mathcal{N}_{A\to B}$ is a positive map if and only if $(\bra{\psi}_A\otimes\bra{\phi}_B)F_{AB}^{\mathcal{N}}(\ket{\psi}_A\otimes\ket{\phi}_B)\geq 0$ for all $\ket{\psi}\in\mathcal{H}_A$ and $\ket{\phi}_B\in\mathcal{H}_B$.
				
			\end{enumerate}
			(\textit{Bibliographic Note}: The representation in \eqref{eq-dePillis_Jam_rep} was defined by \citet{dePillis67}. It is sometimes called the \textit{Jamio\l{}kowski representation} of $\mathcal{N}$ due to the work of \citet{Jam72}, who proved the necessary and sufficient condition on $F_{AB}^{\mathcal{N}}$ in (h) such that $\mathcal{N}_{A\to B}$ is positive.)

%
%

		\item Consider the antisymmetric and symmetric states $\zeta_{AB}$ and $\zeta_{AB}^{\perp}$ defined in \eqref{eq-Werner_zeta} and \eqref{eq-Werner_zeta_perp}, respectively, and recall the channel $\mathcal{N}^{\rho}_{A\to B}$ defined in \eqref{eq-state_to_channel} for a given bipartite state $\rho_{AB}$. 
        Show that
					\begin{align}
						\mathcal{N}_{A\to B}^{\zeta}(X)&=\frac{1}{d-1}\left(\Tr[X]\mathbbm{1}_d-X^{\t}\right),\\[0.1cm]
						\mathcal{N}_{A\to B}^{\zeta^{\perp}}(X)&=\frac{1}{d+1}\left(\Tr[X]\mathbbm{1}_d+X^{\t}\right)
					\end{align}
					for every linear operator $X\in\Lin(\mathcal{H}_A)$. 
					
					(\textit{Bibliographic Note}: The quantum channels $\mathcal{N}^{\zeta}$ and $\mathcal{N}^{\zeta^{\perp}}$ are sometimes called \textit{Werner--Holevo channels}, after \citet{WH02}.)
					
%
				
		
%
%
%
%

	\end{enumerate}

\chapter{Fundamental Quantum Information Processing Tasks}\label{chap-QM_protocols}

	Having studied quantum states, measurements, and channels in detail in the previous two chapters, we are now ready to study three fundamental tasks in quantum information processing: quantum teleportation, quantum super-dense coding, and quantum hypothesis testing. Quantum hypothesis testing has been studied since the late 1960s, with the aim of generalizing (classical) statistical hypothesis testing to the quantum setting. The discovery of quantum teleportation and super-dense coding in the early 1990s demonstrated the practical advantages that entanglement could allow for with respect to communication, and it contributed to the rise of quantum information science as a prominent field of study in both theoretical and experimental physics. 
	
	All of the tasks that we study in this chapter provide us with prototypes of some of the quantum communication scenarios that we consider in Parts~\ref{part:q-comm-prots} and \ref{part-feedback} of this book. In particular, listed below are the tasks and protocols that we study in this chapter and how they are connected to the communication tasks that we study later.
	\begin{itemize}
		\item Quantum teleportation (Section~\ref{sec-teleportation}) is connected to the task of quantum communication (Chapter~\ref{chap-quantum_capacity}), and in particular to LOCC-assisted quantum communication (Chapter~\ref{chap-LOCC-QC}).
		
		\item Quantum super-dense coding (Section~\ref{sec-super_dense_coding}) is connected to the task of entangle\-ment-assisted classical communication (Chapter~\ref{chap-EA_capacity}).
		
		\item Quantum hypothesis testing (Section~\ref{sec-QM_protocols_hypo_testing}), in particular state discrimination in Section~\ref{subsec-state_discrimination} and Section~\ref{sec:QM-over:multiple-state-disc}, is connected to classical communication (Chapter~\ref{chap-classical_capacity}). Furthermore, asymmetric hypothesis testing in Section~\ref{sec-hypo_testing_states_asym} is fundamental to the analysis of every communication scenario that we consider in this book, as it provides us with a method for placing an upper bound on the rate of communication with a finite number of uses of a quantum channel.
	\end{itemize}
	
	Several fundamental quantities used in quantum information theory, particularly in the analysis of quantum communication protocols, arise naturally in the context of quantum hypothesis testing. For example, the \textit{trace distance} arises in terms of the optimal success probability for discriminating between two quantum states in the task of symmetric hypothesis testing (Section~\ref{subsec-state_discrimination}), and similarly the \textit{diamond distance} arises in terms of the optimal success probability for discriminating between two quantum channels in the task of symmetric quantum channel hypothesis testing (Section~\ref{sec-channel_discrimination}). The Chernoff divergence quantifies the optimal error exponent for symmetric hypothesis testing of two quantum states in the asymptotic setting (Section~\ref{sec-hypo_testing_state_symm_asymp}), and this quantity is related to the \textit{Petz--R\'{e}nyi relative entropy}, which we define later in Section~\ref{sec-petz_ren_rel_ent}. With respect to asymmetric hypothesis testing of two quantum states, the optimal error probability defines the so-called \textit{hypothesis testing relative entropy} (Section~\ref{sec-hyp_test_rel_ent}), and in the asymptotic setting the optimal error exponent is given by the \textit{quantum relative entropy} (Section~\ref{sec-rel_ent}). The task of hypothesis testing thus provides several fundamental quantities in quantum information theory with an operational meaning, and we devote Chapters~\ref{chap-QM_dist_meas} and \ref{chap-entropies} to the detailed study of these and other quantities.

\section{Quantum Teleportation}\label{sec-teleportation}

	Quantum teleportation is a remarkable and fundamental protocol in quantum information theory. The simplest version of the protocol allows two parties, Alice and Bob, to transfer the state of a qubit from Alice to Bob while making use of one shared pair of qubits in a maximally entangled state, along with two bits of classical communication.
	
\subsection{Qubit Teleportation Protocol}
	
	Before stating the basic teleportation protocol, let us start by introducing a key element of the protocol, the \textit{Bell measurement}. 

	The Bell measurement is a measurement on two qubits defined by the POVM $\{\ket{\Phi_{z,x}}\bra{\Phi_{z,x}}:z,x\in\{0,1\}\}$, where we recall from \eqref{eq-two_qubit_Bell_states} that the two-qubit \textit{Bell states} are defined as
	\begin{equation}\label{eq-qubit_Bell_compact}
		\ket{\Phi_{z,x}}_{AB}=(\mathbbm{1}_A\otimes X_B^x Z_B^z)\ket{\Phi}_{AB}=(Z_A^zX_A^x\otimes\mathbbm{1}_B)\ket{\Phi}_{AB}
	\end{equation}
	for all $x,z\in\{0,1\}$, so that 
	\begin{align}
		\ket{\Phi_{0,0}}&\coloneqq \frac{1}{\sqrt{2}}(\ket{0,0}+\ket{1,1}),\\
		\ket{\Phi_{1,0}}&\coloneqq \frac{1}{\sqrt{2}}(\ket{0,0}-\ket{1,1}),\\
		\ket{\Phi_{0,1}}&\coloneqq \frac{1}{\sqrt{2}}(\ket{0,1}+\ket{1,0}),\\
		\ket{\Phi_{1,1}}&\coloneqq \frac{1}{\sqrt{2}}(\ket{0,1}-\ket{1,0}).
	\end{align}
	The Bell states are all maximally entangled, and they form an orthonormal basis for the Hilbert space $\mathbb{C}^2\otimes\mathbb{C}^2$ of two qubits. As such, we have that $\sum_{z,x=0}^1\ket{\Phi_{z,x}}\bra{\Phi_{z,x}}=\mathbbm{1}_2\otimes\mathbbm{1}_2$, so that the set $\{\ket{\Phi_{z,x}}\bra{\Phi_{z,x}}:z,x\in\{0,1\}\}$ is indeed a POVM. Furthermore, the classical bits $x$ and $z$ can be viewed as being the outcomes of the measurement. 
	
	We can write the usual computational basis states for two qubits in terms of the Bell states as
	\begin{align}
		\ket{0,0}&=\frac{1}{\sqrt{2}}(\ket{\Phi_{0,0}}+\ket{\Phi_{1,0}}),\label{eq-Bell_to_comp_1}\\
		\ket{0,1}&=\frac{1}{\sqrt{2}}(\ket{\Phi_{0,1}}+\ket{\Phi_{1,1}}),\\
		\ket{1,0}&=\frac{1}{\sqrt{2}}(\ket{\Phi_{0,1}}-\ket{\Phi_{1,1}}),\\
		\ket{1,1}&=\frac{1}{\sqrt{2}}(\ket{\Phi_{0,0}}-\ket{\Phi_{1,0}}).\label{eq-Bell_to_comp_4}
	\end{align} 

	\begin{figure}
		\centering
		\includegraphics[scale=0.8]{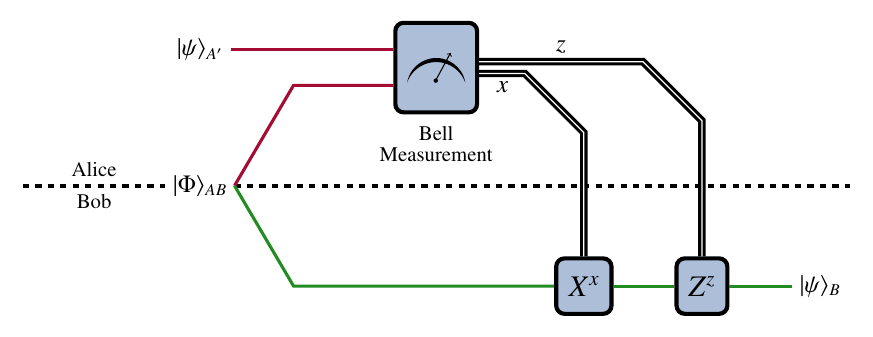}
		\caption{Circuit diagram for the quantum teleportation protocol. The protocol accomplishes the task of sending a quantum state $\psi$ from Alice to Bob using a shared entangled state and two bits of classical communication. The outcomes $(x,z)$ of Alice's Bell measurement on her qubits $A$ and $A'$ are communicated to Bob, who applies the unitary $Z^zX^x$ on his qubit to transform it to the state $\psi$ that Alice wished to send.}\label{fig-teleportation}
	\end{figure}
	
	We now detail the teleportation protocol; see Figure \ref{fig-teleportation} for a circuit diagram depicting the protocol. The protocol starts with Alice and Bob sharing two qubits in the state $\ket{\Phi}_{AB}$. Alice has an additional qubit that she wishes to teleport to Bob; this latter qubit is in the state $\ket{\psi}_{A'}$, where
	\begin{equation}
		\ket{\psi}_{A'}=\alpha\ket{0}_{A'}+\beta\ket{1}_{A'},\quad\alpha,\beta\in\mathbb{C},\quad \abs{\alpha}^2+\abs{\beta}^2=1.
	\end{equation}
	The state $\ket{\psi}_{A'}$ is arbitrary and need not be known to either Alice or Bob. The overall joint state between Alice and Bob at the start of the protocol is therefore
	\begin{equation}\label{eq-teleport_state_start}
		\begin{aligned}
		\ket{\psi}_{A'}\otimes\ket{\Phi}_{AB}&=\frac{1}{\sqrt{2}}(\alpha\ket{0,0,0}_{A'AB}+\alpha\ket{0,1,1}_{A'AB}\\
		&\qquad\qquad+\beta\ket{1,0,0}_{A'AB}+\beta\ket{1,1,1}_{A'AB}).
		\end{aligned}
	\end{equation}
	Alice and Bob then proceed as follows.
	\begin{enumerate}
		\item Alice performs a Bell measurement on her two qubits $A'$ and $A$. To determine the measurement outcomes and their probabilities, it is helpful to write down the initial state \eqref{eq-teleport_state_start} in the Bell basis on Alice's systems. Using \eqref{eq-Bell_to_comp_1}--\eqref{eq-Bell_to_comp_4}, we find that
			\begin{align}
				& \ket{\psi}_{A'}\otimes\ket{\Phi}_{AB}\notag\\
				&=\frac{1}{2}\left(\ket{\Phi_{0,0}}_{A'A}\otimes(\alpha\ket{0}_B+\beta\ket{1}_B)\right.
				\left.+\ket{\Phi_{1,0}}_{A'A}\otimes (\alpha\ket{0}_B-\beta\ket{1}_B)\right.\nonumber\\
				&\quad\quad\left.+\ket{\Phi_{0,1}}_{A'A}\otimes(\alpha\ket{1}_B+\beta\ket{0}_B)\right.
				\left.+\ket{\Phi_{1,1}}_{A'A}\otimes (\alpha\ket{1}_B-\beta\ket{0}_B)\right)\label{eq-qubit_teleportation_step1} \\
				&=\frac{1}{2}\left(\ket{\Phi_{0,0}}_{A'A}\otimes\ket{\psi}_B\right.\left.+\ket{\Phi_{1,0}}_{A'A}\otimes Z_B\ket{\psi}_B\right.\nonumber\\
				&\qquad\qquad\left.+\ket{\Phi_{0,1}}_{A'A}\otimes X_B\ket{\psi}_B\right.\left.+\ket{\Phi_{1,1}}_{A'A}\otimes X_BZ_B\ket{\psi}_B\right). \label{eq-qubit_teleportation_step2}
			\end{align}
			From this, it is clear that each outcome $(x,z)\in\{0,1\}^2$ of the Bell measurement occurs with equal probablity $\frac{1}{4}$ and that the state of Bob's qubit after the measurement is $X_B^xZ_B^z\ket{\psi}_B$.
			
			\begin{exercise}{exer-teleportation_qubit_steps}
				Verify \eqref{eq-qubit_teleportation_step1} and \eqref{eq-qubit_teleportation_step2}.
			\end{exercise}

		\item Alice communicates to Bob the two classical bits $x$ and $z$ resulting from the Bell measurement.
		
		\item Upon receiving the measurement outcomes, Bob performs $X^x$ and then~$Z^z$ on his qubit. The resulting state of Bob's qubit is $\ket{\psi}$. 
		
	\end{enumerate}
	
	Although we have described the teleportation protocol using a pure state~$\ket{\psi}_{A'}$ as the state being teleported, the protocol applies just as well if the state to be teleported is a mixed state $\rho_{A'}$.

\subsection{Qudit Teleportation Protocol}

	The teleportation protocol for qubits described above can be easily generalized to qudits using the Heisenberg--Weyl operators $\{W_{z,x}:0\leq z,x\leq d-1\}$ introduced in Definition~\ref{def-heisenberg-weyl}. Specifically, recall from \eqref{eq-qudit_Bell} that we define the \textit{two-qudit Bell states} in terms of the Heisenberg--Weyl operators as follows:
	\begin{equation}
		\ket{\Phi_{z,x}}_{AB}\coloneqq (W_A^{z,x}\otimes \mathbbm{1}_B)\ket{\Phi}_{AB},
	\end{equation}
	which is a direct generalization of \eqref{eq-qubit_Bell_compact}, where $\ket{\Phi}_{AB}=\frac{1}{\sqrt{d}}\sum_{i=0}^{d-1}\ket{i,i}_{AB}$. Just like the qubit Bell states, the qudit Bell states form an orthonormal basis for $\mathbb{C}^d\otimes\mathbb{C}^d$ (see Exercise~\ref{exer-two_qudit_Bell_ONB}). This means that the set of operators
	\begin{equation}
		\{\ket{\Phi_{z,x}}\!\bra{\Phi_{z,x}}:0\leq z,x\leq d-1\}
	\end{equation}
	constitutes a POVM, which is the POVM corresponding to the Bell measurement on two qudits.
	
	Now we start, like before, with Alice holding two qudits, one shared with Bob and in the joint state $\ket{\Phi}_{AB}$, and the other in the state $\ket{\psi}_{A'}$, where
	\begin{equation}
		\ket{\psi}_{A'}=\sum_{i=0}^{d-1}c_i\ket{i}_{A'},\quad\sum_{i=0}^{d-1}\abs{c_i}^2=1.
	\end{equation}
	The state $\ket{\psi}_{A'}$ is the one to be teleported to Bob's system. The starting joint state on the three qudits $A'$, $A$, and $B$ is
	\begin{equation}
		\ket{\psi}_{A'}\otimes\ket{\Phi}_{AB}=\frac{1}{\sqrt{d}}\sum_{i,j=0}^{d-1} c_i\ket{i,j,j}_{A'AB}.
	\end{equation}
	Alice then performs a Bell measurement on her two qudits. By writing the Bell states $\ket{\Phi_{z,x}}_{A'A}$ as
	\begin{equation}
		\ket{\Phi_{z,x}}_{A'A}=\frac{1}{\sqrt{d}}\sum_{k=0}^{d-1}\e^{\frac{2\pi\I(k+x)z}{d}}\ket{k+x,k}_{A'A},
	\end{equation}
	for each outcome $(z,x)\in\{0,\dotsc,d-1\}^2$, where the sum $k+x$ is understood to be modulo $d$, we use \eqref{eq-post_meas_state_half_proj} to find that the corresponding (unnormalized) post-measurement state of Bob's qudit is
	\begin{align}
		&(\bra{\Phi_{z,x}}_{A'A}\otimes\mathbbm{1}_B)(\ket{\psi}_{A'}\otimes\ket{\Phi}_{AB})\nonumber\\
		&\qquad=\frac{1}{d}\sum_{j,k,\ell=0}^{d-1}c_{\ell}\e^{-\frac{2\pi\I(k+x)z}{d}}\braket{k+x}{\ell}\braket{k}{j}\ket{j}_B\\
		&\qquad=\frac{1}{d}\sum_{k=0}^{d-1}c_{k+x}\e^{-\frac{2\pi\I(k+x)z}{d}}\ket{k}_B\\
		&\qquad=\frac{1}{d}\sum_{k'=0}^{d-1}c_{k'}\e^{-\frac{2\pi\I k'z}{d}}\ket{k'-x}_B\\
		&\qquad=\frac{1}{d}\sum_{k=0}^{d-1}c_{k'} X(-x)Z(-z)\ket{k'}_B\\
		&\qquad=\frac{1}{d}X(-x)Z(-z)\ket{\psi}_B.
	\end{align}
	Therefore, each outcome occurs with probability $\frac{1}{d^2}$, and the corresponding post-measurement state of Bob's qudit is $X(-x)Z(-z)\ket{\psi}_B$. This means that Bob, upon receiving the two classical values corresponding to the outcome of the Bell measurement, can apply the unitary $Z(z)X(x)=W_{z,x}$ in order to transform the state of his qudit to $\ket{\psi}_B$, completing the teleportation protocol.
	
	\begin{figure}
		\centering
		\includegraphics[scale=0.8]{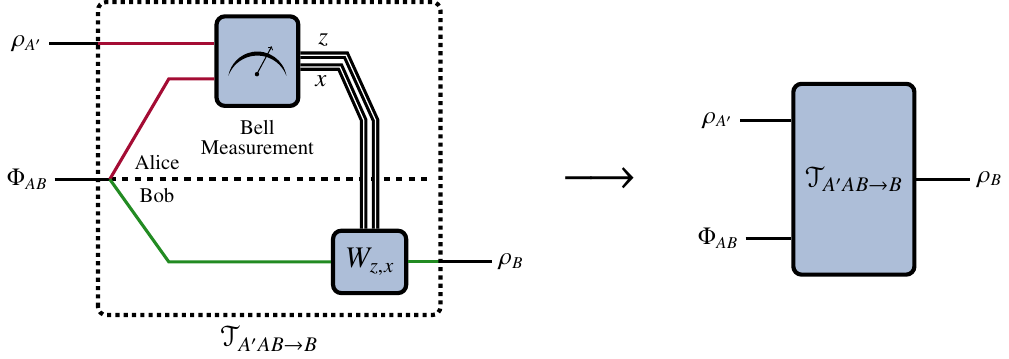}
		\caption{The qudit teleportation protocol, depicted on the left, can be regarded as an LOCC channel $\mathcal{T}_{A'AB\to B}$ with the input states $\rho_{A'}$ and $\ket{\Phi}\!\bra{\Phi}_{AB}$ and the output state $\rho_B$, as shown on the right.}\label{fig-teleportation_chan}
	\end{figure}
	
	Of course, the teleportation protocol works just as well if the state to be teleported is mixed. Also, as shown in Figure \ref{fig-teleportation_chan}, we can write down the entire teleportation protocol as a one-way LOCC channel $\mathcal{T}_{A'AB\to B}$ from Alice to Bob, with the state $\rho_{A'}$ to be teleported and the state $\ket{\Phi}\!\bra{\Phi}_{AB}$ as the inputs. By the analysis above, the output state received by Bob is exactly the same as the original input state:
	\begin{equation}\label{eq-teleportation_channel}
		\mathcal{T}_{A'AB\to B}(\rho_{A'}\otimes\ket{\Phi}\!\bra{\Phi}_{AB})=\rho_B.
	\end{equation}
	This equation can also be interpreted as follows:
	\begin{equation}\label{eq-teleportation_sim_id}
		\mathcal{T}_{A'AB\to B}((\cdot)_{A'}\otimes \ket{\Phi}\!\bra{\Phi}_{AB})=\id_{A'\to B}(\cdot),
	\end{equation}
	i.e., that the teleportation protocol simulates the identity channel.
	
	To see that the teleportation protocol can indeed be viewed as a one-way LOCC channel from Alice to Bob, let us explicitly write down the quantum channel $\mathcal{T}_{A'AB\to B}$ defined above in the form of \eqref{eq-1wayLOCC_decomp}, i.e.,
	\begin{equation}\label{eq-teleportation_LOCC_chan}
		\mathcal{T}_{A'AB\to B}=\mathcal{D}_{BY_1Y_2\to B}\circ\mathcal{C}_{X_1X_2\to Y_1Y_2}\circ\mathcal{E}_{A'A\to X_1X_2}
	\end{equation}
	which we recall is equivalent to the form in \eqref{eq-LOCC_A_to_B}. We have that
	\begin{align}
		\mathcal{E}_{A'A\to X_1X_2}&=\sum_{z,x=0}^{d-1}\mathcal{E}_{A'A\to\varnothing}^{z,x}\otimes\ket{z,x}\!\bra{z,x}_{X_1X_2},\label{eq-teleportation_LOCC_1}\\
		\mathcal{E}_{A'A\to\varnothing}^{z,x}(\cdot)&\coloneqq \Tr_{A'A}[\ket{\Phi_{z,x}}\!\bra{\Phi_{z,x}}_{A'A}(\cdot)],\label{eq-teleportation_LOCC_2}\\
		\mathcal{C}_{X_1X_2\to Y_1Y_2}(\ket{z,x}\!\bra{z,x}_{X_1X_2})&=\ket{z,x}\!\bra{z,x}_{Y_1Y_2},\label{eq-teleportation_LOCC_3}\\
		\mathcal{D}_{BY_1Y_2\to B}((\cdot)_B\otimes\ket{z,x}\!\bra{z,x}_{Y_1Y_2})&=\mathcal{D}_{B\to B}^{z,x}(\cdot),\label{eq-teleportation_LOCC_4}\\
		\mathcal{D}_{B\to B}^{z,x}(\cdot)&\coloneqq W_{z,x}(\cdot)W_{z,x}^\dagger.\label{eq-teleportation_LOCC_5}
	\end{align}
	
	\begin{exercise}{exer-teleportation_LOCC_chan}
		Combine the quantum channels in \eqref{eq-teleportation_LOCC_1}--\eqref{eq-teleportation_LOCC_5} according to \eqref{eq-teleportation_LOCC_chan} and conclude that the channel $\mathcal{T}_{A'AB\to B}$ can be written as 
		\begin{equation}\label{eq-teleportation_LOCC_chan2}
			\mathcal{T}_{A'AB\to B}(\sigma_{A'AB})=\sum_{z,x=0}^{d-1} \Tr_{A'A}\!\left[\Phi_{A'A}^{z,x} W_B^{z,x}(\sigma_{A'AB})(W_B^{z,x})^{\dagger}\right]
		\end{equation}
		for every state  $\sigma_{A'AB}$. Verify that, for the input state $\sigma_{A'AB}=\rho_{A'}\otimes\Phi_{AB}$, we get $\mathcal{T}_{A'AB\to B}(\rho_{A'}\otimes\Phi_{AB})=\rho_B$, as expected.
	\end{exercise}
	
\begin{figure}
		\centering
		\includegraphics[scale=0.8]{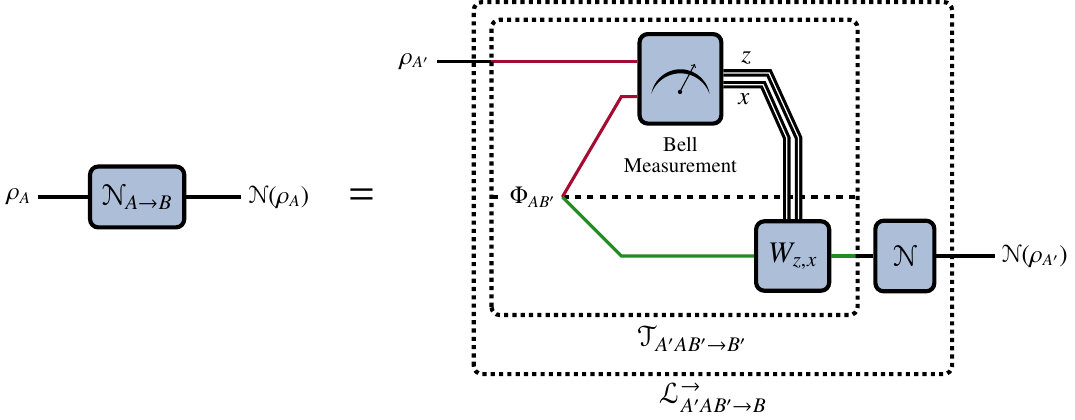}
		\caption{Every quantum channel $\mathcal{N}$ is teleportation simulable, because Alice and Bob can first perform the usual teleportation protocol, and then Bob can apply the channel $\mathcal{N}$ to the state teleported by Alice. The combination of these two steps can be taken as the LOCC channel $\mathcal{L}_{A'AB'\to B}^{\rightarrow}$.}\label{fig-teleportation_sim_arbitrary}
	\end{figure}

	We can also connect with the previously defined notion of LOCC simulation of a quantum channel (Definition~\ref{def-LOCC_sim_chan}). That is, we can understand the teleportation protocol and \eqref{eq-teleportation_channel} as demonstrating that the identity channel is LOCC simulable with associated resource state given by the maximally entangled state. Now, by using the teleportation protocol in conjunction with a quantum channel $\mathcal{N}_{A\to B}$, we find that every quantum channel $\mathcal{N}_{A\to B}$ is LOCC simulable with associated resource state given by the maximally entangled state of an appropriate Schmidt rank. To see this, observe that the channel $\mathcal{N}_{A\to B}$ can be trivially written as $\mathcal{N}_{A\to B}=\mathcal{N}_{B'\to B}\circ\id_{A\to B'}$, where $B'$ is an auxiliary system with the same dimension as $A$. Then, by \eqref{eq-teleportation_sim_id}, we can simulate the identity channel $\id_{A\to B'}$ using the usual teleportation protocol, so that the overall LOCC channel $\mathcal{L}$ is $\mathcal{N}_{B'\to B}\circ\mathcal{T}_{AA'B'\to B'}$ and the resource state is $\ket{\Phi}\!\bra{\Phi}_{A'B'}$, with the dimension of $A'$ equal to the dimension of $A$. This is illustrated in Figure \ref{fig-teleportation_sim_arbitrary}. We can also simulate $\mathcal{N}$ via teleportation in a different manner, in which Alice locally applies the channel $\mathcal{N}$ to her input state $\rho_A$, then teleports the resulting state to Bob. Mathematically, we write this as $\mathcal{N}_{A\to B}=\id_{\hat{A}\to B}\circ\mathcal{N}_{A\to\hat{A}}=\mathcal{T}_{\hat{A}\tilde{A}B\to B}\circ\mathcal{N}_{A\to\hat{A}}$, where $\hat{A},\tilde{A}$ are auxiliary systems with the same dimension as $B$. We thus have have the following two ways to represent the action of the channel $\mathcal{N}$ using teleportation:
	\begin{align}
		\mathcal{N}_{A\to B}(\rho_A)&=\mathcal{N}_{B'\to B}\left(\mathcal{T}_{AA'B'\to B'}(\rho_{A}\otimes\ket{\Phi}\!\bra{\Phi}_{A'B'})\right)\label{eq-chan_tele_sim_1}\\
		&=\mathcal{T}_{\hat{A}\tilde{A}B\to B}\left(\mathcal{N}_{A\to\hat{A}}(\rho_A)\otimes\ket{\Phi}\!\bra{\Phi}_{\tilde{A}B}\right).\label{eq-chan_tele_sim_2}
	\end{align}
	
	Depending on whether the input dimension is smaller than the output dimension of the channel, there can be a more economical way to perform the simulation. If the channel's output dimension is smaller than its input dimension, then the more economical way to simulate the channel is for Alice to apply $\mathcal{N}_{A\to B}$ first  and then for Alice and Bob to perform the teleportation protocol. In this way, they exploit a maximally entangled state of Schmidt rank $d_B$ in order to accomplish the simulation. If the channel's input dimension is smaller than its output dimension, then the more economical way to simulate the channel is for Alice and Bob to perform the teleportation protocol first, and then for Bob to apply the channel locally. In this way, they exploit a maximally entangled state of Schmidt rank $d_A$ in order to accomplish the simulation. As we see in Section~\ref{subsec-tele_sim}, depending on the channel and its symmetries, there can be even more economical methods to simulate a quantum channel via teleportation.

\subsubsection{Qudit Teleportation Protocol With Respect to a Finite Group} \label{sec-QM-fund-prot:TP-finite-group}
	
	The qudit teleportation protocol outlined above is based on the Heisenberg--Weyl operators $\{W_{z,x}:0\leq z,x\leq d-1\}$ acting on a $d$-dimensional system, which are used to form the qudit Bell states and thus the Bell measurement.
	
	More generally, we can consider an arbitrary finite group $G$ and an irreducible unitary representation $\{U^g\}_{g\in G}$ of $G$ acting on a $d$-dimensional Hilbert space, where $d^2\leq|G|$. It then follows from Schur's lemma (see Bibliographic Notes in Section~\ref{sec:qm:bib-notes}) that
	\begin{equation}
		\frac{1}{|G|}\sum_{g\in G}U^g \rho U^{g\dagger}=\Tr[\rho]\frac{\mathbbm{1}_d}{d}
	\end{equation}
	for every state $\rho$. In particular, for bipartite states $\rho_{AB}$ in which the systems $A$ and $B$ are both $d$-dimensional, we find that
	\begin{equation}\label{eq-twirl_G_tensor}
		\frac{1}{|G|}\sum_{g\in G}(U_A^g\otimes \mathbbm{1}_B) \rho_{AB} (U_A^{g\dagger}\otimes\mathbbm{1}_B)=\frac{\mathbbm{1}_A}{d}\otimes\Tr_A[\rho_{AB}].
	\end{equation}
	
	Now, let us take the maximally entangled state $\ket{\Phi}_{AB}$ and define the states
	\begin{equation}\label{eq-qudit_Bell_G}
		\ket{\Phi^g}_{AB}\coloneqq (U_A^g\otimes \mathbbm{1}_B)\ket{\Phi}_{AB}.
	\end{equation}
	We call these states the \textit{generalized Bell states}. We see that these states are a direct generalization of the usual qudit Bell states in \eqref{eq-qudit_Bell}. 
	
	\begin{exercise}{exer-G_Bell_POVM}
		Using \eqref{eq-twirl_G_tensor}, prove that
		\begin{equation}
			\frac{1}{|G|}\sum_{g\in G}\ket{\Phi^g}\!\bra{\Phi^g}_{AB}=\frac{\mathbbm{1}_{AB}}{d^2}.
		\end{equation}
	\end{exercise}
	
	By defining the operators
	\begin{equation}\label{eq-G_Bell_meas}
		M_{AB}^g\coloneqq \frac{d^2}{|G|}\ket{\Phi^g}\!\bra{\Phi^g}_{AB},
	\end{equation}
	we find that
	\begin{equation}
		\sum_{g\in G}M_{AB}^g=\mathbbm{1}_{AB}.
	\end{equation}
	Since the operators $M^g_{AB}$ satisfy $0\leq M_{AB}^g\leq\mathbbm{1}_{AB}$ for all $g\in G$ (the right-most inequality due to the assumption $d^2\leq |G|$), we conclude that the set $\{M^g_{AB}\}_{g\in G}$ constitutes a POVM. This POVM defines the \textit{$G$-Bell measurement}. 
	
	We use the $G$-Bell measurement defined by the POVM $\{M^g\}_{g\in G}$ in order to construct the generalized teleportation protocol. The protocol proceeds as follows. As before, Alice and Bob start by sharing two qudits in the state $\ket{\Phi}_{AB}$, with Alice holding an extra qudit, i.e., in the state $\ket{\psi}_{A'}$, to be teleported to Bob. 
	\begin{enumerate}
		\item Alice performs, on her qudits $A$ and $A'$, the generalized Bell measurement given by the POVM $\{M_{AA'}^g\}_{g\in G}$. For each outcome $g\in G$ of the measurement, according to \eqref{eq-post_meas_state_half_proj} the (unnormalized) post-measurement state of Bob's qudit is
			\begin{align}
				&\left(\frac{d}{\sqrt{|G|}}\bra{\Phi_g}_{A'A}\otimes\mathbbm{1}_B\right)(\ket{\psi}_{A'}\otimes\ket{\Phi}_{AB})\nonumber\\
				&\qquad=\frac{d}{\sqrt{|G|}}\left(\bra{\Phi}_{A'A}(U_{A'}^{g\dagger}\otimes\mathbbm{1}_A)\otimes\mathbbm{1}_B\right)(\ket{\psi}_{A'}\otimes\ket{\Phi}_{AB})\\
				&\qquad=\frac{d}{\sqrt{|G|}}\left(\frac{1}{\sqrt{d}}\sum_{k=1}^d\bra{k,k}_{A'A}(U_{A'}^{g\dagger}\otimes\mathbbm{1}_A)\otimes\mathbbm{1}_B\right)\nonumber\\
				&\qquad\qquad\qquad\qquad\qquad\times \left(\ket{\psi}_{A'}\otimes\frac{1}{\sqrt{d}}\sum_{k'=1}^d\ket{k',k'}_{AB}\right)\\
				&\qquad=\frac{1}{\sqrt{|G|}}\sum_{k,k'=1}^d\bra{k}_{A'}U_{A'}^{g\dagger}\ket{\psi}_{A'}\ket{k}_B\braket{k}{k'}\\
				&\qquad=\frac{1}{\sqrt{|G|}}\sum_{k=1}^d\bra{k}_{A'}U_{A'}^{g\dagger}\ket{\psi}_{A'}\ket{k}_B\\
				&\qquad=\frac{1}{\sqrt{|G|}}\,U_{B}^{g\dagger}\ket{\psi}_B
			\end{align}
			We see that each outcome occurs with probability $\frac{1}{|G|}$, and the post-meas\-urement state of Bob's qudit is $U_{B}^{g\dagger}\ket{\psi}_B$.
		\item Alice communicates the outcome $g$ resulting from the measurement to Bob.
		\item Upon receiving the measurement outcome, Bob applies $U^g$ on his qudit. The resulting state of Bob's qudit is $\ket{\psi}_{B}$.	
	\end{enumerate}
	
	Observe that the original qudit teleportation protocol is a special case of the generalized teleportation protocol outlined above, in which the group $G$ is $\mathbb{Z}_d\times\mathbb{Z}_d$ and its irreducible projective unitary representation $\{U^g\}_{g\in G}$ is taken to be the set of Heisenberg--Weyl operators. Then, the generalized Bell states $\Phi^g$ are precisely the qudit Bell states $\Phi_{z,x}$ defined in \eqref{eq-qudit_Bell}. Furthermore, since $|G|=d^2$, the POVM elements $M^g=\frac{d^2}{|G|}\ket{\Phi^g}\!\bra{\Phi^g}$ are the projections on to the qudit Bell states, exactly as in the qudit teleportation protocol.
	
	\begin{exercise}{exer-G_teleportation_channel}
		By following a development similar to that in \eqref{eq-teleportation_LOCC_chan}--\eqref{eq-teleportation_LOCC_5} and Exercise~\ref{exer-teleportation_LOCC_chan}, verify that the one-way LOCC channel corresponding to the generalized teleportation protocol presented above has the following form analogous to \eqref{eq-teleportation_LOCC_chan2}:
		\begin{equation}
			\mathcal{T}_{A'AB\to B}^G(\sigma_{A'AB})=\sum_{g\in G}\Tr_{A'A}\!\left[M_{A'A}^g U_B^g(\sigma_{A'AB})U_B^{g\dagger}\right]
		\end{equation}
		for every state  $\sigma_{A'AB}$. Conclude that $\mathcal{T}_{A'AB\to B}^{G}(\rho_{A'}\otimes\Phi_{AB})=\rho_B$, as expected.
	\end{exercise}

\subsection{Post-Selected Teleportation}

	Throughout this section, we have described teleportation protocols that involve performing a particular kind of Bell measurement between a system $A'$, whose state is to be teleported, and a system $A$ that is one share of a bipartite system $AB$ in the joint resource state $\ket{\Phi}\!\bra{\Phi}_{AB}$. Based on the outcome of the Bell measurement performed on~$A'A$, Bob applies a particular correction operation in order to obtain the initial state of $A'$ in his system $B$. Thus, although the individual outcomes of the Bell measurement occur with some probability, the overall teleportation protocol is deterministic, due to the correction operations; i.e., it succeeds with probability one.
	
	If Bob does not have the ability to apply correction operations to his system based on the Bell measurement outcomes, then Alice and Bob can perform what is called \textit{post-selected teleportation}. Post-selected teleportation is based on the fact that, in the teleportation protocols that we have considered, Bob does not need to apply a correction operation on his system if the outcome corresponding to $\ket{\Phi}\!\bra{\Phi}_{A'A}$ occurs in the Bell measurement performed on $A'A$. This is due to the fact that the post-measurement state on Bob's system, conditioned on the $\ket{\Phi}\!\bra{\Phi}_{A'A}$ outcome, is precisely the initial state $\rho_{A'}$ to be teleported:
	\begin{equation}\label{eq-post_selected_teleportation}
		\bra{\Phi}_{A'A}(\rho_{A'}\otimes\ket{\Phi}\!\bra{\Phi}_{AB})\ket{\Phi}_{A'A}=\frac{1}{d^2}\,\rho_B,
	\end{equation}
	where $d=d_A=d_B$. This is a special case of \eqref{eq-Choi_state_post_selected_teleportation} in which $\mathcal{N}_{A'\to B}=\id_{A'\to B}$ and $X_{RA'}=\rho_{A'}$. If we thus modify the teleportation protocol so as to consider the $\ket{\Phi}\!\bra{\Phi}_{A'A}$ outcome of the Bell measurement as a ``success'' and the rest of the outcomes as a ``failure'', then we obtain post-selected teleportation. Post-selected teleportation is probabilistic by definition. In particular, from \eqref{eq-post_selected_teleportation}, we see that it succeeds with probability $\frac{1}{d^2}$.

\subsection{Teleportation-Simulable Channels}\label{subsec-tele_sim}

\begin{figure}
		\centering
		\includegraphics[scale=0.8]{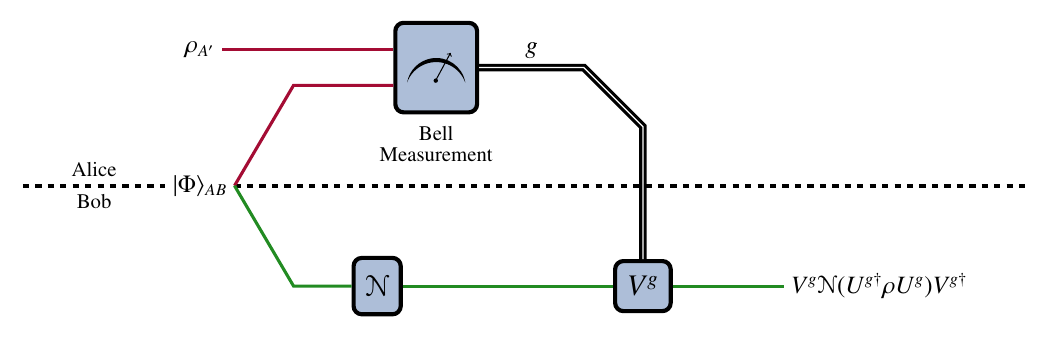}
		\caption{Circuit diagram of a modified teleportation protocol in which Bob applies a channel $\mathcal{N}$ to his qudit before performing a correction operation $V^g$, which is a unitary operation from the set $\{V^g:g\in G\}$ of unitary operators.}\label{fig-teleportation_gen}
	\end{figure}

	Let us now consider the even more general protocol depicted in Figure \ref{fig-teleportation_gen}. Let~$G$ be a finite group. As before, Alice and Bob start with a shared pair of qudits in the state $\ket{\Phi}\!\bra{\Phi}_{AB}$, while Alice holds an extra qudit in the state $\rho_{A'}$ to be teleported to Bob. However, unlike the above teleportation protocol, Bob applies the channel~$\mathcal{N}$ to his qudit before he receives the results of the Bell measurement. Once he receives the measurement results, he applies the unitary operation $V^{g}$ from the set~$\{V^{g}:g\in G\}$ of pre-determined unitary operators constituting a projective unitary representation of $G$. 
	
	The initial tripartite joint state of the protocol is 
	\begin{equation}\label{eq-teleport_gen}
		\rho_{A'}\otimes(\mathbbm{1}_A\otimes\mathcal{N}_B)(\ket{\Phi}\!\bra{\Phi}_{AB}).
	\end{equation}
	Alice performs the same generalized Bell measurement as before on $A$ and $A'$, which we recall has the POVM $\{\Pi_{AA'}^g\}_{g\in G}$ with elements $\Pi_{AA'}^g$ defined in \eqref{eq-G_Bell_meas}. Recall that this POVM corresponds to an irreducible projective unitary representation of $G$ given by $\{U^g\}_{g\in G}$. Since the Bell measurement operates only on the systems $A'$ and $A$, we can bring them inside the action of $\mathcal{N}$ on Bob's share of the state $\ket{\Phi}\!\bra{\Phi}_{AB}$. This means that the analysis for the qudit teleportation protocol from Section~\ref{sec-QM-fund-prot:TP-finite-group} carries over exactly in this case. In other words, each outcome $g\in G$ occurs with an equal probability of $\frac{1}{|G|}$, and the post-measurement state on Bob's qudit corresponding to the outcome $g$ is
	\begin{equation}
		\mathcal{N}(U^{g\dagger}\rho U^g).
	\end{equation}
	After Bob applies the unitary $V^g$, the state of Bob's qudit at the end of the protocol is
	\begin{equation}
		V^g\mathcal{N}(U^{g\dagger}\rho U^g)V^{g\dagger}.
	\end{equation}
	This occurs with probability $\frac{1}{|G|}$ for all $g\in G$.

	\begin{figure}
		\centering
		\includegraphics[scale=0.8]{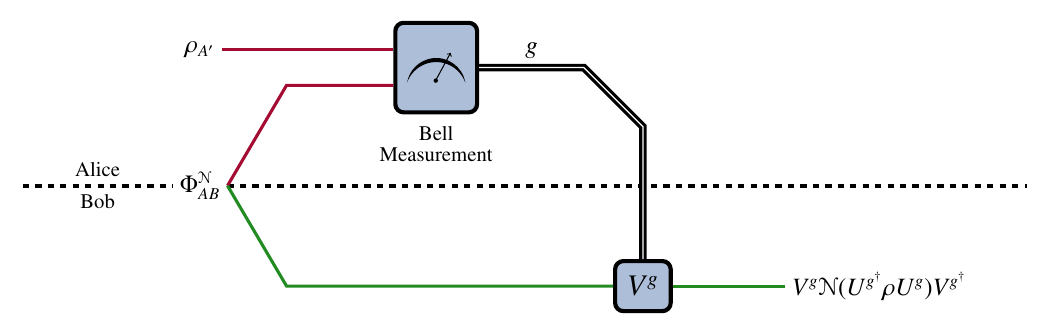}
		\caption{A mathematically equivalent way of describing the protocol in Figure \ref{fig-teleportation_gen}. In this case, Alice and Bob start with the Choi state $\Phi_{AB}^{\mathcal{N}}$ of the channel $\mathcal{N}$ instead of the maximally entangled state $\ket{\Phi}\!\bra{\Phi}_{AB}$.}\label{fig-teleportation_gen2}
	\end{figure}
	
	Now, observe that the state \eqref{eq-teleport_gen} can be written as
	\begin{equation}
		\rho_{A'}\otimes\Phi_{AB}^{\mathcal{N}},
	\end{equation}
	where we recall that $\Phi_{AB}^{\mathcal{N}}=(\id_A\otimes\mathcal{N}_B)(\ket{\Phi}\!\bra{\Phi}_{AB})$ is the Choi state of the channel $\mathcal{N}$. In other words, the protocol depicted in Figure \ref{fig-teleportation_gen} is mathematically equivalent to the teleportation protocol over a group $G$ outlined above, except that instead of starting with the shared maximally entangled state $\ket{\Phi}_{AB}$, Alice and Bob start with the shared state $\Phi_{AB}^{\mathcal{N}}$. This equivalent protocol is depicted in Figure \ref{fig-teleportation_gen2}.
	
		If Bob discards the classical message $g$ at the end of the protocol, then the state of his system is given by
	\begin{equation}\label{eq-twirled-channel}
	   \frac{1}{|G|} \sum_{g \in G} V^g\mathcal{N}(U^{g\dagger}\rho U^g)V^{g\dagger}.
	\end{equation}
	Recall from \eqref{eq-channel_twirl} that this state is simply the output state of the twirl of $\mathcal{N}$ with respect to the unitary representations $\{U^g\}_{g\in G}$ and $\{V^g\}_{g\in G}$, because the twirled channel $\overline{\mathcal{N}}$ is a symmetrized version of the original channel $\mathcal{N}$. Thus, the generalized teleportation protocol gives an explicit procedure for implementing a channel twirl by implementing the teleportation protocol using the Choi state of the channel as the resource state. Specifically, we have
    \begin{equation}
        \mathcal{T}_{A'AB\to B}^{G}(\rho_{A'}\otimes\Phi_{AB}^{\mathcal{N}})=\overline{\mathcal{N}}_{A\to B}(\rho_A).
    \end{equation}

\begin{figure}
		\centering
		\includegraphics[scale=0.8]{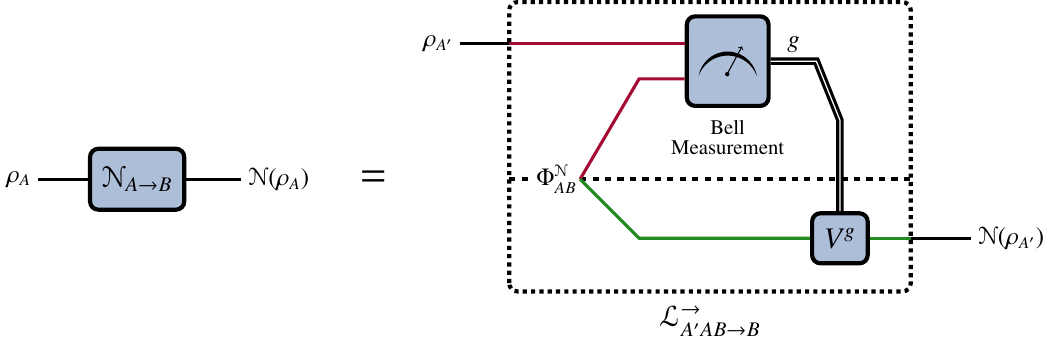}
		\caption{Teleportation simulation of a $G$-covariant channel $\mathcal{N}$, where the operators $\{V^g:g\in G\}$ form a unitary representation of $G$ on the output space of $\mathcal{N}$.}\label{fig-teleportation_sim_covariant}
	\end{figure}

	Suppose now that the channel $\mathcal{N}$ satisfies the group covariance property from Definition~\ref{def-group_cov_chan} for all $g\in G$. In this case, we see that $\mathcal{N}(U^{g\dagger}\rho U^g)=V^{g\dagger}\mathcal{N}(\rho)V^g$ for every outcome $g$ of Alice's generalized Bell measurement. Therefore, after Bob applies~$V^g$, the state of his qudit is precisely~$\mathcal{N}(\rho)$. This generalized teleportation protocol therefore effectively applies the channel $\mathcal{N}$ to the state $\rho_{A'}$ and transfers the resulting state to Bob's qudit; see Figure \ref{fig-teleportation_sim_covariant}. We say that the teleportation protocol \textit{simulates} the action of the channel $\mathcal{N}$ on the input state $\rho_{A'}$. As stated earlier, in this sense, the original teleportation protocol can be regarded as a way to simulate the identity channel from Alice to Bob.

	The notion of simulation of a channel by a teleportation protocol can be extended to a one-way LOCC channel $\mathcal{L}^{\rightarrow}$, as introduced in Definition \ref{def-LOCC}, to obtain the following definition.
	
	\begin{figure}
		\centering
		\includegraphics[scale=0.8]{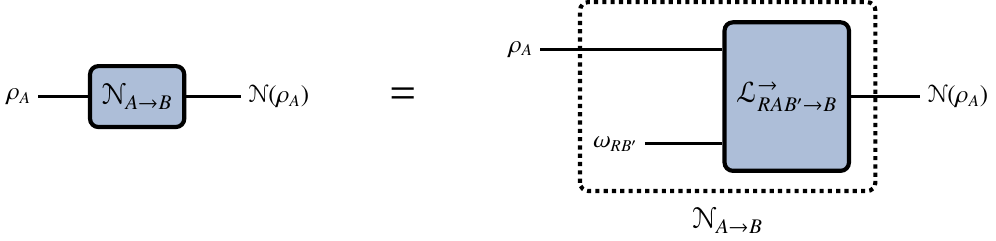}
		\caption{Depiction of a teleportation-simulable channel with associated resource state $\omega_{RB'}$. The teleportation-simulable channel $\mathcal{N}_{A\to B}$ can be realized via the interaction LOCC channel $\mathcal{L}^{\rightarrow}_{RAB\to B}$ and the resource state $\omega_{RB'}$.}\label{fig-teleportation_sim_gen}
	\end{figure}
	
	\begin{definition}{Teleportation-Simulable Channel}{def-teleportation_sim}
		A channel $\mathcal{N}_{A\to B}$ is called \textit{teleportation-simulable} with associated resource state $\omega_{RB}$ if there exists a one-way LOCC channel $\mathcal{L}_{RAB'\to B}^{\rightarrow}$  such that, for every input state $\rho_A$,
		\begin{equation}\label{eq-teleportation_sim}
			\mathcal{N}(\rho_A)=\mathcal{L}_{RAB'\to B}^{\rightarrow}(\rho_A\otimes \omega_{RB'}).
		\end{equation}
        Here, the systems $R$ and $A$ belong to Alice, $B'$ and $B$ to Bob, and the classical communication goes from Alice to Bob.
	\end{definition}
	
	Figure \ref{fig-teleportation_sim_gen} illustrates the concept of a teleportation-simulable channel. Note that in \eqref{eq-teleportation_sim} the resource state $\omega_{RB'}$ is fixed, as well as the LOCC channel $\mathcal{L}^{\rightarrow}$. Both sides of the equation should thus be regarded as functions of $\rho_A$. 

	From the above discussions, we conclude that every group-covariant channel is teleportation-simulable, where the one-way LOCC channel $\mathcal{L}$ is simply the teleportation protocol with respect to the group, and the resource state is the Choi state of the channel.

\section{Quantum Super-Dense Coding}\label{sec-super_dense_coding}

	We now discuss the quantum super-dense coding protocol. This protocol can be viewed as a ``dual'' to the quantum teleportation protocol in the following sense: while in the basic quantum teleportation protocol Alice and Bob make use of two qubits in the entangled state vector $\ket{\Phi^+}=\frac{1}{\sqrt{2}}(\ket{0,0}+\ket{1,1})$ and two bits of classical information to simulate a noiseless qubit channel, in quantum super-dense coding they make use of the shared entangled state $\ket{\Phi^+}$ along with one use of a noiseless qubit channel to communicate \textit{two} bits of classical information. This is remarkable because, without the shared entanglement and only one use of the noiseless qubit channel, they can communicate at most only one bit of classical information. The quantum super-dense coding protocol thus represents one of the simplest examples in which prior shared entanglement provides an advantage for classical communication.
	
	\begin{figure}
		\centering
		\includegraphics[scale=0.8]{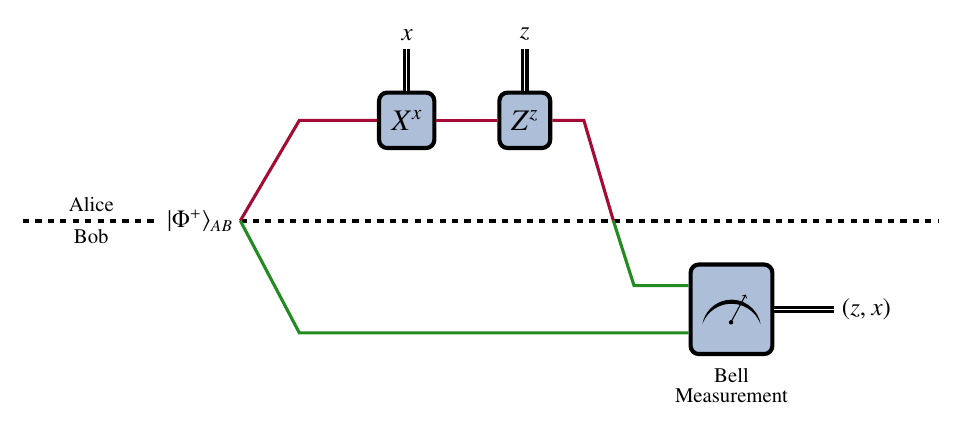}
		\caption{Circuit diagram for the super-dense coding protocol. Using the bits $(z,x)$ that she wishes to send, Alice applies the appropriate Pauli~$X$ and/or~$Z$ operators to her share $A$ of the maximally entangled qubits that are in the state $\ket{\Phi^+}_{AB}$ and sends it through a noiseless qubit channel to Bob. Bob then performs a Bell measurement on the two qubits to recover the encoded bits~$(z,x)$.}\label{fig-super_dense_coding}
	\end{figure}
	
	Let us now go through the quantum super-dense coding protocol. See Figure \ref{fig-super_dense_coding} for a depiction of the protocol. Alice wishes to send two classical bits $(z,x)\in\{0,1\}^2$ to Bob by making use of a shared pair of qubits in the maximally entangled state $\ket{\Phi^+}_{AB}$ and one use of a noiseless qubit channel. Depending on the bits she wishes to send, she performs the following operations on her share of the entangled qubits:
	\begin{itemize}
		\item To send the bits $(0,0)$, she does nothing.
		\item To send the bits $(0,1)$, she applies the Pauli $X$ operator to her qubit, transforming the joint state $\ket{\Phi^+}_{AB}$ to $\ket{\Psi^+}_{AB}$.
		\item To send the bits $(1,0)$, she applies the Pauli $Z$ operator to her qubit, transforming the joint state $\ket{\Phi^+}_{AB}$ to $\ket{\Phi^-}_{AB}$.
		\item To send the bits $(1,1)$, she applies the $X$ operator followed by the $Z$ operator, so that the joint state becomes $\ket{\Psi^-}_{AB}$.
	\end{itemize}
	After applying the appropriate operation, Alice sends her qubit to Bob with the one allowed use of a noiseless qubit channel.
	
	Bob now holds both qubits, and they are in one of the four Bell states
	\begin{equation}
		\ket{\Phi_{z,x}}_{AB}=(Z_A^zX_A^x\otimes\mathbbm{1}_B)\ket{\Phi^+}_{AB}
	\end{equation}
	depending on the bits $(z,x)$ Alice sent. Bob then performs a Bell measurement on his two qubits, and the outcome of this measurement consists precisely of the bits $(z,x)$ that Alice wished to send. 
	
	The super-dense coding protocol has a simple generalization to the qudit case. In this case, Alice and Bob share the qudit Bell state $\ket{\Phi}_{AB}$ before communication begins, and by applying one of the $d^2$ Heisenberg--Weyl operators $W_{z,x}$ from \eqref{eq-Heisenberg_Weyl_operators} on her share of the state, Alice can rotate the global state to one of the $d^2$ qudit Bell states in \eqref{eq-qudit_Bell}. After Alice sends her share of the encoded state over a noiseless qudit channel to Bob, Bob can then perform the qudit Bell measurement to decode which of the $d^2$ messages Alice transmitted.

\section{Quantum Hypothesis Testing}\label{sec-QM_protocols_hypo_testing}

	We now consider the task of quantum hypothesis testing, which is a generalization of classical statistical hypothesis testing to the quantum setting. In the quantum setting, the statistical hypotheses are represented by states of a particular quantum system\footnote{The hypotheses can be represented by quantum channels more generally, as we detail in Section~\ref{sec-channel_discrimination}.}, and the task is to determine which of the hypotheses is ``true'', i.e., to determine the state of the quantum system.

\begin{figure}
		\centering
		\includegraphics[scale=0.8]{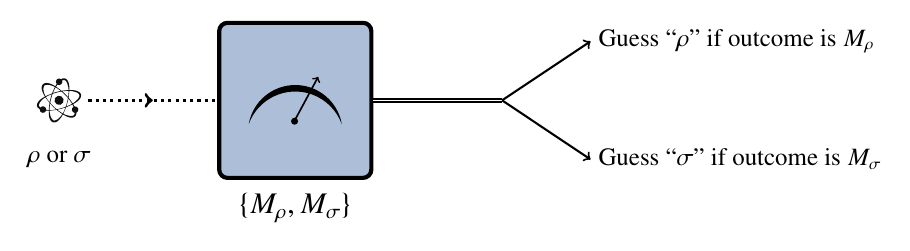}
		\caption{In quantum hypothesis testing, a given quantum system is known to be either in the state $\rho$ or the state $\sigma$. The most general strategy to determine the state of the system consists of measuring it according to a two-outcome POVM $\{M_{\rho},M_{\sigma}\}$. If the outcome corresponding to $M_{\rho}$ occurs, then we guess that the system is in the state $\rho$, and if the outcome corresponding to $M_{\sigma}$ occurs, then we guess that the system is in the state $\sigma$. The goal is to minimize the probability of error of this general strategy.}\label{fig-hypo_testing_single}
	\end{figure}

	To be more specific, consider the following scenario. Bob is given a quantum system by Alice, which is either in the state $\rho$ or in the state $\sigma$, and his task is to determine in which state the system has been prepared. Bob's most general strategy consists of performing a measurement of the system, described by the POVM $\{M_{\rho},M_{\sigma}\}$ (so that $M_{\rho},M_{\sigma}\geq 0$ and $M_{\rho}+M_{\sigma}=\mathbbm{1})$, and then guessing ``$\rho$'' if the outcome corresponds to $M_{\rho}$ and guessing ``$\sigma$'' if the outcome corresponds to $M_{\sigma}$; see Figure~\ref{fig-hypo_testing_single}. Of course, Bob's guess might not always be correct, and there are two types of errors that can occur:
	\begin{enumerate}
		\item\textit{Type-I Error}: Bob guesses ``$\sigma$'', but the system is in the state $\rho$. The probability of this occurring is $\Tr[M_{\sigma}\rho]$.
		\item\textit{Type-II Error}: Bob guesses ``$\rho$'', but the system is in the state $\sigma$. The probability of this occurring is $\Tr[M_{\rho}\sigma]$.
	\end{enumerate}
	
	In order to obtain an optimal strategy for Bob, there are two cases that are typically considered.
	\begin{itemize}
		\item \textit{Symmetric Case}: Also called \textit{quantum state discrimination}; in this setting, Bob has some prior knowledge about the state he is given. Specifically, he knows that the state is $\rho$ with probability $\lambda\in [0,1]$ and $\sigma$ with probability $1-\lambda$. The goal is then to minimize the overall probability of error, that is, the average of the type-I and type-II error probabilities with respect to this probability distribution: 
			\begin{equation}\label{eq-hypo_testing_state_symm_err_prob}
				p_{\text{err}}(\lambda,\rho,\sigma,M)\coloneqq \lambda\Tr[(\mathbbm{1}-M)\rho]+(1-\lambda)\Tr[M\sigma],
			\end{equation}
			where $M\equiv M_{\rho}$. The optimization problem we are interested in is thus:
			\begin{equation}\label{eq-hypo_test_symm_opt_prob}
				\begin{array}{l l}
					\text{minimize} & p_{\text{err}}(\lambda,\rho,\sigma,M) \\ 
					\text{subject to} & 0\leq M\leq\mathbbm{1},
				\end{array}
			\end{equation}
			where the minimization is with respect to operators $0\leq M\leq\mathbbm{1}$ representing the two-outcome POVM $\{M,\mathbbm{1}-M\}$. We discuss symmetric hypothesis testing, and the optimization problem in \eqref{eq-hypo_test_symm_opt_prob}, in detail in Section~\ref{subsec-state_discrimination}.
			
		\item \textit{Asymmetric Case}: In this setting, the goal is to minimize the type-II error probability, given an upper bound on the type-I error probability:
			\begin{equation}\label{eq-hypo_test_asymm_opt_prob}
				\begin{array}{l l}
					\text{minimize} & \Tr[M\sigma] \\
					\text{subject to} & \Tr[(\mathbbm{1}-M)\rho]\leq\varepsilon,\\
					& 0\leq M\leq\mathbbm{1},
				\end{array}
			\end{equation}
			where $\varepsilon\in[0,1]$ is the upper bound on the type-I error probability, and the optimization is with respect to every operator $M\equiv M_\rho$ satisfying $0\leq M\leq\mathbbm{1}$, representing the two-outcome POVM $\{M,\mathbbm{1}-M\}$. We discuss asymmetric hypothesis testing, and the optimization problem in \eqref{eq-hypo_test_asymm_opt_prob}, in detail in Section~\ref{sec-hypo_testing_states_asym}.

	\end{itemize}
	
	\begin{exercise}{exer-hypo_testing_states_simple}
		Consider a very simple hypothesis testing strategy in which Bob discards the state of the quantum system and simply guesses ``$\rho$'' with some probability $q\in[0,1]$ and ``$\sigma$'' with probability $1-q$.
		\begin{enumerate}[topsep=0.3cm]
			\item What is the POVM corresponding to this strategy?
			\item Evaluate the type-I and type-II error probabilities for this strategy.
			\item If, in the symmetric setting, the prior probability for the state $\rho$ is $\lambda\in[0,1]$, then evaluate the error probability in \eqref{eq-hypo_testing_state_symm_err_prob} for this strategy.
		\end{enumerate}
	\end{exercise}
	
	\begin{figure}
		\centering
		\includegraphics[scale=0.8]{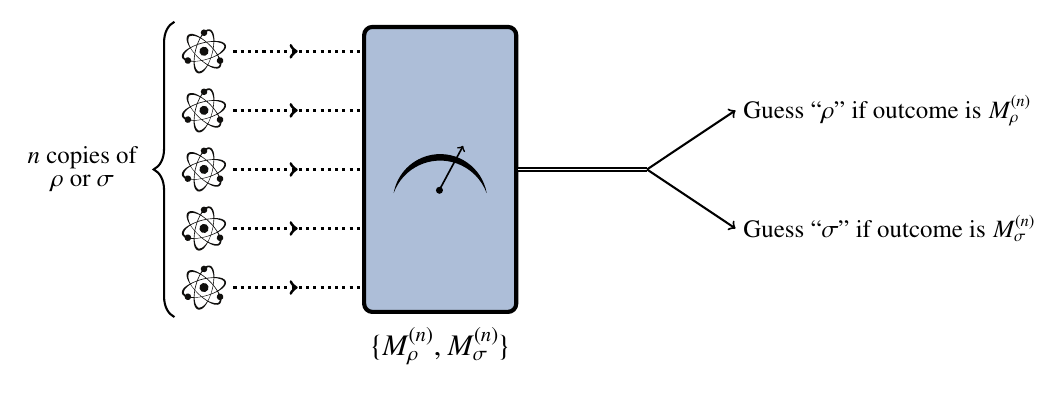}
		\caption{Quantum hypothesis testing with $n\geq 1$ copies of the state. As in the case of one copy shown in Figure~\ref{fig-hypo_testing_single} ($n=1$), the most general decision strategy in this case consists of a measurement of all $n$ copies of the system according to a POVM $\big\{M_{\rho}^{(n)},M_{\sigma}^{(n)}\big\}$. If the outcome corresponding to $M_{\rho}^{(n)}$ occurs, then we guess that each copy of the system is in the state $\rho$; if the outcome corresponding to $M_{\sigma}^{(n)}$ occurs, then we guess that each copy of the system is in the state $\sigma$.}\label{fig-hypo_testing_asymp}
	\end{figure}
	
	Now, suppose that Bob is given several copies, say $n\geq 1$, of a quantum system, each one of which is either in the state $\rho$ or the state $\sigma$. His strategy to determine the state can now make use of these multiple copies in an adaptive manner, for example, and could allow the error probabilities to go below the ``single-shot'' ($n=1$) error probabilities defined above. Since Bob ultimately has to make a decision between $\rho$ and $\sigma$, his strategy is still described by a two-outcome POVM, which we denote by $\big\{M_{\rho}^{(n)},M_{\sigma}^{(n)}\big\}$. This setting of hypothesis testing with multiple copies is depicted in Figure~\ref{fig-hypo_testing_asymp}. The type-I and type-II error probabilities are defined in an analogous manner as before. Specifically, the type-I error is $\Tr\big[M_{\sigma}^{(n)}\rho^{\otimes n}\big]$ and the type-II error is $\Tr\big[M_{\rho}^{(n)}\sigma^{\otimes n}\big]$. In the symmetric case, if $\lambda\in[0,1]$ is the probability that each system is in the state $\rho$, then the error probability is
	\begin{align}
		&\lambda\Tr\big[M_{\sigma}^{(n)}\rho^{\otimes n}\big]+(1-\lambda)\Tr\big[M_{\rho}^{(n)}\sigma^{\otimes n}\big]\nonumber\\
		&\quad=\lambda\Tr\big[\big(\mathbbm{1}^{\otimes n}-M^{(n)}\big)\rho^{\otimes n}\big]+(1-\lambda)\Tr\big[M^{(n)}\sigma^{\otimes n}\big]\\
		&\quad=p_{\text{err}}\big(\lambda,\rho^{\otimes n},\sigma^{\otimes n},M^{(n)}\big),
	\end{align}
	where we have let $M_{\rho}^{(n)}\equiv M^{(n)}$.

	\begin{exercise}{exer-hypo_testing_mult_copies}
		Consider states $\rho$ and $\sigma$ along with a POVM $\{M_0,M_1\}$ representing a strategy for hypothesis testing of a single copy of the quantum system, where the outcome ``0'' corresponds to $\rho$ and the outcome ``1'' corresponds to $\sigma$. Let $\lambda\in[0,1]$ be the prior probability for $\rho$, and let $n\geq 2$. Construct the POVM $\{M_{\rho}^{(n)},M_{\sigma}^{(n)}\}$, and evaluate the type-I and type-II error probabilities for the following two decision strategies for hypothesis testing of $n$ copies of the quantum system.
		\begin{enumerate}[topsep=0.3cm]
			\item The \textit{majority-vote} decision strategy: (1) Measure each system according to the POVM $\{M_0,M_1\}$, and let $N_x$ be the number of times the outcome $x$ occurs. (2) If $N_0>N_1$, guess ``$\rho$'', and if $N_1>N_0$, guess ``$\sigma$''. If $n$ is even and $N_0=N_1$, then guess ``$\rho$'' with probability $q\in[0,1]$ and ``$\sigma$'' with probability $1-q$.
				
			\item The \textit{unanimous-vote} decision strategy: (1) Measure each system according to the POVM $\{M_0,M_1\}$, and let $N_x$ be the number of times the outcome $x$ occurs. (2) If $N_0=n$, then guess ``$\rho$''; otherwise, guess ``$\sigma$''.
		\end{enumerate}
	\end{exercise}

\subsection{Symmetric Case (State Discrimination)}\label{subsec-state_discrimination}

	Given quantum states $\rho$ and $\sigma$, the goal of symmetric hypothesis testing, also known as quantum state discrimination, is to devise a measurement strategy that minimizes the error probability defined in \eqref{eq-hypo_testing_state_symm_err_prob}, where $\lambda\in[0,1]$ is the probability that the state is $\rho$ and $1-\lambda$ is the probability that the state is $\sigma$. The value of the corresponding optimization problem in \eqref{eq-hypo_test_symm_opt_prob} is
	\begin{equation}\label{eq-state_disc_opt_error_def}
	 	p_{\text{err}}^*(\lambda,\rho,\sigma)\coloneqq\inf_{M:0\leq M\leq\mathbbm{1}}\left\{\Tr[(\mathbbm{1}-M)(\lambda\rho)]+\Tr[M(1-\lambda)\sigma]\right\}.
	\end{equation}
	
	\begin{exercise}{exer-hypo_test_states_symm_SDP}
		Show that $p_{\text{err}}^*(\lambda,\rho,\sigma)$ can be evaluated using a semi-definite program. Then, using strong duality, prove that an alternate expression for $p_{\text{err}}^*(\lambda,\rho,\sigma)$ is
		\begin{equation}\label{eq-state_disc_opt_error_dual}
			p_{\text{err}}^*(\lambda,\rho,\sigma)=\sup_{W\text{ Hermitian}}\left\{\Tr[W]:W\leq\lambda\rho,\,W\leq(1-\lambda)\sigma\right\}.
		\end{equation}
		Finally, evaluate the complementary slackness conditions from Proposition~\ref{prop:math-tools:comp-slack}. An optimal operator $W$ is known as the ``greatest lower bound operator''.
	\end{exercise}
	
	\begin{exercise}{exer-hypo_test_states_symm_iso_invar}
		Prove that $p_{\text{err}}^*(\lambda,\rho,\sigma)$ is isometrically invariant: for every isometry $V$, $p_{\text{err}}^*(\lambda,\rho,\sigma)=p_{\text{err}}^*(\lambda,V\rho V^{\dagger},V\sigma V^{\dagger})$.
	\end{exercise}
	
	More generally than isometric invariance, the following \textit{data-processing inequality} holds for the error probability for discriminating two quantum states. The intuition behind the proof of Proposition~\ref{prop-hypo_test_states_symm_data_proc} is as follows: Suppose that we are given a quantum system in an unknown state. Before applying a measurement to determine the state, we could perform a quantum channel $\mathcal{N}$. However, if we do so, this strategy is not necessarily an optimal strategy, and the error probability is never smaller than if we simply apply an optimal measurement to distinguish the states.
	
	\begin{proposition*}{Data-Processing Inequality for State Discrimination}{prop-hypo_test_states_symm_data_proc}
		Consider states $\rho$ and $\sigma$, $\lambda\in[0,1]$, and let $\mathcal{N}$ be a positive and trace preserving superoperator. Then,
		\begin{equation}
			p_{\text{err}}^*(\lambda,\rho,\sigma)\leq p_{\text{err}}^*(\lambda,\mathcal{N}(\rho),\mathcal{N}(\sigma))
		\end{equation}
	\end{proposition*}
	
	\begin{Proof}
		Let $M'$ be an operator satisfying $0\leq M'\leq\mathbbm{1}$, and consider the operator $\mathcal{N}^{\dagger}(M')$ (this is the measurement operator corresponding to performing the channel~$\mathcal{N}$ first and then applying the measurement operator $M'$). Due to the positivity of $\mathcal{N}$, and thus of $\mathcal{N}^{\dagger}$, we have $\mathcal{N}^{\dagger}(M')\geq 0$. The condition $M'\leq\mathbbm{1}$ implies $\mathbbm{1}-M'\geq 0$. Thus, by the positivity of $\mathcal{N}^{\dagger}$, it follows that $\mathcal{N}^{\dagger}(\mathbbm{1}-M')\geq 0$, which implies $\mathcal{N}^{\dagger}(M')\leq\mathcal{N}^{\dagger}(\mathbbm{1})$. Now, $\mathcal{N}^{\dagger}$ is unital, because $\mathcal{N}$ is trace preserving (see Exercise~\ref{exer-adjoint_unital}), which means that $\mathcal{N}^{\dagger}(\mathbbm{1})=\mathbbm{1}$. We thus conclude that $0\leq\mathcal{N}^{\dagger}(M')\leq\mathbbm{1}$. Therefore, $\mathcal{N}^{\dagger}(M')$ is a measurement operator and thus a feasible point in the optimization problem for $p_{\text{err}}^*(\lambda,\rho,\sigma)$, so that
		\begin{align}
			p_{\text{err}}^*(\lambda,\rho,\sigma)&=\inf_{M:0\leq M\leq\mathbbm{1}}\{\Tr[(\mathbbm{1}-M)(\lambda\rho)]+\Tr[M(1-\lambda)\sigma]\}\\
			&\leq \Tr[(\mathbbm{1}-\mathcal{N}^{\dagger}(M'))(\lambda\rho)]+\Tr[\mathcal{N}^{\dagger}(M')(1-\lambda)\sigma]\\
			&=\Tr[(\mathbbm{1}-M')(\lambda\mathcal{N}(\rho))]+\Tr[M'(1-\lambda)\mathcal{N}(\sigma)],
		\end{align}
		where the last line follows from the definition of the adjoint of a superoperator. 
        Finally, because the inequality
		\begin{equation}
			p_{\text{err}}^*(\lambda,\rho,\sigma)\leq \Tr[(\mathbbm{1}-M')(\lambda\mathcal{N}(\rho))]+\Tr[M'(1-\lambda)\mathcal{N}(\sigma)]
		\end{equation}
		holds for all $M'$ satisfying $0\leq M'\leq\mathbbm{1}$, we conclude that
		\begin{align}
			p_{\text{err}}^*(\lambda,\rho,\sigma)&\leq\inf_{M':0\leq M'\leq\mathbbm{1}}\Tr[(\mathbbm{1}-M')(\lambda\mathcal{N}(\rho))]+\Tr[M'(1-\lambda)\mathcal{N}(\sigma)]\\
			&=p_{\text{err}}^*(\lambda,\mathcal{N}(\rho),\mathcal{N}(\sigma)),
		\end{align}
		as required.
	\end{Proof}
	
	It turns out that $p_{\text{err}}^*(\lambda,\rho,\sigma)$ can be written in terms of the trace norm (Section~\ref{sec-math_tools_trace_norm}) as
	\begin{equation}\label{eq-state_disc_opt_error}
		p_{\text{err}}^*(\lambda,\rho,\sigma)=\frac{1}{2}\left(1-\norm{\lambda\rho-(1-\lambda)\sigma}_1\right),
	\end{equation}
	which is an immediate consequence of the following theorem.
	 
	\begin{theorem*}{Helstrom--Holevo Theorem}{thm-Holevo_Helstrom}
		For all positive semi-definite operators $A$ and $B$, 
		\begin{equation}\label{eq-Holevo_Helstrom}
			\inf_{M:0\leq M\leq\mathbbm{1}}\left\{\Tr[(\mathbbm{1}-M)A]+\Tr[MB]\right\}=\frac{1}{2}\left(\Tr[A+B]-\norm{A-B}_1\right).
		\end{equation}
		A measurement operator $M$ is optimal if and only if $M = \Pi_+ + \Lambda_0$, where $\Pi_+$ is the projection onto the strictly  positive part of $A-B$, the operator $\Pi_0$ is the projection onto the zero eigenspace of $A-B$, and $0 \leq \Lambda_0 \leq \Pi_0$. Equivalently, 
		\begin{equation}\label{eq-Holevo_Helstrom-alt}
			\sup_{M:0\leq M\leq\mathbbm{1}}\Tr[M(A-B)] = \frac{1}{2}\left(\Tr[A-B] + \norm{A-B}_1\right),
		\end{equation}
		and the conditions for an optimal $M$ are the same as given above.
	\end{theorem*}
	
	\begin{remark}
		Letting $A=\lambda\rho$ and $B=(1-\lambda)\sigma$ in the statement of Theorem~\ref{thm-Holevo_Helstrom}, we recognize that the objective function on the left-hand side of \eqref{eq-Holevo_Helstrom} is equal to $p_{\text{err}}(\lambda,\rho,\sigma,M)$ as defined in \eqref{eq-hypo_testing_state_symm_err_prob}. We thus obtain \eqref{eq-state_disc_opt_error}. Note that Theorem~\ref{thm-Holevo_Helstrom} also gives us a measurement that achieves the minimal error probability.
	\end{remark}
	
	\begin{Proof}
		Let $M$ be an arbitrary operator satisfying $0\leq M\leq\mathbbm{1}$. Let $\Delta \coloneqq A-B$ and let $\Delta_+$ and $\Delta_-$ be the positive and negative parts, respectively, of $\Delta$, so that $A-B=\Delta_+-\Delta_-$ and $\Delta_+\Delta_-=0$ (recall \eqref{eq-operator_pos_neg_decomp}). We can then write the objective function in \eqref{eq-Holevo_Helstrom} as
		\begin{equation}
			\Tr[(\mathbbm{1}-M)A]+\Tr[MB] = \Tr[A]-\left(\Tr[M\Delta_+]-\Tr[M\Delta_-]\right).
			\label{eq-Holevo_Helstrom-rewrite}
		\end{equation}
		Now, since $\Tr[M\Delta_-]\geq 0$, on account of both $M$ and $\Delta_-$ being positive semi-definite, we find that
		\begin{align}
			\Tr[M\Delta_+]-\Tr[M\Delta_-] \leq \Tr[M\Delta_+] \leq \Tr[\Delta_+],
			\label{eq-up-bnd-trace-norm-1}
		\end{align}
		where  the last inequality follows because $M\leq\mathbbm{1}$. The equality in~\eqref{eq-Holevo_Helstrom-rewrite} and the inequality in~\eqref{eq-up-bnd-trace-norm-1} imply that
		\begin{equation}
		    \Tr[(\mathbbm{1}-M)A]+\Tr[MB] \geq \Tr[A] - \Tr[\Delta_+].
		\end{equation}
		Since
		\begin{equation}
			\norm{A-B}_1=\Tr[|A-B|]=\Tr[\Delta_+]+\Tr[\Delta_-]
		\end{equation}
		and
		\begin{equation}
			\Delta_-=\Delta_++B-A,
		\end{equation}
		we can write $\Tr[\Delta_+]$ as
		\begin{equation}\label{eq-state_disc_pf}
			\Tr[\Delta_+]=\frac{1}{2}\left(\norm{A-B}_1-\Tr[B-A]\right).
		\end{equation}
		This means that the objective function on the left-hand side of \eqref{eq-Holevo_Helstrom-rewrite} can be bounded from below by $\frac{1}{2}\left(\Tr[A\allowbreak+B]-\norm{A-B}_1\right)$. We have thus shown that
		\begin{equation}\label{eq-state_disc_pf2}
			\Tr[(\mathbbm{1}-M)A]+\Tr[MB]\geq \frac{1}{2}\left(\Tr[A+B]-\norm{A-B}_1\right).
		\end{equation}
		for all $M$ such that $0\leq M\leq \mathbbm{1}$, which implies that
		\begin{equation}
			\inf_{M:0\leq M\leq\mathbbm{1}}\{\Tr[(\mathbbm{1}-M)A]+\Tr[MB]\}
			\geq \frac{1}{2}\left(\Tr[A+B]-\norm{A-B}_1\right).
			\end{equation}
		
		To see the reverse inequality, let $M = \Pi_+ + \Lambda_0$, where $\Pi_+$ is the projection onto the strictly positive part of $A-B$ and $\Lambda_0$ satisfies $0 \leq \Lambda_0 \leq \Pi_0$, with $\Pi_0$ the projection onto the zero eigenspace of $A-B$. Then,
		\begin{align}
			\Tr[M(A-B)]&=\Tr[(\Pi_+ + \Lambda_0)(\Delta_+-\Delta_-)] \label{eq-up-bnd-trace-norm-2}\\
			&=\Tr[(\Pi_+ + \Lambda_0)\Delta_+]-\Tr[(\Pi_+ + \Lambda_0)\Delta_-]\\
			&=\Tr[\Delta_+] \label{eq-up-bnd-trace-norm-3},
		\end{align}
		where  the last equality follows because $\Tr[\Pi_+\Delta_+]=\Tr[\Delta_+]$ and $\Tr[\Pi_+\Delta_-]=0$, since $\Pi_+$ and $\Delta_-$ are by definition orthogonal. We also used $\Tr[\Lambda_0 \Delta_+] = \Tr[\Lambda_0 \Delta_-]=0$, with these latter equalities following because $0 \leq \Tr[\Lambda_0 \Delta_{\pm}] \leq \Tr[\Pi_0 \Delta_{\pm}] = 0$. Therefore, using \eqref{eq-state_disc_pf}, we find that
		\begin{equation}
			\Tr[A]-\Tr[(\Pi_+ + \Lambda_0)(A-B)]=\frac{1}{2}\left(\Tr[A+B]-\norm{A-B}_1\right).
		\end{equation}
		The operator $\Pi_+ + \Lambda_0$ thus achieves the bound in \eqref{eq-state_disc_pf2}, which means that
		\begin{equation}
			\inf_{M:0\leq M\leq \mathbbm{1}} \{\Tr[(\mathbbm{1}-M)A]+\Tr[MB]\} = \frac{1}{2}\left(\Tr[A+B]-\norm{A-B}_1\right),
		\end{equation}
		so that \eqref{eq-Holevo_Helstrom} is established.
		
		To see that $\Pi_+ + \Lambda_0$ is the only form for an optimal measurement operator, suppose that $M$ is optimal, i.e., satisfies $0 \leq M \leq \mathbbm{1}$ and saturates \eqref{eq-Holevo_Helstrom} with equality. Then it follows that the two inequalities in \eqref{eq-up-bnd-trace-norm-1} are saturated with equality, so that
		\begin{equation}
		 \Tr[M(\Delta_+ - \Delta_-)] = \Tr[M \Delta_+] = \Tr[ \Delta_+]= \Tr[\Pi_+ \Delta_+].
		\end{equation}
		On the one hand, the leftmost equality implies that $\Tr[M \Delta_-] = 0$, where $\Delta_-$ is the  strictly negative part of $A-B$. Since both $M$ and $\Delta_-$ are positive semi-definite and $\Pi_-$ is the projection onto the strictly negative part of $\Delta_-$, we conclude that $M \Pi_- =  \Pi_- M = 0$. This in turn implies that
		\begin{equation}
		M (\Pi_+ + \Pi_0)= (\Pi_+ + \Pi_0) M = M,
		\end{equation}
		which, after sandwiching $M \leq \mathbbm{1}$ on the left and right by $\Pi_+ + \Pi_0$, implies that $M \leq \Pi_+ + \Pi_0$.
		On the other hand, since
		\begin{align}
		0 & = \Tr[\Delta_+ (\Pi_+  - M)] \\
		& =
		\Tr[\Delta_+ (\Pi_+  - \Pi_+ M \Pi_+ )] \\
		& =\Tr[\Delta_+ \Pi_+ (\mathbbm{1}  -  M  )\Pi_+],
		\end{align}
		we find that $\Pi_+ (\mathbbm{1} - M)\Pi_+ = 0$. Now consider that $\Pi_- (\mathbbm{1} - M)\Pi_+ = 0$ because $\Pi_- \Pi_+ =0 $ and $\Pi_- M  \Pi_+ = \Pi_- (\Pi_+ + \Pi_0) M  \Pi_+ = 0$. So then $\Pi_+ (\mathbbm{1} - M)\Pi_+ = 0$ and $\Pi_- (\mathbbm{1} - M)\Pi_+ = 0$ imply that $(\mathbbm{1} - M)\Pi_+ = 0$. From this equation, we conclude that $\Pi_+ = \Pi_+ M = M \Pi_+$. By sandwiching $\Pi_+ \leq \mathbbm{1}$ by $M$ and remembering that $M^2\leq M$ because $0\leq M \leq \mathbbm{1}$, 
        we conclude that $\Pi_+ \leq M$. Combining this operator inequality with the previous one, we conclude that an optimal $M$ satisfies $\Pi_+ \leq M \leq \Pi_+ + \Pi_0$, which is equivalent to $M$ decomposing as $M = \Pi_+ + \Lambda_0$ for $0 \leq \Lambda_0 \leq \Pi_0$.
		
		The equality in \eqref{eq-Holevo_Helstrom-alt} follows as
		\begin{align}
		\frac{1}{2}\norm{A-B}_1 & = \frac{1}{2}\Tr[A+B] - \inf_{M:0\leq M\leq\mathbbm{1}}\Tr[(\mathbbm{1}-M)A]+\Tr[MB] \\
		&  = \sup_{M:0\leq M\leq\mathbbm{1}}\frac{1}{2}\Tr[A+B] - (\Tr[(\mathbbm{1}-M)A]+\Tr[MB]) \\
		&  = \sup_{M:0\leq M\leq\mathbbm{1}}\frac{1}{2}\Tr[B-A] + \Tr[M(A-B)] \\
		&  = \frac{1}{2}\Tr[B-A] + \sup_{M:0\leq M\leq\mathbbm{1}} \Tr[M(A-B)].
		\end{align}
Rearranging this equality, we arrive at \eqref{eq-Holevo_Helstrom-alt}.
		An optimal $M$ having the form $\Pi_+ + \Lambda_0$  again follows from \eqref{eq-up-bnd-trace-norm-1}, \eqref{eq-up-bnd-trace-norm-2}--\eqref{eq-up-bnd-trace-norm-3}, and the reasoning given above. 
	\end{Proof}
	
	\begin{exercise}{exer-hypo_test_symm_pure_states}
		Let $\rho=\ket{\psi}\!\bra{\psi}\equiv\psi$ and $\sigma=\ket{\phi}\!\bra{\phi}\equiv\phi$ be pure states, and let $\lambda\in[0,1]$. Show that
		\begin{equation}\label{eq-p_err_opt_pure_states}
			p_{\text{err}}^*(\lambda,\psi,\phi)=\frac{1}{2}\left(1-\sqrt{\smash[b]{1-4\lambda(1-\lambda)\abs{\braket{\psi}{\phi}}^{2}}}\right).
		\end{equation}
		What is a measurement that achieves this optimal error probability?
	\end{exercise}
	
	Observe from \eqref{eq-p_err_opt_pure_states} that if $\ket{\psi}$ and $\ket{\phi}$ are two orthogonal pure state vectors, then $p_{\text{err}}^*(\lambda,\psi,\phi)=0$.
	
	\begin{exercise}{exer-hypo_test_states_symm_opt_meas_orthog}
		Let $\rho$ and $\sigma$ be quantum states that are orthogonal, in the sense that $\Pi_{\rho}\Pi_{\sigma}=\Pi_{\sigma}\Pi_{\rho}=0$, where $\Pi_{\rho}$ and $\Pi_{\sigma}$ are the projections onto the support of $\rho$ and $\sigma$, respectively (recall \eqref{eq-proj_support}). Prove that the optimal error probability for discriminating $\rho$ and $\sigma$ vanishes, i.e., that $p_{\text{err}}^*(\lambda,\rho,\sigma)=0$. What is a measurement achieving this optimal error probability?
	\end{exercise}
	
	\begin{exercise}{exer-hypo_test_states_symm_opt_werner_iso}
		Evaluate the optimal error probability $p_{\text{err}}^*\!\left(\lambda,\rho_{AB}^{\text{iso};p_1},\rho_{AB}^{\text{W};p_2}\right)$ for discriminating between the isotropic state $\rho_{AB}^{\text{iso};p_1}$, $p_1\in[0,1]$, and the Werner state $\rho_{AB}^{\text{W};p_2}$, $p_2\in[0,1]$, where $\lambda\in[0,1]$.
	\end{exercise}
	
	The Helstrom--Holevo theorem gives us the lowest possible error probability in distinguishing between two states $\rho$ and $\sigma$, given just one copy of either state. Suppose now that, instead of just one copy, Alice sends Bob $n$~copies of either $\rho$ or $\sigma$. Bob's task is then to discriminate between the states $\rho^{\otimes n}$ and $\sigma^{\otimes n}$. The Helstrom--Holevo theorem still applies in this case, so that
	\begin{equation}
		p_{\text{err}}^*(\lambda,\rho^{\otimes n},\sigma^{\otimes n})=\frac{1}{2}\left(1-\norm{\lambda\rho^{\otimes n}-(1-\lambda)\sigma^{\otimes n}}_1\right)
	\end{equation}
	is the lowest possible error probability. However, because Bob now has $n$ copies of either $\rho$ or $\sigma$, he can perform a  discrimination strategy that involves a collective measurement acting on the $n$ copies of the state. This means that the optimal \textit{error exponent} can generally be lower with $n\geq 2$ copies than with just one copy. 
	
	\begin{exercise}{exer-hypo_test_states_symm_err_decay}
		Prove that the optimal error probability $p_{\text{err}}^*(\lambda,\rho,\sigma)$ for quantum state discrimination is monotonically non-increasing with $n$, i.e., prove that
		\begin{equation}\label{eq-hypo_test_states_symm_err_decay}
			p_{\text{err}}^*\big(\lambda,\rho^{\otimes (n+1)},\sigma^{\otimes (n+1)}\big) \leq p_{\text{err}}^*(\lambda,\rho^{\otimes n},\sigma^{\otimes n})
		\end{equation}
		for all $n\geq 1$.
	\end{exercise}

\subsubsection{Asymptotic Setting}\label{sec-hypo_testing_state_symm_asymp}
	
	
	Given states $\rho$ and $\sigma$ and $\lambda\in(0,1)$, how does the optimal error probability $p_{\text{err}}(\lambda,\rho^{\otimes n},\sigma^{\otimes n})$ behave as the number $n$ of copies of the state increases? If $\rho\equiv\psi$ and $\sigma\equiv\phi$ are pure states, then because $\psi^{\otimes n}$ are $\phi^{\otimes n}$ are both pure states, we can use \eqref{eq-p_err_opt_pure_states} and the expansion $ \frac{1}{2}\left(1-\sqrt{1-4x}\right) = x + O(x^2)$ to see that the following approximation holds as $n$ becomes large:
	\begin{equation}
		p_{\text{err}}^*(\lambda,\psi^{\otimes n},\phi^{\otimes n})\approx \lambda(1-\lambda)\abs{\braket{\psi}{\phi}}^{2n}=\lambda(1-\lambda)2^{-n(-\log_2\abs{\braket{\psi}{\phi}}^2)}.
	\end{equation}
	Now, because $\abs{\braket{\psi}{\phi}}^2\in[0,1]$, we have that $-\log_2\abs{\braket{\phi}{\psi}}^2\geq 0$, which means that, as $n$ becomes large, the optimal error probability decays exponentially to zero. Does the exponential decay hold more generally? In other words, for arbitrary states $\rho$ and $\sigma$, is it true that $p_{\text{err}}^*(\lambda,\rho^{\otimes n},\sigma^{\otimes n})\approx 2^{-n\xi(\lambda,\rho,\sigma)}$ as $n$ becomes large, where $\xi(\lambda,\rho,\sigma)=-\frac{1}{n}\log_2 p_{\text{err}}^*(\lambda,\rho^{\otimes n},\sigma^{\otimes n})$ is a non-negative asymptotic \textit{error exponent} that is independent of $n$? To be more precise, does the limit $\lim_{n\to\infty}-\frac{1}{n}\log_2 p_{\text{err}}^*(\lambda,\rho^{\otimes n},\sigma^{\otimes n})$ exist, and if so, what is its value?
	
	The following theorem provides positive answers to both questions. The characterization given below is useful because the quantity $p_{\text{err}}^*(\lambda,\rho^{\otimes n},\sigma^{\otimes n})$ becomes more and more difficult to calculate as $n$ increases, so that the asymptotic error exponent is a helpful characterization of $p_{\text{err}}^*(\lambda,\rho^{\otimes n},\sigma^{\otimes n})$.
	
		
	\begin{theorem*}{Quantum Chernoff Bound}{thm-quantum_Chernoff}
		For all quantum states $\rho$ and $\sigma$, and $\lambda\in(0,1)$, the following limit exists and is equal to the the \textit{quantum Chernoff divergence} of $\rho$ and $\sigma$:
		\begin{equation}\label{eq-quantum_Chernoff_bound}
			\lim_{n\to\infty}-\frac{1}{n}\log_2 p_{\text{err}}^*(\lambda,\rho^{\otimes n},\sigma^{\otimes n})=C(\rho\Vert\sigma),
		\end{equation}
		where
		\begin{equation}\label{eq:qm:chern-divergence}
			C(\rho\Vert\sigma) \coloneqq \sup_{s\in(0,1)} \left( -\log_2\Tr[\rho^s\sigma^{1-s}]\right).
		\end{equation}
		 That is, $C(\rho\Vert\sigma)$ is the optimal asymptotic error exponent for symmetric hypothesis testing of $\rho$ and $\sigma$.
	\end{theorem*}
	
	
	\begin{remark}
		Theorem~\ref{thm-quantum_Chernoff} tells us that, as $n$ becomes large, the following approximation holds
		\begin{equation}
			p_{\text{err}}^*(\lambda,\rho^{\otimes n},\sigma^{\otimes n}) \approx 2^{-n\,C(\rho\Vert\sigma)},
		\end{equation}
		so that the optimal error probability does indeed decay exponentially with the number $n$ of copies of the state. In particular, the quantum Chernoff divergence is the \textit{optimal asymptotic error exponent} for the exponential decay of the error probability as the number $n$ of copies increases.
		
		Note that the optimal error exponent in \eqref{eq-quantum_Chernoff_bound} is independent of the prior probability distribution. This means that, in the asymptotic limit (i.e., in the limit $n\to\infty$), the prior probability distribution is irrelevant for the optimal error exponent.
	
		We call Theorem~\ref{thm-quantum_Chernoff} the quantum Chernoff bound because it is analogous to the Chernoff bound from (classical) symmetric hypothesis testing, which is the task of discriminating between two hypotheses, each of which has a corresponding probability distribution (please consult the Bibliographic Notes in Section~\ref{sec:qm:bib-notes} for details).
	\end{remark}
	
	An important element of the proof of the quantum Chernoff bound is Lemma~\ref{lemma:spectral-ineq} below, known as ``Audenaert's inequality''.

	\begin{Lemma*}{Audenaert's Inequality}{lemma:spectral-ineq}
		Let $A$ and $B$ be positive semi-definite operators. Then, for all $s\in\left(0,1\right)$,
		\begin{equation} \label{eq:Audenaert_inequality}
			\frac{1}{2}\left(\Tr[A+B]-\norm{A-B}_1\right)  \leq\Tr[A^{s}B^{1-s}].
		\end{equation}
	\end{Lemma*}

Before we prove the above result, it is useful to build some intuition around it. Consider the simple case where $A=a$ and $B=b$ are scalars. Then the left-hand side of~\eqref{eq:Audenaert_inequality} coincides with the minimum of $a$ and $b$, and the inequality simply states that
\begin{equation} \label{eq:scalar_Audenaert_inequality}
\min\{a,b\} \leq a^s b^{1-s} ,
\end{equation}
which becomes clear once one interprets the right-hand side as the weighted geometric mean between $a$ and $b$.

For general operators $A$ and $B$, remembering that $\|X\|_1 = \Tr \abs{X}$ (see Definition~\ref{def-Schatten_norm}), we can formally write
\begin{equation}
\frac{1}{2}\left(\Tr[A+B]-\norm{A-B}_1\right) = \frac12\Tr[A+B - \abs{A-B}] = \Tr [A \wedge B] ,
\end{equation}
where the ``non-commutative minimum'' of $A$ and $B$ is defined as
\begin{equation}
    A \wedge B \coloneqq \frac{A+B - |A-B|}{2}\, .
\end{equation}
Audenaert's inequality~\eqref{eq:Audenaert_inequality} can then be recast as
\begin{equation}
    \Tr [A\wedge B] \leq \Tr [A^s B^{1-s}]\, ,
\end{equation}
which is formally similar to~\eqref{eq:scalar_Audenaert_inequality}, and serves as its operator generalization. We refer the reader who is interested in the properties of the non-commutative minimum to the Bibliographic Notes (Section~\ref{sec-QM_protocols_bib_notes}).

	\paragraph*{Proof of Lemma~\ref{lemma:spectral-ineq}:}
		Let $\Delta\coloneqq A-B$, and let $\Delta_+$ and $\Delta_-$ be the positive and negative parts, respectively, of $\Delta$, so that $\Delta=\Delta_+-\Delta_-$ (recall \eqref{eq-operator_pos_neg_decomp}). Then,
		\begin{equation}
			|\Delta|=|\Delta_+-\Delta_-|=\Delta_++\Delta_-.
		\end{equation}
		Therefore,
		\begin{equation}
			\norm{A-B}_1 = \norm{\Delta}_1=\Tr[|\Delta|]=\Tr[\Delta_+]+\Tr[\Delta_-].
		\end{equation}
		In addition, we write
		\begin{equation}
			A+B=A-B+2B=\Delta_+-\Delta_-+2B,
		\end{equation}
		so that
		\begin{align}
			&\frac{1}{2}\left(\Tr[A+B]-\norm{A-B}_1\right)\nonumber\\
			&=\frac{1}{2}\left(\Tr[\Delta_+]-\Tr[\Delta_-]+2\Tr[B]-\Tr[\Delta_+]-\Tr[\Delta_-]\right)\\
			&=\Tr[B]-\Tr[\Delta_-].
		\end{align}
		So it suffices to prove that the following inequality holds for all $s\in(0,1)$:
		\begin{equation}\label{eq-lem_spec_ineq_pf_0}
			\Tr[B]-\Tr[\Delta_-]\leq\Tr[A^sB^{1-s}].
		\end{equation}
		Using the fact that $\Delta_+\geq 0$, by definition of the positive part of $\Delta$, we obtain
		\begin{equation}\label{eq-lem_spec_ineq_pf_1}
			B+\Delta_+\geq B.
		\end{equation}
		Similarly, using $A-B=\Delta_+-\Delta_-$, we obtain
		\begin{equation}\label{eq-lem_spec_ineq_pf_2}
			A+\Delta_-=B+\Delta_+\geq B.
		\end{equation}
		By the operator monotonicity of $x\mapsto x^s$ for $s\in(0,1)$ (recall Definition~\ref{def-op_conv_conc_mono} and thereafter), the inequality in \eqref{eq-lem_spec_ineq_pf_2} implies that
		\begin{equation}
			B^s\leq (A+\Delta_-)^s.
			\label{eq-QM-fund-prot:key-lemm-chern-proof-1}
		\end{equation}
		Using this, along with the fact that $\Tr[B]=\Tr[B^sB^{1-s}]$, we find that
		\begin{align}
			\Tr[B]-\Tr[A^sB^{1-s}]&=\Tr[(B^s-A^s)B^{1-s}]\\
			&\leq \Tr[((A+\Delta_-)^s-A^s)B^{1-s}]\\
			&\leq \Tr[((A+\Delta_-)^s-A^s)(A+\Delta_-)^{1-s}]\\
			&=\Tr[A]+\Tr[\Delta_-]-\Tr[A^s(A+\Delta_-)^{1-s}] \label{eq-lem_spec_ineq_pf_3}\\
			& \leq \Tr[A]+\Tr[\Delta_-]-\Tr[A] \\
			& = \Tr[\Delta_-].
		\end{align}
		The first inequality follows because $B^{1-s}\geq 0$ and from \eqref{eq-QM-fund-prot:key-lemm-chern-proof-1}. The second inequality follows because $(A+\Delta_{-})^s \geq A^s$ and from \eqref{eq-QM-fund-prot:key-lemm-chern-proof-1} with the substitution $s \to 1-s$. The last inequality follows because
		\begin{equation}
		\Tr[A^s(A+\Delta_-)^{1-s}] \geq \Tr[A^s A^{1-s}] = \Tr[A],
		\end{equation}
		which in turn follows because $A^s \geq 0$ and $(A+\Delta_-)^{1-s} \geq A^{1-s}$.
				Therefore,  we conclude that
		\begin{equation}
			\Tr[B]-\Tr[A^sB^{1-s}]\leq  \Tr[\Delta_-],
		\end{equation}
		establishing the desired inequality in \eqref{eq-lem_spec_ineq_pf_0}. \qed
	
	\begin{exercise}{exer-trace_dist_bound}
		Let $A$ and $B$ be positive semi-definite operators and $s\in(0,1)$. Starting with Lemma~\ref{lemma:spectral-ineq}, prove that
		\begin{equation}\label{eq-trace_dist_bound}
			\frac{1}{2}\norm{A-B}_1\geq\frac{1}{2}\Tr[A+B]-\norm{A^sB^{1-s}}_1.
		\end{equation}
		(\textit{Hint}: Recall Theorem~\ref{prop:math-tools:var-char-t-norm}.)
	\end{exercise}
	
	\begin{remark}
		In the case $s=\frac{1}{2}$, the inequality in \eqref{eq-trace_dist_bound} becomes
		\begin{align}
			\frac{1}{2}\norm{A-B}_1&\geq\frac{1}{2}\Tr[A+B]-\norm{\sqrt{A}\sqrt{B}}_1\\
			&=\frac{1}{2}\Tr[A+B]-\sqrt{F(A,B)},
		\end{align}
		where in the second line we let
		\begin{equation}
			F(A,B)\coloneqq\norm{\sqrt{A}\sqrt{B}}_1^2.
		\end{equation}
		The quantity $F(A,B)$ is called the \textit{fidelity} between $A$ and $B$, and it plays a critical role in the analysis of quantum communication protocols. We study the fidelity function in detail in Chapter~\ref{chap-QM_dist_meas}.
	\end{remark}
		
	We now proceed to the proof of the quantum Chernoff bound.

\paragraph*{Proof of the Quantum Chernoff Bound (Theorem~\ref{thm-quantum_Chernoff}):} Since the limit on the left-hand side of \eqref{eq-quantum_Chernoff_bound} need not exist \textit{a priori}, let us define the following quantities based on the discussion in Section~\ref{sec:math-tools:cont-inf-sup}:
	\begin{align}
		\underline{\xi}(\rho,\sigma) & \coloneqq \liminf_{n\to \infty}-\frac{1}{n}\log_2 p_{\text{err}}^*(\lambda,\rho^{\otimes n},\sigma^{\otimes n}),\label{eq:qm-over:chernoff-def-lower}\\
		\overline{\xi}(\rho,\sigma) & \coloneqq \limsup_{n\to \infty}-\frac{1}{n}\log_2 p_{\text{err}}^*(\lambda,\rho^{\otimes n},\sigma^{\otimes n}).\label{eq:qm-over:chernoff-def-upper}
	\end{align}
	Note that, by definition, we always have $\underline{\xi}(\rho,\sigma)\leq\overline{\xi}(\rho,\sigma)$; see \eqref{eq-liminf_limsup_ineq}. Our goal is to prove that $\underline{\xi}(\rho,\sigma)=\overline{\xi}(\rho,\sigma)=\lim_{n\to\infty}-\frac{1}{n}\log_2 p_{\text{err}}^*(\lambda,\rho^{\otimes n},\sigma^{\otimes n})=C(\rho\Vert\sigma)$.

	Now, if $\lambda$ is the probability with which $\rho$ is chosen, and $1-\lambda$ the probability with which $\sigma$ is chosen, then an application of Lemma~\ref{lemma:spectral-ineq}, with $A=\lambda\rho^{\otimes n}$, $B=(1-\lambda)\sigma^{\otimes n}$, and $s\in(0,1)$, gives the following:
	\begin{align}
		p_{\text{err}}^*(\lambda,\rho^{\otimes n},\sigma^{\otimes n})&=\frac{1}{2}\left(1-\norm{\lambda\rho^{\otimes n}-(1-\lambda)\sigma^{\otimes n}}_1\right)\\
		&\leq \Tr\!\left[\lambda^s(\rho^{\otimes n})^s(1-\lambda)^{1-s}(\sigma^{\otimes n})^{1-s}\right]\\
		&=\lambda^s(1-\lambda)^{1-s}\Tr[(\rho^s)^{\otimes n}(\sigma^{1-s})^{\otimes n}]\\
		&=\lambda^s(1-\lambda)^{1-s}\left(\Tr[\rho^s\sigma^{1-s}]\right)^n\\
		&\leq \left(\Tr[\rho^s\sigma^{1-s}]\right)^n,
	\end{align}
	where  the last line follows from the fact that $\lambda^s(1-\lambda)^{1-s}\leq 1$. By taking a negative logarithm and dividing by $n$, this bound becomes
	\begin{equation}
	-\frac{1}{n}\log_2 p_{\text{err}}^*(\lambda,\rho^{\otimes n},\sigma^{\otimes n}) \geq -\log_2 \Tr[\rho^s\sigma^{1-s}].
	\end{equation}
	Since the bound above holds for all $s\in(0,1)$, we can optimize over $s$, obtaining the following:
	\begin{align}
		\underline{\xi}(\rho,\sigma)&=\liminf_{n\to\infty}-\frac{1}{n}\log_2 p_{\text{err}}^*(\lambda,\rho^{\otimes n},\sigma^{\otimes n})\\
		&\geq \sup_{s\in(0,1)}- \log_2\Tr[\rho^s\sigma^{1-s}]\\
		&=C(\rho\Vert\sigma).\label{eq-quantum_Chernoff_pf}
	\end{align}
	
	For the opposite inequality, we start by observing that it suffices to optimize over projectors when determining the optimal error probability $p_{\text{err}}^*(\lambda,\rho^{\otimes n},\sigma^{\otimes n})$. (This is true because one choice of an optimal measurement is a projective measurement, as shown in the proof of Theorem~\ref{thm-Holevo_Helstrom}, where we can set $\Lambda_0 = 0$.) Next, suppose that $\rho$ and $\sigma$ have the following spectral decompositions:
	\begin{equation}
		\rho=\sum_{i=0}^{d-1} \lambda_i \ket{\psi_i}\!\bra{\psi_i},\quad \sigma=\sum_{j=0}^{d-1} \mu_j \ket{\phi_j}\!\bra{\phi_j},
	\end{equation}
	where $d$ is the dimension of the underlying Hilbert space. Then, for every projection~$\Pi$,
	\begin{align}
		\Tr[(\mathbbm{1}-\Pi)\rho]&=\sum_{i=0}^{d-1} \lambda_i\Tr[(\mathbbm{1}-\Pi)\ket{\psi_i}\!\bra{\psi_i}]\\
		&=\sum_{i=0}^{d-1} \lambda_i\Tr[(\mathbbm{1}-\Pi)\ket{\psi_i}\!\bra{\psi_i}(\mathbbm{1}-\Pi)]\\
		&=\sum_{i,j=0}^{d-1}\lambda_i\Tr[\ket{\phi_j}\!\bra{\phi_j}(\mathbbm{1}-\Pi)\ket{\psi_i}\!\bra{\psi_i}(\mathbbm{1}-\Pi)]\\
		&=\sum_{i,j=0}^{d-1} \lambda_i\abs{\bra{\psi_i}(\mathbbm{1}-\Pi)\ket{\phi_j}}^2,
	\end{align}
	where we have used the fact that $\mathbbm{1}=\sum_{j=0}^{d-1}\ket{\phi_j}\!\bra{\phi_j}$. Similarly, using the fact that $\mathbbm{1}=\sum_{i=0}^{d-1}\ket{\psi_i}\!\bra{\psi_i}$, we have
	\begin{align}
		\Tr[\Pi\sigma]&=\sum_{j=0}^{d-1} \mu_j\Tr[\Pi\ket{\phi_j}\!\bra{\phi_j}]\\
		&=\sum_{j=0}^{d-1}\mu_j\Tr[\Pi\ket{\phi_j}\!\bra{\phi_j}\Pi]\\
		&=\sum_{i,j=0}^{d-1}\mu_j\Tr[\ket{\psi_i}\!\bra{\psi_i}\Pi\ket{\phi_j}\!\bra{\phi_j}\Pi]\\
		&=\sum_{i,j=0}^{d-1}\mu_j\abs{\bra{\psi_i}\Pi\ket{\phi_j}}^2.
	\end{align}
	Then, observing that $\lambda_i\geq\min\{\lambda_i,\mu_j\}$ and $\mu_j\geq\min\{\lambda_i,\mu_j\}$ for all $0\leq i,j\leq d-1$, the error probability under the measurement given by $\Pi$ satisfies
	\begin{align}
		&p_{\text{err}}(\lambda,\rho,\sigma,\Pi)\nonumber\\
		&\quad =\Tr[(\mathbbm{1}-\Pi)(\lambda\rho)]+\Tr[\Pi(1-\lambda)\sigma]\\
		&\quad=\sum_{i,j=0}^{d-1} \left(\lambda\lambda_i\abs{\bra{\psi_i}(\mathbbm{1}-\Pi)\ket{\phi_j}}^2+(1-\lambda)\mu_j\abs{\bra{\psi_i}\Pi\ket{\phi_j}}^2\right)\\
		&\quad \geq \min\{\lambda\lambda_i,(1-\lambda)\mu_j\}\left(\abs{x_{i,j}}^2+\abs{y_{i,j}}^2\right),
	\end{align}
	where we have defined $x_{i,j}\coloneqq\bra{\psi_i}(\mathbbm{1}-\Pi)\ket{\phi_j}$ and $y_{i,j}\coloneqq\bra{\psi_i}\Pi\ket{\phi_j}$ in the last line.  Using the identity $\abs{a}^2+\abs{b}^2\geq\frac{1}{2}\abs{a+b}^2$, which holds for all $a,b\in\mathbb{C}$, we obtain
	\begin{equation}
		p_{\text{err}}(\lambda,\rho,\sigma,\Pi)\geq \frac{1}{2}\sum_{i,j=0}^{d-1} \min\{\lambda\lambda_i,(1-\lambda)\mu_j\}\abs{\braket{\psi_i}{\phi_j}}^2.
	\end{equation}
	Now, let us define two probability distributions, $p,q:\{0,1,\dotsc,d-1\}^2\to[0,1]$ as follows:
	\begin{equation}
		p(i,j)\coloneqq \lambda_i\abs{\braket{\psi_i}{\phi_j}}^2,\quad q(i,j)\coloneqq \mu_j\abs{\braket{\psi_i}{\phi_j}}^2,\quad\forall~0\leq i,j\leq d-1.
	\end{equation}
	It is straightforward to verify that $p$ and $q$ are indeed probability distributions. Then, since the projector $\Pi$ is arbitrary, and it suffices to optimize over projective measurements, as explained above, the following inequality holds
	\begin{equation}
		p_{\text{err}}^*(\lambda,\rho,\sigma)\geq \frac{1}{2}\sum_{i,j=0}^{d-1}\min\{\lambda p(i,j),(1-\lambda)q(i,j)\}.
	\end{equation}
	The expression on the right-hand side of the above inequality is precisely half the optimal error probability for discriminating the two probability distributions $p$ and $q$, with a prior probability of $\lambda$. Indeed, we can see this by an application of Theorem~\ref{thm-Holevo_Helstrom} to the case of commutative $A$ and $B$. To this end, letting $A = \sum_{i=0}^{d-1} a_i \ket{i}\!\bra{i}$ and $B = \sum_{i=0}^{d-1} b_i \ket{i}\!\bra{i}$ where $a_i,b_i\geq 0$ for all $i\in \{0,\dotsc,d-1\}$, it follows that an optimal measurement operator is given by $\Pi_+ = \sum_{i:a_i \geq b_i} \ket{i}\!\bra{i}$, with complement $\Pi_- = \sum_{i:a_i < b_i} \ket{i}\!\bra{i}$; thus,
	\begin{align}
	\inf_{M:0\leq M\leq\mathbbm{1}}\left\{\Tr[(\mathbbm{1}-M)A]+\Tr[MB]\right\} & =
	\Tr[\Pi_- A]+\Tr[\Pi_+ B] \\
	& = \sum_{i : a_i < b_i} a_i + \sum_{i : a_i \geq b_i} b_i \\
	& = \sum_{i=0}^{d-1} \min\{a_i,b_i\}.
	\end{align}
	We denote this optimal error probability by $p_{\text{err}}(\lambda,p,q)$. Therefore,
	\begin{equation}
		p_{\text{err}}^*(\lambda,\rho,\sigma)\geq \frac{1}{2} p_{\text{err}}^*(\lambda,p,q).
	\end{equation}
	Now, for $n\geq 2$, by following the same arguments as above, we obtain
	\begin{equation}\label{eq:qm-c-to-q-chernoff-asymp}
		p_{\text{err}}^*(\lambda,\rho^{\otimes n},\sigma^{\otimes n})\geq \frac{1}{2} p_{\text{err}}^*\left(\lambda,p^{(n)},q^{(n)}\right),
	\end{equation}
	where $p^{(n)},q^{(n)}$ are the $n$-fold i.i.d.~probability distributions corresponding to $p$ and $q$, respectively. Then, by the classical Chernoff bound (please consult the Bibliographic Notes in Section~\ref{sec:qm:bib-notes}), we find that
	\begin{align}
		\overline{\xi}(\rho,\sigma)&=\limsup_{n\to\infty}-\frac{1}{n}\log_2 p_{\text{err}}^*(\lambda,\rho^{\otimes n},\sigma^{\otimes n})\\
		&\leq\limsup_{n\to\infty}-\frac{1}{n}\log_2 p_{\text{err}}^*(\lambda,p^{(n)},q^{(n)})\\
		&=\sup_{s\in(0,1)}-\log_2\sum_{i,j=0}^{d-1} p(i,j)^sq(i,j)^{1-s},
	\end{align}
	where the factor of $\tfrac{1}{2}$ in \eqref{eq:qm-c-to-q-chernoff-asymp} vanishes in the asymptotic limit. 
	Finally, observe that
	\begin{align}
		\sum_{i,j=0}^{d-1} p(i,j)^s q(i,j)^{1-s}&=\sum_{i,j=0}^{d-1} \lambda_i^s\left(\abs{\braket{\psi_i}{\phi_j}}^2\right)^s\mu_j^{1-s}\left(\abs{\braket{\psi_i}{\phi_j}}^2\right)^{1-s}\\
		&=\sum_{i,j=0}^{d-1} \lambda_i^s\mu_j^{1-s}\abs{\braket{\psi_i}{\phi_j}}^2\\
		&=\sum_{i,j=0}^{d-1} \lambda_i^s\mu_j^{1-s}\Tr[\ket{\psi_i}\langle \psi_i| \phi_j \rangle \!\bra{\phi_j}]\\
		&=\Tr\!\left[\left(\sum_{i=0}^{d-1}\lambda_i^s \ket{\psi_i}\!\bra{\psi_i}\right)\left(\sum_{j=0}^{d-1} \mu_j^{1-s}\ket{\phi_j}\!\bra{\phi_j}\right)\right]\\
		&=\Tr[\rho^s\sigma^{1-s}].
	\end{align}
	Therefore,
	\begin{equation}
		\overline{\xi}(\rho,\sigma)\leq \sup_{s\in(0,1)} - \log_2\Tr[\rho^s\sigma^{1-s}]=C(\rho\Vert\sigma),
	\end{equation}
	which, combined with \eqref{eq-quantum_Chernoff_pf} and $\underline{\xi}(\rho,\sigma)\leq\overline{\xi}(\rho,\sigma)$, implies that
	\begin{equation}
		\underline{\xi}(\rho,\sigma) = \overline{\xi}(\rho,\sigma)=\sup_{s\in(0,1)} -\log_2\Tr[\rho^s\sigma^{1-s}]=C(\rho\Vert\sigma),
	\end{equation}
	which is what we set out to prove. \qed

\subsection{Multiple State Discrimination}\label{sec:QM-over:multiple-state-disc}
	
	We now briefly consider state discrimination when there are more than two states. Suppose that Alice prepares a quantum system in a state chosen randomly from a set $\{\rho^x\}_{x\in\mathcal{X}}$ of states. We assume that $\mathcal{X}$ is some finite alphabet with size $|\mathcal{X}|\geq 2$, and that the state $\rho^x$ is chosen with probability $p(x)$, where $p:\mathcal{X}\to[0,1]$ is a probability distribution. Alice sends her chosen state to Bob, whose task is to guess the value of $x$, i.e., which state Alice sent. Bob's knowledge of the system can be described by the ensemble $\{(p(x),\rho_B^x)\}_{x\in\mathcal{X}}$, which has the following associated classical--quantum state:
	\begin{equation}\label{eq-state_disc_cq_state}
		\rho_{X_A B}\coloneqq \sum_{x\in\mathcal{X}}p(x)\ket{x}\!\bra{x}_{X_A}\otimes\rho_B^x,
	\end{equation}
	where $X_A$ is a classical $|\mathcal{X}|$-dimensional register that contains Alice's choice of state. Note that while Bob knows both the prior probability distribution $p$ and the association $x\leftrightarrow\rho^x$ between the labels $x$ and the states $\rho^x$, he does not have access to the register $X_A$. (If Bob did have access to the classical register $X_A$, he could simply measure it in the basis $\{\ket{x}\}_{x\in\mathcal{X}}$ and figure out what state Alice sent.) Therefore, as before, Bob must make a measurement. His strategy is to choose a POVM $\{M_B^x\}_{x\in\mathcal{X}}$ with elements indexed by the elements of~$\mathcal{X}$. If he obtains the outcome corresponding to $x\in\mathcal{X}$, then he guesses that the state sent was $\rho^x$.
	
	The scenario of multiple state discrimination is very similar to the task of classical communication over a quantum channel $\mathcal{N}$ that we consider in Chapter~\ref{chap-classical_capacity}. The classical messages to be sent correspond to the labels $x\in\mathcal{X}$, while the states $\rho^x$ correspond to an encoding of the messages into quantum states, which are then sent through the quantum channel, and $p$ corresponds to the prior probability over the set of messages. The measurement performed to guess the state corresponds to a decoding channel that is applied at the receiving end of the quantum channel in order to determine the message that was sent. The quantity in \eqref{eq-opt_guessing_prob} below, evaluated for the ensemble $\{(p(x),\mathcal{N}(\rho^x))\}_{x \in \mathcal{X}}$, is then the optimal average probability of correctly guessing the message sent, where the optimization is over all POVMs.
	
	Let $\mathcal{M}_{B\to X_B}(\cdot)\coloneqq \sum_{x\in\mathcal{X}}\Tr[M_B^x(\cdot)]\ket{x}\!\bra{x}_{X_B}$ be the measurement channel corresponding to the POVM $\{M_B^x\}_{x\in\mathcal{X}}$, where $X_B$ is a $|\mathcal{X}|$-dimensional classical register. (Recall the definition of a measurement channel from Definition~\ref{def-qc_channel}.) After the measurement, the classical--quantum state in \eqref{eq-state_disc_cq_state} transforms to
	\begin{align}
		\omega_{X_AX_B}&\coloneqq\mathcal{M}_{B\to X_B}(\rho_{X_AB})\\
		&=\sum_{x,x'\in\mathcal{X}}p(x)\Tr[M_B^{x'}\rho_B^x]\ket{x}\!\bra{x}_{X_A}\otimes\ket{x'}\!\bra{x'}_{X_B}.
	\end{align}
	Let
	\begin{equation}
		\Pi_{X_AX_B}^{\text{succ}}\coloneqq\sum_{x\in\mathcal{X}}\ket{x}\!\bra{x}_{X_A}\otimes\ket{x}\!\bra{x}_{X_B},
	\end{equation}
	which is a projector corresponding to the registers $X_A$ and $X_B$ having the same value; this is what we want for state discrimination to be successful. The expected success probability of the strategy given by the POVM $\{M^x\}_{x\in\mathcal{X}}$ is thus
	\begin{align}
		p_{\text{succ}}(\{(p(x),\rho^x)\}_x,\{M^x\}_x)&\coloneqq\Tr[\Pi_{X_AX_B}\omega_{X_AX_B}]\\
		&=\sum_{x\in\mathcal{X}}p(x)\Tr[M^x\rho^x].
	\end{align}
	
	\begin{exercise}{exer-hypo_testing_states_symm_multiple}
		Show that the error probability for discriminating the states in the ensemble $\{(p(x),\rho^x)\}_{x\in\mathcal{X}}$ using the POVM $\{M^x\}_{x\in\mathcal{X}}$ is
		\begin{align}
			p_{\text{err}}(\{(p(x),\rho^x)\}_x,\{M^x\}_x)&\coloneqq 1-p_{\text{succ}}(\{(p(x),\rho^x)\}_x,\{M^x\}_x)\\
			&=\Tr[(\mathbbm{1}_{X_AX_B}-\Pi_{X_AX_B})\omega_{X_AX_B}]\\
			&=\sum_{x\in\mathcal{X}}p(x)\Tr[(\mathbbm{1}-M^x)\rho^x].
		\end{align}
	\end{exercise}

	The optimal success probability for discriminating the states in the ensemble $\{(p(x),\rho^x)\}_{x\in\mathcal{X}}$ is
	\begin{multline}\label{eq-opt_guessing_prob}
		p_{\text{succ}}^*(\{(p(x),\rho^x)\}_{x})\\\coloneqq\sup_{\{M^x\}_x}\left\{\sum_{x\in\mathcal{X}}p(x)\Tr[M^x\rho^x]:0\leq M^x\leq\mathbbm{1}~\forall x,\sum_{x\in\mathcal{X}}M^x=\mathbbm{1}\right\}.
	\end{multline}
	The optimal error probability is then $p_{\text{err}}^*(\{(p(x),\rho^x)\}_x)\coloneqq 1-p_{\text{succ}}^*(\{(p(x),\rho^x)\}_x)$.

	\begin{exercise}{exer-hypo_testing_symm_mult_SDP}
		Given an ensemble $\{(p(x),\rho^x)\}_{x\in\mathcal{X}}$ of quantum states with associated classical--quantum state $\rho_{XB}=\sum_{x\in\mathcal{X}}p(x)\ket{x}\bra{x}_X\otimes\rho_B^x$, show that the optimal success probability $p_{\text{succ}}^*(\{(p(x),\rho^x)\}_x)$ can be evaluated using the following semi-definite program:
		\begin{equation}
			\begin{array}{l l} \text{maximize} & \Tr[M_{XB}\rho_{XB}] \\[0.1cm] \text{subject to} & \Tr_X[M_{XB}]=\mathbbm{1}_B,\\ & M_{XB}\geq 0. \end{array}
		\end{equation}
		In other words, show that
		\begin{equation}
			p^*_{\text{succ}}(\{(p(x),\rho^x)\}_x)= \sup_{ M_{XB}\geq 0 }\left\{\Tr[M_{XB}\rho_{XB}] : \Tr_X[M_{XB}] = \mathbbm{1}_B \right\}.
		\end{equation} 
		Then, using strong duality, prove that
		\begin{align}
			p^*_{\text{succ}}(\{(p(x),\rho^x)\}_x)&=\inf_{Y_B \geq 0} \{\Tr[Y_B] : \mathbbm{1}_X \otimes Y_B \geq \rho_{XB}\},\label{eq:QM-over:multiple-state-disc-dual-SDP}\\
			&=\inf_{Y_B \geq 0} \{\Tr[Y_B] :   Y_B \geq p(x) \rho^x_{B} \ \ \forall x \in \mathcal{X}\}.
		\end{align}
		Finally, evaluate the complementary slackness conditions from Proposition~\ref{prop:math-tools:comp-slack}.
	\end{exercise}
	
	\begin{exercise}{exer-hypo_testing_states_mult_pure_orthog}
		Let $\{\ket{\psi^x}\}_{x\in\mathcal{X}}$ be a finite set of mutually orthogonal vectors, let $\rho^x=\ket{\psi^x}\bra{\psi^x}$ for all $x\in\mathcal{X}$, and consider the ensemble $\{(p(x),\rho^x)\}_{x\in\mathcal{X}}$, where $p:\mathcal{X}\to[0,1]$ is a probability distribution. Prove that $p_{\text{succ}}^*(\{(p(x),\rho^x)\}_x)=1$ and construct the corresponding optimal measurement.
	\end{exercise}
	
	\begin{exercise}{exer-hypo_test_states_symm_mult_data_proc}
		Generalize Proposition~\ref{prop-hypo_test_states_symm_data_proc} to the case of multiple state discrimination. Specifically, for every finite ensemble $\{(p(x),\rho^x)\}_{x\in\mathcal{X}}$ of states and every positive, trace preserving map $\mathcal{N}$, prove that
		\begin{equation}
			p_{\text{succ}}^*(\{(p(x),\rho^x)\}_x)\geq p_{\text{succ}}^*(\{(p(x),\mathcal{N}(\rho^x))\}_x).
		\end{equation}
	\end{exercise}

\subsection{Asymmetric Case}\label{sec-hypo_testing_states_asym}

	False positives and false negatives in medical testing are not, in general, equally consequential; for the same reason, when quantum hypothesis testing is applied to practical scenarios, it may be appropriate to treat type-I and type-II errors differently. This is accomplished by asymmetric hypothesis testing. Given quantum states $\rho$ and $\sigma$, the goal of asymmetric quantum hypothesis testing is to minimize the type-II error probability given an upper bound on the type-I error probability, as per the optimization problem in \eqref{eq-hypo_test_asymm_opt_prob}. The value of that optimization problem is given by
	\begin{equation}\label{eq-entr:beta-quant}
		\beta_{\varepsilon}(\rho\Vert\sigma)\coloneqq\inf\{\Tr[M\sigma]:0\leq M\leq\mathbbm{1},\,\Tr[M\rho]\geq 1-\varepsilon\},
	\end{equation}
	with $\varepsilon\in[0,1]$ being an upper bound on the type-I error probability. Intuitively, we might expect a trade-off between the type-I and type-II error probabilities. In particular, we might expect that we can achieve a lower minimum type-II error probability by increasing our tolerance on the type-I error probability. This is indeed true. Observe that every measurement operator $M$ that satisfies $\Tr[M\rho]\geq 1-\varepsilon$ also satisfies $\Tr[M\rho]\geq 1-\varepsilon'$ for all $\varepsilon'$ greater than $\varepsilon$. All such measurement operators are thus feasible points in the optimization for $\beta_{\varepsilon'}(\rho\Vert\sigma)$. Therefore,
	\begin{equation}
		\beta_{\varepsilon'}(\rho\Vert\sigma)\leq\beta_{\varepsilon}(\rho\Vert\sigma)\quad\forall~\varepsilon'\in[\varepsilon,1].
	\end{equation}
	
	\begin{exercise}{exer-hypo_testing_states_beta_quantity}
		Show that $\beta_{\varepsilon}(\rho\Vert\sigma)$ can be evaluated using a semi-definite program. Then, using strong duality, prove that an alternate expression for $\beta_{\varepsilon}(\rho\Vert\sigma)$ is
		\begin{equation}\label{eq-hypo_test_rel_ent_dual_beta}
			\beta_{\varepsilon}(\rho\Vert\sigma)=\sup_{\mu\geq0,Z\geq 0}\{\mu(1-\varepsilon) - \Tr[Z]: \mu\rho\leq\sigma+Z \}.
		\end{equation}
		Finally, evaluate the complementary slackness conditions from Proposition~\ref{prop:math-tools:comp-slack} and conclude that
		\begin{equation}\label{eq:QEI:CS-HTRE-exercise}
			M (\sigma+Z) = \mu M \rho,\qquad \Tr[M \rho]\mu = (1-\varepsilon)\mu, \qquad MZ = Z.
		\end{equation}
	\end{exercise}
	
	\begin{exercise}{exer-hypo_testing_states_beta_iso_invar}
		Prove that the minimum type-II error probability for asymmetric hypothesis testing of states $\rho$ and $\sigma$, with $\varepsilon\in[0,1]$, is isometrically invariant: for every isometry $V$, we have that $\beta_{\varepsilon}(\rho\Vert\sigma)=\beta_{\varepsilon}(V\rho V^{\dagger}\Vert V\sigma V^{\dagger})$.
	\end{exercise}
	
	As with the minimum error probability for symmetric hypothesis testing, the minimum type-II error probability for asymmetric hypothesis testing obeys the following data-processing inequality.
	
	\begin{proposition*}{Data-Processing Inequality for Asymmetric Hypothesis Testing}{prop-hypo_testing_states_asymm_data_proc}
		Consider states $\rho$ and $\sigma$, a number $\varepsilon\in[0,1]$, and let $\mathcal{N}$ be a positive and trace preserving map. Then,
		\begin{equation}
			\beta_{\varepsilon}(\rho\Vert\sigma)\leq\beta_{\varepsilon}(\mathcal{N}(\rho)\Vert\mathcal{N}(\sigma)).
		\end{equation}
	\end{proposition*}
	
	\begin{Proof}
		The proof is analogous to the proof of Proposition~\ref{prop-hypo_test_states_symm_data_proc}, and the intuition behind it is as well. Let $M'$ be an operator satisfying $0\leq M'\leq\mathbbm{1}$ and $\Tr[M'\mathcal{N}(\rho)]\geq 1-\varepsilon$. Then, due to the positivity of $\mathcal{N}$, and thus of $\mathcal{N}^{\dagger}$, we have that $\mathcal{N}^{\dagger}(M')\geq 0$ and $\mathcal{N}^{\dagger}(\mathbbm{1}-M')\geq 0$, so that $\mathcal{N}^{\dagger}(M')\leq\mathcal{N}^{\dagger}(\mathbbm{1})$. Since $\mathcal{N}$ is trace preserving, the adjoint $\mathcal{N}^{\dagger}$ is unital (see Exercise~\ref{exer-adjoint_unital}), which means that $\mathcal{N}^{\dagger}(\mathbbm{1})=\mathbbm{1}$, entailing that $0\leq\mathcal{N}^{\dagger}(M')\leq\mathbbm{1}$. Furthermore, by definition of the adjoint of a superoperator, the inequality $\Tr[\mathcal{N}^{\dagger}(M')\rho]\geq 1-\varepsilon$ holds. Therefore, $\mathcal{N}^{\dagger}(M')$ is a feasible point in the optimization for $\beta_{\varepsilon}(\rho\Vert\sigma)$, so that 
		\begin{align}
			\beta_{\varepsilon}(\rho\Vert\sigma)&=\inf\{\Tr[M\sigma]:0\leq M\leq\mathbbm{1},\,\Tr[M\rho]\geq 1-\varepsilon\}\\
			&\leq \Tr[\mathcal{N}^{\dagger}(M')\sigma]\\
			&=\Tr[M'\mathcal{N}(\sigma)],
		\end{align}
		where the last line follows by definition of the adjoint of a superoperator. Finally, because the inequality $\beta_{\varepsilon}(\rho\Vert\sigma)\leq\Tr[M'\mathcal{N}(\sigma)]$ holds for every operator $M'$ satisfying $0\leq M'\leq\mathbbm{1}$ and $\Tr[M'\mathcal{N}(\rho)]\geq 1-\varepsilon$, we conclude that
		\begin{align}
			\beta_{\varepsilon}(\rho\Vert\sigma)&\leq\inf\{\Tr[M'\mathcal{N}(\sigma)]:0\leq M'\leq\mathbbm{1},\,\Tr[M'\mathcal{N}(\rho)]\geq 1-\varepsilon\}\\
			&=\beta_{\varepsilon}(\mathcal{N}(\rho)\Vert\mathcal{N}(\sigma)),
		\end{align}
		as required.
	\end{Proof}
	
	\begin{proposition*}{Optimal Measurement for Asymmetric Hypothesis Testing}{prop:QEI:opt-meas-HTRE-protocols}
		For every state $\rho$, positive semi-definite operator $\sigma$, and $\varepsilon\in[0,1]$, the minimum type-II error probability\footnote{Even though this quantity need not be an error probability when $\sigma$ is a general positive semi-definite operator, we still refer to it as such.} $\beta_{\varepsilon}(\rho\Vert\sigma)$ is achieved by the following  measurement operator:
		\begin{equation}
			M(\mu^{\ast},p^{\ast})\coloneqq \Pi_{\mu^{\ast}\rho>\sigma}+p^{\ast}\Pi_{\mu^{\ast}\rho=\sigma},
		\end{equation}
		where $\Pi_{\mu^{\ast}\rho>\sigma}$ is the projection onto the strictly positive part of $\mu^{\ast}\rho-\sigma$, the projection $\Pi_{\mu^{\ast}\rho=\sigma}$ projects onto the zero eigenspace of $\mu^{\ast}\rho-\sigma$, and $\mu^{\ast} \geq 0$ and $p^{\ast} \in [0,1]$ are chosen as follows:
		\begin{align}
			\mu^{\ast} & \coloneqq \sup\left\{  \mu:\Tr[\Pi_{\mu\rho>\sigma}\rho]\leq1-\varepsilon\right\}  ,\\
			p^{\ast}& \coloneqq \frac{1-\varepsilon-\Tr[\Pi_{\mu^{\ast}\rho>\sigma}\rho]}{\Tr[\Pi_{\mu^{\ast}\rho=\sigma}\rho]}.
		\end{align}
	\end{proposition*}
	
	\begin{Proof}
		First, observe that it suffices to optimize with respect to every measurement operator~$M$ that meets the constraint $\Tr[M\rho]\geq 1-\varepsilon$ with equality. This follows because for every measurement operator $M$ such that $\Tr[M\rho]>1-\varepsilon$, there exists a positive number $\lambda\in[0,1)$ such that $\Tr[(\lambda M)\rho]=1-\varepsilon$. Note that $0\leq\lambda M\leq\mathbbm{1}$, so that $\lambda M$ is a measurement operator, and because $\Tr[(\lambda M)\sigma]<\Tr[M\sigma]$, we conclude that
		\begin{equation}
			\beta_{\varepsilon}(\rho\Vert\sigma)=\inf\{\Tr[M\sigma]:0\leq M\leq\mathbbm{1},\,\Tr[M\rho]=1-\varepsilon\}.
		\end{equation}
		Based on this, let $M$ be a measurement operator satisfying $\Tr[M\rho]=1-\varepsilon$ and let $\mu\geq 0$. Then,
		\begin{align}
			\Tr[M\sigma]  & =\Tr[M\sigma]+\mu\left(  1-\varepsilon-\Tr[M\rho]\right)  \label{eq:QM-fund-prot:htre-opt-meas-pf-1}\\
			& =-\mu\varepsilon+\Tr[(\mathbbm{1}-M)\mu\rho]+\Tr[M\sigma]\\
			& \geq-\mu\varepsilon+\frac{1}{2}\left(  \Tr[\mu\rho+\sigma]-\left\Vert \mu\rho-\sigma\right\Vert_{1}\right)  \\
			& =-\mu\varepsilon+\frac{1}{2}\left(  \mu+\Tr[\sigma]-\left\Vert\mu\rho-\sigma\right\Vert_{1}\right).\label{eq:QM-fund-prot:htre-opt-meas-pf-last}
		\end{align}
		The sole inequality follows as an application of Theorem~\ref{thm-Holevo_Helstrom}, with $B=\sigma$ and $A=\mu\rho$. Observe that the final expression is a universal bound independent of $M$.
		
		To determine an optimal measurement operator, we can look to Theorem~\ref{thm-Holevo_Helstrom}. There, it was established that the following measurement operator is an optimal one for $\inf_{M:0\leq M\leq\mathbbm{1}}\left\{\Tr[(\mathbbm{1}-M)\mu\rho]+\Tr[M\sigma]\right\}$:
		\begin{equation}
			M(\mu,p)\coloneqq \Pi_{\mu\rho>\sigma}+p\Pi_{\mu\rho=\sigma},
		\end{equation}
		where $\Pi_{\mu\rho>\sigma}$ is the projection onto the strictly positive part of $\mu\rho-\sigma$, the projection $\Pi_{\mu\rho=\sigma}$ projects onto the zero eigenspace of $\mu\rho-\sigma$, and $p\in\left[  0,1\right]  $. The measurement operator $M(\mu,p)$ is called a \textit{quantum Neyman--Pearson test}. We still need to choose the parameters $\mu\geq0$ and $p\in\left[  0,1\right]$. Let us pick $\mu$ according to the following optimization:
		\begin{equation}
			\mu^{\ast}\coloneqq \sup\left\{  \mu:\Tr[\Pi_{\mu\rho>\sigma}\rho]\leq1-\varepsilon\right\}  .
		\end{equation}
		If it happens that $\mu^{\ast}$ is such that $\Tr[\Pi_{\mu^{\ast}\rho>\sigma}\rho]=1-\varepsilon$, then we are done; we can pick $p=0$. However, if $\mu^{\ast}$ is such that $\Tr[\Pi_{\mu^{\ast}\rho>\sigma}\rho]<1-\varepsilon$, then we pick $p^{\ast}\in\left[0,1\right]  $ such that
		\begin{equation}
			p^{\ast}\coloneqq \frac{1-\varepsilon-\Tr[\Pi_{\mu^{\ast}\rho>\sigma}\rho]}{\Tr[\Pi_{\mu^{\ast}\rho=\sigma}\rho]},
		\end{equation}
		with it following that $p^{\ast}\in\left[  0,1\right]  $ because
		\begin{equation}
			\Tr[\Pi_{\mu^{\ast}\rho>\sigma}\rho]<1-\varepsilon\leq\Tr[\Pi_{\mu^{\ast}\rho\geq\sigma}\rho].
		\end{equation}
		With these choices, we then find that
		\begin{equation}
			\Tr[M(\mu^{\ast},p^{\ast})\rho]=1-\varepsilon.
		\end{equation}
		By the analysis in \eqref{eq:QM-fund-prot:htre-opt-meas-pf-1}--\eqref{eq:QM-fund-prot:htre-opt-meas-pf-last}, it then follows that
		\begin{equation}
			\Tr[M\sigma]\geq\Tr[M(\mu^{\ast},p^{\ast})\sigma]
		\end{equation}
		for every measurement operator~$M$ satisfying $0\leq M\leq\mathbbm{1}$ and $\Tr[M\rho]=1-\varepsilon$.
	\end{Proof}

\subsubsection{Asymptotic Setting}

	Now, suppose that multiple copies, say $n$, of the state ($\rho$ or $\sigma$) are available. Based on the discussion at the beginning of Section~\ref{sec-QM_protocols_hypo_testing}, the optimal type-II error probability, given an upper bound of $\varepsilon$ on the type-I error probability, is $\beta_{\varepsilon}(\rho^{\otimes n}\Vert\sigma^{\otimes n})$ for all $n\geq 1$. Then, as in the symmetric case, we are interested in the behavior of this type-II error probability as $n$ becomes large. Furthermore, based on the earlier discussion of the trade-off between the type-I and type-II error probabilities, we might imagine that as $n$ becomes large it is possible to bring the type-I error probability all the way down to zero, because the states become more distinguishable as~$n$ increases. We might also suspect that this convergence is exponentially fast in $n$, as it is exponentially unlikely to confuse many copies of two different states. In order to investigate this possibility, we consider the \textit{optimal type-II error exponent}, by analogy with the error exponent for state discrimination that we considered in Section~\ref{sec-hypo_testing_state_symm_asymp}. To be precise, we consider the following quantities:
	\begin{align}
		E(\rho,\sigma)&\coloneqq\inf_{\varepsilon\in(0,1)}\liminf_{n\to\infty}-\frac{1}{n}\log_2\beta_{\varepsilon}(\rho^{\otimes n}\Vert\sigma^{\otimes n}),\\
		\widetilde{E}(\rho,\sigma)&\coloneqq\sup_{\varepsilon\in(0,1)}\limsup_{n\to\infty}-\frac{1}{n}\log_2\beta_{\varepsilon}(\rho^{\otimes n}\Vert\sigma^{\otimes n}).
	\end{align}
	We refer to $E(\rho,\sigma)$ as the optimal achievable type-II error exponent and $\widetilde{E}(\rho,\sigma)$ as the optimal strong converse type-II error exponent.
	Note that the following inequality follows immediately from the above definitions:
	\begin{equation}\label{eq-hypo_testing_tII_vs_tII_str_conv-protocols}
		E(\rho,\sigma)\leq\widetilde{E}(\rho,\sigma).
	\end{equation}
	The result below, known as the \textit{quantum Stein's lemma}, provides us with a tractable expression for $E(\rho,\sigma)$ and $\widetilde{E}(\rho,\sigma)$ in terms of the \textit{quantum relative entropy}, an important quantity in quantum information theory that we introduce formally in Chapter~\ref{chap-entropies}. We also delay the proof of the result to Chapter~\ref{chap-entropies}, when all of the required elements of the proof become available to us.
	
	\begin{theorem*}{Quantum Stein's Lemma}{thm-q_Stein_lemma-protocols}
		For all states $\rho$ and $\sigma$, the optimal achievable and strong converse rates are equal to the quantum relative entropy of $\rho$ and $\sigma$, i.e.,
		\begin{equation}
			E(\rho,\sigma)=\widetilde{E}(\rho,\sigma)=D(\rho\Vert\sigma),
		\end{equation}
		where
		\begin{equation}\label{eq-quantum_rel_ent_0}
			D(\rho\Vert\sigma)\coloneqq\left\{\begin{array}{l l} \Tr[\rho(\log_2\rho-\log_2\sigma)] & \text{if }\supp(\rho)\subseteq\supp(\sigma), \\ +\infty & \text{otherwise} \end{array}\right.
		\end{equation}
		is the \textit{quantum relative entropy}.
	\end{theorem*}
	
	\begin{Proof}
		See Section~\ref{subsec-q_Stein_lemma}.
	\end{Proof}
	
	\begin{remark}
		See Section~\ref{sec:math-tools:functions-herm-ops} for the definition of the logarithm of a Hermitian operator. See Section~\ref{sec-rel_ent} for a more detailed explanation of the support conditions in the definition of the quantum relative entropy.
	\end{remark}




\section{Quantum Channel Discrimination}\label{sec-channel_discrimination}

	In Sections~\ref{subsec-state_discrimination} and \ref{sec:QM-over:multiple-state-disc}, we considered symmetric hypothesis testing with respect to quantum states, also known as quantum state discrimination, in which the hypotheses are modelled as quantum states, and the task is to devise a measurement strategy that maximizes the probability of success of correctly identifying the state of a given quantum system. Let us now consider the analogous scenario with quantum channels, which we refer to as \textit{quantum channel discrimination}. We do not discuss asymmetric hypothesis testing with respect to quantum channels in this book; please see the Bibliographic Notes in Section~\ref{sec-QM_protocols_bib_notes} for references.

	The task of quantum channel discrimination is as follows. Consider that a quantum system $A$ undergoes an evolution according to one of two quantum channels, $\mathcal{N}_{A\to B}$ or $\mathcal{M}_{A\to B}$. Furthermore, we suppose that the quantum channel is chosen to be $\mathcal{N}$ with probability $\lambda\in[0,1]$ and $\mathcal{M}$ with probability $1-\lambda$. The goal is to devise a strategy that correctly guesses the unknown quantum channel with the highest probability.
	
	\begin{figure}
		\centering
		\includegraphics[scale=0.8]{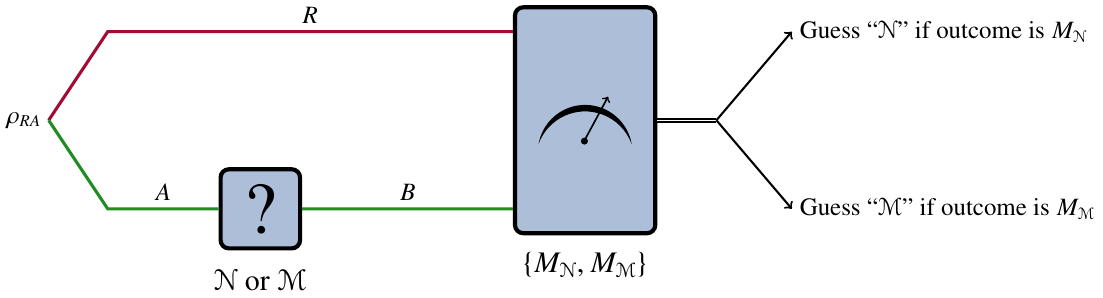}
		\caption{The most general strategy for discriminating two quantum channels $\mathcal{N}$ and $\mathcal{M}$ is to prepare a bipartite state $\rho_{RA}$, with the reference system $R$ having arbitrary dimension, sending the system $A$ through the unknown quantum channel, and then measuring both systems $R$ and $A$ according to a two-outcome POVM $\{M_{\mathcal{N}},M_{\mathcal{M}}\}$. If the outcome corresponding to $M_{\mathcal{N}}$ occurs, then we guess that the unknown channel is $\mathcal{N}$; otherwise, we guess that it is $\mathcal{M}$. The minimum error probability among all such strategies is given by Theorem~\ref{thm-q_channel_disc}.}\label{fig-chan_discrimination}
	\end{figure}

	The most general strategy for quantum channel discrimination is illustrated in Figure~\ref{fig-chan_discrimination}. The strategy consists of a state $\rho_{RA}$ and a two-outcome measurement described by the POVM $\{M_{\mathcal{N}},M_{\mathcal{M}}\}$. The bipartite state $\rho_{RA}$ is such that the system~$R$ can be arbitrarily large in principle, in order try to achieve the lowest error probability. The measurement acts on both the system $R$ and the system $A$ after system $A$ has passed through the unknown channel. The expected error probability of this strategy is analogous to the expected error probability of a strategy for quantum state discrimination: it is the expectation, with respect to the prior probability distribution given by $\lambda$, of the probabilities of the two types of errors that can occur: guessing ``$\mathcal{M}$'' when the channel is $\mathcal{N}$, and guessing ``$\mathcal{N}$'' when the channel is $\mathcal{M}$. In other words, the expected error probability is
	\begin{multline}
		\lambda\Tr[M_{\mathcal{M}}\mathcal{N}_{A\to B}(\rho_{RA})]+(1-\lambda)\Tr[M_{\mathcal{N}}\mathcal{M}_{A\to B}(\rho_{RA})]\label{eq-chan_disc_error_prob}\\
		 =p_{\text{err}}(\lambda,\mathcal{N}_{A\to B}(\rho_{RA}),\mathcal{M}_{A\to B}(\rho_{RA}),M),
	\end{multline}
	where the last line follows by letting $M_{\mathcal{N}}\equiv M$ and from the definition of $p_{\text{err}}$ in \eqref{eq-hypo_testing_state_symm_err_prob}. We see that, given a state $\rho_{RA}$, the task of discriminating $\mathcal{N}$ and $\mathcal{M}$ reduces to the task of discriminating the states $\mathcal{N}_{A\to B}(\rho_{RA})$ and $\mathcal{M}_{A\to B}(\rho_{RA})$. The optimal error probability is obtained by optimizing with respect to every state $\rho_{RA}$ and measurement operator $M$, so that
	\begin{align}
		p_{\text{err}}^*(\lambda,\mathcal{N},\mathcal{M})\coloneqq& \inf_{\substack{\rho_{RA}\\M:0\leq M\leq\mathbbm{1}}} p_{\text{err}}(\lambda,\mathcal{N}_{A\to B}(\rho_{RA}),\mathcal{M}_{A\to B}(\rho_{RA}),M)\\
		=&\ \inf_{\rho_{RA}}p_{\text{err}}^*(\lambda,\mathcal{N}_{A\to B}(\rho_{RA}),\mathcal{M}_{A\to B}(\rho_{RA})),\label{eq-chan_disc_opt_error_def}
	\end{align}
	where the optimization is with respect to every quantum state $\rho_{RA}$, and there is an implicit optimization with respect to the dimension of the system $R$. This gives us a first look into how a quantity defined initially for quantum states can be ``lifted'' to a quantity defined for quantum channels. In particular, the quantity $p_{\text{err}}^*$, initially defined for quantum states as in \eqref{eq-state_disc_opt_error_def}, has been extended to quantum channels by evaluating the state quantity with respect to the states $\mathcal{N}_{A\to B}(\rho_{RA})$ and $\mathcal{M}_{A\to B}(\rho_{RA})$ and then optimizing with respect to both the dimension of $R$ and the state $\rho_{RA}$. Such constructions of channel quantities from state quantities arise throughout the rest of the book.
	
	Just as the optimal error probability for discriminating two states can be expressed using the trace norm (recall \eqref{eq-state_disc_opt_error}), we now show that, analogously, the optimal error probability for discriminating two quantum channels can be expressed in terms of the diamond norm (see Definition~\ref{def-diamond_norm_superop}).

	\begin{theorem*}{Minimum Error Probability for Quantum Channel Discrimination}{thm-q_channel_disc}
		Let $\mathcal{N}_{A\to B}$ and $\mathcal{M}_{A\to B}$ be quantum channels, and $\lambda\in[0,1]$. The optimal error probability for discriminating $\mathcal{N}$ and $\mathcal{M}$ is given by
		\begin{align}
			p_{\text{err}}^*(\lambda,\mathcal{N},\mathcal{M})&=\inf_{\psi_{A'A}}p_{\text{err}}^*(\lambda,\mathcal{N}_{A\to B}(\psi_{A'A}),\mathcal{M}_{A\to B}(\psi_{A'A}))\label{eq-chan_disc_opt_error_0}\\
			&=\frac{1}{2}(1-\norm{\lambda\mathcal{N}-(1-\lambda)\mathcal{M}}_{\diamond}),\label{eq-chan_disc_opt_error}
		\end{align}
		where, in the first line, the optimization is with respect to every pure state $\psi_{A'A}$ with $d_{A'}=d_A$.
	\end{theorem*}
	
	\begin{Proof}
		Using \eqref{eq-state_disc_opt_error}, we have
		\begin{equation}
			p_{\text{err}}^*(\lambda,\mathcal{N},\mathcal{M})=\frac{1}{2}\left(1-\sup_{\rho_{RA}}\norm{\lambda\mathcal{N}_{A\to B}(\rho_{RA})-(1-\lambda)\mathcal{M}_{A\to B}(\rho_{RA})}_1\right).
		\end{equation}
		Now, consider a state $\rho_{RA}$ in the optimization above, and let $\psi_{R'RA}$ be a purification of $\rho_{RA}$. Then,
		\begin{align}
			&\norm{\lambda\mathcal{N}_{A\to B}(\rho_{RA})-(1-\lambda)\mathcal{M}_{A\to B}(\rho_{RA})}_1\nonumber\\
			&\quad=\norm{\lambda\Tr_{R'}[\mathcal{N}_{A\to B}(\psi_{R'RA})]-(1-\lambda)\Tr_{R'}[\mathcal{M}_{A\to B}(\psi_{R'RA})]}_1\\
			&\quad\leq \norm{\lambda\mathcal{N}_{A\to B}(\psi_{R'RA})-(1-\lambda)\mathcal{M}_{A\to B}(\psi_{R'RA})}_1,\label{eq-q_channel_disc_opt_err_prob_pf1}
		\end{align}
		where the inequality follows from \eqref{eq-data_proc_trace_norm_0}, with respect to the partial trace channel~$\Tr_{R'}$. Now, without loss of generality, we can let $d_{R'}\geq d_Rd_A$; see Section~\ref{sec:qm:purification-def}. Then, by the Schmidt decomposition theorem (Theorem~\ref{thm-Schmidt}), in particular \eqref{eq-Schmidt_decomp_gen}, the state vector $\ket{\psi}_{R'RA}$ can be expressed according to the $R'R|A$ bipartition as $\ket{\psi}_{R'RA}=\sum_{k=1}^{d_A}\sqrt{p_k}\ket{u_k}_{R'R}\otimes\ket{v_k}_A$, where  $p_k$ is a probability and the vectors $\ket{u_k}_{R'R}$ form an orthonormal basis for a $d_A$-dimensional vector space. In other words, only a $d_A$-dimensional subspace of $\mathcal{H}_{R'R}$, call it $\mathcal{H}_{A'}$, is relevant for calculating the trace norm in \eqref{eq-q_channel_disc_opt_err_prob_pf1}, and there exists an isometry $V_{A'\to R'R}$ such that $V_{A'\to R'R} \ket{\psi}_{A'A} = \ket{\psi}_{R'RA}$. Adopting the shorthand $V\equiv V_{A'\to R'R}$, it follows that
		\begin{align}
			&\norm{\lambda\mathcal{N}_{A\to B}(\rho_{RA})-(1-\lambda)\mathcal{M}_{A\to B}(\rho_{RA})}_1\nonumber\\
			& \quad \leq \norm{\lambda\mathcal{N}_{A\to B}(V\psi_{A'A}V^\dag)-(1-\lambda)\mathcal{M}_{A\to B}(V\psi_{A'A}V^\dag)}_1 \\
			& \quad = \norm{\lambda\mathcal{N}_{A\to B}(\psi_{A'A})-(1-\lambda)\mathcal{M}_{A\to B}(\psi_{A'A})}_1 \\
			&\quad\leq\sup_{\phi_{A'A}}\norm{\lambda\mathcal{N}_{A\to B}(\phi_{A'A})-(1-\lambda)\mathcal{M}_{A\to B}(\phi_{A'A})}_1\\
			&\quad=\norm{\lambda\mathcal{N}-(1-\lambda)\mathcal{M}}_{\diamond},
		\end{align}
		where the last line follows from \eqref{eq-diamond_norm_Herm_pres}, because the map $\lambda\mathcal{N}-(1-\lambda)\mathcal{M}$ is Hermiticity preserving. Therefore,
		\begin{equation}
			p_{\text{err}}^*(\lambda,\mathcal{N},\mathcal{M})\geq \frac{1}{2}\left(1-\norm{\lambda\mathcal{N}-(1-\lambda)\mathcal{M}}_{\diamond}\right).
		\end{equation}
		The opposite inequality holds simply by restricting the optimization in \eqref{eq-chan_disc_opt_error_def} to every pure state $\psi_{A'A}$ satisfying $d_{A'}=d_A$. We thus obtain \eqref{eq-chan_disc_opt_error}, as required.
	\end{Proof}
	
	\begin{exercise}{exer-hypo_test_chan_symm_SDP}
		Consider quantum channels $\mathcal{N}_{A\to B}$ and $\mathcal{M}_{A\to B}$, and let $\lambda\in[0,1]$. Using \eqref{eq-chan_disc_opt_error_0} and \eqref{eq-chan_disc_error_prob}, show that the optimal error probability $p_{\text{err}}^*(\lambda,\mathcal{N},\mathcal{M})$ can be evaluated using a semi-definite program. Then, using strong duality, prove that an alternate expression for $p_{\text{err}}^*(\lambda,\mathcal{N},\mathcal{M})$ is
		\begin{multline}
			p_{\text{err}}^*(\lambda,\mathcal{N},\mathcal{M})\\=\sup_{\substack{W_{AB}\\\text{Hermitian}}}\!\!\left\{\lambda_{\min}(\Tr_B[W_{AB}]):W_{AB}\leq\lambda\Gamma_{AB}^{\mathcal{N}},\,W_{AB}\leq(1-\lambda)\Gamma_{AB}^{\mathcal{M}}\right\},
		\end{multline}
		where $\lambda_{\min}(\Tr_B[W_{AB}])$ is the smallest eigenvalue of $\Tr_B[W_{AB}]$; see Exercise~\ref{exer-Herm_eigenvals}.
	\end{exercise}
	
	\begin{exercise}{exer-hypo_test_chan_symm_Choi_bounds}
		Consider quantum channels $\mathcal{N}_{A\to B}$ and $\mathcal{M}_{A\to B}$, and let $\lambda\in[0,1]$. Prove the following bounds on the optimal error probability for discriminating $\mathcal{N}$ and $\mathcal{M}$ in terms of the optimal error probability for discriminating the Choi states of $\mathcal{N}$ and $\mathcal{M}$:
		\begin{equation}\label{eq-hypo_test_chan_symm_bounds_Choi}
			d_Ap_{\text{err}}^*(\lambda,\Phi_{AB}^{\mathcal{N}},\Phi_{AB}^{\mathcal{M}})\leq p_{\text{err}}^*(\lambda,\mathcal{N},\mathcal{M})\leq p_{\text{err}}^*(\lambda,\Phi_{AB}^{\mathcal{N}},\Phi_{AB}^{\mathcal{M}}).
		\end{equation}
		(\textit{Hint}: See Exercise~\ref{exer-diamond_norm_Choi_bounds}.)
	\end{exercise}
	
	The upper bound in \eqref{eq-hypo_test_chan_symm_bounds_Choi} corresponds to the strategy that consists of letting the state $\rho_{RA}$ in Figure~\ref{fig-chan_discrimination} be the maximally-entangled state $\Phi_{A'A}=\ket{\Phi}\bra{\Phi}_{A'A}$, with $\ket{\Phi}_{A'A}=\frac{1}{\sqrt{d_A}}\sum_{i=0}^{d_A-1}\ket{i,i}_{A'A}$. The following exercise tells us when this strategy is optimal, i.e., when the upper bound in \eqref{eq-hypo_test_chan_symm_bounds_Choi} is achieved.
	
	\begin{exercise}{exer-hypo_test_chan_symm_covariant}
		Let the quantum channels $\mathcal{N}_{A\to B}$ and $\mathcal{M}_{A\to B}$ be jointly covariant with respect to a group $G$, so that
		\begin{align}
		\mathcal{N}_{A\to B}(U_A^g\rho_A U_A^{g\dagger}) & =V_B^{g}\mathcal{N}_{A\to B}(\rho_A)V_B^{g\dagger},\\
		\mathcal{M}_{A\to B}(U_A^g\rho_A U_A^{g\dagger}) & =V_B^{g}\mathcal{M}_{A\to B}(\rho_A)V_B^{g\dagger},
		\end{align}
		for every $g\in G$ and every state $\rho_A$, where $g\mapsto U_A^g$ and $g\mapsto V_B^g$ are unitary representations of $G$ acting on $\mathcal{H}_A$ and $\mathcal{H}_B$, respectively. Furthermore, let $g\mapsto U_A^g$ be such that
		\begin{equation}
		\mathcal{T}_A^G(\cdot)\coloneqq\frac{1}{|G|}\sum_{g\in G}U_A^g(\cdot)U_A^{g\dagger}  =\Tr[\cdot]\frac{\mathbbm{1}_A}{d_A}.
		\end{equation}
		Prove that $p_{\text{err}}^*(\lambda,\mathcal{N},\mathcal{M})=p_{\text{err}}^*(\lambda,\Phi_{AB}^{\mathcal{N}},\Phi_{AB}^{\mathcal{M}})$. (\textit{Hint}: Use Proposition~\ref{prop-diamond_norm_group_cov}.)
	\end{exercise}

\section{Summary}


\section{Bibliographic Notes}\label{sec-QM_protocols_bib_notes}

	The quantum teleportation protocol for qubits and qudits presented in Section~\ref{sec-teleportation} was devised by \citet{BBC+93}. The generalization to groups that act irreducibly on the input state was presented by \citet{BDMS00,Werner01}. Covariant channels were considered by \citet{Hol02}. For an overview of group- and representation-theoretic concepts for finite groups, see \citet{group_theory_book}.
	
	The notion of teleportation simulation of a quantum channel was introduced by \citet{BDSW96} (see Section~V therein). The LOCC simulation of a quantum channel was given by \citet{HHH99} (see Eq.~(10) therein), and the related PPT simulation of a quantum channel was given by \citet{KW17}.  The generalized teleportation simulation of a quantum channel was developed for groups that act irreducibly on the channel input space by \citet{CDP09} (see Section~VII therein). \citet{PLOB15} played a role in developing the tool of LOCC/teleportation simulation more recently. The fact that the channel twirl can be realized by generalized teleportation simulation was observed by \citet{KW17} (see Appendix~B therein). We also refer to the work of \cite{DHCB05} (particularly Section~II.B) for a closely related result. The teleportation protocol was extended to states of bosonic systems by \citet{prl1998braunstein}.  Teleportation simulation of bosonic Gaussian channels was given by \citet{GI02,NFC09}, and a detailed analysis of convergence in this scenario was presented by \citet{PhysRevA.97.062305}.
	
	The quantum super-dense coding protocol was discovered by \citet{PhysRevLett.69.2881}.
	
	The problem of state discrimination, as described in Section~\ref{subsec-state_discrimination}, was considered by \citet{Hel67} (see also \citet{H69}) and \citet{Hol72}, who determined the optimal success probability in the case of projective measurements and general POVMs, respectively. The proof of the optimal measurement operators in Theorem~\ref{thm-Holevo_Helstrom} is due to \citet{Jen10}. Lemma~\ref{lemma:spectral-ineq} is due to \citet{ACMBMAV07}, with the simple proof presented here attributed by \citet{JOPS12} and \citet{A14} as being due to N.~Ozawa. The notion of non-commutative minimum was studied by~\citet{C23}. The Chernoff bound for probability distributions was given by \citet{Cher52}. The corresponding bound for quantum states in Theorem~\ref{thm-quantum_Chernoff} was established in two works: \citet{ACMBMAV07} determined the upper bound on the optimal error exponent, while \citet{nussbaum2009chernoff} established the lower bound. The semi-definite program and complementary slackness conditions for multiple state discrimination in Section~\ref{sec:QM-over:multiple-state-disc} are due to \citet{YKL75}. For work on measurement strategies in two-state discrimination, including adaptive strategies and strategies that achieve the Chernoff bound, we refer to the work of \citet{BM96,BYH97,ABB+05,HDB+11,BLS+20}.
	
	The operational interpretation of diamond distance in terms of symmetric hypothesis testing of quantum channels (specifically, Theorem~\ref{thm-q_channel_disc}) was given by \citet{RW05,Sacchi05}. The work of \citet{KWerner04,GLN04} provides a different operational interpretation of diamond distance in terms of quantifying the error between an ideal channel and an experimental approximation of it, which we elaborate upon in Chapter~\ref{chap-QM_dist_meas}. ...(references for multiple channel discrimination...references for asymmetric hypothesis testing for channels...)

\section{Problems}


	\begin{enumerate}[left=0cm,itemsep=1cm]
%
		
		\item Let $\rho_{AB}$ be a quantum state with $d_A=d_B=d\geq 2$, and consider the quantity
			\begin{equation}
				q_{\text{corr}}(A|B)_{\rho}\equiv q_{\text{corr}}(\rho_{AB})\coloneqq d\sup_{\mathcal{N}}\bra{\Phi}_{AB}(\id_A\otimes\mathcal{N}_B)(\rho_{AB})\ket{\Phi}_{AB},
			\end{equation}
			where the optimization is with respect to quantum channels $\mathcal{N}_B$ acting on system $B$.
			\begin{enumerate}[itemsep=0.3cm,topsep=0.3cm]
				\item Show that $q_{\text{corr}}(\rho_{AB})$ can be evaluated using a semi-definite program, and determine the corresponding dual program.
				
				\item Suppose that $\rho_{AB}$ is a classical--quantum state, i.e., $\rho_{AB}\equiv\rho_{XB}=\sum_{x\in\mathcal{X}}p(x)\ket{x}\bra{x}_X\otimes\rho_B^x$, where $\mathcal{X}$ is a set of size $d$, $p:\mathcal{X}\to[0,1]$ is a probability distribution, and $\{\rho_B^x\}_{x\in\mathcal{X}}$ is a set of states. Prove that
					\begin{equation}
						q_{\text{corr}}(\rho_{XB})=p_{\text{succ}}^*(\{(p(x),\rho^x)\}_x).
					\end{equation}
					In other words, for classical--quantum states, the quantity $q_{\text{corr}}$ reduces to the optimal success probability for multiple state discrimination of the set $\{\rho_B^x\}_{x\in\mathcal{X}}$. (\textit{Hint}: See Exercise~\ref{exer-POVM_adjoint_Naimark}.)
				
			\end{enumerate}
			(\textit{Bibliographic Note}: The function $q_{\text{corr}}$ was defined by \citet{KRS09} within the context of the min-entropy (a quantity that we encounter in Chapter~\ref{chap-entropies}) and its operational meaning.)
		
%
%

            \item Let $\ket{\psi}$ and $\ket{\phi}$ be two pure state vectors on the same quantum system. Assume that they are distinct and non-orthogonal, so that $0 < |\braket{\psi}{\phi}| < 1$. In what follows, we denote $\psi\coloneqq \ketbra{\psi}{\psi}$ and $\phi\coloneqq \ketbra{\phi}{\phi}$.  
                \begin{enumerate}[itemsep=0.3cm,topsep=0.3cm]
                    \item Using the definition of the quantum relative entropy in \eqref{eq-quantum_rel_ent_0}, argue that $D(\psi\|\phi) = +\infty$.  

                    \item Based on Theorem \ref{thm-q_Stein_lemma-protocols}, this result suggests that given multiple copies of an unknown state, either $\psi$ or $\phi$, one should be able to construct a measurement such that the type-II error probability is \emph{exactly} zero, while the type-I error probability remains nonzero but vanishes asymptotically. Can you describe such a measurement procedure and prove that it behaves as claimed? (\textit{Hint}: Consider the vectors orthogonal to $\ket{\psi}$ and $\ket{\phi}$ within the subspace spanned by these two vectors.)
                \end{enumerate}  
            (\textit{Bibliographic Note}: This protocol, sometimes known as inconclusive state discrimination, was first proposed by \citet{Ivanovic_1987}, \citet{Dieks_1988}, and \citet{Peres_1988}.)
	
	\end{enumerate}


\chapter{Distinguishibility Measures for Quantum States and Channels}\label{chap-QM_dist_meas}

	[in progress]
	
	

\section{Trace Distance}\label{sec-QM-trace-distance}
	
	The trace distance is a distance measure based on the trace norm; see Section~\ref{sec-math_tools_trace_norm}. For two quantum states $\rho$ and $\sigma$, we define the \textit{normalized trace distance between $\rho$ and $\sigma$} as $\frac{1}{2}\norm{\rho-\sigma}_1$.
	
	Using \eqref{exer-trace_distance}, observe that the normalized trace distance between pure states $\ket{\psi}\!\bra{\psi}$ and $\ket{\phi}\!\bra{\phi}$ is given by
	\begin{equation}\label{eq-trace_dist_pure_states}
		\frac{1}{2}\norm{\ket{\psi}\!\bra{\psi}-\ket{\phi}\!\bra{\phi}}_1=\sqrt{1-\abs{\braket{\psi}{\phi}}^2}.
	\end{equation}
	
	
	In Section~\ref{subsec-state_discrimination}, we showed that the trace distance arises in terms of the optimal error probability for discriminating two quantum states. Specifically, for two quantum states $\rho$ and $\sigma$, and for $\lambda=\frac{1}{2}$, we have
	\begin{equation}
		p_{\text{err}}^*\left(1/2,\rho,\sigma\right)=\frac{1}{2}\left(1-\frac{1}{2}\norm{\rho-\sigma}_1\right).
	\end{equation}
	The trace distance thus has an operational meaning as quantifying the optimal error probability for distinguishing two quantum states with equal prior probability.

    A related operational meaning of the trace distance is in terms of assessing the performance of a quantum information processing protocol in which the ideal state to be generated is $\rho$ but the actual state generated is $\sigma$. 
    Namely, suppose that a third party is trying to assess how distinguishable the actual state $\sigma$ is from the ideal state $\rho$. 
    One way to do so is by performing a quantum measurement described by the POVM $\{M_x\}_{x\in\mathcal{X}}$ whose elements are indexed by some finite set $\mathcal{X}$. In the case that $\rho$ was prepared, the probability of obtaining the outcome corresponding to $x$ is $\Tr[M_x\rho]$, and in the case that $\sigma$ was prepared, this probability is $\Tr[M_{x}\sigma]$. In order for $\rho$ and $\sigma$ to be considered ``close,'' what we demand is that the absolute deviation between the actual probability $\Tr[M_{x}\sigma]$ and the ideal probability $\Tr[M_{x}\rho]$ be no larger than some desired tolerance $\varepsilon>0$, so that $\abs{\Tr[M_x\rho]-\Tr[M_x\sigma]}\leq\varepsilon$. We want this condition to hold for \textit{all} possible measurements that one could perform, so what we demand mathematically is that
	\begin{equation}
		\sup_{0\leq M\leq \mathbbm{1}}\abs{\Tr[M\rho] - \Tr[M\sigma]} \leq\varepsilon.
	\end{equation}
	As stated in Theorem~\ref{thm-trace_dist_meas} below, the following identity holds:
	\begin{equation}\label{eq-trace_dist_variational1}
		\sup_{0\leq M\leq \mathbbm{1}}\abs{\Tr[M\rho]-\Tr[M\sigma]} =\frac{1}{2}\norm{\rho-\sigma}_{1},
	\end{equation}
	indicating that if $\frac{1}{2}\norm{\rho-\sigma}_{1}\leq \varepsilon$, then the deviation between probabilities for every possible measurement operator never exceeds $\varepsilon$, so that the approximation between states $\rho$ and $\sigma$ is naturally quantified by the trace distance $\frac{1}{2}\norm{\rho-\sigma}_{1}$.
	
	Let us now prove the variational characterization of the trace distance stated in \eqref{eq-trace_dist_variational1}.
	
	\begin{theorem*}{Trace Distance via Measurement}{thm-trace_dist_meas}
		For two states $\rho$ and $\sigma$, the following equality holds:
		\begin{equation}\label{eq-trace_dist_meas}
			\frac{1}{2}\norm{\rho-\sigma}_1=\sup_{0\leq M\leq\mathbbm{1}}\abs{\Tr\!\left[M(\rho-\sigma)\right]}.
		\end{equation}
        Up to replacing $M$ with $\mathbbm{1}-M$, every optimal measurement operator $M$ is of the form $M=\Pi_+ + \Lambda_0$, where $\Pi_+$ is the projection onto the strictly positive part of $\rho-\sigma$ and $\Lambda_0$ is an operator satisfying $0 \leq \Lambda_0 \leq \Pi_0$, with $\Pi_0$ the projection onto the zero eigenspace of $\rho-\sigma$.
	\end{theorem*}
	
	\begin{Proof}
        Since
        \begin{equation}
            \abs{\Tr\!\left[M(\rho-\sigma)\right]} = \max\left\{\Tr\!\left[M(\rho-\sigma)\right],\, \Tr\!\left[(\mathbbm{1} - M)(\rho-\sigma)\right]\right\},
        \end{equation}
        the equality follows as a direct application of \eqref{eq-Holevo_Helstrom-alt} and the optimality statement following it, with $A=\rho$ and $B=\sigma$.
	\end{Proof}
	
	From Exercise~\ref{exer-trace_norm_Herm_SDP}, we know that, for Hermitian operators, the trace norm can be evaluated using semi-definite programming. The normalized trace distance $\frac{1}{2}\norm{\rho-\sigma}_1$ can therefore be evaluated using semi-definite programming, because $\rho-\sigma$ is Hermitian. Since $\rho$ and $\sigma$ are positive semi-definite, we obtain the following simpler semi-definite programs for their normalized trace distance.
	
	\begin{proposition*}{SDPs for Normalized Trace Distance}{prop-trace_dist_SDP}
		The trace distance between every two quantum states $\rho$ and $\sigma$ can be written as the following semi-definite programs:
		\begin{align}
			\frac{1}{2}\norm{\rho-\sigma}_1&=\sup_{M\geq 0}\{\Tr[M(\rho-\sigma)]: M\leq\mathbbm{1}\}\label{eq-trace_dist_SDP_primal}\\
			&=\inf_{Z\geq 0}\{\Tr[Z]:Z\geq \rho-\sigma\}.\label{eq-trace_dist_SDP_dual}
		\end{align}
	\end{proposition*}
	
	\begin{Proof}
		The expression in \eqref{eq-trace_dist_SDP_primal} is immediate from \eqref{eq-Holevo_Helstrom-alt}, which already provides an expression for $\frac{1}{2}\norm{\rho-\sigma}_1$ as a semi-definite program in the primal form as in \eqref{eq-primal_SDP_def}. Obtaining the expression in \eqref{eq-trace_dist_SDP_dual} is then straightforward; see Exercise~\ref{exer-norm_trace_dist_dual}.
	\end{Proof}
	
	\begin{exercise}{exer-norm_trace_dist_dual}
		Prove \eqref{eq-trace_dist_SDP_dual}.
	\end{exercise}
	
	The trace distance obeys the following data-processing inequality.
	
	\begin{theorem*}{Data-Processing Inequality for Trace Distance}{thm-trace_dist_monotone}
		Let $\rho$ and $\sigma$ be states, and let $\mathcal{N}$ be a positive, trace-non-increasing map. Then, 
        \begin{equation}\label{eq-trace_dist_monotone-PTNI}
			\norm{\rho-\sigma}_1 \geq \norm{\mathcal{N}(\rho)-\mathcal{N}(\sigma)}_1.
		\end{equation}
	\end{theorem*}
	
	\begin{Proof}
		This is immediate from \eqref{eq-data_proc_trace_norm_0}, which tells us that the trace norm is monotone non-increasing under the action of every positive trace-non-increasing superoperator. It is also possible to provide a direct proof using the expression in \eqref{eq-trace_dist_SDP_primal}; see Exercise~\ref{exer-trace_dist_monotone-PTNI}.
	\end{Proof}
	
	\begin{exercise}{exer-trace_dist_monotone-PTNI}
		Provide a direct proof of \eqref{eq-trace_dist_monotone-PTNI} using the expression in \eqref{eq-trace_dist_SDP_primal}. (\textit{Hint}: See the proof of Proposition~\ref{prop-hypo_test_states_symm_data_proc}.)
	\end{exercise}
	
	
	
	
	By combining the results of Theorems~\ref{thm-trace_dist_meas} and \ref{thm-trace_dist_monotone}, we find that the trace distance is achieved by a measurement channel:
	
	\begin{theorem*}{Trace Distance Achieved by Measurement Channel}{thm-trace_dist_ach_by_meas_channel}
		For two states $\rho$ and $\sigma$, the following equality holds:
		\begin{equation}
			\norm{\rho-\sigma}_1=\max_{\{M_x\}_x}\sum_{x\in\mathcal{X}}\abs{\Tr[M_x\rho]-\Tr[M_x\sigma]},
		\end{equation}
		where the optimization is performed with respect to every POVM $\{M_x\}_{x\in\mathcal{X}}$ defined over a finite alphabet $\mathcal{X}$. An optimal POVM is given by $\{\Pi_+, \mathbbm{1} - \Pi_+\}$, where 
        $\Pi_+$ is the projection onto the strictly positive part of $\rho-\sigma$.
	\end{theorem*}
	
	\begin{Proof}
	The inequality
	\begin{equation}
	\norm{\rho-\sigma}_1 \geq\max_{\{M_x\}_x}\sum_{x\in\mathcal{X}}\abs{\Tr[M_x\rho]-\Tr[M_x\sigma]}
	\end{equation}
	follows from Theorem~\ref{thm-trace_dist_monotone} by taking the channel $\mathcal{N}$ to be the quantum--classical channel defined as $\mathcal{N}(\omega)= \sum_{x \in \mathcal{X}} \Tr[M_x \omega] \ket{x}\!\bra{x}$. Then, from Theorem~\ref{thm-trace_dist_meas}, we have the equality
	\begin{equation}
	\frac{1}{2} \norm{\rho-\sigma}_1 = |\Tr[\Pi_+(\rho -\sigma)]|.
	\end{equation}
	Since $|\Tr[\Pi_+(\rho -\sigma)]|=|\Tr[(\mathbbm{1}-\Pi_+)(\rho -\sigma)]|$, we conclude that
	\begin{equation}
	\norm{\rho-\sigma}_1 = |\Tr[\Pi_+(\rho -\sigma)]| + |\Tr[(\mathbbm{1}-\Pi_+)(\rho -\sigma)]|,
	\end{equation}
	so that an optimal POVM is given by $\{\Pi_+, \mathbbm{1} - \Pi_+\}$.
	\end{Proof}

\section{Fidelity}\label{subsec-fidelity}

	In addition to the trace distance, another distinguishability measure for states that we consider in this book is the fidelity (also called Uhlmann fidelity).
	
	\begin{definition}{Fidelity}{def-fidelity}
		For two quantum states $\rho$ and $\sigma$, the \textit{fidelity of $\rho$ and $\sigma$}, denoted by $F(\rho,\sigma)$, is defined as
		\begin{equation}
			F(\rho,\sigma)\coloneqq \norm{\sqrt{\rho}\sqrt{\sigma}}_1^2=\left(\Tr\!\left[\sqrt{\smash[b]{\sqrt{\smash[b]{\sigma}}\rho\sqrt{\smash[b]{\sigma}}}}\right]\right)^2.
		\end{equation}
	\end{definition}
	
	Observe that the fidelity is symmetric in its arguments. We also have that $F(\rho,\sigma)\in[0,1]$ for all states $\rho$ and $\sigma$, a fact that we prove below. 
	
	For a pure state $\ket{\psi}$ and mixed state $\rho$, the fidelity between them is equal to
	\begin{equation}\label{eq-fidelity_pure_mixed}
		F(\rho,\ket{\psi}\!\bra{\psi})=\bra{\psi}\rho\ket{\psi}=\Tr[\ket{\psi}\!\bra{\psi}\rho].
	\end{equation}
	Also, for two pure states $\ket{\psi}$ and $\ket{\phi}$, the fidelity is simply 
	\begin{equation}\label{eq-fidelity_pure_pure}
		F(\ket{\psi}\!\bra{\psi},\ket{\phi}\!\bra{\phi})=\abs{\braket{\psi}{\phi}}^2.
	\end{equation}
	
	The formula in \eqref{eq-fidelity_pure_mixed} gives the fidelity an operational meaning that we employ in later chapters. Suppose that the goal of a quantum information processing protocol is to produce the pure state $\ket{\psi}\!\bra{\psi}$, but it instead produces a mixed state $\rho$. Then the fidelity $F(\rho,\ket{\psi}\!\bra{\psi})$ is equal to the probability that the actual state $\rho$ passes a test for being the ideal state $\ket{\psi}\!\bra{\psi}$, with the test being given by the POVM $\{\ket{\psi}\!\bra{\psi}, \mathbbm{1} -\ket{\psi}\!\bra{\psi}\}$. That is, the probability of obtaining the first outcome of the measurement (i.e., ``success'') is equal to $F(\rho,\ket{\psi}\!\bra{\psi})$. In this way, the fidelity provides another natural way for assessing the performance of quantum information processing protocols.
	
	Like the trace distance, the fidelity can  be computed via a semi-definite program. Specifically, in the following, we show that the square root of the fidelity, which we call the ``root fidelity'' and denote by
	\begin{equation}
		\sqrt{F}(\rho,\sigma)\coloneqq\sqrt{F(\rho,\sigma)}=\norm{\sqrt{\rho}\sqrt{\sigma}}_1,
	\end{equation}
    can be computed via a semi-definite program.

	\begin{proposition*}{SDPs for Root Fidelity of States}{prop:QM-over:SDP-fidelity-states}
	The root fidelity $\sqrt{F}(\rho
,\sigma)\coloneqq\sqrt{F(\rho,\sigma)}=\left\Vert \sqrt{\rho}\sqrt{\sigma}\right\Vert _{1}$\ of quantum
states $\rho$ and $\sigma$ is characterized by the following primal and dual
semi-definite programs:
\begin{align}
\sqrt{F}(\rho,\sigma)  & =\frac{1}{2}\sup_{X\in\Lin(\mathcal{H}
)}\left\{  \operatorname{Tr}[X]+\operatorname{Tr}[X^{\dag}]:
\begin{pmatrix}
\rho & X\\
X^{\dag} & \sigma
\end{pmatrix}
\geq0\right\}  \label{eq:QM-over:fidelity-primal-SDP}\\
& =\frac{1}{2}\inf_{Y,Z\geq 0}\left\{  \operatorname{Tr}[Y\rho]+\operatorname{Tr}
[Z\sigma]:
\begin{pmatrix}
Y & \mathbbm{1}\\
\mathbbm{1} & Z
\end{pmatrix}
\geq0\right\}.  \label{eq:QM-over:fidelity-dual-SDP}
\end{align}
	\end{proposition*}
	
	\begin{Proof}
	See Appendix \ref{app:QM-over:SDP-fid-states}.
	\end{Proof}
	
	\begin{theorem*}{Basic Properties of Fidelity}{thm-fidelity_properties}
		\begin{enumerate}
			\item For two states $\rho$ and $\sigma$, the inequalities $0\leq F(\rho,\sigma)\leq 1$ hold. Furthermore, $F(\rho,\sigma)=1$ if and only if $\rho=\sigma$, and $F(\rho,\sigma)=0$ if and only if $\rho$ and $\sigma$ are supported on orthogonal subspaces.
			
			\item\textit{Isometric invariance}: For all states $\rho$ and $\sigma$, and for every isometry~$V$,
				\begin{equation}
					F(\rho,\sigma)=F(V\rho V^\dagger,V\sigma V^\dagger).
				\end{equation}
			
			\item\textit{Multiplicativity}: The fidelity is multiplicative with respect to tensor-product states, meaning that for all states $\rho_1,\sigma_1,\rho_2,\sigma_2$, we have
				\begin{equation}\label{eq-fidelity_multiplicative}
					F(\rho_1\otimes\rho_2,\sigma_1\otimes\sigma_2)=F(\rho_1,\sigma_1)F(\rho_2,\sigma_2).
				\end{equation}
	
		\end{enumerate}
	\end{theorem*}
	
	\begin{Proof}
		\hfill\begin{enumerate}
			\item The fact that $F(\rho,\sigma)\geq 0$ for all states $\rho$ and $\sigma$ follows from the definition of the fidelity as the squared trace norm and the fact that the trace norm is always non-negative. If $\rho$ and $\sigma$ are supported on orthogonal subspaces, then $\sqrt{\rho}\sqrt{\sigma}=0$, which means that $F(\rho,\sigma)=0$. Conversely, if $F(\rho,\sigma)=0$, then $\norm{\sqrt{\rho}\sqrt{\sigma}}_1=0$, which implies (by definition of a norm) that $\sqrt{\rho}\sqrt{\sigma}=0$, which in turn implies that $\rho$ and $\sigma$ are supported on orthogonal subspaces.
			
			Now, using \eqref{eq-trace_norm_variational}, there exists a unitary $U$ such that
			\begin{equation}
				F(\rho,\sigma)=\norm{\sqrt{\rho}\sqrt{\sigma}}_1^2=\abs{\Tr[U\sqrt{\rho}\sqrt{\sigma}]}^2.
			\end{equation}
			Then, using the Cauchy--Schwarz inequality for the Hilbert--Schmidt inner product (see \eqref{eq-Cauchy_Schwarz_HS}), we find that
			\begin{align}
				F(\rho,\sigma)&=\abs{\Tr[U\sqrt{\rho}\sqrt{\sigma}]}^2\\
				&\leq \Tr[U\sqrt{\rho}\sqrt{\rho}U^\dagger]\Tr[\sqrt{\sigma}\sqrt{\sigma}]\\
				&=\Tr[\sqrt{\rho}\sqrt{\rho}]\Tr[\sqrt{\sigma}\sqrt{\sigma}]\\
				&=\Tr[\rho]\Tr[\sigma]\\
				&=1.
			\end{align}
			If $\rho=\sigma$, then $F(\rho,\sigma)=\norm{\rho}_1^2=\Tr[\rho]^2=1$. On the other hand, if $F(\rho,\sigma)=1$, then the inequality in the Cauchy--Schwarz inquality is saturated. The Cauchy--Schwarz inequality is saturated if and only if the two operators involved are proportional to each other. This means that $\alpha \sqrt{\sigma} = U\sqrt{\rho}$ for some $\alpha\in \mathbb{C}$, which must satisfy $|\alpha|=1$ in order for the Hilbert--Schmidt norm of the left-hand side to be $1$. Evaluating this equality on an orthonormal set $\{\ket{i}\}_i$ of eigenvectors of $\rho$ corresponding to non-negative eigenvalues $\lambda_i>0$ yields
            \begin{equation}
                U\ket{i} = \frac{\alpha}{\sqrt{\lambda_i}} \sqrt{\sigma}\ket{i}.
            \end{equation}
            Taking the inner product of two of these identities gives
            \begin{equation}
                \delta_{i,j} = (\lambda_i\lambda_j)^{-1/2} \bra{i}\sigma\ket{j},
            \end{equation}
            which becomes immediately
            \begin{equation}
                \bra{i}\rho\ket{j} = \bra{i}\sigma\ket{j}.
            \end{equation}
            Noting that the vectors $\{\ket{i}\}_i$ span the support of $\rho$, this immediately implies that $\Tr[\Pi_\rho \sigma ] = 1$, where $\Pi_\rho$ denotes the projector onto the support of $\rho$. In other words, $\supp(\sigma)\subseteq \supp(\rho)$, and therefore the equation above implies that $\rho=\sigma$.
			
			\item \textit{Proof of isometric invariance}: For every isometry $V$ and every two states $\rho$ and $\sigma$, since the action of an isometry does not change the eigenvalues, we have that $\sqrt{V\rho V^\dagger}=V\sqrt{\rho}V^\dagger$ and $\sqrt{V\sigma V^\dagger}=V\sqrt{\sigma}V^\dagger$. Therefore,
				\begin{align}
					F(V\rho V^\dagger,V\sigma V^\dagger)&=\norm{\sqrt{V\rho V^\dagger}\sqrt{V\sigma V^\dagger}}_1^2\\
					&=\norm{V\sqrt{\rho}V^\dagger V\sqrt{\sigma}V^\dagger}_1^2\\
					&=\norm{V\sqrt{\rho}\sqrt{\sigma}V^\dagger}_1^2\\
					&=\norm{\sqrt{\rho}\sqrt{\sigma}}_1^2,
				\end{align}
				as claimed, where the last line is due to the isometric invariance of the Schatten norms, as stated in \eqref{eq-Schatten_norm_iso_invar}.
				
			\item\textit{Proof of multiplicativity}: Using the fact that $\sqrt{\rho_1\otimes\rho_2}=\sqrt{\rho_1}\otimes\sqrt{\rho_2}$, and similarly for $\sqrt{\sigma_1\otimes\sigma_2}$, and using the multiplicativity of the trace norm with respect to the tensor product (see \eqref{eq-Schatten_norm_mult}), we find that
				\begin{align}
					F(\rho_1\otimes\rho_2,\sigma_1\otimes\sigma_2)&=\norm{\sqrt{\rho_1\otimes\rho_2}\sqrt{\sigma_1\otimes\sigma_2}}_1^2\\
					&=\norm{(\sqrt{\rho_1}\otimes\sqrt{\rho_2})(\sqrt{\sigma_1}\otimes\sqrt{\sigma_2})}_1^2\\
					&=\norm{\sqrt{\rho_1}\sqrt{\sigma_1}\otimes\sqrt{\rho_2}\sqrt{\sigma_2}}_1^2\\
					&=\norm{\sqrt{\rho_1}\sqrt{\sigma_1}}_1^2\norm{\sqrt{\rho_2}\sqrt{\sigma_2}}_1^2\\
					&=F(\rho_1,\sigma_1)F(\rho_2,\sigma_2),
				\end{align}
				as claimed. \qedhere 
		\end{enumerate}
	\end{Proof}

	\begin{theorem*}{Uhlmann's Theorem}{thm-Uhlmann_fidelity}
		For two quantum states $\rho_A$ and $\sigma_A$, let $\ket{\psi^{\rho}}\coloneqq(\mathbbm{1}_R\otimes\sqrt{\rho_A})\ket{\Gamma}_{RA}$ and $\ket{\psi^{\sigma}}\coloneqq(\mathbbm{1}_R\otimes\sqrt{\sigma_A})\ket{\Gamma}_{RA}$ be purifications of $\rho$ and $\sigma$, respectively, with the dimension of $R$ equal to the dimension of $A$. Then,
		\begin{equation}\label{eq-Uhlmann_fidelity}
			F(\rho,\sigma)=\max_{U_R}\abs{\bra{\psi^{\rho}}_{RA}(U_R\otimes\mathbbm{1}_A)\ket{\psi^{\sigma}}_{RA}}^2,
		\end{equation}
		where the optimization is over unitaries on $R$.
	\end{theorem*}
	
	\begin{remark}
		Since all purifications are related to each other by isometries on the purifying system (which is the system $R$ as in the statement of the theorem), Uhlmann's theorem tells us that the fidelity between two quantum states is equal to the maximum (squared) overlap between their purifications.
		
		Furthermore, it is straightforward to show that it suffices to take the dimension of $R$ the same as the dimension of $A$, as we have done in the statement of the theorem. In other words, performing an optimization over the dimension of $R$ leads to the same result as in \eqref{eq-Uhlmann_fidelity}.

        Another equivalent interpretation of Uhlmann's theorem is as follows: given an arbitrary fixed purification $\ket{\psi^\rho}_{RA}$ of $\rho$ on some bipartite system $RA$, it holds that
        \begin{equation}
            F(\rho,\sigma) = \max_{\ket{\psi^\sigma}_{RA}} \abs{\braket{\psi^{\rho}}{\psi^{\sigma}}_{RA}}^2 ,
            \label{eq:Uhlmann_fixed_purification}
        \end{equation}
        where the maximization is over all purifications $\ket{\psi^\sigma}_{RA}$ of $\sigma$ on $RA$.
	\end{remark}
	
	\begin{Proof}
		Using the definitions of $\ket{\psi^{\rho}}_{RA}$ and $\ket{\psi^{\sigma}}_{RA}$, we find for every unitary $U_R$ that
		\begin{align}
			&\abs{\bra{\psi^{\rho}}_{RA}(U_R\otimes\mathbbm{1}_A)\ket{\psi^{\sigma}}_{RA}}^2\nonumber\\
			&\quad=\abs{\bra{\Gamma}_{RA}(\mathbbm{1}_R\otimes\sqrt{\rho_A})(U_R\otimes\mathbbm{1}_A)(\mathbbm{1}_R\otimes\sqrt{\sigma_A})\ket{\Gamma}_{RA}}^2\\
			&\quad=\abs{\bra{\Gamma}_{RA}(U_R\otimes\sqrt{\rho_A}\sqrt{\sigma_A})\ket{\Gamma}_{RA}}^2\\
			&\quad=\abs{\bra{\Gamma}_{RA}(\mathbbm{1}_R\otimes\sqrt{\rho_A}\sqrt{\sigma_A}U_A^{\t})\ket{\Gamma_{RA}}}^2,
		\end{align}
		where  the last line follows from the transpose trick in \eqref{eq-transpose_trick}. Then, using \eqref{eq-trace_identity}, we find that
		\begin{equation}
			\abs{\bra{\psi^{\rho}}_{RA}(U_R\otimes\mathbbm{1}_A)\ket{\psi^{\sigma}}_{RA}}^2=\abs{\Tr[\sqrt{\rho_A}\sqrt{\sigma_A}U_A^{\t}]}^2.
		\end{equation}
		Since $U_A$ ranges over all unitaries, the same holds for $U_A^{\t}$; hence, we use \eqref{eq-trace_norm_variational} to obtain
		\begin{align}
			\max_U\abs{\bra{\psi^{\rho}}_{RA}(U_R\otimes\mathbbm{1}_A)\ket{\psi^{\sigma}}_{RA}}^2&=\max_U\abs{\Tr[\sqrt{\rho_A}\sqrt{\sigma_A}U_A^{\t}]}^2\\
			&=\max_U\abs{\Tr[\sqrt{\rho_A}\sqrt{\sigma_A}U_A]}^2\\
			&=\norm{\sqrt{\rho_A}\sqrt{\sigma_A}}_1^2,
		\end{align}
		as claimed.
	\end{Proof}

    \begin{exercise}{ex-QM-dist_meas-fidelity_SDP}
        Let $\rho_A$ and $\sigma_A$ be quantum states, and let $\ket{\psi^{\rho}}_{RA}=(\mathbbm{1}_R\otimes\sqrt{\rho_A})\ket{\Gamma}_{RA}$ be a purification of $\rho_A$, with the dimension of $R$ equal to the dimension of $A$. Using Uhlmann's theorem, show that the fidelity $F(\rho,\sigma)$ can be computed using the following primal-dual pair of semi-definite programs:
        \begin{align}
            F(\rho,\sigma)&=\sup_{W_{RA}\geq 0}\Big\{\bra{\psi^{\rho}}W_{RA}\ket{\psi^{\rho}}:\Tr_R[W_{RA}]\leq\sigma_A\Big\}\label{eq-fidelity_SDP_primal}\\
            &=\inf_{Z_A\geq 0}\Big\{\Tr[\sigma_AZ_A]:\mathbbm{1}_R\otimes Z_A\geq\ketbra{\psi^{\rho}}{\psi^{\rho}}_{RA}\Big\}.\label{eq-fidelity_SDP_dual}
        \end{align}
    \end{exercise}

	\begin{theorem*}{Data-Processing Inequality for Fidelity}{thm-fidelity_monotone}
		Let $\rho$ and $\sigma$ be states, and let $\mathcal{N}$ be a quantum channel. Then, 
		\begin{equation}
			 F(\rho,\sigma) \leq F(\mathcal{N}(\rho),\mathcal{N}(\sigma)).
		\end{equation}
	\end{theorem*}
	
	\begin{Proof}
		Recall that every quantum channel $\mathcal{N}_{A\to B}$ can be written in the Stinespring form as $\mathcal{N}_{A\to B}(\rho_A)=\Tr_E[V\rho_A V^\dagger]$, where $V\equiv V_{A\to BE}$ is some isometric extension of $\mathcal{N}$ and $d_E\leq \rank(\Gamma^{\mathcal{N}}_{AB})$. Since we have shown that the fidelity is invariant under isometric channels, it remains to show that the fidelity is non-decreasing under the action of the partial trace. To this end, consider bipartite states $\rho_{AB}$ and $\sigma_{AB}$, and let $\ket{\psi}_{RAB}$ and $\ket{\phi}_{RAB}$ be arbitrary purifications of $\rho_{AB}$ and $\sigma_{AB}$, respectively; here, $d_R=d_Ad_B$. Observe that $\ket{\psi}_{RAB}$ and $\ket{\phi}_{RAB}$ are also purifications of $\rho_A=\Tr_B[\rho_{AB}]$ and $\sigma_A=\Tr_B[\sigma_{AB}]$, respectively. Then, by Uhlmann's theorem, we have that
		\begin{equation}
			F(\rho_A,\sigma_A)=\max_{U_{RB}}\abs{\bra{\psi}_{RAB}(U_{RB}\otimes\mathbbm{1}_A)\ket{\phi_{RAB}}}^2.
		\end{equation}
		By restricting the maximization above to unitaries of the form $U_R\otimes\mathbbm{1}_B$, we have that
		\begin{equation}
			F(\rho_A,\sigma_A)\geq \abs{\bra{\psi}_{RAB}(U_R\otimes\mathbbm{1}_{AB})\ket{\phi}_{RAB}}^2
		\end{equation}
		for all unitaries $U_R$. Therefore,
		\begin{equation}
			\max_{U_R}\abs{\bra{\psi}_{RAB}(U_R\otimes\mathbbm{1}_{AB})\ket{\phi}_{RAB}}^2 \leq \max_{U_{RB}}\abs{\bra{\psi}_{RAB}(U_{RB}\otimes\mathbbm{1}_A)\ket{\phi}_{RAB}}^2.
		\end{equation}
		But, by Uhlmann's theorem,
		\begin{equation}
			\max_{U_R}\abs{\bra{\psi}_{RAB}(U_R\otimes\mathbbm{1}_{AB})\ket{\phi}_{RAB}}^2=F(\rho_{AB},\sigma_{AB}).
		\end{equation}
		Therefore,
		\begin{equation}
			F(\rho_{AB},\sigma_{AB})\leq F(\rho_A,\sigma_A)=F(\Tr_B[\rho_{AB}],\Tr_B[\sigma_{AB}]).
		\end{equation}
		The fidelity thus satisfies the data-processing inequality with respect to the partial trace.
		
		Using the data-processing inequality for the fidelity with respect to the partial trace, along with its invariance under isometries, we conclude that
		\begin{align}
			F(\mathcal{N}(\rho),\mathcal{N}(\sigma))&=F\!\left(\Tr_E[V\rho V^\dagger],\Tr_E[V\rho V^\dagger]\right)\\
			&\geq F(V\rho V^\dagger,V\sigma V^\dagger)\\
			&=F(\rho,\sigma),
		\end{align}
		as claimed.
	\end{Proof}
	
	With Uhlmann's theorem and the data-processing inequality for the fidelity in hand, we can now establish two more properties of the fidelity.
	
	\begin{theorem*}{Concavity of Fidelity}{thm-fidelity_concave}
		The fidelity is concave in either one of its arguments:
		\begin{equation}\label{eq-fidelity_concave}
			F\!\left(\sum_{x\in\mathcal{X}}p(x)\rho^x,\sigma\right)\geq \sum_{x\in\mathcal{X}}p(x)F(\rho^x,\sigma),
		\end{equation}
		where $\mathcal{X}$ is a finite alphabet, $p:\mathcal{X}\to[0,1]$ is a probability distribution, $\sigma$ is a quantum state, and $\{\rho^x\}_{x\in\mathcal{X}}$ is a set of quantum states.
	\end{theorem*}
	
	\begin{Proof}
		By Uhlmann's theorem, we know that the fidelity is given by the maximum overlap between the purifications of the two states under consideration. Based on this, let $\ket{\psi^{\sigma}}_{RA}$ be a purification of $\sigma_A$. Then, for $x\in\mathcal{X}$, let $\ket{\phi^x}_{RA}$ be a purification of $\rho_A^x$ such that $F(\rho_A^x,\sigma)=\abs{\braket{\phi^x}{\psi^{\sigma}}}^2$. Then,
		\begin{align}
			\sum_{x\in\mathcal{X}}p(x)F(\rho_A^x,\sigma_A)&=\sum_{x\in\mathcal{X}}p(x)\abs{\braket{\phi^x}{\psi^{\sigma}}}^2\\
			&=\bra{\psi^{\sigma}}_{RA}\left(\sum_{x\in\mathcal{X}}p(x)\ket{\phi^x}\!\bra{\phi^x}_{RA}\right)\ket{\psi^{\sigma}}_{RA}\\
			&=F\!\left(\sum_{x\in\mathcal{X}}p(x)\ket{\phi^x}\!\bra{\phi^x}_{RA},\ket{\psi^{\sigma}}\!\bra{\psi^{\sigma}}_{RA}\right),
		\end{align}
		where  the last line follows from the formula in \eqref{eq-fidelity_pure_mixed} for the fidelity between a pure state and a mixed state. Then, using the data-processing inequality for the fidelity with respect to the partial trace, we obtain
		\begin{align}
			\sum_{x\in\mathcal{X}}p(x)F(\rho_A^x,\sigma_A)&=F\!\left(\sum_{x\in\mathcal{X}}p(x)\ket{\phi^x}\!\bra{\phi^x}_{RA},\ket{\psi^{\sigma}}\!\bra{\psi^{\sigma}}_{RA}\right)\\
			&\leq F\!\left(\sum_{x\in\mathcal{X}}p(x)\Tr_R[\ket{\phi^x}\!\bra{\phi^x}_{RA}],\Tr_R[\ket{\psi^{\sigma}}\!\bra{\psi^{\sigma}}_{RA}]\right)\\
			&=F\!\left(\sum_{x\in\mathcal{X}}p(x)\rho_A^x,\sigma_A\right),
		\end{align}
		which is the claimed result.
	\end{Proof}
	
	A more general result than the concavity result proved above, namely \textit{joint concavity}, can be obtained if we consider instead the root fidelity.
	
	\begin{theorem*}{Joint Concavity of Root  Fidelity}{thm-joint_concave_sqrt_fid}
		The root fidelity is jointly concave:
		\begin{equation}
			\sqrt{F}\!\left(\sum_{x\in\mathcal{X}}p(x)\rho^x,\sum_{x\in\mathcal{X}}p(x)\sigma^x\right)\geq\sum_{x\in\mathcal{X}}p(x)\sqrt{F}(\rho^x,\sigma^x),
			\label{eq-joint-concave-root-fidelity}
		\end{equation}
		where $\mathcal{X}$ is a finite alphabet, $p:\mathcal{X}\to [0,1]$ is a probability distribution, and $\{\rho^x\}_{x\in\mathcal{X}}$ and $\{\sigma^x\}_{x\in\mathcal{X}}$ are sets of states.
	\end{theorem*}
	
	\begin{Proof}
		This result follows by defining the classical--quantum states
		\begin{align}
		\rho_{XA} & \coloneqq \sum_{x \in \mathcal{X}} p(x) \ket{x}\!\bra{x}_X \otimes \rho^x_A ,\\
		\sigma_{XA} & \coloneqq \sum_{x \in \mathcal{X}} p(x) \ket{x}\!\bra{x}_X \otimes \sigma^x_A ,
		\end{align}
		and observing that
		\begin{align}
		& \sqrt{F}(\rho_{XA},\sigma_{XA}) \nonumber \\
		& = \norm{\sqrt{\rho_{XA}} \sqrt{\sigma_{XA}} }_1 \label{eq-root_fid_joint_concave_pf1}\\
		& = \norm{\left(\sum_{x \in \mathcal{X}}  \ket{x}\!\bra{x}_X \otimes \sqrt{p(x)\rho^x_A}\right)\left(\sum_{x' \in \mathcal{X}}  \ket{x'}\!\bra{x'}_X \otimes \sqrt{p(x')\sigma^{x'}_A}\right) }_1 \label{eq-root_fid_joint_concave_pf2} \\
		& = \norm{\sum_{x \in \mathcal{X}}  \ket{x}\!\bra{x}_X \otimes p(x) \sqrt{\rho^x_A} \sqrt{\sigma^{x}_A} }_1 \label{eq-root_fid_joint_concave_pf3}\\
		& = \sum_{x \in \mathcal{X}}\norm{   p(x) \sqrt{\rho^x_A} \sqrt{\sigma^{x}_A} }_1 \label{eq-root_fid_joint_concave_pf4}\\
		& = \sum_{x \in \mathcal{X}}p(x)\norm{    \sqrt{\rho^x_A} \sqrt{\sigma^{x}_A} }_1 \label{eq-root_fid_joint_concave_pf5}\\
		 & = \sum_{x \in \mathcal{X}}p(x)\sqrt{F}(\rho^x_A, \sigma^{x}_A ).\label{eq-root_fid_joint_concave_pf6}
		\end{align}
		From the above and the data-processing inequality for the fidelity under partial trace (Theorem~\ref{thm-fidelity_monotone}), we conclude that
		\begin{equation}
		\sum_{x \in \mathcal{X}}p(x)\sqrt{F}(\rho^x_A, \sigma^{x}_A ) =
		\sqrt{F}(\rho_{XA},\sigma_{XA}) \leq \sqrt{F}(\rho_{A},\sigma_{A}),
		\end{equation}
		which is equivalent to \eqref{eq-joint-concave-root-fidelity}.
	\end{Proof}
	
	A calculation analogous to that in \eqref{eq-root_fid_joint_concave_pf1}--\eqref{eq-root_fid_joint_concave_pf6} can be used to demonstrate that the root fidelity satisfies the direct-sum property: for every finite alphabet $\mathcal{X}$, probability distributions $p,q:\mathcal{X}\to[0,1]$, and sets $\{\rho_A^x\}_{x\in\mathcal{X}}$, $\{\sigma_A^x\}_{x\in\mathcal{X}}$ of states, we have
	\begin{multline}\label{eq-direct_sum_root_fid}
		\sqrt{F}\left(\sum_{x\in\mathcal{X}}p(x)\ket{x}\!\bra{x}_X\otimes\rho_A^x,\sum_{x\in\mathcal{X}}q(x)\ket{x}\!\bra{x}_X\otimes\sigma_A^x\right)\\ =\sum_{x\in\mathcal{X}}\sqrt{p(x)q(x)}\sqrt{F}(\rho_A^x,\sigma_A^x).
	\end{multline}
	
	Just as the trace distance can be achieved with a measurement, so it holds that the fidelity can also be achieved with a measurement, as we now show.
	
	\begin{theorem*}{Fidelity via Measurement}{thm-fidelity_meas}
		For two states $\rho$ and $\sigma$, the following equality holds:
		\begin{equation}
			F(\rho,\sigma)=\min_{\{M_x\}_x}\left(\sum_{x\in\mathcal{X}}\sqrt{\Tr[M_x\rho]}\sqrt{\Tr[M_x\sigma]}\right)^2,
		\end{equation}
		where the optimization is with respect to every POVM $\{M_x\}_{x \in \mathcal{X}}$ defined over a finite alphabet $\mathcal{X}$.
	\end{theorem*}
	
	\begin{Proof}
		Let $\mathcal{M}$ be a measurement channel defined as
		\begin{equation}
			\mathcal{M}(\rho)=\sum_{x\in\mathcal{X}}\Tr[M_x\rho]\ket{x}\!\bra{x},
		\end{equation}
		where $\{M_x\}_{x\in\mathcal{X}}$ is a POVM and $\mathcal{X}$ is a finite alphabet. Then, by the data-processing inequality for the fidelity with respect to the channel $\mathcal{M}$, and since the action of $\mathcal{M}$ leads to a state that is diagonal in the orthonormal basis $\{\ket{x}\}_{x\in\mathcal{X}}$, we obtain
		\begin{align}
			F(\rho,\sigma)&\leq F(\mathcal{M}(\rho),\mathcal{M}(\sigma))\\
			&=\norm{\sqrt{\mathcal{M}(\rho)}\sqrt{\mathcal{M}(\sigma)}}_1^2\\
			&=\norm{\sqrt{\sum_{x\in\mathcal{X}}\Tr[M_x\rho]\ket{x}\!\bra{x}}\sqrt{\sum_{x\in\mathcal{X}}\Tr[M_x\sigma]\ket{x}\!\bra{x}}}_1^2\\
			&=\norm{\left(\sum_{x\in\mathcal{X}}\sqrt{\Tr[M_x\rho]}\ket{x}\!\bra{x}\right)\left(\sum_{x\in\mathcal{X}}\sqrt{\Tr[M_x\sigma]}\ket{x}\!\bra{x}\right)}_1^2\\
			&=\norm{\sum_{x\in\mathcal{X}}\sqrt{\Tr[M_x\rho]}\sqrt{\Tr[M_x\sigma]}\ket{x}\!\bra{x}}_1^2\\
			&=\left(\sum_{x\in\mathcal{X}}\sqrt{\Tr[M_x\rho]}\sqrt{\Tr[M_x\sigma]}\right)^2.
		\end{align}
		Since the POVM $\{M_x\}_{x\in\mathcal{X}}$ is arbitrary, we find that
		\begin{equation}
			F(\rho,\sigma)\leq\min_{\{M_x\}_x}\left(\sum_{x\in\mathcal{X}}\sqrt{\Tr[M_x\rho]}\sqrt{\Tr[M_x\sigma]}\right)^2.
		\end{equation}
		
		We now prove the reverse inequality by explicitly constructing a POVM that achieves the fidelity. First, observe that we can write the fidelity $F(\rho,\sigma)$ as
		\begin{equation}
			F(\rho,\sigma)=\norm{\sqrt{\rho}\sqrt{\sigma}}_1^2=\Tr\!\left[\sqrt{\sqrt{\sigma}\rho\sqrt{\sigma}}\right]^2=\Tr[A\sigma]^2,
		\end{equation}
		where
		\begin{equation}
			A\coloneqq \sigma^{-\frac{1}{2}}\left(\sigma^{\frac{1}{2}}\rho\sigma^{\frac{1}{2}}\right)^{\frac{1}{2}}\sigma^{-\frac{1}{2}}.
		\end{equation}
		If $\sigma$ is not invertible, then the inverse is understood to be on the support of $\sigma$, in which case $A$ is supported on the support of $\sigma$. So the fidelity is simply equal to the squared expectation value of the Hermitian operator $A$ with respect to the state $\sigma$. Observe that we can also write
		\begin{align}
			F(\rho,\sigma) & =\left(\Tr\!\left[\sqrt{\sqrt{\sigma}\rho\sqrt{\sigma}}\right]\right)^2 \\
		& =\left(\Tr\!\left[\sqrt{\sqrt{\sigma}\Pi_{\sigma}\rho\Pi_{\sigma}\sqrt{\sigma}}\right]\right)^2\\
		& =F(\Pi_{\sigma}\rho\Pi_{\sigma},\sigma),
		\end{align}
		where $\Pi_{\sigma}$ is the projection onto the support of $\sigma$. This holds because $\sqrt{\sigma}=\Pi_{\sigma}\sqrt{\sigma}=\sqrt{\sigma}\Pi_{\sigma}$.
		
		Now, let us consider a measurement in the eigenbasis of $A$. Let
		\begin{equation}
			A=\sum_{i=0}^{d-1}\lambda_i\ket{\psi_i}\!\bra{\psi_i},
		\end{equation}
            be a spectral decomposition of $A$. Setting $r=\rank(A)$, we can assume that $\lambda_i>0$ for all $i=0,\ldots,r-1$, and $\lambda_i=0$ for $i\geq r$. Then,
		\begin{align}
			\Tr[A\sigma]&=\Tr\!\left[\sum_{i=0}^{d-1}\lambda_i\ket{\psi_i}\!\bra{\psi_i}\sigma\right]\\
			&=\sum_{i=0}^{d-1}\lambda_i\bra{\psi_i}\sigma\ket{\psi_i}\\
			&=\sum_{i=0}^{d-1}\sqrt{\bra{\psi_i}\lambda_i\sigma\lambda_i\ket{\psi_i}}\sqrt{\bra{\psi_i}\sigma\ket{\psi_i}}\\
			&=\sum_{i=0}^{r-1}\sqrt{\bra{\psi_i}A\sigma A\ket{\psi_i}}\sqrt{\bra{\psi_i}\sigma\ket{\psi_i}},
		\end{align}
		where  the last line follows because $A\ket{\psi_i}=\lambda_i\ket{\psi_i}$ for all $0\leq i\leq r-1$, and we have used the fact that $\lambda_i=0$ for all $r\leq i\leq d-1$. Now, it is straightforward to show that $A\sigma A=\Pi_{\sigma}\rho\Pi_{\sigma}$. Therefore,
		\begin{align}
			F(\rho,\sigma)=\Tr[A\sigma]^2&=\left(\sum_{i=0}^{r-1} \sqrt{\bra{\psi_i}\Pi_{\sigma}\rho\Pi_{\sigma}\ket{\psi_i}}\sqrt{\bra{\psi_i}\sigma\ket{\psi_i}}\right)^2\\
			&=\left(\sum_{i=0}^{r-1}\sqrt{\bra{\psi_i}\rho\ket{\psi_i}}\sqrt{\bra{\psi_i}\sigma\ket{\psi_i}}\right)^2,\label{eq-fidelity_meas_pf}
		\end{align}
		where  the last line follows because $A$ is defined on the support of $\sigma$, which means that $\Pi_{\sigma}\ket{\psi_i}=\ket{\psi_i}$ for all $0\leq i\leq r-1$. We thus have
		\begin{align}
			\min_{\{M_x\}_x}\left(\sum_{x\in\mathcal{X}}\sqrt{\Tr[M_x\rho]}\sqrt{\Tr[M_x\sigma]}\right)^2&\leq\left(\sum_{i=0}^{r-1}\sqrt{\bra{\psi_i}\rho\ket{\psi_i}}\sqrt{\bra{\psi_i}\sigma\ket{\psi_i}}\right)^2\\
			&=F(\rho,\sigma),
		\end{align}
		which is precisely the reverse inequality, as desired.
	\end{Proof}
	
	By employing Theorem~\ref{thm-fidelity_meas}, we conclude that the data-processing inequality for fidelity in Theorem~\ref{thm-fidelity_monotone} applies beyond quantum channels to all positive, trace-preserving maps.
	
	\begin{proposition*}{Improved Data Processing for Fidelity}{prop-fidelity-monotone-PTP}
		Let $\rho$ and $\sigma$ be quantum states, and let $\mathcal{N}$ be a positive, trace-preserving map. Then the following inequality holds:
		\begin{equation}
			F(\mathcal{N}(\rho), \mathcal{N}(\sigma))\geq F(\rho, \sigma).
		\end{equation}
	\end{proposition*}
	
	\begin{Proof}
	   Let $\{M_x'\}_{x\in\mathcal{X}}$ be a POVM. Then, consider that
	\begin{align}
	\sum_{x\in \mathcal{X}} \sqrt{\Tr[M_x' \mathcal{N}(\rho)]\Tr[M_x' \mathcal{N}(\sigma)]} & = 
	\sum_{x\in \mathcal{X}} \sqrt{\Tr[\mathcal{N}^\dag(M_x') \rho]\Tr[\mathcal{N}^\dag(M_x') \sigma]} \label{eq:improved_dpi_fidelity_proof_1} \\
	& \geq \min_{\{M_x\}_{x \in \mathcal{X}}}\sum_{x\in \mathcal{X}} \sqrt{\Tr[M_x \rho]\Tr[M_x \sigma]} \label{eq:improved_dpi_fidelity_proof_2} \\
	& = \sqrt{F}(\rho,\sigma), \label{eq:improved_dpi_fidelity_proof_3} 
	\end{align}
    where the last equality follows from Theorem~\ref{thm-fidelity_meas}. The inequality holds because $\{\mathcal{N}^\dag(M_x')\}_{x\in\mathcal{X}}$ is a POVM. Indeed, the adjoint map $\mathcal{N}^{\dagger}$ is positive and unital, because $\mathcal{N}$ is a positive, trace-preserving map. Thus, from $M'_x\geq 0$ we deduce that $\mathcal{N}^\dag(M_x') \geq 0$ for all $x \in \mathcal{X}$; also, $\sum_{x\in\mathcal{X}} \mathcal{N}^\dag(M_x') = \mathcal{N}^\dag(\sum_{x\in\mathcal{X}} M_x') = \mathcal{N}^\dag(\mathbbm{1}) = \mathbbm{1}$. Then, because the inequality~\eqref{eq:improved_dpi_fidelity_proof_1}--\eqref{eq:improved_dpi_fidelity_proof_3} 
    holds for every POVM $\{M_x'\}_{x\in\mathcal{X}}$, we can apply Theorem~\ref{thm-fidelity_meas} again to conclude that
	\begin{align}
	\sqrt{F}(\mathcal{N}(\rho), \mathcal{N}(\sigma)) & = \min_{\{M_x'\}_{x\in\mathcal{X}}}\sum_{x\in \mathcal{X}} \sqrt{\Tr[M_x' \mathcal{N}(\rho)]\Tr[M_x' \mathcal{N}(\sigma)]} \\
	& \geq \sqrt{F}(\rho,\sigma),
	\end{align}
	as claimed.
	\end{Proof}
	
	We now establish a useful relation between trace distance and fidelity, often referred to as the ``Fuchs--van de Graaf inequality.''
	
	\begin{theorem*}{Relation Between Trace Distance and Fidelity}{thm-Fuchs_van_de_graaf}
		For  two states $\rho$ and $\sigma$, the following chain of inequalities relates their trace distance with their fidelity:
		\begin{equation}\label{eq-Fuchs_van_de_graaf}
			1-\sqrt{F(\rho,\sigma)}\leq\frac{1}{2}\norm{\rho-\sigma}_1\leq\sqrt{1-F(\rho,\sigma)}.
		\end{equation}
	\end{theorem*}
	
	\begin{Proof}
		We first prove the upper bound on the trace distance. To do so, recall the formula in \eqref{eq-trace_dist_pure_states} for the trace distance between two pure states. If we let $\ket{\psi^{\rho}}_{RA}$ and $\ket{\psi^{\sigma}}_{RA}$ be purifications of $\rho_A$ and $\sigma_A$, respectively, such that $F(\rho_A,\sigma_A)=\abs{\braket{\psi^{\rho}}{\psi^{\sigma}}}^2$, and if we use the data-processing inequality for the trace distance with respect to the partial trace channel $\Tr_R$, then we obtain
		\begin{align}
			\frac{1}{2}\norm{\rho_A-\sigma_A}_1&=\frac{1}{2}\norm{\Tr_R[\ket{\psi^{\rho}}\!\bra{\psi^{\rho}}_{RA}-\ket{\psi^{\sigma}}\!\bra{\psi^{\sigma}}_{RA}]}_1\\
			&\leq \frac{1}{2}\norm{\ket{\psi^{\rho}}\!\bra{\psi^{\rho}}_{RA}-\ket{\psi^{\sigma}}\!\bra{\psi^{\sigma}}_{RA}}_1\\
			&=\sqrt{1-\abs{\braket{\psi^{\rho}}{\psi^{\sigma}}}^2}\\
			&=\sqrt{1-F(\rho_A,\sigma_A)},
		\end{align}
		as claimed.
		
		For the lower bound on the trace distance, we use the results of Theorems~\ref{thm-fidelity_meas} and~\ref{thm-trace_dist_ach_by_meas_channel}. Theorem~\ref{thm-fidelity_meas} tells us that there exists a POVM $\{M_x\}_{x\in\mathcal{X}}$ such that
		\begin{align}
			F(\rho,\sigma)&=\left(\sum_{x\in\mathcal{X}}\sqrt{\Tr[M_x\rho]}\sqrt{\Tr[M_x\sigma]}\right)^2\\
			&\equiv\left(\sum_{x\in\mathcal{X}}\sqrt{p(x)q(x)}\right)^2,
		\end{align}
		where we have let $p(x)\coloneqq\Tr[M_x\rho]$ and $q(x)\coloneqq\Tr[M_x\sigma]$. Using this, observe that
		\begin{align}
			\sum_{x\in\mathcal{X}}\left(\sqrt{\smash[b]{p(x)}}-\sqrt{\smash[b]{q(x)}}\right)^2&=\sum_{x\in\mathcal{X}}\left(p(x)-2\sqrt{\smash[b]{p(x)q(x)}}+q(x)\right)\\
			&=2-2\sum_{x\in\mathcal{X}}\sqrt{\smash[b]{p(x)q(x)}}\\
			&=2-2\sqrt{F(\rho,\sigma)}.
		\end{align}
		Now, Theorem \ref{thm-trace_dist_ach_by_meas_channel} tells us that
		\begin{equation}
			\norm{\rho-\sigma}_1=\max_{\{M_x'\}_{x}}\sum_{x\in\mathcal{X}}|r(x)-s(x)|,
		\end{equation}
		where $r(x)\coloneqq\Tr[M_x'\rho]$ and $s(x)\coloneqq\Tr[M_x'\sigma]$. In particular, for the POVM $\{M_x\}_{x\in\mathcal{X}}$ that achieves the fidelity, we have
		\begin{equation}
			\sum_{x\in\mathcal{X}}|p(x)-q(x)|\leq\norm{\rho-\sigma}_1.
		\end{equation}
		Using this, and the fact that $\left|\sqrt{p(x)}-\sqrt{q(x)}\right|\leq \left|\sqrt{p(x)}+\sqrt{q(x)}\right|$, we obtain
		\begin{align}
			\sum_{x\in\mathcal{X}}\left(\sqrt{\smash[b]{p(x)}}-\sqrt{\smash[b]{q(x)}}\right)^2&\leq \sum_{x\in\mathcal{X}}\abs{\sqrt{\smash[b]{p(x)}}-\sqrt{\smash[b]{q(x)}}}\abs{\sqrt{\smash[b]{p(x)}}+\sqrt{\smash[b]{q(x)}}}\\
			&=\sum_{x\in\mathcal{X}}|p(x)-q(x)|\\
			&\leq \norm{\rho-\sigma}_1.
		\end{align}
		Therefore,
		\begin{equation}
			2-2\sqrt{F(\rho,\sigma)}=\sum_{x\in\mathcal{X}}\left(\sqrt{\smash[b]{p(x)}}-\sqrt{\smash[b]{q(x)}}\right)^2\leq\norm{\rho-\sigma}_1,
		\end{equation}
		which is the claimed lower bound.
	\end{Proof}

The next result formalizes a physically intuitive feature of quantum theory: if a quantum measurement on a given state $\rho$ yields a specific outcome with probability close to $1$, then the corresponding post-measurement state $\rho^{\prime}$, as defined in~\eqref{eq:gentle-measurement-rho-prime} below, must be very close to the initial state.

\begin{Lemma*}
{Gentle Measurement}{lem-dm:gentle-measurement}
Let $\rho$ be a density
operator. Consider a measurement of $\rho$ with respect to the POVM $\{M,\mathbbm{1}-M\}$, where $0\leq M\leq\mathbbm{1}$. If $\operatorname{Tr}[M\rho]  \geq 1-\varepsilon$, for $\varepsilon\in[0,1]$, then the post-measurement state
\begin{equation}
\rho^{\prime}\coloneqq \frac{\sqrt{M}\rho\sqrt{M}}{\operatorname{Tr}
[  M\rho]  }
\label{eq:gentle-measurement-rho-prime}
\end{equation}
satisfies
\begin{align}
F(\rho,\rho') & \geq 1-\varepsilon , \\
\frac{1}{2} \left\Vert \rho-\rho^{\prime}\right\Vert _{1} & \leq\sqrt{\varepsilon}.
\end{align}
\end{Lemma*}

\begin{Proof}
    We start with the fact that
    \begin{equation}
        F(\rho,\rho')=\frac{1}{\Tr[M\rho]}\left\Vert\sqrt{\rho}\sqrt{\sqrt{M}\rho\sqrt{M}}\right\Vert_1^2.
    \end{equation}
    Then, using the fact that $\sqrt{\rho}\sqrt{M}\sqrt{\rho}$ is positive semi-definite, which implies that 
    \begin{equation}
    \sqrt{\left(\sqrt{\rho}\sqrt{M}\sqrt{\rho}\right)^2}=\sqrt{\rho}\sqrt{M}\sqrt{\rho},
    \end{equation}
    we obtain
    \begin{align}
            \left\Vert\sqrt{\rho}\sqrt{\sqrt{M}\rho\sqrt{M}}\right\Vert_1&=\Tr\!\left[\sqrt{\sqrt{\rho}\sqrt{M}\sqrt{\rho}\sqrt{\rho}\sqrt{M}\sqrt{\rho}}\right]\\
            &=\Tr\!\left[\sqrt{\left(\sqrt{\rho}\sqrt{M}\sqrt{\rho}\right)^2}\right]\\
            &=\Tr\big[\sqrt{M}\rho\big]\\
            &\geq\Tr[M\rho],
        \end{align}
        where the inequality holds because  $\sqrt{M}\geq M$ provided that $M\leq\mathbbm{1}$. Therefore, 
        \begin{equation}
            F(\rho,\rho')\geq\Tr[M\rho]\geq 1-\varepsilon,
        \end{equation}
        where the second inequality holds by assumption. We then obtain the desired bound on the normalized trace distance by applying Theorem~\ref{thm-Fuchs_van_de_graaf}.
\end{Proof}

\subsection{Sine Distance}
	
	Unlike the trace distance, the fidelity is not a distance measure in the mathematical sense because it does not satisfy the triangle inequality. The following distance measure based on the fidelity, however, does satisfy the triangle inequality, along with the other properties that define a distance measure.

	\begin{definition}{Sine Distance}{def-purified_distance}
		For  two states $\rho$ and $\sigma$, we define the \textit{sine distance} between $\rho$ and $\sigma$ as
		\begin{equation}\label{eq-purified_distance}
			P(\rho,\sigma)\coloneqq \sqrt{1-F(\rho,\sigma)}.
		\end{equation}
	\end{definition}
	
	The measure $P(\rho,\sigma)$ is known as the sine distance due to the fact that $F(\rho,\sigma)$ has the interpretation as the largest value of the squared cosine of the angle between two arbitrary purifications of $\rho$ and $\sigma$ (see Theorem~\ref{thm-Uhlmann_fidelity}), 
    which means that $\sqrt{1-F(\rho,\sigma)}$ has the interpretation as being the sine of the same angle. Related to this interpretation, the measure $P(\rho,\sigma)$ is equal to the minimum trace distance between purifications of $\rho$ and $\sigma$:
	\begin{multline}
		\min_{\ket{\psi^\rho}_{RA},\, \ket{\psi^\sigma}_{RA}} \frac{1}{2}\norm{\ket{\psi^\rho}\!\bra{\psi^\rho}_{RA}- \ket{\psi^\sigma}\!\bra{\psi^\sigma}_{RA}}_1 \\= \min_{\ket{\psi^\rho}_{RA},\, \ket{\psi^\sigma}_{RA}} \sqrt{1-\vert \braket{\psi^\rho}{\psi^\sigma}_{RA} \vert^2} = P(\rho,\sigma), \label{eq:sine_distance_min_purifications}
	\end{multline}
	where the optimization is over all purifications $\ket{\psi^\rho}_{RA}$ and $\ket{\psi^\sigma}_{RA}$ of $\rho$ and $\sigma$, respectively. This follows by applying \eqref{eq-trace_dist_pure_states}, as well as Uhlmann's theorem (Theorem~\ref{thm-Uhlmann_fidelity}). In fact, from~\eqref{eq:Uhlmann_fixed_purification} we even deduce a slightly stronger statement: for every fixed purification $\ket{\psi^\rho}_{RA}$ of $\rho$ on a bipartite system $RA$, it holds that
    \begin{equation}
        P(\rho,\sigma) = \min_{\ket{\psi^\sigma}_{RA}} \frac{1}{2}\norm{\ket{\psi^\rho}\!\bra{\psi^\rho}_{RA}- \ket{\psi^\sigma}\!\bra{\psi^\sigma}_{RA}}_1 ,
        \label{eq:sine_distance_one_sided_min_purifications}
    \end{equation}
    where the minimization is over every purification $\ket{\psi^\sigma}_{RA}$ of $\sigma$ on the same system~$RA$. Let us finally note that, due to the aforementioned connection with purifications, the quantity $P(\rho,\sigma)$ is also referred to as \textit{purified distance}.
	
	Since the fidelity satisfies the data-processing inequality with respect to positive, trace-preserving maps (Proposition~\ref{prop-fidelity-monotone-PTP}), so does the sine distance: for  two states $\rho$ and $\sigma$ and a positive, trace-preserving map $\mathcal{N}$, we have that
	\begin{equation}\label{eq-sine_dist_data_proc}
		P(\rho,\sigma)\geq P(\mathcal{N}(\rho),\mathcal{N}(\sigma)).
	\end{equation}
	
	\begin{Lemma*}{Triangle Inequality for Sine Distance}{lem-sine-distance-triangle}
		Let $\rho$, $\sigma$, and $\omega$ be quantum states. Then the triangle inequality holds for the sine distance:
		\begin{equation}
		P(\rho,\sigma) \leq P(\rho,\omega) + P(\omega,\sigma).
		\end{equation}
	\end{Lemma*}

	\begin{Proof}	
		Define the canonical purifications as
		\begin{align}
			\ket{\psi^{\rho}}_{RA}&=(\mathbbm{1}_{R}\otimes\sqrt{\rho_{A}})\ket{\Gamma}_{RA},\\
			\ket{\psi^{\sigma}}_{RA}&=(\mathbbm{1}_{R}\otimes\sqrt{\sigma_{A}})\ket{\Gamma}_{RA},\\
			\ket{\psi^{\omega}}_{RA}&=(\mathbbm{1}_{R}\otimes\sqrt{\omega_{A}})\ket{\Gamma}_{RA},
		\end{align}
		where $\ket{\Gamma}_{RA}$ is the maximally entangled vector from \eqref{eq-max_ent_vector}. Recalling \eqref{eq-trace_dist_pure_states}, for pure states
$|\phi\rangle$ and $|\varphi\rangle$, we have that
\begin{equation}
\frac{1}{2}\left\Vert |\phi\rangle\!\langle\phi|-|\varphi\rangle\!\langle
\varphi|\right\Vert _{1}=\sqrt{1-F(\phi,\varphi)}=\sqrt{1-|\langle\phi
|\varphi\rangle|^{2}}.
\end{equation}
Let $U_{R}$ and $V_{R}$ be arbitrary unitaries acting on the reference system
$R$. From the fact that trace distance obeys the triangle inequality and the
equality given above, we find that
\begin{multline}
\sqrt{1-|\langle\psi^{\sigma}|_{RA}(W_R\otimes \mathbbm{1}_{A})|\psi^{\rho
}\rangle_{RA}|^{2}}\leq\sqrt{1-|\langle\psi^{\sigma}|_{RA}(U_{R}^{\dag}\otimes
\mathbbm{1}_{A})|\psi^{\omega}\rangle_{RA}|^{2}}\\
+\sqrt{1-|\langle\psi^{\omega}|_{RA}(V_{R}\otimes \mathbbm{1}_{A})|\psi^{\rho}\rangle
_{RA}|^{2}},
\end{multline}
where $W_R\coloneqq U_R^\dagger V_R$. By applying
Uhlmann's theorem, we find that
\begin{align}
\sqrt{1-F(\sigma_{A},\rho_{A})} &= \min_{W'_R} \sqrt{1-|\langle\psi^{\sigma}|_{RA}(W'_R\otimes \mathbbm{1}_{A})|\psi^{\rho
}\rangle_{RA}|^{2}} \\
&\leq \sqrt{1-|\langle\psi^{\sigma}|_{RA}(W_R\otimes \mathbbm{1}_{A})|\psi^{\rho
}\rangle_{RA}|^{2}} \\
&\leq \sqrt{1-|\langle\psi^{\sigma}|_{RA}(U_{R}^{\dag}\otimes \mathbbm{1}_{A})|\psi^{\omega}\rangle_{RA}|^{2}} \nonumber \\
& \qquad +\sqrt{1-|\langle
\psi^{\omega}|_{RA}(V_{R}\otimes \mathbbm{1}_{A})|\psi^{\rho}\rangle_{RA}|^{2}}.
\end{align}
Since the inequality holds for arbitrary unitaries $U$ and $V$, it holds for
the minimum of each term on the right, and so this, combined with Uhlmann's
theorem (Theorem~\ref{thm-Uhlmann_fidelity}), implies the desired result:
\begin{equation}
\sqrt{1-F(\sigma_{A},\rho_{A})}\leq\sqrt{1-F(\sigma_{A},\omega_{A})}
+\sqrt{1-F(\omega_{A},\rho_{A})},
\end{equation}
concluding the proof.
%
%
\end{Proof}

\begin{exercise}{exer-sine_dist_triangle_ineq}
    Provide an alternative proof of the triangle inequality for sine distance by making use of \eqref{eq:sine_distance_one_sided_min_purifications} along with the triangle inequality for trace distance.
\end{exercise}

\section{Diamond Distance}
	
	Just as there are measures of distinguishability for quantum states, it is important to develop measures of distinguishability for quantum channels, in order to assess the performance of quantum information-processing protocols that attempt to simulate an ideal process. The measures that we introduce in this section are again motivated by operational concerns, stemming from the ability of an experimenter to distinguish one quantum channel from another when given access to a single use of the channel. In what follows, our discussion mirrors and generalizes the discussion in Section~\ref{sec-QM-trace-distance} that motivated trace distance as a measure of distinguishability for quantum states.
	
	How should we measure the distance between two quantum channels $\mathcal{N}_{A\to B}$ and $\mathcal{M}_{A\to B}$? Related, how should we assess the performance of a quantum information-processing protocol in which the ideal channel to be simulated is $\mathcal{N}_{A\to B}$ but the channel realized in practice is $\mathcal{M}_{A\to B}$? Suppose that a third party is trying to assess how distinguishable the actual channel $\mathcal{M}_{A\to B}$ is from the ideal channel $\mathcal{N}_{A\to B}$. Such an individual, henceforth called ``the distinguisher'', has access to both the input and output ports of the channel, and so the most general strategy that can be implemented is to prepare a state $\rho_{RA}$ of a reference system $R$ and the channel input system $A$. Then the distinguisher transmits the $A$ system of $\rho_{RA}$ via the unknown channel, receiving the corresponding output system $B$. The next step is to perform a measurement, described by the POVM $\{M^x_{RB}\}_x$, on the reference system $R$ and the channel output system~$B$. The probability of obtaining a particular outcome $M^x_{RB}$ is given by the Born rule. In the case that the unknown channel is $\mathcal{N}_{A\to B}$,  this probability is $\Tr[M^x_{RB}\mathcal{N}_{A\to B}(\rho_{RA})]$, and in the case that the unknown channel is $\mathcal{M}_{A\to B}$,  this probability is $\Tr[M^x_{RB}\mathcal{M}_{A\to B}(\rho_{RA})]$. What we demand is that the absolute deviation between the two probabilities $\Tr[M^x_{RB}\mathcal{N}_{A\to B}(\rho_{RA})]$ and $\Tr[M^x_{RB}\mathcal{M}_{A\to B}(\rho_{RA})]$ is no larger than some tolerance $\varepsilon$. Since this should be the case for all possible input states and measurement outcomes, what we demand mathematically is that 
	\begin{equation}
	\sup_{\rho_{RA},\ 0\leq M_{RB} \leq \mathbbm{1}_{RB}} |\Tr[M_{RB}\mathcal{N}_{A\to B}(\rho_{RA})] - 
	\Tr[M_{RB}\mathcal{M}_{A\to B}(\rho_{RA})]| \leq \varepsilon.
	\end{equation}
	As a consequence of the characterization of trace distance from Theorem~\ref{thm-trace_dist_meas}, we have that
	\begin{multline}\label{eq-diamond_dist_meas}
	\sup_{\rho_{RA},\ 0\leq M_{RB} \leq \mathbbm{1}_{RB}} |\Tr[M_{RB}(\mathcal{N}_{A\to B} - 
	\mathcal{M}_{A\to B})(\rho_{RA})]|\\=\sup_{\rho_{RA}}\frac{1}{2}\norm{\mathcal{N}_{A\to B}(\rho_{RA})-\mathcal{M}_{A\to B}(\rho_{RA})}_1\eqqcolon\frac{1}{2}\norm{\mathcal{N}-\mathcal{M}}_{\diamond},
	\end{multline}
	where $\frac{1}{2}\norm{\mathcal{N} - \mathcal{M}}_{\diamond}$ is defined to be the \textit{normalized diamond distance} between $\mathcal{N}$ and $\mathcal{M}$, and we recall the definition of the diamond norm $\norm{\cdot}_{\diamond}$ from Definition~\ref{def-diamond_norm_superop}. 
    Therefore, if $\frac{1}{2}\norm{\mathcal{N} - \mathcal{M}}_{\diamond} \leq \varepsilon$, then the absolute deviation between probabilities for every possible input state and measurement operator never exceeds $\varepsilon$, so that the approximation between quantum channels $\mathcal{N}_{A\to B}$ and $\mathcal{M}_{A\to B}$ is naturally quantified by the normalized diamond distance $\frac{1}{2}\norm{\mathcal{N} - \mathcal{M}}_{\diamond}$. 

	For two quantum channels $\mathcal{N}$ and $\mathcal{M}$, note that the difference $\mathcal{N}-\mathcal{M}$ is a Hermiticity-preserving map. An important simplification of the diamond norm of a Hermiticity-preserving map is given in Theorem~\ref{thm-diamond_norm}, which we recall here for convenience.
	
	\begin{proposition}{prop-diamond_norm_herm_pres}
		The diamond norm of a Hermiticity-preserving map $\mathcal{P}_{A\to B}$ can be calculated as
		\begin{equation}\label{eq-diamond_norm_alt}
			\norm{\mathcal{P}}_{\diamond}=\sup_{\psi_{RA}}\norm{\mathcal{P}_{A\to B}(\psi_{RA})}_1,
		\end{equation}
		where the optimization is over every pure state $\psi_{RA}$, such that the dimension of $R$ is equal to the dimension of the system $A$.
	\end{proposition}

	
	
		The diamond norm of a Hermiticity-preserving map and the normalized diamond distance of two quantum channels can be computed via semi-definite programs (SDPs), which we now show. 

    \begin{proposition*}{SDPs for Diamond Norm and Normalized Diamond Distance}{prop-diamond_dist_SDP}
        The diamond norm of a Hermiticity-preserving map $\mathcal{P}_{A\to B}$, with Choi representation $\Gamma_{AB}^{\mathcal{P}}$, can be written as the following primal-dual pair of semi-definite programs:
        \begin{align}
            \norm{\mathcal{P}}_{\diamond}&=\sup_{\substack{S_{AB}^1,S_{AB}^2\geq 0,\\\rho_A\geq 0}}\Big\{\Tr\big[\big(S_{AB}^1-S_{AB}^2\big)\Gamma_{AB}^{\mathcal{P}}\big]: S_{AB}^1+S_{AB}^2=\rho_A\otimes\mathbbm{1}_B,\,\Tr[\rho_A]=1\Big\}\label{eq-diamond_norm_SDP_primal}\\
            &=\inf_{\mu\geq 0,\,Z_{AB}\geq 0}\Big\{\mu: -Z_{AB}\leq\Gamma_{AB}^{\mathcal{P}}\leq Z_{AB},\,\Tr_B[Z_{AB}]\leq\mu\mathbbm{1}_{A}\Big\}.\label{eq-diamond_norm_SDP_dual}
        \end{align}
         In particular, for quantum channels $\mathcal{N}_{A\to B}$ and $\mathcal{M}_{A\to B}$, their normalized diamond distance is given by
        \begin{align}
			&\frac{1}{2}\norm{\mathcal{N}-\mathcal{M}}_{\diamond}\nonumber\\
			&\quad=\sup_{\substack{\rho_A\geq 0,\\S_{AB}\geq 0}}\Big\{\Tr\big[S_{AB}\big(\Gamma_{AB}^{\mathcal{N}}-\Gamma_{AB}^{\mathcal{M}}\big)\big]:\, S_{AB}\leq\rho_A\otimes\mathbbm{1}_B,\, \Tr[\rho_A]=1 \Big\}\label{eq-diamond_dist_SDP_primal}\\[0.2cm]
			&\quad=\inf_{\substack{\mu\geq 0,\\Z_{AB}\geq 0}}\Big\{ \mu:\, Z_{AB}\geq \Gamma_{AB}^{\mathcal{N}}-\Gamma_{AB}^{\mathcal{M}},\ \mu\mathbbm{1}_A\geq \Tr_B[Z_{AB}]\Big\}.\label{eq-diamond_dist_SDP_dual}
		\end{align}
        The latter expression is equal to
		\begin{equation}
			\inf_{Z_{AB}\geq 0}\Big\{\norm{\Tr_B[Z_{AB}]}_{\infty}:Z_{AB}\geq \Gamma_{AB}^{\mathcal{N}}-\Gamma_{AB}^{\mathcal{M}}\Big\}.
		\end{equation}
    \end{proposition*}

    \begin{Proof}
        See Appendix~\ref{app-trace_dist_fidelity_SDP}.
    \end{Proof}

	As described at the beginning of this section, the diamond distance has an operational meaning in terms of the task of \textit{channel discrimination}, which is a generalization of state discrimination (see Section \ref{subsec-state_discrimination}). Let us now analyze the task of channel discrimination in more detail.
	
	Suppose that Alice gives Bob a device that implements either the channel $\mathcal{N}$ or the channel $\mathcal{M}$, but she does not tell him which one. Bob's task is to decide between these two options. Suppose that the channels $\mathcal{N}$ and $\mathcal{M}$ have prior probabilities $\lambda$ and $1-\lambda$ of being selected, respectively. The only way for Bob to determine which channel the device implements (without guessing randomly) is to pass a quantum system, say in the state $\rho$, through it. He can then perform a measurement on the resulting output state and make a guess as to which channel was implemented. Therefore, in addition to having the freedom to choose any binary measurement (which is the case in state discrimination), in channel discrimination Bob also has the freedom to prepare a system $A$ and a reference system $R$ in any state $\rho_{RA}$ of his choosing, with the system $A$ being passed through the device.

	For every fixed input state $\rho_{RA}$, there are two possible output states, depending on which channel was implemented. This means that, for every fixed input state, the task of channel discrimination reduces to the task of state discrimination. Using the result of Theorem~\ref{thm-Holevo_Helstrom}, for the input state $\rho_{RA}$ the corresponding optimal error probability (i.e., the error probability obtained by optimizing over all measurements) is
	\begin{equation}
		p_{\text{err}}^*(\rho_{RA})=\frac{1}{2}\left(1-\norm{\lambda\mathcal{N}_{A\to B}(\rho_{RA})-(1-\lambda)\mathcal{M}_{A\to B}(\rho_{RA})}_1\right).
	\end{equation}
	Then, optimizing over all input states $\rho_{RA}$ in order to minimize the error probability, we find that
	\begin{align}
		\inf_{\rho_{RA}}p_{\text{err}}^*(\rho_{RA})&=\frac{1}{2}\left(1-\sup_{\rho_{RA}}\norm{(\lambda\mathcal{N}_{A\to B}-(1-\lambda)\mathcal{M}_{A\to B})(\rho_{RA})}_1\right)\\
		&=\frac{1}{2}\left(1-\norm{\lambda\mathcal{N}-(1-\lambda)\mathcal{M}}_{\diamond}\right),
	\end{align}
	where  the last line follows from the definition of the diamond norm. The optimal error probability for the task of channel discrimination is thus a simple function of the diamond norm.

\section{Fidelity Measures for Channels}

\label{sec-QM-DM:fid-meas-chan}

	The diamond distance is a distance measure for channels that is based on the trace distance for states. We now define a fidelity-based quantity for channels that can be used to assess their ability to preserve entanglement. 
		
	\begin{definition}{Entanglement Fidelity of a Channel}{def-ent_fid_chan}
		For a quantum channel $\mathcal{N}_{A}$ with input and output systems of equal dimension $d$, we define its \textit{entanglement fidelity} as
		\begin{equation}
			F_e(\mathcal{N})\coloneqq \bra{\Phi}_{RA} (\id_R\otimes \mathcal{N}_{A})(\Phi_{RA})\ket{\Phi}_{RA},
		\end{equation}
		where
		\begin{equation}
			\ket{\Phi}_{RA}=\frac{1}{\sqrt{d}}\sum_{i=0}^{d-1}\ket{i,i}_{RA}.
		\end{equation}
	\end{definition}

	Notice that the entanglement fidelity of a channel is the fidelity of the maximally entangled state with the Choi state of the channel. Intuitively, then, the entanglement fidelity quantifies how good a channel is at preserving the entanglement between two systems when it acts on one of the two systems.
	
	It turns out that the entanglement fidelity is very closely related to another fidelity-based measure on quantum channels called the \textit{average fidelity}:
	\begin{equation}\label{eq-chan_avg_fid}
		\overline{F}(\mathcal{N})\coloneqq \int_{\psi} \bra{\psi}\mathcal{N}(\psi)\ket{\psi}~\text{d}\psi,
	\end{equation}
	where we integrate over all pure states acting on the input Hilbert space of $\mathcal{N}$ with respect to the Haar measure. The Haar measure is the uniform probability measure on pure quantum states (see the remark after \eqref{eq-sym_proj_integral}). For a quantum channel $\mathcal{N}$ with input system dimension $d$, the following identity holds
	\begin{equation}
		\overline{F}(\mathcal{N})=\frac{dF_e(\mathcal{N})+1}{d+1}.
		\label{eq:QM-over:average-fid-to-ent-fid}
	\end{equation}
	
	Instead of taking the average as in \eqref{eq-chan_avg_fid}, we can take the minimum over all input states to obtain the \textit{minimum fidelity}:
	\begin{equation}\label{eq-minimum_fidelity_chan}
		F_{\min}(\mathcal{N})\coloneqq\inf_{\psi} \bra{\psi}\mathcal{N}(\psi)\ket{\psi},
	\end{equation}
	where the optimization is over all pure states $\psi$ acting on the input Hilbert space of the channel $\mathcal{N}$. By introducing a reference system $R$ and optimizing over all joint states $\ket{\psi}_{RA}$ of $R$ and the input system $A$ of the channel $\mathcal{N}$, we obtain a fidelity measure that generalizes the entanglement fidelity.
	
	\begin{definition}{Fidelity of a Quantum Channel}{eq-worse_case_fid_chan}
		For a quantum channel $\mathcal{N}_A$ with equal input and output system dimension, we define the \textit{fidelity of $\mathcal{N}$} as
		\begin{equation}\label{eq-worst_case_fidelity_chan}
			F(\mathcal{N})\coloneqq \inf_{\psi_{RA}}\bra{\psi}_{RA}(\id_R \otimes \mathcal{N}_{A})(\psi_{RA})\ket{\psi}_{RA},
		\end{equation}
		where we take the infimum over all pure states $\ket{\psi}_{RA}$, with the dimension of $R$ equal to the dimension of $A$.
	\end{definition}
	
	Note that the state $\ket{\psi}_{RA}=\ket{\Phi}_{RA}$ is a special case in the optimization in \eqref{eq-worst_case_fidelity_chan}. This implies that $F(\mathcal{N})\leq F_e(\mathcal{N})$ for every channel $\mathcal{N}$.
	
	More generally, we define the fidelity between two quantum channels $\mathcal{N}_{A\to B}$ and $\mathcal{M}_{A\to B}$ as follows:
	
    \begin{definition}{Fidelity of Quantum Channels}{def:QM-over:fid-channels}
	Let $\mathcal{N}_{A\to B}$ and $\mathcal{M}_{A\to B}$ be quantum channels. Their 
    fidelity is defined as
	\begin{equation}\label{eq-fidelity_chan_distance}
		F(\mathcal{N},\mathcal{M})=\inf_{\rho_{RA}}F(\mathcal{N}_{A\to B}(\rho_{RA}),\mathcal{M}_{A\to B}(\rho_{RA})),
	\end{equation}
	where the infimum is taken over all bipartite states $\rho_{RA}$, with the dimension of $R$ arbitrarily large.
	\end{definition}

	\begin{remark}
	
	Similar to the diamond distance, we define the channel fidelity as above in order to indicate its operational meaning with an infimum over all possible input states, but it is not necessary to take the infimum over all bipartite states. One can instead restrict the infimum to be over pure bipartite states where the reference system $R$ has the same dimension as the channel input system $A$, so that
	\begin{equation}
	F(\mathcal{N},\mathcal{M})=\inf_{\psi_{RA}}F(\mathcal{N}_{A\to B}(\psi_{RA}),\mathcal{M}_{A\to B}(\psi_{RA})),
	\end{equation}
	where $\psi_{RA}$ is a pure bipartite state with system $R$ having the same dimension as system $A$. The same statement thus applies to \eqref{eq-worst_case_fidelity_chan}.
	This can be proved by applying the same reasoning given later on in~\eqref{eq-gen_div_chan_pure_2}--\eqref{eq-gen_div_chan_pure_4} in a more general scenario.
	
	Here, we provide a different argument for this fact. First, we have that
		\begin{align}
			\inf_{\rho_{RA}}F(\mathcal{N}_{A\to B}(\rho_{RA}),\mathcal{M}_{A\to B}(\rho_{RA}))\leq \inf_{\psi_{RA}}F(\mathcal{N}_{A\to B}(\psi_{RA}),\mathcal{M}_{A\to B}(\psi_{RA}))
		\end{align}
		which holds simply by restricting the optimization on the left-hand side to pure states.
		
		Next, given a state $\rho_{RA}$, with the dimension of $R$ not necessarily equal to the dimension of $A$, we can purify it to a state $\psi_{R'RA}$. Then, using the data-processing inequality for the fidelity with respect to the partial trace channel $\Tr_{R'}$ (Proposition~\ref{prop-fidelity-monotone-PTP}), we find that
		\begin{align}
			F(\mathcal{N}_{A\to B}(\rho_{RA}),\mathcal{M}_{A\to B}(\rho_{RA}))&=F\left(\mathcal{N}_{A\to B}(\Tr_{R'}[\psi_{R'RA}]),\mathcal{M}_{A\to B}(\Tr_{R'}[\psi_{R'RA}])\right)\\
			&=F\left(\Tr_{R'}[\mathcal{N}_{A\to B}(\psi_{R'RA})],\Tr_{R'}[\mathcal{M}_{A\to B}(\psi_{R'RA})]\right)\\
			&\geq F(\mathcal{N}_{A\to B}(\psi_{R'RA}),\mathcal{M}_{A\to B}(\psi_{R'RA}))\\
			&\geq \inf_{\psi_{R'RA}}F(\mathcal{N}_{A\to B}(\psi_{R'RA}),\mathcal{M}_{A\to B}(\psi_{R'RA})).
		\end{align}
		Since the state $\rho_{RA}$ is arbitrary, we obtain
		\begin{equation}
			\inf_{\rho_{RA}}F(\mathcal{N}_{A\to B}(\rho_{RA}),\mathcal{M}_{A\to B}(\rho_{RA}))\geq \inf_{\psi_{RA}}F(\mathcal{N}_{A\to B}(\psi_{RA}),\mathcal{M}_{A\to B}(\psi_{RA})).
		\end{equation}
		
		Finally, by the Schmidt decomposition theorem (Theorem~\ref{thm-Schmidt}), for every pure state $\psi_{RA}$, the rank of the reduced state $\psi_R$ need not exceed the dimension of $A$, implying that it suffices to optimize over pure states for which the system $R$ has the same dimension as the system $A$. We thus obtain
		\begin{align}
			\inf_{\rho_{RA}}F(\mathcal{N}_{A\to B}(\rho_{RA}),\mathcal{M}_{A\to B}(\rho_{RA}))&=\inf_{\psi_{RA}}F(\mathcal{N}_{A\to B}(\psi_{RA}),\mathcal{M}_{A\to B}(\psi_{RA})\\
			&=F(\mathcal{N},\mathcal{M}).
		\end{align}
	\end{remark}
	
	We then have that $F(\mathcal{N})=F(\mathcal{N},\id)$. In other words, the fidelity $F(\mathcal{N})$ of a quantum channel $\mathcal{N}$ is equal to the fidelity between $\mathcal{N}$ and the identity channel $\id$.
	
	Similar to the diamond distance, the fidelity of quantum channels can be computed by means of semi-definite programs.
	
%

\begin{proposition*}{SDP for Root Fidelity of Channels}
{prop:QM-over:SDP-fidelity-channels}
Let $\mathcal{N}_{A\rightarrow B}$ and $\mathcal{M}
_{A\rightarrow B}$ be quantum channels with respective Choi operators
$\Gamma_{RB}^{\mathcal{N}}$ and $\Gamma_{RB}^{\mathcal{M}}$. Then their root
channel fidelity
\begin{equation}
\sqrt{F}(\mathcal{N}_{A\rightarrow B},\mathcal{M}_{A\rightarrow B}
)\coloneqq\inf_{\psi_{RA}}\sqrt{F}(\mathcal{N}_{A\rightarrow B}(\psi_{RA}
),\mathcal{M}_{A\rightarrow B}(\psi_{RA}))
\end{equation}
can be calculated by means of the following semi-definite program:
\begin{align}
& \sqrt{F}(\mathcal{N}_{A\rightarrow B},\mathcal{M}_{A\rightarrow B}
) \notag \\
& = \sup_{\lambda\geq0,\, Q_{RB}}\left\{  \lambda:\ \lambda \mathbbm{1}_{R}\leq\operatorname{Re}
[\operatorname{Tr}_{B}[Q_{RB}]],\ \
\begin{pmatrix}
\Gamma_{RB}^{\mathcal{N}} & Q_{RB}^{\dag} \\[.8ex]
Q_{RB} & \Gamma_{RB}^{\mathcal{M}}
\end{pmatrix}
\geq0\right\} \label{eq:SDP-root-fid-ch}\\
& = \frac{1}{2}\inf_{\substack{\rho_{R}\geq 0,\\W_{RB},\,Z_{RB}\text{ Hermitian}}}
\Bigg\{\operatorname{Tr}\!\big[\Gamma
_{RB}^{\mathcal{N}}W_{RB}\big]+\operatorname{Tr}\!\big[\Gamma_{RB}^{\mathcal{M}}Z_{RB}\big]:\ 
\operatorname{Tr}[\rho_{R}]=1,\notag \\
& \qquad\qquad\qquad\qquad\qquad\qquad\quad
\begin{pmatrix}
W_{RB} & \rho_{R}\otimes \mathbbm{1}_{B}\\
\rho_{R}\otimes \mathbbm{1}_{B} & Z_{RB}
\end{pmatrix}
\geq 0\Bigg\}.
\end{align}
The expression in \eqref{eq:SDP-root-fid-ch}\ is equal to
\begin{equation}
\sup_{Q_{RB}}\left\{ \lambda_{\min}\!\left(  \operatorname{Re}
[\operatorname{Tr}_{B}[Q_{RB}]]\right) :\
\begin{pmatrix}
\Gamma_{RB}^{\mathcal{N}} & Q_{RB}^{\dag} \\[.8ex]
Q_{RB} & \Gamma_{RB}^{\mathcal{M}}
\end{pmatrix}
\geq0\right\}  ,
\end{equation}
where $\lambda_{\min}(\cdot)$ denotes the minimum eigenvalue function.
\end{proposition*}
	
	\begin{Proof}
	See Appendix \ref{app:QM-over:SDP-fid-chs}.
	\end{Proof}

	The inequality in \eqref{eq-Fuchs_van_de_graaf} relating the fidelity between two states $\rho$ and $\sigma$ and their trace distance can be used to relate the fidelity-based distance measure $F(\mathcal{N},\mathcal{M})$ on channels and the diamond distance $\frac{1}{2}\norm{\mathcal{N}-\mathcal{M}}_{\diamond}$. It is straightforward to show that
	\begin{equation}\label{eq-fuchs_van_de_graaf_channels}
		1-\sqrt{F(\mathcal{N},\mathcal{M})}\leq\frac{1}{2}\norm{\mathcal{N}-\mathcal{M}}_{\diamond}\leq\sqrt{1-F(\mathcal{N},\mathcal{M})}.
	\end{equation}

    \begin{exercise}{exer-fuchs_van_de_graaf_channels}
        Prove \eqref{eq-fuchs_van_de_graaf_channels}.
    \end{exercise}
    
	The following proposition relates the fidelity $F(\mathcal{N})$ of a channel $\mathcal{N}$ to its minimum fidelity $F_{\min}(\mathcal{N})$ in \eqref{eq-minimum_fidelity_chan}, telling us that if $F_{\min}(\mathcal{N})$ is large then so is $F(\mathcal{N})$.

    \begin{proposition}{prop:min-fid-to-min-ent-fid}
    Let $\mathcal{N}$ be a quantum channel. For all $\varepsilon\in[0,1]$, if $F_{\min}(\mathcal{N})\geq 1-\varepsilon$, then $F(\mathcal{N})\geq 1-\frac{3}{2}\varepsilon$.
    \end{proposition}

\begin{Proof}
From the hypothesis, it follows that  
\begin{equation}
\bra{\psi}\mathcal{N}(\psi)\ket{\psi} \geq 1-\varepsilon
\label{min_fidelities_relation_proof_eq1}
\end{equation}
for every state vector $\ket{\psi}$. Now, consider an arbitrary entangled state vector $\ket{\Psi}\equiv\ket{\Psi}_{RM}$. We can write its Schmidt decomposition as $\ket{\Psi}_{RM} = \sum_j \sqrt{\lambda_j} \ket{e_j}_R \otimes \ket{j}_M$, where $\{\ket{e_j}_R\}_j$ and $\{\ket{j}_M\}_j$ form orthonormal bases of the Hilbert spaces $\mathcal{H}_R$ and $\mathcal{H}_M$, respectively, and the Schmidt coefficients $\lambda_j$ form a probability distribution; that is, $\lambda_j\geq 0$ for all $j$, with $\sum_j \lambda_j = 1$. One can verify that  
\begin{equation}
\bra{\Psi}_{RM}\mathcal{N}_{M\to M}(\Psi_{RM})\ket{\Psi}_{RM} = \sum_{j,k} \lambda_j \lambda_k \bra{j}\mathcal{N}(\ketbra{j}{k})\ket{k}\, ,
\end{equation}
where on the right-hand side we dropped system indices for compactness. The issue is that while we know from \eqref{min_fidelities_relation_proof_eq1} that the terms $\bra{j}\mathcal{N}(\ketbra{j}{k})\ket{k}$ for $j=k$ are close to $1$, we must also bound these terms for $j\neq k$. For this purpose, we use once again \eqref{min_fidelities_relation_proof_eq1}, applying it also to state vectors $\ket{\psi}$ outside of the basis $\{\ket{j}\}_j$. Consider therefore a collection of independent and uniformly distributed random phases $\varphi_j \in [0, 2\pi]$; we can construct the associated random state vector $\ket{\psi_\varphi} \coloneqq \sum_j \sqrt{\lambda_j}\, e^{i\varphi_j} \ket{j}$. We have that
\begin{align}
1-\varepsilon &\leq \mathbb{E}_\varphi \bra{\psi_\varphi}\mathcal{N}\left(\psi_\varphi\right) \ket{\psi_\varphi} \label{min_fidelities_relation_proof_eq3_first} \\
&= \mathbb{E}_\varphi \sum_{j,k,\ell,m} \sqrt{\lambda_j \lambda_k \lambda_\ell \lambda_m}\, e^{i\left(-\varphi_j + \varphi_k - \varphi_\ell + \varphi_m\right)}\bra{j} \mathcal{N}(\ketbra{k}{\ell}) \ket{m} \\
&= \sum_{j,k,\ell,m} \sqrt{\lambda_j \lambda_k \lambda_\ell \lambda_m}\, \left( \mathbb{E}_\varphi\, e^{i\left(-\varphi_j + \varphi_k - \varphi_\ell + \varphi_m\right)} \right) \bra{j} \mathcal{N}(\ketbra{k}{\ell}) \ket{m} \\
&= \sum_{j,k,\ell,m} \sqrt{\lambda_j \lambda_k \lambda_\ell \lambda_m} \left(\delta_{jk} \delta_{\ell m} + \delta_{j m} \delta_{k \ell} - \delta_{jk\ell m} \right) \bra{j} \mathcal{N}(\ketbra{k}{\ell}) \ket{m} \\
&= \sum_{j,\ell} \lambda_j \lambda_\ell \bra{j} \mathcal{N}\left(\ketbra{j}{\ell}\right) \ket{\ell} + \sum_{j,k} \lambda_j \lambda_k \bra{j} \mathcal{N}(\ketbra{k}{k}) \ket{j} \nonumber \\
&\qquad - \sum_j \lambda_j^2 \bra{j} \mathcal{N}\left(\ketbra{j}{j}\right)\ket{j} \\
&= \sum_{j, \ell} \lambda_j \lambda_\ell \bra{j} \mathcal{N}(\ketbra{j}{\ell}) \ket{\ell} + \sum_{j \neq k} \lambda_j \lambda_k \bra{j} \mathcal{N}(\ketbra{k}{k}) \ket{j} \\
&= \bra{\Psi}_{RM}\mathcal{N}_{M\to M}(\Psi_{RM}) \ket{\Psi}_{RM} + \sum_{j \neq k} \lambda_j \lambda_k \bra{j} \mathcal{N}(\ketbra{k}{k}) \ket{j}\, . \label{min_fidelities_relation_proof_eq3_last}
\end{align}
Here, the expectations in the second and third lines are with respect to the random variables $\varphi_j$, while the $\delta$ terms in the fourth line are Kronecker deltas. The term $\delta_{jk\ell m} \coloneqq \delta_{jk} \delta_{j\ell} \delta_{jm}$ equals $1$ if and only if $j = k = \ell = m$.  

The first term in \eqref{min_fidelities_relation_proof_eq3_last} is precisely what we seek to bound from below. To complete the proof, we only need to deal with the second term. Note that  
\begin{align}
\sum_{j \neq k} \lambda_j \lambda_k \bra{j} \mathcal{N}(\ketbra{k}{k}) \ket{j} &= \sum_k \lambda_k \sum_{j:\ j\neq k} \lambda_j \bra{j} \mathcal{N}(\ketbra{k}{k}) \ket{j} \label{min_fidelities_relation_proof_eq4_first} \\
&\leq \sum_k \lambda_k \left( \max_{j:\ j\neq k} \lambda_j \right) \sum_{j:\ j\neq k} \bra{j} \mathcal{N}(\ketbra{k}{k}) \ket{j} \\
&= \sum_k \lambda_k \left( \max_{j:\ j\neq k} \lambda_j \right) \big( 1 - \bra{k} \mathcal{N}(\ketbra{k}{k}) \ket{k} \big) \\
&\leq \varepsilon \sum_k \lambda_k \left( \max_{j:\ j\neq k} \lambda_j \right) \\
&= \varepsilon \left( \lambda_1 \lambda_2 + \lambda_1 \sum_{k\neq 1} \lambda_k \right) \\
&= \varepsilon \left( \lambda_1 \lambda_2 + \lambda_1 (1-\lambda_1) \right) \\
&\leq 2\varepsilon \lambda_1 (1-\lambda_1) \\
&\leq \frac{\varepsilon}{2}\, . \label{min_fidelities_relation_proof_eq4_last} 
\end{align}
Here, in the second line we used the fact that $\bra{j} \mathcal{N}\left(\ketbra{k}{k}\right) \ket{j} \geq 0$, while the equality in the third line follows from the trace-preserving property of $\mathcal{N}$. In the fourth line, we applied \eqref{min_fidelities_relation_proof_eq1}. In the fifth line, we distinguished the two cases $k=1$ and $k\neq 1$, assuming, without loss of generality, that the Schmidt coefficients are sorted in non-increasing order as $\lambda_1 \geq \lambda_2 \geq \ldots$. Finally, the inequality in the last line follows from the fact that $p(1-p)\leq 1/4$ for all $p \in [0,1]$. Combining \eqref{min_fidelities_relation_proof_eq3_first}--\eqref{min_fidelities_relation_proof_eq3_last} and \eqref{min_fidelities_relation_proof_eq4_first}--\eqref{min_fidelities_relation_proof_eq4_last} concludes the proof.
\end{Proof}

\section{Bibliographic Notes}\label{sec-QM_dist_meas_bib_notes}

	The quantum fidelity was defined by \citet{Uhl76}, and Theorem~\ref{thm-Uhlmann_fidelity} is due to \citet{Uhl76}. The semi-definite program for root fidelity in Proposition~\ref{prop:QM-over:SDP-fidelity-states} was established by \citet{Killoran12} and \citet{Wat13}, and we have followed the proof provided in the latter. The semi-definite program for fidelity in \eqref{eq-fidelity_SDP_primal} can be found in \citep[Corollary~5.3]{Wat09}. The fact that the fidelity is achieved by a quantum measurement was realized by \citet{FC95}. The relation between trace distance and fidelity presented in Theorem~\ref{thm-Fuchs_van_de_graaf} was proved by \citet{FG98}. For very closely related inequalities, with the fidelity replaced by the ``Holevo fidelity'' $(\Tr[\sqrt{\rho}\sqrt{\sigma}])^2$, see \citet{Kholevo1972}. The gentle measurement lemma, in a different form, was originally proved by \citet[Lemma~9]{W99}. The fidelity inequality presented in Lemma~\ref{lem-dm:gentle-measurement} is a special case of the inequality proven by \citet[Lemma~7]{BCCRR10}. The sine distance was defined by \citet{R02,R03,GLN04,R06}, and its interpretation in terms of the minimal trace distance of purifications was given by \citet{R06}. The sine distance was generalized to subnormalized states by \citet{TCR10}, where it was given the name ``purified distance.''
	
	The diamond norm was presented and studied by \citet{Kit97}, who applied it to problems in quantum information theory and quantum computation. The operational interpretation of the diamond distance in terms of hypothesis testing of quantum channels was given by \citet{KWerner04,RW05,GLN04}. More properties of the diamond norm can be found in \citet{Wat18}. The SDP in Proposition~\ref{prop-diamond_dist_SDP} for the normalized diamond distance of quantum channels was given by \citet{Wat09}.

\citet{Sch96} introduced the entanglement fidelity of a quantum channel, and \citet{BKN98} made further observations regarding it. \citet{Nielsen2002249} provided a simple proof for the relation between entanglement fidelity and average fidelity in \eqref{eq:QM-over:average-fid-to-ent-fid}. The fidelity of quantum channels was introduced by \citet{GLN04}. A semi-definite program for the root fidelity of channels was given by \citet{Yuan2017}; see also the work of \citet{GRS18}. The particular semi-definite program in Proposition~\ref{prop:QM-over:SDP-fidelity-channels}, for the root fidelity of channels, was presented by~\citet{KW20}. Proposition~\ref{prop:min-fid-to-min-ent-fid} was established by~\citet[Theorem~2]{BKN98}; a qualitatively equivalent statement featuring a worse dependence on $\varepsilon$ was given by~\citet[Proposition~4.3]{KWerner04} (see also~\citet[Theorem~3.56]{Wat18}).

\begin{subappendices}

\section{SDPs for Diamond Norm and Normalized Diamond Distance}\label{app-trace_dist_fidelity_SDP}



	

\paragraph*{Proof of Proposition~\ref{prop-diamond_dist_SDP}:} 

Our starting point is the SDP in \eqref{eq:MT:trace-norm-herm-op_SDP} for the trace norm of an arbitrary Hermitian operator $H$, namely,
\begin{equation}
    \norm{H}_1=\sup_{M_1,M_2\geq 0}\big\{\Tr[H(M_1-M_2)]:M_1+M_2=\mathbbm{1}\big\}.
\end{equation}
Using this, along with \eqref{eq-diamond_dist_meas}, consider that 
\begin{multline}
    \norm{\mathcal{P}}_{\diamond}=\sup_{\substack{\psi_{RA}\geq 0,\\ M_{RB}^1,M_{RB}^2\geq 0}}\Big\{\Tr\big[\big(M_{RB}^1-M_{RB}^2\big)\mathcal{P}_{A\to B}(\psi_{RA})\big]:M_{RB}^1+M_{RB}^2=\mathbbm{1}_{RB},\\[-2ex]\Tr[\psi_{RA}]=1,\,\Tr[\psi_{RA}^2]=1\Big\},
\end{multline}
where $R$ has the same dimension as $A$ and the constraints $\psi_{RA}\geq 0$, $\Tr[\psi_{RA}]=1$, and $\Tr[\psi_{RA}^2]=1$ correspond to $\psi_{RA}$ being a pure bipartite state. Note that the above is equal to
\begin{multline}\label{eq-diamond_dist_SDP_pf_1}
    \norm{\mathcal{P}}_{\diamond}=\sup_{\substack{\psi_{RA}\geq 0,\\ M_{RB}^1,M_{RB}^2\geq 0}}\Big\{\Tr\big[\big(M_{RB}^1-M_{RB}^2\big)\mathcal{P}_{A\to B}(\psi_{RA})\big]:M_{RB}^1+M_{RB}^2=\mathbbm{1}_{RB},\\[-2ex]\psi_R>0,\,\Tr[\psi_{RA}]=1,\,\Tr[\psi_{RA}^2]=1\Big\},
\end{multline}
due to the fact that the set of pure states with reduced state $\psi_R$ positive definite is dense in the set of all pure states. Now, recall from \eqref{eq-pure_state_vec} that every such pure state can be written as $\psi_{RA}=X_R\Gamma_{RA}X_R^\dagger$ for some linear operator $X_R$ such that $\Tr[X_R^\dagger X_R]=1$ and $|X_R|>0$, where $\Gamma_{RA}$ defined in \eqref{eq-max_ent_vector}. Using this, we find that the objective function can be rewritten as
\begin{align}
    &\Tr\big[\big(M_{RB}^1-M_{RB}^2\big)\mathcal{P}_{A\to B}(X_R\Gamma_{RA}X_R^\dagger)\big] \notag \\
    &\quad =\Tr\big[X_R^\dagger \big(M_{RB}^1-M_{RB}^2\big)X_R\mathcal{P}_{A\to B}(\Gamma_{RA})\big]\\
    &\quad =\Tr\big[X_R^\dagger\big(M_{RB}^1-M_{RB}^2\big)X_R\Gamma_{RB}^{\mathcal{P}}\big].
\end{align}
Now, observe the following equivalences:
\begin{align}
    M_{RB}^1\geq 0\quad&\Longleftrightarrow\quad X_R^{\dagger}M_{RB}^1X_R\geq 0,\\
    M_{RB}^2\geq 0\quad&\Longleftrightarrow\quad X_R^{\dagger}M_{RB}^2X_R\geq 0,\\
    M_{RB}^1+M_{RB}^2=\mathbbm{1}_{RB}\quad&\Longleftrightarrow\quad X_{R}^{\dagger}M_{RB}^1X_R+X_R^{\dagger}M_{RB}^2X_R=X_R^{\dagger}X_R\otimes\mathbbm{1}_B.
\end{align}
Thus, defining $S_{RB}^1\coloneqq X_R^{\dagger}M_{RB}^1X_R$, $S_{RB}^2\coloneqq X_R^{\dagger}M_{RB}^2 X_R$, $\rho_R\coloneqq X_R^{\dagger}X_R$, and since~$R$ has the same dimension as $A$, the optimization in \eqref{eq-diamond_dist_SDP_pf_1} is equivalent to the following one:
\begin{equation}
    \sup_{\substack{\rho_A,\\ S_{AB}^1,\,S_{AB}^2\geq 0}}\Big\{\Tr\big[\big(S_{AB}^1-S_{AB}^2\big)\Gamma_{AB}^{\mathcal{P}}\big]: S_{AB}^1+S_{AB}^2=\rho_A\otimes\mathbbm{1}_B,\,\rho_A>0,\,\Tr[\rho_A]=1\Big\},
\end{equation}
giving the equality in \eqref{eq-diamond_norm_SDP_primal}. To determine the dual SDP in \eqref{eq-diamond_norm_SDP_dual}, let us start by casting the above SDP in the form of the following standard form of a primal SDP from \eqref{eq-primal_SDP_def}, namely,
\begin{equation}
    \sup_{Y\geq 0}\{\Tr[DY]:\Phi(Y)\leq C\}.
\end{equation}
By setting
\begin{align}
    Y&\coloneqq\begin{pmatrix} S_{AB}^1 & 0 & 0 \\ 0 & S_{AB}^2 & 0 \\ 0 & 0 & \rho_A \end{pmatrix},\\
    D&\coloneqq\begin{pmatrix} \Gamma_{AB}^{\mathcal{P}} & 0 & 0 \\ 0 & -\Gamma_{AB}^{\mathcal{P}} & 0 \\ 0 & 0 & 0  \end{pmatrix},\\
    \Phi(Y)&\coloneqq\begin{pmatrix} S_{AB}^1+S_{AB}^2-\rho_A\otimes\mathbbm{1}_B & 0 & 0 & 0 \\ 0 & -(S_{AB}^1+S_{AB}^2-\rho_A\otimes\mathbbm{1}_B) & 0 & 0 \\ 0 & 0 & \Tr[\rho_A] & 0 \\ 0 & 0 & 0 & -\Tr[\rho_A] \end{pmatrix},\\
    C&\coloneqq\begin{pmatrix} 0 & 0 & 0 & 0 \\ 0 & 0 & 0 & 0 \\ 0 & 0 & 1 & 0 \\ 0 & 0 & 0 & -1  \end{pmatrix},
\end{align}
we observe that \eqref{eq-diamond_norm_SDP_primal} is indeed in standard form. Now, the corresponding standard form of the dual problem, from \eqref{eq-dual_SDP_def}, is
\begin{equation}
    \inf_{G\geq 0}\{\Tr[CG]:\Phi^\dagger(G)\geq D\}.
\end{equation}
Without loss of generality, we can take the variable $G$ to be of the form
\begin{equation}
    G\coloneqq\begin{pmatrix} Z_{AB}^1 & 0 & 0 & 0 \\ 0 & Z_{AB}^2 & 0 & 0 \\ 0 & 0 & \mu_1 & 0 \\ 0 & 0 & 0 & \mu_2   \end{pmatrix}.
\end{equation}
Then, we find that
\begin{multline}
    \Tr[\Phi(Y)G]=\Tr[S_{AB}^1(Z_{AB}^1-Z_{AB}^2)]+\Tr[S_{AB}^2(Z_{AB}^1-Z_{AB}^2)]\\+\Tr[\rho_A\Tr_B[-Z_{AB}^1+Z_{AB}^2]]+\Tr[\rho_A((\mu_1-\mu_2)\mathbbm{1}_A)],
\end{multline}
from which we conclude that
\begin{equation}
    \Phi^{\dagger}(G)=\begin{pmatrix} Z_{AB}^1-Z_{AB}^2 & 0 & 0 \\ 0 & Z_{AB}^1-Z_{AB}^2 & 0 \\ 0 & 0 & \Tr_B[-Z_{AB}^1+Z_{AB}^2]+(\mu_1-\mu_2)\mathbbm{1}_B  \end{pmatrix}.
\end{equation}
The standard form of the dual SDP then becomes
\begin{multline}
    \inf_{\substack{\mu_1,\,\mu_2\geq 0,\\Z_{AB}^1,\,Z_{AB}^2\geq 0}}\Big\{\mu_1-\mu_2:-(Z_{AB}^1-Z_{AB}^2)\leq\Gamma_{AB}^{\mathcal{P}}\leq Z_{AB}^1-Z_{AB}^2,\\[-3ex]\Tr_B[Z_{AB}^1-Z_{AB}^2]\leq(\mu_1-\mu_2)\mathbbm{1}_A\Big\}.
\end{multline}
Now, observe that the variables $\mu_1$ and $\mu_2$ always appear together in the above optimization as $\mu_1-\mu_2$. Similarly, the variables $Z_{AB}^1$ and $Z_{AB}^2$ always appear together as $Z_{AB}^1-Z_{AB}^2$. Consequently, both pairs of variables can be reduced to a single real variable $\mu\in\mathbb{R}$ and a single Hermitian operator $Z_{AB}$. We therefore have
\begin{equation}
    \inf_{\substack{\mu\in\mathbb{R},\\Z_{AB}\text{ Hermitian}}}\Big\{\mu:-Z_{AB}\leq\Gamma_{AB}^{\mathcal{P}}\leq Z_{AB},\,\Tr_B[Z_{AB}]\leq\mu\mathbbm{1}_A\Big\}.
\end{equation}
Finally, the constraints $-Z_{AB}\leq\Gamma_{AB}^{\mathcal{P}}\leq Z_{AB}$ imply that $Z_{AB}\geq 0$, which in turn implies that $\mu\geq 0$. This leads to the claimed form of the dual SDP in \eqref{eq-diamond_norm_SDP_dual}. Equality of the primal and dual SDPs is due to strong duality, which holds here because the choice $Z_{AB}=\Gamma_{AB}^{\mathcal{P}}+\delta\mathbbm{1}_{AB}$, with $\delta>2\abs{\lambda_{\min}(\Gamma_{AB}^{\mathcal{P}})}$, is a strictly feasible point. Indeed, note that $Z_{AB}-\Gamma_{AB}^{\mathcal{P}}=\delta\mathbbm{1}_{AB}>0$ and $\Gamma_{AB}^{\mathcal{P}}+Z_{AB}=2\Gamma_{AB}^{\mathcal{P}}+\delta\mathbbm{1}_{AB}>0$. Furthermore, the choice $\mu>\abs{\lambda_{\max}(\Tr_B[\Gamma_{AB}^{\mathcal{P}}])}+d_B\delta$ is strictly feasible, because $\mu>0$ and $\Tr_B[Z_{AB}]=\Tr_B[\Gamma_{AB}^{\mathcal{P}}]+d_B\delta\mathbbm{1}_A\leq\lambda_{\max}(\Gamma_{AB}^{\mathcal{P}})\mathbbm{1}_A+d_B\delta\mathbbm{1}_A\leq\abs{\lambda_{\max}(\Gamma_{AB}^{\mathcal{P}})}\mathbbm{1}_A+d_B\delta\mathbbm{1}_A<\mu\mathbbm{1}_A$. 

For the normalized diamond distance, let us observe that we can eliminate the variable $S_{AB}^2$ in \eqref{eq-diamond_norm_SDP_primal} by using the constraint $S_{AB}^1+S_{AB}^2=\rho_A\otimes\mathbbm{1}_B$. Indeed, this constraint implies that $S_{AB}^2=\rho_A\otimes\mathbbm{1}_B-S_{AB}^1$. Moreover, because $S_{AB}^2\geq 0$, we have that $S_{AB}^1\leq\rho_A\otimes\mathbbm{1}_B$. Therefore, we can write the primal SDP in \eqref{eq-diamond_norm_SDP_primal} as follows:
\begin{equation}
    \sup_{S_{AB}\geq 0,\,\rho_A\geq 0}\Big\{2\Tr[\Gamma_{AB}^{\mathcal{P}}S_{AB}]-\Tr[\Tr_B[\Gamma_{AB}^{\mathcal{P}}]\rho_A]:S_{AB}\leq\rho_A\otimes\mathbbm{1}_B,\,\Tr[\rho_A]=1\Big\}.
\end{equation}
Then, by setting $\Gamma_{AB}^{\mathcal{P}}=\Gamma_{AB}^{\mathcal{N}}-\Gamma_{AB}^{\mathcal{M}}$, the second term in the objective function vanishes, leading to \eqref{eq-diamond_dist_SDP_primal}. Furthermore, by following arguments very similar to those above, it is straightforward to determine that the corresponding dual SDP has the claimed form in \eqref{eq-diamond_dist_SDP_dual} and also that strong duality holds. The equality
\begin{multline}
    \inf_{\substack{\mu\geq 0,\\ Z_{AB}\geq 0}}\Big\{\mu:Z_{AB}\geq \Gamma_{AB}^{\mathcal{N}}-\Gamma_{AB}^{\mathcal{M}},\,\mu\mathbbm{1}_A\geq \Tr_B[Z_{AB}]\Big\}\\=\inf_{Z_{AB}\geq 0}\Big\{\norm{\Tr_B[Z_{AB}]}_{\infty}:Z_{AB}\geq \Gamma_{AB}^{\mathcal{N}}-\Gamma_{AB}^{\mathcal{M}}\Big\}
\end{multline}
holds by the expression in \eqref{eq-PSD_largest_eig_SDP} for the Schatten $\infty$-norm for positive semi-definite operators.

\section{SDPs for Fidelity of States and Channels}

\label{app:QM-over:SDPs-fidelity}

\subsection{Proof of Proposition~\ref{prop:QM-over:SDP-fidelity-states}}

\label{app:QM-over:SDP-fid-states}

First, let us verify that strong duality holds for the primal and dual
semi-definite programs in \eqref{eq:QM-over:fidelity-primal-SDP} and
\eqref{eq:QM-over:fidelity-dual-SDP}, respectively. Consider that $X=0$ is a
feasible choice for the primal program, while $Y=Z=2\mathbbm{1}$ is strictly feasible
for the dual program. Thus, strong duality holds according to Theorem~\ref{thm:math-tools:slater-cond}.

In order to prove the equality in \eqref{eq:QM-over:fidelity-primal-SDP}, we
start with the following lemma:

\begin{Lemma}{lem:QM-over:Bhatia-lemma-1}
Let $P$ and $Q$ be positive semi-definite
operators in $\Lin(\mathcal{H})$, and let $X\in\Lin(\mathcal{H})$. Then, the operator
\begin{equation}
\begin{pmatrix}
P & X\\
X^{\dag} & Q
\end{pmatrix}
\end{equation}
is positive semi-definite if and only if there exists $K\in\Lin(\mathcal{H})$ satisfying $\left\Vert K\right\Vert _{\infty}\leq 1$ and
$X=\sqrt{P}K\sqrt{Q}$.
\end{Lemma}

\begin{Proof}
See Theorem~IX.5.9 of \citet{Bha97}.
\end{Proof}

It follows from this lemma that the operator $X$ in the primal optimization
in \eqref{eq:QM-over:fidelity-primal-SDP} can take the form $X=\sqrt{\rho}K\sqrt{\sigma}$,
such that $K\in\Lin(\mathcal{H})$ and $\left\Vert K\right\Vert
_{\infty}\leq1$, so that the optimization with respect to $X$ is reduced to an
optimization with respect to $K$. We then find that the primal optimal value is given by
\begin{align}
& \frac{1}{2}\sup_{X\in\Lin(\mathcal{H})}\left\{  \operatorname{Tr}
[X]+\operatorname{Tr}[X^{\dag}]:
\begin{pmatrix}
\rho & X\\
X^{\dag} & \sigma
\end{pmatrix}
\geq0\right\}  \nonumber\\
& =\frac{1}{2}\sup_{K:\left\Vert K\right\Vert _{\infty}\leq1} \left\{\operatorname{Tr}
[\sqrt{\rho}K\sqrt{\sigma}]+\operatorname{Tr}[\sqrt{\sigma}K^{\dag}\sqrt{\rho
}] \right\} \\
& =\sup_{K:\left\Vert K\right\Vert _{\infty}\leq1}\operatorname{Re}
[\operatorname{Tr}[\sqrt{\rho}K\sqrt{\sigma}]]\\
& =\sup_{K:\left\Vert K\right\Vert _{\infty}\leq1}\left\vert \operatorname{Tr}
[\sqrt{\rho}K\sqrt{\sigma}]\right\vert \\
& =\left\Vert \sqrt{\rho}\sqrt{\sigma}\right\Vert _{1}=\sqrt{F}(\rho,\sigma).
\end{align}
The first equality follows from Lemma~\ref{lem:QM-over:Bhatia-lemma-1}. The
third equality follows because we can use the optimization over $K$ to adjust
a global phase such that the real part is equal to the absolute value. (Here,
one should think of the fact that $\operatorname{Re}[z]=r\cos(\theta)$ for
$z=re^{i\theta}$, and then one can optimize the value of $\theta$ so that
$\operatorname{Re}[z]=r$.) 
The final equality follows by a generalization of
Proposition~\ref{prop:math-tools:var-char-t-norm} (in fact the same proof given there implies that the optimization can
be with respect to $U$ satisfying $\left\Vert U\right\Vert _{\infty}\leq1$,
rather than just with respect to isometries).

We now prove that \eqref{eq:QM-over:fidelity-dual-SDP} is the dual program of
\eqref{eq:QM-over:fidelity-primal-SDP}. We can rewrite the primal SDP\ as
\begin{equation}
\frac{1}{2}\sup\left\{\operatorname{Tr}[X]+\operatorname{Tr}[X^{\dag}]\right\}
\end{equation}
subject to
\begin{equation}
\begin{pmatrix}
\rho & 0\\
0 & \sigma
\end{pmatrix}
\geq
\begin{pmatrix}
0 & X\\
X^{\dag} & 0
\end{pmatrix}
,\qquad
\begin{pmatrix}
R & X\\
X^{\dag} & S
\end{pmatrix}
\geq0
\end{equation}
because $R$ and $S$ are not involved in the objective function and can always
be chosen so that the second operator is PSD. Also, the following equivalences
hold
\begin{align}
\begin{pmatrix}
\rho & X\\
X^{\dag} & \sigma
\end{pmatrix}
\geq0\quad &  \Longleftrightarrow\quad
\begin{pmatrix}
\rho & -X\\
-X^{\dag} & \sigma
\end{pmatrix}
\geq0\\
&  \Longleftrightarrow\quad
\begin{pmatrix}
\rho & 0\\
0 & \sigma
\end{pmatrix}
\geq
\begin{pmatrix}
0 & X\\
X^{\dag} & 0
\end{pmatrix}
.
\end{align}

As given in \eqref{eq-primal_SDP_def} and \eqref{eq-dual_SDP_def}, the standard forms of primal and dual SDPs for Hermitian $A$ and $B$ and
Hermiticity-preserving map $\Phi$ are as follows:
\begin{align}
&  \sup_{G\geq0}\left\{  \operatorname{Tr}[AG]:\Phi(G)\leq B\right\}  ,\\
&  \inf_{Y\geq0}\left\{  \operatorname{Tr}[BY]:\Phi^{\dag}(Y)\geq A\right\}  .
\end{align}

So the SDP\ above is in standard form with
\begin{align}
G &  =
\begin{pmatrix}
R & X\\
X^{\dag} & S
\end{pmatrix}
,\quad A=
\begin{pmatrix}
0 & \mathbbm{1}\\
\mathbbm{1} & 0
\end{pmatrix}
,\\
\Phi(G) &  =
\begin{pmatrix}
0 & X\\
X^{\dag} & 0
\end{pmatrix}
,\quad B=
\begin{pmatrix}
\rho & 0\\
0 & \sigma
\end{pmatrix}
.
\end{align}
Setting
\begin{equation}
Y=
\begin{pmatrix}
W & V\\
V^{\dag} & Z
\end{pmatrix}
,
\end{equation}
the adjoint of $\Phi$ is given by
\begin{align}
\operatorname{Tr}[Y\Phi(X)] &  =\operatorname{Tr}\left[
\begin{pmatrix}
W & V\\
V^{\dag} & Z
\end{pmatrix}
\begin{pmatrix}
0 & X\\
X^{\dag} & 0
\end{pmatrix}
\right]  \\
&  =\operatorname{Tr}\left[
\begin{pmatrix}
VX^{\dag} & WX\\
ZX^{\dag} & V^{\dag}X
\end{pmatrix}
\right]  \\
&  =\operatorname{Tr}[VX^{\dag}]+\operatorname{Tr}[XV^{\dag}]\\
&  =\operatorname{Tr}\left[
\begin{pmatrix}
0 & V\\
V^{\dag} & 0
\end{pmatrix}
\begin{pmatrix}
R & X\\
X^{\dag} & S
\end{pmatrix}
\right]  ,
\end{align}
so that
\begin{equation}
\Phi^{\dag}(Y)=
\begin{pmatrix}
0 & V\\
V^{\dag} & 0
\end{pmatrix}
.
\end{equation}
Then the dual is given by
\begin{equation}
\frac{1}{2}\inf\operatorname{Tr}\left[
\begin{pmatrix}
\rho & 0\\
0 & \sigma
\end{pmatrix}
\begin{pmatrix}
W & V\\
V^{\dag} & Z
\end{pmatrix}
\right]
\end{equation}
subject to
\begin{equation}
\begin{pmatrix}
0 & V\\
V^{\dag} & 0
\end{pmatrix}
\geq
\begin{pmatrix}
0 & \mathbbm{1}\\
\mathbbm{1} & 0
\end{pmatrix}
,\qquad
\begin{pmatrix}
W & V\\
V^{\dag} & Z
\end{pmatrix}
\geq0.
\end{equation}
This simplifies to
\begin{equation}
\frac{1}{2}\inf\left\{\operatorname{Tr}[\rho W]+\operatorname{Tr}[\sigma Z]\right\},
\end{equation}
subject to
\begin{equation}
\begin{pmatrix}
0 & V\\
V^{\dag} & 0
\end{pmatrix}
\geq
\begin{pmatrix}
0 & \mathbbm{1}\\
\mathbbm{1} & 0
\end{pmatrix}
,\qquad
\begin{pmatrix}
W & V\\
V^{\dag} & Z
\end{pmatrix}
\geq0
\end{equation}
Since
\begin{align}
\begin{pmatrix}
W & V\\
V^{\dag} & Z
\end{pmatrix}
\geq0\quad &  \Longleftrightarrow\quad
\begin{pmatrix}
W & -V\\
-V^{\dag} & Z
\end{pmatrix}
\geq0\\
&  \Longleftrightarrow\quad
\begin{pmatrix}
W & 0\\
0 & Z
\end{pmatrix}
\geq
\begin{pmatrix}
0 & V\\
V^{\dag} & 0
\end{pmatrix}
,
\end{align}
we find that there is a single condition
\begin{equation}
\begin{pmatrix}
W & 0\\
0 & Z
\end{pmatrix}
\geq
\begin{pmatrix}
0 & V\\
V^{\dag} & 0
\end{pmatrix}
\geq
\begin{pmatrix}
0 & \mathbbm{1}\\
\mathbbm{1} & 0
\end{pmatrix}
,
\end{equation}
and since $V$ plays no role in the objective function, we can set $V=\mathbbm{1}$. So
the final SDP\ simplifies as follows:
\begin{equation}
\frac{1}{2}\inf_{W,Z} \left\{\operatorname{Tr}[\rho W]+\operatorname{Tr}[\sigma Z]\right\}
\end{equation}
subject to
\begin{equation}
\begin{pmatrix}
W & -\mathbbm{1}\\
-\mathbbm{1} & Z
\end{pmatrix}
\geq0.
\end{equation}
Using the fact that
\begin{equation}
\begin{pmatrix}
W & -\mathbbm{1}\\
-\mathbbm{1} & Z
\end{pmatrix}
\geq0\quad\Longleftrightarrow\quad
\begin{pmatrix}
W & \mathbbm{1}\\
\mathbbm{1} & Z
\end{pmatrix}
\geq0
\end{equation}
we can do one final rewriting as follows:
\begin{equation}
\frac{1}{2}\inf_{W,Z}\left\{\operatorname{Tr}[\rho W]+\operatorname{Tr}[\sigma Z]\right\}
\end{equation}
subject to
\begin{equation}
\begin{pmatrix}
W & \mathbbm{1}\\
\mathbbm{1} & Z
\end{pmatrix}
\geq0.
\end{equation}

\subsection{Proof of Proposition~\ref{prop:QM-over:SDP-fidelity-channels}}

\label{app:QM-over:SDP-fid-chs}

First, strong duality holds, according to Theorem~\ref{thm:math-tools:slater-cond}, because $Q_{RB} = 0$ and $\lambda = 0$ is feasible for the primal program, while $\rho_R = \mathbbm{1}/|R|$ and $W_{RB} = Z_{RB} = 2 \mathbbm{1}_{RB}$ is strictly feasible for the dual.

For a pure bipartite state
$\psi_{RA}$, we use \eqref{eq-pure_state_vec} to conclude that
\begin{equation}
\psi_{RA}=X_{R}\Gamma_{RA}X_{R}^{\dag},
\end{equation}
where $\operatorname{Tr}[X_{R}^{\dag}X_{R}]=1$ to see that
\begin{equation}
\mathcal{N}_{A\rightarrow B}(\psi_{RA})=X_{R}\Gamma_{RB}^{\mathcal{N}}
X_{R}^{\dag},\quad\mathcal{M}_{A\rightarrow B}(\psi_{RA})=X_{R}\Gamma
_{RB}^{\mathcal{M}}X_{R}^{\dag},
\end{equation}
and then plug in to \eqref{eq:QM-over:fidelity-dual-SDP}\ to get that
\begin{equation}
\sqrt{F}(\mathcal{N}_{A\rightarrow B},\mathcal{M}_{A\rightarrow B})=\frac
{1}{2}\inf_{W_{RB},Z_{RB}}\operatorname{Tr}[X_{R}\Gamma_{RB}^{\mathcal{N}
}X_{R}^{\dag}W_{RB}]+\operatorname{Tr}[X_{R}\Gamma_{RB}^{\mathcal{M}}
X_{R}^{\dag}Z_{RB}]
\end{equation}
subject to
\begin{equation}
\begin{pmatrix}
W_{RB} & \mathbbm{1}_{RB}\\
\mathbbm{1}_{RB} & Z_{RB}
\end{pmatrix}
\geq0. \label{eq:QM-over:ineq-const-dual-fid-states-SDP}
\end{equation}
Consider that the objective function can be written as
\begin{equation}
\operatorname{Tr}[\Gamma_{RB}^{\mathcal{N}}W_{RB}^{\prime}]+\operatorname{Tr}
[\Gamma_{RB}^{\mathcal{M}}Z_{RB}^{\prime}],
\end{equation}
with
\begin{equation}
W_{RB}^{\prime}\coloneqq X_{R}^{\dag}W_{RB}X_{R},\quad Z_{RB}^{\prime}\coloneqq X_{R}^{\dag
}Z_{RB}X_{R}
\end{equation}
Now consider that the inequality in
\eqref{eq:QM-over:ineq-const-dual-fid-states-SDP}\ is equivalent to
\begin{equation}
\begin{pmatrix}
X_{R} & 0\\
0 & X_{R}
\end{pmatrix}
^{\dag}
\begin{pmatrix}
W_{RB} & \mathbbm{1}_{RB}\\
\mathbbm{1}_{RB} & Z_{RB}
\end{pmatrix}
\begin{pmatrix}
X_{R} & 0\\
0 & X_{R}
\end{pmatrix}
\geq0 .
\end{equation}
(Here we have assumed that $X_R$ is invertible, but it suffices to do so for this optimization because the set of invertible $X_R$ is dense in the set of all possible~$X_R$.)
Multiplying out the last matrix we find that
\begin{align}
&
\begin{pmatrix}
X_{R} & 0\\
0 & X_{R}
\end{pmatrix}
^{\dag}
\begin{pmatrix}
W_{RB} & \mathbbm{1}_{RB}\\
\mathbbm{1}_{RB} & Z_{RB}
\end{pmatrix}
\begin{pmatrix}
X_{R} & 0\\
0 & X_{R}
\end{pmatrix}
\nonumber\\
&\quad  =
\begin{pmatrix}
X_{R}^{\dag}W_{RB}X_{R} & X_{R}^{\dag}X_{R}\otimes \mathbbm{1}_{B}\\
X_{R}^{\dag}X_{R}\otimes \mathbbm{1}_{B} & X_{R}^{\dag}Z_{RB}X_{R}
\end{pmatrix}
\\
&\quad  =
\begin{pmatrix}
W_{RB}^{\prime} & \rho_{R}\otimes \mathbbm{1}_{B}\\
\rho_{R}\otimes \mathbbm{1}_{B} & Z_{RB}^{\prime}
\end{pmatrix}
,
\end{align}
where we defined $\rho_{R}=X_{R}^{\dag}X_{R}$. Observing that $\rho_{R}\geq0$
and $\operatorname{Tr}[\rho_{R}]=1$, we can write the final SDP\ as follows:
\begin{equation}
\sqrt{F}(\mathcal{N}_{A\rightarrow B},\mathcal{M}_{A\rightarrow B})=\frac
{1}{2}\inf_{\rho_{R},W_{RB},Z_{RB}}\operatorname{Tr}[\Gamma_{RB}^{\mathcal{N}
}W_{RB}]+\operatorname{Tr}[\Gamma_{RB}^{\mathcal{M}}Z_{RB}],
\end{equation}
subject to
\begin{equation}
\rho_{R}\geq0,\quad\operatorname{Tr}[\rho_{R}]=1,\quad
\begin{pmatrix}
W_{RB} & \rho_{R}\otimes \mathbbm{1}_{B}\\
\rho_{R}\otimes \mathbbm{1}_{B} & Z_{RB}
\end{pmatrix}
\geq0. \label{eq:QM-over:ch-fid-SDP-constr}
\end{equation}

Now let us calculate the dual SDP\ to this, using the following standard forms
for primal and dual SDPs, with Hermitian operators $A$ and $B$ and a
Hermiticity-preserving map $\Phi$ (as given in \eqref{eq-primal_SDP_def} and \eqref{eq-dual_SDP_def}):
\begin{equation}
\sup_{X\geq0}\left\{  \operatorname{Tr}[AX]:\Phi(X)\leq B\right\}  ,\qquad
\inf_{Y\geq0}\left\{  \operatorname{Tr}[BY]:\Phi^{\dag}(Y)\geq A\right\}  .
\end{equation}
Consider that the constraint in \eqref{eq:QM-over:ch-fid-SDP-constr} implies
$W_{RB}\geq0$ and $Z_{RB}\geq0$, so that we can set
\begin{align}
Y  &  =
\begin{pmatrix}
W_{RB} & 0 & 0\\
0 & Z_{RB} & 0\\
0 & 0 & \rho_{R}
\end{pmatrix}
,\quad B=
\begin{pmatrix}
\Gamma_{RB}^{\mathcal{N}} & 0 & 0\\
0 & \Gamma_{RB}^{\mathcal{M}} & 0\\
0 & 0 & 0
\end{pmatrix}
,\\
\Phi^{\dag}(Y)  &  =
\begin{pmatrix}
W_{RB} & \rho_{R}\otimes \mathbbm{1}_{B} & 0 & 0\\
\rho_{R}\otimes \mathbbm{1}_{B} & Z_{RB} & 0 & 0\\
0 & 0 & \operatorname{Tr}[\rho_{R}] & 0\\
0 & 0 & 0 & -\operatorname{Tr}[\rho_{R}]
\end{pmatrix}
,\\
A  &  =
\begin{pmatrix}
0 & 0 & 0 & 0\\
0 & 0 & 0 & 0\\
0 & 0 & 1 & 0\\
0 & 0 & 0 & -1
\end{pmatrix}
.
\end{align}
Then with
\begin{equation}
X=
\begin{pmatrix}
P_{RB} & Q_{RB}^{\dag} & 0 & 0\\
Q_{RB} & S_{RB} & 0 & 0\\
0 & 0 & \lambda & 0\\
0 & 0 & 0 & \mu
\end{pmatrix}
\end{equation}
the map $\Phi$ is given by
\begin{align}
&  \operatorname{Tr}[X\Phi^{\dag}(Y)]\nonumber\\
&  =\operatorname{Tr}\left[
\begin{pmatrix}
P_{RB} & Q_{RB}^{\dag} & 0 & 0\\
Q_{RB} & S_{RB} & 0 & 0\\
0 & 0 & \lambda & 0\\
0 & 0 & 0 & \mu
\end{pmatrix}
\begin{pmatrix}
W_{RB} & \rho_{R}\otimes \mathbbm{1}_{B} & 0 & 0\\
\rho_{R}\otimes \mathbbm{1}_{B} & Z_{RB} & 0 & 0\\
0 & 0 & \operatorname{Tr}[\rho_{R}] & 0\\
0 & 0 & 0 & -\operatorname{Tr}[\rho_{R}]
\end{pmatrix}
\right] \\
& =\operatorname{Tr}[P_{RB}W_{RB}]+\operatorname{Tr}[Q_{RB}^{\dag}\left(
\rho_{R}\otimes \mathbbm{1}_{B}\right)  ]+\operatorname{Tr}[Q_{RB}(\rho_{R}\otimes
\mathbbm{1}_{B})] \\
&  \qquad+\operatorname{Tr}[S_{RB}Z_{RB}]+\left(  \lambda-\mu\right)
\operatorname{Tr}[\rho_{R}] \\
&  =\operatorname{Tr}[P_{RB}W_{RB}]+\operatorname{Tr}[S_{RB}Z_{RB}
]+\operatorname{Tr}[(\operatorname{Tr}_{B}[Q_{RB}+Q_{RB}^{\dag}]+\left(
\lambda-\mu\right)  \mathbbm{1}_{R})\rho_{R}] \\
&  =\operatorname{Tr}\left[
\begin{pmatrix}
P_{RB} & 0 & 0\\
0 & S_{RB} & 0\\
0 & 0 & \operatorname{Tr}_{B}[Q_{RB}+Q_{RB}^{\dag}]+\left(  \lambda
-\mu\right)  \mathbbm{1}_{R}
\end{pmatrix}
\begin{pmatrix}
W_{RB} & 0 & 0\\
0 & Z_{RB} & 0\\
0 & 0 & \rho_{R}
\end{pmatrix}
\right]  .
\end{align}
So then
\begin{equation}
\Phi(X)=
\begin{pmatrix}
P_{RB} & 0 & 0\\
0 & S_{RB} & 0\\
0 & 0 & \operatorname{Tr}_{B}[Q_{RB}+Q_{RB}^{\dag}]+\left(  \lambda
-\mu\right)  \mathbbm{1}_{R}
\end{pmatrix}
.
\end{equation}
The primal is then given by
\begin{equation}
\frac{1}{2}\sup\operatorname{Tr}\left[
\begin{pmatrix}
0 & 0 & 0 & 0\\
0 & 0 & 0 & 0\\
0 & 0 & 1 & 0\\
0 & 0 & 0 & -1
\end{pmatrix}
\begin{pmatrix}
P_{RB} & Q_{RB}^{\dag} & 0 & 0\\
Q_{RB} & S_{RB} & 0 & 0\\
0 & 0 & \lambda & 0\\
0 & 0 & 0 & \mu
\end{pmatrix}
\right]  ,
\end{equation}
subject to
\begin{align}
\begin{pmatrix}
P_{RB} & 0 & 0\\
0 & S_{RB} & 0\\
0 & 0 & \operatorname{Tr}_{B}[Q_{RB}+Q_{RB}^{\dag}]+\left(  \lambda
-\mu\right)  \mathbbm{1}_{R}
\end{pmatrix}
&  \leq
\begin{pmatrix}
\Gamma_{RB}^{\mathcal{N}} & 0 & 0\\
0 & \Gamma_{RB}^{\mathcal{M}} & 0\\
0 & 0 & 0
\end{pmatrix}
,\\
\begin{pmatrix}
P_{RB} & Q_{RB}^{\dag} & 0 & 0\\
Q_{RB} & S_{RB} & 0 & 0\\
0 & 0 & \lambda & 0\\
0 & 0 & 0 & \mu
\end{pmatrix}
&  \geq0,
\end{align}
which simplifies to
\begin{equation}
\frac{1}{2}\sup\left(  \lambda-\mu\right)
\end{equation}
subject to
\begin{align}
P_{RB}  &  \leq\Gamma_{RB}^{\mathcal{N}},\\
S_{RB}  &  \leq\Gamma_{RB}^{\mathcal{M}},\\
\operatorname{Tr}_{B}[Q_{RB}+Q_{RB}^{\dag}]+\left(  \lambda-\mu\right)  \mathbbm{1}_{R}
&  \leq0,\\
\begin{pmatrix}
P_{RB} & Q_{RB}^{\dag}\\
Q_{RB} & S_{RB}
\end{pmatrix}
&  \geq0,\\
\lambda,\mu &  \geq0.
\end{align}
We can simplify this even more. We can set $\lambda^{\prime}=\lambda-\mu
\in\mathbb{R}$, and we can substitute $Q_{RB}$ with $-Q_{RB}$ without changing
the value, so then it becomes
\begin{equation}
\frac{1}{2}\sup\lambda^{\prime}
\end{equation}
subject to
\begin{align}
P_{RB}  &  \leq\Gamma_{RB}^{\mathcal{N}},\\
S_{RB}  &  \leq\Gamma_{RB}^{\mathcal{M}},\\
\lambda^{\prime}\mathbbm{1}_{R}  &  \leq\operatorname{Tr}_{B}[Q_{RB}+Q_{RB}^{\dag}],\\
\begin{pmatrix}
P_{RB} & -Q_{RB}^{\dag}\\
-Q_{RB} & S_{RB}
\end{pmatrix}
&  \geq0,\\
\lambda^{\prime}  &  \in\mathbb{R}.
\end{align}
We can rewrite
\begin{align}
\begin{pmatrix}
P_{RB} & -Q_{RB}^{\dag}\\
-Q_{RB} & S_{RB}
\end{pmatrix}
\geq0\quad &  \Longleftrightarrow\quad
\begin{pmatrix}
P_{RB} & Q_{RB}^{\dag}\\
Q_{RB} & S_{RB}
\end{pmatrix}
\geq0\\
&  \Longleftrightarrow\quad
\begin{pmatrix}
P_{RB} & 0\\
0 & S_{RB}
\end{pmatrix}
\geq
\begin{pmatrix}
0 & -Q_{RB}^{\dag}\\
-Q_{RB} & 0
\end{pmatrix}
\end{align}
We then have the simplified condition
\begin{equation}
\begin{pmatrix}
0 & -Q_{RB}^{\dag}\\
-Q_{RB} & 0
\end{pmatrix}
\leq
\begin{pmatrix}
P_{RB} & 0\\
0 & S_{RB}
\end{pmatrix}
\leq
\begin{pmatrix}
\Gamma_{RB}^{\mathcal{N}} & 0\\
0 & \Gamma_{RB}^{\mathcal{M}}
\end{pmatrix}
.
\end{equation}
Since $P_{RB}$ and $S_{RB}$ do not appear in the objective function, we can
set them to their largest value and obtain the following simplification
\begin{equation}
\frac{1}{2}\sup\lambda^{\prime}
\end{equation}
subject to
\begin{equation}
\lambda^{\prime}\mathbbm{1}_{R}\leq\operatorname{Tr}_{B}[Q_{RB}+Q_{RB}^{\dag}],\quad
\begin{pmatrix}
\Gamma_{RB}^{\mathcal{N}} & Q_{RB}^{\dag}\\
Q_{RB} & \Gamma_{RB}^{\mathcal{M}}
\end{pmatrix}
\geq0,\quad\lambda^{\prime}\in\mathbb{R}
\end{equation}
Since a feasible solution is $\lambda^{\prime}=0$ and $Q_{RB}=0$, it is clear
that we can restrict to $\lambda^{\prime}\geq0$. After a relabeling, this
becomes
\begin{multline}
\frac{1}{2}\sup_{\lambda\geq0,Q_{RB}}\left\{  \lambda:\lambda \mathbbm{1}_{R}
\leq\operatorname{Tr}_{B}[Q_{RB}+Q_{RB}^{\dag}],\quad
\begin{pmatrix}
\Gamma_{RB}^{\mathcal{N}} & Q_{RB}^{\dag}\\
Q_{RB} & \Gamma_{RB}^{\mathcal{M}}
\end{pmatrix}
\geq0\right\} \\
=\sup_{\lambda\geq0,Q_{RB}}\left\{  \lambda:\lambda \mathbbm{1}_{R}\leq\operatorname{Re}
[\operatorname{Tr}_{B}[Q_{RB}]],\quad
\begin{pmatrix}
\Gamma_{RB}^{\mathcal{N}} & Q_{RB}^{\dag}\\
Q_{RB} & \Gamma_{RB}^{\mathcal{M}}
\end{pmatrix}
\geq0\right\}  .
\end{multline}
This is equivalent to
\begin{equation}
\sup_{Q_{RB}}\left\{  \lambda_{\min}\left(  \operatorname{Re}
[\operatorname{Tr}_{B}[Q_{RB}]]\right)  :
\begin{pmatrix}
\Gamma_{RB}^{\mathcal{N}} & Q_{RB}^{\dag}\\
Q_{RB} & \Gamma_{RB}^{\mathcal{M}}
\end{pmatrix}
\geq0\right\}  .
\end{equation}
This concludes the proof.

\end{subappendices}

\chapter{Quantum Entropies and Information}\label{chap-entropies}

	In this chapter, we introduce various entropic and information quantities that play a fundamental role in the analysis of quantum communication protocols. Here we see that the notions of entropy and information take many forms. The most basic and fundamental entropic and information measures are members of the ``von Neumann family,'' and these often correspond to optimal communication rates for information-theoretic tasks in the asymptotic regime of many uses of an independent and identically distributed (i.i.d.) resource. More refined entropy measures belong to the ``R\'enyi family,'' and interestingly, in part due to the non-commutativity of quantum states, there are several interesting ways of generalizing the classical R\'enyi relative entropy that are meaningful for understanding information-theoretic tasks. The R\'enyi measures reduce to the von Neumann ones in a particular limit, and they are useful for characterizing optimal rates of information-theoretic tasks in the non-asymptotic regime, in particular when trying to determine how fast an error probability is converging to zero or one (so-called error exponents and strong converse exponents, respectively). Even more broadly, we define entropy measures from the meaningful ``one-shot'' information-theoretic task given by quantum hypothesis testing. Even though one might debate whether such an operationally defined quantity is truly an entropy, our opinion is that this perspective is quite powerful, and so we adopt it in this chapter and the rest of the book. Thus, this chapter explores quite broadly a variety of entropic measures and their mathematical properties, as they are the basis for analyzing a wide variety of quantum information-processing protocols.
	
	We start our development with a brief preview of quantum entropies and information (Section~\ref{sec:qei:preview}) and then proceed to the well-known quantum relative entropy (Section~\ref{sec-rel_ent}), which plays a foundational role in understanding properties of other entropies. We then proceed to defining a generalized divergence, which is a concept that plays an important role in the proofs of strong converse theorems throughout this book (Section~\ref{sec-gen_div}). Of particular focus are several prominent examples of generalized divergences, the Petz--R\'{e}nyi relative entropy (Section~\ref{sec-petz_ren_rel_ent}), the sandwiched R\'{e}nyi relative entropy (Section~\ref{sec-sand_ren_rel_ent}), the geometric R\'{e}nyi relative entropy (Section~\ref{sec-QEI:geometric-renyi}), the Belavkin--Staszewski relative entropy (Section~\ref{sec:QEI:Belavkin--Staszewski}), the (smooth) max-relative entropy (Section~\ref{sec-max_rel_ent}), and the hypothesis testing relative entropy (Section~\ref{sec-hyp_test_rel_ent}). The first of these plays an important role in achievability proofs for quantum channel capacities, and the second plays an important role for strong converses. The hypothesis testing relative entropy is fundamental in establishing bounds on one-shot channel capacities.
	
\section{Preview}\label{sec:qei:preview}
	
	Arguably one of the most important quantities in quantum information theory is the \textit{von Neumann entropy}\index{von Neumann entropy}, which is the quantum generalization of the Shannon entropy\index{Shannon entropy}. We also refer to it as the \textit{quantum entropy}\index{quantum entropy} and do so from now on. For a quantum system $A$ in the state $\rho_A\in\Density(\mathcal{H}_A)$, the von Neumann entropy is defined as
	\begin{equation}\label{eq-quantum_entropy}
		H(\rho_A)\equiv H(A)_\rho\coloneqq -\Tr[\rho_A\log_2\rho_A].
	\end{equation}
	If $\rho_A$ has a spectral decomposition of the form
	\begin{equation}
		\rho_A=\sum_{i=1}^r\lambda_i\ket{\psi_i}\!\bra{\psi_i}_A,
		\label{eq-QEI:spec-decomp-init}
	\end{equation}
	where $r\equiv \rank(\rho_A)$, then we can write $H(\rho_A)$ in terms of the non-zero eigenvalues $\{\lambda_i\}_{i=1}^r$ of $\rho_A$ as
	\begin{equation}\label{eq-quantum_entropy_eig}
		H(\rho_A)=- \sum_{i=1}^r\lambda_i\log_2\lambda_i.
	\end{equation}
	Note that the zero eigenvalues of $\rho_A$ do not contribute to the entropy due to the convention that $0\log_2 0=0$, which is taken because $\lim_{x\to 0^+}x\log_2 x=0$. By viewing $\rho_A$ as a probabilistic mixture of the pure states $\{\ket{\psi_i}_A\}_{i=1}^r$ defined in \eqref{eq-QEI:spec-decomp-init}, the quantum entropy quantifies the uncertainty about which of these pure states the system $A$ is in. In particular, $H(\rho_A)$ is, in a rough sense, the expected information gain upon performing an experiment to determine the state of the system.
	
	Note that the right-hand side of \eqref{eq-quantum_entropy_eig} is the formula for the \textit{Shannon entropy} of the probability distribution $\{\lambda_i\}_i$ corresponding to the eigenvalues of $\rho_A$. The Shannon entropy of the probability distribution $\{p,1-p\}$, for $ p \in [0,1]$, shows up frequently and is denoted by $h_2(p)$, i.e.,
	\begin{equation}\label{eq-ent_bin}
		h_2(p)\coloneqq -p\log_2 p-(1-p)\log_2 (1-p).
	\end{equation}
	It is called the binary entropy function.
	
	Just as the Shannon entropy has an operational meaning as the optimal rate of (classical) data compression, the quantum entropy has an operational interpretation as the optimal rate of quantum data compression. More precisely, given the state $\rho_A^{\otimes n}$, the quantum entropy $H(\rho_A)$ is the minimum number of qubits per copy of the state $\rho_A$ that are needed to faithfully represent $\rho_A^{\otimes n}$, when $n$ becomes large. This task is also called Schumacher compression.
	
	Other fundamental information-theoretic quantities, which are functions of the Shannon entropy, have straightforward generalizations to the quantum setting. Let~$\rho_{AB}$ be a bipartite state, and let $\sigma_{ABC}$ be a tripartite state.
	\begin{enumerate}
		\item The \textit{quantum conditional entropy}\index{quantum conditional entropy} is defined as
			\begin{equation}\label{eq:QEI:cond-ent-def}
				H(A|B)_{\rho}\coloneqq H(AB)_{\rho}-H(B)_{\rho}.
			\end{equation}
			The quantum conditional entropy quantifies the uncertainty about the state of the system $A$ in the presence of additional quantum side information in the form of the quantum system~$B$.
			
		\item The \textit{coherent information}\index{coherent information} is defined as
			\begin{equation}
				I(A\rangle B)_\rho\coloneqq H(B)_\rho-H(AB)_\rho=-H(A|B)_\rho,
				\label{eq:QEI:coh-info-1st-def}
			\end{equation}
			and it arises in the context of communication of quantum information over quantum channels (see Chapter \ref{chap-quantum_capacity}). The coherent information is asymmetric and can be interpreted as having a directionality. 
			We obtain a quantity called the \textit{reverse coherent information} by swapping the systems $A$ and $B$ in \eqref{eq:QEI:coh-info-1st-def}:
			\begin{equation}
				I(B\rangle A)_{\rho}\coloneqq H(A)_{\rho}-H(AB)_{\rho}=-H(B|A)_{\rho}.
			\end{equation}
			This quantity arises when studying feedback-assisted quantum communication.
		\item The \textit{quantum mutual information}\index{quantum mutual information}\index{mutual information|see {quantum mutual information}} is defined as
			\begin{align}
				I(A;B)_{\rho}&\coloneqq H(A)_\rho+H(B)_\rho-H(AB)_\rho,\label{eq-mut_inf_formula}\\
				&=H(A)_{\rho}-H(A|B)_{\rho}\label{eq-mut_inf_formula_2}\\
				&=H(B)_{\rho}-H(B|A)_{\rho}\label{eq-mut_inf_formula_3},
			\end{align}
			and it arises in the context of communicating classical information over quantum channels (see Chapters~\ref{chap-EA_capacity} and \ref{chap-classical_capacity}).
		\item The \textit{quantum conditional mutual information}\index{quantum conditional mutual information}\index{conditional mutual information|see {quantum conditional mutual information}} is defined as
			\begin{equation}\label{eq-QCMI_def}
				I(A;B|C)_\sigma\coloneqq H(A|C)_\sigma+H(B|C)_\sigma-H(AB|C)_\sigma,
			\end{equation}
			and it is the basis for an entanglement measure called squashed entanglement (see Chapter~\ref{chap-ent_measures}). An important result is that the quantum conditional mutual information is non-negative, i.e., $I(A;B|C)_\sigma\geq 0$ for every tripartite state $\sigma_{ABC}$. This inequality goes by the name \textit{strong subadditivity of quantum entropy}, and we show at the end of Section \ref{sec-rel_ent} that it follows from the data-processing inequality for the quantum relative entropy (Theorem~\ref{thm-monotone_rel_ent}).
	\end{enumerate}
		As it turns out, all of these quantities, including the entropy itself, can be derived from a single parent quantity, the quantum relative entropy, which we introduce in the next section.

\section{Quantum Relative Entropy}\label{sec-rel_ent}

	We define the quantum relative entropy as follows:	

	\begin{definition}{Quantum Relative Entropy}{def-rel_ent}
		For every state $\rho$ and  positive semi-definite operator $\sigma$, the \textit{quantum relative entropy of $\rho$ and $\sigma$}, denoted by $D(\rho\Vert\sigma)$, is defined as
		\begin{equation}
			D(\rho\Vert\sigma)=\left\{\begin{array}{l l} \Tr[\rho(\log_2 \rho-\log_2\sigma)] & \text{if }\supp(\rho)\subseteq\supp(\sigma),\\ +\infty & \text{otherwise}. \end{array}\right.
		\end{equation}
	\end{definition}
	
	\begin{remark}
		More generally, we could define the quantum relative entropy exactly as above, but with both arguments being positive semi-definite operators. For our purposes in this book, however, it suffices to restrict the first argument to be a state.
	\end{remark}
	
	The quantum relative entropy is a particular quantum generalization of the \textit{Kullback--Leibler divergence} or classical relative entropy, which for two probability distributions $p,q:\mathcal{X}\to [0,1]$ defined on a finite alphabet $\mathcal{X}$ is given by
	\begin{equation}
		D(p\Vert q)=\sum_{x\in\mathcal{X}} p(x)\log_2\!\left(\frac{p(x)}{q(x)}\right). \label{eq:classical-rel-ent}
	\end{equation}
	The quantum relative entropy has an operational meaning in terms of the task of quantum hypothesis testing, as is shown in Section \ref{subsec-q_Stein_lemma} on the quantum Stein's lemma (Theorem \ref{thm-q_Stein_lemma}). The quantum relative entropy $D(\rho\Vert\sigma)$ can also be interpreted as a distinguishability measure for the quantum states $\rho$ and $\sigma$, in part due to the facts that $D(\rho\Vert\sigma)\geq 0$ and  $D(\rho\Vert\sigma)=0$ if and only if $\rho=\sigma$, which is shown in Proposition \ref{prop-rel_ent} below.
	
	The support condition $\supp(\rho)\subseteq\supp(\sigma)$ in the definition of the quantum relative entropy essentially has to do with the term $\Tr[\rho\log_2\sigma]$ and the fact that the logarithm of an operator is really only well defined for positive definite operators, while we allow for $\sigma$ to be positive semi-definite, which means that it could have some eigenvalues equal to zero. Recall that the expression $\rho\log_2\rho$ is well defined even for states with eigenvalues equal to zero since we set $0\log_2 0=0$. We justified this with the fact that $\lim_{x\to 0^+}x\log_2 x=0$. We can similarly make sense of the support condition in the definition of the quantum relative entropy by using the following fact.
	
	\begin{proposition}{prop-rel_ent_lim}
		For every state $\rho$ and positive semi-definite operator $\sigma$,
		\begin{equation}\label{eq-rel_ent_limit}
			D(\rho\Vert\sigma)=\lim_{\varepsilon\to 0^+} \Tr[\rho(\log_2\rho-\log_2(\sigma+\varepsilon\mathbbm{1}))].
		\end{equation}
		Consequently, whenever $\sigma$ does not have full support (i.e., it is positive semi-definite as opposed to positive definite), we can write $D(\rho\Vert\sigma)$ as the following limit:
		\begin{equation}
			D(\rho\Vert\sigma)=\lim_{\varepsilon\to 0^+}D(\rho\Vert\sigma+\varepsilon\mathbbm{1}).
		\end{equation}
	\end{proposition}
	
	\begin{Proof}
		Observe that, for all $\varepsilon>0$, the operator $\sigma+\varepsilon\mathbbm{1}$ has full support; i.e., $\text{supp}(\sigma+\varepsilon\mathbbm{1})=\mathcal{H}$ for all $\varepsilon>0$, where $\mathcal{H}$ is the underlying Hilbert space. This means that the quantity
		\begin{equation}
			\Tr[\rho\log_2(\sigma+\varepsilon\mathbbm{1})]
		\end{equation}
		is finite for all $\varepsilon>0$. Now, let us decompose the Hilbert space $\mathcal{H}$ into the direct sum of the orthogonal subspaces $\supp(\sigma)$ and $\ker(\sigma)$, so that $\mathcal{H}=\supp(\sigma)\oplus\ker(\sigma)$.  Then with respect to this decomposition, the operators $\rho$ and $\sigma$ can be written as the block matrices
		\begin{equation}\label{eq-q_rel_ent_block}
			\rho= \begin{pmatrix} \rho_{0,0} & \rho_{0,1}\\ \rho_{0,1}^\dagger & \rho_{1,1}\end{pmatrix},\qquad 
			\sigma =\begin{pmatrix} \sigma_{0,0} & 0\\0&0\end{pmatrix}.
		\end{equation}
		
		Now, let us first suppose that $\supp(\rho)\subseteq\supp(\sigma)$.  Then we can write the second term in \eqref{eq-rel_ent_limit} as
		\begin{align}
			\Tr[\rho\log_2(\sigma+\varepsilon\mathbbm{1})]&=\Tr\!\left[\begin{pmatrix}\rho_{0,0} & 0 \\ 0 & 0\end{pmatrix}\log_2\begin{pmatrix} \sigma_{0,0}+\varepsilon\mathbbm{1}&0\\0&\varepsilon\mathbbm{1}\end{pmatrix}\right]\\
			&=\Tr\!\left[\begin{pmatrix}\rho_{0,0}&0\\0&0\end{pmatrix}\begin{pmatrix}\log_2(\sigma_{0,0}+\varepsilon\mathbbm{1})&0\\0&\log_2(\varepsilon\mathbbm{1})\end{pmatrix}\right]\\
			&=\Tr[\rho_{0,0}\log_2(\sigma_{0,0}+\varepsilon\mathbbm{1})].
		\end{align}
		Therefore,
		\begin{equation}
			\lim_{\varepsilon\to 0^+}\Tr[\rho\log_2(\sigma+\varepsilon\mathbbm{1})]=\lim_{\varepsilon\to 0^+} \Tr[\rho_{0,0}\log_2(\sigma_{0,0}+\varepsilon\mathbbm{1})]=\Tr[\rho\log_2 \sigma],
		\end{equation}
		which means that
		\begin{equation}
			\lim_{\varepsilon\to 0^+}\Tr[\rho(\log_2 \rho-\log_2(\sigma+\varepsilon\mathbbm{1}))]=\lim_{\varepsilon\to 0^+}\Tr[\rho(\log_2 \rho-\log_2 \sigma)]
		\end{equation}
		whenever $\supp(\rho)\subseteq\supp(\sigma)$.
		
		If $\supp(\rho)\nsubseteq\supp(\sigma)$, then the block $\rho_{1,1}$ of $\rho$ is non-zero (and the block $\rho_{0,1}$ could be non-zero), and we obtain
		\begin{align}
			\Tr[\rho\log_2(\sigma+\varepsilon\mathbbm{1})]
			& =\Tr\!\left[\begin{pmatrix}\rho_{0,0}&\rho_{0,1}\\\rho_{0,1}^\dagger &\rho_{1,1}\end{pmatrix}\begin{pmatrix}\log_2(\sigma_{0,0}+\varepsilon\mathbbm{1})&0\\0&\log_2(\varepsilon\mathbbm{1})\end{pmatrix}\right]\\
			&=\Tr[\rho_{0,0}\log_2(\sigma_{0,0}+\varepsilon\mathbbm{1})]+\Tr[\rho_{1,1}\log_2(\varepsilon\mathbbm{1})]\\
			&=\Tr[\rho_{0,0}\log_2(\sigma_{0,0}+\varepsilon\mathbbm{1})]+\log_2(\varepsilon)\Tr[\rho_{1,1}].
		\end{align}
		Then, using the fact that $\lim_{\varepsilon\to 0^+}(-\log_2\varepsilon)=+\infty$, we find that
		\begin{multline}
			\lim_{\varepsilon\to 0^+}\Tr[\rho(\log_2 \rho-\log_2(\sigma+\varepsilon\mathbbm{1}))]=
			\Tr[\rho\log_2 \rho]\\
			-\lim_{\varepsilon\to 0^+}\Tr[\rho_{0,0}\log_2(\sigma_{0,0}+\varepsilon\mathbbm{1})]
			-\log_2(\varepsilon)\lim_{\varepsilon\to 0^+}\Tr[\rho_{1,1}] = +\infty.
		\end{multline}
		We thus conclude that
		\begin{equation}
			\begin{aligned}
			&\lim_{\varepsilon\to 0^+}\Tr[\rho(\log_2 \rho-\log_2(\sigma+\varepsilon\mathbbm{1}))]\\
			&\qquad=\left\{\begin{array}{l l} \Tr[\rho(\log_2 \rho-\log_2 \sigma)] & \text{if }\supp(\rho)\subseteq\supp(\sigma),\\+\infty & \text{otherwise.}\end{array}\right. \\
			&\qquad= D(\rho\Vert \sigma),
			\end{aligned}
		\end{equation}
		as required. 
	\end{Proof}

	The proposition above, in particular the fact that we can write the quantum relative entropy as $D(\rho\Vert\sigma)=\lim_{\varepsilon\to 0^+}D(\rho\Vert\sigma+\varepsilon\mathbbm{1})$, allows us to take the logarithm $\log_2(\sigma)$ of $\sigma$  on only the support of $\sigma$ when determining the quantum relative entropy. We can use this fact to write another formula for the quantum relative entropy. We start by writing a spectral decomposition of the state $\rho$ and positive semi-definite operator $\sigma$. In particular, setting $r_\rho \equiv\rank(\rho)$ and $r_\sigma \equiv \rank(\sigma)$, let
	\begin{align}
		\rho&=\sum_{j=1}^d p_j\ket{\psi_j}\!\bra{\psi_j}=\sum_{j=1}^{r_\rho}p_j\ket{\psi_j}\!\bra{\psi_j},\label{eq-spec_decomp_rho}\\
		\sigma&=\sum_{k=1}^d q_k\ket{\phi_k}\!\bra{\phi_k}=\sum_{k=1}^{r_\sigma}q_k\ket{\phi_k}\!\bra{\phi_k},\label{eq-spec_decomp_sigma}
	\end{align}
	be spectral decompositions of $\rho$ and $\sigma$, where $d=\dim(\mathcal{H})$ and in the second equality we have restricted the sum to only those eigenvalues that are non-zero. Then, 
	\begin{align}
		D(\rho\Vert \sigma)&=\sum_{j=1}^{r_\rho} p_j\log_2 p_j-\Tr\!\left[\left(\sum_{j=1}^{r_\rho}p_j\ket{\psi_j}\!\bra{\psi_j}\right)\left(\sum_{k=1}^{r_\sigma}\log_2 q_k \ket{\phi_k}\!\bra{\phi_k}\right)\right]\\
		&=\sum_{j=1}^{r_\rho}p_j\log_2 p_j-\sum_{j=1}^{r_\rho}\sum_{k=1}^{r_\sigma}\abs{\braket{\psi_j}{\phi_k}}^2p_j\log_2 q_k \\
		&=\sum_{j=1}^{r_\rho}\left[p_j\log_2 p_j -\sum_{k=1}^{r_\sigma}\abs{\braket{\psi_j}{\phi_k}}^2p_j\log_2 q_k \right]\label{eq-ren_ent_alt_a}.
	\end{align}
	Now, using the fact that the eigenvectors $\{\ket{\phi_k}:1\leq k\leq d\}$ form a complete orthonormal basis for $\mathcal{H}$, so that $\mathbbm{1}_{\mathcal{H}}=\sum_{k=1}^d\ket{\phi_k}\!\bra{\phi_k}$, we conclude that
	\begin{equation}
		1=\braket{\psi_j}{\psi_j}=\sum_{k=1}^d\braket{\psi_j}{\phi_k}\braket{\phi_k}{\psi_j}=\sum_{k=1}^{r_\sigma}\abs{\braket{\psi_j}{\phi_k}}^2,
        \label{eq:normalisation_for_Nussbaum_Szkola_1}
	\end{equation}
	for $1\leq j\leq r_\rho$, where the last equality follows from the assumption that $\supp(\rho)\subseteq\supp(\sigma)$, which implies that the eigenvectors $\{\ket{\psi_j}:1\leq j\leq r_\rho\}$ of $\rho$ can be expressed as a linear combination of the eigenvectors $\{\ket{\phi_k}:1\leq k\leq r_\sigma\}$ of $\sigma$. Therefore,
	\begin{equation}\label{eq-rel_ent_alt}
		\boxed{\begin{aligned}
		D(\rho\Vert \sigma)=\sum_{j=1}^{r_\rho}\sum_{k=1}^{r_\sigma}\abs{\braket{\psi_j}{\phi_k}}^2 & p_j\log_2\!\left(\frac{p_j}{q_k}\right),
		\end{aligned}}
	\end{equation}
	whenever $\supp(\rho)\subseteq\supp(\sigma)$.
	
	We now detail some important mathematical properties of the quantum relative entropy.
	
	\begin{proposition*}{Basic Properties of the Quantum Relative Entropy}{prop-rel_ent}
		The quantum relative entropy satisfies the following properties for all states $\rho,\rho_1,\rho_2$ and positive semi-definite operators $\sigma,\sigma_1,\sigma_2$:
		\begin{enumerate}
			\item \textit{Isometric invariance}: For every isometry $V$,
				\begin{equation}
					D(V\rho V^\dagger\Vert V\sigma V^\dagger)=D(\rho\Vert \sigma).
				\end{equation}
            
            \item \textit{Additivity}:
				\begin{equation}\label{eq-rel_ent_additivity}
					D(\rho_1\otimes \rho_2\Vert \sigma_1\otimes \sigma_2)=D(\rho_1\Vert \sigma_1)+D(\rho_2\Vert \sigma_2).
				\end{equation}
				As a special case, for all $\beta\in (0,\infty)$,
				\begin{equation}\label{eq-rel_ent_scalar_mult}
					D(\rho\Vert \beta \sigma)=D(\rho\Vert \sigma)+\log_2\!\left(\frac{1}{\beta}\right).
				\end{equation}

			\item 
				\begin{enumerate}
					\item \textit{Klein's inequality}: If $\Tr(\sigma)\leq 1$, then $D(\rho\Vert \sigma)\geq 0$. 
					\item \textit{Faithfulness}: If $\Tr(\sigma)\leq 1$, then $D(\rho\Vert \sigma)=0$ if and only if $\rho=\sigma$.
					\item If $\sigma\leq\sigma'$,  then $D(\rho\Vert \sigma)\geq D(\rho\Vert \sigma')$. In particular, if $\rho\leq\sigma$, then $D(\rho\Vert \sigma)\leq 0$. 
				\end{enumerate}
		
			\item \textit{Direct-sum property}: Let $p\colon \mathcal{X}\rightarrow[0,1]$ be a probability distribution over a finite alphabet $\mathcal{X}$ with associated $|\mathcal{X}|$-dimensional system $X$, and let $q\colon \mathcal{X}\to[0,\infty)$ be a non-negative function on $\mathcal{X}$. Let $\{\rho_A^x\}_{x\in\mathcal{X}}$ be a set of states on a system $A$, and let $\{\sigma_A^x\}_{x\in\mathcal{X}}$ be a set of positive semi-definite operators on $A$. Then,
				\begin{equation}\label{eq-rel_ent_direct_sum}
					D(\rho_{XA}\Vert\sigma_{XA}) =D(p\Vert q)+\sum_{x\in\mathcal{X}}p(x)D(\rho_A^x\Vert\sigma_A^x).
				\end{equation}
				where 
				\begin{align}
				\rho_{XA} & \coloneqq \sum_{x\in\mathcal{X}}p(x)\ket{x}\!\bra{x}_X\otimes\rho_A^x ,\\
				\sigma_{XA} & \coloneqq \sum_{x\in\mathcal{X}}q(x)\ket{x}\!\bra{x}_X\otimes\sigma_A^x.
				\end{align}
		\end{enumerate}
	\end{proposition*}
	
	\begin{remark}
		If we let the first argument of the relative entropy be a general positive semi-definite operator instead of just a state, then  \eqref{eq-rel_ent_scalar_mult}  can be generalized for every $\alpha,\beta\in (0,\infty)$ as 
		\begin{equation}
			D(\alpha\rho\Vert\beta\sigma)=\alpha D(\rho\Vert\sigma)+\alpha\log_2\!\left(\frac{\alpha}{\beta}\right).
		\end{equation}
	\end{remark}

	\begin{Proof}
		\hfill\begin{enumerate}
			\item \textit{Proof of isometric invariance}: When $\supp(\rho)\nsubseteq\supp(\sigma)$, there is nothing to prove because $\supp(V\rho V^\dagger)\nsubseteq\supp(V\sigma V^\dagger)$, which means that both $D(V\rho V^\dagger\Vert V\sigma V^\dagger)$ and $D(\rho\Vert\sigma)$ are equal to $+\infty$.
			
			Suppose that $\supp(\rho)\subseteq\supp(\sigma)$, which implies that also $\supp(V\rho V^\dagger)\subseteq\supp(V \sigma V^\dagger)$. Let $\rho$ and $\sigma$ have the spectral decompositions given in \eqref{eq-spec_decomp_rho} and \eqref{eq-spec_decomp_sigma}, respectively. Using the formula in \eqref{eq-rel_ent_alt}, we find that
			\begin{align}
				D(V\rho V^\dagger\Vert V\sigma V^\dagger)&=\sum_{j=1}^{r_\rho}\sum_{k=1}^{r_\sigma}\abs{\bra{\psi_j}V^\dagger V\ket{\phi_k}}^2 p_j\log_2\!\left(\frac{p_j}{q_k}\right)\\
				&=\sum_{j=1}^{r_\rho}\sum_{k=1}^{r_\sigma}\abs{\braket{\psi_j}{\phi_k}}^2 p_j\log_2\!\left(\frac{p_j}{q_k}\right)\\
				&=D(\rho\Vert\sigma).
			\end{align}
			We used the fact that $V$ is an isometry, i.e., satisfying $V^\dag V = \mathbbm{1}$.

            \item \textit{Proof of additivity}:  Since $\supp(\rho_1\otimes \rho_2)=\supp(\rho_1)\otimes\supp(\rho_2)$ and $\supp(\sigma_1\otimes \sigma_2)=\supp(\sigma_1)\otimes \supp(\sigma_2)$, the condition $\supp(\rho_1\otimes \rho_2)\nsubseteq\supp(\sigma_1\otimes \sigma_2)$ is equivalent to the condition $\supp(\rho_1)\allowbreak\nsubseteq\supp(\sigma_1)$ or $\supp(\rho_2)\nsubseteq\supp(\sigma_2)$. Therefore, $D(\rho_1\otimes \rho_2\Vert \sigma_1\otimes \sigma_2)=+\infty$ and $D(\rho_1\Vert \sigma_1)=+\infty$ or $D( \rho_2\Vert  \sigma_2)=+\infty$ if one of the support conditions is violated. Now suppose that $\supp(\rho_1\otimes \rho_2)\subseteq\supp(\sigma_1\allowbreak\otimes \sigma_2)$. Letting $\rho_1$ and $\rho_2$ have spectral decompositions
				\begin{equation}
					\rho_1=\sum_{j=1}^d p_j^{1}\ket{\psi_j^{1}}\!\bra{\psi_j^{1}},\quad \rho_2=\sum_{k=1}^d p_k^{2}\ket{\psi_k^{2}}\!\bra{\psi_k^{2}},
				\end{equation}
				we find that
				\begin{align}
					\log_2(\rho_1\otimes \rho_2)&=\log_2\!\left(\sum_{j,k=1}^d p_j^{1}\ket{\psi_j^{1}}\!\bra{\psi_j^{1}}\otimes p_k^{2}\ket{\psi_k^{2}}\!\bra{\psi_k^{2}}\right)\\
					&=\sum_{j,k=1}^d\log_2(p_j^{1}p_k^{2})\ket{\psi_j^{1}}\!\bra{\psi_j^{1}}\otimes\ket{\psi_k^{2}}\!\bra{\psi_k^{2}}\\
					&=\sum_{j,k=1}^d\log_2(p_j^{1})\ket{\psi_j^{1}}\!\bra{\psi_j^{1}}\otimes\ket{\psi_k^{2}}\!\bra{\psi_k^{2}}\\
					&\qquad\qquad+\sum_{j,k=1}^d\log_2(p_k^{2})\ket{\psi_j^{1}}\!\bra{\psi_j^{1}}\otimes\ket{\psi_k^{2}}\!\bra{\psi_k^{2}}\\
					&=\log_2(\rho_1)\otimes\mathbbm{1}+\mathbbm{1}\otimes\log_2(\rho_2).
				\end{align}
				Similarly, $\log_2(\sigma_1\otimes \sigma_2)=\log_2(\sigma_1)\otimes\mathbbm{1}+\mathbbm{1}\otimes\log_2(\sigma_2)$. Therefore,
				\begin{align}
					&D(\rho_1\otimes \rho_2\Vert \sigma_1\otimes \sigma_2)\notag \\
					&=\Tr\!\left[(\rho_1\otimes \rho_2)(\log_2(\rho_1)\otimes\mathbbm{1}+\mathbbm{1}\otimes\log_2(\rho_2)\right]\nonumber\\
					&\qquad\qquad -\Tr\!\left[(\rho_1\otimes \rho_2)(\log_2(\sigma_1)\otimes\mathbbm{1}+\mathbbm{1}\otimes\log_2(\sigma_2)\right]\\
					&=\Tr(\rho_2)\left(\Tr[\rho_1\log_2(\rho_1)]-\Tr[\rho_1\log_2(\sigma_1)]\right)\nonumber\\
					&\qquad\qquad -\Tr(\rho_1)\left(\Tr[\rho_2\log_2(\rho_2)]-\Tr[\rho_2\log_2(\sigma_2)]\right)\\
					&=D(\rho_1\Vert \sigma_1)+D(\rho_2\Vert \sigma_2).
				\end{align}
				
				Then, to see \eqref{eq-rel_ent_scalar_mult}, let $\rho=\rho_1$, $\alpha=\rho_2=1$, $\sigma=\sigma_1$, and $\beta=\sigma_2$. Recognizing that the tensor product with a scalar is just multiplication by the scalar, we find that
				\begin{equation}
					\begin{aligned}
					D(\rho\Vert \beta \sigma)&=D(\rho\Vert \sigma)+D(1\Vert\beta)\\
					&=D(\rho\Vert \sigma)+(\log_2(1)-\log_2\beta)\\
					&=D(\rho\Vert \sigma)+\log_2\!\left(\frac{1}{\beta}\right).
					\end{aligned}
				\end{equation}
            
            \item \begin{enumerate}
					\item \textit{Proof of Klein's inequality}: The result is trivial in the case that $\supp(\rho)\nsubseteq\supp(\sigma)$, and so we assume that $\supp(\rho)\subseteq\supp(\sigma)$. We can also assume that $\Tr [\sigma] = 1$, without loss of generality. Indeed, assume that we have proved the inequality in the case where $\Tr [\sigma]=1$. If $\Tr [\sigma] <1$, we can construct the normalized state $\sigma'\coloneqq \sigma / \Tr[\sigma]$. Note that $\Tr [\sigma] > 0$, otherwise we would have $\sigma=0$, contradicting the support condition. We have $D(\rho\|\sigma') \geq 0$ by assumption; therefore, using~\eqref{eq-rel_ent_scalar_mult},
                    \begin{equation}
                        D(\rho\|\sigma) = D(\rho\| \Tr[\sigma] \sigma') = D(\rho\|\sigma') + \log_2 \frac{1}{\Tr [\sigma]} \geq 0,
                        \label{eq:assuming_sigma_normalised}
                    \end{equation}
                    where we observed that both terms are non-negative. In what follows, we will therefore assume that $\Tr [\sigma] = 1$.
                    
                    We can write the relative entropy as in~\eqref{eq-rel_ent_alt},
                    \begin{align}
                        D(\rho\Vert \sigma) &= \sum_{j=1}^{r_\rho}\sum_{k=1}^{r_\sigma}\abs{\braket{\psi_j}{\phi_k}}^2 p_j\log_2\!\left(\frac{p_j}{q_k}\right) \\
                        &= \sum_{j=1}^{r_\rho}\sum_{k=1}^{r_\sigma} \lambda_{jk}\, f\left(\frac{p_j}{q_k}\right),
                        \label{eq:relent_a_la_Nussbaum_Szkola}
                    \end{align}
                    where, in the second line, we defined $\lambda_{jk} \coloneqq \abs{\braket{\psi_j}{\phi_k}}^2 q_k$ and introduced the function $f(x)\coloneqq x \log_2 x$. This function is strictly convex, as one can verify by taking the second derivative:
                    \begin{equation}
                        f^{\prime\prime}(x) = \frac{1}{x \ln 2} > 0,
                        \label{eq:strict_convexity_xlogx}
                    \end{equation}
                    for all $x>0$. Furthermore, the numbers $\lambda_{jk}$, for $1\leq j\leq r_\rho$ and $1\leq k\leq r_{\sigma}$, form a probability distribution. In fact, using~\eqref{eq:normalisation_for_Nussbaum_Szkola_1}, we obtain
                    \begin{equation}
                        \sum_{j=1}^{r_\rho}\sum_{k=1}^{r_\sigma} \lambda_{jk} = \sum_{k=1}^{r_\sigma} q_k \sum_{j=1}^{r_\rho} \abs{\braket{\psi_j}{\phi_k}}^2 = \sum_{k=1}^{r_\sigma} q_k = 1.
                        \label{eq:normalisation_for_Nussbaum_Szkola_2}
                    \end{equation}
                    Therefore, applying the convexity of $f$ to~\eqref{eq:relent_a_la_Nussbaum_Szkola}, we obtain the inequality
                    \begin{align}
                        D(\rho\Vert \sigma) &\geq f\left( \sum_{j=1}^{r_\rho}\sum_{k=1}^{r_\sigma} \lambda_{jk} \frac{p_j}{q_k}\right) \label{eq:Klein_proof_key_inequality} \\
                        &= f\left( \sum_{j=1}^{r_\rho}\sum_{k=1}^{r_\sigma} \abs{\braket{\psi_j}{\phi_k}}^2 p_j \right) \\
                        &= f(1) \label{eq:Klein_proof_key_inequality_f_1} \\
                        &= 0, 
                    \end{align}
                    where, in the third line, we leveraged a calculation analogous to~\eqref{eq:normalisation_for_Nussbaum_Szkola_2}, but with $p_j$ instead of $q_k$.

					\item We are now interested in the case of equality in the statement $D(\rho\Vert\sigma)\allowbreak\geq 0$ that we just proved in (a). 
                    In that proof we made use of two inequalities, both of which have to be saturated if $D(\rho\|\sigma)=0$: the first one is~\eqref{eq:assuming_sigma_normalised}, and its saturation implies that $\Tr [\sigma]=1$, so that $\sigma$ is a normalized state. The second inequality is~\eqref{eq:Klein_proof_key_inequality}, where we employed the convexity of $f$ to state that the convex combination (with coefficients $\lambda_{jk}$) of the values of $f$ at the points $p_j/q_k$ is upper bounded by the value of $f$ at the convex combination of the points (with the same coefficients), which we calculated to be equal to~$1$ in~\eqref{eq:Klein_proof_key_inequality_f_1}. Since $f$ is actually strictly convex, by~\eqref{eq:strict_convexity_xlogx}, equality is only possible if $p_j/q_k=1$ whenever $\lambda_{jk}>0$. This implies that 
                    \begin{equation}
                    p_j \braket{\psi_j}{\phi_k} = q_k \braket{\psi_j}{\phi_k}\quad \forall\ 1\leq j\leq d,\ \ 1\leq k\leq d.
                    \label{eq:equality_Klein_proof_key_identity}
                    \end{equation}
                    To see this, restrict first, without loss of generality, to the case where $1\leq j\leq r_\rho$ and $1\leq k\leq r_\sigma$, and then distinguish the two cases where $\braket{\psi_j}{\phi_k} = 0$ and $\braket{\psi_j}{\phi_k} \neq 0$. From~\eqref{eq:equality_Klein_proof_key_identity} we obtain that
                    \begin{align}
                        \rho &= \sum_{j=1}^{d} p_j \ketbra{\psi_j}{\psi_j} \\
                        &= \sum_{j=1}^{d} p_j \ketbra{\psi_j}{\psi_j} \left(\sum_{k=1}^{d} \ketbra{\phi_k}{\phi_k}\right) \\
                        &= \sum_{j=1}^{d} \sum_{k=1}^{d} p_j \braket{\psi_j}{\phi_k} \ketbra{\psi_j}{\phi_k} \\
                        &= \sum_{j=1}^{d} \sum_{k=1}^{d} q_k \braket{\psi_j}{\phi_k} \ketbra{\psi_j}{\phi_k} \\
                        &= \left( \sum_{j=1}^{d} \ketbra{\psi_j}{\psi_j} \right) \sum_{k=1}^{d} q_k \ketbra{\phi_k}{\phi_k} \\
                        &= \sum_{k=1}^{d} q_k \ketbra{\phi_k}{\phi_k} \\
                        &= \sigma,
                    \end{align}
                    as claimed.

					\item First suppose that $\rho$, $\sigma$, and $\sigma'$ are positive definite. Since the logarithm is operator monotone (see Section~\ref{sec:math-tools:functions-herm-ops}), the operator inequality $\sigma'\geq\sigma$ implies that $\log_2(\sigma')\geq\log_2(\sigma)$, which implies that $\Tr[\rho\log_2(\sigma')]\allowbreak\geq\Tr[\rho\log_2\sigma]$. Therefore,
						\begin{align}
							D(\rho\Vert\sigma)&=\Tr[\rho(\log_2\rho-\log_2\sigma)]\\
							&\geq\Tr[\rho(\log_2\rho-\log_2\sigma')]\\
							&=D(\rho\Vert\sigma'),
						\end{align}
						as required. In the general case that the operators are not positive definite,  we apply the result to the positive definite operators $(1-\delta)\rho+\delta\pi$, $\sigma+\varepsilon\mathbbm{1}$, and $\sigma'+\varepsilon'\mathbbm{1}$, for $\delta,\varepsilon,\varepsilon' >  0$, and then use the following to obtain the result:
                        \begin{equation}\label{eq-q_rel_ent_double_lim}
							\lim_{\varepsilon\to 0^+}\lim_{\delta\to 0^+}D((1-\delta)\rho+\delta\pi\Vert\sigma+\varepsilon\mathbbm{1})=D(\rho\Vert\sigma),
						\end{equation}

                        The second statement follows by substituting $\sigma' \to \sigma$ and $\sigma \to \rho$ and applying faithfulness.
				\end{enumerate}
				
			\item \textit{Proof of the direct-sum property}: Define the classical--quantum operators
				\begin{equation}
					\rho_{XA}=\sum_{x\in\mathcal{X}}p(x)\ket{x}\!\bra{x}_X\otimes\rho_A^x,\quad \sigma_{XA}=\sum_{x\in\mathcal{X}}q(x)\ket{x}\!\bra{x}_X\otimes\sigma_A^x.
				\end{equation}
				If $\supp(\rho_{XA}) \not\subseteq \supp(\sigma_{XA})$, so that $D(\rho_{XA}\|\sigma_{XA}) = +\infty$, then there exists some $x\in\mathcal{X}$ such that $p(x)>0$ and, moreover, either $q(x)=0$ or $\supp(\rho_A^x)\not\subseteq \supp(\sigma_A^x)$. In the former case, we have $D(p\|q)=+\infty$, while, in the latter, we have $D(\rho_A^x\|\sigma_A^x)=+\infty$. In either case, \eqref{eq-rel_ent_direct_sum} is satisfied. Thus, in the rest of the proof we can assume that $\supp(\rho_{XA})\subseteq\supp(\sigma_{XA})$. Observe that
				\begin{align}
					\log_2\rho_{XA}&=\sum_{x\in\mathcal{X}}\ket{x}\!\bra{x}_X\otimes\log_2(p(x)\rho_A^x)\\
					&=\sum_{x\in\mathcal{X}}\ket{x}\!\bra{x}_X\otimes\log_2 p(x)\mathbbm{1}_{A}+\sum_{x\in\mathcal{X}}\ket{x}\!\bra{x}_X\otimes\log_2\rho_A^x,\label{eq-log_cq_state}\\
					\log_2\sigma_{XA}&=\sum_{x\in\mathcal{X}}\ket{x}\!\bra{x}_X\otimes\log_2(q(x)\sigma_A^x)\\
					&=\sum_{x\in\mathcal{X}}\ket{x}\!\bra{x}_X\otimes \log_2 q(x)\mathbbm{1}_{A}+\sum_{x\in\mathcal{X}}\ket{x}\!\bra{x}_X\otimes\log_2\sigma_A^x.
				\end{align}
				Then, 
				\begin{align}
					D(\rho_{XA}\Vert\sigma_{XA})&=\Tr[\rho_{XA}\log_2\rho_{XA}]-\Tr[\rho_{XA}\log_2\sigma_{XA}]\\
					&=\Tr\!\left[\sum_{X\in\mathcal{X}}p(x)\log_2(p(x))\ket{x}\!\bra{x}_X\otimes\rho_A^x\right.\nonumber\\
					&\qquad\qquad\left.+\sum_{x\in\mathcal{X}} p(x)\ket{x}\!\bra{x}_X\otimes\rho_A^x\log_2\rho_A^x\right]\notag\\
					&\qquad -\Tr\!\left[\sum_{x\in\mathcal{X}}p(x)\log_2(q(x))\ket{x}\!\bra{x}_X\otimes\rho_A^x\right.\nonumber\\
					&\qquad\qquad\quad\left.+\sum_{x\in\mathcal{X}}p(x)\ket{x}\!\bra{x}_X\otimes\rho_A^x\log_2\sigma_A^x\right]\\
					&=\sum_{x\in\mathcal{X}}\left[p(x)\log_2 p(x)-p(x)\log_2 q(x)\right]\nonumber\\
					&\qquad +\sum_{x\in\mathcal{X}}p(x)\Tr\!\left[\rho_A^x\log_2\rho_A^x-\rho_A^x\log_2\sigma_A^x\right]\\
					&=D(p\Vert q)+\sum_{x\in\mathcal{X}}p(x) D(\rho_A^x\Vert\sigma_A^x),
				\end{align}
				as required. \qedhere
		\end{enumerate}
	\end{Proof}

	An important consequence of Klein's inequality from the proposition abo\-ve is that
	\begin{equation}\label{eq-Klein_ineq_states}
		\boxed{\begin{aligned}
		D(\rho\Vert\sigma)&\geq 0\text{ for all states }\rho,\sigma,\text{ and }\\
		D(\rho\Vert\sigma)&=0\text{ if and only if }\rho=\sigma.\end{aligned}}
	\end{equation}
	This allows us to use the quantum relative entropy as a distinguishability measure for quantum states. We emphasize, however, that the quantum relative entropy is not a metric in the mathematical sense, since it is neither symmetric in its two arguments nor does it satisfy the triangle inequality.

\begin{figure}
		\centering
		\includegraphics[scale=0.8]{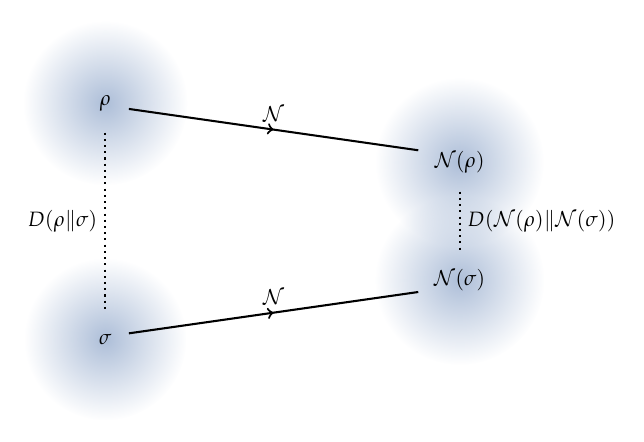}
		\caption{Illustration of the data-processing inequality for the quantum relative entropy (Theorem \ref{thm-monotone_rel_ent}). The quantum states $\rho$, $\sigma$, $\mathcal{N}(\rho)$, and $\mathcal{N}(\sigma)$ are represented by spheres that signify the amount of ``space'' they occupy in the Hilbert space. While the states $\rho$ and $\sigma$ are nearly distinguishable as depicted, since their spheres do not overlap, after processing with the channel~$\mathcal{N}$, the states become much less distinguishable because the spheres overlap significantly.}\label{fig-q_rel_ent_monotone}
	\end{figure}

	We now come to one of the most important properties of the quantum relative entropy that is used frequently throughout this book: the \textit{data-processing inequality}. It is  also called the \textit{monotonicity of the quantum relative entropy}.
\begingroup
\allowdisplaybreaks[0]
	\begin{theorem*}{Data-Processing Inequality for Quantum Relative Entropy}{thm-monotone_rel_ent}
		Let $\rho$ be a state, $\sigma$ a positive semi-definite operator, and $\mathcal{N}$ a quantum channel. Then,
		\begin{equation}
			D(\rho\Vert\sigma)\geq D(\mathcal{N}(\rho)\Vert\mathcal{N}(\sigma)).
		\end{equation}
	\end{theorem*}
\endgroup	
	In other words, the quantum relative entropy $D(\rho\Vert\sigma)$ can only decrease or stay the same if we apply the same quantum channel $\mathcal{N}$ to the states $\rho$ and $\sigma$. When the quantum relative entropy is interpreted as a distinguishability measure on quantum states, the data-processing inequality tells us that the distinguishability of two quantum states cannot increase when we act on them with the same quantum channel; see Figure \ref{fig-q_rel_ent_monotone}. We postpone the proof of the data-processing inequality to later in the chapter, where it follows easily as a consequence of the data-processing inequality for the Petz--R\'{e}nyi and sandwiched R\'{e}nyi relative entropies (see Corollaries~\ref{cor-rel_ent_monotone1} and \ref{cor-rel_ent_monotone2}, respectively).
	
	It is typically interesting and illuminating to investigate the conditions under which an important inequality is saturated. The data-processing inequality for quantum relative entropy is no exception. In the case that the action of the quantum channel can be reversed on $\rho$ and $\sigma$, so that there exists a recovery channel $\mathcal{R}$ satisfying
	\begin{equation}
	\rho = (\mathcal{R} \circ \mathcal{N})(\rho), \qquad \sigma = (\mathcal{R} \circ \mathcal{N})(\sigma),
	\label{eq-reversal-ch-for-rel-ent}
	\end{equation}
	then it follows from an application of Theorem~\ref{thm-monotone_rel_ent} that
	\begin{align}
	D(\mathcal{N}(\rho)\Vert\mathcal{N}(\sigma)) & \geq D((\mathcal{R} \circ \mathcal{N})(\rho)\Vert(\mathcal{R} \circ \mathcal{N})(\sigma))\\
	& = D(\rho\Vert\sigma).
	\end{align}
Thus, by combining with Theorem~\ref{thm-monotone_rel_ent}, we in fact have that
\begin{equation}
D(\rho\Vert\sigma)=  D(\mathcal{N}(\rho)\Vert\mathcal{N}(\sigma)),
\end{equation}
so that the existence of a recovery channel implies saturation of the data-processing inequality in Theorem~\ref{thm-monotone_rel_ent}.

On the other hand, suppose that $\rho$, $\sigma$, and $\mathcal{N}$ are such that the inequality in Theorem~\ref{thm-monotone_rel_ent} is saturated:
\begin{equation}
D(\rho\Vert\sigma)=  D(\mathcal{N}(\rho)\Vert\mathcal{N}(\sigma)),
\end{equation}
Then it is a non-trivial result that there exists a recovery channel $\mathcal{R}$ such that the equality in \eqref{eq-reversal-ch-for-rel-ent} holds. In fact, this channel can be taken as the Petz recovery channel from Definition~\ref{def-Petz_recovery}. We do not provide a proof here and instead point to the Bibliographic Notes in Section~\ref{sec:QEI:bib-notes} for more details.

	One of the remarkable aspects of the data-processing inequality for the qua\-ntum relative entropy is that it alone can be used to prove many of the properties of the quantum relative entropy stated in Proposition~\ref{prop-rel_ent}. For example, Klein's inequality follows by considering the trace channel $\Tr$, so that for every state $\rho$ and positive semi-definite operator $\sigma$ such that $\Tr[\sigma]\leq 1$, we find that
	\begin{align}
		D(\rho\Vert\sigma)&\geq D(\Tr(\rho)\Vert\Tr(\sigma))=\Tr(\rho)\log_2\!\left(\frac{\Tr(\rho)}{\Tr(\sigma)}\right)\\
		& =\log_2\!\left(\frac{1}{\Tr(\sigma)}\right)\geq 0. \label{eq-QEI:alt-klein-proof}
	\end{align}
	Isometric invariance also follows from the data-processing inequality. The inequality $D(\rho\|\sigma) \geq D(V\rho V^\dag \| V \sigma V^\dag)$ follows from data processing because $(\cdot) \to V(\cdot)V^\dag$ is a channel. The reverse inequality also follows from data processing because $D(V\rho V^\dag \| V \sigma V^\dag) \geq D(\mathcal{R}_V(V\rho V^\dag) \| \mathcal{R}_V(V \sigma V^\dag)) = D(\rho  \|  \sigma )$, where $\mathcal{R}_V$ is the reversal channel defined in \eqref{eq-isometry_recover} and we used \eqref{eq-QM-ch:reversal-prop-1}--\eqref{eq-QM-ch:reversal-prop-last}.
	
	Another important fact that follows from the data-processing inequality for quantum relative entropy is its \textit{joint convexity}. This latter property is so central that, together with isometric invariance and additivity (Proposition~\ref{prop-rel_ent}), it actually implies the data-processing inequality.
	
	\begin{proposition*}{Joint Convexity of Quantum Relative Entropy}{prop-rel_ent_joint_convex}
		Let $p\colon \mathcal{X}\rightarrow[0,1]$ be a probability distribution over a finite alphabet $\mathcal{X}$ with associated $|\mathcal{X}|$-dimensional system $X$, let $\{\rho_A^{x}\}_{x\in\mathcal{X}}$ be a set of states on a system~$A$, and let $\{\sigma_A^{x}\}_{x\in\mathcal{X}}$ be a set of positive semi-definite operators on $A$. Then,
		\begin{equation}
			\sum_{x\in\mathcal{X}}p(x)D(\rho_A^x\Vert\sigma_A^x) \geq 
			D\!\left(\sum_{x\in\mathcal{X}}p(x)\rho_A^x\Bigg\Vert\sum_{x\in\mathcal{X}}p(x)\sigma_A^x\right).
		\end{equation}
	\end{proposition*}
	
	\begin{Proof}
		 Define the classical--quantum state and operator, respectively, as
		\begin{equation}
			\rho_{XA} \coloneqq \sum_{x\in\mathcal{X}}p(x)\ket{x}\!\bra{x}_X\otimes\rho_A^{x},\quad\sigma_{XA}\coloneqq\sum_{x\in\mathcal{X}}p(x)\ket{x}\!\bra{x}_X\otimes\sigma_A^{x}.
		\end{equation}
		By the direct-sum property of the quantum relative entropy (Proposition~\ref{prop-rel_ent}), we find that
		\begin{equation}
			D(\rho_{XA}\Vert\sigma_{XA})=\sum_{x\in\mathcal{X}}p(x)D(\rho_A^x\Vert\sigma_A^x).
		\end{equation}
		Then, since we have $\rho_A=\Tr_X[\rho_{XA}]=\sum_{x\in\mathcal{X}}p(x)\rho_A^{x}$ and  $\sigma_A=\Tr_X[\sigma_{XA}]=\sum_{x\in\mathcal{X}}p(x)\sigma_A^{x}$, we apply the data-processing inequality for quantum relative entropy with respect to the partial trace channel $\Tr_X$ to obtain
		\begin{align}
			\sum_{x\in\mathcal{X}}p(x)D(\rho_A^{x}\Vert\sigma_A^{x})&=
			D(\rho_{XA}\Vert\sigma_{XA}) \\
& \geq D(\Tr_X(\rho_{XA})\Vert\Tr_X(\sigma_{XA}))\\
					&=	D\!\left(\sum_{x\in\mathcal{X}}p(x)\rho_A^{x}\Bigg\Vert\sum_{x\in\mathcal{X}}p(x)\sigma_A^{x}\right)
			,
		\end{align}
		which is the desired joint convexity of quantum relative entropy.
	\end{Proof}
	
	\begin{exercise}{ex-QEI:ext-direct-sum-joint-conv}
	Let $\rho_{XA} \coloneqq \sum_{x\in\mathcal{X}}p(x)\ket{x}\!\bra{x}_X\otimes\rho_A^x$ and  $\sigma_{XA}  \coloneqq \sum_{x\in\mathcal{X}}q(x)\ket{x}\!\bra{x}_X\otimes\sigma_A^x$.
	 Prove that
	\begin{equation}
	D(\rho_{XA}\|\sigma_{XA}) \geq D(p\|q) + D\!\left(\sum_{x\in\mathcal{X}}p(x)\rho_A^x\Bigg\Vert\sum_{x\in\mathcal{X}}p(x)\sigma_A^x\right).
	\end{equation}
	\end{exercise}

\subsection{Information Measures from Quantum Relative Entropy}

	As stated at the beginning of this chapter, the quantum relative entropy acts, as in the classical case, as a parent quantity for all of the fundamental information-theoretic quantities based on the quantum entropy. Indeed, using the properties of the quantum relative entropy stated previously, it is straightforward to verify the following:
	\begin{enumerate}
		\item The quantum entropy $H(\rho)$ of a state $\rho$ is given by
			\begin{equation}\label{eq-quantum_entropy_relent}
				H(\rho)=-D(\rho\Vert\mathbbm{1}).
			\end{equation}
		\item The quantum conditional entropy $H(A|B)_\rho$ of a bipartite state $\rho_{AB}$ is given by
			\begin{align}
				H(A|B)_\rho&=-D(\rho_{AB}\Vert\mathbbm{1}_A\otimes \rho_B)\label{eq-cond_ent}\\
				&=-\inf_{\sigma_B\in\Density(\mathcal{H}_B)}D(\rho_{AB}\Vert\mathbbm{1}_A\otimes\sigma_B)\label{eq-cond_ent_opt},
			\end{align}
			and the coherent information $I(A\rangle B)_\rho$ of a bipartite state $\rho_{AB}$ is given by
			\begin{align}
				I(A\rangle B)_\rho&=D(\rho_{AB}\Vert\mathbbm{1}_A\otimes \rho_B)\label{eq-coh_inf}\\
				&=\inf_{\sigma_B\in\Density(\mathcal{H}_B)}D(\rho_{AB}\Vert\mathbbm{1}_A\otimes\sigma_B)\label{eq-coher_inf_opt}.
			\end{align}
			Observe that
			\begin{equation}
			H(A|B)_{\rho} = -I(A\rangle B)_{\rho}
			\label{eq:QEI:coh-info-to-cond-ent}
			\end{equation}
			for every bipartite state $\rho_{AB}$.
			Similarly, we can write the reverse coherent information as
			\begin{align}
				I(B\rangle A)_{\rho}&=D(\rho_{AB}\Vert\rho_A\otimes\mathbbm{1}_B) \label{eq-rev_coh_inf} \\
				&=\inf_{\sigma_A\in\Density(\mathcal{H}_A)}D(\rho_{AB}\Vert\sigma_A\otimes\mathbbm{1}_B)\label{eq-rev_coh_inf_opt}
			\end{align}
			
		\item The quantum mutual information $I(A;B)_\rho$ of a bipartite state $\rho_{AB}$ is given by
			\begin{align}
				I(A;B)_\rho&=D(\rho_{AB}\Vert \rho_A\otimes \rho_B)\label{eq-mut_inf}\\
				&=\inf_{\sigma_B\in\Density(\mathcal{H}_B)}D(\rho_{AB}\Vert\rho_A\otimes\sigma_B)\label{eq-mut_inf_opt}\\
&=\inf_{\tau_A\in\Density(\mathcal{H}_A)}D(\rho_{AB}\Vert\tau_A\otimes\rho_B) \\		&=\inf_{\substack{\tau_A\in\Density(\mathcal{H}_A)\\\sigma_B\in\Density(\mathcal{H}_B)}}D(\rho_{AB}\Vert\tau_A\otimes\sigma_B).\label{eq-mut_inf_double_opt}
			\end{align}
			
		\item The quantum conditional mutual information $I(A;B|C)_\rho$ of a tripartite state $\rho_{ABC}$ is given by
			\begin{equation}\label{eq-QCMI_rel_ent}
				I(A;B|C)_\rho=D(\rho_{ABC}\Vert \sigma_{ABC}),
			\end{equation}
			where
			\begin{equation}
				\sigma_{ABC}\coloneqq 2^{\log_2(\rho_{AC}\otimes\mathbbm{1}_B)+\log_2(\rho_{BC}\otimes\mathbbm{1}_A)-\log_2(\mathbbm{1}_{AB}\otimes \rho_C)}.
			\end{equation}
	\end{enumerate}
	
	The expressions in \eqref{eq-cond_ent_opt}, \eqref{eq-coher_inf_opt}, and \eqref{eq-mut_inf_opt}, in terms of an optimization of the conditional entropy, coherent information, and mutual information, respectively, are useful for defining infor\-mation-theoretic quantities analogous to these ones in the context of generalized divergences, which we introduce in the next section.
	
	\begin{exercise}{ex-QEI:opt-formulas-info-quantities}
	Verify the equalities in \eqref{eq-cond_ent_opt}, \eqref{eq-coher_inf_opt}, and \eqref{eq-mut_inf_opt}. \textit{Hint}: First prove that
	\begin{equation}
		D(\rho_{AB}\Vert\tau_A\otimes\rho_B)+D(\tau_A\otimes\rho_B\Vert\tau_A\otimes\sigma_B)=D(\rho_{AB}\Vert\tau_A\otimes\sigma_B),
	\end{equation}
	for all states $\tau_A$ and $\sigma_B$. Then use the fact that 
	\begin{equation}D(\tau_A\otimes\rho_B\Vert\tau_A\otimes\sigma_B)\geq 0
	\end{equation}
	 which holds for all states $\tau_A$ and $\sigma_B$, by Klein's inequality as stated in \eqref{eq-Klein_ineq_states}. Set $\tau_A=\pi_A$ to prove \eqref{eq-cond_ent_opt} and \eqref{eq-coher_inf_opt}, and set $\tau_A=\rho_A$ to prove \eqref{eq-mut_inf_opt}.
	 \end{exercise}
	
	The properties of the quantum relative entropy, such as the ones stated in Propositions~\ref{prop-rel_ent} and \ref{prop-rel_ent_joint_convex}, can be directly translated to properties of the derived information measures stated above. Some of these properties are used frequently throughout the book, and so we state them here for convenience. They are straightforward to verify using definitions and properties of the quantum relative entropy.
	\begin{itemize}
	\item \textit{Additivity of the quantum entropy for product states $\rho$ and $\tau$}:
				\begin{equation} \label{eq-quantum_entropy_additive}
					H(\rho\otimes\tau) = H(\rho) + H(\tau).
				\end{equation}

	\item \textit{Isometric invariance of the quantum entropy for a state $\rho$ and an isometry $V$}:
				\begin{equation} 
					H(\rho) = H(V \rho V^{\dag}).
				\end{equation}
	
		\item \textit{Concavity of the quantum entropy}: The joint convexity of the quantum relative entropy, as stated in Proposition \ref{prop-rel_ent_joint_convex}, and the identity in \eqref{eq-quantum_entropy_relent} imply that the quantum entropy is concave: if $p\colon \mathcal{X}\to[0,1]$ is a probability distribution over a finite alphabet $\mathcal{X}$ and $\{\rho_A^x\}_{x\in\mathcal{X}}$ is a set of states on a system~$A$, then
			\begin{equation}\label{eq-q_entropy_concave}
				H\!\left(\sum_{x\in\mathcal{X}}p(x)\rho_A^x\right)\geq\sum_{x\in\mathcal{X}}p(x)H(\rho_A^x).
			\end{equation}
			By taking a spectral decomposition of a state $\rho$, applying the concavity inequality above, and the fact that the entropy of a pure state is equal to zero, we conclude that the quantum entropy is non-negative for every state~$\rho$:
			\begin{equation}
			H(\rho)\geq 0.
			\label{eq:QEI:entropy-non-neg}
			\end{equation}
			By employing the mixing property of the Heisenberg--Weyl unitaries from \eqref{eq-HW_twirl}, the invariance of entropy under a unitary, its concavity, and the fact that the entropy of the maximally mixed state $\pi_A$ is equal to $\log_2 d_A$, we conclude the following dimension bound for the entropy of a state of system~$A$:
			\begin{equation}
			H(\rho)\leq \log_2 d_A. \label{eq-entropy-dim-bnd}
			\end{equation}
			
		\item \textit{Direct-sum property of the quantum entropy}: The direct-sum property of the quantum relative entropy (see Proposition \ref{prop-rel_ent}) translates to the following for the quantum entropy: If $p\colon \mathcal{X}\to[0,1]$ is a probability distribution over a finite alphabet $\mathcal{X}$ with associated $|\mathcal{X}|$-dimensional system $X$ and $\{\rho_A^x\}_{x\in\mathcal{X}}$ is a set of states on a system $A$, then
			\begin{equation}\label{eq-q_entropy_direct_sum}
				H\!\left(\sum_{x\in\mathcal{X}}p(x)\ket{x}\!\bra{x}_X\otimes\rho_A^x\right)=H(p)+\sum_{x\in\mathcal{X}}p(x)H(\rho_A^x),
			\end{equation}
			where $H(p)\coloneqq-\sum_{x\in\mathcal{X}}p(x)\log_2p(x)$ is the (classical) Shannon entropy of the probability distribution $p$.
			
			\item \textit{Chain rule for conditional entropy}: For every state $\rho_{ABC}$, the following equality holds:
			\begin{equation}
			H(AB|C)_{\rho} = H(A|C)_{\rho} + H(B|AC)_{\rho} = H(B|C)_\rho + H(A|BC)_\rho.
			\label{eq-QEI:chain-rule-cond-ent}
			\end{equation}
			If the system $C$ is trivial, so that the state is a bipartite state $\rho_{AB}$, then this equality reduces to the following:
			\begin{equation}
			H(AB)_{\rho} = H(A)_{\rho} + H(B|A)_{\rho} = H(B)_\rho + H(A|B)_\rho.
			\end{equation}

		\item \textit{Chain rule for quantum mutual information}: For every state $\rho_{ABC}$, the following equalities hold:
			\begin{align}
				I(A;BC)_\rho &= I(A;B)_\rho + I(A;C|B)_\rho = I(A;C)_\rho + I(A;B|C)_\rho, \label{eq-q_mut_inf_chain_rule} \\
                I(AB;C)_\rho &= I(A;C)_\rho + I(B;C|A)_\rho = I(B;C)_\rho + I(A;C|B)_\rho. \label{eq-q_mut_inf_chain_rule_2}
			\end{align}
			We call \eqref{eq-QEI:chain-rule-cond-ent}, \eqref{eq-q_mut_inf_chain_rule}, and \eqref{eq-q_mut_inf_chain_rule_2} \textit{chain rules}, because they allow us to break down the correlations between a single system and a bipartite system as the sum of two terms, each representing only the correlations between two of these systems. For example, \eqref{eq-q_mut_inf_chain_rule} demonstrates that the correlations between $A$ and $BC$ can be built up by first establishing correlations between $A$ and $B$ (signified by $I(A;B)_\rho$), then establishing correlations between $A$ and $C$, given the correlations with $B$ (signified by $I(A;C|B)_\rho$).
	\end{itemize}
	
	
\begin{exercise}{ex-QEI:entropy-props}
Provide explicit proofs of the properties in \eqref{eq-quantum_entropy_additive}--\eqref{eq-q_entropy_direct_sum}, by following what is stated above.
\end{exercise}

\begin{exercise}{ex-QEI:chain-rules}
Verify the chain rules stated in \eqref{eq-QEI:chain-rule-cond-ent}, \eqref{eq-q_mut_inf_chain_rule}, and \eqref{eq-q_mut_inf_chain_rule_2}. More generally, prove that $I(A;BC|D)_\rho = I(A;B|D)_\rho + I(A;C|BD)_\rho$ for a four-party state $\rho_{ABCD}$.
\end{exercise}
	
\subsection{Quantum Conditional Mutual Information}
	
	We now develop in detail some properties of the quantum conditional mutual information that we use later in the book.
	
	We start with the proof of the strong subadditivity property of the quantum entropy, which we recall from \eqref{eq-QCMI_def} is the statement that
	\begin{equation}
		I(A;B|C)_{\rho}=H(A|C)_{\rho}+H(B|C)_{\rho}-H(AB|C)_{\rho}\geq 0
	\end{equation}
	for every state $\rho_{ABC}$.
	
	\begin{theorem*}{Strong Subadditivity of Quantum Entropy}{thm-SSA}
		For every state $\rho_{ABC}$, the following inequality holds
		\begin{equation}
			I(A;B|C)_{\rho}\geq 0.
			\label{eq:QEI:SSA-ineq-CMI}
		\end{equation}
	\end{theorem*}
	
	\begin{Proof}
		One way to prove this result is by means the data-processing inequality for the quantum relative entropy, along with the expression for the quantum conditional entropy in \eqref{eq-cond_ent}. We start by using the definition of the quantum conditional entropy to rewrite the quantum conditional mutual information defined in \eqref{eq-QCMI_def} as 
		\begin{align}
			I(A;B|C)_\rho&=H(A|C)_\rho+H(B|C)_\rho-H(AB|C)_\rho\\
			&=H(B|C)_\rho-H(B|AC)_\rho.
		\end{align}
		Then, using the expression in \eqref{eq-cond_ent} for the quantum conditional entropy in terms of the quantum relative entropy, we find that
		\begin{equation}
			I(A;B|C)_\rho=D(\rho_{ABC}\Vert\mathbbm{1}_B\otimes\rho_{AC})-D(\rho_{BC}\Vert\mathbbm{1}_B\otimes\rho_C).
		\end{equation}
		Finally, observe that, by the data-processing inequality for the quantum relative entropy with respect to the partial trace channel $\Tr_A$, we obtain
		\begin{align}
			D(\rho_{ABC}\Vert\mathbbm{1}_B\otimes\rho_{AC})&\geq D(\Tr_A[\rho_{ABC}]\Vert\mathbbm{1}_B\otimes\Tr_A[\rho_{AC}])\\
			&=D(\rho_{BC}\Vert\mathbbm{1}_B\otimes\rho_C),
		\end{align}
		which implies that $I(A;B|C)_\rho\geq 0$, as required.
	\end{Proof}
	
	Two direct consequences of strong subadditivity are that the conditional entropy is concave and non-negative for every separable state. We detail these properties below.
	
	\begin{corollary*}{Concavity of Conditional Entropy}{cor-QEI:cond-entr-concave}
	 The conditional entropy is concave, and the coherent information is convex:
	 \begin{align}
	 H(A|B)_{\bar{\rho}} & \geq \sum_{x\in \mathcal{X}} p(x) H(A|B)_{\rho^x},
	 \label{eq:QEI:concavity-cond-ent}\\
	 I(A\rangle B)_{\bar{\rho}} & \leq \sum_{x\in \mathcal{X}} p(x) I(A\rangle B)_{\rho^x},
	  \label{eq:QEI:coh-info-convex}
	 \end{align}
	 where $\mathcal{X}$ is a finite alphabet, $p\colon \mathcal{X}\to [0,1]$ is a probability distribution, $\{\rho^{x}_{AB}\}_{x\in \mathcal{X}}$ is a set of bipartite states, and $\bar{\rho} \coloneqq \sum_{x\in \mathcal{X}} p(x)\rho^{x}_{AB}$.
	 \end{corollary*}
	 
	 \begin{Proof}
	 This follows directly by constructing the classical--quantum state
	 \begin{equation}
	 \rho_{XAB} = \sum_{x\in \mathcal{X}} p(x) |x\rangle \!\langle x| \otimes \rho^{x}_{AB},
	 \end{equation}
	  applying strong subadditivity $H(A|B) - H(A|BX)_{\rho } = I(A;X|B)_{\rho} \geq 0$, and observing that
	  \begin{equation}
	  H(A|BX)_{\rho } = \sum_{x\in \mathcal{X}} p(x) H(A|B)_{\rho^x}.
	  \end{equation}
	  By applying \eqref{eq:QEI:coh-info-to-cond-ent}, we conclude from \eqref{eq:QEI:concavity-cond-ent} that the coherent information is convex.
	  \end{Proof}
	
	\begin{corollary*}{Non-Negativity of Conditional Entropy on Separable States}{cor-QEI:cond-entr-sep-non-neg}
	The conditional entropy $H(A|B)_{\sigma}$ is non-negative for every separable state $\sigma_{AB}$.
	\end{corollary*}
	
	\begin{Proof}
	Recall from Definition~\ref{def-sep_ent_state} that $\sigma_{AB}$ is separable if it can be written as
	\begin{equation}
	\sigma_{AB} = \sum_{x \in\mathcal{X}} p(x) \tau^x_A \otimes \omega^x_B,
	\end{equation}
	 where  $p\colon \mathcal{X}\to[0,1]$ is a probability distribution over a finite alphabet $\mathcal{X}$ and  $\{\tau^x_A\}_{x\in\mathcal{X}}$ and $\{\omega^x_B\}_{x\in\mathcal{X}}$ are sets of states. Defining the extension $\sigma_{XAB}$ of $\sigma_{AB}$ as
	 \begin{equation}
	 \sigma_{XAB} \coloneqq \sum_{x \in\mathcal{X}} p(x) |x\rangle\! \langle x|_X \otimes \tau^x_A \otimes \omega^x_B ,
	 \end{equation}
	 we then conclude that $I(A;X|B)_{\sigma} \geq 0$, which implies the desired inequality:
	 \begin{align}
	 H(A|B)_{\sigma} &
	 \geq H(A|BX)_{\sigma} \\
	 & = \sum_{x\in\mathcal{X}} p(x) H(A|B)_{\tau^x \otimes \omega^x} \\
	 & = \sum_{x\in\mathcal{X}} p(x) H(A)_{\tau^x} \geq 0.
	 \label{eq:QEI:cond-ent-non-neg-sep}
	 \end{align}
	 This concludes the proof.
	 \end{Proof}

	We now prove some other properties of the quantum conditional mutual information that recur throughout the book.
	
	\begin{proposition*}{Properties of Quantum Conditional Mutual Information}{prop-cond_mut_inf_properties}
		The quantum conditional mutual information has the following properties:
		\begin{enumerate}
		\item \textit{Symmetry}: For every state $\rho_{ABC}$, we have
		$
		I(A;B|C)_{\rho} = I(B;A|C)_{\rho}.
		$
			\item \textit{Local entropy and dimension bounds}: For every state $\rho_{ABC}$,
				\begin{align}
					I(A;B|C)_{\rho} & \leq
					2 \min\{H(A)_{\rho},H(B)_{\rho}\} 
					 \leq 2\log_2(\min\{d_A,d_B\}). \label{eq-QCMI_dim_bound}
				\end{align}
				Let $\sigma_{XBC}$ be a classical--quantum state of the form
				\begin{equation}
				\sigma_{XBC} = \sum_{x \in \mathcal{X}} p(x) |x\rangle\!\langle x|_X \otimes \rho^x_{BC},
				\label{eq:QEI:cq-cmi-dim-bnd}
				\end{equation}
				where
				$p\colon \mathcal{X}\to[0,1]$ is a probability distribution over a finite alphabet $\mathcal{X}$ with associated $|\mathcal{X}|$-dimensional system $X$, and  $\{\rho_{BC}^x\}_{x\in\mathcal{X}}$ is a set of states.
				Then the following  bounds hold
				\begin{equation}
				I(X;B|C)_{\sigma} \leq H(X)_{\sigma} \leq \log_2|\mathcal{X}|.
				\label{eq:QEI:dim-bnd-CMI-cq}
				\end{equation}
			\item \textit{Product conditioning system}: For a state $\rho_{ABC}=\sigma_{AB}\otimes\tau_C$, 
				\begin{equation}
					I(A;B|C)_{\rho}=I(A;B)_\sigma.
				\end{equation}
			\item \textit{Additivity}: For states $\rho_{A_1B_1C_1}$ and $\tau_{A_2B_2C_2}$, the following equality holds for the product state $\rho_{A_1B_1C_1}\otimes\tau_{A_2B_2C_2}$:
				\begin{equation}
					I(A_1A_2;B_1B_2|C_1C_2)_{\rho\otimes\tau}=I(A_1;B_1|C_1)_\rho+I(A_2;B_2|C_2)_\tau.
					\label{eq:QEI:CMI-additive}
				\end{equation}
			
			\item \textit{Direct-sum property}: 
For the classical--quantum state
				\begin{equation}
					\sigma_{XABC}=\sum_{x\in\mathcal{X}}p(x)\ket{x}\!\bra{x}_X\otimes\rho_{ABC}^x
				\end{equation}
				the following equality holds: 
				\begin{equation}
					I(A;B|CX)_{\sigma}=\sum_{x\in\mathcal{X}}p(x)I(A;B|C)_{\rho^x}.
					\label{eq:QEI:direct-sum-cmi}
				\end{equation}
			
			\item \textit{Chain rule}: For every state $\rho_{AB_1B_2C}$,
				\begin{align}
					I(A;B_1B_2|C)_\rho&=I(A;B_1|C)_\rho+I(A;B_2|B_1C)_\rho . \label{eq-QCMI_chain_rule}
				\end{align}
			
			\item \textit{Data-processing inequality for local channels}: 
			For every state $\rho_{ABC}$ and all local channels $\mathcal{N}_{A\to A'}$ and $\mathcal{M}_{B\to B'}$, the following inequality holds:
				\begin{equation}
					I(A;B|C)_{\rho}\geq I(A';B'|C)_{\omega},
				\end{equation}
				where $\omega_{A'B'C}\coloneqq(\mathcal{N}_{A\to A'}\otimes\mathcal{M}_{B\to B'})(\rho_{ABC})$.
		\end{enumerate}
	\end{proposition*}

	\begin{Proof}
	 We establish the various properties one by one.
		\begin{enumerate}
		
		\item Symmetry under exchange of $A$ and $B$ follows immediately from the definition.
		
			\item Using the definition of conditional entropy, we can write $I(A;B|C)_{\rho}$ as
				\begin{align}
					I(A;B|C)_{\rho}&=H(AC)_{\rho}-H(C)_{\rho}+H(BC)_{\rho}-H(ABC)_{\rho}\\
					&=H(A|C)_{\rho}-H(A|BC)_{\rho}.
				\end{align}
				The inequality $H(A|C)_{\rho} \leq H(A)_\rho$ is a direct consequence of strong subadditivity, under a relabeling and with trivial conditioning system.
				Then, by the dimension bound from \eqref{eq-entropy-dim-bnd}, it follows that $H(A)_{\rho}\leq\log_2d_A$. Therefore,
				\begin{equation}\label{eq-cond_entr_dim_bound}
					H(A|C)_{\rho}\leq \log_2d_A.
				\end{equation}
				Now, let $\ket{\psi}_{ABCE}$ be a purification of $\rho_{ABC}$. Then, since $\rho_{ABC}$ and $\psi_E$ have the same spectrum, and since $\rho_{BC}$ and $\psi_{AE}$ have the same spectrum, we obtain
				\begin{align}
					H(A|BC)_{\rho}&=H(ABC)_{\rho}-H(BC)_{\rho}\\
					&=H(E)_{\psi}-H(AE)_{\psi}\\
					&=-H(A|E)_{\psi}\\
					&\geq -H(A)\\
					&\geq -\log_2d_A,
				\end{align}
				where the first inequality follows from \eqref{eq-cond_entr_dim_bound}, and the second inequality, as before, from the fact that $H(A)_{\rho}\leq \log_2d_A$ for every state $\rho$. Therefore,
				\begin{equation}
					H(A|BC)_{\rho}\geq -H(A)_{\rho}\geq  -\log_2 d_A, \label{eq:QEI:lower-bnd-cond-ent-4-dim-bnd}
				\end{equation}
				which means that $I(A;B|C)_{\rho}=H(A|C)_{\rho}-H(A|BC)_{\rho}\leq 2 H(A)_{\rho} \leq 2\log_2d_A$. By applying symmetry under exchange of $A$ and $B$ and the same argument, we conclude that $I(A;B|C)_{\rho} \leq 2H(B)_{\rho} \leq 2 \log_2 d_B$.
				%
				Thus, we conclude \eqref{eq-QCMI_dim_bound}.
				
				If the system $A$ is classical (as in \eqref{eq:QEI:cq-cmi-dim-bnd}), then the state is separable with respect to the bipartite cut $A|BC$. As such, the lower bound in \eqref{eq:QEI:lower-bnd-cond-ent-4-dim-bnd} improves to $H(A|BC)_{\rho}\geq 0$ (as a consequence of Corollary~\ref{cor-QEI:cond-entr-sep-non-neg}), implying that the upper bound improves to $I(A;B|C)_{\rho}\leq H(A)_{\rho}\leq \log_2 d_A$ in this case.

			\item For $\rho_{ABC}=\sigma_{AB}\otimes\tau_C$, we have
				\begin{equation}
					I(A;B|C)_{\rho}=H(A|C)_{\sigma\otimes\tau}+H(B|C)_{\sigma\otimes\tau}-H(AB|C)_{\sigma\otimes\tau}.
				\end{equation}
				Now, we use the  fact that
				\begin{equation}\label{eq-cond_entropy_prod}
					H(A|C)_{\sigma\otimes\tau}=H(A)_{\sigma},
				\end{equation}
				which follows from \eqref{eq-quantum_entropy_additive}. Analogously, $H(B|C)_{\sigma\otimes\tau}=H(B)_{\sigma}$ and $H(AB|C)_{\sigma\otimes\tau}=H(AB)_{\sigma}$, and we find that
				\begin{equation}
					I(A;B|C)_\rho=H(A)_{\sigma}+H(B)_{\sigma}-H(AB)_{\sigma}=I(A;B)_{\sigma}.
				\end{equation}

			\item By writing
				\begin{multline}
					I(A_1A_2;B_1B_2|C_1C_2)_{\rho\otimes\tau}=H(A_1A_2|C_1C_2)_{\rho\otimes\tau}+H(B_1B_2|C_1C_2)_{\rho\otimes\tau}\\
					 -H(A_1A_1B_1B_2|C_1C_2)_{\rho\otimes\tau},
				\end{multline}
				and recognizing that each conditional entropy on the right-hand side is evaluated on a product state, we can use \eqref{eq-quantum_entropy_additive} to obtain the desired result. As an example, we evaluate the first term on the right-hand side:
				\begin{align}
					&H(A_1A_2|C_1C_2)_{\rho\otimes\tau}\notag \\
					&\quad =H(A_1A_2C_1C_2)_{\rho\otimes\tau}-H(C_1C_2)_{\rho\otimes\tau}\\
					&\quad=H(A_1C_1)_{\rho}+H(A_2C_2)_{\tau}-H(C_1)_{\rho}-H(C_2)_{\tau}\\
					&\quad=H(A_1|C_1)_{\rho}+H(A_2|C_2)_{\tau}.
				\end{align}
				
			\item By definition, we have that
				\begin{equation}\label{eq-QCMI-direct_sum_pf}
					I(A;B|CX)_{\sigma}=H(A|CX)_{\sigma}+H(B|CX)_{\sigma}-H(AB|CX)_{\sigma}.
				\end{equation}
				Let us consider the first term on the right-hand side. Using the definition of the quantum conditional entropy and using the direct-sum property of the quantum entropy, as stated in \eqref{eq-q_entropy_direct_sum}, we obtain
				\begin{align}
					& H(A|CX)_{\sigma}\notag \\
					&=H(ACX)_{\sigma}-H(CX)_{\sigma}\\
					&=H(p)+\sum_{x\in\mathcal{X}}p(x)H(AC)_{\rho^x}-H(p)-\sum_{x\in\mathcal{X}}p(x)H(C)_{\rho^x}\\
					&=\sum_{x\in\mathcal{X}}p(x)\left(H(AC)_{\rho^x}-H(C)_{\rho^x}\right)\\
					&=\sum_{x\in\mathcal{X}}p(x)H(A|C)_{\rho^x}.
				\end{align}
				The other terms on the right-hand side of \eqref{eq-QCMI-direct_sum_pf} are evaluated similarly, and we ultimately arrive at
				\begin{align}
					\notag I(A;B|CX)_{\sigma}&=\sum_{x\in\mathcal{X}}p(x)\left(H(A|C)_{\rho^x}+H(B|C)_{\rho^x}-H(AB|C)_{\rho^x}\right)\\
					&=\sum_{x\in\mathcal{X}}p(x)I(A;B|C)_{\rho^x}.
				\end{align}
				
			\item This follows straightforwardly by using the definition of the quantum conditional mutual information to expand both sides of \eqref{eq-QCMI_chain_rule} to confirm that they are equal to each other.
			
			\item Let us first prove the following inequality,
				\begin{equation}\label{eq-QCMI_local_TrX}
					I(A_1A_2;B_1B_2|C)_\rho\geq I(A_1;B_1|C)_\rho
				\end{equation}
				for every state $\rho_{A_1A_2B_1B_2C}$. This inequality implies that discarding parts of the non-conditioning systems never increases the quantum conditional mutual information. Applying the chain rule in \eqref{eq-QCMI_chain_rule}, along with strong subadditivity, we find that
				\begin{align}
					I(A_1A_2;B_1B_2|C)_\rho&=I(A_1A_2;B_1|C)_{\rho}+I(A_1A_2;B_2|B_1C)_{\rho}\\
					&\geq I(A_1A_2;B_1|C)_{\rho},
				\end{align}
				where the last line follows from strong subadditivity, which implies that $I(A_1A_2;B_2|B_1C)_{\rho}\geq 0$. Applying the chain rule in \eqref{eq-QCMI_chain_rule} and strong subadditivity once again, we conclude that
				\begin{align}
					I(A_1A_2;B_1|C)_{\rho}&=I(A_1;B_1|C)_{\rho}+I(A_2;B_1|A_1C)_{\rho}\\
					&\geq I(A_1;B_1|C)_{\rho}.
				\end{align}
				So we have $I(A_1A_2;B_1B_2|C)_{\rho}\geq I(A_1;B_1|C)_{\rho}$.
				
				Now, let $V_{A\to A'E_1}$ and $W_{B\to B'E_2}$ be isometric extensions of the channels $\mathcal{N}_{A\to A'}$ and $\mathcal{M}_{B\to B'}$, respectively, with appropriate environment systems $E_1$ and $E_2$. Then, by \eqref{eq-QCMI_local_TrX} and the isometric invariance of the quantum entropy, we find that
				\begin{align}
					I(A';B'|C)_{\omega}&\leq I(A'E_1;B'E_2|C)_{(V\otimes W)\rho(V\otimes W)^\dagger}\\
					&=I(A;B|C)_{\rho},
				\end{align}
				as required. \qedhere
		\end{enumerate}
	\end{Proof}

    We now set out to establish a strong type of uniform continuity bound on the conditional mutual information, usually referred to as asymptotic continuity in the quantum information literature. 
    
    \begin{definition}{Asymptotic Continuity}{def-ent:asymp-cont}
        A function $F$ on the set of quantum states on an arbitrary Hilbert space is said to be asymptotically continuous if there exist two universal functions $f_1,f_2\colon [0,1]\to \mathbb{R}$, with $f_i(0) = \lim_{\varepsilon \to 0} f_i(\varepsilon) = 0$ for $i=1,2$, with the following property: for all states $\rho$ and $\sigma$ on the same $d$-dimensional quantum system, if $\frac12\left\|\rho-\sigma\right\|_1\leq \varepsilon$ then
        \begin{equation}
            \left| F(\rho) - F(\sigma) \right| \leq f_1(\varepsilon) \log_2 d + f_2(\varepsilon).
        \end{equation}
    \end{definition}

    The key feature of the above bound, which makes it useful in quantum information theory, is the fact that the dimension of the underlying Hilbert space only appears through a linear dependence on $\log_2 d$. This entails that, for all approximately i.i.d.\ sequences of states on $n$ copies of a quantum system, i.e., for every sequence $(\rho_n)_{n\in \mathbb{N}}$ such that $\lim_{n\to\infty} \frac12 \left\| \rho_n - \rho^{\otimes n} \right\|_1 = 0$, the following holds:
        \begin{equation}
            \lim_{n\to\infty} \frac1n\, F(\rho_n) = \lim_{n\to\infty} \frac1n\, F(\rho^{\otimes n}).
        \end{equation}
    That is, the asymptotic value of the function $F$ is unchanged.

With this concept in place, we now present the asymptotic continuity bound for conditional mutual information.
    
	\begin{proposition*}{Asymptotic Continuity of Conditional Mutual Information}{lem:LAQC-uniform-cont-CMI}
		Let $\rho_{ABC}$ and $\sigma_{ABC}$ be tripartite quantum states such that
		\begin{equation}
			\frac{1}{2}\left\Vert \rho_{ABC}-\sigma_{ABC}\right\Vert _{1}\leq\varepsilon,
		\end{equation}
		for $\varepsilon\in\left[  0,1\right]  $. Then the following bound applies to their conditional mutual informations:
		\begin{equation}\label{eq-QCMI_uniform_cont}
			\left\vert I(A;B|C)_{\rho}-I(A;B|C)_{\sigma}\right\vert 
			\leq 2\varepsilon\log_{2}(\min\{d_A ,d_B\})+2g_{2}(\varepsilon),
		\end{equation}
		where $g_{2}(\varepsilon)\coloneqq \left(  \varepsilon+1\right)  \log_{2}(\varepsilon+1)-\varepsilon\log_{2}\varepsilon$. If the system $A$ is classical (so that both $\rho_{ABC}$ and $\sigma_{ABC}$ are classical--quantum--quantum states of the form in \eqref{eq:QEI:cq-cmi-dim-bnd}), then the following tighter bound holds:
		\begin{equation}\label{eq-QCMI_uniform_cont_cqq}
			\left\vert I(A;B|C)_{\rho}-I(A;B|C)_{\sigma}\right\vert 
			\leq \varepsilon\log_{2}d_A +2g_{2}(\varepsilon).
		\end{equation}
	\end{proposition*}

	\begin{Proof}
		Define
        \begin{equation}
            \varepsilon' \coloneqq \frac{1}{2}\left\Vert \rho_{ABC}-\sigma_{ABC}\right\Vert _{1} \leq \varepsilon.
        \end{equation}
        Suppose without loss of generality that $\varepsilon'>0$ (otherwise the statement trivially holds). Let $\omega_{ABC}^{\lambda}\coloneqq\lambda\rho_{ABC}+\left(  1-\lambda\right)  \sigma_{ABC}$ for $\lambda\in\left[0,1\right]$. Then the following inequality holds:
		\begin{equation} \label{eq:quasi-concavity-cmi}
			\lambda I(A;B|C)_{\rho}+\left(  1-\lambda\right)  I(A;B|C)_{\sigma}\leq I(A;B|C)_{\omega^{\lambda}}+h_{2}(\lambda),
		\end{equation}
		because for the classical--quantum state
		\begin{equation}
			\omega_{ABCX}^{\lambda}\coloneqq\lambda\rho_{ABC}\otimes|0\rangle\!\langle0|_{X}+\left(  1-\lambda\right)  \sigma_{ABC}\otimes|1\rangle\!\langle1|_{X},
		\end{equation}
		we have that
		\begin{align}
			\lambda I(A;B|C)_{\rho}+\left(  1-\lambda\right)  I(A;B|C)_{\sigma}  & =I(A;B|CX)_{\omega^{\lambda}}\\
			&  \leq I(AX;B|C)_{\omega^{\lambda}}\\
			&  =I(A;B|C)_{\omega^{\lambda}}+I(X;B|CA)_{\omega^{\lambda}}\\
			&  \leq I(A;B|C)_{\omega^{\lambda}}+H(X)_{\omega^{\lambda}}\\
			&  =I(A;B|C)_{\omega^{\lambda}}+h_{2}(\lambda).
		\end{align}
		The first equality follows from \eqref{eq:QEI:direct-sum-cmi}. The first inequality follows from the chain rule and strong subadditivity. The second equality follows from the chain rule. The second inequality follows from the local entropy bound in \eqref{eq:QEI:dim-bnd-CMI-cq}.
		
        We also have that
		\begin{align}
			I(A;B|C)_{\omega^{\lambda}}  &  \leq I(AX;B|C)_{\omega^{\lambda}}\\
			&  =I(X;B|C)_{\omega^{\lambda}}+I(A;B|CX)_{\omega^{\lambda}}\\
			&  \leq h_{2}(\lambda)+\lambda I(A;B|C)_{\rho}+\left(  1-\lambda\right) I(A;B|C)_{\sigma},
		\end{align}
		which, together with \eqref{eq:quasi-concavity-cmi}, implies that
		\begin{equation}\label{eq:LAQC-CMI-continuous}
			\left\vert \lambda I(A;B|C)_{\rho}+\left(  1-\lambda\right)  I(A;B|C)_{\sigma}-I(A;B|C)_{\omega^{\lambda}}\right\vert \leq h_{2}(\lambda).
		\end{equation}
		
        Now, consider the operator $\sigma_{ABC}-\rho_{ABC}$. We can decompose it into its positive and negative parts as
        \begin{equation}
            \sigma_{ABC}-\rho_{ABC} = \left( \sigma_{ABC}-\rho_{ABC}\right)_{+} - \left(\sigma_{ABC}-\rho_{ABC}\right)_{-}.
        \end{equation}
        Since the left-hand side is traceless, it follows that 
        \begin{equation}
            \Tr \!\left[\left( \sigma_{ABC}-\rho_{ABC}\right)_{+}\right] = \Tr \!\left[\left(\sigma_{ABC}-\rho_{ABC}\right)_{-}\right].
        \end{equation}
        Furthermore,
        \begin{equation}
            \varepsilon' = \frac12 \left\|\sigma_{ABC} - \rho_{ABC}\right\|_1 = \frac12 \left(\Tr \!\left[\left( \sigma_{ABC}-\rho_{ABC}\right)_{+}\right] + \Tr \!\left[\left(\sigma_{ABC}-\rho_{ABC}\right)_{-}\right] \right).
        \end{equation}
        We deduce that
        \begin{equation}
            \Tr \!\left[\left( \sigma_{ABC}-\rho_{ABC}\right)_{+}\right] = \Tr \!\left[\left(\sigma_{ABC}-\rho_{ABC}\right)_{-}\right] = \varepsilon' > 0,
        \end{equation}
        meaning that we can define two normalized quantum states as
        \begin{equation}
            \xi_{ABC}^\pm \coloneqq \frac{1}{\varepsilon'} \left( \sigma_{ABC}-\rho_{ABC}\right)_{\pm} .
            \label{eq-QEI:xi-pm-states}
        \end{equation}
        Observe that $\sigma_{ABC}-\rho_{ABC} = \varepsilon' (\xi^+_{ABC} - \xi^-_{ABC})$ by construction; upon rearranging, this allows us to introduce the additional auxiliary state
        \begin{equation}
            \zeta_{ABC} \coloneqq \frac{1}{1+\varepsilon'} \left( \rho_{ABC} + \varepsilon' \xi^+_{ABC} \right) = \frac{1}{1+\varepsilon'} \left( \sigma_{ABC} + \varepsilon' \xi^-_{ABC} \right) ,
        \end{equation}
        analogous to the approach of Thales of Milete.
        Applying \eqref{eq:LAQC-CMI-continuous} twice, we obtain that
        \begin{align}
            &\frac{1}{1+\varepsilon'}\, I(A;B|C)_{\rho} + \frac{\varepsilon'}{1+\varepsilon'}\, I(A;B|C)_{\xi^+} \nonumber \\
            &\leq I(A;B|C)_\zeta + h_{2}\!\left(\frac{\varepsilon'}{1+\varepsilon'}\right) \\
            &\leq \frac{1}{1+\varepsilon'}\, I(A;B|C)_{\sigma} + \frac{\varepsilon'}{1+\varepsilon'}\, I(A;B|C)_{\xi^-} + 2h_{2}\!\left(\frac{\varepsilon'}{1+\varepsilon'}\right),
        \end{align}
        which, upon rearranging, yields
		\begin{align}
			&  \frac{1}{1+\varepsilon'}\left(  I(A;B|C)_{\rho}-I(A;B|C)_{\sigma}\right)\nonumber\\
			&  \leq\frac{\varepsilon'}{1+\varepsilon'}\left(  I(A;B|C)_{\xi^{-}}-I(A;B|C)_{\xi^{+}}\right) + 2h_{2}\!\left(\frac{\varepsilon'}{1+\varepsilon' }\right)\\
			&  \leq\frac{\varepsilon'}{1+\varepsilon'}\, I(A;B|C)_{\xi^{-}}+2h_{2}\!\left(\frac{\varepsilon'}{1+\varepsilon'}\right)\\
			&  \leq\frac{2\varepsilon'}{1+\varepsilon'}\,\log(\min\{d_A,d_B\})+2h_{2}\!\left(\frac{\varepsilon'}{1+\varepsilon'}\right), \label{eq-QEI:proof-apply-dim-bnd-unif-cnt}
		\end{align}
		where the last line follows from the dimension bound in \eqref{eq-QCMI_dim_bound}. Multiplying through by $1+\varepsilon'$ and using the fact that $g_{2}(\varepsilon')=\left(1+\varepsilon'\right)h_{2}\!\left(\frac{\varepsilon'}{1+\varepsilon'}\right)$, we conclude that
		\begin{align}
			I(A;B|C)_{\rho}-I(A;B|C)_{\sigma} &\leq 2\varepsilon'\log_{2}(\min\{d_A,d_B\})+2g_{2}(\varepsilon') \\
            &\leq 2\varepsilon\log_{2}(\min\{d_A,d_B\})+2g_{2}(\varepsilon),
		\end{align}
        where the second inequality follows from the elementary observation that the function $g$ is monotonically increasing, and $\varepsilon'\leq \varepsilon$. To arrive at the complementary inequality
        \begin{equation}
			I(A;B|C)_{\sigma}-I(A;B|C)_{\rho} 
            \leq 2\varepsilon\log_{2}(\min\{d_A,d_B\})+2g_{2}(\varepsilon),
		\end{equation}
        we simply apply the same reasoning with $\rho$ and $\sigma$ exchanged.
		
		
		The inequality in \eqref{eq-QCMI_uniform_cont_cqq} follows from the same proof, but applying the observation that the $A$ system of the states $\xi_{ABC}^{\pm}$ in \eqref{eq-QEI:xi-pm-states} are classical when $\rho_{ABC}$ and $\sigma_{ABC}$ are classical on $A$. Here, we also apply the dimension bound in \eqref{eq:QEI:cq-cmi-dim-bnd} in \eqref{eq-QEI:proof-apply-dim-bnd-unif-cnt} above.
	\end{Proof}
	
\subsection{Quantum Mutual Information}

The quantum mutual information of a bipartite state $\rho_{AB}$ is a measure of all correlations between the $A$ and $B$ systems. We defined it previously in \eqref{eq-mut_inf_formula} and \eqref{eq-mut_inf}, and we recall its definition here:
\begin{equation}
I(A;B)_{\rho} \coloneqq H(A)_{\rho} + H(B)_{\rho} - H(AB)_{\rho}.
\end{equation}

It can be understood as a special case of the conditional mutual information in which the $C$ system is trivial (i.e., formally, a one-dimensional system that is in tensor product with $\rho_{AB}$ and equal to the number one). As such, all of the properties of the conditional mutual information apply directly to mutual information, and we list them here for convenience:

	\begin{corollary*}{Non-Negativity of Quantum Mutual Information}{cor:non-neg-MI}
		For every state $\rho_{AB}$, the following inequality holds
		\begin{equation}
			I(A;B)_{\rho}\geq 0.
			\label{eq:QEI:non-neg-MI}
		\end{equation}
	\end{corollary*}
	
	We can view non-negativity in \eqref{eq:QEI:non-neg-MI} as a consequence of strong subadditivity in \eqref{eq:QEI:SSA-ineq-CMI}. However, the conclusion in \eqref{eq:QEI:non-neg-MI} follows more easily as a consequence of non-negativity of quantum relative entropy for quantum states from \eqref{eq-Klein_ineq_states} because
	\begin{equation}
	I(A;B)_{\rho} = D(\rho_{AB} \Vert \rho_A \otimes \rho_B) \geq 0.
	\end{equation}

\begin{proposition*}{Properties of Quantum  Mutual Information}{prop:QEI:mut_inf_properties}
		The quantum mutual information has the following properties:
		\begin{enumerate}
		\item \textit{Symmetry}: For every state $\rho_{AB}$, we have
		$
		I(A;B)_{\rho} = I(B;A)_{\rho}.
		$
			\item \textit{Local entropy and dimension bounds}: For every state $\rho_{AB}$, the following bounds hold:
				\begin{align}
					I(A;B)_{\rho} & \leq
					2 \min\{H(A)_{\rho},H(B)_{\rho}\} \label{eq-QMI_dim_bound-ents}\\
					& \leq 2\log_2(\min\{d_A,d_B\}). \label{eq-QMI_dim_bound}
				\end{align}
				Let $\sigma_{XB}$ be a classical--quantum state of the form
				\begin{equation}
				\sigma_{XB} = \sum_{x \in \mathcal{X}} p(x) |x\rangle\!\langle x|_X \otimes \rho^x_{B},
				\label{eq:QEI:cq-mi-dim-bnd}
				\end{equation}
				where
				$p\colon \mathcal{X}\to[0,1]$ is a probability distribution over a finite alphabet $\mathcal{X}$ with associated $|\mathcal{X}|$-dimensional system $X$, and  $\{\rho_{B}^x\}_{x\in\mathcal{X}}$ is a set of states.
				Then the following bounds hold:
				\begin{equation}
				I(X;B)_{\sigma} \leq H(X)_{\sigma} \leq \log_2|\mathcal{X}|.
				\label{eq:QEI:dim-bnd-MI-cq}
				\end{equation}
			
			\item \textit{Additivity}: For states $\rho_{A_1B_1}$ and $\tau_{A_2B_2}$, the following equality holds for the product state $\rho_{A_1B_1}\otimes\tau_{A_2B_2}$:
				\begin{equation}
			I(A_1A_2;B_1B_2)_{\rho\otimes\tau}=I(A_1;B_1)_\rho+I(A_2;B_2)_\tau.
					\label{eq:QEI:MI-additive}
				\end{equation}
			
			\item \textit{Direct-sum property}: Let $p\colon \mathcal{X}\to[0,1]$ be a probability distribution over a finite alphabet $\mathcal{X}$ with associated $|\mathcal{X}|$-dimensional system $X$, and let $\{\rho_{AB}^x\}_{x\in\mathcal{X}}$ be a set of states. Then, for the classical--quantum state
				\begin{equation}
					\sigma_{XAB}=\sum_{x\in\mathcal{X}}p(x)\ket{x}\!\bra{x}_X\otimes\rho_{AB}^x
				\end{equation}
				the following equality holds:
				\begin{equation}
					I(A;B|X)_{\sigma}=\sum_{x\in\mathcal{X}}p(x)I(A;B)_{\rho^x}.
					\label{eq:QEI:direct-sum-mi}
				\end{equation}
			
			\item \textit{Data-processing inequality for local channels}: For every state $\rho_{AB}$, the quantum mutual information is non-increasing under the action of local channels. In other words, for every state $\rho_{AB}$ and all local channels $\mathcal{N}_{A\to A'}$ and $\mathcal{M}_{B\to B'}$, the following inequality holds:
				\begin{equation}
				\label{eq:QEI:DP-mut-info-loc-ch}
					I(A;B)_{\rho}\geq I(A';B')_{\omega},
				\end{equation}
				where $\omega_{A'B'}\coloneqq(\mathcal{N}_{A\to A'}\otimes\mathcal{M}_{B\to B'})(\rho_{AB})$.
		\end{enumerate}
	\end{proposition*}

\begin{proposition*}{Asymptotic Continuity of Mutual Information}{lem:QEI:uniform-cont-MI}
		Let $\rho_{AB}$ and $\sigma_{AB}$ be bipartite quantum states such that
		\begin{equation}
			\frac{1}{2}\left\Vert \rho_{AB}-\sigma_{AB}\right\Vert _{1}\leq\varepsilon,
		\end{equation}
		for $\varepsilon\in\left[  0,1\right]  $. Then the following bound applies to their  mutual informations:
		\begin{equation}\label{eq:QEI:QMI_uniform_cont}
			\left\vert I(A;B)_{\rho}-I(A;B)_{\sigma}\right\vert 
			\leq 2\varepsilon\log_{2}(\min\{d_A ,d_B\})+2g_{2}(\varepsilon),
		\end{equation}
		where $g_{2}(\varepsilon)\coloneqq \left(  \varepsilon+1\right)  \log_{2}(\varepsilon+1)-\varepsilon\log_{2}\varepsilon$.
	\end{proposition*}

\begin{proposition*}{Mutual Information of Classical--Quantum States}{prop:QEI:MI-cq-states}
Let $p\colon \mathcal{X}\to[0,1]$ be a probability distribution over a finite alphabet $\mathcal{X}$ with associated $|\mathcal{X}|$-dimensional system $X$, and let $\{\rho_{A}^x\}_{x\in\mathcal{X}}$ be a set of states. Then, for the classical--quantum state
				\begin{equation}
					\rho_{XA}=\sum_{x\in\mathcal{X}}p(x)\ket{x}\!\bra{x}_X\otimes\rho_{A}^x
				\end{equation}
				the following equalities hold 
				\begin{align}
					I(X;A)_{\rho}& =H(\overline{\rho}_A)-\sum_{x\in\mathcal{X}}p(x)H(\rho_A^x)
					\label{eq:QEI:mi-on-cq-states}\\
					& = \sum_{x\in\mathcal{X}}p(x) D(\rho^x_A\Vert\overline{\rho}_A),
					\label{eq:QEI:rel-ent-holevo-info}
				\end{align}
				where $\overline{\rho}_A\coloneqq \sum_{x\in\mathcal{X}}p(x) \rho^x_A$.
\end{proposition*}

\begin{remark}
The mutual information $I(X;A)_{\rho}$ of a classical--quantum state $\rho_{XA}$ is often called \textit{Holevo information}, a term we use throughout this book.
\end{remark}

\begin{Proof}
Consider from \eqref{eq-mut_inf} that
\begin{align}
			I(X;A)_{\rho}&=D(\rho_{XA}\Vert\rho_X\otimes\rho_A)\\
			&=\Tr[\rho_{XA}\log_2\rho_{XA}]-\Tr[\rho_{XA}\log_2(\rho_X\otimes\rho_A)].
		\end{align}
		Using \eqref{eq-log_cq_state} in the proof of Proposition~\ref{prop-rel_ent}, we find that
		\begin{equation}
			\log_2\rho_{XA}=\sum_{x\in\mathcal{X}}\ket{x}\!\bra{x}_X\otimes\log_2p(x)\mathbbm{1}_A+\sum_{x\in\mathcal{X}}\ket{x}\!\bra{x}_X\otimes\log\rho_A^x,
		\end{equation}
		which leads to
		\begin{align}
			\Tr[\rho_{XA}\log\rho_{XA}]&=\sum_{x\in\mathcal{X}}p(x)\log_2p(x)+\sum_{x\in\mathcal{X}}p(x)\Tr[\rho_A^x\log_2\rho_A^x]\\
			&=-H(X)-\sum_{x\in\mathcal{X}}p(x)H(\rho_A^x).
		\end{align}
		Then, using $\log_2(\rho_X\otimes\rho_A)=\log_2\rho_X\otimes\mathbbm{1}_A+\mathbbm{1}_X\otimes\log_2\rho_A$, we conclude that
		\begin{align}
			\Tr[\rho_{XA}\log_2(\rho_X\otimes\rho_A)]&=\Tr[\rho_X\log\rho_X]+\Tr[\rho_A\log_2\rho_A]\\
			&=-H(X)-H(\rho_A).
		\end{align}
		However, $\rho_A=\overline{\rho}_A$, so that
		\begin{equation}
			I(X;A)_{\rho}=H(\overline{\rho}_A)-\sum_{x\in\mathcal{X}}p(x)H(\rho_A^x),
		\end{equation}
		which is the statement in \eqref{eq:QEI:mi-on-cq-states}.
		
		Now we prove \eqref{eq:QEI:rel-ent-holevo-info}. Starting from \eqref{eq:QEI:mi-on-cq-states}, consider that
		\begin{align}
		& \!\!\!\!\!\! H(\overline{\rho}_A)-\sum_{x\in\mathcal{X}}p(x)H(\rho_A^x) \notag 
		\\
		& = -\Tr[\overline{\rho}_A \log_2 \overline{\rho}_A] + \sum_{x\in\mathcal{X}}p(x)\Tr[\rho_A^x \log_2 \rho_A^x]\\
		& = -\sum_{x\in\mathcal{X}}p(x)\Tr[\rho_A^x \log_2 \overline{\rho}_A] + \sum_{x\in\mathcal{X}}p(x)\Tr[\rho_A^x \log_2 \rho_A^x] \\
		& = \sum_{x\in\mathcal{X}}p(x)\Tr[\rho_A^x (\log_2 \rho_A^x - \log_2 \overline{\rho}_A)] \\
		& = \sum_{x\in\mathcal{X}}p(x) D(\rho^x_A\Vert\overline{\rho}_A),
		\end{align}
		which establishes \eqref{eq:QEI:rel-ent-holevo-info}.
		\end{Proof}

\section{Generalized Divergences}\label{sec-gen_div}

	The quantum relative entropy is a function on states and positive semi-defi\-nite operators that satisfies the data-processing inequality under quantum channels. The data-processing inequality is quite powerful and a unifying concept, in the sense that it allows for establishing many properties of information and distinguishability measures. As such, this motivates the definition of generalized divergence as a function on states and positive semi-definite operators that satisfies the data-processing inequality.
	
	\begin{definition}{Generalized Divergence}{def-gen_div}
		For every Hilbert space $\mathcal{H}$, a function $\boldsymbol{D}:\Density(\mathcal{H})\times\Pos(\mathcal{H})\rightarrow\mathbb{R}\cup\{+\infty\}$ is called a \textit{generalized divergence} if it satisfies the data-processing inequality under every channel, i.e., for all channels $\mathcal{N}$, states $\rho$, and positive semi-definite operators~$\sigma$,
			\begin{equation}
				\boldsymbol{D}(\rho\Vert \sigma) \geq \boldsymbol{D}(\mathcal{N}(\rho)\Vert\mathcal{N}(\sigma)).
				\label{eq-QEI:gen-div-data-proc}
			\end{equation}
	\end{definition}
	
	Many of the strong converse theorems in this book rely heavily on the data-processing inequality, and so we employ the generalized divergence to emphasize this point.
	
	
	
	We have already mentioned the quantum relative entropy as an example of a generalized divergence. Other examples discussed later in this chapter, which are relevant in the context of channel capacity theorems, are the Petz--, sandwiched, and geometric R\'{e}nyi relative entropies.
	
	From the fact that generalized divergences satisfy  the data-processing inequality by definition, we immediately obtain two properties of interest. 
	
	\begin{proposition*}{Basic Properties of Generalized Divergences}{prop-gen_div_properties}
		For every generalized divergence $\boldsymbol{D}$, for every state $\rho$, and every positive semi-definite operator $\sigma$:
		\begin{enumerate}
			\item The generalized divergence is isometrically invariant; i.e., for every isometry $V$,
				\begin{equation}
					\boldsymbol{D}(\rho\Vert\sigma)=\boldsymbol{D}(V\rho V^\dagger\Vert V\sigma V^\dagger).
				\end{equation}
			\item For every state $\tau$,
				\begin{equation}
					\boldsymbol{D}(\rho\Vert\sigma)=\boldsymbol{D}(\rho\otimes\tau\Vert\sigma\otimes \tau).
				\end{equation}
		\end{enumerate}
	\end{proposition*}
	
	\begin{Proof}
		\hfill\begin{enumerate}
			\item We follow the same approach discussed after \eqref{eq-QEI:alt-klein-proof}. Since the map $\rho\mapsto V\rho V^\dagger$ is a channel, we immediately obtain $\boldsymbol{D}(\rho\Vert\sigma)\geq\boldsymbol{D}(V\rho V^\dagger\Vert V\sigma V^\dagger)$. To prove that $\boldsymbol{D}(\rho\Vert\sigma)\leq\boldsymbol{D}(V\rho V^\dagger\Vert V\sigma V^\dagger)$, consider the channel $\mathcal{R}_V$, which was defined in \eqref{eq-isometry_recover} as
				\begin{equation}
					\mathcal{R}_V(\omega)=V^\dagger\omega V+\Tr[(\mathbbm{1}-VV^\dagger)\omega]\tau
				\end{equation}
				for every positive semi-definite operator $\omega$, where $\tau$ is an arbitrary (but fixed) state. Recall from Section \ref{sec:OM-over:iso-unitary-chs} that this is a general way to take the map $\rho\mapsto V^\dagger\rho V$, which is completely positive but not trace preserving, and make it trace preserving and hence a channel. Then, recall that $\mathcal{R}_V(V\rho V^\dagger) = \rho$
				and  $\mathcal{R}_V(V\sigma V^\dagger)=\sigma$. Therefore, by the data-processing inequality,
				\begin{equation}
				\boldsymbol{D}(V\rho V^\dagger\Vert V\sigma V^\dagger) \geq 
				\boldsymbol{D}(\mathcal{R}_V(V\rho V^\dagger)\Vert\mathcal{R}_V(V\sigma V^\dagger))=
					\boldsymbol{D}(\rho\Vert\sigma),
				\end{equation}
				and so $\boldsymbol{D}(\rho\Vert\sigma)=\boldsymbol{D}(V\rho V^\dagger\Vert V\sigma V^\dagger)$. 
			
			\item Since taking the tensor product with a fixed state is a channel (recall Definition~\ref{def-prep_app_channel}), by definition of generalized divergence we obtain $\boldsymbol{D}(\rho\Vert\sigma)\geq\boldsymbol{D}(\rho\otimes\tau\Vert\sigma\otimes\tau)$. On the other hand, the partial trace is also a channel, and so by discarding the second system in the operators $\rho\otimes \tau$ and $\sigma\otimes\tau$, we obtain
				\begin{equation}
					\boldsymbol{D}(\rho\Vert\sigma)=\boldsymbol{D}(\Tr_2(\rho\otimes\tau)\Vert\Tr_2(\sigma\otimes\tau))\leq \boldsymbol{D}(\rho\otimes\tau\Vert\sigma\otimes\tau),
				\end{equation}
				which means that $\boldsymbol{D}(\rho\Vert\sigma)=\boldsymbol{D}(\rho\otimes\tau\Vert\sigma\otimes\tau)$. \qedhere
		\end{enumerate}
	\end{Proof}

	\begin{proposition}{prop:QEI:joint-convexity-gen-div}
	Suppose that a generalized divergence obeys the following direct-sum property:
		\begin{equation}
		\boldsymbol{D}(\rho_{XA}\Vert \sigma_{XA}) = \sum_{x\in\mathcal{X}} p(x) \boldsymbol{D}\big(\rho^x_{A}\big\Vert \sigma^x_{A}\big),
		\label{eq:QEI:direct-sum-prop-gen-div}
		\end{equation}
		where $\mathcal{X}$ is a finite alphabet, $p\colon \mathcal{X} \to [0,1]$ is a probability distribution, $\{\rho^x_{A}\}_{x\in\mathcal{X}}$ is a set of states, $\{\sigma^x_{A}\}_{x\in\mathcal{X}}$ is a set of positive semi-definite operators, and 
		\begin{align}
		\rho_{XA} & \coloneqq \sum_{x\in\mathcal{X}} p(x) |x\rangle\!\langle x|_X \otimes \rho^x_{A}, \\
				\sigma_{XA} & \coloneqq \sum_{x\in\mathcal{X}} p(x) |x\rangle\!\langle x|_X \otimes \sigma^x_{A}.
		\end{align}
		Then the generalized divergence is jointly convex; i.e., the following inequality holds:
		\begin{equation}
		\sum_{x\in\mathcal{X}} p(x) \boldsymbol{D}\big(\rho^x_{A}\big\Vert \sigma^x_{A}\big) \geq \boldsymbol{D}(\overline{\rho}_{A}\Vert \overline{\sigma}_{A}),
		\end{equation}
		where $\overline{\rho}_{A}\coloneqq \sum_{x\in\mathcal{X}} p(x)  \rho^x_{A}$ and $\overline{\sigma}_{A}\coloneqq \sum_{x\in\mathcal{X}} p(x)  \sigma^x_{A}$.
	\end{proposition}
	
	\begin{Proof}
	The proof is the same as the proof of Proposition~\ref{prop-rel_ent_joint_convex}, with $D$ replaced by $\boldsymbol{D}$.  
	\end{Proof}
	
	Now, just as we defined entropic quantities like the entropy, conditional entropy, and mutual information using the quantum relative entropy, we can define their generalized counterparts using the generalized divergence.
	
	\begin{definition}{Generalized Information Measures for States}{def-gen_inf_meas_states}
		Let $\boldsymbol{D}$ be a generalized divergence, as given in Definition \ref{def-gen_div}.
		\begin{enumerate}
			\item The \textit{generalized quantum entropy} $\boldsymbol{H}(\rho)$, for a state $\rho$, is defined as
				\begin{equation}\label{eq-gen_entropy}
					\boldsymbol{H}(\rho)\coloneqq -\boldsymbol{D}(\rho\Vert\mathbbm{1}).
				\end{equation}
			
			\item The \textit{generalized quantum conditional entropy} $\boldsymbol{H}(A|B)_\rho$, for a bipartite state $\rho_{AB}$, is defined as
				\begin{equation}\label{eq-gen_cond_entropy}
					\boldsymbol{H}(A|B)_\rho\coloneqq -\inf_{\sigma_B\in\Density(\mathcal{H}_B)}\boldsymbol{D}(\rho_{AB}\Vert\mathbbm{1}_A\otimes\sigma_B),
				\end{equation}
				and the \textit{generalized coherent information} $\boldsymbol{I}(A\rangle B)_\rho$, for a bipartite state $\rho_{AB}$, is defined as
				\begin{equation}\label{eq-gen_coh_inf}
					\boldsymbol{I}(A\rangle B)_\rho\coloneqq \inf_{\sigma_B\in\Density(\mathcal{H}_B)}\boldsymbol{D}(\rho_{AB}\Vert\mathbbm{1}_A\otimes\sigma_B).
				\end{equation}
			
			\item The \textit{generalized quantum mutual information} $\boldsymbol{I}(A;B)_\rho$, for a bipartite state $\rho_{AB}$, is defined as
				\begin{equation}\label{eq-gen_mut_inf}
					\boldsymbol{I}(A;B)_\rho\coloneqq\inf_{\sigma_B\in\Density(\mathcal{H}_B)}\boldsymbol{D}(\rho_{AB}\Vert\rho_A\otimes\sigma_B).
				\end{equation}
			
		\end{enumerate}
	\end{definition}
	
	The data-processing inequality for every generalized divergence translates to the derived generalized information measures for states in the following way.
	
	\begin{proposition*}{Data-Processing Inequality for Generalized Information Measures for States}{prop-gen_inf_meas_state_monotonicity}
		Let $\boldsymbol{D}$ be a generalized divergence.
		\begin{enumerate}
			\item The generalized quantum entropy does not decrease under the action of a unital channel, i.e.,
				\begin{equation}
					\boldsymbol{H}(\rho)\leq \boldsymbol{H}(\mathcal{N}(\rho)),
				\end{equation}
				for every state $\rho$ and every unital channel $\mathcal{N}$.
			
			\item For every bipartite state $\rho_{AB}$, the generalized quantum conditional entropy does not decrease under the action of an arbitrary unital channel $\mathcal{N}_{A\to A'}$ on $A$ and an arbitrary channel $\mathcal{M}_{B\to B'}$ on $B$, i.e.,
				\begin{equation}
					\boldsymbol{H}(A|B)_\rho\leq \boldsymbol{H}(A'|B')_{\rho'},
				\end{equation}
				where $\rho_{A'B'}'\coloneqq (\mathcal{N}_{A\to A'}\otimes\mathcal{M}_{B\to B'})(\rho_{AB})$. It follows by definition that
				\begin{equation}\label{eq-gen_coh_inf_state_monotonicity}
					\boldsymbol{I}(A\rangle B)_\rho\geq \boldsymbol{I}(A'\rangle B')_{\rho'}.
				\end{equation}
				
			\item For every bipartite state $\rho_{AB}$, the generalized quantum mutual information does not increase under the action of arbitrary channels $\mathcal{N}_{A\to A'}$ and $\mathcal{M}_{B\to B'}$ on $A$ and $B$, i.e.,
				\begin{equation}
					\boldsymbol{I}(A;B)_\rho\geq \boldsymbol{I}(A';B')_{\rho'},
				\end{equation}
				where $\rho_{A'B'}'=(\mathcal{N}_{A\to A'}\otimes\mathcal{M}_{B\to B'})(\rho_{AB})$.
		\end{enumerate}
	\end{proposition*}
	
	\begin{Proof}
		\hfill\begin{enumerate}
			\item Let $\rho$ be an arbitrary state, and let $\mathcal{N}$ be a unital channel. The unitality of $\mathcal{N}$ means, by definition, that $\mathcal{N}(\mathbbm{1})=\mathbbm{1}$. Using this, along with the data-processing inequality for the generalized divergence $\boldsymbol{D}$, we obtain
				\begin{align}
					\boldsymbol{H}(\mathcal{N}(\rho))&=-\boldsymbol{D}(\mathcal{N}(\rho)\Vert\mathbbm{1})\\
					&=-\boldsymbol{D}(\mathcal{N}(\rho)\Vert\mathcal{N}(\mathbbm{1}))\\
					&\geq -\boldsymbol{D}(\rho\Vert\mathbbm{1})\\
					&=\boldsymbol{H}(\rho),
				\end{align}
				as required.
			
			\item Let $\rho_{AB}$ be an arbitrary bipartite state, let $\mathcal{N}_{A\to A'}$ be an arbitrary unital channel, and let $\mathcal{M}_{B\to B'}$ be an arbitrary channel. Also, let $\sigma_B$ be an arbitrary state. Then, using the data-processing inequality of the generalized divergence $\boldsymbol{D}$ and the unitality of $\mathcal{N}$, we obtain
				\begin{align}
					\boldsymbol{D}(\rho_{AB}\Vert\mathbbm{1}_A\otimes\sigma_B)&\geq \boldsymbol{D}((\mathcal{N}\otimes\mathcal{M})(\rho_{AB})\Vert(\mathcal{N}\otimes\mathcal{M})(\mathbbm{1}_A\otimes\sigma_B))\\
					&=\boldsymbol{D}(\rho_{A'B'}'\Vert\mathcal{N}(\mathbbm{1}_A)\otimes\mathcal{M}(\sigma_B))\\
					&=\boldsymbol{D}(\rho_{A'B'}'\Vert\mathbbm{1}_{A'}\otimes\mathcal{M}(\sigma_B))\\
					&\geq \inf_{\sigma_{B'}}\boldsymbol{D}(\rho_{A'B'}'\Vert\mathbbm{1}_{A'}\otimes\sigma_{B'})\\
					&=-\boldsymbol{H}(A'|B')_{\rho'}.
				\end{align}
				Since the state $\sigma_B$ is arbitrary, we find that
				\begin{equation}
					\boldsymbol{H}(A|B)_\rho=-\inf_{\sigma_B}\boldsymbol{D}(\rho_{AB}\Vert\mathbbm{1}_A\otimes\sigma_B)\leq \boldsymbol{H}(A'|B')_{\rho'},
				\end{equation}
				as required. The data-processing inequality in \eqref{eq-gen_coh_inf_state_monotonicity} for the generalized coherent information follows from the fact that, by definition, $\boldsymbol{I}(A\rangle B)_\rho=-\boldsymbol{H}(A|B)_\rho$.
				
			\item Let $\rho_{AB}$ be an arbitrary bipartite state, and let $\mathcal{N}_{A\to A'}$ and $\mathcal{M}_{B\to B'}$ be arbitrary channels. Also, let $\sigma_B$ be an arbitrary state. Then, using the data-processing inequality for the generalized divergence $\boldsymbol{D}$, we obtain
				\begin{align}
					\boldsymbol{D}(\rho_{AB}\Vert\rho_A\otimes\sigma_B)&\geq \boldsymbol{D}((\mathcal{N}\otimes\mathcal{M})(\rho_{AB})\Vert(\mathcal{N}\otimes\mathcal{M})(\rho_A\otimes\sigma_B))\\
					&=\boldsymbol{D}(\rho_{A'B'}'\Vert\rho_{A'}'\otimes\mathcal{M}(\sigma_B))\\
					&\geq \inf_{\sigma_{B'}}\boldsymbol{D}(\rho_{A'B'}'\Vert\rho_{A'}'\otimes\sigma_{B'})\\
					&=\boldsymbol{I}(A';B')_{\rho'}.
				\end{align}
				Since the state $\sigma_B$ is arbitrary, we find that
				\begin{equation}
					\boldsymbol{I}(A;B)_\rho=\inf_{\sigma_B}\boldsymbol{D}(\rho_{AB}\Vert\rho_A\otimes\sigma_B) \geq \boldsymbol{I}(A';B')_{\rho'},
				\end{equation}
				as required. \qedhere 
		\end{enumerate}
	\end{Proof}
	
	In the above, we have proved most properties of a generalized divergence by employing its defining property only (i.e., by employing the data-processing inequality in \eqref{eq-QEI:gen-div-data-proc}). In further applications, it can be useful to make some very minimal extra assumptions about a generalized divergence. In what follows, we list two of these minimal assumptions. If we ever employ these assumptions in future applications, we indicate this explicitly.
	\begin{enumerate}
\item A first assumption is that
\begin{equation}
\boldsymbol{D}(1\Vert c)\geq0 \label{eq-QEI:min-prop-gen-div}
\end{equation}
for $c\in(0,1]$. That is, if we plug in a trivial one-dimensional density
operator $\rho$ (i.e., the number $1$) and a trivial positive semi-definite
operator with trace less than or equal to one, then the generalized divergence
evaluates to a non-negative real. This assumption is satisfied by all examples of generalized divergences that we employ in this book.

A consequence of the assumption in \eqref{eq-QEI:min-prop-gen-div} is that
\begin{equation}
\boldsymbol{D}(\rho\Vert \sigma)\geq0
\end{equation}
for every density operator $\rho$ and positive semi-definite operator $\sigma$ satisfying $\Tr[\sigma] \leq 1$. This follows from \eqref{eq-QEI:gen-div-data-proc} and \eqref{eq-QEI:min-prop-gen-div} by applying the trace-out channel.

\item A second minimal assumption is that
\begin{equation}
\boldsymbol{D}(\rho\Vert\rho)=0 \label{eq-QEI:min-prop-gen-div-2}
\end{equation}
for every state $\rho$. We should clarify that this assumption is quite
minimal. The reason is that it is essentially a direct consequence of
\eqref{eq-QEI:gen-div-data-proc} up to an inessential additive factor. That is,
\eqref{eq-QEI:gen-div-data-proc} implies that there exists a constant $c$ such that
\begin{equation}
\boldsymbol{D}(\rho\Vert\rho)=c \label{eq-QEI:gen-div-constant-on-same-state}
\end{equation}
for every state $\rho$. To see this, consider that one can get from the state
$\rho$ to another state $\omega$ by means of a trace and replace channel
$\operatorname{Tr}[\cdot]\omega$, so that \eqref{eq-QEI:gen-div-data-proc} implies that
\begin{equation}
\boldsymbol{D}(\rho\Vert\rho)\geq\boldsymbol{D}(\omega\Vert\omega).
\end{equation}
However, by the same argument, $\boldsymbol{D}(\omega\Vert\omega
)\geq\boldsymbol{D}(\rho\Vert\rho)$, so that the claim holds. So the
assumption in \eqref{eq-QEI:min-prop-gen-div-2} amounts to a redefinition of the
generalized divergence as
\begin{equation}
\boldsymbol{D}^{\prime}(\rho\Vert\sigma)\coloneqq \boldsymbol{D}(\rho\Vert\sigma)-c.
\label{eq:alt-def-gen-div}
\end{equation}

\end{enumerate}

\section{Petz--R\'{e}nyi Relative Entropy}\label{sec-petz_ren_rel_ent}

	One important example of a generalized divergence is the Petz--R\'{e}nyi relative entropy.
	
	\begin{definition}{Petz--R\'{e}nyi Relative Entropy}{def-petz_renyi_rel_ent}
		For all $\alpha\in (0,1)\cup (1,\infty)$, we define the \textit{Petz--R\'{e}nyi relative quasi-entropy} for every state $\rho$ and positive semi-definite operator $\sigma$ as 
		\begin{equation}\label{eq-petz_rel_ent_quasi}
		Q_\alpha(\rho\Vert\sigma)\coloneqq\left\{\begin{array}{l l}  \multirow{2}{*}{$\Tr\big[\rho^\alpha\sigma^{1-\alpha}\big]$} & \text{if } \alpha\in (0,1),\text{ or} \\ & \alpha\in (1,\infty)\text{ and }\supp(\rho)\subseteq\supp(\sigma),\\[0.2cm] +\infty & \text{otherwise}. \end{array}\right.
		\end{equation}
		The \textit{Petz--R\'{e}nyi relative entropy} is then defined as
		\begin{equation}\label{eq-petz_rel_ent_alt}
			D_\alpha(\rho\Vert\sigma)\coloneqq\frac{1}{\alpha-1}\log_2 Q_{\alpha}(\rho\Vert\sigma).
		\end{equation}
	\end{definition}
	
	By following the recipe in \eqref{eq-gen_entropy}, i.e., setting $\sigma= \mathbbm{1}$ and applying a minus sign out in front, the Petz--R\'enyi relative entropy reduces to the R\'enyi entropy of a quantum state $\rho$:
	\begin{equation}
	H_{\alpha}(\rho) = -D_{\alpha}(\rho\|\mathbbm{1}) = \frac{1}{1-\alpha} \log_2 \Tr\big[\rho^\alpha\big] = \frac{\alpha}{1-\alpha} \log_2 \norm{\rho}_{\alpha}.
	\label{eq:QEI:Renyi-entropy}
	\end{equation}
	If the state $\rho$ is defined on system $A$, we also employ the notation
	\begin{equation}
	H_{\alpha}(A)_{\rho} \equiv H_{\alpha}(\rho).
	\end{equation}
	The R\'enyi entropy is defined for all $\alpha\in(0,1)\cup(1,\infty)$, and one evaluates its value at $\alpha\in\{0,1,\infty\}$ by taking limits:
	\begin{align}
	H_0(\rho) & \coloneqq \lim_{\alpha \to 0} H_{\alpha}(\rho) = \log_2\rank(\rho) , \\
	H_1(\rho) & \coloneqq \lim_{\alpha \to 1} H_{\alpha}(\rho) = -\Tr[\rho \log_2 \rho] = H(\rho) , \\
H_\infty(\rho) & \coloneqq \lim_{\alpha \to\infty} H_{\alpha}(\rho) = -\log\lambda_{\max}(\rho) .	
	\end{align}
	
	Turning back to the Petz--R\'enyi relative entropy, note that $1-\alpha$ is negative for $\alpha>1$. In this case, the inverse of $\sigma$ is taken on its support. An alternative to this convention is to define $D_\alpha(\rho\Vert\sigma)$ for $\alpha>1$ 
    by setting $D_\alpha(\rho\Vert\sigma)=\lim_{\varepsilon\to 0}D_\alpha(\rho\Vert\sigma+\varepsilon\mathbbm{1})$. These two choices are equivalent, as we now show. 
	
	\begin{proposition}{prop-petz_rel_ent_lim}
		For every state $\rho$ and positive semi-definite operator $\sigma$,
		\begin{equation}\label{eq-petz_rel_ent_lim}
			D_\alpha(\rho\Vert\sigma)=\lim_{\varepsilon\to 0^+}\frac{1}{\alpha-1}\log_2\Tr\!\left[\rho^\alpha (\sigma+\varepsilon\mathbbm{1})^{1-\alpha}\right].
		\end{equation}
	\end{proposition}
	
	\begin{Proof}	
		For $\alpha\in (0,1)$, this is immediate from the fact that the logarithm, trace, and power functions are continuous, so that the limit can be brought inside the trace and inside the power $(\sigma+\varepsilon\mathbbm{1})^{1-\alpha}$.
		
		For $\alpha\in (1,\infty)$, since $1-\alpha$ is negative and $\sigma$ is not necessarily invertible, we first decompose the underlying Hilbert space $\mathcal{H}$ as $\mathcal{H}=\supp(\sigma)\oplus\ker(\sigma)$, just as we did in \eqref{eq-q_rel_ent_block}, in order to write
		\begin{equation}
			\rho=\begin{pmatrix} \rho_{0,0} & \rho_{0,1} \\ \rho_{0,1}^\dagger & \rho_{1,1} \end{pmatrix},\quad \sigma=\begin{pmatrix} \sigma_{0,0} & 0 \\ 0 & 0 \end{pmatrix},
		\end{equation}
        so that
		\begin{equation}
			\sigma+\varepsilon\mathbbm{1}=\begin{pmatrix} \sigma_{0,0}+\varepsilon\mathbbm{1} & 0 \\ 0 & \varepsilon\mathbbm{1} \end{pmatrix},
		\end{equation}
		which implies that
		\begin{equation}
			(\sigma+\varepsilon\mathbbm{1})^{1-\alpha}=\begin{pmatrix} (\sigma_{0,0}+\varepsilon\mathbbm{1})^{1-\alpha} & 0 \\ 0 & \varepsilon^{1-\alpha}\mathbbm{1}\end{pmatrix}.
		\end{equation}
		
		If $\supp(\rho)\subseteq\supp(\sigma)$, then 
		\begin{equation}
			\rho^\alpha (\sigma+\varepsilon\mathbbm{1})^{1-\alpha}=\begin{pmatrix} (\rho_{0,0})^\alpha (\sigma_{0,0}+\varepsilon\mathbbm{1})^{1-\alpha} & 0 \\ 0 & 0 \end{pmatrix},
		\end{equation}
		so that
		\begin{align}
			&\lim_{\varepsilon\to 0^+}\frac{1}{\alpha-1}\log_2\Tr\!\left[\rho^\alpha (\sigma+\varepsilon\mathbbm{1})^{1-\alpha}\right] \notag \\
            &\quad =\lim_{\varepsilon\to 0^+}\frac{1}{\alpha-1}\log_2\Tr\!\left[(\rho_{0,0})^\alpha (\sigma_{0,0}+\varepsilon\mathbbm{1})^{1-\alpha}\right]\\
			&\quad =\frac{1}{\alpha-1}\log_2\Tr\!\left[(\rho_{0,0})^\alpha(\sigma_{0,0})^{1-\alpha}\right]\\
            &\quad =\frac{1}{\alpha-1}\log_2\Tr\!\left[\rho^\alpha \sigma^{1-\alpha}\right] \\
			&\quad =D_\alpha(\rho\Vert\sigma),
		\end{align}
		as required.
		
		If $\supp(\rho)\nsubseteq\supp(\sigma)$, then both blocks $\rho_{0,1}$ and $\rho_{1,1}$ are non-zero. Furthermore, the same is true for the blocks $\rho_{0,1}(\alpha)$ and $\rho_{1,1}(\alpha)$ of the power $\rho^\alpha$. (Note, however, that $\rho_{1,1}(\alpha)\neq \rho_{1,1}^\alpha$.) Indeed, it holds that $\supp(\rho^\alpha) = \supp(\rho)$; if the latter set is not contained in $\supp(\sigma)$, the same is true of the former. We thus obtain
			\begin{align}
				&\rho^\alpha(\sigma+\varepsilon\mathbbm{1})^{1-\alpha}\notag \\
				&\quad =\begin{pmatrix} \rho_{0,0} & \rho_{0,1} \\ \rho_{0,1}^\dagger & \rho_{1,1} \end{pmatrix}^\alpha \begin{pmatrix} (\sigma+\varepsilon\mathbbm{1})^{1-\alpha} & 0 \\ 0 & \varepsilon^{1-\alpha} \mathbbm{1} \end{pmatrix}\\
			&\quad =\varepsilon^{1-\alpha}\left[\begin{pmatrix} \rho_{0,0}(\alpha) & \rho_{0,1}(\alpha) \\ \rho_{0,1}^\dagger(\alpha) & \rho_{1,1}(\alpha) \end{pmatrix} \begin{pmatrix} \varepsilon^{\alpha-1}(\sigma+\varepsilon\mathbbm{1})^{1-\alpha} & 0 \\ 0 & \mathbbm{1}\end{pmatrix}\right]\label{eq-petz_rel_ent_def_lim_pf}.
		\end{align}
		Due to the fact that $\alpha\in (1,\infty)$, it holds that $\lim_{\varepsilon\to 0^+}\varepsilon^{1-\alpha}=+\infty$, and since the limit $\varepsilon\to 0^+$ of the trace of the matrix inside the square brackets in \eqref{eq-petz_rel_ent_def_lim_pf} is equal to $\Tr[\rho_{1,1}(\alpha)]$, and hence nonzero, we find that
		\begin{equation}
			\lim_{\varepsilon\to 0}\frac{1}{\alpha-1}\log_2\Tr\!\left[\rho^{\alpha}(\sigma+\varepsilon\mathbbm{1})^{1-\alpha}\right]=+\infty=D_\alpha(\rho\Vert\sigma)
		\end{equation}
		for the case $\alpha\in (1,\infty)$ and $\supp(\rho)\nsubseteq\supp(\sigma)$.
	\end{Proof}

	
	The Petz--R\'{e}nyi relative entropy is a natural generalization of the classical R\'{e}nyi relative entropy, which is defined for a probability distribution $p\colon \mathcal{X}\rightarrow[0,1]$ and a positive measure $q\colon \mathcal{X}\rightarrow[0,\infty)$ over the same alphabet $\mathcal{X}$ as 
	\begin{equation}\label{eq-renyi_rel_ent_classical}
		D_\alpha(p\Vert q)=\frac{1}{\alpha-1}\log_2\sum_{x\in\mathcal{X}}p(x)^\alpha q(x)^{1-\alpha}
	\end{equation} 
	for $\alpha\in(0,1)$. For $\alpha\in(1,\infty)$, the same expression defines $D_\alpha(p\Vert q)$ whenever $q(x)=0$ implies $p(x)=0$ for all $x\in\mathcal{X}$; otherwise, $D_\alpha(p\Vert q)=+\infty$.
	
	Recall the quantum Chernoff bound from Theorem~\ref{thm-quantum_Chernoff} in Section~\ref{subsec-state_discrimination}, which states that optimal error exponent for the task of discriminating between the states $\rho$ and $\sigma$ is
	\begin{equation}
		\underline{\xi}(\rho,\sigma) = \overline{\xi}(\rho,\sigma)= C(\rho\Vert\sigma) \coloneqq \sup_{\alpha\in(0,1)} -\log_2\Tr\big[\rho^{\alpha}\sigma^{1-\alpha}\big].
	\end{equation}
	Using the definition of the Petz--R\'{e}nyi relative entropy, we find that
	\begin{equation}
	C(\rho\Vert\sigma)=\sup_{\alpha\in(0,1)}(1-\alpha)D_{\alpha}(\rho\Vert\sigma).
	\end{equation}
	The Petz--R\'{e}nyi relative entropy thus plays a role in providing the optimal error exponent for the task of discriminating between two states (i.e., symmetric hypothesis testing).
	
	
	
	Like the quantum relative entropy, the Petz--R\'{e}nyi relative entropy is faithful, meaning that, for $\alpha\in(0,1)\cup(1,\infty)$ and states $\rho,\sigma$, 
	\begin{equation}
	D_{\alpha}(\rho\Vert\sigma)=0 \qquad \Longleftrightarrow
	\qquad 
	\rho=\sigma.
	\end{equation}
	We prove this in Proposition~\ref{prop-petz_sand_ren_ent_faithful} in the next section, as it requires results from both this section and the next one. Also, like the quantum relative entropy, the Petz--R\'{e}nyi relative entropy is a generalized divergence for certain values of~$\alpha$, which is shown in Theorem \ref{thm-petz_rel_ent_monotone} below.
	
	Before getting to Theorem~\ref{thm-petz_rel_ent_monotone}, we first discuss several important properties of the Petz--R\'{e}nyi relative entropy.  Let us note that, if $\rho$ and $\sigma$ act on a $d$-dimensional Hilbert space and are invertible, then the Petz--R\'{e}nyi relative quasi-entropy can be written as
	\begin{equation}\label{eq-petz_rel_ent_quasi_alt}
		Q_\alpha(\rho\Vert\sigma)=\bra{\varphi^\rho}\big(\rho^{-1}\otimes\sigma^{\t}\big)^{1-\alpha}\ket{\varphi^\rho},
	\end{equation}
	where $\ket{\varphi^\rho}\coloneqq\big(\rho^{\frac{1}{2}}\otimes\mathbbm{1}\big)\ket{\Gamma}$ is a purification of $\rho$ and $\ket{\Gamma}=\sum_{i=1}^{d}\ket{i,i}$. This is due to the transpose trick in \eqref{eq-transpose_trick} and the identity in \eqref{eq-trace_identity}. We can extend the applicability of this expression to states $\rho$ and positive semi-definite operators $\sigma$ that are not invertible by noting that
	\begin{equation}\label{eq-petz_rel_ent_quasi_lim}
		Q_\alpha(\rho\Vert\sigma)=\lim_{\varepsilon\to 0^+}\lim_{\delta\to 0^+}\frac{1}{\alpha-1}\log_2\Tr\!\left[\left((1-\delta)\rho+\delta\pi\right)^\alpha(\sigma+\varepsilon\mathbbm{1})^{1-\alpha}\right].
	\end{equation}
	
	We start by establishing the important fact that the quantum relative entropy is a special case of the Petz--R\'{e}nyi relative entropy in the limit $\alpha\to 1$.

	\begin{proposition}{prop-petz_rel_ent_lim_1}
		Let $\rho$ be a state and $\sigma$ a positive semi-definite operator. Then, in the limit $\alpha\to 1$, the Petz--R\'{e}nyi relative entropy converges to the quantum relative entropy:
		\begin{equation}
			\lim_{\alpha\to 1} D_\alpha(\rho\Vert\sigma)=D(\rho\Vert\sigma).
		\end{equation}
	\end{proposition}
	
	\begin{Proof}
		Let us first consider the case $\alpha\in (1,\infty)$. If $\supp(\rho)\nsubseteq\supp(\sigma)$, then $D_\alpha(\rho\Vert\sigma)=+\infty$, so that $\lim_{\alpha\to 1^+}D_\alpha(\rho\Vert\sigma)=+\infty$, consistent with the definition of the quantum relative entropy in this case (see Definition~\ref{def-rel_ent}). If $\supp(\rho)\subseteq\supp(\sigma)$, then $D_\alpha(\rho\Vert\sigma)$ is finite and we can write
		\begin{equation}
			D_\alpha(\rho\Vert\sigma)=\frac{1}{\alpha-1}\log_2 Q_\alpha(\rho\Vert\sigma).
		\end{equation}
		Now, let us define the function
		\begin{equation}
			Q_{\alpha,\beta}(\rho\Vert\sigma)\coloneqq\Tr\big[\rho^\alpha\sigma^{1-\beta}\big],
		\end{equation}
		so that $Q_\alpha(\rho\Vert\sigma)=Q_{\alpha,\alpha}(\rho\Vert\sigma)$. By noting that $\supp(\rho)\subseteq\supp(\sigma)$ implies $Q_1(\rho\Vert\sigma)=\Tr[\rho\Pi_\sigma]=1$ (where $\Pi_\sigma$ is the projection onto the support of $\sigma$), and since $\log_2 1=0$, we can write $D_\alpha(\rho\Vert\sigma)$ as
		\begin{equation}
			D_\alpha(\rho\Vert\sigma)=\frac{\log_2 Q_\alpha(\rho\Vert\sigma)-\log_2 Q_1(\rho\Vert\sigma)}{\alpha-1},
		\end{equation}
		so that
		\begin{align}
			\lim_{\alpha\to 1}D_\alpha(\rho\Vert\sigma)&=\left.\frac{\D}{\D\alpha}\log_2 Q_\alpha(\rho\Vert\sigma)\right|_{\alpha=1}\\
			&=\frac{1}{\ln(2)}\frac{\left.\frac{\D}{\D\alpha}Q_\alpha(\rho\Vert\sigma)\right|_{\alpha=1}}{Q_1(\rho\Vert\sigma)}\\
			&=\frac{1}{\ln(2)}\left.\frac{\D}{\D\alpha}Q_\alpha(\rho\Vert\sigma)\right|_{\alpha=1},
		\end{align}
		where the first equality follows from the definition of the derivative and the second equality from the derivative of the natural logarithm, along with the chain rule. Using the function $Q_{\alpha,\beta}$ and the chain rule, we write
		\begin{equation}
			\left.\frac{\D}{\D\alpha}Q_\alpha(\rho\Vert\sigma)\right|_{\alpha=1}=\left.\frac{\D}{\D\alpha}Q_{\alpha,1}(\rho\Vert\sigma)\right|_{\alpha=1}+\left.\frac{\D}{\D\beta}Q_{1,\beta}(\rho\Vert\sigma)\right|_{\beta=1}.
		\end{equation}
		Then,
		\begin{equation}
			\frac{\D}{\D\alpha}Q_{\alpha,1}(\rho\Vert\sigma)=\frac{\D}{\D\alpha}\Tr[\rho^\alpha\Pi_\sigma]=\frac{\D}{\D\alpha}\Tr[\rho^{\alpha}]=\Tr[\rho^\alpha\ln\rho],
		\end{equation}
		where we used the fact that $\rho^\alpha\Pi_\sigma=\rho^\alpha$ since $\supp(\rho)\subseteq\supp(\sigma)$. Therefore,
		\begin{equation}
			\left.\frac{\D}{\D\alpha}Q_{\alpha,1}(\rho\Vert\sigma)\right|_{\alpha=1}=\Tr[\rho\ln\rho].
		\end{equation}
		Similarly,
		\begin{equation}
			\frac{\D}{\D\beta}Q_{1,\beta}(\rho\Vert\sigma)=\frac{\D}{\D\beta}\Tr[\rho\sigma^{1-\beta}]=-\Tr[\rho\sigma^{1-\beta}\ln\sigma],
		\end{equation}
		so that
		\begin{equation}
			\left.\frac{\D}{\D\beta}Q_{1,\beta}(\rho\Vert\sigma)\right|_{\beta=1}=-\Tr[\rho\Pi_\sigma\ln\sigma]=-\Tr[\rho\ln\sigma],
		\end{equation}
		where the last equality follows from the fact that the support condition $\supp(\rho)\subseteq\allowbreak\supp(\sigma)$ holds. So we find that
		\begin{equation}\label{eq-petz_rel_ent_lim_pf}
			\lim_{\alpha\to 1}D_\alpha(\rho\Vert\sigma)=\frac{1}{\ln(2)}\left.\frac{\D}{\D\alpha}Q_\alpha(\rho\Vert\sigma)\right|_{\alpha=1}=\Tr[\rho\log_2\rho]-\Tr[\rho\log_2\sigma]
		\end{equation}
		when $\supp(\rho)\subseteq\supp(\sigma)$, which means that, for $\alpha\in (1,\infty)$,
		\begin{equation}
			\begin{aligned}
			\lim_{\alpha\to 1^+}D_\alpha(\rho\Vert\sigma)&=\left\{\begin{array}{l l} \Tr[\rho\log_2\rho]-\Tr[\rho\log_2\sigma] & \text{if }\supp(\rho)\subseteq\supp(\sigma),\\ +\infty & \text{otherwise}\end{array}\right.\\
			&=D(\rho\Vert\sigma).
			\end{aligned}
		\end{equation}
		
		Let us now consider the case $\alpha\in (0,1)$. If $\supp(\rho)\subseteq\supp(\sigma)$, then, since the limit in \eqref{eq-petz_rel_ent_lim_pf} holds from both sides, we find that
		\begin{equation}
		\lim_{\alpha\to 1^-}D_\alpha(\rho\Vert\sigma)=\Tr[\rho\log_2\rho]-\Tr[\rho\log_2\sigma].
		\end{equation}
        If the supports of $\rho$ and $\sigma$ are orthogonal, so that $\Tr[\rho\sigma]= 0$, then $D_\alpha(\rho\|\sigma) = +\infty$ for all $\alpha\in (0,1)$, implying that $\lim_{\alpha\to 1^-} D_\alpha(\rho\|\sigma) = +\infty = D(\rho\|\sigma)$. The last remaining case is when $\supp(\rho)\nsubseteq\supp(\sigma)$ but $\Tr[\rho\sigma]\neq 0$. Then, observe that we can write $D_\alpha$ as
		\begin{equation}
			D_\alpha(\rho\Vert\sigma)=\frac{\log_2 Q_\alpha(\rho\Vert\sigma)-\log_2 Q_1(\rho\Vert\sigma)}{\alpha-1}+\frac{\log_2 Q_1(\rho\Vert\sigma)}{\alpha-1},
		\end{equation}
		so that
		\begin{equation}\label{eq-petz_rel_ent_lim_pf2}
			\lim_{\alpha\to 1^-}D_\alpha(\rho\Vert\sigma)=\left.\frac{\D}{\D\alpha}\log_2 Q_\alpha(\rho\Vert\sigma)\right|_{\alpha=1}+\lim_{\alpha\to 1^-}\frac{\log_2\Tr[\rho\Pi_\sigma]}{\alpha-1},
		\end{equation}
		where we have used $Q_1(\rho\Vert\sigma)=\Tr[\rho\Pi_\sigma]$. Now, since $\Tr[\rho\sigma]\neq 0$ and $\supp(\rho)\allowbreak\nsubseteq\supp(\sigma)$, we have that $0<\Tr[\rho\Pi_\sigma]< 1$, which means that $\log_2\Tr[\rho\Pi_\sigma]<0$. Since $\lim_{\alpha\to 1^-}\frac{1}{\alpha-1}=-\infty$, we get that the second term in \eqref{eq-petz_rel_ent_lim_pf2} is equal to $+\infty$, which means that $\lim_{\alpha\to 1^-}D_\alpha(\rho\Vert\sigma)=+\infty$. Therefore,
		\begin{equation}
			\begin{aligned}
			&\lim_{\alpha\to 1^-}D_\alpha(\rho\Vert\sigma)\\
			&\quad=\left\{\begin{array}{l l} \Tr[\rho\log_2\rho]-\Tr[\rho\log_2\sigma] & \text{if }\supp(\rho)\subseteq\supp(\sigma),\\ +\infty & \text{otherwise} \end{array}\right.\\
			&\quad =D(\rho\Vert\sigma).
			\end{aligned}
		\end{equation}
		To conclude, we have that $\lim_{\alpha\to 1^+}D_\alpha(\rho\Vert\sigma)=\lim_{\alpha\to 1^-}D_\alpha(\rho\Vert\sigma)=D(\rho\Vert\sigma)$, which means that
		\begin{equation}
			\lim_{\alpha\to 1}D_\alpha(\rho\Vert\sigma)=D(\rho\Vert\sigma),
		\end{equation}
		as required.
	\end{Proof}
	
	\begin{proposition*}{Properties of the Petz--R\'{e}nyi Relative Entropy}{prop-Petz_rel_ent}
		For all states $\rho,\rho_1,\rho_2$ and positive semi-definite operators $\sigma,\sigma_1,\sigma_2$, the Petz--R\'{e}nyi relative entropy satisfies the following properties.
		\begin{enumerate}
			\item \textit{Isometric invariance}: For all $\alpha\in (0,1)\cup (1,\infty)$ and for all isometries~$V$,
				\begin{equation}\label{eq-petz_rel_ent_iso_invar}
					D_\alpha(\rho\Vert\sigma)=D_\alpha(V\rho V^\dagger\Vert V\sigma V^\dagger).
				\end{equation}
			\item \textit{Monotonicity in $\alpha$}: For all $\alpha,\beta\in (0,1)\cup(1,\infty)$, $D_\alpha$ is monotonically increasing in $\alpha$, i.e., $\alpha\leq\beta$ implies $ D_\alpha(\rho\Vert\sigma)\leq D_{\beta}(\rho\Vert\sigma)$.
			\item \textit{Additivity}: For all $\alpha\in (0,1)\cup(1,\infty)$,
				\begin{equation}
					D_\alpha(\rho_1\otimes\rho_2\Vert\sigma_1\otimes\sigma_2)=D_\alpha(\rho_1\Vert\sigma_1)+D_\alpha(\rho_2\Vert\sigma_2).
				\end{equation}
			\item \textit{Direct-sum property}: Let $p\colon \mathcal{X}\to[0,1]$ be a probability distribution over a finite alphabet $\mathcal{X}$ with associated $|\mathcal{X}|$-dimensional system $X$, and let $q\colon \mathcal{X}\to[0,\infty)$ be a positive function on $\mathcal{X}$. Let $\{\rho_A^x\}_{x\in\mathcal{X}}$ be a set of states on a system $A$, and let $\{\sigma_A^x\}_{x\in\mathcal{X}}$ be a set of positive semi-definite operators on $A$. Then,
				\begin{equation}
					Q_\alpha(\rho_{XA}\Vert \sigma_{XA})
					 =\sum_{x\in\mathcal{X}}p(x)^{\alpha}q(x)^{1-\alpha} Q_\alpha(\rho_A^x\Vert\sigma_A^x),
					 \label{eq:QEI:direct-sum-petz-renyi}
				\end{equation}
				where 
				\begin{align}
				\rho_{XA} & \coloneqq \sum_{x\in\mathcal{X}}p(x)\ket{x}\!\bra{x}_X\otimes\rho_A^x ,\\
				\sigma_{XA} & \coloneqq \sum_{x\in\mathcal{X}}q(x)\ket{x}\!\bra{x}_X\otimes\sigma_A^x.
				\end{align}

		\end{enumerate}
	\end{proposition*}
	
	\begin{remark}
		Observe that the direct-sum property analogous to that for the quantum relative entropy (see Proposition~\ref{prop-rel_ent}) does not hold for the Petz--R\'{e}nyi relative entropy for every $\alpha\in(0,1)\cup(1,\infty)$. We can instead only make a statement for the Petz--R\'{e}nyi relative quasi-entropy.
	\end{remark}
	
	\begin{Proof}
		\hfill\begin{enumerate}
			\item \textit{Proof of isometric invariance}: Let us start by writing $D_\alpha(\rho\Vert\sigma)$ as in \eqref{eq-petz_rel_ent_lim}:
				\begin{equation}
					D_\alpha(\rho\Vert\sigma)=\lim_{\varepsilon\to 0^+}\frac{1}{\alpha-1}\log_2\Tr\!\left[\rho^{\alpha}(\sigma+\varepsilon\mathbbm{1})^{1-\alpha}\right].
				\end{equation}
				Then, using the fact that $(V\rho V^\dagger)^{\alpha}=V\rho^\alpha V^\dagger$, we find that
				\begin{align}
					&D_\alpha(V\rho V^\dagger\Vert V\sigma V^\dagger)\notag \\
					&\quad =\lim_{\varepsilon\to 0^+}\frac{1}{\alpha-1}\log_2\Tr\!\left[(V\rho V^\dagger)^\alpha (V\sigma V^\dagger+\varepsilon\mathbbm{1})^{1-\alpha}\right]\\
					&\quad =\lim_{\varepsilon\to 0^+}\frac{1}{\alpha-1}\log_2\Tr\!\left[V\rho^\alpha V^\dagger (V\sigma V^\dagger+\varepsilon\mathbbm{1})^{1-\alpha}\right].
				\end{align}
				Now, let $\Pi\coloneqq VV^\dagger$ be the projection onto the image of $V$, and let $\hat{\Pi}\coloneqq\mathbbm{1}-\Pi$. Then, we can write
				\begin{equation}
					V\sigma V^\dagger+\varepsilon\mathbbm{1}=V\sigma V^\dagger+\varepsilon\Pi+\varepsilon\hat{\Pi}=V(\sigma+\varepsilon\mathbbm{1})V^\dagger+\varepsilon\hat{\Pi}.
				\end{equation}
				Since $V(\sigma+\varepsilon\mathbbm{1})V^\dagger$ and $\varepsilon\hat{\Pi}$ are supported on orthogonal subspaces, we obtain
				\begin{equation}
					(V\sigma V^\dagger+\varepsilon\mathbbm{1})^{1-\alpha}=V(\sigma+\varepsilon\mathbbm{1})^{1-\alpha}V^\dagger+\varepsilon^{1-\alpha}\hat{\Pi}.
				\end{equation}
				Therefore,
				\begin{align}
					&\Tr\!\left[V\rho^\alpha V^\dagger (V\sigma V^\dagger+\varepsilon\mathbbm{1})^{1-\alpha}\right]\nonumber\\
					&\quad =\Tr\!\left[V\rho^\alpha V^\dagger V(\sigma+\varepsilon\mathbbm{1})^{1-\alpha} V^\dagger +\varepsilon^{1-\alpha}V\rho^\alpha V^\dagger\hat{\Pi}\right]\\
					&\quad =\Tr\!\left[V\rho^\alpha(\sigma+\varepsilon\mathbbm{1})^{1-\alpha}V^\dagger\right]\\
					&\quad =\Tr\!\left[\rho^\alpha(\sigma+\varepsilon\mathbbm{1})^{1-\alpha}\right],
				\end{align}
				where the second equality follows from the fact that $V^\dagger\hat{\Pi}V=V^\dagger V-V^\dagger VV^\dagger V=\mathbbm{1}-\mathbbm{1}=0$, and  the last equality from cyclicity of the trace. Therefore,
				\begin{align}
					D_\alpha(V\rho V^\dagger \Vert V\sigma V^\dagger)&=\lim_{\varepsilon\to 0^+}\frac{1}{\alpha-1}\log_2\Tr\!\left[\rho^\alpha(\sigma+\varepsilon\mathbbm{1})^{1-\alpha}\right]\\
					&=D_\alpha(\rho\Vert\sigma),
				\end{align}
				as required.
				
			\item \textit{Proof of monotonicity in $\alpha$}: Using the expression in \eqref{eq-petz_rel_ent_alt} for $D_\alpha$ along with the form in \eqref{eq-petz_rel_ent_quasi_alt} for the quasi-entropy $Q_\alpha$, let us write $D_\alpha(\rho\Vert\sigma)$ as
				\begin{equation}
					D_\alpha(\rho\Vert\sigma)=\frac{1}{\alpha-1}\frac{\ln\bra{\varphi^\rho}X^{1-\alpha}\ket{\varphi^\rho}}{\ln(2)}=-\frac{1}{\gamma}\frac{\ln\bra{\varphi^\rho}X^\gamma\ket{\varphi^\rho}}{\ln(2)},
				\end{equation}
				where $X=\rho^{-1}\otimes\sigma^{\t}$, $\gamma\coloneqq 1-\alpha$, and $\ket{\varphi^\rho}=(\rho^{\frac{1}{2}}\otimes\mathbbm{1})\ket{\Gamma}$ is a purification of $\rho$. We first prove the result for $\rho$ invertible, and the proof for non-invertible states $\rho$ follows by \eqref{eq-petz_rel_ent_quasi_lim}. Now, since $\frac{\D}{\D\alpha}=\frac{\D}{\D\gamma}\frac{\D\gamma}{\D\alpha}=-\frac{\D}{\D\gamma}$, we find that
				\begin{align}
					&\frac{\D}{\D\alpha}D_\alpha(\rho\Vert\sigma)\nonumber\\
					&\quad =\frac{1}{\ln(2)}\frac{\D}{\D\gamma}\left(\frac{1}{\gamma}\ln\bra{\varphi^\rho}X^\gamma\ket{\varphi^\rho}\right)\\
					&\quad =\frac{1}{\ln(2)}\left(-\frac{1}{\gamma^2}\ln\bra{\varphi^\rho}X^\gamma\ket{\varphi^\rho}+\frac{1}{\gamma}\frac{\bra{\varphi^\rho}X^\gamma\ln X\ket{\varphi^\rho}}{\bra{\varphi^\rho}X^\gamma\ket{\varphi^\rho}}\right)\\
					&\quad =\frac{1}{\ln(2)}\frac{-\bra{\varphi^\rho}X^\gamma\ket{\varphi^\rho}\ln\bra{\varphi^\rho}X^\gamma\ket{\varphi^\rho}+\gamma\bra{\varphi^\rho}X^\gamma\ln X\ket{\varphi^\rho}}{\gamma^2\bra{\varphi^\rho}X^\gamma\ket{\varphi^\rho}}\\
					&\quad =\frac{1}{\ln(2)}\frac{-\bra{\varphi^\rho}X^\gamma\ket{\varphi^\rho}\ln\bra{\varphi^\rho}X^\gamma\ket{\varphi^\rho}+\bra{\varphi^\rho}X^\gamma\ln X^\gamma\ket{\varphi^\rho}}{\gamma^2\bra{\varphi^\rho}X^\gamma\ket{\varphi^\rho}}.
				\end{align}
				Letting $g(x)\coloneqq x\log_2 x$, it follows that
				\begin{equation}
					\frac{\D}{\D\alpha}D_\alpha(\rho\Vert\sigma)=\frac{\bra{\varphi^\rho}g(X^\gamma)\ket{\varphi^\rho}-g(\bra{\varphi^\rho}X^\gamma\ket{\varphi^\rho})}{\gamma^2\bra{\varphi^\rho}X^\gamma\ket{\varphi^\rho}}.
				\end{equation}
				Then, since $g$ is convex, by Jensen's inequality in \eqref{eq-op_Jensen_alt_2}, we conclude that
				\begin{equation}
					\bra{\varphi^\rho}g(X^\gamma)\ket{\varphi^\rho}\geq g(\bra{\varphi^\rho}X^\gamma\ket{\varphi^\rho}),
				\end{equation}
				which means that $\frac{\D}{\D\alpha}D_\alpha(\rho\Vert\sigma)\geq 0$. Therefore, $D_\alpha(\rho\Vert\sigma)$ is monotonically increasing in $\alpha$, as required.
		
			\item \textit{Proof of additivity}: When all quantities are finite, we have that
				\begin{equation}
					D_\alpha(\rho_1\otimes\rho_2\Vert\sigma_1\otimes\sigma_2)=\frac{1}{\alpha-1}\log_2\Tr\!\left[(\rho_1\otimes\rho_2)^\alpha (\sigma_1\otimes\sigma_2)^{1-\alpha}\right].
				\end{equation}
				Using the fact that $(X\otimes Y)^\beta= X^\beta\otimes Y^\beta$ for all positive semi-definite operators $X,Y$ and all $\beta\in\mathbb{R}$, we obtain
				\begin{align}
					Q_\alpha(\rho_1\otimes\rho_2\Vert\sigma_1\otimes\sigma_2)&=\Tr\!\left[(\rho_1\otimes\rho_2)^\alpha(\sigma_1\otimes\sigma_2)^{1-\alpha}\right]\\
					&=\Tr\!\left[(\rho_1^\alpha\otimes\rho_2^\alpha)(\sigma_1^{1-\alpha}\otimes\sigma_2^{1-\alpha})\right]\\
					&=\Tr\!\left[\rho_1^\alpha\sigma_1^{1-\alpha}\otimes\rho_2^\alpha\sigma_2^{1-\alpha}\right]\\
					&=\Tr[\rho_1^\alpha\sigma_1^{1-\alpha}]\cdot \Tr[\rho_2^\alpha\sigma_2^{1-\alpha}]\\
					&=Q_\alpha(\rho_1\Vert\sigma_1)\cdot Q_\alpha(\rho_2\Vert\sigma_2).
				\end{align}
				Applying $\frac{1}{\alpha-1}\log_2$ and definitions, additivity follows.
			\item \textit{Proof of the direct-sum property}: Define the classical--quantum operators
				\begin{equation}
					\rho_{XA}\coloneqq\sum_{x\in\mathcal{X}}p(x)\ket{x}\!\bra{x}_X\otimes\rho_A^x,\quad\sigma_{XA}\coloneqq\sum_{x\in\mathcal{X}}q(x)\ket{x}\!\bra{x}_X\otimes\sigma_A^x.
				\end{equation}
				Then,
				\begin{align}
					\rho_{XA}^\alpha &=\sum_{x\in\mathcal{X}}\ket{x}\!\bra{x}_X\otimes (p(x)\rho_X^x)^{\alpha}\\
					\sigma_{XA}^{1-\alpha}&=\sum_{x\in\mathcal{X}}\ket{x}\!\bra{x}_X\otimes (q(x)\sigma_A^x)^{1-\alpha},
				\end{align}
				so that
				\begin{align}
					\rho_{XA}^\alpha\sigma_{XA}^{1-\alpha}&=\sum_{x\in\mathcal{X}}\ket{x}\!\bra{x}_X\otimes (p(x)\rho_A^x)^\alpha (q(x)\sigma_A^x)^{1-\alpha}\\
					&=\sum_{x\in\mathcal{X}}p(x)^\alpha q(x)^{1-\alpha}\ket{x}\!\bra{x}_X\otimes (\rho_A^x)^{\alpha} (\sigma_A^x)^{1-\alpha},
				\end{align}
				and
				\begin{align}
					Q_\alpha(\rho_{XA}\Vert\sigma_{XA})&=\Tr[\rho_{XA}^\alpha\sigma_{XA}^{1-\alpha}]\\
					&=\sum_{x\in\mathcal{X}}p(x)^\alpha q(x)^{1-\alpha}\Tr[(\rho_A^x)^\alpha (\sigma_A^x)^{1-\alpha}]\\
					&=\sum_{x\in\mathcal{X}}p(x)^\alpha q(x)^{1-\alpha} Q_\alpha(\rho_A^x\Vert\sigma_A^x),
				\end{align}
				as required. \qedhere
		\end{enumerate}
	\end{Proof}
	
	We now prove the data-processing inequality for the Petz--R\'{e}nyi relative entropy for $\alpha\in [0,1)\cup (1,2]$.
	
	\begin{theorem*}{Data-Processing Inequality for Petz--R\'{e}nyi Relative Entropy}{thm-petz_rel_ent_monotone}
		Let $\rho$ be a state, $\sigma$ a positive semi-definite operator, and $\mathcal{N}$ a quantum channel. Then, for all $\alpha\in[0,1)\cup (1,2]$,
		\begin{equation}
			D_\alpha(\rho\Vert\sigma)\geq D_\alpha(\mathcal{N}(\rho)\Vert\mathcal{N}(\sigma)).
		\end{equation}
	\end{theorem*}
	
	\begin{Proof}
		We prove the statement for $\alpha \in (0,1)\cup(1,2]$. The case of $\alpha = 0$ then follows by taking the limit $\alpha \to 0$. From Stinespring's theorem (Theorem \ref{thm-q_channels}), we know that the action of every channel $\mathcal{N}$ on a linear operator $X$ can be written as
		\begin{equation}
			\mathcal{N}(X)=\Tr_E[VXV^\dagger],
		\end{equation}
		for some $V$, where $V$ is an isometry and $E$ is an auxiliary system with dimension $d_E\geq\rank(\Gamma^{\mathcal{N}})$. As stated in \eqref{eq-petz_rel_ent_iso_invar}, $D_\alpha$ is isometrically invariant. Therefore, it suffices to show the data-processing inequality for $D_\alpha$ under partial trace; i.e., it suffices to show that for every state $\rho_{AB}$ and every positive semi-definite operator $\sigma_{AB}$,
		\begin{equation}\label{eq-petz_rel_ent_monotone_pf_0}
			D_\alpha(\rho_{AB}\Vert\sigma_{AB})\geq D_\alpha(\rho_A\Vert\sigma_A),\quad \alpha\in (0,1)\cup (1,2].
		\end{equation}
		We now proceed to prove this inequality. We prove it for $\rho_{AB}$, and hence $\rho_A$, invertible, as well as for $\sigma_{AB}$ and $\sigma_A$ invertible. The result follows in the general case of $\rho_{AB}$ and/or $\rho_{A}$ non-invertible, as well as $\sigma_{AB}$ and/or $\sigma_A$ non-invertible, by applying the result to the invertible operators $(1-\delta)\rho_{AB}+\delta\pi_{AB}$ and $\sigma_{AB}+\varepsilon\mathbbm{1}_{AB}$, with $\delta,\varepsilon>0$, and taking the limit $\delta\to 0^+$, followed by $\varepsilon\to 0^+$, because
		\begin{align}
			D_\alpha(\rho_{AB}\Vert\sigma_{AB}) & =\lim_{\varepsilon\to 0^+}\lim_{\delta\to 0^+}D_\alpha((1-\delta)\rho_{AB}+\delta\pi_{AB}\Vert\sigma_{AB}+\varepsilon\mathbbm{1}_{AB}),\\
			D_\alpha(\rho_{A}\Vert\sigma_{A}) & =\lim_{\varepsilon\to 0^+}\lim_{\delta\to 0^+}D_\alpha((1-\delta)\rho_{A}+\delta\pi_{A}\Vert\sigma_{A}+d_B\varepsilon\mathbbm{1}_{A}),			
		\end{align}
		which can be verified in a similar manner to the proof of \eqref{eq-petz_rel_ent_lim} in Proposition~\ref{prop-petz_rel_ent_lim}.
		
		Using the quasi-entropy $Q_\alpha$, we can equivalently write \eqref{eq-petz_rel_ent_monotone_pf_0} as
		\begin{equation}\label{eq-petz_rel_ent_monotone_pf}
			\begin{aligned}
			Q_\alpha(\rho_{AB}\Vert\sigma_{AB})&\geq Q_\alpha(\rho_A\Vert\sigma_A),\quad \text{ for } \alpha\in (1,2],\\
			Q_\alpha(\rho_{AB}\Vert\sigma_{AB})&\leq Q_\alpha(\rho_A\Vert\sigma_A),\quad \text{ for } \alpha\in (0,1).
			\end{aligned}
		\end{equation}
		The remainder of this proof is thus devoted to establishing \eqref{eq-petz_rel_ent_monotone_pf}.
		
		Consider that
		\begin{align}
			Q_\alpha(\rho_{AB}\Vert\sigma_{AB})&=\bra{\varphi^{\rho_{AB}}}f\big(\rho_{AB}^{-1}\otimes\sigma_{\hat{A}\hat{B}}^{\t}\big)\ket{\varphi^{\rho_{AB}}},\\
			Q_\alpha(\rho_A\Vert\sigma_A)&=\bra{\varphi^{\rho_A}}f\big(\rho_A^{-1}\otimes\sigma_{\hat{A}}^{\t}\big)\ket{\varphi^{\rho_A}},
		\end{align}
		where we have set
		\begin{equation}\label{eq-petz_rel_ent_monotone_pf_2}
			f(x)\coloneqq x^{1-\alpha}
		\end{equation}
		and
		\begin{align}
			\ket{\varphi^{\rho_{AB}}}&=\big(\rho_{AB}^{\frac{1}{2}}\otimes\mathbbm{1}_{\hat{A}\hat{B}}\big)\ket{\Gamma}_{AB\hat{A}\hat{B}}=\big(\rho_{AB}^{\frac{1}{2}}\otimes\mathbbm{1}_{\hat{A}\hat{B}}\big)\ket{\Gamma}_{A\hat{A}}\ket{\Gamma}_{B\hat{B}},\\
			\ket{\varphi^{\rho_A}}&=\big(\rho_A^{\frac{1}{2}}\otimes\mathbbm{1}_{\hat{A}}\big)\ket{\Gamma}_{A\hat{A}}.
		\end{align}
		Now, let us define the isometry $V_{A\hat{A}\rightarrow AB\hat{A}\hat{B}}$ as\footnote{Observe that the isometry is related to the isometric extension in \eqref{eq-Petz_PT_iso_ext} of the Petz recovery channel for the partial trace, as  discussed in Section~\ref{subsec-Petz_channel}.}
		\begin{equation}\label{eq-petz_rel_ent_monotone_pf_3}
			V_{A\hat{A}\rightarrow AB\hat{A}\hat{B}}=\rho_{AB}^{\frac{1}{2}}\big(\rho_A^{-\frac{1}{2}}\otimes\mathbbm{1}_{\hat{A}}\big)\ket{\Gamma}_{B\hat{B}}.
		\end{equation}
		Observe then that
		\begin{align}
			V_{A\hat{A}\to AB\hat{A}\hat{B}}\ket{\varphi^{\rho_A}}_{A\hat{A}}&=\rho_{AB}^{\frac{1}{2}}\big(\rho_A^{-\frac{1}{2}}\otimes\mathbbm{1}_{\hat{A}}\big)\big(\rho_A^{\frac{1}{2}}\otimes\mathbbm{1}_{\hat{A}}\big)\ket{\Gamma}_{A\hat{A}}\ket{\Gamma}_{B\hat{B}}\\
			&=\big(\rho_{AB}^{\frac{1}{2}}\otimes\mathbbm{1}_{\hat{A}\hat{B}}\big)\ket{\Gamma}_{A\hat{A}}\ket{\Gamma}_{B\hat{B}}\\
			&=\ket{\varphi^{\rho_{AB}}}.
		\end{align}
		We thus obtain, using the operator Jensen inequality (Theorem~\ref{thm-Jensen}),
		\begin{align}
			Q_\alpha(\rho_{AB}\Vert\sigma_{AB})&=\bra{\varphi^{\rho_A}}V^\dagger f\big(\rho_{AB}^{-1}\otimes\sigma_{\hat{A}\hat{B}}^{\t}\big)V\ket{\varphi^{\rho_A}}\\
				&\geq \bra{\varphi^{\rho_A}}f\Big(V^\dagger\big(\rho_{AB}^{-1}\otimes\sigma_{\hat{A}\hat{B}}^{\t}\big)V\Big)\ket{\varphi^{\rho_A}},\quad \text{ for } \alpha\in (1,2],
		\end{align}
		and
		\begin{align}				
			Q_\alpha(\rho_{AB}\Vert\sigma_{AB})&=\bra{\varphi^{\rho_A}}V^\dagger f\big(\rho_{AB}^{-1}\otimes\sigma_{\hat{A}\hat{B}}^{\t}\big)V\ket{\varphi^{\rho_A}}\\
			& \leq \bra{\varphi^{\rho_A}}f\Big(V^\dagger \big(\rho_{AB}^{-1}\otimes\sigma_{\hat{A}\hat{B}}^{\t}\big)V\Big)\ket{\varphi^{\rho_A}}, \quad \text{ for }  \alpha\in [0,1).
		\end{align}
		Note that the operator Jensen inequality is applicable because for $\alpha\in(1,2]$ the function $f$ in \eqref{eq-petz_rel_ent_monotone_pf_2} is operator convex and for $\alpha\in (0,1)$ it is operator concave.\footnote{Indeed, as discussed in Section~\ref{subsec:operator_inequalities}, the function $x^\beta$ is operator convex for $\beta\in[-1,0)\cup[1,2]$ and operator concave for $\beta\in (0,1]$, where here $\beta=1-\alpha$.} Now, consider that
		\begin{align}
			&V^\dagger\big(\rho_{AB}^{-1}\otimes\sigma_{\hat{A}\hat{B}}^{\t}\big)V\nonumber\\
			&\quad =\bra{\Gamma}_{B\hat{B}}\big(\rho_A^{-\frac{1}{2}}\otimes\mathbbm{1}_{\hat{A}}\big)\rho_{AB}^{\frac{1}{2}}\big(\rho_{AB}^{-1}\otimes\sigma_{\hat{A}\hat{B}}^{\t}\big)\rho_{AB}^{\frac{1}{2}}\big(\rho_A^{-\frac{1}{2}}\otimes\mathbbm{1}_{\hat{A}}\big)\ket{\Gamma}_{B\hat{B}}\\
			&\quad =\bra{\Gamma}_{B\hat{B}}\big(\rho_A^{-\frac{1}{2}}\otimes\mathbbm{1}_{\hat{A}}\big)\big(\mathbbm{1}_{AB}\otimes\sigma_{\hat{A}\hat{B}}^{\t}\big)\big(\rho_A^{-\frac{1}{2}}\otimes\mathbbm{1}_{\hat{A}}\big)\ket{\Gamma}_{B\hat{B}}\\			
			&\quad =\rho_A^{-1}\otimes\bra{\Gamma}_{B\hat{B}}\sigma_{\hat{A}\hat{B}}^{\t}\ket{\Gamma}_{B\hat{B}}\\
			&\quad =\rho_A^{-1}\otimes\sigma_{\hat{A}}^{\t},
		\end{align}
		where the last equality follows from the fact that
		\begin{equation}
			\bra{\Gamma}_{B\hat{B}}\sigma_{\hat{A}\hat{B}}^{\t}\ket{\Gamma}_{B\hat{B}}\allowbreak=\Tr_{\hat{B}}[\sigma_{\hat{A}\hat{B}}^{\t}]=\sigma_{\hat{A}}^{\t},
		\end{equation}
		the last equality due to the fact that the transpose is taken on a product basis for $\mathcal{H}_{\hat{A}}\otimes\mathcal{H}_{\hat{B}}$. Therefore, we find that 
		\begin{align}
			Q_\alpha(\rho_{AB}\Vert\sigma_{AB})&\geq Q_\alpha(\rho_A\Vert\sigma_A),\quad \text{ for } \alpha\in(1,2],\\
			Q_\alpha(\rho_{AB}\Vert\sigma_{AB})&\leq Q_\alpha(\rho_A\Vert\sigma_A),\quad \text{ for }\alpha\in(0,1),
		\end{align}
		as required. This establishes the data-processing inequality for $D_\alpha$ under partial trace. Combining this with the isometric invariance of $D_\alpha$ and Stinespring's theorem, we conclude that
		\begin{equation}
			D_\alpha(\rho\Vert\sigma)\geq D_\alpha(\mathcal{N}(\rho)\Vert\mathcal{N}(\sigma)),\quad \alpha\in (0,1)\cup (1,2]
		\end{equation}
		for every state $\rho$, positive semi-definite operator $\sigma$, and channel $\mathcal{N}$. 
	\end{Proof}
	
	
	By taking the limit $\alpha\to 1$ in the statement of data-processing inequality for $D_\alpha$, and using Proposition \ref{prop-petz_rel_ent_lim_1}, we immediately obtain the data-processing inequality for the quantum relative entropy, stated previously as Theorem~\ref{thm-monotone_rel_ent}.
	
	\begin{corollary*}{Data-Processing Inequality for Quantum Relative Entropy}{cor-rel_ent_monotone1}
		Let $\rho$ be a state, $\sigma$ a positive semi-definite operator, and $\mathcal{N}$ a quantum channel. Then,
		\begin{equation}
			D(\rho\Vert\sigma)\geq D(\mathcal{N}(\rho)\Vert\mathcal{N}(\sigma)).
		\end{equation}
	\end{corollary*}
			
	As we showed in the above proof of Theorem~\ref{thm-petz_rel_ent_monotone}, an equivalent form of the data-processing inequality for the Petz--R\'{e}nyi relative entropy is
    \begin{align}
		Q_\alpha(\rho\Vert\sigma)&\geq Q_\alpha(\mathcal{N}(\rho)\Vert\mathcal{N}(\sigma)),\quad\alpha\in(1,2],\label{eq-petz_renyi_quasi_monotone_1}\\
		Q_\alpha(\rho\Vert\sigma)&\leq Q_\alpha(\mathcal{N}(\rho)\Vert\mathcal{N}(\sigma)),\quad\alpha\in [0,1).\label{eq-petz_renyi_quasi_monotone_2}
	\end{align}
	
	With the data-processing inequality 
    for the Petz--R\'{e}nyi relative entropy 
    in hand, it is now straightforward to prove some of the following additional properties.
	
	\begin{proposition*}{Additional Properties of Petz--R\'{e}nyi Relative Entropy}{prop-petz_rel_ent_add_properties}
		The Petz--R\'{e}nyi relative entropy $D_\alpha$ satisfies the following properties for every state $\rho$ and positive semi-definite operator $\sigma$ for $\alpha\in\left(0,1\right)\cup (1,2]$:
		\begin{enumerate}
			\item If $\Tr(\sigma)\leq \Tr(\rho)=1$, then $D_\alpha(\rho\Vert\sigma)\geq 0$.
			\item \textit{Faithfulness}: If $\Tr[\sigma]\leq 1$, we have that $D_\alpha(\rho\Vert\sigma)=0$ if and only if $\rho=\sigma$.
			\item For every positive semi-definite operator $\sigma'$ such that $\sigma'\geq \sigma$, we have $D_\alpha(\rho\Vert\sigma) \geq D_\alpha(\rho\Vert\sigma')$. In particular, if $\rho\leq\sigma$, then $D_\alpha(\rho\Vert\sigma)\leq 0$.
		\end{enumerate}
	\end{proposition*}
	
	\begin{Proof}
		\hfill\begin{enumerate}
			\item By the data-processing inequality for $D_\alpha$ with respect to the trace channel $\Tr$, and letting $x=\Tr(\rho)=1$ and $y=\Tr(\sigma)$, we find that
					\begin{align}
						D_\alpha(\rho\Vert\sigma)\geq D_\alpha(x\Vert y)&=\frac{1}{\alpha-1}\log_2\big(x^\alpha y^{1-\alpha}\big)\\
						&=\frac{1}{\alpha-1}\log_2(y^{1-\alpha})\\
						&=-\log_2 y\\
						&\geq 0,
					\end{align}
					where the last line follows from the assumption that $y=\Tr(\sigma)\leq 1$.
					
			\item \textit{Proof of faithfulness}: If $\rho=\sigma$, then, for all $\alpha\in\left(0,1\right)\cup (1,2)$,
					\begin{align}
						D_\alpha(\rho\Vert\rho)
						&=\frac{1}{\alpha-1}\log_2\Tr[\rho^{\alpha}\rho^{1-\alpha}]\label{eq-petz_rel_ent_faithful_pf3}\\
						&=\frac{1}{\alpha-1}\log_2\Tr(\rho)\label{eq-petz_rel_ent_faithful_pf4}\\
						&=0.\label{eq-petz_rel_ent_faithful_pf5}
					\end{align}
					Next, suppose that $\alpha\in\left(0,1\right)\cup (1,2)$ and  $D_\alpha(\rho\Vert\sigma)=0$. From the above, we conclude that $D_\alpha(\Tr(\rho)\Vert\Tr(\sigma))=-\log_2 y\geq 0$. From the fact that $\log_2 y=0$ if and only if $y=1$, we conclude that $D_\alpha(\rho\Vert\sigma)=0$ implies $\Tr(\sigma)=\Tr(\rho)=1$, so that $\sigma$ is a density operator. Then, for every measurement channel $\mathcal{M}$, 
					\begin{equation}
						D_\alpha(\mathcal{M}(\rho)\Vert\mathcal{M}(\sigma))\leq D_\alpha(\rho\Vert\sigma)=0.
					\end{equation}
					On the other hand, since a measurement channel is trace preserving, non-negativity ensures $D_\alpha(\mathcal{M}(\rho)\Vert\mathcal{M}(\sigma))\geq 0$, 
					so that $ D_\alpha(\mathcal{M}(\rho)\Vert\mathcal{M}(\sigma))=0$ for every measurement channel $\mathcal{M}$. Now, recall that $\mathcal{M}(\rho)$ and $\mathcal{M}(\sigma)$ are effectively probability distributions determined by the measurement. Since the classical R\'{e}nyi relative entropy is equal to zero if and only if its two arguments are equal, we can conclude that $\mathcal{M}(\rho)=\mathcal{M}(\sigma)$. Since this is true for every measurement channel, we conclude from Theorem~\ref{thm-trace_dist_ach_by_meas_channel} and the fact that the trace norm is a norm that $\rho=\sigma$.
					
					So we have that $ D_\alpha(\rho\Vert\sigma)=0$ if and only if $\rho=\sigma$, as required.
					
		
			\item Consider the state $\hat{\rho}\coloneqq\ket{0}\!\bra{0}\otimes\rho$ and the operator $\hat{\sigma}\coloneqq\ket{0}\!\bra{0}\otimes\sigma+\ket{1}\!\bra{1}\otimes (\sigma'-\sigma)$, which is positive semi-definite because $\sigma'\geq \sigma$ by assumption.
			
			Then
					\begin{equation}
						\hat{\rho}^{\alpha}\hat{\sigma}^{1-\alpha} = \ket{0}\!\bra{0}\otimes \rho^{\alpha} \sigma^{1-\alpha},
					\end{equation}
					which implies that
					\begin{equation}
						 D_\alpha(\hat{\rho}\Vert\hat{\sigma})= D_\alpha(\rho\Vert\sigma).
					\end{equation}
					Then, observing that $\Tr_1[\hat{\sigma}]=\sigma'$, and using the data-processing inequality for $ D_\alpha$ with respect to the partial trace channel $\Tr_1$, we conclude that
					\begin{equation}
						 D_\alpha(\rho\Vert\sigma')= D_\alpha(\Tr_1(\hat{\rho})\Vert\Tr_1(\hat{\sigma}))\leq  D_\alpha(\hat{\rho}\Vert\hat{\sigma})=D_\alpha(\rho\Vert\sigma),
					\end{equation}
					as required.
                    
                    The second statement follows by substituting $\sigma' \to \sigma$ and $\sigma \to \rho$ and applying faithfulness.
                    \qedhere

		\end{enumerate}
	\end{Proof}
	
	\begin{proposition*}{Joint Convexity \& Concavity of the Petz--R\'{e}nyi Relative Quasi-Entropy}{prop-petz_renyi_joint_convex}
		Let $p\colon \mathcal{X}\to[0,1]$ be a probability distribution over a finite alphabet $\mathcal{X}$ with associated $|\mathcal{X}|$-dimensional system $X$, let $\{\rho_A^x\}_{x\in\mathcal{X}}$ be a set of states on a system $A$, and let $\{\sigma_A^x\}_{x\in\mathcal{X}}$ be a set of positive semi-definite operators on $A$. Then,
		\begin{align}
			Q_\alpha\!\left(\sum_{x\in\mathcal{X}}p(x)\rho_A^x\middle\Vert\sum_{x\in\mathcal{X}}p(x)\sigma_A^x\right)
			& \leq \sum_{x\in\mathcal{X}}p(x)Q_\alpha(\rho_A^x\Vert\sigma_A^x),\quad\text{ for }\alpha\in(1,2], \label{eq-Petz_Renyi_quasi_convex_pf-other}\\
			Q_\alpha\!\left(\sum_{x\in\mathcal{X}}p(x)\rho_A^x\middle\Vert\sum_{x\in\mathcal{X}}p(x)\sigma_A^x\right) 
			&\geq \sum_{x\in\mathcal{X}}p(x)Q_\alpha(\rho_A^x\Vert\sigma_A^x),\quad\text{ for }\alpha\in[0,1). \label{eq-Petz_Renyi_quasi_convex_pf}
		\end{align}
		Consequently, the Petz--R\'{e}nyi relative entropy $D_\alpha$ is jointly convex for $\alpha\in [0,1)$:
		\begin{equation}
			D_\alpha\!\left(\sum_{x\in\mathcal{X}}p(x)\rho_A^x\middle\Vert\sum_{x\in\mathcal{X}}p(x)\sigma_A^x\right)
			\leq \sum_{x\in\mathcal{X}}p(x)D_\alpha(\rho_A^x\Vert\sigma_A^x),\quad\alpha\in[0,1).
		\end{equation}
	\end{proposition*}
	
	\begin{Proof}
		By the direct-sum property of $Q_\alpha$ and applying \eqref{eq-petz_renyi_quasi_monotone_1}--\eqref{eq-petz_renyi_quasi_monotone_2} and Proposition~\ref{prop:QEI:joint-convexity-gen-div}, we conclude \eqref{eq-Petz_Renyi_quasi_convex_pf-other}--\eqref{eq-Petz_Renyi_quasi_convex_pf}.
		
		For $\alpha\in[0,1)$, applying $\log_2$ to both sides of \eqref{eq-Petz_Renyi_quasi_convex_pf} and multiplying by $\frac{1}{\alpha-1}$, which is negative, we conclude that
		\begin{equation}
			\begin{aligned}
			&\frac{1}{\alpha-1}\log_2 Q_\alpha\!\left(\sum_{x\in\mathcal{X}}p(x)\rho_A^x\middle\Vert\sum_{x\in\mathcal{X}}p(x)\sigma_A^x\right)\\
			&\qquad\qquad\qquad\qquad\leq \frac{1}{\alpha-1}\log_2\!
			\left(\sum_{x\in\mathcal{X}}p(x)Q_\alpha(\rho_A^x\Vert\sigma_A^x)\right).
			\end{aligned}
		\end{equation}
		Then, since $-\log_2$ is a convex function, and using the definition of $D_\alpha$ in terms of $Q_\alpha$, we find that 
		\begin{align}
			D_\alpha\!\left(\sum_{x\in\mathcal{X}}p(x)\rho_A^x\middle\Vert \sum_{x\in\mathcal{X}}p(x)\sigma_A^x\right)&\leq \sum_{x\in\mathcal{X}}p(x)\frac{1}{\alpha-1}\log_2 Q_\alpha(\rho_A^x\Vert\sigma_A^x)\\
			&=\sum_{x\in\mathcal{X}}p(x)D_\alpha(\rho_A^x\Vert\sigma_A^x),
		\end{align}
		as required.
	\end{Proof}

\section{Sandwiched R\'{e}nyi Relative Entropy}

\label{sec-sand_ren_rel_ent}

	A second example of a generalized divergence is the sandwiched R\'{e}nyi relative entropy, which we define as follows.
	
	\begin{definition}{Sandwiched R\'{e}nyi Relative Entropy}{def-sand_rel_ent}
		For all $\alpha\in (0,1)\cup (1,\infty)$, we define the \textit{sandwiched R\'{e}nyi relative quasi-entropy} for every state $\rho$ and positive semi-definite operator $\sigma$ as  
		\begin{equation}
			\begin{aligned}
			&\widetilde{Q}_\alpha(\rho\Vert\sigma)\\
			&\coloneqq \left\{\begin{array}{l l} \multirow{2}{*}{$\Tr\!\left[\left(\sigma^{\frac{1-\alpha}{2\alpha}}\rho\sigma^{\frac{1-\alpha}{2\alpha}}\right)^\alpha\right]$} & \text{if }\alpha\in (0,1),\text{ or}\\ & \alpha\in (1,\infty),~\supp(\rho)\subseteq\supp(\sigma), \\[0.2cm] +\infty & \text{otherwise}. \end{array}\right.
			\end{aligned}
		\end{equation}
		The \textit{sandwiched R\'{e}nyi relative entropy} is then defined as
		\begin{equation}\label{eq-sand_ren_rel_entropy}
			\widetilde{D}_{\alpha}(\rho\Vert\sigma)\coloneqq \frac{1}{\alpha-1}\log_2\widetilde{Q}_\alpha(\rho\Vert\sigma).		\end{equation}
	\end{definition}
	
	Observe that we can use the definition of the Schatten norm from \eqref{eq-Schatten-norm} to write the sandwiched R\'{e}nyi relative entropy $\widetilde{D}_\alpha$ in the following different ways:
	\begin{align}
		\widetilde{D}_\alpha(\rho\Vert\sigma)&=\frac{\alpha}{\alpha-1}\log_2\norm{\sigma^{\frac{1-\alpha}{2\alpha}}\rho\sigma^{\frac{1-\alpha}{2\alpha}}}_{\alpha}\label{eq-sand_rel_ent_Schatten}\\
		&=\frac{\alpha}{\alpha-1}\log_2\norm{\rho^{\frac{1}{2}}\sigma^{\frac{1-\alpha}{\alpha}}\rho^{\frac{1}{2}}}_\alpha\label{eq-sand_rel_ent_Schatten_2}\\
		&=\frac{2\alpha}{\alpha-1}\log_2\norm{\rho^{\frac{1}{2}}\sigma^{\frac{1-\alpha}{2\alpha}}}_{2\alpha}\label{eq-sand_rel_ent_Schatten_3}.
	\end{align}
	The expression in \eqref{eq-sand_rel_ent_Schatten_2} and Proposition \ref{prop-Schatten_pos_var} then lead us to the following variational characterization of $\widetilde{D}_\alpha$ for $\alpha\in(0,1)\cup(1,\infty)$:
	\begin{equation}\label{eq-sand_rel_ent_var}
		 \widetilde{D}_\alpha(\rho\Vert\sigma)=\sup_{\substack{\tau> 0,\\\Tr(\tau)= 1}}\widetilde{D}_\alpha(\rho\Vert\sigma;\tau),
	\end{equation}
	where 
	\begin{equation}\label{eq-sand_rel_ent_opt}
		\begin{aligned}
		&\widetilde{D}_\alpha(\rho\Vert\sigma;\tau)\\
		&\quad \coloneqq\left\{\begin{array}{l l} 
		+\infty & \text{if }\alpha>1,~\supp(\rho)\nsubseteq\supp(\sigma),\\
		\frac{\alpha}{\alpha-1}\log_2\Tr\!\left[\rho^{\frac{1}{2}}\sigma^{\frac{1-\alpha}{\alpha}}\rho^{\frac{1}{2}}\tau^{\frac{\alpha-1}{\alpha}}\right] & \text{otherwise.}\end{array}\right.
		\end{aligned}
	\end{equation}
	Also observe that for $\alpha=\frac{1}{2}$, the sandwiched R\'{e}nyi relative entropy $\widetilde{D}_{\frac{1}{2}}(\rho\Vert\sigma)$ can be expressed as
	\begin{equation}\label{eq-sand_rel_half_fidelity}
		\widetilde{D}_{\frac{1}{2}}(\rho\Vert\sigma)=-\log_2F(\rho,\sigma),
	\end{equation}
	where we recall the definition of the fidelity $F(\rho,\sigma)$ from Definition~\ref{def-fidelity}.
	
	In the case $\alpha\in (1,\infty)$, since $1-\alpha$ is negative, we take the inverse of $\sigma$. In case $\sigma$ is not invertible, we take the inverse of $\sigma$ on its support. An alternative to this convention is to define $\widetilde{D}_\alpha(\rho\Vert\sigma)$ for $\alpha>1$ using only positive definite~$\sigma$, and for positive semi-definite $\sigma$, define
	\begin{equation}
		\widetilde{D}_\alpha(\rho\Vert\sigma)=\lim_{\varepsilon\to 0^+}\widetilde{D}_\alpha(\rho\Vert\sigma+\varepsilon\mathbbm{1}).
	\end{equation}
	Both alternatives are equivalent, as we now show (similar to what we did in the proofs of Propositions~\ref{prop-rel_ent_lim} and \ref{prop-petz_rel_ent_lim}).
	
	\begin{proposition}{prop-sand_rel_ent_lim}
		For every state $\rho$ and positive semi-definite operator $\sigma$,
		\begin{equation}\label{eq-sand_rel_ent_lim}
			\widetilde{D}_\alpha(\rho\Vert\sigma)=\lim_{\varepsilon\to 0^+}\frac{1}{\alpha-1}\log_2\Tr\!\left[\left(\rho^{\frac{1}{2}}(\sigma+\varepsilon\mathbbm{1})^{\frac{1-\alpha}{\alpha}}\rho^{\frac{1}{2}}\right)^\alpha\right].
		\end{equation}
	\end{proposition}
	
	\begin{Proof}
		For $\alpha\in(0,1)$, this is immediate from the fact that the logarithm, trace, and power functions are continuous, so that the limit can be brought inside the trace and inside the power $(\sigma+\varepsilon\mathbbm{1})^{\frac{1-\alpha}{2\alpha}}$.
		
		For $\alpha\in(1,\infty)$, since $1-\alpha$ is negative and $\sigma$ is not necessarily invertible, let us start by decomposing the underlying Hilbert space $\mathcal{H}$ as $\mathcal{H}=\supp(\sigma)\oplus\ker(\sigma)$, as in \eqref{eq-q_rel_ent_block}, so that
		\begin{equation}
			\rho^{\frac{1}{2}}=\begin{pmatrix}(\rho^{\frac{1}{2}})_{0,0} & (\rho^{\frac{1}{2}})_{0,1}\\ (\rho^{\frac{1}{2}})_{0,1}^\dagger & (\rho^{\frac{1}{2}})_{1,1} \end{pmatrix},\quad \sigma=\begin{pmatrix} \sigma_{0,0} & 0 \\ 0 & 0 \end{pmatrix}.
		\end{equation}
        We can then write
		\begin{equation}
			\sigma+\varepsilon\mathbbm{1}=\begin{pmatrix} \sigma_{0,0}+\varepsilon \mathbbm{1} & 0 \\ 0 & \varepsilon\mathbbm{1}\end{pmatrix},
		\end{equation}
		which implies that
		\begin{equation}
			(\sigma+\varepsilon\mathbbm{1})^{\frac{1-\alpha}{\alpha}}=\begin{pmatrix} (\sigma_{0,0}+\varepsilon\mathbbm{1})^{\frac{1-\alpha}{\alpha}} & 0 \\ 0 & (\varepsilon\mathbbm{1})^{\frac{1-\alpha}{\alpha}}\end{pmatrix}.
		\end{equation}
		
		If $\supp(\rho)\subseteq\supp(\sigma)$, then
		\begin{equation}
			\rho^{\frac{1}{2}}(\sigma+\varepsilon\mathbbm{1})^{\frac{1-\alpha}{\alpha}}\rho^{\frac{1}{2}} = \begin{pmatrix} (\rho^{\frac{1}{2}})_{0,0}(\sigma_{0,0}+\varepsilon\mathbbm{1})^{\frac{1-\alpha}{\alpha}}(\rho^{\frac{1}{2}})_{0,0} & 0\\0&0\end{pmatrix},
		\end{equation}
		so that
		\begin{equation}
			\lim_{\varepsilon\to 0^+}\frac{1}{\alpha-1}\log_2\Tr\!\left[\left(\rho^{\frac{1}{2}}(\sigma+\varepsilon\mathbbm{1})^{\frac{1-\alpha}{\alpha}}\rho^{\frac{1}{2}}\right)^\alpha\right]=\widetilde{D}_\alpha(\rho\Vert\sigma),\quad\alpha\in(1,\infty),
		\end{equation}
		as required.
		
		If $\supp(\rho)\nsubseteq\supp(\sigma)$, then $(\rho^{\frac{1}{2}})_{1,1}$ is non-zero. In this case, we use the fact that
		\begin{equation}
			(\sigma+\varepsilon\mathbbm{1})^{\frac{1-\alpha}{\alpha}}=\begin{pmatrix}(\sigma_{0,0}+\varepsilon\mathbbm{1})^{\frac{1-\alpha}{\alpha}}&0 \\ 0&\varepsilon^{\frac{1-\alpha}{\alpha}}\mathbbm{1}\end{pmatrix}\geq\begin{pmatrix}0&0\\0&\varepsilon^{\frac{1-\alpha}{\alpha}}\mathbbm{1}\end{pmatrix}
		\end{equation}
		to conclude that
		\begin{align}
			\rho^{\frac{1}{2}}(\sigma+\varepsilon\mathbbm{1})^{\frac{1-\alpha}{\alpha}}\rho^{\frac{1}{2}} &\geq \rho^{\frac{1}{2}}\begin{pmatrix}0&0\\0&\varepsilon^{\frac{1-\alpha}{\alpha}}\mathbbm{1}\end{pmatrix}\rho^{\frac{1}{2}} \\
			&= \varepsilon^{\frac{1-\alpha}{\alpha}}\begin{pmatrix} (\rho^{\frac{1}{2}})_{0,1}(\rho^{\frac{1}{2}})_{0,1}^\dagger & (\rho^{\frac{1}{2}})_{0,1}(\rho^{\frac{1}{2}})_{1,1} \\ (\rho^{\frac{1}{2}})_{1,1}(\rho^{\frac{1}{2}})_{0,1}^\dagger & (\rho^{\frac{1}{2}})_{1,1}(\rho^{\frac{1}{2}})_{1,1}\end{pmatrix}.
		\end{align}
		Now, since $\alpha\in (1,\infty)$, we have that $\lim_{\varepsilon\to 0^+}\varepsilon^{\frac{1-\alpha}{\alpha}}=+\infty$; therefore, by continuity arguments similar to those given above, we conclude that
		\begin{equation}
			\lim_{\varepsilon\to 0^+}\frac{1}{\alpha-1}\log_2\Tr\!\left[\left(\rho^{\frac{1}{2}}(\sigma+\varepsilon\mathbbm{1})^{\frac{1-\alpha}{\alpha}}\rho^{\frac{1}{2}}\right)^\alpha\right]= +\infty ,
		\end{equation}
		for the case $\alpha\in (1,\infty)$ and $\supp(\rho)\nsubseteq\supp(\sigma)$. This implies that 
		\begin{equation}
			\lim_{\varepsilon\to 0^+}\frac{1}{\alpha-1}\log_2\Tr\!\left[\left(\rho^{\frac{1}{2}}(\sigma+\varepsilon\mathbbm{1})^{\frac{1-\alpha}{\alpha}}\rho^{\frac{1}{2}}\right)^\alpha\right]=\widetilde{D}_{\alpha}(\rho\Vert\sigma) ,
		\end{equation}
		for the case $\alpha\in (1,\infty)$ and $\supp(\rho)\nsubseteq\supp(\sigma)$, as required.
	\end{Proof}

	The Petz--R\'{e}nyi and sandwiched R\'{e}nyi relative entropies are two ways of defining a quantum generalization of the classical R\'{e}nyi relative entropy in \eqref{eq-renyi_rel_ent_classical}. Indeed, if $\rho$ and $\sigma$ are both classical, commuting states (i.e., both are diagonal in the same basis), then both $D_\alpha(\rho\Vert\sigma)$ and $\widetilde{D}_\alpha(\rho\Vert\sigma)$ reduce to the classical R\'{e}nyi relative entropy in \eqref{eq-renyi_rel_ent_classical}. In general, there are often many (in fact, typically infinitely many) ways to generalize classical quantities to the quantum (i.e., non-commutative) case such that we recover the original classical quantity in the special case of commuting operators. What distinguishes one generalization from another is the role that they play in characterizing operational tasks in quantum information theory, which is a theme explored throughout this book.  
	
	
	
	
	We now establish the important fact that the quantum relative entropy is a special case of the sandwiched R\'{e}nyi relative entropy in the limit $\alpha\to 1$. The proof proceeds very similarly to the proof of the same property for the Petz--R\'{e}nyi relative entropy. 
	
	\begin{proposition}{prop-sand_ren_ent_lim}
		Let $\rho$ be a state and $\sigma$ a positive semi-definite operator. Then, in the limit $\alpha\to 1$, the sandwiched R\'{e}nyi relative entropy converges to the quantum relative entropy:
		\begin{equation}
			\lim_{\alpha\to 1}\widetilde{D}_\alpha(\rho\Vert\sigma)=D(\rho\Vert\sigma).
			\label{eq:QEI:sandwiched-Renyi-to-Umegaki}
		\end{equation}
	\end{proposition}
	
	\begin{Proof}
		Let us first consider the case $\alpha\in (1,\infty)$. If $\supp(\rho)\nsubseteq\supp(\sigma)$, then $\widetilde{D}_\alpha(\rho\Vert\sigma)=+\infty$ for all $\alpha\in (1,\infty)$, so that $\lim_{\alpha\to 1^+}\widetilde{D}_\alpha(\rho\Vert\sigma)=+\infty = D(\rho\|\sigma)$. If $\supp(\rho)\subseteq\supp(\sigma)$, then $\widetilde{D}_\alpha(\rho\Vert\sigma)$ is finite; using \eqref{eq-sand_rel_ent_Schatten_2}, we write
		\begin{equation}
			\widetilde{D}_\alpha(\rho\Vert\sigma)=\frac{1}{\alpha-1}\log_2 \widetilde{Q}_\alpha(\rho\Vert\sigma)=\frac{1}{\alpha-1}\log_2\Tr\!\left[\left(\rho^{\frac{1}{2}}\sigma^{\frac{1-\alpha}{\alpha}}\rho^{\frac{1}{2}}\right)^\alpha\right].
		\end{equation}
		Let us define the function
		\begin{equation}
			\widetilde{Q}_{\alpha,\beta}(\rho\Vert\sigma)\coloneqq \Tr\!\left[\left(\rho^{\frac{1}{2}}\sigma^{\frac{1-\alpha}{\alpha}}\rho^{\frac{1}{2}}\right)^\beta\right],
		\end{equation}
		so that $\widetilde{Q}_\alpha(\rho\Vert\sigma)=\widetilde{Q}_{\alpha,\alpha}(\rho\Vert\sigma)$. By noting that $\supp(\rho)\subseteq\supp(\sigma)$ implies $\widetilde{Q}_1(\rho\Vert\sigma)=\Tr[\rho\Pi_\sigma]=1$ (where $\Pi_\sigma$ is the projection onto the support of $\sigma$), and since $\log_2 1=0$, we can write $\widetilde{D}_\alpha(\rho\Vert\sigma)$ as
		\begin{equation}\label{eq-sand_rel_ent_lim_pf2}
			\widetilde{D}_\alpha(\rho\Vert\sigma)=\frac{\log_2\widetilde{Q}_\alpha(\rho\Vert\sigma)-\log_2\widetilde{Q}_1(\rho\Vert\sigma)}{\alpha-1},
		\end{equation}
		so that
		\begin{align}
			\lim_{\alpha\to 1}\widetilde{D}_\alpha(\rho\Vert\sigma)&=\left.\frac{\D}{\D\alpha}\log_2\widetilde{Q}_\alpha(\rho\Vert\sigma)\right|_{\alpha=1}\\
			&=\frac{1}{\ln(2)}\frac{\left.\frac{\D}{\D\alpha}\widetilde{Q}_\alpha(\rho\Vert\sigma)\right|_{\alpha=1}}{\widetilde{Q}_1(\rho\Vert\sigma)}\\
			&=\frac{1}{\ln(2)}\left.\frac{\D}{\D\alpha}\widetilde{Q}_\alpha(\rho\Vert\sigma)\right|_{\alpha=1},
		\end{align}
		where the first equality follows from the definition of the derivative and the second equality from the derivative of the natural logarithm, along with the chain rule. Using the function $\widetilde{Q}_{\alpha,\beta}$ and the chain rule, we write
		\begin{equation}
			\left.\frac{\D}{\D\alpha}\widetilde{Q}_\alpha(\rho\Vert\sigma)\right|_{\alpha=1}=\left.\frac{\D}{\D\alpha}\widetilde{Q}_{\alpha,1}(\rho\Vert\sigma)\right|_{\alpha=1}+\left.\frac{\D}{\D\beta}\widetilde{Q}_{1,\beta}(\rho\Vert\sigma)\right|_{\beta=1}.
		\end{equation}
		Then,
		\begin{equation}
			\frac{\D}{\D\alpha}\widetilde{Q}_{\alpha,1}(\rho\Vert\sigma)=\frac{\D}{\D\alpha}\Tr\!\left[\rho\sigma^{\frac{1-\alpha}{\alpha}}\right]=-\frac{1}{\alpha^2}\Tr\!\left[\rho\sigma^{\frac{1-\alpha}{\alpha}}\ln\sigma\right],
		\end{equation}
		where the logarithm is calculated on the support. Hence,
		\begin{equation}
			\left.\frac{\D}{\D\alpha}\widetilde{Q}_{\alpha,1}(\rho\Vert\sigma)\right|_{\alpha=1}=-\Tr[\rho\Pi_\sigma\ln\sigma]=-\Tr[\rho\ln\sigma].
		\end{equation}
        Similarly, 
		\begin{equation}
			\frac{\D}{\D\beta}\widetilde{Q}_{1,\beta}(\rho\Vert\sigma)=\frac{\D}{\D\beta}\Tr\!\left[\left(\rho^{\frac{1}{2}}\Pi_\sigma\rho^{\frac{1}{2}}\right)^\beta\right]=\frac{\D}{\D\beta}\Tr[\rho^\beta]=\Tr[\rho^\beta\ln\rho],
		\end{equation}
		where we used the fact that $\supp(\rho)\subseteq\supp(\sigma)$ in order to conclude that $\rho^{\frac{1}{2}}\Pi_\sigma\rho^{\frac{1}{2}}=\rho$. Therefore,
		\begin{equation}
			\left.\frac{\D}{\D\beta}\widetilde{Q}_{1,\beta}(\rho\Vert\sigma)\right|_{\beta=1}=\Tr[\rho\ln\rho].
		\end{equation}
		So we find that
		\begin{equation}\label{eq-sand_rel_ent_lim_pf}
			\lim_{\alpha\to 1}\widetilde{D}_\alpha(\rho\Vert\sigma)=\frac{1}{\ln(2)}\left.\frac{\D}{\D\alpha}\widetilde{Q}_\alpha(\rho\Vert\sigma)\right|_{\alpha=1}=\Tr[\rho\log_2\rho]-\Tr[\rho\log_2\sigma]
		\end{equation}
		when $\supp(\rho)\subseteq\supp(\sigma)$. Therefore, for $\alpha\in (1,\infty)$,
		\begin{equation}
			\begin{aligned}
			&\lim_{\alpha\to 1^+}\widetilde{D}_\alpha(\rho\Vert\sigma)\\
			&\quad =\left\{\begin{array}{l l} \Tr[\rho\log_2\rho]-\Tr[\rho\log_2\sigma] & \text{if }\supp(\rho)\subseteq\supp(\sigma),\\ +\infty & \text{otherwise}\end{array}\right.\\
			&\quad =D(\rho\Vert\sigma).
			\end{aligned}
		\end{equation}
		
		Let us now consider the case $\alpha\in (0,1)$. If $\supp(\rho)\subseteq\supp(\sigma)$, then since the limit in \eqref{eq-sand_rel_ent_lim_pf} holds from both sides, we find that
		\begin{equation}
		\lim_{\alpha\to 1^-}\widetilde{D}_\alpha(\rho\Vert\sigma)=\Tr[\rho\log_2\rho]-\Tr[\rho\log_2\sigma].
		\end{equation}
		If $\supp(\rho)\nsubseteq\supp(\sigma)$ (and $\Tr[\rho\sigma]\neq 0$), then observe that we can write $\widetilde{D}_\alpha$ as
		\begin{equation}
			\widetilde{D}_\alpha(\rho\Vert\sigma)=\frac{\log_2\widetilde{Q}_\alpha(\rho\Vert\sigma)-\log_2 \widetilde{Q}_1(\rho\Vert\sigma)}{\alpha-1}+\frac{\log_2\widetilde{Q}_1(\rho\Vert\sigma)}{\alpha-1},
		\end{equation}
		so that
		\begin{equation}\label{eq-sand_rel_ent_lim_pf3}
			\lim_{\alpha\to 1^-}\widetilde{D}_\alpha(\rho\Vert\sigma)=\left.\frac{\D}{\D\alpha}\log_2\widetilde{Q}_\alpha(\rho\Vert\sigma)\right|_{\alpha=1}+\lim_{\alpha\to 1^-}\frac{\log_2\Tr[\rho\Pi_\sigma]}{\alpha-1},
		\end{equation}
		where we have used $\widetilde{Q}_1(\rho\Vert\sigma)=\Tr[\rho\Pi_\sigma]$. Now, since $\supp(\rho)\nsubseteq\supp(\sigma)$ and $\Tr[\rho\sigma]\neq 0$, we have that  $0<\Tr[\rho\Pi_\sigma]< 1$, which means that $\log_2\Tr[\rho\Pi_\sigma]<0$. Since $\lim_{\alpha\to 1^-}\frac{1}{\alpha-1}=-\infty$, we find that the second term in \eqref{eq-sand_rel_ent_lim_pf3} is equal to $+\infty$, which means that $\lim_{\alpha\to 1^-}\widetilde{D}_\alpha(\rho\Vert\sigma)=+\infty$. Therefore,
		\begin{equation}
			\begin{aligned}
			&\lim_{\alpha\to 1^-}\widetilde{D}_\alpha(\rho\Vert\sigma)\\
			&\quad =\left\{\begin{array}{l l} \Tr[\rho\log_2\rho]-\Tr[\rho\log_2\sigma] & \text{if }\supp(\rho)\subseteq\supp(\sigma),\\ +\infty & \text{otherwise} \end{array}\right.\\
			&\quad =D(\rho\Vert\sigma).
			\end{aligned}
		\end{equation}
		To conclude, we have that $\lim_{\alpha\to 1^+}\widetilde{D}_\alpha(\rho\Vert\sigma)=\lim_{\alpha\to 1^-}\widetilde{D}_\alpha(\rho\Vert\sigma)=D(\rho\Vert\sigma)$, which means that \eqref{eq:QEI:sandwiched-Renyi-to-Umegaki} holds.
	\end{Proof}
	
	In the following proposition, we state several basic properties of the sandwiched R\'{e}nyi relative entropy. The proofs of the first four properties are analogous to those of the same properties of the Petz--R\'{e}nyi relative entropy. The last property in the proposition establishes that the sandwiched R\'{e}nyi relative entropy is always less than or equal to the Petz--R\'{e}nyi relative entropy. 
	
	\begin{proposition*}{Properties of Sandwiched R\'{e}nyi Relative Entropy}{prop-sand_rel_ent_properties}
		For all states $\rho,\rho_1,\rho_2$ and positive semi-definite operators $\sigma,\sigma_1,\sigma_2$, the sandwiched R\'{e}nyi relative entropy $\widetilde{D}_\alpha$ satisfies the following properties:
		\begin{enumerate}
			\item \textit{Isometric invariance}: For all $\alpha \in (0,1)\cup (1,\infty)$ and for every isometry~$V$,
				\begin{equation}\label{eq-sand_rel_ent_iso_invar}
					\widetilde{D}_\alpha(\rho\Vert\sigma)=\widetilde{D}_\alpha(V\rho V^\dagger\Vert V\sigma V^\dagger).
				\end{equation}
			\item \textit{Monotonicity in $\alpha$}: For all $\alpha\in (0,1)\cup (1,\infty)$, $\widetilde{D}_\alpha$ is monotonically increasing in $\alpha$, i.e., $\alpha \leq \beta$ implies $\widetilde{D}_\alpha(\rho\Vert\sigma) \leq \widetilde{D}_{\beta}(\rho\Vert\sigma)$.
			\item \textit{Additivity}: For all $\alpha\in(0,1)\cup(1,\infty)$,
				\begin{equation}\label{eq-sand_ren_rel_ent_additive}
					\widetilde{D}_\alpha(\rho_1\otimes\rho_2\Vert\sigma_1\otimes\sigma_2)=\widetilde{D}_\alpha(\rho_1\Vert\sigma_1)+\widetilde{D}_\alpha(\rho_2\Vert\sigma_2).
				\end{equation}
				
			\item \textit{Direct-sum property}: Let $p\colon \mathcal{X}\to[0,1]$ be a probability distribution over a finite alphabet $\mathcal{X}$ with associated $|\mathcal{X}|$-dimensional system $X$, and let $q\colon \mathcal{X}\to [0,\infty)$ be a positive function on $\mathcal{X}$. Let $\{\rho_A^x\}_{x\in\mathcal{X}}$ be a set of states on a system $A$, and let $\{\sigma_A^x\}_{x\in\mathcal{X}}$ be a set of positive semi-definite operators on $A$. Then,
				\begin{equation}\label{eq-sand_quasi_rel_ent_direct_sum}
					\widetilde{Q}_\alpha(\rho_{XA}\Vert \sigma_{XA})
					=\sum_{x\in\mathcal{X}}p(x)^{\alpha}q(x)^{1-\alpha} \widetilde{Q}_\alpha(\rho_A^x\Vert\sigma_A^x).
				\end{equation}
				where 
				\begin{align}
				\rho_{XA} & \coloneqq \sum_{x\in\mathcal{X}}p(x)\ket{x}\!\bra{x}_X\otimes\rho_A^x ,\\
				\sigma_{XA} & \coloneqq \sum_{x\in\mathcal{X}}q(x)\ket{x}\!\bra{x}_X\otimes\sigma_A^x.
				\end{align}

			\item If $\rho_1\leq\gamma\rho_2$ for some $\gamma\geq 1$, then
				\begin{equation}\label{eq-sand_ren_ent_ineq_first_arg}
					\widetilde{D}_{\alpha}(\rho_1\Vert\sigma)\leq\frac{\alpha}{\alpha-1}\log_2\gamma+\widetilde{D}_{\alpha}(\rho_2\Vert\sigma),\quad\alpha>1.
				\end{equation}
				
			\item For all $\alpha\in (0,1)\cup (1,\infty)$, the sandwiched R\'{e}nyi relative entropy $\widetilde{D}_\alpha$ is always less than or equal to the Petz--R\'{e}nyi relative entropy $D_\alpha$, i.e., 
				\begin{equation}
				\label{eq-petz_vs_sandwiched}
					\widetilde{D}_\alpha(\rho\Vert\sigma)\leq D_\alpha(\rho\Vert\sigma).
				\end{equation}
				Furthermore, for $\alpha\in(0,1)$, we have
				\begin{equation}
				\label{eq-petz_vs_sandwiched_2}
					\alpha D_{\alpha}(\rho\Vert\sigma) + (1-\alpha)(-\log_2\Tr[\sigma]) \leq \widetilde{D}_{\alpha}(\rho\Vert\sigma).
				\end{equation}
		\end{enumerate}
	
	\end{proposition*}
	
	\begin{Proof}
		\hfill\begin{enumerate}
			\item \textit{Proof of isometric invariance}: Let us start by writing $\widetilde{D}_\alpha(\rho\Vert\sigma)$ using the function $\norm{\cdot}_{\alpha}$ as in \eqref{eq-sand_rel_ent_Schatten_2}:
				\begin{equation}
					\widetilde{D}_\alpha(\rho\Vert\sigma)=\lim_{\varepsilon\to 0}\frac{\alpha}{\alpha-1}\log_2\norm{\rho^{\frac{1}{2}}(\sigma+\varepsilon\mathbbm{1})^{\frac{1-\alpha}{\alpha}}\rho^{\frac{1}{2}}}_{\alpha},
				\end{equation}
				where we have also made use of the fact that for positive semi-definite operators, $\widetilde{D}_\alpha(\rho\Vert\sigma)$ can be defined as in \eqref{eq-sand_rel_ent_lim}. Now,
				\begin{equation}
					\begin{aligned}
					&\widetilde{D}_\alpha(V\rho V^\dagger\Vert V\sigma V^\dagger)\\
					&\quad =\lim_{\varepsilon\to 0}\frac{\alpha}{\alpha-1}\log_2\norm{(V\rho V^\dagger)^{\frac{1}{2}}(V\sigma V^\dagger+\varepsilon\mathbbm{1})^{\frac{1-\alpha}{\alpha}}(V\rho V^\dagger)^{\frac{1}{2}}}_\alpha.
					\end{aligned}
				\end{equation}
				Since $(V\rho V^\dagger)^{\frac{1}{2}}=V\rho^{\frac{1}{2}}V^\dagger$, we find that
				\begin{align}
					&\norm{(V\rho V^\dagger)^{\frac{1}{2}}(V\sigma V^\dagger+\varepsilon\mathbbm{1})^{\frac{1-\alpha}{\alpha}}(V\rho V^\dagger)^{\frac{1}{2}}}_\alpha\nonumber\\
					&\quad =\norm{V\rho^{\frac{1}{2}}V^\dagger (V\sigma V^\dagger+\varepsilon\mathbbm{1})^{\frac{1-\alpha}{\alpha}}V\rho^{\frac{1}{2}}V^\dagger}_\alpha\\
					&\quad =\norm{\rho^{\frac{1}{2}}V^\dagger(V\sigma V^\dagger+\varepsilon\mathbbm{1})^{\frac{1-\alpha}{\alpha}}V\rho^{\frac{1}{2}}}_\alpha, \label{eq-sand_rel_ent_iso_invar_pf}
				\end{align}
				where the last equality follows from the isometric invariance of the function $\norm{\cdot}_{\alpha}$. Now, let $\Pi\coloneqq VV^\dagger$ be the projection onto the image of $V$, and let $\hat{\Pi}\coloneqq\mathbbm{1}-\Pi$. Then, we write
				\begin{equation}
					V\sigma V^\dagger+\varepsilon\mathbbm{1}=V\sigma V^\dagger+\varepsilon\Pi+\varepsilon\hat{\Pi}=V(\sigma+\varepsilon\mathbbm{1})V^\dagger+\varepsilon\hat{\Pi}.
				\end{equation}
				Since $V(\sigma+\varepsilon\mathbbm{1})V^\dagger$ and $\varepsilon\hat{\Pi}$ are supported on orthogonal subspaces, we obtain
				\begin{equation}
					(V\sigma V^\dagger+\varepsilon\mathbbm{1})^{\frac{1-\alpha}{\alpha}}=V(\sigma+\varepsilon\mathbbm{1})^{\frac{1-\alpha}{\alpha}}V^\dagger+\varepsilon^{\frac{1-\alpha}{\alpha}}\hat{\Pi}.
				\end{equation}
				Continuing from \eqref{eq-sand_rel_ent_iso_invar_pf}, we thus find that
				\begin{align}
					&\norm{\rho^{\frac{1}{2}}V^\dagger(V\sigma V^\dagger+\varepsilon\mathbbm{1})^{\frac{1-\alpha}{\alpha}}V\rho^{\frac{1}{2}}}_\alpha\nonumber\\
					&=\norm{\rho^{\frac{1}{2}}V^\dagger \left(V(\sigma+\varepsilon\mathbbm{1})^{\frac{1-\alpha}{\alpha}}V^\dagger+\varepsilon^{\frac{1-\alpha}{\alpha}}\hat{\Pi}\right)V\rho^{\frac{1}{2}}}_\alpha\\
					&=\norm{\rho^{\frac{1}{2}}V^\dagger V(\sigma+\varepsilon\mathbbm{1})^{\frac{1-\alpha}{\alpha}}V^\dagger V\rho^{\frac{1}{2}}+\varepsilon^{\frac{1-\alpha}{\alpha}}\rho^{\frac{1}{2}}V^\dagger\hat{\Pi}V\rho^{\frac{1}{2}}}_\alpha\\
					&=\norm{\rho^{\frac{1}{2}}(\sigma+\varepsilon\mathbbm{1})^{\frac{1-\alpha}{\alpha}}\rho^{\frac{1}{2}}}_\alpha,
				\end{align}
				where the last equality follows from the fact that $V^\dagger\hat{\Pi}V=V^\dagger V-V^\dagger VV^\dagger V=\mathbbm{1}-\mathbbm{1}=0$. Therefore,
				\begin{equation}
					\begin{aligned}
					\widetilde{D}_\alpha(V\rho V^\dagger\Vert V\sigma V^\dagger)&=\lim_{\varepsilon\to 0}\frac{\alpha}{\alpha-1}\log_2\norm{\rho^{\frac{1}{2}}(\sigma+\varepsilon\mathbbm{1})^{\frac{1-\alpha}{\alpha}}\rho^{\frac{1}{2}}}_{\alpha}\\
					&=\widetilde{D}_\alpha(\rho\Vert\sigma),
					\end{aligned}
				\end{equation}
				as required.
	
			\item \textit{Proof of monotonicity in $\alpha$}: We make use of the function $\widetilde{D}_\alpha(\rho\Vert\sigma;\tau)$ defined in \eqref{eq-sand_rel_ent_opt}, which we can write as
				\begin{equation}
					\widetilde{D}_\alpha(\rho\Vert\sigma;\tau)=-\frac{1}{\gamma}\log_2\bra{\varphi^\rho}X^{\gamma}\ket{\varphi^\rho}=-\frac{1}{\gamma}\frac{\ln\bra{\varphi^\rho}X^{\gamma}\ket{\varphi^\rho}}{\ln(2)},
				\end{equation}
				where $X=\tau^{-1}\otimes\sigma^{\t}$, $\gamma\coloneqq\frac{1-\alpha}{\alpha}$ and $\ket{\varphi^\rho}=(\rho^{\frac{1}{2}}\otimes\mathbbm{1})\ket{\Gamma}$ is a purification of~$\rho$. We prove monotonicity of this quantity by taking its derivative with respect to $\alpha$ and showing that it is non-negative. Since $\frac{\D\gamma}{\D\alpha}=-\frac{1}{\alpha^2}$, we can express the derivative with respect to $\alpha$ in terms of the derivative with respect to $\gamma$ using $\frac{\D}{\D\alpha}=\frac{\D}{\D\gamma}\frac{\D\gamma}{\D\alpha}=-\frac{1}{\alpha^2}\frac{\D}{\D\gamma}$. Therefore,
				\begin{equation}
					\frac{\D}{\D\alpha}\widetilde{D}_\alpha(\rho\Vert\sigma;\tau)=-\frac{1}{\alpha^2}\frac{\D}{\D\gamma}\left(-\frac{1}{\gamma}\frac{\ln\bra{\varphi^\rho}X^{\gamma}\ket{\varphi^\rho}}{\ln(2)}\right)=\frac{1}{\alpha^2}\frac{\D f}{\D\gamma},
				\end{equation}
				where
				\begin{equation}
				f(\gamma)\coloneqq\frac{1}{\gamma}\frac{\ln\bra{\varphi^\rho}X^{\gamma}\ket{\varphi^\rho}}{\ln(2)}.
				\end{equation}
				Now,
				\begin{align}
					\frac{\D f}{\D\gamma}&=\frac{1}{\ln(2)}\left(-\frac{1}{\gamma^2}\ln\bra{\varphi^\rho}X^{\gamma}\ket{\varphi^\rho}+\frac{1}{\gamma}\frac{\bra{\varphi^\rho}X^{\gamma}\ln X\ket{\varphi^\rho}}{\bra{\varphi^\rho}X^{\gamma}\ket{\varphi^\rho}}\right)\\
					&=\frac{-\bra{\varphi^\rho}X^{\gamma}\ket{\varphi^\rho}\log_2\bra{\varphi^\rho}X^{\gamma}\ket{\varphi^\rho}+\bra{\varphi^\rho}X^{\gamma}\log_2 X^{\gamma}\ket{\varphi^\rho}}{\gamma^2\bra{\varphi^\rho}X^\gamma\ket{\varphi^\rho}}.
				\end{align}
				Now, let $g(x)\coloneqq x\log_2 x$. Then, we can write
				\begin{equation}
					\frac{\D f}{\D\gamma}=\frac{\bra{\varphi^\rho}g(X^{\gamma})\ket{\varphi^\rho}-g(\bra{\varphi^\rho}X^{\gamma}\ket{\varphi^\rho})}{\gamma^2\bra{\varphi^\rho}X^{\gamma}\ket{\varphi^\rho}}.
				\end{equation}
				Since $g$ is operator convex,  the operator Jensen inequality in \eqref{eq-op_Jensen_alt} implies that
				\begin{equation}
					\bra{\varphi^\rho}g(X^{\gamma})\ket{\varphi^\rho}\geq g(\bra{\varphi^\rho}X^{\gamma}\ket{\varphi^\rho}),
				\end{equation}
				which implies that $\frac{\D f}{\D\gamma}\geq 0$. Therefore, $\widetilde{D}_\alpha(\rho\Vert\sigma;\tau)$ is monotonically increasing in $\alpha$ for all $\rho,\sigma,\tau$. By \eqref{eq-sand_rel_ent_var}, we conclude that $\widetilde{D}_\alpha(\rho\Vert\sigma)$ is monotonically increasing in $\alpha$, as required.

			\item \textit{Proof of additivity}: When all quantities are finite, we have that
				\begin{equation}
					\begin{aligned}
					&\widetilde{D}_\alpha(\rho_1\otimes\rho_2\Vert\sigma_1\otimes\sigma_2)\\
					&\quad =\frac{1}{\alpha-1}\log_2\Tr\!\left[\left((\sigma_1\otimes\sigma_2)^{\frac{1-\alpha}{2\alpha}}(\rho_1\otimes\rho_2)(\sigma_1\otimes\sigma_2)^{\frac{1-\alpha}{2\alpha}}\right)^\alpha\right].
					\end{aligned}
				\end{equation}
				Using the fact that $(X\otimes Y)^\beta=X^\beta\otimes Y^\beta$ for all positive semi-definite operators $X,Y$ and all $\beta\in\mathbb{R}$, we obtain
				\begin{align}
					&\widetilde{Q}_\alpha(\rho_1\otimes\rho_2\Vert\sigma_1\otimes\sigma_2)\nonumber\\
					&\quad =\Tr\!\left[\left((\sigma_1\otimes\sigma_2)^{\frac{1-\alpha}{2\alpha}}(\rho_1\otimes\rho_2)(\sigma_1\otimes\sigma_2)^{\frac{1-\alpha}{2\alpha}}\right)^\alpha\right]\\
					&\quad =\Tr\!\left[\left(\left(\sigma_1^{\frac{1-\alpha}{2\alpha}}\otimes\sigma_2^{\frac{1-\alpha}{2\alpha}}\right)(\rho_1\otimes\rho_2) \left(\sigma_1^{\frac{1-\alpha}{2\alpha}}\otimes\sigma_2^{\frac{1-\alpha}{2\alpha}}\right)\right)^\alpha\right]\\
					&\quad =\Tr\!\left[\left(\sigma_1^{\frac{1-\alpha}{2\alpha}}\rho_1\sigma_1^{\frac{1-\alpha}{2\alpha}}\otimes\sigma_2^{\frac{1-\alpha}{2\alpha}}\rho_2\sigma_2^{\frac{1-\alpha}{2\alpha}}\right)^\alpha\right]\\
					&\quad =\Tr\!\left[\left(\sigma_1^{\frac{1-\alpha}{2\alpha}}\rho_1\sigma_1^{\frac{1-\alpha}{2\alpha}}\right)^{\alpha}\otimes\left(\sigma_2^{\frac{1-\alpha}{2\alpha}}\rho_2\sigma_2^{\frac{1-\alpha}{2\alpha}}\right)^\alpha\right]\\
					&\quad =\Tr\!\left[\left(\sigma_1^{\frac{1-\alpha}{2\alpha}}\rho_1\sigma_1^{\frac{1-\alpha}{2\alpha}}\right)^\alpha\right] \cdot \Tr\!\left[\left(\sigma_2^{\frac{1-\alpha}{2\alpha}}\rho_2\sigma_2^{\frac{1-\alpha}{2\alpha}}\right)^\alpha\right]\\
					&\quad =\widetilde{Q}_\alpha(\rho_1\Vert\sigma_1)\cdot \widetilde{Q}_\alpha(\rho_2\Vert\sigma_2).
				\end{align}
				Applying $\frac{1}{\alpha-1}\log_2$ and definitions, additivity follows.

                If any of the relative entropies in \eqref{eq-sand_ren_rel_ent_additive} is infinite, then the equality still holds. For example, consider $\alpha\in (1,\infty)$. The only case in which any of the relative entropies in \eqref{eq-sand_ren_rel_ent_additive} can diverge is if $\supp(\rho_1)\not\subseteq \supp(\sigma_1)$ or $\supp(\rho_2)\not\subseteq \supp(\sigma_2)$. In fact, if both inclusions were to hold, we would also have 
                \begin{align}
                \supp(\rho_1\otimes \rho_2) & = \supp(\rho_1) \otimes \supp(\rho_2) \\
                & \subseteq \supp(\sigma_1) \otimes \supp(\sigma_2) \\
                & = \supp(\sigma_1\otimes \sigma_2),
                \end{align}
                implying that all three relative entropies are finite.
                
                If either $\supp(\rho_1)\not\subseteq \supp(\sigma_1)$ or $\supp(\rho_2)\not\subseteq \supp(\sigma_2)$, then it is also possible to show that $\supp(\rho_1\otimes \rho_2)\not\subseteq \supp(\sigma_1\otimes \sigma_2)$. In fact, if this inclusion relation were to hold, then there would exist some $\lambda>0$ with the property that $\rho_1\otimes \rho_2 \leq \lambda\, \sigma_1\otimes \sigma_2$. Tracing away one of the two subsystems, we would conclude that $\rho_i\leq \lambda\,\sigma_i$ for $i=1,2$, implying that $\supp(\rho_i) \subseteq \supp(\sigma_i)$, a contradiction. 
                
                Putting all together, this shows that the left-hand side of \eqref{eq-sand_ren_rel_ent_additive} is equal to $+\infty$ if and only if the right-hand side is also equal to $+\infty$. The case where $\alpha\in (0,1)$ is entirely analogous.

			\item \textit{Proof of the direct-sum property}: Define the classical--quantum state and operator, respectively, as
				\begin{equation}
					\rho_{XA}\coloneqq \sum_{x\in\mathcal{X}}p(x)\ket{x}\!\bra{x}_X\otimes\rho_A^x,\quad\sigma_{XA}\coloneqq \sum_{x\in\mathcal{X}}q(x)\ket{x}\!\bra{x}_X\otimes\sigma_A^x.
				\end{equation}
				Then, since
				\begin{equation}
					\sigma_{XA}^{\frac{1-\alpha}{2\alpha}}=\sum_{x\in\mathcal{X}}\ket{x}\!\bra{x}_X\otimes (q(x)\sigma_A^x)^{\frac{1-\alpha}{2\alpha}},
				\end{equation}
				we find that
				\begin{align}
					\sigma_{XA}^{\frac{1-\alpha}{2\alpha}}\rho_{XA}\sigma_{XA}^{\frac{1-\alpha}{2\alpha}}&=\sum_{x\in\mathcal{X}}\ket{x}\!\bra{x}_X\otimes (q(x)\sigma_A^x)^{\frac{1-\alpha}{2\alpha}}(p(x)\rho_A^x)(q(x)\sigma_A^x)^{\frac{1-\alpha}{2\alpha}}\\
					&=\sum_{x\in\mathcal{X}}p(x)q(x)^{\frac{1-\alpha}{\alpha}}\ket{x}\!\bra{x}_X\otimes (\sigma_A^x)^{\frac{1-\alpha}{2\alpha}}\rho_A^x(\sigma_A^x)^{\frac{1-\alpha}{2\alpha}},
				\end{align}
				which means that
				\begin{equation}
					\begin{aligned}
					&\left(\sigma_{XA}^{\frac{1-\alpha}{2\alpha}}\rho_{XA}\sigma_{XA}^{\frac{1-\alpha}{2\alpha}}\right)^\alpha\\
					&\quad =\sum_{x\in\mathcal{X}}p(x)^\alpha q(x)^{1-\alpha}\ket{x}\!\bra{x}_X\otimes\left((\sigma_A^x)^{\frac{1-\alpha}{2\alpha}}\rho_A^x(\sigma_A^x)^{\frac{1-\alpha}{2\alpha}}\right)^\alpha.
					\end{aligned}
				\end{equation}
				Taking the trace on both sides of this equation, and using the definition of $\widetilde{Q}_\alpha$, we conclude that
				\begin{equation}
					\widetilde{Q}_\alpha(\rho_{XA}\Vert\sigma_{XA})=\sum_{x\in\mathcal{X}}p(x)^\alpha q(x)^{1-\alpha}\widetilde{Q}_\alpha(\rho_A^x\Vert\sigma_A^x),
				\end{equation}
				as required. 

			\item From the assumption that $\rho_1\leq\gamma\rho_2$, we have that
				\begin{equation}
					\sigma^{\frac{1-\alpha}{2\alpha}}\rho_1\sigma^{\frac{1-\alpha}{2\alpha}}\leq \gamma \sigma^{\frac{1-\alpha}{2\alpha}}\rho_2\sigma^{\frac{1-\alpha}{2\alpha}}.
				\end{equation}
				Then, using \eqref{eq-power_func_monotone_PSD}, we obtain
				\begin{equation}
					\Tr\!\left[\left(\sigma^{\frac{1-\alpha}{2\alpha}}\rho_1\sigma^{\frac{1-\alpha}{2\alpha}}\right)^{\alpha}\right]\leq \gamma^\alpha \Tr\!\left[\left(\sigma^{\frac{1-\alpha}{2\alpha}}\rho_2\sigma^{\frac{1-\alpha}{2\alpha}}\right)^{\alpha}\right].
				\end{equation}
				The result follows after applying the logarithm and dividing by $\alpha-1$ on both sides of this inequality.

			\item This follows from the Araki--Lieb--Thirring inequalities, which we state here without proof (see the Bibliographic Notes in Section~\ref{sec:QEI:bib-notes}): for positive semi-definite operators $A$ and $B$ acting on a finite-dimens\-ional Hilbert space, and for $q\geq 0$, the following inequalities hold
				\begin{enumerate}
					\item $\Tr\!\left[\left(B^{\frac{1}{2}}AB^{\frac{1}{2}}\right)^{rq}\right] \geq \Tr\!\left[\left(B^{\frac{r}{2}}A^r B^{\frac{r}{2}}\right)^q\right]$ for all $r \in [0,1]$.
					\item $\Tr\!\left[\left(B^{\frac{1}{2}}A B^{\frac{1}{2}}\right)^{rq}\right]\leq \Tr\!\left[\left(B^{\frac{r}{2}}A^rB^{\frac{r}{2}}\right)^q\right]$ for all $r\geq 1$.
				\end{enumerate}
				For $\alpha\in (0,1)$, we make use of the first of these inequalities. In particular, we set $q=1$, $r=\alpha$, $A=\rho$ and $B=\sigma^{\frac{1-\alpha}{\alpha}}$. Then, letting $\gamma\coloneqq\frac{1-\alpha}{2\alpha}$, we obtain
				\begin{equation}
	\Tr\!\left[\left(\sigma^\gamma\rho\sigma^\gamma\right)^{\alpha}\right]\geq\Tr[\sigma^{\alpha\gamma}\rho^\alpha\sigma^{\alpha\gamma}]=\Tr\!\left[\sigma^{\frac{1-\alpha}{2}}\rho^\alpha\sigma^{\frac{1-\alpha}{2}}\right]=\Tr[\rho^{\alpha}\sigma^{1-\alpha}],
				\end{equation}
				where the last equality holds by cyclicity of the trace. Since the logarithm function is monotonically increasing, this inequality implies that
				\begin{equation}
					\log_2\Tr\!\left[\left(\sigma^{\frac{1-\alpha}{2\alpha}}\rho\sigma^{\frac{1-\alpha}{2\alpha}}\right)^\alpha\right]\geq\log_2\Tr\!\left[\rho^\alpha\sigma^{1-\alpha}\right].
				\end{equation}
				Finally, since $\alpha-1<0$ for all $\alpha\in (0,1)$, we conclude that
				\begin{equation}
					\frac{1}{\alpha-1}\log_2\Tr\!\left[\left(\sigma^{\frac{1-\alpha}{2\alpha}}\rho\sigma^{\frac{1-\alpha}{2\alpha}}\right)^\alpha\right]\leq\frac{1}{\alpha-1}\log_2\Tr[\rho^\alpha\sigma^{1-\alpha}].
				\end{equation}
				That is, $\widetilde{D}_\alpha(\rho\Vert\sigma)\leq D_\alpha(\rho\Vert\sigma)$, as required.
				
				For $\alpha\in (1,\infty)$, we make use of the second Araki--Lieb--Thirring inequality above. As before, we let $q=1$, $r=\alpha$, $A=\rho$, and $B^{\frac{1}{2}}=\sigma^\gamma$. We find that
				\begin{equation}
					\Tr\!\left[\left(\sigma^{\frac{1-\alpha}{2\alpha}}\rho\sigma^{\frac{1-\alpha}{2\alpha}}\right)^\alpha\right]\leq\Tr\!\left[\sigma^{\frac{1-\alpha}{2}}\rho^\alpha\sigma^{\frac{1-\alpha}{2}}\right]=\Tr[\rho^\alpha\sigma^{1-\alpha}].
				\end{equation}
				Then, since the logarithm function is a monotonically increasing function and $\alpha-1>0$ for all $\alpha\in (1,\infty)$, we conclude that
				\begin{equation}
					\frac{1}{\alpha-1}\log_2\Tr\!\left[\left(\sigma^{\frac{1-\alpha}{2\alpha}}\rho\sigma^{\frac{1-\alpha}{2\alpha}}\right)^\alpha\right]\leq\frac{1}{\alpha-1}\log_2\Tr[\rho^\alpha\sigma^{1-\alpha}],
				\end{equation}
				i.e., $\widetilde{D}_\alpha(\rho\Vert\sigma)\leq D_\alpha(\rho\Vert\sigma)$, as required.
				
				For the inequality in \eqref{eq-petz_vs_sandwiched_2}, with $\rho$ a state and $\sigma$ a positive semi-definite operator, we use the following ``reverse'' Araki--Lieb--Thirring inequality, which we state here without proof (see the Bibliographic Notes in Section~\ref{sec:QEI:bib-notes}):
				\begin{equation}
					\Tr\!\left[\left(B^{\frac{1}{2}}AB^{\frac{1}{2}}\right)^{rq}\right]\leq \left(\Tr\!\left[\left(B^{\frac{r}{2}}A^rB^{\frac{r}{2}}\right)^q\right]\right)^r\norm{A^{\frac{1-r}{2}}}_{a}^{2rq}\norm{B^{\frac{1-r}{2}}}_b^{2rq}.
				\end{equation}
				This inequality holds for all positive semi-definite operators $A$ and $B$, as well as for $q>0$, $r\in(0,1]$, $a,b\in(0,\infty]$, and for $\frac{1}{2rq}=\frac{1}{2q}+\frac{1}{a}+\frac{1}{b}$. Taking $q=1$, $r=\alpha$, $A=\rho$, $B=\sigma^{\frac{1-\alpha}{\alpha}}$, $a=\frac{2}{1-\alpha}$, and $b=\frac{2\alpha}{(1-\alpha)^2}$, we obtain
				\begin{multline}
					\Tr\!\left[\left(\sigma^{\frac{1-\alpha}{2\alpha}}\rho\sigma^{\frac{1-\alpha}{2\alpha}}\right)^{\alpha}\right]\\
					\leq \left(\Tr\!\left[\left(\sigma^{\frac{1-\alpha}{2}}\rho\sigma^{\frac{1-\alpha}{2}}\right)\right]\right)^{\alpha}\norm{\rho^{\frac{1-\alpha}{2}}}_{\frac{2}{1-\alpha}}^{2\alpha}\norm{\sigma^{\frac{(1-\alpha)^2}{2\alpha}}}_{\frac{2\alpha}{(1-\alpha)^2}}^{2\alpha}.
				\end{multline}
				Now, because $\rho$ is a state,
				\begin{align}
					\norm{\rho^{\frac{1-\alpha}{2}}}_{\frac{2}{1-\alpha}}^{2\alpha}&=\left(\Tr\!\left[\left|\rho^{\frac{1-\alpha}{2}}\right|^{\frac{2}{1-\alpha}}\right]\right)^{\alpha(1-\alpha)}\\
					&=\left(\Tr\!\left[\left(\rho^{\frac{1-\alpha}{2}}\right)^{\frac{2}{1-\alpha}}\right]\right)^{\alpha(1-\alpha)}\\
					&=\left(\Tr[\rho]\right)^{\alpha(1-\alpha)}\\
					&=1.
				\end{align}
				For $\sigma$, we obtain
				\begin{equation}
					\norm{\sigma^{\frac{(1-\alpha)^2}{2\alpha}}}_{\frac{2\alpha}{(1-\alpha)^2}}^{2\alpha}=(\Tr[\sigma])^{(1-\alpha)^2}.
				\end{equation}
				Therefore,
				\begin{equation}
					\Tr\!\left[\left(\sigma^{\frac{1-\alpha}{2\alpha}}\rho\sigma^{\frac{1-\alpha}{2\alpha}}\right)^{\alpha}\right]\leq \left(\Tr[\rho^{\alpha}\sigma^{1-\alpha}]\right)^{\alpha}\cdot (\Tr[\sigma])^{(1-\alpha)^2}.
				\end{equation}
				Taking the logarithm of both sides and multiplying by $\frac{1}{\alpha-1}$, which is negative for $\alpha\in(0,1)$, we obtain the inequality in \eqref{eq-petz_vs_sandwiched_2}. \qedhere

		\end{enumerate}
	\end{Proof}
	
	The monotonicity in $\alpha$ of the sandwiched R\'enyi relative entropy establishes an inequality relating the quantum relative entropy and the fidelity of quantum states $\rho$ and $\sigma$, by picking $\alpha=1$ and $\alpha=1/2$, respectively, and applying Proposition~\ref{prop-sand_ren_ent_lim}:
	\begin{equation}
	D(\rho\|\sigma) \geq -\log_2 F(\rho,\sigma).
	\label{eq:QEI:pinsker-fidelity}
	\end{equation}
	We can modify the lower bound a bit to establish an inequality relating the quantum relative entropy and the trace distance:
	
	\begin{corollary*}{Quantum Pinsker Inequality}{cor:QEI:pinsker}
	Let $\rho$ and $\sigma$ be quantum states. Then the following inequality holds
	\begin{align}
	D(\rho\|\sigma) \geq \frac{1}{4 \ln 2} \norm{\rho - \sigma}_1^2.
	\end{align}
	\end{corollary*}
	\begin{remark}
	The constant prefactor can be improved from $\frac{1}{4 \ln 2}$ to $\frac{1}{2 \ln 2}$, but we do not give a proof here (please consult the Bibliographic Notes in Section~\ref{sec:QEI:bib-notes}).
	\end{remark}
	\begin{Proof}
	We can rewrite \eqref{eq:QEI:pinsker-fidelity} as follows:
	\begin{align}
	D(\rho\|\sigma) & \geq -\frac{1}{\ln 2} \ln F(\rho,\sigma)\\
	& = - \frac{1}{\ln 2} \ln [1 - (1-F(\rho,\sigma))]\\
	& \geq \frac{1}{\ln 2} (1-F(\rho,\sigma))\\
	& \geq \frac{1}{4 \ln 2} \norm{\rho - \sigma}_1^2.
	\end{align}
	The second inequality follows from $-\ln(1-x) \geq x$, which holds for $x\in[0,1]$. The final inequality follows from Theorem~\ref{thm-Fuchs_van_de_graaf}.
	\end{Proof}
	
	Like the quantum relative entropy and the Petz--R\'{e}nyi relative entropy, the sandwiched R\'{e}nyi relative entropy is faithful, meaning that for all states $\rho,\sigma$ and all $\alpha\in(0,1)\cup(1,\infty)$,
	\begin{equation}
	\widetilde{D}_{\alpha}(\rho\Vert\sigma)=0
	\qquad \Longleftrightarrow \qquad \rho=\sigma .
	\end{equation}
	We prove this in Proposition~\ref{prop-petz_sand_ren_ent_faithful} below.
	
	We now prove the data-processing inequality for the sandwiched R\'{e}nyi relative entropy $\widetilde{D}_\alpha$ for $\alpha\in[\sfrac{1}{2},1)\cup(1,\infty)$. This, along with Proposition \ref{prop-sand_ren_ent_lim}, gives us a different way (apart from using data-processing inequality for the Petz--R\'{e}nyi relative entropy) to prove the data-processing inequality for the quantum relative entropy.

	\begin{theorem*}{Data-Processing Inequality for Sandwiched R\'{e}nyi Relative Entropy}{thm-sand_renyi_monotone}
		Let $\rho$ be a state, $\sigma$ a positive semi-definite operator, and $\mathcal{N}$ a quantum channel. Then, for all $\alpha\in\left[\sfrac{1}{2},1\right)\cup(1,\infty)$,
		\begin{equation}
			\widetilde{D}_\alpha(\rho\Vert\sigma)\geq \widetilde{D}_\alpha(\mathcal{N}(\rho)\Vert\mathcal{N}(\sigma)).
		\end{equation}
	\end{theorem*}
	
	\begin{Proof}
		This proof follows steps very similar to those in the proof of the data-processing inequality for the Petz--R\'{e}nyi relative entropy (Theorem \ref{thm-petz_rel_ent_monotone}), with the key difference being that in this case we make use of the fact that the sandwiched R\'{e}nyi relative entropy can be written as the optimization in \eqref{eq-sand_rel_ent_var}.
		
		From Stinespring's theorem (Theorem \ref{thm-q_channels}), we know that the action of a channel $\mathcal{N}$ on a linear operator $X$ can be written as
		\begin{equation}
			\mathcal{N}(X)=\Tr_E[VXV^\dagger],
		\end{equation}
		for some $V$, where $V$ is an isometry and $E$ is an auxiliary system with dimension $d_E\geq\rank(\Gamma^{\mathcal{N}})$. As stated in \eqref{eq-sand_rel_ent_iso_invar}, $\widetilde{D}_\alpha$ is isometrically invariant. Therefore, it suffices to prove the data-processing inequality for $\widetilde{D}_\alpha$ under partial trace; i.e., it suffices to show that for every state $\rho_{AB}$, every positive semi-definite operator $\sigma_{AB}$, and all $\alpha\in\left[\sfrac{1}{2},1\right)\cup(1,\infty)$:
		\begin{equation}\label{eq-sand_rel_ent_monotone_proof_1}
			\widetilde{D}_\alpha(\rho_{AB}\Vert\sigma_{AB})\geq \widetilde{D}_\alpha(\rho_A\Vert\sigma_A).
		\end{equation}
		We now proceed to prove this inequality. We prove it for $\rho_{AB}$, and hence $\rho_A$, invertible, as well as for $\sigma_{AB}$ and $\sigma_A$ invertible. The result follows in the general case of $\rho_{AB}$ and/or $\rho_{A}$ non-invertible, as well as $\sigma_{AB}$ and/or $\sigma_A$ non-invertible, by applying the result to the invertible operators 
		$(1-\delta)\rho_{AB}+\delta\pi_{AB}$ and $\sigma_{AB}+\varepsilon\mathbbm{1}_{AB}$, with $\delta,\varepsilon>0$, and taking the limits $\varepsilon\to 0^+$ and $\delta\to 0^+$, since
		\begin{align}
			\widetilde{D}_\alpha(\rho_{AB}\Vert\sigma_{AB}) & =\lim_{\varepsilon\to 0^+}\lim_{\delta\to 0^+}\widetilde{D}_\alpha((1-\delta)\rho_{AB}+\delta\pi_{AB}\Vert\sigma_{AB}+\varepsilon\mathbbm{1}_{AB}),\\
			\widetilde{D}_\alpha(\rho_{A}\Vert\sigma_{A}) & =\lim_{\varepsilon\to 0^+}\lim_{\delta\to 0^+}\widetilde{D}_\alpha((1-\delta)\rho_{A}+\delta\pi_{A}\Vert\sigma_{A}+d_B\varepsilon\mathbbm{1}_{A}),			
		\end{align}
		which can be verified in a similar manner to the proof of \eqref{eq-sand_rel_ent_lim} in Proposition~\ref{prop-sand_rel_ent_lim}.
		
		Let us start by defining the quantity $\widetilde{Q}_{\alpha}(\rho\Vert\sigma;\tau)$ as
		\begin{equation}
			\widetilde{Q}_{\alpha}(\rho\Vert\sigma;\tau)\coloneqq\bra{\varphi^{\rho}}(\tau^{-1}\otimes\sigma^{\t})^{\frac{1-\alpha}{\alpha}}\ket{\varphi^{\rho}},
		\end{equation}
		where $\tau$ is a positive definite state and
		\begin{equation}
			\ket{\varphi^{\rho}}\coloneqq (\rho^{\frac{1}{2}}\otimes\mathbbm{1})\ket{\Gamma}
		\end{equation}
		is a purification of $\rho$. We note that
		\begin{equation}
			\widetilde{Q}_\alpha(\rho\Vert\sigma;\tau)=\Tr\!\left[\rho^{\frac{1}{2}}\sigma^{\frac{1-\alpha}{\alpha}}\rho^{\frac{1}{2}}\tau^{\frac{\alpha-1}{\alpha}}\right]
		\end{equation}
		so that
		\begin{equation}\label{eq-sand_rel_ent_monotone_proof_4}
			\widetilde{D}_\alpha(\rho\Vert\sigma;\tau)=\frac{\alpha}{\alpha-1}\log_2\widetilde{Q}_\alpha(\rho\Vert\sigma;\tau),
		\end{equation}
		where we recall the quantity $\widetilde{D}_{\alpha}(\rho\Vert\sigma;\tau)$ defined in \eqref{eq-sand_rel_ent_opt}. Now, to prove \eqref{eq-sand_rel_ent_monotone_proof_1}, we show that for every positive definite state $\omega_A$, there exists a positive definite state $\tau_{AB}$ such that
		\begin{equation}\label{eq-sand_rel_ent_monotone_proof_2}
			\begin{aligned}
			\widetilde{Q}_\alpha(\rho_{AB}\Vert\sigma_{AB};\tau_{AB})&\geq \widetilde{Q}_{\alpha}(\rho_A\Vert\sigma_A;\omega_A),\quad \text{ for } \alpha\in (1,\infty),\\
			\widetilde{Q}_\alpha(\rho_{AB}\Vert\sigma_{AB};\tau_{AB})&\leq \widetilde{Q}_\alpha(\rho_A\Vert\sigma_A;\omega_A),\quad
			\text{ for } \alpha\in\left[\sfrac{1}{2},1 \right).
			\end{aligned}
		\end{equation}
		With these two inequalities, along with \eqref{eq-sand_rel_ent_monotone_proof_4} and \eqref{eq-sand_rel_ent_var}, the result follows.
		
		Consider that
		\begin{align}
			\widetilde{Q}_\alpha(\rho_{AB}\Vert\sigma_{AB};\tau_{AB})&=\bra{\varphi^{\rho_{AB}}}f(\tau_{AB}^{-1}\otimes\sigma_{\hat{A}\hat{B}}^{\t})\ket{\varphi^{\rho_{AB}}},\\
			\widetilde{Q}_\alpha(\rho_A\Vert\sigma_A;\omega_A)&=\bra{\varphi^{\rho_A}}f(\omega_A^{-1}\otimes\sigma_{\hat{A}}^{\t})\ket{\varphi^{\rho_A}},
		\end{align}
		where we have set
		\begin{equation}\label{eq-sand_rel_ent_monotone_proof_3}
			f(x)\coloneqq x^{\frac{1-\alpha}{\alpha}}
		\end{equation}
		and
		\begin{align}
			\ket{\varphi^{\rho_{AB}}}&=(\rho_{AB}^{\frac{1}{2}}\otimes\mathbbm{1}_{\hat{A}\hat{B}})\ket{\Gamma}_{AB\hat{A}\hat{B}}=(\rho_{AB}^{\frac{1}{2}}\otimes\mathbbm{1}_{\hat{A}\hat{B}})\ket{\Gamma}_{A\hat{A}}\ket{\Gamma}_{B\hat{B}},\\
			\ket{\varphi^{\rho_A}}&=(\rho_A^{\frac{1}{2}}\otimes\mathbbm{1}_{\hat{A}})\ket{\Gamma}_{A\hat{A}}.
		\end{align}
		Now, let us use the same isometry $V_{A\hat{A}\to AB\hat{A}\hat{B}}$ from \eqref{eq-petz_rel_ent_monotone_pf_3} that we used in the proof of data-processing inequality for the Petz--R\'{e}nyi relative entropy; that is, let
		\begin{equation}
			V_{A\hat{A}\to AB\hat{A}\hat{B}}\coloneqq \rho_{AB}^{\frac{1}{2}}(\rho_A^{-\frac{1}{2}}\otimes\mathbbm{1}_{\hat{A}})\ket{\Gamma}_{B\hat{B}}.
		\end{equation}
		Recall that
		\begin{align}
			V_{A\hat{A}\to AB\hat{A}\hat{B}}\ket{\varphi^{\rho_A}}_{A\hat{A}}&=\rho_{AB}^{\frac{1}{2}}(\rho_A^{-\frac{1}{2}}\otimes\mathbbm{1}_{\hat{A}})(\rho_A^{\frac{1}{2}}\otimes\mathbbm{1}_{\hat{A}})\ket{\Gamma}_{A\hat{A}}\ket{\Gamma}_{B\hat{B}}\\
			&=(\rho_{AB}^{\frac{1}{2}}\otimes\mathbbm{1}_{\hat{A}\hat{B}})\ket{\Gamma}_{A\hat{A}}\ket{\Gamma}_{B\hat{B}}\\
			&=\ket{\varphi^{\rho_{AB}}}.
		\end{align}
		We thus obtain, for all $\tau_{AB}$,
		\begin{equation}\label{eq-sand_rel_ent_monotone_proof_5}
			\begin{aligned}
			\widetilde{Q}_\alpha(\rho_{AB}\Vert\sigma_{AB};\tau_{AB})&=\bra{\varphi^{\rho_A}}V^\dagger f(\tau_{AB}^{-1}\otimes\sigma_{\hat{A}\hat{B}}^{\t})V\ket{\varphi^{\rho_A}}\\
			&\geq \bra{\varphi^{\rho_A}}f(V^\dagger(\tau_{AB}^{-1}\otimes\sigma_{\hat{A}\hat{B}}^{\t})V)\ket{\varphi^{\rho_A}}
			\end{aligned}
		\end{equation}
		for $\alpha\in (1,\infty)$ and
		\begin{equation}\label{eq-sand_rel_ent_monotone_proof_6}
			\begin{aligned}
			\widetilde{Q}_\alpha(\rho_{AB}\Vert\sigma_{AB};\tau_{AB})&=\bra{\varphi^{\rho_A}}V^\dagger f(\tau_{AB}^{-1}\otimes\sigma_{\hat{A}\hat{B}}^{\t})V\ket{\varphi^{\rho_A}}\\
			&\leq \bra{\varphi^{\rho_A}}f(V^\dagger(\tau_{AB}^{-1}\otimes\sigma_{\hat{A}\hat{B}}^{\t})V)\ket{\varphi^{\rho_A}}
			\end{aligned}
		\end{equation}
		for $\alpha\in\left[\sfrac{1}{2},1\right)$, where to obtain the last inequality in each case we used the operator Jensen inequality (Theorem~\ref{thm-Jensen}), which is applicable since for $\alpha\in (1,\infty)$ the function $f$ in \eqref{eq-sand_rel_ent_monotone_proof_3} is operator convex and for $\alpha\in\left[\sfrac{1}{2},1\right)$ it is operator concave. 
		
		Now, recall that to conclude \eqref{eq-sand_rel_ent_monotone_proof_1}, we should perform an optimization over invertible states $\tau_{AB}$ as per the definition in \eqref{eq-sand_rel_ent_var} of $\widetilde{D}_\alpha(\rho_{AB}\Vert\sigma_{AB};\tau_{AB})$ in order to obtain $\widetilde{D}_{\alpha}(\rho_{AB}\Vert\sigma_{AB})$. Since we only require a lower bound on $\widetilde{D}_\alpha(\rho_{AB}\Vert\sigma_{AB})$, we can obtain the lower bound in \eqref{eq-sand_rel_ent_monotone_proof_1} on $\widetilde{D}_{\alpha}(\rho_{AB}\Vert\sigma_{AB})$ by simply picking a particular state $\tau_{AB}$ in the optimization in \eqref{eq-sand_rel_ent_var}. Let us therefore take 
		\begin{equation}\label{eq-sand_rel_ent_monotone_proof_7}
			\tau_{AB}=\xi_{AB}(\omega_A)\coloneqq \rho_{AB}^{\frac{1}{2}}(\rho_A^{-\frac{1}{2}}\omega_A\rho_A^{-\frac{1}{2}}\otimes\mathbbm{1}_B)\rho_{AB}^{\frac{1}{2}},
		\end{equation}
		where $\omega_A$ is an arbitrary invertible state. Note that this choice of $\tau_{AB}$ is indeed a state because it is the result of applying the Petz recovery channel $\mathcal{P}_{\rho_{AB},\Tr_B}$ defined in \eqref{eq-Petz_channel_partrace} to $\omega_A$. It is also invertible; in particular,
		\begin{equation}
			\tau_{AB}^{-1} = [\xi_{AB}(\omega_A)]^{-1}=\rho_{AB}^{-\frac{1}{2}}(\rho_A^{\frac{1}{2}}\omega_A^{-1}\rho_A^{\frac{1}{2}}\otimes\mathbbm{1}_B)\rho_{AB}^{-\frac{1}{2}}.
		\end{equation}
		With the choice in \eqref{eq-sand_rel_ent_monotone_proof_7} for $\tau_{AB}$, we find that
		\begin{align}
			&V^\dagger(\tau_{AB}^{-1}\otimes\sigma_{\hat{A}\hat{B}}^{\t})V\nonumber\\
			&\quad=\bra{\Gamma}_{B\hat{B}}(\rho_A^{-\frac{1}{2}}\otimes\mathbbm{1}_{\hat{A}})\rho_{AB}^{\frac{1}{2}}(\tau_{AB}^{-1}\otimes\sigma_{\hat{A}\hat{B}}^{\t})\rho_{AB}^{\frac{1}{2}}(\rho_A^{-\frac{1}{2}}\otimes\mathbbm{1}_{\hat{A}})\ket{\Gamma}_{B\hat{B}}\\
			&\quad=\bra{\Gamma}_{B\hat{B}}\left(\rho_A^{-\frac{1}{2}}\rho_{AB}^{\frac{1}{2}}\rho_{AB}^{-\frac{1}{2}}(\rho_A^{\frac{1}{2}}\omega_A^{-1}\rho_A^{\frac{1}{2}}\otimes\mathbbm{1}_B)\rho_{AB}^{-\frac{1}{2}}\rho_{AB}^{\frac{1}{2}}\rho_A^{-\frac{1}{2}}\otimes\sigma_{\hat{A}\hat{B}}^{\t}\right)\ket{\Gamma}_{B\hat{B}}\\
			&\quad=\bra{\Gamma}_{B\hat{B}}\omega_A^{-1}\otimes \mathbbm{1}_B \otimes\sigma_{\hat{A}\hat{B}}^{\t}\ket{\Gamma}_{B\hat{B}}\\
			&\quad=\omega_A^{-1}\otimes\bra{\Gamma}_{B\hat{B}}\sigma_{\hat{A}\hat{B}}^{\t}\ket{\Gamma}_{B\hat{B}}\\
			&\quad=\omega_A^{-1}\otimes\sigma_{\hat{A}}^{\t},
		\end{align}
		where we have used the fact that $\bra{\Gamma}_{B\hat{B}}\sigma_{\hat{A}\hat{B}}^{\t}\ket{\Gamma}_{B\hat{B}}=\Tr_{\hat{B}}[\sigma_{\hat{A}\hat{B}}^{\t}]=\sigma_{\hat{A}}^{\t}$, the last equality due to the fact that the transpose is taken on a product basis for $\mathcal{H}_{\hat{A}}\otimes\mathcal{H}_{\hat{B}}$.
		
		Therefore, for $\alpha\in(1,\infty)$, taking the logarithm on both sides of \eqref{eq-sand_rel_ent_monotone_proof_5} and using the state in \eqref{eq-sand_rel_ent_monotone_proof_7}, we find that
		\begin{align}
			\log_2\widetilde{Q}_\alpha(\rho_{AB}\Vert\sigma_{AB};\xi_{AB}(\omega_A))&\geq\log_2\bra{\varphi^{\rho_A}}f(\omega_A^{-1}\otimes\sigma_{\hat{A}}^{\t})\ket{\varphi^{\rho_A}}\\
			&=\log_2\widetilde{Q}_\alpha(\rho_A\Vert\sigma_A;\omega_A).
		\end{align}
		Multiplying both sides of this inequality by $\frac{\alpha}{\alpha-1}$, we obtain
		\begin{align}
			\widetilde{D}_\alpha(\rho_{AB}\Vert\sigma_{AB})&=\sup_{\substack{\tau_{AB}>0\\\Tr[\tau_{AB}]=1}}\widetilde{D}_\alpha(\rho_{AB}\Vert\sigma_{AB};\tau_{AB})\\
			&\geq \widetilde{D}_\alpha(\rho_{AB}\Vert\sigma_{AB};\xi_{AB}(\omega_A))\\
			&\geq \widetilde{D}_\alpha(\rho_A\Vert\sigma_A;\omega_A)
		\end{align}
		for all invertible states $\omega_A$. Finally, taking the supremum over the set $\{\omega_A:\omega_A>0,~\Tr[\omega_A]=1\}$, we conclude that
		\begin{equation}
			\widetilde{D}_{\alpha}(\rho_{AB}\Vert\sigma_{AB})\geq\widetilde{D}_\alpha(\rho_A\Vert\sigma_A),\quad\text{ for }\alpha\in (1,\infty).
		\end{equation}
		
		For $\alpha\in\left[\sfrac{1}{2},1\right)$, taking the logarithm on both sides of \eqref{eq-sand_rel_ent_monotone_proof_6} and using the state in \eqref{eq-sand_rel_ent_monotone_proof_7}, we conclude that
		\begin{align}
			\log_2\widetilde{Q}_\alpha(\rho_{AB}\Vert\sigma_{AB};\xi_{AB}(\omega_A))&\leq\log_2\bra{\varphi^{\rho_A}}f(\omega_A^{-1}\otimes\sigma_{\hat{A}}^{\t})\ket{\varphi^{\rho_A}}\\
			&=\log_2\widetilde{Q}_\alpha(\rho_A\Vert\sigma_A;\omega_A).
		\end{align}
		Multiplying both sides of this inequality by $\frac{\alpha}{\alpha-1}$, which is negative in this case, so that
		\begin{equation}
			\frac{\alpha}{\alpha-1}\log_2\widetilde{Q}_\alpha(\rho_{AB}\Vert\sigma_{AB};\xi_{AB}(\omega_A))\geq\frac{\alpha}{\alpha-1}\log_2\widetilde{Q}_\alpha(\rho_A\Vert\sigma_A;\omega_A),
		\end{equation}
		we obtain
		\begin{align}
			\widetilde{D}_\alpha(\rho_{AB}\Vert\sigma_{AB})&=\sup_{\substack{\tau_{AB}>0,\\\Tr[\tau_{AB}]=1}}\widetilde{D}_\alpha(\rho_{AB}\Vert\sigma_{AB};\tau_{AB})\\
			&\geq\widetilde{D}_\alpha(\rho_{AB}\Vert\sigma_{AB};\xi_{AB}(\omega_A))\\
			&\geq\widetilde{D}_\alpha(\rho_A\Vert\sigma_A;\omega_A)
		\end{align}
		for all invertible states $\omega_A$. Finally, taking the supremum over the set $\{\omega_A:\omega_A>0,~\Tr[\omega_A]=1\}$, we conclude that
		\begin{equation}
			\widetilde{D}_\alpha(\rho_{AB}\Vert\sigma_{AB})\geq\widetilde{D}_\alpha(\rho_A\Vert\sigma_A),\quad\text{ for }\alpha\in\left[\sfrac{1}{2},1\right).
		\end{equation}
		Having established the data-processing inequality for $\widetilde{D}_\alpha$ under the partial trace channel for $\alpha\in\left[\sfrac{1}{2},1\right)\cup(1,\infty)$, we conclude that
		\begin{equation}
			\widetilde{D}_\alpha(\rho\Vert\sigma)\geq\widetilde{D}_\alpha(\mathcal{N}(\rho)\Vert\mathcal{N}(\sigma)),
		\end{equation}
		for $\alpha\in\left[\sfrac{1}{2},1\right)\cup(1,\infty)$, all states $\rho$, positive semi-definite operators $\sigma$, and all channels $\mathcal{N}$.
	\end{Proof}
	
	
	By taking the limit $\alpha\to 1$ in the statement of the data-processing inequality for $\widetilde{D}_\alpha$, along with Proposition \ref{prop-sand_ren_ent_lim}, we immediately obtain the data-processing inequality for the quantum relative entropy.
	
	\begin{corollary*}{Data-Processing Inequality for Quantum Relative Entropy}{cor-rel_ent_monotone2}
		Let $\rho$ be a state, $\sigma$ a positive semi-definite operator, and $\mathcal{N}$ a quantum channel. Then,
		\begin{equation}
			D(\rho\Vert\sigma)\geq D(\mathcal{N}(\rho)\Vert\mathcal{N}(\sigma)).
		\end{equation}
	\end{corollary*}
	
	With the data-processing inequality for the sandwiched R\'{e}nyi relative entropy in hand, it is now straightforward to prove some of the following additional properties.
	
	\begin{proposition*}{Additional Properties of Sandwiched R\'{e}nyi Relative Entropy}{prop-sand_rel_ent_add_properties}
		The sandwiched R\'{e}nyi relative entropy $\widetilde{D}_\alpha$ satisfies the following properties for every state $\rho$ and positive semi-definite operator $\sigma$ for $\alpha\in\left[\sfrac{1}{2},1\right)\cup (1,\infty)$.
		\begin{enumerate}
			\item If $\Tr(\sigma)\leq \Tr(\rho)=1$, then $\widetilde{D}_\alpha(\rho\Vert\sigma)\geq 0$.
			\item \textit{Faithfulness}: If $\Tr[\sigma]\leq 1$, we have that $\widetilde{D}_\alpha(\rho\Vert\sigma)=0$ if and only if $\rho=\sigma$.
			\item For every positive semi-definite operator $\sigma'$ such that $\sigma'\geq \sigma$, we have $\widetilde{D}_\alpha(\rho\Vert\sigma) \geq \widetilde{D}_\alpha(\rho\Vert\sigma')$. In particular, if $\rho\leq\sigma$, then $\widetilde{D}_\alpha(\rho\Vert\sigma)\leq 0$.
		\end{enumerate}
	\end{proposition*}
	
	\begin{Proof}
		\hfill\begin{enumerate}
			\item By the data-processing inequality for $\widetilde{D}_\alpha$ with respect to the trace channel $\Tr$, and letting $x=\Tr(\rho)=1$ and $y=\Tr(\sigma)$, we find that
					\begin{align}
						\widetilde{D}_\alpha(\rho\Vert\sigma)\geq\widetilde{D}_\alpha(x\Vert y)&=\frac{1}{\alpha-1}\log_2\Tr[(y^{\frac{1-\alpha}{2\alpha}}x y^{\frac{1-\alpha}{2\alpha}})^\alpha]\\
						&=\frac{1}{\alpha-1}\log_2( y^{1-\alpha})\\
						&=\frac{1-\alpha}{\alpha-1}\log_2 y\\
						&=-\log_2 y\\
						&\geq 0,
					\end{align}
					where the last line follows from the assumption that $y=\Tr(\sigma)\leq 1$.
					
			\item \textit{Proof of faithfulness}: If $\rho=\sigma$, then the following equalities hold for all $\alpha\in\left[1/2,1\right)\cup (1,\infty)$:
					\begin{align}
						\widetilde{D}_\alpha(\rho\Vert\rho)&=\frac{1}{\alpha-1}\log_2\Tr\!\left[\left(\rho^{\frac{1-\alpha}{2\alpha}}\rho\rho^{\frac{1-\alpha}{2\alpha}}\right)^\alpha\right]\label{eq-sand_rel_ent_faithful_pf1}\\
						&=\frac{1}{\alpha-1}\log_2\Tr\!\left[\rho^{\frac{1-\alpha}{2}}\rho^\alpha\rho^{\frac{1-\alpha}{2}}\right]\label{eq-sand_rel_ent_faithful_pf2}\\
						&=\frac{1}{\alpha-1}\log_2\Tr[\rho^{1-\alpha}\rho^{\alpha}]\label{eq-sand_rel_ent_faithful_pf3}\\
						&=\frac{1}{\alpha-1}\log_2\Tr(\rho)\label{eq-sand_rel_ent_faithful_pf4}\\
						&=0.\label{eq-sand_rel_ent_faithful_pf5}
					\end{align}
					Next, suppose that $\alpha\in\left[1/2,1\right)\cup (1,\infty)$ and  $\widetilde{D}_\alpha(\rho\Vert\sigma)=0$. From the above, we conclude that $\widetilde{D}_\alpha(\Tr(\rho)\Vert\Tr(\sigma))=-\log_2 y\geq 0$. From the fact that $\log_2 y=0$ if and only if $y=1$, we conclude that $\widetilde{D}_\alpha(\rho\Vert\sigma)=0$ implies $\Tr(\sigma)=\Tr(\rho)=1$, so that $\sigma$ is a density operator. Then, for every measurement channel $\mathcal{M}$, 
					\begin{equation}
						\widetilde{D}_\alpha(\mathcal{M}(\rho)\Vert\mathcal{M}(\sigma))\leq\widetilde{D}_\alpha(\rho\Vert\sigma)=0.
					\end{equation}
					On the other hand, since $\Tr(\sigma)=\Tr(\rho)$,
					\begin{align}
						D(\mathcal{M}(\rho)\Vert\mathcal{M}(\sigma))&\geq\widetilde{D}_\alpha(\Tr(\mathcal{M}(\rho))\Vert\Tr(\mathcal{M}(\sigma)))\\
						&=\widetilde{D}_\alpha(\Tr(\rho)\Vert\Tr(\sigma))\\
						&=0,
					\end{align}
					which means that $\widetilde{D}_\alpha(\mathcal{M}(\rho)\Vert\mathcal{M}(\sigma))=0$ for all measurement channels $\mathcal{M}$. Now, recall that $\mathcal{M}(\rho)$ and $\mathcal{M}(\sigma)$ are effectively probability distributions determined by the measurement. Since the classical R\'{e}nyi relative entropy is equal to zero if and only if its two arguments are equal, we can conclude that $\mathcal{M}(\rho)=\mathcal{M}(\sigma)$. Since this is true for every measurement channel, we conclude from Theorem~\ref{thm-trace_dist_ach_by_meas_channel} and the fact that the trace norm is a norm that $\rho=\sigma$.
					So we have that $\widetilde{D}_\alpha(\rho\Vert\sigma)=0$ if and only if $\rho=\sigma$, as required.

			\item Consider the state $\hat{\rho}\coloneqq\ket{0}\!\bra{0}\otimes\rho$ and the operator $\hat{\sigma}\coloneqq\ket{0}\!\bra{0}\otimes\sigma+\ket{1}\!\bra{1}\otimes (\sigma'-\sigma)$, which is positive semi-definite because $\sigma'\geq \sigma$ by assumption.
			Then
					\begin{equation}
						\hat{\sigma}^{\frac{1-\alpha}{2\alpha}}\hat{\rho}\hat{\sigma}^{\frac{1-\alpha}{2\alpha}}=\ket{0}\!\bra{0}\otimes\sigma^{\frac{1-\alpha}{2\alpha}}\rho\sigma^{\frac{1-\alpha}{2\alpha}},
					\end{equation}
					which implies that
					\begin{equation}
						\widetilde{D}_\alpha(\hat{\rho}\Vert\hat{\sigma})=\widetilde{D}_\alpha(\rho\Vert\sigma).
					\end{equation}
					Then, observing that $\Tr_1[\hat{\sigma}]=\sigma'$, and using the data-processing inequality for $\widetilde{D}_\alpha$ with respect to the partial trace channel $\Tr_1$, we conclude that
					\begin{equation}
						\widetilde{D}_\alpha(\rho\Vert\sigma')=\widetilde{D}_\alpha(\Tr_1(\hat{\rho})\Vert\Tr_1(\hat{\sigma}))\leq \widetilde{D}_\alpha(\hat{\rho}\Vert\hat{\sigma})=\widetilde{D}_\alpha(\rho\Vert\sigma),
					\end{equation}
					as required.

                    The second statement follows by substituting $\sigma \to \rho$ and $\sigma' \to \sigma$ and applying faithfulness.
                    \qedhere

		\end{enumerate}
	\end{Proof}
	
	Let us now prove the faithfulness of both the Petz--R\'{e}nyi and sandwiched R\'{e}nyi relative entropies for the full range of parameters for which they are defined.
	
	\begin{proposition*}{Faithfulness of the Petz--R\'{e}nyi and Sandwiched R\'{e}nyi Relative Entropies}{prop-petz_sand_ren_ent_faithful}
		For all $\alpha\in(0,1)\cup(1,\infty)$ and for all states $\rho,\sigma$, the Petz--R\'{e}nyi and sandwiched R\'{e}nyi relative entropies are faithful, meaning that
		\begin{align}
			D_{\alpha}(\rho\Vert\sigma)&=0\text{ if and only if }\rho=\sigma,\\
			\widetilde{D}_{\alpha}(\rho\Vert\sigma)&=0\text{ if and only if }\rho=\sigma.
		\end{align}
	\end{proposition*}
	
	\begin{Proof}
		Note that the equality $\widetilde{D}_{\alpha}(\rho\Vert\rho)=0$ for all $\alpha\in(0,1)\cup(1,\infty)$ is immediate from the definition (see also \eqref{eq-sand_rel_ent_faithful_pf1}--\eqref{eq-sand_rel_ent_faithful_pf5}). The converse statement has already been established in property 2. of Proposition~\ref{prop-sand_rel_ent_add_properties} for $\alpha\in[\sfrac{1}{2},1)\cup(1,\infty)$. Before getting to the range $\alpha\in(0,\sfrac{1}{2})$, let us consider the Petz--R\'{e}nyi relative entropy.
		
		It is immediately clear from the definition that $D_{\alpha}(\rho\Vert\rho)=0$ for all $\alpha\in(0,1)\cup(1,\infty)$. For $\alpha\in(0,1)\cup(1,2]$, the converse follows from the data-processing inequality, which holds for this parameter range as shown in Theorem~\ref{thm-petz_rel_ent_monotone}, as well as from arguments analogous to those in the proof of property 2.\ in Proposition~\ref{prop-sand_rel_ent_add_properties}. For $\alpha\in(2,\infty)$, we use the fact that $D_{\alpha}(\rho\Vert\sigma)\geq\widetilde{D}_{\alpha}(\rho\Vert\sigma)$ for all $\rho,\sigma$, as shown in Proposition~\ref{prop-sand_rel_ent_properties}. In particular, if $D_{\alpha}(\rho\Vert\sigma)=0$, then $\widetilde{D}_{\alpha}(\rho\Vert\sigma)\leq 0$. However, because $\rho$ and $\sigma$ are states, by property 1.\ of Proposition~\ref{prop-sand_rel_ent_add_properties}, we have that $\widetilde{D}_{\alpha}(\rho\Vert\sigma)\geq 0$, which means that $\widetilde{D}_{\alpha}(\rho\Vert\sigma)=0$. Then, by property 2.\ of Proposition~\ref{prop-sand_rel_ent_add_properties}, we immediately get that $\rho=\sigma$.
		
		Finally, suppose that $\widetilde{D}_{\alpha}(\rho\Vert\sigma)=0$, where $\alpha\in(0,\sfrac{1}{2})$. Then, using \eqref{eq-petz_vs_sandwiched_2}, we have that $\alpha D_{\alpha}(\rho\Vert\sigma)\leq 0$. However, because $\rho$ and $\sigma$ are states, we have that $D_\alpha(\rho\Vert\sigma)\geq 0$, due to Proposition~\ref{prop-petz_rel_ent_add_properties}. 
		Therefore, $D_{\alpha}(\rho\Vert\sigma)=0$, which implies that $\rho=\sigma$ by the faithfulness of the Petz--R\'{e}nyi relative entropy, which we just proved.
	\end{Proof}
	
	The data-processing inequality for the sandwiched R\'{e}nyi relative entropy can be written using the sandwiched R\'{e}nyi relative quasi-entropy $\widetilde{Q}_\alpha$ as
	\begin{equation}
		\frac{1}{\alpha-1}\log_2 \widetilde{Q}_\alpha(\rho\Vert\sigma)\geq\frac{1}{\alpha-1}\log_2 \widetilde{Q}_\alpha(\mathcal{N}(\rho)\vert\mathcal{N}(\sigma)).
	\end{equation}
	Then, since $\alpha-1$ is negative for $\alpha\in\left[\sfrac{1}{2},1\right)$, we can use the monotonicity of the function $\log_2$ to obtain
	\begin{align}
		\widetilde{Q}_\alpha(\rho\Vert\sigma)&\geq \widetilde{Q}_\alpha(\mathcal{N}(\rho)\Vert\mathcal{N}(\sigma)),\quad\text{ for } \alpha\in (1,\infty),\label{eq-sand_quasi_monotone_1}\\
		\widetilde{Q}_\alpha(\rho\Vert\sigma)&\leq \widetilde{Q}_\alpha(\mathcal{N}(\rho)\Vert\mathcal{N}(\sigma)),\quad\text{ for }\alpha\in\left[\sfrac{1}{2},1\right).\label{eq-sand_quasi_monotone_2}
	\end{align}		
	Just as with the Petz--R\'{e}nyi relative entropy, we can use this to prove the joint convexity of the sandwiched R\'{e}nyi relative entropy.
	
	\begin{proposition*}{Joint Convexity \& Concavity of Sandwiched R\'{e}nyi Relative Quasi-Entropy}{prop-sand_rel_ent_joint_convex}
		Let $p\colon \mathcal{X}\to[0,1]$ be a probability distribution over a finite alphabet $\mathcal{X}$ with associated $|\mathcal{X}|$-dimensional system $X$, let $\{\rho_A^x\}_{x\in\mathcal{X}}$ be a set of states on a system $A$, and let $\{\sigma_A^x\}_{x\in\mathcal{X}}$ be a set of positive semi-definite operators on $A$. Then, for $\alpha\in(1,\infty)$,
		\begin{equation}\label{eq-sand_ren_q_ent_joint_convex_ageq1}
			\widetilde{Q}_\alpha\!\left(\sum_{x\in\mathcal{X}}p(x)\rho_A^x\Bigg\Vert\sum_{x\in\mathcal{X}}p(x)\sigma_A^x\right)
			\leq \sum_{x\in\mathcal{X}}p(x)\widetilde{Q}_\alpha(\rho_A^x\Vert\sigma_A^x),
		\end{equation}
		and, for $\alpha\in\left[\sfrac{1}{2},1\right)$,
		\begin{equation}
			\widetilde{Q}_\alpha\!\left(\sum_{x\in\mathcal{X}}p(x)\rho_A^x\Bigg\Vert\sum_{x\in\mathcal{X}}p(x)\sigma_A^x\right)
			\geq \sum_{x\in\mathcal{X}}p(x)\widetilde{Q}_\alpha(\rho_A^x\Vert\sigma_A^x).
			\label{eq-sand_ren_q_ent_joint_convex_aleq1}
		\end{equation}
		Consequently, the sandwiched R\'{e}nyi relative entropy $\widetilde{D}_\alpha$ is jointly convex for $\alpha\in \left[\sfrac{1}{2},1\right)$:
		\begin{equation}
			\widetilde{D}_\alpha\!\left(\sum_{x\in\mathcal{X}}p(x)\rho_A^x\Bigg\Vert\sum_{x\in\mathcal{X}}p(x)\sigma_A^x\right)
			 \leq \sum_{x\in\mathcal{X}}p(x)\widetilde{D}_\alpha(\rho_A^x\Vert\sigma_A^x).
		\end{equation}
	\end{proposition*}
	
	\begin{Proof}
		By the direct-sum property of $\widetilde{Q}_\alpha$ and applying  \eqref{eq-sand_quasi_monotone_1}--\eqref{eq-sand_quasi_monotone_2} and Proposition~\ref{prop:QEI:joint-convexity-gen-div},
		we conclude \eqref{eq-sand_ren_q_ent_joint_convex_ageq1}--\eqref{eq-sand_ren_q_ent_joint_convex_aleq1}.

For $\alpha\in\left[\sfrac{1}{2},1\right)$, taking $\log_2$ of both sides and multiplying by $\frac{1}{\alpha-1}$, which is negative, we find that
		\begin{multline}
			\frac{1}{\alpha-1}\log_2 \widetilde{Q}_\alpha\!\left(\sum_{x\in\mathcal{X}}p(x)\rho_A^x\Bigg\Vert\sum_{x\in\mathcal{X}}p(x)\sigma_A^x\right)\\
			 \leq \frac{1}{\alpha-1}\log_2\!\left(\sum_{x\in\mathcal{X}}p(x)\widetilde{Q}_\alpha(\rho_A^x\Vert\sigma_A^x\right).
		\end{multline}
		Then, since $-\log_2$ is a convex function, and using the definition of $\widetilde{D}_\alpha$ in terms of $\widetilde{Q}_\alpha$, we conclude  that
		\begin{align}
			\widetilde{D}_\alpha\!\left(\sum_{x\in\mathcal{X}}p(x)\rho_A^x\Bigg\Vert \sum_{x\in\mathcal{X}}p(x)\sigma_A^x\right)&\leq \sum_{x\in\mathcal{X}}p(x)\frac{1}{\alpha-1}\log_2 \widetilde{Q}_\alpha(\rho_A^x\Vert\sigma_A^x)\\
			&=\sum_{x\in\mathcal{X}}p(x)\widetilde{D}_\alpha(\rho_A^x\Vert\sigma_A^x),
		\end{align}
		as required.
	\end{Proof}
	
	Although the sandwiched R\'{e}nyi relative entropy is not jointly convex for $\alpha\in(1,\infty)$, it is \textit{jointly quasi-convex}, in the sense that
	\begin{equation}\label{eq-sand_rel_ent_q_convex}
		\widetilde{D}_{\alpha}\!\left(\sum_{x\in\mathcal{X}}p(x)\rho_A^x\Bigg\Vert\sum_{x\in\mathcal{X}} p(x)\sigma_A^x\right)\leq \max_{x\in\mathcal{X}}\widetilde{D}_{\alpha}(\rho_A^x\Vert\sigma_A^x),
	\end{equation}
	for every finite alphabet $\mathcal{X}$, probability distribution $p\colon \mathcal{X}\to[0,1]$, set $\{\rho_A^x\}_{x\in\mathcal{X}}$ of states, and set $\{\sigma_A^x\}_{x\in\mathcal{X}}$ of positive semi-definite operators. Indeed, from \eqref{eq-sand_ren_q_ent_joint_convex_ageq1}, we immediately obtain
	\begin{equation}
		\widetilde{Q}_{\alpha}\!\left(\sum_{x\in\mathcal{X}}p(x)\rho_A^x\Bigg\Vert\sum_{x\in\mathcal{X}}p(x)\sigma_A^x\right)\leq\max_{x\in\mathcal{X}}\widetilde{Q}_{\alpha}(\rho_A^x\Vert\sigma_A^x).
	\end{equation}
	Taking the logarithm and multiplying by $\frac{1}{\alpha-1}$ on both sides of this inequality leads to \eqref{eq-sand_rel_ent_q_convex}.

\section{Geometric R\'enyi Relative Entropy}

\label{sec-QEI:geometric-renyi}

In the previous two sections, we considered two examples of generalized divergences, the Petz-- and sandwiched R\'{e}nyi relative entropies. Both of these are quantum generalizations of the classical R\'{e}nyi relative entropy defined in~\eqref{eq-renyi_rel_ent_classical}.

In this section, we consider another generalization of the classical R\'{e}nyi relative entropy, called the geometric R\'{e}nyi relative entropy. Unlike the two previous R\'{e}nyi relative entropies, the geometric R\'{e}nyi relative entropy does not converge to the quantum relative entropy in the $\alpha\to 1$ limit. Instead, it converges to what is called the Belavkin--Staszewski relative entropy, as shown in Section~\ref{sec:QEI:Belavkin--Staszewski}. This latter quantity represents a different quantum generalization of the classical relative entropy in \eqref{eq:classical-rel-ent}. The main use of the geometric R\'{e}nyi and Belavkin--Staszewski relative entropies is in establishing upper bounds on the rates of feedback-assisted quantum communication protocols, the latter of which is the main focus of Part~\ref{part-feedback} of this book.

\begin{definition}
{Geometric R\'{e}nyi Relative Entropy}{def:geometric-renyi-rel-ent}
Let
$\rho$ be a state, $\sigma$ a positive semi-definite operator, and $\alpha
\in(0,1)\cup(1,\infty)$. The \textit{geometric R\'{e}nyi relative quasi-entropy} is
defined as
\begin{equation}
\label{eq:def-geometric-renyi-rel-quasi-ent}
\widehat{Q}_{\alpha}(\rho\Vert\sigma)\coloneqq\lim_{\varepsilon\rightarrow0^{+}
}\operatorname{Tr}\!\left[  \sigma_{\varepsilon}\!\left(  \sigma_{\varepsilon
}^{-\frac{1}{2}}\rho\sigma_{\varepsilon}^{-\frac{1}{2}}\right)  ^{\alpha
}\right]=\lim_{\varepsilon\to 0^+}\Tr[G_{\alpha}(\sigma_{\varepsilon},\rho)]  ,
\end{equation}
where $\sigma_{\varepsilon}\coloneqq\sigma+\varepsilon\mathbbm{1}$ and
\begin{equation}
	G_{\alpha}(\sigma_{\varepsilon},\rho)\coloneqq \sigma_{\varepsilon}^{\frac{1}{2}}\left(\sigma_{\varepsilon}^{-\frac{1}{2}}\rho\sigma_{\varepsilon}^{-\frac{1}{2}}\right)^{\alpha}\sigma_{\varepsilon}^{\frac{1}{2}}
\end{equation}
is the \textit{weighted operator geometric mean} of $\sigma_{\varepsilon}$ and $\rho$. The \textit{geometric R\'{e}nyi relative entropy} is then defined as
\begin{equation}
\widehat{D}_{\alpha}(\rho\Vert\sigma)\coloneqq\frac{1}{\alpha-1}\log_{2}\widehat
{Q}_{\alpha}(\rho\Vert\sigma). \label{eq:def-geometric-renyi-rel-ent}
\end{equation}

\end{definition}

\begin{remark}
In general, the weighted operator geometric mean of two positive definite operators $X$ and $Y$ is defined as
\begin{equation}\label{eq:weighted-op-geo-mean}
	G_{\beta}(X,Y)\coloneqq X^{\frac{1}{2}}\left(X^{-\frac{1}{2}}YX^{-\frac{1}{2}}\right)^{\beta}X^{\frac{1}{2}},
\end{equation}
where $\beta\in\mathbb{R}$ is the weight parameter. We recover the standard operator geometric mean for $\beta=\frac{1}{2}$. 

An important property of the weighted operator geometric mean is that
\begin{equation}\label{eq:geometric-mean-identity}
	G_{\beta}(X,Y)=G_{1-\beta}(Y,X)
\end{equation}
for all positive definite $X,Y$, and all $\beta\in\mathbb{R}$. To see this, observe that
\begin{align}
G_{1-\beta}(Y,X)  &  =Y^{\frac{1}{2}}\left(  Y^{-\frac{1}{2}}XY^{-\frac{1}{2}
}\right)  ^{1-\beta}Y^{\frac{1}{2}}\\
&  =Y^{\frac{1}{2}}\left(  Y^{-\frac{1}{2}}XY^{-\frac{1}{2}}\right)  \left(
Y^{-\frac{1}{2}}XY^{-\frac{1}{2}}\right)  ^{-\beta}Y^{\frac{1}{2}}\\
&  =X^{\frac{1}{2}}X^{\frac{1}{2}}Y^{-\frac{1}{2}}\left(  Y^{-\frac{1}{2}
}X^{\frac{1}{2}}X^{\frac{1}{2}}Y^{-\frac{1}{2}}\right)  ^{-\beta}Y^{\frac
{1}{2}}
\end{align}
Now we apply Lemma~\ref{lem:sing-val-lemma-pseudo-commute}. Specifically, we set $L=X^{\frac{1}{2}}Y^{-\frac{1}{2}}$ and $f(x)=x^{-\beta}$ therein to conclude that
\begin{align}
G_{1-\beta}(Y,X)  &=X^{\frac{1}{2}}\left(  X^{\frac{1}{2}}Y^{-\frac{1}{2}}Y^{-\frac{1}{2}
}X^{\frac{1}{2}}\right)  ^{-\beta}X^{\frac{1}{2}}Y^{-\frac{1}{2}}Y^{\frac
{1}{2}}\\
&  =X^{\frac{1}{2}}\left(  X^{-\frac{1}{2}}YX^{-\frac{1}{2}}\right)  ^{\beta
}X^{\frac{1}{2}}\\
&  =G_{\beta}(X,Y).
\end{align}
\end{remark}

Definition~\ref{def:geometric-renyi-rel-ent} of the geometric R\'{e}nyi relative entropy involves a limit, which has to do with the possibility that $\sigma$ might not be invertible (i.e., it might not be positive definite). Recall that the same situation arises for the Petz-- and sandwiched R\'{e}nyi relative entropies, which leads to expressions for them in terms of a limit in Propositions~\ref{prop-petz_rel_ent_lim} and \ref{prop-sand_rel_ent_lim}, respectively. For these two quantities, the limits evaluate to a finite value with an explicit expression under the condition $\alpha\in(0,1)$ and $\Tr[\rho\sigma]\neq 0$, or $\alpha\in(1,\infty)$ and $\supp(\rho)\subseteq\supp(\sigma)$. Analogously, for the geometric R\'{e}nyi relative entropy there are several cases for which the limit in \eqref{eq:def-geometric-renyi-rel-quasi-ent} is finite and has an explicit expression. The following proposition outlines some of the simpler cases in which $\sigma$ is positive definite:

\begin{proposition}{prop:alt-rep-geometric-renyi-quasi}
	Let $\rho$ be a state, and let $\sigma$ be a positive definite operator. Then,
	\begin{equation}
		\widehat{Q}_{\alpha}(\rho\Vert\sigma)=\Tr\!\left[\sigma\!\left(\sigma^{-\frac{1}{2}}\rho\sigma^{-\frac{1}{2}}\right)^{\alpha}\right]=\Tr[G_{\alpha}(\sigma,\rho)]
	\end{equation}
for all $\alpha\in(0,1)\cup(1,\infty)$. If $\rho$ is a positive definite state and $\sigma$ a positive definite operator, then
	\begin{align}
		\widehat{Q}_{\alpha}(\rho\Vert\sigma)&=\Tr\!\left[\rho\!\left(\rho^{-\frac{1}{2}}\sigma\rho^{-\frac{1}{2}}\right)^{1-\alpha}\right]=\Tr[G_{1-\alpha}(\rho,\sigma)]\\
		&=\Tr\!\left[\rho\!\left(\rho^{\frac{1}{2}}\sigma^{-1}\rho^{\frac{1}{2}}\right)^{\alpha-1}\right]. \label{eq:alt-op-geo-mean-4-geo-ent}
	\end{align}
	for all $\alpha\in(0,1)\cup(1,\infty)$.
\end{proposition}

\begin{proof}
If $\sigma$ is positive definite, then the support of $\sigma$ is the entire Hilbert space, and so the limit $\varepsilon\to 0^+$ in \eqref{eq:def-geometric-renyi-rel-quasi-ent} simply evaluates to $\widehat{Q}_{\alpha}(\rho\Vert\sigma)=\Tr[G_{\alpha}(\sigma,\rho)]$ for all $\alpha\in(0,1)\cup(1,\infty)$.

If $\rho$ is also positive definite, then by invoking the equality in \eqref{eq:geometric-mean-identity}, we conclude that $\Tr[G_{\alpha}(\sigma,\rho)]=\Tr[G_{1-\alpha}(\rho,\sigma)]$ for all $\alpha\in(0,1)\cup(1,\infty)$. Furthermore, since both $\rho$ and $\sigma$ are positive definite, the following equality holds
\begin{equation}
	\left(\rho^{-\frac{1}{2}}\sigma\rho^{-\frac{1}{2}}\right)^{1-\alpha}=\left(\rho^{\frac{1}{2}}\sigma^{-1}\rho^{\frac{1}{2}}\right)^{\alpha-1}.
\end{equation}
Therefore, the equality in \eqref{eq:alt-op-geo-mean-4-geo-ent} holds.
\end{proof}

We now provide explicit expressions for the geometric R\'enyi relative quasi-entropy $\widehat{Q}_{\alpha}(\rho\Vert\sigma)$ that are consistent with the limit-based definition in~\eqref{eq:def-geometric-renyi-rel-quasi-ent}  whenever $\rho$ and/or $\sigma$ are not  positive definite. The expressions given in \eqref{eq:geometric-rel-quasi-explicit-1} below cover all possible values of $\alpha\in(0,1)\cup(1,\infty)$ and support conditions. Additional expressions are given in \eqref{eq:geometric-rel-quasi-explicit-2}.

\begin{proposition*}{Explicit Expressions for Geometric R\'enyi Relative Quasi-Entropy}{prop:explicit-form-geometric-renyi}For every state $\rho$, positive
semi-definite operator $\sigma$, and $\alpha\in(0,1)\cup(1,\infty)$, the
following equality holds for the geometric R\'enyi relative quasi-entropy:
\begin{equation}
\widehat{Q}_{\alpha}(\rho\Vert\sigma)=\left\{
\begin{array}
[c]{cc}
\operatorname{Tr}\!\left[  \sigma\!\left(  \sigma^{-\frac{1}{2}}\rho
\sigma^{-\frac{1}{2}}\right)  ^{\alpha}\right]  &
\begin{array}
[c]{c}
\text{if }\alpha\in\left(  0,1\right)  \cup(1,\infty)\\
\text{and }\operatorname{supp}(\rho)\subseteq\operatorname{supp}(\sigma)
\end{array}
\\
& \\
\operatorname{Tr}\!\left[  \sigma\!\left(  \sigma^{-\frac{1}{2}}\tilde{\rho
}\sigma^{-\frac{1}{2}}\right)  ^{\alpha}\right]  &
\begin{array}
[c]{c}
\text{if }\alpha\in\left(  0,1\right) \\
\text{and }\operatorname{supp}(\rho)\not \subseteq \operatorname{supp}(\sigma)
\end{array}
\\
& \\
+\infty &
\begin{array}
[c]{c}
\text{if }\alpha\in(1,\infty)\text{ and}\\
\operatorname{supp}(\rho)\not \subseteq \operatorname{supp}(\sigma)\text{,}
\end{array}
\end{array}
\right.   \label{eq:geometric-rel-quasi-explicit-1}
\end{equation}
where
\begin{align}
\tilde{\rho}  &  \coloneqq\rho_{0,0}-\rho_{0,1}^{\vphantom{\dag}}\rho_{1,1}^{-1\vphantom{\dag}}\rho_{0,1}^{\dag}
,\quad\rho=
\begin{pmatrix}
\rho_{0,0} & \rho_{0,1}\\
\rho_{0,1}^{\dag} & \rho_{1,1}
\end{pmatrix}
,
\end{align}
where the block matrix representation above is with respect to the decomposition $\mathcal{H} = \supp(\sigma)\oplus \ker(\sigma)$, 
and the inverses $\sigma^{-\frac{1}{2}}$ and $\rho_{1,1}^{-1}$ are taken on the supports of $\sigma$ and $\rho_{1,1}$, respectively. 
We also have the alternative expressions below for certain cases:
\begin{equation}
\widehat{Q}_{\alpha}(\rho\Vert\sigma)=\left\{
\begin{array}
[c]{cc}
\operatorname{Tr}\!\left[  \rho\!\left(  \rho^{-\frac{1}{2}}\sigma\rho
^{-\frac{1}{2}}\right)  ^{1-\alpha}\right]  &
\begin{array}
[c]{c}
\text{if }\alpha\in\left(  0,1\right) \\
\text{and }\operatorname{supp}(\sigma)\subseteq\operatorname{supp}(\rho)
\end{array}
\\
& \\
\operatorname{Tr}\!\left[  \rho\!\left(  \rho^{\frac{1}{2}}\sigma^{-1}
\rho^{\frac{1}{2}}\right)  ^{\alpha-1}\right]  &
\begin{array}
[c]{c}
\text{if }\alpha\in(1,\infty)\\
\text{and }\operatorname{supp}(\rho)\subseteq\operatorname{supp}(\sigma)
\end{array}
\end{array}
\right.  , \label{eq:geometric-rel-quasi-explicit-2}
\end{equation}
where the inverses $\rho^{-\frac{1}{2}}$ and $\sigma^{-1}$ are taken on the supports of $\rho$ and $\sigma$, respectively.
\end{proposition*}

\begin{Proof}
	The proof is similar in spirit to the proofs of Propositions~\ref{prop-petz_rel_ent_lim} and \ref{prop-sand_rel_ent_lim}, but it is more complicated than these previous proofs. We provide it in Section~\ref{app:QEI:explicit-form-geometric-renyi_pf}.
\end{Proof}

Observe that when $\operatorname{supp}(\rho)\subseteq
\operatorname{supp}(\sigma)$ and $\alpha\in(0,1)$, the expression
$\operatorname{Tr}[  \sigma(  \sigma^{-1/2}\rho
\sigma^{-1/2})  ^{\alpha}]  $ is actually a special case
of $\operatorname{Tr}[  \sigma(  \sigma^{-1/2}
\tilde{\rho}\sigma^{-1/2})  ^{\alpha}]  $, because the
operators $\rho_{0,1}$ and $\rho_{1,1}$ are both equal to zero in this case,
so that $\Pi_{\sigma}\rho=\rho\Pi_{\sigma}=\rho$ and $\tilde{\rho}=\rho_{0,0}
$.

The main intuition behind the first expression in \eqref{eq:geometric-rel-quasi-explicit-1} and those in \eqref{eq:geometric-rel-quasi-explicit-2}  is as follows. If $\rho$
and $\sigma$ are positive definite, then the following equalities hold
\begin{align}
\operatorname{Tr}\!\left[  \sigma\!\left(  \sigma^{-\frac{1}{2}}\rho
\sigma^{-\frac{1}{2}}\right)  ^{\alpha}\right]   &  =\operatorname{Tr}
\!\left[  \rho\!\left(  \rho^{-\frac{1}{2}}\sigma\rho^{-\frac{1}{2}}\right)
^{1-\alpha}\right] \\
&  =\operatorname{Tr}\!\left[  \rho\!\left(  \rho^{\frac{1}{2}}\sigma^{-1}
\rho^{\frac{1}{2}}\right)  ^{\alpha-1}\right]  ,
\end{align}
for all $\alpha\in(0,1)\cup(1,\infty)$, as shown previously in
Proposition~\ref{prop:alt-rep-geometric-renyi-quasi}.
\begin{enumerate}
\item If the support
condition $\operatorname{supp}(\rho)\subseteq\operatorname{supp}(\sigma)$
holds, then we can think of $\operatorname{supp}(\sigma)$ as being the whole
Hilbert space and $\sigma$ being invertible on the whole space. So then
generalized inverses like $\sigma^{-\frac{1}{2}}$ or $\sigma^{-1}$ are true
inverses on $\operatorname{supp}(\sigma)$, and the expression
$\operatorname{Tr}[\sigma(\sigma^{-1/2}\rho\sigma^{-1/2})^{\alpha}]$ is sensible 
 for $\alpha\in(0,1)\cup(1,\infty)$, with the only inverse in the
expression being $\sigma^{-\frac{1}{2}}$; this expression results after taking the limit $\varepsilon \to 0^+$.

\item Similarly, the expression
$\operatorname{Tr}[\rho(\rho^{1/2}\sigma^{-1}\rho^{1/2})^{\alpha-1}]$ is sensible for $\alpha\in(1,\infty)$, with the only inverse in the expression
being $\sigma^{-1}$; this latter expression results after taking the limit $\varepsilon \to 0^+$.

\item If the support condition
$\operatorname{supp}(\sigma)\subseteq\operatorname{supp}(\rho)$ holds, then we
can think of $\operatorname{supp}(\rho)$ as being the whole Hilbert space and
$\rho$ being invertible on the whole space. So then the generalized inverse
$\rho^{-\frac{1}{2}}$ is a true inverse on $\operatorname{supp}(\rho)$, and
the expression $\operatorname{Tr}[\rho(\rho^{-1/2}\sigma\rho^{-1/2}
)^{1-\alpha}]$ is sensible for $\alpha\in(0,1)$, with the only inverse in the
expression being $\rho^{-\frac{1}{2}}$; this expression also results after taking the limit $\varepsilon \to 0^+$.
\end{enumerate}

	In order to understand the first expression in \eqref{eq:geometric-rel-quasi-explicit-2} further, observe that the following identities hold for all $\alpha\in\left(  0,1\right)\cup(1,\infty)    $:
\begin{align}
\label{eq:limit-eq-geometric-Renyi}
\widehat{Q}_{\alpha}(\rho\Vert\sigma) & =\lim_{\varepsilon\rightarrow0^{+}}
\lim_{\delta\rightarrow0^{+}}\operatorname{Tr}\!\left[  \sigma_{\varepsilon
}\!\left(  \sigma_{\varepsilon}^{-\frac{1}{2}}\rho_{\delta}\sigma
_{\varepsilon}^{-\frac{1}{2}}\right)  ^{\alpha}\right]\\
& =\lim_{\varepsilon\to 0^+}\lim_{\delta\to 0^+}\Tr[G_{\alpha}(\sigma_{\varepsilon},\rho_{\delta})],
\end{align}
where
\begin{equation}
\rho_{\delta}\coloneqq\left(  1-\delta\right)  \rho+\delta\pi,
\end{equation}
and $\pi$ is the maximally mixed state. This holds because the expression for the geometric R\'{e}nyi relative quasi-entropy in Definition~\ref{def:geometric-renyi-rel-ent} does not involve an inverse of the state $\rho$.

As it turns out, the order of the limits in \eqref{eq:limit-eq-geometric-Renyi} does not matter for $\alpha\in(0,1)$:

\begin{Lemma*}{Limit Interchange for Geometric R\'enyi Relative Quasi-Entropy}{lem:limit-exchange-geom-renyi-a-0-to1}
Let $\rho$ be a state and
$\sigma$ a positive semi-definite operator. For $\alpha\in(0,1)$, the
following equality holds
\begin{align}
\widehat{Q}_{\alpha}(\rho\Vert\sigma)  &  =\lim_{\varepsilon\rightarrow0^{+}
}\lim_{\delta\rightarrow0^{+}}\operatorname{Tr}\!\left[  \sigma_{\varepsilon
}\!\left(  \sigma_{\varepsilon}^{-\frac{1}{2}}\rho_{\delta}\sigma
_{\varepsilon}^{-\frac{1}{2}}\right)  ^{\alpha}\right] \\
&  =\inf_{\varepsilon,\delta>0}\operatorname{Tr}\!\left[  \sigma_{\varepsilon
}\!\left(  \sigma_{\varepsilon}^{-\frac{1}{2}}\rho_{\delta}\sigma
_{\varepsilon}^{-\frac{1}{2}}\right)  ^{\alpha}\right] \\
&  =\lim_{\delta\rightarrow0^{+}}\lim_{\varepsilon\rightarrow0^{+}
}\operatorname{Tr}\!\left[  \sigma_{\varepsilon}\!\left(  \sigma_{\varepsilon
}^{-\frac{1}{2}}\rho_{\delta}\sigma_{\varepsilon}^{-\frac{1}{2}}\right)
^{\alpha}\right]  ,
\end{align}
where $\rho_{\delta}\coloneqq\left(  1-\delta\right)  \rho+\delta\pi$, $\delta
\in\left(  0,1\right)  $, $\pi$ is the maximally mixed state, $\sigma
_{\varepsilon}\coloneqq\sigma+\varepsilon \mathbbm{1}$, and $\varepsilon>0$.
\end{Lemma*}

\begin{Proof}
See Section~\ref{app:proof:limit-exchange-geom-renyi-a-0-to1}.
\end{Proof}

Now, because both $\sigma_{\varepsilon}$ and $\rho_{\delta}$ are positive definite for $\varepsilon,\delta>0$, we can use the property   in \eqref{eq:geometric-mean-identity}, along with Lemma~\ref{lem:limit-exchange-geom-renyi-a-0-to1}, to obtain the following for $\alpha \in(0,1)$:
\begin{align}
	\widehat{Q}_{\alpha}(\rho\Vert\sigma)&=\lim_{\delta\to 0^+}\lim_{\varepsilon\to 0^+}\Tr[G_{1-\alpha}(\rho_{\delta},\sigma_{\varepsilon})]\\
	&=\lim_{\delta\to 0^+}\lim_{\varepsilon\to 0^+}\Tr\!\left[\rho_{\delta}\!\left(\rho_{\delta}^{-\frac{1}{2}}\sigma_{\varepsilon}\rho_{\delta}^{-\frac{1}{2}}\right)^{1-\alpha}\right]\\
	&=\lim_{\delta\to 0^+}\Tr\!\left[\rho_{\delta}\!\left(\rho_{\delta}^{-\frac{1}{2}}\sigma\rho_{\delta}^{-\frac{1}{2}}\right)^{1-\alpha}\right],\label{eq:limit-eq-geometric-Renyi_2}
\end{align}
where the last equality holds for the analogous reason that \eqref{eq:limit-eq-geometric-Renyi} holds, namely, that the inverse of $\sigma$ is not involved. We are now in a situation that looks like the expression in \eqref{eq:def-geometric-renyi-rel-quasi-ent}, except that the roles of $\rho$ and $\sigma$ are reversed and $\alpha$ is substituted with $1-\alpha$. Then, in the limit $\delta \to 0^+$, if the support condition $\supp(\sigma)\subseteq\supp(\rho)$ holds, the expression converges to $\Tr[\rho(\rho^{-1/2}\sigma\rho^{-1/2})^{1-\alpha}]$.

It is worthwhile to consider the special case of $\alpha=2$. In this case, the geometric R\'{e}nyi relative quasi-entropy collapses to
the Petz--R\'{e}nyi relative quasi-entropy when $\supp(\rho)\subseteq\supp(\sigma)$:
\begin{align}
\widehat{Q}_{2}(\rho\Vert\sigma)  &  =\lim_{\varepsilon\rightarrow0^{+}
}\operatorname{Tr}\!\left[  \sigma_{\varepsilon}\!\left(  \sigma_{\varepsilon
}^{-\frac{1}{2}}\rho\sigma_{\varepsilon}^{-\frac{1}{2}}\right)  ^{2}\right] \\
&  =\lim_{\varepsilon\rightarrow0^{+}}\operatorname{Tr}\!\left[
\sigma_{\varepsilon}\!\left(  \sigma_{\varepsilon}^{-\frac{1}{2}}\rho
\sigma_{\varepsilon}^{-\frac{1}{2}}\right)  \left(  \sigma_{\varepsilon
}^{-\frac{1}{2}}\rho\sigma_{\varepsilon}^{-\frac{1}{2}}\right)  \right] \\
&  =\lim_{\varepsilon\rightarrow0^{+}}\operatorname{Tr}[\rho\sigma
_{\varepsilon}^{-1}\rho]\\
&  =\lim_{\varepsilon\rightarrow0^{+}}\operatorname{Tr}[\rho^{2}
\sigma_{\varepsilon}^{-1}]\\
&  =Q_{2}(\rho\Vert\sigma),
\end{align}
with the last line following from Proposition~\ref{prop-petz_rel_ent_lim}. The
development above implies that the corresponding R\'enyi relative entropies are equal:
\begin{equation}\label{eq-geo_ren_ent_petz_ren_a_2}
\widehat{D}_{2}(\rho\Vert\sigma)=D_{2}(\rho\Vert\sigma).
\end{equation}
The geometric and sandwiched  R\'{e}nyi relative entropies also converge to the same value in the limit $\alpha\to\infty$, as shown in Section~\ref{sec-max_rel_ent}.

A first property of the geometric R\'{e}nyi relative entropy that we establish is
its relation to the sandwiched R\'{e}nyi relative entropy.

\begin{proposition*}{Ordering of Sandwiched and Geometric R\'enyi Relative Entropies}{prop:geometric-to-sandwiched}Let $\rho$ be a state and $\sigma$ a
positive semi-definite operator. The geometric R\'{e}nyi relative entropy is
not smaller than the sandwiched R\'{e}nyi relative entropy for all $\alpha
\in\left(  0,1\right)  \cup\left(  1,\infty\right)  $:
\begin{equation}
\widetilde{D}_{\alpha}(\rho\Vert\sigma)\leq\widehat{D}_{\alpha}(\rho
\Vert\sigma). \label{eq:geometric-renyi-to-sandwiched}
\end{equation}

\end{proposition*}

\begin{Proof}
This is a direct consequence of the Araki--Lieb--Thirring inequality
(Lemma~\ref{lem-ALT_ineq}), which we recall here for convenience. For positive semi-definite operators $X$ and $Y$,
$q\geq0$, and $r\in\left[  0,1\right]  $, it holds that
\begin{equation}
\operatorname{Tr}\!\left[  \left(  Y^{\frac{1}{2}}XY^{\frac{1}{2}}\right)
^{rq}\right]  \geq\operatorname{Tr}\!\left[  \left(  Y^{\frac{r}{2}}
X^{r}Y^{\frac{r}{2}}\right)  ^{q}\right]  . \label{eq:QEI:ALT-1}
\end{equation}
For $r\geq1$, instead,
\begin{equation}
\operatorname{Tr}\!\left[  \left(  Y^{\frac{1}{2}}XY^{\frac{1}{2}}\right)
^{rq}\right]  \leq\operatorname{Tr}\!\left[  \left(  Y^{\frac{r}{2}}
X^{r}Y^{\frac{r}{2}}\right)  ^{q}\right]  . \label{eq:QEI:ALT-2}
\end{equation}
By employing \eqref{eq:QEI:ALT-1} with $q=1$, $r=\alpha\in(0,1)$, $Y=\sigma_{\varepsilon}
^{\frac{1}{\alpha}}$, and $X=\sigma_{\varepsilon}^{-\frac{1}{2}}\rho
\sigma_{\varepsilon}^{-\frac{1}{2}}$, and recalling that $\sigma_{\varepsilon
}\coloneqq\sigma+\varepsilon \mathbbm{1}$, we find that
\begin{align}
\widehat{Q}_{\alpha}(\rho\Vert\sigma_{\varepsilon})  &  =\operatorname{Tr}
\!\left[  \sigma_{\varepsilon}\!\left(  \sigma_{\varepsilon}^{-\frac{1}{2}
}\rho\sigma_{\varepsilon}^{-\frac{1}{2}}\right)  ^{\alpha}\right] \\
&  =\operatorname{Tr}\!\left[  \left(  \sigma_{\varepsilon}^{\frac{1}{2\alpha
}}\right)  ^{\alpha}\left(  \sigma_{\varepsilon}^{-\frac{1}{2}}\rho
\sigma_{\varepsilon}^{-\frac{1}{2}}\right)  ^{\alpha}\left(  \sigma
_{\varepsilon}^{\frac{1}{2\alpha}}\right)  ^{\alpha}\right] \\
&  \leq\operatorname{Tr}\!\left[  \left(  \sigma_{\varepsilon}^{\frac
{1}{2\alpha}}\sigma_{\varepsilon}^{-\frac{1}{2}}\rho\sigma_{\varepsilon
}^{-\frac{1}{2}}\sigma_{\varepsilon}^{\frac{1}{2\alpha}}\right)  ^{\alpha
}\right] \\
&  =\operatorname{Tr}\!\left[  \left(  \sigma_{\varepsilon}^{\frac{1-\alpha
}{2\alpha}}\rho\sigma_{\varepsilon}^{\frac{1-\alpha}{2\alpha}}\right)
^{\alpha}\right] \\
&  =\widetilde{Q}_{\alpha}(\rho\Vert\sigma_{\varepsilon}),
\end{align}
which implies for $\alpha\in(0,1)$, by using definitions, that
\begin{equation}
\widetilde{D}_{\alpha}(\rho\Vert\sigma_{\varepsilon})\leq\widehat{D}_{\alpha
}(\rho\Vert\sigma_{\varepsilon}).
\end{equation}
Now taking the limit as $\varepsilon\rightarrow0^{+}$, employing
Proposition~\ref{prop-sand_rel_ent_lim} and
Definition~\ref{def:geometric-renyi-rel-ent}, we arrive at the inequality in \eqref{eq:geometric-renyi-to-sandwiched}.

Since the Araki--Lieb--Thirring inequality is reversed for $r=\alpha\in\left(
1,\infty\right)  $, we can employ similar reasoning as above, using \eqref{eq:QEI:ALT-2} and definitions,
to arrive at \eqref{eq:geometric-renyi-to-sandwiched} for $\alpha\in
(1,\infty)$.
\end{Proof}

If the state $\rho$ is pure, then the geometric R\'{e}nyi relative entropy
simplifies as follows, such that it is independent of $\alpha$:

\begin{proposition*}{Geometric R\'enyi Relative Entropy for Pure States}{prop:QEI:geo-renyi-pure-states}
Let $\rho=|\psi\rangle\!\langle\psi|$ be a pure state and $\sigma$ a positive
semi-definite operator.\ Then the following equality holds for all $\alpha
\in(0,1)\cup(1,\infty)$:
\begin{equation}
\widehat{D}_{\alpha}(\rho\Vert\sigma)=\left\{
\begin{array}
[c]{cc}
\log_2\langle\psi|\sigma^{-1}|\psi\rangle & \text{if } |\psi\rangle \in \operatorname{supp}(\sigma)\\
+\infty & \text{otherwise}
\end{array}
\right.  ,
\label{eq:geometric-renyi-1st-pure}
\end{equation}
where the inverse $\sigma^{-1}$ is taken on the support of $\sigma$. If $\sigma$ is
also a rank-one operator, so that $\sigma=|\phi\rangle\!\langle\phi|$ and
$\left\Vert |\phi\rangle\right\Vert _{2}>0$, then the following equality holds
for all $\alpha\in(0,1)\cup(1,\infty)$:
\begin{equation}
\widehat{D}_{\alpha}(\rho\Vert\sigma)=\left\{
\begin{array}
[c]{cc}
-\log_2\!\left\Vert |\phi\rangle\right\Vert _{2}^{2} & \text{if }\exists
c\in\mathbb{C}\text{ such that }|\psi\rangle=c|\phi\rangle\\
+\infty & \text{otherwise}
\end{array}
\right.  . \label{eq:geometric-renyi-pure-states-collapse}
\end{equation}
In particular, if $\sigma=|\phi\rangle\!\langle\phi|$ is a state so that
$\left\Vert |\phi\rangle\right\Vert _{2}^{2}=1$, then
\begin{equation}
\widehat{D}_{\alpha}(\rho\Vert\sigma)=\left\{
\begin{array}
[c]{cc}
0 & \text{if $\ket{\psi}= e^{i\varphi} \ket{\phi}$ for some $\varphi\in \mathbb{R}$}\\
+\infty & \text{otherwise}
\end{array}
\right.  .
\label{eq:QEI:pure-state-GRR-degenerate}
\end{equation}
\end{proposition*}

\begin{Proof}
Defining $\sigma_{\varepsilon}\coloneqq\sigma+\varepsilon \mathbbm{1}$, consider that
\begin{align}
\operatorname{Tr}\!\left[  \sigma_{\varepsilon}\!\left(  \sigma_{\varepsilon
}^{-\frac{1}{2}}\rho\sigma_{\varepsilon}^{-\frac{1}{2}}\right)  ^{\alpha
}\right]   &  =\operatorname{Tr}\!\left[  \sigma_{\varepsilon}\!\left(
\sigma_{\varepsilon}^{-\frac{1}{2}}|\psi\rangle\!\langle\psi|\sigma
_{\varepsilon}^{-\frac{1}{2}}\right)  ^{\alpha}\right] \\
&  =\left(  \left\Vert \sigma_{\varepsilon}^{-\frac{1}{2}}|\psi\rangle
\right\Vert _{2}^{2}\right)  ^{\alpha}\operatorname{Tr}\!\left[
\sigma_{\varepsilon}\left(  \frac{\sigma_{\varepsilon}^{-\frac{1}{2}}
|\psi\rangle\!\langle\psi|\sigma_{\varepsilon}^{-\frac{1}{2}}}{\left\Vert
\sigma_{\varepsilon}^{-\frac{1}{2}}|\psi\rangle\right\Vert _{2}^{2}}\right)
^{\alpha}\right] \\
&  =\left(  \left\Vert \sigma_{\varepsilon}^{-\frac{1}{2}}|\psi\rangle
\right\Vert _{2}^{2}\right)  ^{\alpha}\operatorname{Tr}\!\left[
\sigma_{\varepsilon}\frac{\sigma_{\varepsilon}^{-\frac{1}{2}}|\psi
\rangle\!\langle\psi|\sigma_{\varepsilon}^{-\frac{1}{2}}}{\left\Vert
\sigma_{\varepsilon}^{-\frac{1}{2}}|\psi\rangle\right\Vert _{2}^{2}}\right] \\
&  =\left(  \left\Vert \sigma_{\varepsilon}^{-\frac{1}{2}}|\psi\rangle
\right\Vert _{2}^{2}\right)  ^{\alpha-1}\operatorname{Tr}\!\left[
\sigma_{\varepsilon}\sigma_{\varepsilon}^{-\frac{1}{2}}|\psi\rangle\!\langle
\psi|\sigma_{\varepsilon}^{-\frac{1}{2}}\right] \\
&  =\left(  \left\Vert \sigma_{\varepsilon}^{-\frac{1}{2}}|\psi\rangle
\right\Vert _{2}^{2}\right)  ^{\alpha-1}\operatorname{Tr}[|\psi\rangle\!
\langle\psi|]\\
&  =\left[  \langle\psi|\sigma_{\varepsilon}^{-1}|\psi\rangle\right]
^{\alpha-1}.
\end{align}
The third equality follows because $|\varphi\rangle\!\langle\varphi|^{\alpha
}=|\varphi\rangle\!\langle\varphi|$ for all $\alpha>0$ when $\left\Vert |\varphi\rangle\right\Vert
_{2}=1$. Applying the chain of equalities above, we find that
\begin{align}
\frac{1}{\alpha-1}\log_2\operatorname{Tr}\!\left[  \sigma_{\varepsilon}\left(
\sigma_{\varepsilon}^{-\frac{1}{2}}\rho\sigma_{\varepsilon}^{-\frac{1}{2}
}\right)  ^{\alpha}\right]   &  =\frac{1}{\alpha-1}\log_{2}\!\left[
\langle\psi|\sigma_{\varepsilon}^{-1}|\psi\rangle\right]  ^{\alpha-1}\\
&  =\log_2\langle\psi|\sigma_{\varepsilon}^{-1}|\psi\rangle.
\end{align}

Now let a spectral decomposition of $\sigma$ be given by
\begin{equation}
\sigma=\sum_{y}\mu_{y}Q_{y},
\end{equation}
where $\mu_{y}$ are the non-negative eigenvalues, assumed to be distinct, and $Q_{y}$ are the eigenprojections, assumed to add up to the identity. In this decomposition, we include exactly one value of $y$, call it $y_0$, such that $\mu_{y_0}=0$; if $\sigma$ has no zero eigenvalues, we set $Q_{y_0}=0$. Then it follows that
\begin{equation}
\sigma_{\varepsilon}=\sigma+\varepsilon \mathbbm{1}=\sum_{y}\left(  \mu_{y}
+\varepsilon\right)  Q_{y},
\end{equation}
and we find that
\begin{equation}
\sigma_{\varepsilon}^{-1}=\sum_{y}\left(  \mu_{y}+\varepsilon\right)
^{-1}Q_{y}.
\end{equation}
We then conclude that
\begin{align}
\langle\psi|\sigma_{\varepsilon}^{-1}|\psi\rangle &  =
\langle\psi|\sum_{y}\left(  \mu_{y}+\varepsilon\right)  ^{-1}Q_{y}|\psi
\rangle \\
&  =  \sum_{y}\left(  \mu_{y}+\varepsilon\right)  ^{-1}\langle
\psi|Q_{y}|\psi\rangle \\
&  =  \sum_{y:\mu_{y}\neq0}\left(  \mu_{y}+\varepsilon\right)
^{-1}\langle\psi|Q_{y}|\psi\rangle+\varepsilon^{-1}\langle\psi|Q_{y_{0}}
|\psi\rangle  ,
\end{align}
where $y_{0}$ is the value of $y$ for which $\mu_{y}=0$ (if no such value of
$y$ exists, then $Q_{y_{0}}$ is equal to the zero operator). Thus, if
$\langle\psi|Q_{y_{0}}|\psi\rangle\neq0$ (equivalent to $|\psi\rangle$ being
outside the support of $\sigma$), then it follows that
\begin{equation}
\lim_{\varepsilon\rightarrow0^{+}}\log_2\langle\psi|\sigma_{\varepsilon}
^{-1}|\psi\rangle=+\infty.
\end{equation}
Otherwise the expression converges as claimed.

Now suppose that $\sigma$ is a rank-one operator, so that $\sigma=|\phi
\rangle\!\langle\phi|$ and $\left\Vert |\phi\rangle\right\Vert _{2}>0$. By
defining
\begin{equation}
|\phi^{\prime}\rangle   \coloneqq\frac{|\phi\rangle}{\left\Vert |\phi
\rangle\right\Vert _{2}},
\end{equation}
we find that the inverse of $\sigma$ taken on its support (in this case, $\operatorname{span}(\ket{\phi})$) is given by
\begin{equation}
    \sigma^{-1} = \left( \Vert \ket{\phi} \Vert_2^2\, \ketbra{\phi^\prime}{\phi^\prime} \right)^{-1} = \Vert \ket{\phi} \Vert_2^{-2}\, \ketbra{\phi^\prime}{\phi^\prime} .
\end{equation}
Inserting this into \eqref{eq:geometric-renyi-1st-pure} yields \eqref{eq:geometric-renyi-pure-states-collapse}. As a special case, we obtain \eqref{eq:QEI:pure-state-GRR-degenerate}.
\end{Proof}

We note here that, for pure states $\rho$ and $\sigma$ and as indicated by \eqref{eq:QEI:pure-state-GRR-degenerate}, the geometric
R\'{e}nyi relative entropy is either equal to zero or $+\infty$, depending on
whether $\rho=\sigma$ or $\rho\neq \sigma$. This behavior 
is very different from that exhibited by the Petz-- and sandwiched R\'{e}nyi relative entropies. The latter quantities always evaluate to a finite value if $\alpha \in (0,1)$ and the pure states are non-orthogonal.

The geometric R\'{e}nyi relative entropy possesses a number of useful
properties, similar to those for the Petz-- and sandwiched R\'{e}nyi relative
entropies, which we delineate now.

\begin{proposition*}{Properties of Geometric R\'{e}nyi Relative Entropy}
{prop:geometric-renyi-props}
For all states $\rho$, $\rho_{1}$, $\rho
_{2}$ and positive semi-definite operators $\sigma$, $\sigma_{1}$, $\sigma
_{2}$, the geometric R\'{e}nyi relative entropy satisfies the following properties:

\begin{enumerate}
\item Isometric invariance: For all $\alpha\in(0,1)\cup(1,\infty)$ and for every
isometry $V$,
\begin{equation}
\widehat{D}_{\alpha}(\rho\Vert\sigma)=\widehat{D}_{\alpha}(V\rho V^{\dag}\Vert
V\sigma V^{\dag}).
\end{equation}

\item Monotonicity in $\alpha$: For all $\alpha\in(0,1)\cup(1,\infty)$, the
geometric R\'{e}nyi relative entropy $\widehat{D}_{\alpha}$ is monotonically
increasing in $\alpha$; i.e., $\alpha\leq \beta$ implies $\widehat{D}_{\alpha
}(\rho\Vert\sigma)\leq\widehat{D}_{\beta}(\rho\Vert\sigma)$.

\item Additivity:\ For all $\alpha\in(0,1)\cup(1,\infty)$,
\begin{equation}
\widehat{D}_{\alpha}(\rho_{1}\otimes\rho_{2}\Vert\sigma_{1}\otimes\sigma
_{2})=\widehat{D}_{\alpha}(\rho_{1}\Vert\sigma_{1})+\widehat{D}_{\alpha}
(\rho_{2}\Vert\sigma_{2}). \label{eq:additivity-geometric-renyi}
\end{equation}

\item Direct-sum property: Let $p\colon \mathcal{X}\rightarrow\left[  0,1\right]  $
be a probability distribution over a finite alphabet $\mathcal{X}$ with
associated $\left\vert \mathcal{X}\right\vert $-dimensional system $X$, and
let $q\colon \mathcal{X}\rightarrow[0,\infty)$ be a positive function on
$\mathcal{X}$. Let $\left\{  \rho_{A}^{x}\right\}_{x\in\mathcal{X}}  $ be a set
of states on a system $A$, and let $\left\{  \sigma_{A}^{x}\right\}_{x\in
\mathcal{X}}  $ be a set of positive semi-definite operators on $A$.
Then,
\begin{equation}
\widehat{Q}_{\alpha}(\rho_{XA}\Vert\sigma_{XA})=\sum_{x\in\mathcal{X}
}p(x)^{\alpha}q(x)^{1-\alpha}\widehat{Q}_{\alpha}(\rho_{A}^{x}\Vert\sigma
_{A}^{x}), \label{eq:direct-sum-prop-geometric-renyi}
\end{equation}
where
\begin{align}
\rho_{XA}  &  \coloneqq\sum_{x\in\mathcal{X}}p(x)|x\rangle\!\langle x|_{X}\otimes
\rho_{A}^{x},\\
\sigma_{XA}  &  \coloneqq\sum_{x\in\mathcal{X}}q(x)|x\rangle\!\langle x|_{X}
\otimes\sigma_{A}^{x}.
\end{align}

\end{enumerate}
\end{proposition*}

\begin{Proof}
\hfill
\begin{enumerate}
\item \textit{Proof of isometric invariance}: Let us start by writing
$\widehat{D}_{\alpha}(\rho\Vert\sigma)$ as in
\eqref{eq:def-geometric-renyi-rel-quasi-ent}--\eqref{eq:def-geometric-renyi-rel-ent}:
\begin{equation}
\widehat{D}_{\alpha}(\rho\Vert\sigma)=\lim_{\varepsilon\rightarrow0^{+}}
\frac{1}{\alpha-1} \log_2\operatorname{Tr}\!\left[  \sigma_{\varepsilon}\!\left(
\sigma_{\varepsilon}^{-\frac{1}{2}}\rho\sigma_{\varepsilon}^{-\frac{1}{2}
}\right)  ^{\alpha}\right]  .
\end{equation}
where
\begin{equation}
\sigma_{\varepsilon}\coloneqq\sigma+\varepsilon \mathbbm{1}.
\end{equation}
Let $V$ be an isometry. Then, defining
\begin{equation}
\omega_{\varepsilon}\coloneqq V\sigma V^{\dag}+\varepsilon \mathbbm{1},
\end{equation}
we find that
\begin{equation}
\widehat{D}_{\alpha}(V\rho V^{\dag}\Vert V\sigma V^{\dag})=\lim_{\varepsilon
\rightarrow0^{+}}\frac{1}{\alpha-1} \log_2\operatorname{Tr}\!\left[
\omega_{\varepsilon}\!\left(  \omega_{\varepsilon}^{-\frac{1}{2}}V\rho
V^{\dag}\omega_{\varepsilon}^{-\frac{1}{2}}\right)  ^{\alpha}\right]  .
\end{equation}
Now let $\Pi\coloneqq VV^{\dag}$ be the projection onto the image of $V$, so that $\Pi
V=V$, and let $\hat{\Pi}\coloneqq \mathbbm{1}-\Pi$. Then we can write
\begin{equation}
\omega_{\varepsilon}=V\sigma V^{\dag}+\varepsilon\Pi+\varepsilon\hat{\Pi
}=V\sigma_{\varepsilon}V^{\dag}+\varepsilon\hat{\Pi}.
\end{equation}
Since $V\sigma_{\varepsilon}V^{\dag}$ and $\varepsilon\hat{\Pi}$ are supported
on orthogonal subspaces, we obtain
\begin{equation}
\omega_{\varepsilon}^{-\frac{1}{2}}=V\sigma_{\varepsilon}^{^{-\frac{1}{2}}
}V^{\dag}+\varepsilon^{-\frac{1}{2}}\hat{\Pi}.
\end{equation}
Consider then that
\begin{align}
& \omega_{\varepsilon}^{-\frac{1}{2}}V\rho V^{\dag}\omega_{\varepsilon}
^{-\frac{1}{2}}  \notag \\
&  =\left(  V\sigma_{\varepsilon}^{^{-\frac{1}{2}}}V^{\dag
}+\varepsilon^{-\frac{1}{2}}\hat{\Pi}\right)  \Pi V\rho V^{\dag}\Pi\left(
V\sigma_{\varepsilon}^{^{-\frac{1}{2}}}V^{\dag}+\varepsilon^{-\frac{1}{2}}
\hat{\Pi}\right) \\
&  =\left(  V\sigma_{\varepsilon}^{^{-\frac{1}{2}}}V^{\dag}\right)  \Pi V\rho
V^{\dag}\Pi\left(  V\sigma_{\varepsilon}^{^{-\frac{1}{2}}}V^{\dag}\right) \\
&  =V\sigma_{\varepsilon}^{^{-\frac{1}{2}}}\rho\sigma_{\varepsilon}
^{^{-\frac{1}{2}}}V^{\dag},
\end{align}
where the second equality follows because $\hat{\Pi}\Pi=\Pi\hat{\Pi}=0$. Thus,
\begin{equation}
\left(  \omega_{\varepsilon}^{-\frac{1}{2}}V\rho V^{\dag}\omega_{\varepsilon
}^{-\frac{1}{2}}\right)  ^{\alpha}=V\left(  \sigma_{\varepsilon}^{^{-\frac
{1}{2}}}\rho\sigma_{\varepsilon}^{^{-\frac{1}{2}}}\right)  ^{\alpha}V^{\dag},
\end{equation}
and we find that
\begin{align}
& \operatorname{Tr}\!\left[  \omega_{\varepsilon}\!\left(  \omega_{\varepsilon
}^{-\frac{1}{2}}V\rho V^{\dag}\omega_{\varepsilon}^{-\frac{1}{2}}\right)
^{\alpha}\right]   \notag \\
&  \quad=\operatorname{Tr}\!\left[  \left(  V\sigma
_{\varepsilon}V^{\dag}+\varepsilon\hat{\Pi}\right)  V\left(  \sigma
_{\varepsilon}^{-\frac{1}{2}}\rho\sigma_{\varepsilon}^{-\frac{1}{2}}\right)
^{\alpha}V^{\dag}\right] \\
&  \quad=\operatorname{Tr}\!\left[  \sigma_{\varepsilon}\!\left(  \sigma
_{\varepsilon}^{^{-\frac{1}{2}}}\rho\sigma_{\varepsilon}^{^{-\frac{1}{2}}
}\right)  ^{\alpha}\right]  .
\end{align}
Since the equality
\begin{equation}
\operatorname{Tr}\!\left[  \omega_{\varepsilon}\!\left(  \omega_{\varepsilon
}^{-\frac{1}{2}}V\rho V^{\dag}\omega_{\varepsilon}^{-\frac{1}{2}}\right)
^{\alpha}\right]  =\operatorname{Tr}\!\left[  \sigma_{\varepsilon}\!\left(
\sigma_{\varepsilon}^{^{-\frac{1}{2}}}\rho\sigma_{\varepsilon}^{^{-\frac{1}
{2}}}\right)  ^{\alpha}\right]
\end{equation}
holds for all $\varepsilon>0$, we conclude the proof of isometric invariance
by taking the limit $\varepsilon\rightarrow0^{+}$.

\item \textit{Proof of monotonicity in }$\alpha$: We prove this by showing
that the derivative is non-negative for all $\alpha>0$. By applying
\eqref{eq:limit-eq-geometric-Renyi}, we can consider $\rho$ and $\sigma$ to be
positive definite without loss of generality. By applying
\eqref{eq:alt-op-geo-mean-4-geo-ent}, consider that
\begin{align}
\widehat{Q}_{\alpha}(\rho\Vert\sigma)  &  =\operatorname{Tr}\!\left[
\rho\!\left(  \rho^{-\frac{1}{2}}\sigma\rho^{-\frac{1}{2}}\right)  ^{1-\alpha
}\right] \\
&  =\operatorname{Tr}\!\left[  \rho\!\left(  \rho^{\frac{1}{2}}\sigma^{-1}
\rho^{\frac{1}{2}}\right)  ^{\alpha-1}\right]  .
\label{eq:alt-way-for-geometric-in-mono}
\end{align}
Now defining $|\varphi^{\rho}\rangle=(\rho^{\frac{1}{2}}\otimes \mathbbm{1})|\Gamma
\rangle$ as a purification of $\rho$, and setting
\begin{align}
\gamma &  \coloneqq \alpha-1,\\
X  &  \coloneqq \rho^{\frac{1}{2}}\sigma^{-1}\rho^{\frac{1}{2}},
\end{align}
we can write the geometric R\'{e}nyi relative entropy as
\begin{equation}
\widehat{D}_{\alpha}(\rho\Vert\sigma)=\frac{1}{\gamma}\log_2 \langle\varphi^{\rho
}|X^{\gamma}\otimes \mathbbm{1}|\varphi^{\rho}\rangle = 
\frac{1}{\gamma}\frac{\ln \langle\varphi^{\rho
}|X^{\gamma}\otimes \mathbbm{1}|\varphi^{\rho}\rangle}{\ln(2)},
\end{equation}
where we made use of \eqref{eq:alt-way-for-geometric-in-mono}. Then
$\frac{\D}{\D\alpha}=\frac{\D}{\D\gamma}\frac
{\D\gamma}{\D\alpha}=\frac{\D}{\D\gamma}$, and so we
find that
\begin{align}
&  \ln(2) \frac{\D}{\D\alpha}\widehat{D}_{\alpha}(\rho\Vert
\sigma)\nonumber\\
&  =\frac{\D}{\D\gamma}\left[  \frac{1}{\gamma}\ln\langle
\varphi^{\rho}|X^{\gamma}\otimes \mathbbm{1}|\varphi^{\rho}\rangle\right] \\
&  =\left[  -\frac{1}{\gamma^{2}}\ln\langle\varphi^{\rho}|X^{\gamma}\otimes
\mathbbm{1}|\varphi^{\rho}\rangle+\frac{1}{\gamma}\frac{\D}{\D\gamma}
\ln\langle\varphi^{\rho}|X^{\gamma}\otimes \mathbbm{1}|\varphi^{\rho}\rangle\right] \\
&  =\left[  -\frac{1}{\gamma^{2}}\ln\langle\varphi^{\rho}|X^{\gamma}\otimes
\mathbbm{1}|\varphi^{\rho}\rangle+\frac{1}{\gamma}\frac{\langle\varphi^{\rho}|X^{\gamma
}\ln X\otimes \mathbbm{1}|\varphi^{\rho}\rangle}{\langle\varphi^{\rho}|X^{\gamma}\otimes
\mathbbm{1}|\varphi^{\rho}\rangle}\right] \\
&  =\left[  \frac{-\langle\varphi^{\rho}|X^{\gamma}\otimes \mathbbm{1}|\varphi^{\rho
}\rangle\ln\langle\varphi^{\rho}|X^{\gamma}\otimes \mathbbm{1}|\varphi^{\rho}
\rangle+\gamma\langle\varphi^{\rho}|X^{\gamma}\ln X\otimes \mathbbm{1}|\varphi^{\rho
}\rangle}{\gamma^{2}\langle\varphi^{\rho}|X^{\gamma}\otimes \mathbbm{1}|\varphi^{\rho
}\rangle}\right] \\
&  =\left[  \frac{-\langle\varphi^{\rho}|X^{\gamma}\otimes \mathbbm{1}|\varphi^{\rho
}\rangle\ln\langle\varphi^{\rho}|X^{\gamma}\otimes \mathbbm{1}|\varphi^{\rho}
\rangle+\langle\varphi^{\rho}|X^{\gamma}\ln X^{\gamma}\otimes \mathbbm{1}|\varphi^{\rho
}\rangle}{\gamma^{2}\langle\varphi^{\rho}|X^{\gamma}\otimes \mathbbm{1}|\varphi^{\rho
}\rangle}\right]  .
\end{align}
Letting $g(x)\coloneqq x\log_2 x$, we write
\begin{equation}
\frac{\D}{\D\alpha}\widehat{D}_{\alpha}(\rho\Vert\sigma
)=\frac{\langle\varphi^{\rho}|g(X^{\gamma}\otimes \mathbbm{1})|\varphi^{\rho}
\rangle-g(\langle\varphi^{\rho}|(X^{\gamma}\otimes \mathbbm{1})|\varphi^{\rho}\rangle
)}{\gamma^{2}\langle\varphi^{\rho}|X^{\gamma}\otimes \mathbbm{1}|\varphi^{\rho}\rangle}.
\end{equation}
Then, since $g(x)$ is operator convex, by the operator Jensen inequality in \eqref{eq-op_Jensen_alt}, we conclude that
\begin{equation}
\langle\varphi^{\rho}|g(X^{\gamma}\otimes \mathbbm{1})|\varphi^{\rho}\rangle\geq
g(\langle\varphi^{\rho}|(X^{\gamma}\otimes \mathbbm{1})|\varphi^{\rho}\rangle),
\end{equation}
which means that $\frac{\D}{\D\alpha}\widehat{D}_{\alpha}
(\rho\Vert\sigma)\geq0$. Therefore, $\widehat{D}_{\alpha}(\rho\Vert\sigma)$ is
monotonically increasing in $\alpha$, as required.

\item \textit{Proof of additivity}:\ The proof of
\eqref{eq:additivity-geometric-renyi}\ is found by direct evaluation. Consider that
\begin{multline}
\lim_{\varepsilon_{1}\rightarrow0^{+}}\widehat{Q}_{\alpha}(\rho_{1}\Vert
\sigma_{1,\varepsilon_{1}})\cdot\lim_{\varepsilon_{2}\rightarrow0^{+}}
\widehat{Q}_{\alpha}(\rho_{2}\Vert\sigma_{2,\varepsilon_{2}})\\
=\lim_{\varepsilon_{1}\rightarrow0^{+}}\lim_{\varepsilon_{2}\rightarrow0^{+}
}\widehat{Q}_{\alpha}(\rho_{1}\Vert\sigma_{1,\varepsilon_{1}})\cdot\widehat
{Q}_{\alpha}(\rho_{2}\Vert\sigma_{2,\varepsilon_{2}}),
\end{multline}
where $\sigma_{1,\varepsilon_{1}} \coloneqq \sigma_1 + \varepsilon_{1} \mathbbm{1}$ and 
$\sigma_{2,\varepsilon_{2}} \coloneqq \sigma_2 + \varepsilon_{2} \mathbbm{1}$.
We then find that
\begin{align}
& \widehat{Q}_{\alpha}(\rho_{1}\Vert\sigma_{1,\varepsilon_{1}})\cdot
\widehat{Q}_{\alpha}(\rho_{2}\Vert\sigma_{2,\varepsilon_{2}})\notag \\
& =\operatorname{Tr}\!\left[  \sigma_{1,\varepsilon_{1}}\left(  \sigma
_{1,\varepsilon_{1}}^{-\frac{1}{2}}\rho_{1}\sigma_{1,\varepsilon_{1}}
^{-\frac{1}{2}}\right)  ^{\alpha}\right]  \operatorname{Tr}\!\left[
\sigma_{2,\varepsilon_{2}}\left(  \sigma_{2,\varepsilon_{2}}^{-\frac{1}{2}
}\rho_{2}\sigma_{2,\varepsilon_{2}}^{-\frac{1}{2}}\right)  ^{\alpha}\right]
\\
& =\operatorname{Tr}\!\left[  \sigma_{1,\varepsilon_{1}}\left(  \sigma
_{1,\varepsilon_{1}}^{-\frac{1}{2}}\rho_{1}\sigma_{1,\varepsilon_{1}}
^{-\frac{1}{2}}\right)  ^{\alpha}\otimes\sigma_{2,\varepsilon_{2}}\left(
\sigma_{2,\varepsilon_{2}}^{-\frac{1}{2}}\rho_{2}\sigma_{2,\varepsilon_{2}
}^{-\frac{1}{2}}\right)  ^{\alpha}\right]  \\
& =\operatorname{Tr}\!\left[  \left(  \sigma_{1,\varepsilon_{1}}\otimes
\sigma_{2,\varepsilon_{2}}\right)  \left(  \left(  \sigma_{1,\varepsilon_{1}
}^{-\frac{1}{2}}\rho_{1}\sigma_{1,\varepsilon_{1}}^{-\frac{1}{2}}\right)
^{\alpha}\otimes\left(  \sigma_{2,\varepsilon_{2}}^{-\frac{1}{2}}\rho
_{2}\sigma_{2,\varepsilon_{2}}^{-\frac{1}{2}}\right)  ^{\alpha}\right)
\right]  \\
& =\operatorname{Tr}\!\left[  \left(  \sigma_{1,\varepsilon_{1}}\otimes
\sigma_{2,\varepsilon_{2}}\right)  \left(  \sigma_{1,\varepsilon_{1}}
^{-\frac{1}{2}}\rho_{1}\sigma_{1,\varepsilon_{1}}^{-\frac{1}{2}}\otimes
\sigma_{2,\varepsilon_{2}}^{-\frac{1}{2}}\rho_{2}\sigma_{2,\varepsilon_{2}
}^{-\frac{1}{2}}\right)  ^{\alpha}\right]  \\
& =\operatorname{Tr}\!\left[  \left(  \sigma_{1,\varepsilon_{1}}\otimes
\sigma_{2,\varepsilon_{2}}\right)  \left(  \left(  \sigma_{1,\varepsilon_{1}
}\otimes\sigma_{2,\varepsilon_{2}}\right)  ^{-\frac{1}{2}}\left(  \rho
_{1}\otimes\rho_{2}\right)  \left(  \sigma_{1,\varepsilon_{1}}\otimes
\sigma_{2,\varepsilon_{2}}\right)  ^{-\frac{1}{2}}\right)  ^{\alpha}\right]
\\
& =\widehat{Q}_{\alpha}(\rho_{1}\otimes\rho_{2}\Vert\sigma_{1,\varepsilon_{1}
}\otimes\sigma_{2,\varepsilon_{2}}).
\end{align}
By considering that
\begin{equation}
\lim_{\varepsilon_{1}\rightarrow0^{+}}\lim_{\varepsilon_{2}\rightarrow0^{+}
}\sigma_{1,\varepsilon_{1}}\otimes\sigma_{2,\varepsilon_{2}}=\lim
_{\varepsilon\rightarrow0^{+}}\sigma_{1}\otimes\sigma_{2}+\varepsilon \mathbbm{1}\otimes
\mathbbm{1},
\end{equation}
along with continuity of the underlying functions,
we conclude that
\begin{multline}
\lim_{\varepsilon_{1}\rightarrow0^{+}}\lim_{\varepsilon_{2}\rightarrow0^{+}
}\widehat{Q}_{\alpha}(\rho_{1}\otimes\rho_{2}\Vert\sigma_{1,\varepsilon_{1}
}\otimes\sigma_{2,\varepsilon_{2}})\\
=\lim_{\varepsilon\rightarrow0^{+}}\widehat{Q}_{\alpha}(\rho_{1}\otimes
\rho_{2}\Vert\sigma_{1}\otimes\sigma_{2}+\varepsilon \mathbbm{1}\otimes
\mathbbm{1}).
\end{multline}
Finally, by applying the continuous function $\frac{1}{\alpha-1}\log_{2}(\cdot)$ to
all sides of the equalities established, we conclude that additivity holds.

\item \textit{Proof of direct-sum property}: Define the classical--quantum state $\rho_{XA}$ and
operator $\sigma_{XA}$, respectively, as
\begin{equation}
\rho_{XA}\coloneqq\sum_{x\in\mathcal{X}}p(x)|x\rangle\!\langle x|_{X}\otimes\rho
_{A}^{x},
\qquad
\sigma_{XA}\coloneqq\sum_{x\in\mathcal{X}}q(x)|x\rangle\!\langle
x|_{X}\otimes\sigma_{A}^{x}.
\end{equation}
Define
\begin{align}
\sigma_{XA}^{\varepsilon}  & \coloneqq\sum_{x\in\mathcal{X}}q(x)|x\rangle\!\langle
x|_{X}\otimes\sigma_{A}^{x}+\varepsilon \mathbbm{1}_{X}\otimes \mathbbm{1}_{A}\\
& =\sum_{x\in\mathcal{X},q(x)\neq0}q(x)|x\rangle\!\langle x|_{X}\otimes
\sigma_{A}^{x}+\varepsilon\sum_{x\in\mathcal{X}}|x\rangle\!\langle x|_{X}\otimes
\mathbbm{1}_{A}\\
& =\sum_{x\in\mathcal{X},q(x)\neq0}|x\rangle\!\langle x|_{X}\otimes
q(x)\sigma_{A,\varepsilon}^{x}+\sum_{x\in\mathcal{X},q(x)=0}|x\rangle\!\langle
x|_{X}\otimes\varepsilon \mathbbm{1}_{A},
\end{align}
where
\begin{equation}
\sigma_{A,\varepsilon}^{x}\coloneqq\sigma_{A}^{x}+\varepsilon \mathbbm{1}_{A}.
\end{equation}
Then we find that
\begin{multline}
\left(  \sigma_{XA}^{\varepsilon}\right)  ^{-\frac{1}{2}}=\sum_{x\in
\mathcal{X},q(x)\neq0}|x\rangle\!\langle x|_{X}\otimes\left(  q(x)\sigma
_{A,\varepsilon}^{x}\right)  ^{-\frac{1}{2}}+ \\ \sum_{x\in\mathcal{X}
,q(x)=0}|x\rangle\!\langle x|_{X}\otimes\varepsilon^{-\frac{1}{2}}\mathbbm{1}_{A},
\end{multline}
so that (omitting some lines of calculation)
\begin{multline}
\left(  \left(  \sigma_{XA}^{\varepsilon}\right)  ^{-\frac{1}{2}}\rho
_{XA}\left(  \sigma_{XA}^{\varepsilon}\right)  ^{-\frac{1}{2}}\right)
^{\alpha}\\
=\sum_{\substack{x\in\mathcal{X},q(x)\neq0,\\p(x)\neq0}}|x\rangle\!\langle
x|_{X}\otimes\left(  \frac{p(x)}{q(x)}\right)  ^{\alpha}\left(  \left(
\sigma_{A,\varepsilon}^{x}\right)  ^{-\frac{1}{2}}\rho_{A}^{x}\left(
\sigma_{A,\varepsilon}^{x}\right)  ^{-\frac{1}{2}}\right)  ^{\alpha}\\
+\sum_{\substack{x\in\mathcal{X},q(x)=0,\\p(x)\neq0}}|x\rangle\!\langle
x|_{X}\otimes\varepsilon^{-\alpha}(\rho_{A}^{x})^{\alpha}.
\end{multline}
Defining
\begin{equation}
\omega_{A}^{x}\coloneqq\left(  \sigma_{A,\varepsilon}^{x}\right)  ^{-\frac{1}{2}}
\rho_{A}^{x}\left(  \sigma_{A,\varepsilon}^{x}\right)  ^{-\frac{1}{2}},
\end{equation}
it then follows that
\begin{align}
& \widehat{Q}_{\alpha}(\rho_{XA}\Vert\sigma_{XA}^{\varepsilon})\nonumber\\
& =\operatorname{Tr}\!\left[  \sigma_{XA}^{\varepsilon}\left(  \left(
\sigma_{XA}^{\varepsilon}\right)  ^{-\frac{1}{2}}\rho_{XA}\left(  \sigma
_{XA}^{\varepsilon}\right)  ^{-\frac{1}{2}}\right)  ^{\alpha}\right]  \\
& =\operatorname{Tr}\!\left[  \left(  \sum_{\substack{x\in\mathcal{X}
,\\q(x)\neq0}}q(x)|x\rangle\!\langle x|_{X}\otimes\sigma_{A,\varepsilon}
^{x}\right)  \left(  \sum_{\substack{x^{\prime}\in\mathcal{X},\\q(x^{\prime
})\neq0,\\p(x^{\prime})\neq0}}|x^{\prime}\rangle\!\langle x^{\prime}|_{X}
\otimes\left(  \frac{p(x^{\prime})}{q(x^{\prime})}\right)  ^{\alpha}\left(
\omega_{A}^{x^{\prime}}\right)  ^{\alpha}\right)  \right]  \nonumber\\
& +\operatorname{Tr}\!\left[  \left(  \sum_{\substack{x\in\mathcal{X}
,\\q(x)\neq0}}q(x)|x\rangle\!\langle x|_{X}\otimes\sigma_{A,\varepsilon}
^{x}\right)  \left(  \sum_{\substack{x^{\prime}\in\mathcal{X},\\q(x^{\prime
})=0,\\p(x^{\prime})\neq0}}|x^{\prime}\rangle\!\langle x^{\prime}|_{X}
\otimes\varepsilon^{-\alpha}(\rho_{A}^{x^{\prime}})^{\alpha}\right)  \right]
\nonumber\\
& +\operatorname{Tr}\!\left[  \left(  \sum_{\substack{x\in\mathcal{X},\\q(x)=0}}
|x\rangle\!\langle x|_{X}\otimes\varepsilon \mathbbm{1}_{A}\right)  \left(  \sum
_{\substack{x^{\prime}\in\mathcal{X},\\q(x^{\prime})\neq0,\\p(x^{\prime}
)\neq0}}|x^{\prime}\rangle\!\langle x^{\prime}|_{X}\otimes\left(  \frac
{p(x^{\prime})}{q(x^{\prime})}\right)  ^{\alpha}\left(  \omega_{A}^{x^{\prime
}}\right)  ^{\alpha}\right)  \right]  \nonumber\\
& +\operatorname{Tr}\!\left[  \left(  \sum_{\substack{x\in\mathcal{X},\\q(x)=0}}
|x\rangle\!\langle x|_{X}\otimes\varepsilon \mathbbm{1}_{A}\right)  \left(  \sum
_{\substack{x^{\prime}\in\mathcal{X},q(x^{\prime})=0,\\p(x^{\prime})\neq
0}}|x^{\prime}\rangle\!\langle x^{\prime}|_{X}\otimes\varepsilon^{-\alpha}
(\rho_{A}^{x^{\prime}})^{\alpha}\right)  \right]  \\
& =\sum_{x\in\mathcal{X},q(x)\neq0,p(x)\neq0}p(x)^{\alpha}q(x)^{1-\alpha
}\operatorname{Tr}\!\left[  \sigma_{A,\varepsilon}^{x}\left(  \omega_{A}
^{x}\right)  ^{\alpha}\right]  \nonumber\\
& \qquad\qquad\qquad+\sum_{x\in\mathcal{X},q(x)=0,p(x)\neq0}\varepsilon^{1-\alpha}\Tr[(\rho_{A}^{x})^{\alpha}].
\end{align}
Now observing that $\operatorname{Tr}\!\left[  \sigma_{A,\varepsilon}^{x}\left(  \omega_{A}
^{x}\right)  ^{\alpha}\right] = \widehat{Q}_{\alpha}(\rho_{A}^{x}\Vert\sigma_{A,\varepsilon}^{x})$ and taking the limit $\varepsilon\rightarrow0^{+}$ in the last line above, we find that
\begin{multline}
\lim_{\varepsilon\rightarrow0^{+}}\left(  \sum_{\substack{x\in\mathcal{X},\\q(x)\neq
0,\\p(x)\neq0}}p(x)^{\alpha}q(x)^{1-\alpha}\widehat{Q}_{\alpha}(\rho_{A}^{x}
\Vert\sigma_{A,\varepsilon}^{x})+\sum_{\substack{x\in\mathcal{X},\\q(x)=0,\\p(x)\neq
0}}\varepsilon^{1-\alpha} \Tr[(\rho_{A}^{x})^{\alpha}]\right)  \\
=\sum_{x\in\mathcal{X}}p(x)^{\alpha}q(x)^{1-\alpha}\widehat{Q}_{\alpha}
(\rho_{A}^{x}\Vert\sigma_{A}^{x})
\end{multline}
if $\alpha\in(0,1)$ or if $\alpha\in(1,\infty)$, $\operatorname{supp}(\rho
_{A}^{x})\subseteq\operatorname{supp}(\sigma_{A}^{x})$, and there does not
exist a value of $x$ for which $p(x)\neq0$ and $q(x)=0$. The latter support
conditions are precisely the same as $\operatorname{supp}(\rho_{XA}
)\subseteq\operatorname{supp}(\sigma_{XA})$. If $\alpha \in (1,\infty)$ and the support conditions do not hold, then the limit evaluates to $+\infty$, consistent with the right-hand side above. This concludes the proof of the direct-sum property.\qedhere 

\end{enumerate}
\end{Proof}

We now establish the data-processing inequality for the geometric R\'{e}nyi
relative entropy for $\alpha\in\left(  0,1\right)  \cup(1,2]$. 

\begin{theorem*}
{Data-Processing Inequality for Geometric R\'{e}nyi Relative Entropy}
{thm:data-proc-geometric-renyi}Let $\rho$ be a state, $\sigma$ a
positive semi-definite operator, and $\mathcal{N}$ a quantum channel. Then,
for all $\alpha\in\left(  0,1\right)  \cup(1,2]$, 
\begin{equation}
\widehat{D}_{\alpha}(\rho\Vert\sigma)\geq\widehat{D}_{\alpha}(\mathcal{N}
(\rho)\Vert\mathcal{N}(\sigma)).
\end{equation}

\end{theorem*}

\begin{Proof}
From Stinespring's dilation theorem (Theorem~\ref{thm-q_channels}), we know that the action of a
quantum channel $\mathcal{N}$ on every linear operator $X$ can be written as
\begin{equation}
\mathcal{N}(X)=\operatorname{Tr}_{E}[VXV^{\dag}],
\end{equation}
where $V$ is an isometry and $E$ is an auxiliary system with dimension
$d_{E}\geq\ $rank$(\Gamma_{AB}^{\mathcal{N}})$, with $\Gamma_{AB}
^{\mathcal{N}}$ the Choi operator for the channel $\mathcal{N}$. As stated in
Proposition~\ref{prop:geometric-renyi-props}, the geometric R\'{e}nyi relative
entropy $\widehat{D}_{\alpha}$ is isometrically invariant. Therefore, it
suffices to establish the data-processing inequality for $\widehat{D}_{\alpha
}$ under partial trace; i.e., it suffices to show that for every state
$\rho_{AB}$, positive semi-definite operator $\sigma_{AB}$, and for all $\alpha\in\left(  0,1\right) \cup(1,2]$:
\begin{equation}
\widehat{D}_{\alpha}(\rho_{AB}\Vert\sigma_{AB})\geq\widehat{D}_{\alpha}
(\rho_{A}\Vert\sigma_{A}).
\end{equation}
We now proceed to prove this inequality. We prove it for the case where $\rho_{AB}$ and $\sigma_{AB}$, and
hence $\rho_{A}$ and $\sigma_A$, are invertible. 
The general result then follows 
by continuity, i.e., by 
first considering the invertible operators $\left(1-\delta\right) \rho_{AB}+\delta\pi_{AB}$ and $\sigma_{AB}+\varepsilon \mathbbm{1}_{AB}$, with $\delta\in(0,1)$ and $\varepsilon>0$, and then taking the limit $\delta\rightarrow0^{+}$ followed by $\varepsilon\rightarrow0^{+}$, because
\begin{align}
\widehat{D}_{\alpha}(\rho_{AB}\Vert\sigma_{AB})  &  =\lim_{\varepsilon
\rightarrow0^{+}}\lim_{\delta\rightarrow0^{+}}\widehat{D}_{\alpha}(\left(
1-\delta\right)  \rho_{AB}+\delta\pi_{AB}\Vert\sigma_{AB}+\varepsilon
\mathbbm{1}_{AB}),\\
\widehat{D}_{\alpha}(\rho_{A}\Vert\sigma_{A})  &  =\lim_{\varepsilon
\rightarrow0^{+}}\lim_{\delta\rightarrow0^{+}}\widehat{D}_{\alpha}(\left(
1-\delta\right)  \rho_{A}+\delta\pi_{A}\Vert\sigma_{A}+d_{B}\varepsilon
\mathbbm{1}_{A}),
\end{align}
which follows from \eqref{eq:limit-eq-geometric-Renyi} and the fact that the
dimensional factor $d_{B}$ does not affect the limit in the second quantity above.

To establish the data-processing inequality, we make use of the Petz recovery
channel for partial trace (see Section~\ref{sec:petz-recovery-partial-trace}), as well as the operator
Jensen inequality (Theorem~\ref{thm-Jensen}). Recall that the Petz recovery channel
$\mathcal{P}_{\sigma_{AB},\operatorname{Tr}_{B}}$ for partial trace is defined
as
\begin{equation}
\mathcal{P}_{\sigma_{AB},\operatorname{Tr}_{B}}(X_{A})\equiv\mathcal{P}
(X_{A})\coloneqq\sigma_{AB}^{\frac{1}{2}}\left(  \sigma_{A}^{-\frac{1}{2}}X_{A}
\sigma_{A}^{-\frac{1}{2}}\otimes \mathbbm{1}_{B}\right)  \sigma_{AB}^{\frac{1}{2}}.
\end{equation}
The Petz recovery channel satisfies
\begin{equation}
\mathcal{P}(\sigma_{A})=\sigma_{AB}, \label{eq:petz-map-recovers-sig-geo-DP}
\end{equation}
as one can verify by inspection. Since $\mathcal{P}_{\sigma_{AB}
,\operatorname{Tr}_{B}}$ is completely positive and trace preserving, it
follows that its adjoint
\begin{equation}
\mathcal{P}^{\dag}(Y_{AB})\coloneqq\sigma_{A}^{-\frac{1}{2}}\operatorname{Tr}
_{B}\left[\sigma_{AB}^{\frac{1}{2}}Y_{AB}\sigma_{AB}^{\frac{1}{2}}\right]\sigma
_{A}^{-\frac{1}{2}},
\end{equation}
is completely positive and unital. Observe that
\begin{equation}
\mathcal{P}^{\dag}\left(\sigma_{AB}^{-\frac{1}{2}}\rho_{AB}\sigma_{AB}^{-\frac
{1}{2}}\right)=\sigma_{A}^{-\frac{1}{2}}\rho_{A}\sigma_{A}^{-\frac{1}{2}}.
\label{eq:adjoint-petz-funny-prop-geo-DP}
\end{equation}
We then find for $\alpha\in(1,2]$ that
\begin{align}
\widehat{Q}_{\alpha}(\rho_{AB}\Vert\sigma_{AB})  &  =\operatorname{Tr}
\!\left[  \sigma_{AB}\!\left(  \sigma_{AB}^{-\frac{1}{2}}\rho_{AB}\sigma
_{AB}^{-\frac{1}{2}}\right)  ^{\alpha}\right] \\
&  =\operatorname{Tr}\!\left[  \mathcal{P}(\sigma_{A})\!\left(  \sigma
_{AB}^{-\frac{1}{2}}\rho_{AB}\sigma_{AB}^{-\frac{1}{2}}\right)  ^{\alpha
}\right] \\
&  =\operatorname{Tr}\!\left[  \sigma_{A}\mathcal{P}^{\dag}\!\left(\left(
\sigma_{AB}^{-\frac{1}{2}}\rho_{AB}\sigma_{AB}^{-\frac{1}{2}}\right)
^{\alpha}\right)\right] \\
&  \geq\operatorname{Tr}\!\left[  \sigma_{A}\!\left(  \mathcal{P}^{\dag
}\!\left(  \sigma_{AB}^{-\frac{1}{2}}\rho_{AB}\sigma_{AB}^{-\frac{1}{2}
}\right)  \right)  ^{\alpha}\right] \\
&  =\operatorname{Tr}\!\left[  \sigma_{A}\!\left(  \sigma_{A}^{-\frac{1}{2}
}\rho_{A}\sigma_{A}^{-\frac{1}{2}}\right)  ^{\alpha}\right] \\
&  =\widehat{Q}_{\alpha}(\rho_{A}\Vert\sigma_{A}).
\end{align}
The second equality follows from \eqref{eq:petz-map-recovers-sig-geo-DP}. The
sole inequality is a consequence of the operator Jensen inequality and the
fact that $x^{\alpha}$ is operator convex for $\alpha\in(1,2]$. Indeed, for
$\mathcal{M}$ a completely positive unital map, item 2.~of Theorem~\ref{thm-Jensen} implies that
\begin{equation}
f(\mathcal{M}(X))\leq\mathcal{M}(f(X)) \label{eq:Choi-thm-ext-op-jensen}
\end{equation}
for Hermitian $X$ and an operator convex function $f$. The second-to-last
equality follows from \eqref{eq:adjoint-petz-funny-prop-geo-DP}.

Applying the same reasoning as above, but using the fact that $x^{\alpha}$ is
operator concave for $\alpha\in(0,1)$, we find for $\alpha\in(0,1)$ that
\begin{equation}
\widehat{Q}_{\alpha}(\rho_{A}\Vert\sigma_{A})\geq\widehat{Q}_{\alpha}
(\rho_{AB}\Vert\sigma_{AB}).
\end{equation}
Putting together the above and employing definitions, we find that the
following inequality holds for $\alpha\in(0,1)\cup(1,2]$:
\begin{equation}
\widehat{D}_{\alpha}(\rho_{AB}\Vert\sigma_{AB})\geq\widehat{D}_{\alpha}
(\rho_{A}\Vert\sigma_{A}),
\end{equation}
concluding the proof.
\end{Proof}

With the data-processing inequality for the geometric R\'{e}nyi relative
entropy in hand, we can establish some additional properties.

\begin{proposition*}{Additional Properties of Geometric R\'enyi Relative Entropy}{prop:QEI:add-prop-geo-renyi}
The geometric
R\'enyi relative entropy $\widehat{D}_{\alpha}$ satisfies the following
properties for every state $\rho$ and positive semi-definite operator $\sigma$
for $\alpha\in\left(  0,1\right)  \cup(1,2]$:

\begin{enumerate}
\item If $\operatorname{Tr}[\sigma]\leq\operatorname{Tr}[\rho]=1$, then
$\widehat{D}_{\alpha}(\rho\Vert\sigma)\geq0$.

\item Faithfulness: Suppose that $\operatorname{Tr}[\sigma]\leq
\operatorname{Tr}[\rho]=1$ and let $\alpha\in(0,1)\cup(1,\infty)$. Then
$\widehat{D}_{\alpha}(\rho\Vert\sigma)=0$ if and only if $\rho=\sigma$.

\item For every positive semi-definite operator $\sigma^{\prime}$ such that
$\sigma^{\prime}\geq\sigma$, the following inequality holds: $
\widehat{D}_{\alpha}(\rho\Vert
\sigma) \geq \widehat{D}_{\alpha}(\rho\Vert\sigma^{\prime})$. In particular, if $\rho\leq\sigma$, then $\widehat{D}_{\alpha}(\rho\Vert\sigma)\leq 0$.
\end{enumerate}
\end{proposition*}

\begin{Proof}
\hfill
\begin{enumerate}
\item Apply the data-processing inequality with the channel being the full
trace-out channel:
\begin{align}
\widehat{D}_{\alpha}(\rho\Vert\sigma)  &  \geq\widehat{D}_{\alpha
}(\operatorname{Tr}[\rho]\Vert\operatorname{Tr}[\sigma])\\
&  =\frac{1}{\alpha-1}\log_2\!\left[  \left(  \operatorname{Tr}[\rho]\right)
^{\alpha}\left(  \operatorname{Tr}[\sigma]\right)  ^{1-\alpha}\right] \\
&  =-\log_2\operatorname{Tr}[\sigma]\\
&  \geq0.
\end{align}

\item If $\rho=\sigma$, then it follows by direct evaluation that $\widehat
{D}_{\alpha}(\rho\Vert\sigma)=0$ for $\alpha
\in(0,1)\cup(1,\infty)$.

To see the other implication, suppose first that $\left(  0,1\right)
\cup(1,2]$. Then $\widehat{D}_{\alpha}(\rho\Vert\sigma)=0$ implies that equality is achieved in the two inequalities in item~1.~above. So then $\Tr[\sigma]=1$. Furthermore, we conclude from data processing  that $\widehat{D}_{\alpha}(\mathcal{M}(\rho)\Vert\mathcal{M}(\sigma))=0$ for all
measurement channels $\mathcal{M}$. This includes the measurement that achieves the trace distance. By applying the faithfulness of the
classical R\'{e}nyi relative entropy on the distributions that result from the optimal trace-distance measurement, we conclude that $\rho=\sigma$. To get the range outside the
data-processing interval of $\left(  0,1\right)  \cup(1,2]$, note that
$\widehat{D}_{\alpha}(\rho\Vert\sigma)=0$ for $\alpha>2$ implies by
monotonicity (Property~2\ of Proposition~\ref{prop:geometric-renyi-props})
that $\widehat{D}_{\alpha}(\rho\Vert\sigma)=0$ for $\alpha\leq2$. Then it
follows that $\rho=\sigma$.


\item 
The condition $\sigma^{\prime}\geq\sigma$
implies that $\sigma^{\prime}-\sigma\geq0$. Then define the following positive
semi-definite operators:
\begin{align}
\hat{\rho}  &  \coloneqq|0\rangle\!\langle0|\otimes\rho,\\
\hat{\sigma}  &  \coloneqq|0\rangle\!\langle0|\otimes\sigma+|1\rangle\!\langle
1|\otimes\left(  \sigma^{\prime}-\sigma\right)  .
\end{align}
By exploiting the direct-sum property of geometric R\'{e}nyi relative entropy
(Proposition~\ref{prop:geometric-renyi-props}) and the data-processing
inequality (Theorem~\ref{thm:data-proc-geometric-renyi}), we find that
\begin{equation}
\widehat{D}_{\alpha}(\rho\Vert\sigma)=\widehat{D}_{\alpha}(\hat{\rho}\Vert
\hat{\sigma})\geq\widehat{D}_{\alpha}(\rho\Vert\sigma^{\prime}),
\end{equation}
where the inequality follows from data processing with respect to partial
trace over the classical register.

The second statement follows by substituting $\sigma \to \rho$ and $\sigma' \to \sigma$ and applying faithfulness.
\qedhere
\end{enumerate}
\end{Proof}

The data-processing inequality for the geometric R\'{e}nyi relative entropy
can be written using the geometric R\'{e}nyi relative quasi-entropy
$\widehat{Q}_{\alpha}(\rho\Vert\sigma)$ as
\begin{equation}
\frac{1}{\alpha-1} \log_2\widehat{Q}_{\alpha}(\rho\Vert\sigma)\geq\frac{1}
{\alpha-1} \log_2\widehat{Q}_{\alpha}(\mathcal{N}(\rho)\Vert\mathcal{N}(\sigma)).
\end{equation}
Since $\alpha-1$ is negative for $\alpha\in(0,1)$, we obtain:
\begin{align}
\widehat{Q}_{\alpha}(\rho\Vert\sigma)  &  \geq\widehat{Q}_{\alpha}
(\mathcal{N}(\rho)\Vert\mathcal{N}(\sigma)),\qquad\text{for }\alpha\in(1,2],\\
\widehat{Q}_{\alpha}(\rho\Vert\sigma)  &  \leq\widehat{Q}_{\alpha}
(\mathcal{N}(\rho)\Vert\mathcal{N}(\sigma)),\qquad\text{for }\alpha\in(0,1).
\end{align}
We can use this to establish some convexity/concavity statements for the geometric
R\'{e}nyi relative entropy.

\begin{proposition*}{Joint Convexity \& Concavity of the Geometric R\'enyi Relative Quasi-Entropy}{prop:QEI:joint-conv-conc-geometric-renyi}
Let $p\colon \mathcal{X}\rightarrow\left[  0,1\right]  $ be a probability
distribution over a finite alphabet $\mathcal{X}$ with associated $\left\vert
\mathcal{X}\right\vert $-dimensional system $X$, let $\left\{  \rho_{A}
^{x}\right\}_{x\in\mathcal{X}}  $ be a set of states on system $A$, and let
$\left\{  \sigma_{A}^{x}\right\}_{x\in\mathcal{X}}  $ be a set of positive
semi-definite operators on $A$. Then, for $\alpha\in(1,2]$,
\begin{equation}
\widehat{Q}_{\alpha}\!\left(  \sum_{x\in\mathcal{X}}p(x)\rho_{A}
^{x}\middle\Vert\sum_{x\in\mathcal{X}}p(x)\sigma_{A}^{x}\right)  \leq
\sum_{x\in\mathcal{X}}p(x)\widehat{Q}_{\alpha}(\rho_{A}^{x}\Vert\sigma_{A}
^{x}), \label{eq:convex-quasi-geometric-renyi-rel-alpha-higher-1}
\end{equation}
and, for $\alpha\in(0,1)$,
\begin{equation}
\widehat{Q}_{\alpha}\!\left(  \sum_{x\in\mathcal{X}}p(x)\rho_{A}
^{x}\middle\Vert\sum_{x\in\mathcal{X}}p(x)\sigma_{A}^{x}\right)  \geq
\sum_{x\in\mathcal{X}}p(x)\widehat{Q}_{\alpha}(\rho_{A}^{x}\Vert\sigma_{A}
^{x}).
\end{equation}
Consequently, the geometric R\'{e}nyi relative entropy $\widehat{D}_{\alpha}$
is jointly convex for $\alpha\in(0,1)$:
\begin{equation}
\widehat{D}_{\alpha}\!\left(  \sum_{x\in\mathcal{X}}p(x)\rho_{A}
^{x}\middle\Vert\sum_{x\in\mathcal{X}}p(x)\sigma_{A}^{x}\right)  \leq
\sum_{x\in\mathcal{X}}p(x)\widehat{D}_{\alpha}(\rho_{A}^{x}\Vert\sigma_{A}
^{x}). 
\end{equation}

\end{proposition*}

\begin{Proof}
The first two inequalities follow directly from the direct-sum property of the
geometric R\'{e}nyi relative entropy
(Proposition~\ref{prop:geometric-renyi-props}),  the data-processing
inequality (Theorem~\ref{thm:data-proc-geometric-renyi}), and Proposition~\ref{prop:QEI:joint-convexity-gen-div}. The last inequality
follows from the second by applying the logarithm, scaling by $1/\left(
\alpha-1\right)$, and exploiting convexity of the negative logarithm.
\end{Proof}

Although the geometric R\'{e}nyi relative entropy is not jointly convex for
$\alpha\in(1,2]\,$, it is jointly quasi-convex, in the sense that
\begin{equation}
\widehat{D}_{\alpha}\!\left(  \sum_{x\in\mathcal{X}}p(x)\rho_{A}
^{x}\middle\Vert\sum_{x\in\mathcal{X}}p(x)\sigma_{A}^{x}\right)  \leq
\max_{x\in\mathcal{X}}\widehat{D}_{\alpha}(\rho_{A}^{x}\Vert\sigma_{A}^{x}),
\label{eq:quasi-convex-quasi-geometric-renyi-rel-alpha-higher-1}
\end{equation}
for every finite alphabet $\mathcal{X}$, probability distribution $p\colon \mathcal{X}
\rightarrow\left[  0,1\right]  $, set $\left\{  \rho_{A}^{x}\right\}_{x\in \mathcal{X}}  $ of states, and set $\left\{  \sigma_{A}^{x}
\right\}_{x\in \mathcal{X}}  $ of positive semi-definite operators. Indeed, from
\eqref{eq:convex-quasi-geometric-renyi-rel-alpha-higher-1}, we immediately
obtain
\begin{equation}
\widehat{Q}_{\alpha}\!\left(  \sum_{x\in\mathcal{X}}p(x)\rho_{A}
^{x}\middle\Vert\sum_{x\in\mathcal{X}}p(x)\sigma_{A}^{x}\right)  \leq
\max_{x\in\mathcal{X}}\widehat{Q}_{\alpha}(\rho_{A}^{x}\Vert\sigma_{A}^{x}).
\end{equation}
Taking the logarithm and multiplying by $\frac{1}{\alpha-1}$ on both sides of
this inequality leads to \eqref{eq:quasi-convex-quasi-geometric-renyi-rel-alpha-higher-1}.

The geometric R\'{e}nyi relative entropy has another interpretation, which is worth mentioning.

\begin{proposition*}
{Geometric R\'{e}nyi Relative Entropy from Classical Preparations}
{prop:geometric-renyi-from-classical-preps}
Let $\rho$ be a state and $\sigma$ a positive semi-definite operator satisfying $\operatorname{supp} (\rho)\subseteq\operatorname{supp}(\sigma)$. For all $\alpha\in(0,1)\cup (1,2]$, the geometric R\'{e}nyi relative entropy is equal to the smallest R\'{e}nyi relative entropy between any two classical distributions that realize 
$\rho$ and $\sigma$ upon the application of a classical--quantum channel. That is, the following equality holds:
\begin{equation}
\widehat{D}_{\alpha}(\rho\Vert\sigma)=\inf_{\left\{  p,q,\mathcal{P}\right\}}\left\{  D_{\alpha}(p\Vert q):\mathcal{P}(\omega(p))=\rho,\, \mathcal{P}(\omega(q)) = \sigma \right\} , \label{eq:matsumoto-geo-renyi-preps}
\end{equation}
where the classical R\'{e}nyi relative entropy is defined in \eqref{eq-renyi_rel_ent_classical}, the channel $\mathcal{P}$ is a classical--quantum channel, $p\colon \mathcal{X} \rightarrow\left[  0,1\right]$ is a probability distribution over a finite alphabet $\mathcal{X}$, $q\colon \mathcal{X}\rightarrow[0,\infty)$ is a positive function on $\mathcal{X}$, $\omega(p)\coloneqq \sum_{x \in \mathcal{X} } p(x) |x\rangle\!\langle x|$,  and $\omega(q)\coloneqq \sum_{x \in \mathcal{X}} q(x) \ketbra{x}{x}$.
\end{proposition*}

\begin{Proof}
First,
suppose that there exists a quantum channel $\mathcal{P}$ such that
\begin{equation}
\mathcal{P}(\omega(p))=\rho,\qquad\mathcal{P}(\omega(q))=\sigma.
\label{eq:constraint-p-q-geo-renyi}
\end{equation}
Then consider the following chain of inequalities:
\begin{align}
D_{\alpha}(p\Vert q)  &  =\widehat{D}_{\alpha}(\omega(p)\Vert\omega(q))\\
&  \geq\widehat{D}_{\alpha}(\mathcal{P}(\omega(p))\Vert\mathcal{P}
(\omega(q)))\\
&  =\widehat{D}_{\alpha}(\rho\Vert\sigma).
\end{align}
The first equality follows because the geometric R\'{e}nyi relative entropy
reduces to the classical R\'{e}nyi relative entropy for commuting operators.
The inequality is a consequence of the data-processing inequality for the
geometric R\'{e}nyi relative entropy
(Theorem~\ref{thm:data-proc-geometric-renyi}). The final equality follows from
the constraint in \eqref{eq:constraint-p-q-geo-renyi}. Since the inequality
holds for arbitrary $p$, $q$, and $\mathcal{P}$ satisfying
\eqref{eq:constraint-p-q-geo-renyi}, we conclude that
\begin{equation}
\inf_{\left\{  p,q,\mathcal{P}\right\}  }\left\{  D_{\alpha}(p\Vert
q):\mathcal{P}(p)=\rho,\mathcal{P}(q)=\sigma\right\}  \geq\widehat{D}_{\alpha
}(\rho\Vert\sigma). \label{eq:ineq-classical-prep-bigger-geo-ren}
\end{equation}

The equality in \eqref{eq:matsumoto-geo-renyi-preps}\ then follows by demonstrating a specific distribution $p$, positive function $q$, and preparation channel $\mathcal{P}$ that saturate the inequality in
\eqref{eq:ineq-classical-prep-bigger-geo-ren}. Since $\supp(\rho) \subseteq \supp(\sigma)$ by assumption, up to restricting the support we can assume, without loss of generality, that $\sigma>0$ is invertible. The optimal choices of $p$,
$q$, and $\mathcal{P}$ are then given by
\begin{align}
p(x)  &  \coloneqq\lambda_{x}q(x),\label{eq:optimal-choices-classical-preps-1}\\
q(x)  &  \coloneqq\operatorname{Tr}[\Pi_{x}\sigma],\\
\mathcal{P}(\cdot)  &  \coloneqq\sum_{x}\langle x|(\cdot)|x\rangle\frac{\sigma
^{\frac{1}{2}}\Pi_{x}\sigma^{\frac{1}{2}}}{q(x)},
\label{eq:optimal-choices-classical-preps-3}
\end{align}
where the spectral decomposition of the positive semi-definite operator
$\sigma^{-\frac{1}{2}}\rho\sigma^{-\frac{1}{2}}$ is given by
\begin{equation}
\sigma^{-\frac{1}{2}}\rho\sigma^{-\frac{1}{2}}=\sum_{x}\lambda_{x}\Pi_{x}.
\label{eq:def-Delta-geo-renyi}
\end{equation}
The preparation channel $\mathcal{P}$ is a classical--quantum channel that
measures the input in the basis $\{|x\rangle\}_{x}$ and prepares the state
$\frac{\sigma^{\frac{1}{2}}\Pi_{x}\sigma^{\frac{1}{2}}}{q(x)}$ if the
measurement outcome is $x$. We find that
\begin{align}
\mathcal{P}(\omega(p)) &=\sum_{x}\frac{p(x)}{q(x)}\sigma^{\frac{1}{2}}
\Pi_{x}\sigma^{\frac{1}{2}} \\
&= \sum_{x}\frac{\lambda_{x}q(x)}{q(x)}\sigma
^{\frac{1}{2}}\Pi_{x}\sigma^{\frac{1}{2}} \\
&= \sigma^{\frac{1}{2}}\left(
\sum_{x}\lambda_{x}\Pi_{x}\right)  \sigma^{\frac{1}{2}} \\
&= \sigma^{\frac{1}{2}}\left(\sigma^{-\frac{1}{2}}\rho\sigma^{-\frac{1}{2}
}\right) \sigma^{\frac{1}{2}} \\
&= \rho,
\end{align}
and
\begin{align}
\mathcal{P}(\omega(q)) &=\sum_{x}\frac{q(x)}{q(x)}\sigma^{\frac{1}{2}}
\Pi_{x}\sigma^{\frac{1}{2}} \\
&= \sigma^{\frac{1}{2}}\left( \sum_{x}\Pi
_{x}\right) \sigma^{\frac{1}{2}} \\
&= \sigma.
\end{align}
Incidentally, this also shows that $p$ is a probability distribution, because
\begin{align}
\sum_{x}p(x) = \Tr\!\left[\omega(p)\right] = \Tr\!\left[ \mathcal{P}(\omega(p))\right] = \Tr\!\left[\rho\right] = 1. 
\end{align}
Finally, consider the classical R\'{e}nyi relative quasi-entropy:
\begin{align}
\sum_{x}p(x)^{\alpha}q(x)^{1-\alpha}  &  =\sum_{x}\left(  \lambda
_{x}q(x)\right)  ^{\alpha}q(x)^{1-\alpha} \\
&= \sum_{x}\lambda_{x}^{\alpha}q(x) \\
&= \sum_{x}\lambda_{x}^{\alpha}\operatorname{Tr}[\Pi_{x}\sigma] \\
&= \Tr\!\left[  \sigma\!\left(  \sum_{x}\lambda_{x}^{\alpha}
\Pi_{x}\right)  \right] \\
&= \Tr\!\left[  \sigma\!\left(
\sigma^{-\frac{1}{2}}\rho\sigma^{-\frac{1}{2}}\right)  ^{\alpha}\right] \\
&= \widehat{Q}_{\alpha}(\rho\Vert\sigma),
\end{align}
where the second-to-last equality follows from the spectral decomposition in
\eqref{eq:def-Delta-geo-renyi} and the form of the geometric R\'{e}nyi
relative quasi-entropy from
Proposition~\ref{prop:explicit-form-geometric-renyi}. As a consequence of the
equality
\begin{equation}
\sum_{x}p(x)^{\alpha}q(x)^{1-\alpha}=\widehat{Q}_{\alpha}(\rho\Vert\sigma),
\end{equation}
and the fact that these choices of $p$, $q$, and $\mathcal{P}$ satisfy the
constraints $\mathcal{P}(p)=\rho$ and $\mathcal{P}(q)=\sigma$, we conclude
that
\begin{equation}
D_{\alpha}(p\Vert q)=\widehat{D}_{\alpha}(\rho\Vert\sigma).
\end{equation}
Combining this equality with \eqref{eq:ineq-classical-prep-bigger-geo-ren}, we
conclude the equality in \eqref{eq:matsumoto-geo-renyi-preps}.
\end{Proof}

The following proposition establishes the ordering between the sandwiched, Petz--,
and geometric R\'{e}nyi relative entropies for the interval $\alpha
\in(0,1)\cup(1,2]$.  It follows by applying similar reasoning as in the proof of
Proposition~\ref{prop:geometric-renyi-from-classical-preps}.

\begin{proposition}{prop:sand-Petz-geo-ineqs}Let $\rho$ be a state and $\sigma$ a positive
semi-definite operator. For $\alpha\in(0,1)\cup(1,2]$, the following
inequalities hold:
\begin{equation}
\widetilde{D}_{\alpha}(\rho\Vert\sigma)\leq D_{\alpha}(\rho\Vert\sigma
)\leq\widehat{D}_{\alpha}(\rho\Vert\sigma),
\label{eq:sandwiched-Petz-geometric-ineqs}
\end{equation}
for the sandwiched ($\widetilde{D}_{\alpha}$), Petz ($D_{\alpha}$), and
geometric ($\widehat{D}_{\alpha}$) R\'{e}nyi relative entropies.
\end{proposition}

\begin{Proof}
The first inequality was stated as the last property of Proposition~\ref{prop-sand_rel_ent_properties}. So we establish
the proof of the second inequality here. Suppose that $\mathcal{P}$ is a
classical--quantum channel, $p\colon \mathcal{X}\rightarrow\left[  0,1\right]  $ is
a probability distribution over a finite alphabet $\mathcal{X}$, and
$q\colon \mathcal{X}\rightarrow(0,\infty)$ is a positive function on $\mathcal{X}$
satisfying
\begin{equation}
\mathcal{P}(\omega(p))=\rho,\qquad\mathcal{P}(\omega(q))=\sigma,
\label{eq:geo-constraint-up-bnd-Petz-R}
\end{equation}
where
\begin{equation}
\omega(p)\coloneqq\sum_{x\in\mathcal{X}}p(x)|x\rangle\!\langle x|,\qquad\omega
(q)\coloneqq\sum_{x\in\mathcal{X}}q(x)|x\rangle\!\langle x|.
\end{equation}
Then consider the following chain of inequalities:
\begin{align}
D_{\alpha}(p\Vert q)  &  =D_{\alpha}(\omega(p)\Vert\omega(q))\\
&  \geq D_{\alpha}(\mathcal{P}(\omega(p))\Vert\mathcal{P}(\omega(q)))\\
&  =D_{\alpha}(\rho\Vert\sigma).
\end{align}
The first equality follows because the Petz--R\'{e}nyi relative entropy
reduces to the classical R\'{e}nyi relative entropy for commuting operators.
The inequality follows from the data-processing inequality for the
Petz--R\'{e}nyi relative entropy for $\alpha\in(0,1)\cup(1,2]$ (Theorem~\ref{thm-petz_rel_ent_monotone}).
The final equality follows from the constraint in
\eqref{eq:geo-constraint-up-bnd-Petz-R}. Since the inequality above holds for
all $p$, $q$, and $\mathcal{P}$ satisfying
\eqref{eq:geo-constraint-up-bnd-Petz-R}, we conclude that
\begin{equation}
\inf_{\left\{  p,q,\mathcal{P}\right\}  }\left\{  D_{\alpha}(p\Vert
q):\mathcal{P}(p)=\rho,\mathcal{P}(q)=\sigma\right\}  \geq D_{\alpha}
(\rho\Vert\sigma).
\end{equation}
Now applying Proposition~\ref{prop:geometric-renyi-from-classical-preps}, we
conclude the second inequality in \eqref{eq:sandwiched-Petz-geometric-ineqs}.
\end{Proof}

\subsection{Proof of Proposition~\ref{lem:limit-exchange-geom-renyi-a-0-to1}}

\label{app:proof:limit-exchange-geom-renyi-a-0-to1}

First consider that
\begin{equation}
\left(  1-\delta\right)  \rho_{\delta}^{\prime}\leq\rho_{\delta}\leq
\rho_{\delta}^{\prime},
\end{equation}
where
\begin{equation}
\rho_{\delta}^{\prime}\coloneqq\rho+\delta\pi.
\end{equation}
By operator monotonicity of $x^{\alpha}$ for $\alpha\in\left(  0,1\right)  $,
we conclude that
\begin{align}
\left(  1-\delta\right)  ^{\alpha}\operatorname{Tr}\!\left[  \sigma
_{\varepsilon}\!\left(  \sigma_{\varepsilon}^{-\frac{1}{2}}\rho_{\delta
}^{\prime}\sigma_{\varepsilon}^{-\frac{1}{2}}\right)  ^{\alpha}\right]   &
\leq\operatorname{Tr}\!\left[  \sigma_{\varepsilon}\!\left(  \sigma
_{\varepsilon}^{-\frac{1}{2}}\rho_{\delta}\sigma_{\varepsilon}^{-\frac{1}{2}
}\right)  ^{\alpha}\right] \\
&  \leq\operatorname{Tr}\!\left[  \sigma_{\varepsilon}\!\left(  \sigma
_{\varepsilon}^{-\frac{1}{2}}\rho_{\delta}^{\prime}\sigma_{\varepsilon
}^{-\frac{1}{2}}\right)  ^{\alpha}\right]  .
\end{align}
The prefactors in these bounds to the left of the trace expressions are uniform and independent of $\varepsilon$, and so it follows
that
\begin{align}
\lim_{\varepsilon\rightarrow0^{+}}\lim_{\delta\rightarrow0^{+}}
\operatorname{Tr}\!\left[  \sigma_{\varepsilon}\!\left(  \sigma_{\varepsilon
}^{-\frac{1}{2}}\rho_{\delta}\sigma_{\varepsilon}^{-\frac{1}{2}}\right)
^{\alpha}\right]   &  =\lim_{\varepsilon\rightarrow0^{+}}\lim_{\delta
\rightarrow0^{+}}\operatorname{Tr}\!\left[  \sigma_{\varepsilon}\!\left(
\sigma_{\varepsilon}^{-\frac{1}{2}}\rho_{\delta}^{\prime}\sigma_{\varepsilon
}^{-\frac{1}{2}}\right)  ^{\alpha}\right]  ,\\
\lim_{\delta\rightarrow0^{+}}\lim_{\varepsilon\rightarrow0^{+}}
\operatorname{Tr}\!\left[  \sigma_{\varepsilon}\!\left(  \sigma_{\varepsilon
}^{-\frac{1}{2}}\rho_{\delta}\sigma_{\varepsilon}^{-\frac{1}{2}}\right)
^{\alpha}\right]   &  =\lim_{\delta\rightarrow0^{+}}\lim_{\varepsilon
\rightarrow0^{+}}\operatorname{Tr}\!\left[  \sigma_{\varepsilon}\!\left(
\sigma_{\varepsilon}^{-\frac{1}{2}}\rho_{\delta}^{\prime}\sigma_{\varepsilon
}^{-\frac{1}{2}}\right)  ^{\alpha}\right]  ,
\end{align}
provided that the above limits exist. Again from the operator monotonicity of $x^{\alpha}$ for $\alpha\in\left(
0,1\right)  $, we conclude for fixed $\varepsilon>0$ that
\begin{equation}
\delta_{1}\leq\delta_{2}\qquad\Rightarrow\qquad\operatorname{Tr}\!\left[
\sigma_{\varepsilon}\!\left(  \sigma_{\varepsilon}^{-\frac{1}{2}}\rho
_{\delta_{1}}^{\prime}\sigma_{\varepsilon}^{-\frac{1}{2}}\right)  ^{\alpha
}\right]  \leq\operatorname{Tr}\!\left[  \sigma_{\varepsilon}\!\left(
\sigma_{\varepsilon}^{-\frac{1}{2}}\rho_{\delta_{2}}^{\prime}\sigma
_{\varepsilon}^{-\frac{1}{2}}\right)  ^{\alpha}\right]  .
\end{equation}
By exploiting the identity
\begin{equation}
\operatorname{Tr}\!\left[  \sigma_{\varepsilon}\!\left(  \sigma_{\varepsilon
}^{-\frac{1}{2}}\rho_{\delta}^{\prime}\sigma_{\varepsilon}^{-\frac{1}{2}
}\right)  ^{\alpha}\right]  =\operatorname{Tr}\!\left[  \rho_{\delta}^{\prime
}\!\left(  \left(  \rho_{\delta}^{\prime}\right)  ^{-\frac{1}{2}}
\sigma_{\varepsilon}\left(  \rho_{\delta}^{\prime}\right)  ^{-\frac{1}{2}
}\right)  ^{1-\alpha}\right]
\end{equation}
from Proposition~\ref{prop:alt-rep-geometric-renyi-quasi} and operator
monotonicity of $x^{1-\alpha}$ for $\alpha\in\left(  0,1\right)  $, we
conclude for fixed $\delta>0$ that
\begin{equation}
\varepsilon_{1}\leq\varepsilon_{2}\qquad\Rightarrow\qquad\operatorname{Tr}
\!\left[  \sigma_{\varepsilon_{1}}\!\left(  \sigma_{\varepsilon_{1}}
^{-\frac{1}{2}}\rho_{\delta}\sigma_{\varepsilon_{1}}^{-\frac{1}{2}}\right)
^{\alpha}\right]  \leq\operatorname{Tr}\!\left[  \sigma_{\varepsilon_{2}
}\!\left(  \sigma_{\varepsilon_{2}}^{-\frac{1}{2}}\rho_{\delta}^{\prime}
\sigma_{\varepsilon_{2}}^{-\frac{1}{2}}\right)  ^{\alpha}\right]  ,
\end{equation}
where $\varepsilon_{1}>0$. Thus, we find that
\begin{align}
\lim_{\varepsilon\rightarrow0^{+}}\lim_{\delta\rightarrow0^{+}}
\operatorname{Tr}\!\left[  \sigma_{\varepsilon}\!\left(  \sigma_{\varepsilon
}^{-\frac{1}{2}}\rho_{\delta}^{\prime}\sigma_{\varepsilon}^{-\frac{1}{2}
}\right)  ^{\alpha}\right]   &  =\inf_{\varepsilon>0}\inf_{\delta
>0}\operatorname{Tr}\!\left[  \sigma_{\varepsilon}\!\left(  \sigma
_{\varepsilon}^{-\frac{1}{2}}\rho_{\delta}^{\prime}\sigma_{\varepsilon
}^{-\frac{1}{2}}\right)  ^{\alpha}\right]  ,\\
\lim_{\delta\rightarrow0^{+}}\lim_{\varepsilon\rightarrow0^{+}}
\operatorname{Tr}\!\left[  \sigma_{\varepsilon}\!\left(  \sigma_{\varepsilon
}^{-\frac{1}{2}}\rho_{\delta}^{\prime}\sigma_{\varepsilon}^{-\frac{1}{2}
}\right)  ^{\alpha}\right]   &  =\inf_{\delta>0}\inf_{\varepsilon
>0}\operatorname{Tr}\!\left[  \sigma_{\varepsilon}\!\left(  \sigma
_{\varepsilon}^{-\frac{1}{2}}\rho_{\delta}^{\prime}\sigma_{\varepsilon
}^{-\frac{1}{2}}\right)  ^{\alpha}\right]  .
\end{align}
(In particular, the limits on the left-hand side are well defined.) Since infima can be exchanged, we conclude the statement of the proposition.

\subsection{Proof of Proposition~\ref{prop:explicit-form-geometric-renyi}}

\label{app:QEI:explicit-form-geometric-renyi_pf}

First suppose
that $\alpha\in(1,\infty)$ and $\operatorname{supp}(\rho)\not \subseteq
\operatorname{supp}(\sigma)$. Then from 
Propositions~\ref{prop-sand_rel_ent_lim} and \ref{prop:geometric-to-sandwiched} and the fact that the
sandwiched R\'{e}nyi relative quasi-entropy $\widetilde{Q}_{\alpha}(\rho
\Vert\sigma)=+\infty$ in this case, it follows that $\widehat{Q}_{\alpha}
(\rho\Vert\sigma)=+\infty$, thus establishing the third expression in \eqref{eq:geometric-rel-quasi-explicit-1}.

Now suppose that $\alpha\in\left(  0,1\right)  \cup(1,\infty)$ and
$\operatorname{supp}(\rho)\subseteq\operatorname{supp}(\sigma)$. Let us employ
the decomposition of the Hilbert space $\mathcal{H}$ as $\mathcal{H}
=\operatorname{supp}(\sigma)\oplus\ker(\sigma)$.\ Then we can write $\rho$ as
\begin{equation}
\rho=
\begin{pmatrix}
\rho_{0,0} & \rho_{0,1}\\
\rho_{0,1}^{\dag} & \rho_{1,1}
\end{pmatrix}
,\qquad\sigma=
\begin{pmatrix}
\sigma_{0,0} & 0\\
0 & 0
\end{pmatrix}, \label{eq:rho-sig-decompose-for-geometric-formula}
\end{equation}
and we find that
\begin{equation} \label{eq:sigma_epsilon_decomposition}
\sigma_{\varepsilon} = \sigma + \varepsilon\mathbbm{1} =
\begin{pmatrix}
\sigma_{0,0} + \varepsilon\mathbbm{1} & 0\\
0 & \varepsilon\mathbbm{1}
\end{pmatrix}
,
\end{equation}
which implies that
\begin{equation}
\sigma_{\varepsilon}^{-\frac{1}{2}}=
\begin{pmatrix}
\left(\sigma_{0,0}+\varepsilon\mathbbm{1}\right)  ^{-\frac{1}{2}} & 0\\
0 & \varepsilon^{-\frac{1}{2}}\mathbbm{1}
\end{pmatrix}
. \label{eq:rho-sig-decompose-for-geometric-formula-3}
\end{equation}

The condition $\operatorname{supp}(\rho)\subseteq\operatorname{supp}(\sigma)$
implies that $\rho_{0,1}=0$ and $\rho_{1,1}=0$. Then
\begin{equation}
\sigma_{\varepsilon}^{-\frac{1}{2}}\rho\sigma_{\varepsilon}^{-\frac{1}{2}}=
\begin{pmatrix}
\left(  \sigma_{0,0}+\varepsilon\mathbbm{1}\right)  ^{-\frac{1}{2}}\rho
_{0,0}\left(  \sigma_{0,0}+\varepsilon\mathbbm{1}\right)  ^{-\frac{1}{2}} & 0\\
0 & 0
\end{pmatrix}
,
\end{equation}
so that
\begin{align}
&  \operatorname{Tr}\!\left[  \sigma_{\varepsilon}\!\left(  \sigma_{\varepsilon}^{-\frac{1}{2}}\rho\sigma_{\varepsilon}^{-\frac{1}{2}}\right)
^{\alpha}\right] \nonumber\\
&  =\operatorname{Tr}\!\left[
\begin{pmatrix}
\sigma_{0,0}+\varepsilon\mathbbm{1} & 0\\
0 & \varepsilon\mathbbm{1}
\end{pmatrix}
\begin{pmatrix}
\left[  \left(  \sigma_{0,0}+\varepsilon\mathbbm{1}\right)  ^{-\frac{1}{2}}
\rho_{0,0}\left(  \sigma_{0,0}+\varepsilon\mathbbm{1}\right)  ^{-\frac{1}{2}
}\right]  ^{\alpha} & 0\\
0 & 0
\end{pmatrix}
\right] \\
&  =\operatorname{Tr}\!\left[  \left(  \sigma_{0,0}+\varepsilon \mathbbm{1} \right)
\left[  \left(  \sigma_{0,0}+\varepsilon\mathbbm{1}\right)  ^{-\frac{1}{2}}
\rho_{0,0}\left(  \sigma_{0,0}+\varepsilon\mathbbm{1}\right)  ^{-\frac{1}{2}
}\right]  ^{\alpha}\right]  .
\end{align}
Taking the limit $\varepsilon\rightarrow0^{+}$ then leads to
\begin{align}
\lim_{\varepsilon\rightarrow0^{+}}\operatorname{Tr}\!\left[  \sigma
_{\varepsilon}\!\left(  \sigma_{\varepsilon}^{-\frac{1}{2}}\rho\sigma
_{\varepsilon}^{-\frac{1}{2}}\right)  ^{\alpha}\right]   &  =\operatorname{Tr}
\!\left[  \sigma_{0,0}\!\left(  \sigma_{0,0}^{-\frac{1}{2}}\rho_{0,0}\sigma_{0,0}^{-\frac{1}{2}
}\right)  ^{\alpha}\right] \\
&  =\operatorname{Tr}\!\left[  \sigma_{0,0}\!\left(  \sigma_{0,0}^{-\frac{1}{2}}\rho
\sigma_{0,0}^{-\frac{1}{2}}\right)  ^{\alpha}\right]  ,
\label{eq:rho-sig-decompose-for-geometric-formula-last}
\end{align}
thus establishing the first expression in \eqref{eq:geometric-rel-quasi-explicit-1}.

We now establish \eqref{eq:geometric-rel-quasi-explicit-2}. For $\alpha
\in(1,\infty)$ and $\operatorname{supp}(\rho)\subseteq\operatorname{supp}
(\sigma)$, the same analysis implies that
\begin{equation}
\operatorname{Tr}\!\left[  \sigma_{\varepsilon}\!\left(  \sigma_{\varepsilon
}^{-\frac{1}{2}}\rho\sigma_{\varepsilon}^{-\frac{1}{2}}\right)  ^{\alpha
}\right]  =\operatorname{Tr}\!\left[  \hat{\sigma}_{\varepsilon}\!\left(
\hat{\sigma}_{\varepsilon}^{-\frac{1}{2}}\rho_{0,0}\hat{\sigma}_{\varepsilon
}^{-\frac{1}{2}}\right)  ^{\alpha}\right]  ,
\end{equation}
where
\begin{equation}
\hat{\sigma}_{\varepsilon}\coloneqq\sigma_{0,0}+\varepsilon\mathbbm{1},
\end{equation}
and the inverses are taken on the support. Since
\begin{equation}
\left(  \hat{\sigma}_{\varepsilon}^{-\frac{1}{2}}\rho_{0,0}\hat{\sigma
}_{\varepsilon}^{-\frac{1}{2}}\right)  ^{\alpha}=\hat{\sigma}_{\varepsilon
}^{-\frac{1}{2}}\rho_{0,0}\hat{\sigma}_{\varepsilon}^{-\frac{1}{2}}\left(
\hat{\sigma}_{\varepsilon}^{-\frac{1}{2}}\rho_{0,0}\hat{\sigma}_{\varepsilon
}^{-\frac{1}{2}}\right)  ^{\alpha-1}
\end{equation}
for $\alpha>1$, we have that
\begin{align}
&  \operatorname{Tr}\!\left[  \hat{\sigma}_{\varepsilon}\hat{\sigma
}_{\varepsilon}^{-\frac{1}{2}}\rho_{0,0}\hat{\sigma}_{\varepsilon}^{-\frac
{1}{2}}\!\left(  \hat{\sigma}_{\varepsilon}^{-\frac{1}{2}}\rho_{0,0}
\hat{\sigma}_{\varepsilon}^{-\frac{1}{2}}\right)  ^{\alpha-1}\right]
\nonumber\\
&  =\operatorname{Tr}\!\left[  \hat{\sigma}_{\varepsilon}^{\frac{1}{2}}
\rho_{0,0}^{\frac{1}{2}}\rho_{0,0}^{\frac{1}{2}}\hat{\sigma}_{\varepsilon
}^{-\frac{1}{2}}\!\left(  \hat{\sigma}_{\varepsilon}^{-\frac{1}{2}}\rho
_{0,0}^{\frac{1}{2}}\rho_{0,0}^{\frac{1}{2}}\hat{\sigma}_{\varepsilon}
^{-\frac{1}{2}}\right)  ^{\alpha-1}\right] \\
&  =\operatorname{Tr}\!\left[  \hat{\sigma}_{\varepsilon}^{\frac{1}{2}}
\rho_{0,0}^{\frac{1}{2}}\!\left(  \rho_{0,0}^{\frac{1}{2}}\hat{\sigma
}_{\varepsilon}^{-\frac{1}{2}}\hat{\sigma}_{\varepsilon}^{-\frac{1}{2}}
\rho_{0,0}^{\frac{1}{2}}\right)  ^{\alpha-1}\rho_{0,0}^{\frac{1}{2}}
\hat{\sigma}_{\varepsilon}^{-\frac{1}{2}}\right] \\
&  =\operatorname{Tr}\!\left[  \rho_{0,0}\left(  \rho_{0,0}^{\frac{1}{2}}
\hat{\sigma}_{\varepsilon}^{-1}\rho_{0,0}^{\frac{1}{2}}\right)  ^{\alpha
-1}\right]  ,
\end{align}
where we applied Lemma~\ref{lem:sing-val-lemma-pseudo-commute}\ with
$f(x)=x^{\alpha-1}$ and $L=\rho_{0,0}^{\frac{1}{2}}\hat{\sigma}_{\varepsilon
}^{-\frac{1}{2}}$. Now taking the limit $\varepsilon\rightarrow0^{+}$, we
conclude that
\begin{align}
\lim_{\varepsilon\rightarrow0^{+}}\operatorname{Tr}\!\left[  \sigma
_{\varepsilon}\!\left(  \sigma_{\varepsilon}^{-\frac{1}{2}}\rho\sigma
_{\varepsilon}^{-\frac{1}{2}}\right)  ^{\alpha}\right]   &  =\lim
_{\varepsilon\rightarrow0^{+}}\operatorname{Tr}\!\left[  \rho_{0,0}\left(
\rho_{0,0}^{\frac{1}{2}}\hat{\sigma}_{\varepsilon}^{-1}\rho_{0,0}^{\frac{1}
{2}}\right)  ^{\alpha-1}\right] \\
&  =\operatorname{Tr}\!\left[  \rho_{0,0}\left(  \rho_{0,0}^{\frac{1}{2}
}\sigma^{-1}\rho_{0,0}^{\frac{1}{2}}\right)  ^{\alpha-1}\right] \\
&  =\operatorname{Tr}\!\left[  \rho\left(  \rho^{\frac{1}{2}}\sigma^{-1}
\rho^{\frac{1}{2}}\right)  ^{\alpha-1}\right]  ,
\end{align}
for the case $\alpha\in(1,\infty)$ and $\operatorname{supp}(\rho
)\subseteq\operatorname{supp}(\sigma)$, thus establishing \eqref{eq:geometric-rel-quasi-explicit-2}.

For the case that $\alpha\in(0,1)$ and $\operatorname{supp}(\sigma
)\subseteq\operatorname{supp}(\rho)$, we can employ the limit exchange from
Lemma~\ref{lem:limit-exchange-geom-renyi-a-0-to1}\ and a similar argument as
in
\eqref{eq:rho-sig-decompose-for-geometric-formula}--\eqref{eq:rho-sig-decompose-for-geometric-formula-last},
but with respect to the decomposition $\mathcal{H}=\operatorname{supp}
(\rho)\oplus\ker(\rho)$, to conclude that
\begin{equation}
\widehat{Q}_{\alpha}(\rho\Vert\sigma)=\operatorname{Tr}\!\left[  \rho\!\left(
\rho^{-\frac{1}{2}}\sigma\rho^{-\frac{1}{2}}\right)  ^{1-\alpha}\right]  ,
\end{equation}
thus establishing the second expression in
\eqref{eq:geometric-rel-quasi-explicit-1}. This case amounts to the exchange
$\rho\leftrightarrow\sigma$ and $\alpha\leftrightarrow1-\alpha$.

We finally consider the case $\alpha\in(0,1)$ and $\operatorname{supp}
(\rho)\not \subseteq \operatorname{supp}(\sigma)$, which is the most involved
case. 
We define $\sigma_{\varepsilon}$ as in~\eqref{eq:sigma_epsilon_decomposition}. Let us also construct 
\begin{equation}
\rho_{\delta}\coloneqq\left(  1-\delta\right)  \rho+\delta\pi,
\end{equation}
with $\delta\in(0,1)$ and $\pi$ the maximally mixed state. By invoking
Lemma~\ref{lem:limit-exchange-geom-renyi-a-0-to1}, we conclude that the
following exchange of limits is possible for $\alpha\in(0,1)$:
\begin{equation}
\lim_{\varepsilon\rightarrow0^{+}}\widehat{D}_{\alpha}(\rho\Vert\sigma_{\varepsilon
})=\lim_{\varepsilon\rightarrow0^{+}}\lim_{\delta\rightarrow0^{+}}\widehat{D}_{\alpha
}(\rho_{\delta}\Vert\sigma_{\varepsilon})=\lim_{\delta\rightarrow0^{+}}
\lim_{\varepsilon\rightarrow0^{+}}\widehat{D}_{\alpha}(\rho_{\delta}\Vert\sigma
_{\varepsilon}).
\end{equation}
Now write
\begin{equation}
\rho_{\delta}=
\begin{pmatrix}
\rho_{0,0}^{\delta} & \rho_{0,1}^{\delta}\\
(\rho_{0,1}^{\delta})^{\dag} & \rho_{1,1}^{\delta}
\end{pmatrix}
.
\end{equation}
Then
\begin{equation}
\widehat{D}_{\alpha}(\rho_{\delta}\Vert\sigma_{\varepsilon})=\frac{1}{\alpha-1}
\log_2\operatorname{Tr}\!\left[  \sigma_{\varepsilon}\!\left(  \sigma
_{\varepsilon}^{-\frac{1}{2}}\rho_{\delta}\sigma_{\varepsilon}^{-\frac{1}{2}
}\right)  ^{\alpha}\right]  .
\end{equation}
Consider that
\begin{align}
&\sigma_{\varepsilon}^{-\frac{1}{2}}\rho_{\delta}\sigma_{\varepsilon}
^{-\frac{1}{2}} \notag \\
&\ =
\begin{pmatrix}
\sigma_{0,0}+\varepsilon\mathbbm{1} & 0\\
0 & \varepsilon\mathbbm{1}
\end{pmatrix}^{-\frac{1}{2}}
\begin{pmatrix}
\rho_{0,0}^{\delta} & \rho_{0,1}^{\delta}\\
(\rho_{0,1}^{\delta})^{\dag} & \rho_{1,1}^{\delta}
\end{pmatrix}
\begin{pmatrix}
\sigma_{0,0} + \varepsilon\mathbbm{1} & 0\\
0 & \varepsilon\mathbbm{1}
\end{pmatrix}^{-\frac{1}{2}}\\
&\ =
\begin{pmatrix}
(\sigma_{0,0} + \varepsilon\mathbbm{1})^{-\frac{1}{2}} & 0\\
0 & \varepsilon^{-\frac{1}{2}}\mathbbm{1}
\end{pmatrix}
\begin{pmatrix}
\rho_{0,0}^{\delta} & \rho_{0,1}^{\delta}\\
(\rho_{0,1}^{\delta})^{\dag} & \rho_{1,1}^{\delta}
\end{pmatrix}
\begin{pmatrix}
(\sigma_{0,0} + \varepsilon\mathbbm{1})^{-\frac{1}{2}} & 0\\
0 & \varepsilon^{-\frac{1}{2}}\mathbbm{1}
\end{pmatrix}
\\
&\ =
\begin{pmatrix}
(\sigma_{0,0} + \varepsilon\mathbbm{1})^{-\frac{1}{2}}\rho_{0,0}^{\delta}(\sigma_{0,0} + \varepsilon\mathbbm{1})^{-\frac{1}{2}} & \varepsilon^{-\frac{1}{2}}(\sigma_{0,0} + \varepsilon\mathbbm{1})^{-\frac{1}{2}}\rho_{0,1}^{\delta} \\
\varepsilon^{-\frac{1}{2}}(\rho_{0,1}^{\delta})^{\dag}
(\sigma_{0,0} + \varepsilon\mathbbm{1})^{-\frac{1}{2}} & \varepsilon^{-1} \rho_{1,1}^{\delta}
\end{pmatrix}
\\
&\ =
\begin{pmatrix}
(\sigma_{0,0} + \varepsilon\mathbbm{1})^{-\frac{1}{2}}\rho_{0,0}^{\delta}(\sigma_{0,0} + \varepsilon\mathbbm{1})^{-\frac{1}{2}} & \varepsilon^{-\frac{1}{2}}(\sigma_{0,0} + \varepsilon\mathbbm{1})^{-\frac{1}{2}}\rho_{0,1}^{\delta}\\
\varepsilon^{-\frac{1}{2}}(\rho_{0,1}^{\delta})^{\dag}(\sigma_{0,0} + \varepsilon\mathbbm{1})^{-\frac{1}{2}} & \varepsilon^{-1}\rho_{1,1}^{\delta}
\end{pmatrix}
.
\end{align}
So then
\begin{align}
&  \operatorname{Tr}\!\left[  \sigma_{\varepsilon}\!\left(  \sigma
_{\varepsilon}^{-\frac{1}{2}}\rho_{\delta}\sigma_{\varepsilon}^{-\frac{1}{2}
}\right)  ^{\alpha}\right] \nonumber\\
&  =\operatorname{Tr}\!\left[
\begin{pmatrix}
\sigma_{0,0} \!+\! \varepsilon\mathbbm{1} & 0\\
0 & \varepsilon\mathbbm{1}
\end{pmatrix}
\begin{pmatrix}
(\sigma_{0,0} \!+\! \varepsilon\mathbbm{1})^{-\frac{1}{2}}\rho_{0,0}^{\delta}(\sigma_{0,0} \!+\! \varepsilon\mathbbm{1})^{-\frac{1}{2}} & \varepsilon^{-\frac{1}{2}}(\sigma_{0,0} \!+\! \varepsilon\mathbbm{1})^{-\frac{1}{2}}\rho_{0,1}^{\delta} \\
\varepsilon^{-\frac{1}{2}}(\rho_{0,1}^{\delta})^{\dag}(\sigma_{0,0} \!+\! \varepsilon\mathbbm{1})^{-\frac{1}{2}} & \varepsilon^{-1}\rho_{1,1}^{\delta}
\end{pmatrix}^{\alpha}\right] \\
&  =\operatorname{Tr}\!\left[
\begin{pmatrix}
\varepsilon^{-\alpha}(\sigma_{0,0} \!+\! \varepsilon\mathbbm{1}) & 0\\
0 & \varepsilon^{1-\alpha}\mathbbm{1}
\end{pmatrix}
(K(\varepsilon))^{\alpha}\right] ,
\end{align}
where, in the last line, we defined
\begin{equation}
K(\varepsilon)\coloneqq
\begin{pmatrix}
\varepsilon(\sigma_{0,0} + \varepsilon\mathbbm{1})^{-\frac{1}{2}}\rho_{0,0}^{\delta}
(\sigma_{0,0} + \varepsilon\mathbbm{1})^{-\frac{1}{2}} & \varepsilon^{\frac{1}{2}}
(\sigma_{0,0} + \varepsilon\mathbbm{1})^{-\frac{1}{2}}\rho_{0,1}^{\delta}\\
\varepsilon^{\frac{1}{2}}(\rho_{0,1}^{\delta})^{\dag}(\sigma_{0,0} + \varepsilon\mathbbm{1})^{-\frac{1}{2}} & \rho_{1,1}^{\delta}
\end{pmatrix}
.
\end{equation}
Now let us invoke Lemma~\ref{lem:sisi-zhou-lem} with the substitutions
\begin{align}
A  &  \leftrightarrow\rho_{1,1}^{\delta},\\
B  &  \leftrightarrow(\rho_{0,1}^{\delta})^{\dag} (\sigma_{0,0} + \varepsilon\mathbbm{1})^{-\frac{1}{2}},\\
C  &  \leftrightarrow (\sigma_{0,0} + \varepsilon\mathbbm{1})^{-\frac{1}{2}}\rho
_{0,0}^{\delta} (\sigma_{0,0} + \varepsilon\mathbbm{1})^{-\frac{1}{2}},\\
\varepsilon &  \leftrightarrow\varepsilon^{\frac{1}{2}}.
\end{align}
Defining
\begin{align}
L(\varepsilon)  &  \coloneqq
\begin{pmatrix}
\varepsilon S_{\delta, \varepsilon} & 0\\
0 & \rho_{1,1}^{\delta}+\varepsilon R
\end{pmatrix}
,\\
S_{\delta, \varepsilon}  &  \coloneqq (\sigma_{0,0} + \varepsilon\mathbbm{1})^{-\frac{1}{2}}\left(  \rho_{0,0}^{\delta}-\rho_{0,1}^{\delta}(\rho
_{1,1}^{\delta})^{-1}(\rho_{0,1}^{\delta})^{\dag}\right) (\sigma_{0,0} + \varepsilon\mathbbm{1})^{-\frac{1}{2}},\\
R  &  \coloneqq\operatorname{Re}\!\left[(\rho_{1,1}^{\delta})^{-1}(\rho_{0,1}^{\delta
})^{\dag}(\sigma_{0,0} + \varepsilon\mathbbm{1})^{-1}(\rho_{0,1}^{\delta})\right],
\end{align}
we conclude from Lemma~\ref{lem:sisi-zhou-lem}\ that
\begin{equation}
\left\Vert K(\varepsilon)-e^{-i\sqrt{\varepsilon}G}L(\varepsilon
)e^{i\sqrt{\varepsilon}G}\right\Vert _{\infty}\leq o(\varepsilon),
\label{eq:op-norm-bound-geo-support}
\end{equation}
where $G$ in Lemma~\ref{lem:sisi-zhou-lem}\ is defined from $A$ and $B$ above.
The inequality in \eqref{eq:op-norm-bound-geo-support} in turn implies the
following operator inequalities:
\begin{equation}
e^{-i\sqrt{\varepsilon}G}L(\varepsilon)e^{i\sqrt{\varepsilon}G}-o(\varepsilon
)\mathbbm{1}\leq K(\varepsilon)\leq e^{-i\sqrt{\varepsilon}G}L(\varepsilon
)e^{i\sqrt{\varepsilon}G}+o(\varepsilon)\mathbbm{1}.
\label{eq:op-ineq-geo-supp-renyi-bnd}
\end{equation}
Observe that
\begin{equation}
e^{-i\sqrt{\varepsilon}G}L(\varepsilon)e^{i\sqrt{\varepsilon}G}+o(\varepsilon
)\mathbbm{1}=e^{-i\sqrt{\varepsilon}G}\left[  L(\varepsilon)+o(\varepsilon)\mathbbm{1}\right]
e^{i\sqrt{\varepsilon}G}.
\end{equation}
Now invoking these and the operator monotonicity of the function $x^{\alpha}$
for $\alpha\in(0,1)$, we find that
\begin{align}
&  \operatorname{Tr}\!\left[  \sigma_{\varepsilon}\left(  \sigma_{\varepsilon
}^{-\frac{1}{2}}\rho_{\delta}\sigma_{\varepsilon}^{-\frac{1}{2}}\right)
^{\alpha}\right] \label{eq:geo-renyi-sup-arg-exp-to-bnd}\\
&  =\operatorname{Tr}\!\left[
\begin{pmatrix}
\varepsilon^{-\alpha}(\sigma_{0,0} + \varepsilon\mathbbm{1}) & 0\\
0 & \varepsilon^{1-\alpha}\mathbbm{1}
\end{pmatrix}
\left(  K(\varepsilon)\right)  ^{\alpha}\right] \\
&  \leq\operatorname{Tr}\!\left[
\begin{pmatrix}
\varepsilon^{-\alpha}(\sigma_{0,0} + \varepsilon\mathbbm{1}) & 0\\
0 & \varepsilon^{1-\alpha}\mathbbm{1}
\end{pmatrix}
\left(  e^{-i\sqrt{\varepsilon}G}\left[  L(\varepsilon)+o(\varepsilon
)\mathbbm{1}\right]  e^{i\sqrt{\varepsilon}G}\right)  ^{\alpha}\right] \\
&  =\operatorname{Tr}\!\left[
\begin{pmatrix}
\varepsilon^{-\alpha}(\sigma_{0,0} + \varepsilon\mathbbm{1}) & 0\\
0 & \varepsilon^{1-\alpha}\mathbbm{1}
\end{pmatrix}
e^{-i\sqrt{\varepsilon}G}\left(  L(\varepsilon)+o(\varepsilon)\mathbbm{1}\right)
^{\alpha}e^{i\sqrt{\varepsilon}G}\right]  . \label{eq:up-op-bnd-suppo-geo-ren}
\end{align}
Consider that
\begin{align}
&  \left(  L(\varepsilon)+o(\varepsilon)\mathbbm{1}\right)  ^{\alpha}\nonumber\\
&  =
\begin{pmatrix}
\varepsilon S_{\delta, \varepsilon} + o(\varepsilon)\mathbbm{1} & 0\\
0 & \rho_{1,1}^{\delta}+\varepsilon R+o(\varepsilon)\mathbbm{1}
\end{pmatrix}
^{\alpha}\\
&  =
\begin{pmatrix}
\left(  \varepsilon S_{\delta, \varepsilon}+o(\varepsilon
)\mathbbm{1}\right)  ^{\alpha} & 0\\
0 & \left(  \rho_{1,1}^{\delta}+\varepsilon R+o(\varepsilon)\mathbbm{1}\right)
^{\alpha}
\end{pmatrix}
\\
&  =
\begin{pmatrix}
\varepsilon^{\alpha}\left(  S_{\delta, \varepsilon}+o(1)\mathbbm{1}\right)  ^{\alpha} & 0\\
0 & \left(  \rho_{1,1}^{\delta}+\varepsilon R+o(\varepsilon)\mathbbm{1}\right)
^{\alpha}
\end{pmatrix}
.
\end{align}
Now expanding $e^{i\sqrt{\varepsilon}G}$ to first order in order to evaluate
\eqref{eq:up-op-bnd-suppo-geo-ren}\ (higher order terms will end up being
irrelevant), we find that
\begin{align}
&  \operatorname{Tr}\!\left[
\begin{pmatrix}
\varepsilon^{-\alpha}(\sigma_{0,0} + \varepsilon\mathbbm{1}) & 0\\
0 & \varepsilon^{1-\alpha}\mathbbm{1}
\end{pmatrix}
e^{-i\sqrt{\varepsilon}G}\left(  L(\varepsilon)+o(\varepsilon)\mathbbm{1}\right)
^{\alpha}e^{i\sqrt{\varepsilon}G}\right] \nonumber\\
&  =\operatorname{Tr}\!\left[
\begin{pmatrix}
\varepsilon^{-\alpha}(\sigma_{0,0} + \varepsilon\mathbbm{1}) & 0\\
0 & \varepsilon^{1-\alpha}\mathbbm{1}
\end{pmatrix}
\left(  L(\varepsilon)+o(\varepsilon)\mathbbm{1}\right)  ^{\alpha}\right] \nonumber\\
&  \quad+\operatorname{Tr}\!\left[
\begin{pmatrix}
\varepsilon^{-\alpha}(\sigma_{0,0} + \varepsilon\mathbbm{1}) & 0\\
0 & \varepsilon^{1-\alpha}\mathbbm{1}
\end{pmatrix}
\left(  -i\sqrt{\varepsilon}G\right)  \left(  L(\varepsilon)+o(\varepsilon
)\mathbbm{1}\right)  ^{\alpha}\right] \nonumber\\
&  \quad+\operatorname{Tr}\!\left[
\begin{pmatrix}
\varepsilon^{-\alpha}(\sigma_{0,0} + \varepsilon\mathbbm{1}) & 0\\
0 & \varepsilon^{1-\alpha}\mathbbm{1}
\end{pmatrix}
\left(  L(\varepsilon)+o(\varepsilon)\mathbbm{1}\right)  ^{\alpha}\left(  i\sqrt
{\varepsilon}G\right)  \right]  +o(1)\\
&  =\operatorname{Tr}\!\left[
\begin{pmatrix}
(\sigma_{0,0} + \varepsilon\mathbbm{1})\left(  S_{\delta, \varepsilon} +o(1)\mathbbm{1}\right)  ^{\alpha} & 0\\
0 & \varepsilon^{1-\alpha}\left(  \rho_{1,1}^{\delta
}+\varepsilon R+o(\varepsilon)\mathbbm{1}\right)  ^{\alpha}
\end{pmatrix}
\right] \nonumber\\
&  \quad-i\sqrt{\varepsilon}\operatorname{Tr}\!\left[
\begin{pmatrix}
\left(  S_{\delta, \varepsilon}+o(1)\mathbbm{1}\right)  ^{\alpha
}(\sigma_{0,0} + \varepsilon\mathbbm{1}) & 0\\
0 & \varepsilon^{1-\alpha}\left(  \rho_{1,1}^{\delta}+\varepsilon
R+o(\varepsilon)\mathbbm{1}\right)^{\alpha}
\end{pmatrix}
G\right] \nonumber\\
&  \quad+i\sqrt{\varepsilon}\operatorname{Tr}\!\left[
\begin{pmatrix}
(\sigma_{0,0} + \varepsilon\mathbbm{1})\left(  S_{\delta, \varepsilon}+o(1)\mathbbm{1}\right)  ^{\alpha} & 0\\
0 & \varepsilon^{1-\alpha}\left(  \rho_{1,1}^{\delta
}+\varepsilon R+o(\varepsilon)\mathbbm{1}\right)  ^{\alpha}
\end{pmatrix}
G\right]  +o(1)\\
&  =\operatorname{Tr}\!\left[
\begin{pmatrix}
(\sigma_{0,0} + \varepsilon\mathbbm{1})\left(  S_{\delta, \varepsilon} + o(1)\mathbbm{1}\right)  ^{\alpha} & 0\\
0 & \varepsilon^{1-\alpha}\left(\rho_{1,1}^{\delta
}+\varepsilon R+o(\varepsilon)\mathbbm{1}\right)  ^{\alpha}
\end{pmatrix}
\right]  +o(1)\\
&  =\operatorname{Tr}\!\left[  (\sigma_{0,0} + \varepsilon\mathbbm{1})\left(
S_{\delta, \varepsilon}+o(1)\mathbbm{1}\right)  ^{\alpha}\right]
+\varepsilon^{1-\alpha}\operatorname{Tr}\!\left[\left(
\rho_{1,1}^{\delta}+\varepsilon R+o(\varepsilon)\mathbbm{1}\right)^{\alpha}\right]+o(1).
\end{align}
By observing the last line, we see that higher order terms for $e^{i\sqrt
{\varepsilon}G}$ include prefactors of $\varepsilon$ (or higher powers), which
vanish in the $\varepsilon\rightarrow0^{+}$ limit. Now taking the limit
$\varepsilon\rightarrow0^{+}$, we find that
\begin{multline}
\lim_{\varepsilon\rightarrow0^{+}}\operatorname{Tr}\!\left[
\begin{pmatrix}
\varepsilon^{-\alpha}(\sigma_{0,0} + \varepsilon\mathbbm{1}) & 0\\
0 & \varepsilon^{1-\alpha}\mathbbm{1}
\end{pmatrix}
e^{-i\sqrt{\varepsilon}G}\left(  L(\varepsilon)+o(\varepsilon)\mathbbm{1}\right)
^{\alpha}e^{i\sqrt{\varepsilon}G}\right]
\label{eq:eps-final-limit-geo-ren-supp}\\
=\operatorname{Tr}\!\left[  \sigma\left(  \sigma^{-\frac{1}{2}}\left(
\rho_{0,0}^{\delta}-\rho_{0,1}^{\delta}(\rho_{1,1}^{\delta})^{-1}(\rho
_{0,1}^{\delta})^{\dag}\right)  \sigma^{-\frac{1}{2}}\right)  ^{\alpha
}\right]  ,
\end{multline}
where the inverses are taken on the support of $\sigma$. By proceeding in a
similar way, but using the lower bound in
\eqref{eq:op-ineq-geo-supp-renyi-bnd}, we find the following lower bound on
\eqref{eq:geo-renyi-sup-arg-exp-to-bnd}:
\begin{equation}
\operatorname{Tr}\!\left[
\begin{pmatrix}
\varepsilon^{-\alpha}(\sigma_{0,0} + \varepsilon\mathbbm{1}) & 0\\
0 & \varepsilon^{1-\alpha}\mathbbm{1}
\end{pmatrix}
e^{-i\sqrt{\varepsilon}G}\left(  L(\varepsilon)-o(\varepsilon)\mathbbm{1}\right)
^{\alpha}e^{i\sqrt{\varepsilon}G}\right]  .
\end{equation}
Then by the same argument above, the lower bound on
\eqref{eq:geo-renyi-sup-arg-exp-to-bnd} after taking the limit $\varepsilon
\rightarrow0^{+}$ is the same as in \eqref{eq:eps-final-limit-geo-ren-supp}.
So we conclude that
\begin{equation}
\lim_{\varepsilon\rightarrow0^{+}}\operatorname{Tr}\!\left[  \sigma
_{\varepsilon}\left(  \sigma_{\varepsilon}^{-\frac{1}{2}}\rho_{\delta}
\sigma_{\varepsilon}^{-\frac{1}{2}}\right)  ^{\alpha}\right]
=\operatorname{Tr}\!\left[  \sigma\left(  \sigma^{-\frac{1}{2}}\left(
\rho_{0,0}^{\delta}-\rho_{0,1}^{\delta}(\rho_{1,1}^{\delta})^{-1}(\rho
_{0,1}^{\delta})^{\dag}\right)  \sigma^{-\frac{1}{2}}\right)  ^{\alpha
}\right]  .
\end{equation}
Now consider that
\begin{equation}
\lim_{\delta\rightarrow0^{+}} \left(\rho_{0,0}^{\delta}-\rho_{0,1}^{\delta}
(\rho_{1,1}^{\delta})^{-1}(\rho_{0,1}^{\delta})^{\dag}\right) = \rho_{0,0}-\rho
_{0,1}^{\vphantom{dag}}\rho_{1,1}^{-1\vphantom{\dag}}\rho_{0,1}^{\dag},
\end{equation}
where the inverse on the right is taken on the support of $\rho_{1,1}$. This
follows because the image of $\rho_{0,1}^{\dag}$ is contained in the support
of $\rho_{1,1}$. Thus, we take the limit $\delta\rightarrow0^{+}$, and find
that
\begin{equation}
\lim_{\delta\rightarrow0^{+}}\lim_{\varepsilon\rightarrow0^{+}}
\operatorname{Tr}\!\left[  \sigma_{\varepsilon}\left(  \sigma_{\varepsilon
}^{-\frac{1}{2}}\rho_{\delta}\sigma_{\varepsilon}^{-\frac{1}{2}}\right)
^{\alpha}\right]  =\operatorname{Tr}\!\left[  \sigma\left(  \sigma^{-\frac
{1}{2}}\left(  \rho_{0,0}-\rho_{0,1}\rho_{1,1}^{-1}\rho_{0,1}^{\dag}\right)
\sigma^{-\frac{1}{2}}\right)  ^{\alpha}\right]  ,
\end{equation}
where all inverses are taken on the support. This concludes the proof.

\begin{Lemma}{lem:sisi-zhou-lem}
Let $A$ be an invertible Hermitian operator, $B$ a
linear operator, $C$ a Hermitian operator, and let $\varepsilon>0$.\ Then with
\begin{align}
M(\varepsilon)  &  \coloneqq 
\begin{bmatrix}
A & \varepsilon B\\
\varepsilon B^{\dag} & \varepsilon^{2}C
\end{bmatrix}
,\\
D(\varepsilon)  &  \coloneqq 
\begin{bmatrix}
A+\varepsilon^{2}\operatorname{Re}[A^{-1}BB^{\dag}] & 0\\
0 & \varepsilon^{2}\left(  C-B^{\dag}A^{-1}B\right)
\end{bmatrix}
,\\
G  &  \coloneqq 
\begin{bmatrix}
0 & -iA^{-1}B\\
iB^{\dag}A^{-1} & 0
\end{bmatrix}
,
\end{align}
the following inequality holds
\begin{equation}
\left\Vert M(\varepsilon)-e^{-i\varepsilon G}D(\varepsilon)e^{i\varepsilon
G}\right\Vert _{\infty}\leq o(\varepsilon^{2}). \label{eq:sis-zhou-lemm}
\end{equation}

\end{Lemma}

\begin{proof}
Observe that $G$ is Hermitian and consider that
\[
e^{i\varepsilon G}M(\varepsilon)e^{-i\varepsilon G}=\left(  I+i\varepsilon
G-\frac{\varepsilon^{2}}{2}G^{2}\right)  M(\varepsilon)\left(  I-i\varepsilon
G-\frac{\varepsilon^{2}}{2}G^{2}\right)  +o(\varepsilon^{2}).
\]
Then we find that
\begin{multline}
\left(  I+i\varepsilon G-\frac{\varepsilon^{2}}{2}G^{2}\right)  M(\varepsilon
)\left(  I-i\varepsilon G-\frac{\varepsilon^{2}}{2}G^{2}\right)
=M(\varepsilon)+i\varepsilon\left[  GM(\varepsilon)-M(\varepsilon)G\right] \\
+\varepsilon^{2}\left[  GM(\varepsilon)G-\frac{1}{2}G^{2}M(\varepsilon
)-\frac{1}{2}M(\varepsilon)G^{2}\right]  +o(\varepsilon^{2}).
\end{multline}
Now observe that
\begin{align}
GM(\varepsilon)  &  =
\begin{bmatrix}
0 & -iA^{-1}B\\
iB^{\dag}A^{-1} & 0
\end{bmatrix}
\begin{bmatrix}
A & \varepsilon B\\
\varepsilon B^{\dag} & \varepsilon^{2}C
\end{bmatrix}
\\
&  =
\begin{bmatrix}
-i\varepsilon A^{-1}BB^{\dag} & -i\varepsilon^{2}A^{-1}BC\\
iB^{\dag} & i\varepsilon B^{\dag}A^{-1}B
\end{bmatrix}
\\
&  =
\begin{bmatrix}
-i\varepsilon A^{-1}BB^{\dag} & o(\varepsilon)\\
iB^{\dag} & i\varepsilon B^{\dag}A^{-1}B
\end{bmatrix}
,\\
M(\varepsilon)G  &  =\left[  GM(\varepsilon)\right]  ^{\dag}\\
&  =
\begin{bmatrix}
i\varepsilon BB^{\dag}A^{-1} & -iB\\
o(\varepsilon) & -i\varepsilon B^{\dag}A^{-1}B
\end{bmatrix}
,
\end{align}
which implies that
\begin{align}
&  i\varepsilon\left[  GM(\varepsilon)-M(\varepsilon)G\right] \nonumber\\
&  =i\varepsilon\left(
\begin{bmatrix}
-i\varepsilon A^{-1}BB^{\dag} & o(\varepsilon)\\
iB^{\dag} & i\varepsilon B^{\dag}A^{-1}B
\end{bmatrix}
-
\begin{bmatrix}
i\varepsilon BB^{\dag}A^{-1} & -iB\\
o(\varepsilon) & -i\varepsilon B^{\dag}A^{-1}B
\end{bmatrix}
\right) \\
&  =
\begin{bmatrix}
2\varepsilon^{2}\operatorname{Re}[A^{-1}BB^{\dag}] & -\varepsilon
B+o(\varepsilon^{2})\\
-\varepsilon B^{\dag}+o(\varepsilon^{2}) & -2\varepsilon^{2}B^{\dag}A^{-1}B
\end{bmatrix}
.
\end{align}
Also, observe that
\begin{align}
GM(\varepsilon)G  &  =
\begin{bmatrix}
o(1) & o(\varepsilon)\\
iB^{\dag} & o(1)
\end{bmatrix}
\begin{bmatrix}
0 & -iA^{-1}B\\
iB^{\dag}A^{-1} & 0
\end{bmatrix}
\\
&  =
\begin{bmatrix}
o(\varepsilon) & o(1)\\
o(1) & B^{\dag}A^{-1}B
\end{bmatrix}
,\\
G^{2}M(\varepsilon)  &  =G[GM(\varepsilon)]\\
&  =
\begin{bmatrix}
0 & -iA^{-1}B\\
iB^{\dag}A^{-1} & 0
\end{bmatrix}
\begin{bmatrix}
o(1) & o(\varepsilon)\\
iB^{\dag} & o(1)
\end{bmatrix}
\\
&  =
\begin{bmatrix}
A^{-1}BB^{\dag} & o(1)\\
o(1) & o(\varepsilon)
\end{bmatrix}
,\\
M(\varepsilon)G^{2}  &  =\left[  G^{2}M(\varepsilon)\right]  ^{\dag}\\
&  =
\begin{bmatrix}
BB^{\dag}A^{-1} & o(1)\\
o(1) & o(\varepsilon)
\end{bmatrix}
.
\end{align}
So then we find that
\begin{align}
&  \varepsilon^{2}\left[  GM(\varepsilon)G-\frac{1}{2}G^{2}M(\varepsilon
)-\frac{1}{2}M(\varepsilon)G^{2}\right] \nonumber\\
&  =\varepsilon^{2}\left(
\begin{bmatrix}
o(\varepsilon) & o(1)\\
o(1) & B^{\dag}A^{-1}B
\end{bmatrix}
-\frac{1}{2}
\begin{bmatrix}
A^{-1}BB^{\dag} & o(1)\\
o(1) & o(\varepsilon)
\end{bmatrix}
-\frac{1}{2}
\begin{bmatrix}
BB^{\dag}A^{-1} & o(1)\\
o(1) & o(\varepsilon)
\end{bmatrix}
\right) \\
&  =
\begin{bmatrix}
-\varepsilon^{2}\operatorname{Re}[A^{-1}BB^{\dag}]+o(\varepsilon^{3}) &
o(\varepsilon^{2})\\
o(\varepsilon^{2}) & \varepsilon^{2}B^{\dag}A^{-1}B+o(\varepsilon^{3})
\end{bmatrix}
.
\end{align}
So then
\begin{align}
&  \left(  I+i\varepsilon G-\frac{\varepsilon^{2}}{2}G^{2}\right)
M(\varepsilon)\left(  I-i\varepsilon G-\frac{\varepsilon^{2}}{2}G^{2}\right)
\nonumber\\
&  =M(\varepsilon)+i\varepsilon\left[  GM(\varepsilon)-M(\varepsilon)G\right]
\nonumber\\
&  \qquad+\varepsilon^{2}\left[  GM(\varepsilon)G-\frac{1}{2}G^{2}
M(\varepsilon)-\frac{1}{2}M(\varepsilon)G^{2}\right]  +o(\varepsilon^{2})\\
&  =
\begin{bmatrix}
A & \varepsilon B\\
\varepsilon B^{\dag} & \varepsilon^{2}C
\end{bmatrix}
+
\begin{bmatrix}
2\varepsilon^{2}\operatorname{Re}[A^{-1}BB^{\dag}] & -\varepsilon
B+o(\varepsilon^{2})\\
-\varepsilon B^{\dag}+o(\varepsilon^{2}) & -2\varepsilon^{2}B^{\dag}A^{-1}B
\end{bmatrix}
\nonumber\\
&  \qquad+
\begin{bmatrix}
-\varepsilon^{2}\operatorname{Re}[A^{-1}BB^{\dag}]+o(\varepsilon^{3}) &
o(\varepsilon^{2})\\
o(\varepsilon^{2}) & \varepsilon^{2}B^{\dag}A^{-1}B+o(\varepsilon^{3})
\end{bmatrix}
+o(\varepsilon^{2})\\
&  =
\begin{bmatrix}
A+\varepsilon^{2}\operatorname{Re}[A^{-1}BB^{\dag}] & 0\\
0 & \varepsilon^{2}\left(  C-B^{\dag}A^{-1}B\right)
\end{bmatrix}
+o(\varepsilon^{2})\\
&  =D(\varepsilon)+o(\varepsilon^{2}).
\end{align}
So we conclude that
\begin{equation}
e^{i\varepsilon G}M(\varepsilon)e^{-i\varepsilon G}=D(\varepsilon
)+o(\varepsilon^{2}),
\end{equation}
which in turn implies that
\begin{equation}
M(\varepsilon)=e^{-i\varepsilon G}D(\varepsilon)e^{i\varepsilon G}
+o(\varepsilon^{2}),
\end{equation}
from which we conclude the claim in \eqref{eq:sis-zhou-lemm}.
\end{proof}

\section{Belavkin--Staszewski Relative Entropy}

\label{sec:QEI:Belavkin--Staszewski}

A different quantum generalization of the classical relative entropy is given
by the Belavkin--Staszewski\footnote{The name Staszewski is pronounced
Stah$\cdot$shev$\cdot$ski, with emphasis on the second syllable.} relative
entropy:

\begin{definition}
{Belavkin--Staszewski Relative Entropy}{def:belavkin-sta-rel-ent}The
Belavkin--Staszewski relative entropy of a quantum state $\rho$ and a positive
semi-definite operator $\sigma$ is defined as
\begin{equation}
\widehat{D}(\rho\Vert\sigma)\coloneqq\left\{
\begin{array}
[c]{cc}
\operatorname{Tr}\!\left[  \rho\log_2\!\left(  \rho^{\frac{1}{2}}\sigma^{-1}
\rho^{\frac{1}{2}}\right)  \right]  & \text{if }\operatorname{supp}
(\rho)\subseteq\operatorname{supp}(\sigma)\\
+\infty & \text{otherwise}
\end{array}
\right.  ,
\end{equation}
where the inverse $\sigma^{-1}$ is taken on the support of $\sigma$ and the
logarithm is evaluated on the support of $\rho$.
\end{definition}

This quantum generalization of classical relative entropy is not known to
possess an information-theoretic meaning. However, it is quite useful for obtaining
upper bounds on quantum channel capacities and quantum channel discrimination
rates, as considered in Part~\ref{part-feedback} of this book.

An important property of the Belavkin--Staszewski relative entropy is that it
is the limit of the geometric R\'{e}nyi relative entropy as $\alpha
\rightarrow1$.

\begin{proposition}{prop:BS-rel-ent-to-geometric}
Let $\rho$ be a state and $\sigma$ a
positive semi-definite operator.\ Then, in the limit $\alpha\rightarrow1$, the
geometric R\'{e}nyi relative entropy converges to the Belavkin--Staszewski
relative entropy:
\begin{equation}
\lim_{\alpha\rightarrow1}\widehat{D}_{\alpha}(\rho\Vert\sigma)=\widehat
{D}(\rho\Vert\sigma).
\end{equation}

\end{proposition}

\begin{Proof}
Suppose at first that $\operatorname{supp}(\rho)\subseteq\operatorname{supp}
(\sigma)$. Then $\widehat{D}_{\alpha}(\rho\Vert\sigma)$ is finite for all
$\alpha\in(0,1)\cup(1,\infty)$, and we can write the following explicit
formula for the geometric R\'{e}nyi relative entropy by employing
Proposition~\ref{prop:explicit-form-geometric-renyi}:
\begin{align}
\widehat{D}_{\alpha}(\rho\Vert\sigma)  &  =\frac{1}{\alpha-1} \log_2\widehat
{Q}_{\alpha}(\rho\Vert\sigma)\\
&  =\frac{1}{\alpha-1} \log_2\operatorname{Tr}\!\left[  \sigma\left(
\sigma^{-\frac{1}{2}}\rho\sigma^{-\frac{1}{2}}\right)  ^{\alpha}\right]  .
\end{align}
Our assumption implies that $\operatorname{Tr}[\Pi_{\sigma}\rho]=1$, and we
find that
\begin{align}
\widehat{Q}_{1}(\rho\Vert\sigma)  &  =\operatorname{Tr}\!\left[  \sigma\left(
\sigma^{-\frac{1}{2}}\rho\sigma^{-\frac{1}{2}}\right)  \right] \\
&  =\operatorname{Tr}[\Pi_{\sigma}\rho]\\
&  =1.
\end{align}
Since $\log_21=0$, we can write
\begin{equation}
\widehat{D}_{\alpha}(\rho\Vert\sigma)=\frac{ \log_2\widehat{Q}_{\alpha}(\rho
\Vert\sigma)- \log_2\widehat{Q}_{1}(\rho\Vert\sigma)}{\alpha-1},
\end{equation}
so that
\begin{align}
\lim_{\alpha\rightarrow1}\widehat{D}_{\alpha}(\rho\Vert\sigma)  &
=\lim_{\alpha\rightarrow1}\frac{ \log_2\widehat{Q}_{\alpha}(\rho\Vert\sigma)-
\log_2\widehat{Q}_{1}(\rho\Vert\sigma)}{\alpha-1}\\
&  =\left.  \frac{\D}{\D\alpha} \log_2\widehat{Q}_{\alpha}(\rho\Vert\sigma
)\right\vert _{\alpha=1}\\
&  =\frac{1}{\ln(2)}\frac{\left.  \frac{\D}{\D\alpha}\widehat{Q}_{\alpha}
(\rho\Vert\sigma)\right\vert _{\alpha=1}}{\widehat{Q}_{1}(\rho\Vert\sigma)}\\
&  =\frac{1}{\ln(2)}\left.  \frac{\D}{\D\alpha}\widehat{Q}_{\alpha}(\rho
\Vert\sigma)\right\vert _{\alpha=1}.
\end{align}
Then
\begin{align*}
\left.  \frac{\D}{\D\alpha}\widehat{Q}_{\alpha}(\rho\Vert\sigma)\right\vert
_{\alpha=1}  &  =\left.  \frac{\D}{\D\alpha}\operatorname{Tr}\!\left[
\sigma\left(  \sigma^{-\frac{1}{2}}\rho\sigma^{-\frac{1}{2}}\right)  ^{\alpha
}\right]  \right\vert _{\alpha=1}\\
&  =\left.  \operatorname{Tr}\!\left[  \sigma\frac{\D}{\D\alpha}\left(
\sigma^{-\frac{1}{2}}\rho\sigma^{-\frac{1}{2}}\right)  ^{\alpha}\right]
\right\vert _{\alpha=1}.
\end{align*}
For a positive semi-definite operator $X$ with spectral decomposition
\begin{equation}
X=\sum_{z}\nu_{z}\Pi_{z},
\end{equation}
we have
\begin{align}
\left.  \frac{\D}{\D\alpha}X^{\alpha}\right\vert _{\alpha=1}  &  =\left.
\frac{\D}{\D\alpha}\sum_{z}\nu_{z}^{\alpha}\Pi_{z}\right\vert _{\alpha=1}\\
&  =\sum_{z}\left(  \left.  \frac{\D}{\D\alpha}\nu_{z}^{\alpha}\right\vert
_{\alpha=1}\right)  \Pi_{z}\\
&  =\sum_{z}\left(  \left.  \nu_{z}^{\alpha}\ln \nu_{z}^{\alpha}\right\vert
_{\alpha=1}\right)  \Pi_{z}\\
&  =\sum_{z}\left(  \nu_{z}\ln \nu_{z}\right)  \Pi_{z}\\
&  =X\ln_{\ast}X,
\end{align}
where
\begin{equation}
\ln_{\ast}(x)\coloneqq\left\{
\begin{array}
[c]{cc}
\ln(x) & x>0\\
0 & x=0
\end{array}
\right.  . \label{eq:log-star-fnc-geometric-renyi}
\end{equation}
Thus we find that
\begin{align}
&  \left.  \operatorname{Tr}\!\left[  \sigma\frac{\D}{\D\alpha}\left(
\sigma^{-\frac{1}{2}}\rho\sigma^{-\frac{1}{2}}\right)  ^{\alpha}\right]
\right\vert _{\alpha=1}\nonumber\\
&  \qquad =\operatorname{Tr}\!\left[  \sigma\!\left(  \sigma^{-\frac{1}{2}}\rho
\sigma^{-\frac{1}{2}}\right)  \ln_{\ast}\!\left(  \sigma^{-\frac{1}{2}}
\rho\sigma^{-\frac{1}{2}}\right)  \right]
\label{eq:special-log-func-steps-BS-1}\\
&  \qquad =\operatorname{Tr}\!\left[  \sigma^{\frac{1}{2}}\rho^{\frac{1}{2}}
\rho^{\frac{1}{2}}\sigma^{-\frac{1}{2}}\ln_{\ast}\!\left(  \sigma^{-\frac
{1}{2}}\rho^{\frac{1}{2}}\rho^{\frac{1}{2}}\sigma^{-\frac{1}{2}}\right)
\right] \\
& \qquad =\operatorname{Tr}\!\left[  \sigma^{\frac{1}{2}}\rho^{\frac{1}{2}}\ln_{\ast}\!\left(  \rho^{\frac{1}{2}}\sigma^{-\frac{1}{2}}\sigma^{-\frac{1}{2}
}\rho^{\frac{1}{2}}\right)  \rho^{\frac{1}{2}}\sigma^{-\frac{1}{2}}\right] \\
& \qquad =\operatorname{Tr}\!\left[  \rho^{\frac{1}{2}}\Pi_{\sigma}\rho^{\frac{1}
{2}}\ln_{\ast}\!\left(  \rho^{\frac{1}{2}}\sigma^{-\frac{1}{2}}\sigma
^{-\frac{1}{2}}\rho^{\frac{1}{2}}\right)  \right] \\
& \qquad =\operatorname{Tr}\!\left[  \rho\ln\!\left(  \rho^{\frac{1}{2}}\sigma
^{-1}\rho^{\frac{1}{2}}\right)  \right]  .
\label{eq:special-log-func-steps-BS-last}
\end{align}
The third equality follows from Lemma~\ref{lem:sing-val-lemma-pseudo-commute}.
The final equality follows from the assumption $\operatorname{supp}
(\rho)\subseteq\operatorname{supp}(\sigma)$ and by applying the interpretation
of the logarithm exactly as stated in
Definition~\ref{def:belavkin-sta-rel-ent}. Then we find that
\begin{align}
\lim_{\alpha\rightarrow1}\widehat{D}_{\alpha}(\rho\Vert\sigma)  &  = \operatorname{Tr}\!\left[  \rho\log_2\!\left(  \rho^{\frac{1}{2}}
\sigma^{-1}\rho^{\frac{1}{2}}\right)  \right] ,
\end{align}
for the case in which $\operatorname{supp}(\rho)\subseteq\operatorname{supp}
(\sigma)$.

Now suppose that $\alpha\in(1,\infty)$ and $\operatorname{supp}(\rho
)\not \subseteq \operatorname{supp}(\sigma)$. Then $\widehat{D}_{\alpha}
(\rho\Vert\sigma)=+\infty$, so that $\lim_{\alpha\rightarrow1^{+}}\widehat
{D}_{\alpha}(\rho\Vert\sigma)=+\infty$, consistent with the definition of the
Belavkin--Staszewski relative entropy in this case (see
Definition~\ref{def:belavkin-sta-rel-ent}).

Suppose that $\alpha\in(0,1)$ and $\operatorname{supp}(\rho)\not \subseteq
\operatorname{supp}(\sigma)$. Employing
Proposition~\ref{prop:geometric-to-sandwiched}, we have that $\widehat
{D}_{\alpha}(\rho\Vert\sigma)\geq\widetilde{D}_{\alpha}(\rho\Vert\sigma)$ for
all $\alpha\in(0,1)$. Since $\lim_{\alpha\rightarrow1^{-}}\widetilde
{D}_{\alpha}(\rho\Vert\sigma)=+\infty$ in this case, it follows that also $\lim_{\alpha\rightarrow1^{-}}\widehat{D}_{\alpha
}(\rho\Vert\sigma)=+\infty$.

Therefore,
\begin{align}
\lim_{\alpha\rightarrow1^{-}}\widehat{D}_{\alpha}(\rho\Vert\sigma
) &= \left\{
\begin{array}
[c]{cc}
\operatorname{Tr}\!\left[  \rho\log_2\!\left(  \rho^{\frac{1}{2}}\sigma^{-1}
\rho^{\frac{1}{2}}\right)  \right]  & \text{if }\operatorname{supp}
(\rho)\subseteq\operatorname{supp}(\sigma)\\
+\infty & \text{otherwise}
\end{array}
\right. \\
&  = \widehat{D}(\rho\Vert\sigma).
\end{align}
To conclude, we have established that $\lim_{\alpha\rightarrow1^{+}}
\widehat{D}_{\alpha}(\rho\Vert\sigma)=\lim_{\alpha\rightarrow1^{-}}\widehat
{D}_{\alpha}(\rho\Vert\sigma)=\widehat{D}(\rho\Vert\sigma)$, which means that
\begin{equation}
\lim_{\alpha\rightarrow1}\widehat{D}_{\alpha}(\rho\Vert\sigma)=\widehat
{D}(\rho\Vert\sigma),
\end{equation}
as claimed.
\end{Proof}

The following inequality relates the quantum relative entropy to the
Belavkin--Staszewski relative entropy:

\begin{proposition}{cor:BS-to-q-rel-ent}
Let $\rho$ be a state and $\sigma$ a positive
semi-definite operator. Then the quantum relative entropy is never larger than
the Belavkin--Staszewski relative entropy:
\begin{equation}
D(\rho\Vert\sigma)\leq\widehat{D}(\rho\Vert\sigma).
\label{eq:BS-rel-ent-to-usual-rel-ent}
\end{equation}

\end{proposition}

\begin{Proof}
If $\operatorname{supp}(\rho)\not \subseteq \operatorname{supp}(\sigma)$, then
there is nothing to prove in this case because both
\begin{equation}
D(\rho\Vert\sigma)=\widehat{D}(\rho\Vert\sigma)=+\infty,
\end{equation}
and so the inequality in \eqref{eq:BS-rel-ent-to-usual-rel-ent} holds
trivially. Now, let us suppose instead that $\operatorname{supp}
(\rho)\subseteq\operatorname{supp}(\sigma)$. From
Propositions~\ref{prop:geometric-to-sandwiched} and
\ref{prop:explicit-form-geometric-renyi}, we conclude for all $\alpha
\in\left(  0,1\right)  \cup\left(  1,\infty\right)  $ that
\begin{equation}
\widetilde{D}_{\alpha}(\rho\Vert\sigma)\leq\widehat{D}_{\alpha}(\rho
\Vert\sigma). \label{eq:progress-BS-to-rel-ent-ineq}
\end{equation}
From Proposition~\ref{prop-sand_ren_ent_lim}, we know that
\begin{equation}
\lim_{\alpha\rightarrow1}\widetilde{D}_{\alpha}(\rho\Vert\sigma)=D(\rho
\Vert\sigma),
\end{equation}
while, from Proposition~\ref{prop:BS-rel-ent-to-geometric}, we know that
\begin{equation}
\lim_{\alpha\rightarrow1}\widehat{D}_{\alpha}(\rho\Vert\sigma)=\widehat
{D}(\rho\Vert\sigma).
\end{equation}
Thus, applying the limit $\alpha\rightarrow1$ to
\eqref{eq:progress-BS-to-rel-ent-ineq} and the two equalities above, we
conclude \eqref{eq:BS-rel-ent-to-usual-rel-ent}.
\end{Proof}

Similar to what was shown in Proposition~\ref{prop-rel_ent_lim},
Definition~\ref{def:belavkin-sta-rel-ent}\ is consistent with the following limit:

\begin{proposition}{prop:QEI:BS-rel-ent-limits-eps-delta}
For every state $\rho$ and positive semi-definite operator $\sigma$, we have
\begin{equation}
\widehat{D}(\rho\Vert\sigma)=\lim_{\varepsilon\rightarrow0^{+}}\lim
_{\delta\rightarrow0^{+}}\operatorname{Tr}\!\left[  \rho_{\delta}\log
_{2}\!\left(  \rho_{\delta}^{\frac{1}{2}}\sigma_{\varepsilon}^{-1}\rho
_{\delta}^{\frac{1}{2}}\right)  \right]  , \label{eq:limit-formula-BS-entropy}
\end{equation}
where $\delta\in\left(  0,1\right)  $ and
\begin{equation}
\rho_{\delta}\coloneqq\left(  1-\delta\right)  \rho+\delta\pi,\qquad\sigma
_{\varepsilon}\coloneqq\sigma+\varepsilon \mathbbm{1},
\end{equation}
with $\pi$ the maximally mixed state.
\end{proposition}

\begin{Proof}
Suppose first that $\operatorname{supp}(\rho)\subseteq\operatorname{supp}
(\sigma)$. We follow an approach similar to that given in the proof of
Proposition~\ref{prop:explicit-form-geometric-renyi}. Let us employ the
decomposition of the Hilbert space into $\operatorname{supp}(\sigma)\oplus
\ker(\sigma)$.\ Then we can write $\rho$ and $\sigma$ as in
\eqref{eq:rho-sig-decompose-for-geometric-formula}, so that
\begin{equation}
\sigma_{\varepsilon}^{-1}=
\begin{pmatrix}
\left(  \sigma_{0,0} + \varepsilon \mathbbm{1}\right)^{-1} & 0\\
0 & \varepsilon^{-1}\mathbbm{1}
\end{pmatrix}
,
\end{equation}
where we have followed the developments in
\eqref{eq:rho-sig-decompose-for-geometric-formula}--\eqref{eq:rho-sig-decompose-for-geometric-formula-3}.
The condition $\operatorname{supp}(\rho)\subseteq\operatorname{supp}(\sigma)$
implies that $\rho_{0,1}=0$ and $\rho_{1,1}=0$. It thus follows that
\begin{equation}
\lim_{\delta\rightarrow0^{+}}\rho_{\delta}=\begin{pmatrix}\rho_{0,0} & 0\\
0 & 0
\end{pmatrix}. \label{eq:BS-rel-ent-rho-delta-limit}     
\end{equation}
We then find that
\begin{align}
\operatorname{Tr}\!\left[  \rho_{\delta}\log_2\!\left(  \rho_{\delta}^{\frac
{1}{2}}\sigma_{\varepsilon}^{-1}\rho_{\delta}^{\frac{1}{2}}\right)  \right]
&  =\operatorname{Tr}\!\left[  \rho_{\delta}^{\frac{1}{2}}\sigma_{\varepsilon
}^{\frac{1}{2}}\sigma_{\varepsilon}^{-\frac{1}{2}}\rho_{\delta}^{\frac{1}{2}
}\log_2\!\left(  \rho_{\delta}^{\frac{1}{2}}\sigma_{\varepsilon}^{-\frac{1}{2}
}\sigma_{\varepsilon}^{-\frac{1}{2}}\rho_{\delta}^{\frac{1}{2}}\right)
\right] \\
&  =\operatorname{Tr}\!\left[  \rho_{\delta}^{\frac{1}{2}}\sigma_{\varepsilon
}^{\frac{1}{2}}\log_2\!\left(  \sigma_{\varepsilon}^{-\frac{1}{2}}\rho_{\delta
}^{\frac{1}{2}}\rho_{\delta}^{\frac{1}{2}}\sigma_{\varepsilon}^{-\frac{1}{2}
}\right)  \sigma_{\varepsilon}^{-\frac{1}{2}}\rho_{\delta}^{\frac{1}{2}
}\right] \\
&  =\operatorname{Tr}\!\left[  \log_2\!\left(  \sigma_{\varepsilon}^{-\frac{1}
{2}}\rho_{\delta}\sigma_{\varepsilon}^{-\frac{1}{2}}\right)  \left(
\sigma_{\varepsilon}^{-\frac{1}{2}}\rho_{\delta}\sigma_{\varepsilon}
^{-\frac{1}{2}}\right)  \sigma_{\varepsilon}\right] \\
&  =\operatorname{Tr}\!\left[  \sigma_{\varepsilon}\left(  \sigma
_{\varepsilon}^{-\frac{1}{2}}\rho_{\delta}\sigma_{\varepsilon}^{-\frac{1}{2}
}\right)  \log_2\!\left(  \sigma_{\varepsilon}^{-\frac{1}{2}}\rho_{\delta}
\sigma_{\varepsilon}^{-\frac{1}{2}}\right)  \right] \\
&  =\operatorname{Tr}\!\left[  \sigma_{\varepsilon}\eta\!\left(
\sigma_{\varepsilon}^{-\frac{1}{2}}\rho_{\delta}\sigma_{\varepsilon}
^{-\frac{1}{2}}\right)  \right]  ,
\end{align}
where the second equality holds due to
Lemma~\ref{lem:sing-val-lemma-pseudo-commute}, applied with $f=\log_2$ and $L=\rho
_{\delta}^{\frac{1}{2}}\sigma_{\varepsilon}^{-\frac{1}{2}}$, the third by cyclicity of the trace, and the fourth because $\sigma_{\varepsilon}^{-\frac{1}{2}}\rho_{\delta}\sigma_{\varepsilon}^{-\frac{1}{2}}$ commutes with $\log_2
\big(\sigma_{\varepsilon}^{-\frac{1}{2}}\rho_{\delta}\sigma_{\varepsilon}
^{-\frac{1}{2}}\big)$; we also employed once again the cyclicity of trace. In the last line, we
made use of the following function:
\begin{equation}
\eta(x)\coloneqq x\log_2 x,
\end{equation}
defined for all $x\in\lbrack0,\infty)$ with $\eta(0)=0$. By appealing to the
continuity of the function $\eta(x)$ on $x\in\lbrack0,\infty)$ and the fact
in \eqref{eq:BS-rel-ent-rho-delta-limit}, we find that
\begin{equation}
\lim_{\delta\rightarrow0^{+}}\operatorname{Tr}\!\left[  \sigma_{\varepsilon
}\eta\!\left(  \sigma_{\varepsilon}^{-\frac{1}{2}}\rho_{\delta}\sigma
_{\varepsilon}^{-\frac{1}{2}}\right)  \right]  =\operatorname{Tr}\!\left[
\hat{\sigma}_\varepsilon\eta\!\left(  \hat{\sigma}_\varepsilon^{-\frac{1}{2}}
\rho_{0,0}\hat{\sigma}_\varepsilon^{-\frac{1}{2}}\right)  \right]  ,
\end{equation}
where
\begin{equation}
    \hat{\sigma}_\varepsilon \coloneqq \sigma_{0,0} + \varepsilon \mathbbm{1}.
\end{equation}
Now recall the function $\log_{2,\ast}$ defined in
\eqref{eq:log-star-fnc-geometric-renyi}. Using it and again applying 
Lemma~\ref{lem:sing-val-lemma-pseudo-commute}, we can write
\begin{align}
&  \operatorname{Tr}\!\left[  \hat{\sigma}_\varepsilon\eta\!\left(  \hat{\sigma}_\varepsilon^{-\frac{1}{2}}\rho_{0,0}\hat{\sigma}_\varepsilon^{-\frac{1}{2}
}\right)  \right] \nonumber\\
&  =\operatorname{Tr}\!\left[  \hat{\sigma}_\varepsilon\hat{\sigma}_\varepsilon^{-\frac{1}{2}}\rho_{0,0}\hat{\sigma}_\varepsilon^{-\frac{1}{2}}\log_{2,\ast
}\!\left(  \hat{\sigma}_\varepsilon^{-\frac{1}{2}}\rho_{0,0}\hat{\sigma}_\varepsilon^{-\frac{1}{2}}\right)  \right] \\
&  =\operatorname{Tr}\!\left[  \hat{\sigma}_\varepsilon^{\frac{1}{2}}\rho
_{0,0}^{\frac{1}{2}}\rho_{0,0}^{\frac{1}{2}}\hat{\sigma}_\varepsilon^{-\frac{1}
{2}}\log_{2,\ast}\!\left(  \hat{\sigma}_\varepsilon^{-\frac{1}{2}}\rho_{0,0}
^{\frac{1}{2}}\rho_{0,0}^{\frac{1}{2}}\hat{\sigma}_\varepsilon^{-\frac{1}{2}
}\right)  \right] \\
&  =\operatorname{Tr}\!\left[  \hat{\sigma}_\varepsilon^{\frac{1}{2}}\rho
_{0,0}^{\frac{1}{2}}\log_{2,\ast}\!\left(  \rho_{0,0}^{\frac{1}{2}}
\hat{\sigma}_\varepsilon^{-\frac{1}{2}}\hat{\sigma}_\varepsilon^{-\frac{1}{2}}
\rho_{0,0}^{\frac{1}{2}}\right)  \rho_{0,0}^{\frac{1}{2}}\hat{\sigma}_\varepsilon^{-\frac{1}{2}}\right] \\
&  =\operatorname{Tr}\!\left[  \rho_{0,0}\log_{2,\ast}\!\left(  \rho_{0,0}
^{\frac{1}{2}}\hat{\sigma}_\varepsilon^{-1}\rho_{0,0}^{\frac{1}{2}}\right)
\right] \\
&  =\operatorname{Tr}\!\left[  \rho_{0,0}\log_{2,\ast}\!\left(  \rho_{0,0}
^{\frac{1}{2}}\left(  \sigma_{0,0} +\varepsilon \mathbbm{1}\right)  ^{-1}\rho
_{0,0}^{\frac{1}{2}}\right)  \right]  .
\end{align}
Now taking the limit as $\varepsilon\rightarrow0^{+}$, and appealing to
continuity of $\log_{2,\ast}(x)$ and $x^{-1}$ for $x>0$, we find that
\begin{align}
&  \lim_{\varepsilon\rightarrow 0^{+}}\operatorname{Tr}\!\left[  \rho_{0,0}
\log_{2,\ast}\!\left(  \rho_{0,0}^{\frac{1}{2}}\left(  \sigma_{0,0}+\varepsilon
\mathbbm{1}\right)^{-1}\rho_{0,0}^{\frac{1}{2}}\right)  \right] \nonumber\\
&  =\operatorname{Tr}\!\left[  \rho_{0,0}\log_{2,\ast}\!\left(  \rho_{0,0}
^{\frac{1}{2}}\,\sigma_{0,0}^{-1}\,\rho_{0,0}^{\frac{1}{2}}\right)  \right] \\
&  =\operatorname{Tr}\!\left[  \rho\log_2\!\left(  \rho^{\frac{1}{2}}\sigma
^{-1}\rho^{\frac{1}{2}}\right)  \right]
\end{align}
where the formula in the last line is interpreted exactly as stated in
Definition~\ref{def:belavkin-sta-rel-ent}. Thus, we conclude that
\begin{equation}
\lim_{\varepsilon\rightarrow0^{+}}\lim_{\delta\rightarrow0^{+}}
\operatorname{Tr}\!\left[  \rho_{\delta}\log_2\!\left(  \rho_{\delta}^{\frac
{1}{2}}\sigma_{\varepsilon}^{-1}\rho_{\delta}^{\frac{1}{2}}\right)  \right]
=\operatorname{Tr}\!\left[  \rho\log_2\!\left(  \rho^{\frac{1}{2}}\sigma^{-1}
\rho^{\frac{1}{2}}\right)  \right]  .
\end{equation}

Now suppose that $\operatorname{supp}(\rho)\not \subseteq \operatorname{supp}
(\sigma)$. Then applying Proposition~\ref{cor:BS-to-q-rel-ent}, we find that
the following inequality holds for all $\delta\in(0,1)$ and $\varepsilon>0$:
\begin{equation}
\widehat{D}(\rho_{\delta}\Vert\sigma_{\varepsilon})\geq D(\rho_{\delta}
\Vert\sigma_{\varepsilon}).
\end{equation}
Now taking limits and applying Proposition~\ref{prop-rel_ent_lim}, we find that
\begin{align}
\lim_{\varepsilon\rightarrow0^{+}}\lim_{\delta\rightarrow0^{+}}\widehat
{D}(\rho_{\delta}\Vert\sigma_{\varepsilon})  &  \geq\lim_{\varepsilon
\rightarrow0^{+}}\lim_{\delta\rightarrow0^{+}}D(\rho_{\delta}\Vert
\sigma_{\varepsilon})\\
&  =\lim_{\varepsilon\rightarrow0^{+}}D(\rho\Vert\sigma_{\varepsilon})\\
&  =+\infty.
\end{align}
This concludes the proof.
\end{Proof}

By taking the limit $\alpha\rightarrow1$ in the statement of the
data-processing inequality for $\widehat{D}_{\alpha}$, and applying
Proposition~\ref{prop:BS-rel-ent-to-geometric}, we immediately obtain the
data-processing inequality for the Belavkin--Staszewski relative entropy. 

\begin{corollary*}
{Data-Processing Inequality for Belavkin--Staszewski Relative Entropy}
{cor:DP-BS-rel-ent}Let $\rho$ be a state, $\sigma$ a positive
semi-definite operator, and $\mathcal{N}$ a quantum channel. Then
\begin{equation}
\widehat{D}(\rho\Vert\sigma)\geq\widehat{D}(\mathcal{N}(\rho)\Vert
\mathcal{N}(\sigma)).
\end{equation}

\end{corollary*}

Some basic properties of the Belavkin--Staszewski relative entropy are as follows:

\begin{proposition*}{Properties of Belavkin--Staszewski Relative Entropy}{prop:QEI:basic-props-BS-rel-ent}
The
Belavkin--Staszewski relative entropy satisfies the following properties for
states $\rho,\rho_{1},\rho_{2}$ and positive semi-definite operators
$\sigma,\sigma_{1},\sigma_{2}$.

\begin{enumerate}
\item \textit{Isometric invariance}: For every isometry $V$,
\begin{equation}
\widehat{D}(V\rho V^{\dagger}\Vert V\sigma V^{\dagger})=\widehat{D}(\rho
\Vert\sigma).
\end{equation}

\item
\begin{enumerate}
\item If $\operatorname{Tr}[\sigma]\leq1$, then $\widehat{D}(\rho\Vert
\sigma)\geq0$.

\item \textit{Faithfulness}: Suppose that $\operatorname{Tr}[\sigma
]\leq\operatorname{Tr}[\rho]=1$. Then $\widehat{D}(\rho\Vert\sigma)=0$ if and
only if $\rho=\sigma$.

\item If $\rho\leq\sigma$, then $\widehat{D}(\rho\Vert\sigma)\leq0$.

\item If $\sigma\leq\sigma^{\prime}$, then $\widehat{D}(\rho\Vert\sigma
)\geq\widehat{D}(\rho\Vert\sigma^{\prime})$.
\end{enumerate}

\item \textit{Additivity}:
\begin{equation}
\widehat{D}(\rho_{1}\otimes\rho_{2}\Vert\sigma_{1}\otimes\sigma_{2}
)=\widehat{D}(\rho_{1}\Vert\sigma_{1})+D(\rho_{2}\Vert\sigma_{2}).
\label{eq-BS_rel_ent_additivity}
\end{equation}
As a special case, for every $\beta\in(0,\infty)$,
\begin{equation}
\widehat{D}(\rho\Vert\beta\sigma)=\widehat{D}(\rho\Vert\sigma)+\log
_{2}\!\left(  \frac{1}{\beta}\right)  . \label{eq-BS_rel_ent_scalar_mult}
\end{equation}

\item \textit{Direct-sum property}: Let $p\colon \mathcal{X}\rightarrow\lbrack0,1]$
be a probability distribution over a finite alphabet $\mathcal{X}$ with
associated $|\mathcal{X}|$-dimensional system $X$, and let $q\colon \mathcal{X}
\rightarrow\lbrack0,\infty)$ be a positive function on $\mathcal{X}$. Let
$\{\rho_{A}^{x}:x\in\mathcal{X}\}$ be a set of states on a system $A$, and let
$\{\sigma_{A}^{x}:x\in\mathcal{X}\}$ be a set of positive semi-definite
operators on $A$. Then,
\begin{equation}
\widehat{D}(\rho_{XA}\Vert\sigma_{XA})=\widehat{D}(p\Vert q)+\sum
_{x\in\mathcal{X}}p(x)\widehat{D}(\rho_{A}^{x}\Vert\sigma_{A}^{x}).
\label{eq-BS_rel_ent_direct_sum}
\end{equation}
where
\begin{align}
\rho_{XA}  &  \coloneqq\sum_{x\in\mathcal{X}}p(x)|x\rangle\!\langle x|_{X}\otimes
\rho_{A}^{x},\\
\sigma_{XA}  &  \coloneqq\sum_{x\in\mathcal{X}}q(x)|x\rangle\!\langle x|_{X}
\otimes\sigma_{A}^{x}.
\end{align}

\end{enumerate}
\end{proposition*}

\begin{Proof}
\hfill
\begin{enumerate}
    \item Isometric invariance is a direct consequence of Propositions
\ref{prop:geometric-renyi-props}\ and \ref{prop:BS-rel-ent-to-geometric}.

\item All of the properties in the second item follow from data processing
(Corollary~\ref{cor:DP-BS-rel-ent}).
\begin{enumerate}
\item Applying the trace-out channel, we find
that
\begin{align}
\widehat{D}(\rho\Vert\sigma)  &  \geq\widehat{D}(\operatorname{Tr}[\rho
]\Vert\operatorname{Tr}[\sigma])\\
&  =\operatorname{Tr}[\rho] \log_2(\operatorname{Tr}[\rho]/\operatorname{Tr}
[\sigma])\\
&  =- \log_2\operatorname{Tr}[\sigma]\\
&  \geq0.
\end{align}

\item If $\rho=\sigma$, then it follows by direct evalution that $\widehat{D}
(\rho\Vert\sigma)=0$. If $\widehat{D}(\rho\Vert\sigma)=0$ and
$\operatorname{Tr}[\sigma]\leq1$, then $D(\rho\Vert\sigma)=0$ by
Proposition~\ref{cor:BS-to-q-rel-ent}\ and we conclude that $\rho=\sigma$ from
faithfulness of the quantum relative entropy (Proposition~\ref{prop-rel_ent}).

\item If $\rho\leq\sigma$, then $\sigma-\rho$ is positive semi-definite, and the
following operator is positive semi-definite:
\begin{equation}
\hat{\sigma}\coloneqq|0\rangle\!\langle0|\otimes\rho+|1\rangle\!\langle1|\otimes\left(
\sigma-\rho\right)  .
\end{equation}
Defining $\hat{\rho}\coloneqq|0\rangle\!\langle0|\otimes\rho$, by a straightforward calculation we find 
\begin{equation}
0=\widehat{D}(\rho\Vert\rho)=\widehat{D}(\hat{\rho}\Vert\hat{\sigma}
)\geq\widehat{D}(\rho\Vert\sigma),
\end{equation}
where the inequality follows from data processing by tracing out the first
classical register of $\hat{\rho}$ and $\hat{\sigma}$.

\item If $\sigma\leq\sigma^{\prime}$, then the operator $\sigma^{\prime}-\sigma$ is
positive semi-definite and so is the following one:
\begin{equation}
\hat{\sigma}\coloneqq|0\rangle\!\langle0|\otimes\sigma+|1\rangle\!\langle1|\otimes\left(
\sigma^{\prime}-\sigma\right)  .
\end{equation}
Defining $\hat{\rho}\coloneqq|0\rangle\!\langle0|\otimes\rho$, we find from the
direct-sum property that
\begin{equation}
\widehat{D}(\rho\Vert\sigma)=\widehat{D}(\hat{\rho}\Vert\hat{\sigma}
)\geq\widehat{D}(\rho\Vert\sigma^{\prime}),
\end{equation}
where the inequality follows from data processing by tracing out the first
classical register of $\hat{\rho}$ and $\hat{\sigma}$.
\end{enumerate}

\item Additivity follows by direct evaluation.

\item The direct-sum property follows by direct evaluation. \qedhere

\end{enumerate}
\end{Proof}

A statement similar to that made by
Proposition~\ref{prop:geometric-renyi-from-classical-preps} holds for the
Belavkin--Staszewski relative entropy:

\begin{proposition*}{Belavkin--Staszewski Relative Entropy from Classical Preparations}{prop:QEI:BS-ent-from-classical}
Let $\rho$
be a state and $\sigma$ a positive semi-definite operator satisfying
$\operatorname{supp}(\rho)\subseteq\operatorname{supp}(\sigma)$. The
Belavkin--Staszewski relative entropy is equal to the smallest value that the
classical relative entropy can take by minimizing over classical--quantum
channels that realize the state $\rho$ and the positive semi-definite operator
$\sigma$. That is, the following equality holds:
\begin{equation}
\widehat{D}(\rho\Vert\sigma)=\inf_{\left\{  p,q,\mathcal{P}\right\}  }\left\{
D(p\Vert q):\mathcal{P}(\omega(p))=\rho,\, \mathcal{P}(\omega(q))=\sigma\right\}  ,
\label{eq:BS-rel-ent-equality-classical-preps}
\end{equation}
where the classical relative entropy is defined in \eqref{eq:classical-rel-ent},
the channel $\mathcal{P}$ is a classical--quantum channel, $p\colon \mathcal{X}
\rightarrow\left[  0,1\right]  $ is a probability distribution over a finite
alphabet $\mathcal{X}$,  $q\colon \mathcal{X}\rightarrow[0,\infty)$ is a positive
function on $\mathcal{X}$, $\omega(p)\coloneqq \sum_{x \in \mathcal{X} } p(x) |x\rangle\!\langle x|$,  and $\omega(q)\coloneqq \sum_{x \in \mathcal{X} } q(x) |x\rangle\!\langle x|$.
\end{proposition*}

\begin{Proof}
The proof is very similar to the proof of
Proposition~\ref{prop:geometric-renyi-from-classical-preps}, and so we use the
same notation to provide a brief proof. By following the same reasoning that
leads to \eqref{eq:ineq-classical-prep-bigger-geo-ren}, it follows that
\begin{equation}
\inf_{\left\{  p,q,\mathcal{P}\right\}  }\left\{  D(p\Vert q):\mathcal{P}
(p)=\rho,\mathcal{P}(q)=\sigma\right\}  \geq\widehat{D}(\rho\Vert\sigma).
\label{eq:BS-rel-ent-lower-bound-classical-preps}
\end{equation}
The optimal choices of $p$, $q$, and $\mathcal{P}$ saturating the inequality
in \eqref{eq:BS-rel-ent-lower-bound-classical-preps} are again given by
\eqref{eq:optimal-choices-classical-preps-1}--\eqref{eq:optimal-choices-classical-preps-3}.
Consider for those choices that
\begin{align}
\sum_{x}p(x) \log_2\!\left(  \frac{p(x)}{q(x)}\right)   &  =\sum_{x}p(x)
\log_2\!\left(  \lambda_{x}\right) \\
&  =\sum_{x}\lambda_{x}q(x) \log_2\!\left(  \lambda_{x}\right) \\
&  =\sum_{x}\lambda_{x}\operatorname{Tr}[\Pi_{x}\sigma] \log_2\!\left(
\lambda_{x}\right) \\
&  =\operatorname{Tr}\!\left[  \sigma\left(  \sum_{x}\lambda_{x}\log
_{2}\!\left(  \lambda_{x}\right)  \Pi_{x}\right)  \right] \\
&  =\operatorname{Tr}\!\left[  \sigma\left(  \sigma^{-\frac{1}{2}}\rho
\sigma^{-\frac{1}{2}}\right)  \log_2\!\left(  \sigma^{-\frac{1}{2}}\rho
\sigma^{-\frac{1}{2}}\right)  \right] \\
&  =\operatorname{Tr}\!\left[  \rho\log_2\!\left(  \rho^{\frac{1}{2}}\sigma
^{-1}\rho^{\frac{1}{2}}\right)  \right]  ,
\end{align}
where the last equality follows from reasoning similar to that used to justify
\eqref{eq:special-log-func-steps-BS-1}--\eqref{eq:special-log-func-steps-BS-last}.
Then by following the reasoning at the end of the proof of
Proposition~\ref{prop:geometric-renyi-from-classical-preps}, we conclude \eqref{eq:BS-rel-ent-equality-classical-preps}.
\end{Proof}

\section{Max-Relative Entropy}

\label{sec-max_rel_ent}

	An important generalized divergence that appears in the context of placing upper bounds on communication rates of feedback-assisted protocols is the max-relative entropy.

	\begin{definition}{Max-Relative Entropy}{def-max_rel_ent}
		The \textit{max-relative entropy} $D_{\max}(\rho\Vert\sigma)$ of a state $\rho$ and a positive semi-definite operator $\sigma$ is defined as
		\begin{equation}
			D_{\max}(\rho\Vert\sigma)\coloneqq\log_{2}\!\left\Vert \sigma^{-\frac{1}{2}}\rho\sigma^{-\frac{1}{2}}\right\Vert_{\infty},
		\end{equation}
		if $\supp(\rho)\subseteq\supp(\sigma)$; otherwise, $D_{\max}(\rho\Vert\sigma)=+\infty$.
	\end{definition}

	The max-relative entropy has the following equivalent representations:
	\begin{align}
		D_{\max}(\rho\Vert\sigma)  & =\log_{2}\!\left\Vert \rho^{\frac{1}{2}}\sigma^{-1}\rho^{\frac{1}{2}}\right\Vert_{\infty}\\
		& =2\log_{2}\!\left\Vert \rho^{\frac{1}{2}}\sigma^{-\frac{1}{2}}\right\Vert_{\infty}\\
		& =\log_2\inf_{\lambda \geq 0}\{\lambda:\rho\leq \lambda \sigma\} \label{eq-D_max_SDP}\\
		& =\inf_{\lambda \in \mathbb{R}}\{\lambda:\rho\leq 2^{\lambda} \sigma\} 
		\label{eq-D_max_SDP_alt} \\
		& = \log_2 \sup_{M\geq 0}\{\Tr[M\rho]:\Tr[M\sigma]\leq 1\}. \label{eq:QEI:D_max_SDP_dual}
	\end{align}
	The second-to-last equality demonstrates that $D_{\max}(\rho\Vert\sigma)$ can be calculated using a semi-definite program (SDP) (see Section~\ref{sec-SDPs}). Indeed, the optimization in \eqref{eq-D_max_SDP} can be cast in the standard form in \eqref{eq-dual_SDP_def}, i.e.,
	\begin{equation}
		\inf\{\lambda:\rho\leq\lambda\sigma\}=\left\{\begin{array}{l l} \text{infimum} & \Tr[B Y] \\ \text{subject to} & \Phi^\dagger(Y)\geq A, \\ & Y\geq 0, \end{array}\right.
	\end{equation}
	with $Y\equiv\lambda$, $A\equiv \rho$, $B\equiv 1$, and $\Phi^\dagger(Y)= Y\sigma$. (Note that taking the trace on both sides of the constraint $\rho\leq\lambda\sigma$ in \eqref{eq-D_max_SDP} results in $\lambda\geq 1/\Tr[\sigma]$, so that $\lambda\geq 0$.) The final equality in \eqref{eq:QEI:D_max_SDP_dual} results from calculating the SDP dual to that in~\eqref{eq-D_max_SDP}. 
    
    We also observe that the identities in \eqref{eq-D_max_SDP} and \eqref{eq-D_max_SDP_alt} are consistent even when $\supp(\rho) \not\subseteq \supp(\sigma)$. Indeed, when that is the case, there is no real $\lambda$ such that $\rho\leq 2^\lambda \sigma$. The set in \eqref{eq-D_max_SDP_alt} is therefore empty, and, according to standard mathematical conventions, its infimum equals $+\infty$, in accordance with Definition~\ref{def-max_rel_ent}.
	
	The max-relative entropy has the following alternative representation, which, when $\rho$ and $\sigma$ are states, allows for thinking of it as being related to the largest weight that one can place on $\rho$ to realize $\sigma$ as a probabilistic mixture of $\rho$ and some other state.
	
\begin{Lemma}{lem-QEI:max-rel-ent-alt-exp}
The max-relative entropy $D_{\max}(\rho\Vert\sigma)$ of two states $\rho$ and 
$\sigma$ can be written as follows:
\begin{equation}
D_{\max}(\rho\Vert\sigma)=\inf_{\lambda\geq 0,\,\omega\geq 0}\left\{
\lambda:\sigma=2^{-\lambda}\rho+\left(  1-2^{-\lambda}\right)  \omega
,\ \operatorname{Tr}[\omega]=1\right\}
.\label{eq-QEI:alt-dmax-exp-convex-comb}
\end{equation}
\end{Lemma}

\begin{Proof}
Let $\mu\geq 0$ be such that $\rho \leq 2^{\mu}\sigma$. 
If $\mu=0$, then, on the one hand, it must be the case that $\rho=\sigma$, so that $D_{\max}(\rho\Vert\sigma) = 0$. On the other, the right-hand side of \eqref{eq-QEI:alt-dmax-exp-convex-comb} vanishes, because $0$ belongs to the set. The equality in \eqref{eq-QEI:alt-dmax-exp-convex-comb} therefore holds in this case. 

We now assume that $\mu>0$. Since $2^{\mu}\sigma - \rho \geq 0$ has non-zero trace, we can normalize it by defining $\omega\coloneqq\frac{2^{\mu}\sigma-\rho}{2^{\mu}-1}$, which is a valid quantum state. Now, consider that
\begin{align}
2^{-\mu}\rho+\left( 1-2^{-\mu}\right)  \omega & =2^{-\mu}\rho+\left(
1-2^{-\mu}\right)  \frac{2^{\mu}\sigma-\rho}{2^{\mu}-1}\\
& =2^{-\mu}\rho+\sigma-2^{-\mu}\rho\\
& =\sigma.
\end{align}
Thus, $\mu$ and $\omega$ satisfy the constraints for the optimization problem
in \eqref{eq-QEI:alt-dmax-exp-convex-comb}, and we conclude that
\begin{equation}
\mu\geq\inf_{\lambda\geq 0,\,\omega\geq0}\left\{  \lambda:\sigma
=2^{-\lambda}\rho+\left(  1-2^{-\lambda}\right)  \omega,\ \operatorname{Tr}
[\omega]=1\right\}  .
\end{equation}
By taking an infimum over all $\mu$ satisfying $\rho\leq2^{\mu}\sigma$ and
applying \eqref{eq-D_max_SDP_alt}, we conclude that
\begin{equation}
D_{\max}(\rho\Vert\sigma)\geq\inf_{\lambda\geq 0,\, \omega\geq0}\left\{
\lambda:\sigma=2^{-\lambda}\rho+\left(  1-2^{-\lambda}\right)  \omega
,\ \operatorname{Tr}[\omega]=1\right\}  .
\end{equation}

Now we prove the opposite inequality. Let $\lambda\geq 0$ and let $\omega$ be an arbitrary state satisfying $\sigma=2^{-\lambda}\rho+\left(
1-2^{-\lambda}\right) \omega$. Due to the fact that $\left(1-2^{-\lambda}\right)\omega \geq 0$, it follows that
\begin{equation}
\sigma=2^{-\lambda}\rho+\left(  1-2^{-\lambda}\right)  \omega\geq2^{-\lambda
}\rho,
\end{equation}
from which we conclude that $\rho\leq2^{\lambda}\sigma$. It follows that
$\lambda\geq D_{\max}(\rho\Vert\sigma)$. Since $\omega$ and $\lambda$ are
arbitrary, we conclude that
\begin{equation}
\inf_{\lambda\geq 0,\,\omega\geq0}\left\{
\lambda:\sigma=2^{-\lambda}\rho+\left(  1-2^{-\lambda}\right)  \omega
,\ \operatorname{Tr}[\omega]=1\right\} \geq D_{\max}(\rho\Vert\sigma),
\end{equation}
which completes the proof.
\end{Proof}

	\begin{proposition*}{Data-Processing Inequality for Max-Relative Entropy}{prop-max_rel_ent_monotone}
		Let $\rho$ be a state, $\sigma$ a positive semi-definite operator, and $\mathcal{N}$ a quantum channel. Then,
		\begin{equation}
			D_{\max}(\rho\Vert\sigma)\geq D_{\max}(\mathcal{N}(\rho)\Vert\mathcal{N}(\sigma)).
		\end{equation}
	\end{proposition*}
	
	\begin{remark}
		This result holds more generally for positive maps that are not necessarily trace preserving.
	\end{remark}

	\begin{Proof}
		To see this, let $\lambda \in \mathbb{R}$ be such that the operator inequality $\rho\leq2^{\lambda}\sigma$\ holds. Then the operator inequality $\mathcal{N}(\rho)\leq2^{\lambda}\mathcal{N}(\sigma)$ holds because the quantum channel $\mathcal{N}$ is a positive map. Therefore,
		\begin{equation}
			D_{\max}(\mathcal{N}(\rho)\Vert\mathcal{N}(\sigma))=\inf_{\mu \in \mathbb{R}} \left\{\mu:\mathcal{N}(\rho)\leq2^{\mu}\mathcal{N}(\sigma)\right\}  \leq\lambda.
		\end{equation}
		Since the inequality holds for all choices of $\lambda$ such that $\rho \leq 2^{\lambda}\sigma$ holds, we conclude that
		\begin{equation}
			D_{\max}(\mathcal{N}(\rho)\Vert\mathcal{N}(\sigma))\leq\inf_{\lambda \in \mathbb{R}} \{\lambda:\rho\leq2^{\lambda}\sigma\}  =D_{\max}(\rho\Vert\sigma).
		\end{equation}
		This concludes the proof.
	\end{Proof}
	
	It turns out that the max-relative entropy is a limiting case of the sandwiched and geometric R\'{e}nyi relative entropies, as we now show. 

	\begin{proposition}{prop-sand_rel_ent_limit_max}
		The sandwiched and geometric R\'{e}nyi relative entropies converge to the max-relative entropy in the limit $\alpha\rightarrow\infty$:
		\begin{equation}\label{eq-sand_rel_ent_to_max_rel_ent}
			\lim_{\alpha\rightarrow\infty}\widetilde{D}_{\alpha}(\rho\Vert\sigma) = 			\lim_{\alpha\rightarrow\infty}\widehat{D}_{\alpha}(\rho\Vert\sigma) = D_{\max}(\rho\Vert\sigma).
		\end{equation}
	\end{proposition}

	\begin{Proof}
		We begin with the case in which $\operatorname{supp}(\rho)\not \subseteq\operatorname{supp}(\sigma)$. We trivially have $\widetilde{D}_{\alpha}(\rho\Vert\sigma)=\widehat{D}_{\alpha}(\rho\Vert\sigma) = D_{\max}(\rho\Vert\sigma)=+\infty$ for all $\alpha>1$, which implies the equality in \eqref{eq-sand_rel_ent_to_max_rel_ent} in this case.
		
		In the case that $\operatorname{supp}(\rho)\subseteq\operatorname{supp}(\sigma)$, we can consider, without loss of generality, that $\operatorname{supp}(\sigma)=\mathcal{H}$, so that $\lambda_{\min}(\sigma)>0$.
		
		We begin with the sandwiched R\'enyi relative entropy. Consider that
		\begin{align}
			\widetilde{D}_{\alpha}(\rho\Vert\sigma) &  =\frac{1}{\alpha-1}\log_{2}\Tr\big[\big(\rho^{\frac{1}{2}}\sigma^{\frac{1-\alpha}{\alpha}}\rho^{\frac{1}{2}}\big)^{\alpha}\big]\\
			&  =\frac{1}{\alpha-1}\log_{2}\Tr\big[\big(\rho^{\frac{1}{2}}\sigma^{-\frac{1}{2}}\sigma^{\frac{1}{\alpha}}\sigma^{-\frac{1}{2}}\rho^{\frac{1}{2}}\big)^{\alpha}\big].
		\end{align}
		By the operator inequalities $\left[  \lambda_{\min}(\sigma)\right]^{\frac{1}{\alpha}}\mathbbm{1}\leq\sigma^{\frac{1}{\alpha}}\leq\left[  \lambda_{\max}(\sigma)\right]  ^{\frac{1}{\alpha}}\mathbbm{1}$ and the monotonicity $\Tr[X^{\alpha}]\leq\Tr[Y^{\alpha}]$ for positive semi-definite $X$ and $Y$ satisfying $X\leq Y$ (see Lemma~\ref{prop-operator_ineqs}), we find for $\alpha>1$ that
		\begin{align}
			\lambda_{\min}(\sigma)\,\Tr\big[\big(\rho^{\frac{1}{2}}\sigma^{-1}\rho^{\frac{1}{2}}\big)^{\alpha}\big] &  \leq\Tr\big[\big(\rho^{\frac{1}{2}}\sigma^{-\frac{1}{2}}\sigma^{\frac{1}{\alpha}}\sigma^{-\frac{1}{2}}\rho^{\frac{1}{2}}\big)^{\alpha}\big]\\
			&  \leq\lambda_{\max}(\sigma)\,\Tr\big[\big(\rho^{1/2}\sigma^{-1}\rho^{\frac{1}{2}}\big)^{\alpha}\big].
		\end{align}
		Using the fact that
		\begin{equation}
			\Tr\big[\big(\rho^{\frac{1}{2}}\sigma^{-1}\rho^{\frac{1}{2}}\big)^{\alpha}\big]=\left\Vert\rho^{\frac{1}{2}}\sigma^{-1}\rho^{\frac{1}{2}}\right\Vert _{\alpha}^{\alpha},
		\end{equation}
		and taking a logarithm followed by multiplication of $\frac{1}{\alpha-1}$, we find that
		\begin{multline}
			\frac{1}{\alpha-1}\log_{2}\lambda_{\min}(\sigma)+\frac{\alpha}{\alpha-1}\log_2\!\left\Vert \rho^{\frac{1}{2}}\sigma^{-1}\rho^{\frac{1}{2}}\right\Vert _{\alpha}\leq\widetilde{D}_{\alpha}(\rho\Vert\sigma)\\
			\leq\frac{1}{\alpha-1}\log_{2}\lambda_{\max}(\sigma)+\frac{\alpha}{\alpha-1}\log_2\!\left\Vert \rho^{\frac{1}{2}}\sigma^{-1}\rho^{\frac{1}{2}}\right\Vert _{\alpha}.
		\end{multline}
		Now taking the limit $\alpha\rightarrow\infty$ and recalling that $\lim_{\alpha\rightarrow\infty}\left\Vert X\right\Vert _{\alpha}=\left\Vert X\right\Vert _{\infty}$ (Proposition~\ref{prop:math-tools:infty-norm-from-alpha}), we conclude the equality $\lim_{\alpha\rightarrow\infty}\widetilde{D}_{\alpha}(\rho\Vert\sigma) = 			 D_{\max}(\rho\Vert\sigma)$.
		

We now consider the geometric R\'enyi relative entropy. Since we have
that
\begin{equation}
\lambda_{\min}(\sigma)\mathbbm{1}\leq\sigma\leq\lambda_{\max}(\sigma)\mathbbm{1},
\end{equation}
it follows that
\begin{align}
\lambda_{\min}(\sigma)\operatorname{Tr}\!\left[  \left(  \sigma^{-\frac{1}{2}
}\rho\sigma^{-\frac{1}{2}}\right)  ^{\alpha}\right]   &  \leq\operatorname{Tr}
\left[  \sigma\left(  \sigma^{-\frac{1}{2}}\rho\sigma^{-\frac{1}{2}}\right)
^{\alpha}\right] \\
&  \leq\lambda_{\max}(\sigma)\operatorname{Tr}\!\left[  \left(  \sigma
^{-\frac{1}{2}}\rho\sigma^{-\frac{1}{2}}\right)  ^{\alpha}\right]  .
\end{align}
Now taking a logarithm, dividing by $\alpha-1$, and applying definitions, we
find that the following inequalities hold for $\alpha>1$:
\begin{align}
&  \frac{1}{\alpha-1}\log_2\lambda_{\min}(\sigma)+\frac{1}{\alpha-1}\log
_{2}\operatorname{Tr}\!\left[  \left(  \sigma^{-\frac{1}{2}}\rho\sigma
^{-\frac{1}{2}}\right)  ^{\alpha}\right] \nonumber\\
&  \leq\widehat{D}_{\alpha}(\rho\Vert\sigma
)\label{eq:geometric-renyi-to-max-1}\\
&  \leq\frac{1}{\alpha-1}\log_2\lambda_{\max}(\sigma)+\frac{1}{\alpha-1}
\log_2\operatorname{Tr}\!\left[  \left(  \sigma^{-\frac{1}{2}}\rho\sigma^{-\frac
{1}{2}}\right)  ^{\alpha}\right]  . \label{eq:geometric-renyi-to-max-2}
\end{align}
Now, note that
\begin{align}
\frac{1}{\alpha-1}\log_2\operatorname{Tr}\!\left[  \left(  \sigma^{-\frac{1}{2}
}\rho\sigma^{-\frac{1}{2}}\right)  ^{\alpha}\right]   &  =\frac{\alpha}
{\alpha-1}\log_2\!\left(  \operatorname{Tr}\!\left[  \left(  \sigma^{-\frac{1}{2}
}\rho\sigma^{-\frac{1}{2}}\right)  ^{\alpha}\right]  \right)  ^{\frac
{1}{\alpha}}\\
&  =\frac{\alpha}{\alpha-1}\log_2\!\left\Vert \sigma^{-\frac{1}{2}}\rho
\sigma^{-\frac{1}{2}}\right\Vert _{\alpha}.
\end{align}
Then, using again the fact that $\lim_{\alpha\rightarrow\infty}\left\Vert X\right\Vert
_{\alpha}=\left\Vert X\right\Vert _{\infty}$, it follows that
\begin{equation}
\lim_{\alpha\rightarrow\infty}\frac{1}{\alpha-1}\log_2\operatorname{Tr}\!\left[
\left(  \sigma^{-\frac{1}{2}}\rho\sigma^{-\frac{1}{2}}\right)  ^{\alpha
}\right]  =D_{\max}(\rho\Vert\sigma).
\end{equation}
Combining this limit with the inequalities in
\eqref{eq:geometric-renyi-to-max-1} and \eqref{eq:geometric-renyi-to-max-2},
we arrive at the equality $\lim_{\alpha\rightarrow\infty}\widehat{D}_{\alpha}(\rho\Vert\sigma) = D_{\max}(\rho\Vert\sigma)$.
	\end{Proof}
	
	As a consequence of Proposition~\ref{prop-sand_rel_ent_limit_max}, it is customary to use the notation
	\begin{equation}
		\widetilde{D}_{\infty}(\rho\Vert\sigma)\equiv D_{\max}(\rho\Vert\sigma) \equiv \widehat{D}_{\infty}(\rho\Vert\sigma)
	\end{equation}
	to denote the max-relative entropy. It also follows that the max-relative entropy satisfies the properties of isometric invariance and additivity, as stated in Proposition~\ref{prop-sand_rel_ent_properties}. Proposition~\ref{prop-sand_rel_ent_properties} also tells us that the sandwiched R\'{e}nyi relative entropy $\widetilde{D}_{\alpha}$ is monotonically increasing in $\alpha$, which means that the max-relative entropy has the largest value among all sandwiched R\'{e}nyi relative entropies. Due to Proposition~\ref{prop:geometric-renyi-props}, a similar conclusion holds for the geometric R\'enyi relative entropies. The max-relative entropy also satisfies all of the properties stated in Proposition~\ref{prop-sand_rel_ent_add_properties}.
	
	The conditional entropy arising from the max-relative entropy (according to the general definition in \eqref{eq-gen_cond_entropy}) is known as the \textit{conditional min-entropy}:
	\begin{equation}\label{eq-cond_min_entropy}
		H_{\min}(A|B)_{\rho}\coloneqq \widetilde{H}_{\infty}(A|B)_{\rho} = -\inf_{\sigma_B}D_{\max}(\rho_{AB}\Vert\mathbbm{1}_A\otimes\sigma_B)
	\end{equation}
	for all bipartite states $\rho_{AB}$, where the optimization is over states $\sigma_B$. Since the max-relative entropy has the largest value among all the sandwiched R\'{e}nyi relative entropies, the quantity in \eqref{eq-cond_min_entropy} has the smallest value among all conditional sandwiched R\'{e}nyi entropies, which is why it is called the conditional min-entropy. Note that the conditional sandwiched R\'{e}nyi entropy is defined through \eqref{eq-gen_cond_entropy}, with the generalized divergence $\boldsymbol{D}$ therein replaced by the sandwiched R\'enyi relative entropy $\widetilde{D}_\alpha$, the latter defined in \eqref{eq-sand_ren_rel_entropy}.  On the other hand, the quantity
	\begin{equation}\label{eq-cond_max_entropy}
		H_{\max}(A|B)_{\rho}\coloneqq\widetilde{H}_{\frac{1}{2}}(A|B)_{\rho}=-\inf_{\sigma_B}\widetilde{D}_{\frac{1}{2}}(\rho_{AB}\Vert\mathbbm{1}_A\otimes\sigma_B)
	\end{equation}
	is known as the \textit{conditional max-entropy} of the state $\rho_{AB}$, where the optimization is over states $\sigma_B$. This name comes from the fact that $\alpha=\frac{1}{2}$ is the smallest value of $\alpha$ for which the sandwiched R\'{e}nyi relative entropy is known to satisfy the data-processing inequality (recall Theorem~\ref{thm-sand_renyi_monotone}). Since the sandwiched R\'{e}nyi relative entropy $\widetilde{D}_{\alpha}$ is monotonically increasing in $\alpha$, the quantity in \eqref{eq-cond_max_entropy} is known as the conditional max-entropy because it has the largest value among all conditional sandwiched R\'{e}nyi entropies for which the data-processing inequality is known to hold.
	
	\begin{remark}
		Let $\rho_{XB}$ be a classical--quantum state of the form
		\begin{equation}
			\rho_{XB}=\sum_{x\in\mathcal{X}}p(x)\ket{x}\!\bra{x}_X\otimes\rho_B^x,
		\end{equation}
		where $\mathcal{X}$ is a finite alphabet $p\colon \mathcal{X}\to[0,1]$ is a probability distribution, and $\{\rho_B^x\}_{x\in\mathcal{X}}$ is a set of states. Using the duality of semi-definite programs (see Section~\ref{sec-SDPs}), as done in \eqref{eq:QM-over:multiple-state-disc-dual-SDP}, it follows that
		\begin{equation}
			H_{\min}(X|B)_{\rho}=-\log_2 p_{\text{succ}}^*(\{(p(x),\rho_B^x)\}_x),
		\end{equation}
		where $p_{\text{succ}}^*(\{(p(x),\rho_B^x)\})$, defined in \eqref{eq-opt_guessing_prob}, is the optimal success probability for multiple state discrimination. The conditional min-entropy of a classical--quantum state thus has an operational interpretation in terms of the optimal success probability for multiple state discrimination.
	\end{remark}

\subsection{Smooth Max-Relative Entropy}\label{subsec-smooth_max_rel_ent}

	For the analysis of lower bounds on quantum and private communication rates, we require the smooth max-relative entropy, which is an example of a smooth generalized divergence. A smooth generalized divergence, denoted by $\boldsymbol{D}^{\varepsilon}(\rho\Vert\sigma)$, is defined by taking a generalized divergence $\boldsymbol{D}(\rho\Vert\sigma)$, for a given state $\rho$ and positive semi-definite operator $\sigma$, and optimizing the quantity $\boldsymbol{D}(\widetilde{\rho}\Vert\sigma)$ over states $\widetilde{\rho}$ that are within a distance $\varepsilon$ from the given state $\rho$. Specifically, it is defined as follows:
	\begin{equation}
		\boldsymbol{D}^{\varepsilon}(\rho\Vert\sigma) \coloneqq \inf_{\widetilde{\rho}\in\mathcal{B}^{\varepsilon}(\rho)}\boldsymbol{D}(\widetilde{\rho}\Vert\sigma),
	\end{equation}
	where
	\begin{equation}\label{eq-eps_smooth_ball}
		\mathcal{B}^{\varepsilon}(\rho)\coloneqq \{\tau : \tau \geq 0, \, \Tr[\tau] = 1, \, P(\rho,\tau)\leq\varepsilon\}
	\end{equation}
	is the set of states $\tau$ that are $\varepsilon$-close to $\rho$ in terms of the sine distance (Definition~\ref{def-purified_distance}).

	\begin{definition}{Smooth Max-Relative Entropy}{def-smooth_max_rel_ent}
		Let $\rho$ be a state and $\sigma$ a positive semi-definite operator. Then, the \textit{$\varepsilon$-smooth max-relative entropy}, for $\varepsilon\in[0,1)$, is defined as
		\begin{equation}
			D_{\text{max}}^{\varepsilon}(\rho\Vert\sigma)\coloneqq\inf_{\widetilde{\rho}\in\mathcal{B}^{\varepsilon}(\rho)} D_{\text{max}}(\widetilde{\rho}\Vert\sigma).
			\label{eq-QEI:smooth-max-rel-ent-def}
		\end{equation}
	\end{definition}
	
	Just like the max-relative entropy, the smooth max-relative entropy is a generalized divergence,  satisfying the data-processing inequality:

	\begin{proposition*}{Data-Processing Inequality for Smooth Max-Relative Entropy}{prop:QEI:DP-smooth-max-rel-ent}
	Let $\rho$ be a state, $\sigma$ a positive semi-definite operator, and $\mathcal{N}$ a quantum channel. The smooth max-relative entropy obeys the following data-processing inequality  for all $\varepsilon
\in\left(  0,1\right) $:
\begin{equation}
D_{\max}^{\varepsilon}(\rho\Vert\sigma)\geq D_{\max}^{\varepsilon}
(\mathcal{N}(\rho)\Vert\mathcal{N}(\sigma)). \label{eq:QEI:dp-smooth-d-max}
\end{equation}
\end{proposition*}

\begin{Proof}
To see this, let $\widetilde{\rho}$ be an arbitrary state such that
\begin{equation}
P( \widetilde{\rho},\rho)\leq\varepsilon.
\label{eq:QEI:eps-close-d-max}
\end{equation}
Then from the data-processing inequality for the sine distance under
positive trace-preserving maps (see \eqref{eq-sine_dist_data_proc}), it follows that
\begin{equation}
P( \mathcal{N}(\widetilde{\rho}),\mathcal{N}(\rho
))\leq\varepsilon.
\end{equation}
So it follows that
\begin{align}
D_{\max}(\widetilde{\rho}\Vert\sigma)  &  \geq D_{\max}(\mathcal{N}
(\widetilde{\rho})\Vert\mathcal{N}(\sigma))\\
&  \geq D_{\max}^{\varepsilon}(\mathcal{N}(\rho)\Vert\mathcal{N}(\sigma)).
\end{align}
Since the inequality holds for an arbitrary state $\widetilde{\rho}$
satisfying \eqref{eq:QEI:eps-close-d-max}, we conclude \eqref{eq:QEI:dp-smooth-d-max}.
\end{Proof}

\begin{remark}
The proof given above holds more generally when $\mathcal{N}$ is a positive, trace-preserving map, so that \eqref{eq:QEI:dp-smooth-d-max} holds in this more general case.
\end{remark}

	The smooth max-relative entropy can be related to the sandwiched R\'{e}nyi relative entropy as follows:
	
	\begin{proposition*}{Smooth Max- to Sandwiched R\'{e}nyi Relative Entropy}{prop-smooth_max_to_petz_renyi}
		Let $\rho$ be a state, $\sigma$  a positive semi-definite operator, $\alpha\in(1,\infty)$, and $\varepsilon\in(0,1)$. Then,
		\begin{equation}
		\label{eq-smooth_max_to_petz_renyi}
		D_{\max}^{\varepsilon}(\rho\Vert\sigma)\leq\widetilde{D}_{\alpha}(\rho
\Vert\sigma)
+\frac{1}{\alpha-1}\log_{2}\!\left(  \frac{1}{\varepsilon^{2}}\right)
+\log_{2}\!\left(  \frac{1}{1-\varepsilon^{2}}\right)  .
		\end{equation}
	\end{proposition*}
	
	\begin{Proof}
		The statement is trivially true if $\rho=\sigma$ or if $\operatorname{supp}
(\rho)\not \subseteq \operatorname{supp}(\sigma)$. In the former case, $D_{\max}^{\varepsilon}(\rho\Vert\sigma)=\widetilde{D}_{\alpha}(\rho
\Vert\sigma)=0$, and in the latter, $\widetilde{D}_{\alpha}(\rho
\Vert\sigma)=+\infty$.

So going forward, we assume
that $\rho\neq\sigma$ and $\operatorname{supp}(\rho)\subseteq
\operatorname{supp}(\sigma)$. As mentioned in \eqref{eq:QEI:D_max_SDP_dual}, the SDP dual of $D_{\max}(\tau\Vert\omega)$ is
given by
\begin{equation}
D_{\max}(\tau\Vert\omega)=\log_{2}\sup_{\Lambda \geq 0}\left\{  \operatorname{Tr}
[\Lambda\tau]:\operatorname{Tr}[\Lambda\omega]\leq 1\right\}  ,
\label{eq:QEI:app-SDP-dual-dmax}
\end{equation}
implying that
\begin{equation}
D_{\max}^{\varepsilon}(\rho\Vert\sigma)=\log_{2}\inf_{\widetilde{\rho}
:P( \widetilde{\rho},\rho)\leq\varepsilon
}\ \sup_{\substack{ \Lambda\geq0,\operatorname{Tr}[\Lambda
\sigma]\leq1}}\operatorname{Tr}[\Lambda\widetilde{\rho}].
\end{equation}
Since the objective function $\operatorname{Tr}[\Lambda\widetilde{\rho}]$ is
linear in $\Lambda$ and $\widetilde{\rho}$, the set $\{\Lambda:\Lambda
\geq0,\operatorname{Tr}[\Lambda\sigma]\leq1\}$ is compact and concave, and the
set
\begin{equation}
\left\{  \widetilde{\rho}:P( \widetilde{\rho},
\rho)\leq\varepsilon,\ \widetilde{\rho}\geq
0,\ \operatorname{Tr}[\widetilde{\rho}]=1\right\}
\end{equation}
is compact and convex (due to convexity of sine distance), the
minimax theorem (Theorem~\ref{thm-Sion_minimax}) applies and we find that
\begin{equation}
D_{\max}^{\varepsilon}(\rho\Vert\sigma)=\log_{2} \sup_{\substack{\Lambda\geq0, \operatorname{Tr}[\Lambda\sigma]\leq1}}\ \inf_{\widetilde
{\rho}:P( \widetilde{\rho},\rho)
\leq\varepsilon}\operatorname{Tr}[\Lambda\widetilde{\rho}].
\end{equation}
For a fixed operator $\Lambda\geq0$ with spectral decomposition
\begin{equation}
\Lambda=\sum_{i}\lambda_{i}|\phi_{i}\rangle\!\langle\phi_{i}|,
\end{equation}
let us define the following set, for a choice of $\lambda>0$ to be specified
later:
\begin{equation}
\mathcal{S}\coloneqq \left\{  i:\langle\phi_{i}|\rho|\phi_{i}\rangle>2^{\lambda
}\langle\phi_{i}|\sigma|\phi_{i}\rangle\right\}  .
\end{equation}
Let
\begin{equation}
\Pi\coloneqq \sum_{i\in\mathcal{S}}|\phi_{i}\rangle\!\langle\phi_{i}|.
\label{eq:QEI:app-proj-dmaxeps-up-bnd}
\end{equation}
Then from the definition, we find that
\begin{equation}
\operatorname{Tr}[\Pi\rho]>2^{\lambda}\operatorname{Tr}[\Pi\sigma],
\end{equation}
which implies that
\begin{equation}
\frac{\operatorname{Tr}[\Pi\rho]}{\operatorname{Tr}[\Pi\sigma]}>2^{\lambda}.
\label{eq:QEI:app-ratio-probs-ineq}
\end{equation}
Now consider
from the data-processing inequality under the channel
\begin{equation}
\Delta(\omega)\coloneqq \operatorname{Tr}[\Pi\omega]|0\rangle\!\langle
0|+\operatorname{Tr}[\hat{\Pi}\omega]|1\rangle\!\langle1|
\end{equation}
that
\begin{align}
  \widetilde{D}_{\alpha}(\rho\Vert\sigma) & \geq\widetilde{D}_{\alpha}(\Delta(\rho)\Vert\Delta(\sigma))\\
&  =\frac{1}{\alpha-1}\log_{2}\!\left(
\begin{array}
[c]{c}
\left(  \operatorname{Tr}[\Pi\rho]\right)  ^{\alpha}\left(  \operatorname{Tr}
[\Pi\sigma]\right)  ^{1-\alpha}\\
+\left(  \operatorname{Tr}[\hat{\Pi}\rho]\right)  ^{\alpha}\left(
\operatorname{Tr}[\hat{\Pi}\sigma]\right)  ^{1-\alpha}
\end{array}
\right) \\
&  \geq\frac{1}{\alpha-1}\log_{2}\!\left(  \left(  \operatorname{Tr}[\Pi
\rho]\right)  ^{\alpha}\left(  \operatorname{Tr}[\Pi\sigma]\right)
^{1-\alpha}\right) \\
&  =\frac{1}{\alpha-1}\log_{2}\!\left(  \operatorname{Tr}[\Pi\rho]\left(
\frac{\operatorname{Tr}[\Pi\rho]}{\operatorname{Tr}[\Pi\sigma]}\right)
^{\alpha-1}\right) \\
&  =\frac{1}{\alpha-1}\log_{2}\!\left(  \operatorname{Tr}[\Pi\rho]\right)
+\log_{2}\!\left(  \frac{\operatorname{Tr}[\Pi\rho]}{\operatorname{Tr}[\Pi
\sigma]}\right) \\
&  \geq\frac{1}{\alpha-1}\log_{2}\!\left(  \operatorname{Tr}[\Pi\rho]\right)
+\lambda.
\end{align}
Now picking
\begin{equation}
\lambda=\widetilde{D}_{\alpha}(\rho\Vert\sigma)+\frac{1}{\alpha-1}\log
_{2}\!\left(  \frac{1}{\varepsilon^{2}}\right)  ,
\end{equation}
we conclude that
\begin{equation}
\operatorname{Tr}[\Pi\rho]\leq\varepsilon^{2}.
\end{equation}
Defining $\hat{\Pi}\coloneqq \mathbbm{1} - \Pi$, this means that
\begin{equation}
\operatorname{Tr}[\hat{\Pi}\rho]\geq1-\varepsilon^{2}.
\label{eq:QEI:app-final-steps-1}
\end{equation}
Thus, the state
\begin{equation}
\rho^{\prime}\coloneqq \frac{\hat{\Pi}\rho\hat{\Pi}}{\operatorname{Tr}[\hat{\Pi}\rho]}
\label{eq:QEI:app-rho-prime-state}
\end{equation}
is such that 
\begin{equation}
F(\rho,\rho^{\prime})\geq1-\varepsilon^{2},
\end{equation}
by applying Lemma~\ref{lem-dm:gentle-measurement},
and in turn that 
\begin{equation}
P( \rho,\rho^{\prime})\leq\varepsilon.
\end{equation}
We also have that
\begin{equation}
\rho^{\prime}\leq\frac{\hat{\Pi}\rho\hat{\Pi}}{1-\varepsilon^{2}}.
\end{equation}

Now let $\Lambda$ be an arbitrary operator satisfying $\Lambda\geq0$ and
$\operatorname{Tr}[\Lambda\sigma]\leq1$, and let $\Pi$ be the projection
defined in \eqref{eq:QEI:app-proj-dmaxeps-up-bnd}\ for this choice of $\Lambda$.
Then we find that
\begin{align}
\left(  1-\varepsilon^{2}\right)  \operatorname{Tr}[\Lambda\rho^{\prime}]  &
\leq\operatorname{Tr}[\Lambda\hat{\Pi}\rho\hat{\Pi}]\\
&  =\operatorname{Tr}[\hat{\Pi}\Lambda\hat{\Pi}\rho]\\
&  =\sum_{i\notin\mathcal{S}}\lambda_{i}\langle\phi_{i}|\rho|\phi_{i}\rangle\\
&  \leq2^{\lambda}\sum_{i\notin\mathcal{S}}\lambda_{i}\langle\phi_{i}
|\sigma|\phi_{i}\rangle\\
&  \leq2^{\lambda}\operatorname{Tr}[\Lambda\sigma]\\
&  \leq2^{\lambda}.
\end{align}
Thus, we have found the following uniform bound for any operator $\Lambda$
satisfying $\Lambda\geq0$ and $\operatorname{Tr}[\Lambda\sigma]\leq1$, with
$\rho^{\prime}$ the state in \eqref{eq:QEI:app-rho-prime-state} depending on
$\Lambda$ and\ satisfying $P(\rho,\rho^{\prime})\leq\varepsilon$:
\begin{equation}
\operatorname{Tr}[\Lambda\rho^{\prime}]\leq2^{\lambda+\log_{2}\!\left(  \frac
{1}{1-\varepsilon^{2}}\right)  }.
\end{equation}
Then it follows that
\begin{align}
D_{\max}^{\varepsilon}(\rho\Vert\sigma)  &  =\log_{2}\sup_{\substack{\Lambda\geq0, \operatorname{Tr}[\Lambda\sigma]\leq1}}\ \inf_{\widetilde
{\rho}:P( \widetilde{\rho},\rho)
\leq\varepsilon}\operatorname{Tr}[\Lambda\widetilde{\rho}]\\
&  \leq\log_{2}\sup_{\substack{\ \Lambda\geq0,\operatorname{Tr}
[\Lambda\sigma]\leq1}}\operatorname{Tr}[\Lambda\rho^{\prime}]\\
&  \leq\lambda+\log_{2}\!\left(  \frac{1}{1-\varepsilon^{2}}\right)  .
\label{eq:QEI:app-final-steps-last}
\end{align}
This concludes the proof.
	\end{Proof}
	
	A quantity of interest is the \textit{smooth conditional min-entropy}, which is a conditional entropy that we define via the general construction of conditional entropies in \eqref{eq-gen_cond_entropy}. For every bipartite state $\rho_{AB}$, and every $\varepsilon\in[0,1)$, we define it as
	\begin{equation}\label{eq-smooth_cond_min_ent}
		H_{\text{min}}^{\varepsilon}(A|B)_\rho\coloneqq -\inf_{\sigma_B} D_{\max}^{\varepsilon}(\rho_{AB}\Vert\mathbbm{1}_A\otimes\sigma_B),
	\end{equation}
	where we take the infimum over states $\sigma_B$.

	Using the definition of the smooth conditional min-entropy in \eqref{eq-smooth_cond_min_ent} and applying Proposition~\ref{prop-smooth_max_to_petz_renyi}, we conclude that
	\begin{equation}\label{eq-smooth_cond_min_ent_to_petz_renyi}
		H_{\min}^{\varepsilon}(A|B)_{\rho}\geq \widetilde{H}_{\alpha}(A|B)_{\rho}
-\frac{1}{\alpha-1}\log_{2}\!\left(  \frac{1}{\varepsilon^{2}}\right)
-\log_{2}\!\left(  \frac{1}{1-\varepsilon^{2}}\right) ,  
	\end{equation}
	for all $\alpha>1$ and $\varepsilon \in (0,1)$. Note that the conditional sandwiched R\'{e}nyi entropy $\widetilde{H}_{\alpha}(A|B)_{\rho}$ is defined through \eqref{eq-gen_cond_entropy}, with the generalized divergence~$\boldsymbol{D}$ therein replaced by the sandwiched R\'enyi relative entropy $\widetilde{D}_\alpha$, the latter defined in \eqref{eq-sand_ren_rel_entropy}.

\section{Hypothesis Testing Relative Entropy}\label{sec-hyp_test_rel_ent}

	We now explore another important generalized divergence, called the hypothesis testing relative entropy. This particular entropy is defined to be the optimal value of an operationally defined problem in the context of quantum hypothesis testing. As such, it is debatable as to whether such a quantity should be given the name ``entropy.'' However, our perspective is that the advantages of doing so far outweigh this semantic point about nomenclature, and so we adopt this perspective here and throughout the book. At the most fundamental level, the hypothesis testing relative entropy obeys the quantum data-processing inequality, and for this reason and others, it is useful for characterizing the optimal limits of various communication protocols.

	\begin{definition}{$\boldsymbol{\varepsilon}$-Hypothesis Testing Relative Entropy}{def-hypo_testing_rel_ent}
		Given a state $\rho$, a positive semi-definite operator $\sigma$, and $\varepsilon\in\left[0,1\right]$, the \textit{$\varepsilon$-hypothesis testing relative entropy} is defined as
		\begin{equation}
			D_{H}^{\varepsilon}(\rho\Vert\sigma)\coloneqq-\log_{2}\beta_{\varepsilon}(\rho\Vert\sigma),
			\label{eq:QEI:def-hypo-test-rel-ent}
		\end{equation}
		where
		\begin{equation}
			\beta_{\varepsilon}(\rho\Vert\sigma)\coloneqq\inf_{\Lambda}\left\{\Tr[\Lambda\sigma]:0\leq\Lambda\leq \mathbbm{1}, \ \Tr[\Lambda
\rho]\geq 1-\varepsilon\right\}.
		\end{equation}
	\end{definition}
	
	Observe that $D_{H}^{\varepsilon}(\rho\Vert\sigma)$ can be written as
	\begin{equation}\label{eq:hypo-rewrite}
		D_{H}^{\varepsilon}(\rho\Vert\sigma)=\sup_{\Lambda}\{-\log_2\Tr[\Lambda\sigma]:0\leq\Lambda\leq \mathbbm{1}, \, \Tr[\Lambda\rho]=1-\varepsilon\}.
	\end{equation}
	That is, the monotonicity of the $\log_2$ function allows us to bring $-\log_2$ inside the minimization in the definition of $\beta_{\varepsilon}(\rho\Vert\sigma)$, and it suffices to optimize over measurement operators that meet the constraint $\Tr[\Lambda\rho]\geq 1-\varepsilon$ with equality. This follows because for every measurement operator $\Lambda$ such that $\Tr[\Lambda\rho]>1-\varepsilon$, we can modify it by scaling it by a positive number $\lambda\in[0,1)$ such that $\Tr[\left(\lambda\Lambda\right)\rho]=1-\varepsilon$. The new operator $\lambda\Lambda$ is a legitimate measurement operator and the error probability $\Tr[(\lambda\Lambda)\sigma]$ only decreases under this scaling (i.e., $\Tr[(\lambda\Lambda)\sigma]<\Tr[\Lambda\sigma]$), which allows us to conclude \eqref{eq:hypo-rewrite}.
	
	The hypothesis testing relative entropy can be computed using a semi-definite program, as indicated in the following proposition:
		
	\begin{proposition*}{Hypothesis Testing Relative Entropy as an SDP}{prop-hypo_test_rel_ent_dual}
		For every state $\rho$, positive semi-definite operator $\sigma$, and $\varepsilon\in[0,1]$, the $\varepsilon$-hypothesis testing relative entropy can be expressed as the following SDPs:
		\begin{align}
			D_H^{\varepsilon}(\rho\Vert\sigma) & = -\log_2 \inf_{\Lambda\geq 0}\left\{\Tr[\Lambda\sigma]:\Lambda\leq \mathbbm{1}, \ \Tr[\Lambda
\rho]\geq 1-\varepsilon\right\}
\label{eq:QEI:hypo_test_rel_ent_primal}\\
			& = -\log_2\sup_{\mu\geq0,Z\geq 0}\{\mu(1-\varepsilon) - \Tr[Z]: \mu\rho\leq\sigma+Z \}.
					\label{eq-hypo_test_rel_ent_dual}
		\end{align}
		Complementary slackness implies that the following equalities hold for optimal $\Lambda$, $\mu$, and $Z$:
		\begin{equation}
		\Lambda (\sigma+Z) = \mu \Lambda \rho,\qquad \Tr[\Lambda \rho]\mu = (1-\varepsilon)\mu, \qquad \Lambda Z = Z.
		\label{eq:QEI:CS-HTRE}
		\end{equation}
	\end{proposition*}
	
	\begin{Proof}
	The primal formulation in \eqref{eq:QEI:hypo_test_rel_ent_primal} is immediate from Definition~\ref{def-hypo_testing_rel_ent}. Indeed, considering the standard form in \eqref{eq-dual_SDP_def}, we see that we can set
	\begin{equation}
	B  = \sigma, \quad Y = \Lambda, \quad 
	\Phi^{\dag}(Y) =
	\begin{pmatrix}
	\Tr[\Lambda \rho] & 0\\
	0 & - \Lambda
	\end{pmatrix},\quad 
	A  =
	\begin{pmatrix}
	1-\varepsilon & 0\\
	0 & - \mathbbm{1}
	\end{pmatrix}.
	\label{eq:QEI:SDP-HTRE-1}
	\end{equation}
	
		To figure out the dual and having already identified $A$, $B$, and $\Phi^{\dag}$, we need to determine the map $\Phi$ and plug into the standard form in \eqref{eq-primal_SDP_def}. Letting
		\begin{equation}
		X \coloneqq
		\begin{pmatrix}
		\mu & 0 \\
		0 & Z
		\end{pmatrix},
		\end{equation}
		we find that
		\begin{align}
		\Tr[\Phi^{\dag}(Y)X] & = 
		\Tr\!\left[\begin{pmatrix}
	\Tr[\Lambda \rho] & 0\\
	0 & - \Lambda
	\end{pmatrix} \begin{pmatrix}
		\mu & 0 \\
		0 & Z
		\end{pmatrix}\right] \\
		& = \mu \Tr[\Lambda \rho] - \Tr[\Lambda Z] \\
		& = \Tr[\Lambda(\mu\rho - Z)],
		\end{align}
		which implies that
		\begin{equation}
		\Phi(X) = \mu\rho - Z.
		\end{equation}
		Now substituting into \eqref{eq-primal_SDP_def} and simplifying, we conclude that the right-hand side of \eqref{eq-hypo_test_rel_ent_dual} is the dual SDP.
		
		To show that this is equal to $D_H^{\varepsilon}(\rho\Vert\sigma)$, we should demonstrate that the primal and dual SDPs satisfy the strong duality property. It is clear that $\Lambda = \mathbbm{1}$ is a feasible point for the primal SDP. Furthermore, the choices $\mu=1$ and $Z = \mathbbm{1} + [\sigma-\rho]_+$, where $[\sigma-\rho]_+$ is the positive part of $\sigma-\rho$, are strictly feasible for the dual. Thus, we conclude \eqref{eq-hypo_test_rel_ent_dual}  by applying Theorem~\ref{thm:math-tools:slater-cond}.
		
		The complementary slackness conditions in \eqref{eq:QEI:CS-HTRE} follow directly from Proposition~\ref{prop:math-tools:comp-slack}.
	\end{Proof}
	
	\begin{proposition*}{Optimal Measurement for Hypothesis Testing Relative Entropy}{prop:QEI:opt-meas-HTRE}
	For every state $\rho$, positive semi-definite operator $\sigma$, and $\varepsilon\in[0,1]$, the $\varepsilon$-hypothesis testing relative entropy $D_H^{\varepsilon}(\rho\Vert \sigma)$ is achieved by the following  measurement operator:
	\begin{equation}
\Lambda(\mu^{\ast},p^{\ast})\coloneqq \Pi_{\mu^{\ast}\rho>\sigma}+p^{\ast}\Pi_{\mu^{\ast}\rho=\sigma},
\end{equation}
where $\Pi_{\mu^{\ast}\rho>\sigma}$ is the projection onto the strictly positive part
of $\mu^{\ast}\rho-\sigma$, the projection $\Pi_{\mu^{\ast}\rho=\sigma}$ projects onto the
zero eigenspace of $\mu^{\ast}\rho-\sigma$, and $\mu^{\ast} \geq 0$ and $p^{\ast} \in [0,1]$ are chosen as follows:
\begin{align}
\mu^{\ast} & \coloneqq \sup\left\{  \mu:\operatorname{Tr}[\Pi_{\mu\rho>\sigma}\rho
]\leq1-\varepsilon\right\}  ,\\
p^{\ast}& \coloneqq \frac{1-\varepsilon-\operatorname{Tr}[\Pi_{\mu^{\ast}\rho>\sigma
}\rho]}{\operatorname{Tr}[\Pi_{\mu^{\ast}\rho=\sigma}\rho]}.
\end{align}
	\end{proposition*}
	
	\begin{Proof}
	To find the form of an optimal measurement operator for the hypothesis testing
relative entropy, let $\Lambda$ be a measurement operator satisfying
$\operatorname{Tr}[\Lambda\rho]=1-\varepsilon$ and let $\mu\geq0$. Then
\begin{align}
\operatorname{Tr}[\Lambda\sigma]  & =\operatorname{Tr}[\Lambda\sigma
]+\mu\left(  1-\varepsilon-\operatorname{Tr}[\Lambda\rho]\right)  \label{eq:QEI:htre-opt-meas-pf-1}\\
& =-\mu\varepsilon+\operatorname{Tr}[(I-\Lambda)\mu\rho]+\operatorname{Tr}
[\Lambda\sigma]\\
& \geq-\mu\varepsilon+\frac{1}{2}\left(  \operatorname{Tr}[\mu\rho
+\sigma]-\left\Vert \mu\rho-\sigma\right\Vert _{1}\right)  \\
& =-\mu\varepsilon+\frac{1}{2}\left(  \mu+\operatorname{Tr}[\sigma]-\left\Vert
\mu\rho-\sigma\right\Vert _{1}\right)  .
\label{eq:QEI:htre-opt-meas-pf-last}
\end{align}
The sole inequality follows as an application of Theorem~\ref{thm-Holevo_Helstrom}, with $B=\sigma$ and
$A=\mu\rho$. Observe that the final expression is a universal bound
independent of $\Lambda$. To determine an optimal measurement operator, we can
look to Theorem~~\ref{thm-Holevo_Helstrom}. There, it was established that the following measurement
operator is an optimal one for $\inf_{\Lambda:0\leq\Lambda\leq I}\left\{
\operatorname{Tr}[(I-\Lambda)\mu\rho]+\operatorname{Tr}[\Lambda\sigma
]\right\}  $:
\begin{equation}
\Lambda(\mu,p)\coloneqq \Pi_{\mu\rho>\sigma}+p\Pi_{\mu\rho=\sigma},
\end{equation}
where $\Pi_{\mu\rho>\sigma}$ is the projection onto the strictly positive part
of $\mu\rho-\sigma$, the projection $\Pi_{\mu\rho=\sigma}$ projects onto the
zero eigenspace of $\mu\rho-\sigma$, and $p\in\left[  0,1\right]  $. The
measurement operator $\Lambda(\mu,p)$ is called a quantum Neyman--Pearson test.
We still need to choose the parameters $\mu\geq0$ and $p\in\left[  0,1\right]
$. Let us pick $\mu$ according to the following optimization:
\begin{equation}
\mu^{\ast}\coloneqq \sup\left\{  \mu:\operatorname{Tr}[\Pi_{\mu\rho>\sigma}\rho
]\leq1-\varepsilon\right\}  .
\end{equation}
If it so happens that $\mu^{\ast}$ is such that $\operatorname{Tr}[\Pi
_{\mu^{\ast}\rho>\sigma}\rho]=1-\varepsilon$, then we are done; we can pick
$p=0$. However, if $\mu^{\ast}$ is such that $\operatorname{Tr}[\Pi_{\mu
^{\ast}\rho>\sigma}\rho]<1-\varepsilon$, then we pick $p^{\ast}\in\left[
0,1\right]  $ such that
\begin{equation}
p^{\ast}\coloneqq \frac{1-\varepsilon-\operatorname{Tr}[\Pi_{\mu^{\ast}\rho>\sigma
}\rho]}{\operatorname{Tr}[\Pi_{\mu^{\ast}\rho=\sigma}\rho]},
\end{equation}
with it following that $p^{\ast}\in\left[  0,1\right]  $ because
\begin{equation}
\operatorname{Tr}[\Pi_{\mu^{\ast}\rho>\sigma}\rho]<1-\varepsilon
\leq\operatorname{Tr}[\Pi_{\mu^{\ast}\rho\geq\sigma}\rho].
\end{equation}
With these choices, we then find that
\begin{equation}
\operatorname{Tr}[\Lambda(\mu^{\ast},p^{\ast})\rho]=1-\varepsilon.
\end{equation}
By the analysis in \eqref{eq:QEI:htre-opt-meas-pf-1}--\eqref{eq:QEI:htre-opt-meas-pf-last}, it then follows that
\begin{equation}
\operatorname{Tr}[\Lambda\sigma]\geq\operatorname{Tr}[\Lambda(\mu^{\ast
},p^{\ast})\sigma]
\end{equation}
for all measurement operators $\Lambda$ satisfying $0\leq\Lambda\leq I$ and
$\operatorname{Tr}[\Lambda\rho]=1-\varepsilon$.
	\end{Proof}
	
	
	Note that the other generalized divergences we have considered so far satisfy $\boldsymbol{D}(\rho\Vert\rho)=0$ for all states $\rho$. The $\varepsilon$-hypothesis testing relative entropy, however, does not have this property unless $\varepsilon=0$. In fact, it is clear from the definition, along with \eqref{eq:hypo-rewrite}, that
	\begin{equation}
	D_H^{\varepsilon}(\rho\Vert\rho)=-\log_2(1-\varepsilon)
	\end{equation}
	for all states $\rho$ and  $\varepsilon\in[0,1]$.

	Like the quantum relative entropy, the Petz--R\'{e}nyi relative entropy, the sandwiched R\'{e}nyi relative entropy, and the max-relative entropy, the $\varepsilon$-hypo\-thesis testing relative entropy is also a generalized divergence, meaning that is satisfies the data-processing inequality.
	
	\begin{theorem*}{Data-Processing Inequality for Hypothesis Testing Relative Entropy}{thm-hypo_rel_ent_monotone}
		Let $\rho$ be a state, $\sigma$ a positive semi-definite operator, and $\mathcal{N}$ a quantum channel. Then, for all $\varepsilon\in[0,1]$,
		\begin{equation}
			D_H^{\varepsilon}(\rho\Vert\sigma)\geq D_H^{\varepsilon}(\mathcal{N}(\rho)\Vert\mathcal{N}(\sigma)).
		\end{equation}
	\end{theorem*}
	
	\begin{Proof}
		The intuition for this proof is as follows: A measurement operator $\Lambda$ satisfying the constraints $0\leq\Lambda\leq\mathbbm{1}$ and $\Tr[\Lambda\mathcal{N}(\rho)]\geq 1-\varepsilon$ can be understood as a particular measurement strategy for distinguishing $\rho$ from $\sigma$ in which we first apply the channel $\mathcal{N}$ and then apply the measurement operator $\Lambda$. Then this particular measurement strategy cannot perform better than the optimal measurement strategy for distinguishing $\rho$ from $\sigma$.
		
		We start by writing $D_H^{\varepsilon}(\mathcal{N}(\rho)\Vert\mathcal{N}(\sigma))$ as in \eqref{eq:QEI:def-hypo-test-rel-ent}:
		\begin{multline}\label{eq-hypo_monotone_pf}
			D_H^{\varepsilon}(\mathcal{N}(\rho)\Vert\mathcal{N}(\sigma))\\=\sup_{\Lambda}\{-\log_2\Tr[\Lambda\mathcal{N}(\sigma)]:0\leq \Lambda\leq\mathbbm{1},~\Tr[\Lambda\mathcal{N}(\rho)]\geq 1-\varepsilon\}.
		\end{multline}
		Fix $\Lambda$ such that $0\leq\Lambda\leq\mathbbm{1}$ and $\Tr[\Lambda\mathcal{N}(\rho)]\geq 1-\varepsilon$. By definition of the adjoint, we have that
		\begin{equation}
			\Tr[\Lambda\mathcal{N}(\sigma)]=\Tr[\mathcal{N}^\dagger(\Lambda)\sigma],\quad \Tr[\Lambda\mathcal{N}(\rho)]=\Tr[\mathcal{N}^\dagger(\Lambda)\rho].
		\end{equation}
		Also, note that $0\leq\mathcal{N}^\dagger(\Lambda)\leq\mathbbm{1}$. The leftmost inequality is due to the fact that $\mathcal{N}^\dagger$ is a positive map because $\mathcal{N}$ is. The rightmost inequality is due to the fact that $\mathcal{N}^\dagger$ is subunital because $\mathcal{N}$ is trace non-increasing. By the positivity of $\mathcal{N}^\dagger$, we obtain
		\begin{equation}
			\Lambda\leq\mathbbm{1}\Rightarrow \mathbbm{1}-\Lambda\geq 0\Rightarrow\mathcal{N}^\dagger(\mathbbm{1}-\Lambda)\geq 0\Rightarrow \mathcal{N}^\dagger(\Lambda)\leq\mathcal{N}^\dagger(\mathbbm{1})\leq\mathbbm{1}.
		\end{equation}
		Since $\Lambda$ is arbitrary, we bound \eqref{eq-hypo_monotone_pf} as
		\begin{equation}
			\begin{aligned}
			&D_H^{\varepsilon}(\mathcal{N}(\rho)\Vert\mathcal{N}(\sigma))\\
			&\quad \leq\sup_{\Lambda}\{-\log_2\Tr[\mathcal{N}^\dagger(\Lambda)\sigma]:0\leq\mathcal{N}^\dagger(\Lambda)\leq \mathbbm{1},~\Tr[\mathcal{N}^\dagger(\Lambda)\rho]\geq 1-\varepsilon\}.
			\end{aligned}
		\end{equation}
		Now, by enlarging the optimization set from measurement operators $\mathcal{N}^\dagger(\Lambda)$ satisfying $0\leq\mathcal{N}^\dagger(\Lambda)\allowbreak\leq\mathbbm{1}$ and $\Tr[\mathcal{N}^\dagger(\Lambda)\rho]\geq 1-\varepsilon$ to all measurement operators, say $\Lambda'$, satisfying $0\leq\Lambda'\leq\mathbbm{1}$ and $\Tr[\Lambda'\rho]\geq 1-\varepsilon$, we obtain
		\begin{align}
			&D_H^{\varepsilon}(\mathcal{N}(\rho)\Vert\mathcal{N}(\sigma))\nonumber\\
			&\leq\sup_{\Lambda}\{-\log_2\Tr[\mathcal{N}^\dagger(\Lambda)\sigma]:0\leq\mathcal{N}^\dagger(\Lambda)\leq \mathbbm{1},~\Tr[\mathcal{N}^\dagger(\Lambda)\rho]\geq 1-\varepsilon\}\\
			&\leq \sup_{\Lambda'}\{-\log_2\Tr[\Lambda'\sigma]:0\leq\Lambda'\leq\mathbbm{1},~\Tr[\Lambda'\rho]\geq 1-\varepsilon\}\\
			&=D_H^{\varepsilon}(\rho\Vert\sigma),
		\end{align}
		as required.
	\end{Proof}
	\begin{remark}
	Inspection of the proof above reveals that it holds more generally for $\mathcal{N}$ a positive, trace-non-increasing map.
	\end{remark}
	
	\begin{proposition*}{Properties of Hypothesis Testing Relative Entropy}{prop-properties_hyp_rel_ent}
		The $\varepsilon$-hypothesis testing relative entropy satisfies the following properties for all $\varepsilon\in[0,1]$:
		\begin{enumerate}
			\item If $\varepsilon' \in (\varepsilon,1]$, then
			\begin{equation}
			D_H^{\varepsilon}(\rho\Vert \sigma) \leq D_H^{\varepsilon'}(\rho\Vert \sigma).
			\label{eq:QEI:HTRE-mono-eps}
			\end{equation}
					\item The following limit holds:
\begin{equation}
\lim_{\varepsilon\rightarrow0}D_{H}^{\varepsilon}(\rho\Vert\sigma)=D_{0
}(\rho\Vert\sigma), \label{eq:lim-dmineps-to-dmin}
\end{equation}
where $D_{0
}(\rho\Vert\sigma)=-\log_2\Tr[\Pi_{\rho}\sigma]$ is the Petz--R\'enyi relative entropy of order zero and $\Pi_{\rho}$ is the projection onto the support of $\rho$.

			\item For every state $\rho$ and positive semi-definite operators $\sigma,\sigma'$ such that $\sigma'\geq\sigma$, we have that $D_H^{\varepsilon}(\rho\Vert\sigma)\geq D_H^{\varepsilon}(\rho\Vert\sigma')$.
			\item For every state $\rho$, positive semi-definite operator $\sigma$, and $\alpha>0$, we have that $D_H^{\varepsilon}(\rho\Vert\alpha\sigma)=D_H^{\varepsilon}(\rho\Vert\sigma)-\log_2\alpha$.
			\item Let $p,q:\mathcal{X}\to[0,1]$ be two probability distributions over a finite alphabet $\mathcal{X}$ with associated $|\mathcal{X}|$-dimensional system $X$, let $\{\rho_A^x\}_{x\in\mathcal{X}}$ be a set of states on a system~$A$, and let $\sigma_A$ be a state on system~$A$. Then,
				\begin{multline}
					D_H^{\varepsilon}\!\left(\sum_{x\in\mathcal{X}}p(x)\ket{x}\!\bra{x}_X\otimes\rho_A^x\Bigg\Vert\sum_{x\in\mathcal{X}}q(x)\ket{x}\!\bra{x}_X\otimes\sigma^x_A\right)\\\geq \min_{x\in\mathcal{X}}D_H^{\varepsilon}(\rho_A^x\Vert\sigma^x_A).
				\end{multline}
		\end{enumerate}
	\end{proposition*}
	
	\begin{Proof}
		\hfill\begin{enumerate}
			\item 
			Eq.~\eqref{eq:QEI:HTRE-mono-eps} follows from Definition~\ref{def-hypo_testing_rel_ent}: increasing $\varepsilon$ increases the set of measurement operators $\Lambda$ over which we can optimize, and $D_H^{\varepsilon}(\rho\Vert \sigma)$ does not decrease under such a change.
			
			\item Consider that the following inequality
holds for all $\varepsilon\in(0,1)$:
\begin{equation}
D_{H}^{\varepsilon}(\rho\Vert\sigma)\geq D_{0}(\rho\Vert\sigma),
\end{equation}
because the measurement operator $\Pi_{\rho}$ (projection onto support of
$\rho$) satisfies $\operatorname{Tr}[\Pi_{\rho}\rho]\geq1-\varepsilon$ for all
$\varepsilon\in(0,1)$. So we conclude that
\begin{equation}
\liminf_{\varepsilon\rightarrow0}D_{H}^{\varepsilon}(\rho\Vert\sigma)\geq
D_{0}(\rho\Vert\sigma). \label{eq:liminf-dmineps-to-dmin}
\end{equation}
Alternatively, suppose that $\Lambda$ is a measurement operator satisfying
$\operatorname{Tr}[\Lambda\rho]=1-\varepsilon$ (note that when optimizing
$D_{H}^{\varepsilon}$, it suffices to optimize over measurement operators
satisfying the constraint $\operatorname{Tr}[\Lambda\rho]\geq1-\varepsilon$
with equality, as mentioned in \eqref{eq:hypo-rewrite}). Then applying the data-processing inequality for
$D_{\alpha}(\rho\Vert\sigma)$ under the measurement $\left\{  \Lambda
,\mathbbm{1}-\Lambda\right\}  $, which holds for $\alpha\in(0,1)$, we find that
\begin{equation}
D_{\alpha}(\rho\Vert\sigma)\geq
\frac{1}{\alpha-1}\log_{2}\!\left[  \left(  1-\varepsilon\right)  ^{\alpha
}\operatorname{Tr}[\Lambda\sigma]^{1-\alpha}+\varepsilon^{\alpha}\left(
1-\operatorname{Tr}[\Lambda\sigma]\right)  ^{1-\alpha}\right]  .
\end{equation}
Since this bound holds for all measurement operators $\Lambda$ satisfying
$\operatorname{Tr}[\Lambda\rho]=1-\varepsilon$, we conclude the following
bound for all $\alpha\in(0,1)$:
\begin{multline}
D_{\alpha}(\rho\Vert\sigma)\geq\\
\frac{1}{\alpha-1}\log_{2}\!\left[
\left(  1-\varepsilon\right)  ^{\alpha}\left(  2^{-D_{H}^{\varepsilon}
(\rho\Vert\sigma)}\right)  ^{1-\alpha}
+\ \varepsilon^{\alpha}\left(  1-2^{-D_{H}^{\varepsilon}(\rho\Vert\sigma
)}\right)  ^{1-\alpha}
\right]  .
\end{multline}
Now taking the limit of the right-hand side as $\varepsilon\rightarrow0$, we
find that the following bound holds for all $\alpha\in(0,1)$:
\begin{equation}
D_{\alpha}(\rho\Vert\sigma)\geq\limsup_{\varepsilon\rightarrow0}D_{H
}^{\varepsilon}(\rho\Vert\sigma).
\end{equation}
Since the bound holds for all $\alpha\in(0,1)$, we can take the limit on the
left-hand side to arrive at
\begin{equation}
\lim_{\alpha\rightarrow0}D_{\alpha}(\rho\Vert\sigma)=D_{0}(\rho\Vert
\sigma)\geq\limsup_{\varepsilon\rightarrow0}D_{H}^{\varepsilon}(\rho
\Vert\sigma). \label{eq:limsup-dmineps-to-dmin}
\end{equation}
Now putting together \eqref{eq:liminf-dmineps-to-dmin} and
\eqref{eq:limsup-dmineps-to-dmin}, we conclude~\eqref{eq:lim-dmineps-to-dmin}.
			
			\item Fix an operator $\Lambda$ satisfying $0\leq\Lambda\leq\mathbbm{1}$. The assumption $\sigma'\geq \sigma$ implies that
				$\Tr[\Lambda\sigma']\geq\Tr[\Lambda\sigma]$, which in turn implies that
				\begin{equation}
					-\log_2\Tr[\Lambda\sigma']\leq-\log_2\Tr[\Lambda\sigma].
				\end{equation}
				Since the operator $\Lambda$ is arbitrary, we obtain $D_H^{\varepsilon}(\rho\Vert\sigma)\geq D_H^{\varepsilon}(\rho\Vert\sigma')$, as required.
			
			\item This follows immediately from the fact that
				\begin{equation}
					-\log_2\Tr[\Lambda(\alpha\sigma)]=-\log_2\!\left(\alpha\Tr[\Lambda\sigma]\right)=-\log_2\Tr[\Lambda\sigma]-\log_2\alpha
				\end{equation}
				for all operators $\Lambda$ satisfying $0\leq\Lambda\leq\mathbbm{1}$.
			
			\item Since $D_H^{\varepsilon}(\rho\Vert\sigma)=-\log_2\beta_{\varepsilon}(\rho\Vert\sigma)$, where $\beta_{\varepsilon}(\rho\Vert\sigma)$ is defined in \eqref{eq-entr:beta-quant}, we can equivalently show that
				\begin{multline}\label{eq-hyp_rel_ent_quasi_concave_pf}
					\beta_{\varepsilon}\!\left(\sum_{x\in\mathcal{X}}p(x)\ket{x}\!\bra{x}_X\otimes\rho_A^x\Bigg\Vert\sum_{x\in\mathcal{X}}q(x)\ket{x}\!\bra{x}_X\otimes\sigma^x_A\right)\\
					\leq \max_{x\in\mathcal{X}}\beta_{\varepsilon}(\rho_A^x\Vert\sigma^x_A).
				\end{multline}
				Now, in the definition of $\beta_{\varepsilon}$ on the left-hand side of the  inequality above, let us restrict the infimum over all measurement operators to those of the form $\Lambda_{XA}=\sum_{x\in\mathcal{X}}\ket{x}\!\bra{x}_X\otimes M_A^x$ such that $0\leq M_A^x\leq\mathbbm{1}_A$ and $\Tr[M_A^x\rho_A^x]\geq 1-\varepsilon$ for all $x\in\mathcal{X}$. Doing this leads to
				\begin{align}
					&\beta_{\varepsilon}\!\left(\sum_{x\in\mathcal{X}}p(x)\ket{x}\!\bra{x}_X\otimes\rho_A^x\Bigg\Vert\sum_{x\in\mathcal{X}}q(x)\ket{x}\!\bra{x}_X\otimes\sigma_A^x\right)\\
					&\quad\leq \inf_{\{M_A^x\}_{x\in\mathcal{X}}}\bigg\{\sum_{x\in\mathcal{X}}q(x)\Tr[M_A^x\sigma^x_A]:0\leq M_A^x\leq\mathbbm{1}_A,\nonumber\\
					&\qquad\qquad\qquad\qquad\qquad\qquad\qquad\Tr[M_A^x\rho_A^x]\geq 1-\varepsilon~\forall x\in\mathcal{X}\bigg\}\\
					&\quad =\sum_{x\in\mathcal{X}}q(x)\inf_{M_A^x}\left\{\Tr[M_A^x\sigma^x_A]:0\leq M_A^x\leq\mathbbm{1}_A,\right.\nonumber\\
					&\qquad\qquad\qquad\qquad\qquad\qquad\qquad\left.\Tr[M_A^x\rho_A^x]\geq 1-\varepsilon\right\}\\
					&=\sum_{x\in\mathcal{X}}q(x)\beta_{\varepsilon}(\rho_A^x\Vert\sigma^x_A)\\
					&\leq \max_{x\in\mathcal{X}}\beta_{\varepsilon}(\rho_A^x\Vert\sigma^x_A).
				\end{align}
				The last inequality follows because $\beta_{\varepsilon}(\rho_A^x\Vert\sigma^x_A)\leq\max_{x\in\mathcal{X}}\beta_{\varepsilon}(\rho_A^x\Vert\sigma^x_A)$ for all $x\in\mathcal{X}$. So we have shown that the inequality \eqref{eq-hyp_rel_ent_quasi_concave_pf} holds, which completes the proof. \qedhere
		\end{enumerate}
	\end{Proof}

\subsection{Connection to Quantum Relative Entropy}

	We now prove a bound relating the $\varepsilon$-hypothesis testing relative entropy to the quantum relative entropy.

	\begin{proposition*}{Hypothesis Testing to Quantum Relative Entropy}{prop-hypo_to_rel_ent}
		Fix $\varepsilon\in[0,1)$. Let $\rho$ be a state and $\sigma$ a positive semi-definite operator. Then the following bound relates the $\varepsilon$-hypothesis testing relative entropy to the quantum relative entropy:
		\begin{equation}
			D_{H}^{\varepsilon}(\rho\Vert\sigma)\leq\frac{1}{1-\varepsilon}\left(D(\rho\Vert\sigma)+h_{2}(\varepsilon)+\varepsilon\log_{2}\Tr[\sigma]\right),\label{eq:WR-bound-to-rel-ent}
		\end{equation}
		where $h_2(\varepsilon)$ is the binary entropy defined in \eqref{eq-ent_bin}. 
	\end{proposition*}

	\begin{Proof}
		To see this, we use the fact that the optimization in the definition of $D_H^{\varepsilon}(\rho\Vert\sigma)$ can be restricted as in \eqref{eq:hypo-rewrite}, i.e.,
		\begin{equation}
			D_{H}^{\varepsilon}(\rho\Vert\sigma)=\sup_{\Lambda}\{-\log_2\Tr[\Lambda\sigma]:0\leq\Lambda\leq \mathbbm{1},\
			\Tr[\Lambda\rho]=1-\varepsilon\}.
		\end{equation}
		For every measurement operator $\Lambda$ such that $\Tr[\Lambda\rho]=1-\varepsilon$, the data-processing inequality for the quantum relative entropy (Theorem \ref{thm-monotone_rel_ent}) implies that
		\begin{align}
			D(\rho\Vert\sigma) &  \geq D(\{1-\varepsilon,\varepsilon\}\Vert\{\Tr[\Lambda\sigma],\Tr[\sigma]-\Tr[\Lambda\sigma]\})\nonumber\\
			&  =(1-\varepsilon)\log_{2}\!\left(  \frac{1-\varepsilon}{\Tr[\Lambda\sigma]}\right)  +\varepsilon\log_{2}\!\left(  \frac{\varepsilon}{\Tr[\sigma]-\Tr[\Lambda\sigma]}\right)  \\
			&  =(1-\varepsilon)\log_{2}\!\left(  1-\varepsilon\right)  -(1-\varepsilon)\log_{2}\!\left(  \Tr[\Lambda\sigma]\right)  \nonumber\\
			&  \qquad+\varepsilon\log_{2}\!\left(  \varepsilon\right)  +\varepsilon \log_{2}\!\left(  \frac{1}{\Tr[\sigma]\left(
1-\Tr[\Lambda\sigma/\Tr[\sigma]]\right)  }\right)
\\
			&  =-\left(  1-\varepsilon\right)  \log_{2}\Tr[\Lambda
\sigma]-h_{2}(\varepsilon)\nonumber\\
			&   \qquad+\varepsilon\log_{2}\!\left(  \frac{1}{1-\Tr[\Lambda\sigma/\Tr[\sigma]]}\right)-\varepsilon\log_{2}\Tr[\sigma]\\
			&  \geq-\left(  1-\varepsilon\right)  \log_{2}\Tr[\Lambda\sigma]-h_{2}(\varepsilon)-\varepsilon\log_{2}\Tr[\sigma],
		\end{align}
		where the inequality holds because $\varepsilon\log_{2}\!\left(\frac{1}{  1-\Tr[\Lambda\sigma/\Tr[\sigma]]  }\right)\geq 0$. Rewriting this gives
		\begin{equation}
			-\log\Tr[\Lambda\sigma]\leq\frac{1}{1-\varepsilon}\left[D(\rho\Vert\sigma)+h_{2}(\varepsilon)+\varepsilon\log_{2}\Tr[\sigma]\right].
		\end{equation}
		Since this bound holds for all measurement operators $\Lambda$ satisfying $\Tr[\Lambda\rho]=1-\varepsilon$, we conclude \eqref{eq:WR-bound-to-rel-ent}.
	\end{Proof}

\subsection{Connections to Quantum R\'{e}nyi Relative Entropies}


	We now show a connection between the hypothesis testing relative entropy and the Petz-- and sandwiched R\'{e}nyi relative entropies. 
	
	\begin{proposition*}{Hypothesis Testing to Sandwiched R\'enyi Relative Entropy}{prop:sandwich-to-htre}
		Let $\rho$ be a state and $\sigma$ a positive semi-definite operator. Let $\alpha\in(1,\infty)$ and $\varepsilon\in\lbrack 0,1)$. Then the following inequality holds
		\begin{equation}\label{eq:sandwich-to-htre}
			D_{H}^{\varepsilon}(\rho\Vert\sigma)\leq\widetilde{D}_{\alpha}(\rho\Vert \sigma)+\frac{\alpha}{\alpha-1}\log_{2}\!\left(\frac{1}{1-\varepsilon}\right).
		\end{equation}
		In particular, in the limit $\alpha\to\infty$,
		\begin{equation}\label{eq-D_max_to_htre}
			D_H^{\varepsilon}(\rho\Vert\sigma)\leq D_{\max}(\rho\Vert\sigma)+\log_2\!\left(\frac{1}{1-\varepsilon}\right).
		\end{equation}
	\end{proposition*}

	\begin{Proof}
		If the support condition $\supp(\rho)\subseteq
\supp(\sigma)$ does not hold, then the right-hand side of \eqref{eq:sandwich-to-htre} is equal to $+\infty$, and so the result is trivially true. Thus, in what follows, we suppose that the support condition $\supp(\rho)\subseteq \supp(\sigma)$ holds. Let $\Lambda$ denote a measurement operator such that $\Tr[\Lambda\rho]=1-\varepsilon$. Let $q\coloneqq\Tr[\Lambda\sigma]$. By the data-processing inequality for the sandwiched R\'{e}nyi relative entropy for $\alpha>1$ (Theorem \ref{thm-sand_renyi_monotone}), under the measurement channel
		\begin{equation}
			\omega\mapsto\Tr[\Lambda\omega]|0\rangle\!\langle 0|+\Tr[(\mathbbm{1}-\Lambda)\omega]|1\rangle\!\langle1|,
			\label{eq-meas-channel-sandwiched-hypotest}
		\end{equation}
		we find that
		\begin{align}
			\widetilde{D}_{\alpha}(\rho\Vert\sigma) &  \geq\widetilde{D}_{\alpha}(\{1-\varepsilon,\varepsilon\}\Vert\{q,\Tr[\sigma]-q\})\label{eq:sandwich-to-p-and-q-1}\\
			&  =\frac{1}{\alpha-1}\log_{2}\!\left[  \left(  1-\varepsilon\right)^{\alpha}q^{1-\alpha}+\varepsilon^{\alpha}\left(\Tr[\sigma]-q\right)  ^{1-\alpha}\right]  \\
			&  \geq\frac{1}{\alpha-1}\log_{2}[\left(  1-\varepsilon\right)^{\alpha}q^{1-\alpha}]\\
			&  =\frac{\alpha}{\alpha-1}\log_{2}(1-\varepsilon)-\log_{2} q.\label{eq:sandwich-to-p-and-q-last}
		\end{align}
		The statement in \eqref{eq:sandwich-to-htre} follows by taking the supremum over all $\Lambda$ such that $\Tr[\Lambda\rho]=1-\varepsilon$. Furthermore, \eqref{eq-D_max_to_htre} follows because $\lim_{\alpha\to\infty}\widetilde{D}_\alpha(\rho\Vert\sigma)=D_{\max}(\rho\Vert\sigma)$ (as shown in Proposition \ref{prop-sand_rel_ent_limit_max}) and the fact that $\lim_{\alpha\to\infty}\frac{\alpha}{\alpha-1}=1$.
	\end{Proof}

	The following proposition establishes an inequality relating hypothesis testing relative entropy and the Petz--R\'{e}nyi relative entropy, and it represents a counterpart to Proposition \ref{prop:sandwich-to-htre}. After giving its proof, we show how Proposition \ref{prop:sandwich-to-htre} and the following proposition lead to a proof of the quantum Stein's lemma.

	\begin{proposition*}{Hypothesis Testing to Petz--R\'enyi Relative Entropy}{prop:ineq-hypo-renyi}
		Let $\rho$ be a state, and let $\sigma$ be a positive semi-definite operator. Let $\alpha\in(0,1)$ and $\varepsilon \in(0,1]$. Then the following inequality holds:
		\begin{equation}\label{eq-hypo_testing_lower}
			D_{H}^{\varepsilon}(\rho\Vert\sigma)\geq D_{\alpha}(\rho\Vert\sigma)+\frac{\alpha}{\alpha-1}\log_{2}\!\left(  \frac{1}{\varepsilon}\right).
		\end{equation}
	\end{proposition*}

	\begin{Proof}
		The statement is trivially true if $\rho\sigma=0$ because both $D_{H}^{\varepsilon}(\rho\Vert\sigma)=+\infty$ and $D_{\alpha} (\rho\Vert\sigma)=+\infty$ in this case. So we consider the non-trivial case when this equality does not hold. Recall from Lemma \ref{lemma:spectral-ineq} that the following inequality holds for positive semi-definite operators $A$ and $B$ and for $\alpha\in(0,1)$:
		\begin{align}
			\inf_{\Lambda:0\leq \Lambda\leq \mathbbm{1}}\Tr[(\mathbbm{1}-\Lambda)A]+\Tr[\Lambda B] & =\frac{1}{2}\left(  \Tr[A+B]-\norm{A-B}_{1}\right)  \\
			&  \leq\Tr[A^{\alpha}B^{1-\alpha}],
		\end{align}
		where the first equality is the result of Theorem \ref{thm-Holevo_Helstrom}. For $p\in(0,1)$, pick $A=p\rho$ and $B=\left(  1-p\right)  \sigma$. Plugging in to the  inequality above, we find that there exists a measurement operator $\Lambda^*=\Lambda(p,\rho,\sigma)$ such that
		\begin{equation}
			p\Tr[(\mathbbm{1}-\Lambda^*)\rho]+(1-p)\Tr[\Lambda^*\sigma]\leq p^{\alpha}(1-p)^{1-\alpha}\Tr[\rho^{\alpha}\sigma^{1-\alpha}].
		\end{equation}
		This implies that
		\begin{equation}
			p\Tr[(\mathbbm{1}-\Lambda^*)\rho]\leq p^{\alpha}(1-p)^{1-\alpha}\Tr[\rho^{\alpha}\sigma^{1-\alpha}],
		\end{equation}
		and in turn that
		\begin{equation}
			\Tr[(\mathbbm{1}-\Lambda^*)\rho]\leq\left(\frac{1-p}{p}\right)^{1-\alpha}\Tr[\rho^{\alpha}\sigma^{1-\alpha}].
		\end{equation}
		For a given $\varepsilon\in(0,1]$ and $\alpha\in(0,1)$, we pick $p\in(0,1)$ such that
		\begin{equation}
			\left(\frac{1-p}{p}\right)^{1-\alpha}\Tr[\rho^{\alpha}\sigma^{1-\alpha}]=\varepsilon.
		\end{equation}
		This is possible because we can rewrite the equation above as
		\begin{align}
			\varepsilon &  =\left(  \frac{1-p}{p}\right)  ^{1-\alpha}\Tr[\rho^{\alpha}\sigma^{1-\alpha}]\nonumber\\
			&=\left(  \frac{1}{p}-1\right)^{1-\alpha}\Tr[\rho^{\alpha}\sigma^{1-\alpha}]\\
			\Longleftrightarrow\qquad \left(  \frac{1}{p}-1\right)  ^{1-\alpha} &  =\frac{\varepsilon}{\Tr[\rho^{\alpha}\sigma^{1-\alpha}]}\\
			\Longleftrightarrow\qquad\frac{1}{p} &  =\left(\frac{\varepsilon}{\Tr[\rho^{\alpha}\sigma^{1-\alpha}]}\right)^{\frac{1}{1-\alpha}}+1\\
			\Longleftrightarrow\qquad p &  =\frac{1}{\left(\frac{\varepsilon}{\Tr[\rho^{\alpha}\sigma^{1-\alpha}]}\right)^{\frac{1}{1-\alpha}}+1}\in\left(0,1\right)  .
		\end{align}
		This means that $\Lambda^*=\Lambda(p,\rho,\sigma)$, with $p$ selected as above, is a measurement operator such that
		\begin{equation}\label{eq:ineq-hypo-renyi_pf}
			\Tr[(\mathbbm{1}-\Lambda^*)\rho]\leq\varepsilon.
		\end{equation}
		Now, 
		we use the fact that
		\begin{equation}
			(1-p)\Tr[\Lambda^*\sigma]\leq p^{\alpha}(1-p)^{1-\alpha}\Tr[\rho^{\alpha}\sigma^{1-\alpha}]
		\end{equation}
		implies
		\begin{equation}
			\Tr[\Lambda^*\sigma]\leq\left(  \frac{p}{1-p}\right)  ^{\alpha}\Tr[\rho^{\alpha}\sigma^{1-\alpha}].
		\end{equation}
		Considering that
		\begin{equation}
			\varepsilon=\left(  \frac{1-p}{p}\right)  ^{1-\alpha}\Tr[\rho^{\alpha}\sigma^{1-\alpha}]=\left(  \frac{p}{1-p}\right)^{\alpha-1}\Tr[\rho^{\alpha}\sigma^{1-\alpha}]
		\end{equation}
		implies that
		\begin{equation}
			\left(\frac{\varepsilon}{\Tr[\rho^{\alpha}\sigma^{1-\alpha}]}\right)^{\frac{1}{\alpha-1}}=\frac{p}{1-p},
		\end{equation}
		we get that
		\begin{align}
			\Tr[\Lambda^*\sigma] &  \leq\left(  \frac{p}{1-p}\right)^{\alpha}\Tr[\rho^{\alpha}\sigma^{1-\alpha}]\\
			&  =\left(  \left(  \frac{\varepsilon}{\Tr[\rho^{\alpha}\sigma^{1-\alpha}]}\right)^{\frac{1}{\alpha-1}}\right)  ^{\alpha}\Tr[\rho^{\alpha}\sigma^{1-\alpha}]\\
			&  =\varepsilon^{\frac{\alpha}{\alpha-1}}\left(  \Tr[\rho^{\alpha}\sigma^{1-\alpha}]\right)^{\frac{\alpha}{1-\alpha}}\Tr[\rho^{\alpha}\sigma^{1-\alpha}]\\
			&  =\varepsilon^{\frac{\alpha}{\alpha-1}}\left(  \Tr[\rho^{\alpha}\sigma^{1-\alpha}]\right)^{\frac{1}{1-\alpha}}.
		\end{align}
		Then, by taking the negative logarithm and optimizing over all $\Lambda^*$ satisfying \eqref{eq:ineq-hypo-renyi_pf}, we find that
		\begin{align}
			D_H^{\varepsilon}(\rho\Vert\sigma)&\geq -\log_{2}\Tr[\Lambda^*\sigma]\\
			&\geq-\log_{2}\!\left(\varepsilon^{\frac{\alpha}{\alpha-1}}\left(\Tr[\rho^{\alpha}\sigma^{1-\alpha}]\right)^{\frac{1}{1-\alpha}}\right)  \\
			&  =-\frac{\alpha}{\alpha-1}\log_{2}(\varepsilon)+\frac{1}{\alpha-1}\log_{2}\Tr[\rho^{\alpha}\sigma^{1-\alpha}]\\
			&  =-\frac{\alpha}{\alpha-1}\log_{2}(\varepsilon)+D_{\alpha}(\rho\Vert\sigma).
		\end{align}
		Rearranging this inequality leads to \eqref{eq-hypo_testing_lower}, as required.
	\end{Proof}

\section{Quantum Stein's Lemma}\label{subsec-q_Stein_lemma}

	In this section, we show how Propositions \ref{prop:sandwich-to-htre} and \ref{prop:ineq-hypo-renyi} lead to a proof of the quantum Stein's lemma, which is one of the most important results in the asymptotic theory of quantum hypothesis testing. Before doing so, let us discuss the task of asymmetric hypothesis testing.

\begin{figure}
		\centering
		\includegraphics[scale=0.7]{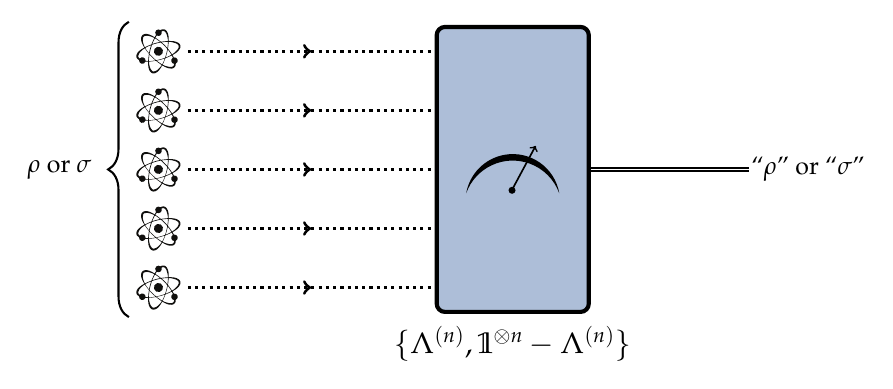}
		\caption{Schematic depiction of hypothesis testing. Many copies, say $n$, of an identical quantum system are each prepared either in the state $\rho$ or the state $\sigma$. A joint binary measurement, described by the POVM $\{\Lambda^{(n)},\mathbbm{1}^{\otimes n}-\Lambda^{(n)}\}$, is made on the overall state, which is either $\rho^{\otimes n}$ or~$\sigma^{\otimes n}$. }\label{fig-hypo_testing}
	\end{figure}

	The general setting of asymmetric i.i.d.~hypothesis testing is illustated in Figure \ref{fig-hypo_testing}. Bob is given $n$ copies of a quantum system, each of which is either in the state $\rho$ or in the state $\sigma$, and his task is to determine in which state the systems have been prepared. Bob's strategy consists of performing a joint measurement on all of the systems at once, described by the POVM $\{\Lambda^{(n)},\mathbbm{1}^{\otimes n}-\Lambda^{(n)}\}$, guessing ``$\rho$'' if the outcome corresponds to $\Lambda^{(n)}$ and guessing ``$\sigma$'' if the outcome corresponds to $\mathbbm{1}^{\otimes n}-\Lambda^{(n)}$. In this case, there are two types of errors that can occur.
	\begin{enumerate}
		\item\textit{Type-I Error}: Bob guesses ``$\sigma$'', but the systems are in the state $\rho^{\otimes n}$. The probability of this occurring is $\Tr[(\mathbbm{1}^{\otimes n}-\Lambda^{(n)})\rho^{\otimes n}]$.
		\item\textit{Type-II Error}: Bob guesses ``$\rho$'', but the systems are in the state $\sigma^{\otimes n}$. The probability of this occurring is $\Tr[\Lambda^{(n)}\sigma^{\otimes n}]$.
	\end{enumerate}
	
	The quantity $\beta_{\varepsilon}(\rho^{\otimes n}\Vert\sigma^{\otimes n})$, defined from \eqref{eq-entr:beta-quant}, can be interpreted as the minimum type-II error probability subject to the constraint $1-\Tr[\Lambda^{(n)}\rho]=\Tr[(\mathbbm{1}^{\otimes n}-\Lambda^{(n)})\rho]\leq \varepsilon$ on the type-I error probability. The \textit{rate} of this protocol, given a type-I error probability constraint of $\varepsilon$, is $\frac{1}{n}D_H^{\varepsilon}(\rho^{\otimes n}\Vert\sigma^{\otimes n})$, which is essentially the minimum type-II error probability exponent per copy of the states $\rho$ and $\sigma$. By increasing the number $n$ of copies, one might imagine that this  normalized minimum type-II error probability exponent can be increased. Also, there is generally a trade-off between the type-I and type-II error probabilities, meaning that both cannot be made arbitrarily small simultaneously. However, by increasing the number $n$ of copies of the states $\rho$ and $\sigma$, we might expect that the type-I error probability can be brought down all the way to zero. We thus define the \textit{maximum rate for hypothesis testing} of the states $\rho$ and $\sigma$ as the largest value of the normalized type-II error probability exponent, as $n\to\infty$, such that the type-I error probability vanishes in this limit, i.e.,  
	\begin{equation}\label{eq-hypo_testing_opt_rate}
		\inf_{\varepsilon\in (0,1)}\liminf_{n\to \infty}\frac{D_H^{\varepsilon}(\rho^{\otimes n}\Vert\sigma^{\otimes n})}{n}.
	\end{equation}
	A tractable expression for this optimal rate is given by the \textit{quantum Stein's lemma}, which we state and prove in Theorem \ref{thm-q_Stein_lemma} below.
	
	The task of asymmetric hypothesis testing is similar to the task of state discrimination that we considered in Section~\ref{subsec-state_discrimination}. While in hypothesis testing we consider two error probabilities, the type-I and type-II error probabilities, in state discrimination we consider only one error probability that is in fact the average of the type-I and type-II error probabilities taken with respect to a prior probability distribution. If $\lambda\in[0,1]$ is the probability of choosing the state $\rho$, then the average of the type-I and type-II error probabilities is
	\begin{multline}
		\lambda\Tr[(\mathbbm{1}^{\otimes n}-\Lambda^{(n)})\rho^{\otimes n}]+(1-\lambda)\Tr[\Lambda^{(n)}\sigma^{\otimes n}]\\
		=p_{\text{err}}(\{\Lambda^{(n)},\mathbbm{1}^{\otimes n}-\Lambda^{(n)}\},\{\rho^{\otimes n},\sigma^{\otimes n}\}),
	\end{multline}
	where we recall the definition of the error probability $p_{\text{err}}(\{\Lambda^{(n)},\mathbbm{1}^{\otimes n}-\Lambda^{(n)}\};\{\rho,\sigma\})$ of state discrimination from \eqref{eq-hypo_testing_state_symm_err_prob}. Due to the fact that we combine both the type-I and type-II error probabilities, the task of state discrimination is often referred to as \textit{symmetric hypothesis testing}, while the task of hypothesis testing that we are considering in this section is referred to as \textit{asymmetric hypothesis testing}.
	
	
	More formally, a \textit{hypothesis testing protocol} is defined by the four elements $(n,\rho,\sigma,\Lambda^{(n)})$, where $n$ is the number of copies of the system, each of which is either in the state $\rho$ or $\sigma$, and $0\leq \Lambda^{(n)}\leq\mathbbm{1}^{\otimes n}$ is the operator defining the POVM $\{\Lambda^{(n)},\mathbbm{1}^{\otimes n}-\Lambda^{(n)}\}$ used to decide the state of the system.
	
	\begin{definition}{$\boldsymbol{(n,\varepsilon_{\text{II}},\varepsilon_{\text{I}})}$ Hypothesis Testing Protocol}{def-neIeII_hypo_testing_protocol}
		Let $(n,\rho,\sigma,\Lambda^{(n)})$ be the elements of a hypothesis testing protocol. The protocol is called an \textit{$(n,\varepsilon_{\text{II}},\varepsilon_{\text{I}})$ protocol for $\rho$ and $\sigma$} if the type-I error probability 
		$
		\Tr[(\mathbbm{1}^{\otimes n}-\Lambda^{(n)})\rho^{\otimes n}]
		$
		 satisfies 
		 \begin{equation}
		 \Tr[(\mathbbm{1}^{\otimes n}-\Lambda^{(n)})\rho^{\otimes n}]\leq\varepsilon_{\text{I}},
		 \end{equation}
		  and the type-II error probability $\Tr[\Lambda^{(n)}\sigma^{\otimes n}]$ satisfies 
		  \begin{equation}
		  \Tr[\Lambda^{(n)}\sigma^{\otimes n}]\leq\varepsilon_{\text{II}}.
		  \end{equation}
	\end{definition}
	
	Observe that, by definition, if there exists an $(n,\varepsilon_{\text{II}},\varepsilon_{\text{I}})$ hypothesis testing protocol for $\rho$ and $\sigma$, then there exists an $(n,\varepsilon_{\text{II}}',\varepsilon_{\text{I}})$ hypothesis testing protocol for all $\varepsilon_{\text{II}}'\geq\varepsilon_{\text{II}}$ because every measurement operator $\Lambda^{(n)}$ for which $\Tr[\Lambda^{(n)}\sigma^{\otimes n}]\leq\varepsilon_{\text{II}}$ clearly satisfies $\Tr[\Lambda^{(n)}\sigma^{\otimes n}]\leq\varepsilon_{\text{II}}'$. Similarly, if there does not exist an $(n,\varepsilon_{\text{II}},\varepsilon_{\text{I}})$ hypothesis testing protocol, then there does not exist an $(n,\varepsilon_{\text{II}}',\varepsilon_{\text{I}})$ hypothesis testing protocol for all $\varepsilon_{\text{II}}'\leq\varepsilon_{\text{II}}$. 
	
	We define the \textit{rate} of a hypothesis testing protocol $(n,\rho,\sigma,\Lambda^{(n)})$ as
	\begin{equation}
		R(n,\rho,\sigma,\Lambda^{(n)})\coloneqq -\frac{1}{n}\log_2 \Tr[\Lambda^{(n)}\sigma^{\otimes n}],
	\end{equation}
	The rate is also called the \textit{(normalized) type-II error exponent} because the type-II error probability $\Tr[\Lambda^{(n)}\sigma^{\otimes n}]$ can be expressed as $\Tr[\Lambda^{(n)}\sigma^{\otimes n}]=2^{-nR}$. The goal of the hypothesis testing protocol is to find the highest rate such that the type-I error probability tends to zero as~$n$ increases.
	
	\begin{definition}{Achievable Rate for Hypothesis Testing}{def-hypo_testing_ach_rate}
		Given quantum states $\rho$ and $\sigma$, a rate $R\in\mathbb{R}^+$ is called an \textit{achievable rate for hypothesis testing of $\rho$ and $\sigma$} if for all $\varepsilon\in(0,1]$, $\delta>0$, and  sufficiently large $n$ there exists an $(n,2^{-n(R-\delta)},\varepsilon)$ hypothesis testing protocol for $\rho$ and $\sigma$.
	\end{definition}
	
	\begin{remark}
		When we say that there exists an $(n,2^{-n(R-\delta)},\varepsilon)$ hypothesis testing protocol for $\rho$ and $\sigma$ for all $\varepsilon\in(0,1]$,  $\delta>0$, and ``sufficiently large $n$'', we mean that for all $\varepsilon\in(0,1]$ and  $\delta>0$, there exists a number $N_{\varepsilon,\delta}\in\mathbb{N}$ such that for all $n\geq N_{\varepsilon,\delta}$, there exists an $(n,2^{-n(R-\delta)},\varepsilon)$ hypothesis testing protocol for $\rho$ and $\sigma$. This convention with the nomenclature ``sufficiently large $n$'' is taken throughout the rest of the book.
	\end{remark}
	
	Note that, by definition, for every achievable rate $R$ there exists a value of $n$ such that the type-I error probability $\varepsilon$ becomes arbitrarily close to zero. 
	
	\begin{definition}{Optimal Achievable Rate for Hypothesis Testing}{def-opt_tII_exp_hypo_testing}
		Given quantum states $\rho$ and $\sigma$, the \textit{optimal achievable rate}, denoted by $E(\rho,\sigma)$, is defined as the supremum of all achievable rates for hypothesis testing of $\rho$ and $\sigma$, i.e.,
		\begin{equation}
			E(\rho,\sigma)\coloneqq\sup\{R: R\text{ is an achievable rate for }\rho,\sigma\}.
		\end{equation}
	\end{definition}

	\begin{definition}{Strong Converse Rate for Hypothesis Testing}{def-hypo_testing_str_conv_rate}
		Given quantum states $\rho$ and $\sigma$, a rate $R\in\mathbb{R}^+$ is called a \textit{strong converse rate for hypothesis testing of $\rho$ and $\sigma$} if for all $\varepsilon\in[0,1)$, $\delta>0$, and sufficiently large $n$, there does not exist an $(n,2^{-n(R+\delta)},\varepsilon)$ hypothesis testing protocol for $\rho$ and $\sigma$.
	\end{definition}
	
	Note that, by definition, for every strong converse rate $R$ there exists a value of $n$ such that the type-I error probability $\varepsilon$ is arbitrarily close to one.
	
	\begin{definition}{Optimal Strong Converse Rate for Hypothesis Testing}{def-opt_str_conv_tII_exp_hypo_testing}
		Given quantum states $\rho$ and $\sigma$, the \textit{optimal strong converse rate}, denoted by $\widetilde{E}(\rho,\sigma)$, is defined as the infimum of all strong converse rates for hypothesis testing of $\rho$ and $\sigma$, i.e.,
		\begin{equation}
			\widetilde{E}(\rho,\sigma)\coloneqq\inf\{R: R\text{ is a strong converse rate for }\rho,\sigma\}.
		\end{equation}
	\end{definition}

	Note that the following inequality is a direct consequence of definitions:\begin{equation}\label{eq-hypo_testing_tII_vs_tII_str_conv}
			E(\rho,\sigma)\leq\widetilde{E}(\rho,\sigma).
		\end{equation}
		Indeed, suppose for a contradiction that this is not true, i.e., that $E(\rho,\sigma)>\widetilde{E}(\rho,\sigma)$ holds. This means that there exists an achievable rate $R$ such that $\widetilde{E}(\rho,\sigma)<R<E(\rho,\sigma)$, so that for all $\varepsilon\in(0,1]$, all $\delta>0$, and all sufficiently large $n$ there exists an $(n,2^{-n(R-\delta)},\varepsilon)$ hypothesis testing protocol. On the other hand, since $\widetilde{E}(\rho,\sigma)$ is the optimal strong converse rate, there exists a $\delta>0$ such that $R-\delta>\widetilde{E}(\rho,\sigma)$ is a strong converse rate. This implies, by definition, that an $(n,2^{-n(R-\delta+\delta')},\varepsilon)$ protocol, with $\delta'\in(0,\delta)$, does not exist. However, since $R$ was claimed to be an achievable rate, such a protocol should exist since $R-\delta+\delta'<R$. We have thus reached a contradiction. The inequality $E(\rho,\sigma)>\widetilde{E}(\rho,\sigma)$ therefore cannot be true, which means that $E(\rho,\sigma)\leq\widetilde{E}(\rho,\sigma)$.
	
	We now state and prove the quantum Stein's lemma:  both the optimal achievable and strong converse rates are equal to the quantum relative entropy. As alluded to at the beginning of Section \ref{sec-rel_ent} on the quantum relative entropy, the quantum Stein's lemma gives the quantum relative entropy its most fundamental operational meaning as the optimal rate in asymmetric quantum hypothesis testing.
	
	\begin{theorem*}{Quantum Stein's Lemma}{thm-q_Stein_lemma}
		For all states $\rho$ and $\sigma$, the optimal achievable and strong converse rates are equal to the quantum relative entropy of $\rho$ and $\sigma$, i.e.,
		\begin{equation}
			E(\rho,\sigma)=\widetilde{E}(\rho,\sigma)=D(\rho\Vert\sigma).
		\end{equation}
	\end{theorem*}
	
	\begin{Proof}
		
		Note that if $\supp(\rho)\nsubseteq\supp(\sigma)$, then $D(\rho\Vert\sigma)=+\infty$ by definition. In this singular case, the optimal strong converse rate $\widetilde{E}(\rho,\sigma)$ is undefined, and so we prove that $E(\rho,\sigma)=+\infty$.
		
		Fix $\varepsilon\in(0,1]$, and let $\Pi_\sigma$ be the projection onto the support of $\sigma$. Note that since $\supp(\rho)\nsubseteq\supp(\sigma)$, we have that $\Tr[\Pi_{\sigma}\rho]<1$. Now, pick $n$ large enough so that $(\Tr[\Pi_{\sigma}\rho])^n\leq\varepsilon$. Define the measurement operator $\Lambda^{(n)}$ as $\Lambda^{(n)}\coloneqq \mathbbm{1}-(\Pi_{\sigma})^{\otimes n}$. Observe that the type-I error probability is
		\begin{equation}
		\Tr[(\mathbbm{1}^{\otimes n}-\Lambda^{(n)})\rho^{\otimes n}]=(\Tr[\Pi_{\sigma}\rho])^n\leq\varepsilon
		\end{equation}
		and the type-II error probability is
		\begin{equation}
		\Tr[\Lambda^{(n)}\sigma^{\otimes n}]=1-(\Tr[\Pi_{\sigma}\sigma])^n=0.
		\end{equation}
		Therefore, for all sufficiently large $n$ such that $(\Tr[\Pi_\sigma\rho])^n\leq\varepsilon$ holds, the elements $(n,\rho,\sigma,\Lambda^{(n)})$ constitute an $(n,0,\varepsilon)$ hypothesis testing protocol for $\rho$ and $\sigma$, the rate of which is $+\infty=D(\rho\Vert\sigma)$. Since $\varepsilon$ is arbitrary, we conclude that, for all $\varepsilon\in(0,1]$ and sufficiently large $n$, there exists a hypothesis testing protocol for $\rho$ and $\sigma$ with rate $R=+\infty$. This implies that $E(\rho,\sigma)=+\infty$ in the singular case of $\supp(\rho)\nsubseteq\supp(\sigma)$. (See the Problems at the end of Chapter~\ref{chap-QM_protocols}.)
			
		For the remainder of the proof, we assume that the support condition $\supp(\rho)\subseteq\supp(\sigma)$ holds, so that $D(\rho\Vert\sigma)$ is finite.
	
		Let us first show that $D(\rho\Vert\sigma)$ is an achievable rate, which establishes that $E(\rho,\sigma)\geq D(\rho\Vert\sigma)$. To this end, fix $\varepsilon\in(0,1]$ and $\delta>0$. Let $\delta_1,\delta_2>0$ be such that
		\begin{equation}\label{eq-stein_lemma_pf1}
			\delta_1+\delta_2=\delta.
		\end{equation}
		Set $\alpha\in(0,1)$ such that
		\begin{equation}\label{eq-stein_lemma_pf2}
			\delta_1\geq D(\rho\Vert\sigma)-D_\alpha(\rho\Vert\sigma),
		\end{equation}
		which is possible because $\lim_{\alpha\to 1}D_\alpha(\rho\Vert\sigma)=D(\rho\Vert\sigma)$ by Proposition \ref{prop-petz_rel_ent_lim_1} and  $D_\alpha$ is monotonically increasing in $\alpha$, as established in Proposition \ref{prop-Petz_rel_ent}. Then, with this choice of $\alpha$, take $n$ large enough so that
		\begin{equation}\label{eq-stein_lemma_pf3}
			\delta_2\geq\frac{\alpha}{n(1-\alpha)}\log_2\!\left(\frac{1}{\varepsilon}\right).
		\end{equation}
		
		Now, let $0\leq\Lambda^{(n)}\leq\mathbbm{1}^{\otimes n}$ be a measurement operator that achieves the $\varepsilon$-hypothesis testing relative entropy $D_H^{\varepsilon}(\rho^{\otimes n}\Vert\sigma^{\otimes n})$, which means that 
		\begin{equation}
			\Tr[(\mathbbm{1}^{\otimes n}-\Lambda^{(n)})\rho^{\otimes n}]=\varepsilon
		\end{equation}
		and
		\begin{equation}
			-\frac{1}{n}\log_2\Tr[\Lambda^{(n)}\sigma^{\otimes n}]=\frac{1}{n}D_H^{\varepsilon}(\rho^{\otimes n}\Vert\sigma^{\otimes n}).
		\end{equation}
		The elements $(n,\rho,\sigma,\Lambda^{(n)})$ thus constitute an $(n,2^{-n\left(\frac{1}{n}D_H^{\varepsilon}(\rho^{\otimes n}\Vert\sigma^{\otimes n})\right)},\varepsilon)$ hypothesis testing protocol for $\rho$ and $\sigma$. We now apply Proposition \ref{prop:ineq-hypo-renyi} and the additivity of the Petz--R\'{e}nyi relative entropy from Proposition \ref{prop-Petz_rel_ent} to find that
		\begin{align}
			\frac{1}{n}D_{H}^{\varepsilon}(\rho^{\otimes n}\Vert\sigma^{\otimes n}) & \geq\frac{1}{n}D_{\alpha}(\rho^{\otimes n}\Vert\sigma^{\otimes n})+\frac{\alpha}{n(\alpha-1)}\log_{2}\!\left(  \frac{1}{\varepsilon}\right)\\
			&  =D_{\alpha}(\rho\Vert\sigma)+\frac{\alpha}{n(\alpha-1)}\log_{2}\!\left(  \frac{1}{\varepsilon}\right).\label{eq:stein-lower}
		\end{align} 
		Rearranging the right-hand side of this inequality and using \eqref{eq-stein_lemma_pf1}--\eqref{eq-stein_lemma_pf3}, we conclude that
		\begin{align}
			&\frac{1}{n}D_H^{\varepsilon}(\rho^{\otimes n}\Vert\sigma^{\otimes n})\nonumber\\
			&\quad \geq D(\rho\Vert\sigma)-\left(D(\rho\Vert\sigma)-D_\alpha(\rho\Vert\sigma)+\frac{\alpha}{n(1-\alpha)}\log_2\!\left(\frac{1}{\varepsilon}\right)\right)\\
			&\quad\geq D(\rho\Vert\sigma)-(\delta_1+\delta_2)\\
			&\quad\geq D(\rho\Vert\sigma)-\delta.
		\end{align}
		We thus have
		\begin{equation}
			D(\rho\Vert\sigma)-\delta\leq \frac{1}{n}D_H^{\varepsilon}(\rho^{\otimes n}\Vert\sigma^{\otimes n}).
		\end{equation}
		The error $2^{-n(D(\rho\Vert\sigma)-\delta)}$ is then greater than or equal to $2^{-n\left(\frac{1}{n}D_H^{\varepsilon}(\rho^{\otimes n}\Vert\sigma^{\otimes n})\right)}$, which means, by the fact stated in the paragraph immediately after Definition \ref{def-neIeII_hypo_testing_protocol}, that there exists an $(n,2^{-n(R-\delta)},\varepsilon)$ hypothesis testing protocol with $R=D(\rho\Vert\sigma)$ for all sufficiently large $n$ such that \eqref{eq-stein_lemma_pf3} holds. Since $\varepsilon$ and $\delta$ are arbitrary, we conclude that for all $\varepsilon\in(0,1]$, $\delta>0$, and sufficiently large $n$, there exists an $(n,2^{-n(R-\delta)},\varepsilon)$ hypothesis testing protocol with $R=D(\rho\Vert\sigma)$. Then $D(\rho\Vert\sigma)$ is an achievable rate, so that 
		\begin{equation}\label{eq-stein_lemma_pf7}
			E(\rho,\sigma)\geq D(\rho\Vert\sigma).
		\end{equation}
		
		Let us now show that the quantum relative entropy $D(\rho\Vert\sigma)$ is a strong converse rate, which establishes that $\widetilde{E}(\rho,\sigma)\leq D(\rho\Vert\sigma)$. Fix $\varepsilon\in[0,1)$ and $\delta>0$. Let $\delta_1,\delta_2>0$ be such that
		\begin{equation}\label{eq-stein_lemma_pf4}
			\delta>\delta_1+\delta_2\eqqcolon\delta'.
		\end{equation}
		Set $\alpha\in(1,\infty)$ such that
		\begin{equation}\label{eq-stein_lemma_pf5}
			\delta_1\geq \widetilde{D}_\alpha(\rho\Vert\sigma)-D(\rho\Vert\sigma),
		\end{equation}
		which is possible because $\lim_{\alpha\to 1}\widetilde{D}_\alpha(\rho\Vert\sigma)=D(\rho\Vert\sigma)$ by Proposition \ref{prop-sand_ren_ent_lim} and  $\widetilde{D}_\alpha$ is monotonically increasing in $\alpha$, as established in Proposition \ref{prop-sand_rel_ent_properties}. With this value of $\alpha$, take $n$ large enough so that
		\begin{equation}\label{eq-stein_lemma_pf6}
			\delta_2\geq\frac{\alpha}{n(\alpha-1)}\log_2\!\left(\frac{1}{1-\varepsilon}\right).
		\end{equation}
		
		Now, consider an arbitrary measurement operator $\Lambda^{(n)}$ such that the hypothesis testing protocol given by $(n,\rho,\sigma,\Lambda^{(n)})$ satisfies $\varepsilon\geq \Tr[(\mathbbm{1}^{\otimes n}-\Lambda^{(n)})\rho^{\otimes n}]$ and $\varepsilon_{\text{II}}\geq \Tr[\Lambda^{(n)}\sigma^{\otimes n}]$. By definition of the hypothesis testing relative entropy, we have that $-\log_2\Tr[\Lambda^{(n)}\sigma^{\otimes n}]\leq D_H^{\varepsilon}(\rho^{\otimes n}\Vert\sigma^{\otimes n})$. Applying Proposition \ref{prop:sandwich-to-htre}, we thus find that
		\begin{align}
			-\frac{1}{n}\log_2\Tr[\Lambda^{(n)}\sigma^{\otimes n}]&\leq \frac{1}{n}D_{H}^{\varepsilon}(\rho^{\otimes n}\Vert\sigma^{\otimes n})\\
			& \leq\frac{\alpha}{n(\alpha-1)}\log_{2}\!\left(  \frac{1}{1-\varepsilon }\right)  +\frac{1}{n}\widetilde{D}_{\alpha}(\rho^{\otimes n}\Vert \sigma^{\otimes n})\\
			&  =\frac{\alpha}{n(\alpha-1)}\log_{2}\!\left(  \frac{1}{1-\varepsilon }\right)  +\widetilde{D}_{\alpha}(\rho\Vert\sigma),\label{eq:stein-upper}
		\end{align}
		where the second line follows from the additivity of the sandwiched R\'{e}nyi relative entropy, as stated in Proposition \ref{prop-sand_rel_ent_properties}. Rearranging the right-hand side of this inequality and using \eqref{eq-stein_lemma_pf4}--\eqref{eq-stein_lemma_pf6}, we obtain
		\begin{align}
			-\frac{1}{n}\log_2\Tr[\Lambda^{(n)}\sigma^{\otimes n}]&\leq D(\rho\Vert\sigma) +\widetilde{D}_\alpha(\rho\Vert\sigma)-D(\rho\Vert\sigma)\nonumber\\
			&\qquad +\frac{\alpha}{n(\alpha-1)}\log_2\!\left(\frac{1}{1-\varepsilon}\right)\\
			&\leq D(\rho\Vert\sigma)+\delta'\\
			&<D(\rho\Vert\sigma)+\delta.
		\end{align}
		We thus have that $ \Tr[\Lambda^{(n)}\sigma^{\otimes n}]> 2^{-n(D(\rho\Vert\sigma)+\delta)}$. Since $\Lambda^{(n)}$ is an arbitrary measurement operator satisfying $\varepsilon\geq \Tr[(\mathbbm{1}^{\otimes n}-\Lambda^{(n)})\rho^{\otimes n}]$, we see that, for all sufficiently large $n$ such that \eqref{eq-stein_lemma_pf6} holds, an $(n,2^{-n(D(\rho\Vert\sigma)+\delta)},\varepsilon)$ hypothesis testing protocol cannot exist, for if it did we would have $\Tr[\Lambda^{(n)}\sigma^{\otimes n}]\leq 2^{-n(D(\rho\Vert\sigma)+\delta)}$ for some $\Lambda^{(n)}$. Since $\varepsilon$ and $\delta$ are arbitrary, we have that for all $\varepsilon\in[0,1)$, $\delta>0$, and sufficiently large $n$, there does not exist an $(n,2^{-n(D(\rho\Vert\sigma)+\delta)},\varepsilon)$ hypothesis testing protocol for $\rho$ and $\sigma$, which means that $D(\rho\Vert\sigma)$ is a strong converse rate, so that
		\begin{equation}\label{eq-stein_lemma_pf8}
			\widetilde{E}(\rho,\sigma)\leq D(\rho\Vert\sigma).
		\end{equation}
		

		Using \eqref{eq-stein_lemma_pf7}, \eqref{eq-stein_lemma_pf8}, and \eqref{eq-hypo_testing_tII_vs_tII_str_conv}, we obtain
		\begin{equation}
			E(\rho,\sigma)\leq\widetilde{E}(\rho,\sigma)\leq D(\rho\Vert\sigma)\leq E(\rho,\sigma),
		\end{equation}
		which means that $E(\rho,\sigma)=\widetilde{E}(\rho,\sigma)=D(\rho\Vert\sigma)$.
			\end{Proof}
	
	We can conclude the main result of Theorem~\ref{thm-q_Stein_lemma} in a different yet related way. Recall that an alternate definition of the optimal type-II error exponent is given in \eqref{eq-hypo_testing_opt_rate}, i.e., 
	\begin{equation}
		E(\rho,\sigma)=\inf_{\varepsilon\in(0,1)}\liminf_{n\to\infty}\frac{D_H^{\varepsilon}(\rho^{\otimes n}\Vert\sigma^{\otimes n})}{n}.
	\end{equation}
	It is straightforward to show that
	\begin{equation}
		\inf_{\varepsilon\in( 0,1)}\liminf_{n\to\infty}\frac{D_H^{\varepsilon}(\rho^{\otimes n}\Vert\sigma^{\otimes n})}{n}=D(\rho\Vert\sigma).
	\end{equation}
	Indeed, using \eqref{eq:stein-lower}, we find that
	\begin{equation}
		\inf_{\varepsilon\in (0,1)}\liminf_{n\to\infty}\frac{1}{n}D_H^{\varepsilon}(\rho^{\otimes n}\Vert\sigma^{\otimes n})\geq D_\alpha(\rho\Vert\sigma)
	\end{equation}
	for all $\alpha\in(0,1)$, so that taking the supremum over $\alpha\in(0,1)$ on the right-hand side leads to
	\begin{equation}
		E(\rho,\sigma)\geq D(\rho\Vert\sigma).
	\end{equation}
	We can also write the optimal strong converse type-II error exponent as 
	\begin{equation}
		\widetilde{E}(\rho,\sigma)=\sup_{\varepsilon\in (0,1)}\limsup_{n\to\infty}\frac{1}{n}D_H^{\varepsilon}(\rho^{\otimes n}\Vert\sigma^{\otimes n}).
	\end{equation}
	Similarly, using \eqref{eq:stein-upper}, we conclude that
	\begin{equation}
		\sup_{\varepsilon\in(0,1) }\limsup_{n\to\infty}\frac{1}{n}D_H^{\varepsilon}(\rho^{\otimes n}\Vert\sigma^{\otimes n})\leq\widetilde{D}_\alpha(\rho\Vert\sigma)
	\end{equation}
	for all $\alpha>1$, so that taking the infimum over $\alpha\in (1,\infty)$ on the right-hand side leads to
	\begin{equation}
		\widetilde{E}(\rho,\sigma)\leq D(\rho\Vert\sigma).
	\end{equation}
	We therefore conclude that $E(\rho,\sigma)=\widetilde{E}(\rho,\sigma)=D(\rho\Vert\sigma)$. Note that in the  arguments above for the lower bound on $E(\rho,\sigma)$ and the upper bound on $\widetilde{E}(\rho,\sigma)$, we did not have to explicitly take the infimum or supremum over $\varepsilon\in (0,1)$, respectively.
	
	
\subsection{Error and Strong Converse Exponents}
	
	Given states $\rho$ and $\sigma$, as well as the number $n$ of copies of the states, we can change our perspective a bit from that given in the previous section and instead determine bounds on the type-I error probability $\varepsilon$. In particular, we can change our focus a bit, such that we are now interested in how fast the type-I error probability converges to zero if the type-II error exponent is equal to a constant smaller than the quantum relative entropy, and we are also interested in how fast the type-I error probability converges to one if the type-II error exponent is equal to a constant larger than the quantum relative entropy. To assist with this analysis, we establish the following propositions, whose proofs are closely related to the proofs of Propositions~\ref{prop:sandwich-to-htre} and \ref{prop:ineq-hypo-renyi}.
	
	\begin{proposition}{prop-fixed-rate-sandwiched-renyi}
Let $\rho$ be a state, and let $\sigma$ be a positive semi-definite operator.
Let $\alpha>1$ and $R\geq0$. Then, for $\Lambda$ a measurement operator
satisfying
\begin{equation}
\operatorname{Tr}[\Lambda\sigma]\leq2^{-R},
\end{equation}
the following bound holds:
\begin{equation}
\operatorname{Tr}[\left(  \mathbbm{1}-\Lambda\right)  \rho]\geq1-2^{-\left(
\frac{\alpha-1}{\alpha}\right)  \left(  R-\widetilde{D}_{\alpha}(\rho
\Vert\sigma)\right)  }.
\end{equation}

\end{proposition}

\begin{remark}
	Note that the second bound is nontrivial only in the case that $R>D(\rho
\Vert\sigma)$ because $\widetilde{D}_{\alpha}(\rho\Vert\sigma)>D(\rho
\Vert\sigma)$ for $\alpha>1$.
\end{remark}

\begin{Proof}

The proof is similar to the proof of Proposition~\ref{prop:sandwich-to-htre}. Let $p\coloneqq\operatorname{Tr}
[\Lambda\rho]$ and $q\coloneqq\operatorname{Tr}[\Lambda\sigma]$. By applying the data-processing inequality for the sandwiched R\'enyi relative entropy along with the
measurement channel from \eqref{eq-meas-channel-sandwiched-hypotest}, we conclude that
\begin{align}
\widetilde{D}_{\alpha}(\rho\Vert\sigma)  & \geq\widetilde{D}_{\alpha}(\left\{
p,1-p\right\}  \Vert\left\{  q,\operatorname{Tr}[\sigma]-q\right\}  )\\
& =\frac{1}{\alpha-1}\log_{2}\!\left[  p^{\alpha}q^{1-\alpha}+\left(
1-p\right)  ^{\alpha}\left(  \operatorname{Tr}[\sigma]-q\right)  ^{1-\alpha
}\right]  \\
& \geq\frac{1}{\alpha-1}\log_{2}\!\left[  p^{\alpha}q^{1-\alpha}\right]  \\
& =\frac{\alpha}{\alpha-1}\log_{2}p-\log_{2}q\\
& \geq\frac{\alpha}{\alpha-1}\log_{2}p+R\\
& =\frac{\alpha}{\alpha-1}\log_{2}\operatorname{Tr}[\Lambda\rho]+R,
\end{align}
which implies that
\begin{equation}
\operatorname{Tr}[\Lambda\rho]\leq2^{-\left(  \frac{\alpha-1}{\alpha}\right)
\left(  R-\widetilde{D}_{\alpha}(\rho\Vert\sigma)\right)  }
\end{equation}
Rewriting this gives the bound in the statement of the proposition.
\end{Proof}
	
	\begin{proposition}{prop-fixed-rate-petz-renyi}
		Let $\rho$ be a state, and let $\sigma$ be a positive semi-definite operator. Let $\alpha\in(0,1)$ and $R\geq0$. Then, there exists a measurement operator $\Lambda$ such that
		\begin{align}
			\operatorname{Tr}[\Lambda\sigma]  & \leq2^{-R},\\
			\operatorname{Tr}[\left(  \mathbbm{1}-\Lambda\right)  \rho]  & \leq2^{-\left(\frac{1-\alpha}{\alpha}\right)  \left(  D_{\alpha}(\rho\Vert\sigma)-R\right)}.
		\end{align}
	\end{proposition}

	\begin{remark}
		Note that the second bound above is nontrivial only in the case that $R<D(\rho\Vert\sigma)$ because $D_{\alpha}(\rho\Vert\sigma)<D(\rho\Vert \sigma)$ for $\alpha\in(0,1)$.
	\end{remark}

\begin{Proof}

The proof is similar to the proof of Proposition~\ref{prop:ineq-hypo-renyi}. Employing the same
measurement operator $\Lambda^{\ast}$ therein, we conclude from the same reasoning
in that proof that
\begin{align}
\operatorname{Tr}[\left(  \mathbbm{1}-\Lambda^{\ast}\right)  \rho]  & \leq\left(  \frac
{1-p}{p}\right)  ^{1-\alpha}\operatorname{Tr}[\rho^{\alpha}\sigma^{1-\alpha
}],\label{eq-err-expon-bnd-type-I}\\
\operatorname{Tr}[\Lambda^{\ast}\sigma]  & \leq\left(  \frac{p}{1-p}\right)
^{\alpha}\operatorname{Tr}[\rho^{\alpha}\sigma^{1-\alpha}].
\end{align}
We then pick $p\in(0,1)$ such that the following equation is satisfied
\begin{align}
2^{-R}  & =\left(  \frac{p}{1-p}\right)  ^{\alpha}\operatorname{Tr}
[\rho^{\alpha}\sigma^{1-\alpha}]\\
& =\left(  \frac{p}{1-p}\right)  ^{\alpha}2^{\left(
\alpha-1\right)  D_{\alpha}(\rho\Vert\sigma)}\\
\Longleftrightarrow \qquad 2^{-R}2^{\left(  1-\alpha\right)  D_{\alpha}(\rho\Vert\sigma)}
& =\left(  \frac{p}{1-p}\right)  ^{\alpha}
\\
\Longleftrightarrow \qquad2^{R/\alpha}2^{\left(  \alpha-1\right)  D_{\alpha}(\rho
\Vert\sigma)/\alpha}  & =\left(  \frac{1-p}{p}\right)  .
\end{align}
We see that picking $p$ in such a way is always possible because one more step
of the development above  leads to the conclusion that
\begin{equation}
p=\frac{1}{1+2^{R/\alpha}2^{\left(  \alpha-1\right)  D_{\alpha}(\rho
\Vert\sigma)/\alpha}}\in\left(  0,1\right)  .
\end{equation}
Substituting into \eqref{eq-err-expon-bnd-type-I}, we find that
\begin{align}
\operatorname{Tr}[\left(  \mathbbm{1}-\Lambda^{\ast}\right)  \rho]  & \leq\left(  2^{R/\alpha
}2^{\left(  \alpha-1\right)  D_{\alpha}(\rho\Vert\sigma)/\alpha}\right)
^{1-\alpha}2^{\left(  \alpha-1\right)  D_{\alpha}(\rho\Vert\sigma)}\\
& =2^{\left(  \frac{1-\alpha}{\alpha}\right)  R}2^{-\frac{\left(
1-\alpha\right)  ^{2}}{\alpha}D_{\alpha}(\rho\Vert\sigma)}2^{\left(
\alpha-1\right)  D_{\alpha}(\rho\Vert\sigma)}\\
& =2^{\left(  \frac{1-\alpha}{\alpha}\right)  R}2^{\frac{\left(
\alpha-1\right)  }{\alpha}D_{\alpha}(\rho\Vert\sigma)}\\
& =2^{-\left(  \frac{1-\alpha}{\alpha}\right)  \left(  D_{\alpha}(\rho
\Vert\sigma)-R\right)  },
\end{align}
concluding the proof.
\end{Proof}

	The inequalities from Propositions~\ref{prop-fixed-rate-sandwiched-renyi} and \ref{prop-fixed-rate-petz-renyi} lead to the following bounds on the type-I error probability $\varepsilon$ for quantum hypothesis testing, when the type-II error probability has a fixed rate $R$:
	\begin{equation}\label{eq-hypo_testing_error_exponent}
		1-2^{-n\left(\frac{\alpha-1}{\alpha}\right)\left(R-\widetilde{D}_\alpha(\rho\Vert\sigma)\right)}\leq \varepsilon
		\leq 2^{-n\left(\frac{1-\alpha}{\alpha}\right)\left(D_\alpha(\rho\Vert\sigma)-R\right)}.
	\end{equation}
	The left inequality holds for all $\alpha>1$, while the right inequality holds for all $\alpha\in(0,1)$. Let us now examine the behavior of $\varepsilon$ above and below the optimal rate $D(\rho\Vert\sigma)$.
	
	\begin{figure}
		\centering
		\includegraphics[scale=0.7]{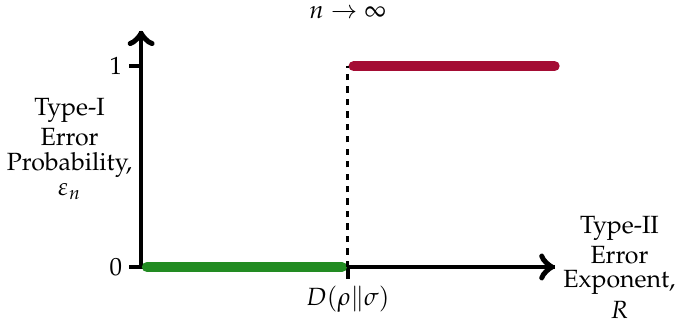}
		\caption{The type-I error probability $\varepsilon_n$ as a function of the rate $R$, i.e., the type-II error exponent, as the number $n$ of copies of the system approaches infinity for the task of asymmetric hypothesis testing for the states $\rho$ and $\sigma$. The optimal rate of $D(\rho\Vert\sigma)$ for this task, as established by the quantum Stein's lemma in Theorem \ref{thm-q_Stein_lemma}, has what is called the \textit{strong converse property}, which means that it is the optimal strong converse rate. Therefore, for every rate above it, the type-I error probability converges to one in the limit of arbitrarily many copies of the system.}\label{fig-hypo_testing_str_converse}
	\end{figure}
	
	\begin{enumerate}
		\item Consider a sequence $\{(n,2^{-nR},\varepsilon_n)\}_{n\in\mathbb{N}}$ of hypothesis testing protocols for $\rho$ and $\sigma$ such that the rate of each protocol has some arbitrary (but fixed) value $R<D(\rho\Vert\sigma)$. By Proposition~\ref{prop-fixed-rate-petz-renyi}, the sequence can be chosen such that each element satisfies the right-most inequality in \eqref{eq-hypo_testing_error_exponent}, so that
		\begin{equation}
			\varepsilon_n\leq 2^{-n\left(\frac{1-\alpha}{\alpha}\right)\left(D_\alpha(\rho\Vert\sigma)-R\right)}.
		\end{equation}
		Since the Petz--R\'{e}nyi relative entropy $D_\alpha(\rho\Vert\sigma)$ is monotonically increasing in $\alpha$, as established in Proposition \ref{prop-Petz_rel_ent}, there exists an $\alpha^*<1$ such that $D_{\alpha^*}(\rho\Vert\sigma)$ lies in between $D(\rho\Vert\sigma)$ and $R$; i.e., we have that $R<D_{\alpha^*}(\rho\Vert\sigma)$. We thus obtain
			\begin{equation}
				\varepsilon_n\leq 2^{-n\left(\frac{1-\alpha^*}{\alpha^*}\right)\left(D_{\alpha^*}(\rho\Vert\sigma)-R\right)}.
			\end{equation}
			Since $R<D_{\alpha^*}(\rho\Vert\sigma)$, taking the limit $n\to\infty$ on both sides of this inequality gives us $\lim_{n\to\infty}\varepsilon_n\leq 0$. However, $\varepsilon_n\geq 0$ for all $n$ because $\varepsilon_n$ is by definition a probability. So we find that
			\begin{equation}
				\varepsilon_n\to 0\quad\text{as}\quad n\to\infty\quad\text{if}\quad R<D(\rho\Vert\sigma).
			\end{equation}
			
		\item Now consider a sequence $\{(n,2^{-nR},\varepsilon_n)\}_{n\in\mathbb{N}}$ of hypothesis testing protocols for $\rho$ and $\sigma$ such that the rate of each protocol has some arbitrary (but fixed) value $R>D(\rho\Vert\sigma)$. In other words, suppose that we would like to perform hypothesis testing at a rate above the optimal rate. Each element of the sequence satisfies the left-most inequality in \eqref{eq-hypo_testing_error_exponent}, so that
			\begin{equation}
				\varepsilon_n\geq 1-2^{-n\left(\frac{\alpha-1}{\alpha}\right)\left(R-\widetilde{D}_\alpha(\rho\Vert\sigma)\right)}.
			\end{equation}
			Then, recalling from Proposition~\ref{prop-sand_rel_ent_properties} that the sandwiched R\'{e}nyi relative entropy $\widetilde{D}_\alpha(\rho\Vert\sigma)$ is monotonically increasing in $\alpha$, there exists an $\alpha^*>1$ such that $\widetilde{D}_{\alpha^*}(\rho\Vert\sigma)$ lies in between $D(\rho\Vert\sigma)$ and $R$; i.e., we have that $R>\widetilde{D}_{\alpha^*}(\rho\Vert\sigma)$. We thus obtain
			\begin{equation}
				\varepsilon_n\geq 1-2^{-n\left(\frac{\alpha^*-1}{\alpha^*}\right)\left(R-\widetilde{D}_{\alpha^*}(\rho\Vert\sigma)\right)}.
			\end{equation}
			Since $R>\widetilde{D}_{\alpha^*}(\rho\Vert\sigma)$, taking the limit $n\to\infty$ on both sides of this inequality gives us $\lim_{n\to\infty}\varepsilon_n\geq 1$. However, $\varepsilon_n\leq 1$ for all $n$ because $\varepsilon_n$ is by definition a probability. So we conclude that
			\begin{equation}
				\varepsilon_n\to 1\quad\text{as}\quad n\to\infty\quad\text{if}\quad R>D(\rho\Vert\sigma).
			\end{equation}
			So the type-I error probability grows exponentially to one in the limit $n\to\infty$ for every rate above the optimal rate.
	\end{enumerate}
	
	The optimal rate $D(\rho\Vert\sigma)$ is therefore a sharp dividing point below which the type-I error probability $\varepsilon_n$ exponentially drops to zero as $n\to\infty$ and above which it exponentially increases to one as $n\to\infty$. This behavior is illustrated in Figure \ref{fig-hypo_testing_str_converse}.

\section{Information Measures for Quantum Channels}\label{sec-inf_meas_chan}

	We conclude this chapter with a discussion of information measures for quantum channels. 
	
	Given an information measure defined for quantum states, we define a corresponding information measure for channels by sending one share of a bipartite state through a given channel and evaluating the information measure on the corresponding output state. Specifically, for every generalized divergence $\boldsymbol{D}:\Density(\mathcal{H})\times\Pos(\mathcal{H})\rightarrow\mathbb{R}\cup\{+\infty\}$, which was given in Definition~\ref{def-gen_div}, we define the \textit{generalized channel divergence} as follows:
	
	\begin{definition}{Generalized Channel Divergence}{def-gen_channel_div}
		Given a generalized divergence $\boldsymbol{D}:\Density(\mathcal{H})\times\Pos(\mathcal{H})\rightarrow\mathbb{R}\cup\{+\infty\}$, a quantum channel $\mathcal{N}_{A\to B}$, and a completely positive map $\mathcal{M}_{A\to B}$, we define the \textit{generalized channel divergence} of $\mathcal{N}$ and $\mathcal{M}$ as
		\begin{equation}\label{eq-gen_chan_div}
			\boldsymbol{D}(\mathcal{N}\Vert\mathcal{M})\coloneqq \sup_{\rho_{RA}}\boldsymbol{D}\!\left(\mathcal{N}_{A\to B}(\rho_{RA})\Vert \mathcal{M}_{A\to B}(\rho_{RA})\right),
		\end{equation}
		where the supremum is over all mixed states $\rho_{RA}$ with an arbitrary reference system $R$.
	\end{definition}
	
	\begin{proposition}{prop:QEI:gen-ch-div-pure-states-opt}
	Let $\boldsymbol{D}$, $\mathcal{N}_{A\to B}$, and $\mathcal{M}_{A\to B}$ be as given in Definition~\ref{def-gen_channel_div}. It suffices to optimize the generalized channel divergence $\boldsymbol{D}(\mathcal{N}\Vert\mathcal{M})$ with respect to pure states $\psi_{RA}$ with the reference system $R$ isomorphic to the input system $A$:
	\begin{align}
	\label{eq-gen_chan_div_pure}
		\boldsymbol{D}(\mathcal{N}\Vert\mathcal{M}) & = \sup_{\psi_{RA}}\boldsymbol{D}(\mathcal{N}_{A\to B}(\psi_{RA})\Vert\mathcal{M}_{A\to B}(\psi_{RA}))\\
		&  = \sup_{\rho_{A}}\boldsymbol{D}(\sqrt{\rho_A}\Gamma^{\mathcal{N}}_{A B} \sqrt{\rho_A}\Vert\sqrt{\rho_A}\Gamma^{\mathcal{M}}_{A B}\sqrt{\rho_A}).
		\label{eq-gen_chan_div_pure-dens-op-opt}
	\end{align}
	In the second line, $\Gamma^{\mathcal{N}}_{A B}$ and $\Gamma^{\mathcal{M}}_{A B}$ denote the Choi operators of $\mathcal{N}_{A\to B}$ and $\mathcal{M}_{A\to B}$, respectively, and the optimization is with respect to every density operator $\rho_A$.
	\end{proposition}
	
	\begin{Proof}
	Observe that if we take a purification $\ket{\phi}_{R'RA}$ of the state $\rho_{RA}$ in the optimization in \eqref{eq-gen_chan_div}, then we find that
	\begin{align}
		&\boldsymbol{D}\!\left(\mathcal{N}_{A\to B}(\rho_{RA})\Vert \mathcal{M}_{A\to B}(\rho_{RA})\right)\notag \\
		&\quad =\boldsymbol{D}\!\left(\mathcal{N}_{A\to B}(\Tr_{R'}[\phi_{R'RA}])\Vert\mathcal{M}_{A\to B}(\Tr_{R'}[\phi_{R'RA}])\right)\label{eq-gen_div_chan_pure_2}\\
		&\quad =\boldsymbol{D}\!\left(\Tr_{R'}\!\left[\mathcal{N}_{A\to B}(\phi_{R'RA})\right]\Vert\Tr_{R'}\!\left[\mathcal{M}_{A\to B}(\phi_{R'RA})\right]\right)\label{eq-gen_div_chan_pure_3}\\
		&\quad\leq \boldsymbol{D}\!\left(\mathcal{N}_{A\to B}(\phi_{R'RA})\Vert\mathcal{M}_{A\to B}(\phi_{R'RA})\right),\label{eq-gen_div_chan_pure_4}
	\end{align}
	where we have used the data-processing inequality for the generalized divergence in the last line. This means that for every state $\rho_{RA}$, the generalized divergence $\boldsymbol{D}\!\left(\mathcal{N}_{A\to B}(\rho_{RA})\Vert \mathcal{M}_{A\to B}(\rho_{RA})\right)$ is never larger than the corresponding generalized divergence evaluated on a purification of $\rho_{RA}$. This means that it suffices in \eqref{eq-gen_chan_div} to optimize over only pure states. Furthermore, by the Schmidt decomposition theorem (Theorem \ref{thm-Schmidt}), the purifying space $\mathcal{H}_{R'R}$ need not have dimension exceeding that of the dimension of $\mathcal{H}_A$. Therefore, the generalized channel divergence can be written as in \eqref{eq-gen_chan_div_pure}.
	
	To see \eqref{eq-gen_chan_div_pure-dens-op-opt}, we first use the fact in \eqref{eq-pure_state_vec}, which implies that for every pure state $\psi_{RA}$, there exists a state $\rho_R$ and a unitary $U_R$ such that
	\begin{equation}
		\ket{\psi}_{RA}=(U_R \sqrt{\rho_R}\otimes\mathbbm{1}_A)\ket{\Gamma}_{RA}. \label{eq:QEI:pure-bi-state-rewrite}
	\end{equation}
	Thus, it follows that
	\begin{align}
	\mathcal{N}_{A\to B}(\psi_{RA}) &  = 
	\mathcal{N}_{A\to B}((U_R \sqrt{\rho_R}\otimes\mathbbm{1}_A)\Gamma_{RA}( \sqrt{\rho_R}U_R^\dag\otimes\mathbbm{1}_A)) \\
		&  = (U_R \sqrt{\rho_R}\otimes\mathbbm{1}_B)\mathcal{N}_{A\to B}(\Gamma_{RA})( \sqrt{\rho_R}U_R^\dag\otimes\mathbbm{1}_B) \\
		&  = (U_R \sqrt{\rho_R}\otimes\mathbbm{1}_B)\Gamma^{\mathcal{N}}_{R B}( \sqrt{\rho_R}U_R^\dag\otimes\mathbbm{1}_B),
	\end{align}
	where we employed the Choi representation $\Gamma^{\mathcal{N}}_{RB}$  of the channel $\mathcal{N}$. Similarly, $\mathcal{M}_{A\to B}(\psi_{RA}) = U_R \sqrt{\rho_R}\Gamma^{\mathcal{M}}_{R B} \sqrt{\rho_R}U_R^\dag$.
	By employing the unitary invariance of a generalized divergence, we conclude that
	\begin{equation}
	\boldsymbol{D}(\mathcal{N}_{A\to B}(\psi_{RA})\Vert\mathcal{M}_{A\to B}(\psi_{RA}))\\
		  = \boldsymbol{D}(\sqrt{\rho_A}\Gamma^{\mathcal{N}}_{A B} \sqrt{\rho_A}\Vert\sqrt{\rho_A}\Gamma^{\mathcal{M}}_{A B}\sqrt{\rho_A}).
		  \label{eq:QEI:gen-ch-div-canon-purif}
	\end{equation}
	Then it suffices to optimize over states $\rho_A$, so that \eqref{eq-gen_chan_div_pure-dens-op-opt} holds.
\end{Proof}

\begin{proposition}{prop:QEI:gen-ch-div-concavity-convexity}
Let $\boldsymbol{D}$, $\mathcal{N}_{A\to B}$, and $\mathcal{M}_{A\to B}$ be as given in Definition~\ref{def-gen_channel_div}, and suppose that $\boldsymbol{D}$ obeys the direct-sum property in \eqref{eq:QEI:direct-sum-prop-gen-div}. Then the function
\begin{equation}
f(\rho_A,\mathcal{M}_{A\to B}) \coloneqq \boldsymbol{D}(\sqrt{\rho_A}\Gamma^{\mathcal{N}}_{AB}\sqrt{\rho_A}\Vert\sqrt{\rho_A}\Gamma^{\mathcal{M}}_{AB}\sqrt{\rho_A})
\label{eq:QEI:abbr-ch-div-func}
\end{equation}
is concave in the first argument and convex in the second. If $\mathfrak{M}$ is a convex set of completely positive maps, then
\begin{multline}
\inf_{\mathcal{M}\in\mathfrak{M}}\sup_{\rho_A}\boldsymbol{D}(\sqrt{\rho_A}\Gamma^{\mathcal{N}}_{AB}\sqrt{\rho_A}\Vert\sqrt{\rho_A}\Gamma^{\mathcal{M}}_{AB}\sqrt{\rho_A})\\
		=\sup_{\rho_A}\inf_{\mathcal{M}\in\mathfrak{M}}\boldsymbol{D}(\sqrt{\rho_A}\Gamma^{\mathcal{N}}_{AB}\sqrt{\rho_A}\Vert\sqrt{\rho_A}\Gamma^{\mathcal{M}}_{AB}\sqrt{\rho_A}).
		\label{eq:QEI:minimax-gen-ch-div-1}
\end{multline}
 Equivalently,
\begin{equation}
\inf_{\mathcal{M} \in \mathfrak{M}}\boldsymbol{D}(\mathcal{N} \Vert \mathcal{M}) = \sup_{\psi_{RA}}\inf_{\mathcal{M}  \in \mathfrak{M}} \boldsymbol{D}(\mathcal{N}_{A\to B}(\psi_{RA})\Vert\mathcal{M}_{A\to B}(\psi_{RA})),
\label{eq:QEI:minimax-gen-ch-div}
\end{equation}
where $\psi_{RA}$ is a pure state with system $R$ isomorphic to system $A$.
\end{proposition}

\begin{Proof}
To see the concavity, let $\psi_{RA}^{0}$ and $\psi_{RA}^{1}$ be pure states
with reduced states $\psi_{A}^{0}$ and $\psi_{A}^{1}$. Let $\psi
_{SRA}^{\lambda}$ denote the following pure state:
\begin{equation}
|\psi^{\lambda}\rangle_{SRA}\coloneqq \sqrt{1-\lambda}|0\rangle_{S}|\psi^{0}
\rangle_{RA}+\sqrt{\lambda}|1\rangle_{S}|\psi^{1}\rangle_{RA}.
\end{equation}
Observe that
\begin{equation}
\psi_{A}^{\lambda}=\left(  1-\lambda\right)  \psi_{A}^{0}+\lambda\psi_{A}^{1},
\end{equation}
so that the reduced state $\psi_{A}^{\lambda}$ is a convex combination of the
reduced states $\psi_{A}^{0}$ and $\psi_{A}^{1}$. Define
\begin{equation}
\overline{\psi}_{SRA}^{\lambda}\coloneqq \left(  1-\lambda\right)  |0\rangle\!
\langle0|_{S}\otimes\psi_{RA}^{0}+\lambda|1\rangle\!\langle1|_{S}\otimes
\psi_{RA}^{1},
\end{equation}
which is the state resulting from the action of a completely dephasing qubit
channel on system $S$. Let $\phi_{RA}^{\lambda}$ be an arbitrary pure state
with reduced state equal to $\psi_{A}^{\lambda}$. Then we find that
\begin{align}
& \!\!\!\!\!\!\!\!\!\boldsymbol{D}(\mathcal{N}_{A\rightarrow B}(\phi_{RA}^{\lambda})\Vert\mathcal{M}
_{A\rightarrow B}(\phi_{RA}^{\lambda}))\nonumber\\
& =\boldsymbol{D}(\mathcal{N}_{A\rightarrow B}(\psi_{SRA}^{\lambda})\Vert\mathcal{M}
_{A\rightarrow B}(\psi_{SRA}^{\lambda}))\\
& \geq \boldsymbol{D}(\mathcal{N}_{A\rightarrow B}(\overline{\psi}_{SRA}^{\lambda}
)\Vert\mathcal{M}_{A\rightarrow B}(\overline{\psi}_{SRA}^{\lambda}))\\
& =\lambda \boldsymbol{D}(\mathcal{N}_{A\rightarrow B}(\psi_{RA}^{0})\Vert\mathcal{M}
_{A\rightarrow B}(\psi_{RA}^{0}))\nonumber\\
& \qquad+\left(  1-\lambda\right)  \boldsymbol{D}(\mathcal{N}_{A\rightarrow B}(\psi
_{RA}^{1})\Vert\mathcal{M}_{A\rightarrow B}(\psi_{RA}^{1})).
\end{align}
The first equality follows because every two purifications of the same state are
related by an isometric channel acting on the reference system, as well as the
isometric invariance of the generalized divergence. The inequality follows
from quantum data processing, by the action of a completely dephasing qubit
channel on the system $S$. The final equality follows from the direct-sum
property and because the generalized divergence is invariant under tensoring in the same state (Proposition~\ref{prop-gen_div_properties}). Finally, we have the following equalities, by employing the isometric invariance of the generalized divergence (Proposition~\ref{prop-gen_div_properties}), the equality in \eqref{eq:QEI:gen-ch-div-canon-purif}, and the definition in \eqref{eq:QEI:abbr-ch-div-func}:
\begin{align}
\boldsymbol{D}(\mathcal{N}_{A\rightarrow B}(\phi_{RA}^{\lambda})\Vert\mathcal{M}
_{A\rightarrow B}(\phi_{RA}^{\lambda})) & = f(\psi_{A}^{\lambda},\mathcal{M}_{A\to B}), \\
\boldsymbol{D}(\mathcal{N}_{A\rightarrow B}(\psi_{RA}^{0})\Vert\mathcal{M}
_{A\rightarrow B}(\psi_{RA}^{0})) & = f(\psi_{A}^{0},\mathcal{M}_{A\to B}), \\
\boldsymbol{D}(\mathcal{N}_{A\rightarrow B}(\psi_{RA}^{1})\Vert\mathcal{M}
_{A\rightarrow B}(\psi_{RA}^{1})) & = f(\psi_{A}^{1},\mathcal{M}_{A\to B}).
\end{align}

	To see the convexity in $\mathcal{M}$, consider that for every $\lambda\in[0,1]$ and completely positive maps $\mathcal{M}_1,\mathcal{M}_2\in\mathfrak{M}$, the joint convexity of the generalized divergence (Proposition~\ref{prop:QEI:joint-convexity-gen-div}) gives that
	\begin{align}
		&f(\rho_A,\lambda\mathcal{M}_1+(1-\lambda)\mathcal{M}_2)\nonumber\\
		&\quad =\boldsymbol{D}(\sqrt{\rho_A}\Gamma^{\mathcal{N}}_{AB}\sqrt{\rho_A}\Vert\sqrt{\rho_A}\Gamma^{\lambda\mathcal{M}_1+(1-\lambda)\mathcal{M}_2}_{AB}\sqrt{\rho_A})\\
		&\quad =\boldsymbol{D}(\lambda\sqrt{\rho_A}\Gamma^{\mathcal{N}}_{AB}\sqrt{\rho_A}+(1-\lambda)\sqrt{\rho_A}\Gamma^{\mathcal{N}}_{AB}\sqrt{\rho_A}\nonumber\\
		&\qquad\qquad \Vert\lambda\sqrt{\rho_A}\Gamma^{\mathcal{M}_1}_{AB}\sqrt{\rho_A}+(1-\lambda)\sqrt{\rho_A}\Gamma^{\mathcal{M}_2}_{AB}\sqrt{\rho_A})\\
		&\quad\leq \lambda \boldsymbol{D}(\sqrt{\rho_A}\Gamma^{\mathcal{N}}_{AB}\sqrt{\rho_A}\Vert\sqrt{\rho_A}\Gamma^{\mathcal{M}_1}_{AB}\sqrt{\rho_A})\nonumber\\
		&\qquad\qquad +(1-\lambda)\boldsymbol{D}(\sqrt{\rho_A}\Gamma^{\mathcal{N}}_{AB}\sqrt{\rho_A}\Vert\sqrt{\rho_A}\Gamma^{\mathcal{M}_2}_{AB}\sqrt{\rho_A})\\
		&=\lambda f(\rho_A,\mathcal{M}_1)+(1-\lambda)f(\rho_A,\mathcal{M}_2).
	\end{align}
	
	The equality in \eqref{eq:QEI:minimax-gen-ch-div-1} follows from what was just shown and Sion's minimax theorem (Theorem~\ref{thm-Sion_minimax}). The equality in \eqref{eq:QEI:minimax-gen-ch-div} follows from \eqref{eq:QEI:gen-ch-div-canon-purif} and Proposition~\ref{prop:QEI:gen-ch-div-pure-states-opt}.
\end{Proof}
	
	The generalized channel divergence takes a simple form if the channel $\mathcal{N}$ and the completely positive map $\mathcal{M}$ both happen to be jointly covariant with respect to a group, as shown in the following proposition.
	
	\begin{proposition*}{Generalized Divergence for Jointly Covariant Channels}{prop-gen_div_group_cov}
		Let $\boldsymbol{D}:\Density(\mathcal{H})\times\Pos(\mathcal{H})\rightarrow\mathbb{R}\cup\{+\infty\}$ be a generalized divergence and $G$ a finite group with unitary representations $\{U_A^g\}_{g\in G}$ and  $\{V_B^g\}_{g\in G}$. Let $\mathcal{N}_{A\to B}$ be a quantum channel and $\mathcal{M}_{A\to B}$ a completely positive map that are both covariant with respect to $G$, i.e., (Definition~\ref{def-group_cov_chan})
		\begin{equation}
			\begin{aligned}
			\mathcal{N}(U_g\rho U_g^\dagger)&=V_g\mathcal{N}(\rho)V_g^\dagger,\\
			\mathcal{M}(U_g\rho U_g^\dagger)&=V_g\mathcal{M}(\rho)V_g^\dagger,
			\end{aligned}
		\end{equation}
		for every $g\in G$ and state $\rho$. Then, for every state $\psi_{A'A}$, with the dimension of $A'$ equal to the dimension of $A$,
		\begin{multline}
		\label{eq-gen_div_chan_cov_ineq}
			\boldsymbol{D}(\mathcal{N}_{A\to B}(\psi_{A'A})\Vert\mathcal{M}_{A\to B}(\psi_{A'A}))\\\leq \boldsymbol{D}(\mathcal{N}_{A\to B}(\phi_{RA}^{\overline{\rho}})\Vert\mathcal{M}_{A\to B}(\phi_{RA}^{\overline{\rho}})),
		\end{multline}
		where $\rho_A\coloneqq\psi_A=\Tr_{A'}[\psi_{A'A}]$,
		\begin{equation}
		\overline{\rho}_A\coloneqq\mathcal{T}_G(\rho_A)\coloneqq\frac{1}{|G|}\sum_{g \in G}U_A^g\rho_AU_A^{g\dagger},
		\end{equation}
		and $\phi_{RA}^{\overline{\rho}}$ is a purification of $\overline{\rho}_A$. Consequently, the generalized channel divergence $\boldsymbol{D}(\mathcal{N}\Vert\mathcal{M})$ is given by optimizing over pure states $\psi_{A'A}$ such that the reduced state $\psi_A$ is invariant under the channel $\mathcal{T}_G$; i.e.,
		\begin{multline}\label{eq-gen_div_chan_cov}
			\boldsymbol{D}(\mathcal{N}\Vert\mathcal{M})\\=\sup_{\psi_{A'A}}\{\boldsymbol{D}(\mathcal{N}_{A\to B}(\psi_{A'A})\Vert\mathcal{M}_{A\to B}(\psi_{A'A})):\psi_A=\mathcal{T}_G(\psi_A)\}.
		\end{multline}
		In particular, if the representation $\{U_A^g\}_{g\in G}$ is irreducible, then the optimal state in \eqref{eq-gen_div_chan_cov} is the maximally entangled state $\Phi_{A'A}$, so that
		\begin{equation}\label{eq-gen_div_cov_irrep}
			\boldsymbol{D}(\mathcal{N}\Vert\mathcal{M})=\boldsymbol{D}(\rho_{AB}^{\mathcal{N}}\Vert\rho_{AB}^{\mathcal{M}}),
		\end{equation}
		where $\rho_{AB}^{\mathcal{N}}$ and $\rho_{AB}^{\mathcal{M}}$ are Choi states of $\mathcal{N}$ and $\mathcal{M}$, respectively.
	\end{proposition*}
	
	
	\begin{Proof}
		The inequality
		\begin{equation}\label{eq-gen_div_chan_cov_pf_1}
			\boldsymbol{D}(\mathcal{N}\Vert\mathcal{M})\geq \sup_{\psi_{A'A}}\{\boldsymbol{D}(\mathcal{N}_{A\to B}(\psi_{A'A})\Vert\mathcal{M}_{A\to B}(\psi_{A'A})):\psi_A=\mathcal{T}_G(\psi_A)\}
		\end{equation}
		is immediate from the fact that the set  $\{\psi_{A'A}:\psi_A=\mathcal{T}_G(\psi_A)\}$ of pure states is a subset of all pure states. The remainder of this proof is devoted to the reverse inequality.
		
		Let $\psi_{A'A}$ be an arbitrary state, let $\rho_A=\psi_A$, and let $\overline{\rho}_A=\mathcal{T}_G(\rho_A)$. Furthermore, let $\phi_{RA}^{\overline{\rho}}$ be a purification of $\overline{\rho}_A$. Let us also consider the following purification of $\overline{\rho}_A$:
		\begin{equation}
			\ket{\psi^{\overline{\rho}}}_{R'A'A}\coloneqq\frac{1}{\sqrt{|G|}}\sum_{g\in G}\ket{g}_{R'}\otimes(\mathbbm{1}_{A'}\otimes U_A^g)\ket{\psi}_{A'A},
			\label{eq:QEI:gen-ch-div-symmetries-1}
		\end{equation}
		where $\{\ket{g}\}_{g\in G}$ is an orthonormal basis for $\mathcal{H}_{R'}$ indexed by the elements of $G$. Since all purifications of a state can be mapped to each other by isometries acting on the purifying systems, there exists an isometry $W_{R\to R'A'}$ such that $\ket{\psi^{\overline{\rho}}}_{R'A'A}=W_{R\to R'A'}\ket{\phi^{\overline{\rho}}}_{RA}$. Using this, we find that
		\begin{align}
			&\boldsymbol{D}(\mathcal{N}_{A\to B}(\psi_{R'A'A}^{\overline{\rho}})\Vert\mathcal{M}_{A\to B}(\psi_{R'A'A}^{\overline{\rho}}))\\
			&\quad=\boldsymbol{D}(\mathcal{N}_{A\to B}(\mathcal{W}_{R\to R'A'}(\phi^{\overline{\rho}}_{RA}))\Vert\mathcal{M}_{A\to B}(\mathcal{W}_{R\to R'A'}(\phi^{\overline{\rho}}_{RA})))\\
			&\quad=\boldsymbol{D}(\mathcal{W}_{R\to R'A'}(\mathcal{N}_{A\to B}(\phi^{\overline{\rho}}_{RA}))\Vert\mathcal{W}_{R\to R'A'}(\mathcal{M}_{A\to B}(\phi^{\overline{\rho}}_{RA})))\\
			&\quad=\boldsymbol{D}(\mathcal{N}_{A\to B}(\phi_{RA}^{\overline{\rho}})\Vert\mathcal{M}_{A\to B}(\phi_{RA}^{\overline{\rho}})),\label{eq-gen_div_chan_cov_pf_2}
		\end{align}
		where the last equality follows from the fact that every generalized divergence is isometrically invariant (recall Proposition~\ref{prop-gen_div_properties}). Now, let us apply the dephasing map $X\mapsto\sum_{g\in G}\ket{g}\!\bra{g}X\ket{g}\!\bra{g}$ to the $R'$ system. Since this map is a channel, by the data-processing inequality for the generalized divergence, we obtain
		\begin{multline}
			\boldsymbol{D}\!\left(\mathcal{N}_{A\to B}(\psi^{\overline{\rho}}_{R'A'A})\Vert\mathcal{M}_{A\to B}(\psi^{\overline{\rho}}_{R'A'A})\right)\\
			\geq \boldsymbol{D}\!\left(\frac{1}{|G|}\sum_{g\in G}\ket{g}\!\bra{g}_{R'}\otimes(\mathcal{N}_{A\to B}\circ\mathcal{U}_A^g)(\psi_{A'A})\Bigg\Vert\right.\\
			\left.\frac{1}{|G|}\sum_{g\in G}\ket{g}\!\bra{g}_{R'}\otimes(\mathcal{M}_{A\to B}\circ\mathcal{U}_A^g)(\psi_{A'A})\right).
		\end{multline}
		Then, because generalized divergences are invariant under unitaries, we can apply the unitary channel given by the unitary $\sum_{g\in G}\ket{g}\!\bra{g}\otimes V_B^{g\dagger}$ at the output of $\mathcal{N}$ and $\mathcal{M}$ to obtain
		\begin{align}
			&\boldsymbol{D}\!\left(\mathcal{N}_{A\to B}(\psi^{\overline{\rho}}_{R'A'A})\Vert\mathcal{M}_{A\to B}(\psi^{\overline{\rho}}_{R'A'A})\right)\notag\\
			&\geq \boldsymbol{D}\!\left(\frac{1}{|G|}\sum_{g\in G}\ket{g}\!\bra{g}_{R'}\otimes((\mathcal{V}_B^g)^\dagger\circ\mathcal{N}_{A\to B}\circ\mathcal{U}_A^g)(\psi_{A'A})\Bigg\Vert\right.\nonumber\\
			&\qquad\qquad\left.\frac{1}{|G|}\sum_{g\in G}\ket{g}\!\bra{g}_{R'}\otimes((\mathcal{V}_B^g)^\dagger\circ\mathcal{M}_{A\to B}\circ\mathcal{U}_A^g)(\psi_{A'A})\right).\label{eq-gen_div_group_cov_pf2}
		\end{align}
		Finally, since the group-covariance of $\mathcal{N}$ and $\mathcal{M}$ with respect to the representations $\{U_A^g\}_{g\in G}$ and $\{V_B^g\}_{g\in G}$ implies that
		\begin{equation}
			(\mathcal{V}_B^g)^\dagger\circ\mathcal{N}\circ\mathcal{U}_A^g=\mathcal{N},\quad \mathcal{V}_B^{g\dagger}\circ\mathcal{M}\circ\mathcal{U}_A^g=\mathcal{M}
		\end{equation}
		for all $g\in G$, and since from Proposition \ref{prop-gen_div_properties} generalized divergences are invariant under tensoring with the same state in both arguments, we obtain
		\begin{align}
			&\boldsymbol{D}(\mathcal{N}_{A\to B}(\phi_{RA}^{\overline{\rho}})\Vert\mathcal{M}_{A\to B}(\phi_{RA}^{\overline{\rho}}))\\
			&=\boldsymbol{D}\!\left(\mathcal{N}_{A\to B}(\psi^{\overline{\rho}}_{R'A'A})\Vert\mathcal{M}_{A\to B}(\psi^{\overline{\rho}}_{R'A'A})\right)\\
			&\geq\boldsymbol{D}\!\left(\frac{1}{|G|}\sum_{g\in G}\ket{g}\!\bra{g}_{R'}\otimes \mathcal{N}_{A\to B}(\psi_{A'A})\Bigg\Vert\frac{1}{|G|}\sum_{g\in\ G}\ket{g}\!\bra{g}_{R'}\otimes\mathcal{M}_{A\to B}(\psi_{A'A})\right)\\
			&=\boldsymbol{D}\!\left(\mathcal{N}_{A\to B}(\psi_{A'A})\Vert(\mathcal{M}_{A\to B}(\psi_{A'A})\right),\label{eq-gen_div_chan_cov_pf_3}
		\end{align}
		which is precisely \eqref{eq-gen_div_chan_cov_ineq}. By definition, the pure state $\phi_{RA}^{\overline{\rho}}$ is such that its reduced state on $A$ is invariant under the channel $\mathcal{T}_G$. Therefore, optimizing over all such pure states, we obtain
		\begin{multline}
			\boldsymbol{D}(\mathcal{N}_{A\to B}(\psi_{A'A})\Vert\mathcal{M}_{A\to B}(\psi_{A'A}))\\ \leq \sup_{\phi_{RA}}\{\boldsymbol{D}(\mathcal{N}_{A\to B}(\phi_{RA})\Vert\mathcal{M}_{A\to B}(\phi_{RA})):\phi_A=\mathcal{T}_G(\phi_A)\}.
		\end{multline}
		Since this inequality holds for all pure states $\psi_{A'A}$, we obtain
		\begin{align}
			\boldsymbol{D}(\mathcal{N}\Vert\mathcal{M})&=\sup_{\psi_{A'A}}\boldsymbol{D}(\mathcal{N}_{A\to B}(\psi_{A'A})\Vert\mathcal{M}_{A\to B}(\psi_{A'A}))\\
			&\leq \sup_{\psi_{A'A}}\{\boldsymbol{D}(\mathcal{N}_{A\to B}(\psi_{A'A})\Vert\mathcal{M}_{A\to B}(\psi_{A'A})):\psi_A=\mathcal{T}_G(\psi_A)\}.
		\end{align}
		Combining this inequality with \eqref{eq-gen_div_chan_cov_pf_1} gives us \eqref{eq-gen_div_chan_cov}.
		
		Finally, if the representation $\{U_A^g\}_{g\in G}$ acting on the input space of the channel $\mathcal{N}$ and the completely positive map $\mathcal{M}$ is irreducible, then for every state $\psi_{A'A}$ such that $\rho_A=\psi_A$, it holds that $\overline{\rho}_A=\mathbbm{1}_A/d_A$. Then, since the maximally entangled state is a purification of the maximally mixed state, we let $\phi_{RA}^{\overline{\rho}}=\Phi_{RA}$, which implies by \eqref{eq-gen_div_chan_cov_pf_2} that 
		\begin{align}
			&\boldsymbol{D}(\mathcal{N}_{A\to B}(\psi_{R'A'A}^{\overline{\rho}})\Vert\mathcal{M}_{A\to B}(\psi_{R'A'A}^{\overline{\rho}}))\\
			&\quad=\boldsymbol{D}(\mathcal{N}_{A\to B}(\Phi_{RA})\Vert\mathcal{M}_{A\to B}(\Phi_{RA}))\\
			&\quad=\boldsymbol{D}(\rho_{RB}^{\mathcal{N}}\Vert\rho_{RB}^{\mathcal{M}}).\label{eq-gen_div_chan_cov_pf_4}
		\end{align}
		Then, by \eqref{eq-gen_div_chan_cov_pf_3}, we have that
		\begin{equation}
			\boldsymbol{D}(\rho_{RB}^{\mathcal{N}}\Vert\rho_{RB}^{\mathcal{M}})\geq \boldsymbol{D}(\mathcal{N}_{A\to B}(\psi_{A'A})\Vert\mathcal{M}_{A\to B}(\psi_{A'A}))
		\end{equation}
		for all states $\psi_{A'A}$, so that $\boldsymbol{D}(\rho_{RB}^{\mathcal{N}}\Vert\rho_{RB}^{\mathcal{M}})\geq \boldsymbol{D}(\mathcal{N}\Vert\mathcal{M})$. Since the reverse inequality trivially holds, we obtain \eqref{eq-gen_div_cov_irrep}. 
	\end{Proof}
	
	We can also define information measures for channels based on information measures for states derived from generalized divergences. Using the generalized coherent information and the generalized mutual information defined in \eqref{eq-gen_coh_inf} and \eqref{eq-gen_mut_inf}, respectively, we make the following definitions.
	
	\begin{definition}{Generalized Information Measures for Quantum Channels}{def-gen_inf_meas_chan}
		Let $\boldsymbol{D}$ be a generalized divergence, as defined in Definition \ref{def-gen_div}, and let $\mathcal{N}_{A\to B}$ be a quantum channel.
		\begin{enumerate}
			\item The \textit{generalized mutual information of $\mathcal{N}$}, denoted by $\boldsymbol{I}(\mathcal{N})$, is defined as
				\begin{equation}\label{eq-gen_mut_inf_chan}
					\begin{aligned}
					\boldsymbol{I}(\mathcal{N})&\coloneqq\sup_{\psi_{RA}}\boldsymbol{I}(R;B)_{\omega}\\
					&=\sup_{\psi_{RA}}\inf_{\sigma_B}\boldsymbol{D}(\mathcal{N}_{A\to B}(\psi_{RA})\Vert\psi_R\otimes\sigma_B),
					\end{aligned}
				\end{equation}
				where $\omega_{RB}=\mathcal{N}_{A\to B}(\psi_{RA})$, $\psi_{RA}$ is a pure state with the dimension of $R$ the same as the dimension of $A$, and the generalized mutual information $\boldsymbol{I}(A;B)_\rho$ of a bipartite state $\rho_{AB}$ has been defined in \eqref{eq-gen_mut_inf}. 
			
			\item The \textit{generalized coherent information of $\mathcal{N}$}, denoted by $\boldsymbol{I}^c(\mathcal{N})$, is defined as
				\begin{equation}
					\begin{aligned}
					\boldsymbol{I}^c(\mathcal{N})&\coloneqq\sup_{\psi_{RA}}\boldsymbol{I}(R\rangle B)_{\omega}\\
					&=\sup_{\psi_{RA}}\inf_{\sigma_B}\boldsymbol{D}(\mathcal{N}_{A\to B}(\psi_{RA})\Vert\mathbbm{1}_R\otimes\sigma_B),
					\end{aligned}
				\end{equation}
				where $\omega_{RB}=\mathcal{N}_{A\to B}(\psi_{RA})$, $\psi_{RA}$ is a pure state with the dimension of $R$ the same as the dimension of $A$, and the generalized coherent information $\boldsymbol{I}^c(A\rangle B)_\rho$ of a bipartite state $\rho_{AB}$ has been defined in~\eqref{eq-gen_coh_inf}.
				
			\item The \textit{generalized Holevo information of $\mathcal{N}$}, denoted by $\boldsymbol{\chi}(\mathcal{N})$, is defined as
				\begin{equation}\label{eq-gen_Hol_inf_chan}
					\begin{aligned}
					\boldsymbol{\chi}(\mathcal{N})&\coloneqq\sup_{\rho_{XA}}\boldsymbol{I}(X;B)_{\omega}\\
					&=\sup_{\rho_{XA}}\inf_{\sigma_B}\boldsymbol{D}(\mathcal{N}_{A\to B}(\rho_{XA})\Vert\rho_X\otimes\sigma_B),
					\end{aligned}
				\end{equation}
				where $\omega_{XA}=\mathcal{N}_{A\to B}(\rho_{XA})$. Here, $\rho_{XA}=\sum_{x\in\mathcal{X}}p(x)\ket{x}\!\bra{x}_X\otimes\rho_A^x$ is a classical--quantum state, with $\mathcal{X}$ a finite alphabet with corresponding $|\mathcal{X}|$-dimensional system $X$, $\{\rho_A^x\}_{x\in\mathcal{X}}$ is a set of states, and $p\colon \mathcal{X}\to[0,1]$ is a probability distribution on $\mathcal{X}$. The supremum is additionally over the classical system $X$; i.e., no restriction is made on the size of $\mathcal{X}$.
		\end{enumerate}
	\end{definition}
	
	The generalized mutual information $\boldsymbol{I}(\mathcal{N})$ and the generalized coherent information $\boldsymbol{I}^c(\mathcal{N})$ both involve an optimization over pure states. It is straigh\-tforward to show that it suffices to optimize over pure states for both quantities. The argument is similar to that in~\eqref{eq-gen_div_chan_pure_2}--\eqref{eq-gen_div_chan_pure_4} above.

	For the generalized mutual information of a covariant channel, we can prove a result that is analogous to Proposition~\ref{prop-gen_div_group_cov}.
	
	\begin{proposition*}{Generalized Mutual Information for Covariant Channels}{prop-gen_mut_inf_cov_chan}
		Let $\mathcal{N}_{A\to B}$ be a $G$-covariant quantum channel for a finite group $G$. Then, for all pure states $\psi_{RA}$, with the dimension of $R$ equal to the dimension of $A$, we have that
		\begin{equation}\label{eq-gen_chan_mut_inf_cov_state}
			\boldsymbol{I}(R;B)_{\omega}\leq \boldsymbol{I}(R;B)_{\overline{\omega}},
		\end{equation}
		where $\omega_{RB}\coloneqq\mathcal{N}_{A\to B}(\psi_{RA})$, $\overline{\omega}_{RB}\coloneqq\mathcal{N}_{A\to B}(\phi_{RA}^{\overline{\rho}})$,
		\begin{equation}\label{eq-psi_RA_group_symmetrize1}
			\overline{\rho}_A=\frac{1}{|G|}\sum_{g\in G}U_A^g\rho_A U_A^{g\dagger}\eqqcolon\mathcal{T}_G(\rho_A),
		\end{equation}
		$\rho_A=\psi_A=\Tr_R[\psi_{RA}]$, and $\phi_{RA}^{\overline{\rho}}$ is a purification of $\overline{\rho}_A$. Consequently,
		\begin{equation}
			\boldsymbol{I}(\mathcal{N})=\sup_{\phi_{RA}}\{\boldsymbol{I}(R;B)_{\omega}:\phi_A=\mathcal{T}_G(\phi_A),\omega_{RB}=\mathcal{N}_{A\to B}(\phi_{RA})\}.
		\end{equation}
		In other words, in order to calculate $\boldsymbol{I}(\mathcal{N})$, it suffices to optimize over pure states $\psi_{RA}$ such that the reduced state $\psi_A$ is invariant under the channel $\mathcal{T}_G$ defined in \eqref{eq-psi_RA_group_symmetrize1}.\\
		
		If the representation $\{U_A^g\}_{g\in G}$ is irreducible, then $\boldsymbol{I}(\mathcal{N})$ is equal to the generalized mutual information of the Choi state of the channel, i.e.,
		\begin{equation}\label{eq-gen_mut_inf_chan_cov_irrep}
			\boldsymbol{I}(\mathcal{N})=\boldsymbol{I}(R;B)_{\rho^{\mathcal{N}}}.
		\end{equation}
	\end{proposition*}
	
	\begin{Proof}
		The proof is similar to the proof of Proposition~\ref{prop-gen_div_group_cov} and uses some steps therein.
		
		The inequality
		\begin{equation}
			\boldsymbol{I}(\mathcal{N})\geq\sup_{\phi_{RA}}\{\boldsymbol{I}(R;B)_{\omega}:\phi_A=\mathcal{T}_G(\phi_A),\omega_{RB}=\mathcal{N}_{A\to B}(\phi_{RA})\}
		\end{equation}
		holds simply by restricting the optimization in the definition of $\boldsymbol{I}(\mathcal{N})$ to pure states $\phi_{RA}$ whose reduced states $\phi_A$ are invariant under the channel $\mathcal{T}_G$. The remainder of the proof is devoted to showing that the reverse inequality holds as well.
		
		Let $\psi_{RA}$ be an arbitrary pure state, let $\rho_A=\psi_A$, and let $\overline{\rho}_A=\mathcal{T}_G(\rho_A)$. Furthermore, let $\phi_{RA}^{\overline{\rho}}$ be a purification of $\overline{\rho}_A$.
		Let $\mathcal{M}_{A\to B}$ be the replacement channel that traces out the $A$ system and replaces it with a state $\sigma_B$. Now following the steps in \eqref{eq:QEI:gen-ch-div-symmetries-1}--\eqref{eq-gen_div_group_cov_pf2}, and incorporating the covariance of the channel~$\mathcal{N}$, we conclude that
		\begin{align}
			&\boldsymbol{D}(\mathcal{N}_{A\to B}(\varphi_{RA}^{\overline{\rho}})\Vert\mathcal{M}_{A\to B}(\varphi_{RA}^{\overline{\rho}}))\nonumber\\
			&\quad = \boldsymbol{D}(\mathcal{N}_{A\to B}(\varphi_{RA}^{\overline{\rho}})\Vert\varphi_{R}^{\overline{\rho}}\otimes\sigma_B)\\
			&\quad\geq \boldsymbol{D}\!\left(\frac{1}{|G|}\sum_{g\in G}\ket{g}\!\bra{g}_{R'}\otimes\mathcal{N}_{A\to B}(\psi_{RA})\Bigg\Vert\frac{1}{|G|}\sum_{g\in G}\ket{g}\!\bra{g}_{R'}\otimes\psi_R\otimes V_B^{g\dagger}\sigma_B V_B^g\right)\\
			&\quad\geq \boldsymbol{D}(\mathcal{N}_{A\to B}(\psi_{RA})\Vert\psi_R\otimes\tau_B),
		\end{align}
		where the last line follows from the data-processing inequality under the partial-trace channel $\Tr_{R'}$ and we let $\tau_B\coloneqq \frac{1}{|G|}\sum_{g\in G}V_B^{g\dagger}\sigma_B V_B^g$. By taking the infimum over all states $\tau_B$ on the right-hand side of the inequality above,
		we find that
		\begin{align}
			\boldsymbol{D}(\mathcal{N}_{A\to B}(\phi_{RA}^{\overline{\rho}})\Vert\phi_R^{\overline{\rho}}\otimes\sigma_B)
			&\geq \boldsymbol{D}(\mathcal{N}_{A\to B}(\psi_{RA})\Vert\psi_R\otimes\tau_B)\\
			&\geq \inf_{\tau_B}\boldsymbol{D}(\mathcal{N}_{A\to B}(\psi_{RA})\Vert\psi_R\otimes\tau_B)\\
			&=\boldsymbol{I}(R;B)_{\omega},
		\end{align}
		where $\omega_{RB}=\mathcal{N}_{A\to B}(\psi_{RA})$. The inequality above holds for all states $\psi_{RA}$ and all states $\sigma_B$. Therefore, optimizing over all states $\sigma_B$ on the left-hand side of the above inequality leads to 
		\begin{equation}\label{eq-gen_mut_inf_chan_cov_pf_3}
			\inf_{\sigma_B}\boldsymbol{D}(\mathcal{N}_{A\to B}(\phi_{RA}^{\overline{\rho}})\Vert\phi_R^{\overline{\rho}}\otimes\sigma_B)=\boldsymbol{I}(R;B)_{\overline{\omega}}\geq \boldsymbol{I}(R;B)_{\omega},
		\end{equation}
		where $\overline{\omega}_{RB}=\mathcal{N}_{A\to B}(\phi_{RA}^{\overline{\rho}})$. Thus, we conclude \eqref{eq-gen_chan_mut_inf_cov_state}.
		
		Next, by construction, the state $\phi_{RA}^{\overline{\rho}}$ is such that its reduced state on $A$ is invariant under the channel $\mathcal{T}_G$. Optimizing over all such states leads to
		\begin{equation}
			\sup_{\phi_{RA}}\{\boldsymbol{I}(R;B)_{\omega}:\phi_A=\mathcal{T}_G(\phi_A),\omega_{RB}=\mathcal{N}_{A\to B}(\phi_{RA})\}\geq \boldsymbol{I}(R;B)_{\omega}.
		\end{equation}
		Since this inequality holds for all pure states $\psi_{RA}$, we finally obtain
		\begin{multline}
			\sup_{\phi_{RA}}\{\boldsymbol{I}(R;B)_{\omega}:\phi_A=\mathcal{T}_G(\phi_A),\omega_{RB}=\mathcal{N}_{A\to B}(\phi_{RA})\}\\ \geq \sup_{\psi_{RA}}\boldsymbol{I}(R;B)_{\omega}=\boldsymbol{I}(\mathcal{N}),
		\end{multline}
		as required.
		
		To prove \eqref{eq-gen_mut_inf_chan_cov_irrep}, note that if $\{U_A^g\}_{g\in G}$ is irreducible, then for every state $\psi_{RA}$, the state $\rho_A=\psi_A$ satisfies $\overline{\rho}_A=\mathcal{T}_G(\rho_A)=\frac{\mathbbm{1}_A}{d_A}$. Then, since the maximally entangled state is a purification of the maximally mixed state, we let $\phi_{RA}^{\overline{\rho}}=\Phi_{RA}$, which implies via \eqref{eq-gen_mut_inf_chan_cov_pf_3} that
		\begin{equation}
			\inf_{\sigma_B}\boldsymbol{D}(\mathcal{N}_{A\to B}(\Phi_{RA})\Vert\pi_R\otimes\sigma_B)=\boldsymbol{I}(R;B)_{\rho^{\mathcal{N}}}\geq \boldsymbol{I}(R;B)_{\omega},
		\end{equation}
		where $\omega_{RB}=\mathcal{N}_{A\to B}(\psi_{RA})$. Since the pure state $\psi_{RA}$ is arbitrary, we have that 
		\begin{equation}
			\boldsymbol{I}(R;B)_{\rho^{\mathcal{N}}}\geq\sup_{\psi_{RA}}\boldsymbol{I}(R;B)_{\omega}=\boldsymbol{I}(\mathcal{N}).
		\end{equation}
		The reverse inequality holds simply by restricting the optimization in the definition of $\boldsymbol{I}(\mathcal{N})$ to the maximally entangled state $\Phi_{RA}$. We thus have \eqref{eq-gen_mut_inf_chan_cov_irrep}, as required.
	\end{Proof}
	
	The following proposition is helpful in simplifying the computation of the generalized Holevo information $\boldsymbol{\chi}(\mathcal{N})$ of a quantum channel $\mathcal{N}$:	

	
	\begin{proposition}{prop-Holevo_inf_pure_states}
		Let $\mathcal{N}$ be a quantum channel. To compute its generalized Holevo information $\boldsymbol{\chi}(\mathcal{N})$, as defined in \eqref{eq-gen_Hol_inf_chan},  it suffices to optimize over ensembles consisting of pure states. If the underlying generalized divergence is continuous, then no more than $d^2$ pure states are needed for the optimization, where $d$ is the dimension of the input space of $\mathcal{N}$.
	\end{proposition}
	
	\begin{Proof}
		Let $\rho_{XA}$ be a classical--quantum state of the form
		\begin{equation}
			\rho_{XA}=\sum_{x\in\mathcal{X}}p(x)\ket{x}\!\bra{x}_X\otimes\rho_A^x,
		\end{equation}
		where $\mathcal{X}$ is a finite alphabet with associated $|\mathcal{X}|$-dimensional system $X$, $p\colon \mathcal{X}\to[0,1]$ is a probability distribution on $\mathcal{X}$, and $\{\rho_A^x\}_{x\in\mathcal{X}}$ is a set of states.
		
		First, by simply restricting the optimization in the definition of the Hol\-evo information to ensembles containing pure states only, we obtain
		\begin{equation}
			\boldsymbol{\chi}(\mathcal{N})=\sup_{\rho_{XA}}\boldsymbol{I}(X;B)_{\mathcal{N}_{A\to B}(\rho_{XA})}\geq\sup_{\tau_{ZA}} \boldsymbol{I}(Z;B)_{\mathcal{N}_{A\to B}(\tau_{ZA})},
		\end{equation}
		where $\tau_{ZA}$ is a classical--quantum state consisting only of pure states, i.e.,
		\begin{equation}
			\tau_{ZA}=\sum_{z\in\mathcal{Z}}p(x)\ket{z}\!\bra{z}_Z\otimes\ket{\psi^z}\!\bra{\psi^z}_A.
		\end{equation}
		
		Now, let each state $\rho_A^x$ in the classical--quantum state $\rho_{XA}$ have a spectral decomposition of the form
		\begin{equation}
			\rho_A^x=\sum_{y=1}^{r_x}\lambda_y^x\ket{\phi_y^x}\!\bra{\phi_y^x},
		\end{equation}
		where $r_x=\rank(\rho_A^x)$. So $\rho_{XA}$ can be written as
		\begin{equation}
			\rho_{XA}=\sum_{x\in\mathcal{X}}\sum_{y=1}^{r_x}p(x)\lambda_y^x\ket{x}\!\bra{x}_X\otimes\ket{\phi_y^x}\!\bra{\phi_y^x}_A.
		\end{equation}
		Now, let us define the state
		\begin{equation}
			\omega_{XYA}=\sum_{x\in\mathcal{X}}\sum_{y=1}^{r_x}p(x)\lambda_y^x\ket{x}\!\bra{x}_X\otimes\ket{y}\!\bra{y}_Y\otimes\ket{\phi_y^x}\!\bra{\phi_y^x}_A.
		\end{equation}
		Then, we have that
		\begin{equation}
			\rho_{XA}=\Tr_Y[\omega_{XYA}].
		\end{equation}
		Therefore, by the data-processing inequality for the generalized divergence, we find that
		\begin{align}
			\boldsymbol{I}(X;B)_{\mathcal{N}(\rho)}&=\inf_{\sigma_B}\boldsymbol{D}(\mathcal{N}_{A\to B}(\rho_{XA})\Vert\rho_X\otimes\sigma_B)\\
			&=\inf_{\sigma_B}\boldsymbol{D}(\Tr_Y[\mathcal{N}_{A\to B}(\omega_{XYA})]\Vert\Tr_Y[\omega_{XY}]\otimes\sigma_B)\\
			&\leq \inf_{\sigma_B}\boldsymbol{D}(\mathcal{N}_{A\to B}(\omega_{XYA})\Vert\omega_{XY}\otimes\sigma_B)\\
			&=\boldsymbol{I}(XY;B)_{\mathcal{N}(\omega)}.
		\end{align}
		Since $\omega_{XYA}$ is a classical--quantum state with pure states, it holds that
		\begin{equation}
			\boldsymbol{I}(XY;B)_{\mathcal{N}(\omega)}\leq\sup_{\tau_{ZA}}\boldsymbol{I}(Z;B)_{\mathcal{N}(\tau)}.
		\end{equation}
		Therefore,
		\begin{equation}
			\boldsymbol{\chi}(\mathcal{N})=\sup_{\rho_{XA}}\boldsymbol{I}(X;B)_{\mathcal{N}_{A\to B}(\rho_{XA})}\leq\sup_{\tau_{ZA}}\boldsymbol{I}(Z;B)_{\mathcal{N}_{A\to B}(\tau_{ZA})}
		\end{equation}
		which means that
		\begin{equation}
			\boldsymbol{\chi}(\mathcal{N})=\sup_{\tau_{ZA}}\boldsymbol{I}(Z;B)_{\mathcal{N}_{A\to B}(\tau_{ZA})},
		\end{equation}
		as required.
		
		When the underlying generalized divergence is continuous, the fact that the alphabet $\mathcal{X}$ of the classical--quantum states $\tau_{ZA}$ need not exceed $d^2$ elements is due to the Fenchel--Eggleston--Carath\'{e}odory Theorem (Theorem~\ref{thm-Caratheodory}) and the fact that dimension of the space of density operators on a $d$-dimensional space is $d^2$.
	\end{Proof}

	The information measures for channels on which we primarily focus in this book are those based on the following generalized divergences: the quantum relative entropy, the Petz--R\'{e}nyi relative entropy, the sandwiched R\'{e}nyi relative entropy, and the hypothesis testing relative entropy. Specifically, giv\-en a channel $\mathcal{N}_{A\to B}$, we are interested in the following mutual information quantities. In each case, $\psi_{RA}$ is a pure state with the dimension of the system~$R$ the same as that of $A$,  the state $\omega_{RB}=\mathcal{N}_{A\to B}(\psi_{RA})$, and $\sigma_B$ is a state.
	\begin{enumerate}
		\item \textit{$\varepsilon$-hypothesis testing mutual information} of $\mathcal{N}$,  defined for $\varepsilon\in[0,1]$ as
			\begin{equation}\label{eq-hypo_testing_mutual_inf_chan}
				I_H^{\varepsilon}(\mathcal{N})\coloneqq\sup_{\psi_{RA}}I_H^{\varepsilon}(R;B)_{\omega},
			\end{equation}
			where 
			\begin{equation}\label{eq-hypo_testing_mutual_inf}
				I_H^{\varepsilon}(A;B)_\rho\coloneqq\inf_{\sigma_B}D_H^{\varepsilon}(\rho_{AB}\Vert\rho_A\otimes\sigma_B)
			\end{equation}
			is the \textit{$\varepsilon$-hypothesis testing mutual information} of the bipartite state $\rho_{AB}$.
			
		\item \textit{Petz--R\'{e}nyi mutual information} of $\mathcal{N}$:
			\begin{equation}\label{eq-petz_renyi_mut_inf_chan}
				I_\alpha(\mathcal{N})\coloneqq\sup_{\psi_{RA}}I_\alpha(R;B)_{\omega}\quad\forall~\alpha\in[0,1)\cup (1,2],
			\end{equation}
			where 
			\begin{equation}
			\label{eq-petz_renyi_mut_inf}
							I_\alpha(A;B)_\rho\coloneqq\inf_{\sigma_B}D_\alpha(\rho_{AB}\Vert\rho_A\otimes\sigma_B)
			\end{equation}
			is the \textit{Petz--R\'{e}nyi mutual information} of the bipartite state $\rho_{AB}$.
		
		\item \textit{Sandwiched R\'{e}nyi mutual information} of $\mathcal{N}$:
			\begin{equation}\label{eq-sand_ren_mut_inf_chan}
				\widetilde{I}_\alpha(\mathcal{N})\coloneqq\sup_{\psi_{RA}}\widetilde{I}_\alpha(R;B)_{\omega}\quad\forall~\alpha\in\left[\sfrac{1}{2},1\right)\cup(1,\infty),
			\end{equation}
			where 
			\begin{equation}
				\widetilde{I}_\alpha(A;B)_\rho\coloneqq\inf_{\sigma_B}\widetilde{D}_\alpha(\rho_{AB}\Vert\rho_A\otimes\sigma_B)
				\label{eq:QEI:sand-Ren-MI-states}
			\end{equation}
			is the \textit{sandwiched R\'{e}nyi mutual information} of the bipartite state $\rho_{AB}$.	
	\end{enumerate}
	
	For each of these quantities, we are also interested in the special case of classical--quantum states. If $\mathcal{X}$ is a finite alphabet with corresponding $|\mathcal{X}|$-dimensional system $X$, $\{\rho_A^x\}_{x\in\mathcal{X}}$ is a set of states, $p\colon \mathcal{X}\to[0,1]$ a probability distribution on $\mathcal{X}$, and $\rho_{XA}=\sum_{x\in\mathcal{X}}p(x)\ket{x}\!\bra{x}_X\otimes\rho_A^x$, then for every channel $\mathcal{N}$ we define the following quantities. In each case, we define $\omega_{XB}\coloneqq \mathcal{N}_{A\to B}(\rho_{XA})$.
	\begin{enumerate}
		\item \textit{$\varepsilon$-hypothesis testing Holevo information} of $\mathcal{N}$,  defined for all $\varepsilon\in[0,1]$ as
			\begin{equation}\label{eq-hypo_testing_Holevo_inf_chan}
				\chi_H^{\varepsilon}(\mathcal{N})\coloneqq\sup_{\rho_{XA}}I_H^{\varepsilon}(X;B)_{\omega}.
			\end{equation}

		\item \textit{Petz--R\'{e}nyi Holevo information} of $\mathcal{N}$:
			\begin{equation}\label{eq-petz_renyi_Hol_inf_chan}
				\chi_\alpha(\mathcal{N})\coloneqq\sup_{\rho_{XA}}I_\alpha(X;B)_{\omega}\quad\forall~\alpha\in[0,1)\cup(1,2].
			\end{equation}

		\item \textit{Sandwiched R\'{e}nyi Holevo information} of $\mathcal{N}$:
			\begin{equation}\label{eq-sand_rel_Holevo_inf_chan}
				\widetilde{\chi}_\alpha(\mathcal{N})\coloneqq\sup_{\rho_{XA}}\widetilde{I}_\alpha(X;B)_{\omega}\quad\forall~\alpha\in\left[\sfrac{1}{2},1\right)\cup(1,\infty).
			\end{equation}

	\end{enumerate}
	
	We are also interested in the corresponding coherent information quantities. In each case, $\psi_{RA}$ is a pure state with the dimension of the system~$R$ the same as that of $A$,  the state $\omega_{RB}=\mathcal{N}_{A\to B}(\psi_{RA})$, and $\sigma_B$ is a state. The coherent information quantities are defined as follows for every quantum channel $\mathcal{N}_{A\to B}$:
	\begin{enumerate}
		\item \textit{$\varepsilon$-hypothesis testing coherent information} of $\mathcal{N}$, defined for all $\varepsilon\in[0,1]$ as
			\begin{equation}\label{eq-hypo_test_coh_inf_chan}
				I_H^{c,\varepsilon}(\mathcal{N})\coloneqq\sup_{\psi_{RA}}I_H^{\varepsilon}(R\rangle B)_{\omega},
			\end{equation}
			where
			\begin{equation}\label{eq-hypo_test_coh_inf_state}
				I_H^{\varepsilon}(A\rangle B)_\rho\coloneqq\inf_{\sigma_B}D_H^{\varepsilon}(\rho_{AB}\Vert\mathbbm{1}_A\otimes\sigma_B)
			\end{equation}
			is the \textit{$\varepsilon$-hypothesis testing coherent information} of the bipartite state $\rho_{AB}$.
			
		\item \textit{Petz--R\'{e}nyi coherent information} of $\mathcal{N}$:
			\begin{equation}
				I_\alpha^c(\mathcal{N})\coloneqq\sup_{\psi_{RA}}I_\alpha(R\rangle B)_{\omega}\quad\forall~\alpha\in[0,1)\cup(1,2],
			\end{equation}
			where 
			\begin{equation}
				I_\alpha(A\rangle B)_\rho\coloneqq\inf_{\sigma_B}D_\alpha(\rho_{AB}\Vert\mathbbm{1}_A\otimes\sigma_B)
			\end{equation}
			is the \textit{Petz--R\'{e}nyi coherent information} of the bipartite state $\rho_{AB}$.
		
		\item \textit{Sandwiched R\'{e}nyi coherent information} of $\mathcal{N}$:
			\begin{equation}\label{eq-sand_renyi_coh_inf_chan}
				\widetilde{I}_\alpha^c(\mathcal{N})\coloneqq\sup_{\psi_{RA}}\widetilde{I}_\alpha(R\rangle B)_{\omega}\quad\forall~\alpha\in\left[\sfrac{1}{2},1\right)\cup(1,\infty),
			\end{equation}
			where 
			\begin{equation}
				\widetilde{I}_\alpha(A\rangle B)_\rho\coloneqq\inf_{\sigma_B}\widetilde{D}_\alpha(\rho_{AB}\Vert\mathbbm{1}_A\otimes\sigma_B)
			\end{equation}
			is the \textit{sandwiched R\'{e}nyi coherent information} of the bipartite state $\rho_{AB}$.		
	\end{enumerate}
	
	For all of the quantities defined above, we define the corresponding quantities based on the quantum relative entropy by taking the limit $\alpha\to 1$. The key such quantities of interest in this book are the following:
	\begin{enumerate}
		\item \textit{Mutual information of $\mathcal{N}$}, denoted by $I(\mathcal{N})$ and defined as
			\begin{equation}\label{eq-mut_inf_chan}
				I(\mathcal{N})\coloneqq \sup_{\psi_{RA}}I(R;B)_{\omega},
			\end{equation}
			where $\omega_{RB}=\mathcal{N}_{A\to B}(\psi_{RA})$, and we recall from \eqref{eq-mut_inf} that the mutual information $I(A;B)_\rho$ of a bipartite state $\rho_{AB}$ is given by
			\begin{align}
				I(A;B)_\rho&=H(A)_{\rho}+H(B)_{\rho}-H(AB)_{\rho}\label{eq-mut_inf_state_2}\\
				&=D(\rho_{AB}\Vert\rho_A\otimes\rho_B)\label{eq-mut_inf_state_3}\\
				&=\inf_{\sigma_B}D(\rho_{AB}\Vert\rho_A\otimes\sigma_B),
			\end{align}
			where the optimization in the last line is over states $\sigma_B$.
		\item \textit{Holevo information of $\mathcal{N}$}, denoted by $\chi(\mathcal{N})$ and defined as
			\begin{equation}\label{eq-Hol_inf_chan}
				\chi(\mathcal{N})\coloneqq \sup_{\rho_{XA}}I(X;B)_{\omega},
			\end{equation}
			where $\omega_{RB}=\mathcal{N}_{A\to B}(\rho_{XA})$, and the supremum is over all classical-quan\-tum states of the form $\rho_{XA}=\sum_{x\in\mathcal{X}}p(x)\ket{x}\!\bra{x}_X\otimes \rho_A^x$, with $\mathcal{X}$ a finite alphabet with associated $|\mathcal{X}|$-dimensional system $X$, $\{\rho_A^x\}_{x\in\mathcal{X}}$ a set of states, and $p\colon \mathcal{X}\to[0,1]$ a probability distribution on $\mathcal{X}$.
		\item \textit{Coherent information of $\mathcal{N}$}, denoted by $I^c(\mathcal{N})$ and defined as
			\begin{equation}\label{eq-coh_inf_chan}
				I^c(\mathcal{N})\coloneqq\sup_{\psi_{RA}}I(R\rangle B)_{\omega},
			\end{equation}
			where $\omega_{RB}=\mathcal{N}_{A\to B}(\psi_{RA})$, and we recall from \eqref{eq-coh_inf} that the coherent information $I(A\rangle B)_\rho$ of a bipartite state $\rho_{AB}$ is given by
			\begin{align}
				I(A\rangle B)_\rho&=H(B)_{\rho}-H(AB)_{\rho}\label{eq-coh_inf_state_def}\\
				&=D(\rho_{AB}\Vert\mathbbm{1}_A\otimes\rho_B)\\
				&=\inf_{\sigma_B}D(\rho_{AB}\Vert\mathbbm{1}_A\otimes\sigma_B),
			\end{align}
			where the optimization in the last line is over states $\sigma_B$.
			
			Using \eqref{eq-coh_inf_state_def}, we can write the coherent information of the channel $\mathcal{N}$ as
			\begin{equation}\label{eq-coh_inf_chan_2}
				I^c(\mathcal{N})=\sup_{\psi_{RA}}\{H(B)_{\omega}-H(RB)_{\omega}\}.
			\end{equation}
			Given a Stinespring representation of $\mathcal{N}$, so that $\mathcal{N}(\rho)=\Tr_E[V\rho V^\dagger]$, where $V:\mathcal{H}_A\to\mathcal{H}_B\otimes\mathcal{H}_E$ is an isometry such that $d_E\geq\rank(\Gamma^{\mathcal{N}})$, observe that
			\begin{equation}
				H(RB)_{\omega}=H(E)_{\tau},
			\end{equation}
			where $\tau_E=\mathcal{N}^c(\psi_A)$. This is due to the fact that the state $\phi_{RBE}\coloneqq V\psi_{RA}V^\dagger$ is pure, meaning that $\omega_{RB}$ and $\tau_E=\Tr_{RB}[V\psi_{RA}V^\dagger]=\mathcal{N}^c(\psi_A)$ have the same spectrum. Furthermore, the state on which $H(B)_{\omega}$ is evaluated is equal to $\mathcal{N}(\psi_A)$. Therefore, we have that
			\begin{equation}\label{eq-coh_inf_chan_alt}
				I^c(\mathcal{N})=\sup_{\rho}\{H(\mathcal{N}(\rho))-H(\mathcal{N}^c(\rho))\},
			\end{equation}
			where the optimization is over all states $\rho$.
	\end{enumerate}

	\subsection{Simplified Formulas for R\'enyi Information Measures}

In this section, we provide some simplified formulas for the Petz--R\'enyi
information quantities for general bipartite states and for all R\'enyi
information quantities for pure bipartite states.

\begin{proposition*}{Quantum Sibson Identities}{prop:QEI:sibson-identities}
Let $\rho_{AB}$ be a bipartite state. Then the Petz--R\'enyi mutual and coherent
informations simplify as follows for all $\alpha\in(0,1)\cup(1,\infty)$:
\begin{align}
I_{\alpha}(A;B)_{\rho}  & =\frac{\alpha}{\alpha-1}\log_{2}\operatorname{Tr}\!\left[  \left(  \operatorname{Tr}_{A}[\rho_{AB}^{\alpha}\rho_{A}^{1-\alpha
}]\right)  ^{\frac{1}{\alpha}}\right]  ,\label{eq:QEI:sibson-id-MI}\\
I_{\alpha}(A\rangle B)_{\rho}  & =\frac{\alpha}{\alpha-1}\log_{2}
\operatorname{Tr}\!\left[  \left(  \operatorname{Tr}_{A}[\rho_{AB}^{\alpha
}]\right)  ^{\frac{1}{\alpha}}\right]  .\label{eq:QEI:sibson-id-CI}
\end{align}
\end{proposition*}

\begin{Proof}
We show both identities with a unified approach. Let $\tau_{A}$ be a positive
semi-definite operator, let $\sigma_{B}$ be a state, and let $\omega
_{B}(\alpha)$ denote the following state:
\begin{equation}
\omega_{B}(\alpha)\coloneqq \frac{\left(  \operatorname{Tr}_{A}[\rho_{AB}^{\alpha}
\tau_{A}^{1-\alpha}]\right)  ^{\frac{1}{\alpha}}}{\operatorname{Tr}\!\left[
\left(  \operatorname{Tr}_{A}[\rho_{AB}^{\alpha}\tau_{A}^{1-\alpha}]\right)
^{\frac{1}{\alpha}}\right]  },
\end{equation}
so that
\begin{equation}
\left(  \operatorname{Tr}_{A}[\rho_{AB}^{\alpha}\tau_{A}^{1-\alpha}]\right)
^{\frac{1}{\alpha}}=\operatorname{Tr}\!\left[  \left(  \operatorname{Tr}
_{A}[\rho_{AB}^{\alpha}\tau_{A}^{1-\alpha}]\right)  ^{\frac{1}{\alpha}
}\right]  \cdot\omega_{B}(\alpha)
\end{equation}
We first prove that
\begin{align}
D_{\alpha}(\rho_{AB}\Vert\tau_{A}\otimes\sigma_{B})  & =D_{\alpha}(\rho
_{AB}\Vert\tau_{A}\otimes\omega_{B}(\alpha))+D_{\alpha}(\omega_{B}
(\alpha)\Vert\sigma_{B})\label{eq:QEI:chain-rule-identity-sibson-MI}\\
& \geq D_{\alpha}(\rho_{AB}\Vert\tau_{A}\otimes\omega_{B}(\alpha
)),\label{eq:QEI:chain-rule-identity-sibson-MI-lower}
\end{align}
where the inequality follows because $D_{\alpha}(\omega_{B}(\alpha)\Vert
\sigma_{B})\geq0$ for all states. Consider that
\begin{align}
& Q_{\alpha}(\rho_{AB}\Vert\tau_{A}\otimes\sigma_{B})\nonumber\\
& =\operatorname{Tr}[\rho_{AB}^{\alpha}\left(  \tau_{A}\otimes\sigma
_{B}\right)  ^{1-\alpha}]\\
& =\operatorname{Tr}[\rho_{AB}^{\alpha}(  \tau_{A}^{1-\alpha}
\otimes\sigma_{B}^{1-\alpha})  ]\\
& =\operatorname{Tr}_{B}[\operatorname{Tr}_{A}[\rho_{AB}^{\alpha}(
\tau_{A}^{1-\alpha}\otimes\sigma_{B}^{1-\alpha})  ]]\\
& =\operatorname{Tr}[\operatorname{Tr}_{A}[\rho_{AB}^{\alpha}\tau
_{A}^{1-\alpha}]\sigma_{B}^{1-\alpha}]\\
& =\operatorname{Tr}\!\left[  \left(  \operatorname{Tr}\!\left[  \left(
\operatorname{Tr}_{A}[\rho_{AB}^{\alpha}\tau_{A}^{1-\alpha}]\right)
^{\frac{1}{\alpha}}\right]  \cdot\omega_{B}(\alpha)\right)  ^{\alpha}
\sigma_{B}^{1-\alpha}\right]  \\
& =\left(  \operatorname{Tr}\!\left[  \left(  \operatorname{Tr}_{A}[\rho
_{AB}^{\alpha}\tau_{A}^{1-\alpha}]\right)  ^{\frac{1}{\alpha}}\right]
\right)  ^{\alpha}\cdot\operatorname{Tr}\!\left[  \omega_{B}(\alpha)^{\alpha
}\sigma_{B}^{1-\alpha}\right]  .
\end{align}
Applying the function $(\cdot)\rightarrow\frac{1}{\alpha-1}\log_{2}(\cdot)$ to
both sides, we conclude that
\begin{multline}
D_{\alpha}(\rho_{AB}\Vert\tau_{A}\otimes\sigma_{B})=\\
\frac{\alpha}{\alpha-1}\log_{2}\operatorname{Tr}\!\left[  \left(
\operatorname{Tr}_{A}[\rho_{AB}^{\alpha}\tau_{A}^{1-\alpha}]\right)
^{\frac{1}{\alpha}}\right]  +D_{\alpha}(\omega_{B}(\alpha)\Vert\sigma_{B}).
\end{multline}
Now consider that
\begin{align}
& Q_{\alpha}(\rho_{AB}\Vert\tau_{A}\otimes\omega_{B}(\alpha))\nonumber\\
& =\operatorname{Tr}[\rho_{AB}^{\alpha}(\tau_{A}^{1-\alpha}\otimes\omega
_{B}(\alpha)^{1-\alpha})]\\
& =\operatorname{Tr}[\operatorname{Tr}_{A}[\rho_{AB}^{\alpha}\tau
_{A}^{1-\alpha}]\omega_{B}(\alpha)^{1-\alpha}]\\
& =\operatorname{Tr}\!\left[  \operatorname{Tr}_{A}[\rho_{AB}^{\alpha}\tau
_{A}^{1-\alpha}]\left(  \frac{\left(  \operatorname{Tr}_{A}[\rho_{AB}^{\alpha
}\tau_{A}^{1-\alpha}]\right)  ^{\frac{1}{\alpha}}}{\operatorname{Tr}\!\left[
\left(  \operatorname{Tr}_{A}[\rho_{AB}^{\alpha}\tau_{A}^{1-\alpha}]\right)
^{\frac{1}{\alpha}}\right]  }\right)  ^{1-\alpha}\right]  \\
& =\left(  \operatorname{Tr}\!\left[  \left(  \operatorname{Tr}_{A}[\rho
_{AB}^{\alpha}\tau_{A}^{1-\alpha}]\right)  ^{\frac{1}{\alpha}}\right]
\right)  ^{\alpha-1}\operatorname{Tr}\!\left[  \operatorname{Tr}_{A}[\rho
_{AB}^{\alpha}\tau_{A}^{1-\alpha}]\left(  \operatorname{Tr}_{A}[\rho
_{AB}^{\alpha}\tau_{A}^{1-\alpha}]\right)  ^{\frac{1-\alpha}{\alpha}}\right]
\\
& =\left(  \operatorname{Tr}\!\left[  \left(  \operatorname{Tr}_{A}[\rho
_{AB}^{\alpha}\tau_{A}^{1-\alpha}]\right)  ^{\frac{1}{\alpha}}\right]
\right)  ^{\alpha-1}\operatorname{Tr}\!\left[  \left(  \operatorname{Tr}
_{A}[\rho_{AB}^{\alpha}\tau_{A}^{1-\alpha}]\right)  ^{\frac{1}{\alpha}
}\right]  \\
& =\left(  \operatorname{Tr}\!\left[  \left(  \operatorname{Tr}_{A}[\rho
_{AB}^{\alpha}\tau_{A}^{1-\alpha}]\right)  ^{\frac{1}{\alpha}}\right]
\right)  ^{\alpha}.
\end{align}
Now applying the function $(\cdot)\rightarrow\frac{1}{\alpha-1}\log_{2}
(\cdot)$ to both sides, we conclude that
\begin{equation}
D_{\alpha}(\rho_{AB}\Vert\tau_{A}\otimes\omega_{B}(\alpha))=\frac{\alpha
}{\alpha-1}\log_{2}\operatorname{Tr}\!\left[  \left(  \operatorname{Tr}_{A}
[\rho_{AB}^{\alpha}\tau_{A}^{1-\alpha}]\right)  ^{\frac{1}{\alpha}}\right]  .
\end{equation}
So this establishes \eqref{eq:QEI:chain-rule-identity-sibson-MI}. We then
conclude from \eqref{eq:QEI:chain-rule-identity-sibson-MI-lower} that
\begin{equation}
\inf_{\sigma_{B}}D_{\alpha}(\rho_{AB}\Vert\tau_{A}\otimes\sigma_{B}
)=D_{\alpha}(\rho_{AB}\Vert\tau_{A}\otimes\omega_{B}(\alpha)),
\end{equation}
because the lower bound in \eqref{eq:QEI:chain-rule-identity-sibson-MI-lower}
is achieved by picking $\sigma_{B}=\omega_{B}(\alpha)$. So this establishes
that
\begin{equation}
\inf_{\sigma_{B}}D_{\alpha}(\rho_{AB}\Vert\tau_{A}\otimes\sigma_{B}
)=\frac{\alpha}{\alpha-1}\log_{2}\operatorname{Tr}\!\left[  \left(
\operatorname{Tr}_{A}[\rho_{AB}^{\alpha}\tau_{A}^{1-\alpha}]\right)
^{\frac{1}{\alpha}}\right]  .
\end{equation}
We conclude the formula in \eqref{eq:QEI:sibson-id-MI} by setting $\tau_{A}=\rho_{A}$, and we
conclude the formula in \eqref{eq:QEI:sibson-id-CI} by setting $\tau_{A}=\mathbbm{1}_{A}$.
\end{Proof}

\begin{proposition*}{R\'enyi Information Measures for Pure Bipartite States}{prop:QEI:renyi-infos-pure-bipartite}
Let $\psi_{AB}$ be a pure bipartite state, and let $\alpha\in(0,1)\cup
(1,\infty)$. Then the Petz--, sandwiched, and geometric R\'enyi mutual
informations simplify as follows:
\begin{align}
I_{\alpha}(A;B)_{\psi}  & =2H_{\frac{2-\alpha}{\alpha}}(A)_{\psi
},\label{eq:QEI:Petz-Renyi-MI-pure-bi}\\
\widetilde{I}_{\alpha}(A;B)_{\psi}  & =2H_{\frac{1}{2\alpha-1}}(A)_{\psi
},\label{eq:QEI:SW-Renyi-MI-pure-bi}\\
\widehat{I}_{\alpha}(A;B)_{\psi}  & =2H_{0}(A)_{\psi}
,\label{eq:QEI:geo-Renyi-MI-pure-bi}
\end{align}
where the  R\'enyi entropy $H_{\alpha}(A)$ is defined in \eqref{eq:QEI:Renyi-entropy}. 
The Petz--, sandwiched, and geometric R\'enyi coherent informations simplify as
follows:
\begin{align}
I_{\alpha}(A\rangle B)_{\psi}  & =H_{\frac{1}{\alpha}}(A)_{\psi}
,\label{eq:QEI:Petz-Renyi-CI-pure-bi}\\
\widetilde{I}_{\alpha}(A\rangle B)_{\psi}  & =H_{\frac{\alpha}{2\alpha-1}
}(A)_{\psi},\label{eq:QEI:SW-Renyi-CI-pure-bi}\\
\widehat{I}_{\alpha}(A\rangle B)_{\psi}  & =H_{\frac{1}{2}}(A)_{\psi
}.\label{eq:QEI:geo-Renyi-CI-pure-bi}
\end{align}
\end{proposition*}

\begin{Proof}
We start by proving \eqref{eq:QEI:Petz-Renyi-MI-pure-bi}. We assume that
$\psi_{AB}$ is in its Schmidt form, so that without loss of generality the
Hilbert spaces for systems $A$ and $B$ are isomorphic and each have dimension
equal to the Schmidt rank of $\psi_{AB}$. With $\Gamma_{AB}$ the maximally
entangled operator with local bases chosen to match those from the Schmidt
decomposition, we have that $\psi_{AB}=\psi_{A}^{\frac{1}{2}}\Gamma_{AB}
\psi_{A}^{\frac{1}{2}}$, where $\psi_{A}=\operatorname{Tr}_{B}[\psi_{AB}]$. We
apply the Sibson identity in \eqref{eq:QEI:sibson-id-MI} to find that
\begin{align}
2^{\frac{\alpha-1}{\alpha}I_{\alpha}(A;B)_{\psi}}  & =\operatorname{Tr}\!\left[
\left(  \operatorname{Tr}_{A}[\psi_{AB}^{\alpha}\psi_{A}^{1-\alpha}]\right)
^{\frac{1}{\alpha}}\right]  
 =\operatorname{Tr}\!\left[  \left(  \operatorname{Tr}_{A}[\psi_{AB}\psi
_{A}^{1-\alpha}]\right)  ^{\frac{1}{\alpha}}\right]  \\
& =\operatorname{Tr}\!\left[  \left(  \operatorname{Tr}_{A}[\psi_{A}^{\frac
{1}{2}}\Gamma_{AB}\psi_{A}^{\frac{1}{2}}\psi_{A}^{1-\alpha}]\right)
^{\frac{1}{\alpha}}\right]  \\
& =\operatorname{Tr}\!\left[  \left(  \operatorname{Tr}_{A}[\Gamma_{AB}\psi
_{A}^{\frac{1}{2}}\psi_{A}^{1-\alpha}\psi_{A}^{\frac{1}{2}}]\right)
^{\frac{1}{\alpha}}\right]  \\
& =\operatorname{Tr}\!\left[  \left(  \operatorname{Tr}_{A}[\Gamma_{AB}\psi
_{A}^{2-\alpha}]\right)  ^{\frac{1}{\alpha}}\right]  \\
& =\operatorname{Tr}\!\left[  \left(  \operatorname{Tr}_{A}[\Gamma_{AB}(\psi
_{B}^{T})^{2-\alpha}]\right)  ^{\frac{1}{\alpha}}\right]  
 =\operatorname{Tr}\!\left[  \left(  (\psi_{B}^{T})^{2-\alpha}\right)
^{\frac{1}{\alpha}}\right]  \\
& =\operatorname{Tr}\!\left[  \psi_{B}^{\frac{2-\alpha}{\alpha}}\right]  
 =\operatorname{Tr}\!\left[  \psi_{A}^{\frac{2-\alpha}{\alpha}}\right]  .
\end{align}
The fourth equality follows from cyclicity of partial trace and the sixth
follows from the transpose trick in \eqref{eq-transpose_trick}. Rearranging the first and last lines gives
\begin{align}
I_{\alpha}(A;B)_{\psi}  & =\frac{\alpha}{\alpha-1}\log_{2}\operatorname{Tr}\!\left[  \psi_{A}^{\frac{2-\alpha}{\alpha}}\right]  \\
& =2\left(  \frac{1}{1-\frac{2-\alpha}{\alpha}}\right)  \log_{2}
\operatorname{Tr}\!\left[  \psi_{A}^{\frac{2-\alpha}{\alpha}}\right]  \\
& =2H_{\frac{2-\alpha}{\alpha}}(A)_{\psi}.
\end{align}

We now prove \eqref{eq:QEI:Petz-Renyi-CI-pure-bi}. It
 follows from the Sibson identity in
\eqref{eq:QEI:sibson-id-CI}:
\begin{align}
2^{\frac{\alpha-1}{\alpha}I_{\alpha}(A\rangle B)_{\psi}}  & =\operatorname{Tr}\!\left[  \left(  \operatorname{Tr}_{A}[\psi_{AB}^{\alpha}]\right)  ^{\frac
{1}{\alpha}}\right]  
 =\operatorname{Tr}\!\left[  \left(  \operatorname{Tr}_{A}[\psi_{AB}]\right)
^{\frac{1}{\alpha}}\right]  \\
& =\operatorname{Tr}\!\left[  \left(  \psi_{B}\right)  ^{\frac{1}{\alpha}
}\right]  
 =\operatorname{Tr}\!\left[  \psi_{A}^{\frac{1}{\alpha}}\right]  .
\end{align}
Rearranging this gives
\begin{align}
I_{\alpha}(A\rangle B)_{\psi}  & =\frac{\alpha}{\alpha-1}\log_{2}
\operatorname{Tr}\!\left[  \psi_{A}^{\frac{1}{\alpha}}\right]  
 =\frac{1}{1-\frac{1}{\alpha}}\log_{2}\operatorname{Tr}\!\left[  \psi
_{A}^{\frac{1}{\alpha}}\right]  
 =H_{\frac{1}{\alpha}}(A)_{\psi}.
\end{align}

We now prove \eqref{eq:QEI:SW-Renyi-MI-pure-bi}. Consider, for an arbitrary
state $\sigma_{B}$, that
\begin{align}
& \widetilde{Q}_{\alpha}(\psi_{AB}\Vert\psi_{A}\otimes\sigma_{B})\nonumber\\
& =\operatorname{Tr}\!\left[  \left(  \psi_{AB}^{\frac{1}{2}}\left(  \psi
_{A}\otimes\sigma_{B}\right)  ^{\frac{1-\alpha}{\alpha}}\psi_{AB}^{\frac{1}
{2}}\right)  ^{\alpha}\right]  \\
& =\operatorname{Tr}\!\left[  \left(  |\psi\rangle\!\langle\psi|_{AB}\left(
\psi_{A}\otimes\sigma_{B}\right)  ^{\frac{1-\alpha}{\alpha}}|\psi
\rangle\!\langle\psi|_{AB}\right)  ^{\alpha}\right]  \\
& =\left(  \langle\psi|_{AB}\left(  \psi_{A}\otimes\sigma_{B}\right)
^{\frac{1-\alpha}{\alpha}}|\psi\rangle_{AB}\right)  ^{\alpha}\operatorname{Tr}\!\left[  |\psi\rangle\!\langle\psi|_{AB}^{\alpha}\right]  \\
& =\left(  \langle\psi|_{AB}\left(  \psi_{A}\otimes\sigma_{B}\right)
^{\frac{1-\alpha}{\alpha}}|\psi\rangle_{AB}\right)  ^{\alpha}\\
& =\left(  \langle\Gamma|_{AB}\psi_{A}^{\frac{1}{2}}\left(  \psi_{A}
^{\frac{1-\alpha}{\alpha}}\otimes\sigma_{B}^{\frac{1-\alpha}{\alpha}}\right)
\psi_{A}^{\frac{1}{2}}|\Gamma\rangle_{AB}\right)  ^{\alpha}\\
& =\left(  \langle\Gamma|_{AB}\left(  \psi_{A}^{\frac{1}{\alpha}}\otimes
\sigma_{B}^{\frac{1-\alpha}{\alpha}}\right)  |\Gamma\rangle_{AB}\right)
^{\alpha}\\
& =\left(  \langle\Gamma|_{AB}\left(  \psi_{A}^{\frac{1}{\alpha}}\left[
\T_{A}(\sigma_{A})\right]  ^{\frac{1-\alpha}{\alpha}}\otimes \mathbbm{1}_{B}\right)
|\Gamma\rangle_{AB}\right)  ^{\alpha}\\
& =\left(  \operatorname{Tr}\!\left[  \psi_{A}^{\frac{1}{\alpha}}\left[
\T_{A}(\sigma_{A})\right]  ^{\frac{1-\alpha}{\alpha}}\right]  \right)
^{\alpha}.
\end{align}
Now applying the function $(\cdot)\rightarrow\frac{1}{\alpha-1}\log_{2}
(\cdot)$ to both sides, we conclude that
\begin{equation}
\widetilde{D}_{\alpha}(\psi_{AB}\Vert\psi_{A}\otimes\sigma_{B})=\frac{\alpha
}{\alpha-1}\log_{2}\operatorname{Tr}\!\left[  \psi_{A}^{\frac{1}{\alpha}}\left[
\T_{A}(\sigma_{A})\right]  ^{\frac{1-\alpha}{\alpha}}\right]  ,
\end{equation}
and applying Proposition~\ref{prop-Schatten_pos_var}, we conclude that
\begin{align}
I_{\alpha}(A;B)_{\psi}  & =\inf_{\sigma_{B}}\widetilde{D}_{\alpha}(\psi
_{AB}\Vert\psi_{A}\otimes\sigma_{B})\\
& =\inf_{\sigma_{A}}\frac{\alpha}{\alpha-1}\log_{2}\operatorname{Tr}\!\left[
\psi_{A}^{\frac{1}{\alpha}}\left[  \T_{A}(\sigma_{A})\right]  ^{\frac{1-\alpha
}{\alpha}}\right]  \\
& =\frac{\alpha}{\alpha-1}\log_{2}\left\Vert \psi_{A}^{\frac{1}{\alpha}
}\right\Vert _{\frac{\alpha}{2\alpha-1}}\\
& =\frac{\alpha}{\alpha-1}\log_{2}\left(  \operatorname{Tr}\!\left[  \left(
\psi_{A}^{\frac{1}{\alpha}}\right)  ^{\frac{\alpha}{2\alpha-1}}\right]
\right)  ^{\frac{2\alpha-1}{\alpha}}\\
& =\frac{\alpha}{\alpha-1}\frac{2\alpha-1}{\alpha}\log_{2}\operatorname{Tr}\!\left[  \left(  \psi_{A}^{\frac{1}{\alpha}}\right)  ^{\frac{\alpha}{2\alpha
-1}}\right]  \\
& =\frac{2\alpha-1}{\alpha-1}\log_{2}\operatorname{Tr}\!\left[  \psi_{A}
^{\frac{1}{2\alpha-1}}\right]  \\
& =2\left(  \frac{1}{1-\frac{1}{2\alpha-1}}\right)  \log_{2}\operatorname{Tr}\!\left[  \psi_{A}^{\frac{1}{2\alpha-1}}\right]  \\
& =2H_{\frac{1}{2\alpha-1}}(A)_{\psi}.
\end{align}
The third equality follows from Proposition~\ref{prop-Schatten_pos_var}.

We now prove \eqref{eq:QEI:SW-Renyi-CI-pure-bi}:
\begin{align}
& \widetilde{Q}_{\alpha}(\psi_{AB}\Vert \mathbbm{1}_{A}\otimes\sigma_{B})\nonumber\\
& =\operatorname{Tr}\!\left[  \left(  \psi_{AB}^{\frac{1}{2}}\left(
\mathbbm{1}_{A}\otimes\sigma_{B}\right)  ^{\frac{1-\alpha}{\alpha}}\psi_{AB}^{\frac
{1}{2}}\right)  ^{\alpha}\right]  \\
& =\operatorname{Tr}\!\left[  \left(  |\psi\rangle\!\langle\psi|_{AB}\left(
\mathbbm{1}_{A}\otimes\sigma_{B}^{\frac{1-\alpha}{\alpha}}\right)  |\psi\rangle
\langle\psi|_{AB}\right)  ^{\alpha}\right]  \\
& =\left(  \langle\psi|_{AB}\left(  \mathbbm{1}_{A}\otimes\sigma_{B}^{\frac{1-\alpha
}{\alpha}}\right)  |\psi\rangle_{AB}\right)  ^{\alpha}\operatorname{Tr}\!\left[
\left(  |\psi\rangle\!\langle\psi|_{AB}\right)  ^{\alpha}\right]  \\
& =\left(  \langle\psi|_{AB}\left(  \mathbbm{1}_{A}\otimes\sigma_{B}^{\frac{1-\alpha
}{\alpha}}\right)  |\psi\rangle_{AB}\right)  ^{\alpha}\\
& =\left(  \langle\Gamma|_{AB}\left(  \psi_{A}\otimes\sigma_{B}^{\frac
{1-\alpha}{\alpha}}\right)  |\Gamma\rangle_{AB}\right)  ^{\alpha}\\
& =\left(  \langle\Gamma|_{AB}\left(  \psi_{A}\left[  \T_{A}(\sigma
_{A})\right]  ^{\frac{1-\alpha}{\alpha}}\otimes \mathbbm{1}_{B}\right)  |\Gamma
\rangle_{AB}\right)  ^{\alpha}\\
& =\left(  \operatorname{Tr}\!\left[  \psi_{A}\left[  \T_{A}(\sigma_{A})\right]
^{\frac{1-\alpha}{\alpha}}\right]  \right)  ^{\alpha}.
\end{align}
Then consider that
\begin{align}
\widetilde{I}_{\alpha}(A\rangle B)_{\psi}  & =\inf_{\sigma_{B}}\frac{1}
{\alpha-1}\log_{2}\widetilde{Q}_{\alpha}(\psi_{AB}\Vert \mathbbm{1}_{A}\otimes\sigma
_{B})\\
& =\inf_{\sigma_{A}}\frac{1}{\alpha-1}\log_{2}\left(  \operatorname{Tr}\!\left[
\psi_{A}\left[  \T_{A}(\sigma_{A})\right]  ^{\frac{1-\alpha}{\alpha}}\right]
\right)  ^{\alpha}\\
& =\frac{\alpha}{\alpha-1}\log_{2}\left\Vert \psi_{A}\right\Vert
_{\frac{\alpha}{2\alpha-1}}\\
& =\frac{\alpha}{\alpha-1}\log_{2}\left(  \operatorname{Tr}\!\left[  \psi
_{A}^{\frac{\alpha}{2\alpha-1}}\right]  \right)  ^{\frac{2\alpha-1}{\alpha}
}\\
& =\frac{\alpha}{\alpha-1}\frac{2\alpha-1}{\alpha}\log_{2}\operatorname{Tr}\!\left[  \psi_{A}^{\frac{\alpha}{2\alpha-1}}\right]  \\
& =\frac{2\alpha-1}{\alpha-1}\log_{2}\operatorname{Tr}\!\left[  \psi_{A}
^{\frac{\alpha}{2\alpha-1}}\right]  \\
& =\frac{1}{1-\frac{\alpha}{2\alpha-1}}\log_{2}\operatorname{Tr}\!\left[
\psi_{A}^{\frac{\alpha}{2\alpha-1}}\right]  \\
& =H_{\frac{\alpha}{2\alpha-1}}(A)_{\psi}.
\end{align}
The third equality follows from Proposition~\ref{prop-Schatten_pos_var}.

We prove \eqref{eq:QEI:geo-Renyi-MI-pure-bi}. Let $\sigma_{B}$ be a state with the same support as
$\psi_{B}$. Recall the formula in Proposition~\ref{prop:QEI:geo-renyi-pure-states} for the geometric R\'enyi relative
entropy when the state $\rho$ is pure. We use this to conclude that
\begin{align}
\widehat{D}_{\alpha}(\psi_{AB}\Vert\psi_{A}\otimes\sigma_{B})  & =\log
_{2}\langle\psi|_{AB}\left(  \psi_{A}\otimes\sigma_{B}\right)  ^{-1}
|\psi\rangle_{AB}\\
& =\log_{2}\langle\psi|_{AB}\left(  \psi_{A}^{-1}\otimes\sigma_{B}
^{-1}\right)  |\psi\rangle_{AB}\\
& =\log_{2}\langle\Gamma|_{AB}\psi_{A}^{\frac{1}{2}}\left(  \psi_{A}
^{-1}\otimes\sigma_{B}^{-1}\right)  \psi_{A}^{\frac{1}{2}}|\Gamma\rangle
_{AB}\\
& =\log_{2}\langle\Gamma|_{AB}\left(  \mathbbm{1}_{A}\otimes\sigma_{B}^{-1}\right)
|\Gamma\rangle_{AB}\\
& =\log_{2}\operatorname{Tr}\!\left[  \sigma_{B}^{-1}\right]  .
\end{align}
Now consider that the minimum value of $\inf_{\sigma_{B}}\operatorname{Tr}\!\left[  \sigma_{B}^{-1}\right]  $ occurs when $\sigma_{B}$ is the maximally
mixed state $\pi_B$. This follows from using the Lagrange multiplier method (or alternatively $\inf_{\sigma_{B}}\operatorname{Tr}\!\left[  \sigma_{B}^{-1}\right]$ can be evaluated as $d_B^2$ by applying Proposition~\ref{prop-Schatten_pos_var} again, with an implicit identity operator acting on the support of $\psi_B$). We then
conclude that
\begin{align}
\widehat{I}_{\alpha}(A;B)_{\psi}  & =\inf_{\sigma_{B}}\log_{2}
\operatorname{Tr}\!\left[  \sigma_{B}^{-1}\right]  
 =\log_{2}\operatorname{Tr}\!\left[  \pi_{B}^{-1}\right] \\ 
& =2\log_{2}\text{rank}(\psi_{A})
 =2H_{0}(A)_{\psi}.
\end{align}

We finally prove \eqref{eq:QEI:geo-Renyi-CI-pure-bi}:
\begin{align}
\widehat{D}_{\alpha}(\psi_{AB}\Vert \mathbbm{1}_{A}\otimes\sigma_{B})  & =\log
_{2}\langle\psi|_{AB}\left(  \mathbbm{1}_{A}\otimes\sigma_{B}\right)  ^{-1}|\psi
\rangle_{AB}\\
& =\log_{2}\langle\psi|_{AB}\left(  \mathbbm{1}_{A}\otimes\sigma_{B}^{-1}\right)
|\psi\rangle_{AB}\\
& =\log_{2}\langle\Gamma|_{AB}\left(  \psi_{A}\otimes\sigma_{B}^{-1}\right)
|\Gamma\rangle_{AB}\\
& =\log_{2}\langle\Gamma|_{AB}\left(  \mathbbm{1}_{A}\otimes\sigma_{B}^{-1}\T_{B}
(\psi_{B})\right)  |\Gamma\rangle_{AB}\\
& =\log_{2}\operatorname{Tr}\!\left[  \sigma_{B}^{-1}\T_{B}(\psi_{B})\right]  .
\end{align}
Now applying Proposition~\ref{prop-Schatten_pos_var}, we conclude that
\begin{align}
\widehat{I}_{\alpha}(A\rangle B)_{\psi}  & =\inf_{\sigma_{B}}\widehat
{D}_{\alpha}(\psi_{AB}\Vert \mathbbm{1}_{A}\otimes\sigma_{B})
 =\inf_{\sigma_{B}}\log_{2}\operatorname{Tr}\!\left[  \sigma_{B}^{-1}\T_{B}
(\psi_{B})\right]  \\
& =\log_{2}\left\Vert \T_{B}(\psi_{B})\right\Vert _{\frac{1}{2}}
 =\log_{2}\left\Vert \psi_{B}\right\Vert _{\frac{1}{2}}\\
& =\log_{2}\left\Vert \psi_{A}\right\Vert _{\frac{1}{2}}
 =H_{\frac{1}{2}}(A)_{\psi}.
\end{align}
This concludes the proof.
\end{Proof}

\subsection{Remarks on Defining Channel Quantities from State Quantities}\label{sec-chan_inf_measures}

	Observe that all of the generalized information measures for quantum channels given in Definition~\ref{def-gen_inf_meas_chan}, as well as all of the channel information measures given above for specific generalized divergences, are defined in a common manner. Specifically, given a function $f:\Density(\mathcal{H}_{AB})\to\mathbb{R}$ for bipartite states, the corresponding function $f$ for quantum channels\footnote{In a slight abuse of notation, we use the same letter to denote the channel quantity as the state quantity.} is defined as
	\begin{equation}\label{eq-state_to_chan_quantity}
		f(\mathcal{N})\coloneqq\sup_{\rho_{RA}}f(R;B)_{\omega},
	\end{equation}
	where $\mathcal{N}_{A\to B}$ is a quantum channel, $\rho_{RA}$ is a quantum state, with the dimension of $R$ unbounded, and $\omega_{RB}=\mathcal{N}_{A\to B}(\rho_{RA})$. In other words, we define the channel quantity by taking a quantum state $\rho_{RA}$ of a bipartite system consisting of the input system $A$ of the channel and a reference system $R$ (whose dimension is in general unbounded), passing $A$ through the channel, then evaluating the state quantity on the output state $\mathcal{N}_{A\to B}(\rho_{RA})$. We then optimize over all states $\rho_{RA}$. 
	
	A similar principle as in \eqref{eq-state_to_chan_quantity} has been used in Definition~\ref{def-gen_channel_div} for the generalized divergence between two quantum channels. In particular, if we have a function $f:\Density(\mathcal{H})\times\Lin_+(\mathcal{H})\to\mathbb{R}\cup\{+\infty\}$ on two quantum states,\footnote{We allow for the second argument to be a positive semi-definite operator more generally.} then we define a corresponding quantity $f$ for two quantum channels as
	\begin{equation}\label{eq-state_to_chan_quantity_2}
		f(\mathcal{N},\mathcal{M})\coloneqq\sup_{\rho_{RA}}f(\mathcal{N}_{A\to B}(\rho_{RA}),\mathcal{M}_{A\to B}(\rho_{RA})),
	\end{equation}
	where $\mathcal{N}_{A\to B}$ and $\mathcal{M}_{A\to B}$ are quantum channels and $\rho_{RA}$ is a quantum state, with the dimension of $R$ unbounded. In other words, we define the channel quantity by evaluating the state quantity on the states $\mathcal{N}_{A\to B}(\rho_{RA})$ and $\mathcal{M}_{A\to B}(\rho_{RA})$ and optimizing over all states $\rho_{RA}$. We have already seen this principle being used in Chapter~\ref{chap-QM_dist_meas} to define the diamond distance (Definition~\ref{eq-diamond_dist_meas}) and fidelity (Definition~\ref{def:QM-over:fid-channels}) of two quantum channels, in which case the state quantity $f$ is the trace distance or fidelity.
	
	In both \eqref{eq-state_to_chan_quantity} and \eqref{eq-state_to_chan_quantity_2}, properties of the underlying state quantity (namely, the data-processing inequality), as well as the Schmidt decomposition theorem, allow us to restrict the optimizations in \eqref{eq-state_to_chan_quantity} and \eqref{eq-state_to_chan_quantity_2} to pure states $\ket{\psi}_{RA}$ without loss of generality, with the dimension of $R$ equal to the dimension of $A$. If the underlying state quantity $f$ in \eqref{eq-state_to_chan_quantity} is invariant with respect to local unitaries, then we can use this simplification to write $f(\mathcal{N})$ as
	\begin{equation}\label{eq-state_to_chan_quantity_alt}
		f(\mathcal{N})=\sup_{\rho_A}f(\rho_A,\mathcal{N}_{A\to B}),
	\end{equation}
	where the optimization is now only over states $\rho_A$ for the input system $A$ for the channel $\mathcal{N}$, and
	\begin{equation}
		f(\rho_A,\mathcal{N}_{A\to B})\coloneqq f(A;B)_{\omega},\quad \omega_{AB}=\sqrt{\rho_A}\Gamma_{AB}^{\mathcal{N}}\sqrt{\rho_A}.
	\end{equation}
	This holds due to \eqref{eq-pure_state_vec}, which states for every purification $\ket{\psi^{\rho}}_{AA'}$ of $\rho_A$ there exists an operator $Y_A$ such that $(Y_A\otimes\mathbbm{1}_{A'})\ket{\Gamma}_{AA'}=\ket{\psi^{\rho}}_{AA'}$. Then, by the polar decomposition (Theorem~\ref{thm-polar_decomposition}), and the fact that $Y_AY_A^\dagger=\rho_A$, it holds that $Y_A=U_A\sqrt{\rho_A}$ for some unitary $U_A$. Finally, using the definition of the Choi representation $\Gamma_{AB}^{\mathcal{N}}$ and the unitary invariance of $f$, we obtain \eqref{eq-state_to_chan_quantity_alt}. This equivalent formulation of the channel quantity $f(\mathcal{N})$ has been used in \eqref{eq-coh_inf_chan_alt} for the coherent information of a channel.
	
	If the underlying state quantity in \eqref{eq-state_to_chan_quantity_2} is unitarily invariant, then by using the same reasoning as above we can write $f(\mathcal{N},\mathcal{M})$ in a form analogous to \eqref{eq-state_to_chan_quantity_alt}:
	\begin{equation}\label{eq-state_to_chan_quantity_2_alt}
		f(\mathcal{N},\mathcal{M})=\sup_{\rho_A}f(\sqrt{\rho_A}\Gamma_{AB}^{\mathcal{N}}\sqrt{\rho_A},\sqrt{\rho_A}\Gamma_{AB}^{\mathcal{M}}\sqrt{\rho_A}).
	\end{equation}

\section{Summary}

	In this chapter, we studied various entropic quantities, starting with quantum relative entropy. We proved many of its most important properties, and we saw that it acts as a parent quantity for well-known quantities such as von Neumann entropy, quantum conditional entropy, quantum mutual information and conditional mutual information, and coherent information. We then studied the Petz--R\'{e}nyi, sandwiched R\'{e}nyi, geometric R\'{e}nyi, and hypothesis testing relative entropies, and we proved many of their most important properties.
	
	The unifying concept of this chapter is that of generalized divergence. A generalized divergence is a function $\boldsymbol{D}\colon \Density(\mathcal{H})\times\Lin_+(\mathcal{H})\to\mathbb{R}$ that satisifes the data-processing inequality: for every state $\rho$, positive semi-definite operator $\sigma$, and quantum channel $\mathcal{N}$,
	\begin{equation}
		\boldsymbol{D}(\rho\Vert\sigma)\geq \boldsymbol{D}(\mathcal{N}(\rho)\Vert\mathcal{N}(\sigma)).
	\end{equation}
	This inequality holds for all of the quantum relative entropies that we considered in this chapter, and many of their important properties (such as joint convexity) can be derived using it. The data-processing inequality is a core concept in information theory, and it underlies virtually all of the results that we present in this book.
	
	At the end of the chapter, we defined information measures for quantum channels. Given a state information measure (or, more generally, a generalized divergence), we define an information measure for channels in a manner analogous to the way that the diamond norm (a generalized divergence for channels) is defined from the trace norm (a generalized divergence for states): we send one share of a bipartite pure state through the channel, evaluate the state measure, and then optimize over all input states. All of the upper and lower bounds on communication capacities that we present in this book are given in terms of channel information measures defined in this way.

\section{Bibliographic Notes}

\label{sec:QEI:bib-notes}

	The von Neumann quantum entropy was originally defined by \citet{Neu27entropy}. 
The quantum conditional entropy was considered implicitly by \citet{LR73,LR73b}, proposed as a quantum-information theoretic quantity of interest by \citet{CA97}, and the coherent information thereafter by \citet{SN96}. The modern definition of the quantum mutual information was proposed by \citet{Stratonovich1965}. Its non-negativity was proved by \citet{MD35,LR68}. The quantum conditional mutual information was considered implicitly by \citet{LR68,LR73,LR73b} and was proposed as a quantum infor\-mation-theoretic quantity of interest by \citet{CA97,CA98}.  Chain rules for quantum conditional entropy and information were employed by \citet{CA98}.

	The definition of the quantum relative entropy as presented in Definition~\ref{def-rel_ent} is due to \citet{U62}. It took many years after this until the paper by \citet{HP91} was published, which solidified the operational interpretation of the ``Umegaki quantum relative entropy'' in terms of quantum hypothesis testing and the quantum Stein's lemma. The strong converse for the quantum Stein's lemma was established by \citet{ON00}.
	
	The non-negativity of quantum relative entropy follows as a consequence of an inequality by \citet{O31}, and its data-processing inequality for quantum channels was established by \citet{Lin75}. The data-processing inequality for the quantum relative entropy under partial trace was established by \citet{LR73}, from which its joint convexity follows. The expression in \eqref{eq-QCMI_rel_ent} for the quantum conditional mutual information was given implicitly by \citet{LR73}. Strong subadditivity of quantum entropy (or equivalently, non-negativity of quantum conditional mutual information) was established by \citet{LR73,LR73b}. The data-processing inequality for the quantum conditional mutual information under local channels was established by \citet{CW04}. The notion of asymptotic continuity was introduced by \citet{DH1999} and formalized by \citet{SRH2006}. The asymptotic continuity bound for quantum conditional mutual information, as presented in Proposition~\ref{lem:LAQC-uniform-cont-CMI}, was established by \citet{Shirokov17}.
	
	The notion of generalized divergence in the classical case was proposed by \citet{PV10}, and in the quantum case by \citet{SW12}. Proposition~\ref{prop-gen_div_properties} was established by \citet{WWY14,TWW17}. The generalized mutual information and conditional quantum entropy were proposed by \citet{SW12}.
	
	The Petz--R\'enyi relative entropy was proposed by \citet{P85,P86}, wherein its data-processing inequality was established for the case of partial trace. The general definition of Petz--R\'enyi relative entropy incorporating support conditions and its data-processing inequality for general channels was established by \citet{TCR09}, along with several of its other properties such as monotonicity in $\alpha$. The expression in \eqref{eq-petz_rel_ent_quasi_alt} for the Petz--R\'enyi relative entropy was presented by \citet{TCR09,Sharma10}.
	
	The sandwiched R\'enyi relative entropy was independently proposed by \citet{MDSFT13} and \citet{WWY14}, and the alternative expression in \eqref{eq-sand_rel_ent_Schatten_3} was given by \citet{Dupuis2016}. The variational expression in \eqref{eq-sand_rel_ent_var}, as well as Proposition~\ref{prop-sand_rel_ent_lim}, are due to \citet{MDSFT13}. Proposition~\ref{prop-sand_ren_ent_lim} was established by \citet{MDSFT13,WWY14}. The fact that sandwiched R\'enyi relative entropy is monotone with respect to $\alpha$ is due to \citet{MDSFT13}, with an independent proof for $\alpha >1$ by \citet{Bei13}. The inequality in \eqref{eq-petz_vs_sandwiched} for $\alpha > 1$ is due to \citet{WWY14}, as a direct consequence of the inequality by \citet{LT76}. The inequality in \eqref{eq-petz_vs_sandwiched} for $\alpha \in (0,1)$ is due to \citet{DL14limit}, as a direct consequence of the Araki--Lieb--Thirring inequalities by \citet{LT76,Araki1990}. The ``reverse'' Araki--Lieb--Thirring inequality, which leads to the inequality in \eqref{eq-petz_vs_sandwiched_2}, was proved by \citet{IRS17}. The data-processing inequality for the sandwiched R\'enyi relative entropy was established in a number of papers for various parameter ranges of $\alpha$: \citet{MDSFT13,WWY14,FL13,Bei13,Mosonyi2015}, being established for the full range $\alpha \in [\sfrac{1}{2},\infty]$ by \citet{FL13}. The proof that we presented here is due to \citet{W18opt}. Counterexamples to data processing for the sandwiched R\'{e}nyi relative entropy in the range $\alpha\in \left(0,\sfrac{1}{2}\right)$ were given by \citet{BFT15}.
	
	The geometric R\'enyi relative entropy has its roots in work of \citet{PR98}, and it was further
developed by \citet{M13,Matsumoto2018}. See also \citep{T15book,HM17} for other expositions. It was given the name ``geometric R\'enyi relative entropy'' by \citet{Fang2019a} because it is a function
of the matrix geometric mean of its arguments. See, e.g., \citet{LL01} for a review of matrix geometric means.
	Proposition~\ref{prop:explicit-form-geometric-renyi} was established by \citet{KW20}, with roots in the earlier work of \citet{Mat14,Mat14condconv}. In particular, the expression $\operatorname{Tr}[  \sigma(  \sigma
^{-1/2}\tilde{\rho}\sigma^{-1/2})  ^{\alpha}]  $
for $\alpha=1/2$ and $\operatorname{supp}(\rho)\not \subseteq
\operatorname{supp}(\sigma)$ was identified in \citet[Section~3]{Mat14} and
later generalized to all $\alpha\in(0,1)$ in \citet[Section~2]{Mat14condconv}.
Lemma~\ref{lem:limit-exchange-geom-renyi-a-0-to1} and Proposition~\ref{prop:QEI:geo-renyi-pure-states} were also established by \citet{KW20}, along with monotoniticity in $\alpha$ for all $\alpha\in (0,1)\cup(1,\infty)$ (in Proposition~\ref{prop:geometric-renyi-props}).
	The inequality in Proposition~\ref{prop:geometric-to-sandwiched} was established for the interval $\alpha\in
(0,1)\cup(1,2]$ in \citet{T15book} (by making use of a general result in
\citet{M13,Matsumoto2018}) and for the full interval $\alpha\in(1,\infty)$
in \citet{WWW19}. Here, we have followed the approach of \citet{WWW19} and offered a
unified proof in terms of the Araki--Lieb--Thirring inequality
\citet{Araki1990,LT76}.
		The first inequality in
Proposition~\ref{prop:sand-Petz-geo-ineqs} was established for $\alpha
\in(1,2]$ in \citet{WWY14}\ and for $\alpha\in(0,1)$ in \citet{DL14limit}, by
employing the Araki--Lieb--Thirring inequality \citet{Araki1990,LT76}. The
second inequality was established by \citet{M13,Matsumoto2018} and reviewed
by \citet{T15book}.
	Data processing was
established by an operator-theoretic approach in \citet{PR98} and by an
operational method in \citet{M13,Matsumoto2018}. The operator-theoretic
approach taken here has its roots in \citet[Proposition~2.5]{HP91} and was reviewed in
\citet[Corollary 3.31]{HM17}. 
The interpretation of geometric R\'{e}nyi relative entropy given in Proposition~\ref{prop:geometric-renyi-from-classical-preps} was
discovered by \citet{M13,Matsumoto2018}. Lemma~\ref{lem:sisi-zhou-lem} was presented by \citet{KW20} and is based on \citet[Lemma~3]{Zhou2019a}.

\citet{Belavkin1982} discovered the quantum generalization of the classical relative entropy given in Section~\ref{sec:QEI:Belavkin--Staszewski}, now known as the Belavkin--Staszewski relative
entropy.
\citet{M13,Matsumoto2018} showed that
 the Belavkin--Staszewski relative entropy is  the limit of the geometric R\'{e}nyi relative entropy as $\alpha
\rightarrow1$. The proof given here was presented by \citet{KW20}.
\citet{HP91} found the inequality in Proposition~\ref{cor:BS-to-q-rel-ent} relating the quantum relative entropy to the
Belavkin--Staszewski relative entropy (the proof given here is due to \citet{KW20}).
\citet{HP91} established the data-processing inequality for the Belavkin--Staszewski relative
entropy, by a method  different from that given here.
\citet{M13,Matsumoto2018} found the interpretation of the Belavkin--Staszewski relative entropy given in Proposition~\ref{prop:QEI:BS-ent-from-classical}.

	The max-relative entropy was proposed by \citet{Datta2009b} as a quantum information-theoretic quantity of interest. Datta also established many basic information processing properties of the max-relative entropy and studied its role in quantum hypothesis testing and entanglement theory \citet{Datta2009b,Dat09}. Proposition~\ref{prop-sand_rel_ent_limit_max} is due to \citet{MDSFT13,KW20}. Proposition~\ref{prop-smooth_max_to_petz_renyi} is due to \citet{Wang2019b} and \citet{ABJT19}.
	
	The conditional min-entropy was defined by \citet{RennerThesis}, as well as the smooth conditional min-entropy. The operational interpretations of the conditional min- and max-entropies were examined by \citet{KRS09}.

	The hypothesis testing relative entropy was studied implicitly by a variety of authors for a long time in the context of quantum hypothesis testing: \citet{HP91,ON00,hayashi2003generalCapacity,N06,Hay07}. It was proposed as a quantum information-theoretic quantity of interest by \citet{BD10} (in the context of operator smoothing of a one-shot entropic quantity), and given the name hypothesis testing relative entropy by \citet{WR12}, wherein its connection to classical communication was explored (see also \citet{hayashi2003generalCapacity}). An alternate proof of the data-processing inequality for quantum relative entropy in terms of hypothesis testing was given by \citet{BS12}. Various properties of the hypothesis testing relative entropy were established by \citet{DKFRR13} and \citet{DTW14}, including the fact that it can be written as an SDP (see also \citet{Wang2019b} in this context). Eq.~\eqref{eq:lim-dmineps-to-dmin} was established by \citet{Wang2019b}. Proposition~\ref{prop:QEI:opt-meas-HTRE} was considered by \citet{Hel76} and discussed more recently by \citet{GV16}. A special case of Proposition~\ref{prop-hypo_to_rel_ent} was established by \citet{WR12,MW12}. Proposition~\ref{prop:sandwich-to-htre} was established by \citet{CMW14}. Proposition~\ref{prop:ineq-hypo-renyi} is essentially due to \citet{Hay07}, with a refinement by \citet{AMV12} and a later rediscovery of it, formulated in a different way, by \citet{QWW17}. Proposition~\ref{prop-fixed-rate-petz-renyi} is essentially due to \citet{Hay07}.
	
	The generalized channel divergence of Definition~\ref{def-gen_channel_div} was proposed by \citet{LKDW18}. The argument for concavity in Proposition~\ref{prop:QEI:gen-ch-div-concavity-convexity} is based on the construction and argument given for Proposition~13 of \citet{Wang2019magic}.  Proposition~\ref{prop-gen_div_group_cov} was established by \citet{LKDW18}. The various generalized channel information measures can be found in the papers of \citet{WWY14,GW15}, and the related channel information measures based on hypothesis testing, Petz--R\'enyi, and sandwiched R\'enyi relative entropy are from \citet{KW09,SW12,WWY14,GW15,DTW14}. The channel mutual information was defined by \citet{AC97}, the channel Holevo information by \citet{schumacher1997sending} (based on the Holevo quantity for ensembles \citet{Holevo73}), and the channel coherent information by \citet{L97}.	These papers together thus developed a general concept of promoting a measure of correlations in a quantum state to a measure of a channel's ability to create the same correlations, by optimizing the state measure with respect to a (subset of) all of the states that can be generated by means of the channel.
	
	The review by \citet{Ruskai02} is helpful not only for understanding entropy inequalities in quantum information, but also for understanding the history of developments with respect to quantum entropy and information. The book of \citet{T15book} provides an exposition of R\'enyi relative entropies and their properties (see also \citet{Leditzky16}). The book of \citet{Wbook17} provides an overview of entropies in the von Neumann family, their properties, and the derived channel information measures.

\chapter{Information Measures for Quantum Channels}\label{chap-entropies_chan}


\chapter{Entanglement Measures}\label{chap-ent_measures}

	In the previous chapter, we laid the foundation for analyzing quantum communication protocols by defining entropic quantities, such as the Petz-- and sandwiched R\'{e}nyi relative entropies, as well as information measures for quantum states and  channels derived from these relative entropies. We now use these information measures to define entanglement measures for quantum states and channels. Given quantum systems $A$ and $B$, an entanglement measure is a function $E:\Density(\mathcal{H}_{AB})\to\mathbb{R}$ that  quantifies the amount of entanglement present in a state $\rho_{AB}$ of these systems. The notion of ``quantifying entanglement'' is explained in Section~\ref{sec-ent_measures_def} below, with the defining requirement of an entanglement measure being that it does not increase under channels realized by local operations and classical communication (Definition~\ref{def-LOCC}). We can think of this requirement of ``LOCC monotonicity'' as a restricted form of the data-processing inequality, but now applied to a single bipartite state rather than to a pair of states. The data-processing inequality indicates that the distinguishability of two states does not increase under the action of the same quantum channel (Definition~\ref{def-gen_div}), whereas LOCC monotonicity indicates that entanglement does not increase under the action of an LOCC channel on a bipartite state.
	
	Given an entanglement measure $E$ for states, the corresponding entanglement measure for channels is defined using the general principle in Section~\ref{sec-chan_inf_measures}, which is to optimize the state measure with respect to all bipartite states that can be shared between the sender and receiver of the channel by making use of the channel. We develop entanglement measures for channels in Chapter~\ref{chap-ent_measures_chan}, and these naturally quantify how much entanglement can be generated by a channel connecting a sender to a receiver.
	
	Entanglement measures feature prominently in the analysis of optimal rates for distillation and communication protocols. In particular, entanglement measures for states arise as upper bounds on their distillable entanglement and secret key, which we examine in Chapters~\ref{chap-ent_distill} and \ref{chap-secret_key_distill}, respectively. Entanglement measures for quantum channels arise as upper bounds on the rates of quantum and private communication over a quantum channel, which we consider in Chapters~\ref{chap-quantum_capacity} and \ref{chap-private_capacity}, respectively, as well as for their feedback-assisted counterparts that we consider in Part~\ref{part-feedback} of this book.

	Being a uniquely quantum-mechanical property, it is perhaps not surprising that entanglement features prominently in quantum communication protocols, both in the encoding and decoding of messages, as well as in the analysis of their optimal rates. In fact, any state that is not entangled (i.e., separable) is useless for entanglement and secret key distillation, and similarly, entanglement breaking channels are useless for quantum and private communication. The distinguishability of a given state from the set of separable states, which we show in this chapter is an entanglement measure for states, can thus give an indication of how much entanglement or secret key can be distilled from it. Similarly, for communication tasks over quantum channels, the distinguishability of a given quantum channel from the set of entanglement breaking channels, which we show in this chapter is an entanglement measure for channels, can be used to determine how good the channel is for quantum or private communication.
	
	
	The rest of this chapter proceeds as follows. We start in Section~\ref{sec-ent_measures_def} by formally defining what it means for a function $E$ to be an entanglement measure for bipartite states and by providing examples of entanglement measures. In Section~\ref{sec-ent_measures_sep_distance}, we consider entanglement measures that quantify the distinguishability of a given state $\rho_{AB}$ from the set $\SEP(A\!:\!B)$ of separable states, with the measure given by some generalized divergence $\boldsymbol{D}$ (see Definition~\ref{def-gen_div}). We also consider in Section~\ref{sec-ent_measures_Rains_dist} a class of entanglement measures based on the distinguishability of a given state from the larger set $\PPT'\supset\SEP$. In Section~\ref{sec-LAQC:sq-ent-and-props}, we consider a different kind of entanglement measure for states called squashed entanglement. 

\section{Definition and Basic Properties}\label{sec-ent_measures_def}

	Recall from Definition~\ref{def-sep_ent_state} that a bipartite state $\rho_{AB}$ is called entangled if it is not separable, meaning that it \textit{cannot} be written in the following form
	\begin{equation}
		\sum_{x\in\mathcal{X}}p(x)\tau_A^x\otimes\omega_B^x,
	\end{equation}
	for some finite alphabet $\mathcal{X}$, probability distribution $p:\mathcal{X}\to[0,1]$, and sets $\{\tau_A^x\}_{x\in\mathcal{X}}$ and $\{\omega_B^x\}_{x\in\mathcal{X}}$ of states. Determining whether a given quantum state $\rho_{AB}$ is entangled is a fundamental problem in quantum information theory. In Section~\ref{subsec-bipartite_state}, in the discussion after Definition~\ref{def-sep_ent_state}, we listed the following criteria for the entanglement of pure and mixed states:
	\begin{itemize}
		\item \textit{Schmidt rank criterion}: A pure bipartite state $\psi_{AB}$ is entangled if and only if its Schmidt rank is strictly greater than one.
		\item \textit{PPT criterion}: If a bipartite mixed state $\rho_{AB}$ has negative partial transpose (i.e., the partial transpose $\rho_{AB}^{\t_B}$ has at least one negative eigenvalue), then it is entangled. If both systems $A$ and $B$ are qubit systems or if one of the systems is a qubit and the other a qutrit, then $\rho_{AB}$ is entangled if and only if $\rho_{AB}^{\t_B}$ has negative partial transpose.
	\end{itemize}
	In the case of mixed states, there is generally not a simple necessary and sufficient criterion to determine whether a given bipartite state is entangled, and in fact it is known that it is computationally difficult, in a precise sense, to decide if a state is entangled (please consult the Bibliographic Notes in Section~\ref{sec:E-meas:bib-notes}).
	
	In addition to determining whether or not a given quantum state is entangled, we are interested in quantifying the amount of entanglement present in a quantum state. Doing so allows us to compare quantum states based on the amount of entanglement present in them. An \textit{entanglement measure} is a function $E:\Density(\mathcal{H}_{AB})\to\mathbb{R}$ from the set of density operators acting on the Hilbert space of a bipartite system to the set of real numbers, and it  quantifies the entanglement of a quantum state $\rho_{AB}\in\Density(\mathcal{H}_{AB})$. (The formal definition of an entanglement measure is given in Definition~\ref{def-LAQC:ent-measure}.) To  indicate the partitioning of the subsystems explicitly, we often write $E(A;B)_{\rho}$ instead of $E(\rho_{AB})$.
	
	How exactly do we quantify entanglement? Suppose that we have a bipartite state $\rho_{AB}$ and we would like to quantify the entanglement between the systems $A$ and $B$. One fundamental observation is that the entanglement of $\rho_{AB}$ cannot increase under the action of a local operations and classical communication (LOCC) channel (recall Definition~\ref{def-LOCC}). This is intuitive because entanglement is a non-local property of a bipartite state, and so local operations alone do not increase it. Similarly, classical communication should only affect the \textit{classical} correlations between the two systems $A$ and $B$, and not the quantum correlations, i.e., the entanglement. 
			
			Given the reasoning above, the defining property of an entanglement measure $E:\Density(\mathcal{H}_{AB})\to\mathbb{R}$ for the quantum systems $A$ and $B$ is that it does not increase under the action of an LOCC channel:
	\begin{definition}{Entanglement Measure}{def-LAQC:ent-measure}
		We say that $E(A;B)_{\rho}$ is an entanglement measure
for a bipartite state $\rho_{AB}$ if the following inequality holds for every bipartite state $\rho_{AB}$ and every LOCC\ channel $\mathcal{L}_{AB\rightarrow A'B'}$ that acts on $\rho_{AB}$:%
		\begin{equation}
			E(A;B)_{\rho}\geq E(A';B')_{\omega},
			\label{eq:E-meas:LOCC-mono}
		\end{equation}
		where $\omega_{A'B'}\coloneqq \mathcal{L}_{AB\rightarrow
A'B'}(\rho_{AB})$.
	\end{definition}

	This property of LOCC monotonicity is  the most operational property of an entanglement measure, in the sense that it relates directly to the distillation or communication tasks that we consider in this book, which involve LOCC between the two parties sharing a state $\rho_{AB}$ or between sender and receiver at the terminals of a quantum channel, respectively. It is also analogous to the data-processing inequality for generalized divergences. 
	
	The defining requirement of LOCC monotonicity implies that an entanglement measure $E$ takes on its minimum value on the set of separable states. To see this, recall from the discussion after Definition~\ref{def-sep_ent_state} that a separable state can be prepared by LOCC. Thus, starting from an arbitrary state $\rho_{AB}$, Alice and Bob can trace out their local systems $A$ and $B$ and perform LOCC to prepare a separable state $\sigma_{A'B'}$. The serial concatenation of these two actions is itself an LOCC channel. Thus, it follows from the definition above that
	\begin{equation}
		E(A;B)_{\rho}\geq E(A';B')_{\sigma}
	\end{equation}
	for every separable state $\sigma_{A'B'}$. Now, given another separable state $\sigma_{A''B''}'$, it is possible to transform between $\sigma_{A'B'}$ and $\sigma_{A''B''}'$ using LOCC, meaning that $E(A';B')_{\sigma}\geq E(A'';B'')_{\sigma'}$ and $E(A';B')_{\sigma}\leq E(A'';B'')_{\sigma'}$. Therefore,
	\begin{equation}
		E(A';B')_{\sigma}=E(A'';B'')_{\sigma'},
	\end{equation}
	for all separable states $\sigma_{AB}$ and $\sigma_{A'B'}'$. As a consequence, an entanglement measure $E$ takes on its minimum value and is equal to a constant $c\in\mathbb{R}$ for all separable states. It is often convenient and simpler if an entanglement measure $E$ is equal to zero for all separable states. If this is not the case, then we can simplify redefine the entanglement measure as $E'(A;B)_{\rho}=E(A;B)_{\rho}-c$. By this reasoning and adjustment (if needed), every entanglement measure (as per Definition~\ref{def-LAQC:ent-measure})  satisfies the following two properties of non-negativity on all states and vanishing on separable states:
	\begin{enumerate}
		\item \textit{Non-negativity}: $E(\rho_{AB})\geq 0$ for every state $\rho_{AB}$.
		\item \textit{Vanishing for separable states}: $E(\sigma_{AB})=0$ for every separable state $\sigma_{AB}$.
		\end{enumerate}

	Other properties that are desirable for an entanglement measure $E$ are as follows:
	\begin{enumerate}
		\item \textit{Faithfulness}: $E(\sigma_{AB})=0$ if and only if $\sigma_{AB}$ is separable, so that $E(\rho_{AB})>0$ if and only if $\rho_{AB}$ is entangled.
		\item \textit{Invariance under classical communication}: For every finite alphabet $\mathcal{X}$, probability distribution $p:\mathcal{X}\to[0,1]$ on $\mathcal{X}$, and set $\{\rho_{AB}^x\}_{x\in\mathcal{X}}$ of states, define the following classical--quantum state:
		\begin{equation}
		\rho_{X AB } \coloneqq \sum_{x\in\mathcal{X}} p(x) |x\rangle\!\langle x|_{X} \otimes \rho_{AB}^x.
		\label{eq:E-meas:cq-state-inv-CC}
		\end{equation}
		Then the entanglement measure $E$ satisfies invariance under classical communication if
		\begin{equation}
		 E(X A; B )_{\rho} = E( A; B X)_{\rho} = \sum_{x\in\mathcal{X}} p(x) E(A;B)_{\rho^x}.
		\label{eq:E-meas:invar-class-comm}
		\end{equation}
		This property has the interpretation of invariance under classical communication because the equality $E(X A; B )_{\rho} = E( A; B X)_{\rho}$ indicates that the classical value $x$ in register $X$ can be communicated classically to Bob and discarded locally, and the entanglement measure does not change under this action. Furthermore, the value of the entanglement is simply the expected entanglement, where the expectation is calculated with respect to the probability distribution $p(x)$.
		
		This property is also known as the ``flags'' property in the research literature.
		
		\item \textit{Convexity}: For every finite alphabet $\mathcal{X}$, probability distribution $p:\mathcal{X}\to[0,1]$ on $\mathcal{X}$, and set $\{\rho_{AB}^x\}_{x\in\mathcal{X}}$ of states,
			\begin{equation}
			\label{eq:E-meas:convexity-e-meas}
		\sum_{x\in\mathcal{X}}p(x)E(\rho_{AB}^x) \geq
				E\!\left(\sum_{x\in\mathcal{X}}p(x)\rho_{AB}^x\right).
			\end{equation}
			Convexity is an intuitive property of entanglement, and for entanglement measures that are invariant under classical communication (obeying \eqref{eq:E-meas:invar-class-comm}), it captures the idea that entanglement should not increase on average if classical information about the identity of a state is lost. (In fact, convexity is an immediate consequence of LOCC monotonicity and invariance under classical communication.)
		\item \textit{Additivity}: The entanglement of a tensor-product state $\rho_{A_1A_2B_1B_2}=\tau_{A_1B_1}\otimes\omega_{A_2B_2}$ is the sum of the entanglement of the individual states in the tensor product:
			\begin{equation}
				\label{eq:E-meas:ent-meas-additive}				E(A_1A_2;B_1B_2)_{\tau\otimes\omega}=E(A_1;B_1)_{\tau}+E(A_2;B_2)_{\omega}.
			\end{equation}
			If instead we have only that
			\begin{equation}\label{def-ent_meas_subadditive}
				E(A_1A_2;B_1B_2)_{\tau\otimes\omega}\leq E(A_1;B_1)_{\tau}+E(A_2;B_2)_{\omega}
			\end{equation}
			for all states $\tau_{A_1B_1}$ and $\omega_{A_2B_2}$, then the entanglement measure $E$ is \textit{subadditive}.
			
		\item \textit{Selective LOCC monotonicity}: 
A property stronger than LOCC monotonicity is that $E$ is non-increasing on average under an \textit{LOCC instrument}. In more detail, let $\rho_{AB}$ be a bipartite state, and let $\{\mathcal{L}^x_{AB\to A'B'}\}_{x\in\mathcal{X}}$ be a collection of maps, such that $\mathcal{L}^{\leftrightarrow}_{AB\to A'B'}$ is an LOCC channel of the form:
\begin{equation}
\mathcal{L}^{\leftrightarrow}_{AB\to A'B'} = \sum_{x\in\mathcal{X}} \mathcal{L}^x_{AB\to A'B'},
\label{eq:E-meas:LOCC-instrument}
\end{equation}
for some finite alphabet $\mathcal{X}$ and where each map $\mathcal{L}^x_{AB\to A'B'}$ is completely positive such that the sum map $\mathcal{L}^{\leftrightarrow}_{AB\to A'B'}$ is trace preserving (i.e., a quantum channel). Furthermore, each map $\mathcal{L}^x_{AB\to A'B'}$ can be written in the form of \eqref{eq-LOCC_chan_gen}, as follows:
\begin{equation}
\mathcal{L}^x_{AB\to A'B'} = \sum_{y \in \mathcal{Y}}
\mathcal{E}^{x,y}_{A\to A'} \otimes  \mathcal{F}^{x,y}_{B\to B'},
\label{eq:E-meas:LOCC-instrument-indiv-maps}
\end{equation}
where
$\{\mathcal{E}^{x,y}_{A\to A'}\}_{x\in \mathcal{X}}$ and $\{\mathcal{F}^{x,y}_{B\to B'}\}_{x\in \mathcal{X}}$ are sets of completely positive maps.  Set
\begin{equation}
p(x)\coloneqq \Tr[\mathcal{L}^x_{AB\to A'B'}(\rho_{AB})],
\end{equation}
and for $x\in\mathcal{X}$ such that $p(x) \neq 0$, set
\begin{equation}
\omega_{AB}^x \coloneqq \frac{1}{p(x)}\mathcal{L}^x_{AB\to A'B'}(\rho_{AB}).
\end{equation}
If the classical value of $x$ is not discarded, then the given state $\rho_{AB}$ is transformed to the ensemble $\{(p(x),\omega_{AB}^x)\}_{x\in\mathcal{X}}$ via LOCC. 

The entanglement measure $E$ satisfies \textit{selective LOCC monotonicity} if
		\begin{equation}
			E(\rho_{AB})\geq \sum_{x\in\mathcal{X}:p(x)\neq 0}p(x)E(\omega_{AB}^x),
			\label{eq:E-meas:strong-LOCC-mono}
		\end{equation}
		for every ensemble $\{(p(x),\omega_{AB}^x)\}_{x\in\mathcal{X}}$ that arises from $\rho_{AB}$ via LOCC as specified above. Selective LOCC monotonicity indicates that entanglement does not increase on average under the action of LOCC.
		Many entanglement measures satisfy this stronger property.
		
		Observe that selective LOCC monotonicity in \eqref{eq:E-meas:strong-LOCC-mono} implies LOCC monotonicity in \eqref{eq:E-meas:LOCC-mono}, simply because the alphabet $\mathcal{X}$ in \eqref{eq:E-meas:LOCC-instrument} can consist of only one letter.
		\end{enumerate}
	The entanglement measures that we consider in this chapter satisfy many of the properties listed above.

	Given that we would like to quantify entanglement, it makes sense to ask what the basic \textit{unit} of entanglement should be. We take as our unit of entanglement the two-qubit maximally entangled Bell state $\ket{\Phi}=\frac{1}{\sqrt{2}}(\ket{0,0}+\ket{1,1})$, and we thus say that the state $\ket{\Phi}$ represents ``one ebit.'' A maximally entangled state of Schmidt rank $d$ is then referred to as having ``$\log_2 d$ ebits.'' All of the entanglement measures that we consider in this chapter are equal to one for a two-qubit maximally entangled state, which is another justification for using it as the unit of entanglement.\footnote{This ``normalization'' condition is sometimes taken to be a requirement for an entanglement measure.} Similarly, for a maximally entangled state of Schmidt rank $d$, all of the entanglement measures that we consider in this chapter are equal to $\log_2 d$.
	 
	 
	 To close out this introductory section, we prove a lemma that helps to reduce the difficulty
in determining whether a given function is an entanglement measure. 

\begin{Lemma}{lem:E-meas:convex-strong-LOCC-mono}
Let $E:D(\mathcal{H}_{AB})\rightarrow\mathbb{R}$ be a function that, for every
bipartite state $\rho_{AB}$, is
\begin{enumerate}
\item
 invariant under classical communication,
as defined in \eqref{eq:E-meas:invar-class-comm}, and 
\item obeys data processing under local channels, in
the sense that%
\begin{equation}
E(A;B)_{\rho}\geq E(A^{\prime};B^{\prime})_{\omega},
\end{equation}
for all channels $\mathcal{N}_{A\rightarrow A^{\prime}}$ and $\mathcal{M}%
_{B\rightarrow B^{\prime}}$, where%
\begin{equation}
\omega_{A^{\prime}B^{\prime}}\coloneqq (\mathcal{N}_{A\rightarrow A^{\prime}}%
\otimes\mathcal{M}_{B\rightarrow B^{\prime}})(\rho_{AB}).
\end{equation}
\end{enumerate}
Then $E$ is convex, as defined in \eqref{eq:E-meas:convexity-e-meas}, and a selective LOCC\ monotone, as defined in \eqref{eq:E-meas:strong-LOCC-mono}.
\end{Lemma}

\begin{Proof}
We first prove convexity. Let
\begin{equation}
\rho_{XAB}\coloneqq \sum_{x\in\mathcal{X}}p(x)|x\rangle\!\langle x|_{X}\otimes\rho
_{AB}^{x},
\end{equation}
where $\mathcal{X}$ is a finite alphabet, $p:\mathcal{X}\rightarrow\left[
0,1\right]  $ is a probability distribution, and $\{\rho_{AB}^{x}%
\}_{x\in\mathcal{X}}$ is a set of states. Then%
\begin{align}
\sum_{x\in\mathcal{X}}p(x)E(A;B)_{\rho^{x}}  & =E(XA;B)_{\rho}\\
& \geq E(A;B)_{\rho},
\end{align}
where the entanglement $E(A;B)$ in the last line is evaluated with respect to
the reduced state $\rho_{AB}=\sum_{x\in\mathcal{X}}p(x)\rho_{AB}^{x}$. The
equality follows from the assumption of invariance under classical communication, as defined in \eqref{eq:E-meas:invar-class-comm}, and
the inequality follows because the partial trace channel $\operatorname{Tr}%
_{X}$ is a local channel that discards the classical system $X$.

To establish selective LOCC\ monotonicity, consider that an LOCC\ channel
$\mathcal{L}_{AB\rightarrow A^{\prime}B^{\prime}}^{\leftrightarrow}$ of the
form in \eqref{eq:E-meas:LOCC-instrument}, by Definition~\ref{def-LOCC}, can be built up as a serial concatenation of one-way
LOCC\ channels. Furthermore, each one-way LOCC channel can keep some classical
information and discard some of it. It is helpful
conceptually to think of the retained classical information as part of the
variable $x$ in \eqref{eq:E-meas:LOCC-instrument} and the discarded classical information as part of the
variable $y$ in \eqref{eq:E-meas:LOCC-instrument-indiv-maps}. In more detail, each such one-way LOCC\ channel (from
Alice to Bob in the case discussed below) has the following form:%
\begin{equation}
\sum_{k,\ell}\mathcal{E}_{A\rightarrow A^{\prime}}^{k,\ell}\otimes
\mathcal{F}_{B\rightarrow B^{\prime}}^{k,\ell}%
,\label{eq:E-meas:one-way-LOCC-double-ind}%
\end{equation}
where $\{\mathcal{E}_{A}^{k,\ell}\}_{k,\ell}$ is a collection of completely
positive maps such that the sum map $\sum_{k,\ell}\mathcal{E}_{A}^{k,\ell}$ is
trace preserving, and $\{\mathcal{F}_{B}^{k,\ell}\}_{k,\ell}$ is a collection
of quantum channels. For now and for simplicity, let use the superindex
$m\coloneqq \left(  k,\ell\right)  $. We should think of the classical information $k$
as that which is being \textit{kept} and that in $\ell$ as that which is being
\textit{lost} or \textit{let go}. The one-way LOCC\ channel in
\eqref{eq:E-meas:one-way-LOCC-double-ind}\ can be implemented using the
following steps:

\begin{enumerate}
\item Alice applies the following local quantum channel:%
\begin{equation}
\tau_{AB}\mapsto\sum_{m}\mathcal{E}_{A\rightarrow A^{\prime}}^{m}%
(\tau_{AB})\otimes|m\rangle\!\langle m|_{M_{A}}%
.\label{eq:E-meas:1-way-step-1-local}%
\end{equation}

\item Alice employs a classical communication channel%
\begin{equation}
(\cdot)_{M_{A}}\mapsto\sum_{m}|m\rangle_{M_{B}}\langle m|_{M_{A}}%
(\cdot)_{M_{A}}|m\rangle_{M_{A}}\langle m|_{M_{B}}%
\label{eq:E-meas:1-way-step-2-CC}%
\end{equation}
to communicate the value in $M_{A}$ to Bob:%
\begin{equation}
\sum_{m}\mathcal{E}_{A\rightarrow A^{\prime}}^{m}(\tau_{AB})\otimes
|m\rangle\!\langle m|_{M_{A}}\mapsto\sum_{m}\mathcal{E}_{A\rightarrow
A^{\prime}}^{m}(\tau_{AB})\otimes|m\rangle\!\langle m|_{M_{B}}.
\end{equation}

\item Bob performs the local channel 
\begin{equation}
(\cdot)_{B M_B} \mapsto \sum_{m}
\mathcal{F}_{B\rightarrow
B^{\prime}}^{m}(\cdot) \otimes
|m\rangle\!\langle m|_{M_{B}
}(\cdot)_{M_{B}}|m\rangle\!\langle m|_{M_{B}} ,
\end{equation}
which can be understood as \textquotedblleft looking
in the classical register $M_{B}$\textquotedblright\ to determine the value
$m$ and performing the local quantum channel $\mathcal{F}_{B\rightarrow
B^{\prime}}^{m}$ based on the value $m$ found. Under this local channel, the
global state becomes as follows:%
\begin{equation}
\sum_{m}\mathcal{E}_{A\rightarrow A^{\prime}}^{m}(\tau_{AB})\otimes
|m\rangle\!\langle m|_{M_{B}}\mapsto\sum_{m}(\mathcal{E}_{A\rightarrow
A^{\prime}}^{m}\otimes\mathcal{F}_{B\rightarrow B^{\prime}}^{m})(\tau
_{AB})\otimes|m\rangle\!\langle m|_{M_{B}}.\label{eq:E-meas:1-way-step-3-local}%
\end{equation}

\item Bob then discards the $\ell$ part of the classical information $m$, as
follows:%
\begin{align}
& \sum_{m}(\mathcal{E}_{A\rightarrow A^{\prime}}^{m}\otimes\mathcal{F}%
_{B\rightarrow B^{\prime}}^{m})(\tau_{AB})\otimes|m\rangle\!\langle m|_{M_{B}%
}\nonumber\\
& =\sum_{k,\ell}(\mathcal{E}_{A\rightarrow A^{\prime}}^{k,\ell}\otimes
\mathcal{F}_{B\rightarrow B^{\prime}}^{k,\ell})(\tau_{AB})\otimes
|k,\ell\rangle\!\langle k,\ell|_{K_{B}L_{B}}\\
& \mapsto\sum_{k,\ell}(\mathcal{E}_{A\rightarrow A^{\prime}}^{k,\ell
}\otimes\mathcal{F}_{B\rightarrow B^{\prime}}^{k,\ell})(\tau_{AB}%
)\otimes|k\rangle\!\langle k|_{K_{B}}\label{eq:E-meas:1-way-step-4-local}\\
& =\sum_{k}p(k)\omega_{A^{\prime}B^{\prime}}^{k}\otimes|k\rangle\!\langle
k|_{K_{B}},
\end{align}
where%
\begin{align}
p(k)  & \coloneqq \operatorname{Tr}\!\left[  \sum_{\ell}(\mathcal{E}_{A\rightarrow
A^{\prime}}^{k,\ell}\otimes\mathcal{F}_{B\rightarrow B^{\prime}}^{k,\ell
})(\tau_{AB})\right]  ,\\
\omega_{A^{\prime}B^{\prime}}^{k}  & \coloneqq \frac{1}{p(k)}\sum_{\ell}%
(\mathcal{E}_{A\rightarrow A^{\prime}}^{k,\ell}\otimes\mathcal{F}%
_{B\rightarrow B^{\prime}}^{k,\ell})(\tau_{AB}).
\end{align}
Bob could, if desired, finally discard the classical register $K_{B}$ to
implement the one-way LOCC channel in
\eqref{eq:E-meas:one-way-LOCC-double-ind}. However, it is helpful to hold on
to it for our analysis below.
\end{enumerate}

Now we analyze how the entanglement changes under each of these steps,
omitting the state subscripts at each step except for the first and last
lines, because those not shown are clear from the context:%
\begin{align}
E(A;B)_{\tau}  & \geq E(A^{\prime}M_{A};B)\\
& =E(A^{\prime};BM_{B})\\
& \geq E(A^{\prime};B^{\prime}M_{B})\\
& =E(A^{\prime};B^{\prime}K_{B}L_{B})\\
& \geq E(A^{\prime};B^{\prime}K_{B})\\
& =\sum_{k}p(k)E(A^{\prime};B^{\prime})_{\omega^{k}}.
\end{align}
The first inequality follows from data processing under the local channel in
\eqref{eq:E-meas:1-way-step-1-local}. The first equality follows from the
assumption of invariance of classical communication, i.e., invariance under
the action of the classical channel in \eqref{eq:E-meas:1-way-step-2-CC}. The
second inequality follows from data processing under the local channel in
\eqref{eq:E-meas:1-way-step-3-local}. The second equality is trivial,
following because $M_{B}=(K_{B},L_{B})$ by definition. The third inequality
follows again from data processing under the local channel in
\eqref{eq:E-meas:1-way-step-4-local}. The final equality follows again from
invariance under classical communication.

Thus, we have shown selective one-way LOCC monotonicity (from Alice to Bob) in
the following sense:%
\begin{equation}
E(A;B)_{\tau}\geq\sum_{k}p(k)E(A^{\prime};B^{\prime})_{\omega^{k}},
\end{equation}
where the ensemble $\{(p(k),\omega_{A^{\prime}B^{\prime}}^{k})\}_{k}$ arises
from the state $\tau_{AB}$ by means of one-way LOCC from Alice to Bob. By the
same argument, but flipping the role of Alice and Bob, it follows that selective one-way LOCC monotonicity from Bob to
Alice holds for the function $E$. Since every LOCC channel is built up as a serial concatenation of
one-way LOCC\ channels and since we have proven that selective monotonicity holds
for the function $E$ for each of them, it follows that $E$ obeys selective LOCC\ monotonicity.
\end{Proof}

\subsection{Examples}\label{sec-ent_meas_examples}

	Let us now consider some examples of entanglement measures. The first two entanglement measures that we consider are related to the Schmidt rank criterion and the PPT criterion, respectively, stated in Section~\ref{subsec-bipartite_state} and reiterated at the beginning of this chapter. They are known as the entanglement of formation and the log-negativity, respectively, and are some of the simplest and earliest entanglement measures defined. They are also conceptually linked to more complex entanglement measures like squashed entanglement and the Rains relative entropy, the latter of which are the best known upper bounds on distillable entanglement (studied in Chapter~\ref{chap-ent_distill}).

\subsubsection{Entanglement of Formation}
	
	Given a pure bipartite state $\psi_{AB}=\ket{\psi}\!\bra{\psi}_{AB}$, there exists a Schmidt decomposition of $\ket{\psi}_{AB}$ such that
	\begin{equation}
		\ket{\psi}_{AB}=\sum_{k=1}^r\sqrt{\lambda_k}\ket{e_k}_A\otimes\ket{f_k}_B,
	\end{equation}
	where $r$ is the Schmidt rank, $\lambda_k>0$ are the Schmidt coefficients, and $\{\ket{e_k}_A\}_{k=1}^r$, $\{\ket{f_k}_B\}_{k=1}^r$ are orthonormal sets. Observe that the reduced states $\psi_A\coloneqq\Tr_B[\psi_{AB}]$ and $\psi_B\coloneqq\Tr_A[\psi_{AB}]$ have the same non-zero eigenvalues, which means that their entropies are equal, i.e., $H(\psi_A)=H(\psi_B)$. Furthermore, $H(\psi_A)=0$ if and only if $r=1$, and $r=1$ if and only if $\psi_{AB}$ is separable, by the Schmidt rank criterion. Therefore, the entropy of the reduced state of a pure bipartite state provides us with a signature of entanglement for pure bipartite states:
	\begin{equation}\label{eq-Schmidt_rank_ent_meas}
		\psi_{AB}\text{ entangled }\qquad \Longleftrightarrow \qquad H(\psi_A)>0.
	\end{equation}
	We let
	\begin{equation}\label{eq-ent_formation_pure_state}
		E_F(\psi_{AB})\coloneqq H(\Tr_B[\psi_{AB}])=-\sum_{k=1}^r\lambda_k\log_2\lambda_k
	\end{equation}
	for every pure state $\psi_{AB}$.
	
	The function $E_F$ is an entanglement measure, as proven in Proposition~\ref{def:E-meas:EoF} below. When evaluated on pure bipartites as above, it is known as the \textit{entropy of entanglement} or \textit{entanglement entropy}, and it is often simply denoted by $E(\psi_{AB})$ in the research literature. By \eqref{eq-Schmidt_rank_ent_meas}, it is also a faithful entanglement measure on pure states, i.e., $E_F(\psi_{AB})=0$ if and only if $\psi_{AB}$ is a separable state. Recall from Section~\ref{subsubsec-entanglement} that a maximally entangled state is defined by having $\lambda_k=\frac{1}{r}$ for all $1\leq k\leq r$. For such states, $E_F(\psi_{AB})=\log_2 r$, which justifies calling them maximally entangled because $\log_2 r$ is the largest value of the quantum entropy for states supported on an $r$-dimensional space.
	
	The definition in \eqref{eq-ent_formation_pure_state} for the entanglement measure $E_F$, so far, has been defined only for pure states. To extend the definition to mixed states, we use the fact that a mixed state $\rho_{AB}$ can be decomposed into a convex combination of pure states as follows:
	\begin{equation}
		\rho_{AB}=\sum_{x\in\mathcal{X}}p(x)\psi_{AB}^x,
	\end{equation}
	where $\mathcal{X}$ is a finite alphabet, $p:\mathcal{X}\to[0,1]$ is a probability distribution, and $\{\psi^x_{AB}\}_{x\in\mathcal{X}}$ is a set of pure states. We can then measure the entanglement of $\rho_{AB}$ by taking the expected entanglement of the pure states involved in the decomposition of $\rho_{AB}$, i.e., by $\sum_{x\in\mathcal{X}}p(x)H(\psi_A^x)$, where $\psi_A^x\coloneqq\Tr_B[\psi_{AB}^x]$. However, this strategy can lead to different values for the entanglement of $\rho_{AB}$, depending on the chosen decomposition, because the decomposition of mixed states into a convex combination of pure states is generally not unique. We can address this issue by minimizing over all possible decompositions, leading to the following definition:
	
	\begin{definition}{Entanglement of Formation}{def:E-meas:EoF}
The entanglement of formation of a bipartite state $\rho_{AB}$ is defined as
	\begin{equation}
		E_F(\rho_{AB})\coloneqq\inf_{\{(p(x),\psi_{AB}^x)\}_{x\in\mathcal{X}}}\left\{\sum_{x\in\mathcal{X}}p(x)H(\psi_A^x):\rho_{AB}=\sum_{x\in\mathcal{X}}p(x)\psi^x_{AB}\right\},
		\label{eq:E-meas:EoF-def}
	\end{equation}
	where $\rho_{AB}=\sum_{x\in\mathcal{X}}p(x)\psi^x_{AB}$ is a pure-state decomposition of $\rho_{AB}$.
	\end{definition}
	
	It suffices to take $|\mathcal{X}|\leq \dim(\mathcal{H}_{AB})^2$ in the optimization in \eqref{eq:E-meas:EoF-def}, and furthermore, the infimum is achieved by at least one pure-state decomposition. The fact that the alphabet $\mathcal{X}$ need not exceed $\dim(\mathcal{H}_{AB})^2$ elements is due to the entropy being a continuous function and the Fenchel--Eggleston--Carath\'{e}odory Theorem (Theorem~\ref{thm-Caratheodory}) and the fact that dimension of the space of density operators on a $\dim(\mathcal{H}_{AB})$-dimensional space is $\dim(\mathcal{H}_{AB})^2$. The fact that the infimum is achieved follows because the optimization is with respect to a compact space and the function being optimized is continuous.

	For every pure bipartite state $\psi_{AB}$, the equality in \eqref{eq-ent_formation_pure_state} holds because it is not possible to decompose a pure state $\psi_{AB}$ as a mixture of other pure states different from $\psi_{AB}$. As mentioned above, the entanglement of formation is also known as the \textit{entropy of entanglement} for the special case of a pure bipartite state, due to the fact that it is equal to the entropy of the reduced state.
	
		It is a direct consequence of Definition~\ref{def:E-meas:EoF} and the non-negativity of quantum entropy in \eqref{eq:QEI:entropy-non-neg} that the entanglement of formation is non-negative, i.e.,
	\begin{equation}
	E_F(\rho_{AB}) \geq 0
	\end{equation}
	for every bipartite state $\rho_{AB}$.
	
		Not only is the entanglement of formation non-negative, but it is also faithful; and we can  make an even more refined statement about approximate faithfulness (when $\rho_{AB}$ is close to separable or when $E_F(\rho_{AB})$ is close to zero).

		Before establishing this statement in Proposition~\ref{prop:E-meas:EoF-faithful} below, we first prove an asymptotic continuity bound for the entanglement of formation. \textit{Asymptotic continuity} is a desirable property of an entanglement measure: if two bipartite states $\rho_{AB}$ and $\sigma_{AB}$ are not very distinguishable from each other (according to some distinguishability measure), then the difference of their entanglement should be small. 	(See Definition~\ref{def-ent:asymp-cont} for a definition of asymptotic continuity.)
	
	\begin{proposition*}{Asymptotic Continuity of Entanglement of Formation}{prop:E-meas:unif-cont-EoF}
Let $\rho_{AB}$ and $\sigma_{AB}$ be states satisfying%
\begin{equation}
\frac{1}{2}\left\Vert \rho_{AB}-\sigma_{AB}\right\Vert _{1}\leq\varepsilon
,\label{eq:E-meas:unif-cont-EoF}%
\end{equation}
where $\varepsilon\in\left[  0,1\right]  $. Then the entanglement of
formations of $\rho_{AB}$ and $\sigma_{AB}$ satisfy%
\begin{equation}
\left\vert E_{F}(\rho_{AB})-E_{F}(\sigma_{AB})\right\vert \leq\delta\log
_{2}\min\left\{  d_{A},d_{B}\right\}  +g_{2}(\delta
),\label{eq:E-meas:unif-cont-EoF-statement}%
\end{equation}
where $\delta\coloneqq \sqrt{\varepsilon\left(  2-\varepsilon\right)  }$ and
$g_{2}(x)\coloneqq \left(  x+1\right)  \log_{2}(x+1)-x\log_{2}x$.
\end{proposition*}

\begin{Proof}
By applying Theorem~\ref{thm-Fuchs_van_de_graaf} to \eqref{eq:E-meas:unif-cont-EoF}, we find that%
\begin{equation}
\sqrt{F}(\rho_{AB},\sigma_{AB})\geq1-\varepsilon,
\end{equation}
which implies that%
\begin{equation}
\sqrt{1-F(\rho_{AB},\sigma_{AB})}\leq\sqrt{1-\left(  1-\varepsilon\right)
^{2}}=\sqrt{\varepsilon\left(  2-\varepsilon\right)  }%
.\label{eq:E-meas:cont-EoF-pf-1}%
\end{equation}
Let%
\begin{equation}
|\psi^{\rho}\rangle_{RAB}\coloneqq \sum_{x}\sqrt{p(x)}|x\rangle_{R}|\psi^{x}%
\rangle_{AB}.
\end{equation}
be a purification of $\rho_{AB}$, with $\rho_{AB}=\sum
_{x}p(x)\psi_{AB}^{x}$ a pure-state decomposition of $\rho_{AB}$. By
applying Uhlmann's theorem (Theorem~\ref{thm-Uhlmann_fidelity}), there exists a purification $\psi
_{RAB}^{\sigma}$ of $\sigma_{AB}$ such that $F(\psi_{RAB}^{\rho},\psi
_{RAB}^{\sigma})=F(\rho_{AB},\sigma_{AB})$. By combining this observation with
\eqref{eq:E-meas:cont-EoF-pf-1}, and the fact that the sine distance of two
pure states is equal to the normalized trace distance (see \eqref{eq-trace_dist_pure_states}), we conclude that%
\begin{equation}
\frac{1}{2}\left\Vert \psi_{RAB}^{\rho}-\psi_{RAB}^{\sigma}\right\Vert
_{1}\leq\sqrt{\varepsilon\left(  2-\varepsilon\right)  }.
\end{equation}
We now apply the measurement channel $\mathcal{M}_{R\rightarrow X}%
(\cdot)\coloneqq \sum_{x}|x\rangle_{X}\langle x|_{R}(\cdot)|x\rangle_{R}\langle
x|_{X}$ to the $R$ systems, as well as the data-processing inequality for the trace distance, to
conclude that%
\begin{equation}
\frac{1}{2}\left\Vert \mathcal{M}_{R\rightarrow X}(\psi_{RAB}^{\rho
})-\mathcal{M}_{R\rightarrow X}(\psi_{RAB}^{\sigma})\right\Vert _{1}\leq
\sqrt{\varepsilon\left(  2-\varepsilon\right)  }.
\end{equation}
Consider that%
\begin{equation}
\rho_{XAB}\coloneqq \mathcal{M}_{R\rightarrow X}(\psi_{RAB}^{\rho})=\sum
_{x}p(x)|x\rangle\!\langle x|_{X}\otimes\psi_{AB}^{x},
\end{equation}
and there exists a probability distribution $q(x)$ and a set $\{\varphi
_{AB}^{x}\}_{x}$, satisfying%
\begin{equation}
\sigma_{AB}=\sum_{x}q(x)\varphi_{AB}^{x}%
,\label{eq:E-meas:sigma-decomp-EoF-cont-pf}%
\end{equation}
such that%
\begin{equation}
\sigma_{XAB}\coloneqq \mathcal{M}_{R\rightarrow X}(\psi_{RAB}^{\sigma})=\sum
_{x}q(x)|x\rangle\!\langle x|_{X}\otimes\varphi_{AB}^{x}.
\end{equation}
Now applying the asymptotic continuity of conditional mutual information (Proposition~\ref{lem:LAQC-uniform-cont-CMI}), we conclude that%
\begin{equation}
\frac{1}{2}I(A;B|X)_{\sigma}\leq\frac{1}{2}I(A;B|X)_{\rho}+\delta\log_{2}%
\min\{d_{A},d_{B}\}+g_{2}(\delta),
\end{equation}
with $\delta=\sqrt{\varepsilon\left(  2-\varepsilon\right)  }$. Since the
states of systems $AB$ are pure when conditioned on the classical system $X$,
for both $\rho_{XAB}$ and $\sigma_{XAB}$, consider that%
\begin{equation}
\frac{1}{2}I(A;B|X)_{\sigma}=H(A|X)_{\sigma},\qquad\frac{1}{2}I(A;B|X)_{\rho
}=H(A|X)_{\rho}.\label{eq:E-meas:EoF-rewrite-CMI}%
\end{equation}
So we conclude that%
\begin{equation}
E_{F}(\sigma_{AB})\leq H(A|X)_{\sigma}\leq H(A|X)_{\rho}+\delta\log_{2}%
\min\{d_{A},d_{B}\}+g_{2}(\delta),
\end{equation}
where the first inequality follows from Definition~\ref{def:E-meas:EoF} and
\eqref{eq:E-meas:sigma-decomp-EoF-cont-pf}. Since the pure-state decomposition
of $\rho_{AB}$ is arbitrary, we conclude that%
\begin{equation}
E_{F}(\sigma_{AB})\leq E_{F}(\rho_{AB})+\delta\log_{2}\min\{d_{A}%
,d_{B}\}+g_{2}(\delta).\label{eq:E-meas:cont-EoF-last-step-pf}%
\end{equation}
Running the argument again, but starting from an arbitrary pure-state
decomposition of $\sigma_{AB}$, we conclude the inequality%
\begin{equation}
E_{F}(\rho_{AB})\leq E_{F}(\sigma_{AB})+\delta\log_{2}\min\{d_{A}%
,d_{B}\}+g_{2}(\delta),
\end{equation}
which, together with \eqref{eq:E-meas:cont-EoF-last-step-pf}, implies \eqref{eq:E-meas:unif-cont-EoF-statement}.
\end{Proof}

\begin{proposition*}{Faithfulness of Entanglement of Formation}{prop:E-meas:EoF-faithful}
The entanglement of formation is faithful, so that $E_F(\rho_{AB})=0$ if and only if the state $\rho_{AB}$ is separable. More quantitatively, for a state
$\rho_{AB}$ and $\varepsilon\in\left[  0,1\right]  $, if%
\begin{equation}
\inf_{\sigma_{AB}\in\operatorname{SEP}(A:B)}\frac{1}{2}\left\Vert \rho
_{AB}-\sigma_{AB}\right\Vert _{1}\leq\varepsilon
,\label{eq:E-meas:approx-faithful-EoF-1-1}%
\end{equation}
then%
\begin{equation}
E_{F}(\rho_{AB})\leq\delta\log_{2}\min\{d_{A},d_{B}\}+g_{2}(\delta
),\label{eq:E-meas:approx-faithful-EoF-1-2}%
\end{equation}
where $\delta\coloneqq \sqrt{\varepsilon\left(  2-\varepsilon\right)  }$. Conversely, for $\varepsilon\geq0$,
if%
\begin{equation}
E_{F}(\rho_{AB})\leq\varepsilon,\label{eq:E-meas:approx-faithful-EoF-2-1}
\end{equation}
 then%
\begin{equation}
\inf_{\sigma_{AB}\in\operatorname{SEP}(A:B)}\frac{1}{2}\left\Vert \rho
_{AB}-\sigma_{AB}\right\Vert _{1}\leq\sqrt{\varepsilon \ln2}%
.\label{eq:E-meas:approx-faithful-EoF-2-2}%
\end{equation}
\end{proposition*}

\begin{Proof}
We begin by proving the first statement. Suppose that the state $\sigma_{AB}$
is separable. Then by the remark after Definition~\ref{def-sep_ent_state}, there exists a pure-state decomposition of
$\sigma_{AB}$ as%
\begin{equation}
\sigma_{AB}=\sum_{x}p(x)\phi_{A}^{x}\otimes\varphi_{B}^{x}.
\end{equation}
For this decomposition, we have that $\sum_{x}p(x)H(\phi_{A}^{x})=0$ because
the quantum entropy is equal to zero for a pure state. This implies by
definition that $E_{F}(\sigma_{AB})=0$. The statement in
\eqref{eq:E-meas:approx-faithful-EoF-1-1}--\eqref{eq:E-meas:approx-faithful-EoF-1-2}
then follows by combining this observation with \eqref{eq:E-meas:unif-cont-EoF}--\eqref{eq:E-meas:unif-cont-EoF-statement}, as well as the fact that the function on the right-hand side of \eqref{eq:E-meas:approx-faithful-EoF-1-2} is monotone in $\varepsilon$.

To see the second statement, let%
\begin{equation}
\rho_{AB}=\sum_{x}p(x)\psi_{AB}^{x}%
\end{equation}
be an arbitrary pure-state decomposition of $\rho_{AB}$. By applying the
observation in \eqref{eq:E-meas:EoF-rewrite-CMI}, we find that%
\begin{align}
\sum_{x}p(x)H(\psi_{A}^{x})  & =\frac{1}{2}\sum_{x}p(x)I(A;B)_{\psi^{x}}\\
& =\frac{1}{2}\sum_{x}p(x)D(\psi_{AB}^{x}\Vert\psi_{A}^{x}\otimes\psi_{B}%
^{x})\\
& \geq\frac{1}{4\ln2}\sum_{x}p(x)\left\Vert \psi_{AB}^{x}-\psi_{A}^{x}%
\otimes\psi_{B}^{x}\right\Vert _{1}^{2}\\
& \geq\frac{1}{4\ln2}\left\Vert \sum_{x}p(x)\psi_{AB}^{x}-\sum_{x}p(x)\psi
_{A}^{x}\otimes\psi_{B}^{x}\right\Vert _{1}^{2}\\
& =\frac{1}{4\ln2}\left\Vert \rho_{AB}-\sum_{x}p(x)\psi_{A}^{x}\otimes\psi
_{B}^{x}\right\Vert _{1}^{2}\\
& \geq\frac{1}{\ln2}\left(  \inf_{\sigma_{AB}\in\operatorname{SEP}(A:B)}%
\frac{1}{2}\left\Vert \rho_{AB}-\sigma_{AB}\right\Vert _{1}\right)  ^{2}.
\end{align}
The second equality follows from rewriting the mutual information in terms of
relative entropy (see \eqref{eq-mut_inf}). The first inequality follows from the quantum Pinsker
inequality (Corollary~\ref{cor:QEI:pinsker} and the remark thereafter). The second inequality follows from convexity of the square
function and the trace norm. Since the inequality holds for an arbitrary
pure-state decomposition of $\rho_{AB}$, we conclude that%
\begin{equation}
E_{F}(\rho_{AB})\geq\frac{1}{\ln2}\left(  \inf_{\sigma_{AB}\in
\operatorname{SEP}(A:B)}\frac{1}{2}\left\Vert \rho_{AB}-\sigma_{AB}\right\Vert
_{1}\right)  ^{2}.
\end{equation}
From this and \eqref{eq:E-meas:approx-faithful-EoF-2-1}, we conclude \eqref{eq:E-meas:approx-faithful-EoF-2-2}.

Finally, to conclude exact faithfulness from approximate faithfulness, we
argue that the infimum in \eqref{eq:E-meas:approx-faithful-EoF-1-1} and
\eqref{eq:E-meas:approx-faithful-EoF-2-2} is achieved (i.e., can be replaced
with a minimum). This follows because the trace distance is continuous in
$\sigma_{AB}$ and the set of separable states is compact.
\end{Proof}
	
	As the following proposition indicates, the entanglement of formation is an entanglement measure according to Definition~\ref{def-LAQC:ent-measure}. The proof detailed below is also a good opportunity to apply Lemma~\ref{lem:E-meas:convex-strong-LOCC-mono} to a simple example.
	
\begin{proposition}{prop:E-meas:EoF-convex-str-LOCC-mono}	
The entanglement of formation is convex, so that \eqref{eq:E-meas:convexity-e-meas} holds with $E$ set to $E_F$, and it is a selective LOCC monotone, so that \eqref{eq:E-meas:strong-LOCC-mono} holds with $E$ set to $E_F$.
\end{proposition}

\begin{Proof}
By Lemma~\ref{lem:E-meas:convex-strong-LOCC-mono}, we only need to show that the entanglement of formation does not
increase under the action of a local channel and is invariant under classical
communication. We begin with the first one. Let $\rho_{AB}$ be a bipartite
state, and let $\mathcal{N}_{A\rightarrow A^{\prime}}$ be a local quantum
channel. Let $\{(p(x),\psi_{AB}^{x})\}_{x}$ be a pure-state decomposition of
$\rho_{AB}$, i.e., satisfying $\sum_{x}p(x)\psi_{AB}^{x}=\rho_{AB}$. Let
$\mathcal{N}_{A\rightarrow A^{\prime}}$ have the Kraus representation
$\{N_{A\rightarrow A^{\prime}}^{y}\}_{y}$. Then%
\begin{equation}
\omega_{A^{\prime}B}   \coloneqq \mathcal{N}_{A\rightarrow A^{\prime}}(\rho_{AB})
 =\sum_{x}p(x)\mathcal{N}_{A\rightarrow A^{\prime}}(\psi_{AB}^{x}),
\end{equation}
and%
\begin{equation}
\mathcal{N}_{A\rightarrow A^{\prime}}(\psi_{AB}^{x})   =\sum_{y}%
N_{A\rightarrow A^{\prime}}^{y}\psi_{AB}^{x}(N_{A\rightarrow A^{\prime}}%
^{y})^{\dag}
 =\sum_{y}p(y|x)\varphi_{A^{\prime}B}^{x,y},
\end{equation}
where%
\begin{align}
p(y|x)  & \coloneqq \operatorname{Tr}[N_{A\rightarrow A^{\prime}}^{y}\psi_{AB}%
^{x}(N_{A\rightarrow A^{\prime}}^{y})^{\dag}],\\
\varphi_{A^{\prime}B}^{x,y}  & \coloneqq \frac{1}{p(y|x)}N_{A\rightarrow A^{\prime}%
}^{y}\psi_{AB}^{x}(N_{A\rightarrow A^{\prime}}^{y})^{\dag}.
\end{align}
Thus, $\{(p(x)p(y|x),\varphi_{A^{\prime}B}^{x,y})\}_{x,y}$ is a pure-state
decomposition of $\omega_{A^{\prime}B}$. Also, observe that%
\begin{align}
\psi_{B}^{x}  & =\operatorname{Tr}_{A}[\psi_{AB}^{x}]\\
& =\operatorname{Tr}_{A^{\prime}}[\mathcal{N}_{A\rightarrow A^{\prime}}%
(\psi_{AB}^{x})]\\
& =\sum_{y}p(y|x)\operatorname{Tr}_{A^{\prime}}[\varphi_{A^{\prime}B}%
^{x,y}]\\
& =\sum_{y}p(y|x)\varphi_{B}^{x,y}.
\end{align}
Then we have that%
\begin{align}
\sum_{x}p(x)H(\psi_{B}^{x})  & \geq\sum_{x,y}p(x)p(y|x)H(\varphi_{B}^{x,y})\\
& \geq E_{F}(A^{\prime};B)_{\omega},
\end{align}
where the first inequality follows from the concavity of entropy (see \eqref{eq-q_entropy_concave}) and the
second from the definition of entanglement of formation. Since the inequality
holds for all pure-state decompositions of $\rho_{AB}$, we conclude the
desired inequality:%
\begin{equation}
E_{F}(A;B)_{\rho}\geq E_{F}(A^{\prime};B)_{\omega}.
\end{equation}
By flipping the role of Alice and Bob in the analysis above, we conclude that
the entanglement of formation does not increase under the action of a local
channel on Bob's system.

Now we prove that $E_{F}$ is invariant under classical communication. Let
$\rho_{XAB}$ be the classical--quantum state defined in \eqref{eq:E-meas:cq-state-inv-CC}. A pure-state
decomposition of $\rho_{XAB}$ has the form $\{(p(x)p(y|x),|x\rangle\!\langle
x|_{X}\otimes\psi_{AB}^{x,y})\}_{x,y}$, where
\begin{equation}
\rho_{AB}^{x}=\sum_{y}p(y|x)\psi_{AB}^{x,y}.
\end{equation}
This ensemble serves as a decomposition of $\rho_{XAB}$ for $E_{F}%
(XA;B)_{\rho}$. Then%
\begin{equation}
\sum_{x,y}p(x)p(y|x)H(\psi_{B}^{x,y})\geq\sum_{x}p(x)E_{F}(A;B)_{\rho^{x}}.
\end{equation}
Since the inequality holds for all pure-state decompositions of $\rho_{XAB}$,
we conclude that%
\begin{equation}
E_{F}(XA;B)_{\rho}\geq\sum_{x}p(x)E_{F}(A;B)_{\rho^{x}}%
.\label{eq:E-meas:EoF-inv-class-comm-1}%
\end{equation}
Now let $\{(p(y|x),\psi_{AB}^{x,y})\}_{y}$ be a pure-state decomposition of
$\rho_{AB}^{x}$. Then we find that%
\begin{equation}
\sum_{x,y}p(x)p(y|x)H(\psi_{B}^{x,y})\geq E_{F}(XA;B)_{\rho}%
\end{equation}
because the ensemble $\{(p(x)p(y|x),|x\rangle\!\langle x|_{X}\otimes\psi
_{AB}^{x,y})\}_{x,y}$ is a particular pure-state decomposition of $\rho_{XAB}%
$. Since the inequality holds for all pure-state decompositions of $\rho
_{AB}^{x}$, we conclude that%
\begin{equation}
\sum_{x}p(x)E_{F}(A;B)_{\rho^{x}}\geq E_{F}(XA;B)_{\rho}%
.\label{eq:E-meas:EoF-inv-class-comm-2}%
\end{equation}
Putting together \eqref{eq:E-meas:EoF-inv-class-comm-1} and
\eqref{eq:E-meas:EoF-inv-class-comm-2}, we conclude that%
\begin{equation}
\sum_{x}p(x)E_{F}(A;B)_{\rho^{x}}=E_{F}(XA;B)_{\rho}.
\end{equation}
By the same argument, but exchanging the roles of Alice and Bob, we conclude
that%
\begin{equation}
\sum_{x}p(x)E_{F}(A;B)_{\rho^{x}}=E_{F}(A;BX)_{\rho}.
\end{equation}
This concludes the proof.
\end{Proof}

\begin{proposition*}{Subadditivity of Entanglement of Formation}{prop:E-meas:subadditivity-EoF}
The entanglement of formation is subadditive; i.e., \eqref{def-ent_meas_subadditive} holds with $E$ set to $E_F$.
\end{proposition*}

\begin{Proof}
Let $\sum_x p(x) \psi^x_{A_1B_1}$ and $\sum_y q(y) \phi^y_{A_2B_2}$ be respective pure-state decompositions of $\tau_{A_1B_1}$ and  $\omega_{A_2B_2}$. Then $\sum_{x,y} p(x)q(y) \psi^x_{A_1B_1} \otimes \phi^y_{A_2B_2}$ is a pure-state decomposition of $\tau_{A_1B_1}\otimes \omega_{A_2B_2}$. It follows that 
\begin{align}
E_F(A_1A_2;B_1B_2)_{\tau \otimes \omega}
& \leq \sum_{x,y}p(x)q(y) H(\psi^x_{A_1} \otimes \phi^y_{A_2}) \\
& = \sum_{x,y}p(x)q(y) [H(\psi^x_{A_1}) +H(\phi^y_{A_2})] \\
& = \sum_{x}p(x) H(\psi^x_{A_1}) + \sum_{y}q(y) H(\phi^y_{A_2}).
\end{align}
Since the inequality holds for arbitrary pure-state decompositions of $\tau_{A_1B_1}$ and $\omega_{A_2B_2}$, we conclude that subadditivity holds.
\end{Proof}

The opposite inequality, superadditivity of entanglement of formation, is known not to hold in general. Thus, the entanglement of formation is non-additive in general. The proof of this statement is highly nontrivial (please consult the Bibliographic Notes in Section~\ref{sec:E-meas:bib-notes}). 

The entanglement of formation is connected to an information-theoretic task: $E_F(\rho_{AB})$ is an achievable rate for the task of preparing the state $\rho_{AB}$ from many copies of the two-qubit maximally entangled state $\ket{\Phi}_{AB}$ when allowing LOCC for free (please consult the Bibliographic Notes in Section~\ref{sec:E-meas:bib-notes}).
	
	For two-qubit states $\rho_{AB}$, the entanglement of formation has the following analytic expression:
	\begin{equation}\label{eq-ent_formation_two_qubits}
		E_F(\rho_{AB})=h_2\!\left(\frac{1+\sqrt{1-C(\rho_{AB})^2}}{2}\right)\quad \text{(two-qubit states)}.
	\end{equation}
	(please consult the Bibliographic Notes in Section~\ref{sec:E-meas:bib-notes}.) Here,
	\begin{equation}
		C(\rho_{AB})=\max\{0,\lambda_1-\lambda_2-\lambda_3-\lambda_4\},
	\end{equation}
	where $\lambda_1,\lambda_2,\lambda_3,\lambda_4$ are the eigenvalues of $\sqrt{\sqrt{\rho_{AB}}\, \widetilde{\rho}_{AB}\sqrt{\rho_{AB}}}$  in decreasing order. The operator $\widetilde{\rho}_{AB}$ is defined as $\widetilde{\rho}_{AB}\coloneqq (Y\otimes Y)\conj{\rho}_{AB}(Y\otimes Y)$, with $Y$ being the Pauli-$Y$ operator (see \eqref{eq-Pauli_operators}) and $\conj{\rho}_{AB}$ being the complex conjugate of $\rho_{AB}$ in the standard basis.

\subsubsection{Negativity and Logarithmic Negativity}
	
	Similar to how we motivated the entanglement of formation from the Schmidt rank criterion for pure states, we can also motivate an entanglement measure from the PPT criterion. The PPT criterion states that if the partial transpose $\T_B(\rho_{AB})$ of a given state $\rho_{AB}$ has a negative eigenvalue, then $\rho_{AB}$ is entangled.\footnote{Note that it does not matter in which basis the transpose is defined.} We use this fact to define the \textit{negativity} of $\rho_{AB}$ as
	\begin{equation}
\label{eq:E-meas:def-negativity}
		N(\rho_{AB})\coloneqq\frac{\norm{\T_B(\rho_{AB})}_1-1}{2},
	\end{equation}
	and the \textit{logarithmic negativity} (often written simply as \textit{log-negativity}) of $\rho_{AB}$ as
	\begin{equation}
		E_N(\rho_{AB})\coloneqq \log_2\norm{\T_B(\rho_{AB})}_1.
		\label{eq:E-meas:def-log-negativity}
	\end{equation}
	
	Both the negativity and the log-negativity quantify the extent to which the partial transpose $\T_B(\rho_{AB})$ has negative eigenvalues. In particular, suppose that $\T_B(\rho_{AB})$ has the following Jordan--Hahn decomposition:
	\begin{equation}
		\T_B(\rho_{AB})=P-N,
	\end{equation}
	where $P$ and $N$ are the positive and negative parts of $\T_B(\rho_{AB})$, satisfying $P,N\geq 0$ and $PN=0$, and we have used \eqref{eq-operator_pos_neg_decomp} and \eqref{eq-MT:Jordan-Hahn}. By definition of the trace norm,
	\begin{equation}
		\norm{\T_B(\rho_{AB})}_1=\Tr[P+N].
		\label{eq:E-meas:Jordan-Hahn-pos-sum-neg}
	\end{equation}
	On the other hand, observe that $\Tr[\T_B(\rho_{AB})]=\Tr[\rho_{AB}]=1$, so that
	\begin{equation}
		1=\Tr[\T_B(\rho_{AB})]=\Tr[P-N].
		\label{eq:E-meas:Jordan-Hahn-pos-sum-neg-other}
	\end{equation}
	Therefore,
	\begin{equation}
	N(\rho_{AB})=\frac{\norm{\T_B(\rho_{AB})}_1-1}{2}=\frac{\norm{\T_B(\rho_{AB})}_1-\Tr[\T_B(\rho_{AB})]}{2}=\Tr[N].
	\end{equation}
	So, according to \eqref{eq-MT:Jordan-Hahn}, the negativity is the sum of the absolute values of the negative eigenvalues of $\rho_{AB}^{\t_B}$.
	
	By utilizing H\"older duality and semi-definite programming duality, it is possible to write $\norm{\T_B(\rho_{AB})}_1$ as the
following primal and dual semi-definite programs:
\begin{align}
\!\! \norm{\T_B(\rho_{AB})}_1 & =\sup_{R_{AB}}\left\{  \operatorname{Tr}[R_{AB}\rho_{AB}%
]:-\mathbbm{1}_{AB}\leq \T_{B}(R_{AB})\leq \mathbbm{1}_{AB}\right\}  ,
\label{eq:E-meas:SDP-primal-log-neg}
\\
& =\inf_{K_{AB},L_{AB}\geq0}\left\{  \operatorname{Tr}%
[K_{AB}+L_{AB}]: \T_B(K_{AB}-L_{AB})=\rho_{AB}\right\}
.\label{eq:E-meas:SDP-dual-log-neg}
\end{align}
where the optimization in the first line is with respect to Hermitian $R_{AB}$. We give a proof of \eqref{eq:E-meas:SDP-dual-log-neg}--\eqref{eq:E-meas:SDP-primal-log-neg} in Appendix~\ref{app:E-meas:SDP-negativity}.

\begin{proposition}{prop:E-meas}
The log-negativity is non-negative for all bipartite states, and it is faithful on the set of PPT states (i.e., it is equal to zero if and only if a state is~PPT). 
\end{proposition}

\begin{Proof}
To see the first statement, we note that the choice $R_{AB} = \mathbbm{1}_{AB}$ is feasible for the primal SDP in \eqref{eq:E-meas:SDP-primal-log-neg}, so that $\norm{\T_B(\rho_{AB})}_1 \geq 1$, and hence $E_N(\rho_{AB}) \geq 0$, for every bipartite state $\rho_{AB}$.

Suppose that $\rho_{AB}$ is a PPT state. Then $\norm{\T_B(\rho_{AB})}_1=\Tr[\T_B(\rho_{AB})] = \Tr[\rho_{AB}]=  1$ due to the assumption that $\T_B(\rho_{AB})\geq 0$, implying that $E_N(\rho_{AB}) = 0$ for every PPT state.

Finally, suppose that $E_N(\rho_{AB}) = 0$. Then $\norm{\T_B(\rho_{AB})}_1 = 1$, and employing the notation of \eqref{eq:E-meas:Jordan-Hahn-pos-sum-neg}--\eqref{eq:E-meas:Jordan-Hahn-pos-sum-neg-other}, we conclude that $1=\Tr[P+N]=\Tr[P-N]$, which implies that $\Tr[N]=0$. Since $N \geq 0$, this implies that $N=0$. Thus, $\T_B(\rho_{AB})$ has no negative part and $\rho_{AB}$ is thus a PPT state. 
\end{Proof}

\begin{definition}{Selective PPT Monotonicity}{def:E-meas:strong-ppt-mono}
As a generalization of selective LOCC monotonicity defined in \eqref{eq:E-meas:strong-LOCC-mono}, we say that a
function $E:\Density(\mathcal{H}_{AB})\rightarrow\mathbb{R}$\ is a selective PPT
monotone if it satisfies%
\begin{equation}
E(\rho_{AB})\geq\sum_{x\in\mathcal{X}:p(x)>0}p(x)E(\omega_{A^{\prime}%
B^{\prime}}^{x}),\label{eq:E-meas:strong-PPT-monotone}%
\end{equation}
for every bipartite state $\rho_{AB}$ and C-PPT-P instrument $\{\mathcal{P}%
_{AB\rightarrow A^{\prime}B^{\prime}}^{x}\}_{x\in\mathcal{X}}$, with%
\begin{align}
p(x)  & \coloneqq \operatorname{Tr}[\mathcal{P}_{AB\rightarrow A^{\prime}B^{\prime}%
}^{x}(\rho_{AB})],\label{eq:E-meas:post-PPT-ensemble-1}\\
\omega_{A^{\prime}B^{\prime}}^{x}  & \coloneqq \frac{1}{p(x)}\mathcal{P}%
_{AB\rightarrow A^{\prime}B^{\prime}}^{x}(\rho_{AB}%
).\label{eq:E-meas:post-PPT-ensemble-2}%
\end{align}
A C-PPT-P instrument is such that every map $\mathcal{P}_{AB\rightarrow
A^{\prime}B^{\prime}}^{x}$ is completely positive and $\T_{B^{\prime}}%
\circ\mathcal{P}_{AB\rightarrow A^{\prime}B^{\prime}}^{x}\circ \T_{B}$ is
completely positive, and the sum map $\sum_{x\in\mathcal{X}}\mathcal{P}_{AB\rightarrow
A^{\prime}B^{\prime}}^{x}$ is trace preserving.

It follows that $E$ is a PPT monotone if it is a selective PPT monotone, because the former is a special case of the latter in which the alphabet $\mathcal{X}$ has only one letter.
\end{definition}

An LOCC instrument in \eqref{eq:E-meas:LOCC-instrument}
is a C-PPT-P instrument because every map $\mathcal{L}_{AB\rightarrow
A^{\prime}B^{\prime}}^{x}$ in an LOCC instrument satisfies the requirements
for a C-PPT-P instrument.

The negativity and the log-negativity are entanglement measures, as shown in Proposition~\ref{prop:E-meas:neg-log-neg-E-mono} below. Interestingly, the method of proof does not involve making use of Lemma~\ref{lem:E-meas:convex-strong-LOCC-mono} because it is impossible to do so for the log-negativity, as the latter is not convex. In any case, we prove a stronger result than selective LOCC monotonicity for the log-negativity: we prove that it is a selective PPT monotone. 

\begin{proposition}{prop:E-meas:neg-log-neg-E-mono}
The negativity and the log-negativity are selective PPT monotones, satisfying
\eqref{eq:E-meas:strong-PPT-monotone}. The negativity is convex, satisfying
\eqref{eq:E-meas:convexity-e-meas}, but the log-negativity is not.
\end{proposition}

\begin{Proof}
Define $\rho_{AB}$ and $\{(p(x),\omega_{A^{\prime}B^{\prime}}^{x}%
)\}_{x\in\mathcal{X}}$ as in
\eqref{eq:E-meas:post-PPT-ensemble-1}--\eqref{eq:E-meas:post-PPT-ensemble-2},
and let $\{\mathcal{P}_{AB\rightarrow A^{\prime}B^{\prime}}^{x}\}_{x\in
\mathcal{X}}$ be a C-PPT-P instrument. Let $K_{AB}$ and $L_{AB}$ be arbitrary
positive semi-definite operators satisfying%
\begin{equation}
\T_{B}(K_{AB}-L_{AB})=\rho_{AB}.\label{eq:E-meas:neg-K-L-constr}%
\end{equation}
Then we find that%
\begin{align}
p(x)\omega_{A^{\prime}B^{\prime}}^{x}  & =\mathcal{P}_{AB\rightarrow
A^{\prime}B^{\prime}}^{x}(\rho_{AB})\\
& =\mathcal{P}_{AB\rightarrow A^{\prime}B^{\prime}}^{x}(\T_{B}(K_{AB}%
-L_{AB}))\\
& =\T_{B^{\prime}}(\T_{B^{\prime}}\circ\mathcal{P}_{AB\rightarrow A^{\prime
}B^{\prime}}^{x}\circ \T_{B})(K_{AB}-L_{AB})).
\end{align}
Let us define%
\begin{align}
K_{A^{\prime}B^{\prime}}^{x}  & \coloneqq \frac{1}{p(x)}( \T_{B^{\prime}}\circ
\mathcal{P}_{AB\rightarrow A^{\prime}B^{\prime}}^{x}\circ \T_{B})(K_{AB}%
),\label{eq:E-meas:defs-ensemble-ops-neg-mono-pf-1}\\
L_{A^{\prime}B^{\prime}}^{x}  & \coloneqq \frac{1}{p(x)}( \T_{B^{\prime}}\circ
\mathcal{P}_{AB\rightarrow A^{\prime}B^{\prime}}^{x}\circ \T_{B})(L_{AB}%
),\label{eq:E-meas:defs-ensemble-ops-neg-mono-pf-2}%
\end{align}
so that%
\begin{equation}
\omega_{A^{\prime}B^{\prime}}^{x}=\T_{B^{\prime}}(K_{A^{\prime}B^{\prime}}%
^{x}-L_{A^{\prime}B^{\prime}}^{x}).
\label{eq:E-meas:eq-constr-omega-x-log-neg-mono}
\end{equation}
Furthermore, since $ \T_{B^{\prime}}\circ\mathcal{P}_{AB\rightarrow A^{\prime
}B^{\prime}}^{x}\circ \T_{B}$ is completely positive, it follows that
$K_{A^{\prime}B^{\prime}}^{x},L_{A^{\prime}B^{\prime}}^{x}\geq0$. Thus,
$K_{A^{\prime}B^{\prime}}^{x}$ and $L_{A^{\prime}B^{\prime}}^{x}$ are feasible
for the SDP\ in \eqref{eq:E-meas:SDP-dual-log-neg}\ for $\norm{\T_B(\omega_{A^{\prime
}B^{\prime}}^{x})}_1$, and we conclude that%
\begin{equation}
\operatorname{Tr}[K_{A^{\prime}B^{\prime}}^{x}+L_{A^{\prime}B^{\prime}}%
^{x}]\geq \norm{\T_B(\omega_{A^{\prime
}B^{\prime}}^{x})}_1%
.\label{eq:E-meas:neg-lower-feasible-SDP}%
\end{equation}
Then consider that%
\begin{align}
& \operatorname{Tr}[K_{AB}+L_{AB}]\nonumber\\
& =\operatorname{Tr}[\T_{B}(K_{AB}+L_{AB})]\\
& =\sum_{x\in\mathcal{X}:p(x)>0}\operatorname{Tr}[\mathcal{P}_{AB\rightarrow
A^{\prime}B^{\prime}}^{x}( \T_{B}(K_{AB}+L_{AB}))]\\
& =\sum_{x\in\mathcal{X}:p(x)>0}\operatorname{Tr}[( \T_{B^{\prime}}%
\circ\mathcal{P}_{AB\rightarrow A^{\prime}B^{\prime}}^{x}\circ \T_{B}%
)(K_{AB}+L_{AB})]\\
& =\sum_{x\in\mathcal{X}:p(x)>0}p(x)\operatorname{Tr}[K_{A^{\prime}B^{\prime}%
}^{x}+L_{A^{\prime}B^{\prime}}^{x}]\\
& \geq\sum_{x\in\mathcal{X}:p(x)>0}p(x)\norm{\T_B(\omega_{A^{\prime
}B^{\prime}}^{x})}_1.
\end{align}
The first and third equalities hold because the trace is invariant under a
partial transpose. The second equality follows because the sum map $\sum
_{x}\mathcal{P}_{AB\rightarrow A^{\prime}B^{\prime}}^{x}$ is trace preserving.
The fourth equality follows from the definitions in
\eqref{eq:E-meas:defs-ensemble-ops-neg-mono-pf-1}--\eqref{eq:E-meas:defs-ensemble-ops-neg-mono-pf-2}.
The inequality follows from \eqref{eq:E-meas:neg-lower-feasible-SDP}. Since
the inequality holds for all $K_{AB},L_{AB}\geq0$ satisfying
\eqref{eq:E-meas:neg-K-L-constr}, we conclude that%
\begin{equation}
\norm{\T_B(\rho_{AB})}_1 \geq\sum_{x\in\mathcal{X}:p(x)>0}p(x)\norm{\T_B(\omega_{A^{\prime
}B^{\prime}}^{x})}_1.
\label{eq:E-meas:log-neg-PPT-mono-key-ineq}
\end{equation}

By applying the definition in \eqref{eq:E-meas:def-negativity}, we conclude that the negativity is a selective PPT monotone.
Now considering \eqref{eq:E-meas:log-neg-PPT-mono-key-ineq} and taking the logarithm, and using its monotonicity and concavity, we
conclude that the log-negativity is a selective PPT monotone:
\begin{align}
E_{N}(\rho_{AB})  & =\log_{2}\norm{\T_B(\rho_{AB})}_1\\
& \geq\log_{2}\!\left[  \sum_{x\in\mathcal{X}:p(x)>0}p(x)\norm{\T_B(\omega_{A^{\prime
}B^{\prime}}^{x})}_1\right]  \\
& \geq\sum_{x\in\mathcal{X}:p(x)>0}p(x)\log_{2}\norm{\T_B(\omega_{A^{\prime
}B^{\prime}}^{x})}_1\\
& =\sum_{x\in\mathcal{X}:p(x)>0}p(x)E_{N}(\omega_{A^{\prime}B^{\prime}}^{x}).
\end{align}

That the negativity is convex follows directly from the definition, convexity of the trace
norm, and linearity of the partial transpose.

The lack of convexity of log-negativity follows from direct evaluation for the
states $\Phi_{AB}\coloneqq \frac{1}{2}\sum_{i,j\in\left\{  0,1\right\}  }%
|ii\rangle\!\langle jj|_{AB}$, $\sigma_{AB}\coloneqq \frac{1}{2}\sum_{i\in\left\{
0,1\right\}  }|ii\rangle\!\langle ii|_{AB}$, and $\overline{\rho}_{AB}\coloneqq \frac
{1}{2}\left(  \Phi_{AB}+\sigma_{AB}\right)  $, for which we have%
\begin{equation}
E_{N}(\Phi_{AB})=1,\qquad E_{N}(\sigma_{AB})=0,\qquad E_{N}(\overline{\rho
}_{AB})=\log_{2}\frac{3}{2},
\end{equation}
so that%
\begin{equation}
E_{N}(\overline{\rho}_{AB})>\frac{1}{2}\left(  E_{N}(\Phi_{AB})+E_{N}%
(\sigma_{AB})\right)  .
\end{equation}
This concludes the proof.
\end{Proof}
	
	\begin{proposition*}{Additivity of Log-Negativity}{prop:E-meas:additivity-log-neg}
	The logarithmic negativity is additive; i.e.,  \eqref{eq:E-meas:ent-meas-additive} holds with $E$ set to $E_N$.
	\end{proposition*}
	
	\begin{Proof}
For every two states $\tau_{A_1B_1}$ and $\omega_{A_2B_2}$, consider that
	\begin{align}
		E_N(A_1A_2;B_1B_2)_{\tau\otimes\omega}&=\log_2\norm{\T_{B_1 B_2}(\tau_{A_1B_1}\otimes\omega_{A_2B_2})}_1\label{eq-ent_meas_log_neg_additive_1}\\
&=\log_2\norm{\T_{B_1 }(\tau_{A_1B_1})\otimes\T_{B_2}(\omega_{A_2B_2})}_1\\	
		&=\log_2\!\left(\norm{\T_{B_1 }(\tau_{A_1B_1})}_1\cdot\norm{\T_{B_2}(\omega_{A_2B_2})}_1\right)\label{eq-ent_meas_log_neg_additive_2}\\
		&=\log_2\norm{\T_{B_1}(\tau_{A_1B_1})}_1+\log_2\norm{\T_{B_2}(\omega_{A_2B_2})}_1\label{eq-ent_meas_log_neg_additive_3}\\
		&=E_N(A_1;B_1)_{\tau}+E_N(A_2;B_2)_{\omega}.\label{eq-ent_meas_log_neg_additive_4}
	\end{align}
	This concludes the proof.
	\end{Proof}
	
	\begin{proposition*}{Log-Negativity of Pure Bipartite States}{prop:E-meas:log-neg-pure-states}
	For a pure bipartite state $\psi_{AB}$, the log-negativity is equal to the R\'enyi entropy of order $\frac{1}{2}$ of the reduced state~$\psi_{A}$:
	\begin{equation}
	E_N(\psi_{AB}) = H_{\frac{1}{2}}(\psi_A).
	\end{equation}
	\end{proposition*}
	
	\begin{Proof}
	For every pure state $\psi_{AB}=\ket{\psi}\!\bra{\psi}_{AB}$ such that
	\begin{equation}
	\ket{\psi}_{AB}=\sum_{k=1}^r\sqrt{\lambda_k}\ket{e_k}_A\otimes\ket{f_k}_B
	\end{equation}
	is a Schmidt decomposition, we have that
	\begin{equation}
		\ket{\psi}\!\bra{\psi}_{AB}^{\t_B}=\sum_{k,k'=1}^r\sqrt{\lambda_k\lambda_{k'}}\ket{e_k}\!\bra{e_{k'}}_A\otimes\ket{f_{k'}}\!\bra{f_k}_B,
	\end{equation}
	where we have taken the partial transpose with respect to the orthonormal set $\{\ket{f_k}_B\}_{k=1}^r$. Observe that
	\begin{equation}
		\ket{\psi}\!\bra{\psi}_{AB}^{\t_B}=F_{AB}\left(\sum_{k'=1}^r\sqrt{\lambda_{k'}}\ket{e_{k'}}\!\bra{e_{k'}}_A\otimes\sum_{k=1}^r\sqrt{\lambda_k}\ket{f_k}\!\bra{f_k}_B\right),
	\end{equation}
	where $F_{AB}=\sum_{k,k'=1}^r\ket{e_{k'}}\!\bra{e_k}_A\otimes\ket{f_k}\!\bra{f_{k'}}_B$ is a unitary swap operator. Thus, by unitary invariance of the trace norm, we obtain
	\begin{equation}
		E_N(\psi_{AB})=\log_2\!\left(\sum_{k=1}^r\sqrt{\lambda_k}\right)^2=2\log_2\!\left(\sum_{k=1}^r\sqrt{\lambda_k}\right).
	\end{equation}
	By comparing with \eqref{eq:QEI:Renyi-entropy}, we conclude the statement of the proposition.
		\end{Proof}
		
	For a maximally entangled state ($\lambda_k=\frac{1}{r}$ for all $1\leq k\leq r$), we find that $E_N(\psi_{AB})=\log_2 r$, exactly as with the entanglement of formation.

\subsubsection{Divergence-Based Measures}
	
\begin{figure}
		\centering
		\includegraphics[scale=0.9]{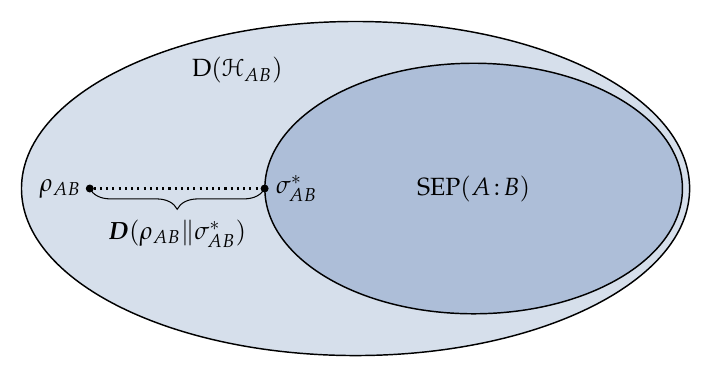}
		\caption{A simple way to measure the entanglement of a bipartite state $\rho_{AB}$ is to calculate its divergence with the set of separable states. If we use a generalized divergence $\boldsymbol{D}$ as our measure, then the measure of the entanglement in $\rho_{AB}$ is given by the smallest value of $\boldsymbol{D}(\rho_{AB}\Vert\sigma_{AB})$, where $\sigma_{AB}\in\SEP(A\!:\!B)$ is a separable state, i.e., by the quantity $\boldsymbol{D}(\rho_{AB}\Vert\sigma_{AB}^*)=\inf_{\sigma_{AB}\in\SEP(A:B)}\boldsymbol{D}(\rho_{AB}\Vert\sigma_{AB})$.}\label{fig-dist_based_ent_measure}
	\end{figure}

	The two entanglement measures considered above are based on specific mathematical properties of entanglement. However, using the fact that entangled states are, by definition, not separable, we can construct a broad class of entanglement measures by finding the divergence of a given state $\rho_{AB}$ with the set of separable states. This idea is illustrated in Figure~\ref{fig-dist_based_ent_measure}. We primarily consider such \textit{divergence-based} entanglement measures in this book (in the research literature, these are also called ``distance-based'' entanglement measures, even though divergences that are not distances, such as relative entropy, are used in this approach).
	
	As an example of a divergence-based entanglement measure, let us consider a concrete divergence, the normalized trace distance, which we defined in Section~\ref{sec-QM-trace-distance} as $\frac{1}{2}\norm{\rho-\sigma}_1$ for every two states $\rho$ and $\sigma$. Mathematically, the distance of a point to a set is defined by finding the element of that set that is closest to the given point. With this idea, we define the \textit{trace distance of entanglement} of a state $\rho_{AB}$ as the normalized trace distance from $\rho_{AB}$ to the closest state $\sigma_{AB}\in\SEP(A\!:\!B)$:
	\begin{equation}\label{eq-trace_dist_ent}
		E_T(A;B)_{\rho}\coloneqq \inf_{\sigma_{AB}\in\SEP(A:B)}\frac{1}{2}\norm{\rho_{AB}-\sigma_{AB}}_1.
	\end{equation}
	Note that the infimum is indeed achieved, because $\SEP(A\!:\!B)$ is a compact set and the trace norm is continuous in $\sigma_{AB}$, so that there always exists a closest separable state to the given state $\rho_{AB}$. Recall that we implicitly introduced the trace distance of entanglement in Proposition~\ref{prop:E-meas:EoF-faithful}, when considering approximate faithfulness of the entanglement of formation.
	
	The quantity $E_T$ is indeed an entanglement measure. To see this, we use the data-processing inequality for the trace distance (Theorem~\ref{thm-trace_dist_monotone}), and the fact that separable states are preserved under LOCC channels (which follows immediately from the definition of LOCC channels). Then, for every state $\rho_{AB}$, every LOCC channel $\mathcal{L}_{AB\to A'B'}$, and letting $\omega_{A'B'}=\mathcal{L}_{AB\to A'B'}(\rho_{AB})$, we obtain
	\begin{align}
	E_T(A;B)_{\rho} & = \inf_{\sigma_{AB}\in\SEP(A:B)}\frac{1}{2}\norm{\rho_{AB}-\sigma_{AB}}_1 \\ 
	& \geq \inf_{\sigma_{AB}\in\SEP(A:B)}\frac{1}{2}\norm{\mathcal{L}_{AB\to A'B'}(\rho_{AB})-\mathcal{L}_{AB\to A'B'}(\sigma_{AB})}_1 \\
	& \geq \inf_{\tau_{A'B'}\in\SEP(A':B')}\frac{1}{2}\norm{\mathcal{L}_{AB\to A'B'}(\rho_{AB})-\tau_{A'B'}}_1\\
		&= E_T(A';B')_{\omega}.
	\end{align}
	Although the simple proof above makes it clear that the trace distance of entanglement is an LOCC monotone, it is known that the trace distance of entanglement is not a selective LOCC monotone, as defined in \eqref{eq:E-meas:strong-LOCC-mono} (please consult the Bibliographic Notes in Section~\ref{sec:E-meas:bib-notes}).
	
	The trace distance of entanglement is also faithful, which is due to the fact that the trace distance is a metric in the mathematical sense: $\frac{1}{2}\norm{\rho_{AB}-\sigma_{AB}}_1\geq 0$ for all states $\rho_{AB},\sigma_{AB}$, and $\frac{1}{2}\norm{\rho_{AB}-\sigma_{AB}}_1=0$ if and only if $\rho_{AB}=\sigma_{AB}$.
	
	Beyond the trace distance, we can take any distinguishability measure and define an entanglement measure analogous to the one in \eqref{eq-trace_dist_ent}. That is, we can take any \textit{generalized divergence} $\boldsymbol{D}$ as our divergence. Recall from Definition~\ref{def-gen_div} that a generalized divergence is a function $\boldsymbol{D}:\Density(\mathcal{H})\times\Lin_+(\mathcal{H})\to\mathbb{R}\cup\{+\infty\}$ that obeys the data-processing inequality. We then define the \textit{generalized divergence of entanglement of $\rho_{AB}$} as follows:
	\begin{equation}\label{eq-gen_div_ent_0}
		\boldsymbol{E}(A;B)_{\rho}\coloneqq\inf_{\sigma_{AB}\in\SEP(A:B)}\boldsymbol{D}(\rho_{AB}\Vert\sigma_{AB}).
	\end{equation}
	See Figure~\ref{fig-dist_based_ent_measure} for a visual depiction of the idea behind this quantity.
	If the generalized divergence $\boldsymbol{D}$ is continuous in its second argument, then the infimum in \eqref{eq-gen_div_ent_0} is achieved. We study this entanglement measure in much more detail in Section~\ref{sec-ent_measures_sep_distance}. By the data-processing inequality for the generalized divergence, as well as the fact that separable states are preserved under LOCC channels, it follows that $\boldsymbol{E}(A;B)_{\rho}$ is an entanglement measure. We prove this and other properties of the generalized divergence of entanglement in Proposition~\ref{prop-gen_div_ent_properties}. As was the case for the trace distance of entanglement, it does not necessarily follow that the generalized divergence of entanglement is a selective LOCC monotone, as defined in \eqref{eq:E-meas:strong-LOCC-mono}, but it does hold for some important cases.


	Although the generalized divergence of entanglement of a bipartite state is conceptually simple, it is in general difficult to optimize over the set of separable states because it does not have a simple characterization (except in low dimensions). This means that the generalized divergence of entanglement is difficult to compute in most cases.
	
	To obtain an entanglement measure that is simpler to compute, one idea is to relax the optimization in \eqref{eq-gen_div_ent_0} from the set of separable states to some other set of states that contains the set of separable states. It is ideal if this other set is easier to characterize than the set of separable states. As a first step, let us recall the PPT criterion from Section~\ref{subsec-PPT}, which states that if a bipartite state is separable, then it is PPT, meaning that it has positive partial transpose (recall Definition~\ref{def-PPT}). This fact immediately leads to the containment $\SEP(A\!:\!B)\subseteq\PPT(A\!:\!B)$. (As stated in Section~\ref{subsec-PPT}, if both $A$ and $B$ are qubits, or if one of them is a qubit and the other a qutrit, then $\PPT(A\!:\!B)=\SEP(A\!:\!B)$.) We can thus define the \textit{PPT generalized divergence} of a bipartite state $\rho_{AB}$ as
	\begin{equation}\label{eq-PPT_divergence}
		\boldsymbol{E}_{\PPT}(A;B)_{\rho}\coloneqq\inf_{\sigma_{AB}\in\operatorname{PPT}(A:B)}\boldsymbol{D}(\rho_{AB}\Vert\sigma_{AB}).
	\end{equation}
	If the generalized divergence $\boldsymbol{D}$ is continuous in its second argument, then the infimum is achieved. Also, note that $\boldsymbol{E}_{\PPT}(A;B)_{\rho}=\boldsymbol{E}(A;B)_{\rho}$ when both $A$ and $B$ are qubits or when one of them is a qubit and the other a qutrit.
	
	Like the generalized divergence of entanglement, the PPT divergence is an entanglement measure. In fact, the PPT divergence is monotone under C-PPT-P channels, as defined in Definition~\ref{def-PPT_pres_chan}.
	This is due to the data-processing inequality for the generalized divergence and the fact that the set of PPT states is closed under C-PPT-P channels (see Proposition~\ref{prop:QM:C-PPT-P-preserves-PPT}).
	Since the set of LOCC channels is contained in the set of C-PPT-P channels (Propositions~\ref{prop:QM-over:sep-ch-contains-LOCC} and \ref{prop-LOCC_PPT_preserving}), it follows that the PPT divergence is an entanglement measure.
	
	Unlike the generalized divergence of entanglement, the PPT divergence is  not a faithful entanglement measure. It is true that $\boldsymbol{E}_{\PPT}(A;B)_{\rho}=0$ for all separable states $\rho_{AB}$ due to the containment
	\begin{equation}
	\SEP(A\!:\!B)\subseteq\PPT(A\!:\!B).
	\end{equation}
	However, the converse statement is not true because the infimum in \eqref{eq-PPT_divergence}, if achieved, need not be achieved by a separable state. In other words, there exist PPT entangled states $\rho_{AB}$ for which $\boldsymbol{E}_{\PPT}(A;B)_{\rho}=0$. 
	
	\begin{figure}
		\centering
		\includegraphics[scale=0.9]{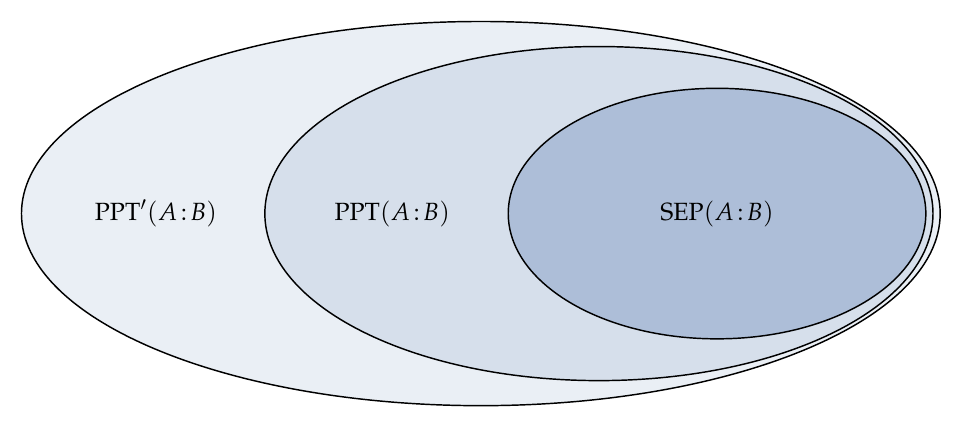}
		\caption{The set $\SEP(A\!:\!B)$ of separable states acting on the Hilbert space $\mathcal{H}_{AB}$ is contained in the set $\PPT(A\!:\!B)$ of positive partial transpose (PPT) states, which in turn is contained in the set $\PPT'(A\!:\!B)$ of operators defined in \eqref{eq:PPT-prime-set}. The sets $\PPT$ and $\PPT'$ are relaxations of the set of separable states that can be easily characterized in terms of semi-definite constraints.}\label{fig-sep_ppt_ppt_prime}
	\end{figure}

	It turns out to be useful to relax  the set of PPT states further:
	
	\begin{definition}{PPT${}^{\prime}$}{def:E-meas:PPT-prime-set}
	Let $\text{PPT}'(A\!:\!B)$ denote the following convex set of positive semi-definite operators:
	\begin{equation}\label{eq:PPT-prime-set}
		\PPT'(A\!:\!B)\coloneqq \left\{  \sigma_{AB}:\sigma_{AB}%
\geq 0,\, \left\Vert \T_{B}(\sigma_{AB})\right\Vert_{1}\leq 1\right\}.%
	\end{equation}
	\end{definition}
	Convexity of the set $\PPT'(A\!:\!B)$ follows from convexity of the trace norm.
	Furthermore, the set $\PPT'(A\!:\!B)$ contains the set of PPT states because every PPT state $\sigma_{AB}$ satisfies $\left\Vert \T_{B}(\sigma_{AB})\right\Vert_{1}= 1$. Furthermore, every operator $\sigma_{AB} \in \PPT'(A\!:\!B)$ is subnormalized, satisfying $\Tr[\sigma_{AB}]\leq 1$, which follows because 
	\begin{equation}
	\Tr[\sigma_{AB}] = \Tr[\T_B(\sigma_{AB})]\leq \left\Vert \T_{B}(\sigma_{AB})\right\Vert_{1}\leq 1.
	\end{equation}	
	The set $\text{PPT}'(A\!:\!B)$ can be written equivalently as 
	\begin{equation}
		\PPT'(A\!:\!B)\coloneqq \left\{  \sigma_{AB}:\sigma_{AB}%
\geq 0,\, E_N(\sigma_{AB})\leq 0\right\},
	\end{equation}
	by inspecting the formula for log-negativity in \eqref{eq:E-meas:def-log-negativity}.
	By comparing with \eqref{eq:PPT-states-set}, we clearly have the containment
	\begin{equation}\label{eq-SEP_PPT_Rains_containment}
		\SEP(A\!:\!B)\subseteq\PPT(A\!:\!B)\subseteq\PPT'(A\!:\!B).
	\end{equation}
	See Figure~\ref{fig-sep_ppt_ppt_prime} for a visual depiction of this containment.

	We define the \textit{generalized Rains divergence} of a bipartite state $\rho_{AB}$ as
	\begin{equation}\label{eq-gen_Rains_div_0}
		\boldsymbol{R}(A;B)_{\rho}\coloneqq \inf_{\sigma_{AB}\in\PPT'(A:B)}\boldsymbol{D}(\rho_{AB}\Vert\sigma_{AB}).
	\end{equation}
	If the underlying generalized divergence $\boldsymbol{D}$ is continuous in its second argument, then the infimum is achieved in \eqref{eq-gen_Rains_div_0}. Like the entanglement measures $\boldsymbol{E}(A;B)_{\rho}$ and $\boldsymbol{E}_{\PPT}(A;B)_{\rho}$, the generalized Rains divergence is an entanglement measure. In fact, it is monotone under C-PPT-P channels, which follows from  the data-processing inequality for the generalized divergence and because the set $\PPT'$ is preserved under C-PPT-P channels (a consequence of Lemma~\ref{prop-PPT_prime_properties} below). Since the set of LOCC channels is contained in the set of C-PPT-P channels (Propositions~\ref{prop:QM-over:sep-ch-contains-LOCC} and \ref{prop-LOCC_PPT_preserving}), it follows that the generalized Rains divergence is an entanglement measure.
	
	As with the PPT divergence, the generalized Rains divergence is not a faithful entanglement measure. It is true that $\boldsymbol{R}(A;B)_{\rho}=0$ for all separable states $\rho_{AB}$, due to the containment $\SEP(A\!:\!B)\subseteq\PPT'(A\!:\!B)$. However, the converse statement is not true because the infimum in \eqref{eq-gen_Rains_div_0} need not be achieved by a separable state.
	
	Depending on the form of the generalized divergence $\boldsymbol{D}$, the relaxation from the set $\SEP$ to the set $\PPT'$ leads to an entanglement measure that can be computed efficiently via semi-definite programming (Section~\ref{sec-SDPs}). We investigate one such example of an entanglement measure in Section~\ref{subsec-max_Rains_rel_ent_state}. Also, due to the containments in \eqref{eq-SEP_PPT_Rains_containment}, we have that
	\begin{equation}\label{eq-SEP_PPT_Rains_ineq}
		\boldsymbol{E}(A;B)_{\rho}\geq\boldsymbol{E}_{\PPT}(A;B)_{\rho}\geq\boldsymbol{R}(A;B)_{\rho},
	\end{equation}
	for every bipartite state $\rho_{AB}$. Thus, as we show later in the book, the relaxation from $\SEP$ to $\PPT'$ via the generalized Rains divergence not only allows for the possibility of efficiently computable entanglement measures, but due to the inequality in \eqref{eq-SEP_PPT_Rains_ineq}, it also allows for the possibility of obtaining a tighter upper bound on communication rates in certain scenarios. We investigate the properties of the generalized Rains divergence in detail in Section~\ref{sec-ent_measures_Rains_dist}.
	
	Before proceeding, let us state some properties of the set $\PPT'$.
	
	\begin{Lemma*}{Properties of the Set $\text{PPT}'$}{prop-PPT_prime_properties}
		The set $\text{PPT}'(A\!:\!B)$ defined in \eqref{eq:PPT-prime-set} has the following properties:
		\begin{enumerate}
			\item It is closed under completely PPT-preserving channels (recall Definition \ref{def-PPT_pres_chan}). In more detail, let $\mathcal{P}_{AB\to A'B'}$ be a completely PPT-preserving channel. Then, for every state $\rho_{AB}\in\text{PPT}'(A\!:\!B)$, we have that $\mathcal{P}_{AB\to A'B'}(\rho_{AB})\in\text{PPT}'(A'\!:\!B')$.
			\item It is closed under LOCC channels (recall Definition \ref{def-LOCC}). In more detail, let $\mathcal{L}_{AB\to A'B'}$ be an LOCC channel. Then, for every state $\rho_{AB}\in\text{PPT}'(A\!:\!B)$, it holds that $\mathcal{L}_{AB\to A'B'}(\rho_{AB})\in\text{PPT}'(A'\!:\!B')$.
		\end{enumerate}
	\end{Lemma*}
	
	\begin{remark}
		We emphasize that not all operators in the set $\PPT'$ are quantum states, meaning that not all operators $\sigma_{AB}\in\PPT'(A\!:\!B)$ satisfy $\Tr[\sigma_{AB}]=1$.
	\end{remark}
	
	\begin{Proof}
		\hfill\begin{enumerate}
			\item Let $\sigma_{A'B'}=\mathcal{P}_{AB\to A'B'}(\rho_{AB})$. Since $\mathcal{P}_{AB\to A'B'}$ is a channel, and $\rho_{AB}$ is a state, we have that $\sigma_{A'B'}\geq 0$. Then,
				\begin{align}
					\T_{B'}(\sigma_{A'B'})&=(\T_{B'}\circ\mathcal{P}_{AB\to A'B'})(\rho_{AB})\\
					&=(\T_{B'}\circ\mathcal{P}_{AB\to A'B'}\circ\T_{B}\circ\T_{B})(\rho_{AB})\\
					&=(\T_{B'}\circ\mathcal{P}_{AB\to A'B'}\circ\T_B)(\T_B(\rho_{AB})).
				\end{align}
				Now, consider that the induced trace norm of $\T_{B'}\circ\mathcal{P}_{AB\to A'B'}\circ\T_B$ satisfies $\norm{\T_{B'}\circ\mathcal{P}_{AB\to A'B'}\circ\T_B}_1=1$, which follows from  \eqref{eq-induced_trace_norm}--\eqref{eq-induced_trace_norm_2} and the fact that $\T_{B'}\circ\mathcal{P}_{AB\to A'B'}\circ\T_B$ is a channel by definition of $\mathcal{P}_{AB\to A'B'}$. Furthermore, we have that $\norm{\T_B(\rho_{AB})}_1\leq 1$ because $\rho_{AB}\in\text{PPT}'(A\!:\!B)$. Putting these observations together, we find that
				\begin{align}
					\norm{\T_{B'}(\sigma_{A'B'})}_1&=\norm{(\T_{B'}\circ\mathcal{P}_{AB\to A'B'}\circ\T_B)(\T_B(\rho_{AB}))}_1\\
					&\leq \norm{\T_{B'}\circ\mathcal{P}_{AB\to A'B'}\circ\T_{B}}_1\norm{\T_{B}(\rho_{AB})}_1\\
					&\leq 1.
				\end{align}
				Therefore, $\sigma_{A'B'}\in\text{PPT}'(A\!:\!B)$.
			
			\item Since every LOCC channel is a completely PPT-preserving channel (Propositions~\ref{prop:QM-over:sep-ch-contains-LOCC} and \ref{prop-LOCC_PPT_preserving}), the result immediately follows from the proof above. \qedhere
		\end{enumerate}
	\end{Proof}

\subsubsection{Squashed Entanglement}

\label{sec:E-meas:sq-ent-intro}

	In the previous example, we considered the generalized divergence of entanglement, which is simply the generalized divergence of a given bipartite state with the set of separable states. If we restrict ourselves to \textit{product states} only, i.e., states of the form $\tau_A\otimes\sigma_B$, and we consider the quantum relative entropy, then we obtain
	\begin{equation}
		\inf_{\tau_A,\sigma_B}D(\rho_{AB}\Vert\tau_A\otimes\sigma_B)=I(A;B)_{\rho},
	\end{equation}
	where we recall the expression in \eqref{eq-mut_inf_double_opt} for the mutual information $I(A;B)_{\rho}$ of the state $\rho_{AB}$. Thus, the mutual information is the minimal relative entropy between the state of interest and the set of product states. We can thus view the mutual information as a measure of the correlations contained in the bipartite state $\rho_{AB}$. However, the mutual information quantifies both classical and quantum correlations because it detects any correlation whatsoever. Thus, it cannot be the case that the mutual information is an entanglement measure, because it is strictly positive even for some separable states that have only classical correlations, and so it can in general increase under LOCC channels.

	Regardless, we can still use the mutual information in a meaningful way to quantify entanglement. For example, suppose that $\sigma_{AB}$ is a separable state, so that
	\begin{equation}\label{eq-sq_ent_motivation_0}
		\sigma_{AB}=\sum_{x\in\mathcal{X}}p(x)\rho_{A}^{x}\otimes\tau_{B}^{x},
	\end{equation}
	for a finite alphabet $\mathcal{X}$, probability distribution $p:\mathcal{X}\to[0,1]$, and sets $\{\rho_{A}^{x}\}_{x\in\mathcal{X}}$, $\{\tau_{B}^{x}\}_{x\in\mathcal{X}}$ of states. Let us form the following extension of $\sigma_{AB}$ to a state $\omega_{ABX}$, with $X$ a classical register:
	\begin{equation}\label{eq-sq_ent_motivation_1}
		\omega_{ABX}=\sum_{x\in\mathcal{X}}p(x)\rho_A^x\otimes\tau_B^x\otimes\ket{x}\!\bra{x}_X.
	\end{equation}
	This is indeed an extension because $\Tr_X[\omega_{ABX}]=\sigma_{AB}$. Let us now consider the \textit{conditional} mutual information $I(A;B|X)_{\omega}$ of this extension (recall the definition of the quantum conditional mutual information in \eqref{eq-QCMI_def}). Since $\omega_{ABX}$ is a classical--quantum state, it follows that
	\begin{equation}\label{eq:LAQC-mutual-info-zero-sep}%
		I(A;B|X)_{\omega}=\sum_{x\in\mathcal{X}}p(x)I(A;B)_{\rho^{x}\otimes\tau^{x}}=0.
	\end{equation}
	Therefore, while the mutual information of a separable state can in general be non-zero, the conditional mutual information is always zero. Intuitively, this is due to the fact that the classical system acts as a ``probe,'' which, when measured, reveals a value $x\in\mathcal{X}$ for the classical system $X$. Conditioned on this value, the joint state $\rho_A^x\otimes\tau_B^x$ is product.
	
	Thus, for every separable state $\rho_{AB}$, there exists a classical extension $\omega_{ABX}$ such that the conditional mutual information $I(A;B|X)_{\omega}$ is equal to zero. Using this idea, we could propose a potential measure of entanglement as follows:
	\begin{equation}\label{eq:LAQC-classical-squashed}%
		\frac{1}{2}\inf_{\omega_{ABX}}\left\{  I(A;B|X)_{\omega}:\Tr_{X}[\omega_{ABX}]=\rho_{AB}\right\}  , 
	\end{equation}
	where the optimization is with respect to extensions of $\rho_{AB}$ having a classical extension system $X$ of arbitrary (finite) dimension $|\mathcal{X}|$, so that
	\begin{equation}
		\omega_{ABX}=\sum_{x\in\mathcal{X}}p(x)\rho_{AB}^{x}\otimes\ket{x}\!\bra{x}_X
	\end{equation}
	for some set $\{\rho_{AB}^x\}_{x\in\mathcal{X}}$ of states and a probability distribution $p(x)$ satisfying $\rho_{AB} = \sum_{x\in\mathcal{X}}p(x)\rho^x_{AB}$. The normalization factor of $\frac{1}{2}$ is there for reasons that become apparent later. If we require that every state $\rho^{x}_{AB}$ in the extension $\omega_{ABX}$ should be pure, then the measure in \eqref{eq:LAQC-classical-squashed} reduces to the entanglement of formation (this was actually used in \eqref{eq:E-meas:EoF-rewrite-CMI} in the proof of Proposition~\ref{prop:E-meas:unif-cont-EoF}).
	
	The quantity proposed in \eqref{eq:LAQC-classical-squashed} is non-negative for every state $\rho_{AB}$, due to the non-negativity of mutual information and the fact that conditional mutual information with a classical conditioning system is equal to a convex combination of mutual informations. It is already clear that the quantity proposed in \eqref{eq:LAQC-classical-squashed} is equal to zero for every separable state---if a state is separable, then the optimization in \eqref{eq:LAQC-classical-squashed} finds the separable decomposition and the value of the quantity is zero, as discussed just after~\eqref{eq:LAQC-mutual-info-zero-sep}. The converse is also true, which follows from the same proof given for \eqref{eq:E-meas:approx-faithful-EoF-2-1}--\eqref{eq:E-meas:approx-faithful-EoF-2-2}. It is actually also possible to show that the quantity in  \eqref{eq:LAQC-classical-squashed} is an entanglement measure. However, we do not make further use of this quantity in this book, because there is an entanglement measure more suitable for our purposes, as introduced below.
	
	Instead of taking a classical extension of the separable state $\sigma_{AB}$ in \eqref{eq-sq_ent_motivation_0}, as we did in \eqref{eq-sq_ent_motivation_1}, we can take a ``quantum extension,'' in the sense that we could define an extension $\omega_{ABE}$ in which the system $E$ is some finite-dimensional quantum system. Optimizing over all such extensions, we obtain the \textit{squashed entanglement}:
	\begin{equation}\label{eq-squashed_entanglement_0}
		E_{\text{sq}}(A;B)_{\rho}\coloneqq\frac{1}{2}\inf_{\omega_{ABE}}\{I(A;B|E)_{\omega}:\Tr_E[\omega_{ABE}]=\rho_{AB}\},
	\end{equation}
	which can only be smaller than the quantity proposed in \eqref{eq:LAQC-classical-squashed}. Note that we optimize with respect to extensions $\omega_{ABE}$ for which the extension system $E$ can have \textit{arbitrary} finite dimension. It is not known whether the optimization can be restricted to extension systems of a certain fixed dimension. In general, therefore, it is not known whether the infimum in \eqref{eq-squashed_entanglement_0} can be replaced by a minimum.
	
	The squashed entanglement is indeed an entanglement measure, as we show in Section~\ref{sec-LAQC:sq-ent-and-props}. It is also a faithful entanglement measure. That it vanishes for separable states follows from the arguments presented above. For a proof of the converse direction, please consult the Bibliographic Notes in Section~\ref{sec:E-meas:bib-notes}. We establish other important properties of the squashed entanglement in Section~\ref{sec-LAQC:sq-ent-and-props}.

\section{Generalized Divergence of Entanglement}\label{sec-ent_measures_sep_distance}

	In this section, we investigate properties of the generalized divergence of entanglement, which is a general construction of an entanglement measure. Let us recall from \eqref{eq-gen_div_ent_0} above that the generalized divergence of entanglement of a bipartite state is the generalized divergence between that state and the set of separable states.
	
	\begin{definition}{Generalized Divergence of Entanglement}{def-gen_div_ent}
		Let $\boldsymbol{D}$ be a generalized divergence (see Definition~\ref{def-gen_div}). For every bipartite state $\rho_{AB}$, we define the \textit{generalized divergence of entanglement of $\rho_{AB}$} as
		\begin{equation}\label{eq-gen_div_ent}
			\boldsymbol{E}(A;B)_{\rho}\coloneqq\inf_{\sigma_{AB}\in\SEP(A:B)}\boldsymbol{D}(\rho_{AB}\Vert\sigma_{AB}).
		\end{equation}
		If the underlying generalized divergence $\boldsymbol{D}$ is continuous in its second argument, then the infimum can be replaced by a minimum.
	\end{definition}
	
	We are particularly interested throughout the rest of this book in the following generalized divergences of entanglement for every state $\rho_{AB}$:
	\begin{enumerate}
		\item The \textit{relative entropy of entanglement of $\rho_{AB}$},
			\begin{equation}
				E_R(A;B)_{\rho}\coloneqq \inf_{\sigma_{AB}\in\SEP(A:B)}D(\rho_{AB}\Vert\sigma_{AB}),
				\label{eq-EM:rel-entr-enta-def}
			\end{equation}
			where $D(\rho_{AB}\Vert\sigma_{AB})$ is the quantum relative entropy of $\rho_{AB}$ and $\sigma_{AB}$ (Definition~\ref{def-rel_ent}).
			
		\item The \textit{$\varepsilon$-hypothesis testing relative entropy of entanglement of $\rho_{AB}$},
			\begin{equation}\label{eq-hypo_test_rel_ent_entanglement}
				E_R^{\varepsilon}(A;B)_{\rho}\coloneqq\inf_{\sigma_{AB}\in\SEP(A:B)}D_H^{\varepsilon}(\rho_{AB}\Vert\sigma_{AB}),
			\end{equation}
			where $D_H^{\varepsilon}(\rho_{AB}\Vert\sigma_{AB})$ is the $\varepsilon$-hypothesis testing relative entropy of $\rho_{AB}$ and $\sigma_{AB}$ (Definition~\ref{def-hypo_testing_rel_ent}).
			
		\item The \textit{sandwiched R\'{e}nyi relative entropy of entanglement of $\rho_{AB}$},
			\begin{equation}\label{eq-sand_Ren_rel_entropy_entanglement}
				\widetilde{E}_{\alpha}(A;B)_{\rho}\coloneqq\inf_{\sigma_{AB}\in\SEP(A:B)}\widetilde{D}_{\alpha}(\rho_{AB}\Vert\sigma_{AB}),
			\end{equation}
			where $\widetilde{D}_{\alpha}(\rho_{AB}\Vert\sigma_{AB})$, $\alpha\in\left[\sfrac{1}{2},1\right)\cup(1,\infty)$, is the sandwiched R\'{e}nyi relative entropy of $\rho_{AB}$ and $\sigma_{AB}$ (Definition~\ref{def-sand_rel_ent}). Note that $\widetilde{E}_{\alpha}(A;B)_{\rho}$ is monotonically increasing in $\alpha$ for all $\rho_{AB}$ (see Proposition~\ref{prop-sand_rel_ent_properties}). This fact, along with the fact that $\lim_{\alpha\to 1}\widetilde{D}_{\alpha}=D$ (see Proposition~\ref{prop-sand_ren_ent_lim}), leads to
			\begin{equation}
				\lim_{\alpha\to 1}\widetilde{E}_{\alpha}(A;B)_{\rho}=E_R(A;B)_{\rho}
			\end{equation}
			for every state $\rho_{AB}$. See Appendix~\ref{app-sand_ren_inf_limit} for details of the proof.
			
		\item The \textit{max-relative entropy of entanglement of $\rho_{AB}$},
			\begin{equation}
				E_{\max}(A;B)_{\rho}\coloneqq\inf_{\sigma_{AB}\in\SEP(A:B)}D_{\max}(\rho_{AB}\Vert\sigma_{AB}),
			\end{equation}
			where $D_{\max}(\rho_{AB}\Vert\sigma_{AB})$ is the max-relative entropy of $\rho_{AB}$ and $\sigma_{AB}$ (Definition~\ref{def-max_rel_ent}). Using the fact that $\lim_{\alpha\to\infty}\widetilde{D}_{\alpha}=D_{\max}$ (see Proposition~\ref{prop-sand_rel_ent_limit_max}), we find that
			\begin{equation}
				E_{\max}(A;B)_{\rho}=\lim_{\alpha\to\infty}\widetilde{E}_{\alpha}(A;B)_{\rho}
			\end{equation}
			for every state $\rho_{AB}$. See Appendix~\ref{app-sand_ren_inf_limit} for details of the proof. As a consequence of this fact, and the fact that $\widetilde{E}_{\alpha}(A;B)_{\rho}$ is monotonically increasing in $\alpha$ for all $\rho_{AB}$, we have that
			\begin{equation}
				E_{\max}(A;B)_{\rho}\geq \widetilde{E}_{\alpha}(A;B)_{\rho} 
			\end{equation}
			for all $\alpha\in(1,\infty)$ and every state $\rho_{AB}$.	
	\end{enumerate}
	
	In addition to the quantities above, we also can define the Petz-- and geometric R\'enyi relative entropies of entanglement in a similar way, but based on Definitions~\ref{def-petz_renyi_rel_ent} and \ref{def:geometric-renyi-rel-ent}, respectively, for the range of $\alpha$ for which data processing holds. These are denoted by $E_{\alpha}(A;B)_{\rho}$ and $\widehat{E}_{\alpha}(A;B)_{\rho}$, respectively.
	

	\begin{proposition*}{Properties of Generalized Divergence of Entanglement}{prop-gen_div_ent_properties}
		Let $\boldsymbol{D}$ be a generalized divergence that is continuous in its second argument, and consider the generalized divergence of entanglement $\boldsymbol{E}(A;B)_{\rho}$ of a state $\rho_{AB}$, as defined in \eqref{eq-gen_div_ent}.
		
		\begin{enumerate}
			\item \textit{Separable monotonicity}: For every separable channel $\mathcal{S}_{AB\to A'B'}$, the generalized divergence of entanglement is monotone non-increasing:
				\begin{equation}\label{eq:SKA-sand-REE-monotone}
					\boldsymbol{E}(A;B)_{\rho}\geq \boldsymbol{E}(A';B')_{\omega},
				\end{equation}
				where $\omega_{A'B'}=\mathcal{S}_{AB\to A'B'}(\rho_{AB})$. Since every LOCC channel is a separable channel, the generalized divergence of entanglement is also monotone non-increasing under LOCC channels. It is therefore an entanglement measure as per Definition~\ref{def-LAQC:ent-measure}.
			
			\item \textit{Faithfulness}: If $\boldsymbol{D}$ satisfies $\boldsymbol{D}(\rho\Vert\sigma)\geq 0$ and $\boldsymbol{D}(\rho\Vert\sigma)=0$ if and only if $\rho=\sigma$ (for all states $\rho$ and $\sigma$), then $\boldsymbol{E}(A;B)_{\rho}=0$ if and only if $\rho_{AB}\in\SEP(A\!:\!B)$. The generalized divergence of entanglement is then a faithful entanglement measure.
				
			\item \textit{Subadditivity}: If $\boldsymbol{D}$ is additive for product positive semi-definite operators, i.e., $\boldsymbol{D}(\rho\otimes\omega\Vert\sigma\otimes\tau)=\boldsymbol{D}(\rho\Vert\sigma)+\boldsymbol{D}(\omega\Vert\tau)$, then for every two quantum states $\rho_{A_1B_1}$ and $\omega_{A_2B_2}$, the generalized divergence of entanglement is sub-additive:
				\begin{equation}\label{eq:SKA-Renyi-REE-subadditivity}
					\boldsymbol{E}(A_1A_2;B_1B_2)_{\rho\otimes\omega}\leq \boldsymbol{E}(A_1;B_1)_{\rho}+\boldsymbol{E}(A_2;B_2)_{\omega}.
				\end{equation}
			
			\item \textit{Convexity}: If $\boldsymbol{D}$ is jointly convex, meaning that
				\begin{equation}
					\boldsymbol{D}\!\left(\sum_{x\in\mathcal{X}}p(x)\rho_{AB}^x\Bigg\Vert\sum_{x\in\mathcal{X}}p(x)\sigma_{AB}^x\right)\leq\sum_{x\in\mathcal{X}}p(x)\boldsymbol{D}(\rho_{AB}^x\Vert\sigma_{AB}^x),
				\end{equation}
				for every finite alphabet $\mathcal{X}$, probability distribution $p:\mathcal{X}\to[0,1]$, and sets $\{\rho_{AB}^x\}_{x\in\mathcal{X}}$, $\{\sigma_{AB}^x\}_{x\in\mathcal{X}}$ of states, then the generalized divergence of entanglement is convex:
				\begin{equation}
					\boldsymbol{E}(A;B)_{\overline{\rho}}\leq\sum_{x\in\mathcal{X}}p(x)\boldsymbol{E}(A;B)_{\rho^x},
				\end{equation}
				where $\overline{\rho}_{AB}=\sum_{x\in\mathcal{X}}p(x)\rho_{AB}^x$.

		\end{enumerate}
		
		Properties 1., 2., 3. are satisfied when the generalized divergence is the quantum relative entropy, the Petz--, sandwiched, and geometric R\'{e}nyi relative entropy, and the max-relative entropy. Property 4. is satisfied when the generalized divergence is the quantum relative entropy and the Petz--, sandwiched, and geometric  R\'{e}nyi relative entropies for the range of $\alpha < 1 $ for which data processing holds.
	\end{proposition*}
	
	\begin{Proof}
		\hfill\begin{enumerate}
			\item For $\omega_{A'B'}=\mathcal{S}_{AB\to A'B'}(\rho_{AB})$, we have by definition,
				\begin{align}
					\boldsymbol{E}(A';B')_{\omega}&=\inf_{\tau_{A'B'}\in\SEP(A':B')}\boldsymbol{D}(\omega_{A'B'}\Vert\tau_{A'B'})\\
					&=\inf_{\tau_{A'B'}\in\SEP(A':B')}\boldsymbol{D}(\mathcal{S}_{AB\to A'B'}(\rho_{AB})\Vert\tau_{A'B'}).\label{eq-gen_div_ent_sep_monotone_pf}
				\end{align}
				Now, recall that every separable channel $\mathcal{S}_{AB\rightarrow A'B'}$ takes $\sigma_{AB}\in\SEP(A\!:\!B)$ to a state in $\SEP(A'\!:\!B')$, as shown already in \eqref{eq:QM-over:sep-preserves-sep-1}--\eqref{eq:QM-over:sep-preserves-sep-2}.
				Therefore, restricting the optimization in \eqref{eq-gen_div_ent_sep_monotone_pf} leads to
				\begin{align}
					\boldsymbol{E}(A';B')_{\omega}&\leq \inf_{\sigma_{AB}\in\SEP(A:B)}\boldsymbol{D}(\mathcal{S}_{AB\to A'B'}(\rho_{AB})\Vert\mathcal{S}_{AB\to A'B'}(\sigma_{AB}))\\
					&\leq \inf_{\sigma_{AB}\in\SEP(A:B)}\boldsymbol{D}(\rho_{AB}\Vert\sigma_{AB})\\
					&=\boldsymbol{E}(A;B)_{\rho},
				\end{align}
				as required, where we used the data-processing inequality for the generalized divergence to obtain the second inequality.
				
			\item We have
				\begin{equation}\label{eq-gen_div_ent_2}
					\boldsymbol{E}(A;B)_{\rho}=\inf_{\sigma_{AB}\in\SEP(A:B)}\boldsymbol{D}(\rho_{AB}\Vert\sigma_{AB}).
				\end{equation}
				If $\rho_{AB}\in\SEP(A\!:\!B)$, then the state $\rho_{AB}$ itself achieves the minimum in \eqref{eq-gen_div_ent_2} because $\boldsymbol{D}(\rho_{AB}\Vert\rho_{AB})=0$. We thus have $\boldsymbol{E}(A;B)_{\rho}=0$. On the other hand, if $\boldsymbol{E}(A;B)_{\rho}=0$, then there exists a separable state $\sigma_{AB}^*$ such that $\boldsymbol{D}(\rho_{AB}\Vert\sigma_{AB}^*)=0$, which by assumption implies that $\rho_{AB}=\sigma_{AB}^*$, i.e., that $\rho_{AB}$ is separable.
		
			\item By definition, the optimization in the definition of $\boldsymbol{E}(A_1A_2;B_1B_2)_{\rho\otimes\omega}$ is over the set $\SEP(A_1A_2\!:\!B_1B_2)$. It is straightforward to see that this set contains states of the form $\xi_{A_1B_1}\otimes\tau_{A_2B_2}$, where $\xi_{A_1B_1}\in\SEP(A_1\!:\!B_1)$ and $\tau_{A_2B_2}\in\SEP(A_2\!:\!B_2)$. By restricting the optimization to such states, and by using additivity of the generalized divergence $\boldsymbol{D}$, we obtain
				\begin{align}
					\boldsymbol{E}(A_1A_2;B_1B_2)_{\rho\otimes\omega}&\leq \boldsymbol{D}(\rho_{A_1B_1}\otimes\omega_{A_2B_2}\Vert\xi_{A_1B_1}\otimes\tau_{A_2B_2})\\
					&=\boldsymbol{D}(\rho_{A_1B_1}\Vert\xi_{A_1B_1})+\boldsymbol{D}(\omega_{A_2B_2}\Vert\tau_{A_2B_2}).
				\end{align}
				Since $\xi_{A_1B_1}\in\SEP(A_1\!:\!B_1)$ and $\tau_{A_2B_2}\in\SEP(A_2\!:\!B_2)$ are arbitrary, the inequality in \eqref{eq:SKA-Renyi-REE-subadditivity} follows. 
			
			\item We have
				\begin{equation}
					\boldsymbol{E}(A;B)_{\overline{\rho}}=\inf_{\sigma_{AB}\in\SEP(A:B)}\boldsymbol{D}\!\left(\sum_{x\in\mathcal{X}}p(x)\rho_{AB}^x\Bigg\Vert\sigma_{AB}\right).
				\end{equation}
				Let us restrict the optimization over all separable states to an optimization over sets $\{\sigma_{AB}^x\}_{x\in\mathcal{X}}$ of separable states indexed by the alphabet $\mathcal{X}$. Then, 
				$\sum_{x\in\mathcal{X}}p(x)\sigma_{AB}^x$ is a separable state because the set of separable states is convex. Therefore, using the joint convexity of $\boldsymbol{D}$, we obtain
				\begin{align}
					\boldsymbol{E}(A;B)_{\overline{\rho}}&\leq\inf_{\{\sigma_{AB}^x\}_x\subset\SEP(A:B)}\boldsymbol{D}\!\left(\sum_{x\in\mathcal{X}}p(x)\rho_{AB}^x\Bigg\Vert\sum_{x\in\mathcal{X}}p(x)\sigma_{AB}^x\right)\\
					&\leq \inf_{\{\sigma_{AB}^x\}_x\subset\SEP(A:B)}\sum_{x\in\mathcal{X}}p(x)\boldsymbol{D}(\rho_{AB}^x\Vert\sigma_{AB}^x)\\
					&\leq\sum_{x\in\mathcal{X}}p(x)\inf_{\sigma_{AB}^x\in\SEP(A:B)}\boldsymbol{D}(\rho_{AB}^x\Vert\sigma_{AB}^x)\\
					&=\sum_{x\in\mathcal{X}}p(x)\boldsymbol{E}(A;B)_{\rho^x},
				\end{align}
				as required. \qedhere
		\end{enumerate}
	\end{Proof}
	
	We now delve a bit more into particular examples of the generalized divergence of entanglement, which are based on the relative entropy and the Petz--, sandwiched, and geometric R\'enyi relative entropies.
	
	\begin{proposition}{prop:E-meas:REE-inv-cc}
	The relative entropy of entanglement is invariant under classical communication; i.e., \eqref{eq:E-meas:invar-class-comm} holds with $E$ set to $E_{R}$.
\end{proposition}

\begin{Proof}
Let $\rho_{XAB}$ be a classical--quantum state of the form in \eqref{eq:E-meas:cq-state-inv-CC}. Let
$\{\sigma_{AB}^{x}\}_{x\in\mathcal{X}}$ be an arbitrary set of separable
states, and set%
\begin{equation}
\sigma_{XAB}\coloneqq \sum_{x\in\mathcal{X}}p(x)|x\rangle\!\langle x|_{X}\otimes
\sigma_{AB}^{x}.
\end{equation}
Consider that%
\begin{align}
E_{R}(XA;B)_{\rho}  & =\inf_{\sigma_{XAB}\in\operatorname{SEP}(XA:B)}%
D(\rho_{XAB}\Vert\sigma_{XAB})\\
& \leq D(\rho_{XAB}\Vert\sigma_{XAB})\\
& =\sum_{x\in\mathcal{X}}p(x)D(\rho_{AB}^{x}\Vert\sigma_{AB}^{x}),
\end{align}
where the last equality follows from the direct-sum property of relative entropy in \eqref{eq-rel_ent_direct_sum}. Since the inequality
holds for every set $\{\sigma_{AB}^{x}\}_{x\in\mathcal{X}}$ of separable
states, we conclude that%
\begin{equation}
E_{R}(XA;B)_{\rho}\leq\sum_{x\in\mathcal{X}}p(x)E_{R}(A;B)_{\rho^{x}%
}.\label{eq:E-meas:REE-inv-CC-1}%
\end{equation}

Now suppose that $\sigma_{XAB}$ is an arbitrary separable state of the systems
$XA|B$ (here we assume that the system $X$ is not necessarily classical). After performing
the completely dephasing channel $\Delta_{X}(\cdot)\coloneqq \sum_{x\in\mathcal{X}%
}|x\rangle\!\langle x|_{X}(\cdot)|x\rangle\!\langle x|_{X}$ on the $X$ system of
$\sigma_{XAB}$, the resulting state is a classical--quantum state of the
following form:%
\begin{equation}
\Delta_{X}(\sigma_{XAB})=\sum_{x\in\mathcal{X}}q(x)|x\rangle\!\langle
x|_{X}\otimes\sigma_{AB}^{x},
\end{equation}
where $q(x)$ is a probability distribution and $\{\sigma_{AB}^{x}%
\}_{x\in\mathcal{X}}$ is a set of separable states. Then consider that%
\begin{align}
D(\rho_{XAB}\Vert\sigma_{XAB})  & \geq D(\Delta_{X}(\rho_{XAB})\Vert\Delta
_{X}(\sigma_{XAB}))\\
& =\sum_{x\in\mathcal{X}}p(x)D(\rho_{AB}^{x}\Vert\sigma_{AB}^{x})+D(p\Vert
q)\\
& \geq\sum_{x\in\mathcal{X}}p(x)D(\rho_{AB}^{x}\Vert\sigma_{AB}^{x})\\
& \geq\sum_{x\in\mathcal{X}}p(x)E_{R}(A;B)_{\rho^{x}}.
\end{align}
The first inequality follows from the data-processing inequality for relative
entropy (Theorem~\ref{thm-monotone_rel_ent}). The equality follows from the direct-sum property of relative entropy in \eqref{eq-rel_ent_direct_sum}. The second inequality follows from
the non-negativity of the classical relative entropy $D(p\Vert q)$. The final
inequality follows from the definition of the relative entropy of entanglement
and the fact that $\sigma_{AB}^{x}$ is separable. Since the chain of
inequalities holds for every separable state $\sigma_{XAB}$, we conclude that%
\begin{equation}
E_{R}(XA;B)_{\rho}\geq\sum_{x\in\mathcal{X}}p(x)E_{R}(A;B)_{\rho^{x}%
}.\label{eq:E-meas:REE-inv-CC-2}%
\end{equation}
Putting together \eqref{eq:E-meas:REE-inv-CC-1}\ and
\eqref{eq:E-meas:REE-inv-CC-2}, and noting that the same argument applies when
exchanging the roles of Alice and Bob, we conclude the statement of the proposition.
\end{Proof}

As an immediate corollary of Proposition~\ref{prop:E-meas:REE-inv-cc}, Lemma~\ref{lem:E-meas:convex-strong-LOCC-mono}, and
Property~1.~of Proposition~\ref{prop-gen_div_ent_properties}, we conclude that the relative entropy of
entanglement is a selective LOCC monotone. However, we can conclude something
stronger, which is what we prove in Proposition~\ref{prop:E-meas:strong-sep-mono-rel-ents} below after defining selective separable monotonicity.

\begin{definition}{Selective Separable Monotonicity}{def:E-meas:strong-separable-mono}
As a generalization of selective LOCC monotonicity defined in \eqref{eq:E-meas:strong-LOCC-mono} and in the spirit of the selective PPT monotonicity from Definition~\ref{def:E-meas:strong-ppt-mono}, we say that a
function $E:\Density(\mathcal{H}_{AB})\rightarrow\mathbb{R}$\ is a selective separable
monotone if it satisfies%
\begin{equation}
E(\rho_{AB})\geq\sum_{x\in\mathcal{X}:p(x)>0}p(x)E(\omega_{A^{\prime}%
B^{\prime}}^{x}),\label{eq:E-meas:strong-separable-monotone}%
\end{equation}
for every bipartite state $\rho_{AB}$ and separable instrument $\{\mathcal{S}%
_{AB\rightarrow A^{\prime}B^{\prime}}^{x}\}_{x\in\mathcal{X}}$, with%
\begin{align}
p(x)  & \coloneqq \operatorname{Tr}[\mathcal{S}_{AB\rightarrow A^{\prime}B^{\prime}%
}^{x}(\rho_{AB})],\label{eq:E-meas:post-separable-ensemble-1}\\
\omega_{A^{\prime}B^{\prime}}^{x}  & \coloneqq \frac{1}{p(x)}\mathcal{S}%
_{AB\rightarrow A^{\prime}B^{\prime}}^{x}(\rho_{AB}%
).\label{eq:E-meas:post-separable-ensemble-2}%
\end{align}
A separable instrument is such that every map $\mathcal{S}_{AB\rightarrow
A^{\prime}B^{\prime}}^{x}$ is completely positive and separable (with Kraus operators of the form in \eqref{eq:QM-over:kraus-form-sep-ch}), and the sum map $\sum_{x\in\mathcal{X}}\mathcal{S}_{AB\rightarrow
A^{\prime}B^{\prime}}^{x}$ is trace preserving.

It follows that $E$ is a separable monotone if it is a selective separable monotone, because the former is a special case of the latter in which the alphabet $\mathcal{X}$ has only one letter.
\end{definition}

\begin{proposition*}{Selective Separable Monotonicity of Relative Entropies of Entanglement}{prop:E-meas:strong-sep-mono-rel-ents}
The relative entropy of entanglement is a selective separable monotone; i.e.,  \eqref{eq:E-meas:strong-separable-monotone} holds with $E$ set to $E_R$. The
Petz--, sandwiched, and geometric R\'enyi relative entropies of entanglement are
selective separable monotones for the range $\alpha>1$ for which data processing holds.
\end{proposition*}

\begin{Proof}
Let us begin with the relative entropy of entanglement. Let $\rho_{AB}$ be an
arbitrary bipartite state, let $\{\mathcal{S}_{AB\rightarrow A^{\prime
}B^{\prime}}^{x}\}_{x\in\mathcal{X}}$ be a separable instrument, and let
$\mathcal{S}_{AB\rightarrow XA^{\prime}B^{\prime}}$ denote the following
quantum channel:%
\begin{equation}
\mathcal{S}_{AB\rightarrow XA^{\prime}B^{\prime}}(\omega_{AB})\coloneqq \sum
_{x\in\mathcal{X}}|x\rangle\!\langle x|_{X}\otimes\mathcal{S}_{AB\rightarrow
A^{\prime}B^{\prime}}^{x}(\omega_{AB}).
\end{equation}
Let $\tau_{XA^{\prime}B^{\prime}}\coloneqq \mathcal{S}_{AB\rightarrow XA^{\prime
}B^{\prime}}(\rho_{AB})$, and note that%
\begin{equation}
\tau_{XA^{\prime}B^{\prime}}=\sum_{x\in\mathcal{X}}p(x)|x\rangle\!\langle
x|_{X}\otimes\tau_{A^{\prime}B^{\prime}}^{x},
\end{equation}
for some probability distribution $p(x)$ and set $\{\tau_{A^{\prime}B^{\prime
}}^{x}\}_{x\in\mathcal{X}}$ of states. Then consider that%
\begin{align}
E_{R}(A;B)_{\rho}  & \geq E_{R}(XA^{\prime};B^{\prime})_{\tau}\\
& =\sum_{x\in\mathcal{X}}p(x)E_{R}(A^{\prime};B^{\prime})_{\tau^{x}}.
\end{align}
The inequality follows because $E_{R}$ is monotone under separable channels
(Property~1.~of Proposition~\ref{prop-gen_div_ent_properties}), and the equality follows from Proposition~\ref{prop:E-meas:REE-inv-cc}.

Let us now consider proving the statement for the Petz--R\'enyi relative entropy
of entanglement for $\alpha\in(1,2]$. Consider the same channel $\mathcal{S}%
_{AB\rightarrow XA^{\prime}B^{\prime}}$ and state $\tau_{XA^{\prime}B^{\prime
}}$ defined above. Let $\sigma_{AB}$ be an arbitrary separable state, and
consider that%
\begin{equation}
\mathcal{S}_{AB\rightarrow XA^{\prime}B^{\prime}}(\sigma_{AB})=\sum
_{x\in\mathcal{X}}q(x)|x\rangle\!\langle x|_{X}\otimes\sigma_{A^{\prime
}B^{\prime}}^{x},
\end{equation}
for some probability distribution $q(x)$ and set $\{\sigma_{A^{\prime
}B^{\prime}}^{x}\}_{x\in\mathcal{X}}$ of separable states. Then we find that%
\begin{align}
Q_{\alpha}(\rho_{AB}\Vert\sigma_{AB})  & \geq Q_{\alpha}(\mathcal{S}%
_{AB\rightarrow XA^{\prime}B^{\prime}}(\rho_{AB})\Vert\mathcal{S}%
_{AB\rightarrow XA^{\prime}B^{\prime}}(\sigma_{AB}))\\
& =\sum_{x\in\mathcal{X}:p(x)>0}p(x)^{\alpha}q(x)^{1-\alpha}Q_{\alpha}%
(\tau_{A^{\prime}B^{\prime}}^{x}\Vert\sigma_{A^{\prime}B^{\prime}}^{x})\\
& =\sum_{x\in\mathcal{X}:p(x)>0}p(x)\left(  \frac{p(x)}{q(x)}\right)
^{\alpha-1}Q_{\alpha}(\tau_{A^{\prime}B^{\prime}}^{x}\Vert\sigma_{A^{\prime
}B^{\prime}}^{x}).
\end{align}
The first inequality follows from the data-processing inequality for the
Petz--R\'enyi relative quasi-entropy (Theorem~\ref{thm-petz_rel_ent_monotone}), and the first equality
follows from its direct-sum property (see \eqref{eq:QEI:direct-sum-petz-renyi}). Now applying the
monotonicity and concavity of the function $(\cdot)\rightarrow\frac{1}%
{\alpha-1}\log_{2}(\cdot)$ for $\alpha\in(1,2]$, we find that%
\begin{align}
& D_{\alpha}(\rho_{AB}\Vert\sigma_{AB})\nonumber\\
& =\frac{1}{\alpha-1}\log_{2}Q_{\alpha}(\rho_{AB}\Vert\sigma_{AB})\\
& \geq\frac{1}{\alpha-1}\log_{2}\!\left[  \sum_{x\in\mathcal{X}:p(x)>0}%
p(x)\left(  \frac{p(x)}{q(x)}\right)  ^{\alpha-1}Q_{\alpha}(\tau_{A^{\prime
}B^{\prime}}^{x}\Vert\sigma_{A^{\prime}B^{\prime}}^{x})\right]  \\
& \geq\sum_{x\in\mathcal{X}:p(x)>0}p(x)\frac{1}{\alpha-1}\log_{2}\!\left[
\left(  \frac{p(x)}{q(x)}\right)  ^{\alpha-1}Q_{\alpha}(\tau_{A^{\prime
}B^{\prime}}^{x}\Vert\sigma_{A^{\prime}B^{\prime}}^{x})\right]  \\
& =\sum_{x\in\mathcal{X}:p(x)>0}p(x)\left[  \log_{2}\!\left(  \frac{p(x)}%
{q(x)}\right)  +D_{\alpha}(\tau_{A^{\prime}B^{\prime}}^{x}\Vert\sigma
_{A^{\prime}B^{\prime}}^{x})\right]  \\
& =D(p\Vert q)+\sum_{x\in\mathcal{X}:p(x)>0}p(x)D_{\alpha}(\tau_{A^{\prime
}B^{\prime}}^{x}\Vert\sigma_{A^{\prime}B^{\prime}}^{x})\\
& \geq\sum_{x\in\mathcal{X}:p(x)>0}p(x)E_{\alpha}(A^{\prime};B^{\prime}%
)_{\tau^{x}}.
\end{align}
The final two equalities follow by direct evaluation and applying definitions.
The final inequality follows because $D(p\Vert q)\geq0$ for probability
distributions $p$ and $q$, and it also follows from the definition of the
Petz--R\'enyi relative entropy of entanglement and the fact that the state $\sigma^x_{A'B'}$ is separable. Since the inequality holds for
every separable state $\sigma_{AB}$, we conclude the desired inequality:%
\begin{equation}
E_{\alpha}(A;B)_{\rho}\geq\sum_{x\in\mathcal{X}:p(x)>0}p(x)E_{\alpha
}(A^{\prime};B^{\prime})_{\tau^{x}}.
\end{equation}

By applying the same method of proof for the sandwiched and geometric R\'enyi
relative entropies for the range of $\alpha>1$ for which data processing
holds, along with their data processing and direct-sum properties, we conclude
the same inequality for the sandwiched and geometric R\'enyi relative entropies
of entanglement.
\end{Proof}
	
	The following additional facts are known specifically about the relative entropy of entanglement and the Petz--, sandwiched, and geometric R\'enyi relative entropies of entanglement.
	
	\begin{proposition}{prop-ent_meas_rel_ent_entanglement}
		\begin{enumerate}
			\item For every bipartite state $\rho_{AB}$,
				\begin{align}
					E_R(A;B)_{\rho} & \geq\max\{I(A\rangle B)_{\rho},I(B\rangle A)_{\rho}\} , \\
					E_{\alpha}(A;B)_{\rho} & \geq\max\{I_{\alpha}(A\rangle B)_{\rho},I_{\alpha}(B\rangle A)_{\rho}\} \label{eq:E-meas:REE-greater-than-CI-1}, \\
										\widetilde{E}_{\alpha}(A;B)_{\rho} & \geq\max\{\widetilde{I}_{\alpha}(A\rangle B)_{\rho},\widetilde{I}_{\alpha}(B\rangle A)_{\rho}\} \\
															\widehat{E}_{\alpha}(A;B)_{\rho} & \geq\max\{\widehat{I}_{\alpha}(A\rangle B)_{\rho},\widehat{I}_{\alpha}(B\rangle A)_{\rho}\} ,
															\label{eq:E-meas:REE-greater-than-CI-3}
				\end{align}
				where the last three inequalities hold for the range of $\alpha$ for which data processing holds.
			
			\item For every pure bipartite state $\psi_{AB}$,
				\begin{align}
				\label{eq-rel_ent_entanglement_pure_states}
					E_R(A;B)_{\psi}& =H(A)_{\psi},\\									E_{\alpha}(A;B)_{\psi}& =H_{\frac{1}{\alpha}}(A)_{\psi},\label{eq:E-meas:REE-equal-to-CI-1}\\
	\widetilde{E}_{\alpha}(A;B)_{\psi}& =H_{\frac{\alpha}{2\alpha-1}}(A)_{\psi},\\
	\widehat{E}_{\alpha}(A;B)_{\psi}& =H_{\frac{1}{2}}(A)_{\psi},\label{eq:E-meas:REE-equal-to-CI-3}
				\end{align}
				where the last three equalities hold for the range of $\alpha$ for which data processing holds.
		\end{enumerate}
	\end{proposition}
	
	\begin{remark}
		Observe that for pure states, the relative entropy of entanglement is equal to the entanglement of formation (see \eqref{eq-ent_formation_pure_state}).
	\end{remark}
	
	\begin{Proof}
		\hfill\begin{enumerate}
			\item Let $\sigma_{AB}$ be an arbitrary separable state, which can be written as
				\begin{equation}
					\sigma_{AB}=\sum_{x\in\mathcal{X}}p(x)\omega_A^x\otimes\tau_B^x,
				\end{equation}
				where $\mathcal{X}$ is some finite alphabet $p:\mathcal{X}\to[0,1]$ is a probability distribution, and $\{\omega_A^x\}_{x\in\mathcal{X}}$, $\{\tau_B^x\}_{x\in\mathcal{X}}$ are sets of states. Since every $\omega_A^x$ is a state (in particular, since all of its eigenvalues are bounded from above by one), the inequality $\omega_A^x\leq\mathbbm{1}_A$ holds, which implies that $\mathbbm{1}_A\otimes\tau_B^x\geq \omega_A^x\otimes\tau_B^x$ for all $x\in\mathcal{X}$, which thus implies that
				\begin{equation}
					\mathbbm{1}_A\otimes\sigma_B=\sum_{x\in\mathcal{X}}p(x)\mathbbm{1}_A\otimes\tau_B^x\geq \sum_{x\in\mathcal{X}}p(x)\omega_A^x\otimes\tau_B^x=\sigma_{AB}.
					\label{eq:E-meas:key-ineq-REE-to-CI}
				\end{equation}
				Therefore, using property 2.(d) in Proposition~\ref{prop-rel_ent}, we have that
				\begin{align}
					D(\rho_{AB}\Vert\sigma_{AB})&\geq D(\rho_{AB}\Vert\mathbbm{1}_A\otimes\sigma_B)\\
					&\geq\inf_{\sigma_B}D(\rho_{AB}\Vert\mathbbm{1}_A\otimes\sigma_B)\\
					&=I(A\rangle B)_{\rho}
				\end{align}
				for all separable states, where the optimization is with respect to every state $\sigma_B$ on the right-hand side, and where we have used the expression in \eqref{eq-coher_inf_opt} for coherent information. We thus have
				\begin{equation}\label{eq-rel_ent_entanglement_coh_inf_LB}
					E_R(A;B)_{\rho}=\inf_{\sigma_{AB}\in\SEP(A:B)}D(\rho_{AB}\Vert\sigma_{AB})\geq I(A\rangle B)_{\rho}.
				\end{equation}
				
				By the same argument, but flipping the roles of Alice and Bob, we conclude that
				\begin{equation}
					E_R(A;B)_{\rho}\geq I(B\rangle A)_{\rho}.
				\end{equation}
				Combining this inequality with the one in \eqref{eq-rel_ent_entanglement_coh_inf_LB} leads to  the desired result.
				
				The same proof, but using \eqref{eq:E-meas:key-ineq-REE-to-CI} and Property~4.~of Propositions~\ref{prop-petz_rel_ent_add_properties}, \ref{prop-sand_rel_ent_add_properties}, and \ref{prop:QEI:add-prop-geo-renyi}, leads to the inequalities in \eqref{eq:E-meas:REE-greater-than-CI-1}--\eqref{eq:E-meas:REE-greater-than-CI-3}.

			\item Let
				\begin{equation}\label{eq-rel_ent_entangle_pure_states_pf}
					\ket{\psi}_{AB}=\sum_{k=1}^r\sqrt{\lambda_k}\ket{e_k}_A\otimes\ket{f_k}_B
				\end{equation}
				be the Schmidt decomposition of $\ket{\psi}_{AB}$, where $r$ is the Schmidt rank, $\lambda_k>0$ for all $1\leq k\leq r$, and $\{\ket{e_k}_A\}_{k=1}^r$, $\{\ket{f_k}_B\}_{k=1}^r$ are orthonormal sets of vectors. 

			Since the entropy $H(AB)_{\psi}$ vanishes for all pure states, we immediately have
				\begin{equation}
					I(A\rangle B)_{\psi}=H(B)_{\psi}=H(A)_{\psi}=I(B\rangle A)_{\psi},
				\end{equation}
				where the equality $H(B)_{\psi}=H(A)_{\psi}$ follows from the Schmidt decomposition in \eqref{eq-rel_ent_entangle_pure_states_pf}, which tells us that the reduced states $\psi_A$ and $\psi_B$ have the same non-zero eigevalues. Based on the fact that $E_R(A;B)_{\psi}\geq I(A\rangle B)_{\psi}$, which we just proved above, we thus have the lower bound
				\begin{equation}\label{eq-rel_ent_entanglement_pure_states_pf2}
					E_R(A;B)_{\psi}\geq H(A)_{\psi}.
				\end{equation}
				The same reasoning, but using the lower bounds in \eqref{eq:E-meas:REE-greater-than-CI-1}--\eqref{eq:E-meas:REE-greater-than-CI-3}, as well as \eqref{eq:QEI:Petz-Renyi-CI-pure-bi}--\eqref{eq:QEI:geo-Renyi-CI-pure-bi},  leads to the inequalities
				\begin{align}
					E_{\alpha}(A;B)_{\psi}& \geq H_{\frac{1}{\alpha}}(A)_{\psi},
					\label{eq:E-meas:RREE-greater-CI-1}\\
	\widetilde{E}_{\alpha}(A;B)_{\psi}& \geq H_{\frac{\alpha}{2\alpha-1}}(A)_{\psi},\\
	\widehat{E}_{\alpha}(A;B)_{\psi}& \geq H_{\frac{1}{2}}(A)_{\psi} .\label{eq:E-meas:RREE-greater-CI-3}
	\end{align}

				For the reverse inequality, let
				\begin{equation}
					\Pi\coloneqq\sum_{k=1}^r \ket{e_k}\!\bra{e_k}_A\otimes\ket{f_k}\!\bra{f_k}_B
				\end{equation}
				be the projection onto the $r^2$-dimension subspace of $\mathcal{H}_{AB}$ on which $\psi_{AB}$ is supported. Also, define a channel $\mathcal{N}$ as
				\begin{equation}
					\mathcal{N}(X_{AB})\coloneqq \Pi X_{AB}\Pi + (\mathbbm{1}_{AB}-\Pi)X_{AB}(\mathbbm{1}_{AB}-\Pi).
				\end{equation}
				Note that, by definition, $\mathcal{N}(\psi_{AB})=\psi_{AB}$. Also, let $\sigma_B$ be a state, and, for $p(k)\coloneqq \bra{f_k}\sigma_B\ket{f_k}_B$, set
				\begin{equation}
					\overline{\sigma}_{AB}\coloneqq \sum_{k=1}^r p(k)\ket{e_k}\!\bra{e_k}_A\otimes\ket{f_k}\!\bra{f_k}_B,
				\end{equation}
				which is a separable state. It is straightforward to show that
				\begin{equation}
					\Pi(\mathbbm{1}_A\otimes\sigma_B)\Pi=\overline{\sigma}_{AB}.
				\end{equation}
				Also,  we have that
				\begin{align}
					&D(\mathcal{N}(\psi_{AB})\Vert\mathcal{N}(\mathbbm{1}_A\otimes\sigma_B))\nonumber\\
					&\qquad =D(\psi_{AB}\Vert\overline{\sigma}_{AB}+(\mathbbm{1}_{AB}-\Pi)(\mathbbm{1}_A\otimes\sigma_B)(\mathbbm{1}_{AB}-\Pi))\\
					&\qquad =D(\psi_{AB}\Vert\overline{\sigma}_{AB}),
				\end{align}
				because the operator $(\mathbbm{1}_{AB}-\Pi)(\mathbbm{1}_A\otimes\sigma_B)(\mathbbm{1}_{AB}-\Pi)$ is supported on the space orthogonal to the support of $\psi_{AB}$.
				Therefore, by the data-processing inequality for quantum relative entropy, we obtain
				\begin{align}
					E_R(A;B)_{\psi}&=\inf_{\sigma_{AB}\in\SEP(A:B)}D(\psi_{AB}\Vert\sigma_{AB})\\
					&\leq D(\psi_{AB}\Vert\overline{\sigma}_{AB})\\
					&=D(\mathcal{N}(\psi_{AB})\Vert\mathcal{N}(\mathbbm{1}_A\otimes\sigma_B))\\
					&\leq D(\psi_{AB}\Vert\mathbbm{1}_A\otimes\sigma_B).
				\end{align}
				Since the inequality holds for every state $\sigma_B$, we conclude that
				\begin{equation}
		E_R(A;B)_{\psi}\leq \inf_{\sigma_B}D(\psi_{AB}\Vert\mathbbm{1}_A\otimes\sigma_B)							=I(A\rangle B)_{\psi}
					=H(A)_{\psi}.
					\end{equation}
				Combining this with \eqref{eq-rel_ent_entanglement_pure_states_pf2} gives us $E_R(A;B)_{\psi}=H(A)_{\psi}$, as required.
				
				Applying the same method of proof, but using the properties of the Petz--, sandwiched, and geometric R\'enyi relative entropies, as well as \eqref{eq:QEI:Petz-Renyi-CI-pure-bi}--\eqref{eq:QEI:geo-Renyi-CI-pure-bi}, we conclude the inequalities
				\begin{align}
					E_{\alpha}(A;B)_{\psi}& \leq H_{\frac{1}{\alpha}}(A)_{\psi},\\
	\widetilde{E}_{\alpha}(A;B)_{\psi}& \leq H_{\frac{\alpha}{2\alpha-1}}(A)_{\psi},\\
	\widehat{E}_{\alpha}(A;B)_{\psi}& \leq H_{\frac{1}{2}}(A)_{\psi} ,
	\end{align}
which, combined with \eqref{eq:E-meas:RREE-greater-CI-1}--\eqref{eq:E-meas:RREE-greater-CI-3}, leads to \eqref{eq:E-meas:REE-equal-to-CI-1}--\eqref{eq:E-meas:REE-equal-to-CI-3}. \qedhere 
		\end{enumerate}
	\end{Proof}

\subsection{Cone Program Formulations}\label{subsec-gen_div_ent_cone_prog}

	Computing a generalized divergence of entanglement involves an optimization over the set of separable states, which is known to be hard (please consult the Bibliographic Notes in Section~\ref{sec:E-meas:bib-notes}). The optimization is made more complicated by the fact that most of the generalized divergences we consider in this book are non-linear functions of the input state (for example, sandwiched R\'{e}nyi relative entropy). However, both the max-relative entropy and hypothesis testing relative entropy can be formulated as semi-definite programs (SDPs). Indeed, recall from \eqref{eq-D_max_SDP} that
	\begin{equation}\label{eq-D_max_SDP_2}
		D_{\max}(\rho\Vert\sigma)=\log_2\inf\{\lambda:\rho\leq\lambda\sigma\},
	\end{equation}
	and recall from \eqref{eq-entr:beta-quant} that
	\begin{equation}
		D_H^{\varepsilon}(\rho\Vert\sigma) = -\log_2\inf\{\Tr[\Lambda\sigma]:0\leq\Lambda\leq\mathbbm{1},\, \Tr[\Lambda\rho]\geq 1-\varepsilon\}.
		\label{eq:E-meas:HTRE-SDP}
	\end{equation}
	As discussed earlier, both of these generalized divergences can be cast into the SDP standard forms in Definition~\ref{def-SDPs}, and thus their corresponding generalized divergence of entanglement can be formulated as a \textit{cone program}. A cone program is an optimization problem over a convex cone\footnote{A subset $C$ of a vector space is called a cone if $\alpha x\in C$ for every $x\in C$ and $\alpha>0$. A convex cone is one for for which $\alpha x+\beta y\in C$ for all $\alpha,\beta>0$ and $x,y\in C$.} with a convex objective function. An SDP is a special case of a cone program in which the convex cone is the set of positive semi-definite operators.
	
	The convex cone of interest here is the set $\widehat{\SEP}(A\!:\!B)$ of all separable operators, which we define as follows: $X_{AB}\in\widehat{\SEP}(A\!:\!B)$ if there exists a positive integer $\ell$ and positive semi-definite operators $\{P_{A}^{x}\}_{x=1}^{\ell}$ and $\{Q_{B}^{x}\}_{x=1}^{\ell}$ such that%
	\begin{equation}
		X_{AB}=\sum_{x=1}^{\ell }P_{A}^{x}\otimes Q_{B}^{x}.
	\end{equation}
	In what follows, we sometimes employ the shorthands $\SEP$ and $\widehat{\SEP}$ when the bipartition is clear from the context.
	
	We now show that the max-relative entropy of entanglement can be written as a cone program.

	\begin{proposition*}{Cone Program for  Max-Relative Entropy of Entanglement}{lem:SKA-alt-emax}
		Let $\rho_{AB}$ be a bipartite state. Then,
		\begin{equation}\label{eq-E_max_cone_prog_1}
			E_{\max}(A;B)_{\rho}=\log_2 G_{\max}(A;B)_{\rho},
		\end{equation}
		where
		\begin{equation}\label{eq-E_max_cone_prog_2}
			G_{\max}(A;B)_{\rho}\coloneqq \inf\left\{\Tr[X_{AB}]:\rho_{AB}\leq X_{AB},X_{AB}\in\widehat{\SEP}\right\}.
		\end{equation}
	\end{proposition*}

	\begin{Proof}
		Employing the expression in \eqref{eq-D_max_SDP_2}, we find that
		\begin{align}
			E_{\max}(A;B)_{\rho} &  \inf_{\sigma_{AB}\in\SEP(A:B)}D_{\max}(\rho_{AB}\Vert\sigma_{AB})\nonumber\\
			&  =\log_{2}\inf\{\mu:\rho_{AB}\leq\mu\sigma_{AB},\ \sigma_{AB}\in\SEP\}\\
			&  =\log_{2}\inf\left\{\Tr[X_{AB}]:\rho_{AB}\leq X_{AB},\ X_{AB}\in\widehat{\SEP}\right\},
		\end{align}
		as required, where in the last step we made the change of variable $\mu\sigma_{AB}\equiv X_{AB}$. Since $\sigma_{AB}\in\SEP(A\!:\!B)$ and $\mu\geq 0$, we have that $X_{AB}\in\widehat{\SEP}(A\!:\!B)$.
	\end{Proof}

	Next, we show that the hypothesis testing relative entropy of entanglement can be written as a cone program.
	
	\begin{proposition*}{Cone Program for  Hypothesis Testing Relative Entropy of Entanglement}{prop-hypo_test_rel_entropy_ent}
		Let $\rho_{AB}$ be a bipartite state. Then, for all $\varepsilon\in[0,1]$,
		\begin{multline}
			E_R^{\varepsilon}(A;B)_{\rho}= -\log_2\sup\{\mu(1-\varepsilon) - \Tr[Z_{AB}]: \mu\geq 0, Z_{AB}\geq 0, \\
			\sigma_{AB}\in\widehat{\SEP} ,\, \mu \rho_{AB} \leq \sigma_{AB} + Z_{AB}, \, \Tr[\sigma_{AB}]=1\}.
			\label{eq:E-meas:HTRE-REE-cone}
		\end{multline}
	\end{proposition*}
	
	\begin{Proof}
		This follows from the definition in \eqref{eq-hypo_test_rel_ent_entanglement} and		
		the dual formulation of the hypothesis testing relative entropy stated in \eqref{eq-hypo_test_rel_ent_dual}.
	\end{Proof}
	
Recall that $\SEP=\PPT$ in the case of qubit-qubit and qubit-qutrit states, which means that the optimizations in \eqref{eq-E_max_cone_prog_2} and \eqref{eq:E-meas:HTRE-REE-cone} are SDPs when $\rho_{AB}$ is either a two-qubit state or a qubit-qutrit state.

\section{Generalized Rains Divergence}\label{sec-ent_measures_Rains_dist}
	
	As explained in Section~\ref{sec-ent_meas_examples}, the generalized divergence of entanglement is in general complicated to compute because the set of separable states does not admit a simple characterization, making optimization over separable states difficult. By relaxing the set of separable states to the set $\PPT'$ defined in \eqref{eq:PPT-prime-set}, which has a simple characterization in terms of semi-definite constraints, we defined the generalized Rains divergence in \eqref{eq-gen_Rains_div_0}. Recall that
	\begin{equation}
		\PPT'(A\!:\!B)=\{\sigma_{AB}:\sigma_{AB}\geq 0,~\norm{\T_B(\sigma_{AB})}_1\leq 1\}.
	\end{equation}
	
	In this section, we investigate properties of the generalized Rains divergence. We start by recalling its definition.
	
	\begin{definition}{Generalized Rains Divergence of a Bipartite State}{def-gen_Rains_rel_ent}
		Let $\boldsymbol{D}$ be a generalized divergence (see Definition~\ref{def-gen_div}). For every bipartite state $\rho_{AB}$, we define the \textit{generalized Rains divergence of $\rho_{AB}$} as
		\begin{equation}\label{eq-gen_Rains_rel_ent_state}
			\boldsymbol{R}(A;B)_{\rho}\coloneqq \inf_{\sigma_{AB}\in\operatorname{PPT}'(A:B)}\boldsymbol{D}(\rho_{AB}\Vert\sigma_{AB}).
		\end{equation}
		If $\boldsymbol{D}$ is continuous in its second argument, then the infimum can be replaced by a minimum.
	\end{definition}
	
	Since $\SEP\subseteq\PPT'$, optimizing over states in $\PPT'$ can never lead to a value that is greater than the value obtained by optimizing over separable states. Therefore, as stated in \eqref{eq-SEP_PPT_Rains_ineq},
	\begin{equation}\label{eq-ent_meas_Rains_vs_sep}
		\boldsymbol{R}(A;B)_{\rho}\leq \boldsymbol{E}(A;B)_{\rho}
	\end{equation}
	for every state $\rho_{AB}$.
	
	We are particularly interested throughout the rest of this book in the following generalized Rains divergences for every state $\rho_{AB}$:
	\begin{enumerate}
		\item The \textit{Rains relative entropy of $\rho_{AB}$},
			\begin{equation}\label{eq-Rains_rel_ent}
				R(A;B)_{\rho}\coloneqq \inf_{\sigma_{AB}\in\operatorname{PPT}'(A:B)}D(\rho_{AB}\Vert\sigma_{AB}),
			\end{equation}
			where $D(\rho_{AB}\Vert\sigma_{AB})$ is the quantum relative entropy of $\rho_{AB}$ and $\sigma_{AB}$ (Definition~\ref{def-rel_ent}). For two-qubit states, it is known that the infimum in \eqref{eq-Rains_rel_ent} is achieved by a separable state, which means that $E_R(A;B)_{\rho}=R(A;B)_{\rho}$ for two-qubit states $\rho_{AB}$ (please consult the Bibliographic Notes in Section~\ref{sec:E-meas:bib-notes}).
		
		\item The \textit{$\varepsilon$-hypothesis testing Rains relative entropy of $\rho_{AB}$},
			\begin{equation}\label{def:eps-Rains-rel-entr}
				R_H^{\varepsilon}(A;B)_{\rho}\coloneqq\inf_{\sigma_{AB}\in\operatorname{PPT}'(A:B)}D_{H}^{\varepsilon}(\rho_{AB}\Vert\sigma_{AB}),
			\end{equation}
			where $D_H^{\varepsilon}(\rho_{AB}\Vert\sigma_{AB})$ is the $\varepsilon$-hypothesis testing relative entropy of $\rho_{AB}$ and $\sigma_{AB}$ (Definition~\ref{def-hypo_testing_rel_ent}).
			
		\item The \textit{sandwiched R\'{e}nyi Rains relative entropy of $\rho_{AB}$},
			\begin{equation}\label{eq-sand_renyi_Rains_rel_ent}
				\widetilde{R}_{\alpha}(A;B)_{\rho}\coloneqq\inf_{\sigma_{AB}\in\operatorname{PPT}'(A:B)}\widetilde{D}_{\alpha}(\rho_{AB}\Vert\sigma_{AB}),
			\end{equation}
			where $\widetilde{D}_{\alpha}(\rho_{AB}\Vert\sigma_{AB})$, $\alpha\in[\sfrac{1}{2},1)\cup(1,\infty)$, is the sandwiched R\'{e}nyi relative entropy of $\rho_{AB}$ and $\sigma_{AB}$ (Definition~\ref{def-sand_rel_ent}). Note that $\widetilde{R}_{\alpha}(A;B)_{\rho}$ is monotonically increasing in $\alpha$ for all $\rho_{AB}$ (see Proposition~\ref{prop-sand_rel_ent_properties}). This fact, along with the fact that $\lim_{\alpha\to 1}\widetilde{D}_{\alpha}=D$ (see Proposition~\ref{prop-sand_ren_ent_lim}), leads to
			\begin{equation}
				\lim_{\alpha\to 1}\widetilde{R}_{\alpha}(A;B)_{\rho}=R(A;B)_{\rho}
				\label{eq:E-meas:Rains-limit-alph-1}
			\end{equation}
			for every state $\rho_{AB}$. See Appendix~\ref{app-sand_ren_inf_limit} for details of the proof.
			
		\item The \textit{max-Rains relative entropy of $\rho_{AB}$},
			\begin{equation}
				R_{\max}(A;B)_{\rho}\coloneqq \inf_{\sigma_{AB}\in\operatorname{PPT}'(A:B)}D_{\max}(\rho_{AB}\Vert\sigma_{AB}),
				\label{eq:E-meas:max-Rains-def}
			\end{equation}
			where $D_{\max}(\rho_{AB}\Vert\sigma_{AB})$ is the max-relative entropy of $\rho_{AB}$ and $\sigma_{AB}$ (Definition~\ref{def-max_rel_ent}). Using the fact that $\lim_{\alpha\to\infty}\widetilde{D}_{\alpha}=D_{\max}$ (see Proposition~\ref{prop-sand_rel_ent_limit_max}), we find that
			\begin{equation}
				R_{\max}(A;B)_{\rho}=\lim_{\alpha\to\infty}\widetilde{R}_{\alpha}(A;B)_{\rho}
				\label{eq:E-meas:R-max-limit-alph-infty}
			\end{equation}
			for every state $\rho_{AB}$. See Appendix~\ref{app-sand_ren_inf_limit} for details of the proof. Due to this fact, along with the fact that $\widetilde{R}_{\alpha}(A;B)_{\rho}$ is monotonically increasing in $\alpha$ for all $\rho_{AB}$, we have that
			\begin{equation}
				R_{\max}(A;B)_{\rho}\geq \widetilde{R}_{\alpha}(A;B)_{\rho}
			\end{equation}
			for all $\alpha\in(1,\infty)$ and every state $\rho_{AB}$.
	\end{enumerate}
	
	The following inequalities relate the log-negativity, as defined in \eqref{eq:E-meas:def-log-negativity}, to the Rains relative entropy, the sandwiched R\'enyi Rains relative entropy, and the max-Rains relative entropy:
	\begin{proposition*}{Log-Negativity to Rains Relative Entropies}{prop:E-meas:log-neg-to-sand-Rains}
	For a bipartite state $\rho_{AB}$, the following inequalities hold:
	\begin{equation}
	R(A;B)_{\rho} \leq R_{\max}(A;B)_{\rho}  \leq E_N(A;B)_{\rho}.
	\label{eq:E-meas:log-neg-to-sand-Rains}
	\end{equation}
	Furthermore, for all $\alpha,\beta\in[\sfrac{1}{2},1)\cup(1,\infty)$ such that $\alpha < \beta$, we have that
	\begin{equation}
	\widetilde{R}_{\alpha}(A;B)_{\rho} \leq \widetilde{R}_{\beta}(A;B)_{\rho}.
	\label{eq:E-meas:ordering-sand-Rains}
	\end{equation}
	\end{proposition*}
	
	\begin{Proof}
	The inequality $R(A;B)_{\rho} \leq R_{\max}(A;B)_{\rho}$ and the inequality in \eqref{eq:E-meas:ordering-sand-Rains} are a direct consequence of the monotonicity in $\alpha$ of the sandwiched R\'enyi relative entropy (Proposition~\ref{prop-sand_rel_ent_properties}), as well as \eqref{eq:E-meas:Rains-limit-alph-1} and \eqref{eq:E-meas:R-max-limit-alph-infty}. 
	
	To see the inequality $R_{\max}(A;B)_{\rho}  \leq E_N(A;B)_{\rho}$, consider picking
	\begin{equation}
	\sigma_{AB}=\frac{\rho_{AB}}{  \norm{\T_B(\rho_{AB})}_1}
	\end{equation}
	in \eqref{eq:E-meas:max-Rains-def}. For this choice, we have that $\sigma_{AB}\geq 0$ and $\norm{\T_B(\sigma_{AB})}_1\leq 1$, so that $\sigma_{AB}\in \text{PPT}'(A\!:\!B)$. Thus 
	\begin{align}
	R_{\max}(A;B)_{\rho} & \leq D_{\max}(\rho_{AB}\Vert \sigma_{AB}) \\
	& = D_{\max}(\rho_{AB}\Vert \rho_{AB}) + \log_2\norm{\T_B(\rho_{AB})}_1 \\
	& = E_N(A;B)_{\rho}.
	\end{align}
	The first equality follows by direct evaluation using Definition~\ref{def-max_rel_ent}.
	The last equality follows because $D_{\max}(\rho_{AB}\Vert \rho_{AB})=0$ and from the definition in \eqref{eq:E-meas:def-log-negativity}.
	\end{Proof}

	\begin{proposition*}{Properties of Generalized Rains Divergence}{prop-gen_Rains_rel_ent_properties}
		Let $\boldsymbol{D}$ be a generalized divergence, and consider the generalized Rains divergence $\boldsymbol{R}(A;B)_{\rho}$ of a state $\rho_{AB}$ as defined in \eqref{eq-gen_Rains_rel_ent_state}.
		
		\begin{enumerate}
			\item \textit{PPT monotonicity}: For every completely PPT-preserving channel $\mathcal{P}_{AB\to A'B'}$,
				\begin{equation}
					\boldsymbol{R}(A;B)_{\rho}\geq\boldsymbol{R}(A';B')_{\omega},
				\end{equation}
				where $\omega_{A'B'}=\mathcal{P}_{AB\to A'B'}(\rho_{AB})$. In other words, the generalized Rains divergence is monotonically non-increasing under completely PPT-preserving channels. Since every LOCC channel is a completely PPT-preserving channel (Propositions~\ref{prop:QM-over:sep-ch-contains-LOCC} and \ref{prop-LOCC_PPT_preserving}), the generalized Rains divergence is monotone non-increasing under LOCC channels, and thus it is an entanglement measure as per Definition~\ref{def-LAQC:ent-measure}.
	
			\item \textit{Subadditivity}: If $\boldsymbol{D}$ is additive for product positive semi-definite operators, i.e., $\boldsymbol{D}(\rho\otimes\omega\Vert\sigma\otimes\tau)=\boldsymbol{D}(\rho\Vert\sigma)+\boldsymbol{D}(\omega\Vert\tau)$, then for every two quantum states $\rho_{A_1B_1}$ and $\omega_{A_2B_2}$ the generalized Rains relative entropy is subadditive:
				\begin{equation}\label{eq:LAQC-Renyi-Rains-subadditivity}
					\boldsymbol{R}(A_1A_2;B_1B_2)_{\rho\otimes\omega}\leq\boldsymbol{R}(A_1;B_1)_{\rho}+\boldsymbol{R}(A_2;B_2)_{\omega}.
				\end{equation}
			
			\item \textit{Convexity}: If $\boldsymbol{D}$ is jointly convex, meaning that
				\begin{equation}
					\boldsymbol{D}\!\left(\sum_{x\in\mathcal{X}}p(x)\rho_{AB}^x\Bigg\Vert\sum_{x\in\mathcal{X}}p(x)\sigma_{AB}^x\right)\leq\sum_{x\in\mathcal{X}}p(x)\boldsymbol{D}(\rho_{AB}^x\Vert\sigma_{AB}^x)
				\end{equation}
				for every finite alphabet $\mathcal{X}$, probability distribution $p:\mathcal{X}\to[0,1]$, set $\{\rho_{AB}^x\}_{x\in\mathcal{X}}$ of states, set $\{\sigma_{AB}^x\}_{x\in\mathcal{X}}$ of positive semi-definite operators, then the generalized Rains divergence is convex:
				\begin{equation}
					\boldsymbol{R}(A;B)_{\overline{\rho}}\leq\sum_{x\in\mathcal{X}}p(x)\boldsymbol{R}(A;B)_{\rho^x},
				\end{equation}
				where $\overline{\rho}_{AB}=\sum_{x\in\mathcal{X}}p(x)\rho_{AB}^x$.
		\end{enumerate}
		
		Properties 1.~and 2.~are satisfied when the generalized divergence is the quantum relative entropy, the Petz--, sandwiched, and geometric R\'{e}nyi relative entropies, and the max-relative entropy. Property 3.~is satisfied when the generalized divergence is the quantum relative entropy and the Petz--, sandwiched, and geometric R\'{e}nyi relative entropies for the range of $ \alpha <1$ for which data processing holds.
	\end{proposition*}
	
	\begin{remark}
		Note that the generalized Rains divergence is generally not a faithful entanglement measure. Although $\boldsymbol{R}(A;B)_{\rho}=0$ for all separable states $\rho_{AB}$ due to the containment $\SEP(A\!:\!B)\subseteq\PPT'(A\!:\!B)$, the converse statement is not generally true because the infimum in the definition of $\boldsymbol{R}(A;B)_{\rho}$ is not generally achieved by a separable state.
	\end{remark}
	
	\begin{Proof}
		\hfill\begin{enumerate}
			\item For $\omega_{A'B'}=\mathcal{P}_{AB\to A'B'}(\rho_{AB})$, we have by definition,
				\begin{align}
					\boldsymbol{R}(A';B')_{\omega}&=\inf_{\tau_{A'B'}\in\text{PPT}'(A':B')}\boldsymbol{D}(\omega_{A'B'}\Vert\tau_{A'B'})\\
					&=\inf_{\tau_{A'B'}\in\text{PPT}'(A':B')}\boldsymbol{D}(\mathcal{P}_{AB\to A'B'}(\rho_{AB})\Vert\tau_{A'B'}).\label{eq-eps_Rains_state_PPT_monotone_pf}
				\end{align}
				Now, recall from Lemma \ref{prop-PPT_prime_properties} that the set $\text{PPT}'$ is closed under completely PPT-preserving channels. Based on this, it follows that the output operators of the completely PPT-preserving channel $\mathcal{P}_{AB\to A'B'}$ are in the set $\text{PPT}'(A'\!:\!B')$. In other words, we have
				\begin{equation}
					\{\mathcal{P}_{AB\to A'B'}(\sigma_{AB}):\sigma_{AB}\in\text{PPT}'(A\!:\!B)\}\subseteq \text{PPT}'(A'\!:\!B').
				\end{equation}
				Therefore, restricting the optimization in \eqref{eq-eps_Rains_state_PPT_monotone_pf} to this set leads to
				\begin{align}
					\boldsymbol{R}(A';B')_{\omega}&\leq \inf_{\sigma_{AB}\in\text{PPT}'(A:B)}\boldsymbol{D}(\mathcal{P}_{AB\to A'B'}(\rho_{AB})\Vert\mathcal{P}_{AB\to A'B'}(\sigma_{AB}))\\
					&\leq \inf_{\sigma_{AB}\in\text{PPT}'(A:B)}\boldsymbol{D}(\rho_{AB}\Vert\sigma_{AB})\\
					&=\boldsymbol{R}(A;B)_{\rho},
				\end{align}
				as required, where to obtain the second inequality we used the data-processing inequality for the generalized divergence.
				
			\item By definition, the optimization in the definition of $\boldsymbol{R}(A_1A_2;B_1B_2)_{\rho\otimes\omega}$ is over the set
				\begin{multline}
					\operatorname{PPT}'(A_1A_2\!:\!B_1B_2)\\=\{\sigma_{A_1A_2B_1B_1}:\sigma_{A_1A_2B_1B_2}\geq 0,\norm{\T_{B_1B_2}(\sigma_{A_1A_2B_1B_2})}_1\leq 1\},
				\end{multline}
				which contains operators of the form $\sigma_{A_1A_2B_1B_2}=\xi_{A_1B_1}\otimes\tau_{A_2B_2}$, where $\xi_{A_1B_1}\in\operatorname{PPT}'(A_1\!:\!B_1)$ and $\tau_{A_2B_2}\in\operatorname{PPT}'(A_2\!:\!B_2)$. By restricting the optimization to such tensor product operators, and by using the additivity of the generalized divergence $\boldsymbol{D}$, we obtain
		\begin{align}
			\boldsymbol{R}(A_{1}A_{2};B_{1}B_{2})_{\rho\otimes\omega}  & \leq\boldsymbol{D}(\rho_{A_{1}B_{1}}\otimes\omega_{A_{2}B_{2}}\Vert\xi_{A_{1}B_{1}}\otimes\tau_{A_{2}B_{2}})\\
			&  =\boldsymbol{D}(\rho_{A_{1}B_{1}}\Vert\xi_{A_{1}B_{1}})+\boldsymbol{D}(\omega_{A_{2}B_{2}}\Vert\tau_{A_{2}B_{2}}).
		\end{align}
		Since $\xi_{A_{1}B_{1}}\in\operatorname{PPT}'(A_{1}\!:\!B_{1})$ and $\tau_{A_{2}B_{2}}\in\operatorname{PPT}'(A_{2}\!:\!B_{2})$ are arbitrary, the inequality in \eqref{eq:LAQC-Renyi-Rains-subadditivity} follows.
		
		\item We have
			\begin{equation}
				\boldsymbol{R}(A;B)_{\overline{\rho}}=\inf_{\sigma_{AB}\in\PPT'(A:B)}\boldsymbol{D}\!\left(\sum_{x\in\mathcal{X}}p(x)\rho_{AB}^x\Bigg\Vert\sigma_{AB}\right).
			\end{equation}
			Let us restrict the optimization over all $\PPT'$ operators to an optimization over sets $\{\sigma_{AB}^x\}_{x\in\mathcal{X}}$ of $\PPT'$ operators indexed by the alphabet $\mathcal{X}$. Then, because $\PPT'(A\!:\!B)$ is a convex set, we have that $\sum_{x\in\mathcal{X}}p(x)\sigma_{AB}^x\in\PPT'(A\!:\!B)$. Therefore, using the joint convexity of $\boldsymbol{D}$, we obtain
			\begin{align}
				\boldsymbol{R}(A;B)_{\overline{\rho}}&\leq \inf_{\{\sigma_{AB}^x\}_x\subset\PPT'(A:B)}\boldsymbol{D}\!\left(\sum_{x\in\mathcal{X}}p(x)\rho_{AB}^x\Bigg\Vert\sum_{x\in\mathcal{X}}p(x)\sigma_{AB}^x\right)\\
				&\leq \inf_{\{\sigma_{AB}^x\}_x\subset\PPT'(A:B)}\sum_{x\in\mathcal{X}}p(x)\boldsymbol{D}(\rho_{AB}^x\Vert\sigma_{AB}^x)\\
				&\leq\sum_{x\in\mathcal{X}}p(x)\inf_{\sigma_{AB}^x\in\PPT'(A:B)}\boldsymbol{D}(\rho_{AB}^x\Vert\sigma_{AB}^x)\\
				&=\sum_{x\in\mathcal{X}}p(x)\boldsymbol{R}(A;B)_{\rho^x},
			\end{align}
			as required. \qedhere
		\end{enumerate}
	\end{Proof}
	
	By Proposition~\ref{prop-gen_Rains_rel_ent_properties}, the sandwiched R\'{e}nyi Rains relative entropy $\widetilde{R}_{\alpha}$ is convex for $\alpha\in\left[\sfrac{1}{2},1\right)$. Although convexity does not hold for $\alpha>1$, we do have quasi-convexity, which we now prove.

	\begin{proposition*}{Quasi-Convexity of R\'{e}nyi Rains Relative Entropy}{prop-ren_rains_rel_ent_q_conv}
		Let $p:\mathcal{X}\to[0,1]$ be a probability distribution over a finite alphabet $\mathcal{X}$, and let $\{\rho_{AB}^x\}_{x\in\mathcal{X}}$ be a set of states. Then, for all $\alpha\in(1,\infty)$,
		\begin{equation}
			\widetilde{R}_{\alpha}(A;B)_{\overline{\rho}}\leq \max_{x\in\mathcal{X}} \widetilde{R}_{\alpha}(A;B)_{\rho^x},
		\end{equation}
		where $\overline{\rho}_{AB}=\sum_{x\in\mathcal{X}}p(x)\rho_{AB}^x$.
	\end{proposition*}
	
	\begin{Proof}
Let us recall the definition of the quasi-sandwiched R\'enyi relative entropy from Definition~\ref{def-sand_rel_ent}:
\begin{equation}
\widetilde{Q}_{\alpha}(\rho\|\sigma)\coloneqq\Tr\!\left[\left(\sigma^{\frac{1-\alpha}{2\alpha}}\rho\sigma^{\frac{1-\alpha}{2\alpha}}\right)^{\alpha}\right].
\end{equation}
This function is jointly convex in $\rho$ and $\sigma$
for $\alpha>1$ (recall Proposition~\ref{prop-sand_rel_ent_joint_convex}). We now prove that the function
\begin{equation}
\rho_{AB}\mapsto\widetilde{Q}_{\alpha,\PPT'}(\rho_{AB})\coloneqq\inf_{\sigma_{AB}\in\PPT'(A:B)}\widetilde{Q}_{\alpha}(\rho_{AB}\|\sigma_{AB})
\end{equation}
is convex. To see this, pick arbitrary $\sigma_{AB}^{1},\sigma_{AB}^{2}\in\PPT'(A:B)$
and consider that
\begin{align}
& \widetilde{Q}_{\alpha,\PPT'}(\lambda\rho_{AB}^{1}+\left(1-\lambda\right)\rho_{AB}^{2}) \notag \\
& \leq\widetilde{Q}_{\alpha}(\lambda\rho_{AB}^{1}+\left(1-\lambda\right)\rho_{AB}^{2}\|\lambda\sigma_{AB}^{1}+\left(1-\lambda\right)\sigma_{AB}^{2})\\
 & \leq\lambda\widetilde{Q}_{\alpha}(\rho_{AB}^{1}\|\sigma_{AB}^{1})+\left(1-\lambda\right)\widetilde{Q}_{\alpha}(\rho_{AB}^{2}\|\sigma_{AB}^{2}),
\end{align}
where the second inequality follows from joint convexity.
Since the inequality holds for all $\sigma_{AB}^{1},\sigma_{AB}^{2}\in\PPT'(A:B)$,
we conclude that
\begin{equation}
\widetilde{Q}_{\alpha,\PPT'}(\lambda\rho_{AB}^{1}+\left(1-\lambda\right)\rho_{AB}^{2})\leq\lambda\widetilde{Q}_{\alpha,\PPT'}(\rho_{AB}^{1})+\left(1-\lambda\right)\widetilde{Q}_{\alpha,\PPT'}(\rho_{AB}^{2}),
\end{equation}
thus justifying that $\rho_{AB}\mapsto\widetilde{Q}_{\alpha,\PPT'}(\rho_{AB}) $ is convex. Now consider that
\begin{align}
& \widetilde{R}_{\alpha}(\lambda\rho_{AB}^{1}+\left(1-\lambda\right)\rho_{AB}^{2}) \notag \\
& =\frac{1}{\alpha-1}\log_{2}\widetilde{Q}_{\alpha,\PPT'}(\lambda\rho_{AB}^{1}+\left(1-\lambda\right)\rho_{AB}^{2})\\
 & \leq\frac{1}{\alpha-1}\log_{2}\left[\lambda\widetilde{Q}_{\alpha,\PPT'}(\rho_{AB}^{1})+\left(1-\lambda\right)\widetilde{Q}_{\alpha,\PPT'}(\rho_{AB}^{2})\right]\\
 & \leq\frac{1}{\alpha-1}\log_{2}\left[\max_{x\in\left\{ 1,2\right\} }\widetilde{Q}_{\alpha,\PPT'}(\rho_{AB}^{x})\right]\\
 & \leq\max_{x\in\left\{ 1,2\right\} }\frac{1}{\alpha-1}\log_{2}\widetilde{Q}_{\alpha,\PPT'}(\rho_{AB}^{x})\\
 & =\max_{x\in\left\{ 1,2\right\} }\widetilde{R}_{\alpha}(\rho_{AB}^{x}),
\end{align}
thus establishing the claim.
	\end{Proof}

\subsection{Semi-Definite Program Formulations}\label{subsec-max_Rains_rel_ent_state}

	One of the advantages of using the generalized Rains divergence as an entanglement measure is that the set $\PPT'$ involved in its definition has a simple characterization in terms of semi-definite constraints. Indeed, let us recall that
	\begin{equation}\label{eq-PPT_prime_set_max_Rains}
		\PPT'(A\!:\!B)=\{\sigma_{AB}:\sigma_{AB}\geq 0,~\norm{\T_B(\sigma_{AB})}_1\leq 1\}.
	\end{equation}
	Using the expression in \eqref{eq:E-meas:SDP-dual-log-neg} for $\norm{\T_B(\sigma_{AB})}_1$, this set can equivalently be written as follows:
	\begin{multline}
		\PPT'(A\!:\!B)=\{\sigma_{AB}:\sigma_{AB}\geq 0,~\exists K_{AB}, L_{AB}\geq 0 \text{ such that } \\
		\Tr[K_{AB}+ L_{AB}]\leq 1 ,\, \T_B(K_{AB}-L_{AB}) = \sigma_{AB}\},
	\end{multline}
	which can be further simplified to
	\begin{multline}
		\PPT'(A\!:\!B)=\{\T_B(K_{AB}-L_{AB}):\T_B(K_{AB}-L_{AB})\geq 0, \\K_{AB}, L_{AB}\geq 0, \,
		\Tr[K_{AB}+ L_{AB}]\leq 1 \},
		\label{eq:E-meas:SDP-PPT-prime}
	\end{multline}	
	In this section, we show how these  characterizations of the set $\PPT'$ allow us to compute both the max-Rains relative entropy and the hypothesis testing Rains relative entropy via semi-definite programs (Section~\ref{sec-SDPs}).

	We first consider the max-Rains relative entropy, which we recall is defined as
	\begin{equation}
		R_{\max}(A;B)_{\rho}=\inf_{\sigma_{AB}\in\PPT'(A:B)}D_{\max}(\rho_{AB}\Vert\sigma_{AB}).
	\end{equation}
	Let us also recall from \eqref{eq-D_max_SDP} that $D_{\max}(\rho_{AB}\Vert\sigma_{AB})$ can be written as follows:
	\begin{equation}
		D_{\max}(\rho_{AB}\Vert\sigma_{AB})=\log_2\inf\{\lambda:\rho_{AB}\leq\lambda\sigma_{AB}\}.
	\end{equation}
	As shown in the discussion after \eqref{eq-D_max_SDP}, the optimization in the  equation above is a semi-definite program (SDP). This, along with the definition in \eqref{eq-PPT_prime_set_max_Rains} of the set $\PPT'(A\!:\!B)$, leads to the following SDP formulation for~$R_{\max}$.

	\begin{proposition*}{SDPs for the Max-Rains Relative Entropy}{prop-max_Rains_rel_ent_SDP}
		Let $\rho_{AB}$ be a bipartite state. Then the max-Rains relative entropy can be written as%
		\begin{equation}\label{eq:LAQC-Rains-to-W}%
			R_{\max}(A;B)_{\rho}=\log_{2}W_{\max}(A;B)_{\rho}, 
		\end{equation}
		where
				\begin{align}
			&\!\!\!\! W_{\max}(A;B)_{\rho}\notag \\
			&=
			\inf_{K_{AB},L_{AB}\geq 0} \{ \Tr[K_{AB}+L_{AB}] : \T_B[K_{AB}-L_{AB}]\geq\rho_{AB}\},
			\label{eq:LAQC-primal-form-W-Rains} \\
			&=\sup_{Y_{AB}\geq 0} \{\Tr[Y_{AB}\rho_{AB}] : \norm{\T_B[Y_{AB}]}_{\infty}\leq 1\} .\label{eq:LAQC-dual-form-W-Rains}
		\end{align}
	\end{proposition*}

	\begin{Proof}
		First, we establish the equality for $W(A;B)_{\rho}$ in \eqref{eq:LAQC-primal-form-W-Rains}. Due to the fact that the infimum over $\PPT'$ operators in the definition of $R_{\max}$ can be achieved, we have that
		\begin{align}
			R_{\max}(A;B)_{\rho}&=\inf_{\sigma_{AB}\in\PPT'(A:B)}D_{\max}(\rho_{AB}\Vert\sigma_{AB})\\
			&=\inf_{\sigma_{AB}\in\PPT'(A:B)}\log_2\inf\{\lambda:\rho_{AB}\leq\lambda\sigma_{AB}\}\\
			&=\log_2\inf\{\lambda:\rho_{AB}\leq\lambda\sigma_{AB},~\sigma_{AB}\geq 0,~\norm{\T_B[\sigma_{AB}]}_1\leq 1\}.
		\end{align}
		The constraint $\rho_{AB}\leq\lambda\sigma_{AB}$ implies (by taking the trace on both sides of the inequality) that $\lambda\geq 1$. This in turn implies that $\lambda\sigma_{AB}\geq 0$ and that $\norm{\T_B[\lambda\sigma_{AB}]}_1\leq\lambda$. So we have
		\begin{align}
			R_{\max}(A;B)_{\rho}&=\log_2\inf\{\lambda:\rho_{AB}\leq\lambda\sigma_{AB},~\lambda\sigma_{AB}\geq 0,~\norm{\T_B[\lambda\sigma_{AB}]}_1\leq\lambda\}
		\end{align}
		Let us now make the change of variable $S_{AB}\equiv\lambda\sigma_{AB}$. We then have
		\begin{align}
			R_{\max}(A;B)_{\rho}&=\log_2\inf\{\lambda :\rho_{AB}\leq S_{AB},~S_{AB}\geq 0,~\norm{\T_B[S_{AB}]}_1\leq\lambda\}\\
			&=\log_2\inf\{\norm{\T_B[S_{AB}]}_1:\rho_{AB}\leq S_{AB}\},
		\end{align}
		where to obtain the last line we eliminated the constraint $S_{AB}\geq 0$ (because it is implied by $\rho_{AB}\leq S_{AB}$) and we used the fact that $\mu\geq\norm{\T_B[S_{AB}]}_1$, meaning that the smallest value of $\lambda$ is $\norm{\T_B[S_{AB}]}_1$.

		Now, for an arbitrary operator $S_{AB}$ satisfying $\rho_{AB}\leq S_{AB}$, recall from \eqref{eq-MT:Jordan-Hahn} that the Jordan--Hahn decomposition of $\T_{B}[S_{AB}]$ is given by $\T_{B}(S_{AB})=K_{AB}-L_{AB}$, with $K_{AB},L_{AB}\geq0$ and $K_{AB}L_{AB}=0$. We then find that%
		\begin{align}
			\norm{\T_{B}[S_{AB}]}_{1}&=\Tr[K_{AB}+L_{AB}],\\
			S_{AB}&=\T_{B}[K_{AB}-L_{AB}].
		\end{align}
		The following inequality thus holds:
		\begin{multline}\label{eq:LAQC-max-Rains-to-dual-1}
			\inf\{\Tr[K_{AB}+L_{AB}]:\rho_{AB}\leq \T_{B}[K_{AB}-L_{AB}],K_{AB},L_{AB}\geq0\}\\\leq \inf\{\left\Vert\T_{B}(S_{AB})\right\Vert _{1}:\rho_{AB}\leq S_{AB}\}.
		\end{multline}
		To see the opposite inequality, let $K_{AB}$ and $L_{AB}$ be arbitrary operators such that $K_{AB},L_{AB}\geq 0$ and $\rho_{AB}\leq\T_{B}[K_{AB}-L_{AB}]$. Then, setting $S_{AB}=\T_{B}[K_{AB}-L_{AB}]$, we find that $\rho_{AB}\leq S_{AB}$ and%
		\begin{align}
			\norm{\T_{B}(S_{AB})}_{1}&=\norm{K_{AB}-L_{AB}}_{1}\\
			&  \leq\norm{K_{AB}}_{1}+\norm{L_{AB}}_{1}\\
			&  =\Tr[K_{AB}+L_{AB}],
		\end{align}
		where we used the triangle inequality. This implies that%
		\begin{multline}\label{eq:LAQC-max-Rains-to-dual-2}
			\inf\{\left\Vert\T_{B}(S_{AB})\right\Vert _{1}:\rho_{AB}\leq S_{AB}\}\leq\\
		\inf\{\Tr[K_{AB}+L_{AB}]:\rho_{AB}\leq\T_{B}(K_{AB}-L_{AB}),K_{AB},L_{AB}\geq0\}.
		\end{multline}
		Putting together \eqref{eq:LAQC-max-Rains-to-dual-1} and
\eqref{eq:LAQC-max-Rains-to-dual-2}, we conclude that $W_{\max}(A;B)_{\rho}$ is given by the equality in \eqref{eq:LAQC-primal-form-W-Rains}.

		To see the equality in \eqref{eq:LAQC-dual-form-W-Rains}, we employ semi-definite programming duality (see Section~\ref{sec-SDPs}). We first put \eqref{eq:LAQC-primal-form-W-Rains} into standard form (see Definition~\ref{def-SDPs}) as follows:%
		\begin{equation}
			\inf_{X\geq0}\{\Tr[CX]:\Phi(X)\geq D\},
		\end{equation}
		with%
		\begin{align}
			X  &  =%
			\begin{pmatrix}
			K_{AB} & 0\\
			0 & L_{AB}%
			\end{pmatrix}
			,\quad C=%
			\begin{pmatrix}
			\mathbbm{1}_{AB} & 0\\
			0 & \mathbbm{1}_{AB}%
			\end{pmatrix},\\
			\Phi(X)  &  =\T_{B}[K_{AB}-L_{AB}],\quad D=\rho_{AB}.
		\end{align}
		The dual program is then given by%
		\begin{equation}\label{eq:LAQC-dual-max-Rains-standard-form}
			\sup_{Y\geq0} \{\Tr[DY]:\Phi^{\dag}(Y)\leq C\}.
		\end{equation}
		In order to determine the adjoint $\Phi^{\dagger}$ of $\Phi$, consider that
		\begin{align}
			\Tr[\Phi(X)Y]&=\Tr[\T_{B}[K_{AB}-L_{AB}]Y_{AB}]\\
			&  =\Tr[(K_{AB}-L_{AB})\T_{B}[Y_{AB}]]\\
			&  =\Tr\left[\begin{pmatrix}
				K_{AB} & 0\\
				0 & L_{AB}%
				\end{pmatrix}%
				\begin{pmatrix}
				\T_{B}[Y_{AB}] & 0\\
				0 & -\T_{B}[Y_{AB}]
				\end{pmatrix} \right]  .
		\end{align}
		Therefore,
		\begin{equation}
			\Phi^{\dag}(Y)=\begin{pmatrix} \T_{B}[Y_{AB}] & 0\\ 0 & -\T_{B}[Y_{AB}]\end{pmatrix}
		\end{equation}
		so that $\Phi^{\dagger}(Y)\leq C$ is equivalent to
		\begin{equation}%
			\begin{pmatrix}
			\T_{B}[Y_{AB}] & 0\\
			0 & -\T_{B}[Y_{AB}]
			\end{pmatrix}
			\leq%
			\begin{pmatrix}
			\mathbbm{1}_{AB} & 0\\
			0 & \mathbbm{1}_{AB}%
			\end{pmatrix}.
		\end{equation}
		This is equivalent to the condition $-\mathbbm{1}_{AB}\leq \T_{B}[Y_{AB}]\leq \mathbbm{1}_{AB}$, which is equivalent to $\norm{\T_B[Y_{AB}]}_{\infty}\leq 1$ (see \eqref{eq:math-tools:example-SDP-sd-eq}). We thus conclude that \eqref{eq:LAQC-dual-max-Rains-standard-form} is equal to \eqref{eq:LAQC-dual-form-W-Rains}.

	To arrive at the equality in \eqref{eq:LAQC-primal-form-W-Rains}--\eqref{eq:LAQC-dual-form-W-Rains}, we need to verify that strong duality holds, and so we check the conditions of Theorem~\ref{thm:math-tools:slater-cond}. For the dual program in \eqref{eq:LAQC-dual-form-W-Rains}, pick $Y_{AB}=\frac{1}{2}\mathbbm{1}_{AB}$. Then the constraints in \eqref{eq:LAQC-primal-form-W-Rains} are strict, because $Y_{AB}>0$ and $\norm{\T_{B}(Y_{AB})}_{\infty}<1$ for this choice. Furthermore, for the primal in \eqref{eq:LAQC-primal-form-W-Rains}, pick $K_{AB}$ and $L_{AB}$ equal to the positive and negative parts of $\T_B(\rho_{AB})$, respectively. These are positive semi-definite by definition and satisfy $\rho_{AB} = \T_B(K_{AB}-L_{AB})$, so that they are feasible for the primal.
	\end{Proof}

It is worthwhile to observe the close connection of the SDP in \eqref{eq:LAQC-primal-form-W-Rains}, related to the max-Rains relative entropy, to that in \eqref{eq:E-meas:SDP-dual-log-neg}, related to the log-negativity. In fact, the SDP in \eqref{eq:LAQC-primal-form-W-Rains} is a relaxation of that in \eqref{eq:E-meas:SDP-dual-log-neg}, and this leads to an alternate proof of the rightmost inequality from \eqref{eq:E-meas:log-neg-to-sand-Rains}:
\begin{equation}
R_{\max}(A;B)_{\rho} \leq E_N(A;B)_{\rho},
\end{equation}
holding for every bipartite state $\rho_{AB}$.

Furthermore, the form of the SDP in \eqref{eq:LAQC-primal-form-W-Rains}, along with the same proof given for Proposition~\ref{prop:E-meas:neg-log-neg-E-mono}, except with \eqref{eq:E-meas:neg-K-L-constr} and \eqref{eq:E-meas:eq-constr-omega-x-log-neg-mono} replaced with inequality constraints, leads to the following conclusion:

\begin{proposition*}{Max-Rains Relative Entropy is a Selective PPT Monotone}{prop:E-meas:max-rains-strong-ppt-mono}
The max-Rains relative entropy is a selective PPT monotone; i.e., \eqref{eq:E-meas:strong-PPT-monotone} holds with $E$ set to $R_{\max}$.
\end{proposition*}

	Due to the additivity of $D_{\max}$,  Proposition~\ref{prop-gen_Rains_rel_ent_properties} implies that $R_{\max}$ is subadditive:
	\begin{equation}\label{eq-R_max_subadditivity}
		R_{\max}(A_1A_2;B_1B_2)_{\rho\otimes\omega}\leq R_{\max}(A_1;B_1)_{\rho}+R_{\max}(A_2;B_2)_{\omega},
	\end{equation}
	where the inequality holds for all states $\rho_{A_1B_1}$ and $\omega_{A_2B_2}$. 
	Using the dual formulation in \eqref{eq:LAQC-primal-form-W-Rains} for $R_{\max}$, we find that the reverse inequality also holds, implying that $R_{\max}$ is an additive entanglement measure.
	
	\begin{proposition*}{Additivity of Max-Rains Relative Entropy}{prop:LAQC-add-max-Rains-rel-ent}
		Let $\rho_{A_{1}B_{1}}$ and $\omega_{A_{2}B_{2}}$ be quantum states. Then,
		\begin{equation}
			R_{\max}(A_{1}A_{2};B_{1}B_{2})_{\rho\otimes\omega}=R_{\max}(A_{1};B_{1})_{\rho}+R_{\max}(A_{2};B_{2})_{\omega}.
		\end{equation}
	\end{proposition*}

	\begin{Proof}
		We prove the inequality reverse to the one in \eqref{eq-R_max_subadditivity}. To this end, we employ the dual formulation of $R_{\max}$ in \eqref{eq:LAQC-dual-form-W-Rains}. Let $Y_{A_{1}B_{1}}$ and $S_{A_{2}B_{2}}$ be arbitrary operators satisfying%
		\begin{align}
			\norm{\T_{B_{1}}[Y_{A_{1}B_{1}}]}_{\infty}  &  \leq 1,\quad  Y_{A_{1}B_{1}}\geq 0,\label{eq:LAQC-constraints-max-Rains-superadd-1}\\
			\norm{\T_{B_{2}}[S_{A_{2}B_{2}}]}_{\infty}  &  \leq 1,\quad S_{A_{2}B_{2}}\geq 0. \label{eq:LAQC-constraints-max-Rains-superadd-2}%
		\end{align}
		Then it follows from multiplicativity of the Schatten $\infty$-norm under tensor products (see \eqref{eq-Schatten_norm_mult}) that%
		\begin{align}
			\norm{\T_{B_{1}B_{2}}[Y_{A_{1}B_{1}}\otimes S_{A_{2}B_{2}}]}_{\infty} &=\norm{\T_{B_{1}}[Y_{A_{1}B_{1}}]\otimes \T_{B_{2}}[S_{A_{2}B_{2}}]}_{\infty}\\
			&=\norm{\T_{B_{1}}[Y_{A_{1}B_{1}}]}_{\infty}\norm{\T_{B_{2}}[S_{A_{2}B_{2}}]}_{\infty}\\
			&\leq 1.
		\end{align}
		Furthermore, we have that $Y_{A_{1}B_{1}}\otimes S_{A_{2}B_{2}}\geq0$. So it follows that%
		\begin{align}
			&  \log_{2}\Tr[Y_{A_{1}B_{1}}\rho_{A_{1}B_{1}}]+\log
_{2}\Tr[S_{A_{2}B_{2}}\omega_{A_{2}B_{2}}]\nonumber\\
			&  =\log_{2}\!\left(  \Tr[Y_{A_{1}B_{1}}\rho_{A_{1}B_{1}}] \Tr[S_{A_{2}B_{2}}\omega_{A_{2}B_{2}}]\right) \\
			&  =\log_{2}\!\left(  \Tr[\left(  Y_{A_{1}B_{1}}\otimes S_{A_{2}B_{2}}\right)\left(  \rho_{A_{1}B_{1}}\otimes \omega_{A_{2}B_{2}}\right)    ]\right) \\
			&  \leq R_{\max}(A_{1}A_{2};B_{1}B_{2})_{\rho\otimes\omega}.
		\end{align}
		The inequality follows because $Y_{A_{1}B_{1}}\otimes S_{A_{2}B_{2}}$ is a particular operator satisfying the constraints in \eqref{eq:LAQC-primal-form-W-Rains}\ for $R_{\max}(A_{1}A_{2};B_{1}B_{2})_{\rho\otimes\omega}$. Since the inequality holds for all $Y_{A_{1}B_{1}}$ and $S_{A_{2}B_{2}}$ satisfying \eqref{eq:LAQC-constraints-max-Rains-superadd-1}--\eqref{eq:LAQC-constraints-max-Rains-superadd-2}, the superadditivity inequality%
		\begin{equation}
			R_{\max}(A_{1}A_{2};B_{1}B_{2})_{\rho\otimes\omega}\geq R_{\max}(A_{1};B_{1})_{\rho}+R_{\max}(A_{2};B_{2})_{\omega}%
		\end{equation}
		follows.
	\end{Proof}

	We now consider the hypothesis testing Rains relative entropy, which is defined as
	\begin{equation}
		R_H^{\varepsilon}(A;B)_{\rho}=\inf_{\sigma_{AB}\in\PPT'(A:B)}D_H^{\varepsilon}(\rho_{AB}\Vert\sigma_{AB}),
	\end{equation}
	for $\varepsilon\in[0,1]$. Recall the primal and dual formulations of $D_H^{\varepsilon}$ from  Proposition~\ref{prop-hypo_test_rel_ent_dual}. 
	Using the dual formulation, we obtain the following:
	
	\begin{proposition*}{SDP for Hypothesis Testing Rains Relative Entropy}{prop-hypo_test_Rains_rel_ent_SDP}
		Let $\rho_{AB}$ be a bipartite state. Then, the hypothesis testing Rains relative entropy can be written as
		\begin{equation}
			R_H^{\varepsilon}(A;B)_{\rho}=-\log_2 W_H^{\varepsilon}(A;B)_{\rho},
		\end{equation}
		for all $\varepsilon\in[0,1]$, where
		\begin{align}
		\label{eq-hypo_test_Rains_rel_ent_SDP}
			& W_H^{\varepsilon}(A;B)_{\rho} \notag \\
			& \coloneqq \sup_{\substack{\mu\geq 0, Z_{AB},  \\K_{AB}, L_{AB} \geq 0}} \{ \mu(1-\varepsilon) - \Tr[Z_{AB}] : 
			\mu \rho_{AB} \leq \T_B(K_{AB} -L_{AB}) + Z_{AB}, \notag \\
			& \qquad\qquad\qquad\qquad \T_B(K_{AB} -L_{AB}) \geq 0,\,
			 \Tr[K_{AB} + L_{AB}] \leq 1
			  \} \\
			  & = \inf_{M_{AB},N_{AB}\geq0}\{\left\Vert \T_{B}(M_{AB}+N_{AB})\right\Vert
_{\infty}:\operatorname{Tr}[M_{AB}\rho_{AB}]\geq1-\varepsilon,\ M_{AB}\leq
\mathbbm{1}_{AB}\}.
		\end{align}
	\end{proposition*}
	
	\begin{Proof}
		Using the dual SDP formulation of the hypothesis testing relative entropy from Proposition~\ref{prop-hypo_test_rel_ent_dual}, we obtain
		\begin{multline}
			W_H^{\varepsilon}(A;B)_{\rho}= \\ \sup_{\mu\geq 0, Z_{AB}\geq 0, \sigma_{AB} \in\operatorname{PPT}'(A:B)} \{ \mu(1-\varepsilon) - \Tr[Z_{AB}] : 
			\mu \rho_{AB} \leq \sigma_{AB} + Z_{AB} \} 
		\end{multline}
		Now, combining with the characterization of $\operatorname{PPT}'(A\!:\!B)$ from \eqref{eq:E-meas:SDP-PPT-prime}, we conclude \eqref{eq-hypo_test_Rains_rel_ent_SDP}.
		
		The SDP\ for $W_{H}^{\varepsilon}(A;B)_{\rho}$ in \eqref{eq-hypo_test_Rains_rel_ent_SDP} can be written in the standard form of \eqref{eq-primal_SDP_def} 
as%
\begin{equation}
\sup_{X\geq0}\left\{  \operatorname{Tr}[AX]:\Phi(X)\leq B\right\}  ,
\end{equation}
where%
\begin{align}
X  & =%
\begin{pmatrix}
\mu & 0 & 0 & 0\\
0 & Z_{AB} & 0 & 0\\
0 & 0 & K_{AB} & 0\\
0 & 0 & 0 & L_{AB}%
\end{pmatrix}
,\quad A=%
\begin{pmatrix}
1-\varepsilon & 0 & 0 & 0\\
0 & -\mathbbm{1}_{AB} & 0 & 0\\
0 & 0 & 0 & 0\\
0 & 0 & 0 & 0
\end{pmatrix}
,\\
\Phi(X)  & =%
\begin{pmatrix}
\mu\rho_{AB}-\T_{B}(K_{AB}-L_{AB})-Z_{AB} & 0 & 0\\
0 & -\T_{B}(K_{AB}-L_{AB}) & 0\\
0 & 0 & \operatorname{Tr}[K_{AB}+L_{AB}]
\end{pmatrix}
,\\
B  & =%
\begin{pmatrix}
0 & 0 & 0\\
0 & 0 & 0\\
0 & 0 & 1
\end{pmatrix}
.
\end{align}
Setting%
\begin{equation}
Y=%
\begin{pmatrix}
M_{AB} & 0 & 0\\
0 & N_{AB} & 0\\
0 & 0 & \lambda
\end{pmatrix}
,
\end{equation}
we compute the dual map $\Phi^{\dag}$ as follows:%
\begin{align}
& \operatorname{Tr}[Y\Phi(X)] \notag \\
& =\operatorname{Tr}[M_{AB}\left(  \mu\rho_{AB}-\T_{B}(K_{AB}-L_{AB}%
)-Z_{AB}\right)  ]-\operatorname{Tr}[N_{AB}\T_{B}(K_{AB}-L_{AB})]\notag \\
& \qquad+\lambda\operatorname{Tr}[K_{AB}+L_{AB}]\\
& =\mu\operatorname{Tr}[M_{AB}\rho_{AB}]-\operatorname{Tr}[M_{AB}%
Z_{AB}]+\operatorname{Tr}[\left(  \lambda \mathbbm{1}_{AB}-\T_{B}(M_{AB}+N_{AB})\right)
K_{AB}]\notag\\
& \qquad +\operatorname{Tr}[\left(  \lambda \mathbbm{1}_{AB}+\T_{B}(M_{AB}+N_{B})\right)
L_{AB}],
\end{align}
which implies that%
\begin{multline}
\Phi^{\dag}(Y)=\\%
\begin{pmatrix}
\operatorname{Tr}[M_{AB}\rho_{AB}] & 0 & 0 & 0\\
0 & -M_{AB} & 0 & 0\\
0 & 0 & \lambda \mathbbm{1}_{AB}-\T_{B}(M_{AB}+N_{AB}) & 0\\
0 & 0 & 0 & \lambda \mathbbm{1}_{AB}+\T_{B}(M_{AB}+N_{B})
\end{pmatrix}
.
\end{multline}
Then using the standard form of the dual program in \eqref{eq-dual_SDP_def}, i.e.,%
\begin{equation}
\inf_{Y\geq0}\left\{  \operatorname{Tr}[BY]:\Phi^{\dag}(Y)\geq A\right\}  ,
\end{equation}
we find that the dual SDP\ is given by%
\begin{multline}
\inf_{\lambda,M_{AB},N_{AB}\geq0}\{\lambda:\operatorname{Tr}[M_{AB}\rho
_{AB}]\geq1-\varepsilon,\ M_{AB}\leq \mathbbm{1}_{AB},\\
\lambda \mathbbm{1}_{AB}\pm \T_{B}(M_{AB}+N_{AB})\geq0\}.
\end{multline}
This can alternatively be written as%
\begin{equation}
\inf_{M_{AB},N_{AB}\geq0}\{\left\Vert \T_{B}(M_{AB}+N_{AB})\right\Vert
_{\infty}:\operatorname{Tr}[M_{AB}\rho_{AB}]\geq1-\varepsilon,\ M_{AB}\leq
\mathbbm{1}_{AB}\}.
\end{equation}

Finally, we verify that strong duality holds, by applying Theorem~\ref{thm:math-tools:slater-cond}. A strictly feasible choice for the primal variables is $\mu=1$, $Z_{AB} = \mathbbm{1}_{AB}$, $K_{AB} = \pi_{AB}/2$, $L_{AB} = \pi_{AB}/3$. A feasible choice for the dual variables is $M_{AB} = \mathbbm{1}_{AB}$ and $N_{AB} = \mathbbm{1}_{AB}$.  
	\end{Proof}

\section{Squashed Entanglement}\label{sec-LAQC:sq-ent-and-props}

	In this section, we investigate the properties of the squashed entanglement, which we introduced as an entanglement measure in Section~\ref{sec-ent_meas_examples}. We first recall the definition of squashed entanglement from \eqref{eq-squashed_entanglement_0}.
	
	\begin{definition}{Squashed Entanglement}{def:LAQC-sq-ent}
		Let $\rho_{AB}$ be a bipartite state. Then, the squashed entanglement is defined as%
		\begin{equation}\label{eq-squashed_entanglement}
			E_{\operatorname{sq}}(A;B)_{\rho}\coloneqq \frac{1}{2}\inf_{\omega_{ABE}}\left\{  I(A;B|E)_{\omega}:\Tr_{E}[\omega_{ABE}]=\rho_{AB}\right\},
		\end{equation}
		where the quantum conditional mutual information $I(A;B|E)_{\omega}$ is defined in \eqref{eq-QCMI_def} as
		\begin{align}
			I(A;B|E)_{\omega}&=H(A|E)_{\omega}+H(B|E)_{\omega}-H(AB|E)_{\omega}\\
			&=H(AE)_{\omega}-H(E)_{\omega}+H(BE)_{\omega}-H(ABE)_{\omega}\\
			&=H(B|E)_{\omega}-H(B|AE)_{\omega}\label{eq-sq_ent_def_QCMI}
		\end{align}
	\end{definition}
	
	The optimization in the definition of squashed entanglement is with respect to all extensions $\omega_{ABE}$ of $\rho_{AB}$, with the extension system $E$ having arbitrarily large, yet finite dimension, which means that the infimum cannot in general be replaced by a minimum. The fact that the extension system can have arbitrarily large, yet finite dimension also means that computing the squashed entanglement is in general difficult; however, we can always place an upper bound on it by calculating the quantum conditional mutual information of a specific extension.
	
	To understand why the quantity in \eqref{eq-squashed_entanglement} is called squashed entanglement, it is helpful to think in terms of a cryptographic scenario. This cryptographic perspective turns out to be useful in the context of secret key agreement, which we consider in Chapters~\ref{chap-secret_key_distill} and \ref{chap-SKA}. Consider three parties, the protagonists Alice and Bob, as well as an eavesdropper. Alice and Bob possess the quantum systems $A$ and $B$, respectively, and the eavesdropper possesses a system $E$, such that the global state $\omega_{ABE}$ is consistent with the reduced state $\rho_{AB}$ of Alice and Bob. The conditional mutual information $I(A;B|E)_{\omega}$ corresponds to the correlations between Alice and Bob from the perspective of the eavesdropper. The squashed entanglement $E_{\text{sq}}(A;B)_{\rho}$ is then an optimization over all possible global states $\omega_{ABE}$ such that $\Tr_E[\omega_{ABE}]=\rho_{AB}$, and this optimization corresponds to the worst possible scenario in which the eavesdropper attempts to ``squash down'' the correlations of Alice and Bob, i.e., to reduce the value of $I(A;B|E)_{\omega}$ as much as possible. This cryptographic perspective actually allows us to write the squashed entanglement in an alternative way, which we do in Proposition~\ref{prop:LAQC-sq-chann-form} below.
	
	We begin by establishing some basic properties of squashed entanglement. As we will see, the squashed entanglement possesses \textit{all} of the desired properties of an entanglement measure stated at the beginning of Section~\ref{sec-ent_measures_def}.

	\begin{proposition*}{Properties of Squashed Entanglement}{prop-squashed_ent_properties}
		The squashed entanglement $E_{\text{sq}}(A;B)_{\rho}$ has the following properties.
		\begin{enumerate}
			\item \textit{Non-negativity}: For every bipartite state $\rho_{AB}$,
				\begin{equation}\label{prop:LAQC-squashed-sep}
					E_{\text{sq}}(A;B)_{\rho}\geq 0.
				\end{equation}
			
			\item \textit{Faithfulness}: We have $E_{\text{sq}}(A;B)_{\sigma}=0$ if and only if $\sigma_{AB}$ is a separable state.
			
			\item \textit{Convexity}: Let $\mathcal{X}$ be a finite alphabet, $p:\mathcal{X}\to[0,1]$ a probability distribution on $\mathcal{X}$, and $\{\rho_{AB}^x\}_{x\in\mathcal{X}}$ a set of states. Then,
				\begin{equation}\label{prop-LOCC-QC:convexity-sq-ent}
					E_{\operatorname{sq}}(A;B)_{\overline{\rho}}\leq\sum_{x\in\mathcal{X}}p(x)E_{\operatorname{sq}}(A;B)_{\rho^{x}},
				\end{equation}
				where $\overline{\rho}_{AB}=\sum_{x\in\mathcal{X}}p(x)\rho_{AB}^x$.
			
			\item \textit{Monogamy}: For every  state $\rho_{A_1A_2B_1B_2}$, the overall squashed entanglement is not smaller than the sum of the individual squashed entanglements:
				\begin{multline}\label{eq:LAQC-sq-ent-superadd}
					E_{\operatorname{sq}}(A_{1}A_{2};B_{1}B_{2})_{\rho}\geq E_{\operatorname{sq}}(A_{1};B_{1})_{\rho}+ E_{\operatorname{sq}}(A_{1};B_{2})_{\rho} + E_{\operatorname{sq}}(A_{2};B_{1})_{\rho} \\+ E_{\operatorname{sq}}(A_{2};B_{2})_{\rho}.
				\end{multline}
\item \textit{Additivity}:	For a tensor-product state $\sigma_{A_1A_2B_1B_2}=\omega_{A_1B_1}\otimes\tau_{A_2B_2}$, the following additivity identity holds:
				\begin{equation}\label{eq:LAQC-sq-additive-states}%
					E_{\operatorname{sq}}(A_{1}A_{2};B_{1}B_{2})_{\sigma}=E_{\operatorname{sq}}(A_{1};B_{1})_{\omega}+E_{\operatorname{sq}}(A_{2};B_{2})_{\tau}.
				\end{equation}
		\end{enumerate}
	\end{proposition*}

	\begin{Proof}
		\hfill\begin{enumerate}
			\item This follows immediately from the fact that the conditional mutual information of an arbitrary state is non-negative (Theorem~\ref{thm-SSA}). 
			
			\item The statement ``if $\sigma_{AB}$ is a separable state, then $E_{\text{sq}}(A;B)_{\sigma}=0$'' follows from the line of reasoning in \eqref{eq-sq_ent_motivation_0}--\eqref{eq:LAQC-mutual-info-zero-sep} used to motivate the definition of squashed entanglement. For a proof of the converse statement, please consult the Bibliographic Notes in Section~\ref{sec:E-meas:bib-notes}.
				 
			
			\item Let $\omega_{ABE}^{x}$ denote an arbitrary extension of $\rho_{AB}^{x}$. Then%
				\begin{equation}
					\omega_{ABEX}\coloneqq \sum_{x\in\mathcal{X}}p(x)\omega_{ABE}^{x}\otimes|x\rangle\!\langle x|_{X}%
				\end{equation}
				is a particular extension of $\overline{\rho}_{AB}$. It follows that%
				\begin{align}
					2\cdot E_{\operatorname{sq}}(A;B)_{\overline{\rho}} &  \leq I(A;B|EX)_{\omega }\\
					&  =\sum_{x\in\mathcal{X}}p(x)I(A;B|E)_{\omega^{x}},
				\end{align}
				where to obtain the equality we used the direct-sum property of conditional mutual information (see Proposition~\ref{prop-cond_mut_inf_properties}). Since the inequality holds for arbitrary extensions of $\rho_{AB}^{x}$, we conclude the desired inequality.
			
			\item To see the inequality in \eqref{eq:LAQC-sq-ent-superadd}, let $\omega_{A_{1}A_{2}B_{1}B_{2}E}$ be an arbitrary extension of $\rho_{A_{1}A_{2}B_{1}B_{2}}$. Then by two applications of the chain rule for conditional mutual information, we find that%
				\begin{align}
					& I(A_{1}A_{2};B_{1}B_{2}|E)_{\omega}  \notag \\
					&  =I(A_{1};B_{1}B_{2}|E)_{\omega}+I(A_{2};B_{1}B_{2}|EA_{1})_{\omega}\\
					&  =I(A_{1};B_{1}|E)_{\omega}+I(A_{1};B_{2}|EB_{1})_{\omega}\nonumber\\
					&  \qquad+I(A_{2};B_{2}|EA_{1})_{\omega}+I(A_{2};B_{1}|EA_{1}B_{2})_{\omega}\\
					&  \geq 2\left[  E_{\operatorname{sq}}(A_{1};B_{1})_{\rho}+ E_{\operatorname{sq}}(A_{1};B_{2})_{\rho} + E_{\operatorname{sq}}(A_{2};B_{1})_{\rho} + E_{\operatorname{sq}}(A_{2};B_{2})_{\rho}\right]  .
				\end{align}
				The inequality follows because $\omega_{A_{1}B_{1}E}$ is a particular extension of the reduced state $\rho_{A_{1}B_{1}}$, the state $\omega_{A_{1}B_1 B_{2}E}$ is a particular extension of the reduced state $\rho_{A_{1}B_{2}}$, the state $\omega_{A_{1}A_{2} B_{2}E}$ is a particular extension of the reduced state $\rho_{A_{2}B_{2}}$, and $\omega_{A_{1}A_{2}B_{1}B_2 E}$ is a particular extension of the reduced state $\rho_{A_{2}B_{1}}$. Since the extension $\omega_{A_1A_2B_1B_2E}$ is arbitrary, optimizing over all such extensions on the left-hand side of the inequality above gives \eqref{eq:LAQC-sq-ent-superadd}.

				\item To see the equality in \eqref{eq:LAQC-sq-additive-states}, first consider that for a tensor-product state $\omega_{A_1B_1}\otimes\tau_{A_2B_2}$, the reduced state on systems $A_1B_2$ is the product state  $\omega_{A_1}\otimes\tau_{B_2}$, and the reduced state on systems $A_2B_1$ is the product state $\omega_{A_2}\otimes\tau_{B_1}$. Thus, faithfulness implies that $E_{\operatorname{sq}}(A_{1};B_{2})_{\sigma} = E_{\operatorname{sq}}(A_{2};B_{1})_{\sigma}= 0 $, and then the monogamy inequality in \eqref{eq:LAQC-sq-ent-superadd} implies that 
				\begin{equation}
				E_{\operatorname{sq}}(A_{1}A_{2};B_{1}B_{2})_{\sigma}\geq E_{\operatorname{sq}}(A_{1};B_{1})_{\omega}+ E_{\operatorname{sq}}(A_{2};B_{2})_{\tau}.
				\label{eq-EM:super-add-SE}
				\end{equation}
				Now let $\omega_{A_{1}B_{1}E_{1}}$ be an extension of $\omega_{A_{1}B_{1}}$, and let $\tau_{A_{2}B_{2}E_{2}}$ be an extension of $\tau_{A_{2}B_{2}}$. Then $\omega_{A_{1}B_{1}E_{1}}\otimes\tau_{A_{2}B_{2}E_{2}}$ is an extension of $\omega_{A_{1}B_{1}}\otimes\tau_{A_{2}B_{2}}$. We then have that%
				\begin{align}
					2\cdot E_{\operatorname{sq}}(A_{1}A_{2};B_{1}B_{2})_{\sigma}  &  \leq I(A_{1}A_{2};B_{1}B_{2}|E_{1}E_{2})_{\omega\otimes\tau}\\
					&  =I(A_{1};B_{1}|E_{1})_{\omega}+I(A_{2};B_{2}|E_{2})_{\tau},
				\end{align}
				where the equality follows from the additivity of conditional mutual information with respect to tensor-product states (Proposition~\ref{prop-cond_mut_inf_properties}). Since the extensions $\omega_{A_{1}B_{1}E_{1}}$ and $\tau_{A_{2}B_{2}E_{2}}$ are arbitrary, the following inequality holds:
				\begin{equation}
				E_{\operatorname{sq}}(A_{1}A_{2};B_{1}B_{2})_{\sigma}\leq E_{\operatorname{sq}}(A_{1};B_{1})_{\omega}+ E_{\operatorname{sq}}(A_{2};B_{2})_{\tau}.
								\label{eq-EM:sub-add-SE}
				\end{equation}
				We then conclude \eqref{eq:LAQC-sq-additive-states} by combining \eqref{eq-EM:super-add-SE} and \eqref{eq-EM:sub-add-SE}. \qedhere
		\end{enumerate}
	\end{Proof}
	
	Let us now prove that the squashed entanglement is indeed an entanglement measure, as per Definition~\ref{def-LAQC:ent-measure}.
	
	\begin{theorem*}{Selective LOCC Monotonicity  of Squashed Entanglement}{thm:LAQC-mono-LOCC-sq}
The squashed entanglement is a selective LOCC monotone, so that \eqref{eq:E-meas:strong-LOCC-mono} holds with $E$ set to $E_{\operatorname{sq}}$.
	\end{theorem*}

	\begin{Proof}
	We prove that the conditions of Lemma~\ref{lem:E-meas:convex-strong-LOCC-mono} hold and then we apply it.
		The first part of the proof follows  from the fact that conditional mutual information does not increase under the action of local channels (see Proposition~\ref{prop-cond_mut_inf_properties}): for a tripartite state $\xi_{ABE}$ and local channels $\mathcal{N}_{A\rightarrow A'}$ and $\mathcal{M}_{B\rightarrow B'}$,
		\begin{equation}
			I(A;B|E)_{\xi}\geq I(A';B'|E)_{\zeta},
		\end{equation}
		where $\zeta_{A'B'E}\coloneqq (\mathcal{N}_{A\rightarrow
A'}\otimes\mathcal{M}_{B\rightarrow B'})(\xi_{ABE})$. After incorporating optimizations, it then follows immediately that squashed entanglement does not increase under the action of local channels:%
		\begin{equation}
			E_{\operatorname{sq}}(A;B)_{\rho}\geq E_{\operatorname{sq}}(A';B')_{\kappa},
			\label{eq-LOCC-QC:sq-ent-mono-local-ch}
		\end{equation}
		where $\kappa_{A'B'}\coloneqq (\mathcal{N}_{A\rightarrow
A'}\otimes\mathcal{M}_{B\rightarrow B'})(\rho_{AB})$.

Also, the squashed entanglement is invariant under classical communication, which
				 follows from Lemma~\ref{lem:LAQC-sq-inv-classical-comm} below. Now applying Lemma~\ref{lem:E-meas:convex-strong-LOCC-mono}, we conclude the statement of the theorem.
	\end{Proof}

		\begin{Lemma*}{Invariance of Squashed Entanglement Under Classical Communication}{lem:LAQC-sq-inv-classical-comm}
		Let $\rho_{X A B}$ be a classical--quantum state of the following form:%
		\begin{equation}
			\rho_{XAB}\coloneqq \sum_{x\in\mathcal{X}}p(x)|x\rangle\!\langle x|_{X}\otimes\rho_{AB}^{x},
		\end{equation}
		where $\mathcal{X}$ is a finite alphabet, $p:\mathcal{X}\to[0,1]$ is a probability distribution, and $\{\rho_{AB}^{x}\}_{x\in\mathcal{X}}$ is a set of states. Then%
		\begin{equation}
			E_{\operatorname{sq}}(XA;B)_{\rho}  
			  =E_{\operatorname{sq}}(A;BX)_{\rho}
			  =\sum_{x\in\mathcal{X}}p(x)E_{\operatorname{sq}}(A;B)_{\rho^{x}}.
		\end{equation}
	\end{Lemma*}

	\begin{Proof}
		Note that discarding or appending a classical state $|x\rangle\!\langle x|_{X_{A}}$ is a local channel, so that $E_{\operatorname{sq}}(A;B)_{\rho^{x}}=E_{\operatorname{sq}}(XA;B)_{|x\rangle\!\langle x|\otimes\rho^{x}}$. In Proposition~\ref{prop-squashed_ent_properties}, we proved that the squashed entanglement is a convex function. Using this fact, it follows that
		\begin{equation}
			\sum_{x\in\mathcal{X}}p(x)E_{\operatorname{sq}}(A;B)_{\rho^{x}} = \sum_{x\in\mathcal{X}}p(x)E_{\operatorname{sq}}(XA;B)_{|x\rangle\!\langle x|\otimes\rho^{x}} \geq E_{\operatorname{sq}}(XA;B)_{\rho}.
		\end{equation}
		Similarly, we have that%
		\begin{equation}
			\sum_{x\in\mathcal{X}}p(x)E_{\operatorname{sq}}(A;B)_{\rho^{x}}\geq E_{\operatorname{sq}}(A;BX)_{\rho}
			.
		\end{equation}
		
		So it suffices to establish the following inequality:%
		\begin{equation}
			\sum_{x\in\mathcal{X}}p(x)E_{\operatorname{sq}}(A;B)_{\rho^{x}}\leq \min\{E_{\operatorname{sq}}(XA;B)_{\rho},\, E_{\operatorname{sq}}(A;BX)_{\rho}\}.
		\end{equation}
		To this end, let $\omega_{XABE}$ be an arbitrary extension of $\rho_{XAB}$. After the action of a local completely dephasing channel $\Delta_X(\cdot) \coloneqq \sum_{x\in \mathcal{X}} |x\rangle \!\langle x|_X (\cdot) |x\rangle \!\langle x|_X$, it follows that the state $\theta_{XABE}\coloneqq\Delta_X(\omega_{XABE})$ has the following form:%
		\begin{equation}
			\theta_{XABE} = \sum_{x\in\mathcal{X}}p(x)|x\rangle\!\langle x|_{X_{A}}\otimes \theta_{ABE}^{x},
			\label{eq-EM:extension-for-SE-inv-cc}
		\end{equation}
		where $\theta_{ABE}^{x}$ is an extension of $\rho_{AB}^x$. To see this, let $\ket{\phi^x}_{ABR}$ purify $\rho^x_{AB}$ for each $x\in\mathcal{X}$, and consider that
		\begin{equation}
		\ket{\varphi}_{XX_E ABR} \coloneqq \sum_{x \in \mathcal{X} } \sqrt{p(x)} \ket{x}_X \ket{x}_{X_E} \ket{\phi^x}_{ABR}
		\end{equation}
		is a purification of $\rho_{XAB}$ with purifying systems $X_E R$. By applying Proposition~\ref{prop-extension_purif}, we conclude that the extension $\omega_{XABE}$ can be realized by the action of a quantum channel $\mathcal{N}_{X_E R \to E}$ on systems $X_E R$ of $\varphi_{XX_E ABR}$:
		\begin{equation}
		\omega_{XABE} = \mathcal{N}_{X_E R \to E}(\varphi_{XX_E ABR}).
		\end{equation}
		The conclusion stated in \eqref{eq-EM:extension-for-SE-inv-cc} then follows because
		\begin{equation}
		\theta_{XABE} = (\Delta_X \otimes \mathcal{N}_{X_E R \to E})(\varphi_{XX_E ABR})
		= \sum_{x \in \mathcal{X} } p(x) \ket{x}\!\bra{x}_X \otimes \theta_{ABE}^{x},
		\end{equation}
		where $\theta_{ABE}^{x} \coloneqq \mathcal{N}_{X_E R \to E}(\ket{x}\!\bra{x}_{X_E} \otimes  \phi^x_{ABR})$.
		We then find that
		\begin{align}
			2\sum_{x\in\mathcal{X}}p(x)E_{\operatorname{sq}}(A;B)_{\rho^{x}}  &  \leq\sum_{x}p(x)I(A;B|E)_{\theta^{x}}\\
			&  =I(A;B|EX)_{\theta}\\
			& = I(XA;B|E)_{\theta}-I(X;B|E)_{\theta}\\
			&  \leq I(XA;B|E)_{\theta}\\
			& \leq I(XA;B|E)_{\omega}.
		\end{align}
		The last equality follows from the chain rule for conditional mutual information and the second-to-last inequality from non-negativity of conditional mutual information: $I(X;B|E)_{\rho}\geq0$. The final inequality holds because the conditional mutual information does not increase under the action of a local channel on system $X$ (see Proposition~\ref{prop-cond_mut_inf_properties}). Since the inequality holds for an arbitrary extension of $\rho_{XAB}$, we conclude that%
		\begin{equation}
			\sum_{x\in\mathcal{X}}p(x)E_{\operatorname{sq}}(A;B)_{\rho^{x}}\leq E_{\operatorname{sq}}(XA;B)_{\rho}.
		\end{equation}
		A proof for the other inequality%
		\begin{equation}
			\sum_{x\in\mathcal{X}}p(x)E_{\operatorname{sq}}(A;B)_{\rho^{x}}\leq E_{\operatorname{sq}}(A;BX)_{\rho}%
		\end{equation}
		follows similarly.
	\end{Proof}

	In Section~\ref{sec-ent_meas_examples}, we introduced the entanglement of formation as an entanglement measure. The following inequality relates the entanglement of formation to the squashed entanglement:
	\begin{proposition*}{Squashed Entanglement and Entanglement of Formation}{prop:E-meas:sq-ent-to-EoF}
	The entanglement of formation is never smaller than the squashed entanglement:
	\begin{equation}
	E_{\operatorname{sq}}(A;B)_{\rho} \leq 
	E_F(A;B)_{\rho},
	\label{eq:E-meas:sq-to-EoF}
	\end{equation}
	for every bipartite state $\rho_{AB}$.
	\end{proposition*}
	
	\begin{Proof}
	We already alluded to this inequality when introducing squashed entanglement in Section~\ref{sec:E-meas:sq-ent-intro}. If the extension $\omega_{ABE}$ is restricted to be of the form
	\begin{equation}
	\omega_{ABE} = \sum_{x}p(x)\phi^x_{AB} \otimes |x\rangle\!\langle x|,
	\end{equation}
	where $p(x)$ is a probability distribution and $\{\phi^x_{AB}\}_x$ is a set of pure states satisfying $\rho_{AB} = \sum_{x}p(x)\phi^x_{AB}$, then it follows that
	\begin{equation}
	E_{\operatorname{sq}}(A;B)_{\rho} \leq \frac{1}{2} I(A;B|X)_{\omega} = H(A|X)_{\omega}.
	\end{equation}
	Since the inequality holds for all such extensions, the inequality in \eqref{eq:E-meas:sq-to-EoF} follows by applying the definition of entanglement of formation in \eqref{eq:E-meas:EoF-def}.
	\end{Proof}
	
	For pure bipartite states, the entanglement of formulation is simply the entropy of the reduced state of one of the subsystems. It turns out that the squashed entanglement reduces to the same quantity for pure bipartite states.

	\begin{proposition*}{Squashed Entanglement for Pure States}{prop:LAQC-reduction-for-pure-squashed}
		Let $\psi_{AB}$ be a pure bipartite state. Then, the squashed entanglement of $\psi_{AB}$ is equal to the entropy of its reduced state on $A$:
		\begin{equation}
			E_{\operatorname{sq}}(A;B)_{\psi}=H(A)_{\psi}.
		\end{equation}
	\end{proposition*}

	\begin{Proof}
		Every extension $\omega_{ABE}$ of the pure bipartite state $\psi_{AB}$ is a product state of the form $\omega_{ABE}=\psi_{AB}\otimes\rho_E$ for some state $\rho_{E}$. By Proposition~\ref{prop-cond_mut_inf_properties}, it follows that%
		\begin{equation}
			I(A;B|E)_{\psi\otimes\rho}=I(A;B)_{\psi}=2H(A)_{\psi},
		\end{equation}
		where the second equality holds because $H(AB)_{\psi}=0$ and $H(A)_{\psi}=H(B)_{\psi}$ for a pure bipartite state. We thus have $E_{\text{sq}}(A;B)_{\psi}=H(A)_{\psi}$.
	\end{Proof}
	
	An immediate consequence of Proposition~\ref{prop:LAQC-reduction-for-pure-squashed} is that, for the maximally entangled state $\ket{\Phi}_{AB}=\frac{1}{\sqrt{d}}\sum_{i=0}^{d-1}\ket{i,i}_{AB}$ of Schmidt rank $d$, the squashed entanglement is equal to $\log_{2}d$:%
	\begin{equation}\label{eq:LAQC-sq-max-ent}%
		E_{\operatorname{sq}}(A;B)_{\Phi}=\log_{2}d. 
	\end{equation}

	Let us now return to the discussion after Definition~\ref{def:LAQC-sq-ent} on the cryptographic interpretation of squashed entanglement. We stated that the quantum conditional mutual information $I(A;B|E)_{\omega}$ can be interpreted as the amount of correlations between Alice ($A$) and Bob ($B$) from the point of view of an eavesdropper ($E$), where $\omega_{ABE}$ is the joint state shared by all three parties. If the eavesdropper wants to reduce, or ``squash'' these correlations as much as possible, then we optimize with respect to every state $\omega_{ABE}$ that is consistent with the state $\rho_{AB}$ shared by Alice and Bob, leading to the squashed entanglement $E_{\text{sq}}(A;B)_{\rho}$. Now, recall Proposition~\ref{prop-extension_purif}, which states that for every extension $\omega_{ABE}$ of a given state $\rho_{AB}$ there exists a quantum channel $\mathcal{S}_{E'\to E}$ such that $\mathcal{S}_{E'\to E}(\psi_{ABE'}^{\rho})=\omega_{ABE}$, where $\psi_{ABE'}^{\rho}$ is a purification of $\rho_{AB}$. We therefore immediately have the following.
	
	\begin{proposition*}{Other Representations of Squashed Entanglement}{prop:LAQC-sq-chann-form}
		Let $\rho_{AB}$ be a bipartite quantum state, and let $\psi_{ABE'}^{\rho}$ be a purification of it. Then the squashed entanglement $E_{\text{sq}}(A;B)_{\rho}$ can be written as
		\begin{equation}\label{eq:LAQC-sq-ch-rep-1}
			E_{\text{sq}}(A;B)_{\rho}=\frac{1}{2}\inf_{\mathcal{S}_{E'\to E}}\{I(A;B|E)_{\omega}:\omega_{ABE}=\mathcal{S}_{E'\to E}(\psi_{ABE'}^{\rho})\},
		\end{equation}
		where the infimum is with respect to every quantum channel $\mathcal{S}_{E'\to E}$.
		Another representation of squashed entanglement is
		\begin{multline}\label{eq:LAQC-sq-ch-rep-2}%
			E_{\operatorname{sq}}(A;B)_{\rho}= \\ \frac{1}{2}\inf_{\mathcal{V}_{E'\rightarrow EF}}\left\{  H(B|E)_{\theta}+H(B|F)_{\theta}: \theta_{BEF}\coloneqq\mathcal{V}_{E'\rightarrow EF}(\psi_{ABE'}^{\rho}) \right\}  ,
		\end{multline}
		where the infimum is with respect to every isometric channel $\mathcal{V}_{E'\rightarrow EF}$.
	\end{proposition*}
	
	The act of ``squashing'' the correlations between Alice and Bob can thus be thought of explicitly in terms of an eavesdropper applying the channel $\mathcal{S}_{E'\to E}$ to their purifying system $E'$ of $\rho_{AB}$. For this reason, we call $\mathcal{S}_{E'\to E}$ a \textit{squashing channel}.
	

	\begin{Proof}
		The equality \eqref{eq:LAQC-sq-ch-rep-1} follows immediately from Proposition~\ref{prop-extension_purif}.

		In order to establish \eqref{eq:LAQC-sq-ch-rep-2}, let $\mathcal{S}_{E'\rightarrow E}$ be an arbitrary squashing channel, and let $\mathcal{V}_{E'\rightarrow EF}$ be an isometric extension of $\mathcal{S}_{E'\to E}$, so that $\mathcal{S}_{E'\to E}=\Tr_F\circ\mathcal{V}_{E'\to EF}$, where the system $F$ satisfies $d_F\geq\rank(\Gamma_{EE'}^{\mathcal{S}})$. Consider that $\theta_{ABEF}\coloneqq\mathcal{V}_{E'\rightarrow EF}(\psi_{ABE'}^{\rho})$ is a purification of both $\omega_{ABE}$ and $\theta_{BEF}$, i.e., $\omega_{ABE}=\mathcal{S}_{E'\rightarrow E}(\psi_{ABE'}^{\rho})=\Tr_{F}[\theta_{ABEF}]$ and $\theta_{BEF}=\Tr_{A}[\theta_{ABEF}]$. Then, using \eqref{eq-sq_ent_def_QCMI}, it follows that%
		\begin{align}
			I(A;B|E)_{\omega}  &  =H(B|E)_{\omega}-H(B|AE)_{\omega}\\
			&  =H(B|E)_{\theta}-H(B|AE)_{\theta}.
		\end{align}
		Now, since $\theta_{ABEF}$ is a pure state, we have from duality of conditional entropy   that
		\begin{equation}
			-H(B|AE)_{\theta}
			=H(B|F)_{\theta}.
		\end{equation}
		Therefore,
		\begin{equation}
			I(A;B|E)_{\omega}=H(B|E)_{\theta}+H(B|F)_{\theta}.
		\end{equation}
		We conclude that \eqref{eq:LAQC-sq-ch-rep-2} holds because the squashing channel $\mathcal{S}_{E'\to E}$ is arbitrary in the development above.
	\end{Proof}
	
	 We now establish an asymptotic continuity bound for the squashed entanglement:
	
	\begin{proposition*}{Asymptotic Continuity of Squashed Entanglement}{prop:LAQC-cont-sq-unif}
		Let $\rho_{AB}$ and $\sigma_{AB}$ be bipartite states such that the following fidelity bound holds%
		\begin{equation}
			F(\rho_{AB},\sigma_{AB})\geq1-\varepsilon,
		\end{equation}
		for $\varepsilon\in\left[  0,1\right]  $. Then the following bound applies to their squashed entanglements:%
		\begin{equation}\label{eq-sq_ent_uniform_cont}
			\left\vert E_{\operatorname{sq}}(A;B)_{\rho}-E_{\operatorname{sq}}(A;B)_{\sigma}\right\vert \leq\sqrt{\varepsilon}\log_{2}\min\left\{d_A,d_B\right\}+g_{2}(\sqrt{\varepsilon}),
		\end{equation}
		where%
		\begin{equation}
			g_{2}(\delta)\coloneqq \left(  \delta+1\right)  \log_{2}(\delta+1)-\delta \log_{2}\delta.
		\end{equation}
	\end{proposition*}

	\begin{Proof}
		Due to Uhlmann's theorem (Theorem~\ref{thm-Uhlmann_fidelity}) and Proposition~\ref{prop-extension_purif}, for an arbitrary extension $\rho_{ABE}$ of $\rho_{AB}$, there exists an extension $\sigma_{ABE}$ of $\sigma_{AB}$ such that%
		\begin{equation}
			F(\rho_{ABE},\sigma_{ABE})=F(\rho_{AB},\sigma_{AB})\geq1-\varepsilon.
		\end{equation}
		By the relation between trace distance and fidelity (Theorem~\ref{thm-Fuchs_van_de_graaf}), it follows that%
		\begin{equation}
			\frac{1}{2}\left\Vert \rho_{ABE}-\sigma_{ABE}\right\Vert _{1}\leq \sqrt{\varepsilon}.
		\end{equation}
		Then, applying the asymptotic continuity of conditional mutual information (Proposition \ref{lem:LAQC-uniform-cont-CMI}), we find that
		\begin{align}
			2E_{\operatorname{sq}}(A;B)_{\sigma}  &  \leq I(A;B|E)_{\sigma}\\
			&  \leq I(A;B|E)_{\rho}+2\sqrt{\varepsilon}\log_{2}\min\left\{d_A,d_B\right\}+2g_{2}(\sqrt{\varepsilon}).
		\end{align}
		Since the extension $\rho_{ABE}$ is arbitrary, it follows that%
		\begin{equation}
			E_{\operatorname{sq}}(A;B)_{\sigma}\leq E_{\operatorname{sq}}(A;B)_{\rho}+\sqrt{\varepsilon}\log_{2}\min\left\{d_A,d_B\right\}+g_{2}(\sqrt{\varepsilon}).
		\end{equation}
		The other bound,
		\begin{equation}
			E_{\operatorname{sq}}(A;B)_{\rho}\leq E_{\operatorname{sq}}(A;B)_{\sigma}+\sqrt{\varepsilon}\log_{2}\min\left\{d_A,d_B\right\}  + g_{2}(\sqrt{\varepsilon}),
		\end{equation}
		follows from a similar proof.
	\end{Proof}


\section{Summary}

	In this chapter, we studied entanglement measures for quantum states and quantum channels. The defining property of an entanglement measure for states is monotonicity under local operations and classical communication (LOCC): a function $E:\mathcal{H}_{AB}\to\mathbb{R}$ is an entanglement measure if $E(\rho_{AB})\geq E(\mathcal{L}(\rho_{AB}))$ for every bipartite state $\rho_{AB}$ and every LOCC channel $\mathcal{L}$. LOCC monotonicity can be thought of as a special kind of data-processing inequality, and it is a core concept in entanglement theory in the same way that the data-processing inequality is the core concept behind generalized divergence.
	
	An important type of state entanglement measure for our purposes in this book is a divergence-based measure, in which the entanglement in a given bipartite quantum state is quantified by its divergence with the set of separable states. As our divergence, we take a generalized divergence $\boldsymbol{D}:\Density(\mathcal{H})\times\Lin_+(\mathcal{H})\to\mathbb{R}$, and we call the resulting quantity \textit{generalized divergence of entanglement}. Due to the data-processing inequality (which holds for a generalized divergence by definition), we immediately obtain LOCC monotonicity for the generalized divergence of entanglement, thus making it an entanglement measure. We also consider the divergence with the larger set of $\PPT'$ operators that contains all separable states, and call the resulting quantity \textit{generalized Rains divergence}.

%
%
%
%
%
%
%
%
%

\section{Bibliographic Notes}	

\label{sec:E-meas:bib-notes}

	Entanglement theory has a long history, with one of the seminal papers being that of \citet{BDSW96} (see also \citet{BBPSSW96EPP,BBPS96} for earlier works). \citet{BDSW96} introduced the resource theory of entanglement, with the separable states as the free states and LOCC channels as the free channels, along with the related operational notions of distillable entanglement and entanglement cost.  The review of \citet{HHHH09} is a useful resource for gaining an understanding of notable accomplishments in the area. See also the review of \citet{PV07}, which focuses specifically on entanglement measures.
	
	The axiomatic approach to defining an entanglement measure that we have taken in this chapter was proposed by \citet{BDSW96,VPRK97,VP98,Vid00}, with LOCC monotonicity emerging as the defining property of an entanglement measure \citet{HHHH09}. \citet{Vid00} and \citet{MH05} established simplified conditions for proving that a function is a selective LOCC monotone. Lemma~\ref{lem:E-meas:convex-strong-LOCC-mono} is in this spirit.
	
	Entanglement of formation was defined by \citet{BDSW96}, and they showed that it is a selective LOCC monotone. They also showed that it is an upper bound on entanglement cost, and thus an upper bound on distillable entanglement (we did not define these concepts here, but we will define distillable entanglement in detail in Chapter~\ref{chap-ent_distill}). The asymptotic continuity bound for entanglement of formation in Proposition~\ref{prop:E-meas:unif-cont-EoF} was given by \citet{Win16}. The faithfulness bounds for entanglement of formation in Proposition~\ref{prop:E-meas:EoF-faithful} were given by \citet{LW14} (see also \citet{Nielsen2000}). The non-additivity of entanglement of formation was established by \citet{Hastings09}, which built upon an earlier result of \citet{Shor04} connecting various additivity conjectures in quantum information theory. The formula in \eqref{eq-ent_formation_two_qubits} for the entanglement of formation of two-qubit states was determined by \citet{Woot98}.
	
	Negavitity and log-negativity were defined by \citet{ZHS+98} (see also \citep{VW02} and \citep{Ple05}). \citet{VW02} showed that log-negativity is monotone under LOCC. They also proved that it is additive. See also \citep{Ple05} for a proof of selective LOCC monotonicity of log-negativity, and for a proof of the fact that log-negativity is not convex. The semi-definite programming approach, which we have taken here to prove selective LOCC monotonicity of log-negativity, is based on \citet{WD16a}.

	The relative entropy of entanglement was defined by \citet{VPRK97,VP98}, who also proved many of its properties, such as LOCC monotonicity and convexity. They also proposed concepts closely related to the generalized divergence of entanglement. The fact that trace distance of entanglement is not a selective LOCC monotone was shown by \citet{PhysRevA.98.052351}. Optimization over the set of separable states has been shown by \citet{Gur04,Ghar10} to be NP-hard in general (see also \citep{Ioa07,SW12b}). A closed-form formula for the relative entropy of entanglement for two-qubit states was derived by \citet{MI08}. The max-relative entropy of entanglement was defined by \citet{Datta2009b}, the hypothesis testing relative entropy of entanglement by \citet{BD11}, and the sandwiched R\'enyi relative entropy of entanglement by \citet{WTB16}. The fact that the relative entropy of entanglement is invariant under classical communication (Proposition~\ref{prop:E-meas:REE-inv-cc}) was stated by \citet{MH05} and proven by \citet{KW17}. Our proof of selective separable monotonicity in Proposition~\ref{prop:E-meas:strong-sep-mono-rel-ents}, for the R\'enyi relative entropies, is based on the approach from \citet{WW19alog}.
	The proof of Property~1.~of Proposition~\ref{prop-ent_meas_rel_ent_entanglement} follows the approach from \citet{PVP00}, and the proof of Property~2.~follows the approach of \citet[Proposition~10]{TWW17}. 	Property~2.~of Proposition~\ref{prop-ent_meas_rel_ent_entanglement} was first established by \citet{VP98}. The cone program formulations of max-relative entropy of entanglement of a bipartite state, as well as max-relative entropy of entanglement of a quantum channel, were given by \citet{BW17}.

	The relative entropy  with the set of PPT states was defined by \citet{R99} in the context of entanglement distillation (see Chapter~\ref{chap-ent_distill}). \citet{R01} then modified the quantity to obtain a tighter bound on the distillable entanglement. After this development, \citet{AdMVW02} defined the set $\PPT'$ and showed that this improved bound can be written as the relative entropy with the set of $\PPT'$ operators, which we refer to here as the Rains relative entropy. The fact that the set PPT' is preserved by completely PPT preserving channels (Property~1.~of Lemma~\ref{prop-PPT_prime_properties}) was shown by \citet{TWW17}. The sandwiched Rains relative entropy of a bipartite state was defined by \citet{TWW17}, as well as the generalized Rains divergence, both as generalizations of the Rains relative entropy of \citet{R01,AdMVW02}. The semi-definite program formulation of the max-Rains relative entropy of a bipartite state was given by \citet{WD16a,WFD17}, who also recognized that the semi-definite programming bound from \citet{WD16a} is equal to the max-Rains relative entropy. \citet{WD16a} proved that the max-Rains relative entropy is a selective PPT monotone (Proposition~\ref{prop:E-meas:max-rains-strong-ppt-mono}) and that it is additive (Proposition~\ref{prop:LAQC-add-max-Rains-rel-ent}).
	\citet{MI08} proved that the Rains relative entropy is equal to the relative entropy of entanglement for two-qubit states.

	The squashed entanglement of a bipartite state was defined by \citet{CW04}, who established several of its key properties mentioned in Propositions~\ref{prop-squashed_ent_properties}, \ref{thm:LAQC-mono-LOCC-sq}, and \ref{prop:LAQC-reduction-for-pure-squashed}, including non-negativity, vanishing on separable states, convexity, superadditivity in general, additivity for tensor-product states, LOCC monotonicity, reduction for pure states, and the squashing channel representation in \eqref{eq:LAQC-sq-ch-rep-1}. The faithfulness of squashed entanglement was established by \citet{BCY11}. A function related to squashed entanglement was discussed by \citet{Tuc99,Tuc02}. Our discussions motivating squashed entanglement are related to those presented by \citet{Tuc99,Tuc02}. The representation of squashed entanglement in \eqref{eq:LAQC-sq-ch-rep-2} is due to \citet{TGW14IEEE}. Asymptotic continuity of the squashed entanglement of a bipartite state was established by \citet{AF04}, and the explicit bound given here is due to \citet{Shirokov17}.


\begin{subappendices}

\section{Semi-Definite Programs for Negativity}

\label{app:E-meas:SDP-negativity}

Here we prove that the quantity $\norm{\T_B(\rho_{AB})}_1$ has the
following primal and dual SDP\ formulations:%
\begin{align}
\norm{\T_B(\rho_{AB})}_1  & =\sup_{R_{AB}}\left\{  \operatorname{Tr}[R_{AB}\rho
_{AB}]:-\mathbbm{1}_{AB}\leq \T_B(R_{AB})\leq \mathbbm{1}_{AB}\right\}  ,\\
& =\inf_{K_{AB},L_{AB}\geq0}\left\{  \operatorname{Tr}[K_{AB}+L_{AB}%
]: \T_B(K_{AB}-L_{AB})=\rho_{AB}\right\},
\end{align}
where the optimization in the first line is with respect to Hermitian $R_{AB}%
$. Since this is the core quantity underlying both the negativity and the log-negativity, it follows that these entanglement measures can be computed by means of semi-definite programs. To see the first equality, consider that%
\begin{align}
\norm{\T_B(\rho_{AB})}_1   & =\left\Vert \T_B(\rho_{AB})\right\Vert _{1}\\
& =\sup_{\left\Vert R_{AB}\right\Vert _{\infty}\leq1}\operatorname{Tr}%
[R_{AB}\T_{B}(\rho_{AB})]\\
& =\sup_{\left\Vert R_{AB}\right\Vert _{\infty}\leq1}\operatorname{Tr}%
[ \T_B(R_{AB})\rho_{AB}]\\
& =\sup_{\left\Vert \T_B(R_{AB})\right\Vert _{\infty}\leq1}\operatorname{Tr}%
[R_{AB}\rho_{AB}]\\
& =\sup_{R_{AB}}\left\{  \operatorname{Tr}[R_{AB}\rho_{AB}]:-\mathbbm{1}_{AB}\leq
\T_{B}(R_{AB})\leq \mathbbm{1}_{AB}\right\}  .
\end{align}
The second equality follows from H\"older duality (see \eqref{eq-Schatten_norm_var}), and since $\T_{B}(\rho_{AB})$
is Hermitian, it suffices to optimize over Hermitian $R_{AB}$. The third
equality follows because the partial transpose is its own Hilbert--Schmidt
adjoint. The fourth equality follows from the substitution $R_{AB}\rightarrow
\T_{B}(R_{AB})$. The final equality follows because the inequality $\left\Vert
\T_{B}(R_{AB})\right\Vert _{\infty}\leq1$ is equivalent to $-\mathbbm{1}_{AB}\leq
\T_{B}(R_{AB})\leq \mathbbm{1}_{AB}$ for a Hermitian operator $\T_{B}(R_{AB})$. 

Now consider that the set of Hermitian operators is equivalent to the set of
operators formed as differences of positive semi-definite operators. So this
implies that%
\begin{multline}
\norm{\T_B(\rho_{AB})}_1 =\\
\sup_{P_{AB},Q_{AB}\geq0}\left\{  \operatorname{Tr}[\left(
P_{AB}-Q_{AB}\right)  \rho_{AB}]:-\mathbbm{1}_{AB}\leq \T_B(P_{AB}-Q_{AB})\leq
\mathbbm{1}_{AB}\right\}  .
\end{multline}
Then by setting%
\begin{align}
X  & =%
\begin{pmatrix}
P_{AB} & 0\\
0 & Q_{AB}%
\end{pmatrix}
,\quad A=%
\begin{pmatrix}
\rho_{AB} & 0\\
0 & -\rho_{AB}%
\end{pmatrix}
,\quad
B=%
\begin{pmatrix}
\mathbbm{1}_{AB} & 0\\
0 & \mathbbm{1}_{AB}%
\end{pmatrix},\\
\Phi(X)  & =%
\begin{pmatrix}
\T_{B}(P_{AB}-Q_{AB}) & 0\\
0 & -\T_{B}(P_{AB}-Q_{AB})
\end{pmatrix}
,
\end{align}
this primal SDP is now in the standard form of \eqref{eq-primal_SDP_def}. Then, setting
\begin{equation}
Y=%
\begin{pmatrix}
K_{AB} & 0\\
0 & L_{AB}%
\end{pmatrix}
,
\end{equation}
we can calculate the Hilbert--Schmidt adjoint of $\Phi$ as%
\begin{align}
& \operatorname{Tr}[Y\Phi(X)]\nonumber\\
& =\operatorname{Tr}\left[
\begin{pmatrix}
K_{AB} & 0\\
0 & L_{AB}%
\end{pmatrix}%
\begin{pmatrix}
\T_{B}(P_{AB}-Q_{AB}) & 0\\
0 & -\T_{B}(P_{AB}-Q_{AB})
\end{pmatrix}
\right]  \\
& =\operatorname{Tr}[K_{AB}( \T_B(P_{AB}-Q_{AB}))]-\operatorname{Tr}%
[L_{AB}( \T_B(P_{AB}-Q_{AB}))]\\
& =\operatorname{Tr}[ \T_B(K_{AB})(P_{AB}-Q_{AB}))]-\operatorname{Tr}%
[ \T_B(L_{AB})(P_{AB}-Q_{AB})]\\
& =\operatorname{Tr}[ \T_B(K_{AB}-L_{AB})P_{AB}]+\operatorname{Tr}%
[ \T_B(L_{AB}-K_{AB})Q_{AB}]\\
& =\operatorname{Tr}\left[
\begin{pmatrix}
\T_{B}(K_{AB}-L_{AB}) & 0\\
0 & -\T_{B}(K_{AB}-L_{AB})
\end{pmatrix}%
\begin{pmatrix}
P_{AB} & 0\\
0 & Q_{AB}%
\end{pmatrix}
\right]  ,
\end{align}
so that%
\begin{equation}
\Phi^{\dag}(Y)=%
\begin{pmatrix}
\T_{B}(K_{AB}-L_{AB}) & 0\\
0 & -\T_{B}(K_{AB}-L_{AB})
\end{pmatrix}
.
\end{equation}
Then, plugging into the standard form for the dual SDP in \eqref{eq-dual_SDP_def} and
simplifying a bit, we find that it is given by%
\begin{multline}
\inf_{K_{AB},L_{AB}\geq0}\left\{
\begin{array}
[c]{c}%
\operatorname{Tr}[K_{AB}+L_{AB}]: \T_B(K_{AB}-L_{AB})\geq\rho_{AB},\\
-\T_{B}(K_{AB}-L_{AB})\geq-\rho_{AB}%
\end{array}
\right\}  \\
=\inf_{K_{AB},L_{AB}\geq0}\left\{  \operatorname{Tr}[K_{AB}+L_{AB}%
]: \T_B(K_{AB}-L_{AB})=\rho_{AB}\right\}  .
\end{multline}

Strong duality holds according to Theorem~\ref{thm:math-tools:slater-cond}. Indeed, setting $K_{AB}$ and $L_{AB}$
to the respective positive and negative parts of $\T_{B}(\rho_{AB})$ is
feasible for the dual, while setting $R_{AB}=\mathbbm{1}_{AB}/2$ is strictly feasible
for the primal.

\end{subappendices}

\chapter{Entanglement Measures for Quantum Channels}\label{chap-ent_measures_chan}

So far we have considered entanglement measures for quantum states. We now consider entanglement measures for channels. Using the general principle discussed in Section~\ref{sec-chan_inf_measures} for constructing channel quantities out of state quantitites, we arrive at the following definition for the entanglement of a channel.

...

\section{Definition and Basic Properties}
	
	...

	\begin{definition}{Entanglement of a Quantum Channel}{def:LAQC-ent-channel}
		From an entanglement measure $E$ defined on bipartite quantum states, we define the corresponding entanglement of a quantum channel $\mathcal{N}_{A\rightarrow B}$ as follows:
		\begin{equation}\label{eq:LAQC-ent-channel-pure}
			E(\mathcal{N})\coloneqq \sup_{\rho_{RA}}E(R;B)_{\omega},
		\end{equation}
		where $\omega_{RB}\coloneqq \mathcal{N}_{A\rightarrow B}(\rho_{RA})$, and the optimization is with respect to every bipartite state $\rho_{RA}$, with system $R$ arbitrarily large, yet finite.
	\end{definition}
	
	\begin{remark}
		Note that it suffices to optimize \eqref{eq:LAQC-ent-channel-pure} with respect to pure states $\psi_{RA}$, with the dimension of $R$ equal to the dimension of $A$, when calculating the entanglement of a channel, so that
		\begin{equation}
		E(\mathcal{N})\coloneqq \sup_{\psi_{RA}}E(R;B)_{\omega},
		\end{equation}
		where $\omega_{RB}\coloneqq \mathcal{N}_{A\rightarrow B}(\psi_{RA})$.
		This follows from the fact that an entanglement measure for states is, by definition, monotone under LOCC channels. It is therefore monotone under a local partial trace channel. In particular, consider a mixed state $\rho_{RA}$, with the dimension of $R$ not necessarily equal to the dimension of $A$. Let $\omega_{RB}=\mathcal{N}_{A\to B}(\omega_{RA})$. Then, if we take a purification $\phi_{R'RA}$ of $\rho_{RA}$, we obtain
		\begin{align}
			E(R;B)_{\omega}&=E(\mathcal{N}_{A\to B}(\rho_{RA}))\label{eq-ent_meas_chan_pure_pf1}\\
			&=E(\mathcal{N}_{A\to B}(\Tr_{R'}[\phi_{R'RA}]))\label{eq-ent_meas_chan_pure_pf2}\\
			&=E(\Tr_{R'}[\mathcal{N}_{A\to B}(\phi_{R'RA})])\label{eq-ent_meas_chan_pure_pf3}\\
			&\leq E(\mathcal{N}_{A\to B}(\phi_{R'RA}))\label{eq-ent_meas_chan_pure_pf4}\\
			&=E(R'R;B)_{\tau},
		\end{align}
		where $\tau_{R'RB}=\mathcal{N}_{A\to B}(\phi_{R'RA})$ and to obtain the inequality we used the fact that $E$ is monotone under the partial trace channel $\Tr_{R'}$. This demonstrates that it suffices to optimize with respect to pure states when calculating the entanglement of a channel. Furthermore, by the Schmidt decomposition theorem (Theorem~\ref{thm-Schmidt}), the dimension of the purifying system $R'R$ need not exceed the dimension of $A$.
	\end{remark}
	
	Note that in the  definition above the channel $\mathcal{N}_{A\to B}$ acts locally on the state $\psi_{RA}$ to produce the state $\omega_{RB}=\mathcal{N}_{A\to B}(\psi_{RA})$. We can thus view $\mathcal{N}_{A\to B}$ as an LOCC channel, which means that $E(R;B)_{\omega}\leq E(R;A)_{\psi}$, by the definition of an entanglement measure for states. In other words, by sending one share of a bipartite state through the channel $\mathcal{N}$, the entanglement can only stay the same or go down. The quantity $E(\mathcal{N})$ thus indicates how well  entanglement is preserved when one share of it is sent through the channel $\mathcal{N}$.

	Let us consider three examples of entanglement measures for quantum channels, defined using entanglement measures for bipartite quantum states.
	\begin{enumerate}
		\item The \textit{generalized divergence of entanglement} of a channel $\mathcal{N}_{A\to B}$, defined for every generalized divergence $\boldsymbol{D}$ as
			\begin{align}
				\boldsymbol{E}(\mathcal{N})&\coloneqq\sup_{\psi_{RA}}\boldsymbol{E}(R;B)_{\omega}\\
				&=\sup_{\psi_{RA}}\inf_{\sigma_{RB}\in\SEP(R:B)}\boldsymbol{D}(\mathcal{N}_{A\to B}(\psi_{RA})\Vert\sigma_{RB}),\label{eq-gen_div_ent_channel}
			\end{align}
			where $\omega_{RB}=\mathcal{N}_{A\to B}(\psi_{RA})$, and the optimization is with respect to pure states $\psi_{RA}$, with the dimension of $R$ equal to the dimension of $A$. We investigate this entanglement measure in Section~\ref{subsec-gen_div_ent_channel}.
			
		\item The \textit{generalized Rains divergence} of a channel $\mathcal{N}_{A\to B}$, defined for every generalized divergence $\boldsymbol{D}$ as
			\begin{align}
				\boldsymbol{R}(\mathcal{N})&\coloneqq\sup_{\psi_{SA}}\boldsymbol{R}(A;B)_{\omega}\\
				&=\sup_{\psi_{SA}}\inf_{\sigma_{SB}\in\PPT'(S:B)}\boldsymbol{D}(\mathcal{N}_{A\to B}(\psi_{SA})\Vert\sigma_{SB}),
			\end{align}
			where $\omega_{SB}=\mathcal{N}_{A\to B}(\psi_{SA})$, and the optimization is with respect to pure states $\psi_{SA}$, with the dimension of $S$ equal to the dimension of $A$. We investigate this entanglement measure in Section~\ref{subsec-Rains_gen_div_ent_channel}.
			
		\item The \textit{squashed entanglement} of a channel $\mathcal{N}_{A\to B}$, defined as
			\begin{equation}
				E_{\text{sq}}(\mathcal{N})\coloneqq\sup_{\psi_{RA}}E_{\text{sq}}(R;B)_{\omega},
			\end{equation}
			where $\omega_{RB}=\mathcal{N}_{A\to B}(\psi_{RA})$, and the optimization is with respect to pure states $\psi_{RA}$, with the dimension of $R$ equal to the dimension of $A$. We investigate this entanglement measure in Section~\ref{subsec-sq_ent_channel}.
	\end{enumerate}
	
	\begin{remark}
		Instead of defining an entanglement measure for channels via an entanglement measure for states, consider that the channel analogue of a separable state is an entanglement-breaking channel, which follows from the discussion in Section~\ref{subsec-ent_break_channels}. Another way to construct an entanglement measure for quantum channels is through the generalized channel divergence (Definition~\ref{def-gen_channel_div}) between the channel and the set of entanglement-breaking channels:
		\begin{equation}
			\boldsymbol{E}'(\mathcal{N})\coloneqq\inf_{\mathcal{M} \in \operatorname{EB}(A\to B)}\boldsymbol{D}(\mathcal{N}\Vert\mathcal{M}),
		\end{equation}
		where $\operatorname{EB}(A\to B)$ denotes the set of entanglement-breaking channels taking system $A$ to system $B$.
		Now, using the expression for the generalized channel divergence in \eqref{eq-gen_chan_div_pure}, we obtain
		\begin{equation}
			\boldsymbol{E}'(\mathcal{N})=\inf_{\mathcal{M}\in\text{EB}(A\to B)}\sup_{\psi_{RA}}\boldsymbol{D}(\mathcal{N}_{A\to B}(\psi_{RA})\Vert\mathcal{M}_{A\to B}(\psi_{RA})),
		\end{equation}
		where the optimization is with respect to pure states $\psi_{RA}$ is such that the dimension of $R$ is equal to the dimension of $A$.
		
		Now, because entanglement-breaking channels and separable states (with maximally mixed reduced state) are in one-to-one correspondence (see Section~\ref{subsec-ent_break_channels}), we find that the generalized divergence of entanglement of $\mathcal{N}$ is bounded from above as follows: 
		\begin{equation}
			\boldsymbol{E}(\mathcal{N})\leq\sup_{\psi_{RA}}\inf_{\mathcal{M}\in\text{EB}(A\to B)}\boldsymbol{D}(\mathcal{N}_{A\to B}(\psi_{RA})\Vert\mathcal{M}_{A\to B}(\psi_{RA})).
		\end{equation}
		The right-hand side of the above inequality and the quantity $\boldsymbol{E}'(\mathcal{N})$ differ in the order of the infimum and supremum. From the discussion in Section~\ref{sec-analysis_probability}, in particular \eqref{eq-minimax_gen_inequality}, we conclude that $\boldsymbol{E}(\mathcal{N})\leq\boldsymbol{E}'(\mathcal{N})$ for all quantum channels $\mathcal{N}$. For the rest of this chapter, and throughout the rest of this book, we thus stick with the definition of a channel entanglement measure given in Definition~\ref{def:LAQC-ent-channel}.
	\end{remark}
	
	Many properties of state entanglement measures carry over, or have an analogue, to the corresponding channel entanglement measure.
	
	\begin{proposition*}{Properties of Entanglement Measures for Channels}{prop-chan_ent_measure_properties}
		Let $E$ be an entanglement measure for states, as defined in Definition~\ref{def-LAQC:ent-measure}, and consider the corresponding channel entanglement measure defined in Definition~\ref{def:LAQC-ent-channel}.
		\begin{enumerate}
			\item\textit{Faithfulness}: If $E$ vanishes for all separable states, then $E(\mathcal{N})=0$ for all entanglement breaking channels. If $E$ is faithful (vanishing if and only if the input state is separable), then $E(\mathcal{N})=0$ implies that $\mathcal{N}$ is entanglement breaking.
			
			\item\textit{Convexity}: If $E$ is a convex entanglement measure for states, then the corresponding channel entanglement measure is convex: for every finite alphabet $\mathcal{X}$, probability distribution $p:\mathcal{X}\to[0,1]$, and set $\{\mathcal{N}_{A\to B}^x\}_{x\in\mathcal{X}}$ of quantum channels,
				\begin{equation}
					E\!\left(\sum_{x\in\mathcal{X}}p(x)\mathcal{N}^x\right)\leq\sum_{x\in\mathcal{X}} p(x)E(\mathcal{N}^x).
				\end{equation}
			
			\item\textit{Superadditivity}: If $E$ is superadditive, meaning that
				\begin{equation}
					E(A_1A_2;B_1B_2)_{\rho\otimes\tau}\geq E(A_1;B_1)_{\rho}+E(A_2;B_2)_{\tau}
				\end{equation}
				for all states $\rho_{A_1B_1}$ and $\tau_{A_2B_2}$, then the channel entanglment measure is also superadditive: for every two channels $\mathcal{N}_{A_1\to B_1}$ and $\mathcal{M}_{A_2\to B_2}$,
				\begin{equation}
					E(\mathcal{N}\otimes\mathcal{M})\geq E(\mathcal{N})+E(\mathcal{M}).
				\end{equation}
		\end{enumerate}
	\end{proposition*}
	
	\begin{Proof}
		\hfill\begin{enumerate}
			\item Let $\mathcal{N}$ be an entanglement breaking channel. This means that $\omega_{RB}=\mathcal{N}_{A\to B}(\psi_{RA})$ is separable for every pure state $\psi_{RA}$. Therefore, since $E$ vanishes for all separable states, we have $E(R;B)_{\omega}=0$ for every pure state $\psi_{RA}$, so that $E(\mathcal{N})=0$.
				
				Now, suppose that $E$ is a faithful state entanglement measure, and let $E(\mathcal{N})=0$. Since $E$ is faithful, it is non-negative for all input states, which implies that  $E(R;B)_{\omega}=0$ for every state $\omega_{RB}=\mathcal{N}_{A\to B}(\psi_{RA})$, i.e., for every pure state $\psi_{RA}$. Furthermore, by faithfulness of $E$ for states, it holds that $\mathcal{N}_{A\to B}(\psi_{RA})$ is separable for every pure state $\psi_{RA}$. Therefore, $\mathcal{N}$ is entanglement breaking.
				
			\item Let $\psi_{RA}$ be an arbitrary pure state, and let
				\begin{equation}
					\omega_{RB}=\left(\sum_{x\in\mathcal{X}}\mathcal{N}_{A\to B}^x\right)(\psi_{RA})				
					=\sum_{x\in\mathcal{X}}p(x)\omega_{RB}^x,
				\end{equation}
				where $\omega_{RB}^x=\mathcal{N}_{A\to B}^x(\psi_{RA})$. Then, by the convexity of $E$ for states,
				\begin{equation}
					E(R;B)_{\omega} \leq \sum_{x\in\mathcal{X}}p(x)E(R;B)_{\omega^x}
					\leq \sum_{x\in\mathcal{X}}p(x)E(\mathcal{N}^x),
				\end{equation}
				where for the last inequality we used the definition of the channel entanglement measure. Therefore, for every pure state $\psi_{RA}$, we have
				\begin{equation}
					E(R;B)_{\omega}\leq\sum_{x\in\mathcal{X}}p(x)E(\mathcal{N}^x).
				\end{equation}
				Thus,
				\begin{equation}
					E\!\left(\sum_{x\in\mathcal{X}}p(x)\mathcal{N}^x\right)=\sup_{\psi_{RA}}E(R;B)_{\omega}\leq\sum_{x\in\mathcal{X}}p(x)E(\mathcal{N}^x),
				\end{equation}
				as required.
				
			\item By restricting the optimization in the definition of $E(\mathcal{N}\otimes\mathcal{M})$ to pure product states $\phi_{R_1A_1}\otimes\varphi_{R_2A_2}$, letting $\xi_{R_1B_1}=\mathcal{N}_{A_1\to B_1}(\varphi_{R_1A_1})$, $\tau_{R_2B_2}=\mathcal{M}_{A_2\to B_2}(\varphi_{R_2A_2})$, and using superadditivity of the state entanglement measure $E$, we obtain
				\begin{align}
					E(\mathcal{N}\otimes\mathcal{M})&=\sup_{\psi_{RA_1A_2}}E(R;B_1B_2)_{\omega}\\
					&\geq \sup_{\phi_{R_1A_2}\otimes\varphi_{R_2A_2}}E(R_1R_2;B_1B_2)_{\xi\otimes\tau}\\
					&\geq \sup_{\phi_{R_1A_1}}E(R_1;B_1)_{\xi}+\sup_{\varphi_{R_2A_2}}E(R_2;B_2)_{\tau}\\
					&=E(\mathcal{N})+E(\mathcal{M}),
				\end{align}
				as required. \qedhere
		\end{enumerate}
	\end{Proof}

\section{Amortized Entanglement}

\begin{figure}
		\centering
		\includegraphics[scale=0.9]{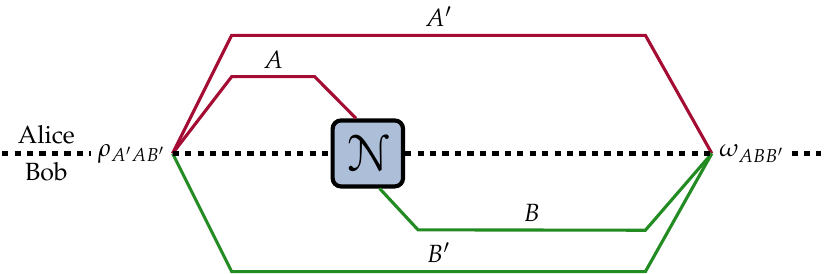}
		\caption{Starting from a state $\rho_{A'AB'}$, Alice sends the system $A$ through the channel $\mathcal{N}_{A\to B}$ to Bob, resulting in the state $\omega_{A'BB'}=\mathcal{N}_{A\to B}(\rho_{A'AB'})$. The difference between the final and initial entanglement (as quantified by an entanglement measure for states), optimized over all initial states $\rho_{A'AB'}$, is equal to the amortized entanglement of $\mathcal{N}$; see Definition~\ref{def:LAQC-amortized-ent}.}\label{fig-amort_ent}
	\end{figure}

	There is another way to define the entanglement of a quantum channel from an entanglement measure on quantum states, and this method turns out to be useful in the feedback-assisted communciation protocols that we consider in Part~\ref{part-feedback}. To see how this measure is defined, consider the situation shown in Figure~\ref{fig-amort_ent}. In this setup, Alice and Bob each have access to systems $A'$ and $B'$, respectively. Alice also possesses the system $A$, and she passes it through the channel $\mathcal{N}_{A\to B}$ to Bob. The initial joint state $\rho_{A'AB'}$ then becomes $\omega_{ABB'}=\mathcal{N}_{A\to B}(\rho_{A'AB'})$. The systems $A'$ and $B'$ can be thought of as ``memory systems'' that hold auxiliary information. In feedback-assisted quantum communication protocols, these memory systems can hold the results of previous rounds of communication between Alice and Bob for the purpose of deciding what local operations to perform in subsequent rounds. By taking the difference between the final entanglement in the state $\omega_{A'BB'}$ and the initial entanglement in the state $\rho_{A'AB'}$, and optimizing over all initial states $\rho_{A'AB'}$, we arrive at what is called the \textit{amortized entanglement}.

	\begin{definition}{Amortized Entanglement of a Quantum Channel}{def:LAQC-amortized-ent}%
		From an entanglement measure $E$ defined on bipartite quantum states, we define the \textit{amortized entanglement} of a quantum channel $\mathcal{N}_{A\rightarrow B}$ as follows:%
		\begin{equation}
			E^{\mathcal{A}}(\mathcal{N})\coloneqq \sup_{\rho_{A'AB'}}\left\{E(A';BB')_{\omega}-E(A'A;B')_{\rho}\right\},
		\end{equation}
		where $\omega_{A'BB'}\coloneqq \mathcal{N}_{A\rightarrow
B}(\rho_{A'AB'})$ and the optimization is with respect to
states $\rho_{A'AB'}$. The systems $A'$ and
$B'$ have arbitrarily large, yet finite dimensions.
	\end{definition}

	Due to the fact that the systems $A'$ and $B'$ can be
arbitrarily large, it is not necessarily the case that the supremum above can be achieved. Thus, in general, it might be difficult to compute a channel's amortized entanglement.

	For every entanglement measure $E$ that is equal to zero for all separable states, we always have that the entanglement of the channel never exceeds the amortized entanglement of the channel:

	\begin{Lemma}{lem:LAQC-amort-to-unamort}
		For a given quantum channel $\mathcal{N}$ and entanglement measure $E$ that is equal to zero for all separable states, the channel's entanglement does not exceed its amortized entanglement:%
		\begin{equation}\label{eq-ent_measure_chan_vs_amort}
			E(\mathcal{N})\leq E^{\mathcal{A}}(\mathcal{N}).
		\end{equation}
	\end{Lemma}

	\begin{Proof}
		By choosing the input state $\rho_{A'AB'}$ in the optimization for amortized entanglement to have a trivial (one-dimensional) system $B'$ (so that $\rho_{A'AB'}$ is trivially a separable state between Alice and Bob), we find that $E(A';BB')_{\omega}=E(A';B)_{\omega}$ and
$E(AA';B')_{\rho}=0$. Since such a state is an arbitrary state
to consider for optimizing the channel's entanglement, the inequality follows.
	\end{Proof}
	
	Whether the inequality reverse to the one in \eqref{eq-ent_measure_chan_vs_amort} holds, which would imply that $E(\mathcal{N})=E^{\mathcal{A}}(\mathcal{N})$, depends on the entanglement measure $E$. In Section~\ref{sec-amort_collapses} below, we show that this so-called ``amortization collapse'' occurs for some entanglement measures.

	The amortized entanglement of a channel has several interesting properties, which we list in some detail in this section. These include convexity, faithfulness, and (sub)additivity.

	\begin{proposition*}{Properties of Amortized Entanglement of a Quantum Channel}{prop-chan_amort_ent_properties}
		
		\begin{enumerate}
			\item \textit{Dimension bound}: Let $E$ be a subadditive entanglement measure for states, and let $\mathcal{N}_{A\rightarrow B}$ be a quantum channel. Then,
			\begin{equation}\label{prop:LAQC-tp-dim-upper-bound}
				E^{\mathcal{A}}(\mathcal{N})\leq\min\{E(A;A')_{\Phi},E(B;B')_{\Phi}\},
			\end{equation}
			where $A'$ has the same dimension as $A$, $B'$ has the same dimension as $B$, and $\Phi^+$ is a maximally entangled state of systems $AA'$ or $BB'$.
			
			\item \textit{Faithfulness}: Let $E$ be an entanglement measure that is equal to zero for all separable states. If a channel $\mathcal{N}$ is entanglement-breaking, then its amortized entanglement $E^{\mathcal{A}}(\mathcal{N})$ is equal to zero. If the entanglement measure $E$ is faithful (equal to zero if and only if the state is separable) and the amortized entanglement $E^{\mathcal{A}}(\mathcal{N})$ of a channel $\mathcal{N}$ is equal to zero, then the channel
$\mathcal{N}$ is entanglement breaking.
			
			\item \textit{Convexity}: Let $E$ be a convex entanglement measure for states. Then, the amortized entanglement $E^{\mathcal{A}}$ of a channel is convex: for every finite alphabet $\mathcal{X}$, probability distribution $p:\mathcal{X}\to[0,1]$, and set $\{\mathcal{N}_{A\to B}^x\}_{x\in\mathcal{X}}$ of quantum channels,
			\begin{equation}\label{eq:LAQC-amortized-ent-convex}
				E^{\mathcal{A}}\!\left(\sum_{x\in\mathcal{X}}p(x)\mathcal{N}^x\right)\leq \sum_{x\in\mathcal{X}}p(x)E^{\mathcal{A}}(\mathcal{N}^x).
			\end{equation}
			
			\item \textit{Subdditivity and additivity}: For every entanglement measure $E$, the amortized entanglement $E^{\mathcal{A}}$\ of a channel is subadditive, meaning that for every two quantum channels $\mathcal{N}$ and $\mathcal{M}$,
				\begin{equation}\label{eq:LAQC-amortized-subadd-gen}%
					E^{\mathcal{A}}(\mathcal{N}\otimes\mathcal{M})\leq E^{\mathcal{A}}(\mathcal{N})+E^{\mathcal{A}}(\mathcal{M}).
				\end{equation}
				
				If $E$ is an additive entanglement measure, then the amortized entanglement $E^{\mathcal{A}}$ is additive, meaning that
				\begin{equation} \label{eq:LAQC-additivity-amortized}%
					E^{\mathcal{A}}(\mathcal{N}\otimes\mathcal{M})=E^{\mathcal{A}}(\mathcal{N})+E^{\mathcal{A}}(\mathcal{M})
				\end{equation}
				for every two quantum channels $\mathcal{N}$ and $\mathcal{M}$.
		\end{enumerate}
	
	\end{proposition*}
	
	\begin{Proof}
		\hfill\begin{enumerate}
			\item To prove \eqref{prop:LAQC-tp-dim-upper-bound}, we use the fact that $\mathcal{N}$ can be simulated via teleporation. Specifically, from \eqref{eq-chan_tele_sim_1} and \eqref{eq-chan_tele_sim_2}, we can represent the action of $\mathcal{N}_{A\to B}$ on every state $\rho_{A'AB'}$ in the following two ways:
				\begin{align}
					\mathcal{N}_{A\to B}(\rho_{A'AB'})&=\mathcal{N}_{B_i'\to B}\left(\mathcal{T}_{AA_iB_i\to B_i'}(\rho_{A'AB'}\otimes \Phi_{A_iB_i})\right),\\
					\mathcal{N}_{A\to B}(\rho_{A'AB'})&=\mathcal{T}_{A_o'A_oB_o\to B}\left(\mathcal{N}_{A\to A_o'}(\rho_{A'AB'})\otimes\Phi_{A_oB_o}^+\right),
				\end{align}
				where $A_i,B_i,B_i'$ are auxiliary systems such that $d_{A_i}=d_{B_i}=d_{B_i'}=d_A$ (i.e., systems with the same dimension as the input system $A$ of the channel) and $A_o,A_o',B_o$ are auxiliary systems such that $d_{A_o}=d_{A_o'}=d_{B_o}=d_{B}$ (i.e., systems with the same dimension as the output system $B$ of the channel). The teleportation channel $\mathcal{T}$ is given by \eqref{eq-teleportation_LOCC_chan}. Since $\mathcal{T}$ is an LOCC channel, and $\mathcal{N}$ is a local channel, by LOCC monotonicity of the entanglement measure $E$ we obtain
				\begin{align}
					E(A';BB')_{\omega}&\leq E(A'AA_i;B'B_i)_{\rho\otimes\Phi},\\
					E(A';BB')_{\omega}&\leq E(A'AA_o;B'B_o)_{\rho\otimes\Phi},
				\end{align}
				where $\omega_{A'BB'}=\mathcal{N}_{A\to B}(\rho_{A'AB'})$. Then, by subadditivity of $E$, we find that
				\begin{align}
					E(A';BB')_{\omega}&\leq E(A'A;B')_{\rho}+E(A_i;B_i)_{\Phi},\\
					E(A';BB')_{\omega}&\leq E(A'A;B')_{\rho}+E(A_o;B_o)_{\Phi}.
				\end{align}
				Since the state $\rho_{A'AB'}$ is arbitrary, we conclude that
				\begin{align}
					E^{\mathcal{A}}(\mathcal{N})&\leq E(A_i;B_i)_{\Phi}=E(A;A')_{\Phi},\\
					E^{\mathcal{A}}(\mathcal{N})&\leq E(A_o;B_o)_{\Phi}=E(B;B')_{\Phi},
				\end{align}
				where for the last equality in each case we used $d_{A_i}=d_{B_i}=d_{A'}=d_A$ and $d_{A_o}=d_{B_o}=d_{B'}=d_B$. We thus have
				\begin{equation}
					E^{\mathcal{A}}(\mathcal{N})\leq\min\{E(A;A')_{\Phi},E(B;B')_{\Phi}\},
				\end{equation}
				as required.
			
			\item Let $\mathcal{N}$ be an entanglement breaking channel. For every state $\rho_{A'AB'}$, let $\omega_{ABB'}=\mathcal{N}_{A\to B}(\rho_{A'AB'})$. Recall from Section~\ref{subsec-ent_break_channels}, specifically Theorem~\ref{thm-ent_break_meas_reprepare}, that every entanglement breaking channel can be represented as a measurement of the input system followed by the preparation of a state on the output system conditioned on the outcome of the measurement. As such, every entanglement breaking channel can be simulated by an LOCC channel. Therefore, by the monotonicity of the entanglement measure $E$ under LOCC,
			\begin{equation}
				E(A';BB')_{\omega}\leq E(A'A;B')_{\rho},
			\end{equation}
			which means that
			\begin{equation}
				E(A';BB')_{\omega}-E(A'A;B')_{\rho}\leq E(A'A;B')_{\rho}-E(A'A;B')_{\rho}=0
			\end{equation}
			for every state $\rho_{A'AB'}$. Therefore, $E^{\mathcal{A}}(\mathcal{N})\leq 0$. On the other hand, because $E$ vanishes for all separable states, it holds that $E(\mathcal{N})\geq 0$. Therefore, by Lemma~\ref{lem:LAQC-amort-to-unamort}, $E^{\mathcal{A}}(\mathcal{N})\geq 0$, and we conclude that $E^{\mathcal{A}}(\mathcal{N})=0$.

				Now, let $E$ be a faithful entanglement measure, meaning that it vanishes if and only if the input state is separable, and suppose that $E^{\mathcal{A}}(\mathcal{N})=0$. By Lemma~\ref{lem:LAQC-amort-to-unamort}, we have that $0=E^{\mathcal{A}}(\mathcal{N})\geq E(\mathcal{N})$, which in turn implies that $E(\mathcal{N})=0$ because $E(\mathcal{N})\geq 0$ for all channels $\mathcal{N}$. Therefore, by Proposition~\ref{prop-chan_ent_measure_properties}, we conclude that $\mathcal{N}$ is entanglement breaking. 
				
			\item Let $\rho_{A'AB'}$ be an arbitrary state, and let
				\begin{equation}
					\omega_{A'BB'}=\left(\sum_{x\in\mathcal{X}}p(x)\mathcal{N}_{A\to B}^x\right)(\rho_{A'AB'})=\sum_{x\in\mathcal{X}}p(x)\omega_{A'BB'}^x,
				\end{equation}
				where $\omega_{A'BB'}^x=\mathcal{N}_{A\to B}^x(\rho_{A'AB'})$ for all $x\in\mathcal{X}$. Then, by convexity of the entanglement measure $E$, we obtain
				\begin{equation}
					E(A';BB')_{\omega}\leq \sum_{x\in\mathcal{X}}p(x)E(A';BB')_{\omega^x}.
				\end{equation}
				Also, $E(A'A;B')_{\rho}=\sum_{x\in\mathcal{X}}p(x)E(A'A;B')_{\rho}$. Therefore, by the definition of amortized entanglement, we find that
				\begin{align}
					& E(A';BB')_{\omega}-E(A'A;B')_{\rho}\\
					&\qquad \leq \sum_{x\in\mathcal{X}}p(x)\left(E(A';BB')_{\omega^x}-E(A'A;B')_{\rho}\right)\\
					&\qquad \leq \sum_{x\in\mathcal{X}}p(x)E^{\mathcal{A}}(\mathcal{N}^x).
				\end{align}
				Since the state $\rho_{A'AB'}$ is arbitrary, by optimizing over every state $\rho_{A'AB'}$ on the left-hand side of the  inequality above, we obtain
				\begin{equation}
					E^{\mathcal{A}}\!\left(\sum_{x\in\mathcal{X}}p(x)\mathcal{N}^x\right)\leq\sum_{x\in\mathcal{X}}p(x)E^{\mathcal{A}}(\mathcal{N}^x),
				\end{equation}
				as required.

			\item Let $A_{1}$ and $B_{1}$ denote the respective input and output systems for the quantum channel $\mathcal{N}$, and let $A_{2}$ and $B_{2}$ denote the respective input and output quantum systems for the quantum channel $\mathcal{M}$. Let $\rho_{A'A_{1}A_{2}B'}$ be an arbitrary state. Let
				\begin{align}
					\omega_{A'B_{1}B_{2}B'}&=(\mathcal{N}_{A_{1}\rightarrow B_{1}}\otimes\mathcal{M}_{A_{2}\rightarrow B_{2}})(\rho_{A'A_{1}A_{2}B'})\\
					&=\mathcal{N}_{A_1\to B_1}(\tau_{A'A_1B_2B'}),
				\end{align}
				where
				\begin{equation}
					\tau_{A'A_1B_2B'}\coloneqq \mathcal{M}_{A_2\to B_2}(\rho_{A'A_1A_2B'}).
				\end{equation}
				Observe that the state $\omega_{A'B_1B_2B'}$ is both an example of an output state in the optimization defining $E^{\mathcal{A}}(\mathcal{N}\otimes\mathcal{M})$ and in the optimization defining $E^{\mathcal{A}}(\mathcal{N})$ (with an appropriate identification of the $A'$, $B$, and $B'$ systems for the latter). Observe also that $\tau_{A'A_1B_2B'}$ is an example of an output state in the optimization defining $E^{\mathcal{A}}(\mathcal{M})$. Therefore,
				\begin{align}
					&  E(A';B_{1}B_{2}B')_{\omega}-E(A'A_{1}A_{2};B')_{\rho}\nonumber\\
					&  =E(A';B_{1}B_{2}B')_{\omega}-E(A'A_{1};B_{2}B')_{\tau}\nonumber\\
					&  \qquad+E(A'A_{1};B_{2}B')_{\tau}-E(A'A_{1}A_{2};B')_{\rho}\\
					&  \leq E^{\mathcal{A}}(\mathcal{N})+E^{\mathcal{A}}(\mathcal{M}).\label{eq:LAQC-amortized-subadd-critical-ineq}%
				\end{align}
				Since the state $\rho_{A'A_1A_2B'}$ is arbitrary, we can optimize over all such states on the left-hand side of the  inequality above to obtain
				\begin{equation}\label{eq-ent_meas_amort_ent_subadditive_pf}
					E^{\mathcal{A}}(\mathcal{N}\otimes\mathcal{M})\leq E^{\mathcal{A}}(\mathcal{N})+E^{\mathcal{A}}(\mathcal{M}).
				\end{equation}
				
				Now, suppose that $E$ is an additive entanglement measure. Let us restrict the optimization over states $\rho_{A'A_1A_2B'}$ in the definition of $E^{\mathcal{A}}(\mathcal{N}\otimes\mathcal{M})$ such that $A'\equiv A_1'A_2'$, $B'\equiv B_1'B_2'$, and $\rho_{A'A_1A_2B'}=\rho_{A_1'A_1B_1'}^1\otimes\rho_{A_2'A_2B_2'}^2$, for states $\rho_{A_1'A_1B_1'}^1$ and $\rho_{A_2'A_2B_2'}^2$. Then,
				\begin{align}
					\omega_{A'B_1B_2B'}&=(\mathcal{N}_{A_1\to B_1}\otimes\mathcal{M}_{A_2\to B_2})(\rho_{A'A_1A_2B'})\\
					&=\mathcal{N}_{A_1\to B_1}(\rho_{A_1'A_1B_1'}^1)\otimes\mathcal{M}_{A_2\to B_2}(\rho_{A_2'A_2B_2'}^2)\\
					&=:\omega_{A_1'B_1B_1'}^1\otimes\omega_{A_2B_2B_2'}^2.
				\end{align}
				Therefore, using additivity of $E$, we obtain
				\begin{align}
					E^{\mathcal{A}}(\mathcal{N}\otimes\mathcal{M})&=\sup_{\rho_{A'A_1A_2B'}}\{E(A';B_1B_2B')_{\omega}-E(A'A_1A_2;B')_{\rho}\}\\
					&\geq \sup_{\rho_{A_1'A_1B_1'}^1\otimes\rho_{A_2'A_2B_2'}^2}\{E(A_1'A_2';B_1B_2B_1'B_2')_{\omega^1\otimes\omega^2}\nonumber\\
					&\qquad\qquad\qquad\qquad-E(A_1'A_2'A_1A_2;B_1'B_2')_{\rho^1\otimes\rho^2}\}\\
					&\geq \sup_{\rho_{A_1'A_1B_1'}^1\otimes\rho_{A_2'A_2B_2'}^2}\{E(A_1';B_1B_1')_{\omega^1}+E(A_2';B_2B_2')_{\omega^2} \nonumber \\
					&\qquad\qquad\qquad -E(A_1'A_1;B_1')_{\rho^1}-E(A_2'A_2;B_2')_{\rho^2}\}\\
					&=\sup_{\rho_{A_1'A_1B_1'}^1}\{E(A_1';B_1B_1')_{\omega^1}-E(A_1'A_1;B_1')_{\rho^1}\} \nonumber\\
					&\qquad\qquad+\sup_{\rho_{A_2'A_2B_2'}^2}\{E(A_2';B_2B_2')_{\omega^2}-E(A_2'A_2;B_2')_{\rho^2}\}\\
					&=E^{\mathcal{A}}(\mathcal{N})+E^{\mathcal{A}}(\mathcal{M}).
				\end{align}
				Combining this with \eqref{eq-ent_meas_amort_ent_subadditive_pf}, we conclude that
				\begin{equation}
					E^{\mathcal{A}}(\mathcal{N}\otimes\mathcal{M})=E^{\mathcal{A}}(\mathcal{N})+E^{\mathcal{A}}(\mathcal{M}),
				\end{equation}
				as required. \qedhere
		\end{enumerate}
	\end{Proof}

	An immediate consequence of \eqref{eq:LAQC-amortized-subadd-gen} is the following inequality:%
	\begin{equation}
		\sup_{\mathcal{M}}\left[  E^{\mathcal{A}}(\mathcal{N}\otimes\mathcal{M})-E^{\mathcal{A}}(\mathcal{M})\right]  \leq E^{\mathcal{A}}(\mathcal{N}),
	\end{equation}
	where the supremum is with respect to quantum channels $\mathcal{M}$. This inequality demonstrates that no other channel can help to enhance the amortized entanglement of a quantum channel.

\subsection{Amortized Entanglement and Teleportation Simulation}\label{sec:LAQC-TP-simulation-def}

	Teleportation (or, more generally, LOCC, separable, or PPT) simulation of a quantum channel is a key tool that we can use to establish upper bounds on capacities of certain quantum channels when they are assisted by LOCC. Recalling Section~\ref{subsec-tele_sim}, the basic idea behind this tool is that a quantum channel can be simulated by the action of a teleportation protocol, with a maximally entangled resource state shared between the sender $A$ and receiver~$B$. More generally, recalling Definition~\ref{def-LOCC_sim_chan}, a channel $\mathcal{N}_{A\rightarrow B}$ with input system $A$ and output system $B$ is defined to be LOCC-simulable with associated resource state $\omega_{RB'}$ if the following equality holds for all input states $\rho_{A}$:%
	\begin{equation}
		\mathcal{N}_{A\rightarrow B}(\rho_{A})=\mathcal{L}_{ARB'\rightarrow B}(\rho_{A}\otimes\omega_{RB'}),
	\end{equation}
	where $\mathcal{L}_{ARB'\rightarrow B}$ is a quantum channel consisting of LOCC\ between the sender, who has systems $A$ and $R$, and the receiver, who has system $B'$. Whenever the underlying state entanglement measure is subadditive, the amortized entanglement of an LOCC-simulable channel can be bounded from above by the entanglement of the resource state. In fact, this is precisely what we did when proving the dimension bound in Proposition~\ref{prop-chan_amort_ent_properties} above. We can therefore understand the dimension bound as being a consequence of the fact that all channels are teleportation simulable, and hence LOCC simulable, by using a maximally entangled resource state. 

	\begin{proposition}{prop:LAQC-tp-upper-bound}
		Let $E$ be a subadditive state entanglement measure (recall \eqref{def-ent_meas_subadditive}). If a quantum channel $\mathcal{N}_{A\rightarrow B}$ is LOCC-simulable with associated resource state $\omega_{RB'}$, i.e.,
		\begin{equation}
			\mathcal{N}_{A\to B}(\rho_A)=\mathcal{L}_{ARB'\to B}(\rho_A\otimes\omega_{RB'}),
		\end{equation}
		where $\mathcal{L}_{ARB'\to B}$ is an LOCC channel, then the amortized entanglement $E^{\mathcal{A}}(\mathcal{N})$ of $\mathcal{N}$ is bounded from above by the entanglement of the resource state:
		\begin{equation}\label{eq:LAQC-tp-upper-bound}
			E^{\mathcal{A}}(\mathcal{N})\leq E(R;B')_{\omega}.
		\end{equation}
	\end{proposition}

	\begin{Proof}
		For every state $\rho_{A'AB^{\prime\prime}}$, we use monotonicity of the state entanglement measure under LOCC, as well as subadditivity of the measure, to obtain
		\begin{align}
			&  E(A';BB^{\prime\prime})_{\mathcal{L}(\rho\otimes\omega)}-E(A'A;B^{\prime\prime})_{\rho}\nonumber\\
			&  \leq E(A'AR;B^{\prime\prime}B')_{\rho\otimes\omega}-E(A'A;B^{\prime\prime})_{\rho}\\
			&  \leq E(A'A;B^{\prime\prime})_{\rho}+E(R;B')_{\omega}-E(A'A;B^{\prime\prime})_{\rho}\\
			&  =E(R;B')_{\omega},
		\end{align}
		where for the first inequality we made use of LOCC monotonicity and for the second inequality we made use of the assumption of subadditivity. Since the state $\rho_{A'AB''}$ was arbitrary, we conclude \eqref{eq:LAQC-tp-upper-bound}.
	\end{Proof}
	
	If it happens that a channel $\mathcal{N}_{A\to B}$ is LOCC-simulable with resource state $\omega_{RB'}=\mathcal{N}_{A\to B'}(\rho_{RA})$ for some state $\rho_{RA}$, then the inequality in \eqref{eq:LAQC-tp-upper-bound} becomes an equality. In Section~\ref{subsec-tele_sim}, we saw an example in which such a situation arises, namely, when the channel $\mathcal{N}$ is group covariant. In this case, the resource state is simply the Choi state of the channel.

	\begin{proposition}{prop:LAQC-TP-sim-amortized-equal}
		Let $E$ be an entanglement measure that is subadditive with respect to states and zero  on separable states, and let $E^{\mathcal{A}}$ denote its amortized version. If a channel $\mathcal{N}_{A\rightarrow B}$\ is LOCC-simulable with associated resource state $\omega_{RB}=\mathcal{N}_{A\rightarrow B}(\rho_{RA})$ for some input state $\rho_{RA}$, then the following equality holds:
		\begin{equation}
			E^{\mathcal{A}}(\mathcal{N})=E(R;B)_{\omega}.
		\end{equation}
	\end{proposition}

	\begin{Proof}
		From Proposition~\ref{prop:LAQC-tp-upper-bound}, we have that $E^{\mathcal{A}}(\mathcal{N})\leq E(R;B)_{\omega}$. For the reverse inequality, we take $\rho_{A'AB''}=\rho_{RA}$ in the optimization that defines $E^{\mathcal{A}}(\mathcal{N})$, where we identify $A'\equiv R$ and $B''\equiv\varnothing$ (i.e., $B''$ is a trivial one-dimensional system). Then, $\mathcal{N}_{A\to B}(\rho_{A'AB''})=\mathcal{N}_{A\to B}(\rho_{RA})$, which is the resource state. Furthermore, since $B''$ is a one-dimensional system, the state $\rho_{A'AB''}$ is trivially separable, so that $E(A'A;B'')_{\rho}=0$. Therefore, $E^{\mathcal{A}}(\mathcal{N})\geq E(A';B)_{\omega}\equiv E(R;B)_{\omega}$.
	\end{Proof}

\section{Generalized Divergence of Entanglement}\label{subsec-gen_div_ent_channel}

	In this section, we examine the generalized divergence of entanglement of quantum channels, which is a channel entanglement measure that arises from the generalized divergence of entanglement for quantum states that we considered in Section~\ref{sec-ent_measures_sep_distance}.
	
	\begin{definition}{Generalized Divergence of Entanglement}{def-gen_div_ent_chan}
		Let $\boldsymbol{D}$ be a generalized divergence (see Definition~\ref{def-gen_div}). For every quantum channel $\mathcal{N}_{A\to B}$, we define the \textit{generalized divergence of entanglement of $\mathcal{N}$} as
		\begin{align}
			\boldsymbol{E}(\mathcal{N})&\coloneqq\sup_{\psi_{RA}} \boldsymbol{E}(R;B)_{\omega}\\
			&=\sup_{\psi_{RA}}\inf_{\sigma_{RB}\in\SEP(R:B)}\boldsymbol{D}(\mathcal{N}_{A\to B}(\psi_{RA})\Vert\sigma_{RB}),
		\end{align}
		where $\omega_{RB}\coloneqq\mathcal{N}_{A\to B}(\psi_{RA})$. The supremum is with respect to every pure state $\psi_{RA}$, with the dimension of $R$ equal to the dimension of $A$.
	\end{definition}
	
	In the remark immediately after Definition~\ref{def:LAQC-ent-channel}, we stated how it suffices to optimize with respect to pure bipartite states (with equal dimension for each subsystem) when calculating the generalized divergence of entanglement of a quantum channel. 
	
	We can write the generalized divergence of entanglement of $\mathcal{N}_{A\to B}$ in the following alternate form:
	\begin{align}
		\boldsymbol{E}(\mathcal{N})&=\sup_{\rho_A}\boldsymbol{E}(\mathcal{N}_{A\to B},\rho_A),\label{eq-gen_div_ent_chan_alt}\\
		\boldsymbol{E}(\mathcal{N}_{A\to B},\rho_A)&\coloneqq \inf_{\sigma_{AB}\in\SEP(A:B)}\boldsymbol{D}(\sqrt{\rho_A}\Gamma_{AB}^{\mathcal{N}}\sqrt{\rho_A}\Vert\sigma_{AB}).
	\end{align}
	This is indeed true because  we can write every purification of $\rho_A$ as $(V_{A'}\sqrt{\rho_{A'}}\otimes\mathbbm{1}_A)\ket{\Gamma}_{A'A}$ for some isometry $V_{A'}$ (see \eqref{eq-pure_state_vec} and Theorem~\ref{thm-polar_decomposition}), that the set of separable states is invariant under local isometries, and that generalized divergences are invariant under local isometries (see Proposition~\ref{prop-gen_div_properties}).

	
	As with the state quantities, we are interested in the following generalized divergences of entanglement of a quantum channel $\mathcal{N}_{A\to B}$:
	\begin{enumerate}
		\item The \textit{relative entropy of entanglement of $\mathcal{N}$},
			\begin{align}
				E_R(\mathcal{N})&\coloneqq\sup_{\psi_{SA}}E_R(S;B)_{\omega}\\
				&= \sup_{\psi_{SA}}\inf_{\sigma_{SB}\in\SEP(S:B)}D(\mathcal{N}_{A\to B}(\psi_{SA})\Vert\sigma_{SB}),
				\label{eq-EM:REE-channel-def}
			\end{align}
			where $\omega_{SB}=\mathcal{N}_{A\to B}(\psi_{SA})$.
		
		\item The \textit{$\varepsilon$-hypothesis testing relative entropy of entanglement of $\mathcal{N}$},
			\begin{align}
				E_R^{\varepsilon}(\mathcal{N})&\coloneqq\sup_{\psi_{RA}}E_R^{\varepsilon}(R;B)_{\omega}\\
				&=\sup_{\psi_{RA}}\inf_{\sigma_{RB}\in\SEP(R:B)}D_H^{\varepsilon}(\mathcal{N}_{A\to B}(\psi_{RA})\Vert\sigma_{RB}),
				\label{eq-EM:eps-REE-channel}
			\end{align}
			where $\omega_{RB}=\mathcal{N}_{A\to B}(\psi_{RA})$.
			
		\item The \textit{sandwiched R\'{e}nyi relative entropy of entanglement of $\mathcal{N}$},
			\begin{align}
				\widetilde{E}_{\alpha}(\mathcal{N})&\coloneqq\sup_{\psi_{RA}}\widetilde{E}_{\alpha}(R;B)_{\omega}\\
				&=\sup_{\psi_{RA}}\inf_{\sigma_{RB}\in\SEP(R:B)}\widetilde{D}_{\alpha}(\mathcal{N}_{A\to B}(\psi_{RA})\Vert\sigma_{RB}),
				\label{eq-EM:sandwiched-REE-channel}
			\end{align}
			where $\omega_{RB}\in\mathcal{N}_{A\to B}(\psi_{RA})$ and $\alpha\in\left[\sfrac{1}{2},1\right)\cup(1,\infty)$. It follows from Proposition~\ref{prop-sand_rel_ent_properties} that $\widetilde{E}_{\alpha}(\mathcal{N})$ is monotonically increasing in $\alpha$. Also, in Appendix~\ref{app-sand_ren_inf_limit}, we prove that
			\begin{equation}
				E_R(\mathcal{N})=\lim_{\alpha\to 1^+}\widetilde{E}_{\alpha}(\mathcal{N})
			\end{equation}
			for every quantum channel $\mathcal{N}$.
			
		\item The \textit{max-relative entropy of entanglement of $\mathcal{N}$},
			\begin{align}
				E_{\max}(\mathcal{N})&\coloneqq\sup_{\psi_{RA}}E_{\max}(R;B)_{\omega}\\
				&=\sup_{\psi_{RA}}\inf_{\sigma_{RB}\in\SEP(R:B)}D_{\max}(\mathcal{N}_{A\to B}(\psi_{RA})\Vert\sigma_{RB}),
			\end{align}
			where $\omega_{RB}=\mathcal{N}_{A\to B}(\psi_{RA})$. In Appendix~\ref{app-sand_ren_inf_limit}, we prove that
			\begin{equation}
				E_{\max}(\mathcal{N})=\lim_{\alpha\to\infty}\widetilde{E}_{\alpha}(\mathcal{N})
			\end{equation}
			for every quantum channel $\mathcal{N}$.
	\end{enumerate}
	
	The generalized divergence of entanglement of a quantum channel satisfies all of the general properties of a channel entanglement measure shown in Proposition~\ref{prop-chan_ent_measure_properties} except for the superadditivity property, which holds when the generalized divergence of a bipartite state is superadditive. However, as shown in Proposition~\ref{prop-gen_div_ent_properties}, the generalized divergence of entanglement of a bipartite state is generally only subadditive. Thus, neither the superadditivity nor the subadditivity of the generalized divergence of entanglement of a channel immediately follows. In Section~\ref{sec-amort_collapses} below, we show that the max-relative entropy of entanglement of a quantum channel is subadditive, meaning that
	\begin{equation}\label{eq-E_max_additivity_0}
		E_{\max}(\mathcal{N}\otimes\mathcal{M})\leq E_{\max}(\mathcal{N})+E_{\max}(\mathcal{M})
	\end{equation}
	for all quantum channels $\mathcal{N}$ and $\mathcal{M}$.
	
	The amortized generalized divergence of entanglement, defined according to Definition~\ref{def:LAQC-amortized-ent} as
	\begin{equation}
		\boldsymbol{E}^{\mathcal{A}}(\mathcal{N})\coloneqq\sup_{\rho_{A'AB'}}\{\boldsymbol{E}(A';BB')_{\omega}-\boldsymbol{E}(A'A;B)_{\rho}\},
	\end{equation}
	where $\omega_{A'BB'}=\mathcal{N}_{A\to B}(\rho_{A'AB'})$, satisfies all of the properties stated in Proposition~\ref{prop-chan_amort_ent_properties}. In particular, it is subadditive. We show in Section~\ref{sec-amort_collapses} below that $E_{\max}(\mathcal{N})=E_{\max}^{\mathcal{A}}(\mathcal{N})$ for every quantum channel $\mathcal{N}$, and this is what leads to the subadditivity statement in \eqref{eq-E_max_additivity_0}.
	
	For covariant channels, the optimization over pure input states in the generalized divergence of entanglement can be simplified, as we now show. This simplification is similar to the simplification that occurs for the generalized channel divergence for jointly covariant channels (see Proposition~\ref{prop-gen_div_group_cov}).
	
	\begin{proposition*}{Generalized Divergence of Entanglement for Covariant Channels}{prop-gen_div_ent_chan_cov}
		Let $\mathcal{N}_{A\to B}$ be a $G$-covariant quantum channel for a finite group $G$ (recall Definition~\ref{def-group_cov_chan}). Then, for every pure state $\psi_{RA}$, with the dimension of $R$ equal to the dimension of $A$, we have that
		\begin{equation}\label{eq-gen_div_ent_chan_cov_state_symm}
			\boldsymbol{E}(R;B)_{\omega}\leq \boldsymbol{E}(R;B)_{\overline{\omega}},
		\end{equation}
		where $\omega_{RB}\coloneqq\mathcal{N}_{A\to B}(\psi_{RA})$, $\overline{\omega}_{RB}\coloneqq\mathcal{N}_{A\to B}(\phi^{\overline{\rho}}_{RA})$,
		\begin{equation}\label{eq-psi_RA_group_symmetrize0}
			\overline{\rho}_{A}=\frac{1}{|G|}\sum_{g\in G}U_A^g\psi_{A}U_A^{g\dagger}\eqqcolon\mathcal{T}_G(\psi_A),
		\end{equation}
		 and $\phi_{RA}^{\overline{\rho}}$ is a purification of $\overline{\rho}_A$. Consequently,
		\begin{equation}\label{eq-gen_div_ent_chan_cov}
			\boldsymbol{E}(\mathcal{N})=\sup_{\phi_{RA}}\{\boldsymbol{E}(R;B)_{\omega}:\phi_A=\mathcal{T}_G(\phi_A)\},
		\end{equation}
		where $\omega_{RB}=\mathcal{N}_{A\to B}(\phi_{RA})$.
		In other words, in order to calculate $\boldsymbol{E}(\mathcal{N})$, it suffices to optimize with respect to pure states $\phi_{RA}$ such that the reduced state $\phi_A$ is invariant under the channel $\mathcal{T}_G$ defined in \eqref{eq-psi_RA_group_symmetrize0}.
	\end{proposition*}
	
	\begin{remark}
		Using \eqref{eq-gen_div_ent_chan_alt}, we can write \eqref{eq-gen_div_ent_chan_cov_state_symm} as
		\begin{equation}
			\boldsymbol{E}(\mathcal{N},\rho)\leq\boldsymbol{E}(\mathcal{N},\mathcal{T}_G(\rho)),
		\end{equation}
		which holds for every state $\rho$ acting on the input space of the channel $\mathcal{N}$. We can write \eqref{eq-gen_div_ent_chan_cov} as
		\begin{equation}
			\boldsymbol{E}(\mathcal{N})=\sup_{\rho}\{\boldsymbol{E}(\mathcal{N},\rho):\rho=\mathcal{T}_G(\rho)\}.
		\end{equation}
	\end{remark}
	
	\begin{Proof}
		The proof is similar to the proof of Proposition~\ref{prop-gen_div_group_cov}. Let $\psi_{RA}$ be an arbitrary pure state,  and let $\overline{\rho}_A=\mathcal{T}_G(\psi_A)$. Furthermore, let $\phi_{RA}^{\overline{\rho}}$ be a purification of $\overline{\rho}_A$. Let us also consider the following purification of $\overline{\rho}_A$:
		\begin{equation}
			\ket{\psi^{\overline{\rho}}}_{R'RA}\coloneqq\frac{1}{\sqrt{|G|}}\sum_{g\in G}\ket{g}_{R'}\otimes (\mathbbm{1}_{R}\otimes U_A^g)\ket{\psi}_{RA},
		\end{equation}
		where $\{\ket{g}_{R'}\}_{g\in G}$ is an orthonormal basis for $\mathcal{H}_{R'}$ indexed by the elements of $G$. Since all purifications of a state can be mapped to each other by isometries on the purifying systems, there exists an isometry $W_{R\to R'R}$ such that $\ket{\psi^{\overline{\rho}}}_{R'RA}=W_{R\to R'R}\ket{\phi^{\overline{\rho}}}_{RA}$. Then, because the set $\SEP$ of separable states is invariant under local isometries, for every state $\sigma_{RB}\in\SEP(R\!:\!B)$ we have that $\tau_{R'RB}\coloneqq\mathcal{W}_{R\to R'R}(\sigma_{RB})\in\SEP(R'R\!:\!B)$. Therefore,
		\begin{align}
			&\boldsymbol{D}(\mathcal{N}_{A\to B}(\psi_{R'RA}^{\overline{\rho}})\Vert\tau_{R'RB})\\
			&\quad=\boldsymbol{D}(\mathcal{N}_{A\to B}(\mathcal{W}_{R\to R'R}(\phi_{RA}^{\overline{\rho}}))\Vert\mathcal{W}_{R\to R'A'}(\sigma_{RB}))\\
			&\quad=\boldsymbol{D}(\mathcal{W}_{R\to R'R}(\mathcal{N}_{A\to B}(\phi_{RA}^{\overline{\rho}}))\Vert\mathcal{W}_{R\to R'R}(\sigma_{RB}))\\
			&\quad=\boldsymbol{D}(\mathcal{N}_{A\to B}(\phi_{RA}^{\overline{\rho}})\Vert\sigma_{RB}),
		\end{align}
		where, to obtain the last equality, we used the fact that any generalized divergence is isometrically invariant (recall Proposition~\ref{prop-gen_div_properties}). Now, if we apply the dephasing channel $X\mapsto\sum_{g\in G}\ket{g}\!\bra{g}X\ket{g}\!\bra{g}$ to the $R'$ system, then by the data-processing inequality for the generalized divergence $\boldsymbol{D}$, we obtain
		\begin{align}
			&\boldsymbol{D}(\mathcal{N}_{A\to B}(\psi_{R'RA}^{\overline{\rho}})\Vert\tau_{R'RB})\nonumber\\
			&\quad \geq \boldsymbol{D}\!\left(\frac{1}{|G|}\sum_{g\in G}\ket{g}\!\bra{g}_{R'}\otimes(\mathcal{N}_{A\to B}\circ\mathcal{U}_A^g)(\psi_{RA})\Bigg\Vert\sum_{g\in G}p(g)\ket{g}\!\bra{g}_{R'}\otimes \tau_{RB}^g\right)\\
			&\quad =\boldsymbol{D}\!\left(\frac{1}{|G|}\sum_{g\in G}\ket{g}\!\bra{g}_{R'}\otimes ((\mathcal{V}_B^g)^\dagger\circ\mathcal{N}_{A\to B}\circ\mathcal{U}_A^g)(\psi_{RA})\Bigg\Vert\right.\nonumber\\
			&\qquad\qquad\qquad\qquad\qquad\qquad\qquad\left. \sum_{g\in G}p(g)\ket{g}\!\bra{g}_{R'}\otimes V_B^{g\dagger}\tau_{RB}^gV_B^g\right),
		\end{align}
		where to obtain the last line we applied the unitary channel given by the unitary $\sum_{g\in G}\ket{g}\!\bra{g}_{R'}\otimes V_B^{g\dagger}$ and we used the fact that generalized divergences are invariant under unitaries. Furthermore, we wrote the action of the dephasing channel on $\tau_{R'RB}$ as $\sum_{g\in G}p(g)\ket{g}\!\bra{g}_{R'}\otimes \tau_{RB}^g$, where $p:G\to[0,1]$ is a probability distribution and $\{\tau_{RB}^g\}_{g\in G}$ is a set of states. This operator is in the set $\SEP(R'R:B)$ because the set $\SEP$ is closed under local channels. Next, due to the covariance of $\mathcal{N}$, we have that $(\mathcal{V}_B^g)^\dagger\circ\mathcal{N}\circ\mathcal{U}_A^g=\mathcal{N}$, so that
		\begin{align}
			&\boldsymbol{D}(\mathcal{N}_{A\to B}(\psi_{R'RA}^{\overline{\rho}})\Vert\tau_{R'RB})\\
			&\quad \geq \boldsymbol{D}\!\left(\frac{1}{|G|}\sum_{g\in G}\ket{g}\!\bra{g}_{R'}\otimes\mathcal{N}_{A\to B}(\psi_{RA})\Bigg\Vert \sum_{g\in G}p(g)\ket{g}\!\bra{g}_{R'}\otimes V_B^{g\dagger}\tau_{RB}^g V_B^g \right)\\
			&\geq \boldsymbol{D}\!\left(\mathcal{N}_{A\to B}(\psi_{RA})\Bigg\Vert\sum_{g\in G}p(g)V_B^{g\dagger}\tau_{RB}^g V_B^g\right),\label{eq-gen_div_ent_inf_chan_cov_pf}
		\end{align}
		where to obtain the last inequality we used the data-processing inequality for $\boldsymbol{D}$ under the channel $\Tr_{R'}$. Now, observe that the state $\sum_{g\in G}p(g)\ket{g}\!\bra{g}_{R'}\otimes V_B^{g\dagger}\tau_{RB}^gV_B^g$ is in the set $\SEP(R'R:B)$. This is due to the fact that $\sum_{g\in G}\ket{g}\!\bra{g}_{R'}\allowbreak\otimes V_B^{g\dagger}$ is a controlled unitary, and since register $R'$ is classical, this controlled unitary can be implemented as an LOCC channel. Also, the set $\SEP$ is closed under LOCC channels. It follows then that  $\sum_{g\in G}p(g) V_B^{g\dagger}\tau_{RB}V_B^g\in\SEP(R\!:\!B)$ because we obtain it from the previous separable state by applying a local partial trace over $R'$. By taking the infimum over every state $\tau_{RB}\in\SEP(R\!:\!B)$ in \eqref{eq-gen_div_ent_inf_chan_cov_pf}, we have that
		\begin{align}
			\boldsymbol{D}(\mathcal{N}_{A\to B}(\phi_{RA}^{\overline{\rho}})\Vert\sigma_{RB})&=\boldsymbol{D}(\mathcal{N}_{A\to B}(\psi_{R'RA}^{\overline{\rho}})\Vert\tau_{R'RB})\\
			&\geq \inf_{\tau_{RB}\in\SEP(R:B)}\boldsymbol{D}(\mathcal{N}_{A\to B}(\psi_{RA})\Vert\tau_{RB})\\
			&=\boldsymbol{E}(R;B)_{\omega},
		\end{align}
		where $\omega_{RB}=\mathcal{N}_{A\to B}(\psi_{RA})$. This inequality holds for every state $\psi_{RA}$ and every state $\sigma_{RB}\in\SEP(R\!:\!B)$. Therefore, optimizing over all $\sigma_{RB}\in\SEP(R\!:\!B)$ leads to
		\begin{equation}
			\inf_{\sigma_{RB}\in\SEP(R:B)}\boldsymbol{D}(\mathcal{N}_{A\to B}(\phi_{RA}^{\overline{\rho}})\Vert\sigma_{RB})=\boldsymbol{E}(R;B)_{\overline{\omega}}\geq \boldsymbol{E}(R;B)_{\omega},
		\end{equation}
		where $\overline{\omega}_{RB}=\mathcal{N}_{A\to B}(\phi_{RA}^{\overline{\rho}})$. This is precisely the inequality in \eqref{eq-gen_div_ent_chan_cov_state_symm}.
		
		Next, by construction, the state $\phi_{RA}^{\overline{\rho}}$ is such that its reduced state on $A$ is invariant under the channel $\mathcal{T}_G$. Optimizing over all such states leads to
		\begin{equation}
			\sup_{\phi_{RA}}\{\boldsymbol{E}(R;B)_{\overline{\omega}}:\phi_A=\mathcal{T}_G(\phi_A),\overline{\omega}_{RB}=\mathcal{N}_{A\to B}(\phi_{RA})\}\geq \boldsymbol{E}(R;B)_{\omega}.
		\end{equation}
		Since this inequality holds for every pure state $\psi_{RA}$, we finally obtain
		\begin{equation}
			\sup_{\phi_{RA}}\{\boldsymbol{E}(R;B)_{\omega}:\phi_A=\mathcal{T}_G(\phi_A),\omega_{RB}=\mathcal{N}_{A\to B}(\phi_{RA})\}\geq \boldsymbol{E}(\mathcal{N}).
		\end{equation}
		Since the reverse inequality trivially holds, we obtain \eqref{eq-gen_div_ent_chan_cov}.
	\end{Proof}

	
	We saw in Section~\ref{subsec-gen_div_ent_cone_prog} that both the max-relative entropy of entanglement and the hypothesis testing relative entropy of entanglement can be formulated as cone programs. We now show that the max-relative entropy of entanglement for channels can also be formulated as a cone program.
	
	\begin{proposition*}{Cone Program for the Max-Relative Entropy of Entanglement of a Quantum Channel}{lem:SKA-alt-e-max-channel}
		Let $\mathcal{N}_{A\rightarrow B}$ be a quantum channel. Then%
		\begin{equation}
			E_{\max}(\mathcal{N})=\log_{2}\Sigma_{\max}(\mathcal{N}),
		\end{equation}
		where%
		\begin{equation}
			\Sigma_{\max}(\mathcal{N})\coloneqq \inf_{Y_{SB}\in\widehat{\SEP}}\left\{  \left\Vert\Tr_{B}[Y_{SB}]\right\Vert _{\infty}:\Gamma_{SB}^{\mathcal{N}}\leq Y_{SB}\right\},
		\end{equation}
		and $\Gamma_{SB}^{\mathcal{N}}$ is the Choi operator of the channel $\mathcal{N}_{A\rightarrow B}$.
	\end{proposition*}

	\begin{Proof}
		Using the definition of a channel's max-relative entropy of entanglement, and the cone program formulation of the max-relative entropy of entanglement for states from Proposition~\ref{lem:SKA-alt-emax}, we have
		\begin{align}
			E_{\max}(\mathcal{N})  &  =\sup_{\psi_{SA}}E_{\max}(S;B)_{\omega}\\&  =\sup_{\psi_{SA}}\inf_{X_{SB}\in\widehat{\SEP}}\log_{2}\!\left\{\Tr[X_{SB}]:\mathcal{N}_{A\rightarrow B}(\psi_{SA})\leq X_{SB}\right\},
		\end{align}
		where $\omega_{SB}=\mathcal{N}_{A\rightarrow B}(\psi_{SA})$. Now, recall from \eqref{eq-pure_state_vec} that an arbitrary pure bipartite state $\psi_{SA}$ can be written as $Z_{S}\Gamma_{SA}Z_{S}^{\dag}$, where $\Gamma_{SA}=\ket{\Gamma}\!\bra{\Gamma}$, $\ket{\Gamma}_{SA}=\sum_{i=0}^{d_A-1}\ket{i,i}_{SA}$, and $Z_{S}$ is an operator satisfying $\Tr[Z_{S}^{\dagger}Z_{S}]=1$. Then%
		\begin{align}
			\mathcal{N}_{A\rightarrow B}(\psi_{SA})  &  =\mathcal{N}_{A\rightarrow B}(Z_{S}\Gamma_{SA}Z_{S}^{\dag})\\
			&  =Z_{S}\mathcal{N}_{A\rightarrow B}(\Gamma_{SA})Z_{S}^{\dag}\\
			&  =Z_{S}\Gamma_{SB}^{\mathcal{N}}Z_{S}^{\dag},
		\end{align}
		where $\Gamma_{SB}^{\mathcal{N}}=\mathcal{N}_{A\rightarrow B}(\Gamma_{SA})$ is the Choi operator of $\mathcal{N}_{A\rightarrow B}$. Since the set of operators $Z_{S}$ satisfying $Z_{S}^{\dag}Z_{S}>0$ and $\Tr[Z_{S}^{\dag}Z_{S}]=1$ is dense in the set of all operators satisfying $\Tr[Z_{S}^{\dag}Z_{S}]=1$, we find that%
		\begin{multline}
			E_{\max}(\mathcal{N})\\
				=\log_2\sup_{Z_{S}}\inf_{X_{SB}\in\widehat{\SEP}}\left\{\Tr[X_{SB}]:Z_{S}\Gamma_{SB}^{\mathcal{N}}Z_{S}^{\dag}\leq X_{SB},\ Z_{S}^{\dag}Z_{S}>0,\ \Tr[Z_{S}^{\dag}Z_{S}]=1\right\}.
				\label{eq:E-meas:Emax-ch-proof-cone-1}
		\end{multline}
		Let us now make a change of variable, defining the variable $Y_{SB}$ according to the relation $X_{SB}=Z_{S}Y_{SB}Z_{S}^{\dagger}$. Then, since
		\begin{align}
			Z_{S}\Gamma_{SB}^{\mathcal{N}}Z_{S}^{\dagger}\leq X_{SB}=Z_SY_{SB}Z_S^\dagger\qquad & \Longleftrightarrow\qquad   \Gamma_{SB}^{\mathcal{N}}\leq Y_{SB},\\
X_{SB}    \in\widehat{\SEP} \qquad &\Longleftrightarrow \qquad Y_{SB}\in\widehat{\SEP},
		\end{align}
		we find that%
		\begin{align}
			&  \text{Eq. }\eqref{eq:E-meas:Emax-ch-proof-cone-1} \nonumber\\
			&  =\sup_{Z_{S}}\inf_{Y_{SB}\in\widehat{\SEP}}\left\{\Tr[Z_{S}Y_{SB}Z_{S}^{\dag}]:\Gamma_{SB}^{\mathcal{N}}\leq Y_{SB},\ Z_{S}^{\dagger}Z_{S}>0,\ \Tr[Z_{S}^{\dag}Z_{S}]=1\right\}\nonumber\\
			&  =\sup_{Z_{S}}\inf_{Y_{SB}\in\widehat{\SEP}}\left\{\Tr[Z_{S}^{\dagger}Z_{S}Y_{SB}]:\Gamma_{SB}^{\mathcal{N}}\leq Y_{SB},\ Z_{S}^{\dagger}Z_{S}>0,\ \Tr[Z_{S}^{\dagger}Z_{S}]=1\right\}\nonumber\\
			&  =\sup_{\rho_{S}}\inf_{Y_{SB}\in\widehat{\SEP}}\left\{\Tr[\rho_{S}Y_{SB}]:\Gamma_{SB}^{\mathcal{N}}\leq Y_{SB}\right\}  ,
		\end{align}
		where in the last line we made the substitution $\rho_{S}=Z_{S}^{\dag}Z_{S}$, so that the optimization is with respect to density operators. Furthermore, we have employed the fact that the set of density operators satisfying $\rho_{S}>0$ is dense in the set of all density operators. Now observing that the objective function is linear in $\rho_{S}$ and $Y_{SB}$, the set of density operators is compact and convex, and the set of separable operators is convex, the Sion minimax theorem (Theorem~\ref{thm-Sion_minimax}) applies, such that we can exchange the optimizations to find that%
		\begin{align}
			&  \sup_{\rho_{S}}\inf_{Y_{SB}\in\widehat{\SEP}}\left\{\Tr[\rho_{S}Y_{SB}]:\Gamma_{SB}^{\mathcal{N}}\leq Y_{SB}\right\}
\nonumber\\
			&  =\inf_{Y_{SB}\in\widehat{\SEP}}\sup_{\rho_{S}}\left\{\Tr[\rho_{S}Y_{SB}]:\Gamma_{SB}^{\mathcal{N}}\leq Y_{SB}\right\} \\
			&  =\inf_{Y_{SB}\in\widehat{\SEP}}\sup_{\rho_{S}}\left\{\Tr[\rho_{S}\Tr_{B}[Y_{SB}]]:\Gamma_{SB}^{\mathcal{N}%
}\leq Y_{SB}\right\} \\
			&  =\inf_{Y_{SB}\in\widehat{\SEP}}\left\{  \left\Vert \Tr_{B}[Y_{SB}]\right\Vert_{\infty}:\Gamma_{SB}^{\mathcal{N}}\leq Y_{SB}\right\} \\
&  =\Sigma_{\max}(\mathcal{N}).
		\end{align}
		The second equality follows from partial trace, and the third follows because $\norm{X}_{\infty}=\sup_{\rho}\Tr[X\rho]$ for positive semi-definite operators $X$, where the optimization is with respect to density operators (see \eqref{eq-X_PSD_largest_eig}).
	\end{Proof}

	

\section{Generalized Rains Divergence}\label{subsec-Rains_gen_div_ent_channel}

	We now examine the generalized Rains divergence of quantum channels, which is a channel entanglement measure that arises from the generalized Rains divergence for bipartite quantum states that we considered in Section~\ref{sec-ent_measures_Rains_dist}.
	
	\begin{definition}{Generalized Rains Information of a Quantum Channel}{def-gen_Rains_inf_chan}
		Let $\boldsymbol{D}$ be a generalized divergence (see Definition~\ref{def-gen_div}. For every quantum channel $\mathcal{N}_{A\to B}$, we define the \textit{generalized Rains information of $\mathcal{N}$} as
		\begin{align}
			\boldsymbol{R}(\mathcal{N})&\coloneqq\sup_{\psi_{SA}}\boldsymbol{R}(S;B)_{\omega}\\
			&=\sup_{\psi_{SA}}\inf_{\sigma_{SB}\in\operatorname{PPT}'(S:B)}\boldsymbol{D}(\mathcal{N}_{A\to B}(\psi_{SA})\Vert\sigma_{SB}),
		\end{align}
		where $\omega_{SB}\coloneqq\mathcal{N}_{A\to B}(\psi_{SA})$. The supremum is with respect to every pure state $\psi_{SA}$, with the dimension of $S$ the same as the dimension of $A$.
	\end{definition}
	
	In the remark immediately after Definition~\ref{def:LAQC-ent-channel} we show that it suffices to optimize with respect to pure bipartite states (with equal dimension for each subsystem) when calculating the generalized Rains divergence of a quantum channel. 
	
	We can write the generalized Rains divergence of $\mathcal{N}_{A\to B}$ in the following alternate form:
	\begin{align}
		\boldsymbol{R}(\mathcal{N})&=\sup_{\rho_A}\boldsymbol{R}(\mathcal{N}_{A\to B},\rho_A),\label{eq-Rains_gen_div_chan_alt}\\
		\boldsymbol{R}(\mathcal{N}_{A\to B},\rho_A)&\coloneqq \inf_{\sigma_{AB}\in\PPT'(A:B)}\boldsymbol{D}(\sqrt{\rho_A}\Gamma_{AB}^{\mathcal{N}}\sqrt{\rho_A}\Vert\sigma_{AB}).
	\end{align}
	This is indeed true because we can write every purification of $\rho_A$ as $(V_{S}\sqrt{\rho_{S}}\otimes\mathbbm{1}_A)\ket{\Gamma}_{SA}$ for some isometry $V_{S}$ (see \eqref{eq-pure_state_vec} and Theorem~\ref{thm-polar_decomposition}),  the set of $\PPT'$ operators is invariant under local isometries, and the generalized divergences are invariant under local isometries (see Proposition~\ref{prop-gen_div_properties}).

	
	As with the state quantities, we are interested in the following generalized Rains information quantities for every quantum channel $\mathcal{N}_{A\to B}$. For every case below, we define $\omega_{SB}=\mathcal{N}_{A\to B}(\psi_{SA})$.
	\begin{enumerate}
		\item The \textit{Rains information of $\mathcal{N}$},
			\begin{align}
				R(\mathcal{N})&\coloneqq \sup_{\psi_{SA}}R(S;B)_{\omega}\\
				&=\sup_{\psi_{SA}}\inf_{\sigma_{SB}\in\operatorname{PPT}'(S:B)}D(\mathcal{N}_{A\to B}(\psi_{SA})\Vert\sigma_{SB})\label{eq-Rains_inf_chan}.
			\end{align}

		\item The \textit{$\varepsilon$-hypothesis testing Rains information of $\mathcal{N}$},
			\begin{align}
				R_H^{\varepsilon}(\mathcal{N})&\coloneqq\sup_{\psi_{SA}}R_H^{\varepsilon}(S;B)_{\omega}\\
				&=\sup_{\psi_{SA}}\inf_{\sigma_{SB}\in\operatorname{PPT}'(S:B)}D_H^{\varepsilon}(\mathcal{N}_{A\to B}(\psi_{SA})\Vert\sigma_{SB}) \label{eq-hypo_test_rains_inf_chan},
			\end{align}

		\item The \textit{sandwiched R\'{e}nyi Rains information of $\mathcal{N}$},
			\begin{align}
				\widetilde{R}_{\alpha}(\mathcal{N})&\coloneqq\sup_{\psi_{SA}}\widetilde{R}_{\alpha}(S;B)_{\omega}\\
				&=\sup_{\psi_{SA}}\inf_{\sigma_{SB}\in\operatorname{PPT}'(S:B)}\widetilde{D}_{\alpha}(\mathcal{N}_{A'\to B}(\psi_{SA})\Vert\sigma_{SB})\label{eq-sand_Renyi_Rains_inf_chan},
			\end{align}
			where $\alpha\in[\sfrac{1}{2},1)\cup(1,\infty)$. It follows from Proposition~\ref{prop-sand_rel_ent_properties} that $\widetilde{R}_{\alpha}(\mathcal{N})$ is monotonically increasing in $\alpha$. Also, in Appendix~\ref{app-sand_ren_inf_limit}, we prove that
			\begin{equation}
				R(\mathcal{N})=\lim_{\alpha\to 1}\widetilde{R}_{\alpha}(\mathcal{N})
			\end{equation}
			for every quantum channel $\mathcal{N}$.
			
		\item The \textit{max-Rains information of $\mathcal{N}$},
			\begin{align}
				R_{\max}(\mathcal{N})&\coloneqq\sup_{\psi_{SA}}R_{\max}(S;B)_{\omega}\\
				&=\sup_{\psi_{SA}}\inf_{\sigma_{SB}\in\PPT'(S:B)}D_{\max}(\mathcal{N}_{A\to B}(\psi_{SA})\Vert\sigma_{SB}).
			\end{align}
			 In Appendix~\ref{app-sand_ren_inf_limit}, we prove that
			\begin{equation}
				R_{\max}(\mathcal{N})=\lim_{\alpha\to\infty}\widetilde{R}_{\alpha}(\mathcal{N})
			\end{equation}
			for every quantum channel $\mathcal{N}$.
	\end{enumerate}
	
	The generalized Rains divergence of a quantum channel satisfies all of the properties of a channel entanglement measure laid out in Proposition~\ref{prop-chan_ent_measure_properties}, except for faithfulness and superadditivity. Faithfulness generally does not hold because the generalized Rains divergence of a bipartite quantum state is not faithful. Superadditivity does not hold in general because the Rains divergence of a bipartite quantum state is generally only subadditive. The max-Rains information of a quantum channel, however, is additive, meaning that
	\begin{equation}\label{eq-max_Rains_inf_additive_0}
		R_{\max}(\mathcal{N}\otimes\mathcal{M})=R_{\max}(\mathcal{N})+R_{\max}(\mathcal{M})
	\end{equation}
	for all quantum channels $\mathcal{N}$ and $\mathcal{M}$. We defer a proof of this to Section~\ref{sec-amort_collapses} below.
	
	The amortized generalized Rains divergence $\boldsymbol{R}^{\mathcal{A}}(\mathcal{N})$, defined according to Definition~\ref{def:LAQC-amortized-ent} as
	\begin{equation}
		\boldsymbol{R}^{\mathcal{A}}(\mathcal{N})\coloneqq\sup_{\rho_{A'AB'}}\{\boldsymbol{R}(A';BB')_{\omega}-\boldsymbol{R}(A'A;B)_{\rho}\},
	\end{equation}
	where $\omega_{A'BB'}=\mathcal{N}_{A\to B}(\rho_{A'AB'})$, satisfies all of the properties stated in Proposition~\ref{prop-chan_amort_ent_properties} except for faithfulness, because the generalized Rains divergence of a bipartite quantum state is not faithful. In particular, due to additivity of the max-Rains relative entropy (Proposition~\ref{prop:LAQC-add-max-Rains-rel-ent}), we immediately obtain additivity of the amortized max-Rains information of a quantum channel, i.e.,
	\begin{equation}
		R_{\max}^{\mathcal{A}}(\mathcal{N}\otimes\mathcal{M})=R_{\max}^{\mathcal{A}}(\mathcal{N})+R_{\max}^{\mathcal{A}}(\mathcal{M})
	\end{equation}
	for all quantum channels $\mathcal{N}$ and $\mathcal{M}$. We show in Section~\ref{sec-amort_collapses} below that
	\begin{equation}
	R_{\max}(\mathcal{N})=R_{\max}^{\mathcal{A}}(\mathcal{N})
	\end{equation}
	 for every quantum channel $\mathcal{N}$, and it is this fact that leads to the additivity statement in \eqref{eq-max_Rains_inf_additive_0}.
	
	For covariant channels, the optimization over pure input states in the generalized Rains divergence simplifies in the same way as it does for the generalized divergence of entanglement.
	
	\begin{proposition*}{Generalized Rains Information for Covariant Channels}{prop-gen_Rains_inf_chan_cov}
		Let $\mathcal{N}_{A\to B}$ be a $G$-covariant quantum channel for a finite group $G$ (recall Definition~\ref{def-group_cov_chan}). Then, for every pure state $\psi_{RA}$, with the dimension of $R$ equal to the dimension of $A$, we have that
		\begin{equation}\label{eq-Rains_inf_chan_cov_state_symm}
			\boldsymbol{R}(S;B)_{\omega}\leq \boldsymbol{R}(S;B)_{\overline{\omega}},
		\end{equation}
		where $\omega_{RB}\coloneqq\mathcal{N}_{A\to B}(\psi_{RA})$, $\overline{\omega}_{RB}\coloneqq\mathcal{N}_{A\to B}(\phi^{\overline{\rho}}_{RA})$,
		\begin{equation}\label{eq-psi_RA_group_symmetrize}
			\overline{\rho}_{A}=\frac{1}{|G|}\sum_{g\in G}U_A^g\rho_{A}U_A^{g\dagger}\eqqcolon\mathcal{T}_G(\rho_A),
		\end{equation}
		$\rho_A=\psi_A=\Tr_{S}[\psi_{SA}]$, and $\phi_{SA}^{\overline{\rho}}$ a purification of $\overline{\rho}_A$. Consequently,
		\begin{equation}\label{eq-Rains_inf_chan_cov}
			\boldsymbol{R}(\mathcal{N})=\sup_{\phi_{SA}}\{\boldsymbol{R}(S;B)_{\omega}:\phi_A=\mathcal{T}_G(\phi_A),\omega_{SB}=\mathcal{N}_{A\to B}(\phi_{SA})\}.
		\end{equation}
		In other words, in order to calculate $\boldsymbol{R}(\mathcal{N})$, it suffices to optimize with respect to pure states $\phi_{SA}$ such that the reduced state $\phi_A$ is invariant under the channel $\mathcal{T}_G$ defined in \eqref{eq-psi_RA_group_symmetrize}.
	\end{proposition*}
	
	\begin{remark}
		Using \eqref{eq-Rains_gen_div_chan_alt}, we can write \eqref{eq-Rains_inf_chan_cov_state_symm} as
		\begin{equation}
			\boldsymbol{R}(\mathcal{N},\rho)\leq\boldsymbol{R}(\mathcal{N},\mathcal{T}_G(\rho)),
		\end{equation}
		which holds for every state $\rho$ acting on the input space of the channel $\mathcal{N}$. We can write \eqref{eq-Rains_inf_chan_cov} as
		\begin{equation}
			\boldsymbol{R}(\mathcal{N})=\sup_R\{\boldsymbol{R}(\mathcal{N},\rho):\rho=\mathcal{T}_G(\rho)\}.
		\end{equation}
	\end{remark}
	
	\begin{Proof}
		The proof is identical to the proof of Proposition~\ref{prop-gen_div_ent_chan_cov}, with the exception that the set PPT' is involved rather than the set SEP. The LOCC channels discussed there preserve the set PPT', and this is the main reason why the same proof applies.
	\end{Proof}
	
	
	Using the SDP formulation of max-Rains relative entropy in Proposition~\ref{prop-max_Rains_rel_ent_SDP}, we arrive at an SDP formulation for the max-Rains information of a quantum channel.

	\begin{proposition*}{SDPs for Max-Rains Information of a Quantum Channel}{prop-SDP_max_rains}
		Let $\mathcal{N}_{A\rightarrow B}$ be a quantum channel. Then%
		\begin{equation}\label{eq:LAQC-max-Rains-Gamma}
			R_{\max}(\mathcal{N})=\log_{2}\Gamma_{\max}(\mathcal{N}),
		\end{equation}
		where%
		\begin{align}
			& \Gamma_{\max}(\mathcal{N}) \notag \\ 
			& =\inf_{Y_{SB},V_{SB}\geq0}\{\left\Vert \Tr_{B}[V_{SB}+Y_{SB}]\right\Vert _{\infty} :  \T_{B}(V_{SB}-Y_{SB})\geq \Gamma_{SB}^{\mathcal{N}}\}\label{eq:LAQC-max-Rains-Gamma-primal}\\
			&  =\sup_{\rho_{S}\geq0}\{\Tr[\Gamma_{SB}^{\mathcal{N}}X_{SB}]: \Tr[\rho_{S}]\leq 1,\, -\rho_{S}\otimes \mathbbm{1}_{B}\leq \T_{B}(X_{SB})\leq\rho_{S}\otimes \mathbbm{1}_{B}\}. \label{eq:LAQC-max-Rains-Gamma-dual}%
		\end{align}
	\end{proposition*}

	\begin{Proof}
		To arrive at \eqref{eq:LAQC-max-Rains-Gamma-dual}, consider that, with $\omega_{SB}=\mathcal{N}_{A\rightarrow B}(\psi_{SA})$,%
		\begin{align}
			&  2^{R_{\max}(\mathcal{N})}\nonumber\\
			&  =\sup_{\psi_{SA}}2^{R_{\max}(S;B)_{\omega}}\\
			&  =\sup\left\{\Tr[\mathcal{N}_{A\rightarrow B}(\psi_{SA})X_{SB}]:\norm{\T_{B}(X_{SB})}_{\infty}\leq 1, X_{SB}\geq0\right\}  ,
		\end{align}
		where the last equality follows from \eqref{eq:LAQC-Rains-to-W}--\eqref{eq:LAQC-dual-form-W-Rains}. Recall from \eqref{eq-pure_state_vec} that an arbitrary pure bipartite state $\psi_{SA}$ can be written as $Z_{S}\Gamma_{SA}Z_{S}^{\dagger}$, where $\Gamma_{SA}=|\Gamma\rangle\!\langle\Gamma|_{SA}$, $\ket{\Gamma}_{SA}=\sum_{i=0}^{d_A=1}\ket{i,i}_{SA}$, and $Z_{S}$ is an operator satisfying $\Tr[Z_{S}^{\dag}Z_{S}]=1$. Then%
		\begin{align}
			\Tr[\mathcal{N}_{A\rightarrow B}(\psi_{SA})X_{SB}] & =\Tr[\mathcal{N}_{A\rightarrow B}(Z_{S}\Gamma_{SA}Z_{S}^{\dag})X_{SB}]\\
			&  =\Tr[Z_{S}\mathcal{N}_{A\rightarrow B}(\Gamma_{SA})Z_{S}^{\dag}X_{SB}]\\
			&  =\Tr[\Gamma_{SB}^{\mathcal{N}}Z_{S}^{\dag}X_{SB}Z_{S}],
		\end{align}
		where $\Gamma_{SB}^{\mathcal{N}}=\mathcal{N}_{A\rightarrow B}(\Gamma_{SA})$ denotes the Choi operator of the channel $\mathcal{N}_{A\rightarrow B}$. Since the set of operators $Z_{S}$ satisfying $Z_{S}^{\dag}Z_{S}>0$ and $\Tr[Z_{S}^{\dag}Z_{S}]=1$ is dense in the set of all operators satisfying $\Tr[Z_{S}^{\dag}Z_{S}]=1$, we find that%
		\begin{multline}
			2^{R_{\max}(\mathcal{N})}=\sup\{\Tr[\Gamma_{SB}^{\mathcal{N}} Z_{S}^{\dag}X_{SB}Z_{S}]:\left\Vert \T_{B}(X_{SB})\right\Vert _{\infty}\leq1,\\
			X_{SB}\geq0,Z_{S}^{\dag}Z_{S}>0,\ \Tr[Z_{S}^{\dag}Z_{S}]=1\}.
		\end{multline}
		Consider that $X_{SB}\geq0\Leftrightarrow Z_{S}^{\dag}X_{SB}Z_{S}\geq0$ and
		\begin{align}
			&  \left\Vert \T_{B}(X_{SB})\right\Vert _{\infty}\leq1\nonumber\\
			  \Longleftrightarrow\quad & -\mathbbm{1}_{SB}\leq \T_{B}(X_{SB})\leq \mathbbm{1}_{SB}\\
	\Longleftrightarrow\quad & -Z_{S}^{\dag}Z_{S}\otimes \mathbbm{1}_{B}\leq Z_{S}^{\dag}\T_{B}(X_{SB})Z_{S}\leq Z_{S}^{\dag}Z_{S}\otimes \mathbbm{1}_{B}\\
			\Longleftrightarrow\quad &  -Z_{S}^{\dag}Z_{S}\otimes \mathbbm{1}_{B}\leq \T_{B}(Z_{S}^{\dag}X_{SB}Z_{S})\leq Z_{S}^{\dag}Z_{S}\otimes \mathbbm{1}_{B}.
		\end{align}
		We now set $X_{SB}'\coloneqq Z_{S}^{\dag}X_{SB}Z_{S}$ and $\rho_{S}=Z_{S}^{\dag}Z_{S}>0$ and rewrite as follows:
		\begin{multline}
			2^{R_{\max}(\mathcal{N})}=\sup\{\Tr[\Gamma_{SB}^{\mathcal{N}}X_{SB}']:-\rho_{S}\otimes \mathbbm{1}_{B}\leq \T_{B}(X_{SB}')\leq\rho
_{S}\otimes \mathbbm{1}_{B},\\
			X_{SB}'\geq0,\ \rho_{S}>0,\Tr[\rho_{S}]=1\},
		\end{multline}
		which is the equality in \eqref{eq:LAQC-max-Rains-Gamma} and \eqref{eq:LAQC-max-Rains-Gamma-dual}, after observing that the set $\{\rho_{S}:\rho_{S}>0,\Tr[\rho_{S}]=1\}$ is dense in the set $\{\rho_{S}:\rho_{S}\geq0,\Tr[\rho_{S}]=1\}$.

		To arrive at the equality in \eqref{eq:LAQC-max-Rains-Gamma-primal}, we employ the dual formulation of the max-Rains relative entropy in \eqref{eq:LAQC-dual-form-W-Rains}. Consider that%
		\begin{align}
			& 2^{R_{\max}(\mathcal{N})}  \notag \\
			&  =\sup_{\psi_{SA}}2^{R_{\max}(S;B)_{\omega}}\\
			&  =\sup_{\psi_{SA}}\inf_{K_{SB},L_{SB}\geq 0}\{\Tr[K_{SB}%
+L_{SB}]: \T_{B}\!\left(  K_{SB}-L_{SB}\right)  \geq\mathcal{N}_{A\rightarrow B}(\psi_{SA})\}.
		\end{align}
		Making the same observations as we did previously, we have that $\mathcal{N}_{A\rightarrow B}(\psi_{SA})=Z_{S}\Gamma_{SB}^{\mathcal{N}}Z_{S}^{\dag}$, as well as%
		\begin{equation}
			\T_{B}\!\left(  K_{SB}-L_{SB}\right)  \geq\mathcal{N}_{A\rightarrow B}(\psi _{SA})\quad\Longleftrightarrow\quad \T_{B}\!\left(  K_{SB}'-L_{SB}'\right)  \geq \Gamma_{SB}^{\mathcal{N}},
		\end{equation}
		where $K_{SB}'$ and $L_{SB}'$ are such that $K_{SB}=Z_{S}K_{SB}'Z_{S}^{\dag}$ and $L_{SB}=Z_{S}L_{SB}'Z_{S}^{\dag}$, respectively. Then $K_{SB},L_{SB}\geq0\Longleftrightarrow K_{SB}',L_{SB}'\geq0$, and we find that%
		\begin{multline}
			2^{R_{\max}(\mathcal{N})}=\sup_{Z_{S}}\inf_{K_{SB}',L_{SB}'\geq 0}\{\Tr[Z_{S}K_{SB}'Z_{S}^{\dag}+Z_{S}L_{SB}'Z_{S}^{\dag}]:
			\T_{B}\!\left(  K_{SB}'-L_{SB}'\right)  \geq \Gamma_{SB}^{\mathcal{N}},\\
			Z_{S}^{\dag}Z_{S}>0,\ \Tr[Z_{S}^{\dag}Z_{S}]=1\}.
		\end{multline}
		Employing cyclicity of trace, setting $\rho_{S}=Z_{S}^{\dag}Z_{S}$, and exploiting the fact that the set $\{\rho_{S}:\rho_{S}>0,\Tr[\rho_{S}]=1\}$ is dense in the set $\{\rho_{S}:\rho_{S}\geq 0,\Tr[\rho_{S}]=1\}$, we find that%
		\begin{multline}
			2^{R_{\max}(\mathcal{N})}=\sup_{\rho_{S}}\inf_{K_{SB}',L_{SB}'}\{\Tr[\rho_{S}(K_{SB}'+L_{SB}')]:K_{SB}%
',L_{SB}'\geq0,\\
			\T_{B}\!\left(  K_{SB}'-L_{SB}'\right)  \geq \Gamma_{SB}^{\mathcal{N}},\ \rho_{S}\geq0,\ \Tr[\rho_{S}]=1\}.
		\end{multline}
		The function that we are optimizing is linear in $\rho_{S}$ and jointly convex in $K_{SB}'$ and $L_{SB}'$ (the set with respect to which the infimum is performed is also compact), so that the minimax theorem (Theorem~\ref{thm-Sion_minimax}) applies and we can exchange $\sup$ with $\inf$ to find that
		\begin{multline}
			2^{R_{\max}(\mathcal{N})}=\inf_{K_{SB}',L_{SB}'}\sup_{\rho_{S}}\{\Tr[\rho_{S}(K_{SB}'+L_{SB}')]:K_{SB}',L_{SB}'\geq0,\\
			\T_{B}\!\left(  K_{SB}'-L_{SB}'\right)  \geq \Gamma_{SB}^{\mathcal{N}},\ \rho_{S}\geq0,\ \Tr[\rho_{S}]=1\}.
		\end{multline}
		For fixed $K_{SB}'$ and $L_{SB}'$, consider that%
		\begin{align}
			&  \sup_{\rho_{S}}\{\Tr[\rho_{S}(K_{SB}'+L_{SB}')]:\rho_{S}\geq0,\ \Tr[\rho_{S}]=1\}\nonumber\\
			&  =\sup_{\rho_{S}}\{\Tr[\rho_{S}\Tr_{B}[K_{SB}'+L_{SB}']]:\rho_{S}\geq0,\ \Tr[\rho_{S}]=1\}\\
			&  =\left\Vert \Tr_{B}[K_{SB}'+L_{SB}']\right\Vert _{\infty},
		\end{align}
		where for the last line we used \eqref{eq-X_PSD_largest_eig}. Substituting back in, we find that%
		\begin{multline}
			2^{R_{\max}(\mathcal{N})}=\inf_{K_{SB}',L_{SB}'}\{\left\Vert \Tr_{B}[K_{SB}'+L_{SB}']\right\Vert _{\infty}:K_{SB}',L_{SB}'\geq0,\\
			\T_{B}\!\left(  K_{SB}'-L_{SB}'\right)  \geq \Gamma_{SB}^{\mathcal{N}}\},
		\end{multline}
		as claimed in \eqref{eq:LAQC-max-Rains-Gamma-dual}.
		
		According to Theorem~\ref{thm:math-tools:slater-cond}, strong duality holds by picking $V_{SB}$ and $Y_{SB}$ equal to the positive and negative parts of $\T_B(\Gamma_{SB}^{\mathcal{N}})$, respectively, which are feasible for \eqref{eq:LAQC-max-Rains-Gamma-dual}. Furthermore, we can pick $\rho_{S} = \mathbbm{1}_S/(2d_S)$ and $X_{SB} = \mathbbm{1}_{SB}/(3d_S)$, which are strictly feasible for \eqref{eq:LAQC-max-Rains-Gamma-primal}.
	\end{Proof}

\section{Squashed Entanglement}\label{subsec-sq_ent_channel}

	We now move on to the squashed entanglement of a quantum channel, which is a channel entanglement measure that arises from the squashed entanglement of a bipartite state, the latter  defined in Section~\ref{sec-LAQC:sq-ent-and-props}.
	
	\begin{definition}{Squashed Entanglement of a Quantum Channel}{def-sq_ent_channel}
		For every quantum channel $\mathcal{N}_{A\to B}$, we define the \textit{squashed entanglement of $\mathcal{N}$} as
		\begin{align}
			E_{\text{sq}}(\mathcal{N})&=\sup_{\psi_{RA}}E_{\text{sq}}(R;B)_{\omega}\\
			&=\frac{1}{2}\sup_{\psi_{RA}}\inf_{\tau_{RBE}}\{I(R;B|E)_{\tau}:\Tr_E[\tau_{RBE}]=\omega_{RB}\},
		\end{align}
		where $\omega_{RB}=\mathcal{N}_{A\to B}(\psi_{RA})$ and the quantum conditional mutual information $I(R;B|E)_{\tau}$ is defined as
		\begin{equation}
			I(R;B|E)_{\tau}=H(R|E)_{\tau}+H(B|E)_{\tau}-H(RB|E)_{\tau}.
		\end{equation}
	\end{definition}
	
	We can write the squashed entanglement of a quantum channel in the following alternate form:
	\begin{align}
		E_{\text{sq}}(\mathcal{N})&=\sup_{\rho_{A}}E_{\text{sq}}(\mathcal{N},\rho), \label{eq:LAQC-sq-ch-opt-with-no-purify}\\
		E_{\text{sq}}(\mathcal{N},\rho)&\coloneqq E_{\text{sq}}(R;B)_{\omega}. \label{eq:LAQC-sq-ch-def-with-no-purify}
	\end{align}
	where $\omega_{RB}=\mathcal{N}_{A\to B}(\psi_{RA}^{\rho})$, with $\psi_{RA}^{\rho}$ some purification of $\rho_A$. This is indeed true because  the squashed entanglement of a bipartite state is invariant under isometries and because all purifications of a given state are related to each other by local isometries acting on the purifying system.
	
	Let us now recall the following alternate expression for the squashed entanglement of a bipartite state from \eqref{eq:LAQC-sq-ch-rep-2}:
	\begin{multline}
		E_{\text{sq}}(A;B)_{\rho}=\frac{1}{2}\inf_{\mathcal{V}_{E'\rightarrow EF}}\left\{  H(B|E)_{\theta}+H(B|F)_{\theta}: \theta_{BEF}=\mathcal{V}_{E'\rightarrow EF}(\psi_{ABE'}^{\rho}), \right\},
	\end{multline}
	where $\psi_{ABE'}^{\rho}$ is some purification of $\rho_{AB}$. Now, given an input state $\rho_A$ of a quantum channel $\mathcal{N}_{A\to B}$, let $\phi_{RA}$ be a purification of $\rho_A$. Then, for the state $\omega_{RB}\coloneqq\mathcal{N}_{A\to B}(\phi_{RA})$, we can take a purification to be $\psi_{RBE'}=\mathcal{U}_{A\to BE'}^{\mathcal{N}}(\phi_{RA})$, where $\mathcal{U}^{\mathcal{N}}$ is an isometric channel that extends $\mathcal{N}$. Then, for every isometric channel $\mathcal{V}_{E'\to EF}$, we can define the state
	\begin{equation}
	\theta_{BEF}=\mathcal{V}_{E'\to EF}(\psi_{BE'})=(\mathcal{V}_{E'\to EF}\circ\mathcal{U}^{\mathcal{N}}_{A\to BE'})(\rho_A),
	\end{equation}
	where $\psi_{BE'}=\Tr_R[\psi_{RBE'}]$. We then have
	\begin{multline}
		E_{\text{sq}}(\mathcal{N},\rho_A)=E(R;B)_{\omega}
		=\\ \frac{1}{2}\inf_{\mathcal{V}_{E'\to EF}}\left\{H(B|E)_{\theta}+H(B|F)_{\theta}:\theta_{BEF}=(\mathcal{V}_{E'\to EF}\circ\mathcal{U}_{A\to BE'}^{\mathcal{N}})(\rho_A)\right\}\label{eq-sq_ent_state_alt_expr}
	\end{multline}
	for every state $\rho_A$. We thus obtain the following expression for the squashed entanglement of a channel:
	\begin{multline}\label{eq-sq_ent_chan_alt_expr}
		E_{\text{sq}}(\mathcal{N})= \\
		\frac{1}{2}\sup_{\rho_A}\inf_{\mathcal{V}_{E'\to EF}}\left\{H(B|E)_{\theta}+H(B|F)_{\theta}:\theta_{BEF}=(\mathcal{V}_{E'\to EF}\circ\mathcal{U}_{A\to BE'}^{\mathcal{N}})(\rho_A)\right\}.
	\end{multline}

	The squashed entanglement of a quantum channel, as well as its amortized version $E_{\text{sq}}^{\mathcal{A}}$ defined as
	\begin{equation}
		E_{\text{sq}}^{\mathcal{A}}(\mathcal{N})\coloneqq\sup_{\rho_{A'AB'}}\{E_{\text{sq}}(A';BB')_{\omega}-E_{\text{sq}}(A'A;B')_{\rho}\},
	\end{equation}
	with $\omega_{A'BB'}=\mathcal{N}_{A\to B}(\rho_{A'AB'})$, satisfy all of the properties stated in Proposition~\ref{prop-chan_ent_measure_properties} and Proposition~\ref{prop-chan_amort_ent_properties}, respectively. In particular, because the squashed entanglement for states is additive (see \eqref{eq:LAQC-sq-additive-states}), we immediately have that the amortized squashed entanglement of a channel is additive, i.e.,
	\begin{equation}
		E_{\text{sq}}^{\mathcal{A}}(\mathcal{N}\otimes\mathcal{M})=E_{\text{sq}}^{\mathcal{A}}(\mathcal{N})+E_{\text{sq}}^{\mathcal{A}}(\mathcal{M})
	\end{equation}
	for all quantum channels $\mathcal{N}$ and $\mathcal{M}$. In Section~\ref{sec-amort_collapses} below, we prove that $E_{\text{sq}}(\mathcal{N})=E_{\text{sq}}^{\mathcal{A}}(\mathcal{N})$ for every quantum channel $\mathcal{N}$, which then implies the additivity of the squashed entanglement of a channel, i.e.,
	\begin{equation}
		E_{\text{sq}}(\mathcal{N}\otimes\mathcal{M})=E_{\text{sq}}(\mathcal{N})+E_{\text{sq}}(\mathcal{M})
	\end{equation}
	for all channels $\mathcal{N}$ and $\mathcal{M}$.
	
	\begin{proposition}{lem:concavity-sq-ent-ch-input}
		Let $\mathcal{N}_{A\rightarrow B}$ be a quantum channel. The function $\rho\mapsto E_{\text{sq}}(\mathcal{N},\rho)$, where $E_{\text{sq}}(\mathcal{N},\rho)$ is defined in \eqref{eq:LAQC-sq-ch-def-with-no-purify}, is concave: for $\mathcal{X}$ a finite alphabet, $p:\mathcal{X}\to[0,1]$ a probability distribution on $\mathcal{X}$, and $\{\rho_A^x\}_{x\in\mathcal{X}}$ a set of states, the following inequality holds
		\begin{equation}
			E_{\text{sq}}\!\left(\mathcal{N},\sum_{x\in\mathcal{X}}p(x)\rho_A^x\right)\geq \sum_{x\in\mathcal{X}}p(x)E_{\text{sq}}(\mathcal{N},\rho_A^x).
		\end{equation}
	\end{proposition}

	\begin{Proof}
		In order to prove this, we make use of the expression for $E_{\text{sq}}(\mathcal{N},\rho_A)$ in \eqref{eq-sq_ent_state_alt_expr}.
	
		For every state $\rho_A^x$, with $x\in\mathcal{X}$, define the state
		\begin{equation}
			\theta_{BEF}^x\coloneqq(\mathcal{V}_{E'\to EF}\circ\mathcal{U}_{A\to BE'}^{\mathcal{N}})(\rho_A^x),
		\end{equation}
		where $\mathcal{V}_{E'\to EF}$ is an arbitrary isometric channel. Then, using \eqref{eq-sq_ent_state_alt_expr}, we have
		\begin{equation}
			E_{\text{sq}}(\mathcal{N},\rho_A^x)=E(R;B)_{\omega^x}\leq \frac{1}{2}(H(B|E)_{\theta^x}+H(B|F)_{\theta^x},)
		\end{equation}
		
		Now, let
		\begin{align}
			\overline{\rho}_A&\coloneqq\sum_{x\in\mathcal{X}}p(x)\rho_A^x,\\
			\overline{\theta}_{BEF}&\coloneqq\sum_{x\in\mathcal{X}}p(x)\theta_{BEF}^x\\
			&=\sum_{x\in\mathcal{X}}p(x)(\mathcal{V}_{E'\to EF}\circ\mathcal{U}_{A\to BE'}^{\mathcal{N}})(\rho_A^x)\\
			&=(\mathcal{V}_{E'\to EF}\circ\mathcal{U}_{A\to BE'}^{\mathcal{N}})(\overline{\rho}_A)\label{eq-sq_ent_chan_no_purif_concave_pf}.
		\end{align}
		Using concavity of conditional entropy (see \eqref{eq:QEI:concavity-cond-ent}), we obtain
		\begin{align}
			\sum_{x\in\mathcal{X}}p(x)E_{\text{sq}}(\mathcal{N},\rho_A^x)&\leq \frac{1}{2}\sum_{x\in\mathcal{X}}p(x)\left(H(B|E)_{\theta^x}+H(B|F)_{\theta^x}\right)\\
			&\leq \frac{1}{2}(H(B|E)_{\overline{\theta}}+H(B|F)_{\overline{\theta}}).
		\end{align}
		Finally, since the isometric channel $\mathcal{V}_{E'\to EF}$ is arbitrary, taking the infimum over all such channels on the right-hand side of the inequality above and using \eqref{eq-sq_ent_chan_no_purif_concave_pf} gives us
		\begin{align}
			\sum_{x\in\mathcal{X}}p(x)E_{\text{sq}}(\mathcal{N},\rho_A^x)&\leq \frac{1}{2}\inf_{\mathcal{V}_{E'\to EF}}\{H(B|E)_{\overline{\theta}}+H(B|F)_{\overline{\theta}}\}\\
			&=E_{\text{sq}}(\mathcal{N},\overline{\rho}_A),
		\end{align}
		which is what we set out to prove.
	\end{Proof}

\section{Amortization Collapses}\label{sec-amort_collapses}

	In Lemma~\ref{lem:LAQC-amort-to-unamort}, we proved the following relation between the entanglement $E(\mathcal{N})$ of a channel $\mathcal{N}$ and its amortized entanglement $E^{\mathcal{A}}(\mathcal{N})$:
	\begin{equation}
		E(\mathcal{N})\leq E^{\mathcal{A}}(\mathcal{N}).
	\end{equation}
	In general, therefore, amortization can yield a larger value for the entanglement of a channel than the usual channel entanglement measure.
	
	For which entanglement measures does the reverse inequality hold? In this section, we investigate this question, and we prove that three of the channel entanglement measures that we have considered in this chapter --- max-relative entropy of entanglement, max-Rains information, and squashed entanglement --- satisfy the reverse inequality. Thus, for these three entanglement measures, amortization does not yield a higher entanglement value than the usual channel entanglement measure. This so-called ``amortization collapse'' is important because it immediately implies additivity of the usual channel entanglement measure.

\subsection{Max-Relative Entropy of Entanglement}

	We start by proving that the amortization collapse occurs for max-relative entropy of entanglement. The key tools in the proof are Propositions~\ref{lem:SKA-alt-emax} and \ref{lem:SKA-alt-e-max-channel}, which provide cone programs for both the max-relative entropy of entanglement for bipartite states and the max-relative entropy of entanglement for quantum channels. Let us recall these now:
	\begin{align}
		E_{\max}(A;B)_{\rho}&=\log_2 G_{\max}(A;B)_{\rho}\label{eq-max_rel_ent_state_cone_prog_0}\\
		&=\log_2\inf_{X_{AB}\in\widehat{\SEP}}\left\{\Tr[X_{AB}]:\rho_{AB}\leq X_{AB}\right\}\label{eq-max_rel_ent_state_cone_prog_1},\\[0.2cm]
		E_{\max}(\mathcal{N})&=\log_2\Sigma_{\max}(\mathcal{N})\label{eq-max_rel_ent_chan_cone_prog_0}\\
		&=\log_2\inf_{Y_{SB}\in\widehat{\SEP}}\left\{\norm{\Tr_B[Y_{SB}]}_{\infty}:\Gamma_{SB}^{\mathcal{N}}\leq Y_{SB}\right\},\label{eq-max_rel_ent_chan_cone_prog_1}
	\end{align}
	where $\rho_{AB}$ is a bipartite state and $\mathcal{N}$ is a quantum channel with Choi operator $\Gamma_{SB}^{\mathcal{N}}$.

	\begin{theorem*}{Amortization Collapse for Max-Relative Entropy of Entanglement}{prop:SKA-amort-doesnt-help-e-max}
		Let $\mathcal{N}_{A\rightarrow B}$ be a quantum channel. For every state $\rho_{A'AB'}$,
		\begin{equation}\label{eq-max_rel_ent_amort_collapse}
			E_{\max}(A';BB')_{\omega}\leq E_{\max}(\mathcal{N})+E_{\max}(A'A;B')_{\rho},
		\end{equation}
		where $\omega_{A'BB'}\coloneqq \mathcal{N}_{A\rightarrow B}(\rho_{A'AB'})$. Consequently,
		\begin{equation}
			E_{\max}^{\mathcal{A}}(\mathcal{N})\leq E_{\max}(\mathcal{N}),
		\end{equation}
		and thus (by Lemma~\ref{lem:LAQC-amort-to-unamort}), we have that
		\begin{equation}\label{eq-max_rel_ent_amort_no_help}
			E_{\max}^{\mathcal{A}}(\mathcal{N})=E_{\max}(\mathcal{N})
		\end{equation}
		for every quantum channel $\mathcal{N}$.
	\end{theorem*}

	\begin{Proof}
		Using the cone program formulations in \eqref{eq-max_rel_ent_state_cone_prog_0}--\eqref{eq-max_rel_ent_chan_cone_prog_1}, we find that the inequality in \eqref{eq-max_rel_ent_amort_collapse} is equivalent to
		\begin{equation}\label{eq:SKA-no-logs-amort-no-help-e-max}
			G_{\max}(A';BB')_{\omega}=\Sigma_{\max}(\mathcal{N})\cdot G_{\max}(A'A;B')_{\rho}.
		\end{equation}
		We now set out to prove this inequality.
		
		Using \eqref{eq-max_rel_ent_state_cone_prog_1}, we find that%
		\begin{equation}
			G_{\max}(A'A;B')_{\rho}=\inf\Tr[C_{A'AB'}],
		\end{equation}
		subject to the constraints%
		\begin{align}
			C_{A'AB'}  & \in\widehat{\SEP}(A'A\!:\!B'), \label{eq:SKA-C-e-max-constr-1} \\
			C_{A'AB'}  &  \geq\rho_{A'AB'}, \label{eq:SKA-C-e-max-constr-2}%
		\end{align}
		and
		\begin{equation}
			G_{\max}(A';BB')_{\omega}=\inf\Tr[D_{A'BB'}],
		\end{equation}
		subject to the constraints%
		\begin{align}
			D_{A'BB'}  &  \in\widehat{\SEP}(A'\!:\!BB'),\label{eq:SKA-sep-constr-1}\\
			D_{A'BB'}  &  \geq\mathcal{N}_{A\rightarrow B}(\rho_{A'AB'}). \label{eq:SKA-sep-constr-2}
		\end{align}
		Furthermore, \eqref{eq-max_rel_ent_chan_cone_prog_1} gives that
		\begin{equation}
			\Sigma_{\max}(\mathcal{N})=\inf\left\Vert \Tr_{B}[Y_{SB}]\right\Vert_{\infty},
		\end{equation}
		subject to the constraints%
		\begin{align}
			Y_{SB}  &  \in\widehat{\SEP}(S\!:\!B),\label{eq:SKA-Y-e-max-chan-constr-1}\\
			Y_{SB}  &  \geq \Gamma_{SB}^{\mathcal{N}}. \label{eq:SKA-Y-e-max-chan-constr-2}%
		\end{align}

		With these optimizations in place, we can now establish the inequality in \eqref{eq:SKA-no-logs-amort-no-help-e-max} by making a judicious choice for $D_{A'BB'}$. Let $C_{A'AB'}$ be an arbitrary operator to consider in the optimization for $G_{\max}(A'A;B')_{\rho}$ (i.e., satisfying \eqref{eq:SKA-C-e-max-constr-1}--\eqref{eq:SKA-C-e-max-constr-2}), and let $Y_{SB}$ be an arbitrary operator to consider in the optimization for $\Sigma_{\max}(\mathcal{N})$ (i.e., satisfying \eqref{eq:SKA-Y-e-max-chan-constr-1}--\eqref{eq:SKA-Y-e-max-chan-constr-2}). Let $\ket{\Gamma}_{SA}=\sum_{i=0}^{d_A-1}\ket{i,i}_{SA}$. Pick
		\begin{equation}
			D_{A'BB'}=\langle\Gamma|_{SA}C_{A'AB'}\otimes
Y_{SB}|\Gamma\rangle_{SA}.
		\end{equation}
		We need to prove that $D_{A'BB'}$ is feasible for $G_{\max}(A';BB')_{\omega}$. To this end, consider that%
		\begin{align}
			\langle\Gamma|_{SA}C_{A'AB'}\otimes Y_{SB}|\Gamma\rangle_{SA} &  \geq\langle\Gamma|_{SA}\rho_{A'AB'}\otimes \Gamma_{SB}^{\mathcal{N}}|\Gamma\rangle_{SA}\\
			&  =\mathcal{N}_{A\rightarrow B}(\rho_{A'AB'}),
		\end{align}
		which follows from \eqref{eq:SKA-C-e-max-constr-2},
\eqref{eq:SKA-Y-e-max-chan-constr-2}, and \eqref{eq-QM:post-selected-TP-Choi-op}. Now, since $C_{A'AB'}\in\widehat{\SEP}(A'A\!:\!B')$, it can be written as $\sum_{x\in\mathcal{X}}P_{A'A}^{x}\otimes Q_{B'}^{x}$ for some finite alphabet $\mathcal{X}$ and for sets $\{P_{A'A}^{x}\}_{x\in\mathcal{X}}$, $\{Q_{B'}^{x}\}_{x\in\mathcal{X}}$ of positive semi-definite operators. Similarly, since $Y_{SB}\in\widehat{\SEP}\left( S\!:\!B\right) $, it can be written as $\sum_{y\in\mathcal{Y}}L_{S}^{y}\otimes M_{B}^{y}$, for some finite alphabet $\mathcal{Y}$ and sets $\{L_{S}^{y}\}_{y\in\mathcal{Y}}$, $\{M_{B}^{y}\}_{y\in\mathcal{Y}}$ of positive semi-definite operators. Then, using \eqref{eq-transpose_trick} and \eqref{eq-trace_identity}, we have that%
		\begin{align}
			&  \!\!\!\!\bra{\Gamma}_{SA}C_{A'AB'}\otimes Y_{SB}\ket{\Gamma}_{SA}\nonumber\\
			&  =\sum_{\substack{x\in\mathcal{X}\\y\in\mathcal{Y}}}\bra{\Gamma}_{SA}P_{A'A}^{x}\otimes Q_{B'}^{x}\otimes L_{S}^{y}\otimes M_{B}^{y}\ket{\Gamma}_{SA}\\
			&  =\sum_{\substack{x\in\mathcal{X}\\y\in\mathcal{Y}}}\bra{\Gamma}_{SA}P_{A'A}^{x}\T_{A}(L_{A}^{y})\otimes Q_{B'}^{x}\otimes \mathbbm{1}_{S}\otimes M_{B}^{y}\ket{\Gamma}_{SA}\\
			&  =\sum_{\substack{x\in\mathcal{X}\\y\in\mathcal{Y}}}\Tr_{A}[P_{A'A}^{x}\T_{A}(L_{A}^{y})]\otimes Q_{B'}^{x}\otimes M_{B}^{y}\\
			&  \in\widehat{\SEP}(A'\!:\!BB').
		\end{align}
		For the second equality, we used the transpose trick from \eqref{eq-transpose_trick}, and for the third, we used \eqref{eq-trace_identity}.
		The last statement follows because
		\begin{equation}
			\Tr_{A}[P_{A'A}^{x}\T_{A}(L_{A}^{y})] =\Tr_{A}\!\left[\sqrt{\T_{A}(L_{A}^{y})}P_{A'A}^{x}\sqrt{\T_{A}(L_{A}^{y})}\right]
		\end{equation}
		is positive semi-definite for each $x\in\mathcal{X}$ and $y\in\mathcal{Y}$. Thus, $D_{A'BB'}$ is feasible for $G_{\max}(A';BB')_{\omega}$. Finally, using \eqref{eq-trace_identity} again, consider that%
		\begin{align}
			G_{\max}(A';BB')_{\omega} & \leq \Tr[D_{A'BB'}]  \\
&  =\Tr[\bra{\Gamma}_{SA}C_{A'AB'}\otimes Y_{SB}\ket{\Gamma}_{SA}]\\
			&  =\Tr[C_{A'AB'}\T_{A}(Y_{AB})]\\
			&  =\Tr[C_{A'AB'}\T_{A}(\Tr_{B}[Y_{AB}])].
		\end{align}
		For the second equality, we used the transpose trick  from \eqref{eq-transpose_trick}.
		Since $C_{A'AB'}$ and $Y_{SB}$ are positive semi-definite (this follows from \eqref{eq:SKA-C-e-max-constr-2} and \eqref{eq:SKA-Y-e-max-chan-constr-2}, respectively), using \eqref{eq-Schatten_norm_var} we have that
		\begin{align}
			\Tr[C_{A'AB'}\T_{A}(\Tr_{B}[Y_{AB}])]&=\abs{\Tr[C_{A'AB'}\T_{A}(\Tr_{B}[Y_{AB}])]}\\
			&\leq \norm{C_{A'AB'}}_1\norm{\T_{A}(\Tr_{B}[Y_{AB}])}_{\infty}\\
			&=\Tr[C_{A'AB'}]\norm{\T_{A}(\Tr_{B}[Y_{AB}])}_{\infty}\\
			&=\Tr[C_{A'AB'}]\norm{\Tr_{B}[Y_{AB}]}_{\infty},
		\end{align}
		where for the last equality we used the fact that the spectrum of an operator is invariant under the action of a full transpose (note, in this case, that $\T_A$ is a full transpose because the operator $\Tr_B[Y_{AB}]$ acts only on $A$).
		Therefore,
		\begin{align}
			G_{\max}(A';BB')_{\omega}& \leq \Tr[C_{A'AB'}]\norm{\Tr_{B}[Y_{AB}]}_{\infty}.
		\end{align}
		Since this inequality holds for all $C_{A'AB'}$ satisfying \eqref{eq:SKA-C-e-max-constr-1}--\eqref{eq:SKA-C-e-max-constr-2} and for all $Y_{SB}$ satisfying \eqref{eq:SKA-Y-e-max-chan-constr-1}--\eqref{eq:SKA-Y-e-max-chan-constr-2}, we conclude \eqref{eq:SKA-no-logs-amort-no-help-e-max} after taking an infimum with respect to all such operators.
		
		Having shown that
		\begin{multline}
			 E_{\max}(A';BB')_{\omega}\leq E_{\max}(\mathcal{N})+E_{\max}(A'A;B')_{\rho}
			\\
			\Longrightarrow \quad  E_{\max}(A';BB')_{\omega}-E_{\max}(A'A;B')_{\rho}\leq E_{\max}(\mathcal{N})
		\end{multline}
		for every state $\rho_{A'AB'}$, it follows immediately from the definition of $E_{\max}^{\mathcal{A}}(\mathcal{N})$ that $E_{\max}^{\mathcal{A}}(\mathcal{N})\leq E_{\max}(\mathcal{N})$. Combined with the result of Lemma~\ref{lem:LAQC-amort-to-unamort}, which is the reverse inequality, we obtain $E_{\max}^{\mathcal{A}}(\mathcal{N})=E_{\max}(\mathcal{N})$.
	\end{Proof}

	With the equality $E_{\max}^{\mathcal{A}}(\mathcal{N})=E_{\max}(\mathcal{N})$ in hand, the subadditivity of max-relative entropy of entanglement of quantum channels immediately follows.

	\begin{corollary*}{Subadditivity of Max-Relative Entropy of Entanglement of a Quantum Channel}{cor-chan_max_rel_subadditive}
		The max-relative entropy of entanglement of a quantum channel is subadditive: for every two quantum channels $\mathcal{N}$ and $\mathcal{M}$,
		\begin{equation}
			E_{\max}(\mathcal{N}\otimes\mathcal{M})\leq E_{\max}(\mathcal{N})+E_{\max}(\mathcal{M}).
		\end{equation}
	\end{corollary*}
	
	\begin{Proof}
		Given that the amortized entanglement of a quantum channel is always subadditive, regardless of whether or not the underlying state entanglement measure is additive (see Proposition~\ref{prop-chan_amort_ent_properties}), we have that
		\begin{equation}
			E_{\max}^{\mathcal{A}}(\mathcal{N}\otimes\mathcal{M})\leq E_{\max}^{\mathcal{A}}(\mathcal{N})+E_{\max}^{\mathcal{A}}(\mathcal{M})
		\end{equation}
		for all quantum channels $\mathcal{N}$ and $\mathcal{M}$. Using \eqref{eq-max_rel_ent_amort_no_help}, we immediately obtain the desired result.
	\end{Proof}

\subsection{Max-Rains Information}

	Let us now prove that the amortization collapse also occurs for the max-Rains information of a quantum channel. The key tools needed for the proof are Propositions~\ref{prop-max_Rains_rel_ent_SDP} and \ref{prop-SDP_max_rains}, which provide semi-definite programs for both the max-Rains relative entropy for bipartite states and the max-Rains information for quantum channels. Let us recall these now:
	\begin{align}
		2^{R_{\max}(A;B)_{\rho}}
		& = W_{\max}(A;B)_{\rho}\label{eq-max_Rains_rel_ent_SDP_0}\\
		& =\inf_{K_{AB},L_{AB}\geq 0}\{\Tr[K_{AB}+L_{AB}]:\T_B[K_{AB}-L_{AB}]\geq\rho_{AB}\},\label{eq-max_Rains_rel_ent_SDP_1}\\
		2^{R_{\max}(\mathcal{N})}
		& =\Gamma_{\max}(\mathcal{N})\label{eq-max_Rains_inf_chan_SDP_0}\\
		&=\inf_{Y_{SB},V_{SB}\geq 0}\{\norm{\Tr_B[V_{SB}+Y_{SB}]}_{\infty}:\T_B[V_{SB}-Y_{SB}]\geq\Gamma_{SB}^{\mathcal{N}}\},\label{eq-max_Rains_inf_chan_SDP_1}
	\end{align}
	where $\rho_{AB}$ is a bipartite state and $\mathcal{N}$ is a quantum channel with Choi operator~$\Gamma_{SB}^{\mathcal{N}}$.
	
	\begin{theorem*}{Amortization Collapse for Max-Rains Information}{prop:LAQC-amort-doesnt-help-max-Rains}
		Let $\mathcal{N}_{A\rightarrow B}$ be a quantum channel. For every state $\rho_{A'AB'}$,
		\begin{equation}\label{eq:r-max-amort-no-help-1}%
			R_{\max}(A';BB')_{\omega}\leq R_{\max}(\mathcal{N})+R_{\max}(A'A;B')_{\rho}, 
		\end{equation}
		where $\omega_{A'BB'}\coloneqq\mathcal{N}_{A\rightarrow B}(\rho_{A'AB'})$. Consequently,
		\begin{equation}
			R_{\max}^{\mathcal{A}}(\mathcal{N})\leq R_{\max}(\mathcal{N}),
		\end{equation}
		and thus (by Lemma~\ref{lem:LAQC-amort-to-unamort}), we have that
		\begin{equation}\label{eq-max_Rains_chan_inf_amort_no_help}
			R_{\max}^{\mathcal{A}}(\mathcal{N})=R_{\max}(\mathcal{N})
		\end{equation}
		for every quantum channel $\mathcal{N}$.
	\end{theorem*}

	\begin{Proof}
	The proof given below is conceptually similar to the proof of Theorem~\ref{prop:SKA-amort-doesnt-help-e-max}, but it has some key differences. 
	
		Using the semi-definite program formulations in \eqref{eq-max_Rains_rel_ent_SDP_0}--\eqref{eq-max_Rains_inf_chan_SDP_1}, we find that the inequality in \eqref{eq:r-max-amort-no-help-1} is equivalent to
		\begin{equation}\label{eq:LAQC-amortized-ineq}%
			W_{\max}(A';BB')_{\omega}\leq\Gamma_{\max}(\mathcal{N})\cdot W_{\max}(A'A;B')_{\rho}, 
		\end{equation}
		and so we aim to prove this one.
		
		Using \eqref{eq-max_Rains_rel_ent_SDP_1}, we have that
		\begin{equation}
			W(A'A;B')_{\rho}=\inf\Tr[C_{A'AB'}+D_{A'AB'}],
		\end{equation}
		subject to the constraints%
		\begin{align}
			C_{A'AB'},\ D_{A'AB'}  &  \geq 0,\label{eq:LAQC-W-constraint_0}\\
			\T_{B'}(C_{A'AB'}-D_{A'AB'})  & \geq\rho_{A'AB'}, \label{eq:LAQC-W-constraint}%
		\end{align}
		and we also have
		\begin{equation}
			W(A';BB')_{\omega}=\inf\Tr[G_{A'BB'}+F_{A'BB'}],
		\end{equation}
		subject to the constraints%
		\begin{align}
			G_{A'BB'},F_{A'BB'}  &  \geq 0,\label{eq:LAQC-W-input-const-1}\\
			\mathcal{N}_{A\rightarrow B}(\rho_{A'AB'})  &  \leq \T_{BB'}[G_{A'BB'}-F_{A'BB'}].\label{eq:LAQC-W-input-const-2}%
		\end{align}
		Using \eqref{eq-max_Rains_inf_chan_SDP_1}, we have that
		\begin{equation}
			\Gamma_{\max}(\mathcal{N})=\inf\left\Vert \Tr_{B}[V_{SB}+Y_{SB}]\right\Vert _{\infty},
		\end{equation}
		subject to the constraints%
		\begin{align}
			Y_{SB},V_{SB}  &  \geq 0,\label{eq:LAQC-Gamma-constraint_0}\\
			\T_{B}[V_{SB}-Y_{SB}]  &  \geq \Gamma_{SB}^{\mathcal{N}}.\label{eq:LAQC-Gamma-constraint}%
		\end{align}
		
		With these SDP formulations in place, we can now establish the inequality in \eqref{eq:LAQC-amortized-ineq} by making judicious choices for $G_{A'BB'}$ and $F_{A'BB'}$. Let $C_{A'AB'}$ and $D_{A'AB'}$ be arbitrary operators in the optimization for $W_{\max}(A'A;B')_{\rho}$, and let $Y_{SB}$ and $V_{SB}$ be arbitrary operators in the optimization for $\Gamma_{\max}(\mathcal{N})$. Let $\ket{\Gamma}_{SA}=\sum_{i=0}^{d_A-1}\ket{i,i}_{SA}$. Pick
		\begin{align}
			G_{A'BB'} &  =\bra{\Gamma}_{SA}C_{A'AB'}\otimes V_{SB}+D_{A'AB'}\otimes Y_{SB}\ket{\Gamma}_{SA},\\
			F_{A'BB'} &  =\bra{\Gamma}_{SA}C_{A'AB'}\otimes Y_{SB}+D_{A'AB'}\otimes V_{SB}\ket{\Gamma}_{SA}.
		\end{align}
		Note that $G_{A'BB'},F_{A'BB'}\geq 0$
because $C_{A'AB'}$, $D_{A'AB'}$, $Y_{SB}$, $V_{SB}\geq0$. Using \eqref{eq:LAQC-W-constraint} and \eqref{eq:LAQC-Gamma-constraint}, consider that
		\begin{align}
			&  \T_{BB'}[G_{A'BB'}-F_{A'BB'}]\nonumber\\
			&  =\T_{BB'}\left[\bra{\Gamma}_{SA}(C_{A'AB'}-D_{A'AB'})\otimes(V_{SB}-Y_{SB})\ket{\Gamma}_{SA}\right] \\
			&  =\bra{\Gamma}_{SA}\T_{B'}[C_{A'AB'}-D_{A'AB'}]\otimes \T_{B}[V_{SB}-Y_{SB}]\ket{\Gamma}_{SA}\\
			&  \geq \bra{\Gamma}_{SA} \rho_{A'AB'}\otimes \Gamma_{SB}^{\mathcal{N}}\ket{\Gamma}_{SA}\\
			&  =\mathcal{N}_{A\rightarrow B}(\rho_{A'AB'}),
		\end{align}
		where the last equality follows from \eqref{eq-QM:post-selected-TP-Choi-op}. Our choices of $G_{A'BB'}$ and $F_{A'BB'}$ are thus feasible points for $W_{\max}(A';BB')_{\omega}$. Using this, along with \eqref{eq-transpose_trick} and \eqref{eq-trace_identity}, we obtain
		\begin{align}
			&  W_{\max}(A';BB')_{\omega}\nonumber\\
			&  \leq\Tr[G_{A'BB'}+F_{A'BB'}]
			\label{eq-LAQC-amort-max-Rains-core-part-1}\\
			&  =\Tr[\bra{\Gamma}_{SA}(C_{A'AB'}+D_{A'AB'})\otimes(V_{SB}+Y_{SB})\ket{\Gamma}_{SA}]\\
			&  =\Tr[(C_{A'AB'}+D_{A'AB'})\T_{A}(V_{AB}+Y_{AB})]\\
			&  =\Tr[(C_{A'AB'}+D_{A'AB'})\T_{A}(\Tr_{B}[V_{AB}+Y_{AB}])].
		\end{align}
		The second equality follows from the transpose trick from \eqref{eq-transpose_trick}.
		Now, since $C_{A'AB'},D_{A'AB'}\geq 0$ (recall \eqref{eq:LAQC-W-constraint_0}), and $V_{AB},Y_{AB}\geq 0$ (recall \eqref{eq:LAQC-Gamma-constraint_0}), we can use \eqref{eq-Schatten_norm_var} to conclude that
		\begin{align}
			&\Tr[(C_{A'AB'}+D_{A'AB'})\T_A(\Tr_B[V_{AB}+Y_{AB}])]\\
			&\quad=\abs{\Tr[(C_{A'AB'}+D_{A'AB'})\T_A(\Tr_B[V_{AB}+Y_{AB}])]}\\
			&\quad \leq \norm{C_{A'AB'}+D_{A'AB'}}_1\norm{\T_A(\Tr_B[V_{AB}+Y_{AB}])}_{\infty}\\
			&\quad =\Tr[C_{A'AB'}+D_{A'AB'}]\norm{\T_A(\Tr_B[V_{AB}+Y_{AB}])}_{\infty}\\
			&\quad =\Tr[C_{A'AB'}+D_{A'AB'}]\norm{\Tr_B[V_{AB}+Y_{AB}]}_{\infty},
		\end{align}
		where the final equality follows because the spectrum of an operator is invariant under the action of a (full) transpose (note, in this case, that $\T_{A}$ is a full transpose because the operator $\Tr_{B}[V_{AB}+Y_{AB}]$ acts only on system $A$). We thus have
		\begin{align}
			W_{\max}(A';BB')_{\omega} &\leq\Tr[C_{A'AB'}+D_{A'AB'}]\norm{\T_{A}[\Tr_{B}[V_{AB}+Y_{AB}]]}_{\infty}
		\end{align}
		Since this inequality holds for all $C_{A'AB'}$ and $D_{A'AB'}$ satisfying \eqref{eq:LAQC-W-constraint} and for all $V_{AB}$ and $Y_{AB}$ satisfying \eqref{eq:LAQC-Gamma-constraint}, we conclude the inequality in \eqref{eq:LAQC-amortized-ineq}.
		
		Having shown that
		\begin{multline}
			R_{\max}(A';BB')_{\omega}\leq R_{\max}(\mathcal{N})+R_{\max}(A'A;B')_{\rho},\\
			\Longrightarrow \quad R_{\max}(A';BB')_{\omega}-R_{\max}(A'A;B')_{\rho}\leq R_{\max}(\mathcal{N})
		\end{multline}
		for every state $\rho_{A'AB'}$, it immediately follows from the definition of amortized entanglement of a channel that
		\begin{equation}
			R_{\max}^{\mathcal{A}}(\mathcal{N})\leq R_{\max}(\mathcal{N}).
		\end{equation}
		Thus, by Lemma~\ref{lem:LAQC-amort-to-unamort}, which proves the reverse inequality, we obtain
		\begin{equation}
			R_{\max}^{\mathcal{A}}(\mathcal{N})=R_{\max}(\mathcal{N})
		\end{equation}
		for every quantum channel $\mathcal{N}$.
	\end{Proof}

	With the equality $R_{\max}^{\mathcal{A}}(\mathcal{N})=R_{\max}(\mathcal{N})$ in hand, along with additivity of max-Rains relative entropy for bipartite states, additivity of max-Rains information of a quantum channel immediately follows.
	
	\begin{corollary*}{Additivity of Max-Rains Information of a Channel}{cor-max_Rains_inf_chan_additive}
		The max-Rains information of a quantum channel is additive: for every two quantum channels $\mathcal{N}$ and $\mathcal{M}$,
		\begin{equation}
			R_{\max}(\mathcal{N}\otimes\mathcal{M})=R_{\max}(\mathcal{N})+R_{\max}(\mathcal{M}).
		\end{equation}
	\end{corollary*}
	
	\begin{Proof}
		The additivity of max-Rains relative entropy for bipartite states (see Proposition~\ref{prop:LAQC-add-max-Rains-rel-ent}), along with Proposition~\ref{prop-chan_amort_ent_properties}, implies that the amortized max-Rains information of a quantum channel is additive, meaning that
		\begin{equation}
			R_{\max}^{\mathcal{A}}(\mathcal{N}\otimes\mathcal{M})=R_{\max}^{\mathcal{A}}(\mathcal{N})+R_{\max}^{\mathcal{A}}(\mathcal{M})
		\end{equation}
		for all quantum channels $\mathcal{N}$ and $\mathcal{M}$. Then, from \eqref{eq-max_Rains_chan_inf_amort_no_help}, we obtain the desired result.
	\end{Proof}

\subsection{Squashed Entanglement}
	
	Finally, let us prove that the amortization collapse occurs for the squashed entanglement of a quantum channel.

	\begin{theorem*}{Amortization Collapse for Squashed Entanglement}{thm:LAQC-amort-collapse-squashed}
		The squashed entanglement of a channel does not increase under amortization, i.e.,
		\begin{equation}\label{eq-sq_ent_chan_amort_no_help}
			E_{\text{sq}}(\mathcal{N})=E_{\text{sq}}^{\mathcal{A}}(\mathcal{N})
		\end{equation}
		for every quantum channel $\mathcal{N}$.
	\end{theorem*}

	\begin{Proof}
		The inequality $E_{\text{sq}}(\mathcal{N})\leq E_{\text{sq}}^{\mathcal{A}}(\mathcal{N})$ has already been shown in Lemma~\ref{lem:LAQC-amort-to-unamort}. We thus prove the reverse inequality.

		Let $\rho_{A'AB'}$ be an arbitrary state, and let $\omega_{A'BB'}=\mathcal{N}_{A\rightarrow B}(\rho_{A'AB'})$. Let $\mathcal{U}_{A\rightarrow BE}^{\mathcal{N}}$ be an isometric channel extending $\mathcal{N}_{A\rightarrow B}$, and let $\varphi_{A'AB'R}$ be a purification of $\rho_{A'AB'}$. Then, the state $\theta_{A'BB'ER}\coloneqq \mathcal{U}_{A\rightarrow BE}^{\mathcal{N}}(\varphi_{A'AB'R})$ is a purification of $\omega_{A'BB'}$. As we show in Lemma~\ref{lem:LAQC-sq-subadd-lem} below, the following inequality holds
		\begin{equation}
			E_{\text{sq}}(A';BB')_{\omega}=E_{\text{sq}}(A';BB')_{\theta}\leq E_{\text{sq}}(A'B'R;B)_{\theta}+E_{\text{sq}}(A'BE;B')_{\theta}.
		\end{equation}
		Now, squashed entanglement is invariant under the action of an isometric channel, in particular $\mathcal{U}_{A\to BE}^{\mathcal{N}}$, so that $E_{\text{sq}}(A'BE;B^{\prime })_{\theta}=E_{\text{sq}}(A'A;B')_{\rho}$. Also, the state $\theta_{A'BB'ER}$ can be viewed as a particular state in the optimization over  states that defines $E_{\text{sq}}(\mathcal{N})$ because $\theta_{A'BB'ER}$ is a purification of $\omega_{A'BB'}$. Therefore,
		\begin{equation}
			E_{\text{sq}}(A'B'R;B)_{\theta}\leq E_{\text{sq}}(\mathcal{N}).
		\end{equation}
		Altogether, we thus have
		\begin{multline}
			 E_{\text{sq}}(A';BB')_{\omega}\leq E_{\text{sq}}(\mathcal{N})+E_{\text{sq}}(A'A;B')_{\rho}\\
			\Longrightarrow \quad E_{\text{sq}}(A';BB')_{\omega}-E_{\text{sq}}(A'A;B')_{\rho}\leq E_{\text{sq}}(\mathcal{N})
		\end{multline}
		for every state $\rho_{A'AB'}$, which means by definition of amortized entanglement that
		\begin{equation}
			E_{\text{sq}}^{\mathcal{A}}(\mathcal{N})\leq E_{\text{sq}}(\mathcal{N}),
		\end{equation}
		which is what we sought to prove.
	\end{Proof}
	
	With the equality $E_{\text{sq}}^{\mathcal{A}}(\mathcal{N})=E_{\text{sq}}(\mathcal{N})$ in hand, along with additivity of squashed entanglement for bipartite states (Property 4.~of Proposition~\ref{prop-squashed_ent_properties}), additivity of squashed entanglement for channels immediately follows.

	\begin{corollary*}{Additivity of Squashed Entanglement of a Channel}{cor-sq_ent_additive}
		The squashed entanglement of a quantum channel is additive: for every two quantum channels $\mathcal{N}$ and $\mathcal{M}$,
		\begin{equation}
			E_{\text{sq}}(\mathcal{N}\otimes\mathcal{M})=E_{\text{sq}}(\mathcal{N})+E_{\text{sq}}(\mathcal{M}).
		\end{equation}
	\end{corollary*}
	
	\begin{Proof}
		The additivity of the squashed entanglement for bipartite states (see Proposition~\ref{prop-squashed_ent_properties}), along with Proposition~\ref{prop-chan_amort_ent_properties}, implies that the amortized squashed entanglement of a quantum channel is additive, meaning that
		\begin{equation}
			E_{\text{sq}}^{\mathcal{A}}(\mathcal{N}\otimes\mathcal{M})=E_{\text{sq}}^{\mathcal{A}}(\mathcal{N})+E_{\text{sq}}^{\mathcal{A}}(\mathcal{M})
		\end{equation}
		for all quantum channels $\mathcal{N}$ and $\mathcal{M}$. Then, from \eqref{eq-sq_ent_chan_amort_no_help}, we obtain the desired result.
	\end{Proof}

	\begin{Lemma}{lem:LAQC-sq-subadd-lem}
		Let $\psi_{KL_{1}L_{2}M_{1}M_{2}}$ be a pure state. Then%
		\begin{equation}
			E_{\operatorname{sq}}(K;L_{1}L_{2})_{\psi}\leq E_{\operatorname{sq}}(KL_{2}M_{2};L_{1})_{\psi}+E_{\operatorname{sq}}(KL_{1}M_{1};L_{2})_{\psi}.
		\end{equation}
	\end{Lemma}

	\begin{Proof}
		We make use of the squashing channel representation of squashed entanglement in \eqref{eq:LAQC-sq-ch-rep-1}, namely,
		\begin{equation}\label{eq:LAQC-sq-ch-rep-1_2}
			E_{\text{sq}}(A;B)_{\rho}=\frac{1}{2}\inf_{\mathcal{S}_{E'\to E}}\{I(A;B|E)_{\omega}:\omega_{ABE}=\mathcal{S}_{E'\to E}(\psi_{ABE'}^{\rho})\},
		\end{equation}
		where $\psi_{ABE'}^{\rho}$ is a purification of $\rho$ and the infimum is with respect to every quantum channel $\mathcal{S}_{E'\to E}$. Let us also recall that
		\begin{align}
			I(A;B|E)_{\omega}&=H(A|E)_{\omega}+H(B|E)_{\omega}-H(AB|E)_{\omega}\label{eq-qcmi_cond_ent_expr_0}\\
			&=H(B|E)_{\omega}-H(B|AE)_{\omega}\label{eq-qcmi_cond_ent_expr},
		\end{align}
		and that strong subadditivity (Theorem~\ref{thm-SSA}) is the statement that $I(A;B|E)_{\omega}\geq 0$. From this we obtain the following two inequalitites:
		\begin{align}
			H(AB|E)_{\omega}&\leq H(A|E)_{\omega}+H(B|E)_{\omega},\label{eq-SSA_1}\\
			H(B|AE)_{\omega}&\leq H(B|E)_{\omega}.\label{eq-SSA_2}
		\end{align}
		
		Now, the given pure state $\psi_{KL_1L_2M_1M_2}$ can be thought of as a purification of the reduced state $\psi_{KL_1L_2}$ for which the squashed entanglement $E_{\text{sq}}(K;L_1L_2)_{\psi}$ is evaluated, with the purifying systems being $M_1$ and $M_2$. Then, considering an arbitrary product squashing channel $\mathcal{S}_{M_1\to M_1'}^1\otimes\mathcal{S}_{M_2\to M_2'}^2$, and letting
		\begin{equation}
			\omega_{KL_1L_2M_1'M_2'}=(\mathcal{S}_{M_1\to M_1'}^1\otimes\mathcal{S}_{M_2\to M_2'}^2)(\psi_{KL_1L_2M_1M_2}),
		\end{equation}
		we find from \eqref{eq:LAQC-sq-ch-rep-1_2} that
		\begin{equation}
			2 E_{\text{sq}}(K;L_1L_2)_{\psi}\leq I(K;L_1L_2|M_1'M_2')_{\omega}.
		\end{equation}
		Expanding the quantum conditional mutual information using \eqref{eq-qcmi_cond_ent_expr}, we have that 
		\begin{equation}
			I(K;L_1L_2|M_1'M_2')_{\omega}=H(L_1L_2|M_1'M_2')_{\omega}-H(L_1L_2|M_1'M_2'K)_{\omega}.
		\end{equation}
		Now, let $\phi_{KL_1L_2M_1'M_2'R}$ be a purification of $\omega_{KL_1L_2M_1'M_2'}$ with purifying system $R$. Then, by definition of conditional entropy, and using the fact that $\phi_{KL_1L_2M_1'M_2'R}$ is pure, we obtain\footnote{The steps in \eqref{eq-sq_ent_subadd_lem_pf2}--\eqref{eq-sq_ent_subadd_lem_pf4} establish a general fact called \textit{duality of conditional entropy}: for every pure state $\psi_{ABE}$, the following equality holds $H(A|E)_{\psi}+H(B|E)_{\psi}=0$.}
		\begin{align}
			H(L_1L_2|M_1'M_2'K)_{\omega}&=H(L_1L_2M_1'M_2'K)_{\omega}-H(M_1'M_2'K)_{\omega}\label{eq-sq_ent_subadd_lem_pf2}\\
			&=H(R)_{\phi}-H(L_1L_2R)_{\phi}\label{eq-sq_ent_subadd_lem_pf3}\\
			&=-H(L_1L_2|R)_{\phi}.\label{eq-sq_ent_subadd_lem_pf4}
		\end{align}
		Therefore,
		\begin{equation}\label{eq-sq_ent_subadd_lem_pf1}
			2E_{\text{sq}}(K;L_1L_2)_{\psi}\leq H(L_1L_2|M_1'M_2')_{\phi}+H(L_1L_2|R)_{\phi}.
		\end{equation}
		Using the inequality in \eqref{eq-SSA_1} followed by two applications of \eqref{eq-SSA_2} (with appropriate identification of subsystems in all three cases), we obtain
		\begin{align}
			H(L_1L_2|M_1'M_2')_{\phi}&\leq H(L_1|M_1'M_2')_{\phi}+H(L_2|M_1'M_2')_{\phi}\\
			&\leq H(L_1|M_1')_{\phi}+H(L_2|M_2')_{\phi}.
		\end{align}
		Next, using \eqref{eq-SSA_1}, we conclude that
		\begin{equation}
			H(L_1L_2|R)_{\phi}\leq H(L_1|R)_{\phi}+H(L_2|R)_{\phi}.
		\end{equation}
		Therefore, continuing from \eqref{eq-sq_ent_subadd_lem_pf1}, we have 
		\begin{equation}
			2 E_{\text{sq}}(K;L_1L_2)_{\psi}\leq H(L_1|M_1')_{\phi}+H(L_2|M_2')_{\phi}					+H(L_1|R)_{\phi}+H(L_2|R)_{\phi}.
		\end{equation}
		Now, applying reasoning analogous to that in \eqref{eq-sq_ent_subadd_lem_pf2}--\eqref{eq-sq_ent_subadd_lem_pf4} for the last two terms, we find that
		\begin{align}
			2E_{\text{sq}}(K;L_1L_2)_{\psi}&\leq H(L_1|M_1')_{\omega}-H(L_2|KL_2M_1'M_2')_{\omega}\notag \\
			&\qquad +H(L_2|M_2')_{\omega}-H(L_2|KL_1M_1'M_2')_{\omega}\\
			&=I(KL_2M_2';L_1|M_1')_{\omega}+I(KL_1M_1';L_2|M_2')_{\omega},\label{eq-sq_ent_subadd_lem_pf5}
		\end{align}
		where in the last line we used the expression in \eqref{eq-qcmi_cond_ent_expr} for the conditional mutual information. 
		
		Now, let us regard $\psi_{KL_1L_2M_1M_2}$ as a purification of the reduced state $\psi_{KL_2M_2L_1}$, for which the squashed entanglement $E_{\text{sq}}(KL_2M_2;L_1)_{\psi}$ is evaluated, with the purifying system being $M_1$. Then, the state
		\begin{equation}
			\tau_{KL_2M_2L_1M_1'}\coloneqq\mathcal{S}_{M_1\to M_1'}^1(\psi_{KL_2M_1L_1M_1})
		\end{equation}
		is a particular extension of $\psi_{KL_2M_2L_1}$ in the optimization for $E_{\text{sq}}(KL_2M_2;L_1)_{\psi}$. Similarly, we can regard $\psi_{KL_1L_2M_1M_2}$ as a purification of the reduced state $\psi_{KL_1M_1L_2}$, for which the squashed entanglement $E_{\text{sq}}(KL_1M_1;L_2)_{\psi}$ is evaluated, with the purifying system being $M_2$. Then, the state
		\begin{equation}
			\sigma_{KL_1M_1L_2M_2'}\coloneqq\mathcal{S}_{M_2\to M_2'}^2(\psi_{KL_1M_1L_2M_2})
		\end{equation}
		is a particular extension of $\psi_{KL_1M_1L_2}$ in the optimization for $E_{\text{sq}}(KL_1M_1;L_2)_{\psi}$. Using all of this, we proceed from \eqref{eq-sq_ent_subadd_lem_pf5} to obtain
		\begin{align}
			2E_{\text{sq}}(K;L_1L_2)_{\psi}&\leq I(KL_2M_2';L_1|M_1')_{\omega}+I(KL_1M_1';L_2|M_2')_{\omega}\\
			&\leq I(KL_2M_2;L_1|M_1')_{\tau}+I(KL_1M_1;L_2|M_2')_{\sigma},
		\end{align}
		where for the second inequality we used the data-processing inequality for conditional mutual information (Proposition~\ref{prop-cond_mut_inf_properties}). Since the squashing channels $\mathcal{S}_{M_1\to M_1'}^1$ and $\mathcal{S}_{M_2\to M_2'}^2$ are arbitrary, optimizing over all such channels on the right-hand side of the inequality above leads us to
		\begin{equation}
			E_{\text{sq}}(K;L_1L_2)_{\psi}\leq E_{\text{sq}}(KL_2M_2;L_1)_{\psi}+E_{\text{sq}}(KL_1M_1;L_2)_{\psi},
		\end{equation}
		as required.
	\end{Proof}

\section{Summary}

	...	We considered two types of channel entanglement measures. The first type quantifies the entanglement of a bipartite state after one share of it is sent through the given quantum channel, in a manner analogous to the channel information measures defined in Chapter~\ref{chap-entropies}. The second type of channel entanglement measure is called \textit{amortized entanglement}, which essentially quantifies the difference in entanglement between a bipartite state and the state obtained after sending one share of it through the given channel. The concept of amortized entanglement turns out to play an important role in feedback-assisted communciation scenarios (as considered in Part~\ref{part-feedback}), as it can be used to prove important properties  of entanglement measures of the first kind....

\section{Bibliographic Notes}

	...The entanglement of a quantum channel was presented by \citet{TGW14IEEE,TWW17}, by employing the squashed entanglement and the Rains relative entropy entanglement measures, respectively. The amortized entanglement of a quantum channel has its roots in early work by \citet{BHLS03,LHL03}, and it was formally defined and its various properties established by \citet{KW17}. The work by \citet{RKBKMA17} is related to the notion of amortized entanglement. The connection of amortized entanglement to teleportation simulation of quantum channels was elucidated by \citet{KW17}. A channel's relative entropy of entanglement was defined by \citet{PLOB15}, max-relative entropy of entanglement by \citet{CM17}, and the hypothesis testing and sandwiched R\'enyi relative entropy of entanglement by \citet{WTB16}. Proposition~\ref{prop-gen_div_ent_chan_cov} is based on \citep[Proposition~2]{TWW17} and \citep[Proposition~14]{WTB16}. The generalized Rains information of a quantum channel was defined by \citet{TWW17}, who explicitly considered the Rains information and the sandwiched R\'enyi Rains information as special cases. \citet{TBR15} focused on the hypothesis testing Rains information of a quantum channel. \citet{WFD17} observed  that the quantity  defined by \citet{WD16b} is equal to the max-Rains information of a quantum channel. The semi-definite program formulation of the max-Rains information of a quantum channel in Proposition~\ref{prop-SDP_max_rains} is due to \citet{WD16b,WFD17}. Proposition~\ref{prop-gen_Rains_inf_chan_cov} is due to \citet{TWW17}. Concavity of a channel's unoptimized squashed entanglement (Proposition~\ref{lem:concavity-sq-ent-ch-input}) is due to \citet{TGW14IEEE}. The amortization collapse of max-relative entropy of entanglement and of max-Rains information (Theorems~\ref{prop:SKA-amort-doesnt-help-e-max} and \ref{prop:LAQC-amort-doesnt-help-max-Rains}, respectively) were shown by \citet{BW17}, with the amortization collapse of max-relative entropy implicitly considered by \citet{CM17}. Lemma~\ref{lem:LAQC-sq-subadd-lem} is due to \citet{TGW14IEEE}, and the explicit observation that Lemma~\ref{lem:LAQC-sq-subadd-lem} implies that amortization does not increase the squashed entanglement of a quantum channel (Theorem~\ref{thm:LAQC-amort-collapse-squashed}) was realized by \citet{KW17}. Corollary~\ref{cor-sq_ent_additive} was established by \citet{TGW14IEEE}. ...

\section{Problems}

\begin{subappendices}

\section[The \texorpdfstring{$\alpha\to 1$}{a to 1} and \texorpdfstring{$\alpha\to\infty$}{a to inf} Limits of the Sandwic\-hed R\'{e}nyi Entanglement Measures]{The \texorpdfstring{$\alpha\to 1$}{a to 1} and \texorpdfstring{$\alpha\to\infty$}{a to inf} Limits of the Sandwic\-hed R\'{e}nyi Entanglement\\Measures}\label{app-sand_ren_inf_limit}

	In this section, we prove the $\alpha\to 1$ and $\alpha\to\infty$ limits of the sandwiched R\'{e}nyi state and channel entanglement measures that we have considered in this chapter. Specifically, we consider the limits of
	\begin{align}
		\widetilde{E}_{\alpha}(A;B)_{\rho}&=\inf_{\sigma_{AB}\in\SEP(A:B)}\widetilde{D}_{\alpha}(\rho_{AB}\Vert\sigma_{AB}),\\
		\widetilde{E}_{\alpha}(\mathcal{N})&=\sup_{\psi_{RA}}\inf_{\sigma_{RB}\in\SEP(R:B)}\widetilde{D}_{\alpha}(\mathcal{N}_{A\to B}(\psi_{RA})\Vert\sigma_{RB}),\\
		\widetilde{R}_{\alpha}(A;B)_{\rho}&=\inf_{\sigma_{AB}\in\PPT'(A:B)}\widetilde{D}_{\alpha}(\rho_{AB}\Vert\sigma_{AB}),\\
		\widetilde{R}_{\alpha}(\mathcal{N})&=\sup_{\psi_{RA}}\inf_{\sigma_{RB}\in\PPT'(R:B)}\widetilde{D}_{\alpha}(\mathcal{N}_{A\to B}(\psi_{RA})\Vert\sigma_{RB}),
	\end{align}
	where $\alpha\in[\sfrac{1}{2},1)\cup(1,\infty)$.
	
	We make consistent use throughout this section of the fact that the sandwiched R\'{e}nyi relative entropy $\widetilde{D}_{\alpha}$ is monotonically increasing in $\alpha$ for all $\alpha\in(0,1)\cup(1,\infty)$ (see Proposition~\ref{prop-sand_rel_ent_properties}), as well as the fact that $\lim_{\alpha\to 1}\widetilde{D}_{\alpha}=D$, where $D$ is the quantum relative entropy (see Proposition~\ref{prop-sand_ren_ent_lim}). We also use the fact that $\lim_{\alpha\to\infty}\widetilde{D}_{\alpha}=D_{\max}$ (see Proposition~\ref{prop-sand_rel_ent_limit_max}).

\subsection{\texorpdfstring{$\alpha\to 1$}{a to 1} Limits}

	We start by showing that
	\begin{equation}\label{eq-sand_rel_ent_of_ent_lim_a1_pf}
		\lim_{\alpha\to 1}\widetilde{E}_{\alpha}(A;B)_{\rho}=E_R(A;B)=\inf_{\sigma_{AB}\in\SEP(A:B)}D(\rho_{AB}\Vert\sigma_{AB}).
	\end{equation}
	The proof of
	\begin{equation}
		\lim_{\alpha\to 1}\widetilde{R}_{\alpha}(A;B)_{\rho}=R(A;B)_{\rho}=\inf_{\sigma_{AB}\in\PPT'(A:B)}D(\rho_{AB}\Vert\sigma_{AB})
	\end{equation}
	is analogous, and so we omit it.
	
	When approaching one from above, due to monotonicity in $\alpha$ of $\widetilde{D}_{\alpha}$ (Proposition~\ref{prop-sand_rel_ent_properties}), we have that
	\begin{equation}
		\lim_{\alpha\to 1^+}\widetilde{D}_{\alpha}=\inf_{\alpha\in(1,\infty)}\widetilde{D}_{\alpha}=D.
	\end{equation}
	Therefore, we readily obtain
	\begin{align}
		\lim_{\alpha\to 1^+}\widetilde{E}_{\alpha}(A;B)_{\rho}&=\inf_{\alpha\in(1,\infty)}\inf_{\sigma_{AB}\in\SEP(A:B)}\widetilde{D}_{\alpha}(\rho_{AB}\Vert\sigma_{AB})\\
		&=\inf_{\sigma_{AB}\in\SEP(A:B)}\inf_{\alpha\in(1,\infty)}\widetilde{D}_{\alpha}(\rho_{AB}\Vert\sigma_{AB})\\
		&=\inf_{\sigma_{AB}\in\SEP(A:B)}D(\rho_{AB}\Vert\sigma_{AB})\\
		&=E_R(A;B)_{\rho}.
	\end{align}
	Now, when approaching one from below, we have
	\begin{equation}
		\lim_{\alpha\to 1^-}\widetilde{D}_{\alpha}=\sup_{\alpha\in(0,1)}\widetilde{D}_{\alpha}=D.
	\end{equation}
	Therefore,
	\begin{equation}
		\lim_{\alpha\to 1^-}\widetilde{E}_{\alpha}(A;B)_{\rho}=\sup_{\alpha\in(\sfrac{1}{2},1)}\inf_{\sigma_{AB}\in\SEP(A:B)}\widetilde{D}_{\alpha}(\rho_{AB}\Vert\sigma_{AB}).
	\end{equation}
	Now, we apply Theorem~\ref{thm-Mosonyi_minimax}. Specifically, we can apply the theorem in order to exchange the order of the infimum and supremum because the function
	\begin{equation}
		(\sigma_{AB},\alpha)\mapsto \widetilde{D}_{\alpha}(\rho_{AB}\Vert\sigma_{AB})
	\end{equation}
	is continuous in the first argument and montonically increasing in the second argument (also, the set of separable states is compact). We thus obtain
	\begin{align}
		\lim_{\alpha\to 1^-}\widetilde{E}_{\alpha}(A;B)_{\rho}&=\sup_{\alpha\in(\sfrac{1}{2},1)}\inf_{\sigma_{AB}\in\SEP(A:B)}\widetilde{D}_{\alpha}(\rho_{AB}\Vert\sigma_{AB})\\
		&=\inf_{\sigma_{AB}\in\SEP(A:B)}\sup_{\alpha\in(\sfrac{1}{2},1)}\widetilde{D}_{\alpha}(\rho_{AB}\Vert\sigma_{AB})\\
		&=\inf_{\sigma_{AB}\in\SEP(A:B)}D(\rho_{AB}\Vert\sigma_{AB})\\
		&=E_R(A;B)_{\rho}.
	\end{align}
	This concludes the proof of \eqref{eq-sand_rel_ent_of_ent_lim_a1_pf}.

	Now, for the channel measure, we show that
	\begin{equation}\label{eq-sand_rel_ent_chan_lim_a1_pf}
		\lim_{\alpha\to 1}\widetilde{E}_{\alpha}(\mathcal{N})=E_R(\mathcal{N})=\sup_{\psi_{RA}}\inf_{\sigma_{RB}\in\SEP(R:B)}D(\mathcal{N}_{A\to B}(\psi_{RA})\Vert\sigma_{RB}).
	\end{equation}
	The proof of
	\begin{equation}
		\lim_{\alpha\to 1}\widetilde{R}_{\alpha}(\mathcal{N})=R(\mathcal{N})=\sup_{\psi_{RA}}\inf_{\sigma_{RB}\in\PPT'(R:B)}D(\mathcal{N}_{A\to B}(\psi_{RA})\Vert\sigma_{RB})
	\end{equation}
	is analogous, and so we omit it.
	
	When approaching one from below, we use exactly the same arguments as above to exchange the infimum and supremum, in order to conclude that
	\begin{align}
		\lim_{\alpha\to 1^-}\widetilde{E}_{\alpha}(\mathcal{N})&=\sup_{\alpha\in(\sfrac{1}{2},1)}\sup_{\psi_{RA}}\inf_{\sigma_{RB}\in\SEP(R:B)}\widetilde{D}_{\alpha}(\mathcal{N}_{A\to B}(\psi_{RA})\Vert\sigma_{RB})\\
		&=\sup_{\psi_{RA}}\sup_{\alpha\in(\sfrac{1}{2},1)}\inf_{\sigma_{RB}\in\SEP(R:B)}\widetilde{D}_{\alpha}(\mathcal{N}_{A\to B}(\psi_{RA})\Vert\sigma_{RB})\\
		&=\sup_{\psi_{RA}}\inf_{\sigma_{RB}\in\SEP(R:B)}\sup_{\alpha\in(\sfrac{1}{2},1)}\widetilde{D}_{\alpha}(\mathcal{N}_{A\to B}(\psi_{RA})\Vert\sigma_{RB})\\
		&=\sup_{\psi_{RA}}\inf_{\sigma_{RB}\in\SEP(R:B)}D(\mathcal{N}_{A\to B}(\psi_{RA})\Vert\sigma_{RB})\\
		&=E_R(\mathcal{N}).
	\end{align}
	Next, when approaching from above, we again use Theorem~\ref{thm-Mosonyi_minimax}. This time, since the function
	\begin{equation}
		(\psi_{RA},\alpha)\mapsto \inf_{\sigma_{RB}\in\SEP(R:B)}\widetilde{D}_{\alpha}(\mathcal{N}_{A\to B}(\psi_{RA})\Vert\sigma_{RB})
	\end{equation}
	is continuous in the first argument and monotonically increasing in the second argument, we can exchange the order of the infimum and supremum to obtain
	\begin{align}
		\lim_{\alpha\to 1^+}\widetilde{E}_{\alpha}(\mathcal{N})&=\inf_{\alpha\in(1,\infty)}\sup_{\psi_{RA}}\inf_{\sigma_{RB}\in\SEP(R:B)}\widetilde{D}_{\alpha}(\mathcal{N}_{A\to B}(\psi_{RA})\Vert\sigma_{RB})\\
		&=\sup_{\psi_{RA}}\inf_{\alpha\in(1,\infty)}\inf_{\sigma_{RB}\in\SEP(R:B)}\widetilde{D}_{\alpha}(\mathcal{N}_{A\to B}(\psi_{RA})\Vert\sigma_{RB})\\
		&=\sup_{\psi_{RA}}\inf_{\sigma_{RB}\in\SEP(R:B)}\inf_{\alpha\in(1,\infty)}\widetilde{D}_{\alpha}(\mathcal{N}_{A\to B}(\psi_{RA})\Vert\sigma_{RB})\\
		&=\sup_{\psi_{RA}}\inf_{\sigma_{RB}\in\SEP(R:B)}D(\mathcal{N}_{A\to B}(\psi_{RA})\Vert\sigma_{RB})\\
		&=E_R(\mathcal{N}).
	\end{align}
	This concludes the proof of \eqref{eq-sand_rel_ent_chan_lim_a1_pf}.

\subsection{\texorpdfstring{$\alpha\to \infty$}{a to inf} Limits}

	We now move on to the $\alpha\to\infty$ limits. We first show that
	\begin{equation}\label{eq-sand_rel_ent_state_lim_aInf_pf}
		\lim_{\alpha\to\infty}\widetilde{E}_{\alpha}(A;B)_{\rho}=E_{\max}(A;B)=\inf_{\sigma_{AB}\in\SEP(A:B)}D_{\max}(\rho_{AB}\Vert\sigma_{AB}).
	\end{equation}
	The proof of
	\begin{equation}
		\lim_{\alpha\to\infty}\widetilde{R}_{\alpha}(A;B)_{\rho}=R_{\max}(A;B)_{\rho}=\inf_{\sigma_{AB}\in\PPT'(A:B)}D_{\max}(\rho_{AB}\Vert\sigma_{AB})
	\end{equation}
	is analogous, and so we omit it.
	
	Note that due to monotonicity in $\alpha$ of $\widetilde{D}_{\alpha}$, the following equality holds
	\begin{equation}
		\lim_{\alpha\to\infty}\widetilde{D}_{\alpha}=\sup_{\alpha\in(1,\infty)}\widetilde{D}_{\alpha}=D_{\max}.
	\end{equation}
	Therefore,
	\begin{equation}
		\lim_{\alpha\to\infty}\widetilde{E}_{\alpha}(A;B)_{\rho}=\sup_{\alpha\in(1,\infty)}\inf_{\sigma_{AB}\in\SEP(A:B)}\widetilde{D}_{\alpha}(\rho_{AB}\Vert\sigma_{AB}).
	\end{equation}
	Now, since the function
	\begin{equation}
		(\sigma_{AB},\alpha)\mapsto \widetilde{D}_{\alpha}(\rho_{AB}\Vert\sigma_{AB})
	\end{equation}
	is continuous in the first argument, monotonically increasing in the second argument, and because the set of separable states is compact, we can use Theorem~\ref{thm-Mosonyi_minimax} to change the order of the supremum and infimum to obtain
	\begin{align}
		\lim_{\alpha\to\infty}\widetilde{E}_{\alpha}(A;B)_{\rho}&=\sup_{\alpha\in(1,\infty)}\inf_{\sigma_{AB}\in\SEP(A:B)}\widetilde{D}_{\alpha}(\rho_{AB}\Vert\sigma_{AB})\\
		&=\inf_{\sigma_{AB}\in\SEP(A:B)}\sup_{\alpha\in(1,\infty)}\widetilde{D}_{\alpha}(\rho_{AB}\Vert\sigma_{AB})\\
		&=\inf_{\sigma_{AB}\in\SEP(A:B)}D_{\max}(\rho_{AB}\Vert\sigma_{AB})\\
		&=E_{\max}(A;B)_{\rho}.
	\end{align}
	This completes the proof of \eqref{eq-sand_rel_ent_state_lim_aInf_pf}.
	
	Finally, we prove that
	\begin{equation}\label{eq-sand_rel_ent_chan_lim_aInf_pf}
		\lim_{\alpha\to\infty}\widetilde{E}_{\alpha}(\mathcal{N})=E_{\max}(\mathcal{N})=\sup_{\psi_{RA}}\inf_{\sigma_{RB}\in\SEP(R:B)}D_{\max}(\mathcal{N}_{A\to B}(\psi_{RA})\Vert\sigma_{RB}).
	\end{equation}
	The proof of 
	\begin{equation}
		\lim_{\alpha\to\infty}\widetilde{R}_{\alpha}(\mathcal{N})=R_{\max}(\mathcal{N})=\sup_{\psi_{RA}}\inf_{\sigma_{RB}\in\PPT'(R:B)}D_{\max}(\mathcal{N}_{A\to B}(\psi_{RA})\Vert\sigma_{RB})
	\end{equation}
	is analogous, and so we omit it.
	
	Using exactly the same argument as in the proof of \eqref{eq-sand_rel_ent_state_lim_aInf_pf} above, we obtain
	\begin{align}
		\lim_{\alpha\to\infty}\widetilde{E}_{\alpha}(\mathcal{N})&=\sup_{\alpha\in(1,\infty)}\sup_{\psi_{RA}}\inf_{\sigma_{RB}\in\SEP(R:B)}\widetilde{D}_{\alpha}(\mathcal{N}_{A\to B}(\psi_{RA})\Vert\sigma_{RB})\\
		&=\sup_{\psi_{RA}}\sup_{\alpha\in(1,\infty)}\inf_{\sigma_{RB}\in\SEP(R:B)}\widetilde{D}_{\alpha}(\mathcal{N}_{A\to B}(\psi_{RA})\Vert\sigma_{RB})\\
		&=\sup_{\psi_{RA}}\inf_{\sigma_{RB}\in\SEP(R:B)}\sup_{\alpha\in(1,\infty)}\widetilde{D}_{\alpha}(\mathcal{N}_{A\to B}(\psi_{RA})\Vert\sigma_{RB})\\
		&=\sup_{\psi_{RA}}\inf_{\sigma_{RB}\in\SEP(R:B)}D_{\max}(\mathcal{N}_{A\to B}(\psi_{RA})\Vert\sigma_{RB})\\
		&=E_{\max}(\mathcal{N}),
	\end{align}
	where in the third line we used Theorem~\ref{thm-Mosonyi_minimax} in order to exchange the infimum and supremum based on exactly the same arguments used in the proof of \eqref{eq-sand_rel_ent_state_lim_aInf_pf} above. This concludes the proof of \eqref{eq-sand_rel_ent_chan_lim_aInf_pf}.

\end{subappendices}

\part{Quantum Communication Protocols}[	We now begin our study of quantum communication protocols. In this part of the book, we focus on point-to-point communication protocols, most of which do not make use of feedback between the sender and receiver. The settings include classical communication, entanglement-assisted classical communication, entanglement distillation, quantum co\-mmunication, secret key distillation, and private communication. These point-to-point protocols are the most basic communication models in quantum information, and in many cases, they are relevant from a practical perspective. Furthermore, these protocols serve as benchmarks for the usefulness of the feedback-assisted protocols that we consider in Part~\ref{part-feedback}, in the sense that the optimal communication rates of any feedback-assist\-ed protocol should not be smaller than the corresponding point-to-point protocol, in order for the feedback-assisted protocol to be deemed useful or advantageous.]\label{part:q-comm-prots}

\chapter{Entanglement-Assisted Classical Communication}\label{chap-EA_capacity}

	The first communication task that we consider is entanglement-assi\-sted classical communication. In this scenario, Alice and Bob are allowed to share an unlimited amount of entanglement prior to communication, and the goal is for Alice to transmit the maximum possible amount of classical information over a given channel $\mathcal{N}$, by using this prior shared entanglement as a resource. We consider this particular setting before all other communication settings because, perhaps unexpectedly, the main information-theoretic results in this setting are much simpler than those in all other communication settings that we consider in this book. 
	
	Entanglement is a uniquely quantum phenomenon, and it is natural to ask, when communicating over quantum channels, whether it can be used to provide an advantage for sending classical information. The  super-dense coding protocol, described in Section~\ref{sec-super_dense_coding}, is an example of such an advantage. Recall that in this protocol, Alice and Bob share a pair of quantum systems in the maximally entangled state $\ket{\Phi^+}=\frac{1}{\sqrt{2}}(\ket{0,0}+\ket{1,1})$, and they are connected by a noiseless qubit channel. With this shared entanglement, along with only one use of the channel, Alice can communicate \textit{two} bits of classical information to Bob. In the case of qudits, using the maximally entangled state $\frac{1}{\sqrt{d}}\sum_{i=0}^{d-1}\ket{i,i}$, Alice can communication $2\log_2 d$ bits to Bob with only one use of a noiseless qudit quantum channel. Does this kind of advantage exist in general? Specifically, supposing that we allow Alice and Bob unlimited shared entanglement, what is the maximum amount of classical information that can be communicated over a given quantum channel $\mathcal{N}$?
	
	The answer to this question is provided by Theorem~\ref{thm-ea_classical_capacity}, which tells us that the entanglement-assisted classical capacity of a channel $\mathcal{N}$ is equal to the mutual information $I(\mathcal{N})$ of the channel (see \eqref{eq-mut_inf_chan}). The strength of this  result is that it holds for \textit{all} channels. Entanglement-assisted classical communication is one of the few scenarios in which such a profoundly simple statement---applying to all channels---can be made. Furthermore, the fact that the mutual information $I(\mathcal{N})$ is the optimal rate for entangl\-ement-assisted classical communication for all channels $\mathcal{N}$ makes this communication scenario formally analogous to communication over classical channels. Indeed, the famous result of Shannon from 1948 is that the capacity of a classical channel described by a conditional probability distribution $p_{Y|X}(y|x)$ with input and output random variables $X$ and $Y$, respectively, is equal to $\max_{p_X}I(X;Y)$, where $I(X;Y)$ is the mutual information between the random variables $X$ and $Y$ and the optimization is performed over all probability distributions $p_X$ corresponding to the input $X$. Entangl\-ement-assisted classical communication can thus be viewed as a ``natural'' analogue of classical communication in the quantum setting. 
	
	
	

\section{One-Shot Setting}\label{sec-eacc_one_shot}

\begin{figure}
		\centering
		\includegraphics[scale=0.8]{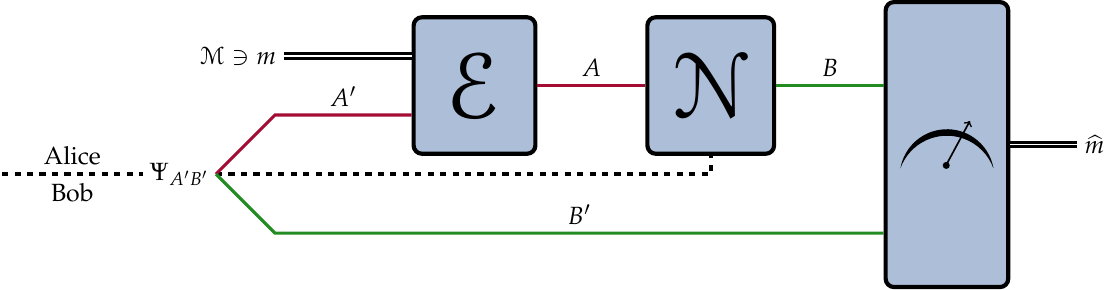}
		\caption{Depiction of a protocol for entanglement-assisted classical communication over one use of the quantum channel $\mathcal{N}$. Alice and Bob initially share a pair of quantum systems in the state $\Psi_{A'B'}$. Alice, who wishes to send a message $m$ selected from a set $\mathcal{M}$ of messages, first encodes the message into a quantum state on a quantum system $A$ by using an encoding channel $\mathcal{E}$. She then sends the quantum system $A$ through the channel $\mathcal{N}$. After Bob receives the system $B$, he performs a measurement on both of his systems $BB'$, using the outcome of the measurement to give an estimate $\widehat{m}$ of the message sent by Alice.}\label{fig-ea_classical_comm_oneshot}
	\end{figure}

	We begin by considering the one-shot setting for entanglement-assisted classical communication over $\mathcal{N}$, with such a protocol depicted in Figure~\ref{fig-ea_classical_comm_oneshot}. We call this the ``one-shot setting'' because the channel $\mathcal{N}$ is used only once. This is in contrast to the ``asymptotic setting'' that we consider in the next section, in which the channel may be used an arbitrarily large number of times. 
	
	The protocol depicted in Figure~\ref{fig-ea_classical_comm_oneshot} is defined by the four elements $(\mathcal{M},\allowbreak\Psi_{A'B'},\allowbreak\mathcal{E}_{M'A'\to A},\mathcal{D}_{BB'\to\widehat{M}})$, in which $\mathcal{M}$ is a message set, $\Psi_{A'B'}$ is an entangled state shared by Alice and Bob, $\mathcal{E}_{M'A'\to A}$ is an encoding channel, and $\mathcal{D}_{BB'\to\widehat{M}}$ is a decoding channel. The triple $(\Psi, \mathcal{E},\mathcal{D})$, consisting of the resource state and encoding and decoding channels, is called an entanglement-assisted code or, more simply, a code. In what follows, we employ the abbreviation
	\begin{equation}
	\mathcal{P}\equiv(\Psi, \mathcal{E},\mathcal{D}),
	\end{equation}
	where the notation $\mathcal{P}$ indicates \textit{protocol}.
	
	Given that there are $|\mathcal{M}|$ messages in the message set, it holds that each message can be uniquely associated with a bit string of size at least $\log_2|\mathcal{M}|$. The quantity $\log_2|\mathcal{M}|$ thus represents the number of bits that are communicated in the protocol. One of the goals of this section is to obtain upper and lower bounds, in terms of information measures for channels, on the maximum number $\log_2|\mathcal{M}|$ of bits that can be communicated in an entanglement-assisted classical communication protocol.
	
	The protocol proceeds as follows: let $p:\mathcal{M}\to[0,1]$ be a probability distribution over the message set. Alice starts by preparing two systems $M$ and $M'$ in the following classically correlated state:
	\begin{equation}\label{eq-ea_classical_comm_initial_state}
		\overline{\Phi}_{MM'}^p\coloneqq\sum_{m\in\mathcal{M}}p(m)\ket{m}\!\bra{m}_M\otimes\ket{m}\!\bra{m}_{M'}.
	\end{equation}
	Note that if Alice wishes to send a particular message $m$ deterministically, then she can choose the distribution $p$ to be the degenerate distribution, equal to one for $m$ and zero for all other messages.
	Alice and Bob share the state $\Psi_{A'B'}$ before communication begins, so that the global state shared between them is
	\begin{equation}
		\overline{\Phi}_{MM'}^p\otimes\Psi_{A'B'}.
	\end{equation}
	
	Alice then sends the $M'$ and $A'$ registers through the encoding channel $\mathcal{E}_{M'A'\to A}$. Due to the fact that the system $M'$ is classical, this encoding channel realizes a set $\{\mathcal{E}_{A'\to A}^m\}_{m\in\mathcal{M}}$ of quantum channels as follows:
	\begin{equation}
		\mathcal{E}_{A'\rightarrow A}^{m}(\tau_{A'})\coloneqq\mathcal{E}_{M'A'\to A}(\ket{m}\!\bra{m}_{M'}\otimes\tau_{A'})
	\end{equation}
	for every state $\tau_{A'}$. The global state after the encoding channel is therefore
	\begin{equation}
		\mathcal{E}_{M^{\prime}A^{\prime}\rightarrow A}(\overline{\Phi}^p_{MM^{\prime}}\otimes\Psi_{A^{\prime}B^{\prime}})=\sum_{m\in\mathcal{M}}p(m)\ket{m}\!\bra{m}_{M}\otimes\mathcal{E}_{A^{\prime}\rightarrow A}^{m}(\Psi_{A^{\prime}B^{\prime}}).
	\end{equation}

	Alice then transmits the $A$ system through the channel $\mathcal{N}_{A\rightarrow B}$, leading to the state
	\begin{equation}\label{eq-eac:state-after-channel}
		\begin{aligned}
		&(\mathcal{N}_{A\rightarrow B}\circ\mathcal{E}_{M^{\prime}A^{\prime}\rightarrow A})(\overline{\Phi}^p_{MM^{\prime}}\otimes\Psi_{A^{\prime}B^{\prime}})\\
		&\quad=\sum_{m\in\mathcal{M}}p(m)\ket{m}\!\bra{m}_{M}\otimes(\mathcal{N}_{A\rightarrow B}\circ\mathcal{E}_{A^{\prime}\rightarrow A}^{m})(\Psi_{A^{\prime}B^{\prime}}).\\
		&\quad=\sum_{m\in\mathcal{M}}p(m)\ket{m}\!\bra{m}_M\otimes\tau_{BB'}^m,
		\end{aligned}
	\end{equation}
	where
	\begin{equation}\label{eq-eac:state-after-channel_Bob_cond}
		\tau_{BB'}^m\coloneqq (\mathcal{N}_{A\to B}\circ\mathcal{E}_{A'\to A}^m)(\Psi_{A'B'})\quad\forall~m\in\mathcal{M}.
	\end{equation}
	
	Bob, whose task is to determine which message Alice sent, applies a decoding channel $\mathcal{D}_{BB'\to\widehat{M}}$ on his system $B'$ and the system $B$ received through the channel $\mathcal{N}$. The decoding channel is a quantum-classical channel (Definition~\ref{def-qc_channel}) associated with a POVM $\{\Lambda_{BB'}^m\}_{m\in\mathcal{M}}$, so that
	\begin{equation}
		\mathcal{D}_{BB'\rightarrow\widehat{M}}(\tau_{BB'}^m)\coloneqq\sum_{\widehat{m}\in\mathcal{M}}\Tr[\Lambda_{BB'}^{\widehat{m}}\tau_{BB'}^m]\ket{\widehat{m}}\!\bra{\widehat{m}}_{\widehat{M}},
	\end{equation}
	for all $m\in\mathcal{M}$. The global state in \eqref{eq-eac:state-after-channel} thus becomes
	\begin{align}
		\omega_{M\widehat{M}}^p&\coloneqq(\mathcal{D}_{BB'\rightarrow\widehat{M}}\circ\mathcal{N}_{A\rightarrow B}\circ\mathcal{E}_{M'A'\rightarrow A})(\overline{\Phi}^p_{MM'}\otimes\Psi_{A'B'})\label{eq-eac:end-EAC-state}\\
		&=\sum_{m,\widehat{m}\in\mathcal{M}}p(m)\ket{m}\!\bra{m}_M\otimes\Tr[\Lambda_{BB'}^{\widehat{m}}\mathcal{N}_{A\rightarrow B}(\mathcal{E}_{A'\rightarrow A}^{m}(\Psi_{A'B'}))]\ket{\widehat{m}}\!\bra{\widehat{m}}_{\widehat{M}}.\label{eq-eac:end-EAC-state_2}
	\end{align}
	
	The final decoding measurement by Bob induces the conditional probability distribution $q:\mathcal{M}\times\mathcal{M}\to[0,1]$ defined by
	\begin{align}
		q(\widehat{m}|m) & \coloneqq\Pr[\widehat{M}=\widehat{m}|M=m]\\
		&=\Tr[\Lambda_{BB'}^{\widehat{m}}\mathcal{N}_{A\rightarrow B}(\mathcal{E}_{A'\rightarrow A}^{m}(\Psi_{A'B'}))].
	\end{align}
	Bob's strategy is such that if the outcome $\widehat{m}$ occurs from his measurement, then he declares that the message sent was $\widehat{m}$.
	
	The probability that Bob correctly identifies a given message $m$ is then given by $q(m|m)$. The \textit{message error probability of the code $\mathcal{P}\equiv (\Psi, \mathcal{E},\mathcal{D})$ and message $m$} is then given by
	\begin{equation}\label{eq-eac-mess_error_prob}
		\begin{aligned}
		p_{\text{err}}(m,\mathcal{P};\mathcal{N})&\coloneqq 1-q(m|m)\\
		&=\Tr[(\mathbbm{1}_{BB'}-\Lambda_{BB'}^m)\mathcal{N}_{A\to B}(\mathcal{E}_{A'\to A}^m(\Psi_{A'B'}))]\\
		&=\sum_{\widehat{m}\in\mathcal{M}\setminus\{m\}}q(\widehat{m}|m).
		\end{aligned}
	\end{equation}
	The \textit{average error probability of the code} is
	\begin{equation}\label{eq-eac-avg_error_prob}
		\begin{aligned}
		\overline{p}_{\text{err}}(\mathcal{P};p,\mathcal{N})&\coloneqq\sum_{m\in\mathcal{M}}p(m)p_{\text{err}}(m;\mathcal{P})\\
		&=\sum_{m\in\mathcal{M}}p(m)(1-q(m|m))\\
		&=\sum_{m\in\mathcal{M}}\sum_{\widehat{m}\in\mathcal{M}\setminus\{m\}}p(m)q(\widehat{m}|m).
		\end{aligned}
	\end{equation}
	The \textit{maximal error probability of the code} is
	\begin{equation}\label{eq-eac-maximal_error_prob}
		p_{\text{err}}^*(\mathcal{P};\mathcal{N})\coloneqq\max_{m\in\mathcal{M}}p_{\text{err}}(m,\mathcal{P};\mathcal{N}).
	\end{equation}
	Each of these three error probabilities can be used to assess the \textit{reliability} of the protocol, i.e., how well the encoding and decoding allow Alice to transmit her message to Bob.
	
	\begin{definition}{$\boldsymbol{(|\mathcal{M}|,\varepsilon)}$ Entanglement-Assisted Classical Communication Protocol}{def-Me-eacc-protocol}
		Let $(\mathcal{M},\Psi_{A'B'},\mathcal{E}_{M'A'\to A},\mathcal{D}_{BB'\to\widehat{M}})$ be the elements of an entanglement-assisted classical communication protocol over the channel $\mathcal{N}_{A\to B}$. The protocol is called an \textit{$(|\mathcal{M}|,\varepsilon)$ protocol}, with $\varepsilon\in[0,1]$, if $p_{\text{err}}^*(\mathcal{P};\mathcal{N})\leq\varepsilon$.
	\end{definition}
	
	\begin{Lemma}{lem:EA-comm:error-criterion-analysis}
	The following equalities hold
	\begin{align}
	 \overline{p}_{\text{err}}(\mathcal{P};p,\mathcal{N}) &  = \frac{1}{2}\norm{\overline{\Phi}_{MM'}^p-\omega_{M\widehat{M}}^p}_1,\\
	\label{eq-ea_classical_comm_reliability}
		p_{\text{err}}^*(\mathcal{P};\mathcal{N}) & = \max_{p:\mathcal{M}\to[0,1]}\frac{1}{2}\norm{\overline{\Phi}_{MM'}^p-\omega_{M\widehat{M}}^p}_1,
	\end{align}
	where $\overline{\Phi}_{MM'}^p$ and $\omega_{M\widehat{M}}^p$ are defined in \eqref{eq-ea_classical_comm_initial_state} and \eqref{eq-eac:end-EAC-state}, respectively. Thus, the error criterion $p_{\text{err}}^*(\mathcal{P};\mathcal{N})\leq\varepsilon$ is equivalent to $\max_{p:\mathcal{M}\to[0,1]}\frac{1}{2}\norm{\overline{\Phi}_{MM'}^p-\omega_{M\widehat{M}}^p}_1\leq\varepsilon$.
	\end{Lemma}
	
	\begin{remark}
	The final criterion above states that the normalized trace distance between the initial and final states of the protocol, maximized over all possible prior probability distributions, does not exceed $\varepsilon$. 
	\end{remark}
	
	\begin{Proof}
	To see this, let us first note that the normalized trace distance in \eqref{eq-ea_classical_comm_reliability} is equal to the average error probability of the code. Indeed,
	\begin{align}
		&\frac{1}{2}\norm{\overline{\Phi}_{MM'}^p-\omega_{M\widehat{M}}^p}_1\nonumber\\
		&\quad=\frac{1}{2}\left\lVert\sum_{m\in\mathcal{M}}p(m)\ket{m,m}\!\bra{m,m}_{MM'}\right.\nonumber\\
		&\qquad\qquad\qquad\left.-\sum_{m,\widehat{m}\in\mathcal{M}}p(m)q(\widehat{m}|m)\ket{m,\widehat{m}}\!\bra{m,\widehat{m}}_{M\widehat{M}}\right\rVert_1\label{eq-eacc_trace_dist_avg_error_1}\\
		&\quad=\frac{1}{2}\left\lVert\sum_{m\in\mathcal{M}}p(m)\ket{m}\!\bra{m}_M\right.\nonumber\\
		&\qquad\qquad\qquad\left.\otimes\left(\ket{m}\!\bra{m}_{M'}-\sum_{\widehat{m}\in\mathcal{M}}q(\widehat{m}|m)\ket{\widehat{m}}\!\bra{\widehat{m}}_{\widehat{M}}\right)\right\rVert_1\\
		&\quad=\frac{1}{2}\sum_{m\in\mathcal{M}}p(m)\norm{\ket{m}\!\bra{m}_{M'}-\sum_{\widehat{m}\in\mathcal{M}}q(\widehat{m}|m)\ket{\widehat{m}}\!\bra{\widehat{m}}_{\widehat{M}}}_1\\
		&\quad=\frac{1}{2}\sum_{m\in\mathcal{M}}p(m)\left\lVert(1-q(m|m))\ket{m}\!\bra{m}\right.\nonumber\\
		&\qquad\qquad\qquad\left.-\sum_{\widehat{m}\in\mathcal{M}\setminus\{m\}}q(\widehat{m}|m)\ket{\widehat{m}}\!\bra{\widehat{m}}_{\widehat{M}}\right\rVert_1\\
		&\quad=\frac{1}{2}\sum_{m\in\mathcal{M}}p(m)\left((1-q(m|m))+\sum_{\widehat{m}\in\mathcal{M}\setminus\{m\}}q(\widehat{m}|m)\right)\\
		&\quad=\frac{1}{2}\underbrace{\sum_{m\in\mathcal{M}}p(m)(1-q(m|m))}_{\overline{p}_{\text{err}}(\mathcal{P};p,\mathcal{N})}+\frac{1}{2}\underbrace{\sum_{m\in\mathcal{M}}\sum_{\widehat{m}\in\mathcal{M}\setminus\{m\}}p(m)q(\widehat{m}|m)}_{\overline{p}_{\text{err}}(\mathcal{P};p,\mathcal{N})}\\
		&\quad=\overline{p}_{\text{err}}(\mathcal{P};p,\mathcal{N})\label{eq-eacc_trace_dist_avg_error},
	\end{align}
	where the third and fifth equalities follow from \eqref{eq-trace_norm_blkdiag} with $\alpha=1$. Then, if $m^*\in\mathcal{M}$ is the message attaining the maximum error probability $p_{\text{err}}^*$,  let $\tilde{p}:\mathcal{M}\to[0,1]$ be the probability distribution such that $\tilde{p}(m^*)=1$ and $\tilde{p}(m)=0$ for all $m\neq m^*$. Using this probability distribution, we obtain
	\begin{align}
		p_{\text{err}}^*(\mathcal{P};\mathcal{N})&=\max_{m\in\mathcal{M}}p_{\text{err}}(m,\mathcal{P};\mathcal{N}) \label{eq-EAC:connect-err-probs-1}\\
		&= \sum_{m\in\mathcal{M}}\tilde{p}(m)p_{\text{err}}(m,\mathcal{P};\mathcal{N})\\
		&=\overline{p}_{\text{err}}(\mathcal{P};\tilde{p},\mathcal{N})\\
		&=\frac{1}{2}\norm{\overline{\Phi}_{MM'}^{\tilde{p}}-\omega_{M\widehat{M}}^{\tilde{p}}}_1\\
		&\leq\max_{p:\mathcal{M}\to[0,1]}\frac{1}{2}\norm{\overline{\Phi}_{MM'}^p-\omega_{M\widehat{M}}^p}_1.
	\end{align}
	Furthermore, letting $p^*$ be the distribution attaining the maximum average error probability, we find that
	\begin{align}
		\max_{p:\mathcal{M}\to[0,1]}\frac{1}{2}\norm{\overline{\Phi}_{MM'}^p-\omega_{M\widehat{M}}^p}_1&=\overline{p}_{\text{err}}(\mathcal{P};p^*,\mathcal{N})\\
		&=\sum_{m\in\mathcal{M}}p^*(m)p_{\text{err}}(m,\mathcal{P};\mathcal{N})\\
		&\leq\sum_{m\in\mathcal{M}}p^*(m)p_{\text{err}}^*(\mathcal{P};\mathcal{N})\\
		&=p_{\text{err}}^*(\mathcal{P};\mathcal{N}),
	\end{align}
	where the inequality follows from the fact that
	\begin{equation}
		p_{\text{err}}(m,\mathcal{P};\mathcal{N})\leq \max_{m\in\mathcal{M}}p_{\text{err}}(m,\mathcal{P};\mathcal{N})=p_{\text{err}}^*(\mathcal{P};\mathcal{N})
		\label{eq-EAC:connect-err-probs-last}
	\end{equation}
	for all $m\in\mathcal{M}$. So we find that
	\begin{align}
		p_{\text{err}}^*(\mathcal{P};\mathcal{N}) &= \max_{p:\mathcal{M}\to[0,1]}\overline{p}_{\text{err}}(\mathcal{P};p)\\
		&=\max_{p:\mathcal{M}\to[0,1]}\frac{1}{2}\norm{\overline{\Phi}_{MM'}^p-\omega_{M\widehat{M}}^p}_1.\label{eq-eacc_max_err_prob_trace_distance}
	\end{align}
	This concludes the proof.
	\end{Proof}
	
	Another way to define the error criterion of an $(|\mathcal{M}|,\varepsilon)$ protocol, which is equivalent to the average error probability, is through what is called the \textit{comparator test}. The comparator test is a measurement defined by the two-element POVM $\{\Pi_{M\widehat{M}},\mathbbm{1}-\Pi_{M\widehat{M}}\}$, where $\Pi_{M\widehat{M}}$ is the projection defined as
	\begin{equation}\label{eq-eacc_comparator_test}
		\Pi_{M\widehat{M}}\coloneqq\sum_{m\in\mathcal{M}}\ket{m}\!\bra{m}_M\otimes\ket{m}\!\bra{m}_{\widehat{M}}.
	\end{equation}
	Note that $\Tr[\Pi_{M\widehat{M}}\omega_{M\widehat{M}}^p]$ is equal to the probability that the classical registers $M$ and $\widehat{M}$ in the state $\omega_{M\widehat{M}}$ have the same values. In particular, observe that
	\begin{align}
		&\Tr\!\left[\Pi_{M\widehat{M}}\omega_{M\widehat{M}}^p\right]\nonumber\\
		&\quad =\Tr\!\left[\left(\sum_{m\in\mathcal{M}}\ket{m,m}\!\bra{m,m}_{M\widehat{M}}\right)\right.\nonumber\\
		&\qquad\qquad\times \left.\left(\sum_{m',m\in\mathcal{M}}p(m')q(\widehat{m}|m')\ket{m',\widehat{m}}\!\bra{m',\widehat{m}}_{M\widehat{M}}\right)\right]\label{eq-ea_classical_comm_comparator_succ_prob_1}\\
		&\quad =\sum_{m,m',\widehat{m}\in\mathcal{M}}p(m')q(\widehat{m}|m')\delta_{m,m'}\delta_{m,\widehat{m}}\label{eq-ea_classical_comm_comparator_succ_prob_2}\\
		&\quad =\sum_{m\in\mathcal{M}}p(m)q(m|m)\label{eq-ea_classical_comm_comparator_succ_prob_3}\\
		&\quad =1-\overline{p}_{\text{err}}(\mathcal{P};p,\mathcal{N}).\label{eq-ea_classical_comm_comparator_succ_prob}
	\end{align}
	We can interpret this expression as the \textit{average success probability of the code} $\mathcal{P} \equiv (\Psi, \mathcal{E},\mathcal{D})$, which we denote by
	\begin{equation}
	\overline{p}_{\text{succ}}(\mathcal{P};p,\mathcal{N}) \coloneqq 1-\overline{p}_{\text{err}}(\mathcal{P};p,\mathcal{N}).
	\end{equation}

	As mentioned at the beginning of this chapter, our goal is to bound (from above and below) the maximum number $\log_2|\mathcal{M}|$ of transmitted bits in any entanglement-assisted classical communication protocol over $\mathcal{N}$. Given an error probability tolerance of $\varepsilon$, we call the maximum bits of transmitted bits the \textit{one-shot entanglement-assisted classical capacity} of $\mathcal{N}$.
	
	\begin{definition}{One-Shot Entanglement-Assisted Classical Capacity}{def-ea_classical_comm_one_shot_capacity}
		Given a quantum channel $\mathcal{N}_{A\to B}$ and $\varepsilon\in[0,1]$, the \textit{one-shot $\varepsilon$-error entanglement-assisted classical capacity of $\mathcal{N}$}, denoted by $C_{\operatorname{EA}}^{\varepsilon}(\mathcal{N})$, is defined to be the maximum number $\log_2|\mathcal{M}|$ of transmitted bits among all $(|\mathcal{M}|,\varepsilon)$ entanglement-assisted classical communication protocols over $\mathcal{N}$. In other words,
		\begin{equation}
			C_{\operatorname{EA}}^{\varepsilon}(\mathcal{N})\coloneqq \sup_{(\mathcal{M},\Psi,\mathcal{E},\mathcal{D})}\{\log_2|\mathcal{M}|:p_{\text{err}}^*((\Psi,\mathcal{E},\mathcal{D});\mathcal{N})\leq\varepsilon\},
			\label{eq:EA-comm:ea-capacity}
		\end{equation}
		where the optimization is with respect to all protocols $(\mathcal{M},\Psi_{A'B'},\mathcal{E}_{M'A'\to A},\mathcal{D}_{BB'\to\widehat{M}})$ such that $d_{M'}=d_{\widehat{M}}=|\mathcal{M}|$.
	\end{definition}
	
	In addition to finding, for a given $\varepsilon\in[0,1]$, the maximum number of transmitted bits among all $(|\mathcal{M}|,\varepsilon)$ classical communication protocols over $\mathcal{N}_{A\to B}$, we can consider the following complementary problem: for a given number of messages $|\mathcal{M}|$, find the smallest possible error among all $(|\mathcal{M}|,\varepsilon)$ entanglement-assisted classical communication protocols, which we denote by $\varepsilon_{\operatorname{EA}}^*(|\mathcal{M}|;\mathcal{N})$. In other words, the problem is to determine
	\begin{equation}\label{eq-EA_comm_opt_error_one_shot}
		\varepsilon_{\operatorname{EA}}^*(|\mathcal{M}|;\mathcal{N})\coloneqq\inf_{(\Psi,\mathcal{E},\mathcal{D})}\{p_{\text{err}}^*((\Psi,\mathcal{E},\mathcal{D});\mathcal{N}): d_{M'}=d_{\widehat{M}}=|\mathcal{M}|\},
	\end{equation}
	where the optimization is over every state $\Psi_{A'B'}$, encoding channel $\mathcal{E}_{M'A'\to A}$, and decoding channel $\mathcal{D}_{BB'\to\widehat{M}}$, such that $d_{M'}=d_{\widehat{M}}=|\mathcal{M}|$. In this chapter, we focus primarily on the problem of optimizing the number of transmitted bits rather than the error, and so our primary quantity of interest is the one-shot capacity $C_{\operatorname{EA}}^{\varepsilon}(\mathcal{N})$.

\subsection{Protocol Over a Useless Channel}\label{sec-EAC:useless-protocol}

\begin{figure}
		\centering
		\includegraphics[scale=0.8]{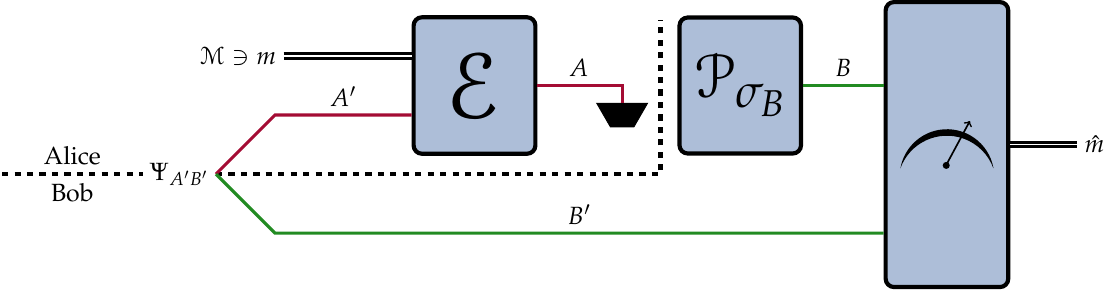}
		\caption{Depiction of a protocol that is useless for entanglement-assisted classical communication. The state encoding the message $m$ via $\mathcal{E}$ is discarded and replaced by an arbitrary (but fixed) state $\sigma_{B}$.}\label{fig-ea_classical_comm_useless_oneshot}
	\end{figure}

Our first goal is to obtain an upper bound on the one-shot entanglement-assisted classical capacity defined in \eqref{eq:EA-comm:ea-capacity}. To do so, along with the entanglement-assisted classical communication protocol over the actual channel $\mathcal{N}$ described above, we also consider the same protocol but over the \textit{useless} channel depicted in Figure~\ref{fig-ea_classical_comm_useless_oneshot}. This useless channel  discards the quantum state encoded with the message and replaces it with some arbitrary (but fixed) state $\sigma_B$. This replacement channel is useless for communication because the state $\sigma_B$ does not contain any information about the message $m$. Intuitively, we can say that such a channel corresponds to ``cutting the communication line.'' As we show in Lemma~\ref{lem-eac-meta_conv}, comparing this protocol over the useless channel with the actual protocol allows us to obtain an upper bound on the quantity $\log_2|\mathcal{M}|$, which we recall represents the number of bits that are transmitted over the channel.
	
	The definition of the useless channel implies that, for every message $m\in\mathcal{M}$,
	\begin{equation}\label{eq-eacc_useless_channel}
		\mathcal{E}_{A'\to A}^m(\Psi_{A'B'})\mapsto \mathcal{P}_{\sigma_B}\circ\Tr_A\circ\mathcal{E}_{A'\to A}^m(\Psi_{A'B'})=\sigma_B\otimes\Psi_{B'},
	\end{equation}
	where $\Psi_{B'}\coloneqq\Tr_{A'}[\Psi_{A'B'}]$. Making use of the definition of the replacement channel $\mathcal{R}_{A\to B}^{\sigma_B}$ in Definition~\ref{def-replace_channel}, we can write this as
	\begin{equation}
		\mathcal{E}_{A'\to A}^m(\Psi_{A'B'})\mapsto \mathcal{R}_{A\to B}^{\sigma_B}(\mathcal{E}_{A'\to A}(\Psi_{A'B'})).
	\end{equation}
	The state at the end of the protocol over the useless channel is then
	\begin{equation}\label{eq-eacc_useless_final_state}
		\begin{aligned}
		\tau_{M\widehat{M}}^p&\coloneqq\sum_{m,\widehat{m}\in\mathcal{M}}p(m)\Tr[\Lambda_{BB'}^{\widehat{m}}(\sigma_B\otimes\Psi_{B'})]\ket{m}\!\bra{m}_M\otimes\ket{\widehat{m}}\!\bra{\widehat{m}}_{\widehat{M}}\\
		&=\pi_M^p\otimes \sum_{\widehat{m}\in\mathcal{M}}\Tr[\Lambda_{BB'}^{\widehat{m}}(\sigma_B\otimes\Psi_{B'})]\ket{\widehat{m}}\!\bra{\widehat{m}}_{\widehat{M}},
		\end{aligned}
	\end{equation}
	where
	\begin{equation}
		\pi_M^p\coloneqq\sum_{m\in\mathcal{M}}p(m)\ket{m}\!\bra{m}_M.
	\end{equation}
	Now, recall from \eqref{eq-eac:end-EAC-state_2} that the state $\omega_{M\widehat{M}}^p$ at the end of the actual protocol over the channel $\mathcal{N}$ is given by
	\begin{equation}
		\omega_{M\widehat{M}}^p=\sum_{m,\widehat{m}\in\mathcal{M}}p(m)\ket{m}\!\bra{m}_M\otimes\Tr[\Lambda_{BB'}^{\widehat{m}}\mathcal{N}_{A\rightarrow B}(\mathcal{E}_{A'\rightarrow A}^{m}(\Psi_{A'B'}))]\ket{\widehat{m}}\!\bra{\widehat{m}}_{\widehat{M}}.
	\end{equation}
	It is helpful in what follows to let
	\begin{equation}\label{eq-ea_classical_comm_initial_state_uniform}
			\overline{\Phi}_{MM'}\coloneqq\frac{1}{|\mathcal{M}|}\sum_{m\in\mathcal{M}}\ket{m}\!\bra{m}_M\otimes\ket{m}\!\bra{m}_{M'}
	\end{equation}
	be the state in \eqref{eq-ea_classical_comm_initial_state} in which the probability distribution $p$ is the uniform distribution over $\mathcal{M}$.
	
	We now state a lemma that is helpful for placing an upper bound on the number $\log_2|\mathcal{M}|$ of bits communicated in an entanglement-assisted classical communication protocol. This lemma can also be used for the same purpose for unassisted classical communication protocols, as discussed in the next chapter.
	
	\begin{Lemma}{lem-eac-meta_conv}
		Let $\overline{\Phi}_{MM'}$ be the state defined in \eqref{eq-ea_classical_comm_initial_state_uniform}, and let $\omega_{MM'}$ be a state on the two classical registers $M$ and $M'$ such that $\omega_M=\Tr_{M'}[\omega_{MM'}]=\pi_M=\frac{\mathbbm{1}_M}{|\mathcal{M}|}$. If the probability $\Tr[\Pi_{MM'}\omega_{MM'}]$ that the state $\omega_{MM'}$ passes the comparator test defined by the POVM $\{\Pi_{MM'},\mathbbm{1}-\Pi_{MM'}\}$, where $\Pi_{MM'}$ is the projection defined in \eqref{eq-eacc_comparator_test}, satisfies
		\begin{equation}
			\Tr[\Pi_{MM'}\omega_{MM'}]\geq 1-\varepsilon,
		\end{equation}
		for some $\varepsilon\in[0,1]$,
		then
		\begin{equation}\label{eq-eqcc_meta_conv_upper_bound}
			\log_2|\mathcal{M}|\leq I_{H}^{\varepsilon}(M;M')_{\omega},
		\end{equation}
		where the $\varepsilon$-hypothesis testing mutual information $I_{H}^{\varepsilon}(M;M')_{\omega}$ is defined in \eqref{eq-hypo_testing_mutual_inf}.
	\end{Lemma}

	\begin{Proof}
		By assumption, we have that
		\begin{equation}\label{eq-ea_classical_comm_reliability_uniform_alt}
			\Tr[\Pi_{MM'}\omega_{MM'}]\geq 1-\varepsilon,
		\end{equation}
		Now, consider a state $\tau_{MM'}$ of the form $\tau_{MM'}=\omega_{M}\otimes\sigma_{M'}=\pi_{M}\otimes\sigma_{M'}$, where $\sigma_{M'}$ is some state.\footnote{Note that the state in \eqref{eq-eacc_useless_final_state} at the end of the entanglement-assisted classical communication protocol over the useless channel has precisely this form (when $p$ is taken to be the uniform distribution over the message set $\mathcal{M}$).} Then, 
		\begin{align}
			\Tr[\Pi_{MM'}\tau_{MM'}]&=\Tr[\Pi_{MM'}(\pi_M\otimes\sigma_{M'})]\label{lem-eac-meta_conv_pf1}\\
			&=\frac{1}{|\mathcal{M}|}\Tr[\Pi_{MM'}(\mathbbm{1}_M\otimes\sigma_{M'})]\label{lem-eac-meta_conv_pf2}\\
			&=\frac{1}{|\mathcal{M}|}\Tr[\Tr_M[\Pi_{MM'}]\sigma_{M'}]\label{lem-eac-meta_conv_pf3}\\
			&=\frac{1}{|\mathcal{M}|}\Tr[\mathbbm{1}_{M'}]\sigma_{M'}]\label{lem-eac-meta_conv_pf3-1}\\			
			&=\frac{1}{|\mathcal{M}|}\label{lem-eac-meta_conv_pf4},
		\end{align}
		We thus obtain
		\begin{align}
			\log_2|\mathcal{M}|&=-\log_2\Tr[\Pi_{MM'}\tau_{MM'}]\\
			&\leq D_H^{\varepsilon}(\omega_{MM'}\Vert\tau_{MM'})\\
			&=D_H^{\varepsilon}(\omega_{MM'}\Vert\omega_M\otimes\sigma_{M'}),\label{eq-eacc_strong_conv_lem_pf_2}
		\end{align}
		where the inequality follows from the definition of the hypothesis testing relative entropy in \eqref{eq:QEI:def-hypo-test-rel-ent} (i.e., $\Pi_{MM'}$ is a particular measurement operator satisfying \eqref{eq-ea_classical_comm_reliability_uniform_alt}, but $D_H^{\varepsilon}(\omega_{MM'}\Vert\tau_{MM'})$ involves an optimization over all such operators).
		Since the state $\sigma_{M'}$ is arbitrary, we conclude that
		\begin{equation}
			\log_2|\mathcal{M}|\leq \inf_{\sigma_{M'}}D_H^{\varepsilon}(\omega_{MM'}\Vert\omega_M\otimes\sigma_{M'})=I_H^{\varepsilon}(M;M')_{\omega},
		\end{equation}
		which is \eqref{eq-eqcc_meta_conv_upper_bound}, as required.
	\end{Proof}
	
	The right-hand side of \eqref{eq-eqcc_meta_conv_upper_bound} is an upper bound on the number $\log_2|\mathcal{M}|$ of bits communicated using an $(|\mathcal{M}|,\varepsilon)$ entanglement-assisted classical communication protocol over the channel $\mathcal{N}$. Indeed, since the error criterion $p_{\text{err}}^*(\mathcal{P})\leq\varepsilon$ holds by definition of an $(|\mathcal{M}|,\varepsilon)$ protocol, using \eqref{eq-eacc_max_err_prob_trace_distance} and \eqref{eq-eacc_trace_dist_avg_error} we obtain
	\begin{align}
		\overline{p}_{\text{err}}(\mathcal{P};p,\mathcal{N})&
		\leq \max_{p:\mathcal{M}\to[0,1]}\overline{p}_{\text{err}}(\mathcal{P};p,\mathcal{N})\label{eq-ea_classical_comm_arb_to_uniform_reduction_1}\\
		&=\max_{p:\mathcal{M}\to[0,1]}\frac{1}{2}\norm{\overline{\Phi}_{MM'}^p-\omega_{M\widehat{M}}^p}_1\label{eq-ea_classical_comm_arb_to_uniform_reduction_2}\\
		&=p_{\text{err}}^*(\mathcal{P};\mathcal{N})\\
		&\leq\varepsilon \label{eq-ea_classical_comm_arb_to_uniform_reduction}
	\end{align}
	for every probability distribution $p$ on $\mathcal{M}$. In particular, the  inequality above holds with $p$ being the uniform distribution on $\mathcal{M}$, so that $p(m)=\frac{1}{|\mathcal{M}|}$ for all $m\in\mathcal{M}$. Let us define the state 
	\begin{equation}\label{eq-ea_classical_comm_final_state_uniform}
		\omega_{M\widehat{M}}\coloneqq\frac{1}{|\mathcal{M}|}\sum_{m,\widehat{m}\in\mathcal{M}}\Tr[\Lambda_{BB'}^{\widehat{m}}\mathcal{N}_{A\to B}(\mathcal{E}_{A'\to A}^m(\Psi_{A'B'}))]\ket{m,\widehat{m}}\!\bra{m,\widehat{m}}_{M\widehat{M}},
	\end{equation}
	which is the state $\omega_{M\widehat{M}}^p$ with $p$ being the uniform distribution over $\mathcal{M}$. Observe that 
	\begin{align}
		\Tr_{\widehat{M}}[\omega_{M\widehat{M}}]&=\frac{1}{|\mathcal{M}|}\sum_{m\in\mathcal{M}}\Tr\!\left[\left(\sum_{\widehat{m}\in\mathcal{M}}\Lambda_{BB'}^{\widehat{m}}\right)\mathcal{N}_{A\to B}(\mathcal{E}_{A'\to A}^m(\Psi_{A'B'}))\right]\ket{m}\!\bra{m}_M\\
		&=\frac{1}{|\mathcal{M}|}\sum_{m\in\mathcal{M}}\Tr[\mathcal{N}_{A\to B}(\mathcal{E}_{A'\to A}^m(\Psi_{A'B'}))]\ket{m}\!\bra{m}_{M}\\
		&=\pi_M,
	\end{align}
	where  the last equality follows because the channels $\mathcal{N}_{A\to B}$ and $\mathcal{E}_{A'\to A}^m$ are trace preserving. Finally, since the probability of passing the comparator test is given by \eqref{eq-ea_classical_comm_comparator_succ_prob}, i.e., 
	\begin{equation}
		\Tr\!\left[\Pi_{M\widehat{M}}\omega_{M\widehat{M}}^p\right]=1-\overline{p}_{\text{err}}(\mathcal{P};p,\mathcal{N})
	\end{equation}
	for every probability distribution $p$, we find that $\Tr[\Pi_{M\widehat{M}}\omega_{M\widehat{M}}]\geq 1-\varepsilon$. The state $\omega_{M\widehat{M}}$ thus satisfies the condition of Lemma~\ref{lem-eac-meta_conv}. We conclude that
	\begin{equation}\label{eq-eqcc_meta_conv_upper_bound_2}
		\log_2|\mathcal{M}|\leq I_H^{\varepsilon}(M;\widehat{M})_{\omega}
	\end{equation}
	for every $(|\mathcal{M}|,\varepsilon)$ entanglement-assisted classical communication protocol.
	
	Recall from Section~\ref{sec-hyp_test_rel_ent} that the hypothesis testing relative entropy has an operational meaning as the optimal type-II error exponent in asymmetric hypothesis testing. The quantity $D_H^{\varepsilon}(\omega_{M\widehat{M}}\Vert\omega_M\otimes\sigma_{\widehat{M}})$ thus represents the optimal type-II error exponent, subject to the upper bound of $\varepsilon$ on the type-I error exponent, for distinguishing between the state resulting from the actual entanglement-assisted classical communication protocol over $\mathcal{N}$ and the state resulting from an entanglement-assisted classical communication protocol over a useless channel, which discards the state encoded with the message and replaces it with the state $\sigma_{\widehat{M}}$. By taking an infimum over all states $\sigma_{\widehat{M}}$, the quantity $I_H^{\varepsilon}(M;\widehat{M})_\omega$ represents the smallest possible minimum type-II error exponent. The bound in \eqref{eq-eqcc_meta_conv_upper_bound_2} thus establishes a close link between the tasks of reliable communication and hypothesis testing.
	
	Given a particular choice of the encoding and decoding channels, as well as a particular choice of the shared state $\Psi_{A'B'}$, if $p_{\text{err}}^*(\mathcal{P};\mathcal{N})\leq\varepsilon$, then the quantity $I_H^{\varepsilon}(M;\widehat{M})_{\omega}$ in \eqref{eq-eqcc_meta_conv_upper_bound_2} is an upper bound on the maximum number of bits that can be transmitted over the channel $\mathcal{N}$. The optimal value of this upper bound can be realized by finding the state $\sigma_{\widehat{M}}$ defining the useless channel that optimizes the quantity $I_H^{\varepsilon}(M;\widehat{M})_{\omega}$ in addition to the measurement that achieves the $\varepsilon$-hypothesis testing relative entropy. Importantly, a different choice of encoding, decoding, and of the state $\Psi_{A'B'}$ produces a different value for this upper bound. We would thus like to find an upper bound that applies regardless of which specific protocol is chosen. In other words, we would like an upper bound that is a function of the channel $\mathcal{N}$ only, and this is the topic of the next section.

\subsection{Upper Bound on the Number of Transmitted Bits}

	We now give a general upper bound on the number of transmitted bits possible for an arbitrary one-shot entanglement-assisted classical communication protocol for a channel $\mathcal{N}$. This result is stated in Theorem~\ref{cor-eacc_meta_str_weak_conv}. The upper bound obtained therein holds independently of the encoding and decoding channels used in the protocol and depends only on the given communication channel $\mathcal{N}$.
	
	Let us start with an arbitrary $(|\mathcal{M}|,\varepsilon)$ entanglement-assisted classical communication protocol over $\mathcal{N}$ corresponding to, as described at the beginn\-ing of this chapter, a message set $\mathcal{M}$, a prior shared entangled state $\Psi_{A'B'}$, an encoding channel $\mathcal{E}$, and a decoding channel $\mathcal{D}$. The error criterion $p_{\text{err}}^*(\mathcal{P};\mathcal{N})\leq\varepsilon$ holds by the definition of an $(|\mathcal{M}|,\varepsilon)$ protocol. Then, by the arguments at the end of the previous section, Lemma~\ref{lem-eac-meta_conv} implies that the inequality $\log_2|\mathcal{M}|\leq I_H^{\varepsilon}(M;\widehat{M})_{\omega}$ holds. Using this bound on the number $\log_2|\mathcal{M}|$ of transmitted bits, we obtain the following result:
	
	\begin{proposition*}{Upper Bound on One-Shot Entanglement-Assisted Classical Capacity}{prop-eac:one-shot-bound-meta}
		Let $\mathcal{N}_{A\rightarrow B}$ be a quantum channel. For every $(|\mathcal{M}|,\varepsilon)$ entanglement-assisted classical communication protocol over $\mathcal{N}_{A\rightarrow B}$, with $\varepsilon\in[0,1]$, the number of bits transmitted over $\mathcal{N}$ is bounded from above by the $\varepsilon$-hypothesis testing mutual information of $\mathcal{N}$, defined in \eqref{eq-hypo_testing_mutual_inf_chan}, i.e.,
		\begin{equation}\label{eq-eac:one-shot-bound-meta}
			\log_{2}|\mathcal{M}|\leq I_H^{\varepsilon}(\mathcal{N}).
		\end{equation}
		Consequently, for the one-shot entanglement-assisted classical capacity, we have
		\begin{equation}
			C_{\operatorname{EA}}^{\varepsilon}(\mathcal{N})\leq I_H^{\varepsilon}(\mathcal{N})
		\end{equation}
		for all $\varepsilon\in[0,1]$.
	\end{proposition*}

	\begin{Proof}
		First, let us apply Lemma~\ref{lem-eac-meta_conv} to conclude that%
		\begin{equation}
			\log_{2}|\mathcal{M}| \leq I_{H}^{\varepsilon}(M;\widehat{M})_{\omega},
		\end{equation}
		where $\omega_{M\widehat{M}}$ is defined in \eqref{eq-ea_classical_comm_final_state_uniform}. Using the data-processing inequality for the hypothesis testing mutual information under the action of the decoding channel $\mathcal{D}_{BB'\to\widehat{M}}$ (from Proposition~\ref{prop-gen_inf_meas_state_monotonicity}), we find that
		\begin{equation}\label{eq-eac:one-shot-bound-meta_pf1}
			I_{H}^{\varepsilon}(M;\widehat{M})_{\omega}\leq I_{H}^{\varepsilon}(M;BB')_{\theta},
		\end{equation}
		where the state $\theta_{MBB^{\prime}}$ is the same as that in \eqref{eq-eac:state-after-channel}:
		\begin{equation}
			\theta_{MBB^{\prime}}\coloneqq(\mathcal{N}_{A\rightarrow B}\circ\mathcal{E}_{M^{\prime}A^{\prime}\rightarrow A})(\overline{\Phi}_{MM^{\prime}}\otimes \Psi_{A^{\prime}B^{\prime}}).
		\end{equation}
		Observe that the reduced state $\theta_{MB^{\prime}}$ is a product state because the channel $\mathcal{N}_{A\rightarrow B}$ and encoding $\mathcal{E}_{M^{\prime}A^{\prime}\rightarrow A}$ are trace preserving:
		\begin{align}
			\theta_{MB^{\prime}}  & =\Tr_{B}[(\mathcal{N}_{A\rightarrow B}\circ\mathcal{E}_{M^{\prime}A^{\prime}\rightarrow A})(\overline{\Phi}_{MM^{\prime}}\otimes\Psi_{A^{\prime}B^{\prime}})]\\
			&=\Tr_{M^{\prime}A^{\prime}}[\overline{\Phi}_{MM^{\prime}}\otimes\Psi_{A^{\prime}B^{\prime}}]\\
			&=\overline{\Phi}_{M}\otimes\Psi_{B^{\prime}}\\
			&=\theta_{M}\otimes\theta_{B^{\prime}}.\label{eq-eac:initial-reduced-state-product}%
		\end{align}
		Now, by definition, we have that
		\begin{equation}\label{eq-eac:one-shot-bound-meta_pf2}
			I_{H}^{\varepsilon}(M;BB^{\prime})_{\theta}=\inf_{\sigma_{BB^{\prime}}}D_{H}^{\varepsilon}(\theta_{MBB^{\prime}}\Vert\theta_{M}\otimes\sigma_{BB^{\prime}})
		\end{equation}
		Choosing $\sigma_{BB^{\prime}}$ to be $\sigma_{B}\otimes\theta_{B^{\prime}}$ and optimizing over $\sigma_{B}$ only, we find that
		\begin{align}
			I_H^{\varepsilon}(M;BB')_{\theta}&\leq\inf_{\sigma_{B}}D_{H}^{\varepsilon}(\theta_{MBB^{\prime}}\Vert\theta_{M}\otimes\sigma_{B}\otimes\theta_{B^{\prime}})\label{eq-eac:one-shot-bound-meta_pf3}\\
			&=\inf_{\sigma_{B}}D_{H}^{\varepsilon}(\theta_{MBB^{\prime}}\Vert\theta_{MB^{\prime}}\otimes\sigma_{B})\label{eq-eac:one-shot-bound-meta_pf4}\\
			&=I_{H}^{\varepsilon}(MB^{\prime};B)_{\theta}\label{eq-eac:one-shot-bound-meta_pf5},
		\end{align}
		where the first equality follows from the observation in \eqref{eq-eac:initial-reduced-state-product} and the second equality follows by definition. Now, observe that the state $\theta_{MBB'}$ has the form $\mathcal{N}_{A\rightarrow B}(\rho_{SA})$, where $S\equiv MB'$ and $\rho_{SA}\equiv\mathcal{E}_{M^{\prime}A^{\prime}\rightarrow A}(\overline{\Phi}_{MM^{\prime}}\otimes\Psi_{A^{\prime}B^{\prime}})$. This means that we can optimize over every state $\rho_{SA}$ to obtain
		\begin{equation}\label{eq-eac:one-shot-bound-meta_pf6}
			I_{H}^{\varepsilon}(MB^{\prime};B)_{\theta}\leq\sup_{\rho_{SA}}I_{H}^{\varepsilon}(S;B)_{\xi},
		\end{equation}
		where $\xi_{SB}=\mathcal{N}_{A\rightarrow B}(\rho_{SA})$. Note that this optimization over states $\rho_{SA}$ is effectively an optimization over all possible encoding channels $\mathcal{E}_{M'A'\to A}$ that define the $(|\mathcal{M}|,\varepsilon)$ protocol. Now, since it suffices to take pure states when optimizing the $\varepsilon$-hypothesis testing mutual information of bipartite states, following from the same reasoning in \eqref{eq-gen_div_chan_pure_2}--\eqref{eq-gen_div_chan_pure_4} in the context of generalized divergences, we find that
		\begin{align}
			I_H^{\varepsilon}(MB';B)_\theta&\leq\sup_{\psi_{SA}} I_H^{\varepsilon}(S;B)_{\zeta}\label{eq-eac:one-shot-bound-meta_pf7}\\
			&=I_H^{\varepsilon}(\mathcal{N})\label{eq-eac:one-shot-bound-meta_pf8},
		\end{align}
		where $\psi_{SA}$ is a pure state, with the dimension of $S$ the same as that of $A$, and $\zeta_{SB}=\mathcal{N}_{A\to B}(\psi_{SA})$. So we have
		\begin{equation}
			\log_2|\mathcal{M}|\leq I_H^{\varepsilon}(M;\widehat{M})_{\omega}\leq I_H^{\varepsilon}(MB';B)_\theta\leq I_H^{\varepsilon}(\mathcal{N}),
		\end{equation}
		as required.
	\end{Proof}
	
	The result of Proposition~\ref{prop-eac:one-shot-bound-meta} can be written explicitly as
	\begin{align}
		\log_2|\mathcal{M}|&\leq \sup_{\psi_{RA}}\inf_{\sigma_B}D_H^{\varepsilon}(\mathcal{N}_{A\to B}(\psi_{RA})\Vert\psi_R\otimes\sigma_B)\\
		&=\sup_{\psi_{RA}}\inf_{\sigma_B}D_H^{\varepsilon}(\mathcal{N}_{A\to B}(\psi_{RA})\Vert\mathcal{R}_{A\to B}^{\sigma_B}(\psi_{RA})),\label{eq-eac:one-shot-bound-meta_2}
	\end{align}
	in which we explictly see the comparison, via the hypothesis testing relative entropy, between the actual entanglement-assisted classical communication protocol and the protocols over the useless channels $\mathcal{R}_{A\to B}^{\sigma_B}$, with each labeled by the state $\sigma_B$. The state $\psi_{RA}$ corresponds to the state after the encoding channel, and optimizing over these states is effectively an optimization over all encoding channels and all shared entangled states. The latter is true since the preparation of the shared state $\Psi_{A'B'}$ can always be incorporated into a larger encoding channel. Specifically, the encoding $\mathcal{E}_{M'A'\to A}(\overline{\Phi}_{MM'}\otimes\Psi_{A'B'})$ can be written as $\mathcal{E}'_{M'\to A}(\overline{\Phi}_{MM'})$, where $\mathcal{E}'_{M'\to A}\coloneqq \mathcal{E}_{M'A'\to A}\circ\mathcal{P}_{\Psi_{A'B'}}$.

	
	As an immediate consequence of Propositions~\ref{prop-eac:one-shot-bound-meta}, \ref{prop-hypo_to_rel_ent}, and \ref{prop:sandwich-to-htre}, we have the following two bounds:

	\begin{theorem*}{One-Shot Upper Bounds for Entanglement-Assisted Classical Communication}{cor-eacc_meta_str_weak_conv}
		Let $\mathcal{N}_{A\rightarrow B}$ be a quantum channel, and let $\varepsilon\in [0,1)$. For all $(|\mathcal{M}|,\varepsilon)$ entanglement-assisted classical communication protocols over the channel $\mathcal{N}$, the following bounds hold,
		\begin{align}
			\log_{2}|\mathcal{M}|  &  \leq\frac{1}{1-\varepsilon}(I(\mathcal{N})+h_{2}(\varepsilon)),\label{eq-eacc_weak_conv_one_shot_1}\\
			\log_{2}|\mathcal{M}|  &  \leq \widetilde{I}_{\alpha}(\mathcal{N})+\frac{\alpha}{\alpha-1}\log_{2}\!\left(  \frac{1}{1-\varepsilon}\right)\quad\forall~\alpha>1,\label{eq-eacc_str_conv_one_shot_1}
		\end{align}
		where $I(\mathcal{N})$ is the mutual information of $\mathcal{N}$, as defined in \eqref{eq-mut_inf_chan}, and $\widetilde{I}_{\alpha}(\mathcal{N})$ is the sandwiched R\'{e}nyi mutual information of $\mathcal{N}$, as defined in~\eqref{eq-sand_ren_mut_inf_chan}. 
	\end{theorem*}
	
	Since the bounds in \eqref{eq-eacc_weak_conv_one_shot_1} and \eqref{eq-eacc_str_conv_one_shot_1} hold for every $(|\mathcal{M}|,\varepsilon)$ entanglement-assisted classical communication protocol over $\mathcal{N}$, we have that
	\begin{align}
		C_{\operatorname{EA}}^{\varepsilon}(\mathcal{N})&\leq\frac{1}{1-\varepsilon}(I(\mathcal{N})+h_2(\varepsilon)),\\
		C_{\operatorname{EA}}^{\varepsilon}(\mathcal{N})&\leq\widetilde{I}_{\alpha}(\mathcal{N})+\frac{\alpha}{\alpha-1}\log_2\!\left(\frac{1}{1-\varepsilon}\right)\quad\forall~\alpha>1,
	\end{align}
	for all $\varepsilon\in[0,1)$.

	Let us recap the steps we took to arrive at the bounds in \eqref{eq-eacc_weak_conv_one_shot_1} and \eqref{eq-eacc_str_conv_one_shot_1}.
	\begin{enumerate}
		\item We first compared the entanglement-assisted classical communication protocol over $\mathcal{N}$ with the same protocol over a useless channel, by using the hypothesis testing relative entropy. Lemma~\ref{lem-eac-meta_conv} plays a role in bounding the maximum number of transmitted bits for a particular protocol.
		
		\item We then used the data-processing inequality for the hypothesis testing relative entropy to obtain a quantity that is independent of the decoding channel, as well as being minimized over all useless protocols when compared to the actual protocol. This is done in \eqref{eq-eac:one-shot-bound-meta_pf1} and \eqref{eq-eac:one-shot-bound-meta_pf2}--\eqref{eq-eac:one-shot-bound-meta_pf5} in the proof of Proposition~\ref{prop-eac:one-shot-bound-meta}. 
		\item Finally, we optimized over all encoding channels (and, effectively, over all shared states $\Psi_{A'B'}$) in \eqref{eq-eac:one-shot-bound-meta_pf6}--\eqref{eq-eac:one-shot-bound-meta_pf8} to obtain Proposition~\ref{prop-eac:one-shot-bound-meta}, in which the bound is a function solely of the channel and the error probability. 
		\item Using Propositions \ref{prop-hypo_to_rel_ent} and \ref{prop:sandwich-to-htre}, whi\-ch relate the hypothesis testing relative entropy to the quantum relative entropy and the sandwiched R\'{e}nyi relative entropy, respectively, we arrived at Theorem~\ref{cor-eacc_meta_str_weak_conv}.
		
	\end{enumerate}

	The bounds in \eqref{eq-eacc_weak_conv_one_shot_1} and \eqref{eq-eacc_str_conv_one_shot_1} are fundamental upper bounds on the number of transmitted bits for \textit{every} entanglement-assisted classical communication protocol. 
	A natural question to ask now is whether the upper bounds in \eqref{eq-eacc_weak_conv_one_shot_1} and \eqref{eq-eacc_str_conv_one_shot_1} can be achieved. In other words, is it possible to devise protocols such that the number of transmitted bits is equal to the right-hand side of either \eqref{eq-eacc_weak_conv_one_shot_1} or \eqref{eq-eacc_str_conv_one_shot_1}? We do not know how to, especially if we demand that we exactly attain the right-hand side of either \eqref{eq-eacc_weak_conv_one_shot_1} or \eqref{eq-eacc_str_conv_one_shot_1}. However, when given many uses of a channel (in the asymptotic setting), we can come close to achieving these upper bounds. This motivates finding lower bounds on the number of transmitted bits.

\subsection{Lower Bound on the Number of Transmitted Bits via Position-Based Coding and Sequential Decoding}

\label{subsec-pos_coding}
	
	To obtain lower bounds on the number of transmitted bits, as discussed in Appendix~\ref{chap-str_conv}, we should devise particular coding schemes. Concretely, we should devise, for all $\varepsilon\in (0,1)$, an entanglement-assisted classical communication protocol $(\mathcal{M},\Psi_{A'B'},\mathcal{E},\mathcal{D})$ that is an $(|\mathcal{M}|,\varepsilon)$ protocol, meaning that the maximal error probability $p_{\text{err}}^*(\mathcal{P};\mathcal{N})$ satisfies $p_{\text{err}}^*(\mathcal{P};\mathcal{N})\leq\varepsilon$. Recall from \eqref{eq-eac-maximal_error_prob} that the maximal error probability is defined as
	\begin{equation}
		p_{\text{err}}^*(\mathcal{P};\mathcal{N})=\max_{m\in\mathcal{M}}p_{\text{err}}(m,\mathcal{P};\mathcal{N}),
	\end{equation}
	where for $m\in\mathcal{M}$ the message error probability $p_{\text{err}}(m,\mathcal{P};\mathcal{N})$ is defined in \eqref{eq-eac-mess_error_prob} as
	\begin{equation}
		p_{\text{err}}(m,\mathcal{P};\mathcal{N})=1-q(m|m),
	\end{equation}
	with $q(\widehat{m}|m)$ being the probability of identifying the message sent as $\widehat{m}$, given that the message $m$ was sent.
	
	We make use of a technique called \textit{position-based coding} along with \textit{sequential decoding} to establish the lower bound \eqref{eq-eac:one-shot-lower_bound} in Proposition~\ref{prop-eac:one-shot-lower_bound} below, which is analogous to the upper bound \eqref{eq-eac:one-shot-bound-meta} in Proposition~\ref{prop-eac:one-shot-bound-meta}. We now give a brief description of position-based coding and sequential decoding, while leaving the details to the proof of Proposition~\ref{prop-eac:one-shot-lower_bound}.
	
\begin{figure}
		\centering
		\includegraphics[scale=0.8]{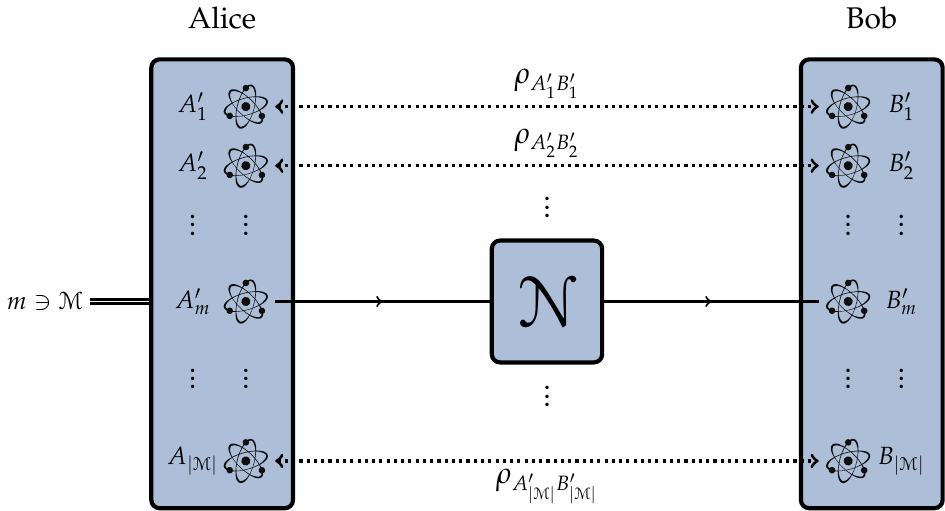}
		\caption{Schematic depiction of position-based coding. Alice and Bob start with $|\mathcal{M}|$ copies of a state $\rho_{A'B'}$. If Alice wants to send the message $m\in\mathcal{M}$, she sends the $m{\text{th}}$ system through the channel $\mathcal{N}$ to Bob.}\label{fig-pos_based_coding}
	\end{figure}

	Let us consider an entanglement-assisted classical communication protocol defined by the four elements $(\mathcal{M},\rho_{A'B'}^{\otimes|\mathcal{M}|},\mathcal{E},\mathcal{D})$ and depicted in Figure~\ref{fig-pos_based_coding}. The state shared by Alice and Bob prior to communication is $|\mathcal{M}|$ copies of a state $\rho_{A'B'}$. The encoding $\mathcal{E}$ is defined such that if Alice wishes to send a message $m\in\mathcal{M}$, then she sends her $m{\text{th}}$ $A'$ system through the channel. Specifically, the encoding channels $\mathcal{E}_{(A')^{|\mathcal{M}|}\to A}^m$ are defined as
	\begin{equation}\label{eq-eacc-pos_encode_AB}
		\mathcal{E}_{(A')^{|\mathcal{M}|}\to A}^m(\rho_{A'B'}^{\otimes|\mathcal{M}|})=\rho_{B_1'}\otimes\dotsb\otimes\rho_{AB_m'}\otimes\dotsb\otimes\rho_{B_m'}=\Tr_{\bar{A}_m}\!\left[\rho_{A'B'}^{\otimes|\mathcal{M}|}\right],
	\end{equation}
	where $\bar{A}_m$ indicates all systems $A_k$ except for $A_m$. This encoding procedure is called position-based coding because the message is encoded into the particular $A'$ system that is sent to Bob. In other words, the message is encoded into the ``position'' of the $A'$ systems.\footnote{In practice, it would be wasteful for the sender to discard so much entanglement by explicitly using the encoding procedure in \eqref{eq-eacc-pos_encode_AB}. The explicit encoding given should thus be considered a conceptual tool for understanding that the $m$th system is sent through the channel, and in practice it can be realized simply by sending the $m$th system through the channel.}
	
	The state held by the receiver Bob after Alice sends the $A$ system in \eqref{eq-eacc-pos_encode_AB} through the channel $\mathcal{N}_{A\to B}$ is
	\begin{equation}\label{eq-eacc-pos_encode_B}
		\tau_{B_1'\dotsb B_m'\dotsb B_{|\mathcal{M}|}'B}^m\coloneqq\rho_{B_1'}\otimes\dotsb\otimes\mathcal{N}_{A\to B}(\rho_{AB_m'})\otimes\dotsb\otimes\rho_{B_{|\mathcal{M}|}'}.
	\end{equation}
	
	Bob, whose task is to determine the message $m$ sent to him, should apply a decoding channel that ideally succeeds with high probability. The sequential decoding strategy consists of Bob performing a sequence of measurements on systems $B_i'$ and $B$. Each of these measurements is defined by the POVM $\{\Pi_{B'BR},\mathbbm{1}_{B'BR}-\Pi_{B'BR}\}$, where $R$ is an arbitrary (finite-dimensional) reference system held by Bob that he makes use of to help with the decoding. In particular, Bob has identical reference systems $R_1,\dotsc, R_{|\mathcal{M}|}$, each associated with his $B'$ systems. The projectors defining the sequential decoding strategy are then
	\begin{equation}\label{eq-eacc-seq_decode}
		P_i\coloneqq \mathbbm{1}_{B_1'R_1}\otimes\dotsb\otimes\mathbbm{1}_{B_{i-1}'R_{i-1}}\otimes\Pi_{B_i'BR_i}\otimes\mathbbm{1}_{B_{i+1}'R_{i+1}}\otimes\dotsb\otimes\mathbbm{1}_{B_{|\mathcal{M}|}'R_{|\mathcal{M}|}}
	\end{equation}
	for all $1\leq i\leq |\mathcal{M}|$, and they correspond to measuring systems $B_i'BR_i$ with the POVM $\{\Pi_{B'BR},\mathbbm{1}_{B'BR}-\Pi_{B'BR}\}$. This measurement can be thought of intuitively as asking the question ``Was the $i{\text{th}}$ message sent?'', with the outcome corresponding to $P_i$ being ``yes'' and the outcome corresponding to
	\begin{equation}
	\widehat{P}_i\coloneqq\mathbbm{1}-P_i
	\end{equation}
	 being ``no.'' Bob performs a measurement on the systems $B_1'BR_1$, followed by a measurement on $B_2'BR_2$, followed by a measurement on $B_3'BR_3$, etc., until he obtains the outcome corresponding to ``yes.'' The system number corresponding to this outcome is then his guess for the message. The probability $q(\widehat{m}|m)$ of guessing $\widehat{m}$ given that the message $m$ was sent is, therefore,
	\begin{equation}
		q(\widehat{m}|m)=\Tr[P_{\widehat{m}}\widehat{P}_{\widehat{m}-1}\dotsb \widehat{P}_1\omega_{B_1'\dotsb B_{|\mathcal{M}|}'BR_1\dotsb R_{|\mathcal{M}|}}^m\widehat{P}_1\dotsb\widehat{P}_{\widehat{m}-1}P_{\widehat{m}}],
	\end{equation}
	where
	\begin{equation}\label{eq-eacc-seq_decode_state_B_ext}
		\omega_{B_1'\dotsb B_{|\mathcal{M}|}'BR_1\dotsb R_{|\mathcal{M}|}}^m\coloneqq\tau_{B_1'\dotsb B_{|\mathcal{M}|}'B}^m\otimes\ket{0,\cdots,0}\bra{0,\cdots,0}_{R_1\dotsb R_{|\mathcal{M}|}}.
	\end{equation}
	The message error probability is then
	\begin{equation}\label{eq-eacc_seq_coding_mess_err_prob}
		p_{\text{err}}(m;\mathcal{P})
		=1-\Tr[P_m\widehat{P}_{m-1}\dotsb \widehat{P}_1\omega_{B_1'\dotsb B_{|\mathcal{M}|}'BR_1\cdots R_{|\mathcal{M}|}}^m \widehat{P}_1\cdots \widehat{P}_{m-1}P_m]
	\end{equation}
	for all $m\in\mathcal{M}$.
	
	Recall that our goal is to place an upper bound on the maximal error probability $p_{\text{err}}^*(\mathcal{P};\mathcal{N})$ of this position-based coding and sequential decoding protocol. We obtain an upper bound on $p_{\text{err}}(m,\mathcal{P};\mathcal{N})$ for each message $m$ from applying the following theorem, called the \textit{quantum union bound}, whose proof can be found in Appendix~\ref{sec-q_union_bd_pf}. This theorem can be thought of as a quantum generalization of the union bound from probability theory. Indeed, if $A_1$, \ldots, $A_N$ is a sequence of events, then the union bound is as follows:
	\begin{equation}
	\operatorname{Pr}[(A_1 \cap \cdots \cap A_N)^c] = \operatorname{Pr}[A_1^c \cup \cdots \cup A_N^c] \leq \sum_{i=1}^N \operatorname{Pr}[A_i^c],
	\end{equation}
	where the superscript $c$ denotes the complement of an event.
	
	\begin{theorem*}{Quantum Union Bound}{thm-q_union_bd}
		Let $\{P_i\}_{i=1}^N$ be a set of projectors. For every state $\rho$ and  $c>0$,
		\begin{equation}\label{eq-q_union_bd}
			\begin{aligned}
			1-&\Tr[P_NP_{N-1}\dotsb P_1\rho P_1\dotsb P_{N-1}P_N]\\
			&\leq (1+c)\Tr[(\mathbbm{1}-P_N)\rho]+(2+c+c^{-1})\sum_{i=1}^{N-1}\Tr[(\mathbbm{1}-P_i)\rho].
			\end{aligned}
		\end{equation}
	\end{theorem*}
	
	\begin{Proof}
		See Appendix~\ref{sec-q_union_bd_pf}.
	\end{Proof}
	
	Using this theorem, we place the following upper bound on the message error probability $p_{\text{err}}(m,\mathcal{P};\mathcal{N})$:
	\begin{equation}\label{eq-eacc_seq_decode_err_prob}
		\begin{aligned}
		p_{\text{err}}(m,\mathcal{P};\mathcal{N})&\leq (1+c)\Tr[\widehat{P}_m\omega_{B_1'\dotsb B_{|\mathcal{M}|}'BR_1\dotsb R_{|\mathcal{M}|}}^m]\\
		&\qquad +(2+c+c^{-1})\sum_{i=1}^{m-1}\Tr[P_i\omega_{B_1'\dotsb B_{|\mathcal{M}|}'BR_1\dotsb R_{|\mathcal{M}|}}^m],
		\end{aligned}
	\end{equation}
	which holds for all $c>0$. By making a particular choice for the projectors $P_1,\dotsc,P_{|\mathcal{M}|}$, and a particular choice for the constant $c$, we obtain the following.
	
	\begin{proposition*}{Lower Bound on One-Shot Entanglement-Assisted Classical Capacity}{prop-eac:one-shot-lower_bound}
		Let $\mathcal{N}_{A\to B}$ be a quantum channel. For  $\varepsilon\in (0,1)$ and $\eta\in (0,\varepsilon)$, there exists an $(|\mathcal{M}|,\varepsilon)$ entanglement-assisted classical communication protocol over $\mathcal{N}_{A\to B}$ such that
		\begin{equation}\label{eq-eac:one-shot-lower_bound}
			\log_2|\mathcal{M}|= \overline{I}_H^{\,\varepsilon-\eta}(\mathcal{N})-\log_2\!\left(\frac{4\varepsilon}{\eta^2}\right)
		\end{equation}
		Consequently, for all $\varepsilon\in(0,1)$ and for all $\eta\in(0,\varepsilon)$,
		\begin{equation}
			C_{\operatorname{EA}}^{\varepsilon}(\mathcal{N})\geq \overline{I}_H^{\varepsilon-\eta}(\mathcal{N})-\log_2\!\left(\frac{4\varepsilon}{\eta^2}\right).
		\end{equation}
		Here,
		\begin{equation}
			\overline{I}_H^{\varepsilon}(\mathcal{N})\coloneqq \sup_{\psi_{RA}}\overline{I}_H^{\varepsilon}(R;B)_{\omega},
		\end{equation}
		where $\omega_{RB}=\mathcal{N}_{A\to B}(\psi_{RA})$, $\psi_{RA}$ is a pure state with the dimension of $R$ the same as that of $A$, and
		\begin{equation}
			\overline{I}_H^{\varepsilon}(A;B)_\rho\coloneqq D_H^{\varepsilon}(\rho_{AB}\Vert\rho_A\otimes\rho_B).
		\end{equation}
	\end{proposition*}
	
	\begin{remark}
		The quantity $\overline{I}_H^\varepsilon(A;B)_\rho$ defined in the statement of Proposition~\ref{prop-eac:one-shot-lower_bound} above is similar to the quantity $I_H^{\varepsilon}(A;B)_\rho$ defined in \eqref{eq-hypo_testing_mutual_inf}, except that we do not perform an optimization over states $\sigma_B$. The resulting channel quantity $\overline{I}_H^{\varepsilon}(\mathcal{N})$ is then similar to the quantity $I_H^{\varepsilon}(\mathcal{N})$ defined in \eqref{eq-hypo_testing_mutual_inf_chan}. The fact that it suffices to optimize over pure states $\psi_{RA}$ in $\overline{I}_H^{\varepsilon}(\mathcal{N})$, with the dimension of $R$ the same as that of $A$, follows from arguments analogous to those presented in Section~\ref{sec-inf_meas_chan}.
	\end{remark}
	
	\begin{Proof}
		Fix $\varepsilon\in(0,1)$ and $\eta\in(0,\varepsilon)$. Starting with the encoded state on Bob's systems as defined in \eqref{eq-eacc-pos_encode_B}, observe that the state of the systems $B$ and $B_{\widehat{m}}'$ is given by
		\begin{equation}
			\tau_{BB_{\widehat{m}}'}^m=\left\{\begin{array}{c l} \mathcal{N}_{A'\to B}(\rho_{A'B'}) & \text{if }\widehat{m}=m,\\ \mathcal{N}_{A'\to B}(\rho_{A'})\otimes\rho_{B'} & \text{if }\widehat{m}\neq m. \end{array}\right.
		\end{equation}
		Recall that the system $B$ results from the system $A_m$ being sent through the channel by Alice. If, along with system $B$, he measures the system $B_m'$, then Bob is performing a measurement on the state $\mathcal{N}_{A'\to B}(\rho_{A'B'})$. On the other hand, if Bob measures a system $B_{\widehat{m}}'$, with $\widehat{m}\neq m$, then Bob is performing a measurement on the state $\mathcal{N}_{A'\to B}(\rho_{A'})\otimes\rho_{B'}$. Bob, knowing the states $\rho_{A'B'}$ as well as the channel $\mathcal{N}$, can devise the measurement $\{\Lambda_{B'B},\mathbbm{1}_{B'B}-\Lambda_{B'B}\}$ that achieves the minimum value of the probability $\Tr[\Lambda_{B'B}(\rho_{B'}\otimes\mathcal{N}_{A'\to B}(\rho_{A'}))]$ while satisfying
		\begin{equation}\label{eq-eac:one-shot-lower_bound_pf}
			\Tr[\Lambda_{B'B}\mathcal{N}_{A'\to B}(\rho_{A'B'})]\geq 1-(\varepsilon-\eta).
		\end{equation}
		Equivalently, by recalling Definition~\ref{def-hypo_testing_rel_ent}, the measurement achieves the $\varepsilon$-hypothesis testing relative entropy
		\begin{equation}
			D_H^{\varepsilon-\eta}(\mathcal{N}_{A'\to B}(\rho_{A'B'})\Vert\rho_{B'}\otimes\mathcal{N}_{A'\to B}(\rho_{A'})),
		\end{equation}
		meaning that
		\begin{equation}\label{eq-eac:one-shot-lower_bound_pf2}
			\begin{aligned}
			&-\log_2\Tr[\Lambda_{B'B}(\rho_{B'}\otimes\mathcal{N}_{A'\to B}(\rho_{A'}))]\\
			&\quad=D_H^{\varepsilon-\eta}(\mathcal{N}_{A'\to B}(\rho_{A'B'})\Vert\rho_{B'}\otimes\mathcal{N}_{A'\to B}(\rho_{A'}))\\
			&\quad=\overline{I}_H^{\varepsilon-\eta}(B';B)_{\xi},
			\end{aligned}
		\end{equation}
		where $\xi_{B'B}\coloneqq\mathcal{N}_{A'\to B}(\rho_{A'B'})$.
	
		The measurement with POVM $\{\Lambda_{B'B},\mathbbm{1}_{B'B}-\Lambda_{B'B}\}$ forms one part of Bob's decoding strategy. The other part of the decoding strategy is based on the fact that Bob does not know the position corresponding to the system $B$ he receives through the channel from Alice. He therefore does not know which of the systems $B_1',\dotsc,B_{|\mathcal{M}|}'$ to measure along with $B$. As described before the statement of the proposition, the sequential decoding strategy consists of Bob performing a sequence of projective measurements on the systems $B_i'BR$ corresponding to the question ``Was the $i{\text{th}}$ message sent?''. Let us define the projectors $\{\Pi_{B'BR},\mathbbm{1}_{B'BR}-\Pi_{B'BR}\}$ on which this measurement is based as follows:
		\begin{equation}\label{eq-eacc_one_shot_lower_bound_pf0a}
			\Pi_{B'BR}\coloneqq U_{B'BR}^\dagger(\mathbbm{1}_{B'B}\otimes\ket{1}\!\bra{1}_R)U_{B'BR},
		\end{equation}
		where $R$ is a qubit system and the unitary $U_{B'BR}$ is defined as
		\begin{multline}\label{eq-eacc_one_shot_lower_bound_pf0b}
			U_{B'BR}\coloneqq\sqrt{\mathbbm{1}_{B'B}-\Lambda_{B'B}}\otimes\left(\ket{0}\!\bra{0}_R+\ket{1}\!\bra{1}_R\right)\\
			+\sqrt{\Lambda_{B'B}}\otimes\left(\ket{1}\!\bra{0}_R-\ket{0}\!\bra{1}_R\right).
		\end{multline}
		Then, it follows that
		\begin{align}
			&\Tr[\Pi_{B'BR}(\mathcal{N}_{A'\to B}(\rho_{A'B'})\otimes\ket{0}\!\bra{0}_R)]\nonumber\\
			\qquad&=\Tr[(\mathbbm{1}_{B'B}\otimes\bra{0}_R)\Pi_{B'BR}(\mathbbm{1}_{B'B}\otimes\ket{0}_R)\mathcal{N}_{A'\to B}(\rho_{A'B'})]\\
			\qquad&=\Tr[\Lambda_{B'B}\mathcal{N}_{A'\to B}(\rho_{A'B'})]\\
			\qquad&\geq 1-(\varepsilon-\eta),
		\end{align}
		where the second equality holds by the definition of $\Pi_{B'BR}$ in \eqref{eq-eacc_one_shot_lower_bound_pf0a} and the fact that $(\mathbbm{1}_{B'B}\otimes\bra{1}_R)U_{B'BR}(\mathbbm{1}_{B'B}\otimes\ket{0}_R)=\sqrt{\Lambda_{B'B}}$, which can be seen from \eqref{eq-eacc_one_shot_lower_bound_pf0b}. To obtain the inequality, we used \eqref{eq-eac:one-shot-lower_bound_pf}. Defining the projection operators $\{P_i\}_{i=1}^{|\mathcal{M}|}$ as in \eqref{eq-eacc-seq_decode}, and defining the state $\omega$ as in \eqref{eq-eacc-seq_decode_state_B_ext}, it also holds that
		\begin{align}
			\Tr[P_i\omega_{B_1'\dotsb B_{|\mathcal{M}|}'BR_1\dotsb R_{|\mathcal{M}|}}^m]&=\Tr[\Lambda_{B'B}(\rho_{B'}\otimes\mathcal{N}_{A'\to B}(\rho_{A'}))]\label{eq-eacc_one_shot_lower_bound_pf1}
		\end{align}
		for all $i<m$, and
		\begin{align}
			\Tr[\widehat{P}_m\omega_{B_1'\dotsb B_{|\mathcal{M}|}'BR_1\dotsb R_{|\mathcal{M}|}}^m]&=\Tr[(\mathbbm{1}_{B'B}-\Lambda_{B'B})\mathcal{N}_{A'\to B}(\rho_{B'A'})]\label{eq-eacc_one_shot_lower_bound_pf2}.
		\end{align}
		Now, recall that the message error probability $p_{\text{err}}(m,\mathcal{P};\mathcal{N})$ is defined as in \eqref{eq-eacc_seq_coding_mess_err_prob}, i.e.,
		\begin{equation}
			\begin{aligned}
			&p_{\text{err}}(m,\mathcal{P};\mathcal{N})\\
			&\qquad=1-\Tr[P_m\widehat{P}_{m-1}\dotsb\widehat{P}_1\omega_{B_1'\dotsb B_{|\mathcal{M}|}'BR_1\dotsb R_{|\mathcal{M}|}}\widehat{P}_1\dotsb\widehat{P}_{m-1}P_m],
			\end{aligned}
		\end{equation}
		and that we can use the quantum union bound (Theorem~\ref{thm-q_union_bd}) to place an upper bound on this quantity as in \eqref{eq-eacc_seq_decode_err_prob}, i.e.,
		 \begin{equation}
		 	\begin{aligned}
			p_{\text{err}}(m;\mathcal{P})&\leq (1+c)\Tr[\widehat{P}_m\omega_{B_1'\dotsb B_{|\mathcal{M}|}'BR_1\dotsb R_{|\mathcal{M}|}}^m]\\
			&\qquad +(2+c+c^{-1})\sum_{i=1}^{m-1}\Tr[P_i\omega_{B_1'\dotsb B_{|\mathcal{M}|}'BR_1\dotsb R_{|\mathcal{M}|}}^m]
			\end{aligned}
		\end{equation}
		for all $c>0$. Using \eqref{eq-eacc_one_shot_lower_bound_pf1} and \eqref{eq-eacc_one_shot_lower_bound_pf2}, the inequality in \eqref{eq-eac:one-shot-lower_bound_pf}, and the equality in \eqref{eq-eac:one-shot-lower_bound_pf2}, the upper bound can be simplified so that
		\begin{align}
			&p_{\text{err}}(m,\mathcal{P};\mathcal{N})\nonumber\\
			&\quad \leq (1+c)\Tr[(\mathbbm{1}_{B'B}-\Lambda_{B'B})\mathcal{N}_{A'\to B}(\rho_{B'A'})]\nonumber\\
			&\quad\qquad +(2+c+c^{-1})(m-1)\Tr[\Lambda_{B'B}(\rho_{B'}\otimes\mathcal{N}_{A'\to B}(\rho_{A'})]\\
			&\quad\leq(1+c)(\varepsilon-\eta)+(2+c+c^{-1})|\mathcal{M}|2^{-\overline{I}_H^{\varepsilon-\eta}(B';B)_{\xi}}\label{eq-eac:one-shot-lower_bound_pf3}
		\end{align}
		for all $c>0$, where the second inequality follows because $m-1\leq |\mathcal{M}|$. The inequality in \eqref{eq-eac:one-shot-lower_bound_pf3} holds for all $m\in\mathcal{M}$, which means that for all $c>0$,
		\begin{equation}
			 p_{\text{err}}^*(\mathcal{P};\mathcal{N})\leq (1+c)(\varepsilon-\eta)+(2+c+c^{-1})|\mathcal{M}|2^{-\overline{I}_H^{\varepsilon-\eta}(B';B)_{\xi}}.
		\end{equation}
		Let us set $\gamma \equiv \overline{I}_H^{\varepsilon-\eta}(B';B)_{\xi}$ and solve for the value of $|\mathcal{M}|$ such that
		\begin{equation}
		(1+c)(\varepsilon-\eta)+(2+c+c^{-1})|\mathcal{M}|2^{-\gamma} = \varepsilon.
		\end{equation}
		We find that
		\begin{equation}
		|\mathcal{M}| = 2^\gamma b (\eta - b \varepsilon),
		\label{eq-EAC:solving-for-M-size}
		\end{equation}
		where $b \equiv \frac{c}{1+c}$. Since $b$ is a variable and our goal is to make $|\mathcal{M}|$ as large as possible for fixed $\varepsilon$ and $\eta$, let us maximize $|\mathcal{M}|$ with respect to $b$. Solving $\frac{\partial |\mathcal{M}|}{\partial b} = 0$, we find that $b = \frac{\eta}{2 \varepsilon}$. This is a permissible value of $b$ since it is required that $b>0$ and $\eta - b \varepsilon \geq 0$. Plugging back into \eqref{eq-EAC:solving-for-M-size}, we find that
		\begin{equation}
		|\mathcal{M}| = 2^\gamma \frac{\eta^2}{4\varepsilon} = 2^{\overline{I}_H^{\varepsilon-\eta}(B';B)_{\xi}-\log_2\left(\frac{4\varepsilon}{\eta^2}\right)}.
		\label{eq-eac:one-shot-lower_bound_pf4}
		\end{equation}
		Thus, with $|\mathcal{M}|$ given by \eqref{eq-eac:one-shot-lower_bound_pf4}, we conclude that
		\begin{equation}\label{eq-eac:one-shot-lower_bound_pf6}
			p_{\text{err}}^*(\mathcal{P})\leq\varepsilon,
		\end{equation}
		and this proves the existence of an $(|\mathcal{M}|,\varepsilon)$ protocol with $|\mathcal{M}|$ given by \eqref{eq-eac:one-shot-lower_bound_pf4}.
		
		However, \eqref{eq-eac:one-shot-lower_bound_pf4} holds for every state $\rho_{A'B'}$, which means that we can take
		\begin{equation}\label{eq-eac:one-shot-lower_bound_pf5}
			\begin{aligned}
			\log_2|\mathcal{M}|&=\sup_{\rho_{A'B'}}\overline{I}_H^{\varepsilon-\eta}(B';B)_{\xi}-\log_2\!\left(\frac{4\varepsilon}{\eta^2}\right)\\
			&=\overline{I}_H^{\varepsilon-\eta}(\mathcal{N})-\log_2\!\left(\frac{4\varepsilon}{\eta^2}\right),
			\end{aligned}
		\end{equation}
		and have \eqref{eq-eac:one-shot-lower_bound_pf6} hold. This is precisely \eqref{eq-eac:one-shot-lower_bound}, and since $\varepsilon\in(0,1)$ and $\eta\in(0,\varepsilon)$ are arbitrary, the proof is complete.
	\end{Proof}
	
	Let us take note of the following two facts from the  proof of Proposition~\ref{prop-eac:one-shot-lower_bound} given above:
	\begin{enumerate}
		\item Given a particular $\varepsilon\in(0,1)$ and an $\eta\in(0,\varepsilon)$, we can construct a position-based coding and sequential decoding protocol achieving a maximal error probability of $p_{\text{err}}^*(\mathcal{P})\leq\varepsilon$ by taking
			\begin{equation}
				\log_2|\mathcal{M}|=\widehat{I}_{H}^{\varepsilon-\eta}(B';B)_{\xi}-\log_2\!\left(\frac{4\varepsilon}{\eta^2}\right),
			\end{equation}
			where $\xi_{B'B}=\mathcal{N}_{A'\to B}(\rho_{A'B'})$ and $\rho_{A'B'}$ is the shared state that Alice and Bob start with (see \eqref{eq-eac:one-shot-lower_bound_pf4}). Note that this holds for \textit{every} state $\rho_{A'B'}$. We take the supremum at the end of the proof in order to obtain the highest possible number of transmitted bits, and also to obtain a quantity that is a function of the channel $\mathcal{N}$ only. The right-hand side of \eqref{eq-eac:one-shot-lower_bound} can thus be achieved by determining the optimal shared state $\rho_{A'B'}$.
		\item Since it suffices to optimize over pure states in order to obtain $\widehat{I}_{H}^{\varepsilon-\eta}(\mathcal{N})$, we see that the shared state $\rho_{A'B'}$ can be taken to be pure, with the dimension of $B'$ the same as that of $A'$.
	\end{enumerate}
	
	An immediate consequence of Propositions~\ref{prop-eac:one-shot-lower_bound} and \ref{prop:ineq-hypo-renyi} is the following theorem.
	
	\begin{theorem*}{One-Shot Lower Bounds for Entanglement-Assisted Classical Communication}{thm-eacc_one_shot_lower_bound}
		Let $\mathcal{N}_{A\to B}$ be a quantum channel. For all $\varepsilon\in (0,1)$, $\eta\in(0,\varepsilon)$, and $\alpha\in (0,1)$, there exists an $(|\mathcal{M}|,\varepsilon)$ entanglement-assisted classical communication protocol over $\mathcal{N}_{A\to B}$ such that
		\begin{equation}\label{eq-eacc_one_shot_lower_bound}
			\begin{aligned}
			\log_2|\mathcal{M}|&\geq\overline{I}_\alpha(\mathcal{N})-\frac{\alpha}{1-\alpha}\log_2\!\left(\frac{1}{\varepsilon-\eta}\right)-\log_2\!\left(\frac{4\varepsilon}{\eta^2}\right).
			\end{aligned}
		\end{equation}
		Here,
		\begin{equation}\label{eq-petz_renyi_chan_mut_inf_noopt}
			\overline{I}_\alpha(\mathcal{N})\coloneqq\sup_{\psi_{RA}} \overline{I}_\alpha(R;B)_{\omega},
		\end{equation}
		where $\omega_{RB}=\mathcal{N}_{A\to B}(\psi_{RA})$, $\psi_{RA}$ is a pure state, $R$ has dimension equal to that of $A$, and
		\begin{equation}\label{eq-petz_renyi_mut_inf_noopt}
			\overline{I}_\alpha(A;B)_\rho\coloneqq D_\alpha(\rho_{AB}\Vert\rho_A\otimes\rho_B).
		\end{equation}
	\end{theorem*}
	
	\begin{remark}
		The quantity $\overline{I}_\alpha(A;B)_\rho$ defined in the statement of Theorem~\ref{thm-eacc_one_shot_lower_bound} above is similar to the quantity $I_\alpha(A;B)_\rho$ defined in \eqref{eq-petz_renyi_mut_inf}, except that we do not perform an optimization over states $\sigma_B$. The resulting channel quantity $\overline{I}_\alpha(\mathcal{N})$ is then similar to the quantity $I_\alpha(\mathcal{N})$ defined in \eqref{eq-petz_renyi_mut_inf_chan}. The fact that it suffices to optimize over pure states $\psi_{RA}$ in $\overline{I}_\alpha(\mathcal{N})$, with the dimension of $R$ the same as that of $A$, follows from arguments analogous to those presented in Section~\ref{sec-inf_meas_chan}.
	\end{remark}
	
	\begin{Proof}
		From Proposition~\ref{prop-eac:one-shot-lower_bound}, we know that for all $\varepsilon\in (0,1)$ and $\eta\in(0,\varepsilon)$ there exists an $(|\mathcal{M}|,\varepsilon)$ entanglement-assisted classical communication protocol such that
		\begin{equation}\label{eq-eacc_one_shot_lower_bound_pf}
			\log_2|\mathcal{M}|=\overline{I}_H^{\varepsilon-\eta}(\mathcal{N})-\log_2\!\left(\frac{4\varepsilon}{\eta^2}\right).
		\end{equation}
		Proposition~\ref{prop:ineq-hypo-renyi} relates the hypothesis testing relative entropy to the Petz--R\'{e}nyi relative entropy according to
		\begin{equation}
			D_H^{\varepsilon}(\rho\Vert\sigma)\geq D_\alpha(\rho\Vert\sigma)+\frac{\alpha}{\alpha-1}\log_2\!\left(\frac{1}{\varepsilon}\right)
		\end{equation}
		for all $\alpha\in (0,1)$, which implies that
		\begin{equation}
			\overline{I}_H^{\varepsilon}(\mathcal{N})\geq \overline{I}_\alpha(\mathcal{N})+\frac{\alpha}{\alpha-1}\log_2\!\left(\frac{1}{\varepsilon}\right).
		\end{equation}
		Combining this inequality with \eqref{eq-eacc_one_shot_lower_bound_pf}, we obtain the desired result.
	\end{Proof}
	
	Since the inequality in \eqref{eq-eacc_one_shot_lower_bound} holds for all $(|\mathcal{M}|,\varepsilon)$ entanglement-assisted classical communication protocols, we have that
	\begin{equation}
		C_{\operatorname{EA}}^{\varepsilon}(\mathcal{N})\geq \overline{I}_\alpha(\mathcal{N})+\frac{\alpha}{\alpha-1}\log_2\!\left(\frac{1}{\varepsilon-\eta}\right)-\log_2\!\left(\frac{4\varepsilon}{\eta^2}\right)
	\end{equation}
	for all $\alpha\in(0,1)$, where $\varepsilon\in(0,1)$ and $\eta\in(0,\varepsilon)$.

\section[Entanglement-Assisted Classical Capacity of a \\ Quantum Channel]{Entanglement-As\-sisted Classical Capacity of a Quantum Channel}

\label{sec-eacc_asymp_setting}

\begin{figure}
		\centering
		\includegraphics[scale=0.65]{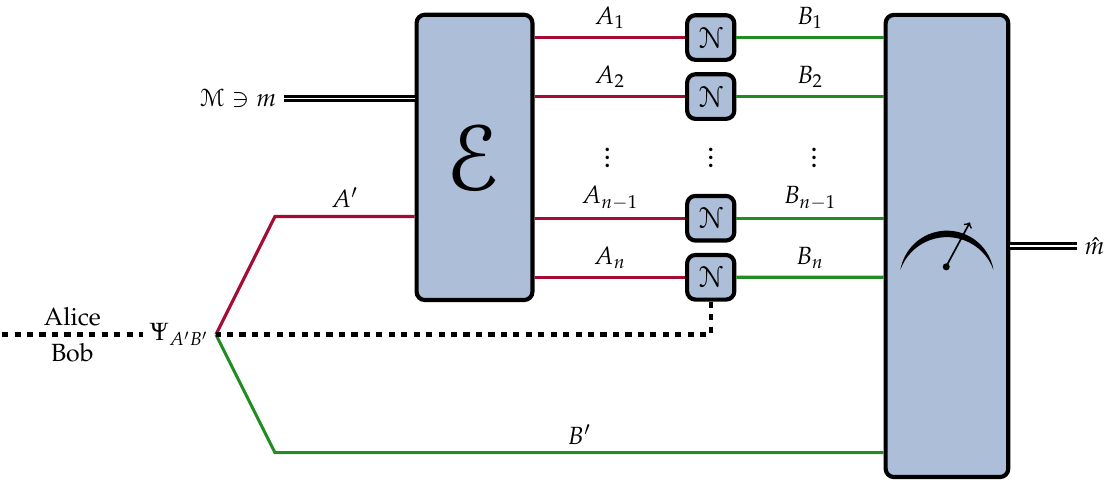}
		\caption{The most general entanglement-assisted classical communication protocol over a multiple number $n\geq 1$ uses of a quantum channel $\mathcal{N}$. Alice and Bob intially share a pair of quantum systems in the state $\Psi_{A'B'}$. Alice, who wishes to send a message $m$ from a set $\mathcal{M}$ of messages, first encodes the message into a quantum state on $n$ quantum systems using an encoding channel $\mathcal{E}$. She then sends each quantum system through the channel $\mathcal{N}$. After Bob receives the systems, he performs a joint measurement on them and the system $B'$, using the outcome of the measurement to give an estimate $\widehat{m}$ of the message $m$ sent to him by Alice.}\label{fig-ea_classical_comm}
	\end{figure}

	Let us now consider the asymptotic setting of entanglement-assisted classical communication. In this scenario, depicted in Figure~\ref{fig-ea_classical_comm}, instead of encoding the message into one quantum system and consequently using the channel $\mathcal{N}$ only once, Alice encodes the message into $n\geq 1$ quantum systems $A_1,\dotsc, A_n$, all with the same dimension as that of $A$, and sends each one of these through the channel $\mathcal{N}$. We call this the asymptotic setting because the number $n$ can be arbitrarily large.
	
	
	The analysis of the asymptotic setting is almost exactly the same as that of the one-shot setting. This is due to the fact that $n$ independent uses of the channel $\mathcal{N}$ can be regarded as a single use of the tensor-product channel $\mathcal{N}^{\otimes n}$. So the only change that needs to be made is to replace $\mathcal{N}$ with $\mathcal{N}^{\otimes n}$ and to define the states and POVM elements as acting on $n$ systems instead of just one. In particular, the state at the end of the protocol is
	\begin{equation}
		\omega_{M\widehat{M}}^p=(\mathcal{D}_{B^nB'\to\widehat{M}}\circ\mathcal{N}_{A\to B}^{\otimes n}\circ\mathcal{E}_{M'A'\to A^n})(\overline{\Phi}^p_{MM'}\otimes\Psi_{A'B'}),
	\end{equation}
	where $p$ is the prior probability distribution over the set of messages $\mathcal{M}$, the encoding channel $\mathcal{E}_{M'A'\to A^n}$ defines a set $\{\mathcal{E}_{M'\to A^n}^m\}_{m\in\mathcal{M}}$ of channels so that
	\begin{equation}
		\mathcal{E}_{M'A'\to A^n}(\overline{\Phi}^p_{MM'}\otimes\Psi_{A'B'})=\sum_{m\in\mathcal{M}}p(m)\ket{m}\!\bra{m}_M\otimes\mathcal{E}_{A'\to A^n}^m(\Psi_{A'B'}),
	\end{equation}
	and the decoding channel $\mathcal{D}_{B^nB'\to\widehat{M}}$, with associated POVM $\{\Lambda_{B^nB'}^m\}_{m\in\mathcal{M}}$, is defined as
	\begin{equation}
		\mathcal{D}_{B^nB'\to\widehat{M}}(\tau_{B^nB'})=\sum_{m\in\mathcal{M}}\Tr[\Lambda_{B^nB'}^m\tau_{B^nB'}]\ket{m}\!\bra{m}_{\widehat{M}}.
	\end{equation}
	Then, for a given code specified by the encoding and decoding channels, the definitions of the message error probability of the code, the average error probability of the code, and the maximal error probability of the code all follow analogously from their definitions in \eqref{eq-eac-mess_error_prob}, \eqref{eq-eac-avg_error_prob}, and \eqref{eq-eac-maximal_error_prob}, respectively, from the one-shot setting.
	
	\begin{definition}{$\boldsymbol{(n,|\mathcal{M}|,\varepsilon)}$ Entanglement-Assisted Classical Communication Protocol}{def-eacc-nMe-protocol}
		Let $(\mathcal{M},\Psi_{A'B'},\mathcal{E}_{M'A'\to A^n},\mathcal{D}_{B^nB'\to\widehat{M}})$ be the elements of an entanglement-assisted classical communication protocol over $n$ uses of the channel $\mathcal{N}_{A\to B}$. The protocol is called an \textit{$(n,|\mathcal{M}|,\varepsilon)$ protocol}, with $\varepsilon\in[0,1]$, if $p_{\text{err}}^*(\mathcal{P};\mathcal{N}^{\otimes n})\leq\varepsilon$.
	\end{definition}
	
	Note that if there exists an $(n,|\mathcal{M}|,\varepsilon)$ entanglement-assist\-ed classical communication protocol, then there exists an $(n,|\mathcal{M}'|,\varepsilon)$ entangle\-ment-assisted classical communication protocol for all $\mathcal{M}'$ satisfying $|\mathcal{M}'|\leq|\mathcal{M}|$. Indeed, simply take a subset $\mathcal{M}'\subseteq\mathcal{M}$ of size $|\mathcal{M}'|$ and define the encoding and decoding channels $\mathcal{E}'$ and $\mathcal{D}'$ as the restrictions of the original channels $\mathcal{E},\mathcal{D}$ to the set $\mathcal{M}'$. Then, using the shorthand $\mathcal{P}'\equiv (\Psi, \mathcal{E}',\mathcal{D}')$,
	\begin{align}
		p_{\text{err}}^*(\mathcal{P}';\mathcal{N})&=\max_{m'\in\mathcal{M}'}p_{\text{err}}(m',\mathcal{P}';\mathcal{N})\\
		&\leq \max_{m\in\mathcal{M}}p_{\text{err}}(m,\mathcal{P};\mathcal{N})\\
		&=p_{\text{err}}^*(\mathcal{P};\mathcal{N})\\
		&\leq\varepsilon,
	\end{align}
	where the first inequality holds because $\mathcal{M}'$ is a subset of $\mathcal{M}$. So we have an $(n,|\mathcal{M}'|,\varepsilon)$ protocol. Similarly, if there does not exist an $(n,|\mathcal{M}|,\varepsilon)$ entanglement-assisted classical communication protocol, then there does not exist an $(n,|\mathcal{M}'|,\varepsilon)$ entanglement-assisted classical communication protocol for all $\mathcal{M}'$ satisfying $|\mathcal{M}'|\geq|\mathcal{M}|$.
	
	The \textit{rate} of an entanglement-assisted classical communication protocol ov\-er $n$ uses of a channel is equal to the number of bits that can be transmitted per channel use, i.e.,
	\begin{equation}
		R(n,|\mathcal{M}|)\coloneqq\frac{1}{n}\log_2|\mathcal{M}|.
	\end{equation}
	Observe that the rate depends only on the number of messages in the set and on the number of uses of the channel. In particular, it does not directly depend on the communication channel nor on the encoding and decoding channels. Given a channel $\mathcal{N}_{A\to B}$ and $\varepsilon\in(0,1)$, the maximum rate of entanglement-assisted classical communication over $\mathcal{N}$ among all $(n,|\mathcal{M}|,\varepsilon)$ protocols is
	\begin{align}
		C_{\operatorname{EA}}^{n,\varepsilon}(\mathcal{N})&\coloneqq \frac{1}{n}C_{\operatorname{EA}}^{\varepsilon}(\mathcal{N}^{\otimes n})\\
		&=\sup_{(\mathcal{M},\Psi,\mathcal{E},\mathcal{D})}\left\{\frac{1}{n}\log_2|\mathcal{M}|:p_{\text{err}}^*((\Psi,\mathcal{E},\mathcal{D});\mathcal{N}^{\otimes n})\leq\varepsilon\right\},
	\end{align}
	where the optimization is over all entanglement-assisted classical communication protocols $(\mathcal{M},\Psi_{A'B'},\mathcal{E}_{M'A'\to A^n},\mathcal{D}_{B^nB'\to\widehat{M}})$ over $\mathcal{N}^{\otimes n}$, with $d_{M'}=d_{\widehat{M}}=|\mathcal{M}|$.
	
	
	The goal of an entanglement-assisted classical communication protocol in the asymptotic setting is to maximize the rate while at the same time keeping the maximal error probability low, using the number $n$ of channel uses as a tunable parameter. Ideally, we would want the error probability to vanish, and since we want to determine the highest possible rate, we are not necessarily concerned about the practical question regarding how many channel uses might be required, at least in the asymptotic setting. In particular, as indicated by definitions, 
	it might take an arbitrarily large number of channel uses to obtain the highest rate with a vanishing error probability. 
	
	\begin{definition}{Achievable Rate for Entanglement-Assisted Classical Communication}{def-eacc_ach_rate}
		Given a quantum channel $\mathcal{N}$, a rate $R\in\mathbb{R}^+$ is called an \textit{achievable rate for entanglement-assisted classical communication over $\mathcal{N}$} if for all $\varepsilon\in (0,1]$, $\delta>0$, and sufficiently large $n$, there exists an $(n,2^{n(R-\delta)},\varepsilon)$ entanglement-assisted classical communication protocol.
	\end{definition}
	
	As we prove in Appendix~\ref{chap-str_conv},
	\begin{equation}
		R\text{ achievable rate }\Longleftrightarrow \lim_{n\to\infty}\varepsilon_{\operatorname{EA}}^*(2^{n(R-\delta)};\mathcal{N}^{\otimes n})=0\quad\forall~\delta>0.
	\end{equation}
	In other words, a rate $R$ is achievable if the optimal error probability for a sequence of protocols with rate $R-\delta$, $\delta>0$, vanishes as the number $n$ of uses of $\mathcal{N}$ increases.
	
	
	\begin{definition}{Entanglement-Assisted Classical Capacity of a Quantum Channel}{def-eacc-cap}
		The \textit{entanglement-assisted classical capacity} of a quantum channel $\mathcal{N}$, denoted by $C_{\operatorname{EA}}(\mathcal{N})$, is defined as the supremum of all achievable rates, i.e.,
		\begin{align}
			C_{\operatorname{EA}}(\mathcal{N})&\coloneqq\sup\{R:R\text{ is an achievable rate for }\mathcal{N}\}.
		\end{align}
	\end{definition}
	
	The entanglement-assisted classical capacity can also be written as
	\begin{equation}
		C_{\operatorname{EA}}(\mathcal{N})=\inf_{\varepsilon\in(0, 1]}\liminf_{n\to\infty}\frac{1}{n}C_{\operatorname{EA}}^{\varepsilon}(\mathcal{N}^{\otimes n}).
	\end{equation}
	See Appendix~\ref{chap-str_conv} for a proof.
	
	\begin{definition}{Weak Converse Rate for Entanglement-Assisted Classical Communication}{def-eacc_weak_conv_rate}
		Given a quantum channel $\mathcal{N}$, a rate $R\in\mathbb{R}^+$ is called a \textit{weak converse rate for entanglement-assisted classical communication over $\mathcal{N}$} if every $R'>R$ is not an achievable rate for $\mathcal{N}$.
	\end{definition}
	
	We show in Appendix~\ref{chap-str_conv} in that
	\begin{equation}\label{eq-EA_comm_weak_conv_rate_alt}
		R\text{ weak converse rate }\Longleftrightarrow\lim_{n\to\infty}\varepsilon_{\operatorname{EA}}^*(2^{n(R-\delta)};\mathcal{N}^{\otimes n})>0\quad\forall~\delta>0.
	\end{equation}
	In other words, a weak converse rate is a rate above which the optimal error probability cannot be made to vanish in the limit of a large number of channel uses.
	
	\begin{definition}{Strong Converse Rate for Entanglement-Assisted Classical Communication}{def-eacc_str_conv_rate}
		Given a quantum channel $\mathcal{N}$, a rate $R\in\mathbb{R}^+$ is called a \textit{strong converse rate for entanglement-assisted classical communication over $\mathcal{N}$} if for all $\varepsilon\in[0,1)$, $\delta>0$, and sufficiently large $n$, there does not exist an $(n,2^{n(R+\delta)},\varepsilon)$ entanglement-assisted classical communication protocol over $\mathcal{N}$.
	\end{definition}
	
	We show in Appendix~\ref{chap-str_conv} that
	\begin{equation}\label{eq-EA_comm_strong_conv_rate_alt}
		R\text{ strong converse rate }\Longleftrightarrow\lim_{n\to\infty}\varepsilon_{\operatorname{EA}}^*(2^{n(R+\delta)};\mathcal{N}^{\otimes n})=1\quad\forall~\delta>0.
	\end{equation}
	In other words, unlike the weak converse, in which the optimal error is required to simply be bounded away from zero as the number $n$ of channel uses increases, in order to have a strong converse rate the optimal error has to converge to one as $n$ increases. By comparing \eqref{eq-EA_comm_weak_conv_rate_alt} and \eqref{eq-EA_comm_strong_conv_rate_alt}, it is clear that every strong converse rate is a weak converse rate.
	


	\begin{definition}{Strong Converse Entanglement-Assisted Classical Capacity of a Quantum Channel}{def-eacc_str_conv_cap}
		The \textit{strong converse entanglement-assisted classical capacity} of a quantum channel $\mathcal{N}$, denoted by $\widetilde{C}_{\operatorname{EA}}(\mathcal{N})$, is defined as the infimum of all strong converse rates, 
		i.e.,
		\begin{align}
			\widetilde{C}_{\operatorname{EA}}(\mathcal{N})&\coloneqq \inf\{R:R\text{ is a strong converse rate for }\mathcal{N}\}.
		\end{align}
	\end{definition}
	
	As shown in general 
	in Appendix~\ref{chap-str_conv}, we have that
	\begin{equation}\label{eq-eacc_str_conv_rate_lower_bound}
		C_{\operatorname{EA}}(\mathcal{N})\leq \widetilde{C}_{\operatorname{EA}}(\mathcal{N})
	\end{equation}
	for every quantum channel $\mathcal{N}$. We can also write the strong converse entanglement-assisted classical capacity as
	\begin{equation}
		\widetilde{C}_{\operatorname{EA}}(\mathcal{N})=\sup_{\varepsilon\in[0, 1)}\limsup_{n\to\infty}\frac{1}{n}C_{\operatorname{EA}}^{\varepsilon}(\mathcal{N}^{\otimes n}).
	\end{equation}
	See Appendix~\ref{chap-str_conv} for a proof.

	Having defined the entanglement-assisted classical capacity of a quantum channel, as well as the strong converse capacity, we now state the main theorem of this chapter, which gives an expression for the entanglement-assisted classical capacity of a quantum channel.

	\begin{theorem*}{Entanglement-Assisted Classical Capacity}{thm-ea_classical_capacity}
		For every quantum channel $\mathcal{N}$, its entanglement-assisted classical capacity $C_{\operatorname{EA}}(\mathcal{N})$ and its strong converse entanglement-assisted classical capacity $\widetilde{C}_{\operatorname{EA}}(\mathcal{N})$ are both equal to the mutual information $I(\mathcal{N})$, i.e.,
		\begin{equation}\label{eq-ea_classical_capacity}
			C_{\operatorname{EA}}(\mathcal{N})=\widetilde{C}_{\operatorname{EA}}(\mathcal{N})=I(\mathcal{N}),
		\end{equation}
		where $I(\mathcal{N})$ is defined in \eqref{eq-mut_inf_chan}.
	\end{theorem*}

	There are two ingredients to proving Theorem~\ref{thm-ea_classical_capacity}:
	\begin{enumerate}
	
		\item \textit{Achievability}: We show that $I(\mathcal{N})$ is an achievable rate. In general, to show that $R\in\mathbb{R}^+$ is achievable, we define the shared entangled state $\Psi_{A'B'}$ and construct encoding and decoding channels such that for all $\varepsilon\in(0,1]$ and sufficiently large $n$, the encoding and decoding channels correspond to $(n,2^{nr},\varepsilon)$ protocols, as per Definition~\ref{def-eacc-nMe-protocol}, with rates $r< R$. Thus, if $R$ is an achievable rate, then, for every error probability $\varepsilon$, it is possible to find an $n$ large enough, along with encoding and decoding channels, such that the resulting protocol has rate arbitrarily close to $R$ and maximal error probability bounded from above by $\varepsilon$.
		
			The achievability part of the proof establishes that $C_{\operatorname{EA}}(\mathcal{N})\geq I(\mathcal{N})$.
		
		\item \textit{Strong Converse}: We show that $I(\mathcal{N})$ is a strong converse rate, from which it follows that $\widetilde{C}_{\operatorname{EA}}(\mathcal{N})\leq I(\mathcal{N})$. In general, to show that $R\in\mathbb{R}^+$ is a strong converse rate, we show that, given any shared entangled state $\Psi_{A'B'}$ and any encoding and decoding channels, for every rate $r> R$,  $\varepsilon\in[0,1)$, and sufficiently large $n$, the communication protocol defined by the encoding and decoding channels is not an $(n,2^{nr},\varepsilon)$ protocol.	
	\end{enumerate}
	
	After showing the achievability and strong converse parts, we can use the inequality in \eqref{eq-eacc_str_conv_rate_lower_bound} to conclude that
	\begin{equation}
		I(\mathcal{N})\leq C_{\operatorname{EA}}(\mathcal{N})\leq\widetilde{C}_{\operatorname{EA}}(\mathcal{N})\leq I(\mathcal{N}),
	\end{equation}
	which immediately implies that $C_{\operatorname{EA}}(\mathcal{N})=\widetilde{C}_{\operatorname{EA}}(\mathcal{N})=I(\mathcal{N})$.
	
	We first establish in Section~\ref{subsec-eacc_achievability} that the rate $I(\mathcal{N})$ is achievable for entan\-glement-assisted classical communication over $\mathcal{N}$. We then address the additivity of the mutual information of a channel, in particular of the sandwiched R\'{e}nyi mutual information of a channel, in Section~\ref{app-sand_ren_mut_inf_additivity}. Finally, we prove that $I(\mathcal{N})$ is a strong converse rate in Section~\ref{sec-eacc_str_conv}. This implies that $I(\mathcal{N})$ is a weak converse rate; however, in Section~\ref{subsec-eacc_weak_conv}, we provide an independent proof of this fact, as the technique used in the proof is useful for alternate communication scenarios (besides entanglement-assisted communication) for which a strong converse theorem is not known to hold.


\subsection{Proof of Achievability}\label{subsec-eacc_achievability}

	In this section, we prove that $I(\mathcal{N})$ is an achievable rate for entanglement-assisted classical communication over $\mathcal{N}$.
	
	First, recall from Theorem~\ref{thm-eacc_one_shot_lower_bound} that for all $\varepsilon\in (0,1)$, $\eta\in(0,\varepsilon)$, and $\alpha\in (0,1)$, there exists an $(|\mathcal{M}|,\varepsilon)$ entanglement-assisted classical communication protocol over $\mathcal{N}$ such that
	\begin{equation}\label{eq-eacc_achieve_pf}
		\begin{aligned}
			\log_2|\mathcal{M}|&\geq \overline{I}_\alpha(\mathcal{N})+\frac{\alpha}{\alpha-1}\log_2\!\left(\frac{1}{\varepsilon-\eta}\right)-\log_2\!\left(\frac{4\varepsilon}{\eta^2}\right),
		\end{aligned}
	\end{equation}
	where $\overline{I}_\alpha(\mathcal{N})$ is defined in  \eqref{eq-petz_renyi_chan_mut_inf_noopt}.
	We obtained this result through a position-based coding strategy along with sequential decoding. A simple corollary of this result is the following.
	
	\begin{corollary*}{Lower Bound for Entanglement-Assisted Classical Communication in Asymptotic Setting}{cor-eacc_asymp_lower_bound}
		Let $\mathcal{N}$ be a quantum channel. For all $\varepsilon\in(0,1]$,  $n\in\mathbb{N}$, and $\alpha\in (0,1)$, there exists an $(n,|\mathcal{M}|,\varepsilon)$ entanglement-assisted classical communication protocol over $n$ uses of $\mathcal{N}$ such that
		\begin{equation}\label{eq-eacc_achieve_pf3}
			\frac{1}{n}\log_2|\mathcal{M}|\geq \overline{I}_\alpha(\mathcal{N})-\frac{1}{n(1-\alpha)}\log_2\!\left(\frac{2}{\varepsilon}\right)-\frac{3}{n}.
		\end{equation}
	\end{corollary*}
	
	\begin{Proof}
		Fix $\varepsilon\in(0,1]$. The inequality \eqref{eq-eacc_achieve_pf} holds for every channel $\mathcal{N}$, which means that it holds for $\mathcal{N}^{\otimes n}$. Applying the inequality in \eqref{eq-eacc_achieve_pf} to $\mathcal{N}^{\otimes n}$ and dividing both sides by $n$, we obtain
		\begin{equation}
			\begin{aligned}
			\frac{\log_2|\mathcal{M}|}{n}&\geq\frac{1}{n}\overline{I}_\alpha(\mathcal{N}^{\otimes n})+\frac{\alpha}{n(\alpha-1)}\log_2\!\left(\frac{1}{\varepsilon-\eta}\right)-\frac{1}{n}\log_2\!\left(\frac{4\varepsilon}{\eta^2}\right)
			\end{aligned}
		\end{equation}
		for all $\alpha\in (0,1)$. By restricting the optimization in the definition of $\overline{I}_\alpha(\mathcal{N}^{\otimes n})$ to tensor-power states, we conclude that $\overline{I}_\alpha(\mathcal{N}^{\otimes n})\geq n\overline{I}_\alpha(\mathcal{N})$. This follows from the additivity of the Petz--R\'{e}nyi relative entropy under tensor-product states (see Proposition~\ref{prop-Petz_rel_ent}). So we obtain
		\begin{equation}\label{eq-eacc_achieve_pf2}
			\frac{\log_2|\mathcal{M}|}{n}\geq\overline{I}_\alpha(\mathcal{N})+\frac{\alpha}{n(\alpha-1)}\log_2\!\left(\frac{1}{\varepsilon-\eta}\right)-\frac{1}{n}\log_2\!\left(\frac{4\varepsilon}{\eta^2}\right)
		\end{equation}
		for all $\alpha\in (0,1)$. Letting $\eta=\frac{\varepsilon}{2}$, and using the fact that $\alpha-1$ is negative for $\alpha\in(0,1)$, this inequality becomes
		\begin{equation}
			\frac{\log_2|\mathcal{M}|}{n}\geq \overline{I}_\alpha(\mathcal{N})-\frac{1}{n(1-\alpha)}\log_2\!\left(\frac{2}{\varepsilon}\right)-\frac{3}{n}
		\end{equation}
		for all $\alpha\in(0,1)$. Since $\varepsilon$ is arbitrary, we find that for all $\varepsilon\in(0,1]$, there exists an $(n,|\mathcal{M}|,\varepsilon)$ protocol such that \eqref{eq-eacc_achieve_pf3} is satisfied, as required.
	\end{Proof}
	
	The inequality in \eqref{eq-eacc_achieve_pf3} gives us, for every $\varepsilon\in(0,1]$ and  $n\in\mathbb{N}$, a lower bound on the size $|\mathcal{M}|$ of the message set that we can take for a corresponding $(n,|\mathcal{M}|,\varepsilon)$ entanglement-assisted classical communication protocol defined by position-based coding and sequential decoding. If instead we fix a particular communication rate $R$ by letting $|\mathcal{M}|=2^{nR}$, then we can rearrange the inequality in \eqref{eq-eacc_achieve_pf3} to obtain an upper bound on the maximal error probability of the corresponding $(n,2^{nR},\varepsilon)$ entanglement-assisted classical communication protocol. Specifically, we conclude that
	\begin{equation}\label{eq-eacc_achieve_error_bound}
		\varepsilon\leq 2\cdot 2^{-n(1-\alpha)\left(\overline{I}_\alpha(\mathcal{N})-R-\frac{3}{n}\right)}
	\end{equation}
	for all $\alpha\in(0,1)$.
	
	The inequality in \eqref{eq-eacc_achieve_pf3} implies that
	\begin{equation}
		C_{\operatorname{EA}}^{n,\varepsilon}(\mathcal{N})\geq \overline{I}_\alpha(\mathcal{N})-\frac{1}{n(\alpha-1)}\log_2\!\left(\frac{2}{\varepsilon}\right)-\frac{3}{n}
	\end{equation}
	for all $\varepsilon\in(0,1]$ and $\alpha\in(0,1)$.
	
	We can now use \eqref{eq-eacc_achieve_pf3} to prove that the mutual information $I(\mathcal{N})$ is an achievable rate for entanglement-assisted classical communication over $\mathcal{N}$.
	
\subsubsection*{Proof of the Achievability Part of Theorem~\ref{thm-ea_classical_capacity}}
	
	
	Fix $\varepsilon\in(0,1]$ and $\delta>0$. Let $\delta_1,\delta_2>0$ be such that
	\begin{equation}\label{eq-eacc_asymp_achieve_pf1}
		\delta=\delta_1+\delta_2.
	\end{equation}
	Set $\alpha\in(0,1)$ such that
	\begin{equation}\label{eq-eacc_asymp_achieve_pf2}
		\delta_1\geq I(\mathcal{N})-\overline{I}_\alpha(\mathcal{N}),
	\end{equation}
	which is possible since $\overline{I}_\alpha(\mathcal{N})$ is monotonically increasing in $\alpha$ (this follows from Proposition~\ref{prop-Petz_rel_ent}), and since $\lim_{\alpha\to 1^-}\overline{I}_\alpha(\mathcal{N})=I(\mathcal{N})$ (see Appendix~\ref{app-sand_ren_mut_inf_chan_limit} for a proof). With this value of $\alpha$, take $n$ large enough so that
	\begin{equation}\label{eq-eacc_asymp_achieve_pf3}
		\delta_2\geq\frac{1}{n(1-\alpha)}\log_2\!\left(\frac{2}{\varepsilon}\right)+\frac{3}{n}.
	\end{equation}
		
	Now, making use of the inequality in \eqref{eq-eacc_achieve_pf3} of Corollary~\ref{cor-eacc_asymp_lower_bound}, there exists an $(n,|\mathcal{M}|,\varepsilon)$ protocol, with $n$ and $\varepsilon$ chosen as above, such that
	\begin{equation}
		\frac{\log_2|\mathcal{M}|}{n}\geq\overline{I}_\alpha(\mathcal{N})-\frac{1}{n(1-\alpha)}\log_2\!\left(\frac{2}{\varepsilon}\right)-\frac{3}{n}.
	\end{equation}
	Rearranging the right-hand side of this inequality, and using \eqref{eq-eacc_asymp_achieve_pf1}--\eqref{eq-eacc_asymp_achieve_pf3}, we find that
	\begin{align}
		\frac{\log_2|\mathcal{M}|}{n}&\geq I(\mathcal{N})-\left(I(\mathcal{N})-\overline{I}_\alpha(\mathcal{N})+\frac{1}{n(1-\alpha)}\log_2\!\left(\frac{2}{\varepsilon}\right)+\frac{3}{n}\right)\\
		&\geq I(\mathcal{N})-(\delta_1+\delta_2)\\
		&=I(\mathcal{N})-\delta.
	\end{align}
	We thus have $I(\mathcal{N})-\delta\leq \frac{1}{n}\log_2|\mathcal{M}|$. By the fact stated immediately after Definition~\ref{def-eacc-nMe-protocol}, we conclude that there exists an $(n,2^{n(R-\delta)},\varepsilon)$ entanglement-assisted classical communication protocol with $R=I(\mathcal{N})$ for all sufficiently large $n$ such that \eqref{eq-eacc_asymp_achieve_pf3} holds. Since $\varepsilon$ and $\delta$ are arbitrary, we have that for all $\varepsilon\in(0,1]$, $\delta>0$, and sufficiently large $n$, there exists an $(n,2^{n(I(\mathcal{N})-\delta)},\allowbreak\varepsilon)$ entanglement-assisted classical communication protocol. This means that $I(\mathcal{N})$ is an achievable rate, and thus that $C_{\operatorname{EA}}(\mathcal{N})\geq I(\mathcal{N})$. See Appendix~\ref{subsec-EA_comm_ach_diff_POV} for a discussion of a different way of seeing the achievability proof.

\subsection{Additivity of the San\-dwiched R\'{e}nyi Mutual Information of a Channel}\label{app-sand_ren_mut_inf_additivity}

We now turn our attention to establishing converse bounds for entanglement-assisted classical communication in the asymptotic setting. Recall from Theorem~\ref{cor-eacc_meta_str_weak_conv} that, for every quantum channel $\mathcal{N}$,  $\varepsilon\in[0,1)$, and all $(|\mathcal{M}|,\varepsilon)$ entanglement-assisted classical communication protocols over $\mathcal{N}$,
	\begin{align}
		\log_2|\mathcal{M}|&\leq\frac{1}{1-\varepsilon}[I(\mathcal{N})+h_2(\varepsilon)],\label{eq-eacc_weak_conv_one_shot_1a}\\
		\log_2|\mathcal{M}|&\leq\widetilde{I}_\alpha(\mathcal{N})+\frac{\alpha}{\alpha-1}\log_2\!\left(\frac{1}{1-\varepsilon}\right)\quad\forall~\alpha>1 \label{eq-eacc_str_conv_one_shot_1a}.
	\end{align}
	To obtain these inequalities, we considered an entanglement-assisted classical communication protocol over a useless channel and used the hypothesis testing relative entropy to compare this protocol with the actual protocol over the channel $\mathcal{N}$. The useless channel in the asymptotic setting is analogous to the one in Figure~\ref{fig-ea_classical_comm_useless_oneshot} and is shown in Figure~\ref{fig-ea_classical_comm_useless}. A simple corollary of Theorem~\ref{cor-eacc_meta_str_weak_conv}, which is relevant for the asymptotic setting, is the following:
	
	\begin{figure}
		\centering
		\includegraphics[scale=0.8]{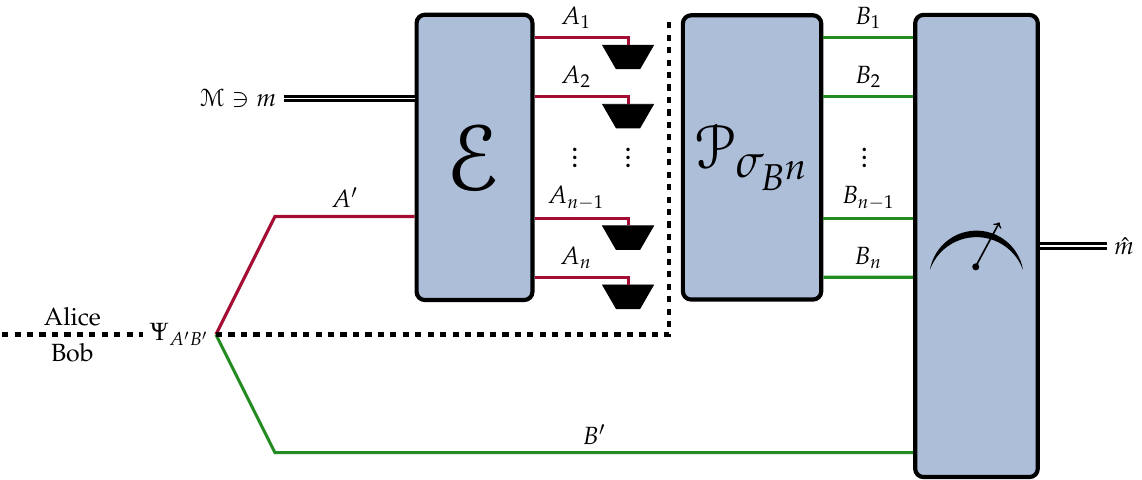}
		\caption{Depiction of a protocol that is useless for entanglement-assisted classical communication in the asymptotic setting. The state encoding the message $m$ via $\mathcal{E}$ is discarded and replaced by an arbitrary (but fixed) state $\sigma_{B^n}$.}\label{fig-ea_classical_comm_useless}
	\end{figure}
	
	\begin{corollary*}{Upper Bounds for Entanglement-Assisted Classical Communication in Asymptotic Setting}{cor-eacc_str_weak_conv_upper}
		Let $\mathcal{N}$ be a quantum channel. For all $\varepsilon\in[0,1)$,  $n\in\mathbb{N}$, and $(n,|\mathcal{M}|,\varepsilon)$ entanglement-assisted classical communication protocols over $n$ uses of $\mathcal{N}$, the rate of transmitted bits is bounded from above as follows:
		\begin{align}
			\frac{1}{n}\log_{2}|\mathcal{M}|  &  \leq\frac{1}{1-\varepsilon}\left(\frac{1}{n}I(\mathcal{N}^{\otimes n})+\frac{1}{n}h_{2}(\varepsilon)\right),\label{eq-eacc_weak_conv_one_shot_2}\\
			\frac{1}{n}\log_{2}|\mathcal{M}|  &  \leq \frac{1}{n}\widetilde{I}_{\alpha}(\mathcal{N}^{\otimes n})+\frac{\alpha}{n(\alpha-1)}\log_{2}\!\left(  \frac{1}{1-\varepsilon}\right)\quad\forall~\alpha>1.\label{eq-eacc_str_conv_one_shot_2}
		\end{align}
	\end{corollary*}
	
	\begin{Proof}
		Since the inequalities \eqref{eq-eacc_weak_conv_one_shot_1a} and \eqref{eq-eacc_str_conv_one_shot_1a} of Theorem~\ref{cor-eacc_meta_str_weak_conv} hold for every channel $\mathcal{N}$, they hold for the channel $\mathcal{N}^{\otimes n}$. Therefore, applying \eqref{eq-eacc_weak_conv_one_shot_1a} and \eqref{eq-eacc_str_conv_one_shot_1a} to $\mathcal{N}^{\otimes n}$ and dividing both sides by $n$, we immediately obtain the desired result.
	\end{Proof}
	
	The inequalities in the corollary above give us, for all $\varepsilon\in[0,1)$ and  $n\in\mathbb{N}$, an upper bound on the size $|\mathcal{M}|$ of the message set we can take for every corresponding $(n,|\mathcal{M}|,\varepsilon)$ entanglement-assisted classical communication protocol. If instead we fix a particular communication rate $R$ by letting $|\mathcal{M}|=2^{nR}$, then we can obtain a lower bound on the maximal error probability of the corresponding $(n,2^{nR},\varepsilon)$ entanglement-assisted classical communication protocol. Specifically, using \eqref{eq-eacc_str_conv_one_shot_2}, we find that
	\begin{equation}
		\varepsilon\geq 1-2^{-n\left(\frac{\alpha-1}{\alpha}\right)\left(R-\frac{1}{n}\widetilde{I}_\alpha(\mathcal{N}^{\otimes n})\right)}
	\end{equation}
	for all $\alpha>1$. 
	
	The inequality in \eqref{eq-eacc_weak_conv_one_shot_2} can be  simplified by using the fact that the mutual information of a channel is additive, as stated in the following theorem:
	
	\begin{theorem*}{Additivity of  Mutual Information of a Quantum Channel}{thm-chan_mut_inf_additive}
		For every two quantum channels $\mathcal{N}_1$ and $\mathcal{N}_2$, the mutual information of $\mathcal{N}_1\otimes\mathcal{N}_2$ is equal to the sum of mutual informations of $\mathcal{N}_1$ and $\mathcal{N}_2$, i.e.,
		\begin{equation}\label{eq-chan_mut_inf_additivity}
			I(\mathcal{N}_1\otimes\mathcal{N}_2)=I(\mathcal{N}_1)+I(\mathcal{N}_2).
		\end{equation}
	\end{theorem*}
	
	\begin{Proof}
		We first recall from \eqref{eq-mut_inf_chan} that the mutual information $I(\mathcal{N})$ of the channel $\mathcal{N}$ is defined as
		\begin{equation}
			\begin{aligned}
			I(\mathcal{N})&=\sup_{\psi_{RA}}I(R;B)_{\omega}\\
			&=\sup_{\psi_{RA}}D(\mathcal{N}_{A\to B}(\psi_{RA})\Vert\psi_R\otimes\mathcal{N}_{A\to B}(\psi_A)),
			\end{aligned}
		\end{equation}
		where $\omega_{RB}=\mathcal{N}_{A\to B}(\psi_{RA})$, $\psi_{RA}$ is a pure state, and $R$ is a system with the same dimension as $A$. Now that, as shown in Section~\ref{sec-inf_meas_chan} in the context of generalized divergences, optimizing over all states $\rho_{RA}$ is not required, since
		\begin{equation}\label{eq-mut_inf_additive_pf2}
			\sup_{\rho_{RA}}I(R;B)_{\omega}=\sup_{\psi_{RA}}I(R;B)_{\omega}.
		\end{equation}
		We also recall that the mutual information $I(A;B)_\rho$ of a bipartite state $\rho_{AB}$ is a special case of conditional mutual information $I(A;B|C)$ with trivial conditioning system $C$. By applying the additivity of conditional mutual information (see \eqref{eq:QEI:CMI-additive}), we thus conclude additivity of mutual information
		for product states $\tau_{A_1B_1}\otimes \omega_{A_2B_2}$:
		\begin{equation}			I(A_1A_2;B_1B_2)_{\tau\otimes\omega}			=I(A_1;B_1)_\tau+I(A_2;B_2)_\omega.
		\label{eq-mut_inf_additive_1}
		\end{equation}
		
		Using these facts, the inequality
		\begin{equation}\label{eq-mut_inf_additive_pf3}
			I(\mathcal{N}_1\otimes\mathcal{N}_2)\geq I(\mathcal{N}_1)+I(\mathcal{N}_2),
		\end{equation}
		is straightforward to establish. Letting
		\begin{align}
			\rho_{R_1R_2B_1B_2}&\coloneqq((\mathcal{N}_1)_{A_1\to B_1}\otimes(\mathcal{N}_2)_{A_2\to B_2})(\psi_{R_1R_2A_1A_2}),\\
			\tau_{R_1B_1}&\coloneqq(\mathcal{N}_1)_{A_1\to B_1}(\phi_{R_1A_1}),\\
			\omega_{R_2B_2}&\coloneqq(\mathcal{N}_2)_{A_2\to B_2}(\varphi_{R_2A_2}),
		\end{align}
		and restricting the optimization in the mutual information of $\mathcal{N}$ to pure product states $\phi\otimes\varphi$, we get that
		\begin{align}
			I(\mathcal{N}_1\otimes\mathcal{N}_2)&=\sup_{\psi}I(R_1R_2;B_1B_2)_{\rho}\label{eq-mut_inf_chan_additive_1}\\
			&\geq\sup_{\phi\otimes\varphi}I(R_1R_2;B_1B_2)_{\tau\otimes\omega}\label{eq-mut_inf_chan_additive_2}\\
			&=\sup_{\phi\otimes\varphi}\left\{I(R_1;B_1)_\tau+I(R_2;B_2)_\omega\right\}\label{eq-mut_inf_chan_additive_3}\\
			&=\sup_\phi I(R_1;B_1)_\tau+\sup_\varphi I(R_2;B_2)_\omega\label{eq-mut_inf_chan_additive_4}\\
			&=I(\mathcal{N}_1)+I(\mathcal{N}_2).\label{eq-mut_inf_chan_additive_5}
		\end{align}
		
		To prove the reverse inequality,  let $\rho_{RB_1B_1}\coloneqq((\mathcal{N}_1)_{A_1\to B_1}\otimes(\mathcal{N}_2)_{A_2\to B_2})\allowbreak(\psi_{RA_1A_2})$. Then, using the formula in \eqref{eq-mut_inf_formula} for the mutual information in terms of the quantum entropy, it is straightforward to verify that
		\begin{align}
			I(R;B_1B_2)_\rho 
			&=I(R;B_1)_\rho+I(RB_1;B_2)_\rho-I(B_1;B_2)_\rho.\label{eq-mut_inf_additive_pf}
		\end{align}
		Now, Klein's inequality in Proposition~\ref{prop-rel_ent}, implies that the mutual information is non-negative. Using this fact on the last term in \eqref{eq-mut_inf_additive_pf}, we find that
		\begin{equation}
			I(R;B_1B_2)_\rho\leq I(R;B_1)_\rho+I(RB_1;B_2)_\rho.
		\end{equation}
		Since $\mathcal{N}_2$ is trace preserving, we have that
		\begin{equation}
			\rho_{RB_1}=\Tr_{B_2}[\rho_{RB_1B_2}]=(\mathcal{N}_1)_{A_1\to B_1}(\psi_{RA_1}).
		\end{equation}
		Therefore,
		\begin{equation}
			I(R;B_1)_\rho\leq\sup_{\rho_{RA_1}}I(R;B_1)_{\tau}=I(\mathcal{N}_1),
		\end{equation}
		where the equality follows from \eqref{eq-mut_inf_additive_pf2}. Similarly, by writing $\rho_{RB_1B_2}$ as
		\begin{equation}
			\rho_{RB_1B_2}=(\mathcal{N}_2)_{A_2\to B_2}(\omega_{RB_1A_2}),\quad\omega_{RB_1A_2}\coloneqq(\mathcal{N}_1)_{A_1\to B_1}(\psi_{RA_1A_2}),
		\end{equation}
		we get that
		\begin{equation}
			I(RB_1;B_2)_\rho\leq I(\mathcal{N}_2).
		\end{equation}
		Therefore, 
		\begin{equation}
			I(R;B_1B_1)_\rho\leq I(\mathcal{N}_1)+I(\mathcal{N}_2).
		\end{equation}
		Since the state $\psi_{RA_1A_2}$ that we started with is arbitrary, we obtain
		\begin{equation}
			I(\mathcal{N}_1\otimes\mathcal{N}_2)=\sup_{\psi_{RA_1A_2}}I(R;B_1B_2)_\rho\leq I(\mathcal{N}_1)+I(\mathcal{N}_2),
		\end{equation}
		as required. Combining this inequality with that in \eqref{eq-mut_inf_additive_pf3}, we have the required equality, $I(\mathcal{N}_1\otimes\mathcal{N}_2)=I(\mathcal{N}_1)+I(\mathcal{N}_2)$.
	\end{Proof}

	Using the additivity of the mutual information of a channel, the inequality in \eqref{eq-eacc_weak_conv_one_shot_2} can be rewritten as
	\begin{equation}\label{eq-eacc_weak_conv_one_shot_3}
		\frac{\log_2|\mathcal{M}|}{n}\leq\frac{1}{1-\varepsilon}\left(I(\mathcal{N})+\frac{1}{n}h_2(\varepsilon)\right),
	\end{equation}
	which implies that
	\begin{equation}
		C_{\operatorname{EA}}^{n,\varepsilon}(\mathcal{N})\leq\frac{1}{1-\varepsilon}\left(I(\mathcal{N})+\frac{1}{n}h_2(\varepsilon)\right)
	\end{equation}
	for all $n\geq 1$ and  $\varepsilon\in(0,1)$. Using this inequality, it is straightforward to conclude that $I(\mathcal{N})$ is a weak converse rate for entanglement-assisted classical communication, and the interested reader can jump ahead to Section~\ref{subsec-eacc_weak_conv} to see this. 

	We now show that the sandwiched R\'{e}nyi mutual information $\widetilde{I}_\alpha(\mathcal{N})$ of a channel $\mathcal{N}$ is also additive, i.e., that
	\begin{equation}
		\widetilde{I}_\alpha(\mathcal{N}_1\otimes\mathcal{N}_2)=\widetilde{I}_\alpha(\mathcal{N}_1)+\widetilde{I}_\alpha(\mathcal{N}_2)
	\end{equation}
	for all $\alpha>1$. One ingredient of the proof is the additivity of the sandwiched R\'{e}nyi mutual information of bipartite states with respect to tensor-product states, i.e.,
	\begin{equation}\label{eq-sand_rel_ent_mut_inf_additive}
		\widetilde{I}_\alpha(A_1A_2;B_1B_2)_{\xi\otimes\omega}=\widetilde{I}_\alpha(A_1;B_1)_\xi+\widetilde{I}_\alpha(A_2;B_2)_\omega,
	\end{equation}
	where $\xi_{A_1B_1}$ and $\omega_{A_2B_2}$ are states. To show this, let us first recall the definition of the sandwiched R\'{e}nyi mutual information of a bipartite state $\rho_{AB}$ from \eqref{eq:QEI:sand-Ren-MI-states}:
	\begin{equation}
		\widetilde{I}_\alpha(A;B)_\rho=\inf_{\sigma_B}\widetilde{D}_\alpha(\rho_{AB}\Vert\rho_A\otimes\sigma_B),
	\end{equation}
	where the optimization is over states $\sigma_B$. This quantity, as well as the sandwiched R\'{e}nyi mutual information of a channel, can be written in an alternate way, as we show in the following lemma, the proof of which can be found in Appendix~\ref{lem-sand_rel_mut_inf_alt_pf}.
	
	\begin{Lemma}{lem-sand_rel_mut_inf_alt}
		For every bipartite state $\rho_{AB}$ and $\alpha>1$, the sandwiched R\'{e}nyi mutual information $\widetilde{I}_\alpha(A;B)_\rho$ can be written as
		\begin{equation}\label{eq-sand_rel_mut_inf_alt}
			\begin{aligned}
			&\widetilde{I}_{\alpha}(A;B)_\rho\\
			&=\frac{\alpha}{\alpha-1}\log_2\sup_{\tau_C}\norm{\Tr_{AC}\!\left[\left(\rho_A^{\frac{1-\alpha}{\alpha}}\otimes\tau_C^{\frac{\alpha-1}{\alpha}}\right)\ket{\psi}\!\bra{\psi}_{ABC}\right]}_{\frac{\alpha}{2\alpha-1}},
			\end{aligned}
		\end{equation}
		where $\ket{\psi}_{ABC}$ is a purification of $\rho_{AB}$ and $\tau_C$ is a state. The sandwiched R\'{e}nyi mutual information $I(\mathcal{N})$ of a channel $\mathcal{N}$ can be written as
		\begin{equation}\label{eq-sand_rel_mut_inf_alt_second}
			\widetilde{I}_\alpha(\mathcal{N})=\frac{\alpha}{\alpha-1}\inf_{\sigma_B}\log_2\norm{\mathcal{S}_{\sigma_B}^{(\alpha)}\circ\mathcal{N}}_{\text{CB},\,1\to\alpha},
		\end{equation}
		where $\mathcal{S}_{\sigma_B}^{(\alpha)}(\cdot)\coloneqq\sigma_B^{\frac{1-\alpha}{2\alpha}}(\cdot)\sigma_B^{\frac{1-\alpha}{2\alpha}}$ and
		\begin{equation}\label{eq-operator_CB_alpha_norm}
			\begin{aligned}
			\norm{\mathcal{M}}_{\text{CB},\,1\to\alpha}&\coloneqq\sup_{\substack{Y_R>0,\\\Tr[Y_R]\leq 1}}\norm{\mathcal{M}_{A\to B}\!\left(Y_R^{\frac{1}{2\alpha}}\ket{\Gamma}\!\bra{\Gamma}_{RA}Y_R^{\frac{1}{2\alpha}}\right)}_{\alpha},
			\end{aligned}
		\end{equation}
		with $\mathcal{M}$ a completely positive map. (See Appendix~\ref{app-CB_alpha_norm_alt} for an alternate expression for $\norm{\cdot}_{\text{CB},1\to\alpha}$.)
	\end{Lemma}
	
	\begin{Proof}
		See Appendix~\ref{lem-sand_rel_mut_inf_alt_pf}.
	\end{Proof}
	
	Using the alternate expression in \eqref{eq-sand_rel_mut_inf_alt}, we establish the additivity statement in \eqref{eq-sand_rel_ent_mut_inf_additive} of the sandwiched R\'{e}nyi mutual information of a bipartite state. 

	\begin{proposition*}{Additivity of  Sandwiched R\'{e}nyi Mutual Information of Bipartite States}{prop-sand_rel_mut_inf_additive}
		For every product state $\xi_{A_1B_1}\otimes\omega_{A_2B_2}$ and  $\alpha>1$, the sandwiched R\'{e}nyi mutual information $\widetilde{I}_\alpha(A_1A_2;B_1B_2)_{\xi\otimes\omega}$ is additive, i.e.,
		\begin{equation}\label{eq-sand_rel_ent_mut_inf_additive_2}
			\widetilde{I}_{\alpha}(A_1A_2;B_1B_2)_{\xi\otimes\omega}=\widetilde{I}_\alpha(A_1;B_1)_\xi+\widetilde{I}_\alpha(A_2;B_2)_\omega.
		\end{equation}
	\end{proposition*}
	
	\begin{Proof}
		By definition, we have that
		\begin{align}
			\widetilde{I}_{\alpha}(A_1A_2;B_1B_2)_{\xi\otimes\omega}=\inf_{\sigma_{B_1B_2}}\widetilde{D}_{\alpha}(\xi_{A_1B_1}\otimes\omega_{A_2B_2}\Vert\xi_{A_1}\otimes\omega_{A_2}\otimes\sigma_{B_1B_2})
		\end{align}
		If we restrict the optimization to product states $\sigma_{B_1}^1\otimes\sigma_{B_2}^2$, then we find that
		\begin{align}
			&\widetilde{I}_{\alpha}(A_1A_2;B_1B_2)_{\xi\otimes\omega}\nonumber\\
			&\quad\leq\inf_{\sigma^1\otimes\sigma^2}\widetilde{D}_\alpha(\xi_{A_1B_1}\otimes\omega_{A_2B_2}\Vert\xi_{A_1}\otimes\omega_{A_2}\otimes\sigma_{B_1}^1\otimes\sigma_{B_2}^2)\\
			&\quad=\inf_{\sigma^1,\sigma^2}\left\{\widetilde{D}_{\alpha}(\xi_{A_1B_1}\Vert\xi_{A_1}\otimes\sigma_{B_1}^1)+\widetilde{D}_{\alpha}(\omega_{A_2B_2}\Vert\omega_{A_2}\otimes\sigma_{B_2}^2)\right\}\\
			&\quad=\widetilde{I}_{\alpha}(A_1;B_1)_{\xi}+\widetilde{I}_{\alpha}(A_2;B_2)_{\omega}.
		\end{align}
		So
		\begin{equation}
			\widetilde{I}_{\alpha}(A_1A_2;B_1B_2)_{\xi\otimes\omega}\leq\widetilde{I}_{\alpha}(A_1;B_1)_{\xi}+\widetilde{I}_{\alpha}(A_2;B_2)_{\omega}.
		\end{equation}
		
		To show the reverse inequality, we use the alternate expression \eqref{eq-sand_rel_mut_inf_alt} in Lemma~\ref{lem-sand_rel_mut_inf_alt} for the sandwiched R\'{e}nyi mutual information of a bipartite state. 
In this expression, if we take a product purification $\ket{\psi_1}_{A_1B_1C_1}\otimes\ket{\psi_2}_{A_2B_2C_2}$ of $\xi_{A_1B_1}\otimes\omega_{A_2B_2}$ and restrict the optimization to product states, we obtain
		\begin{align}
			&\widetilde{I}_{\alpha}(A_1A_2;B_1B_2)_{\xi\otimes\omega}\nonumber\\
			&=\frac{\alpha}{\alpha-1}\log_2\sup_{\tau_{C_1C_2}}\norm{\Tr_{A_1A_2C_1C_2}\!\left[\left(\left(\xi_{A_1}\otimes\omega_{A_2}\right)^{\frac{1-\alpha}{\alpha}}\otimes\tau_{C_1C_2}^{\frac{\alpha-1}{\alpha}}\right)\right.\right.\nonumber\\
			&\qquad\qquad\qquad\qquad\qquad\times\left.\left.\ket{\psi_1}\!\bra{\psi_1}_{A_1B_1C_1}\otimes\ket{\psi_2}\!\bra{\psi_2}_{A_2B_2C_2}\right]}_{\frac{\alpha}{2\alpha-1}}\\
			&\geq \frac{\alpha}{\alpha-1}\log_2\sup_{\tau_{C_1}\otimes\tau_{C_2}}\norm{\Tr_{A_1A_2C_1C_2}\!\left[\left(\xi_{A_1}^{\frac{1-\alpha}{\alpha}}\otimes\omega_{A_2}^{\frac{1-\alpha}{\alpha}}\otimes\tau_{C_1}^{\frac{\alpha-1}{\alpha}}\otimes\tau_{C_2}^{\frac{\alpha-1}{\alpha}}\right)\right.\right.\nonumber\\
			&\qquad\qquad\qquad\qquad\qquad\left.\left.\times\ket{\psi_1}\!\bra{\psi_1}_{A_1B_1C_1}\otimes\ket{\psi_2}\!\bra{\psi_2}_{A_2B_2C_2}\right]}_{\frac{\alpha}{2\alpha-1}}\\
			&=\frac{\alpha}{\alpha-1}\log_2\sup_{\tau_{C_1},\tau_{C_2}}\norm{\Tr_{A_1C_1}\!\left[\left(\xi_{A_1}^{\frac{1-\alpha}{\alpha}}\otimes\tau_{C_1}^{\frac{\alpha-1}{\alpha}}\right)\ket{\psi_1}\!\bra{\psi_1}_{A_1B_1C_1}\right]\right.\nonumber\\
			&\qquad\qquad\qquad \left.\otimes\Tr_{A_2C_2}\!\left[\left(\omega_{A_2}^{\frac{1-\alpha}{\alpha}}\otimes\tau_{C_2}^{\frac{\alpha-1}{\alpha}}\right)\ket{\psi_2}\!\bra{\psi_2}_{A_2B_2C_2}\right]}_{\frac{\alpha}{2\alpha-1}}\\
			&=\frac{\alpha}{\alpha-1}\log_2\sup_{\tau_{C_1},\tau_{C_2}}\left\{\norm{\Tr_{A_1C_1}\!\left[\left(\xi_{A_1}^{\frac{1-\alpha}{\alpha}}\otimes\tau_{C_1}^{\frac{\alpha-1}{\alpha}}\right)\ket{\psi_1}\!\bra{\psi_1}_{A_1B_1C_1}\right]}_{\frac{\alpha}{2\alpha-1}}\right.\nonumber\\
			&\qquad\qquad\left. \times \norm{\Tr_{A_2C_2}\!\left[\left(\omega_{A_2}^{\frac{1-\alpha}{\alpha}}\otimes\tau_{C_2}^{\frac{\alpha-1}{\alpha}}\right)\ket{\psi_2}\!\bra{\psi_2}_{A_2B_2C_2}\right]}_{\frac{\alpha}{2\alpha-1}}\right\}\\
			&=\frac{\alpha}{\alpha-1}\log_2\sup_{\tau_{C_1}}\norm{\Tr_{A_1C_1}\!\left[\left(\xi_{A_1}^{\frac{1-\alpha}{\alpha}}\otimes\tau_{C_1}^{\frac{\alpha-1}{\alpha}}\right)\ket{\psi_1}\!\bra{\psi_1}_{A_1B_1C_1}\right]}_{\frac{\alpha}{2\alpha-1}}\nonumber\\
			&\quad+\frac{\alpha}{\alpha-1}\log_2\sup_{\tau_{C_2}}\norm{\Tr_{A_2C_2}\!\left[\left(\omega_{A_2}^{\frac{1-\alpha}{\alpha}}\otimes\tau_{C_2}^{\frac{\alpha-1}{\alpha}}\right)\ket{\psi_2}\!\bra{\psi_2}_{A_2B_2C_2}\right]}_{\frac{\alpha}{2\alpha-1}}\\
			&=\widetilde{I}_{\alpha}(A_1;B_1)_{\xi}+\widetilde{I}_{\alpha}(A_2;B_2)_{\omega}.
		\end{align}
		We have thus shown \eqref{eq-sand_rel_ent_mut_inf_additive_2},
		as required.
	\end{Proof}

	\begin{theorem*}{Additivity of  Sandwiched R\'{e}nyi Mutual Information of a Channel}{thm-sand_rel_ent_additivity}
		For every two channels $\mathcal{N}_1$ and $\mathcal{N}_2$, and for all $\alpha>1$,
		\begin{equation}
			\widetilde{I}_{\alpha}(\mathcal{N}_1\otimes\mathcal{N}_2)=\widetilde{I}_\alpha(\mathcal{N}_1)+\widetilde{I}_\alpha(\mathcal{N}_2).
			\label{eq:EA-comm:add-I-sand-alpha-ch}
		\end{equation}
	\end{theorem*}
	
	\begin{Proof}
		Recall that
		\begin{equation}\label{eq-sand_rel_mut_inf_chan_1}
			\widetilde{I}_\alpha(\mathcal{N})=\sup_{\psi_{RA}}\widetilde{I}_\alpha(R;B)_\omega,
		\end{equation}
		where $\omega_{RB}\coloneqq\mathcal{N}_{A\to B}(\psi_{RA})$, and the supremum is taken over every pure state $\psi_{RA}$, with $R$ having the same dimension as $A$. The superadditivity of the sandwiched R\'{e}nyi mutual information of a channel, namely,
		\begin{equation}
			\widetilde{I}_\alpha(\mathcal{N}_1\otimes\mathcal{N}_2)\geq\widetilde{I}_\alpha(\mathcal{N}_1)+\widetilde{I}_\alpha(\mathcal{N}_2)
			\label{eq:EA-comm:superadd-I-sand-alpha-ch}
		\end{equation}
		follows immediately by restricting the optimization in the definition \eqref{eq-sand_rel_mut_inf_chan_1} to product states and using the additivity of the sandwiched R\'{e}nyi mutual information of bipartite states, as proven in Proposition~\ref{prop-sand_rel_mut_inf_additive}. An explicit proof of this statement goes along the same lines as \eqref{eq-mut_inf_chan_additive_1}--\eqref{eq-mut_inf_chan_additive_5}, with $I$ replaced by $\widetilde{I}_\alpha$.
		
		To prove the reverse inequality, namely, the subadditivity of the sandwiched R\'{e}nyi mutual information of a channel, we use the expression in \eqref{eq-sand_rel_mut_inf_alt_second}. By restricting the infimum in \eqref{eq-sand_rel_mut_inf_alt_second} to product states, we find that
		\begin{align}
			&\widetilde{I}_\alpha(\mathcal{N}_1\otimes\mathcal{N}_2)\nonumber\\
			&\quad=\frac{\alpha}{\alpha-1}\inf_{\sigma_{B_1B_2}}\log_2\norm{\mathcal{S}_{\sigma_{B_1B_2}}^{(\alpha)}\circ\left(\mathcal{N}_1\otimes \mathcal{N}_2\right)}_{\text{CB},\,1\to\alpha}\\
			&\quad\leq\frac{\alpha}{\alpha-1}\inf_{\sigma_{B_1}^1\otimes\sigma_{B_2}^2}\log_2\norm{\mathcal{S}^{(\alpha)}_{\sigma_{B_1}^1\otimes\sigma_{B_2}^2}\circ\left(\mathcal{N}_1\otimes \mathcal{N}_2\right)}_{\text{CB},\,1\to\alpha}\\
			&\quad=\frac{\alpha}{\alpha-1}\inf_{\sigma_{B_1}^1,\sigma_{B_2}^2}\log_2\norm{\left(\mathcal{S}_{\sigma_{B_1}^1}^{(\alpha)}\circ \mathcal{N}_1\right)\otimes\left(\mathcal{S}_{\sigma_{B_2}^2}^{(\alpha)}\circ \mathcal{N}_2\right)}_{\text{CB},\,1\to\alpha},
		\end{align}
		where the last equality follows because
		\begin{equation}
			\mathcal{S}_{\sigma_{B_1}^1\otimes\sigma_{B_2}^2}^{(\alpha)}=\left(\mathcal{S}_{\sigma_{B_1}^1}^{(\alpha)}\otimes\id_{B_2}\right)\circ\left(\id_{B_1}\otimes\mathcal{S}_{\sigma_{B_2}^2}^{(\alpha)}\right).
		\end{equation}
		Now, consider that the norm $\norm{\cdot}_{\text{CB},\,1\to\alpha}$ is multiplicative with respect to tensor products of completely positive maps, i.e.,
		\begin{equation}\label{eq-CB_alpha_norm_mult}
			\norm{\mathcal{M}_1\otimes\mathcal{M}_2}_{\text{CB},\,1\to\alpha}=\norm{\mathcal{M}_1}_{\text{CB},\,1\to\alpha}\norm{\mathcal{M}_2}_{\text{CB},\,1\to\alpha}
		\end{equation}
		for every two completely positive maps $\mathcal{M}_1,\mathcal{M}_2$ and all $\alpha>1$ (see Appendix~\ref{app-CB_norm_mult_pf} for a proof). Using this, we find that
		\begin{align}
			&\!\!\!\!\!\!\!\widetilde{I}_{\alpha}(\mathcal{N}_1\otimes\mathcal{N}_2)\nonumber\\
			& \leq \frac{\alpha}{\alpha-1}\inf_{\sigma_{B_1}^1,\sigma_{B_2}^2}\log_2\norm{\left(\mathcal{S}_{\sigma_{B_1}^1}^{(\alpha)}\circ \mathcal{N}_1\right)\otimes\left(\mathcal{S}_{\sigma_{B_2}^2}^{(\alpha)}\circ \mathcal{N}_2\right)}_{\text{CB},\,1\to\alpha}\\
			&=\frac{\alpha}{\alpha-1}\inf_{\sigma_{B_1}^1}\log_2\norm{\mathcal{S}_{\sigma_{B_1}^1}^{(\alpha)}\circ\mathcal{N}_1}_{\text{CB},\,1\to\alpha} +\frac{\alpha}{\alpha-1}\inf_{\sigma_{B_2}^2}\log_2\norm{\mathcal{S}_{\sigma_{B_2}^2}^{(\alpha)}\circ\mathcal{N}_2}_{\text{CB},\,1\to\alpha}\\
			&=\widetilde{I}_\alpha(\mathcal{N}_1)+\widetilde{I}_\alpha(\mathcal{N}_2).
		\end{align}
		We have thus shown that $\widetilde{I}_\alpha(\mathcal{N}_1\otimes\mathcal{N}_2)\leq\widetilde{I}_\alpha(\mathcal{N}_1)+\widetilde{I}_\alpha(\mathcal{N}_2)$, and by combining this with \eqref{eq:EA-comm:superadd-I-sand-alpha-ch}, we conclude \eqref{eq:EA-comm:add-I-sand-alpha-ch}.
	\end{Proof}
	
	Note that the additivity of the mutual information of a channel, i.e., Theorem~\ref{thm-chan_mut_inf_additive}, follows straightforwardly from the  theorem above by taking the limit $\alpha\to 1$ (see Appendix~\ref{app-sand_ren_mut_inf_chan_limit} for a proof).
	
	Using the additivity of the sandwiched R\'{e}nyi mutual information of a channel (Theorem~\ref{prop-sand_rel_mut_inf_additive}), the inequality in \eqref{eq-eacc_str_conv_one_shot_2} can be written as
	\begin{equation}\label{eq-eacc_str_conv_one_shot_3}
		\frac{1}{n}\log_2|\mathcal{M}| \leq \widetilde{I}_{\alpha}(\mathcal{N})+\frac{\alpha}{n(\alpha-1)}\log_2\!\left(\frac{1}{1-\varepsilon}\right)
	\end{equation}
	for all $\alpha>1$. This implies that
	\begin{equation}
		C_{\operatorname{EA}}^{n,\varepsilon}(\mathcal{N})\leq \widetilde{I}_{\alpha}(\mathcal{N})+\frac{\alpha}{n(\alpha-1)}\log_2\!\left(\frac{1}{1-\varepsilon}\right)
		\label{eq:EA-comm:str-conv-bound-EA-cap-finite-n}
	\end{equation}
	for all $n\geq 1$, $\varepsilon\in(0,1)$, and $\alpha>1$.

\subsection{Proof of the Strong Converse}\label{sec-eacc_str_conv}

	With the inequality in \eqref{eq:EA-comm:str-conv-bound-EA-cap-finite-n} in hand, we can now prove that the mutual information $I(\mathcal{N})$ is a strong converse rate for entanglement-assisted classical communication over $\mathcal{N}$ and establish that $\widetilde{C}_{\operatorname{EA}}(\mathcal{N})=I(\mathcal{N})$.


\subsubsection*{Proof of the Strong Converse Part of Theorem~\ref{thm-ea_classical_capacity}}
	
	Fix $\varepsilon\in[0,1)$ and $\delta>0$. Let $\delta_1,\delta_2>0$ be such that
	\begin{equation}\label{eq-eacc_str_conv_pf1}
		\delta>\delta_1+\delta_2\eqqcolon\delta'.
	\end{equation}
	Set $\alpha\in(1,\infty)$ such that
	\begin{equation}\label{eq-eacc_str_conv_pf2}
		\delta_1\geq\widetilde{I}_\alpha(\mathcal{N})-I(\mathcal{N}),
	\end{equation}
	which is possible since $\widetilde{I}_\alpha(\mathcal{N})$ is monotonically increasing with $\alpha$ (following from Proposition~\ref{prop-sand_rel_ent_properties}), and since $\lim_{\alpha\to 1^+}\widetilde{I}_\alpha(\mathcal{N})=I(\mathcal{N})$ (see Appendix~\ref{app-sand_ren_mut_inf_chan_limit} for a proof). With this value of $\alpha$, take $n$ large enough so that
	\begin{equation}\label{eq-eacc_str_conv_pf3}
		\delta_2\geq \frac{\alpha}{n(\alpha-1)}\log_2\!\left(\frac{1}{1-\varepsilon}\right).
	\end{equation}
		
	Now, with the values of $n$ and $\varepsilon$ as above, every $(n,|\mathcal{M}|,\varepsilon)$ entanglement-assisted classical communication protocol satisfies \eqref{eq-eacc_str_conv_one_shot_3}.
	Rearranging the right-hand side of this inequality, and using \eqref{eq-eacc_str_conv_pf1}--\eqref{eq-eacc_str_conv_pf3}, we obtain
	\begin{align}
		\frac{\log_2|\mathcal{M}|}{n}&\leq I(\mathcal{N})+\widetilde{I}_\alpha(\mathcal{N})-I(\mathcal{N})+\frac{\alpha}{n(\alpha-1)}\log_2\!\left(\frac{1}{1-\varepsilon}\right)\\
		&\leq I(\mathcal{N})+\delta_1+\delta_2\\
		&=I(\mathcal{N})+\delta'\\
		&<I(\mathcal{N})+\delta.
	\end{align}
	So we have that $I(\mathcal{N})+\delta>\frac{\log_2|\mathcal{M}|}{n}$ for all $(n,|\mathcal{M}|,\varepsilon)$ entanglement-assisted classical communication protocols and sufficiently large $n$. Due to this strict inequality, it follows that there cannot exist an $(n,2^{n(I(\mathcal{N})+\delta)},\varepsilon)$ entanglement-assisted classical communication protocol for all sufficiently large $n$ such that \eqref{eq-eacc_str_conv_pf3} holds, for if it did there would exist a set $\mathcal{M}$ such that $|\mathcal{M}|= 2^{n(I(\mathcal{N})+\delta)}$, which we have just seen is not possible. Since $\varepsilon$ and $\delta$ are arbitrary, we conclude that for all $\varepsilon\in[0,1)$, $\delta>0$, and sufficiently large $n$, there does not exist an $(n,2^{n(I(\mathcal{N})+\delta)},\varepsilon)$ entanglement-assisted classical communication protocol. This means that $I(\mathcal{N})$ is a strong converse rate, and thus that $\widetilde{C}_{\operatorname{EA}}(\mathcal{N})\leq I(\mathcal{N})$. See Appendix~\ref{app:EA-comm:strong-conv-diff-POV} for a different way of understanding the strong converse.

\subsection{Proof of the Weak Converse}

\label{subsec-eacc_weak_conv}

	We now conclude Section~\ref{sec-eacc_asymp_setting} by providing an independent proof of the fact that the mutual information $I(\mathcal{N})$ of a channel $\mathcal{N}$ is a weak converse rate.\footnote{Recall that any strong converse rate is also a weak converse rate, so that by the proof of the strong converse part of Theorem~\ref{thm-ea_classical_capacity} we can immediately conclude that $I(\mathcal{N})$ is a weak converse rate.}
	
	\begin{theorem*}{Weak Converse for Entanglement-Assisted Classical Communication}{thm-ea_classical_comm_weak_converse}
		For every quantum channel $\mathcal{N}$, the mutual information $I(\mathcal{N})$ is a weak converse rate for entanglement-assisted classical communication over~$\mathcal{N}$.
	\end{theorem*}
	
	\begin{Proof}
		Suppose that $R$ is an achievable rate. Then, by definition, for all $\varepsilon\in(0,1]$,  $\delta>0$, and sufficiently large $n$, there exists an $(n,2^{n(R-\delta)},\varepsilon)$ entanglement-assisted classical communication protocol over $\mathcal{N}$. For all such protocols, the inequality \eqref{eq-eacc_weak_conv_one_shot_3} holds by Corollary~\ref{cor-eacc_str_weak_conv_upper} and the additivity of the mutual information of a channel, i.e.,
		\begin{equation}
			R-\delta \leq \frac{1}{1-\varepsilon}\left(I(\mathcal{N})+\frac{1}{n}h_2(\varepsilon)\right).
		\end{equation}
		Since this bound holds for all sufficiently large $n$, it holds in the limit $n\to\infty$, so that
		\begin{equation}
			R\leq\frac{1}{1-\varepsilon}I(\mathcal{N})+\delta.
		\end{equation}
		Then, since this inequality holds for all $\varepsilon\in(0,1]$ and $\delta>0$, we obtain
		\begin{equation}
			R\leq \lim_{\varepsilon,\delta\to 0}\left\{\frac{1}{1-\varepsilon}I(\mathcal{N})+\delta\right\}=I(\mathcal{N}).
		\end{equation}
		We have thus shown that if $R$ is an achievable rate, then $R\leq I(\mathcal{N})$. The contrapositive of this statement is that if $R>I(\mathcal{N})$, then $R$ is not an achievable rate. By definition, therefore, $I(\mathcal{N})$ is a weak converse rate.
	\end{Proof}

\section{Examples}\label{sec-eacc_examples}

	In this section, we determine the entanglement-assisted classical capacity of some of the channels that we introduced in Chapter~\ref{chap-QM_channels}. Recall that  Theorem~\ref{thm-ea_classical_capacity} states that the entanglement-assisted classical capacity $C_{\operatorname{EA}}(\mathcal{N})$ is given by the mutual information of the channel $\mathcal{N}$, i.e.,
	\begin{equation}
		C_{\operatorname{EA}}(\mathcal{N})=I(\mathcal{N})=\sup_{\psi_{RA}}I(R;B)_{\omega},
	\end{equation}
	where $\omega_{RB}=\mathcal{N}_{A\to B}(\psi_{RA})$ and  the optimization is over every pure state $\psi_{RA}$, with the dimension of $R$ the same as the dimension of the input system~$A$ of the channel. The mutual information $I(R;B)_{\omega}$ can be calculated using either the quantum relative entropy or the quantum entropy via
	\begin{align}
		I(R;B)_{\omega}&=D(\mathcal{N}_{A\to B}(\psi_{RA})\Vert\psi_R\otimes\mathcal{N}_{A\to B}(\psi_A))\\
		&=H(R)_{\omega}+H(B)_{\omega}-H(AB)_{\omega}.
	\end{align}

	\begin{figure}
		\centering
		\includegraphics[scale=1]{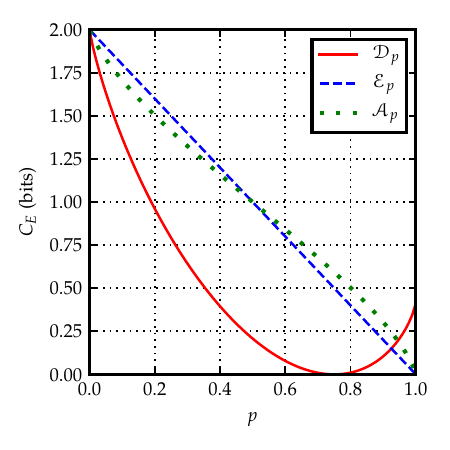}
		\caption{
		The entanglement-assisted classical capacity $C_{\operatorname{EA}}$ of the depolarizing channel $\mathcal{D}_p$ (expressed in \eqref{eq-ea_cap_dep}), the erasure channel $\mathcal{E}_p$ (expressed in \eqref{eq-ea_cap_erasure}), and the amplitude damping channel $\mathcal{A}_p$ (expressed in \eqref{eq-ea_cap_amp_damp}), all of which are defined for the parameter $p\in[0,1]$.}\label{fig-EA_capacities}
	\end{figure}
	
	In what follows, we consider the entanglement-assisted classical capacity of channels that are covariant with respect to a group $G$, and then we provide an explicit expression for the entanglement-assisted classical capacity of the depolarizing, erasure, and generalized amplitude damping channels (see Section~\ref{subsec-qubit_channel}). See Figure~\ref{fig-EA_capacities} for a plot of these capacities.

\subsection{Covariant Channels}

	Let us start with covariant channels. Recall from Definition~\ref{def-group_cov_chan} that a channel $\mathcal{N}$ is covariant with respect to a group $G$ if there exist projective unitary representations $\{U_A^g\}_{g\in G}$ and $\{V_B^g\}_{g\in G}$ such that
	\begin{equation}\label{eq-group_cov_channel}
		\mathcal{N}(U_A^g\rho (U_A^g)^\dagger)=V_B^g\mathcal{N}(\rho)(V_B^g)^\dagger
	\end{equation}
	for every state $\rho$ and all $g\in G$.
	
	Suppose that a channel $\mathcal{N}$ is covariant with respect to a group $G$ such that the representation $\{U_A^g\}_{g\in G}$ of $G$ acting on the input space of $\mathcal{N}$ satisfies
	\begin{equation}\label{eq-G_irrep_twirl}
		\frac{1}{|G|}\sum_{g\in G}U_A^g\rho_A (U_A^g)^\dagger=\Tr[\rho_A]\frac{\mathbbm{1}}{d},
	\end{equation}
	for all $\rho_A$, where $d$ is the dimension of the input space of the channel $\mathcal{N}$. Such channels are called \textit{irreducibly covariant}.
	
	Let us now recall Proposition~\ref{prop-gen_mut_inf_cov_chan}, which tells us that the generalized mutual information for every covariant channel is given as follows:
	\begin{equation}\label{eq-gen_mut_inf_chan_cov_2}
		\boldsymbol{I}(\mathcal{N})=\sup_{\phi_{RA}}\{\boldsymbol{I}(R;B)_{\omega}:\phi_A=\mathcal{T}_G(\phi_A)\},
	\end{equation}
	where $\omega_{RB}=\mathcal{N}_{A\to B}(\phi_{RA})$ and $\mathcal{T}_G(\cdot)\coloneqq\frac{1}{|G|}\sum_{g\in G}U^g(\cdot)U^{g\dagger}$. In other words, for covariant channels, it suffices to optimize over pure states $\phi_{RA}$ for which the reduced state $\phi_A$ is invariant under the channel $\mathcal{T}_G$. For irreducibly covariant channels, the expression in \eqref{eq-gen_mut_inf_chan_cov_2} simplifies to
	\begin{equation}\label{eq-gen_mut_inf_chan_cov_irrep_2}
		\boldsymbol{I}(\mathcal{N})=\boldsymbol{I}(A;B)_{\rho^{\mathcal{N}}},
	\end{equation}
	where $\rho_{AB}^{\mathcal{N}}=\mathcal{N}_{A'\to B}(\Phi_{AA'})$ is the Choi state of $\mathcal{N}$.
	
	Using \eqref{eq-gen_mut_inf_chan_cov_2}, we immediately obtain the following theorem.
	
	\begin{theorem*}{Entanglement-Assisted Classical Capacity of Covariant Channels}{thm-EA_cap_covariant}
		If a channel $\mathcal{N}_{A\to B}$ is covariant with respect to a group $G$ as in \eqref{eq-group_cov_channel}, then its entanglement-assisted classical capacity is given by
		\begin{equation}\label{eq-EA_cap_cov_chan}
			C_{\operatorname{EA}}(\mathcal{N})=\sup_{\phi_{RA}}\{I(R;B)_{\omega}:\phi_A=\mathcal{T}_G(\phi_A)\},
		\end{equation}
		where $\omega_{RB}=\mathcal{N}_{A\to B}(\phi_{RA})$.
		If the representation $\{U_A^g\}_{g\in G}$ is irreducible, then the entanglement-assisted classical capacity of $\mathcal{N}$ is given by the mutual information of its Choi state $\rho_{AB}^{\mathcal{N}}$, i.e.,
		\begin{equation}
			C_{\operatorname{EA}}(\mathcal{N})=I(A;B)_{\rho^{\mathcal{N}}}.
		\end{equation} 
	\end{theorem*}

\subsubsection{Depolarizing Channel}\label{sec-eacc_depolarizing}

	In Section~\ref{subsec-qubit_channel}, specifically in \eqref{eq-qubit_depolarizing_channel}, we defined the depolarizing channel on qubits as
	\begin{equation}
		\mathcal{D}_{p}(\rho)\coloneqq (1-p)\rho+\frac{p}{3}(X\rho X+Y\rho Y+Z\rho Z)
	\end{equation}
	From this, it follows that the channel is covariant with respect to the Pauli operators, meaning that
	\begin{align}
		\mathcal{D}_{p}(X\rho X)&=X\mathcal{D}_{p}(\rho)X,\\
		\mathcal{D}_{p}(Y\rho Y)&=Y\mathcal{D}_{p}(\rho)Y,\\
		\mathcal{D}_{p}(Z\rho Z)&=Z\mathcal{D}_{p}(\rho)Z.
	\end{align}
	Furthermore, we have the identity in \eqref{eq-Pauli_twirl}, which asserts that for every state~$\rho$,
	\begin{equation}
		\frac{1}{4}\rho+\frac{1}{4}X\rho X+\frac{1}{4}Y\rho Y+\frac{1}{4}Z\rho Z=\frac{\mathbbm{1}}{2}.
	\end{equation}
	This means that the operators $\{\mathbbm{1},X,Y,Z\}$ satisfy the property in \eqref{eq-G_irrep_twirl}.\footnote{The operators $\{\mathbbm{1},X,Y,Z\}$ form a projective unitary representation of the group $\mathbb{Z}_2\times\mathbb{Z}_2=\{(0,0),(0,1),(1,0),(1,1)\}$, where $\mathbb{Z}_2$ is the group consisting of the set $\{0,1\}$ with addition modulo two. Specficially, we have $U_{(0,0)}=\mathbbm{1}$, $U_{(0,1)}=X$, $U_{(1,0)}=Z$, and $U_{(1,1)}=Y$.} By Theorem~\ref{thm-EA_cap_covariant}, we thus have that the entanglement-assisted classical capacity of the depolarizing channel is simply the mutual information of its Choi state.
	
	It is straightforward to see that the Choi state $\rho_{AB}^{\mathcal{D}_p}$ of the depolarizing channel is
	\begin{multline}
		\rho_{AB}^{\mathcal{D}_p}=(1-p)\ket{\Phi^+}\!\bra{\Phi^+}_{AB}\\
		+\frac{p}{3}\left(\ket{\Psi^+}\!\bra{\Psi^+}_{AB}+\ket{\Psi^-}\!\bra{\Psi^-}_{AB}+\ket{\Phi^-}\!\bra{\Phi^-}_{AB}\right).
	\end{multline}
	Since $H(A)_{\rho^{\mathcal{D}_p}}=H(B)_{\rho^{\mathcal{D}_p}}=\log_2(2)=1$ and
	\begin{equation}
		H(AB)_{\rho^{\mathcal{D}_p}}=-(1-p)\log_2(1-p)-p\log_2\!\left(\frac{p}{3}\right),
	\end{equation}
	we find that
	\begin{align}
		C_{\operatorname{EA}}(\mathcal{D}_p)&=I(A;B)_{\rho^{\mathcal{D}_p}}=H(A)_{\rho^{\mathcal{D}_p}}+H(B)_{\rho^{\mathcal{D}_p}}-H(AB)_{\rho^{\mathcal{D}_p}}\\
		&=2+(1-p)\log_2(1-p)+p\log_2\!\left(\frac{p}{3}\right)\\
		& = 2-h_2(p) - p\log_2(3)\label{eq-ea_cap_dep}
	\end{align}
	for all $p\in[0,1]$. See Figure~\ref{fig-EA_capacities} above for a plot of the capacity.
	
	\begin{figure}
		\centering
		\includegraphics[scale=0.85]{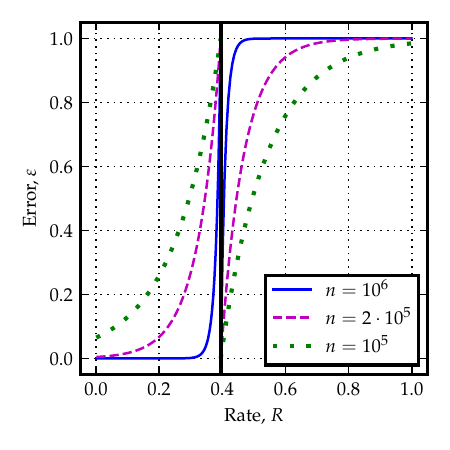}~
		\includegraphics[scale=0.85]{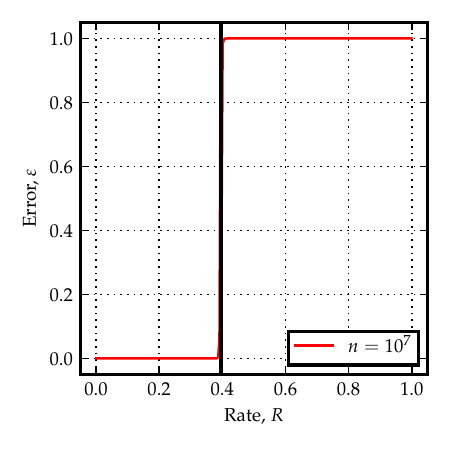}
		\caption{Plot of the error bounds in \eqref{eq-eacc_dep_errorVrate_pf2} and \eqref{eq-eacc_dep_errorVrate_pf3} for the depolarizing channel $\mathcal{D}_p$ with $p=0.4$. By increasing the number $n$ of channel uses, it is possible to communicate at rates closer to the capacity (indicated by the black vertical line) with vanishing error probability. Furthermore, for every rate above the capacity, as $n$ increases, the error probability approaches one at an exponential rate, consistent with the fact that the mutual information $I(\mathcal{D}_p)$ is a strong converse rate.}\label{fig-eacc_dep_errorVrate}
	\end{figure}

	Let us also briefly analyze the lower and upper bounds obtained in Corollaries~\ref{cor-eacc_asymp_lower_bound} and \ref{cor-eacc_str_weak_conv_upper}, respectively. Specifically, let us consider the following bounds on the maximal error probability that results from these bounds, i.e.,
	\begin{align}
		\varepsilon&\leq 2\cdot 2^{-n(1-\alpha)\left(\overline{I}_\alpha(\mathcal{D}_p)-R-\frac{3}{n}\right)},\\
		\varepsilon&\geq 1-2^{-n\left(\frac{\alpha-1}{\alpha}\right)\left(R-\widetilde{I}_{\alpha}(\mathcal{D}_p)\right)},
	\end{align}
	which are rearrangements of \eqref{eq-eacc_achieve_pf3} and \eqref{eq-eacc_str_conv_one_shot_2} and are discussed further in Appendices~\ref{subsec-EA_comm_ach_diff_POV} and \ref{app:EA-comm:strong-conv-diff-POV}, respectively.
	Now, by \eqref{eq-gen_mut_inf_chan_cov_irrep_2}, we have
	\begin{equation}\label{eq-eacc_dep_errorVrate_pf}
		\widetilde{I}_{\alpha}(\mathcal{D}_p)=\widetilde{I}_{\alpha}(R;B)_{\rho^{\mathcal{D}_p}}\leq \overline{\widetilde{I}}_{\alpha}(R;B)_{\rho^{\mathcal{D}_p}},
	\end{equation}
	where
	\begin{equation}
		\overline{\widetilde{I}}_{\alpha}(A;B)_\rho\coloneqq\widetilde{D}_{\alpha}(\rho_{AB}\Vert\rho_A\otimes\rho_B).
	\end{equation}
	For simplicity, let us use the quantity in \eqref{eq-eacc_dep_errorVrate_pf}, which does not involve an optimization over states $\sigma_B$, to place a lower bound on $\varepsilon$, so that
	\begin{equation}\label{eq-eacc_dep_errorVrate_pf2}
		\varepsilon\geq 1-2^{-n\left(\frac{\alpha-1}{\alpha}\right)\left(R-\overline{\widetilde{I}}_{\alpha}(R;B)_{\omega}\right)},
	\end{equation}
	where $\omega_{RB}=\rho^{\mathcal{D}_p}_{RB}$ is the Choi state of $\mathcal{D}_p$. Similarly, for simplicity, let us take the quantity $\overline{I}_\alpha(\mathcal{N})$, which by definition involves an optimization over all pure states $\psi_{RA}$, and let $\psi_{RA}$ be the maximally entangled state $\Phi_{RA}^+$. So we take the upper bound on the error probability to be
	\begin{equation}\label{eq-eacc_dep_errorVrate_pf3}
		\varepsilon\leq 2\cdot 2^{-n(1-\alpha)\left(\overline{I}_\alpha(R;B)_{\omega}-R-\frac{3}{n}\right)},
	\end{equation}
	where $\omega_{RB}=\rho^{\mathcal{D}_p}_{RB}$ is the Choi state of $\mathcal{D}_p$. Then, for $p=0.4$, we plot in Figure~\ref{fig-eacc_dep_errorVrate} the bounds in \eqref{eq-eacc_dep_errorVrate_pf2} and \eqref{eq-eacc_dep_errorVrate_pf3} (with $\alpha=1.0001$ and $\alpha=0.9999$, respectively) to obtain plots that are analogous to the generic plot in Figure~\ref{fig-ea_classical_comm_str_converse} in Appendix~\ref{app:EA-comm:strong-conv-diff-POV}. As portrayed in Figure~\ref{fig-ea_classical_comm_str_converse}, we indeed see that, as the number $n$ of channel uses increases, the capacity $C_{\operatorname{EA}}(\mathcal{D}_p)$ becomes a clearer dividing point between reliable communication---with nearly-vanish\-ing error probability---and unreliable communication---with error probability approaching one at an exponential rate.
	
	Let us consider the qudit depolarizing channel, as defined in \eqref{eq:QM-over:qudit-depolarizing} in terms of the Heisenberg--Weyl operators $\{W_{z,x}\}_{z,x}$ from \eqref{eq-Heisenberg_Weyl_operators}.
	From the definition in \eqref{eq:QM-over:qudit-depolarizing} and the properties stated in \eqref{eq-Heisenberg_Weyl_prop1}--\eqref{eq-Heisenberg_Weyl_prop3}, it follows that the qudit depolarizing channel is covariant with respect to the Heisenberg--Weyl operators. Furthermore, the Heisenberg--Weyl operators form an irreducible projective unitary representation of the group $\mathbb{Z}_d\times\mathbb{Z}_d$, and thus satisfy
	\begin{equation}
		\frac{1}{d^2}\sum_{z,x=0}^{d-1} W_{z,x}\rho W_{z,x}^\dagger=\Tr[\rho]\frac{\mathbbm{1}}{d}
		\label{eq:EA-comm:qudit-depol-cov}
	\end{equation}
	for all $\rho$. Therefore, by Theorem~\ref{thm-EA_cap_covariant}, we have that
	\begin{equation}
		C_{\operatorname{EA}}(\mathcal{D}_p^{(p)})=I(A;B)_{\rho^{\mathcal{D}_p^{(d)}}}.
	\end{equation}
	By calculations analogous to those above, we obtain
	\begin{equation}
		C_{\operatorname{EA}}(\mathcal{D}_p^{(d)})=2\log_2 d - h_2(p) - p\log_2(d^2-1).
	\end{equation}

\subsubsection{Erasure Channel}\label{subsec-eacc_erasure_channel}

	In Section~\ref{subsec-qubit_channel}, specifically in \eqref{eq-erasure_channel}, we defined the qubit erasure channel as
	\begin{equation}
		\mathcal{E}_p(\rho)\coloneqq (1-p)\rho+p\Tr[\rho]\ket{e}\!\bra{e}
	\end{equation}
	for $p\in[0,1]$, where $\ket{e}$ is a state that is orthogonal to all states in the input qubit system. Recall that if we let the output space simply be a qutrit system with the orthonormal basis $\{\ket{0},\ket{1},\ket{2}\}$, then the input qubit space can be naturally embedded into the subspace of the qutrit system spanned by $\ket{0}$ and $\ket{1}$, so that we can let the erasure state simply be $\ket{2}$. This means that we can write the action of the erasure channel as
	\begin{equation}
		\mathcal{E}_p(\rho)=(1-p)\rho+p\ket{2}\!\bra{2}.
	\end{equation}
	
	Observe that, like the depolarizing channel, the erasure channel is covariant with respect to the group $\mathbb{Z}_2\times\mathbb{Z}_2$, with the representation $\{\mathbbm{1},X,Y,Z\}$ on the input qubit space and the representation $\{\mathbbm{1}\oplus\ket{2}\!\bra{2},X\oplus\ket{2}\!\bra{2},Y\oplus\ket{2}\!\bra{2},Z\oplus\ket{2}\!\bra{2}\}$ on the output space. Then, by Theorem~\ref{thm-EA_cap_covariant}, the entangle\-ment-assisted classical capacity of the erasure channel is equal to the mutual information of its Choi state.
	
	The Choi state $\rho_{AB}^{\mathcal{E}_p}$ of the erasure channel is
	\begin{equation}
		\rho_{AB}^{\mathcal{E}_p}=(1-p)\ket{\Phi^+}\!\bra{\Phi^+}_{AB}+p\frac{\mathbbm{1}_A}{2}\otimes\ket{2}\!\bra{2}.
	\end{equation}
	It is straightforward to verify that
	\begin{align}
		H(A)_{\rho^{\mathcal{E}_p}}&=1,\\
		H(B)_{\rho^{\mathcal{E}_p}}&=-(1-p)\log_2\!\left(\frac{1-p}{2}\right)-p\log_2p,\\
		H(AB)_{\rho^{\mathcal{E}_p}}&=-(1-p)\log_2(1-p)-p\log_2\!\left(\frac{p}{2}\right),
	\end{align}
	so that
	\begin{equation}\label{eq-ea_cap_erasure}
		C_{\operatorname{EA}}(\mathcal{E}_p)=I(A;B)_{\rho^{\mathcal{E}_p}}=2(1-p).
	\end{equation}
	
	In general, as discussed in Section~\ref{sec:QM-over:qudit-erasure}, we can consider an erasure channel $\mathcal{E}_p^{(d)}$ acting on a qudit system with dimension $d$ and orthonormal basis $\{\ket{1},\dotsc,\ket{d}\}$, so that the output of the erasure channel is a state on a $(d+1)$-dimensional system, i.e.,
	\begin{equation}
		\mathcal{E}_p^{(d)}(\rho)=(1-p)\rho+p\Tr[\rho]\ket{d+1}\!\bra{d+1}.
	\end{equation}
	Then, it is straightforward to see that the qudit erasure channel is irreducibly covariant with respect to the group $\mathbb{Z}_d\times\mathbb{Z}_d$, with the corresponding representation on the input space being $\{W_{z,x}:0\leq z,x\leq d-1\}$ and the representation on the output space being $\{W_{z,x}\oplus\ket{d+1}\!\bra{d+1}:0\leq z,x\leq d-1\}$. Therefore, by reasoning analogous to the above, we obtain
	\begin{equation}
		C_{\operatorname{EA}}(\mathcal{E}_p^{(d)})=2(1-p)\log_2d.
	\end{equation}

\subsection{Generalized Amplitude Damping Channel}\label{subsec-eacc_GADC}

	In \eqref{eq-gen_amp_damp}, we defined the generalized amplitide damping channel $\mathcal{A}_{\gamma,N}$ as the channel with the four Kraus operators in \eqref{eq-gen_amp_damp_Kraus1} and \eqref{eq-gen_amp_damp_Kraus2}, i.e.,
	\begin{equation}
		\mathcal{A}_{\gamma,N}(\rho)=A_1\rho A_1^\dagger+A_2\rho A_2^\dagger+A_3\rho A_3^\dagger +A_4\rho A_4^\dagger,
	\end{equation}
	where
	\begin{align}
		A_1=\sqrt{1-N}\begin{pmatrix}1&0\\0&\sqrt{1-\gamma}\end{pmatrix},&\quad A_2=\sqrt{1-N}\begin{pmatrix}0&\sqrt{\gamma}\\0&0\end{pmatrix},\\
		A_3=\sqrt{N}\begin{pmatrix}\sqrt{1-\gamma}&0\\0&1\end{pmatrix},&\quad A_4=\sqrt{N}\begin{pmatrix}0&0\\\sqrt{\gamma}&0\end{pmatrix}.
	\end{align}
	
	Now, it is straightforward to verify that $ZA_1=A_1Z$, $ZA_2=-A_2Z$, $ZA_3=A_3Z$, and $ZA_4=-ZA_4$. Therefore,
	\begin{equation}
		\mathcal{A}_{\gamma,N}(Z\rho Z)=Z\mathcal{A}_{\gamma,N}(\rho)Z
	\end{equation}
	for every state $\rho$. The generalized amplitude damping channel is thus covariant with respect to $\{\mathbbm{1},Z\}$, which can be viewed as a representation of the group $G=\mathbb{Z}_2$ on both the input and output spaces of the channel (see Exercise~\ref{exer-GADC}). Note that this representation does not satisfy the property in \eqref{eq-G_irrep_twirl}.
	
	Nevertheless, we can use the expression in \eqref{eq-EA_cap_cov_chan} to determine the entan\-glement-assisted classical capacity of the generalized amplitude damping ch\-annel. First, observe that the channel $\rho\mapsto\frac{1}{|G|}\sum_{g\in G}U_g\rho U_g^\dagger$ is same as the completely dephasing channel defined in \eqref{eq-completely_dephasing_channel}. Since the output of the completely dephasing channel is always diagonal in the basis $\{\ket{0},\ket{1}\}$, by \eqref{eq-EA_cap_cov_chan} it suffices to optimize over pure states $\psi_{RA}$ such that the reduced state $\psi_A$ is diagonal in the basis $\{\ket{0},\ket{1}\}$. Thus, up to an (irrelevant) unitary on the system $R$, the pure states $\psi_{RA}$ in \eqref{eq-EA_cap_cov_chan} can be taken to have the form
	\begin{equation}
		\ket{\psi}_{RA}=\sqrt{1-p}\ket{0,0}_{RA}+\sqrt{p}\ket{1,1}_{RA}
	\end{equation}
	for $p\in[0,1]$. For every such state, it is straightforward to show that the corresponding output state $\omega_{RB}=(\mathcal{A}_{\gamma,N})_{A\to B}(\psi_{RA})$ has four eigenvalues: $\left(1-p\right)\gamma N$, $
		\left(1-N\right)p\gamma$, and
	\begin{align}
		\lambda_{\pm} & \coloneqq \frac{1}{2}\left(1-\gamma(N+p-2Np)\pm f(N,p,\gamma) \right),\\
		f(N,p,\gamma) & \coloneqq \sqrt{\gamma^2(N-p)^2-2\gamma(N+p-2Np)+1},
	\end{align}
	which means that
	\begin{multline}
		H(RB)_{\rho^{\mathcal{A}_{\gamma,N}}} =-(1-p)\gamma N\log_2((1-p)\gamma N)\\
		 -(1-N)p\gamma\log_2((1-N)p\gamma)\\
		 -\lambda_+\log_2\lambda_+-\lambda_-\log_2\lambda_-.
	\end{multline}
	Since $\omega_R=\psi_R=(1-p)\ket{0}\!\bra{0}+p\ket{1}\!\bra{1}$, we have that $H(R)_{\rho^{\mathcal{A}_{\gamma,N}}}=h_2(p)$. Finally, we have
	\begin{equation}
		H(B)_{\rho^{\mathcal{A}_{\gamma,N}}} =h_2(\gamma(p-N)+1-p)).
	\end{equation}
	Therefore,
	\begin{multline}\label{eq-EA_cap_gen_amp_damp}
		I(\mathcal{A}_{\gamma,N})
		=\max_{p\in[0,1]}\{h_2(p) +h_2(\gamma(p-N)+1-p))
		 \\
		 +(1-p)\gamma N\log_2((1-p)\gamma N)\\
		 +(1-N)p\gamma\log_2((1-N)p\gamma)\\
		 +\lambda_+\log_2\lambda_++\lambda_-\log_2\lambda_-\}.
	\end{multline}
	
	In the case $N=0$, the channel $\mathcal{A}_{\gamma,0}$ is the amplitude damping channel $\mathcal{A}_{\gamma}$ (see \eqref{eq-amplitude_damping_channel}). Using \eqref{eq-EA_cap_gen_amp_damp}, we find that
	\begin{equation}\label{eq-ea_cap_amp_damp}
		I(\mathcal{A}_{\gamma})=\max_{p\in[0,1]}\left\{h_2(p(1-\gamma))+h_2(p)-h_2(\gamma p)\right\}
	\end{equation}
	for all $\gamma\in[0,1]$.

\section{Summary}

In this chapter, we formally defined and studied entanglement-assisted classical communication. We began with the fundamental one-shot setting, in which a quantum channel is used just once for entanglement-assisted classical communication, and we defined the one-shot entanglement-assisted classical capacity in \eqref{eq:EA-comm:ea-capacity}. We then derived upper and lower bounds on the one-shot capacity (Propositions~\ref{prop-eac:one-shot-bound-meta} and \ref{prop-eac:one-shot-lower_bound}), making a fundamental link between communication and hypothesis testing for both bounds. To derive the upper bound, the main conceptual point was to compare an actual protocol for entanglement-assisted communication with a useless one. This approach led to the hypothesis testing mutual information as an upper bound. To derive the lower bound, we employed the combined approach of sequential decoding and position-based coding, which at its core, is about how well a correlated state can be distinguished from a product state. Stepping back a bit, this is conceptually similar to the idea behind the converse upper bound, which ultimately features a comparison between a correlated state and a product state. We can consider the one-shot setting to contain the fundamental information theoretic argument for entanglement-assisted communication.

With the one-shot setting in hand, we moved on the asymptotic setting, in which the channel is allowed to be used multiple times (as a model of how communication channels would be used in practice). We defined various notions of communication rates, including achievable rates, capacity, weak converse rates, strong converse rates, and strong converse capacity. With the fundamental one-shot bounds in place, we then substituted one use of the channel $\mathcal{N}$ with $n$ uses (the tensor-product channel $\mathcal{N}^{\otimes n}$) and applied various technical arguments to prove that the mutual information of a channel is equal to both its capacity and strong converse capacity  for entanglement-assisted communication. As a main step to establish the capacity, we proved that the mutual information of a channel is additive, and as a main step to establish the strong converse capacity, we proved that the sandwiched R\'enyi mutual information of a channel is additive.

Finally, we calculated the entanglement-assisted classical for several key channels, including the depolarizing, erasure, and generalized amplitude damping channels, in order to illustrate the theory on some concrete examples.

As it turns out, the strongest results known in quantum information theory are for the entanglement-assisted capacity. The results stated above hold for all quantum channels, and thus can be viewed from the physics perspective as universal physical laws delineating the ultimate limits of entanglement-assisted classical communication for any physical process (i.e., described by a quantum channel). In this sense, shared entanglement simplifies quantum information theory immensely.

Going forward from here, the same concepts such as capacity, achievable rate, etc.~can be defined for different communication tasks (i.e., unassisted classical communication, quantum communication, private communication, etc.). What changes is that the known results  are not as strong as they are for the entanglement-assisted setting. We know the capacity of these other communication tasks only for certain subclasses of channels. This might be considered unfortunate, but a different perspective is that it is exciting, because rather exotic phenomena such as superadditivity and superactivation can occur.

\section{Bibliographic Notes}

Entanglement-assisted classical communication was devised by \citet{PhysRevLett.83.3081}, as an information-theoretic extension of super-dense coding. The entanglement-assisted classical capacity theorem was proven by \citet{bennett2002entanglement}, and \citet{Hol01a} gave a different proof for this theorem.

Entanglement-assisted classical communication in the one-shot setting was considered by \citet{datta2013one-EA,MW12,DTW14,AJW17b,QWW17,AJW19}. Proposition~\ref{prop-eac:one-shot-bound-meta} is due to \citet{MW12}, and the proof given here was found independently by \citet{QWW17} and \citet{AJW19}.

The position-based coding method was introduced by \citet{AJW17b}. It can be understood as a quantum generalization of pulse position modulation \citep{verdu1990channel,eltit03}. Sequential decoding was considered by \citet{giovannetti2012achieving,Sen11,Wilde20130259,Gao15,OMW19}, and Theorem~\ref{thm-q_union_bd} is due to \citet{OMW19}. Proposition~\ref{prop-eac:one-shot-lower_bound} is due to \citet{QWW17}, and the proof given here was found by \citet{OMW19}.

The strong converse for entanglement-assisted classical capacity was established by \citet{BDHSW12} and \citet{GW15}, with the latter paper employing the R\'enyi entropic method used in this chapter. Eq.~\eqref{eq-sand_rel_mut_inf_alt} and the additivity of sandwiched R\'enyi mutual information of bipartite states (Proposition~\ref{prop-sand_rel_mut_inf_additive}) were established by \citet{Bei13}. Eq.~\eqref{eq-sand_rel_mut_inf_alt_second} was established by \citet{GW15}, and the completely-bound $1\to \alpha$ norm was studied in depth by \citet{DJKR06}. Theorem~\ref{thm-sand_rel_ent_additivity} was proven by \citet{GW15}, by employing the multiplicativity of completely bounded norms (Eq.~\eqref{eq-CB_alpha_norm_mult}) found by \citet{DJKR06}.

The entanglement-assisted classical capacity of the depolarizing and erasure channels was evaluated by \citet{PhysRevLett.83.3081},  the same capacity for the amplitude damping channel was evaluated by \citet{GF05}, and  the same capacity for the generalized amplitude damping channel was evaluated by \citet{LM07}.

The proofs in Appendix~\ref{app-sand_ren_mut_inf_chan_limit} are due to \citet{CMW14}, and the proofs in Appendices~\ref{app-CB_alpha_norm_alt} and \ref{app-CB_norm_mult_pf} are due to \citet{Jencova06} (with the proofs in this book containing some slight variations). The Lieb concavity theorem (Theorem~\ref{thm:lieb-concavity}) is due to \citet{L73}.

\begin{subappendices}

\section{Proof of Theorem~\ref{thm-q_union_bd}}\label{sec-q_union_bd_pf}

	Theorem~\ref{thm-q_union_bd} is a consequence of the following more general result:

	\begin{theorem}{thm-T1}
		Let $\{P_i\}_{i=1}^{N}$ be a finite set of projectors. For every vector $\ket{\psi}$ and $c>0$,
		\begin{equation}
			\begin{aligned}
			&\norm{\ket{\psi}}_2^2-\norm{P_NP_{N-1}\dotsb P_1\ket{\psi}}_2^2\leq (1+c)\norm{(\mathbbm{1}-P_N)\ket{\psi}}_2^2\\
			&\qquad\qquad\qquad\qquad+(2+c+c^{-1})\sum_{i=2}^{N-1}\norm{(\mathbbm{1}-P_i)\ket{\psi}}_2^2\\
			&\qquad\qquad\qquad\qquad+(2+c^{-1})\norm{(\mathbbm{1}-P_1)\ket{\psi}}_2^2.
			\end{aligned}
		\end{equation}
	\end{theorem}

	Indeed, recall from Theorem~\ref{thm-spectral_theorem} that every density operator $\rho$ has a spectral decomposition of the following form:
	\begin{equation}
		\rho=\sum_{j\in\mathcal{J}}p(j)\ket{\psi_j}\!\bra{\psi_j},
	\end{equation}
	where $p:\mathcal{J}\to[0,1]$ is a probability distribution, and $\{\ket{\psi_j}\}_{j\in\mathcal{J}}$ is an orthonormal set of eigenvectors. Applying Theorem~\ref{thm-T1}, we find that
	\begin{align}
		& 1-\Tr[P_{N}P_{N-1}\cdots P_{1}\ket{\psi_j}\!\bra{\psi_j}P_{1}\cdots P_{N-1}]\nonumber\\
		& =\norm{\ket{\psi_j}}_{2}^{2}-\norm{P_{N}P_{N-1}\cdots P_{1}\ket{\psi_j}}_{2}^{2}\\
		&  \leq(1+c)\norm{\left(\mathbbm{1}-P_{N}\right)\ket{\psi_j}}_{2}^{2}+(2+c+c^{-1})\sum_{i=2}^{N-1}\norm{\left(\mathbbm{1}-P_{i}\right)\ket{\psi_j}}_{2}^{2}\nonumber\\
		&  \qquad\qquad+(2+c^{-1})\norm{\left(\mathbbm{1}-P_{1}\right)\ket{\psi_j}}_{2}^{2}\\
		&  =\left(1+c\right)  \Tr[\left(\mathbbm{1}-P_{N}\right)\ket{\psi_j}\!\bra{\psi_j}]+(2+c+c^{-1}) \sum_{i=2}^{N-1}\Tr[(\mathbbm{1}-P_{i})\ket{\psi_j}\!\bra{\psi_j}]\nonumber\\
		&  \qquad\qquad +(2+c^{-1})\Tr[\left(\mathbbm{1}-P_{1}\right) \ket{\psi_j}\!\bra{\psi_j}].\label{eq-q_union_bd_pf1}
	\end{align}
	The reduction from Theorem~\ref{thm-T1} to Theorem~\ref{thm-q_union_bd} follows by averaging the above inequality over the probability distribution $p:\mathcal{J}\to[0,1]$ and from the fact that the right-hand side of the resulting inequality can be bounded from above by
	\begin{equation}
		(1+c)\Tr[(\mathbbm{1}-P_N)\rho]+(2+c+c^{-1})\sum_{i=1}^{N-1}\Tr[(\mathbbm{1}-P_i)\rho],
	\end{equation}
	so that
	\begin{equation}
		\begin{aligned}
		&1-\Tr[P_NP_{N-1}\dotsb P_1\rho P_1\dotsb P_{N-1}P_N]\\
		&\qquad\leq (1+c)\Tr[(\mathbbm{1}-P_N)\rho]+(2+c+c^{-1})\sum_{i=1}^{N-1}\Tr[(\mathbbm{1}-P_i)\rho].
		\end{aligned}
	\end{equation}	
	We therefore shift our focus to proving Theorem~\ref{thm-T1}, and we do so with the
aid of several lemmas. To simplify the notation, we hereafter employ the following shorthands:%
	\begin{align}
		\norm{\dotsb}_{\psi}&\equiv\norm{\cdots\ket{\psi}}_{2},\label{eq:shorthand-1}\\
		\langle\dotsb\rangle_{\psi} &  \equiv\bra{\psi}\dotsb\ket{\psi},\label{eq:shorthand_2}\\
		\widehat{P}_{i}  &  \equiv \mathbbm{1}-P_{i}, \label{eq:q_i}%
	\end{align}
	where for a given operator $A$ the expression $\langle A\rangle_{\psi}=\bra{\psi}A\ket{\psi}$ is assumed to mean $\bra{\psi}(A\ket{\psi})$. Furthermore, we also assume without loss of generality that the vector $\ket{\psi}$ in Theorem~\ref{thm-T1} is a unit vector. This assumption can be dropped by scaling the resulting inequality by an arbitrary positive number corresponding to the norm of $\ket{\psi}$.

	First recall that, due to the fact that $P^2=P$ for every projector $P$, we have the following identities holding for all $i\in\{1,2,\dotsc,N\}$:
	\begin{equation}\label{projection}
		\begin{aligned}
		\langle \widehat{P}_{i}P_{i-1}\cdots P_{1}\rangle_{\psi}&=\langle \widehat{P}_{i}\widehat{P}_{i}P_{i-1}\dotsb P_{1}\rangle_{\psi},\\
		\langle P_{1}\cdots P_{i}\rangle_{\psi}&=\langle P_{1}\cdots P_{i}P_{i}\rangle_{\psi},
		\end{aligned} 
	\end{equation}
	under the convention that $P_{i-1}\dotsb P_{1}=P_{1}\dotsb P_{i-1}=\mathbbm{1}$ for $i=1$.

	\begin{Lemma}{lem-L1}
		For a set $\{P_{i}\}_{i=1}^{N}$, a unit vector $\ket{\psi}$, and employing the shorthand in \eqref{eq:shorthand-1}--\eqref{eq:q_i}, we have the following identities and inequality:
		\begin{align}
			\sum_{i=1}^{N}\langle \widehat{P}_{i}P_{i-1}\dotsb P_{1}\rangle_{\psi} &  =1-\langle P_{N}\dotsb P_{1}\rangle_{\psi},\label{pro1}\\
			\sum_{i=1}^{N}\langle P_{1}\dotsb P_{i-1}\widehat{P}_{i}\rangle_{\psi} &  =1-\langle P_{1}\dotsb P_{N}\rangle_{\psi},\label{pro2}\\
			\sum_{i=1}^{N}\langle P_{1}\dotsb P_{i-1}\widehat{P}_{i}P_{i-1}\dotsb P_{1}\rangle_{\psi} & =1-\langle P_{1}\dotsb P_{N}\dotsb P_{1}\rangle_{\psi},\label{pro3}\\
			1-\sqrt{\langle P_{N}\rangle_{\psi}}\sqrt{\langle P_{1}\dotsb P_{N}\dotsb P_{1}\rangle_{\psi}}  \nonumber\\
			&  \!\!\!\!\!\!\!\!\!\!\!\!\!\!\!\!\!\!\!\!\!\leq\sum_{i=1}^{N}\sqrt{\langle \widehat{P}_{i}\rangle_{\psi}}\sqrt{\langle P_{1}\cdots P_{i-1}\widehat{P}_{i}P_{i-1}\cdots P_{1}\rangle_{\psi}}, \label{pro4}%
		\end{align}
		under the convention that $P_{i-1}\cdots P_{1}=P_{1}\cdots P_{i-1}=\mathbbm{1}$ for $i=1$.
	\end{Lemma}

	\begin{Proof}
		The following identities are straightforward to verify:
		\begin{align}
			1  &  =\langle \widehat{P}_{1}\rangle_{\psi}+\langle \widehat{P}_{2}P_{1}\rangle_{\psi}+\dotsb+\langle \widehat{P}_{N-1}P_{N-2}\dotsb P_{1}\rangle_{\psi}+\langle \widehat{P}_{N}P_{N-1}\dotsb P_{1} \rangle_{\psi}\nonumber\\
			&\qquad+\langle P_{N}P_{N-1}\dotsb P_{1}\rangle_{\psi},\label{rel1}\\
			1  &  =\langle \widehat{P}_{1}\rangle_{\psi}+\langle P_{1}\widehat{P}_{2}\rangle_{\psi}+\dotsb+\langle P_{1}\dotsb P_{N-2}\widehat{P}_{N-1}\rangle_{\psi}+\langle P_{1}\dotsb P_{N-1}\widehat{P}_{N}\rangle_{\psi}\nonumber\\
			&\qquad+\langle P_{1}\dotsb P_{N-1}P_{N}\rangle_{\psi},\label{rel2}\\
			1  &  =\langle \widehat{P}_{1}\rangle_{\psi}+\langle P_{1}\widehat{P}_{2}P_{1}\rangle_{\psi}+\dotsb+\langle P_{1}\dotsb P_{N-2}\widehat{P}_{N-1}P_{N-2}\dotsb P_{1}\rangle_{\psi}\nonumber\\
			&  \qquad+\langle P_{1}\dotsb P_{N-1}\widehat{P}_{N}P_{N-1}\dotsb P_{1} \rangle_{\psi}+\langle P_{1}\dotsb P_{N-1}P_{N}P_{N-1}\dotsb P_{1}\rangle_{\psi}. \label{rel3}%
		\end{align}
		Consequently, from the equalities in \eqref{rel1}, \eqref{rel2}, and \eqref{rel3}, we obtain \eqref{pro1}, \eqref{pro2}, and \eqref{pro3}, respectively. The following equality is a direct consequence of \eqref{rel1} and \eqref{projection}:
		\begin{multline}
			1=\langle \widehat{P}_{1}\rangle_{\psi}+\langle \widehat{P}_{2}\widehat{P}_{2}P_{1}\rangle_{\psi}+\dotsb+\langle \widehat{P}_{N-1}\widehat{P}_{N-1}P_{N-2}\dotsb P_{1}\rangle_{\psi}\label{rel4}\\
			+\langle \widehat{P}_{N}\widehat{P}_{N}P_{N-1}\dotsb P_{1}\rangle_{\psi}+\langle P_{N}P_{N}P_{N-1}\cdots P_{1}\rangle_{\psi}.
		\end{multline}
		By applying the Cauchy--Schwarz inequality from \eqref{eq-Cauchy_Schwarz_HS}  to \eqref{rel4}, we find that%
		\begin{multline}
			1\leq\langle \widehat{P}_{1}\rangle_{\psi}+\sqrt{\langle \widehat{P}_{2}\rangle_{\psi}}\sqrt{\langle P_{1} \widehat{P}_{2}P_{1}\rangle_{\psi}}\\
			+\dotsb+\sqrt{\langle \widehat{P}_{N}\rangle_{\psi}}\sqrt{\langle P_{1}\dotsb P_{N-1}\widehat{P}_{N}P_{N-1}\dotsb P_{1}\rangle_{\psi}}\\
			+\sqrt{\langle P_{N}\rangle_{\psi}}\sqrt{\langle P_{1}\dotsb P_{N-1}P_{N} P_{N-1}\dotsb P_{1}\rangle_{\psi}},
		\end{multline}
		from which \eqref{pro4} immediately follows.
	\end{Proof}

	\begin{Lemma}{lem-L3}
		For a set $\{P_{i}\}_{i=1}^{N}$ of projectors, a unit vector $\ket{\psi}$, and employing the shorthand in \eqref{eq:shorthand-1}--\eqref{eq:q_i}, the following inequality holds for $N\geq2$:%
		\begin{equation}
			\sum_{i=1}^{N}\Norm{\widehat{P}_{i}(\mathbbm{1}-P_{i-1}\dotsb P_{1})}_{\psi}^{2}\leq \sum_{i=1}^{N-1}\Norm{\widehat{P}_{i}}_{\psi}^{2},
		\end{equation}
		under the convention that $P_{i-1}\dotsb P_{1}=P_{1}\dotsb P_{i-1}=\mathbbm{1}$ for $i=1$. Equivalently,%
		\begin{equation}
			\sum_{i=2}^{N}\Norm{\widehat{P}_{i}(\mathbbm{1}-P_{i-1}\dotsb P_{1})}_{\psi}^{2}\leq \sum_{i=1}^{N-1}\Norm{\widehat{P}_{i}}_{\psi}^{2},
		\end{equation}
		due to the aforementioned convention.
	\end{Lemma}

	\begin{Proof}
		Consider the following chain of equalities:%
		\begin{align}
			&  \sum_{i=1}^{N}\Norm{\widehat{P}_{i}(\mathbbm{1}-P_{i-1}\dotsb P_{1})}_{\psi}^{2}\nonumber\\
			&  =\sum_{i=1}^{N}\Norm{\widehat{P}_{i}-\widehat{P}_{i}P_{i-1}\dotsb P_{1}}_{\psi}^{2}\\
			&  =\sum_{i=1}^{N}\left(\Norm{\widehat{P}_{i}}_{\psi}^{2}-\langle
\widehat{P}_{i}P_{i-1}\dotsb P_{1}\rangle_{\psi}-\langle P_{1}\dotsb P_{i-1}\widehat{P}_{i}%
\rangle_{\psi}\right.\nonumber\\
			&\qquad\qquad\qquad\qquad\qquad\qquad\left.+\langle P_{1}\dotsb P_{i-1}\widehat{P}_{i}P_{i-1}\dotsb P_{1}\rangle_{\psi}\right) \label{apro1}\\
			&  =\left(\sum_{i=1}^{N}\Norm{\widehat{P}_{i}}_{\psi}^{2}\right)-1+\langle P_{N}\dotsb P_{1}\rangle_{\psi}-1+\langle P_{1}\dotsb P_{N}\rangle_{\psi}\nonumber\\
			&\qquad\qquad\qquad\qquad\qquad\qquad\qquad+1-\langle P_{1}\dotsb P_{N}\dotsb P_{1}\rangle_{\psi}\label{apro2}\\
			&  =\left(\sum_{i=1}^{N}\Norm{\widehat{P}_{i}}_{\psi}^{2}\right)-1+\langle P_{N}P_{N}P_{N-1}\dotsb P_{1}\rangle_{\psi}+\langle P_{1}\dotsb P_{N-1}P_{N}P_{N}\rangle_{\psi}\nonumber\\
			&\qquad\qquad\qquad\qquad\qquad\qquad\qquad-\langle P_{1}\dotsb P_{N}\dotsb P_{1}\rangle_{\psi}. \label{eq:connector}%
		\end{align}
		To obtain \eqref{apro1}, we used the identities in \eqref{projection}. Next, to get \eqref{apro2}, the identities in~\eqref{pro1}, \eqref{pro2}, and \eqref{pro3} of Lemma~\ref{lem-L1} were used. Continuing, we have that%
		\begin{align}
			\text{Eq.}~\eqref{eq:connector}  &  \leq\left(\sum_{i=1}^{N}\Norm{\widehat{P}_{i}}_{\psi}^{2}\right)-1-\langle P_{1}\cdots P_{N}\dotsb P_{1}\rangle_{\psi}\nonumber\\
			&\qquad\qquad\qquad\qquad+2\sqrt{\langle P_{N}\rangle_{\psi}}\sqrt{\langle P_{1}\dotsb P_{N}\dotsb P_{1}\rangle_{\psi}}\label{apro3}\\
			&  =\left(\sum_{i=1}^{N}\Norm{\widehat{P}_{i}}_{\psi}^{2}\right)-1+\langle P_{N}\rangle_{\psi}\nonumber\\
			&\qquad\qquad\qquad\qquad-\left(\sqrt{\langle P_{N}\rangle_{\psi}}-\sqrt{\langle P_{1}\dotsb P_{N}\dotsb P_{1}\rangle_{\psi}}\right)^{2}\\
			&  \leq\left(\sum_{i=1}^{N}\Norm{\widehat{P}_{i}}_{\psi}^{2}\right)-\Norm{\widehat{P}_{N}}_{\psi}^{2}=\sum_{i=1}^{N-1}\Norm{\widehat{P}_{i}}_{\psi}^{2}.
		\end{align}
		To obtain \eqref{apro3}, the Cauchy-Schwarz inequality was employed.
	\end{Proof}

	We are now in a position to prove Theorem~\ref{thm-T1}.

\subsubsection*{Proof of Theorem~\ref{thm-T1}}

	Consider that%
	\begin{align}
		1-\Norm{P_{N}\dotsb P_{1}}_{\psi}^{2}  &  =1-\langle P_{1}\dotsb P_{N}\dotsb P_{1}\rangle_{\psi} \nonumber\\
		&  \qquad\qquad+2\left(  1-\sqrt{\langle P_{N}\rangle_{\psi}}\sqrt{\langle P_{1}\dotsb P_{N}\dotsb P_{1}\rangle_{\psi}}\right)\nonumber\\
		&\qquad\qquad-2\left(  1-\sqrt{\langle P_{N}\rangle_{\psi}}\sqrt{\langle P_{1}\dotsb P_{N}\dotsb P_{1}\rangle_{\psi}}\right) \\
		&  =2\left(  1-\sqrt{\langle P_{N}\rangle_{\psi}}\sqrt{\langle P_{1}\cdots P_{N}\dotsb P_{1}\rangle_{\psi}}\right) \nonumber\\
		&  \qquad\qquad-\left(\sqrt{\langle P_{N}\rangle_{\psi}}-\sqrt{\langle P_{1}\cdots P_{N}\dotsb P_{1}\rangle_{\psi}}\right)^{2}\nonumber\\
		&\qquad\qquad -1+\langle P_{N}\rangle_{\psi}. \label{eq:connector-3}%
	\end{align}
	Continuing, we have that%
	\begin{align}
		&\text{Eq.}~\eqref{eq:connector-3} \nonumber\\
		&\quad  \leq-\Norm{\widehat{P}_{N}}_{\psi}^{2}+2\left(1-\sqrt{P_{N}}\sqrt{\langle P_{1}\dotsb P_{N}\dotsb P_{1}\rangle_{\psi}}\right) \label{mpro1}\\
		&\quad  \leq-\Norm{\widehat{P}_{N}}_{\psi}^{2}+2\sum_{i=1}^{N}\sqrt{\langle
\widehat{P}_{i}\rangle_{\psi}}\sqrt{\langle P_{1}\dotsb P_{i-1}\widehat{P}_{i}P_{i-1}\dotsb P_{1}\rangle_{\psi}}\label{mpro2}\\
		&\quad  \leq-\Norm{\widehat{P}_{N}}_{\psi}^{2}+2\sum_{i=1}^{N}\sqrt{\langle
\widehat{P}_{i}\rangle_{\psi}}\left(\Norm{\widehat{P}_{i}}_{\psi} +\Norm{\widehat{P}_{i}(\mathbbm{1}-P_{i-1}\dotsb P_{1})}_{\psi}\right). \label{mpro3}%
	\end{align}
	First, \eqref{mpro1} is obtained by observing that
	\begin{align}
		-\left(\sqrt{\langle P_{N}\rangle_{\psi}}-\sqrt{\langle P_{1}\dotsb P_{N}\dotsb P_{1}\rangle_{\psi}}\right)^{2}-1+\langle P_{N}\rangle_{\psi}&\leq-1+\langle P_{N}\rangle_{\psi}\nonumber\\
		&=-\Norm{\widehat{P}_{N}}_{\psi}^{2}.
	\end{align}
	Next, \eqref{mpro2} follows from \eqref{pro4} of Lemma~\ref{lem-L1}. Then, \eqref{mpro3} is a consequence of the triangle inequality:
	\begin{align}
		\sqrt{\langle P_1\dotsb P_{i-1}\widehat{P}_iP_{i-1}\dotsb P_1\rangle_{\psi}}&=\sqrt{\langle P_1\dotsb P_{i-1}\widehat{P}_i\widehat{P}_iP_{i-1}\dotsb P_1\rangle_{\psi}}\\
		&=\Norm{\widehat{P}_{i}P_{i-1}\dotsb P_1}_{\psi}\\
		&=\Norm{\widehat{P}_i(-\mathbbm{1}+\mathbbm{1}-P_{i-1}\dotsb P_1)}_{\psi}\\
		&=\Norm{-\widehat{P}_i+\widehat{P}_i(\mathbbm{1}-P_{i-1}\dotsb P_1)}_{\psi}\\
		&\leq \Norm{\widehat{P}_i}_{\psi}+\Norm{\widehat{P}_i(\mathbbm{1}-P_{i-1}\dotsb P_1)}_{\psi}.
	\end{align}
	Continuing, we have that%
	\begin{align}
		&\text{Eq.}~\eqref{mpro3}  \nonumber\\
		&\quad=-\Norm{\widehat{P}_{N}}_{\psi}^{2}+2\sum
_{i=1}^{N}\Norm{\widehat{P}_{i}}_{\psi}^{2}+2\sum_{i=1}^{N}\left(\Norm{\widehat{P}_{i}}_{\psi} \Norm{\widehat{P}_{i}(\mathbbm{1}-P_{i-1}\dotsb P_{1})}_{\psi} \right) \\
		&\quad  =-\Norm{\widehat{P}_{N}}_{\psi}^{2}+2\sum_{i=1}^{N}\Norm{\widehat{P}_{i}}_{\psi}^{2}+2\sum_{i=2}^{N}\left(\Norm{\widehat{P}_{i}}_{\psi} \Norm{\widehat{P}_{i}(\mathbbm{1}-P_{i-1}\dotsb P_{1})}_{\psi}\right) \label{eq:apply-conv}\\
		&\quad  \leq-\Norm{\widehat{P}_{N}}_{\psi}^{2}+2\sum_{i=1}^{N}\Norm{\widehat{P}_{i}}_{\psi}^{2}\nonumber\\
		&\qquad\qquad\qquad\quad+\sum_{i=2}^{N}\left(c\Norm{\widehat{P}_{i}}_{\psi}^{2}+c^{-1}\Norm{\widehat{P}_{i}(\mathbbm{1}-P_{i-1}\dotsb P_{1})}_{\psi}^{2}\right) \label{mpro3bis}\\
		&\quad  \leq-\Norm{\widehat{P}_{N}}_{\psi}^{2}+2\sum_{i=1}^{N}\Norm{\widehat{P}_{i}}_{\psi}^{2}+c\sum_{i=2}^{N}\Norm{\widehat{P}_{i}}_{\psi}^{2}+c^{-1}\sum_{i=1}^{N-1}\Norm{\widehat{P}_{i}}_{\psi}^{2}\label{mpro4}\\
		&\quad  \leq(1+c)\Norm{\widehat{P}_{N}}_{\psi}^{2}+(2+c^{-1})\Norm{\widehat{P}_{1}}_{\psi}^{2}+(2+c+c^{-1})\sum_{i=2}^{N-1}\Norm{\widehat{P}_{i}}_{\psi}^{2}. \label{mpro5}%
\end{align}
	Eq. \eqref{eq:apply-conv} follows from the convention that $P_{i-1}\dotsb P_{1}=\mathbbm{1}$ for $i=1$. Eq. \eqref{mpro3bis} is a consequence of the inequality $2xy\leq cx^{2}+c^{-1}y^{2}$, holding for $x,y\in\mathbb{R}$ and $c>0$. Finally, \eqref{mpro4} is obtained by using Lemma~\ref{lem-L3}.
	
	\section{The \texorpdfstring{$\alpha\to 1$}{a to 1} Limit of the Sandwic\-hed R\'{e}nyi Mutual Information of a Channel}\label{app-sand_ren_mut_inf_chan_limit}

	In this appendix, we show that
	\begin{equation}
		\lim_{\alpha\to 1^-}\overline{I}_\alpha(\mathcal{N})=\lim_{\alpha\to 1^+}\widetilde{I}_\alpha(\mathcal{N})=I(\mathcal{N}),
	\end{equation}
	where we recall that
	\begin{align}
		\overline{I}_\alpha(\mathcal{N})&=\sup_{\psi_{RA}}D_\alpha(\mathcal{N}_{A\to B}(\psi_{RA})\Vert\psi_R\otimes\mathcal{N}_{A\to B}(\psi_A)),\\
		\widetilde{I}_{\alpha}(\mathcal{N})&=\sup_{\psi_{RA}}\inf_{\sigma_B}\widetilde{D}_{\alpha}(\mathcal{N}_{A\to B}(\psi_{RA})\Vert\psi_R\otimes\sigma_B).
	\end{align}
	In these definitions, $\psi_{RA}$ is a pure state, with the dimension of $R$ equal to the dimension of $A$, and in the definition of $\widetilde{I}_\alpha(\mathcal{N})$ the infimum is over states $\sigma_B$.
	
	All of the arguments presented here are similar to those in Appendix~\ref{app-sand_ren_inf_limit}, which we refer to for additional details.
	
	As a consequence of the fact that $\overline{I}_\alpha(\mathcal{N})$ increases monotonically with $\alpha$ (see Proposition~\ref{prop-Petz_rel_ent}), as well as the fact that $\lim_{\alpha\to 1}D_\alpha(\rho\Vert\sigma)=D(\rho\Vert\sigma)$ (see Proposition~\ref{prop-petz_rel_ent_lim_1}), we find that
	\begin{align}
		\lim_{\alpha\to 1^-}\overline{I}_\alpha(\mathcal{N})&=\sup_{\alpha\in (0,1)}\sup_{\psi_{RA}}D_\alpha(\mathcal{N}_{A\to B}(\psi_{RA})\Vert\psi_R\otimes\mathcal{N}_{A\to B}(\psi_A))\\
		&=\sup_{\psi_{RA}}\sup_{\alpha\in(0,1)}D_\alpha(\mathcal{N}_{A\to B}(\psi_{RA})\Vert\psi_R\otimes\mathcal{N}_{A\to B}(\psi_A))\\
		&=\sup_{\psi_{RA}}D(\mathcal{N}_{A\to B}(\psi_{RA})\Vert\psi_R\otimes\mathcal{N}_{A\to B}(\psi_A))\\
		&=I(\mathcal{N}),
	\end{align}
	as required.
	
	Similarly, for the sandwiched R\'{e}nyi mutual information, we use the fact that it increases monotonically with $\alpha$ (see Proposition~\ref{prop-sand_rel_ent_properties}), along with the fact that $\lim_{\alpha\to 1}\widetilde{D}_\alpha(\rho\Vert\sigma)=D(\rho\Vert\sigma)$ (see Proposition~\ref{prop-sand_ren_ent_lim}), to obtain
	\begin{align}
		\lim_{\alpha\to 1^+}\widetilde{I}_\alpha(\mathcal{N})&=\inf_{\alpha\in(1,\infty)}\sup_{\psi_{RA}}\inf_{\sigma_B}\widetilde{D}_\alpha(\mathcal{N}_{A\to B}(\psi_{RA})\Vert\psi_R\otimes\sigma_B)\\
		&=\sup_{\psi_{RA}}\inf_{\alpha\in(1,\infty)}\inf_{\sigma_B}\widetilde{D}_\alpha(\mathcal{N}_{A\to B}(\psi_{RA})\Vert\psi_R\otimes\sigma_B)\\
		&=\sup_{\psi_{RA}}\inf_{\sigma_B}\inf_{\alpha\in(1,\infty)}\widetilde{D}_\alpha(\mathcal{N}_{A\to B}(\psi_{RA})\Vert\psi_R\otimes\sigma_B)\\
		&=\sup_{\psi_{RA}}\inf_{\sigma_B}D(\mathcal{N}_{A\to B}(\psi_{RA})\Vert\psi_R\otimes\sigma_B)\\
		&=\sup_{\psi_{RA}}D(\mathcal{N}_{A\to B}(\psi_{RA})\Vert\psi_R\otimes\mathcal{N}_{A\to B}(\psi_A))\\
		&=I(\mathcal{N}).
	\end{align}
To obtain the second equality, we made use of the minimax theorem in Theorem~\ref{thm-Mosonyi_minimax} to exchange $\inf_{\alpha\in(1,\infty)}$ and $\sup_{\psi_{RA}}$. Specifically, we applied that theorem to the function 
	\begin{equation}
		(\alpha,\psi_{RA})\mapsto \inf_{\sigma_B}\widetilde{D}_{\alpha}(\mathcal{N}_{A\to B}(\psi_{RA})\Vert\psi_R\otimes\sigma_B),
	\end{equation}
	which is monotonically increasing in the first argument and continuous in the second argument.

\section{Achievability from a Different Point of View}

\label{subsec-EA_comm_ach_diff_POV}

	Here we show that the mutual information $I(\mathcal{N})$ is an achievable rate based on the alternate definition given in Appendix~\ref{chap-str_conv}.
	According to that definition, a rate $R\in\mathbb{R}^+$ is an achievable rate for entanglement-asssisted classical communication over $\mathcal{N}$ if there exists a sequence $\{(n,|\mathcal{M}_n|,\allowbreak\varepsilon_n)\}_{n\in\mathbb{N}}$ of $(n,|\mathcal{M}|,\allowbreak\varepsilon)$ entanglement-assisted classical communication protocols over $n$ uses of $\mathcal{N}$ such that 
	\begin{equation}
	\liminf_{n\to\infty}\frac{1}{n}\log_2|\mathcal{M}_n|\geq R \quad\text{ and } \quad\lim_{n\to\infty}\varepsilon_n=0.
	\end{equation}
	
	To start, let us recall Corollary~\ref{cor-eacc_asymp_lower_bound}, which states that for all $\varepsilon \in (0,1]$, $n \in \mathbb{N}$, and $\alpha \in (0,1)$, there exists an $(n,|\mathcal{M}|,\varepsilon) $ protocol satisfying
	\begin{equation}
	\frac{1}{n}\log_2|\mathcal{M}|\geq \overline{I}_\alpha(\mathcal{N})-\frac{1}{n(1-\alpha)}\log_2\!\left(\frac{2}{\varepsilon}\right)-\frac{3}{n}.
	\label{eq:EA-comm:ach-starting-point-app}
	\end{equation}
	Fix constants $\delta_1, \delta_2$ satisfying $0<\delta_2 < \delta_1 <1$. Pick $\alpha_n \in (0,1)$ and $\varepsilon_n \in (0,1]$ as follows:
	\begin{equation}
	\alpha_n \coloneqq 1-n^{-(1-\delta_1)}, \qquad \varepsilon_n \coloneqq 2^{-n^{\delta_2}}.
	\end{equation}
	Plugging in to \eqref{eq:EA-comm:ach-starting-point-app}, we find that there exists a sequence of $\{(n,|\mathcal{M}_n|,\varepsilon_n)\}_{n\in\mathbb{N}}$ protocols satisfying
	\begin{align}
	\frac{1}{n}\log_2|\mathcal{M}_n| & \geq \overline{I}_{\alpha_n}(\mathcal{N})-\frac{1}{n(1-\alpha_n)}\log_2\!\left(\frac{2}{\varepsilon_n}\right)-\frac{3}{n} \\
&  = \overline{I}_{\alpha_n}(\mathcal{N}) - \frac{1 + n^{\delta_2}}{n^{\delta_1}}-\frac{3}{n}\\
&  = \overline{I}_{\alpha_n}(\mathcal{N}) - \frac{1}{n^{\delta_1}} - \frac{1}{n^{\delta_1-\delta_2}}-\frac{3}{n}.
	\end{align}
	Now taking the limit $n\to \infty$, we find that
	\begin{align}
	\liminf_{n \to \infty} \frac{1}{n}\log_2|\mathcal{M}_n| & \geq \liminf_{n \to \infty}\left[\overline{I}_{\alpha_n}(\mathcal{N}) - \frac{1}{n^{\delta_1}} - \frac{1}{n^{\delta_1-\delta_2}}-\frac{3}{n} \right]\\
	& = I(\mathcal{N}),	
	\end{align}
	and $\lim_{n\to \infty}\varepsilon_n = 0$. The equality above follows because $\lim_{n \to \infty}\alpha_n = 1$ and $\lim_{\alpha\to 1}\overline{I}_\alpha(\mathcal{N})=I(\mathcal{N})$ (see Appendix~\ref{app-sand_ren_mut_inf_chan_limit} for a proof). Thus, it follows that the mutual information rate $I(\mathcal{N})$ is achievable according to the alternate definition given in Appendix~\ref{chap-str_conv}.
	 
	In the approach detailed above, the error probability decays subexponentially to zero (i.e., slower than an exponential decay) and the rate increases to $I(\mathcal{N})$ with increasing $n$. If we would like to have exponential decay of the error probability, then we can instead fix the rate $R$ to be a constant satisfying $R<I(\mathcal{N})$ and reconsider the analysis.
Rearranging the inequality in \eqref{eq-eacc_achieve_pf3} in order to get a bound on $\varepsilon_n$, we find that for all $\alpha\in(0,1)$, there exists a sequence of $\{(n,|\mathcal{M}_n|,\varepsilon_n)\}_{n\in\mathbb{N}}$ protocols satisfying
	\begin{equation}\label{eq-eacc_ach_alt_pf1}
		\varepsilon_n\leq 2\cdot 2^{-n(1-\alpha)\left(\overline{I}_\alpha(\mathcal{N})-R-\frac{3}{n}\right)}.
	\end{equation}
 Since $R<I(\mathcal{N})$, $\lim_{\alpha\to 1}\overline{I}_\alpha(\mathcal{N})=I(\mathcal{N})$, and since $\overline{I}_\alpha(\mathcal{N})$ is monotonically increasing in $\alpha$ (this follows from Proposition~\ref{prop-sand_rel_ent_properties}), there exists an $\alpha^*<1$ such that $\overline{I}_{\alpha^*}(\mathcal{N})>R$. Applying the bound in \eqref{eq-eacc_ach_alt_pf1} to this value of $\alpha$, we find that
	\begin{equation}
		\varepsilon_n\leq 2\cdot 2^{-n(1-\alpha^*)\left(\overline{I}_{\alpha^*}(\mathcal{N})-R-\frac{3}{n}\right)}.
	\end{equation}
	Then, taking the limit $n\to\infty$ on both sides of this inequality, we conclude that $\lim_{n\to\infty}\varepsilon_n\allowbreak= 0$ exponentially fast. 
	Thus, by choosing $R$ as a constant satisfying $R<I(\mathcal{N})$ it follows that there exists a sequence of $\{(n,2^{nR},\varepsilon_n)\}_{n\in\mathbb{N}}$ protocols such that the error probability $\varepsilon_n$ decays exponentially fast to zero.

\section{Proof of Lemma~\ref{lem-sand_rel_mut_inf_alt}}\label{lem-sand_rel_mut_inf_alt_pf}

	We start by writing the definition of $\widetilde{I}_{\alpha}(A;B)_\rho$ as
	\begin{align}
		\widetilde{I}_{\alpha}(A;B)_\rho&=\inf_{\sigma_B}\frac{\alpha}{\alpha-1}\log_2\norm{\left(\rho_A^{\frac{1-\alpha}{2\alpha}}\otimes\sigma_B^{\frac{1-\alpha}{2\alpha}}\right)\rho_{AB}\left(\rho_A^{\frac{1-\alpha}{2\alpha}}\otimes\sigma_B^{\frac{1-\alpha}{2\alpha}}\right)}_{\alpha}\\
		&=\frac{\alpha}{\alpha-1}\log_2\inf_{\sigma_B}\norm{\left(\rho_A^{\frac{1-\alpha}{2\alpha}}\otimes\sigma_B^{\frac{1-\alpha}{2\alpha}}\right)\rho_{AB}\left(\rho_A^{\frac{1-\alpha}{2\alpha}}\otimes\sigma_B^{\frac{1-\alpha}{2\alpha}}\right)}_{\alpha},\label{eq-sand_rel_mut_inf_alt2}
	\end{align}
	where $\alpha>1$ and the optimization is over states $\sigma_B$. Then, for every purification $\ket{\psi}_{ABC}$ of $\rho_{AB}$, we have
	\begin{multline}
		\left(\rho_A^{\frac{1-\alpha}{2\alpha}}\otimes\sigma_B^{\frac{1-\alpha}{2\alpha}}\right)\rho_{AB}\left(\rho_A^{\frac{1-\alpha}{2\alpha}}\otimes\sigma_B^{\frac{1-\alpha}{2\alpha}}\right)\\
		 =\Tr_C\!\left[\left(\rho_A^{\frac{1-\alpha}{2\alpha}}\otimes\sigma_B^{\frac{1-\alpha}{2\alpha}}\right)\ket{\psi}\!\bra{\psi}_{ABC}\left(\rho_A^{\frac{1-\alpha}{2\alpha}}\otimes\sigma_B^{\frac{1-\alpha}{2\alpha}}\right)\right].
	\end{multline}
	Now, the operator inside $\Tr_C$ on the last line in the  equation above is rank one, which means that
	\begin{equation*}
		\begin{aligned}
		&\Tr_C\!\left[\left(\rho_A^{\frac{1-\alpha}{2\alpha}}\otimes\sigma_B^{\frac{1-\alpha}{2\alpha}}\right)\ket{\psi}\!\bra{\psi}_{ABC}\left(\rho_A^{\frac{1-\alpha}{2\alpha}}\otimes\sigma_B^{\frac{1-\alpha}{2\alpha}}\right)\right]\quad\text{and}\\
		&\quad \Tr_{AB}\!\left[\left(\rho_A^{\frac{1-\alpha}{2\alpha}}\otimes\sigma_B^{\frac{1-\alpha}{2\alpha}}\right)\ket{\psi}\!\bra{\psi}_{ABC}\left(\rho_A^{\frac{1-\alpha}{2\alpha}}\otimes\sigma_B^{\frac{1-\alpha}{2\alpha}}\right)\right]
		\end{aligned}
	\end{equation*}
	have the same non-zero eigenvalues. This means that their Schatten norms are equal, so that 
	\begin{align}
		&\widetilde{I}_{\alpha}(A;B)_\rho\nonumber\\
		&=\frac{\alpha}{\alpha-1}\log_2\inf_{\sigma_B}\norm{\Tr_C\!\left[\left(\rho_A^{\frac{1-\alpha}{2\alpha}}\otimes\sigma_B^{\frac{1-\alpha}{2\alpha}}\right)\ket{\psi}\!\bra{\psi}_{ABC}\left(\rho_A^{\frac{1-\alpha}{2\alpha}}\otimes\sigma_B^{\frac{1-\alpha}{2\alpha}}\right)\right]}_{\alpha}\\
		&=\frac{\alpha}{\alpha-1}\log_2\inf_{\sigma_B}\norm{\Tr_{AB}\!\left[\left(\rho_A^{\frac{1-\alpha}{2\alpha}}\otimes\sigma_B^{\frac{1-\alpha}{2\alpha}}\right)\ket{\psi}\!\bra{\psi}_{ABC}\left(\rho_A^{\frac{1-\alpha}{2\alpha}}\otimes\sigma_B^{\frac{1-\alpha}{2\alpha}}\right)\right]}_{\alpha}\label{eq-sand_rel_mut_inf_alt_pf2}
	\end{align}
	Now, we use the variational characterization of the Schatten norm in \eqref{eq-Schatten_norm_var}, which states that for every operator $X$,
	\begin{equation}\label{eq-sand_rel_mut_inf_alt_pf}
		\norm{X}_p=\sup_{\norm{Y}_{p'}=1}\abs{\Tr[Y^\dagger X]}
	\end{equation}
	for all $1\leq p\leq\infty$, where $p'$ satisfies $\frac{1}{p}+\frac{1}{p'}=1$. Note that if $X$ is positive semi-definite, then we can restrict the optimization to positive semi-definite operators $Y$. Using \eqref{eq-sand_rel_mut_inf_alt_pf}, with $p=\alpha$ and $p'=\frac{\alpha}{\alpha-1}$, on the expression in \eqref{eq-sand_rel_mut_inf_alt_pf2}, and since the argument of the norm in that expression is positive semi-definite, we can optimize over positive semi-definite operators $\tau_C$ to obtain
	\begin{align}
		&\widetilde{I}_{\alpha}(A;B)_\rho\nonumber\\
		&=\frac{\alpha}{\alpha-1}\log_2\inf_{\sigma_B}\sup_{\tau_C}\Tr\!\left[\tau_C^{\frac{\alpha-1}{\alpha}}\Tr_{AB}\!\left[\left(\rho_A^{\frac{1-\alpha}{2\alpha}}\otimes\sigma_B^{\frac{1-\alpha}{2\alpha}}\right)\times\right.\right.\nonumber\\
		&\qquad\qquad\qquad\qquad\qquad\qquad\qquad\qquad\left.\left.\ket{\psi}\!\bra{\psi}_{ABC}\left(\rho_A^{\frac{1-\alpha}{2\alpha}}\otimes\sigma_B^{\frac{1-\alpha}{2\alpha}}\right)\right]\right]\\
		&=\frac{\alpha}{\alpha-1}\log_2\inf_{\sigma_B}\sup_{\tau_C}\Tr\!\left[\left(\rho_A^{\frac{1-\alpha}{2\alpha}}\otimes\sigma_B^{\frac{1-\alpha}{2\alpha}}\otimes\tau_C^{\frac{\alpha-1}{2\alpha}}\right)\right.\nonumber\\
		&\qquad\qquad\qquad\qquad\qquad\qquad\left.\times \ket{\psi}\!\bra{\psi}_{ABC}\left(\rho_A^{\frac{1-\alpha}{2\alpha}}\otimes\sigma_B^{\frac{1-\alpha}{2\alpha}}\otimes\tau_C^{\frac{\alpha-1}{2\alpha}}\right)\right]\\
		&=\frac{\alpha}{\alpha-1}\log_2\inf_{\sigma_B}\sup_{\tau_C}\Tr\!\left[\left(\rho_A^{\frac{1-\alpha}{\alpha}}\otimes\sigma_B^{\frac{1-\alpha}{\alpha}}\otimes\tau_C^{\frac{\alpha-1}{\alpha}}\right)\ket{\psi}\!\bra{\psi}_{ABC}\right]\\
		&=\frac{\alpha}{\alpha-1}\log_2\sup_{\tau_C}\inf_{\sigma_B}\Tr\!\left[\left(\rho_A^{\frac{1-\alpha}{\alpha}}\otimes\sigma_B^{\frac{1-\alpha}{\alpha}}\otimes\tau_C^{\frac{\alpha-1}{\alpha}}\right)\ket{\psi}\!\bra{\psi}_{ABC}\right]\label{eq-chi_alpha_superadditive_pf_1},
	\end{align}
	where the last line follows by applying Sion's minimax theorem (Theorem~\ref{thm-Sion_minimax}) to the function
	\begin{equation}
		(\tau_C,\sigma_B)\mapsto \Tr\!\left[\left(\rho_A^{\frac{1-\alpha}{\alpha}}\otimes\sigma_B^{\frac{1-\alpha}{\alpha}}\otimes\tau_C^{\frac{\alpha-1}{\alpha}}\right)\ket{\psi}\!\bra{\psi}_{ABC}\right],
	\end{equation}
	which is convex in the first argument because $\sigma_B\mapsto\sigma_B^{\frac{1-\alpha}{\alpha}}$ is operator convex and concave in the second argument because $\tau_C\mapsto\tau_C^{\frac{\alpha-1}{\alpha}}$ is operator concave.
		
	Finally, we use Proposition~\ref{prop-Schatten_pos_var}, which is that 
	\begin{equation}
		\norm{X}_p=\inf_{\substack{Y\geq 0,\\\norm{Y}_{p'}=1}}\Tr[XY]
	\end{equation}
	for all $0<p<1$, where $\frac{1}{p}+\frac{1}{p'}=1$. Applying this to \eqref{eq-chi_alpha_superadditive_pf_1} with $p'=\frac{\alpha}{1-\alpha}$, so that $p=\frac{\alpha}{2\alpha-1}$, we conclude that
	\begin{align}
		&\widetilde{I}_{\alpha}(A;B)_\rho\nonumber\\
		&=\frac{\alpha}{\alpha-1}\log_2\sup_{\tau_C}\inf_{\sigma_B}\Tr\!\left[\left(\rho_A^{\frac{1-\alpha}{\alpha}}\otimes\sigma_B^{\frac{1-\alpha}{\alpha}}\otimes\tau_C^{\frac{\alpha-1}{\alpha}}\right)\ket{\psi}\!\bra{\psi}_{ABC}\right]\\
		&=\frac{\alpha}{\alpha-1}\log_2\sup_{\tau_C}\inf_{\sigma_B}\Tr\!\left[\sigma_B^{\frac{1-\alpha}{\alpha}}\Tr_{AC}\!\left[\left(\rho_A^{\frac{1-\alpha}{\alpha}}\otimes\tau_C^{\frac{\alpha-1}{\alpha}}\right)\ket{\psi}\!\bra{\psi}_{ABC}\right]\right]\\
		&=\frac{\alpha}{\alpha-1}\log_2\sup_{\tau_C}\norm{\Tr_{AC}\!\left[\left(\rho_A^{\frac{1-\alpha}{\alpha}}\otimes\tau_C^{\frac{\alpha-1}{\alpha}}\right)\ket{\psi}\!\bra{\psi}_{ABC}\right]}_{\frac{\alpha}{2\alpha-1}},
	\end{align}
	the last line of which is \eqref{eq-sand_rel_mut_inf_alt}, as required. 

	To prove \eqref{eq-sand_rel_mut_inf_alt_second}, we use the fact that the definition of the sandwiched R\'{e}nyi mutual information of a bipartite state can be written as in \eqref{eq-sand_rel_mut_inf_alt2}, i.e.,
	\begin{equation}
		\widetilde{I}_\alpha(R;B)_\rho=\frac{\alpha}{\alpha-1}\log_2\inf_{\sigma_B}\norm{\left(\rho_R^{\frac{1-\alpha}{2\alpha}}\otimes\sigma_B^{\frac{1-\alpha}{2\alpha}}\right)\rho_{RB}\left(\rho_R^{\frac{1-\alpha}{2\alpha}}\otimes\sigma_B^{\frac{1-\alpha}{2\alpha}}\right)}_{\alpha},
	\end{equation}
	which means that the definition in \eqref{eq-sand_rel_mut_inf_chan_1} can be written as
	\begin{equation}
		\begin{aligned}
		&\widetilde{I}_{\alpha}(\mathcal{N})\\
		&=\frac{\alpha}{\alpha-1}\sup_{\psi_{RA}}\log_2\inf_{\sigma_B}\norm{\left(\psi_R^{\frac{1-\alpha}{2\alpha}}\otimes\sigma_B^{\frac{1-\alpha}{2\alpha}}\right)\mathcal{N}_{A\to B}(\psi_{RA})\left(\psi_R^{\frac{1-\alpha}{2\alpha}}\otimes\sigma_B^{\frac{1-\alpha}{2\alpha}}\right)}_{\alpha}.
		\end{aligned}
	\end{equation}
	Now, we use the fact mentioned in \eqref{eq-pure_state_vec2}, which is that for every pure state $\psi_{RA}$, with the systems $R$ and $A$ having the same dimensions, there exists an operator $X_R$ such that
	\begin{equation}
		\ket{\psi}_{RA}=(X_R\otimes\mathbbm{1}_A)\ket{\Gamma}_{RA},
	\end{equation}
	and $\Tr[X_R^\dag X_R] = 1$, with this latter equality following from \eqref{eq-trace_identity}. By taking a polar decomposition of $X_R$ as $X_R = U_R \sqrt{\tau}_R$ for a unitary $U_R$ and a state $\tau_R$ (see Theorem~\ref{thm-polar_decomposition}), we can then write
	\begin{equation}
		\ket{\psi}_{RA}=(U_R \sqrt{\tau_R}\otimes\mathbbm{1}_A)\ket{\Gamma}_{RA}.
	\end{equation}
	This implies that
	\begin{align}
		\psi_R
		&=\Tr_B[\mathcal{N}_{A\to B}(\psi_{RA})]\\
		&=\Tr_B[(U_R \sqrt{\tau_R}\otimes\mathbbm{1}_B)\mathcal{N}_{A\to B}(\Gamma_{RA})(\sqrt{\tau_R}U_R^\dag \otimes\mathbbm{1}_B)]\\
		&=U_R \tau_R U_R^\dag,
	\end{align}
	where the last equality follows because $\mathcal{N}$ is trace preserving and $\Tr_A[\ket{\Gamma}\!\bra{\Gamma}_{RA}]=\mathbbm{1}_R$. Using this, we find that
	\begin{align}
	& \left(\psi_R^{\frac{1-\alpha}{2\alpha}}\otimes\sigma_B^{\frac{1-\alpha}{2\alpha}}\right)
	\mathcal{N}_{A\to B}(\psi_{RA})\left(\psi_R^{\frac{1-\alpha}{2\alpha}}\otimes\sigma_B^{\frac{1-\alpha}{2\alpha}}\right) \notag \\
	& = \left(U_R \tau_R^{\frac{1-\alpha}{2\alpha} }U_R^\dag \otimes\sigma_B^{\frac{1-\alpha}{2\alpha}}\right)
	U_R \sqrt{\tau_R}\mathcal{N}_{A\to B}(\Gamma_{RA})\sqrt{\tau_R}U_R^\dag \left(U_R \tau_R^{\frac{1-\alpha}{2\alpha}}U_R^\dag \otimes\sigma_B^{\frac{1-\alpha}{2\alpha}}\right)\notag \\
		& = \left(U_R \tau_R^{\frac{1}{2\alpha} } \otimes\sigma_B^{\frac{1-\alpha}{2\alpha}}\right)
	\mathcal{N}_{A\to B}(\Gamma_{RA}) \left( \tau_R^{\frac{1}{2\alpha}}U_R^\dag \otimes\sigma_B^{\frac{1-\alpha}{2\alpha}}\right).
	\end{align}
	Therefore, by exploiting unitary invariance of the $\alpha$-Schatten norm, we can write $\widetilde{I}_\alpha(\mathcal{N})$ as
	\begin{align}
		&\widetilde{I}_\alpha(\mathcal{N})\nonumber\\
		&=\frac{\alpha}{\alpha-1}\sup_{\rho_R}\log_2\inf_{\sigma_B}\norm{\left(\rho_R^{\frac{1}{2\alpha}}\otimes\sigma_B^{\frac{1-\alpha}{2\alpha}}\right)\mathcal{N}_{A\to B}(\Gamma_{RA})\left(\rho_R^{\frac{1}{2\alpha}}\otimes\sigma_B^{\frac{1-\alpha}{2\alpha}}\right)}_{\alpha}\\
		&=\frac{\alpha}{\alpha-1}\sup_{\rho_R}\log_2\inf_{\sigma_B}\norm{\mathcal{N}_{A\to B}(\Gamma_{RA})^{\frac{1}{2}}\left(\rho_R^{\frac{1}{\alpha}}\otimes\sigma_B^{\frac{1-\alpha}{\alpha}}\right)\mathcal{N}_{A\to B}(\Gamma_{RA})^{\frac{1}{2}}}_{\alpha}.
	\end{align}
	Now, the function
	\begin{equation}
		(\rho_R,\sigma_B)\mapsto \norm{\mathcal{N}_{A\to B}(\Gamma_{RA})^{\frac{1}{2}}\left(\rho_R^{\frac{1}{\alpha}}\otimes\sigma_B^{\frac{1-\alpha}{\alpha}}\right)\mathcal{N}_{A\to B}(\Gamma_{RA})^{\frac{1}{2}}}_{\alpha}
	\end{equation}
	is concave in the first argument (this follows from Lemma~\ref{lem:EA-comm-nontrivial-concavity} in Appendix~\ref{app-CB_norm_mult_pf} below) and convex in the second argument (this follows from the operator convexity of $\sigma_B\mapsto\sigma_B^{\frac{1-\alpha}{\alpha}}$ for $\alpha>1$ and convexity of the Schatten norm). Thus, by the Sion minimax theorem (Theorem~\ref{thm-Sion_minimax}), we can exchange $\sup_{\rho_R}$ and $\inf_{\sigma_B}$. Also, we define the completely positive map $\mathcal{S}_{\sigma_B}^{(\alpha)}$ by $\mathcal{S}_{\sigma_B}^{(\alpha)}(\cdot)\coloneqq \sigma_B^{\frac{1-\alpha}{2\alpha}}(\cdot)\sigma_B^{\frac{1-\alpha}{2\alpha}}$. We can then further rewrite $\widetilde{I}_\alpha(\mathcal{N})$ as
	\begin{align}
		&\widetilde{I}_\alpha(\mathcal{N})\nonumber\\
		&=\frac{\alpha}{\alpha-1}\inf_{\sigma_B}\log_2\sup_{\rho_R}\norm{(\mathcal{S}_{\sigma_B}^{(\alpha)}\circ\mathcal{N}_{A\to B})\left(\rho_R^{\frac{1}{2\alpha}}\ket{\Gamma}\!\bra{\Gamma}_{RA}\rho_R^{\frac{1}{2\alpha}}\right)}_{\alpha}\\
		&=\frac{\alpha}{\alpha-1}\inf_{\sigma_B}\log_2\norm{\mathcal{S}_{\sigma_B}^{(\alpha)}\circ\mathcal{N}}_{\text{CB},~1\to\alpha},
	\end{align}
	where, to arrive at the last line, we used the definition in \eqref{eq-operator_CB_alpha_norm}. Also, consider that the optimum in \eqref{eq-operator_CB_alpha_norm} is achieved when $\Tr[Y_R]=1$. Therefore,
	\begin{equation}
		\widetilde{I}_\alpha(\mathcal{N})=\frac{\alpha}{\alpha-1}\inf_{\sigma_B}\log_2\norm{\mathcal{S}_{\sigma_B}^{(\alpha)}\circ\mathcal{N}}_{\text{CB},~1\to\alpha},
	\end{equation}
	as required.

\section{Alternate Expression for the \texorpdfstring{$1\to\alpha$}{1 to a} CB Norm}\label{app-CB_alpha_norm_alt}

	In this section, we show that
	\begin{equation}\label{eq-eacc_CB_1alpha_norm_alt}
		\norm{\mathcal{M}}_{\text{CB},1\to\alpha}=\sup_{Y_{RA}> 0}\frac{\norm{\mathcal{M}_{A\to B}(Y_{RA})}_{\alpha}}{\norm{\Tr_A[Y_{RA}]}_{\alpha}}
	\end{equation}
	for every completely positive map $\mathcal{M}$. We start with the expression in \eqref{eq-operator_CB_alpha_norm} and write it alternatively as follows:
	\begin{align}
		\norm{\mathcal{M}}_{\text{CB},1\to\alpha}&=\sup_{\substack{Y_R>0,\\\Tr[Y_R]\leq 1}}\norm{\mathcal{M}_{A\to B}\!\left(Y_R^{\frac{1}{2\alpha}}\ket{\Gamma}\!\bra{\Gamma}_{RA}Y_R^{\frac{1}{2\alpha}}\right)}_{\alpha}\\
		&=\sup_{\substack{Y_R>0,\\\norm{Y_R}_{\alpha}\leq 1}}\norm{\mathcal{M}_{A\to B}\!\left(Y_R^{\frac{1}{2}}\ket{\Gamma}\!\bra{\Gamma}_{RA}Y_R^{\frac{1}{2}}\right)}_{\alpha}\\
		&=\sup_{Y_R>0}\frac{\norm{\mathcal{M}_{A\to B}\!\left(Y_R^{\frac{1}{2}}\ket{\Gamma}\!\bra{\Gamma}_{RA}Y_R^{\frac{1}{2}}\right)}_{\alpha}}{\norm{Y_R}_{\alpha}}.\label{eq-CB_alpha_norm_alt_pf1}
	\end{align}
	Now, we use the fact that there is a one-to-one correspondence between the operators $Y_R$ and the vectors
	\begin{equation}
		\ket{\Gamma^Y}_{RA}\coloneqq (Y_R^{\frac{1}{2}}\otimes\mathbbm{1}_A)\ket{\Gamma}_{RA}.
	\end{equation}
	This allows us to rewrite the optimization in \eqref{eq-CB_alpha_norm_alt_pf1} in terms of such vectors. Then, by employing isometric invariance of the norms with respect to an isometry acting on the reference system $R$, we can restrict the optimization to arbitrary vectors $\ket{\psi}_{RA}$. Therefore, we have that
	\begin{equation}
		\norm{\mathcal{M}}_{\text{CB},1\to\alpha}=\sup_{\psi_{RA}}\frac{\norm{\mathcal{M}_{A\to B}(\psi_{RA})}_{\alpha}}{\norm{\Tr_A[\psi_{RA}]}_{\alpha}},
		\label{eq-EAC:rewrite-CB-1-alph-norm}
	\end{equation}
	where $\psi_{RA} \equiv \ket{\psi}\!\bra{\psi}_{RA}$.
	Since the optimization in \eqref{eq-EAC:rewrite-CB-1-alph-norm} is over a subset of the positive semi-definite operators $Y_{RA}$ (and by approximation), we conclude the inequality
	\begin{equation}
	\norm{\mathcal{M}}_{\text{CB},1\to\alpha} \leq 
	\sup_{Y_{RA}> 0}\frac{\norm{\mathcal{M}_{A\to B}(Y_{RA})}_{\alpha}}{\norm{\Tr_A[Y_{RA}]}_{\alpha}}.
	\end{equation}
	
	It remains to show the opposite inequality.
	Consider a vector $\ket{\phi}_{SRA}$ that purifies $Y_{RA}>0$, in the sense that $\Tr_{S}[\phi_{SRA}] = Y_{RA}$. Then we have that
	\begin{align}
	\frac{\norm{\mathcal{M}_{A\to B}(Y_{RA})}_{\alpha}}{\norm{\Tr_A[Y_{RA}]}_{\alpha}} &= 
	\frac{\norm{(\mathcal{M}_{A\to B} \otimes \Tr_S) (\phi_{SRA})}_{\alpha}}{\norm{\Tr_{SA}[\phi_{SRA}]}_{\alpha}}\\
	& \leq \sup_{\ket{\phi}_{SRA}} \frac{\norm{(\mathcal{M}_{A\to B} \otimes \Tr_S) (\phi_{SRA})}_{\alpha}}{\norm{\Tr_{SA}[\phi_{SRA}]}_{\alpha}} \\
	& =  \norm{\mathcal{M} \otimes \Tr }_{\text{CB},1\to\alpha} \\
	 & = \norm{\mathcal{M}  }_{\text{CB},1\to\alpha} \norm{ \Tr }_{\text{CB},1\to\alpha} \\
	 & = \norm{\mathcal{M}  }_{\text{CB},1\to\alpha} 
	\end{align}
	The third-to-last equality follows from \eqref{eq-EAC:rewrite-CB-1-alph-norm}. The second-to-last equality follows from \eqref{eq-CB_alpha_norm_mult}, as shown in Appendix~\ref{app-CB_norm_mult_pf}. The final inequality follows because $\norm{ \Tr }_{\text{CB},1\to\alpha}=1$, as can be readily verified.
	

\section{Proof of the Multiplicativity of the \texorpdfstring{$1\to\alpha$}{1 to a} CB Norm}\label{app-CB_norm_mult_pf}

	In this appendix, we prove the statement in \eqref{eq-CB_alpha_norm_mult}, which is that
	\begin{equation}
		\norm{\mathcal{M}_1\otimes\mathcal{M}_2}_{\text{CB},1\to\alpha}=\norm{\mathcal{M}_1}_{\text{CB},1\to\alpha}\norm{\mathcal{M}_2}_{\text{CB},1\to\alpha}
	\end{equation}
	for every two completely positive maps $\mathcal{M}_1$ and $\mathcal{M}_2$ and all $\alpha>1$. Recall from \eqref{eq-operator_CB_alpha_norm} that 
	\begin{equation}
		\norm{\mathcal{M}}_{\text{CB},1\to\alpha}=\sup_{\substack{Y_{R}>0,\\\operatorname{Tr}[Y_{R}]\leq 1}}\norm{Y_{R}^{\frac{1}{2\alpha}}\Gamma_{RB}^{\mathcal{M}}Y_{R}^{\frac{1}{2\alpha}}}_{\alpha},
	\end{equation}
	where $\Gamma_{RB}^{\mathcal{M}}\coloneqq\mathcal{M}_{A\rightarrow B}(\Gamma_{RA})$ is the Choi representation of $\mathcal{M}$, and the dimension of $R$ is the same as the dimension of $A$. For a completely positive map $\mathcal{P}_{C\rightarrow D}$, let us also define
	\begin{equation}\label{eq:p-to-p-norm-EA-comm}
		\left\Vert \mathcal{P}\right\Vert _{\alpha\rightarrow\alpha}\coloneqq\sup_{Z_{C}>0}\frac{\left\Vert \mathcal{P}_{C\rightarrow D}(Z_{C})\right\Vert _{\alpha}}{\left\Vert Z_{C}\right\Vert _{\alpha}}.
	\end{equation}
	Note that
	\begin{align}
		\norm{\mathcal{P}}_{\alpha\to\alpha}&=\sup_{Z_C>0}\frac{\norm{\mathcal{P}_{C\to D}(Z_C)}_{\alpha}}{\norm{Z_C}_\alpha}\label{eq:EA-comm-CB-to-comp-1}\\
		&=\sup_{\substack{Z_C>0,\\\norm{Z_C}_\alpha\leq 1}}\norm{\mathcal{P}_{C\to D}(Z_C)}_{\alpha}\label{eq:EA-comm-CB-to-comp-2}\\
		&=\sup_{\substack{Y_C>0,\\\Tr[Y_C]\leq 1}}\norm{\mathcal{P}_{C\to D}(Y_C^{\frac{1}{\alpha}})}_{\alpha},\label{eq:EA-comm-CB-to-comp-3}
	\end{align}
	where the last equality follows from the substitution $Y_C = Z_C^{\alpha}$ so that $\Tr[Y_C]= \Tr[Z_C^\alpha] = \norm{Z_C}_{\alpha}^{\alpha}$.

	Now, it immediately follows that%
	\begin{equation}
		\norm{\mathcal{M}_{1}\otimes\mathcal{M}_{2}\right\Vert
_{\operatorname{CB},1\rightarrow\alpha}\geq\left\Vert \mathcal{M}%
_{1}}_{\text{CB},1\to\alpha}\norm{\mathcal{M}_{2}}_{\text{CB},1\to\alpha}.
	\end{equation}
	Indeed, due to the fact that the Choi representation of $\mathcal{M}_{1}\otimes\mathcal{M}_{2}$ has a tensor-product form (see \eqref{eq-Choi_rep_tensor_prod}), we can restrict the optimization in the definition of the norm $\norm{\mathcal{M}_1\otimes\mathcal{M}_2}_{\text{CB},1\to\alpha}$ to tensor-product operators $Y_{R_1}\otimes Y_{R_2}$ to obtain
	\begin{align}
		&\norm{\mathcal{M}_{1}\otimes\mathcal{M}_{2}}_{\text{CB},1\to\alpha}\nonumber\\
		&=\sup_{\substack{Y_{R_{1}R_{2}}>0,\\ \operatorname{Tr}[Y_{R_{1}R_{2}}]\leq 1}}\norm{Y_{R_{1}R_{2}}^{\frac{1}{2\alpha}}(\Gamma_{R_{1}B_{1}}^{\mathcal{M}_{1}}\otimes\Gamma_{R_{2}B_{2}}^{\mathcal{M}_{2}})Y_{R_{1}R_{2}}^{\frac{1}{2\alpha}}}_{\alpha}\\
		&\geq\sup_{\substack{Y_{R_{1}}>0,Y_{R_{2}}>0,\\\operatorname{Tr}[Y_{R_{1}}]\leq 1,\operatorname{Tr}[Y_{R_{2}}]\leq 1}}\norm{(Y_{R_{1}}^{\frac{1}{2\alpha}}\otimes Y_{R_{2}}^{\frac{1}{2\alpha}})(\Gamma_{R_{1}B_{1}}^{\mathcal{M}_{1}}\otimes\Gamma_{R_{2}B_{2}}^{\mathcal{M}_{2}})(Y_{R_{1}}^{\frac{1}{2\alpha}}\otimes
Y_{R_{2}}^{\frac{1}{2\alpha}})}_{\alpha}\\
		&=\sup_{\substack{Y_{R_{1}}>0,Y_{R_{2}}>0,\\\operatorname{Tr}[Y_{R_{1}}]\leq 1,\operatorname{Tr}[Y_{R_{2}}]\leq1}}\norm{Y_{R_{1}}^{\frac{1}{2\alpha}}\Gamma_{R_{1}B_{1}}^{\mathcal{M}_{1}}Y_{R_{1}}^{\frac{1}{2\alpha}}\otimes Y_{R_{2}}^{\frac{1}{2\alpha}}\Gamma_{R_{2}B_{2}}^{\mathcal{M}_{2}}Y_{R_{2}}^{\frac{1}{2\alpha}}}_{\alpha}\\
		&=\sup_{\substack{Y_{R_{1}}>0,Y_{R_{2}}>0,\\\operatorname{Tr}[Y_{R_{1}}]\leq 1,\operatorname{Tr}[Y_{R_{2}}]\leq1}}\norm{Y_{R_{1}}^{\frac{1}{2\alpha}}\Gamma_{R_{1}B_{1}}^{\mathcal{M}_{1}}Y_{R_{1}}^{\frac{1}{2\alpha}}}_{\alpha}\norm{Y_{R_{2}}^{\frac{1}{2\alpha}}\Gamma_{R_{2}B_{2}}^{\mathcal{M}_{2}}Y_{R_{2}}^{\frac{1}{2\alpha}}}_{\alpha}\\
		&=\sup_{\substack{Y_{R_{1}}>0,\\\operatorname{Tr}[Y_{R_{1}}]\leq 1}}\norm{Y_{R_{1}}^{\frac{1}{2\alpha}}\Gamma_{R_{1}B_{1}}^{\mathcal{M}_{1}}Y_{R_{1}}^{\frac{1}{2\alpha}}}_{\alpha}\sup_{\substack{Y_{2}>0,\\\operatorname{Tr}[Y_{R_{2}}]\leq1}}\norm{Y_{R_{2}}^{\frac{1}{2\alpha}}\Gamma_{R_{2}B_{2}}^{\mathcal{M}_{2}}Y_{R_{2}}^{\frac{1}{2\alpha}}}_{\alpha}\\
		&=\norm{\mathcal{M}_{1}}_{\text{CB},1\to\alpha}\norm{\mathcal{M}_{2}}_{\text{CB},1\to\alpha}.
	\end{align}

	Now we establish the opposite inequality. Let $\mathcal{U}_{A\to BE}^{\mathcal{M}}$ be a linear map that extends $\mathcal{M}_{A\to B}$, in the sense that there is a linear operator $U_{A\to BE}^{\mathcal{M}}$ such that%
	\begin{align}
		\mathcal{U}_{A\rightarrow BE}^{\mathcal{M}}(Y_{A}) &  =U_{A\to BE}^{\mathcal{M}}Y_{A}(U_{A\rightarrow BE}^{\mathcal{M}})^{\dagger},\\
		\Tr_{E}[\mathcal{U}_{A\to BE}^{\mathcal{M}}(Y_{A})]&=\mathcal{M}_{A\to B}(Y_{A}).
	\end{align}
	Due to the fact that $Y_{R}^{\frac{1}{2\alpha}}\mathcal{U}_{A\to BE}^{\mathcal{M}}(\Gamma_{RA})Y_{R}^{\frac{1}{2\alpha}}$ is a rank-one operator, and from an application of a generalization of the Schmidt decomposition (Theorem~\ref{thm-Schmidt}), the following operators have the same non-zero eigenvalues:%
	\begin{align}
		\Tr_{E}[Y_{R}^{\frac{1}{2\alpha}}\mathcal{U}_{A\to BE}^{\mathcal{M}}(\Gamma_{RA})Y_{R}^{\frac{1}{2\alpha}}] & =Y_{R}^{\frac{1}{2\alpha}}\mathcal{M}_{A\to B}(\Gamma_{RA})Y_{R}^{\frac{1}{2\alpha}}
	\end{align}
	and
	\begin{align}
		\Tr_{RB}[Y_{R}^{\frac{1}{2\alpha}}\mathcal{U}_{A\to BE}^{\mathcal{M}}(\Gamma_{RA})Y_{R}^{\frac{1}{2\alpha}}] & = \Tr_{R}[Y_{R}^{\frac{1}{2\alpha}}\mathcal{M}_{A\rightarrow B}^{c}(\Gamma_{RA})Y_{R}^{\frac{1}{2\alpha}}]\\
		&=\mathcal{M}_{A\rightarrow E}^{c}((Y_{A}^{\t})^{\frac{1}{\alpha}}),
	\end{align}
	where $\mathcal{M}_{A\rightarrow E}^{c}=\Tr_{B}\circ
\mathcal{U}_{A\to BE}^{\mathcal{M}}$ denotes the complementary map and the last equality follows by applying the transpose trick in \eqref{eq-transpose_trick} and the fact that $\Tr_{R}[\Gamma_{RA}]=\mathbbm{1}_{A}$. Then we find that%
	\begin{align}
		\norm{\mathcal{M}}_{\text{CB},1\to\alpha}&=\sup_{\substack{Y_{A}>0,\\\Tr[Y_{A}]\leq 1}}\norm{\mathcal{M}_{A\to E}^{c}((Y_{A}^{\t})^{\frac{1}{\alpha}})}_{\alpha}\\
		&=\sup_{\substack{Y_{A}>0,\\\Tr[Y_{A}]\leq 1}}\norm{\mathcal{M}_{A\to E}^{c}(Y_{A}^{\frac{1}{\alpha}})}_{\alpha}\\
		&=\sup_{Y_{A}>0}\frac{\norm{\mathcal{M}_{A\to E}^{c}(Y_{A})}_{\alpha}}{\norm{Y_{A}}_{\alpha}}\\
		&=\norm{\mathcal{M}^c}_{\alpha\to\alpha},
	\end{align}
	where the second equality follows from \eqref{eq:EA-comm-CB-to-comp-1}--\eqref{eq:EA-comm-CB-to-comp-3}. So we have that%
	\begin{equation}\label{eq:EA-comm-CB-to-comp}
		\norm{\mathcal{M}_{A\to B}}_{\text{CB},1\to\alpha}=\norm{\mathcal{M}_{A\to E}^{c}}_{\alpha\to\alpha}.
	\end{equation}
	Finally, for an arbitrary operator $Y_{A_{1}A_{2}}$ satisfying $Y_{A_{1}A_{2}}>0$, and setting $X_{A_{1}E_{2}}\coloneqq\left(\mathcal{M}_{2}^c\right)_{A_{2}\to E_{2}}(Y_{A_{1}A_{2}})$, we can write
	\begin{align}
		&\left((\mathcal{M}_1^c)_{A_1\to E_1}\otimes(\mathcal{M}_2^c)_{A_2\to E_2}\right)(Y_{A_1A_2})\nonumber\\
		&\quad=(\mathcal{M}_1^c)_{A_1\to E_1}\left((\mathcal{M}_2^c)_{A_2\to E_2}(Y_{A_1A_2})\right)\\
		&\quad=(\mathcal{M}_1^c)_{A_1\to E_1}(X_{A_1E_2}).
	\end{align}
	Then, multiplying and dividing by $\norm{X_{A_1E_2}}_{\alpha}=\norm{(\mathcal{M}_2^c)_{A_2\to E_2}(Y_{A_1A_2})}_{\alpha}$ gives
	\begin{align}
		&\frac{\norm{\left(\left(\mathcal{M}_{1}^c\right)_{A_{1}\to E_{1}}\otimes\left(\mathcal{M}_{2}^c\right)_{A_{2}\to E_{2}}\right)(Y_{A_{1}A_{2}})}_{\alpha}}{\norm{Y_{A_{1}A_{2}}}_{\alpha}}\nonumber\\
		&=\frac{\norm{\left(\mathcal{M}_{1}^c\right)_{A_{1}\to E_{1}}\left(X_{A_{1}E_{2}}\right)}_{\alpha}}{\norm{X_{A_{1}E_{2}}}_{\alpha}}\frac{\norm{\left(\mathcal{M}_{2}^c\right)_{A_{2}\to E_{2}}(Y_{A_{1}A_{2}})}_{\alpha}}{\norm{Y_{A_{1}A_{2}}}_{\alpha}}\\
		&\leq\left(\sup_{X_{A_{1}E_{2}}>0}\frac{\norm{\left(\mathcal{M}_{1}^c\right)_{A_{1}\to E_{1}}(X_{A_{1}E_{2}})}_{\alpha}}{\norm{X_{A_{1}E_{2}}}_{\alpha}}\right)\nonumber\\
		&\qquad\qquad\qquad\times \left(\sup_{Y_{A_{1}A_{2}}>0}\frac{\norm{\left(\mathcal{M}_{2}^c\right)_{A_{2}\to E_{2}}(Y_{A_{1}A_{2}})}_{\alpha}}{\norm{Y_{A_{1}A_{2}}}_{\alpha}}\right)  \\
		&=\norm{\left(\mathcal{M}_{1}^c\right)_{A_{1}\to E_{1}}\otimes\id_{E_{2}}}_{\alpha\to\alpha}\norm{\id_{A_{1}}\otimes\left(\mathcal{M}_{2}^c\right)_{A_{2}\rightarrow E_{2}}}_{\alpha\to\alpha}\\
		&=\norm{\left(\mathcal{M}_{1}^c\right)_{A_{1}\to E_{1}}}_{\alpha\to\alpha}\norm{\left(\mathcal{M}_{2}^c\right)_{A_{2}\to E_{2}}}_{\alpha\to\alpha}\\
		&=\norm{\mathcal{M}_{1}}_{\text{CB},1\to\alpha}\norm{\mathcal{M}_{2}}_{\text{CB},1\rightarrow\alpha}.
	\end{align}
	The third equality follows from Le\-mma~\ref{lem:EA-comm-collapse-of-CB-norm-for-CP} below. The final equality holds by \eqref{eq:EA-comm-CB-to-comp}. Since $Y_{A_1A_2}$ is arbitrary, we find that
	\begin{align}
		\norm{\mathcal{M}_1\otimes\mathcal{M}_2}_{\text{CB},1\to\alpha}&=\norm{\mathcal{M}_1^c\otimes\mathcal{M}_2^c}_{\alpha\to \alpha}\\
		&=\sup_{Y_{A_1A_2}>0}\frac{\norm{(\mathcal{M}_1^c\otimes\mathcal{M}_2^c)(Y_{A_1A_2})}_{\alpha}}{\norm{Y_{A_1A_2}}_{\alpha}}\\
		&\leq \norm{\mathcal{M}_1}_{\text{CB},1\to\alpha}\norm{\mathcal{M}_2}_{\text{CB},1\to\alpha}.
	\end{align}
	So we have that $\norm{\mathcal{M}_1\otimes\mathcal{M}_2}_{\text{CB},1\to \alpha}=\norm{\mathcal{M}_1}_{\text{CB},1\to\alpha}\norm{\mathcal{M}_2}_{\text{CB},1\to\alpha}$.

	\begin{Lemma}{lem:EA-comm-collapse-of-CB-norm-for-CP}
		Let $\mathcal{M}$ be a completely positive map. Then, for $\id$ an arbitrary identity map, the following equality holds,
		\begin{equation}
			\norm{\id\otimes\mathcal{M}}_{\alpha\to\alpha}=\norm{\mathcal{M}}_{\alpha\to\alpha}.
		\end{equation}
	\end{Lemma}

	\begin{Proof}
		The inequality $\norm{\id\otimes\mathcal{M}}_{\alpha\to\alpha}\geq\norm{\mathcal{M}}_{\alpha\to\alpha}$ immediately follows by restricting the optimization on the left-hand side of the inequality. So we now establish the non-trivial inequality $\norm{\id\otimes\mathcal{M}}_{\alpha\to\alpha}\leq\norm{\mathcal{M}}_{\alpha\rightarrow\alpha}$. Letting the identity map act on a reference system$~R$, consider from \eqref{eq:EA-comm-CB-to-comp-3} that%
		\begin{equation}
			\norm{\id\otimes\mathcal{M}}_{\alpha\to\alpha}=\sup_{\substack{Y_{RA}>0,\\\operatorname{Tr}[Y_{RA}]\leq 1}}\norm{\mathcal{M}_{A\to B}(Y_{RA}^{\frac{1}{\alpha}})}_{\alpha}.
	\end{equation}
		Let $\{V^{i}\}_{i=1}^{d_R^{2}}$ denote a set of Heisenberg--Weyl operators acting on the reference system $R$ (see \eqref{eq-Heisenberg_Weyl_operators}), so that
		\begin{equation}
			\frac{1}{d_R^2}\sum_{i=1}^{d_R^2}V_R^i(\cdot)(V_R^i)^\dagger=\Tr[\cdot]\frac{\mathbbm{1}}{d_R}.
		\end{equation}
		Then, for an arbitrary $Y_{RA}>0$ satisfying $\Tr[Y_{RA}]\leq 1$, we use the unitary invariance of the Schatten norm to obtain
		\begin{align}
			\norm{\mathcal{M}_{A\to B}(Y_{RA}^{\frac{1}{\alpha}})}_{\alpha}&=\frac{1}{d_R^{2}}\sum_{i=1}^{d_R^{2}}\norm{V_{R}^{i}\mathcal{M}_{A\rightarrow B}(Y_{RA}^{\frac{1}{\alpha}})(V_{R}^{i})^{\dagger}}_{\alpha}\\
			&=\frac{1}{d_R^{2}}\sum_{i=1}^{d_R^{2}}\norm{\mathcal{M}_{A\to B}((V_{R}^{i}Y_{RA}(V_{R}^{i})^{\dagger})^{\frac{1}{\alpha}})}_{\alpha}\\
			&\leq\norm{\mathcal{M}_{A\to B}\!\left(\left[\frac{1}{d_R^{2}}\sum_{i=1}^{d_R^{2}}V_{R}^{i}Y_{RA}(V_{R}^{i})^{\dagger}\right]^{\frac{1}{\alpha}}\right)}_{\alpha}\\
			&=\norm{\mathcal{M}_{A\rightarrow B}\left(\left[\pi_{R}\otimes Y_{A}\right]^{\frac{1}{\alpha}}\right)}_{\alpha},
		\end{align}
		where the inequality follows from Lemma~\ref{lem:EA-comm-nontrivial-concavity} below, which states that the function $X\mapsto \norm{\mathcal{M}(X^{\frac{1}{\alpha}})}_{\alpha}$ is concave for all $\alpha>1$. The last equality follows from
\eqref{eq-HW_twirl}, with $\pi_{R}=\frac{\mathbbm{1}_{R}}{|R|}$ the maximally mixed state and $Y_{A}=\operatorname{Tr}_{R}[Y_{RA}]$. Continuing, we find that
		\begin{align}
			\norm{\mathcal{M}_{A\to B}\!\left(\left[\pi_{R}\otimes Y_{A}\right]^{\frac{1}{\alpha}}\right)}_{\alpha}&=\norm{\mathcal{M}_{A\to B}\!\left(\pi_{R}^{\frac{1}{\alpha}}\otimes Y_{A}^{\frac{1}{\alpha}}\right)}_{\alpha}\\
			&=\norm{\pi_{R}^{\frac{1}{\alpha}}\otimes\mathcal{M}_{A\rightarrow B}(Y_{A}^{\frac{1}{\alpha}})}_{\alpha}\\
			&=\norm{\pi_{R}^{\frac{1}{\alpha}}}_{\alpha}\norm{\mathcal{M}_{A\to B}(Y_{A}^{\frac{1}{\alpha}})}_{\alpha}\\
			&=\norm{\mathcal{M}_{A\to B}(Y_{A}^{\frac{1}{\alpha}})}_{\alpha}\\
			&\leq\sup_{\substack{Y_{A}>0,\\\operatorname{Tr}[Y_{A}]\leq 1}}\norm{\mathcal{M}_{A\to B}(Y_{A}^{\frac{1}{\alpha}})}_{\alpha}\\
			&=\norm{\mathcal{M}}_{\alpha\to\alpha}.
		\end{align}
		Since the inequality holds for arbitrary $Y_{RA}>0$ satisfying
$\Tr[Y_{RA}]\leq 1$, we find that%
		\begin{equation}
			\norm{\id\otimes\mathcal{M}}_{\alpha\to\alpha}\leq\norm{\mathcal{M}}_{\alpha\to\alpha},
		\end{equation}
		concluding the proof.
	\end{Proof}

	\begin{Lemma}{lem:EA-comm-nontrivial-concavity}
		Let $X$ be a positive semi-definite operator, and let $\mathcal{M}$ be a completely positive map. For $\alpha>1$, the following function is concave:
		\begin{equation}
			X\mapsto\norm{\mathcal{M}(X^{\frac{1}{\alpha}})}_{\alpha}.
		\end{equation}
	\end{Lemma}

	\begin{Proof}
		Since $\mathcal{M}$ is completely positive, it has a Kraus representation as%
		\begin{equation}
			\mathcal{M}(Z)=\sum_{i}M_{i}ZM_{i}^{\dagger}.
		\end{equation}
		From Proposition~\ref{prop-Schatten_pos_var}, consider that
		\begin{align}
			\norm{\mathcal{M}(X^{\frac{1}{\alpha}})}_{\alpha}&=\sup_{\substack{Y>0,\\\norm{Y}_{\frac{\alpha}{\alpha-1}}\leq
1}}\Tr\!\left[\mathcal{M}(X^{\frac{1}{\alpha}})Y\right]\\
			&=\sup_{\substack{Y>0,\\\Tr[Y]\leq 1}}\Tr\!\left[\mathcal{M}(X^{\frac{1}{\alpha}})Y^{\frac{\alpha-1}{\alpha}}\right]\\
			&=\sup_{\substack{Y>0,\\\Tr[Y]\leq 1}}\sum_{i}\Tr\!\left[M_{i}X^{\frac{1}{\alpha}}M_{i}^{\dagger}Y^{\frac{\alpha-1}{\alpha}}\right].\label{eq:holder-duality-EA-comm-mult}
		\end{align}
		The Lieb concavity theorem (see Theorem~\ref{thm:lieb-concavity} below) is the statement that the following function is jointly concave with respect to positive semi-definite $R$ and $S$ for arbitrary $t\in(0,1)$ and an arbitrary operator $K$:%
		\begin{equation}
			(R,S)\mapsto \Tr[KR^{t}K^{\dag}S^{1-t}].
		\end{equation}
		Let $X_{0},X_{1}\geq 0$ and let $Y_{0},Y_{1}>0$ be such that $\Tr[Y_{0}],\Tr[Y_{1}]\leq 1$. Then for $\lambda\in(0,1]$, and
defining%
		\begin{equation}
			X_{\lambda}\coloneqq\lambda X_{0}+\left(1-\lambda\right)X_{1},\qquad Y_{\lambda}\coloneqq\lambda Y_{0}+\left(1-\lambda\right)Y_{1},
		\end{equation}
		we find that%
		\begin{align}
			&\lambda\Tr\!\left[\mathcal{M}(X_{0}^{\frac{1}{\alpha}})Y_{0}^{\frac{\alpha-1}{\alpha}}\right]+\left(1-\lambda\right)\Tr\!\left[\mathcal{M}(X_{1}^{\frac{1}{\alpha}})Y_{1}^{\frac{\alpha-1}{\alpha}}\right]\nonumber\\
			&=\sum_{i}\lambda\Tr\!\left[M_{i}X_{0}^{\frac{1}{\alpha}}M_{i}^{\dagger}Y_{0}^{\frac{\alpha-1}{\alpha}}\right]+\sum_{i}(1-\lambda)\Tr\!\left[M_{i}X_{1}^{\frac{1}{\alpha}}M_{i}^{\dagger}Y_{1}^{\frac{\alpha-1}{\alpha}}\right]  \\
			&=\sum_{i}\lambda\Tr\!\left[M_{i}X_{0}^{\frac{1}{\alpha}}M_{i}^{\dagger}Y_{0}^{\frac{\alpha-1}{\alpha}}\right]+\left(1-\lambda\right)\Tr\!\left[M_{i}X_{1}^{\frac{1}{\alpha}}M_{i}^{\dagger}Y_{1}^{\frac{\alpha-1}{\alpha}}\right]\\
			&\leq\sum_{i}\left(\Tr\!\left[M_{i}X_{\lambda}^{\frac{1}{\alpha}}M_{i}^{\dagger}Y_{\lambda}^{\frac{\alpha-1}{\alpha}}\right]\right)\\
			&=\Tr\!\left[\mathcal{M}(X_{\lambda}^{\frac{1}{\alpha}})Y_{\lambda}^{\frac{\alpha-1}{\alpha}}\right]\\
			&\leq \sup_{\substack{Y_A>0,\\\Tr[Y_A]\leq 1}}\Tr\!\left[\mathcal{M}(X_{\lambda}^{\frac{1}{\alpha}})Y_A^{\frac{\alpha-1}{\alpha}}\right]
			=\norm{\mathcal{M}(X_{\lambda}^{\frac{1}{\alpha}})}_{\alpha},
		\end{align}
		where the first inequality follows from an application of the Lieb concavity theorem, and the second inequality follows from applying \eqref{eq:holder-duality-EA-comm-mult} and because $Y_{\lambda}$ is a particular operator satisfying $Y_{\lambda}>0$ and $\Tr[Y_{\lambda}]\leq 1$. Since the chain of inequalities holds for arbitary $Y_{0},Y_{1}>0$ such that $\Tr[Y_{0}],\Tr[Y_{1}]\leq 1$, we conclude that%
		\begin{equation}
			 \lambda\norm{\mathcal{M}(X_{0}^{\frac{1}{\alpha}})}_{\alpha}+\left(1-\lambda\right)\norm{\mathcal{M}(X_{1}^{\frac{1}{\alpha}})}_{\alpha} \leq
			 \norm{\mathcal{M}(X_{\lambda}^{\frac{1}{\alpha}})}_{\alpha},
		\end{equation}
		which concludes the proof.
	\end{Proof}

	\begin{theorem*}{Lieb Concavity}{thm:lieb-concavity}
		The following function is jointly concave with respect to positive semi-definite operators $R$ and $S$ for arbitrary $t\in(0,1)$ and an arbitrary operator $K$:%
		\begin{equation}\label{eq-lieb_function}
			(R,S)\mapsto \Tr[KR^{t}K^{\dag}S^{1-t}].
		\end{equation}
	\end{theorem*}

	\begin{Proof}
		We begin by restricting the first argument of the function in \eqref{eq-lieb_function} to positive definite operators. Defining $\ket{K}_{RA}=K_{A}^{\dagger}\ket{\Gamma}_{RA}$, consider that%
		\begin{align}
			\Tr[KR^{t}K^{\dagger}S^{1-t}]&=\bra{\Gamma}_{RA}\mathbbm{1}_{R}\otimes\left(KR^{t}K^{\dagger}S^{1-t}\right)_{A}\ket{\Gamma}_{RA}\\
			&=\bra{\Gamma}_{RA}(S_R^{\t})^{1-t}\otimes K_AR_A^{t}K_A^{\dagger}\ket{\Gamma}_{RA}\\
			&=\bra{\Gamma}_{RA}K_{A}\left[(S_R^{\t})^{1-t}\otimes R_A^{t}\right]K_{A}^{\dagger}\ket{\Gamma}_{RA}\\
			&=\bra{K}_{RA}R_{A}^{\frac{1}{2}}\left[(S_R^{\t})^{1-t}\otimes R_A^{t-1}\right]R_{A}^{\frac{1}{2}}\ket{K}_{RA}\\
			&=\bra{K}_{RA}R_{A}^{\frac{1}{2}}\left[S_R^{\t}\otimes R_A^{-1}\right]^{1-t}R_{A}^{\frac{1}{2}}\ket{K}_{RA}\\
			&=\bra{K}_{RA}R_{A}^{\frac{1}{2}}g(S_R^{\t}\otimes R_A^{-1})R_{A}^{\frac{1}{2}}\ket{K}_{RA},\label{eq-Lieb_concave_pf_1}
		\end{align}
		where the fourth equality holds by the positive definiteness of $R$, and where $g(x)\coloneqq x^{1-t}$ is an operator concave function. For $\lambda\in[0,1]$, let
		\begin{equation}
			R_{\lambda}\coloneqq\lambda R_{0}+\left(1-\lambda\right)  R_{1},\qquad S_{\lambda}\coloneqq\lambda S_{0}+\left(1-\lambda\right)S_{1},
		\end{equation}
		where $R_{0}$ and $R_{1}$ are positive definite and $S_{0}$ and $S_{1}$ are positive semi-definite. Also, let
		\begin{align}
			G_{0}&\coloneqq \mathbbm{1}\otimes\sqrt{\lambda R_{0}}\left(R_{\lambda}\right)^{-\frac{1}{2}},\\
			G_{1}&\coloneqq \mathbbm{1}\otimes\sqrt{\left(  1-\lambda\right)  R_{1}}\left(  R_{\lambda}\right)^{-\frac{1}{2}}.
		\end{align}
		Then
		\begin{align}
			G_{0}^{\dagger}G_{0}+G_{1}^{\dagger}G_{1}&=\mathbbm{1}\otimes\left(R_{\lambda}\right)^{-\frac{1}{2}}\lambda R_{0}\left(R_{\lambda}\right)^{-\frac{1}{2}}\nonumber\\
			&\qquad+\mathbbm{1}\otimes\left(R_{\lambda}\right)^{-\frac{1}{2}}\left(1-\lambda\right) R_{1}\left(R_{\lambda}\right)^{-\frac{1}{2}}\\
			&=\mathbbm{1}\otimes\left(R_{\lambda}\right)^{-\frac{1}{2}}R_{\lambda}\left(  R_{\lambda}\right)^{-\frac{1}{2}}\\
			&=\mathbbm{1}\otimes\mathbbm{1}.
		\end{align}
		A variation of the operator Jensen inequality (Theorem~\ref{thm-Jensen}) is that the following inequality holds for an operator concave function $f$, a finite set $\{X_{i}\}_{i}$ of Hermitian operators, and a finite set $\{A_{i}\}_{i}$ of operators satisfying $\sum_{i}A_{i}^{\dagger}A_{i}=\mathbbm{1}$:%
		\begin{equation}
			\sum_{i}A_{i}^{\dag}f(X_{i})A_{i}\leq f\!\left(\sum_{i}A_{i}^{\dag}X_{i}A_{i}\right).
		\end{equation}
		Then from the operator Jensen inequality and \eqref{eq-Lieb_concave_pf_1}, we conclude that%
		\begin{align}
			&\lambda\Tr[KR_{0}^{t}K^{\dagger}S_{0}^{1-t}]+\left(1-\lambda\right)\Tr[KR_{1}^{t}K^{\dagger}S_{1}^{1-t}]\nonumber\\
			&=\lambda\bra{K}_{RA}\left(R_{0}\right)_{A}^{\frac{1}{2}}g(S_{0}^{\t}\otimes R_{0}^{-1})\left(R_{0}\right)_{A}^{\frac{1}{2}}\ket{K}_{RA}\nonumber\\
			&\qquad+\left(1-\lambda\right)\bra{K}_{RA}\left(R_{1}\right)_{A}^{\frac{1}{2}}g(S_{1}^{\t}\otimes R_{1}^{-1})\left(R_{1}\right)_{A}^{\frac{1}{2}}\ket{K}_{RA}\\
			&=\lambda\bra{K}_{RA}\left(R_{0}\right)_{A}^{\frac{1}{2}}g(\lambda S_{0}^{\t}\otimes\left(\lambda R_{0}\right)^{-1}) \left(R_{0}\right)_{A}^{\frac{1}{2}}\ket{K}_{RA}\nonumber\\
			&\quad+\left(1-\lambda\right)\bra{K}_{RA}\left(R_{1}\right)_{A}^{\frac{1}{2}}g(\left(  1-\lambda\right)S_{1}^{\t}\otimes\left(\left(1-\lambda\right)R_{1}\right)^{-1})\left(R_{1}\right)_{A}^{1/2}\ket{K}_{RA}\\
			&=\bra{K}_{RA}\left(R_{\lambda}\right)_{A}^{\frac{1}{2}}G_{0}^{\dagger}g(\lambda S_{0}^{\t}\otimes\left(\lambda R_{0}\right)^{-1})G_{0}\left(R_{\lambda}\right)_{A}^{\frac{1}{2}}\ket{K}_{RA}\nonumber\\
			&\quad+\bra{K}_{RA}\left(R_{\lambda}\right)_{A}^{\frac{1}{2}}G_{1}^{\dagger}g(\left(1-\lambda\right)S_{1}^{\t}\otimes\left(\left(1-\lambda\right)R_{1}\right)^{-1})G_{1}\left(R_{\lambda}\right)_{A}^{\frac{1}{2}}\ket{K}_{RA}\\
			&\leq\bra{K}_{RA}\left(R_{\lambda}\right)_{A}^{\frac{1}{2}}g(L)\left(R_{\lambda}\right)_{A}^{\frac{1}{2}}\ket{K}_{RA},
		\end{align}
		where the third equality follows because $\mathbbm{1}_R\otimes\sqrt{\lambda}R_0^{\frac{1}{2}}=\mathbbm{1}_R\otimes(R_\lambda)^{\frac{1}{2}}G_0^\dagger$. In the last line, we have let
		\begin{multline}
			L\coloneqq G_{0}^{\dagger}\left(\lambda S_{0}^{\t}\otimes\left(\lambda R_{0}\right)^{-1}\right)G_{0}\\
			+G_{1}^{\dagger}\left(\left(1-\lambda\right)S_{1}^{\t}\otimes\left(\left(1-\lambda\right)R_{1}\right)^{-1}\right)G_{1}.
		\end{multline}
		Consider that%
		\begin{align}
			L&=G_{0}^{\dagger}\left(\lambda S_{0}^{\t}\otimes\left(\lambda R_{0}\right)^{-1}\right)G_{0}+G_{1}^{\dagger}\left(\left(1-\lambda\right)S_{1}^{\t}\otimes\left(\left(1-\lambda\right)R_{1}\right)^{-1}\right)G_{1}\nonumber\\
			&=\left(\mathbbm{1}\otimes\left(R_{\lambda}\right)^{-\frac{1}{2}}\sqrt{\lambda R_{0}}\right)\left(\lambda S_{0}^{\t}\otimes\left(\lambda R_{0}\right)^{-1}\right)\left(\mathbbm{1}\otimes\sqrt{\lambda R_{0}}\left(R_{\lambda}\right)^{-\frac{1}{2}}\right)  \nonumber\\
			&\qquad+\left(\mathbbm{1}\otimes (R^{\lambda})^{-\frac{1}{2}}\sqrt{\left(1-\lambda\right)R_{1}}\right)\nonumber\\
			&\qquad\quad\times\left(\left(1-\lambda\right)S_{1}^{\t}\otimes\left(\left(1-\lambda\right)R_{1}\right)^{-1}\right)\left(\mathbbm{1}\otimes\sqrt{\left(1-\lambda\right)R_{1}}\left(R_{\lambda}\right)^{-\frac{1}{2}}\right)  \\
			&=\lambda S_{0}^{\t}\otimes\left(R_{\lambda}\right)^{-1}+\left(1-\lambda\right)S_{1}^{\t}\otimes\left(R_{\lambda}\right)^{-1}\\
			&=S_{\lambda}^{\t}\otimes\left(R_{\lambda}\right)^{-1}.
		\end{align}
		Continuing, we find that%
		\begin{align}
			&\lambda\Tr[KR_{0}^{t}K^{\dagger}S_{0}^{1-t}]+\left(1-\lambda\right)\Tr[KR_{1}^{t}K^{\dagger}S_{1}^{1-t}]\nonumber\\
			&\leq\bra{K}_{RA}\left(R_{\lambda}\right)_{A}^{\frac{1}{2}}g\left(L\right)\left(R_{\lambda}\right)_{A}^{\frac{1}{2}}\ket{K}_{RA}\\
			&=\bra{K}_{RA}\left(R_{\lambda}\right)_{A}^{\frac{1}{2}}g(S_{\lambda}^{\t}\otimes\left(R_{\lambda}\right)^{-1})\left(R_{\lambda}\right)_{A}^{\frac{1}{2}}\ket{K}_{RA}\\
			&=\Tr[KR_{\lambda}^{t}K^{\dag}S_{\lambda}^{1-t}].
		\end{align}
		So the function $(R,S)\mapsto \Tr[KR^tK^\dagger S^{1-t}]$ is jointly concave when the first argument is restricted to be a positive definite operator. The more general case of positive semi-definite operators in the first argument can be established by adding $\varepsilon\mathbbm{1}$ to any positive semi-definite operator to ensure that it is positive definite, applying the above inequality, and then taking the limit $\varepsilon\to 0$ at the end. This concludes the proof.
	\end{Proof}

\section{The Strong Converse from a Different Point of View}

\label{app:EA-comm:strong-conv-diff-POV}
	
	Here we show that the mutual information $I(\mathcal{N})$ is a strong converse rate based on the 
	 the alternate definition given in Appendix~\ref{chap-str_conv}.
	According to that definition,  a rate $R\in\mathbb{R}^+$ is a strong converse rate for entanglement-assisted classical communication over a channel $\mathcal{N}$ if for every sequence $\{(n,|\mathcal{M}_n|,\varepsilon_n)\}_{n\in\mathbb{N}}$ of $(n,|\mathcal{M}|,\varepsilon)$ entanglement-assisted classical communication protocols over $n$ uses of $\mathcal{N}$, we have that $\liminf_{n\to\infty}\frac{1}{n}\log_2|\mathcal{M}_n|>R\Rightarrow \lim_{n\to\infty}\varepsilon_n=1$.
	
	Let us show that the mutual information $I(\mathcal{N})$ of the channel $\mathcal{N}$ is a strong converse rate under this alternate definition. 
	Let $\{(n,\left\vert \mathcal{M}_{n}\right\vert ,\varepsilon_{n})\}_{n\in \mathbb{N}}$
be a sequence of protocols satisfying $\liminf_{n\rightarrow\infty}\frac{1}%
{n}\log_{2}\vert \mathcal{M}_{n}\vert >I(\mathcal{N})$. Due to this
strict inequality, the fact that $\lim_{\alpha\rightarrow1}\widetilde
{I}_{\alpha}(\mathcal{N})=I(\mathcal{N})$,  and since the sandwiched R\'{e}nyi mutual information $\widetilde{I}_\alpha(\mathcal{N})$ is monotonically increasing in $\alpha$ (this follows from Proposition~\ref{prop-sand_rel_ent_properties}), there exists a value $\alpha^{\ast
}>1$ such that%
\begin{equation}
\liminf_{n\rightarrow\infty}\frac{1}{n}\log_{2}\vert \mathcal{M}%
_{n}\vert >\widetilde{I}_{\alpha^{\ast}}(\mathcal{N}%
).\label{eq:EA-comm:str-conv-alt-view-step-1}%
\end{equation}
Now recall the following bound from \eqref{eq-eacc_str_conv_one_shot_3}, which holds for all $\alpha>1$ and
for every $(n,\left\vert \mathcal{M}\right\vert ,\varepsilon)$ protocol:%
\begin{equation}
\frac{1}{n}\log_{2}\vert \mathcal{M}\vert \leq\widetilde{I}%
_{\alpha}(\mathcal{N})+\frac{\alpha}{n\left(  \alpha-1\right)  }\log
_{2}\!\left(  \frac{1}{1-\varepsilon}\right)  .
\end{equation}
We can apply it in our case to conclude that%
\begin{equation}
\frac{1}{n}\log_{2}\!\left\vert \mathcal{M}_{n}\right\vert \leq\widetilde
{I}_{\alpha^{\ast}}(\mathcal{N})+\frac{\alpha^{\ast}}{n\left(  \alpha^{\ast
}-1\right)  }\log_{2}\!\left(  \frac{1}{1-\varepsilon_{n}}\right)  .
\end{equation}
Now suppose that%
\begin{equation}
\liminf_{n\rightarrow\infty}\varepsilon_{n}=c\in\lbrack
0,1).\label{eq:EA-comm:str-conv-alt-view-step-2}%
\end{equation}
Then it follows that%
\begin{align}
\liminf_{n\rightarrow\infty}\frac{1}{n}\log_{2}\!\left\vert \mathcal{M}%
_{n}\right\vert  & \leq\liminf_{n\rightarrow\infty}\left[  \widetilde
{I}_{\alpha^{\ast}}(\mathcal{N})+\frac{\alpha^{\ast}}{n\left(  \alpha^{\ast
}-1\right)  }\log_{2}\!\left(  \frac{1}{1-\varepsilon_{n}}\right)  \right]  \\
& =\widetilde{I}_{\alpha^{\ast}}(\mathcal{N})+\liminf_{n\rightarrow\infty
}\left[  \frac{\alpha^{\ast}}{n\left(  \alpha^{\ast}-1\right)  }\log
_{2}\!\left(  \frac{1}{1-\varepsilon_{n}}\right)  \right]  \\
& =\widetilde{I}_{\alpha^{\ast}}(\mathcal{N}),
\end{align}
where the last equality follows because $\alpha^{\ast}>1$ is a constant and
the sequence $\left\{  \varepsilon_{n}\right\}  _{n\in N}$ converges to a
constant $c\in\lbrack0,1)$. However, this contradicts
\eqref{eq:EA-comm:str-conv-alt-view-step-1}. Thus,
\eqref{eq:EA-comm:str-conv-alt-view-step-2} cannot hold, and so we conclude
that $\liminf_{n\rightarrow\infty}\varepsilon_{n}=1$.

	The argument given above makes no statement about how fast the error probability converges to one in the large $n$ limit. If we fix the rate $R$ of communication to be a constant satisfying $R>I(\mathcal{N})$, then we can argue that the error probability converges exponentially fast to one. To this end, consider a sequence $\{(n,2^{nR},\varepsilon_n)\}_{n\in\mathbb{N}}$ of $(n,|\mathcal{M}|,\varepsilon)$ protocols, with each element of the sequence having an arbitrary (but fixed) rate $R>I(\mathcal{N})$. For each element of the sequence, the inequality in \eqref{eq-eacc_str_conv_one_shot_3} holds, which means that
	\begin{equation}
		R\leq \widetilde{I}_\alpha(\mathcal{N})+\frac{\alpha}{n(\alpha-1)}\log_2\!\left(\frac{1}{1-\varepsilon_n}\right)
	\end{equation}
	for all $\alpha>1$. Rearranging this inequality leads to the following lower bound on the error probabilities $\varepsilon_n$:
	\begin{equation}\label{eq-eacc_str_conv_alt_pf1}
		\varepsilon_n\geq 1-2^{-n\left(\frac{\alpha-1}{\alpha}\right)\left(R-\widetilde{I}_\alpha(\mathcal{N})\right)}
	\end{equation}
	for all $\alpha>1$. Now, since $R>I(\mathcal{N})$, $\lim_{\alpha\to 1}\widetilde{I}_\alpha(\mathcal{N})=I(\mathcal{N})$, and since the sandwiched R\'{e}nyi mutual information $\widetilde{I}_\alpha(\mathcal{N})$ is monotonically increasing in $\alpha$ (this follows from Proposition~\ref{prop-sand_rel_ent_properties}), there exists an $\alpha^*>1$ such that $R>\widetilde{I}_{\alpha^*}(\mathcal{N})$. Applying the inequality in \eqref{eq-eacc_str_conv_alt_pf1} to this value of $\alpha$, we find that
	\begin{equation}
		\varepsilon_n\geq 1-2^{-n\left(\frac{\alpha^*-1}{\alpha^*}\right)\left(R-\widetilde{I}_{\alpha^*}(\mathcal{N})\right)}.
	\end{equation}
	Then, taking the limit $n\to\infty$ on both sides of this inequality, we conclude that $\lim_{n\to\infty}\varepsilon_n= 1$ and the convergence to one is exponentially fast. 
	\begin{figure}
		\centering
		\includegraphics[scale=0.7]{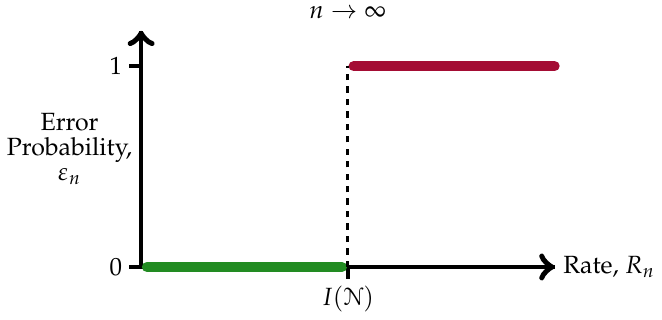}
		\caption{The error probability $\varepsilon_n$ as a function of the rate $R_n$ for  entanglement-assisted classical communication  over a quantum channel~$\mathcal{N}$. As $n\to\infty$, for every rate below the mutual information $I(\mathcal{N})$, there exists a sequence of protocols with error probability converging to zero. For every rate above the mutual information $I(\mathcal{N})$, the error probability converges to one for all possible protocols.}\label{fig-ea_classical_comm_str_converse}
	\end{figure}
		
	From the  arguments above, we find not only that $I(\mathcal{N})$ is a strong converse rate according to the alternate definition provided in Appendix~\ref{chap-str_conv}, but also that the maximal error probability of every sequence of $(n,|\mathcal{M}|,\varepsilon)$ entanglement-assisted classical communication protocols with fixed rate strictly above the mutual information $I(\mathcal{N})$ approaches one at an exponential rate.

	In Section~\ref{subsec-EA_comm_ach_diff_POV}, we showed that the error probability vanishes in the limit $n\to\infty$ for every fixed rate $R<I(\mathcal{N})$. We thus see that, as $n\to\infty$, the mutual information $I(\mathcal{N})$ is a sharp dividing point between reliable, error-free communication and communication with error probability approaching one exponentially fast. This situation is depicted in Figure~\ref{fig-ea_classical_comm_str_converse}.

\end{subappendices}

\chapter{Classical Communication}\label{chap-classical_capacity}

	
	We now move on to classical communication over quantum channels. Unlike the previous chapter, here we suppose that Alice and Bob do not have access to shared entanglement prior to communication. Thus, the scenario considered in this chapter is more practical than the entanglement-assisted setting---in the previous chapter, we made the simplifying assumption that shared entanglement is available for free to the sender and receiver. However, without widespread entanglement-sharing networks available, this assumption is not really practical, and so the entanglement-assisted capacity is mostly of academic interest at the moment.
	
	Without shared entanglement available to the sender and receiver, is it still advantageous to use a quantum strategy to send classical information over a quantum channel?
	At first glance, it may seem that, without prior shared entanglement, there might not be any point in using a quantum strategy to send classical information over a quantum channel. However, when using a channel multiple times, there is still the possibility of encoding a message into a state at the encoder that is entangled across multiple channel uses and then performing a collective measurement at the decoder. For many examples of channels, it is known that collective measurements can enhance communication capacity, and it is known that in principle, there exists a channel for which  entangled states at the encoder provides a further enhancement to communication capacity.
	
	 Although it may seem that determining the maximum amount of classical information that can be communicated using a given quantum channel, i.e., determining the classical capacity of a quantum channel, might be easier than its entanglement-assisted counterpart, this problem turns out to be one of the most challenging problems in quantum Shannon theory. 
	 This is due to the fact that, as we discuss in this chapter, the relevant quantity in calculating the classical capacity of a quantum channel $\mathcal{N}$ is related to its Holevo information $\chi(\mathcal{N})$, and this quantity is not known to be additive for all quantum channels. This means that the best we can say for a given channel is that its Holevo information is an achievable rate for classical communication---we cannot necessarily say that it is the highest possible achievable rate. This is in stark contrast to the case of entanglement-assisted classical communication, for which we know that the mutual information $I(\mathcal{N})$ of a channel is additive for all channels and thus is equal to the entanglement-assisted classical capacity of any quantum channel. 
	 	
	We start in the next brief section by considering some simple motivating examples of communication. Then we consider the one-shot setting. This setting for classical communication is similar to the one-shot setting of entanglement-assisted classical communication from the previous chapter. The only difference is that, in this case, there is no entanglement assistance. We then move on to the asymptotic setting, for which we prove Theorem~\ref{thm-classical_capacity}, which states that the classical capacity of a quantum channel is equal to the \textit{regularized} Holevo information of the channel. This is a quantity that in general requires computing the Holevo information for an arbitrarily large number of uses of the given channel, and it is therefore intractable unless the Holevo information happens to be additive for the channel. From here, we consider various classes of channels for which the Holevo information is additive or for which we can establish the strong converse. We also consider various methods for bounding the classical capacity from above. Finally, we calculate the classical capacity for some examples of channels.

\subsubsection*{Simple Example of Classical Communication Over a Quantum Channel}

	At the beginning of the previous chapter, we stated that super-dense coding is a simple example of an entanglement-assisted classical communication protocol over a noiseless quantum channel. The essense of that protocol is the encoding of $d^2$ messages into $d^2$ mutually orthogonal pure states (the maximally entangled states defined in \eqref{eq-qudit_Bell}). The $d^2$ messages contain $\log_2 d^2=2\log_2 d$ bits of classical information, which can be communicated without error from Alice to Bob, with just one use of a noiseless qudit quantum channel.
	
	Now, without the assistance of prior shared entanglement, one result of  this chapter is that the maximum amount of classical information that can be communicated over a noiseless quantum channel without error is $\log_2 d$, where $d$ is the dimension of the channel. Let us  describe a simple protocol that achieves this number of communicated bits.	Consider a discrete set $\mathcal{M}$ of messages, and suppose that Alice encodes each message $m\in\mathcal{M}$ into a quantum state $\ket{m}$, such that the set $\{\ket{m}\}_{m\in\mathcal{M}}$, is orthonormal, i.e., $\braket{m}{m'}=\delta_{m,m'}$ for all $m,m'\in\mathcal{M}$. Bob, knowing Alice's encoding of the messages, devises a measurement to extract the message described by the POVM $\{\ket{m}\!\bra{m}\}_{m\in\mathcal{M}}$. His strategy is to guess that the message sent was ``$m$'' if the outcome of his measurement is $m\in\mathcal{M}$. If Alice sends the state $\ket{m}\!\bra{m}$ through a noiseless quantum channel, then Bob is guaranteed to receive the state $\ket{m}\!\bra{m}$ unaltered, so that his guess will always be correct. Alice can thus send $\log_2|\mathcal{M}|$ bits of classical information to Bob without error.
	
	Now, if the channel is noisy, the initially orthogonal states in general become non-orthogonal, so that if Alice sends the state $\ket{m}\!\bra{m}$ through the channel then Bob generally receive a mixed state $\rho^m$ instead. As a consequence of using a noisy quantum channel, Bob's decoding strategy will not always succeed, meaning that there will be errors. In order to mitigate the effects of noise, Alice can choose a more clever encoding of the message, and similarly Bob can devise a more clever decoding strategy.\footnote{We assume, as in all communication tasks considered in this book, that Alice and Bob know the channel connecting them, so that they can use this knowledge to develop their encoding and decoding.} Alice and Bob can also use the channel multiple times, which can decrease the error in general, while also  allowing for the messages to be encoded into higher-dimensional entangled states.
	
	Observe that the task of classical communication over a quantum channel is closely related to the task of state discrimination (see Section~\ref{subsec-state_discrimination}). Recall that the goal of state discrimination is to minimize the error probability for a given set $\{\rho^m\}_{m\in\mathcal{M}}$ of states corresponding to the message set $\mathcal{M}$ and a particular decoding POVM $\{\Lambda_B^m\}_{m\in\mathcal{M}}$ indexed by the messages. In classical communication, we focus primarily on maximizing the rate $\frac{1}{n}\log_2|\mathcal{M}|$ of communication for a given error probability $\varepsilon$, and we are interested in determining the maximum rate $R$ for which $\varepsilon$ vanishes as the number $n$ of channel uses increases.

\section{One-Shot Setting}\label{sec-cc_one_shot}

\begin{figure}
		\centering
		\includegraphics[scale=0.8]{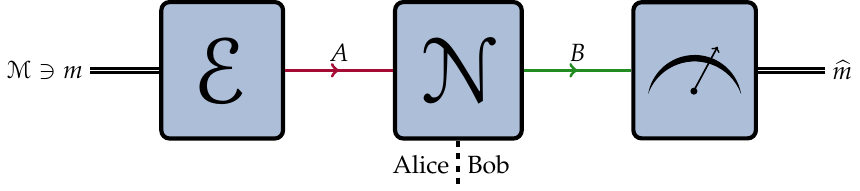}
		\caption{Depiction of a protocol for classical communication over one use of the quantum channel $\mathcal{N}$. Alice, who wishes to send a message $m$ chosen from a set $\mathcal{M}$ of messages, first encodes the message into a quantum state on a quantum system $A$, using a classical--quantum encoding channel $\mathcal{E}$. She then sends the quantum system $A$ through the channel $\mathcal{N}_{A\to B}$. After Bob receives the system $B$, he performs a measurement on it, using the outcome of the measurement to give an estimate $\widehat{m}$ of the message sent by Alice.}\label{fig-classical_comm_oneshot}
	\end{figure}

	In the one-shot setting, we start by considering a classical communication protocol over a quantum channel $\mathcal{N}$, as depicted in Figure~\ref{fig-classical_comm_oneshot}. The protocol is defined by the triple $(\mathcal{M},\mathcal{E}_{M\to A},\allowbreak\mathcal{D}_{B\to\widehat{M}})$, consisting of a message set $\mathcal{M}$, an \textit{encoding channel} $\mathcal{E}_{M\to A}$, and a \textit{decoding channel} $\mathcal{D}_{B\to\widehat{M}}$. The pair $(\mathcal{E},\mathcal{D})$ of encoding and decoding channels is often called a \textit{code} and denoted by $\mathcal{C}=(\mathcal{E},\mathcal{D})$. The encoding channel is a classical--quantum channel (see Definition~\ref{def-cq_channel}), and the decoding channel is a quantum--classical or measurement channel (see Definition~\ref{def-qc_channel}).
	
	Now, given that there are $|\mathcal{M}|$ messages in the message set, it follows that each message can be uniquely associated with a bit string of size at least $\log_2|\mathcal{M}|$. The quantity $\log_2|\mathcal{M}|$ thus represents the number of bits communicated in the protocol. One of the goals of this section is to obtain upper and lower bounds on maximum number of $\log_2|\mathcal{M}|$ of bits that can be communicated in any classical communication protocol.
	
	The protocol proceeds as follows: let $p:\mathcal{M}\to[0,1]$ be a probability distribution over the message set. With probability $p(m)$, Alice picks a message $m\in\mathcal{M}$ and makes a local copy of it. Letting $\{\ket{m}\}_{m\in\mathcal{M}}$ be an orthonormal basis indexed by the messages, 
her initial state is described by the following classically correlated state:
	\begin{equation}\label{eq-classical_comm_initial_state}
		\overline{\Phi}_{MM'}^p\coloneqq \sum_{m\in \mathcal{M}}p(m)\ket{m}\!\bra{m}_{M}\otimes\ket{m}\!\bra{m}_{M'}.
	\end{equation}
	Note that if Alice wishes to send a particular message $m$ deterministically, then she can choose the distribution $p$ to be the degenerate distribution, equal to one for $m$ and zero for all other messages.
	
	She then
	uses an encoding channel $\mathcal{E}_{M\to A}$ to map the message to a quantum state $\rho_A^m$. We can explicitly define the encoding channel $\mathcal{E}_{M'\to A}$ as
	\begin{equation}\label{eq-classical_comm_encoding}
		\mathcal{E}_{M'\to A}(\ket{m}\!\bra{m'}_{M'})=\delta_{m,m'}\rho_{A}^m\quad\forall~m,m'\in\mathcal{M}.
	\end{equation}
	Note that this channel has the form of a classical--quantum channel (recall Definition~\ref{def-cq_channel}). The action of the encoding channel on the initial state in \eqref{eq-classical_comm_initial_state} is as follows:
	\begin{align}
		\mathcal{E}_{M'\to A}(\overline{\Phi}_{MM'}^p)&=\sum_{m\in\mathcal{M}}p(m)\ket{m}\!\bra{m}_M\otimes\mathcal{E}_{M'\to A}(\ket{m}\!\bra{m})\\
		&=\sum_{m\in\mathcal{M}}p(m)\ket{m}\!\bra{m}_M\otimes\rho_{A}^m  \\
		&=: \rho_{MA}^p. \label{eq-classical_comm_cq_state}
	\end{align}
	

	Alice then sends the system $A$ through the channel $\mathcal{N}_{A\to B}$, resulting in the state 
	\begin{equation}\label{eq-cc:state-after-channel}
		\mathcal{N}_{A\to B}(\rho_{MA})=\sum_{m\in\mathcal{M}}p(m)\ket{m}\!\bra{m}_M\otimes\mathcal{N}_{A\to B}(\rho_{A}^m).
	\end{equation}
	Bob, whose task is to determine which message Alice sent, performs a decoding measurement on his received system $B$, which has the corresponding POVM $\{\Lambda_{B}^m\}_{m\in\mathcal{M}}$. The measurement is associated with the decoding channel $\mathcal{D}_{B\to \widehat{M}}$, which is simply a quantum--classical channel as given in Definition~\ref{def-qc_channel}, i.e.,
	\begin{equation}
		 \mathcal{D}_{B\to\widehat{M}}(\tau_{B})\coloneqq \sum_{m\in\mathcal{M}}\Tr[\Lambda_{B}^m\tau_{B}]\ket{m}\!\bra{m}_{\widehat{M}}
	\end{equation}
	for every state $\tau_{B}$. So the final state of the protocol is
	\begin{align}
		\omega_{M\widehat{M}}^p &\coloneqq (\mathcal{D}_{B\to\widehat{M}}\circ\mathcal{N}_{A\to B}\circ\mathcal{E}_{M'\to A})(\overline{\Phi}_{MM'}^p)\label{eq-classical_comm_final_state_1}\\
		&=\sum_{m,\widehat{m}\in\mathcal{M}}p(m)\ket{m}\!\bra{m}_M\otimes\Tr[\Lambda_{B}^{\widehat{m}}\mathcal{N}_{A\to B}(\rho^m_{A})]\ket{\widehat{m}}\!\bra{\widehat{m}}_{\widehat{M}}.\label{eq-classical_comm_final_state_2}
	\end{align}
	
	The measurement by Bob induces the conditional probability distribution $q:\mathcal{M}\times\mathcal{M}\to[0,1]$ defined by
	\begin{equation}
		q(\widehat{m}|m)\coloneqq\Pr[\widehat{M}=\widehat{m}|M=m]=\Tr[\Lambda_{B}^{\widehat{m}}\mathcal{N}_{A\to B}(\rho_{A}^m)].
	\end{equation}
	Bob's strategy is such that if the outcome $\widehat{m}$ occurs from his measurement, then he guesses that the message sent was $\widehat{m}$. The probability that Bob correctly identifies a given message $m$ is then equal to $q(m|m)$. The \textit{message error probability of the code} is given by
	\begin{equation}\label{eq-mess_error_prob}
		\begin{aligned}
		p_{\text{err}}(m,(\mathcal{E},\mathcal{D});\mathcal{N})&\coloneqq 1-q(m|m)\\
		&=\Tr[(\mathbbm{1}_B-\Lambda_{B}^m)\mathcal{N}_{A\to B}(\rho_{A}^m)]\\
		&=\sum_{\widehat{m}\in\mathcal{M}\setminus\{m\}}q(\widehat{m}|m).
		\end{aligned}
	\end{equation}
	The \textit{average error probability of the code} is
	\begin{align}
		\overline{p}_{\text{err}}((\mathcal{E},\mathcal{D});p,\mathcal{N})&\coloneqq\sum_{m\in\mathcal{M}}p(m)p_{\text{err}}(m,(\mathcal{E},\mathcal{D});\mathcal{N})\\
		&=\sum_{m\in\mathcal{M}}p(m)(1-q(m|m))\label{eq-avg_error_prob}\\
		&=\sum_{m\in\mathcal{M}}\sum_{\widehat{m}\in\mathcal{M}\setminus\{m\}}p(m)q(\widehat{m}|m).
	\end{align}
	The \textit{maximal error probability of the code} is
	\begin{equation}\label{eq-maximal_error_prob}
		p_{\text{err}}^*(\mathcal{E},\mathcal{D};\mathcal{N})\coloneqq\max_{m\in\mathcal{M}}p_{\text{err}}(m,(\mathcal{E},\mathcal{D});\mathcal{N}).
	\end{equation}
	Just as in the case of entanglement-assisted classical communication in Chapter~\ref{chap-EA_capacity}, each of these three error probabilities can be used to assess the reliability of the protocol, i.e., how well the encoding and decoding allows Alice to transmit her message to Bob.

	\begin{definition}{$\boldsymbol{(|\mathcal{M}|,\varepsilon)}$ Classical Communication Protocol}{def-cc_Me_protocol}
		A classical communication protocol $(\mathcal{M},\mathcal{E}_{M\to A},\mathcal{D}_{B\to \widehat{M}})$ over the channel $\mathcal{N}_{A\to B}$ is called an \textit{$(|\mathcal{M}|,\varepsilon)$ protocol}, with $\varepsilon\in[0,1]$, if $p_{\text{err}}^*(\mathcal{E},\mathcal{D};\mathcal{N})\leq\varepsilon$.
	\end{definition}
	
	As with entanglement-assisted classical communication, the error criterion $p_{\text{err}}(\mathcal{E},\mathcal{D};\mathcal{N})^*\leq\varepsilon$ is equivalent to 
	\begin{equation}\label{eq-classical_comm_reliability}
		\max_{p:\mathcal{M}\to[0,1]}\frac{1}{2}\norm{\overline{\Phi}_{MM'}^p-\omega_{M\widehat{M}}^p}_1\leq\varepsilon,
	\end{equation}
	and the steps to show this are the same as those shown in the proof of Lemma~\ref{lem:EA-comm:error-criterion-analysis}. In particular, the following equality holds
	\begin{equation}\label{eq-cc_trace_dist_avg_error}
		\frac{1}{2}\norm{\overline{\Phi}_{MM'}^p-\omega_{M\widehat{M}}^p}_1=\overline{p}_{\text{err}}((\mathcal{E},\mathcal{D});p),
	\end{equation}
	which leads to
	\begin{equation}
		\begin{aligned}
		p_{\text{err}}^*(\mathcal{E},\mathcal{D};\mathcal{N})&=\max_{p:\mathcal{M}\to[0,1]}\overline{p}_{\text{err}}((\mathcal{E},\mathcal{D});p,\mathcal{N})\\
		&=\max_{p:\mathcal{M}\to[0,1]}\frac{1}{2}\norm{\overline{\Phi}_{MM'}^p-\omega_{M\widehat{M}}^p}_1.
		\end{aligned}
	\end{equation}

	Also, as in Chapter~\ref{chap-EA_capacity}, another way to define the error criterion of the protocol is through a comparator test. Recall that the comparator test is a measurement defined by the two-element POVM $\{\Pi_{M\widehat{M}},\mathbbm{1}_{M\widehat{M}} - \Pi_{M\widehat{M}}\}$, where $\Pi_{M\widehat{M}}$ is the projection defined as
	\begin{equation}\label{eq-comparator_test}
		\Pi_{M\widehat{M}}\coloneqq\sum_{m\in\mathcal{M}}\ket{m}\!\bra{m}_M\otimes\ket{m}\!\bra{m}_{\widehat{M}}.
	\end{equation}
	Note that $\Tr[\Pi_{M\widehat{M}}\omega_{M\widehat{M}}^p]$ is simply the probability that the classical registers $M$ and $\widehat{M}$ in the state $\omega_{M\widehat{M}}^p$ have the same values. In particular, following the same steps as in \eqref{eq-ea_classical_comm_comparator_succ_prob_1}--\eqref{eq-ea_classical_comm_comparator_succ_prob_3}, we have
	\begin{equation}\label{eq-classical_comm_comparator_succ_prob}
		\Tr\!\left[\Pi_{M\widehat{M}}\omega_{M\widehat{M}}^p\right]=1-\overline{p}_{\text{err}}((\mathcal{E},\mathcal{D});p,\mathcal{N}) =: \overline{p}_{\text{succ}}((\mathcal{E},\mathcal{D});p,\mathcal{N}),
	\end{equation}
	where we have acknowledged that the expression on the left-hand side can be interpreted as the average success probability of the code~$(\mathcal{E},\mathcal{D})$ and denoted it by $\overline{p}_{\text{succ}}((\mathcal{E},\mathcal{D});p,\mathcal{N})$.
	
	As mentioned at the beginning of this chapter, our goal is to bound (from above and below) the maximum number $\log_2|\mathcal{M}|$ of transmitted bits for every classical communication protocol over $\mathcal{N}$. Given an error probability threshold of $\varepsilon$, we call the maximum number of transmitted bits the \textit{one-shot classical capacity} of $\mathcal{N}$.

	\begin{definition}{One-Shot Classical Capacity of a Quantum Channel}{def-classical_comm_one_shot_capacity}
		Given a quantum channel $\mathcal{N}_{A\to B}$ and $\varepsilon\in[0,1]$, the \textit{one-shot $\varepsilon$-error classical capacity of $\mathcal{N}$}, denoted by $C^{\varepsilon}(\mathcal{N})$, is defined to be the maximum number $\log_2|\mathcal{M}|$ of transmitted bits among all $(|\mathcal{M}|,\varepsilon)$ classical communication protocols over $\mathcal{N}$. In other words,
		\begin{equation}
			C^{\varepsilon}(\mathcal{N})\coloneqq\sup_{(\mathcal{M},\mathcal{E},\mathcal{D})}\{\log_2|\mathcal{M}| : p_{\text{err}}^*(\mathcal{E},\mathcal{D};\mathcal{N})\leq\varepsilon\},
		\end{equation}
		where the optimization is over all protocols $(\mathcal{M},\mathcal{E}_{M'\to A},\mathcal{D}_{B\to \widehat{M}})$ satisfying $d_{M'}=d_{\widehat{M}}=|\mathcal{M}|$.
	\end{definition}

	In addition to finding, for a given $\varepsilon\in[0,1]$, the maximum number of transmitted bits among all $(|\mathcal{M}|,\varepsilon)$ classical communication protocols over $\mathcal{N}_{A\to B}$, we can consider the following complementary problem: for a given number of messages $|\mathcal{M}|$, find the smallest possible error probability among all $(|\mathcal{M}|,\varepsilon)$ classical communication protocols, which we denote by $\varepsilon_C^*(|\mathcal{M}|;\mathcal{N})$. In other words, to problem is to determine
	\begin{equation}\label{eq-classical_comm_one_shot_opt_error}
		\varepsilon_C^*(|\mathcal{M}|;\mathcal{N})\coloneqq \inf_{\mathcal{E},\mathcal{D}}\{p_{\text{err}}^*(\mathcal{E},\mathcal{D};\mathcal{N}): d_{M'}=d_{\widehat{M}}=|\mathcal{M}|\},
	\end{equation}
	where the optimization is over encoding channels $\mathcal{E}$ with input space dimension $|\mathcal{M}|$ and decoding channels $\mathcal{D}$ with output space dimension $|\mathcal{M}|$. In this book, we focus primarily on the problem of optimizing the number of transmitted bits rather than the error probability, and so our primary quantity of interest is the one-shot capacity $C^{\varepsilon}(\mathcal{N})$.

\subsection{Protocol Over a Useless Channel}\label{subsec-cc_useless_channel}
	
\begin{figure}
		\centering
		\includegraphics[scale=0.8]{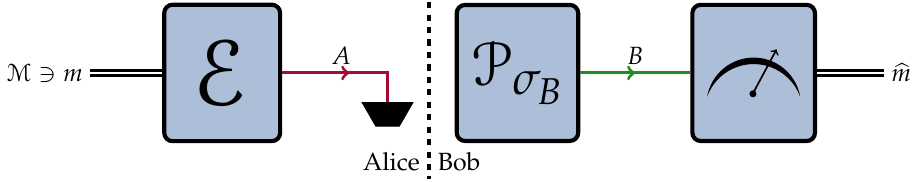}
		\caption{Depiction of a protocol that is useless for classical communication. The state encoding the message $m$ via $\mathcal{E}$ is discarded and replaced with an arbitrary (but fixed) state $\sigma_{B}$.}\label{fig-classical_comm_useless_oneshot}
	\end{figure}

	We now turn to establishing an upper bound on the one-shot classical capacity, and our approach is similar to the approach outlined in Section~\ref{sec-EAC:useless-protocol}. With this goal in mind, along with the actual classical communication protocol, we also consider the same protocol but performed over a \textit{useless} channel as depicted in Figure~\ref{fig-classical_comm_useless_oneshot}. This useless channel discards the state encoded with the message and replaces it with some arbitrary (but fixed) state~$\sigma_{B}$. In other words,
	\begin{equation}\label{eq-cc_useless_channel}
		\rho_{A}^m\mapsto \sigma_{B}\eqqcolon(\mathcal{P}_{\sigma_{B}}\circ\Tr)(\rho_{A}^m)\quad\forall~m\in\mathcal{M},
	\end{equation}
	where the encoded states $\rho_{A}^m$ are defined in \eqref{eq-classical_comm_encoding}. This channel is useless because the state $\sigma_{B}$ does not contain any information about the message. As with entanglement-assisted classical communication, comparing this protocol over the useless channel with the actual protocol allows us to obtain an upper bound on the quantity $\log_2|\mathcal{M}|$, which we recall represents the number of bits that are transmitted over the channel.
	
	The state at the end of the protocol over the useless channel is the following tensor-product state:
	\begin{equation}\label{eq-cc_useless_final_state}
		\tau_{M\widehat{M}}^p\coloneqq \sum_{m\in\mathcal{M}}p(m)\ket{m}\!\bra{m}_M\otimes\sum_{\widehat{m}\in\mathcal{M}}\Tr[\Lambda_{B}^{\widehat{m}}\sigma_{B}]\ket{\widehat{m}}\!\bra{\widehat{m}}_{\widehat{M}},
	\end{equation}
	which indicates that the decoded message system $\widehat{M}$ is independent of the message system $M$ in this case.
	Now, recall from \eqref{eq-classical_comm_final_state_2} that the state $\omega_{M\widehat{M}}^p$ at the end of the actual protocol over the channel $\mathcal{N}$ is given by
	\begin{equation}\label{eq-classical_comm_final_state_3}
		\omega_{M\widehat{M}}^p=\sum_{m,\widehat{m}\in\mathcal{M}}p(m)\Tr[\Lambda_{B}^{\widehat{m}}\mathcal{N}_{A\to B}(\rho_{A}^m)]\ket{m}\!\bra{m}_M\otimes \ket{\widehat{m}}\!\bra{\widehat{m}}_{\widehat{M}}.
	\end{equation}
	Similar to the notation from Chapter~\ref{chap-EA_capacity}, we let 
	\begin{equation}\label{eq-classical_comm_final_state_uniform}
		\omega_{M\widehat{M}}\coloneqq\frac{1}{|\mathcal{M}|}\sum_{m,\widehat{m}\in\mathcal{M}}\Tr[\Lambda_{B}^{\widehat{m}}\mathcal{N}_{A\to B}(\rho_{A}^m)]\ket{m}\!\bra{m}_M\otimes \ket{\widehat{m}}\!\bra{\widehat{m}}_{\widehat{M}},
	\end{equation}
	be the state $\omega_{M\widehat{M}}^p$ with the probability distribution $p$ over the message set equal to the uniform distribution, i.e., $p(m)=\frac{1}{|\mathcal{M}|}$. We also let
	\begin{equation}\label{eq-classical_comm_initial_state_uniform}
		\overline{\Phi}_{MM'}\coloneqq\frac{1}{|\mathcal{M}|}\sum_{m\in\mathcal{M}}\ket{m}\!\bra{m}_M\otimes\ket{m}\!\bra{m}_{M'}
	\end{equation}
	be the state in \eqref{eq-classical_comm_initial_state} with $p$ being the uniform distribution over $\mathcal{M}$.
	
	Now, observe that $\Tr_{\widehat{M}}[\omega_{M\widehat{M}}]=\pi_M$. Also, for every $(|\mathcal{M}|,\varepsilon)$ classical communication protocol, the condition $p_{\text{err}}^*(\mathcal{E},\mathcal{D};\mathcal{N})\leq\varepsilon$ holds. By following the same steps as in \eqref{eq-ea_classical_comm_arb_to_uniform_reduction_1}--\eqref{eq-ea_classical_comm_arb_to_uniform_reduction}, this condition implies that $\overline{p}_{\text{err}}((\mathcal{E},\mathcal{D});p,\mathcal{N})\leq\varepsilon$ for the uniform distribution $p$. Then, by \eqref{eq-classical_comm_comparator_succ_prob}, we find that
	\begin{equation}
		\Tr[\Pi_{M\widehat{M}}\omega_{M\widehat{M}}]\geq 1-\varepsilon.
	\end{equation}
	We can therefore use Lemma~\ref{lem-eac-meta_conv} to conclude that
	\begin{equation}\label{eq-cc_meta_conv_upper_bound}
		\log_2|\mathcal{M}|\leq I_H^{\varepsilon}(M;\widehat{M})_{\omega}
	\end{equation}
	for every $(|\mathcal{M}|,\varepsilon)$ classical communication protocol. This means that, given a particular choice of the encoding and decoding channels, if $p_{\text{err}}^*(\mathcal{E},\mathcal{D};\mathcal{N})\leq\varepsilon$, then the upper bound in \eqref{eq-cc_meta_conv_upper_bound} is the maximum number of bits that can be transmitted over the channel $\mathcal{N}$. The optimal value of this upper bound is realized by finding the state $\sigma_{\widehat{M}}$ defining the useless channel that optimizes the quantity $I_H^{\varepsilon}(M;\widehat{M})_{\omega}$ in addition to the measurement that achieves the $\varepsilon$-hypothesis testing relative entropy in \eqref{eq-eacc_strong_conv_lem_pf_2}. Importantly, a different choice of encoding and decoding produces a different value for this upper bound. We would thus like to find an upper bound that applies regardless of which specific protocol is chosen. In other words, we would like an upper bound that is a function of the channel $\mathcal{N}$ only.

\subsection{Upper Bound on the Number of Transmitted Bits}

	We now give a general upper bound on the number of transmitted bits that can be communicated in any classical communication protocol. This result is stated in Theorem~\ref{cor-cc_meta_str_weak_conv}, and the upper bound obtained therein holds independently of the encoding and decoding channels used in the protocol and depends only on the given communication channel $\mathcal{N}$.
	
	Let us start with an arbitrary $(|\mathcal{M}|,\varepsilon)$ classical communication protocol over the channel $\mathcal{N}$, corresponding to, as described at the beginning of this chapter, a message set $\mathcal{M}$, an encoding channel $\mathcal{E}$, and a decoding channel~$\mathcal{D}$. The error criterion $p_{\text{err}}^*(\mathcal{E},\mathcal{D};\mathcal{N})\leq\varepsilon$ holds by definition of an $(|\mathcal{M}|,\varepsilon)$ protocol, which implies the upper bound in \eqref{eq-cc_meta_conv_upper_bound} for the number $\log_2|\mathcal{M}|$ of transmitted bits in any $(|\mathcal{M}|,\varepsilon)$ classical communication protocol. Using this upper bound, we obtain the following:

	\begin{proposition*}{Upper Bound on  One-Shot Classical Capacity}{prop-cc:one-shot-bound-meta}
		Let $\mathcal{N}$ be a quantum channel. For every $(|\mathcal{M}|,\varepsilon)$ classical communication protocol over $\mathcal{N}$, with $\varepsilon\in[0,1]$, the number of bits transmitted over $\mathcal{N}$ is bounded from above by the $\varepsilon$-hypothesis testing Holevo information of $\mathcal{N}$, as defined in \eqref{eq-hypo_testing_Holevo_inf_chan}, i.e., 
		\begin{equation}\label{eq-cc:one-shot-bound-meta}
			\log_{2}|\mathcal{M}| \leq\chi_H^{\varepsilon}(\mathcal{N}).
		\end{equation}
		Therefore,
		\begin{equation}
			C^{\varepsilon}(\mathcal{N})\leq\chi_H^{\varepsilon}(\mathcal{N}).
		\end{equation}
	\end{proposition*}

	\begin{Proof}
		We start with the upper bound in \eqref{eq-cc_meta_conv_upper_bound}, i.e.,
		\begin{equation}
			\log_2|\mathcal{M}|\leq I_H^{\varepsilon}(M;\widehat{M})_{\omega},
		\end{equation}
		where the state $\omega_{M\widehat{M}}$ is defined in \eqref{eq-classical_comm_final_state_uniform}. Recall that this bound follows from Lemma~\ref{lem-eac-meta_conv}. Note that the state $\omega_{M\widehat{M}}$ can be written as
		\begin{equation}\label{eq-cc:one-shot-bound-meta_pf_2}
			\omega_{M\widehat{M}}=\mathcal{D}_{B\to\widehat{M}}(\theta_{MB}),
		\end{equation}
		where
		\begin{equation}
			\theta_{MB}\coloneqq \frac{1}{|\mathcal{M}|}\sum_{m\in\mathcal{M}}\ket{m}\!\bra{m}_M\otimes\mathcal{N}_{A\to B}(\rho_{A}^m).
		\end{equation}
		
		Now, from the data-processing inequality for the hypothesis testing relative entropy under the action of the decoding channel $\mathcal{D}_{B\to\widehat{M}}$, we find that
		\begin{equation}
			I_H^{\varepsilon}(M;\widehat{M})_\omega=\inf_{\sigma_{\widehat{M}}'}D_H^{\varepsilon}(\omega_{M\widehat{M}}\Vert\omega_M\otimes\sigma_{\widehat{M}}')\leq I_H^{\varepsilon}(M;B)_{\theta}\label{prop-cc:one-shot-bound-meta_pf4},
		\end{equation}
		where we have used the fact that $\theta_M=\pi_M=\omega_M$. Note that
		\begin{equation}
			\theta_{MB}=\mathcal{N}_{A\to B}(\rho_{MA}),
		\end{equation}
		where $\rho_{MA}$ is the classical--quantum state $\rho_{MA}^p$ defined in \eqref{eq-classical_comm_cq_state} with $p$ equal to the uniform probability distribution. Optimizing over all classical--quan\-tum states $\xi_{MA}$ then leads to
		\begin{equation}
			I_H^{\varepsilon}(M;B)_{\theta}\leq\sup_{\xi_{MA}}I_H^{\varepsilon}(M;B)_{\zeta}=\chi_{H}^{\varepsilon}(\mathcal{N}),
		\end{equation}
		where $\zeta_{MB}=\mathcal{N}_{A\to B}(\xi_{MA})$ and we have used the definition in \eqref{eq-hypo_testing_Holevo_inf_chan} for the $\varepsilon$-hypothesis testing Holevo information of a channel. Note that this optimization over all classi\-cal--quantum states is effectively an optimization over all possible encoding channels $\mathcal{E}_{M'\to A}$ that define the $(|\mathcal{M}|,\varepsilon)$ protocol. Putt\-ing everything together, we obtain
		\begin{equation}
			\log_2|\mathcal{M}|\leq I_H^{\varepsilon}(M;\widehat{M})_{\omega}\leq I_H^{\varepsilon}(M;B)_{\theta}\leq \chi_H^{\varepsilon}(\mathcal{N}),
		\end{equation}
		as required.
	\end{Proof}
	
	The result of Proposition~\ref{prop-cc:one-shot-bound-meta} can be written explicitly as
	\begin{align}
		\log_2|\mathcal{M}|&\leq \sup_{\rho_{MA}}\inf_{\sigma_B}D_H^{\varepsilon}(\mathcal{N}_{A\to B}(\rho_{MA})\Vert\rho_M\otimes\sigma_B)\\
		&=\sup_{\rho_{MA}}\inf_{\sigma_B}D_H^{\varepsilon}(\mathcal{N}_{A\to B}(\rho_{MA})\Vert\mathcal{R}_{A\to B}^{\sigma_B}(\rho_{MA})),\label{eq-cc:one-shot-bound-meta_2a}
	\end{align}
	where $\rho_{MA}$ is a classical--quantum state. By doing so, we explictly see here the comparison, via the hypothesis testing relative entropy, between the actual classical communication protocol and the protocol over the useless channels $\mathcal{R}_{A\to B}^{\sigma_B}$, labeled by the states $\sigma_B$. The state $\rho_{MA}$ corresponds to the state after the encoding channel, and optimizing over these states is effectively an optimization over all encoding channels.

	As an immediate consequence of Propositions~\ref{prop-cc:one-shot-bound-meta}, \ref{prop-hypo_to_rel_ent}, and \ref{prop:sandwich-to-htre}, we have the following two bounds:

	\begin{theorem*}{One-Shot Upper Bounds for Classical Communication}{cor-cc_meta_str_weak_conv}
		Let $\mathcal{N}$ be a quantum channel, let $\varepsilon \in [0,1)$, and let $\alpha>1$. For every $(|\mathcal{M}|,\varepsilon)$ classical communication protocol over $\mathcal{N}$, the following bounds hold
		\begin{align}
			\log_{2}|\mathcal{M}|  &  \leq\frac{1}{1-\varepsilon}\left(\chi(\mathcal{N})+h_2(\varepsilon)\right),\label{eq-cc_weak_conv_one_shot_1}\\
			\log_{2}|\mathcal{M}|  &  \leq\widetilde{\chi}_{\alpha}(\mathcal{N})+\frac{\alpha}{\alpha-1}\log_{2}\!\left(  \frac{1}{1-\varepsilon}\right),\label{eq-cc_str_conv_one_shot_1}
		\end{align}
		where $\chi(\mathcal{N})$ is the Holevo information of $\mathcal{N}$, as defined in \eqref{eq-Hol_inf_chan}, and $\widetilde{\chi}_{\alpha}(\mathcal{N})$ is the sandwiched R\'{e}nyi Holevo information of $\mathcal{N}$, as defined in~\eqref{eq-sand_rel_Holevo_inf_chan}.
	\end{theorem*}
	
	Since the bounds in \eqref{eq-cc_weak_conv_one_shot_1} and \eqref{eq-cc_str_conv_one_shot_1} hold for every $(|\mathcal{M}|,\varepsilon)$ classical communication protocol over $\mathcal{N}$, we have that
	\begin{align}
		C^{\varepsilon}(\mathcal{N})&\leq\frac{1}{1-\varepsilon}(\chi(\mathcal{N})+h_2(\varepsilon)),\\
		C^{\varepsilon}(\mathcal{N})&\leq\widetilde{\chi}_{\alpha}(\mathcal{N})+\frac{\alpha}{\alpha-1}\log_2\!\left(\frac{1}{1-\varepsilon}\right)\quad\forall~\alpha>1,
	\end{align}
	for all $\varepsilon\in[0,1)$.
	
	Let us recap the steps that we took to arrive at the bounds in \eqref{eq-cc_weak_conv_one_shot_1} and~\eqref{eq-cc_str_conv_one_shot_1}.
	\begin{enumerate}
		\item We first compared the classical communication protocol over the channel~$\mathcal{N}$ with a protocol over a useless channel, by using the $\varepsilon$-hypothesis testing relative entropy. This led us to the upper bound in \eqref{eq-cc_meta_conv_upper_bound}.
		\item We then used the data-processing inequality for the hypothesis testing relative entropy to obtain a quantity that is independent of the decoding channel, and also optimized over all useless protocols. This is done in \eqref{prop-cc:one-shot-bound-meta_pf4} in the proof of Proposition~\ref{prop-cc:one-shot-bound-meta}. 
		\item Finally, to obtain a bound that is a function solely of the channel $\mathcal{N}$ and the error probability, we optimized over all encoding channels to obtain Proposition~\ref{prop-cc:one-shot-bound-meta}. 
		\item Using Propositions \ref{prop-hypo_to_rel_ent} and \ref{prop:sandwich-to-htre}, which relate the hypothesis testing relative entropy to the quantum relative entropy and the sandwiched-R\'{e}nyi relative entropy, respectively, we arrived at Theorem~\ref{cor-cc_meta_str_weak_conv}.
		
	\end{enumerate} 
	
	The bounds in \eqref{eq-cc_weak_conv_one_shot_1} and \eqref{eq-cc_str_conv_one_shot_1} are fundamental upper bounds on the number of transmitted bits for \textit{every} classical communication protocol. 
	A natural question to ask is whether the upper bounds in \eqref{eq-cc_weak_conv_one_shot_1} and \eqref{eq-cc_str_conv_one_shot_1} can be achieved. In other words, is it possible to devise protocols such that the number of transmitted bits is equal to the right-hand side of either \eqref{eq-cc_weak_conv_one_shot_1} or \eqref{eq-cc_str_conv_one_shot_1}? We do not know how to, especially if we demand that we exactly attain the right-hand side of either \eqref{eq-cc_weak_conv_one_shot_1} or \eqref{eq-cc_str_conv_one_shot_1}. However, when given many uses of a channel (in the asymptotic setting), we can come close to achieving these upper bounds. This motivates finding lower bounds on the number of transmitted bits.

\subsection{Lower Bound on the Number of Transmitted Bits}

	Having obtained upper bounds on the number transmitted bits in the previous section, let us now determine lower bounds. The key result of this section is Proposition~\ref{prop-cc_one-shot_lower_bound}, resulting in Theorem~\ref{thm-cc_one_shot_lower_bound}, which contains a lower bound on the number of transmitted bits for every $(|\mathcal{M}|,\varepsilon)$ classical communication protocol.
	
	As we saw in the previous chapter on entanglement-assisted classical co\-mmunication, in order to obtain a lower bound on the number of transmitted bits, we should devise an explicit classical communication protocol $(\mathcal{M},\mathcal{E},\mathcal{D})$ such that the maximal error probability satisfies $p_{\text{err}}^*(\mathcal{E},\mathcal{D};\mathcal{N})\leq\varepsilon$ for $\varepsilon\in[0,1]$. Recall from \eqref{eq-maximal_error_prob} that the maximal error probability is defined as
	\begin{equation}
		p_{\text{err}}^*(\mathcal{E},\mathcal{D};\mathcal{N})=\max_{m\in\mathcal{M}}p_{\text{err}}(m,(\mathcal{E},\mathcal{D});\mathcal{N}),
	\end{equation}
	where, for all $m\in\mathcal{M}$, the message error probability $p_{\text{err}}(m;(\mathcal{E},\mathcal{D}))$ is defined in \eqref{eq-mess_error_prob} as
	\begin{equation}
		p_{\text{err}}(m,(\mathcal{E},\mathcal{D});\mathcal{N})=1-q(m|m),
	\end{equation}
	with $q(\widehat{m}|m)$ being the probability of identifying the message sent as $\widehat{m}$ given that the message $m$ was sent.
	
	The classical communication protocol discussed here is related to the enta\-nglement-assisted classical communication protocol in Section~\ref{subsec-pos_coding}. We suppose at first that Alice and Bob have some shared randomness prior to communication. This shared randomness is strictly speaking not part of the classical communication protocol as outlined at the beginning of Section~\ref{sec-cc_one_shot}, but the advantage of using it is that we can directly employ all of the developments for the position-based coding and sequential decoding strategy from Section~\ref{subsec-pos_coding}. We then perform what is called \textit{derandomization} and \textit{expurgation} (both of which we outline below) ultimately to remove this shared randomness from the protocol and thus obtain the desired lower bound on the number of transmitted bits for the true unassisted classical communication protocol.
	
	\begin{proposition*}{Lower Bound on  One-Shot Classical Capacity}{prop-cc_one-shot_lower_bound}
		Let $\mathcal{N}_{A\to B}$ be a quantum channel. For all $\varepsilon\in (0,1)$ and $\eta\in\left(0,\frac{\varepsilon}{2}\right)$, there exists an $(|\mathcal{M}|,\varepsilon)$ classical communication protocol over $\mathcal{N}_{A\to B}$ such that
		\begin{equation}\label{eq-cc_one-shot_lower_bound}
			\log_2|\mathcal{M}|=\overline{\chi}_H^{\frac{\varepsilon}{2}-\eta}(\mathcal{N})-\log_2\!\left(\frac{4\varepsilon}{\eta^2}\right).
		\end{equation}
		 Consequently, for all $\varepsilon\in(0,1)$ and $\eta\in\left(0,\frac{\varepsilon}{2}\right)$, 
		\begin{equation}
			C^{\varepsilon}(\mathcal{N})\geq\overline{\chi}_H^{\frac{\varepsilon}{2}-\eta}(\mathcal{N})-\log_2\!\left(\frac{4\varepsilon}{\eta^2}\right).
		\end{equation}
		Here,
		\begin{equation}
			\overline{\chi}_H^{\varepsilon}(\mathcal{N})\coloneqq\sup_{\rho_{XA}}\overline{I}_H^{\varepsilon}(X;B)_{\omega},
		\end{equation}
		where $\omega_{XB}=\mathcal{N}_{A\to B}(\rho_{XA})$, the state $\rho_{XA}$ is a classical--quantum state, and
		\begin{equation}
			\overline{I}_H^{\varepsilon}(X;B)_\omega=D_H^{\varepsilon}(\omega_{XB}\Vert\omega_X\otimes\omega_B).
		\end{equation}
	\end{proposition*}
	
	\begin{remark}
		The quantity $\overline{\chi}_H^{\varepsilon}(\mathcal{N})$ defined in the statement of Proposition~\ref{prop-cc_one-shot_lower_bound} above is similar to the quantity $\chi_H^{\varepsilon}(\mathcal{N})$ defined in \eqref{eq-hypo_testing_Holevo_inf_chan}, except that it is defined with respect to the mutual information $\overline{I}_H^{\varepsilon}(X;B)_\rho$ that we encountered in Proposition~\ref{prop-eac:one-shot-lower_bound}, which does not involve an optimization over states $\sigma_B$. 
	\end{remark}
	
	\begin{Proof}
		As described before the statement of the proposition, we start with a protocol based on randomness-assisted classical communication, in which Alice and Bob have shared randomness prior to communication via the state $\rho_{XB'}^{\otimes|\mathcal{M}'|}$, where $\mathcal{M}'$ is a message set and $\rho_{XB'}$ is the following classically correlated state:
		\begin{equation}\label{eq-cc_shared_randomness_state}
			\rho_{XB'}\coloneqq\sum_{x\in\mathcal{X}}r(x)\ket{x}\!\bra{x}_{X}\otimes\ket{x}\!\bra{x}_{B'}.
		\end{equation}
		Here, $\mathcal{X}$ is a finite alphabet and $r:\mathcal{X}\rightarrow [0,1]$ is a probability distribution on $\mathcal{X}$. The system $X$ is held by Alice and the system $B'$ is held by Bob. We denote the encoding and decoding channels for this protocol by $\mathcal{E}'$ and $\mathcal{D}'$, respectively, and they correspond to the position-based coding and sequential decoding strategies developed in Section~\ref{subsec-pos_coding}. The goal is to use this protocol to determine the existence of encoding and decoding channels $\mathcal{E}$ and $\mathcal{D}$
for an $(|\mathcal{M}|,\varepsilon)$ classical communication protocol.

		As a first step, Alice processes each of her $X$ systems with a classical--quantum channel $x\mapsto\rho_{A}^{x}$ (see Definition~\ref{def-cq_channel}), so that the state shared by them becomes $\rho_{A^{\prime}B^{\prime}}^{\otimes|\mathcal{M}'|}$, where%
		\begin{equation}
			\rho_{A'B'}=\sum_{x\in\mathcal{X}}r(x)\rho_{A'}^{x}\otimes\ket{x}\!\bra{x}_{B^{\prime}}.
		\end{equation}
		Just as in Section~\ref{subsec-pos_coding}, the rest of the encoding channel $\mathcal{E}'$ is defined such that if Alice wishes to send the message $m\in\mathcal{M}'$, then she sends the $m{\text{th}}$ $A$ system through the channel. Thus, the state shared by Alice and Bob becomes
		\begin{equation}\label{eq-cc_one-shot_lower_bound_pf1}
			\rho_{A_{1}^{\prime}B_{1}^{\prime}}\otimes\dotsb\otimes\mathcal{N}_{A\rightarrow B}(\rho_{AB_{m}^{\prime}}%
)\otimes\dotsb\otimes\rho_{A_{|\mathcal{M}^{\prime}|}^{\prime}B_{|\mathcal{M}^{\prime}|}^{\prime}}
		\end{equation}
		for all $m\in\mathcal{M}^{\prime}$, where
		\begin{equation}\label{eq-cc_one-shot_lower_bound_pf1a}
			\rho_{A_{i}^{\prime}B_{i}^{\prime}}\coloneqq \sum_{x\in\mathcal{X}}r(x)\rho_{A'_{i}}^{x}\otimes\ket{x}\!\bra{x}_{B_{i}^{\prime}}%
.
		\end{equation}
		The reduced state on Bob's systems is then
		\begin{equation}
			\tau^m_{B_{1}^{\prime}\dotsb B_{m}^{\prime}\dotsb B_{|\mathcal{M}'|}^{\prime}B}=\rho_{B_{1}^{\prime}}\otimes\dotsb\otimes\mathcal{N}_{A\rightarrow B}(\rho_{AB_{m}^{\prime}})\otimes\dotsb\otimes \rho_{B_{|\mathcal{M}^{\prime}|}^{\prime}}.
			\label{eq:CC-comm:reduced-state-Bob}
		\end{equation}
		In particular, we have that
		\begin{equation}
			\tau^m_{B_{\widehat{m}}^{\prime}B}=\left\{\begin{array}[c]{cl} \mathcal{N}_{A^{\prime}\rightarrow B}(\rho_{A^{\prime}B^{\prime}}) & \text{if }\widehat{m}=m,\\
			\rho_{B^{\prime}}\otimes\mathcal{N}_{A^{\prime}\rightarrow B}(\rho_{A^{\prime}}) & \text{if }\widehat{m}\neq m,
			\end{array}\right.
		\end{equation}
		for all $m\in\mathcal{M}^{\prime}$. Bob's task is to perform a test to guess which of the two states $\mathcal{N}_{A^{\prime}\rightarrow B}(\rho_{A^{\prime}B^{\prime}})$ and $\rho_{B^{\prime}}\otimes\mathcal{N}_{A^{\prime}\rightarrow B}(\rho_{A^{\prime}})$ he has on $B$ and the $|\mathcal{M}%
^{\prime}|$ systems $B_{1}^{\prime}\dotsb B_{|\mathcal{M}^{\prime}|}^{\prime}$. Since the shared state in the proof of Proposition~\ref{prop-eac:one-shot-lower_bound} is arbitrary, we can apply all the arguments in that proof with the state $\rho_{A^{\prime}B^{\prime}}$ defined in \eqref{eq-cc_one-shot_lower_bound_pf1a} above to conclude immediately via
\eqref{eq-eac:one-shot-lower_bound_pf3} that
		\begin{equation}\label{eq-cc_one-shot_lower_bound_pf1b}
			p_{\text{err}}(m,(\mathcal{E}^{\prime},\mathcal{D}^{\prime});\mathcal{N})\leq \varepsilon
		\end{equation}
		for all $m\in\mathcal{M}^{\prime}$, provided that
		\begin{equation}
		\log_{2}|\mathcal{M}^{\prime}|=\overline{I}_{H}^{\varepsilon-\eta}(B^{\prime};B)_{\xi}-\log_{2}\!\left(  \frac{4\varepsilon}{\eta^{2}}\right),
		\end{equation}
		where $\xi_{B^{\prime}B}\coloneqq\mathcal{N}_{A^{\prime}\rightarrow B}(\rho_{A^{\prime}B^{\prime}})$. Note that $\xi_{B^{\prime}B}$ is a classical--quantum state, which means that $\overline{I}_{H}^{\varepsilon-\eta}(B^{\prime};B)_{\xi}=\overline{\chi}_{H}^{\varepsilon-\eta}(B^{\prime};B)_{\xi}$. Furthermore, by taking a supremum over every input ensemble $\{(r(x),\rho_{A}^{x})\}_{x\in\mathcal{X}}$,
we find that
		\begin{equation}\label{eq-cc_one-shot_lower_bound_pf4}
			\log_{2}|\mathcal{M}^{\prime}|=\overline{\chi}_{H}^{\varepsilon-\eta}(\mathcal{N})-\log_{2}\!\left(  \frac{4\varepsilon}{\eta^{2}}\right).
		\end{equation}
		Combining \eqref{eq-cc_one-shot_lower_bound_pf1b} and \eqref{eq-cc_one-shot_lower_bound_pf4}, we can already conclude that, when shared randomness is available for free, a lower bound on the number of transmitted bits is given by \eqref{eq-cc_one-shot_lower_bound_pf4}. The condition in \eqref{eq-cc_one-shot_lower_bound_pf1b} on the message error probability implies that the average error probability $\overline{p}_{\text{err}}((\mathcal{E}^{\prime},\mathcal{D}^{\prime});p)$, with $p$ the uniform distribution over $\mathcal{M}^{\prime}$, satisfies
		\begin{equation}\label{eq-cc_one-shot_lower_bound_pf2}%
			\overline{p}_{\text{err}}((\mathcal{E}^{\prime},\mathcal{D}^{\prime});p,\mathcal{N})= \frac{1}{|\mathcal{M}^{\prime}|} \sum_{m\in\mathcal{M}'}p_{\text{err}}(m,(\mathcal{E}^{\prime},\mathcal{D}^{\prime}),\mathcal{N})\leq\varepsilon.
		\end{equation}

		Let us now use the expression in \eqref{eq-eacc_seq_coding_mess_err_prob} to derive an exact expression for the average error probability. The expression in \eqref{eq-eacc_seq_coding_mess_err_prob} is
		\begin{equation}
			\begin{aligned} 
			&p_{\text{err}}(m,(\mathcal{E},\mathcal{D});\mathcal{N})\\ &\qquad\qquad=1-\Tr[P_m\widehat{P}_{m-1}\dotsb \widehat{P}_1\omega_{B_1'\dotsb B_{|\mathcal{M}'|}'BR_1\dotsb R_{|\mathcal{M}'|}}^m \widehat{P}_1\dotsb \widehat{P}_{m-1}P_m] \end{aligned}\label{eq-cc_seq_coding_mess_err_prob}%
		\end{equation}
		for all $m\in\mathcal{M}^{\prime}$. First, observe that both $\mathcal{N}_{A^{\prime}\rightarrow B}(\rho_{A^{\prime}B^{\prime}})$ and $\rho_{B^{\prime}}\otimes\mathcal{N}_{A^{\prime}\rightarrow B}(\rho_{A^{\prime}})$ are classical--quantum states. This means that, for every measurement operator $\Lambda_{BB^{\prime}}$, we find that
		\begin{align}
			\Tr[\Lambda_{BB'}\mathcal{N}_{A'\to B}(\rho_{B'A'})]&=\sum_{x\in\mathcal{X}}r(x)\Tr[\Lambda_{B'B}(\ket{x}\!\bra{x}_{B'}\otimes\rho_B^x)]\\
			&=\sum_{x\in\mathcal{X}}r(x)\Tr[(\bra{x}_{B'}\otimes\mathbbm{1}_B)\Lambda_{B'B}(\ket{x}_{B'}\otimes\mathbbm{1}_B)\rho_B^x]\\
			&=\sum_{x\in\mathcal{X}}r(x)\Tr[M_B^x\rho_B^x],
		\end{align}
		where $\rho_{B}^{x}=\mathcal{N}_{A^{\prime}\rightarrow B}(\rho_{A^{\prime}}^{x})$, and we have defined the operators
		\begin{equation}
			M_{B}^{x}\coloneqq\Tr_{B^{\prime}}[(\ket{x}\!\bra{x}_{B^{\prime}}\otimes\mathbbm{1}_{B^{\prime}})\Lambda_{B^{\prime}B}].
		\end{equation}
		Similarly, letting $\overline{\rho}_{B}\coloneqq\sum_{x\in\mathcal{X}}r(x)\rho_{B}^{x}$, we find that
		\begin{align}
			\Tr[\Lambda_{B'B}(\rho_{B'}\otimes\mathcal{N}_{A'\to B}(\rho_{A'}))] & =\sum_{x\in\mathcal{X}}r(x)\Tr[\Lambda_{B'B}(\ket{x}\!\bra{x}_{B'}\otimes\overline{\rho}_B)]\\
			&  =\sum_{x\in\mathcal{X}}r(x)\Tr[M_B^x\overline{\rho}_B].
		\end{align}
		This implies that the measurement operator $\Lambda_{B^{\prime}B}^{\ast}$ that achieves the optimal value for the quantity $D_{H}^{\varepsilon-\eta}(\mathcal{N}_{A^{\prime}\rightarrow B}(\rho_{A^{\prime}B^{\prime}})\Vert\rho_{B^{\prime}}\otimes\mathcal{N}_{A^{\prime}\rightarrow B}(\rho_{A^{\prime}}))$ can be taken to have the form
		\begin{equation}\label{eq-cc_pos_coding_pf1}%
			\Lambda_{B^{\prime}B}^{\ast}=\sum_{x\in\mathcal{X}}\ket{x}\!\bra{x}_{B^{\prime}}\otimes M_{B}^{x}.
		\end{equation}
		Now using the fact that%
		\begin{equation}
			\sqrt{\Lambda_{B^{\prime}B}^{\ast}}=\sum_{x\in\mathcal{X}}\ket{x}\!\bra{x}_{B^{\prime}}\otimes\sqrt{M_{B}^{x}},
		\end{equation}
		the projectors $\Pi_{B^{\prime}BR}$ defined in \eqref{eq-eacc_one_shot_lower_bound_pf0a} have the following form:%
		\begin{equation}
			\Pi_{B^{\prime}BR}=\sum_{x\in\mathcal{X}}\ket{x}\!\bra{x}_{B^{\prime}}\otimes \Pi_{BR}^{x},
		\end{equation}
		where $R$ is a reference system held by Bob to help with the decoding, and the projector $\Pi_{BR}^{x}$ is given by
		\begin{align}
			\Pi_{BR}^{x}&\coloneqq (U_{BR}^{x})^{\dagger}\left(\mathbbm{1}_{B}\otimes\ket{1}\!\bra{1}_{R}\right)U_{BR}^{x},\\
			U_{BR}^{x}&\coloneqq\sqrt{\mathbbm{1}_{B}-M_{B}^{x}}\otimes\left(\ket{0}\!\bra{0}_{R}+\ket{1}\!\bra{1}_{R}\right)\nonumber\\
			&\qquad+\sqrt{M_{B}^{x}}\otimes\left(\ket{1}\bra{0}_{R}-\ket{0}\!\bra{1}_{R}\right)  .
		\end{align}
		This in turn implies that the measurement operators $P_{i}$, which are used for the sequential decoding and are defined in \eqref{eq-eacc-seq_decode}, have the form
		\begin{equation}
			P_{i}=\sum_{x_{1},\dotsc,x_{|\mathcal{M}^{\prime}|}\in\mathcal{X}}\ket{\underline{x}}\bra{\underline{x}}_{B_{1}^{\prime}\dotsb B_{|\mathcal{M}^{\prime}|}^{\prime}}\otimes P_{i}^{x_{i}},
		\end{equation}
		where $\ket{\underline{x}}\equiv\ket{x_1,\dotsc,x_{|\mathcal{M}'|}}$ and
		\begin{equation}
			P_{i}^{x_{i}}\coloneqq\mathbbm{1}_{R_{1}}\otimes\dotsb\otimes\mathbbm{1}_{R_{i-1}}\otimes\Pi_{BR_{i}}^{x_{i}}\otimes\mathbbm{1}_{R_{i+1}}\otimes\dotsb\otimes\mathbbm{1}_{R_{|\mathcal{M}^{\prime}|}}.
		\end{equation}
		Finally, since we can write the state $\tau^m_{B_{1}^{\prime}\dotsb  B_{|\mathcal{M}^{\prime}|}^{\prime}B}$ in \eqref{eq:CC-comm:reduced-state-Bob} 
		as
		\begin{equation}
			\tau^m_{B_{1}^{\prime}\dotsb B_{|\mathcal{M}^{\prime}|}^{\prime}B}=\sum_{x_{1},\dotsc,x_{|\mathcal{M}^{\prime}|}\in\mathcal{X}}r(x_{1})\dotsb r(x_{|\mathcal{M}^{\prime}|})\ket{\underline{x}}\bra{\underline{x}}_{B_{1}^{\prime}\dotsb B_{|\mathcal{M}^{\prime}|}^{\prime}}\otimes\rho_{B}^{x_{m}},
		\end{equation}
		we find that
		\begin{align}
			&\omega^m_{B_{1}^{\prime}\dotsb B_{|\mathcal{M}^{\prime}|}^{\prime}BR_{1}\dotsb R_{|\mathcal{M}^{\prime}|}}\nonumber\\
			&\coloneqq\tau^m_{B_{1}^{\prime}\dotsb B_{|\mathcal{M}^{\prime}|}^{\prime}B}\otimes\ket{\underline{0}}\bra{\underline{0}}_{R_{1}\dotsb R_{|\mathcal{M}^{\prime}|}}\\
			&  =\sum_{x_{1},\dotsc,x_{|\mathcal{M}^{\prime}|}\in\mathcal{X}}r(x_{1})\dotsb r(x_{|\mathcal{M}^{\prime}|})\ket{\underline{x}}\bra{\underline{x}}_{B_{1}^{\prime}\dotsb B_{|\mathcal{M}^{\prime}|}^{\prime}}\otimes(\rho_{B}^{x_{m}}\otimes\ket{\underline{0}}\bra{\underline{0}}_{R_{1}\dotsb R_{|\mathcal{M}^{\prime}|}}),
		\end{align}
		where $\ket{\underline{0}}\equiv\ket{0,\dotsc,0}$. Therefore, by definition,
		\begin{align}
			&p_{\text{err}}(m;(\mathcal{E}^{\prime},\mathcal{D}^{\prime}))\nonumber\\
			&\quad =1-\Tr[P_m\widehat{P}_{m-1}\dotsb \widehat{P}_1\omega_{B_1'\dotsb B_{|\mathcal{M}'|}'BR_1\dotsb R_{|\mathcal{M}'|}}\widehat{P}_1\dotsb\widehat{P}_{m-1}P_m]\\
			&  \quad=\sum_{x_{1},\dotsc,x_{|\mathcal{M}^{\prime}|}\in\mathcal{X}}\left[r(x_{1})\dotsb r(x_{|\mathcal{M}^{\prime}|})\right.  \nonumber\\
			&  \qquad\qquad\qquad\qquad\left.  \times\left(
1-\Tr[\Omega_m^{x_m}(\rho_B^{x_m}\otimes\ket{\underline{0}}\bra{\underline{0}}_{R_1\dotsb R_{|\mathcal{M}'|}})]\right)\right]  ,
		\end{align}
		for all $m\in\mathcal{M}^{\prime}$, where
		\begin{equation}
			\Omega_{m}^{x_{m}}\coloneqq\widehat{P}_{1}^{x_{1}}\dotsb\widehat{P}_{m-1}^{x_{m-1}}P_{m}^{x_{m}}\widehat{P}_{m-1}^{x_{m-1}}\dotsb\widehat{P}_{1}^{x_{1}}.
		\end{equation}
		Therefore, the average error probability is bounded as
		\begin{align}
			&  \overline{p}_{\text{err}}((\mathcal{E}^{\prime},\mathcal{D}^{\prime});p)\nonumber\\
			&  =\sum_{m\in\mathcal{M}^{\prime}}\frac{1}{|\mathcal{M}^{\prime}|}\sum_{x_{1},\dotsc,x_{|\mathcal{M}^{\prime}|}\in\mathcal{X}}\left[  r(x_{1})\dotsb r(x_{|\mathcal{M}^{\prime}|})\right.  \nonumber\\
			&  \qquad\qquad\qquad\qquad\left.  \times\left(1-\Tr[\Omega_m^{x_m}(\rho_B^{x_m}\otimes\ket{\underline{0}}\bra{\underline{0}}_{R_1\dotsb R_{|\mathcal{M}'|}})]\right)\right]  \\
			&  \leq\sum_{m\in\mathcal{M}^{\prime}}\frac{1}{|\mathcal{M}^{\prime}|}\sum_{x_{1},\dotsc,x_{|\mathcal{M}^{\prime}|}\in\mathcal{X}}\left[  r(x_{1})\dotsb r(x_{|\mathcal{M}^{\prime}|})\right.  \nonumber\\
			&  \qquad\qquad\left.  \times\left(  \gamma_{\text{I}}\Tr[(I_{B}-M_{B}^{x_{m}})\rho_{B}^{x_{m}}]+\gamma_{\text{II}}\sum_{i=1}^{m-1}\Tr[M_{B}^{x_{i}}\rho_{B}^{x_{m}}]\right)  \right]  \\
			&  \leq\varepsilon,
		\end{align}
		where $\gamma_{\text{I}}\coloneqq 1+c$ and $\gamma_{\text{II}}\coloneqq 2+c+c^{-1}$, with $c=\frac{\eta}{2\varepsilon-\eta}$, and the inequality on the last line holds due to \eqref{eq-cc_one-shot_lower_bound_pf2}. Exchanging the sum over $\mathcal{M}^{\prime}$ with the sum over the elements $x_{1},\dotsc,x_{|\mathcal{M}^{\prime}|}$ of $\mathcal{X}$ (in the spirit of the famous trick of Shannon), we find that%
		\begin{multline}
			\sum_{x_{1},\dotsc,x_{|\mathcal{M}^{\prime}|}\in\mathcal{X}}r(x_{1})\dotsb r(x_{|\mathcal{M}^{\prime}|})\overline{p}_{\text{err}}(\mathcal{C};p)\label{eq-cc_one-shot_lower_bound_pf3}\\
			 \leq\sum_{x_{1},\dotsc,x_{|\mathcal{M}^{\prime}|}\in\mathcal{X}}r(x_{1})\dotsb r(x_{|\mathcal{M}^{\prime}|})\overline{u}_{\text{err}}(\mathcal{C};p)\leq\varepsilon,
		\end{multline}
		where
		\begin{equation}
			\overline{p}_{\text{err}}(\mathcal{C};p,\mathcal{N})\coloneqq\sum_{m\in\mathcal{M}^{\prime}}\frac{1}{|\mathcal{M}^{\prime}|}\left(1-\Tr[\Omega_m^{x_m}(\rho_B^{x_m}\otimes\ket{\underline{0}}\bra{\underline{0}}_{R_1\dotsb R_{|\mathcal{M}'|}})]\right)
		\end{equation}
		is the average error probability under a code $\mathcal{C}$ in which each message $m$ is encoded as $m\mapsto x_{m}\mapsto\rho_{A}^{x_{m}}$ and%
		\begin{equation}
			\overline{u}_{\text{err}}(\mathcal{C};p,\mathcal{N})\coloneqq\sum_{m\in\mathcal{M}^{\prime}}\frac{1}{|\mathcal{M}^{\prime}|}\left(
\gamma_{\text{I}}\operatorname{Tr}[(I_{B}-M_{B}^{x_{m}})\rho_{B}^{x_{m}}]+\gamma_{\text{II}}\sum_{i=1}^{m-1}\operatorname{Tr}[M_{B}^{x_{i}}\rho_{B}^{x_{m}}]\right)
		\end{equation}
		is an upper bound on the average error probability $\overline{p}_{\text{err}}(\mathcal{C};p,\mathcal{N})$. The decoding is defined by the measurement operators $\{\Omega_{m}^{x_{m}}\}_{m\in\mathcal{M}^{\prime}}$. Note that the code $\mathcal{C}$ is a random variable, in the sense that the string $x_{1},\dotsc,x_{|\mathcal{M}^{\prime}|}$ of length $|\mathcal{M}^{\prime}|$ is used for the encoding and decoding with probability $r(x_{1})\dotsb r(x_{|\mathcal{M}^{\prime}|})$.

		Since the minimum does not exceed the average, the inequality in \eqref{eq-cc_one-shot_lower_bound_pf3} implies that there exists a code $\mathcal{C}^{\ast}$, with corresponding string $x_{1}^{\ast},\dotsc,x_{|\mathcal{M}^{\prime}|}^{\ast}$, such that
		\begin{equation}
			\overline{u}_{\text{err}}(\mathcal{C}^{\ast};p,\mathcal{N})\leq\varepsilon,\label{eq-cc_one-shot_lower_bound_pf5}%
		\end{equation}
		and in turn, via Theorem~\ref{thm-q_union_bd}, that%
		\begin{equation}
			\overline{p}_{\text{err}}(\mathcal{C}^{\ast};p,\mathcal{N})\leq\overline{u}_{\text{err}}(\mathcal{C}^{\ast};p,\mathcal{N})\leq\varepsilon.
		\end{equation}
		By choosing this particular code, we can now follow through the entire argument above \textit{without} the shared randomness (in the form of the state $\rho_{A^{\prime}B^{\prime}}$) in order to conclude that with the code $\mathcal{C}^{\ast}$, the number of transmitted bits is given by \eqref{eq-cc_one-shot_lower_bound_pf4}, and the average error probability of
the code is bounded from above by $\varepsilon$. This completes the
\textit{derandomization} part of the proof.

		Finally, we are interested in a code, call it $(\mathcal{E},\mathcal{D})$, satisfying the maximal error probability criterion $p_{\text{err}}^{\ast}(\mathcal{E},\mathcal{D};\mathcal{N})\leq\varepsilon$ instead of the average error probability criterion. To find such a code, we can apply \textit{expurgation} to the code $\mathcal{C}^{\ast}$ defined above. Formally, this means the
following: since we have a code satisfying \eqref{eq-cc_one-shot_lower_bound_pf5}, by Markov's inequality (see \eqref{eq-MT:Markov-ineq}), half of the codewords in $\mathcal{C}^{\ast}$  (call them $c_{1},\dotsc,c_{\frac{|\mathcal{M}^{\prime}|}{2}}$) satisfy
		\begin{equation}
			u_{\text{err}}(m;\mathcal{C}^{\ast})\leq2\varepsilon ,
		\end{equation}
		for all $m\in\mathcal{M}^{\prime}$ corresponding to the codewords $c_{1},\dotsc,c_{\frac{|\mathcal{M}^{\prime}|}{2}}$, where%
		\begin{equation}
			u_{\text{err}}(m,\mathcal{C}^{\ast};\mathcal{N})\coloneqq\gamma_{\text{I}}\Tr[(I_{B}-M_{B}^{x_{m}})\rho_{B}^{x_{m}}]+\gamma_{\text{II}}\sum_{i=1}^{m-1}\Tr[M_{B}^{x_{i}}\rho_{B}^{x_{m}}].
		\end{equation}
		We thus define a new message set $\mathcal{M}\subset\mathcal{M}^{\prime}$, with $|\mathcal{M}|=\frac{|\mathcal{M}^{\prime}|}{2}$, by removing all but those messages in $\mathcal{M}^{\prime}$ whose encodings are given by $c_{1},\dotsc,c_{\frac{|\mathcal{M}^{\prime}|}{2}}$. Let $\mathcal{C}$ denote
the expurgated code. Due to the fact that all of the terms in $u_{\text{err}}(m,\mathcal{C}^{\ast};\mathcal{N})$ are non-negative, we find for all $m\in\mathcal{M}$ that%
		\begin{equation}
			u_{\text{err}}(m,\mathcal{C};\mathcal{N})\leq u_{\text{err}}(m,\mathcal{C}^{\ast};\mathcal{N}),
		\end{equation}
		where%
		\begin{equation}
			u_{\text{err}}(m,\mathcal{C};\mathcal{N})\coloneqq\gamma_{\text{I}}\Tr[(I_{B}-M_{B}^{c_{m}})\rho_{B}^{c_{m}}]+\gamma_{\text{II}}\sum_{i=1}^{m-1}\Tr[M_{B}^{c_{i}}\rho_{B}^{c_{m}}].
		\end{equation}
		Again applying the quantum union bound (Theorem~\ref{thm-q_union_bd}), we then find that%
		\begin{equation}
			p_{\text{err}}(m,\mathcal{C};\mathcal{N})\leq u_{\text{err}}(m,\mathcal{C};\mathcal{N}),
		\end{equation}
		where%
		\begin{multline}
			p_{\text{err}}(m,\mathcal{C};\mathcal{N})\coloneqq 1-\\
			\Tr[P_{m}^{c_{m}}\widehat{P}_{m-1}^{c_{m-1}}\dotsb\widehat{P}_{1}^{c_{1}}(\rho_{B}^{c_{m}}\otimes\ket{\underline{0}}\bra{\underline{0}}_{R_{1}\dotsb R_{|\mathcal{M}^{\prime}|}})\widehat{P}_{1}^{c_{1}}\dotsb\widehat{P}_{m-1}^{c_{m-1}}P_{m}^{c_{m}}].
		\end{multline}
		We thus have a code $(\mathcal{E},\mathcal{D})$ satisfying
		\begin{equation}
			p_{\text{err}}^{\ast}(\mathcal{E},\mathcal{D};\mathcal{N})\leq 2\varepsilon.
		\end{equation}
		Specifically, the encoding is given by $m\mapsto c_{m}$ for all $m\in \mathcal{M}$, and the decoding is given by the sequential decoding procedure
consisting of sequentially applying the binary measurements $\{P_{m}^{c_{m}},\widehat{P}_{m}^{c_{m}}\}$ for all $m\in\mathcal{M}$ and decoding as message $m$ as soon as the outcome $P_{m}^{c_{m}}$ occurs.

		Therefore, we can use \eqref{eq-cc_one-shot_lower_bound_pf4} to obtain the following for the number $\log_{2}|\mathcal{M}|$ of transmitted bits with the reduced message set:
		\begin{align}
			\log_{2}|\mathcal{M}|&=\log_{2}\!\left(  \frac{|\mathcal{M}^{\prime}|}{2}\right)=\log_{2}|\mathcal{M}^{\prime}|-\log_{2}(2)\\
			&=\overline{\chi}_{H}^{\varepsilon-\eta}(\mathcal{N})-\log_{2}\!\left(\frac{4\varepsilon}{\eta^{2}}\right)  -\log_{2}(2)\\
			&=\overline{\chi}_{H}^{\varepsilon-\eta}(\mathcal{N})-\log_{2}\!\left(\frac{8\varepsilon}{\eta^{2}}\right)  .
		\end{align}
		Since $\varepsilon$ and $\eta$ are arbitrary, we have shown that for all $\varepsilon\in(0,1)$ and $\eta\in(0,\varepsilon)$, there exists an $(|\mathcal{M}|,2\varepsilon)$ classical communication protocol satisfying $\log_{2}|\mathcal{M}| = \overline{\chi}_{H}^{\varepsilon-\eta}(\mathcal{N})-\log_{2}\!\left(
\frac{8\varepsilon}{\eta^{2}}\right)$. By the substitution $2\varepsilon\to \varepsilon$, we can finally say that for all
$\varepsilon\in(0,1)$ and  $\eta\in\left(0,\frac{\varepsilon}{2}\right)$, there exists an $(|\mathcal{M}|,\varepsilon)$ classical communication protocol satisfying
		\begin{equation}
			\log_{2}|\mathcal{M}|=\overline{\chi}_{H}^{\frac{\varepsilon}{2}-\eta}(\mathcal{N})-\log_{2}\!\left(\frac{4\varepsilon}{\eta^{2}}\right).
		\end{equation}
		This concludes the proof.
	\end{Proof}

	An immediate consequence of Propositions~\ref{prop-cc_one-shot_lower_bound} and \ref{prop:ineq-hypo-renyi} is the following theorem.
	
	\begin{theorem*}{One-Shot Lower Bounds for Classical Communication}{thm-cc_one_shot_lower_bound}
		Let $\mathcal{N}_{A\to B}$ be a quantum channel. For all $\varepsilon\in(0,1)$, $\eta\in\left(0,\frac{\varepsilon}{2}\right)$, and $\alpha\in(0,1)$, there exists an $(|\mathcal{M}|,\varepsilon)$ classical communication protocol over $\mathcal{N}_{A\to B}$ such that
		\begin{equation}\label{eq-cc_one_shot_lower_bound}
			\log_2|\mathcal{M}|\geq\overline{\chi}_\alpha(\mathcal{N})+\frac{\alpha}{\alpha-1}\log_2\!\left(\frac{1}{\frac{\varepsilon}{2}-\eta}\right)-\log_2\!\left(\frac{4\varepsilon}{\eta^2}\right).
		\end{equation}
		Here,
		\begin{equation}\label{eq-petz_renyi_Hol_inf_chan_noopt}
			\overline{\chi}_\alpha(\mathcal{N})\coloneqq\sup_{\rho_{XA}}\overline{I}_\alpha(X;B)_{\omega},
		\end{equation}
		where $\omega_{XB}=\mathcal{N}_{A\to B}(\rho_{XA})$, the state $\rho_{XA}$ is a classical--quantum state, and 
		\begin{equation}
			\overline{I}_\alpha(X;B)_{\omega}\coloneqq D_\alpha(\omega_{XB}\Vert\omega_X\otimes\omega_B).
		\end{equation}
	\end{theorem*}
	
	\begin{remark}
		The quantity $\overline{\chi}_\alpha(\mathcal{N})$ defined in the statement of Theorem~\ref{thm-cc_one_shot_lower_bound} above is similar to the quantity $\chi_{\alpha}(\mathcal{N})$ defined in \eqref{eq-petz_renyi_Hol_inf_chan}, except that it is defined with respect to the mutual information $\overline{I}_\alpha(X;B)_\omega$ that we encountered in Theorem~\ref{thm-eacc_one_shot_lower_bound}, which does not involve an optimization over states $\sigma_B$.
	\end{remark}
	
	\begin{Proof}
		From Proposition~\ref{prop-cc_one-shot_lower_bound}, we know that for all $\varepsilon\in(0,1)$ and $\eta\in\left(0,\frac{\varepsilon}{2}\right)$, there exists an $(|\mathcal{M}|,\varepsilon)$ classical communication protocol such that
		\begin{equation}\label{eq-cc_one_shot_lower_bound_pf}
			\log_2|\mathcal{M}|=\overline{\chi}_H^{\frac{\varepsilon}{2}-\eta}(\mathcal{N})-\log_2\!\left(\frac{4\varepsilon}{\eta^2}\right).
		\end{equation}
		Proposition~\ref{prop:ineq-hypo-renyi} relates the hypothesis testing relative entropy to the Petz--R\'{e}nyi relative entropy according to
		\begin{equation}
			D_H^{\varepsilon}(\rho\Vert\sigma)\geq D_\alpha(\rho\Vert\sigma)+\frac{\alpha}{\alpha-1}\log_2\!\left(\frac{1}{\varepsilon}\right)
		\end{equation}
		for all $\alpha\in(0,1)$, which implies that
		\begin{equation}
			\overline{\chi}_H^{\varepsilon}(\mathcal{N})\geq\overline{\chi}_{\alpha}(\mathcal{N})+\frac{\alpha}{\alpha-1}\log_2\!\left(\frac{1}{\varepsilon}\right).
		\end{equation}
		Combining this inequality with \eqref{eq-cc_one_shot_lower_bound_pf}, we immediately get the desired result.		
	\end{Proof}
	
	Since the inequality in \eqref{eq-cc_one_shot_lower_bound} holds for every $(|\mathcal{M}|,\varepsilon)$ classical communication protocol, we have that
	\begin{equation}
		C^{\varepsilon}(\mathcal{N})\geq\overline{\chi}_\alpha(\mathcal{N})+\frac{\alpha}{\alpha-1}\log_2\!\left(\frac{1}{\frac{\varepsilon}{2}-\eta}\right)-\log_2\!\left(\frac{4\varepsilon}{\eta^2}\right)
	\end{equation}
	for all $\alpha\in(0,1)$, $\varepsilon\in(0,1)$, and $\eta\in\left(0,\frac{\varepsilon}{2}\right)$.

\section{Classical Capacity of a Quantum Channel}

\begin{figure}
		\centering
		\includegraphics[scale=0.8]{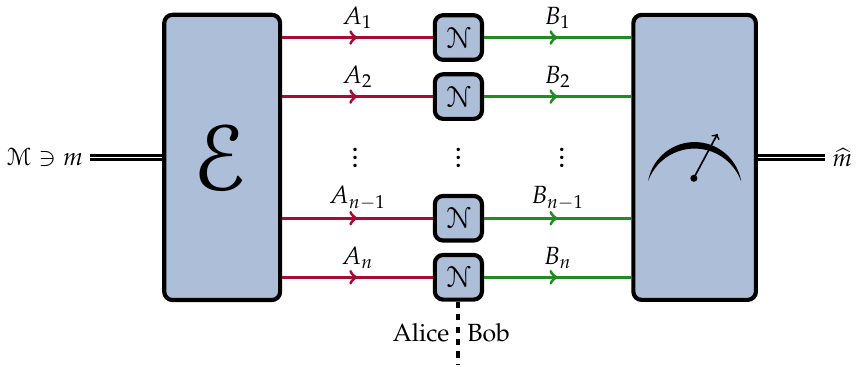}
		\caption{The most general classical communication protocol over a multiple number $n\geq 1$ uses of a quantum channel $\mathcal{N}$. Alice, who wishes to send a message $m$ selected from a set $\mathcal{M}$, first encodes the message into a quantum state on $n$ quantum systems using a classical--quantum encoding channel $\mathcal{E}$. She then sends each quantum system through the channel $\mathcal{N}$. After Bob receives the systems, he performs a collective measurement on them, using the outcome of the measurement to give an estimate~$\widehat{m}$ of the message $m$ sent to him by Alice.}\label{fig-classical_comm}
	\end{figure}

	Let us now consider the asymptotic setting of classical communication, as depicted in Figure~\ref{fig-classical_comm}. Similar to entangle\-ment-assisted classical communication, instead of encoding the message into one quantum system and consequently using the channel $\mathcal{N}$ only once, Alice encodes the message into $n\geq 1$ quantum systems $A_1,\dotsc, A_n$, all with the same dimension as $A$, and sends each one of these through the channel $\mathcal{N}$. We call this the asymptotic setting because the number $n$ of channel uses can be arbitrarily large. 

	Recall that in the case of entanglement-assisted classical communication, we  showed that encoding channels that entangle the $n$ systems $A_1,\dotsc, A_n$ do not help to achieve higher rates in the asymptotic setting. This is due to the additivity of the mutual information and the additivity of the sandwiched R\'{e}nyi mutual information of a channel for all channels and $\alpha>1$. In the case of classical communication that we consider in this chapter, it turns out that, so far, such a statement is known to be generally false for the Holevo information of a quantum channel (please consult the Bibliographic Notes in Section~\ref{sec:CC-comm:bib-notes}). That is, in principle there exists a channel for which the Holevo information is not additive.
	Therefore, unlike entanglement-assisted classical communication, concrete expressions for the classical capacity exist only for specific classes of channels.
	
	The analysis of the classical communication protocol in the asymptotic setting is almost exactly the same as in the one-shot setting. This is due to the fact that $n$ independent uses of the channel $\mathcal{N}$ can be regarded as a single use of the channel $\mathcal{N}^{\otimes n}$. So the only change that needs to be made is to replace $\mathcal{N}$ with $\mathcal{N}^{\otimes n}$ and to define the states and POVM elements as acting on $n$ systems instead of just one. In particular, the state at the end of the protocol presented in \eqref{eq-classical_comm_final_state_1}--\eqref{eq-classical_comm_final_state_2} at the beginning of Section~\ref{sec-cc_one_shot} is
	\begin{equation}
		\omega_{M\widehat{M}}^p= (\mathcal{D}_{B^n\to\widehat{M}}\circ\mathcal{N}_{A\to B}^{\otimes n}\circ\mathcal{E}_{M'\to A^n})(\overline{\Phi}_{MM'}^p),
	\end{equation}
	where $p$ is the prior probability distribution over the message set $\mathcal{M}$, the encoding channel $\mathcal{E}_{M'\to A^n}$ is defined as
	\begin{equation}
		\mathcal{E}_{M'\to A^n}(\ket{m}\!\bra{m}_{M'})=\rho_{A^n}^m\quad\forall~m\in\mathcal{M},
	\end{equation}
	and the decoding channel $\mathcal{D}_{B^n\to\widehat{M}}$, with associated POVM $\{\Lambda_{B^n}^m\}_{m\in\mathcal{M}}$, is defined as
	\begin{equation}
		\mathcal{D}_{B^n\to\widehat{M}}(\tau_{B^n})=\sum_{m\in\mathcal{M}}\Tr[\Lambda_{B^n}^m\tau_{B^n}]\ket{m}\!\bra{m}_{\widehat{M}}.
	\end{equation}
	Then, for every given code specified by the encoding and decoding channels, the definitions of the message error probability of the code, the average error probability of the code, and the maximal error probability of the code all follow analogously from their definitions in \eqref{eq-mess_error_prob}, \eqref{eq-avg_error_prob}, and \eqref{eq-maximal_error_prob}, respectively, in the one-shot setting.
	
	\begin{definition}{$\boldsymbol{(n,|\mathcal{M}|,\varepsilon)}$ Classical Communication Protocol}{def-cc_nMe_protocol}
		A classical communication protocol $(\mathcal{M},\mathcal{E}_{M\to A^n},\mathcal{D}_{B^n\to \widehat{M}})$ over $n$ uses of the channel $\mathcal{N}_{A\to B}$ is called an \textit{$(n,|\mathcal{M}|,\varepsilon)$ protocol}, with $\varepsilon\in[0,1]$, if $p_{\text{err}}^*(\mathcal{E},\mathcal{D})\leq\varepsilon$.
	\end{definition}
	
	Just as in the case of entanglement-assisted classical communication, the \textit{rate} of a classical communication protocol over $n$ uses of a channel is simply the number of bits that can transmitted per channel use, i.e.,
	\begin{equation}
		R(n,|\mathcal{M}|)\coloneqq\frac{1}{n}\log_2|\mathcal{M}|.
	\end{equation}
	Given a channel $\mathcal{N}_{A\to B}$ and $\varepsilon\in[0,1]$, the maximum rate of classical communication over $\mathcal{N}$ among all $(n,|\mathcal{M}|,\varepsilon)$ protocols is
	\begin{equation}
		C^{n,\varepsilon}(\mathcal{N}) \coloneqq\frac{1}{n}C^{\varepsilon}(\mathcal{N}^{\otimes n})
		 =\sup_{(\mathcal{M},\mathcal{E},\mathcal{D})}\left\{\frac{1}{n}\log_2|\mathcal{M}|:p_{\text{err}}^*(\mathcal{E},\mathcal{D};\mathcal{N}^{\otimes n})\leq\varepsilon\right\},
	\end{equation}
	where the optimization is with respect to every classical communication protocol $(\mathcal{M},\allowbreak\mathcal{E}_{M'\to A},\mathcal{D}_{B\to\widehat{M}})$ over $\mathcal{N}^{\otimes n}$, with $d_{M'}=d_{\widehat{M}}=|\mathcal{M}|$.
	
	 	
	As with entanglement-assisted classical communication, the goal of a classical communication protocol is to maximize the rate while at the same time keeping the maximal error probability low. Ideally, we would want the error probability to vanish, and since we want to determine the highest possible rate, we are not concerned about the practical question regarding how many channel uses might be required, at least in the asymptotic setting. In particular, as we will see below, it might take an arbitrarily large number of channel uses to obtain the highest rate with a vanishing error probability. 
	
	
	\begin{definition}{Achievable Rate for Classical Communication}{def-cc_ach_rate}
		Given a quantum channel $\mathcal{N}$, a rate $R\in\mathbb{R}^+$ is called an \textit{achievable rate for classical communication over $\mathcal{N}$} if for all $\varepsilon\in (0,1]$, $\delta>0$, and sufficiently large $n$, there exists an $(n,2^{n(R-\delta)},\varepsilon)$ classical communication protocol.
	\end{definition}
	
	As we prove in Appendix~\ref{chap-str_conv},
	\begin{equation}
		R\text{ acheivable rate }\Longleftrightarrow \lim_{n\to\infty}\varepsilon_C^*(2^{n(R-\delta)};\mathcal{N}^{\otimes n})=0\quad\forall~\delta>0.
	\end{equation}
	In other words, a rate $R$ is achievable if the optimal error probability for a sequence of protocols with rate $R-\delta$, $\delta>0$, vanishes as the number $n$ of uses of $\mathcal{N}$ increases.
	
	
	\begin{definition}{Classical Capacity of a Quantum Channel}{def-cc_cap}
		The \textit{classical capacity} of a quantum channel $\mathcal{N}$, denoted by $C(\mathcal{N})$, is defined as the supremum of all achievable rates, i.e.,
		\begin{align}
			C(\mathcal{N})&\coloneqq\sup\{R:R\text{ is an achievable rate for $\mathcal{N}$}\}.
		\end{align}
	\end{definition}
	
	The classical capacity can also be written as
	\begin{equation}
		C(\mathcal{N})=\inf_{\varepsilon\in( 0,1]}\liminf_{n\to\infty}\frac{1}{n}C^{\varepsilon}(\mathcal{N}^{\otimes n}).
	\end{equation}
	See Appendix~\ref{chap-str_conv} for a proof.
	
	\begin{definition}{Weak Converse Rate for Classical Communication}{def-cc_weak_conv_rate}
		Given a quantum channel $\mathcal{N}$, a rate $R\in\mathbb{R}^+$ is called a \textit{weak converse rate for classical communication over $\mathcal{N}$} if every $R'>R$ is not an achievable rate for $\mathcal{N}$.
	\end{definition}
	
	We show in Appendix~\ref{chap-str_conv} that
	\begin{equation}\label{eq-classical_comm_weak_conv_rate_alt}
		R\text{ weak converse rate }\Longleftrightarrow \lim_{n\to\infty}\varepsilon_C^*(2^{n(R-\delta)};\mathcal{N}^{\otimes n})>0\quad\forall~\delta>0.
	\end{equation}
	In other words, a weak converse rate is a rate above which the optimal error probability cannot be made to vanish in the limit of a large number of channel uses.
	
	\begin{definition}{Strong Converse Rate for Classical Communication}{def-cc_str_conv_rate}
		Given a quantum channel $\mathcal{N}$, a rate $R\in\mathbb{R}^+$ is called a \textit{strong converse rate for classical communication over $\mathcal{N}$} if for all $\varepsilon\in[0,1)$, $\delta>0$, and  sufficiently large $n$, there does not exist an $(n,2^{n(R+\delta)},\varepsilon)$ classical communication protocol over $\mathcal{N}$.
	\end{definition}
	
	We show in Appendix~\ref{chap-str_conv} that
	\begin{equation}\label{eq-classical_comm_str_conv_rate_alt}
		R\text{ strong converse rate }\Longleftrightarrow\lim_{n\to\infty}\varepsilon_C^*(2^{n(R+\delta)};\mathcal{N}^{\otimes n})=1\quad\forall~\delta>0.
	\end{equation}
	In other words, unlike the weak converse, in which the optimal error probability is required to simply be bounded away from zero as the number $n$ of channel uses increases, in order to have a strong converse rate the optimal error has to converge to one as $n$ increases. By comparing \eqref{eq-classical_comm_weak_conv_rate_alt} and \eqref{eq-classical_comm_str_conv_rate_alt}, it is clear that every strong converse rate is a weak converse rate.
	

	
	\begin{definition}{Strong Converse Classical Capacity of a Quantum Channel}{def-cc_str_conv_capacity}
		The \textit{strong converse classical capacity} of a quantum channel $\mathcal{N}$, denoted by $\widetilde{C}(\mathcal{N})$, is defined as the infimum of all strong converse rates,
		i.e.,
		\begin{align}
			\widetilde{C}(\mathcal{N})&\coloneqq\inf\{R:R\text{ is a strong converse rate for }\mathcal{N}\}.
		\end{align}
	\end{definition}
	
	As shown in general 
	in Appendix~\ref{chap-str_conv}, the following inequality holds
	\begin{equation}\label{eq-cc_str_conv_rate_lower_bound}
		C(\mathcal{N})\leq \widetilde{C}(\mathcal{N})
	\end{equation}
	for every quantum channel $\mathcal{N}$. We can also write the strong converse classical capacity as
	\begin{equation}
		\widetilde{C}(\mathcal{N})=\sup_{\varepsilon\in [0,1) }\limsup_{n\to\infty}\frac{1}{n}C^{\varepsilon}(\mathcal{N}^{\otimes n}).
	\end{equation}
	See Appendix~\ref{chap-str_conv} for a proof.

	Having defined the classical capacity of a quantum channel, as well as the strong converse capacity, we now state one of the main theorems of this chapter, which gives us a formal expression for the classical capacity of every quantum channel.

	\begin{theorem*}{Classical Capacity of a Quantum Channel}{thm-classical_capacity}
		The classical capacity of a quantum channel $\mathcal{N}$ is equal to its \textit{regularized Holevo information} $\chi_{\text{reg}}(\mathcal{N})$ of $\mathcal{N}$, i.e.,
		\begin{equation}\label{eq-classical_capacity}
			C(\mathcal{N})=\chi_{\text{reg}}(\mathcal{N})\coloneqq\lim_{n\to\infty}\frac{1}{n}\chi(\mathcal{N}^{\otimes n}).
		\end{equation}
	\end{theorem*}
	
	\begin{remark}
		The quantity $\chi_{\text{reg}}(\mathcal{N})\coloneqq\lim_{n\to\infty}\frac{1}{n}\chi(\mathcal{N}^{\otimes n})$ is called the \textit{regularization} of the Holevo information.  It can be shown that the limit in the definition of $\chi_{\text{reg}}(\mathcal{N})$ does indeed exist (please consult the Bibliographic Notes in Section~\ref{sec:CC-comm:bib-notes}).
	\end{remark}
	
	Note that, unlike the case of entanglement-assisted classical communication in Chapter~\ref{chap-EA_capacity}, the right-hand side of \eqref{eq-classical_capacity} does not depend on only a single use of the channel $\mathcal{N}$. Rather, the capacity formula involves a limit over an arbitrarily large number of uses of the channel and is essentially impossible to compute in general, firstly because of the difficulty of computing the Holevo information of a channel and secondly due to the limit over an arbitrarily large number of uses of the channel. The issue is essentially that the Holevo information of a channel is not known to be additive for all channels, while the mutual information of a channel is known to have this property (we show this in Theorem~\ref{thm-chan_mut_inf_additive}). However, as we show below in Section~\ref{sec-cc_additivity_question}, the Holevo information is \textit{superadditive}, meaning that $\chi(\mathcal{N}^{\otimes n})\geq n\chi(\mathcal{N})$. This implies that the Holevo information is always a lower bound on the quantum capacity of every channel $\mathcal{N}$:
	\begin{equation}
		C(\mathcal{N})\geq \chi(\mathcal{N})\text{ for all channels }\mathcal{N}.
		\label{eq:CC-comm:Holevo-info-lower-bnd-cap}
	\end{equation}
	Channels for which the Holevo information is known to be additive include the following:
	\begin{enumerate}
		\item All entanglement-breaking channels. (See Definition~\ref{def-ent_break_chan}.)
		\item All Hadamard channels. (See Definition~\ref{def-Hadamard_chan}.)
		\item The depolarizing channel. (See \eqref{eq-qubit_depolarizing_channel}.)
		\item The erasure channel. (See \eqref{eq-erasure_channel}.)
	\end{enumerate}
	For all of these channels, we thus have that $C(\mathcal{N})=\chi(\mathcal{N})$.
	
	Also notice that, unlike Theorem~\ref{thm-ea_classical_capacity} for entanglement-assisted classical communication, Theorem~\ref{thm-classical_capacity} only makes a statement about the classical capacity $C(\mathcal{N})$ of all channels, not about the strong converse classical capacity $\widetilde{C}(\mathcal{N})$. In the case of entanglement-assisted classical communication, proving that the mutual information of a channel is a strong converse rate involved proving that the sandwiched R\'{e}nyi mutual information is additive for all channels, which we established in Theorem~\ref{thm-sand_rel_ent_additivity}. Similarly, in the case of classical communication and attempting to follow a similar approach, the relevant quantity is the sandwiched R\'{e}nyi Holevo information $\widetilde{\chi}_\alpha$, defined in \eqref{eq-sand_rel_Holevo_inf_chan}. Unlike the sandwiched R\'{e}nyi mutual information, the sandwiched R\'{e}nyi Holevo information is not known to be additive for all channels. However, as we show in Section~\ref{sec-classical_capacity_ent_break}, it is additive for all entanglement-breaking channels. It is also additive for Hadamard and depolarizing channels (please consult the Bibliographic Notes in Section~\ref{sec:CC-comm:bib-notes}). The best we can do, at the moment, is to say that the regularized Holevo information is a \textit{weak} converse rate for all channels.

	There are two ingredients to the proof of Theorem~\ref{thm-classical_capacity}:
	\begin{enumerate}
		\item \textit{Achievability}: We show that $\chi_{\text{reg}}(\mathcal{N})$ is an achievable rate. In general, to show that $R\in\mathbb{R}^+$ is achievable, we define encoding and decoding channels such that for all $\varepsilon\in(0,1]$ and sufficiently large $n$, the encoding and decoding channels correspond to $(n,2^{nr},\varepsilon)$ protocols with rates $r< R$, as per Definition~\ref{def-cc_ach_rate}. Thus, if $R$ is an achievable rate, then, given an error probability $\varepsilon$, it is possible to find an $n$ large enough, along with encoding and decoding channels, such that the resulting protocol has rate arbitrarily close to $R$ and maximal error probability bounded from above by~$\varepsilon$.
		
		The achievability part of the proof establishes that $C(\mathcal{N})\geq\chi_{\text{reg}}(\mathcal{N})$.
		
		\item \textit{Weak Converse}: We show that $\chi_{\text{reg}}(\mathcal{N})$ is a weak converse rate, from which it follows that $C(\mathcal{N})\leq\chi_{\text{reg}}(\mathcal{N})$. To show that $\chi_{\text{reg}}(\mathcal{N})$ is a weak converse rate, we show that every achievable rate $r$ satisfies $r\leq\chi_{\text{reg}}(\mathcal{N})$.
	\end{enumerate}
	
	The achievability and weak converse proofs establish that the classical capacity is equal to the regularized Holevo information: $C(\mathcal{N})=\chi_{\text{reg}}(\mathcal{N})$. Theorem~\ref{thm-classical_capacity} and the inequality in \eqref{eq-cc_str_conv_rate_lower_bound} allow us to conclude that
	\begin{equation}
		\widetilde{C}(\mathcal{N})\geq \lim_{n\to\infty}\frac{1}{n}\chi(\mathcal{N}^{\otimes n}).
	\end{equation}
	
	We first establish in Section~\ref{sec-cc_achievability} that the rate $\chi_{\text{reg}}(\mathcal{N})$ is achievable for classical communication over $\mathcal{N}$. Then, in Section~\ref{sec-cc_weak_conv}, we prove that $\chi_{\text{reg}}(\mathcal{N})$ is a weak converse rate. We prove that the sandwiched R\'{e}nyi Holevo information of a entanglement-breaking channel is additive in Section~\ref{sec-cc_additivity_question}. With this additivity result, we prove in Section~\ref{sec-cc_strong_conv} that $C(\mathcal{N})=\widetilde{C}(\mathcal{N})=\chi(\mathcal{N})$ for all entanglement-breaking channels. 
	

\subsection{Proof of Achievability}\label{sec-cc_achievability}

	In this section, we prove that $\chi_{\text{reg}}(\mathcal{N})$ is an achievable rate for classical communication over $\mathcal{N}$.
	
	First, recall from Theorem~\ref{thm-cc_one_shot_lower_bound} that for all $\varepsilon\in(0,1)$ and $\eta\in(0,\frac{\varepsilon}{2})$, there exists an $(|\mathcal{M}|,\varepsilon)$ classical communication protocol over $\mathcal{N}$ such that
	\begin{equation}\label{eq-cc_one_shot_lower_bound_2}
		\log_2|\mathcal{M}|\geq\overline{\chi}_{\alpha}(\mathcal{N})+\frac{\alpha}{\alpha-1}\log_2\!\left(\frac{1}{\frac{\varepsilon}{2}-\eta}\right)-\log_2\!\left(\frac{4\varepsilon}{\eta^2}\right)
	\end{equation}
	for all $\alpha\in(0,1)$, where we recall from \eqref{eq-petz_renyi_Hol_inf_chan_noopt} that
	\begin{align}
		\overline{\chi}_{\alpha}(\mathcal{N}) & = \sup_{\rho_{XA}}\overline{I}_\alpha(X;B)_{\omega}\\
		& = \sup_{\rho_{XA}}D_\alpha(\mathcal{N}_{A\to B}(\rho_{XA})\Vert\rho_X\otimes\mathcal{N}_{A\to B}(\rho_A)),
	\end{align}
	where $\omega_{XB} \coloneqq \mathcal{N}_{A\to B}(\rho_{XA})$ and the optimization is over all classical--quantum states $\rho_{XA}$. A simple corollary of this result is the following.
	
	\begin{corollary*}{Lower Bound for Classical Communication in Asymptotic Setting}{cor-cc_asymp_lower_bound}
		Let $\mathcal{N}$ be a quantum channel. For all $\varepsilon\in(0,1]$,  $n\in\mathbb{N}$, and $\alpha\in (0,1)$, there exists an $(n,|\mathcal{M}|,\varepsilon)$ classical communication protocol over $n$ uses of $\mathcal{N}$ such that
		\begin{equation}\label{eq-cc_lower_bound_asymp}
			\frac{1}{n}\log_2|\mathcal{M}|\geq \overline{\chi}_{\alpha}(\mathcal{N})-\frac{1}{n(1-\alpha)}\log_2\!\left(\frac{4}{\varepsilon}\right)-\frac{4}{n}.
		\end{equation}
	\end{corollary*}
	
	\begin{Proof}
		The inequality \eqref{eq-cc_one_shot_lower_bound_2} holds for every channel $\mathcal{N}$, which means that it holds for $\mathcal{N}^{\otimes n}$. Applying the inequality in \eqref{eq-cc_one_shot_lower_bound_2} to $\mathcal{N}^{\otimes n}$ and dividing both sides by $n$, we obtain
		\begin{equation}
			\frac{1}{n}\log_2|\mathcal{M}|\geq\frac{1}{n}\overline{\chi}_{\alpha}(\mathcal{N}^{\otimes n})+\frac{\alpha}{n(\alpha-1)}\log_2\!\left(\frac{1}{\frac{\varepsilon}{2}-\eta}\right)-\frac{1}{n}\log_2\!\left(\frac{4\varepsilon}{\eta^2}\right)
		\end{equation}
		for all $\alpha\in(0,1)$. By restricting the optimization in the definition of $\overline{\chi}_{\alpha}(\mathcal{N}^{\otimes n})$ to tensor-power states, we find that $\overline{\chi}_{\alpha}(\mathcal{N}^{\otimes n})\geq n\overline{\chi}_{\alpha}(\mathcal{N})$. This follows from the additivity of the Petz--R\'{e}nyi relative entropy under tensor-product states (see Proposition~\ref{prop-Petz_rel_ent}). So we obtain
		\begin{equation}
			\frac{1}{n}\log_2|\mathcal{M}|\geq \overline{\chi}_{\alpha}(\mathcal{N})+\frac{\alpha}{n(\alpha-1)}\log_2\!\left(\frac{1}{\frac{\varepsilon}{2}-\eta}\right)-\frac{1}{n}\log_2\!\left(\frac{4\varepsilon}{\eta^2}\right)
		\end{equation}
		for all $\alpha\in(0,1)$. Now, letting $\eta=\frac{\varepsilon}{4}$ and using the fact that $\alpha-1$ is negative for $\alpha\in(0,1)$, the following inequality holds for all $\alpha\in(0,1)$:
		\begin{equation}
			\frac{1}{n}\log_2|\mathcal{M}|\geq \overline{\chi}_{\alpha}(\mathcal{N})-\frac{1}{n(1-\alpha)}\log_2\!\left(\frac{4}{\varepsilon}\right)-\frac{4}{n}.
		\end{equation}
		 In other words, for all $\varepsilon\in(0,1]$, there exists an $(n,|\mathcal{M}|,\allowbreak\varepsilon)$ classical communication protocol such that \eqref{eq-cc_lower_bound_asymp} is satisfied. This concludes the proof.
	\end{Proof}
	
	The inequality in \eqref{eq-cc_lower_bound_asymp} gives us, for every $\varepsilon\in(0,1]$ and  $n\in\mathbb{N}$, a lower bound on the rate of a corresponding $(n,|\mathcal{M}|,\varepsilon)$ classical communication protocol, which is known to exist due to Proposition~\ref{prop-cc_one-shot_lower_bound}. If instead we fix a particular communication rate $R$ by letting $|\mathcal{M}|=2^{nR}$, then we can rearrange the inequality in \eqref{eq-cc_lower_bound_asymp} to obtain an exponentially decaying upper bound on the maximal error probability of the corresponding $(n,2^{nR},\varepsilon)$ classical communication protocol. Specifically, we find that
	\begin{equation}
		\varepsilon\leq 4\cdot 2^{-(1-\alpha)\left(\overline{\chi}_{\alpha}(\mathcal{N})-R-\frac{4}{n}\right)}
	\end{equation}
	for all $\alpha\in(0,1)$.
	
	The inequality in \eqref{eq-cc_lower_bound_asymp} implies that
	\begin{equation}
		C^{n,\varepsilon}(\mathcal{N})\geq \overline{\chi}_{\alpha}(\mathcal{N})-\frac{1}{n(\alpha-1)}\log_2\!\left(\frac{4}{\varepsilon}\right)-\frac{4}{n}
	\end{equation}
	for all $n\geq 1$, $\varepsilon\in(0,1)$, and $\alpha\in(0,1)$.
	
	We can now use \eqref{eq-cc_lower_bound_asymp} to prove that $\chi_{\text{reg}}(\mathcal{N})$ is an achievable rate for classical communication over $\mathcal{N}$.

\subsubsection*{Proof of the Achievability Part of Theorem~\ref{thm-classical_capacity}}

	Fix $\varepsilon\in(0,1]$ and $\delta>0$. Let $\delta_1,\delta_2>0$ be such that
	\begin{equation}\label{eq-cc_achieve_pf_1}
		\delta=\delta_1+\delta_2.
	\end{equation}
	Set $\alpha\in(0,1)$ such that
	\begin{equation}\label{eq-cc_achieve_pf_2}
		\delta_1\geq\chi(\mathcal{N})-\overline{\chi}_{\alpha}(\mathcal{N}),
	\end{equation}
	which is possible because $\overline{\chi}_{\alpha}(\mathcal{N})$ is monotonically increasing in $\alpha$ (this follows from Proposition~\ref{prop-Petz_rel_ent}), because $\lim_{\alpha\to 1^-}\overline{\chi}_{\alpha}(\mathcal{N})=\chi(\mathcal{N})$ (the proof of this is analogous to the one presented in Appendix~\ref{app-sand_ren_mut_inf_chan_limit}). With this value of $\alpha$, take $n$ large enough so that
	\begin{equation}\label{eq-cc_achieve_pf_3}
		\delta_2\geq\frac{1}{n(1-\alpha)}\log_2\!\left(\frac{4}{\varepsilon}\right)+\frac{4}{n}.
	\end{equation}
	
	Now, making use of the inequality in \eqref{eq-cc_lower_bound_asymp} of Corollary~\ref{cor-cc_asymp_lower_bound}, there exists an $(n,|\mathcal{M}|,\varepsilon)$ protocol, with $n$ and $\varepsilon$ chosen as above, such that
	\begin{equation}
		\frac{1}{n}\log_2|\mathcal{M}|\geq\overline{\chi}_{\alpha}(\mathcal{N})-\frac{1}{n(1-\alpha)}\log_2\!\left(\frac{4}{\varepsilon}\right)-\frac{4}{n}.
	\end{equation}
	Rearranging the right-hand side of this inequality, and using \eqref{eq-cc_achieve_pf_1}--\eqref{eq-cc_achieve_pf_3}, we find that
	\begin{align}
		\frac{1}{n}\log_2|\mathcal{M}|&\geq \chi(\mathcal{N})-\left(\chi(\mathcal{N})-\overline{\chi}_{\alpha}(\mathcal{N})+\frac{1}{n(1-\alpha)}\log_2\!\left(\frac{4}{\varepsilon}\right)+\frac{4}{n}\right)\\
		&\geq \chi(\mathcal{N})-(\delta_1+\delta_2)\\
		&=\chi(\mathcal{N})-\delta.
	\end{align}
	We thus have $\chi(\mathcal{N})-\delta\leq\frac{1}{n}\log_2|\mathcal{M}|$. Recall that if an $(n,|\mathcal{M}|,\varepsilon)$ protocol exists, then an $(n,|\mathcal{M}'|,\varepsilon)$ also exists for all $\mathcal{M}'$ satisfying $|\mathcal{M}'|\leq|\mathcal{M}|$. We thus conclude that there exists an $(n,2^{n(R-\delta)},\varepsilon)$ classical communication with $R=\chi(\mathcal{N})$ for all sufficiently large $n$ such that \eqref{eq-cc_achieve_pf_3} holds. Since $\varepsilon$ and $\delta$ are arbitrary, we conclude that, for all $\varepsilon\in(0,1]$, $\delta>0$, and  sufficiently large $n$, there exists an $(n,2^{n(\chi(\mathcal{N})-\delta)},\varepsilon)$ classical communication protocol. This means that $\chi(\mathcal{N})$ is an achievable rate, and thus that $C(\mathcal{N})\geq\chi(\mathcal{N})$.
	
	Now, we can repeat the arguments above  for the tensor-power channel~$\mathcal{N}^{\otimes k}$ with $k\geq 1$, and we conclude that $\frac{1}{k}\chi(\mathcal{N}^{\otimes k})$ is an achievable rate. Since this holds for all $k$, we conclude that $\lim_{k\to\infty}\frac{1}{k}\chi(\mathcal{N}^{\otimes k}) = \chi_{\text{reg}}(\mathcal{N})$ is an achievable rate. Therefore, $C(\mathcal{N})\geq \chi_{\text{reg}}(\mathcal{N})$.

\subsubsection{Achievability from a Different Point of View}

	Using arguments similar to those given in Appendix~\ref{subsec-EA_comm_ach_diff_POV}, we can make the following statement: there exists a sequence $\{(n,2^{nR_n},\varepsilon_n)\}_{n\in\mathbb{N}}$ of $(n,|\mathcal{M}|,\varepsilon)$ classical communication protocols over $\mathcal{N}$, such that $\liminf_{n\to\infty}R_n\geq\chi(\mathcal{N})$ and $\lim_{n\to\infty}\varepsilon_n = 0$. If we consider a sequence $\{(n,2^{nR},\varepsilon_n)\}_{n\in\mathbb{N}}$ of $(n,|\mathcal{M}|,\varepsilon)$ classical communication protocols, this time keeping the rate at an arbitrary (but fixed) value $R<\chi(\mathcal{N})$ and varying the error probability, we conclude that there exists a sequence of protocols for which the error probabilities $\varepsilon_n$ approach zero exponentially fast as $n\to\infty$.

\subsection{Proof of the Weak Converse}\label{sec-cc_weak_conv}

	We now show that the regularized Holevo information $\chi_{\text{reg}}(\mathcal{N})$ is a weak converse rate. The result is to establish that $C(\mathcal{N})\leq\chi_{\text{reg}}(\mathcal{N})$ and therefore that $C(\mathcal{N})\allowbreak=\chi_{\text{reg}}(\mathcal{N})$, completing the proof of Theorem~\ref{thm-classical_capacity}.
	
\begin{figure}
		\centering
		\includegraphics[scale=0.8]{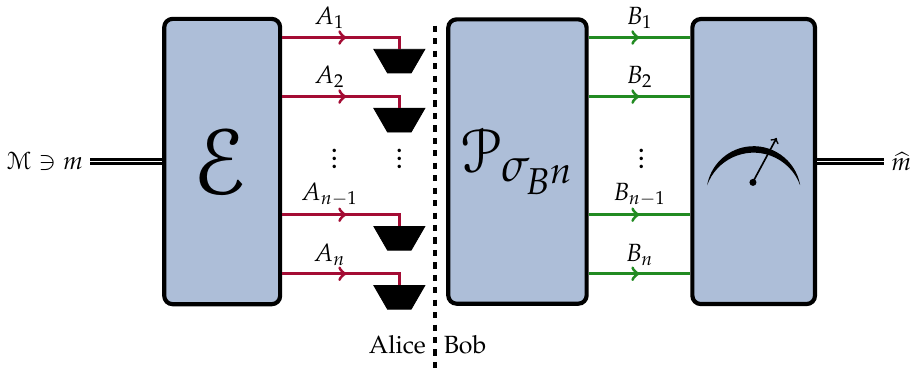}
		\caption{Depiction of a protocol that is useless for classical communication in the asymptotic setting. The state encoding the message $m$ via $\mathcal{E}$ is discarded and replaced by an arbitrary (but fixed) state $\sigma_{B^n}$.}\label{fig-classical_comm_useless}
	\end{figure}

	Let us first recall from Theorem~\ref{cor-cc_meta_str_weak_conv} that for every quantum channel $\mathcal{N}$ we have the following: for all $\varepsilon\in[0,1)$ and $(|\mathcal{M}|,\varepsilon)$ classical communication protocols over $\mathcal{N}$,
	\begin{align}
		\log_2|\mathcal{M}|&\leq\frac{1}{1-\varepsilon}\left(\chi(\mathcal{N})+h_2(\varepsilon)\right),\label{eq-cc_weak_conv_one_shot_2}\\
		\log_2|\mathcal{M}|&\leq\widetilde{\chi}_{\alpha}(\mathcal{N})+\frac{\alpha}{\alpha-1}\log_2\!\left(\frac{1}{1-\varepsilon}\right),\quad \forall~\alpha>1.\label{eq-cc_strong_conv_one_shot_2}
	\end{align}
	To obtain these inequalities, we considered a classical communication protocol over a useless channel and used the hypothesis testing relative entropy to compare this protocol with the actual protocol over the channel $\mathcal{N}$. The useless channel in the asymptotic setting is analogous to the one in Figure~\ref{fig-classical_comm_useless_oneshot} and is shown in Figure~\ref{fig-classical_comm_useless}. A simple corollary of Theorem~\ref{cor-cc_meta_str_weak_conv}, which is relevant for the asymptotic setting, is the following.
	
	\begin{corollary*}{Upper Bounds for Classical Communication in Asymptotic Setting}{cor-cc_str_weak_conv_upper}
		Let $\mathcal{N}$ be a quantum channel. For all $\varepsilon\in[0,1)$,  $n\in\mathbb{N}$, and  $(n,|\mathcal{M}|,\varepsilon)$ classical communication protocols over $n$ uses of $\mathcal{N}$, the rate of transmitted bits is bounded from above as follows:
		\begin{align}
			\frac{\log_{2}|\mathcal{M}|}{n}  &  \leq\frac{1}{1-\varepsilon}\left(\frac{1}{n}\chi(\mathcal{N}^{\otimes n})+\frac{1}{n}h_{2}(\varepsilon)\right),\label{eq-cc_weak_conv_1}\\
			\frac{\log_{2}|\mathcal{M}|}{n}  & \leq \frac{1}{n}\widetilde{\chi}_{\alpha}(\mathcal{N}^{\otimes n})+\frac{\alpha}{n(\alpha-1)}\log_{2}\!\left(\frac{1}{1-\varepsilon}\right)\quad\forall~\alpha>1.\label{eq-cc_str_conv_1}
		\end{align}
	\end{corollary*}
	
	\begin{Proof}
		Since the inequalities in \eqref{eq-cc_weak_conv_one_shot_2} and \eqref{eq-cc_strong_conv_one_shot_2} of Theorem~\ref{cor-cc_meta_str_weak_conv} hold for every channel $\mathcal{N}$, they hold for the channel $\mathcal{N}^{\otimes n}$. Therefore, applying \eqref{eq-cc_weak_conv_one_shot_2} and \eqref{eq-cc_strong_conv_one_shot_2} to $\mathcal{N}^{\otimes n}$ and dividing both sides by $n$, we immediately obtain the desired result.
	\end{Proof}
	
	The inequalities in the  corollary above give us, for every $\varepsilon\in[0,1)$ and  $n\in\mathbb{N}$, an upper bound on the size $|\mathcal{M}|$ of the message set we can take for an arbitrary $(n,|\mathcal{M}|,\varepsilon)$ classical communication protocol. If instead we fix a particular communication rate $R$ by letting $|\mathcal{M}|=2^{nR}$, then we can obtain a lower bound on the maximal error probability of an arbitrary $(n,2^{nR},\varepsilon)$ classical communication protocol. Specifically, using \eqref{eq-cc_str_conv_1}, we find that
	\begin{equation}
		\varepsilon\geq 1-2^{-n\left(\frac{\alpha-1}{\alpha}\right)\left(R-\frac{1}{n}\widetilde{\chi}_\alpha(\mathcal{N}^{\otimes n})\right)}
	\end{equation}
	for all $\alpha>1$.
	
	The inequalities in \eqref{eq-cc_weak_conv_1} and \eqref{eq-cc_str_conv_1} imply that
	\begin{align}
		C^{n,\varepsilon}(\mathcal{N})&\leq \frac{1}{1-\varepsilon}\left(\frac{1}{n}\chi(\mathcal{N}^{\otimes n})+\frac{1}{n}h_2(\varepsilon)\right),\\
		C^{n,\varepsilon}(\mathcal{N})&\leq\frac{1}{n}\widetilde{\chi}_{\alpha}(\mathcal{N}^{\otimes n})+\frac{\alpha}{n(\alpha-1)}\log_2\!\left(\frac{1}{1-\varepsilon}\right)\quad\forall~\alpha>1,
	\end{align}
	where $n\geq 1$ and $\varepsilon\in(0,1)$.
	
	Using \eqref{eq-cc_weak_conv_1}, we can now prove the weak converse part of Theorem~\ref{thm-classical_capacity}.
	

\subsubsection*{Proof of the Weak Converse Part of Theorem~\ref{thm-classical_capacity}}

	Suppose that $R$ is an achievable rate. Then, by definition, for all $\varepsilon\in(0,1]$, $\delta>0$, and sufficiently large $n$, there exists an $(n,2^{n(R-\delta)},\varepsilon)$ classical communication protocol over $\mathcal{N}$. For all such protocols, the inequality \eqref{eq-cc_weak_conv_1} in Corollary~\ref{cor-cc_str_weak_conv_upper} holds, so that
	\begin{equation}
		R-\delta\leq\frac{1}{1-\varepsilon}\left(\frac{1}{n}\chi(\mathcal{N}^{\otimes n})+\frac{1}{n}h_2(\varepsilon)\right).
	\end{equation}
	Since this bound holds for all  $n$, it holds in the limit $n\to\infty$, so that
	\begin{align}
		R&\leq\lim_{n\to\infty}\frac{1}{1-\varepsilon}\left(\frac{1}{n}\chi(\mathcal{N}^{\otimes n})+\frac{1}{n}h_2(\varepsilon)\right)+\delta\\
		&=\frac{1}{1-\varepsilon}\lim_{n\to\infty}\frac{1}{n}\chi(\mathcal{N}^{\otimes n})+\delta.
	\end{align}
	Then, since this inequality holds for all $\varepsilon,\delta>0$, we then conclude that
	\begin{equation}
		R\leq \lim_{\varepsilon,\delta\to 0}\left\{\frac{1}{1-\varepsilon}\lim_{n\to\infty}\frac{1}{n}\chi(\mathcal{N}^{\otimes n})+\delta\right\}=\lim_{n\to\infty}\frac{1}{n}\chi(\mathcal{N}^{\otimes n}).
		\label{eq-CC:weak-converse-pf-last-step}
	\end{equation}
	We have thus shown that if $R$ is an achievable rate, then $R\leq\chi_{\text{reg}}(\mathcal{N})$. The contrapositive of this statement is that if $R>\chi_{\text{reg}}(\mathcal{N})$, then $R$ is not an achievable rate. By definition, therefore, $\chi_{\text{reg}}(\mathcal{N})$ is a weak converse rate.

	Recall that Theorem~\ref{thm-classical_capacity} only gives an expression for the capacity $C(\mathcal{N})$, and not for the strong converse capacity $\widetilde{C}(\mathcal{N})$. The sandwiched R\'{e}nyi Holevo information $\widetilde{\chi}_{\alpha}(\mathcal{N})$ of a channel $\mathcal{N}$ can be used to obtain the upper bound in \eqref{eq-cc_str_conv_1}, holding for every $(n,|\mathcal{M}|,\varepsilon)$ protocol. This inequality then leads to an expression for the strong converse capacity in the case that $\widetilde{\chi}_{\alpha}(\mathcal{N})$ happens to be additive for $\mathcal{N}$. We now, therefore, address this question regarding the additivity of the sandwiched R\'{e}nyi Holevo information.

\subsection{The Additivity Question}\label{sec-cc_additivity_question}

	Although we have shown that the classical capacity $C(\mathcal{N})$ of a channel $\mathcal{N}$ is given by the regularized Holevo information $\chi_{\text{reg}}(\mathcal{N})=\lim_{n\to\infty}\frac{1}{n}\chi(\mathcal{N}^{\otimes n})$, as mentioned earlier, without the additivity of $\chi(\mathcal{N})$ this result is not particularly helpful since it is not known how to compute the regularized Holevo information  in general.
	
	Note, however, that for all channels $\mathcal{N}_1$ and $\mathcal{N}_2$ we always have the \textit{superadditivity} of the Holevo information, i.e.,
	\begin{equation}\label{eq-chi-superadditive}
		\chi(\mathcal{N}_1\otimes\mathcal{N}_2)\geq \chi(\mathcal{N}_1)+\chi(\mathcal{N}_2).
	\end{equation}
	This follows by performing exactly the same steps in \eqref{eq-mut_inf_additive_pf3}--\eqref{eq-mut_inf_chan_additive_5}, but with the systems $R_1$ and $R_2$ therein taken to be classical systems.
	Therefore, to prove the additivity of $\chi$ for a channel $\mathcal{N}$, it suffices to show that $\chi(\mathcal{N}\otimes\mathcal{M})\leq \chi(\mathcal{N})+\chi(\mathcal{M})$.
	
	Similarly, the sandwiched R\'{e}nyi Holevo information is superadditive; i.e., for all $\alpha\geq 1$ and all channels $\mathcal{N}_1$ and $\mathcal{N}_2$, it holds that
	\begin{equation}\label{eq-chi_alpha_superadditive}
		\widetilde{\chi}_{\alpha}(\mathcal{N}_1\otimes\mathcal{N}_2)\geq\widetilde{\chi}_{\alpha}(\mathcal{N}_1)+\widetilde{\chi}_{\alpha}(\mathcal{N}_2).
	\end{equation}
	First, recall that
	\begin{equation}
		\widetilde{\chi}_\alpha(\mathcal{N})=\sup_{\rho_{XA}}\widetilde{I}_{\alpha}(X;B)_{\omega},
	\end{equation}
	where $\omega_{XB}=\mathcal{N}_{A\to B}(\rho_{XA})$, and where we optimize over classical--quantum states $\rho_{XA}=\sum_{x\in\mathcal{X}}p(x)\ket{x}\!\bra{x}_X\otimes\rho_A^x$, with $\mathcal{X}$ a finite alphabet with associated $|\mathcal{X}|$-dimensional system $X$ and $\{\rho_A^x\}_{x\in\mathcal{X}}$ a set of states. Also, recall that for every state $\rho_{AB}$,
	\begin{equation}
		\widetilde{I}_\alpha(A;B)_\rho=\inf_{\sigma_B}\widetilde{D}_\alpha(\rho_{AB}\Vert\rho_A\otimes\sigma_B).
	\end{equation}
		
	Now, the proof of \eqref{eq-chi_alpha_superadditive} proceeds similarly to the proof of the corresponding inequality \eqref{eq-chi-superadditive} for the Holevo information. By restricting the optimization in the definition of $\widetilde{\chi}_{\alpha}(\mathcal{N}_1\otimes\mathcal{N}_2)$ to product states, and letting $\rho_{X_1X_2B_1B_2}'$ be defined as
	\begin{equation}
	\rho_{X_1X_2B_1B_2}'\coloneqq ((\mathcal{N}_1)_{A_1\to B_1}\otimes(\mathcal{N}_2)_{A_2\to B_2})(\rho_{X_1X_2A_1A_2}),
	\end{equation}
	for a classical--quantum state $\rho_{X_1X_2A_1A_2}$ ($X$ systems classical and $A$ systems quantum), we find that
	\begin{align}
		\widetilde{\chi}_{\alpha}(\mathcal{N}_1\otimes\mathcal{N}_2)&=\sup_{\rho}\widetilde{I}_\alpha(X_1X_2;B_1B_2)_{\rho'}\\
		&\geq\sup_{\tau\otimes\omega}\widetilde{I}_\alpha(X_1X_2;B_1B_2)_{\xi'\otimes\omega'},
	\end{align}
	where $\tau_{X_1B_1}'\coloneqq(\mathcal{N}_1)_{A_1\to B_1}(\xi_{X_1A_2})$ and $\omega_{X_2B_2}'\coloneqq(\mathcal{N}_2)_{A_2\to B_2}(\omega_{X_2A_2})$. Proposition~\ref{prop-sand_rel_mut_inf_additive} states that the sandwiched R\'{e}nyi mutual information $\widetilde{I}_{\alpha}$ is additive for product states, meaning that 
	\begin{equation}
		\widetilde{I}_{\alpha}(A_1A_2;B_1B_2)_{\tau\otimes\omega}=\widetilde{I}_{\alpha}(A_1;B_1)_{\tau}+\widetilde{I}_{\alpha}(A_2;B_2)_{\omega}
	\end{equation}
	for every  state $\tau_{A_1B_1}\otimes\omega_{A_2B_2}$. Using this, we find that
	\begin{align}
		\widetilde{\chi}_{\alpha}(\mathcal{N}_1\otimes\mathcal{N}_2)&\geq\sup_{\tau,\omega}\left\{\widetilde{I}_{\alpha}(X_1;B_1)_{\tau'}+\widetilde{I}_{\alpha}(X_2;B_2)_{\omega'}\right\}\\
		&=\sup_{\tau}\widetilde{I}_{\alpha}(X_1;B_1)_{\tau'}+\sup_{\omega}\widetilde{I}_{\alpha}(X_2;B_2)_{\omega'}\\
		&=\widetilde{\chi}_{\alpha}(\mathcal{N}_1)+\widetilde{\chi}_{\alpha}(\mathcal{N}_2),
	\end{align}
	i.e.,
	\begin{equation}
		\widetilde{\chi}_{\alpha}(\mathcal{N}_1\otimes\mathcal{N}_2)\geq\widetilde{\chi}_{\alpha}(\mathcal{N}_1)+\widetilde{\chi}_{\alpha}(\mathcal{N}_2),
	\end{equation}
	as required.
	
	We see that in order to show the additivity of the sandwiched R\'{e}nyi Holevo information for $\mathcal{N}$, it suffices to show \textit{subadditivity for $\mathcal{N}$}, i.e.,
	\begin{equation}
		\widetilde{\chi}_\alpha(\mathcal{N}^{\otimes n})\leq n\widetilde{\chi}_{\alpha}(\mathcal{N})\quad\forall ~n\geq 1.
	\end{equation}
	We now show that subadditivity, and thus additivity, of the sandwiched R\'{e}nyi Holevo information holds for all entanglement-breaking channels.

\subsubsection{Entanglement-Breaking Channels}\label{sec-classical_capacity_ent_break}

	In this section, we prove that the sandwiched R\'{e}nyi Holevo information is additive for all entanglement-breaking channels.
	
	\begin{theorem*}{Additivity of $\boldsymbol{\widetilde{\chi}_\alpha}$ for Entanglement-Breaking Channels}{thm-classical_comm_ent_break_additive}
		For an entanglement-breaking channel $\mathcal{N}$ and an arbitrary channel $\mathcal{M}$, the following equality holds for all $\alpha> 1$:
		\begin{equation}\label{eq-chi_alpha_subadditive}
			\widetilde{\chi}_{\alpha}(\mathcal{N}\otimes\mathcal{M})=\widetilde{\chi}_{\alpha}(\mathcal{N})+\widetilde{\chi}_{\alpha}(\mathcal{M}).
		\end{equation}
	\end{theorem*}
	

%
%
%
	
	The proof of this theorem relies on two lemmas, the first of which states that the sandwiched R\'{e}nyi Holevo information $\widetilde{\chi}_\alpha(\mathcal{N})$ of a channel $\mathcal{N}$ is equal to a quantity $\widetilde{K}_\alpha(\mathcal{N})$, called the \textit{sandwiched R\'{e}nyi information radius of $\mathcal{N}$}.
	
	\begin{Lemma}{lem-ent_break_mult_1}
		For every quantum channel $\mathcal{N}$ and $\alpha>1$, the following equality holds
		\begin{equation}
			\widetilde{\chi}_\alpha(\mathcal{N})=\inf_\sigma\sup_\rho\widetilde{D}_\alpha(\mathcal{N}(\rho)\Vert\sigma)\eqqcolon \widetilde{K}_\alpha(\mathcal{N}),
			\label{eq-CC:info-radius-K-chan}
		\end{equation}
		where the optimizations are over states $\rho$ and $\sigma$. The quantity $\widetilde{K}_\alpha(\mathcal{N})$ is called the \textit{sandwiched R\'{e}nyi information radius of $\mathcal{N}$}.
	\end{Lemma}
	
	\begin{Proof}
		To prove this lemma, we show that $\widetilde{\chi}_\alpha(\mathcal{N})\leq\allowbreak\widetilde{K}_\alpha(\mathcal{N})$ and $\widetilde{\chi}_\alpha(\mathcal{N})\allowbreak\geq\widetilde{K}_\alpha(\mathcal{N})$.
		
		First, using the definition in \eqref{eq-sand_rel_Holevo_inf_chan} of $\widetilde{\chi}_\alpha(\mathcal{N})$, we find that for every state~$\tau_B$,
		\begin{align}
			&\widetilde{\chi}_\alpha(\mathcal{N})\nonumber\\
			&\quad=\sup_{\rho_{XA}}\widetilde{I}_{\alpha}(X;B)_{\omega}\\
			&\quad=\sup_{\rho_{XA}}\inf_{\sigma_B}\widetilde{D}_{\alpha}\!\left(\sum_{x\in\mathcal{X}}p(x)\ket{x}\!\bra{x}_X\otimes\mathcal{N}(\rho_A^x)\Bigg\Vert\sum_{x\in\mathcal{X}}p(x)\ket{x}\!\bra{x}_X\otimes\sigma_B\right)\\
			&\quad\leq \sup_{\rho_{XA}}\widetilde{D}_\alpha\!\left(\sum_{x\in\mathcal{X}}p(x)\ket{x}\!\bra{x}_X\otimes\mathcal{N}(\rho_A^x)\Bigg\Vert\sum_{x\in\mathcal{X}}p(x)\ket{x}\!\bra{x}_X\otimes\tau_B\right),
		\end{align}
		where $\omega_{XB}=\mathcal{N}_{A\to B}(\rho_{XA})$ and the supremum is over all classical--quantum states $\rho_{XA}$ of the form $\rho_{XA}=\sum_{x\in\mathcal{X}}p(x)\ket{x}\!\bra{x}_X\allowbreak\otimes\rho_A^x$, with $\mathcal{X}$ a finite alphabet with associated $|\mathcal{X}|$-dimensional quantum system $X$ and $\{\rho_A^x\}_{x\in\mathcal{X}}$ is a set of states. Now, recall from \eqref{eq-sand_rel_ent_q_convex} that the sandwiched R\'enyi relative entropy is jointly quasi-convex for $\alpha > 1$ and invariant under tensoring in the same state $|x\rangle\!\langle x|$,
		which implies that
		\begin{align}
			\widetilde{\chi}_\alpha(\mathcal{N})&\leq \sup_{\rho_{XA}}\widetilde{D}_\alpha\!\left(\sum_{x\in\mathcal{X}}p(x)\ket{x}\!\bra{x}_X\otimes\mathcal{N}(\rho_A^x)\Bigg\Vert\sum_{x\in\mathcal{X}}p(x)\ket{x}\!\bra{x}_X\otimes\tau_B\right)\\
			&\leq \sup_{\rho_{XA}}\max_{x\in\mathcal{X}}\widetilde{D}_\alpha(\mathcal{N}(\rho_A^x)\Vert\tau_B)\\
			&\leq \sup_{\rho_A}\widetilde{D}_\alpha(\mathcal{N}(\rho_A)\Vert\tau_B).
		\end{align}
		The final inequality above holds for every state $\tau_B$, which implies that
		\begin{equation}
			\widetilde{\chi}_\alpha(\mathcal{N})\leq\inf_{\tau_B}\sup_{\rho_A}\widetilde{D}_\alpha(\mathcal{N}(\rho_A)\Vert\tau_B)=\widetilde{K}_\alpha(\mathcal{N}),
		\end{equation}
		i.e., $\widetilde{\chi}_\alpha(\mathcal{N})\leq\widetilde{K}_\alpha(\mathcal{N})$. 
		
		We now show that $\widetilde{K}_{\alpha}(\mathcal{N})\leq\widetilde{\chi}_{\alpha}(\mathcal{N})$. First, consider that
		\begin{align}
			\widetilde{K}_{\alpha}(\mathcal{N})&=\inf_{\sigma_B}\sup_{\rho_A}\widetilde{D}_\alpha(\mathcal{N}(\rho_A)\Vert\sigma_B)\\
			&=\frac{1}{\alpha-1}\inf_{\sigma_B}\sup_{\rho_A}\log_2\widetilde{Q}_\alpha(\mathcal{N}(\rho_A)\Vert\sigma_B)\\
			&=\frac{1}{\alpha-1}\log_2\inf_{\sigma_B}\sup_{\rho_A}\widetilde{Q}_\alpha(\mathcal{N}(\rho_A)\Vert\sigma_B).
		\end{align}
		Now, by taking a supremum over all probability measures $\mu$ on the set of all states $\rho_A$, we find that
		\begin{equation}
			\sup_{\rho_A}\widetilde{Q}_\alpha(\mathcal{N}(\rho_A)\Vert\sigma_B)\leq\sup_\mu\int \widetilde{Q}_\alpha(\mathcal{N}(\rho_A)\Vert\sigma_B)~\text{d}\mu(\rho_A).
		\end{equation}
		So we have that
		\begin{equation}
			\widetilde{K}_\alpha(\mathcal{N})\leq\frac{1}{\alpha-1}\log_2\inf_{\sigma_B}\sup_\mu\int \widetilde{Q}_\alpha(\mathcal{N}(\rho_A)\Vert\sigma_B)~\text{d}\mu(\rho_A).
		\end{equation}
		We now apply the Sion minimax theorem (Theorem~\ref{thm-Sion_minimax}) to exchange $\inf_{\sigma_B}$ and $\sup_\mu$. This theorem is applicable because the function
		\begin{equation}
			(\mu,\sigma_B)\mapsto \int \widetilde{Q}_\alpha(\mathcal{N}(\rho_A)\Vert\sigma_B)~\text{d}\mu(\rho_A)
		\end{equation}
		is linear in the measure $\mu$ and convex in the states $\sigma_B$. The latter is indeed true because
		\begin{equation}
			\widetilde{Q}_\alpha(\mathcal{N}(\rho_A)\Vert\sigma_B)=\norm{\mathcal{N}(\rho)^{\frac{1}{2}}\sigma_B^{\frac{1-\alpha}{\alpha}}\mathcal{N}(\rho_A)^{\frac{1}{2}}}_{\alpha}^\alpha,
		\end{equation}
		for all $\alpha>1$ the function $\sigma_B\mapsto \sigma_B^{\frac{1-\alpha}{\alpha}}$ is operator convex, the Schatten norm $\norm{\cdot}_{\alpha}$ is convex, and the function $x\mapsto x^{\alpha}$ is convex  for all $x\geq 0$. So we find that
		\begin{equation}
			\widetilde{K}_{\alpha}(\mathcal{N})\leq\frac{1}{\alpha-1}\log_2\sup_{\mu}\inf_{\sigma_B}\int \widetilde{Q}_{\alpha}(\mathcal{N}(\rho_A)\Vert\sigma_B)~\text{d}\mu(\rho_A).
		\end{equation}
		Now, by Carath\'{e}odory's theorem (Theorem~\ref{thm-Caratheodory}), if $\rho$ is a density operator acting on a $d$-dimensional space, then there exists an alphabet $\mathcal{X}$ of size no more than $d^2$, a probability distribution $p:\mathcal{X}\to[0,1]$ on $\mathcal{X}$, and an ensemble $\{(p(x),\rho_A^x)\}_{x\in\mathcal{X}}$ of states such that
		\begin{equation}
			\int \widetilde{Q}_\alpha(\mathcal{N}(\rho_A)\Vert\sigma_B)~\text{d}\mu(\rho)=\sum_{x\in\mathcal{X}}p(x)\widetilde{Q}_\alpha(\mathcal{N}(\rho_A^x)\Vert\sigma_B).
		\end{equation}
		Therefore,
		\begin{align}
			\widetilde{K}_{\alpha}(\mathcal{N})&\leq \frac{1}{\alpha-1}\log_2\sup_{\{(p(x),\rho_A^x)\}_x}\inf_{\sigma_B} \sum_{x\in\mathcal{X}}p(x)\widetilde{Q}_\alpha(\mathcal{N}(\rho_A^x)\Vert\sigma_B)\\
			&=\frac{1}{\alpha-1}\log_2\sup_{\{(p(x),\rho_A^x)\}_x}\inf_{\sigma_B}\widetilde{Q}_{\alpha}(\mathcal{N}_{A\to B}(\rho_{XB})\Vert\rho_X\otimes\sigma_B)\\
			&=\sup_{\rho_{XA}}\inf_{\sigma_B}\widetilde{D}_{\alpha}(\rho_{XA}\Vert\rho_X\otimes\sigma_B)\\
			&=\sup_{\rho_{XA}}\widetilde{I}_\alpha(X;B)_\omega\\
			&=\widetilde{\chi}_\alpha(\mathcal{N}),
		\end{align}
		where $\omega_{XB}=\mathcal{N}_{A\to B}(\rho_{XA})$, and to obtain the first equality we used the direct-sum property of $\widetilde{Q}_\alpha$ in \eqref{eq-sand_quasi_rel_ent_direct_sum}. So we have $\widetilde{K}_\alpha(\mathcal{N})\leq\widetilde{\chi}_\alpha(\mathcal{N})$ in addition to $\widetilde{K}_\alpha(\mathcal{N})\geq\widetilde{\chi}_\alpha(\mathcal{N})$, which means that $\widetilde{K}_\alpha(\mathcal{N})=\widetilde{\chi}_\alpha(\mathcal{N})$, as required.
	\end{Proof}
	
	Using the fact that $\lim_{\alpha\to 1^+}\widetilde{\chi}_{\alpha}(\mathcal{N})=\chi(\mathcal{N})$ (the proof of this analogous to the one presented in Appendix~\ref{app-sand_ren_mut_inf_chan_limit}), we obtain the following alternate formula for the Holevo information of a quantum channel $\mathcal{N}$:
	\begin{equation}\label{eq-Hol_inf_chan_SW}
		\chi(\mathcal{N})=\inf_{\sigma}\sup_{\rho}D(\mathcal{N}(\rho)\Vert\sigma),
	\end{equation}
	where the optimizations are over states $\rho$ and $\sigma$. 
	
	Before stating the following lemma, let us define an entanglement-breaking map $\mathcal{N}_{A\to B}$ to be a completely positive map such that $\mathcal{N}_{A\to B}(X_{RA})$ is a separable operator of systems $R$ and $B$ for every positive semi-definite input operator $X_{RA}$. We can think of it as a generalization of an entanglement-breaking channel (Definition~\ref{def-ent_break_chan}) in which there is no requirement of trace preservation.

	\begin{Lemma}{lem-ent_break_mult_2}
Let $\mathcal{M}_{A\rightarrow B}$ be a completely positive map, and let
$P_{RA}$ be a positive semi-definite separable operator, i.e., such that it
can be written in the following form:%
\begin{equation}
P_{RA}=\sum_{x\in\mathcal{X}}C_{R}^{x}\otimes D_{A}^{x},
\label{eq-CC:P_RA_op}
\end{equation}
where $\mathcal{X}$ is a finite alphabet and $C_{R}^{x},D_{A}^{x}\geq0$ for
all $x\in\mathcal{X}$. Let $P_{R}=\operatorname{Tr}_{A}[P_{RA}]=\sum
_{x\in\mathcal{X}}\operatorname{Tr}[D_{A}^{x}]C_{R}^{x}$. For all $\alpha
\geq1$, the following inequality holds%
\begin{equation}
\left\Vert \mathcal{M}_{A\rightarrow B}(P_{RA})\right\Vert _{\alpha}\leq
\nu_{\alpha}(\mathcal{M}_{A\rightarrow B})\cdot\left\Vert P_{R}\right\Vert
_{\alpha},\label{eq-CC:mult-a-norm-EB-main-step}%
\end{equation}
where%
\begin{equation}
\nu_{\alpha}(\mathcal{P})\coloneqq\sup_{\rho}\left\Vert \mathcal{P}(\rho)\right\Vert
_{\alpha},\label{eq-max_output_norm}%
\end{equation}
$\mathcal{P}$ is a completely positive map, and the supremum is taken over
every density operator in the domain of $\mathcal{P}$. As a consequence, if
$\mathcal{N}_{A^{\prime}\rightarrow B^{\prime}}$ is an entanglement-breaking
map, then the following equality holds for all $\alpha\geq1$:%
\begin{equation}
\nu_{\alpha}(\mathcal{M}_{A\rightarrow B}\otimes\mathcal{N}_{A^{\prime
}\rightarrow B^{\prime}})=\nu_{\alpha}(\mathcal{M}_{A\rightarrow B})\cdot
\nu_{\alpha}(\mathcal{N}_{A^{\prime}\rightarrow B^{\prime}}%
).\label{eq-CC:mult-a-norm-EB}%
\end{equation}

\end{Lemma}

\begin{Proof}
Without loss of generality, we can suppose that each $D_{A}^{x}$ is
normalized, in the sense that $\operatorname{Tr}[D_{A}^{x}]=1$. If it is not
the case, then we can redefine $C_{R}^{x}$ as $C_{R}^{x}\operatorname{Tr}%
[D_{A}^{x}]$ and $D_{A}^{x}$ as $D_{A}^{x}/\operatorname{Tr}[D_{A}^{x}]$
without changing the separable operator~$P_{RA}$. Next, we observe that%
\begin{align}
\mathcal{M}_{A\rightarrow B}(P_{RA}) &  =\sum_{x\in\mathcal{X}}C_{R}%
^{x}\otimes\mathcal{M}_{A\rightarrow B}(D_{A}^{x})\\
&  =VTV^{\dag},
\end{align}
where%
\begin{align}
V &  \coloneqq\sum_{x\in\mathcal{X}}\langle x|\otimes\sqrt{C_{R}^{x}}\otimes \mathbbm{1}_{B},\\
T &  \coloneqq\sum_{x\in\mathcal{X}}|x\rangle\!\langle x|\otimes \mathbbm{1}_{R}\otimes
\mathcal{M}_{A\rightarrow B}(D_{A}^{x}).
\end{align}
This implies that%
\begin{equation}
\operatorname{Tr}[(\mathcal{M}_{A\rightarrow B}(P_{RA}))^{\alpha
}]=\operatorname{Tr}[(VTV^{\dag})^{\alpha}].
\end{equation}
Now let us apply the Araki--Lieb--Thirring inequality (Lemma~\ref{lem-ALT_ineq}), which
states that for all positive semi-definite operators $X$ and $Y$,%
\begin{equation}
\operatorname{Tr}\!\left[  \left(  Y^{\frac{1}{2}}XY^{\frac{1}{2}}\right)
^{rq}\right]  \leq\operatorname{Tr}\!\left[  \left(  Y^{\frac{r}{2}}%
X^{r}Y^{\frac{r}{2}}\right)  ^{q}\right]
\end{equation}
for all $q\geq0$ and $r\geq1$. For $q=1$, we obtain%
\begin{equation}
\operatorname{Tr}\!\left[  \left(  Y^{\frac{1}{2}}XY^{\frac{1}{2}}\right)
^{r}\right]  \leq\operatorname{Tr}\!\left[  X^{r}Y^{r}\right]
.\label{eq-CC:ALT-recall-twist}%
\end{equation}
Now, for every operator $Z$, note that $ZXZ^{\dag}$ has the same non-zero
eigenvalues as $\left(  Z^{\dag}Z\right)  ^{\frac{1}{2}}X\left(  Z^{\dag
}Z\right)  ^{\frac{1}{2}}$ (this follows by considering the polar
decomposition of~$Z$). In addition, since $Z^{\dag}Z$ is positive
semi-definite, applying \eqref{eq-CC:ALT-recall-twist} with $Y=Z^{\dag}Z$
gives us%
\begin{align}
\operatorname{Tr}\!\left[  \left(  ZXZ^{\dag}\right)  ^{r}\right]   &
=\operatorname{Tr}\!\left[  \left(  \left(  Z^{\dag}Z\right)  ^{\frac{1}{2}%
}X\left(  Z^{\dag}Z\right)  ^{\frac{1}{2}}\right)  ^{r}\right]  \\
&  \leq\operatorname{Tr}\!\left[  X^{r}(Z^{\dag}Z)^{r}\right]  .
\end{align}
Substituting $r=\alpha$, $Z=V$, and $X=T$ into this inequality gives us%
\begin{align}
\operatorname{Tr}[(\mathcal{M}_{A\rightarrow B}(P_{RA}))^{\alpha}] &
=\operatorname{Tr}[(VTV^{\dag})^{\alpha}]\\
&  \leq\operatorname{Tr}[(V^{\dag}V)^{\alpha}T^{\alpha}].
\end{align}
Letting%
\begin{equation}
S\coloneqq\sum_{x\in\mathcal{X}}\langle x|\otimes\sqrt{C_{R}^{x}},
\end{equation}
observe that%
\begin{equation}
V^{\dag}V=S^{\dag}S\otimes \mathbbm{1}_{B},
\end{equation}
which implies that%
\begin{equation}
(V^{\dag}V)^{\alpha}=(S^{\dag}S)^{\alpha}\otimes \mathbbm{1}_{B}.
\end{equation}
Therefore, since $T$, and thus $T^{\alpha}$, is block diagonal, we find that%
\begin{align}
& \operatorname{Tr}[(V^{\dag}V)^{\alpha}T^{\alpha}]\notag \\
& =\operatorname{Tr}\!\left[  \left(  (S^{\dag}S)^{\alpha}\otimes
\mathbbm{1}_{B}\right)  \left(  \sum_{x\in\mathcal{X}}|x\rangle\!\langle x|\otimes
\mathbbm{1}_{R}\otimes(\mathcal{M}_{A\rightarrow B}(D_{A}^{x}))^{\alpha}\right)
\right]  \\
& =\sum_{x\in\mathcal{X}}\operatorname{Tr}\!\left[  \left(  (S^{\dag
}S)^{\alpha}\right)  _{x}\right]  \operatorname{Tr}[(\mathcal{M}_{A\rightarrow
B}(D_{A}^{x}))^{\alpha}],
\end{align}
where%
\begin{equation}
\left(  (S^{\dag}S)^{\alpha}\right)  _{x}\coloneqq(\langle x|\otimes \mathbbm{1}_{R})(S^{\dag
}S)^{\alpha}(|x\rangle\otimes \mathbbm{1}_{R}).
\end{equation}
Now,%
\begin{align}
\operatorname{Tr}[(\mathcal{M}_{A\rightarrow B}(D_{A}^{x}))^{\alpha}%
]^{\frac{1}{\alpha}}  & =\left\Vert \mathcal{M}_{A\rightarrow B}(D_{A}%
^{x})\right\Vert _{\alpha}\leq\nu_{\alpha}(\mathcal{M}_{A\rightarrow B})\\
& \Rightarrow\operatorname{Tr}[(\mathcal{M}_{A\rightarrow B}(D_{A}%
^{x}))^{\alpha}]\leq\nu_{\alpha}(\mathcal{M}_{A\rightarrow B})^{\alpha}.
\end{align}
By taking the partial trace over system $A$ of $P_{RA}$, we find that%
\begin{equation}
P_{R}=\operatorname{Tr}_{A}[P_{RA}]=\sum_{x\in\mathcal{X}}C_{R}^{x}=SS^{\dag}.
\end{equation}
Using this, we find that%
\begin{align}
\sum_{x\in\mathcal{X}}\operatorname{Tr}\!\left[  \left(  (S^{\dag}S)^{\alpha
}\right)  _{x}\right]    & =\operatorname{Tr}[(S^{\dag}S)^{\alpha}]\\
& =\operatorname{Tr}[(SS^{\dag})^{\alpha}]\\
& =\operatorname{Tr}[P_{R}^{\alpha}]\\
& =\left\Vert P_{R}\right\Vert _{\alpha}^{\alpha}.
\end{align}
Putting everything together, we conclude that%
\begin{align}
 \left\Vert \mathcal{M}_{A\rightarrow B}(P_{RA})\right\Vert _{\alpha}
& =\left(  \operatorname{Tr}[(\mathcal{M}_{A\rightarrow B}(P_{RA}))^{\alpha
}]\right)  ^{\frac{1}{\alpha}}\\
& =\left(  \operatorname{Tr}[(VTV^{\dag})^{\alpha}]\right)  ^{\frac{1}{\alpha
}}\\
& \leq\left(  \operatorname{Tr}[(V^{\dag}V)^{\alpha}T^{\alpha}]\right)
^{\frac{1}{\alpha}}\\
& \leq\nu_{\alpha}(\mathcal{M}_{A\rightarrow B})\cdot\left\Vert P_{R}%
\right\Vert _{\alpha}.
\end{align}

To see the equality in \eqref{eq-CC:mult-a-norm-EB}, we prove it in two steps.
First, consider that the following inequality holds for all completely
positive maps $\mathcal{M}_{A\rightarrow B}$ and $\mathcal{N}_{A^{\prime
}\rightarrow B^{\prime}}$:%
\begin{equation}
\nu_{\alpha}(\mathcal{M}_{A\rightarrow B}\otimes\mathcal{N}_{A^{\prime
}\rightarrow B^{\prime}})\geq\nu_{\alpha}(\mathcal{M}_{A\rightarrow B}%
)\cdot\nu_{\alpha}(\mathcal{N}_{A^{\prime}\rightarrow B^{\prime}}).
\end{equation}
This follows simply by restricting the optimization in the definition of
$\nu_{\alpha}(\mathcal{M}_{A\rightarrow B}\otimes\mathcal{N}_{A^{\prime
}\rightarrow B^{\prime}})$ to tensor-product states. Specifically,%
\begin{align}
\nu_{\alpha}(\mathcal{M}_{A\rightarrow B}\otimes\mathcal{N}_{A^{\prime
}\rightarrow B^{\prime}})  & =\sup_{\rho_{AA^{\prime}}}\left\Vert
(\mathcal{M}_{A\rightarrow B}\otimes\mathcal{N}_{A^{\prime}\rightarrow
B^{\prime}})(\rho_{AA^{\prime}})\right\Vert _{\alpha}\\
& \geq\sup_{\sigma_{A},\omega_{A^{\prime}}}\left\Vert (\mathcal{M}%
_{A\rightarrow B}\otimes\mathcal{N}_{A^{\prime}\rightarrow B^{\prime}}%
)(\sigma_{A}\otimes\omega_{A^{\prime}})\right\Vert _{\alpha}\\
& =\sup_{\sigma_{A},\omega_{A^{\prime}}}\left\Vert (\mathcal{M}_{A\rightarrow
B}(\sigma_{A})\otimes\mathcal{N}_{A^{\prime}\rightarrow B^{\prime}}%
(\omega_{A^{\prime}})\right\Vert _{\alpha}\\
& =\sup_{\sigma_{A}}\left\Vert (\mathcal{M}_{A\rightarrow B}(\sigma
_{A})\right\Vert _{\alpha}\cdot\sup_{\omega_{A^{\prime}}}\left\Vert
\mathcal{N}_{A^{\prime}\rightarrow B^{\prime}}(\omega_{A^{\prime}})\right\Vert
_{\alpha}\\
& =\nu_{\alpha}(\mathcal{M}_{A\rightarrow B})\cdot\nu_{\alpha}(\mathcal{N}%
_{A^{\prime}\rightarrow B^{\prime}}).
\end{align}
The following reverse inequality%
\begin{equation}
\nu_{\alpha}(\mathcal{M}_{A\rightarrow B}\otimes\mathcal{N}_{A^{\prime
}\rightarrow B^{\prime}})\leq\nu_{\alpha}(\mathcal{M}_{A\rightarrow B}%
)\cdot\nu_{\alpha}(\mathcal{N}_{A^{\prime}\rightarrow B^{\prime}%
})\label{eq-CC:rev-ineq-EB-mult}%
\end{equation}
holds when $\mathcal{N}_{A^{\prime}\rightarrow B^{\prime}}$ is an
entanglement-breaking map. Indeed, considering an arbitrary input state
$\rho_{AA^{\prime}}$, the output state $\omega_{AB^{\prime}}\coloneqq\mathcal{N}%
_{A^{\prime}\rightarrow B^{\prime}}(\rho_{AA^{\prime}})$ is a separable
operator. Applying \eqref{eq-CC:mult-a-norm-EB-main-step} to the separable
operator $\omega_{AB^{\prime}}$ and identifying system $B^{\prime}$ with $R$
in \eqref{eq-CC:mult-a-norm-EB-main-step}, we conclude that%
\begin{align}
\left\Vert (\mathcal{M}_{A\rightarrow B}\otimes\mathcal{N}_{A^{\prime
}\rightarrow B^{\prime}})(\rho_{AA^{\prime}})\right\Vert _{\alpha}  &
=\left\Vert \mathcal{M}_{A\rightarrow B}(\omega_{AB^{\prime}})\right\Vert
_{\alpha}\\
& \leq\nu_{\alpha}(\mathcal{M}_{A\rightarrow B})\cdot\left\Vert \omega
_{B^{\prime}}\right\Vert _{\alpha}\\
& =\nu_{\alpha}(\mathcal{M}_{A\rightarrow B})\cdot\left\Vert \mathcal{N}%
_{A^{\prime}\rightarrow B^{\prime}}(\rho_{A^{\prime}})\right\Vert _{\alpha}\\
& \leq\nu_{\alpha}(\mathcal{M}_{A\rightarrow B})\cdot\nu_{\alpha}%
(\mathcal{N}_{A^{\prime}\rightarrow B^{\prime}}).
\end{align}
Since the inequality holds for every input state $\rho_{AA^{\prime}}$, we
conclude the inequality in \eqref{eq-CC:rev-ineq-EB-mult}.
\end{Proof}
	
	With Lemmas \ref{lem-ent_break_mult_1} and \ref{lem-ent_break_mult_2} in hand, we can now prove Theorem~\ref{thm-classical_comm_ent_break_additive}.

\subsubsection*{Proof of Theorem~\ref{thm-classical_comm_ent_break_additive}}\label{subsubsec-classical_comm_ent_break_additive}
	
	We start by using \eqref{eq-sand_rel_ent_Schatten} to write the definition of $\widetilde{K}_{\alpha}(\mathcal{N})$ as
	\begin{align}
		\widetilde{K}_\alpha(\mathcal{N})&=\inf_{\sigma_B}\sup_{\rho_A}\widetilde{D}_\alpha(\mathcal{N}(\rho_A)\Vert\sigma_B)\\
		&=\inf_{\sigma_B}\sup_{\rho_A}\frac{\alpha}{\alpha-1}\log_2\norm{\sigma_B^{\frac{1-\alpha}{2\alpha}}\mathcal{N}(\rho_A)\sigma_B^{\frac{1-\alpha}{2\alpha}}}_{\alpha}.
	\end{align}
	Then, for every channel $\mathcal{M}$,
	\begin{align}
		& \widetilde{K}_{\alpha}(\mathcal{N}\otimes\mathcal{M})\notag \\
		&=\inf_{\sigma_{A'B'}}\sup_{\rho_{AB}}\widetilde{D}_\alpha((\mathcal{N}\otimes\mathcal{M})(\rho_{AB})\Vert\sigma_{A'B'})\\
		&=\frac{\alpha}{\alpha-1}\inf_{\sigma_{A'B'}}\sup_{\rho_{AB}}\log_2\norm{\sigma_{A'B'}^{\frac{1-\alpha}{2\alpha}}\left((\mathcal{N}\otimes\mathcal{M})(\rho_{AB})\right)\sigma_{A'B'}^{\frac{1-\alpha}{2\alpha}}}_{\alpha}\\
		&=\frac{\alpha}{\alpha-1}\inf_{\sigma_{A'B'}}\log_2\sup_{\rho_{AB}}\norm{\sigma_{A'B'}^{\frac{1-\alpha}{2\alpha}}\left((\mathcal{N}\otimes\mathcal{M})(\rho_{AB})\right)\sigma_{A'B'}^{\frac{1-\alpha}{2\alpha}}}_{\alpha}\label{eq-classical_comm_ent_break_additive_pf}\\
		&\leq\frac{\alpha}{\alpha-1}\inf_{\sigma_{A'},\tau_{B'}}\log_2\sup_{\rho_{AB}}\Bigg\lVert\left(\sigma_{A'}^{\frac{1-\alpha}{2\alpha}}\otimes\tau_{B'}^{\frac{1-\alpha}{2\alpha}}\right)\left((\mathcal{N}\otimes\mathcal{M})(\rho_{AB})\right)\left(\sigma_{A'}^{\frac{1-\alpha}{2\alpha}}\otimes\tau_{B'}^{\frac{1-\alpha}{2\alpha}}\right)\Bigg\rVert_{\alpha},
	\end{align}
	where to obtain the inequality we have restricted the infimum to tensor product states. Now, observe that since $\mathcal{N}$ is entanglement-breaking, then sandwiching the output of the channel by the positive semi-definite operator $\sigma_{A'}^{\frac{1-\alpha}{2\alpha}}$ leads to a new map $\mathcal{N}'$ that is a completely positive entanglement-breaking map (though not necessarily trace preserving). Similarly, sandwiching the output of $\mathcal{M}$ by the positive semi-definite operator $\tau_{B'}^{\frac{1-\alpha}{2\alpha}}$ leads to a new completely positive map $\mathcal{M}'$. Therefore, using Lemma~\ref{lem-ent_break_mult_2}, we obtain
	\begin{align}
		&\widetilde{K}_{\alpha}(\mathcal{N}\otimes\mathcal{M})\notag \\
		&\leq \frac{\alpha}{\alpha-1}\inf_{\sigma_{A'},\tau_{B'}}\log_2\!\left(\sup_{\rho_{AB}}\norm{(\mathcal{N}'\otimes\mathcal{M}')(\rho_{AB})}_{\alpha}\right)\\
		&=\frac{\alpha}{\alpha-1}\inf_{\sigma_{A'},\tau_{B'}}\log_2\nu_{\alpha}(\mathcal{N}'\otimes\mathcal{M}')\\
		&=\frac{\alpha}{\alpha-1}\inf_{\sigma_{A'},\tau_{B'}}\log_2\!\left(\nu_{\alpha}(\mathcal{N}')\nu_{\alpha}(\mathcal{M}')\right)\\
		&=\frac{\alpha}{\alpha-1}\inf_{\sigma_{A'},\tau_{B'}}\left[\log_2\nu_{\alpha}(\mathcal{N}')+\log_2\nu_{\alpha}(\mathcal{M}')\right]\\
		&=\inf_{\sigma_{A'}}\frac{\alpha}{\alpha-1}\log_2\sup_{\rho_A}\norm{\mathcal{N}'(\rho_A)}_{\alpha}+\inf_{\tau_{B'}}\frac{\alpha}{\alpha-1}\log_2\sup_{\omega_B}\norm{\mathcal{M}'(\omega_B)}_{\alpha}\\
		&=\inf_{\sigma_{A'}}\sup_{\rho_A}\frac{\alpha}{\alpha-1}\log_2\norm{\mathcal{N}'(\rho_A)}_{\alpha}+\inf_{\tau_{B'}}\sup_{\omega_B}\frac{\alpha}{\alpha-1}\log_2\norm{\mathcal{M}'(\omega_B)}_{\alpha}\\
		&=\widetilde{K}_{\alpha}(\mathcal{N})+\widetilde{K}_{\alpha}(\mathcal{M}).
	\end{align}
	So we have that $\widetilde{K}_{\alpha}(\mathcal{N}\otimes\mathcal{M})\leq \widetilde{K}_\alpha(\mathcal{N})+\widetilde{K}_{\alpha}(\mathcal{M})$ for every channel $\mathcal{M}$. Using Lemma~\ref{lem-ent_break_mult_1}, we obtain the desired result.
	
	Note that the additivity of the Holevo information of every entanglement-breaking channel follows from the additivity of the sandwiched R\'{e}nyi Holevo information of such channels by taking the limit $\alpha\to 1^{+}$ (the proof is analogous to the one presented in Appendix~\ref{app-sand_ren_mut_inf_chan_limit}).

\subsection{Proof of the Strong Converse for Entanglement-Break\-ing Channels}\label{sec-cc_strong_conv}
	
	Having shown that the sandwiched R\'{e}nyi Holevo information is additive for all entanglement-breaking channels, we can now proceed further from \eqref{eq-cc_str_conv_1} to prove a strong converse theorem for all entanglement-breaking channels. Moreover, since the sandwiched R\'{e}nyi Holevo information $\widetilde{\chi}_{\alpha}(\mathcal{N})$ satisfies $\lim_{\alpha\to 1^+}\widetilde{\chi}_{\alpha}(\mathcal{N})=\chi(\mathcal{N})$ (the proof of this is analogous to the one presented in Appendix~\ref{app-sand_ren_mut_inf_chan_limit}), we can go beyond the statement of Theorem~\ref{thm-classical_capacity} and say that $C(\mathcal{N})=\chi(\mathcal{N})$ for all entanglement-breaking channels $\mathcal{N}$.

	\begin{theorem*}{Classical Capacity of Entanglement-Break\-ing Channels}{thm-str_conv_additive}
		For every entanglement-breaking channel $\mathcal{N}$,
		\begin{equation}\label{eq-cc_capacity_ent_breaking}
			C(\mathcal{N})=\widetilde{C}(\mathcal{N})=\chi(\mathcal{N}).
		\end{equation}
	\end{theorem*}
	
	\begin{remark} Note that this theorem holds more generally for every channel $\mathcal{N}$ for which the sandwiched R\'{e}nyi Holevo information $\widetilde{\chi}_{\alpha}(\mathcal{N})$ is additive.
	\end{remark}
	
	\begin{Proof}
		Since $\lim_{\alpha\to 1^+}\widetilde{\chi}_{\alpha}(\mathcal{N})=\chi(\mathcal{N})$ (the proof of this is analogous to the one presented in Appendix~\ref{app-sand_ren_mut_inf_chan_limit}), we find that the Holevo information is additive for all entanglement-breaking channels. The equality $C(\mathcal{N})=\chi(\mathcal{N})$ then follows from Theorem~\ref{thm-classical_capacity}.
		
		The remainder of the proof is devoted to establishing that $\chi(\mathcal{N})$ is a strong converse rate for classical communication over $\mathcal{N}$, from which it follows that $\widetilde{C}(\mathcal{N})\leq\chi(\mathcal{N})$, which in turn implies, via \eqref{eq-cc_str_conv_rate_lower_bound}, that $\widetilde{C}(\mathcal{N})=\chi(\mathcal{N})$.
		
		Fix $\varepsilon\in[0,1)$ and $\delta>0$. Let $\delta_1,\delta_2>0$ be such that
		\begin{equation}\label{eq-cc_str_conv_ent_break_pf1}
			\delta>\delta_1+\delta_2\eqqcolon\delta'.
		\end{equation}
		Set $\alpha\in(1,\infty)$ such that
		\begin{equation}\label{eq-cc_str_conv_ent_break_pf2}
			\delta_1\geq\widetilde{\chi}_{\alpha}(\mathcal{N})-\chi(\mathcal{N}),
		\end{equation}
		which is possible since $\widetilde{\chi}_{\alpha}(\mathcal{N})$ is monotonically increasing with $\alpha$ (this follows from Proposition~\ref{prop-sand_rel_ent_properties}), and since $\lim_{\alpha\to 1^+}\widetilde{\chi}_{\alpha}(\mathcal{N})=\chi(\mathcal{N})$. With this value of $\alpha$, take $n$ large enough so that
		\begin{equation}\label{eq-cc_str_conv_ent_break_pf3}
			\delta_2\geq\frac{\alpha}{n(\alpha-1)}\log_2\!\left(\frac{1}{1-\varepsilon}\right).
		\end{equation}
		
		Now, with the values of $n$ and $\varepsilon$ chosen as above, every $(n,|\mathcal{M}|,\varepsilon)$ classical communication protocol satisfies \eqref{eq-cc_str_conv_1} in Corollary~\ref{cor-cc_str_weak_conv_upper}. In particular, using the additivity of the sandwiched R\'{e}nyi Holevo information for all $\alpha>1$, we can write \eqref{eq-cc_str_conv_1} as
		\begin{equation}
			\frac{1}{n}\log_2|\mathcal{M}|\leq\widetilde{\chi}_{\alpha}(\mathcal{N})+\frac{\alpha}{n(\alpha-1)}\log_2\!\left(\frac{1}{1-\varepsilon}\right).
		\end{equation}
		Rearranging the right-hand side of this inequality, and using the assumptions in \eqref{eq-cc_str_conv_ent_break_pf1}--\eqref{eq-cc_str_conv_ent_break_pf3}, we obtain
		\begin{align}
			\frac{1}{n}\log_2|\mathcal{M}|&\leq\chi(\mathcal{N})+\widetilde{\chi}_{\alpha}(\mathcal{N})-\chi(\mathcal{N})+\frac{\alpha}{n(\alpha-1)}\log_2\!\left(\frac{1}{1-\varepsilon}\right)\\
			&\leq \chi(\mathcal{N})+\delta_1+\delta_2\\
			&=\chi(\mathcal{N})+\delta'\\
			&<\chi(\mathcal{N})+\delta.
		\end{align}
		So we have that $\chi(\mathcal{N})+\delta>\frac{1}{n}\log_2|\mathcal{M}|$ for all $(n,|\mathcal{M}|,\varepsilon)$ classical communication protocols with $n$ sufficiently large. Due to this strict inequality, it follows that there cannot exist an $(n,2^{n(\chi(\mathcal{N})+\delta)},\varepsilon)$ classical communication protocol for all sufficiently large $n$ such that \eqref{eq-cc_str_conv_ent_break_pf3} holds, for if it did there would exist some message set $\mathcal{M}$ such that $\frac{1}{n}\log_2|\mathcal{M}|=\chi(\mathcal{N})+\delta$, which we have just seen is not possible. Since $\varepsilon$ and $\delta$ are arbitrary, we conclude that for all $\varepsilon\in[0,1)$, $\delta>0$, and sufficiently large $n$, there does not exist an $(n,2^{n(\chi(\mathcal{N})+\delta)},\varepsilon)$ classical communication protocol. This means that $\chi(\mathcal{N})$ is a strong converse rate, which completes the proof.
	\end{Proof}

\subsubsection{The Strong Converse from a Different Point of View}
	
\begin{figure}
		\centering
		\includegraphics[scale=0.8]{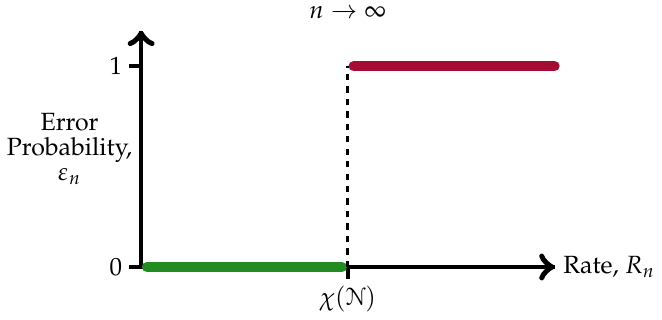}
		\caption{The error probability $\varepsilon_n$ as a function of the rate $R_n$ for classical communication  over a quantum channel $\mathcal{N}$ for which the sandwiched R\'{e}nyi Holevo information $\widetilde{\chi}_\alpha(\mathcal{N})$ is additive. As $n\to\infty$, for every rate below the Holevo information $\chi(\mathcal{N})$, there exists a sequence of protocols with error probability converging to zero. For every rate above the Holevo information $\chi(\mathcal{N})$, the error probability converges to one for all possible protocols.}\label{fig-classical_comm_str_converse}
	\end{figure}

	Just as we did in the case of entanglement-assisted classical communication in Appendix~\ref{app:EA-comm:strong-conv-diff-POV}, we can use the alternative definitions of classical capacity and strong converse classical capacity (stated in Appendix~\ref{chap-str_conv}) to see that $C(\mathcal{N})=\widetilde{C}(\mathcal{N})=\chi(\mathcal{N})$ for all channels $\mathcal{N}$ for which the sandwiched R\'{e}nyi Holevo information is additive and that, as shown in Figure~\ref{fig-classical_comm_str_converse}, the quantity $\chi(\mathcal{N})$ is a sharp dividing point between reliable, error-free communication and communication with error approaching one exponentially fast. Specifically, by following the arguments in Appendix~\ref{app:EA-comm:strong-conv-diff-POV}, we obtain the following: for every sequence $\{(n,2^{nR},\varepsilon_n)\}_{n\in\mathbb{N}}$ of $(n,|\mathcal{M}|,\varepsilon)$ protocols, with each element of the sequence having an arbitrary (but fixed) rate $R>\chi(\mathcal{N})$, the sequence $\{\varepsilon_n\}_{n\in\mathbb{N}}$ of error probabilities approaches one at an exponential rate.
	
\subsection{General Upper Bounds on the Strong Converse Classical Capacity}\label{subsec-cc_general_upper_bounds}

	The difficulty in proving the additivity of the Holevo information for a general channel, and thus obtaining an upper bound on its classical capacity, has motivated the study of other, more tractable upper bounds on the classical capacity of a quantum channel. In this section, we present two upper bounds on the strong converse classical capacity of a quantum channel.

\subsubsection{\texorpdfstring{$\Upsilon$}{Y}-Information Upper Bound}\label{subsubsec-cc_Upsilon_inf}

	Recall from Proposition~\ref{prop-cc:one-shot-bound-meta} that the following upper bound on the number of transmitted bits holds for every $(|\mathcal{M}|,\varepsilon)$ classical communication protocol:
	\begin{equation}\label{eq-cc:one-shot-bound-meta_2}
		\log_2|\mathcal{M}|\leq\chi_H^{\varepsilon}(\mathcal{N}),
	\end{equation}
	where the $\varepsilon$-hypothesis testing Holevo information $\chi_H^{\varepsilon}(\mathcal{N})$ of the quantum channel $\mathcal{N}$ is defined in \eqref{eq-hypo_testing_Holevo_inf_chan} as
	\begin{align}
		\chi_H^{\varepsilon}(\mathcal{N})& =
		\sup_{\rho_{XA}} I_H^{\varepsilon}(X;B)_{\omega}\\
		& = \sup_{\rho_{XA}}\inf_{\sigma_B}D_H^{\varepsilon}(\mathcal{N}_{A\to B}(\rho_{XA})\Vert\rho_X\otimes\sigma_B).
	\end{align}
	Here, $\omega_{XB} = \mathcal{N}_{A\to B}(\rho_{XA})$, and the optimization is over classical--quantum states $\rho_{XA}$ of the form $\rho_{XA}=\sum_{x\in\mathcal{X}}p(x)\ket{x}\!\bra{x}_X\otimes\rho_A^x$, where $\mathcal{X}$ is a finite alphabet, $p:\mathcal{X}\to[0,1]$ is a probability distribution, and $\{(p(x),\rho_A^x)\}_{x\in\mathcal{X}}$ is an ensemble of states.
	
	We derived the inequality in \eqref{eq-cc:one-shot-bound-meta_2} by comparing the actual classical communication protocol over the channel $\mathcal{N}$ with a classical communication protocol over the replacement channel $\mathcal{R}$ (see \eqref{eq-cc_useless_channel} in Section~\ref{subsec-cc_useless_channel}). The replacement channel is useless for classical communication  because it discards the state encoded with the message and replaces it with a fixed state. We can make the comparison between the channel $\mathcal{N}$ and the channel $\mathcal{R}$ for classical communication more explicit by writing the quantity $\chi_H^{\varepsilon}(\mathcal{H})$ as
	\begin{equation}
		\chi_H^{\varepsilon}(\mathcal{N})=\sup_{\rho_{XA}}\inf_{\mathcal{R}_{A\to B}}D_H^{\varepsilon}(\mathcal{N}_{A\to B}(\rho_{XA})\Vert\mathcal{R}_{A\to B}(\rho_{XA})),
	\end{equation}
	where $\mathcal{R}_{A\to B}$ has the action $\mathcal{R}_{A\to B}(\rho_{XA})=\rho_X\otimes\sigma_B$.
	
	Intuitively, an approach for obtaining an alternative upper bound on the strong converse classical capacity is to expand the set of useless channels from the replacement channels to the following set of completely positive trace non-increasing maps:
	\begin{equation}\label{eq-cc_gen_useless_maps}
		\mathfrak{F}\coloneqq \{\mathcal{F}_{A\to B}:\exists~\sigma_B\geq 0,\, \Tr[\sigma_B]\leq 1,  \, \mathcal{F}_{A\to B}(\rho_A)\leq \sigma_B\, \forall\rho_A\in \Density(\mathcal{H}_A)\},
	\end{equation}
	This set of maps, even though they contain completely positive, non-trace-preserving maps, can be thought of intuitively as also being useless for classical communication, and it contains the set of replacement channels. Using this set, we define the generalized $\Upsilon$-information as follows:
	
	\begin{definition}{Generalized $\Upsilon$-Information}{def_gen_upsilon_inf}
		Let $\boldsymbol{D}$ be a generalized divergence (see Definition~\ref{def-gen_div}). For every quantum channel $\mathcal{N}_{A\to B}$, we define the \textit{generalized $\Upsilon$-information of $\mathcal{N}$} as
		\begin{equation}
			\boldsymbol{\Upsilon}(\mathcal{N})\coloneqq \sup_{\psi_{RA}}\inf_{\mathcal{F}\in\mathfrak{F}}\boldsymbol{D}(\mathcal{N}_{A\to B}(\psi_{RA})\Vert\mathcal{F}_{A\to B}(\psi_{RA})),
		\end{equation}
		where the supremum is over pure states $\psi_{RA}$, with the dimension of $R$ the same as the dimension of $A$.
	\end{definition}
	
	
	\begin{remark}
		Note that it suffices to optimize over pure states $\psi_{RA}$, with the dimension of $R$ equal to the dimension of $A$, when calculating the generalized $\Upsilon$-information of a channel, i.e., for general states $\rho_{RA}$ (with the dimension of $R$ not necessarily equal to the dimension of $A$),
		\begin{equation}
			\sup_{\rho_{RA}}\inf_{\mathcal{F}\in\mathfrak{F}}\boldsymbol{D}(\mathcal{N}_{A\to B}(\rho_{RA})\Vert\mathcal{F}_{A\to B}(\rho_{RA}))=\boldsymbol{\Upsilon}(\mathcal{N}).
		\end{equation}
		The proof of this proceeds analogously to the steps in \eqref{eq-gen_div_chan_pure_2}--\eqref{eq-gen_div_chan_pure_4} for proving that it suffices to optimize over pure states when calculating the generalized  channel divergence.
	\end{remark}
	
	In this section, we are interested in the following generalized $\Upsilon$-informa\-tion channel quantities:
	\begin{enumerate}
		\item The \textit{$\Upsilon$-information of $\mathcal{N}$},
			\begin{equation}
				\Upsilon(\mathcal{N})\coloneqq\sup_{\psi_{RA}}\inf_{\mathcal{F}\in\mathfrak{F}}D(\mathcal{N}_{A\to B}(\psi_{RA})\Vert\mathcal{F}_{A\to B}(\psi_{RA})).
			\end{equation} 
		
		\item The \textit{$\varepsilon$-hypothesis testing $\Upsilon$-information of $\mathcal{N}$},
			\begin{equation}
				\Upsilon_H^{\varepsilon}(\mathcal{N})\coloneqq \sup_{\psi_{RA}}\inf_{\mathcal{F}\in\mathfrak{F}}D_H^{\varepsilon}(\mathcal{N}_{A\to B}(\psi_{RA})\Vert\mathcal{F}_{A\to B}(\psi_{RA})).
			\end{equation}
			
		\item The \textit{sandwiched R\'{e}nyi $\Upsilon$-information of $\mathcal{N}$},
			\begin{equation}
				\widetilde{\Upsilon}_{\alpha}(\mathcal{N})\coloneqq\sup_{\psi_{RA}}\inf_{\mathcal{F}\in\mathfrak{F}}\widetilde{D}_{\alpha}(\mathcal{N}_{A\to B}(\psi_{RA})\Vert\mathcal{F}_{A\to B}(\psi_{RA})),
			\end{equation}
			where $\alpha\in \left[1/2,1\right)\cup(1,\infty)$.
	\end{enumerate}
	
	\begin{proposition*}{Holevo Information and $\Upsilon$-Information}{prop:CC-comm:holevo-to-upsilon}
	The $\Upsilon$-information $\Upsilon(\mathcal{N})$ of a quantum channel $\mathcal{N}$ is greater than or equal to its Holevo information $\chi(\mathcal{N})$:
	\begin{equation}
	\Upsilon(\mathcal{N}) \geq \chi(\mathcal{N}).
	\end{equation}
	\end{proposition*}
	
	\begin{Proof}
	To see this,
%
%
we apply Proposition~\ref{prop:QEI:gen-ch-div-concavity-convexity} to obtain
	\begin{align}
		\Upsilon(\mathcal{N})&=\sup_{\rho_A}\inf_{\mathcal{F}\in\mathfrak{F}}D(\sqrt{\rho_A}\Gamma^{\mathcal{N}}_{AB}\sqrt{\rho_A}\Vert\sqrt{\rho_A}\Gamma^{\mathcal{F}}_{AB}\sqrt{\rho_A})\\
		&=\inf_{\mathcal{F}\in\mathfrak{F}}\sup_{\rho_A}D(\sqrt{\rho_A}\Gamma^{\mathcal{N}}_{AB}\sqrt{\rho_A}\Vert\sqrt{\rho_A}\Gamma^{\mathcal{F}}_{AB}\sqrt{\rho_A}).
		\label{eq-CC:minimax-upsilon-info}
	\end{align}
		Then, by employing \eqref{eq-CC:minimax-upsilon-info}, we find that
	\begin{align}
		\Upsilon(\mathcal{N})&=\inf_{\mathcal{F}\in\mathfrak{F}}\sup_{\rho_A}D(\sqrt{\rho_A}\Gamma^{\mathcal{N}}_{AB}\sqrt{\rho_A}\Vert\sqrt{\rho_A}\Gamma^{\mathcal{F}}_{AB}\sqrt{\rho_A})\\
		&\geq \inf_{\mathcal{F}\in\mathfrak{F}}\sup_{\rho_A}D(\mathcal{N}_{A\to B}(\rho_A)\Vert\mathcal{F}_{A\to B}(\rho_A)),
	\end{align}
	where, to obtain the inequality, we used the fact that
	\begin{align}
		\sqrt{\rho_A}\Gamma^{\mathcal{N}}_{AB}\sqrt{\rho_A}&=(\sqrt{\rho_{A'}}\otimes\mathbbm{1}_B)\mathcal{N}_{A\to B}(\Gamma_{A'A})(\sqrt{\rho_{A'}}\otimes\mathbbm{1}_B)\\
		&=\mathcal{N}_{A\to B}\left((\sqrt{\rho_{A'}}\otimes\mathbbm{1}_A)\Gamma_{A'A}(\sqrt{\rho_{A'}}\otimes\mathbbm{1}_A)\right)\\
		&=\mathcal{N}_{A\to B}\left((\mathbbm{1}_{A'}\otimes\sqrt{\smash[b]{\rho_A^{\t}}})\Gamma_{AA}(\mathbbm{1}_{A'}\otimes\sqrt{\smash[b]{\rho_A^{\t}}})\right).
	\end{align}
	In the last line we used the transpose trick (see \eqref{eq-transpose_trick}. Then, we applied the monotonicity of the quantum relative entropy with respect to the partial trace $\Tr_A$. Finally, we used the fact that $\rho_A^{\t}$ is a state for every $\rho_A$, so that the optimization over states remains unchanged. 
	
	Continuing, we have that
	\begin{align}
		\Upsilon(\mathcal{N})&\geq \inf_{\mathcal{F}\in\mathfrak{F}}\sup_{\rho_A}D(\mathcal{N}_{A\to B}(\rho_A)\Vert\sigma_{\mathcal{F}})\\
		&\geq \inf_{\sigma_B}\sup_{\rho_A}D(\mathcal{N}_{A\to B}(\rho_A)\Vert\sigma_B)\\
		&=\chi(\mathcal{N}).
	\end{align}
	To obtain the first inequality, we first used the fact that for every map $\mathcal{F}\in\mathfrak{F}$ there exists a state, which we call $\sigma_{\mathcal{F}}$, such that $\mathcal{F}_{A\to B}(\rho_A)\leq \sigma_{\mathcal{F}}$ for all input states $\rho_A$. We then used 2.(d) in Proposition~\ref{prop-rel_ent}. To obtain the last inequality, we simply enlarged the set over which the infimum is performed to include all states. Then, to obtain the equality on the last line, we used the expression in \eqref{eq-Hol_inf_chan_SW} for the Holevo information. 
	\end{Proof}
	
	We now prove an analogue of Proposition~\ref{prop-cc:one-shot-bound-meta} involving the $\varepsilon$-hypothesis testing $\Upsilon$-information, $\Upsilon_H^{\varepsilon}(\mathcal{N})$, in place of the $\varepsilon$-hypothesis testing Holevo information $\chi_H^{\varepsilon}(\mathcal{N})$.
	
	\begin{proposition}{prop-cc_upsilon_meta_converse}
		Let $\mathcal{N}$ be a quantum channel. For every $(|\mathcal{M}|,\varepsilon)$ classical communication protocol over $\mathcal{N}$, the number of bits transmitted over $\mathcal{N}$ is bounded from above by the $\varepsilon$-hypothesis testing $\Upsilon$-information of $\mathcal{N}$, i.e.,
		\begin{equation}
			\log_2|\mathcal{M}|\leq \Upsilon_H^{\varepsilon}(\mathcal{N}).
		\end{equation}
	\end{proposition}
	
	\begin{Proof}
		For every $(|\mathcal{M}|,\varepsilon)$ classical communication protocol, with encoding and decoding channel given by $\mathcal{E}$ and $\mathcal{D}$, respectively, the maximal error probability criterion $p_{\text{err}}^*(\mathcal{E},\mathcal{D};\mathcal{N})\leq \varepsilon$ holds. This implies $\overline{p}_{\text{err}}((\mathcal{E},\mathcal{D});p,\mathcal{N})\leq \varepsilon$  for the average probability,  where $p:\mathcal{M}\to[0,1]$ is the uniform prior probability distribution over the messages in $\mathcal{M}$. If the encoding channel $\mathcal{E}$ is defined such that we obtain the set $\{\rho_A^m\}_{m\in\mathcal{M}}$ of states associated to each message $m\in\mathcal{M}$ (see \eqref{eq-classical_comm_encoding}), and the decoding channel $\mathcal{D}$ is defined by the POVM $\{\Lambda_B^{\widehat{m}}\}_{\widehat{m}\in\mathcal{M}}$, then we can write the average success probability $\overline{p}_{\text{succ}}((\mathcal{E},\mathcal{D});p,\mathcal{N})$ of the code $(\mathcal{E},\mathcal{D})$ as
		\begin{align}
			\overline{p}_{\text{succ}}((\mathcal{E},\mathcal{D});p,\mathcal{N})&=1-\overline{p}_{\text{err}}((\mathcal{E},\mathcal{D});p,\mathcal{N})\\
			&=\frac{1}{|\mathcal{M}|}\sum_{m\in\mathcal{M}}\Tr[\Lambda_B^{m}\mathcal{N}_{A\to B}(\rho_A^m)],
		\end{align}
		and we have that $\overline{p}_{\text{succ}}((\mathcal{E},\mathcal{D});p,\mathcal{N})\geq 1-\varepsilon$. Now, recall from \eqref{eq-Choi_rep_action} that we can write the action of $\mathcal{N}_{A\to B}$ in terms of its Choi representation $\Gamma^{\mathcal{N}}_{AB}$ as
		\begin{equation}
			\mathcal{N}_{A\to B}(\rho_A^m)=\Tr_A\!\left[\left((\rho_A^m)^{\t}\otimes\mathbbm{1}_B\right)\Gamma^{\mathcal{N}}_{AB}\right]
		\end{equation}
		for all $m\in\mathcal{M}$. Also, let us define the average state
		\begin{equation}
			\overline{\rho}_A\coloneqq\frac{1}{|\mathcal{M}|}\sum_{m\in\mathcal{M}}\rho_A^m,
		\end{equation}
		and a purification of it
		\begin{equation}
			\ket{\overline{\phi}}_{AA'}\coloneqq(\mathbbm{1}_{A}\otimes\sqrt{\overline{\rho}_{A'}})\ket{\Gamma}_{AA'}=\left(\sqrt{\smash[b]{\overline{\rho}_{A}^{\t}}}\otimes\mathbbm{1}_{A'}\right)\ket{\Gamma}_{A'A},
		\end{equation}
		where we used the transpose trick in \eqref{eq-transpose_trick} to obtain the last equality. Then, observe that
		\begin{align}
			\mathcal{N}_{A'\to B}(\overline{\phi}_{AA'})&=\sqrt{\smash[b]{\overline{\rho}_{A}^{\t}}}\mathcal{N}_{A'\to B}(\Gamma_{AA'})\sqrt{\smash[b]{\overline{\rho}_{A}^{\t}}}\\
			&=\sqrt{\smash[b]{\overline{\rho}_A^{\t}}}\Gamma^{\mathcal{N}}_{AB}\sqrt{\smash[b]{\overline{\rho}_A^{\t}}},
		\end{align}
		which implies that the Choi operator $\Gamma^{\mathcal{N}}_{AB}$ can be written as\footnote{Note that if $\overline{\rho}_A^{\t}$ is not invertible, then the inverse is understood to be on the support of~$\overline{\rho}_A^{\t}$.}
		\begin{equation}\label{eq-cc_upsilon_meta_conv_pf_1}
			\Gamma^{\mathcal{N}}_{AB} = (\overline{\rho}_{A}^{\t})^{-\frac{1}{2}}\mathcal{N}_{A'\to B}(\overline{\phi}_{AA'})(\overline{\rho}_A^{\t})^{-\frac{1}{2}}.
		\end{equation}
		Therefore,
		\begin{align}
			&\overline{p}_{\text{succ}}((\mathcal{E},\mathcal{D});p,\mathcal{N})\nonumber\\
			&\qquad=\frac{1}{|\mathcal{M}|}\sum_{m\in\mathcal{M}}\Tr[\Lambda_B^m\mathcal{N}_{A\to B}(\rho_A^m)]\\
			&\qquad=\frac{1}{|\mathcal{M}|}\sum_{m\in\mathcal{M}}\Tr\!\left[\left((\rho_A^m)^{\t}\otimes\Lambda_B^m\right)\Gamma^{\mathcal{N}}_{AB}\right]\\
			&\qquad=\frac{1}{|\mathcal{M}|}\sum_{m\in\mathcal{M}}\Tr\!\left[\left((\rho_A^m)^{\t}\otimes\Lambda_B^m\right)(\overline{\rho}_A^{\t})^{-\frac{1}{2}}\mathcal{N}_{A'\to B}(\overline{\phi}_{AA'})(\overline{\rho}_A^{\t})^{-\frac{1}{2}}\right]\\
			&\qquad=\Tr\!\left[(\overline{\rho}_A^{\t})^{-\frac{1}{2}}\left(\frac{1}{|\mathcal{M}|}\sum_{m\in\mathcal{M}}(\rho_A^m)^{\t}\otimes\Lambda_B^m\right)(\overline{\rho}_A^{\t})^{-\frac{1}{2}}\mathcal{N}_{A'\to B}(\overline{\phi}_{AA'})\right]\\
			&\qquad=\Tr[\Omega_{AB}\mathcal{N}_{A'\to B}(\overline{\phi}_{AA'})],
		\end{align}
		where
		\begin{equation}
			\Omega_{AB}\coloneqq (\overline{\rho}_A^{\t})^{-\frac{1}{2}}\left(\frac{1}{|\mathcal{M}|}\sum_{m\in\mathcal{M}}(\rho_A^m)^{\t}\otimes\Lambda_B^m\right)(\overline{\rho}_A^{\t})^{-\frac{1}{2}}.
		\end{equation}
		Note that $\Omega_{AB}$ is positive semi-definite, i.e., $\Omega_{AB}\geq 0$. Also, observe that since $\Lambda_B^m\leq\mathbbm{1}_B$ for all $m\in\mathcal{M}$, we have that
		\begin{equation}
			\Omega_{AB}\leq (\overline{\rho}_A^{\t})\left(\frac{1}{|\mathcal{M}|}\sum_{m\in\mathcal{M}}(\rho_A^m)^{\t}\otimes\mathbbm{1}_B\right)(\overline{\rho}_A^{\t})^{-\frac{1}{2}}=\mathbbm{1}_{AB}.
		\end{equation}
		Together with $\Omega_{AB}\geq 0$, this means that $\Omega_{AB}$ is a measurement operator. So we have that
		\begin{equation}
			p_{\text{succ}}((\mathcal{E},\mathcal{D});p,\mathcal{N})=\Tr[\Omega_{AB}\mathcal{N}_{A\to B}(\overline{\phi}_{AA'})]\geq 1-\varepsilon.
		\end{equation}
		
		Now, let $\mathcal{F}\in\mathfrak{F}$. This means that there exists a state, call it $\sigma_B$, such that $\mathcal{F}(\rho_A)\leq\sigma_B$ for all states $\rho_A$. We find that
		\begin{align}
			&\Tr[\Omega_{AB}\mathcal{F}_{A'\to B}(\overline{\phi}_{AA'})]\nonumber\\
			&\quad=\Tr\!\left[(\overline{\rho}_{A}^{\t})^{-\frac{1}{2}}\left(\frac{1}{|\mathcal{M}|}\sum_{m\in\mathcal{M}}(\rho_A^m)^{\t}\otimes\Lambda_B^m\right)(\overline{\rho}_A^{\t})^{-\frac{1}{2}}\mathcal{F}_{A'\to B}(\overline{\phi}_{AA'})\right]\\
			&\quad=\frac{1}{|\mathcal{M}|}\sum_{m\in\mathcal{M}}\Tr\!\left[\left((\rho_A^m)^{\t}\otimes\Lambda_B^m\right)\Gamma^{\mathcal{F}}_{AB}\right]\\
			&\quad=\frac{1}{|\mathcal{M}|}\sum_{m\in\mathcal{M}}\Tr[\Lambda_B^m\mathcal{F}_{A\to B}(\rho_A^m)]\\
			&\quad\leq \frac{1}{|\mathcal{M}|}\sum_{m\in\mathcal{M}}\Tr[\Lambda_B^m\sigma_B]\\
			&\quad\leq\frac{1}{|\mathcal{M}|},
		\end{align}
		where we used \eqref{eq-cc_upsilon_meta_conv_pf_1} to obtain the second equality and we used the fact that $\mathcal{F}(\rho_A)\leq\sigma_B$ for every input state $\rho_A$ to obtain the second-to-last inequality.
		
		Now, by optimizing the quantity $\Tr[\Omega_{AB}\mathcal{F}_{A'\to B}(\overline{\phi}_{AA'})]$ over all measurement operators, subject to the constraint $\Tr[\Omega_{AB}\mathcal{N}_{A'\to B}(\overline{\phi}_{AA'})]\geq 1-\varepsilon$, we get that 
		\begin{align}
			\log_2|\mathcal{M}|&\leq D_H^{\varepsilon}(\mathcal{N}_{A'\to B}(\overline{\phi}_{AA'})\Vert\mathcal{F}_{A'\to B}(\overline{\phi}_{AA'})).
		\end{align}
		Since this holds for every $\mathcal{F}\in\mathfrak{F}$, we have that
		\begin{equation}
			\log_2|\mathcal{M}|\leq \inf_{\mathcal{F}\in\mathfrak{F}}D_H^{\varepsilon}(\mathcal{N}_{A'\to B}(\overline{\phi}_{AA'})\Vert\mathcal{F}_{A'\to B}(\overline{\phi}_{AA'})).
		\end{equation}
		Finally, optimizing over all pure states $\overline{\phi}_{AA'}$, we conclude that
		\begin{equation}
			\log_2|\mathcal{M}|\leq\sup_{\psi_{RA}}\inf_{\mathcal{F}\in\mathfrak{F}}D_H^{\varepsilon}(\mathcal{N}_{A\to B}(\psi_{RA})\Vert\mathcal{F}_{A\to B}(\psi_{RA}))=\Upsilon_H^{\varepsilon}(\mathcal{N}),
		\end{equation}
		as required.
	\end{Proof}

	As an immediate consequence of Propositions~\ref{prop-cc_upsilon_meta_converse}, \ref{prop-hypo_to_rel_ent}, and \ref{prop:sandwich-to-htre}, we have the following two bounds:
	
	\begin{proposition}{prop-cc_upsilon_upper_bounds}
		Let $\mathcal{N}$ be a quantum channel, let $\varepsilon\in[0,1)$, and let $\alpha>1$. For every $(|\mathcal{M}|,\varepsilon)$ classical communication protocol over $\mathcal{N}$, the following bounds hold
		\begin{align}
			\log_2|\mathcal{M}|&\leq\frac{1}{1-\varepsilon}\left(\Upsilon(\mathcal{N})+h_2(\varepsilon)\right),\\
			\log_2|\mathcal{M}|&\leq\widetilde{\Upsilon}_\alpha(\mathcal{N})+\frac{\alpha}{\alpha-1}\log_2\!\left(\frac{1}{1-\varepsilon}\right).
		\end{align}
	\end{proposition}
	
	In the asymptotic setting, the bounds above become the following:
	\begin{align}
		\frac{1}{n}\log_2|\mathcal{M}|&\leq\frac{1}{1-\varepsilon}\left(\frac{1}{n}\Upsilon(\mathcal{N}^{\otimes n})+\frac{1}{n}h_2(\varepsilon)\right),\\
		\frac{1}{n}\log_2|\mathcal{M}|&\leq\frac{1}{n}\widetilde{\Upsilon}_\alpha(\mathcal{N}^{\otimes n})+\frac{\alpha}{n(\alpha-1)}\log_2\!\left(\frac{1}{1-\varepsilon}\right)\quad\forall~\alpha>1.\label{eq-cc_upsilon_str_conv_bound}
	\end{align}
	These upper bounds hold for every $(n,|\mathcal{M}|,\varepsilon)$ classical communication protocol over a quantum channel $\mathcal{N}$, where $n\in\mathbb{N}$ and $\varepsilon\in[0,1)$.
	
	Now, as with the Holevo information and the sandwiched R\'{e}nyi Holevo information, we are faced with the additivity of the $\Upsilon$-information  and the sandwiched R\'{e}nyi $\Upsilon$-information. Our primary focus is on the latter, since we would like to make a statement about the strong converse for channels more general than entanglement-breaking channels. It turns out that the sandwiched R\'{e}nyi $\Upsilon$-information is additive for irreducibly-covariant channels.
	
	Recall from Definition~\ref{def-group_cov_chan} that a channel $\mathcal{N}_{A\to B}$ is covariant with respect to a group $G$ if there exist projective unitary representations $\{U_A^g\}_{g\in G}$ and $\{V_B^g\}_{g\in G}$ such that
	\begin{equation}
		\mathcal{N}(U_A^g\rho_A (U_A^g)^\dagger)=V_B^g\mathcal{N}(\rho_A)(V_B^g)^\dagger
	\end{equation}
	for all states $\rho_A$ and all $g\in G$. The channel $\mathcal{N}$ is called irreducibly covariant if the representation $\{U_A^g\}_{g\in G}$ acting on the input space of the channel is irreducible, which means that it satisfies
	\begin{equation}
		\frac{1}{|G|}\sum_{g\in G}U_A^g\rho_A(U_A^g)^\dagger=\frac{\mathbbm{1}}{d_A}
	\end{equation}
	for every state $\rho_A$.
	
	\begin{proposition*}{Generalized $\boldsymbol{\Upsilon}$-Information for Irreducibly-Covariant Channels}{prop-gen_upsilon_inf_cov}
		Let $\mathcal{N}_{A\to B}$ be an irreducibly-covariant quantum channel. Then the generalized $\Upsilon$-information of $\mathcal{N}$ can be calculated using the maximally entangled state $\Phi_{RA}$, i.e.,
		\begin{equation}\label{eq-cc_gen_up_inf_cov}
			\begin{aligned}
			\boldsymbol{\Upsilon}(\mathcal{N})&=\inf_{\mathcal{F}\in\mathfrak{F}}\boldsymbol{D}(\mathcal{N}_{A\to B}(\Phi_{RA})\Vert\mathcal{F}_{A\to B}(\Phi_{RA}))\\
			&=\inf_{\mathcal{F}\in\mathfrak{F}}\boldsymbol{D}(\rho_{RB}^{\mathcal{N}}\Vert\rho_{RB}^{\mathcal{F}}).
			\end{aligned}
		\end{equation}
	\end{proposition*}
	
	\begin{Proof}
		By simply restricting the optimization over states $\psi_{RA}$ in the definition of the generalized $\Upsilon$-information to the maximally entangled state $\Phi_{RA}$, we obtain
		\begin{align}
			\boldsymbol{\Upsilon}(\mathcal{N})&=\sup_{\psi_{RA}}\inf_{\mathcal{F}\in\mathfrak{F}}\boldsymbol{D}(\mathcal{N}_{A\to B}(\psi_{RA})\Vert\mathcal{F}_{A\to B}(\psi_{RA}))\\
			&\geq \inf_{\mathcal{F}\in\mathfrak{F}}\boldsymbol{D}(\mathcal{N}_{A\to B}(\Phi_{RA})\Vert\mathcal{F}_{A\to B}(\Phi_{RA})).
		\end{align}
		To prove the reverse inequality, let us recall Proposition~\ref{prop-gen_div_group_cov}, specifically its proof. Let $G$ be the group with respect to which $\mathcal{N}$ is irreducibly covariant, let $\{U_A^g\}_{g\in G}$ be the irreducible representation of $G$ acting on the input space of $\mathcal{N}_{A\to B}$, and let $\{V_B^g\}_{g\in G}$ be the representation of $G$ acting on the output space of $\mathcal{N}$. Since the maps $\mathcal{F}_{A\to B}$ in $\mathfrak{F}$, in particular the map achieving the infimum in the definition of $\boldsymbol{\Upsilon}(\mathcal{N})$, need not be irreducibly covariant, we cannot use Proposition~\ref{prop-gen_div_group_cov} directly. Instead, we consider \eqref{eq-gen_div_group_cov_pf2} in its proof. For every $\mathcal{F}_{A\to B}\in\mathfrak{F}$, by using \eqref{eq-gen_div_chan_cov_pf_4}, the inequality in \eqref{eq-gen_div_group_cov_pf2} becomes
		\begin{align}
			&\boldsymbol{D}(\mathcal{N}_{A\to B}(\Phi_{RA})\Vert\mathcal{F}_{A\to B}(\Phi_{RA}))\nonumber\\
			&\quad\geq\boldsymbol{D}\!\left(\frac{1}{|G|}\sum_{g\in G}\ket{g}\!\bra{g}_{R'}\otimes\left((\mathcal{V}_B^g)^\dagger\circ\mathcal{N}_{A\to B}\circ \mathcal{U}_A^g\right)(\psi_{RA})\right.\Bigg\Vert\\
			&\qquad\qquad\quad \left.\frac{1}{|G|}\sum_{g\in G}\ket{g}\!\bra{g}_{R'}\otimes\left((\mathcal{V}_B^g)^\dagger\circ\mathcal{F}_{A\to B}\circ \mathcal{U}_A^g\right)(\psi_{RA})\right)
		\end{align}
		for every pure state $\psi_{RA}$, with the dimension of $R$ equal to the dimension of $A$. Using the data-processing inequality for the generalized divergence with respect to the partial trace $\Tr_{R}$, and using the fact that $\mathcal{N}$ is covariant, we find that
		\begin{align}
			&\boldsymbol{D}(\mathcal{N}_{A\to B}(\Phi_{RA})\Vert\mathcal{F}_{A\to B}(\Phi_{RA}))\nonumber\\
			&\quad \geq \boldsymbol{D}\!\left(\mathcal{N}_{A\to B}(\psi_{RA})\Bigg\Vert\frac{1}{|G|}\sum_{g\in G}\left((\mathcal{V}_B^g)^\dagger\circ\mathcal{F}_{A\to B}\circ\mathcal{U}_A^g\right)(\psi_{RA})\right).
		\end{align}
		Now, observe that the map $\mathcal{F}_{A\to B}'\coloneqq\frac{1}{|G|}\sum_{g\in G}(\mathcal{V}_B^g)^\dagger\circ\mathcal{F}_{A\to B}\circ\mathcal{U}_A^g$ is in the set $\mathfrak{F}$ since $\mathcal{F}$ is. Therefore, optimizing over all $\mathcal{F}'\in\mathfrak{F}$, we find that
		\begin{multline}
			\boldsymbol{D}(\mathcal{N}_{A\to B}(\Phi_{RA})\Vert\mathcal{F}_{A\to B}(\Phi_{RA}))\\
			\geq\inf_{\mathcal{F}'\in\mathfrak{F}}\boldsymbol{D}(\mathcal{N}_{A\to B}(\psi_{RA})\Vert\mathcal{F}_{A\to B}'(\psi_{RA})).
		\end{multline}
		Since the map $\mathcal{F}\in\mathfrak{F}$ is arbitrary, we conclude that
		\begin{multline}
			\inf_{\mathcal{F}\in\mathfrak{F}}\boldsymbol{D}(\mathcal{N}_{A\to B}(\Phi_{RA})\Vert\mathcal{F}_{A\to B}(\Phi_{RA}))\\
			\geq \inf_{\mathcal{F}'\in\mathfrak{F}}\boldsymbol{D}(\mathcal{N}_{A\to B}(\psi_{RA})\Vert\mathcal{F}_{A\to B}'(\psi_{RA})).
		\end{multline}
		Finally, since the state $\psi_{RA}$ is arbitrary, we obtain
		\begin{align}
			&\inf_{\mathcal{F}\in\mathfrak{F}}\boldsymbol{D}(\mathcal{N}_{A\to B}(\Phi_{RA})\Vert\mathcal{F}_{A\to B}(\Phi_{RA}))\nonumber\\
			&\qquad\qquad\qquad\geq \sup_{\psi_{RA}}\inf_{\mathcal{F}'\in\mathfrak{F}}\boldsymbol{D}(\mathcal{N}_{A\to B}(\psi_{RA})\Vert\mathcal{F}_{A\to B}'(\psi_{RA}))\\
			&\qquad\qquad\qquad =\boldsymbol{\Upsilon}(\mathcal{N}).
		\end{align}
		We thus conclude \eqref{eq-cc_gen_up_inf_cov}, as required.
	\end{Proof}
	
	By Proposition~\ref{prop-gen_upsilon_inf_cov}, we have that for irreducibly-covariant channels the sandwiched R\'{e}nyi $\Upsilon$-information can be calculated without an optimization over all pure states $\psi_{RA}$---we simply set $\psi_{RA}=\Phi_{RA}$. Using this fact, we obtain the following:
	
	\begin{proposition*}{Subadditivity of Sandwiched R\'{e}nyi $\Upsilon$-Information for Irreducibly-Covariant Channels}{prop-sand_ren_up_inf_cov_additive}
		Let $\mathcal{N}_{A\to B}$ be an irreducibly covariant quantum channel. Then, for all $\alpha\in [1/2,1)\cup(1,\infty)$, the sandwiched R\'{e}nyi $\Upsilon$-information $\widetilde{\Upsilon}_\alpha(\mathcal{N})$ is subadditive, i.e.,
		\begin{equation}
			\widetilde{\Upsilon}_{\alpha}(\mathcal{N}^{\otimes n})\leq n\widetilde{\Upsilon}_{\alpha}(\mathcal{N})
		\end{equation}
		for all $n\geq 1$.
	\end{proposition*}
	
	
	\begin{Proof}
		Since $\mathcal{N}$ is irreducibly covariant, we use Proposition~\ref{prop-gen_upsilon_inf_cov} to conclude that
		\begin{equation}
			\widetilde{\Upsilon}_\alpha(\mathcal{N}^{\otimes n})=\inf_{\mathcal{F}\in\mathfrak{F}_n}\widetilde{D}_\alpha(\mathcal{N}_{A\to B}^{\otimes n}(\Phi_{R^nA^n})\Vert\mathcal{F}_{A^n\to B^n}(\Phi_{R^nA^n})),
		\end{equation}
		where $\mathfrak{F}_n$ is the set of completely positive maps in \eqref{eq-cc_gen_useless_maps} acting on the space of the system $A^n$. Now, the maximally entangled state $\Phi_{R^nA^n}$ on $n$ identical copies $R_1\dotsb R_n$ and $A_1\dotsb A_n$ of the systems $R$ and $A$ splits into a tensor product in the following way:
		\begin{equation}
			\Phi_{R^nA^n}=\Phi_{R_1A_1}\otimes\dotsb\otimes\Phi_{R_nA_n}.
		\end{equation}
		Furthermore, if we restrict the optimization over maps $\mathcal{F}_{A^n\to B^n}\in\mathfrak{F}_n$ to a tensor product of identical maps $\mathcal{G}$ in the  set $\mathfrak{F}$ such that
		\begin{equation}
			\mathcal{F}_{A^n\to B^n}=\mathcal{G}_{A_1\to B_1}\otimes\dotsb\otimes\mathcal{G}_{A_n\to B_n},
		\end{equation}
		then, we arrive at the following bound:
		\begin{align}
			&\widetilde{\Upsilon}_{\alpha}(\mathcal{N}^{\otimes n})\nonumber\\
			&\quad \leq \inf_{\mathcal{G}\in\mathfrak{F}}\widetilde{D}_{\alpha}(\mathcal{N}_{A_1\to B_1}(\Phi_{R_1A_1})\otimes\dotsb\otimes\mathcal{N}_{A_n\to B_n}(\Phi_{R_nA_n})\Vert\nonumber\\
			&\qquad\qquad\qquad\qquad\qquad\mathcal{G}_{A_1\to B_1}(\Phi_{R_1A_1})\otimes\dotsb\otimes\mathcal{G}_{A_n\to B_n}(\Phi_{R_nA_n}))\\
			&\quad=\inf_{\mathcal{G}\in\mathfrak{F}_1}n\widetilde{D}_\alpha(\mathcal{N}_{A\to B}(\Phi_{RA})\Vert\mathcal{G}_{A\to B}(\Phi_{RA}))\\
			&\quad=n\widetilde{\Upsilon}_\alpha(\mathcal{N}),
		\end{align}
		as required, where  the first equality follows from  the additivity of the sandwiched R\'{e}nyi relative entropy for tensor-product states (see \eqref{eq-sand_ren_rel_ent_additive} in Proposition~\ref{prop-sand_rel_ent_properties}).
	\end{Proof}

	With the subadditivity of the sandwiched R\'{e}nyi $\Upsilon$-information for irreducibly covariant channels, we can now state the following strong converse theorem.
	
	\begin{theorem*}{$\Upsilon$-Information Upper Bound on the Strong Converse Classical Capacity of Irreducibly-Covariant Channels}{thm-cc_str_conv_upsilon_cov}
The $\Upsilon$-information $\Upsilon(\mathcal{N})$ of an irreducibly-covariant quantum channel~$\mathcal{N}$ is a strong converse rate for classical communication over $\mathcal{N}$; i.e.,
		\begin{equation}
		\widetilde{C}(\mathcal{N}) \leq \Upsilon(\mathcal{N}).
		\end{equation}
	\end{theorem*}
	
	\begin{Proof}
		Fix $\varepsilon\in[0,1)$ and $\delta>0$. Let $\delta_1,\delta_2>0$ be such that
		\begin{equation}\label{eq-cc_str_conv_upsilon_cov_pf1}
			\delta>\delta_1+\delta_2\eqqcolon\delta'.
		\end{equation}
		Set $\alpha\in(1,\infty)$ such that
		\begin{equation}\label{eq-cc_str_conv_upsilon_cov_pf2}
			\delta_1\geq\widetilde{\Upsilon}_{\alpha}(\mathcal{N})-\Upsilon(\mathcal{N}),
		\end{equation}
		which is possible because $\widetilde{\Upsilon}_{\alpha}(\mathcal{N})$ is monotonically increasing with $\alpha$ (this follows from Proposition~\ref{prop-sand_rel_ent_properties}) and $\lim_{\alpha\to 1^+}\widetilde{\Upsilon}_{\alpha}(\mathcal{N})=\Upsilon(\mathcal{N})$ (see Appendix~\ref{app-sand_ren_upsilon_inf_chan_limit} for a proof). With this value of $\alpha$, take $n$ large enough so that
		\begin{equation}\label{eq-cc_str_conv_upsilon_cov_pf3}
			\delta_2\geq\frac{\alpha}{n(\alpha-1)}\log_2\!\left(\frac{1}{1-\varepsilon}\right).
		\end{equation}
		
		Now, with the values of $n$ and $\varepsilon$ chosen as above, every $(n,|\mathcal{M}|,\varepsilon)$ classical communication protocol satisfies \eqref{eq-cc_upsilon_str_conv_bound}. In particular, using the subadditivity of the sandwiched R\'{e}nyi $\Upsilon$-information for all $\alpha>1$, we can write \eqref{eq-cc_upsilon_str_conv_bound} as
		\begin{equation}
			\frac{1}{n}\log_2|\mathcal{M}|\leq\widetilde{\Upsilon}_{\alpha}(\mathcal{N})+\frac{\alpha}{n(\alpha-1)}\log_2\!\left(\frac{1}{1-\varepsilon}\right).
		\end{equation}
		Rearranging the right-hand side of this inequality, and using the assumptions in \eqref{eq-cc_str_conv_upsilon_cov_pf1}--\eqref{eq-cc_str_conv_upsilon_cov_pf3}, we obtain
		\begin{align}
			\frac{1}{n}\log_2|\mathcal{M}|&\leq\Upsilon(\mathcal{N})+\widetilde{\Upsilon}_{\alpha}(\mathcal{N})-\Upsilon(\mathcal{N})+\frac{\alpha}{n(\alpha-1)}\log_2\!\left(\frac{1}{1-\varepsilon}\right)\\
			&\leq \Upsilon(\mathcal{N})+\delta_1+\delta_2\\
			&=\Upsilon(\mathcal{N})+\delta'\\
			&<\Upsilon(\mathcal{N})+\delta.
		\end{align}
		So we have that $\Upsilon(\mathcal{N})+\delta>\frac{1}{n}\log_2|\mathcal{M}|$ for all $(n,|\mathcal{M}|,\varepsilon)$ classical communication protocols with $n$ sufficiently large. Due to this strict inequality, it follows that there cannot exist an $(n,2^{n(\Upsilon(\mathcal{N})+\delta)},\varepsilon)$ classical communication protocol for all sufficiently large $n$ such that \eqref{eq-cc_str_conv_upsilon_cov_pf3} holds, for if it did there would exist some message set $\mathcal{M}$ such that $\frac{1}{n}\log_2|\mathcal{M}|=\Upsilon(\mathcal{N})+\delta$, which we have just seen is not possible. Since $\varepsilon$ and $\delta$ are arbitrary, we have that for all $\varepsilon\in[0,1)$, $\delta>0$, and sufficiently large $n$, there does not exist an $(n,2^{n(\Upsilon(\mathcal{N})+\delta)},\varepsilon)$ classical communication protocol. This means that $\Upsilon(\mathcal{N})$ is a strong converse rate, which completes the proof.
	\end{Proof}
	
	Theorem~\ref{thm-cc_str_conv_upsilon_cov} thus gives us an upper bound on the strong converse classical capacity $\widetilde{C}(\mathcal{N})$ of any irreducibly-covariant channel $\mathcal{N}$, namely,
	\begin{equation}
		\widetilde{C}(\mathcal{N})\leq \Upsilon(\mathcal{N}),
	\end{equation}
	which in turn implies, via \eqref{eq-cc_str_conv_rate_lower_bound}, that
	\begin{equation}
		C(\mathcal{N})\leq \Upsilon(\mathcal{N})
	\end{equation}
	for every irreducibly-covariant channel $\mathcal{N}$. Recall that the Holevo information $\chi(\mathcal{N})$ is an achievable rate for classical communication over any quantum channel, which implies that $C(\mathcal{N})\geq \chi(\mathcal{N})$. (This in fact gives another way for concluding that $\Upsilon(\mathcal{N})\geq \chi(\mathcal{N})$.)
	
	It turns out that the $\Upsilon$-information $\Upsilon(\mathcal{N})$ is \textit{equal} to the Holevo information in the case of the erasure channel $\mathcal{E}_p^{(d)}$. Recall from Section~\ref{subsec-eacc_erasure_channel} that the erasure channel is irreducibly covariant. This fact, along with other reasoning, allows us to conclude that
	\begin{equation}
		C(\mathcal{E}_p^{(d)})=\widetilde{C}(\mathcal{E}_p^{(d)})=\chi(\mathcal{E}_p^{(d)})=(1-p)\log_2 d
	\end{equation}
	for all dimensions $d\geq 2$ and all $p\in[0,1]$. We provide a proof of this chain of equalities in Section~\ref{subsec-cc_erasure_channel} below.

\subsubsection{SDP Upper Bound}

	While the $\Upsilon$-information gives us an upper bound on the strong converse classical capacity of any irreducibly-covariant channel, computing it is relatively challenging due to the minimization over the set $\mathfrak{F}$. In this section, we define a subset of $\mathfrak{F}$, denoted by $\mathfrak{F}_\beta$, that allows us to obtain a quantity that can be computed using a semi-definite program (SDP). Furthermore, this quantity turns out to be additive for \textit{all} channels, which means that it is an upper bound on the strong converse classical capacity for all channels.
	
	The set $\mathfrak{F}_\beta$ is defined as the following set of completely positive maps:
	\begin{equation}
		\mathfrak{F}_\beta\coloneqq\{\mathcal{F}\text{ completely positive}:\beta(\mathcal{F})\leq 1\},
	\end{equation}
	where $\beta(\mathcal{F})$ is defined as the solution to the following optimization problem:
	\begin{equation}\label{eq-cc_beta_SDP}
		\beta(\mathcal{F})\coloneqq \left\{\begin{array}{l l} \text{infimum} & \Tr[S_B] \\
		\text{subject to} & -R_{AB}\leq (\Gamma^{\mathcal{F}}_{AB})^{\t_B}\leq R_{AB},\\[0.2cm]
		& -\mathbbm{1}_A\otimes S_B\leq R_{AB}^{\t_B}\leq \mathbbm{1}_A\otimes S_B
		\end{array}\right.
	\end{equation}
	Note that the optimization occurs over the operators $S_B$ and $R_{AB}$, and that the optimization problem as a whole is an SDP. Indeed, recalling the general form of an SDP from Section~\ref{sec-SDPs}, we can write it as
	\begin{equation}\label{eq-cc_beta_SDP_alt}
		\beta(\mathcal{F})=\left\{\begin{array}{l l}\text{infimum} & \Tr[CX]\\
		\text{subject to} & \Phi(X)\geq D,\\
		& X\geq 0, \end{array}\right.
	\end{equation}
	where
	\begin{align}
		X&=\begin{pmatrix} S_B & 0\\0 & R_{AB} \end{pmatrix},\quad C=\begin{pmatrix} \mathbbm{1}_B & 0 \\ 0 & 0_{AB}\end{pmatrix},\\
		\Phi(X)&=\begin{pmatrix} R_{AB} & 0 & 0 & 0 \\ 0 & R_{AB} & 0 & 0 \\0 & 0 & \mathbbm{1}_R\otimes S_B-R_{AB}^{\t_B} & 0 \\ 0 & 0 & 0 & \mathbbm{1}_R\otimes S_B+R_{AB}^{\t_B}\end{pmatrix},\\
		D&=\begin{pmatrix} (\Gamma^{\mathcal{F}}_{AB})^{\t_B} & 0 & 0 & 0 \\ 0 & -(\Gamma^{\mathcal{F}}_{AB})^{\t_B} & 0 & 0 \\ 0 & 0 & 0 & 0 \\ 0 & 0 & 0 & 0\end{pmatrix}.
	\end{align}
	Note that the constraints in \eqref{eq-cc_beta_SDP} imply that $S_B$ and $R_{AB}$ are positive semi-definite, since the constraint $-R_{AB}\leq (\Gamma^{\mathcal{F}}_{AB})^{\t_B}\leq R_{AB}$ implies that
	\begin{align}
		R_{AB}-(\Gamma^{\mathcal{F}}_{AB})^{\t_B}&\geq 0,\\
		R_{AB}+(\Gamma^{\mathcal{F}}_{AB})^{\t_B}&\geq 0.
	\end{align}
	Adding the two inequalities leads to $R_{AB}\geq 0$. Similarly, the constraint $-\mathbbm{1}_A\otimes S_B\leq R_{AB}^{\t_B}\leq \mathbbm{1}_A\otimes S_B$ implies that
$S_B\geq 0$.
	
	It is straightforward to show that $\mathfrak{F}_\beta$ is a subset of $\mathfrak{F}$, i.e., $\mathfrak{F}_\beta\subseteq \mathfrak{F}$. Indeed, suppose that for a given completely positive map $\mathcal{F}\in\mathfrak{F}_\beta$, the quantity $\beta(\mathcal{F})$ in \eqref{eq-cc_beta_SDP} is achieved by the operators $(R_{AB}^*,S_B^*)$. Note that since $\mathcal{F}\in\mathfrak{F}_\beta$, by definition we have that $\Tr[S_B^*]\leq 1$, and we also have that $(\Gamma^{\mathcal{F}}_{AB})^{\t_B}\leq R_{AB}^*$ and $(R_{AB}^*)^{\t_B}\leq\mathbbm{1}_A\otimes S_B^*$. Letting $\sigma_B\equiv S_B^*$, for every state $\rho_A$ we find that
	\begin{align}
		\mathcal{F}_{A\to B}(\rho_A)&=\Tr_A\!\left[(\rho_A^{\t}\otimes\mathbbm{1}_B)\Gamma^{\mathcal{F}}_{AB}\right]\\
		&=\left(\Tr_A\!\left[(\rho_A^{\t}\otimes\mathbbm{1}_B)(\Gamma^{\mathcal{F}}_{AB})^{\t_B}\right]\right)^{\t}\\
		&=\left(\Tr_A\!\left[(\sqrt{\smash[b]{\rho_A^{\t}}}\otimes\mathbbm{1}_B)(\Gamma^{\mathcal{F}}_{AB})^{\t_B}(\sqrt{\smash[b]{\rho_A^{\t}}}\otimes\mathbbm{1}_B)\right]\right)^{\t}\\
		&\leq \left(\Tr_A\!\left[(\sqrt{\smash[b]{\rho_A^{\t}}}\otimes\mathbbm{1}_B)R_{AB}^*(\sqrt{\smash[b]{\rho_A^{\t}}}\otimes\mathbbm{1}_B)\right]\right)^{\t}\\
		&=\Tr_A\!\left[(\sqrt{\smash[b]{\rho_A^{\t}}}\otimes\mathbbm{1}_B)(R_{AB}^*)^{\t_B}(\sqrt{\smash[b]{\rho_A^{\t}}}\otimes\mathbbm{1}_B)\right]\\
		&\leq \Tr_A\!\left[(\rho_A^{\t}\otimes\mathbbm{1}_B)(\mathbbm{1}_A\otimes\sigma_B)\right]\\
		&=\sigma_B,
	\end{align}
	where to obtain the first inequality we used $(\Gamma^{\mathcal{F}}_{AB})^{\t_B}\leq R_{AB}^*$ and to obtain the second inequality we used $(R_{AB}^*)^{\t_B}\leq\mathbbm{1}_A\otimes\sigma_B$. Therefore, $\mathcal{F}_{A\to B}(\rho_A)\leq\sigma_B$ for all $\rho_A$, which means that $\mathcal{F}_{A\to B}\in\mathfrak{F}$. Since $\mathcal{F}\in\mathfrak{F}_\beta$ is arbitrary, we conclude that $\mathfrak{F}_\beta\subset\mathfrak{F}$.
	 
	By replacing the set $\mathfrak{F}$ in the definition of the generalized $\Upsilon$-information of a channel $\mathcal{N}$ with the set $\mathfrak{F}_\beta$, we obtain the following quantity:
	\begin{equation}
		\boldsymbol{\Upsilon}^\beta(\mathcal{N})\coloneqq\sup_{\psi_{RA}}\inf_{\mathcal{F}\in\mathfrak{F}_\beta}\boldsymbol{D}(\mathcal{N}_{A\to B}(\psi_{RA})\Vert\mathcal{F}_{A\to B}(\psi_{RA})),
	\end{equation}
	which we call the \textit{$\Upsilon^\beta$-information of $\mathcal{N}$}. When we take the generalized divergence $\boldsymbol{D}$ to be the quantum relative entropy, the hypothesis testing relative entropy, and the sandwiched R\'{e}nyi relative entropy, we have
	\begin{align}
	\Upsilon^\beta(\mathcal{N})&\coloneqq\sup_{\psi_{RA}}\inf_{\mathcal{F}\in\mathfrak{F}_\beta}D(\mathcal{N}_{A\to B}(\psi_{RA})\Vert\mathcal{F}_{A\to B}(\psi_{RA})),\\
		\Upsilon_H^{\beta,\varepsilon}(\mathcal{N})&\coloneqq\sup_{\psi_{RA}}\inf_{\mathcal{F}\in\mathfrak{F}_\beta}D_H^{\varepsilon}(\mathcal{N}_{A\to B}(\psi_{RA})\Vert\mathcal{F}_{A\to B}(\psi_{RA})),\\
		\widetilde{\Upsilon}_\alpha^\beta(\mathcal{N})&\coloneqq \sup_{\psi_{RA}}\inf_{\mathcal{F}\in\mathfrak{F}_\beta}\widetilde{D}_\alpha(\mathcal{N}_{A\to B}(\psi_{RA})\Vert\mathcal{F}_{A\to B}(\psi_{RA})).
	\end{align}
	 
	Since the set $\mathfrak{F}_\beta$ is a subset of $\mathfrak{F}$, minimizing over $\mathfrak{F}_\beta$ can never lead to a smaller value compared to minimizing over $\mathfrak{F}$, which means that
	\begin{equation}
		\boldsymbol{\Upsilon}(\mathcal{N})\leq \boldsymbol{\Upsilon}^\beta(\mathcal{N}).
	\end{equation}
	Therefore, using the $\Upsilon^\beta$-information, the bound in Proposition~\ref{prop-cc_upsilon_meta_converse} on every $(|\mathcal{M}|,\varepsilon)$ classical communication protocol thus becomes
	\begin{equation}\label{prop-cc_upsilon_meta_converse_beta}
		\log_2|\mathcal{M}|\leq \Upsilon_H^{\beta,\varepsilon}(\mathcal{N}).
	\end{equation}
	This bound is looser than the one in Proposition~\ref{prop-cc_upsilon_meta_converse}, but it has the advantange that it can be computed using an SDP. This is due to the fact that the hypothesis testing relative entropy can itself be computed via an SDP. 
	 
	Although we get an efficiently computable upper bound in the one-shot setting via the $\Upsilon^\beta$-information, in the asymptotic setting this bound is not known to be additive, making its evaluation computationally prohibitive as the number $n$ of channel uses increases. Instead, for the purpose of obtaining an efficiently computable upper bound in the asymptotic setting, we define the following quantity for every quantum channel $\mathcal{N}$:
	\begin{equation}\label{eq-cc_C_beta}
		C_\beta(\mathcal{N})=\log_2\beta(\mathcal{N}),
	\end{equation}
	Since $\beta(\mathcal{N})$ can be computed using an SDP (in particular, via the optimization problem in \eqref{eq-cc_beta_SDP}), we have that $C_\beta(\mathcal{N})$ can also be computed using an SDP.
	
	A useful fact about the quantity $C_\beta(\mathcal{N})$ is the fact that it is additive, i.e.,
	\begin{equation}
		C_\beta(\mathcal{N}_1\otimes\mathcal{N}_2)=C_\beta(\mathcal{N}_1)+C_\beta(\mathcal{N}_2)
	\end{equation}
	for all channels $\mathcal{N}_1$ and $\mathcal{N}_2$, as proved in Appendix~\ref{app-cc-C_beta_additive_pf} by employing semi-definite programming duality. We use this fact to prove that $C_\beta(\mathcal{N})$ is a strong converse rate for classical communication over a channel $\mathcal{N}$ in Theorem~\ref{thm-cc-str_conv_upper_bnd_SDP} below. However, first we establish the following proposition:

	\begin{proposition}{prop-upsilon_vs_C_beta}
		For a quantum channel $\mathcal{N}$, the following inequalities hold for all $\alpha>1$:
		\begin{equation}
		\Upsilon(\mathcal{N})\leq C_\beta(\mathcal{N}),\qquad  \widetilde{\Upsilon}_\alpha(\mathcal{N})\leq C_{\beta}(\mathcal{N}).
		\end{equation}
	\end{proposition}
	
	\begin{Proof}
		Let $\mathcal{F}=\frac{1}{\beta(\mathcal{N})}\mathcal{N}$. Then, $\beta(\mathcal{F})=\frac{1}{\beta(\mathcal{N})}\beta(\mathcal{N})=1$, which means that $\mathcal{F}\in\mathfrak{F}_\beta$. Then, since $\mathfrak{F}_\beta\subseteq \mathfrak{F}$, we can  choose $\mathcal{F}$ as above when performing the infimum in the definition of $\Upsilon(\mathcal{N})$. This leads to
		\begin{align}
			\Upsilon(\mathcal{N})&=\sup_{\psi_{RA}}\inf_{\mathcal{F}\in\mathfrak{F}}D(\mathcal{N}_{A\to B}(\psi_{RA})\Vert\mathcal{F}_{A\to B}(\psi_{RA}))\\
			&\leq \sup_{\psi_{RA}}D\!\left(\mathcal{N}_{A\to B}(\psi_{RA})\Bigg\Vert\frac{1}{\beta(\mathcal{N})}\mathcal{N}_{A\to B}(\psi_{RA})\right)\\
			&=\sup_{\psi_{RA}}D(\mathcal{N}_{A\to B}(\psi_{RA})\Vert\mathcal{N}_{A\to B}(\psi_{RA}))+\log_2\beta(\mathcal{N})\\
			&=C_\beta(\mathcal{N}),
		\end{align}
		as required, where  the second equality follows because $D(\rho\Vert\frac{1}{x}\sigma)=D(\rho\Vert\sigma)+\log_2x$ for every state $\rho$, positive semi-definite operator $\sigma$, and $x>0$ (see \eqref{eq-rel_ent_scalar_mult}). The last equality follows because $D(\rho\Vert\rho)=0$.
		
		Similarly, using the fact that $\widetilde{D}_\alpha(\rho\Vert\frac{1}{x}\sigma)=\widetilde{D}_\alpha(\rho\Vert\sigma)+\log_2x$, and using the same choice for the map $\mathcal{F}$ as above, we find that $\widetilde{\Upsilon}_\alpha(\mathcal{N})\leq C_\beta(\mathcal{N})$.
	\end{Proof}

	The additivity of $C_\beta$, along with Proposition~\ref{prop-upsilon_vs_C_beta}, leads us to the following:
	
	\begin{theorem*}{SDP Upper Bound on  Strong Converse Classical Capacity}{thm-cc-str_conv_upper_bnd_SDP}
		For a quantum channel $\mathcal{N}$, the quantity $C_\beta(\mathcal{N})$ is a strong converse rate for classical communication over $\mathcal{N}$.
	\end{theorem*}
	
	\begin{Proof}
		We start by observing that, using Proposition~\ref{prop-upsilon_vs_C_beta}, the inequality in \eqref{eq-cc_upsilon_str_conv_bound} can be written as
		\begin{equation}
			\frac{1}{n}\log_2|\mathcal{M}|\leq \frac{1}{n}C_\beta(\mathcal{N}^{\otimes n})+\frac{\alpha}{n(\alpha-1)}\log_2\!\left(\frac{1}{1-\varepsilon}\right)
		\end{equation}
		for all $\alpha>1$. This inequality holds for all $(n,|\mathcal{M}|,\varepsilon)$ classical communication protocols. Using the additivity of $C_\beta$, we find that
		\begin{equation}
			\frac{1}{n}\log_2|\mathcal{M}|\leq C_\beta(\mathcal{N})+\frac{\alpha}{n(\alpha-1)}\log_2\!\left(\frac{1}{1-\varepsilon}\right).
		\end{equation}
		Now, let us fix $\varepsilon\in[0,1)$ and $\delta>0$. Let $\delta'$ be such that $\delta>\delta'$. Since $C_\beta(\mathcal{N})$ does not depend on $\alpha$, let us choose $\alpha$ such that the right-hand side of the above inequality is as small as possible, which occurs as $\alpha\to\infty$. With this choice of $\alpha$, take $n$ large enough so that
		\begin{equation}\label{eq-cc_C_beta_str_conv_pf}
			\delta'\geq \frac{1}{n}\log_2\!\left(\frac{1}{1-\varepsilon}\right).
		\end{equation}
		Then, we obtain
		\begin{align}
			\frac{1}{n}\log_2|\mathcal{M}|&\leq C_\beta(\mathcal{N})+\frac{1}{n}\log_2\!\left(\frac{1}{1-\varepsilon}\right)\label{eq-CC:cbeta-exponential-decay}\\
			&\leq C_\beta(\mathcal{N})+\delta'\\
			&<C_\beta(\mathcal{N})+\delta.
		\end{align}
		So we have that $C_\beta(\mathcal{N})+\delta>\frac{1}{n}\log_2|\mathcal{M}|$ for all $(n,|\mathcal{M}|,\varepsilon)$ classical communication protocols with $n$ sufficiently large. Due to this strict inequality, it follows that there cannot exist an $(n,2^{n(C_\beta(\mathcal{N})+\delta)},\varepsilon)$ classical communication protocol for all sufficiently large $n$ such that \eqref{eq-cc_C_beta_str_conv_pf} holds, for if it did there would exist some message set $\mathcal{M}$ such that $\frac{1}{n}\log_2|\mathcal{M}|=C_\beta(\mathcal{N})+\delta$, which we have just seen is not possible. Since $\varepsilon$ and $\delta$ are arbitrary, we have that for all $\varepsilon\in[0,1)$, $\delta>0$, and sufficiently large $n$, there does not exist an $(n,2^{n(C_\beta(\mathcal{N})+\delta)},\varepsilon)$ classical communication protocol. This means that $C_\beta(\mathcal{N})$ is a strong converse rate, which completes the proof.
	\end{Proof}

	By examining \eqref{eq-CC:cbeta-exponential-decay} in the above proof, we see that the following bound holds for an arbitrary $(n,|\mathcal{M}|,\varepsilon)$ classical communication protocol:
	\begin{equation}
		\frac{1}{n}\log_2|\mathcal{M}| \leq C_\beta(\mathcal{N})+\frac{1}{n}\log_2\!\left(\frac{1}{1-\varepsilon}\right).
	\end{equation}
	If we fix the rate $R = \frac{1}{n}\log_2|\mathcal{M}|$, then this bound can be rewritten as follows:
	\begin{equation}
		1-\varepsilon \leq 2^{-n(R - C_\beta(\mathcal{N}))},
	\end{equation}
	which indicates that communicating at a rate $R > C_\beta(\mathcal{N})$ implies the success probability $1-\varepsilon$ of every sequence of such protocols decays exponentially fast to zero.

\section{Examples}\label{sec-Holevo_inf}

	In this section, we present various examples of channels with known formulas for the Holevo information and/or known results on additivity of the Holevo information.
	
	Let us start by making some observations about the Holevo information $\chi(\mathcal{N})$ of a channel $\mathcal{N}$. First, by expanding the definition of the Holevo information using the expression for the mutual information in terms of the relative entropy, we arrive at the following:
	
	\begin{proposition*}{Alternate Forms for Channel Holevo Information}{prop-Holevo_inf_alt}
		For a channel $\mathcal{N}$, the following equalities hold
		\begin{align}
			\chi(\mathcal{N})&=\sup_{\{(p(x),\psi_A^x)\}_{x}}\left[H(\mathcal{N}(\overline{\rho}_A))-\sum_{x\in\mathcal{X}}p(x)H(\mathcal{N}(\psi_A^x))\right]\label{eq-Holevo_inf_alt},\\
			&=\sup_{\{(p(x),\psi_A^x)\}_x}\left[H(\mathcal{N}(\overline{\rho}_A))-\sum_{x\in\mathcal{X}}p(x)H(\mathcal{N}^c(\psi_A^x))\right]\label{eq-Holevo_inf_alt_2}.
		\end{align}
		where $\{(p(x),\psi_A^x)\}_{x\in\mathcal{X}}$ is an ensemble of pure states, $\overline{\rho}_A\coloneqq \sum_{x\in\mathcal{X}}p(x)\psi_A^x$, and we recall from \eqref{eq-quantum_entropy} and \eqref{eq-quantum_entropy_relent} that $H(\rho)=-\Tr[\rho\log\rho]$ is the quantum entropy of $\rho$.
	\end{proposition*}
	
	\begin{Proof}
		We start by recalling the definition of the Holevo information $\chi(\mathcal{N})$ of $\mathcal{N}$ from \eqref{eq-Hol_inf_chan}:
		\begin{equation}
			\chi(\mathcal{N})\coloneqq \sup_{\rho_{XA}}I(X;B)_{\omega},
		\end{equation}
		where $\omega_{RB}=\mathcal{N}_{A\to B}(\rho_{XA})$, and the supremum is over all classical-quan\-tum states of the form $\rho_{XA}=\sum_{x\in\mathcal{X}}p(x)\ket{x}\!\bra{x}_X\otimes \rho_A^x$, with $\mathcal{X}$ a finite alphabet with associated $|\mathcal{X}|$-dimensional system $X$, $\{\rho_A^x\}_{x\in\mathcal{X}}$ a set of states, and $p:\mathcal{X}\to[0,1]$ a probability distribution on $\mathcal{X}$. Recall from Proposition~\ref{prop-Holevo_inf_pure_states} that to compute the Holevo information it suffices to take ensembles consisting only of pure states.
		
		Defining the classical--quantum state $\rho_{XA}$ as
		\begin{equation}
		\rho_{XA}=\sum_{x\in\mathcal{X}}p(x)\ket{x}\!\bra{x}_X\otimes\psi_{A}^x,
		\end{equation}
		where $\{\psi_{A}^x\}_{x\in \mathcal{X}}$ is a set of pure states,
	and defining  $\rho_{XB}'=\mathcal{N}_{A\to B}(\rho_{XA})$ is another classical--quantum state, it follows from Proposition~\ref{prop:QEI:MI-cq-states} that
		\begin{equation}
			I(X;B)_{\rho'}=H\!\left(\sum_{x\in\mathcal{X}}p(x)\mathcal{N}(\psi_A^x)\right)-\sum_{x\in\mathcal{X}}p(x)H(\mathcal{N}(\psi_A^x))
		\end{equation}
		for all states $\rho_{XA}$. Since optimizing over classical--quantum states is equivalent to optimizing over ensembles $\{(p(x),\psi_A^x)\}_{x\in\mathcal{X}}$, we obtain \eqref{eq-Holevo_inf_alt}.
		
		To prove \eqref{eq-Holevo_inf_alt_2}, let $V_{A\to BE}$ be an isometric extension of $\mathcal{N}$. Then, for every pure state $\psi$ on the channel input system $A$, we can write
		\begin{equation}
			\mathcal{N}(\psi)=\Tr_{E}[V\psi V^\dagger].
		\end{equation}
		Since $V\ket{\psi}$ is a pure state, it follows that $\Tr_{E}[V\psi V^\dagger]$ and $\Tr_{B}[V\psi V^\dagger]$ have the same (non-zero) eigenvalues. The latter state is equal to $\mathcal{N}^c(\psi_{A})$ by definition, which means that
		\begin{equation}
			H(\mathcal{N}(\psi_{A}))=H(\mathcal{N}^c(\psi_{A})).
		\end{equation}
		Therefore, \eqref{eq-Holevo_inf_alt_2} follows.
	\end{Proof}

\subsection{Covariant Channels}

	For irreducibly-covariant channels (see Definition~\ref{def-group_cov_chan}), the Holevo information takes a particularly simple form.

	\begin{theorem*}{Holevo Information of Irreducibly-Covariant Channels}{thm-Holevo_inf_covariant}
		Suppose $\mathcal{N}_{A\to B}$ is a covariant channel with respect to a finite group $G$, with an irreducible representation $\{U_A^g\}_{g\in G}$ of $G$ acting on the input space of the channel and another representation $\{V_B^g\}_{g\in G}$ of $G$ acting on the output space of the channel. Then,
		\begin{equation}
			\chi(\mathcal{N})=H\!\left(\mathcal{N}(\pi_A)\right)-H_{\text{min}}(\mathcal{N}),
		\end{equation}
		where $\pi_A=\frac{\mathbbm{1}_A}{d_A}$ and
		\begin{equation}
			H_{\text{min}}(\mathcal{N})\coloneqq\min_{\rho_A}H(\mathcal{N}(\rho_A))
		\end{equation}
		is called the \textit{minimum output entropy} of $\mathcal{N}$. The minimization is with respect to all input states $\rho_A$ in the domain of $\mathcal{N}$.
	\end{theorem*}
	
	\begin{Proof}
		We have
		\begin{align}
			\chi(\mathcal{N})&=\sup_{\{(p(x),\rho_A^x)\}_{x}}\left[H\!\left(\sum_{x\in\mathcal{X}}p(x)\mathcal{N}(\rho_A^x)\right)-\sum_{x\in\mathcal{X}}p(x)H(\mathcal{N}(\rho_A^x))\right]\\
			&\leq \sup_{\{(p(x),\rho_A^x)\}_{x}}\left\{H\!\left(\sum_{x\in\mathcal{X}}p(x)\mathcal{N}(\rho_A^x)\right)\right\}\nonumber\\
			&\qquad+\sup_{\{(p(x),\rho_A^x)\}_{x}}\left\{-\sum_{x\in\mathcal{X}}p(x)H(\mathcal{N}(\rho_A^x))\right\}\\
			&\leq \sup_{\rho_A}H(\mathcal{N}(\rho_A))+\sup_{\rho_A} \left\{-H(\mathcal{N}(\rho_A))\right\}\\
			&=\sup_{\rho_A}H(\mathcal{N}(\rho_A))-\inf_{\rho_A} H(\mathcal{N}(\rho_A)).\label{eq-Holevo_inf_covariant_pf_1}
		\end{align}
		Now, by the unitary invariance of the quantum entropy, for every state $\rho_A$ we obtain
		\begin{equation}
			H(\mathcal{N}(\rho_A))=H(V_B^g\mathcal{N}(\rho_A)(V_B^g)^\dagger)=H(\mathcal{N}(U_A^g\rho_A (U_A^g)^\dagger))
		\end{equation}
		for all $g\in G$. This implies that
		\begin{align}
			H(\mathcal{N}(\rho_A))&=\frac{1}{|G|}\sum_{g\in G}H(\mathcal{N}(U_A^g\rho_A (U_A^g)^\dagger))\\
			&\leq H\!\left(\sum_{g\in G}\frac{1}{|G|}\mathcal{N}(U_A^g\rho_A (U_A^g)^\dagger)\right)\\
			&=H\!\left(\mathcal{N}\left(\frac{1}{|G|}\sum_{g\in G}U_A^g \rho_A (U_A^g)^\dagger\right)\right)\\
			&=H\!\left(\mathcal{N}(\pi_A)\right),
		\end{align}
		where the inequality follows from concavity of the quantum entropy and the last equality follows because $\{U^g\}_{g\in G}$ is an irreducible representation, which implies that
		\begin{equation}\label{eq-group_irrep_twirl}
			\frac{1}{|G|}\sum_{g\in G}U_A^g\rho_A (U_A^g)^\dagger=\frac{\mathbbm{1}_A}{d_A}
		\end{equation}
		for every state $\rho_A$. Then, since we are optimizing a continuous function over a compact and convex set, the infimum in \eqref{eq-Holevo_inf_covariant_pf_1} can be achieved, meaning that we can replace the infimum in \eqref{eq-Holevo_inf_covariant_pf_1} with a minimum, which means that
		\begin{equation}
			\chi(\mathcal{N})\leq H\!\left(\mathcal{N}(\pi_A)\right)-H_{\text{min}}(\mathcal{N}).
		\end{equation}
		
		To show the reverse inequality, let $\rho_A^*$ be a state for which $H(\mathcal{N}(\rho_A^*))=H_{\text{min}}(\mathcal{N})$. Then, we consider the ensemble $\left\{\left(\frac{1}{|G|},U_A^g\rho_A^* (U_A^g)^\dagger\right)\right\}_{g\in G}$ and obtain
		\begin{align}
			\chi(\mathcal{N})&\geq H\!\left(\sum_{g\in G}\frac{1}{|G|}\mathcal{N}(U_A^g\rho_A^* (U_A^g)^\dagger)\right)-\sum_{g\in G}\frac{1}{|G|}H(\mathcal{N}(U_A^g\rho_A^* (U_A^g)^\dagger))\\
			&=H\!\left(\mathcal{N}(\pi_A)\right)-\sum_{g\in G}\frac{1}{|G|}H(V_B^g\rho_A^* (V_B^g)^\dagger)\\
			&=H\!\left(\mathcal{N}(\pi_A)\right)-\sum_{g\in G}\frac{1}{|G|}H(\mathcal{N}(\rho_A^*))\\
			&=H\!\left(\mathcal{N}(\pi_A)\right)-H(\mathcal{N}(\rho_A^*))\\
			&=H\!\left(\mathcal{N}(\pi_A)\right)-H_{\text{min}}(\mathcal{N}).
		\end{align}
		Therefore,
		\begin{equation}
			\chi(\mathcal{N})\geq H\!\left(\mathcal{N}(\pi_A)\right)-H_{\text{min}}(\mathcal{N}),
		\end{equation}
		and the proof is complete.
	\end{Proof}
	
	We note that to compute the minimum output entropy of any channel $\mathcal{N}$, it suffices to optimize over pure states. Indeed, by restricting the optimization to pure states $\psi$ in the definition of $H_{\text{min}}(\mathcal{N})$, we find that
	\begin{equation}
		H_{\text{min}}(\mathcal{N})=\min_\rho H(\mathcal{N}(\rho))\leq\min_{\psi}H(\mathcal{N}(\psi)).
	\end{equation}
	On the other hand, since every state $\rho$ can be written as a convex combination of pure states, so that $\rho=\sum_{x\in\mathcal{X}} p(x)\psi^x$, we see that
	\begin{align}
		H(\mathcal{N}(\rho))&=H\!\left(\mathcal{N}\left(\sum_{x\in\mathcal{X}}p(x)\psi^x\right)\right)\\
		&=H\!\left(\sum_{x\in\mathcal{X}}p(x)\mathcal{N}(\psi^x)\right)\\
		&\geq \sum_{x\in\mathcal{X}}p(x)H(\mathcal{N}(\psi^x))\\
		&\geq \min_{x\in\mathcal{X}}H(\mathcal{N}(\psi^x))\\
		&\geq \min_{\psi}H(\mathcal{N}(\psi)),
	\end{align}
	where the first inequality follows from concavity of the quantum entropy. So we have
	\begin{equation}
		H_{\text{min}}(\mathcal{N})=\min_{\psi}H(\mathcal{N}(\psi)).
	\end{equation}
	
\begin{figure}
		\centering
		\includegraphics[scale=1]{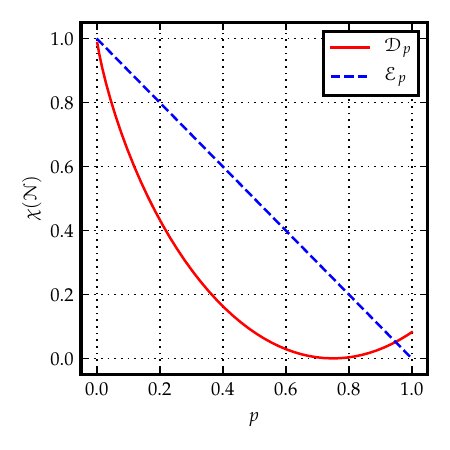}
		\caption{The Holevo information of the depolarizing channel $\mathcal{D}_p$ (expressed in \eqref{eq-Hol_inf_dep}) and the erasure channel $\mathcal{E}_p$ (expressed in \eqref{eq-Hol_inf_erasure}), both of which are defined for the parameter $p\in[0,1]$.}\label{fig-Hol_inf_cov}
	\end{figure}

	We now look at two irreducibly covariant channels, the depolarizing channel and the erasure channel. A plot of the Holevo information for these channels is given in Figure~\ref{fig-Hol_inf_cov}.
	
\subsubsection{Depolarizing Channel}

	In Section~\ref{subsec-qubit_channel}, specifically in \eqref{eq-qubit_depolarizing_channel}, we defined the qubit depolarizing channel as
	\begin{equation}
		\mathcal{D}_{p}(\rho) \coloneqq (1-p)\rho+\frac{p}{3}(X\rho X+Y\rho Y+Z\rho Z)
	\end{equation}
	for all $p\in[0,1]$. From the arguments in Section~\ref{sec-eacc_depolarizing}, it follows that this channel is covariant with respect to the Pauli operators on both the input and output spaces. Furthermore, the Pauli operators $\{\mathbbm{1},X,Y,Z\}$ satisfy the property in \eqref{eq-group_irrep_twirl}. We thus conclude that
	\begin{equation}
		\chi(\mathcal{D}_{p})=H\!\left(\mathcal{D}_{p}(\pi)\right)-H_{\text{min}}(\mathcal{D}_{p})=1-H_{\text{min}}(\mathcal{D}_{p}),
	\end{equation}
	where we used the fact that the depolarizing channel is unital, i.e., $\mathcal{D}_p(\mathbbm{1})=\mathbbm{1}$, and that $H(\pi)\allowbreak=\log_22=1$. To compute the minimum output entropy, we use the fact that it suffices to minimize over pure states. It is straightforward to show that the two eigenvalues of $\mathcal{D}_{p}(\psi)$ are $(\sfrac{2p}{3},1-\sfrac{2p}{3})$ for \textit{every} pure state~$\psi$. Therefore, 
	\begin{equation}
		H_{\text{min}}(\mathcal{D}_{p})=h_2\!\left(\frac{2p}{3}\right)
	\end{equation}
	so that
	\begin{equation}\label{eq-Hol_inf_dep}
		\chi(\mathcal{D}_{p})=1-h_2\!\left(\frac{2p}{3}\right)
	\end{equation}
	for $p\in[0,1]$. Since the Holevo information is known to be additive for the depolarizing channel (please consult the Bibliographic Notes in Section~\ref{sec:CC-comm:bib-notes}), it follows that $\chi(\mathcal{D}_{p})$ is equal to the classical capacity of the depolarizing channel. See Figure~\ref{fig-Hol_inf_cov} for a plot of the Holevo information $\chi(\mathcal{D}_p)$ of the depolarizing channel.
	
	For the qudit depolarizing channel $\mathcal{D}_p^{(d)}$, recall from the discussion around \eqref{eq:EA-comm:qudit-depol-cov} that it is irreducibly covariant. Therefore, by Theorem~\ref{thm-Holevo_inf_covariant}, we obtain
	\begin{equation}
		\chi(\mathcal{D}_{p}^{(d)})=\log_2 d-H_{\text{min}}(\mathcal{D}_{p}^d).
	\end{equation}
	The calculation of the minimum output entropy for the qudit depolarizing channel is analogous to the calculation of the minimum output entropy of the qubit depolarizing channel. In particular, for every pure state $\psi$, the eigenvalues of $\mathcal{D}_{p}^{(d)}(\psi)$ are $1-\frac{d}{d+1}p$ (with multiplicity one) and $\frac{d}{d^2-1}p$ (with multiplicity $d-1$). Indeed, by using the parameterization in \eqref{eq:QM-over:qudit-depol-reparam} with $q=\frac{pd^{2}}{d^{2}-1}$, consider that%
\begin{align}
\mathcal{D}_{p}^{(d)}(\psi)  & =\left(  1-q\right)  \psi+q\frac{I}{d}\\
& =\left(  1-q+\frac{q}{d}\right)  \psi+\frac{q}{d}\left(  I-\psi\right)  \\
& =\left(  1-\frac{pd}{d+1}\right)  \psi+\frac{pd}{d^{2}-1}\left(
I-\psi\right)  .
\end{align}
Therefore,
	\begin{equation}
		H_{\text{min}}(\mathcal{D}_{p}^{(d)})=-\left(1-\frac{dp}{d+1}\right)\log_2\!\left(1-\frac{dp}{d+1}\right)
		-\frac{dp}{d+1}\log_2\!\left(\frac{dp}{d^2-1}\right),
	\end{equation}
	so that
	\begin{equation}\label{eq-Holevo_inf_depolarizing}
		\chi(\mathcal{D}_{p}^{(d)})=\log_2d+\left(1-\frac{dp}{d+1}\right)\log_2\!\left(1-\frac{dp}{d+1}\right) +\frac{dp}{d+1}\log_2\!\left(\frac{dp}{d^2-1}\right)
	\end{equation}
	for $d\geq 2$ and $p\in[0,1]$.
	
	The Holevo information is also known to be additive for the qudit depolarizing channel, which means that the expression in \eqref{eq-Holevo_inf_depolarizing} is equal to its classical capacity.
	
	\begin{theorem*}{Additivity of the Holevo Information for the Depolarizing Channel}{thm-Hol_inf_additive_depolarizing}
		For every channel $\mathcal{M}$,
		\begin{equation}
			\chi(\mathcal{D}_{p}^{(d)}\otimes\mathcal{M})=\chi(\mathcal{D}_{p}^{(d)})+\chi(\mathcal{M})
		\end{equation}
		for all $d\geq 2$ and all $p\in[0,1]$. Consequently,
		\begin{equation}
			C(\mathcal{D}_p^{(d)})=\chi(\mathcal{D}_p^{(d)}).
		\end{equation}
	\end{theorem*}
	
	\begin{Proof}
		Please consult the Bibliographic Notes in Section~\ref{sec:CC-comm:bib-notes}.
	\end{Proof}
	
	It also holds that the Holevo information is the strong converse classical capacity of the qudit depolarizing channel, i.e.,
	\begin{equation}
		\widetilde{C}(\mathcal{D}_p^{(d)})=\chi(\mathcal{D}_p^{(d)})
	\end{equation}
	for all $d\geq 2$ and all $p\in[0,1]$. Please consult the Bibliographic Notes in Section~\ref{sec:CC-comm:bib-notes} for a reference to the proof.

\subsubsection{Erasure Channel}\label{subsec-cc_erasure_channel}

	Let us now consider the erasure channel. Recall from \eqref{eq-erasure_channel} that the erasure channel $\mathcal{E}_p$, with $p\in[0,1]$, is defined as
	\begin{equation}
		\mathcal{E}_p(\rho)=(1-p)\rho+p\Tr[\rho]\ket{e}\!\bra{e},
	\end{equation}
	where $\ket{e}$ is called the erasure state and is not in the Hilbert space of the input system $A$. In other words, the state $\ket{e}\!\bra{e}$ is supported on the space orthogonal to the input space. As argued in Section~\ref{subsec-eacc_erasure_channel}, we can consider the output space of the channel to be a qutrit system with the orthonormal basis $\{\ket{0},\ket{1},\ket{2}\}$, and we can let the state $\ket{2}$ be the erasure state. Then,
	\begin{equation}
		\mathcal{E}_p(\rho)=(1-p)\rho+p\ket{2}\!\bra{2}
	\end{equation}
	for every state $\rho$.
	
	We also argued in Section~\ref{subsec-eacc_erasure_channel} that the erasure channel is irreducibly covariant. Therefore, by Theorem~\ref{thm-Holevo_inf_covariant}, we have that
	\begin{equation}
		\chi(\mathcal{E}_p)=H\!\left(\mathcal{E}_p(\pi)\right)-H_{\text{min}}(\mathcal{E}_p).
	\end{equation}
	Now, on the input qubit space, we have $\pi=\frac{1}{2}\ket{0}\!\bra{0}+\frac{1}{2}\ket{1}\!\bra{1}$. Therefore,
	\begin{equation}
		\mathcal{E}_p(\pi)=\frac{1-p}{2}\ket{0}\!\bra{0}+\frac{1-p}{2}\ket{1}\!\bra{1}+p\ket{2}\!\bra{2},
	\end{equation}
	which means that
	\begin{align}
		H\!\left(\mathcal{E}_p(\pi)\right) & =-(1-p)\log_2\!\left(\frac{1-p}{2}\right)-p\log_2p \\
		& = 1 - p + h_2(p).
	\end{align}
	In addition, for every pure state $\psi$, we have
	\begin{align}
		H(\mathcal{E}_p(\psi))&=H\!\left((1-p)\psi+p\ket{2}\!\bra{2}\right)\\
		&=-(1-p)\log_2(1-p)-p\log_2 p,\\
		& = h_2(p),
	\end{align}
	where  the second equality follows because the state $\ket{2}\!\bra{2}$ is orthogonal to $\psi$. Therefore,
	\begin{equation}
		H_{\text{min}}(\mathcal{E}_p)=h_2(p),
	\end{equation}
	which means that the Holevo information of the erasure channel is
	\begin{equation}\label{eq-Hol_inf_erasure}
		\chi(\mathcal{E}_p)=1-p.
	\end{equation}
	This is consistent with what one might expect intuitively because communication over the erasure channel is only possible with probability $1-p$, when no erasure occurs, and conditioned on this outcome, the erasure channel is simply the identity channel.
	
	In general, for the qudit erasure channel $\mathcal{E}_p^{(d)}$, whose action can be defined on the $d$-dimensional space with orthonormal basis $\{\ket{1},\dots,\ket{d}\}$ such that the state $\ket{d+1}$ is the erasure state, we have that it is irreducibly covariant (see Section~\ref{subsec-eacc_erasure_channel}). Using this fact, which implies that 
	\begin{equation}
		\chi(\mathcal{E}_p^{(d)})=H\!\left(\mathcal{E}_p^{(d)}(\pi)\right)-H_{\text{min}}(\mathcal{E}_p^{(d)}),
	\end{equation}
	along with arguments analogous to those presented above, we obtain
	\begin{equation}
		\chi(\mathcal{E}_p^{(d)})=(1-p)\log_2 d. 
		\label{eq:CC-comm:holevo-info-erasure-ch}
	\end{equation}
	
	\begin{proposition*}{$\Upsilon$-Information of the Erasure Channel}{prop:CC-comm:ups-erasure}
	The $\Upsilon$-information of the qudit erasure channel $\mathcal{E}_p^{(d)}$ is given by
	\begin{equation}
	\Upsilon(\mathcal{E}_p^{(d)}) = \chi(\mathcal{E}_p^{(d)}) = (1-p)\log_2 d.
	\end{equation}
	\end{proposition*}
	
	\begin{Proof}
	By combining \eqref{eq:CC-comm:holevo-info-erasure-ch} and Proposition~\ref{prop:CC-comm:holevo-to-upsilon}, we conclude that $\Upsilon(\mathcal{E}_p^{(d)}) \geq \chi(\mathcal{E}_p^{(d)}) = (1-p)\log_2 d$. So we establish the opposite inequality.
	
		Since the erasure channel is irreducibly covariant,  Theorem~\ref{prop-gen_upsilon_inf_cov} implies that the optimization over states $\psi_{RA}$ in the definition of the $\Upsilon$-information is unnecessary, and we have that
		\begin{equation}
			\Upsilon(\mathcal{E}_p^{(d)})=\inf_{\mathcal{F}\in\mathfrak{F}}D((\mathcal{E}_p^{(d)})_{A\to B}(\Phi_{RA})\Vert\mathcal{F}_{A\to B}(\Phi_{RA})),
		\end{equation}
		where $\Phi_{RA}$ is the maximally entangled state with Schmidt rank $d$. The Choi state $\rho_{RB}^{\mathcal{E}_p^{(d)}}=(\mathcal{E}_p^{(d)})_{A\to B}(\Phi_{RA})$ of the erasure channel is
		\begin{equation}
			\rho_{RB}^{\mathcal{E}_p^{(d)}}=(1-p)\Phi_{RB} + p\frac{\mathbbm{1}_R}{d}\otimes\ket{e}\!\bra{e}.
		\end{equation}
		Now, let us make a particular choice of the map $\mathcal{F}$ in the minimization over the completely positive maps in $\mathfrak{F}$. Suppose that $\mathcal{F}_{A\to B}$ is such that
		\begin{equation}
			\mathcal{F}_{A\to B}(\Phi_{RA})=\frac{1-p}{d}\Phi_{RB}+p\frac{\mathbbm{1}_R}{d}\otimes\ket{e}\!\bra{e}\eqqcolon \sigma_{RB}^{\mathcal{F}},
		\end{equation} 
		which implies via \eqref{eq-Choi_state_post_selected_teleportation} that its action on a general input state $\rho_A$ is as follows:
		\begin{equation}
		\mathcal{F}_{A\to B}(\rho_A) = \frac{1-p}{d} \rho_A + p \ket{e}\!\bra{e} \leq \frac{1-p}{d} \mathbbm{1}_A + p \ket{e}\!\bra{e}. \label{eq-CC:erasure-F-map-indeed-F-map}
		\end{equation}
		Note that the map $\mathcal{F}_{A\to B}$ defined in this way is indeed in the set $\mathfrak{F}$, as demonstrated by the inequality in \eqref{eq-CC:erasure-F-map-indeed-F-map} and the fact that the operator on the right-hand side of \eqref{eq-CC:erasure-F-map-indeed-F-map} is a quantum state. Then,
		\begin{equation}
			\Upsilon(\mathcal{E}_p^{(d)})\leq D(\rho_{RB}^{\mathcal{E}_p^{(d)}}\Vert\sigma_{RB}^{\mathcal{F}}).
		\end{equation}
		It is straightforward to show that
		\begin{equation}
			\Tr\!\left[\rho_{RB}^{\mathcal{E}_p^{(d)}}\log_2\rho_{RB}^{\mathcal{E}_p^{(d)}}\right]=(1-p)\log_2(1-p)+p\log_2\!\left(\frac{p}{d}\right),
		\end{equation}
		and
		\begin{equation}
			\Tr\!\left[\rho_{RB}^{\mathcal{E}_p^{(d)}}\log_2\sigma_{RB}^{\mathcal{F}}\right]=(1-p)\log_2\!\left(\frac{1-p}{d}\right)+p\log_2\!\left(\frac{p}{d}\right),
		\end{equation}
		which means that
		\begin{equation}
			\Upsilon(\mathcal{E}_p^{(d)})\leq (1-p)\log_2d=\chi(\mathcal{E}_p^{(d)}).
		\end{equation}
		This concludes the proof.
	\end{Proof}

	The classical capacity of the quantum erasure channel and its strong converse now follow as a direct corollary of \eqref{eq:CC-comm:holevo-info-erasure-ch}, \eqref{eq:CC-comm:Holevo-info-lower-bnd-cap},  Proposition~\ref{prop:CC-comm:ups-erasure}, the irreducible covariance of the erasure channel, and Theorem~\ref{thm-cc_str_conv_upsilon_cov}.
	
	\begin{theorem*}{Classical Capacity of the Erasure Channel}{thm-cc_erasure_chan_cap}
		For  $d\geq 2$ and $p\in[0,1]$, the following equality holds for the classical capacity and strong converse classical capacity of the quantum erasure channel $\mathcal{E}_p^{(d)}$:
		\begin{equation}
			C(\mathcal{E}_p^{(d)})=\widetilde{C}(\mathcal{E}_p^{(d)})=(1-p)\log_2d.
		\end{equation}
	\end{theorem*}
	
%
%
%

%

\subsection{Amplitude Damping Channel}

	Recall from \eqref{eq-amplitude_damping_channel} that the amplitude damping channel is defined as
	\begin{equation}
		\mathcal{A}_\gamma(\rho)=A_1\rho A_1^\dagger+A_2\rho A_2^\dagger,
	\end{equation}
	where
	\begin{equation}\label{eq-amplitude_damping_channel_2}
		A_1=\sqrt{\gamma}\ket{0}\!\bra{1},\quad A_2=\ket{0}\!\bra{0}+\sqrt{1-\gamma}\ket{1}\!\bra{1}.
	\end{equation}
	
	It can be shown that
	\begin{equation}\label{eq-Hol_inf_AD}
		\chi(\mathcal{A}_\gamma)=\frac{1}{2}\left(f(r^*)-\log_2(1-q^2)-qf'(q)\right),
	\end{equation}
	where
	\begin{align}
		f(x)&\coloneqq (1+x)\log_2(1+x)-(1-x)\log_2(1-x),\\
		f'(x)&=\log_2\!\left(\frac{1+x}{1-x}\right),\\
		r^*&\coloneqq\sqrt{1-\gamma-\frac{(q-\gamma)^2}{1-\gamma}+q^2},
	\end{align}
	and $q$ is determined via
	\begin{equation}
		(\gamma q-\gamma^2-\gamma(1-\gamma))f'(r^*)=-r^*(1-\gamma)f'(q).
	\end{equation}
	(Please consult the Bibliographic Notes in Section~\ref{sec:CC-comm:bib-notes}.) It is worth noting that neither the additivity of the Holevo information for nor the classical capacity of the amplitude damping channel are known. 
	
\begin{figure}
		\centering
		\includegraphics[scale=1]{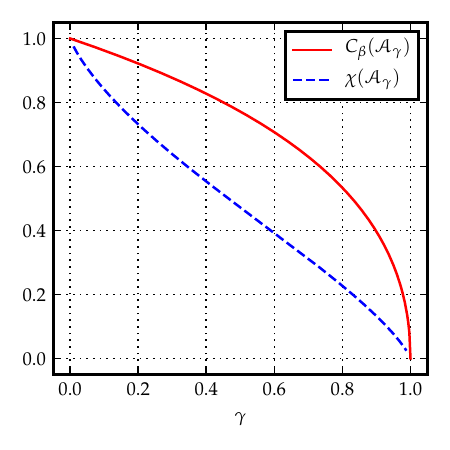}
		\caption{The Holevo information $\chi(\mathcal{A}_\gamma)$ in \eqref{eq-Hol_inf_AD} of the amplitude damping channel, which represents a lower bound on its classical capacity. Also shown is the upper bound $C_\beta(\mathcal{A}_\gamma)$ on the classical capacity, which is defined in \eqref{eq-cc_C_beta} via the SDP in \eqref{eq-cc_beta_SDP}, and for the amplitude damping channel it is given by the expression in \eqref{eq-cc_C_beta_AD}.
		}\label{fig-cc_C_beta_Hol_inf_AD}
	\end{figure}

	As we have seen, the quantity $C_\beta(\mathcal{A}_\gamma)$ is an upper bound on the classical capacity of the amplitude damping channel, and it can be shown (please consult the Bibliographic Notes in Section~\ref{sec:CC-comm:bib-notes}) that
	\begin{equation}\label{eq-cc_C_beta_AD}
		C_\beta(\mathcal{A}_\gamma)=\log_2(1+\sqrt{1-\gamma}).
	\end{equation}
	See Figure~\ref{fig-cc_C_beta_Hol_inf_AD} for a plot of both $\chi(\mathcal{A}_\gamma)$ and $C_\beta(\mathcal{A}_\gamma)$.
	
\subsection{Hadamard Channels}

	In this section, we prove that the Holevo information is additive for all Hada\-mard channels.
	
	\begin{theorem*}{Additivity of Holevo Information for Hadamard Channels}{thm-Hol_inf_additive_Hadamard}
		For a Hadamard channel $\mathcal{N}$ and an arbitrary channel $\mathcal{M}$, the following additivity relation holds
		\begin{equation}\label{eq-Holevo_additive_Hadamard}
			\chi(\mathcal{N}\otimes\mathcal{M})=\chi(\mathcal{N})+\chi(\mathcal{M}).
		\end{equation}
	\end{theorem*}
	
	\begin{Proof}
		Using the expression in \eqref{eq-Holevo_inf_alt_2}, we have that 
		\begin{equation}\label{eq-Hadamard_Holevo_inf_additive_pf_1}
			\begin{aligned}
		 	&\chi(\mathcal{N}\otimes\mathcal{M})\\
		 	&\quad=\sup_{\{(p(x),\psi^x)\}_{x\in\mathcal{X}}}\left[H\!\left(\sum_{x\in\mathcal{X}}p(x)(\mathcal{N}\otimes\mathcal{M})(\psi_{A_1A_2}^x)\right)\right. \\
		 	&\qquad\qquad\qquad\qquad\qquad\qquad\left. -\sum_{x\in\mathcal{X}}p(x)H((\mathcal{N}^c\otimes\mathcal{M}^c)(\psi_{A_1A_2}^x))\right].
		 	\end{aligned}
		\end{equation}
		 
		Now, for every bipartite state $\rho_{AB}$, it follows from strong subadditivity that $H(\rho_{AB})\leq H(\rho_A)+H(\rho_B)$ a fact known as the subadditivity of the quantum entropy (consider \eqref{eq-SSA_1} with system $C$ trivial). Using this for the first term in \eqref{eq-Hadamard_Holevo_inf_additive_pf_1}, we find that
		\begin{equation}
			\begin{aligned}
			&H\!\left(\sum_{x\in\mathcal{X}}p(x)(\mathcal{N}\otimes\mathcal{M})(\psi_{A_1A_2}^x)\right)\\
			&\qquad\qquad\leq H\!\left(\sum_{x\in\mathcal{X}}p(x)\mathcal{N}(\psi_{A_1}^x)\right)+H\!\left(\sum_{x\in\mathcal{X}}p(x)\mathcal{M}(\psi_{A_2}^x)\right).
			\end{aligned}
		\end{equation}
		We now make use of the following identity, which is straightforward to verify: for every finite alphabet $\mathcal{X}$ and  ensemble $\{(p(x),\rho_{AB}^x)\}$,
		\begin{multline}\label{eq-Hadamard_Holevo_inf_additive_pf}
			\sum_{x\in\mathcal{X}}p(x)D(\rho_{AB}^x\Vert\rho_A^x\otimes\rho_B^x)=\sum_{x\in\mathcal{X}}p(x)H(\rho_A^x)+\sum_{x\in\mathcal{X}}p(x)H(\rho_B^x)\\
			\quad-\sum_{x\in\mathcal{X}}p(x)H(\rho_{AB}^x).
		\end{multline}
		We use this for the second term in \eqref{eq-Hadamard_Holevo_inf_additive_pf_1} to conclude that
		\begin{equation}
			\begin{aligned}
		 	&\sum_{x\in\mathcal{X}}p(x)H((\mathcal{N}^c\otimes\mathcal{M}^c)(\psi_{A_1A_2}^x))\\
		 	&\quad=\sum_{x\in\mathcal{X}}p(x)H(\mathcal{N}^c(\psi_{A_1}^x))+\sum_{x\in\mathcal{X}}p(x)H(\mathcal{M}^c(\psi_{A_2}^x))\\
		 	&\qquad\qquad-\sum_{x\in\mathcal{X}}p(x)D((\mathcal{N}^c\otimes\mathcal{M}^c)(\psi_{A_1A_2}^x)\Vert\mathcal{N}^c(\psi_{A_1}^x)\otimes\mathcal{M}^c(\psi_{A_2}^x)).
			\end{aligned}
		\end{equation}
		Now, let us focus on the relative entropy term in the expression above. Since $\mathcal{N}$ is a Hadamard channel, by Proposition~\ref{prop-Hadamard_chan_comp} we know that the complementary channel $\mathcal{N}^c$ is entanglement-breaking. Then, from Theorem~\ref{thm-ent_break_meas_reprepare}, we know that every entanglement-breaking channel can be written as the composition of a measurement channel followed by a preparation channel. This means that we can write $\mathcal{N}^c$ as $\mathcal{N}^c=\mathcal{P}\circ\mathcal{M}_{\text{qc}}$, where $\mathcal{M}_{\text{qc}}$ is the measurement (or quantum--classical) channel, and $\mathcal{P}$ is the preparation channel. Using the data-processing inequality for the quantum relative entropy, for all $x\in\mathcal{X}$ we obtain
		\begin{align}
			&D((\mathcal{N}^c\otimes\mathcal{M}^c)(\psi_{A_1A_2}^x)\Vert\mathcal{N}^c(\psi_{A_1}^x)\otimes\mathcal{M}(\psi_{A_2}^x))\nonumber\\
			&\qquad=D((\mathcal{P}\circ\mathcal{M}^c_{\text{qc}}\otimes\mathcal{M}^c)(\psi_{A_1A_2}^x)\Vert(\mathcal{P}\circ\mathcal{M}_{\text{qc}})(\psi_{A_1}^x)\otimes\mathcal{M}^c(\psi_{A_2}^x))\\
			&\qquad\leq D((\mathcal{M}_{\text{qc}}\otimes\mathcal{M})(\psi_{A_1A_2}^x)\Vert\mathcal{M}^c_{\text{qc}}(\psi_{A_1}^x)\otimes\mathcal{M}^c(\psi_{A_2}^x)).
		\end{align}
		Then, using the identity \eqref{eq-Hadamard_Holevo_inf_additive_pf} once again, we obtain
		\begin{align}
			&\sum_{x\in\mathcal{X}}p(x)D((\mathcal{M}_{\text{qc}}\otimes\mathcal{M}^c)(\psi_{A_1A_2}^x)\Vert\mathcal{M}_{\text{qc}}(\psi_{A_1}^x)\otimes\mathcal{M}^c(\psi_{A_2}^x))\nonumber\\
			&\qquad =\sum_{x\in\mathcal{X}}p(x)H(\mathcal{M}_{\text{qc}}(\psi_{A_1}^x))+\sum_{x\in\mathcal{X}}p(x)H(\mathcal{M}^c(\psi_{A_2}^x))\nonumber\\
			&\qquad\qquad -\sum_{x\in\mathcal{X}}p(x)H((\mathcal{M}_{\text{qc}}\otimes\mathcal{M}^c)(\psi_{A_1A_2}^x)).
		\end{align}
		Now, let the measurement channel $\mathcal{M}_{\text{qc}}$ have the associated POVM $\{M_{A_1}^y\}_{y\in\mathcal{Y}}$ for some finite alphabet $\mathcal{Y}$. Then, letting $q(y|x)\coloneqq \Tr[M_y\psi_{A_1}^x]$, we have that
		\begin{equation}
			H(\mathcal{M}_{\text{qc}}(\psi_{A_1}^x))=H(Y|X=x)
		\end{equation}
		for all $x\in\mathcal{X}$. Also, for every $x\in\mathcal{X}$, we find that
		\begin{equation}
			(\mathcal{M}_{\text{qc}}\otimes\mathcal{M}^c)(\psi^x_{A_1A_2})=\sum_{y\in\mathcal{Y}} q(y|x)\ket{y}\!\bra{y}_Y\otimes \rho_{A_2}^{x,y},
		\end{equation}
		where $\rho_{A_2}^{x,y}\coloneqq\frac{1}{q(y|x)}\Tr_{A_1}[(M^y_{A_1}\otimes\mathbbm{1}_{A_2})\psi^x_{A_1A_2}]$. Note that
		\begin{equation}\label{eq-Hadamard_Holevo_inf_additive_pf_2}
			\begin{aligned}
			\sum_{y\in\mathcal{Y}}q(y|x)\rho_{A_2}^{x,y}&=\sum_{y\in\mathcal{Y}}\Tr_{A_1}[(M^y_{A_1}\otimes\mathbbm{1}_{A_2})\psi^x_{A_1A_2}]\\
			&=\Tr_{A_1}[\psi^x_{A_1A_2}]\\
			&=\psi_{A_2}^x
			\end{aligned}
		\end{equation}
		for all $x\in\mathcal{X}$. Therefore, for all $x\in\mathcal{X}$,
		\begin{align}
			&H((\mathcal{M}_{\text{qc}}\otimes\mathcal{M})(\psi^x_{A_1A_2}))\nonumber\\
			&\quad=H\!\left(\sum_{y\in\mathcal{Y}}q(y|x)\ket{y}\!\bra{y}_Y\otimes\mathcal{M}^c(\rho_{A_2}^{x,y})\right)\\
			&\quad=H(Y|X=x)+\sum_{y\in\mathcal{Y}}q(y|x)H(\mathcal{M}^c(\rho_{A_2}^{x,y})),
		\end{align}
		where the last equality follows from the direct-sum property of the quantum entropy. Putting everything together, we obtain
		\begin{align}
			&H\!\left(\sum_{x\in\mathcal{X}}p(x)(\mathcal{N}\otimes\mathcal{M})(\psi_{A_1A_2}^x)\right)-\sum_{x\in\mathcal{X}}p(x)H((\mathcal{N}^c\otimes\mathcal{M}^c)(\psi_{A_1A_2}^x))\nonumber\\
			&\quad \leq H\!\left(\sum_{x\in\mathcal{X}}p(x)\mathcal{N}(\psi_{A_1}^x)\right)-\sum_{x\in\mathcal{X}}p(x)H(\mathcal{N}^c(\psi_{A_1}^x))+H\!\left(\sum_{x\in\mathcal{X}}p(x)\mathcal{M}(\psi_{A_2}^x)\right)\nonumber\\
			&\qquad\qquad -\sum_{x\in\mathcal{X}}p(x)H(\mathcal{M}^c(\psi_{A_2}^x))+\sum_{x\in\mathcal{X}}p(x)H(Y|X=x)\nonumber\\
			&\qquad\qquad+\sum_{x\in\mathcal{X}}p(x)H(\mathcal{M}^c(\psi_{A_2}^x))-\sum_{x\in\mathcal{X}}p(x)H(Y|X=x)\nonumber\\
			&\qquad\qquad-\sum_{x\in\mathcal{X}}\sum_{y\in\mathcal{Y}}p(x)q(y|x)H(\mathcal{M}^c(\psi_{A_2}^{x,y}))\\
			&=H\!\left(\sum_{x\in\mathcal{X}}p(x)\mathcal{N}(\psi_{A_1}^x)\right)-\sum_{x\in\mathcal{X}}p(x)H(\mathcal{N}^c(\psi_{A_1}^x))\\
			&\qquad\qquad+H\!\left(\sum_{x\in\mathcal{X}}\sum_{y\in\mathcal{Y}}p(x)q(y|x)\mathcal{M}(\rho_{A_2}^{x,y})\right)\nonumber\\
			&\qquad\qquad-\sum_{x\in\mathcal{X}}\sum_{y\in\mathcal{Y}}p(x)q(y|x)H(\mathcal{M}^c(\rho_{A_2}^{x,y}))\nonumber\\
			&\leq \chi(\mathcal{N})+\chi(\mathcal{M})\label{eq-Hadamard_Holevo_inf_additive_pf_3},
		\end{align}
		where we have used \eqref{eq-Hadamard_Holevo_inf_additive_pf_2} and, to obtain the last inequality, the fact that the first two terms in \eqref{eq-Hadamard_Holevo_inf_additive_pf_3} are of the form of the objective function in the expression in \eqref{eq-Holevo_inf_alt} for the Holevo information $\chi(\mathcal{N})$, and similarly for the last two terms, in which the ensemble is $\{(p(x)q(y|x),\rho_{A_2}^{x,y}):x\in\mathcal{X},~y\in\mathcal{Y}\}$.
		 
		Since the ensemble $\{(p(x),\psi^x_{A_1A_2})\}_{x\in\mathcal{X}}$ used to obtain \eqref{eq-Hadamard_Holevo_inf_additive_pf_3} is arbitrary, we conclude that
		\begin{equation}
			\chi(\mathcal{N}\otimes\mathcal{M})\leq\chi(\mathcal{M})+\chi(\mathcal{N}),
		\end{equation}
		which implies, via the superadditivity in \eqref{eq-chi-superadditive} that $\chi(\mathcal{N}\otimes\mathcal{M})=\chi(\mathcal{N})+\chi(\mathcal{M})$, as required.
	\end{Proof}
	
	\begin{exercise}{exer-classical_cap_dephasing}
		Prove that the classical capacity of the $d$-dimensional dephasing channel, $d\geq 2$, is $\log_2 d$.
	\end{exercise}

\section{Summary}
	
	In this chapter, we developed the theory of classical communication over a quantum channel, adopting a similar structure to that of the previous chapter. We began with the one-shot setting of classical communication, and we defined the one-shot classical capacity of a quantum channel in Definition~\ref{def-classical_comm_one_shot_capacity}. We then derived  upper (Proposition~\ref{prop-cc:one-shot-bound-meta}) and lower (Proposition~\ref{prop-cc_one-shot_lower_bound}) bounds on the one-shot classical capacity in terms of the hypothesis testing Holevo information of a quantum channel. The approaches to doing so are conceptually similar to those from the previous chapter. However, there are extra steps involved in deriving the lower bound, called derandomization and expurgation, that establish the existence of a code with maximum error probability no larger than a given threshold and number of bits transmitted roughly equal to the one-shot Holevo information.
	
	With the fundamental information-theoretic arguments established in the one-shot setting, we then moved on to the asymptotic setting of classical communication. One of the main results is that the regularized Holevo information of a channel is equal to its classical capacity (Theorem~\ref{thm-classical_capacity}). We then considered some special cases: for entanglement-breaking, Hadamard, depolarizing, and erasue channels, the Holevo information is not only equal to the classical capacity but also equal to the strong converse classical capacity (we showed the proofs in full for entanglement-breaking and erasure channels, but deferred to the literature for the others). We discussed general upper bounds on the classical capacity, including the $\Upsilon$-information and $C_{\beta}$ semi-definite programming bound.
	
	Going forward from here, the methods of position-based coding and sequential decoding are useful for the tasks of secret key distillation (Chapter~\ref{chap-secret_key_distill}) and private communication (Chapter~\ref{chap-private_capacity}), and the concept of derandomization appears again in the context of private communication. The Holevo information will also play a role in achievable rates for these tasks.
	
\section{Bibliographic Notes}
	
	\label{sec:CC-comm:bib-notes}
	
Classical communication over quantum channels is one of the earliest settings considered in quantum information theory. A key early work on the topic includes \citet{Holevo73}, in which the Holevo upper bound on classical capacity was established. Many years later, after the advent of quantum computing, the Holevo information lower bound on classical capacity was established by \citet{holevo1998capacity} and \citet{schumacher1997sending}. Prior to these works, \citet{PhysRevA.54.1869} proved the same lower bound for the special case of a channel that accepts classical inputs and outputs pure quantum states.

Classical communication in the one-shot setting has been considered by a number of authors, including \citet{hayashi2003generalCapacity,Hay07,MD09,MH11,RR11,WR12,MW12,SW12,DMHB13,TH12,Wilde20130259,AJW17b,QWW17,OMW19}. Proposition~\ref{prop-cc:one-shot-bound-meta} is due to \citet{MW12}. The second part of Theorem~\ref{cor-cc_meta_str_weak_conv} is due to \citet{WWY14}. A variation of Theorem~\ref{prop-cc_one-shot_lower_bound} (for average error probability with uniformly random message) is due to \citet{WR12}. The proof given here is due to \citet{OMW19}, however with some variations given in this book to account for maximal error probability. At the same time, the proof uses the method of position-based coding \citep{AJW17b}, with the derandomization argument as given by \citet{QWW17}.

Additivity of Holevo information for entanglement-breaking channels was established by \citet{S02}, for Hadamard channels by \citet{KMNR07}, for the depolarizing channel by \citet{King03}, and for the erasure channel by \citet{PhysRevLett.78.3217}. The fact that the Holevo information is the strong converse classical capacity of the depolarizing channel was proven by \citet{KW09}. Additivity of the sandwiched R\'enyi-Holevo information for entanglement-breaking channels (Theorem~\ref{thm-classical_comm_ent_break_additive}) was established by \citet{WWY14}, by building upon earlier seminal results of \citet{K03} subsequently generalized by \citet{Hol06}. That is, Lemma~\ref{lem-ent_break_mult_2} is due to \citet{K03,Hol06}. Lemma~\ref{lem-ent_break_mult_1} is due to \citet{WWY14}.

The $\Upsilon$-information of a quantum channel and its variants were defined by \citet{WFT18}. The same authors established bounds on classical capacity involving $\Upsilon$-information. The strong converse for the classical capacity of the quantum erasure channel is due to \citet{WW14}, but here we have followed the approach of \citet{WFT18}. The semi-definite programming upper bound $C_{\beta}(\mathcal{N})$ for the classical capacity of a quantum channel $\mathcal{N}$ was established by \citet{WXD18}.

The Holevo information of covariant channels was studied by \citet{Hol02}.
	
	A proof of the fact that the limit in the definition of the regularized Holevo information of a channel exists was given by \citet{BNS98}.
	
	The formula in \eqref{eq-Hol_inf_AD} for the Holevo information of the amplitude damping channel was derived by \citet{ZF07}, using the techniques of \citet{Cor02} and \citet{Ber05}. The formula in \eqref{eq-cc_C_beta_AD} for the quantity $C_\beta$ for the same channel was determined by \citet{WXD18} (see also \citet{KSW19}).

\begin{subappendices}

\section{The \texorpdfstring{$\alpha\to 1$}{a to 1} Limit of the Sandwic\-hed R\'{e}nyi \texorpdfstring{$\Upsilon$}{Y}-Information of a Channel}\label{app-sand_ren_upsilon_inf_chan_limit}

	In this section, we show that
	\begin{equation}
		\lim_{\alpha\to 1^+}\widetilde{\Upsilon}_\alpha(\mathcal{N})=\Upsilon(\mathcal{N}),
	\end{equation}
	where we recall that
	\begin{align}
		\widetilde{\Upsilon}_\alpha(\mathcal{N})&=\sup_{\psi_{RA}}\inf_{\mathcal{F}\in\mathfrak{F}}\widetilde{D}_\alpha(\mathcal{N}_{A\to B}(\psi_{RA})\Vert\mathcal{F}_{A\to B}(\psi_{RA})),\\
		\Upsilon(\mathcal{N})&=\sup_{\psi_{RA}}\inf_{\mathcal{F}\in\mathfrak{F}}D(\mathcal{N}_{A\to B}(\psi_{RA})\Vert\mathcal{F}_{A\to B}(\psi_{RA})).
	\end{align}
	Here, $\psi_{RA}$ is a pure state, with the dimension of $R$ equal to the dimension of $A$, and the infimum is over the set $\mathfrak{F}$ of completely positive maps defined as
	\begin{equation}
		\mathfrak{F}=\{\mathcal{F}_{A\to B}:\exists ~\sigma_B\geq 0,\, \Tr[\sigma_B]\leq 1,\, \mathcal{F}_{A\to B}(\rho_A)\leq\sigma_B~\forall~\rho_A\in \Density(\mathcal{H}_A)\}.
	\end{equation}
	
	Now, since the sandwiched R\'{e}nyi relative entropy increases monotonically with $\alpha$ (see Proposition~\ref{prop-sand_rel_ent_properties}), and since $\lim_{\alpha\to 1}\widetilde{D}_\alpha(\rho\Vert\sigma)=D(\rho\Vert\sigma)$ (see Proposition~\ref{prop-sand_ren_ent_lim}), we obtain
	\begin{align}
		\lim_{\alpha\to 1^+}\widetilde{\Upsilon}_\alpha(\mathcal{N})&=\inf_{\alpha\in(1,\infty)}\sup_{\psi_{RA}}\inf_{\mathcal{F}\in\mathfrak{F}}\widetilde{D}_\alpha(\mathcal{N}_{A\to B}(\psi_{RA})\Vert\mathcal{F}_{A\to B}(\psi_{RA}))\\
		&=\sup_{\psi_{RA}}\inf_{\alpha\in(1,\infty)}\inf_{\mathcal{F}\in\mathfrak{F}}\widetilde{D}_\alpha(\mathcal{N}_{A\to B}(\psi_{RA})\Vert\mathcal{F}_{A\to B}(\psi_{RA}))\\
		&=\sup_{\psi_{RA}}\inf_{\mathcal{F}\in\mathfrak{F}}\inf_{\alpha\in(1,\infty)}\widetilde{D}_\alpha(\mathcal{N}_{A\to B}(\psi_{RA})\Vert\mathcal{F}_{A\to B}(\psi_{RA}))\\
		&=\sup_{\psi_{RA}}\inf_{\mathcal{F}\in\mathfrak{F}}D(\mathcal{N}_{A\to B}(\psi_{RA})\Vert\mathcal{F}_{A\to B}(\psi_{RA}))\\
		&=\Upsilon(\mathcal{N}),
	\end{align}
	as required, where to obtain the second equality we made use of the minimax theorem in Theorem~\ref{thm-Mosonyi_minimax} to exchange $\inf_{\alpha\in(1,\infty)}$ and $\sup_{\psi_{RA}}$. Specifically, we applied that theorem to the function
	\begin{equation}
		(\alpha,\psi_{RA})\mapsto \inf_{\mathcal{F}\in\mathfrak{F}}\widetilde{D}_\alpha(\mathcal{N}_{A\to B}(\psi_{RA})\Vert\mathcal{F}_{A\to B}(\psi_{RA})),
	\end{equation}
	which is monotonically increasing in the first argument and continuous in the second argument.

%

\section{Proof of the Additivity of \texorpdfstring{$C_\beta(\mathcal{N})$}{Cb}}\label{app-cc-C_beta_additive_pf}

	In this section, we prove that
	\begin{equation}
		C_\beta(\mathcal{N}_1\otimes\mathcal{N}_2)=C_\beta(\mathcal{N}_1)+C_\beta(\mathcal{N}_2).
	\end{equation}
	Noting that $C_\beta(\mathcal{N})=\log_2\beta(\mathcal{N})$, with $\beta(\mathcal{N})$ defined in \eqref{eq-cc_beta_SDP}, the expression above for the additivity of $C_\beta(\mathcal{N})$ is equivalent to the multiplicativity of $\beta(\mathcal{N})$, i.e.,
	\begin{equation}\label{eq-cc-beta_multiplicative}
		\beta(\mathcal{N}_1\otimes\mathcal{N}_2)=\beta(\mathcal{N}_1)\cdot\beta(\mathcal{N}_2).
	\end{equation}
	We now prove that this equality holds. We start with the following lemma.
	
	\begin{Lemma}{lem-tensor_op_ineq}
		Let $A$ and $B$ be Hermitian operators such that $-A\leq B\leq A$, and let $C$ and $D$ be Hermitian operators such that $-C\leq D\leq C$. Then,
		\begin{equation}\label{eq-tensor_op_ineq}
			-A\otimes C\leq B\otimes D\leq A\otimes C.
		\end{equation}
	\end{Lemma}
	
	\begin{Proof}
		The condition $-A\leq B\leq A$ is equivalent to $A-B\geq 0$ and $A+B\geq 0$, and the condition $-C\leq D\leq C$ is equivalent to $C-D\geq 0$ and $C+D\geq 0$. These inequalities imply that
		\begin{align}
			(A-B)\otimes (C-D)&\geq 0,\\
			(A+B)\otimes (C+D)&\geq 0,\\
			(A-B)\otimes (C+D)&\geq 0,\\
			(A+B)\otimes (C-D)&\geq 0.
		\end{align}
		Expanding the left-hand side of these inequalities gives
		\begin{align}
			A\otimes C-B\otimes C-A\otimes D+B\otimes D&\geq 0,\\
			A\otimes C+B\otimes C+A\otimes D+B\otimes D&\geq 0,\\
			A\otimes C-B\otimes C+A\otimes D-B\otimes D&\geq 0,\\
			A\otimes C+B\otimes C-A\otimes D-B\otimes D&\geq 0.
		\end{align}
		Now, adding the first two of these inequalities implies that $A\otimes C+B\otimes D\geq 0$, which is equivalent to the left-hand side of \eqref{eq-tensor_op_ineq}. Adding the last two inequalities implies that $A\otimes C-B\otimes D\geq 0$, which is equivalent to the right-hand side of \eqref{eq-tensor_op_ineq}.
	\end{Proof}
	
	An immediate corollary of the  lemma above is the following: for all Hermitian operators $A,B,C,D$ such that $0\leq B\leq A$ and $0\leq D\leq C$, it holds that
	\begin{equation}\label{eq-tensor_op_ineq_cor}
		0\leq B\otimes D\leq A\otimes C.
	\end{equation}
	Indeed, the condition $0\leq B\leq A$ implies that $A\geq 0$, which is equivalent to $-A\leq 0$, which means that $-A\leq B$ holds. Similarly, we get that $-C\leq D$. So we have that $-A\leq B\leq A$ and $-C\leq D\leq C$. The result then follows by applying the  lemma above.
	
	Now, let us start the proof of \eqref{eq-cc-beta_multiplicative} by showing
	\begin{equation}
		\beta(\mathcal{N}_1\otimes\mathcal{N}_2)\leq\beta(\mathcal{N}_1)\cdot\beta(\mathcal{N}_2).
	\end{equation}
	Recall from \eqref{eq-cc_beta_SDP} that
	\begin{equation}\label{eq-cc_beta_SDP2}
		\beta(\mathcal{N})=\left\{\begin{array}{l l} \text{infimum} & \Tr[S_B] \\
		\text{subject to} & -R_{AB}\leq \T_B[\Gamma^{\mathcal{N}}_{AB}]\leq R_{AB},\\
		& -\mathbbm{1}_A\otimes S_B\leq \T_B[R_{AB}]\leq\mathbbm{1}_A\otimes S_B. \end{array}\right.
	\end{equation}
	Now, let $(R_{AB}^1,S_B^1)$ be a feasible point in the SDP for $\beta(\mathcal{N}_1)$, and let $(R_{AB}^2,S_B^2)$ be a feasible point in the SDP for $\beta(\mathcal{N}_2)$. Each pair thus satisfies the constraints in \eqref{eq-cc_beta_SDP2}. Using Lemma~\ref{lem-tensor_op_ineq}, the first of these constraints implies that
	\begin{equation}
		-R_{A_1B_1}^1\otimes R_{A_2B_2}^2\leq \T_{B_1}[\Gamma^{\mathcal{N}_1}_{A_1B_1}]\otimes \T_{B_2}[\Gamma^{\mathcal{N}_2}_{A_2B_2}]
		\leq R_{A_1B_1}^1\otimes R_{A_2B_2}^2.
	\end{equation}
	Furthermore, observe that
	\begin{align}
		\T_{B_1}[\Gamma^{\mathcal{N}_1}_{A_1B_1}]\otimes \T_{B_2}[\Gamma^{\mathcal{N}_2}_{A_2B_2}]&=\T_{B_1B_2}[\Gamma^{\mathcal{N}_1}_{A_1B_1}\otimes \Gamma^{\mathcal{N}_2}_{A_2B_2}]\\
		&=\T_{B_1B_2}[\Gamma^{\mathcal{N}_1\otimes\mathcal{N}_2}_{A_1A_2B_1B_2}].\label{eq-cc_beta_submultiplicative_pf1}
	\end{align}
	Using this, along with Lemma~\ref{lem-tensor_op_ineq}, the second constraint in \eqref{eq-cc_beta_SDP2} implies that
	\begin{equation}\label{eq-cc_beta_submultiplicative_pf2}
		-\mathbbm{1}_{A_1A_2}\otimes S_{B_1}^1\otimes S_{B_2}^2\leq \T_{B_1B_2}[R_{A_1B_1}^1\otimes R_{A_2B_2}^2]\leq \mathbbm{1}_{A_1A_2}\otimes S_{B_1}^1\otimes S_{B_2}^2.
	\end{equation}
	Now, the inequalities in \eqref{eq-cc_beta_submultiplicative_pf1} and \eqref{eq-cc_beta_submultiplicative_pf2} imply that $(R_{A_1B_1}^1\otimes R_{A_2B_2}^2,S_{B_1}^1\otimes S_{B_2}^2)$ is a feasible point in the SDP for $\beta(\mathcal{N}_1\otimes\mathcal{N}_2)$. This means that
	\begin{equation}\label{eq-cc_beta_submultiplicative_pf3}
		\beta(\mathcal{N}_1\otimes\mathcal{N}_2)\leq \Tr[S_{B_1}^1\otimes S_{B_2}^2]=\Tr[S_{B_1}^1]\Tr[S_{B_2}^2].
	\end{equation}
	Since $(R_{A_1B_1}^1,S_{B_1}^1)$ and $(R_{A_2B_2}^2,S_{B_2}^2)$ are arbitrary feasible points in the SDPs for $\beta(\mathcal{N}_1)$ and $\beta(\mathcal{N}_2)$, respectively, the inequality in \eqref{eq-cc_beta_submultiplicative_pf3} holds for the feasible points achieving $\beta(\mathcal{N}_1)$ and $\beta(\mathcal{N}_2)$. This means that
	\begin{equation}
		\beta(\mathcal{N}_1\otimes\mathcal{N}_2)\leq \beta(\mathcal{N}_1)\cdot \beta(\mathcal{N}_2),
	\end{equation}
	as required.
	
	The prove the reverse inequality, i.e.,
	\begin{equation}\label{eq-cc-beta_supermultiplicative}
		\beta(\mathcal{N}_1\otimes\mathcal{N}_2)\geq \beta(\mathcal{N}_1)\beta(\mathcal{N}_2),
	\end{equation}
	we turn to the SDP dual to the one in \eqref{eq-cc_beta_SDP2}. 
	
	\begin{Lemma}{lem-cc_beta_dual}
		For every quantum channel $\mathcal{N}$, the SDP dual to the SDP in \eqref{eq-cc_beta_SDP2} for $\beta(\mathcal{N})$ is given by
		\begin{equation}
			 \widehat{\beta}(\mathcal{N})\coloneqq\left\{\begin{array}{l l}\text{supremum} & \Tr\!\left[\T_B[\Gamma^{\mathcal{N}}_{AB}](K_{AB}-M_{AB})\right],\\[0.1cm]
			 \text{subject to} & K_{AB}+M_{AB}\leq \T_B[E_{AB}+F_{AB}],\\
			 & E_B+F_B\leq\mathbbm{1}_B,\\
			 & K_{AB},M_{AB},E_{AB},F_{AB}\geq 0.
			 \end{array}\right.
		\end{equation}
		Furthermore, it holds that $\widehat{\beta}(\mathcal{N})=\beta(\mathcal{N})$.
	\end{Lemma}
	
	\begin{Proof}
		Using the formulation of the SDP for $\beta(\mathcal{N})$ as in \eqref{eq-cc_beta_SDP_alt}, the dual to the SDP for $\beta(\mathcal{N})$ is simply
		\begin{equation}
			\widehat{\beta}(\mathcal{N})=\left\{\begin{array}{l l} \text{supremum} & \Tr[DY]\\
			\text{subject to} & \Phi^\dagger(Y)\leq C,\\
			& Y\geq 0,
			\end{array}\right.
		\end{equation}
		where
		\begin{equation}
			C=\begin{pmatrix} \mathbbm{1}_B & 0 \\ 0 & 0_{AB}\end{pmatrix},\quad D=\begin{pmatrix} \T_B[\Gamma^{\mathcal{N}}_{AB}] & 0 & 0 & 0 \\ 0 & -\T_B[\Gamma^{\mathcal{N}}_{AB}] & 0 & 0 \\ 0 & 0 & 0 & 0 \\ 0 & 0 & 0 & 0 \end{pmatrix},
		\end{equation}
		and the map $\Phi$ is defined as
		\begin{align}
			\Phi(X)&=\begin{pmatrix} R_{AB} & 0 & 0 & 0 \\ 0 & R_{AB} & 0 & 0 \\ 0 & 0 & \mathbbm{1}_R\otimes S_B-\T_B[R_{AB}] & 0 \\ 0 & 0 & 0 & \mathbbm{1}_R\otimes S_B+\T_B[R_{AB}] \end{pmatrix},\\
			X&=\begin{pmatrix} S_B & 0\\ 0 & R_{AB}\end{pmatrix}.
		\end{align}
		To determine the adjoint $\Phi^\dagger$, we first observe that, since the operators $C$ and $D$ are block diagonal, the objective function $\Tr[DY]$ of the dual problem involves only the diagonal blocks of $Y$. Furthermore, the fact that $\Phi(X)$ and $X$ are block diagonal means that the condition $\Tr[\Phi(X)Y]=\Tr[X\Phi^\dagger(Y)]$ defining the adjoint map $\Phi^\dagger$ involves only the diagonal blocks of $Y$. Therefore, if the dual problem is feasible, then there is always a feasible point $Y$ that is block diagonal. This means that, without loss of generality, we can let
		\begin{equation}
			Y=\begin{pmatrix} Y_{AB}^1 & 0 & 0 & 0 \\ 0 & Y_{AB}^2 & 0 & 0 \\ 0 & 0 & Y_{AB}^3 & 0 \\ 0 & 0 & 0 & Y_{AB}^4\end{pmatrix},
		\end{equation}
		with $Y_{AB}^1,Y_{AB}^2,Y_{AB}^3,Y_{AB}^4\geq 0$. Then,
		\begin{align}
			&\Tr[\Phi(X)Y]\nonumber\\
			&\quad =\Tr\!\left[\begin{pmatrix} R_{AB} & 0 & 0 & 0 \\ 0 & R_{AB} & 0 & 0 \\ 0 & 0 & \mathbbm{1}_R\otimes S_B-\T_B[R_{AB}] & 0 \\ 0 & 0 & 0 & \mathbbm{1}_R\otimes S_B+\T_B[R_{AB}] \end{pmatrix}\right.\nonumber\\
			&\qquad\qquad\qquad\qquad\qquad\left.\times\begin{pmatrix} Y_{AB}^1 & 0 & 0 & 0 \\ 0 & Y_{AB}^2 & 0 & 0 \\ 0 & 0 & Y_{AB}^3 & 0 \\ 0 & 0 & 0 & Y_{AB}^4 \end{pmatrix}\right]\\
			&\quad =\Tr\!\left[R_{AB}Y_{AB}^1+R_{AB}Y_{AB}^2+(\mathbbm{1}_R\otimes S_B-\T_B[R_{AB}])Y_{AB}^3\right.\nonumber\\
			&\qquad\qquad\qquad\left.+(\mathbbm{1}_R\otimes S_B+\T_B[R_{AB}])Y_{AB}^4\right]\\
			&\quad =\Tr[S_B(Y_B^3+Y_B^4)]+\Tr[R_{AB}(Y_{AB}^1+Y_{AB}^2+\T_B[Y_{AB}^4-Y_{AB}^3])]\\
			&=\Tr\!\left[\begin{pmatrix}S_B & 0 \\ 0 & R_{AB}\end{pmatrix}\begin{pmatrix}Y_B^3+Y_B^4 & 0 \\ 0 & Y_{AB}^1+Y_{AB}^2+\T_B[Y_{AB}^4-Y_{AB}^3]\end{pmatrix}\right].
		\end{align}
		For the last line to be equal to $\Tr[X\Phi^\dagger(Y)]$, we must have
		\begin{equation}
			\Phi^\dagger(Y)=\begin{pmatrix} Y_B^3+Y_B^4 & 0 \\ 0 & Y_{AB}^1+Y_{AB}^2+\T_B[Y_{AB}^4-Y_{AB}^3]\end{pmatrix}.
		\end{equation}
		Then, the condition $\Phi^\dagger(Y)\leq C$ is given by
		\begin{equation}
			\begin{pmatrix} Y_B^3+Y_B^4 & 0 \\ 0 & Y_{AB}^1+Y_{AB}^2+\T_B[Y_{AB}^4-Y_{AB}^3]\end{pmatrix}\leq\begin{pmatrix} \mathbbm{1}_B & 0 \\ 0 & 0_{RB}\end{pmatrix},
		\end{equation}
		which implies that
		\begin{align}
			Y_B^3+Y_B^4&\leq\mathbbm{1}_B,\\
			Y_{AB}^1+Y_{AB}^2&\leq \T_B[Y_{AB}^3-Y_{AB}^4].
		\end{align}
		Then, 
		\begin{align}
			\Tr[DY]&=\Tr[\T_B[\Gamma^{\mathcal{N}}_{AB}]Y_{AB}^1]-\Tr[\T_B[\Gamma^{\mathcal{N}}_{AB}]Y_{AB}^2]\\
			&=\Tr[\T_B[\Gamma^{\mathcal{N}}_{AB}](Y_{AB}^1-Y_{AB}^2)].
		\end{align}
		Therefore, the dual is given by
		\begin{equation}\label{eq-cc-beta_SDP_dual}
			\widehat{\beta}(\mathcal{N})=\left\{\begin{array}{l l} \text{supremum} & \Tr\!\left[\T_B[\Gamma^{\mathcal{N}}_{AB}](K_{AB}-M_{AB})\right]\\
			\text{subject to} & K_{AB}+M_{AB}\leq \T_B[E_{AB}-F_{AB}],\\
			& E_B+F_B\leq\mathbbm{1}_B,\\
			& K_{AB},M_{AB},E_{AB},F_{AB}\geq 0,
			\end{array}\right.
		\end{equation}
		as required.
		
		To show that $\widehat{\beta}(\mathcal{N})=\beta(\mathcal{N})$, we need to check that Slater's condition holds (Theorem~\ref{thm:math-tools:slater-cond}). We can pick $E_{AB}=\frac{\mathbbm{1}_{AB}}{3d_A}$, $F_{AB}=\frac{\mathbbm{1}_{AB}}{6d_A}$, and $K_{AB}=M_{AB}=\frac{\mathbbm{1}_{AB}}{24d_A}$, where $d_A$ is the dimension of the space of the system $A$. Then we have strict inequalities for all of the constraints of the dual problem, which means that Slater's condition holds. The primal $\beta(\mathcal{N})$ and dual $\widehat{\beta}(\mathcal{N})$ are thus equal.
	\end{Proof}
	
	With the dual problem in hand, we can now prove \eqref{eq-cc-beta_supermultiplicative}. Let $(K_{A_1B_1}^1,\allowbreak M_{A_1B_1}^1,\allowbreak E_{A_1B_1}^1,F_{A_1B_1}^1)$ be a feasible point for the dual SDP for $\mathcal{N}_1$, and let $(K_{A_2B_2}^2,\allowbreak M_{A_2B_2}^2,\allowbreak E_{A_2B_2}^2,\allowbreak F_{A_2B_2}^2)$ be a feasible point for the dual SDP for $\mathcal{N}_2$. Then, pick
	\begin{align}
		K_{A_1B_1A_2B_2}&=K_{A_1B_1}^1\otimes K_{A_2B_2}^2+M_{A_1B_1}^1\otimes M_{A_2B_2}^2,\\
		M_{A_1B_1A_2B_2}&=K_{A_1B_1}^1\otimes M_{A_2B_2}^2+M_{A_1B_1}^1\otimes K_{A_2B_2}^2,\\
		E_{A_1B_1A_2B_2}&=E_{A_1B_1}^1\otimes E_{A_2B_2}^2+F_{A_1B_1}^1\otimes F_{A_2B_2}^2,\\
		F_{A_1B_1A_2B_2}&=E_{A_1B_1}^1\otimes F_{A_2B_2}^2+F_{A_1B_1}^1\otimes E_{A_2B_2}^2.
	\end{align}
	Note that $K_{A_1B_1A_2B_2},M_{A_1B_1A_2B_2},E_{A_1B_1A_2B_2},F_{A_1B_1A_2B_2}\geq 0$. Then,
	\begin{align}
		K_{A_1B_1A_2B_2}-M_{A_1B_1A_2B_2}&=(K_{A_1B_1}^1-M_{A_1B_1}^1)\otimes (K_{A_2B_2}^2-M_{A_2B_2}^2),\\
		K_{A_1B_1A_2B_2}+M_{A_1B_1A_2B_2}&=(K_{A_1B_1}^1+M_{A_1B_1}^1)\otimes (K_{A_2B_2}^2+M_{A_2B_2}^2),\\
		E_{A_1B_1A_2B_2}-F_{A_1B_1A_2B_2}&=(E_{A_1B_1}^1-F_{A_1B_1}^1)\otimes (E_{A_2B_2}^2-F_{A_2B_2}^2),\\
		E_{A_1B_1A_2B_2}+F_{A_1B_1A_2B_2}&=(E_{A_1B_1}^1+F_{A_1B_1}^1)\otimes (E_{A_2B_2}^2+F_{A_2B_2}^2).
	\end{align}
	Consider that
	\begin{align}
		& K_{A_1B_1A_2B_2}+M_{A_1B_1A_2B_2}\nonumber\\
		&\quad =(K_{A_1B_1}^1+M_{A_1B_1}^1)\otimes (K_{A_2B_2}^2+M_{A_2B_2}^2)\\
		&\quad\leq \T_{B_1}[E_{A_1B_1}^1-F_{A_1B_1}^1]\otimes\T_{B_2}[E_{A_2B_2}^2-F_{A_2B_2}^2]\\
		&\quad=\T_{B_1B_2}[(E_{A_1B_1}^1-F_{A_1B_1}^1)\otimes (E_{A_2B_2}^2-F_{A_2B_2}^2)]\\
		&\quad=\T_{B_1B_2}[E_{A_1B_1A_2B_2}-F_{A_1B_1A_2B_2}],
	\end{align}
	where the inequality follows from the constraints $K_{A_iB_i}^i,M_{A_iB_i}^i\geq 0$ and $K_{A_iB_i}^i\allowbreak+M_{A_iB_i}^i\leq\T_{B_i}[E_{A_iB_i}^i-F_{A_iB_i}^i]$ for $i\in\{1,2\}$ and from an application of \eqref{eq-tensor_op_ineq_cor}. Furthermore, we have that
	\begin{align}
		E_{B_1B_2}+F_{B_1B_2}&=(E_{B_1}^1+F_{B_1}^1)\otimes (E_{B_2}^2+F_{B_2}^2)\\
		&\leq \mathbbm{1}_{B_1}\otimes\mathbbm{1}_{B_2}\\
		&=\mathbbm{1}_{B_1B_2},
	\end{align}
	where the inequality follows from the constraints $E_{B_i}^i,F_{B_i}^i\geq 0$ and $E_{B_i}^i+F_{B_i}^i\leq\mathbbm{1}_{B_i}$ for $i\in\{1,2\}$ and from an application of \eqref{eq-tensor_op_ineq_cor}. The collection
	\begin{equation}(K_{A_1B_1A_2B_2},M_{A_1B_1A_2B_2},E_{A_1B_1A_2B_2},F_{A_1B_1A_2B_2})
	\end{equation}
	thus constitutes a feasible point for the SDP in \eqref{eq-cc-beta_SDP_dual}. By restricting the optimization in the SDP to this point, we find that
	\begin{align}
		&\beta(\mathcal{N}_1\otimes\mathcal{N}_2)\\
		&\quad\geq \Tr\!\left[\T_{B_1B_2}[\Gamma^{\mathcal{N}_1\otimes\mathcal{N}_2}_{A_1A_2B_1B_2}](K_{A_1B_1A_2B_2}-M_{A_1B_1A_2B_2})\right]\\
		&\quad=\Tr\!\left[\left(\T_{B_1}[\Gamma^{\mathcal{N}_1}_{A_1B_1}]\otimes\T_{B_2}[\Gamma^{\mathcal{N}_2}_{A_2B_2}]\right)(K_{A_1B_1A_2B_2}-M_{A_1B_1A_2B_2})\right]\\
		&=\Tr\!\left[\left(\T_{B_1}[\Gamma^{\mathcal{N}_1}_{A_1B_1}]\otimes\T_{B_2}[\Gamma^{\mathcal{N}_2}_{A_2B_2}]\right)\right.\nonumber\\
		&\qquad\qquad\left.\times \left((K_{A_1B_1}^1-M_{A_1B_1}^1)\otimes(K_{A_2B_2}^2-M_{A_2B_2}^2)\right)\right]\\
		&=\Tr\!\left[\T_{B_1}[\Gamma^{\mathcal{N}_1}_{A_1B_1}](K_{A_1B_1}^1-M_{A_1B_1}^1)\right]\nonumber\\
		&\qquad\qquad\times \Tr\!\left[\T_{B_2}[\Gamma^{\mathcal{N}_2}_{A_2B_2}](K_{A_2B_2}^2-M_{A_2B_2}^2)\right].\label{eq-cc_beta_supermultiplicative_pf}
	\end{align}
	Now, since $(K_{AB}^1,M_{AB}^1,E_{AB}^1,F_{AB}^1)$ $(K_{AB}^2,M_{AB}^2,E_{AB}^2,F_{AB}^2)$ were arbitrary feasible points in the SDPs for $\widehat{\beta}(\mathcal{N}_1)=\beta(\mathcal{N}_1)$ and $\widehat{\beta}(\mathcal{N}_2)=\beta(\mathcal{N}_2)$, respectively, the inequality in \eqref{eq-cc_beta_supermultiplicative_pf} holds for the feasible points achieving $\beta(\mathcal{N}_1)$ and $\beta(\mathcal{N}_2)$. Therefore,
	\begin{equation}
		\beta(\mathcal{N}_1\otimes\mathcal{N}_2)\geq \beta(\mathcal{N}_1)\cdot\beta(\mathcal{N}_2).
	\end{equation}
	We have thus shown that $\beta(\mathcal{N}_1\otimes\mathcal{N}_2)=\beta(\mathcal{N}_1)\cdot\beta(\mathcal{N}_2)$.

\end{subappendices}

\chapter{Entanglement Distillation}\label{chap-ent_distill}

	In the last two chapters, we explored classical communication over quantum channels, in which classical information is encoded into a quantum state, transmitted over a quantum channel, and decoded at the receiving end. In this chapter, we begin our exploration of quantum communication. The goal here is to send quantum information between two spatially separated parties. By ``quantum information,'' we mean that a particular quantum state is transmitted, which is carried physically by some quantum system. As was the case in previous chapters, the particular information carrier is unimportant to us when developing the theoretical results; however, the most common physical manifestation is a photonic encoding, which is useful for long-distance quantum communication.
	
	A basic quantum communication protocol is teleportation, which we developed in Section~\ref{sec-teleportation}. In this protocol, the sender, Alice, initially shares a maximally entangled state with the receiver, Bob. This shared entanglement, along with classical communication, can be used to transmit an arbitrary quantum state perfectly from Alice to Bob. Specifically, if Alice and Bob share a maximally entangled state of Schmidt rank $d\geq 2$, then using this entanglement along with $2\log_2 d$ bits of classical communication, Alice can perfectly transmit an arbitrary state of $\log_2 d$ qubits to Bob. Thus, the quantum teleportation protocol realizes a noiseless quantum channel between Alice and Bob without having to physically transport the particles carrying the quantum information. Of course, this achievement comes at the cost of having a pre-shared maximally entangled state.
	
	How do we obtain maximally entangled states in the first place? In practice, due to noise and other device imperfections, physical sources of entanglement often only produce mixed entangled states, not the pure, maximally entangled states that are needed for quantum teleportation. The purpose of this chapter is to show that many copies of a mixed entangled state can be used to extract, or \textit{distill}, some smaller number of pure maximally entangled states. These distilled maximally entangled states can then be used for quantum communication via the teleportation protocol. This is a basic strategy for quantum communication that we consider in more detail in Chapter~\ref{chap-LOCC-QC}, in order to obtain achievable rates for quantum communication over a quantum channel.
	
\begin{figure}
		\centering
		\includegraphics[scale=0.9]{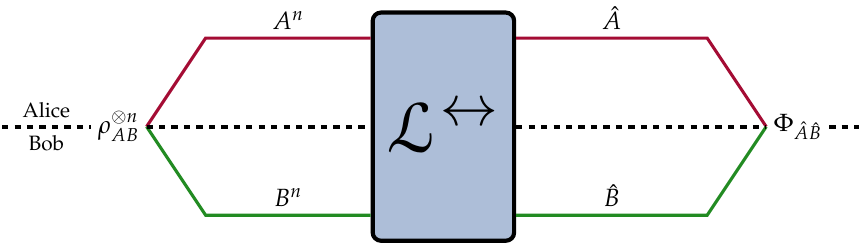}
		\caption{Given a bipartite state $\rho_{AB}$ shared by Alice and Bob, the task of entanglement distillation is to find the largest $d$ for which a maximally entangled state $\ket{\Phi}_{\hat{A}\hat{B}}$ of Schmidt rank $d$ can be extracted from $n$ copies of $\rho_{AB}$ with the smallest possible error, given a two-way LOCC channel $\mathcal{L}_{A^nB^n\to\hat{A}\hat{B}}^{\leftrightarrow}$ between Alice and Bob.}\label{fig-ent_distill}
	\end{figure}

	Similar to quantum teleportation, in which the allowed resources are local operations by Alice and Bob and one-way classical communication from Alice to Bob, in entanglement distillation we allow Alice and Bob local operations with \textit{two-way} classical communication (that is, communication from Alice to Bob and from Bob to Alice); see Figure~\ref{fig-ent_distill}. The goal is to determine, given many copies of a quantum state $\rho_{AB}$, the maximum rate at which maximally entangled states (i.e., ebits) can be distilled approximately from $\rho_{AB}$, where the rate is defined as the ratio $\frac{1}{n}\log_2 d$ between the number $\log_2 d$ of approximate ebits extracted and the initial number $n$ of copies of $\rho_{AB}$. In the asymptotic setting, this maximum rate of entanglement distillation is called the \textit{distillable entanglement of $\rho_{AB}$}, and we denote it by $E_D(\rho_{AB})$. We often write $E_D(\rho_{AB})$ as $E_D(A;B)_{\rho}$ in order to explicitly indicate the bipartition between the subsystems.

	The shared resource state $\rho_{AB}$ for entanglement distillation has to be entangled to begin with in order for entanglement distillation to be successful.  If $\rho_{AB}$ is separable to begin with, then it stays separable after the application of an LOCC channel, and it is not possible to distill high fidelity maximally entangled states from a separable state. This intuitive reasoning becomes formalized in this chapter: some of the entanglement measures from Chapter~\ref{chap-ent_measures} serve as upper bounds on the distillable entanglement, in both the one-shot (Section~\ref{sec-ent_distill_one_shot}) and asymptotic (Section~\ref{sec-ent_distill_asymptotic}) settings. In particular, the Rains relative entropy and squashed entanglement are upper bounds on distillable entanglement. These entanglement measures are currently the best known upper bounds on distillable entanglement, and so we focus exclusively on them for this purpose in this chapter. It is then a trivial consequence of Proposition~\ref{prop:E-meas:log-neg-to-sand-Rains}, \eqref{eq-SEP_PPT_Rains_ineq}, and Proposition~\ref{prop:E-meas:sq-ent-to-EoF} that log-negativity, relative entropy of entanglement, and entanglement of formation are upper bounds on distillable entanglement, and so we do not focus on these entanglement measures in this chapter.
	
	We also consider lower bounds on distillable entanglement in this chapter: the lower bound on distillable entanglement in the one-shot setting in Section~\ref{subsec-ent_distill_one_shot_lower_bound} is based on the concept of \textit{decoupling}, which is an important concept that we discuss later. This lower bound, when applied in the asymptotic setting, leads to the coherent information lower bound $E_D(\rho_{AB})\geq I(A\rangle B)_{\rho}$ on distillable entanglement.


\section{One-Shot Setting}\label{sec-ent_distill_one_shot}

	The one-shot setting for entanglement distillation begins with Alice and Bob sharing the state $\rho_{AB}$. An \textit{entanglement distillation protocol for $\rho_{AB}$} is defined by the pair $(d,\mathcal{L}_{AB\to\hat{A}\hat{B}}^{\leftrightarrow})$, where $d\in\mathbb{N}$, $d\geq 1$, and $\mathcal{L}_{AB\to\hat{A}\hat{B}}^{\leftrightarrow}$ is an LOCC channel (Definition~\ref{def-LOCC}), with $d_{\hat{A}}=d_{\hat{B}}=d$. The \textit{distillation error} $p_{\text{err}}(\mathcal{L}^{\leftrightarrow};\rho_{AB})$ of the protocol is given by the \textit{infidelity}, defined as
	\begin{align}
		p_{\text{err}}(\mathcal{L}^{\leftrightarrow};\rho_{AB})&\coloneqq 1-F(\Phi_{\hat{A}\hat{B}},\mathcal{L}_{AB\to\hat{A}\hat{B}}^{\leftrightarrow}(\rho_{AB}))\\
		&=1-\bra{\Phi}_{\hat{A}\hat{B}}\mathcal{L}_{AB\to\hat{A}\hat{B}}^{\leftrightarrow}(\rho_{AB})\ket{\Phi}_{\hat{A}\hat{B}},\label{eq-ent_distill_one_shot_error}
	\end{align}
	where $\Phi_{\hat{A}\hat{B}}$ is the maximally entangled state of Schmidt rank $d$, defined as
	\begin{equation}\label{eq-max_ent_state_Schmidt_rank_M}
		\ket{\Phi}_{\hat{A}\hat{B}}=\frac{1}{\sqrt{d}}\sum_{i=0}^{d-1}\ket{i}_{\hat{A}}\otimes\ket{i}_{\hat{B}},
	\end{equation}
	and $F$ is the fidelity (see Section~\ref{subsec-fidelity}). To obtain \eqref{eq-ent_distill_one_shot_error}, we used the formula in \eqref{eq-fidelity_pure_mixed} for the fidelity between a pure state and a mixed state. 
	
	The figure of merit in \eqref{eq-ent_distill_one_shot_error} is sensible: the error probability $p_{\text{err}}(\mathcal{L}^{\leftrightarrow};\rho_{AB})$ is equal to the probability that the state $\omega_{\hat{A}\hat{B}}\coloneqq\mathcal{L}^{\leftrightarrow}_{AB\to\hat{A}\hat{B}}(\rho_{AB})$ fails an ``entanglement test,'' which is a measurement defined by the POVM
	\begin{equation}
	\{\Phi_{\hat{A}\hat{B}},\mathbbm{1}_{\hat{A}\hat{B}}-\Phi_{\hat{A}\hat{B}}\}.
	\label{eq:ED:entanglement-test}
	\end{equation}
	 Passing the test corresponds to the measurement operator $\Phi_{\hat{A}\hat{B}}$ and failing corresponds to $\mathbbm{1}_{\hat{A}\hat{B}}-\Phi_{\hat{A}\hat{B}}$. If $1-\Tr[\omega_{\hat{A}\hat{B}}\Phi_{\hat{A}\hat{B}}]\leq\varepsilon\in[0,1]$, and $d_{\hat{A}}=d_{\hat{B}}=d\geq 1$, then we say that the final state $\omega_{\hat{A}\hat{B}}$ contains $\log_2 d$ \textit{$\varepsilon$-approximate ebits}.

	\begin{definition}{$(d,\varepsilon)$ Entanglement Distillation Protocol}{def-Me_ent_dist_protocol}
		An entanglement distillation protocol $(d,\mathcal{L}_{AB\to \hat{A}\hat{B}}^{\leftrightarrow})$ for the state $\rho_{AB}$ is called a \textit{$(d,\varepsilon)$ protocol}, with $\varepsilon\in[0,1]$, if $p_{\text{err}}(\mathcal{L}^{\leftrightarrow};\rho_{AB})\leq\varepsilon$.
	\end{definition}
	
	Given $\varepsilon\in[0,1]$, the largest number $\log_2 d$ of $\varepsilon$-approximate ebits that can be extracted from a state $\rho_{AB}$ among all $(d,\varepsilon)$ entanglement distillation protocols is called the \textit{one-shot $\varepsilon$-distillable entanglement of $\rho_{AB}$}.
	
	\begin{definition}{One-Shot Distillable Entanglement}{def-ent_distill_eps_distillable_ent}
		Given a bipartite state $\rho_{AB}$ and $\varepsilon\in[0,1]$, the \textit{one-shot $\varepsilon$-distillable entanglement of $\rho_{AB}$}, denoted by $E_D^{\varepsilon}(\rho_{AB})\equiv E_D^{\varepsilon}(A;B)_{\rho}$, is defined as
		\begin{equation}\label{eq-ent_distill_one_shot}
			E_D^{\varepsilon}(A;B)_{\rho}\coloneqq \sup_{(d,\mathcal{L}^{\leftrightarrow})}\{\log_2 d: p_{\text{err}}(\mathcal{L}^{\leftrightarrow};\rho_{AB})\leq\varepsilon\},
		\end{equation}
		where the optimization is over all $d\geq 1$ and every LOCC channel $\mathcal{L}^{\leftrightarrow}_{AB\to\hat{A}\hat{B}}$ with $d_{\hat{A}}=d_{\hat{B}}=d$.
	\end{definition}
	
	In addition to finding the largest number $\log_2 d$ of $\varepsilon$-approximate ebits that can be extracted from all $(d,\varepsilon)$ entanglement distilltion protocols for a given $\varepsilon\in[0,1]$, we can consider the following complementary question: for a given $d\geq 1$, what is the lowest value of $\varepsilon$ that can be attained among all $(d,\varepsilon)$ entanglement distillation protocols? In other words, what is the value of
	\begin{equation}\label{eq-ent_distill_one_shot_opt_error}
		\varepsilon^*_D(d;\rho_{AB})\coloneqq\inf_{\mathcal{L}^{\leftrightarrow}_{AB\to\hat{A}\hat{B}}\in \text{LOCC}}\{p_{\text{err}}(\mathcal{L}^{\leftrightarrow};\rho_{AB}): d_{\hat{A}}=d_{\hat{B}}=d\},
	\end{equation}
	where the optimization is over every LOCC channel $\mathcal{L}^{\leftrightarrow}_{AB\to\hat{A}\hat{B}}$, with $d_{\hat{A}}=d_{\hat{B}}=d$? In this book, we focus primarily on the problem of optimizing the number of extracted (approximate) ebits rather than the error, and so our primary quantity of interest is the one-shot distillable entanglement $E_D^{\varepsilon}(\rho_{AB})$.

	Calculating the one-shot distillable entanglement is generally a difficult task, because it involves optimizing over every Schmidt rank $d\geq 1$ of the maximally entangled state $\Phi_{\hat{A}\hat{B}}$ and over every LOCC channel $\mathcal{L}_{AB\to\hat{A}\hat{B}}$, with $d_{\hat{A}}=d_{\hat{B}}=d$. We therefore try to estimate the one-shot distillable entanglement by devising upper and lower bounds. We begin in the next section with upper bounds.

\subsection{Upper Bounds on the Number of Ebits}\label{subsec-ent_distill_one_shot_UB}

In this section, we provide three different upper bounds on one-shot distillable entanglement, based on coherent information, Rains relative entropy, and squashed entanglement.
	Our study of upper bounds on one-shot distillable entanglement begins with coherent information and the following lemma. This can be understood as a fully quantum generalization of Lemma~\ref{lem-eac-meta_conv}, in which we are performing the entanglement test in \eqref{eq:ED:entanglement-test} rather than the comparator test from \eqref{eq-eacc_comparator_test}.
	
	\begin{Lemma}{prop-qc_meta_conv}
		Let $A$ and $B$ be quantum systems with the same dimension $d\geq 1$. Let $\Phi_{AB}$ be a maximally entangled state of Schmidt rank $d$, and let $\omega_{AB}$ be an arbitrary bipartite state. If the probability $\Tr[\Phi_{AB}\omega_{AB}]$ that the state $\omega_{AB}$ passes the entanglement test defined by the POVM $\{\Phi_{AB},\mathbbm{1}_{AB}-\Phi_{AB}\}$ satisfies
		\begin{equation}\label{eq-qc_meta_conv_cond}
			\Tr[\Phi_{AB}\omega_{AB}]\geq 1-\varepsilon
		\end{equation}
		for some $\varepsilon\in[0,1]$, then 
		\begin{equation}\label{eq-entr:coh-inf-ent-test-ineq}
			\log_2 d\leq I_H^{\varepsilon}(A\rangle B)_{\omega},
		\end{equation}
		where $I_H^{\varepsilon}(A\rangle B)_{\omega}$ is the $\varepsilon$-hypothesis testing coherent information (see \eqref{eq-hypo_test_coh_inf_state}).\\
		
		If, in addition to \eqref{eq-qc_meta_conv_cond}, we have that $\Tr_B[\omega_{AB}]=\pi_A=\frac{\mathbbm{1}_A}{d}$, then
		\begin{equation}\label{eq-entr:mut-inf-ent-test-ineq}
			2\log_2 d\leq I_H^{\varepsilon}(A;B)_{\omega}.
		\end{equation}
	\end{Lemma}
	
	\begin{Proof}
		By assumption, we have that
		\begin{equation}\label{eq-qc_meta_conv_pf1}
			F(\Phi_{AB},\rho_{AB})=\Tr[\Phi_{AB}\omega_{AB}]\geq 1-\varepsilon.
		\end{equation}
		Now, for every state $\sigma_B$, we have that
		\begin{align}
			\Tr[\Phi_{AB}(\mathbbm{1}_A\otimes\sigma_{B})]&=\frac{1}{d}\bra{\Gamma}_{AB}(\mathbbm{1}_A\otimes\sigma_{B})\ket{\Gamma}_{AB}\label{eq-qc_meta_conv_pf2}\\
			&=\frac{1}{d}\Tr[\sigma_{B}]\label{eq-qc_meta_conv_pf3}\\
			&=\frac{1}{d},\label{eq-qc_meta_conv_pf4}
		\end{align}
		where we used \eqref{eq-trace_identity}.
		Next, recall from \eqref{eq-hypo_test_coh_inf_state} that
		\begin{equation}
			I_H^{\varepsilon}(A\rangle B)_{\omega}=\inf_{\sigma_{B}}D_H^{\varepsilon}(\omega_{AB}\Vert\mathbbm{1}_{A}\otimes\sigma_{B}),
		\end{equation}
		where 
		\begin{multline}\label{eq-ent_distill_hypo_rel_ent_def}
			D_H^{\varepsilon}(\rho_{AB}\Vert\mathbbm{1}_A\otimes\sigma_B)=-\log_2\inf_{\Lambda_{AB}}\{\Tr[\Lambda_{AB}(\mathbbm{1}_A\otimes\sigma_B)]:0\leq\Lambda_{AB}\leq\mathbbm{1}_{AB},\\\Tr[\Lambda_{AB}\omega_{AB}]\geq 1-\varepsilon\}.
		\end{multline}
		Based on \eqref{eq-qc_meta_conv_pf1}, we see that $\Phi_{AB}$ is a measurement operator satisfying the constraints for the optimization in the definition of $D_H^{\varepsilon}(\omega_{AB}\Vert\mathbbm{1}_A\otimes\sigma_{B})$. Therefore,
		\begin{equation}
			\Tr[\Phi_{AB}(\mathbbm{1}_A\otimes\sigma_{B})]=\frac{1}{d}\geq 2^{-D_H^{\varepsilon}(\omega_{AB}\Vert\mathbbm{1}_A\otimes\sigma_{B})},
		\end{equation}
		which implies that
		\begin{equation}
			\log_2 d\leq D_H^{\varepsilon}(\omega_{AB}\Vert\mathbbm{1}_A\otimes\sigma_{B})
		\end{equation}
		for every state $\sigma_{B}$. Optimizing over $\sigma_{B}$ leads to
		\begin{equation}
			\log_2 d\leq \inf_{\sigma_{B}}D_H^{\varepsilon}(\omega_{AB}\Vert\mathbbm{1}_A\otimes\sigma_{B})=I_H^{\varepsilon}(A\rangle B)_{\omega},
		\end{equation}
		which is precisely \eqref{eq-entr:coh-inf-ent-test-ineq}.
		
		Similarly, 
		\begin{align}
			\Tr[\Phi_{AB}(\pi_A\otimes \sigma_{B})]&=\frac{1}{d}\Tr[\Phi_{AB}(\mathbbm{1}_{A}\otimes\sigma_{B})]\label{eq-entr:mut-inf-2nd-state-1}\\
			&=\frac{1}{d^2}\bra{\Gamma}_{AB}(\mathbbm{1}_A\otimes\sigma_{B})\ket{\Gamma}_{AB}\label{eq-entr:mut-inf-2nd-state-2} \\
			&=\frac{1}{d^2},\label{eq-entr:mut-inf-2nd-state-4}
		\end{align}
		where the last line follows from the same reasoning for \eqref{eq-qc_meta_conv_pf2}--\eqref{eq-qc_meta_conv_pf4}. Next, recall that
		\begin{align}
			I_H^{\varepsilon}(A;B)_{\omega}&=\inf_{\sigma_{B}}D_H^{\varepsilon}(\omega_{AB}\Vert\omega_{A}\otimes\sigma_{B}).
		\end{align}
		Therefore, by definition of the hypothesis testing relative entropy, 
		\begin{equation}
			\Tr[\Phi_{AB}(\pi_A\otimes\sigma_{B})]=\frac{1}{d^2}\geq 2^{-D_H^{\varepsilon}(\omega_{AB}\Vert\pi_A\otimes\sigma_{B})},
		\end{equation}
		which implies that
		\begin{equation}
			2\log_2 d\leq D_H^{\varepsilon}(\omega_{AB}\Vert\pi_A\otimes\sigma_{B}).
		\end{equation}
		Since the state $\sigma_{B}$ is arbitrary, we obtain
		\begin{equation}
			2\log_2 d\leq \inf_{\sigma_{B}}D_H^{\varepsilon}(\omega_{AB}\Vert\pi_A\otimes\sigma_{B})=I_H^{\varepsilon}(A;B)_{\omega},
		\end{equation}
		which is precisely \eqref{eq-entr:mut-inf-ent-test-ineq}. To obtain the last equality, we made use of the assumption $\Tr_{B}[\omega_{AB}]=\pi_A$.
	\end{Proof}
	
	Note that the result of Lemma~\ref{prop-qc_meta_conv} is general and applies to every bipartite state that is close in fidelity to a maximally entangled state. Applying it to the state $\omega_{\hat{A}\hat{B}}=\mathcal{L}_{AB\to\hat{A}\hat{B}}(\rho_{AB})$ at the output of a $(d,\varepsilon)$ entanglement distillation protocol for a state $\rho_{AB}$, we obtain the following result:
	
	\begin{theorem*}{Upper Bound on One-Shot Distillable Entanglement}{prop-ent_distill_one_shot_UB_0}
		Let $\rho_{AB}$ be a bipartite state. For every $(d,\varepsilon)$ entanglement distillation protocol $(d,\mathcal{L}_{AB\to\hat{A}\hat{B}})$ for $\rho_{AB}$, with $\varepsilon\in(0,1]$ and $d_{\hat{A}}=d_{\hat{B}}=d$, the number of $\varepsilon$-approximate ebits extracted at the end of the protocol is bounded from above by the LOCC-optimized $\varepsilon$-hypothesis testing coherent information of $\rho_{AB}$, i.e.,
		\begin{equation}\label{eq-ent_distill_one_shot_UB_0}
			\log_2 d\leq \sup_{\mathcal{L}} I_H^{\varepsilon}(A'\rangle B')_{\mathcal{L}(\rho)},
		\end{equation}
				where the optimization is over every LOCC channel $\mathcal{L}_{AB\to A' B'}$.
		Consequently, for the one-shot $\varepsilon$-distillable entanglement, we obtain
		\begin{equation}\label{eq-ent_distill_one_shot_UB}
			E_D^{\varepsilon}(A;B)_{\rho}\leq \sup_{\mathcal{L}} I_H^{\varepsilon}(A'\rangle B')_{\mathcal{L}(\rho)}.
		\end{equation}
	\end{theorem*}
	
	\begin{Proof}
		For a $(d,\varepsilon)$ entanglement distillation protocol $(d,\mathcal{L}_{AB\to\hat{A}\hat{B}})$ for $\rho_{AB}$, by definition the state $\omega_{\hat{A}\hat{B}}=\mathcal{L}_{AB\to\hat{A}\hat{B}}(\rho_{AB})$ satisfies $\Tr[\Phi_{\hat{A}\hat{B}}\omega_{\hat{A}\hat{B}}]\geq 1-\varepsilon$. Therefore, using \eqref{eq-entr:coh-inf-ent-test-ineq}, we conclude that $\log_2 d\leq I_H^{\varepsilon}(\hat{A}\rangle\hat{B})_{\mathcal{L}(\rho)}$. Since $\mathcal{L}_{AB\to\hat{A}\hat{B}}$ is a particular LOCC channel, we conclude that
		\begin{equation}
		I_H^{\varepsilon}(\hat{A}\rangle\hat{B})_{\mathcal{L}(\rho)} \leq 
		\sup_{\mathcal{L}} I_H^{\varepsilon}(A'\rangle B')_{\mathcal{L}(\rho)}.
		\end{equation}
We thus conclude \eqref{eq-ent_distill_one_shot_UB_0}. Now using the definition of $E_D^{\varepsilon}(A;B)_{\rho}$ in \eqref{eq-ent_distill_one_shot}, we obtain the inequality in \eqref{eq-ent_distill_one_shot_UB}.
	\end{Proof}
	
	
	We now consider an upper bound based on the Rains relative entropy. In order to place an upper bound on the one-shot distillable entanglement $E_D^{\varepsilon}(\rho_{AB})$ for a given state $\rho_{AB}$ and  $\varepsilon\in[0,1]$,  we consider states that are useless for entanglement distillation. This is entirely analogous conceptually to what is done for classical and entanglement-assisted classical communication in the previous two chapters, in which we used the set of replacement channels (which are useless for both of these communication tasks) to place an upper bound on the number of transmitted bits in an $(|\mathcal{M}|,\varepsilon)$ protocol, such  that the upper bound depends only on the channel $\mathcal{N}$ being used for communication.
	
	What states are useless for entanglement distillation? Note that an intuitive necessary condition for successful entanglement distillation is that the initial state $\rho_{AB}$ should be entangled:  if $\rho_{AB}$ is separable, then the output state $\mathcal{L}_{AB\to\hat{A}\hat{B}}(\rho_{AB})$ of an arbitrary entanglement distillation protocol is still a separable state. This suggests that \textit{separable states are useless for entanglement distillation}. To be more precise, separable states are useless for entanglement distillation because they have a very small probability of passing the entanglement test. As we show in Lemma~\ref{lem:fail-ent-test} below, the following bound holds for every separable state $\sigma_{AB}$:
	\begin{equation}\label{eq-ent_distill_sep_useless_0}
		\Tr[\Phi_{AB}\sigma_{AB}]\leq\frac{1}{d}\, ,
	\end{equation}
	where $d$ is the Schmidt rank of $\Phi_{AB}$.
	More generally, operators in the set $\PPT'$, defined as
		\begin{equation}\label{eq:PPT-prime-set_2}
		\PPT'(A\!:\!B)=\{\sigma_{AB}:\sigma_{AB}\geq 0,~\norm{\T_B(\sigma_{AB})}_1\leq 1\},
	\end{equation}
	are also useless for entanglement distillation, in the sense that a statement analgous to \eqref{eq-ent_distill_sep_useless_0} can be made for them. We now prove this statement.
	
	\begin{Lemma}{lem:fail-ent-test}
		Let $A$ and $B$ be quantum systems with the same dimension $d\geq 1$. Let $\Phi_{AB}$ be a maximally entangled state of Schmidt rank $d$. If $\sigma_{AB}\in\PPT'\!\left(A\!:\!B\right)$, then $\Tr[\Phi_{AB}\sigma_{AB}]\leq\frac{1}{d}$.
	\end{Lemma}
	
	\begin{remark}
		Note that  Lemma~\ref{lem:fail-ent-test} implies that $\Tr[\Phi_{AB}\sigma_{AB}]\leq\frac{1}{d}$ for every separable state $\sigma_{AB}$ because $\SEP(A\!:\!B)\subseteq\PPT'(A\!:\!B)$ (recall Figure~\ref{fig-sep_ppt_ppt_prime}).
	\end{remark}
	
	\begin{Proof}
		Using the fact that the partial transpose is self-inverse and self-adjoint, as discussed in \eqref{eq:partial-transpose-self-inverse} and \eqref{eq:partial-transpose-self-adjoint}, respectively, we find that 
		\begin{align}
			\Tr[\Phi_{AB}\sigma_{AB}]  & =\Tr[\T_{B}(\Phi_{AB})\T_{B}(\sigma_{AB})]\\
			& =\frac{1}{d}\Tr[U_B F_{AB} U_B^\dag \T_{B}(\sigma_{AB})],
		\end{align}
		where $F_{AB}$ is the unitary swap operator and $U_B$ is a local unitary acting on system~$B$. Here we applied the identity $\T_{B}(\Phi_{AB}) = \frac{1}{d} U_B F_{AB} U_B^\dag$ from \eqref{eq-partial_transpose_gamma_different_bases}. Since $U_B F_{AB} U_B^\dag$ is a unitary operator, by the variational characterization of the trace norm (see \eqref{eq-trace_norm_variational}), we obtain
		\begin{align}
			\Tr[\Phi_{AB}\sigma_{AB}]& \leq\frac{1}{d}\norm{\T_{B}(\sigma_{AB})}_{1}\\
			& \leq\frac{1}{d},
		\end{align}
		where the last line follows from the definition of the set $\PPT'(A\!:\!B)$ in \eqref{eq:PPT-prime-set_2}.
	\end{Proof}
	
	Due to the fact that $\SEP\subseteq\PPT\subseteq\PPT'$ (see Figure~\ref{fig-sep_ppt_ppt_prime}), Lemma~\ref{lem:fail-ent-test} tells us that both separable and PPT states are useless for entanglement distillation. However, due to the fact that separable states are strictly contained in the set of PPT states for all bipartite states except for qubit-qubit and qubit-qutrit states, it follows that there are PPT entangled states that are useless for entanglement distillation. We elaborate upon this point further in Section~\ref{subsec-bound_ent} below, and we show that the distillable entanglement (in the asymptotic setting) vanishes for all PPT  states.
	
	The steps followed in the proof of Lemma~\ref{lem:fail-ent-test} above are completely analogous to the steps in \eqref{eq-qc_meta_conv_pf2}--\eqref{eq-qc_meta_conv_pf4} and in \eqref{eq-entr:mut-inf-2nd-state-1}--\eqref{eq-entr:mut-inf-2nd-state-4} of the proof of Lemma~\ref{prop-qc_meta_conv}. Therefore, just as Lemma~\ref{prop-qc_meta_conv} was used to establish Proposition~\ref{prop-ent_distill_one_shot_UB_0}, we can use Lemma~\ref{lem:fail-ent-test} to place an upper bound on the number $\log_2 d$ of approximate ebits in a bipartite state $\omega_{AB}$.
	
	\begin{proposition}{prop:core-meta-converse-privacy_a}
		Fix $\varepsilon\in [0,1]$, and let $A$ and $B$ be quantum systems with the same dimension $d\geq 1$. Fix a
maximally entangled state $\Phi_{AB}$ of Schmidt rank $d$. Let $\omega_{AB}$ be an $\varepsilon$-approximate maximally entangled state, in the sense that%
		\begin{equation}\label{eq:approx-max-ent}
			F(\Phi_{AB},\omega_{AB})=\Tr[\Phi_{AB}\omega_{AB}]\geq 1-\varepsilon.
		\end{equation}
		Then, the number $\log_{2}d$ of $\varepsilon$-approximate ebits in $\omega_{AB}$ is bounded from above as follows:
		\begin{equation}\label{eq:eps-Rains-rel-ent-approx-max-ent}
			\log_2 d\leq R_H^{\varepsilon}(A;B)_{\omega},
		\end{equation}
		where $R_H^{\varepsilon}(A;B)_{\omega}$ is the $\varepsilon$-hypothesis testing Rains relative entropy of $\omega_{AB}$ (see \eqref{def:eps-Rains-rel-entr}).
	\end{proposition}

	\begin{Proof}
		Let $\sigma_{AB}$ be an arbitrary operator in $\operatorname{PPT}'(A\!:\!B)$. The inequality $\Tr[\Phi_{AB}\omega_{AB}]\geq1-\varepsilon$ guarantees that $\omega_{AB}$ passes the entanglement test with probability greater than $1-\varepsilon$. Thus, we conclude that $\Phi_{AB}$ is a particular measurement operator satisfying the constraints for $2^{-D_H^{\varepsilon}(\omega_{AB}\Vert\sigma_{AB})}$. Applying Lemma~\ref{lem:fail-ent-test} and the definition of $D_H^{\varepsilon}(\omega_{AB}\Vert\sigma_{AB})$, we conclude that%
		\begin{equation}
			2^{-D_H^{\varepsilon}(\omega_{AB}\Vert\sigma_{AB})}\leq\Tr[\Phi_{AB}\sigma_{AB}]\leq\frac{1}{d}.
		\end{equation}
		Rearranging this leads to
		\begin{equation}
			\log_2 d\leq D_H^{\varepsilon}(\omega_{AB}\Vert\sigma_{AB})
		\end{equation}
		Since this inequality holds for every operator $\sigma_{AB}\in\PPT'(A\!:\!B)$, we conclude that%
		\begin{equation}
			\log_2 d\leq\inf_{\sigma_{AB}\in\PPT'(A:B)}D_H^{\varepsilon}(\omega_{AB}\Vert\sigma_{AB})=R_H^{\varepsilon}(A;B)_{\omega} ,
		\end{equation}
		where we used the definition of $R_H^{\varepsilon}(A;B)_{\omega}$ in \eqref{def:eps-Rains-rel-entr} to obtain the last equality.
			\end{Proof}%
		%
		

	A consequence of Proposition~\ref{prop:core-meta-converse-privacy_a} is the following upper bound on the one-shot distillable entanglement of $\rho_{AB}$. 
	
	\begin{theorem*}{Rains Upper Bound on One-Shot Distillable Entanglement}{cor-ent_distill_one-shot_UB_alt}
		Let $\rho_{AB}$ be a bipartite state. For every $(d,\varepsilon)$ entanglement distillation protocol for $\rho_{AB}$, with $\varepsilon\in[0,1]$, we have that 
		\begin{equation}\label{eq-ent_distill_one_shot_UB_alt_0}
			\log_2 d\leq R_H^{\varepsilon}(A;B)_{\rho}.
		\end{equation}
		Consequently, for the one-shot distillable entanglement, we have
		\begin{equation}\label{eq-ent_distill_one_shot_UB_alt}
			E_D^{\varepsilon}(A;B)_{\rho}\leq R_H^{\varepsilon}(A;B)_{\rho}
		\end{equation}
		for every state $\rho_{AB}$ and all $\varepsilon\in[0,1]$.
	\end{theorem*}
	
	
	\begin{Proof}
		Consider a $(d,\varepsilon)$ entanglement distillation protocol for $\rho_{AB}$ with the corresponding LOCC channel $\mathcal{L}_{AB\to\hat{A}\hat{B}}$. Then, by definition, we have that
		\begin{equation}
			p_{\text{err}}(\mathcal{L};\rho_{AB})=1-\Tr[\Phi_{\hat{A}\hat{B}}\mathcal{L}_{AB\to\hat{A}\hat{B}}(\rho_{AB})]\leq \varepsilon.
		\end{equation}
		Letting $\omega_{\hat{A}\hat{B}}=\mathcal{L}_{AB\to\hat{A}\hat{B}}(\rho_{AB})$, we have that $\Tr[\Phi_{\hat{A}\hat{B}}\omega_{\hat{A}\hat{B}}]\geq 1-\varepsilon$. The output state $\omega_{\hat{A}\hat{B}}$ of the entanglement distillation protocol therefore satisfies the conditions of Proposition~\ref{prop:core-meta-converse-privacy_a}, which means that
		\begin{equation}
			\log_2 d\leq R_H^{\varepsilon}(\hat{A};\hat{B})_{\omega} .
		\end{equation}
		Now,  it follows from Proposition~\ref{prop-gen_Rains_rel_ent_properties} that $R_H^{\varepsilon}(\hat{A};\hat{B})$ is an entanglement measure. Thus, it satisfies the data-processing inequality under LOCC channels, which means that $R_H^{\varepsilon}(\hat{A};\hat{B})_{\omega}\leq R_H^{\varepsilon}(A;B)_{\rho}$. We thus have $\log_2 d\leq R_H^{\varepsilon}(A;B)_{\rho}$. Since this inequality holds for all $d\geq 1$ and every LOCC channel $\mathcal{L}_{AB\to\hat{A}\hat{B}}$, by definition of one-shot $\varepsilon$-distillable entanglement, we obtain $E_D^{\varepsilon}(A;B)_{\rho}\leq R_H^{\varepsilon}(A;B)_{\rho}$, as required. 
	\end{Proof}
	
	
	The main step that allows us to conclude the bound in \eqref{eq-ent_distill_one_shot_UB_alt_0} in terms of the state $\rho_{AB}$ alone is the fact that $R_H^{\varepsilon}$ is an entanglement measure, meaning that it is monotone non-increasing under LOCC channels. In other words, the set of $\PPT'$ operators is preserved under LOCC channels. This fact is not true for the set $\{\mathbbm{1}_A\otimes\sigma_B:\sigma_B\in\Density(\mathcal{H})\}$ appearing in the optimization that defines the $\varepsilon$-hypothesis testing coherent information, meaning that the operator $\mathcal{L}_{AB\to\hat{A}\hat{B}}(\mathbbm{1}_A\otimes\sigma_B)$ (where $\mathcal{L}_{AB\to\hat{A}\hat{B}}$ is an LOCC channel) is not in general of the form $\mathbbm{1}_{\hat{B}}\otimes\tau_{\hat{B}}$ for some state $\tau_{\hat{B}}$. We therefore cannot  use the data-processing inequality for the bound in \eqref{prop-ent_distill_one_shot_UB_0} in order to reduce it to $\log_2 d\leq I_H^{\varepsilon}(A\rangle B)_{\rho}$.

	Combining Theorems~\ref{prop-ent_distill_one_shot_UB_0} and \ref{cor-ent_distill_one-shot_UB_alt} with Propositions~\ref{prop-hypo_to_rel_ent} and \ref{prop:sandwich-to-htre} immediately leads to the following upper bounds.

	\begin{corollary}{cor-ent_distill_one_shot_UB_all}
		Let $\rho_{AB}$ be a bipartite state, and let $\varepsilon\in[0,\sfrac{1}{2})$. For every $(d,\varepsilon)$ entanglement distillation protocol $(d,\mathcal{L}_{AB\to\hat{A}\hat{B}})$, with $d_{\hat{A}}=d_{\hat{B}}=d$, we have that
		\begin{equation}
			\log_2 d \leq \frac{1}{1-2\varepsilon}\left(\sup_{\mathcal{L}}I(A'\rangle B')_{\mathcal{L}(\rho)}+h_2(\varepsilon)\right),\label{eq-ent_distill_one_shot_UB_weak_conv} 
		\end{equation}
		where $I( A'\rangle B')_{\mathcal{L}(\rho)}$
		is the coherent information of $\mathcal{L}_{AB\to A' B'}(\rho_{AB})$ (see \eqref{eq-coher_inf_opt}). For $\varepsilon\in[0,1)$,
		\begin{equation}
		\log_2 d \leq \widetilde{R}_{\alpha}(A;B)_{\rho}+\frac{\alpha}{\alpha-1}\log_2\!\left(\frac{1}{1-\varepsilon}\right)\quad\forall~\alpha>1,\label{eq-ent_distill_one_shot_UB_str_conv}
		\end{equation}
		where
		\begin{equation}
			\widetilde{R}_{\alpha}(A;B)_{\rho}=\inf_{\sigma_{AB}\in\PPT'(A:B)}\widetilde{D}_{\alpha}(\rho_{AB}\Vert\sigma_{AB})
		\end{equation}
		is the sandwiched R\'{e}nyi Rains relative entropy of $\rho_{AB}$ (see \eqref{eq-sand_renyi_Rains_rel_ent}).
	\end{corollary}
	
	
	\begin{Proof}
		Combining the upper bound in \eqref{eq-ent_distill_one_shot_UB_0} from Theorem~\ref{prop-ent_distill_one_shot_UB_0} with the upper bound in \eqref{eq:WR-bound-to-rel-ent} from Proposition~\ref{prop-hypo_to_rel_ent}, we obtain
		\begin{align}
			\log_2 d&\leq \sup_{ \mathcal{L}}I_H^{\varepsilon}(A'\rangle B')_{\mathcal{L}(\rho)}\\
			&\leq \frac{1}{1-\varepsilon}\left(\sup_{ \mathcal{L}}I(A'\rangle B')_{\mathcal{L}(\rho)}+h_2(\varepsilon)\right)+\frac{\varepsilon}{1-\varepsilon}\log_2 d.
		\end{align}
		Rearranging this and simplifying leads to
		\begin{equation}
			\log_2 d\leq\frac{1}{1-2\varepsilon}\left(\sup_{ \mathcal{L}}I(A'\rangle B')_{\mathcal{L}(\rho)}+h_2(\varepsilon)\right),
		\end{equation}
		which is the inequality in \eqref{eq-ent_distill_one_shot_UB_weak_conv}. The inequality in \eqref{eq-ent_distill_one_shot_UB_str_conv} follows from Theorem~\ref{cor-ent_distill_one-shot_UB_alt} and \eqref{eq:sandwich-to-htre} in Proposition~\ref{prop:sandwich-to-htre}.
	\end{Proof}
	
	Since the upper bounds in \eqref{eq-ent_distill_one_shot_UB_weak_conv} and \eqref{eq-ent_distill_one_shot_UB_str_conv} hold for all $(d,\varepsilon)$ entanglement distillation protocols, we conclude the following upper bounds on distillable entanglement: 
	\begin{align}
		E_D^{\varepsilon}(A;B)_{\rho}&\leq\frac{1}{1-2\varepsilon}\left(\sup_{\mathcal{L}}I(A'\rangle B')_{\mathcal{L}(\rho)}+h_2(\varepsilon)\right),\\
		E_D^{\varepsilon}(A;B)_{\rho}&\leq \widetilde{R}_{\alpha}(A;B)_{\rho}+\frac{\alpha}{\alpha-1}\log_2\!\left(\frac{1}{1-\varepsilon}\right)\quad\forall~\alpha>1.
	\end{align}
	
	We finally turn to squashed entanglement and establish it as an upper bound on one-shot distillable entanglement:

\begin{theorem*}{Squashed Entanglement Upper Bound on One-Shot Distillable Entanglement}{thm:ED:squashed-ent-upp-1-shot-DE}
Let $\rho_{AB}$ be a bipartite state. For every $(d, \varepsilon)$ entanglement distillation protocol for $\rho_{AB}$, with $\varepsilon\in[0,1)$, we have that
\begin{equation}
\log_2 d \leq \frac{1}{1-\sqrt{\varepsilon}}\left(E_{\operatorname{sq}}(A;B)_{\rho} +g_2(\sqrt{\varepsilon})\right),
\end{equation}
where $E_{\operatorname{sq}}(A;B)_{\rho}$ is the squashed entanglement of $\rho_{AB}$ (see \eqref{eq-squashed_entanglement_0}) and  $g_2(\delta)\coloneqq (\delta+1)\log_2(\delta+1) - \delta \log_2 \delta$.
		Consequently, for the one-shot distillable entanglement, we have
		\begin{equation}\label{eq-ent_distill_one_shot_UB_alt_squashed}
			E_D^{\varepsilon}(A;B)_{\rho}\leq \frac{1}{1-\sqrt{\varepsilon}}\left(E_{\operatorname{sq}}(A;B)_{\rho} +g_2(\sqrt{\varepsilon})\right)
		\end{equation}
		for every state $\rho_{AB}$ and  $\varepsilon\in[0,1)$.

\end{theorem*}

\begin{Proof}
Consider a $(d,\varepsilon)$ entanglement distillation protocol for $\rho_{AB}$ with the corresponding LOCC channel $\mathcal{L}_{AB\to\hat{A}\hat{B}}$. From the LOCC monotonicity of squashed entanglement (Theorem~\ref{thm:LAQC-mono-LOCC-sq}), we have that
\begin{equation}
E_{\operatorname{sq}}(\hat{A};\hat{B})_{\omega} \leq E_{\operatorname{sq}}(A;B)_{\rho},
\end{equation}
where $\omega_{\hat{A}\hat{B}}=\mathcal{L}_{AB\to\hat{A}\hat{B}}(\rho_{AB})$.
Continuing, by definition, the following inequality holds
		\begin{equation}
			p_{\text{err}}(\mathcal{L};\rho_{AB})=1-\Tr[\Phi_{\hat{A}\hat{B}}\mathcal{L}_{AB\to\hat{A}\hat{B}}(\rho_{AB})]\leq \varepsilon.
		\end{equation}
		It follows that $\Tr[\Phi_{\hat{A}\hat{B}}\omega_{\hat{A}\hat{B}}]\geq 1-\varepsilon$, which is the same as
		\begin{equation}
		F(\Phi_{\hat{A}\hat{B}},\omega_{\hat{A}\hat{B}})\geq 1-\varepsilon.
		\end{equation}
				As a consequence of Proposition~\ref{prop:LAQC-cont-sq-unif}, we find that%
		\begin{align}
			  E_{\operatorname{sq}}(\hat{A};\hat{B})_{\omega}
			&  \geq E_{\operatorname{sq}}(\hat{A};\hat{B})_{\Phi}-\left(  \sqrt{\varepsilon}\log_{2}\min\left\{  \left\vert \hat{A}\right\vert ,\left\vert \hat{B}\right\vert\right\}  +g_{2}(\sqrt{\varepsilon})\right) \\
			&  =\log_{2}d-\left(  \sqrt{\varepsilon}\log_{2}d+g_{2}(\sqrt{\varepsilon })\right) \\
			&  =(1-\sqrt{\varepsilon})\log_{2}d-g_{2}(\sqrt{\varepsilon}).
		\end{align}
		The first equality follows from Proposition~\ref{prop:LAQC-reduction-for-pure-squashed}.\ We can finally rearrange the established inequality $E_{\operatorname{sq}}(A;B)_{\rho}\geq(1-\sqrt{\varepsilon})\log_{2}d-g_{2}(\sqrt{\varepsilon})$ to be in the form stated in the theorem.
\end{Proof}

\subsection{Lower Bound on the Number of Ebits via Decoupling}\label{subsec-ent_distill_one_shot_lower_bound}

\begin{figure}
		\centering
		\includegraphics[scale=0.9]{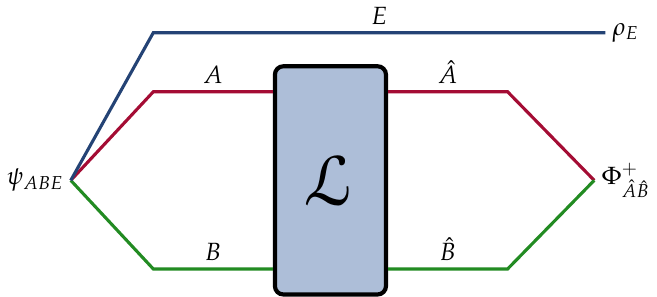}
		\caption{The task of entanglement distillation can be understood from the perspective of decoupling: given a bipartite state $\rho_{AB}$ with purification $\psi_{ABE}$, the entanglement distillation protocol given by the LOCC channel $\mathcal{L}$ should result in the pure maximally entangled state $\Phi_{\hat{A}\hat{B}}$, which by definition is in tensor product with the environment, so that the joint state is $\Phi_{\hat{A}\hat{B}}\otimes\rho_E$, with $\rho_E=\Tr_{AB}[\psi_{ABE}]$.}\label{fig-ent_distill_decoupling}
	\end{figure}

	Having found upper bounds on one-shot distillable entanglement, we now focus on lower bounds. In order to find a lower bound on distillable entanglement, we have to find an explicit entanglement distillation protocol that works for an arbitrary bipartite state $\rho_{AB}$ and an arbitrary error $\varepsilon\in(0,1)$. Recall that the goal of entanglement distillation is for two parties, Alice and Bob, to make use of LOCC to transform their shared bipartite state $\rho_{AB}$ to the maximally entangled state $\Phi_{\hat{A}\hat{B}}$, for some $d_{\hat{A}}=d_{\hat{B}}=d\geq 1$. Now, the initial state $\rho_{AB}$ has some purification $\ket{\psi^{\rho}}_{ABE}$, with the purifying system $E$ in general correlated with $A$ and $B$. However, because the maximally entangled state is pure, every purification of it must be of the form $\Phi_{\hat{A}\hat{B}}\otimes\phi_{E'}$, with the system $E'$ in tensor product with systems $\hat{A}$ and $\hat{B}$. Since the goal of entanglement distillation is to distill a maximally entangled state and the maximally entangled state has this property, we can thus think of entanglement distillation as the task of \textit{decoupling} $A$ and $B$ from their environment $E$; see Figure~\ref{fig-ent_distill_decoupling}. Our lower bound on one-shot distillable entanglement tells us what dimension $d$ of $\hat{A}$ and $\hat{B}$ is sufficient in order to achieve this decoupling up to error $\varepsilon$.
	
	The lower bound on one-shot distillable entanglement that we determine in this section is expressed in terms of an information measure that is derived from a \textit{smoothed} version of the max-relative entropy $D_{\text{max}}$, which we briefly cover in Section~\ref{subsec-smooth_max_rel_ent}. Recall from Definition~\ref{def-max_rel_ent} that the max-relative entropy of a state $\rho$ and a positive semi-definite operator $\sigma$ is defined as
	\begin{equation}
		D_{\text{max}}(\rho\Vert\sigma)=\log_2\norm{\sigma^{-\frac{1}{2}}\rho\sigma^{-\frac{1}{2}}}_{\infty}.
	\end{equation}
	Using this, we define the conditional min-entropy as
	\begin{equation}
		H_{\min}(A|B)_{\rho}\coloneqq -\inf_{\sigma_B}D_{\max}(\rho_{AB}\Vert\mathbbm{1}_A\otimes\sigma_B)
	\end{equation}
	where the optimization is with respect to every state $\sigma_B$.
	
	From Definition~\ref{def-smooth_max_rel_ent}, the smooth max-relative entropy is defined as
	\begin{equation}
		D_{\max}^{\varepsilon}(\rho\Vert\sigma)=\inf_{\widetilde{\rho}\in\mathcal{B}^{\varepsilon}(\rho)}D_{\max}(\widetilde{\rho}\Vert\sigma),
	\end{equation}
	where we recall from \eqref{eq-eps_smooth_ball} that
	\begin{equation}
		\mathcal{B}^{\varepsilon}(\rho)=\{\widetilde{\rho}:P(\rho,\widetilde{\rho})\leq\varepsilon\},
	\end{equation}
	and the sine distance $P(\rho,\widetilde{\rho})$ is given by (see Definition~\ref{def-purified_distance})
	\begin{equation}
		P(\rho,\widetilde{\rho})=\sqrt{1-F(\rho,\widetilde{\rho})}.
	\end{equation}
	Using the smooth max-relative entropy, we define the smooth conditional min-entropy of $\rho_{AB}$ as
	\begin{align}
		H_{\min}^{\varepsilon}(A|B)_{\rho}&=-\inf_{\sigma_B}D_{\max}^{\varepsilon}(\rho_{AB}\Vert\mathbbm{1}_A\otimes\sigma_B)\\
		&=\sup_{\widetilde{\rho}\in\mathcal{B}^{\varepsilon}(\rho)}H_{\min}(A|B)_{\widetilde{\rho}}
	\end{align}
	for all $\varepsilon\in(0,1)$, where the optimization in the first line is with respect to states $\sigma_B$.
	
	We also need the \textit{smooth conditional max-entropy} of $\rho_{AB}$, which is defined as
	\begin{equation}
		H_{\max}^{\varepsilon}(A|B)_{\rho}\coloneqq\inf_{\widetilde{\rho}\in\mathcal{B}^{\varepsilon}(\rho)}H_{\max}(A|B)_{\rho}
	\end{equation}
	for all $\varepsilon\in(0,1)$, where
	\begin{align}
		H_{\max}(A|B)_{\rho}&\coloneqq \widetilde{H}_{\frac{1}{2}}(A|B)_{\rho}\\
		&=-\inf_{\sigma_B}\widetilde{D}_{\frac{1}{2}}(\rho_{AB}\Vert\mathbbm{1}_A\otimes\sigma_B)\\
		&=\sup_{\sigma_B}\log_2F(\rho_{AB},\mathbbm{1}_A\otimes\sigma_B).
	\end{align}
	The obtain the last equality, we made use of \eqref{eq-sand_rel_half_fidelity}, and the optimization therein is with respect to states $\sigma_B$.
	
	For every state $\rho_{AB}$, the conditional min- and max-entropies are related as follows:
	\begin{align}
		H_{\max}(A|B)_{\rho}&=-H_{\min}(A|E)_{\psi}\label{eq-Hmax_Hmin_pure}\\
		H_{\max}^{\varepsilon}(A|B)_{\rho}&=-H_{\min}^{\varepsilon}(A|E)_{\psi},\label{eq-Hmax_Hmin_pure_smooth}
	\end{align}
	for all $\varepsilon\in(0,1)$, where $\psi_{ABE}$ is a purification of $\rho_{AB}$.
	
	Both the conditional min-entropy and the smooth conditional min-entropy can be formulated as semi-definite programs. The same is true for the conditional max-entropy and the smooth conditional max-entropy. Please consult the  Bibliographic Notes in Section~\ref{sec:ed:bib-notes} for details.
	
	Finally, we need the quantity
	\begin{align}
		\widetilde{H}_2(A|B)_{\rho}&\coloneqq -\inf_{\sigma_B}\widetilde{D}_2(\rho_{AB}\Vert\mathbbm{1}_A\otimes\sigma_B)\\
		&=-\inf_{\sigma_B}\log_2\Tr\!\left[\left(\sigma_B^{-\frac{1}{4}}\rho_{AB}\sigma_B^{-\frac{1}{4}}\right)^2\right],
	\end{align}
	which is known as the ``conditional collision entropy'' of $\rho_{AB}$, where the optimization is with respect to states $\sigma_B$. Due to monotonicity in $\alpha$ of the sandwiched R\'{e}nyi relative entropy $\widetilde{D}_{\alpha}$ (see Proposition~\ref{prop-sand_rel_ent_properties}), we have that
	\begin{equation}\label{eq-ent_distill_Hmin_vs_H2}
		H_{\min}(A|B)_{\rho}\leq\widetilde{H}_2(A|B)_{\rho}
	\end{equation}
	for every bipartite state $\rho_{AB}$.
	
	We are now ready to state a lower bound on one-shot distillable entanglement.

	\begin{theorem*}{Lower Bound on One-Shot Distillable Entanglement}{prop-ent_dist_lower_bound}
		Let $\rho_{AB}$ be a quantum state. For all $\varepsilon\in(0,1]$ and $\eta\in[0,\sqrt{\varepsilon})$, there exists a $(d,\varepsilon)$ one-way entanglement distillation protocol for $\rho_{AB}$ with
		\begin{equation}
			\log_2 d=-H_{\max}^{\sqrt{\varepsilon}-\eta}(A|B)_{\rho}+4\log_2\eta.
			\label{eq:ED:lower-bnd-one-shot-ed}
		\end{equation}
		Consequently, for the one-shot distillable entanglement of $\rho_{AB}$, we have
		\begin{equation}
			E_D^{\varepsilon}(A;B)_{\rho}\geq \sup_{\mathcal{L}}\left(-H_{\max}^{\sqrt{\varepsilon}-\eta}(A'|B')_{\mathcal{L}(\rho)}\right)+4\log_2\eta
		\end{equation}
		for all $\varepsilon\in[0,1]$ and $\eta\in[0,\sqrt{\varepsilon})$, where the optimization is over  every LOCC channel $\mathcal{L}_{AB\to A'B'}$.
	\end{theorem*}
	
	In order to prove Theorem~\ref{prop-ent_dist_lower_bound}, we exhibit an entanglement distillation protocol $(d,\mathcal{L}_{AB\to\hat{A}\hat{B}})$, with $d_{\hat{A}}=d_{\hat{B}}=d=2^{-H_{\max}^{\sqrt{\varepsilon}-\eta}(A|B)_{\rho}+4\log_2\eta}$, such that $p_{\text{err}}(\mathcal{L};\rho_{AB})\leq\varepsilon$ for all $\varepsilon\in(0,1]$. To this end, we construct a \textit{one-way} LOCC channel $\mathcal{L}_{AB\to\hat{A}\hat{B}}^{\rightarrow}$ of the form
	\begin{equation}\label{eq-ent_distill_one_shot_lower_bound_pf4_0}
		\mathcal{L}_{AB\to\hat{A}\hat{B}}^{\rightarrow}=\sum_{x\in\mathcal{X}}\mathcal{E}^x_{A\to\hat{A}}\otimes\mathcal{D}_{B\to\hat{B}}^x,
	\end{equation}
	where $\mathcal{X}$ is a finite alphabet, $\{\mathcal{E}_{A\to\hat{A}}^x\}_{x\in\mathcal{X}}$ is a set of completely positive maps such that $\sum_{x\in\mathcal{X}}\mathcal{E}_{A\to\hat{A}}^x$ is trace preserving, and $\{\mathcal{D}_{B\to\hat{B}}^x\}_{x\in\mathcal{X}}$ is a set of channels. Recall from Section~\ref{subsec-LOCC_channels} that every one-way Alice-to-Bob LOCC channel can be written as
	\begin{equation}\label{eq-ent_distill_one_shot_lower_bound_pf4}
		\mathcal{L}_{AB\to\hat{A}\hat{B}}^{\rightarrow}=\mathcal{D}_{BX_B\to\hat{B}}\circ\mathcal{C}_{X_A\to X_B}\circ\mathcal{E}_{A\to\hat{A}X_A},
	\end{equation}
	where $\mathcal{E}_{A\to\hat{A}X_A}$ is a local channel for Alice that corresponds to the quantum instrument given by the maps $\{\mathcal{E}_{A\to\hat{A}}^x\}_{x\in\mathcal{X}}$, i.e., (see also \eqref{eq-instrument_output})
	\begin{equation}
		\mathcal{E}_{A\to\hat{A}X_A}(\rho_A)=\sum_{x\in\mathcal{X}}\mathcal{E}_{A\to\hat{A}}^x(\rho_A)\otimes\ket{x}\!\bra{x}_{X_A}.
	\end{equation}
	The map $\mathcal{C}_{X_A\to X_B}$ is a noiseless classical channel that transforms the classical register $X_A$, held by Alice, to the classical register $X_B$ (which is simply a copy of $X_A$), held by Bob. The final channel $\mathcal{D}_{BX_B\to\hat{B}}$ is a local channel for Bob defined as
	\begin{equation}\label{eq-ent_distill_one_shot_lower_bound_pf7}
		\mathcal{D}_{BX_B\to\hat{B}}(\rho_B \otimes \ket{x}\!\bra{x}_{X_B})=\mathcal{D}_{B\to\hat{B}}^x(\rho_B)
	\end{equation}
	for all $x\in\mathcal{X}$. In the proof below, we explicitly construct the CP maps $\{\mathcal{E}_{A\to\hat{A}}\}_{x\in\mathcal{X}}$ and the channels $\{\mathcal{D}_{B\to\hat{B}}^x\}_{x\in\mathcal{X}}$.
	
	In addition to providing explicit forms for the channels $\mathcal{E}_{A\to\hat{A}X_A}$ and $\mathcal{D}_{BX_B\to\hat{B}}$ involved in the LOCC channel $\mathcal{L}_{AB\to\hat{A}\hat{B}}^{\rightarrow}$ in \eqref{eq-ent_distill_one_shot_lower_bound_pf4_0}, we  prove that $p_{\text{err}}(\mathcal{L};\rho_{AB})\leq\varepsilon$. To do this, we make use of the following general decoupling result, which we explain and prove in Appendix~\ref{app-ent_distill_decoupling_pf}.
	
	\begin{theorem}{thm-ent_distill_decoupling}
		Given a subnormalized state $\rho_{AE}$ (i.e., $\Tr[\rho_{AE}]\leq 1$), and a completely positive map $\mathcal{N}_{A\to A'}$, the following bound holds 
		\begin{multline}\label{eq-ent_distill_decoupling}
			\int_{U_A}\norm{\mathcal{N}_{A\to A'}(U_A\rho_{AE}U_A^\dagger)-\Phi_{A'}^{\mathcal{N}}\otimes\rho_E}_1~\text{d}U_A\\\leq 2^{-\frac{1}{2}\widetilde{H}_2(A|E)_{\rho}-\frac{1}{2}\widetilde{H}_2(A|A')_{\Phi^{\mathcal{N}}}},
		\end{multline}
		where $\Phi_{A'}^{\mathcal{N}}\coloneqq\Tr_{A}[\Phi_{AA'}^{\mathcal{N}}]$, $\Phi_{AA'}^{\mathcal{N}}$ is given by $\mathcal{N}_{A\to A'}(\Phi_{AA})$, $\rho_E\coloneqq\Tr_A[\rho_{AE}]$, and the integral is over unitaries $U_A$ acting on system $A$, taken with respect to the Haar measure.
	\end{theorem}
	
	\begin{Proof}
		See Appendix~\ref{app-ent_distill_decoupling_pf}.
	\end{Proof}
	
	\begin{remark}
		The integral in \eqref{eq-ent_distill_decoupling} with respect to the Haar measure should be thought of as a uniform average over the continuous set of all unitaries $U_A$ acting on the system $A$. In other words, the integral is analogous to a uniform average over a discrete set of unitaries. In fact, for every dimension $d\geq 1$, there exists a set $\{U_x\}_{x\in\mathcal{X}}$ of unitaries, called a \textit{unitary one-design}, such that
		\begin{equation}\label{eq-unitary_1_design}
			\int_U UXU^\dagger~\text{d}U=\frac{1}{|\mathcal{X}|}\sum_{x\in\mathcal{X}}U_x X U_x^\dagger=\Tr[X]\frac{\mathbbm{1}}{d}
		\end{equation}
		for every operator $X$. An example of a unitary one-design is the Heisenberg-Weyl operators $\{W_{z,x}:0\leq z,x\leq d-1\}$, which are defined in \eqref{eq-Heisenberg_Weyl_operators}--\eqref{eq-gen_X_pauli}. Please consult the Bibliographic Notes in Section~\ref{sec:ed:bib-notes} for more information about integration over unitaries with respect to the Haar measure and about unitary designs. A simple argument for   the right-most equality in \eqref{eq-unitary_1_design} goes as follows. First, it follows for a unitary $V$ that
		\begin{equation}
		V \left(\int_U UXU^\dagger~\text{d}U\right) V^\dag =  \int_U VU X(VU)^\dagger~\text{d}U = \int_U U X U^\dagger~\text{d}U,
		\end{equation}
		where the final equality follows because the Haar measure is a unitarily invariant measure. So it follows that the operator $\int_U UXU^\dagger~\text{d}U$ commutes with all unitaries. The only operator that does so is the identity operator, which implies that $\int_U UXU^\dagger~\text{d}U \propto \mathbbm{1}$. The normalization factor of $\Tr[X]/d$ follows by taking a trace of the left-hand side, using its cyclicity, and the fact that $\text{d}U$ is a probability measure.
	\end{remark}

\subsubsection*{Proof of Theorem~\ref{prop-ent_dist_lower_bound}}

	Fix $\eta \in (0,\sqrt{\varepsilon})$, and  let $\psi_{ABE}$ be a purification of $\rho_{AB}$, with $\rho_{AE}\coloneqq\Tr_B[\psi_{ABE}]$. Fix $d$ such that \eqref{eq:ED:lower-bnd-one-shot-ed} holds. Then, starting from the expression in \eqref{eq:ED:lower-bnd-one-shot-ed} and using \eqref{eq-Hmax_Hmin_pure_smooth}, we find that
	\begin{equation}
		\log_2 d=H_{\min}^{\sqrt{\varepsilon}-\eta}(A|E)_{\rho}+4\log_2\eta.
	\end{equation}
	Now, pick a state $\widetilde{\rho}_{AE}\in\mathcal{B}^{\sqrt{\varepsilon}-\eta}(\rho_{AE})$ such that
	\begin{equation}
	H_{\min}(A|E)_{\widetilde{\rho}} = H_{\min}^{\sqrt{\varepsilon}-\eta}(A|E)_{\rho}.
	\label{eq:ED:non-smooth-equal-smooth-decoupling}
	\end{equation}
	Then, using \eqref{eq-ent_distill_Hmin_vs_H2}, we find that
	\begin{align}
		\log_2 d &=H_{\min}(A|E)_{\widetilde{\rho}}+4\log_2\eta\\
		&\leq \widetilde{H}_2(A|E)_{\widetilde{\rho}}+4\log_2\eta\\
		&=\widetilde{H}_2(A|E)_{\widetilde{\rho}}-2\log_2\!\left(\frac{1}{\eta^2}\right),
	\end{align}
	where
	\begin{equation}
		\widetilde{H}_2(A|E)_{\widetilde{\rho}}=-\inf_{\sigma_E}\log_2\Tr\!\left[\left(\sigma_E^{-\frac{1}{4}}\widetilde{\rho}_{AE}\sigma_E^{-\frac{1}{4}}\right)^2\right]
	\end{equation}

	We now define a channel $\mathcal{E}_{A\to\hat{A}X_A}$ as follows:
	\begin{equation}\label{eq-ent_distill_one_shot_lower_bound_pf5}
		\mathcal{E}_{A\to\hat{A}X_A}(\cdot)\coloneqq\sum_{x\in\mathcal{X}}V_{A\to\hat{A}}^x\Pi_A^x(\cdot)\Pi_A^xV_{A\to\hat{A}}^{x\dagger}\otimes\ket{x}\!\bra{x}_{X_A},
	\end{equation}
	where $d_{\hat{A}}=d$, $\mathcal{X}$ is a finite alphabet with\footnote{We assume that $d$ divides $d_A$ without loss of generality. If it is not the case, then we can repeat the whole analysis with the system $A$ embedded in a larger Hilbert space that is divided by $d$. We would also need to start with a state $\widetilde{\rho}_{AE}$ such that \eqref{eq:ED:non-smooth-equal-smooth-decoupling} holds with the definition $H_{\min}^{\sqrt{\varepsilon}-\eta}(A|E)_{\rho} \coloneqq -\inf_{\sigma_E} D^{\sqrt{\varepsilon}-\eta}_{\max}(\rho_{AE}\Vert \Pi_A \otimes \sigma_B) $, where $\Pi_A$ is the projection onto the support of $\Tr_E[\rho_{AE}]$. Then we would repeat the whole analysis with such a $\widetilde{\rho}_{AE}$. We do not go into further details here.} $|\mathcal{X}|=d_{X_A}=\frac{d_A}{d}$, $\{\Pi_A^x\}_{x\in\mathcal{X}}$ is a set of projectors such that $\sum_{x\in\mathcal{X}}\Pi_A^x=\mathbbm{1}_A$, and $\{V_{A\to\hat{A}}^x\}_{x\in\mathcal{X}}$ is a set of isometries. So we have that
	\begin{equation}
		\mathcal{E}_{A\to\hat{A}}^x(\cdot)\coloneqq V_{A\to\hat{A}}^x\Pi_A^x(\cdot)\Pi_A^xV_{A\to\hat{A}}^{x\dagger}
	\end{equation}
	for all $x\in\mathcal{X}$. Each isometry $V_{A\to\hat{A}}^x$ takes the subspace of $\mathcal{H}_A$ onto which $\Pi_A^x$ projects and embeds it into the fixed $d$-dimensional space $\mathcal{H}_{\hat{A}}$, i.e., $\text{im}(V_{A\to\hat{A}}^x)=\mathcal{H}_{\hat{A}}$ for all $x\in\mathcal{X}$. The projectors $\{\Pi_A^x\}_{x\in\mathcal{X}}$ correspond to a measurement of the input state, with $\Pi_A^x(\cdot)\Pi_A^x$ the (unnormalized) post-measurement state, and the isometries $\{V_{A\to\hat{A}}^x\}_{x\in\mathcal{X}}$ can be thought of as encodings of the initial system $A$ into the system $\hat{A}$ on which one share of the desired maximally entangled state $\Phi_{\hat{A}\hat{B}}$ is to be generated. We have
	\begin{align}
		\mathcal{E}_{A\to\hat{A}X_A}(\rho_{AB})&=\sum_{x\in\mathcal{X}}\mathcal{E}_{A\to\hat{A}}(\rho_{AB})\otimes\ket{x}\!\bra{x}_{X_A}\\
		&=\sum_{x\in\mathcal{X}}\mathcal{V}_{A\to\hat{A}}^x(\Pi_A^x\rho_{AB}\Pi_A^x)\otimes\ket{x}\!\bra{x}_{X_A}\\
		&=\sum_{x\in\mathcal{X}}p(x)\omega_{\hat{A}B}^x\otimes\ket{x}\!\bra{x}_{X_A},
	\end{align}
	where
	\begin{align}
		p(x)&\coloneqq\Tr[\Pi_A\rho_A],\\
		\omega_{\hat{A}B}^x&\coloneqq\frac{1}{p(x)}\mathcal{V}_{A\to\hat{A}}^x(\Pi_A^x\rho_{AB}\Pi_A^x).\label{eq-ent_distill_one_shot_lower_bound_pf8}
	\end{align}
	
	Now, by Theorem~\ref{thm-ent_distill_decoupling}, the following inequality holds
	\begin{multline}
		\int_{U_A}\norm{\mathcal{E}_{A\to\hat{A}X_A}(U_A\widetilde{\rho}_{AE}U_A^\dagger)-\Phi_{\hat{A}X_A}^{\mathcal{E}}\otimes\widetilde{\rho}_E}_1~\text{d}U_A\\ \leq 2^{-\frac{1}{2}\widetilde{H}_2(A|E)_{\widetilde{\rho}}-\frac{1}{2}\widetilde{H}_2(A|\hat{A}X_A)_{\Phi^{\mathcal{E}}}},
	\end{multline}
	where $\Phi^{\mathcal{E}}_{A\hat{A}X_A}$ is the Choi state of $\mathcal{E}_{A\to\hat{A}X_A}$.
	Given that $\log_2 d\leq \widetilde{H}_2(A|E)_{\widetilde{\rho}}-2\log_2\!\left(\frac{1}{\eta^2}\right)$, we obtain
	\begin{equation}
		2^{-\frac{1}{2}\widetilde{H}_2(A|E)_{\widetilde{\rho}}}\leq \frac{1}{\sqrt{d}}\eta^2.
	\end{equation}
	We also have that
	\begin{align}
		\widetilde{H}_2(A|\hat{A}X_A)_{\Phi^{\mathcal{E}}}&=\sup_{\sigma_{\hat{A}X_A}}\left\{-\log_2\Tr\!\left[\left(\sigma_{\hat{A}X_A}^{-\frac{1}{4}}\Phi_{A\hat{A}X_A}^{\mathcal{E}}\sigma_{\hat{A}X_A}^{-\frac{1}{4}}\right)^2\right]\right\}\\
		&\geq -\log_2 d.
	\end{align}
	Indeed, in the optimization over $\sigma_{\hat{A}X_{A}}$, take%
\begin{align}
\sigma_{\hat{A}X_{A}}  & =\frac{1}{d_{X_{A}}}\sum_{x\in\mathcal{X}}\pi
_{\hat{A}}\otimes|x\rangle\!\langle x|_{X_{A}}\\
& =\pi_{\hat{A}}\otimes\pi_{X_{A}}\\
& =\frac{1}{d_{A}}\mathbbm{1}_{\hat{A}}\otimes \mathbbm{1}_{X_{A}}.
\end{align}

With this choice of $\sigma_{\hat{A}X_{A}}$, we find that%
\begin{align}
& \operatorname{Tr}\!\left[  \left(  \sigma_{\hat{A}X_{A}}^{-\frac{1}{4}}%
\Phi_{A\hat{A}X_{A}}^{\mathcal{E}}\sigma_{\hat{A}X_{A}}^{-\frac{1}{4}}\right)
^{2}\right]  \notag \\
& =d_{A}\operatorname{Tr}\!\left[  \left(  \Phi_{A\hat{A}X_{A}}^{\mathcal{E}%
}\right)  ^{2}\right]  \\
& =d_{A}\operatorname{Tr}\!\left[  \left(  \sum_{x\in\mathcal{X}}V_{A\rightarrow
\hat{A}}^{x}\Pi_{A}^{x}\Phi_{AA}\Pi_{A}^{x}(V_{A\rightarrow\hat{A}}^{x}%
)^{\dag}\otimes|x\rangle\!\langle x|_{X_{A}}\right)  ^{2}\right]  \\
& =d_{A}\sum_{x\in\mathcal{X}}\operatorname{Tr}\!\left[  \left(  V_{A\rightarrow
\hat{A}}^{x}\Pi_{A}^{x}\Phi_{AA}\Pi_{A}^{x}(V_{A\rightarrow\hat{A}}^{x}%
)^{\dag}\right)  ^{2}\right]  \\
& =d_{A}\sum_{x\in\mathcal{X}}\operatorname{Tr}\!\left[  \left(  V_{A\rightarrow
\hat{A}}^{x}\Pi_{A}^{x}\frac{1}{d_{A}}\sum_{i,j=0}^{d_{A}-1}|i\rangle\!\langle
j|_{A}\otimes|i\rangle\!\langle j|_{A}\Pi_{A}^{x}(V_{A\rightarrow\hat{A}}%
^{x})^{\dag}\right)  ^{2}\right]  \\
& =d_{A}\sum_{x\in\mathcal{X}}\operatorname{Tr}\!\left[  \left(  \frac{1}{d_{A}%
}\sum_{i,j=0}^{d-1}|i\rangle\!\langle j|_{\hat{A}}\otimes|i\rangle\!\langle
j|_{\hat{A}}\right)  ^{2}\right]  \\
& =d_{A}\sum_{x\in\mathcal{X}}\operatorname{Tr}\!\left[  \left(  \frac{d}{d_{A}%
}\frac{1}{d}\sum_{i,j=0}^{d-1}|i\rangle\!\langle j|_{\hat{A}}\otimes
|i\rangle\!\langle j|_{\hat{A}}\right)  ^{2}\right]  \\
& =\frac{d^{2}}{d_{A}}\sum_{x\in\mathcal{X}}\operatorname{Tr}\!\left[  \left(
\frac{1}{d}\sum_{i,j=0}^{d-1}|i\rangle\!\langle j|_{\hat{A}}\otimes
|i\rangle\!\langle j|_{\hat{A}}\right)  ^{2}\right]  \\
& =\frac{d^{2}}{d_{A}}\left\vert \mathcal{X}\right\vert \\
& =d,
\end{align}
	where we recall that $d_{X_A}=\left\vert \mathcal{X}\right\vert=\frac{d_A}{d}$.
We thus have
	\begin{equation}
		2^{-\frac{1}{2}\widetilde{H}_2(A|\hat{A}X_A)_{\Phi^{\mathcal{E}}}}\leq \sqrt{d},
	\end{equation}
	which means that
	\begin{equation}
		\int_{U_A}\norm{\mathcal{E}_{A\to\hat{A}X_A}(U_A\widetilde{\rho}_{AE}U_A^\dagger)-\Phi_{\hat{A}X_A}^{\mathcal{E}}\otimes\widetilde{\rho}_E}_1~\text{d}U_A\leq \eta^2.
	\end{equation}
	Note that
	\begin{align}
		\Phi_{\hat{A}X_A}^{\mathcal{E}}&=\frac{1}{d_A}\sum_{x\in\mathcal{X}}V_{A\to\hat{A}}^x\Pi_A^x\mathbbm{1}_A\Pi_A^xV_{A\to\hat{A}}^{x\dagger}\otimes\ket{x}\!\bra{x}_{X_A}\\
		&=\frac{1}{d_A}\sum_{x\in\mathcal{X}}V_{A\to\hat{A}}^x\Pi_A^xV_{A\to\hat{A}}^{x\dagger}\otimes\ket{x}\!\bra{x}_{X_A}\\
		&=\frac{1}{d_A}\mathbbm{1}_{\hat{A}}\otimes\mathbbm{1}_{X_A}\\
		&=\pi_{\hat{A}}\otimes\pi_{X_A},
	\end{align}
	where 
	the last equality follows because $d_{X_A}=\frac{d_A}{d}$. So we have
	\begin{equation}
		\int_{U_A}\norm{\mathcal{E}_{A\to\hat{A}X_A}(U_A\widetilde{\rho}_{AE}U_A^\dagger)-\pi_{\hat{A}}\otimes\pi_{X_A}\otimes\widetilde{\rho}_E}_1~\text{d}U_A\leq\eta^2.
	\end{equation}
	Now, since the average over a set of elements is never less than the minimum over the same set, we have that
	\begin{multline}
		\int_{U_A}\norm{\mathcal{E}_{A\to\hat{A}X_A}(U_A\widetilde{\rho}_{AE}U_A^\dagger)-\pi_{\hat{A}}\otimes\pi_{X_A}\otimes\widetilde{\rho}_E}_1~\text{d}U_A\\
		\geq \min_{U_A}\norm{\mathcal{E}_{A\to\hat{A}X_A}(U_A\widetilde{\rho}_{AE}U_A^\dagger)-\pi_{\hat{A}}\otimes\pi_{X_A}\otimes\widetilde{\rho}_E}_1
	\end{multline}
	This implies that there exists a unitary $U_A$ (in particular, one that achieves the minimum on the right-hand side of the above inequality) such that
	\begin{multline}
		\eta^2\geq \int_{U_A}\norm{\mathcal{E}_{A\to\hat{A}X_A}(U_A\widetilde{\rho}_{AE}U_A^\dagger)-\pi_{\hat{A}}\otimes\pi_{X_A}\otimes\widetilde{\rho}_E}_1~\text{d}U_A\\
		\geq \norm{\mathcal{E}_{A\to\hat{A}X_A}(U_A\widetilde{\rho}_{AE}U_A^\dagger)-\pi_{\hat{A}}\otimes\pi_{X_A}\otimes\widetilde{\rho}_E}_1
	\end{multline}
	Now, let
	\begin{align}
		\widetilde{\omega}_{\hat{A}X_AE}=\mathcal{E}_{A\to\hat{A}X_A}(U_A\widetilde{\rho}_{AE}U_A^\dagger),&\quad \widetilde{\tau}_{\hat{A}X_AE}=\pi_{\hat{A}}\otimes\pi_{X_A}\otimes\widetilde{\rho}_E,\\
		\omega_{\hat{A}X_AE}=\mathcal{E}_{A\to\hat{A}X_A}(U_A\rho_{AE}U_A^\dagger),&\quad \tau_{\hat{A}X_AE}=\pi_{\hat{A}}\otimes\pi_{X_A}\otimes \rho_E.
	\end{align}
	Then, by the Fuchs--van de Graaf inequality (see \eqref{eq-Fuchs_van_de_graaf}), and by the definition of the sine distance (see Definition~\ref{def-purified_distance}), we have that
	\begin{align}
		\eta^2&\geq \norm{\widetilde{\omega}_{\hat{A}X_AE}-\widetilde{\tau}_{\hat{A}X_AE}}_1\\
		&\geq 2-2\sqrt{F(\widetilde{\omega}_{\hat{A}X_AE},\widetilde{\tau}_{\hat{A}X_AE})}\\
		&\geq 2-2\sqrt{1-P(\widetilde{\omega}_{\hat{A}X_AE},\widetilde{\tau}_{\hat{A}X_AE})^2},
	\end{align}
	which implies that
	\begin{align}
		P(\widetilde{\omega}_{\hat{A}X_AE},\widetilde{\tau}_{\hat{A}X_AE})&\leq \sqrt{1-\left(1-\frac{\eta^2}{2}\right)^2}\\
		&= \eta\sqrt{1-\frac{\eta^2}{4}}\\
		&\leq\eta.\label{eq-ent_distill_one_shot_lower_bound_pf2}
	\end{align}
	Then, by the triangle inequality for sine distance (Lemma~\ref{lem-sine-distance-triangle}), we have
	\begin{align}
		P(\omega_{\hat{A}X_AE},\widetilde{\tau}_{\hat{A}X_AE})&\leq P(\omega_{\hat{A}X_AE},\widetilde{\omega}_{\hat{A}X_AE})+P(\widetilde{\omega}_{\hat{A}X_AE},\widetilde{\tau}_{\hat{A}X_AE})\\
		&\leq P(\rho_{AE},\widetilde{\rho}_{AE})+\eta\\
		&\leq \sqrt{\varepsilon}-\eta+\eta\\
		&=\sqrt{\varepsilon},\label{eq-ent_distill_one_shot_lower_bound_pf3}
	\end{align}
	where the second inequality follows from the data-processing inequality for the sine distance, unitary invariance of the sine distance, and the inequality in \eqref{eq-ent_distill_one_shot_lower_bound_pf2}. To obtain the last inequality, we used the definition of the state $\widetilde{\rho}_{AE}$ as one that is $(\sqrt{\varepsilon}-\eta)$-close to $\rho_{AE}$ in sine distance. We can write the inequality in \eqref{eq-ent_distill_one_shot_lower_bound_pf3} in terms of fidelity as
	\begin{equation}
		F(\omega_{\hat{A}X_AE},\widetilde{\tau}_{\hat{A}X_AE})\geq 1-\varepsilon.
	\end{equation}
	Note that both $\omega_{\hat{A}X_AE}$ and $\widetilde{\tau}_{\hat{A}X_AE}$ are classical-quantum states because $X_A$ is a classical register. In particular,
	\begin{align}
		\omega_{\hat{A}X_AE}&=\sum_{x\in\mathcal{X}}p(x)\ket{x}\!\bra{x}_{X_A}\otimes\omega_{\hat{A}E}^x,\\
		p(x)&=\Tr[\Pi_A^x U_A\rho_A U_A^\dag],\\
		\omega_{\hat{A}E}^x&=\frac{1}{p(x)}\mathcal{V}_{A\to\hat{A}}^x(\Pi_A^x U_A\rho_{AE} U_A^\dag \Pi_A^x).
	\end{align}
	Also,
	\begin{equation}
		\widetilde{\tau}_{\hat{A}X_AE}=\pi_{\hat{A}}\otimes\frac{1}{d_{X_A}}\sum_{x\in\mathcal{X}}\ket{x}\!\bra{x}_{X_A}\otimes\widetilde{\rho}_E.
	\end{equation}
	Then, using the direct-sum property of the root fidelity (see \eqref{eq-direct_sum_root_fid}), we obtain
	\begin{align}
		F(\omega_{\hat{A}X_AE},\widetilde{\tau}_{\hat{A}X_AE})&=\left(\sqrt{F}(\omega_{\hat{A}X_AE},\widetilde{\tau}_{\hat{A}X_AE})\right)^2\\
		&=\left(\sum_{x\in\mathcal{X}}\sqrt{\frac{p(x)}{d_{X_A}}}\sqrt{F}(\omega_{\hat{A}E}^x,\pi_{\hat{A}}\otimes\widetilde{\rho}_E)\right)^2.
	\end{align}
	Now, let
	\begin{equation}
		\psi_{\hat{A}BE}^x\coloneqq \frac{1}{p(x)}\mathcal{V}_{A\to\hat{A}}^x(\Pi_A^x U_A\psi_{ABE}U_A^\dag \Pi_A^x)
	\end{equation}
	be a purification of $\omega_{\hat{A}E}^x$ for all $x\in\mathcal{X}$, and let $\Phi_{\hat{A}\hat{B}}\otimes\widetilde{\phi}_{EB'}$ be a purification of $\pi_{\hat{A}}\otimes\widetilde{\rho}_E$. Then, by Uhlmann's theorem (Theorem~\ref{thm-Uhlmann_fidelity}), for every $x\in\mathcal{X}$ there exists an isometric channel $\mathcal{W}_{B\to\hat{B}B'}^x$ such that 
	\begin{equation}
		\sqrt{F}(\omega_{\hat{A}E}^x,\pi_{\hat{A}}\otimes\widetilde{\rho}_E)=\sqrt{F}(\mathcal{W}_{B\to\hat{B}B'}^x(\psi_{\hat{A}BE}^x),\Phi_{\hat{A}\hat{B}}\otimes\widetilde{\phi}_{EB'})
	\end{equation}
	for all $x\in\mathcal{X}$. Using the set $\{\mathcal{W}_{B\to\hat{B}B'}^x\}_{x\in\mathcal{X}}$, we define the quantum channels $\{\mathcal{D}_{B\to\hat{B}}^x\}_{x\in\mathcal{X}}$ as follows:
	\begin{equation}\label{eq-ent_distill_one_shot_lower_bound_pf6}
		\mathcal{D}_{B\to\hat{B}}^x\coloneqq\Tr_{B'}\circ\mathcal{W}_{B\to\hat{B}B'}^x.
	\end{equation}
	By the data-processing inequality for fidelity (see Theorem~\ref{thm-fidelity_monotone}) under the partial trace channel $\Tr_{EB'}$, we obtain
	\begin{align}
		&\sqrt{F}(\omega_{\hat{A}E}^x,\pi_{\hat{A}}\otimes\widetilde{\rho}_E)\nonumber\\
		&\qquad=\sqrt{F}(\mathcal{W}_{B\to\hat{B}B'}^x(\psi_{\hat{A}BE}^x),\Phi_{\hat{A}\hat{B}}\otimes\widetilde{\phi}_{EB'})\\
		&\qquad\leq\sqrt{F}(\Tr_{EB'}[\mathcal{W}_{B\to\hat{B}B'}^x(\psi_{\hat{A}BE}^x)],\Tr_{EB'}[\Phi_{\hat{A}\hat{B}}\otimes\widetilde{\phi}_{EB'}])\\
		&\qquad= \sqrt{F}(\mathcal{D}_{B\to\hat{B}}^x(\omega_{\hat{A}B}^x),\Phi_{\hat{A}\hat{B}}),\label{eq-ent_distill_one_shot_lower_bound_pf9}
	\end{align}
	for all $x\in\mathcal{X}$, where we recall the definition of $\omega_{\hat{A}B}^x$ from \eqref{eq-ent_distill_one_shot_lower_bound_pf8}. Since the inequality in \eqref{eq-ent_distill_one_shot_lower_bound_pf9} holds for all $x\in\mathcal{X}$, we have that
	\begin{multline}\label{eq-ent_distill_one_shot_lower_bound_pf10}
		\left(\sum_{x\in\mathcal{X}}\sqrt{\frac{p(x)}{d_{X_A}}}\sqrt{F}(\omega_{\hat{A}E}^x,\pi_{\hat{A}}\otimes\widetilde{\rho}_E)\right)^2\\
		\leq \left(\sum_{x\in\mathcal{X}}\sqrt{\frac{p(x)}{d_{X_A}}}\sqrt{F}(\mathcal{D}_{B\to\hat{B}}^x(\omega_{\hat{A}B}^x),\Phi_{\hat{A}\hat{B}})\right)^2.
	\end{multline}
	
	Now, the final state of Alice and Bob after executing the LOCC channel defined by \eqref{eq-ent_distill_one_shot_lower_bound_pf4}, with $\mathcal{E}_{A\to\hat{A}X_A}$ defined by \eqref{eq-ent_distill_one_shot_lower_bound_pf5} and $\mathcal{D}_{BX_B\to\hat{B}}$ defined by \eqref{eq-ent_distill_one_shot_lower_bound_pf7} and \eqref{eq-ent_distill_one_shot_lower_bound_pf6}, is
	\begin{align}
		\omega_{\hat{A}\hat{B}}&=(\mathcal{D}_{X_BB\to\hat{B}}\circ\mathcal{C}_{X_A\to X_B}\circ\mathcal{E}_{A\to\hat{A}X_A})(\rho_{AB})\\
		&=\sum_{x\in\mathcal{X}}\mathcal{D}_{X_BB\to\hat{B}}(\mathcal{E}_{A\to\hat{A}}^x(\rho_{AB})\otimes\ket{x}\!\bra{x}_{X_B})\\
		&=\sum_{x\in\mathcal{X}}(\mathcal{E}_{A\to\hat{A}}^x\otimes\mathcal{D}_{B\to\hat{B}}^x)(\rho_{AB})\\
		&=\Tr_{X_B}\!\left[\sum_{x\in\mathcal{X}}p(x)\ket{x}\!\bra{x}_{X_B}\otimes\mathcal{D}_{B\to\hat{B}}^x(\omega_{\hat{A}B}^x)\right],
	\end{align}
	where in the third equality we recognize the required form in \eqref{eq-ent_distill_one_shot_lower_bound_pf4_0} for a one-way Alice-to-Bob LOCC channel, and in the last inequality we made use of \eqref{eq-ent_distill_one_shot_lower_bound_pf8}. Using the form of $\omega_{\hat{A}\hat{B}}$ in the last equality, along with all of the developments above, we finally obtain
	\begin{align}
		&F(\omega_{\hat{A}\hat{B}},\Phi_{\hat{A}\hat{B}})\nonumber\\
		&\quad=F\!\left(\Tr_{X_B}\!\left[\sum_{x\in\mathcal{X}}p(x)\ket{x}\!\bra{x}_{X_B}\otimes\mathcal{D}_{B\to\hat{B}}^x(\omega_{\hat{A}B}^x)\right],\Tr_{X_B}[\pi_{X_B}\otimes\Phi_{\hat{A}\hat{B}}]\right)\\
		&\quad\geq F\!\left(\sum_{x\in\mathcal{X}}p(x)\ket{x}\!\bra{x}_{X_B}\otimes\mathcal{D}_{B\to\hat{B}}^x(\omega_{\hat{A}B}^x),\frac{1}{d_{X_A}}\sum_{x\in\mathcal{X}}\ket{x}\!\bra{x}_{X_A}\otimes\Phi_{\hat{A}\hat{B}}\right)\\
		&\quad=\left(\sum_{x\in\mathcal{X}}\sqrt{\frac{p(x)}{d_{X_A}}}\sqrt{F}(\mathcal{D}_{B\to\hat{B}}^x(\omega_{\hat{A}B}^x),\Phi_{\hat{A}\hat{B}})\right)^2\\
		&\quad=\left(\sum_{x\in\mathcal{X}}\sqrt{\frac{p(x)}{d_{X_A}}}\sqrt{F}((\Tr_{EB'}\circ\mathcal{W}_{B\to\hat{B}B'}^x)(\psi_{\hat{A}BE}^x),\Tr_{EB'}[\Phi_{\hat{A}\hat{B}}\otimes\widetilde{\phi}_{EB'}])\right)^2\\
		&\quad\geq\left(\sum_{x\in\mathcal{X}}\sqrt{\frac{p(x)}{d_{X_A}}}\sqrt{F}(\mathcal{W}_{B\to\hat{B}B'}^x(\psi_{\hat{A}BE}^x),\Phi_{\hat{A}\hat{B}}\otimes\widetilde{\phi}_{EB'})\right)^2\\
		&\quad=\left(\sum_{x\in\mathcal{X}}\sqrt{\frac{p(x)}{d_{X_A}}}\sqrt{F}(\omega_{\hat{A}E}^x,\pi_{\hat{A}}\otimes\widetilde{\rho}_{E})\right)^2\\
		&\quad\geq 1-\varepsilon.
	\end{align}
	Therefore,
	\begin{equation}
		p_{\text{err}}(\mathcal{L};\rho_{AB})=1-F(\omega_{\hat{A}\hat{B}},\Phi_{\hat{A}\hat{B}})\leq\varepsilon.
	\end{equation}
	
	To summarize, we have shown that, given a state $\rho_{AB}$ and $\varepsilon\in(0,1)$, there exists a $(d,\varepsilon)$ one-way entanglement distillation protocol $\mathcal{L}_{AB\to\hat{A}\hat{B}}$ if the dimension $d=d_{\hat{A}}=d_{\hat{B}}$ satisfies $\log_2 d=-H_{\max}^{\sqrt{\varepsilon}-\eta}(A|B)_{\rho}+4\log_2 \eta$, where $\eta\in[0,\sqrt{\varepsilon})$. Although we explicitly constructed the encoding channels $\{\mathcal{E}_{A\to\hat{A}}^x\}_{x\in\mathcal{X}}$ on Alice's side, on Bob's side we relied on Uhlmann's theorem to guarantee the existence of a set of decoding channels $\{\mathcal{D}_{B\to\hat{B}}^x\}_{x\in\mathcal{X}}$ such that the overall LOCC channel $\mathcal{L}_{AB\to\hat{A}\hat{B}}$ satisfies $p_{\text{err}}(\mathcal{L};\rho_{AB})\leq\varepsilon$.  \qedsymbol
	
	Combining Theorem~\ref{prop-ent_dist_lower_bound} with \eqref{eq-smooth_cond_min_ent_to_petz_renyi}, and using \eqref{eq-Hmax_Hmin_pure_smooth}, leads to the following lower bound on the one-shot distillable entanglement:
	
	\begin{corollary}{cor-ent_distill_one_shot_UB_alt}
		Let $\rho_{AB}$ be a bipartite quantum state with purification $\psi_{ABE}$. For all $\varepsilon\in(0,1)$,  $\eta\in[0,\sqrt{\varepsilon})$, and $\alpha>1$, there exists a $(d,\varepsilon)$ one-way entanglement distillation protocol for $\rho_{AB}$ satisfying
		\begin{multline}\label{eq-ent_distill_one_shot_UB_renyi}
			\log_2 d \geq \widetilde{H}_{\alpha}(A|E)_{\psi}-\frac{1}{\alpha-1}\log_2\!\left(\frac{1}{(\sqrt{\varepsilon}-\eta)^2}\right) \\
			- \log_2\!\left(\frac{1}{1-(\sqrt{\varepsilon}-\eta)^2}\right) +4\log_2\eta .
		\end{multline}
	\end{corollary}
	
	\begin{Proof}
		The inequality follows from taking the results of Theorem~\ref{prop-ent_dist_lower_bound}, using \eqref{eq-Hmax_Hmin_pure_smooth}, and applying the inequality in \eqref{eq-smooth_cond_min_ent_to_petz_renyi}.
	\end{Proof}
	
	Since the inequality in \eqref{eq-ent_distill_one_shot_UB_renyi} holds for all $(d,\varepsilon)$ entanglement distillation protocols, we obtain the following bound for all $\varepsilon\in(0,1)$, $\eta\in[0,\sqrt{\varepsilon})$, and $\alpha>1$:
		\begin{multline}
			E_D^{\varepsilon}(A;B)_{\rho} \geq \sup_{\mathcal{L}}\widetilde{H}_{\alpha}(A'|E')_{\phi}-\frac{1}{\alpha-1}\log_2\!\left(\frac{1}{(\sqrt{\varepsilon}-\eta)^2}\right) \\
			- \log_2\!\left(\frac{1}{1-(\sqrt{\varepsilon}-\eta)^2}\right) + 4\log_2\eta ,
		\end{multline}
		where the optimization is with respect to every LOCC channel $\mathcal{L}_{AB\to A'B'}$, such that $\phi_{A'B'E'}$ is a purification of $\mathcal{L}_{AB\to A' B'}(\rho_{AB})$. This comes about by first applying the LOCC channel $\mathcal{L}_{AB\to A' B'}$ to $\rho_{AB}$ for free, applying Corollary~\ref{cor-ent_distill_one_shot_UB_alt} to the state $\mathcal{L}_{AB\to A' B'}(\rho_{AB})$, and finally optimizing over every LOCC channel $\mathcal{L}_{AB\to A'B'}$.

\section[Distillable Entanglement of a Quantum State]{Distillable Entanglement of a Quantum State}\label{sec-ent_distill_asymptotic}

	Having found upper and lower bounds on the one-shot distillable entanglement $E_D^{\varepsilon}(A;B)_{\rho}$ of a bipartite quantum state $\rho_{AB}$, let us now move on to the asymptotic setting. In this setting, we allow Alice and Bob to make use of an arbitrarily large number $n$ of copies of the state $\rho_{AB}$ in order to obtain a maximally entangled state. An \textit{entanglement distillation protocol for $n$ copies of $\rho_{AB}$} is defined by the triple $(n,d,\mathcal{L}_{A^nB^n\to \hat{A}\hat{B}})$, consisting of the number $n$ of copies of $\rho_{AB}$, an integer $d\geq 1$, and an LOCC channel $\mathcal{L}_{A^nB^n\to\hat{A}\hat{B}}$ with $d_{\hat{A}}=d_{\hat{B}}=d$. Observe that an entanglement distillation protocol for $n$ copies of $\rho_{AB}$ is equivalent to a (one-shot) entanglement distillation protocol for the state $\rho_{AB}^{\otimes n}$. All of the results of Section~\ref{sec-ent_distill_one_shot} thus carry over to the asymptotic setting simply by replacing $\rho_{AB}$ with $\rho_{AB}^{\otimes n}$. In particular, the error probability for an entanglement distillation protocol for $\rho_{AB}$ defined by $(n,d,\mathcal{L}_{A^nB^n\to\hat{A}\hat{B}})$ is equal to
	\begin{equation}
		p_{\text{err}}(\mathcal{L};\rho_{AB}^{\otimes n})=1-\bra{\Phi}_{\hat{A}\hat{B}}\mathcal{L}_{A^nB^n\to\hat{A}\hat{B}}(\rho_{AB}^{\otimes n})\ket{\Phi}_{\hat{A}\hat{B}}.
	\end{equation}
	
	\begin{definition}{$(n,d,\varepsilon)$ Entanglement Distillation Protocol}{def-ent_distill_n_d_eps_protocol}
		An entanglement distillation protocol $(n,d,\mathcal{L}_{A^nB^n\to\hat{A}\hat{B}})$ for $n$ copies of $\rho_{AB}$, with $d_{\hat{A}}=d_{\hat{B}}=d$, is called an \textit{$(n,d,\varepsilon)$ protocol}, with $\varepsilon\in[0,1]$, if $p_{\text{err}}(\mathcal{L};\rho_{AB}^{\otimes n})\leq\varepsilon$.
	\end{definition}
	
	Based on the discussion above, we note that an $(n,d,\varepsilon)$ entanglement distillation protocol for $\rho_{AB}$ is a $(d,\varepsilon)$ entanglement distillation protocol for $\rho_{AB}^{\otimes n}$.
	
	The \textit{rate} $R(n,d)$ of an $(n,d,\varepsilon)$ entanglement distillation protocol for $n$ copies of a given state is 
	\begin{equation}
		R(n,d)\coloneqq\frac{\log_2 d}{n},
	\end{equation}
	which can be thought of as the number of $\varepsilon$-approximate ebits contained in the final state of the protocol, per copy of the given initial state. Given a state $\rho_{AB}$ and $\varepsilon\in[0,1]$, the maximum rate of entanglement distillation among all $(n,d,\varepsilon)$ entanglement distillation protocols for $\rho_{AB}$ is given by
	\begin{align}
		E_D^{n,\varepsilon}(\rho_{AB})\equiv E_D^{n,\varepsilon}(A;B)_{\rho}&\coloneqq\frac{1}{n}E_D^{\varepsilon}(\rho_{AB}^{\otimes n})\\
		&=\sup_{(d,\mathcal{L})}\left\{\frac{\log_2 d}{n}: p_{\text{err}}(\mathcal{L};\rho_{AB}^{\otimes n})\leq\varepsilon\right\},\label{eq-ent_distill_n_shot}
	\end{align}
	where the optimization is with respect to all $d\geq 1$ and every LOCC channel $\mathcal{L}_{A^nB^n\to\hat{A}\hat{B}}$ with $d_{\hat{A}}=d_{\hat{B}}=d$.

	\begin{definition}{Achievable Rate for Entanglement Distillation}{def-ent_distill_ach_rate}
		Given a bipartite quantum state $\rho_{AB}$, a rate $R\in\mathbb{R}^+$ is called an \textit{achievable rate for entanglement distillation for $\rho_{AB}$} if for all $\varepsilon\in(0,1]$, $\delta>0$, and sufficiently large $n$, there exists an $(n,2^{n(R-\delta)},\varepsilon)$ entanglement distillation protocol for $\rho_{AB}$.
	\end{definition}
	
	As we prove in Appendix~\ref{chap-str_conv},
	\begin{equation}
		R\text{ achievable rate }\Longleftrightarrow \lim_{n\to\infty}\varepsilon_D(2^{n(R-\delta)};\rho_{AB}^{\otimes n})=0\quad\forall~\delta>0.
	\end{equation}
	In other words, a rate $R$ is achievable if the optimal error probability for a sequence of protocols with rate $R-\delta$, $\delta>0$, vanishes as the number $n$ of copies of $\rho_{AB}$ increases.
	
	\begin{definition}{Distillable Entanglement of a Quantum State}{def-ent_distill_disill_ent}
		The \textit{distillable entanglement of a bipartite state $\rho_{AB}$}, denoted by $E_D(A;B)_{\rho}$, is defined to be the supremum of all achievable rates for entanglement distillation for $\rho_{AB}$, i.e.,
		\begin{align}
			E_D(A;B)_{\rho}&\coloneqq\sup\{R:R\text{ is an achievable rate for }\rho_{AB}\}.
		\end{align}
	\end{definition}
	
	The distillable entanglement can also be written as
	\begin{equation}\label{eq-distillable_entanglement_alt_def}
		E_D(A;B)_{\rho}=\inf_{\varepsilon\in( 0,1]}\liminf_{n\to\infty}\frac{1}{n}E_D^{\varepsilon}(\rho_{AB}^{\otimes n}).
	\end{equation}
	See Appendix~\ref{chap-str_conv} for a proof.
	
	\begin{definition}{Weak Converse Rate for Entanglement Distillation}{def-ent_distill_weak_conv_rate}
		Given a bipartite state $\rho_{AB}$, a rate $R\in\mathbb{R}^+$ is called a \textit{weak converse rate for entanglement distillation for $\rho_{AB}$} if every $R'>R$ is not an achievable rate for $\rho_{AB}$.
	\end{definition}
	
	As we show in Appendix~\ref{chap-str_conv},
	\begin{equation}\label{eq-ent_distill_weak_conv_rate_alt}
		R\text{ weak converse rate }\Longleftrightarrow \lim_{n\to\infty}\varepsilon_D(2^{n(R-\delta)};\rho_{AB}^{\otimes n})>0\quad\forall~\delta>0.
	\end{equation}
	
	\begin{definition}{Strong Converse Rate for Entanglement Distillation}{def-ent_distill_str_conv_rate}
		Given a bipartite state $\rho_{AB}$, a rate $R\in\mathbb{R}^+$ is called a \textit{strong converse rate for entanglement distillation for $\rho_{AB}$} if for all $\varepsilon\in[0,1)$, $\delta>0$, and sufficiently large $n$, there does not exist an $(n,2^{n(R+\delta)},\varepsilon)$ entanglement distillation protocol for $\rho_{AB}$.
	\end{definition}
	
	We show in Appendix~\ref{chap-str_conv} that
	\begin{equation}\label{eq-ent_distill_str_conv_rate_alt}
		R\text{ strong converse rate }\Longleftrightarrow \lim_{n\to\infty}\varepsilon_D(2^{n(R+\delta)};\rho_{AB})=1\quad\forall~\delta>0.
	\end{equation}
	
	\begin{definition}{Strong Converse Distillable Entanglement of a Quantum State}{def-ent_distill_str_conv_distill_ent}
		The \textit{strong converse distillable entanglement} of a bipartite state $\rho_{AB}$, denoted by $\widetilde{E}_D(A;B)_{\rho}$, is defined as the infimum of all strong converse rates, 
		i.e.,
		\begin{align}
			\widetilde{E}_D(A;B)_{\rho}&\coloneqq\inf\{R:R\text{ is a strong converse rate for }\rho_{AB}\}.
		\end{align}
	\end{definition}

	Note that
	\begin{equation}\label{eq-ent_distill_cap_vs_strong_conv}
		E_D(A;B)_{\rho}\leq \widetilde{E}_D(A;B)_{\rho}
	\end{equation}
	for all bipartite states $\rho_{AB}$. We can also write the strong converse distillable entanglement as
	\begin{equation}
		\widetilde{E}_D(A;B)_{\rho}=\sup_{\varepsilon\in[0,1) }\limsup_{n\to\infty}\frac{1}{n}E_D^{\varepsilon}(\rho_{AB}^{\otimes n}).
	\end{equation}
	See Appendix~\ref{chap-str_conv} for a proof.

	We are now ready to present a general expression for the distillable entanglement of a bipartite quantum state, as well as two upper bounds on it.
	
	\begin{theorem*}{Distillable Entanglement of a Bipartite State}{thm-distillable_entanglement}
		The distillable entanglement of a bipartite state $\rho_{AB}$ is given by
		\begin{equation}\label{eq-distillable_entanglement}
			E_D(A;B)_{\rho}=\lim_{n\to\infty}\frac{1}{n}\sup_{\mathcal{L}^{(n)}}I(A'\rangle B')_{\mathcal{L}^{(n)}(\rho^{\otimes n})},
		\end{equation}
		where the optimization is with respect to  (two-way) LOCC channels $\mathcal{L}^{(n)}_{A^nB^n\to A'B'}$. Furthermore, the Rains relative entropy $R(A;B)_{\rho}$ from \eqref{eq-Rains_rel_ent} is a strong converse rate for distillable entanglement, in the sense that
		\begin{equation}\label{eq-distillable_entanglement_str_conv_UB}
			\widetilde{E}_D(A;B)_{\rho}\leq R(A;B)_{\rho},
		\end{equation}
		and the squashed entanglement from \eqref{eq-squashed_entanglement} is a weak converse rate, in the sense that
		\begin{equation}
		E_D(A;B)_{\rho}\leq E_{\operatorname{sq}}(A;B)_{\rho}.
		\label{eq:ED:squashed-ent-weak-conv-ED}
		\end{equation}
	\end{theorem*}
	
	If we define
	\begin{equation}
		D^{\leftrightarrow}(\rho_{AB})\equiv D^{\leftrightarrow}(A;B)_{\rho}\coloneqq\sup_{\mathcal{L}}I(A'\rangle B')_{\mathcal{L}(\rho)},
	\end{equation}
	then we can write \eqref{eq-distillable_entanglement} as
	\begin{equation}
		E_D(A;B)_{\rho}=\lim_{n\to\infty}\frac{1}{n}D^{\leftrightarrow}(\rho_{AB}^{\otimes n})\eqqcolon D_{\text{reg}}^{\leftrightarrow}(\rho_{AB}),
		\label{eq-ED:dist-ent-alt-exp-1}
	\end{equation}
	so that the distillable entanglement can be viewed as the regularized version of $D^{\leftrightarrow}$, and it is reminiscent of the regularized Holevo information that gives the classical capacity of a quantum channel.

	Let us make the following observations about  Theorem~\ref{thm-distillable_entanglement}.
	\begin{itemize}
		\item The coherent information of a bipartite state $\rho_{AB}$ is an achievable rate for entanglement distillation, i.e.,
			\begin{equation}\label{eq-ent_distill_coh_inf_LB}
				E_D(A;B)_{\rho}\geq I(A\rangle B)_{\rho}=H(B)_{\rho}-H(AB)_{\rho}.
			\end{equation}
			This follows immediately from \eqref{eq-distillable_entanglement} by dropping the optimization over two-way LOCC channels and due to the fact that the coherent information is additive for product states, meaning that $I(A^n\rangle B^n)_{\rho^{\otimes n}}= nI(A\rangle B)_{\rho}$. As we show in Section~\ref{sec-ent_distill_ach} below, the strategy to attain the coherent information rate is essentially the one-way entanglement distillation protocol considered in Section~\ref{subsec-ent_distill_one_shot_lower_bound} for the one-shot lower bound, and it is sometimes called the ``hashing protocol.'' For this reason, the inequality in \eqref{eq-ent_distill_coh_inf_LB} is known as the \textit{hashing bound} (please consult the Bibliographic Notes in Section~\ref{sec:ed:bib-notes} for pointers to the research literature).
			
		\item In order to obtain a higher entanglement distillation rate than $I(A\rangle B)_{\rho}$, one strategy is to use $n\geq 2$ copies of $\rho_{AB}$ along with a two-way LOCC channel $\mathcal{L}_{A^nB^n\to A'B'}$ in order to obtain a state $\omega_{A'B'}\coloneqq\mathcal{L}_{A^nB^n\to A'B'}(\rho_{AB}^{\otimes n})$ whose coherent information is potentially larger than that of $\rho_{AB}$. Then, we can apply the hashing protocol to the state $\omega_{A'B'}$. The overall rate of this strategy (the two-way LOCC channel followed by the hashing protocl) is then $\frac{1}{n}I(A'\rangle B')_{\omega}$, and Theorem~\ref{thm-distillable_entanglement} tells us that such a strategy is optimal in the large $n$ limit. With increasingly more copies of $\rho_{AB}$ to start with, it might be possible to obtain a better rate, which is why we need to regularize in general.
	\end{itemize}

	As with the proof of the entanglement-assisted classical capacity and classical capacity theorems in Chapters~\ref{chap-EA_capacity} and \ref{chap-classical_capacity}, respectively, we prove Theorem~\ref{thm-distillable_entanglement} in two steps:
	\begin{enumerate}
		\item\textit{Achievability}: We show that the right-hand side of \eqref{eq-distillable_entanglement} is an achievable rate for entanglement distillation for $\rho_{AB}$. Doing so involves exhibiting an explicit entanglement distillation protocol. The protocol we use is based on the one we used in Section~\ref{subsec-ent_distill_one_shot_lower_bound} to obtain a lower bound on the one-shot distillable entanglement. 
		
		The achievability part of the proof establishes that
		\begin{equation}
			E_D(A;B)_{\rho}\geq\lim_{n\to\infty}\frac{1}{n}\sup_{\mathcal{L}}I(A'\rangle B')_{\mathcal{L}(\rho^{\otimes n})}.
		\end{equation}
		
		\item\textit{Weak converse}: We show that the right-hand side of \eqref{eq-distillable_entanglement} is a weak converse rate for entanglement distillation for $\rho_{AB}$, from which it follows that $E_D(A;B)_{\rho}\leq\lim_{n\to\infty}\frac{1}{n}\sup_{\mathcal{L}}I(A'\rangle B')_{\mathcal{L}(\rho^{\otimes n})}$. In order to show this, we use the one-shot upper bounds from Section~\ref{subsec-ent_distill_one_shot_UB} to prove that every achievable rate $R$ satisfies $R \leq \lim_{n\to\infty}\frac{1}{n}\sup_{\mathcal{L}}I(A'\rangle B')_{\mathcal{L}(\rho^{\otimes n})}$.
	\end{enumerate}
	
	We go through the achievability part of the proof of Theorem~\ref{thm-distillable_entanglement} in Section~\ref{sec-ent_distill_ach}. We then proceed with the weak converse part in Section~\ref{sec-ent_distill_weak_conv}.
	
	The expression in \eqref{eq-distillable_entanglement} for the distillable entanglement involves both a limit over an unbounded number of copies of the state $\rho_{AB}$, as well as an optimization over all two-way LOCC channels. Computing the distillable entanglement is therefore intractable in general. After establishing a proof of \eqref{eq-distillable_entanglement}, we proceed to establish upper bounds on distillable entanglement that depend only on the given state $\rho_{AB}$. Specifically, in Section~\ref{sec-ent_distill_strong_converse}, we use the one-shot results in Section~\ref{subsec-ent_distill_one_shot_UB} to show that the Rains relative entropy is a strong converse rate for entanglement distillation. We also show that the squashed entanglement is a weak converse rate for entanglement distillation.
	
\subsubsection{Bound Entanglement}\label{subsec-bound_ent}

	The inequality in \eqref{eq-distillable_entanglement_str_conv_UB} implies that $E_D(A;B)_{\rho}\leq 0$ for every PPT state $\rho_{AB}$, because, by definition, the Rains relative entropy vanishes for all PPT states. On the other hand, we always have $E_D(A;B)_{\rho}\geq 0$ for every state $\rho_{AB}$. Therefore,
	\begin{equation}
		E_D(A;B)_{\rho}=0\text{ for all PPT states.}
	\end{equation}
	Recall from the discussion after Lemma~\ref{lem:fail-ent-test} (see also Section~\ref{subsec-PPT}) that there exist PPT entangled states in higher dimensional bipartite systems because $\SEP(A\!:\!B)\neq\PPT(A\!:\!B)$ except for when $A$ and $B$ are both qubits or when one is a qubit and the other is a qutrit. All of these entangled states have zero distillable entanglement, and thus we refer to them as \textit{bound entangled}. Remarkably, therefore, except for qubit-qubit and qubit-qutrit states, prior entanglement is only necessary, but not sufficient, for distilling pure maximally entangled states. Please consult the Bibliographic Notes in Section~\ref{sec:ed:bib-notes} for more information about bound entanglement.
	
	\begin{figure}
		\centering
		\includegraphics[width=\textwidth]{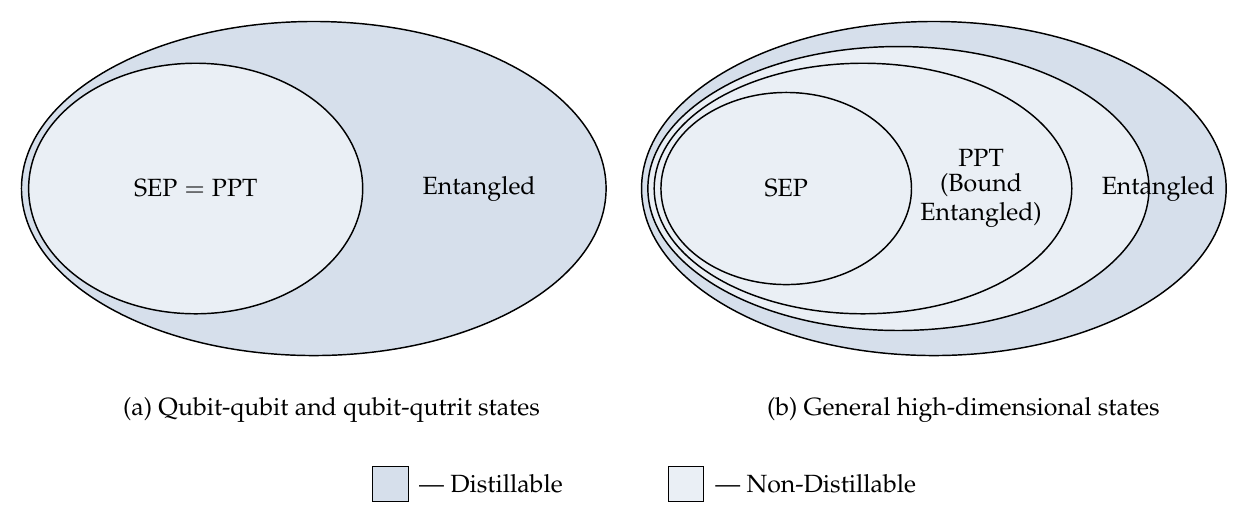}
		\caption{The set of all bipartite states can be split into distillable and non-distillable sets. (a) For qubit-qubit and qubit-qutrit states entanglement and distillability are in one-to-one correspondence because all PPT states are separable. (b) In higher dimensions, there are entangled states belonging to the set PPT, which we call bound entangled states. Distillability and entanglement are thus not synonymous in general for high-dimensional quantum systems.}\label{fig-dist_nondist_gen}
	\end{figure}

	As shown in Figure~\ref{fig-dist_nondist_gen}, we can use entanglement distillation to split up the set of all bipartite states into distillable and non-distillable states. For two-qubit states and qubit-qutrit states, non-distillable states are exactly equal to the set of separable states by the PPT criterion. For higher dimensions, as stated above, this is not the case. Also in higher dimensions, it is in general possible to have states with negative partial transpose (NPT) that are nonetheless non-distillable. These \textit{NPT bound entangled} states are shown in Figure~\ref{fig-dist_nondist_gen}(b) as the region between the PPT bound entangled states and the distillable entangled states. It is not known whether NPT bound entangled states exist, but since they have not been ruled out, we nevertheless depict them in the figure.

\subsection{Proof of Achievability}\label{sec-ent_distill_ach}

	As the first step in proving the achievability part of Theorem~\ref{thm-distillable_entanglement}, let us recall Corollary~\ref{cor-ent_distill_one_shot_UB_alt}: given a bipartite state $\rho_{AB}$ with purification $\psi_{ABE}$, for all $\varepsilon\in(0,1)$, $\eta\in[0,\sqrt{\varepsilon})$, and $\alpha>1$, there exists a $(d,\varepsilon)$ entanglement distillation protocol for $\rho_{AB}$ such that
	\begin{multline}
	\label{eq-ent_distill_coh_inf_ach_pf0}
		\log_2 d\geq \widetilde{H}_{\alpha}(A|E)_{\psi}-\frac{1}{\alpha-1}\log_2\!\left(\frac{1}{(\sqrt{\varepsilon}-\eta)^2}\right) \\
			- \log_2\!\left(\frac{1}{1-(\sqrt{\varepsilon}-\eta)^2}\right) +4\log_2\eta ,
	\end{multline}
	where
	\begin{equation}
		\widetilde{H}_{\alpha}(A|E)_{\psi}=-\inf_{\sigma_E}\widetilde{D}_{\alpha}(\psi_{AE}\Vert\mathbbm{1}_{A}\otimes\sigma_E)
	\end{equation}
	is the sandwiched R\'{e}nyi conditional entropy. Applying this inequality to the state $\rho_{AB}^{\otimes n}$ for all $n\geq 1$ leads to the following:
	
	\begin{proposition}{prop-ent_distill_asymp_LB}
		For every state $\rho_{AB}$ and $\varepsilon\in(0,1)$, there exists an $(n,d,\varepsilon)$ entanglement distillation protocol for $\rho_{AB}$ such that the rate $\frac{\log_2 d}{n}$ satisfies
		\begin{equation}\label{eq-ent_distill_coh_inf_ach_pf5}
			\frac{\log_2 d}{n}\geq \widetilde{H}_{\alpha}(A|E)_{\psi}-\frac{2\alpha-1}{n(\alpha-1)} \log_2\!\left(\frac{4}{\varepsilon}\right)
			-\frac{1}{n}\log_2\!\left(\frac{1}{1-\frac{\varepsilon}{4}}\right),
		\end{equation}
		for all $n\geq 1$ and $\alpha>1$, where $\psi_{ABE}$ is a purification of $\rho_{AB}$. In general,
		\begin{multline}\label{eq-ent_distill_coh_inf_ach_pf1a}
			E_D^{n,\varepsilon}(A;B) \geq \sup_{\mathcal{L}}\frac{1}{n}\widetilde{H}_{\alpha}(A'|E')_{\phi}-\frac{2\alpha-1}{n(\alpha-1)} \log_2\!\left(\frac{4}{\varepsilon}\right)
			\\-\frac{1}{n}\log_2\!\left(\frac{1}{1-\frac{\varepsilon}{4}}\right), 
		\end{multline}
		for all $n\geq 1$ and $\alpha>1$, where the optimization is over every LOCC channel $\mathcal{L}_{A^nB^n\to A'B'}$, such that $\phi_{A'B'E'}$ is a purification of $\mathcal{L}_{A^nB^n\to A'B'}(\rho_{AB}^{\otimes n})$.
	\end{proposition}
	
	\begin{Proof}
		Let $\psi_{ABE}$ be a purification of $\rho_{AB}$, and use the tensor-product purification $\psi_{ABE}^{\otimes n}$ for $\rho_{AB}^{\otimes n}$. Also, let $\eta=\frac{\sqrt{\varepsilon}}{2}$. Substituting all of this into the inequality in \eqref{eq-ent_distill_coh_inf_ach_pf0} and simplifying leads to 
		\begin{multline}\label{eq-ent_distill_coh_inf_ach_pf1}
		\frac{\log_2 d}{n}\geq \frac{1}{n} \widetilde{H}_{\alpha}(A^n|E^n)_{\psi^{\otimes n}}-\frac{1}{n(\alpha-1)} \log_2\!\left(\frac{4}{\varepsilon}\right)
			\\-\frac{1}{n}\log_2\!\left(\frac{1}{1-\frac{\varepsilon}{4}}\right) -\frac{2}{n} \log_2\!\left(\frac{4}{\varepsilon}\right),
		\end{multline}
		Then, optimizing over every LOCC channel $\mathcal{L}_{A^nB^n\to A'B'}$, and using the definition of $E_D^{n,\varepsilon}(A;B)_{\rho}$ in \eqref{eq-ent_distill_n_shot}, we obtain \eqref{eq-ent_distill_coh_inf_ach_pf1a}.
		
		By employing additivity of the sandwiched R\'{e}nyi conditional entropy for all $\alpha > 1$, we have that
		\begin{equation}
			\widetilde{H}_{\alpha}(A^n|E^n)_{\psi^{\otimes n}}= n \widetilde{H}_{\alpha}(A|E)_{\psi}.
		\end{equation}
		Note that the proof of additivity follows similarly to the proof of Proposition~\ref{prop-sand_rel_mut_inf_additive}.
		This leads to \eqref{eq-ent_distill_coh_inf_ach_pf5}.
	\end{Proof}
	
	With the inequality in \eqref{eq-ent_distill_coh_inf_ach_pf5}, we can prove that the coherent information is an achievable rate for entanglement distillation.
	
	\begin{theorem*}{Achievability of Coherent Information for Entanglement Distillation}{thm-ent_distill_coh_inf_ach}
		The coherent information $I(A\rangle B)_{\rho}$ of a bipartite state $\rho_{AB}$ is an achievable rate for entanglement distillation for $\rho_{AB}$. In other words, $E_D(A;B)_{\rho}\geq I(A\rangle B)_{\rho}$ for every bipartite state $\rho_{AB}$.
	\end{theorem*}
	
	\begin{Proof}
		Let $\psi_{ABE}$ be a purification of $\rho_{AB}$. Fix $\varepsilon\in(0,1]$ and $\delta>0$. Let $\delta_1,\delta_2>0$ be such that
		\begin{equation}\label{eq-ent_distill_coh_inf_ach_pf3}
			\delta=\delta_1+\delta_2.
		\end{equation}
		Set $\alpha\in(1,\infty)$ such that
		\begin{equation}\label{eq-ent_distill_coh_inf_ach_pf2}
			\delta_1\geq I(A\rangle B)_{\rho}-\widetilde{H}_{\alpha}(A|E)_{\psi}.
		\end{equation}
		Note that this is possible because $\widetilde{H}_{\alpha}(A|E)_{\psi}$ increases monotonically with decreasing $\alpha$ (this follows from Proposition~\ref{prop-Petz_rel_ent}), so that
		\begin{equation}
			\lim_{\alpha\to 1^+}\widetilde{H}_{\alpha}(A|E)_{\psi}=\sup_{\alpha\in(1,\infty)}\widetilde{H}_{\alpha}(A|E)_{\psi}.
		\end{equation}
		Also,
		\begin{align}
			\lim_{\alpha\to 1^+}\widetilde{H}_{\alpha}(A|E)_{\psi}&=\sup_{\alpha\in(1,\infty)}\widetilde{H}_{\alpha}(A|E)_{\psi}\\
			&=\sup_{\alpha\in(1,\infty)}\left(-\inf_{\sigma_E}\widetilde{D}_{\alpha}(\psi_{AE}\Vert\mathbbm{1}_A\otimes\sigma_E)\right)\\
			&=-\inf_{\alpha\in(1,\infty)}\inf_{\sigma_E}\widetilde{D}_{\alpha}(\psi_{AE}\Vert\mathbbm{1}_A\otimes\sigma_E)\\
			&=-\inf_{\sigma_E}\inf_{\alpha\in(1,\infty)}\widetilde{D}_{\alpha}(\psi_{AE}\Vert\mathbbm{1}_A\otimes\sigma_E)\\
			&=-\inf_{\sigma_E}D(\psi_{AE}\Vert\mathbbm{1}_A\otimes\sigma_E)\\
			&=H(A|E)_{\psi},
		\end{align}
		where the fifth equality follows from Proposition~\ref{prop-sand_ren_ent_lim}. Then, by definition of conditional entropy, and the fact that $\psi_{ABE}$ is a pure state, we find that
		\begin{equation}
			H(A|E)_{\psi}=H(AE)_{\psi}-H(E)_{\psi}=H(B)_{\psi}-H(AB)_{\psi}=I(A\rangle B)_{\rho}.
		\end{equation}
		
		With $\alpha\in(1,\infty)$ chosen such that \eqref{eq-ent_distill_coh_inf_ach_pf2} holds, take $n$ large enough so that
		\begin{equation}\label{eq-ent_distill_coh_inf_ach_pf4}
			\delta_2\geq \frac{2\alpha-1}{n(\alpha-1)} \log_2\!\left(\frac{4}{\varepsilon}\right)
			+\frac{1}{n}\log_2\!\left(\frac{1}{1-\frac{\varepsilon}{4}}\right).
		\end{equation}
		Now, we use the fact that for the $n$ and $\varepsilon$ chosen above there exists an $(n,d,\varepsilon)$ protocol such that 
		\begin{equation}
			\frac{\log_2 d}{n}\geq \widetilde{H}_{\alpha}(A|E)_{\psi} - \frac{2\alpha-1}{n(\alpha-1)} \log_2\!\left(\frac{4}{\varepsilon}\right)
			-\frac{1}{n}\log_2\!\left(\frac{1}{1-\frac{\varepsilon}{4}}\right).
		\end{equation}
		(This follows from Proposition~\ref{prop-ent_distill_asymp_LB} above.) Rearranging the right-hand side of this inequality, and using \eqref{eq-ent_distill_coh_inf_ach_pf3}, \eqref{eq-ent_distill_coh_inf_ach_pf2}, and \eqref{eq-ent_distill_coh_inf_ach_pf4}, we find that
		\begin{align}
			\frac{\log_2 d}{n}&\geq I(A\rangle B)_{\rho}-\left(I(A\rangle B)_{\rho}-\widetilde{H}_{\alpha}(A|E)_{\psi}+ \frac{2\alpha-1}{n(\alpha-1)} \log_2\!\left(\frac{4}{\varepsilon}\right)
			\right.\nonumber\\
			&\qquad\qquad\qquad\qquad\qquad\qquad \left.+\frac{1}{n}\log_2\!\left(\frac{1}{1-\frac{\varepsilon}{4}}\right) \right)\\
			&\geq I(A\rangle B)_{\rho}-(\delta_1+\delta_2)\\
			&=I(A\rangle B)_{\rho}-\delta.
		\end{align}
		We thus have that there exists an $(n,d,\varepsilon)$ entanglement distillation protocol with rate $\frac{\log_2 d}{n}\geq I(A\rangle B)_{\rho}-\delta$. Therefore, there exists an $(n,2^{n(R-\delta)},\varepsilon)$ entanglement distillation protocol with $R=I(A\rangle B)_{\rho}$ for all sufficiently large $n$ such that \eqref{eq-ent_distill_coh_inf_ach_pf4} holds. Since $\varepsilon$ and $\delta$ are arbitrary, we conclude that for all $\varepsilon\in(0,1]$, $\delta>0$, and  sufficiently large $n$, there exists an $(n,2^{n(I(A\rangle B)_{\rho}-\delta)},\varepsilon)$ entanglement distillation protocol. This means that, by definition, $I(A\rangle B)_{\rho}$ is an achievable rate.
	\end{Proof}

\subsubsection*{Proof of the Achievability Part of Theorem~\ref{thm-distillable_entanglement}}

	Let $\mathcal{L}_{A^kB^k\to A'B'}$ be an arbitrary LOCC channel with $k\geq 1$, let
	\begin{equation}
	\omega_{A'B'}=\mathcal{L}_{A^kB^k\to A'B'}(\rho_{AB}^{\otimes k}),
	\end{equation}
	 and let $\phi_{A'B'E'}$ be a purification of $\omega_{A'B'}$. Fix $\varepsilon\in(0,1]$ and $\delta>0$. Let $\delta_1,\delta_2>0$ be such that
	\begin{equation}\label{eq-ent_distill_ach_pf1}
		\delta=\delta_1+\delta_2.
	\end{equation}
	Set $\alpha\in(1,\infty)$ such that
	\begin{equation}\label{eq-ent_distill_ach_pf2}
		\delta_1\geq \frac{1}{k}I(A'\rangle B')_{\omega}-\frac{1}{k}\widetilde{H}_{\alpha}(A'|E')_{\phi},
	\end{equation}
	which is possible based on the arguments given in the proof of Theorem~\ref{thm-ent_distill_coh_inf_ach} above. Then, with this choice of $\alpha$, take $n$ large enough so that
	\begin{equation}\label{eq-ent_distill_ach_pf3}
		\delta_2\geq \frac{2\alpha-1}{kn(\alpha-1)} \log_2\!\left(\frac{4}{\varepsilon}\right)
			+\frac{1}{kn}\log_2\!\left(\frac{1}{1-\frac{\varepsilon}{4}}\right).
	\end{equation}
	Now, we use the fact that, for the chosen $n$ and $\varepsilon$, there exists an $(n,d,\varepsilon)$ entanglement distillation protocol such that \eqref{eq-ent_distill_coh_inf_ach_pf5} holds, i.e., 
	\begin{equation}
		\frac{\log_2 d}{n}\geq \widetilde{H}_{\alpha}(A'|E')_{\phi} - \frac{2\alpha-1}{n(\alpha-1)} \log_2\!\left(\frac{4}{\varepsilon}\right)
			-\frac{1}{n}\log_2\!\left(\frac{1}{1-\frac{\varepsilon}{4}}\right).
	\end{equation}
	Dividing both sides by $k$ gives
	\begin{equation}
		\frac{\log_2 d}{kn}\geq \frac{1}{k} \widetilde{H}_{\alpha}(A'|E')_{\phi} - \frac{2\alpha-1}{kn(\alpha-1)} \log_2\!\left(\frac{4}{\varepsilon}\right)
			-\frac{1}{kn}\log_2\!\left(\frac{1}{1-\frac{\varepsilon}{4}}\right).
	\end{equation}
	Rearranging the right-hand side of this inequality, and using \eqref{eq-ent_distill_ach_pf1}--\eqref{eq-ent_distill_ach_pf3}, we find that
	\begin{align}
		\frac{\log_2 d}{kn}&\geq \frac{1}{k}I(A'\rangle B')_{\omega}-\left(\frac{1}{k}I(A'\rangle B')_{\omega}-\frac{1}{k}\widetilde{H}_{\alpha}(A'|E')_{\phi}\right.\nonumber\\
		&\qquad\qquad \left. + \frac{2\alpha-1}{kn(\alpha-1)} \log_2\!\left(\frac{4}{\varepsilon}\right)
			+\frac{1}{kn}\log_2\!\left(\frac{1}{1-\frac{\varepsilon}{4}}\right) \right)\\
		&\geq\frac{1}{k}I(A'\rangle B')_{\omega}-(\delta_1+\delta_2)\\
		&=\frac{1}{k}I(A'\rangle B')_{\omega}-\delta.
	\end{align}
	Thus, there exists a $(kn,d,\varepsilon)$ entanglement distillation protocol with rate $\frac{\log_2 d}{kn}\geq \frac{1}{k}I(A'\rangle B')_{\omega}-\delta$. Therefore, letting $n'\equiv kn$, there exists an $(n',2^{n'(R-\delta)},\varepsilon)$ entanglement distillation protocol with $R=\frac{1}{k}I( A'\rangle B')_{\omega}$ for all sufficiently large $n$ such that \eqref{eq-ent_distill_ach_pf3} holds. Since $\varepsilon$ and $\delta$ are arbitrary, we conclude that for all $\varepsilon\in(0,1]$,  $\delta>0$, and sufficiently large $n$, there exists an $(n,2^{n(\frac{1}{k}I( A'\rangle B')_{\omega}-\delta)},\varepsilon)$ entanglement distillation protocol. This means that $\frac{1}{k}I( A'\rangle B')_{\omega}$ is an achievable rate. (Recall that $\omega_{ A' B'}=\mathcal{L}_{A^kB^k\to A'B'}(\rho_{AB}^{\otimes k})$.)
	
	Now, since in the  arguments above  the LOCC channel $\mathcal{L}_{A^kB^k\to A' B'}$ is arbitrary, we conclude that
	\begin{equation}
		\frac{1}{k}\sup_{\mathcal{L}}I( A'\rangle B')_{\mathcal{L}(\rho^{\otimes k})}
	\end{equation}
	is an achievable rate. Finally, since the number $k$ of copies of $\rho_{AB}$ is arbitrary, we conclude that
	\begin{equation}
		\lim_{k\to\infty}\frac{1}{k}\sup_{\mathcal{L}}I(A'\rangle B')_{\mathcal{L}(\rho^{\otimes k})}
	\end{equation}
	is an achievable rate. 


\subsection{Proof of the Weak Converse}\label{sec-ent_distill_weak_conv}

	In order to prove the weak converse part of Theorem~\ref{thm-distillable_entanglement}, we make use of Corollary~\ref{cor-ent_distill_one_shot_UB_all}, specifically \eqref{eq-ent_distill_one_shot_UB_weak_conv}: given a bipartite state $\rho_{AB}$, for every $(d,\varepsilon)$ entanglement distillation protocol for $\rho_{AB}$, with $\varepsilon\in[0,\sfrac{1}{2})$, it holds that
	\begin{equation}
		\log_2 d\leq\frac{1}{1-2\varepsilon}\left(\sup_{\mathcal{L}} I(A'\rangle B')_{\mathcal{L}(\rho)}+h_2(\varepsilon)\right),
	\end{equation}
	where the optimization is with respect to every LOCC channel $\mathcal{L}_{AB\to A' B'}$. Applying this inequality to the state $\rho_{AB}^{\otimes n}$ immediately leads to the following.
	
	\begin{proposition}{prop-ent_distill_asymp_UB}
		Let $\rho_{AB}$ be a bipartite state, and let $n\geq 1$ and $\varepsilon\in[0,\sfrac{1}{2})$. For an $(n,d,\varepsilon)$ entanglement distillation protocol for $\rho_{AB}$ with corresponding LOCC channel $\mathcal{L}_{A^nB^n\to A' B'}$, the rate $\frac{\log_2 d}{n}$ satisfies
		\begin{equation}\label{eq-ent_distill_n_shot_UB_weak_conv}
			\frac{\log_2 d}{n}\leq \frac{1}{1-2\varepsilon}\left(\sup_{\mathcal{L}} \frac{1}{n}I(A'\rangle B')_{\mathcal{L}(\rho^{\otimes n})}+\frac{1}{n}h_2(\varepsilon)\right).
		\end{equation}
		Consequently,
		\begin{equation}\label{eq-ent_distill_distill_ent_n_shot}
			E_D^{n,\varepsilon}(A;B)_{\rho}\leq\frac{1}{1-2\varepsilon}\left(\frac{1}{n}\sup_{\mathcal{L}}I( A'\rangle B')_{\mathcal{L}(\rho^{\otimes n})}+\frac{1}{n} h_2(\varepsilon)\right),
		\end{equation}
		where the optimization is over every LOCC channel $\mathcal{L}_{A^nB^n\to A'B'}$.
	\end{proposition}
	
	\begin{Proof}
		The inequality in \eqref{eq-ent_distill_n_shot_UB_weak_conv} is immediate from \eqref{eq-ent_distill_one_shot_UB_weak_conv} in Corollary~\ref{cor-ent_distill_one_shot_UB_all} by applying that inequality to the state $\rho_{AB}^{\otimes n}$ and dividing both sides by $n$. The inequality in \eqref{eq-ent_distill_distill_ent_n_shot} follows immediately by definition of $E_D^{n,\varepsilon}$ in \eqref{eq-ent_distill_n_shot}.
	\end{Proof}

\subsubsection*{Proof of the Weak Converse Part of Theorem~\ref{thm-distillable_entanglement}}

	Suppose that $R$ is an achievable rate for entanglement distillation for the bipartite state $\rho_{AB}$. Then, by definition, for all $\varepsilon\in(0,1]$,  $\delta>0$, and sufficiently large $n$, there exists an $(n,2^{n(R-\delta)},\varepsilon)$ entanglement distillation protocol for $\rho_{AB}$. For all such protocols for which $\varepsilon\in(0,\sfrac{1}{2})$,  the inequality in \eqref{eq-ent_distill_n_shot_UB_weak_conv} holds, so that
	\begin{equation}
		R-\delta \leq\frac{1}{1-2\varepsilon}\left(\frac{1}{n}\sup_{\mathcal{L}}I(A'\rangle B')_{\mathcal{L}(\rho^{\otimes n})}+\frac{1}{n}h_2(\varepsilon)\right).
	\end{equation}
	 Since the inequality holds for all sufficiently large $n$, it holds in the limit $n\to\infty$, so that
	\begin{align}
		R&\leq \lim_{n\to\infty}\frac{1}{1-2\varepsilon}\left(\frac{1}{n}\sup_{\mathcal{L}}I(A'\rangle B')_{\mathcal{L}(\rho^{\otimes n})}+\frac{1}{n}h_2(\varepsilon)\right)+\delta\\
		&=\frac{1}{1-2\varepsilon}\lim_{n\to\infty}\frac{1}{n}\sup_{\mathcal{L}}I(A'\rangle B')_{\mathcal{L}(\rho^{\otimes n})}+\delta.
	\end{align}
	Then, since this inequality holds for all $\varepsilon \in (0,\sfrac{1}{2}),\delta>0$, we conclude that
	\begin{align}
		R&\leq \lim_{\varepsilon,\delta\to 0}\left(\frac{1}{1-2\varepsilon}\lim_{n\to\infty}\frac{1}{n}\sup_{\mathcal{L}}I( A'\rangle B')_{\mathcal{L}(\rho^{\otimes n})}+\delta\right)\\
		&=\lim_{n\to\infty}\frac{1}{n}\sup_{\mathcal{L}}I(A'\rangle B')_{\mathcal{L}(\rho^{\otimes n})}.
	\end{align}
	We have thus shown that the quantity $\lim_{n\to\infty}\frac{1}{n}\sup_{\mathcal{L}}I(A'\rangle B')_{\mathcal{L}(\rho^{\otimes n})}$ is a weak converse rate for entanglement distillation for $\rho_{AB}$.


\subsection{Rains Relative Entropy Strong Converse Upper Bound}\label{sec-ent_distill_strong_converse}

	As stated previously, the expression in \eqref{eq-distillable_entanglement} for distillable entanglement involves both a limit over an unbounded number of copies of the initial state $\rho_{AB}$, as well as an optimization over all two-way LOCC channels. Computing the distillable entanglement is therefore intractable in general. In this section, we use the one-shot upper bound established in Section~\ref{subsec-ent_distill_one_shot_UB} to show that the Rains relative entropy is a strong converse upper bound  on the distillable entanglement of a bipartite state.  
	
	We start by recalling the upper bound in \eqref{eq-ent_distill_one_shot_UB_str_conv}, which tells us that
	\begin{align}
		\log_2 d&\leq \widetilde{R}_{\alpha}(A;B)_{\rho}+\frac{\alpha}{\alpha-1}\log_2\!\left(\frac{1}{1-\varepsilon}\right)\quad\forall~\alpha>1,\label{eq-ent_distill_one_shot_UB_str_conv2}
	\end{align}
	for an arbitrary $(d,\varepsilon)$ entanglement distillation protocol, where $\varepsilon\in(0,1)$. Recall that
	\begin{align}
		\widetilde{R}_{\alpha}(A;B)_{\rho}&=\inf_{\sigma_{AB}\in\PPT'(A:B)}\widetilde{D}_{\alpha}(\rho_{AB}\Vert\sigma_{AB}).
	\end{align}
	The upper bound above is a consequence of the fact that $\PPT'$ operators are useless for entanglement distillation, in the sense that for every $\sigma_{AB}\in\PPT'(A\!:\!B)$, the bound  $\Tr[\Phi_{AB}\sigma_{AB}]\leq\frac{1}{d}$ holds.
	
	Applying the upper bound in \eqref{eq-ent_distill_one_shot_UB_str_conv2} to the state $\rho_{AB}^{\otimes n}$ leads to the following result:
	
	\begin{corollary}{cor-ent_distill_n_shot_UB_str_conv}
		Let $\rho_{AB}$ be a bipartite state, let $n\geq 1$, $\varepsilon\in [0,1)$, and $\alpha>1$. For an $(n,d,\varepsilon)$ entanglement distillation protocol, the following bound holds
		\begin{align}
			\frac{\log_2 d}{n}&\leq \widetilde{R}_{\alpha}(A;B)_{\rho}+\frac{\alpha}{n(\alpha-1)}\log_2\!\left(\frac{1}{1-\varepsilon}\right)\label{eq-ent_distill_n_shot_UB_str_conv}	.
		\end{align}
Consequently,
		\begin{align}
			E_D^{n,\varepsilon}(A;B)_{\rho}&\leq \widetilde{R}_{\alpha}(A;B)_{\rho}+\frac{\alpha}{n(\alpha-1)}\log_2\!\left(\frac{1}{1-\varepsilon}\right).\label{eq-ent_distill_distill_ent_n_shot_UB_str_conv}
		\end{align}
	\end{corollary}
	
	\begin{Proof}
		An $(n,d,\varepsilon)$ entanglement distillation protocol for $\rho_{AB}$ is a $(d,\varepsilon)$ entanglement distillation protocol for $\rho_{AB}^{\otimes n}$. Therefore, applying the inequality in \eqref{eq-ent_distill_one_shot_UB_str_conv2} to the state $\rho_{AB}^{\otimes n}$ and dividing both sides by $n$ leads to
		\begin{equation}
			\frac{\log_2 d}{n}\leq \frac{1}{n}\widetilde{R}_{\alpha}(A^n;B^n)_{\rho^{\otimes n}}+\frac{\alpha}{n(\alpha-1)}\log_2\!\left(\frac{1}{1-\varepsilon}\right).
		\end{equation}
		Now, by subadditivity of the sandwiched R\'{e}nyi Rains relative entropy (see \eqref{eq:LAQC-Renyi-Rains-subadditivity}), we have that
		\begin{equation}
			\widetilde{R}_{\alpha}(A^n;B^n)_{\rho^{\otimes n}}\leq n\widetilde{R}_{\alpha}(A;B)_{\rho}.
		\end{equation}
		Therefore,
		\begin{equation}
			\frac{\log_2 d}{n}\leq \widetilde{R}_{\alpha}(A;B)_{\rho}+\frac{\alpha}{n(\alpha-1)}\log_2\!\left(\frac{1}{1-\varepsilon}\right),
		\end{equation}
		as required. Since this inequality holds for all $(n,d,\varepsilon)$ protocols, we obtain \eqref{eq-ent_distill_distill_ent_n_shot_UB_str_conv} by optimizing over all protocols $(d,\mathcal{L}_{AB\to\hat{A}\hat{B}})$, with $d_{\hat{A}}=d_{\hat{B}}=d\geq 1$.
	\end{Proof}

	Given an $\varepsilon\in(0,1)$, the inequality in \eqref{eq-ent_distill_n_shot_UB_str_conv} gives us a bound on the rate of an arbitrary $(n,d,\varepsilon)$ entanglement distillation protocol for a state $\rho_{AB}$. If instead we fix the rate to be $r$, so that $d=2^{nr}$, then the inequality in \eqref{eq-ent_distill_n_shot_UB_str_conv} is as follows:
	\begin{equation}\label{eq-ent_distill_n_shot_UB_str_conv_2}
		r\leq \widetilde{R}_{\alpha}(A;B)_{\rho}+\frac{\alpha}{n(\alpha-1)}\log_2\!\left(\frac{1}{1-\varepsilon}\right)
	\end{equation}
	for all $\alpha>1$. Rearranging this inequality gives us the following lower bound on $\varepsilon$:
	\begin{equation}
		\varepsilon\geq 1-2^{-n\left(\frac{\alpha-1}{\alpha}\right)\left(r-\widetilde{R}_{\alpha}(A;B)_{\rho}\right)}
	\end{equation}
	for all $\alpha>1$.

	\begin{theorem*}{Strong Converse Upper Bound on Distillable Entanglement}{thm-distillable_ent_str_conv_UB}
		Let $\rho_{AB}$ be  a bipartite state. The Rains relative entropy $R(A;B)_{\rho}$ is a strong converse rate for entanglement distillation for $\rho_{AB}$, i.e.,
		\begin{equation}\label{eq-ent_distill_strong_conv_UBs}
			 \widetilde{E}_D(A;B)_{\rho}\leq R(A;B)_{\rho},
		\end{equation}
		where we recall that $R(A;B)_{\rho}$ is defined as
		\begin{equation}
		R(A;B)_{\rho}=\inf_{\sigma_{AB}\in\PPT'(A:B)}D(\rho_{AB}\Vert\sigma_{AB}).		
		\end{equation}
	\end{theorem*}
	
	
	\begin{Proof}
		Let $\varepsilon\in[0,1)$ and $\delta>0$. Let $\delta_1,\delta_2>0$ be such that
		\begin{equation}\label{eq-ent_distill_Rains_str_conv_pf1}
			\delta>\delta_1+\delta_2\eqqcolon\delta'.
		\end{equation}
		Set $\alpha\in(1,\infty)$ such that
		\begin{equation}\label{eq-ent_distill_Rains_str_conv_pf2}
			\delta_1\geq \widetilde{R}_{\alpha}(A;B)_{\rho}-R(A;B)_{\rho},
		\end{equation}
		which is possible because $\widetilde{R}_{\alpha}(A;B)_{\rho}$ is monotonically increasing in $\alpha$ (which follows from Proposition~\ref{prop-sand_rel_ent_properties}) and because $\lim_{\alpha\to 1^+}\widetilde{R}_{\alpha}(A;B)_{\rho}=R(A;B)_{\rho}$ (see Appendix~\ref{app-sand_ren_inf_limit} for a proof). With this value of $\alpha$, take $n$ large enough so that
		\begin{equation}\label{eq-ent_distill_Rains_str_conv_pf3}
			\delta_2\geq \frac{\alpha}{n(\alpha-1)}\log_2\!\left(\frac{1}{1-\varepsilon}\right).
		\end{equation}
		
		Now, with the values of $n$ and $\varepsilon$ as above, an arbitrary $(n,d,\varepsilon)$ entanglement distillation protocol for $\rho_{AB}$ satisfies \eqref{eq-ent_distill_n_shot_UB_str_conv}, i.e.,
		\begin{equation}
			\frac{\log_2 d}{n}\leq \widetilde{R}_{\alpha}(A;B)_{\rho}+\frac{\alpha}{n(\alpha-1)}\log_2\!\left(\frac{1}{1-\varepsilon}\right)
		\end{equation}
		for all $\alpha\in(1,\infty)$. Rearranging the right-hand side of this inequality, and using \eqref{eq-ent_distill_Rains_str_conv_pf1}--\eqref{eq-ent_distill_Rains_str_conv_pf3}, we obtain
		\begin{align}
			\frac{\log_2 d}{n}&\leq R(A;B)_{\rho}+\widetilde{R}_{\alpha}(A;B)_{\rho}-R(A;B)_{\rho}+\frac{\alpha}{n(\alpha-1)}\log_2\!\left(\frac{1}{1-\varepsilon}\right)\\
			&\leq R(A;B)_{\rho}+\delta_1+\delta_2\\
			&=R(A;B)_{\rho}+\delta'\\
			&< R(A;B)_{\rho}+\delta.
		\end{align}
		So we have that $\frac{\log_2 d}{n}<R(A;B)_{\rho}+\delta$ for all $(n,d,\varepsilon)$ entanglement distillation protocols for $\rho_{AB}$ with sufficiently large $n$ such that \eqref{eq-ent_distill_Rains_str_conv_pf3} holds. Due to this strict inequality, it follows that there cannot exist an $(n,2^{n(R(A;B)_{\rho}+\delta)},\varepsilon)$ entanglement distillation protocol for $\rho_{AB}$ for all sufficiently large $n$ such that \eqref{eq-ent_distill_Rains_str_conv_pf3} holds. For if it were to exist, there would be a $d\geq 1$ such that $\log_2 d=n(R(A;B)_{\rho}+\delta)$, which we have just seen is not possible. Since $\varepsilon$ and $\delta$ are arbitrary, we conclude that for all $\varepsilon\in[0,1)$,  $\delta>0$, and sufficiently large $n$, there does not exist an $(n,2^{n(R(A;B)_{\rho}+\delta)},\varepsilon)$ entanglement distillation protocol for $\rho_{AB}$. This means that $R(A;B)_{\rho}$ is a strong converse rate, so that $\widetilde{E}_D(A;B)_{\rho}\leq R(A;B)_{\rho}$.
	\end{Proof}
	
	Given that the Rains relative entropy is a strong converse rate for distillable entanglement, by following arguments analogous to those in the proof above, we can conclude that $\frac{1}{k}R(A^k;B^k)_{\rho^{\otimes k}}$ is a strong converse rate for all $k\geq 2$. Therefore, the regularized quantity
	\begin{align}
		R^{\text{reg}}(A;B)_{\rho}&\coloneqq \lim_{n\to\infty}\frac{1}{n}R(A^n;B^n)_{\rho^{\otimes n}},	\end{align}
	is a strong converse rate for entanglement distillation for $\rho_{AB}$, so that 
	\begin{equation}
		 \widetilde{E}_D(A;B)_{\rho}\leq R^{\text{reg}}(A;B)_{\rho}.
	\end{equation}
	By subadditivity of Rains relative entropy (see \eqref{eq:LAQC-Renyi-Rains-subadditivity}), 
	\begin{equation}
		R^{\text{reg}}(A;B)_{\rho}\leq R(A;B)_{\rho},
	\end{equation}
	so that the regularized quantity in general gives a tighter upper bound on distillable entanglement.

\subsubsection{The Strong Converse from a Different Point of View}

	We now show that the Rains relative entropy is a strong converse rate using the equivalent characterization of a strong converse rate in \eqref{eq-ent_distill_str_conv_rate_alt}. In other words, given a bipartite state $\rho_{AB}$, we show that for an arbitrary sequence $\{(n,2^{nr},\varepsilon_n)\}_{n\in\mathbb{N}}$ of $(n,d,\varepsilon)$ protocols with rates $r>R(A;B)_{\rho}$, it holds that $\lim_{n\to\infty} \varepsilon_n=1$. Indeed, for every element of the sequence, the inequality in \eqref{eq-ent_distill_n_shot_UB_str_conv_2} applies; namely,
	\begin{equation}
		r\leq\widetilde{R}_{\alpha}(A;B)_{\rho}+\frac{\alpha}{n(\alpha-1)}\log_2\!\left(\frac{1}{1-\varepsilon_n}\right)
	\end{equation}
	for all $\alpha>1$. Rearranging this inequality gives us the following lower bound on $\varepsilon_n$:
	\begin{equation}\label{eq-ent_distill_n_shot_UB_str_conv_3}
		\varepsilon_n\geq 1-2^{-n\left(\frac{\alpha-1}{\alpha}\right)\left(r-\widetilde{R}_{\alpha}(A;B)_{\rho}\right)}
	\end{equation}
	for all $\alpha>1$. Now, since $r>R(A;B)_{\rho}$, $\lim_{\alpha\to 1^+}\widetilde{R}_{\alpha}(A;B)_{\rho}=R(A;B)_{\rho}$ (see Appendix~\ref{app-sand_ren_inf_limit}), and because $\widetilde{R}_{\alpha}(A;B)_{\rho}$ is monotonically increasing in $\alpha$ (this follows from Proposition~\ref{prop-sand_rel_ent_properties}), there exists an $\alpha^*>1$ such that $r>\widetilde{R}_{\alpha^*}(A;B)_{\rho}$. Applying the inequality in \eqref{eq-ent_distill_n_shot_UB_str_conv_3} to this value of $\alpha$, we find that
	\begin{equation}\label{eq-ent_distill_n_shot_UB_str_conv_4}
		\varepsilon_n\geq 1-2^{-n\left(\frac{\alpha^*-1}{\alpha^*}\right)\left(r-\widetilde{R}_{\alpha^*}(A;B)_{\rho}\right)}.
	\end{equation}
	Then, taking the limit $n\to\infty$ on both sides of this inequality, we conclude that $\lim_{n\to\infty}\varepsilon_n\geq 1$. But $\varepsilon_n\leq 1$ for all $n$ because $\varepsilon_n$ is  a probability by definition. So we obtain $\lim_{n\to\infty}\varepsilon_n =1$. Since the rate $r>R(A;B)_{\rho}$ is arbitrary, we conclude that $R(A;B)_{\rho}$ is a strong converse rate for entanglement distillation for $\rho_{AB}$. We also see from \eqref{eq-ent_distill_n_shot_UB_str_conv_4} that the sequence $\{\varepsilon_n\}_{n\in\mathbb{N}}$ approaches one at an exponential rate.

	\subsection{Squashed Entanglement Weak Converse Upper Bound}
	
	In this section, we establish the squashed entanglement of a bipartite state as a weak converse upper bound on its distillable entanglement. The main idea is to apply the one-shot bound from Theorem~\ref{thm:ED:squashed-ent-upp-1-shot-DE} and the additivity of the squashed entanglement (Proposition~\ref{prop-squashed_ent_properties}) in order to arrive at this conclusion.  
	
	\begin{corollary}{cor:ED:sq-ent-n-bound}
	Let $\rho_{AB}$ be a bipartite state, let $n\geq 1$, and let $\varepsilon \in [0,1)$. For an $(n,d,\varepsilon)$ entanglement distillation protocol, the following bound holds
	\begin{equation}
	\frac{1}{n} \log_2 d \leq \frac{1}{1-\sqrt{\varepsilon}}\left(E_{\operatorname{sq}}(A;B)_{\rho} + \frac{g_2(\sqrt{\varepsilon})}{n}\right).
	\label{eq:ED:squashed-ent-n-shot-bnd-DE}
	\end{equation}
	\end{corollary}
	
	\begin{Proof}
	An $(n,d,\varepsilon)$ entanglement distillation protocol for $\rho_{AB}$ is a $(d,\varepsilon)$ entanglement distillation protocol for $\rho_{AB}^{\otimes n}$. Therefore, applying the inequality in Theorem~\ref{thm:ED:squashed-ent-upp-1-shot-DE} to the state $\rho_{AB}^{\otimes n}$ and dividing both sides by $n$ leads to
		\begin{equation}
			\frac{1}{n} \log_2 d \leq \frac{1}{1-\sqrt{\varepsilon}}\left(\frac{1}{n} E_{\operatorname{sq}}(A^n;B^n)_{\rho^{\otimes n}} + \frac{g_2(\sqrt{\varepsilon})}{n}\right).
		\end{equation}
		Now, by additivity of the squashed entanglement (see \eqref{eq:LAQC-sq-additive-states}), we have that
		\begin{equation}
			E_{\operatorname{sq}}(A^n;B^n)_{\rho^{\otimes n}} = n E_{\operatorname{sq}}(A;B)_{\rho}.
		\end{equation}
		This concludes the proof.
	\end{Proof}

	We now provide a proof of \eqref{eq:ED:squashed-ent-weak-conv-ED}, the statement that the squashed entanglement is a weak converse rate for entanglement distillation. 
		Suppose that $R$ is an achievable rate for entanglement distillation for the bipartite state $\rho_{AB}$. Then, by definition, for all $\varepsilon\in(0,1]$,  $\delta>0$, and sufficiently large $n$, there exists an $(n,2^{n(R-\delta)},\varepsilon)$ entanglement distillation protocol for $\rho_{AB}$. For all such protocols,  the inequality in \eqref{eq:ED:squashed-ent-n-shot-bnd-DE} holds, so that
	\begin{equation}
		R-\delta \leq\frac{1}{1-\sqrt{\varepsilon}}\left(E_{\operatorname{sq}}(A;B)_{\rho}+\frac{1}{n}g_2(\sqrt{\varepsilon})\right).
	\end{equation}
	 Since the inequality holds for all sufficiently large $n$, it holds in the limit $n\to\infty$, so that
	\begin{align}
		R&\leq \lim_{n\to\infty}\frac{1}{1-\sqrt{\varepsilon}}\left(E_{\operatorname{sq}}(A;B)_{\rho}+\frac{1}{n}g_2(\sqrt{\varepsilon})\right)+\delta\\
		&=\frac{1}{1-\sqrt{\varepsilon}}E_{\operatorname{sq}}(A;B)_{\rho}+\delta.
	\end{align}
	Then, since this inequality holds for all $\varepsilon\in(0,1]$ and $\delta>0$, we conclude that
	\begin{align}
		R&\leq \lim_{\varepsilon,\delta\to 0}\left(\frac{1}{1-\sqrt{\varepsilon}}E_{\operatorname{sq}}(A;B)_{\rho}+\delta\right)\\
		&=E_{\operatorname{sq}}(A;B)_{\rho}.
	\end{align}
	We have thus shown that the squashed entanglement $E_{\operatorname{sq}}(A;B)_{\rho}$ is a weak converse rate for entanglement distillation.

\subsection{One-Way Entanglement Distillation}\label{sec-one_way_ent_distill}

	In Section~\ref{subsec-ent_distill_one_shot_lower_bound}, we considered a one-way entanglement distillation protocol to derive a lower bound on the one-shot distillable entanglement of a bipartite state. In the asymptotic setting, this leads to the coherent information lower bound on distillable entanglement, i.e.,
	\begin{equation}
		E_D(A;B)_{\rho}\geq I(A\rangle B)_{\rho}=H(B)_{\rho}-H(AB)_{\rho},
	\end{equation}
	which holds for every bipartite state $\rho_{AB}$. By simply reversing the roles of Alice and Bob in the protocol, it follows that 
	\begin{equation}
		E_D(A;B)_{\rho}\geq I(B\rangle A)_{\rho}=H(A)_{\rho}-H(AB)_{\rho}.
	\end{equation}
	The quantity on the right-hand side of the above inequality is sometimes called \textit{reverse coherent information}. Thus, in general, we have the following lower bound on distillable entanglement:
	\begin{equation}
		E_D(A;B)_{\rho}\geq\max\{I(A\rangle B)_{\rho},I(B\rangle A)_{\rho}\},
	\end{equation}
	which holds for every bipartite state $\rho_{AB}$.
	
	The coherent information lower bound can be improved by first applying a two-way LOCC channel to $n$ copies of the given state, and then performing a one-way entanglement distillation protocol at the coherent information rate. This leads to
	\begin{equation}\label{eq-distillable_entanglement_2}
		E_D(A;B)_{\rho}=\lim_{n\to\infty}\frac{1}{n}\sup_{\mathcal{L}}I(A'\rangle B')_{\mathcal{L}(\rho^{\otimes n})}=\lim_{n\to\infty}\frac{1}{n}D^{\leftrightarrow}(\rho_{AB}^{\otimes n})
	\end{equation}
	where the optimization is over every two-way LOCC channel $\mathcal{L}_{A^nB^n\to A' B'}$.
	
	If we restrict the optimization in \eqref{eq-distillable_entanglement_2} above to one-way LOCC channels of the form $\mathcal{L}_{A^nB^n\to A' B'}$, then we obtain what is called the \textit{one-way distillable entanglement}, denoted by $E_{D}^{\rightarrow}(A;B)_{\rho}$, and defined operationally in a similar way to the distillable entanglement $E_{D}(A;B)_{\rho}$, but with the free operations allowed restricted to one-way LOCC. A key result is the following equality:
	\begin{equation}\label{eq-one_way_distillable_entanglement}
		E_{D}^{\rightarrow}(A;B)_{\rho}=\lim_{n\to\infty}\frac{1}{n}\sup_{\mathcal{L}^{\rightarrow} }I( A'\rangle B')_{\mathcal{L}^{\rightarrow}(\rho^{\otimes n})}=\lim_{n\to\infty}\frac{1}{n}D^{\rightarrow}(\rho_{AB}^{\otimes n}),
	\end{equation}
	where
	\begin{equation}
		D^{\rightarrow}(\rho_{AB})\coloneqq\sup_{\mathcal{L}^{\rightarrow} }I( A'\rangle B')_{\mathcal{L}^{\rightarrow}(\rho)}.
	\end{equation}
	Like the distillable entanglement, the one-way distillable entanglement is an operational quantity of interest in entanglement theory. Furthermore, the equality in \eqref{eq-one_way_distillable_entanglement} can be proved along similar lines to how we proved~\eqref{eq-distillable_entanglement}.
	
	In what follows, we show that this expression for one-way distillable entanglement can be simplified.
	
	\begin{theorem*}{One-Way Distillable Entanglement of a Bipartite State}{thm-one_way_distillable_entanglement}
		The one-way distillable entanglement of a bipartite state $\rho_{AB}$ is given by
		\begin{equation}\label{eq-one_way_distillable_entanglement_thm}
			E_{D}^{\rightarrow}(A;B)_{\rho}=\lim_{n\to\infty}\frac{1}{n}\sup_{V}I( A'\rangle X B^n)_{V\rho^{\otimes n}V^\dagger},
		\end{equation}
		where $V_{A^n\to A' X E}$ is an isometry of the form
		\begin{equation}\label{eq-one_way_distill_ent_isometries}
			V_{A^n\to A' X E}=\sum_{x\in\mathcal{X}}K_{A^n\to A'}^x\otimes\ket{x}_{X}\otimes\ket{x}_E,
		\end{equation}
		with $\sum_{x\in\mathcal{X}}(K_{A^n}^x)^\dagger K_{A^n}^x=\mathbbm{1}_{A^n}$, $d_{A'}=d_A^n$, $d_{X}\leq d_A^2$, and $d_E=|\mathcal{X}|=d_X$. Additionally,
		\begin{equation}\label{eq-one_way_distill_ent_one_copy}
			D^{\rightarrow}(\rho_{AB})=\sup_V I(A'\rangle X B)_{V\rho V^\dagger},
		\end{equation}
		with the optimization over isometries $V$ as in \eqref{eq-one_way_distill_ent_isometries}, with $n=1$.
	\end{theorem*}
	
	This theorem tells us that, to determine the one-way distillable entanglement of a bipartite state, it suffices to optimize over one-way LOCC channels that consist of only a quantum instrument for Alice, with each of the corresponding maps containing just one Kraus operator. Furthermore, it suffices to take $A'=A^n$ and $B'=B^n$.
	
	\begin{Proof}
		We start by recalling from Section~\ref{subsec-LOCC_channels} (see also the beginning of Section~\ref{subsec-ent_distill_one_shot_lower_bound}) that every one-way LOCC channel $\mathcal{L}_{A^nB^n\to A'B'}^{\rightarrow}$ can be expressed as
		\begin{align}
			\omega_{A'B'}&\coloneqq\mathcal{L}_{A^nB^n\to A'B'}^{\rightarrow}(\rho_{A^nB^n})\label{eq-one_way_ent_distill_pf_1}\\
			&=\sum_{x\in\mathcal{X}}(\mathcal{E}_{A^n\to A'}^x\otimes\mathcal{D}_{B^n\to B'}^x)(\rho_{A^nB^n})\label{eq-one_way_ent_distill_pf_2}\\
			&=\left(\mathcal{D}_{X_B B^n\to B'}\circ\mathcal{C}_{X_A\to X_B}\circ\mathcal{E}_{A^n\to A' X_A}\right)(\rho_{A^nB^n}),\label{eq-one_way_ent_distill_pf_3}
		\end{align}
		where $\mathcal{X}$ is some finite alphabet, $d_{X_A}=d_{X_B}=|\mathcal{X}|$, $\{\mathcal{E}_{A^n\to A'}^x\}_{x\in\mathcal{X}}$ is a set of completely positive maps such that $\sum_{x\in\mathcal{X}}\mathcal{E}_{A^n\to A'}^x$ is trace preserving, and $\{\mathcal{D}_{B^n\to B'}^x\}_{x\in\mathcal{X}}$ is a set of channels. In particular,
		\begin{align}
			\mathcal{E}_{A^n\to A' X_A}(\rho_{A^nB^n})&=\sum_{x\in\mathcal{X}}\mathcal{E}_{A^n\to A'}^x(\rho_{A^nB^n})\otimes\ket{x}\!\bra{x}_{X_A},\\
			\mathcal{D}_{X_BB^n\to B'}(\ket{x}\!\bra{x}_{X_B}\otimes\rho_{A^nB^n})&=\mathcal{D}_{B^n\to B'}^x(\rho_{A^nB^n}).
		\end{align}
		
		For every $n\geq 1$, if we restrict the optimization in \eqref{eq-one_way_distillable_entanglement} such that $|\mathcal{X}|=d_{A^n}^2=d_A^{2n}$, $d=d_{ A'}=d_A^n$, $\mathcal{D}_{B^n\to B'}^x=\id_{B^n}$ for all $x\in\mathcal{X}$, and $\mathcal{E}_{A^n\to A'}^x(\cdot)=K_{A^n}^x(\cdot)K_{A^n}^x$ for all $x\in\mathcal{X}$ such that $\sum_{x\in\mathcal{X}}(K_{A^n}^x)^\dagger K_{A^n}^x=\mathbbm{1}_{A^n}$, then the LOCC channel $\mathcal{L}_{A^nB^n\to A' B'}^{\rightarrow}$ reduces to
		\begin{equation}
			\mathcal{L}_{A^nB^n\to A' B'}^{\rightarrow}(\rho_{A^nB^n})=\sum_{x\in\mathcal{X}}K_{A^n}^x\rho_{A^nB^n}(K_{A^n}^x)^\dagger\otimes\ket{x}\!\bra{x}_{X}
		\end{equation}
		for every state $\rho_{A^nB^n}$, and it has an isometric extension of the form
		\begin{equation}
			V_{A^n\to A^nXE}=\sum_{x\in\mathcal{X}}K_{A^n}^x\otimes\ket{x}_{X}\otimes\ket{x}_E.
		\end{equation}
		We thus obtain
		\begin{equation}\label{eq-one_way_ent_distill_pf_6}
			E_{D}^{\rightarrow}(A;B)_{\rho}\geq \lim_{n\to\infty}\frac{1}{n}\sup_{V}I(A^n\rangle X B^n)_{\omega}
		\end{equation}
		
		The rest of the proof is devoted to proving the reverse inequality.
		Let $\mathcal{L}_{A^nB^n\to A'B'}^{\rightarrow}$ be an arbitrary one-way LOCC channel of the form in \eqref{eq-one_way_ent_distill_pf_1}--\eqref{eq-one_way_ent_distill_pf_3}. For every state $\rho_{A^nB^n}$, let
		\begin{align}
			p_x&\coloneqq\Tr[\mathcal{E}_{A^n\to A'}^x(\rho_{A^nB^n})],\\
			\rho_{A'B^n}^x&\coloneqq\frac{1}{p_x}\mathcal{E}_{A^n\to A'}^x(\rho_{A^nB^n}),\\
			\rho_{B^n}^x&\coloneqq\frac{1}{p_x}\Tr_{ A'}[\mathcal{E}_{A^n\to A'}^x(\rho_{A^nB^n})],
		\end{align}
		for all $x\in\mathcal{X}$. Then, using the data-processing inequality for coherent information in \eqref{eq-gen_coh_inf_state_monotonicity} and the direct-sum property for quantum relative entropy, we obtain
		\begin{align}
			I( A'\rangle B')_{\omega}
			& \leq I( A'\rangle B^nX_B)_{\mathcal{E}(\rho)}\\
			& =\sum_{x\in\mathcal{X}}p_xD(\rho_{ A' B^n}^x\Vert\mathbbm{1}_{A'}\otimes\rho_{B^n}^x)\\
			& =\sum_{x\in\mathcal{X}}p_x I( A' \rangle B^n)_{\rho^x}.\label{eq-one_way_ent_distill_pf_7}
		\end{align}
		Now,
		\begin{align}
			\mathcal{E}_{A^n\to A' X_A}(\rho_{A^nB^n})&=\sum_{x\in\mathcal{X}}\mathcal{E}_{A^n\to A'}^x(\rho_{AB})\otimes\ket{x}\!\bra{x}_{X_A}\\
			&=\sum_{x\in\mathcal{X}}p_x\rho_{A'B}^x\otimes\ket{x}\!\bra{x}_{X_A}.
		\end{align}
		Suppose that $\mathcal{E}_{A^n\to A'}^x$ has the following Kraus representation:
		\begin{equation}
			\mathcal{E}_{A^n\to A'}^x(\cdot)=\sum_{y\in\mathcal{Y}} K_{A^n\to A'}^{x,y}(\cdot)(K_{A^n\to A'}^{x,y})^\dagger,
		\end{equation}
		where $\mathcal{Y}$ is some finite alphabet. Let
		\begin{equation}
			q_{x,y}\coloneqq \Tr[K_{A^n\to A'}^{x,y}\rho_{ A^n B^n}(K_{A^n\to A'}^{x,y})^\dagger],
		\end{equation}
		and observe that $p_x=\sum_{y\in\mathcal{Y}}q_{x,y}$. Therefore, for each $x\in\mathcal{X}$, the values $r_{y|x} \coloneqq \frac{q_{x,y}}{p_x}$ constitute a probability distribution on $\mathcal{Y}$, in the sense that $r_{y|x}\geq 0$ for all $y\in\mathcal{Y}$, and $\sum_{y\in\mathcal{Y}}r_{y|x}=1$. Using this, and letting
		\begin{equation}
			\rho_{ A'B^n}^{x,y}\coloneqq \frac{1}{q_{x,y}}K_{A^n\to A'}^{x,y}\rho_{A^nB^n}(K_{A^n\to A'}^{x,y})^\dagger,
		\end{equation}
		so that
		\begin{equation}
			p_x\rho_{ A' B^n}^x=\sum_{y\in\mathcal{Y}}q_{x,y}\rho_{A' B^n}^{x,y},
		\end{equation}
		we find that
		\begin{align}
			p_xI( A'\rangle B^n)_{\rho^x}&
			 \leq p_x\sum_{y\in\mathcal{Y}}r_{y|x}I(A'\rangle B^n)_{\rho^{x,y}},
		\end{align}
		for all $x\in\mathcal{X}$, where  the  inequality follows from  convexity of coherent information (see \eqref{eq:QEI:coh-info-convex}). Without loss of generality, we can take $A'\equiv A^n$:  if $A'$ has a dimension smaller than that of $A^n$, we can always first isometrically embed $A'$ into $A^n$. The coherent information remains unchanged under this isometric embedding.
		
		Combining the last inequality above with the one in \eqref{eq-one_way_ent_distill_pf_7}, we conclude that
		\begin{equation}
		I(A'\rangle B')_{\omega} \leq \sum_{x \in \mathcal{X}}p_x\sum_{y\in\mathcal{Y}}r_{y|x}I(A'\rangle B^n)_{\rho^{x,y}} = \sum_{x \in \mathcal{X}, y\in\mathcal{Y}} q_{x,y}I(A'\rangle B^n)_{\rho^{x,y}}.
		\end{equation}
		Then assigning the superindex $z=(x,y)$ with $\mathcal{Z} = \mathcal{X} \times \mathcal{Y}$, we finally have
		\begin{equation}
			I(A'\rangle B')_{\omega}\leq \sup_{\mathcal{E}} I(A'\rangle B^n Z_B)_{\mathcal{E}(\rho^{\otimes n})},
		\end{equation}
		where the optimization is over channels $\mathcal{E}_{A^n\to A'Z_B}$ of the form
		\begin{equation}
			\mathcal{E}_{A^n\to A'Z_B}(\cdot)\coloneqq\sum_{z\in\mathcal{X}} K_{A^n\to A'}^z(\cdot)(K_{A^n\to A'}^z)^\dagger\otimes\ket{z}\!\bra{z}_{Z_B},
		\end{equation}
		such that $\sum_{z\in\mathcal{Z}}(K_{A^n\to A'}^z)^\dagger K_{A^n\to A'}^z=\mathbbm{1}_{A^n}$. (This is effectively an optimization over operators $\{K_{A^n\to A'}^z\}_{z\in\mathcal{Z}}$ such that $\sum_{z\in\mathcal{Z}}(K_{A^n\to A'}^z)^\dagger K_{A^n\to A'}^z=\mathbbm{1}_{A^n}$.) The channel $\mathcal{E}_{A^n\to A' Z_B}$ has an isometric extension of the form
		\begin{equation}\label{eq-one_way_ent_distill_pf_5}
			V_{A^n\to A' Z_BE}=\sum_{z\in\mathcal{Z}} K_{A^n\to A'}^z\otimes\ket{z}_{Z_B}\otimes\ket{z}_E,
		\end{equation}
		where $d_E=d_{Z_B}=|\mathcal{Z}|$. Since the number of Kraus operators need not exceed $d_{A^n}^2=d_A^{2n}$ (see Theorem~\ref{thm-q_channels}), we can  take  $d_{Z}=d_A^{2n}$ without loss of generality. We can thus optimize over all isometries of the form in \eqref{eq-one_way_ent_distill_pf_5}. Altogether, we have that
		\begin{equation}
			I( A'\rangle B')_{\omega}\leq \sup_V I(A'\rangle Z B^n)_{V\rho^{\otimes n}V^\dagger}
		\end{equation}
		for every one-way LOCC channel $\mathcal{L}_{A^nB^n\to A' B'}^{\rightarrow}$ and all $n\geq 1$. Optimizing over all one-way LOCC channels on the left-hand side of the  inequality above, and taking the limit $n\to\infty$ leads us to conclude that
		\begin{equation}
			E_{D}^{\rightarrow}(A;B)_{\rho}\leq \lim_{n\to\infty}\frac{1}{n}\sup_V I(A'\rangle Z B^n)_{V\rho^{\otimes n}V^\dagger}.
		\end{equation}
		Combining this with \eqref{eq-one_way_ent_distill_pf_6} and reassigning $Z$ as $X$ finishes the proof.
	\end{Proof}

	\begin{Lemma}{lem-one_way_ent_distill_one_copy}
		For every bipartite state $\rho_{AB}$, the optimized coherent information lower bound on distillable entanglement is non-negative, i.e., $D^{\rightarrow}(\rho_{AB})\geq 0$.
	\end{Lemma}
	
	\begin{Proof}
		Let $\psi_{ABR}=\ket{\psi}\!\bra{\psi}_{ABR}$ be a purification of $\rho_{AB}$, and consider the following Schmidt decomposition of $\ket{\psi}_{ABR}$:
		\begin{equation}
			\ket{\psi}_{ABR}=\sum_{k=0}^{r-1} \sqrt{\lambda_k}\ket{\phi_k}_A\otimes\ket{\varphi_k}_{BR}.
		\end{equation}
		Then, let
		\begin{equation}
			V_{A\to A'XE}\coloneqq \sum_{k=0}^{r-1}\ket{k}_{A'} \bra{\phi_k}_A\otimes \ket{k}_X\otimes\ket{k}_E.
		\end{equation}
		It is then straightforward to show that $I(A' \rangle XB)_{V\rho V^\dagger}=0$. Since $V$ is an example of an isometry in the optimization for $D^{\rightarrow}(\rho_{AB})$, we conclude that $D^{\rightarrow}(\rho_{AB})\geq I(A' \rangle XB)_{V\rho V^\dagger}=0$.
	\end{Proof}

\section{Examples}

	We now consider classes of bipartite states and evaluate the upper and lower bounds on their distillable entanglement that we have established in this chapter. In some cases, the distillable entanglement can be determined exactly because the upper and lower bounds coincide.

\subsection{Pure States}

\label{sec-ED:pure-states}

	The simplest example for which distillable entanglement can be determined exactly is the class of pure bipartite states. In this case, the coherent information lower bound and the Rains relative entropy upper bound coincide and are equal to the entropy of the reduced state. Indeed, for the coherent information, the joint entropy $H(AB)_{\psi}=0$ for every pure state $\psi_{AB}$, so that
	\begin{equation}\label{eq-coh_inf_pure_state}
		I(A\rangle B)_{\psi}=H(B)_{\psi}-H(AB)_{\psi}=H(B)_{\psi}=H(A)_{\psi},
	\end{equation}
	where the last equality follows from the Schmidt decomposition theorem (Theorem~\ref{thm-Schmidt}) to see that the reduced states $\psi_A$ and $\psi_B$ have the same non-zero eigenvalues, and thus the same value for the entropy. On the other hand, Proposition~\ref{prop-ent_meas_rel_ent_entanglement} states that the relative entropy of entanglement $E_R(A;B)_{\psi}=H(A)_{\psi}$, and we also know that $E_R(A;B)_{\psi} \geq R(A;B)_{\psi}$ (see~\eqref{eq-SEP_PPT_Rains_ineq}). We thus have the following:
	
	\begin{theorem*}{Distillable Entanglement for Pure States}{thm-ent_distill_pure_states}
		The distillable entanglement of a pure bipartite state $\psi_{AB}$ is equal to the entropy of the reduced state on $A$, i.e.,
		\begin{equation}
			E_D(A;B)_{\psi} = H(A)_{\psi}.
		\end{equation}
	\end{theorem*}
	

\subsection{Degradable and Anti-Degradable States}

\label{subsec-deg_antideg_states}

	In this section, we define two classes of states for which the one-way distillable entanglement takes on a simple form.
	
	\begin{definition}{Degradable and Anti-Degradable Bipartite States}{def-deg_antideg_states}
		Given a bipartite state $\rho_{AB}$ with purification $\psi_{ABE}$, we call it \textit{degradable} if there exists a quantum channel $\mathcal{D}_{B\to E'}$ such that the state $\tau_{AE'E}\coloneqq\mathcal{D}_{B\to E'}(\psi_{ABE})$ satisfies
		\begin{equation}
			\tau_{AE'}=\tau_{AE}=\psi_{AE}.
		\end{equation}
		
		We call $\rho_{AB}$ \textit{anti-degradable} if there exists a quantum channel $\mathcal{A}_{E\to B'}$ such that the state $\omega_{ABB'}\coloneqq \mathcal{A}_{E\to B'}(\psi_{ABE})$ satisfies
		\begin{equation}
			\omega_{AB'}=\omega_{AB}=\rho_{AB}.
		\end{equation}
	\end{definition}
	
	\begin{remark}
		Degradable and anti-degradable states are the state counterparts of degradable and anti-degradable channels; see Definition~\ref{def-deg_antideg_chan}. In fact, observe that the Choi state of a degradable channel is a degradable state, and the Choi state of an anti-degradable channel is an anti-degradable state.
		
		Anti-degradable states are also sometimes called symmetrically extendible states or two-extendible states (please consult the Bibliographic Notes in Section~\ref{sec:ed:bib-notes}).
	\end{remark}
	
	Intuitively, a degradable state is one for which the system $B$ can be used to simulate (via a quantum channel $\mathcal{D}_{B\to E'}$) the correlations between $A$ and $E$. Analogously, an anti-degrdable state is one for which the system $E$ can be used to simulate (via a quantum channel $\mathcal{A}_{E\to B'}$) the correlations between $A$ and $B$.
	
	An anti-degradable state $\rho_{AB}$ is one for which the environment~$E$ (corresponding to the purifying system of $\rho_{AB}$) \textit{cannot} be decoupled from $A$ and $B$ through LOCC from $A$ to $B$ alone. Indeed, recall the task of decoupling from Section~\ref{subsec-ent_distill_one_shot_lower_bound} (in particular, see Figure~\ref{fig-ent_distill_decoupling}). Since a channel can always be applied to $E$ in order to simulate the correlations between $A$ and $B$, from the point of view of $A$, the systems $B$ and $E$ become indistinguishable, so that $A$ and $B$ cannot be (perfectly) decoupled from $E$. Given that decoupling is not possible for anti-degradable states, we might expect that anti-degradable states have zero one-way distillable entanglement. This is indeed true, as we now show.
	
	\begin{theorem*}{One-Way Distillable Entanglement for Anti-Degradable States}{thm-distill_ent_anti_deg_states}
		For an anti-degradable state $\rho_{AB}$, the one-way distillable entanglement is equal to zero, i.e., $E_D^{\rightarrow}(A;B)_{\rho}=0$.
	\end{theorem*}
	
	\begin{Proof}
		Let $V_{A\to A'XE}$ be an arbitrary isometry in the optimization for $D^{\rightarrow}(\rho_{AB})$. Also, let $\psi_{ABR}$ be a purification of $\rho_{AB}$. Then, because $\rho_{AB}$ is anti-degradable, there exists a channel $\mathcal{A}_{R\to B}$ such that
		\begin{equation}
			\rho_{AB}=\mathcal{A}_{R\to B}(\psi_{AR}).
		\end{equation}
		Now, let
		\begin{equation}
			\omega_{A'XEBR}=V_{A\to A'XE}\psi_{ABR}V_{A\to A'XE}^\dagger,
		\end{equation}
		which is a pure state. Then, using the fact that
		\begin{align}
			\omega_{ A'XB}&=\Tr_E[V_{A\to A'XE}\rho_{AB}V_{A\to A'XE}^\dagger]\\
			&=(\mathcal{A}_{R\to B}\circ\Tr_E)(V_{A\to A'XE}\psi_{AR}V_{A\to A'XE}^\dagger)\\
			&=\mathcal{A}_{R\to B}(\omega_{ A'XR}),
		\end{align}
		and that $\omega_{XB}=\mathcal{A}_{R\to B}(\omega_{XR})$, we find that	
		\begin{align}
			I( A'\rangle XB)_{\omega}&
			\leq I(A'\rangle RX)_{\omega}\\
			&=H(RX)_{\omega}-H(A'RX)_{\omega}\\
			&=H(A'EB)_{\omega}-H(EB)_{\omega}\\
			&=H(A'XB)_{\omega}-H(XB)_{\omega}\\
			&=-I(A'\rangle XB)_{\omega},
		\end{align}
		where we used the data-processing inequality in \eqref{eq-gen_coh_inf_state_monotonicity}, and for the subsequent equalities we used the fact that $\omega_{A'XEBR}$ is a pure state that is symmetric in $X$ and $E$. We thus have $I(A'\rangle XB)_{V\rho V^\dagger}\leq 0$ for every isometry $V$ used in the optimization for $D^{\rightarrow}(\rho_{AB})$, implying that $D^{\rightarrow}(\rho_{AB})\leq 0$. However, since $D^{\rightarrow}(\rho_{AB})\geq 0$ by Lemma~\ref{lem-one_way_ent_distill_one_copy}, we obtain $D^{\rightarrow}(\rho_{AB})=0$. The statement that $E_D^{\rightarrow}(A;B)_{\rho}=0$ follows by repeating the same argument for $n$ copies of $\rho_{AB}$ and using the fact that $\rho_{AB}^{\otimes n}$ is an anti-degradable state if $\rho_{AB}$ is.
	\end{Proof}

	\begin{theorem*}{One-Way Distillable Entanglement for Degradable States}{thm-dist_ent_one_way_deg_states}
		For a degradable state $\rho_{AB}$, we have
		\begin{equation}
			D^{\rightarrow}(\rho_{AB})=I(A\rangle B)_{\rho}.
		\end{equation}
		Consequently, $D^{\rightarrow}(\rho_{AB}^{\otimes n})=nD^{\rightarrow}(\rho_{AB})$, and thus the one-way distillable entanglement of a degradable state $\rho_{AB}$ is equal to its coherent information:
		\begin{equation}
			E_D^{\rightarrow}(A;B)_{\rho}=I(A\rangle B)_{\rho}.
		\end{equation}
	\end{theorem*}
	
	\begin{Proof}
		First, observe that if we pick the isometry $V$ in  \eqref{eq-one_way_distill_ent_one_copy} to be $V_{A\to A'XE}=\mathbbm{1}_{A}\otimes\ket{0,0}_{XE}$ (so that $X$ and $E$ are one-dimensional systems), then we obtain $D^{\rightarrow}(\rho_{AB})\geq I(A\rangle B)_{\rho}$. Then, we conclude that $D^{\rightarrow}(\rho_{AB}^{\otimes n})\geq nI(A\rangle B)_{\rho}$ because coherent information is additive for product states; thus,  $E_D(A;B)_{\rho}=\lim_{n\to\infty}\frac{1}{n}D^{\rightarrow}(\rho_{AB}^{\otimes n})\geq I(A\rangle B)_{\rho}$.
		
		We now prove the reverse inequality.
		Let $V_{A\to A'XE}$ be an arbitrary isometry in the optimization for $D^{\rightarrow}(\rho_{AB})$. Also, let $\psi_{ABR}$ be a purification of $\rho_{AB}$. Then, since $\rho_{AB}$ is degradable, there exists a channel $\mathcal{D}_{B\to R'}$ such that the state $\tau_{AR'R}\coloneqq\mathcal{D}_{B\to R'}(\psi_{ABR})$ satisfies
		\begin{equation}
			\tau_{AR'}=\tau_{AR}=\psi_{AR}.
		\end{equation}
		Let $W_{B\to R'F}$ be an isometric extension of $\mathcal{D}_{B\to R'}$, and let
		\begin{align}
			\ket{\varphi}_{AR'FR}&=W_{B\to R'F}\ket{\psi}_{ABR},\\
			\ket{\omega}_{A'XEBR}&=V_{A\to A'XE}\ket{\psi}_{ABR},\\
			\ket{\phi}_{A'XER'FR}&=W_{B\to R'F}\ket{\omega}_{A'XEBR}=V_{A\to A'XE}\ket{\varphi}_{AR'FR}.
		\end{align}
		Then, by invariance of entropy under the isometry $W_{B\to R'F}$,
		\begin{align}
			I(A'\rangle XB)_{\omega}&=H(XB)_{\omega}-H(A'XB)_{\omega}\\
			&=H(XR'F)_{\phi}-H(A'XR'F)_{\phi}\\
			&=H(XR'F)_{\phi}-H(ER)_{\phi}\\
			&=H(XR'F)_{\phi}-H(XR)_{\phi},
		\end{align}
		where the second-to-last line follows because $\phi_{A'XER'FR}$ is a pure state, so that $H(A'XR'F)_{\phi}=H(ER)_{\phi}$, and then for the last line we used the fact that $\phi_{A'XER'FR}$ is symmetric in $X$ and $E$ by definition of $V_{A\to A'XE}$. Next, due to the fact that $\tau_{AR'}=\tau_{AR}$, it holds that $\phi_{XR'}=\phi_{XR}$. We thus obtain
		\begin{align}
			I(A'\rangle XB)_{\omega}&=H(XR'F)_{\phi}-H(XR')_{\phi}\\
			&=-I(F\rangle XR')_{\phi}\\
			&\leq -I(F\rangle R')_{\phi}\\
			&=H(FR')_{\phi}-H(R')_{\phi}\\
			&=H(FR')_{\phi}-H(R)_{\phi},
		\end{align}
		where the inequality follows from the data-processing inequality with the partial trace channel $\Tr_X$, and the last equality follows because $\phi_R=\phi_{R'}$, due to the degradability of $\rho_{AB}$. Finally, observe that 
		\begin{equation}
			\phi_{R'F}=W_{B\to R'F}\rho_B W_{B\to R'F}^\dagger,
		\end{equation}
		so that $H(R'F)_{\phi}=H(B)_{\rho}$ by isometric invariance of entropy. Also, $\phi_R=\Tr_{AB}[\psi_{ABR}]$, which implies that $H(R)_{\phi}=H(AB)_{\rho}$. We thus obtain
		\begin{equation}
			I(A'\rangle XB)_{\omega}\leq I(A\rangle B)_{\rho}
		\end{equation}
		for every isometry $V_{A\to A'XE}$. This implies that $D^{\rightarrow}(\rho_{AB})\leq I(A\rangle B)_{\rho}$, i.e.,
		\begin{equation}
			D^{\rightarrow}(\rho_{AB})= I(A\rangle B)_{\rho}.
		\end{equation}
		In other words, the trivial isometry $V_{A\to A'XE}=\mathbbm{1}_A\otimes\ket{0,0}_{XE}$ is optimal for $D^{\rightarrow}(\rho_{AB})$ when $\rho_{AB}$ is a degradable state. Thus, by additivity of coherent information for product states, we obtain $E_D(A;B)_{\rho}=I(A\rangle B)_{\rho}$, as required.
	\end{Proof}

\section{Summary}

	In this chapter, we studied the task of entanglement distillation, in which the goal is for Alice and Bob to convert many copies of a shared entangled state $\rho_{AB}$ to some (smaller) number of ebits, i.e., copies of a two-qubit maximally entangled state using local operations and classical communication. The largest rate at which this can be done, given arbitrarily many copies of $\rho_{AB}$ and such that the error vanishes, is called distillable entanglement, and we denote it by $E_D(A;B)_{\rho}$. We started with the one-shot setting, in which we allow for some error in the distillation protocol, and we determined both upper and lower bounds on the number of (approximate) ebits that can be obtained. Then, in the asymptotic setting, we showed that the coherent information $I(A\rangle B)_{\rho}$ is a lower bound on distillable entanglement for every bipartite state $\rho_{AB}$. We also found that two different entanglement measures are upper bounds on distillable entanglement, namely, the Rains relative entropy and squashed entanglement. These are the best known upper bounds on distillable entanglement.  
	
	By first performing entanglement distillation to transform their mixed entangled states to approximately pure maximally entangled states and then performing the quantum teleportation protocol, Alice can transmit any quantum state to Bob. In this sense, entanglement distillation can be used to realize a near-ideal quantum channel between Alice and Bob. This fact underlies achievable strategies for quantum communication, which is the task of perfectly transmitting an arbitrary quantum state from Alice to Bob when the resource is a quantum channel $\mathcal{N}_{A\to B}$ connecting Alice and Bob rather than a shared bipartite state~$\rho_{AB}$. Quantum communication is the subject of the next chapter.





\section{Bibliographic Notes}

\label{sec:ed:bib-notes}

	Although our focus in this book is on communication, maximally entangled states are useful resources for many other quantum information processing tasks, (see, e.g., \citet{HHHH09} for applications of entanglement in quantum computing), which makes entanglement distillation a relevant topic in its own right.

	The concept of entanglement distillation was initially developed by \citet{BBPSSW96EPP,BDSW96}, who also provided one-way and two-way protocols for distillation from two-qubit Bell-diagonal states. The precise mathematical definition of distillable entanglement was given by \citet{Rains1998,R99,R99b,HHH00,PV07}. Relaxing the set of allowed operations from LOCC channels to separable and completely PPT-preserving channels as a means to obtain upper bounds on distillable entanglement was considered by \citet{Rains1998,R99,R99b}.

	\citet{Berta08,BD10a,BD11,WTB16} have considered lower bounds on distillable entanglement in the one-shot setting. The lower bound that we present in Proposition \ref{prop-ent_dist_lower_bound} is the one given by \citet[Proposition~21]{WTB16}, which makes use of the one-shot decoupling results obtained by \citet{Dupuis2014}. In particular, the proof of Theorem~\ref{thm-ent_distill_decoupling} provided in Appendix~\ref{app-ent_distill_decoupling_pf} comes directly from the proof of \citep[Theorem~3.3]{Dupuis2014}. The notion of decoupling has played an important role in the development  of quantum information theory. It was originally proposed by \citet{SW02} in the context of understanding approximate quantum error correction and quantum communication. It was then developed in much more detail by \citet{Horodecki:2005:673,Horodecki:2007:107} in the context of state merging and by  \citet{HHWY08} for understanding the coherent information lower bound on quantum capacity. \citet{D10} developed the method in more detail in his PhD thesis for a variety of information-processing tasks, and this culminated in the general decoupling theorem presented as \citep[Theorem~3.3]{Dupuis2014}.
	
	For the SDP formulations of conditional min- and max-entropy, as well as their smoothed variants, see \citep[Chapter 6]{T15book}. For more information about unitary designs and about Haar measure integration over unitaries, we refer to \citep{CS06,RS09}. The one-shot upper bound that we present in Proposition~\ref{prop:core-meta-converse-privacy_a} and Theorem~\ref{cor-ent_distill_one-shot_UB_alt} based on the fact that PPT' operators are useless for entanglement distillation was determined by \citet{TBR15}.
	
	In the asymptotic setting, \citet{DW05} used random coding arguments to establish the coherent information lower bound (also called the ``hashing inequality'') on the distillable entanglement of a bipartite quantum state. The corresponding hashing protocol was presented by \citet{BDSW96} for two-qubit Bell-diagonal states. \citet{DW05} also determined that the general expression in Theorem~\ref{thm-distillable_entanglement} is an achievable rate for entanglement distillation from a bipartite state, and they also proved the converse. \citet{HHH00} conjectured this formula earlier, conditioned on the hashing inequality being true. 
	
	Theorem~\ref{thm-one_way_distillable_entanglement} is due to \citet{DW05}. The other results in Sections~\ref{sec-one_way_ent_distill} and \ref{subsec-deg_antideg_states} on one-way entanglement distillation were obtained by \citet{LDS18}, who in the same work used the concepts of approximate degradablity and approximate anti-degradability of bipartite states to derive upper bounds on distillable entanglement. We note that anti-degradable quantum states, as defined by \citet{LDS18}, are also known as \textit{symmetrically extendible} states, \textit{two-extendible} states, or \textit{two-shareable} states \citep{W89a,DPS04,Yang06}.
	
	PPT entangled states (i.e., bound entangled states) were discovered by \citet{PH97}, and \citet{HHH98} showed that PPT states are useless for entanglement distillation. A major open and challenging question, which is alluded to in Section~\ref{subsec-bound_ent} (see also Figure~\ref{fig-dist_nondist_gen}) is whether there exist NPT (negative partial transpose) bound entangled states. For discussions concerning NPT bound entanglement, we refer to \citep{PhysRevA.59.4206,DSS+00,DCLB00}.

\begin{subappendices}

\section{One-Shot Decoupling and Proof of Theorem~\ref{thm-ent_distill_decoupling}}\label{app-ent_distill_decoupling_pf}

	\begin{figure}
		\centering
		\includegraphics[scale=0.9]{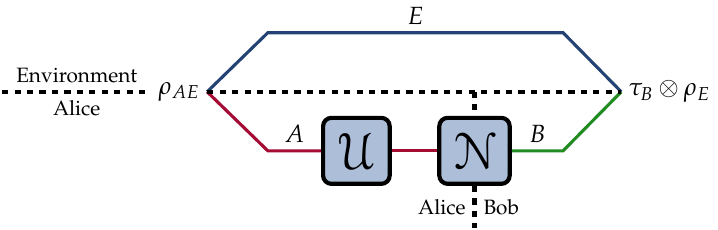}
		\caption{Given a bipartite state $\rho_{AE}$ and a quantum channel $\mathcal{N}_{A\to B}$, the goal of decoupling is to obtain a state $\tau_B\otimes\rho_E$ that is in tensor product with the environment $E$, where $\rho_E=\Tr_A[\rho_{AE}]$. To assist with the task, Alice is allowed to apply an arbitrary unitary $U$ to her system $A$.}\label{fig-decoupling}
	\end{figure}

	The key insight needed to obtain the lower bound on one-shot distillable entanglement in Theorem~\ref{prop-ent_dist_lower_bound} is that entanglement distillation can be thought of in terms of decoupling. The general scenario of decoupling is depicted in Figure~\ref{fig-decoupling}. Given a quantum channel $\mathcal{N}_{A\to B}$ and a bipartite state $\rho_{AE}$, the goal of decoupling is to obtain a state $\mathcal{N}_{A\to B}(\rho_{AE})$ that is decoupled from the system $E$, i.e., a state approximately of the form $\tau_B\otimes\rho_E$, where $\rho_E=\Tr_A[\rho_{AE}]$ and $\tau_B$ is some state. Note that  the reduced state of $E$ at the output is the same as the reduced state of $E$ at the input because we apply a channel only to the system $A$ and such a channel is trace preserving. We also allow Alice an arbitrary unitary that she can apply to her system before sending it through the channel $\mathcal{N}_{A\to B}$, so that we require
	\begin{equation}
		\mathcal{N}_{A\to B}(U_A\rho_{AE}U_A^\dagger) \approx_{\varepsilon}\tau_B\otimes\rho_E,
		\label{eq-ent_distill_decoupling_goal}
	\end{equation}
	for some state $\tau_B$,  up to some error $\varepsilon$. 
	Of course, exact equality in \eqref{eq-ent_distill_decoupling_goal} cannot be obtained in general, and we make the goal instead to obtain an output state $\mathcal{N}_{A\to B}(U_A\rho_{AE}U_A^\dagger)$ that is as close as possible to the state $\Phi_B^{\mathcal{N}}\otimes\rho_E$, where $\Phi_B^{\mathcal{N}} = \operatorname{Tr}_A[\Phi_{AB}^{\mathcal{N}}]$ and $\Phi_{AB}^{\mathcal{N}}$ is the Choi state of $\mathcal{N}_{A\to B}$. The choice for $\tau_B$ given by the reduced Choi state might seem arbitrary, but it is taken for analytical considerations and leads to a good bound for our purposes. By using trace distance as our measure of closeness, the goal is to determine an upper bound on the following quantity:
	\begin{equation}
		\min_{U_A}\norm{\mathcal{N}_{A\to B}(U_A\rho_{AE}U_A^\dagger)-\Phi_B^{\mathcal{N}}\otimes\rho_E}_1
	\end{equation}
	Note that the minimum over all unitaries never exceeds the average, meaning that
	\begin{multline}
		\min_{U_A}\norm{\mathcal{N}_{A\to B}(U_A\rho_{AE}U_A^\dagger)-\Phi_B^{\mathcal{N}}\otimes\rho_E}_1\\
		\leq \int_{U_A} \norm{\mathcal{N}_{A\to B}(U_A\rho_{AE}U_A^\dagger)-\Phi_B^{\mathcal{N}}\otimes\rho_E}_1~\text{d}U_A,
	\end{multline}
	where the integral over all unitaries $U_A$ is with respect to the Haar measure, and it can be thought of as a uniform average over the continuous set of all unitaries $U_A$ acting on the system $A$. Theorem~\ref{thm-ent_distill_decoupling} provides an upper bound on the right-hand side of the inequality above, and we restate the result here for convenience:
	\begin{equation}
		\int_{U_A} \norm{\mathcal{N}_{A\to B}(U_A\rho_{AE}U_A^\dagger)-\Phi_B^{\mathcal{N}}\otimes\rho_E}_1~\text{d}U_A \leq 2^{-\frac{1}{2}\widetilde{H}_2(A|E)_{\rho}-\frac{1}{2}\widetilde{H}_2(A|B)_{\Phi^{\mathcal{N}}}},
	\end{equation}
	where we recall that the sandwiched R\'enyi conditional entropy of order two of a bipartite state $\omega_{CD}$ is defined as
	\begin{align}
		\widetilde{H}_2(C|D)_{\omega}&=-\inf_{\sigma_D}\widetilde{D}_2(\omega_{CD}\Vert\mathbbm{1}_C\otimes\sigma_D)\\
		&=-\inf_{\sigma_D}\log_2\Tr\!\left[\left(\sigma_D^{-\frac{1}{4}}\omega_{CD}\sigma_D^{-\frac{1}{4}}\right)^2\right],
	\end{align}
	and the optimization is over every state $\sigma_D$.

\subsubsection*{Proof of Theorem~\ref{thm-ent_distill_decoupling}}

	Let
	\begin{align}
		M_{BE}&\coloneqq\mathcal{N}_{A\to B}(U_A\rho_{AE}U_A^\dagger)-\Phi_B^{\mathcal{N}}\otimes\rho_E,\\
		\sigma_{BE}&\coloneqq\tau_B\otimes\zeta_E,
	\end{align}
	with $\tau_B$ and $\zeta_E$ arbitrary positive definite states. By the variational characterization of the trace norm in \eqref{eq-trace_norm_variational}, we have that
	\begin{equation}
		\norm{M_{BE}}_1=\max_{U_{BE}}\abs{\Tr[U_{BE}M_{BE}]},
	\end{equation}
	where the optimization is over every unitary $U_{BE}$. Using the Cauchy--Schwarz inequality (see \eqref{eq-Cauchy_Schwarz_HS}), and suppressing system labels for brevity, we obtain
	\begin{align}
		\norm{M}_1&=\max_U\abs{\Tr[UM]}\\
		&=\max_U\abs{\Tr\!\left[\left(\sigma^{\frac{1}{4}}U\sigma^{\frac{1}{4}}\right)\left(\sigma^{-\frac{1}{4}}M\sigma^{-\frac{1}{4}}\right)\right]}\\
		&\leq \max_U\sqrt{\Tr\!\left[\left(\sigma^{\frac{1}{4}}U\sigma^{\frac{1}{4}}\right)\left(\sigma^{\frac{1}{4}}U^\dagger\sigma^{\frac{1}{4}}\right)\right]\Tr\!\left[\sigma^{-\frac{1}{4}}M\sigma^{-\frac{1}{2}}M^\dagger\sigma^{-\frac{1}{4}}\right]}\\
		&=\sqrt{\max_U\Tr\!\left[\sigma^{\frac{1}{2}}U\sigma^{\frac{1}{2}}U^\dagger\right]\Tr\!\left[\sigma^{-\frac{1}{4}}M\sigma^{-\frac{1}{2}}M^\dagger\sigma^{-\frac{1}{4}}\right]}.
	\end{align}
	Since $\sigma^{\frac{1}{2}}$ and $U\sigma^{\frac{1}{2}}U^\dagger$ are positive definite for every unitary $U$, by the Cauchy--Schwarz inequality, we conclude that
	\begin{align}
		\Tr\!\left[\sigma^{\frac{1}{2}}U\sigma^{\frac{1}{2}}U^\dagger\right]&=\abs{\Tr\!\left[\sigma^{\frac{1}{2}}U\sigma^{\frac{1}{2}}U^\dagger\right]}\\
		&\leq \sqrt{\Tr\!\left[\sigma^{\frac{1}{2}}\sigma^{\frac{1}{2}}\right]\Tr\!\left[U\sigma^{\frac{1}{2}}U^\dagger U\sigma^{\frac{1}{2}}U^\dagger\right]}\\
		&=\Tr[\sigma]\\
		&=1
	\end{align}
	for every unitary $U$, which implies that
	\begin{equation}
		\max_U\Tr\!\left[\sigma^{\frac{1}{2}}U\sigma^{\frac{1}{2}}U^\dagger\right]\leq \Tr[\sigma]=1.
	\end{equation}
	On the other hand, by taking $U=\mathbbm{1}$ in the optimization over $U$, we obtain
	\begin{equation}
		\max_U\Tr\!\left[\sigma^{\frac{1}{2}}U\sigma^{\frac{1}{2}}U^\dagger\right]\geq \Tr[\sigma]=1,
	\end{equation}
	which means that
	\begin{equation}
		\max_U\Tr\!\left[\sigma^{\frac{1}{2}}U\sigma^{\frac{1}{2}}U^\dagger\right]=\Tr[\sigma]=1.
	\end{equation}
	Therefore,
	\begin{align}
		\norm{M}_1&\leq \sqrt{\Tr\!\left[\sigma^{-\frac{1}{4}}M\sigma^{-\frac{1}{2}}M^\dagger \sigma^{-\frac{1}{4}}\right]}\\
		&=\sqrt{\Tr\!\left[\left(\sigma^{-\frac{1}{4}}M\sigma^{-\frac{1}{4}}\right)^2\right]},
	\end{align}
	where the last line follows because $M$ is Hermitian. So we have that
	\begin{multline}
		\norm{\mathcal{N}_{A\to B}(U_A\rho_{AE}U_A^\dagger)-\Phi_B^{\mathcal{N}}\otimes\rho_E}_1\\
		\leq\sqrt{\Tr\!\left[\left((\tau_B\otimes\zeta_E)^{-\frac{1}{4}}(\mathcal{N}_{A\to B}(U_A\rho_{AE}U_A^\dagger)-\Phi_B^{\mathcal{N}}\otimes\rho_E)(\tau_B\otimes\zeta_E)^{-\frac{1}{4}}\right)^2\right]}.
	\end{multline}
	Now, define
	\begin{align}
		\widetilde{\mathcal{N}}_{A\to B}(\cdot)&\coloneqq \tau_B^{-\frac{1}{4}}\mathcal{N}_{A\to B}(\cdot)\tau_B^{-\frac{1}{4}},\label{eq-decoupling_pf3}\\
		\widetilde{\rho}_{AE}&\coloneqq\zeta_E^{-\frac{1}{4}}\rho_{AE}\zeta_E^{-\frac{1}{4}}.\label{eq-decoupling_pf4}
	\end{align}
	Using these definitions, we can write the inequality above  as
	\begin{multline}
		\norm{\mathcal{N}_{A\to B}(U_A\rho_{AE}U_A^\dagger)-\Phi_B^{\mathcal{N}}\otimes\rho_E}_1\\ \leq \sqrt{\Tr\!\left[\left(\widetilde{\mathcal{N}}_{A\to B}(U_A\widetilde{\rho}_{AE}U_A^\dagger)-\Phi_B^{\widetilde{\mathcal{N}}}\otimes\widetilde{\rho}_E\right)^2\right]}.
	\end{multline}
	Taking the integral over unitaries $U_A$ on both sides of this inequality, and using Jensen's inequality (see \eqref{eq-MT:Jensen_ineq}), which  applies because the square root function is concave, we obtain
	\begin{align}
		&\int_{U_A}\norm{\mathcal{N}_{A\to B}(U_A\rho_{AE}U_A^\dagger)-\Phi_B^{\mathcal{N}}\otimes\rho_E}_1~\text{d}U_A\nonumber\\
		&\qquad \leq \int_{U_A}\sqrt{\Tr\!\left[\left(\widetilde{\mathcal{N}}_{A\to B}(U_A\widetilde{\rho}_{AE}U_A^\dagger)-\Phi_B^{\widetilde{\mathcal{N}}}\otimes\widetilde{\rho}_E\right)^2\right]}~\text{d}U_A\\
		&\qquad \leq \sqrt{\int_{U_A}\Tr\!\left[\left(\widetilde{\mathcal{N}}_{A\to B}(U_A\widetilde{\rho}_{AE}U_A^\dagger)-\Phi_B^{\widetilde{\mathcal{N}}}\otimes\widetilde{\rho}_E\right)^2\right]~\text{d}U_A}.\label{eq-decoupling_pf2}
	\end{align}
	Expanding the integral on the right-hand side of the inequality above  leads to
	\begin{align}
		&\int_{U_A}\Tr\!\left[\left(\widetilde{\mathcal{N}}_{A\to B}(U_A\widetilde{\rho}_{AE}U_A^\dagger)-\Phi_B^{\widetilde{\mathcal{N}}}\otimes\widetilde{\rho}_E\right)^2\right]~\text{d}U_A\nonumber\\
		&\quad =\int_{U_A}\Tr\!\left[\left(\widetilde{\mathcal{N}}_{A\to B}(U_A\widetilde{\rho}_{AE}U_A^\dagger)\right)^2\right]~\text{d}U_A\\
		&\qquad -2\int_{U_A}\Tr\!\left[\widetilde{\mathcal{N}}_{A\to B}(U_A\widetilde{\rho}_{AE}U_A^\dagger)(\Phi_B^{\widetilde{\mathcal{N}}}\otimes\widetilde{\rho}_E)\right]~\text{d}U_A+\Tr[(\Phi_B^{\widetilde{\mathcal{N}}}\otimes\widetilde{\rho}_E)^2]\\
		&\quad =\int_{U_A}\Tr\!\left[\left(\widetilde{\mathcal{N}}_{A\to B}(U_A\widetilde{\rho}_{AE}U_A^\dagger)\right)^2\right]~\text{d}U_A\nonumber\\
		&\qquad -2~\Tr\!\left[\widetilde{\mathcal{N}}_{A\to B}\left(\int_{U_A}U_A\widetilde{\rho}_{AE}U_A^\dagger~\text{d}U_A\right)(\Phi_B^{\widetilde{\mathcal{N}}}\otimes\widetilde{\rho}_E)\right]+\Tr[(\Phi_B^{\widetilde{\mathcal{N}}}\otimes\widetilde{\rho}_E)^2].
	\end{align}
	Now, using \eqref{eq-unitary_1_design}, we have that
	\begin{equation}
		\widetilde{\mathcal{N}}_{A\to B}\left(\int_{U_A} U_A\widetilde{\rho}_{AE}U_A^\dagger~\text{d}U_A\right)=\widetilde{\mathcal{N}}_{A\to B}(\pi_A \otimes \Tr_A[\widetilde{\rho}_{AE}])=\Phi_B^{\widetilde{\mathcal{N}}}\otimes\widetilde{\rho}_E.
	\end{equation}
	Therefore,
	\begin{multline}\label{eq-decoupling_pf1}
		\int_{U_A}\Tr\!\left[\left(\widetilde{\mathcal{N}}_{A\to B}(U_A\widetilde{\rho}_{AE}U_A^\dagger)-\Phi_B^{\widetilde{\mathcal{N}}}\otimes\widetilde{\rho}_E\right)^2\right]~\text{d}U_A\\ = \int_{U_A}\Tr\!\left[\left(\widetilde{\mathcal{N}}_{A\to B}(U_A\widetilde{\rho}_{AE}U_A^\dagger)\right)^2\right]~\text{d}U_A-\Tr[(\Phi_B^{\widetilde{\mathcal{N}}})^2]\Tr[\widetilde{\rho}_E^2].
	\end{multline}
	We now use the fact that
	\begin{equation}\label{eq-swap_trick}
		\Tr[X^2]=\Tr[X^{\otimes 2}F]
	\end{equation}
	for every operator $X$, where $F$ is the swap operator defined in \eqref{eq-swap_op_standard_0}. In \eqref{eq-decoupling_pf1} above, we have $X\equiv X_{BE}=\widetilde{\mathcal{N}}_{A\to B}(U_A\widetilde{\rho}_{AE}U_A^\dagger)$, which is a bipartite operator. The corresponding swap operator is $F_{BE}=F_B\otimes F_E$, with $F_B$ the swap operator acting on two copies of $\mathcal{H}_B$ and $F_E$ the swap operator acting on two copies of $\mathcal{H}_E$. Therefore,
	\begin{align}
		&\Tr\!\left[\left(\widetilde{\mathcal{N}}_{A\to B}(U_A\widetilde{\rho}_{AE}U_A^\dagger)\right)^2\right]\nonumber\\
		&\qquad=\Tr\!\left[\left(\widetilde{\mathcal{N}}_{A\to B}(U_A\widetilde{\rho}_{AE}U_A^\dagger)\right)^{\otimes 2}(F_B\otimes F_E)\right]\\
		&\qquad=\Tr\!\left[\left(\widetilde{\mathcal{N}}_{A\to B}^{\otimes 2}(U_A^{\otimes 2}\widetilde{\rho}_{AE}^{\otimes 2}(U_A^\dagger)^{\otimes 2})\right)(F_B\otimes F_E)\right]\\
		&\qquad=\Tr\!\left[\widetilde{\rho}_{AE}^{\otimes 2}\left(\left((U_A^\dagger)^{\otimes 2}(\widetilde{\mathcal{N}}_{A\to B}^{\otimes 2})^\dagger(F_B)U_A^{\otimes 2}\right)\otimes F_E\right)\right],
	\end{align}
	where  the last line follows from the definition of the adjoint of a channel. We thus have
	\begin{multline}
		\int_{U_A}\Tr\!\left[\left(\widetilde{\mathcal{N}}_{A\to B}(U_A\widetilde{\rho}_{AE}U_A^\dagger)\right)^2\right]~\text{d}U_A \\
		=\Tr\!\left[\widetilde{\rho}_{AE}^{\otimes 2}\left(\left(\int_{U_A}(U_A^\dagger)^{\otimes 2}(\widetilde{\mathcal{N}}_{A\to B}^{\otimes 2})^\dagger(F_B)U_A^{\otimes 2}~\text{d}U_A\right)\otimes F_E\right)\right].
	\end{multline}
	Now, we use the following known fact (a standard result in Schur--Weyl duality): for every operator $X$ acting on $\mathbb{C}^d\otimes\mathbb{C}^d$, with $d\geq 1$,
	\begin{equation}
		\int_U (U^\dagger)^{\otimes 2} X U^{\otimes 2}~\text{d}U=\alpha\mathbbm{1}+\beta F,
	\end{equation}
	where $F$ is again the swap operator, and
	\begin{align}
		\alpha&=\frac{\Tr[X]}{d^2-1}-\frac{\Tr[XF]}{d(d^2-1)},\\
		\beta&=\frac{\Tr[XF]}{d^2-1}-\frac{\Tr[X]}{d(d^2-1)}.
	\end{align}
	Taking $X\equiv (\widetilde{\mathcal{N}}_{A\to B}^{\otimes 2})^\dagger(F_B)$, which is an operator acting on two copies of $\mathcal{H}_A$, we obtain
	\begin{align}
		\Tr[(\widetilde{\mathcal{N}}_{A\to B}^{\otimes 2})^\dagger(F_B)]&=\Tr[\widetilde{\mathcal{N}}_{A\to B}^{\otimes 2}(\mathbbm{1}_A^{\otimes 2})F_B]\\
		&=d_A^2\Tr[(\Phi_B^{\widetilde{\mathcal{N}}})^{\otimes 2}F_B]\\
		&=d_A^2\Tr[(\Phi_B^{\widetilde{\mathcal{N}}})^2],
	\end{align}
	where the last line follows from \eqref{eq-swap_trick}. By similar reasoning, we obtain
	\begin{align}
		\Tr[F_A(\widetilde{\mathcal{N}}_{A\to B}^{\otimes 2})^\dagger(F_B)]&=\Tr[\widetilde{\mathcal{N}}_{A\to B}^{\otimes 2}(F_A)F_B]\\
		&=d_A^2\Tr[(F_A\otimes F_B)(\Phi_{AB}^{\widetilde{\mathcal{N}}})^{\otimes 2}]\\
		&=d_A^2\Tr[(\Phi_{AB}^{\widetilde{\mathcal{N}}})^2],
	\end{align}
	where the second equality follows by expressing the action of $\widetilde{\mathcal{N}}_{A\to B}^{\otimes 2}$ on $F_A$ with the Choi state $\Phi_{AB}^{\widetilde{\mathcal{N}}}$, using \eqref{eq-Choi_rep_action}. To obtain the last equality, we again used \eqref{eq-swap_trick}. We thus have
	\begin{align}
		\alpha&=\frac{\Tr[(\Phi_B^{\widetilde{\mathcal{N}}})^2]}{d_A^2-1}\left(d_A^2-d_A\frac{\Tr[(\Phi_{AB}^{\widetilde{\mathcal{N}}})^2]}{\Tr[(\Phi_B^{\widetilde{\mathcal{N}}})^2]}\right),\\
		\beta&=\frac{\Tr[(\Phi_{AB}^{\widetilde{\mathcal{N}}})^2]}{d_A^2-1}\left(d_A^2-d_A\frac{\Tr[(\Phi_B^{\widetilde{\mathcal{N}}})^2]}{\Tr[(\Phi_{AB}^{\widetilde{\mathcal{N}}})^2]}\right).
	\end{align}
	We now make use of the following general fact, whose proof we provide below in Lemma~\ref{lem-marginal_purity_ratio}: for every non-zero positive semi-definite operator $P_{AB}$ with $P_B\coloneqq\Tr_A[P_{AB}]$, the following inequalities hold
	\begin{equation}
		\frac{1}{d_A}\leq\frac{\Tr[P_{AB}^2]}{\Tr[P_B^2]}\leq d_A.
	\end{equation}
	Applying these inequalities to the expressions for $\alpha$ and $\beta$ above, we obtain
	\begin{equation}
		\alpha\leq \Tr[(\Phi_B^{\widetilde{\mathcal{N}}})^2],\quad \beta\leq \Tr[(\Phi_{AB}^{\widetilde{\mathcal{N}}})^2].
	\end{equation}
	We thus have that
	\begin{align}
		&\int_{U_A}\Tr\!\left[\left(\widetilde{\mathcal{N}}_{A\to B}(U_A\widetilde{\rho}_{AE}U_A^\dagger)\right)^2\right]~\text{d}U_A\\
		&\quad =  \Tr\!\left[\widetilde{\rho}_{AE}^{\otimes 2}\left((\alpha \mathbbm{1}_A^{\otimes 2}+\beta F_A)\otimes F_E\right)\right]\\
	&\quad =\alpha\Tr[\widetilde{\rho}_E^2]+\beta \Tr[\widetilde{\rho}_{AE}^2] \\
	&\quad \leq \Tr[(\Phi_B^{\widetilde{\mathcal{N}}})^2]\Tr[\widetilde{\rho}_E^2]+\Tr[(\Phi_{AB}^{\widetilde{\mathcal{N}}})^2]\Tr[\widetilde{\rho}_{AE}^2].
	\end{align}
	Combining this with \eqref{eq-decoupling_pf1}, we find that
	\begin{multline}
		\int_{U_A}\Tr\!\left[\left(\widetilde{\mathcal{N}}_{A\to B}(U_A\widetilde{\rho}_{AE}U_A^\dagger)-\Phi_B^{\widetilde{\mathcal{N}}}\otimes\widetilde{\rho}_E\right)^2\right]~\text{d}U_A \\ \leq \Tr[(\Phi_{AB}^{\widetilde{\mathcal{N}}})^2]\Tr[\widetilde{\rho}_{AE}^2].
	\end{multline}
	Then, by \eqref{eq-decoupling_pf2}, and recalling the definitions in \eqref{eq-decoupling_pf3} and \eqref{eq-decoupling_pf4}, we obtain
	\begin{align}
		&\int_{U_A}\norm{\mathcal{N}_{A\to B}(U_A\rho_{AE}U_A^\dagger)-\Phi_B^{\mathcal{N}}\otimes\rho_E}_1~\text{d}U_A\nonumber\\
		&\quad \leq \sqrt{\Tr[(\Phi_{AB}^{\widetilde{\mathcal{N}}})^2]\Tr[\widetilde{\rho}_{AE}^2]}\\
		&\quad=\left(\Tr\!\left[\left(\tau_B^{-\frac{1}{4}}\Phi_{AB}^{\mathcal{N}}\tau_B^{-\frac{1}{4}}\right)^2\right]\right)^{\frac{1}{2}}\left(\Tr\!\left[\left(\zeta^{-\frac{1}{4}}\rho_{AE}\zeta_E^{-\frac{1}{4}}\right)^2\right]\right)^{\frac{1}{4}}.
	\end{align}
	This inequality holds for all states $\tau_B$ and $\zeta_E$, which means that
	\begin{align}
		&\int_{U_A}\norm{\mathcal{N}_{A\to B}(U_A\rho_{AE}U_A^\dagger)-\Phi_B^{\mathcal{N}}\otimes\rho_E}_1~\text{d}U_A\nonumber\\
		&\quad \leq \left(\inf_{\tau_B}\Tr\!\left[\left(\tau_B^{-\frac{1}{4}}\Phi_{AB}^{\mathcal{N}}\tau_B^{-\frac{1}{4}}\right)^2\right]\right)^{\frac{1}{2}}\left(\inf_{\zeta_E}\Tr\!\left[\left(\zeta_E^{-\frac{1}{4}}\rho_{AE}\zeta_E^{-\frac{1}{4}}\right)^2\right]\right)^{\frac{1}{2}}\\
		&\quad =2^{-\frac{1}{2}H_2(A|B)_{\Phi^{\mathcal{N}}}-\frac{1}{2}\widetilde{H}_2(A|E)_{\rho}}\\
		&\quad =2^{-\frac{1}{2}\widetilde{H}_2(A|E)_{\rho}-\frac{1}{2}H_2(A|B)_{\Phi^{\mathcal{N}}}},
	\end{align}
	which completes the proof. \qedsymbol

	\begin{Lemma}{lem-marginal_purity_ratio}
		For every non-zero positive semi-definite operator $P_{AB}$, with $P_B=\Tr_A[P_{AB}]$, it holds that
		\begin{equation}
			\frac{1}{d_A}\leq\frac{\Tr[P_{AB}^2]}{\Tr[P_B^2]}\leq d_A.
		\end{equation}
	\end{Lemma}
	
	\begin{Proof}
		Letting $A'$ denote a copy of $A$, and applying the Cauchy--Schwarz inequality (see \eqref{eq-Cauchy_Schwarz_HS}), we find that
		\begin{align}
			\Tr[P_B^2]&=\Tr[(P_{AB}\otimes\mathbbm{1}_{A'})(P_{A'B}\otimes\mathbbm{1}_{A})]\\
			&\leq \sqrt{\Tr[(P_{AB}\otimes\mathbbm{1}_{A'})^2]\Tr[(P_{A'B}\otimes\mathbbm{1}_{A})^2]}\\
			&=\Tr[P_{AB}^2\otimes\mathbbm{1}_{A'}]\\
			&=d_A\Tr[P_{AB}^2].
		\end{align}
		
		The other inequality follows from the operator inequality $P_{AB} \leq d_A \mathbbm{1}_A \otimes P_B$, after sandwiching it by $P_{AB}^{\frac{1}{2}}$ and taking a full trace. This operator inequality follows in turn because $\frac{1}{d_A^2} \sum_{i=0}^{d^2-1}U^i_A P_{AB}(U^i_A)^\dag = \pi_A \otimes P_B$ (Lemma~\ref{lem:QM-over:HW-twirl}), for $\{U^i_A\}_i$ the set of Heisenberg--Weyl operators, and by noticing that all terms in the sum are positive semi-definite and one term in the sum is $\frac{1}{d_A^2} P_{AB}$.
	\end{Proof}

\end{subappendices}

\chapter{Quantum Communication}\label{chap-quantum_capacity}

	In the previous chapter, we considered entanglement distillation, which is the task of taking many copies of a mixed entangled state $\rho_{AB}$ shared by Alice and Bob and transforming them to a maximally entangled state $\Phi_{\hat{A}\hat{B}}$ of Schmidt rank $d\geq 2$. Using the quantum teleportation protocol, the maximally entangled state resulting from entanglement distillation can be used for quantum communication, in the sense that Alice can transfer an arbitrary state of $\log_2 d$ qubits   to Bob.
	
	Now, if Alice and Bob are distantly separated, then how do they obtain many copies of the shared entangled state $\rho_{AB}$ in the first place? Typically, one of the parties, say Alice, prepares two quantum systems in an entangled state and sends one of them through a quantum channel $\mathcal{N}_{A\to B}$ to Bob, thereby establishing the shared entangled state. Rather than use the shared entangled state as the resource for communication, it is more natural to use the quantum channel itself as the resource, as it could in principle lead to better strategies and higher rates. This is the scenario that we consider in this chapter.

\begin{figure}
		\centering
		\includegraphics[scale=0.8]{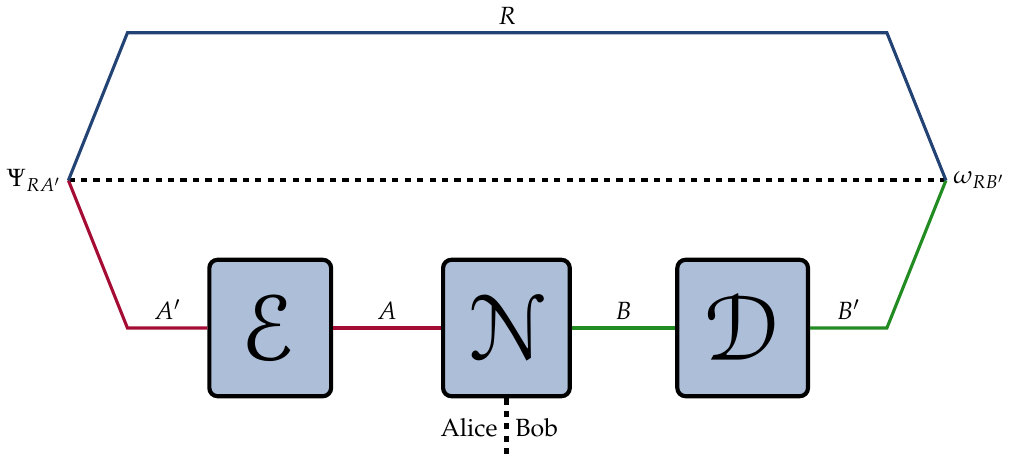}
		\caption{Depiction of a quantum communication protocol for one use of the quantum channel $\mathcal{N}$. Alice shares the maximally entangled state $\Psi_{RA'}$ with an inaccessible reference system $R$. She uses the channel $\mathcal{E}_{A'\to A}$ to encode her system $A'$ into a system $A$, which is sent through the channel $\mathcal{N}_{A\to B}$. Bob then applies the decoding channel $\mathcal{D}_{B\to B'}$, such that the final state is $\omega_{RB'}$ shared between Bob and the reference system.}\label{fig-q_comm_one_shot}
	\end{figure}

	Recall that in the case of classical communication from Chapter~\ref{chap-classical_capacity}, we considered messages from a set $\mathcal{M}$, and the goal was to find upper and lower bounds on the maximum number $\log_2|\mathcal{M}|$ of transmitted bits over a quantum channel for a given error $\varepsilon$. Now, in the case of quantum communication, the goal is to transmit a given number of \textit{qubits}, rather than bits, for a given error $\varepsilon$. Formally, suppose that the sender, Alice, holds a quantum system $A'$ with dimension $d\geq 1$ that she would like to transmit over the channel $\mathcal{N}$ to Bob, the receiver. In general, the state of this system could be entangled with the state of some other system $R$ (of arbitrary dimension) to which Alice does not have access, and so we suppose that the joint state is a pure state $\Psi_{RA'}$ with Schmidt rank $d$. Note that, by the Schmidt decomposition theorem (Theorem~\ref{thm-Schmidt}), the dimension of $R$ need not exceed the dimension of $A'$, which is $d$. The goal is to determine the largest value of $\log_2 d$ (which can be thought of as the number of qubits in the system $A'$) for which the $A'$ part of an arbitrary entangled state $\Psi_{RA'}$ can be transmitted with error at most $\varepsilon$. This general quantum communication scenario is known as \textit{strong subspace transmission}. As usual, Alice and Bob are allowed local encoding and decoding channels, respectively, to help with this task; see Figure~\ref{fig-q_comm_one_shot} for a depiction of a one-shot protocol for quantum communication. In the asymptotic setting, they are also allowed as many uses of the channel $\mathcal{N}$ as desired. The quantum capacity of~$\mathcal{N}$, denoted by $Q(\mathcal{N})$, is then the largest value of $\frac{1}{n}\log_2 d$ such that the $A'$ part of an arbitrary pure state $\Psi_{RA'}$ can be transmitted to Bob with error that vanishes as the number~$n$ of channel uses increases.
	
	Note that the notion of quantum communication presented above (strong subspace transmission) is completely general and includes as special cases the following information-processing tasks:
	\begin{enumerate}
		\item \textit{Entanglement transmission}: Here, Alice's system $A'$ is in the maximally entangled state $\Phi_{RA'}$ with the reference system $R$, and the goal is to transmit the system $A'$ to Bob. This is a special case of strong subspace transmission in which $\Psi_{RA'}\allowbreak=\Phi_{RA'}$.
		
		\item \textit{Entanglement generation}: Alice prepares a pure entangled state $\Psi_{A'A}$, with $d_{A'}=d\geq 1$ and $A$ the input system to the channel $\mathcal{N}$. The goal is to transmit the system $A$ to Bob such that the resulting state shared by Alice and Bob is the maximally entangled state of Schmidt rank $d$. We show in Appendix~\ref{app-q_comm_alt_notions} how entanglement generation is related to the notion of quantum communication that we consider here.
		
		Note that entanglement generation is similar to entanglement distillation, and we elaborate more upon this similarity in Section~\ref{subsec-qcomm_one_shot_lower_bound}. 
				
		\item \textit{Subspace transmission}: In this scenario, Alice wishes to send a system $A'$, in an arbitrary pure state $\varphi_{A'}$, to Bob. This is a special case of the protocol above  in which the system $R$ is not entangled with Alice's system, so that $\Psi_{RA'}=\phi_R\otimes\varphi_{A'}$ for some pure state $\phi_R$ on $R$.
		
		Note that subspace transmission can be accomplished by first performing an entanglement transmission or entanglement generation protocol and then performing the quantum teleportation protocol (however, this approach requires the use of a forward classical channel).
	\end{enumerate}
	We discuss these alternative notions of quantum communication in more detail, as well as prove some relationships between them, in Appendix~\ref{app-q_comm_alt_notions}.
	
	We start our development of quantum communication with the one-shot setting. Recall that mutual-information channel measures appear in the one-shot upper bounds for entanglement-assisted classical communication and that Holevo-information channel measures appear in the one-shot upper bounds for classical communication. In the case of quantum communication, we find that coherent-information channel measures appear in the one-shot upper bounds. The one-shot lower bound that we obtain is based on the one-shot, one-way entanglement distillation protocol in Section~\ref{subsec-ent_distill_one_shot_lower_bound} of the previous chapter, along with an argument for the removal of the classical communication used in that protocol. We then move on to the asymptotic setting, and we find that the quantum capacity of a quantum channel $\mathcal{N}$ is equal to its regularized coherent information, i.e.,
	\begin{equation}
	Q(\mathcal{N})=I_{\text{reg}}^c(\mathcal{N})\coloneqq\lim_{n\to\infty}\frac{1}{n}I^c(\mathcal{N}^{\otimes n}),
	\end{equation}
	where we recall the definition of the coherent information $I^c(\mathcal{N})$ of a channel from \eqref{eq-coh_inf_chan}. Thus, as with the classical capacity, the quantum capacity is difficult to compute in general. We then find tractable upper bounds on the quantum capacity, and for this purpose, the channel entanglement measures defined in Chapter~\ref{chap-ent_measures_chan} play an important role.

\section{One-Shot Setting}\label{sec-qcomm_one_shot}

	A (strong subspace) quantum communication protocol for a quantum channel $\mathcal{N}$ in the one-shot setting is illustrated in Figure~\ref{fig-q_comm_one_shot}. It is defined by the three elements $(d,\allowbreak\mathcal{E}_{A'\to A},\mathcal{D}_{B\to B'})$, in which $d$ is the  dimension of the system~$A'$, $\mathcal{E}_{A'\to A}$ is an encoding channel with $d_{A'}=d$, and $\mathcal{D}_{B\to B'}$ is a decoding channel with $d_{B'}=d_{A'}=d$. We call the pair $(\mathcal{E},\mathcal{D})$ of encoding and decoding channels a \textit{quantum communication code}\footnote{The quantum communication codes that we consider in this chapter are essentially equivalent to codes for performing \textit{approximate quantum error correction}. Please consult the Bibliographic Notes in Section~\ref{sec:Q-cap:bib-notes} for more information.} for $\mathcal{N}$. 
	
	\begin{remark}
		In a strong subspace transmission protocol, the goal is to transmit one share of a pure state $\Psi_{RA'}$, with corresponding state vector $\ket{\Psi}_{RA'}$. Note that the state vector $\ket{\Psi}_{RA'}$ has a Schmidt decomposition of the form 
		\begin{equation}\label{eq-q_comm_starting_state_Schmidt}
			\ket{\Psi}_{RA'}=\sum_{x=1}^d\sqrt{\smash[b]{p(x)}}\ket{\xi_x}_R\otimes\ket{\zeta_x}_{A'},
		\end{equation}
		where $\{p(x)\}_{x=1}^d$ are the Schmidt coefficients and $\{\ket{\xi_x}_R\}_{x=1}^d$, $\{\ket{\zeta_x}_{A'}\}_{x=1}^d$ are orthonormal sets of vectors for $R$ and $A'$, respectively. When written in this form, the state vector~$\ket{\Psi}_{RA'}$ can be understood as a coherent version of the initial state $\Phi^p_{MM'}$ for classical and entanglement-assisted classical communication (see, e.g., \eqref{eq-ea_classical_comm_initial_state}). The key difference in the classical-communication case is that there is a fixed orthonormal basis $\{\ket{m}\}_{m\in\mathcal{M}}$ corresponding to the messages $m$ in the message set $\mathcal{M}$. 
		In quantum communication, the goal is to transmit a state of a quantum system, which means that there is no particular basis used for communication. The encoding and decoding channels should thus be defined so that they can reliably transmit states of the system in an \textit{arbitrary} basis.
	\end{remark}
	
	The protocol proceeds as follows: we start with the entangled state $\Psi_{RA'}$, where the system $A'$ belongs to Alice and the system $R$ is an arbitrary reference system inaccessible to Alice. Alice then sends the system $A'$ through the encoding channel $\mathcal{E}_{A'\to A}$ and sends $A$ through the channel $\mathcal{N}_{A\to B}$. Once Bob receives the system $B$, he applies the decoding channel $\mathcal{D}_{B\to B'}$ to it. The final state of the protocol is therefore
	\begin{equation}\label{eq-qcomm_one_shot_final_state}
		\omega_{RB'}\coloneqq (\mathcal{D}_{B\to B'}\circ\mathcal{N}_{A\to B}\circ\mathcal{E}_{A'\to A})(\Psi_{RA'}).
	\end{equation}
	
	Let us now quantify the reliability of the protocol described above, i.e., how close the final state $\omega_{RB'}$ is to the initial state $\Psi_{RA'}$. In Chapter~\ref{chap-QM_dist_meas}, we discussed two measures of closeness for states:
	\begin{itemize}
		\item Normalized trace distance, using which the distance between the initial and final states is $\frac{1}{2}\norm{\Psi_{RA'}-\omega_{RB'}}_1$. The lower the normalized trace distance, the more reliable the protocol is.
		\item Fidelity, in which case we have
		\begin{equation}
		F(\Psi_{RA'},\omega_{RB'})=\norm{\sqrt{\Psi_{RA'}}\sqrt{\omega_{RB'}}}_1^2=\bra{\Psi}_{RA'}\omega_{RB'}\ket{\Psi}_{RA'}
		\end{equation}
		as the closeness measure between the initial and final states of the protocol. The higher the fidelity, the more reliable the protocol is.
	\end{itemize}
	These two measures of closeness are arguably equivalent to each other, in the sense that one can be used to bound the other via the inequality \eqref{eq-Fuchs_van_de_graaf} shown in Theorem \ref{thm-Fuchs_van_de_graaf}, which we restate here: for all states $\rho$ and $\sigma$,
	\begin{equation}\label{eq-Fuchs_van_de_graaf_2}
		1-\sqrt{F(\rho,\sigma)}\leq\frac{1}{2}\norm{\rho-\sigma}_1\leq\sqrt{1-F(\rho,\sigma)}.
	\end{equation}

	Now, our figure of merit for the quantum communication protocol should not be based on just one particular initial state, in this case $\Psi_{RA'}$. Recall that the task of quantum communication is to reliably transmit one share of an \textit{arbitrary} pure state through the channel $\mathcal{N}_{A\to B}$. Intuitively, therefore, the closer the overall channel $\mathcal{D}_{B\to B'}\circ\mathcal{N}_{A\to B}\circ\mathcal{E}_{A'\to A}$ is to the identity channel $\id_{A'\to B'}$, the better the code $(\mathcal{E},\mathcal{D})$ is at the quantum communication task, and so our figure of merit should quantify this distance. One method to determine this distance is to calculate how well a given code can transmit one share of a state in the worst case, i.e., by either the highest value of the trace distance or by the lowest value of the fidelity. If a code can be designed such that, in the worst case, the fidelity (trace distance) is high (low), then by definition any other state will do just as well or better. We are thus led to define the following two figures of merit:
	\begin{enumerate}
		\item\textit{Worst-case trace distance}: We define this as
			\begin{equation}
				\sup_{\Psi_{RA'}}\frac{1}{2}\norm{\Psi_{RA'}-(\mathcal{D}_{B\to B'}\circ\mathcal{N}_{A\to B}\circ\mathcal{E}_{A'\to A})(\Psi_{RA'})}_1.
			\end{equation}
			Recalling Definition~\ref{eq-diamond_dist_meas}, we see that worse-case trace distance is equal to the \textit{diamond distance} between the identity channel $\id_{A'\to B'}$ and the channel $\mathcal{D}_{B\to B'}\circ\mathcal{N}_{A\to B}\circ\mathcal{E}_{A'\to A}$:
			\begin{equation}
				\frac{1}{2}\norm{\id_{A'\to B'}-\mathcal{D}_{B\to B'}\circ\mathcal{N}_{A\to B}\circ\mathcal{E}_{A'\to A}}_{\diamond}.
			\end{equation}
		\item\textit{Worst-case fidelity}: We define this as
			\begin{equation}
				\inf_{\Psi_{RA'}}\!\bra{\Psi}_{RA'}(\mathcal{D}_{B\to B'}\circ\mathcal{N}_{A\to B}\circ\mathcal{E}_{A'\to A})(\Psi_{RA'})\ket{\Psi}_{RA'},
			\end{equation}
			which is the same as the channel fidelity $F(\mathcal{D}\circ\mathcal{N}\circ\mathcal{E})$ from  Definition~\ref{eq-worse_case_fid_chan}.
	\end{enumerate}
	These two figures of merit are arguably equivalent, as mentioned before, due to the inequality in \eqref{eq-Fuchs_van_de_graaf_2} relating the trace distance and the fidelity. For the rest of this chapter, we  exclusively use the worst-case fidelity of the code as the figure of merit, and we define the \textit{error probability of the quantum communication code $(\mathcal{E},\mathcal{D})$} for $\mathcal{N}$ as
	\begin{align}
		p_{\text{err}}^*(\mathcal{E},\mathcal{D};\mathcal{N})&\coloneqq 1-F(\mathcal{D}\circ\mathcal{N}\circ\mathcal{E})\\
		&=\sup_{\Psi_{RA'}}\left\{1-\bra{\Psi}_{RA'}(\mathcal{D}_{B\to B'}\circ\mathcal{N}_{A\to B}\circ\mathcal{E}_{A'\to A})(\Psi_{RA'})\ket{\Psi}_{RA'}\right\}.\label{eq-q_comm_max_err_prob}
	\end{align}
	One can view this quantity as the quantum analogue of the maximum error probability for the classical communication tasks of Chapters \ref{chap-EA_capacity} and \ref{chap-classical_capacity}, which is why we use the same notation for it as in those chapters.
	
	\begin{definition}{$(d,\varepsilon)$ Quantum Communication Protocol}{def-q_comm_Me_protocol}
		Let $(d,\mathcal{E},\mathcal{D})$ be the elements of a quantum communication protocol for the quantum channel $\mathcal{N}$. The protocol is called a \textit{$(d,\varepsilon)$ protocol}, with $\varepsilon\in[0,1]$, if $p_{\text{err}}^*(\mathcal{E},\mathcal{D};\mathcal{N})\leq\varepsilon$.
	\end{definition}
	
	As alluded to at the beginning of this chapter, a special case of interest in quantum communication is entanglement transmission, which is when the state $\Psi_{RA'}$ is fixed to be the maximally entangled state $\Phi_{RA'}=\ket{\Phi}\!\bra{\Phi}_{RA'}$, where we recall that
	\begin{equation}\label{eq-max_ent_message}
		\ket{\Phi}_{RA'}=\frac{1}{\sqrt{d}}\sum_{i=0}^{d-1}\ket{i,i}_{RA'}.
	\end{equation}
	Now, since this state is a particular state in the optimization in \eqref{eq-q_comm_max_err_prob} for the error probability $p_{\text{err}}^*(\mathcal{E},\mathcal{D};\mathcal{N})$, we conclude that
	\begin{align}
		&p_{\text{err}}^*(\mathcal{E},\mathcal{D};\mathcal{N})\nonumber\\
		&\quad\geq 1-\bra{\Phi}_{RA'}(\mathcal{D}_{B\to B'}\circ\mathcal{N}_{A\to B}\circ\mathcal{E}_{A'\to A})(\Phi_{RA'})\ket{\Phi}_{RA'}\label{eq-q_comm_max_error_to_avg_error_1}\\
		&\quad=1-F_e(\mathcal{D}_{B\to B'}\circ\mathcal{N}_{A\to B}\circ\mathcal{E}_{A'\to A})\label{eq-q_comm_max_error_to_avg_error_2}\\
		&\quad\eqqcolon \overline{p}_{\text{err}}(\mathcal{E},\mathcal{D};\mathcal{N}),\label{eq-q_comm_max_error_to_avg_error_3}
	\end{align}
	where in the second line we have identified the entanglement fidelity of the channel $\mathcal{D}_{B\to B'}\circ\mathcal{N}_{A\to B}\circ\mathcal{E}_{A'\to A}$, as stated in Definition \ref{def-ent_fid_chan}. In the last line, we have defined the quantity $\overline{p}_{\text{err}}(\mathcal{E},\mathcal{D};\mathcal{N})$. As the notation suggests, this quantity is a quantum analogue of the average error probability for classical and entanglement-assisted classical communication. In classical and entanglement-assisted classical communication, the average error probability corresponds to taking a uniform distribution over the messages being sent. Similarly, in quantum communication, the average error probability can be thought of as taking a uniform distribution for the Schmidt coefficients in \eqref{eq-q_comm_starting_state_Schmidt}, which by definition gives a maximally entangled state.
	
	Another way of writing the average error probability for a quantum communication code is via what is known as the \textit{entanglement test}, which we introduced in the previous chapter. It is analogous to the comparator test that we defined in Chapters \ref{chap-EA_capacity} and \ref{chap-classical_capacity} in the context of classical communication. The entanglement test is defined by the POVM $\{\Phi_{RB'},\mathbbm{1}_{RB'}-\Phi_{RB'}\}$. The outcomes of the entanglement test tell us whether  the state being measured is the maximally entangled state $\Phi_{RB'}$. Since the state $\Phi_{RB'}$ is pure, using \eqref{eq-fidelity_pure_mixed}, the probability that the state $\omega_{RB'}$ at the end of the protocol is in the maximally entangled state, i.e., the probability that the state ``passes the entanglement test,'' is
	\begin{align}
		\Tr[\Phi_{RB'}\omega_{RB'}]&=\bra{\Phi}_{RB'}\omega_{RB'}\ket{\Phi}_{RB'}\\
		&=F(\Phi_{RB'},\omega_{RB'})\\
		&=1-\overline{p}_{\text{err}}(\mathcal{E},\mathcal{D};\mathcal{N}).
	\end{align}
	
	As stated at the beginning of this chapter, the goal of quantum communication is to determine the maximum number $\log_2 d$ of qubits that can be transmitted over a quantum channel $\mathcal{N}$, in the sense that the $A'$ part of an arbitrary pure state $\Psi_{RA'}$, with $d_{A'}=d$, can be transmitted over the channel with error at most $\varepsilon\in(0,1]$. We call this maximum number of transmitted qubits the \textit{one-shot quantum capacity of $\mathcal{N}$}.
	
	\begin{definition}{One-Shot Quantum Capacity of a Quantum Channel}{def-q_comm_one_shot_capacity}
		Given a quantum channel $\mathcal{N}$ and $\varepsilon\in(0,1]$, the \textit{one-shot $\varepsilon$-error quantum capacity of $\mathcal{N}$}, denoted by $Q^{\varepsilon}(\mathcal{N})$, is defined to be the maximum number $\log_2 d$ of transmitted qubits among all $(d,\varepsilon)$ quantum communication protocols over $\mathcal{N}$. In other words,
		\begin{equation}\label{eq-q_comm_one_shot_capacity}
			Q^{\varepsilon}(\mathcal{N})\coloneqq\sup_{(d,\mathcal{E},\mathcal{D})}\{\log_2 d : p_{\text{err}}^*(\mathcal{E},\mathcal{D};\mathcal{N})\leq\varepsilon\},
		\end{equation}
		where the optimization is with respect to $d\in\mathbb{N}$, $d\geq 1$, encoding channels $\mathcal{E}$ with input system dimension $d$, and decoding channels $\mathcal{D}$ with output system dimension $d$.
	\end{definition}
	
	In addition to finding, for a given $\varepsilon\in(0,1]$, the maximum number of transmitted qubits among all $(d,\varepsilon)$ quantum communication protocols over $\mathcal{N}_{A\to B}$, we can consider the following complementary problem: for a given dimension $d\geq 1$, find the smallest possible error among all $(d,\varepsilon)$ quantum communication protocols for $\mathcal{N}_{A\to B}$, which we denote by $\varepsilon_Q^*(d;\mathcal{N})$. In other words, the complementary problem is to determine
	\begin{equation}\label{eq-q_comm_one_shot_opt_error}
		\varepsilon_Q^*(d;\mathcal{N})\coloneqq\inf_{\mathcal{E},\mathcal{D}}\{p_{\text{err}}^*(\mathcal{E},\mathcal{D};\mathcal{N}):d_{A'}=d_{B'}=d\},
	\end{equation}
	where the optimization is with respect to all encoding channels $\mathcal{E}_{A'\to A}$ and decoding channels $\mathcal{D}_{B\to B'}$ such that $d_{A'}=d_{B'}=d$. In this book, we focus primarily on the problem of optimizing the number of transmitted qubits rather than the error, and so our primary quantity of interest is the one-shot quantum capacity~$Q^{\varepsilon}(\mathcal{N})$.

	\begin{figure}
		\centering
		\includegraphics[scale=0.8]{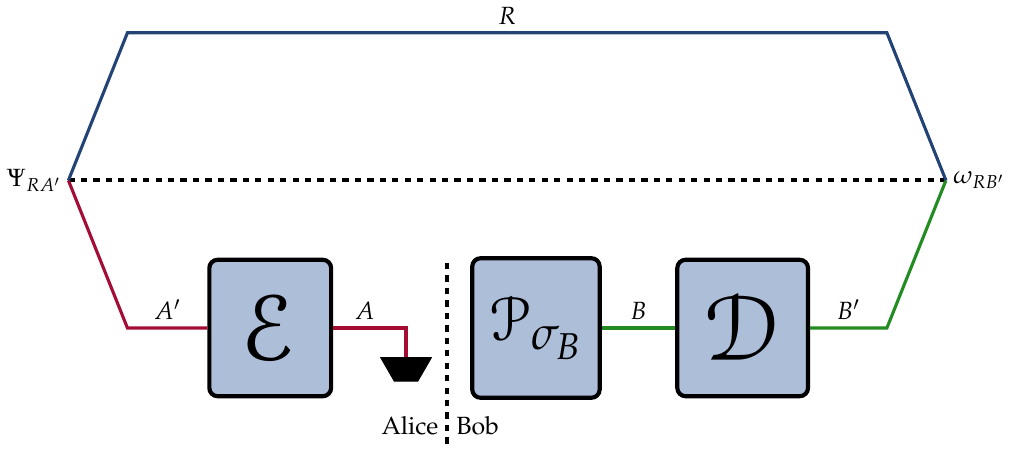}
		\caption{Depiction of a protocol that is useless for entanglement transmission. The encoded half of Alice's share of the pure state $\Psi_{RA'}$ is discarded and replaced by an arbitrary (but fixed) state $\sigma_B$.}\label{fig-q_comm_oneshot_useless}
	\end{figure}

\subsection{Protocol for a Useless Channel}\label{subsec-q_comm_useless_chan}

	Consider an arbitrary $(d,\varepsilon)$ quantum communication protocol for a channel~$\mathcal{N}_{A\to B}$, with $\varepsilon\in(0,1]$ and with encoding and decoding channels $\mathcal{E}$ and $\mathcal{D}$, respectively. This means that $p_{\text{err}}^*(\mathcal{E},\mathcal{D};\mathcal{N})\leq\varepsilon$. By the arguments in \eqref{eq-q_comm_max_error_to_avg_error_1}--\eqref{eq-q_comm_max_error_to_avg_error_3}, this protocol realizes a $(d,\varepsilon)$ entanglement transmission protocol, in the sense that 
	\begin{equation}\label{eq-q_comm_avg_error_meta_conv_bd_pf}
		\Tr[\Phi_{RB'}\rho_{RB'}]\geq 1-\varepsilon,
	\end{equation}
	where $\Phi_{RA'}$ is the maximally entangled state defined in \eqref{eq-max_ent_message} and
	\begin{equation}\label{eq-q_comm_ent_trans_final_state}
		\rho_{RB'}\coloneqq (\mathcal{D}_{B\to B'}\circ\mathcal{N}_{A\to B}\circ\mathcal{E}_{A'\to A})(\Phi_{RA'}).
	\end{equation}
	Consider now the same protocol but over the useless channel depicted in Figure \ref{fig-q_comm_oneshot_useless}. This useless channel is exactly the same as the one considered in Chapters \ref{chap-EA_capacity} and \ref{chap-classical_capacity}; namely, it is the replacement channel for some state $\sigma_B$. For the initial state $\Phi_{RA'}$, the state at the end of the protocol for the replacement channel is
	\begin{align}
		\tau_{RB'}&=(\mathcal{D}_{B\to B'}\circ\mathcal{R}_{A\to B}^{\sigma_B}\circ\mathcal{E}_{A'\to A})(\Phi_{RA'})=\pi_{R}\otimes\mathcal{D}_{B\to B'}(\sigma_B).
	\end{align}
	As in classical communication and entanglement-assisted classical communication, we now use the hypothesis testing relative entropy to compare the state $\rho_{RB'}$ obtained at the end of the quantum communication protocol over the channel $\mathcal{N}$ with the state $\tau_{RB'}$ obtained at the end of the quantum communication protocol for the replacement channel $\mathcal{R}^{\sigma_B}$. In particular, we make use of Lemma~\ref{prop-qc_meta_conv}, because the state $\tau_{RB'}$ satisfies $\Tr_{B'}[\tau_{RB}]=\pi_R$ and due to \eqref{eq-q_comm_avg_error_meta_conv_bd_pf}. Therefore, using \eqref{eq-entr:mut-inf-ent-test-ineq} in Lemma~\ref{prop-qc_meta_conv}, we conclude that 
	\begin{equation}
		\log_2 d\leq \frac{1}{2}I_H^{\varepsilon}(R;B')_{\rho}.
	\end{equation}
	Another bound from Lemma~\ref{prop-qc_meta_conv}, namely, the one in \eqref{eq-entr:coh-inf-ent-test-ineq}, is a more general upper bound that  requires only the assumption in \eqref{eq-q_comm_avg_error_meta_conv_bd_pf} and does not have the interpretation of being a comparison between a quantum communication protocol for $\mathcal{N}$ and a quantum communication protocol for $\mathcal{R}_{A\to B}^{\sigma_B}$. Applying this bound gives
	\begin{equation}\label{eq-entr:coh-inf-ent-test-ineq_2}
		\log_2 d\leq I_H^{\varepsilon}(R\rangle B')_{\rho}.
	\end{equation}
	This latter bound is the one that we employ in this chapter because it leads to a formula for the quantum capacity of some channels of interest in applications.  This inequality tells us that, given an arbitrary $(d,\varepsilon)$ quantum communication protocol with corresponding code $(\mathcal{E},\mathcal{D})$, the $\varepsilon$-hypothesis testing coherent information $I_H^{\varepsilon}(R\rangle B')_{\rho}$, with $\rho_{RB'}$ given by \eqref{eq-q_comm_ent_trans_final_state}, is an upper bound on the maximum number of qubits that can be transmitted over the channel with error at most $\varepsilon$. Note that a different choice for the encoding and decoding generally produces a different value for the upper bound. We would like an upper bound that applies regardless of the specific protocol. In other words, we would like an upper bound that is a function of the channel $\mathcal{N}_{A\to B}$ only.

\subsection{Upper Bound on the Number of Transmitted Qubits}\label{sec-q_comm_one_shot_upper_bounds}

	We now establish a general upper bound on the number of transmitted qubits in an arbitrary quantum communication protocol. This bound holds independently of the encoding and decoding channels used in the protocol and depends only on the given communication channel $\mathcal{N}_{A\to B}$ and the error $\varepsilon$.
	
	\begin{theorem*}{Upper Bound on One-Shot Quantum Capacity}{prop-qcomm:one-shot-bound-meta}
		Let $\mathcal{N}_{A\to B}$ be a quantum channel. For a $(d,\varepsilon)$ quantum communication protocol for $\mathcal{N}_{A\to B}$, with $\varepsilon\in(0,1]$, the number of qubits transmitted over $\mathcal{N}$ is bounded from above by the $\varepsilon$-hypothesis testing coherent information of $\mathcal{N}$ defined in \eqref{eq-hypo_test_coh_inf_chan}, i.e.,
		\begin{equation}
			\log_2 d\leq I_H^{c,\varepsilon}(\mathcal{N}).
		\end{equation}
		Consequently, for the one-shot quantum capacity of $\mathcal{N}$,
		\begin{equation}
			Q^{\varepsilon}(\mathcal{N})\leq I_H^{c,\varepsilon}(\mathcal{N}).
		\end{equation}
	\end{theorem*} 
	
	\begin{Proof}
		Let $\mathcal{E}$ and $\mathcal{D}$ be the encoding and decoding channels, respectively, for a $(d,\varepsilon)$ quantum communication protocol for $\mathcal{N}$. Then, by \eqref{eq-entr:coh-inf-ent-test-ineq_2}, we have that
		\begin{equation}
			\log_2 d\leq I_H^{\varepsilon}(R\rangle B')_{\rho}=\inf_{\sigma_{B'}}D_H^{\varepsilon}(\rho_{RB'}\Vert\mathbbm{1}_R\otimes\sigma_{B'}),
		\end{equation}
		where $\rho_{RB'}=(\mathcal{D}_{B\to B'}\circ\mathcal{N}_{A\to B}\circ\mathcal{E}_{A'\to A})(\Phi_{RA'})$ is the state defined in \eqref{eq-q_comm_ent_trans_final_state}. By restricting the optimization in the definition of $I_H^{\varepsilon}(R\rangle B')_{\rho}$ over every state~$\sigma_{B'}$ to the set $\{\mathcal{D}_{B\to B'}(\tau_B):\tau_B\in\Density(\mathcal{H}_B)\}$, we obtain
		\begin{align}
			&I_H^{\varepsilon}(R\rangle B')_{\rho}\nonumber\\
			&\quad =\inf_{\sigma_{B'}}D_H^{\varepsilon}(\rho_{RB'}\Vert\mathbbm{1}_R\otimes\sigma_{B'})\\
			&\quad\leq\inf_{\tau_B}D_H^{\varepsilon}((\mathcal{D}_{B\to B'}\circ\mathcal{N}_{A\to B}\circ\mathcal{E}_{A'\to A})(\Phi_{RA'})\Vert\mathbbm{1}_{R}\otimes\mathcal{D}_{B\to B'}(\tau_B))\\
			&\quad\leq\inf_{\tau_B}D_H^{\varepsilon}(\mathcal{N}_{A\to B}(\rho_{RA})\Vert\mathbbm{1}_R\otimes\tau_B)\label{eq-one-shot-bound-meta_pf1}
		\end{align}
		where the second inequality follows from the data-processing inequality for hypothesis testing relative entropy and we let $\rho_{RA}\coloneqq\mathcal{E}_{A'\to A}(\Phi_{RA'})$. We now take the supremum over every state $\rho_{RA}$, which effectively corresponds to taking the supremum over all encoding channels, and since it suffices to consider only pure states when optimizing the coherent information (see the arguments after Definition \ref{def-gen_inf_meas_chan}), we conclude that
		\begin{align}
			I_H^{\varepsilon}(R\rangle B')_{\rho}&\leq \inf_{\tau_B}D_H^{\varepsilon}(\mathcal{N}_{A\to B}(\rho_{RA})\Vert\mathbbm{1}_R\otimes\tau_B)\label{eq-one-shot-bound-meta_pf2}\\
			&\leq\sup_{\psi_{RA}} \inf_{\tau_B}D_H^{\varepsilon}(\mathcal{N}_{A\to B}(\psi_{RA})\Vert\mathbbm{1}_R\otimes\tau_B)\label{eq-one-shot-bound-meta_pf3}\\
			&=I_H^{c,\varepsilon}(\mathcal{N}),\label{eq-one-shot-bound-meta_pf4}
		\end{align}
		as required.
	\end{Proof}
	
	As an immediate consequence of Theorem~\ref{prop-qcomm:one-shot-bound-meta} and Propositions \ref{prop-hypo_to_rel_ent} and \ref{prop:sandwich-to-htre}, we obtain the following:
	
	\begin{corollary}{thm-qcomm_meta_str_weak_conv}
		Let $\mathcal{N}_{A\to B}$ be a quantum channel, and let $\varepsilon\in[0,1)$. For all $(d,\varepsilon)$ quantum communication protocols for $\mathcal{N}$, the following bounds hold:
		\begin{align}
			(1-2\varepsilon) \log_2 d&\leq I^c(\mathcal{N})+h_2(\varepsilon),\label{eq-qcomm_weak_conv_one_shot_1}\\
			\log_2 d&\leq \widetilde{I}_\alpha^c(\mathcal{N})+\frac{\alpha}{\alpha-1}\log_2\!\left(\frac{1}{1-\varepsilon}\right)\quad\forall~\alpha>1,\label{eq-qcomm_str_conv_one_shot_1}
		\end{align}
		where $I^c(\mathcal{N})$ is the coherent information of $\mathcal{N}$, as defined in \eqref{eq-coh_inf_chan}, and $\widetilde{I}_\alpha^c(\mathcal{N})$ is the sandwiched R\'{e}nyi coherent information of $\mathcal{N}$, as defined in \eqref{eq-sand_renyi_coh_inf_chan}. 
	\end{corollary}
	
	The proof of \eqref{eq-qcomm_weak_conv_one_shot_1} is analogous to the proof of \eqref{eq-ent_distill_one_shot_UB_weak_conv}. The proof of \eqref{eq-qcomm_str_conv_one_shot_1} follows by combining Theorem~\ref{prop-qcomm:one-shot-bound-meta} with Proposition~\ref{prop:sandwich-to-htre}.
	
	Since the bounds in \eqref{eq-qcomm_weak_conv_one_shot_1} and \eqref{eq-qcomm_str_conv_one_shot_1} hold for an arbitrary $(d,\varepsilon)$ quantum communication protocol for $\mathcal{N}$, we have that
	\begin{align}
		(1-2\varepsilon) Q^{\varepsilon}(\mathcal{N})&\leq I^c(\mathcal{N})+h_2(\varepsilon),\\
		Q^{\varepsilon}(\mathcal{N})&\leq\widetilde{I}_{\alpha}^c(\mathcal{N})+\frac{\alpha}{\alpha-1}\log_2\!\left(\frac{1}{1-\varepsilon}\right)\quad\forall~\alpha>1,
	\end{align}
	for all $\varepsilon\in[0,1)$.
	
	Let us summarize the steps that we took to arrive at the bounds in \eqref{eq-qcomm_weak_conv_one_shot_1} and \eqref{eq-qcomm_str_conv_one_shot_1}:
	\begin{enumerate}
		\item We first compared a quantum communication protocol for $\mathcal{N}$ with the same protocol for a useless channel by using the hypothesis testing relative entropy. This led us to Lemma~\ref{prop-qc_meta_conv}, and the resulting upper bound in~\eqref{eq-entr:coh-inf-ent-test-ineq_2}.
		\item We then used the data-processing inequality for the hypothesis testing relative entropy to remove the decoding channel from the bound in \eqref{eq-entr:coh-inf-ent-test-ineq_2}. This is done in \eqref{eq-one-shot-bound-meta_pf1} in the proof of Theorem~\ref{prop-qcomm:one-shot-bound-meta}.
		\item Finally, we optimized over all encoding channels in \eqref{eq-one-shot-bound-meta_pf2}--\eqref{eq-one-shot-bound-meta_pf4} to obtain Theorem~\ref{prop-qcomm:one-shot-bound-meta}, in which the bound is a function solely of the channel and the error probability. Using Propositions \ref{prop-hypo_to_rel_ent} and \ref{prop:sandwich-to-htre}, which relate hypothesis testing relative entropy to quantum relative entropy and sandwiched R\'{e}nyi relative entropy, we arrived at Corollary~\ref{thm-qcomm_meta_str_weak_conv}.
	\end{enumerate}

\subsection{Lower Bound on the Number of Transmitted Qubits via Entanglement Distillation}\label{subsec-qcomm_one_shot_lower_bound}

	Having derived upper bounds on the number of transmitted qubits for an arbitrary quantum communication protocol, let us now determine a lower bound on the number of transmitted qubits. As with the other communication scenarios that we have considered so far, in order to obtain a lower bound on the number qubits that can be transmitted, we need to devise an explicit $(d,\varepsilon)$ quantum communication protocol for all $\varepsilon\in(0,1)$. The protocol we consider is based on the one-shot, one-way entanglement distillation protocol from Proposition~\ref{prop-ent_dist_lower_bound} in Chapter~\ref{chap-ent_distill}, which establishes that, for an arbitrary bipartite state $\rho_{AB}$ and for all $\varepsilon\in(0,1]$ and $\eta\in[0,\sqrt{\varepsilon})$, there exists a $(d,\varepsilon)$ one-way entanglement distillation protocol for $\rho_{AB}$ such that
	\begin{equation}\label{eq-q_comm_ent_distill_one_shot_LB}
		\log_2 d =-H_{\max}^{\sqrt{\varepsilon}-\eta}(A|B)_{\rho}+4\log_2\eta.
	\end{equation}
	The goal in this section is to show that entanglement distillation can be used to develop a quantum communication strategy. Specifically, we show that the existence of a $(d,\varepsilon)$ one-way entanglement distillation protocol for the bipartite state $\omega_{AB}=\mathcal{N}_{A'\to B}(\psi_{AA'})$ implies the existence of a $(d',\varepsilon')$ quantum communication protocol, with $d'$ and $\varepsilon'$ being functions of $d$ and $\varepsilon$. The claim is as follows:
	
	\begin{theorem*}{Lower Bound on One-Shot Quantum Capacity}{prop-qcomm_one-shot_lower_bound}
		Let $\mathcal{N}_{A\to B}$ be a quantum channel. For all $\varepsilon\in(0,1)$, $\eta\in[0,\sfrac{\varepsilon\sqrt{\delta}}{4})$, and $\delta\in(0,1)$, there exists a $(d,\varepsilon)$ quantum communication protocol for $\mathcal{N}_{A\to B}$ such that
		\begin{equation}\label{eq-q_comm_one_shot_UB}
			\log_2 d=\sup_{\psi_{AA'}}\left(-H_{\max}^{\frac{\varepsilon\sqrt{\delta}}{4}-\eta}(A|B)_{\omega}\right)+\log_2(\eta^4(1-\delta)),
		\end{equation}
		where $\omega_{AB}=\mathcal{N}_{A'\to B}(\psi_{AA'})$. Consequently,
		\begin{equation}
			Q^{\varepsilon}(\mathcal{N})\geq \sup_{\psi_{AA'}}\left(-H_{\max}^{\frac{\varepsilon\sqrt{\delta}}{4}-\eta}(A|B)_{\omega}\right)+\log_2(\eta^4(1-\delta))
		\end{equation}
		for all $\eta\in[0,\sfrac{\varepsilon\sqrt{\delta}}{4})$ and $\delta\in(0,1)$, where $\omega_{AB}=\mathcal{N}_{A'\to B}(\psi_{AA'})$.
	\end{theorem*}
	
	The first step in the proof of Theorem~\ref{prop-qcomm_one-shot_lower_bound} is to observe that one-way entanglement distillation is an example of entanglement generation, albeit with forward (i.e., sender to receiver) classical communication, which we introduced at the beginning of this chapter and formally define below. We then show that forward classical communication does not help for entanglement generation, even in the non-asymptotic setting. One-way entanglement distillation thus implies entanglement generation. We then show that entanglement generation implies entanglement transmission, which we defined at the beginning of this chapter. Finally, we show that entanglement transmission implies quantum communication.
	
	Before proceeding with the proof of Theorem~\ref{prop-qcomm_one-shot_lower_bound}, let us formally define entanglement generation (with and without one-way LOCC assistance) and entanglement transmission.
	\begin{itemize}
		\item\textit{Entanglement generation}: An entanglement generation protocol for $\mathcal{N}_{A\to B}$ is defined by the three elements $(d,\Psi_{A'A},\mathcal{D}_{B\to B'})$, where $\Psi_{A'A}$ is a pure state with $d_{A'}=d$, and $\mathcal{D}_{B\to B'}$ is a decoding channel with $d_{B'}=d$. The goal of the protocol is to transmit the system $A$ such that the final state
			\begin{equation}
				\sigma_{A'B'}\coloneqq(\mathcal{D}_{B\to B'}\circ\mathcal{N}_{A\to B})(\Psi_{A'A})
			\end{equation}
			is close in fidelity to a maximally entangled state of Schmidt rank $d$. The \textit{entanglement generation error} of the protocol is given by
			\begin{align}
				p_{\text{err}}^{\text{(EG)}}(\Psi_{A'A},\mathcal{D};\mathcal{N})&\coloneqq 1-\bra{\Phi}_{A'B'}\sigma_{A'B'}\ket{\Phi}_{A'B'}\\
				&=1-F(\Phi_{A'B'},\sigma_{A'B'}).
			\end{align}
			We call the protocol $(d,\Psi_{A'A},\mathcal{D}_{B\to B'})$ a \textit{$(d,\varepsilon)$ protocol}, with $\varepsilon\in[0,1]$, if $p_{\text{err}}^{\text{(EG)}}(\Psi_{A'A},\mathcal{D};\mathcal{N})\leq\varepsilon$.
			
			Note that an entanglement generation protocol $(d,\Psi_{A'A},\mathcal{D}_{B\to B'})$ over $\mathcal{N}_{A\to B}$ is an example of an entanglement distillation protocol $(d,\mathcal{L}_{AB\to\hat{A}\hat{B}})$ for the state $\rho_{A'B}=\mathcal{N}_{A\to B}(\Psi_{A'A})$, with $\hat{A}\equiv A'$, $\hat{B}\equiv B'$, and $\mathcal{L}_{AB\to\hat{A}\hat{B}}\equiv \mathcal{D}_{B\to B'}$.
			
		\item \textit{Entanglement generation assisted by one-way LOCC}: An entanglement generation protocol for $\mathcal{N}_{A\to B}$ assisted by one-way LOCC from $A$ to $B$ is defined by $(d,\Psi_{A'A},\{\mathcal{E}_{A'A\to A'A}^x\}_x,\{\mathcal{D}_{B\to B'}^x\}_x)$, where $d\geq 1$, $\Psi_{A'A}$ is a pure state with $d_{A'}=d$, $\{\mathcal{E}_{A'A\to A'A}^x\}_{x\in\mathcal{X}}$ is a set of completely positive maps indexed by a finite alphabet $\mathcal{X}$ such that $\sum_{x\in\mathcal{X}}\mathcal{E}_{A'A\to A'A}^x$ is trace preserving, and $\{\mathcal{D}_{B\to B'}^x\}_{x\in\mathcal{X}}$ is a set of quantum channels indexed by $\mathcal{X}$, with $d_{B'}=d$. The goal of the protocol is to transmit the system $A$ such that the final state
			\begin{equation}\label{eq-q_comm_ent_gen_1WLOCC_output}
				\sigma_{A'B'}^{\rightarrow}\coloneqq\sum_{x\in\mathcal{X}}(\mathcal{D}_{B\to B'}^x\circ\mathcal{N}_{A\to B}\circ\mathcal{E}_{A'A\to A'A}^x)(\Psi_{A'A})
			\end{equation}
			is close in fidelity to a maximally entangled state of Schmidt rank $d$. The error of the protocol is given by
			\begin{equation}
				p_{\text{err}}^{\text{(EG)},\rightarrow}(\Psi_{A'A},\{\mathcal{E}^x\}_x,\{\mathcal{D}^x\}_x;\mathcal{N})=1-F(\Phi_{A'B'},\sigma_{A'B'}^{\rightarrow}).
			\end{equation}
			We call the protocol $(d,\Psi_{A'A},\{\mathcal{E}_{A'A\to A'A}^x\}_x,\{\mathcal{D}_{B\to B'}^x\}_x)$ a \textit{$(d,\varepsilon)$ protocol}, with $\varepsilon\in[0,1]$, if $p_{\text{err}}^{\text{(EG)},\rightarrow}(\Psi_{A'A},\{\mathcal{E}^x\}_x,\{\mathcal{D}^x\}_x;\mathcal{N})\leq\varepsilon$.
	
		\item\textit{Entanglement transmission}: An entanglement transmission protocol for $\mathcal{N}_{A\to B}$ consists of the three elements $(d,\mathcal{E},\mathcal{D})$, where $d\geq 1$, $\mathcal{E}_{A'\to A}$ is an encoding channel with $d_{A'}=d$, and $\mathcal{D}_{B\to B'}$ is a decoding channel with $d_{B'}=d$. The goal of the protocol is to transmit the $A'$ system of a maximally entangled state $\Phi_{RA'}$ of Schmidt rank $d$ such that the final state
			\begin{equation}
				\omega_{RB'}\coloneqq (\mathcal{D}_{B\to B'}\circ\mathcal{N}_{A\to B}\circ\mathcal{E}_{A'\to A})(\Phi_{RA'})
			\end{equation}
			is close to the initial maximally entangled state. The \textit{entanglement transmission error} of the protocol is
			\begin{align}
				p_{\text{err}}^{\text{(ET)}}(\mathcal{E},\mathcal{D};\mathcal{N})&\coloneqq 1-\bra{\Phi}_{RB'}\omega_{RB'}\ket{\Phi}_{RB'}\\
				&=1-F_e(\mathcal{D}\circ\mathcal{N}\circ\mathcal{E}),
			\end{align}
			where we recall the entanglement fidelity of a channel from Definition~\ref{def-ent_fid_chan}. We call the protocol $(d,\mathcal{E},\mathcal{D})$ a \textit{$(d,\varepsilon)$ protocol}, with $\varepsilon\in[0,1]$, if $p_{\text{err}}^{\text{(ET)}}(\mathcal{E},\mathcal{D};\mathcal{N})\allowbreak\leq\varepsilon$.
			
			Observe that the error criterion for entanglement transmission is the same as the average error criterion for quantum communication (see \eqref{eq-q_comm_max_error_to_avg_error_1}--\eqref{eq-q_comm_max_error_to_avg_error_3}). This means that the existence of an arbitrary $(d,\varepsilon)$ quantum communication protocol implies the existence of a $(d,\varepsilon)$ entanglement transmission protocol. Also observe that the existence of an arbitrary $(d,\varepsilon)$ entanglement transmission protocol implies the existence of a $(d,\varepsilon)$ entanglement generation protocol with respect to the state $\Psi_{RA}\equiv\mathcal{E}_{A'\to A}(\Phi_{RA'})$ (with the systems $R$, $A$, and $A'$ belonging to Alice).

	\end{itemize}

\subsubsection{Proof of Theorem~\ref{prop-qcomm_one-shot_lower_bound}}

	We start by showing that an arbitrary entanglement distillation protocol for the state $\mathcal{N}_{A\to B}(\psi_{A'A})$, with $\psi_{A'A}$ a pure state, has the same performance parameters  as an entanglement generation protocol with one-way LOCC assistance.
	
	Consider an arbitrary $(d,\varepsilon)$ entanglement distillation protocol for $\mathcal{N}_{A\to B}(\psi_{A'A})$ given by a one-way LOCC channel $\mathcal{L}_{A'B\to \hat{A}\hat{B}}$, with $d_{\hat{A}}=d_{\hat{B}}=d$. In general, this LOCC channel has the form $\mathcal{L}_{A'B\to\hat{A}\hat{B}}=\sum_{x\in\mathcal{X}}\mathcal{E}_{A'\to\hat{A}}^x\otimes\mathcal{D}_{B\to\hat{B}}^x$, where $\mathcal{X}$ is some finite alphabet, $\{\mathcal{E}_{A'\to\hat{A}}^x\}_{x\in\mathcal{X}}$ is a set of completely positive maps such that $\sum_{x\in\mathcal{X}}\mathcal{E}_{A'\to\hat{A}}^x$ is trace preserving, and $\{\mathcal{D}_{B\to\hat{B}}^x\}_{x\in\mathcal{X}}$ is a set of channels. The output state of the entanglement distillation protocol is
	\begin{multline}
		\sum_{x\in\mathcal{X}}(\mathcal{E}_{A'\to\hat{A}}^x\otimes\mathcal{D}_{B\to\hat{B}}^x)(\mathcal{N}_{A\to B}(\psi_{A'A}))\\=\sum_{x\in\mathcal{X}}(\mathcal{D}_{B\to\hat{B}}^x\circ\mathcal{N}_{A\to B}\circ\mathcal{E}_{A'\to\hat{A}}^x)(\psi_{A'A}),
	\end{multline}
	which has the form of a state at the output of an entanglement generation protocol with one-way LOCC assistance. We thus have that a $(d,\varepsilon)$ entanglement distillation protocol for $\mathcal{N}_{A\to B}(\psi_{A'A})$ is equivalent to a $(d,\varepsilon)$ entanglement generation protocol for $\mathcal{N}$ with one-way LOCC assistance. We now show that one-way LOCC assistance does not help for entanglement generation.
	
	\begin{Lemma}{lem-q_comm_ent_gen_1WLOCC_no_help}
		Given a $(d,\varepsilon)$ entanglement generation protocol for a channel $\mathcal{N}$, assisted by one-way LOCC, with $d\geq 1$ and $\varepsilon\in[0,1]$, there exists a $(d,\varepsilon)$ entanglement generation protocol for $\mathcal{N}$ (without one-way LOCC assistance).
	\end{Lemma}
	
	\begin{Proof}
		Consider an arbitrary $(d,\varepsilon)$ entanglement generation protocol assisted by one-way LOCC. The output state of such a protocol has the form in \eqref{eq-q_comm_ent_gen_1WLOCC_output}, i.e.,
		\begin{equation}
			\sigma_{A'B'}^{\rightarrow}=\sum_{x\in\mathcal{X}}(\mathcal{D}_{B\to B'}^x\circ\mathcal{N}_{A\to B}\circ\mathcal{E}_{A'A\to A'A}^x)(\Psi_{A'A}),
		\end{equation}
		and by definition we have
		\begin{equation}
			F(\Phi_{A'B'},\sigma_{A'B'}^{\rightarrow})=\Tr[\Phi_{A'B'}\sigma_{A'B'}^{\rightarrow}]\geq 1-\varepsilon.
		\end{equation}
		Now, let
		\begin{align}
			p(x)&\coloneqq \Tr[\mathcal{E}_{A'A\to A'A}^x(\Psi_{A'A})],\\
			\rho_{A'A}^x&\coloneqq \frac{1}{p(x)}\mathcal{E}_{A'A\to A'A}^x(\Psi_{A'A}).
		\end{align}
		Using this, we can write $\sigma_{A'B'}^{\rightarrow}$ as
		\begin{equation}
			\sigma_{A'B'}^{\rightarrow}=\sum_{x\in\mathcal{X}}p(x)(\mathcal{D}_{B\to B'}^x\circ\mathcal{N}_{A\to B})(\rho_{A'A}^x).
		\end{equation}
		For every $x\in\mathcal{X}$, let $\rho_{A'A}^x$ have the following spectral decomposition:
		\begin{equation}
			\rho_{A'A}^x=\sum_{k=1}^{r_x}q(k|x)\phi_{A'A}^{x,k},
		\end{equation}
		where $r_x=\rank(\rho_{A'A}^x)$. We thus have that
		\begin{equation}
			\sigma_{A'B'}^{\rightarrow}=\sum_{x\in\mathcal{X}}\sum_{k=1}^{r_x}p(x)q(k|x)(\mathcal{D}_{B\to B'}^x\circ\mathcal{N}_{A\to B})(\phi_{A'A}^{x,k}).
		\end{equation}
		Then, letting $\sigma_{A'B'}^{x,k}\coloneqq(\mathcal{D}_{B\to B'}^x\circ\mathcal{N}_{A\to B})(\phi_{A'A}^{x,k})$, we conclude that
		\begin{align}
			\Tr[\Phi_{A'B'}\sigma_{A'B'}^{\rightarrow}]&=\sum_{x\in\mathcal{X}}\sum_{k=1}^{r_x}p(x)q(k|x)\Tr[\Phi_{A'B'}\sigma_{A'B'}^{x,k}]\\
			&\leq \max_{\substack{x\in\mathcal{X},\\1\leq k\leq r_x}}\Tr[\Phi_{A'B'}\sigma_{A'B'}^{x,k}]\\
			&=\max_{\substack{x\in\mathcal{X},\\1\leq k\leq r_x}}\Tr[\Phi_{A'B'}(\mathcal{D}_{B\to B'}^x\circ\mathcal{N}_{A\to B})(\phi_{A'A}^{x,k})].
		\end{align}
		In other words, there exists a pair $(\phi_{A'A}^{x,k},\mathcal{D}_{B\to B'}^x)$ (namely, the one that achieves the maximum on the right-hand side of the inequality above) such that 
		\begin{equation}
			\Tr[\Phi_{A'B'}\sigma_{A'B'}^{x,k}]\geq \Tr[\Phi_{A'B'}\sigma_{A'B'}]\geq 1-\varepsilon.
		\end{equation}
		Therefore, the triple $(d,\phi_{A'A}^{k,x},\mathcal{D}_{B\to B'}^x)$ constitutes a $(d,\varepsilon)$ entanglement generation protocol (without one-way LOCC assistance).
	\end{Proof}
	
	As we have seen in Chapter~\ref{chap-ent_distill}, forward classical communication certainly helps in general for entanglement distillation. However, it does not help for entanglement generation because the resource for entanglement generation is a quantum channel, whereas for entanglement distillation the resource is a bipartite quantum state. Having a quantum channel as the resource is more powerful than having a bipartite quantum state because, when using a quantum channel, there is an extra degree of freedom in the input state to the channel. The proof of Lemma~\ref{lem-q_comm_ent_gen_1WLOCC_no_help} demonstrates that the forward classical communication in an arbitrary $(d,\varepsilon)$ is not needed, and the proof essentially relies on convexity of the entanglement fidelity performance criterion.
	
	We now show that entanglement generation implies entanglement transmission, up to a transformation of the performance parameters.
	
	\begin{Lemma*}{Entanglement Generation to Entanglement Transmission}{lem-q_comm_ent_gen_to_ent_trans}
		Given a $(d,\varepsilon)$ entanglement generation protocol for a channel $\mathcal{N}_{A\to B}$, with $d\geq 1$ and $\varepsilon\in[0,1]$, there exists a $(d,4\varepsilon)$ entanglement transmission protocol for $\mathcal{N}$.
	\end{Lemma*}
	
	\begin{Proof}
		Let $(d,\Psi_{A'A},\mathcal{D}_{B\to B'})$ be the elements of a $(d,\varepsilon)$ entanglement generation protocol for $\mathcal{N}_{A\to B}$, with $d_{A'}=d_{B'}=d$. This implies that the output state
		\begin{equation}
			\sigma_{A'B'}=(\mathcal{D}_{B\to B'}\circ\mathcal{N}_{A\to B})(\Psi_{A'A})
		\end{equation}
		satisfies
		\begin{equation}
			F(\Phi_{A'B'},\sigma_{A'B'})\geq 1-\varepsilon.
		\end{equation}
		We now construct an entanglement transmission protocol. To this end, let $A'\equiv R$ be a reference system inaccessible to both Alice and Bob. By the data-processing inequality for fidelity (Theorem~\ref{thm-fidelity_monotone}) with respect to the partial trace channel $\Tr_{B'}$, we have that 
		\begin{equation}
			F(\Phi_{R},\Psi_{R})=F(\Tr_{B'}[\Phi_{RB'}],\Tr_{B'}[\sigma_{RB'}])\geq F(\Phi_{RB'},\sigma_{RB'})\geq 1-\varepsilon.
		\end{equation}
		Next, by Uhlmann's theorem (Theorem~\ref{thm-Uhlmann_fidelity}), there exists an isometric channel $\mathcal{U}_{A'\to A}$ such that
		\begin{equation}\label{eq-q_comm_ent_gen_to_ent_trans_pf}
			F(\Phi_R,\Psi_R)=F(\mathcal{U}_{A'\to A}(\Phi_{RA'}),\Psi_{RA'})\geq 1-\varepsilon.
		\end{equation}
		We let this isometric channel $\mathcal{U}_{A'\to A}$ be the encoding channel for the entanglement transmission protocol, and we let
		\begin{equation}
			\omega_{RB'}=(\mathcal{D}_{B\to B'}\circ\mathcal{N}_{A\to B}\circ\mathcal{U}_{A'\to A})(\Phi_{RA'}).
		\end{equation}
		Next, using the sine distance (Definition~\ref{def-purified_distance}), by definition of the $(d,\varepsilon)$ entanglement generation protocol, we have that $P(\Phi_{RB'},\sigma_{RB'})\leq\sqrt{\varepsilon}$. Similarly, from \eqref{eq-q_comm_ent_gen_to_ent_trans_pf} we have that
		\begin{equation}\label{eq-q_comm_ent_gen_to_ent_trans_pf2}
			P(\Psi_{RA},\mathcal{U}_{A'\to A}(\Phi_{RA'}))\leq\sqrt{\varepsilon}.
		\end{equation}
		Therefore, by the triangle inequality for the sine distance (Lemma~\ref{lem-sine-distance-triangle}), we conclude that
		\begin{align}
			P(\Phi_{RB'},\omega_{RB'})&\leq P(\Phi_{RB'},\sigma_{RB'})+P(\sigma_{RB'},\omega_{RB'})\\
			&\leq \sqrt{\varepsilon}+\sqrt{\varepsilon}\\
			&\leq 2\sqrt{\varepsilon},
		\end{align}
		where the second inequality follows from the data-processing inequality for sine distance (see \eqref{eq-sine_dist_data_proc}) and \eqref{eq-q_comm_ent_gen_to_ent_trans_pf2} to see that
		\begin{align}
			&P(\sigma_{RB'},\omega_{RB'}) \nonumber\\
			&\quad=P((\mathcal{D}_{B\to B'}\circ\mathcal{N}_{A\to B})(\Psi_{RA}),(\mathcal{D}_{B\to B'}\circ\mathcal{N}_{A\to B}\circ\mathcal{U}_{A'\to A})(\Phi_{RA'}))\\
			&\quad\leq P(\Psi_{RA},\mathcal{U}_{A'\to A}(\Phi_{RA'}))\\
			&\quad\leq \sqrt{\varepsilon}.
		\end{align}
		Therefore, by definition of the sine distance, we conclude that
		\begin{equation}
			1-F(\Phi_{RB'},\omega_{RB'})\leq 4\varepsilon,
		\end{equation}
		so that $(d,\mathcal{U}_{A'\to A},\mathcal{D}_{B\to B'})$ constitutes a $(d,4\varepsilon)$ entanglement transmission protocol, as required.
	\end{Proof}
	
	Finally, we show that entanglement transmission implies quantum communication, up to a transformation of the performance parameters. We could alternatively call this statement ``quantum expurgation,'' because the arguments in the proof are analogous to the expurgation arguments applied in the proof of the lower bound for one-shot classical communication in Proposition~\ref{prop-cc_one-shot_lower_bound}.

	\begin{Lemma*}{Entanglement Transmission to Quantum Communication}{prop-ent_trans_code_to_q_comm_code}
		Given a $(d,\varepsilon)$ entanglement transmission protocol for a channel $\mathcal{N}_{A\to B}$, with $d\geq 1$ and $\varepsilon\in[0,1]$, for all $\delta\in(0,1)$, there exists a $(\floor{(1-\delta)d},2\sqrt{\sfrac{\varepsilon}{\delta}})$ quantum communication protocol for $\mathcal{N}$.
	\end{Lemma*}
	
	\begin{Proof}
		Suppose that a $(d,\varepsilon)$ entanglement transmission code for $\mathcal{N}_{A\to B}$ exists, and let $\mathcal{E}_{A'\to A}$ and $\mathcal{D}_{B\to B'}$ be the corresponding encoding and decoding channels, respectively, with $d_{A'}=d_{B'}=d$. The condition $\overline{p}_{\text{err}}(\mathcal{E},\mathcal{D};\mathcal{N})\leq\varepsilon$ then holds, namely,
		\begin{equation}\label{eq-q_comm_ent_trans_to_q_comm_pf1}
			1-\Tr[\Phi_{RB'}\omega_{RB'}]\leq\varepsilon,
		\end{equation}
		where
		\begin{equation}
			\omega_{RB'}=(\mathcal{D}_{B\to B'}\circ\mathcal{N}_{A\to B}\circ\mathcal{E}_{A'\to A})(\Phi_{RA'}).
		\end{equation}
		Let
		\begin{equation}
			\mathcal{C}_{A'\rightarrow B'}\coloneqq\mathcal{D}_{B\rightarrow B'}\circ\mathcal{N}_{A\rightarrow B}\circ\mathcal{E}_{A'\rightarrow A}.
		\end{equation}
		We proceed with the following algorithm:
		\begin{enumerate}
			\item Set $k=d$ and $\mathcal{H}_d=\mathcal{H}_{A'}$. Suppose for now that $(1-\delta)d$ is a positive integer.

			\item Set $\ket{\phi_{k}}\in\mathcal{H}_{k}$ to be a state vector that achieves the minimum fidelity of $\mathcal{C}_{A'\to B'}$:
				\begin{equation}
					\ket{\phi_{k}}\coloneqq \argmin_{\ket{\phi}\in\mathcal{H}_{k}}\bra{\phi}\mathcal{C}_{A'\rightarrow B'}(\ket{\phi}\!\bra{\phi})\ket{\phi},
				\end{equation}
				and set the fidelity $F_{k}$ of $|\phi_{k}\rangle$ as follows:%
				\begin{align}
					F_{k}&\coloneqq\min_{\ket{\phi}\in\mathcal{H}_{k}}\bra{\phi}\mathcal{C}_{A'\rightarrow B'}(\ket{\phi}\!\bra{\phi})\ket{\phi}\\
					&=\bra{\phi_{k}}\mathcal{C}_{A'\rightarrow B'}(\ket{\phi_k}\!\bra{\phi_k})\ket{\phi_k}.
				\end{align}

			\item Set
				\begin{equation}
					\mathcal{H}_{k-1}\coloneqq\text{span}\{\ket{\psi}\in\mathcal{H}_{k}:\abs{\braket{\psi}{\phi_k}}=0\}.
					\end{equation}
				That is, $\mathcal{H}_{k-1}$ is set to the orthogonal complement of $\ket{\phi_{k}}$ in $\mathcal{H}_{k}$, so that $\mathcal{H}_{k}=\mathcal{H}_{k-1}\oplus\text{span}\{\ket{\phi_{k}}\}$. Set $k\to k-1$.

			\item Repeat steps 2-3 until $k=(1-\delta)d$ after step 3.
		\end{enumerate}

		The idea behind this algorithm is to successively remove minimum fidelity states from $\mathcal{H}_{A'}$ until $k=(1-\delta)d$. By the structure of the algorithm and some analysis given below, we are then guaranteed that for this $k$ and lower that%
		\begin{equation}
			1-\min_{\ket{\phi}\in\mathcal{H}_{k}}\bra{\phi}\mathcal{C}(\ket{\phi}\!\bra{\phi})\ket{\phi}\leq\sfrac{\varepsilon}{\delta}.
		\end{equation}
		That is, the subspace $\mathcal{H}_{k}$ is good for quantum communication of states at the channel input with fidelity at least $1-\varepsilon/\delta$ (to be precise, the subspace $\mathcal{H}_{k}$ is good for subspace transmission as defined in the introduction of this chapter). Furthermore, the algorithm implies that%
		\begin{align}
			F_{d}&\leq F_{d-1}\leq \dotsb \leq F_{(1-\delta)d},\label{eq:fidelity-ordering}\\
			\mathcal{H}_{d} & \supseteq\mathcal{H}_{d-1}\supseteq\dotsb \supseteq\mathcal{H}_{(1-\delta)d}.
		\end{align}
		Also, $\{\ket{\phi_{k}}\}_{k=1}^{\ell}$ is an orthonormal basis for $\mathcal{H}_{\ell}$, where $\ell\in\{1,\dotsc,d\}$. Note that the unit vectors $\ket{\phi_{k}}$, $k\in\{(1-\delta)d-1,\dotsc,1\}$ can be generated by repeating the algorithm above exhaustively.

		We now analyze the claims above by employing Markov's inequality and some other tools. From \eqref{eq-q_comm_ent_trans_to_q_comm_pf1}, we have that
		\begin{equation}
			F(\Phi_{RB'},\mathcal{C}_{A'\rightarrow B'}(\Phi_{RA'}))\geq1-\varepsilon.
		\end{equation}
		Since $\{\ket{\phi_{k}}\}_{k=1}^{d}$ is an orthonormal basis for $\mathcal{H}_{d}$, we can write%
		\begin{equation}
			\ket{\Phi}_{RA'}=\frac{1}{\sqrt{d}}\sum_{k=1}^{d}\ket{\overline{\phi_{k}}}_{R}\otimes\ket{\phi_{k}}_{A'},
		\end{equation}
		where complex conjugation is taken with respect to the basis $\{\ket{i}\}_{i=0}^{d-1}$ used in \eqref{eq-max_ent_message}. A consequence of the data-processing inequality for fidelity under the dephasing channel $\omega_{R}\mapsto\sum_{k=1}^d\ket{\overline{\phi_{k}}}\!\bra{\overline{\phi_{k}}}_{R}\omega_{R}\ket{\overline{\phi_{k}}}\!\bra{\overline{\phi_{k}}}_{R}$ and convexity of the square function is that%
		\begin{align}
			F(\Phi_{RB'},(\id_{R}\otimes\mathcal{C}_{A'\rightarrow B'})(\Phi_{RA'}))&\leq\frac{1}{d}\sum_{k=1}^d \bra{\phi_k}\mathcal{C}_{A'\rightarrow B'}(\ket{\phi_k}\!\bra{\phi_k})\ket{\phi_k}\\
			&=\frac{1}{d}\sum_{k=1}^d F_{k}.
		\end{align}
		This means that 
		\begin{equation}
			\frac{1}{d}\sum_{k=1}^d F_{k}\geq 1-\varepsilon \Longleftrightarrow\frac{1}{d}\sum_{k=1}^d(1-F_{k})\leq\varepsilon.
		\end{equation}
		Now, taking $K$ to be a uniform random variable with realizations $k\in\{1,\dotsc,d\}$ and applying Markov's inequality (see \eqref{eq-MT:Markov-ineq}), we find that%
		\begin{equation}
			\Pr[1-F_{K}\geq\sfrac{\varepsilon}{\delta}]\leq\frac{\mathbb{E}[1-F_{K}]}{\sfrac{\varepsilon}{\delta}}\leq\frac{\varepsilon}{\sfrac{\varepsilon}{\delta}}=\delta.
		\end{equation}
		So this implies that $(1-\delta)d$ of the $F_{k}$ values are such that $F_{k}\geq 1-\sfrac{\varepsilon}{\delta}$. Since they are ordered as given in \eqref{eq:fidelity-ordering}, we conclude that $\mathcal{H}_{(1-\delta)d}$, which by definition has dimension $(1-\delta)d$, is a good subspace for quantum communication in the following sense (subspace transmission):%
		\begin{equation}\label{eq:min-fid-condition-almost-done}
			\min_{\ket{\phi}\in\mathcal{H}_{(1-\delta)d}}\bra{\phi}\mathcal{C}_{A'\rightarrow B'}(\ket{\phi}\!\bra{\phi})\ket{\phi}\geq 1-\sfrac{\varepsilon}{\delta}.
		\end{equation}
		Now, applying Proposition~\ref{prop:min-fid-to-min-ent-fid} to \eqref{eq:min-fid-condition-almost-done}, we conclude that
		\begin{equation}\label{eq:resulting-code-fid-1}
			\min_{\ket{\psi}\in\mathcal{H}_{(1-\delta)d}^{\prime}\otimes\mathcal{H}_{(1-\delta)d}}\bra{\psi}(\id_{\mathcal{H}_{(1-\delta)d}'}\otimes\mathcal{C}_{A'\rightarrow B'})(\ket{\psi}\!\bra{\psi})\ket{\psi}\geq 1-2\sqrt{\sfrac{\varepsilon}{\delta}},
		\end{equation}
		i.e., $p_{\text{err}}^*(\mathcal{E},\mathcal{D};\mathcal{N})\leq 2\sqrt{\sfrac{\varepsilon}{\delta}}$, which is the criterion for strong subspace transmission (the strongest notion of quantum communication).

		To finish off the proof, suppose that $(1-\delta)d$ is not an integer. Then there exists a $\delta'>\delta$ such that $(1-\delta')d=\floor{(1-\delta)d}$ is a positive integer. By the above reasoning, there exists a code satisfying \eqref{eq:resulting-code-fid-1}, except with $\delta$ replaced by $\delta'$, and with the code dimension equal to $\floor{(1-\delta)d}$. We also have that $1-2\sqrt{\sfrac{\varepsilon}{\delta'}}>1-2\sqrt{\sfrac{\varepsilon}{\delta}}$. This concludes the proof.
	\end{Proof}
	
	We now return to the proof of Theorem~\ref{prop-qcomm_one-shot_lower_bound}. To finish it off, we combine the results of Lemmas~\ref{lem-q_comm_ent_gen_1WLOCC_no_help}, \ref{lem-q_comm_ent_gen_to_ent_trans}, and \ref{prop-ent_trans_code_to_q_comm_code} to conclude that the existence of a $(d,\varepsilon)$ entanglement distillation protocol for $\omega_{AB}=\mathcal{N}_{A'\to B}(\psi_{AA'})$ implies the existence of a $(d',\varepsilon')$ quantum communication protocol, where
	\begin{equation}
		d'=(1-\delta)d,\quad \varepsilon'=4\sqrt{\frac{\varepsilon}{\delta}}\ ,\quad\delta\in(0,1).
	\end{equation}
	Recalling that $d$ is given by \eqref{eq-q_comm_ent_distill_one_shot_LB}, we conclude that
	\begin{equation}
		\log_2 d'=-H_{\max}^{\frac{\varepsilon'\sqrt{\delta}}{4}-\eta}(A|B)_{\omega}+\log_2(\eta^4(1-\delta)).
		\label{eq:Q-cap:proof-ent-trans-to-q-comm-step-mid}
	\end{equation}
	Then, since the pure state $\psi_{AA'}$ used in \eqref{eq:Q-cap:proof-ent-trans-to-q-comm-step-mid} is arbitrary, we conclude that there exists a $(d',\varepsilon')$ quantum communication protocol satisfying
	\begin{equation}
		\log_2 d'=\sup_{\psi_{AA'}}\left(-H_{\max}^{\frac{\varepsilon'\sqrt{\delta}}{4}-\eta}(A|B)_{\omega}\right)+\log_2(\eta^4(1-\delta))
	\end{equation}
	for all $\eta\in[0,\sfrac{\varepsilon'\sqrt{\delta}}{4})$ and $\delta\in(0,1)$. This is precisely the statement in \eqref{eq-q_comm_one_shot_UB}, and so the proof of Theorem~\ref{prop-qcomm_one-shot_lower_bound} is complete. 
	
	Applying the relation between smooth conditional min- and max-entropy in \eqref{eq-Hmax_Hmin_pure_smooth} to the result of Theorem~\ref{prop-qcomm_one-shot_lower_bound}, and combining it with \eqref{eq-smooth_cond_min_ent_to_petz_renyi}, we obtain the following.
	
	\begin{corollary}{cor-q_comm_one_shot_LB_alt}
		Let $\mathcal{N}_{A\to B}$ be a quantum channel, and let $\mathcal{V}^{\mathcal{N}}_{A\to BE}$ be an isometric channel extending $\mathcal{N}_{A\to B}$. For all $\varepsilon\in(0,1)$, $\delta\in(0,1)$, $\eta\in[0,\sfrac{\varepsilon\sqrt{\delta}}{4})$, and $\alpha>1$, there exists a $(d,\varepsilon)$ quantum communication protocol for $\mathcal{N}$ with
		\begin{multline}\label{eq-q_comm_one_shot_LB_alt}
			\log_2 d\geq\sup_{\psi_{AA'}}\widetilde{H}_{\alpha}(A|E)_{\phi} -\frac{1}{\alpha-1}\log_2\!\left(\frac{1}{f(\varepsilon,\delta,\eta)}\right)\\
			-\log_2\!\left(\frac{1}{1-f(\varepsilon,\delta,\eta)}\right)+\log_2(\eta^4(1-\delta)),
		\end{multline}
		 where $f(\varepsilon,\delta,\eta)\coloneqq \left(\frac{\varepsilon\sqrt{\delta}}{4}-\eta\right)^2$, $\phi_{AE}=\Tr_B[\mathcal{V}^{\mathcal{N}}_{A'\to BE}(\psi_{AA'})]=\mathcal{N}_{A'\to E}^c(\psi_{AA'})$, and $\psi_{AA'}$ is a pure state with the dimension of $A'$ equal to the dimension of~$A$.
	\end{corollary}
	
	Since the inequality in \eqref{eq-q_comm_one_shot_LB_alt} holds for all $(d,\varepsilon)$ quantum communication protocols, we have that
	\begin{multline}
		Q^{\varepsilon}(\mathcal{N})\geq\sup_{\psi_{AA'}}\widetilde{H}_{\alpha}(A|E)_{\phi} -\frac{1}{\alpha-1}\log_2\!\left(\frac{1}{f(\varepsilon,\delta,\eta)}\right)\\
			-\log_2\!\left(\frac{1}{1-f(\varepsilon,\delta,\eta)}\right)+\log_2(\eta^4(1-\delta)),
	\end{multline}
	where
	\begin{equation}
	\phi_{AE}=\Tr_B[\mathcal{V}^{\mathcal{N}}_{A'\to BE}(\psi_{AA'})]=\mathcal{N}_{A'\to E}^c(\psi_{AA'}),
	\end{equation}
	$f(\varepsilon,\delta,\eta)$ is defined just above, 
	 $\eta\in[0,\sfrac{\varepsilon\sqrt{\delta}}{4})$,  $\delta\in(0,1)$, and $\alpha>1$.

\subsubsection{Remark on Forward Classical Communication}

	To summarize what we did in this section, we used the result from Proposition~\ref{prop-ent_dist_lower_bound} on one-shot entanglement distillation to prove the existence of a quantum communication protocol in the one-shot setting. Note that the entanglement distillation protocol of Proposition~\ref{prop-ent_dist_lower_bound} involves one-way classical communication, while quantum communication (as we defined it at the beginning of this chapter) does not. In other words, in this section we managed to remove the one-way classical communication from the entanglement distillation protocol and thereby argue for the existence of a quantum communication protocol. More generally, it holds that forward classical communication (i.e., from the sender to the receiver) does not enhance the corresponding quantum capacity of the channel. In other words, if $Q^{\rightarrow}(\mathcal{N})$ denotes the quantum capacity of the channel $\mathcal{N}$ when classical communication from the sender to the receiver is allowed as part of the protocol, then $Q^{\rightarrow}(\mathcal{N})=Q(\mathcal{N})$. This is a direct consequence of the chain of reasoning given in Lemmas~\ref{lem-q_comm_ent_gen_1WLOCC_no_help}, \ref{lem-q_comm_ent_gen_to_ent_trans}, and \ref{prop-ent_trans_code_to_q_comm_code}, and we return to this point in Chapter~\ref{chap-LOCC-QC} when we consider LOCC-assisted quantum communication.

\section{Quantum Capacity of a Quantum Channel}\label{sec-q_comm_asymptotic}

\begin{figure}
		\centering
		\includegraphics[scale=0.8]{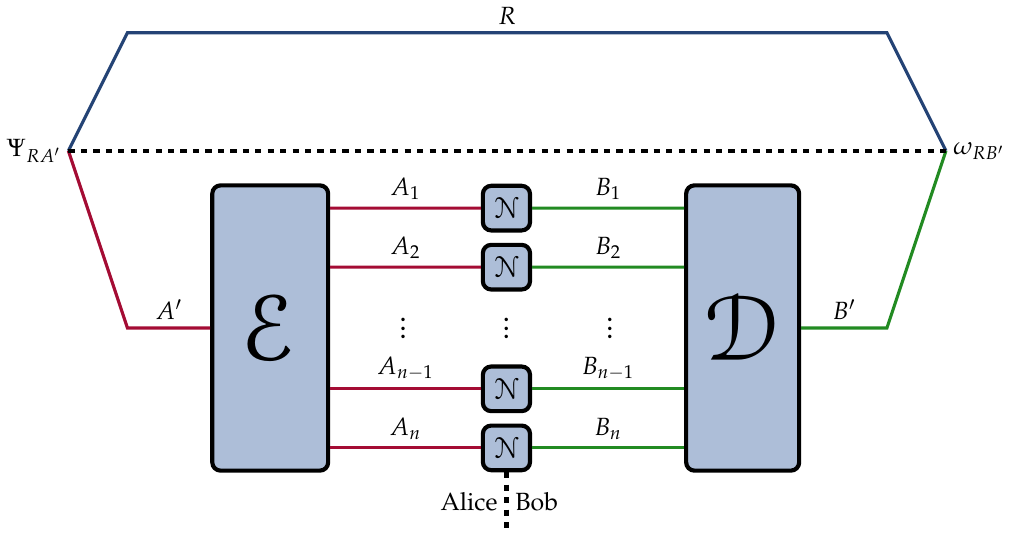}
		\caption{A general quantum communication protocol for $n\geq 1$ memoryless/unassisted uses of a quantum channel $\mathcal{N}$. Alice uses the channel~$\mathcal{E}$ to encode her share $A'$ of the pure state $\Psi_{RA'}$ into $n$ quantum systems $A_1,A_2,\dotsc,A_n$. She then sends each one of these through the channel $\mathcal{N}$. Bob finally applies a joint decoding channel $\mathcal{D}$ on the systems $B_1,B_2,\dotsc,B_n$, resulting in the state $\omega_{RB'}$ given by \eqref{eq-qcomm_final_state_asymptotic}.}\label{fig-qcomm_asymptotic}
	\end{figure}

	We now consider the asymptotic setting. In this scenario, depicted in Figure~\ref{fig-qcomm_asymptotic}, the quantum system $A'$ to be transmitted to Bob is encoded into $n$~copies $A_1,\dotsc, A_n$ of a quantum system $A$, for $n\geq 1$. Each of these systems is then sent independently through the channel $\mathcal{N}$. We call this the asymptotic setting because the number $n$ can be arbitrarily large.
	
	Analysis of the asymptotic setting is almost exactly the same as that of the one-shot setting. This is due to the fact that $n$ independent uses of the channel $\mathcal{N}$ can be regarded as a single use of the channel $\mathcal{N}^{\otimes n}$. So the only change that needs to be made is to replace $\mathcal{N}$ with $\mathcal{N}^{\otimes n}$ and to define the encoding and decoding channels as acting on $n$ systems instead of just one. In particular, the state at the end of the protocol becomes
	\begin{equation}\label{eq-qcomm_final_state_asymptotic}
		\omega_{RB'}=(\mathcal{D}_{B^n\to B'}\circ\mathcal{N}_{A\to B}^{\otimes n}\circ\mathcal{E}_{A'\to A^n})(\Psi_{RA'}).
	\end{equation}
	Then, just as in the one-shot setting, we define the error probability of the code $(\mathcal{E},\mathcal{D})$ for $n$ independent uses of $\mathcal{N}$ as
	\begin{equation}
		p_{\text{err}}^*(\mathcal{E},\mathcal{D};\mathcal{N}^{\otimes n})=1-F(\mathcal{D}\circ\mathcal{N}^{\otimes n}\circ\mathcal{E}).
	\end{equation}
	
	\begin{definition}{$(n,d,\varepsilon)$ Quantum Communication Protocol}{def-qcomm_nMe_protocol}
		Let $(d,\mathcal{E}_{A'\to A^n},\mathcal{D}_{B^n\to B'})$ be the elements of a quantum communication protocol for $n$ independent uses of the channel $\mathcal{N}_{A\to B}$, where $d_{A'}=d_{B'}=d$. The protocol is called an \textit{$(n,d,\varepsilon)$ protocol}, with $\varepsilon\in[0,1]$, if $p_{\text{err}}^*(\mathcal{E},\mathcal{D};\mathcal{N}^{\otimes n})\leq\varepsilon$.
	\end{definition}
	
	The \textit{rate} of an $(n,d,\varepsilon)$ quantum communication protocol is defined as the number of qubits transmitted per channel use, i.e.,
	\begin{equation}
		R(n,d)\coloneqq\frac{\log_2 d}{n}.
	\end{equation}
	Observe that the rate depends only on the dimension $d$ of the system $A'$ of the pure state~$\Psi_{RA'}$ to be transmitted and on the number of channel uses. In particular, it does not directly depend on the communication channel nor on the encoding and decoding channels. For a given $\varepsilon\in[0,1]$ and $n\geq 1$, the highest rate among all $(n,d,\varepsilon)$ protocols is denoted by $Q^{n,\varepsilon}(\mathcal{N})$, and it is defined as
	\begin{equation}
		Q^{n,\varepsilon}(\mathcal{N})\coloneqq\frac{1}{n}Q^{\varepsilon}(\mathcal{N}^{\otimes n})=\sup_{(d,\mathcal{E},\mathcal{D})}\left\{\frac{\log_2 d}{n}:p_{\text{err}}^*(\mathcal{E},\mathcal{D};\mathcal{N}^{\otimes n})\leq\varepsilon \right\},
	\end{equation}
	where in the second equality we use the definition of the one-shot quantum capacity $Q^{\varepsilon}$ given in \eqref{eq-q_comm_one_shot_capacity}, and the supremum is over all $d\geq 1$, encoding channels $\mathcal{E}$ with input system dimension $d$, and decoding channels $\mathcal{D}$ with output system dimension $d$.
	
	\begin{definition}{Achievable Rate for Quantum Communication}{def-qcomm_ach_rate}
		Given a quantum channel $\mathcal{N}$, a rate $R\in\mathbb{R}^+$ is called an \textit{achievable rate for quantum communication over $\mathcal{N}$} if for all $\varepsilon\in(0,1]$, $\delta>0$, and sufficiently large $n$, there exists an $(n,2^{n(R-\delta)},\varepsilon)$ quantum communication protocol for $\mathcal{N}$.
	\end{definition}
	
	As we prove in Appendix~\ref{chap-str_conv},
	\begin{equation}
		R\text{ achievable rate }\Longleftrightarrow \lim_{n\to\infty}\varepsilon_Q^*(2^{n(R-\delta)};\mathcal{N}^{\otimes n})=0\quad\forall~\delta>0.
	\end{equation}
	In other words, a rate $R$ is achievable if for all $\delta>0$, the optimal error probability for a sequence of protocols with rate $R-\delta$ vanishes as the number $n$ of uses of $\mathcal{N}$ increases.
		
	\begin{definition}{Quantum Capacity of a Quantum Channel}{def-qcomm_quantum_cap}
		The \textit{quantum capacity of a quantum channel $\mathcal{N}$}, denoted by $Q(\mathcal{N})$, is defined to be the supremum of all achievable rates, i.e.,
		\begin{align}
			Q(\mathcal{N})&\coloneqq\sup\{R:R\text{ is an achievable rate for }\mathcal{N}\}
		\end{align}
	\end{definition}
	
	An equivalent definition of quantum capacity is
		\begin{equation}
		Q(\mathcal{N}) = \inf_{\varepsilon\in(0,1] }\liminf_{n\to\infty}\frac{1}{n}Q^{\varepsilon}(\mathcal{N}^{\otimes n}).
	\end{equation}
	We prove this in Appendix~\ref{chap-str_conv}.
	
	\begin{definition}{Weak Converse Rate for Quantum Communication}{def-qcomm_weak_conv_rate}
		Given a quantum channel $\mathcal{N}$, a rate $R\in\mathbb{R}^+$ is called a \textit{weak converse rate for quantum communication over $\mathcal{N}$} if every $R'>R$ is not an achievable rate for $\mathcal{N}$.
	\end{definition}
	
	We show in Appendix~\ref{chap-str_conv} that
	\begin{equation}\label{eq-q_comm_weak_conv_rate_alt}
		R\text{ weak converse rate }\Longleftrightarrow \lim_{n\to\infty}\varepsilon_Q^*(2^{n(R-\delta)};\mathcal{N}^{\otimes n})>0\quad\forall~\delta>0.
	\end{equation}
	In other words, a weak converse rate is a rate for which the optimal error probability cannot be made to vanish, even in the limit of a large number of channel uses.
	
	\begin{definition}{Strong Converse Rate for Quantum Communication}{def-qcomm_str_conv_rate}
		Given a quantum channel $\mathcal{N}$, a rate $R\in\mathbb{R}^+$ is called a \textit{strong converse rate for quantum communication over $\mathcal{N}$} if for all $\varepsilon\in[0,1)$, $\delta>0$, and sufficiently large $n$, there does not exist an $(n,2^{n(R+\delta)},\varepsilon)$ quantum communication protocol for $\mathcal{N}$.
	\end{definition}
	
	We show in Appendix~\ref{chap-str_conv} that
	\begin{equation}\label{eq-q_comm_str_conv_rate_alt}
		R\text{ strong converse rate }\Longleftrightarrow \lim_{n\to\infty}\varepsilon_Q^*(2^{n(R+\delta)};\mathcal{N}^{\otimes n})=1\quad\forall~\delta>0.
	\end{equation}
	Unlike the weak converse, in which the optimal error is required to simply be bounded away from zero as the number $n$ of channel uses increases, in order to have a strong converse rate, the optimal error has to converge to one as $n$ increases. By comparing \eqref{eq-q_comm_weak_conv_rate_alt} and \eqref{eq-q_comm_str_conv_rate_alt}, we conclude that every strong converse rate is a weak converse rate. 
	
	\begin{definition}{Strong Converse Quantum Capacity of a Quantum Channel}{def-qcomm_str_conv_cap}
		The \textit{strong converse quantum capacity} of a quantum channel $\mathcal{N}$, denoted by $\widetilde{Q}(\mathcal{N})$, is defined as the infimum of all strong converse rates, 
		i.e.,
		\begin{align}
			\widetilde{Q}(\mathcal{N})&\coloneqq\inf\{R:\text{ $R$ is a strong converse rate for }\mathcal{N}\}
		\end{align}
	\end{definition}
	
	We can also write the strong converse quantum capacity as
	\begin{equation}
		\widetilde{Q}(\mathcal{N})=\sup_{\varepsilon\in [0,1)}\limsup_{n\to\infty}\frac{1}{n}Q^{\varepsilon}(\mathcal{N}^{\otimes n}).
	\end{equation}
	See Appendix~\ref{chap-str_conv} for a proof. We also show in Appendix~\ref{chap-str_conv} that
	\begin{equation}\label{eq-qcomm_str_conv_rate_lower_bound}
		Q(\mathcal{N})\leq \widetilde{Q}(\mathcal{N})
	\end{equation}
	for every quantum channel $\mathcal{N}$.
	
	We now state the main theorem of this chapter, which gives an expression for the quantum capacity of a quantum channel.
	
	\begin{theorem*}{Quantum Capacity}{thm-qcomm_capacity}
		The quantum capacity of a quantum channel $\mathcal{N}_{A\to B}$ is equal to the regularized coherent information $I_{\text{reg}}^c(\mathcal{N})$ of $\mathcal{N}$, i.e.,
		\begin{equation}\label{eq-quantum_capacity}
			Q(\mathcal{N})=I_\text{reg}^c(\mathcal{N})\coloneqq\lim_{n\to\infty}\frac{1}{n}I^c(\mathcal{N}^{\otimes n}).
		\end{equation}
	\end{theorem*}
	
	Recall from \eqref{eq-coh_inf_chan} that the channel coherent information is defined as
	\begin{equation}
		I^c(\mathcal{N})=\sup_{\psi_{RA}}I(R\rangle B)_{\omega},
	\end{equation}
	where $\omega_{RB}=\mathcal{N}_{A\to B}(\psi_{RA})$ and $\psi_{RA}$ is a pure state with the dimension of $R$ equal to the dimension of $A$.
	
	Observe that the expression in \eqref{eq-quantum_capacity} for the quantum capacity of a quantum channel is somewhat similar to the expression in \eqref{eq-classical_capacity} for the classical capacity of a quantum channel, in the sense that both capacities involve a regularization of a corresponding channel measure. In the case of quantum communication, we obtain the regularization of the channel's coherent information, whereas in the case of classical communication, we obtain the regularization of Holevo information. Due to the regularization, which involves a limit of an arbitrarily large number of uses of the channel, the quantum capacity is in general difficult to compute.
	
	We show below in Section~\ref{sec-qcomm_additivity} that the coherent information is always \textit{superadditive}, meaning that $I^c(\mathcal{N}^{\otimes n})\geq nI^c(\mathcal{N})$ for every channel $\mathcal{N}$. This means that the coherent information is always a lower bound on the quantum capacity of a channel $\mathcal{N}$:
	\begin{equation}
		Q(\mathcal{N})\geq I^c(\mathcal{N})\text{ for every channel }\mathcal{N}.
	\end{equation}
	If the coherent information happens to be additive for a particular channel, then the regularization in \eqref{eq-quantum_capacity} is not required. The coherent information is known to be additive for
 all degradable and anti-degradable channels. (See Definition \ref{def-deg_antideg_chan}.) In fact, for anti-degradable channels, the coherent information is equal to zero, a fact that we prove in Section~\ref{subsec-qcomm_anti_deg} below. Examples of degradable and anti-degradable channels include the amplitude damping channel (Section~\ref{sec:QM-over:amp-damp-ch}) and the quantum erasure channel (Section~\ref{sec:QM-over:qudit-erasure}).
	For all such channels, we thus have $Q(\mathcal{N})=I^c(\mathcal{N})$.
	
	Also, just as with classical communication, in this case Theorem \ref{thm-qcomm_capacity} only makes a statement about the quantum capacity and not about the strong converse quantum capacity. One way of attempting to prove that the coherent information of a channel is equal to its strong converse quantum capacity involves proving that the sandwiched R\'{e}nyi coherent information is additive for the channel. Unfortunately, this quantity has not been shown to be additive for \textit{any} quantum channel thus far, which means that this approach is not known to be useful for obtaining strong converse quantum capacities. We consider another approach to strong converse quantum capacities in Section~\ref{sec:Q-cap:Rains-info-SC-bnd}, which leads to a strong converse theorem for dephasing channels (see \eqref{eq:QM-over:d-dim-dephase}). In terms of a general statement about the converse, the best we can  say generally is that the regularized coherent information is a weak converse rate for all quantum channels. 
	
	There are two ingredients to the proof of Theorem \ref{thm-qcomm_capacity}:
	\begin{enumerate}
		\item\textit{Achievability}: We show that $I_{\text{reg}}^c(\mathcal{N})$ is an achievable rate, which involves explicitly constructing a quantum communication protocol. The developments in Section~\ref{subsec-qcomm_one_shot_lower_bound} on a lower bound for one-shot quantum capacity can be used (via the substitution $\mathcal{N}\to\mathcal{N}^{\otimes n}$) to argue for the existence of a quantum communication protocol for $\mathcal{N}$ in the asymptotic setting at the rate $I_{\text{reg}}^c(\mathcal{N})$.
		
			The achievability part of the proof establishes that $Q(\mathcal{N})\geq I_{\text{reg}}^c(\mathcal{N})$.
		
		\item\textit{Weak Converse}: We show that $I_{\text{reg}}^c(\mathcal{N})$ is a weak converse rate, from which it follows that $Q(\mathcal{N})\leq I_{\text{reg}}^c(\mathcal{N})$. To show that $I_{\text{reg}}^c(\mathcal{N})$ is a weak converse rate, we use the one-shot upper bounds from Section~\ref{sec-q_comm_one_shot_upper_bounds} to conclude that every achievable rate $R$ satisfies $R\leq I_{\text{reg}}^c(\mathcal{N})$.
		
	\end{enumerate}
	
	We first establish in Section \ref{sec-qcomm_ach} that the quantity $I_{\text{reg}}^c(\mathcal{N})$ is an achievable rate for quantum communication over $\mathcal{N}$. Then, in Section \ref{sec-qcomm_weak_conv}, we prove that $I_{\text{reg}}^c(\mathcal{N})$ is a weak converse rate. 

\subsection{Proof of Achievability}\label{sec-qcomm_ach}

	In this section, we prove that $I_{\text{reg}}^c(\mathcal{N})$ is an achievable rate for quantum communication over a channel $\mathcal{N}$.
	
	First, recall from Corollary~\ref{cor-q_comm_one_shot_LB_alt} that for all $\varepsilon\in(0,1)$, $\delta\in(0,1)$, $\eta\in[0,\sfrac{\varepsilon\sqrt{\delta}}{4})$, and $\alpha>1$, there exists a $(d,\varepsilon)$ quantum communication protocol for $\mathcal{N}_{A\to B}$ with 
	\begin{multline}\label{eq-q_comm_one_shot_LB_renyi_2}
		\log_2 d\geq \sup_{\psi_{AA'}}\widetilde{H}_{\alpha}(A|E)_{\phi} -\frac{1}{\alpha-1}\log_2\!\left(\frac{1}{f(\varepsilon,\delta,\eta)}\right)\\
			-\log_2\!\left(\frac{1}{1-f(\varepsilon,\delta,\eta)}\right)+\log_2(\eta^4(1-\delta)),
		\end{multline}
		 where $f(\varepsilon,\delta,\eta)\coloneqq \left(\frac{\varepsilon\sqrt{\delta}}{4}-\eta\right)^2$,
	 $\phi_{AE}=\mathcal{N}_{A'\to E}^c(\psi_{AA'})$, and $\psi_{AA'}$ is a pure state with the dimension of $A'$ equal to the dimension of $A$. Note that
	\begin{equation}\label{eq-q_comm_petz_cond_entr}
		\widetilde{H}_{\alpha}(A|E)_{\phi}=-\inf_{\sigma_E}\widetilde{D}_{\alpha}(\phi_{AE}\Vert\mathbbm{1}_A\otimes\sigma_E),
	\end{equation}
	where the optimization is with respect to states $\sigma_E$. A simple corollary of \eqref{eq-q_comm_one_shot_LB_renyi_2} is the following.
	
	\begin{corollary*}{Lower Bound for Quantum Communication in the Asymptotic Setting}{cor-q_comm_asymp_UB}
		Let $\mathcal{N}_{A\to B}$ be a quantum channel. For all $n\in\mathbb{N}$, $\varepsilon\in(0,1)$, and $\alpha>1$, there exists an $(n,d,\varepsilon)$ quantum communication protocol for $\mathcal{N}$ with
		\begin{multline}\label{eq-q_comm_asymp_UB_renyi}
			\frac{\log_2 d}{n}\geq \sup_{\psi_{AA'}}\widetilde{H}_{\alpha}(A|E)_{\phi}-\frac{1}{n(\alpha-1)}\log_2\!\left(\frac{128}{\varepsilon^2}\right)\\-\frac{1}{n}\log_2\!\left(\frac{1}{1-\frac{\varepsilon^2}{128}}\right)-\frac{4}{n} \log_2\!\left(\frac{1}{\varepsilon}\right)-\frac{15}{n},
		\end{multline}
		where $\phi_{AE}=\mathcal{N}_{A'\to E}^c(\psi_{AA'})$ and $\psi_{AA'}$ is a pure state with the dimension of $A'$ equal to the dimension of $A$.
	\end{corollary*}

	\begin{Proof}
		Applying the inequality in \eqref{eq-q_comm_one_shot_LB_renyi_2} to the channel $\mathcal{N}^{\otimes n}$, letting $\delta=\sfrac{1}{2}$, and letting $\eta=\frac{\varepsilon}{8\sqrt{2}}$ leads to
		\begin{multline}\label{eq-q_comm_asymp_UB_renyi_pf1}
			\frac{\log_2 d}{n}\geq \sup_{\Psi_{A^nA^{'n}}}H(A^n|E^n)_{\Phi}-\frac{1}{n(\alpha-1)}\log_2\!\left(\frac{128}{\varepsilon^2}\right)\\-\frac{1}{n}\log_2\!\left(\frac{1}{1-\frac{\varepsilon^2}{128}}\right)-\frac{4}{n} \log_2\!\left(\frac{1}{\varepsilon}\right)-\frac{15}{n},
		\end{multline}
		where $\Phi_{A^nE^n}=(\mathcal{N}_{A'\to E}^c)^{\otimes n}(\Psi_{A^nA^{'n}})$, and the optimization is with respect to pure states $\Psi_{A^nA^{'n}}$ with $d_A=d_{A'}$. Now, let $\psi_{AA'}$ be an arbitrary pure state, and let $\phi_{AE}=\mathcal{N}_{A'\to E}^c(\psi_{AA'})$. Then, restricting the optimization in the definition of $\widetilde{H}_{\alpha}(A^n|E^n)_{\phi^{\otimes n}}$ to product states (see \eqref{eq-q_comm_petz_cond_entr}) leads to
		\begin{equation}
			\widetilde{H}_{\alpha}(A^n|E^n)_{\phi^{\otimes n}}\geq n \widetilde{H}_{\alpha}(A|E)_{\phi}.
		\end{equation}
		In other words,
		\begin{equation}
			\sup_{\Psi_{A^nA^{'n}}}\widetilde{H}_{\alpha}(A^n|E^n)_{\Phi}\geq \widetilde{H}_{\alpha}(A^n|E^n)_{\phi^{\otimes n}}\geq n\widetilde{H}_{\alpha}(A|E)_{\phi}
		\end{equation}
		for every pure state $\psi_{AA'}$, which implies that
		\begin{equation}
			\sup_{\Psi_{A^nA^{'n}}}\widetilde{H}_{\alpha}(A^n|E^n)_{\Phi}\geq n\sup_{\psi_{AA'}}\widetilde{H}_{\alpha}(A|E)_{\phi}.
		\end{equation}
		Therefore, the inequality in \eqref{eq-q_comm_asymp_UB_renyi_pf1} simplifies to \eqref{eq-q_comm_asymp_UB_renyi}, as required.
	\end{Proof}
	
	The inequality in \eqref{eq-q_comm_asymp_UB_renyi} implies that
	\begin{multline}
		Q^{n,\varepsilon}(\mathcal{N})\geq\sup_{\psi_{AA'}}\widetilde{H}_{\alpha}(A|E)_{\phi}-\frac{1}{n(\alpha-1)}\log_2\!\left(\frac{128}{\varepsilon^2}\right)\\-\frac{1}{n}\log_2\!\left(\frac{1}{1-\frac{\varepsilon^2}{128}}\right)-\frac{4}{n} \log_2\!\left(\frac{1}{\varepsilon}\right)-\frac{15}{n},
	\end{multline}
	for all $n\geq 1$, $\varepsilon\in(0,1)$, and $\alpha>1$, where $d_{A'}=d_{A}$ and $\phi_{AE}=\mathcal{N}_{A'\to E}^c(\psi_{AA'})$.
	
	We can now use \eqref{eq-q_comm_asymp_UB_renyi} to prove that the regularized coherent information $I_{\text{reg}}^c(\mathcal{N})$ is an achievable rate for quantum communication over $\mathcal{N}$.

\subsubsection*{Proof of the Achievability Part of Theorem~\ref{thm-qcomm_capacity}}

	Fix $\varepsilon\in(0,1]$ and $\delta>0$. Let $\delta_1,\delta_2>0$ be such that
	\begin{equation}\label{eq-q_comm_ach_pf1}
		\delta=\delta_1+\delta_2.
	\end{equation}
	Set $\alpha\in(1,\infty)$ such that
	\begin{equation}\label{eq-q_comm_ach_pf2}
		\delta_1\geq I^c(\mathcal{N})-\sup_{\psi_{AA'}}\widetilde{H}_{\alpha}(A|E)_{\phi},
	\end{equation}
	where $\psi_{AA'}$ is a pure state with the dimension of~$A'$ equal to the dimension of~$A$ and $\phi_{AE}=\mathcal{N}_{A'\to E}^c(\psi_{AA'})$. Note that this is possible because $\widetilde{H}_{\alpha}(A|E)_{\phi}$ increases monotonically with decreasing $\alpha$ (this follows from Proposition~\ref{prop-Petz_rel_ent}), so that
	\begin{align}
		\lim_{\alpha\to 1^+}\sup_{\psi_{AA'}}\widetilde{H}_{\alpha}(A|E)_{\phi}&=\sup_{\alpha\in(1,\infty)}\sup_{\psi_{AA'}}\widetilde{H}_{\alpha}(A|E)_{\phi}\\
		&=\sup_{\alpha\in(1,\infty)}\sup_{\psi_{AA'}}\left(-\inf_{\sigma_E}\widetilde{D}_{\alpha}(\phi_{AE}\Vert\mathbbm{1}_A\otimes\sigma_E)\right)\\
		&=-\inf_{\alpha\in(1,\infty)}\inf_{\psi_{AA'}}\inf_{\sigma_E}\widetilde{D}_{\alpha}(\phi_{AE}\Vert\mathbbm{1}_A\otimes\sigma_E)\\
		&=-\inf_{\psi_{AA'}}\inf_{\sigma_E}\inf_{\alpha\in(1,\infty)}\widetilde{D}_{\alpha}(\phi_{AE}\Vert\mathbbm{1}\otimes\sigma_E)\\
		&=-\inf_{\psi_{AA'}}\inf_{\sigma_E}D(\phi_{AE}\Vert\mathbbm{1}_A\otimes\sigma_E)\\
		&=\sup_{\psi_{AA'}}\left(-\inf_{\sigma_E}D(\phi_{AE}\Vert\mathbbm{1}_A\otimes\sigma_E)\right)\\
		&=\sup_{\psi_{AA'}}H(A|E)_{\phi},
	\end{align}
	where the fifth equality follows from Proposition~\ref{prop-petz_rel_ent_lim_1}. Now, let $\mathcal{V}^{\mathcal{N}}_{A'\to BE}$ be an isometric channel extending $\mathcal{N}_{A'\to B}$ such that $\phi_{AE}=\mathcal{N}_{A'\to E}^c(\psi_{AA'})=\Tr_B[\mathcal{V}^{\mathcal{N}}_{A'\to BE}(\psi_{AA'})]$. Since the state $\mathcal{V}^{\mathcal{N}}_{A'\to BE}(\psi_{AA'})$ is pure, we can view it as a purification of $\omega_{AB}=\mathcal{N}_{A'\to B}(\psi_{AA'})$, so that
	\begin{align}
		H(A|E)_{\phi}&=H(AE)_{\phi}-H(E)_{\phi}\\
		&=H(B)_{\omega}-H(AB)_{\omega}\\
		&=I(A\rangle B)_{\omega}
	\end{align}
	for every pure state $\psi_{AA'}$. Therefore,
	\begin{equation}
		\sup_{\psi_{AA'}}H(A|E)_{\phi}=\sup_{\psi_{AA'}}I(A\rangle B)_{\omega}=I^c(\mathcal{N}).
	\end{equation}
	
	With $\alpha\in(1,\infty)$ chosen such that \eqref{eq-q_comm_ach_pf2} holds, take $n$ large enough so that
	\begin{equation}\label{eq-q_comm_ach_pf3}
		\delta_2\geq \frac{1}{n(\alpha-1)}\log_2\!\left(\frac{128}{\varepsilon^2}\right)+\frac{1}{n}\log_2\!\left(\frac{1}{1-\frac{\varepsilon^2}{128}}\right)+\frac{4}{n} \log_2\!\left(\frac{1}{\varepsilon}\right)+\frac{15}{n}.
	\end{equation}
	Now, we use the fact that for the $n$ and $\varepsilon$ chosen above there exists an $(n,d,\varepsilon)$ protocol such that
	\begin{multline}
		\frac{\log_2 d}{n}\geq \sup_{\psi_{AA'}}\widetilde{H}_{\alpha}(A|E)_{\phi}-\frac{1}{n(\alpha-1)}\log_2\!\left(\frac{128}{\varepsilon^2}\right)\\-\frac{1}{n}\log_2\!\left(\frac{1}{1-\frac{\varepsilon^2}{128}}\right)-\frac{4}{n} \log_2\!\left(\frac{1}{\varepsilon}\right)-\frac{15}{n},
	\end{multline}
	which holds due to Corollary~\ref{cor-q_comm_asymp_UB}. Rearranging the right-hand side of this inequality, and using \eqref{eq-q_comm_ach_pf1}, \eqref{eq-q_comm_ach_pf2}, and \eqref{eq-q_comm_ach_pf3}, we find that
	\begin{align}
		\frac{\log_2 d}{n}&\geq I^c(\mathcal{N})-\left(I^c(\mathcal{N})-\sup_{\psi_{AA'}}\widetilde{H}_{\alpha}(A|E)_{\phi}+\frac{1}{n(\alpha-1)}\log_2\!\left(\frac{128}{\varepsilon^2}\right)
		\right.\nonumber\\
		&\qquad\qquad\qquad\qquad\left.+\frac{1}{n}\log_2\!\left(\frac{1}{1-\frac{\varepsilon^2}{128}}\right)+\frac{4}{n} \log_2\!\left(\frac{1}{\varepsilon}\right)+\frac{15}{n}\right)\\
		&\geq I^c(\mathcal{N})-(\delta_1+\delta_2)\\
		&=I^c(\mathcal{N})-\delta.
	\end{align}
	Thus, there exists an $(n,d,\varepsilon)$ quantum communication protocol with rate $\frac{\log_2 d}{n}\geq I^c(\mathcal{N})-\delta$. Therefore, there exists an $(n,2^{n(R-\delta)},\varepsilon)$ quantum communication protocol with $R=I^c(\mathcal{N})$ for all sufficiently large $n$ such that \eqref{eq-q_comm_ach_pf3} holds. Since $\varepsilon$ and $\delta$ are arbitrary, we conclude that for all $\varepsilon\in(0,1]$, $\delta>0$, and sufficiently large $n$, there exists an $(n,2^{n(I^c(\mathcal{N})-\delta)},\varepsilon)$ quantum communication protocol. This means, by definition, that $I^c(\mathcal{N})$ is an achievable rate for quantum communication over $\mathcal{N}$.
	
	Now, we can repeat the arguments above for the tensor-power channel $\mathcal{N}^{\otimes k}$ for all $k\geq 1$, and so we conclude that $\frac{1}{k}I^c(\mathcal{N}^{\otimes k})$ is an achievable rate (the arguments are similar to the arguments in the proof of the achievability part of Proposition~\ref{thm-distillable_entanglement}). Since $k$ is arbitrary, we conclude that $\lim_{k\to\infty}\frac{1}{k}I^c(\mathcal{N}^{\otimes k})=I_{\text{reg}}^c(\mathcal{N})$ is an achievable rate for quantum communication over $\mathcal{N}$.  \qedsymbol


\subsection{Proof of the Weak Converse}\label{sec-qcomm_weak_conv}

	We now show that the regularized coherent information $I_{\text{reg}}^c(\mathcal{N})$ is a weak converse rate for quantum communication over $\mathcal{N}$. This establishes that $Q(\mathcal{N})\allowbreak\leq I_{\text{reg}}^c(\mathcal{N})$ and therefore that $Q(\mathcal{N})=I_{\text{reg}}^c(\mathcal{N})$, completing the proof of Theorem \ref{thm-qcomm_capacity}.
	
\begin{figure}
		\centering
		\includegraphics[scale=0.8]{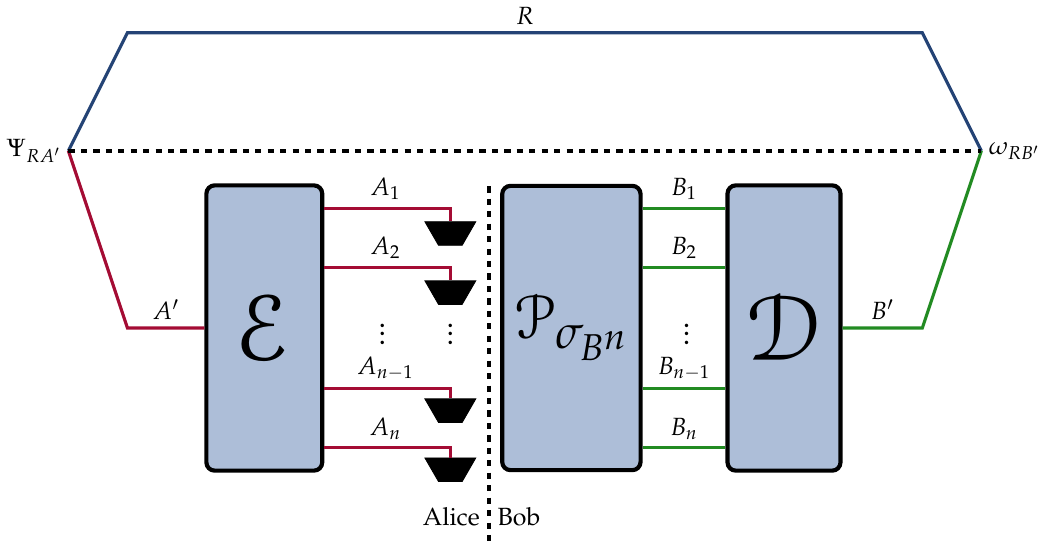}
		\caption{Depiction of a protocol that is useless for quantum communication in the asymptotic setting. The state encoding Alice's share of the pure state $\Psi_{RA'}$ is discarded and replaced by an arbitrary (but fixed) state~$\sigma_{B^n}$.}\label{fig-qcomm_asymptotic_useless}
	\end{figure}

	Let us first recall from Theorem \ref{thm-qcomm_meta_str_weak_conv} that for every quantum channel $\mathcal{N}$, we have the following: for all $\varepsilon\in[0,1)$ and $(d,\varepsilon)$ quantum communication protocols for $\mathcal{N}$,
	\begin{align}
		(1-2\varepsilon) \log_2 d&\leq I^c(\mathcal{N})+h_2(\varepsilon),\label{eq-qcomm_weak_conv_one_shot_2}\\
		\log_2 d&\leq \widetilde{I}_\alpha^c(\mathcal{N})+\frac{\alpha}{\alpha-1}\log_2\!\left(\frac{1}{1-\varepsilon}\right)\quad\forall~\alpha>1.\label{eq-qcomm_str_conv_one_shot_2}
	\end{align}
	To obtain these inequalities, we considered a quantum communication protocol for a useless channel. The useless channel in the asymptotic setting is analogous to the one in Figure \ref{fig-q_comm_oneshot_useless} and is shown in Figure \ref{fig-qcomm_asymptotic_useless}. Applying \eqref{eq-qcomm_weak_conv_one_shot_2} and \eqref{eq-qcomm_str_conv_one_shot_2} to the channel $\mathcal{N}^{\otimes n}$ leads to the following.
	
	\begin{corollary*}{Upper Bounds for Quantum Communication in the Asymptotic Setting}{cor-qcomm_str_weak_conv_upper}
		Let $\mathcal{N}$ be a quantum channel. For all $\varepsilon\in[0,1)$, $n\in\mathbb{N}$, and $(n,d,\varepsilon)$ quantum communication protocols over $n$ uses of $\mathcal{N}$, the number of transmitted qubits is bounded from above as follows:
		\begin{align}
			(1-2\varepsilon) \frac{\log_2 d}{n}&\leq \frac{1}{n}I^c(\mathcal{N}^{\otimes n})+\frac{1}{n}h_2(\varepsilon),\label{eq-qcomm_weak_conv_n_shot}\\
			\frac{\log_2 d}{n}&\leq \frac{1}{n}\widetilde{I}_{\alpha}^c(\mathcal{N}^{\otimes n})+\frac{\alpha}{n(\alpha-1)}\log_2\!\left(\frac{1}{1-\varepsilon}\right)\quad\forall~\alpha>1.\label{eq-qcomm_str_conv_n_shot}
		\end{align}
	\end{corollary*}
	
	\begin{Proof}
		Since the inequalities in \eqref{eq-qcomm_weak_conv_one_shot_2} and \eqref{eq-qcomm_str_conv_one_shot_2} of Theorem~\ref{thm-qcomm_meta_str_weak_conv} hold for every channel $\mathcal{N}$, they hold for the channel $\mathcal{N}^{\otimes n}$. Therefore, applying \eqref{eq-qcomm_weak_conv_one_shot_2} and \eqref{eq-qcomm_str_conv_one_shot_2} to $\mathcal{N}^{\otimes n}$ and dividing both sides by $n$, we obtain the desired result.
	\end{Proof}
	
	The inequalities in the  corollary above give us, for all $\varepsilon\in[0,1)$ and $n\in\mathbb{N}$, an upper bound on the rate of an arbitrary $(n,d,\varepsilon)$ quantum communication protocol. If instead we fix a particular rate $R$ by letting $d=2^{nR}$, then we can obtain a lower bound on the error probability of an $(n,2^{nR},\varepsilon)$ quantum communication protocol. Specifically, using \eqref{eq-qcomm_str_conv_n_shot}, we find that
	\begin{equation}
		\varepsilon\geq 1-2^{-n\left(\frac{\alpha-1}{\alpha}\right)\left(R-\frac{1}{n}\widetilde{I}_{\alpha}^c(\mathcal{N}^{\otimes n})\right)}
	\end{equation}
	for all $\alpha>1$.
	
	The inequalities in \eqref{eq-qcomm_weak_conv_n_shot} and \eqref{eq-qcomm_str_conv_n_shot} imply that 
	\begin{align}
		(1-2\varepsilon) Q^{n,\varepsilon}(\mathcal{N})&\leq\frac{1}{n}I^c(\mathcal{N}^{\otimes n})+\frac{1}{n}h_2(\varepsilon),\\
		Q^{n,\varepsilon}(\mathcal{N})&\leq\frac{1}{n}\widetilde{I}_{\alpha}^c(\mathcal{N}^{\otimes n})+\frac{\alpha}{n(\alpha-1)}\log_2\!\left(\frac{1}{1-\varepsilon}\right)\quad\forall~\alpha>1,
	\end{align}
	with $n\geq 1$ and $\varepsilon\in[0,1)$.
	
	Using \eqref{eq-qcomm_weak_conv_n_shot}, we can now prove the weak converse part of Theorem \ref{thm-qcomm_capacity}.

\subsubsection*{Proof of the Weak Converse Part of Theorem \ref{thm-qcomm_capacity}}

	Suppose that $R$ is an achievable rate. Then, by definition, for all $\varepsilon\in(0,1]$, $\delta>0$, and  sufficiently large $n$, there exists an $(n,2^{n(R-\delta)},\varepsilon)$ quantum communication protocol for $\mathcal{N}$. For all such protocols, the inequality \eqref{eq-qcomm_weak_conv_n_shot} in Corollary~\ref{cor-qcomm_str_weak_conv_upper} holds, so that
	\begin{equation}
		(1-2\varepsilon) (R-\delta)\leq\frac{1}{n}I^c(\mathcal{N}^{\otimes n})+\frac{1}{n}h_2(\varepsilon).
	\end{equation}
	Since this bound holds for all sufficiently large $n$, it holds in the limit $n\to\infty$, so that
	\begin{align}
		(1-2\varepsilon)R&\leq \lim_{n\to\infty}\left(\frac{1}{n}I^c(\mathcal{N}^{\otimes n})+\frac{1}{n}h_2(\varepsilon)\right)+(1-2\varepsilon)\delta,\\
		&= \lim_{n\to\infty}\frac{1}{n}I^c(\mathcal{N}^{\otimes n})+(1-2\varepsilon)\delta.
	\end{align}
	Then, since this inequality holds for all $\varepsilon\in(0,\sfrac{1}{2})$ and $\delta>0$, we obtain
	\begin{equation}
		R\leq \lim_{\varepsilon,\delta\to 0}\left\{\frac{1}{1-2\varepsilon}\lim_{n\to\infty}\frac{1}{n}I^c(\mathcal{N})+\delta\right\}=\lim_{n\to\infty}\frac{1}{n}I^c(\mathcal{N}^{\otimes n})=I_{\text{reg}}^c(\mathcal{N}).
	\end{equation}
	We have thus shown that if $R$ is an achievable rate, then $R\leq I_{\text{reg}}^c(\mathcal{N})$. The contrapositive of this statement is that if $R>I_{\text{reg}}^c(\mathcal{N})$, then $R$ is not an achievable rate. By definition, therefore, $I_{\text{reg}}^c(\mathcal{N})$ is a weak converse rate.

\subsection{The Additivity Question}\label{sec-qcomm_additivity}

	Although we have shown that the quantum capacity $Q(\mathcal{N})$ of a channel $\mathcal{N}$ is given by its regularized coherent information $I_{\text{reg}}^c(\mathcal{N})=\lim_{n\to\infty}\frac{1}{n}I^c(\mathcal{N}^{\otimes n})$, without the additivity of $I^c(\mathcal{N})$, this result is not particularly helpful since it is not clear whether the regularized coherent information can be computed in general.
	
	The coherent information is always \textit{superadditive}, meaning that for two arbitrary quantum channels $\mathcal{N}_1$ and $\mathcal{N}_2$,
	\begin{equation}\label{eq-coh_inf_chan_superadditive}
		I^c(\mathcal{N}_1\otimes\mathcal{N}_2)\geq I^c(\mathcal{N}_1)+I^c(\mathcal{N}_2).
	\end{equation}
	This follows from the fact that coherent information is additive for product states $\tau_{A_1B_1}\otimes\omega_{A_2B_2}$:
	\begin{equation}
		I(A_1A_2\rangle B_1B_2)_{\tau\otimes\omega}=I(A_1\rangle B_1)_{\tau}+I(A_2\rangle B_2)_{\omega},
	\end{equation}
	which is a consequence of \eqref{eq:QEI:coh-info-1st-def} and the additivity of entropy for product states (see \eqref{eq-quantum_entropy_additive}).
	
	Now, let $\psi_{R_1R_2A_1A_2}$, $\phi_{R_1A_1}$, $\varphi_{R_2A_2}$ be arbitrary pure states, where $A_1$ and $A_2$ are input systems to the channels $\mathcal{N}_1$ and $\mathcal{N}_2$, respectively, and $d_{R_1}=d_{A_1}$ and $d_{R_2}=d_{A_2}$. Then, letting
	\begin{align}
		\rho_{R_1R_2B_1B_2}\coloneqq((\mathcal{N}_1)_{A_1\to B_1}\otimes(\mathcal{N}_2)_{A_2\to B_2})(\psi_{R_1R_2A_1A_2}),\\
		\tau_{R_1B_1}\coloneqq (\mathcal{N}_1)_{A_1\to B_1}(\phi_{R_1A_1}),\\
		\omega_{R_2B_2}\coloneqq(\mathcal{N}_2)_{A_2\to B_2}(\varphi_{R_2A_2}),
	\end{align}
	and restricting the optimization in the definition of coherent information of a channel to pure product states, we find that
	\begin{align}
		I^c(\mathcal{N}_1\otimes\mathcal{N}_2)&=\sup_{\psi_{R_1R_2A_1A_2}}I(R_1R_2\rangle B_1B_2)_{\rho} \label{eq:Q-cap:coh-info-superadd-1}\\
		&\geq \sup_{\phi_{R_1A_1}\otimes\varphi_{R_2A_2}}I(R_1R_2\rangle B_1B_2)_{\tau\otimes\omega}\\
		&=\sup_{\phi_{R_1A_1}\otimes\varphi_{R_2A_2}}\{I(R_1\rangle A_1)_{\tau}+I(R_2\rangle B_2)_{\omega}\}\\
		&=\sup_{\phi_{R_1A_1}}I(R_1\rangle B_1)_{\tau}+\sup_{\varphi_{R_2A_2}}I(R_2\rangle B_2)_{\omega}\\
		&=I^c(\mathcal{N}_1)+I^c(\mathcal{N}_2),
		\label{eq:Q-cap:coh-info-superadd-last}
	\end{align}
	which is precisely \eqref{eq-coh_inf_chan_superadditive}. The reverse inequality does not hold in general, but it does for degradable channels (see Section~\ref{subsec-qcomm_deg} below).
	
	For the sandwiched R\'{e}nyi coherent information of a bipartite state $\rho_{AB}$, which is defined as
	\begin{equation}
		\widetilde{I}_{\alpha}^c(A\rangle B)_{\rho}=\inf_{\sigma_B}\widetilde{D}_{\alpha}(\rho_{AB}\Vert\mathbbm{1}_A\otimes\sigma_B),
	\end{equation}
	where the optimization is over states $\sigma_B$, the following additivity equality holds for all product states $\tau_{A_1B_1}\otimes\omega_{A_2B_2}$ and $\alpha \in (1,\infty)$:
	\begin{equation}
		\widetilde{I}_{\alpha}(A_1A_2\rangle B_1B_2)_{\tau\otimes\omega}= \widetilde{I}_{\alpha}(A_1\rangle B_1)_{\tau}+\widetilde{I}_{\alpha}(A_2\rangle B_2)_{\omega}.
	\end{equation}
	This equality follows by reasoning similar to that given for the proof of Proposition~\ref{prop-sand_rel_mut_inf_additive}. 
	By the same reasoning given in \eqref{eq:Q-cap:coh-info-superadd-1}--\eqref{eq:Q-cap:coh-info-superadd-last}, we conclude that
	\begin{equation}
	I^c_{\alpha}(\mathcal{N}_1\otimes\mathcal{N}_2) \geq 
	I^c_{\alpha}(\mathcal{N}_1)+I^c_{\alpha}(\mathcal{N}_2)
	\end{equation}
	for all $\alpha \in (1,\infty)$, where $I^c_{\alpha}(\mathcal{N})$ is the sandwiched R\'enyi coherent information of the channel $I^c_{\alpha}(\mathcal{N})$.
	Whether the reverse inequality holds, even for particular classes of channels, is an open question.

\subsection{Rains Information Strong Converse Upper Bound}\label{sec:Q-cap:Rains-info-SC-bnd}

	Except for channels for which the coherent information is known to be additive (such as the class of degradable channels; see Section~\ref{subsec-qcomm_deg} below), the quantum capacity of a channel is difficult to compute. This prompts us to find tractable upper bounds on quantum capacity. This search for tractable upper bounds is entirely analogous to what was done in Section \ref{subsec-cc_general_upper_bounds} for classical communication in order to obtain tractable strong converse upper bounds on classical capacity.

	Recall that in the previous chapter on entanglement distillation, our approach to obtaining strong converse upper bounds on distillable entanglement consisted of comparing the state at the output of an entanglement distillation protocol with one that is useless for entanglement distillation. We considered the set of  $\PPT'$ operators as the useless set, and we obtained state entanglement measures as upper bounds in the one-shot and asymptotic settings. Now, observe that entanglement transmission is similar to entanglement distillation in the sense that, like entanglement distillation, the error criterion for entanglement transmission involves comparing the output state of the protocol to the maximally entangled state. This suggests that the state entanglement measures defined in Section~\ref{sec-ent_measures_Rains_dist}, and in particular the results of Proposition~\ref{prop:core-meta-converse-privacy_a} and Corollary~\ref{cor-ent_distill_one-shot_UB_alt}, are relevant. However, the main resource that we are considering in this chapter is a quantum channel and not a quantum state, and so we have an extra degree of freedom in the input state to the channel, which we can optimize. This suggests that the channel entanglement measures from Chapter~\ref{chap-ent_measures_chan} are relevant, and this is indeed what we find. 
	
	\begin{proposition}{prop-qcomm:one-shot-bound-meta_Rains}
		Let $\mathcal{N}_{A\to B}$ be a quantum channel. For an arbitrary $(d,\varepsilon)$ quantum communication protocol for $\mathcal{N}_{A\to B}$, the number $\log_2 d$ of qubits transmitted over $\mathcal{N}$ satisfies
		\begin{equation}
			\log_2 d\leq R_H^{\varepsilon}(\mathcal{N}),
		\end{equation}
		where 
		\begin{align}
			R_H^{\varepsilon}(\mathcal{N})&=\sup_{\psi_{SA}}R_H^{\varepsilon}(S;B)_{\omega}\\
			&=\sup_{\psi_{SA}}\inf_{\sigma_{SB}\in\PPT'(S:B)}D_H^{\varepsilon}(\mathcal{N}_{A\to B}(\psi_{SA})\Vert\sigma_{SB})
		\end{align}
		is the $\varepsilon$-hypothesis testing Rains information of the channel $\mathcal{N}$, defined in \eqref{eq-hypo_test_rains_inf_chan}. Consequently, the one-shot quantum capacity does not exceed the $\varepsilon$-hypothesis testing Rains information of the channel $\mathcal{N}$:
		\begin{equation}
			Q^{\varepsilon}(\mathcal{N})\leq R_H^{\varepsilon}(\mathcal{N}).
		\end{equation}
	\end{proposition}
	
	\begin{remark}
		Note that in the expression for $R_H^{\varepsilon}(\mathcal{N})$ above it suffices to optimize over pure states $\psi_{RA}$, with the dimension of $S$ equal to the dimension of $A$. We showed this in \eqref{eq-ent_meas_chan_pure_pf1}--\eqref{eq-ent_meas_chan_pure_pf4} immediately after Definition~\ref{def:LAQC-ent-channel}.
	\end{remark}
	
	\begin{Proof}
		By the arguments in \eqref{eq-q_comm_max_error_to_avg_error_1}--\eqref{eq-q_comm_max_error_to_avg_error_3}, a $(d,\varepsilon)$ quantum communication protocol is a $(d,\varepsilon)$ entanglement transmission protocol. As such, we conclude that the state $\rho_{RB'}$ defined in \eqref{eq-q_comm_avg_error_meta_conv_bd_pf} satisfies $\Tr[\Phi_{RB'}\rho_{RB'}]\geq 1-\varepsilon$. We can therefore apply Proposition \ref{prop:core-meta-converse-privacy_a} to conclude that
		\begin{equation}\label{eq-qcomm:one-shot-bound-meta_Rains_pf}
			\log_2 d\leq R_H^{\varepsilon}(S;B')_{\rho}.
		\end{equation}
		Note that
		\begin{align}
			&R_H^{\varepsilon}(S;B')_{\rho}\nonumber\\
			&\qquad=\inf_{\sigma_{SB'}\in\text{PPT}'(S:B')}D_H^{\varepsilon}(\rho_{SB'}\Vert\sigma_{SB'})\\
			&\qquad=\inf_{\sigma_{SB'}\in\text{PPT}'(S:B')}D_H^{\varepsilon}((\mathcal{D}_{B\to B'}\circ\mathcal{N}_{A\to B}\circ\mathcal{E}_{A'\to A})(\Phi_{SA'})\Vert\sigma_{SB'}).
		\end{align}
		Now, since every local channel is completely PPT preserving (this follows immediately from Proposition~\ref{prop-LOCC_PPT_preserving} and Lemma~\ref{lemma-transpose-CPTP}), we conclude that the channel $\mathcal{D}_{B\to B'}\equiv\id_{S}\otimes\mathcal{D}_{B\to B'}$ is completely PPT preserving, so that the set
		\begin{equation}
			\{\mathcal{D}_{B\to B'}(\tau_{SB}):\tau_{SB}\in\PPT'(S\!:\!B)\}
		\end{equation}
		is a subset of $\operatorname{PPT}'(S\!:\!B')$. Thus, by restricting the optimization over all operators $\sigma_{SB'}\in\PPT'(S\!:\!B')$ to the outputs $\mathcal{D}_{B\to B'}(\tau_{SB})$ of the decoding channel $\mathcal{D}_{B\to B'}$ acting on operators $\tau_{SB}\in\PPT'(S\!:\!B)$, we obtain
		\begin{align}
			&R_H^{\varepsilon}(S;B')_{\omega}\nonumber\\
			&\quad\leq\inf_{\tau_{SB}\in\PPT'(S:B)}D_H^{\varepsilon}((\mathcal{D}_{B\to B'}\circ\mathcal{N}_{A\to B}\circ\mathcal{E}_{A'\to A})(\Phi_{SA'})\Vert\mathcal{D}_{B\to B'}(\tau_{SB}))\\
			&\quad\leq \inf_{\tau_{SB}\in\PPT'(S:B)}D_H^{\varepsilon}((\mathcal{N}_{A\to B}\circ\mathcal{E}_{A'\to A})(\Phi_{SA'})\Vert\tau_{SB})\\
			&\quad=\inf_{\tau_{SB}\in\PPT'(S:B)}D_H^{\varepsilon}(\mathcal{N}_{A\to B}(\rho_{SA})\Vert\tau_{SB}),
		\end{align}
		where the second inequality follows from the data-processing inequality for hypothesis testing relative entropy, and the equality follows by letting $\rho_{SA}=\mathcal{E}_{A'\to A}(\Phi_{SA'})$. Finally, after optimizing over all states $\rho_{SA}$, we obtain
		\begin{align}
			R_H^{\varepsilon}(S;B')_{\rho}&\leq \sup_{\rho_{SA}}\inf_{\tau_{SB}\in\operatorname{PPT}'(S:B)}D_H^{\varepsilon}(\mathcal{N}_{A\to B}(\rho_{SA})\Vert\tau_{SB})\\
			&=R_H^{\varepsilon}(\mathcal{N}),
		\end{align}
		so that, by \eqref{eq-qcomm:one-shot-bound-meta_Rains_pf}, we conclude that
		\begin{equation}
			\log_2 d\leq R_H^{\varepsilon}(\mathcal{N}),
		\end{equation}
		as required.
	\end{Proof}
	
	The result of Proposition~\ref{prop-qcomm:one-shot-bound-meta_Rains} is analogous to the result of Theorem~\ref{prop-qcomm:one-shot-bound-meta}. Combining it with Proposition \ref{prop:sandwich-to-htre}  leads to the following:
	
	\begin{corollary*}{One-Shot Rains Upper Bound for Quantum Communication}{thm-qcomm_meta_str_weak_conv_Rains}
		Let $\mathcal{N}_{A\to B}$ be a quantum channel, and let $\varepsilon\in[0,1)$. For all $(d,\varepsilon)$ quantum communication protocols over $\mathcal{N}$, we have that
		\begin{equation}\label{eq-qcomm_str_conv_one_shot_Rains}
			\log_2 d\leq \widetilde{R}_{\alpha}(\mathcal{N})+\frac{\alpha}{\alpha-1}\log_2\!\left(\frac{1}{1-\varepsilon}\right)\quad\forall ~\alpha>1,
		\end{equation}
		where 
		\begin{align}
			\widetilde{R}_{\alpha}(\mathcal{N})&=\sup_{\psi_{SA}}\widetilde{R}_{\alpha}(S;B)_{\omega}\\
			&=\sup_{\psi_{SA}}\inf_{\sigma_{SB}\in\PPT'(S:B)}\widetilde{D}_{\alpha}(\mathcal{N}_{A\to B}(\psi_{SA})\Vert\sigma_{SB})
		\end{align}
		is the \textit{sandwiched R\'{e}nyi Rains information of $\mathcal{N}$}, defined in \eqref{eq-sand_Renyi_Rains_inf_chan}.
	\end{corollary*}
	
	\begin{remark}
		Note that in the expression for $\widetilde{R}_{\alpha}(\mathcal{N})$ above it suffices to optimize over pure states $\psi_{SA}$, with the dimension of $S$ equal to the dimension of $A$. We showed this in \eqref{eq-ent_meas_chan_pure_pf1}--\eqref{eq-ent_meas_chan_pure_pf4} immediately after Definition~\ref{def:LAQC-ent-channel}.
	\end{remark}
	
	Since the inequality in \eqref{eq-qcomm_str_conv_one_shot_Rains} holds for all $(d,\varepsilon)$ quantum communication protocols, we conclude the following upper bound on the one-shot quantum capacity:
	\begin{equation}
		Q^{\varepsilon}(\mathcal{N})\leq\widetilde{R}_{\alpha}(\mathcal{N})+\frac{\alpha}{\alpha-1}\log_2\!\left(\frac{1}{1-\varepsilon}\right)
	\end{equation}
	for all $\alpha>1$. 
	
	For $n$ channel uses, the bound in \eqref{eq-qcomm_str_conv_one_shot_Rains} becomes
	\begin{equation}\label{eq-qcomm_str_conv_n_shot_Rains}
		\frac{\log_2 d}{n}\leq \frac{1}{n}\widetilde{R}_\alpha(\mathcal{N}^{\otimes n})+\frac{\alpha}{n(\alpha-1)}\log_2\!\left(\frac{1}{1-\varepsilon}\right)\quad\forall~\alpha>1,
	\end{equation}
	which holds for an arbitrary $(n,d,\varepsilon)$ quantum communication protocol that employs  $n$ uses of the channel $\mathcal{N}$, where $n\geq 1$ and $\varepsilon\in[0,1)$. We can simplify this inequality by making use of the following fact.
	
	\begin{proposition*}{Weak Subadditivity of R\'{e}nyi Rains Information of a Channel}{prop-Renyi_Rains_inf_chan_weak_additive}
		Let $\mathcal{N}_{A\to B}$ be a quantum channel, with $d_A$ the dimension of the input system $A$. For all $\alpha>1$ and $n\in\mathbb{N}$, we have
		\begin{equation}\label{eq-ren_Rains_inf_chan_subadd}
			\widetilde{R}_{\alpha}(\mathcal{N}^{\otimes n})\leq n\widetilde{R}_{\alpha}(\mathcal{N})+\frac{\alpha (d_A^2-1)}{\alpha-1}\log_2(n+1).
		\end{equation}
	\end{proposition*}
	
	\begin{Proof}
		Throughout this proof, for convenience we make use of the alternate notation
		\begin{equation}
			\widetilde{R}_{\alpha}(\rho_{AB})\equiv\widetilde{R}_{\alpha}(A;B)_{\rho}
		\end{equation}
		where $\rho_{AB}$ is a bipartite state.
		
		Let $\psi_{SA^n}$ be an arbitrary pure state, with the dimension of $S$ equal to the dimension of $A^n$, and let $\rho_{A^n}\coloneqq \Tr_{S}[\psi_{SA^n}]$. We start by observing that the channel $\mathcal{N}^{\otimes n}$ is covariant with respect to the symmetric group $\mathcal{S}_n$. In particular, if we let $\{W_{A^n}^{\pi}\}_{\pi\in\mathcal{S}_n}$ and $\{W_{B^n}^{\pi}\}_{\pi\in\mathcal{S}_n}$ be the unitary representations of $\mathcal{S}_n$, defined in \eqref{eq-permutation_rep}, acting on $\mathcal{H}_A^{\otimes n}$ and $\mathcal{H}_B^{\otimes n}$, respectively, then for every state $\rho_{A^n}$, we have that
		\begin{equation}
			\mathcal{N}^{\otimes n}(W_{A^n}^{\pi}\rho_{A^n}W_{A^n}^{\pi\dagger})=W_{B^n}^{\pi}\mathcal{N}^{\otimes n}(\rho_{A^n})W_{B^n}^{\pi\dagger}
		\end{equation}
		for all $\pi\in\mathcal{S}_n$. Consequently, by Proposition~\ref{prop-gen_Rains_inf_chan_cov}, in particular by \eqref{eq-Rains_inf_chan_cov_state_symm}, we find that
		\begin{equation}
			\widetilde{R}_{\alpha}(\mathcal{N}_{A\to B}^{\otimes n}(\psi_{SA^n}))\leq \widetilde{R}_{\alpha}(\mathcal{N}_{A\to B}^{\otimes n}(\psi_{SA^n}^{\overline{\rho}})),
		\end{equation}
		where the state $\overline{\rho}_{A^n}$ is defined as
		\begin{equation}
			\overline{\rho}_{A^n}=\frac{1}{n!}\sum_{\pi\in\mathcal{S}_n}W_{A^n}^{\pi}\rho_{A^n} W_{A^n}^{\pi\dagger},
		\end{equation}
		and $\psi_{SA^n}^{\overline{\rho}}$ is a purification of $\overline{\rho}_{A^n}$.
		
		Since the state $\overline{\rho}_{A^n}$ is permutation invariant by definition, by Lemma~\ref{lem-symm_purif}, it has a purification $\ket{\phi^{\overline{\rho}}}_{\hat{A}^nA^n}\in\text{Sym}_n(\mathcal{H}_{\hat{A}A})$, where the dimension of $\hat{A}$ is equal to the dimension of $A$. Consequently, there exists an isometry $V_{S\to\hat{A}^n}$ such that
		\begin{equation}
			V_{S\to\hat{A}^n}\ket{\psi^{\overline{\rho}}}_{SA^n}=\ket{\phi^{\overline{\rho}}}_{\hat{A}^nA^n}.
		\end{equation}
		Therefore, by isometric invariance of the sandwiched R\'{e}nyi Rains relative entropy, we obtain
		\begin{equation}
			\widetilde{R}_{\alpha}(\mathcal{N}_{A\to B}^{\otimes n}(\psi_{SA^n}^{\overline{\rho}}))=\widetilde{R}_{\alpha}(\mathcal{N}_{A\to B}^{\otimes n}(\phi^{\overline{\rho}}_{\hat{A}^nA^n})).
		\end{equation}
		
		Now, since $\phi^{\overline{\rho}}_{\hat{A}^nA^n}$ is a state, the operator inequality $\phi^{\overline{\rho}}_{\hat{A}^nA^n}\leq\mathbbm{1}_{\hat{A}^nA^n}$ holds. Multiplying both sides of this inequality from the left and right by the projection $\Pi_{\text{Sym}_n(\mathcal{H}_{\hat{A}A})}\equiv\Pi_{\hat{A}^nA^n}$ onto the symmetric subspace of $\mathcal{H}_{\hat{A}A}^{\otimes n}$, we obtain
		\begin{equation}
			\Pi_{\hat{A}^nA^n}\ket{\phi^{\overline{\rho}}}\!\bra{\phi^{\overline{\rho}}}_{\hat{A}^nA^n}\Pi_{\hat{A}^nA^n}\leq \Pi_{\hat{A}^nA^n}^2=\Pi_{\hat{A}^nA^n}.
		\end{equation}
		But $\ket{\phi^{\overline{\rho}}}_{\hat{A}^nA^n}\in\text{Sym}_n(\mathcal{H}_{\hat{A}A})$, which means that
		\begin{equation}
			\Pi_{\hat{A}^nA^n}\ket{\phi^{\overline{\rho}}}\!\bra{\phi^{\overline{\rho}}}_{\hat{A}^nA^n}\Pi_{\hat{A}^nA^n}=\ket{\phi^{\overline{\rho}}}\!\bra{\phi^{\overline{\rho}}}_{\hat{A}^nA^n}.
		\end{equation}
		Therefore,
		\begin{equation}
			\phi^{\overline{\rho}}_{\hat{A}^nA^n}\leq \Pi_{\hat{A}^nA^n}=\binom{d_A^2+n-1}{n}\int\phi_{\hat{A}A}^{\otimes n}~\text{d}\phi,
		\end{equation}
		where the equality follows from \eqref{eq-sym_proj_integral}. Now, note that
		\begin{align}
			\binom{n + d_A^2-1}{n}&=\frac{(n+d_A^2-1)(n+d_A^2-2)\dotsb (n+2)(n+1)}{(d_A^2-1)(d_A^2-2)\dotsb 2\cdot 1}\\
			&=\frac{n+d_A^2-1}{d_A^2-1}\cdot\frac{n+d_A^2-2}{d_A^2-2}\dotsb\frac{n+2}{2}\cdot\frac{n+1}{1}.
		\end{align}
		Then, using the fact that $\frac{n+k}{k}\leq n+1$ for all $k\geq 1$, and applying this inequality to each factor on the right-hand side of the above equation, we obtain
		\begin{equation}
			\binom{n + d_A^2-1}{n}\leq (n+1)^{d_A^2-1}
		\end{equation}
		Therefore,
		\begin{equation}
			\phi^{\overline{\rho}}_{\hat{A}^nA^n}\leq (n+1)^{d_A^2-1}\int \phi_{\hat{A}A}^{\otimes n}~\text{d}\phi\equiv (n+1)^{d_A^2-1}\xi_{\hat{A}^nA^n}.
		\end{equation}
		Next, we use \eqref{eq-sand_ren_ent_ineq_first_arg} to obtain
		\begin{equation}
			\widetilde{R}_{\alpha}(\mathcal{N}_{A\to B}^{\otimes n}(\phi_{\hat{A}^nA^n}^{\overline{\rho}})\leq\widetilde{R}_{\alpha}(\mathcal{N}_{A\to B}^{\otimes n}(\xi_{\hat{A}^nA^n}))+\frac{\alpha}{\alpha-1}\log_2 (n+1)^{d_A^2-1}.
		\end{equation}
		Then, by quasi-convexity of the R\'{e}nyi Rains relative entropy (Proposition~\ref{prop-ren_rains_rel_ent_q_conv}), we obtain
		\begin{equation}
			\widetilde{R}_{\alpha}(\mathcal{N}_{A\to B}^{\otimes n}(\xi_{\hat{A}^nA^n}))\leq \sup_{\phi_{\hat{A}A}}\widetilde{R}_{\alpha}(\mathcal{N}_{A\to B}^{\otimes n}(\phi_{\hat{A}A}^{\otimes n})).
		\end{equation}
		Then, using subadditivity of the sandwiched R\'{e}nyi Rains relative entropy for tensor-product states, as given by \eqref{eq:LAQC-Renyi-Rains-subadditivity}, we find that
		\begin{equation}
			\widetilde{R}_{\alpha}(\mathcal{N}_{A\to B}^{\otimes n}(\phi_{\hat{A}A}^{\otimes n}))\leq n\widetilde{R}_{\alpha}(\mathcal{N}_{A\to B}(\phi_{\hat{A}A})).
		\end{equation}
		Putting everything together, we finally obtain
		\begin{align}
			& \!\!\!\!\!\widetilde{R}_{\alpha}(\mathcal{N}_{A\to B}^{\otimes n}(\psi_{SA^n})) \notag
			\\&\leq n\sup_{\phi_{\hat{A}A}}\widetilde{R}_{\alpha}(\mathcal{N}_{A\to B}(\phi_{\hat{A}A}))+\frac{\alpha (d_A^2-1)}{\alpha-1}\log_2(n+1)\\
			&=n\widetilde{R}_{\alpha}(\mathcal{N})+\frac{\alpha (d_A^2-1)}{\alpha-1}\log_2(n+1).
		\end{align}
		Since the pure state $\psi_{SA^n}$ is arbitrary, we conclude that
		\begin{equation}
			\widetilde{R}_{\alpha}(\mathcal{N}^{\otimes n})=\sup_{\psi_{SA^n}}\widetilde{R}_{\alpha}(\mathcal{N}_{A\to B}^{\otimes n}(\psi_{SA^n}))\leq n\widetilde{R}_{\alpha}(\mathcal{N})+\frac{\alpha (d_A^2-1)}{\alpha-1}\log_2(n+1),
		\end{equation}
		as required.
	\end{Proof}
	
	Combining \eqref{eq-ren_Rains_inf_chan_subadd} with \eqref{eq-qcomm_str_conv_n_shot_Rains} leads to the following upper bound on the rate of an arbitrary $(n,d,\varepsilon)$ quantum communication protocol for a quantum channel $\mathcal{N}_{A\to B}$:
	\begin{equation}\label{eq-qcomm_str_conv_n_shot_Rains_2}
		\frac{\log_2 d}{n}\leq\widetilde{R}_{\alpha}(\mathcal{N})+\frac{\alpha}{n(\alpha-1)}\log_2\!\left(\frac{(n+1)^{d_A^2-1}}{1-\varepsilon}\right)
	\end{equation}
	for all $\alpha>1$. Consequently, the following bound holds for the $n$-shot quantum capacity:
	\begin{equation}
		Q^{n,\varepsilon}(\mathcal{N})\leq\widetilde{R}_{\alpha}(\mathcal{N})+\frac{\alpha}{n(\alpha-1)}\log_2\!\left(\frac{(n+1)^{d_A^2-1}}{1-\varepsilon}\right)
	\end{equation}
	for all $\alpha>1$.
	
	With this bound, we are now ready to state the main result of this section, which is that the Rains information of a channel is an upper bound on the strong converse capacity of an arbitrary quantum channel $\mathcal{N}$.
	
	\begin{theorem*}{Strong Converse Upper Bound on Quantum Capacity}{thm-Rains_inf_strong_conv_upper_bound}
		The Rains information $R(\mathcal{N})$ of a  quantum channel $\mathcal{N}_{A\to B}$ is a strong converse rate for quantum communication over $\mathcal{N}$. In other words, $\widetilde{Q}(\mathcal{N})\leq R(\mathcal{N})$ for every quantum channel $\mathcal{N}$.
	\end{theorem*}
	
	Recall from \eqref{eq-Rains_inf_chan} that
	\begin{equation}
		R(\mathcal{N})=\sup_{\psi_{SA}}\inf_{\sigma_{SB}\in\PPT'(S:B)}D(\mathcal{N}_{A\to B}(\psi_{SA})\Vert\sigma_{SB}),
	\end{equation}
	where the supremum is with respect to pure states $\psi_{SA}$ with $d_S=d_A$.
	
	\begin{Proof}
		Let $\varepsilon\in[0,1)$ and $\delta>0$. Let $\delta_1,\delta_2>0$ be such that
		\begin{equation}\label{eq-qcomm_Rains_inf_str_conv_pf_1}
			\delta>\delta_1+\delta_2\eqqcolon\delta'.
		\end{equation}
		Set $\alpha\in(1,\infty)$ such that
		\begin{equation}\label{eq-qcomm_Rains_inf_str_conv_pf_2}
			\delta_1\geq \widetilde{R}_{\alpha}(\mathcal{N})-R(\mathcal{N}),
		\end{equation}
		which is possible because $\widetilde{R}_{\alpha}(\mathcal{N})$ is monotonically increasing in $\alpha$ (which follows from Proposition~\ref{prop-sand_rel_ent_properties}) and because $\lim_{\alpha\to 1^+}\widetilde{R}_{\alpha}(\mathcal{N})=R(\mathcal{N})$ (see Appendix~\ref{app-sand_ren_inf_limit} for a proof). With this value of $\alpha$, take $n$ large enough so that
		\begin{equation}\label{eq-qcomm_Rains_inf_str_conv_pf_3}
			\delta_2\geq \frac{\alpha}{n(\alpha-1)}\log_2\!\left(\frac{(n+1)^{d_A^2-1}}{1-\varepsilon}\right),
		\end{equation}
		where $d_A$ is the dimension of the input space of the channel $\mathcal{N}$.
		
		Now, with the values of $n$ and $\varepsilon$ as above, an $(n,d,\varepsilon)$ quantum communication protocol satisfies \eqref{eq-qcomm_str_conv_n_shot_Rains_2}, i.e.,
		\begin{equation}
			\frac{\log_2 d}{n}\leq\widetilde{R}_{\alpha}(\mathcal{N})+\frac{\alpha}{n(\alpha-1)}\log_2\!\left(\frac{(n+1)^{d_A^2-1}}{1-\varepsilon}\right),
		\end{equation}
		for all $\alpha>1$. Rearranging the right-hand side of this inequality, and using \eqref{eq-qcomm_Rains_inf_str_conv_pf_1}--\eqref{eq-qcomm_Rains_inf_str_conv_pf_3}, we obtain
		\begin{align}
			\frac{\log_2 d}{n}&\leq R(\mathcal{N})+\widetilde{R}_{\alpha}(\mathcal{N})-R(\mathcal{N})+\frac{\alpha}{n(\alpha-1)}\log_2\!\left(\frac{(n+1)^{d_A^2-1}}{1-\varepsilon}\right)\\
			&\leq R(\mathcal{N})+\delta_1+\delta_2\\
			&=R(\mathcal{N})+\delta'\\
			&<R(\mathcal{N})+\delta.
		\end{align}
		So we have that $R(\mathcal{N})+\delta>\frac{\log_2 d}{n}$ for all $(n,d,\varepsilon)$ quantum communication protocols with sufficiently large $n$. Due to this strict inequality, it follows that there cannot exist an $(n,2^{n(R(\mathcal{N})+\delta)},\varepsilon)$ quantum communication protocol for all sufficiently large $n$ such that \eqref{eq-qcomm_Rains_inf_str_conv_pf_3} holds, for if it did there would exist a $d$ such that $\log_2 d=n(R(\mathcal{N})+\delta)$, which we have just seen is not possible. Since $\varepsilon$ and $\delta$ are arbitrary, we conclude that for all $\varepsilon\in[0,1)$, $\delta>0$, and sufficiently large $n$, there does not exist an $(n,2^{n(R(\mathcal{N})+\delta)},\varepsilon)$ quantum communication protocol. This means that $R(\mathcal{N})$ is a strong converse rate, and thus that $\widetilde{Q}(\mathcal{N})\leq R(\mathcal{N})$.
	\end{Proof}

\subsubsection{The Strong Converse from a Different Point of View}

	Let us now show that the Rains relative entropy of a quantum channel $\mathcal{N}$ is a strong converse rate according to the definition of a strong converse rate in Appendix~\ref{chap-str_conv}. To this end, consider a sequence $\{(n,2^{nr},\varepsilon_n)\}_{n\in\mathbb{N}}$ of $(n,d,\varepsilon)$ quantum communication protocols, with each element of the sequence having an arbitrary (but fixed) rate $r>R(\mathcal{N})$. For each element of the sequence, the inequality in \eqref{eq-qcomm_str_conv_n_shot_Rains_2} holds, which means that
	\begin{equation}
		r\leq \widetilde{R}_{\alpha}(\mathcal{N})+\frac{\alpha}{n(\alpha-1)}\log_2\!\left(\frac{(n+1)^{d_A^2-1}}{1-\varepsilon_n}\right)
	\end{equation}
	for all $\alpha>1$. Rearranging this inequality leads to the following lower bound on the error probabilities $\varepsilon_n$:
	\begin{equation}\label{eq-q_comm_Rains_str_conv_pf_alt}
		\varepsilon_n\geq 1-(n+1)^{d_A^2-1}\cdot 2^{-n\left(\frac{\alpha-1}{\alpha}\right)\left(r-\widetilde{R}_{\alpha}(\mathcal{N})\right)}.
	\end{equation}
	Now, since $r>R(\mathcal{N})$, $\lim_{\alpha\to 1^+}\widetilde{R}_{\alpha}(\mathcal{N})=R(\mathcal{N})$ (see Appendix~\ref{app-sand_ren_inf_limit} for a proof), and since the sandwiched R\'{e}nyi Rains relative entropy is monotonically increasing in $\alpha$ (see Proposition~\ref{prop-sand_rel_ent_properties}), there exists an $\alpha^*>1$ such that $R>\widetilde{R}_{\alpha^*}(\mathcal{N})$. Applying the inequality in \eqref{eq-q_comm_Rains_str_conv_pf_alt} to this value of $\alpha$, we find that
	\begin{equation}\label{eq-q_comm_Rains_str_conv_pf_alt_2}
		\varepsilon_n\geq 1-(n+1)^{d_A^2-1}\cdot 2^{-n\left(\frac{\alpha^*-1}{\alpha^*}\right)\left(r-\widetilde{R}_{\alpha^*}(\mathcal{N})\right)}.
	\end{equation}
	Then, taking the limit $n\to\infty$ on both sides of this inequality, we conclude that $\lim_{n\to\infty}\varepsilon_n\geq 1$. However, $\varepsilon_n\leq 1$ for all $n$ since $\varepsilon_n$ is a probability  by definition. So we conclude that $\lim_{n\to\infty}\varepsilon_n=1$. Since the rate $r>R(\mathcal{N})$ is arbitrary, we conclude that $R(\mathcal{N})$ is a strong converse rate. We also see from \eqref{eq-q_comm_Rains_str_conv_pf_alt_2} that the sequence $\{\varepsilon_n\}_{n\in\mathbb{N}}$ approaches one at an exponential rate.
	
\subsection{Squashed Entanglement Weak Converse Bound}

One of the results of Chapter~\ref{chap-LOCC-QC} is that the squashed entanglement of a quantum channel (see Definition~\ref{def-sq_ent_channel}) is a weak converse rate for quantum communication assisted by LOCC. Since the LOCC-assisted quantum capacity is an upper bound on the unassisted quantum capacity considered in this chapter, we conclude that the squashed entanglement of a channel is a weak converse rate for unassisted quantum communication.

We present some statements along these lines briefly here, indicating that the squashed entanglement gives an upper bound on the one-shot quantum capacity, the $n$-shot quantum capacity, as well as the asymptotic quantum capacity. Complete proofs of these statements are available in Chapter~\ref{chap-LOCC-QC}, and they follow from the fact that the assistance of LOCC can only increase rates of quantum communication. Propositions~\ref{prop:Q-cap:sq-ent-bnd-q-cap} and \ref{prop:Q-cap:sq-ent-bnd-q-cap-n-shot} below are a direct consequence of Theorem~\ref{thm:LAQC-sq-ent-bnd-LOCC-as-cap}, and Theorem~\ref{thm-squashed_ent_weak_conv_upper_bound} below is a consequence of Theorem~\ref{prop-LOCC_q_comm_sq_ent_weak_conv}.

\begin{proposition*}{One-Shot Squashed Entanglement Upper Bound}
{prop:Q-cap:sq-ent-bnd-q-cap}
		Let $\mathcal{N}_{A\rightarrow B}$ be a quantum channel, and let $\varepsilon\in[0,1)$. For all $(d,\varepsilon)$  quantum communication protocols over the channel $\mathcal{N}_{A\rightarrow B}$, the following bound holds%
		\begin{equation}
			\log_{2}d\leq\frac{1}{1-\sqrt{\varepsilon}}\left(   E_{\operatorname{sq}}(\mathcal{N})+g_{2}(\sqrt{\varepsilon})\right)  ,
		\end{equation}
		where $E_{\operatorname{sq}}(\mathcal{N})$ is the squashed entanglement of the channel $\mathcal{N}$ (see Definition~\ref{def-sq_ent_channel}) and  $g_2(\delta)\coloneqq (\delta+1)\log_2(\delta+1) - \delta \log_2 \delta$. 
		Consequently, for the one-shot quantum capacity of $\mathcal{N}$,
		\begin{equation}
			Q^{\varepsilon}(\mathcal{N})\leq \frac{1}{1-\sqrt{\varepsilon}}\left(   E_{\operatorname{sq}}(\mathcal{N})+g_{2}(\sqrt{\varepsilon})\right).
		\end{equation}
	\end{proposition*}

\begin{proposition*}{$n$-Shot Squashed Entanglement Upper Bound}
{prop:Q-cap:sq-ent-bnd-q-cap-n-shot}
		Let $\mathcal{N}_{A\rightarrow B}$ be a quantum channel, and let $\varepsilon\in[0,1)$. For all $(n,d,\varepsilon)$  quantum communication protocols over the channel $\mathcal{N}_{A\rightarrow B}$, the following bound holds%
		\begin{equation}
			\frac{1}{n} \log_{2}d\leq\frac{1}{1-\sqrt{\varepsilon}}\left(   E_{\operatorname{sq}}(\mathcal{N})+\frac{g_{2}(\sqrt{\varepsilon})}{n}\right)  .
		\end{equation}
		Consequently, for the $n$-shot quantum capacity of $\mathcal{N}$,
		\begin{equation}
			Q^{n,\varepsilon}(\mathcal{N})\leq \frac{1}{1-\sqrt{\varepsilon}}\left(   E_{\operatorname{sq}}(\mathcal{N})+\frac{g_{2}(\sqrt{\varepsilon})}{n}\right).
		\end{equation}

	\end{proposition*}

	\begin{theorem*}{Weak Converse Upper Bound on Quantum Capacity}{thm-squashed_ent_weak_conv_upper_bound}
		The squashed entanglement $E_{\operatorname{sq}}(\mathcal{N})$ of a  quantum channel $\mathcal{N}_{A\to B}$ is a weak converse rate for quantum communication over $\mathcal{N}$. In other words, $Q(\mathcal{N})\leq E_{\operatorname{sq}}(\mathcal{N})$ for every quantum channel $\mathcal{N}$.
	\end{theorem*}

\section{Examples}\label{sec-q_comm_examples}

	We now consider the quantum capacity for particular classes of quantum channels. As remarked earlier, computing the quantum capacity of an arbitrary channel is a difficult task. This task is made more difficult by the fact that, in some cases, the coherent information is known to be \textit{strictly superadditive}, meaning that
	\begin{equation}
		I^c(\mathcal{N}^{\otimes n})>n I^c(\mathcal{N}).
	\end{equation}
	This fact confirms that regularization of the coherent information really is needed in general in order to compute the quantum capacity, and that additivity of coherent information is simply not true for all channels. Another interesting phenomenon related to quantum capacity is \textit{superactivation}, which is when two channels $\mathcal{N}_1$ and $\mathcal{N}_2$, each with zero quantum capacity, i.e., $Q(\mathcal{N}_1)=Q(\mathcal{N}_2)=0$, can combine to have non-zero quantum capacity, i.e., $Q(\mathcal{N}_1\otimes\mathcal{N}_2)>0$. Please consult the Bibliographic Notes in Section~\ref{sec:Q-cap:bib-notes} for more information about strict superadditivity and superactivation.
	
	In this section, we show that coherent information is additive for all degradable channels, which means that regularization is not needed in order to compute their capacities. The same turns out to be true for generalized dephasing channels, and we prove this by showing that the Rains relative entropy of those channels coincides with their coherent information. We also show that anti-degradable channels have zero quantum capacity. Finally, we evaluate the upper and lower bounds established in this chapter for the generalized amplitude damping channel.
	
	Before starting, let us first recall the definition of coherent information of a channel:
	\begin{equation}\label{eq-coh_inf_chan_alt_2}
		I^c(\mathcal{N})=\sup_{\psi_{RA}}I(R\rangle B)_{\omega}=\sup_{\rho}\{H(\mathcal{N}(\rho))-H(\mathcal{N}^c(\rho))\},
	\end{equation}
	where $\omega_{RB}=\mathcal{N}_{A\to B}(\psi_{RA})$ and the second equality is explained in \eqref{eq-coh_inf_chan_2}--\eqref{eq-coh_inf_chan_alt}. We let
	\begin{equation}\label{eq-coh_inf_chan_state}
		I^c(\rho,\mathcal{N})\coloneqq H(\mathcal{N}(\rho))-H(\mathcal{N}^c(\rho)).
	\end{equation}

\subsection{Degradable Channels}\label{subsec-qcomm_deg}

	Recall from Definition~\ref{def-deg_antideg_chan} that a channel $\mathcal{N}_{A\to B}$ is degradable if there exists a channel $\mathcal{D}_{B\to E}$ such that
	\begin{equation}
		\mathcal{N}^c=\mathcal{D}\circ\mathcal{N},
	\end{equation}
	where $\mathcal{N}^c$ is a channel complementary to $\mathcal{N}$ (see Definition~\ref{def-complementary_chan}) and $d_E\geq\rank(\Gamma_{AB}^{\mathcal{N}})$. In particular, if $V_{A\to BE}$ is an isometric extension of $\mathcal{N}$, so that
	\begin{equation}
		\mathcal{N}(\rho)=\Tr_E[V\rho V^\dagger]
	\end{equation}
	for every state $\rho$, then
	\begin{equation}
		\mathcal{N}^c(\rho)=\Tr_B[V\rho V^\dagger].
	\end{equation}
	
	We now show that the coherent information is additive for degradable quantum channels, meaning that
	\begin{equation}
		I^c(\mathcal{N}\otimes\mathcal{M})=I^c(\mathcal{N})+I^c(\mathcal{M})
	\end{equation}
	for all degradable quantum channels $\mathcal{N}$ and $\mathcal{M}$. Consequently, regularization is unnecessary, and we conclude that the quantum capacity of a degradable channel is equal to its coherent information:
	\begin{equation}
		Q(\mathcal{N})=I^c(\mathcal{N})\text{ for every degradable channel }\mathcal{N}.
	\end{equation}
	
	\begin{proposition*}{Additivity of Coherent Information for Degradable Channels}{prop-coh_inf_chan_additive}
		Let $\mathcal{N}$ and $\mathcal{M}$ be degradable channels. Then, the coherent information is additive, i.e.,
		\begin{equation}
			I^c(\mathcal{N}\otimes\mathcal{M})=I^c(\mathcal{N})+I^c(\mathcal{M}).
		\end{equation}
	\end{proposition*}
	
	\begin{Proof}
		As shown in Section~\ref{sec-qcomm_additivity}, we always have superadditivity of coherent information, so that $I^c(\mathcal{N}\otimes\mathcal{M})\geq I^c(\mathcal{N})+I^c(\mathcal{M})$. So we prove that the reverse inequality also holds for the case of degradable channels.
		
		Let $\mathcal{D}_1$ and $\mathcal{D}_2$ be the degrading channels for $\mathcal{N}$ and $\mathcal{M}$, respectively, meaning that
		\begin{equation}\label{eq-deg_chans_additive_coh_inf_pf}
			\mathcal{N}^c=\mathcal{D}_1\circ\mathcal{N},\quad \mathcal{M}^c=\mathcal{D}_2\circ\mathcal{M}.
		\end{equation}
		
		Now, let $\rho_{A_1A_2}$ be an arbitrary state on the input systems $A_1$ and $A_2$ of the channels $\mathcal{N}$ and $\mathcal{M}$, respectively. Using \eqref{eq-mut_inf_state_2} and \eqref{eq-mut_inf_state_3}, along with the fact that $(\mathcal{N}\otimes\mathcal{M})^c=\mathcal{N}^c\otimes\mathcal{M}^c$, we find that
		\begin{align}
			&H(\mathcal{N}^c(\rho_{A_1}))+H(\mathcal{M}^c(\rho_{A_2})-H((\mathcal{N}\otimes\mathcal{M})^c(\rho_{A_1A_2})\\
			&\quad =D((\mathcal{N}\otimes\mathcal{M})^c(\rho_{A_1A_2})\Vert\mathcal{N}^c(\rho_{A_1})\otimes\mathcal{M}^c(\rho_{A_2}))\\
			&\quad =D((\mathcal{D}_1\circ\mathcal{N}\otimes\mathcal{D}_2\circ\mathcal{M})(\rho_{A_1A_2})\Vert(\mathcal{D}_1\circ\mathcal{N})(\rho_{A_1})\otimes(\mathcal{D}_2\circ\mathcal{M})(\rho_{A_2}))\\
			&\quad \leq D((\mathcal{N}\otimes\mathcal{M})(\rho_{A_1A_2})\Vert\mathcal{N}(\rho_{A_1})\otimes\mathcal{M}(\rho_{A_2}))\\
			&\quad=H(\mathcal{N}(\rho_{A_1}))+H(\mathcal{M}(\rho_{A_2}))-H((\mathcal{N}\otimes\mathcal{M})(\rho_{A_1A_2})),
		\end{align}
		where the third equality follows from \eqref{eq-deg_chans_additive_coh_inf_pf}, the inequality follows from the data-processing inequality for quantum relative entropy, and the last equality from \eqref{eq-mut_inf_state_2} and \eqref{eq-mut_inf_state_3}. Rearranging this inequality and applying subadditivity of the entropy $H((\mathcal{N}\otimes\mathcal{M})(\rho_{A_1A_2}))$ gives
		\begin{align}
			&H((\mathcal{N}\otimes\mathcal{M})(\rho_{A_1A_2}))-H((\mathcal{N}\otimes\mathcal{M})^c(\rho_{A_1A_2}))\\
			&\quad \leq H(\mathcal{N}(\rho_{A_1}))-H(\mathcal{N}^c(\rho_{A_2}))+H(\mathcal{M}(\rho_{A_2}))-H(\mathcal{M}^c(\rho_{A_2}))\\
			&\quad \leq \sup_{\rho_{A_1}}\{H(\mathcal{N}(\rho_{A_1}))-H(\mathcal{N}^c(\rho_{A_1}))\}\\
			&\qquad\qquad+\sup_{\rho_{A_2}}\{H(\mathcal{M}(\rho_{A_2}))-H(\mathcal{M}^c(\rho_{A_2}))\}\\
			&\quad =I^c(\mathcal{N})+I^c(\mathcal{M})
		\end{align}
		Since the state $\rho_{A_1A_2}$ is arbitrary, we conclude that
		\begin{align}
			I^c(\mathcal{N}\otimes\mathcal{M})&=\sup_{\rho_{A_1A_2}}\{H((\mathcal{N}\otimes\mathcal{M})(\rho_{A_1A_2}))-H((\mathcal{N}\otimes\mathcal{N})^c(\rho_{A_1A_2}))\}\\
			&\leq I^c(\mathcal{N})+I^c(\mathcal{M}),
		\end{align}
		as required.
	\end{Proof}
	
	Another useful fact about a degradable channel $\mathcal{N}$ is that the coherent information $I^c(\rho,\mathcal{N})$ defined in \eqref{eq-coh_inf_chan_state} is concave in the input state $\rho$.
	
	\begin{Lemma}{lem-chan_coh_inf_deg_concave}
		For a degradable channel $\mathcal{N}$, the function $\rho\mapsto I^c(\rho,\mathcal{N})$ is concave in the input state $\rho$. In other words, for every finite alphabet $\mathcal{X}$, probability distribution $p:\mathcal{X}\to[0,1]$, and set $\{\rho_A^x\}_{x\in\mathcal{X}}$ of states,
		\begin{equation}\label{eq-chan_coh_inf_deg_concave}
			I^c\!\left(\sum_{x\in\mathcal{X}}p(x)\rho_A^x,\mathcal{N}\right)\geq \sum_{x\in\mathcal{X}}p(x)I^c(\rho_A^x,\mathcal{N}).
		\end{equation}
	\end{Lemma}
	
	\begin{Proof}
		Let $\mathcal{D}$ be a degrading channel corresponding to $\mathcal{N}$, so that $\mathcal{N}^c=\mathcal{D}\circ\mathcal{N}$. Next, define the following states:
		\begin{align}
			\omega_{XB}&\coloneqq\sum_{x\in\mathcal{X}}p(x)\ket{x}\!\bra{x}_X\otimes\mathcal{N}_{A\to B}(\rho_A^x),\\
			\tau_{XE}&\coloneqq\sum_{x\in\mathcal{X}}p(x)\ket{x}\!\bra{x}_X\otimes\mathcal{N}^c(\rho_A^x)=\mathcal{D}_{B\to E}(\omega_{XB})
		\end{align}
		Then, by noting that $\tau_{XE} = \mathcal{D}_{B\to E}(\omega_{XB})$ and applying the data-processing inequality for quantum mutual information (see \eqref{eq:QEI:DP-mut-info-loc-ch}), we obtain
		\begin{equation}
			I(X;E)_{\tau} \leq I(X;B)_{\omega}\ .
		\end{equation}
		Then, using \eqref{eq-mut_inf_formula_3} and rearranging leads to
		\begin{equation}\label{eq-coh_inf_chan_deg_concave_pf}
			H(B)_{\omega}-H(E)_{\tau}\geq H(B|X)_{\omega}-H(E|X)_{\tau},
		\end{equation}
		which is the desired inequality in \eqref{eq-chan_coh_inf_deg_concave}. Indeed, the left-hand side of the inequality above  is simply $I^c\!\left(\sum_{x\in\mathcal{X}} p(x)\rho_A^x,\mathcal{N}\right)$. For the right-hand side, we find that
		\begin{align}
			H(B|X)_{\omega}&=\sum_{x\in\mathcal{X}}p(x)H(\mathcal{N}(\rho_A^x)),\\
			H(E|X)_{\tau}&=\sum_{x\in\mathcal{X}}p(x)H(\mathcal{N}^c(\rho_A^x)),
		\end{align}
		because $\omega_{XB}$ and $\tau_{XE}$ are classical-quantum states.
		Therefore, the right-hand side of \eqref{eq-coh_inf_chan_deg_concave_pf} is equal to $\sum_{x\in\mathcal{X}}p(x)I^c(\rho_A^x,\mathcal{N})$.
	\end{Proof}

\subsubsection{Generalized Dephasing Channels}
	
	\label{sec-QC:gen-deph-chs}
	
	While additivity of coherent information for degradable channels allows for a tractable formula for their quantum capacity, the question about the \textit{strong converse} quantum capacity $\widetilde{Q}(\mathcal{N})$ remains. In other words, is it the case that $\widetilde{Q}(\mathcal{N})=I^c(\mathcal{N})$ for all degradable quantum channels $\mathcal{N}$? We answer this question here for a particular class of degradable channels.
	
	We consider the class of degradable channels $\mathcal{N}$ called \textit{generalized dephasing channels}. Such channels are defined by the following isometric extension:
	\begin{equation}\label{eq-gen_dephase_iso_ext}
		V^{\mathcal{N}}_{A\to BE}=\sum_{i=0}^{d-1}\ket{i}_B\bra{i}_A\otimes\ket{\psi_i}_E,
	\end{equation}
	where $d\geq 1$ and where the state vectors $\{\ket{\psi_i}\}_{i=0}^{d-1}$ are arbitrary (not necessarily orthonormal). Recalling the discussion in Section~\ref{subsec-Hadamard_chan} on Hadamard channels, in particular \eqref{eq-Hadamard_chan_iso_ext}, we see that generalized dephasing channels are Hadamard channels, as in \eqref{eq-Hadamard_channel}, with $V$ therein set to $\mathbbm{1}$.
	
	For a state $\rho_A$, we have that
	\begin{equation}
		\mathcal{N}_{A\to B}(\rho_A)=\Tr_E[V_{A\to BE}^{\mathcal{N}}\rho_AV_{A\to BE}^{\mathcal{N}\dagger}]=\sum_{i,j=0}^{d-1}\bra{i}\rho_A\ket{j}\braket{\psi_i}{\psi_j}\ket{i}\!\bra{j}_B,
	\end{equation}
	and
	\begin{equation}
		\mathcal{N}^c_{A\to E}(\rho_A)=\Tr_B[V_{A\to BE}^{\mathcal{N}}\rho_A V_{A\to BE}^{\mathcal{N}\dagger}]=\sum_{i=0}^{d-1}\bra{i}\rho_A\ket{i}\ket{\psi_i}\!\bra{\psi_i}_E.
	\end{equation}
	Then, it is straightforward to see that
	\begin{equation}
		\mathcal{N}^c\circ\mathcal{N}(\rho)=\mathcal{N}^c(\rho)
	\end{equation}
	for every state $\rho$. This implies that generalized dephasing channels $\mathcal{N}$ are degradable, with $\mathcal{N}^c$ being the degrading channel.
	
	We now show that $\widetilde{Q}(\mathcal{N})=I^c(\mathcal{N})$ for every generalized dephasing channel~$\mathcal{N}$. We do this by showing that the Rains information $R(\mathcal{N})$ of a generalized dephasing channel is equal to its coherent information.
	
	\begin{theorem*}{Quantum Capacity of Generalized Dephasing Channels}{prop-gen_dephase_q_cap}
		For every generalized dephasing channel $\mathcal{N}$ (defined by the isometric extension in \eqref{eq-gen_dephase_iso_ext}), the following equalities hold
		\begin{equation}
			Q(\mathcal{N})=\widetilde{Q}(\mathcal{N})=R(\mathcal{N})=I^c(\mathcal{N}),
		\end{equation}
		which establish the coherent information as the quantum capacity and strong converse quantum capacity.
	\end{theorem*}
	
	\begin{Proof}
		It suffices to show that $R(\mathcal{N})=I^c(\mathcal{N})$. Note that the inequality $I^c(\mathcal{N})\leq R(\mathcal{N})$ holds for every quantum channel $\mathcal{N}$ by combining the result of Theorem~\ref{thm-qcomm_capacity} with the result of Theorem~\ref{thm-Rains_inf_strong_conv_upper_bound}. We now show that the reverse inequality holds for all generalized dephasing channels.
		
		We start by observing that every generalized dephasing channel $\mathcal{N}$ is covariant with respect to the operators $\{Z(j)\}_{j=0}^{d-1}$ defined in \eqref{eq-gen_Z_Pauli}:
		\begin{equation}
			\mathcal{N}(Z(j)\rho Z(j)^\dagger)=Z(j)\mathcal{N}(\rho)Z(j)^\dagger
		\end{equation}
		for every state $\rho$ and for all $0\leq j\leq d-1$, where
		\begin{equation}
			Z(j)=\sum_{k=0}^{d-1}\e^{\frac{2\pi\I kj}{d}}\ket{k}\!\bra{k}.
		\end{equation}
		Then, for every state $\rho$, the average state
		\begin{equation}
			\overline{\rho}\coloneqq\frac{1}{d}\sum_{j=0}^{d-1} Z(j)\rho Z(j)^\dagger=\sum_{k=0}^{d-1}\bra{k}\rho\ket{k}\ket{k}\!\bra{k}
		\end{equation}
		is diagonal in the basis $\{\ket{i}\}_{i=0}^{d-1}$. Since the quantities $\bra{k}\rho\ket{k}$ are probabilities, we conclude that for every state $\rho$, its corresponding average state $\overline{\rho}$ has a purification of the following form:
		\begin{equation}\label{eq-gen_dephase_q_cap_pf}
			\ket{\phi^{\overline{\rho}}}_{RA}=\sum_{i=0}^{d-1}\sqrt{\smash[b]{p(i)}}\ket{i}_R\otimes\ket{i}_A\eqqcolon \ket{\psi^p}_{RA},
		\end{equation}
		where $p:\{0,1,\dotsc,d-1\}\to[0,1]$ is a probability distribution. Therefore, by Proposition~\ref{prop-gen_Rains_inf_chan_cov}, when calculating the Rains information $R(\mathcal{N})$, it suffices to optimize over the pure states $\psi^p_{RA}$:
		\begin{equation}
			R(\mathcal{N})=\sup_{\psi^p_{RA}}\inf_{\sigma_{RB}\in\PPT'(R:B)} D(\mathcal{N}_{A\to B}(\psi_{RA}^p)\Vert\sigma_{RB}).
		\end{equation}
		
		Now, restricting the optimization in the definition of the coherent information~$I^c(\mathcal{N})$ to the pure states in \eqref{eq-gen_dephase_q_cap_pf}, we obtain
		\begin{align}
			I^c(\mathcal{N})&=\sup_{\psi_{RA}}\inf_{\sigma_B}D(\mathcal{N}_{A\to B}(\psi_{RA})\Vert\mathbbm{1}_R\otimes\sigma_B)\\
			&\geq \sup_{\psi_{RA}^p}\inf_{\sigma_B}D(\mathcal{N}_{A\to B}(\psi_{RA}^p)\Vert\mathbbm{1}_R\otimes\sigma_B)\\
			&\geq \sup_{\psi_{RA}^p}\inf_{\sigma_B}D(\Delta(\mathcal{N}_{A\to B}(\psi_{RA}^p))\Vert\Delta(\mathbbm{1}_R\otimes\sigma_B)),
		\end{align}
		where the last line follows from the data-processing inequality for quantum relative entropy, and we introduced the following channel:
		\begin{equation}
			\Delta(\rho)\coloneqq \Pi\rho\Pi+(\mathbbm{1}-\Pi)\rho(\mathbbm{1}-\Pi),\quad\Pi=\sum_{i=0}^{d-1}\ket{i}\!\bra{i}_R\otimes\ket{i}\!\bra{i}_B.
		\end{equation}
		Now, it is straightforward to check that
		\begin{equation}
			\Pi\mathcal{N}_{A\to B}(\psi^p_{RA})\Pi=\mathcal{N}_{A\to B}(\psi_{RA}^p),
		\end{equation}
		which implies that
		\begin{equation}
			\Delta(\mathcal{N}_{A\to B}(\psi_{RA}^p))=\Pi\mathcal{N}_{A\to B}(\psi_{RA}^p)\Pi.
		\end{equation}
		Therefore, because $\Pi$ and $\mathbbm{1}-\Pi$ project onto orthogonal subspaces, we obtain
		\begin{align}
			&D(\Delta(\mathcal{N}_{A\to B}(\psi_{RA}^p))\Vert\Delta(\mathbbm{1}_R\otimes\sigma_B))\notag \\
			&\quad=D(\Pi\mathcal{N}_{A\to B}(\psi_{RA}^p)\Pi\Vert\Pi(\mathbbm{1}_{R}\otimes\sigma_B)\Pi)\\
			&\quad =D\!\left(\Pi\mathcal{N}_{A\to B}(\psi_{RA}^p)\Pi\Bigg\Vert\sum_{i=0}^{d-1}q(i)\ket{i}\!\bra{i}_R\otimes\ket{i}\!\bra{i}_B\right),
		\end{align}
		where  the last line follows because
		\begin{equation}
			\Pi(\mathbbm{1}_R\otimes\sigma_B)\Pi=\sum_{i=0}^{d-1}q(i)\ket{i}\!\bra{i}_R\otimes\ket{i}\!\bra{i}_B,
			\label{eq-QC:sep-state-gen-deph-pf}
		\end{equation}
		with the probability distribution $q(i)\coloneqq\bra{i}\sigma_B\ket{i}$. Note that the right-hand side of the equation above  is a state in $\PPT'(R\!:\!B)$. Therefore, we have
		\begin{align}
			I^c(\mathcal{N})&\geq \sup_{\psi_{RA}^p}\inf_{\sigma_B}D(\Delta(\mathcal{N}_{A\to B}(\psi_{RA}^p))\Vert\Delta(\mathbbm{1}_R\otimes\sigma_B))\\
			&=\sup_{\psi_{RA}^p}\inf_{\sigma_B}D(\Pi\mathcal{N}_{A\to B}(\psi_{RA}^p)\Pi\Vert\Pi(\mathbbm{1}_R\otimes\sigma_B)\Pi)\\
			&=\sup_{\psi_{RA}^p}\inf_{\sigma_B} D\!\left(\mathcal{N}_{A\to B}(\psi_{RA}^p)\Bigg\Vert\sum_{i=0}^{d-1}\bra{i}\sigma_B\ket{i}\ket{i}\!\bra{i}_R\otimes\ket{i}\!\bra{i}_B\right)\\
			&\geq \sup_{\psi_{RA}^p}\inf_{\sigma_{RB}\in\PPT'(R:B)}D(\mathcal{N}_{A\to B}(\psi_{RA}^p)\Vert\sigma_{RB})\\
			&=R(\mathcal{N}),
		\end{align}
		completing the proof.
	\end{Proof}

\subsection{Anti-Degradable Channels}\label{subsec-qcomm_anti_deg}

	Let us now consider anti-degradable channels. Recall from Definition~\ref{def-deg_antideg_chan} that a channel $\mathcal{N}_{A\to B}$ is anti-degradable if there exists an anti-degrading channel $\mathcal{A}_{E\to B}$ such that
	\begin{equation}
		\mathcal{N}=\mathcal{A}\circ\mathcal{N}^c,
	\end{equation}
	where $\mathcal{N}^c$ is a channel complementary to $\mathcal{N}$ and $d_E\geq\rank(\Gamma_{AB}^{\mathcal{N}})$.
	

	\begin{proposition*}{Coherent Information for Anti-Degradable Channels}{prop-coh_inf_chan_antidegradable}
		The coherent information vanishes for all anti-degradable channels, i.e., $I^c(\mathcal{N})=0$ for every anti-degradable channel $\mathcal{N}$. Therefore, $Q(\mathcal{N})=0$ for all anti-degradable channels.
	\end{proposition*}
	
	\begin{Proof}
		Let $\mathcal{N}_{A\to B}$ have the following Stinespring representation: $\mathcal{N}(\rho_A)=\Tr_E[V_{A\to BE}\rho_A V_{A\to BE}^\dagger]$. Then, for every pure state $\psi_{RA}$, the state vector
		\begin{equation}
			\ket{\phi}_{RBE}\coloneqq V_{A\to BE}\ket{\psi}_{RA}
		\end{equation}
		is such that
		\begin{equation}\label{eq-coh_inf_chan_antidegrade_pf}
			I(R\rangle B)_{\phi}=\frac{1}{2}\left(I(R;B)_{\phi}-I(R;E)_{\phi}\right).
		\end{equation}
		To see this, let us first note that $H(E)_{\phi}=H(RB)_{\phi}$ and $H(RE)_{\phi}=H(B)_{\phi}$. These identities hold  because $\phi_{RBE}$ is a pure state, implying that the reduced states $\phi_{E}$ and $\phi_{RB}$ have the same spectrum and the reduced states $\phi_{RE}$ and $\phi_B$ have the same spectrum. This, along with \eqref{eq-mut_inf_state_2}, leads to
		\begin{align}
			&\frac{1}{2}\left(I(R;B)_{\phi}-I(R;E)_{\phi}\right)\\
			&\quad =\frac{1}{2}\left(H(R)_{\phi}+H(B)_{\phi}-H(RB)_{\phi}-H(R)_{\phi}-H(E)_{\phi}+H(RE)_{\phi}\right)\\
			&=H(B)_{\phi}-H(RB)_{\phi}\\
			&=I(R\rangle B)_{\phi}.
		\end{align}
		Next, using \eqref{eq-mut_inf_state_3}, noting that $\phi_{RB}=\mathcal{N}_{A\to B}(\psi_{RA})\eqqcolon\omega_{RB}$, and using the fact that $\mathcal{N}$ is anti-degradable, so that $\mathcal{N}=\mathcal{A}\circ\mathcal{N}^c$, we have
				\begin{equation}
			I(R;B)_{\phi} \leq I(R;E)_{\phi},
		\end{equation}
		where  the inequality follows from the data-processing inequality for mutual information under local channels (see \eqref{eq:QEI:DP-mut-info-loc-ch}) and  the facts that $\mathcal{N}_{A\to B}(\psi_{RA})= \mathcal{A}_{E\to B}\circ\mathcal{N}_{A\to E}^c(\psi_{RA})$ and the reduced state $\phi_{RE}=\mathcal{N}_{A\to E}^c(\psi_{RA})$. Therefore, from \eqref{eq-coh_inf_chan_antidegrade_pf} we conclude that
		\begin{equation}
			I(R\rangle B)_{\omega}\leq 0
		\end{equation}
		for every pure state $\psi_{RA}$. This implies that
		\begin{equation}
			I^c(\mathcal{N})=\sup_{\psi_{RA}}I(R\rangle B)_{\omega}=0,
		\end{equation}
		as required.
	\end{Proof}


\subsection{Generalized Amplitude Damping Channel}

	Let us recall the definition of the generalized amplitude damping channel (GADC) from \eqref{eq-gen_amp_damp}:
	\begin{equation}
		\mathcal{A}_{\gamma,N}(\rho)=A_1\rho A_1^\dagger +A_2\rho A_2^\dagger+ A_3\rho A_3^\dagger +A_4\rho A_4^\dagger,
	\end{equation}
	where
	\begin{align}
		A_1=\sqrt{1-N}\begin{pmatrix} 1 & 0 \\ 0 & \sqrt{1-\gamma} \end{pmatrix},&\quad A_2=\sqrt{1-N}\begin{pmatrix} 0 & \sqrt{\gamma} \\ 0 & 0 \end{pmatrix},\\ A_3=\sqrt{N}\begin{pmatrix} \sqrt{1-\gamma} & 0 \\ 0 & 1 \end{pmatrix},& \quad A_4=\sqrt{N}\begin{pmatrix} 0 & 0 \\ \sqrt{\gamma} & 0 \end{pmatrix},
	\end{align}
	and $\gamma,N\in[0,1]$. It is straightforward to show that
	\begin{equation}\label{eq-GADC_pre_post_N}
		\mathcal{A}_{\gamma,N}(\rho)=X\mathcal{A}_{\gamma,1-N}(X\rho X)X
	\end{equation}
	for every state $\rho$ and all $\gamma,N\in[0,1]$. In other words, the GADC $\mathcal{A}_{\gamma,N}$ is related to the GADC $\mathcal{A}_{\gamma,1-N}$ via a simple pre- and post-processing by the Pauli unitary $X=\ket{0}\!\bra{1}+\ket{1}\!\bra{0}$. The information-theoretic aspects of the GADC are thus invariant under the interchange $N\leftrightarrow 1-N$, which means that we can, without loss of generality, restrict the parameter $N$ to the interval~$\left[0,\sfrac{1}{2}\right]$.
	
	For $N=0$, the GADC reduces to the amplitude damping channel $\mathcal{A}_{\gamma}$ defined in \eqref{eq-amplitude_damping_channel}, which is degradable. Indeed, we first note that
	\begin{equation}
		\mathcal{A}_{\gamma,0}^c=\mathcal{A}_{1-\gamma,0},
	\end{equation}
	where the complementary channel $\mathcal{A}_{\gamma,0}^c$ (recall Definition~\ref{def-complementary_chan}) is defined via the following isometric extension:
	\begin{equation}
		V^{\gamma,N}\coloneqq A_1\otimes\ket{0}+A_2\otimes\ket{1}+A_3\otimes\ket{2}+A_4\otimes\ket{3}.
	\end{equation}
	We now use the fact that, for all $\gamma_1,\gamma_2,N_1,N_2\in[0,1]$,
	\begin{equation}\label{eq-GADC_composition}
		\mathcal{A}_{\gamma,N}=\mathcal{A}_{\gamma_2,N_2}\circ\mathcal{A}_{\gamma_1,N_1},
	\end{equation}
	where $\gamma=\gamma_1+\gamma_2-\gamma_1\gamma_2$ and $N=\frac{\gamma_1(1-\gamma_2)N_1+\gamma_2N_2}{\gamma_1+\gamma_2-\gamma_1\gamma_2}$. From this fact, it follows that the defining condition for degradability, namely, $\mathcal{D}_{\gamma,0}\circ\mathcal{A}_{\gamma,0}=\mathcal{A}_{\gamma,0}^c=\mathcal{A}_{1-\gamma,0}$, is satisfied by the quantum channel $\mathcal{D}_{\gamma,0}\coloneqq\mathcal{A}_{\frac{1-2\gamma}{1-\gamma},0}$. It can be shown that for $N>0$, the GADC $\mathcal{A}_{\gamma,N}$ is not degradable for all $\gamma\in(0,1]$ (please consult the Bibliographic Notes in Section~\ref{sec:Q-cap:bib-notes}).
	
\begin{figure}
		\centering
		\includegraphics[scale=1]{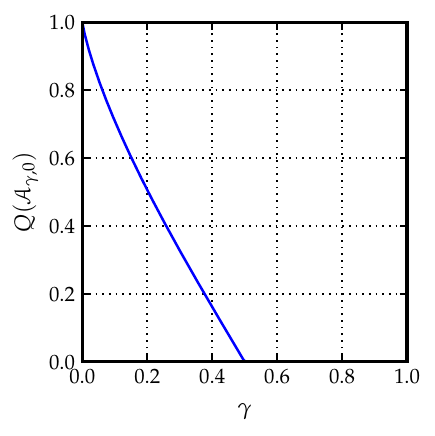}
		\caption{Quantum capacity of the amplitude damping channel, as given by \eqref{eq-QCap_amplitude_damping}. The capacity is equal to zero for $\gamma\geq\frac{1}{2}$ because in this parameter range the channel is anti-degradable.}\label{fig-AD_QCap}
	\end{figure}

	Since $\mathcal{A}_{\gamma,0}$ is degradable, its coherent information is additive, which means that its quantum capacity is equal to its coherent information, i.e.,
	\begin{align}
		Q(\mathcal{A}_{\gamma,0})&=I^c(\mathcal{A}_{\gamma,0})=\sup_{\rho}\left\{H(\mathcal{A}_{\gamma,0}(\rho))-H(\mathcal{A}_{\gamma,0}^c(\rho))\right\}\\
		&=\sup_{\rho} I^c(\rho,\mathcal{A}_{\gamma,0}),
	\end{align}
	where we have used the expression in \eqref{eq-coh_inf_chan_alt_2}. Now, as explained in Section~\ref{subsec-eacc_GADC}, the GADC is covariant with respect to the Pauli operator $Z$. Furthermore, by Lemma~\ref{lem-chan_coh_inf_deg_concave}, the function $\rho\mapsto I^c(\rho,\mathcal{A}_{\gamma,0})$ is concave. Therefore, for every state $\rho$,
	\begin{equation}
		I^c\!\left(\frac{1}{2}\rho+\frac{1}{2}Z\rho Z,\mathcal{A}_{\gamma,0}\right)\geq \frac{1}{2}I^c(\rho,\mathcal{A}_{\gamma,0})+\frac{1}{2}I^c(Z\rho Z,\mathcal{A}_{\gamma,0}).
	\end{equation}
	Now, using the fact that $\mathcal{A}_{\gamma,0}$ is covariant with respect to $Z$, and the fact that $\mathcal{A}_{\gamma,0}^c=\mathcal{A}_{1-\gamma,0}$, we obtain
	\begin{align}
		I^c(Z\rho Z,\mathcal{A}_{\gamma,0})&=H(\mathcal{A}_{\gamma,0}(Z\rho Z))-H(\mathcal{A}_{\gamma,0}^c(Z\rho Z))\\
		&=H(Z\mathcal{A}_{\gamma,0}(\rho)Z)-H(Z\mathcal{A}_{1-\gamma,0}(\rho)Z)\\
		&=H(\mathcal{A}_{\gamma,0}(\rho))-H(\mathcal{A}_{1-\gamma,0}(\rho))\\
		&=H(\mathcal{A}_{\gamma,0}(\rho))-H(\mathcal{A}_{\gamma,0}^c(\rho))\\
		&=I^c(\rho,\mathcal{A}_{\gamma,0}).
	\end{align}
	Therefore,
	\begin{equation}
		I^c\!\left(\frac{1}{2}\rho+\frac{1}{2}Z\rho Z,\mathcal{A}_{\gamma,0}\right)\geq I^c(\rho,\mathcal{A}_{\gamma,0})
	\end{equation}
	for every state $\rho$. Recalling from \eqref{eq-completely_dephasing_channel} that the state $\frac{1}{2}\rho+\frac{1}{2}Z\rho Z$ results from the action of the completely dephasing channel on $\rho$, which means that it is diagonal in the standard basis, we find that
	\begin{equation}
		\max_{p\in[0,1]}I^c((1-p)\ket{0}\!\bra{0}+p\ket{1}\!\bra{1},\mathcal{A}_{\gamma,0})\geq I^c(\rho,\mathcal{A}_{\gamma,0})
	\end{equation}
	for every state $\rho$, which means that
	\begin{align}
		Q(\mathcal{A}_{\gamma,0})&=I^c(\mathcal{A}_{\gamma,0})=\max_{p\in[0,1]}I^c((1-p)\ket{0}\!\bra{0}+p\ket{1}\!\bra{1},\mathcal{A}_{\gamma,0})\\
		&=\max_{p\in[0,1]}\{h_2((1-\gamma)p)-h_2(\gamma p)\}\label{eq-QCap_amplitude_damping},
	\end{align}
	where in the last line we have evaluated $I^c((1-p)\ket{0}\!\bra{0}+p\ket{1}\!\bra{1},\mathcal{A}_{\gamma,0})$. See Figure~\ref{fig-AD_QCap} for a plot of the quantum capacity of the amplitude damping channel $\mathcal{A}_{\gamma,0}$. Note that the capacity vanishes at $\gamma=\frac{1}{2}$, which is due to the fact that for $\gamma\geq\frac{1}{2}$ the amplitude damping channel $\mathcal{A}_{\gamma,0}$ (and more generally the GADC $\mathcal{A}_{\gamma,N}$ for $N\in[0,1]$) is anti-degradable. From Proposition~\ref{prop-coh_inf_chan_antidegradable}, we thus have that $Q(\mathcal{A}_{\gamma,N})=0$ for all $N\in[0,1]$ and $\gamma\geq\frac{1}{2}$. 
	
	Let us now consider the coherent information of the GADC $\mathcal{A}_{\gamma,N}$ for $N>0$. In this case, the coherent information $I^c(\mathcal{A}_{\gamma,N})$ is a lower bound on the quantum capacity of the GADC. As with the amplitude damping channel, it can be shown that for the GADC $\mathcal{A}_{\gamma,N}$ with $N>0$ it suffices to optimize over states diagonal in the standard basis in order to compute the coherent information:
	\begin{equation}\label{eq-GADC_coh_inf}
		I^c(\mathcal{A}_{\gamma,N})=\max_{p\in[0,1]}I^c((1-p)\ket{0}\!\bra{0}+p\ket{1}\!\bra{1},\mathcal{A}_{\gamma,N}),
	\end{equation}
	for all $\gamma\in(0,1)$ and all $N>0$. The proof of this is more involved, since for $N>0$ the GADC is not degradable, meaning that we cannot use Lemma~\ref{lem-chan_coh_inf_deg_concave}. Please consult the Bibliographic Notes in Section~\ref{sec:Q-cap:bib-notes} for a source of the proof.
	
\begin{figure}
		\centering
		\includegraphics[scale=0.8]{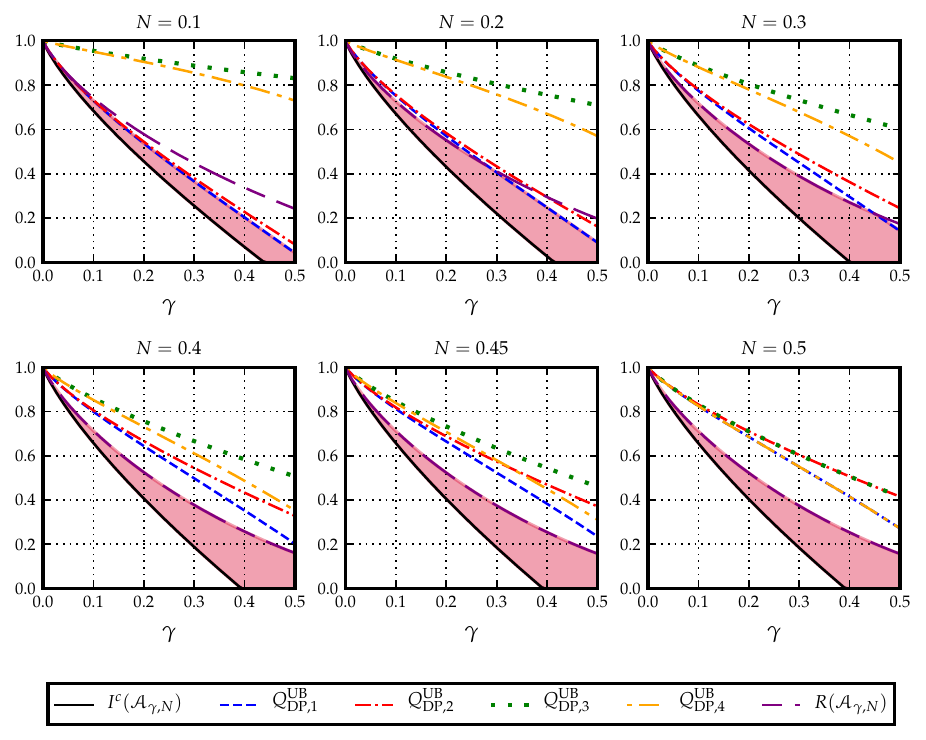}
		\caption{The coherent information lower bound $I^c(\mathcal{A}_{\gamma,N})$ and four upper bounds on the quantum capacity of the generalized amplitude damping channel $\mathcal{A}_{\gamma,N}$. The quantum capacity lies within the shaded region.}\label{fig-GADC_QCap_bounds}
	\end{figure}

	In Figure~\ref{fig-GADC_QCap_bounds}, we plot the coherent information lower bound given by \eqref{eq-GADC_coh_inf}. We also plot the Rains information upper bound $R(\mathcal{A}_{\gamma,N})$ as well as four other upper bounds that are based on the following identities, which follow from \eqref{eq-GADC_composition}:
	\begin{align}
		\mathcal{A}_{\gamma,N}&=\mathcal{A}_{\gamma N,1}\circ\mathcal{A}_{\frac{\gamma(1-N)}{1-\gamma N},0},\\
		\mathcal{A}_{\gamma,N}&=\mathcal{A}_{\gamma(1-N),0}\circ\mathcal{A}_{\frac{\gamma N}{1-\gamma(1-N)},1}.
	\end{align}
	It then follows that
	\begin{align}
		Q(\mathcal{A}_{\gamma,N})&\leq Q(\mathcal{A}_{\frac{\gamma(1-N)}{1-\gamma N},0})\eqqcolon Q_{\text{DP},1}^{\text{UB}}(\gamma,N),\\
		Q(\mathcal{A}_{\gamma,N})&\leq Q(\mathcal{A}_{\gamma(1-N),0})\eqqcolon Q_{\text{DP},2}^{\text{UB}}(\gamma,N),\\
		Q(\mathcal{A}_{\gamma,N})&\leq Q(\mathcal{A}_{\gamma N,0})\eqqcolon Q_{\text{DP},3}^{\text{UB}}(\gamma,N),\\
		Q(\mathcal{A}_{\gamma,N})&\leq Q(\mathcal{A}_{\frac{\gamma N}{1-\gamma(1-N)},0})\eqqcolon Q_{\text{DP},4}^{\text{UB}}(\gamma,N).
	\end{align}
	
	\noindent Note that the right-hand side of each inequality can be calculated using \eqref{eq-QCap_amplitude_damping}. We have also made use of \eqref{eq-GADC_pre_post_N}, which implies that $Q(\mathcal{A}_{\gamma,1})=Q(\mathcal{A}_{\gamma,0})$. These inequalities hold due to the fact that, for the composition of two quantum channels $\mathcal{N}$ and $\mathcal{M}$,
	\begin{equation}
		Q(\mathcal{N}\circ\mathcal{M})\leq Q(\mathcal{M})\quad\text{and}\quad Q(\mathcal{N}\circ\mathcal{M})\leq Q(\mathcal{N}).
	\end{equation}
	The first inequality holds by the data-processing inequality. The second inequality can be viewed as a lower bound on the quantum capacity of the channel $\mathcal{N}$ that arises from a coding strategy consisting of some encoding followed by many uses of the channel $\mathcal{M}$.

\section{Summary}

	In this chapter, we studied quantum communication. Given a quantum channel $\mathcal{N}_{A\to B}$ connecting Alice and Bob, the goal in quantum communication is to determine the highest rate, called the quantum capacity and denoted by $Q(\mathcal{N})$, at which the $A'$ part of an arbitrary pure state $\Psi_{RA'}$ can be transmitted to Bob without error. At the disposal of Alice and Bob are local encoding and decoding channels, as well as an arbitrary number of (unassisted) uses of the channel $\mathcal{N}_{A\to B}$. By unassisted, we mean that Alice and Bob are not allowed to communicate with each other between channel uses. We found that the coherent information $I^c(\mathcal{N})$ of $\mathcal{N}$ is always a lower bound on its quantum capacity, and that, in general, computing the exact value of the capacity involves a regularization, so that $Q(\mathcal{N})=I_{\text{reg}}^c(\mathcal{N})$.
	
	Starting with the one-shot setting, in which only one use of the channel is allowed and there is some tolerable non-zero error, we determined both upper and lower bounds on the number of qubits that can be transmitted. The one-shot upper bound involves the hypothesis testing relative entropy in a way similar to how it is involved in classical communication and entanglement distillation. Specifically, we establish the hypothesis testing coherent information as an upper bound. This leads to the coherent information (hence regularized coherent information) weak converse upper bound in the asymptotic setting. To obtain a lower bound, we used the results of Chapter~\ref{chap-ent_distill} on entanglement distillation. We found that we could take the entanglement distillation protocol developed in that chapter and convert it to a suitable quantum communication protocol. We proved that this lower bound is optimal when applied to the asymptotic setting, in the sense that it leads to the coherent information (hence regularized coherent information) as an achievable rate, which matches the upper bound. For degradable channels, we showed that the coherent information is additive, meaning that $Q(\mathcal{N})=I^c(\mathcal{N})$ for all degradable channels. We also showed that anti-degradable channels have zero quantum capacity.
	
	With the goal of obtaining tractable estimates of quantum capacity for general channels, we found that the Rains information $R(\mathcal{N})$ of $\mathcal{N}$ is a strong converse upper bound on the quantum capacity of $\mathcal{N}$. This allowed us to conclude that the quantum capacity of the generalized dephasing channel is equal to its coherent information, because its Rains information and coherent information coincide. We also looked ahead to Chapter~\ref{chap-LOCC-QC} and concluded from the results there that the squashed entanglement of a quantum channel is an upper bound on quantum capacity.

\section{Bibliographic Notes}

\label{sec:Q-cap:bib-notes}

	The problem of determining the capacity of a quantum channel for transmitting quantum information, in a manner analogous to Shannon's channel capacity theorem, was proposed by \citet{Sho95}. The notion of quantum communication that we consider in this chapter, as well as the notion of entanglement transmission, was defined by \citet{Sch96}. The notion of subspace transmission was defined by \citet{BKN98} (see also \citep{PhysRevLett.78.3217}), and the notion of entanglement generation was defined by \citet{D05}. These different notions of quantum communication, and the connections between them, have been examined by \citet{KWerner04}, where they also proved that the capacities for these variations are all equal to each other.
	
	Upper and lower bounds on one-shot quantum capacity have been established by \citet{BD10,datta2013one-EA,BDL15,TBR15,AJW17b,WFD17}. The approach of using hypothesis testing relative entropy for obtaining an upper bound on one-shot quantum capacity (specifically, Theorem~\ref{prop-qcomm:one-shot-bound-meta}) comes from work by \citet{MW12}. The lower bound on the one-shot quantum capacity in Theorem~\ref{prop-qcomm_one-shot_lower_bound} comes from work on one-shot decoupling  \citep{Dupuis2014}, which was then used by \citet[Proposition~21]{WTB16} to obtain a lower bound on the one-shot distillable entanglement. \textcolor{red}The various code conversions in Lemmas~\ref{lem-q_comm_ent_gen_1WLOCC_no_help}, \ref{lem-q_comm_ent_gen_to_ent_trans}, and \ref{prop-ent_trans_code_to_q_comm_code} are available in a number of works, including \citet{BKN98,KWerner04,Kles07,Wat18} (see also \citet{WQ18}). The one-shot Rains upper bound in Corollary~\ref{thm-qcomm_meta_str_weak_conv_Rains} was obtained by \citet{TWW17}.
	
	In the asymptotic setting, \citet{Sch96,SN96,BNS98,BKN98} established coherent information as an upper bound on quantum capacity, and \citet{L97,capacity2002shor,D05} established the lower bound. (See also the proofs of \citet{qcap2008second,qcap2008fourth}.) Decoupling as a method for understanding quantum capacity was initially studied by \citet{SW02} and developed in further detail by \citet{HHWY08}. The Rains information strong converse upper bound (Theorem~\ref{thm-Rains_inf_strong_conv_upper_bound}) was established by \citet{TWW17}. Weak subadditivity of R\'{e}nyi Rains information of a channel (Proposition~\ref{prop-Renyi_Rains_inf_chan_weak_additive}) is also due to \citet{TWW17}. We also mention that upper bounds on quantum capacity based on approximate degradability and approximate anti-degradability of channels have been established by \citet{SSWR14} (see also \citep{LDS18}). 

	Additivity of coherent information for degradable channels was shown by \citet{DS05}. \citet{science2008smith,SSY11} demonstrated the phenomenon of superactivation of quantum capacity, and \citet{DDPS98,SS07,CEM+15,ES15} demonstrated superadditivity of coherent information. Lemma~\ref{lem-chan_coh_inf_deg_concave} was presented in \citep[Lemma~5]{YHD05MQAC}. The quantum capacity of generalized dephasing channels was established by \citet{DS05} and the strong converse by \citet{TWW17}. See \citep{MW13} for the pretty-strong converse for the quantum capacity of degradable channels. The generalized amplitude damping channel (GADC) has been studied in detail by \citet{KSW19}. The fact that this channel is not degradable for $N\in(0,1)$ and $\gamma\in(0,1]$ follows from \citep[Theorem 4]{CRS08}. The expression in \eqref{eq-QCap_amplitude_damping} for the quantum capacity of the amplitude damping channel was given by \citet{GF05}. For a proof of \eqref{eq-GADC_coh_inf}, see \citep[Appendix]{GPLS09}. A proof of anti-degradability of the GADC $\mathcal{A}_{\gamma,N}$ for all $\gamma\geq\frac{1}{2}$ can be found in \citep[Proposition~2]{KSW19}.

\begin{subappendices}

\section[Alternative Notions of Quantum Communication]{Alternative Notions of\\Quantum Communication}\label{app-q_comm_alt_notions}
	
	At the beginning of this chapter, we considered three alternative notions of quantum communication, and we described how they are implied by the notion of quantum communication as defined at the beginning of Section~\ref{sec-qcomm_one_shot}. We now precisely define these other notions of quantum communication, and we show how the notion of quantum communication considered in the chapter (strong subspace transmission) implies all three alternatives.
	
	\begin{definition}{Entanglement Transmission}{def-q_comm_ent_trans}
		An \textit{entanglement transmission protocol} for $\mathcal{N}_{A\to B}$ consists of the three elements $(d,\mathcal{E},\mathcal{D})$, where $d\geq 1$, $\mathcal{E}_{A'\to A}$ is an encoding channel with $d_{A'}=d$, and $\mathcal{D}_{B\to B'}$ is a decoding channel with $d_{B'}=d$. The goal of the protocol is to transmit the $A'$ system of a maximally entangled state $\Phi_{RA'}$ of Schmidt rank $d$ such that the final state
			\begin{equation}
				\omega_{RB'}\coloneqq (\mathcal{D}_{B\to B'}\circ\mathcal{N}_{A\to B}\circ\mathcal{E}_{A'\to A})(\Phi_{RA'})
			\end{equation}
			is close to the initial maximally entangled state. The \textit{entanglement transmission error} of the protocol is
			\begin{align}
				p_{\text{err}}^{\text{(ET)}}(\mathcal{E},\mathcal{D};\mathcal{N})&\coloneqq 1-\bra{\Phi}_{RB'}\omega_{RB'}\ket{\Phi}_{RB'}\\
				&=1-F_e(\mathcal{D}\circ\mathcal{N}\circ\mathcal{E}),
			\end{align}
			where we recall the entanglement fidelity of a channel from Definition~\ref{def-ent_fid_chan}. We call the protocol $(d,\mathcal{E},\mathcal{D})$ a \textit{$(d,\varepsilon)$ protocol}, with $\varepsilon\in[0,1]$, if $p_{\text{err}}^{\text{(ET)}}(\mathcal{E},\mathcal{D};\mathcal{N})\allowbreak\leq\varepsilon$.
	\end{definition}
	
	It is straightforward to see that if there exists a $(d,\varepsilon)$ quantum communication protocol for a quantum channel $\mathcal{N}$ (as per Definition~\ref{def-q_comm_Me_protocol}), then there exists a $(d,\varepsilon)$ entanglement transmission protocol. Indeed, for a $(d,\varepsilon)$ quantum communication protocol with encoding and decoding channel $\mathcal{E}$ and $\mathcal{D}$, we have that
	\begin{align}
		&p_{\text{err}}^*(\mathcal{E},\mathcal{D};\mathcal{N})\nonumber\\
		&\quad=\max_{\ket{\Psi}_{RA'}}\{1-\bra{\Psi}_{RA'}(\mathcal{D}_{B\to B'}\circ\mathcal{N}_{A\to B}\circ\mathcal{E}_{A'\to A})(\Psi_{RA'})\ket{\Psi}_{RA'}\}\\
		&\quad=1-F(\mathcal{D}\circ\mathcal{N}\circ\mathcal{E})\\
		&\quad\leq\varepsilon.
	\end{align}
	However, since the maximally entangled state $\Phi_{AB'}$ is a particular pure state in the optimization for $F(\mathcal{D}\circ\mathcal{N}\circ\mathcal{E})$, we conclude that
	\begin{equation}
		p_{\text{err}}^{\text{(ET)}}(\mathcal{E},\mathcal{D};\mathcal{N})=1-F_e(\mathcal{D}\circ\mathcal{N}\circ\mathcal{E})\leq 1-F(\mathcal{D}\circ\mathcal{N}\circ\mathcal{E})\leq\varepsilon.
	\end{equation}
	So the elements $(d,\mathcal{E},\mathcal{D})$ form a $(d,\varepsilon)$ entanglement transmission protocol.

	\begin{definition}{Entanglement Generation}{def-q_comm_ent_gen}
		An \textit{entanglement generation protocol} for $\mathcal{N}_{A\to B}$ is defined by the three elements $(d,\Psi_{A'A},\mathcal{D}_{B\to B'})$, where $\Psi_{A'A}$ is a pure state with $d_{A'}=d$, and $\mathcal{D}_{B\to B'}$ is a decoding channel with $d_{B'}=d$. The goal of the protocol is to transmit the system $A$ such that the final state
			\begin{equation}
				\sigma_{A'B'}\coloneqq(\mathcal{D}_{B\to B'}\circ\mathcal{N}_{A\to B})(\Psi_{A'A})
			\end{equation}
			is close in fidelity to a maximally entangled state of Schmidt rank $d$. The \textit{entanglement generation error} of the protocol is given by
			\begin{align}
				p_{\text{err}}^{\text{(EG)}}(\Psi_{A'A},\mathcal{D};\mathcal{N})&\coloneqq 1-\bra{\Phi}_{A'B'}\sigma_{A'B'}\ket{\Phi}_{A'B'}\\
				&=1-F(\Phi_{A'B'},\sigma_{A'B'}).
			\end{align}
			We call the protocol $(d,\Psi_{A'A},\mathcal{D}_{B\to B'})$ a \textit{$(d,\varepsilon)$ protocol}, with $\varepsilon\in[0,1]$, if $p_{\text{err}}^{\text{(EG)}}(\Psi_{A'A},\mathcal{D};\mathcal{N})\leq\varepsilon$.
	\end{definition}
	
	Consider a $(d,\varepsilon)$ quantum communication protocol for $\mathcal{N}_{A\to B}$ given by the elements $(d,\mathcal{E}_{A'\to A},\mathcal{D}_{B\to B'})$, where $d_{A'}=d_{B'}=d$. Then, by the arguments above, the same elements constitute a $(d,\varepsilon)$ entanglement transmission protocol, so that
	\begin{equation}
		\bra{\Phi}_{RB'}(\mathcal{D}_{B\to B'}\circ\mathcal{N}_{A\to B}\circ\mathcal{E}_{A'\to A})(\Phi_{RA'})\ket{\Phi}_{RB'}\geq 1-\varepsilon.
	\end{equation}
	Now, let the system $R\equiv\tilde{A}$ belong to Alice, and let $\Psi_{\tilde{A}A}\coloneqq\mathcal{E}_{A'\to A}(\Phi_{\tilde{A}A})$. Then,
	\begin{equation}
		\bra{\Phi}_{\tilde{A}B'}(\mathcal{D}_{B\to B'}\circ\mathcal{N}_{A\to B})(\Psi_{\tilde{A}A})\ket{\Phi}_{RB'}\geq 1-\varepsilon.
	\end{equation}
	Therefore, by definition, the elements $(d,\Psi_{\tilde{A}A},\mathcal{D}_{B\to B'})$ constitute a $(d,\varepsilon)$ entanglement generation protocol.
	
	\begin{definition}{Subspace Transmission}{def-q_comm_state_trans}
		A \textit{subspace transmission protocol} over the quantum channel $\mathcal{N}_{A\to B}$ consists of the three elements $(d,\mathcal{E},\mathcal{D})$, where $d\geq 1$ and $\mathcal{E}$ and $\mathcal{D}$ are encoding and decoding channels. The goal of the protocol is to transmit an arbitrary pure state $\psi_{A'}$ such that the final state
		\begin{equation}
			\omega_{B'}\coloneqq (\mathcal{D}_{B\to B'}\circ\mathcal{N}_{A\to B}\circ\mathcal{E}_{A'\to A})(\psi_{A'})
		\end{equation}
		is close in fidelity to the initial state. The \textit{state transmission error} of the protocol is
		\begin{equation}
			p_{\text{err}}^{\text{(ST)}}(\mathcal{E},\mathcal{D};\mathcal{N})\coloneqq 1-\min_{\psi}\bra{\psi}\mathcal{N}(\psi)\ket{\psi}=1-F_{\min}(\mathcal{D}\circ\mathcal{N}\circ\mathcal{E}),
		\end{equation}
		where we recall the minimum fidelity of a channel defined in \eqref{eq-minimum_fidelity_chan}. We call the protocol $(d,\mathcal{E},\mathcal{D})$ a $(d,\varepsilon)$ protocol, with $\varepsilon\in[0,1]$, if $p_{\text{err}}^{\text{(ST)}}(\mathcal{E},\mathcal{D};\mathcal{N})\leq\varepsilon$.
	\end{definition}

	\begin{remark}
		An alternative way to define the error criterion for a subspace transmission code would be to use the average fidelity, defined in \eqref{eq-chan_avg_fid}; please consult the Bibliographic Notes in Section~\ref{sec:Q-cap:bib-notes}.
	\end{remark}
	
	Given a $(d,\varepsilon)$ quantum communication protocol for the channel $\mathcal{N}$ with the elements $(d,\mathcal{E},\mathcal{D})$, the equality $p_{\text{err}}^*(\mathcal{E},\mathcal{D};\mathcal{N})=1-F(\mathcal{D}\circ\mathcal{N}\circ\mathcal{E})$ holds, where $F(\cdot)$ is the channel fidelity defined in \eqref{eq-worst_case_fidelity_chan}. Then, restricting the optimization in $F(\mathcal{D}\circ\mathcal{N}\circ\mathcal{E})$ to pure states $\Psi_{RA'}=\ket{\Psi}\!\bra{\Psi}_{RA'}$ such that $\ket{\Psi}_{RA'}=\ket{\phi}_R\otimes\ket{\psi}_{A'}$, we obtain
	\begin{align}
		&p_{\text{err}}^{\text{(ST)}}(\mathcal{E},\mathcal{D};\mathcal{N})\nonumber\\
		&\quad=1-F_{\min}(\mathcal{D}\circ\mathcal{N}\circ\mathcal{E})\\
		&\quad=1-\min_{\ket{\psi}} \bra{\psi}\mathcal{N}(\psi)\ket{\psi}\\
		&\quad=1-\min_{\ket{\phi},\ket{\psi}}(\bra{\phi}_{R}\otimes\bra{\psi}_{A'})(\phi_{R}\otimes\mathcal{N}(\psi_{A'}))(\ket{\phi}_{R}\otimes\ket{\psi}_{A'})\\
		&\quad\leq 1-F(\mathcal{D}\circ\mathcal{N}\circ\mathcal{E})\\
		&\quad\leq\varepsilon.
	\end{align}
	So the elements $(d,\mathcal{E},\mathcal{D})$ form a $(d,\varepsilon)$ subspace transmission protocol.

\end{subappendices}

\chapter{Secret Key Distillation}\label{chap-secret_key_distill}

	This chapter considers the task of secret key distillation. The setting of
this task is that Alice and Bob share a bipartite quantum state $\rho_{AB}$,
and the goal is for them  to perform local operations and public communication in
order to transform $\rho_{AB}$ to a state that approximates an ideal secret key. Some questions are in order: What
is an ideal secret key and for whom is it secret?\ How much secret key can
they extract from this state?\ These are the main questions addressed in this chapter.

The information-theoretic model we assume is that the physical laboratories of
Alice and Bob are secure, so that system $A$ of the state $\rho_{AB}$ is physically secured in
Alice's laboratory and system $B$ is physically secured in Bob's. We suppose that an eavesdropper
Eve possesses a system $E$ that purifies $\rho_{AB}$. That is, if $\psi_{ABE}$
is a purification of $\rho_{AB}$, then we suppose that system $E$ of
$\psi_{ABE}$ is in Eve's possession. This model gives the eavesdropper a lot
of power. Indeed, if $\omega_{ABE^{\prime}}$ is an arbitrary extension of the state
$\rho_{AB}$, then as a consequence of Proposition~\ref{prop-extension_purif}, Eve can transform $\psi_{ABE}$
to $\omega_{ABE^{\prime}}$ by means of a channel acting on her system $E$. We
also assume that any classical data transmitted between Alice and Bob is
public, so that Eve has access to all of it.

An ideal secret key of $\log_{2}K$ secret bits is a tripartite state of the
following form:%
\begin{equation}
\overline{\Phi}_{AB}\otimes\sigma_{E}, \label{eq-SKD:ideal-key-intro}%
\end{equation}
where%
\begin{equation}
\overline{\Phi}_{AB}\coloneqq \frac{1}{K}\sum_{i=0}^{K-1}|i\rangle\!\langle
i|_{A}\otimes|i\rangle\!\langle i|_{B}.
\end{equation}
There are three salient aspects of such a tripartite key state:

\begin{enumerate}
\item The key value is uniformly random and thus hard to guess.

\item The key values in the registers of Alice and Bob are perfectly correlated. That is, if Alice
measures the key value to be $i\in\left\{  0,\ldots,K-1\right\}  $, then Bob
is guaranteed to measure the same value.

\item The overall state is a product state between systems $AB$ and $E$. This
means that Eve's system $E$ is of no use in guessing the key value.
\end{enumerate}

\noindent The goal of a secret-key distillation protocol is for Alice and Bob to
transform the initial state $\psi_{ABE}$, by means of local operations and
public classical communication, to a state that approximates an ideal key
state of the form in~\eqref{eq-SKD:ideal-key-intro}.

A secret key is useful in a communication task called the one-time pad
protocol (also known as the Vernam cipher). In this protocol, we suppose that Alice has a message $m\in
\left\{0,\ldots,K-1\right\}$ that she would like to send to Bob. By making use of the
key, Alice can calculate $\tilde{m}\coloneqq m\oplus i$, where $i$ is the key value
and the addition is modulo $K$, and then send the encrypted message $\tilde
{m}$ over a public classical channel. Since the key is ideal, no one else
besides Alice and Bob knows the precise key value $i$, and the encrypted message $\tilde{m}$ is
uniformly random, which means that it is hard to guess (i.e., there is a $1/K$
chance that an eavesdropper could guess it, which becomes small as $K$ becomes
large). When Bob receives the encrypted message $\tilde{m}$, he can calculate
$m=\tilde{m}\ominus i$ and decrypt the message $m$ because he knows the key
value $i$. This is one of the main uses of a secret key and in turn why we are
interested in secret key distillation.

It turns out that there are strong connections between entanglement
distillation from Chapter~\ref{chap-ent_distill} and secret key distillation. They are not
precisely the same tasks but there are strong links, and the structure of this
chapter follows the structure of Chapter~\ref{chap-ent_distill} quite closely. The main reason
for the strong connection is that the maximally entangled state $\Phi
_{AB}=\frac{1}{K}\sum_{i,j=0}^{K-1}|i\rangle\!\langle j|_{A}\otimes
|i\rangle\!\langle j|_{B}$ can be used to generate an ideal key state. To see
this, consider that the state $\Phi_{AB}$ is unextendible, so that the only
possible extension of it is a tensor-product extension of the form $\Phi
_{AB}\otimes\sigma_{E}$. Then, if Alice and Bob perform local measurement
channels on their systems $A$ and $B$, with respect to the computational
basis, they can realize the ideal tripartite key state of the form in~\eqref{eq-SKD:ideal-key-intro}. Thus, if one can generate maximally entangled
states, then one can generate key states. However, the converse is not true in
general, and this is what distinguishes secret key distillation from
entanglement distillation.

Similar to what we have done in previous chapters, here we establish lower and
upper bounds on the number of secret key bits that can be distilled from a
bipartite state $\rho_{AB}$. The lower bounds are given in terms of the
private information of the state, and the upper bounds are given in terms of
not only the private information but also the squashed entanglement and the
relative entropy of entanglement. The fact that we can use entanglement
measures as bounds further highlights the connection between secret key
distillation and entanglement.

\section{One-Shot Setting}

\label{sec-SKD:one-shot-setting}The one-shot setting for secret key
distillation begins with Alice and Bob sharing a state $\rho_{AB}$, and we
assume that the eavesdropper Eve has access to a system $E$ of a purification
of $\rho_{AB}$. For concreteness, let $\psi_{ABE}$ denote the purification of
$\rho_{AB}$, with system $A$ of $\psi_{ABE}$ held by Alice, $B$ by Bob, and
$E$ by Eve. Keep in mind that all purifications of $\rho_{AB}$ are related by
an isometric channel acting on the $E$ system, so that Eve can reach all
purifications easily by performing an isometric channel on her system$~E$. The
model we assume is that the laboratory of Alice is physically secure and the
quantum system$~A$ is fully contained in it. Similarly, we assume that the
laboratory of Bob is physically secure and contains the system $B$. However,
if the state $\rho_{AB}$ is mixed, then the purifying degrees of freedom in
$E$ are available to Eve (if, on the other hand, $\rho_{AB}$ is pure, then an arbitrary
purification $\psi_{ABE}$ is always a tensor-product state of the systems $AB$
and $E$ and, in this sense, it is understood that Eve does not really have
access to purifying degrees of freedom). This approach gives the most power to
the eavesdropper for the setting of secret key distillation.

In a secret-key distillation protocol, Alice and Bob are allowed to use local
operations and public classical communication (abbreviated as LOPC). An
LOPC\ channel is similar to an LOCC\ channel (as discussed in Section~\ref{subsec-LOCC_channels}), but
the critical difference is that Eve gets a copy of all of the classical data
exchanged. Recall that a generic LOCC channel $\mathcal{L}_{AB\rightarrow
A^{\prime}B^{\prime}}$\ can be written as follows, as discussed in Definition~\ref{def-LOCC}:
\begin{equation}
\mathcal{L}_{AB\rightarrow A^{\prime}B^{\prime}}=\sum_{z\in\mathcal{Z}%
}\mathcal{E}_{A\rightarrow A^{\prime}}^{z}\otimes\mathcal{F}_{B\rightarrow
B^{\prime}}^{z},
\end{equation}
where $\mathcal{Z}$ is a finite alphabet and $\{\mathcal{E}_{A\rightarrow
A^{\prime}}^{z}\}_{z\in\mathcal{Z}}$ and $\{\mathcal{F}_{B\rightarrow
B^{\prime}}^{z}\}_{z\in\mathcal{Z}}$ are sets of completely positive maps such
that the sum map $\mathcal{L}_{AB\rightarrow A^{\prime}B^{\prime}}$ is trace
preserving. Then an LOPC channel is the following enlargement of
$\mathcal{L}_{AB\rightarrow A^{\prime}B^{\prime}}$:%
\begin{equation}
\mathcal{L}_{AB\rightarrow A^{\prime}B^{\prime}Z}=\sum_{z\in\mathcal{Z}%
}\mathcal{E}_{A\rightarrow A^{\prime}}^{z}\otimes\mathcal{F}_{B\rightarrow
B^{\prime}}^{z}\otimes|z\rangle\!\langle z|_{Z}, \label{eq-SKD:LOPC-channel}%
\end{equation}
such that Eve has access to the system $Z$, which contains all of the
classical data exchanged to realize $\mathcal{L}_{AB\rightarrow A^{\prime
}B^{\prime}}$.

The goal of a secret-key distillation protocol is for Alice and Bob to produce
an approximation of an ideal secret-key state, which is defined as follows:

\begin{definition}
{Tripartite Key State}
{def:tripartite-key-state}A state $\gamma_{ABE}$
is a tripartite key state of size $K$, or containing $\log_{2}K$ bits of
secrecy, if local measurements of the $A$ and $B$ systems lead to the same
uniformly random outcome and the system $E$ is product with the measurement
outcomes. That is, after Alice and Bob send their systems through local
dephasing (measurement) channels
\begin{equation}
\mathcal{M}_{A,B}(\cdot)\coloneqq \sum_{i=0}^{K-1}|i\rangle\!\langle i|_{A,B}%
(\cdot)|i\rangle\!\langle i|_{A,B}, \label{eq:local-dephasing-key-state}%
\end{equation}
the resulting state on $AB$ and $E$ is as follows:
\begin{equation}
(\mathcal{M}_{A}\otimes\mathcal{M}_{B})(\gamma_{ABE})=\overline{\Phi}%
_{AB}\otimes\sigma_{E},
\end{equation}
for some state $\sigma_{E}$ and where $\overline{\Phi}_{AB}$ is the maximally
classically correlated state
\begin{equation}
\overline{\Phi}_{AB}\coloneqq \frac{1}{K}\sum_{i=0}^{K-1}|i\rangle\!\langle
i|_{A}\otimes|i\rangle\!\langle i|_{B}.
\end{equation}

\end{definition}

As stated in Definition~\ref{def:tripartite-key-state}, the defining aspect of
an ideal tripartite key state is that the systems $A$ and $B$ of Alice and Bob
are perfectly correlated and uniformly random. This property makes the actual
key value, which ends up being observed by both Alice and Bob, hard to guess
if there are many key values. Furthermore, the fact that the overall state is
such that it is tensor product between $AB$ and $E$ implies that Eve's system
cannot provide any help at all in guessing the key value.

With the notions above in place, we can now formally define a secret-key
distillation protocol. Such a protocol for the state $\rho_{AB}$ is defined by
the pair $(K,\mathcal{L}_{AB\rightarrow K_{A}K_{B}Z}^{\leftrightarrow})$,
where $K\in\mathbb{N}$ and $\mathcal{L}_{AB\rightarrow K_{A}K_{B}%
Z}^{\leftrightarrow}$ is an LOPC\ channel as defined in
\eqref{eq-SKD:LOPC-channel}, with $d_{K_{A}}=d_{K_{B}}=K$. The \textit{key
distillation error} $p_{\text{err}}(\mathcal{L}^{\leftrightarrow};\rho_{AB})$
of the protocol is given by the infidelity, defined as%
\begin{equation}
p_{\text{err}}(\mathcal{L}^{\leftrightarrow};\rho_{AB})\coloneqq \inf_{\gamma
_{K_{A}K_{B}EZ}}\left(  1-F(\gamma_{K_{A}K_{B}EZ},\mathcal{L}_{AB\rightarrow
K_{A}K_{B}Z}^{\leftrightarrow}(\psi_{ABE}))\right)  ,
\label{eq-SKD:error-criterion}%
\end{equation}
where the optimization is with respect to every tripartite key state
$\gamma_{K_{A}K_{B}EZ}$ of size $K$, which is of the form in
Definition~\ref{def:tripartite-key-state} under the identifications
$K_{A}\leftrightarrow A$, $K_{B}\leftrightarrow B$, and $EZ\leftrightarrow E$.
Furthermore, $\psi_{ABE}$ is a purification of $\rho_{AB}$. Note that the key
distillation error $p_{\text{err}}(\mathcal{L}^{\leftrightarrow};\rho_{AB})$
is invariant under the choice of a purification $\psi_{ABE}$ because it
involves an optimization over every tripartite key state $\gamma_{K_{A}%
K_{B}EZ}$, purifications are related by isometric channels, the fidelity is
invariant under isometric channels, and $\mathcal{V}_{E}(\gamma_{K_{A}K_{B}%
EZ})$ is an ideal tripartite key state if $\gamma_{K_{A}K_{B}EZ}$ is, where
$\mathcal{V}_{E}$ is an isometric channel. The optimization in
\eqref{eq-SKD:error-criterion} guarantees the existence of at least one state
$\sigma_{EZ}$ of the eavesdropper such that the actual state $\mathcal{L}%
_{AB\rightarrow K_{A}K_{B}Z}^{\leftrightarrow}(\psi_{ABE})$ of the protocol
approximates an ideal tripartite key state, in the following sense%
\begin{align}
(\mathcal{M}_{K_{A}}\otimes\mathcal{M}_{K_{B}})(\mathcal{L}_{AB\rightarrow
K_{A}K_{B}Z}^{\leftrightarrow}(\psi_{ABE}))  &  \approx(\mathcal{M}_{K_{A}%
}\otimes\mathcal{M}_{K_{B}})(\gamma_{K_{A}K_{B}EZ})\\
&  =\overline{\Phi}_{K_{A}K_{B}}\otimes\sigma_{EZ},
\end{align}
if the key distillation error $p_{\text{err}}(\mathcal{L}^{\leftrightarrow
};\rho_{AB})$ is small.

At this point, it might not be clear why we employ the infidelity error
criterion in \eqref{eq-SKD:error-criterion} rather than the normalized trace
distance. We did so in Chapter~\ref{chap-ent_distill} in the context of entanglement
distillation because it corresponded to the operational notion of an
entanglement test (see \eqref{eq:ED:entanglement-test}). We later show how the infidelity error
criterion corresponds to the operational notion of a \textquotedblleft privacy
test,\textquotedblright\ which justifies its use in the context of secret key distillation.

\begin{definition}
{$(K,\varepsilon)$ secret-key distillation protocol}{} A secret key distillation
protocol $(K,\mathcal{L}_{AB\rightarrow K_{A}K_{B}Z}^{\leftrightarrow})$ for
the state $\rho_{AB}$ is called a $(K,\varepsilon)$ \textit{protocol}, with
$\varepsilon\in\lbrack0,1]$, if $p_{\text{err}}(\mathcal{L}^{\leftrightarrow
};\rho_{AB})\leq\varepsilon$.
\end{definition}

Given $\varepsilon\in\lbrack0,1]$, the largest number $\log_{2}K$ of
$\varepsilon$-approximate secret-key bits that can be extracted from a state
$\rho_{AB}$ among all $(K,\varepsilon)$ secret-key distillation protocols is
called the \textit{one-shot }$\varepsilon$\textit{-distillable key of}
$\rho_{AB}$.

\begin{definition}
{One-Shot Distillable Key}{} Given a bipartite state $\rho_{AB}$ and
$\varepsilon\in\lbrack0,1]$, the \textit{one-shot distillable key of}
$\rho_{AB}$, denoted by $K_{D}^{\varepsilon}(\rho_{AB})\equiv K_{D}%
^{\varepsilon}(A;B)_{\rho}$, is defined as%
\begin{equation}
K_{D}^{\varepsilon}(A;B)_{\rho}\coloneqq \sup_{(K,\mathcal{L}^{\leftrightarrow}%
)}\left\{  \log_{2}K:p_{\text{err}}(\mathcal{L}^{\leftrightarrow};\rho
_{AB})\leq\varepsilon\right\}  , \label{eq-SKD:one-shot-distillable-key-def}%
\end{equation}
where the optimization is over all $K\in\mathbb{N}$ and every LOPC\ channel
$\mathcal{L}_{AB\rightarrow K_{A}K_{B}Z}^{\leftrightarrow}$ with $d_{K_{A}%
}=d_{K_{B}}=K$.
\end{definition}

Calculating the one-shot distillable key is difficult computationally because
it involves optimizing over the key size $K$ and over every LOPC\ channel
$\mathcal{L}_{AB\rightarrow K_{A}K_{B}Z}^{\leftrightarrow}$, with $d_{K_{A}%
}=d_{K_{B}}=K$. We thus try to estimate the one-shot distillable key by
determining upper and lower bounds on it. Section~\ref{sec-SKD:upper-bounds}
introduces upper bounds on the one-shot distillable key. Before doing so, we
first clarify how secret-key distillation protocols can be thought of from a
different perspective as bipartite private-state distillation protocols.

\subsection{Tripartite Key States and Bipartite Private States}

\label{sec-SKD:tri-to-bi-relate}

An important insight for secret key distillation is that there is a way to
describe the whole theory exclusively in terms of a bipartite scenario. This
is related to the assumption that the eavesdropper Eve possesses a full
purification $\psi_{ABE}$\ of the original state $\rho_{AB}$, along with the
structure of quantum mechanics.

To motivate this concept, consider that an approximate tripartite state
$\gamma_{ABE}$ (as described in Definition~\ref{def:tripartite-key-state})\ is
generated at the end of a key distillation protocol, and it is such that all
that the eavesdropper possesses is only available in the system $E$ (in this
context, let us make the same identifications  $K_{A}%
\leftrightarrow A$, $K_{B}\leftrightarrow B$, and $EZ\leftrightarrow E$
discussed around \eqref{eq-SKD:error-criterion}). As such, we can consider a
purification of the state $\gamma_{ABE}$ of the form $\gamma_{AA^{\prime
}BB^{\prime}E}$, in which the joint system $A^{\prime}B^{\prime}$ constitutes
the purifying system. Since a secret-key distillation protocol involves only
three parties, and we already argued that the system $E$ is all that Eve
possesses, it follows that Alice and Bob jointly possess the purifying system,
which can be split among them as $A^{\prime}B^{\prime}$. The reduced state
$\gamma_{AA^{\prime}BB^{\prime}}=\operatorname{Tr}_{E}[\gamma_{AA^{\prime
}BB^{\prime}E}]$ is then a bipartite state because all systems involved are in
possession of Alice and Bob. If the original state $\gamma_{ABE}$ is a
tripartite key state according to Definition~\ref{def:tripartite-key-state},
then by constructing $\gamma_{AA^{\prime}BB^{\prime}}$ according to this
procedure, the resulting state is called a bipartite private state, and it has
a particular structure. Conversely, if $\gamma_{AA^{\prime}BB^{\prime}}$ is a
state with the structure of a bipartite private state, then it follows that by
purifying this state to $\gamma_{AA^{\prime}BB^{\prime}E}$ with an $E$ system
and tracing over systems $A^{\prime}$ and $B^{\prime}$, we arrive at a
tripartite key state. So there is an equivalence between these two viewpoints
(tripartite picture of key distillation and bipartite picture of private state
distillation). We develop this correspondence in detail in what follows.

Before starting, we briefly mention that the equivalence between the
tripartite and bipartite pictures of key distillation implies that we can
bring the tools of entanglement theory (Chapter~\ref{chap-ent_measures}) to
bear on the problem of establishing upper bounds on the number of approximate
secret-key bits that can be generated in a key distillation protocol. This is
one of the main applications of this correspondence, and we note here that it
has led to other insights in quantum information theory.

\begin{definition}
{Bipartite Private State}{def:private-state-bi}A state $\gamma
_{ABA^{\prime}B^{\prime}}$\ is a bipartite private state of size $K$,
containing $\log_{2}K$ bits of secrecy, if after purifying $\gamma
_{ABA^{\prime}B^{\prime}}$ to a pure state $\gamma_{ABA^{\prime}B^{\prime}E}$
with purifying system $E$ and tracing over the systems $A^{\prime}B^{\prime}$,
the resulting state $\gamma_{ABE}$ is a tripartite key state of size $K$. The
systems $A$ and $B$ are called key systems, and the systems $A^{\prime}$ and
$B^{\prime}$ are called shield systems.
\end{definition}

\begin{theorem}
{thm-private_state} A state $\gamma_{ABA^{\prime}B^{\prime}}$ is a
bipartite private state if and only if it has the following form:%
\begin{equation}
\gamma_{ABA^{\prime}B^{\prime}}=U_{ABA^{\prime}B^{\prime}}\left(  \Phi
_{AB}\otimes\theta_{A^{\prime}B^{\prime}}\right)  U_{ABA^{\prime}B^{\prime}%
}^{\dag}, \label{eq:form-for-private-states}%
\end{equation}
where $\Phi_{AB}$ is a maximally entangled state of Schmidt rank $K$:%
\begin{equation}
\Phi_{AB}\coloneqq \frac{1}{K}\sum_{i,j=0}^{K-1}|i\rangle\!\langle j|_{A}%
\otimes|i\rangle\!\langle j|_{B},
\end{equation}
$\theta_{A^{\prime}B^{\prime}}$ is some state, and $U_{ABA^{\prime}B^{\prime}%
}$ is a global twisting unitary of the following form:%
\begin{equation}
U_{ABA^{\prime}B^{\prime}}\coloneqq \sum_{i,j=0}^{K-1}|i\rangle\!\langle i|_{A}%
\otimes|j\rangle\!\langle j|_{B}\otimes U_{A^{\prime}B^{\prime}}^{ij}.
\label{eq:twisting-unitary}%
\end{equation}
In the above, $U_{A^{\prime}B^{\prime}}^{ij}$ is a unitary operator for all
$i,j\in\{0,\ldots,K-1\}$.
\end{theorem}

\begin{Proof}
Suppose that $\gamma_{ABA^{\prime}B^{\prime}}$ has the form in
\eqref{eq:form-for-private-states}. A particular purification of
$\gamma_{ABA^{\prime}B^{\prime}}$ is%
\begin{align}
&  |\phi^{\gamma}\rangle_{ABA^{\prime}B^{\prime}E}\nonumber\\
&  =U_{ABA^{\prime}B^{\prime}}|\Phi\rangle_{AB}\otimes|\psi^{\theta}%
\rangle_{A^{\prime}B^{\prime}E}\\
&  =\left(  \sum_{i,j=0}^{K-1}|i\rangle\!\langle i|_{A}\otimes|j\rangle
\!\langle j|_{B}\otimes U_{A^{\prime}B^{\prime}}^{ij}\right)  \left(  \frac
{1}{\sqrt{K}}\sum_{k=0}^{K-1}|k\rangle_{A}|k\rangle_{B}\otimes|\psi^{\theta
}\rangle_{A^{\prime}B^{\prime}E}\right) \\
&  =\frac{1}{\sqrt{K}}\sum_{k=0}^{K-1}|k\rangle_{A}|k\rangle_{B}\otimes
U_{A^{\prime}B^{\prime}}^{kk}|\psi^{\theta}\rangle_{A^{\prime}B^{\prime}E},
\end{align}
where $|\psi^{\theta}\rangle_{A^{\prime}B^{\prime}E}$ purifies $\theta
_{A^{\prime}B^{\prime}}$. The local dephasing channels in
\eqref{eq:local-dephasing-key-state}\ lead to the following state%
\begin{multline}
(\mathcal{M}_{A}\otimes\mathcal{M}_{B})(|\phi^{\gamma}\rangle\!\langle
\phi^{\gamma}|_{ABA^{\prime}B^{\prime}E})\\
=\frac{1}{K}\sum_{k=0}^{K-1}|k\rangle\!\langle k|_{A}\otimes|k\rangle\!\langle
k|_{B}\otimes U_{A^{\prime}B^{\prime}}^{kk}|\psi^{\theta}\rangle\!\langle
\psi^{\theta}|_{A^{\prime}B^{\prime}E}\left(  U_{A^{\prime}B^{\prime}}%
^{kk}\right)  ^{\dag}.
\end{multline}
Taking a partial trace over the $A^{\prime}B^{\prime}$ systems leads to%
\begin{align}
&  \operatorname{Tr}_{A^{\prime}B^{\prime}}\!\left[  \frac{1}{K}\sum
_{k=0}^{K-1}|k\rangle\!\langle k|_{A}\otimes|k\rangle\!\langle k|_{B}\otimes
U_{A^{\prime}B^{\prime}}^{kk}|\psi^{\theta}\rangle\!\langle\psi^{\theta
}|_{A^{\prime}B^{\prime}E}\left(  U_{A^{\prime}B^{\prime}}^{kk}\right)
^{\dag}\right] \nonumber\\
&  =\frac{1}{K}\sum_{k=0}^{K-1}|k\rangle\!\langle k|_{A}\otimes|k\rangle
\!\langle k|_{B}\otimes\operatorname{Tr}_{A^{\prime}B^{\prime}}\!\left[
U_{A^{\prime}B^{\prime}}^{kk}|\psi^{\theta}\rangle\!\langle\psi^{\theta
}|_{A^{\prime}B^{\prime}E}\left(  U_{A^{\prime}B^{\prime}}^{kk}\right)
^{\dag}\right] \\
&  =\frac{1}{K}\sum_{k=0}^{K-1}|k\rangle\!\langle k|_{A}\otimes|k\rangle
\!\langle k|_{B}\otimes\operatorname{Tr}_{A^{\prime}B^{\prime}}\!\left[
\left(  U_{A^{\prime}B^{\prime}}^{kk}\right)  ^{\dag}U_{A^{\prime}B^{\prime}%
}^{kk}|\psi^{\theta}\rangle\!\langle\psi^{\theta}|_{A^{\prime}B^{\prime}%
E}\right] \\
&  =\overline{\Phi}_{AB}\otimes\rho_{E}.
\end{align}
Thus, the particular purification $|\phi^{\gamma}\rangle_{ABA^{\prime
}B^{\prime}E}$ leads to a tripartite key state on systems $ABE$. Now, in the
development above, we chose a particular purification of $\gamma_{ABA^{\prime
}B^{\prime}}$. However, given that all purifications are related by isometries
acting on the purifying system, every purification can be written as
$V_{E\rightarrow E^{\prime}}|\phi^{\gamma}\rangle_{ABA^{\prime}B^{\prime}E}$
for some isometry $V_{E\rightarrow E^{\prime}}$. Then repeating the
calculation above gives that the reduced state on $ABE^{\prime}$ after local
dephasing channels on $A$ and $B$ is%
\begin{equation}
\overline{\Phi}_{AB}\otimes V_{E\rightarrow E^{\prime}}\rho_{E}(V_{E\rightarrow E^{\prime}})^{\dag},
\end{equation}
so that there is no correlation between the measurement outcomes of Alice and
Bob and the system $E^{\prime}$. Furthermore, the measurement outcomes are
perfectly correlated and uniformly random. So we conclude that a state
$\gamma_{ABA^{\prime}B^{\prime}}$ of the form in
\eqref{eq:form-for-private-states} is a bipartite private state.

Conversely, suppose now that $\gamma_{ABA^{\prime}B^{\prime}}$ is a bipartite
private state held by Alice and Bob, and let $|\phi^{\gamma}\rangle
_{ABA^{\prime}B^{\prime}E}$ be a purification of it, with $E$ the purifying
system. Expanding the state in the basis of the local measurements of Alice
and Bob gives%
\begin{equation}
|\phi^{\gamma}\rangle_{ABA^{\prime}B^{\prime}E}=\sum_{i,j=0}^{K-1}\alpha
_{i,j}|i\rangle_{A}|j\rangle_{B}|\phi_{i,j}\rangle_{A^{\prime}B^{\prime}E},
\end{equation}
for some states $|\phi_{i,j}\rangle_{A^{\prime}B^{\prime}E}$ and probability
amplitudes $\{\alpha_{i,j}\}_{i,j}$. However, in order for the measurement
outcomes of Alice and Bob to be perfectly correlated and uniformly random, it
is necessary that%
\begin{equation}
|\alpha_{i,j}|^{2}=\left\{
\begin{array}
[c]{cc}%
\frac{1}{K} & \text{if }i=j\\
0 & \text{if }i\neq j
\end{array}
\right.  .
\end{equation}
(Any other values for the amplitudes $\alpha_{i,j}$ would lead to a different
distribution upon measurement of the $A$ and $B$ systems.) So the global state
should have the following form:%
\begin{equation}
|\phi^{\gamma}\rangle_{ABA^{\prime}B^{\prime}E}=\sum_{i=0}^{K-1}\frac{1}%
{\sqrt{K}}|i\rangle_{A}|i\rangle_{B}e^{i\varphi_{i}}|\phi_{i,i}\rangle
_{A^{\prime}B^{\prime}E}.
\end{equation}
In order for the reduced density operator on $E$ to be independent of the
measurement outcomes of Alice and Bob, it is necessary for it to be a fixed
state with no dependence on $i$:%
\begin{equation}
\text{Tr}_{A^{\prime}B^{\prime}}[|\phi_{i,i}\rangle\!\langle\phi
_{i,i}|_{A^{\prime}B^{\prime}E}]=\sigma_{E}.
\end{equation}
In such a case, then all of the states $|\phi_{i,i}\rangle_{A^{\prime
}B^{\prime}E}$ are purifications of the same state $\sigma_{E}$, so that there
exists a unitary $U_{A^{\prime}B^{\prime}}^{i}$ relating each $|\phi
_{i,i}\rangle_{A^{\prime}B^{\prime}E}$ to a fixed purification $|\phi^{\sigma
}\rangle_{A^{\prime}B^{\prime}E}$\ of $\sigma$:%
\begin{equation}
e^{i\varphi_{i}}|\phi_{i,i}\rangle_{A^{\prime}B^{\prime}E}=U_{A^{\prime
}B^{\prime}}^{i}|\phi^{\sigma}\rangle_{A^{\prime}B^{\prime}E}.
\end{equation}
Thus, we can write the global state as%
\begin{equation}
\frac{1}{\sqrt{K}}\sum_{i=0}^{K-1}|i\rangle_{A}|i\rangle_{B}U_{A^{\prime
}B^{\prime}}^{i,i}|\phi^{\sigma}\rangle_{A^{\prime}B^{\prime}E},
\end{equation}
which is equivalent to%
\begin{equation}
\left(  \sum_{i,j=0}^{K-1}|i\rangle\!\langle i|_{A}\otimes|j\rangle\!\langle
j|_{B}\otimes U_{A^{\prime}B^{\prime}}^{i,j}\right)  |\Phi\rangle_{AB}%
\otimes|\phi^{\sigma}\rangle_{A^{\prime}B^{\prime}E},
\end{equation}
after setting $U_{A^{\prime}B^{\prime}}^{i,j}=U_{A^{\prime}B^{\prime}}^{i}$
for all $j\in\{0,\ldots,K-1\}$ (there is in fact full freedom in how the
unitary $U_{A^{\prime}B^{\prime}}^{i,j}$ is chosen for $i\neq j$). One can now
deduce that the reduced state on systems $ABA^{\prime}B^{\prime}$ has the form
in \eqref{eq:form-for-private-states}.
\end{Proof}

\begin{definition}
{$\boldsymbol{\varepsilon}$-Approximate Tripartite Key State}%
{def:approx-priv} Fix $\varepsilon\in\left[  0,1\right]  $. A state
$\rho_{ABE}$ is an $\varepsilon$-approximate tripartite key state if there
exists a tripartite key state $\gamma_{ABE}$, as in
Definition~\ref{def:tripartite-key-state}, such that%
\begin{equation}
F(\rho_{ABE},\gamma_{ABE})\geq1-\varepsilon. \label{eq:approx-tri-priv-state}%
\end{equation}
Similarly, a state $\rho_{ABA^{\prime}B^{\prime}}$ is an $\varepsilon
$-approximate bipartite private state if there exists a bipartite private
state $\gamma_{ABA^{\prime}B^{\prime}}$, as in
Definition~\ref{def:private-state-bi}, such that%
\begin{equation}
F(\rho_{ABA^{\prime}B^{\prime}},\gamma_{ABA^{\prime}B^{\prime}})\geq
1-\varepsilon. \label{eq:approx-bi-priv-state}%
\end{equation}

\end{definition}

Approximate tripartite key states are in one-to-one correspondence with
approximate bipartite private states, as summarized below:

\begin{proposition}
{prop-approx_key_state} If $\rho_{ABA^{\prime}B^{\prime}}$ is an
$\varepsilon$-approximate bipartite key state with $K$ key values, then the
state $\rho_{ABE}$ is an $\varepsilon$-approximate tripartite key state with
$K$ key values, where $\rho_{ABE}=\operatorname{Tr}_{A^{\prime}B^{\prime}%
}[\psi_{ABA^{\prime}B^{\prime}E}^{\rho}]$ and $\psi_{ABA^{\prime}B^{\prime}%
E}^{\rho}$ is an arbitrary purification of $\rho_{ABA^{\prime}B^{\prime}}$.
The converse statement is true as well.
\end{proposition}

\begin{Proof}
Suppose that the inequality in \eqref{eq:approx-tri-priv-state} is satisfied.
Let $\psi_{ABA^{\prime}B^{\prime}E}^{\rho}$ be a purification of $\rho_{ABE}$.
Then by applying Uhlmann's theorem (Theorem~\ref{thm-Uhlmann_fidelity}), there exists a
purification $\gamma_{ABA^{\prime}B^{\prime}E}$\ of $\gamma_{ABE}$ such that%
\begin{equation}
F(\rho_{ABE},\gamma_{ABE})=F(\psi_{ABA^{\prime}B^{\prime}E}^{\rho}%
,\gamma_{ABA^{\prime}B^{\prime}E}).
\end{equation}
Tracing over the $E$ system and applying the data-processing inequality for
fidelity (Theorem~\ref{thm-fidelity_monotone}), we conclude that%
\begin{equation}
F(\psi_{ABA^{\prime}B^{\prime}}^{\rho},\gamma_{ABA^{\prime}B^{\prime}}%
)\geq1-\varepsilon.
\end{equation}
Since $\gamma_{ABE}$ is an ideal tripartite key state and the state
$\gamma_{ABA^{\prime}B^{\prime}}$ arises from it via purification and tracing
over system $E$, it follows from Definition~\ref{def:private-state-bi} that
$\gamma_{ABA^{\prime}B^{\prime}}$ is an ideal bipartite private state. In
turn, according to Definition~\ref{def:approx-priv}, it follows that
$\psi_{ABA^{\prime}B^{\prime}}^{\rho}$ is an $\varepsilon$-approximate
bipartite private state.

For the other implication, suppose that the inequality in
\eqref{eq:approx-bi-priv-state} is satisfied. Let $\psi_{ABA^{\prime}%
B^{\prime}E}^{\rho}$ be a purification of $\rho_{ABA^{\prime}B^{\prime}}$. By
applying Uhlmann's theorem (Theorem~\ref{thm-Uhlmann_fidelity}), there exists a purification $\gamma_{ABA^{\prime
}B^{\prime}E}$ of the ideal bipartite private state $\gamma_{ABA^{\prime
}B^{\prime}}$ such that%
\begin{equation}
F(\rho_{ABA^{\prime}B^{\prime}},\gamma_{ABA^{\prime}B^{\prime}})=F(\psi
_{ABA^{\prime}B^{\prime}E}^{\rho},\gamma_{ABA^{\prime}B^{\prime}E})
\end{equation}
Tracing over the $A^{\prime}B^{\prime}$ systems and applying the
data-processing inequality for fidelity, we conclude that%
\begin{equation}
F(\psi_{ABE}^{\rho},\gamma_{ABE})\geq1-\varepsilon.
\end{equation}
Since $\gamma_{ABA^{\prime}B^{\prime}}$ is an ideal bipartite private state
and the state $\gamma_{ABE}$ arises from it via purification and tracing over
systems $A^{\prime}B^{\prime}$, it follows from
Definition~\ref{def:private-state-bi} that $\gamma_{ABE}$ is an ideal
tripartite key state. In turn, according to Definition~\ref{def:approx-priv},
it follows that $\psi_{ABE}^{\rho}$ is an $\varepsilon$-approximate tripartite
key state.
\end{Proof}

\subsection{Equivalence of Tripartite Key Distillation and Bipartite Private
State Distillation}

\label{sec-SKD:tri-bi-equivalence}

The equivalence between ideal and
approximate tripartite key states and bipartite private states extends
further, and it is a correspondence that allows us to consider secret key
distillation in the bipartite picture. To this end, we define a
bipartite private-state distillation protocol, and then we prove the equivalence.

A bipartite private-state distillation protocol for the state $\rho_{AB}$ is
defined by the pair $(K,\mathcal{L}_{AB\rightarrow K_{A}K_{B}A^{\prime
}B^{\prime}}^{\leftrightarrow})$, where $K\in\mathbb{N}$ and $\mathcal{L}%
_{AB\rightarrow K_{A}K_{B}A^{\prime}B^{\prime}}^{\leftrightarrow}$ is an LOCC
channel with $d_{K_{A}}=d_{K_{B}}=K$. The key distillation error
$p_{\text{err}}^{b}(\mathcal{L}^{\leftrightarrow};\rho_{AB})$ of the protocol
is given in terms of the infidelity, defined as%
\begin{equation}
p_{\text{err}}^{b}(\mathcal{L}^{\leftrightarrow};\rho_{AB})\coloneqq \inf
_{\gamma_{K_{A}K_{B}A^{\prime}B^{\prime}}}\left(  1-F(\gamma_{K_{A}%
K_{B}A^{\prime}B^{\prime}},\mathcal{L}_{AB\rightarrow K_{A}K_{B}A^{\prime
}B^{\prime}}^{\leftrightarrow}(\rho_{AB}))\right)  ,
\end{equation}
where the optimization is with respect to every bipartite private state
$\gamma_{K_{A}K_{B}A^{\prime}B^{\prime}}$ such that $d_{K_{A}}=d_{K_{B}}=K$.

\begin{definition}
{$(K,\varepsilon)$ Private-State Distillation Protocol}{} A bipartite
private-state distillation protocol $(K,\mathcal{L}_{AB\rightarrow K_{A}%
K_{B}A^{\prime}B^{\prime}}^{\leftrightarrow})$ for the state $\rho_{AB}$ is
called a $(K,\varepsilon)$ protocol, with $\varepsilon\in\lbrack0,1]$, if
$p_{\text{err}}^{b}(\mathcal{L}^{\leftrightarrow};\rho_{AB})\leq\varepsilon$.
\end{definition}

We now establish the main result of this section, which is the equivalence of
tripartite key distillation and bipartite private-state distillation:

\begin{theorem}
{thm-SKD:tri-bi-equivalence}Let $K\in\mathbb{N}$ and $\varepsilon
\in\left[  0,1\right]  $. Let $\rho_{AB}$ be a bipartite state. There exists a
$(K,\varepsilon)$ tripartite key distillation protocol for $\rho_{AB}$ if and
only if there exists a $(K,\varepsilon)$\ bipartite private-state distillation
protocol for $\rho_{AB}$.
\end{theorem}

\begin{Proof}
We start by proving that there exists a $(K,\varepsilon)$ bipartite
private-state distillation protocol if there exists a $(K,\varepsilon)$
tripartite key distillation protocol. Let $\psi_{ABE}$ be a purification of
$\rho_{AB}$, let $\mathcal{L}_{AB\rightarrow K_{A}K_{B}Z}^{\leftrightarrow}$
be the LOPC\ channel realizing the key distillation, and let $\gamma_{K_{A}K_{B}%
EZ}$ be a tripartite key state such that%
\begin{equation}
1-F(\gamma_{K_{A}K_{B}EZ},\mathcal{L}_{AB\rightarrow K_{A}K_{B}Z}%
^{\leftrightarrow}(\psi_{ABE}))\leq\varepsilon.
\end{equation}
The LOPC channel $\mathcal{L}_{AB\rightarrow K_{A}K_{B}Z}^{\leftrightarrow}%
$\ has the form in \eqref{eq-SKD:LOPC-channel}, so that%
\begin{equation}
\mathcal{L}_{AB\rightarrow K_{A}K_{B}Z}^{\leftrightarrow}=\sum_{z\in
\mathcal{Z}}\mathcal{E}_{A\rightarrow K_{A}}^{z}\otimes\mathcal{F}%
_{B\rightarrow K_{B}}^{z}\otimes|z\rangle\!\langle z|_{Z}.
\end{equation}
An isometric extension $U_{AB\rightarrow K_{A}K_{B}A^{\prime}B^{\prime}%
Z}^{\mathcal{L}^{\leftrightarrow}}$\ of this LOPC\ channel is as follows:%
\begin{equation}
U_{AB\rightarrow K_{A}K_{B}A^{\prime}B^{\prime}Z}^{\mathcal{L}%
^{\leftrightarrow}}\coloneqq \sum_{z\in\mathcal{Z}}V_{A\rightarrow K_{A}A^{\prime}%
}^{\mathcal{E}^{z}}\otimes V_{B\rightarrow K_{B}B^{\prime}}^{\mathcal{F}^{z}%
}\otimes|z\rangle_{Z},
\end{equation}
where $\{V_{A\rightarrow K_{A}A^{\prime}}^{\mathcal{E}^{z}}\}_{z\in
\mathcal{Z}}$ and $\{V_{B\rightarrow K_{B}B^{\prime}}^{\mathcal{F}^{z}%
}\}_{z\in\mathcal{Z}}$ are sets of linear operators such that
$U_{AB\rightarrow K_{A}K_{B}A^{\prime}B^{\prime}Z}^{\mathcal{L}%
^{\leftrightarrow}}$ is an isometry and%
\begin{equation}
\operatorname{Tr}_{A^{\prime}B^{\prime}}\circ\mathcal{U}_{AB\rightarrow
K_{A}K_{B}A^{\prime}B^{\prime}Z}^{\mathcal{L}^{\leftrightarrow}}%
=\mathcal{L}_{AB\rightarrow K_{A}K_{B}Z}^{\leftrightarrow},
\end{equation}
with%
\begin{equation}
\mathcal{U}_{AB\rightarrow K_{A}K_{B}A^{\prime}B^{\prime}Z}^{\mathcal{L}%
^{\leftrightarrow}}(\cdot)\coloneqq U_{AB\rightarrow K_{A}K_{B}A^{\prime}B^{\prime}%
Z}^{\mathcal{L}^{\leftrightarrow}}(\cdot)(U_{AB\rightarrow K_{A}K_{B}%
A^{\prime}B^{\prime}Z}^{\mathcal{L}^{\leftrightarrow}})^{\dag}.
\end{equation}
To meet these requirements, note that it is necessary for each
$V_{A\rightarrow K_{A}A^{\prime}}^{\mathcal{E}^{z}}$ and $V_{B\rightarrow
K_{B}B^{\prime}}^{\mathcal{F}^{z}}$ to be a contraction, i.e., satisfying%
\begin{equation}
\left\Vert V_{A\rightarrow K_{A}A^{\prime}}^{\mathcal{E}^{z}}\right\Vert
_{\infty},\left\Vert V_{B\rightarrow K_{B}B^{\prime}}^{\mathcal{F}^{z}%
}\right\Vert _{\infty}\leq1.
\end{equation}
It then follows that the state $\mathcal{U}_{AB\rightarrow K_{A}K_{B}%
A^{\prime}B^{\prime}Z}^{\mathcal{L}^{\leftrightarrow}}(\psi_{ABE})$ purifies
$\mathcal{L}_{AB\rightarrow K_{A}K_{B}Z}^{\leftrightarrow}(\psi_{ABE})$, and
by applying Uhlmann's theorem (Theorem~\ref{thm-Uhlmann_fidelity}), there exists a pure state $\gamma_{K_{A}%
K_{B}A^{\prime}B^{\prime}EZ}$ satisfying%
\begin{multline}
F(\gamma_{K_{A}K_{B}EZ},\mathcal{L}_{AB\rightarrow K_{A}K_{B}Z}%
^{\leftrightarrow}(\psi_{ABE}))\\
=F(\gamma_{K_{A}K_{B}A^{\prime}B^{\prime}EZ},\mathcal{U}_{AB\rightarrow
K_{A}K_{B}A^{\prime}B^{\prime}Z}^{\mathcal{L}^{\leftrightarrow}}(\psi_{ABE})).
\end{multline}
Now applying the same reasoning given in
Proposition~\ref{prop-approx_key_state}, we conclude that the following
inequality holds%
\begin{equation}
1-F(\gamma_{K_{A}K_{B}A^{\prime}B^{\prime}},(\operatorname{Tr}_{Z}%
\circ\mathcal{U}_{AB\rightarrow K_{A}K_{B}A^{\prime}B^{\prime}Z}%
^{\mathcal{L}^{\leftrightarrow}})(\rho_{AB}))\leq\varepsilon,
\end{equation}
where $\gamma_{K_{A}K_{B}A^{\prime}B^{\prime}}$ is an ideal bipartite private
state of size $K$. Note that the channel $\operatorname{Tr}_{Z}\circ
\mathcal{U}_{AB\rightarrow K_{A}K_{B}A^{\prime}B^{\prime}Z}^{\mathcal{L}%
^{\leftrightarrow}}$ is an LOCC channel, because it has the following form:%
\begin{equation}
\operatorname{Tr}_{Z}\circ\mathcal{U}_{AB\rightarrow K_{A}K_{B}A^{\prime
}B^{\prime}Z}^{\mathcal{L}^{\leftrightarrow}}=\sum_{z\in\mathcal{Z}%
}\mathcal{V}_{A\rightarrow K_{A}A^{\prime}}^{\mathcal{E}^{z}}\otimes
\mathcal{V}_{B\rightarrow K_{B}B^{\prime}}^{\mathcal{F}^{z}}.
\end{equation}
Thus, there exists a $(K,\varepsilon)$ bipartite private-state distillation
protocol if there exists a $(K,\varepsilon)$\ tripartite key distillation protocol.

We now prove the opposite implication. Suppose that there exists a
$(K,\varepsilon)$ bipartite private-state distillation protocol. Let
$\mathcal{L}_{AB\rightarrow K_{A}K_{B}A^{\prime}B^{\prime}}^{\leftrightarrow}$
be the LOCC channel realizing the private-state distillation, and let
$\gamma_{K_{A}K_{B}A^{\prime}B^{\prime}}$ be an ideal bipartite private state
satisfying%
\begin{equation}
1-F(\gamma_{K_{A}K_{B}A^{\prime}B^{\prime}},\mathcal{L}_{AB\rightarrow
K_{A}K_{B}A^{\prime}B^{\prime}}^{\leftrightarrow}(\rho_{AB}))\leq\varepsilon.
\end{equation}
Let $\psi_{ABE}$ be a purification of $\rho_{AB}$. Suppose that the LOCC
channel $\mathcal{L}_{AB\rightarrow K_{A}K_{B}A^{\prime}B^{\prime}%
}^{\leftrightarrow}$ has the following form:%
\begin{equation}
\mathcal{L}_{AB\rightarrow K_{A}K_{B}A^{\prime}B^{\prime}}^{\leftrightarrow
}=\sum_{z\in\mathcal{Z}}\mathcal{E}_{A\rightarrow K_{A}A^{\prime}}^{z}%
\otimes\mathcal{F}_{B\rightarrow K_{B}B^{\prime}}^{z},
\end{equation}
where $\mathcal{Z}$ is a finite alphabet and $\{\mathcal{E}_{A\rightarrow
K_{A}A^{\prime}}^{z}\}_{z\in\mathcal{Z}}$ and $\{\mathcal{F}_{B\rightarrow
K_{B}B^{\prime}}^{z}\}_{z\in\mathcal{Z}}$ are sets of completely positive maps
such that $\mathcal{L}_{AB\rightarrow K_{A}K_{B}A^{\prime}B^{\prime}%
}^{\leftrightarrow}$ is trace preserving. Without loss of generality, we can
suppose that each completely positive map $\mathcal{E}_{A\rightarrow
K_{A}A^{\prime}}^{z}$ consists of a single Kraus operator, and we can suppose
the same for $\mathcal{F}_{B\rightarrow K_{B}B^{\prime}}^{z}$ (the reasoning here is similar to that given in the remark after Definition~\ref{def-sep_ent_state}). Let us denote
these as $E_{A\rightarrow K_{A}A^{\prime}}^{z}$ and $F_{B\rightarrow
K_{B}B^{\prime}}^{z}$, respectively. Then an isometric extension of this
LOCC\ channel is as follows:%
\begin{equation}
U_{AB\rightarrow K_{A}K_{B}A^{\prime}B^{\prime}Z}^{\mathcal{L}%
^{\leftrightarrow}}\coloneqq \sum_{z\in\mathcal{Z}}E_{A\rightarrow K_{A}A^{\prime}%
}^{z}\otimes F_{B\rightarrow K_{B}B^{\prime}}^{z}\otimes|z\rangle_{Z}.
\end{equation}
Thus, the state $\mathcal{U}_{AB\rightarrow K_{A}K_{B}A^{\prime}B^{\prime}%
Z}^{\mathcal{L}^{\leftrightarrow}}(\psi_{ABE})$ is a purification of
$\mathcal{L}_{AB\rightarrow K_{A}K_{B}A^{\prime}B^{\prime}}^{\leftrightarrow
}(\rho_{AB})$. Applying Uhlmann's theorem (Theorem~\ref{thm-Uhlmann_fidelity}), it follows that there exists a
purification $\gamma_{K_{A}K_{B}A^{\prime}B^{\prime}EZ}$ of $\gamma
_{K_{A}K_{B}A^{\prime}B^{\prime}}$ satisfying%
\begin{equation}
F(\gamma_{K_{A}K_{B}A^{\prime}B^{\prime}EZ},\mathcal{U}_{AB\rightarrow
K_{A}K_{B}A^{\prime}B^{\prime}Z}^{\mathcal{L}^{\leftrightarrow}}(\psi
_{ABE}))\geq1-\varepsilon.
\end{equation}
Tracing over systems $A^{\prime}B^{\prime}$ and applying the data-processing
inequality for fidelity, we conclude that%
\begin{equation}
F(\gamma_{K_{A}K_{B}EZ},(\operatorname{Tr}_{A^{\prime}B^{\prime}}%
\circ\mathcal{U}_{AB\rightarrow K_{A}K_{B}A^{\prime}B^{\prime}Z}%
^{\mathcal{L}^{\leftrightarrow}})(\psi_{ABE}))\geq1-\varepsilon.
\end{equation}
Now by applying the same reasoning in Proposition~\ref{prop-approx_key_state},
we conclude that the state $\gamma_{K_{A}K_{B}EZ}$ is an ideal tripartite key
state. However, the channel
\begin{equation}
\operatorname{Tr}_{A^{\prime}B^{\prime}}%
\circ\mathcal{U}_{AB\rightarrow K_{A}K_{B}A^{\prime}B^{\prime}Z}%
^{\mathcal{L}^{\leftrightarrow}}
\end{equation}
 is not necessarily an LOPC channel due to
the coherence of the $Z$ system with the other systems. We can apply
a completely dephasing channel $\overline{\Delta}_{Z}(\cdot)\coloneqq \sum
_{z\in\mathcal{Z}}|z\rangle\!\langle z|_{Z}(\cdot)|z\rangle\!\langle z|_{Z}$
to the $Z$ system, and the fidelity does not decrease under the action of this
channel, implying that%
\begin{equation}
F(\overline{\Delta}_{Z}(\gamma_{K_{A}K_{B}EZ}),(\overline{\Delta}_{Z}%
\circ\operatorname{Tr}_{A^{\prime}B^{\prime}}\circ\mathcal{U}_{AB\rightarrow
K_{A}K_{B}A^{\prime}B^{\prime}Z}^{\mathcal{L}^{\leftrightarrow}})(\psi
_{ABE}))\geq1-\varepsilon.
\end{equation}
The state $\overline{\Delta}_{Z}(\gamma_{K_{A}K_{B}EZ})$ is a tripartite key
state, and the channel
\begin{equation}
\overline{\Delta}_{Z}\circ\operatorname{Tr}%
_{A^{\prime}B^{\prime}}\circ\mathcal{U}_{AB\rightarrow K_{A}K_{B}A^{\prime
}B^{\prime}Z}^{\mathcal{L}^{\leftrightarrow}}
\end{equation}
 is an LOPC channel, being
explicitly written as follows:%
\begin{equation}
\overline{\Delta}_{Z}\circ\operatorname{Tr}_{A^{\prime}B^{\prime}}%
\circ\mathcal{U}_{AB\rightarrow K_{A}K_{B}A^{\prime}B^{\prime}Z}%
^{\mathcal{L}^{\leftrightarrow}}=\sum_{z\in\mathcal{Z}}\mathcal{E}%
_{A\rightarrow K_{A}}^{z}\otimes\mathcal{F}_{B\rightarrow K_{B}}^{z}%
\otimes|z\rangle\!\langle z|_{Z},
\end{equation}
where%
\begin{align}
\mathcal{E}_{A\rightarrow K_{A}}^{z}(\cdot)  &  \coloneqq \operatorname{Tr}%
_{A^{\prime}}[E_{A\rightarrow K_{A}A^{\prime}}^{z}(\cdot)(E_{A\rightarrow
K_{A}A^{\prime}}^{z})^{\dag}],\\
\mathcal{F}_{B\rightarrow K_{B}}^{z}(\cdot)  &  \coloneqq \operatorname{Tr}%
_{B^{\prime}}[F_{B\rightarrow K_{B}B^{\prime}}^{z}(\cdot)(F_{B\rightarrow
K_{B}B^{\prime}}^{z})^{\dag}].
\end{align}
Thus, we have proven that the existence of a $(K,\varepsilon)$ bipartite
private-state distillation protocol for $\rho_{AB}$ implies the existence of a
$(K,\varepsilon)$ tripartite key distillation protocol.
\end{Proof}

\subsubsection{Entanglement Distillation and Secret Key Distillation}

The equivalence between tripartite key distillation and bipartite private-state distillation allows us to relate secret key distillation to entanglement distillation. Indeed, a maximally entangled state $\Phi_{AB}$ is a particular kind of bipartite private state in which the shield systems $A'B'$ are trivial and the twisting unitary $U_{ABA'B'}$ is the identity. This and Theorem~\ref{thm-SKD:tri-bi-equivalence} imply that a $(K,\varepsilon)$ entanglement distillation protocol is a $(K,\varepsilon)$ secret-key distillation protocol. However, the converse is not necessarily true because it is not generally possible to convert a bipartite private state of size $K$ to a maximally entangled state of Schmidt rank~$K$.

As a consequence of the discussion above, it follows that the one-shot distillable entanglement of a bipartite state $\rho_{AB}$ is a lower bound on the one-shot distillable key of $\rho_{AB}$:
\begin{equation}
E_D^{\varepsilon}(A;B)_{\rho} \leq K_D^{\varepsilon}(A;B)_{\rho}
\end{equation}
for all $\varepsilon \in [0,1]$. Since this relationship holds on the fundamental one-shot level, it also holds for the asymptotic quantities as well:
\begin{align}
E_D(A;B)_{\rho} & \leq K_D(A;B)_{\rho},\\
\widetilde{E}_D(A;B)_{\rho} & \leq \widetilde{K}_D(A;B)_{\rho},
\end{align}
where $E_D(A;B)_{\rho}$ is the distillable entanglement of $\rho_{AB}$ (Definition~\ref{def-ent_distill_disill_ent}), $K_D(A;B)_{\rho}$ is the distillable key of $\rho_{AB}$ (given later in Definition~\ref{def-SKD:dist-key-def}), and $\widetilde{E}_D(A;B)_{\rho}$ and $ \widetilde{K}_D(A;B)_{\rho}$ are the strong converse quantities.

\subsection{Upper Bounds on the Number of Secret-Key Bits}

\label{sec-SKD:upper-bounds}In this section, we provide three different upper
bounds on one-shot distillable key, based on private information, relative
entropy of entanglement, and squashed entanglement.

\subsubsection{Private Information Upper Bound}

Our study of upper bounds on one-shot distillable key begins with the private
information and the following lemma.

\begin{Lemma}
{lemma-SKD:approx-key-state-one-shot-private-info}Let $A$ and $B$ be
quantum systems with the same dimension $K\in \mathbb{N}$, let $E$ be another quantum
system of arbitrary dimension, and let $\varepsilon\in\left(  0,1\right)  $.
Let $\omega_{ABE}$ be an $\varepsilon$-approximate tripartite key state of
size$~K$, as specified in Definition~\ref{def:approx-priv}, and let
$\omega_{ABE}^{\mathcal{M}}\coloneqq (\mathcal{M}_{A}\otimes\mathcal{M}_{B}%
)(\omega_{ABE})$, where $\mathcal{M}_{A}$ and $\mathcal{M}_{B}$ are the
measurement channels in Definition~\ref{def:approx-priv}. Then the following
inequality holds%
\begin{equation}
\log_{2}K\leq I_{H}^{\sqrt{\varepsilon}+\delta}(A;B)_{\omega^{\mathcal{M}}%
}-I_{\max}^{\sqrt{\varepsilon}}(A;E)_{\omega^{\mathcal{M}}}+\log_{2}\!\left(
\frac{1}{\delta}\right)  , \label{eq-SKD:tripartite-key-bound-priv-info}%
\end{equation}
where $\delta\in(0,1-\sqrt{\varepsilon})$ and%
\begin{equation}
I_{\max}^{\sqrt{\varepsilon}}(A;E)_{\omega^{\mathcal{M}}}\coloneqq \inf_{\widetilde
{\omega}_{AE}:P(\widetilde{\omega}_{AE},\omega_{AE}^{\mathcal{M}})\leq
\sqrt{\varepsilon}}\inf_{\tau_{E}}D_{\max}(\widetilde{\omega}_{AE}%
\Vert\widetilde{\omega}_{A}\otimes\tau_{E}). \label{eq-SKD:smooth-max-MI-1}%
\end{equation}

\end{Lemma}

\begin{Proof}
Consider that the following condition holds from
Definition~\ref{def:approx-priv}:%
\begin{equation}
F(\gamma_{ABE},\omega_{ABE})\geq1-\varepsilon,
\end{equation}
where $\gamma_{ABE}$ is an ideal tripartite key state. Applying the
measurement channels $\mathcal{M}_{A}$ and $\mathcal{M}_{B}$ from
Definition~\ref{def:tripartite-key-state} and the data-processing inequality
for fidelity, we conclude that%
\begin{equation}
F(\overline{\Phi}_{AB}\otimes\sigma_{E},(\mathcal{M}_{A}\otimes\mathcal{M}%
_{B})(\omega_{ABE}))\geq1-\varepsilon. \label{eq-SKD:err-crit-proof-up-bnd}%
\end{equation}
Now tracing over system $E$ and again applying the data-processing inequality
for fidelity, we conclude that%
\begin{equation}
F(\overline{\Phi}_{AB},\omega_{AB}^{\mathcal{M}})=F(\overline{\Phi}%
_{AB},(\mathcal{M}_{A}\otimes\mathcal{M}_{B})(\omega_{AB}))\geq1-\varepsilon.
\label{eq-SKD:fidelity-to-max-class-corr}%
\end{equation}
Observe that the state $\omega_{AB}^{\mathcal{M}}$ is a classical state and
can be written as%
\begin{equation}
\omega_{AB}^{\mathcal{M}}=\sum_{i,j=0}^{K-1}p(i)q(j|i)|i\rangle\!\langle
i|_{A}\otimes|j\rangle\!\langle j|_{B},
\end{equation}
for a probability distribution $p(i)$ and a conditional probability
distribution $q(j|i)$. If we perform the comparator test $\{\Pi_{AB}%
,I_{AB}-\Pi_{AB}\}$ on systems $AB$ of $\omega_{AB}^{\mathcal{M}}$, where%
\begin{equation}
\Pi_{AB}\coloneqq \sum_{i=0}^{K-1}|i\rangle\!\langle i|_{A}\otimes|i\rangle\!\langle
i|_{B},
\end{equation}
then the probability of passing it is given by%
\begin{align}
&  \operatorname{Tr}[\Pi_{AB}\omega_{AB}^{\mathcal{M}}]\nonumber\\
&  =\operatorname{Tr}\!\left[  \left(  \sum_{i^{\prime}=0}^{K-1}|i^{\prime
}\rangle\!\langle i^{\prime}|_{A}\otimes|i^{\prime}\rangle\!\langle i^{\prime
}|_{B}\right)  \left(  \sum_{i,j=0}^{K-1}p(i)q(j|i)|i\rangle\!\langle
i|_{A}\otimes|j\rangle\!\langle j|_{B}\right)  \right] \\
&  =\sum_{i=0}^{K-1}p(i)q(i|i)
\end{align}
Now consider the following channel:%
\begin{equation}
\mathcal{T}_{AB}(\tau_{AB})\coloneqq \operatorname{Tr}[\Pi_{AB}\tau_{AB}%
]|1\rangle\!\langle1|+\operatorname{Tr}[\left(  I_{AB}-\Pi_{AB}\right)
\tau_{AB}]|0\rangle\!\langle0|,
\end{equation}
which outputs a classical flag register indicating if the comparator test is
successful or not. Consider that%
\begin{align}
\mathcal{T}_{AB}(\overline{\Phi}_{AB})  &  =|1\rangle\!\langle1|,\\
\mathcal{T}_{AB}(\omega_{AB}^{\mathcal{M}})  &  =\left(  \sum_{i=0}%
^{K-1}p(i)q(i|i)\right)  |1\rangle\!\langle1|+\left(  1-\sum_{i=0}%
^{K-1}p(i)q(i|i)\right)  |0\rangle\!\langle0|.
\end{align}
Employing the data-processing inequality for the fidelity and the findings
above, we conclude that%
\begin{align}
1-\varepsilon &  \leq F(\overline{\Phi}_{AB},\omega_{AB}^{\mathcal{M}})\\
&  \leq F(\mathcal{T}_{AB}(\overline{\Phi}_{AB}),\mathcal{T}_{AB}(\omega
_{AB}^{\mathcal{M}}))\\
&  =\sum_{i=0}^{K-1}p(i)q(i|i).
\end{align}
Thus, we conclude that the probability of passing the comparator test
satisfies%
\begin{equation}
\operatorname{Tr}[\Pi_{AB}\omega_{AB}^{\mathcal{M}}]\geq1-\varepsilon.
\end{equation}
Now let $\Pi_{A}^{\delta}$ be the projection onto the positive eigenspace of
$\frac{1}{\delta}\overline{\Phi}_{A}-\omega_{A}^{\mathcal{M}}$, where
$\delta\in(0,1)$. Consider that%
\begin{equation}
\Pi_{A}^{\delta}\left(  \frac{1}{\delta}\overline{\Phi}_{A}-\omega
_{A}^{\mathcal{M}}\right)  \Pi_{A}^{\delta}\geq0\qquad\Longrightarrow\qquad
\Pi_{A}^{\delta}\omega_{A}^{\mathcal{M}}\Pi_{A}^{\delta}\leq\frac{1}{\delta
}\Pi_{A}^{\delta}\overline{\Phi}_{A}\Pi_{A}^{\delta},
\end{equation}
and%
\begin{align}
\left(  I_{A}-\Pi_{A}^{\delta}\right)  \left(  \frac{1}{\delta}\overline{\Phi
}_{A}-\omega_{A}^{\mathcal{M}}\right)  \left(  I_{A}-\Pi_{A}^{\delta}\right)
&  \leq0\\
\Longrightarrow\qquad\operatorname{Tr}[(I_{A}-\Pi_{A}^{\delta})\overline{\Phi
}_{A}]  &  \leq\delta\operatorname{Tr}[(I_{A}-\Pi_{A}^{\delta})\omega
_{A}^{\mathcal{M}}]\leq\delta.
\end{align}
The latter inequality can be rewritten as%
\begin{equation}
\operatorname{Tr}[\Pi_{A}^{\delta}\overline{\Phi}_{A}]\geq1-\delta.
\end{equation}
Also, let $\sigma_{B}$ be an arbitrary state, and consider that%
\begin{align}
\operatorname{Tr}[\Pi_{A}^{\delta}\Pi_{AB}\Pi_{A}^{\delta}(\omega
_{A}^{\mathcal{M}}\otimes\sigma_{B})]  &  =\operatorname{Tr}[\Pi_{AB}(\Pi
_{A}^{\delta}\omega_{A}^{\mathcal{M}}\Pi_{A}^{\delta}\otimes\sigma_{B})]\\
&  \leq\frac{1}{\delta}\operatorname{Tr}[\Pi_{AB}(\Pi_{A}^{\delta}%
\overline{\Phi}_{A}\Pi_{A}^{\delta}\otimes\sigma_{B})]\\
&  \leq\frac{1}{\delta}\operatorname{Tr}[\Pi_{AB}(\overline{\Phi}_{A}%
\otimes\sigma_{B})]\\
&  =\frac{1}{\delta K}\operatorname{Tr}[\Pi_{AB}(I_{A}\otimes\sigma_{B})]\\
&  =\frac{1}{\delta K},
\end{align}
where the second inequality follows because $\Pi_{A}^{\delta}$ and $\overline{\Phi}_{A}$ commute.
Then consider that%
\begin{align}
&  \operatorname{Tr}[(I_{AB}-\Pi_{A}^{\delta}\Pi_{AB}\Pi_{A}^{\delta}%
)\omega_{AB}^{\mathcal{M}}]\nonumber\\
&  \leq\operatorname{Tr}[(I_{AB}-\Pi_{A}^{\delta}\Pi_{AB}\Pi_{A}^{\delta
})\overline{\Phi}_{AB}]+\frac{1}{2}\left\Vert \overline{\Phi}_{AB}-\omega
_{AB}^{\mathcal{M}}\right\Vert _{1}\\
&  \leq\operatorname{Tr}[(I_{AB}-\Pi_{AB})\overline{\Phi}_{AB}%
]+\operatorname{Tr}[(I_{AB}-\Pi_{A}^{\delta}\otimes I_{B})\overline{\Phi}%
_{AB}]+\frac{1}{2}\left\Vert \overline{\Phi}_{AB}-\omega_{AB}^{\mathcal{M}%
}\right\Vert _{1}\\
&  =\operatorname{Tr}[(I_{A}-\Pi_{A}^{\delta})\overline{\Phi}_{A}]+\frac{1}%
{2}\left\Vert \overline{\Phi}_{AB}-\omega_{AB}^{\mathcal{M}}\right\Vert _{1}\\
&  \leq\delta+\sqrt{1-F(\overline{\Phi}_{AB},\omega_{AB}^{\mathcal{M}})}\\
&  \leq\delta+\sqrt{\varepsilon}.
\end{align}
The first inequality is a consequence of the variational characterization of
the normalized trace distance from Theorem~\ref{thm-trace_dist_meas}. The second inequality is a consequence
of the following union bound for commuting projectors $P$ and $Q$:%
\begin{equation}
I-PQP\leq I-P+I-Q,
\end{equation}
which in turn follows from $\left(  I-P\right)  \left(  I-Q\right)  \geq0$.
The third inequality follows from Theorem~\ref{thm-Fuchs_van_de_graaf}, and the last from
\eqref{eq-SKD:fidelity-to-max-class-corr}. As such, the measurement operator
$\Pi_{A}^{\delta}\Pi_{AB}\Pi_{A}^{\delta}$ is a particular measurement
operator satisfying the contraints given in the optimization for the
hypothesis testing relative entropy $D_{H}^{\sqrt{\varepsilon}+\delta}%
(\omega_{AB}^{\mathcal{M}}\Vert\omega_{A}^{\mathcal{M}}\otimes\sigma_{B})$,
and we thus conclude that%
\begin{align}
\log_{2}\delta+\log_{2}K  &  =\log_{2}\delta K\\
&  \leq-\log_{2}\operatorname{Tr}[\Pi_{A}^{\delta}\Pi_{AB}\Pi_{A}^{\delta
}(\omega_{A}^{\mathcal{M}}\otimes\sigma_{B})]\\
&  \leq D_{H}^{\sqrt{\varepsilon}+\delta}(\omega_{AB}^{\mathcal{M}}\Vert
\omega_{A}^{\mathcal{M}}\otimes\sigma_{B}).
\end{align}
Since the bound holds for every state $\sigma_{B}$, we conclude that%
\begin{equation}
\log_{2}K\leq I_{H}^{\sqrt{\varepsilon}+\delta}(A;B)_{\omega^{\mathcal{M}}%
}+\log_{2}\!\left(  \frac{1}{\delta}\right)  .
\label{eq-SKD:intermed-proof-1-decode-error}%
\end{equation}

Now we aim to show that%
\begin{equation}
I_{\max}^{\sqrt{\varepsilon}}(A;E)_{\omega}\leq0.
\label{eq-SKD:smooth-max-MI-less-than-zero}
\end{equation}
Consider that \eqref{eq-SKD:err-crit-proof-up-bnd} implies that%
\begin{equation}
F(\overline{\Phi}_{A}\otimes\sigma_{E},\omega_{AE}^{\mathcal{M}}%
)\geq1-\varepsilon.
\end{equation}
Thus, the state $\overline{\Phi}_{A}\otimes\sigma_{E}$ is such that%
\begin{equation}
P(\overline{\Phi}_{A}\otimes\sigma_{E},\omega_{AE}^{\mathcal{M}})\leq
\sqrt{\varepsilon}. \label{eq-SKD:sine-dist-to-product}%
\end{equation}
Then%
\begin{align}
I_{\max}^{\sqrt{\varepsilon}}(A;E)_{\omega^{\mathcal{M}}}  &  =\inf
_{\widetilde{\omega}_{AE}:P(\widetilde{\omega}_{AE},\omega_{AE}^{\mathcal{M}%
})\leq\sqrt{\varepsilon}}\inf_{\tau_{E}}D_{\max}(\widetilde{\omega}_{AE}%
\Vert\widetilde{\omega}_{A}\otimes\tau_{E})\\
&  \leq D_{\max}(\overline{\Phi}_{A}\otimes\sigma_{E}\Vert\overline{\Phi}%
_{A}\otimes\sigma_{E})\\
&  =0, \label{eq-SKD:intermed-proof-2}%
\end{align}
where the inequality follows from the choices $\widetilde{\omega}%
_{AE}=\overline{\Phi}_{A}\otimes\sigma_{E}$, $\tau_{E}=\sigma_{E}$,
\eqref{eq-SKD:sine-dist-to-product}, and the fact that $D_{\max}(\rho\Vert
\rho)=0$ for every state $\rho$.
\end{Proof}

Note that the result of
Lemma~\ref{lemma-SKD:approx-key-state-one-shot-private-info} is general and
applies to every tripartite state that is close in fidelity to an ideal
tripartite key state. Applying it to the state $\omega_{K_{A}K_{B}%
EZ}=\mathcal{L}_{AB\rightarrow K_{A}K_{B}EZ}^{\leftrightarrow}(\psi_{ABE})$
that is the final output of a $(K,\varepsilon)$ tripartite key distillation
protocol for a state $\rho_{AB}$ with purification $\psi_{ABE}$, we obtain the
following result:

\begin{theorem*}{Upper Bound on One-Shot Distillable Key}{}Let $\rho_{AB}$ be a bipartite state
with purification $\psi_{ABE}$. For every $(K,\varepsilon)$ tripartite key
distillation protocol $(K,\mathcal{L}_{AB\rightarrow K_{A}K_{B}EZ}%
^{\leftrightarrow})$ for $\psi_{ABE}$, with $\varepsilon\in(0,1)$ and
$d_{K_{A}}=d_{K_{B}}=K$, the number of $\varepsilon$-approximate secret-key
bits extracted at the end of the protocol is bounded from above by the
LOPC-optimized private information of $\rho_{AB}$, i.e.,%
\begin{equation}
\log_{2}K\leq\sup_{\mathcal{L}}\left(  I_{H}^{\sqrt{\varepsilon}+\delta
}(X;B^{\prime})_{\mathcal{L}(\psi)}-I_{\max}^{\sqrt{\varepsilon}%
}(X;EZ)_{\mathcal{L}(\psi)}\right)  +\log_{2}\!\left(  \frac{1}{\delta
}\right)  , \label{eq-SKD:private-info-bound-one-shot-key}%
\end{equation}
where $\delta\in(0,1-\sqrt{\varepsilon})$ and the optimization is over every
LOPC\ channel $\mathcal{L}_{AB\rightarrow XB^{\prime}Z}^{\leftrightarrow}$,
where $X$ and $Z$ are classical systems. Consequently, for the one-shot
$\varepsilon$-distillable key, the following bound holds%
\begin{equation}
K_{D}^{\varepsilon}(A;B)_{\rho}\leq\sup_{\mathcal{L}}\left(  I_{H}%
^{\sqrt{\varepsilon}+\delta}(X;B^{\prime})_{\mathcal{L}(\psi)}-I_{\max}%
^{\sqrt{\varepsilon}}(X;EZ)_{\mathcal{L}(\psi)}\right)  +\log_{2}\!\left(
\frac{1}{\delta}\right)  . \label{eq-SKD:private-info-bound-one-shot-key-2}%
\end{equation}

\end{theorem*}

\begin{Proof}
For a $(K,\varepsilon)$ tripartite key distillation protocol $(K,\mathcal{L}%
_{AB\rightarrow K_{A}K_{B}Z}^{\leftrightarrow})$ for $\psi_{ABE}$, by
definition the state $\omega_{K_{A}K_{B}EZ}=\mathcal{L}_{AB\rightarrow
K_{A}K_{B}Z}^{\leftrightarrow}(\psi_{ABE})$ satisfies%
\begin{equation}
F(\gamma_{K_{A}K_{B}EZ},\omega_{K_{A}K_{B}EZ})\geq1-\varepsilon,
\end{equation}
where $\gamma_{K_{A}K_{B}EZ}$ is an ideal tripartite key state. Upon
performing the local measurements $\mathcal{M}_{K_{A}}$ and $\mathcal{M}%
_{K_{B}}$ mentioned in Definition~\ref{def:tripartite-key-state}, we conclude
that%
\begin{equation}
F(\overline{\Phi}_{K_{A}K_{B}}\otimes\sigma_{EZ},(\mathcal{M}_{K_{A}}%
\otimes\mathcal{M}_{K_{B}})(\omega_{K_{A}K_{B}EZ}))\geq1-\varepsilon.
\end{equation}
Set%
\begin{equation}
\omega_{K_{A}K_{B}EZ}^{\mathcal{M}}\coloneqq (\mathcal{M}_{K_{A}}\otimes
\mathcal{M}_{K_{B}})(\omega_{K_{A}K_{B}EZ}).
\end{equation}
Therefore, using \eqref{eq-SKD:tripartite-key-bound-priv-info}, we conclude
that%
\begin{equation}
\log_{2}K\leq I_{H}^{\sqrt{\varepsilon}+\delta}(K_{A};K_{B})_{\omega
^{\mathcal{M}}}-I_{\max}^{\sqrt{\varepsilon}}(K_{A};EZ)_{\omega^{\mathcal{M}}%
}+\log_{2}\!\left(  \frac{1}{\delta}\right)  ,
\label{eq-SKD:intermediate-bnd-private-info-help}%
\end{equation}
where $\delta\in(0,1-\sqrt{\varepsilon})$. Since $(\mathcal{M}_{K_{A}}%
\otimes\mathcal{M}_{K_{B}})\circ\mathcal{L}_{AB\rightarrow K_{A}K_{B}%
Z}^{\leftrightarrow}$ is a particular LOPC channel of the form $\mathcal{L}%
_{AB\rightarrow XB^{\prime}Z}^{\leftrightarrow}$, with $X$ and $Z$ classical
systems, we conclude that%
\begin{multline}
I_{H}^{\sqrt{\varepsilon}+\delta}(K_{A};K_{B})_{\omega^{\mathcal{M}}}-I_{\max
}^{\sqrt{\varepsilon}}(K_{A};EZ)_{\omega^{\mathcal{M}}}\\
\leq\sup_{\mathcal{L}}\left(  I_{H}^{\sqrt{\varepsilon}+\delta}(X;B^{\prime
})_{\mathcal{L}(\psi)}-I_{\max}^{\sqrt{\varepsilon}}(X;EZ)_{\mathcal{L}(\psi
)}\right)  .
\end{multline}
We thus conclude \eqref{eq-SKD:private-info-bound-one-shot-key}. Now employing
the definition of the one-shot $\varepsilon$-distillable key in
\eqref{eq-SKD:one-shot-distillable-key-def}, we conclude \eqref{eq-SKD:private-info-bound-one-shot-key-2}.
\end{Proof}

\subsubsection{Relative Entropy of Entanglement Upper Bound}

\label{sec-SKD:rel-ent-up-bnd}

We now consider an upper bound based on the
relative entropy of entanglement (Section~\ref{sec-ent_measures_sep_distance}). In order to place an upper
bound on the one-shot distillable key $K_{D}^{\varepsilon}(A;B)_{\rho}$ for a
given state $\rho_{AB}$ and $\varepsilon\in\left[  0,1\right]  $, we consider
state that are useless for key distillation. This approach is analogous to
what we did previously for entanglement distillation in Section~\ref{subsec-ent_distill_one_shot_UB}.

Which states are useless for key distillation? Suppose that a state
$\sigma_{AB}$ is separable, so that it can be written as%
\begin{equation}
\sigma_{AB}=\sum_{x\in\mathcal{X}}p(x)\psi_{A}^{x}\otimes\varphi_{B}^{x},
\end{equation}
where $\mathcal{X}$ is a finite alphabet, $p:\mathcal{X}\rightarrow\left[
0,1\right]  $ is a probability distribution, and $\{\psi_{A}^{x}%
\}_{x\in\mathcal{X}}$ and $\{\varphi_{B}^{x}\}_{x\in\mathcal{X}}$ are sets of
pure states. Consistent with the model of key distillation that we have discussed
so far, the eavesdropper Eve is allowed to have access to the purifying system
$E$ of a purification of $\sigma_{AB}$, which in this case can be chosen as
follows:%
\begin{align}
\psi_{ABE}  &  =|\psi\rangle\!\langle\psi|_{ABE},\\
|\psi\rangle_{ABE}  &  \coloneqq \sum_{x\in\mathcal{X}}\sqrt{p(x)}|\psi^{x}\rangle
_{A}\otimes|\varphi^{x}\rangle_{B}\otimes|x\rangle_{E}.
\end{align}
Then Eve can measure the system $E$ and obtain an outcome $x\in\mathcal{X}$
with probability $p(x)$, and the resulting state of Alice and Bob is the
product state $\psi_{A}^{x}\otimes\varphi_{B}^{x}$. Being a product state, the
resulting state $\psi_{A}^{x}\otimes\varphi_{B}^{x}$ of Alice and Bob has no
correlation whatsoever and so cannot be used to generate a secret key. If
Alice and Bob attempt to process this state using LOCC, the same problem
arises. In the model of key distillation that we assume, Eve gets a copy of
all classical data exchanged between Alice and Bob, and so the resulting state
is still a product state and is useless for generating a secret key.\ As such,
\textit{all separable states are useless for key distillation}.

The intuition above is useful for reasoning about key distillation, but there
is a way to make it precise by means of a construct called the
\textquotedblleft privacy test.\textquotedblright\ In doing so, we exploit the
equivalence between tripartite key distillation and bipartite private-state
distillation discussed in Section~\ref{sec-SKD:tri-bi-equivalence}\ and
identified in Theorem~\ref{thm-SKD:tri-bi-equivalence}. The privacy test is
analogous to the entanglement test used in Chapter~\ref{chap-ent_distill},
which we used to establish upper bounds on the number of approximate ebits
that can be generated in an entanglement distillation protocol. Here we define
a \textquotedblleft privacy test\textquotedblright\ as a method for testing
whether a given bipartite state is private. It forms an essential component in
Proposition \ref{prop:core-meta-converse-privacy}, which states that the
$\varepsilon$-relative entropy of entanglement is an upper bound on the number
of private bits in an $\varepsilon$-approximate bipartite private state.

\begin{definition}
{Privacy Test}{def:privacy-test}Let $\gamma_{ABA^{\prime}B^{\prime}}$ be
a bipartite private state as given in Definition~\ref{def:private-state-bi}. A
privacy test corresponding to $\gamma_{ABA^{\prime}B^{\prime}}$ (a $\gamma
$-privacy test)\ is defined as the following measurement:%
\begin{equation}
\left\{  \Pi_{ABA^{\prime}B^{\prime}},\mathbbm{1}_{ABA^{\prime}B^{\prime}}%
-\Pi_{ABA^{\prime}B^{\prime}}\right\}  ,
\end{equation}
where%
\begin{equation}
\Pi_{ABA^{\prime}B^{\prime}}\coloneqq U_{ABA^{\prime}B^{\prime}}\left(  \Phi
_{AB}\otimes \mathbbm{1}_{A^{\prime}B^{\prime}}\right)  U_{ABA^{\prime}B^{\prime}}%
^{\dag} \label{eq:gamma-privacy-test}%
\end{equation}
and $U_{ABA^{\prime}B^{\prime}}$ is the twisting unitary specified in \eqref{eq:twisting-unitary}.
\end{definition}

If one has access to the systems $ABA^{\prime}B^{\prime}$ of a bipartite state
$\rho_{ABA^{\prime}B^{\prime}}$ and has a description of $\gamma_{ABA^{\prime
}B^{\prime}}$ satisfying \eqref{eq:approx-bi-priv-state}, then the $\gamma
$-privacy test decides whether $\rho_{ABA^{\prime}B^{\prime}}$ is a private
state with respect to $\gamma_{ABA^{\prime}B^{\prime}}$. The first outcome
corresponds to the decision \textquotedblleft yes, it is a $\gamma$-private
state,\textquotedblright\ and the second outcome corresponds to
\textquotedblleft no.\textquotedblright\ Physically, this test is just
untwisting the purported private state and projecting onto a maximally
entangled state. The following lemma states that the probability for an
$\varepsilon$-approximate bipartite private state to pass the $\gamma$-privacy
test is not smaller than $1-\varepsilon$:

\begin{Lemma}
{lem:pass-privacy-test}Let $\varepsilon\in\left[  0,1\right]  $ and let
$\rho_{ABA^{\prime}B^{\prime}}$ be an $\varepsilon$-approximate private state
as given in Definition~\ref{def:approx-priv}, with $\gamma_{ABA^{\prime
}B^{\prime}}$ satisfying \eqref{eq:approx-bi-priv-state}. The probability for
$\rho_{ABA^{\prime}B^{\prime}}$ to pass the $\gamma$-privacy test is never
smaller than $1-\varepsilon$:%
\begin{equation}
\operatorname{Tr}[\Pi_{ABA^{\prime}B^{\prime}}\rho_{ABA^{\prime}B^{\prime}%
}]\geq1-\varepsilon, \label{eq:priv-state-pass-test}%
\end{equation}
where $\Pi_{ABA^{\prime}B^{\prime}}$ is defined in \eqref{eq:gamma-privacy-test}.
\end{Lemma}

\begin{Proof}
One can see this bound explicitly by inspecting the following steps:
\begin{align}
&  \operatorname{Tr}[\Pi_{ABA^{\prime}B^{\prime}}\rho_{ABA^{\prime}B^{\prime}%
}]\nonumber\\
&  =\operatorname{Tr}[U_{ABA^{\prime}B^{\prime}}\left(  \Phi_{AB}\otimes
I_{A^{\prime}B^{\prime}}\right)  U_{ABA^{\prime}B^{\prime}}^{\dag}%
\rho_{ABA^{\prime}B^{\prime}}]\\
&  =\operatorname{Tr}[\left(  \Phi_{AB}\otimes I_{A^{\prime}B^{\prime}%
}\right)  U_{ABA^{\prime}B^{\prime}}^{\dag}\rho_{ABA^{\prime}B^{\prime}%
}U_{ABA^{\prime}B^{\prime}}]\\
&  =\langle\Phi|_{AB}\operatorname{Tr}_{A^{\prime}B^{\prime}}[U_{ABA^{\prime
}B^{\prime}}^{\dag}\rho_{ABA^{\prime}B^{\prime}}U_{ABA^{\prime}B^{\prime}%
}]|\Phi\rangle_{AB}\\
&  =F(\Phi_{AB},\operatorname{Tr}_{A^{\prime}B^{\prime}}[U_{ABA^{\prime
}B^{\prime}}^{\dag}\rho_{ABA^{\prime}B^{\prime}}U_{ABA^{\prime}B^{\prime}}])\\
&  \geq F(\Phi_{AB}\otimes\theta_{A^{\prime}B^{\prime}},U_{ABA^{\prime
}B^{\prime}}^{\dag}\rho_{ABA^{\prime}B^{\prime}}U_{ABA^{\prime}B^{\prime}})\\
&  =F(U_{ABA^{\prime}B^{\prime}}(\Phi_{AB}\otimes\theta_{A^{\prime}B^{\prime}%
})U_{ABA^{\prime}B^{\prime}}^{\dag},\rho_{ABA^{\prime}B^{\prime}})\\
&  =F(\gamma_{ABA^{\prime}B^{\prime}},\rho_{ABA^{\prime}B^{\prime}})\\
&  \geq1-\varepsilon.
\end{align}
The third equality follows because $\Phi_{AB}$ is pure and by taking applying
the definition of partial trace (over $A^{\prime}B^{\prime}$). The fourth
equality follows from the expression in \eqref{eq-fidelity_pure_mixed}, for the fidelity between a pure
state and a mixed state. The first inequality follows from the data-processing
inequality for fidelity. The second-to-last equality follows from the unitary
invariance of the fidelity, and the last equality follows because
$\gamma_{ABA^{\prime}B^{\prime}}$ is an ideal private state, written as
$\gamma_{ABA^{\prime}B^{\prime}}=U_{ABA^{\prime}B^{\prime}}(\Phi_{AB}%
\otimes\theta_{A^{\prime}B^{\prime}})U_{ABA^{\prime}B^{\prime}}^{\dag}$.
\end{Proof}

On the other hand, a separable state $\sigma_{ABA^{\prime}B^{\prime}}%
\in\operatorname{SEP}(AA^{\prime}\!:\!BB^{\prime})$ of the key and shield
systems has a small chance of passing an arbitrary $\gamma$-privacy test:

\begin{Lemma}
{lem:fail-privacy-test}For a separable state $\sigma_{ABA^{\prime
}B^{\prime}}\in\operatorname{SEP}(AA^{\prime}\!:\!BB^{\prime})$, the
probability of passing an arbitrary $\gamma$-privacy test is not larger than
$\frac{1}{K}$:%
\begin{equation}
\operatorname{Tr}[\Pi_{ABA^{\prime}B^{\prime}}\sigma_{ABA^{\prime}B^{\prime}%
}]\leq\frac{1}{K}\ , \label{eq:priv-test-separable}%
\end{equation}
where $K$ is the number of values that the secret key can take (i.e.,
$K=d_{A}=d_{B}$).
\end{Lemma}

\begin{Proof}
The idea is to begin by establishing the bound for an arbitrary pure product
state $|\phi\rangle_{AA^{\prime}}\otimes|\varphi\rangle_{BB^{\prime}}$, i.e.,
to show that%
\begin{equation}
\operatorname{Tr}[\Pi_{ABA^{\prime}B^{\prime}}|\phi\rangle\!\langle
\phi|_{AA^{\prime}}\otimes|\varphi\rangle\!\langle\varphi|_{BB^{\prime}}%
]\leq\frac{1}{K}. \label{eq:meta-conv-privacy-bnd-goal}%
\end{equation}
We can expand these states with respect to the standard bases of $A$ and $B$
as follows:%
\begin{equation}
|\phi\rangle_{AA^{\prime}}\otimes|\varphi\rangle_{BB^{\prime}}=\left[
\sum_{i=1}^{K}\alpha_{i}|i\rangle_{A}\otimes|\phi_{i}\rangle_{A^{\prime}%
}\right]  \otimes\left[  \sum_{j=1}^{K}\beta_{j}|j\rangle_{B}\otimes
|\varphi_{j}\rangle_{B^{\prime}}\right]  ,
\end{equation}
where $\sum_{i=1}^{K}\left\vert \alpha_{i}\right\vert ^{2}=\sum_{j=1}%
^{K}\left\vert \beta_{j}\right\vert ^{2}=1$. We then find that%
\begin{align}
&  \!\!\!\!\!\!\operatorname{Tr}[\Pi_{ABA^{\prime}B^{\prime}}|\phi
\rangle\!\langle\phi|_{AA^{\prime}}\otimes|\varphi\rangle\!\langle
\varphi|_{BB^{\prime}}]\nonumber\\
&  =\operatorname{Tr}[U_{ABA^{\prime}B^{\prime}}\left(  \Phi_{AB}\otimes
I_{A^{\prime}B^{\prime}}\right)  U_{ABA^{\prime}B^{\prime}}^{\dag}|\phi
\rangle\!\langle\phi|_{AA^{\prime}}\otimes|\varphi\rangle\!\langle
\varphi|_{BB^{\prime}}]\\
&  =\left\Vert \left(  \langle\Phi|_{AB}\otimes I_{A^{\prime}B^{\prime}%
}\right)  U_{ABA^{\prime}B^{\prime}}^{\dag}|\phi\rangle_{AA^{\prime}}%
\otimes|\varphi\rangle_{BB^{\prime}}\right\Vert _{2}^{2}\\
&  =\left\Vert
\begin{array}
[c]{c}%
\frac{1}{\sqrt{K}}\left(  \sum_{i=1}^{K}\langle i|_{A}\otimes\langle
i|_{B}\otimes U_{A^{\prime}B^{\prime}}^{ii\dag}\right)  \times\\
\left(  \sum_{i^{\prime},j^{\prime}=1}^{K}\alpha_{i^{\prime}}\beta_{j^{\prime
}}|i^{\prime}\rangle_{A}\otimes|j^{\prime}\rangle_{B}\otimes|\phi_{i^{\prime}%
}\rangle_{A^{\prime}}|\varphi_{j^{\prime}}\rangle_{B^{\prime}}\right)
\end{array}
\right\Vert _{2}^{2}\\
&  =\frac{1}{K}\left\Vert \sum_{i,i^{\prime},j^{\prime}=1}^{K}\alpha
_{i^{\prime}}\beta_{j^{\prime}}\langle i|i^{\prime}\rangle_{A}\otimes\langle
i|j^{\prime}\rangle_{B}\otimes U_{A^{\prime}B^{\prime}}^{ii\dag}%
|\phi_{i^{\prime}}\rangle_{A^{\prime}}|\varphi_{j^{\prime}}\rangle_{B^{\prime
}}\right\Vert _{2}^{2}\\
&  =\frac{1}{K}\left\Vert \sum_{i=1}^{K}\alpha_{i}\beta_{i}U_{A^{\prime
}B^{\prime}}^{ii\dag}|\phi_{i}\rangle_{A^{\prime}}|\varphi_{i}\rangle
_{B^{\prime}}\right\Vert _{2}^{2}\\
&  =\frac{1}{K}\left\Vert \sum_{i=1}^{K}\alpha_{i}\beta_{i}|\xi_{i}%
\rangle_{A^{\prime}B^{\prime}}\right\Vert _{2}^{2}\\
&  =\frac{1}{K}\sum_{i,j=1}^{K}\alpha_{i}\beta_{i}\alpha_{j}^{\ast}\beta
_{j}^{\ast}\langle\xi_{j}|\xi_{i}\rangle_{A^{\prime}B^{\prime}}.
\end{align}
where $|\xi_{i}\rangle_{A^{\prime}B^{\prime}}\coloneqq (U_{A^{\prime}B^{\prime}}%
^{ii})^{\dag}|\phi_{i}\rangle_{A^{\prime}}|\varphi_{i}\rangle_{B^{\prime}}$ is
a quantum state. The desired bound in
\eqref{eq:meta-conv-privacy-bnd-goal}\ is then equivalent to%
\begin{equation}
\sum_{i,j=1}^{K}\alpha_{i}\beta_{i}\alpha_{j}^{\ast}\beta_{j}^{\ast}\langle
\xi_{j}|\xi_{i}\rangle_{A^{\prime}B^{\prime}}\leq1.
\end{equation}
Setting $\alpha_{i}=\sqrt{p_{i}}e^{i\theta_{i}}$ and $\beta_{i}=\sqrt{q_{i}%
}e^{i\eta_{i}}$, we find that%
\begin{align}
\sum_{i,j=1}^{K}\alpha_{i}\beta_{i}\alpha_{j}^{\ast}\beta_{j}^{\ast}\langle
\xi_{j}|\xi_{i}\rangle_{A^{\prime}B^{\prime}}  &  =\left\vert \sum_{i,j=1}%
^{K}\sqrt{p_{i}q_{i}p_{j}q_{j}}e^{i\left(  \theta_{i}+\eta_{i}-\theta_{j}%
-\eta_{j}\right)  }\langle\xi_{j}|\xi_{i}\rangle_{A^{\prime}B^{\prime}%
}\right\vert \\
&  \leq\sum_{i,j=1}^{K}\sqrt{p_{i}q_{i}p_{j}q_{j}}\left\vert \langle\xi
_{j}|\xi_{i}\rangle_{A^{\prime}B^{\prime}}\right\vert \\
&  \leq\sum_{i,j=1}^{K}\sqrt{p_{i}q_{i}p_{j}q_{j}}\\
&  =\left[  \sum_{i=1}^{K}\sqrt{p_{i}q_{i}}\right]  ^{2}\leq1,
\end{align}
where the last inequality holds for all probability distributions (this is
just the statement that the classical fidelity cannot exceed one). The above
reasoning thus establishes \eqref{eq:priv-test-separable} for pure product
states, and the bound for general separable states follows because every such
state can be written as a convex combination of pure product states.
\end{Proof}

Recall from \eqref{eq-hypo_test_rel_ent_entanglement} that the $\varepsilon$-relative entropy of
entanglement of a bipartite state $\rho_{AB}$ is defined as
\begin{equation}
E_{R}^{\varepsilon}(A;B)_{\rho}\coloneqq \inf_{\sigma_{AB}\in\operatorname{SEP}%
(A:B)}D_{H}^{\varepsilon}(\rho_{AB}\Vert\sigma_{AB}).
\label{eq:hypo-rel-ent-ent}%
\end{equation}
This quantity is an LOCC\ monotone, meaning that
\begin{equation}
E_{R}^{\varepsilon}(A;B)_{\rho}\geq E_{R}^{\varepsilon}(A^{\prime};B^{\prime
})_{\omega},
\end{equation}
for $\omega_{A^{\prime}B^{\prime}}\coloneqq \mathcal{L}_{AB\rightarrow A^{\prime
}B^{\prime}}(\rho_{AB})$, with $\mathcal{L}_{AB\rightarrow A^{\prime}%
B^{\prime}}$ an LOCC\ channel.

\begin{proposition}
{prop:core-meta-converse-privacy} Fix $\varepsilon\in\lbrack0,1]$. Let
$\rho_{ABA^{\prime}B^{\prime}}$ be an $\varepsilon$-approximate bipartite
private state, as given in Definition~\ref{def:approx-priv}. Then the number
$\log_{2}K$ of private bits in such a state is bounded from above by the
$\varepsilon$-relative entropy of entanglement of $\rho_{ABA^{\prime}%
B^{\prime}}$:
\begin{equation}
\log_{2}K\leq E_{R}^{\varepsilon}(AA^{\prime};BB^{\prime})_{\rho}.
\label{eq:eps-rel-ent-approx-priv-key}%
\end{equation}

\end{proposition}

\begin{Proof}
Let $\sigma_{ABA^{\prime}B^{\prime}}$ be an arbitrary separable state in
$\operatorname{SEP}(AA^{\prime}\!:\!BB^{\prime})$. From
Definition~\ref{def:approx-priv} and Lemma~\ref{lem:pass-privacy-test}, we
conclude that the $\gamma$-privacy test $\Pi_{ABA^{\prime}B^{\prime}}$\ from
\eqref{eq:gamma-privacy-test}\ is a particular measurement operator satisfying
the constraint $\operatorname{Tr}[\Pi_{ABA^{\prime}B^{\prime}}\rho
_{ABA^{\prime}B^{\prime}}]\geq1-\varepsilon$ for $\beta_{\varepsilon}%
(\rho_{ABA^{\prime}B^{\prime}}\Vert\sigma_{ABA^{\prime}B^{\prime}})$. Applying
Lemma~\ref{lem:fail-privacy-test} and the definition of $\beta_{\varepsilon}$,
we conclude that
\begin{equation}
\beta_{\varepsilon}(\rho_{ABA^{\prime}B^{\prime}}\Vert\sigma_{ABA^{\prime
}B^{\prime}})\leq\operatorname{Tr}[\Pi_{ABA^{\prime}B^{\prime}}\sigma
_{ABA^{\prime}B^{\prime}}]\leq\frac{1}{K}.
\end{equation}
Since the inequality holds for all separable states $\sigma_{ABA^{\prime
}B^{\prime}}\in\operatorname{SEP}(AA^{\prime}\!:\!BB^{\prime})$, we conclude
that
\begin{equation}
\sup_{\sigma_{ABA^{\prime}B^{\prime}}\in\operatorname{SEP}(AA^{\prime
}:BB^{\prime})}\beta_{\varepsilon}(\rho_{ABA^{\prime}B^{\prime}}%
\Vert\sigma_{ABA^{\prime}B^{\prime}})\leq\frac{1}{K}.
\end{equation}
Applying a negative logarithm and the definition in \eqref{eq:hypo-rel-ent-ent}, we
arrive at the inequality in \eqref{eq:eps-rel-ent-approx-priv-key}.
\end{Proof}

A consequence of Proposition~\ref{prop:core-meta-converse-privacy} is the
following upper bound on the one-shot distillable key of $\rho_{AB}$:

\begin{theorem*}
{Relative Entropy of Entanglement Upper Bound on One-Shot Distillable
Key}{thm-SKD:one-shot-rel-ent-key-up-bound}Let $\rho_{AB}$ be a
bipartite state. For every $(K,\varepsilon)$ secret-key distillation protocol
for $\rho_{AB}$, with $\varepsilon\in\left[  0,1\right]  $, we have that%
\begin{equation}
\log_{2}K\leq E_{R}^{\varepsilon}(A;B)_{\rho}.
\end{equation}
Consequently, for the one-shot distillable key, we have%
\begin{equation}
K_{D}^{\varepsilon}(A;B)_{\rho}\leq E_{R}^{\varepsilon}(A;B)_{\rho},
\end{equation}
for every state $\rho_{AB}$ and $\varepsilon\in\left[  0,1\right]  $.
\end{theorem*}

\begin{Proof}
By Theorem~\ref{thm-SKD:tri-bi-equivalence}, we can work in the picture of
bipartite private-state distillation. Consider a $(K,\varepsilon)$
private-state distillation protocol for $\rho_{AB}$ with the corresponding
LOCC channel $\mathcal{L}_{AB\rightarrow K_{A}K_{B}A^{\prime}B^{\prime}}$.
Then, by definition, we have that%
\begin{equation}
1-F(\gamma_{K_{A}K_{B}A^{\prime}B^{\prime}},\mathcal{L}_{AB\rightarrow
K_{A}K_{B}A^{\prime}B^{\prime}}(\rho_{AB}))\leq\varepsilon
\end{equation}
for some ideal bipartite private state $\gamma_{K_{A}K_{B}A^{\prime}B^{\prime
}}$ of size $K$. Letting $\omega_{K_{A}K_{B}A^{\prime}B^{\prime}}%
\coloneqq \mathcal{L}_{AB\rightarrow K_{A}K_{B}A^{\prime}B^{\prime}}(\rho_{AB})$, we
have that $1-F(\gamma_{K_{A}K_{B}A^{\prime}B^{\prime}},\omega_{K_{A}%
K_{B}A^{\prime}B^{\prime}})\leq\varepsilon$. The output state $\omega
_{K_{A}K_{B}A^{\prime}B^{\prime}}$ of the private-state distillation protocol
therefore satisfies the conditions of
Proposition~\ref{prop:core-meta-converse-privacy}, which means that%
\begin{equation}
\log_{2}K\leq E_{R}^{\varepsilon}(K_{A}A^{\prime};K_{B}B^{\prime})_{\omega}.
\end{equation}
Now, as mentioned above and discussed in Section~\ref{sec-ent_measures_sep_distance}, $E_{R}^{\varepsilon}$ is an
entanglement measure. Thus, it satisfies the data-processing inequality under
LOCC channels, which means that $E_{R}^{\varepsilon}(K_{A}A^{\prime}%
;K_{B}B^{\prime})_{\omega}\leq E_{R}^{\varepsilon}(A;B)_{\rho}$. We thus have
$\log_{2}K\leq E_{R}^{\varepsilon}(A;B)_{\rho}$. Since this inequality holds
for all $K\in\mathbb{N}$ and for every LOCC channel $\mathcal{L}_{AB\rightarrow
K_{A}K_{B}A^{\prime}B^{\prime}}$, by definition of the one-shot $\varepsilon
$-distillable key, we obtain $K_{D}^{\varepsilon}(A;B)_{\rho}\leq
E_{R}^{\varepsilon}(A;B)_{\rho}$, as required.
\end{Proof}

We then find the following upper bounds on the distillable key available in
$(K,\varepsilon)$ key distillation protocols:

\begin{corollary}
{cor-SKD:private-info-upper-bnd-weak-conv-1-shot}Let $\rho_{AB}$ be a
bipartite state, and let $\varepsilon\in\lbrack0,1)$. For every
$(K,\varepsilon)$ secret-key distillation protocol for $\rho_{AB}$, we have
that%
\begin{multline}
\left(  1-2\sqrt{\varepsilon}-\delta\right)  \log_{2}K\leq\sup_{\mathcal{L}^{\leftrightarrow}%
}\left(  I(X;B^{\prime})_{\mathcal{L}^{\leftrightarrow}(\psi)}-I(X;EZ)_{\mathcal{L}^{\leftrightarrow}(\psi
)}\right) \label{eq-SKD:private-info-upper-bnd-weak-conv-1-shot}\\
+h_{2}(\sqrt{\varepsilon}+\delta)+\left(  1-\sqrt{\varepsilon}-\delta\right)
\log_{2}\!\left(  \frac{1}{\delta}\right)  +2g_{2}(\sqrt{\varepsilon}),
\end{multline}
where $\delta\in\left(  0,1-\sqrt{\varepsilon}\right)  $, $\psi_{ABE}$ is a
purification of $\rho_{AB}$, the information quantities are evaluated on
the state $\mathcal{L}^{\leftrightarrow}_{AB\rightarrow XB^{\prime}Z}(\psi_{ABE})$, and the optimization is over every LOPC channel $\mathcal{L}^{\leftrightarrow}_{AB\rightarrow XB^{\prime}Z}$ with classical systems $X$ and $Z$. The
following bound holds for all $\alpha>1$:%
\begin{equation}
\log_{2}K\leq\widetilde{E}_{\alpha}(A;B)_{\rho}+\frac{\alpha}{\alpha-1}%
\log_{2}\!\left(  \frac{1}{1-\varepsilon}\right)  ,
\label{eq-SKD:rel-ent-alpha-bnd}%
\end{equation}
where%
\begin{equation}
\widetilde{E}_{\alpha}(A;B)_{\rho}=\inf_{\sigma_{AB}\in\operatorname{SEP}%
(A;B)}\widetilde{D}_{\alpha}(\rho_{AB}\Vert\sigma_{AB})
\end{equation}
is the sandwiched R\'enyi relative entropy of entanglement (see \eqref{eq-sand_Ren_rel_entropy_entanglement}).
\end{corollary}

\begin{Proof}
Employing the same reasoning that led to
\eqref{eq-SKD:intermed-proof-1-decode-error} and
\eqref{eq-SKD:intermed-proof-2}, consider that the following bounds hold for a
given $(K,\varepsilon)$ secret-key distillation protocol:%
\begin{align}
\log_{2}K  &  \leq I_{H}^{\sqrt{\varepsilon}+\delta}(K_{A};K_{B}%
)_{\omega^{\mathcal{M}}}+\log_{2}\!\left(  \frac{1}{\delta}\right)
,\label{eq-SKD:vN-key-bound-1}\\
I_{\max}^{\sqrt{\varepsilon}}(K_{A};EZ)_{\omega^{\mathcal{M}}}  &  \leq0.
\label{eq-SKD:vN-key-bound-2}%
\end{align}
where $\delta\in(0,1-\sqrt{\varepsilon})$. Consider from Proposition~\ref{prop-hypo_to_rel_ent} that%
\begin{equation}
I_{H}^{\sqrt{\varepsilon}+\delta}(K_{A};K_{B})_{\omega^{\mathcal{M}}}\leq
\frac{1}{1-\sqrt{\varepsilon}-\delta}\left(  I(K_{A};K_{B})_{\omega
^{\mathcal{M}}}+h_{2}(\sqrt{\varepsilon}+\delta)\right)  .
\label{eq-SKD:vN-key-bound-3}%
\end{equation}
Combining \eqref{eq-SKD:vN-key-bound-1} and \eqref{eq-SKD:vN-key-bound-3}, we
obtain%
\begin{multline}
\left(  1-\sqrt{\varepsilon}-\delta\right)  \log_{2}K\leq I(K_{A}%
;K_{B})_{\omega^{\mathcal{M}}}\label{eq-SKD:vN-key-bound-5}\\
+h_{2}(\sqrt{\varepsilon}+\delta)+\left(  1-\sqrt{\varepsilon}-\delta\right)
\log_{2}\!\left(  \frac{1}{\delta}\right)  .
\end{multline}
Also, we have that%
\begin{align}
&  I_{\max}^{\sqrt{\varepsilon}}(K_{A};EZ)_{\omega^{\mathcal{M}}}\notag \\
&  =\inf_{\widetilde{\omega}_{K_{A}EZ}:P(\widetilde{\omega}_{K_{A}EZ}%
,\omega^{\mathcal{M}})\leq\sqrt{\varepsilon}}\inf_{\tau_{EZ}}D_{\max
}(\widetilde{\omega}_{K_{A}EZ}\Vert\widetilde{\omega}_{K_{A}}\otimes\tau
_{EZ})\\
&  \geq\inf_{\widetilde{\omega}_{K_{A}EZ}:P(\widetilde{\omega}_{K_{A}%
EZ},\omega^{\mathcal{M}})\leq\sqrt{\varepsilon}}\inf_{\tau_{EZ}}%
D(\widetilde{\omega}_{K_{A}EZ}\Vert\widetilde{\omega}_{K_{A}}\otimes\tau
_{EZ})\\
&  =\inf_{\widetilde{\omega}_{K_{A}EZ}:P(\widetilde{\omega}_{K_{A}EZ}%
,\omega^{\mathcal{M}})\leq\sqrt{\varepsilon}}I(K_{A};EZ)_{\widetilde{\omega}%
}\\
&  \geq I(K_{A};EZ)_{\omega^{\mathcal{M}}}-\sqrt{\varepsilon}\log_{2}%
K-2g_{2}(\sqrt{\varepsilon}).
\end{align}
The first inequality follows because $D_{\max}(\rho\Vert\sigma) \geq D(\rho\Vert\sigma)$, and the second inequality is a consequence of Theorem~\ref{thm-Fuchs_van_de_graaf} and \eqref{eq-QCMI_uniform_cont_cqq}.
We then find that
\begin{equation}
I_{\max}^{\sqrt{\varepsilon}}(K_{A};EZ)_{\omega^{\mathcal{M}}}\geq
I(K_{A};EZ)_{\omega^{\mathcal{M}}}-\sqrt{\varepsilon}\log_{2}K-2g_{2}%
(\sqrt{\varepsilon}). \label{eq-SKD:vN-key-bound-4}%
\end{equation}
Combining \eqref{eq-SKD:vN-key-bound-2} and \eqref{eq-SKD:vN-key-bound-4}, we
conclude that%
\begin{equation}
-\sqrt{\varepsilon}\log_{2}K\leq-I(K_{A};EZ)_{\omega^{\mathcal{M}}}%
+2g_{2}(\sqrt{\varepsilon}). \label{eq-SKD:vN-key-bound-6}%
\end{equation}
Adding \eqref{eq-SKD:vN-key-bound-5} and \eqref{eq-SKD:vN-key-bound-6} gives%
\begin{multline}
\left(  1-2\sqrt{\varepsilon}-\delta\right)  \log_{2}K\leq I(K_{A}%
;K_{B})_{\omega^{\mathcal{M}}}-I(K_{A};EZ)_{\omega^{\mathcal{M}}}\\
+h_{2}(\sqrt{\varepsilon}+\delta)+\left(  1-\sqrt{\varepsilon}-\delta\right)
\log_{2}\!\left(  \frac{1}{\delta}\right)  +2g_{2}(\sqrt{\varepsilon}).
\end{multline}
Now by optimizing over every LOPC\ channel $\mathcal{L}^{\leftrightarrow}_{AB\rightarrow
XB^{\prime}Z}$ with $X$ and $Z$ classical systems and observing that the state
$\omega_{K_{A}K_{B}EZ}^{\mathcal{M}}$ results from the action of a particular
LOPC channel on $\psi_{ABE}$,\ we conclude that%
\begin{equation}
I(K_{A};K_{B})_{\omega^{\mathcal{M}}}-I(K_{A};EZ)_{\omega^{\mathcal{M}}}%
\leq\sup_{\mathcal{L}}\left(  I(X;B^{\prime})_{\mathcal{L}(\psi)}%
-I(X;EZ)_{\mathcal{L}(\psi)}\right)  ,
\end{equation}
thus giving \eqref{eq-SKD:private-info-upper-bnd-weak-conv-1-shot}.

The inequality in \eqref{eq-SKD:rel-ent-alpha-bnd} follows from
Theorem~\ref{thm-SKD:one-shot-rel-ent-key-up-bound}\ and \eqref{eq:sandwich-to-htre} in Proposition~\ref{prop:sandwich-to-htre}.
\end{Proof}

Since the upper bounds in
\eqref{eq-SKD:private-info-upper-bnd-weak-conv-1-shot} and
\eqref{eq-SKD:rel-ent-alpha-bnd} hold for all $(K,\varepsilon)$ secret-key
distillation protocols, we conclude the following upper bounds on one-shot
$\varepsilon$-distillable key:%
\begin{multline}
\left(  1-2\sqrt{\varepsilon}-\delta\right)  K_{D}^{\varepsilon}(A;B)_{\rho
}\leq\sup_{\mathcal{L}}\left(  I(X;B^{\prime})_{\mathcal{L}(\psi
)}-I(X;EZ)_{\mathcal{L}(\psi)}\right)
\label{eq-SKD:private-info-up-=bnd-1-shot-dist-key}\\
+h_{2}(\sqrt{\varepsilon}+\delta)+\left(  1-\sqrt{\varepsilon}-\delta\right)
\log_{2}\!\left(  \frac{1}{\delta}\right)  +2g_{2}(\sqrt{\varepsilon}),
\end{multline}%
\begin{equation}
\log_{2}K\leq\widetilde{E}_{\alpha}(A;B)_{\rho}+\frac{\alpha}{\alpha-1}%
\log_{2}\!\left(  \frac{1}{1-\varepsilon}\right)  ,\qquad\forall\alpha>1,
\end{equation}
where $\delta\in(0,1-\sqrt{\varepsilon})$ and the optimization in
\eqref{eq-SKD:private-info-up-=bnd-1-shot-dist-key} is over every
LOPC\ channel $\mathcal{L}_{AB\rightarrow XB^{\prime}Z}$.

\subsubsection{Squashed Entanglement Upper\ Bound}

We now turn to squashed entanglement and establish it as an upper bound on
one-shot distillable key. Before doing so, we establish some preparatory lemmas.

We begin by establishing Lemma~\ref{thm-SKD:SKD-bipartite-bound}, which is an
upper bound on the logarithm of the dimension $K$ of a key system of an
$\varepsilon$-approximate private state, as given in
Definition~\ref{def:approx-priv}, in terms of its squashed entanglement, plus
another term depending only on $\varepsilon$ and $\log_{2}K$. In what follows,
we suppose that $\gamma_{AA^{\prime}BB^{\prime}}$ is a private state with key
systems $AB$ and shield systems $A^{\prime}B^{\prime}$. Recall from
Theorem~\ref{thm-private_state} that a private state of $\log_{2}K$ private
bits can be written in the following form:
\begin{equation}
\gamma_{ABA^{\prime}B^{\prime}}=U_{ABA^{\prime}B^{\prime}}\left(  \Phi
_{AB}\otimes\sigma_{A^{\prime}B^{\prime}}\right)  U_{ABA^{\prime}B^{\prime}%
}^{\dagger}, \label{eq:SKD-private-1}%
\end{equation}
where $\Phi_{AB}$ is a maximally entangled state of Schmidt rank $K$
\begin{equation}
\Phi_{AB}\coloneqq\frac{1}{K}\sum_{i,j}|i\rangle\!\langle j|_{A}%
\otimes|i\rangle\!\langle j|_{B}, \label{eq:SKD-max-ent-state}%
\end{equation}
and
\begin{equation}
U_{ABA^{\prime}B^{\prime}}=\sum_{i,j}|i\rangle\!\langle i|_{A}\otimes
|j\rangle\!\langle j|_{B}\otimes U_{A^{\prime}B^{\prime}}^{ij}%
\end{equation}
is a controlled unitary known as a \textquotedblleft twisting
unitary,\textquotedblright\ with each $U_{A^{\prime}B^{\prime}}^{ij}$ a
unitary operator.\ Due to the fact that the maximally entangled state
$\Phi_{AB}$ is unextendible, an arbitrary extension $\gamma_{AA^{\prime
}BB^{\prime}E}$ of a private state $\gamma_{AA^{\prime}BB^{\prime}}$
necessarily has the following form:
\begin{equation}
\gamma_{AA^{\prime}BB^{\prime}E}=U_{AA^{\prime}BB^{\prime}}\left(  \Phi
_{AB}\otimes\sigma_{A^{\prime}B^{\prime}E}\right)  U_{AA^{\prime}BB^{\prime}%
}^{\dag}, \label{eq:SKD-ext-private-state}%
\end{equation}
where $\sigma_{A^{\prime}B^{\prime}E}$ is an extension of $\sigma_{A^{\prime
}B^{\prime}}$. We start with the following lemma, which applies to an
arbitrary extension of a bipartite private state:

\begin{Lemma}
{lem-SKD:SKD-log-K-to-info-measures} Let $\gamma_{AA^{\prime}BB^{\prime}}$
be a bipartite private state, and let $\gamma_{AA^{\prime}BB^{\prime}E}$ be an
extension of it, as given above. Then the following identity holds for every
such extension:
\begin{equation}
2\log_{2}K=I(A;BB^{\prime}|E)_{\gamma}+I(A^{\prime};B|AB^{\prime}E)_{\gamma}.
\label{eq-SKD:SKD-logK-to-info-measures}%
\end{equation}

\end{Lemma}

\begin{Proof}
First consider that the following identity holds as a consequence of two
applications of the chain rule for conditional quantum mutual information (see
\eqref{eq-QCMI_chain_rule}):%
\begin{align}
I(AA^{\prime};BB^{\prime}|E)_{\gamma}  &  =I(A;BB^{\prime}|E)_{\gamma
}+I(A^{\prime};BB^{\prime}|AE)_{\gamma}\nonumber\\
&  =I(A;BB^{\prime}|E)_{\gamma}+I(A^{\prime};B^{\prime}|AE)_{\gamma
}+I(A^{\prime};B|B^{\prime}AE)_{\gamma}. \label{eq-SKD:SKD-chain-rule-CMI}%
\end{align}
Combined with the following identity, which holds for an arbitrary extension
$\gamma_{AA^{\prime}BB^{\prime}E}$ of a private state $\gamma_{AA^{\prime
}BB^{\prime}}$,
\begin{equation}
I(AA^{\prime};BB^{\prime}|E)_{\gamma}=2\log_{2}K+I(A^{\prime};B^{\prime
}|AE)_{\gamma}, \label{eq-SKD:SKD-christandl-thesis}%
\end{equation}
we recover the statement in \eqref{eq-SKD:SKD-logK-to-info-measures}. So it
remains to prove \eqref{eq-SKD:SKD-christandl-thesis}.

By definition, we have that
\begin{equation}
I(AA^{\prime};BB^{\prime}|E)_{\gamma}=H(AA^{\prime}E)_{\gamma}+H(BB^{\prime
}E)_{\gamma}-H(E)_{\gamma}-H(AA^{\prime}BB^{\prime}E)_{\gamma}.
\label{eq:SKD-def-CMI-for-PS}%
\end{equation}
By applying \eqref{eq:SKD-max-ent-state}--\eqref{eq:SKD-ext-private-state}, we
can write $\gamma_{AA^{\prime}BB^{\prime}E}$ as follows:
\begin{equation}
\gamma_{AA^{\prime}BB^{\prime}E}=\frac{1}{K}\sum_{i,j}|i\rangle\!\langle
j|_{A}\otimes|i\rangle\!\langle j|_{B}\otimes U_{A^{\prime}B^{\prime}}%
^{ii}\sigma_{A^{\prime}B^{\prime}E}(U_{A^{\prime}B^{\prime}}^{jj})^{\dagger}.
\end{equation}
Tracing over system $B$ leads to the following state:
\begin{equation}
\gamma_{AA^{\prime}B^{\prime}E}=\frac{1}{K}\sum_{i}|i\rangle\!\langle
i|_{A}\otimes\gamma_{A^{\prime}B^{\prime}E}^{i}, \label{eq-SKD:SKD-trace-out-B}%
\end{equation}
where
\begin{equation}
\gamma_{A^{\prime}B^{\prime}E}^{i}\coloneqq U_{A^{\prime}B^{\prime}}%
^{ii}\sigma_{A^{\prime}B^{\prime}E}(U_{A^{\prime}B^{\prime}}^{ii})^{\dag}.
\end{equation}
Similarly, tracing over system $A$ of $\gamma_{AA^{\prime}BB^{\prime}E}$ leads
to
\begin{equation}
\gamma_{BA^{\prime}B^{\prime}E}=\frac{1}{K}\sum_{i}|i\rangle\!\langle
i|_{B}\otimes\gamma_{A^{\prime}B^{\prime}E}^{i}. \label{eq:SKD-trace-out-A}%
\end{equation}
So these and the chain rule for conditional entropy (see \eqref{eq-QEI:chain-rule-cond-ent}) imply that
\begin{equation}
H(AA^{\prime}E)_{\gamma}=H(A)_{\gamma}+H(A^{\prime}E|A)_{\gamma}=\log
_{2}K+H(A^{\prime}E|A)_{\gamma}. \label{eq-SKD:SKD-approx-priv-squashed-step-1}%
\end{equation}
Similarly, we have that
\begin{equation}
H(BB^{\prime}E)_{\gamma}=\log_{2}K+H(B^{\prime}E|B)_{\gamma}=\log
_{2}K+H(B^{\prime}E|A)_{\gamma}, \label{eq-SKD:SKD-approx-priv-squashed-step-2}%
\end{equation}
where we have used the symmetries in
\eqref{eq-SKD:SKD-trace-out-B}--\eqref{eq:SKD-trace-out-A}. Since $\gamma
_{E}=\gamma_{E}^{i}$ for all $i$ (this is a consequence of $\gamma
_{ABA^{\prime}B^{\prime}}$ being an ideal private state), we find that
\begin{equation}
H(E)_{\gamma}=\frac{1}{K}\sum_{i}H(E)_{\gamma^{i}}=H(E|A)_{\gamma}.
\label{eq-SKD:SKD-approx-priv-squashed-step-3}%
\end{equation}
Finally, we have that
\begin{align}
H(AA^{\prime}BB^{\prime}E)_{\gamma}  &  =H(ABA^{\prime}B^{\prime}%
E)_{\Phi\otimes\sigma}\\
&  =H(AB)_{\Phi}+H(A^{\prime}B^{\prime}E)_{\sigma}\\
&  =\frac{1}{K}\sum_{i}H(A^{\prime}B^{\prime}E)_{\gamma^{i}}\\
&  =H(A^{\prime}B^{\prime}E|A)_{\gamma}.
\label{eq-SKD:SKD-approx-priv-squashed-step-4}%
\end{align}
The first equality follows from unitary invariance of quantum entropy. The
second equality follows because the entropy is additive for tensor-product
states. The third equality follows because $H(AB)_{\Phi}=0$ since $\Phi_{AB}$
is a pure state, and $\sigma_{A^{\prime}B^{\prime}E}$ is related to
$\gamma_{A^{\prime}B^{\prime}E}^{i}$ by the unitary $U_{A^{\prime}B^{\prime}%
}^{ii}$. The final equality follows by applying \eqref{eq-SKD:SKD-trace-out-B},
and the fact that conditional entropy is a convex combination of entropies for
a classical-quantum state where the conditioning system is classical.
Combining \eqref{eq:SKD-def-CMI-for-PS},
\eqref{eq-SKD:SKD-approx-priv-squashed-step-1},
\eqref{eq-SKD:SKD-approx-priv-squashed-step-2},
\eqref{eq-SKD:SKD-approx-priv-squashed-step-3},
\eqref{eq-SKD:SKD-approx-priv-squashed-step-4}, and the fact that
\begin{equation}
I(A^{\prime};B^{\prime}|AE)_{\gamma}=H(A^{\prime}E|A)_{\gamma}+H(B^{\prime
}E|A)_{\gamma}-H(E|A)_{\gamma}-H(A^{\prime}B^{\prime}E|A)_{\gamma},
\end{equation}
we recover \eqref{eq-SKD:SKD-christandl-thesis}.
\end{Proof}

We now establish the squashed entanglement upper bound for an approximate
bipartite private state:

\begin{proposition}
{thm-SKD:SKD-bipartite-bound} Let $\gamma_{AA^{\prime}BB^{\prime}}$ be a
private state, with key systems $AB$ and shield systems $A^{\prime}B^{\prime}%
$, and let $\omega_{AA^{\prime}BB^{\prime}}$ be an $\varepsilon$-approximate
private state, in the sense that
\begin{equation}
F(\gamma_{AA^{\prime}BB^{\prime}},\omega_{AA^{\prime}BB^{\prime}}%
)\geq1-\varepsilon
\end{equation}
for $\varepsilon\in\left[  0,1\right]  $. Suppose that $d_{A}=d_{B}=K$. Then
\begin{equation}
(1-2\sqrt{\varepsilon})\log_{2}K\leq E_{\operatorname{sq}}(AA^{\prime
};BB^{\prime})_{\omega}+2g_{2}(\sqrt{\varepsilon}),
\end{equation}
where
\begin{equation}
g_{2}(\delta)\coloneqq\left(  \delta+1\right)  \log_{2}(\delta+1)-\delta
\log_{2}\delta.
\end{equation}

\end{proposition}

\begin{Proof}
By applying Uhlmann's theorem for fidelity (Theorem~\ref{thm-Uhlmann_fidelity}%
) and the inequalities relating trace distance and fidelity from
Theorem~\ref{thm-Fuchs_van_de_graaf}, for a given extension $\omega
_{AA^{\prime}BB^{\prime}E}$ of $\omega_{AA^{\prime}BB^{\prime}}$, there exists
an extension $\gamma_{AA^{\prime}BB^{\prime}E}$ of $\gamma_{AA^{\prime
}BB^{\prime}}$ such that
\begin{equation}
\frac{1}{2}\left\Vert \gamma_{AA^{\prime}BB^{\prime}E}-\omega_{AA^{\prime
}BB^{\prime}E}\right\Vert _{1}\leq\sqrt{\varepsilon}.
\end{equation}
Defining $f_{1}(\delta,K)\coloneqq 2\delta\log_{2}K+2g_{2}(\delta)$, we then
find that
\begin{align}
2\log_{2}K  &  =I(A;BB^{\prime}|E)_{\gamma}+I(A^{\prime};B|AB^{\prime
}E)_{\gamma}\\
&  \leq I(A;BB^{\prime}|E)_{\omega}+I(A^{\prime};B|AB^{\prime}E)_{\omega
}+2f_{1}(\sqrt{\varepsilon},K)\\
&  \leq I(A;BB^{\prime}|E)_{\omega}+I(A^{\prime};B|AB^{\prime}E)_{\omega
}\nonumber\\
&  \qquad\qquad+I(A^{\prime};B^{\prime}|AE)_{\omega}+2f_{1}(\sqrt{\varepsilon
},K)\\
&  =I(AA^{\prime};BB^{\prime}|E)_{\omega}+2f_{1}(\sqrt{\varepsilon},K).
\end{align}
The first equality follows from Lemma~\ref{lem-SKD:SKD-log-K-to-info-measures}.
The first inequality follows from two applications of
Proposition~\ref{lem:LAQC-uniform-cont-CMI} (asymptotic continuity of conditional mutual information). The second inequality follows because $I(A^{\prime};B^{\prime}|AE)_{\omega}\geq0$ (this is strong subadditivity from
Theorem~\ref{thm-SSA}). The last equality is a consequence of the chain rule
for conditional mutual information, as used in \eqref{eq-SKD:SKD-chain-rule-CMI}.
Since the inequality
\begin{equation}
2\log_{2}K\leq I(AA^{\prime};BB^{\prime}|E)_{\omega}+2f_{1}(\sqrt{\varepsilon
},K)
\end{equation}
holds for an arbitrary extension of $\omega_{AA^{\prime}BB^{\prime}}$, the
statement of the proposition follows.
\end{Proof}

We now put these statements together and arrive at the following
squashed-entanglement upper bound on one-shot distillable key:

\begin{theorem*}
{Squashed Entanglement Upper Bound on One-Shot Distillable Key}
{thm-SKD:sq-ent-one-shot-up-bnd}Let $\rho_{AB}$ be a bipartite state.
For every $(K,\varepsilon)$ secret-key distillation protocol for $\rho_{AB}$,
with $\varepsilon\in\lbrack0,1)$, we have that%
\begin{equation}
\left(  1-2\sqrt{\varepsilon}\right)  \log_{2}K\leq E_{\operatorname{sq}%
}(A;B)_{\rho}+2g_{2}(\sqrt{\varepsilon}),
\label{eq-SKD:sq-ent-bound-one-shot-key}%
\end{equation}
where $E_{\operatorname{sq}}(A;B)_{\rho}$ is the squashed entanglement of
$\rho_{AB}$ (see Section~\ref{sec-LAQC:sq-ent-and-props}) and $g_{2}(\delta)\coloneqq \left(  \delta+1\right)  \log
_{2}(\delta+1)-\delta\log_{2}\delta$. Consequently, for the one-shot
distillable key, we have%
\begin{equation}
\left(  1-2\sqrt{\varepsilon}\right)  K_{D}^{\varepsilon}(A;B)_{\rho}\leq
E_{\operatorname{sq}}(A;B)_{\rho}+2g_{2}(\sqrt{\varepsilon}),
\label{eq-SKD:sq-ent-one-shot-key-bnd-def}%
\end{equation}
for every state $\rho_{AB}$ and $\varepsilon\in\lbrack0,1)$.
\end{theorem*}

\begin{Proof}
We exploit Theorem~\ref{thm-SKD:tri-bi-equivalence}\ and work in the bipartite
picture of private-state distillation, instead of the tripartite picture of
key distillation. With this in mind, consider a $(K,\varepsilon)$ bipartite
private-state distillation protocol for $\rho_{AB}$ with the corresponding
LOCC channel $\mathcal{L}_{AB\rightarrow K_{A}K_{B}A^{\prime}B^{\prime}}$.
From the LOCC\ monotonicity of squashed entanglement (Theorem~\ref{thm:LAQC-mono-LOCC-sq}), we have
that%
\begin{equation}
E_{\operatorname{sq}}(K_{A}A^{\prime};K_{B}B^{\prime})_{\omega}\leq
E_{\operatorname{sq}}(A;B)_{\rho}, \label{eq-SKD:sq-ent-bound-LOCC-mono}%
\end{equation}
where $\omega_{K_{A}A^{\prime}K_{B}B^{\prime}}\coloneqq \mathcal{L}_{AB\rightarrow
K_{A}K_{B}A^{\prime}B^{\prime}}(\rho_{AB})$. Continuing, by the definition in
\eqref{eq-SKD:error-criterion}\ and applying
Theorem~\ref{thm-SKD:tri-bi-equivalence}, the following inequality holds%
\begin{equation}
p_{\text{err}}(\mathcal{L};\rho_{AB})=1-F(\gamma_{K_{A}A^{\prime}%
K_{B}B^{\prime}},\omega_{K_{A}A^{\prime}K_{B}B^{\prime}})\leq\varepsilon
\end{equation}
for some ideal bipartite private state $\gamma_{K_{A}A^{\prime}K_{B}B^{\prime
}}$. As a consequence of Proposition~\ref{thm-SKD:SKD-bipartite-bound}, we find
that%
\begin{equation}
(1-2\sqrt{\varepsilon})\log_{2}K\leq E_{\operatorname{sq}}(K_{A}A^{\prime
};K_{B}B^{\prime})_{\omega}+2g_{2}(\sqrt{\varepsilon}).
\label{eq-SKD:sq-ent-bound-private-state}%
\end{equation}
Combining \eqref{eq-SKD:sq-ent-bound-LOCC-mono} and
\eqref{eq-SKD:sq-ent-bound-private-state}, we conclude
\eqref{eq-SKD:sq-ent-bound-one-shot-key}. Since this bound holds for all
$(K,\varepsilon)$ key distillation protocols, the bound in
\eqref{eq-SKD:sq-ent-one-shot-key-bnd-def} follows after applying the
definition in \eqref{eq-SKD:one-shot-distillable-key-def}.
\end{Proof}

\subsection{Lower Bound on the Number of Secret-Key Bits via Position-Based
Coding and Convex Splitting}

\label{sec-SKD:low-bound-1-shot-dist-key}

Having found upper bounds on one-shot
distillable key, we now turn to establishing a lower bound. In order to
establish a lower bound on distillable key, we have to find an explicit
secret-key distillation protocol that works for an arbitrary bipartite state
$\rho_{AB}$ and an arbitrary error $\varepsilon\in(0,1)$. Recall that the goal
of secret key distillation is for two parties, Alice and Bob, to make use of
LOPC\ to transform a purification $\psi_{ABE}$ of their shared state
$\rho_{AB}$ to an ideal key state of the form in
Definition~\ref{def:tripartite-key-state}, with the key size $K$ as large as
possible, subject to the constraint that the error not exceed $\varepsilon$.
Furthermore, we allow them to make use of public classical communication for free.

Before we get into the details, let us first slightly modify the model of
secret key distillation, and we discuss later how the model we have already
discussed can fit together with this alternative model. The alternative model
consists of supposing that the state shared by Alice, Bob, and Eve, is a
classical--quantum--quantum state $\rho_{XBE}$ of the following form:%
\begin{equation}
\rho_{XBE}\coloneqq \sum_{x\in\mathcal{X}}p(x)|x\rangle\!\langle x|_{X}\otimes
\rho_{BE}^{x}, \label{eq-SKD:alt-model-1}%
\end{equation}
where $\mathcal{X}$ is a finite alphabet, $p:\mathcal{X}\rightarrow
\lbrack0,1]$ is a probability distribution, and $\{\rho_{BE}^{x}%
\}_{x\in\mathcal{X}}$ is a set of states. The goal of a secret-key
distillation protocol in this setting is to perform an LOPC channel
$\mathcal{L}_{XB\rightarrow K_{A}K_{B}Z}^{\leftrightarrow}$ such that the
final state $\omega_{K_{A}K_{B}EZ}\coloneqq \mathcal{L}_{XB\rightarrow K_{A}K_{B}%
EZ}^{\leftrightarrow}(\rho_{XBE})$ satisfies%
\begin{equation}
p_{\text{err}}(\mathcal{L}^{\leftrightarrow};\rho_{XBE})\coloneqq \inf_{\sigma_{EZ}%
}\left(  1-F(\overline{\Phi}_{K_{A}K_{B}}\otimes\sigma_{EZ},\omega_{K_{A}%
K_{B}EZ})\right)  \leq\varepsilon,
\end{equation}
where the error $\varepsilon\in\left[  0,1\right]  $, the infimum is with
respect to every state $\sigma_{EZ}$, and $\overline{\Phi}_{K_{A}K_{B}}$ is a
maximally classically correlated state of size $K$%
\begin{equation}
\overline{\Phi}_{K_{A}K_{B}}\coloneqq \frac{1}{K}\sum_{i=0}^{K-1}|i\rangle\!\langle
i|_{K_{A}}\otimes|i\rangle\!\langle i|_{K_{B}}.
\end{equation}
We then define the one-shot distillable key of $\rho_{XBE}$ as follows:%
\begin{equation}
K_{D}^{\varepsilon}(\rho_{XBE})\coloneqq \sup_{(K,\mathcal{L}^{\leftrightarrow}%
)}\left\{  \log_{2}K:p_{\text{err}}(\mathcal{L}^{\leftrightarrow};\rho
_{XBE})\leq\varepsilon\right\}  , \label{eq-SKD:alt-model-4}%
\end{equation}
where the optimization is over all $K\in\mathbb{N}$ and every LOPC\ channel
$\mathcal{L}_{XB\rightarrow K_{A}K_{B}Z}^{\leftrightarrow}$ with $d_{K_{A}%
}=d_{K_{B}}=K$.

The main idea behind the lower bound is to exhibit a particular protocol that
accomplishes the task of secret key distillation. The protocol we devise is
simple to describe but more involved to analyze. Additionally, it really does
take advantage of the fact that free public classical communication is
allowed, in the sense that a large amount of public classical communication is employed. The protocol begins with Alice, Bob, and Eve sharing the state
$\rho_{XBE}$, with Alice possessing system $X$, Bob $B$, and Eve $E$. Alice
picks a value~$k$ uniformly at random from the set $\left[  K\right]
\coloneqq \left\{  1,\ldots,K\right\}  $. This will end up being the value of the key.
She also picks a value $r$ uniformly at random from the set $\left[  R\right]
\coloneqq \left\{  1,\ldots,R\right\}  $, where $R\in\mathbb{N}$. This variable plays
the role of randomness that is used to confuse Eve about which key value $k$
was chosen. Once $k$ and $r$ have been selected, Alice labels her system $X$
of the state $\rho_{XBE}$ by the pair $(k,r)$, as $X_{k,r}$. Alice then
prepares $KR-1$ independent instances of the classical state%
\begin{equation}
\sum_{x\in\mathcal{X}}p(x)|x\rangle\!\langle x|,
\label{eq-SKD:key-dist-prot-1}%
\end{equation}
and labels the resulting systems as $X_{1,1}$, \ldots, $X_{k,r-1}$,
$X_{k,r+1}$, \ldots, $X_{KR}$. Alice then sends the classical registers
$X_{1,1}$, \ldots, $X_{K,R}$ in lexicographic order over a public classical
communication channel, so that both Bob and Eve receive copies of them. At
this point, for fixed values of $k$ and $r$, the global shared state of Alice,
Bob, and Eve is as follows:%
\begin{multline}
\rho_{X^{KR}X^{\prime KR}X^{\prime\prime KR}BE}^{k,r}\coloneqq \rho_{X_{1,1}%
X_{1,1}^{\prime}X_{1,1}^{\prime\prime}}\otimes\cdots\otimes\rho_{X_{k,r-1}%
X_{k,r-1}^{\prime}X_{k,r-1}^{\prime\prime}}\otimes
\label{eq-SKD:global-state-key-dist-prot}\\
\rho_{X_{k,r}X_{k,r}^{\prime}X_{k,r}^{\prime\prime}BE}\otimes\rho
_{X_{k,r+1}X_{k,r+1}^{\prime}X_{k,r+1}^{\prime\prime}}\otimes\cdots\otimes
\rho_{X_{K,R}X_{K,R}^{\prime}X_{K,R}^{\prime\prime}},
\end{multline}
where Bob possesses all systems labeled as $X^{\prime}$ (in addition to his
$B$ system) and Eve possesses all systems labeled as $X^{\prime\prime}$ (in
addition to her $E$ system). Furthermore,%
\begin{align}
\rho_{X_{1,1}X_{1,1}^{\prime}X_{1,1}^{\prime\prime}}  &  =\cdots
=\rho_{X_{k,r-1}X_{k,r-1}^{\prime}X_{k,r-1}^{\prime\prime}}\\
&  =\rho_{X_{k,r+1}X_{k,r+1}^{\prime}X_{k,r+1}^{\prime\prime}}=\cdots
=\rho_{X_{K,R}X_{K,R}^{\prime}X_{K,R}^{\prime\prime}}\\
&  =\sum_{x\in\mathcal{X}}p(x)|xxx\rangle\!\langle xxx|,
\end{align}
and%
\begin{equation}
\rho_{X_{k,r}X_{k,r}^{\prime}X_{k,r}^{\prime\prime}BE}=\sum_{x\in\mathcal{X}%
}p(x)|xxx\rangle\!\langle xxx|\otimes\rho_{BE}^{x}.
\end{equation}
Thus, it is only the $X_{k,r}$ classical system that has correlation with Bob
and Eve's systems $BE$ and all others have no correlation whatsoever. The
objective of the key distillation protocol is for Bob to identify the
$X_{k,r}$ system that has correlation with his (and in this way, identify the
key value), while the randomness variable $r$ should have sufficient size $R$
to severely reduce the chance that Eve can guess which $X^{\prime\prime}$
system is correlated with hers. The reduced state of Bob, for fixed $k$ and
$r$, is as follows:%
\begin{equation}
\rho_{X^{\prime KR}B}^{k,r}=\rho_{X_{1,1}^{\prime}}\otimes\cdots\otimes
\rho_{X_{k,r-1}^{\prime}}\otimes\rho_{X_{k,r}^{\prime}B}\otimes\rho
_{X_{k,r+1}^{\prime}}\otimes\cdots\otimes\rho_{X_{K,R}^{\prime}},
\label{eq-SKD:Bobs-state-for-PBC}%
\end{equation}
while the reduced state of Eve, for a fixed value of $k$, is as follows:%
\begin{equation}
\rho_{X^{\prime\prime KR}E}^{k}\coloneqq \frac{1}{R}\sum_{r=1}^{R}\rho_{X_{1,1}%
^{\prime\prime}}\otimes\cdots\otimes\rho_{X_{k,r-1}^{\prime\prime}}\otimes
\rho_{X_{k,r}^{\prime\prime}E}\otimes\rho_{X_{k,r+1}^{\prime\prime}}%
\otimes\cdots\otimes\rho_{X_{K,R}^{\prime\prime}}.
\label{eq-SKD:eves-state-for-convex-split}%
\end{equation}
The idea behind confusing Eve is that if $R$ is large enough, then it becomes
difficult for Eve to determine which $X^{\prime\prime}$ system is correlated
with her system$~E$. What we show later is that if $R$ is large enough, then
her reduced state, for all key values $k$, is essentially indistinguishable
from the following product state:%
\begin{equation}
\rho_{X_{1,1}^{\prime\prime}}\otimes\cdots\otimes\rho_{X_{K,R}^{\prime\prime}%
}\otimes\rho_{E}, \label{eq-SKD:eves-state-for-convex-split-ideal-product}%
\end{equation}
where $\rho_{E}=\operatorname{Tr}_{XB}[\rho_{XBE}]$. If that is the case, then
she can figure out essentially nothing about the key value $k$, leaving her no
strategy other than to try and randomly guess it.

To analyze this protocol in detail, we employ two methods: position-based
coding, as used previously in Section~\ref{subsec-pos_coding} in the context of classical
communication, and another idea known as convex splitting. Looking at Bob's
state in \eqref{eq-SKD:Bobs-state-for-PBC} and comparing it with that in \eqref{eq-eacc-pos_encode_B},
it is natural to employ position-based coding to figure out the value of $k$
and $r$. Indeed, invoking Proposition~\ref{prop-eac:one-shot-lower_bound} (in particular, \eqref{eq-eac:one-shot-lower_bound_pf4}--\eqref{eq-eac:one-shot-lower_bound_pf6}), if the following condition holds%
\begin{equation}
\log_{2}KR=\overline{I}_{H}^{\varepsilon-\eta}(X;B)_{\rho}-\log_{2}\!\left(
\frac{4\varepsilon}{\eta^{2}}\right)  , \label{eq-SKD:PBC-rate}%
\end{equation}
for $\eta\in\left(  0,\varepsilon\right)  $, and where $\overline{I}%
_{H}^{\varepsilon-\eta}(X;B)_{\rho}$ is the hypothesis testing mutual
information defined in \eqref{eq-hypo_testing_mutual_inf}, then Bob can decode $k$ and $r$ with error
probability no larger than $\varepsilon$. We would also like to guarantee that
Eve's state in \eqref{eq-SKD:eves-state-for-convex-split} is close to the
product state in \eqref{eq-SKD:eves-state-for-convex-split-ideal-product}.
This is where the convex-split lemma is useful, which states the following: If%
\begin{equation}
\log_{2}R=\overline{I}_{\max}^{\sqrt{\varepsilon}-\eta}(E;X)_{\rho}+\log
_{2}\!\left(  \frac{2}{\eta^{2}}\right)  , \label{eq-SKD:convex-split-rate}%
\end{equation}
then there exists a state $\widetilde{\rho}_{E}$ satisfying%
\begin{equation}
1-F(\rho_{X^{\prime KR}E}^{k},\rho_{X_{1,1}^{\prime\prime}}\otimes
\cdots\otimes\rho_{X_{K,R}^{\prime\prime}}\otimes\widetilde{\rho}_{E}%
)\leq\varepsilon, \label{eq-SKD:security-parameter-infidelity}%
\end{equation}
and $P(\widetilde{\rho}_{E},\rho_{E})\leq\sqrt{\varepsilon}-\eta$. Observe
that the inequality above holds for all key values $k$. In the above,
$\overline{I}_{\max}^{\delta}(E;X)_{\rho}$ is a smooth max-mutual information
quantity defined for $\delta\in(0,1)$ as%
\begin{equation}
\overline{I}_{\max}^{\delta}(E;X)_{\rho}\coloneqq \inf_{\widetilde{\rho}%
_{XE}:P(\widetilde{\rho}_{XE},\rho_{XE})\leq\delta}D_{\max}(\widetilde{\rho
}_{XE}\Vert\rho_{X}\otimes\widetilde{\rho}_{E}).
\label{eq-SKD:alt-smooth-max-MI-def}%
\end{equation}
Observe that $\overline{I}_{\max}^{\delta}(E;X)_{\rho}$ is different from the
smooth max-mutual information quantity defined previously in
\eqref{eq-SKD:smooth-max-MI-1}. By suitably combining position-based coding
with convex splitting and subtracting
\textit{\eqref{eq-SKD:convex-split-rate}}\ from \eqref{eq-SKD:PBC-rate}, we
thus arrive at the conclusion that Alice and Bob can distill a key $K$ of size%
\begin{equation}
\log_{2}K=\overline{I}_{H}^{\varepsilon-\eta}(X;B)_{\rho}-\overline{I}_{\max
}^{\sqrt{\varepsilon}-\eta}(E;X)_{\rho}-\log_{2}\!\left(  \frac{4\varepsilon
}{\eta^{2}}\right)  -\log_{2}\!\left(  \frac{2}{\eta^{2}}\right)  ,
\end{equation}
and be guaranteed that

\begin{enumerate}
\item Bob can decode the key value $k$ with error probability no larger than
$\varepsilon$ and

\item the key value is secure from Eve with security parameter $\varepsilon$
(as given in \eqref{eq-SKD:security-parameter-infidelity}).
\end{enumerate}

Having discussed the protocol for key distillation and some intuition
justifying why the scheme works, we now formally state a lower bound on the one-shot
distillable key of a state $\rho_{XBE}$:

\begin{theorem}
{thm-SKD:one-shot-key-lower-bnd}Let $\rho_{XBE}$ be a
classical--quantum--quantum state, with system $X$ held by Alice, $B$ by Bob,
and $E$ by Eve. For all $\varepsilon\in(0,1]$, $\varepsilon^{\prime}%
=1-\sqrt{1-\varepsilon}$, $\delta\in(0,\varepsilon^{\prime})$, $\eta
\in(0,\varepsilon^{\prime}-\delta)$, and $\zeta\in(0,\delta)$, there exists a
$(K,\varepsilon)$ one-way key distillation protocol for $\rho_{XBE}$ with%
\begin{multline}
\log_{2}K=\overline{I}_{H}^{\varepsilon^{\prime}-\delta-\eta}(X;B)_{\rho
}-\overline{I}_{\max}^{\delta-\zeta}(E;X)_{\rho}\\
-\log_{2}\!\left(
\frac{4(\varepsilon^{\prime}-\delta)}{\eta^{2}}\right)  -\log_{2}\!\left(
\frac{2}{\zeta^{2}}\right)  , \label{eq-SKD:key-bits-lower-bound-priv-info}%
\end{multline}
where the hypothesis testing mutual information $\overline{I}_{H}%
^{\varepsilon^{\prime}-\delta-\eta}(X;B)_{\rho}$ is defined in \eqref{eq-hypo_testing_mutual_inf} and the
smooth max-mutual information $\overline{I}_{\max}^{\delta-\zeta}(E;X)_{\rho}$
is defined in \eqref{eq-SKD:alt-smooth-max-MI-def}.
\end{theorem}

As discussed above, one of the main tools that we employ to prove this theorem
is the smooth convex-split lemma, which we state here and prove in
Appendix~\ref{app-SKD:convex-split}.

\begin{Lemma*}
{Smooth convex split}{lem-SKD:smooth-convex-split}Let $\rho_{AE}$ be a
state, and let $R\in\mathbb{N}$. Let $\tau_{A_{1}\cdots A_{R}E}$ denote the
following state:%
\begin{equation}
\tau_{A_{1}\cdots A_{R}E}\coloneqq \frac{1}{R}\sum_{r=1}^{R}\rho_{A_{1}}\otimes
\cdots\otimes\rho_{A_{r-1}}\otimes\rho_{A_{r}E}\otimes\rho_{A_{r+1}}%
\otimes\cdots\otimes\rho_{A_{R}}.
\end{equation}
Let $\varepsilon\in(0,1)$ and $\eta\in(0,\varepsilon)$. If%
\begin{equation}
\log_{2}R\geq\overline{I}_{\max}^{\varepsilon-\eta}(E;A)_{\rho}+\log
_{2}\!\left(  \frac{2}{\eta^{2}}\right)  ,
\end{equation}
then there exists a state $\widetilde{\rho}_{E}$ satisfying%
\begin{equation}
P(\tau_{A_{1}\cdots A_{R}E},\rho_{A_{1}}\otimes\cdots\otimes\rho_{A_{R}%
}\otimes\widetilde{\rho}_{E})\leq\varepsilon,
\end{equation}
and $P(\widetilde{\rho}_{E},\rho_{E})\leq\varepsilon-\eta$.
\end{Lemma*}

We now prove Theorem~\ref{thm-SKD:one-shot-key-lower-bnd}.

\begin{Proof}
[Proof of Theorem~\ref{thm-SKD:one-shot-key-lower-bnd}]Fix $\varepsilon
\in(0,1]$, $\delta\in(0,\varepsilon)$, $\eta\in(0,\varepsilon-\delta)$, and
$\zeta\in(0,\delta)$. Alice performs the key distillation protocol discussed
in the paragraph surrounding
\eqref{eq-SKD:key-dist-prot-1}--\eqref{eq-SKD:eves-state-for-convex-split-ideal-product}.
The global state, for fixed $k$ and $r$ is as given in
\eqref{eq-SKD:global-state-key-dist-prot}; the reduced state of Bob, for fixed
$k$ and $r$ is as given in \eqref{eq-SKD:Bobs-state-for-PBC}; and the reduced
state of Eve, for fixed $k$, is given by
\eqref{eq-SKD:eves-state-for-convex-split}. The overall global state,
including Alice's classical registers that hold the key value $k$ and the
randomness value $r$, is as follows:%
\begin{multline}
\rho_{K_{A}R_{A}X^{KR}X^{\prime KR}X^{\prime\prime KR}BE}%
\coloneqq \label{eq-SKD:total-global-state-key-dist-prot}\\
\frac{1}{KR}\sum_{k=1}^{K}\sum_{r=1}^{R}|k\rangle\!\langle k|_{K_{A}}%
\otimes|r\rangle\!\langle r|_{R_{A}}\otimes\rho_{X^{KR}X^{\prime KR}%
X^{\prime\prime KR}BE}^{k,r},
\end{multline}
where $\rho_{X^{KR}X^{\prime KR}X^{\prime\prime KR}BE}^{k,r}$ is defined in
\eqref{eq-SKD:global-state-key-dist-prot}. Tracing over the $R_{A}$ and
$X^{KR}$ systems, the state becomes%
\begin{equation}
\rho_{K_{A}X^{\prime KR}X^{\prime\prime KR}BE}\coloneqq \frac{1}{K}\sum_{k=1}%
^{K}|k\rangle\!\langle k|_{K_{A}}\otimes\frac{1}{R}\sum_{r=1}^{R}%
\rho_{X^{\prime KR}X^{\prime\prime KR}BE}^{k,r}%
\end{equation}
By Proposition~\ref{prop-eac:one-shot-lower_bound}, we conclude that if%
\begin{equation}
\log_{2}KR=\overline{I}_{H}^{\varepsilon-\delta-\eta}(X;B)_{\rho}-\log
_{2}\!\left(  \frac{4(\varepsilon-\delta)}{\eta^{2}}\right),
\end{equation}
then there exists a POVM\ $\{\Lambda_{X^{\prime KR}B}^{k,r}\}_{k\in\left[
K\right]  ,r\in\left[  R\right]  }$ such that%
\begin{equation}
\operatorname{Tr}[\Lambda_{X^{\prime KR}B}^{k,r}\rho_{X^{\prime KR}B}%
^{k,r}]\geq1-(\varepsilon-\delta)\qquad\forall k\in\left[  K\right]
,r\in\left[  R\right]  .
\end{equation}
Let us define the measurement channel $\mathcal{M}_{X^{\prime KR}B\rightarrow
K_{B}R_{B}}^{\prime}$ as follows:%
\begin{equation}
\mathcal{M}_{X^{\prime KR}B\rightarrow K_{B}R_{B}}^{\prime}(\tau_{X^{\prime
KR}B})\coloneqq \sum_{k=1}^{K}\sum_{r=1}^{R}\operatorname{Tr}[\Lambda_{X^{\prime KR}%
B}^{k,r}\tau_{X^{\prime KR}B}]|k\rangle\!\langle k|_{K_{B}}\otimes
|r\rangle\!\langle r|_{R_{B}},
\end{equation}
and the reduced measurement channel $\mathcal{M}_{X^{\prime KR}B\rightarrow
K_{B}}$ as%
\begin{align}
\label{eq-SKD:reduced-meas-ch}
\mathcal{M}_{X^{\prime KR}B\rightarrow K_{B}}(\tau_{X^{\prime KR}B})  &
\coloneqq (\operatorname{Tr}_{R_{B}}\circ\mathcal{M}_{X^{\prime KR}B\rightarrow
K_{B}R_{B}}^{\prime})(\tau_{X^{\prime KR}B})\\
&  =\sum_{k=1}^{K}\operatorname{Tr}[\Lambda_{X^{\prime KR}B}^{k,r}%
\tau_{X^{\prime KR}B}]|k\rangle\!\langle k|_{K_{B}}.
\end{align}
Observe that%
\begin{multline}
\frac{1}{2}\left\Vert \mathcal{M}_{X^{\prime KR}B\rightarrow K_{B}R_{B}%
}^{\prime}(\rho_{X^{\prime KR}B}^{k,r})-|k\rangle\!\langle k|_{K_{B}}%
\otimes|r\rangle\!\langle r|_{R_{B}}\right\Vert _{1}\\
=1-\operatorname{Tr}[\Lambda_{X^{\prime KR}B}^{k,r}\rho_{X^{\prime KR}B}%
^{k,r}]\leq\varepsilon-\delta,
\end{multline}
which follows from the same calculation given in \eqref{eq-eacc_trace_dist_avg_error_1}--\eqref{eq-eacc_trace_dist_avg_error}. By the data-processing
inequality for the trace distance, we conclude that%
\begin{equation}
\frac{1}{2}\left\Vert \mathcal{M}_{X^{\prime KR}B\rightarrow K_{B}}%
(\rho_{X^{\prime KR}B}^{k,r})-|k\rangle\!\langle k|_{K_{B}}\right\Vert
_{1}\leq\varepsilon-\delta.
\end{equation}
From convexity of trace distance, we conclude that%
\begin{equation}
\frac{1}{2}\left\Vert \mathcal{M}_{X^{\prime KR}B\rightarrow K_{B}}\!\left(
\frac{1}{R}\sum_{r=1}^{R}\rho_{X^{\prime KR}B}^{k,r}\right)  -|k\rangle
\!\langle k|_{K_{B}}\right\Vert _{1}\leq\varepsilon-\delta.
\end{equation}

The actual state at the end of the protocol is as follows:%
\begin{equation}
\mathcal{M}_{X^{\prime KR}B\rightarrow K_{B}}(\rho_{K_{A}X^{\prime
KR}X^{\prime\prime KR}BE}),
\end{equation}
and the ideal state to generate is%
\begin{equation}
\overline{\Phi}_{K_{A}K_{B}}\otimes\widetilde{\rho}_{X^{\prime\prime KR}%
E}=\frac{1}{K}\sum_{k=1}^{K}|k\rangle\!\langle k|_{K_{A}}\otimes
|k\rangle\!\langle k|_{K_{B}}\otimes\widetilde{\rho}_{X^{\prime\prime KR}E},
\end{equation}
where $\widetilde{\rho}_{X^{\prime\prime KR}E}$ is some state of the
eavesdropper Eve's systems $X^{\prime\prime KR}E$. Thus, our goal is to find
an upper bound on the following quantity%
\begin{equation}
\frac{1}{2}\left\Vert \mathcal{M}_{X^{\prime KR}B\rightarrow K_{B}}%
(\rho_{K_{A}X^{\prime KR}X^{\prime\prime KR}BE})-\overline{\Phi}_{K_{A}K_{B}%
}\otimes\widetilde{\rho}_{X^{\prime\prime KR}E}\right\Vert _{1},
\end{equation}
which we will convert at the end to an upper bound on%
\begin{equation}
1-F(\mathcal{M}_{X^{\prime KR}B\rightarrow K_{B}}(\rho_{K_{A}X^{\prime
KR}X^{\prime\prime KR}BE}),\overline{\Phi}_{K_{A}K_{B}}\otimes\widetilde{\rho
}_{X^{\prime\prime KR}E}).
\end{equation}

To this end, let us first consider bounding the following intermediate
quantity:%
\begin{equation}
\frac{1}{2}\left\Vert \mathcal{M}_{X^{\prime KR}B\rightarrow K_{B}}%
(\rho_{K_{A}X^{\prime KR}X^{\prime\prime KR}BE})-\frac{1}{K}\sum_{k=1}%
^{K}|k\rangle\!\langle k|_{K_{A}}\otimes|k\rangle\!\langle k|_{K_{B}}%
\otimes\frac{1}{R}\sum_{r=1}^{R}\rho_{X^{\prime\prime KR}E}^{k,r}\right\Vert
_{1}.
\end{equation}
We find that%
\begin{align}
&  \frac{1}{2}\left\Vert \mathcal{M}_{X^{\prime KR}B\rightarrow K_{B}}%
(\rho_{K_{A}X^{\prime KR}X^{\prime\prime KR}BE})-\frac{1}{K}\sum_{k=1}%
^{K}|k\rangle\!\langle k|_{K_{A}}\otimes|k\rangle\!\langle k|_{K_{B}}%
\otimes\frac{1}{R}\sum_{r=1}^{R}\rho_{X^{\prime\prime KR}E}^{k,r}\right\Vert
_{1}\nonumber\\
&  =\frac{1}{2}\left\Vert
\begin{array}
[c]{c}%
\frac{1}{K}\sum_{k=1}^{K}|k\rangle\!\langle k|_{K_{A}}\otimes\mathcal{M}%
_{X^{\prime KR}B\rightarrow K_{B}}\!\left(  \frac{1}{R}\sum_{r=1}^{R}%
\rho_{X^{\prime KR}X^{\prime\prime KR}BE}^{k,r}\right) \\
\qquad-\frac{1}{K}\sum_{k=1}^{K}|k\rangle\!\langle k|_{K_{A}}\otimes
|k\rangle\!\langle k|_{K_{B}}\otimes\frac{1}{R}\sum_{r=1}^{R}\rho
_{X^{\prime\prime KR}E}^{k,r}%
\end{array}
\right\Vert _{1}\\
&  =\frac{1}{K}\sum_{k=1}^{K}\frac{1}{2}\left\Vert \mathcal{M}_{X^{\prime
KR}B\rightarrow K_{B}}\!\left(  \frac{1}{R}\sum_{r=1}^{R}\rho_{X^{\prime
KR}X^{\prime\prime KR}BE}^{k,r}\right)  -|k\rangle\!\langle k|_{K_{B}}%
\otimes\frac{1}{R}\sum_{r=1}^{R}\rho_{X^{\prime\prime KR}E}^{k,r}\right\Vert
_{1}.
\end{align}
Now let us define the state%
\begin{align}
\omega_{X^{\prime\prime KR}E}^{k^{\prime},k}  &  \coloneqq \frac{1}{q(k^{\prime}%
|k)}\left(  \frac{1}{R}\sum_{r,r^{\prime}=1}^{R}\operatorname{Tr}_{X^{\prime
KR}B}[\Lambda_{X^{\prime KR}B}^{k^{\prime},r^{\prime}}\rho_{X^{\prime
KR}X^{\prime\prime KR}BE}^{k,r}]\right)  ,\\
q(k^{\prime}|k)  &  \coloneqq \frac{1}{R}\sum_{r,r^{\prime}=1}^{R}\operatorname{Tr}%
[\Lambda_{X^{\prime KR}B}^{k^{\prime},r^{\prime}}\rho_{X^{\prime KR}%
X^{\prime\prime KR}BE}^{k,r}].
\end{align}
Consider that%
\begin{equation}
\sum_{k^{\prime}=1}^{K}q(k^{\prime}|k)\omega_{X^{\prime\prime KR}E}%
^{k^{\prime},k}=\frac{1}{R}\sum_{r=1}^{R}\rho_{X^{\prime\prime KR}E}^{k,r}.
\end{equation}
Then we can write%
\begin{equation}
\mathcal{M}_{X^{\prime KR}B\rightarrow K_{B}}\!\left(  \frac{1}{R}\sum
_{r=1}^{R}\rho_{X^{\prime KR}X^{\prime\prime KR}BE}^{k,r}\right)
=\sum_{k^{\prime}=1}^{K}q(k^{\prime}|k)|k^{\prime}\rangle\!\langle k^{\prime
}|_{K_{B}}\otimes\omega_{X^{\prime\prime KR}E}^{k^{\prime},k},
\end{equation}
so that%
\begin{equation}
\mathcal{M}_{X^{\prime KR}B\rightarrow K_{B}}\!\left(  \frac{1}{R}\sum
_{r=1}^{R}\rho_{X^{\prime KR}B}^{k,r}\right)  =\sum_{k^{\prime}=1}%
^{K}q(k^{\prime}|k)|k^{\prime}\rangle\!\langle k^{\prime}|_{K_{B}}.
\end{equation}
Using these observations, we can finally write%
\begin{align}
&  \frac{1}{K}\sum_{k=1}^{K}\frac{1}{2}\left\Vert \mathcal{M}_{X^{\prime
KR}B\rightarrow K_{B}}\!\left(  \frac{1}{R}\sum_{r=1}^{R}\rho_{X^{\prime
KR}X^{\prime\prime KR}BE}^{k,r}\right)  -|k\rangle\!\langle k|_{K_{B}}%
\otimes\frac{1}{R}\sum_{r=1}^{R}\rho_{X^{\prime\prime KR}E}^{k,r}\right\Vert
_{1}\nonumber\\
&  =\frac{1}{K}\sum_{k=1}^{K}\frac{1}{2}\left\Vert \sum_{k^{\prime}=1}%
^{K}q(k^{\prime}|k)|k^{\prime}\rangle\!\langle k^{\prime}|_{K_{B}}%
\otimes\omega_{X^{\prime\prime KR}E}^{k^{\prime},k}-|k\rangle\!\langle
k|_{K_{B}}\otimes\sum_{k^{\prime}=1}^{K}q(k^{\prime}|k)\omega_{X^{\prime\prime
KR}E}^{k^{\prime},k}\right\Vert _{1}\\
&  \leq\frac{1}{K}\sum_{k=1}^{K}\sum_{k^{\prime}=1}^{K}q(k^{\prime}|k)\left(
\frac{1}{2}\left\Vert |k^{\prime}\rangle\!\langle k^{\prime}|_{K_{B}}%
\otimes\omega_{X^{\prime\prime KR}E}^{k^{\prime},k}-|k\rangle\!\langle
k|_{K_{B}}\otimes\omega_{X^{\prime\prime KR}E}^{k^{\prime},k}\right\Vert
_{1}\right) \\
&  =\frac{1}{K}\sum_{k=1}^{K}\sum_{k^{\prime}=1}^{K}q(k^{\prime}|k)\left(
\frac{1}{2}\left\Vert |k^{\prime}\rangle\!\langle k^{\prime}|_{K_{B}%
}-|k\rangle\!\langle k|_{K_{B}}\right\Vert _{1}\right) \\
&  =\frac{1}{K}\sum_{k=1}^{K}\sum_{\substack{k^{\prime}=1,\\k^{\prime}\neq
k}}^{K}q(k^{\prime}|k)\\
&  =\frac{1}{K}\sum_{k=1}^{K}\frac{1}{2}\left\Vert \mathcal{M}_{X^{\prime
KR}B\rightarrow K_{B}}\!\left(  \frac{1}{R}\sum_{r=1}^{R}\rho_{X^{\prime KR}%
B}^{k,r}\right)  -|k\rangle\!\langle k|_{K_{B}}\right\Vert _{1}\\
&  \leq\varepsilon-\delta.
\end{align}
We thus conclude that%
\begin{multline}
\frac{1}{2}\left\Vert \mathcal{M}_{X^{\prime KR}B\rightarrow K_{B}}%
(\rho_{K_{A}X^{\prime KR}X^{\prime\prime KR}BE})-\frac{1}{K}\sum_{k=1}%
^{K}|k\rangle\!\langle k|_{K_{A}}\otimes|k\rangle\!\langle k|_{K_{B}}%
\otimes\frac{1}{R}\sum_{r=1}^{R}\rho_{X^{\prime\prime KR}E}^{k,r}\right\Vert
_{1}\label{eq-SKD:final-err-prob-step-key-dist-prot}\\
\leq\varepsilon-\delta.
\end{multline}

We now turn to the analysis of privacy. Starting from the overall global state
\eqref{eq-SKD:total-global-state-key-dist-prot}, and fixing a value of $k$,
the reduced state of Eve's systems is as follows:%
\begin{multline}
\rho_{X^{\prime\prime KR}E}^{k}=\frac{1}{R}\sum_{r=1}^{R}\rho_{X^{\prime\prime
KR}E}^{k,r}=\rho_{X_{1,1}^{\prime\prime}}\otimes\cdots\otimes\rho
_{X_{k-1,R}^{\prime\prime}}\\
\otimes\frac{1}{R}\sum_{r=1}^{R}\rho_{X_{k,1}^{\prime\prime}}\otimes
\cdots\otimes\rho_{X_{k,r-1}^{\prime\prime}}\otimes\rho_{X_{k,r}^{\prime
\prime}E}\otimes\rho_{X_{k,r+1}^{\prime\prime}}\otimes\cdots\otimes
\rho_{X_{k,R}^{\prime\prime}}\\
\otimes\rho_{X_{k+1,1}^{\prime\prime}}\otimes\cdots\otimes\rho_{X_{K,R}%
^{\prime\prime}.}.
\end{multline}
Our goal is to show that%
\begin{equation}
\frac{1}{2}\left\Vert \rho_{X^{\prime\prime KR}E}^{k}-\rho_{X^{\prime\prime
KR}}\otimes\widetilde{\rho}_{E}\right\Vert _{1}\leq\delta,
\end{equation}
for some state $\widetilde{\rho}_{E}$. By the invariance of the trace distance
with respect to tensor-product states, i.e., $\left\Vert \sigma\otimes
\tau-\omega\otimes\tau\right\Vert _{1}=\left\Vert \sigma-\omega\right\Vert
_{1}$, we find that%
\begin{align}
&  \frac{1}{2}\left\Vert \rho_{X^{\prime\prime KR}E}^{k}-\rho_{X^{\prime\prime
KR}}\otimes\widetilde{\rho}_{E}\right\Vert _{1}\\
&  =\frac{1}{2}\left\Vert \rho_{X_{k,1}^{\prime\prime}\cdots X_{k,R}%
^{\prime\prime}E}^{k}-\rho_{X_{k,1}^{\prime\prime}\cdots X_{k,R}^{\prime
\prime}}\otimes\widetilde{\rho}_{E}\right\Vert _{1}\\
&  =\frac{1}{2}\left\Vert \frac{1}{R}\sum_{r=1}^{R}\rho_{X_{k,1}^{\prime
\prime}}\otimes\cdots\otimes\rho_{X_{k,r-1}^{\prime\prime}}\otimes\left(
\rho_{X_{k,r}^{\prime\prime}E}-\rho_{X_{k,r}^{\prime\prime}}\otimes
\widetilde{\rho}_{E}\right)  \otimes\rho_{X_{k,r+1}^{\prime\prime}}%
\otimes\cdots\otimes\rho_{X_{k,R}^{\prime\prime}}\right\Vert _{1}.
\end{align}
By invoking the smooth convex-split lemma
(Lemma~\ref{lem-SKD:smooth-convex-split}) and the inequality relating
normalized trace distance and sine distance (see \eqref{eq-Fuchs_van_de_graaf}), we find that if we
pick $R$ such that%
\begin{equation}
\log_{2}R=\overline{I}_{\max}^{\delta-\zeta}(E;X)_{\rho}+\log_{2}\!\left(
\frac{2}{\zeta^{2}}\right)  ,
\end{equation}
then we are guaranteed that%
\begin{equation}
\frac{1}{2}\left\Vert \rho_{X^{\prime\prime KR}E}^{k}-\rho_{X^{\prime\prime
KR}}\otimes\widetilde{\rho}_{E}\right\Vert _{1}\leq\delta,
\label{eq-SKD:final-sec-step-key-dist-prot}%
\end{equation}
where $\widetilde{\rho}_{E}$ is some state such that $P(\widetilde{\rho}%
_{E},\rho_{E})\leq\delta-\zeta$. Now combining
\eqref{eq-SKD:final-err-prob-step-key-dist-prot} and
\eqref{eq-SKD:final-sec-step-key-dist-prot} with the triangle inequality, we
conclude the desired statement:%
\begin{equation}
\frac{1}{2}\left\Vert \mathcal{M}_{X^{\prime KR}B\rightarrow K_{B}}%
(\rho_{K_{A}X^{\prime KR}X^{\prime\prime KR}BE})-\overline{\Phi}_{K_{A}K_{B}%
}\otimes\rho_{X^{\prime\prime KR}}\otimes\widetilde{\rho}_{E}\right\Vert
_{1}\leq\varepsilon.
\end{equation}
We finally conclude that%
\begin{equation}
1-F(\mathcal{M}_{X^{\prime KR}B\rightarrow K_{B}}(\rho_{K_{A}X^{\prime
KR}X^{\prime\prime KR}BE}),\overline{\Phi}_{K_{A}K_{B}}\otimes\rho
_{X^{\prime\prime KR}}\otimes\widetilde{\rho}_{E})\leq\varepsilon\left(
2-\varepsilon\right)
\end{equation}
by exploiting the inequality in \eqref{eq-Fuchs_van_de_graaf} relating fidelity and trace distance. Now using
the fact that the inverse function of $\varepsilon(2-\varepsilon)$, with domain and range given by $[0,1]$, is
$1-\sqrt{1-\varepsilon}$ and reassigning $\varepsilon\left(  2-\varepsilon
\right)  $ as $\varepsilon$, we conclude the desired statement in \eqref{eq-SKD:key-bits-lower-bound-priv-info}.
\end{Proof}

The result of Theorem~\ref{thm-SKD:one-shot-key-lower-bnd} applies to the
model of secret key distillation outlined in the paragraph containing
\eqref{eq-SKD:alt-model-1}--\eqref{eq-SKD:alt-model-4}. To extend it to the
main model considered in this chapter (and outlined in
Section~\ref{sec-SKD:one-shot-setting}), we can allow Alice and Bob to perform
an LOPC channel $\mathcal{L}_{AB\rightarrow XB^{\prime}Z}^{\leftrightarrow}$
to obtain the following state:%
\begin{equation}
\rho_{XB^{\prime}EZ}\coloneqq \mathcal{L}_{AB\rightarrow XB^{\prime}Z}%
^{\leftrightarrow}(\psi_{ABE}), \label{eq-SKD:state-after-LOPC-cor}%
\end{equation}
where $\psi_{ABE}$ is a purification of the state $\rho_{AB}$ of interest and
$\mathcal{L}_{AB\rightarrow XB^{\prime}Z}^{\leftrightarrow}$ is an
LOPC\ channel with classical output system $X^{\prime}$ and quantum output
system~$B^{\prime}$. Then we obtain the following by applying
Theorem~\ref{thm-SKD:one-shot-key-lower-bnd}:

\begin{corollary}
{cor-SKD:one-shot-lower-bnd-standard-model-SKD}Let $\rho_{AB}$ be a
bipartite state, with system $X$ held by Alice, $B$ by Bob, and $E$ by Eve.
For all $\varepsilon\in(0,1]$, $\varepsilon^{\prime}\coloneqq 1-\sqrt{1-\varepsilon}$,
$\delta\in(0,\varepsilon^{\prime})$, $\eta\in(0,\varepsilon^{\prime}-\delta)$,
and $\zeta\in(0,\delta)$, there exists a $(K,\varepsilon)$ one-way key
distillation protocol for $\rho_{AB}$ with%
\begin{multline}
\log_{2}K=\overline{I}_{H}^{\varepsilon^{\prime}-\delta-\eta}(X;B^{\prime
})_{\rho}-\overline{I}_{\max}^{\delta-\zeta}(EZ;X)_{\rho}\\
-\log_{2}\!\left(
\frac{4(\varepsilon^{\prime}-\delta)}{\eta^{2}}\right)  -\log_{2}\!\left(
\frac{2}{\zeta^{2}}\right)  ,
\end{multline}
where the hypothesis testing mutual information $\overline{I}_{H}%
^{\varepsilon^{\prime}-\delta-\eta}(X;B^{\prime})_{\rho}$ is defined in \eqref{eq-hypo_testing_mutual_inf}, the smooth max-mutual information $\overline{I}_{\max}^{\delta-\zeta
}(EZ;X)_{\rho}$ is defined in \eqref{eq-SKD:alt-smooth-max-MI-def}, and these
quantities are evaluated with respect to the state in
\eqref{eq-SKD:state-after-LOPC-cor}, with $\mathcal{L}_{AB\rightarrow
XB^{\prime}Z}^{\leftrightarrow}$ an LOPC\ channel with classical output system
$X$ and quantum output system $B^{\prime}$. Consequently, for the one-shot
distillable key of $\rho_{AB}$, we have%
\begin{multline}
K_{D}^{\varepsilon}(A;B)_{\rho}\geq\sup_{\substack{\mathcal{L}%
^{\leftrightarrow},\delta\in(0,\varepsilon^{\prime}),\\\eta\in(0,\varepsilon
^{\prime}-\delta),\zeta\in(0,\delta)}}\overline{I}_{H}^{\varepsilon^{\prime
}-\delta-\eta}(X;B^{\prime})_{\rho}-\overline{I}_{\max}^{\delta-\zeta
}(EZ;X)_{\rho}\\
-\log_{2}\!\left(  \frac{4(\varepsilon^{\prime}-\delta)}{\eta^{2}}\right)
-\log_{2}\!\left(  \frac{2}{\zeta^{2}}\right)  ,
\end{multline}
where $\varepsilon^{\prime}\coloneqq 1-\sqrt{1-\varepsilon}$ and the optimization is
over every LOPC\ channel $\mathcal{L}_{AB\rightarrow XB^{\prime}%
Z}^{\leftrightarrow}$.
\end{corollary}

Now combining Corollary~\ref{cor-SKD:one-shot-lower-bnd-standard-model-SKD}
with Propositions~\ref{prop-smooth_max_to_petz_renyi} and \ref{prop:ineq-hypo-renyi}, we conclude the following lower bound on one-shot
distillable key:

\begin{corollary}
{cor-SKD:secret-key-rates-R\'enyi}Let $\rho_{AB}$ be a bipartite state
with purification $\psi_{ABE}$. For all $\varepsilon\in(0,1)$, $\varepsilon
^{\prime}=1-\sqrt{1-\varepsilon}$, $\delta\in(0,\varepsilon^{\prime})$,
$\eta\in(0,\varepsilon^{\prime}-\delta)$, $\zeta\in(0,\delta)$, $\nu
\in(0,\delta-\zeta)$, $\alpha\in(0,1),$ and $\beta>1$, there exists a
$(K,\varepsilon)$ one-way key distillation protocol for $\rho_{AB}$ satisfying%
\[
\log_{2}K\geq\overline{I}_{\alpha}(X;B^{\prime})_{\rho}-\widetilde{I}_{\beta
}^{\prime}(X;EZ)_{\rho}-f(\varepsilon^{\prime},\delta,\eta,\nu,\zeta
,\alpha,\beta)
\]
where
\begin{equation}
\rho_{XB^{\prime}EZ}\coloneqq \mathcal{L}_{AB\rightarrow XB^{\prime}Z}%
^{\leftrightarrow}(\psi_{ABE}),
\end{equation}
$\mathcal{L}_{AB\rightarrow XB^{\prime}Z}^{\leftrightarrow}$ is an
LOPC\ channel with classical output system $X$ and quantum output system
$B^{\prime}$,%
\begin{equation}
\widetilde{I}_{\beta}^{\prime}(X;EZ)_{\rho}\coloneqq \widetilde{D}_{\beta}(\rho
_{XEZ}\Vert\rho_{X}\otimes\rho_{EZ}), \label{eq-SKD:R\'enyi-MI-quantity}%
\end{equation}
and%
\begin{multline}
f(\varepsilon^{\prime},\delta,\eta,\nu,\zeta,\alpha,\beta)\coloneqq \frac{\alpha
}{1-\alpha}\log_{2}\!\left(  \frac{1}{\varepsilon^{\prime}-\delta-\eta
}\right)  +\log_{2}\!\left(  \frac{8}{\nu^{2}}\right)
\label{eq-SKD:complic-function}\\
+\frac{1}{\beta-1}\log_{2}\!\left(  \frac{1}{\left(  \delta-\zeta-\nu\right)
^{2}}\right)  +\log_{2}\!\left(  \frac{1}{1-\left(  \delta-\zeta-\nu\right)
^{2}}\right) \\
+\log_{2}\!\left(  \frac{4(\varepsilon^{\prime}-\delta)}{\eta^{2}}\right)
+\log_{2}\!\left(  \frac{2}{\zeta^{2}}\right)  .
\end{multline}

\end{corollary}

\begin{Proof}
The main idea here is to convert the smooth mutual information quantities
$\overline{I}_{H}^{\varepsilon^{\prime}-\delta-\eta}(X;B^{\prime})_{\rho}$ and
$\overline{I}_{\max}^{\delta-\zeta}(EZ;X)_{\rho}$ from
Corollary~\ref{cor-SKD:one-shot-lower-bnd-standard-model-SKD} to R\'{e}nyi
mutual information quantities with correction terms related to the smoothing
parameters. Let us first invoke Proposition~\ref{prop:ineq-hypo-renyi} to conclude the following
lower bound:%
\begin{equation}
\overline{I}_{H}^{\varepsilon^{\prime}-\delta-\eta}(X;B^{\prime})_{\rho}%
\geq\overline{I}_{\alpha}(X;B^{\prime})_{\rho}-\frac{\alpha}{1-\alpha}\log
_{2}\left(  \frac{1}{\varepsilon^{\prime}-\delta-\eta}\right)  .
\end{equation}
Next, we invoke Lemma~\ref{lem-SKD:smooth-max-MI-relations}\ below to conclude
that%
\begin{equation}
\overline{I}_{\max}^{\delta-\zeta}(EZ;X)_{\rho}\leq D_{\max}^{\delta-\zeta
-\nu}(\rho_{XEZ}\Vert\rho_{X}\otimes\rho_{EZ})+\log_{2}\!\left(  \frac{8}%
{\nu^{2}}\right)  ,
\end{equation}
where $\nu\in(0,\delta-\zeta)$. Then we invoke Proposition~\ref{prop-smooth_max_to_petz_renyi} to conclude
that%
\begin{multline}
D_{\max}^{\delta-\zeta-\nu}(\rho_{XEZ}\Vert\rho_{X}\otimes\rho_{EZ}%
)\leq\widetilde{D}_{\beta}(\rho_{XEZ}\Vert\rho_{X}\otimes\rho_{EZ})\\
+\frac{1}{\beta-1}\log_{2}\!\left(  \frac{1}{\left(  \delta-\zeta-\nu\right)
^{2}}\right)  +\log_{2}\!\left(  \frac{1}{1-\left(  \delta-\zeta-\nu\right)
^{2}}\right)  .
\end{multline}
Considering that%
\begin{equation}
\widetilde{D}_{\beta}(\rho_{XEZ}\Vert\rho_{X}\otimes\rho_{EZ})=\widetilde
{I}_{\beta}^{\prime}(X;EZ)_{\rho}.
\end{equation}
Putting all of the above together with
Corollary~\ref{cor-SKD:one-shot-lower-bnd-standard-model-SKD}, we conclude the proof.
\end{Proof}

\begin{Lemma}
{lem-SKD:smooth-max-MI-relations}Let $\rho_{AE}$ be a bipartite state,
and let $\varepsilon,\delta>0$ be such that $\varepsilon+\delta<1$. Then%
\begin{equation}
\overline{I}_{\max}^{\varepsilon+\delta}(E;A)_{\rho}\leq D_{\max
}^{\varepsilon}(\rho_{AE}\Vert\rho_{A}\otimes\rho_{E})+\log_{2}\!\left(
\frac{8}{\delta^{2}}\right)  ,
\label{eq-SKD:alt-sm-imax-to-sm-imax}
\end{equation}
where $\overline{I}_{\max}^{\varepsilon+\delta}(E;A)_{\rho}$ is defined in
\eqref{eq-SKD:alt-smooth-max-MI-def} and $D_{\max}^{\varepsilon}\!\left(  \rho_{AE}%
\Vert\rho_{A}\otimes\rho_{E}\right)  $ in~\eqref{eq-QEI:smooth-max-rel-ent-def}.
\end{Lemma}

\begin{Proof}
See Appendix~\ref{sec-SKD:two-smooth-maxes}.
\end{Proof}

\section{Distillable Key of a Quantum State}

Having found upper and lower bounds on the one-shot distillable key
$K_{D}^{\varepsilon}(A;B)_{\rho}$ of a bipartite state $\rho_{AB}$, let us now
move on to the asymptotic setting. In this setting, we allow Alice and Bob to
make use of an arbitrarily large number $n$ of copies of the state $\rho_{AB}$
in order to obtain a secret-key state. A \textit{secret key distillation
protocol for }$n$\textit{ copies of }$\rho_{AB}$ is defined by the triple
$(n,K,\mathcal{L}_{A^{n}B^{n}\rightarrow K_{A}K_{B}Z}^{\leftrightarrow})$,
consisting of the number $n$ of copies of $\rho_{AB}$, an integer
$K\in\mathbb{N}$, and an LOPC\ channel $\mathcal{L}_{A^{n}B^{n}\rightarrow
K_{A}K_{B}Z}^{\leftrightarrow}$ with $d_{K_{A}}=d_{K_{B}}=K$. Observe that a
secret-key distillation protocol for $n$ copies of $\rho_{AB}$ is equivalent
to a one-shot secret-key distillation protocol for the state $\rho
_{AB}^{\otimes n}$. All of the results of
Section~\ref{sec-SKD:one-shot-setting} thus carry over to the asymptotic
setting simply by replacing $\rho_{AB}$ with $\rho_{AB}^{\otimes n}$. In
particular, the error probability for a secret-key distillation protocol for
$\rho_{AB}$ defined by $(n,K,\mathcal{L}_{A^{n}B^{n}\rightarrow K_{A}K_{B}%
Z}^{\leftrightarrow})$ is equal to%
\begin{equation}
p_{\text{err}}(\mathcal{L}^{\leftrightarrow};\rho_{AB}^{\otimes n}%
)=\inf_{\gamma_{K_{A}K_{B}EZ}}\left(  1-F(\gamma_{K_{A}K_{B}EZ},\mathcal{L}%
_{A^{n}B^{n}\rightarrow K_{A}K_{B}Z}^{\leftrightarrow}(\psi_{ABE}^{\otimes
n}))\right)  , \label{eq-SKD:n-shot-err-def}%
\end{equation}
where the infimum is with respect to every ideal tripartite key state
$\gamma_{K_{A}K_{B}EZ}$ and $\psi_{ABE}$ is a purification of $\rho_{AB}$. The
definition in \eqref{eq-SKD:n-shot-err-def} is thus the same as that in
\eqref{eq-SKD:error-criterion}, but for the tensor-power state $\rho
_{AB}^{\otimes n}$.

\begin{definition}
{$(n,K,\varepsilon)$ Secret-Key Distillation Protocol}{}A secret-key
distillation protocol $(n,K,\mathcal{L}_{A^{n}B^{n}\rightarrow K_{A}K_{B}%
Z}^{\leftrightarrow})$ for $n$ copies of $\rho_{AB}$, with $d_{K_{A}}%
=d_{K_{B}}=K$, is called an $(n,K,\varepsilon)$ \textit{protocol}, with
$\varepsilon\in\left[  0,1\right]  $, if $p_{\text{err}}(\mathcal{L}%
^{\leftrightarrow};\rho_{AB}^{\otimes n}) \leq \varepsilon$.
\end{definition}

Based on the discussion above, we note here that an $(n,K,\varepsilon)$
secret-key distillation protocol for $\rho_{AB}$ is a $(K,\varepsilon)$
secret-key distillation protocol for $\rho_{AB}^{\otimes n}$.

The rate $R(n,K)$ of an $(n,K,\varepsilon)$ secret-key distillation protocol
for $n$ copies of a given state is%
\begin{equation}
R(n,K)\coloneqq \frac{\log_{2}K}{n},
\end{equation}
which can be thought of as the number of $\varepsilon$-approximate secret-key
bits contained in the final state of the protocol, per copy of the given
initial state. Given a state $\rho_{AB}$ and $\varepsilon\in\left[
0,1\right]  $, the maximum rate of secret key distillation among all
$(n,K,\varepsilon)$ secret-key distillation protocols for $\rho_{AB}$ is%
\begin{align}
K_{D}^{n,\varepsilon}(\rho_{AB})  &  \equiv K_{D}^{n,\varepsilon}(A;B)_{\rho
}\coloneqq \frac{1}{n}K_{D}^{\varepsilon}(\rho_{AB}^{\otimes n}%
)\label{eq-SKD:distillable-key-finite-n-eps}\\
&  =\sup_{(K,\mathcal{L}^{\leftrightarrow})}\left\{  \frac{\log_{2}K}%
{n}:p_{\text{err}}(\mathcal{L}^{\leftrightarrow};\rho_{AB}^{\otimes n}%
)\leq\varepsilon\right\}  ,
\end{align}
where the optimization is with respect to all $K\in\mathbb{N}$ and every
LOPC\ channel $\mathcal{L}_{A^{n}B^{n}\rightarrow K_{A}K_{B}Z}%
^{\leftrightarrow}$ with $d_{K_{A}}=d_{K_{B}}=K$.

\begin{definition}
{Achievable Rate for Secret Key Distillation}{}Given a bipartite state
$\rho_{AB}$, a rate $R\in\mathbb{R}^{+}$ is called an \textit{achievable rate
for secret key distillation for }$\rho_{AB}$ if for all $\varepsilon\in(0,1]$,
$\delta>0$, and sufficiently large $n$, there exists an $(n,2^{n(R-\delta
)},\varepsilon)$ secret-key distillation protocol for~$\rho_{AB}$.
\end{definition}

As we prove in Appendix~\ref{chap-str_conv},%
\begin{equation}
R\text{ achievable rate\quad}\Longleftrightarrow\quad\lim_{n\rightarrow\infty
}\varepsilon_{D}(2^{n(R-\delta)};\rho_{AB}^{\otimes n})=0\qquad\forall
\delta>0.
\end{equation}
In other words, a rate $R$ is achievable if the optimal error probability for
a sequence of protocols with rate $R-\delta$ vanishes as the number $n$ of
copies of $\rho_{AB}$ increases.

\begin{definition}
{Distillable Key of a Quantum State}{def-SKD:dist-key-def} The \textit{distillable key of a bipartite
state }$\rho_{AB}$, denoted by $K_{D}(A;B)_{\rho}$, is defined to be the
supremum of all achievable rates for secret key distillation for $\rho_{AB}$,
i.e.,%
\begin{equation}
K_{D}(A;B)_{\rho}\coloneqq \sup\left\{  R:R\text{ is an achievable rate for }\rho
_{AB}\right\}  .
\end{equation}

\end{definition}

The distillable key can also be written as%
\begin{equation}
K_{D}(A;B)_{\rho}=\inf_{\varepsilon\in(0,1]}\liminf_{n\rightarrow\infty}%
\frac{1}{n}K_{D}^{\varepsilon}(\rho_{AB}^{\otimes n}).
\end{equation}
See Appendix~\ref{chap-str_conv}\ for a proof.

\begin{definition}
{Weak Converse Rate for Secret Key Distillation}{} Given a bipartite state
$\rho_{AB}$, a rate $R\in\mathbb{R}^{+}$ is called a \textit{weak converse
rate for secret key distillation for} $\rho_{AB}$ if every $R^{\prime}>R$ is
not an achievable rate for $\rho_{AB}$.
\end{definition}

As we show in Appendix~\ref{chap-str_conv},%
\begin{equation}
R\text{ weak converse rate\quad}\Longleftrightarrow\quad\lim_{n\rightarrow
\infty}\varepsilon_{D}(2^{n(R-\delta)};\rho_{AB}^{\otimes n})>0\qquad
\forall\delta>0.
\end{equation}

\begin{definition}
{Strong Converse Rate for Secret Key Distillation}{}Given a bipartite state
$\rho_{AB}$, a rate $R\in\mathbb{R}^{+}$ is called a \textit{strong converse
rate for secret key distillation for }$\rho_{AB}$ if for all $\varepsilon
\in\lbrack0,1)$, $\delta>0$, and sufficiently large $n$, there does not exist
an $(n,2^{n\left(  R+\delta\right)  },\varepsilon)$ secret key distillation
protocol for $\rho_{AB}$.
\end{definition}

We show in Appendix~\ref{chap-str_conv} that%
\begin{equation}
R\text{ strong converse rate\quad}\Longleftrightarrow\quad\lim_{n\rightarrow
\infty}\varepsilon_{D}(2^{n(R-\delta)};\rho_{AB}^{\otimes n})=1\qquad
\forall\delta>0.
\end{equation}

\begin{definition}
{Strong Converse Distillable Entanglement of a Quantum State}{}The
\textit{strong converse distillable key} of a bipartite state $\rho_{AB}$,
denoted by $\widetilde{K}_{D}(A;B)_{\rho}$, is defined as the infimum of all
strong converse rates, i.e.,%
\begin{equation}
\widetilde{K}_{D}(A;B)_{\rho}\coloneqq \inf\left\{  R:R\text{ is a strong converse
rate for }\rho_{AB}\right\}  .
\end{equation}

\end{definition}

Note that%
\begin{equation}
K_{D}(A;B)_{\rho}\leq\widetilde{K}_{D}(A;B)_{\rho}%
\end{equation}
for every bipartite state $\rho_{AB}$. We can also write the strong converse
distillable key as%
\begin{equation}
\widetilde{K}_{D}(A;B)_{\rho}=\sup_{\varepsilon\in\lbrack0,1)}\limsup
_{n\rightarrow\infty}\frac{1}{n}K_{D}^{\varepsilon}(\rho_{AB}^{\otimes n}).
\end{equation}
See Appendix~\ref{chap-str_conv} for a proof.

We are now ready to present a general expression for the distillable key of a
bipartite state, as well as two upper bounds on it.

\begin{theorem*}
{Distillable Key of a Bipartite State}{thm-SKD:distillable-key}The
distillable key of a bipartite state $\rho_{AB}$ is given by%
\begin{equation}
K_{D}(A;B)_{\rho}=\lim_{n\rightarrow\infty}\frac{1}{n}\sup_{\mathcal{L}^{(n)}%
}\left(  I(X;B^{\prime})_{\mathcal{L}^{(n)}(\psi^{\otimes n})}%
-I(X;EZ)_{\mathcal{L}^{(n)}(\psi^{\otimes n})}\right)  ,
\label{eq-SKD:distillable-key-theorem}%
\end{equation}
where the optimization is with respect to every two-way LOPC\ channel
$\mathcal{L}_{A^{n}B^{n}\rightarrow XB^{\prime}Z}^{(n)}$ with classical output
system $X$ and quantum output system~$B^{\prime}$. The information quantities
are evaluated with respect to the state $\mathcal{L}_{A^{n}B^{n}\rightarrow
XB^{\prime}Z}^{(n)}(\psi_{ABE}^{\otimes n})$, where $\psi_{ABE}$ is a
purification of $\rho_{AB}$. Furthermore, the relative entropy of entanglement
$E_{R}(A;B)_{\rho}$ from \eqref{eq-EM:rel-entr-enta-def} is a strong converse rate for distillable key,
in the sense that%
\begin{equation}
\widetilde{K}_{D}(A;B)_{\rho}\leq E_{R}(A;B)_{\rho},
\end{equation}
and the squashed entanglement from \eqref{eq-squashed_entanglement} is a weak converse rate, in the sense
that%
\begin{equation}
K_{D}(A;B)_{\rho}\leq E_{\operatorname{sq}}(A;B)_{\rho}.
\label{eq-SKD:sq-ent-asympt-dist-key-up-bnd}%
\end{equation}

\end{theorem*}

If we define
\begin{equation}
D_{K}^{\longleftrightarrow}(\rho_{AB})\equiv D_{K}^{\longleftrightarrow
}(A;B)_{\rho}\coloneqq \sup_{\mathcal{L}}\left(  I(X;B^{\prime})_{\mathcal{L}(\psi
)}-I(X;EZ)_{\mathcal{L}(\psi)}\right)  ,
\end{equation}
where the entropic quantities are evaluated with respect to $\mathcal{L}%
_{A^{n}B^{n}\rightarrow XB^{\prime}Z}(\psi_{ABE})$, with $\psi_{ABE}$ a
purification of $\rho_{AB}$, then we can write
\eqref{eq-SKD:distillable-key-theorem} as%
\begin{equation}
K_{D}(A;B)_{\rho}=\lim_{n\rightarrow\infty}\frac{1}{n}D_{K}%
^{\longleftrightarrow}(\rho_{AB}^{\otimes n})=:D_{\text{reg},K}%
^{\longleftrightarrow}(\rho_{AB}).
\end{equation}
Thus, the distillable key can be viewed as the regularization of
$D_{K}^{\longleftrightarrow}$, similar to what we found in \eqref{eq-ED:dist-ent-alt-exp-1} in Chapter~\ref{chap-ent_distill} for distillable entanglement.

Let us make the following observations about
Theorem~\ref{thm-SKD:distillable-key}.

\begin{itemize}
\item The private information is an achievable rate for secret key
distillation, i.e.,%
\begin{equation}
K_{D}(A;B)_{\rho}\geq\max\{I(X;B)_{\tau}-I(X;E)_{\tau},I(A;Y)_{\omega
}-I(Y;E)_{\omega}\}, \label{eq-SKD:unoptimized-private-info}%
\end{equation}
where%
\begin{align}
\tau_{XBE}  &  \coloneqq \sum_{x}|x\rangle\!\langle x|_{X}\otimes\operatorname{Tr}%
_{A}[\Lambda_{A}^{x}\psi_{ABE}],\label{eq-SKD:ccq-st-1}\\
\omega_{YAE}  &  \coloneqq \sum_{y}|y\rangle\!\langle y|_{Y}\otimes\operatorname{Tr}%
_{B}[\Gamma_{B}^{y}\psi_{ABE}], \label{eq-SKD:ccq-st-2}%
\end{align}
$\psi_{ABE}$ is a purification of the bipartite state $\rho_{AB}$, and
$\{\Lambda_{A}^{x}\}_{x}$ and $\{\Gamma_{B}^{y}\}_{y}$ are POVMs. The idea
behind the first achievable rate $I(X;B)_{\tau}-I(X;E)_{\tau}$ is that Alice
performs the measurement $\{\Lambda_{A}^{x}\}_{x}$ on her system $A$, and this
produces the classical--quantum--quantum state $\tau_{XBE}$. Alice and Bob
then execute the protocol from Theorem~\ref{thm-SKD:one-shot-key-lower-bnd}%
\ on many copies of the state $\tau_{XBE}$. Alternatively, the idea behind the
second achievable rate $I(A;Y)_{\omega}-I(Y;E)_{\omega}$ is similar, but with
the roles of Alice and Bob swapped and distilling a key from many copies of
the state $\omega_{YAE}$.\newline\newline We can also consider these
conclusions to be immediate consequences of
\eqref{eq-SKD:distillable-key-theorem}, which follow from dropping the
optimization over two-way LOPC channels and from the fact that the unoptimized
private informations in \eqref{eq-SKD:unoptimized-private-info} are additive
for product states.

\item In order to obtain a higher key distillation rate than the private
informations in \eqref{eq-SKD:unoptimized-private-info}, one strategy is to
use $n\geq2$ copies of $\psi_{ABE}$ along with a two-way LOPC\ channel
$\mathcal{L}_{A^{n}B^{n}\rightarrow XB^{\prime}Z}$, in order to obtain a state
$\omega_{A^{\prime}B^{\prime}E^{n}Z}\coloneqq \mathcal{L}_{A^{n}B^{n}\rightarrow
XB^{\prime}Z}(\psi_{ABE}^{\otimes n})$. The normalized private informations of
this latter state are potentially higher than that of $\tau_{XBE}$ and
$\omega_{YAE}$ in \eqref{eq-SKD:ccq-st-1}--\eqref{eq-SKD:ccq-st-2}. The
overall rate of this strategy is then
\begin{equation}
\frac{1}{n}\left(  I(X;B^{\prime
})_{\mathcal{L}^{(n)}(\psi^{\otimes n})}-I(X;EZ)_{\mathcal{L}^{(n)}%
(\psi^{\otimes n})}\right)  ,
\end{equation}
and Theorem~\ref{thm-SKD:distillable-key} tells
us that such a strategy is optimal in the large $n$ limit. With increasingly
more copies of $\psi_{ABE}$ to start with, it might be possible to obtain a
better rate, which is why we need to regularize in general.
\end{itemize}

As with other previous capacity theorem proofs in this book, we prove
Theorem~\ref{thm-SKD:distillable-key} in two steps:

\begin{enumerate}
\item \textit{Achievability}: We show that the right-hand side of
\eqref{eq-SKD:distillable-key-theorem} is an achievable rate for secret key
distillation for $\rho_{AB}$. Doing so involves exhibiting an explicit secret
key distillation protocol. The protocol we use is based on the one we
introduced in Section~\ref{sec-SKD:low-bound-1-shot-dist-key}\ to obtain a
lower bound on the one-shot distillable secret key.\newline\newline The
achievability part of the proof establishes that%
\begin{equation}
K_{D}(A;B)_{\rho}\geq\lim_{n\rightarrow\infty}\frac{1}{n}\sup_{\mathcal{L}%
^{(n)}}\left(  I(X;B^{\prime})_{\mathcal{L}^{(n)}(\psi^{\otimes n}%
)}-I(X;EZ)_{\mathcal{L}^{(n)}(\psi^{\otimes n})}\right)  .
\end{equation}

\item \textit{Weak converse}: We show that the right-hand side of
\eqref{eq-SKD:distillable-key-theorem} is a weak converse rate for secret key
distillation for $\rho_{AB}$, from which it follows that%
\begin{equation}
K_{D}(A;B)_{\rho}\leq\lim_{n\rightarrow\infty}\frac{1}{n}\sup_{\mathcal{L}%
^{(n)}}\left(  I(X;B^{\prime})_{\mathcal{L}^{(n)}(\psi^{\otimes n}%
)}-I(X;EZ)_{\mathcal{L}^{(n)}(\psi^{\otimes n})}\right)  .
\end{equation}
In order to show this, we use the one-shot upper bounds from Section~\ref{sec-SKD:upper-bounds} to
prove that every achievable rate $R$ satisfies%
\begin{equation}
R\leq\lim_{n\rightarrow\infty}\frac{1}{n}\sup_{\mathcal{L}^{(n)}}\left(
I(X;B^{\prime})_{\mathcal{L}^{(n)}(\psi^{\otimes n})}-I(X;EZ)_{\mathcal{L}%
^{(n)}(\psi^{\otimes n})}\right)  .
\end{equation}

\end{enumerate}

We go through the achievability part of the proof of
Theorem~\ref{thm-SKD:distillable-key}\ in Section~\ref{sec-SKD:ach-proof}. We then proceed with
the weak converse part in Section~\ref{sec-SKD:weak-conv-proof}.

The expression in \eqref{eq-SKD:distillable-key-theorem}\ for the distillable
key involves both a limit over an unbounded number of copies of the state
$\rho_{AB}$, as well as an optimization over all two-way LOPC channels.
Computing the distillable key is therefore intractable in general. After
establishing a proof of \eqref{eq-SKD:distillable-key-theorem}, we proceed to
establish upper bounds on distillable entanglement that depend only on the
given state $\rho_{AB}$. Specifically, in Section~\ref{sec-SKD:rel-entr-enta-str-conv}, we use the one-shot
results in Section~\ref{sec-SKD:rel-ent-up-bnd} to show that the 
relative entropy of entanglement is a strong converse rate for secret key distillation. We
also show that the squashed entanglement is a weak converse rate for
secret key distillation.

\subsection{Proof of Achievability}

\label{sec-SKD:ach-proof}

As the first step in proving the achievability part of
Theorem~\ref{thm-SKD:distillable-key}, let us recall
Corollary~\ref{cor-SKD:secret-key-rates-R\'enyi}:\ given a bipartite state
$\rho_{AB}$ with purification $\psi_{ABE}$, for all $\varepsilon\in(0,1)$,
$\varepsilon^{\prime}=1-\sqrt{1-\varepsilon}$, $\delta\in(0,\varepsilon
^{\prime})$, $\eta\in(0,\varepsilon^{\prime}-\delta)$, $\zeta\in(0,\delta)$,
$\nu\in(0,\delta-\zeta)$, $\alpha\in(0,1),$ and $\beta>1$, there exists a
$(K,\varepsilon)$ one-way key distillation protocol for $\rho_{AB}$ satisfying%
\begin{equation}
\log_{2}K\geq\overline{I}_{\alpha}(X;B^{\prime})_{\rho}-\widetilde{I}_{\beta
}^{\prime}(X;EZ)_{\rho}-f(\varepsilon^{\prime},\delta,\eta,\nu,\zeta
,\alpha,\beta) \label{eq-SKD:recall-one-shot-renyi-lb}%
\end{equation}
where
\begin{equation}
\rho_{XB^{\prime}EZ}\coloneqq \mathcal{L}_{AB\rightarrow XB^{\prime}Z}%
^{\leftrightarrow}(\psi_{ABE}), \label{eq-SKD:state-for-key-dist-5}%
\end{equation}
$\mathcal{L}_{AB\rightarrow XB^{\prime}Z}^{\leftrightarrow}$ is an
LOPC\ channel with classical output system $X$ and quantum output system
$B^{\prime}$, the R\'enyi mutual information $\widetilde{I}_{\beta}^{\prime
}(X;EZ)_{\rho}$ is defined in \eqref{eq-SKD:R\'enyi-MI-quantity}, and the
function $f(\varepsilon^{\prime},\delta,\eta,\nu,\zeta,\alpha,\beta)$ in
\eqref{eq-SKD:complic-function}. Applying this inequality to the state
$\rho_{AB}^{\otimes n}$ for all $n\in \mathbb{N}$ leads to the following:

\begin{proposition}
{prop-SKD:ach-rate-priv-info-renyi}For every state $\rho_{AB}$ and
$\varepsilon\in(0,1)$, there exists an $(n,K,\varepsilon)$ key distillation
protocol for $\rho_{AB}$ such that the rate $\frac{\log_{2}K}{n}$ satisfies%
\begin{equation}
\frac{\log_{2}K}{n}\geq\overline{I}_{\alpha}(X;B^{\prime})_{\rho}%
-\widetilde{I}_{\beta}^{\prime}(X;EZ)_{\rho}-\frac{1}{n}f\!\left(
\varepsilon^{\prime},\frac{\varepsilon^{\prime}}{2},\frac{\varepsilon^{\prime
}}{4},\frac{\varepsilon^{\prime}}{4},\frac{\varepsilon^{\prime}}{2}%
,\alpha,\beta\right)  , \label{eq-SKD:lb-dist-key-tensor-power-renyi}%
\end{equation}
for all $n\in \mathbb{N}$, $\alpha\in(0,1)$, $\beta>1$, where the information
quantities are with respect to the state in
\eqref{eq-SKD:state-for-key-dist-5}. More generally, we have the following
lower bound on the finite-length distillable key:%
\begin{multline}
K_{D}^{n,\varepsilon}(A;B)_{\rho}\geq\frac{1}{n}\sup_{\mathcal{L}%
^{\leftrightarrow}}\left(  \overline{I}_{\alpha}(X;B^{\prime})_{\tau
}-\widetilde{I}_{\beta}^{\prime}(X;E^{n}Z)_{\tau}\right)
\label{eq-SKD:distillable-key-finite-n-eps-lb}\\
-\frac{1}{n}f\!\left(  \varepsilon^{\prime},\frac{\varepsilon^{\prime}}%
{2},\frac{\varepsilon^{\prime}}{4},\frac{\varepsilon^{\prime}}{4}%
,\frac{\varepsilon^{\prime}}{2},\alpha,\beta\right)  ,
\end{multline}
for all $n\in \mathbb{N}$, $\alpha\in(0,1)$, $\beta>1$, where the optimization is over
every LOPC\ channel $\mathcal{L}_{A^{n}B^{n}\rightarrow XB^{\prime}%
Z}^{\leftrightarrow}$ and the information quantities are with respect to the
following state:%
\begin{equation}
\tau_{XB^{\prime}E^{n}Z}\coloneqq \mathcal{L}_{A^{n}B^{n}\rightarrow XB^{\prime}%
Z}^{\leftrightarrow}(\psi_{ABE}^{\otimes n}).
\end{equation}

\end{proposition}

\begin{Proof}
Let $\psi_{ABE}$ be a purification of $\rho_{AB}$, and use the tensor-product
purification $\psi_{ABE}^{\otimes n}$ for $\rho_{AB}^{\otimes n}$. Also, let
$\delta=\frac{\varepsilon^{\prime}}{2}$, $\eta=\frac{\varepsilon^{\prime}}{4}%
$, $\nu=\frac{\varepsilon^{\prime}}{4}$, and $\zeta=\frac{\varepsilon^{\prime
}}{2}$. Substituting all of this into the inequality in
\eqref{eq-SKD:recall-one-shot-renyi-lb} and simplifying leads to the
inequality%
\begin{equation}
\frac{\log_{2}K}{n}\geq\frac{1}{n}\left(  \overline{I}_{\alpha}(X;B^{\prime
})_{\tau}-\widetilde{I}_{\beta}^{\prime}(X;E^{n}Z)_{\tau}-f\!\left(
\varepsilon^{\prime},\frac{\varepsilon^{\prime}}{2},\frac{\varepsilon^{\prime
}}{4},\frac{\varepsilon^{\prime}}{4},\frac{\varepsilon^{\prime}}{2}%
,\alpha,\beta\right)  \right)  ,
\end{equation}
where $\tau_{XB^{\prime}E^{n}Z}\coloneqq \mathcal{L}_{A^{n}B^{n}\rightarrow
XB^{\prime}Z}^{\leftrightarrow}(\psi_{ABE}^{\otimes n})$. Then, optimizing
over every LOPC\ channel $\mathcal{L}_{AB\rightarrow XB^{\prime}%
Z}^{\leftrightarrow}$, and using the definition of $K_{D}^{n,\varepsilon
}(A;B)_{\rho}$ in \eqref{eq-SKD:distillable-key-finite-n-eps}, we obtain \eqref{eq-SKD:distillable-key-finite-n-eps-lb}.

By restricting the state $\tau_{XB^{\prime}E^{n}Z}$ to have the form
$\tau_{XB^{\prime}EZ}^{\otimes n}=\left(  \mathcal{L}_{AB\rightarrow
XB^{\prime}Z}^{\leftrightarrow}(\psi_{ABE})\right)  ^{\otimes n}$ (i.e., using
a tensor-power LOPC\ strategy) and employing additivity of $\overline
{I}_{\alpha}(X^{n};B^{\prime n})_{\tau^{\otimes n}}$ and $\widetilde{I}%
_{\beta}^{\prime}(X^{n};E^{n}Z^{n})_{\tau^{\otimes n}}$, we conclude that%
\begin{equation}
\overline{I}_{\alpha}(X^{n};B^{\prime n})_{\tau^{\otimes n}}-\widetilde
{I}_{\beta}^{\prime}(X^{n};E^{n}Z^{n})_{\tau^{\otimes n}}=n\left(
\overline{I}_{\alpha}(X;B^{\prime})_{\tau}-\widetilde{I}_{\beta}^{\prime
}(X;EZ)_{\tau}\right)  .
\end{equation}
The lower bound in \eqref{eq-SKD:lb-dist-key-tensor-power-renyi} then follows.
\end{Proof}

Using the inequality in \eqref{eq-SKD:lb-dist-key-tensor-power-renyi}, we can
prove the following lower bound on distillable key:

\begin{theorem*}
{Achievability of Private Information for Secret Key Distillation}
{thm-SKD:ach-priv-info-SKD}The private information $I(X;B^{\prime
})_{\tau}-I(X;EZ)_{\tau}$ is an achievable rate for secret key distillation
for $\rho_{AB}$, where the private information is evaluated with respect to
the following state:%
\begin{equation}
\tau_{XB^{\prime}EZ}\coloneqq \mathcal{L}_{AB\rightarrow XB^{\prime}Z}%
^{\leftrightarrow}(\psi_{ABE}),
\end{equation}
and $\psi_{ABE}$ is a purification of $\rho_{AB}$. In other words,%
\begin{equation}
K_{D}(A;B)_{\rho}\geq I(X;B^{\prime})_{\tau}-I(X;EZ)_{\tau}%
\end{equation}
for every bipartite state $\rho_{AB}$.
\end{theorem*}

\begin{Proof}
Let $\psi_{ABE}$ be a purification of $\rho_{AB}$. Fix $\varepsilon\in(0,1]$
and $\delta>0$. Let $\delta_{1},\delta_{2}>0$ be such that $\delta=\delta
_{1}+\delta_{2}$. Set $\alpha\in(0,1)$ and $\beta>1$ such that%
\begin{equation}
\delta_{1}\geq I(X;B^{\prime})_{\tau}-I(X;EZ)_{\tau}-\left(  \overline
{I}_{\alpha}(X;B^{\prime})_{\tau}-\widetilde{I}_{\beta}^{\prime}(X;EZ)_{\tau
}\right)  . \label{eq-SKD:to-asymp-ach-delta1}%
\end{equation}
Note that this is possible because $\overline{I}_{\alpha}(X;B^{\prime})_{\tau
}$ increases monotonically with increasing $\alpha\in(0,1)$ (see Proposition~\ref{prop-Petz_rel_ent}) and $\widetilde
{I}_{\beta}^{\prime}(X;EZ)_{\tau}$ decreases monotonically with decreasing
$\beta$ (see Proposition~\ref{prop-sand_rel_ent_properties}), so that%
\begin{align}
\lim_{\alpha\rightarrow1^{-}}\overline{I}_{\alpha}(X;B^{\prime})_{\tau}  &
=\sup_{\alpha\in(0,1)}\overline{I}_{\alpha}(X;B^{\prime})_{\tau},\\
\lim_{\beta\rightarrow1^{+}}\widetilde{I}_{\beta}^{\prime}(X;EZ)_{\tau}  &
=\inf_{\beta\in(1,\infty)}\widetilde{I}_{\beta}^{\prime}(X;EZ)_{\tau}.
\end{align}
Also,%
\begin{align}
I(X;B^{\prime})_{\tau}  &  =\lim_{\alpha\rightarrow1^{-}}\overline{I}_{\alpha
}(X;B^{\prime})_{\tau},\\
I(X;EZ)_{\tau}  &  =\lim_{\beta\rightarrow1^{+}}\widetilde{I}_{\beta}^{\prime
}(X;EZ)_{\tau}.
\end{align}
With $\alpha$ and $\beta$ chosen such that \eqref{eq-SKD:to-asymp-ach-delta1}
holds, take $n$ large enough so that%
\begin{equation}
\delta_{2}\geq\frac{1}{n}f\!\left(  \varepsilon^{\prime},\frac{\varepsilon
^{\prime}}{2},\frac{\varepsilon^{\prime}}{4},\frac{\varepsilon^{\prime}}%
{4},\frac{\varepsilon^{\prime}}{2},\alpha,\beta\right)  .
\label{eq-SKD:ach-rate-proof-delta2}%
\end{equation}
Now, we use the fact that for the $n$ and $\varepsilon$ chosen above, there
exists an $(n,K,\varepsilon)$ protocol such that%
\begin{equation}
\frac{\log_{2}K}{n}\geq\overline{I}_{\alpha}(X;B^{\prime})_{\rho}%
-\widetilde{I}_{\beta}^{\prime}(X;EZ)_{\rho}-\frac{1}{n}f\!\left(
\varepsilon^{\prime},\frac{\varepsilon^{\prime}}{2},\frac{\varepsilon^{\prime
}}{4},\frac{\varepsilon^{\prime}}{4},\frac{\varepsilon^{\prime}}{2}%
,\alpha,\beta\right)  .
\label{eq-SKD:ach-proof-one-shot-to-asymp-55}
\end{equation}
(This follows from Proposition~\ref{prop-SKD:ach-rate-priv-info-renyi}%
\ above.) Rearranging the right-hand side of this inequality, and using \eqref{eq-SKD:to-asymp-ach-delta1}, \eqref{eq-SKD:ach-rate-proof-delta2}, and \eqref{eq-SKD:ach-proof-one-shot-to-asymp-55}, we
find that%
\begin{align}
\frac{\log_{2}K}{n}  &  \geq I(X;B^{\prime})_{\tau}-I(X;EZ)_{\tau}\nonumber\\
&  \quad - \left(
\begin{array}
[c]{c}%
I(X;B^{\prime})_{\tau}-I(X;EZ)_{\tau}-\left(  \overline{I}_{\alpha
}(X;B^{\prime})_{\tau}-\widetilde{I}_{\beta}^{\prime}(X;EZ)_{\tau}\right) \\
+\frac{1}{n}f\!\left(  \varepsilon^{\prime},\frac{\varepsilon^{\prime}}%
{2},\frac{\varepsilon^{\prime}}{4},\frac{\varepsilon^{\prime}}{4}%
,\frac{\varepsilon^{\prime}}{2},\alpha,\beta\right)
\end{array}
\right) \\
&  \geq I(X;B^{\prime})_{\tau}-I(X;EZ)_{\tau}-\left(  \delta_{1}+\delta
_{2}\right) \\
&  =I(X;B^{\prime})_{\tau}-I(X;EZ)_{\tau}-\delta.
\end{align}
We thus have shown that there exists an $(n,K,\varepsilon)$ secret key
distillation protocol with rate $\frac{\log_{2}K}{n}\geq I(X;B^{\prime}%
)_{\tau}-I(X;EZ)_{\tau}-\delta$. Therefore, there exists an $(n,2^{n\left(
R-\delta\right)  },\varepsilon)$ secret-key distillation protocol with
$R=I(X;B^{\prime})_{\tau}-I(X;EZ)_{\tau}$ for all sufficiently large $n$ such
that \eqref{eq-SKD:ach-rate-proof-delta2} holds.\ Since $\varepsilon$ and
$\delta$ are arbitrary, we conclude that for all $\varepsilon\in(0,1]$,
$\delta>0$, and sufficiently large $n$, there exists an $(n,2^{n\left(
I(X;B^{\prime})_{\tau}-I(X;EZ)_{\tau}-\delta\right)  },\varepsilon)$ secret
key distillation protocol. This means that, by definition, $I(X;B^{\prime
})_{\tau}-I(X;EZ)_{\tau}$ is an achievable rate.
\end{Proof}

\subsubsection*{Proof of the Achievability Part of
Theorem~\ref{thm-SKD:distillable-key}}

Let $\mathcal{L}_{A^{k}B^{k}\rightarrow XB^{\prime}Z}^{\leftrightarrow}$ be an
arbitrary LOPC channel with $k\in \mathbb{N}$, let%
\begin{equation}
\tau_{XB^{\prime}E^{k}Z}\coloneqq \mathcal{L}_{A^{k}B^{k}\rightarrow XB^{\prime}%
Z}^{\leftrightarrow}(\psi_{ABE}^{\otimes k}),
\end{equation}
where $\psi_{ABE}$ is a purification of $\rho_{AB}$. Fix $\varepsilon\in(0,1]$
and $\delta>0$. Let $\delta_{1},\delta_{2}>0$ be such that $\delta=\delta
_{1}+\delta_{2}$. Set $\alpha\in(0,1)$ and $\beta\in(1,\infty)$ such that%
\begin{equation}
\delta_{1}\geq\frac{1}{k}\left(  I(X;B^{\prime})_{\tau}-I(X;E^{k}Z)_{\tau
}\right)  -\frac{1}{k}\left(  \overline{I}_{\alpha}(X;B^{\prime})_{\tau
}-\widetilde{I}_{\beta}^{\prime}(X;E^{k}Z)_{\tau}\right)  ,
\label{eq-SKD:regularized-ach-proof-1}
\end{equation}
which is possible based on the arguments given in the proof of
Theorem~\ref{thm-SKD:ach-priv-info-SKD}\ above.\ Then, with this choice of
$\alpha$ and $\beta$, take $n$ large enough so that%
\begin{equation}
\delta_{2}\geq\frac{1}{kn}f\!\left(  \varepsilon^{\prime},\frac{\varepsilon
^{\prime}}{2},\frac{\varepsilon^{\prime}}{4},\frac{\varepsilon^{\prime}}%
{4},\frac{\varepsilon^{\prime}}{2},\alpha,\beta\right)  .
\label{eq-SKD:delta-2-ach-reg-priv-info}
\end{equation}
Now, we use the fact that, for the chosen $n$ and $\varepsilon$, there exists
an $(n,K,\varepsilon)$ secret-key distillation protocol such that
\eqref{eq-SKD:lb-dist-key-tensor-power-renyi} holds, i.e.,%
\begin{equation}
\frac{\log_{2}K}{n}\geq\overline{I}_{\alpha}(X;B^{\prime})_{\tau}%
-\widetilde{I}_{\beta}^{\prime}(X;E^{k}Z)_{\tau}-\frac{1}{n}f\!\left(
\varepsilon^{\prime},\frac{\varepsilon^{\prime}}{2},\frac{\varepsilon^{\prime
}}{4},\frac{\varepsilon^{\prime}}{4},\frac{\varepsilon^{\prime}}{2}%
,\alpha,\beta\right)  .
\end{equation}
Dividing both sides by $k$ gives%
\begin{equation}
\frac{\log_{2}K}{kn}\geq\frac{1}{k}\left(  \overline{I}_{\alpha}(X;B^{\prime
})_{\tau}-\widetilde{I}_{\beta}^{\prime}(X;E^{k}Z)_{\tau}\right)  -\frac
{1}{kn}f\!\left(  \varepsilon^{\prime},\frac{\varepsilon^{\prime}}{2}%
,\frac{\varepsilon^{\prime}}{4},\frac{\varepsilon^{\prime}}{4},\frac
{\varepsilon^{\prime}}{2},\alpha,\beta\right)  .
\label{eq-SKD:regularized-ach-proof-last}
\end{equation}
Rearranging the right-hand side of this inequality, and using \eqref{eq-SKD:regularized-ach-proof-1}--\eqref{eq-SKD:regularized-ach-proof-last}, we find
that%
\begin{align}
\frac{\log_{2}K}{kn}  &  \geq\frac{1}{k}\left(  I(X;B^{\prime})_{\tau
}-I(X;E^{k}Z)_{\tau}\right) \nonumber\\
&  \qquad -\left(  \frac{1}{k}\left(  I(X;B^{\prime})_{\tau}-I(X;E^{k}Z)_{\tau
}\right)  -\frac{1}{k}\left(  \overline{I}_{\alpha}(X;B^{\prime})_{\tau
}-\widetilde{I}_{\beta}^{\prime}(X;E^{k}Z)_{\tau}\right)  \right) \nonumber\\
&  \qquad -\frac{1}{kn}f\!\left(  \varepsilon^{\prime},\frac{\varepsilon^{\prime}}%
{2},\frac{\varepsilon^{\prime}}{4},\frac{\varepsilon^{\prime}}{4}%
,\frac{\varepsilon^{\prime}}{2},\alpha,\beta\right) \\
&  \geq\frac{1}{k}\left(  I(X;B^{\prime})_{\tau}-I(X;E^{k}Z)_{\tau}\right)
-\left(  \delta_{1}+\delta_{2}\right) \\
&  =\frac{1}{k}\left(  I(X;B^{\prime})_{\tau}-I(X;E^{k}Z)_{\tau}\right)
-\delta.
\end{align}
Thus, there exists a $(kn,K,\varepsilon)$ secret-key distillation protocol
with rate $\frac{\log_{2}K}{kn}\geq\frac{1}{k}\left(  I(X;B^{\prime})_{\tau
}-I(X;E^{k}Z)_{\tau}\right)  -\delta$. Therefore, letting $n^{\prime}\equiv
kn$, we conclude that there exists an $(n^{\prime},2^{n^{\prime}(R-\delta
)},\varepsilon)$ secret-key distillation protocol with 
\begin{equation}
R=\frac{1}{k}\left(
I(X;B^{\prime})_{\tau}-I(X;E^{k}Z)_{\tau}\right)  
\end{equation}
 for all sufficiently large
$n$ such that \eqref{eq-SKD:delta-2-ach-reg-priv-info} holds. Since $\varepsilon$ and $\delta$ are arbitrary, we
conclude that for all $\varepsilon\in(0,1]$, $\delta>0$, and sufficiently
large $n$, there exists an $(n,2^{n(\frac{1}{k}\left(  I(X;B^{\prime})_{\tau
}-I(X;E^{k}Z)_{\tau}\right)  -\delta)},\varepsilon)$ secret key distillation
protocol. This means that $\frac{1}{k}\left(  I(X;B^{\prime})_{\tau}%
-I(X;E^{k}Z)_{\tau}\right)  $ is an achievable rate.

Now, since in the arguments above the LOPC\ channel $\mathcal{L}_{A^{k}%
B^{k}\rightarrow XB^{\prime}Z}^{\leftrightarrow}$ is arbitrary, we conclude
that%
\begin{equation}
\frac{1}{k}\sup_{\mathcal{L}^{\leftrightarrow}}\left(  I(X;B^{\prime})_{\tau
}-I(X;E^{k}Z)_{\tau}\right)
\end{equation}
is an achievable rate. Finally, since the number $k$ of copies of $\rho_{AB}$
is arbitrary, we conclude that%
\begin{equation}
\lim_{k\rightarrow\infty}\frac{1}{k}\sup_{\mathcal{L}^{\leftrightarrow}%
}\left(  I(X;B^{\prime})_{\tau}-I(X;E^{k}Z)_{\tau}\right)
\end{equation}
is an achievable rate.

\subsection{Proof of the Weak Converse}

\label{sec-SKD:weak-conv-proof}

In order to prove the weak converse part of
Theorem~\ref{thm-SKD:distillable-key}, we make use of
Corollary~\ref{cor-SKD:private-info-upper-bnd-weak-conv-1-shot}, specifically
\eqref{eq-SKD:private-info-upper-bnd-weak-conv-1-shot}:\ given a bipartite
state $\rho_{AB}$, for every $(K,\varepsilon)$ secret key distillation
protocol for $\rho_{AB}$, with $\varepsilon\in\lbrack0,1)$, the following
bound holds%
\begin{multline}
\left(  1-2\sqrt{\varepsilon}-\delta\right)  \log_{2}K\leq\sup_{\mathcal{L}%
}\left(  I(X;B^{\prime})_{\mathcal{L}(\psi)}-I(X;EZ)_{\mathcal{L}(\psi
)}\right) \\
+h_{2}(\sqrt{\varepsilon}+\delta)+\left(  1-\sqrt{\varepsilon}-\delta\right)
\log_{2}\!\left(  \frac{1}{\delta}\right)  +2g_{2}(\sqrt{\varepsilon}),
\end{multline}
where $\delta\in\left(  0,1-\sqrt{\varepsilon}\right)  $, $\psi_{ABE}$ is a
purification of $\rho_{AB}$, and the information quantities are evaluated on
the state $\mathcal{L}_{AB\rightarrow XB^{\prime}Z}(\psi_{ABE})$. Applying
this inequality to the state $\rho_{AB}^{\otimes n}$ leads to the following.

\begin{proposition}{}
Let $\rho_{AB}$ be a bipartite state, let $n\in\mathbb{N}$, $\varepsilon
\in\lbrack0,1)$, and $\delta\in(0,1-\sqrt{\varepsilon})$. For an
$(n,K,\varepsilon)$ secret-key distillation protocol for $\rho_{AB}$ with
corresponding LOPC\ channel $\mathcal{L}_{A^{n}B^{n}\rightarrow XB^{\prime}Z}%
$, with classical systems $X$ and $Z$, the rate $\frac{\log_{2}K}{n}$
satisfies%
\begin{multline}
\left(  1-2\sqrt{\varepsilon}-\delta\right)  \frac{\log_{2}K}{n}\leq\frac
{1}{n}\sup_{\mathcal{L}}\left(  I(X;B^{\prime})_{\mathcal{L}(\psi^{\otimes
n})}-I(X;EZ)_{\mathcal{L}(\psi^{\otimes n})}\right)
\label{eq-SKD:up-bnd-asymp-dist-key}\\
+\frac{1}{n}\left(  h_{2}(\sqrt{\varepsilon}+\delta)+\left(  1-\sqrt
{\varepsilon}-\delta\right)  \log_{2}\!\left(  \frac{1}{\delta}\right)
+2g_{2}(\sqrt{\varepsilon})\right)  .
\end{multline}
Consequently,%
\begin{multline}
\left(  1-2\sqrt{\varepsilon}-\delta\right)  K_{D}^{n,\varepsilon}(A;B)_{\rho
}\leq\frac{1}{n}\sup_{\mathcal{L}}\left(  I(X;B^{\prime})_{\mathcal{L}%
(\psi^{\otimes n})}-I(X;EZ)_{\mathcal{L}(\psi^{\otimes n})}\right) \\
+\frac{1}{n}\left(  h_{2}(\sqrt{\varepsilon}+\delta)+\left(  1-\sqrt
{\varepsilon}-\delta\right)  \log_{2}\!\left(  \frac{1}{\delta}\right)
+2g_{2}(\sqrt{\varepsilon})\right)  ,
\end{multline}
where the optimization is over every LOCC channel $\mathcal{L}_{A^{n}%
B^{n}\rightarrow XB^{\prime}Z}$.
\end{proposition}

\subsubsection*{Proof of the Weak Converse Part of
Theorem~\ref{thm-SKD:distillable-key}}

Suppose that $R$ is an achievable rate for secret key distillation for the
bipartite state $\rho_{AB}$. Then, by definition, for all $\varepsilon
\in(0,1]$, $\delta>0$, and sufficiently large $n$, there exists an
$(n,2^{n\left(  R-\delta\right)  },\varepsilon)$ secret-key distillation
protocol for $\rho_{AB}$. For all such protocols, the inequality in
\eqref{eq-SKD:up-bnd-asymp-dist-key} holds, so that%
\begin{multline}
\left(  1-2\sqrt{\varepsilon}-\delta^{\prime}\right)  \left(  R-\delta\right)
\leq\frac{1}{n}\sup_{\mathcal{L}}\left(  I(X;B^{\prime})_{\mathcal{L}%
(\psi^{\otimes n})}-I(X;EZ)_{\mathcal{L}(\psi^{\otimes n})}\right) \\
+\frac{1}{n}\left(  h_{2}(\sqrt{\varepsilon}+\delta^{\prime})+\left(
1-\sqrt{\varepsilon}-\delta^{\prime}\right)  \log_{2}\!\left(  \frac{1}%
{\delta^{\prime}}\right)  +2g_{2}(\sqrt{\varepsilon})\right)  .
\end{multline}
Since the inequality holds for all sufficiently large $n$, it holds in the
limit $n\rightarrow\infty$, so that%
\begin{align}
&\left(  1-2\sqrt{\varepsilon}-\delta^{\prime}\right)  \left(  R-\delta\right)\notag \\
&  \leq\lim_{n\rightarrow\infty}\Bigg(\frac{1}{n}\sup_{\mathcal{L}}\left(
I(X;B^{\prime})_{\mathcal{L}(\psi^{\otimes n})}-I(X;EZ)_{\mathcal{L}%
(\psi^{\otimes n})}\right) \notag \\
& \qquad  +\frac{1}{n}\left(  h_{2}(\sqrt{\varepsilon}+\delta^{\prime})+\left(
1-\sqrt{\varepsilon}-\delta^{\prime}\right)  \log_{2}\!\left(  \frac{1}%
{\delta}\right)  +2g_{2}(\sqrt{\varepsilon})\right)  \Bigg)\\
&  =\lim_{n\rightarrow\infty}\frac{1}{n}\sup_{\mathcal{L}}\left(
I(X;B^{\prime})_{\mathcal{L}(\psi^{\otimes n})}-I(X;EZ)_{\mathcal{L}%
(\psi^{\otimes n})}\right)  .
\end{align}
Then since this inequality holds for all $\varepsilon\in(0,1)$, $\delta>0$, it
holds in particular for $\delta^{\prime}=\sqrt{\varepsilon}$, $\varepsilon
\in(0,\frac{1}{9})$, which gives%
\begin{equation}
R\leq\frac{1}{\left(  1-3\sqrt{\varepsilon}\right)  }\lim_{n\rightarrow\infty
}\frac{1}{n}\sup_{\mathcal{L}}\left(  I(X;B^{\prime})_{\mathcal{L}%
(\psi^{\otimes n})}-I(X;EZ)_{\mathcal{L}(\psi^{\otimes n})}\right)  +\delta,
\end{equation}
and we thus conclude that%
\begin{align}
R  &  \leq\lim_{\varepsilon,\delta\rightarrow0}\frac{1}{\left(  1-3\sqrt
{\varepsilon}\right)  }\lim_{n\rightarrow\infty}\frac{1}{n}\sup_{\mathcal{L}%
}\left(  I(X;B^{\prime})_{\mathcal{L}(\psi^{\otimes n})}-I(X;EZ)_{\mathcal{L}%
(\psi^{\otimes n})}\right)  +\delta\\
&  =\lim_{n\rightarrow\infty}\frac{1}{n}\sup_{\mathcal{L}}\left(
I(X;B^{\prime})_{\mathcal{L}(\psi^{\otimes n})}-I(X;EZ)_{\mathcal{L}%
(\psi^{\otimes n})}\right)  .
\end{align}
We have thus shown that the quantity $\lim_{n\rightarrow\infty}\frac{1}{n}%
\sup_{\mathcal{L}}\Big(  I(X;B^{\prime})_{\mathcal{L}(\psi^{\otimes n}%
)}-I(X;EZ)_{\mathcal{L}(\psi^{\otimes n})}\Big)  $ is a weak converse rate
for secret key distillation for $\rho_{AB}$.

\subsection{Relative Entropy of Entanglement Strong Converse Upper Bound}

\label{sec-SKD:rel-entr-enta-str-conv}

As indicated previously, the expression in
\eqref{eq-SKD:distillable-key-theorem} for distillable key involves both a
limit over an unbounded number of copies of the initial state $\rho_{AB}$, as
well as an optimization over all two-way LOPC\ channels. Computing the
distillable key is therefore intractable in general. In this section, we use
the one-shot upper bound established in Section~\ref{sec-SKD:rel-ent-up-bnd} to show that the relative
entropy of entanglement is a strong converse upper bound on the distillable
key of a bipartite state $\rho_{AB}$.

We start by recalling the upper bound in \eqref{eq-SKD:rel-ent-alpha-bnd},
which tells us that%
\begin{equation}
\log_{2}K\leq\widetilde{E}_{\alpha}(A;B)_{\rho}+\frac{\alpha}{\alpha-1}%
\log_{2}\!\left(  \frac{1}{1-\varepsilon}\right)  \qquad\forall\alpha>1,
\label{eq-SKD:rel-ent-alpha-bnd-after}%
\end{equation}
for an arbitrary $(K,\varepsilon)$ secret-key distillation protocol, where
$\varepsilon\in(0,1)$. Recall that%
\begin{equation}
\widetilde{E}_{\alpha}(A;B)_{\rho}=\inf_{\sigma_{AB}\in\operatorname{SEP}%
(A:B)}\widetilde{D}_{\alpha}(\rho_{AB}\Vert\sigma_{AB}).
\end{equation}
Recall that the upper bound above is a consequence of the fact that separable
states are useless for secret key distillation.

Applying the upper bound in \eqref{eq-SKD:rel-ent-alpha-bnd-after} to the
state $\rho_{AB}^{\otimes n}$ leads to the following result:

\begin{corollary}{}
Let $\rho_{AB}$ be a bipartite state, let $n\in\mathbb{N}$, $\varepsilon
\in\lbrack0,1)$, and $\alpha>1$. For an $(n,K,\varepsilon)$ secret-key
distillation protocol, the following bound holds%
\begin{equation}
\frac{\log_{2}K}{n}\leq\widetilde{E}_{\alpha}(A;B)_{\rho}+\frac{\alpha
}{n\left(  \alpha-1\right)  }\log_{2}\!\left(  \frac{1}{1-\varepsilon}\right)
. \label{eq-SKD:key-dist-bnd-renyi-REE-1}%
\end{equation}
Consequently,
\begin{equation}
K_{D}^{n,\varepsilon}(A;B)_{\rho}\leq\widetilde{E}_{\alpha}(A;B)_{\rho}%
+\frac{\alpha}{n\left(  \alpha-1\right)  }\log_{2}\!\left(  \frac
{1}{1-\varepsilon}\right)  . \label{eq-SKD:key-dist-bnd-renyi-REE}%
\end{equation}

\end{corollary}

\begin{Proof}
An $(n,K,\varepsilon)$ secret-key distillation protocol for $\rho_{AB}$ is a
$(K,\varepsilon)$ secret-key distillation protocol for $\rho_{AB}^{\otimes n}%
$. Therefore, applying the inequality in
\eqref{eq-SKD:rel-ent-alpha-bnd-after} to the state $\rho_{AB}^{\otimes n}$
and dividing both sides by $n$ leads to%
\begin{equation}
\frac{\log_{2}K}{n}\leq\frac{1}{n}\widetilde{E}_{\alpha}(A^{n};B^{n}%
)_{\rho^{\otimes n}}+\frac{\alpha}{n\left(  \alpha-1\right)  }\log
_{2}\!\left(  \frac{1}{1-\varepsilon}\right)  .
\end{equation}
Now, by subadditivity of the sandwiched R\'enyi relative entropy of entanglement
(see \eqref{eq:SKA-Renyi-REE-subadditivity}), we have that%
\begin{equation}
\widetilde{E}_{\alpha}(A^{n};B^{n})_{\rho^{\otimes n}}\leq n\widetilde
{E}_{\alpha}(A;B)_{\rho}.
\end{equation}
Therefore,%
\begin{equation}
\frac{\log_{2}K}{n}\leq\widetilde{E}_{\alpha}(A;B)_{\rho}+\frac{\alpha
}{n\left(  \alpha-1\right)  }\log_{2}\!\left(  \frac{1}{1-\varepsilon}\right)
,
\end{equation}
as required. Since this inequality holds for all $(n,K,\varepsilon)$
protocols, we obtain \eqref{eq-SKD:key-dist-bnd-renyi-REE} by optimizing over
all key distillation protocols.
\end{Proof}

Given an $\varepsilon\in(0,1)$, the inequality in
\eqref{eq-SKD:key-dist-bnd-renyi-REE-1} gives us a bound on the rate of an
arbitrary $(n,K,\varepsilon)$ secret-key distillation protocol for a state
$\rho_{AB}$. If we instead fix the rate to be $r$, so that $K=2^{nr}$, then
the inequality in \eqref{eq-SKD:key-dist-bnd-renyi-REE-1} is as follows:%
\begin{equation}
r\leq\widetilde{E}_{\alpha}(A;B)_{\rho}+\frac{\alpha}{n\left(  \alpha
-1\right)  }\log_{2}\!\left(  \frac{1}{1-\varepsilon}\right)
\end{equation}
for all $\alpha>1$. Rearranging this inequality gives us the following lower
bound on $\varepsilon$:%
\begin{equation}
\varepsilon\geq1-2^{-n\left(  \frac{\alpha-1}{\alpha}\right)  \left(
r-\widetilde{E}_{\alpha}(A;B)_{\rho}\right)  }%
\end{equation}
for all $\alpha>1$.

\begin{theorem*}
{Strong Converse Upper Bound on Distillable Key}{}
Let $\rho_{AB}$ be a bipartite
state. The relative entropy of entanglement $E_{R}(A;B)_{\rho}$ is a strong
converse rate for secret key distillation for $\rho_{AB}$, i.e.,%
\begin{equation}
\widetilde{K}_{D}(A;B)_{\rho}\leq E_{R}(A;B)_{\rho},
\end{equation}
where we recall that $E_{R}(A;B)_{\rho}$ is defined as%
\begin{equation}
\inf_{\sigma_{AB}\in\operatorname{SEP}(A:B)}D(\rho_{AB}\Vert\sigma_{AB}).
\end{equation}

\end{theorem*}

\begin{Proof}
The proof is identical that given for Theorem~\ref{thm-distillable_ent_str_conv_UB}, except we make use of \eqref{eq-SKD:key-dist-bnd-renyi-REE-1}.
\end{Proof}

Given that the relative entropy of entanglement is a strong converse rate for
distillable key, by following arguments analogous to those in the referenced
proof, we conclude that $\frac{1}{k}E_{R}(A^{k};B^{k})_{\rho^{\otimes k}}$ is
a strong converse rate for all $k\in\mathbb{N}$. Therefore, the regularized
quantity%
\begin{equation}
E_{R}^{\text{reg}}(A;B)_{\rho}\coloneqq \lim_{n\rightarrow\infty}\frac{1}{n}%
E_{R}(A^{n};B^{n})_{\rho^{\otimes n}}%
\end{equation}
is a strong converse rate for secret key distillation for $\rho_{AB}$, so that%
\begin{equation}
\widetilde{K}_{D}(A;B)_{\rho}\leq E_{R}(A;B)_{\rho}.
\end{equation}
By the subaddivity of relative entropy of entanglement (see  \eqref{eq:SKA-Renyi-REE-subadditivity}),%
\begin{equation}
E_{R}^{\text{reg}}(A;B)_{\rho}\leq E_{R}(A;B)_{\rho},
\end{equation}
so that the regularized quantity in general gives a tighter upper bound on
distillable key.

\subsection{Squashed Entanglement Weak Converse Upper Bound}

\label{sec-SKD:squashed-ent-weak-conv}

In this section, we establish the squashed entanglement of a bipartite state
as a weak converse upper bound on its distillable key. The main idea is to
apply the one-shot bound from Theorem~\ref{thm-SKD:sq-ent-one-shot-up-bnd} and
the additivity of the squashed entanglement (Proposition~\ref{prop-squashed_ent_properties}) in order to
arrive at this conclusion.

\begin{corollary}{}
Let $\rho_{AB}$ be a bipartite state, let $n\in \mathbb{N}$, and let $\varepsilon
\in\lbrack0,1)$. For an $(n,K,\varepsilon)$ secret-key distillation protocol,
the following bound holds%
\begin{equation}
\left(  1-2\sqrt{\varepsilon}\right)  \frac{1}{n}\log_{2}K\leq
E_{\operatorname{sq}}(A;B)_{\rho}+\frac{2}{n}g_{2}(\sqrt{\varepsilon}).
\label{eq-SKD:sq-ent-finite-n-up-bnd}%
\end{equation}

\end{corollary}

\begin{Proof}
An $(n,K,\varepsilon)$ secret-key distillation protocol for $\rho_{AB}$ is a
$(K,\varepsilon)$ secret-key distillation protocol for $\rho_{AB}^{\otimes n}%
$. Therefore, applying the inequality in
\eqref{eq-SKD:sq-ent-bound-one-shot-key} to the state $\rho_{AB}^{\otimes n}$
and dividing both sides by $n$ leads to%
\begin{equation}
\left(  1-2\sqrt{\varepsilon}\right)  \frac{1}{n}\log_{2}K\leq\frac{1}%
{n}E_{\operatorname{sq}}(A^{n};B^{n})_{\rho^{\otimes n}}+\frac{2}{n}%
g_{2}(\sqrt{\varepsilon}).
\end{equation}
Now, by additivity of the squashed entanglement (Proposition~\ref{prop-squashed_ent_properties}), we have that%
\begin{equation}
E_{\operatorname{sq}}(A^{n};B^{n})_{\rho^{\otimes n}}=nE_{\operatorname{sq}%
}(A;B)_{\rho}.
\end{equation}
This concludes the proof.
\end{Proof}

We now provide a proof of \eqref{eq-SKD:sq-ent-asympt-dist-key-up-bnd}, the
statement that the squashed entanglement is a weak converse rate for secret
key distillation. Suppose that $R$ is an achievable rate for secret key
distillation for the bipartite state $\rho_{AB}$. Then, by definition, for all
$\varepsilon\in(0,1]$, $\delta>0$, and sufficiently large $n$, there exists an
$(n,2^{n(R-\delta)},\varepsilon)$ secret-key distillation protocol for
$\rho_{AB}$. For all such protocols, the inequality in
\eqref{eq-SKD:sq-ent-finite-n-up-bnd} holds, so that%
\begin{equation}
\left(  1-2\sqrt{\varepsilon}\right)  \left(  R-\delta\right)  \leq
E_{\operatorname{sq}}(A;B)_{\rho}+\frac{2}{n}g_{2}(\sqrt{\varepsilon}).
\end{equation}
Since the inequality holds for all sufficiently large $n$, it holds in the
limit $n\rightarrow\infty$, so that%
\begin{align}
\left(  1-2\sqrt{\varepsilon}\right)  \left(  R-\delta\right)   &  \leq
\lim_{n\rightarrow\infty}\left(  E_{\operatorname{sq}}(A;B)_{\rho}+\frac{2}%
{n}g_{2}(\sqrt{\varepsilon})\right) \\
&  =E_{\operatorname{sq}}(A;B)_{\rho}.
\end{align}
Then, since this inequality holds for all $\varepsilon\in(0,1]$ and $\delta
>0$, it holds in particular for all $\varepsilon\in(0,\frac{1}{4})$ and
$\delta>0$, implying that%
\begin{equation}
R\leq\frac{1}{1-2\sqrt{\varepsilon}}E_{\operatorname{sq}}(A;B)_{\rho}+\delta,
\end{equation}
and furthermore, that%
\begin{align}
R  &  \leq\lim_{\varepsilon,\delta\rightarrow0}\left(  \frac{1}{1-2\sqrt
{\varepsilon}}E_{\operatorname{sq}}(A;B)_{\rho}+\delta\right) \\
&  =E_{\operatorname{sq}}(A;B)_{\rho}.
\end{align}
We have thus shown that the squashed entanglement is a weak converse rate for
secret key distillation.

\section{One-Way Secret Key Distillation}

In Section~\ref{sec-SKD:low-bound-1-shot-dist-key}, we considered a one-way secret-key distillation protocol to
derive a lower bound on the one-shot distillable key of a bipartite state. In
the asymptotic setting, this leads to the private information lower bound on
the distillable key of a bipartite state $\rho_{AB}$, i.e.,%
\begin{equation}
K_{D}(A;B)_{\rho}\geq I(X;B)_{\tau}-I(X;E)_{\tau},
\end{equation}
where%
\begin{equation}
\tau_{XBE}\coloneqq \sum_{x}|x\rangle\!\langle x|_{X}\otimes\operatorname{Tr}%
_{A}[\Lambda_{A}^{x}\psi_{ABE}],
\end{equation}
$\psi_{ABE}$ is a purification of $\rho_{AB}$, and $\{\Lambda_{A}^{x}\}_{x}$
is a POVM. By reversing the roles of Alice and Bob in the protocol, we find
that%
\begin{equation}
K_{D}(A;B)_{\rho}\geq I(A;Y)_{\omega}-I(Y;E)_{\omega},
\end{equation}
where%
\begin{equation}
\omega_{YAE}\coloneqq \sum_{y}|y\rangle\!\langle y|_{Y}\otimes\operatorname{Tr}%
_{B}[\Gamma_{B}^{y}\psi_{ABE}],
\end{equation}
where $\{\Gamma_{B}^{y}\}_{y}$ is a POVM. Then, in general, we have the
following lower bound on distillable key:%
\begin{equation}
K_{D}(A;B)_{\rho}\geq\max\{I(X;B)_{\tau}-I(X;E)_{\tau},I(A;Y)_{\omega
}-I(Y;E)_{\omega}\}.
\end{equation}

This private information lower bound can be improved by first applying a
two-way LOPC\ channel to $n$ copies of the given state, and then performing a
one-way secret-key distillation protocol at the private information rate. This
leads to%
\begin{equation}
K_{D}(A;B)_{\rho}=\lim_{n\rightarrow\infty}\frac{1}{n}\sup_{\mathcal{L}%
^{\leftrightarrow}}I(X;B^{\prime})_{\mathcal{L}^{\leftrightarrow}%
(\psi^{\otimes n})}-I(X;E^{n}Z)_{\mathcal{L}^{\leftrightarrow}(\psi^{\otimes
n})}, \label{eq-SKD:dist-key-review}%
\end{equation}
where the information quantities are evaluated on the state $\mathcal{L}%
_{A^{n}B^{n}\rightarrow XB^{\prime}Z}^{\leftrightarrow}(\psi_{ABE}^{\otimes
n})$, $\mathcal{L}_{A^{n}B^{n}\rightarrow XB^{\prime}Z}^{\leftrightarrow}$ is
an LOPC\ channel with classical systems $X$ and $Z$, and $\psi_{ABE}$ is a
purification of $\rho_{AB}$.

If we restrict the optimization in \eqref{eq-SKD:dist-key-review} above to
one-way LOPC channels of the form $\mathcal{L}_{A^{n}B^{n}\rightarrow
XB^{\prime}Z}^{\rightarrow}$, then we obtain what is called the one-way
distillable key of $\rho_{AB}$, denoted by $K_{D}^{\rightarrow}(A;B)_{\rho}$,
and defined operationally in a similar way to the distillable key
$K_{D}(A;B)_{\rho}$, but with the free operations allowed restricted to
one-way LOPC. A key result is the following equality:%
\begin{align}
K_{D}^{\rightarrow}(A;B)_{\rho}  &  =\lim_{n\rightarrow\infty}\frac{1}{n}%
\sup_{\mathcal{L}^{\rightarrow}}I(X;B^{\prime})_{\mathcal{L}^{\rightarrow
}(\psi^{\otimes n})}-I(X;E^{n}Z)_{\mathcal{L}^{\rightarrow}(\psi^{\otimes n}%
)}\label{eq-SKD:one-way-dist-key-intro}\\
&  =\lim_{n\rightarrow\infty}\frac{1}{n}D_{K}^{\rightarrow}(\rho_{AB}^{\otimes
n}),
\end{align}
where%
\begin{equation}
D_{K}^{\rightarrow}(\rho_{AB})\coloneqq \sup_{\mathcal{L}^{\rightarrow}}I(X;B^{\prime
})_{\mathcal{L}^{\rightarrow}(\psi)}-I(X;EZ)_{\mathcal{L}^{\rightarrow}(\psi
)}.
\end{equation}
Like the distillable key, the one-way distillable key is an operational
quantity of interest. Furthermore, the equality in
\eqref{eq-SKD:one-way-dist-key-intro} can be proved similarly to how we proved \eqref{eq-SKD:distillable-key-theorem}.

In what follows, we show that this expression for one-way distillable key can
be simplified.

\begin{theorem*}
{One-Way Distillable Key of a Bipartite State}{thm-SKD:1-way-dist-key}%
The one-way distillable key of a bipartite state $\rho_{AB}$ is given by%
\begin{equation}
K_{D}^{\rightarrow}(A;B)_{\rho}=\lim_{n\rightarrow\infty}\frac{1}{n}%
\sup_{\{\Lambda_{A^{n}}^{x,z}\}_{x\in\mathcal{X},z\in\mathcal{Z}}}%
I(X;B^{n}|Z)_{\tau}-I(X;E^{n}|Z)_{\tau},
\end{equation}
where%
\begin{equation}
\tau_{XZB^{n}E^{n}}\coloneqq \sum_{x\in\mathcal{X},z\in\mathcal{Z}}|x\rangle\!\langle
x|_{X}\otimes|z\rangle\!\langle z|_{Z}\otimes\operatorname{Tr}_{A^{n}}%
[\Lambda_{A^{n}}^{x,z}\psi_{ABE}^{\otimes n}],
\end{equation}
and the optimization is over every POVM\ $\{\Lambda_{A^{n}}^{x,z}%
\}_{x\in\mathcal{X},z\in\mathcal{Z}}$ with output alphabets $\mathcal{X}$ and
$\mathcal{Z}$.
\end{theorem*}

This theorem tells us that, to determine the one-way distillable key of a
bipartite state, it suffices to optimize over one-way LOPC\ channels that
consist of a POVM\ conducted on Alice's systems.

\begin{Proof}
Let us start by recalling from Definition~\ref{def-LOCC} and the discussion around \eqref{eq-SKD:LOPC-channel} that every one-way LOPC channel
$\mathcal{L}_{A^{n}B^{n}\rightarrow XB^{\prime}Z}^{\rightarrow}$ can be
expressed as%
\begin{align}
\omega_{XB^{\prime}Z}  &  \coloneqq \mathcal{L}_{A^{n}B^{n}\rightarrow XB^{\prime}%
Z}^{\rightarrow}(\xi_{A^{n}B^{n}}) \label{eq-SKD:arbitrary-1-way-lopc-proof-1}%
\\
&  =\sum_{z\in\mathcal{Z}}(\mathcal{E}_{A^{n}\rightarrow X}^{z}\otimes
\mathcal{D}_{B^{n}\rightarrow B^{\prime}}^{z})(\xi_{A^{n}B^{n}})\otimes
|z\rangle\!\langle z|_{Z},\\
&  =(\mathcal{D}_{Z_{B}B^{n}\rightarrow B^{\prime}}\circ\mathcal{C}%
_{Z_{A}\rightarrow Z_{B}Z}\circ\mathcal{E}_{A^{n}\rightarrow XZ_{A}}%
)(\xi_{A^{n}B^{n}}),
\end{align}
where $\mathcal{Z}$ is some finite alphabet, $\{\mathcal{E}_{A^{n}\rightarrow
X}^{z}\}_{z\in\mathcal{Z}}$ is a set of completely positive maps such that
$\sum_{z\in\mathcal{Z}}\mathcal{E}_{A^{n}\rightarrow X}^{z}$ is trace
preserving, and $\{\mathcal{D}_{B^{n}\rightarrow B^{\prime}}^{z}%
\}_{z\in\mathcal{Z}}$ is a set of channels. Furthermore,%
\begin{align}
\mathcal{E}_{A^{n}\rightarrow XZ_{A}}(\xi_{A^{n}B^{n}})  &  =\sum
_{z\in\mathcal{Z}}\mathcal{E}_{A^{n}\rightarrow X}^{z}(\xi_{A^{n}B^{n}%
})\otimes|z\rangle\!\langle z|_{Z},\\
\mathcal{D}_{Z_{B}B^{n}\rightarrow B^{\prime}}(|z\rangle\!\langle z|_{Z_{B}%
}\otimes\xi_{A^{n}B^{n}})  &  =\mathcal{D}_{B^{n}\rightarrow B^{\prime}}%
^{z}(\xi_{A^{n}B^{n}}),
\end{align}
and since the map $\mathcal{E}_{A^{n}\rightarrow X}^{z}$ has a classical
output $X$, it can be written as%
\begin{equation}
\mathcal{E}_{A^{n}\rightarrow X}^{z}(\xi_{A^{n}B^{n}})=\sum_{x\in\mathcal{X}%
}\operatorname{Tr}_{A^{n}}[\Lambda_{A^{n}}^{x,z}\xi_{A^{n}B^{n}}%
]|x\rangle\!\langle x|_{X}, \label{eq-SKD:arbitrary-1-way-lopc-proof-last}%
\end{equation}
where $\{\Lambda_{A^{n}}^{x,z}\}_{x\in\mathcal{X},z\in\mathcal{Z}}$ is a POVM.

For every $n\in\mathbb{N}$, if we restrict the optimization in
\eqref{eq-SKD:one-way-dist-key-intro} to $\mathcal{D}_{B^{n}\rightarrow
B^{\prime}}^{z}=\operatorname{id}_{B^{n}}$ for all $z\in\mathcal{Z}$ and
$\mathcal{E}_{A^{n}\rightarrow X}^{z}(\cdot)=\sum_{x\in\mathcal{X}%
}\operatorname{Tr}_{A^{n}}[\Lambda_{A^{n}}^{x,z}(\cdot)]$ for all
$z\in\mathcal{Z}$, then the LOPC\ channel $\mathcal{L}_{A^{n}B^{n}\rightarrow
XB^{\prime}Z}^{\rightarrow}$ reduces to%
\begin{equation}
\mathcal{L}_{A^{n}B^{n}\rightarrow XB^{n}Z}^{\rightarrow}(\xi_{A^{n}B^{n}%
})=\sum_{x\in\mathcal{X},z\in\mathcal{Z}}\operatorname{Tr}_{A^{n}}%
[\Lambda_{A^{n}}^{x,z}\xi_{A^{n}B^{n}}]\otimes|x\rangle\!\langle x|_{X}%
\otimes|z\rangle\!\langle z|_{Z}%
\end{equation}
for every input state $\xi_{A^{n}B^{n}}$. We thus conclude that%
\begin{equation}
K_{D}^{\rightarrow}(A;B)_{\rho}\geq\lim_{n\rightarrow\infty}\frac{1}{n}%
\sup_{\{\Lambda_{A^{n}}^{x,z}\}_{x\in\mathcal{X},z\in\mathcal{Z}}}%
I(X;B^{n}|Z)_{\tau}-I(X;E^{n}|Z)_{\tau}.
\end{equation}

The rest of the proof is devoted to proving the reverse inequality. Let
$\mathcal{L}_{A^{n}B^{n}\rightarrow XB^{\prime}Z}^{\rightarrow}$ be an
arbitrary LOPC\ channel of the form in
\eqref{eq-SKD:arbitrary-1-way-lopc-proof-1}--\eqref{eq-SKD:arbitrary-1-way-lopc-proof-last}.
Consider that%
\begin{align}
&  I(X;B^{\prime})_{\mathcal{L}^{\rightarrow}(\psi^{\otimes n})}%
-I(X;E^{n}Z)_{\mathcal{L}^{\rightarrow}(\psi^{\otimes n})}\nonumber\\
&  \leq I(X;B^{n}Z)_{\mathcal{L}^{\prime}(\psi^{\otimes n})}-I(X;E^{n}%
Z)_{\mathcal{L}^{\prime}(\psi^{\otimes n})}\\
&  =I(X;Z)_{\mathcal{L}^{\prime}(\psi^{\otimes n})}+I(X;B^{n}|Z)_{\mathcal{L}%
^{\prime}(\psi^{\otimes n})}-I(X;Z)_{\mathcal{L}^{\prime}(\psi^{\otimes n}%
)}-I(X;E^{n}|Z)_{\mathcal{L}^{\prime}(\psi^{\otimes n})}\\
&  =I(X;B^{n}|Z)_{\mathcal{L}^{\prime}(\psi^{\otimes n})}-I(X;E^{n}%
|Z)_{\mathcal{L}^{\prime}(\psi^{\otimes n})}%
\end{align}
where%
\begin{equation}
\mathcal{L}_{A^{n}B^{n}\rightarrow XB^{n}Z}^{\prime}(\xi_{A^{n}B^{n}}%
)\coloneqq \sum_{x\in\mathcal{X},z\in\mathcal{Z}}\operatorname{Tr}_{A^{n}}%
[\Lambda_{A^{n}}^{x,z}\xi_{A^{n}B^{n}}]\otimes|x\rangle\!\langle x|_{X}%
\otimes|z\rangle\!\langle z|_{Z}.
\end{equation}
The inequality follows from data-processing with respect to the decoding
channel $\mathcal{D}_{Z_{B}B^{n}\rightarrow B^{\prime}}$ of Bob. This
concludes the proof.
\end{Proof}

\begin{Lemma}
{lem-SKD:1-way-dist-key-non-neg}For every bipartite state $\rho_{AB}$,
the optimized private information lower bound on distillable key is
non-negative, i.e., $D_{K}^{\rightarrow}(\rho_{AB})\geq0$.
\end{Lemma}

\begin{Proof}
Let $\psi_{ABE}$ be a purification of $\rho_{AB}$, and consider the following
Schmidt decomposition of $\psi_{ABE}$:%
\begin{equation}
|\psi\rangle_{ABE}=\sum_{k=0}^{r-1}\sqrt{\lambda_{k}}|\phi_{k}\rangle
_{A}\otimes|\varphi_{k}\rangle_{BE}.
\end{equation}
Then let $\Lambda_{A}^{x,z}=|\phi_{x}\rangle\!\langle\phi_{x}|\delta_{z,x}$ so
that the POVM\ measures in the local Schmidt basis of $A$ and broadcasts the
measurement result through $x$ and $z$. It is then straightforward to show
that the private information $I(X;B|Z)-I(X;E|Z)=0$. Since the POVM\ we chose
is a particular choice in the optimization for $D_{K}^{\rightarrow}(\rho
_{AB})\geq0$, we conclude that $D_{K}^{\rightarrow}(\rho_{AB})\geq
I(X;B|Z)-I(X;E|Z)=0$.
\end{Proof}

\begin{Lemma}
{lem-SKD:coh-info-lb-1-way-dist-key}For every bipartite state $\rho
_{AB}$, the optimized private information lower bound on distillable key is
not smaller than the coherent information of $\rho_{AB}$, i.e., $D_{K}%
^{\rightarrow}(\rho_{AB})\geq I(A\rangle B)_{\rho}$. Thus, the coherent
information is a lower bound for one-way distillable key:%
\begin{equation}
K_{D}^{\rightarrow}(\rho_{AB})\geq I(A\rangle B)_{\rho}.
\end{equation}

\end{Lemma}

\begin{Proof}
Let $\Lambda_{A}^{x,z}=|\varphi_{x}\rangle\!\langle\varphi_{x}|_{A}$ be a
rank-one POVM for which there is no output $z$. Let the state after the
measurement be as follows:%
\begin{align}
\tau_{XBE}  &  \coloneqq \sum_{x\in\mathcal{X}}|x\rangle\!\langle x|_{X}%
\otimes\operatorname{Tr}_{A}[|\varphi_{x}\rangle\!\langle\varphi_{x}|_{A}%
\psi_{ABE}]\\
&  =\sum_{x\in\mathcal{X}}p(x)|x\rangle\!\langle x|_{X}\otimes\psi_{BE}^{x},
\end{align}
where%
\begin{align}
p(x)  &  \coloneqq \operatorname{Tr}[|\varphi_{x}\rangle\!\langle\varphi_{x}|_{A}%
\psi_{ABE}],\\
\psi_{BE}^{x}  &  \coloneqq \frac{1}{p(x)}\operatorname{Tr}_{A}[|\varphi_{x}%
\rangle\!\langle\varphi_{x}|_{A}\psi_{ABE}].
\end{align}
Note that each $\psi_{BE}^{x}$ is a pure state. Then it follows that%
\begin{align}
I(X;B|Z)_{\tau}-I(X;E|Z)_{\tau}  &  =I(X;B)_{\tau}-I(X;E)_{\tau}\\
&  =H(B)_{\tau}-H(B|X)_{\tau}-H(E)_{\tau}+H(E|X)_{\tau}\\
&  =H(B)_{\tau}-H(E)_{\tau}\\
&  =H(B)_{\rho}-H(E)_{\rho}\\
&  =I(A\rangle B)_{\rho}.
\end{align}
The first equality follows because the $Z$ system is trivial. The third
equality follows because $H(B|X)_{\tau}=H(E|X)_{\tau}$, which in turn follows
because each state $\psi_{BE}^{x}$ is pure.
\end{Proof}

\section{Examples}

We now consider classes of bipartite states and evaluate the upper and lower
bounds on their distillable key that we have established in this chapter. In
some cases, the distillable key can be determined exactly because the upper
and lower bounds coincide.

\subsection{Pure States}

The simplest example for which distillable key can be determined exactly is
the class of pure bipartite states. In this case, the coherent information
lower bound from Lemma~\ref{lem-SKD:coh-info-lb-1-way-dist-key} and the
relative entropy of entanglement upper bound from
Theorem~\ref{thm-SKD:distillable-key}\ coincide and are equal to the entropy
of the reduced state. Thus, applying this same reasoning from Section~\ref{sec-ED:pure-states}, we conclude the following:

\begin{theorem*}
{Distillable Key for Pure States}{} The distillable key of a pure bipartite state
$\psi_{AB}$ is equal to the entropy of the reduced state on $A$, i.e.,%
\begin{equation}
K_{D}(A;B)_{\psi}=H(A)_{\psi}.
\end{equation}

\end{theorem*}

\subsection{Degradable and Anti-Degradable States}

In Section~\ref{subsec-deg_antideg_states}, we defined degradable and anti-degradable states, and we proved
that the one-way distillable entanglement of a degradable state is equal to
its coherent information. Also, we proved that the one-way distillable
entanglement of an anti-degradable state vanishes. It turns out that the same
results hold for one-way distillable key.

\begin{theorem*}
{One-Way Distillable Key for Anti-Degradable States}
{thm-SKD:antideg-key-rate}For an anti-degradable state $\rho_{AB}$, the
one-way distillable key is equal to zero, i.e., $K_{D}^{\rightarrow
}(A;B)_{\rho}=0$.
\end{theorem*}

\begin{Proof}
This is a direct consequence of the definition of an anti-degradable state and
the result in Theorem~\ref{thm-SKD:1-way-dist-key}. Indeed, for an
anti-degradable state $\rho_{AB}$ with purification $\psi_{ABE}$, there exists
an anti-degrading channel $\mathcal{A}_{E\rightarrow B}$ such that $\rho
_{AB}=\mathcal{A}_{E\rightarrow B}(\psi_{AE})$. A similar statement holds for
$\rho_{AB}^{\otimes n}$, i.e., $\rho_{AB}^{\otimes n}=[\mathcal{A}%
_{E\rightarrow B}(\psi_{AE})]^{\otimes n}$. Applying this fact and the
data-processing inequality to the expression $I(X;B^{n}|Z)_{\tau}%
-I(X;E^{n}|Z)_{\tau}$ from Theorem~\ref{thm-SKD:1-way-dist-key}, we conclude
that%
\begin{equation}
I(X;B^{n}|Z)_{\tau}-I(X;E^{n}|Z)_{\tau}\leq0.
\end{equation}
So we conclude that $K_{D}^{\rightarrow}(A;B)_{\rho}\leq0$. Combined with the
general lower bound from Lemma~\ref{lem-SKD:1-way-dist-key-non-neg}, we
conclude that $K_{D}^{\rightarrow}(A;B)_{\rho}=0$ for an anti-degradable state
$\rho_{AB}$.
\end{Proof}

\begin{theorem*}
{One-Way Distillable Key for Degradable States}{thm-SKD:deg-key-rate}For
a degradable state $\rho_{AB}$, we have%
\begin{equation}
D_{K}^{\rightarrow}(\rho_{AB})=I(A\rangle B)_{\rho}.
\end{equation}
Consequently, $D_{K}^{\rightarrow}(\rho_{AB}^{\otimes n})=nD_{K}^{\rightarrow
}(\rho_{AB})$, and thus the one-way distillable key of a degradable state
$\rho_{AB}$ is equal to its coherent information:%
\begin{equation}
K_{D}^{\rightarrow}(A;B)_{\rho}=I(A\rangle B)_{\rho}.
\end{equation}

\end{theorem*}

\begin{Proof}
It suffices to prove the upper bound $D_{K}^{\rightarrow}(\rho_{AB})\leq
I(A\rangle B)_{\rho}$ because Lemma~\ref{lem-SKD:coh-info-lb-1-way-dist-key}
established the lower bound $D_{K}^{\rightarrow}(\rho_{AB})\geq I(A\rangle
B)_{\rho}$ in general. Recall that the defining property of a degradable state
$\rho_{AB}$ with purification $\psi_{ABE}$ is that there exists a degrading
channel $\mathcal{D}_{B\rightarrow E}$ such that $\psi_{AE}=\mathcal{D}%
_{B\rightarrow E}(\rho_{AB})$. Then the same is true for the tensor-power
states, i.e., $\psi_{AE}^{\otimes n}=(\mathcal{D}_{B\rightarrow E}(\rho
_{AB}))^{\otimes n}$. Then consider that%
\begin{align}
&  I(X;B^{n}|Z)_{\tau}-I(X;E^{n}|Z)_{\tau}\nonumber\\
&  =I(XZ;B^{n})_{\tau}-I(Z;B^{n})_{\tau}-\left[  I(XZ;E^{n})_{\tau}%
-I(Z;E^{n})_{\tau}\right] \\
&  =I(XZ;B^{n})_{\tau}-I(XZ;E^{n})_{\tau}-\left[  I(Z;B^{n})_{\tau}%
-I(Z;E^{n})_{\tau}\right] \\
&  \leq I(XZ;B^{n})_{\tau}-I(XZ;E^{n})_{\tau},
\end{align}
where%
\begin{equation}
\tau_{XZB^{n}E^{n}}=\sum_{x\in\mathcal{X},z\in\mathcal{Z}}|x\rangle\!\langle
x|_{X}\otimes|z\rangle\!\langle z|_{Z}\otimes\operatorname{Tr}_{A^{n}}%
[\Lambda_{A^{n}}^{x,z}\psi_{ABE}^{\otimes n}].
\end{equation}
The sole inequality above follows from the data-processing inequality for mutual information and the fact that there is a degrading channel from $B^n$ to $E^n$.
Now let $\Lambda_{A^{n}}^{x,z}=\sum_{y}|\varphi^{x,y,z}\rangle\!\langle
\varphi^{x,y,z}|_{A^{n}}$ be a rank-one decomposition of the POVM\ $\{\Lambda
_{A^{n}}^{x,z}\}_{x,z}$ and define the following extension of the state
$\tau_{XZB^{n}E^{n}}$:%
\begin{equation}
\tau_{XZYB^{n}E^{n}}=\sum_{x\in\mathcal{X},z\in\mathcal{Z}}|x\rangle\!\langle
x|_{X}\otimes|z\rangle\!\langle z|_{Z}\otimes|y\rangle\!\langle y|_{Y}%
\otimes\operatorname{Tr}_{A^{n}}[|\varphi^{x,y,z}\rangle\!\langle
\varphi^{x,y,z}|_{A^{n}}\psi_{ABE}^{\otimes n}].
\end{equation}
Then consider that%
\begin{align}
&  I(XZ;B^{n})_{\tau}-I(XZ;E^{n})_{\tau}\nonumber\\
&  =I(XZY;B^{n})_{\tau}-I(Y;B^{n}|XZ)_{\tau}-\left[  I(XZY;E^{n})_{\tau
}-I(Y;E^{n}|XZ)_{\tau}\right] \\
&  =I(XZY;B^{n})_{\tau}-I(XZY;E^{n})_{\tau}-\left[  I(Y;B^{n}|XZ)_{\tau
}-I(Y;E^{n}|XZ)_{\tau}\right] \\
&  \leq I(XZY;B^{n})_{\tau}-I(XZY;E^{n})_{\tau}\\
&  =H(B^{n})_{\tau}-H(E^{n})_{\tau}-\left[  H(B^{n}|XZY)_{\tau}-H(E^{n}%
|XZY)_{\tau}\right] \\
&  =H(B^{n})_{\tau}-H(E^{n})_{\tau}\\
&  =H(B^{n})_{\psi}-H(E^{n})_{\psi}\\
&  =n\left(  H(B)_{\psi}-H(E)_{\psi}\right) \\
&  =nI(A\rangle B)_{\rho}.
\end{align}
The sole inequality above follows from the data-processing inequality for conditional mutual information and the fact that there is a degrading channel from $B^n$ to $E^n$.
This concludes the proof.
\end{Proof}

\section{Summary}

In this chapter, we considered the task of secret key distillation, in which
the goal is for Alice and Bob to convert a bipartite state to an approximate
tripartite key state with as many secret key bits as possible. In doing so,
they are allowed to perform local operations and public classical
communication, in which an eavesdropper obtains a copy of all of the classical
communication exchanged. The highest rate at which this can be accomplished is
called the distillable key of the state. We began with the one-shot setting,
in which we allow some error in the distillation protocol, and we determined
lower and upper bounds on the number of approximate secret key bits that can
be distilled. In the asymptotic setting, we proved that the private
information of the state is an achievable rate, and we proved that the
squashed entanglement and the relative entropy of entanglement are upper
bounds. These latter quantities are the best known upper bounds on distillable key.

By performing secret key distillation and then the one-time pad protocol
(described in the introduction of this chapter), Alice can transmit a
classical message privately to Bob. This process thus induces an ideal private
classical channel from Alice to Bob. If Alice and Bob are connected by a
quantum channel, then they can use it to share a bipartite state, from which they
can induce a private classical channel in the aforementioned manner. This is
one way to communicate privately over a quantum channel. In the next chapter,
we discuss other, more direct approaches for private communication, which give
an optimal private communication strategy for some quantum channels.

\section{Bibliographic Notes}

The task of secret key distillation, like many tasks in quantum information
theory, has its roots in classical information theory. \citet{M93} and
\citet{AC93} developed the theory of secret key distillation in the classical
case. There, the assumption is that Alice, Bob, and Eve share a tripartite
distribution $p_{XYZ}$, from which they are trying to extract an approximation
of an ideal secret key by means of local operations and public classical
communication. The quantum case considered here is thus a generalization of
this scenario, with a tripartite pure state $\psi_{ABE}$ replacing the
classical distribution~$p_{XYZ}$. Recall that the eavesdropper sharing the purifying system of a purification of Alice and Bob's state gives Eve more power, because she can realize any possible extension of Alice and Bob's state by acting on the purifying system.

The one-time pad protocol traces its roots much further back. It was invented
by \citet{V26}, and its security was established by \citet{S49}. As discussed in
the introduction of this chapter, the main application of secret key
distillation is to distill a secret key that can be used in conjunction with
the one-time pad protocol in order to transmit a message privately.

Much of the technical work on secret key distillation was motivated by the
development of quantum key distribution \citep{BB84,E91}. An early paper on the
topic is about privacy amplification \citep{BBCM95}, which is a component of a
key distillation protocol. Secret key distillation from a bipartite quantum state was then studied by a number of researchers, including \citep{DW05,HHHO05,C06,HHHLO08,HHHO09,CEHHOR07,CSW12}.

The one-shot setting of secret key distillation was studied by \citet{RR12} and
\citet{KKGW19}. We follow the approach of \citet{KKGW19} closely in this chapter.

The connection between the tripartite picture of secret key distillation and
the bipartite picture of private state distillation was identified by
\citet{HHHO05,HHHO09}. This work led to the understanding of the difference
between entanglement and secret key, and it allowed for using the tools of
entanglement theory (such as entanglement measures) in the context of secret
key distillation. In the context of asymptotic secret key distillation, the
relative entropy of entanglement upper bound on distillable key was
established by \citet{HHHO05,HHHO09} and the squashed entanglement upper bound
by \citet{C06,CEHHOR07,CSW12} (see also \citep{Wilde2016a}\ in this context).

The privacy test was defined by \citet{PhysRevLett.100.110502,HHHLO08}, and its
use in establishing one-shot converse bound was established by \citet{WTB16}.
Lemma~\ref{lem:pass-privacy-test}\ is due to \citet{WTB16}.
Lemma~\ref{lem:fail-privacy-test}\ is implicit in the work of \citet{HHHO09} and was
explicitly proved by \citet{WTB16}.
Proposition~\ref{prop:core-meta-converse-privacy}\ and
Theorem~\ref{thm-SKD:one-shot-rel-ent-key-up-bound}\ were established by
\citet{WTB16}. Lemma~\ref{lem-SKD:SKD-log-K-to-info-measures},
Proposition~\ref{thm-SKD:SKD-bipartite-bound}, and
Theorem~\ref{thm-SKD:sq-ent-one-shot-up-bnd}\ were established by
\citet{Wilde2016a}.

Theorem~\ref{thm-SKD:one-shot-key-lower-bnd} was established by \citet{KKGW19}.
The convex split method was introduced by \citet{ADJ17}, and the smooth variant
in Lemma~\ref{lem-SKD:smooth-convex-split} is due to \citet{KKGW19}, making use
of methods in the appendix of \citet{LW19}.
Lemma~\ref{lem-SKD:smooth-max-MI-relations} is a variant of a result in \citet{AJW17b} and was established by \citet{KKGW19} (see also \citet{wilde2017position}).

The expression for distillable key in
\eqref{eq-SKD:distillable-key-theorem}\ of\ Theorem~\ref{thm-SKD:distillable-key}
is due to \citet{DW05}. Eq.~\eqref{eq-SKD:key-dist-bnd-renyi-REE-1} is due to
\citet{WTB16}, and Eq.~\eqref{eq-SKD:sq-ent-finite-n-up-bnd} to
\citet{Wilde2016a}. One-way secret key distillation was also considered by
\citet{DW05}. Theorems~\ref{thm-SKD:antideg-key-rate} and
\ref{thm-SKD:deg-key-rate} were established by \citet{Led20}.

\begin{subappendices}

\section{Proof of Smooth Convex Split Lemma}

\label{app-SKD:convex-split}

In this appendix, we prove Lemma~\ref{lem-SKD:smooth-convex-split}.

Let $\widetilde{\rho}_{AE}$ be an arbitrary state
satisfying $P(\widetilde{\rho}_{AE},\rho_{AE})\leq\sqrt{\varepsilon}-\eta$ and
such that%
\begin{equation}
\rho_{A}\otimes\widetilde{\rho}_{E}=p\widetilde{\rho}_{AE}+\left(  1-p\right)
\omega_{AE}, \label{eq:dmax-condition-cs}%
\end{equation}
for some $p\in(0,1)$ and $\omega_{AE}$ some state. We define the following
state, which we think of as an approximation to $\tau_{A_{1}\cdots A_{R}E}$:%
\begin{equation}
\widetilde{\tau}_{A_{1}\cdots A_{R}E}\coloneqq \frac{1}{R}\sum_{r=1}^{R}\rho_{A_{1}%
}\otimes\cdots\otimes\rho_{A_{r-1}}\otimes\widetilde{\rho}_{A_{r}E}\otimes
\rho_{A_{r+1}}\otimes\cdots\otimes\rho_{A_{R}}.
\end{equation}
It is a good approximation if $\sqrt{\varepsilon}-\eta$ is small, because%
\begin{align}
&  \sqrt{F}(\tau_{A_{1}\cdots A_{R}E},\widetilde{\tau}_{A_{1}\cdots A_{R}%
E})\nonumber\\
&  \geq\frac{1}{R}\sum_{r=1}^{R}\sqrt{F}(\rho_{A}^{\otimes r-1}\otimes
\rho_{A_{r}E}\otimes\rho_{A}^{\otimes R-r},\rho_{A}^{\otimes r-1}%
\otimes\widetilde{\rho}_{A_{r}E}\otimes\rho_{A}^{\otimes R-r})\\
&  =\frac{1}{R}\sum_{r=1}^{R}\sqrt{F}(\rho_{A_{r}E},\widetilde{\rho}_{A_{r}%
E})\\
&  =\sqrt{F}(\rho_{AE},\widetilde{\rho}_{AE}),
\end{align}
where the inequality follows from the concavity of the root fidelity (Theorem~\ref{thm-joint_concave_sqrt_fid}).
This in turn implies that
\begin{equation}
\sqrt{F}(\tau_{A_{1}\cdots A_{R}E},\widetilde{\tau}_{A_{1}\cdots A_{R}E}%
)\geq\sqrt{F}(\rho_{AE},\widetilde{\rho}_{AE}). \label{eq:cs-state-approx}%
\end{equation}
So the inequality in \eqref{eq:cs-state-approx}, the definition of the sine
distance (Definition~\ref{def-purified_distance}), and the fact that $P(\widetilde{\rho}_{AE},\rho_{AE})\leq
\sqrt{\varepsilon}-\eta$, imply that%
\begin{equation}
P(\tau_{A_{1}\cdots A_{R}E},\widetilde{\tau}_{A_{1}\cdots A_{R}E})\leq
\sqrt{\varepsilon}-\eta. \label{eq:smoothed-state-cs}%
\end{equation}
Now, let us define the following states:%
\begin{align}
\beta_{AE}  &  \coloneqq \rho_{A}\otimes\widetilde{\rho}_{E},\\
\alpha_{AE}  &  \coloneqq \widetilde{\rho}_{AE},\\
\widetilde{\tau}_{A^{R}E^{R}}  &  \coloneqq \frac{1}{R}\sum_{r=1}^{R}\beta_{A_{1}%
E_{1}}\otimes\cdots\otimes\beta_{A_{r-1}E_{r-1}}\otimes\alpha_{A_{r}E_{r}%
}\otimes\beta_{A_{r+1}E_{r+1}}\otimes\cdots\otimes\beta_{A_{R}E_{R}},
\end{align}
and observe that%
\begin{align}
\operatorname{Tr}_{E_{2}^{R}}[(\beta_{AE})^{\otimes R}]  &  =\rho_{A_{1}%
}\otimes\cdots\otimes\rho_{A_{R}}\otimes\widetilde{\rho}_{E},\\
\operatorname{Tr}_{E_{2}^{R}}[\widetilde{\tau}_{A^{R}E^{R}}]  &
=\widetilde{\tau}_{A_{1}\cdots A_{R}E}.
\end{align}
Thus, it follows from the data-processing inequality for the sine distance that
\begin{equation}
P(\widetilde{\tau}_{A_{1}\cdots A_{R}E},\rho_{A_{1}}\otimes\cdots\otimes
\rho_{A_{R}}\otimes\widetilde{\rho}_{E})\leq P(\widetilde{\tau}_{A^{R}E^{R}%
},(\beta_{AE})^{\otimes R}). \label{eq:c-split-partial-trace-mono}%
\end{equation}
Now consider that%
\begin{align}
(\beta_{AE})^{\otimes R}  &  =\left(  p\widetilde{\rho}_{AE}+\left(
1-p\right)  \omega_{AE}\right)  ^{\otimes R}\\
&  =\sum_{S\subset\left[  R\right]  }p^{\left\vert S\right\vert }\left(
1-p\right)  ^{R-\left\vert S\right\vert }\widetilde{\rho}_{AE}^{\otimes
S}\otimes \omega_{AE}^{\otimes\left[  R\right]  \backslash S}\\
&  =\sum_{k=0}^{R}\binom{R}{k}p^{k}\left(  1-p\right)  ^{n-k}\theta_{k}%
\end{align}
where $\theta_{k}$ is the following state:%
\begin{equation}
\theta_{k}\coloneqq \frac{1}{\binom{R}{k}}\sum_{\left\vert S\right\vert =k}%
\widetilde{\rho}_{AE}^{\otimes S} \otimes \omega_{AE}^{\otimes\left[  R\right]
\backslash S}.
\end{equation}
Also, consider that%
\begin{align}
\widetilde{\tau}_{A^{R}E^{R}} 
&  =\frac{1}{R}\sum_{r=1}^{R}\beta_{A_{1}E_{1}%
}\otimes\cdots\otimes\beta_{A_{r-1}E_{r-1}}\otimes\alpha_{A_{r}E_{r}}\notag \\
& \qquad\qquad\qquad \otimes\beta_{A_{r+1}E_{r+1}}\otimes\cdots\otimes\beta_{A_{R}E_{R}}\\
&  =\sum_{\emptyset\neq S\subset\left[  R\right]  }p^{\left\vert S\right\vert
-1}\left(  1-p\right)  ^{R-\left\vert S\right\vert }\widetilde{\rho}%
_{AE}^{\otimes S}\otimes\omega_{AE}^{\otimes\left[  R\right]  \backslash S}\\
&  =\sum_{k=1}^{R}\binom{R-1}{k-1}p^{k-1}\left(  1-p\right)  ^{R-k}\theta
_{k}\\
&  =\sum_{k=0}^{R}\frac{k}{Rp}\binom{R}{k}p^{k}\left(  1-p\right)
^{R-k}\theta_{k}.
\end{align}
In the last line, we used the identity $\binom{R-1}{k-1}=\frac{k}{Rp}\binom
{R}{k}$. Defining the following classical--quantum states:%
\begin{align}
\beta_{A^{R}E^{R}K}  &  \coloneqq \sum_{k=0}^{R}\binom{R}{k}p^{k}\left(  1-p\right)
^{n-k}\theta_{k}\otimes|k\rangle\!\langle k|_{K},\\
\widetilde{\tau}_{A^{R}E^{R}K}  &  \coloneqq \sum_{k=0}^{R}\frac{k}{Rp}\binom{R}%
{k}p^{k}\left(  1-p\right)  ^{R-k}\theta_{k}\otimes|k\rangle\!\langle k|_{K},
\end{align}
consider that%
\begin{align}
&  \sqrt{F}((\beta_{AE})^{\otimes R},\widetilde{\tau}_{A^{R}E^{R}})\nonumber\\
&  \geq\sqrt{F}(\beta_{A^{R}E^{R}K},\widetilde{\tau}_{A^{R}E^{R}K})\\
&  =\sum_{k=0}^{R}\sqrt{\binom{R}{k}p^{k}\left(  1-p\right)  ^{n-k}}%
\sqrt{\frac{k}{Rp}\binom{R}{k}p^{k}\left(  1-p\right)  ^{R-k}}\sqrt{F}%
(\theta_{k},\theta_{k})\\
&  =\sum_{k=0}^{R}\binom{R}{k}p^{k}\left(  1-p\right)  ^{R-k}\sqrt{\frac
{k}{Rp}}\\
&  =\sqrt{\frac{1}{Rp}}\sum_{k=0}^{R}\binom{R}{k}p^{k}\left(  1-p\right)
^{R-k}\sqrt{k}\\
&  =\sqrt{\frac{1}{Rp}}\mathbb{E}_{K}\!\left[  \sqrt{K}\right]  ,
\end{align}
where $\mathbb{E}_{K}$ denotes the expectation with respect to the binomial
random variable $K$. The first inequality follows from the data-processing inequality for fidelity with respect to partial trace. The other steps follow by direct evaluation. Let $\mu=Rp$ (i.e., the mean of a binomial random
variable). Consider that the following inequality holds for all $k\geq0$ and
$\mu>0$:%
\begin{equation}
\sqrt{k}\geq\sqrt{\mu}+\frac{k-\mu}{2\sqrt{\mu}}-\frac{\left(  k-\mu\right)
^{2}}{2\mu^{3/2}}.
\end{equation}
Then we find that%
\begin{align}
\sqrt{\frac{1}{Rp}}\mathbb{E}_{K}\!\left[  \sqrt{K}\right]   &  \geq\sqrt
{\frac{1}{Rp}}\mathbb{E}_{K}\!\left[  \sqrt{\mu}+\frac{K-\mu}{2\sqrt{\mu}}%
-\frac{\left(  K-\mu\right)  ^{2}}{2\mu^{3/2}}\right] \\
&  =\sqrt{\frac{1}{Rp}}\left(  \sqrt{\mu}-\frac{\text{Var}(K)}{2\mu^{3/2}%
}\right) \\
&  =\sqrt{\frac{1}{Rp}}\left(  \sqrt{Rp}-\frac{\text{Var}(K)}{2\left(
Rp\right)  ^{3/2}}\right) \\
&  =1-\frac{Rp\left(  1-p\right)  }{\left(  Rp\right)  ^{2}}\\
&  =1-\frac{\left(  1-p\right)  }{Rp}\\
&  \geq1-\frac{1}{Rp}.
\end{align}
Thus it follows that%
\begin{equation}
\sqrt{F}((\beta_{AE})^{\otimes R},\widetilde{\tau}_{A^{R}E^{R}})\geq
1-\frac{\eta^{2}}{2}%
\end{equation}
if%
\begin{equation}
\log_{2}R\geq\log_{2}(1/p)+\log_{2}\!\left(  \frac{2}{\eta^{2}}\right)  .
\end{equation}
This implies that%
\begin{equation}
P((\beta_{AE})^{\otimes R},\widetilde{\tau}_{A^{R}E^{R}})\leq\eta.
\end{equation}
For the same choice of $R$, it follows from
\eqref{eq:c-split-partial-trace-mono}\ that%
\begin{equation}
P(\widetilde{\tau}_{A_{1}\cdots A_{R}E},\rho_{A_{1}}\otimes\cdots\otimes
\rho_{A_{R}}\otimes\widetilde{\rho}_{E})\leq\eta. \label{eq:cs-result}%
\end{equation}
Applying the triangle inequality to \eqref{eq:smoothed-state-cs} and
\eqref{eq:cs-result}, we find that%
\begin{equation}
P(\tau_{A_{1}\cdots A_{R}E},\rho_{A_{1}}\otimes\cdots\otimes\rho_{A_{R}%
}\otimes\widetilde{\rho}_{E})\leq\sqrt{\varepsilon}.
\end{equation}
The whole argument above holds for an arbitrary state $\widetilde{\rho}_{AE}$
satisfying $P(\widetilde{\rho}_{AE},\rho_{AE})\leq\sqrt{\varepsilon}-\eta$ and
\eqref{eq:dmax-condition-cs}, and so taking an infimum of $\log_{2}(1/p)$ over $p$ and 
all states satisfying these conditions, and applying the definition in
\eqref{eq-SKD:alt-smooth-max-MI-def}, as well as Lemma~\ref{lem-QEI:max-rel-ent-alt-exp}, we find that%
\begin{equation}
P(\tau_{A_{1}\cdots A_{R}E},\rho_{A_{1}}\otimes\cdots\otimes\rho_{A_{R}%
}\otimes\widetilde{\rho}_{E})\leq\sqrt{\varepsilon}%
\end{equation}
if%
\begin{equation}
\log_{2}R\geq\overline{I}_{\max}^{\sqrt{\varepsilon}-\eta}(E;A)_{\rho}%
+\log_{2}\!\left(  \frac{2}{\eta^{2}}\right)  .
\end{equation}
This concludes the proof.

\section{Relating Two Variants of Smooth-Max Mutual Information}

\label{sec-SKD:two-smooth-maxes}

In this appendix, we prove Lemma~\ref{lem-SKD:smooth-max-MI-relations}. The steps consist of constructing a state $\widehat{\rho}_{AE}$ such that
\begin{equation}
D_{\max}(\widehat{\rho}_{AE}\Vert\rho_{A} \otimes \widehat{\rho
}_{E})
\leq
D_{\max}(\widetilde{\rho}_{AE}\Vert\rho_{A}\otimes\rho_{E})+\log_{2}\!\left(
\frac{8}{\delta^{2}}\right)
\label{eq:dmax-inequality-smooth-max-relate-prelim}
\end{equation}
and $P(\widehat{\rho}_{AE},\rho
_{AE})\leq\varepsilon+\delta$. Using these inequalities, we can apply the definition of $\overline{I}_{\max}^{\varepsilon+\delta}(E;A)_{\rho}$ to conclude the desired inequality in \eqref{eq-SKD:alt-sm-imax-to-sm-imax}. We begin by showing the first inequality, and after that, we establish the second one.

We begin by establishing some preparatory facts. Let $\widetilde{\rho}_{AE}$ be a state
satisfying $P(\widetilde{\rho}_{AE},\rho_{AE})\leq\varepsilon$. Let
$\gamma=\delta^{2}/8$, and set $\Pi_{E}^{\gamma}$ to be the projection onto
the positive eigenspace of $\frac{1}{\gamma}\widetilde{\rho}_{E}-\rho_{E}$.
Then it follows that%
\begin{equation}
\Pi_{E}^{\gamma}\left(  \frac{1}{\gamma}\widetilde{\rho}_{E}-\rho_{E}\right)
\Pi_{E}^{\gamma}\geq0\quad\Rightarrow\quad\Pi_{E}^{\gamma}\rho_{E}\Pi
_{E}^{\gamma}\leq\frac{1}{\gamma}\Pi_{E}^{\gamma}\widetilde{\rho}_{E}\Pi
_{E}^{\gamma}=\frac{8}{\delta^{2}}\Pi_{E}^{\gamma}\widetilde{\rho}_{E}\Pi
_{E}^{\gamma}, \label{eq:op-ineq-rho-tilde-rho}%
\end{equation}
and%
\begin{multline}
\left(  I-\Pi_{E}^{\gamma}\right)  \left(  \frac{1}{\gamma}\widetilde{\rho
}_{E}-\rho_{E}\right)  \left(  I-\Pi_{E}^{\gamma}\right)  \leq
0\label{eq:trace-ineq-delta-gamma}\\
\Rightarrow\operatorname{Tr}[\left(  I-\Pi_{E}^{\gamma}\right)  \widetilde
{\rho}_{E}]\leq\gamma\operatorname{Tr}[\left(  I-\Pi_{E}^{\gamma}\right)
\rho_{E}]\leq\gamma=\frac{\delta^{2}}{8},
\end{multline}
where the last inequality follows because $\operatorname{Tr}[\left(  I-\Pi
_{E}^{\gamma}\right)  \rho_{E}]\leq1$. The inequality in
\eqref{eq:trace-ineq-delta-gamma} can be rewritten as%
\begin{equation}
\operatorname{Tr}[\Pi_{E}^{\gamma}\widetilde{\rho}_{E}]\geq1-\frac{\delta^{2}%
}{8}. \label{eq:gamma-project-onto-smooth-rho}%
\end{equation}

We now establish \eqref{eq:dmax-inequality-smooth-max-relate-prelim}. Let us define the following states:
\begin{equation}
\overline{\rho}_{AEX}\coloneqq \Pi_{E}^{\gamma}\widetilde{\rho}_{AE}\Pi_{E}^{\gamma
}\otimes|0\rangle\!\langle0|_{X}+\left(  I-\Pi_{E}^{\gamma}\right)
\widetilde{\rho}_{AE}\left(  I-\Pi_{E}^{\gamma}\right)  \otimes|1\rangle
\!\langle1|_{X},
\end{equation}%
\begin{equation}
\widehat{\rho}_{AEX}\coloneqq \left(  \Pi_{E}^{\gamma}\widetilde{\rho}_{AE}\Pi
_{E}^{\gamma}+\rho_{A} \otimes \widetilde{\rho}_{E}^{1/2}\left(  I-\Pi_{E}^{\gamma}\right)
\widetilde{\rho}_{E}^{1/2}\right)  \otimes|0\rangle
\!\langle0|_{X},
\end{equation}
so that%
\begin{align}
\widehat{\rho}_{AE}  &  =\operatorname{Tr}_{X}[\widehat{\rho}_{AEX}]\\
&  =\Pi_{E}^{\gamma}\widetilde{\rho}_{AE}\Pi_{E}^{\gamma}+
\rho_{A} \otimes
\widetilde{\rho}%
_{E}^{1/2}\left(  I-\Pi_{E}^{\gamma}\right)  \widetilde{\rho}_{E}^{1/2}.
\end{align}
Then, using the inequality $\widetilde{\rho}_{AE}\leq\mu\rho_{A}\otimes
\rho_{E}$, with%
\begin{equation}
\mu\coloneqq 2^{D_{\max}(\widetilde{\rho}_{AE}\Vert\rho_{A}\otimes\rho_{E})},
\end{equation}
and the fact that $\mu\frac{8}{\delta^{2}}\geq1$ (which holds because
$D_{\max}(\widetilde{\rho}_{AE}\Vert\rho_{A}\otimes\rho_{E})\geq0$ and
$8\geq\delta^{2}$), we find that%
\begin{align}
\widehat{\rho}_{AE}  &  \leq\mu
\rho_{A} \otimes \Pi_{E}^{\gamma}\rho_{E}\Pi_{E}^{\gamma}+\rho_{A} \otimes \widetilde{\rho}_{E}^{1/2}\left(  I-\Pi_{E}^{\gamma}\right)
\widetilde{\rho}_{E}^{1/2}\\
&  \leq\mu\frac{8}{\delta^{2}} \rho_{A} \otimes \Pi_{E}^{\gamma}\widetilde{\rho}_{E}\Pi
_{E}^{\gamma} + \rho_{A} \otimes \widetilde{\rho}_{E}^{1/2}\left(  I-\Pi
_{E}^{\gamma}\right)  \widetilde{\rho}_{E}^{1/2}\\
&  \leq\mu\frac{8}{\delta^{2}}
\left[  \rho_{A} \otimes \Pi_{E}^{\gamma}\widetilde{\rho}_{E}
\Pi_{E}^{\gamma}+
\rho_{A} \otimes \widetilde{\rho}_{E}^{1/2}\left(  I-\Pi
_{E}^{\gamma}\right)  \widetilde{\rho}_{E}^{1/2}\right]
\\
&  =\mu\frac{8}{\delta^{2}}\rho_{A} \otimes \left[  \Pi_{E}^{\gamma}\widetilde{\rho}_{E}\Pi
_{E}^{\gamma}+\widetilde{\rho}_{E}^{1/2}\left(  I-\Pi_{E}^{\gamma}\right)
\widetilde{\rho}_{E}^{1/2}\right]  \\
&  =\mu\frac{8}{\delta^{2}}\rho_{A} \otimes \widehat{\rho}_{E}.
\end{align}
The second inequality above follows from \eqref{eq:op-ineq-rho-tilde-rho}.
Applying the definition of $D_{\max}(\widehat{\rho}_{AE}\Vert\rho_{A} \otimes \widehat{\rho
}_{E})$, we conclude that%
\begin{equation}
D_{\max}(\widehat{\rho}_{AE}\Vert\rho_{A} \otimes \widehat{\rho
}_{E})
\leq
D_{\max}(\widetilde{\rho}_{AE}\Vert\rho_{A}\otimes\rho_{E})+\log_{2}\!\left(
\frac{8}{\delta^{2}}\right)  . \label{eq:dmax-inequality-smooth-max-relate}%
\end{equation}

We can conclude the statement of the lemma if $P(\widehat{\rho}_{AE},\rho
_{AE})\leq\varepsilon+\delta$, and so it is our aim to show this now. Consider
that%
\begin{equation}
P(\widehat{\rho}_{AEX},\overline{\rho}_{AEX})=\sqrt{1-F(\widehat{\rho}%
_{AEX},\overline{\rho}_{AEX})}.
\end{equation}
The following chain of inequalities holds%
\begin{align}
&  \sqrt{F(\widehat{\rho}_{AEX},\overline{\rho}_{AEX})}\nonumber\\
&  =\operatorname{Tr}\!\left[  \left(  \sqrt{\Pi_{E}^{\gamma}\widetilde{\rho
}_{AE}\Pi_{E}^{\gamma}}\widehat{\rho}_{AE}\sqrt{\Pi_{E}^{\gamma}%
\widetilde{\rho}_{AE}\Pi_{E}^{\gamma}}\right)  ^{1/2}\right] \\
&  \geq\operatorname{Tr}\!\left[  \left(  \sqrt{\Pi_{E}^{\gamma}%
\widetilde{\rho}_{AE}\Pi_{E}^{\gamma}}\left(  \Pi_{E}^{\gamma}\widetilde{\rho
}_{AE}\Pi_{E}^{\gamma}\right)  \sqrt{\Pi_{E}^{\gamma}\widetilde{\rho}_{AE}%
\Pi_{E}^{\gamma}}\right)  ^{1/2}\right] \\
&  =\operatorname{Tr}\!\left[  \Pi_{E}^{\gamma}\widetilde{\rho}_{AE}\Pi
_{E}^{\gamma}\right] \\
&  =\operatorname{Tr}[\Pi_{E}^{\gamma}\widetilde{\rho}_{E}]\\
&  \geq1-\frac{\delta^{2}}{8},
\end{align}
where the inequality follows from operator monotonicity of the square root and
the fact that%
\begin{align}
\widehat{\rho}_{AE}  &  =\Pi_{E}^{\gamma}\widetilde{\rho}_{AE}\Pi_{E}^{\gamma
}+\rho_{A} \otimes \widetilde{\rho}_{E}^{1/2}\left(  I-\Pi_{E}^{\gamma}\right)  \widetilde
{\rho}_{E}^{1/2}\\
&  \geq\Pi_{E}^{\gamma}\widetilde{\rho}_{AE}\Pi_{E}^{\gamma}%
\end{align}
From the above and \eqref{eq:gamma-project-onto-smooth-rho}, we conclude that
$F(\widehat{\rho}_{AEX},\overline{\rho}_{AEX})\geq1-\frac{\delta^{2}}{4}$,
which implies that%
\begin{equation}
P(\widehat{\rho}_{AEX},\overline{\rho}_{AEX})\leq\frac{\delta}{2}.
\label{eq:purified-bar-hat-delta-2}%
\end{equation}
Now consider that%
\begin{align}
&  P(\overline{\rho}_{AEX},\rho_{AE}\otimes|0\rangle\!\langle0|_{X}%
)\nonumber\\
&  \leq P(\overline{\rho}_{AEX},\widetilde{\rho}_{AE}\otimes|0\rangle
\!\langle0|_{X})\nonumber\\
&  \qquad+P(\widetilde{\rho}_{AE}\otimes|0\rangle\!\langle0|_{X},\rho
_{AE}\otimes|0\rangle\!\langle0|_{X})\\
&  =\sqrt{1-F(\overline{\rho}_{AEX},\widetilde{\rho}_{AE}\otimes
|0\rangle\!\langle0|_{X})}+P(\widetilde{\rho}_{AE},\rho_{AE})\\
&  =\sqrt{1-\left\Vert \sqrt{\Pi_{E}^{\gamma}\widetilde{\rho}_{AE}\Pi
_{E}^{\gamma}}\sqrt{\widetilde{\rho}_{AE}}\right\Vert _{1}^{2}}+P(\widetilde
{\rho}_{AE},\rho_{AE})\\
&  =\sqrt{1-\left\Vert \sqrt{\Pi_{E}^{\gamma}\widetilde{\rho}_{AE}\Pi
_{A}^{\gamma}}\sqrt{\Pi_{E}^{\gamma}\widetilde{\rho}_{AE}\Pi_{A}^{\gamma}%
}\right\Vert _{1}^{2}}+P(\widetilde{\rho}_{AE},\rho_{AE})\\
&  =\sqrt{1-\left(  \operatorname{Tr}[\Pi_{E}^{\gamma}\widetilde{\rho}%
_{AE}]\right)  ^{2}}+P(\widetilde{\rho}_{AE},\rho_{AE})\\
&  \leq\frac{\delta}{2}+\varepsilon,
\end{align}
where we applied the triangle inequality of the sine distance
(Lemma~\ref{lem-sine-distance-triangle}) for the first inequality and the fact that
$\left\Vert \sqrt{\Pi\omega\Pi}\sqrt{\tau}\right\Vert _{1}=\left\Vert
\sqrt{\Pi\omega\Pi}\sqrt{\Pi\tau\Pi}\right\Vert _{1}$ for a projector $\Pi$
and states $\omega$ and $\tau$. Combining this with
\eqref{eq:purified-bar-hat-delta-2}, we find that%
\begin{align}
  P(\widehat{\rho}_{AE},\rho_{AE})
&  =P(\widehat{\rho}_{AEX},\rho_{AE}\otimes|0\rangle\!\langle0|_{X})\\
&  \leq P(\widehat{\rho}_{AEX},\overline{\rho}_{AEX})+P(\overline{\rho}%
_{AEX},\rho_{AE}\otimes|0\rangle\!\langle0|_{X})\\
&  =\varepsilon+\delta.
\end{align}
Since we have found a state $\widehat{\rho}_{AE}$ satisfying $P(\widehat{\rho
}_{AE},\rho_{AE})\leq\varepsilon+\delta$ and
\eqref{eq:dmax-inequality-smooth-max-relate}, we conclude that%
\begin{equation}
\widetilde{I}_{\max}^{\varepsilon+\delta}(E;A)_{\rho}\leq D_{\max}%
(\widetilde{\rho}_{AE}\Vert\rho_{A}\otimes\rho_{E})+\log_{2}\!\left(  \frac
{8}{\delta^{2}}\right)  .
\end{equation}
Since this inequality has been shown for all states $\widetilde{\rho}_{AE}$
satisfying $P(\widetilde{\rho}_{AE},\rho_{AE})\leq\varepsilon$, we conclude
the statement of the lemma.

\end{subappendices}

\chapter{Private Communication}\label{chap-private_capacity}

	This chapter focuses on the task of private communication, in which the goal
is for a sender to communicate classical information privately over a quantum
channel to a receiver, such that the environment of the channel gains essentially no
information about the message transmitted. There are connections between this
task and secret key distillation from Chapter~\ref{chap-secret_key_distill}, as well as with quantum
communication from Chapter~\ref{chap-quantum_capacity}. Private communication can be considered a
dynamic version of the general problem of establishing secret correlations
between two parties, whereas secret key distillation is a static version of
the same problem. Indeed, the resource shared between the two parties in the former
task is a quantum channel (a dynamic resource), whereas the resource shared in the latter is a
bipartite quantum state (a static resource). The cryptographic models are similar as well:\ in key
distillation, we assumed that an eavesdropper possesses the purifying system
of a purification of the shared state, whereas, in this chapter, we assume
that an eavesdropper possesses the purifying system of a purification of the
channel connecting the sender to receiver (i.e., the eavesdropper possesses the environment of the
channel). The connection of private communication to quantum communication is
as follows: if two parties can
communicate some amount of quantum information with some error, then the
amount of private information that they can communicate is related to this amount by an inequality. This inequality in turn implies that the private capacity of a quantum channel is not smaller than its quantum capacity.

As with other communication tasks that we have considered in previous
chapters, there are multiple ways to define how communication can be private,
based on various error criteria. In this chapter, we define two such criteria that lead to two different but related communication tasks,
one that we call secret-key transmission and another that we call private
communication. The criterion for the former task is most similar to an average error
criterion, in which the goal is for the sender to use the channel to transmit one
share of a secret key to the receiver, and the criterion for the latter task is a maximal
infidelity criterion, in which all messages transmitted over the channel are
required to meet a particular error criterion, which captures both the
decoding error probability of the receiver, as well as the security of the
message transmitted.

As usual by now, we begin our development in the one-shot setting, with the goal of
establishing lower and upper bounds on the one-shot private capacity. We find
several upper bounds on the one-shot private capacity, in terms of the
one-shot private information of the channel, the hypothesis testing relative
entropy of entanglement, and the squashed entanglement. The lower bound that
we establish is related to a different variation of the one-shot private
information of the channel (not the same quantity as in the upper bound), and
we juxtapose the methods of position-based coding and convex splitting to
prove the achievability of this one-shot private information. Some of the mathematical steps in the proof of the lower bound are similar to those that we used in the previous chapter, in which we established a lower bound on the one-shot distillable key. Moving on to the
asymptotic setting, we prove that the private capacity of a quantum channel is
equal to its regularized private information. This quantity is difficult to
compute in general, and so we then establish some upper bounds on it in terms of
the relative entropy of entanglement and squashed entanglement.

\section{One-Shot Setting}

\label{sec-PC:one-shot-setting}

Let $\mathcal{N}_{A\rightarrow B}$ be a quantum
channel connecting a sender Alice to a receiver Bob, and let $\mathcal{U}%
_{A\rightarrow BE}^{\mathcal{N}}$ be an isometric channel extending
$\mathcal{N}_{A\rightarrow B}$, in the sense that $\mathcal{N}_{A\rightarrow
B}=\operatorname{Tr}_{E}\circ\mathcal{U}_{A\rightarrow BE}^{\mathcal{N}}$. The
goal of a private communication protocol is for Alice to communicate a
classical message to Bob reliably, in the sense that Bob can decode it with
high probability, and such that it is secure from anyone who possesses the
environment system $E$ (we personify the environment as the eavesdropper Eve). A private
communication protocol in the one-shot setting is illustrated in Figure~[REF].
It is defined by the three elements $(\mathcal{M},\mathcal{E}_{M^{\prime
}\rightarrow A},\mathcal{D}_{B\rightarrow\hat{M}})$, in which $\mathcal{M}$ is
a message set, $\mathcal{E}_{M^{\prime}\rightarrow A}$ is an encoding channel,
and $\mathcal{D}_{B\rightarrow\hat{M}}$ is a decoding channel. The pair
$(\mathcal{E}_{M^{\prime}\rightarrow A},\mathcal{D}_{B\rightarrow\hat{M}})$,
consisting of the encoding and decoding channels, is called a private
communication code or, more simply, a code. The encoding channel is a
classical--quantum channel, and the decoding channel is a quantum--classical
or measurement channel.

The steps of the protocol proceed similarly to those of the classical
communication protocol discussed in Section~\ref{sec-cc_one_shot}, with the key difference that the
message transmitted should be kept private from Eve. Let us employ notation
similar to that discussed in \eqref{eq-classical_comm_initial_state}--\eqref{eq-classical_comm_final_state_2}, in which Alice's probability for
selecting message $m\in\mathcal{M}$ is denoted by $p(m)$, the initial state is
denoted by
\begin{equation}
\overline{\Phi}_{MM^{\prime}}^{p}\coloneqq \sum_{m\in\mathcal{M}}p(m)|m\rangle
\!\langle m|_{M}\otimes|m\rangle\!\langle m|_{M^{\prime}},
\end{equation}
the state after the encoding channel by%
\begin{align}
\rho_{MA}^{p}  &  \coloneqq \mathcal{E}_{M^{\prime}\rightarrow A}(\overline{\Phi
}_{MM^{\prime}}^{p})\\
&  =\sum_{m\in\mathcal{M}}p(m)|m\rangle\!\langle m|_{M}\otimes\rho_{A}^{m},
\end{align}
where we have defined
\begin{equation}
\rho_{A}^{m} \coloneqq \mathcal{E}_{M^{\prime}\rightarrow A}(|m\rangle\!\langle m|_{M'}),
\end{equation}
the state before the decoding channel by%
\begin{equation}
\mathcal{U}_{A\rightarrow BE}^{\mathcal{N}}(\rho_{MA}^{p}),
\end{equation}
and the final state of the protocol by%
\begin{align}
\omega_{M\hat{M}E}^{p}  &  \coloneqq (\mathcal{D}_{B\rightarrow\hat{M}}\circ
\mathcal{U}_{A\rightarrow BE}^{\mathcal{N}}\circ\mathcal{E}_{M^{\prime
}\rightarrow A})(\overline{\Phi}_{MM^{\prime}}^{p})\\
&  =\sum_{m,\hat{m}\in\mathcal{M}}p(m)|m\rangle\!\langle m|_{M}\otimes|\hat
{m}\rangle\!\langle\hat{m}|_{\hat{M}}\otimes\operatorname{Tr}_{B}[\Lambda_{B}^{\hat{m}%
}\mathcal{U}_{A\rightarrow BE}^{\mathcal{N}}(\rho_{A}^{m})],
\end{align}
where we have used the fact that the decoding channel is a measurement channel
and thus can be written in terms of a POVM\ $\{\Lambda_{B}^{m}\}_{m\in
\mathcal{M}}$ as%
\begin{equation}
\mathcal{D}_{B\rightarrow\hat{M}}(\tau_{B})\coloneqq \sum_{\hat{m}\in\mathcal{M}%
}\operatorname{Tr}[\Lambda_{B}^{\hat{m}}\tau_{B}]\ |\hat{m}\rangle
\!\langle\hat{m}|_{\hat{M}}.
\end{equation}
If we define the following states
\begin{align}
\omega^{m,\hat{m}}_E & \coloneqq \frac{\operatorname{Tr}_{B}[\Lambda_{B}^{\hat{m}%
}\mathcal{U}_{A\rightarrow BE}^{\mathcal{N}}(\rho_{A}^{m})]}{q(\hat{m}|m)},\\
q(\hat{m}|m) & \coloneqq \operatorname{Tr}[\Lambda_{B}^{\hat{m}%
}\mathcal{U}_{A\rightarrow BE}^{\mathcal{N}}(\rho_{A}^{m})] = \operatorname{Tr}[\Lambda_{B}^{\hat{m}%
}\mathcal{N}_{A\rightarrow B}(\rho_{A}^{m})],
\end{align}
then we can write the final state of the protocol alternatively as follows:
\begin{equation}
\omega_{M\hat{M}E}^{p} = \sum_{m,\hat{m}\in\mathcal{M}}p(m)q(\hat{m}|m) |m\rangle\!\langle m|_{M}\otimes|\hat
{m}\rangle\!\langle\hat{m}|_{\hat{M}}\otimes \omega^{m,\hat{m}}_E.
\end{equation}

The difference between a protocol for (public) communication, as discussed in Section~\ref{sec-cc_one_shot},
and one for private communication is the metric used for characterizing
performance. Here, we demand that the message remain private from the
eavesdropper in addition to being decodable by the receiver. We combine these
constraints into a single metric, which we define in terms of the infidelity.
We delineate two different cases, similar to how we did in Section~\ref{sec-cc_one_shot}:
\begin{enumerate}
\item 
 The
average infidelity of the code is given by%
\begin{align}
&  \overline{p}_{\text{err}}(\mathcal{E},\mathcal{D};p,\mathcal{N})\nonumber\\
&  \coloneqq \inf_{\sigma_{E}}\left(  1-F(\overline{\Phi}_{M\hat{M}}^{p}\otimes
\sigma_{E},\mathcal{P}_{M^{\prime}\rightarrow\hat{M}E}(\overline{\Phi
}_{MM^{\prime}}^{p}))\right) \\
&  =\inf_{\sigma_{E}}\left(  1-\left(  \sum_{m\in\mathcal{M}}p(m)\sqrt
{F}(|m\rangle\!\langle m|_{\hat{M}}\otimes\sigma_{E},\mathcal{P}_{M^{\prime
}\rightarrow\hat{M}E}(|m\rangle\!\langle m|_{M^{\prime}}))\right)
^{2}\right)  ,
\end{align}
where%
\begin{equation}
\mathcal{P}_{M^{\prime}\rightarrow\hat{M}E}\coloneqq \mathcal{D}_{B\rightarrow\hat{M}%
}\circ\mathcal{U}_{A\rightarrow BE}^{\mathcal{N}}\circ\mathcal{E}_{M^{\prime
}\rightarrow A}%
\end{equation}
and the infimum is taken over every state $\sigma_{E}$ of the eavesdropper's
system~$E$. Also, we employed Proposition~\ref{prop-sand_rel_ent_properties} with $\alpha = \frac{1}{2}$ in the last line above. If the prior probability distribution $p(m)$ is the uniform
distribution (i.e., $p(m)=1/\left\vert \mathcal{M}\right\vert $), then the
communication task is called \textit{secret-key transmission}, because the
goal is for Alice to transmit one share of a secret key to the receiver Bob.

\item 
An alternative error criterion is the maximal infidelity of the code, defined
as%
\begin{equation}
p_{\text{err}}^{\ast}(\mathcal{E},\mathcal{D};\mathcal{N})\coloneqq \inf_{\sigma_{E}%
}\max_{m\in\mathcal{M}}\left(  1-F(|m\rangle\!\langle m|_{\hat{M}}%
\otimes\sigma_{E},\mathcal{P}_{M^{\prime}\rightarrow\hat{M}E}(|m\rangle
\!\langle m|_{M^{\prime}}))\right)  .
\end{equation}
When the communication task employs this error criterion, we refer to it as
\textit{private communication}.
\end{enumerate}

The interpretation of the average infidelity obeying the inequality
\begin{equation}
\overline{p}_{\text{err}}(\mathcal{E},\mathcal{D};p,\mathcal{N}%
)\leq\varepsilon
\end{equation}
 is that there exists a state $\sigma_{E}$ of the
eavesdropper's system $E$ such that the state of systems $\hat{M}$ and $E$ is
close to the product state $|m\rangle\!\langle m|_{\hat{M}}\otimes\sigma_{E}$,
on average. This means that not only can Bob can decode well, but also, that the
state of Eve's system is close to the constant state $\sigma_{E}$, such that
her system is not useful for figuring out the message transmitted (on
average). Indeed, by applying the data-processing inequality with respect to
partial trace of system $E$ and letting $\sigma_{E}$ be the state that
achieves $\overline{p}_{\text{err}}(\mathcal{E},\mathcal{D};p,\mathcal{N})$,
we conclude that%
\begin{align}
\varepsilon &  \geq\overline{p}_{\text{err}}(\mathcal{E},\mathcal{D}%
;p,\mathcal{N})\label{eq-PC:avg-infid-implies-reliability-1}\\
&  =1-\left(  \sum_{m\in\mathcal{M}}p(m)\sqrt{F}(|m\rangle\!\langle
m|_{\hat{M}}\otimes\sigma_{E},\mathcal{P}_{M^{\prime}\rightarrow\hat{M}%
E}(|m\rangle\!\langle m|_{M^{\prime}}))\right)  ^{2}\\
&  \geq1-\sum_{m\in\mathcal{M}}p(m)F(|m\rangle\!\langle m|_{\hat{M}}%
\otimes\sigma_{E},\mathcal{P}_{M^{\prime}\rightarrow\hat{M}E}(|m\rangle
\!\langle m|_{M^{\prime}}))\\
&  \geq\sum_{m\in\mathcal{M}}p(m)\left(  1-F(|m\rangle\!\langle m|_{\hat{M}%
},(\mathcal{D}_{B\rightarrow\hat{M}}\circ\mathcal{N}_{A\rightarrow B}%
\circ\mathcal{E}_{M^{\prime}\rightarrow A})(|m\rangle\!\langle m|_{M^{\prime}%
}))\right) \\
&  =\sum_{m\in\mathcal{M}}p(m)\left(  1-\langle m|_{\hat{M}}(\mathcal{D}%
_{B\rightarrow\hat{M}}\circ\mathcal{N}_{A\rightarrow B}\circ\mathcal{E}%
_{M^{\prime}\rightarrow A})(|m\rangle\!\langle m|_{M^{\prime}})|m\rangle
_{\hat{M}}\right) \\
&  =\sum_{m\in\mathcal{M}}p(m)\left(  1-\operatorname{Tr}[\Lambda_{B}%
^{m}\mathcal{N}_{A\rightarrow B}(\rho_{A}^{m})]\right)  .
\label{eq-PC:avg-infid-implies-reliability-last}%
\end{align}
The second inequality follows from convexity of the square function and the
third from the data-processing inequality for fidelity. The latter expression is the same as
the average error probability from \eqref{eq-avg_error_prob}. Now applying the data-processing
inequality with respect to partial trace over $\hat{M}$, we conclude that%
\begin{align}
\varepsilon &  \geq\overline{p}_{\text{err}}(\mathcal{E},\mathcal{D}%
;p,\mathcal{N})\label{eq-PC:avg-infid-implies-security-1}\\
&  =1-F(\overline{\Phi}_{M\hat{M}}^{p}\otimes\sigma_{E},(\mathcal{D}%
_{B\rightarrow\hat{M}}\circ\mathcal{U}_{A\rightarrow BE}^{\mathcal{N}}%
\circ\mathcal{E}_{M^{\prime}\rightarrow A})(\overline{\Phi}_{MM^{\prime}}%
^{p}))\\
&  \geq1-F(\operatorname{Tr}_{\hat{M}}[\overline{\Phi}_{M\hat{M}}^{p}%
\otimes\sigma_{E}],\operatorname{Tr}_{\hat{M}}[(\mathcal{D}_{B\rightarrow
\hat{M}}\circ\mathcal{U}_{A\rightarrow BE}^{\mathcal{N}}\circ\mathcal{E}%
_{M^{\prime}\rightarrow A})(\overline{\Phi}_{MM^{\prime}}^{p})])\\
&  =1-F(\pi_{M}^{p}\otimes\sigma_{E},(\mathcal{N}^c_{A\rightarrow
E}\circ\mathcal{E}_{M^{\prime}\rightarrow A})(\overline{\Phi}_{MM^{\prime}%
}^{p}))\\
&  =1-\left(  \sum_{m\in\mathcal{M}}p(m)\sqrt{F}(\sigma_{E},
\mathcal{N}^c_{A\rightarrow E}(\rho_{A}^{m}))\right)  ^{2},
\label{eq-PC:avg-infid-implies-security-last}%
\end{align}
which indicates that the state of Eve's system $E$ is close to the constant
state $\sigma_{E}$ on average. In the above, $\mathcal{N}
^c_{A\rightarrow E}$ is a complementary channel of $\mathcal{N}_{A\rightarrow
B}$, as defined in Section~\ref{sec-QM:complementary-chs}, and is given by $\mathcal{N}^c_{A\rightarrow
E}=\operatorname{Tr}_{B}\circ\mathcal{U}_{A\rightarrow BE}^{\mathcal{N}}$. Also, in the last line above, we employed Proposition~\ref{prop-sand_rel_ent_properties} with $\alpha = \frac{1}{2}$.

The interpretation of the maximum infidelity obeying the constraint
\begin{equation}
p_{\text{err}}^{\ast}(\mathcal{E},\mathcal{D};\mathcal{N})\leq\varepsilon
\end{equation}
is
similar. If this condition holds, then there exists a state $\sigma_{E}$ of
the eavesdropper's system $E$ such that the state of systems $\hat{M}$ and $E$
is close to the product state $|m\rangle\!\langle m|_{\hat{M}}\otimes
\sigma_{E}$, for every message $m\in\mathcal{M}$. So this is a much stronger
constraint in general and the one we aim to achieve for private communication.
By applying the data-processing inequality to $p_{\text{err}}^{\ast
}(\mathcal{E},\mathcal{D};\mathcal{N})\leq\varepsilon$ and letting $\sigma
_{E}$ be the state that achieves $p_{\text{err}}^{\ast}(\mathcal{E}%
,\mathcal{D};\mathcal{N})$, we conclude by similar reasoning as given above
that%
\begin{equation}
\varepsilon   \geq p_{\text{err}}^{\ast}(\mathcal{E},\mathcal{D}%
;\mathcal{N})
  \geq\max_{m\in\mathcal{M}}\left(  1-\operatorname{Tr}[\Lambda_{B}%
^{m}\mathcal{N}_{A\rightarrow B}(\rho_{A}^{m})]\right)  ,
\end{equation}
and%
\begin{equation}
\varepsilon   \geq p_{\text{err}}^{\ast}(\mathcal{E},\mathcal{D}%
;\mathcal{N})
  \geq\max_{m\in\mathcal{M}}\left(  1-F(\sigma_{E},\mathcal{N}^c_{A\rightarrow E}(\rho_{A}^{m}))\right)  .
\end{equation}
Thus, if $p_{\text{err}}^{\ast}(\mathcal{E},\mathcal{D};\mathcal{N}%
)\leq\varepsilon$ holds, then Bob can reliably decode every message
$m\in\mathcal{M}$, in the sense that%
\begin{equation}
\operatorname{Tr}[\Lambda_{B}^{m}\mathcal{N}_{A\rightarrow B}(\rho_{A}%
^{m})]\geq1-\varepsilon\qquad\forall m\in\mathcal{M},
\end{equation}
and Eve's system $E$ is not useful for determining any of the messages, in the
sense that%
\begin{equation}
F(\sigma_{E},\mathcal{N}^c_{A\rightarrow E}(\rho_{A}^{m}%
))\geq1-\varepsilon\qquad\forall m\in\mathcal{M}.
\end{equation}

These two different infidelity criteria can be used to assess the performance of a
protocol, i.e., how well Bob can decode the message and how secure it is from Eve.

\begin{definition}
{$(\left\vert \mathcal{M}\right\vert ,\varepsilon)$ Private Communication
Protocol}{}
A private communication protocol $(\mathcal{M},\mathcal{E}%
_{M^{\prime}\rightarrow A},\mathcal{D}_{B\rightarrow\hat{M}})$ over the
channel $\mathcal{N}_{A\rightarrow B}$ is called an $(\left\vert
\mathcal{M}\right\vert ,\varepsilon)$ protocol, with $\varepsilon\in\left[
0,1\right]  $, if $p_{\text{err}}^{\ast}(\mathcal{E},\mathcal{D}%
;\mathcal{N})\leq\varepsilon$.
\end{definition}

Similar to the case of entanglement-assisted and unassisted classical
communication, the infidelity criterion $p_{\text{err}}^{\ast}(\mathcal{E}%
,\mathcal{D};\mathcal{N})\leq\varepsilon$ is equivalent to%
\begin{equation}
\inf_{\sigma_{E}}\max_{p:\mathcal{M}\rightarrow\left[  0,1\right]  }\left(
1-F(\overline{\Phi}_{M\hat{M}}^{p}\otimes\sigma_{E},(\mathcal{D}%
_{B\rightarrow\hat{M}}\circ\mathcal{U}_{A\rightarrow BE}^{\mathcal{N}}%
\circ\mathcal{E}_{M^{\prime}\rightarrow A})(\overline{\Phi}_{MM^{\prime}}%
^{p}))\right)  \leq\varepsilon, \label{eq-PC:max-infid-alt-exp}%
\end{equation}
where the optimization is over every probability distribution $p(m)$ for the
messages in $\mathcal{M}$. This follows because%
\begin{multline}
F(\overline{\Phi}_{M\hat{M}}^{p}\otimes\sigma_{E},(\mathcal{D}_{B\rightarrow
\hat{M}}\circ\mathcal{U}_{A\rightarrow BE}^{\mathcal{N}}\circ\mathcal{E}%
_{M^{\prime}\rightarrow A})(\overline{\Phi}_{MM^{\prime}}^{p}))=\\
\left[  \sum_{m\in\mathcal{M}}p(m)\sqrt{F}(|m\rangle\!\langle m|_{\hat{M}%
}\otimes\sigma_{E},(\mathcal{D}_{B\rightarrow\hat{M}}\circ\mathcal{U}%
_{A\rightarrow BE}^{\mathcal{N}}\circ\mathcal{E}_{M^{\prime}\rightarrow
A})(|m\rangle\!\langle m|_{M^{\prime}})))\right]  ^{2},
\end{multline}
as a consequence of Proposition~\ref{prop-sand_rel_ent_properties} with $\alpha = \frac{1}{2}$, and then one can employ arguments
similar to those in \eqref{eq-EAC:connect-err-probs-1}--\eqref{eq-EAC:connect-err-probs-last} to conclude \eqref{eq-PC:max-infid-alt-exp}.

The one-shot private capacity of the channel $\mathcal{N}$ is equal to the
maximum number of private bits that can be transmitted for a fixed infidelity
threshold~$\varepsilon$:

\begin{definition}
{One-Shot Private Capacity of a Quantum Channel}{}Given a quantum channel
$\mathcal{N}_{A\rightarrow B}$ and $\varepsilon\in\left[  0,1\right]  $, the
one-shot $\varepsilon$-error private capacity of $\mathcal{N}$, denoted by
$P^{\varepsilon}(\mathcal{N})$, is defined to be the maximum number $\log
_{2}\left\vert \mathcal{M}\right\vert $ of private bits among all $(\left\vert
\mathcal{M}\right\vert ,\varepsilon)$ private communication protocols over
$\mathcal{N}$. In other words,%
\begin{equation}
P^{\varepsilon}(\mathcal{N})\coloneqq \sup_{(\mathcal{M},\mathcal{E},\mathcal{D}%
)}\left\{  \log_{2}\!\left\vert \mathcal{M}\right\vert :p_{\text{err}}^{\ast
}(\mathcal{E},\mathcal{D};\mathcal{N})\leq\varepsilon\right\}  ,
\label{eq-PC:1-shot-priv-cap-def}%
\end{equation}
where the optimization is over all protocols $(\mathcal{M},\mathcal{E}%
_{M^{\prime}\rightarrow A},\mathcal{D}_{B\rightarrow\hat{M}})$ satisfying
$d_{M^{\prime}}=d_{\hat{M}}=\left\vert \mathcal{M}\right\vert $.
\end{definition}

\subsection{Private Communication and Quantum Communication}

\label{sec-PC:private-to-quantum-rel}This subsection establishes that a
quantum communication protocol can always be converted to one for private
communication, such that there is negligible loss with respect to code parameters. This result then implies an inequality relating the one-shot quantum capacity to the one-shot private capacity.

\begin{proposition}{thm-PC:QC-to-PC}
The existence of an $(M,\varepsilon)$ quantum
communication protocol for a quantum channel $\mathcal{N}_{A\rightarrow B}$
implies the existence of an $(\left\lfloor M/2\right\rfloor ,\min
\{1,2\varepsilon\})$ private communication protocol for $\mathcal{N}%
_{A\rightarrow B}$.
\end{proposition}

\begin{Proof}
Starting from an $(M,\varepsilon)$ quantum communication protocol, we can use
it to transmit one share of a maximally entangled state%
\begin{equation}
\Phi_{RS}\coloneqq \frac{1}{M}\sum_{m,m^{\prime}=1}^{M}|m\rangle\!\langle m^{\prime
}|_{R}\otimes|m\rangle\!\langle m^{\prime}|_{S}%
\end{equation}
of Schmidt rank $M$ faithfully, by definition (see Definition~\ref{def-q_comm_Me_protocol}):%
\begin{equation}
F(\Phi_{RS},(\mathcal{D}_{B\rightarrow S}\circ\mathcal{N}_{A\rightarrow
B}\circ\mathcal{E}_{S^{\prime}\rightarrow A})(\Phi_{RS^{\prime}}%
))\geq1-\varepsilon. \label{eq-PC:fid-crit-q-to-priv}%
\end{equation}
Consider that the state%
\begin{equation}
\sigma_{RSE}\coloneqq (\mathcal{D}_{B\rightarrow S}\circ\mathcal{U}_{A\rightarrow
BE}^{\mathcal{N}}\circ\mathcal{E}_{S^{\prime}\rightarrow A})(\Phi_{RS^{\prime
}})
\end{equation}
extends the state output from the actual protocol. By Uhlmann's theorem (Theorem~\ref{thm-Uhlmann_fidelity}), there exists an extension of $\Phi_{RS}$ such that the fidelity
between this extension and the state $\sigma_{RSE}$ is equal to the fidelity
in \eqref{eq-PC:fid-crit-q-to-priv}. However, the maximally entangled state
$\Phi_{RS}$ is unextendible\ in the sense that the only possible extension is
a tensor-product state $\Phi_{RS}\otimes\omega_{E}$ for some state $\omega
_{E}$. So, putting these statements together, we find that%
\begin{equation}
F(\Phi_{RS}\otimes\omega_{E},(\mathcal{D}_{B\rightarrow S}\circ\mathcal{U}%
_{A\rightarrow BE}^{\mathcal{N}}\circ\mathcal{E}_{S^{\prime}\rightarrow
A})(\Phi_{RS^{\prime}}))\geq1-\varepsilon.
\end{equation}
Furthermore, measuring the $R$ and $S$ systems locally in the Schmidt basis of
$\Phi_{RS}$ only increases the fidelity, so that%
\begin{equation}
F(\overline{\Phi}_{RS}\otimes\omega_{E^{n}},(\overline{\mathcal{D}%
}_{B\rightarrow S}\circ\mathcal{U}_{A\rightarrow BE}^{\mathcal{N}}%
\circ\mathcal{E}_{S^{\prime}\rightarrow A})(\overline{\Phi}_{RS}%
))\geq1-\varepsilon,
\end{equation}
where $\overline{\mathcal{D}}_{B\rightarrow S}$ denotes the concatenation of
the original decoder $\mathcal{D}_{B\rightarrow S}$ followed by the local
measurement:%
\begin{align}
\overline{\mathcal{D}}_{B\rightarrow S}(\cdot)  &  \coloneqq \sum_{m}|m\rangle
\!\langle m|\mathcal{D}_{B\rightarrow S}(\cdot)|m\rangle\!\langle m|\\
&  =\sum_{m}\operatorname{Tr}[(\mathcal{D}_{B\rightarrow S})^{\dag}%
[|m\rangle\!\langle m|](\cdot)]|m\rangle\!\langle m|_{S}.
\end{align}
Observe that $\{(\mathcal{D}_{B\rightarrow S})^{\dag}[|m\rangle\!\langle
m|]\}_{m}$ is a valid POVM. Using the direct-sum property of the fidelity (Proposition~\ref{prop-sand_rel_ent_properties} with $\alpha = \frac{1}{2}$) and defining $\rho_{A}^{m}\coloneqq \mathcal{E}_{S^{\prime}\rightarrow
A}(|m\rangle\!\langle m|_{S'})$, we can then rewrite this as%
\begin{equation}
\left(  \frac{1}{M}\sum_{m=1}^{M}\sqrt{F}(|m\rangle\!\langle m|_{S}%
\otimes\omega_{E},(\overline{\mathcal{D}}_{B\rightarrow S}\circ\mathcal{U}%
_{A\rightarrow BE}^{\mathcal{N}})(\rho_{A}^{m}))\right)  ^{2}\geq
1-\varepsilon.
\end{equation}
We can in turn rewrite this inequality as%
\begin{equation}
\frac{1}{M}\sum_{m=1}^{M}\sqrt{F}(|m\rangle\!\langle m|_{S}\otimes\omega
_{E},(\overline{\mathcal{D}}_{B\rightarrow S}\circ\mathcal{U}_{A\rightarrow
BE}^{\mathcal{N}})(\rho_{A}^{m}))\geq\sqrt{1-\varepsilon}%
\end{equation}
and again as%
\begin{equation}
\frac{1}{M}\sum_{m=1}^{M}\left(  1-\sqrt{F}(|m\rangle\!\langle m|_{S}%
\otimes\omega_{E},(\overline{\mathcal{D}}_{B\rightarrow S}\circ\mathcal{U}%
_{A\rightarrow BE}^{\mathcal{N}})(\rho_{A}^{m}))\right)  \leq1-\sqrt
{1-\varepsilon}%
\end{equation}
Markov's inequality then guarantees that there exists a subset $\mathcal{M}%
^{\prime}$\ of the set $\{1,\ldots,M\}$ of size $\left\lfloor M/2\right\rfloor
$ such that the following condition holds for all $m\in\mathcal{M}^{\prime}$:%
\begin{equation}
1-\sqrt{F}(|m\rangle\!\langle m|_{S}\otimes\omega_{E},(\overline{\mathcal{D}%
}_{B\rightarrow S}\circ\mathcal{U}_{A\rightarrow BE}^{\mathcal{N}})(\rho
_{A}^{m}))\leq2\left(  1-\sqrt{1-\varepsilon}\right)  .
\label{eq:q-to-p-good-condition}%
\end{equation}
We can rewrite this condition as%
\begin{align}
F(|m\rangle\!\langle m|_{S}\otimes\omega_{E},(\overline{\mathcal{D}%
}_{B\rightarrow S}\circ\mathcal{U}_{A\rightarrow BE}^{\mathcal{N}})(\rho
_{A}^{m}))  &  \geq\left(  1-2\left(  1-\sqrt{1-\varepsilon}\right)  \right)
^{2}\\
&  =\left(  1-2\sqrt{1-\varepsilon}\right)  ^{2}\\
&  \geq1-2\varepsilon.
\end{align}
We now define the private communication protocol to consist of codewords
$\{\rho_{A}^{m}\coloneqq \mathcal{E}_{S\rightarrow A}(|m\rangle\!\langle
m|_{S})\}_{m\in\mathcal{M}^{\prime}}$ and the decoding POVM\ to be%
\begin{equation}
\{\Lambda_{B}^{m}\equiv(\mathcal{D}_{B\rightarrow S})^{\dag}(|m\rangle
\!\langle m|)\}_{m\in\mathcal{M}^{\prime}}\cup\left\{  \Lambda_{B^{n}}%
^{0}\coloneqq (\mathcal{D}_{B\rightarrow S})^{\dag}\!\left(  \sum_{m\not \in
\mathcal{M}^{\prime}}|m\rangle\!\langle m|\right)  \right\}  .
\end{equation}
Thus, we have shown that from an $(M,\varepsilon)$ quantum communication
protocol, one can realize an $(\left\lfloor M/2\right\rfloor ,2\varepsilon)$
protocol for private communication.
\end{Proof}

Proposition~\ref{thm-PC:QC-to-PC} then implies the following for the one-shot capacities:

\begin{theorem}{}
For a quantum channel $\mathcal{N}_{A\rightarrow B}$ and $\varepsilon\in
(0,1)$, the following inequality relates the one-shot quantum capacity
$Q^{\frac{\varepsilon}{2}}(\mathcal{N})$\ to the one-shot private capacity
$P^{\varepsilon}(\mathcal{N})$:%
\begin{equation}
Q^{\frac{\varepsilon}{2}}(\mathcal{N})\leq P^{\varepsilon}(\mathcal{N})+1.
\end{equation}

\end{theorem}

\begin{Proof}
Given an arbitrary $(M,\varepsilon/2)$ quantum communication protocol, by
Proposition~\ref{thm-PC:QC-to-PC}, we can realize an arbitrary
$(M/2,\varepsilon)$ private communication protocol. Letting the protocol be
one that achieves the one-shot quantum capacity $Q^{\frac{\varepsilon}{2}%
}(\mathcal{N})$ (i.e., $\log_{2}M=Q^{\frac{\varepsilon}{2}}(\mathcal{N})$), we
conclude that there exists an $(M/2,\varepsilon)$ private communication
protocol. Since this is a particular $(M/2,\varepsilon)$ private communication
protocol, we conclude that%
\begin{equation}
\log_{2}(M/2)\leq P^{\varepsilon}(\mathcal{N}),
\end{equation}
which follows from the definition of the one-shot private capacity
$P^{\varepsilon}(\mathcal{N})$. We finally use the fact that $\log
_{2}(M/2)=Q^{\frac{\varepsilon}{2}}(\mathcal{N})-1$.
\end{Proof}

\subsection{Secret-Key Transmission and Bipartite Private-State Transmission}

\label{sec-PC:bi-to-tri-key}

In this section, we establish a connection between secret-key transmission and
bipartite private-state transmission. Before doing so, we first define what is
meant by a bipartite private-state transmission protocol. To do so, we follow
the same spirit in Section~\ref{sec-SKD:tri-to-bi-relate}, and we purify each step of a secret-key
transmission protocol and trace out the system possessed by the eavesdropper
Eve. In this case, it is only the environment $E$ of the isometric channel
$\mathcal{U}_{A\rightarrow BE}^{\mathcal{N}}$ that belongs to the
eavesdropper, and tracing it out leads to the original channel~$\mathcal{N}_{A\rightarrow B}$.

A bipartite private-state transmission protocol is defined by the triple
\begin{equation}
(\mathcal{M},\mathcal{U}_{M^{\prime}\rightarrow AA^{\prime}}^{\mathcal{E}%
},\mathcal{U}_{B\rightarrow\hat{M}B^{\prime}}^{\mathcal{D}}),
\end{equation}
where
$\mathcal{M}$ is a message set, $\mathcal{U}_{M^{\prime}\rightarrow
AA^{\prime}}^{\mathcal{E}}$ is an isometric encoding channel, and
$\mathcal{U}_{B\rightarrow\hat{M}B^{\prime}}^{\mathcal{D}}$ is an isometric
decoding channel. The protocol begins with Alice preparing a GHZ\ state
$\Phi_{M^{\prime\prime}MM^{\prime}}$ of the following form:%
\begin{equation}
\Phi_{M^{\prime\prime}MM^{\prime}}\coloneqq |\Phi\rangle\!\langle\Phi|_{M^{\prime
\prime}MM^{\prime}},
\end{equation}
where%
\begin{equation}
|\Phi\rangle_{M^{\prime\prime}MM^{\prime}}\coloneqq \frac{1}{\sqrt{\left\vert
\mathcal{M}\right\vert }}\sum_{m\in\mathcal{M}}|m\rangle_{M^{\prime\prime}%
}|m\rangle_{M}|m\rangle_{M^{\prime}}.
\end{equation}
She transmits the $M^{\prime}$ system through the isometric encoding channel
$\mathcal{U}_{M^{\prime}\rightarrow AA^{\prime}}^{\mathcal{E}}$, leading to
the state $\mathcal{U}_{M^{\prime}\rightarrow AA^{\prime}}^{\mathcal{E}}%
(\Phi_{M^{\prime\prime}MM^{\prime}})$. She transmits the $A$ system through
the channel $\mathcal{N}_{A\rightarrow B}$, leading to the state%
\begin{equation}
\mathcal{N}_{A\rightarrow B}(\mathcal{U}_{M^{\prime}\rightarrow AA^{\prime}%
}^{\mathcal{E}}(\Phi_{M^{\prime\prime}MM^{\prime}})).
\end{equation}
Bob finally performs the isometric decoding channel $\mathcal{U}%
_{B\rightarrow\hat{M}B^{\prime}}^{\mathcal{D}}$.\ The final state of the
protocol is then as follows:%
\begin{equation}
\omega_{M^{\prime\prime}MA^{\prime}\hat{M}B^{\prime}}\coloneqq (\mathcal{U}%
_{B\rightarrow\hat{M}B^{\prime}}^{\mathcal{D}}\circ\mathcal{N}_{A\rightarrow
B}\circ\mathcal{U}_{M^{\prime}\rightarrow AA^{\prime}}^{\mathcal{E}}%
)(\Phi_{M^{\prime\prime}MM^{\prime}}),
\end{equation}
where the systems $M^{\prime\prime}MA^{\prime}$ are in possession of Alice and
systems $\hat{M}B^{\prime}$ are in possession of Bob.

Observe that each step of the protocol involves a purification of the steps in
a secret-key transmission protocol, as outlined in
Section~\ref{sec-PC:one-shot-setting}. The initial GHZ\ state is a
purification of the maximally classically correlated state $\overline{\Phi
}_{MM^{\prime}}$. The isometric encoding channel $\mathcal{U}_{M^{\prime
}\rightarrow AA^{\prime}}^{\mathcal{E}}$ purifies the encoding channel
$\mathcal{E}_{M^{\prime}\rightarrow A}$, and the isometric decoding channel
$\mathcal{U}_{B\rightarrow\hat{M}B^{\prime}}^{\mathcal{D}}$ purifies the
decoding channel $\mathcal{D}_{B\rightarrow\hat{M}}$.

The infidelity of a bipartite private-state transmission protocol of the form
above is then defined as follows:%
\begin{equation}
p_{\text{err}}^{b}(\mathcal{U}^{\mathcal{E}},\mathcal{U}^{\mathcal{D}%
};\mathcal{N})\coloneqq \inf_{\gamma_{M^{\prime\prime}MA^{\prime}\hat{M}B^{\prime}}%
}\left(  1-F(\gamma_{M^{\prime\prime}MA^{\prime}\hat{M}B^{\prime}}%
,\omega_{M^{\prime\prime}MA^{\prime}\hat{M}B^{\prime}})\right)  ,
\end{equation}
where the optimization is with respect to every ideal bipartite private state
$\gamma_{M^{\prime\prime}MA^{\prime}\hat{M}B^{\prime}}$, with key system $M$
held by Alice, shield systems $M^{\prime\prime}A^{\prime}$ by Alice, key
system $\hat{M}$ by Bob, and shield system $B^{\prime}$\ by Bob (see Section~\ref{sec-SKD:tri-to-bi-relate}).

\begin{definition}
{$(\left\vert \mathcal{M}\right\vert ,\varepsilon)$ Private-State Transmission
Protocol}{def-PC:private-state-trans-prot}
A bipartite private-state
transmission protocol $(\mathcal{M},\mathcal{U}_{M^{\prime}\rightarrow
AA^{\prime}}^{\mathcal{E}},\mathcal{U}_{B\rightarrow\hat{M}B^{\prime}%
}^{\mathcal{D}})$ for the channel $\mathcal{N}_{A\rightarrow B}$ is called an
$\left(  \left\vert \mathcal{M}\right\vert ,\varepsilon\right)  $ protocol,
with $\varepsilon\in\left[  0,1\right]  $, if $p_{\text{err}}^{b}%
(\mathcal{U}^{\mathcal{E}},\mathcal{U}^{\mathcal{D}};\mathcal{N}%
)\leq\varepsilon$.
\end{definition}

We now establish the main result of this section, which is the equivalence of
secret-key transmission and bipartite private-state transmission:

\begin{theorem}{thm-PC:sec-trans-bi-priv-trans}Let $\mathcal{M}$ be a message set, and
let $\varepsilon\in\left[  0,1\right]  $. Let $\mathcal{N}_{A\rightarrow B}$
be a quantum channel. There exists an $(\left\vert \mathcal{M}\right\vert
,\varepsilon)$ secret-key transmission protocol for $\mathcal{N}_{A\rightarrow
B}$ if and only if there exists an $(\left\vert \mathcal{M}\right\vert
,\varepsilon)$ bipartite private-state transmission protocol for
$\mathcal{N}_{A\rightarrow B}$.
\end{theorem}

\begin{Proof}
We start by proving that there exists an $(\left\vert \mathcal{M}\right\vert
,\varepsilon)$ bipartite private-state transmission protocol if there exists
an $(\left\vert \mathcal{M}\right\vert ,\varepsilon)$ secret-key transmission
protocol. Let $\mathcal{U}_{A\rightarrow BE}^{\mathcal{N}}$ be an isometric
channel extending $\mathcal{N}_{A\rightarrow B}$. Let $\mathcal{E}_{M^{\prime
}\rightarrow A}$ be the encoding channel, and let $\mathcal{D}_{B\rightarrow
\hat{M}}$ be the decoding channel. The final state of the protocol is as
follows:%
\begin{equation}
\omega_{M\hat{M}E}\coloneqq (\mathcal{D}_{B\rightarrow\hat{M}}\circ\mathcal{N}%
_{A\rightarrow B}\circ\mathcal{E}_{M^{\prime}\rightarrow A})(\overline{\Phi
}_{MM^{\prime}})
\end{equation}
and satisfies the inequality%
\begin{equation}
1-F(\overline{\Phi}_{M\hat{M}}\otimes\sigma_{E},\omega_{M\hat{M}E}%
)\leq\varepsilon.
\end{equation}
Observe that $\overline{\Phi}_{M\hat{M}}\otimes\sigma_{E}$ is an ideal
tripartite key state, according to Definition~\ref{def:tripartite-key-state}. Now let $\mathcal{U}%
_{M^{\prime}\rightarrow AA^{\prime}}^{\mathcal{E}}$ be an isometric channel
that extends the encoding channel $\mathcal{E}_{M^{\prime}\rightarrow A}$, and
let $\mathcal{U}_{B\rightarrow\hat{M}B^{\prime}}^{\mathcal{D}}$ be an
isometric channel that extends the decoding channel $\mathcal{D}%
_{B\rightarrow\hat{M}}$. Let $\Phi_{M^{\prime\prime
}MM^{\prime}}$ be a GHZ\ state that purifies $\overline{\Phi}_{MM^{\prime}}$.
Then the following state is a purification of $\omega_{M\hat{M}E}$:%
\begin{equation}
\omega_{MM^{\prime\prime}A^{\prime}\hat{M}B^{\prime}E}\coloneqq (\mathcal{U}%
_{B\rightarrow\hat{M}B^{\prime}}^{\mathcal{D}}\circ\mathcal{U}_{A\rightarrow
BE}^{\mathcal{N}}\circ\mathcal{U}_{M^{\prime}\rightarrow AA^{\prime}%
}^{\mathcal{E}})(\Phi_{M^{\prime\prime}MM^{\prime}}).
\label{eq-PC:final-omega-purification-tri-to-bi}%
\end{equation}
Applying Uhlmann's theorem, we conclude that there is a purification of
$\overline{\Phi}_{M\hat{M}}\otimes\sigma_{E}$, call it $\gamma_{MM^{\prime
\prime}A^{\prime}\hat{M}B^{\prime}E}$, such that%
\begin{equation}
F(\overline{\Phi}_{M\hat{M}}\otimes\sigma_{E},\omega_{M\hat{M}E}%
)=F(\gamma_{MM^{\prime\prime}A^{\prime}\hat{M}B^{\prime}E},\omega
_{MM^{\prime\prime}A^{\prime}\hat{M}B^{\prime}E}).
\end{equation}
Tracing over the $E$ system, we conclude from the data-processing inequality
for fidelity that%
\begin{equation}
1-F(\gamma_{MM^{\prime\prime}A^{\prime}\hat{M}B^{\prime}},\omega
_{MM^{\prime\prime}A^{\prime}\hat{M}B^{\prime}})\leq\varepsilon,
\end{equation}
where%
\begin{equation}
\omega_{MM^{\prime\prime}A^{\prime}\hat{M}B^{\prime}}=(\mathcal{U}%
_{B\rightarrow\hat{M}B^{\prime}}^{\mathcal{D}}\circ\mathcal{N}_{A\rightarrow
B}\circ\mathcal{U}_{M^{\prime}\rightarrow AA^{\prime}}^{\mathcal{E}}%
)(\Phi_{M^{\prime\prime}MM^{\prime}}).
\label{eq-PC:final-omega-state-tri-to-bi}%
\end{equation}
Furthermore, by Definition~\ref{def:private-state-bi}, the state $\gamma_{MM^{\prime\prime}A^{\prime}%
\hat{M}B^{\prime}}$ is an ideal bipartite private state with key system
$M$ held by Alice, shield systems $M^{\prime\prime}A^{\prime}$ held by Alice,
key system $\hat{M}$ held by Bob, and shield system $B^{\prime}$ held by Bob.
Thus, we have shown the first claim.

Now we establish the opposite implication (which follows essentially by
running the argument above backwards). To this end, let $\mathcal{U}%
_{M^{\prime}\rightarrow AA^{\prime}}^{\mathcal{E}}$ be an isometric encoding
channel, and let $\mathcal{U}_{B\rightarrow\hat{M}B^{\prime}}^{\mathcal{D}}$
be an isometric decoding channel for a bipartite private-state transmission
protocol. The initial state of the protocol is the GHZ\ state $\Phi
_{M^{\prime\prime}MM^{\prime}}$, and the final state is $\omega_{MM^{\prime
\prime}A^{\prime}\hat{M}B^{\prime}}$, as given in
\eqref{eq-PC:final-omega-state-tri-to-bi}. For an $(\left\vert \mathcal{M}%
\right\vert ,\varepsilon)$ bipartite private-state transmission protocol, the
following inequality holds%
\begin{equation}
1-F(\gamma_{MM^{\prime\prime}A^{\prime}\hat{M}B^{\prime}},\omega
_{MM^{\prime\prime}A^{\prime}\hat{M}B^{\prime}})\leq\varepsilon,
\end{equation}
where $\gamma_{MM^{\prime\prime}A^{\prime}\hat{M}B^{\prime}}$ is an ideal
bipartite private state. The state $\omega_{MM^{\prime\prime}A^{\prime}\hat
{M}B^{\prime}E}$ in \eqref{eq-PC:final-omega-purification-tri-to-bi}\ is a
purification of $\omega_{MM^{\prime\prime}A^{\prime}\hat{M}B^{\prime}}$, and
by Uhlmann's theorem, there exists a purification $\gamma_{MM^{\prime\prime
}A^{\prime}\hat{M}B^{\prime}E}$\ of $\gamma_{MM^{\prime\prime}A^{\prime}%
\hat{M}B^{\prime}}$ such that%
\begin{equation}
F(\gamma_{MM^{\prime\prime}A^{\prime}\hat{M}B^{\prime}},\omega_{MM^{\prime
\prime}A^{\prime}\hat{M}B^{\prime}})=F(\gamma_{MM^{\prime\prime}A^{\prime}%
\hat{M}B^{\prime}E},\omega_{MM^{\prime\prime}A^{\prime}\hat{M}B^{\prime}E}).
\end{equation}
Tracing over the systems $M^{\prime\prime}$, $A^{\prime}$, and $B^{\prime}$,
the following inequality holds%
\begin{equation}
1-F(\gamma_{M\hat{M}E},\omega_{M\hat{M}E})\leq\varepsilon.
\end{equation}
By the definition of an ideal private state (see Definition~\ref{def:private-state-bi}) and since
the state $\gamma_{MM^{\prime\prime}A^{\prime}\hat{M}B^{\prime}}$ is an ideal bipartite
private state, it follows that $\gamma_{M\hat{M}E}$ is an ideal tripartite key
state. Thus, we have proven the second claim.
\end{Proof}

\subsection{Upper Bounds on the Number of Transmitted Private Bits}

\label{sec-PC:upp-bnd-one-shot-priv-cap}We now establish some general upper
bounds on the number of private bits that can be communicated in an arbitrary
private communication protocol. The results are stated in
Proposition~\ref{prop-PC:one-shot-upper-bnd} and Theorems~\ref{thm-PC:REE-upp-bnd} and \ref{thm-PC:sq-ent-up-bnd-one-shot}, and, like the
upper bounds established in previous chapters, they hold independently of the
encoding and decoding channels used in the protocol and depends only on the
given communication channel $\mathcal{N}$. The first upper bound is in terms
of the one-shot private information of the channel, and the others are in terms of the
channel's $\varepsilon$-relative entropy of entanglement and squashed entanglement.

\subsubsection{Private Information Upper Bound}

\begin{proposition*}
{Upper Bound on One-Shot Private Capacity}{prop-PC:one-shot-upper-bnd}%
Let $\mathcal{N}_{A\rightarrow B}$ be a quantum channel. For every
$(\left\vert \mathcal{M}\right\vert ,\varepsilon)$ private communication
protocol over $\mathcal{N}$, with $\varepsilon\in\left[  0,1\right]  $, the
number of private bits transmitted over $\mathcal{N}$ is bounded from above by
the one-shot private information of $\mathcal{N}$:%
\begin{equation}
\log_{2}\!\left\vert \mathcal{M}\right\vert \leq\sup_{\{p(x),\rho_{A}%
^{x}\}_{x\in\mathcal{X}}}\left(  I_{H}^{\varepsilon}(X;B)_{\rho}-I_{\max
}^{\sqrt{\varepsilon}}(X;E)_{\rho}\right)  ,
\end{equation}
where the optimization is over every ensemble $\{p(x),\rho_{A}^{x}%
\}_{x\in\mathcal{X}}$ and the state $\rho_{XBE}$ is given by%
\begin{equation}
\rho_{XBE}\coloneqq \sum_{x\in\mathcal{X}}p(x)|x\rangle\!\langle x|_{X}\otimes
\mathcal{U}_{A\rightarrow BE}^{\mathcal{N}}(\rho_{A}^{x}),
\end{equation}
with $\mathcal{U}_{A\rightarrow BE}^{\mathcal{N}}$ an isometric channel
extending $\mathcal{N}_{A\rightarrow B}$. The hypothesis testing mutual
information $I_{H}^{\varepsilon}(X;B)_{\rho}$ is defined in \eqref{eq-hypo_testing_mutual_inf} and the
smooth max-mutual information $I_{\max}^{\sqrt{\varepsilon}}(X;E)_{\rho}$ in
\eqref{eq-SKD:smooth-max-MI-1}. Therefore,%
\begin{equation}
P^{\varepsilon}(\mathcal{N})\leq\sup_{\{p(x),\rho_{A}^{x}\}_{x\in\mathcal{X}}%
}\left(  I_{H}^{\varepsilon}(X;B)_{\rho}-I_{\max}^{\sqrt{\varepsilon}%
}(X;E)_{\rho}\right)  .
\end{equation}

\end{proposition*}

\begin{Proof}
The proof has some similarities with the proof of Lemma~\ref{lemma-SKD:approx-key-state-one-shot-private-info}. Since $p_{\text{err}}^{\ast}(\mathcal{E},\mathcal{D}%
;\mathcal{N})\geq\overline{p}_{\text{err}}(\mathcal{E},\mathcal{D}%
;p,\mathcal{N})$ for every probability distribution $p(m)$ over the messages,
it follows by definition that%
\begin{equation}
P^{\varepsilon}(\mathcal{N})\leq\sup_{(\mathcal{M},\mathcal{E},\mathcal{D}%
)}\left\{  \log_{2}\!\left\vert \mathcal{M}\right\vert :\overline{p}%
_{\text{err}}(\mathcal{E},\mathcal{D};p,\mathcal{N})\leq\varepsilon\right\}  ,
\end{equation}
with $p$ set to the uniform distribution over messages. So we bound the
right-hand side instead (note that it is equal to the one-shot secret-key transmission capacity). Let $(\mathcal{M},\mathcal{E}_{M^{\prime}\rightarrow
A},\mathcal{D}_{B\rightarrow\hat{M}})$ be an arbitrary private communication
protocol. By the reasoning in
\eqref{eq-PC:avg-infid-implies-reliability-1}--\eqref{eq-PC:avg-infid-implies-reliability-last},
it follows that%
\begin{equation}
\frac{1}{\left\vert \mathcal{M}\right\vert }\sum_{m\in\mathcal{M}%
}\operatorname{Tr}[\Lambda_{B}^{m}\mathcal{N}_{A\rightarrow B}(\rho_{A}%
^{m})]\geq1-\varepsilon.
\end{equation}
By the same reasoning given in the proof of Proposition~\ref{prop-cc:one-shot-bound-meta}, we conclude
that%
\begin{equation}
\log_{2}\!\left\vert \mathcal{M}\right\vert \leq I_{H}^{\varepsilon}(M;B)_{\tau
}, \label{eq-PC:reliability-bnd}%
\end{equation}
where the state $\tau_{MBE}$ is defined as%
\begin{equation}
\tau_{MBE}\coloneqq \frac{1}{\left\vert \mathcal{M}\right\vert }\sum_{m\in\mathcal{M}%
}|m\rangle\!\langle m|_{M}\otimes\mathcal{U}_{A\rightarrow BE}^{\mathcal{N}%
}(\rho_{A}^{m}). \label{eq-PC:tau-state-proof}%
\end{equation}
Observe that%
\begin{align}
\tau_{MBE}  &  =\frac{1}{\left\vert \mathcal{M}\right\vert }\sum
_{m\in\mathcal{M}}|m\rangle\!\langle m|_{M}\otimes(\mathcal{U}_{A\rightarrow
BE}^{\mathcal{N}}\circ\mathcal{E}_{M^{\prime}\rightarrow A})(|m\rangle
\!\langle m|_{M^{\prime}})\\
&  =(\mathcal{U}_{A\rightarrow BE}^{\mathcal{N}}\circ\mathcal{E}_{M^{\prime
}\rightarrow A})(\overline{\Phi}_{MM^{\prime}}).
\end{align}

From
\eqref{eq-PC:avg-infid-implies-security-1}--\eqref{eq-PC:avg-infid-implies-security-last},
we know that there exists a state $\sigma_{E}$ such that
\begin{align*}
\varepsilon &  \geq1-F(\pi_{M}\otimes\sigma_{E},(\widehat{\mathcal{N}%
}_{A\rightarrow E}\circ\mathcal{E}_{M^{\prime}\rightarrow A})(\overline{\Phi
}_{MM^{\prime}}))\\
&  =1-F(\tau_{M}\otimes\sigma_{E},\tau_{ME}),
\end{align*}
which, by applying the same reasoning in \eqref{eq-SKD:smooth-max-MI-less-than-zero}--\eqref{eq-SKD:intermed-proof-2}, allows us to conclude that%
\begin{equation}
I_{\max}^{\sqrt{\varepsilon}}(M;E)_{\tau}\leq0.
\label{eq-PC:smooth-max-MI-bnd}%
\end{equation}

Putting together \eqref{eq-PC:reliability-bnd} and
\eqref{eq-PC:smooth-max-MI-bnd} implies that
\begin{align}
\log_{2}\!\left\vert \mathcal{M}\right\vert  &  \leq I_{H}^{\varepsilon
}(M;B)_{\tau}-I_{\max}^{\sqrt{\varepsilon}}(M;E)_{\tau}\\
&  \leq\sup_{\{p(x),\rho_{A}^{x}\}_{x\in\mathcal{X}}}\left(  I_{H}%
^{\varepsilon}(X;B)_{\rho}-I_{\max}^{\sqrt{\varepsilon}}(X;E)_{\rho}\right)  ,
\end{align}
where the final inequality follows by noting that $\left\{  \frac
{1}{\left\vert \mathcal{M}\right\vert },\rho_{A}^{m}\right\}  _{m\in
\mathcal{M}}$ is a particular input ensemble and the one-shot private
information in the last line involves an optimization over all input ensembles.
\end{Proof}

As a consequence of the reasoning behind
Proposition~\ref{prop-PC:one-shot-upper-bnd}, along with \eqref{eq-mut_inf}, Proposition~\ref{prop-hypo_to_rel_ent}, and \eqref{eq-SKD:vN-key-bound-4},
we obtain the following:

\begin{corollary}{cor-PC:weak-conv-one-shot-bnd}
Let $\mathcal{N}_{A\rightarrow B}$ be a
quantum channel, and let $\varepsilon\in\lbrack0,1)$. For all $(\left\vert
\mathcal{M}\right\vert ,\varepsilon)$ private communication protocols for
$\mathcal{N}$, the following bound holds%
\begin{multline}
\left(  1-\varepsilon-\sqrt{\varepsilon}\right)  \log_{2}\!\left\vert
\mathcal{M}\right\vert \leq\label{eq-PC:weak-conv-one-shot-bnd}\\
\sup_{\left\{  p(x),\rho_{A}^{x}\right\}  _{x\in\mathcal{X}}}\left(
I(X;B)_{\rho}-I(X;E)_{\rho}\right)  +h_{2}(\varepsilon)+2g(\sqrt{\varepsilon
}).
\end{multline}
Consequently, the following bound holds for the one-shot private capacity of a
channel $\mathcal{N}$:%
\begin{multline}
\left(  1-\varepsilon-\sqrt{\varepsilon}\right)  P^{\varepsilon}%
(\mathcal{N})\leq\\
\sup_{\left\{  p(x),\rho_{A}^{x}\right\}  _{x\in\mathcal{X}}}\left(
I(X;B)_{\rho}-I(X;E)_{\rho}\right)  +h_{2}(\varepsilon)+2g(\sqrt{\varepsilon
}).
\end{multline}

\end{corollary}

\begin{Proof}
Employing the same reasoning that led to \eqref{eq-PC:reliability-bnd} and
\eqref{eq-PC:smooth-max-MI-bnd}, consider that the following bounds hold for a
given $(\left\vert \mathcal{M}\right\vert ,\varepsilon)$ private communication
protocol:%
\begin{align}
\log_{2}\!\left\vert \mathcal{M}\right\vert  &  \leq I_{H}^{\varepsilon
}(M;B)_{\tau},\\
I_{\max}^{\sqrt{\varepsilon}}(M;E)_{\tau}  &  \leq0,
\end{align}
where the state $\tau_{MBE}$ is defined in \eqref{eq-PC:tau-state-proof}. Now
we apply \eqref{eq-mut_inf} and Proposition~\ref{prop-hypo_to_rel_ent} to conclude that%
\begin{equation}
I_{H}^{\varepsilon}(M;B)_{\tau}\leq\frac{1}{1-\varepsilon}\left(
I(M;B)_{\tau}+h_{2}(\varepsilon)\right)  ,
\end{equation}
which implies that%
\begin{equation}
\left(  1-\varepsilon\right)  \log_{2}\!\left\vert \mathcal{M}\right\vert \leq
I(M;B)_{\tau}+h_{2}(\varepsilon), \label{eq-PC:reliability-bnd-to-vN}%
\end{equation}
and the same reasoning that led to \eqref{eq-SKD:vN-key-bound-4} to conclude that%
\begin{equation}
I_{\max}^{\sqrt{\varepsilon}}(M;E)_{\tau}\geq I(M;E)_{\tau}-\sqrt{\varepsilon
}\log_{2}\!\left\vert \mathcal{M}\right\vert -2g_{2}(\sqrt{\varepsilon}).
\label{eq-PC:security-bnd-to-vN}%
\end{equation}
Combining \eqref{eq-PC:reliability-bnd-to-vN} and
\eqref{eq-PC:security-bnd-to-vN}, we conclude that%
\begin{align}
&  \left(  1-\varepsilon-\sqrt{\varepsilon}\right)  \log_{2}\!\left\vert
\mathcal{M}\right\vert \nonumber\\
&  \leq I(M;B)_{\tau}-I(M;E)_{\tau}+h_{2}(\varepsilon)+2g(\sqrt{\varepsilon
})\\
&  \leq\sup_{\left\{  p(x),\rho_{A}^{x}\right\}  _{x\in\mathcal{X}}}\left(
I(X;B)_{\rho}-I(X;E)_{\rho}\right)  +h_{2}(\varepsilon)+2g(\sqrt{\varepsilon
}),
\end{align}
where the last inequality follows by optimizing over all input ensembles.
\end{Proof}

\subsubsection{Relative Entropy of Entanglement Upper Bound}

We now consider an upper bound based on the channel's relative entropy of
entanglement. In order to do so, we exploit the equivalence between secret-key
transmission and bipartite private-state transmission established in
Section~\ref{sec-PC:bi-to-tri-key}. We also make use of Proposition~\ref{prop:core-meta-converse-privacy}, which gives an upper
bound on the number $\log_{2}\!\left\vert \mathcal{M}\right\vert $ of private
bits contained in the final state of a bipartite private-state transmission protocol.

In the previous chapter on secret key distillation, our approach to obtaining
upper bounds on distillable key consisted of 1)\ establishing a connection
between a tripartite key distillation protocol and a bipartite private state
distillation protocol (see Section~\ref{sec-SKD:tri-bi-equivalence}) and 2)\ comparing the state at the
output of a bipartite private-state distillation protocol with one that is
useless for this task. We considered the set of separable states as the
useless set, and we proved that certain state entanglement measures are upper
bounds on distillable key in the one-shot and asymptotic settings.

In Section~\ref{sec-PC:bi-to-tri-key}, we established a similar correspondence between
secret-key transmission, as defined in Section~\ref{sec-PC:one-shot-setting},
and bipartite private-state transmission.\ Here, we observe that bipartite
private-state transmission is similar to bipartite private-state distillation
in the sense that, like private-state distillation, the error criterion for
private-state transmission involves comparing the output state to an ideal
private state (see Definition~\ref{def-PC:private-state-trans-prot}). This suggests that the state
entanglement measures defined in Section~\ref{sec-ent_measures_sep_distance} are relevant (in
particular the result of Proposition~\ref{prop:core-meta-converse-privacy}). However, as
was the case for quantum communication in Chapter~\ref{chap-quantum_capacity}, the main resource
that we are considering in this chapter is a quantum channel and not a quantum
state, and so we have an extra degree of freedom in the input state to the
channel, which we can optimize. This suggests that the channel entanglement
measures from Chapter~\ref{chap-ent_measures_chan} are relevant, and it is indeed what we find.

\begin{theorem*}
{Relative Entropy of Entanglement Upper Bound on One-Shot Private
Capacity}{thm-PC:REE-upp-bnd}
Let $\mathcal{N}_{A\rightarrow B}$ be a
quantum channel, and let $\varepsilon\in\lbrack0,1)$. For all $(\left\vert
\mathcal{M}\right\vert ,\varepsilon)$ private communication protocols for
$\mathcal{N}$, the following bound holds%
\begin{equation}
\log_{2}\!\left\vert \mathcal{M}\right\vert \leq E_{R}^{\varepsilon}%
(\mathcal{N}),
\end{equation}
where $E_{R}^{\varepsilon}(\mathcal{N})$ is the $\varepsilon$-relative entropy
of entanglement of $\mathcal{N}$, defined in \eqref{eq-EM:eps-REE-channel} as%
\begin{equation}
E_{R}^{\varepsilon}(\mathcal{N})\coloneqq \sup_{\psi_{SA}}\inf_{\sigma_{SB}%
\in\operatorname{SEP}(S:B)}D_{H}^{\varepsilon}(\mathcal{N}_{A\rightarrow
B}(\psi_{SA})\Vert\sigma_{SB}).
\end{equation}
Consequently, we have the following bound on the one-shot private capacity:%
\begin{equation}
P^{\varepsilon}(\mathcal{N})\leq E_{R}^{\varepsilon}(\mathcal{N}).
\end{equation}

\end{theorem*}

\begin{Proof}
By definition, it follows that the one-shot secret-key transmission capacity
is an upper bound on $P^{\varepsilon}(\mathcal{N})$. Applying
Theorem~\ref{thm-PC:sec-trans-bi-priv-trans}, we conclude that the one-shot
bipartite private-state transmission capacity is equal to the one-shot
secret-key transmission capacity. So let us bound the one-shot bipartite
private-state transmission capacity. Consider an arbitrary $(M,\varepsilon)$
bipartite private-state transmission protocol. The final state of such a
protocol satisfies the condition in
Definition~\ref{def-PC:private-state-trans-prot}, which means that there is
an ideal private state $\gamma_{M^{\prime\prime}MA^{\prime}\hat{M}B^{\prime}}$
such that%
\begin{equation}
1-F(\gamma_{M^{\prime\prime}MA^{\prime}\hat{M}B^{\prime}},\omega
_{M^{\prime\prime}MA^{\prime}\hat{M}B^{\prime}})\leq\varepsilon.
\end{equation}
As such, Proposition~\ref{prop:core-meta-converse-privacy} applies, and we conclude that%
\begin{align}
\log_{2}\!\left\vert \mathcal{M}\right\vert  &  \leq E_{R}^{\varepsilon
}(M^{\prime\prime}MA^{\prime};\hat{M}B^{\prime})_{\omega}\\
&  \leq E_{R}^{\varepsilon}(M^{\prime\prime}MA^{\prime};B)_{\rho}\\
&  \leq E_{R}^{\varepsilon}(\mathcal{N}).
\end{align}
The second inequality follows from the data-processing inequality for
$D_{H}^{\varepsilon}$ under the action of the isometric decoding channel
$\mathcal{U}_{B\rightarrow\hat{M}B^{\prime}}^{\mathcal{D}}$ and where the
state $\rho_{M^{\prime\prime}MA^{\prime}B}$ is defined as%
\begin{equation}
\rho_{M^{\prime\prime}MA^{\prime}B}\coloneqq (\mathcal{N}_{A\rightarrow B}%
\circ\mathcal{U}_{M^{\prime}\rightarrow AA^{\prime}}^{\mathcal{E}}%
)(\Phi_{M^{\prime\prime}MM^{\prime}}).
\end{equation}
The systems $M^{\prime\prime}MA^{\prime}$ extend the system $A$ of the state
$\mathcal{U}_{M^{\prime}\rightarrow AA^{\prime}}^{\mathcal{E}}(\Phi
_{M^{\prime\prime}MM^{\prime}})$, with $A$ being the input to the channel
$\mathcal{N}_{A\rightarrow B}$. As such, we can optimize over all such input
states, and then conclude the final inequality above (here, we need to apply
the remark after Definition~\ref{def:LAQC-ent-channel} as well).
\end{Proof}

We then have the following bound as a direct application of Proposition~\ref{prop:sandwich-to-htre}:

\begin{corollary}{}
Let $\mathcal{N}_{A\rightarrow B}$ be a quantum channel, and let
$\varepsilon\in\lbrack0,1)$. For all $(\left\vert \mathcal{M}\right\vert
,\varepsilon)$ private communication protocols for $\mathcal{N}$, the
following bound holds for all $\alpha > 1$:
\begin{equation}
\log_{2}\!\left\vert \mathcal{M}\right\vert \leq\widetilde{E}_{\alpha
}(\mathcal{N})+\frac{\alpha}{\alpha-1}\log_{2}\!\left(  \frac{1}{1-\varepsilon
}\right)  , \label{eq-PC:renyi-REE-one-shot-bnd}%
\end{equation}
where $\widetilde{E}_{\alpha}(\mathcal{N})$ is the sandwiched Renyi relative
entropy of entanglement of $\mathcal{N}$, defined in \eqref{eq-EM:sandwiched-REE-channel} as%
\begin{equation}
\widetilde{E}_{\alpha}(\mathcal{N})\coloneqq \sup_{\psi_{SA}}\inf_{\sigma_{SB}%
\in\operatorname{SEP}(S:B)}\widetilde{D}_{\alpha}(\mathcal{N}_{A\rightarrow
B}(\psi_{SA})\Vert\sigma_{SB}).
\end{equation}

\end{corollary}

\subsubsection{Squashed Entanglement Upper Bound}

\label{eq-PC:squashed-ent-one-shot-bnd}

We now turn to squashed entanglement and establish it as an upper bound on
one-shot private capacity. The reasoning behind this result is very similar to that
given in the proof of Proposition~\ref{thm-PC:REE-upp-bnd}, except that we
employ Proposition~\ref{thm-SKD:SKD-bipartite-bound} instead:

\begin{theorem*}
{Squashed Entanglement Upper Bound on One-Shot Private Capacity}
{thm-PC:sq-ent-up-bnd-one-shot}Let $\mathcal{N}_{A\rightarrow B}$ be a
quantum channel, and let $\varepsilon\in\lbrack0,1)$. For all $(\left\vert
\mathcal{M}\right\vert ,\varepsilon)$ private communication protocols for
$\mathcal{N}$, the following bound holds%
\begin{equation}
\left(  1-2\sqrt{\varepsilon}\right)  \log_{2}\!\left\vert \mathcal{M}%
\right\vert \leq E_{\text{sq}}(\mathcal{N})+2g_{2}(\sqrt{\varepsilon}),
\end{equation}
where $E_{\text{sq}}(\mathcal{N})$ is the squashed entanglement of the channel
$\mathcal{N}$, given in Definition~\ref{def-sq_ent_channel} as%
\begin{equation}
E_{\text{sq}}(\mathcal{N})\coloneqq \sup_{\psi_{SA}}E_{\text{sq}}(S;B)_{\tau},
\end{equation}
and $\tau_{SB}\coloneqq \mathcal{N}_{A\rightarrow B}(\psi_{SA})$. Consequently, we
have the following bound on the one-shot private capacity:%
\begin{equation}
\left(  1-2\sqrt{\varepsilon}\right)  P^{\varepsilon}(\mathcal{N})\leq
E_{\text{sq}}(\mathcal{N})+2g_{2}(\sqrt{\varepsilon}).
\end{equation}

\end{theorem*}

\begin{Proof}
As indicated above, the argument is precisely the same as in the proof of
Proposition~\ref{thm-PC:REE-upp-bnd}, except that we apply the following bound
from Proposition~\ref{thm-SKD:SKD-bipartite-bound} instead:%
\begin{equation}
\left(  1-2\sqrt{\varepsilon}\right)  \log_{2}\!\left\vert \mathcal{M}%
\right\vert \leq E_{\text{sq}}(M^{\prime\prime}MA^{\prime};\hat{M}B^{\prime
})_{\omega}+2g_{2}(\sqrt{\varepsilon}).
\end{equation}
After this step, we apply the data-processing inequality for $E_{\text{sq}}$
and optimize over channel input states.
\end{Proof}

\subsection{Lower Bound on the Number of Transmitted Private Bits via
Position-Based Coding and Convex Splitting}

\label{sec-PC:lower-bnd-priv-cap-1-shot}

Having derived upper bounds on the
number of private bits that can be transmitted in an arbitrary private
communication protocol, let us now determine a lower bound. Here we use the
methods of position-based coding and convex splitting to derive an explicit
$(\left\vert \mathcal{M}\right\vert ,\varepsilon)$ protocol for all
$\varepsilon\in(0,1)$.

To derive this lower bound, let us consider a slightly different model of
communication, in which there is a one-input, two-output classical--quantum
channel connecting the sender Alice to the legitimate receiver Bob and the
eavesdropper Eve:%
\begin{equation}
x\rightarrow\rho_{BE}^{x}, \label{eq-PC:cq-wiretap}%
\end{equation}
where $x\in\mathcal{X}$ is the classical input symbol and $\rho_{BE}^{x}$ is
the bipartite quantum state that appears at the output when $x$ is input. Bob
has access to the system~$B$ of the output and Eve to $E$.\ The channel can
alternatively be written as a quantum channel as follows:%
\begin{equation}
\mathcal{N}_{X\rightarrow BE}(\omega)\coloneqq \sum_{x\in\mathcal{X}}\langle
x|_{X}\omega|x\rangle_{X}\rho_{BE}^{x},
\end{equation}
where $\{|x\rangle_{X}\}_{x\in\mathcal{X}}$ is an orthonormal basis. In this
way, a private communication protocol for $\mathcal{N}_{X\rightarrow BE}$ is
defined exactly as we did in Section~\ref{sec-PC:one-shot-setting}, with $\mathcal{N}_{X\rightarrow BE}$
replacing the isometric channel $\mathcal{U}_{A\rightarrow BE}$ therein.
Furthermore, the notions of code infidelity, an $(\left\vert \mathcal{M}%
\right\vert ,\varepsilon)$ private communication protocol, and one-shot
private capacity are defined in the same way, but with $\mathcal{N}%
_{X\rightarrow BE}$ replacing the isometric channel $\mathcal{U}_{A\rightarrow
BE}$.

The main result of this section is the following lower bound on the one-shot
private capacity $P^{\varepsilon}(\mathcal{N})$ of a classical--quantum
wiretap channel $\mathcal{N}_{X\rightarrow BE}$:%
\begin{multline}
P^{\varepsilon}(\mathcal{N})\geq\overline{I}_{H}^{\varepsilon^{\prime}%
-\delta-\eta}(X;B)_{\rho}-\overline{I}_{\max}^{\delta-\zeta}(E;X)_{\rho
}\label{eq-PC:1-shot-private-bnd}\\
-\log_{2}\!\left(  \frac{8\left(  \varepsilon^{\prime}-\delta\right)  }{\eta
^{2}}\right)  -\log_{2}\!\left(  \frac{2}{\zeta^{2}}\right)  ,
\end{multline}
where $\varepsilon^{\prime}=1-\sqrt{1-\varepsilon/2}$, $\delta\in
(0,\varepsilon^{\prime})$, $\eta\in(0,\varepsilon^{\prime}-\delta)$, and
$\zeta\in(0,\delta)$, and the information measures are evaluated with respect
to the state%
\begin{equation}
\rho_{XBE}\coloneqq \sum_{x\in\mathcal{X}}p(x)|x\rangle\!\langle x|_{X}\otimes
\rho_{BE}^{x}.
\label{eq-PC:state-rho-cqq-wiretap}
\end{equation}
In the above, $P^{\varepsilon}(\mathcal{N})$ represents the maximum number of
bits that can be sent from Alice to Bob, using a classical--quantum wiretap channel once, such
that the infidelity does not exceed $\varepsilon\in(0,1)$. The quantities on
the right-hand side of the inequality in \eqref{eq-PC:1-shot-private-bnd}\ are
particular one-shot generalizations of the Holevo information to Bob and Eve,
which are defined in \eqref{eq-hypo_testing_mutual_inf} and~\eqref{eq-SKD:alt-smooth-max-MI-def}, respectively.

To prove the one-shot bound in \eqref{eq-PC:1-shot-private-bnd}, we
employ\ position-based coding (Section~\ref{subsec-pos_coding})\ and convex splitting
(Section~\ref{sec-SKD:low-bound-1-shot-dist-key}). The main idea of position-based coding is conceptually simple
and we review it briefly here. To communicate a classical
message from Alice to Bob, we allow them to share a quantum state $\rho
_{RA}^{\otimes M}$ before communication begins, where $M$ is the number of
messages, Bob possesses the $R$ systems, and Alice the $A$ systems. If Alice
wishes to communicate message $m$, then she sends the $m$th $A$ system through
the channel. The reduced state of Bob's systems is then%
\begin{equation}
\rho_{R_{1}}\otimes\cdots\otimes\rho_{R_{m-1}}\otimes\rho_{R_{m}B}\otimes
\rho_{R_{m+1}}\otimes\cdots\otimes\rho_{R_{M}},
\label{eq-PC:position-based-decoding}%
\end{equation}
where $\rho_{R_{m}B}=\mathcal{N}_{A_{m}\rightarrow B}(\rho_{R_{m}A_{m}})$ and
$\mathcal{N}_{A_{m}\rightarrow B}$ is the quantum channel. For all $m^{\prime
}\neq m$, the reduced state for systems $R_{m^{\prime}}$ and $B$ is the
product state $\rho_{R_{m^{\prime}}}\otimes\rho_{B}$. However, the reduced
state of systems $R_{m}B$ is the (generally)\ correlated state $\rho_{R_{m}B}%
$. So if Bob has a binary measurement that can distinguish the joint state
$\rho_{RB}$ from the product state $\rho_{R}\otimes\rho_{B}$ sufficiently
well, he can base a decoding strategy off of this, and the scheme is reliable
as long as the number of bits $\log_{2}M$ to be communicated is chosen to be
roughly equal to the hypothesis testing
mutual information. This is exactly what is used in position-based coding,
thus forging a transparent and intuitive link between quantum hypothesis
testing and communication for the case of entanglement-assisted communication.

Convex splitting is rather intuitive as well and can be thought of as dual to
the coding scenario mentioned above. Suppose instead that Alice and Bob have a
means of generating the state in \eqref{eq-PC:position-based-decoding},
perhaps by the strategy mentioned above. But now suppose that Alice chooses
the variable $m$ uniformly at random, so that the state, from the perspective
of someone ignorant of the choice of $m$, is the following mixture:%
\begin{equation}
\frac{1}{M}\sum_{m=1}^{M}\rho_{R_{1}}\otimes\cdots\otimes\rho_{R_{m-1}}%
\otimes\rho_{R_{m}B}\otimes\rho_{R_{m+1}}\otimes\cdots\otimes\rho_{R_{M}}.
\end{equation}
The convex-split lemma (Lemma~\ref{lem-SKD:smooth-convex-split}) guarantees that as long as $\log_{2}M$ is roughly equal
to the smooth max-mutual information in \eqref{eq-SKD:alt-smooth-max-MI-def}, then the state above is nearly
indistinguishable from the product state $\rho_{R}^{\otimes M}\otimes\rho_{B}$.

Here we use the approaches of position-based coding and convex splitting in
conjunction to construct codes for the classical--quantum wiretap channel. The
main underlying idea is to have a message variable $m\in\{1,\ldots,M\}$ and a
local randomness variable $r\in\{1,\ldots,R\}$, the latter of which is
selected uniformly at random and used to confuse the eavesdropper Eve. Before
communication begins, Alice, Bob, and Eve are allowed share to $MR$ copies of
the common randomness state%
\begin{equation}
\rho_{XX^{\prime}X^{\prime\prime}}\coloneqq \sum_{x\in\mathcal{X}}p_{X}(x)|xxx\rangle\!\langle xxx|_{XX^{\prime}X^{\prime\prime}}.
\end{equation}
We can think of the $MR$ copies of $\rho_{XX^{\prime}X^{\prime\prime}}$ as
being partitioned into $M$ blocks, each of which contain $R$ copies of the
state $\rho_{XX^{\prime}X^{\prime\prime}}$. If Alice wishes to send message
$m$, then she picks $r$ uniformly at random and sends the  $X_{A}$
system labeled by $(m,r)$ through the classical--quantum wiretap channel in \eqref{eq-PC:cq-wiretap}. As long as
$\log_{2}MR$ is roughly equal to the hypothesis testing mutual information
$\overline{I}_{H}^{\varepsilon}(X;B)$, then Bob can use a position-based
decoder to figure out both $m$ and $r$. As long as $\log_{2}R$ is roughly
equal to the smooth max-mutual information $\overline{I}_{\max}^{\varepsilon
}(E;X)$, then the convex-split lemma guarantees that the overall state of
Eve's systems, regardless of which message $m$ was chosen, is nearly
indistinguishable from the product state $\rho_{X_{E}}^{\otimes MR}\otimes
\rho_{E}$. Thus, in such a scheme, Bob can figure out $m$ while Eve cannot
figure out anything about~$m$. This is the intuition behind the coding scheme
and gives a sense of why $\log_{2}M=\log_{2}MR-\log_{2}R\approx\overline
{I}_{H}^{\varepsilon}(X;B)-\overline{I}_{\max}^{\varepsilon}(E;X)$ is an
achievable number of bits that can be sent privately from Alice to Bob. The
main purpose of this section is to develop the details of this argument and
furthermore to show how the scheme can be derandomized, so that the $MR$ copies
of the common randomness state $\rho_{XX^{\prime}X^{\prime\prime}}$ are in
fact not necessary.

There are strong connections between the approach for establishing a lower bound on one-shot distillable key detailed in Section~\ref{sec-SKD:low-bound-1-shot-dist-key}, and the approach we have outlined above and detail below. In fact, there is a point in the analysis below at which it becomes precisely the same, and at that point, we simply invoke the proof of Theorem~\ref{thm-SKD:one-shot-key-lower-bnd} to complete the analysis.

We now state the main theorem of this section:

\begin{theorem}{thm-PC:lower-bnd-one-shot-priv-cap}
Let $\mathcal{N}_{X\rightarrow BE}:x\rightarrow\rho_{BE}^{x}$ be a
classical--quantum wiretap channel, in which Alice has access to the
input, Bob to the output system $B$, and Eve to the output system $E$. For all
$\varepsilon\in(0,1]$, $\varepsilon^{\prime}=1-\sqrt{1-\varepsilon/2}$,
$\delta\in(0,\varepsilon^{\prime})$, $\eta\in(0,\varepsilon^{\prime}-\delta)$,
and $\zeta\in(0,\delta)$, there exists an $(\left\vert \mathcal{M}\right\vert
,\varepsilon)$ private communication protocol for $\mathcal{N}_{X\rightarrow
BE}$, such that%
\begin{multline}
\log_{2}\!\left\vert \mathcal{M}\right\vert =\overline{I}_{H}^{\varepsilon
^{\prime}-\delta-\eta}(X;B)_{\rho}-\overline{I}_{\max}^{\delta-\zeta
}(E;X)_{\rho}\nonumber\\
-\log_{2}\!\left(  \frac{8\left(  \varepsilon^{\prime}-\delta\right)  }{\eta
^{2}}\right)  -\log_{2}\!\left(  \frac{2}{\zeta^{2}}\right)  .
\end{multline}
where the hypothesis testing mutual information $\overline{I}_{H}%
^{\varepsilon^{\prime}-\delta-\eta}(X;B)_{\rho}$ is defined in \eqref{eq-hypo_testing_mutual_inf} and the
smooth max-mutual information $\overline{I}_{\max}^{\delta-\zeta}(E;X)_{\rho}$
in \eqref{eq-SKD:alt-smooth-max-MI-def}, and they are evaluated with respect to the state $\rho_{XBE}$ in~\eqref{eq-PC:state-rho-cqq-wiretap}.
\end{theorem}

\begin{Proof}
We first exhibit a public shared randomness assisted protocol for private
communication and then show later how to derandomize it. The protocol proceeds
exactly as discussed above. We suppose that Alice, Bob, and Eve share the
state $\rho_{XX^{\prime}X^{\prime\prime}}^{\otimes MR}$ before communication
begins, where $M=\left\vert \mathcal{M}\right\vert $. If Alice wants to send
the message $m$, she picks $r$ uniformly at random from $\left\{
1,\ldots,R\right\}  $ and transmits a classical copy $X^{\prime\prime\prime}$
of the $X$ system labeled by $(m,r)$ through the channel $\mathcal{N}%
_{X\rightarrow BE}$. The resulting state of Alice, Bob, and Eve, for fixed $m$
and $r$, is then as follows:%
\begin{multline}
\rho_{X^{MR}X^{\prime MR}X^{\prime\prime MR}BE}^{m,r}\coloneqq \rho_{X_{1,1}%
X_{1,1}^{\prime}X_{1,1}^{\prime\prime}}\otimes\cdots\otimes\rho_{X_{m,r-1}%
X_{m,r-1}^{\prime}X_{m,r-1}^{\prime\prime}}\otimes\\
\rho_{X_{m,r}X_{m,r}^{\prime}X_{m,r}^{\prime\prime}BE}\otimes\rho
_{X_{m,r+1}X_{m,r+1}^{\prime}X_{m,r+1}^{\prime\prime}}\otimes\cdots\otimes
\rho_{X_{M,R}X_{M,R}^{\prime}X_{M,R}^{\prime\prime}},
\end{multline}
where%
\begin{align}
\rho_{X_{1,1}X_{1,1}^{\prime}X_{1,1}^{\prime\prime}}  &  =\cdots
=\rho_{X_{m,r-1}X_{m,r-1}^{\prime}X_{m,r-1}^{\prime\prime}}\\
&  =\rho_{X_{m,r+1}X_{m,r+1}^{\prime}X_{m,r+1}^{\prime\prime}}=\cdots
=\rho_{X_{M,R}X_{M,R}^{\prime}X_{M,R}^{\prime\prime}}\\
&  =\sum_{x\in\mathcal{X}}p_{X}(x)|xxx\rangle\!\langle xxx|_{XX^{\prime
}X^{\prime\prime}},
\end{align}
and%
\begin{align}
\rho_{X_{m,r}X_{m,r}^{\prime}X_{m,r}^{\prime\prime}BE}  &  =\sum
_{x\in\mathcal{X}}p_{X}(x)|xxx\rangle\!\langle xxx|_{XX^{\prime}%
X^{\prime\prime}}\otimes\mathcal{N}_{X^{\prime\prime\prime}\rightarrow
BE}(|x\rangle\!\langle x|_{X^{\prime\prime\prime}})\\
&  =\sum_{x\in\mathcal{X}}p_{X}(x)|xxx\rangle\!\langle xxx|_{XX^{\prime
}X^{\prime\prime}}\otimes\rho_{BE}^{x}.
\end{align}

At this point, the state here is precisely the same as that given in \eqref{eq-SKD:global-state-key-dist-prot},
and the goal from here is the same as well. Thus, we can apply the same
reasoning given there to conclude that the following infidelity condition
holds%
\begin{equation}
1-F(\mathcal{M}_{X^{\prime MR}B\rightarrow M_{B}}(\rho_{M_{A}X^{\prime
MR}X^{\prime\prime MR}BE}),\overline{\Phi}_{M_{A}M_{B}}\otimes\rho
_{X^{\prime\prime MR}}\otimes\widetilde{\rho}_{E})\leq\varepsilon
\label{eq-PC:infid-condition-priv-comm}%
\end{equation}
if%
\begin{multline}
\log_{2}\!\left\vert \mathcal{M}\right\vert =\overline{I}_{H}^{\varepsilon
^{\prime}-\delta-\eta}(X;B)_{\tau}-\overline{I}_{\max}^{\delta-\zeta
}(E;X)_{\tau}\label{eq-PC:private-comm-shared-rand}\\
-\log_{2}\!\left(  \frac{4\left(  \varepsilon^{\prime}-\delta\right)  }{\eta
^{2}}\right)  -\log_{2}\!\left(  \frac{2}{\zeta^{2}}\right)  ,
\end{multline}
where
\begin{equation}
\tau_{XBE} \coloneqq \sum_{x \in \mathcal{X}} p(x) |x\rangle\!\langle x|_X \otimes \rho^x_{BE}
\end{equation}
and
$\rho_{M_{A}X^{\prime MR}X^{\prime\prime MR}BE}$ is the reduction of the
following state:%
\begin{multline}
\rho_{M_{A}R_{A}X^{MR}X^{\prime MR}X^{\prime\prime MR}BE}\coloneqq \\
\frac{1}{MR}\sum_{m=1}^{M}\sum_{r=1}^{R}|m\rangle\!\langle m|_{M_{A}}%
\otimes|r\rangle\!\langle r|_{R_{A}}\otimes\rho_{X^{MR}X^{\prime MR}%
X^{\prime\prime MR}BE}^{m,r}.
\end{multline}
That is,%
\begin{equation}
\rho_{M_{A}X^{\prime MR}X^{\prime\prime MR}BE}=\operatorname{Tr}_{R_{A}X^{MR}%
}[\rho_{M_{A}R_{A}X^{\prime MR}X^{\prime\prime MR}BE}].
\end{equation}
Furthermore, $\mathcal{M}_{X^{\prime MR}B\rightarrow M_{B}}$ is a measurement
channel of the form in \eqref{eq-SKD:reduced-meas-ch}, and $\widetilde{\rho}_{E}$ is a state satisfying%
\begin{equation}
P(\rho_{E},\widetilde{\rho}_{E})\leq\delta-\zeta.
\end{equation}
Thus, Bob's strategy is to decode both $m$ and $r$ (as before), and he can do
so as long as $\log_{2}MR=\overline{I}_{H}^{\varepsilon^{\prime}-\delta-\eta
}(X;B)_{\tau}-\log_{2}\!\left(  \frac{4\left(  \varepsilon^{\prime}%
-\delta\right)  }{\eta^{2}}\right)  $. At the same time, the message $m$
should be private from Eve, and this is possible as long as $\log
_{2}R=\overline{I}_{\max}^{\delta-\zeta}(E;X)_{\tau}+\log_{2}\!\left(  \frac
{2}{\zeta^{2}}\right)  $. Calculating $\log_{2}M=\log_{2}MR-\log_{2}R$ gives
\eqref{eq-PC:private-comm-shared-rand}. Then the analysis in the proof of Theorem~\ref{thm-SKD:one-shot-key-lower-bnd} guarantees
that the condition in \eqref{eq-PC:infid-condition-priv-comm} holds.

We now discuss how to derandomize the protocol. First, let us define the
following measurement channels%
\begin{equation}
\mathcal{M}_{B\rightarrow M_{B}}^{x_{1,1},\ldots,x_{M,R}}(\omega
_{B})\coloneqq \mathcal{M}_{X^{\prime MR}B\rightarrow M_{B}}(|x_{1,1},\ldots
,x_{M,R}\rangle\!\langle x_{1,1},\ldots,x_{M,R}|_{X^{\prime MR}}\otimes
\omega_{B}),
\end{equation}
where $\omega_{B}$ is an input state. Also, consider that the state
$\rho_{M_{A}X^{\prime MR}X^{\prime\prime MR}BE}$ can be written as%
\begin{multline}
\rho_{M_{A}X^{\prime MR}X^{\prime\prime MR}BE}=\frac{1}{M}\sum_{m=1}%
^{M}|m\rangle\!\langle m|_{M_{A}}\otimes\\
\sum_{x_{1,1},\ldots,x_{M,R}}p(x_{1,1})\cdots p(x_{M,R})|x_{1,1}%
,\ldots,x_{M,R}\rangle\!\langle x_{1,1},\ldots,x_{M,R}|_{X^{\prime MR}}%
\otimes\\
|x_{1,1},\ldots,x_{M,R}\rangle\!\langle x_{1,1},\ldots,x_{M,R}|_{X^{\prime
\prime MR}}\otimes\frac{1}{R}\sum_{r=1}^{R}\rho_{BE}^{x_{m,r}}.
\end{multline}
With this in mind, the state $\mathcal{M}_{X^{\prime MR}B\rightarrow M_{B}%
}(\rho_{M_{A}X^{\prime MR}X^{\prime\prime MR}BE})$ can be written as follows:%
\begin{multline}
\mathcal{M}_{X^{\prime MR}B\rightarrow M_{B}}(\rho_{M_{A}X^{\prime
MR}X^{\prime\prime MR}BE})=\\
\frac{1}{M}\sum_{m=1}^{M}\sum_{x_{1,1},\ldots,x_{M,R}}p(x_{1,1})\cdots
p(x_{M,R})|m\rangle\!\langle m|_{M_{A}}\otimes\\
|x_{1,1},\ldots,x_{M,R}\rangle\!\langle x_{1,1},\ldots,x_{M,R}|_{X^{\prime
\prime MR}}\otimes\mathcal{M}_{B\rightarrow M_{B}}^{x_{1,1},\ldots,x_{M,R}%
}\left(  \frac{1}{R}\sum_{r=1}^{R}\rho_{BE}^{x_{m,r}}\right)  ,
\end{multline}
and the condition in \eqref{eq-PC:infid-condition-priv-comm} as%
\begin{align}
&  1-\varepsilon\nonumber\\
&  \leq F(\mathcal{M}_{X^{\prime MR}B\rightarrow M_{B}}(\rho_{M_{A}X^{\prime
MR}X^{\prime\prime MR}BE}),\overline{\Phi}_{M_{A}M_{B}}\otimes\rho
_{X^{\prime\prime MR}}\otimes\widetilde{\rho}_{E})\\
&  =\left[
\begin{array}
[c]{c}%
\frac{1}{M}\sum_{m=1}^{M}\sum_{x_{1,1},\ldots,x_{M,R}}p(x_{1,1})\cdots
p(x_{M,R})\times\\
\sqrt{F}\left(  \mathcal{M}_{B\rightarrow M_{B}}^{x_{1,1},\ldots,x_{M,R}%
}\left(  \frac{1}{R}\sum_{r=1}^{R}\rho_{BE}^{x_{m,r}}\right)  ,|m\rangle\!\langle m|_{M_{B}}\otimes\widetilde{\rho}_{E}\right)
\end{array}
\right]  ^{2},
\end{align}
which is the same as%
\begin{multline}
\frac{1}{M}\sum_{m=1}^{M}\sum_{x_{1,1},\ldots,x_{M,R}}p(x_{1,1})\cdots
p(x_{M,R})\times\\
\sqrt{F}\left(  \mathcal{M}_{B\rightarrow M_{B}}^{x_{1,1},\ldots,x_{M,R}%
}\left(  \frac{1}{R}\sum_{r=1}^{R}\rho_{BE}^{x_{m,r}}\right)  ,|m\rangle\!\langle m|_{M_{B}}\otimes\widetilde{\rho}_{E}\right) \\
\geq\sqrt{1-\varepsilon}.
\end{multline}
We can now exploit the \textquotedblleft Shannon trick\textquotedblright\ of
exchanging the sum over the messages $m$ and the sum over the codewords to
rewrite this inequality as%
\begin{multline}
\sum_{x_{1,1},\ldots,x_{M,R}}p(x_{1,1})\cdots p(x_{M,R})\times\\
\left(  \frac{1}{M}\sum_{m=1}^{M}\sqrt{F}\left(  \mathcal{M}_{B\rightarrow
M_{B}}^{x_{1,1},\ldots,x_{M,R}}\left(  \frac{1}{R}\sum_{r=1}^{R}\rho
_{BE}^{x_{m,r}}\right)  ,|m\rangle\!\langle m|_{M_{B}}\otimes\widetilde{\rho
}_{E}\right)  \right) \\
\geq\sqrt{1-\varepsilon}.
\end{multline}
Since the average does not exceed the maximum, we conclude that there
exists some choice of codewords $x_{1,1},\ldots,x_{M,R}$ such that the
following inequality holds%
\begin{equation}
\frac{1}{M}\sum_{m=1}^{M}\sqrt{F}\left(  \mathcal{M}_{B\rightarrow M_{B}%
}^{x_{1,1},\ldots,x_{M,R}}\left(  \frac{1}{R}\sum_{r=1}^{R}\rho_{BE}^{x_{m,r}%
}\right)  ,|m\rangle\!\langle m|_{M_{B}}\otimes\widetilde{\rho}_{E}\right)
\geq\sqrt{1-\varepsilon}.
\end{equation}
Let us then use the shorthand $\mathcal{M}_{B\rightarrow M_{B}}\equiv
\mathcal{M}_{B\rightarrow M_{B}}^{x_{1,1},\ldots,x_{M,R}}$, so that we can
rewrite the above as%
\begin{equation}
\frac{1}{M}\sum_{m=1}^{M}\sqrt{F}\left(  \mathcal{M}_{B\rightarrow M_{B}}\!\left(  \frac{1}{R}\sum_{r=1}^{R}\rho_{BE}^{x_{m,r}}\right)  ,|m\rangle\!\langle m|_{M_{B}}\otimes\widetilde{\rho}_{E}\right)  \geq\sqrt{1-\varepsilon
}.
\end{equation}
This completes the \textit{derandomization} part of the proof.

Finally, we are interested in a code that satisfies the maximal infidelity
criterion $p_{\text{err}}^{\ast}(\mathcal{E},\mathcal{D};\mathcal{N}%
)\leq\varepsilon$. To find such a code, we can apply \textit{expurgation} to
the code found above. Since the square root function is concave, after
bringing the average inside the square root and squaring both sides of the
inequality, we conclude that%
\begin{equation}
\frac{1}{M}\sum_{m=1}^{M}F\!\left(  \mathcal{M}_{B\rightarrow M_{B}}\!\left(
\frac{1}{R}\sum_{r=1}^{R}\rho_{BE}^{x_{m,r}}\right)  ,|m\rangle\!\langle
m|_{M_{B}}\otimes\widetilde{\rho}_{E}\right)  \geq1-\varepsilon,
\end{equation}
which we can rewrite one more time as%
\begin{equation}
\frac{1}{M}\sum_{m=1}^{M}\left(  1-F\!\left(  \mathcal{M}_{B\rightarrow M_{B}}\!\left(  \frac{1}{R}\sum_{r=1}^{R}\rho_{BE}^{x_{m,r}}\right)  ,|m\rangle\!\langle m|_{M_{B}}\otimes\widetilde{\rho}_{E}\right)  \right)  \leq
\varepsilon.
\end{equation}
Now applying Markov's inequality, we conclude that at least half of the
messages are such that the following inequality holds%
\begin{equation}
1-F\!\left(  \mathcal{M}_{B\rightarrow M_{B}}\!\left(  \frac{1}{R}\sum_{r=1}%
^{R}\rho_{BE}^{x_{m,r}}\right)  ,|m\rangle\!\langle m|_{M_{B}}\otimes
\widetilde{\rho}_{E}\right)  \leq2\varepsilon.
\end{equation}
Thus, these messages and the corresponding codewords are retained as the final
code. To be clear, suppose without loss of generality, that messages $1$,
\ldots, $\left\lfloor M/2\right\rfloor $ are retained and messages
$\left\lfloor M/2\right\rfloor +1$, \ldots, $M$ are expurgated. Then this
means that the corresponding codewords retained are $x_{1,1}$, \ldots,
$x_{1,R}$, $x_{2,1}$, \ldots, $x_{2,R}$, \ldots, $x_{\left\lfloor
M/2\right\rfloor ,1}$, \ldots, $x_{\left\lfloor M/2\right\rfloor ,R}$, and the
ones discarded are $x_{\left\lfloor M/2\right\rfloor +1,1}$, \ldots,
$x_{\left\lfloor M/2\right\rfloor +1,R}$, $x_{\left\lfloor M/2\right\rfloor
+2,1}$, \ldots, $x_{\left\lfloor M/2\right\rfloor +2,R}$, \ldots, $x_{M,1}$,
\ldots, $x_{M,R}$. After the expurgation, the rate of the code is given by%
\begin{align}
\log_{2}\!\left\vert \mathcal{M}\right\vert /2  &  =\log_{2}\!\left\vert
\mathcal{M}\right\vert -\log_{2}(2)\\
&  =\overline{I}_{H}^{\varepsilon^{\prime}-\delta-\eta}(X;B)_{\rho}%
-\overline{I}_{\max}^{\delta-\zeta}(E;X)_{\rho}\nonumber\\
&  \qquad-\log_{2}\!\left(  \frac{4\left(  \varepsilon^{\prime}-\delta\right)
}{\eta^{2}}\right)  -\log_{2}\!\left(  \frac{2}{\zeta^{2}}\right)  -\log
_{2}(2)\\
&  =\overline{I}_{H}^{\varepsilon^{\prime}-\delta-\eta}(X;B)_{\rho}%
-\overline{I}_{\max}^{\delta-\zeta}(E;X)_{\rho}\nonumber\\
&  \qquad-\log_{2}\!\left(  \frac{8\left(  \varepsilon^{\prime}-\delta\right)
}{\eta^{2}}\right)  -\log_{2}\!\left(  \frac{2}{\zeta^{2}}\right)  .
\end{align}
By a final substitution of $2\varepsilon$ with $\varepsilon$ and rewriting, we
arrive at the claim of the theorem.
\end{Proof}

We can induce a classical--quantum wiretap channel from an isometric channel
$\mathcal{U}_{A\rightarrow BE}^{\mathcal{N}}$ extending $\mathcal{N}%
_{A\rightarrow B}$ by the following pre-processing:%
\begin{equation}
x\rightarrow\rho_{A}^{x}\rightarrow\mathcal{U}_{A\rightarrow BE}^{\mathcal{N}%
}(\rho_{A}^{x}).
\end{equation}
That is, based on the value of a letter $x$, Alice inputs the state $\rho
_{A}^{x}$ into the isometric channel $\mathcal{U}_{A\rightarrow BE}%
^{\mathcal{N}}$. Optimizing over all such preprocessings and applying
Theorem~\ref{thm-PC:lower-bnd-one-shot-priv-cap}, we arrive at the following lower bound on the one-shot private capacity of a quantum channel $\mathcal{N}_{A\rightarrow B}$ (according to the definition given in Section~\ref{sec-PC:one-shot-setting}):

\begin{corollary}{cor-PC:one-shot-priv-comm-ach}Let $\mathcal{N}_{A\rightarrow B}$ be a
quantum channel that is extended by the isometric channel $\mathcal{U}%
_{A\rightarrow BE}^{\mathcal{N}}$. For all $\varepsilon\in(0,1]$,
$\varepsilon^{\prime}=1-\sqrt{1-\varepsilon/2}$, $\delta\in(0,\varepsilon
^{\prime})$, $\eta\in(0,\varepsilon^{\prime}-\delta)$, and $\zeta\in
(0,\delta)$, there exists an $(\left\vert \mathcal{M}\right\vert
,\varepsilon)$ private communication protocol for $\mathcal{N}_{A\rightarrow
B}$, such that%
\begin{multline}
\log_{2}\!\left\vert \mathcal{M}\right\vert =\sup_{\left\{  p(x),\rho_{A}%
^{x}\right\}  _{x\in\mathcal{X}}}\overline{I}_{H}^{\varepsilon^{\prime}%
-\delta-\eta}(X;B)_{\rho}-\overline{I}_{\max}^{\delta-\zeta}(E;X)_{\rho
}\nonumber\\
-\log_{2}\!\left(  \frac{8\left(  \varepsilon^{\prime}-\delta\right)  }{\eta
^{2}}\right)  -\log_{2}\!\left(  \frac{2}{\zeta^{2}}\right)  .
\end{multline}
where the hypothesis testing mutual information $\overline{I}_{H}%
^{\varepsilon^{\prime}-\delta-\eta}(X;B)_{\rho}$ is defined in 
\eqref{eq-hypo_testing_mutual_inf} 
 and the
smooth max-mutual information $\overline{I}_{\max}^{\delta-\zeta}(E;X)_{\rho}$
in  \eqref{eq-SKD:alt-smooth-max-MI-def}, and the information quantities are evaluated with respect to the
following state:%
\begin{equation}
\rho_{XBE}\coloneqq \sum_{x\in\mathcal{X}}p(x)|x\rangle\!\langle x|_{X}\otimes
\mathcal{U}_{A\rightarrow BE}^{\mathcal{N}}(\rho_{A}^{x}).
\end{equation}

\end{corollary}

Now applying Propositions~\ref{prop:ineq-hypo-renyi} and \ref{prop-smooth_max_to_petz_renyi}, we conclude the following bound:

\begin{corollary}{cor-PC:one-shot-to-renyi-ach}Let $\mathcal{N}_{A\rightarrow B}$ be a
quantum channel that is extended by the isometric channel $\mathcal{U}%
_{A\rightarrow BE}^{\mathcal{N}}$. For all $\varepsilon\in(0,1]$,
$\varepsilon^{\prime}=1-\sqrt{1-\varepsilon/2}$, $\delta\in(0,\varepsilon
^{\prime})$, $\eta\in(0,\varepsilon^{\prime}-\delta)$, $\zeta\in(0,\delta)$,
$\nu\in(0,\delta-\zeta)$, $\alpha\in(0,1)$, and $\beta>1$, there exists an
$(\left\vert \mathcal{M}\right\vert ,\varepsilon)$ private communication
protocol for $\mathcal{N}_{A\rightarrow B}$, such that%
\begin{equation}
\log_{2}\!\left\vert \mathcal{M}\right\vert \geq\sup_{\left\{  p(x),\rho_{A}%
^{x}\right\}  _{x\in\mathcal{X}}}\left(  \overline{I}_{\alpha}(X;B)_{\rho
}-\widetilde{I}_{\beta}^{\prime}(X;E)_{\rho}\right)  -f(\varepsilon^{\prime
},\delta,\eta,\nu,\zeta,\alpha,\beta).
\label{eq-PC:rate-ach-one-shot-renyi-priv-comm}%
\end{equation}
where the Petz--Renyi mutual information $\overline{I}_{\alpha}(X;B)_{\rho}$
is defined in \eqref{eq-petz_renyi_mut_inf_noopt}, the sandwiched Renyi mutual information $\widetilde
{I}_{\beta}^{\prime}(X;E)_{\rho}$ as%
\begin{equation}
\widetilde{I}_{\beta}^{\prime}(X;E)_{\rho}\coloneqq \widetilde{D}_{\beta}(\rho
_{XE}\Vert\rho_{X}\otimes\rho_{E}),
\end{equation}
and the information quantities are evaluated with respect to the following
state:%
\begin{equation}
\rho_{XBE}\coloneqq \sum_{x\in\mathcal{X}}p(x)|x\rangle\!\langle x|_{X}\otimes
\mathcal{U}_{A\rightarrow BE}^{\mathcal{N}}(\rho_{A}^{x}).
\end{equation}
Furthermore,%
\begin{multline}
f(\varepsilon^{\prime},\delta,\eta,\nu,\zeta,\alpha,\beta)\coloneqq \frac{\alpha
}{1-\alpha}\log_{2}\!\left(  \frac{1}{\varepsilon^{\prime}-\delta-\eta}\right)
+\log_{2}\!\left(  \frac{8}{\nu^{2}}\right)
\label{eq-PC:f-func-priv-lower-renyi}\\
+\frac{1}{\beta-1}\log_{2}\!\left(  \frac{1}{\left(  \delta-\zeta-\nu\right)
^{2}}\right)  +\log_{2}\!\left(  \frac{1}{1-\left(  \delta-\zeta-\nu\right)
^{2}}\right) \\
+\log_{2}\!\left(  \frac{8\left(  \varepsilon^{\prime}-\delta\right)  }{\eta
^{2}}\right)  +\log_{2}\!\left(  \frac{2}{\zeta^{2}}\right)  .
\end{multline}

\end{corollary}

\begin{Proof}
The reasoning here is precisely the same as that given in the proof of
Corollary~\ref{cor-SKD:secret-key-rates-R\'enyi}. The only difference is that we optimize over every ensemble
$\left\{  p(x),\rho_{A}^{x}\right\}  _{x\in\mathcal{X}}$.
\end{Proof}

\section{Private Capacity of a Quantum Channel}

We now consider the asymptotic setting. In this scenario, depicted in Figure~[REF],
the classical message system $M^{\prime}$ to be transmitted to Bob is encoded
into $n$~copies $A_{1}$, \ldots, $A_{n}$ of a quantum system $A$, for
$n\in\mathbb{N}$. Each of the systems is then sent through an independent use
of the isometric channel $\mathcal{U}_{A\rightarrow BE}^{\mathcal{N}}$, which
extends the point-to-point channel $\mathcal{N}_{A\rightarrow B}$.
As before, this is the asymptotic setting because $n$ can be arbitrarily large.

Due to the fact that $n$ independent uses of the channel $\mathcal{U}%
_{A\rightarrow BE}^{\mathcal{N}}$ is no different from one use of the
tensor-power channel $(\mathcal{U}_{A\rightarrow BE}^{\mathcal{N}})^{\otimes
n}$, the information theory underlying the asymptotic setting is no different
from that in the one-shot setting, and the main task we accomplish here is to
analyze performance of the protocols in the large $n$ limit. Indeed, the only
change that we make here is to replace $\mathcal{U}_{A\rightarrow
BE}^{\mathcal{N}}$ with $(\mathcal{U}_{A\rightarrow BE}^{\mathcal{N}%
})^{\otimes n}$ and define the encoding and decoding channels as acting on $n$
systems instead of one. If Alice transmits message $m$, then the final state
of the protocol is%
\begin{equation}
(\mathcal{D}_{B^{n}\rightarrow\hat{M}}\circ(\mathcal{U}_{A\rightarrow
BE}^{\mathcal{N}})^{\otimes n}\circ\mathcal{E}_{M^{\prime}\rightarrow A^{n}%
})(|m\rangle\!\langle m|_{M^{\prime}}),
\end{equation}
where $\mathcal{E}_{M^{\prime}\rightarrow A^{n}}$ is the encoding channel and
$\mathcal{D}_{B^{n}\rightarrow\hat{M}}$ the decoding channel. Just as in the
one-shot setting, we define the maximal infidelity of the code as%
\begin{multline}
p_{\text{err}}^{\ast}(\mathcal{E},\mathcal{D};\mathcal{N}^{\otimes n})=\\
\inf_{\sigma_{E^{n}}}\max_{m\in\mathcal{M}}(1-F(|m\rangle\!\langle m|_{\hat
{M}}\otimes\sigma_{E^{n}},(\mathcal{D}_{B^{n}\rightarrow\hat{M}}%
\circ(\mathcal{U}_{A\rightarrow BE}^{\mathcal{N}})^{\otimes n}\circ
\mathcal{E}_{M^{\prime}\rightarrow A^{n}})(|m\rangle\!\langle m|_{M^{\prime}%
}))),
\end{multline}
where the infimum is with respect to every state $\sigma_{E^{n}}$ of the
eavesdropper's system~$E$.

\begin{definition}
{$(n,\left\vert \mathcal{M}\right\vert ,\varepsilon)$ Private Communication
Protocol}{}Let $(\mathcal{M},\mathcal{E}_{M^{\prime}\rightarrow A^{n}%
},\mathcal{D}_{B^{n}\rightarrow\hat{M}})$ be the elements of a private
communication protocol for $n$ independent uses of the channel $\mathcal{N}%
_{A\rightarrow B}$, where $d_{M^{\prime}}=d_{\hat{M}}=\left\vert
\mathcal{M}\right\vert $. The protocol is called an $(n,\left\vert
\mathcal{M}\right\vert ,\varepsilon)$ protocol, with $\varepsilon\in\left[
0,1\right]  $, if $p_{\text{err}}^{\ast}(\mathcal{E},\mathcal{D}%
;\mathcal{N}^{\otimes n})\leq\varepsilon$.
\end{definition}

The rate of an $(n,\left\vert \mathcal{M}\right\vert ,\varepsilon)$ private
communication protocol is defined as the number of private bits transmitted
per channel use, i.e.,%
\begin{equation}
R(n,\left\vert \mathcal{M}\right\vert )\coloneqq \frac{\log_{2}\!\left\vert
\mathcal{M}\right\vert }{n}.
\end{equation}
The rate depends only on the size $\left\vert \mathcal{M}\right\vert $ of the
message set $\mathcal{M}$ and on the number of channel uses. In particular, it
does not depend on the communication channel nor on the encoding and decoding
channels. For a given $\varepsilon\in\left[  0,1\right]  $ and $n\in
\mathbb{N}$, the highest rate among all $(n,\left\vert \mathcal{M}\right\vert
,\varepsilon)$ protocols is denoted by $P^{n,\varepsilon}(\mathcal{N})$, and
it is defined as%
\begin{equation}
P^{n,\varepsilon}(\mathcal{N})\coloneqq \frac{1}{n}P^{\varepsilon}(\mathcal{N}%
^{\otimes n})=\sup_{\left(  \mathcal{M},\mathcal{E},\mathcal{D}\right)
}\left\{  \frac{\log_{2}\!\left\vert \mathcal{M}\right\vert }{n}:p_{\text{err}%
}^{\ast}(\mathcal{E},\mathcal{D};\mathcal{N}^{\otimes n})\leq\varepsilon
\right\}  ,
\end{equation}
where, in the second equality, we use the definition of the one-shot private
capacity $P^{\varepsilon}$ given in \eqref{eq-PC:1-shot-priv-cap-def}, and the
supremum is over every message set~$\mathcal{M}$, encoding channel~$\mathcal{E}$ with input dimension $\left\vert \mathcal{M}\right\vert $, and
decoding channel~$\mathcal{D}$ with output dimension $\left\vert
\mathcal{M}\right\vert $.

We now provide several definitions related to private capacity and its associated concepts.

\begin{definition}
{Achievable Rate for Private Communication}{} Given a quantum channel
$\mathcal{N}$, a rate $R\in\mathbb{R}^{+}$ is called an achievable rate for
private communication over $\mathcal{N}$ if for all $\varepsilon\in(0,1]$,
$\delta>0$, and sufficiently large $n$, there exists an $(n,2^{n\left(
R-\delta\right)  },\varepsilon)$ private communication protocol for
$\mathcal{N}$.
\end{definition}

\begin{definition}
{Private Capacity of a Quantum Channel}{} The private capacity of a quantum
channel $\mathcal{N}$, denoted by $P(\mathcal{N})$, is defined to be the
supremum of all achievable rates, i.e.,%
\begin{equation}
P(\mathcal{N})\coloneqq \sup\left\{  R:R\text{ is an achievable rate for }%
\mathcal{N}\right\}  .
\end{equation}

\end{definition}

An equivalent definition of private capacity is
\begin{equation}
P(\mathcal{N})=\inf_{\varepsilon\in(0,1]}\liminf_{n\rightarrow\infty}\frac
{1}{n}P^{\varepsilon}(\mathcal{N}^{\otimes n}),
\end{equation}
which we prove in Appendix~\ref{chap-str_conv}.

\begin{definition}
{Weak Converse Rate for Private Communication}{} Given a quantum channel
$\mathcal{N}$, a rate $R\in\mathbb{R}^{+}$ is called a weak converse rate for
private communication over $\mathcal{N}$ if every $R^{\prime}>R$ is not an
achievable rate for $\mathcal{N}$.
\end{definition}

\begin{definition}
{Strong Converse Rate for Private Communication}{} Given a quantum channel
$\mathcal{N}$, a rate $R\in\mathbb{R}^{+}$ is called a strong converse rate
for private communication over $\mathcal{N}$ if for all $\varepsilon\in
\lbrack0,1)$, $\delta>0$, and sufficiently large $n$, there does not exist an
$(n,2^{n\left(  R+\delta\right)  },\varepsilon)$ private communication
protocol for $\mathcal{N}$.
\end{definition}

\begin{definition}
{Strong Converse Private Capacity of a Quantum Channel}{} The strong converse
private capacity of a quantum channel $\mathcal{N}$, denoted by $\widetilde
{P}(\mathcal{N})$, is defined as the infimum of all strong converse rates,
i.e.,%
\begin{equation}
\widetilde{P}(\mathcal{N})\coloneqq \inf\left\{  R:R\text{ is a strong converse rate
for }\mathcal{N}\right\}  .
\end{equation}

\end{definition}

We can also write the strong converse private capacity as%
\begin{equation}
\widetilde{P}(\mathcal{N})=\sup_{\varepsilon\in\lbrack0,1)}\limsup
_{n\rightarrow\infty}\frac{1}{n}P^{\varepsilon}(\mathcal{N}^{\otimes n}).
\end{equation}
See Appendix~\ref{chap-str_conv} for a proof. We also show in Appendix~\ref{chap-str_conv} that%
\begin{equation}
P(\mathcal{N})\leq\widetilde{P}(\mathcal{N})
\end{equation}
for every quantum channel $\mathcal{N}$.

We now state one of the main theorems of this chapter, which gives an
experssion for the private capacity of a quantum channel.

\begin{theorem*}{Private Capacity}{thm-PC:private-cap}
The private capacity of a quantum
channel $\mathcal{N}_{A\rightarrow B}$ is equal to the regularized private
information $P_{\text{reg}}(\mathcal{N})$ of $\mathcal{N}$, i.e.,%
\begin{equation}
P(\mathcal{N})=I_{\text{reg}}^{p}(\mathcal{N})\coloneqq \lim_{n\rightarrow\infty}%
\frac{1}{n}I^{p}(\mathcal{N}^{\otimes n}), \label{eq-PC:reg-exp-priv-cap}%
\end{equation}
where the private information of a channel is defined as%
\begin{equation}
I^{p}(\mathcal{N})\coloneqq \sup_{\left\{  p(x),\rho_{A}^{x}\right\}  _{x\in
\mathcal{X}}}I(X;B)_{\rho}-I(X;E)_{\rho}, \label{eq-PC:priv-info-def}%
\end{equation}
and the information quantities are evaluated with respect to the state%
\begin{equation}
\rho_{XBE}\coloneqq \sum_{x\in\mathcal{X}}p(x)|x\rangle\!\langle x|_{X}\otimes
\mathcal{U}_{A\rightarrow BE}^{\mathcal{N}}(\rho_{A}^{x}),
\end{equation}
with $\mathcal{U}_{A\rightarrow BE}^{\mathcal{N}}$ an isometric channel
extending $\mathcal{N}_{A\rightarrow B}$.
\end{theorem*}

Observe that the expression in \eqref{eq-PC:reg-exp-priv-cap} for the private
capacity involves a regularization of the private information. Thus, in
general, it is difficult to compute because the optimization is over an
arbitrarily large number of channel uses.

By following an argument similar to that given in Section~\ref{sec-qcomm_additivity}, it follows
that the private information is always superadditive, meaning that
$I^{p}(\mathcal{N}^{\otimes n})\geq nI^{p}(\mathcal{N})$ for every
$n\in\mathbb{N}$ and channel $\mathcal{N}$. This means that the private
information is always a lower bound on the private capacity of a channel
$\mathcal{N}$:%
\begin{equation}
P(\mathcal{N})\geq I^{p}(\mathcal{N})\text{ for every channel }\mathcal{N}.
\end{equation}
If the private information happens to be additive for a particular channel,
then the regularization in \eqref{eq-PC:reg-exp-priv-cap} is not required. For
example, the private information is known to be additive for all degradable
and anti-degradable channels (see Definition~\ref{def-deg_antideg_chan}).  Furthermore, for degradable channels, the private information
is equal to the coherent information and so there is no difference between the
quantum capacity and the private capacity for these channels. That is, for
degradable channels, we have that%
\begin{equation}
P(\mathcal{N})=Q(\mathcal{N})=I^{c}(\mathcal{N}),
\end{equation}
where the coherent information $I^{c}(\mathcal{N})$ of a channel $\mathcal{N}$
is defined in \eqref{eq-coh_inf_chan}. For anti-degradable
channels, the private information is equal to zero, which is what we prove in
Section~\ref{sec-PC:anti-deg-zero-priv-cap}.

Theorem~\ref{thm-PC:private-cap} only makes a statement about the private
capacity and not about the strong converse private capacity. Later on, we
prove that a channel's relative entropy of entanglement is a strong converse
rate for private communication, and for some channels, it coincides with the
private information, thus leading to the strong converse property holding for
these channels. More generally, however, the best statement we can make is that
the regularized private information is a weak converse rate for all quantum channels.

There are two ingredients to the proof of Theorem~\ref{thm-PC:private-cap}:

\begin{enumerate}
\item \textit{Achievability}:\ We prove that $I_{\text{reg}}^{p}(\mathcal{N})$
is an achievable rate, which involves explicitly constructing a private
communication protocol. The developments in
Section~\ref{sec-PC:lower-bnd-priv-cap-1-shot} on a lower bound for one-shot
private capacity can be used, via the substitution $\mathcal{N}\rightarrow
\mathcal{N}^{\otimes n}$, to argue for the existence of a private
communication protocol for $\mathcal{N}$ in the asymptotic setting at the rate
$I_{\text{reg}}^{p}(\mathcal{N})$.\newline\newline The achievability part of
the proof establishes that $P(\mathcal{N})\geq I_{\text{reg}}^{p}%
(\mathcal{N})$.

\item \textit{Weak Converse}: We prove that $I_{\text{reg}}^{p}(\mathcal{N})$
is a weak converse rate, from which it follows that $P(\mathcal{N})\leq
I_{\text{reg}}^{p}(\mathcal{N})$. To show that $I_{\text{reg}}^{p}%
(\mathcal{N})$ is a weak converse rate, we use the upper bounds on one-shot
private capacity from Section~\ref{sec-PC:upp-bnd-one-shot-priv-cap} to
conclude that every achievable rate $R$ satisfies $R\leq I_{\text{reg}}%
^{p}(\mathcal{N})$.
\end{enumerate}

We first establish in Section~\ref{sec-PC:achievable-proof-asymp} that the
quantity $I_{\text{reg}}^{p}(\mathcal{N})$ is an achievable rate for private
communication over $\mathcal{N}$. Then, in
Section~\ref{sec-PC:weak-conv-proof-asymp}, we prove that $I_{\text{reg}}%
^{p}(\mathcal{N})$ is a weak converse rate.

Before proceeding, we establish a relationship between the private information
of a quantum channel and its coherent information, which mirrors the
operational relationship established in
Section~\ref{sec-PC:private-to-quantum-rel}. This relationship gives another
way to arrive at the conclusion that the private capacity of a quantum channel
is not smaller than its quantum capacity.

\begin{theorem}{thm-PC:generic-rel-priv-info-c-info}For a quantum channel
$\mathcal{N}_{A\rightarrow B}$, its private information is not smaller than
its coherent information:%
\begin{equation}
I^{c}(\mathcal{N})\leq I^{p}(\mathcal{N}),
\label{eq-PC:coh-info-priv-info-rel}%
\end{equation}
where the coherent information is defined in \eqref{eq-coh_inf_chan} and the private information
in \eqref{eq-PC:priv-info-def}. As a consequence, the private capacity is not
smaller than the quantum capacity:%
\begin{equation}
Q(\mathcal{N})\leq P(\mathcal{N}). \label{eq-PC:priv-cap-q-cap-rel}%
\end{equation}

\end{theorem}

\begin{Proof}
Picking a pure-state ensemble in \eqref{eq-PC:priv-info-def}, i.e.,
$\{p(x),\psi_{A}^{x}\}_{x\in\mathcal{X}}$, and setting%
\begin{equation}
\rho_{XBE}\coloneqq \sum_{x\in\mathcal{X}}p(x)|x\rangle\!\langle x|_{X}\otimes
\mathcal{U}_{A\rightarrow BE}^{\mathcal{N}}(\psi_{A}^{x}),
\end{equation}
with $\mathcal{U}_{A\rightarrow BE}^{\mathcal{N}}$ an isometric channel
extending $\mathcal{N}_{A\rightarrow B}$, we find that%
\begin{align}
I^{p}(\mathcal{N})  &  \geq I(X;B)_{\rho}-I(X;E)_{\rho}\\
&  =H(B)_{\rho}-H(B|X)_{\rho}-\left(  H(E)_{\rho}-H(E|X)_{\rho}\right) \\
&  =H(B)_{\rho}-H(E)_{\rho}.
\end{align}
The first equality follows from rewriting the mutual information, and the
second follows because the conditional entropies can be written as%
\begin{align}
H(B|X)_{\rho}  &  =\sum_{x\in\mathcal{X}}p(x)H(\operatorname{Tr}%
_{E}[\mathcal{U}_{A\rightarrow BE}^{\mathcal{N}}(\psi_{A}^{x})]),\\
H(E|X)_{\rho}  &  =\sum_{x\in\mathcal{X}}p(x)H(\operatorname{Tr}%
_{B}[\mathcal{U}_{A\rightarrow BE}^{\mathcal{N}}(\psi_{A}^{x})]).
\end{align}
They are equal because the entropies of the marginal states of a pure bipartite
state are equal. Now consider that the reduced state of the $BE$
systems is%
\begin{equation}
\rho_{BE}=\sum_{x\in\mathcal{X}}p(x)\mathcal{U}_{A\rightarrow BE}%
^{\mathcal{N}}(\psi_{A}^{x})=\mathcal{U}_{A\rightarrow BE}^{\mathcal{N}%
}\!\left(  \sum_{x\in\mathcal{X}}p(x)\psi_{A}^{x}\right)  .
\end{equation}
Since we can realize an arbitrary input density operator by taking convex
combinations of pure states, and by applying \eqref{eq-coh_inf_chan_alt}, we conclude the claim in \eqref{eq-PC:coh-info-priv-info-rel}.

The conclusion in \eqref{eq-PC:priv-cap-q-cap-rel} follows by applying \eqref{eq-PC:coh-info-priv-info-rel} and
Theorems~\ref{thm-qcomm_capacity} and \ref{thm-PC:private-cap}.
\end{Proof}

\subsection{Proof of Achievability}

\label{sec-PC:achievable-proof-asymp}In this section, we prove that
$I_{\text{reg}}^{p}(\mathcal{N})$ is an achievable rate for private
communication over a channel $\mathcal{N}$.

First, recall Corollary~\ref{cor-PC:one-shot-to-renyi-ach}. Applying it, we
find the following:

\begin{corollary*}
{Lower Bound for Private Communication in the Asymptotic Setting}
{cor-PC:one-shot-to-asymp-renyi}Let $\mathcal{N}_{A\rightarrow B}$ be a
quantum channel, and let $\mathcal{U}_{A\rightarrow BE}^{\mathcal{N}}$ be an
isometric channel extending it. For all $n\in\mathbb{N}$, $\varepsilon
\in(0,1)$, $\alpha\in\left(  0,1\right)  $, and $\beta>1$, there exists an
$(n,\left\vert \mathcal{M}\right\vert ,\varepsilon)$ private communication
protocol for $\mathcal{N}_{A\rightarrow B}$ such that the rate $\frac{\log
_{2}\left\vert \mathcal{M}\right\vert }{n}$ satisfies%
\begin{multline}
\frac{\log_{2}\!\left\vert \mathcal{M}\right\vert }{n}\geq\sup_{\left\{
p(x),\rho_{A}^{x}\right\}  _{x\in\mathcal{X}}}\overline{I}_{\alpha}%
(X;B)_{\rho}-\widetilde{I}_{\beta}^{\prime}(X;E)_{\rho}%
\label{eq-PC:lower-bnd-iid-asymp-renyi}\\
-\frac{1}{n}f\!\left(  \varepsilon^{\prime},\frac{\varepsilon^{\prime}}{2}%
,\frac{\varepsilon^{\prime}}{4},\frac{\varepsilon^{\prime}}{4},\frac
{\varepsilon^{\prime}}{2},\alpha,\beta\right)  ,
\end{multline}
where the information quantities are evaluated with respect to the state%
\begin{equation}
\rho_{XBE}\coloneqq \sum_{x\in\mathcal{X}}p(x)|x\rangle\!\langle x|_{X}\otimes
\mathcal{U}_{A\rightarrow BE}^{\mathcal{N}}(\rho_{A}^{x}),
\label{eq-PC:rho-state-ach-renyi}%
\end{equation}
and the function $f$ is defined in \eqref{eq-PC:f-func-priv-lower-renyi}.
\end{corollary*}

\begin{Proof}
We evaluate the quantities in Corollary~\ref{cor-PC:one-shot-to-renyi-ach} with respect to the tensor-power
isometric channel $(\mathcal{U}_{A\rightarrow BE}^{\mathcal{N}})^{\otimes n}$
and choose the input ensemble to be a tensor-power ensemble $\{p(x_{1})\cdots
p(x_{n}),\rho_{A_{1}}^{x_{1}}\otimes\cdots\otimes\rho_{A_{n}}^{x_{n}}%
\}_{x_{1},\ldots,x_{n}\in\mathcal{X}^{\times n}}$. This implies that the state
being evaluated for the Renyi information quantities is the tensor-power state
$\rho_{XBE}^{\otimes n}$, where $\rho_{XBE}$ is defined in
\eqref{eq-PC:rho-state-ach-renyi}. Let $\delta=\frac{\varepsilon^{\prime}}{2}%
$, $\eta=\frac{\varepsilon^{\prime}}{4}$, $\nu=\frac{\varepsilon^{\prime}}{4}%
$, and $\zeta=\frac{\varepsilon^{\prime}}{2}$. Exploiting the additivity of
$\overline{I}_{\alpha}$ and $\widetilde{I}_{\beta}^{\prime}$, substituting
into the inequality in \eqref{eq-PC:rate-ach-one-shot-renyi-priv-comm}, and
simplifying leads to the inequality%
\begin{multline}
\frac{\log_{2}\!\left\vert \mathcal{M}\right\vert }{n}\geq\sup_{\left\{
p(x),\rho_{A}^{x}\right\}  _{x\in\mathcal{X}}}\left(  \overline{I}_{\alpha
}(X;B)_{\rho}-\widetilde{I}_{\beta}^{\prime}(X;E)_{\rho}\right) \\
-\frac{1}{n}f\!\left(  \varepsilon^{\prime},\frac{\varepsilon^{\prime}}{2}%
,\frac{\varepsilon^{\prime}}{4},\frac{\varepsilon^{\prime}}{4},\frac
{\varepsilon^{\prime}}{2},\alpha,\beta\right)  .
\end{multline}
This concludes the proof.
\end{Proof}

Using the inequality in \eqref{eq-PC:lower-bnd-iid-asymp-renyi}, we conclude
the following lower bound on the private capacity of a quantum channel:

\begin{theorem*}
{Achievability of Private Information for Private Communication}
{thm-PC:ach-priv-info}The private information $I^{p}(\mathcal{N})$ of a
quantum channel $\mathcal{N}$, as defined in \eqref{eq-PC:priv-info-def}, is
an achievable rate for private communication over $\mathcal{N}$. In other
words,%
\begin{equation}
P(\mathcal{N})\geq I^{p}(\mathcal{N}),
\label{eq-PC:priv-cap-greater-than-priv-info}
\end{equation}
for every quantum channel $\mathcal{N}$.
\end{theorem*}

\begin{Proof}
Let $\mathcal{U}_{A\rightarrow BE}^{\mathcal{N}}$ be an isometric channel
extending the channel $\mathcal{N}_{A\rightarrow B}$ of interest. Fix
$\varepsilon\in(0,1]$ and $\delta>0$. Let $\delta_{1},\delta_{2}>0$ be such
that $\delta=\delta_{1}+\delta_{2}$. Set $\alpha\in(0,1)$ and $\beta>1$ such
that%
\begin{equation}
\delta_{1}\geq I(X;B)_{\rho}-I(X;E)_{\rho}-\left(  \overline{I}_{\alpha
}(X;B)_{\rho}-\widetilde{I}_{\beta}^{\prime}(X;E)_{\rho}\right)  ,
\label{eq-PC:delta-1-requirement-pf}%
\end{equation}
where the information quantities are evaluated with respect to the state
$\rho_{XBE}$ in \eqref{eq-PC:rho-state-ach-renyi}. Note that this is possible
because $\overline{I}_{\alpha}(X;B)_{\rho}$ increases monotonically with
increasing $\alpha\in(0,1)$ (see Proposition~\ref{prop-Petz_rel_ent}) and $\widetilde{I}_{\beta
}^{\prime}(X;E)_{\rho}$ decreases monotonically with decreasing $\beta$ (see
Proposition~\ref{prop-sand_rel_ent_properties}), so that%
\begin{align}
\lim_{\alpha\rightarrow1^{-}}\overline{I}_{\alpha}(X;B)_{\rho}  &
=\sup_{\alpha\in(0,1)}\overline{I}_{\alpha}(X;B)_{\rho},\\
\lim_{\beta\rightarrow1^{+}}\widetilde{I}_{\beta}^{\prime}(X;E)_{\rho}  &
=\inf_{\beta\in\left(  1,\infty\right)  }\widetilde{I}_{\beta}^{\prime
}(X;E)_{\rho}.
\end{align}
Also,%
\begin{align}
I(X;B)_{\rho}  &  =\lim_{\alpha\rightarrow1^{-}}\overline{I}_{\alpha
}(X;B)_{\rho},\\
I(X;E)_{\rho}  &  =\lim_{\beta\rightarrow1^{+}}\widetilde{I}_{\beta}^{\prime
}(X;E)_{\rho}.
\end{align}
With $\alpha$ and $\beta$ chosen such that
\eqref{eq-PC:delta-1-requirement-pf} holds, take $n$ large enough so that%
\begin{equation}
\delta_{2}\geq\frac{1}{n}f\!\left(  \varepsilon^{\prime},\frac{\varepsilon
^{\prime}}{2},\frac{\varepsilon^{\prime}}{4},\frac{\varepsilon^{\prime}}%
{4},\frac{\varepsilon^{\prime}}{2},\alpha,\beta\right)  .
\label{eq-PC:delta-2-requirement-pf}%
\end{equation}
Now, we use the fact that for the $n$ and $\varepsilon$ chosen above, there
exists an $(n,\left\vert \mathcal{M}\right\vert ,\varepsilon)$ protocol such
that%
\begin{equation}
\frac{\log_{2}\!\left\vert \mathcal{M}\right\vert }{n}\geq\overline{I}_{\alpha
}(X;B)_{\rho}-\widetilde{I}_{\beta}^{\prime}(X;E)_{\rho}-\frac{1}{n}f\!\left(
\varepsilon^{\prime},\frac{\varepsilon^{\prime}}{2},\frac{\varepsilon^{\prime
}}{4},\frac{\varepsilon^{\prime}}{4},\frac{\varepsilon^{\prime}}{2}%
,\alpha,\beta\right)  , \label{eq-PC:pf-protocol-existence-renyi}%
\end{equation}
which follows from Corollary~\ref{cor-PC:one-shot-to-asymp-renyi} above.
Rearranging the right-hand side of this inequality, and using
\eqref{eq-PC:delta-1-requirement-pf}, \eqref{eq-PC:delta-2-requirement-pf},
and \eqref{eq-PC:pf-protocol-existence-renyi}, we find that%
\begin{align}
\frac{\log_{2}\!\left\vert \mathcal{M}\right\vert }{n}  &  \geq I(X;B)_{\rho
}-I(X;E)_{\rho} \notag\\
& \qquad -\left(
\begin{array}
[c]{c}%
I(X;B)_{\rho}-I(X;E)_{\rho}-\left(  \overline{I}_{\alpha}(X;B)_{\rho
}-\widetilde{I}_{\beta}^{\prime}(X;E)_{\rho}\right) \\
+\frac{1}{n}f\!\left(  \varepsilon^{\prime},\frac{\varepsilon^{\prime}}{2}%
,\frac{\varepsilon^{\prime}}{4},\frac{\varepsilon^{\prime}}{4},\frac
{\varepsilon^{\prime}}{2},\alpha,\beta\right)
\end{array}
\right) \\
&  \geq I(X;B)_{\rho}-I(X;E)_{\rho}-\left(  \delta_{1}+\delta_{2}\right) \\
&  =I(X;B)_{\rho}-I(X;E)_{\rho}-\delta.
\end{align}
We thus have shown that there exists an $(n,\left\vert \mathcal{M}\right\vert
,\varepsilon)$ private communication protocol with rate $\frac{\log
_{2}\left\vert \mathcal{M}\right\vert }{n}\geq I(X;B)_{\rho}-I(X;E)_{\rho
}-\delta$. Therefore, there exists an $(n,2^{n\left(  R-\delta\right)
},\varepsilon)$ private communication protocol with $R=I(X;B)_{\rho
}-I(X;E)_{\rho}$ for all sufficiently large $n$ such that
\eqref{eq-PC:delta-2-requirement-pf} holds. Since $\varepsilon$ and $\delta$
are arbitrary, we conclude that for all $\varepsilon\in(0,1]$, $\delta>0$, and
sufficiently large $n$, there exists an $(n,2^{n\left(  R-\delta\right)
},\varepsilon)$ private communication protocol. This means that, by
definition, $I(X;B)_{\rho}-I(X;E)_{\rho}$ is an achievable rate. Since this is
true for all input ensembles, we can finally take a supremum over all input
ensembles to arrive at the conclusion in \eqref{eq-PC:priv-cap-greater-than-priv-info}.
\end{Proof}

\subsubsection{Proof of the Achievability Part of
Theorem~\ref{thm-PC:private-cap}}

Let $\{p(x),\rho_{A^{k}}^{x}\}_{x\in\mathcal{X}}$ be an arbitrary ensemble
over $k$ channel input systems, with $k\in\mathbb{N}$. Let%
\begin{equation}
\tau_{XB^{k}E^{k}}\coloneqq \sum_{x\in\mathcal{X}}p(x)|x\rangle\!\langle x|_{X}%
\otimes(\mathcal{U}_{A\rightarrow BE}^{\mathcal{N}})^{\otimes k}(\rho_{A^{k}%
}^{x}).
\end{equation}
Fix $\varepsilon\in(0,1]$ and $\delta>0$. Let $\delta_{1},\delta_{2}>0$ be
such that $\delta=\delta_{1}+\delta_{2}$. Set $\alpha\in(0,1)$ and $\beta>1$
such that%
\begin{equation}
\delta_{1}\geq\frac{1}{k}\left(  I(X;B^{k})_{\tau}-I(X;E^{k})_{\tau}\right)
-\frac{1}{k}\left(  \overline{I}_{\alpha}(X;B^{k})_{\tau}-\widetilde{I}%
_{\beta}^{\prime}(X;E^{k})_{\tau}\right)  ,
\label{eq-PC:delta-1-reg-priv-inf-ach}%
\end{equation}
which is possible based on the arguments given in the proof of
Theorem~\ref{thm-PC:ach-priv-info} above. Then, with this choice of $\alpha$
and $\beta$, take $n$ large enough so that%
\begin{equation}
\delta_{2}\geq\frac{1}{kn}f\!\left(  \varepsilon^{\prime},\frac{\varepsilon
^{\prime}}{2},\frac{\varepsilon^{\prime}}{4},\frac{\varepsilon^{\prime}}%
{4},\frac{\varepsilon^{\prime}}{2},\alpha,\beta\right)  .
\label{eq-PC:delta-2-reg-priv-info-ach}%
\end{equation}
Now, we use the fact that, for the chosen $n$ and $\varepsilon$, there exists
an $(n,\left\vert \mathcal{M}\right\vert ,\varepsilon)$ private communication
protocol such that \eqref{eq-PC:lower-bnd-iid-asymp-renyi} holds, i.e.,%
\begin{equation}
\frac{\log_{2}\!\left\vert \mathcal{M}\right\vert }{n}\geq\overline{I}_{\alpha
}(X;B^{k})_{\tau}-\widetilde{I}_{\beta}^{\prime}(X;E^{k})_{\tau}-\frac{1}%
{n}f\!\left(  \varepsilon^{\prime},\frac{\varepsilon^{\prime}}{2},\frac
{\varepsilon^{\prime}}{4},\frac{\varepsilon^{\prime}}{4},\frac{\varepsilon
^{\prime}}{2},\alpha,\beta\right)  .
\end{equation}
Dividing both sides by $k$ gives%
\begin{equation}
\frac{\log_{2}\!\left\vert \mathcal{M}\right\vert }{kn}\geq\frac{1}{k}\left(
\overline{I}_{\alpha}(X;B^{k})_{\tau}-\widetilde{I}_{\beta}^{\prime}%
(X;E^{k})_{\tau}\right)  -\frac{1}{kn}f\!\left(  \varepsilon^{\prime}%
,\frac{\varepsilon^{\prime}}{2},\frac{\varepsilon^{\prime}}{4},\frac
{\varepsilon^{\prime}}{4},\frac{\varepsilon^{\prime}}{2},\alpha,\beta\right)
. \label{eq-PC:lower-bnd-reg-priv-inf-ach}%
\end{equation}
Rearranging the right-hand side of this inequality, and using
\eqref{eq-PC:delta-1-reg-priv-inf-ach}--\eqref{eq-PC:lower-bnd-reg-priv-inf-ach},
we find that%
\begin{align}
\frac{\log_{2}\!\left\vert \mathcal{M}\right\vert }{kn}  &  \geq\frac{1}%
{k}\left(  I(X;B^{k})_{\tau}-I(X;E^{k})_{\tau}\right) \nonumber\\
&  \qquad -\left(  \frac{1}{k}\left(  I(X;B^{k})_{\tau}-I(X;E^{k})_{\tau}\right)
-\overline{I}_{\alpha}(X;B^{k})_{\tau}-\widetilde{I}_{\beta}^{\prime}%
(X;E^{k})_{\tau}\right) \nonumber\\
&  \qquad -\frac{1}{kn}f\!\left(  \varepsilon^{\prime},\frac{\varepsilon^{\prime}}%
{2},\frac{\varepsilon^{\prime}}{4},\frac{\varepsilon^{\prime}}{4}%
,\frac{\varepsilon^{\prime}}{2},\alpha,\beta\right) \\
&  \geq\frac{1}{k}\left(  I(X;B^{k})_{\tau}-I(X;E^{k})_{\tau}\right)  -\left(
\delta_{1}+\delta_{2}\right) \\
&  =\frac{1}{k}\left(  I(X;B^{k})_{\tau}-I(X;E^{k})_{\tau}\right)  -\delta.
\end{align}
Thus, there exists a $(kn,\left\vert \mathcal{M}\right\vert ,\varepsilon)$
private communication protocol with rate $\frac{\log_{2}\!\left\vert
\mathcal{M}\right\vert }{kn}\geq\frac{1}{k}\left(  I(X;B^{k})_{\tau}%
-I(X;E^{k})_{\tau}\right)  -\delta$. Therefore, letting $n^{\prime}\equiv kn$,
we conclude that there exists an $(n^{\prime},2^{n^{\prime}(R-\delta
)},\varepsilon)$ private communication protocol with%
\begin{equation}
R=\frac{1}{k}\left(  I(X;B^{k})_{\tau}-I(X;E^{k})_{\tau}\right)
\end{equation}
for all sufficiently large $n$ such that
\eqref{eq-PC:delta-2-reg-priv-info-ach}\ holds. Since $\varepsilon$ and
$\delta$ are arbitrary, we conclude that for all $\varepsilon\in(0,1]$,
$\delta>0$, and sufficiently large $n$, there exists an $(n,2^{n\left(
\frac{1}{k}\left(  I(X;B^{k})_{\tau}-I(X;E^{k})_{\tau}\right)  -\delta\right)
},\varepsilon)$ private communication protocol. This means that $\frac{1}%
{k}\left(  I(X;B^{k})_{\tau}-I(X;E^{k})_{\tau}\right)  $ is an achievable rate.

Now, since the input ensemble is arbitrary in the arguments above, we
conclude that%
\begin{equation}
\frac{1}{k}I^{p}(\mathcal{N}^{\otimes k})=\sup_{\{p(x),\rho_{A^{k}}%
^{x}\}_{x\in\mathcal{X}}}\frac{1}{k}\left(  I(X;B^{k})_{\tau}-I(X;E^{k}%
)_{\tau}\right)
\end{equation}
is an achievable rate. Finally, since the number $k$ of instances of the
channel $\mathcal{N}$ is arbitrary, we conclude that the regularized private information
$
\lim_{k\rightarrow\infty}\frac{1}{k}I^{p}(\mathcal{N}^{\otimes k})
$
is an achievable rate.

\subsection{Proof of the Weak Converse}

\label{sec-PC:weak-conv-proof-asymp}In order to prove the weak converse part
of Theorem~\ref{thm-PC:private-cap}, we make use of
Corollary~\ref{cor-PC:weak-conv-one-shot-bnd}, specifically
\eqref{eq-PC:weak-conv-one-shot-bnd}. Applying this inequality to the
tensor-power channel $\mathcal{N}_{A\rightarrow B}^{\otimes n}$ leads to the following:

\begin{proposition}{}
Let $\mathcal{N}_{A\rightarrow B}$ be a quantum channel, and let
$\mathcal{U}_{A\rightarrow BE}^{\mathcal{N}}$ be an isometric channel
extending it. Let $n\in\mathbb{N}$ and $\varepsilon\in\lbrack0,1)$. For an
$(n,\left\vert \mathcal{M}\right\vert ,\varepsilon)$ private communication
protocol for $\mathcal{N}_{A\rightarrow B}$, the rate $\frac{\log
_{2}\left\vert \mathcal{M}\right\vert }{n}$ satisfies%
\begin{multline}
\left(  1-\varepsilon-\sqrt{\varepsilon}\right)  \frac{\log_{2}\!\left\vert
\mathcal{M}\right\vert }{n}\leq\sup_{\left\{  p(x),\rho_{A^{n}}^{x}\right\}
_{x\in\mathcal{X}}}\frac{1}{n}\left(  I(X;B^{n})_{\rho}-I(X;E^{n})_{\rho
}\right) \label{eq-PC:weak-conv-n-shot-bnd}\\
+\frac{1}{n}\left(  h_{2}(\varepsilon)+2g(\sqrt{\varepsilon})\right)  ,
\end{multline}
where the information quantities are evaluated with respect to the state%
\begin{equation}
\rho_{XB^{n}E^{n}}\coloneqq \sum_{x\in\mathcal{X}}p(x)|x\rangle\!\langle x|_{X}%
\otimes(\mathcal{U}_{A\rightarrow BE}^{\mathcal{N}})^{\otimes n}(\rho_{A^{n}%
}^{x}),
\end{equation}
with $\mathcal{U}_{A\rightarrow BE}^{\mathcal{N}}$ an isometric channel
extending $\mathcal{N}_{A\rightarrow B}$. Consequently,%
\begin{multline}
\left(  1-\varepsilon-\sqrt{\varepsilon}\right)  P^{n,\varepsilon}%
(\mathcal{N})\leq\sup_{\left\{  p(x),\rho_{A^{n}}^{x}\right\}  _{x\in
\mathcal{X}}}\frac{1}{n}\left(  I(X;B^{n})_{\rho}-I(X;E^{n})_{\rho}\right) \\
+\frac{1}{n}\left(  h_{2}(\varepsilon)+2g(\sqrt{\varepsilon})\right)  .
\end{multline}

\end{proposition}

\subsubsection{Proof of the Weak Converse Part of
Theorem~\ref{thm-PC:private-cap}}

Suppose that $R$ is an achievable rate for private communication over the
channel $\mathcal{N}_{A\rightarrow B}$. Then, by definition, for all
$\varepsilon\in(0,1]$, $\delta>0$, and sufficiently large $n$, there exists an
$(n,2^{n(R-\delta)},\varepsilon)$ private communication protocol for
$\mathcal{N}_{A\rightarrow B}$. For all such protocols, the inequality in
\eqref{eq-PC:weak-conv-n-shot-bnd} holds, so that%
\begin{multline}
\left(  1-\varepsilon-\sqrt{\varepsilon}\right)  \left(  R-\delta\right)
\leq\sup_{\left\{  p(x),\rho_{A^{n}}^{x}\right\}  _{x\in\mathcal{X}}}\frac
{1}{n}\left(  I(X;B^{n})_{\rho}-I(X;E^{n})_{\rho}\right) \\
+\frac{1}{n}\left(  h_{2}(\varepsilon)+2g(\sqrt{\varepsilon})\right)  .
\end{multline}
Since the inequality holds for all sufficiently large $n$, it holds in the
limit $n\rightarrow\infty$, so that%
\begin{align}
\left(  1-\varepsilon-\sqrt{\varepsilon}\right)  \left(  R-\delta\right)   &
\leq\lim_{n\rightarrow\infty}\Bigg(\sup_{\left\{  p(x),\rho_{A^{n}}%
^{x}\right\}  _{x\in\mathcal{X}}}\frac{1}{n}\left(  I(X;B^{n})_{\rho
}-I(X;E^{n})_{\rho}\right) \nonumber\\
&  \qquad+\frac{1}{n}\left(  h_{2}(\varepsilon)+2g(\sqrt{\varepsilon})\right)
\Bigg)\\
&  =\lim_{n\rightarrow\infty}\sup_{\left\{  p(x),\rho_{A^{n}}^{x}\right\}
_{x\in\mathcal{X}}}\frac{1}{n}\left(  I(X;B^{n})_{\rho}-I(X;E^{n})_{\rho
}\right)  .
\end{align}
Then since this inequality holds for all $\varepsilon\in(0,1)$, $\delta>0$, it
holds in particular for $\varepsilon$ satisfying $\varepsilon+\sqrt
{\varepsilon}<1$, which gives%
\begin{equation}
R\leq\frac{1}{1-\varepsilon-\sqrt{\varepsilon}}\lim_{n\rightarrow\infty}%
\sup_{\left\{  p(x),\rho_{A^{n}}^{x}\right\}  _{x\in\mathcal{X}}}\frac{1}%
{n}\left(  I(X;B^{n})_{\rho}-I(X;E^{n})_{\rho}\right)  +\delta,
\end{equation}
and we thus conclude that%
\begin{align}
R  &  \leq\lim_{\varepsilon,\delta\rightarrow0}\frac{1}{1-\varepsilon
-\sqrt{\varepsilon}}\lim_{n\rightarrow\infty}\sup_{\left\{  p(x),\rho_{A^{n}%
}^{x}\right\}  _{x\in\mathcal{X}}}\frac{1}{n}\left(  I(X;B^{n})_{\rho
}-I(X;E^{n})_{\rho}\right)  +\delta\\
&  =\lim_{n\rightarrow\infty}\sup_{\left\{  p(x),\rho_{A^{n}}^{x}\right\}
_{x\in\mathcal{X}}}\frac{1}{n}\left(  I(X;B^{n})_{\rho}-I(X;E^{n})_{\rho
}\right) \\
&  =I_{\text{reg}}^{p}(\mathcal{N}).
\end{align}
We have thus shown that the quantity $I_{\text{reg}}^{p}(\mathcal{N})$ is a
weak converse rate for private communication over $\mathcal{N}$.

\subsection{Relative Entropy of Entanglement Strong Converse Bound}

Except for channels for which the private information is known to be additive
(such as the class of degradable channels; see Section~\ref{sec-PC:degradable-chs} below), the private
capacity of a channel is difficult to compute. This prompts us to find upper
bounds on the private capacity. In this section, we do so in terms of the
channel's relative entropy of entanglement, and in terms of the channel's
squashed entanglement in the next section.

We begin by recalling the bound from \eqref{eq-PC:renyi-REE-one-shot-bnd},
which holds for all $(\left\vert \mathcal{M}\right\vert ,\varepsilon)$ private
communication protocols and for all $\alpha>1$:%
\begin{equation}
P^{\varepsilon}(\mathcal{N})\leq\widetilde{E}_{\alpha}(\mathcal{N}%
)+\frac{\alpha}{\alpha-1}\log_{2}\!\left(  \frac{1}{1-\varepsilon}\right)  .
\end{equation}
For $n$ channel uses, the bound in \eqref{eq-PC:renyi-REE-one-shot-bnd}
becomes%
\begin{equation}
\frac{\log_{2}\!\left\vert \mathcal{M}\right\vert }{n}\leq\frac{1}{n}%
\widetilde{E}_{\alpha}(\mathcal{N}^{\otimes n})+\frac{\alpha}{n\left(
\alpha-1\right)  }\log_{2}\!\left(  \frac{1}{1-\varepsilon}\right)  ,
\label{eq-PC:n-shot-bnd-Renyi-REE}%
\end{equation}
which holds for all $\alpha>1$ and for all $(n,\left\vert \mathcal{M}%
\right\vert ,\varepsilon)$ private communication protocols, with
$n\in\mathbb{N}$ and $\varepsilon\in\lbrack0,1)$. We can simplify this
inequality by making use of the following fact:

\begin{proposition*}{Weak Subadditivity of a Channel's Renyi Relative Entropy of Entanglement}{} Let
$\mathcal{N}_{A\rightarrow B}$ be a quantum channel, with $d_{A}$ the
dimension of the input system $A$. For all $\alpha>1$ and $n\in\mathbb{N}$, we
have%
\begin{equation}
\widetilde{E}_{\alpha}(\mathcal{N}^{\otimes n})\leq n\widetilde{E}_{\alpha
}(\mathcal{N})+\frac{\alpha\left(  d_{A}^{2}-1\right)  }{\alpha-1}\log
_{2}(n+1). \label{eq-PC:weak-subadd-REE-Renyi}%
\end{equation}

\end{proposition*}

\begin{Proof}
The proof is identical to the proof of Proposition~\ref{prop-Renyi_Rains_inf_chan_weak_additive}, but making use of
Proposition~\ref{prop-gen_div_ent_chan_cov} at the beginning instead of Proposition~\ref{prop-gen_Rains_inf_chan_cov}.
\end{Proof}

Combining \eqref{eq-PC:weak-subadd-REE-Renyi} with
\eqref{eq-PC:n-shot-bnd-Renyi-REE}, we conclude the following upper bound on
the rate of an arbitrary $(n,\left\vert \mathcal{M}\right\vert ,\varepsilon)$
private communication protocol:%
\begin{equation}
\frac{\log_{2}\!\left\vert \mathcal{M}\right\vert }{n}\leq\widetilde{E}_{\alpha
}(\mathcal{N})+\frac{\alpha}{n\left(  \alpha-1\right)  }\log_{2}\!\left(
\frac{\left(  n+1\right)  ^{d_{A}^{2}-1}}{1-\varepsilon}\right)  ,
\end{equation}
which holds for all $\alpha>1$. Consequently, the following upper bound holds
for the $n$-shot private capacity:%
\begin{equation}
P^{n,\varepsilon}(\mathcal{N})\leq\widetilde{E}_{\alpha}(\mathcal{N}%
)+\frac{\alpha}{n\left(  \alpha-1\right)  }\log_{2}\!\left(  \frac{\left(
n+1\right)  ^{d_{A}^{2}-1}}{1-\varepsilon}\right)  ,
\end{equation}
for all $\alpha>1$.

With this bound, we are now ready to state the main result of this section,
which is that a channel's relative entropy of entanglement is an upper bound
on the strong converse private capacity of an arbitrary quantum channel
$\mathcal{N}$.

\begin{theorem*}
{Strong Converse Upper Bound on Private Capacity}
{thm-PC:REE-strong-converse}A channel's relative entropy of
entanglement, denoted by $E_{R}(\mathcal{N})$, is a strong converse rate for
private communication over $\mathcal{N}$. In other words, $\widetilde
{P}(\mathcal{N})\leq E_{R}(\mathcal{N})$ for every quantum channel
$\mathcal{N}$.
\end{theorem*}

Recall from \eqref{eq-EM:REE-channel-def} that%
\begin{equation}
E_{R}(\mathcal{N})\coloneqq \sup_{\psi_{SA}}\inf_{\sigma_{SB}\in\operatorname{SEP}%
(S:B)}D(\mathcal{N}_{A\rightarrow B}(\psi_{SA})\Vert\sigma_{SB}),
\end{equation}
where the supremum is with respect to every pure state $\psi_{SA}$ with
$d_{S}=d_{A}$.

\begin{Proof}
The proof here is identical to that given for Theorem~\ref{thm-Rains_inf_strong_conv_upper_bound}, but using the
relative entropy of entanglement $E_{R}(\mathcal{N})$ instead of the Rains
information~$R(\mathcal{N})$.
\end{Proof}

\subsection{Squashed Entanglement Weak Converse Bound}

We showed earlier in Section~\ref{eq-PC:squashed-ent-one-shot-bnd} that the squashed entanglement gives an upper
bound on the one-shot private capacity. Here we extend these results to the
$n$-shot setting, establishing a bound on the $n$-shot private capacity and we
conclude from it that the squashed entanglement of a quantum channel is a weak
converse rate for private communication over it. Later on in the book, in Chapter~\ref{chap-SKA}, we prove that the squashed entanglement of a channel is an upper bound on its secret-key-agreement capacity, which generally can be much larger than its private capacity. Thus,  the squashed entanglement bound is generally a loose upper bound on its (unassisted) private capacity.

\begin{theorem*}
{Squashed Entanglement Upper Bound on $n$-Shot Private Capacity}
{thm-PC:sq-ent-n-shot-bnd}
Let $\mathcal{N}_{A\rightarrow B}$ be a
quantum channel, and let $\varepsilon\in\lbrack0,1)$. For all $(n,\left\vert
\mathcal{M}\right\vert ,\varepsilon)$ private communication protocols for
$\mathcal{N}$, the following bound holds%
\begin{equation}
\left(  1-2\sqrt{\varepsilon}\right)  \frac{\log_{2}\!\left\vert \mathcal{M}%
\right\vert }{n}\leq E_{\text{sq}}(\mathcal{N})+\frac{2}{n}g_{2}%
(\sqrt{\varepsilon}),
\end{equation}
where $E_{\text{sq}}(\mathcal{N})$ is the squashed entanglement of the channel
$\mathcal{N}$, defined in Section~\ref{subsec-sq_ent_channel}.
\end{theorem*}

\begin{Proof}
Plugging the tensor-power channel $\mathcal{N}^{\otimes n}$ into the bound
from Theorem~\ref{thm-PC:sq-ent-up-bnd-one-shot}, we conclude the following
bound%
\begin{equation}
\left(  1-2\sqrt{\varepsilon}\right)  \frac{\log_{2}\!\left\vert \mathcal{M}%
\right\vert }{n}\leq\frac{1}{n}E_{\text{sq}}(\mathcal{N}^{\otimes n})+\frac
{2}{n}g_{2}(\sqrt{\varepsilon}).
\end{equation}
The desired statement then follows from the additivity of squashed
entanglement of a channel (Corollary~\ref{cor-sq_ent_additive}), which implies that $\frac{1}%
{n}E_{\text{sq}}(\mathcal{N}^{\otimes n})=E_{\text{sq}}(\mathcal{N})$.
\end{Proof}

\begin{theorem*}
{Weak Converse Upper Bound on Private Capacity}
{thm-PC:sq-ent-weak-converse}The squashed entanglement $E_{\text{sq}%
}(\mathcal{N})$ of a quantum channel $\mathcal{N}_{A\rightarrow B}$ is a weak
converse rate for private communication over $\mathcal{N}$. In other words,
$P(\mathcal{N})\leq E_{\text{sq}}(\mathcal{N})$ for every quantum channel
$\mathcal{N}$.
\end{theorem*}

\begin{Proof}
We exploit the bound from Theorem~\ref{thm-PC:sq-ent-n-shot-bnd} and an
argument similar to that from Section~\ref{sec-SKD:squashed-ent-weak-conv} to conclude the desired statement.
\end{Proof}

\section{Examples}

We now consider the private capacity for particular classes of quantum
channels. As we indicated earlier, computing the private capacity of an
arbitrary channel is a difficult task. This task is made more difficult by the
fact that, in some cases, the private information is known to be strictly
superadditive in the following sense:%
\begin{equation}
I^{p}(\mathcal{N}^{\otimes n})\geq nI^{p}(\mathcal{N}).
\end{equation}
This fact confirms that regularization of the private information is really
needed in general in order to compute the private capacity, and that
additivity of private information does not hold for all channels. Please
consult the Bibliographic Notes in Section~\ref{sec-PC:bib-notes}\ for more
information about strict superadditivity of private information for certain quantum channels.

Before starting the development below, recall that the private information of
a channel $\mathcal{N}_{A\rightarrow B}$\ is defined as%
\begin{equation}
I^{p}(\mathcal{N})=\sup_{\{p(x),\rho_{A}^{x}\}_{x\in\mathcal{X}}}\left(
I(X;B)_{\rho}-I(X;E)_{\rho}\right)  , \label{eq-PC:private-info-recall}%
\end{equation}
where the state $\rho_{XBE}$ is defined as%
\begin{equation}
\rho_{XBE}\coloneqq \sum_{x\in\mathcal{X}}p(x)|x\rangle\!\langle x|_{X}\otimes
\mathcal{U}_{A\rightarrow BE}^{\mathcal{N}}(\rho_{A}^{x}),
\label{eq-PC:state-4-priv-info}%
\end{equation}
with $\mathcal{U}_{A\rightarrow BE}^{\mathcal{N}}$ an isometric channel
extending $\mathcal{N}_{A\rightarrow B}$ and the optimization over every
ensemble $\{p(x),\rho_{A}^{x}\}_{x\in\mathcal{X}}$.

\subsection{Degradable Channels}

\label{sec-PC:degradable-chs}

Recall from Definition~\ref{def-deg_antideg_chan} that a channel
$\mathcal{N}_{A\rightarrow B}$ is degradable if there exists a degrading channel
$\mathcal{D}_{B\rightarrow E}$ such that
\begin{equation}
\mathcal{N}^{c}=\mathcal{D}\circ\mathcal{N},
\end{equation}
where $\mathcal{N}^{c}$ is a channel complementary to $\mathcal{N}$ (see
Definition~\ref{def-complementary_chan}) and $d_{E}\geq$ rank$(\Gamma
_{AB}^{\mathcal{N}})$. In particular, if $V_{A\rightarrow BE}$ is an isometric
extension of $\mathcal{N}$, so that
\begin{equation}
\mathcal{N}(\rho)=\operatorname{Tr}_{E}[V\rho V^{\dagger}]
\end{equation}
for every state $\rho$, then
\begin{equation}
\mathcal{N}^{c}(\rho)=\operatorname{Tr}_{B}[V\rho V^{\dagger}].
\end{equation}

We now show that the private information is equal to the coherent information
for every degradable channel, i.e.,%
\begin{equation}
I^{p}(\mathcal{N})=I^{c}(\mathcal{N}),
\end{equation}
where $I^{c}(\mathcal{N})$ is defined in \eqref{eq-coh_inf_chan}. As a consequence of this
observation and the fact that a tensor product of degradable channels is also
degradable, it follows that the private capacity of a degradable channel is
equal to its coherent information, and there is no difference between the
private capacity and the quantum capacity in this case, i.e.,%
\begin{equation}
Q(\mathcal{N})=P(\mathcal{N})=I^{p}(\mathcal{N})=I^{c}(\mathcal{N})\text{ for
every degradable channel }\mathcal{N}\text{.} \label{eq-PC:priv-info-deg}%
\end{equation}

\begin{proposition*}
{Private Information of Degradable Channels}{prop-PC:priv-info-deg}
Let
$\mathcal{N}_{A\rightarrow B}$ be a degradable channel. Then its private
information is equal to its coherent information:%
\begin{equation}
I^{p}(\mathcal{N})=I^{c}(\mathcal{N}),
\end{equation}
where the channel's private information $I^{p}(\mathcal{N})$ is defined in \eqref{eq-PC:private-info-recall} and
its coherent information in~\eqref{eq-coh_inf_chan}. As a consequence, \eqref{eq-PC:priv-info-deg} holds.
\end{proposition*}

\begin{Proof}
By Theorem~\ref{thm-PC:generic-rel-priv-info-c-info}, we only need to prove
the inequality $I^{p}(\mathcal{N})\leq I^{c}(\mathcal{N})$ for the case of a
degradable channel. Let $\mathcal{U}_{A\rightarrow BE}^{\mathcal{N}}$ be an
isometric channel extending $\mathcal{N}_{A\rightarrow B}$. Let $\rho_{A}%
^{x}=\sum_{y}p(y|x)\psi_{A}^{x,y}$ be a spectral decomposition of the input
state $\rho_{A}^{x}$, and define the following extension of the state
$\rho_{XBE\,}$ in \eqref{eq-PC:state-4-priv-info}:%
\begin{equation}
\rho_{XYBE}=\sum_{x,y}p(x)p(y|x)|x\rangle\!\langle x|_{X}\otimes
|y\rangle\!\langle y|_{Y}\otimes\mathcal{U}_{A\rightarrow BE}^{\mathcal{N}%
}(\psi_{A}^{x,y}).
\end{equation}
Consider that%
\begin{align}
&  I(X;B)_{\rho}-I(X;E)_{\rho}\nonumber\\
&  =I(XY;B)_{\rho}-I(Y;B|X)_{\rho}-\left(  I(XY;E)_{\rho}-I(Y;E|X)_{\rho
}\right) \\
&  =I(XY;B)_{\rho}-I(XY;E)_{\rho}-\left(  I(Y;B|X)_{\rho}-I(Y;E|X)_{\rho
}\right) \\
&  \leq I(XY;B)_{\rho}-I(XY;E)_{\rho}\\
&  =H(B)_{\rho}-H(B|XY)_{\rho}-\left(  H(E)_{\rho}-H(E|XY)_{\rho}\right) \\
&  =H(B)_{\rho}-H(E)_{\rho}\\
&  \leq I^{c}(\mathcal{N}).
\end{align}
The first equality follows by applying the chain rule for conditional mutual
information. The first inequality follows by applying the data-processing
inequality for conditional mutual information and the fact that there is a
degrading channel $\mathcal{D}_{B\rightarrow E}$ such that $\rho
_{XYE}=\mathcal{D}_{B\rightarrow E}(\rho_{XYB})$. The last few steps follow
the same reasoning given in the proof of
Theorem~\ref{thm-PC:generic-rel-priv-info-c-info}.
\end{Proof}

\subsubsection{Generalized Dephasing Channels}

Recall the definition of a generalized dephasing channel from the discussion
surrounding \eqref{eq-gen_dephase_iso_ext}. Similar to what was found for generalized dephasing
channels in Section~\ref{sec-QC:gen-deph-chs}, we can also consider the question of whether the
strong converse property holds for the private capacity of these channels.
Indeed, it is the case, and the reasoning is essentially the same as that
given in the proof of Theorem~\ref{prop-gen_dephase_q_cap}, except that we use the strong converse
bound on private capacity given by the relative entropy of entanglement. The
main observation to make while examining the proof of Theorem~\ref{prop-gen_dephase_q_cap} is that
the state in \eqref{eq-QC:sep-state-gen-deph-pf} is a separable state.

\begin{theorem*}
{Private Capacity of Generalized Dephasing Channels}{}
For every generalized
dephasing channel $\mathcal{N}$ (defined by the isometric extension in \eqref{eq-gen_dephase_iso_ext}),
the following equalities hold%
\begin{align}
P(\mathcal{N})  &  =\widetilde{P}(\mathcal{N})=E_{R}(\mathcal{N}%
)=I^{p}(\mathcal{N})\\
&  =Q(\mathcal{N})=\widetilde{Q}(\mathcal{N})=R(\mathcal{N})=I^{c}%
(\mathcal{N}).
\end{align}

\end{theorem*}

\begin{Proof}
The following inequalities hold in general%
\begin{align}
I^{c}(\mathcal{N})  &  \leq Q(\mathcal{N})\leq P(\mathcal{N})\leq\widetilde
{P}(\mathcal{N})\leq E_{R}(\mathcal{N}),\\
I^{c}(\mathcal{N})  &  \leq I^{p}(\mathcal{N})\leq P(\mathcal{N}),\\
I^{c}(\mathcal{N})  &  \leq Q(\mathcal{N})\leq\widetilde{Q}(\mathcal{N})\leq
R(\mathcal{N})\leq E_{R}(\mathcal{N}),
\end{align}
and the reasoning given above establishes that $I^{c}(\mathcal{N}%
)=E_{R}(\mathcal{N})$ for generalizing dephasing channels.
\end{Proof}

\subsection{Anti-Degradable Channels}

\label{sec-PC:anti-deg-zero-priv-cap}

Let us consider the private capacity for anti-degradable channels. Recall from
Definition~\ref{def-deg_antideg_chan} that a channel $\mathcal{N}_{A\rightarrow B}$ is
anti-degradable if there exists an anti-degrading channel $\mathcal{A}%
_{E\rightarrow B}$ such that%
\begin{equation}
\mathcal{N}_{A\rightarrow B}=\mathcal{A}_{E\rightarrow B}\circ\mathcal{N}%
_{A\rightarrow E}^{c},
\end{equation}
where $\mathcal{N}_{A\rightarrow E}^{c}$ is a channel complementary to
$\mathcal{N}_{A\rightarrow B}$ and $d_{E}\geq$ rank$(\Gamma_{AB}^{\mathcal{N}%
})$.

\begin{proposition*}
{Private Information for Anti-Degradable Channels}{}
The private information
vanishes for all anti-degradable channels, i.e., $I^{p}(\mathcal{N})=0$ for
every anti-degradable channel $\mathcal{N}$. Therefore, the private capacity
of an anti-degradable channel is equal to zero, i.e., $P(\mathcal{N})=0$ for
every anti-degradable channel $\mathcal{N}$.
\end{proposition*}

\begin{Proof}
The first claim is a direct consequence of the definition of the private
information in \eqref{eq-PC:private-info-recall}, the fact that there is an
anti-degrading channel $\mathcal{A}_{E\rightarrow B}$ such that $\rho
_{XB}=\mathcal{A}_{E\rightarrow B}(\rho_{XE})$, where the state $\rho_{XBE}$
is defined in \eqref{eq-PC:state-4-priv-info}, and the data-processing
inequality for mutual information. The second claim follows from the
regularized expression for private capacity from
Theorem~\ref{thm-PC:private-cap} and the fact that a tensor product of anti-degradable channels is anti-degradable.
\end{Proof}

\section{Summary}

In this chapter, we studied private communication over a quantum channel~$\mathcal{N}_{A\rightarrow B}$. The communication model that we employed is
that a sender has access to the input system $A$, a legitimate receiver has
access to the output system $B$, and an eavesdropper has access to the system
$E$ of an isometric channel $\mathcal{U}_{A\rightarrow BE}^{\mathcal{N}}$
extending $\mathcal{N}_{A\rightarrow B}$. This model gives the most power to
the eavesdropper, subject to the constraints that the systems $A$ and $B$ are
physically secure in the laboratories of the sender and receiver,
respectively. The goal of a private communication protocol is for the sender
to transmit a classical message such that the receiver can decode it with high
probability and the eavesdropper cannot determine which message was
transmitted (i.e., her system $E$ should be essentially useless for figuring
out the transmitted message). The private capacity is defined as the largest
rate at which private communication is possible, such that the decoding error
probability tends to zero and the eavesdropper's system becomes decoupled with
the message system. In our definitions, we combined these requirements into a
single constraint. We found that the private information $I^{p}(\mathcal{N})$
of quantum channel $\mathcal{N}$ is a lower bound on its private capacity, and
that, in general, computing the exact value of the private capacity involves a
regularization, i.e., $P(\mathcal{N})=I_{\text{reg}}^{p}(\mathcal{N})$.

Following the same course as in previous chapters, we began with the one-shot
setting for private communication, in which only one use of the channel is
allowed, along with some non-zero error. We then determined upper and lower
bounds on the number of private bits that can be transmitted. We established
three upper bounds on the one-shot private capacity, involving the one-shot
private information, the hypothesis testing relative entropy of entanglement,
as well as the squashed entanglement. These in turn led to upper bounds on the
asymptotic private capacity. To obtain a lower bound on the one-shot private
capacity, we employed the methods of position-based coding and convex
splitting, similar to how we did in the previous chapter on secret key
distillation (Chapter~\ref{chap-secret_key_distill}). This lower bound is optimal when employed in the
asymptotic setting because it leads to the regularized private information as
an achievable rate for private communication, and this matches the upper
bound. For degradable channels, there is no difference between the private
information and the coherent information, and this implies that there is no
difference between the private capacity and quantum capacity for these
channels. We also proved that the private capacity of anti-degradable channels
is equal to zero.

Since the regularized private information is difficult to compute, we
established other upper bounds on private capacity, in terms of relative
entropy of entanglement (strong converse upper bound) and squashed
entanglement (weak converse upper bound). We then concluded that the strong
converse property holds for all generalized dephasing channels and their
private capacity is equal to their coherent information.

\section{Bibliographic Notes}

\label{sec-PC:bib-notes}

\citet{S49} studied the information-theoretic security of communication systems.
Some years later, the private capacity of a classical channel (also known as
secrecy capacity)\ was introduced and studied by \citet{W75} and some years
later by \citet{CK78}, who established a general formula for the private
capacity of a classical channel.

\citet{BB84} devised the first protocol for sending private classical
information over a quantum channel, which is known as quantum key
distribution. The private capacity of a quantum channel was studied by
\citet{D05,1050633}, who independently established the regularized expression
for it in Theorem~\ref{thm-PC:private-cap}.

Private communication was studied from the one-shot perspective by
\citet{RR11,WTB16,wilde2017position,radhakrishnan2017one}.
Proposition~\ref{thm-PC:QC-to-PC} was established by \citet{WQ18}. The
connection between secret-key transmission and bipartite private-state
transmission is a direct consequence of the insights of \citet{HHHO05,HHHO09}%
\ and was discussed by \citet{WTB16}. The upper bound in
Proposition~\ref{prop-PC:one-shot-upper-bnd}\ is similar to that established
by \citet{Qi_2018}. The upper bound in Theorem~\ref{thm-PC:REE-upp-bnd}\ is due
to \citet{WTB16} and the upper bound in
Theorem~\ref{thm-PC:sq-ent-up-bnd-one-shot}\ to \citet{TGW14IEEE}. The lower
bound in Section~\ref{sec-PC:lower-bnd-priv-cap-1-shot}\ is due to
\citet{wilde2017position}.

As mentioned above, the asymptotic theory of private communication was
developed by \citet{D05,1050633}. \citet{D05} proved
Theorem~\ref{thm-PC:generic-rel-priv-info-c-info}, relating coherent and
private information and the private to quantum capacity. 
Strict superadditivity of the private information of a quantum channel was established by \citet{smith:170502}, and this result was strengthened by \citet{ES15}.
The relative entropy
of entanglement strong converse bound on private capacity in
Theorem~\ref{thm-PC:REE-strong-converse} was proven by \citet{WTB16}. The
squashed entanglement weak converse bound on private capacity in
Theorem~\ref{thm-PC:sq-ent-weak-converse}\ was proven by \citet{TGW14IEEE}. The
private capacity of degradable channels (i.e.,
Theorem~\ref{prop-PC:priv-info-deg}\ and \eqref{eq-PC:priv-info-deg})\ was
established by \citet{S08}. The strong converse property for the private
capacity of generalized dephasing channels was established by \citet{WTB16}.

\part{Quantum Communication Protocols With Feedback Assistance}[We now delve into interactive quantum communication protocols. Such protocols involve interaction between the sender and receiver of a quantum channel, beyond the quantum channel that connects them, and this interaction can potentially increase a given communication capacity because it represents an additional resource that the sender and receiver have at their disposal. Such protocols are richer than the non-interactive protocols that we considered previously, and as such, their analysis is more involved.

One objective of the following chapters is to understand what role this interaction plays and whether it can increase capacity. For the most part, what we accomplish is the establishment of limitations on the ability of feedback to increase capacity. In some cases, such as the case presented in the first chapter, a surprising conclusion is that interaction does not increase capacity at all, so that the theory simplifies.
]\label{part-feedback}

\chapter{Quantum-Feedback-Assisted Communication}\label{chap-QFACC}



	In this chapter, we begin our foray into interactive quantum communication by analyzing communication protocols in which the goal is for the sender to communicate a classical message to the receiver, with the assistance of a free noiseless quantum feedback channel. By a quantum feedback channel, we mean a quantum channel from the receiver to the sender that is separate from the channel from the sender to the receiver being used to communicate the message. We thus call this communication scenario ``quantum-feedback-assisted communication.''
	
	One simple (yet effective) way to make use of this free noiseless quantum feedback channel is for the receiver to transmit one share of a bipartite quantum state to the sender. By doing so, they can establish shared entanglement, and the rates of classical communication that are achievable with such a strategy are given by the limits on entanglement-assisted communication that we studied previously in Chapter~\ref{chap-EA_capacity}.

	Perhaps surprisingly, we show here that the same non-asymptotic converse bounds established in \eqref{eq-eacc_weak_conv_one_shot_3} and \eqref{eq-eacc_str_conv_one_shot_3} apply to protocols assisted by noiseless quantum feedback. These non-asymptotic converse bounds imply that the quantum-feedback-assisted classical capacity of a channel is no larger than its entanglement-assisted capacity. Furthermore, the strong converse property holds for the quantum-feedback-assisted capacity, so that the strong converse capacity is equal to the mutual information of a quantum channel.

	This result demonstrates that the entanglement-assisted capacity of a quantum channel is a rather robust communication capacity. Not only is the mutual information of a channel equal to the strong converse entanglement-assisted capacity for all channels, but it is also equal to the strong converse quantum feedback-assisted capacity for all channels. Thus, the theory of entanglement-assisted and quantum-feedback-assisted communication simplifies immensely.

	It is worth remarking that Shannon proved that a similar result holds for classical channels, and the strong converse property was later demonstrated as well. In this sense, the entanglement-assisted capacity of a quantum channel represents the fully quantum generalization of the classical capacity of a classical channel. Related, the quantum mutual information of a quantum
channel represents the fully quantum generalization of the classical mutual information of a classical channel.

\section{$n$-Shot Quantum Feedback-Assisted Communication Protocols}

\label{sec-QFACC:n-shot-prot}

	\begin{figure}
		\centering
		\includegraphics[width=6in]{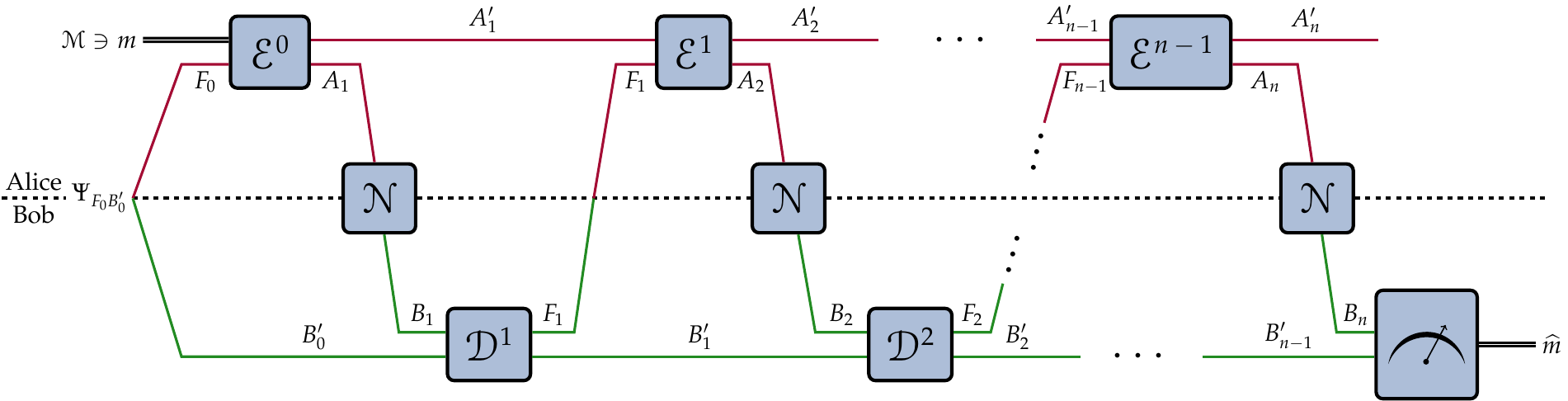}
		\caption{A general quantum-feedback-assisted communication protocol for the channel $\mathcal{N}$, which uses it $n$ times.}
		\label{fig:Feedback-assisted-class-prot}%
	\end{figure}

	We begin by defining the most general form for an $n$-shot classical communication protocol assisted by a noiseless quantum feedback channel, where $n\in\mathbb{N}$. Such a protocol is depicted in Figure~\ref{fig:Feedback-assisted-class-prot}, and it is defined by the following elements:%
	\begin{equation}
		(\mathcal{M},\Psi_{F_{0}B_{0}'},\mathcal{E}_{M'F_{0} \rightarrow A_{1}'A_{1}}^{0},\{\mathcal{E}_{A_{i}'F_{i}\rightarrow A_{i+1}'A_{i+1}}^{i},\mathcal{D}_{B_{i}B_{i-1}'\rightarrow F_{i}B_{i}'}^{i}\}_{i=1}^{n-1},\mathcal{D}_{B_{n}B_{n-1}'\rightarrow\widehat{M}}^{n}),
	\end{equation}
	where $\mathcal{M}$ is the message set, $\Psi_{F_{0}B_{0}'}$ denotes a bipartite quantum state, the objects denoted by $\mathcal{E}$ are encoding channels, and the objects denoted by $\mathcal{D}$ are decoding channels. Let $\mathcal{C}$ denote all of these elements, which together constitute the quantum-feedback-assisted code. The quantum systems labeled by $F$ represent the feedback systems that Bob sends back to Alice. The primed systems
$A_{i}'$ and $B_{i}'$ represent local quantum memory or ``scratch'' registers that Alice and Bob can exploit in the feedback-assisted protocol.

	In such an $n$-round feedback-assisted protocol, the protocol
proceeds as follows: let $p:\mathcal{M}\rightarrow\left[0,1\right]$ be a probability distribution over the message set. Alice starts by preparing two classical registers $M$ and $M'$ in the following state:
	\begin{equation} \label{eq:QFBA-classically-correlated-Phi}
		\overline{\Phi}_{MM'}^{p}\coloneqq\sum_{m\in\mathcal{M}}p(m)\ket{m}\!\bra{m}_{M}\otimes\ket{m}\!\bra{m}_{M'}.
	\end{equation}
	Furthermore, Alice and Bob also initially share a quantum state $\Psi_{F_{0}B_{0}'}$ on Alice's system $F_{0}$ and Bob's system $B_{0}'$. This state is prepared by Bob locally, and then he transmits the system $F_{0}$ to Alice via the noiseless quantum feedback channel. The initial global state shared between them is%
	\begin{equation}
		\overline{\Phi}_{MM'}^{p}\otimes\Psi_{F_{0}B_{0}'}.
	\end{equation}

	Alice then sends the $M'$ and $F_{0}$ registers through the first encoding channel $\mathcal{E}_{M'F_{0}\rightarrow A_{1}'A_{1}}^{0}$. This encoding channel realizes a set $\{\mathcal{E}_{F_{0}\rightarrow A_{1}'A_{1}}^{0,m}\}_{m\in\mathcal{M}}$ of quantum channels as follows:%
	\begin{equation}
		\mathcal{E}_{F_{0}\rightarrow A_{1}'A_{1}}^{0,m}(\tau_{F_{0}})\coloneqq\mathcal{E}_{M'F_{0}\rightarrow A_{1}'A_{1}}^{0}(\ket{m}\!\bra{m}_{M'}\otimes\tau_{F_{0}}),
	\end{equation}
	for all input states $\tau_{F_{0}}$. The global state after the first encoding channel is then as follows:%
	\begin{equation}
		\mathcal{E}_{M'F_{0}\rightarrow A_{1}'A_{1}}^{0}(\overline{\Phi}_{MM'}^{p}\otimes\Psi_{F_{0}B_{0}'}).
	\end{equation}
	Note that the scratch system $A_{1}'$ can contain a classical copy of the particular message $m$ that is being communicated, and the same is true for all of the later scratch systems $A_{i}'$, for $i \in \{2,\dotsc,n\}$. In fact, this is necessary in order for the communication protocol to be effective. Alice then transmits the $A_{1}$ system through the channel $\mathcal{N}_{A_{1}\rightarrow B_{1}}$, leading to the state%
	\begin{equation}\label{eq:QFBA-rho-1-state}
		\rho_{MA_{1}'B_{1}B_{0}'}^{1}\coloneqq(\mathcal{N}_{A_{1}\rightarrow B_{1}}\circ\mathcal{E}_{M'F_{0}\rightarrow A_{1}'A_{1}}^{0})(\overline{\Phi}_{MM'}^{p}\otimes\Psi_{F_{0}B_{0}'}).
	\end{equation}
	After receiving the $B_{1}$ system, Bob performs the  decoding channel $\mathcal{D}_{B_{1}B_{0}'\rightarrow F_{1}B_{1}'}^{1}$, such that the state is then%
	\begin{equation}
		\mathcal{D}_{B_{1}B_{0}'\rightarrow F_{1}B_{1}'}^{1}(\rho_{MA_{1}'B_{1}B_{0}'}^{1}),
	\end{equation}
	with it being understood that the system $B_{1}'$ is Bob's new scratch register and the feedback system $F_{1}$ gets sent over the noiseless quantum feedback channel back to Alice.

	In the next round, Alice processes the $A_{1}'F_{1}$ systems with the encoding channel $\mathcal{E}_{A_{1}'F_{1}\rightarrow A_{2}'A_{2}}^{1}$, and she sends system $A_{2}$ over the channel $\mathcal{N}_{A_{2}\rightarrow B_{2}}$, leading to the state
	\begin{equation}
		\rho_{MA_{2}'B_{2}B_{1}'}^{2}\coloneqq(\mathcal{N}_{A_{2}\rightarrow B_{2}}\circ\mathcal{E}_{A_{1}'F_{1}\rightarrow A_{2}'A_{2}}^{1}\circ\mathcal{D}_{B_{1}B_{0}'\rightarrow F_{1}B_{1}'}^{1})(\rho_{MA_{1}'B_{1}B_{0}'}^{1}).
	\end{equation}
	Bob then applies the second decoding channel $\mathcal{D}_{B_{2}B_{1}'\rightarrow F_{2}B_{2}'}^{2}$. This process then iterates $n-2$ more times, and the state after each use of the channel is as follows:
	\begin{multline}\label{eq:QFBA-rho-i-states}
		\rho_{MA_{i}'B_{i}B_{i-1}'}^{i}\coloneqq\\
		(\mathcal{N}_{A_{i}\rightarrow B_{i}}\circ\mathcal{E}_{A_{i-1}'F_{i-1}\rightarrow A_{i}'A_{i}}^{i-1}\circ\mathcal{D}_{B_{i-1}B_{i-2}'\rightarrow F_{i-1}B_{i-1}'}^{i-1})(\rho_{MA_{i-1}'B_{i-1}B_{i-2}'}^{i-1}),
	\end{multline}
	for $i\in\left\{  3,\dotsc,n\right\}$.

	In the final round, Bob performs the decoding channel $\mathcal{D}_{B_{n}B_{n-1}'\rightarrow\widehat{M}}^{n}$, which is a
quantum-to-classical channel that finally decodes the transmitted message. The final classical--classical state of the protocol is then as follows:
	\begin{equation}
		\omega_{M\widehat{M}}^{p}\coloneqq\mathcal{D}_{B_{n}B_{n-1}'\rightarrow\widehat{M}}^{n}(\Tr_{A_{n}'}[\rho_{MA_{n}'B_{n}B_{n-1}'}^{n}]).\label{eq:QFBA-final-state-omega}%
	\end{equation}

	Now, just as we did in Chapter~\ref{chap-EA_capacity} in the case of entanglement-assisted classical communication, we can define the message error probability, average error probability, and maximal error probability as in \eqref{eq-eac-mess_error_prob}, \eqref{eq-eac-avg_error_prob}, and \eqref{eq-eac-maximal_error_prob}, respectively. Using the alternative expression in \eqref{eq-eacc_trace_dist_avg_error} for the average error probability, we have that the average error probability for the quantum-feedback-assisted code $\mathcal{C}$ is given by
	\begin{equation}
		\overline{p}_{\text{err}}(\mathcal{C};p)=\frac{1}{2}\left\Vert \overline {\Phi}_{MM'}^{p}-\omega_{M\widehat{M}}^{p}\right\Vert _{1}.
	\end{equation}
	Using the alternative expression in \eqref{eq-eacc_max_err_prob_trace_distance} for the maximal error probability, we have that the maximal error probability of the quantum-feedback-assisted code $\mathcal{C}$ is given by%
	\begin{equation}
		p_{\text{err}}^{*}(\mathcal{C})=\max_{p:\mathcal{M}\rightarrow\left[0,1\right]}\frac{1}{2}\left\Vert \overline{\Phi}_{MM'}^{p}-\omega_{M\widehat{M}}^{p}\right\Vert _{1}.
	\end{equation}

	\begin{definition}{$(n,|\mathcal{M}|,\varepsilon)$ Quantum-Feedback-Assisted Classical
Communication Protocol}{def-qfac-nMe_protocol}
		Let $(\mathcal{M},\Psi_{F_{0}B_{0}'},\mathcal{E}_{M'F_{0}\rightarrow A_{1}'A_{1}}^{0},\{\mathcal{E}_{A_{i}'F_{i}\rightarrow A_{i+1}'A_{i+1}}^{i},\mathcal{D}_{B_{i}B_{i-1}'\rightarrow F_{i}B_{i}'}^{i}\}_{i=1}^{n-1},\mathcal{D}_{B_{n}B_{n-1}'\rightarrow\widehat{M}}^{n})$ be the elements of an $n$-shot quantum-feedback-assisted classical communication protocol over the channel $\mathcal{N}_{A\rightarrow B}$. The protocol is called an $(n,|\mathcal{M}|,\varepsilon)$ protocol, with $\varepsilon\in\left[  0,1\right]  $, if $p_{\text{err}}^{*}(\mathcal{C})\leq\varepsilon$.
	\end{definition}

\subsection{Protocol over a Useless Channel}

\label{sec:QFBA-replacer-useless}

	As before, when determining converse bounds on the rate at which classical messages can be communicated reliably using such feedback-assisted protocols, it is helpful to consider a useless channel. Our plan is again to use relative entropy (or some generalized divergence) to compare the states at each time step of the actual protocol with those resulting from employing a useless channel instead of the actual channel. As before, a useless channel that conveys no information at all is one in which the input state is discarded and replaced with some state at the output:%
	\begin{equation}
		\mathcal{R}_{A\rightarrow B}\coloneqq\mathcal{P}_{\sigma_{B}}\circ\Tr_{A},\label{eq:QFBA-replacer-ch}%
	\end{equation}
	where $\mathcal{P}_{\sigma_{B}}$ denotes a preparation channel that prepares the arbitrary (but fixed) state $\sigma_{B}$ at the output.
	
	\begin{figure}
		\centering
		\includegraphics[width=6in]{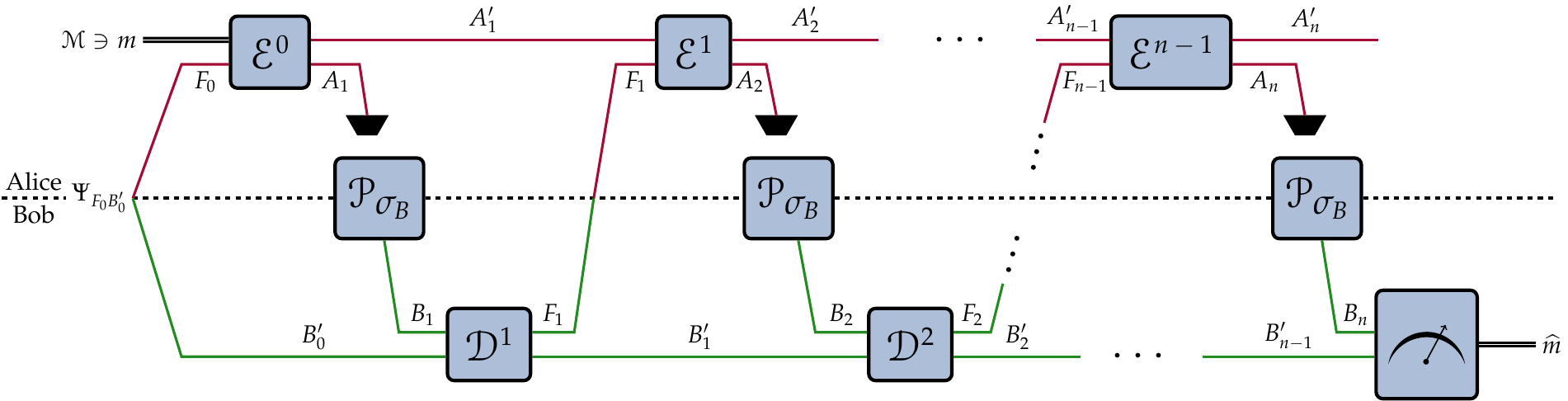}
		\caption{Depiction of a protocol that is useless for quantum-feedback-assisted classical communication. In each round, the encoded state is discarded and replaced with an arbitrary (but fixed) state $\sigma_B$.}\label{fig-eaqf_classical_comm_n_shot_useless}
	\end{figure}

	We can modify the $i^{\text{th}}$ step of the protocol discussed in the previous section, such that instead of the actual channel $\mathcal{N}_{A_{i}\rightarrow B_{i}}$\ being applied, the replacement channel $\mathcal{R}_{A_{i}\rightarrow B_{i}}$ is applied; see Figure \ref{fig-eaqf_classical_comm_n_shot_useless}.
	
	The state after the first round in this protocol over the useless channel is
	\begin{align}
		\tau_{MA_{1}'B_{1}B_{0}'}^{1}  & \coloneqq(\mathcal{R}_{A_{1}\rightarrow B_{1}}\circ\mathcal{E}_{M'F_{0}\rightarrow A_{1}'A_{1}}^{0})(\overline{\Phi}_{MM'}^{p}\otimes\Psi_{F_{0}B_{0}'})\\
		& =\Tr_{A_{1}}[\mathcal{E}_{M'F_{0}\rightarrow A_{1}'A_{1}}^{0}(\overline{\Phi}_{MM'}^{p}\otimes\Psi_{F_{0}B_{0}'})]\otimes\sigma_{B_{1}},
		\label{eq-qfacc:1st-state-useless-prot}
	\end{align}
	where we observe that%
	\begin{equation}\label{eq-qfba-n_shot_useless_analysis}
		\tau_{MA_{1}'B_{1}B_{0}'}^{1}=\tau_{MA_{1}'B_{0}'}^{1}\otimes\sigma_{B_{1}},
	\end{equation}
	and furthermore that%
	\begin{align}
		\tau_{MB_{1}B_{0}'}^{1}  & =\Tr_{A_{1}'A_{1}}[\mathcal{E}_{M'F_{0}\rightarrow A_{1}'A_{1}}^{0}(\overline{\Phi}_{MM'}^{p}\otimes\Psi_{F_{0}B_{0}'})]\otimes\sigma_{B_{1}}\\
		& =\Tr_{M'F_{0}}[\overline{\Phi}_{MM'}^{p}\otimes\Psi_{F_{0}B_{0}'}]\otimes\sigma_{B_{1}}\\
		& =\pi_{M}^{p}\otimes\Psi_{B_{0}'}\otimes\sigma_{B_{1}}\label{eq-qfba-n_shot_useless_analysis2},
	\end{align}
	where the second equality holds due to the fact that the first encoding channel $\mathcal{E}_{M'F_{0}\rightarrow A_{1}'A_{1}}^{0}$ is trace preserving, and where%
	\begin{equation}
		\pi_{M}^{p} \coloneqq \sum_{m\in\mathcal{M}}p(m)|m\rangle\!\langle m|_{M}.
	\end{equation}
	Thus, there is no correlation whatsover between the message system $M$ and Bob's systems $B_{1}B_{0}'$ after tracing over all of Alice's systems. Intuitively, this is a consequence of the fact that the ``communication line has been cut'' when employing the replacement channel.

	The state after the second replacement channel $\mathcal{R}_{A_{2}\rightarrow B_{2}}$\ is then given by%
	\begin{align}
		&  \tau_{MA_{2}'B_{2}B_{1}'}^{2}\nonumber\\
		&  \coloneqq(\mathcal{R}_{A_{2}\rightarrow B_{2}}\circ\mathcal{E}_{A_{1}'F_{1}\rightarrow A_{2}'A_{2}}^{1}\circ\mathcal{D}_{B_{1}B_{0}'\rightarrow F_{1}B_{1}'}^{1})(\tau_{MA_{1}'B_{1}B_{0}'}^{1})\\
		&  =\Tr_{A_{2}}[(\mathcal{E}_{A_{1}'F_{1}\rightarrow A_{2}'A_{2}}^{1}\circ\mathcal{D}_{B_{1}B_{0}'\rightarrow F_{1}B_{1}'}^{1})(\tau_{MA_{1}'B_{1}B_{0}'}^{1})]\otimes\sigma_{B_{2}}\\
		&  =\Tr_{A_{2}}[(\mathcal{E}_{A_{1}'F_{1}\rightarrow A_{2}'A_{2}}^{1}\circ\mathcal{D}_{B_{1}B_{0}'\rightarrow F_{1}B_{1}'}^{1})(\tau_{MA_{1}'B_{0}'}^{1}\otimes\sigma_{B_{1}})]\otimes\sigma_{B_{2}},
	\end{align}
	where we used \eqref{eq-qfba-n_shot_useless_analysis} to obtain the last line. If we take the partial trace over system $A_{2}'$, then the fact that the encoding channel $\mathcal{E}_{A_{1}'F_{1}\rightarrow A_{2}'A_{2}}^{1}$ is trace preserving implies that%
	\begin{align}
		& \tau_{MB_{2}B_{1}'}^{2}\nonumber\\
		& =\Tr_{A_{2}'A_{2}}[(\mathcal{E}_{A_{1}'F_{1}\rightarrow A_{2}'A_{2}}^{1}\circ\mathcal{D}_{B_{1}B_{0}'\rightarrow F_{1}B_{1}'}^{1})(\tau_{MA_{1}'B_{0}'}^{1}\otimes\sigma_{B_{1}})]\otimes\sigma_{B_{2}}\\
		& =\Tr_{A_{1}'F_{1}}[\mathcal{D}_{B_{1}B_{0}'\rightarrow F_{1}B_{1}'}^{1}(\tau_{MA_{1}'B_{0}'}^{1}\otimes\sigma_{B_{1}})]\otimes\sigma_{B_{2}}\\
		& =\Tr_{F_{1}}[\mathcal{D}_{B_{1}B_{0}'\rightarrow F_{1}B_{1}'}^{1}(\tau_{MB_{0}'}^{1}\otimes\sigma_{B_{1}})]\otimes\sigma_{B_{2}}.
	\end{align}
	Then, using \eqref{eq-qfba-n_shot_useless_analysis2}, which implies that $\tau_{MB_0'}^1=\pi_M^p\otimes\Psi_{B_0'}$, we find that
	\begin{align}
		\tau_{MB_2B_1'}^2& =\Tr_{F_{1}}[\mathcal{D}_{B_{1}B_{0}'\rightarrow F_{1}B_{1}'}^{1}(\pi_{M}^{p}\otimes\Psi_{B_{0}'}\otimes \sigma_{B_{1}})]\otimes\sigma_{B_{2}}\\
		& =\pi_{M}^{p}\otimes\Tr_{F_{1}}[\mathcal{D}_{B_{1}B_{0}'\rightarrow F_{1}B_{1}'}^{1}(\Psi_{B_{0}'} \otimes\sigma_{B_{1}})]\otimes\sigma_{B_{2}}\\
		& =\pi_{M}^{p}\otimes\tau_{B_{1}'}^{2}\otimes\sigma_{B_{2}}.
	\end{align}
	Thus, we find again that there is no correlation whatsoever between the message system $M$ and Bob's systems $B_{2}B_{1}'$ after tracing over all of Alice's systems.

	The states for the other rounds $i\in\left\{3,\dotsc,n\right\}  $ are given by%
	\begin{align}
		&  \tau_{MA_{i}'B_{i}B_{i-1}'}^{i}\nonumber\\
		&  \coloneqq(\mathcal{R}_{A_{i}\rightarrow B_{i}}\circ\mathcal{E}_{A_{i-1}'F_{i-1}\rightarrow A_{i}'A_{i}}^{i-1}\circ\mathcal{D}_{B_{i-1} B_{i-2}'\rightarrow F_{i-1}B_{i-1}'}^{i-1})(\tau_{MA_{i-1}'B_{i-1}B_{i-2}'}^{i-1})\\
		&  =\Tr_{A_{i}}[\mathcal{E}_{A_{i-1}'F_{i-1}\rightarrow A_{i}'A_{i}}^{i-1}\circ\mathcal{D}_{B_{i-1}B_{i-2}'\rightarrow F_{i-1}B_{i-1}'}^{i-1})(\tau_{MA_{i-1}'B_{i-1}B_{i-2}'}^{i-1})]\otimes\sigma_{B_{i}}\\
		&  =\Tr_{A_{i}}[\mathcal{E}_{A_{i-1}'F_{i-1}\rightarrow A_{i}'A_{i}}^{i-1}\circ\mathcal{D}_{B_{i-1}B_{i-2}'\rightarrow F_{i-1}B_{i-1}'}^{i-1})(\tau_{MA_{i-1}'B_{i-2}'}^{i-1}\otimes\sigma_{B_{i-1}})]\otimes\sigma_{B_{i}}.
	\end{align}
	Repeating a calculation similar to the above leads to a similar conclusion as above:%
	\begin{equation}\label{eq-qfba_useless_states_i}
		\tau_{MB_{i}B_{i-1}'}^{i}=\pi_{M}^{p}\otimes\tau_{B_{i-1}'}^{i}\otimes\sigma_{B_{i}},
	\end{equation}
	for all $i\in\{3,\dotsc,n\}$. That is, there is no correlation whatsoever between the message system $M$ and Bob's systems $B_{i}B_{i-1}'$ after tracing over all of Alice's systems. Again, this is intuitively a consequence of the fact that the ``communication line has been cut'' when employing the replacement channel.

	Bob's final decoding channel $\mathcal{D}_{B_{n}B_{n-1}'\rightarrow\widehat{M}}^{n}$ therefore leads to the following classical--classical state:%
	\begin{equation}\label{eq:QFBA-final-state-replacer}
		\tau_{M\widehat{M}}\coloneqq\pi_{M}^p \otimes\mathcal{D}_{B_{n}B_{n-1}'\rightarrow\widehat{M}}^{n}(\tau_{B_{n-1}'}^{2}\otimes\sigma_{B_{n}})=\pi_{M}^{p}\otimes\tau_{\widehat{M}},%
	\end{equation}
	where $\tau_{\widehat{M}}\coloneqq\sum_{\widehat{m}\in\mathcal{M}}t(\widehat{m})\ket{\widehat{m}}\!\bra{\widehat{m}}_{\widehat{M}}$ for some probability distribution $t:\mathcal{M}\to[0,1]$, which corresponds to Bob's measurement.

\subsection{Upper Bound on the Number of Transmitted Bits}

	We now give a general upper bound on the number transmitted bits in any quantum-feedback-assisted classical communication protocol. This result is stated in Theorem \ref{thm:QFBA-upper-bnds-finite-length}, and it holds independently of the encoding and decoding channels used in the protocol and depends only on the given communication channel $\mathcal{N}$. Recall from the previous section that $\log_{2}|\mathcal{M}|$ represents the number of bits that are transmitted over the channel $\mathcal{N}$.

	\begin{theorem*}{$n$-Shot Upper Bounds for Quantum-Feedback-Assisted Classical Communication}{thm:QFBA-upper-bnds-finite-length}
		Let $\mathcal{N}_{A\rightarrow B}$ be a quantum channel, and let $\varepsilon\in\lbrack0,1)$. For all $(n,|\mathcal{M}|,\varepsilon)$ quantum-feedback-assisted classical communication protocols over the channel $\mathcal{N}_{A\rightarrow B}$, the following bounds hold,
		\begin{align}
			\frac{\log_{2}|\mathcal{M}|}{n}  & \leq\frac{1}{1-\varepsilon}\left(I(\mathcal{N})+\frac{1}{n}h_{2}(\varepsilon)\right)  ,\label{eq:QFBA-weak-converse-n-shot-bnd}\\
			\frac{\log_{2}|\mathcal{M}|}{n}  & \leq \widetilde{I}_{\alpha}(\mathcal{N})+\frac{\alpha}{n(\alpha-1)}\log_{2}\!\left(  \frac{1}{1-\varepsilon}\right)\qquad\forall~\alpha>1,\label{eq:QFBA-strong-converse-n-shot-bnd}%
		\end{align}
		where $I(\mathcal{N})$ is the mutual information of $\mathcal{N}$, as defined in \eqref{eq-mut_inf_chan}, and $\widetilde{I}_{\alpha}(\mathcal{N})$ is the sandwiched R\'enyi mutual information of $\mathcal{N}$, as defined in~\eqref{eq-sand_ren_mut_inf_chan}.
	\end{theorem*}

	\begin{Proof}
		Let us start with an arbitrary $(n,|\mathcal{M}|,\varepsilon)$ quantum-feedback-assisted classical communication protocol over a channel	$\mathcal{N}_{A\rightarrow B}$, corresponding to, as described earlier, a message set $\mathcal{M}$, a shared quantum state $\Psi_{F_{0}B_{0}'}$, the encoding channels $\mathcal{E}_{M'F_{0}\rightarrow A_{1}'A_{1}}^{0}$ and $\{\mathcal{E}_{A_{i}'F_{i}\rightarrow A_{i+1}'A_{i+1}}^{i}\}_{i=1}^{n-1}$, and the decoding channels $\{\mathcal{D}_{B_{i}B_{i-1}'\rightarrow F_{i}B_{i}'}^{i}\}_{i=1}^{n-1}$ and $\mathcal{D}_{B_{n}B_{n-1}'\rightarrow\widehat{M}}^{n}$. Recall that we refer to all of these objects collectively as the code $\mathcal{C}$. The error criterion $p_{\text{err}}^{*}(\mathcal{C})\leq\varepsilon$ holds by the definition of an $(n,|\mathcal{M}|,\varepsilon)$ protocol, which implies that for all probability distributions $p:\mathcal{M}\rightarrow\left[  0,1\right]  $ on the message set $\mathcal{M}$:%
		\begin{equation}
			\overline{p}_{\text{err}}(\mathcal{C};p)\leq p_{\text{err}}^{*}(\mathcal{C})\leq\varepsilon.
		\end{equation}
		(The reasoning for this is analogous to that in \eqref{eq-ea_classical_comm_arb_to_uniform_reduction_1}--\eqref{eq-ea_classical_comm_arb_to_uniform_reduction}.) In particular, the above inequality holds with $p$ being the uniform distribution on $\mathcal{M}$, so that $p(m)=\frac{1}{|\mathcal{M}|}$ for all $m\in\mathcal{M}$. 
	
		Now, let $\overline{\Phi}_{M\widehat{M}}$ be the state defined in \eqref{eq:QFBA-classically-correlated-Phi} with $p$ the uniform distribution, and similarly let $\omega_{M\widehat{M}}$, defined in \eqref{eq:QFBA-final-state-omega}, be the state at the end of the protocol such that $p$ is the uniform prior probability distribution. Observe that $\Tr[\omega_{M\widehat{M}}]=\pi_M$. Also, letting
		\begin{equation}
			\Pi_{M\widehat{M}}=\sum_{m\in\mathcal{M}}\ket{m}\!\bra{m}_M\otimes\ket{m}\!\bra{m}_{\widehat{M}}
		\end{equation}
		be the projection defining the comparator test, as in \eqref{eq-eacc_comparator_test}, observe that
		\begin{equation}
			1-\Tr[\Pi_{M\widehat{M}}\omega_{M\widehat{M}}]=\frac{1}{2}\left\Vert \overline{\Phi}_{M\widehat{M}}-\omega_{M\widehat{M}}\right\Vert_{1}\leq\varepsilon,
		\end{equation}
		where the first equality follows by combining \eqref{eq-eacc_trace_dist_avg_error} with \eqref{eq-ea_classical_comm_comparator_succ_prob}. This means that
		\begin{equation}
			\Tr[\Pi_{M\widehat{M}}\omega_{M\widehat{M}}]\geq 1-\varepsilon.
		\end{equation}
		We thus have all of the ingredients to apply Lemma \ref{lem-eac-meta_conv}. Doing so gives the following critical first bound:
		\begin{equation}\label{eq:QFBA-init-bnd-hypo}
			\log_{2}|\mathcal{M}|\leq I_{H}^{\varepsilon}(M;\widehat{M})_{\omega}.%
		\end{equation}

		Invoking Proposition~\ref{prop-hypo_to_rel_ent}, the definition of $I_{H}^{\varepsilon}(M;\widehat{M})$ from \eqref{eq-hypo_testing_mutual_inf}, and the expression for  mutual information from \eqref{eq-mut_inf_opt}, we find that
		\begin{equation}\label{eq:QFBA-init-bnd-hypo-to-MI}
			I_{H}^{\varepsilon}(M;\widehat{M})_{\omega}\leq\frac{1}{1-\varepsilon}\left(I(M;\widehat{M})_{\omega}+h_{2}(\varepsilon)\right).%
		\end{equation}
		Now, using the data-processing inequality for the mutual information (see Proposition \ref{prop-gen_inf_meas_state_monotonicity}) with respect to the last decoding channel, $\mathcal{D}_{B_{n}B_{n-1}'\rightarrow\widehat{M}}^{n}$, we find that
		\begin{equation}\label{eq:QFBA-weak-bnd-chain-1}
			I(M;\widehat{M})_{\omega}\leq I(M;B_{n}B_{n-1}')_{\rho^{n}}.
		\end{equation}
		Then, using the chain rule for mutual information in \eqref{eq-q_mut_inf_chain_rule}, we obtain
		\begin{align}
			I(M;B_{n}B_{n-1}')_{\rho^{n}}& = I(M;B_{n}|B_{n-1}')_{\rho^{n}}+I(M;B_{n-1}')_{\rho^{n}}\\
			& \leq I(MB_{n-1}';B_{n})_{\rho^{n}}+I(M;B_{n-1}')_{\rho^{n}},
					\label{eq:QFBA-weak-bnd-step-simple-2}
		\end{align}
		where the second line is a consequence of the chain rule, as well as non-negativity of mutual information:
		\begin{align}
			I(M;B_{n}|B_{n-1}')_{\rho^{n}}&=I(MB_{n-1}';B_{n})_{\rho^{n}}-I(B_{n-1}';B_{n})_{\rho^{n}}\\
			&\leq I(MB_{n-1}';B_{n})_{\rho^{n}}.
		\end{align}
		Finally, observe that the state $\rho_{MB_{n}B_{n-1}'}^{n}$ has the following form:%
		\begin{equation}
			\rho_{MB_{n}B_{n-1}'}^{n}=\mathcal{N}_{A_{n}\rightarrow B_{n}}(\zeta_{MB_{n-1}'A_{n}}^{n}),
		\end{equation}
		where%
		\begin{multline}
			\zeta_{MB_{n-1}'A_{n}}^{n}\coloneqq\label{eq:QFBA-zeta-state-n}\\
			\Tr_{A_{n}'}[(\mathcal{E}_{A_{n-1}'F_{n-1}\rightarrow A_{n}'A_{n}}^{n-1}\circ\mathcal{D}_{B_{n-1}B_{n-2}'\rightarrow F_{n-1}B_{n-1}'}^{n-1})(\rho_{MA_{n-1}'B_{n-1}B_{n-2}'}^{n-1})].
		\end{multline}
		That is, the state $\zeta_{MB_{n-1}'A_{n}}^{n}$ is a particular state to consider in the optimization of the mutual information of a channel (with the channel input system being $A_{n}$ and the external correlated systems being $MB_{n-1}'$), whereas the definition of the mutual information of a channel involves an optimization over all such states. This means that 
		\begin{equation}
		I(MB_{n-1}';B_{n})_{\rho^{n}} \leq I(\mathcal{N}).
		\label{eq:QFBA-weak-bnd-step-simple-3}
		\end{equation}
		Putting together \eqref{eq:QFBA-weak-bnd-chain-1}, \eqref{eq:QFBA-weak-bnd-step-simple-2}, and \eqref{eq:QFBA-weak-bnd-step-simple-3}, we find that
		\begin{equation}\label{eq:QFBA-weak-bnd-chain-last}
			I(M;\widehat{M})_{\omega} \leq I(\mathcal{N})+I(M;B_{n-1}')_{\rho^{n}}.
		\end{equation}

		The quantity $I(M;B_{n-1}')_{\rho^n}$ can be bounded using steps analogous to the above. In particular, using the data-processing inequality for the mutual information with respect to the second-to-last decoding channel $\mathcal{D}_{B_{n-1}B_{n-2}'\rightarrow F_{n-1}B_{n-1}'}^{n-1}$, then employing the same steps as above, we conclude that
		\begin{align}
			I(M;B_{n-1}')_{\rho^{n}}  & \leq I(M;B_{n-1}B_{n-2}')_{\rho^{n-1}}\label{eq:QFBA-weak-bnd-chain-2}\\
			& =I(M;B_{n-1}|B_{n-2}')_{\rho^{n-1}}+I(M;B_{n-2}')_{\rho^{n-1}}\\
			& \leq I(MB_{n-2}';B_{n-1})_{\rho^{n-1}}+I(M;B_{n-2}')_{\rho^{n-1}}\\
			& \leq I(\mathcal{N})+I(M;B_{n-2}')_{\rho^{n-1}},\label{eq:QFBA-weak-bnd-chain-2-last}%
		\end{align}
		Overall, this leads to
		\begin{equation}
			I(M;\widehat{M})_{\omega}\leq 2I(\mathcal{N})+I(M;B_{n-2}')_{\rho^{n-1}}.
		\end{equation}
		Then, bounding $I(M;B_{n-2}')$ in the same manner as above, and continuing this process $n-3$ more times such that we completely ``unwind'' the protocol, we obtain
		\begin{align}
			I(M;\widehat{M})_{\omega}&\leq 2I(\mathcal{N})+I(M;B_{n-1}')_{\rho^{n-1}}\label{eq:QFBA-upper-bnds-finite-length_pf1}\\
			&\leq 3I(\mathcal{N})+I(M;B_{n-3}')_{\rho^{n-2}}\label{eq:QFBA-upper-bnds-finite-length_pf2}\\
			&\vdots\\
			&\leq nI(\mathcal{N})+I(M;B_0')_{\rho^1}\label{eq:QFBA-upper-bnds-finite-length_pf3}.
		\end{align}
		However, from \eqref{eq:QFBA-rho-1-state}, we have that $\rho_{MB_0'}^1=\overline{\Phi}_{M}\otimes \Psi_{B_0'}$, which means that $I(M;B_0')_{\rho^1}=0$. Therefore, putting together \eqref{eq:QFBA-init-bnd-hypo}, \eqref{eq:QFBA-init-bnd-hypo-to-MI}, and	\eqref{eq:QFBA-upper-bnds-finite-length_pf1}--\eqref{eq:QFBA-upper-bnds-finite-length_pf3}, we get		
		\begin{align}
			\log_2|\mathcal{M}|&\leq I_H^{\varepsilon}(M;\widehat{M})_{\omega}\\
			&\leq \frac{1}{1-\varepsilon}\left(I(M;\widehat{M})_{\omega}+h_2(\varepsilon)\right)\\
			&\leq \frac{1}{1-\varepsilon}\left(nI(\mathcal{N})+h_2(\varepsilon)\right),
		\end{align}
		and the last line is equivalent to \eqref{eq:QFBA-weak-converse-n-shot-bnd}, as required.

		We now establish the bound in \eqref{eq:QFBA-strong-converse-n-shot-bnd}. Combining \eqref{eq:QFBA-init-bnd-hypo} with Proposition~\ref{prop:sandwich-to-htre}, we conclude that the following bound holds for all $\alpha>1$:%
		\begin{equation}\label{eq:QFBA-str-conv-init-bnd}%
			\log_{2}|\mathcal{M}|\leq\widetilde{I}_{\alpha}(M;\widehat{M})_{\omega}+\frac{\alpha}{\alpha-1}\log_{2}\!\left(  \frac{1}{1-\varepsilon}\right).
		\end{equation}
		Recall that the sandwiched R\'enyi mutual information $\widetilde{I}_{\alpha}(M;\widehat{M})_{\omega}$ is defined as%
		\begin{align}
			\widetilde{I}_{\alpha}(M;\widehat{M})_{\omega}  & =\inf_{\xi_{\widehat{M}}}\widetilde{D}_{\alpha}(\omega_{M\widehat{M}}\Vert\omega_{M}\otimes\xi_{\widehat{M}})\\
			& =\inf_{\xi_{\widehat{M}}}\widetilde{D}_{\alpha}(\omega_{M\widehat{M}}\Vert\pi_{M}\otimes\xi_{\widehat{M}}).
		\end{align}
		Our goal now is to compare the actual protocol with one that results from employing a useless, replacement channel. To this end, let $\mathcal{R}_{A\rightarrow B}^{\sigma_B}$ be the replacement channel defined in \eqref{eq:QFBA-replacer-ch}, with $\sigma_{B}$ an arbitrary (but fixed) state. Then as discussed in Section~\ref{sec:QFBA-replacer-useless} (in particular, in \eqref{eq:QFBA-final-state-replacer}), the final state of the protocol conducted with the replacement channel is given by $\tau_{M\widehat{M}}=\pi_{M} \otimes\tau_{\widehat{M}}$. Then, we find that%
		\begin{align}
			\widetilde{I}_{\alpha}(M;\widehat{M})_{\omega}&=\inf_{\xi_{\widehat{M}}}\widetilde{D}_{\alpha}(\omega_{M\widehat{M}}\Vert\pi_{M}\otimes\xi_{\widehat{M}})  \\
			& \leq\widetilde{D}_{\alpha}(\omega_{M\widehat{M}}\Vert\pi_{M}\otimes\tau_{\widehat{M}})\\
			& =\widetilde{D}_{\alpha}(\omega_{M\widehat{M}}\Vert\tau_{M\widehat{M}}).\label{eq:QFBA-upper-bnds-finite-length_pf5}
		\end{align}
		We now proceed with a similar method considered in the proof of the bound in \eqref{eq:QFBA-weak-converse-n-shot-bnd}, but using the sandwiched R\'enyi relative entropy as our main tool for analysis. By applying the data-processing inequality for the sandwiched R\'{e}nyi relative entropy with respect to the last decoding channel, and using \eqref{eq-qfba_useless_states_i}, we find that
		\begin{align}
			\widetilde{D}_{\alpha}(\omega_{M\widehat{M}}\Vert\tau_{M\widehat{M}})  & \leq\widetilde{D}_{\alpha}(\rho_{MB_{n}B_{n-1}'}^{n}\Vert\tau_{MB_{n}B_{n-1}'}^{n})\\
			& =\widetilde{D}_{\alpha}(\rho_{MB_{n}B_{n-1}'}^{n}\Vert\pi_{M}\otimes\tau_{B_{n-1}'}^{n}\otimes\sigma_{B_{n}})\\
			& =\frac{\alpha}{\alpha-1}\log_{2}\widetilde{Q}_{\alpha}(\rho_{MB_{n}B_{n-1}'}^{n}\Vert\pi_{M}\otimes\tau_{B_{n-1}'}^{n}\otimes\sigma_{B_{n}})^{\frac{1}{\alpha}}\label{eq:QFBA-conv-renyi-last-decoder},
		\end{align}
		where in the last line we used the definition in \eqref{eq-sand_ren_rel_entropy} of the sandwiched R\'{e}nyi relative entropy. Now, recalling that $\rho_{MB_{n}B_{n-1}'}^{n}=\mathcal{N}_{A_{n}\rightarrow B_{n}}(\zeta_{MB_{n-1}'A_{n}}^{n})$ with the state $\zeta_{MB_{n-1}'A_{n}}^{(n)}$ defined in \eqref{eq:QFBA-zeta-state-n}, and defining the positive semi-definite operator%
		\begin{equation}
			X_{MB_{n-1}'A_{n}}^{(\alpha)}\coloneqq\left(  \pi_{M}\otimes\tau_{B_{n-1}'}^{n}\right)  ^{\frac{1-\alpha}{2\alpha}}\zeta_{MB_{n-1}'A_{n}}^{n}\left(  \pi_{M}\otimes\tau_{B_{n-1}'}^{n}\right)  ^{\frac{1-\alpha}{2\alpha}},
		\end{equation}
		as well as the completely positive map
		\begin{equation}
		\mathcal{S}_{\sigma_{B}}^{(\alpha)}(\cdot)\coloneqq\sigma_{B_{n}}^{\frac{1-\alpha}{2\alpha}}(\cdot)\sigma_{B_{n}}^{\frac{1-\alpha}{2\alpha}},
		\label{eq-QFBA:S-map-alpha}
		\end{equation}
		we use the definition of $\widetilde{Q}_{\alpha}$ in \eqref{eq-sand_rel_ent_Schatten} to obtain
		\begin{align}
			& \widetilde{Q}_{\alpha}(\rho_{MB_{n}B_{n-1}'}^{n}\Vert\pi_{M}\otimes\tau_{B_{n-1}'}^{n}\otimes\sigma_{B_{n}})^{\frac{1}{\alpha}}\nonumber\\
			& =\left\Vert \left(  \pi_{M}\otimes\tau_{B_{n-1}'}^{n}\otimes\sigma_{B_{n}}\right)  ^{\frac{1-\alpha}{2\alpha}}\rho_{MB_{n}B_{n-1}'}^{n}\left(  \pi_{M}\otimes\tau_{B_{n-1}'}^{n}\otimes\sigma_{B_{n}}\right)  ^{\frac{1-\alpha}{2\alpha}}\right\Vert _{\alpha}\label{eq:QFBA-str-peel-1}\\
			& =\left\Vert \mathcal{S}_{\sigma_{B}}^{(\alpha)}\left(  \left(  \pi_{M}\otimes\tau_{B_{n-1}'}^{n}\right)  ^{\frac{1-\alpha}{2\alpha}}\mathcal{N}_{A_{n}\rightarrow B_{n}}(\zeta_{MB_{n-1}'A_{n}}^{n})\left(  \pi_{M}\otimes\tau_{B_{n-1}'}^{n}\right)  ^{\frac{1-\alpha}{2\alpha}}\right)  \right\Vert _{\alpha}\\
			& =\left\Vert (\mathcal{S}_{\sigma_{B}}^{(\alpha)}\circ\mathcal{N}_{A_{n}\rightarrow B_{n}})(X_{MB_{n-1}'A_{n}}^{(\alpha)})\right\Vert_{\alpha}.
		\end{align}
		Multiplying and dividing by $\norm{X^{(\alpha)}_{MB_{n-1}'}}_{\alpha}$ leads to
		\begin{align}
			& \left\Vert (\mathcal{S}_{\sigma_{B}}^{(\alpha)}\circ\mathcal{N}_{A_{n}\rightarrow B_{n}})(X_{MB_{n-1}'A_{n}}^{(\alpha)})\right\Vert_{\alpha}\nonumber\\
			& =\frac{\left\Vert (\mathcal{S}_{\sigma_{B}}^{(\alpha)}\circ\mathcal{N}_{A_{n}\rightarrow B_{n}})(X_{MB_{n-1}'A_{n}}^{(\alpha)})\right\Vert_{\alpha}}{\left\Vert X_{MB_{n-1}'}^{(\alpha)}\right\Vert _{\alpha}}\left\Vert X_{MB_{n-1}'}^{(\alpha)}\right\Vert _{\alpha}\\
			& =\frac{\left\Vert (\mathcal{S}_{\sigma_{B}}^{(\alpha)}\circ\mathcal{N}_{A_{n}\rightarrow B_{n}})(X_{MB_{n-1}'A_{n}}^{(\alpha)})\right\Vert_{\alpha}}{\left\Vert X_{MB_{n-1}'}^{(\alpha)}\right\Vert _{\alpha}}\times\nonumber\\
			& \qquad\left\Vert \left(  \pi_{M}^{\frac{1-\alpha}{2\alpha}}\otimes\left(\tau_{B_{n-1}'}^{n}\right)  ^{\frac{1-\alpha}{2\alpha}}\right)\zeta_{MB_{n-1}'}^{n}\left(  \pi_{M}^{\frac{1-\alpha}{2\alpha}}\otimes\left(  \tau_{B_{n-1}'}^{n}\right)  ^{\frac{1-\alpha}{2\alpha}}\right)  \right\Vert _{\alpha}\\
			& =\frac{\left\Vert (\mathcal{S}_{\sigma_{B}}^{(\alpha)}\circ\mathcal{N}_{A_{n}\rightarrow B_{n}})(X_{MB_{n-1}'A_{n}}^{(\alpha)})\right\Vert_{\alpha}}{\left\Vert X_{MB_{n-1}'}^{(\alpha)}\right\Vert _{\alpha}}\cdot \widetilde{Q}_{\alpha}(\zeta_{MB_{n-1}'}^{n}\Vert \pi_{M}\otimes\tau_{B_{n-1}'}^{n})^{\frac{1}{\alpha}}\\
			& \leq\sup_{Y_{MB_{n-1}'A_{n}}\geq0}\frac{\left\Vert (\mathcal{S}_{\sigma_{B}}^{(\alpha)}\circ\mathcal{N}_{A_{n}\rightarrow B_{n}})(Y_{MB_{n-1}'A_{n}})\right\Vert _{\alpha}}{\left\Vert Y_{MB_{n-1}'}\right\Vert _{\alpha}}\cdot\widetilde{Q}_{\alpha}(\zeta_{MB_{n-1}'}^{n}\Vert\pi_{M}\otimes\tau_{B_{n-1}'}^{n})^{\frac{1}{\alpha}}\\
			& =\left\Vert \mathcal{S}_{\sigma_{B}}^{(\alpha)}\circ\mathcal{N}_{A_{n}\rightarrow B_{n}}\right\Vert _{\text{CB},1\rightarrow\alpha}\cdot \widetilde{Q}_{\alpha}(\zeta_{MB_{n-1}'}^{n}\Vert \pi_{M}\otimes\tau_{B_{n-1}'}^{n})^{\frac{1}{\alpha}},\label{eq:QFBA-str-peel-last}%
		\end{align}
		where to obtain the inequality we performed an optimization with respect to all positive semi-definite operators $Y_{MB_{n-1}'A_{n}}$. To obtain the last line, we have used the norm $\norm{\cdot}_{\text{CB},1\to\alpha}$ defined in \eqref{eq-operator_CB_alpha_norm}, which we show in Appendix \ref{app-CB_alpha_norm_alt} can be written as
		\begin{equation}
			\norm{\mathcal{M}}_{\text{CB},1\to\alpha}=\sup_{Y_{RA}\geq 0}\frac{\norm{\mathcal{M}_{A\to B}(Y_{RA})}_{\alpha}}{\norm{\Tr_A[Y_{RA}]}_{\alpha}}
		\end{equation}
		for any completely positive map $\mathcal{M}$. Plugging \eqref{eq:QFBA-str-peel-last} back in to \eqref{eq:QFBA-conv-renyi-last-decoder}, we conclude the following bound:%
		\begin{multline}
			\widetilde{D}_{\alpha}(\rho_{MB_{n}B_{n-1}'}^{n}\Vert\tau_{MB_{n}B_{n-1}'}^{n})\\
			\leq\frac{\alpha}{\alpha-1}\log_{2}\left\Vert \mathcal{S}_{\sigma_{B}}^{(\alpha)}\circ\mathcal{N}_{A_{n}\rightarrow B_{n}}\right\Vert_{\text{CB},1\rightarrow\alpha}+\widetilde{D}_{\alpha}(\zeta_{MB_{n-1}'}^{n}\Vert\pi_{M}\otimes\tau_{B_{n-1}'}^{n}).
		\end{multline}
		As in the proof of \eqref{eq:QFBA-weak-converse-n-shot-bnd}, we now iterate the above by successively bounding the sandwiched R\'{e}nyi relative entropy terms $\widetilde{D}_{\alpha}(\zeta_{MB_{i-1}'}^i\Vert\pi_M\otimes\tau_{B_{i-1}'}^i)$ for $i\in\{1,\dotsc, n\}$. Starting with the term $\widetilde{D}_\alpha(\zeta_{MB_{n-1}'}^n\Vert\pi_M\otimes\tau_{B_{n-1}'}^n)$, we use the data-processing inequality for the sandwiched R\'{e}nyi relative entropy under the second-to-last decoding channel $\mathcal{D}_{B_{n-1}B_{n-2}'\rightarrow F_{n-1}B_{n-1}'}^{n-1}$, then apply the same reasoning as in \eqref{eq:QFBA-str-peel-1}--\eqref{eq:QFBA-str-peel-last} to obtain
		\begin{align}
			& \widetilde{D}_{\alpha}(\zeta_{MB_{n-1}'}^{n}\Vert\pi_{M}\otimes \tau_{B_{n-1}'}^{n})\nonumber\\
			& \leq\widetilde{D}_{\alpha}(\rho_{MB_{n-1}B_{n-2}'}^{n-1}\Vert\pi_{M}\otimes\tau_{B_{n-1}B_{n-2}'}^{n-1})\\
			& =\widetilde{D}_{\alpha}(\rho_{MB_{n-1}B_{n-2}'}^{n-1}\Vert\pi_M\otimes\tau_{B_{n-2}'}^{n-1}\otimes\sigma_B)\\
			& \leq\frac{\alpha}{\alpha-1}\log_{2}\left\Vert \mathcal{S}_{\sigma_{B}}^{(\alpha)}\circ\mathcal{N}_{A_{n-1}\rightarrow B_{n-1}}\right\Vert_{\text{CB},1\rightarrow\alpha}+\widetilde{D}_{\alpha}(\zeta_{MB_{n-2}'}^{n-1}\Vert\pi_{M}\otimes\tau_{B_{n-2}'}^{n-1}).
		\end{align}
		Iterating this reasoning $n-2$ more times, we end up with the following bound:%
		\begin{align}
			&  \widetilde{D}_{\alpha}(\omega_{M\widehat{M}}\Vert\tau_{M\widehat{M}})\nonumber\\
			&  \leq n\frac{\alpha}{\alpha-1}\log_{2}\left\Vert \mathcal{S}_{\sigma_{B}}^{(\alpha)}\circ\mathcal{N}_{A\rightarrow B}\right\Vert _{\text{CB},1\rightarrow\alpha}+\widetilde{D}_{\alpha}(\rho_{MB_{0}'}^{1}\Vert \pi_{M}\otimes\Psi_{B_{0}'})\nonumber\\
			&  =n\frac{\alpha}{\alpha-1}\log_{2}\left\Vert \mathcal{S}_{\sigma_{B}}^{(\alpha)}\circ\mathcal{N}_{A\rightarrow B}\right\Vert _{\text{CB},1\rightarrow\alpha},\label{eq:QFBA-final-bnd-str-conv}%
		\end{align}
		where the equality holds because $\rho_{MB_{0}'}^{1}=\pi_{M}\otimes\Psi_{B_{0}'}$. Putting together \eqref{eq:QFBA-str-conv-init-bnd}, \eqref{eq:QFBA-upper-bnds-finite-length_pf5}, and \eqref{eq:QFBA-final-bnd-str-conv}, we finally obtain%
		\begin{multline}
			\log_{2}|\mathcal{M}|\leq\\
			n\frac{\alpha}{\alpha-1}\log_{2}\left\Vert \mathcal{S}_{\sigma_{B}}^{(\alpha)}\circ\mathcal{N}_{A\rightarrow B}\right\Vert _{\text{CB},1\rightarrow\alpha}+\frac{\alpha}{\alpha-1}\log_{2}\!\left(  \frac{1}{1-\varepsilon}\right)  .
		\end{multline}
		Since we proved that this bound holds for any choice of the state $\sigma_{B}$, we conclude that%
		\begin{align}
			& \log_{2}|\mathcal{M}|\nonumber\\
			& \leq n\frac{\alpha}{\alpha-1}\inf_{\sigma_{B}}\log_{2}\left\Vert\mathcal{S}_{\sigma_{B}}^{(\alpha)}\circ\mathcal{N}_{A\rightarrow B}\right\Vert _{\text{CB},1\rightarrow\alpha}+\frac{\alpha}{\alpha-1}\log_{2}\!\left(  \frac{1}{1-\varepsilon}\right)  \\
			& =n\widetilde{I}_{\alpha}(\mathcal{N})+\frac{\alpha}{\alpha-1}\log_{2}\!\left(\frac{1}{1-\varepsilon}\right)  ,
		\end{align}
		where the last equality follows from Lemma~\ref{lem-sand_rel_mut_inf_alt}, and it implies \eqref{eq:QFBA-strong-converse-n-shot-bnd}.
	\end{Proof}

\subsection{The Amortized Perspective}

\label{sec-QFBA:amortized-persp}

In this section, we revisit the proofs above for Theorem~\ref{thm:QFBA-upper-bnds-finite-length} that establish bounds on non-asymptotic quantum feedback-assisted capacity. In particular, we adopt a different perspective, which we call the amortized perspective and which turns out to be useful in establishing bounds for all kinds of feedback-assisted protocols other than the ones considered in this chapter. 

For the case of quantum feedback-assisted protocols, the amortized perspective consists of defining the amortized mutual information of a quantum channel as the largest net difference of the output and input mutual information that can be realized by the channel, when Alice and Bob are allowed to share an arbitrary state before the use of the channel. A key property of the amortized mutual information of a channel is that it is more readily seen to lead to an upper bound on the non-asymptotic quantum feedback-assisted capacity. Furthermore, another key property is that the amortized mutual information of a channel collapses to the usual mutual information of a channel, and so this leads to an alternative way of understanding the previous results. Furthermore, as indicated above, this perspective becomes quite useful in later chapters when we analyze LOCC-assisted quantum communication protocols and LOPC-assisted private communication protocols.

\subsubsection{Quantum Mutual Information}

	We begin by defining the following key concept, the amortized mutual information of a quantum channel:

	\begin{definition}{Amortized Mutual Information of a Channel}{def-qfacc-mut_inf_amort}
		The \textit{amortized mutual information of a quantum channel} $\mathcal{N}_{A\rightarrow B}$\ is defined as
		\begin{equation}
			I^{\mathcal{A}}(\mathcal{N})\coloneqq\sup_{\rho_{A'AB'}}\left[I(A';BB')_{\omega}-I(A'A;B')_{\rho}\right],
		\end{equation}
		where
		\begin{equation}
		\omega_{A'BB'}\coloneqq\mathcal{N}_{A\rightarrow B}(\rho_{A'AB'})
		\end{equation}
		and the optimization is over states~$\rho_{A'AB'}$.
	\end{definition}

	Intuitively, the amortized mutual information is equal to the largest net mutual information that can be realized by the channel, if we allow Alice and Bob to share an arbitrary state before communication begins. As mentioned above, this concept turns out to be useful for understanding the feedback-assisted protocols presented previously.

	We have the following simple relationship between mutual information and amortized mutual information:

	\begin{Lemma}{lem:QFBA-trivial-amort-ineq-MI}
		The mutual information of any channel $\mathcal{N}_{A\rightarrow B}$\ does not exceed its amortized mutual information:%
		\begin{equation}
			I(\mathcal{N})\leq I^{\mathcal{A}}(\mathcal{N}).
		\end{equation}
	\end{Lemma}

	\begin{Proof}
		Let us restrict the optimization in the definition of the amortized mutual information to states $\rho_{A'AB'}$ that have a trivial $B'$ system. This means that $\rho_{A'AB'}$ is of the form $\rho_{A'AB'}=\rho_{A'A}\otimes\ket{0}\!\bra{0}_{B'}$. Therefore, $I(A'A;B')_{\rho}=0$ and $I(A';BB')_{\omega}=I(A';B)_{\omega}$, where $\omega_{A'BB'}=\mathcal{N}_{A\to B}(\rho_{A'AB'})$, so that%
		\begin{align}
			I^{\mathcal{A}}(\mathcal{N})&=\sup_{\rho_{A'AB'}}\left[I(A';BB')_{\omega}-I(A'A;B)_{\omega}\right]\\
			&\geq \sup_{\rho_{A'A}}I(A';B)_{\omega}\\
			&=I(\mathcal{N}),
		\end{align}
		i.e., $I^{\mathcal{A}}(\mathcal{N})\geq I(\mathcal{N})$, as required.
	\end{Proof}

	An important question is whether the opposite inequality, i.e., $I^{\mathcal{A}}\leq I(\mathcal{N})$, holds, which would allow us to conclude that $I^{\mathcal{A}}=I(\mathcal{N})$. In this case, we find that it does. Specifically, we have the following.

	\begin{proposition}{prop:QFBA-amortization-collapse-MI}
		Given an arbitrary quantum channel $\mathcal{N}$, amortization does not increase its mutual information:%
		\begin{equation}\label{eq:QFBA-amort-does-not-increase}%
			I(\mathcal{N})=I^{\mathcal{A}}(\mathcal{N}).
		\end{equation}
	\end{proposition}

	\begin{Proof}
		To see this, consider that for an arbitrary input state $\rho_{A'AB'}$, we can use the chain rules for mutual information in \eqref{eq-q_mut_inf_chain_rule}--\eqref{eq-q_mut_inf_chain_rule_2} to obtain
		\begin{align}
			& I(A';BB')_{\omega}-I(A'A;B')_{\rho}\nonumber\\
			& =I(A';B)_{\omega}+I(A';B'|B)_{\omega}-I(A'A;B')_{\rho}\\
			& \leq I(A';B)_{\omega}+I(A'B;B')_{\omega}-I(A'A;B')_{\rho}.
		\end{align}
		In particular, to obtain the third line, note that \eqref{eq-q_mut_inf_chain_rule_2} implies
		\begin{equation}
			I(A'B;B')_{\omega}=I(B;B')_{\omega}+I(A';B'|B)_{\omega}\geq I(A';B'|B)_{\omega}
		\end{equation}
		since $I(B;B')_{\omega}\geq 0$. Continuing, we apply the data-processing inequality for the mutual information under the channel $\mathcal{N}$, which implies that $I(A'A;B)_\rho\geq I(A'B;B')_{\omega}$. Therefore,
		\begin{align}
			&I(A';BB')_{\omega}-I(A'A;B')_\rho\\
			& \leq I(A';B)_{\omega}+I(A'B;B')_{\omega}-I(A'B;B')_{\omega}\\
			& =I(A';B)_{\omega}\\
			& \leq I(\mathcal{N}),
		\end{align}
		where the last line follows because the state $\omega_{A'B}=\mathcal{N}_{A\rightarrow B}(\rho_{A'A})$ has the form of states that we consider when performing the optimization in the definition of the mutual information of a channel. Since the inequality
		\begin{equation}
			I(A';BB')_{\omega}-I(A'A;B')_\rho\leq I(\mathcal{N})
		\end{equation}
		holds for an arbitrary input state $\rho_{A'AB'}$, we conclude the bound in \eqref{eq:QFBA-amort-does-not-increase}.
	\end{Proof}

	We note here that the equality in \eqref{eq:QFBA-amort-does-not-increase} is stronger than the additivity of mutual information shown in Chapter~\ref{chap-EA_capacity} (in particular, that shown in Theorem~\ref{thm-chan_mut_inf_additive}). Indeed, the equality in \eqref{eq:QFBA-amort-does-not-increase} actually implies the additivity relation discussed previously. To see this, consider that the equality in \eqref{eq:QFBA-amort-does-not-increase} implies that%
	\begin{equation}
		I(A';BB')_{\omega}-I(A'A;B')_{\rho}\leq I(\mathcal{N})\label{eq:QFBA-amort-VN}%
	\end{equation}
	for an arbitrary input state $\rho_{A'AB'}$, where $\omega_{A'BB'}=\mathcal{N}_{A\to B}(\rho_{A'AB'})$. Now let $\rho_{A'AB'}=\mathcal{M}_{A^{\prime\prime}\rightarrow B'}(\sigma_{A'AA^{\prime\prime}})$ for some channel $\mathcal{M}_{A^{\prime\prime}\rightarrow B'}$ and some state $\sigma_{A'AA^{\prime\prime}}$. Then it follows that%
	\begin{equation}
		\omega_{A'BB'}=(\mathcal{N}_{A\rightarrow B}\otimes \mathcal{M}_{A^{\prime\prime}\rightarrow B'})(\sigma_{A'AA^{\prime\prime}}),
	\end{equation}
	and applying \eqref{eq:QFBA-amort-VN}, we have that%
	\begin{align}
		I(A';BB')_{\omega}  & \leq I(\mathcal{N})+I(A'A;B')_{\rho}\\
		&\leq I(\mathcal{N})+\sup_{\sigma_{A'AA''}}I(A'A;B')_\rho\\
		& =I(\mathcal{N})+I(\mathcal{M}),
	\end{align}
	where the inequality follows because the state $\sigma_{A'AA^{\prime\prime}}$ is a particular state to consider for the optimization in the definition of the mutual information of the channel $\mathcal{M}_{A^{\prime\prime}\rightarrow B'}$. Since the inequality holds for all input states $\sigma_{A'AA^{\prime\prime}}$ to $\mathcal{N}_{A\rightarrow B}\otimes\mathcal{M}_{A^{\prime\prime}\rightarrow B'}$, we conclude that%
	\begin{equation}
		I(\mathcal{N}\otimes\mathcal{M})\leq I(\mathcal{N})+I(\mathcal{M}),\label{eq:QFBA-amort-VN-last}%
	\end{equation}
	which is the non-trivial inequality needed in the proof of the additivity of the mutual information of a channel (see the proof of Theorem~\ref{thm-chan_mut_inf_additive}).

	How is the amortized mutual information relevant for analyzing a feedback-assisted protocol? Consider that the bound in \eqref{eq:QFBA-weak-bnd-chain-1}\ involves the mutual information $I(M;B_{n}B_{n-1}')_{\rho^{n}}$, so that%
	\begin{align}
		& I(M;B_{n}B_{n-1}')_{\rho^{n}}\nonumber\\
		& =I(M;B_{n}B_{n-1}')_{\rho^{n}}-I(M;B_{0}')_{\rho^{1}}\label{eq:QFBA-amort-feedback-analyze}\\
		& =I(M;B_{n}B_{n-1}')_{\rho^{n}}-I(M;B_{0}')_{\rho^{1}}+\sum_{i=1}^{n-1}I(M;B_{i}')_{\rho^{i}}-I(M;B_{i}')_{\rho^{i}}\\
		& \leq I(M;B_{n}B_{n-1}')_{\rho^{n}}-I(M;B_{0}')_{\rho^{1}}+\sum_{i=1}^{n-1}I(M;B_{i}B_{i-1}')_{\rho^{i}}-I(M;B_{i}')_{\rho^{i}}\\
		& =\sum_{i=1}^{n}I(M;B_{i}B_{i-1}')_{\rho^{i}}-I(M;B_{i-1}')_{\rho^{i}}\\
		& \leq n\cdot\sup_{\rho_{A'AB'}}I(A';BB')_{\omega}-I(A'A;B')_{\rho}\\
		& =n\cdot I^{\mathcal{A}}(\mathcal{N})=n\cdot I(\mathcal{N}).\label{eq:QFBA-amort-feedback-analyze-last}%
	\end{align}
	The first equality follows because the state $\rho_{MB_{0}'}^{1}$ is a product state. The second equality follows by adding and subtracting the mutual information of the state of the message system $M$ and Bob's memory system $B_{i}'$. The inequality is a consequence of data processing under the action of the decoding channels. The third equality follows from collecting terms. The final inequality follows because the state $\rho_{MB_{i}B_{i-1}'}^{i}$ is a particular state to consider in the optimization of the amortized mutual information, and the final equality follows from the amortization collapse in Proposition~\ref{prop:QFBA-amortization-collapse-MI}.

	Thus, we observe that the bound in \eqref{eq:QFBA-weak-converse-n-shot-bnd}, at a fundamental level, is a consequence of the amortization collapse from Proposition~\ref{prop:QFBA-amortization-collapse-MI}.

\subsubsection{Sandwiched R\'enyi Mutual Information}

	We can also consider the concept of amortization for the sandwiched R\'enyi mutual information, and in this subsection, we revisit the bound in \eqref{eq:QFBA-strong-converse-n-shot-bnd} to understand it from this perspective.

	\begin{definition}{Amortized Sandwiched Mutual Information}{def-qafb_amort_sand_ren_mut_inf}
		The amortized sandwiched R\'enyi mutual information of a quantum channel $\mathcal{N}_{A\rightarrow B}$ is defined for $\alpha\in(0,1)\cup(1,\infty)$ as follows:%
		\begin{equation}
			\widetilde{I}_{\alpha}^{\mathcal{A}}(\mathcal{N})\coloneqq\sup_{\rho_{A'AB'}}\left[\widetilde{I}_{\alpha}(A';BB')_{\omega}-\widetilde{I}_{\alpha}(A'A;B')_{\rho}\right],
		\end{equation}
		where $\omega_{A'BB'}\coloneqq\mathcal{N}_{A\rightarrow B}(\rho_{A'AB'})$ and the optimization is over states $\rho_{A'AB'}$.
	\end{definition}

	Just as with the mutual information of a channel, we find that for all $\alpha\in(0,1)\cup(1,\infty)$,
	\begin{equation}\label{eq:QFBA-trivial-amort-sandwiched}%
		\widetilde{I}_{\alpha}(\mathcal{N})\leq\widetilde{I}_{\alpha}^{\mathcal{A}}(\mathcal{N}),
	\end{equation}
	and the proof of this analogous to the proof of Lemma \ref{lem:QFBA-trivial-amort-ineq-MI}, which establishes the corresponding inequality for the mutual information. So the question is to determine whether the opposite inequality holds. Indeed, we find again that it is the case, at least for $\alpha>1$.

	\begin{proposition}{prop:QFBA-amort-collapse-sandwiched}
		Amortization does not increase the sandwiched R\'enyi mutual information of a quantum channel $\mathcal{N}$ for all $\alpha>1$:%
		\begin{equation}
			\widetilde{I}_{\alpha}(\mathcal{N})=\widetilde{I}_{\alpha}^{\mathcal{A}}(\mathcal{N}).
		\end{equation}
	\end{proposition}

	\begin{Proof}
		Let $\rho_{A'AB'}$ be an arbitrary input state, and let $\sigma_{B}$ and $\tau_{B'}$ be arbitrary states. Then, letting $\omega_{A'BB'}=\mathcal{N}_{A\to B}(\rho_{A'AB'})$, we find that
		\begin{align}
			& \widetilde{I}_{\alpha}(A';BB')_{\omega}\nonumber\\
			& =\inf_{\xi_{BB'}}\widetilde{D}_{\alpha}(\omega_{A'BB'}\Vert\omega_{A'}\otimes\xi_{BB'})\\
			& \leq\widetilde{D}_{\alpha}(\omega_{A'BB'}\Vert\omega_{A'}\otimes\sigma_{B}\otimes\tau_{B'})\\
			& =\widetilde{D}_{\alpha}(\mathcal{N}_{A\rightarrow B}(\rho_{A'AB'})\Vert\rho_{A'}\otimes\sigma_{B}\otimes\tau_{B'})\\
			& =\frac{\alpha}{\alpha-1}\log_{2}\left\Vert \left(\rho_{A'}\otimes\sigma_{B}\otimes\tau_{B'}\right)^{\frac{1-\alpha}{2\alpha}}\mathcal{N}_{A\rightarrow B}(\rho_{A'AB'})\left(\rho_{A'}\otimes\sigma_{B}\otimes\tau_{B'}\right)^{\frac{1-\alpha}{2\alpha}}\right\Vert _{\alpha},\label{eq:QFBA-amort-collapse-sandwiched_pf1}
		\end{align}
		where to obtain the last equality we used the alternate expression in \eqref{eq-sand_rel_ent_Schatten} for the sandwiched R\'{e}nyi relative entropy. Defining 
\begin{equation}		
X_{_{A'AB'}}^{(\alpha)}\coloneqq\left(  \rho_{A'}\otimes\tau_{B'}\right)^{\frac{1-\alpha}{2\alpha}}\rho_{A'AB'}\left(\rho_{A^{\prime}}\otimes\tau_{B'}\right)^{\frac{1-\alpha}{2\alpha}},
\end{equation}
and making use of the completely positive map $\mathcal{S}_{\sigma_{B}}^{(\alpha)}$ from \eqref{eq-QFBA:S-map-alpha},
we find that%
		\begin{align}
			& \left\Vert \left(\rho_{A'}\otimes\sigma_{B}\otimes\tau_{B'}\right)^{\frac{1-\alpha}{2\alpha}}\mathcal{N}_{A\rightarrow B}(\rho_{A'AB'})\left(\rho_{A'}\otimes\sigma_{B}\otimes\tau_{B'}\right)^{\frac{1-\alpha}{2\alpha}}\right\Vert_{\alpha}\nonumber\\
			& =\left\Vert (\mathcal{S}_{\sigma_{B}}^{(\alpha)}\circ\mathcal{N}_{A\rightarrow B})(X_{A'AB'}^{(\alpha)})\right\Vert _{\alpha}\\
			& =\frac{\left\Vert (\mathcal{S}_{\sigma_{B}}^{(\alpha)}\circ\mathcal{N}_{A\rightarrow B})(X_{A'AB'}^{(\alpha)})\right\Vert _{\alpha}}{\left\Vert X_{A'B'}^{(\alpha)}\right\Vert _{\alpha}}\left\Vert X_{A'B'}^{(\alpha)}\right\Vert _{\alpha}\\
			& \leq\sup_{Y_{A'AB'}\geq0}\frac{\left\Vert (\mathcal{S}_{\sigma_{B}}^{(\alpha)}\circ\mathcal{N}_{A\rightarrow B})(Y_{A'AB'})\right\Vert _{\alpha}}{\left\Vert Y_{A'B'}\right\Vert _{\alpha}}\times\nonumber\\
			& \qquad\qquad\left\Vert \left[  \rho_{A'}\otimes\tau_{B'}\right]  ^{\frac{1-\alpha}{2\alpha}}\rho_{A'B'}\left[\rho_{A'}\otimes\tau_{B'}\right]  ^{\frac{1-\alpha}{2\alpha}}\right\Vert _{\alpha}\\
			& =\left\Vert \mathcal{S}_{\sigma_{B}}^{(\alpha)}\circ\mathcal{N}_{A\rightarrow B}\right\Vert _{\text{CB},1\rightarrow\alpha}\cdot\widetilde{Q}_{\alpha}(\rho_{A'B'}\Vert\rho_{A'}\otimes\tau_{B'})^{\frac{1}{\alpha}},\label{eq:QFBA-amort-collapse-sandwiched_pf2}
		\end{align}
		where in the last line we have used the expression in \eqref{eq-eacc_CB_1alpha_norm_alt} for the norm $\norm{\cdot}_{\text{CB},1\to\alpha}$. Plugging \eqref{eq:QFBA-amort-collapse-sandwiched_pf2} back into \eqref{eq:QFBA-amort-collapse-sandwiched_pf1}, we find that%
		\begin{multline}
			\widetilde{I}_{\alpha}(A';BB')_{\omega}\leq\frac{\alpha}{\alpha-1}\log_{2}\left\Vert \mathcal{S}_{\sigma_{B}}^{(\alpha)}\circ\mathcal{N}_{A\rightarrow B}\right\Vert _{\text{CB},1\rightarrow\alpha}\\
			+\widetilde{D}_{\alpha}(\rho_{A'B'}\Vert\rho_{A'}\otimes\tau_{B'}).
		\end{multline}
		Since the inequality holds for arbitrary states $\sigma_{B}$ and $\tau_{B'}$, we conclude that%
		\begin{align}
			& \widetilde{I}_{\alpha}(A';BB')_{\omega}\nonumber\\
			& \leq\frac{\alpha}{\alpha-1}\inf_{\sigma_{B}}\log_{2}\left\Vert \mathcal{S}_{\sigma_{B}}^{(\alpha)}\circ\mathcal{N}_{A\rightarrow B}\right\Vert _{\text{CB},1\rightarrow\alpha}+\inf_{\tau_{B'}}\widetilde{D}_{\alpha}(\rho_{A'B'}\Vert\rho_{A'}\otimes\tau_{B'})\\
			& =\widetilde{I}_{\alpha}(\mathcal{N})+\widetilde{I}_{\alpha}(A';B')_{\rho}\\
			& \leq\widetilde{I}_{\alpha}(\mathcal{N})+\widetilde{I}_{\alpha}(A'A;B')_{\rho},
		\end{align}
		where the equality follows from Lemma~\ref{lem-sand_rel_mut_inf_alt} and the final inequality from the data-processing inequality for the mutual information under the partial trace $\Tr_A$. Since we have shown that the following inequality holds for an arbitrary input state $\rho_{A'AB'}$:
		\begin{equation}
			\widetilde{I}_{\alpha}(A';BB')_{\omega}-\widetilde{I}_{\alpha}(A'A;B')_{\rho}\leq\widetilde{I}_{\alpha}(\mathcal{N}),
		\end{equation}
		we conclude that $\widetilde{I}_{\alpha}^{\mathcal{A}}(\mathcal{N})\leq\widetilde{I}_{\alpha}(\mathcal{N})$, which leads to $\widetilde{I}_{\alpha}^{\mathcal{A}}(\mathcal{N})=\widetilde{I}_{\alpha}(\mathcal{N})$ after combining with \eqref{eq:QFBA-trivial-amort-sandwiched}.
	\end{Proof}

	By following exactly the same steps in \eqref{eq:QFBA-amort-VN}--\eqref{eq:QFBA-amort-VN-last}, but replacing $I$ with $\widetilde{I}_{\alpha}$, we can conclude that the amortization collapse in Proposition~\ref{prop:QFBA-amort-collapse-sandwiched} implies the additivity relation (Theorem~\ref{thm-sand_rel_ent_additivity}) for sandwiched R\'enyi mutual information of quantum
channels for all $\alpha>1$:%
	\begin{equation}
		\widetilde{I}_{\alpha}(\mathcal{N})=\widetilde{I}_{\alpha}^{\mathcal{A}}(\mathcal{N})\quad\forall~\alpha>1\quad\Longrightarrow\quad\widetilde{I}_{\alpha}(\mathcal{N}\otimes\mathcal{M})\leq\widetilde{I}_{\alpha}(\mathcal{N})+\widetilde{I}_{\alpha}(\mathcal{M}),
	\end{equation}
	where $\mathcal{N}$ and $\mathcal{M}$ are quantum channels. Furthermore, by following exactly the same steps as in \eqref{eq:QFBA-amort-feedback-analyze}--\eqref{eq:QFBA-amort-feedback-analyze-last}, but replacing $I$ with $\widetilde{I}_{\alpha}$, and employing Proposition~\ref{prop:QFBA-amort-collapse-sandwiched}, we conclude the following bound%
	\begin{equation}
		\widetilde{I}_{\alpha}(M;B_{n}B_{n-1}')_{\rho^{n}}\leq n\cdot \widetilde{I}_{\alpha}(\mathcal{N}),
	\end{equation}
	which in turn implies the bound in \eqref{eq:QFBA-strong-converse-n-shot-bnd}. Thus, we can alternatively analyze feedback-assisted protocols and arrive at the bound in \eqref{eq:QFBA-strong-converse-n-shot-bnd} by utilizing the concept of amortization.

\section{Quantum Feedback-Assisted Classical Capacity of a Quantum Channel}

	In this section, we analyze the asymptotic case, in which we allow for an arbitrarily large number of rounds of feedback. This task is now rather straightforward, given the bounds that we have established in the previous section. So we keep this section brief, only stating some definitions and then some theorems that follow as a direct consequence of definitions and developments in previous chapters.

	\begin{definition}{Achievable Rate for Quantum-Feedback-Assisted Classical Communication}{def-qfacc-ach_rate}
	Given a quantum channel $\mathcal{N}$, a rate $R\in\mathbb{R}^{+}$ is called an achievable rate for quantum-feedback-assisted classical communication over $\mathcal{N}$ if for all $\varepsilon\in(0,1]$, all $\delta>0$, and all sufficiently large $n$, there exists an $(n,2^{n(R-\delta)},\varepsilon)$ quantum-feedback-assisted classical communication protocol.
	\end{definition}

	\begin{definition}{Quantum-Feedback-Assisted Classical Capacity of a Quantum Channel}{def-qfacc-cap}
		The quantum-feedback-assisted classical capacity of a quantum channel $\mathcal{N}$, denoted by $C_{\operatorname{QFB}}(\mathcal{N})$, is defined as the supremum of all achievable rates, i.e.,%
		\begin{equation}
			C_{\operatorname{QFB}}(\mathcal{N})\coloneqq\sup\{R:R\text{ is an achievable rate for }\mathcal{N}\}.
		\end{equation}
	\end{definition}

	\begin{definition}{Strong Converse Rate for Quantum-Feedback-Assisted Classical Communication}{def-qfacc-strong_conv_rate}
		Given a quantum channel $\mathcal{N}$, a rate $R\in\mathbb{R}^{+}$ is called a strong converse rate for quantum-feedback-assisted classical communication over $\mathcal{N}$ if for all $\varepsilon\in\lbrack0,1)$, all $\delta>0$, and all sufficiently large $n$, there does not exist an $(n,2^{n(R+\delta)},\varepsilon)$ quantum-feedback-assisted classical communication protocol.
	\end{definition}

	\begin{definition}{Strong Converse Quantum-Feedback-Assisted Classical Capacity of a Quantum Channel}{def-qfacc-strong_conv_cap}
		The strong converse quantum-feedback-assisted classical capacity of a quantum channel $\mathcal{N}$, denoted by $\widetilde{C}_{\operatorname{QFB}}(\mathcal{N})$, is defined as the infimum of all strong converse rates, i.e.,%
		\begin{equation}
			\widetilde{C}_{\operatorname{QFB}}(\mathcal{N})\coloneqq\inf\{R:R\text{ is a strong converse rate for }\mathcal{N}\}.
		\end{equation}
	\end{definition}

	The main result of this section is the following capacity theorem:

	\begin{theorem*}{Quantum-Feedback-Assisted Classical Capacity}{thm-qfacc-cap}
		For any quantum channel $\mathcal{N}$, its quantum-feedback-assisted classical capacity $C_{\operatorname{QFB}}(\mathcal{N})$ and its strong converse quantum-feedback-assisted classical capacity are both equal to its mutual information $I(\mathcal{N})$, i.e.,%
		\begin{equation}
			C_{\operatorname{QFB}}(\mathcal{N})=\widetilde{C}_{\operatorname{QFB}}(\mathcal{N})=I(\mathcal{N}),
		\end{equation}
		where $I(\mathcal{N})$ is defined in \eqref{eq-mut_inf_chan}.
	\end{theorem*}

	\begin{Proof}
		By previous reasoning, we have that%
		\begin{equation}
			C_{\operatorname{QFB}}(\mathcal{N})\leq\widetilde{C}_{\operatorname{QFB}}(\mathcal{N}),
		\end{equation}
		and by Theorem~\ref{thm-ea_classical_capacity}  and the fact that any entanglement-assisted classical communication protocol is a particular kind of
quantum-feedback-assisted classical communication protocol, we have that%
		\begin{equation}
			I(\mathcal{N})\leq C_{\operatorname{QFB}}(\mathcal{N})\leq\widetilde{C}_{\operatorname{QFB}}(\mathcal{N}).
		\end{equation}
		The upper bound $\widetilde{C}_{\operatorname{QFB}}(\mathcal{N})\leq I(\mathcal{N})$ follows from \eqref{eq:QFBA-strong-converse-n-shot-bnd} and the same reasoning given in the proof detailed in Section~\ref{sec-eacc_str_conv}.
	\end{Proof}

\section{Bibliographic Notes}

\citet{Shannon56} proved that feedback does not increase the classical capacity of a classical channel (his result is a weak-converse bound).
The strong converse for the feedback-assisted classical capacity of a classical channel was established independently by Kemperman and Kesten. Kesten's proof appeared in \citep[Chapter~4]{Wolfowitz1964} and Kemperman's proof appeared later in \citep{Kemp71}. See \citep{Ulrey76} for a discussion of this history.
\citet{PV10} employed a R\'enyi-entropic method to extend Shannon's result to a strong converse statement, i.e., that the mutual information of a classical channel is equal to the strong converse feedback-assisted classical capacity.

\citet{Bowen04}  proved that a quantum feedback channel does not increase the entanglement-assisted classical capacity of a quantum channel (his result is a weak-converse bound). \citet{BDHSW12} proved that the mutual information of a quantum channel is equal to the strong converse quantum-feedback-assisted classical capacity. Their approach was to employ the quantum reverse Shannon theorem to do so. \citet{CMW14} used a R\'enyi-entropic method to prove this same result, and this is the approach that we have followed in this book. As far as we are aware, the concept of amortized mutual information of a quantum channel and the fact that it reduces to the mutual information of a quantum channel are original to this book.

\chapter{Classical-Feedback-Assisted Communication}

In this chapter, we continue with our study of feedback-assisted capacities.
The class of protocols that we consider in this chapter are very similar to
those from the previous chapter (Chapter~\ref{chap-QFACC}), with the exception that the feedback channel
is a classical channel instead of a quantum channel. The resulting
communication task is then called classical communication assisted by a
classical channel (or classical-feedback-assisted communication for short).

Interestingly, this slight change has the effect of complicating the theory
quite a bit:\ a general expression for the capacity is not known. It is
only known for certain channels such as entanglement-breaking channels and erasure
channels. Additionally, there are examples of channels for which classical
feedback can increase the classical capacity significantly, due to the
interplay between classical feedback and entanglement that can be generated by
the channel. We do not discuss this example in this chapter and instead point
to the Bibliographic Notes for details (Section~\ref{sec-CF:bib-notes}). All
of the above implies that the increase of capacity due to classical feedback
is a truly quantum-mechanical phenomenon that separates the classical and
quantum theories of communication. Indeed, it is necessary for a channel to
have the ability to generate entanglement in order for classical feedback to
give a boost to capacity.

Our main focus in this chapter is on establishing upper bounds on the
classical-feedback-assisted capacity. First, we prove that classical feedback
does not increase the capacity of entanglement-breaking channels. The main
tools here are similar to those employed in Section~\ref{sec-classical_capacity_ent_break}. Next, we establish
that the average output entropy of a channel is an upper bound on the
feedback-assisted capacity. Finally, we establish that the $\Upsilon
$-information of a channel, introduced in Section~\ref{subsubsec-cc_Upsilon_inf}, is actually an upper bound
on the feedback-assisted capacity. We close out the chapter by discussing some
example channels and summarizing the main concepts presented.

\section{$n$-Shot Classical Feedback-Assisted Communication Protocols}

\label{sec-CF:FB-assisted-protocol}In this section, we briefly summarize what
is meant by an $n$-shot protocol for classical communication assisted by a
classical feedback channel, where $n\in\mathbb{N}$. This section is brief
because such a protocol is defined exactly as in Section~\ref{sec-QFACC:n-shot-prot}, with the
exception that every feedback channel acting on system~$F_{i}$, sent from the
receiver to the sender, for all $i\in\left\{  0,\ldots,n\right\}  $, is a
classical feedback channel of the following form:%
\begin{equation}
\overline{\Delta}_{F_{i}}(\rho_{F_{i}})\coloneqq \sum_{j=0}^{d_{F_{i}}-1}%
|j\rangle\!\langle j|_{F_{i}}\rho_{F_{i}}|j\rangle\!\langle j|_{F_{i}},
\label{eq-CF:classical-FB-channel}%
\end{equation}
where $\{|j\rangle\}_{j=0}^{d_{F_{i}}-1}$ is some orthonormal basis known to
both the sender and the receiver.

In short, every such protocol has the form given in Figure~\ref{fig:Feedback-assisted-class-prot}, with the
aforementioned exception that every channel acting on $F_{i}$, for
$i\in\left\{  0,\ldots,n\right\}  $, is a classical feedback channel as in
\eqref{eq-CF:classical-FB-channel}. Every such protocol is defined by the
following elements:%
\begin{equation}
(\mathcal{M},\Psi_{F_{0}B_{0}^{\prime}},\mathcal{E}_{M^{\prime}F_{0}%
\rightarrow A_{1}^{\prime}A_{1}}^{0},\{\mathcal{E}_{A_{i}^{\prime}%
F_{i}\rightarrow A_{i+1}^{\prime}A_{i+1}}^{i},\mathcal{D}_{B_{i}%
B_{i-1}^{\prime}\rightarrow F_{i}B_{i}^{\prime}}^{i}\}_{i=1}^{n-1}%
,\mathcal{D}_{B_{n}B_{n-1}^{\prime}\rightarrow\hat{M}}^{n}),
\end{equation}
where $\mathcal{M}$ is the message set, $\Psi_{F_{0}B_{0}^{\prime}}$ is a
bipartite quantum state, the objects denoted by $\mathcal{E}$ are encoding
channels, and those denoted by $\mathcal{D}$ decoding channels. Let $\mathcal{C}$ denote
all of these elements, which together constitute the
classical-feedback-assisted code. The systems labeled by $F$ are feedback
systems, the $i$th of which is sent by the receiver Bob to the sender Alice
through the classical feedback channel in \eqref{eq-CF:classical-FB-channel}.
The initial state of such a protocol, prepared by Alice, is $\overline{\Phi
}_{MM^{\prime}}^{p}$, as defined in \eqref{eq:QFBA-classically-correlated-Phi}. The states throughout the protocol
are the same as defined in Section~\ref{sec-QFACC:n-shot-prot}, with the exception that every state with
an $F$ label is replaced by the same state succeeded by the completely
dephasing channel $\overline{\Delta}_{F_{i}}$. That is, the initial state is%
\begin{equation}
\overline{\Phi}_{MM^{\prime}}^{p}\otimes\overline{\Delta}_{F_{0}}(\Psi
_{F_{0}B_{0}^{\prime}}), \label{eq-CF:initial-state}%
\end{equation}
and the other states are%
\begin{align}
\rho_{MA_{1}^{\prime}B_{1}B_{0}^{\prime}}^{1}  &  \coloneqq (\mathcal{N}%
_{A_{1}\rightarrow B_{1}}\circ\mathcal{E}_{M^{\prime}F_{0}\rightarrow
A_{1}^{\prime}A_{1}}^{0})(\overline{\Phi}_{MM^{\prime}}^{p}\otimes
\overline{\Delta}_{F_{0}}(\Psi_{F_{0}B_{0}^{\prime}}%
)),\label{eq-CF:rho-1-state}\\
\rho_{MA_{2}^{\prime}B_{2}B_{1}^{\prime}}^{2}  &  \coloneqq (\mathcal{N}%
_{A_{2}\rightarrow B_{2}}\circ\mathcal{E}_{A_{1}^{\prime}F_{1}\rightarrow
A_{2}^{\prime}A_{2}}^{1}\circ\overline{\Delta}_{F_{1}}\circ\mathcal{D}%
_{B_{1}B_{0}^{\prime}\rightarrow F_{1}B_{1}^{\prime}}^{1})(\rho_{MA_{1}%
^{\prime}B_{1}B_{0}^{\prime}}^{1}),
\end{align}%
\begin{multline}
\rho_{MA_{i}^{\prime}B_{i}B_{i-1}^{\prime}}^{i}\coloneqq \\
(\mathcal{N}_{A_{i}\rightarrow B_{i}}\circ\mathcal{E}_{A_{i-1}^{\prime}%
F_{i-1}\rightarrow A_{i}^{\prime}A_{i}}^{i-1}\circ\overline{\Delta}_{F_{i-1}%
}\circ\mathcal{D}_{B_{i-1}B_{i-2}^{\prime}\rightarrow F_{i-1}B_{i-1}^{\prime}%
}^{i-1})(\rho_{MA_{2}^{\prime}B_{2}B_{1}^{\prime}}^{2}),
\end{multline}
where $i\in\left\{  3,\ldots,n\right\}  $. The final state of the protocol is
then as follows:%
\begin{equation}
\omega_{M\hat{M}}^{p}\coloneqq \mathcal{D}_{B_{n}B_{n-1}^{\prime}\rightarrow\hat{M}%
}^{n}(\operatorname{Tr}_{A_{n}^{\prime}}[\rho_{MA_{n}^{\prime}B_{n}%
B_{n-1}^{\prime}}^{n}]). \label{eq-CF:final-omega-state}%
\end{equation}

Consider that the initial state of the protocol, as given in
\eqref{eq-CF:initial-state}, has the following form:%
\begin{multline}
\overline{\Phi}_{MM^{\prime}}^{p}\otimes\overline{\Delta}_{F_{0}}(\Psi
_{F_{0}B_{0}^{\prime}})=\\
\sum_{m\in\mathcal{M}}p(m)|m\rangle\!\langle m|_{M}\otimes|m\rangle\!\langle
m|_{M^{\prime}}\otimes\sum_{f_{0}}p(f_{0})|f_{0}\rangle\!\langle f_{0}%
|_{F_{0}}\otimes\Psi_{B_{0}^{\prime}}^{f_{0}},
\label{eq-CF:initial-state-detailed}
\end{multline}
where $p(f_{0})$ is a probability distribution over the possible classical
values sent through the feedback channel and each $\Psi_{B_{0}^{\prime}%
}^{f_{0}}$ is a state of the system $B_{0}^{\prime}$. After the encoding
channel $\mathcal{E}_{M^{\prime}F_{0}\rightarrow A_{1}^{\prime}A_{1}}^{0}$
acts, the state becomes as follows:%
\begin{multline}
\mathcal{E}_{M^{\prime}F_{0}\rightarrow A_{1}^{\prime}A_{1}}^{0}%
(\overline{\Phi}_{MM^{\prime}}^{p}\otimes\overline{\Delta}_{F_{0}}(\Psi
_{F_{0}B_{0}^{\prime}}))=\label{eq-CF:after-encoding-EB}\\
\sum_{m\in\mathcal{M}}\sum_{f_{0}}p(m)p(f_{0})|m\rangle\!\langle m|_{M}%
\otimes\varsigma_{A_{1}^{\prime}A_{1}}^{0,m,f_{0}}\otimes\Psi_{B_{0}^{\prime}%
}^{f_{0}},
\end{multline}
where the state $\varsigma_{A_{1}^{\prime}A_{1}}^{0,m,f_{0}}$ is defined as%
\begin{equation}
\varsigma_{A_{1}^{\prime}A_{1}}^{0,m,f_{0}}\coloneqq \mathcal{E}_{M^{\prime}%
F_{0}\rightarrow A_{1}^{\prime}A_{1}}^{0}(|m\rangle\!\langle m|_{M^{\prime}%
}\otimes|f_{0}\rangle\!\langle f_{0}|_{F_{0}}).
\end{equation}
Then one can proceed from here, defining states of the protocol conditioned on
the value of the message and the classical feedback.

Just as we did in Chapter~\ref{chap-EA_capacity}, we define the message error probability,
average error probability, and maximal error probability, as in \eqref{eq-eac-mess_error_prob}, \eqref{eq-eac-avg_error_prob}, and \eqref{eq-eac-maximal_error_prob}, respectively. Using the expression in \eqref{eq-eacc_trace_dist_avg_error}, the average error
probability for the classical-feedback-assisted code $\mathcal{C}$ is given by%
\begin{equation}
\overline{p}_{\text{err}}(\mathcal{C};p)\coloneqq \frac{1}{2}\left\Vert \overline
{\Phi}_{M\hat{M}}^{p}-\omega_{M\hat{M}}^{p}\right\Vert _{1},
\end{equation}
and using the expression in \eqref{eq-eacc_max_err_prob_trace_distance}, the maximal error probability for the
classical-feedback-assisted code $\mathcal{C}$ is given by%
\begin{equation}
p_{\text{err}}^{\ast}(\mathcal{C})\coloneqq \max_{p:\mathcal{M}\rightarrow\left[
0,1\right]  }\frac{1}{2}\left\Vert \overline{\Phi}_{M\hat{M}}^{p}%
-\omega_{M\hat{M}}^{p}\right\Vert _{1}, \label{eq-CF:maximal-error}%
\end{equation}
where the maximization is over every probability distribution $p$.

\begin{definition}
{$(n,\left\vert \mathcal{M}\right\vert ,\varepsilon)$
Classical-Feedback-Assisted Classical Communication Protocol}{def-CF:cfacc-protocol}
Let
$(\mathcal{M},\Psi_{F_{0}B_{0}^{\prime}},\mathcal{E}_{M^{\prime}%
F_{0}\rightarrow A_{1}^{\prime}A_{1}}^{0},\{\mathcal{E}_{A_{i}^{\prime}%
F_{i}\rightarrow A_{i+1}^{\prime}A_{i+1}}^{i},\mathcal{D}_{B_{i}%
B_{i-1}^{\prime}\rightarrow F_{i}B_{i}^{\prime}}^{i}\}_{i=1}^{n}%
,\mathcal{D}_{B_{n}B_{n-1}^{\prime}\rightarrow\hat{M}}^{n})$ be the elements
of an $n$-shot classical-feedback-assisted classical communication protocol
over the channel $\mathcal{N}_{A\rightarrow B}$. The protocol is called an
$(n,\left\vert \mathcal{M}\right\vert ,\varepsilon)$ protocol, with
$\varepsilon\in\left[  0,1\right]  $, if $p_{\text{err}}^{\ast}(\mathcal{C}%
)\leq\varepsilon$.
\end{definition}

\section{Protocol over a Useless Channel}

\label{sec-CF:useless-protocol}A common theme in this book has been that we
can derive converse bounds by using a generalized divergence to compare the
output of the actual protocol with one that is useless for the task. We did
exactly this in Section~\ref{sec:QFBA-replacer-useless} of the previous chapter, and the only change
that we make here, as in the previous section, is to replace every quantum
feedback channel with a classical one. Figure~\ref{fig-eaqf_classical_comm_n_shot_useless} applies again and we
briefly define the steps and states exactly as before, except with this key
difference for both the figure and the states involved. A useless channel is
one that traces out the input and replaces it with some state at the output:%
\begin{equation}
\mathcal{R}_{A\rightarrow B}\coloneqq \mathcal{P}_{\sigma_{B}}\circ\operatorname{Tr}%
_{A}, \label{eq-CF:replacement-channel}%
\end{equation}
where $\mathcal{P}_{\sigma_{B}}$ denotes a preparation channel that prepares
the state $\sigma_{B}$ at the output. The initial state of this protocol is%
\begin{equation}
\overline{\Phi}_{MM^{\prime}}^{p}\otimes\overline{\Delta}_{F_{0}}(\Psi
_{F_{0}B_{0}^{\prime}}),
\end{equation}
and the others are as follows:%
\begin{align}
\tau_{MA_{1}^{\prime}B_{1}B_{0}^{\prime}}^{1}  &  \coloneqq (\mathcal{R}%
_{A_{1}\rightarrow B_{1}}\circ\mathcal{E}_{M^{\prime}F_{0}\rightarrow
A_{1}^{\prime}A_{1}}^{0})(\overline{\Phi}_{MM^{\prime}}^{p}\otimes
\overline{\Delta}_{F_{0}}(\Psi_{F_{0}B_{0}^{\prime}})),\\
\tau_{MA_{2}^{\prime}B_{2}B_{1}^{\prime}}^{2}  &  \coloneqq (\mathcal{R}%
_{A_{2}\rightarrow B_{2}}\circ\mathcal{E}_{A_{1}^{\prime}F_{1}\rightarrow
A_{2}^{\prime}A_{2}}^{1}\circ\overline{\Delta}_{F_{1}}\circ\mathcal{D}%
_{B_{1}B_{0}^{\prime}\rightarrow F_{1}B_{1}^{\prime}}^{1})(\rho_{MA_{1}%
^{\prime}B_{1}B_{0}^{\prime}}^{1}),
\end{align}%
\begin{multline}
\tau_{MA_{i}^{\prime}B_{i}B_{i-1}^{\prime}}^{i}\coloneqq \\
(\mathcal{R}_{A_{i}\rightarrow B_{i}}\circ\mathcal{E}_{A_{i-1}^{\prime}%
F_{i-1}\rightarrow A_{i}^{\prime}A_{i}}^{i-1}\circ\overline{\Delta}_{F_{i-1}%
}\circ\mathcal{D}_{B_{i-1}B_{i-2}^{\prime}\rightarrow F_{i-1}B_{i-1}^{\prime}%
}^{i-1})(\rho_{MA_{2}^{\prime}B_{2}B_{1}^{\prime}}^{2}),
\end{multline}
where $i\in\left\{  3,\ldots,n\right\}  $. The final state of the protocol is
then as follows:%
\begin{equation}
\omega_{M\hat{M}}^{p}\coloneqq \mathcal{D}_{B_{n}B_{n-1}^{\prime}\rightarrow\hat{M}%
}^{n}(\operatorname{Tr}_{A_{n}^{\prime}}[\rho_{MA_{n}^{\prime}B_{n}%
B_{n-1}^{\prime}}^{n}]).
\end{equation}

Going through calculations similar to those in \eqref{eq-qfacc:1st-state-useless-prot}--\eqref{eq-qfba_useless_states_i}, we arrive at the
following conclusions:%
\begin{align}
\tau_{MA_{1}^{\prime}B_{1}B_{0}^{\prime}}^{1}  &  =\tau_{MA_{1}^{\prime}%
B_{0}^{\prime}}^{1}\otimes\sigma_{B_{1}},\\
\tau_{MB_{1}B_{0}^{\prime}}^{1}  &  =\pi_{M}^{p}\otimes\Psi_{B_{0}^{\prime}%
}\otimes\sigma_{B_{1}},\\
\tau_{MB_{2}B_{1}^{\prime}}^{2}  &  =\pi_{M}^{p}\otimes\tau_{B_{1}^{\prime}%
}^{2}\otimes\sigma_{B_{2}},\\
\tau_{MB_{i}B_{i-1}^{\prime}}^{i}  &  =\pi_{M}^{p}\otimes\tau_{B_{i-1}%
^{\prime}}^{i}\otimes\sigma_{B_{i}},
\label{eq-CF:tau-state-useless-protocol-i}%
\end{align}
where $i\in\left\{  3,\ldots,n\right\}  $ and%
\begin{equation}
\pi_{M}^{p}\coloneqq \sum_{m\in\mathcal{M}}p(m)|m\rangle\!\langle m|_{M}.
\end{equation}
Thus, there is no correlation whatsoever between the message system $M$ and
Bob's systems $B_{i}B_{i-1}^{\prime}$, for each $i\in\left\{  1,\ldots
,n\right\}  $, after tracing over Alice's systems. As before, this is a
consequence of the fact that the \textquotedblleft communication line has been
cut\textquotedblright\ when employing the replacement channel.

Bob's final decoding channel $\mathcal{D}_{B_{n}B_{n-1}^{\prime}%
\rightarrow\hat{M}}^{n}$ thus leads to the following classical--classical
state:%
\begin{equation}
\tau_{M\hat{M}}\coloneqq \pi_{M}^{p}\otimes\tau_{\hat{M}},
\label{eq-CF:final-state-useless-prot}%
\end{equation}
where $\tau_{\hat{M}}\coloneqq \sum_{\hat{m}\in\mathcal{M}}t(\hat{m})|\hat{m}%
\rangle\!\langle\hat{m}|_{\hat{M}}$, for some probability distribution
$t:\mathcal{M}\rightarrow\left[  0,1\right]  $, which corresponds to Bob's measurement.

\section{Upper Bounds on the Number of Transmitted Bits}

We now provide several upper bounds on the number of transmitted bits for
classical communication protocols assisted by classical feedback. The first
bounds that we discuss apply only to entanglement-breaking channels (recall
Definition~\ref{def-ent_break_chan}), and they imply that classical feedback increases neither
the asymptotic classical capacity of entanglement-breaking channels nor their
strong converse classical capacity. Then the next two bounds apply to all
quantum channels, and they are known as the entropy bound and the geometric
$\Upsilon$-information bound.

\subsection{Upper Bounds for Entanglement-Breaking Channels}

Before stating the main theorem of this section, we discuss particular aspects
of a classical-feedback-assisted protocol for classical communication over an
entanglement-breaking channel. Indeed, suppose that $\mathcal{N}_{A\rightarrow
B}$ is an entanglement-breaking channel. We begin our analysis by inspecting
the state in \eqref{eq-CF:after-encoding-EB}. This state is fully
separable with respect to the cut $M:A_{1}^{\prime}A_{1}:B_{0}^{\prime}$. That is, it can be written as follows:
\begin{equation}
\sum_z q(z) \tau^z_M \otimes \sigma^z_{A_{1}^{\prime}A_{1}} \otimes \omega^z_{B_0},
\end{equation}
for $q$ a probability distribution and $\{\tau^z_M\}_z$, $\{\sigma^z_{A_{1}^{\prime}A_{1}}\}_z$, and $\{\omega^z_{B_0}\}_z$ sets of states.
Since the channel $\mathcal{N}_{A\rightarrow B}$ is entanglement breaking,
when it acts on system $A_{1}$ of the state $\varsigma_{A_{1}^{\prime}A_{1}%
}^{0,m,f_{0}}$ in \eqref{eq-CF:after-encoding-EB}, the resulting state is a separable state of the following
form:%
\begin{equation}
\mathcal{N}_{A_{1}\rightarrow B_{1}}(\varsigma_{A_{1}^{\prime}A_{1}%
}^{0,m,f_{0}})=\sum_{y}p(y|m,f_{0})\varsigma_{A_{1}^{\prime}}^{y,m,f_{0}%
}\otimes\varsigma_{B_{1}}^{y,m,f_{0}}.
\end{equation}
So this implies that the state $\rho_{MA_{1}^{\prime}B_{1}B_{0}^{\prime}}^{1}%
$, as defined in \eqref{eq-CF:rho-1-state} and with $\mathcal{N}%
_{A_{1}\rightarrow B_{1}}$ entanglement breaking, is fully separable across
all systems (i.e., with respect to the cut $M:A_{1}^{\prime}:B_{1}%
:B_{0}^{\prime}$).

Bob then applies the decoding channel $\overline{\Delta}_{F_{1}}%
\circ\mathcal{D}_{B_{1}B_{0}^{\prime}\rightarrow F_{1}B_{1}^{\prime}}^{1}$ and
the state at this point is as follows:%
\begin{multline}
(\overline{\Delta}_{F_{1}}\circ\mathcal{D}_{B_{1}B_{0}^{\prime}\rightarrow
F_{1}B_{1}^{\prime}}^{1})(\rho_{MA_{1}^{\prime}B_{1}B_{0}^{\prime}}^{1})=\\
\sum_{m\in\mathcal{M}}\sum_{f_{0},y}p(y|m,f_{0})p(m)p(f_{0})|m\rangle\!\langle
m|_{M}\otimes\\
\varsigma_{A_{1}^{\prime}}^{y,m,f_{0}}\otimes(\overline{\Delta}_{F_{1}}%
\circ\mathcal{D}_{B_{1}B_{0}^{\prime}\rightarrow F_{1}B_{1}^{\prime}}%
^{1})(\varsigma_{B_{1}}^{y,m,f_{0}}\otimes\Psi_{B_{0}^{\prime}}^{f_{0}}).
\end{multline}
Since the $F_{1}$ system is classical, the state $(\overline{\Delta}_{F_{1}%
}\circ\mathcal{D}_{B_{1}B_{0}^{\prime}\rightarrow F_{1}B_{1}^{\prime}}%
^{1})(\varsigma_{B_{1}}^{y,m,f_{0}}\otimes\Psi_{B_{0}^{\prime}}^{f_{0}})$ can
be written as%
\begin{equation}
(\overline{\Delta}_{F_{1}}\circ\mathcal{D}_{B_{1}B_{0}^{\prime}\rightarrow
F_{1}B_{1}^{\prime}}^{1})(\varsigma_{B_{1}}^{y,m,f_{0}}\otimes\Psi
_{B_{0}^{\prime}}^{f_{0}})=\sum_{f_{1}}p(f_{1}|y,m,f_{0})|f_{1}\rangle
\!\langle f_{1}|_{F_{1}}\otimes\varsigma_{B_{1}^{\prime}}^{f_{1},y,m,f_{0}}.
\end{equation}
This means that the state $(\overline{\Delta}_{F_{1}}\circ\mathcal{D}%
_{B_{1}B_{0}^{\prime}\rightarrow F_{1}B_{1}^{\prime}}^{1})(\rho_{MA_{1}%
^{\prime}B_{1}B_{0}^{\prime}}^{1})$ is fully separable with respect to
the cut $M:A_{1}^{\prime}:F_{1}:B_{1}^{\prime}$.

This process continues, and since the channel $\mathcal{N}_{A\rightarrow B}$
is entanglement breaking, by following an analysis similar to that given
above, we observe that the state of the message system $M$,  Alice's, and
Bob's is always fully separable throughout the protocol. This is the
key reason that we obtain the bounds given in the following theorem:

\begin{theorem*}
{$n$-Shot Upper Bounds for Classical Feedback Assisted Classical Communication
over Entanglement Breaking Channels}{thm-CF:ent-break-upp-bnd}Let
$\mathcal{N}_{A\rightarrow B}$ be an entanglement-breaking channel, and let
$\varepsilon\in\lbrack0,1)$. For all $(n,\left\vert \mathcal{M}\right\vert
,\varepsilon)$ classical-feedback-assisted classical communication protocols
over the channel $\mathcal{N}_{A\rightarrow B}$, the following bounds hold%
\begin{align}
\frac{\log_{2}\!\left\vert \mathcal{M}\right\vert }{n}  &  \leq\frac
{1}{1-\varepsilon}\left(  \chi(\mathcal{N})+\frac{1}{n}h_{2}(\varepsilon
)\right)  ,\label{eq-CF:EB-bnd-weak-converse-n-shot}\\
\frac{\log_{2}\!\left\vert \mathcal{M}\right\vert }{n}  &  \leq\widetilde
{\chi}_{\alpha}(\mathcal{N})+\frac{\alpha}{n\left(  \alpha-1\right)  }\log
_{2}\!\left(  \frac{1}{1-\varepsilon}\right)  \qquad\forall\alpha>1,
\label{eq-CF:EB-bnd-renyi-n-shot}%
\end{align}
where $\chi(\mathcal{N})$ is the Holevo information of $\mathcal{N}%
_{A\rightarrow B}$, as defined in \eqref{eq-Hol_inf_chan}, and $\widetilde{\chi}_{\alpha
}(\mathcal{N})$ is the sandwiched R\'enyi Holevo information of $\mathcal{N}%
_{A\rightarrow B}$, as defined in \eqref{eq-sand_rel_Holevo_inf_chan}.
\end{theorem*}

\begin{Proof}
Applying precisely the same reasoning as in the beginning of the proof of
Theorem~\ref{thm:QFBA-upper-bnds-finite-length}, we conclude the following bound:%
\begin{equation}
\log_{2}\!\left\vert \mathcal{M}\right\vert \leq I_{H}^{\varepsilon}(M;\hat
{M})_{\omega}, \label{eq-CF:hypo-test-bnd-ent-break}%
\end{equation}
where $\omega_{M\hat{M}}$ is the final state of the protocol when the
distribution $p$ is set to the uniform distribution.

Invoking Proposition~\ref{prop-hypo_to_rel_ent}, the definition of $I_{H}^{\varepsilon}(M;\hat{M})$
from \eqref{eq-hypo_testing_mutual_inf}, and the expression for the mutual information from \eqref{eq-mut_inf_opt}, we find
that%
\begin{equation}
I_{H}^{\varepsilon}(M;\hat{M})_{\omega}\leq\frac{1}{1-\varepsilon}\left(
I(M;\hat{M})_{\omega}+h_{2}(\varepsilon)\right)  .
\label{eq-CF:hypo-test-bnd-ent-break-2}%
\end{equation}
Now employing the data-processing inequality for the mutual information with
respect to the last decoding channel $\mathcal{D}_{B_{n}B_{n-1}^{\prime
}\rightarrow\hat{M}}^{n}$, we find that%
\begin{equation}
I(M;\hat{M})_{\omega}\leq I(M;B_{n}B_{n-1}^{\prime})_{\rho^{n}}.
\label{eq-CF:apply-holevo-bnd-ent-break}%
\end{equation}
Then using the chain for the mutual information in \eqref{eq-q_mut_inf_chain_rule}, we obtain%
\begin{align}
I(M;B_{n}B_{n-1}^{\prime})_{\rho^{n}}  &  =I(M;B_{n}|B_{n-1}^{\prime}%
)_{\rho^{n}}+I(M;B_{n-1}^{\prime})_{\rho^{n}}%
\label{eq-CF:chain-rule-ent-break-1}\\
&  \leq I(MB_{n-1}^{\prime};B_{n})_{\rho^{n}}+I(M;B_{n-1}^{\prime})_{\rho^{n}%
}. \label{eq-CF:chain-rule-ent-break-2}%
\end{align}
As mentioned above, the state shared between Alice and Bob, at any point
during the protocol, is a separable state. Thus, the global state before the
$n$th channel use can be written as follows:%
\begin{equation}
\rho_{MA_{n}^{\prime}A_{n}B_{n-1}^{\prime}}^{n}=\frac{1}{\left\vert
\mathcal{M}\right\vert }\sum_{m\in\mathcal{M}}|m\rangle\!\langle m|_{M}%
\otimes\sum_{y}p(y|m)\varsigma_{A_{n}^{\prime}A_{n}}^{m,y}\otimes
\varsigma_{B_{n-1}^{\prime}}^{m,y}.
\end{equation}
Then the state after the $n$th channel acts is as follows:%
\begin{align}
\rho_{MA_{n}^{\prime}B_{n}B_{n-1}^{\prime}}^{n}  &  =\mathcal{N}%
_{A_{n}\rightarrow B_{n}}(\rho_{MA_{n}^{\prime}A_{n}B_{n-1}^{\prime}}^{n})\\
&  =\frac{1}{\left\vert \mathcal{M}\right\vert }\sum_{m\in\mathcal{M}%
}|m\rangle\!\langle m|_{M}\otimes\sum_{y}p(y|m)\mathcal{N}_{A_{n}\rightarrow
B_{n}}(\varsigma_{A_{n}^{\prime}A_{n}}^{m,y})\otimes\varsigma_{B_{n-1}%
^{\prime}}^{m,y}.
\end{align}
An extension of the state above is as follows:%
\begin{multline}
\rho_{MYA_{n}^{\prime}B_{n}B_{n-1}^{\prime}}^{n}=\\
\frac{1}{\left\vert \mathcal{M}\right\vert }\sum_{m\in\mathcal{M}}\sum
_{y}p(y|m)|m\rangle\!\langle m|_{M}\otimes|y\rangle\!\langle y|_{Y}%
\otimes\mathcal{N}_{A_{n}\rightarrow B_{n}}(\varsigma_{A_{n}^{\prime}A_{n}%
}^{m,y})\otimes\varsigma_{B_{n-1}^{\prime}}^{m,y},
\end{multline}
and tracing over the system $A_{n}^{\prime}$ leads to the following state:%
\begin{align}
&  \rho_{MYB_{n}B_{n-1}^{\prime}}^{n}\nonumber\\
&  =\operatorname{Tr}_{A_{n}^{\prime}}[\rho_{MYA_{n}^{\prime}B_{n}%
B_{n-1}^{\prime}}^{n}]\\
&  =\frac{1}{\left\vert \mathcal{M}\right\vert }\sum_{m\in\mathcal{M}}\sum
_{y}p(y|m)|m\rangle\!\langle m|_{M}\otimes|y\rangle\!\langle y|_{Y}%
\otimes\mathcal{N}_{A_{n}\rightarrow B_{n}}(\varsigma_{A_{n}}^{m,y}%
)\otimes\varsigma_{B_{n-1}^{\prime}}^{m,y}. \label{eq-CF:nth-ch-use-state-EB}%
\end{align}
Then consider that%
\begin{align}
I(MB_{n-1}^{\prime};B_{n})_{\rho^{n}}  &  \leq I(MYB_{n-1}^{\prime}%
;B_{n})_{\rho^{n}}\label{eq-CF:Holevo-info-bnd-ent-break-1}\\
&  =I(MY;B_{n})_{\rho^{n}}+I(B_{n-1}^{\prime};B_{n}|MY)_{\rho^{n}}\\
&  =I(MY;B_{n})_{\rho^{n}}\\
&  \leq\chi(\mathcal{N}). \label{eq-CF:Holevo-info-bnd-ent-break}%
\end{align}
The first inequality follows from the data-processing inequality for mutual
information. The first equality follows from the chain rule. The second
equality follows because the state in \eqref{eq-CF:nth-ch-use-state-EB} is
product when conditioning on the systems $M$ and $Y$. The last inequality
follows because the state $\rho_{MYB_{n}}^{n}$ is a classical--quantum state
of the following form:%
\begin{align}
\rho_{MYB_{n}}^{n}  &  =\operatorname{Tr}_{B_{n-1}^{\prime}}[\rho
_{MYB_{n}B_{n-1}^{\prime}}^{n}]\\
&  =\frac{1}{\left\vert \mathcal{M}\right\vert }\sum_{m\in\mathcal{M}}\sum
_{y}p(y|m)|m\rangle\!\langle m|_{M}\otimes|y\rangle\!\langle y|_{Y}%
\otimes\mathcal{N}_{A_{n}\rightarrow B_{n}}(\varsigma_{A_{n}}^{m,y}).
\end{align}
Thus, the definition of the Holevo information in \eqref{eq-Hol_inf_chan} implies the last
inequality in \eqref{eq-CF:Holevo-info-bnd-ent-break}. Putting together
\eqref{eq-CF:apply-holevo-bnd-ent-break},
\eqref{eq-CF:chain-rule-ent-break-1}--\eqref{eq-CF:chain-rule-ent-break-2},
and
\eqref{eq-CF:Holevo-info-bnd-ent-break-1}--\eqref{eq-CF:Holevo-info-bnd-ent-break},
we conclude that%
\begin{align}
I(M;\hat{M})_{\omega}  &  \leq\chi(\mathcal{N})+I(M;B_{n-1}^{\prime}%
)_{\rho^{n}}\\
&  \leq\chi(\mathcal{N})+I(M;B_{n-1}B_{n-2}^{\prime})_{\rho^{n-1}},
\end{align}
where the last inequality follows from the data-processing inequality for
mutual information.

Now, we recognize the term $I(M;B_{n-1}B_{n-2}^{\prime})_{\rho^{n-1}}$ as
being of the same form as $I(M;B_{n}B_{n-1}^{\prime})_{\rho^{n}}$ in
\eqref{eq-CF:chain-rule-ent-break-1}. Thus, we iterate through the same
sequence of arguments to conclude that%
\begin{equation}
I(M;B_{n-1}B_{n-2}^{\prime})_{\rho^{n-1}}\leq\chi(\mathcal{N})+I(M;B_{n-2}%
B_{n-3}^{\prime})_{\rho^{n-2}},
\end{equation}
which in turn implies that%
\begin{equation}
I(M;\hat{M})_{\omega}\leq2\chi(\mathcal{N})+I(M;B_{n-2}B_{n-3}^{\prime}%
)_{\rho^{n-2}}.
\end{equation}
Continuing all the way back to the first channel use, we find that%
\begin{equation}
I(M;\hat{M})_{\omega}\leq n\chi(\mathcal{N})
\label{eq-CF:MI-bnd-all-ch-uses-ent-break}%
\end{equation}
because $I(M;B_{0}^{\prime})=0$ (the systems $M$ and $B_{0}^{\prime}$ are in a
product state at the start of the protocol). Putting together
\eqref{eq-CF:hypo-test-bnd-ent-break},
\eqref{eq-CF:hypo-test-bnd-ent-break-2}, and
\eqref{eq-CF:MI-bnd-all-ch-uses-ent-break}, we conclude that%
\begin{equation}
\log_{2}\!\left\vert \mathcal{M}\right\vert \leq\frac{1}{1-\varepsilon}\left(
n\chi(\mathcal{N})+h_{2}(\varepsilon)\right)  ,
\end{equation}
which implies the claim in \eqref{eq-CF:EB-bnd-weak-converse-n-shot}.

We now prove the inequality in \eqref{eq-CF:EB-bnd-renyi-n-shot}. Our starting
point is again \eqref{eq-CF:hypo-test-bnd-ent-break}, but from there, we
instead apply Proposition~\ref{prop:sandwich-to-htre}, the definition of $I_{H}^{\varepsilon}%
(M;\hat{M})_{\omega}$ from \eqref{eq-hypo_testing_mutual_inf}, and the expression for the sandwiched R\'enyi
mutual information from \eqref{eq:QEI:sand-Ren-MI-states} to find that the following holds for all
$\alpha>1$:%
\begin{equation}
I_{H}^{\varepsilon}(M;\hat{M})_{\omega}\leq\widetilde{I}_{\alpha}(M;\hat
{M})_{\omega}+\frac{\alpha}{\alpha-1}\log_{2}\!\left(  \frac{1}{1-\varepsilon
}\right)  . \label{eq-CF:htre-to-renyi-bnd-EB}%
\end{equation}
Recall that the sandwiched R\'enyi mutual information $\widetilde{I}_{\alpha
}(M;\hat{M})_{\omega}$ is defined as%
\begin{align}
\widetilde{I}_{\alpha}(M;\hat{M})_{\omega}  &  =\inf_{\xi_{\hat{M}}}%
\widetilde{D}_{\alpha}(\omega_{M\hat{M}}\Vert\omega_{M}\otimes\xi_{\hat{M}})\\
&  =\inf_{\xi_{\hat{M}}}\widetilde{D}_{\alpha}(\omega_{M\hat{M}}\Vert\pi
_{M}\otimes\xi_{\hat{M}}).
\end{align}
We adopt a similar approach to that given for the proof of \eqref{eq:QFBA-strong-converse-n-shot-bnd}. Our goal is
thus to compare the actual protocol with one that results from employing a
useless, replacement channel (of the form discussed in
Section~\ref{sec-CF:useless-protocol}). To this end, let $\mathcal{R}%
_{A\rightarrow B}^{\sigma_{B}}$ be the replacement channel defined in
\eqref{eq-CF:replacement-channel}, with $\sigma_{B}$ an arbitrary state. Then
as discussed in Section~\ref{sec-CF:useless-protocol} (in particular, in
\eqref{eq-CF:final-state-useless-prot}), the final state of the protocol
conducted with the replacement channel is given by $\tau_{M\hat{M}}=\pi
_{M}\otimes\tau_{\hat{M}}$. Then, we find that%
\begin{align}
\widetilde{I}_{\alpha}(M;\hat{M})_{\omega}  &  =\inf_{\xi_{\hat{M}}}%
\widetilde{D}_{\alpha}(\omega_{M\hat{M}}\Vert\pi_{M}\otimes\xi_{\hat{M}%
})\label{eq-CF:Renyi-mi-relax-EB}\\
&  \leq\widetilde{D}_{\alpha}(\omega_{M\hat{M}}\Vert\pi_{M}\otimes\tau
_{\hat{M}})\\
&  =\widetilde{D}_{\alpha}(\omega_{M\hat{M}}\Vert\tau_{M\hat{M}}).
\label{eq-CF:Renyi-mi-relax-EB-3}%
\end{align}
By applying the data-processing inequality for the sandwiched R\'enyi relative
entropy with respect to the last decoding channel, and using
\eqref{eq-CF:tau-state-useless-protocol-i}, we find that%
\begin{align}
\widetilde{D}_{\alpha}(\omega_{M\hat{M}}\Vert\tau_{M\hat{M}})  &
\leq\widetilde{D}_{\alpha}(\rho_{MB_{n}B_{n-1}^{\prime}}^{n}\Vert\tau
_{MB_{n}B_{n-1}^{\prime}}^{n})\label{eq-CF:sandiwched-equality-1}\\
&  =\widetilde{D}_{\alpha}(\rho_{MB_{n}B_{n-1}^{\prime}}^{n}\Vert
\tau_{MB_{n-1}^{\prime}}^{n}\otimes\sigma_{B_{n}}),
\label{eq-CF:sandiwched-equality-3}%
\end{align}
It is our goal to bound this last term. To do so, consider
that%
\begin{multline}
\widetilde{D}_{\alpha}(\rho_{MB_{n}B_{n-1}^{\prime}}^{n}\Vert\tau
_{MB_{n-1}^{\prime}}^{n}\otimes\sigma_{B_{n}})=\\
\frac{\alpha}{\alpha-1}\log_{2}\!\left\Vert \left(  \Theta_{\sigma_{B_{n}%
}^{\frac{1-\alpha}{\alpha}}}\circ\mathcal{N}_{A_{n}\rightarrow B_{n}}\right)
\left(  \left(  \tau_{MB_{n-1}^{\prime}}^{n}\right)  ^{\frac{1-\alpha}%
{2\alpha}}\rho_{MA_{n}B_{n-1}^{\prime}}^{n}\left(  \tau_{MB_{n-1}^{\prime}%
}^{n}\right)  ^{\frac{1-\alpha}{2\alpha}}\right)  \right\Vert _{\alpha},
\end{multline}
where we define the completely positive map $\Theta_{X}$ by%
\begin{equation}
\Theta_{X}(\rho)\coloneqq X^{\frac{1}{2}}\rho X^{\frac{1}{2}}.
\end{equation}
We now employ the key observation from before:\ if the channel $\mathcal{N}$
is entanglement breaking, then Alice and Bob's systems are always separable
throughout the protocol. Thus, the state $\rho_{MA_{n}B_{n-1}^{\prime}}^{n}$
is fully separable with respect to the cut $M:A_{n}:B_{n-1}^{\prime}$. 
It is in turn separable with respect to the bipartite cut $A_{n}:MB_{n-1}^{\prime}$
and can be written as%
\begin{equation}
\rho_{MA_{n}B_{n-1}^{\prime}}^{n}=\sum_{j}p(j)\rho_{A_{n}}^{j}\otimes
\rho_{MB_{n-1}^{\prime}}^{j},
\end{equation}
which implies that%
\begin{align}
&  \left(  \tau_{MB_{n-1}^{\prime}}^{n}\right)  ^{\frac{1-\alpha}{2\alpha}%
}\rho_{MA_{n}B_{n-1}^{\prime}}^{n}\left(  \tau_{MB_{n-1}^{\prime}}^{n}\right)
^{\frac{1-\alpha}{2\alpha}}\nonumber\\
&  =\left(  \tau_{MB_{n-1}^{\prime}}^{n}\right)  ^{\frac{1-\alpha}{2\alpha}%
}\left(  \sum_{j}p(j)\rho_{A_{n}}^{j}\otimes\rho_{MB_{n-1}^{\prime}}%
^{j}\right)  \left(  \tau_{MB_{n-1}^{\prime}}^{n}\right)  ^{\frac{1-\alpha
}{2\alpha}}\\
&  =\sum_{j}p(j)\rho_{A_{n}}^{j}\otimes\left(  \tau_{MB_{n-1}^{\prime}}%
^{n}\right)  ^{\frac{1-\alpha}{2\alpha}}\rho_{MB_{n-1}^{\prime}}^{j}\left(
\tau_{MB_{n-1}^{\prime}}^{n}\right)  ^{\frac{1-\alpha}{2\alpha}}.
\end{align}
Since conjugation by a positive semi-definite operator is a completely
positive map, we can apply Lemma~\ref{lem-ent_break_mult_2} to conclude that%
\begin{multline}
\left\Vert \left(  \Theta_{\sigma_{B_{n}}^{\frac{1-\alpha}{\alpha}}}%
\circ\mathcal{N}_{A_{n}\rightarrow B_{n}}\right)  \left(  \left(
\tau_{MB_{n-1}^{\prime}}^{n}\right)  ^{\frac{1-\alpha}{2\alpha}}\rho
_{MA_{n}B_{n-1}^{\prime}}^{n}\left(  \tau_{MB_{n-1}^{\prime}}^{n}\right)
^{\frac{1-\alpha}{2\alpha}}\right)  \right\Vert _{\alpha}%
\label{eq-CF:key-ineq-str-conv-EB-nu-alpha}\\
\leq\nu_{\alpha}\!\left(  \Theta_{\sigma_{B_{n}}^{\frac{1-\alpha}{\alpha}}%
}\circ\mathcal{N}_{A_{n}\rightarrow B_{n}}\right)  \cdot\left\Vert \left(
\tau_{MB_{n-1}^{\prime}}^{n}\right)  ^{\frac{1-\alpha}{2\alpha}}\rho
_{MB_{n-1}^{\prime}}^{n}\left(  \tau_{MB_{n-1}^{\prime}}^{n}\right)
^{\frac{1-\alpha}{2\alpha}}\right\Vert _{\alpha},
\end{multline}
where $\nu_{\alpha}(\Theta_{\sigma_{B_{n}}^{\frac{1-\alpha}{\alpha}}}%
\circ\mathcal{N}_{A_{n}\rightarrow B_{n}})$ is defined from \eqref{eq-CC:mult-a-norm-EB-main-step} and we have
identified $MB_{n-1}^{\prime}$ with system $R$ of $P_{RA}$ in \eqref{eq-CC:P_RA_op} and
$A_{n}$ with system $A$ of $P_{RA}$. We then have the following chain of
inequalities:%
\begin{align}
&  \widetilde{D}_{\alpha}(\rho_{MB_{n}B_{n-1}^{\prime}}^{n}\Vert\tau
_{MB_{n}B_{n-1}^{\prime}}^{n})\nonumber\\
&  \leq\frac{\alpha}{\alpha-1}\log_{2}\nu_{\alpha}\!\left(  \Theta
_{\sigma_{B_{n}}^{\frac{1-\alpha}{\alpha}}}\circ\mathcal{N}_{A_{n}\rightarrow
B_{n}}\right)  +\widetilde{D}_{\alpha}(\rho_{MB_{n-1}^{\prime}}^{n}\Vert
\tau_{MB_{n-1}^{\prime}}^{n})\label{eq-CF:iterate-renyi-FB-EB-1}\\
&  \leq\frac{\alpha}{\alpha-1}\log_{2}\nu_{\alpha}\!\left(  \Theta
_{\sigma_{B_{n}}^{\frac{1-\alpha}{\alpha}}}\circ\mathcal{N}_{A_{n}\rightarrow
B_{n}}\right)  +\widetilde{D}_{\alpha}(\rho_{MB_{n-1}B_{n-2}^{\prime}}%
^{n}\Vert\tau_{MB_{n-1}B_{n-2}^{\prime}}^{n})\\
&  \leq n\frac{\alpha}{\alpha-1}\log_{2}\nu_{\alpha}\!\left(  \Theta
_{\sigma_{B}^{\frac{1-\alpha}{\alpha}}}\circ\mathcal{N}_{A\rightarrow
B}\right)  +\widetilde{D}_{\alpha}(\rho_{MB_{0}^{\prime}}^{n}\Vert\tau
_{MB_{0}^{\prime}}^{n})\\
&  =n\frac{\alpha}{\alpha-1}\log_{2}\nu_{\alpha}\!\left(  \Theta_{\sigma
_{B}^{\frac{1-\alpha}{\alpha}}}\circ\mathcal{N}_{A\rightarrow B}\right)
\label{eq-CF:iterate-renyi-FB-EB-last}%
\end{align}
The first inequality follows by combining
\eqref{eq-CF:sandiwched-equality-1}--\eqref{eq-CF:sandiwched-equality-3} and
\eqref{eq-CF:key-ineq-str-conv-EB-nu-alpha}. The second inequality follows
from the data-processing inequality for the sandwiched R\'enyi relative entropy,
with respect to the channel $\operatorname{Tr}_{F_{n-1}}\circ\overline{\Delta
}_{F_{n-1}}\circ\mathcal{D}_{B_{n-1}B_{n-2}^{\prime}\rightarrow F_{n-1}%
B_{n-1}^{\prime}}^{n-1}$. The third inequality follows by recognizing that%
\begin{equation}
\widetilde{D}_{\alpha}(\rho_{MB_{n-1}B_{n-2}^{\prime}}^{n}\Vert\tau
_{MB_{n-1}B_{n-2}^{\prime}}^{n})
\end{equation}
is the sandwiched R\'enyi relative entropy at round $n-1$ of the protocol, which
allows us to apply the argument inductively. The first equality follows
because $\rho_{MB_{0}^{\prime}}^{n}=\tau_{MB_{0}^{\prime}}^{n}$ because no
channels have been applied at this point in the protocol. Putting together
\eqref{eq-CF:hypo-test-bnd-ent-break}, \eqref{eq-CF:htre-to-renyi-bnd-EB},
\eqref{eq-CF:Renyi-mi-relax-EB}--\eqref{eq-CF:Renyi-mi-relax-EB-3},
\eqref{eq-CF:sandiwched-equality-1}, and
\eqref{eq-CF:iterate-renyi-FB-EB-1}--\eqref{eq-CF:iterate-renyi-FB-EB-last},
we conclude that%
\begin{equation}
\log_{2}\!\left\vert \mathcal{M}\right\vert \leq n\frac{\alpha}{\alpha-1}%
\log_{2}\nu_{\alpha}\!\left(  \Theta_{\sigma_{B}^{\frac{1-\alpha}{\alpha}}%
}\circ\mathcal{N}_{A\rightarrow B}\right)  +\frac{\alpha}{\alpha-1}\log
_{2}\!\left(  \frac{1}{1-\varepsilon}\right)  .
\end{equation}
Since this upper bound holds for every state $\sigma_{B}$, we can take an
infimum over all such states and conclude that%
\begin{align}
&  \log_{2}\!\left\vert \mathcal{M}\right\vert \nonumber\\
&  \leq n\frac{\alpha}{\alpha-1}\inf_{\sigma_{B}}\log_{2}\nu_{\alpha}\!\left(
\Theta_{\sigma_{B}^{\frac{1-\alpha}{\alpha}}}\circ\mathcal{N}_{A\rightarrow
B}\right)  +\frac{\alpha}{\alpha-1}\log_{2}\!\left(  \frac{1}{1-\varepsilon
}\right) \\
&  =n\widetilde{K}_{\alpha}(\mathcal{N}_{A\rightarrow B})+\frac{\alpha}%
{\alpha-1}\log_{2}\!\left(  \frac{1}{1-\varepsilon}\right) \\
&  =n\widetilde{\chi}_{\alpha}(\mathcal{N}_{A\rightarrow B})+\frac{\alpha
}{\alpha-1}\log_{2}\!\left(  \frac{1}{1-\varepsilon}\right)  .
\end{align}
The first equality follows from the definition of $\nu_{\alpha}$ in \eqref{eq-CC:mult-a-norm-EB-main-step} and
the definition of $\widetilde{K}_{\alpha}$ in \eqref{eq-CC:info-radius-K-chan}. The last equality follows
from Lemma~\ref{lem-ent_break_mult_1}.
\end{Proof}

\subsection{Entropy Upper Bound on the Number of Transmitted Bits}

We now establish an upper bound that holds for an arbitrary quantum channel.
It is equal to the maximum output entropy of the channel (Theorem~\ref{thm-CF:entropy-bound-finite-length}). A refinement of this
upper bound leads to an upper bound equal to the maximum expected output
entropy of the channel (Theorem~\ref{cor-CF:avg-ent-upper-bnd}), by writing it as a convex combination of
other channels.

We begin by establishing the first upper bound. The main idea for doing so is
to consider a protocol that simulates the general protocol detailed in
Section~\ref{sec-CF:FB-assisted-protocol}. The simulation is a purified
protocol, in which every step of the original protocol is purified. Each state
of the purified protocol, when conditioned on the message being transmitted
and the values of the classical feedback, is in a pure state. We now detail
the form of this purified protocol. In order to simplify notation, we let
$\hat{A}$ denote a joint system throughout, referring to both the original
system $A^{\prime}$ and a purifying reference system, and we take the same
convention when using the notation $\hat{B}$. By inspecting \eqref{eq-CF:initial-state-detailed}, the initial
state of Bob in the purified protocol is as follows:%
\begin{equation}
\sigma_{F_{0}F_{0}^{\prime}\hat{B}_{0}}\coloneqq \sum_{f_{0}}p(f_{0})|f_{0}%
\rangle\!\langle f_{0}|_{F_{0}}\otimes|f_{0}\rangle\!\langle f_{0}%
|_{F_{0}^{\prime}}\otimes\psi_{\hat{B}_{0}}^{f_{0}},
\end{equation}
where the state $\psi_{\hat{B}_{0}}^{f_{0}}$ purifies Bob's state $\Psi
_{B_{0}^{\prime}}^{f_{0}}$, such that tracing over a subsystem of $\psi
_{\hat{B}_{0}}^{f_{0}}$ gives $\Psi_{B_{0}^{\prime}}^{f_{0}}$.\ Additionally,
Bob keeps an extra copy $F_{0}^{\prime}$\ of the classical data transmitted
over the classical feedback channel. Let $\mathcal{U}_{M^{\prime}%
F_{0}\rightarrow\hat{A}_{1}A_{1}}^{0}$ denote an isometric channel extending
the encoding channel $\mathcal{E}_{M^{\prime}F_{0}\rightarrow A_{1}^{\prime
}A_{1}}^{0}$. After $\mathcal{U}_{M^{\prime}F_{0}\rightarrow\hat{A}_{1}A_{1}%
}^{0}$ acts, the global state is as follows:%
\begin{equation}
\omega_{M\hat{A}_{1}A_{1}F_{0}^{\prime}\hat{B}_{0}}^{1}\coloneqq \sum_{m\in
\mathcal{M}}\sum_{f_{0}}p(m)p(f_{0})|m\rangle\!\langle m|_{M}\otimes
\varphi_{\hat{A}_{1}A_{1}}^{0,m,f_{0}}\otimes|f_{0}\rangle\!\langle
f_{0}|_{F_{0}^{\prime}}\otimes\psi_{\hat{B}_{0}}^{f_{0}},
\end{equation}
where%
\begin{equation}
\mathcal{U}_{M^{\prime}F_{0}\rightarrow\hat{A}_{1}A_{1}}^{0}(|m\rangle
\!\langle m|_{M^{\prime}}\otimes|f_{0}\rangle\!\langle f_{0}|_{F_{0}}).
\end{equation}
We perform this purification for each step of the protocol. Let $\mathcal{U}%
_{A_{i}^{\prime}F_{i}\rightarrow\hat{A}_{i+1}A_{i+1}}^{i}$ denote an isometric
channel extending the encoding channel $\mathcal{E}_{A_{i}^{\prime}%
F_{i}\rightarrow A_{i+1}^{\prime}A_{i+1}}^{i}$ for each $i\in\left\{
1,\ldots,n-1\right\}  $. Since the system $F_{i}$ is classical, for each
$i\in\left\{  1,\ldots,n-1\right\}  $, the decoding channel $\overline{\Delta
}_{F_{i}}\circ\mathcal{D}_{B_{i}B_{i-1}^{\prime}\rightarrow F_{i}B_{i}%
^{\prime}}^{i}$ can be written explicitly as%
\begin{equation}
\overline{\Delta}_{F_{i}}\circ\mathcal{D}_{B_{i}B_{i-1}^{\prime}\rightarrow
F_{i}B_{i}^{\prime}}^{i}=\sum_{f_{i}}\mathcal{D}_{B_{i}B_{i-1}^{\prime
}\rightarrow B_{i}^{\prime}}^{i,f_{i}}\otimes|f_{i}\rangle\!\langle
f_{i}|_{F_{i}},
\end{equation}
where $\{\mathcal{D}_{B_{i}B_{i-1}^{\prime}\rightarrow B_{i}^{\prime}%
}^{i,f_{i}}\}_{f_{i}}$ is a collection of completely positive maps such that
the sum map $\sum_{f_{i}}\mathcal{D}_{B_{i}B_{i-1}^{\prime}\rightarrow
B_{i}^{\prime}}^{i,f_{i}}$ is trace preserving. Let $V_{B_{i}B_{i-1}^{\prime
}\rightarrow\hat{B}_{i}}^{i,f_{i}}$ be a linear map such that tracing over a
subsystem of $V_{B_{i}B_{i-1}^{\prime}\rightarrow\hat{B}_{i}}^{i,f_{i}}%
(\cdot)(V_{B_{i}B_{i-1}^{\prime}\rightarrow\hat{B}_{i}}^{i,f_{i}})^{\dag}$
gives the original map $\mathcal{D}_{B_{i}B_{i-1}^{\prime}\rightarrow
B_{i}^{\prime}}^{i,f_{i}}$, and define the map%
\begin{equation}
\mathcal{V}_{B_{i}B_{i-1}^{\prime}\rightarrow\hat{B}_{i}}^{i,f_{i}}%
(\tau_{B_{i}B_{i-1}^{\prime}})\coloneqq V_{B_{i}B_{i-1}^{\prime}\rightarrow\hat{B}%
_{i}}^{i,f_{i}}\tau_{B_{i}B_{i-1}^{\prime}}(V_{B_{i}B_{i-1}^{\prime
}\rightarrow\hat{B}_{i}}^{i,f_{i}})^{\dag}.
\end{equation}
Then we define the extended decoding channel $\mathcal{V}_{B_{i}%
B_{i-1}^{\prime}\rightarrow F_{i}\hat{B}_{i}F_{i}^{\prime}}^{i}$ for each
$i\in\left\{  1,\ldots,n-1\right\}  $ as%
\begin{equation}
\mathcal{V}_{B_{i}B_{i-1}^{\prime}\rightarrow F_{i}\hat{B}_{i}F_{i}^{\prime}%
}^{i}(\tau_{B_{i}B_{i-1}^{\prime}})\coloneqq \sum_{f_{i}}\mathcal{V}_{B_{i}%
B_{i-1}^{\prime}\rightarrow\hat{B}_{i}}^{i,f_{i}}(\tau_{B_{i}B_{i-1}^{\prime}%
})\otimes|f_{i}\rangle\!\langle f_{i}|_{F_{i}}\otimes|f_{i}\rangle\!\langle
f_{i}|_{F_{i}^{\prime}}.
\end{equation}
This extended decoding channel keeps an extra copy of the classical feedback
value $f_{i}$ for Bob in the classical register $F_{i}^{\prime}$. The final
decoding channel in the original protocol is a measurement channel and thus
can be written as%
\begin{equation}
\mathcal{D}_{B_{n}B_{n-1}^{\prime}\rightarrow\hat{M}}^{n}(\tau_{B_{n}%
B_{n-1}^{\prime}})=\sum_{m\in\mathcal{M}}\operatorname{Tr}[\Lambda
_{B_{n}B_{n-1}^{\prime}}^{m}\tau_{B_{n}B_{n-1}^{\prime}}]|m\rangle\!\langle
m|_{\hat{M}},
\end{equation}
where $\{\Lambda_{B_{n}B_{n-1}^{\prime}}^{m}\}_{m\in\mathcal{M}}$ is a POVM.
We enlarge it as follows in the simulation protocol:%
\begin{equation}
\mathcal{V}_{B_{n}B_{n-1}^{\prime}\rightarrow\hat{B}_{n}\hat{M}}^{n}%
(\tau_{B_{n}B_{n-1}^{\prime}})\coloneqq \sum_{m\in\mathcal{M}}\sqrt{\Lambda
_{B_{n}B_{n-1}^{\prime}}^{m}}\tau_{B_{n}B_{n-1}^{\prime}}\sqrt{\Lambda
_{B_{n}B_{n-1}^{\prime}}^{m}}\otimes|m\rangle\!\langle m|_{\hat{M}},
\end{equation}
where the system $\hat{B}_{n}$ is isomorphic to the systems $B_{n}%
B_{n-1}^{\prime}$. In the simulation protocol, we also consider an isometric
channel $\mathcal{U}_{A\rightarrow BE}^{\mathcal{N}}$ that simulates the
original channel $\mathcal{N}_{A\rightarrow B}$ as follows: $\mathcal{N}%
_{A\rightarrow B}=\operatorname{Tr}_{E}\circ\mathcal{U}_{A\rightarrow
BE}^{\mathcal{N}}$.

Thus, the various states involved in the purified protocol are as follows. The
global initial state is $\overline{\Phi}_{MM^{\prime}}^{p}\otimes\sigma
_{F_{0}F_{0}^{\prime}\hat{B}_{0}}$. Alice performs the extended encoding
channel $\mathcal{U}_{M^{\prime}F_{0}\rightarrow\hat{A}_{1}A_{1}}^{0}$ and the
state becomes as follows:%
\begin{equation}
\omega_{M\hat{A}_{1}A_{1}F_{0}^{\prime}\hat{B}_{0}}^{1}=\mathcal{U}%
_{M^{\prime}F_{0}\rightarrow\hat{A}_{1}A_{1}}^{0}(\overline{\Phi}_{MM^{\prime
}}^{p}\otimes\sigma_{F_{0}F_{0}^{\prime}\hat{B}_{0}}).
\end{equation}
Alice transmits system $A_{1}$ through the first use of the extended channel
$\mathcal{U}_{A_{1}\rightarrow B_{1}E_{1}}^{\mathcal{N}}$, resulting in the
following state:%
\begin{equation}
\rho_{M\hat{A}_{1}B_{1}E_{1}F_{0}^{\prime}\hat{B}_{0}}^{1}\coloneqq \mathcal{U}%
_{A_{1}\rightarrow B_{1}E_{1}}^{\mathcal{N}}(\omega_{M\hat{A}_{1}A_{1}%
F_{0}^{\prime}\hat{B}_{0}}^{1}).
\end{equation}
Bob processes his systems $B_{1}$ and $B_{0}^{\prime}$ with the extended
decoding channel $\mathcal{V}_{B_{1}B_{0}^{\prime}\rightarrow F_{1}\hat{B}%
_{1}F_{1}^{\prime}}^{1}$, and Alice acts with the extended encoding channel
$\mathcal{U}_{A_{1}^{\prime}F_{1}\rightarrow\hat{A}_{2}A_{2}}^{1}$, resulting
in the state%
\begin{equation}
\omega_{M\hat{A}_{2}A_{2}\hat{B}_{1}E_{1}F_{0}^{\prime}F_{1}^{\prime}}%
^{2}\coloneqq (\mathcal{U}_{A_{1}^{\prime}F_{1}\rightarrow\hat{A}_{2}A_{2}}^{1}%
\circ\mathcal{V}_{B_{1}B_{0}^{\prime}\rightarrow F_{1}\hat{B}_{1}F_{1}%
^{\prime}}^{1})(\rho_{M\hat{A}_{1}B_{1}E_{1}F_{0}^{\prime}\hat{B}_{0}}^{1}).
\end{equation}
This process iterates $n-2$ more times, resulting in the following states:%
\begin{align}
\rho_{M\hat{A}_{i}B_{i}\hat{B}_{i-1}E_{1}^{i}[F_{0}^{i-1}]^{\prime}}^{i}  &
\coloneqq \mathcal{U}_{A_{i}\rightarrow B_{i}E_{i}}^{\mathcal{N}}(\omega_{M\hat{A}%
_{i}A_{i}\hat{B}_{i-1}E_{1}^{i-1}[F_{0}^{i-1}]^{\prime}}^{i}),\\
\omega_{M\hat{A}_{i+1}A_{i+1}\hat{B}_{i}E_{1}^{i}[F_{0}^{i}]^{\prime}}^{i+1}
&  \coloneqq (\mathcal{U}_{A_{i}^{\prime}F_{i}\rightarrow\hat{A}_{i+1}A_{i+1}}%
^{i}\circ\mathcal{V}_{B_{i}B_{i-1}^{\prime}\rightarrow F_{i}\hat{B}_{i}%
F_{i}^{\prime}}^{i})(\rho_{M\hat{A}_{i}B_{i}\hat{B}_{i-1}E_{1}^{i}[F_{0}%
^{i-1}]^{\prime}}^{i}),
\end{align}
for $i\in\left\{  2,\ldots,n-1\right\}  $. The final extended decoding channel
results in the following state:%
\begin{equation}
\rho_{M\hat{A}_{n}\tilde{B}\hat{M}E_{1}^{n}[F_{0}^{n-1}]^{\prime}}%
^{n}\coloneqq \mathcal{V}_{B_{n}B_{n-1}^{\prime}\rightarrow\hat{B}_{n}\hat{M}}%
^{n}(\rho_{M\hat{A}_{n}B_{n}\hat{B}_{n-1}E_{1}^{n}[F_{0}^{n-1}]^{\prime}}%
^{n}),
\end{equation}
where the $\tilde{B}$ system encompasses all systems in Bob's possession at
the end. Note that we recover each state of the original protocol described in
Section~\ref{sec-CF:FB-assisted-protocol}\ by performing particular partial traces.

Before stating the main theorem of this section, we prove two lemmas that play
an important role in its proof. Both lemmas involve the following information
measure:%
\begin{equation}
I(X;CY)_{\tau}+H(C|XY)_{\tau}, \label{eq-CF:info-quant-entr-bound}%
\end{equation}
where the information quantities are evaluated with respect to the following
classical--quantum state:%
\begin{equation}
\tau_{XYC}\coloneqq \sum_{x,y}p(x,y)|x\rangle\!\langle x|_{X}\otimes|y\rangle\!\langle
y|_{Y}\otimes\tau_{C}^{x,y}. \label{eq-CF:tau-state-entropy-bnd-lemmas}%
\end{equation}
In the above, $p(x,y)$ is a probability distribution and $\tau_{C}^{x,y}$ is a
quantum state for all $x$ and $y$.

\begin{Lemma}{lem-CF:1W-locc-mono-key-info}
Let $\tau_{XYAB}$ be a classical--quantum
state, with classical systems $XY$ and quantum systems $AB$ pure when
conditioned on $XY$. Let $\mathcal{L}_{AB\rightarrow A^{\prime}B^{\prime}Z}$
be a one-way LOCC channel of the following form:%
\begin{equation}
\mathcal{L}_{AB\rightarrow A^{\prime}B^{\prime}Z}\coloneqq \sum_{z}\mathcal{U}%
_{A\rightarrow A^{\prime}}^{z}\otimes\mathcal{V}_{B\rightarrow B^{\prime}}%
^{z}\otimes|z\rangle\!\langle z|_{Z},
\end{equation}
where $\{\mathcal{V}_{B\rightarrow B^{\prime}}^{z}\}_{z}$ is a collection of
completely positive trace-non-increasing maps with $\mathcal{V}_{B\rightarrow
B^{\prime}}^{z}(\cdot)\coloneqq V_{B\rightarrow B^{\prime}}^{z}(\cdot)(V_{B\rightarrow
B^{\prime}}^{z})^{\dag}$ and $\{\mathcal{U}_{A\rightarrow A^{\prime}}%
^{z}\}_{z}$ is a collection of isometric channels. Then the following
inequality holds%
\begin{equation}
I(X;BY)_{\tau}+H(B|XY)_{\tau}\geq I(X;B^{\prime}YZ)_{\omega}+H(B^{\prime
}|XYZ)_{\omega},
\end{equation}
where $\omega_{XYZA^{\prime}B^{\prime}}\coloneqq \mathcal{L}_{AB\rightarrow A^{\prime
}B^{\prime}Z^{\prime}}(\tau_{XYAB})$.
\end{Lemma}

\begin{Proof}
The inequality $I(X;BY)_{\tau}\geq I(X;B^{\prime}YZ)_{\omega}$ follows from
the data-processing inequality for mutual information. In more detail,
consider that $\omega_{XYZB^{\prime}}$ is equal to%
\begin{align}
\omega_{XYZB^{\prime}}  &  =\operatorname{Tr}_{A^{\prime}}[\omega
_{XYZA^{\prime}B^{\prime}}]\\
&  =\operatorname{Tr}_{A^{\prime}}\!\left[  \sum_{z}(\mathcal{U}_{A\rightarrow
A^{\prime}}^{z}\otimes\mathcal{V}_{B\rightarrow B^{\prime}}^{z})(\tau
_{XYAB})\otimes|z\rangle\!\langle z|_{Z}\right] \\
&  =\sum_{z}((\operatorname{Tr}_{A^{\prime}}\circ\mathcal{U}_{A\rightarrow
A^{\prime}}^{z})\otimes\mathcal{V}_{B\rightarrow B^{\prime}}^{z})(\tau
_{XYAB})\otimes|z\rangle\!\langle z|_{Z}\\
&  =\sum_{z}\mathcal{V}_{B\rightarrow B^{\prime}}^{z}(\operatorname{Tr}%
_{A}[\tau_{XYAB}])\otimes|z\rangle\!\langle z|_{Z}\\
&  =\sum_{z}\mathcal{V}_{B\rightarrow B^{\prime}}^{z}(\tau_{XYB}%
)\otimes|z\rangle\!\langle z|_{Z}.
\end{align}
The fourth equality follows because $\mathcal{U}_{A\rightarrow A^{\prime}}%
^{z}$ is an isometric channel for all $z$. Thus, the state $\omega
_{XYZB^{\prime}}$ can be understood as arising from the action of the quantum
instrument $\sum_{z}\mathcal{V}_{B\rightarrow B^{\prime}}^{z}\otimes
|z\rangle\!\langle z|_{Z}$ on the state $\tau_{XYB}$, and since this a channel
taking system $B$ to $B^{\prime}Z$, the data-processing inequality for mutual
information applies. The inequality $H(B|XY)_{\tau}\geq H(B^{\prime
}|XYZ)_{\omega}$ is a consequence of the LOCC monotonicity of the entanglement
of formation (see Proposition~\ref{prop:E-meas:EoF-convex-str-LOCC-mono}). Indeed, consider that%
\begin{align}
H(B|XY)_{\tau}  &  =E_{F}(A;BXY)_{\tau},\\
H(B^{\prime}|XYZ)_{\omega}  &  =E_{F}(A^{\prime};B^{\prime}XYZ)_{\omega},
\end{align}
which follows from the direct-sum property of the entanglement of formation
(see the proof of Proposition~\ref{prop:E-meas:EoF-convex-str-LOCC-mono})\ and its reduction to entropy of entanglement for pure states (see
\eqref{eq-ent_formation_pure_state}). Thus, we apply these equalities and the LOCC\ monotonicity of
entanglement of formation (i.e., $E_{F}(A;BXY)_{\tau}\geq E_{F}(A^{\prime
};B^{\prime}XYZ)_{\omega}$).
\end{Proof}

The following lemma places an entropic upper bound on the amount by which the
information quantity in \eqref{eq-CF:info-quant-entr-bound}\ can increase by
the action of a channel $\mathcal{N}_{A\rightarrow B}$:

\begin{Lemma}{lem-CF:entropy-bound-on-amortized-quant}Let $\mathcal{N}_{A\rightarrow
B}$ be a quantum channel, and let $\tau_{XYAB^{\prime}}$ be a
classical--quantum state of the following form:%
\begin{equation}
\tau_{XYAB^{\prime}}\coloneqq \sum_{x,y}p(x,y)|x\rangle\!\langle x|_{X}\otimes
|y\rangle\!\langle y|_{Y}\otimes\tau_{AB^{\prime}}^{x,y}.
\end{equation}
Then%
\begin{equation}
I(X;BB^{\prime}Y)_{\omega}+H(BB^{\prime}|XY)_{\omega}-\left[  I(X;B^{\prime
}Y)_{\tau}+H(B^{\prime}|XY)_{\tau}\right]  \leq H(B)_{\omega},
\end{equation}
where $\omega_{XYBB^{\prime}}\coloneqq \mathcal{N}_{A\rightarrow B}(\tau
_{XYAB^{\prime}})$.
\end{Lemma}

\begin{Proof}
Consider that%
\begin{align}
&  I(X;BB^{\prime}Y)_{\omega}+H(BB^{\prime}|XY)_{\omega}-\left[
I(X;B^{\prime}Y)_{\tau}+H(B^{\prime}|XY)_{\tau}\right] \nonumber\\
&  =I(X;BB^{\prime}Y)_{\omega}+H(BB^{\prime}|XY)_{\omega}-\left[
I(X;B^{\prime}Y)_{\omega}+H(B^{\prime}|XY)_{\omega}\right] \\
&  =I(X;B|B^{\prime}Y)_{\omega}+H(B|B^{\prime}XY)_{\omega}\\
&  =H(B|B^{\prime}Y)_{\omega}-H(B|B^{\prime}XY)_{\omega}+H(B|B^{\prime
}XY)_{\omega}\\
&  =H(B|B^{\prime}Y)_{\omega}\\
&  \leq H(B)_{\omega}.
\end{align}
All equalities follow from applying definitions and chain rules for mutual
information and entropy. The final inequality follows because conditioning
does not increase entropy.
\end{Proof}

The key properties of the information quantity in \eqref{eq-CF:info-quant-entr-bound} is that it does not increase
under the action of a one-way LOCC channel from Bob to Alice (i.e., the
decoding channel of Bob, the classical feedback channel, and the encoding
channel of Alice) and it cannot increase by more than the output entropy of a
channel under its action. We can use these properties to establish the
following entropy bound on the number of bits that can be transmitted by a
feedback-assisted communication protocol:

\begin{theorem}{thm-CF:entropy-bound-finite-length}
Let $\mathcal{N}_{A\rightarrow B}$
be a quantum channel, and let $\varepsilon\in\lbrack0,1)$. For an
$(n,\left\vert \mathcal{M}\right\vert ,\varepsilon)$ protocol for classical
communication over a quantum channel $\mathcal{N}_{A\rightarrow B}$ assisted
by classical feedback, as described in
Section~\ref{sec-CF:FB-assisted-protocol}, the following bound holds%
\begin{equation}
\frac{\log_{2}\left\vert \mathcal{M}\right\vert }{n}\leq\frac{1}%
{1-\varepsilon}\left(  \sup_{\rho_{A}}H(\mathcal{N}_{A\rightarrow B}(\rho
_{A}))+\frac{h_{2}(\varepsilon)}{n}\right)  .
\label{eq-CF:entropy-bound-n-shot}%
\end{equation}

\end{theorem}

\begin{Proof}
Our starting point is the general bounds in
\eqref{eq-CF:hypo-test-bnd-ent-break}--\eqref{eq-CF:hypo-test-bnd-ent-break-2},
which imply that%
\begin{equation}
\log_{2}\left\vert \mathcal{M}\right\vert \leq\frac{1}{1-\varepsilon}\left(
I(M;\hat{M})_{\omega}+h_{2}(\varepsilon)\right)  ,
\label{eq-CF:entropy-bound-starting-point}%
\end{equation}
where $\omega_{M\hat{M}}$ is the final state of the protocol, as given in
\eqref{eq-CF:final-omega-state}, with $p$ therein set to the uniform
distribution over the set $\mathcal{M}$ of messages. Continuing, and
considering the purified protocol outlined above, we find that%
\begin{align}
&  I(M;\hat{M})_{\omega}\nonumber\\
&  \leq I(M;B_{n}\hat{B}_{n-1}[F_{0}^{n-1}]^{\prime})_{\rho^{n}}+H(B_{n}%
\hat{B}_{n-1}|[F_{0}^{n-1}]^{\prime}M)_{\rho^{n}}%
\label{eq-CF:entropy-bnd-pf-1}\\
&  =I(M;B_{n}\hat{B}_{n-1}[F_{0}^{n-1}]^{\prime})_{\rho^{n}}+H(B_{n}\hat
{B}_{n-1}|[F_{0}^{n-1}]^{\prime}M)_{\rho^{n}}\nonumber\\
&  \qquad-\left[  I(M;\hat{B}_{0}F_{0}^{\prime})_{\omega^{1}}+H(\hat{B}%
_{0}|F_{0}^{\prime}M)_{\omega^{1}}\right] \\
&  =I(M;B_{n}\hat{B}_{n-1}[F_{0}^{n-1}]^{\prime})_{\rho^{n}}+H(B_{n}\hat
{B}_{n-1}|[F_{0}^{n-1}]^{\prime}M)_{\rho^{n}}\nonumber\\
&  \qquad-\left[  I(M;\hat{B}_{0}F_{0}^{\prime})_{\omega^{1}}+H(\hat{B}%
_{0}|F_{0}^{\prime}M)_{\omega^{1}}\right] \nonumber\\
&  \qquad+\sum_{i=2}^{n}I(M;\hat{B}_{i-1}[F_{0}^{i-1}]^{\prime})_{\omega^{i}%
}+H(\hat{B}_{i-1}|[F_{0}^{i-1}]^{\prime}M)_{\omega^{i}}\nonumber\\
&  \qquad\qquad-\left[  I(M;\hat{B}_{i-1}[F_{0}^{i-1}]^{\prime})_{\omega^{i}%
}+H(\hat{B}_{i-1}|[F_{0}^{i-1}]^{\prime}M)_{\omega^{i}}\right]  .
\end{align}
The first inequality follows from non-negativity of quantum entropy and data
processing under the action of the final decoding channel. The first equality
follows because $I(M;\hat{B}_{0}F_{0}^{\prime})_{\omega^{1}}+H(\hat{B}%
_{0}|F_{0}^{\prime}M)_{\omega^{1}} = 0$ for the initial state $\omega_{M\hat
{A}_{1}A_{1}F_{0}^{\prime}\hat{B}_{0}}^{1}$ (indeed, the systems $M$ and
$F_{0}^{\prime}\hat{B}_{0}$ of the reduced state $\omega_{MF_{0}^{\prime}%
\hat{B}_{0}}^{1}$ are product, and the state on system $\hat{B}_{0}$ is pure
when conditioned on $F_{0}^{\prime}M$). The last equality follows by adding
and subtracting the same term. Continuing, we find that the quantity in the
last line above is bounded as%
\begin{align}
&  \leq I(M;B_{n}\hat{B}_{n-1}[F_{0}^{n-1}]^{\prime})_{\rho^{n}}+H(B_{n}%
\hat{B}_{n-1}|[F_{0}^{n-1}]^{\prime}M)_{\rho^{n}}\nonumber\\
&  \qquad-\left[  I(M;\hat{B}_{0}F_{0}^{\prime})_{\omega^{1}}+H(\hat{B}%
_{0}|F_{0}^{\prime}M)_{\omega^{1}}\right] \nonumber\\
&  \qquad+\sum_{i=2}^{n}I(M;B_{i-1}\hat{B}_{i-2}[F_{0}^{i-2}]^{\prime}%
)_{\rho^{i-1}}+H(B_{i-1}\hat{B}_{i-2}|[F_{0}^{i-2}]^{\prime}M)_{\rho^{i-1}%
}\nonumber\\
&  \qquad\qquad-\left[  I(M;\hat{B}_{i-1}[F_{0}^{i-1}]^{\prime})_{\omega^{i}%
}+H(\hat{B}_{i-1}|[F_{0}^{i-1}]^{\prime}M)_{\omega^{i}}\right] \\
&  =\sum_{i=1}^{n}I(M;B_{i}\hat{B}_{i-1}[F_{0}^{i-1}]^{\prime})_{\rho^{i}%
}+H(B_{i}\hat{B}_{i-1}|[F_{0}^{i-1}]^{\prime}M)_{\rho^{i}}\nonumber\\
&  \qquad-\left[  I(M;\hat{B}_{i-1}[F_{0}^{i-1}]^{\prime})_{\omega^{i}}%
+H(\hat{B}_{i-1}|[F_{0}^{i-1}]^{\prime}M)_{\omega^{i}}\right] \\
&  \leq\sum_{i=1}^{n}H(B_{i})_{\rho^{i}}\\
&  \leq n\sup_{\rho_{A}}H(\mathcal{N}_{A\rightarrow B}(\rho_{A}))
\label{eq-CF:entropy-bnd-pf-last}%
\end{align}
The first inequality follows from Lemma~\ref{lem-CF:1W-locc-mono-key-info}%
\ and the second from Lemma~\ref{lem-CF:entropy-bound-on-amortized-quant}. So
we conclude that%
\begin{equation}
I(M;\hat{M})_{\omega}\leq n\sup_{\rho_{A}}H(\mathcal{N}_{A\rightarrow B}%
(\rho_{A})).
\end{equation}
Putting together this inequality and
\eqref{eq-CF:entropy-bound-starting-point}, we conclude the inequality in \eqref{eq-CF:entropy-bound-n-shot}.
\end{Proof}

\subsubsection{Maximum Average Output Entropy Upper Bound for Probabilistic
Mixtures of Channels}

\label{sec-CF:convex-comb-bnd}

In this section, we provide a brief proof of the following theorem, which
generalizes Theorem~\ref{thm-CF:entropy-bound-finite-length} to the maximum
average output entropy of a quantum channel:

\begin{theorem}{cor-CF:avg-ent-upper-bnd} Let $\mathcal{N}_{A\rightarrow B}=\sum
_{x}p_{X}(x)\mathcal{N}_{A\rightarrow B}^{x}$, where $p_{X}$ is a probability
distribution and $\{\mathcal{N}_{A\rightarrow B}^{x}\}_{x}$ is a set of
channels. For an $(n,|\mathcal{M}|,\varepsilon)$ protocol for classical
communication over the channel $\mathcal{N}_{A\rightarrow B}$ assisted by
classical feedback, of the form described in
Section~\ref{sec-CF:FB-assisted-protocol}, the following bound applies%
\begin{equation}
\left(  1-\varepsilon\right)  \log_{2}|\mathcal{M}|\leq n\cdot\sup_{\rho_{A}%
}\sum_{x}p_{X}(x)H(\mathcal{N}_{A\rightarrow B}^{x}(\rho_{A}))+h_{2}%
(\varepsilon).\nonumber
\end{equation}

\end{theorem}

\begin{Proof}
The main idea behind the proof is to observe that an arbitrary
feedback-assisted protocol of the form discussed in
Section~\ref{sec-CF:FB-assisted-protocol}, which is for communication over a
probabilistic mixture channel $\mathcal{N}_{A\rightarrow B}=\sum_{z}%
p_{Z}(z)\mathcal{N}_{A\rightarrow B}^{z}$, has a simulation of the following form:

\begin{enumerate}
\item Before the $i$th use of the channel $\mathcal{N}_{A\rightarrow B}$ in
the feedback-assisted protocol, Bob selects a random variable $Z_{i}$
independently according to the distribution $p_{Z}$. He transmits $Z_{i}$ over
the classical feedback channel to Alice.

\item Each channel use $\mathcal{N}_{A\rightarrow B}$ from the original
protocol is replaced by a simulation in terms of another channel
$\mathcal{M}_{AZ^{\prime}\to B}$, which accepts a quantum input on system $A$
and a classical input on system $Z^{\prime}$. Conditioned on the value $z$ in
system $Z^{\prime}$, the channel $\mathcal{M}_{AZ^{\prime}\to B}$ applies
$\mathcal{N}^{z}_{A\rightarrow B}$ to the quantum system $A$. Thus, if the
random variable $Z \sim p_{Z}$ is fed into the input system $Z^{\prime}$ of
$\mathcal{M}_{AZ^{\prime}\to B}$, then the channel $\mathcal{M}_{AZ^{\prime
}\to B}$ is indistinguishable from the original channel $\mathcal{N}%
_{A\rightarrow B}$.

\item Alice feeds a copy of the classical random variable $Z_{i}$ into the
$i$th use of the channel $\mathcal{M}_{AZ^{\prime}\to B}$.

\item All other aspects of the protocol are executed in the same way as
before. Namely, even though it would be an advantage to Alice to modify her
encodings and Bob to modify later decodings based on the realizations of
$Z_{i}$, they do not do so, and they instead blindly operate all other aspects
of the simulation protocol as they are in the original protocol.
\end{enumerate}

Our goal now is to establish the inequality in
Theorem~\ref{cor-CF:avg-ent-upper-bnd}, relating the $n$, $\left\vert
\mathcal{M}\right\vert $, and $\varepsilon$ parameters of the original
$(n,\left\vert \mathcal{M}\right\vert ,\varepsilon)$ protocol by using the
above simulation.

The main observation to make from here is that the same proof from
Lemma~\ref{lem-CF:entropy-bound-on-amortized-quant} gives the following
bound:
\begin{multline}
I(X;BB^{\prime}YZ)_{\omega}+H(BB^{\prime}|XYZ)_{\omega}%
\label{eq-CF:new-ineq-amortize}\\
-\left[  I(X;B^{\prime}YZ)_{\tau}+H(B^{\prime}|XYZ)_{\tau}\right]  \leq
H(B|Z)_{\omega},
\end{multline}
where $\omega_{XYZBB^{\prime}}$ is the following state:%
\begin{align}
\omega_{XYZBB^{\prime}}  &  \coloneqq \mathcal{M}_{AZ^{\prime}\rightarrow B}%
(\tau_{XYZZ^{\prime\prime}}),\\
\tau_{XYZZ^{\prime\prime}AB^{\prime}}  &  \coloneqq \sum_{x,y,z}%
p(x,y,z)|x,y,z,z\rangle\!\langle x,y,z,z|_{X,Y,Z,Z^{\prime}}\otimes
\tau_{AB^{\prime}}^{x,y,z}.
\end{align}
This follows by grouping $Z$ with $Y$, but then discarding only $Y$ and
$B^{\prime}$ at the end of the proof. We then apply this bound, and the same
reasoning in the proof of Theorem~\ref{thm-CF:entropy-bound-finite-length},
except that the variables $Z_{0},\ldots,Z_{i}$ are grouped together with the
feedback variables $[F_{0}^{i-1}]^{\prime}$ and then the same reasoning in
\eqref{eq-CF:entropy-bnd-pf-1}--\eqref{eq-CF:entropy-bnd-pf-last} applies. At
this point, we invoke \eqref{eq-CF:new-ineq-amortize} and find that
\begin{equation}
(1-\varepsilon)\log_{2}|\mathcal{M}|\leq\sum_{i=1}^{n}H(B_{i}|Z_{i}%
)_{\rho^{(i)}}+h_{2}(\varepsilon).
\end{equation}
We can then bound the sum over entropies as follows:
\begin{align}
\sum_{i=1}^{n}H(B_{i}|Z_{i})_{\rho^{(i)}}  &  \leq nH(B|Z)_{\overline{\rho}}\\
&  =n\sum_{z}p_{Z}(z)H(\mathcal{N}^{z}(\overline{\omega}))\\
&  \leq n\sup_{\rho}\sum_{z}p_{Z}(z)H(\mathcal{N}^{z}(\rho)).
\end{align}
The first inequality is by concavity of conditional entropy, and the
conditional entropy is defined on the averaged channel output state over $n$
uses, defined as%
\begin{align}
\overline{\rho}_{BZ}  &  \coloneqq \sum_{z}p_{Z}(z)|z\rangle\!\langle z|\otimes
\mathcal{N}^{z}(\overline{\omega}),\\
\overline{\omega}_{A}  &  \coloneqq \frac{1}{n}\sum_{i=1}^{n}\omega_{A_{i}}^{(i)}.
\end{align}
The second equality follows from the definition of conditional entropy. The
third inequality follows from optimizing over all states.
\end{Proof}

\subsection{Geometric $\Upsilon$-Information Upper Bound on the Number of
Transmitted Bits}

In this section, we prove that the $\Upsilon$-information bound from
Section~\ref{subsubsec-cc_Upsilon_inf} is actually an upper bound on the classical capacity assisted by
classical feedback. The main idea behind the approach detailed in this section
is to establish a correlation measure for bipartite channels, which is
non-increasing under the action of one-way LOCC channels and measures the
forward classical communication that can be generated by the bipartite channel
for which it is evaluated. Such a measure is relevant in the context of a
feedback-assisted protocol because, in such a protocol, Alice and Bob employ a
one-way LOCC channel from Bob to Alice. In particular, local channels are
allowed for free, as well as the use of a classical feedback channel. Both of
these actions can be considered as particular kinds of bipartite channels and
both of them fall into the class of bipartite channels that are non-signaling
from Alice to Bob and C-PPT-P (call this class NS$_{A\not \rightarrow B}~\cap
$~PPT). Recall the definition of non-signaling channels from Section~\ref{sec-QM:non-sig-chs}
and C-PPT-P channels from Section~\ref{sec-QM:C-PPT-P-chs}. As such, if we employ a measure of
bipartite channels that involves a comparison between a bipartite channel of
interest to all bipartite channels in NS$_{A\not \rightarrow B}~\cap$~PPT,
then the two kinds of free channels would have zero value and the measure
would indicate how different the channel of interest is from this set (i.e.,
how different it is from a channel that has no ability to send quantum
information and no ability to signal from Alice to Bob). This is the main idea
behind the measure that we define below in
Definition~\ref{def:beta-measure-bipartite-ch}, but one should keep in mind
that the measure below does not follow this reasoning precisely.

In Definition~\ref{def:beta-measure-bipartite-ch}, although we motivated the
measure for bipartite channels, we define it more generally for completely
positive bipartite maps, as it turns out to be useful to do so when we define
other measures later.

\begin{definition}{$\beta$-Measure of Classical Communication for Bipartite Channels}
{def:beta-measure-bipartite-ch}
Let $\mathcal{M}_{AB\rightarrow
A^{\prime}B^{\prime}}$ be a completely positive bipartite map. Then we define%
\begin{align}
&  C_{\beta}(\mathcal{M}_{AB\rightarrow A^{\prime}B^{\prime}})\coloneqq\log
_{2}\beta(\mathcal{M}_{AB\rightarrow A^{\prime}B^{\prime}}),\\
&  \beta(\mathcal{M}_{AB\rightarrow A^{\prime}B^{\prime}})\coloneqq\inf
_{\substack{S_{AA^{\prime}BB^{\prime}},\\V_{AA^{\prime}BB^{\prime}}%
\in\operatorname{Herm}}}\left\{
\begin{array}
[c]{c}%
\left\Vert \operatorname{Tr}_{A^{\prime}B^{\prime}}[S_{AA^{\prime}BB^{\prime}%
}]\right\Vert _{\infty}:\\
T_{BB^{\prime}}(V_{AA^{\prime}BB^{\prime}}\pm\Gamma_{AA^{\prime}BB^{\prime}%
}^{\mathcal{M}})\geq0,\\
S_{AA^{\prime}BB^{\prime}}\pm V_{AA^{\prime}BB^{\prime}}\geq0,\\
\operatorname{Tr}_{A^{\prime}}[S_{AA^{\prime}BB^{\prime}}]=\pi_{A}%
\otimes\operatorname{Tr}_{AA^{\prime}}[S_{AA^{\prime}BB^{\prime}}]
\end{array}
\right\}  , \label{eq-CF:basic-measure-NS-PPT-bi}%
\end{align}
where $\operatorname{Herm}$ denotes the set of Hermitian operators and
$\Gamma_{AA^{\prime}BB^{\prime}}^{\mathcal{M}}$ is the Choi operator of
$\mathcal{M}_{AB\rightarrow A^{\prime}B^{\prime}}$:%
\begin{equation}
\Gamma_{AA^{\prime}BB^{\prime}}^{\mathcal{M}}\coloneqq\mathcal{M}_{\hat{A}%
\hat{B}\rightarrow A^{\prime}B^{\prime}}(\Gamma_{A\hat{A}}\otimes\Gamma
_{B\hat{B}}).
\end{equation}
In the above, $\hat{A}$ is isomorphic to $A$, system $\hat{B}$ is isomorphic
to $B$,%
\begin{equation}
\Gamma_{A\hat{A}}\coloneqq\sum_{i,j=0}^{d_{A}-1}|i\rangle\!\langle
j|_{A}\otimes|i\rangle\!\langle j|_{\hat{A}},\qquad\Gamma_{B\hat{B}%
}\coloneqq\sum_{i,j=0}^{d_{B}-1}|i\rangle\!\langle j|_{B}\otimes
|i\rangle\!\langle j|_{\hat{B}},
\end{equation}
and $\pi_{A}\coloneqq I_{A}/d_{A}$.
\end{definition}

Since $S_{AA^{\prime}BB^{\prime}}\pm V_{AA^{\prime}BB^{\prime}}\geq0$ implies
that $S_{AA^{\prime}BB^{\prime}}\geq0$, we can also express $\beta
(\mathcal{M}_{AB\rightarrow A^{\prime}B^{\prime}})$ as follows:%
\begin{equation}
\inf_{\substack{S_{AA^{\prime}BB^{\prime}},\\V_{AA^{\prime}BB^{\prime}}%
\in\operatorname{Herm}}}\left\{
\begin{array}
[c]{c}%
\lambda:\\
\operatorname{Tr}_{A^{\prime}B^{\prime}}[S_{AA^{\prime}BB^{\prime}}%
]\leq\lambda I_{AB}\\
T_{BB^{\prime}}(V_{AA^{\prime}BB^{\prime}}\pm\Gamma_{AA^{\prime}BB^{\prime}%
}^{\mathcal{M}})\geq0,\\
S_{AA^{\prime}BB^{\prime}}\pm V_{AA^{\prime}BB^{\prime}}\geq0,\\
\operatorname{Tr}_{A^{\prime}}[S_{AA^{\prime}BB^{\prime}}]=\pi_{A}%
\otimes\operatorname{Tr}_{AA^{\prime}}[S_{AA^{\prime}BB^{\prime}}]
\end{array}
\right\}
\end{equation}

By exploiting the equality constraint $\operatorname{Tr}_{A^{\prime}%
}[S_{AA^{\prime}BB^{\prime}}]=\pi_{A}\otimes\operatorname{Tr}_{AA^{\prime}%
}[S_{AA^{\prime}BB^{\prime}}]$, we find that%
\begin{align}
\left\Vert \operatorname{Tr}_{A^{\prime}B^{\prime}}[S_{AA^{\prime}BB^{\prime}%
}]\right\Vert _{\infty}  &  =\left\Vert \operatorname{Tr}_{B^{\prime}%
}[\operatorname{Tr}_{A^{\prime}}[S_{AA^{\prime}BB^{\prime}}]]\right\Vert
_{\infty}\\
&  =\left\Vert \operatorname{Tr}_{B^{\prime}}[\pi_{A}\otimes\operatorname{Tr}%
_{AA^{\prime}}[S_{AA^{\prime}BB^{\prime}}]]\right\Vert _{\infty}\\
&  =\left\Vert \pi_{A}\otimes\operatorname{Tr}_{AA^{\prime}B^{\prime}%
}[S_{AA^{\prime}BB^{\prime}}]\right\Vert _{\infty}\\
&  =\frac{1}{d_{A}}\left\Vert \operatorname{Tr}_{AA^{\prime}B^{\prime}%
}[S_{AA^{\prime}BB^{\prime}}]\right\Vert _{\infty}.
\end{align}
Then we find that%
\begin{equation}
\beta(\mathcal{M}_{AB\rightarrow A^{\prime}B^{\prime}})\coloneqq\inf
_{\substack{S_{AA^{\prime}BB^{\prime}},\\V_{AA^{\prime}BB^{\prime}}%
\in\operatorname{Herm}}}\left\{
\begin{array}
[c]{c}%
\frac{1}{d_{A}}\left\Vert \operatorname{Tr}_{AA^{\prime}B^{\prime}%
}[S_{AA^{\prime}BB^{\prime}}]\right\Vert _{\infty}:\\
T_{BB^{\prime}}(V_{AA^{\prime}BB^{\prime}}\pm\Gamma_{AA^{\prime}BB^{\prime}%
}^{\mathcal{M}})\geq0,\\
S_{AA^{\prime}BB^{\prime}}\pm V_{AA^{\prime}BB^{\prime}}\geq0,\\
\operatorname{Tr}_{A^{\prime}}[S_{AA^{\prime}BB^{\prime}}]=\pi_{A}%
\otimes\operatorname{Tr}_{AA^{\prime}}[S_{AA^{\prime}BB^{\prime}}]
\end{array}
\right\}  .
\end{equation}
Since $S_{AA^{\prime}BB^{\prime}}\pm V_{AA^{\prime}BB^{\prime}}\geq0$ implies
that $S_{AA^{\prime}BB^{\prime}}\geq0$, we can also rewrite $\beta
(\mathcal{M}_{AB\rightarrow A^{\prime}B^{\prime}})$ as%
\begin{equation}
\beta(\mathcal{M}_{AB\rightarrow A^{\prime}B^{\prime}})\coloneqq\inf
_{\substack{\lambda,S_{AA^{\prime}BB^{\prime}}\geq0,\\V_{AA^{\prime}%
BB^{\prime}}\in\operatorname{Herm}}}\left\{
\begin{array}
[c]{c}%
\lambda:\\
\frac{1}{d_{A}}\operatorname{Tr}_{AA^{\prime}B^{\prime}}[S_{AA^{\prime
}BB^{\prime}}]\leq\lambda I_{B},\\
T_{BB^{\prime}}(V_{AA^{\prime}BB^{\prime}}\pm\Gamma_{AA^{\prime}BB^{\prime}%
}^{\mathcal{M}})\geq0,\\
S_{AA^{\prime}BB^{\prime}}\pm V_{AA^{\prime}BB^{\prime}}\geq0,\\
\operatorname{Tr}_{A^{\prime}}[S_{AA^{\prime}BB^{\prime}}]=\pi_{A}%
\otimes\operatorname{Tr}_{AA^{\prime}}[S_{AA^{\prime}BB^{\prime}}]
\end{array}
\right\}  . \label{eq-CF:basic-measure-NS-PPT-bi-rewrite}%
\end{equation}

\subsubsection{Properties of the basic
measure\label{sec-CF:props-basic-measure}}

We now establish several properties of $C_{\beta}(\mathcal{N}_{AB\rightarrow
A^{\prime}B^{\prime}})$, which are basic properties that we might expect of a
measure of forward classical communication for a bipartite channel. These
include the following:

\begin{enumerate}
\item non-negativity (Proposition~\ref{prop-CF:lower-bound-ch-non-neg}),

\item stability under tensoring with identity channels
(Proposition~\ref{prop-CF:stability}),

\item zero value for classical feedback channels
(Proposition~\ref{prop-CF:zero-feedback}),

\item zero value for a tensor product of local channels
(Proposition~\ref{prop-CF:zero-local-chs}),

\item subadditivity under serial composition
(Proposition~\ref{prop-CF:subadditivity-bipartite-CP-maps}),

\item data processing under pre- and post-processing by local channels
(Corollary~\ref{cor:DP-LC-beta-p-bipartite-ch}),

\item invariance under local unitary channels
(Corollary~\ref{cor:I-LUC-beta-p-bipartite-ch}),

\item convexity of $\beta$ (Proposition~\ref{prop-CF:convexity-beta}).
\end{enumerate}

All of the properties above hold for bipartite channels, while the second and
fifth through eighth hold more generally for completely positive bipartite maps.

\begin{proposition*}
{Non-Negativity}{prop-CF:lower-bound-ch-non-neg}Let $\mathcal{N}%
_{AB\rightarrow A^{\prime}B^{\prime}}$ be a bipartite channel. Then%
\begin{equation}
C_{\beta}(\mathcal{N}_{AB\rightarrow A^{\prime}B^{\prime}})\geq0.
\end{equation}

\end{proposition*}

\begin{Proof}
We prove the equivalent statement $\beta(\mathcal{N}_{AB\rightarrow A^{\prime
}B^{\prime}})\geq1$. Let $\lambda$, $S_{AA^{\prime}BB^{\prime}}$, and
$V_{AA^{\prime}BB^{\prime}}$ be arbitrary Hermitian operators satisfying the
constraints in \eqref{eq-CF:basic-measure-NS-PPT-bi-rewrite}. Then consider
that%
\begin{align}
\lambda d_{B}  &  =\lambda\operatorname{Tr}_{B}[I_{B}]\\
&  \geq\frac{1}{d_{A}}\operatorname{Tr}_{AA^{\prime}BB^{\prime}}%
[S_{AA^{\prime}BB^{\prime}}]\\
&  \geq\frac{1}{d_{A}}\operatorname{Tr}_{AA^{\prime}BB^{\prime}}%
[V_{AA^{\prime}BB^{\prime}}]\\
&  =\frac{1}{d_{A}}\operatorname{Tr}_{AA^{\prime}BB^{\prime}}[T_{BB^{\prime}%
}(V_{AA^{\prime}BB^{\prime}})]\\
&  \geq\operatorname{Tr}_{AA^{\prime}BB^{\prime}}[\Gamma_{AA^{\prime
}BB^{\prime}}^{\mathcal{N}}]\\
&  =\frac{1}{d_{A}}\operatorname{Tr}_{AB}[I_{AB}]\\
&  =d_{B}.
\end{align}
This implies that $\lambda\geq1$. Since the inequality holds for all $\lambda
$, $S_{AA^{\prime}BB^{\prime}}$, and $V_{AA^{\prime}BB^{\prime}}$ satisfying
the constraints in \eqref{eq-CF:basic-measure-NS-PPT-bi-rewrite}, we conclude
the statement above.
\end{Proof}

\begin{proposition*}
{Stability}{prop-CF:stability}Let $\mathcal{M}_{AB\rightarrow A^{\prime
}B^{\prime}}$ be a completely positive bipartite map.\ Then%
\begin{equation}
C_{\beta}(\operatorname{id}_{\bar{A}\rightarrow\tilde{A}}\otimes
\mathcal{M}_{AB\rightarrow A^{\prime}B^{\prime}}\otimes\operatorname{id}%
_{\bar{B}\rightarrow\tilde{B}})=C_{\beta}(\mathcal{M}_{AB\rightarrow
A^{\prime}B^{\prime}}).
\end{equation}
\end{proposition*}

\begin{Proof}
Let $S_{AA^{\prime}BB^{\prime}}$ and $V_{AA^{\prime}BB^{\prime}}$ be arbitrary
Hermitian operators satisfying the constraints in
\eqref{eq-CF:basic-measure-NS-PPT-bi} for $\mathcal{M}_{AB\rightarrow
A^{\prime}B^{\prime}}$. The Choi operator of $\operatorname{id}_{\bar
{A}\rightarrow\tilde{A}}\otimes\mathcal{M}_{AB\rightarrow A^{\prime}B^{\prime
}}\otimes\operatorname{id}_{\bar{B}\rightarrow\tilde{B}}$ is given by%
\begin{equation}
\Gamma_{\bar{A}\tilde{A}}\otimes\Gamma_{AA^{\prime}BB^{\prime}}^{\mathcal{M}%
}\otimes\Gamma_{\bar{B}\tilde{B}}.
\end{equation}
Let us show that $\Gamma_{\bar{A}\tilde{A}}\otimes S_{AA^{\prime}BB^{\prime}%
}\otimes\Gamma_{\bar{B}\tilde{B}}$ and $\Gamma_{\bar{A}\tilde{A}}\otimes
V_{AA^{\prime}BB^{\prime}}\otimes\Gamma_{\bar{B}\tilde{B}}$ satisfy the
constraints in \eqref{eq-CF:basic-measure-NS-PPT-bi} for $\operatorname{id}%
_{\bar{A}\rightarrow\tilde{A}}\otimes\mathcal{M}_{AB\rightarrow A^{\prime
}B^{\prime}}\otimes\operatorname{id}_{\bar{B}\rightarrow\tilde{B}}$. Consider
that%
\begin{align}
T_{BB^{\prime}}(V_{AA^{\prime}BB^{\prime}}\pm\Gamma_{AA^{\prime}BB^{\prime}%
}^{\mathcal{M}})  &  \geq0\\
\Longleftrightarrow\quad T_{BB^{\prime}}(\Gamma_{\bar{A}\tilde{A}}\otimes
V_{AA^{\prime}BB^{\prime}}\otimes\Gamma_{\bar{B}\tilde{B}}\pm\Gamma_{\bar
{A}\tilde{A}}\otimes\Gamma_{AA^{\prime}BB^{\prime}}^{\mathcal{M}}\otimes
\Gamma_{\bar{B}\tilde{B}})  &  \geq0\\
\Longleftrightarrow\quad T_{BB^{\prime}\bar{B}\tilde{B}}(\Gamma_{\bar{A}%
\tilde{A}}\otimes V_{AA^{\prime}BB^{\prime}}\otimes\Gamma_{\bar{B}\tilde{B}%
}\pm\Gamma_{\bar{A}\tilde{A}}\otimes\Gamma_{AA^{\prime}BB^{\prime}%
}^{\mathcal{M}}\otimes\Gamma_{\bar{B}\tilde{B}})  &  \geq0
\end{align}%
\begin{align}
S_{AA^{\prime}BB^{\prime}}\pm V_{AA^{\prime}BB^{\prime}}  &  \geq0\\
\Longleftrightarrow\quad\Gamma_{\bar{A}\tilde{A}}\otimes S_{AA^{\prime
}BB^{\prime}}\otimes\Gamma_{\bar{B}\tilde{B}}\pm\Gamma_{\bar{A}\tilde{A}%
}\otimes V_{AA^{\prime}BB^{\prime}}\otimes\Gamma_{\bar{B}\tilde{B}}  &  \geq0
\end{align}%
\[
\operatorname{Tr}_{A^{\prime}}[S_{AA^{\prime}BB^{\prime}}]=\pi_{A}%
\otimes\operatorname{Tr}_{AA^{\prime}}[S_{AA^{\prime}BB^{\prime}}],
\]
the latter equivalent to%
\begin{align}
&  \operatorname{Tr}_{A^{\prime}\tilde{A}}[\Gamma_{\bar{A}\tilde{A}}\otimes
S_{AA^{\prime}BB^{\prime}}\otimes\Gamma_{\bar{B}\tilde{B}}]\nonumber\\
&  =I_{\bar{A}}\otimes\pi_{A}\otimes\operatorname{Tr}_{AA^{\prime}%
}[S_{AA^{\prime}BB^{\prime}}\otimes\Gamma_{\bar{B}\tilde{B}}]\\
&  =\pi_{\bar{A}}\otimes\pi_{A}\otimes\operatorname{Tr}_{AA^{\prime}\bar
{A}\tilde{A}}[\Gamma_{\bar{A}\tilde{A}}\otimes S_{AA^{\prime}BB^{\prime}%
}\otimes\Gamma_{\bar{B}\tilde{B}}].
\end{align}
Also, consider that%
\begin{align}
&  \frac{1}{d_{A}d_{\bar{A}}}\left\Vert \operatorname{Tr}_{AA^{\prime}\bar
{A}\tilde{A}B^{\prime}\tilde{B}}[\Gamma_{\bar{A}\tilde{A}}\otimes
S_{AA^{\prime}BB^{\prime}}\otimes\Gamma_{\bar{B}\tilde{B}}]\right\Vert
_{\infty}\nonumber\\
&  =\frac{1}{d_{A}d_{\bar{A}}}\left\Vert d_{\bar{A}}\operatorname{Tr}%
_{AA^{\prime}B^{\prime}}[S_{AA^{\prime}BB^{\prime}}\otimes I_{\bar{B}%
}]\right\Vert _{\infty}\\
&  =\frac{1}{d_{A}}\left\Vert \operatorname{Tr}_{AA^{\prime}B^{\prime}%
}[S_{AA^{\prime}BB^{\prime}}]\otimes I_{\bar{B}}\right\Vert _{\infty}\\
&  =\frac{1}{d_{A}}\left\Vert \operatorname{Tr}_{AA^{\prime}B^{\prime}%
}[S_{AA^{\prime}BB^{\prime}}]\right\Vert _{\infty}.
\end{align}
Thus, it follows that%
\begin{equation}
\beta(\mathcal{M}_{AB\rightarrow A^{\prime}B^{\prime}})\geq\beta
(\operatorname{id}_{\bar{A}\rightarrow\tilde{A}}\otimes\mathcal{M}%
_{AB\rightarrow A^{\prime}B^{\prime}}\otimes\operatorname{id}_{\bar
{B}\rightarrow\tilde{B}}).
\end{equation}

Now let us show the opposite inequality. Let $S_{\bar{A}\tilde{A}AA^{\prime
}BB^{\prime}\bar{B}\tilde{B}}$ and $V_{\bar{A}\tilde{A}AA^{\prime}BB^{\prime
}\bar{B}\tilde{B}}$ be arbitrary Hermitian operators satisfying the
constraints in \eqref{eq-CF:basic-measure-NS-PPT-bi} for $\operatorname{id}%
_{\bar{A}\rightarrow\tilde{A}}\otimes\mathcal{M}_{AB\rightarrow A^{\prime
}B^{\prime}}\otimes\operatorname{id}_{\bar{B}\rightarrow\tilde{B}}$. Set%
\begin{align}
S_{AA^{\prime}BB^{\prime}}^{\prime}  &  \coloneqq\frac{1}{d_{\bar{A}}%
d_{\bar{B}}}\operatorname{Tr}_{\bar{A}\tilde{A}\bar{B}\tilde{B}}[S_{\bar
{A}\tilde{A}AA^{\prime}BB^{\prime}\bar{B}\tilde{B}}],\\
V_{AA^{\prime}BB^{\prime}}^{\prime}  &  \coloneqq\frac{1}{d_{\bar{A}}%
d_{\bar{B}}}\operatorname{Tr}_{\bar{A}\tilde{A}\bar{B}\tilde{B}}[V_{\bar
{A}\tilde{A}AA^{\prime}BB^{\prime}\bar{B}\tilde{B}}].
\end{align}
Consider that%
\begin{equation}
\Gamma_{\bar{A}\tilde{A}AA^{\prime}BB^{\prime}\bar{B}\tilde{B}}%
^{\operatorname{id}\otimes\mathcal{N}\otimes\operatorname{id}}=\Gamma_{\bar
{A}\tilde{A}}\otimes\Gamma_{AA^{\prime}BB^{\prime}}^{\mathcal{M}}\otimes
\Gamma_{\bar{B}\tilde{B}}.
\end{equation}
Then%
\begin{align}
&  T_{BB^{\prime}\bar{B}\tilde{B}}(V_{\bar{A}\tilde{A}AA^{\prime}BB^{\prime
}\bar{B}\tilde{B}}\pm\Gamma_{\bar{A}\tilde{A}}\otimes\Gamma_{AA^{\prime
}BB^{\prime}}^{\mathcal{M}}\otimes\Gamma_{\bar{B}\tilde{B}})\geq0\\
\Longrightarrow\quad &  \operatorname{Tr}_{\bar{A}\tilde{A}\bar{B}\tilde{B}%
}[T_{BB^{\prime}\bar{B}\tilde{B}}(V_{\bar{A}\tilde{A}AA^{\prime}BB^{\prime
}\bar{B}\tilde{B}}\pm\Gamma_{\bar{A}\tilde{A}}\otimes\Gamma_{AA^{\prime
}BB^{\prime}}^{\mathcal{M}}\otimes\Gamma_{\bar{B}\tilde{B}})]\geq0\\
\Longleftrightarrow\quad &  T_{BB^{\prime}}(V_{AA^{\prime}BB^{\prime}}\pm
d_{\bar{A}}d_{\bar{B}}\Gamma_{AA^{\prime}BB^{\prime}}^{\mathcal{M}})\geq0\\
\Longleftrightarrow\quad &  T_{BB^{\prime}}(V_{AA^{\prime}BB^{\prime}}%
^{\prime}\pm\Gamma_{AA^{\prime}BB^{\prime}}^{\mathcal{M}})\geq0.
\end{align}
Also%
\begin{align}
S_{\bar{A}\tilde{A}AA^{\prime}BB^{\prime}\bar{B}\tilde{B}}\pm V_{\bar{A}%
\tilde{A}AA^{\prime}BB^{\prime}\bar{B}\tilde{B}}  &  \geq0\\
\Longrightarrow\quad\operatorname{Tr}_{\bar{A}\tilde{A}\bar{B}\tilde{B}%
}[S_{\bar{A}\tilde{A}AA^{\prime}BB^{\prime}\bar{B}\tilde{B}}\pm V_{\bar
{A}\tilde{A}AA^{\prime}BB^{\prime}\bar{B}\tilde{B}}]  &  \geq0\\
\Longleftrightarrow\quad S_{AA^{\prime}BB^{\prime}}^{\prime}\pm V_{AA^{\prime
}BB^{\prime}}^{\prime}  &  \geq0,
\end{align}
and%
\begin{align}
\operatorname{Tr}_{\tilde{A}A^{\prime}}[S_{\bar{A}\tilde{A}AA^{\prime
}BB^{\prime}\bar{B}\tilde{B}}]  &  =\pi_{\bar{A}A}\otimes\operatorname{Tr}%
_{\bar{A}\tilde{A}AA^{\prime}}[S_{\bar{A}\tilde{A}AA^{\prime}BB^{\prime}%
\bar{B}\tilde{B}}]\\
\Longrightarrow\quad\operatorname{Tr}_{\bar{A}\tilde{A}A^{\prime}\bar{B}%
\tilde{B}}[S_{\bar{A}\tilde{A}AA^{\prime}BB^{\prime}\bar{B}\tilde{B}}]  &
=\operatorname{Tr}_{\bar{A}\bar{B}\tilde{B}}[\pi_{\bar{A}A}\otimes
\operatorname{Tr}_{\bar{A}\tilde{A}AA^{\prime}}[S_{\bar{A}\tilde{A}AA^{\prime
}BB^{\prime}\bar{B}\tilde{B}}]]\\
&  =\pi_{A}\otimes\operatorname{Tr}_{\bar{A}\tilde{A}AA^{\prime}\bar{B}%
\tilde{B}}[S_{\bar{A}\tilde{A}AA^{\prime}BB^{\prime}\bar{B}\tilde{B}}]\\
\Longleftrightarrow\quad\operatorname{Tr}_{A^{\prime}}[S_{AA^{\prime
}BB^{\prime}}^{\prime}]  &  =\pi_{A}\otimes\operatorname{Tr}_{AA^{\prime}%
}[S_{AA^{\prime}BB^{\prime}}^{\prime}].
\end{align}
Finally, let $\lambda$ be such that%
\begin{equation}
\frac{1}{d_{A}d_{\bar{A}}}\operatorname{Tr}_{\bar{A}\tilde{A}AA^{\prime
}B^{\prime}\tilde{B}}[S_{\bar{A}\tilde{A}AA^{\prime}BB^{\prime}\bar{B}%
\tilde{B}}]\leq\lambda I_{B\bar{B}}.
\end{equation}
Then it follows that%
\begin{align}
\operatorname{Tr}_{\bar{B}}\left[  \frac{1}{d_{A}d_{\bar{A}}}\operatorname{Tr}%
_{\bar{A}\tilde{A}AA^{\prime}B^{\prime}\tilde{B}}[S_{\bar{A}\tilde
{A}AA^{\prime}BB^{\prime}\bar{B}\tilde{B}}]\right]   &  \leq\operatorname{Tr}%
_{\bar{B}}[\lambda I_{B\bar{B}}]\\
\Longleftrightarrow\quad\frac{1}{d_{A}d_{\bar{A}}}\operatorname{Tr}_{\bar
{A}\tilde{A}AA^{\prime}B^{\prime}\bar{B}\tilde{B}}[S_{\bar{A}\tilde
{A}AA^{\prime}BB^{\prime}\bar{B}\tilde{B}}]  &  \leq d_{\bar{B}}\lambda
I_{B}\\
\Longleftrightarrow\quad\frac{1}{d_{A}}\operatorname{Tr}_{AA^{\prime}%
B^{\prime}}[S_{AA^{\prime}BB^{\prime}}^{\prime}]  &  \leq\lambda I_{B}.
\end{align}
Thus, we conclude that%
\begin{equation}
\beta(\mathcal{M}_{AB\rightarrow A^{\prime}B^{\prime}})\leq\beta
(\operatorname{id}_{\bar{A}\rightarrow\tilde{A}}\otimes\mathcal{M}%
_{AB\rightarrow A^{\prime}B^{\prime}}\otimes\operatorname{id}_{\bar
{B}\rightarrow\tilde{B}}).
\end{equation}
This concludes the proof.
\end{Proof}

\begin{proposition*}
{Zero on Classical Feedback Channels}{prop-CF:zero-feedback}Let
$\overline{\Delta}_{B\rightarrow A^{\prime}}$ be a classical feedback channel:%
\begin{equation}
\overline{\Delta}_{B\rightarrow A^{\prime}}(\cdot)\coloneqq\sum_{i=0}%
^{d-1}|i\rangle_{A^{\prime}}\langle i|_{B}(\cdot)|i\rangle_{B}\langle
i|_{A^{\prime}},
\end{equation}
where system $A^{\prime}$ is isomorphic to $ B$ and $d=d_{A^{\prime}}=d_{B}$. Then%
\begin{equation}
C_{\beta}(\overline{\Delta}_{B\rightarrow A^{\prime}})=0.
\label{eq-CF:value-beta-p-feedback-ch}%
\end{equation}

\end{proposition*}

\begin{Proof}
We prove the equivalent statement that $\beta(\overline{\Delta}_{B\rightarrow
A^{\prime}})=1$. In this case, the $A$ and $B^{\prime}$ systems are trivial,
so that $d_{A}=1$, and the Choi operator of $\overline{\Delta}_{B\rightarrow
A^{\prime}}$ is given by%
\begin{equation}
\Gamma_{BA^{\prime}}^{\overline{\Delta}}=\overline{\Gamma}_{BA^{\prime}},
\end{equation}
where%
\begin{equation}
\overline{\Gamma}_{BA^{\prime}}\coloneqq\sum_{i=0}^{d_{B}-1}|i\rangle\!\langle
i|_{B}\otimes|i\rangle\!\langle i|_{A^{\prime}}.
\end{equation}
Pick $S_{BA^{\prime}}=V_{BA^{\prime}}=\overline{\Gamma}_{BA^{\prime}}$. Then
we need to check that the constraints in \eqref{eq-CF:basic-measure-NS-PPT-bi}
are satisfied for these choices. Consider that%
\begin{align}
T_{B}(V_{BA^{\prime}}\pm\Gamma_{BA^{\prime}}^{\overline{\Delta}})  &  \geq0\\
\Longleftrightarrow\quad T_{B}(\overline{\Gamma}_{BA^{\prime}}\pm
\overline{\Gamma}_{BA^{\prime}})  &  \geq0\\
\Longleftrightarrow\quad\overline{\Gamma}_{BA^{\prime}}\pm\overline{\Gamma
}_{BA^{\prime}}  &  \geq0,
\end{align}
and the last inequality is trivially satisfied. Also,%
\begin{align}
S_{BA^{\prime}}\pm V_{BA^{\prime}}  &  \geq0\\
\Longleftrightarrow\quad\overline{\Gamma}_{BA^{\prime}}\pm\overline{\Gamma
}_{BA^{\prime}}  &  \geq0,
\end{align}
and the no-signaling condition $\operatorname{Tr}_{A^{\prime}}[S_{AA^{\prime
}BB^{\prime}}]=\pi_{A}\otimes\operatorname{Tr}_{AA^{\prime}}[S_{AA^{\prime
}BB^{\prime}}]$ is trivially satisfied because the $A$ system is trivial,
having dimension equal to one. Finally, let us evaluate the objective function
for these choices:%
\begin{align}
\frac{1}{d_{A}}\left\Vert \operatorname{Tr}_{AA^{\prime}B^{\prime}%
}[S_{AA^{\prime}BB^{\prime}}]\right\Vert _{\infty}  &  =\left\Vert
\operatorname{Tr}_{A^{\prime}}[S_{A^{\prime}B}]\right\Vert _{\infty}\\
&  =\left\Vert \operatorname{Tr}_{A^{\prime}}[\overline{\Gamma}_{BA^{\prime}%
}]\right\Vert _{\infty}\\
&  =\left\Vert I_{B}\right\Vert _{\infty}\\
&  =1.
\end{align}
Combined with the general lower bound from
Proposition~\ref{prop-CF:lower-bound-ch-non-neg}, we conclude \eqref{eq-CF:value-beta-p-feedback-ch}.
\end{Proof}

\begin{proposition*}
{Zero on Tensor Products of Local Channels}{prop-CF:zero-local-chs}Let
$\mathcal{E}_{A\rightarrow A^{\prime}}$ and $\mathcal{F}_{B\rightarrow
B^{\prime}}$ be quantum channels. Then%
\begin{equation}
C_{\beta}(\mathcal{E}_{A\rightarrow A^{\prime}}\otimes\mathcal{F}%
_{B\rightarrow B^{\prime}})=0. \label{eq-CF:local-ch-beta-1}%
\end{equation}

\end{proposition*}

\begin{Proof}
We prove the equivalent statement that $\beta(\mathcal{E}_{A\rightarrow
A^{\prime}}\otimes\mathcal{F}_{B\rightarrow B^{\prime}})=1$. Set
$S_{AA^{\prime}BB^{\prime}}=V_{AA^{\prime}BB^{\prime}}=\Gamma_{AA^{\prime}%
}^{\mathcal{E}}\otimes\Gamma_{BB^{\prime}}^{\mathcal{F}}$, where
$\Gamma_{AA^{\prime}}^{\mathcal{E}}$ and $\Gamma_{BB^{\prime}}^{\mathcal{F}}$
are the Choi operators of $\mathcal{E}_{A\rightarrow A^{\prime}}$ and
$\mathcal{F}_{B\rightarrow B^{\prime}}$, respectively. We need to check that
the constraints in \eqref{eq-CF:basic-measure-NS-PPT-bi} are satisfied for
these choices. Consider that%
\begin{align}
T_{BB^{\prime}}(V_{AA^{\prime}BB^{\prime}}\pm\Gamma_{AA^{\prime}}%
^{\mathcal{E}}\otimes\Gamma_{BB^{\prime}}^{\mathcal{F}})  &  \geq0\\
\Longleftrightarrow\quad T_{BB^{\prime}}(\Gamma_{AA^{\prime}}^{\mathcal{E}%
}\otimes\Gamma_{BB^{\prime}}^{\mathcal{F}}\pm\Gamma_{AA^{\prime}}%
^{\mathcal{E}}\otimes\Gamma_{BB^{\prime}}^{\mathcal{F}})  &  \geq0\\
\Longleftrightarrow\quad\Gamma_{AA^{\prime}}^{\mathcal{E}}\otimes
T_{BB^{\prime}}(\Gamma_{BB^{\prime}}^{\mathcal{F}})\pm\Gamma_{AA^{\prime}%
}^{\mathcal{E}}\otimes T_{BB^{\prime}}(\Gamma_{BB^{\prime}}^{\mathcal{F}})  &
\geq0,
\end{align}
and the last inequality trivially holds because $T_{BB^{\prime}}$ acts as a
positive map on $\Gamma_{BB^{\prime}}^{\mathcal{F}}$. Also,%
\begin{align}
S_{AA^{\prime}BB^{\prime}}\pm V_{AA^{\prime}BB^{\prime}}  &  \geq0\\
\Longleftrightarrow\quad\Gamma_{AA^{\prime}}^{\mathcal{E}}\otimes
\Gamma_{BB^{\prime}}^{\mathcal{F}}\pm\Gamma_{AA^{\prime}}^{\mathcal{E}}%
\otimes\Gamma_{BB^{\prime}}^{\mathcal{F}}  &  \geq0,
\end{align}
and%
\begin{align}
\operatorname{Tr}_{A^{\prime}}[S_{AA^{\prime}BB^{\prime}}]  &
=\operatorname{Tr}_{A^{\prime}}[\Gamma_{AA^{\prime}}^{\mathcal{E}}%
\otimes\Gamma_{BB^{\prime}}^{\mathcal{F}}]\\
&  =I_{A}\otimes\Gamma_{BB^{\prime}}^{\mathcal{F}}\\
&  =\pi_{A}\otimes\operatorname{Tr}_{AA^{\prime}}[\Gamma_{AA^{\prime}%
}^{\mathcal{E}}\otimes\Gamma_{BB^{\prime}}^{\mathcal{F}}]\\
&  =\pi_{A}\otimes\operatorname{Tr}_{AA^{\prime}}[S_{AA^{\prime}BB^{\prime}}].
\end{align}
Finally, consider that the objective function evaluates to%
\begin{align}
\left\Vert \operatorname{Tr}_{A^{\prime}B^{\prime}}[S_{AA^{\prime}BB^{\prime}%
}]\right\Vert _{\infty}  &  =\left\Vert \operatorname{Tr}_{A^{\prime}%
B^{\prime}}[\Gamma_{AA^{\prime}}^{\mathcal{E}}\otimes\Gamma_{BB^{\prime}%
}^{\mathcal{F}}]\right\Vert _{\infty}\\
&  =\left\Vert I_{AB}\right\Vert _{\infty}\\
&  =1.
\end{align}
Combined with the general lower bound from
Proposition~\ref{prop-CF:lower-bound-ch-non-neg}, we conclude \eqref{eq-CF:local-ch-beta-1}.
\end{Proof}

\begin{proposition*}
{Subadditivity under Composition}
{prop-CF:subadditivity-bipartite-CP-maps}Let $\mathcal{M}_{AB\rightarrow
A^{\prime}B^{\prime}}^{1},\mathcal{M}_{A^{\prime}B^{\prime}\rightarrow
A^{\prime\prime}B^{\prime\prime}}^{2}$ be  completely positive bipartite maps,
and define%
\begin{equation}
\mathcal{M}_{AB\rightarrow A^{\prime\prime}B^{\prime\prime}}^{3}%
\coloneqq\mathcal{M}_{A^{\prime}B^{\prime}\rightarrow A^{\prime\prime
}B^{\prime\prime}}^{2}\circ\mathcal{M}_{AB\rightarrow A^{\prime}B^{\prime}%
}^{1}.
\end{equation}
Then%
\begin{equation}
C_{\beta}(\mathcal{M}_{AB\rightarrow A^{\prime\prime}B^{\prime\prime}}%
^{3})\leq C_{\beta}(\mathcal{M}_{A^{\prime}B^{\prime}\rightarrow
A^{\prime\prime}B^{\prime\prime}}^{2})+C_{\beta}(\mathcal{M}_{AB\rightarrow
A^{\prime}B^{\prime}}^{1}). \label{eq-CF:subadd-serial-comp-maps}%
\end{equation}

\end{proposition*}

\begin{Proof}
We prove the equivalent statement that%
\begin{equation}
\beta(\mathcal{M}_{AB\rightarrow A^{\prime\prime}B^{\prime\prime}}^{3}%
)\leq\beta(\mathcal{M}_{A^{\prime}B^{\prime}\rightarrow A^{\prime\prime
}B^{\prime\prime}}^{2})\cdot\beta(\mathcal{M}_{AB\rightarrow A^{\prime
}B^{\prime}}^{1}).
\end{equation}
Let $S_{AA^{\prime}BB^{\prime}}^{1}$ and $V_{AA^{\prime}BB^{\prime}}^{1}$
satisfy%
\begin{align}
T_{BB^{\prime}}(V_{AA^{\prime}BB^{\prime}}^{1}\pm\Gamma_{AA^{\prime}%
BB^{\prime}}^{\mathcal{M}^{1}})  &  \geq0,\label{eq-CF:map-1-constraint-1}\\
S_{AA^{\prime}BB^{\prime}}^{1}\pm V_{AA^{\prime}BB^{\prime}}^{1}  &
\geq0,\label{eq-CF:map-1-constraint-2}\\
\operatorname{Tr}_{A^{\prime}}[S_{AA^{\prime}BB^{\prime}}^{1}]  &  =\pi
_{A}\otimes\operatorname{Tr}_{AA^{\prime}}[S_{AA^{\prime}BB^{\prime}}^{1}],
\label{eq-CF:map-1-constraint-3}%
\end{align}
and let $S_{A^{\prime}A^{\prime\prime}B^{\prime}B^{\prime\prime}}^{2}$ and
$V_{A^{\prime}A^{\prime\prime}B^{\prime}B^{\prime\prime}}^{2}$ satisfy%
\begin{align}
T_{B^{\prime}B^{\prime\prime}}(V_{A^{\prime}A^{\prime\prime}B^{\prime
}B^{\prime\prime}}^{2}\pm\Gamma_{A^{\prime}A^{\prime\prime}B^{\prime}%
B^{\prime\prime}}^{\mathcal{M}^{2}})  &  \geq
0,\label{eq-CF:map-2-constraint-1}\\
S_{A^{\prime}A^{\prime\prime}B^{\prime}B^{\prime\prime}}^{2}\pm V_{A^{\prime
}A^{\prime\prime}B^{\prime}B^{\prime\prime}}^{2}  &  \geq
0,\label{eq-CF:map-2-constraint-2}\\
\operatorname{Tr}_{A^{\prime\prime}}[S_{A^{\prime}A^{\prime\prime}B^{\prime
}B^{\prime\prime}}^{2}]  &  =\pi_{A^{\prime}}\otimes\operatorname{Tr}%
_{A^{\prime}A^{\prime\prime}}[S_{A^{\prime}A^{\prime\prime}B^{\prime}%
B^{\prime\prime}}^{2}]. \label{eq-CF:map-2-constraint-3}%
\end{align}
Then it follows that%
\begin{align}
T_{BB^{\prime}B^{\prime}B^{\prime\prime}}(V_{AA^{\prime}BB^{\prime}}%
^{1}\otimes V_{A^{\prime}A^{\prime\prime}B^{\prime}B^{\prime\prime}}^{2}%
\pm\Gamma_{AA^{\prime}BB^{\prime}}^{\mathcal{M}^{1}}\otimes\Gamma_{A^{\prime
}A^{\prime\prime}B^{\prime}B^{\prime\prime}}^{\mathcal{M}^{2}})  &
\geq0,\label{eq-CF:subadd-proof-intermed-1}\\
S_{AA^{\prime}BB^{\prime}}^{1}\otimes S_{A^{\prime}A^{\prime\prime}B^{\prime
}B^{\prime\prime}}^{2}\pm V_{AA^{\prime}BB^{\prime}}^{1}\otimes V_{A^{\prime
}A^{\prime\prime}B^{\prime}B^{\prime\prime}}^{2}  &  \geq0.
\label{eq-CF:subadd-proof-intermed-2}%
\end{align}
This latter statement is a consequence of the general fact that if $A$, $B$,
$C$, and $D$ are Hermitian operators satisfying $A\pm B\geq0$ and $C\pm
D\geq0$, then $A\otimes C\pm B\otimes D\geq0$. To see this, consider that the
original four operator inequalities imply the four operator inequalities
$\left(  A\pm B\right)  \otimes\left(  C\pm D\right)  \geq0$, and then summing
these four different operator inequalities in various ways leads to $A\otimes
C\pm B\otimes D\geq0$.

Now apply the following positive map to
\eqref{eq-CF:subadd-proof-intermed-1}--\eqref{eq-CF:subadd-proof-intermed-2}:%
\begin{equation}
(\cdot)\rightarrow(\langle\Gamma|_{A^{\prime}A^{\prime}}\otimes\langle
\Gamma|_{B^{\prime}B^{\prime}})(\cdot)(|\Gamma\rangle_{A^{\prime}A^{\prime}%
}\otimes|\Gamma\rangle_{B^{\prime}B^{\prime}}),
\end{equation}
where%
\begin{align}
|\Gamma\rangle_{A^{\prime}A^{\prime}}  &  \coloneqq\sum_{i}|i\rangle
_{A^{\prime}}|i\rangle_{A^{\prime}},\\
|\Gamma\rangle_{B^{\prime}B^{\prime}}  &  \coloneqq\sum_{i}|i\rangle
_{B^{\prime}}|i\rangle_{B^{\prime}}.
\end{align}
This gives%
\begin{align}
T_{BB^{\prime\prime}}(V_{AA^{\prime\prime}BB^{\prime\prime}}^{3}\pm
\Gamma_{AA^{\prime\prime}BB^{\prime\prime}}^{\mathcal{M}^{2}\circ
\mathcal{M}^{1}})  &  \geq0,\label{eq-CF:concat-constraint-1}\\
S_{AA^{\prime\prime}BB^{\prime\prime}}^{3}\pm V_{AA^{\prime\prime}%
BB^{\prime\prime}}^{3}  &  \geq0, \label{eq-CF:concat-constraint-2}%
\end{align}
where%
\begin{align}
V_{AA^{\prime\prime}BB^{\prime\prime}}^{3}  &  \coloneqq(\langle
\Gamma|_{A^{\prime}A^{\prime}}\otimes\langle\Gamma|_{B^{\prime}B^{\prime}%
})(V_{AA^{\prime}BB^{\prime}}^{1}\otimes V_{A^{\prime}A^{\prime\prime
}B^{\prime}B^{\prime\prime}}^{2})(|\Gamma\rangle_{A^{\prime}A^{\prime}}%
\otimes|\Gamma\rangle_{B^{\prime}B^{\prime}}),\\
\Gamma_{AA^{\prime\prime}BB^{\prime\prime}}^{\mathcal{M}^{2}\circ
\mathcal{M}^{1}}  &  \coloneqq(\langle\Gamma|_{A^{\prime}A^{\prime}}%
\otimes\langle\Gamma|_{B^{\prime}B^{\prime}})(\Gamma_{AA^{\prime}BB^{\prime}%
}^{\mathcal{M}^{1}}\otimes\Gamma_{A^{\prime}A^{\prime\prime}B^{\prime
}B^{\prime\prime}}^{\mathcal{M}^{2}})(|\Gamma\rangle_{A^{\prime}A^{\prime}%
}\otimes|\Gamma\rangle_{B^{\prime}B^{\prime}}),\\
S_{AA^{\prime\prime}BB^{\prime\prime}}^{3}  &  \coloneqq(\langle
\Gamma|_{A^{\prime}A^{\prime}}\otimes\langle\Gamma|_{B^{\prime}B^{\prime}%
})(S_{AA^{\prime}BB^{\prime}}^{1}\otimes S_{A^{\prime}A^{\prime\prime
}B^{\prime}B^{\prime\prime}}^{2})(|\Gamma\rangle_{A^{\prime}A^{\prime}}%
\otimes|\Gamma\rangle_{B^{\prime}B^{\prime}}),
\end{align}
and we applied \eqref{eq-QM:serial-comp-choi} to conclude that%
\begin{equation}
(\langle\Gamma|_{A^{\prime}A^{\prime}}\otimes\langle\Gamma|_{B^{\prime
}B^{\prime}})(\Gamma_{AA^{\prime}BB^{\prime}}^{\mathcal{M}^{1}}\otimes
\Gamma_{A^{\prime}A^{\prime\prime}B^{\prime}B^{\prime\prime}}^{\mathcal{M}%
^{2}})(|\Gamma\rangle_{A^{\prime}A^{\prime}}\otimes|\Gamma\rangle_{B^{\prime
}B^{\prime}})=\Gamma_{AA^{\prime\prime}BB^{\prime\prime}}^{\mathcal{M}%
^{2}\circ\mathcal{M}^{1}}.
\end{equation}
Also, consider that%
\begin{align}
&  \operatorname{Tr}_{A^{\prime\prime}}[S_{AA^{\prime\prime}BB^{\prime\prime}%
}^{3}]\nonumber\\
&  =\operatorname{Tr}_{A^{\prime\prime}}[(\langle\Gamma|_{A^{\prime}A^{\prime
}}\otimes\langle\Gamma|_{B^{\prime}B^{\prime}})(S_{AA^{\prime}BB^{\prime}}%
^{1}\otimes S_{A^{\prime}A^{\prime\prime}B^{\prime}B^{\prime\prime}}%
^{2})(|\Gamma\rangle_{A^{\prime}A^{\prime}}\otimes|\Gamma\rangle_{B^{\prime
}B^{\prime}})]\\
&  =(\langle\Gamma|_{A^{\prime}A^{\prime}}\otimes\langle\Gamma|_{B^{\prime
}B^{\prime}})(S_{AA^{\prime}BB^{\prime}}^{1}\otimes\operatorname{Tr}%
_{A^{\prime\prime}}[S_{A^{\prime}A^{\prime\prime}B^{\prime}B^{\prime\prime}%
}^{2}])(|\Gamma\rangle_{A^{\prime}A^{\prime}}\otimes|\Gamma\rangle_{B^{\prime
}B^{\prime}})\\
&  =(\langle\Gamma|_{A^{\prime}A^{\prime}}\otimes\langle\Gamma|_{B^{\prime
}B^{\prime}})(S_{AA^{\prime}BB^{\prime}}^{1}\otimes\pi_{A^{\prime}}%
\otimes\operatorname{Tr}_{A^{\prime}A^{\prime\prime}}[S_{A^{\prime}%
A^{\prime\prime}B^{\prime}B^{\prime\prime}}^{2}])(|\Gamma\rangle_{A^{\prime
}A^{\prime}}\otimes|\Gamma\rangle_{B^{\prime}B^{\prime}})\\
&  =\frac{1}{d_{A^{\prime}}}(\langle\Gamma|_{A^{\prime}A^{\prime}}%
\otimes\langle\Gamma|_{B^{\prime}B^{\prime}})(S_{AA^{\prime}BB^{\prime}}%
^{1}\otimes I_{A^{\prime}}\otimes\operatorname{Tr}_{A^{\prime}A^{\prime\prime
}}[S_{A^{\prime}A^{\prime\prime}B^{\prime}B^{\prime\prime}}^{2}])(|\Gamma
\rangle_{A^{\prime}A^{\prime}}\otimes|\Gamma\rangle_{B^{\prime}B^{\prime}})\\
&  =\frac{1}{d_{A^{\prime}}}\langle\Gamma|_{B^{\prime}B^{\prime}%
}(\operatorname{Tr}_{A^{\prime}}[S_{AA^{\prime}BB^{\prime}}^{1}]\otimes
\operatorname{Tr}_{A^{\prime}A^{\prime\prime}}[S_{A^{\prime}A^{\prime\prime
}B^{\prime}B^{\prime\prime}}^{2}])|\Gamma\rangle_{B^{\prime}B^{\prime}}\\
&  =\frac{1}{d_{A^{\prime}}}\langle\Gamma|_{B^{\prime}B^{\prime}}(\pi
_{A}\otimes\operatorname{Tr}_{AA^{\prime}}[S_{AA^{\prime}BB^{\prime}}%
^{1}]\otimes\operatorname{Tr}_{A^{\prime}A^{\prime\prime}}[S_{A^{\prime
}A^{\prime\prime}B^{\prime}B^{\prime\prime}}^{2}])|\Gamma\rangle_{B^{\prime
}B^{\prime}}\\
&  =\pi_{A}\otimes\frac{1}{d_{A^{\prime}}}\langle\Gamma|_{B^{\prime}B^{\prime
}}(\operatorname{Tr}_{AA^{\prime}}[S_{AA^{\prime}BB^{\prime}}^{1}%
]\otimes\operatorname{Tr}_{A^{\prime}A^{\prime\prime}}[S_{A^{\prime}%
A^{\prime\prime}B^{\prime}B^{\prime\prime}}^{2}])|\Gamma\rangle_{B^{\prime
}B^{\prime}}.
\end{align}
Now consider that%
\begin{equation}
\operatorname{Tr}_{AA^{\prime\prime}}[S_{AA^{\prime\prime}BB^{\prime\prime}%
}^{3}]=\frac{1}{d_{A^{\prime}}}\langle\Gamma|_{B^{\prime}B^{\prime}%
}(\operatorname{Tr}_{AA^{\prime}}[S_{AA^{\prime}BB^{\prime}}^{1}%
]\otimes\operatorname{Tr}_{A^{\prime}A^{\prime\prime}}[S_{A^{\prime}%
A^{\prime\prime}B^{\prime}B^{\prime\prime}}^{2}])|\Gamma\rangle_{B^{\prime
}B^{\prime}}.
\end{equation}
So we conclude that%
\begin{equation}
\operatorname{Tr}_{A^{\prime\prime}}[S_{AA^{\prime\prime}BB^{\prime\prime}%
}^{3}]=\pi_{A}\otimes\operatorname{Tr}_{AA^{\prime\prime}}[S_{AA^{\prime
\prime}BB^{\prime\prime}}^{3}]. \label{eq-CF:concat-constraint-3}%
\end{equation}
Finally, consider that%
\begin{align}
&  \left\Vert \operatorname{Tr}_{A^{\prime\prime}B^{\prime\prime}%
}[S_{AA^{\prime\prime}BB^{\prime\prime}}^{3}]\right\Vert _{\infty}\nonumber\\
&  =\left\Vert \operatorname{Tr}_{A^{\prime\prime}B^{\prime\prime}}[\left(
\langle\Gamma|_{A^{\prime}A^{\prime}}\otimes\langle\Gamma|_{B^{\prime
}B^{\prime}}\right)  \left(  S_{AA^{\prime}BB^{\prime}}^{1}\otimes
S_{A^{\prime}A^{\prime\prime}B^{\prime}B^{\prime\prime}}^{2}\right)
(|\Gamma\rangle_{A^{\prime}A^{\prime}}\otimes|\Gamma\rangle_{B^{\prime
}B^{\prime}})]\right\Vert _{\infty}\\
&  =\left\Vert [\left(  \langle\Gamma|_{A^{\prime}A^{\prime}}\otimes
\langle\Gamma|_{B^{\prime}B^{\prime}}\right)  \left(  S_{AA^{\prime}%
BB^{\prime}}^{1}\otimes\operatorname{Tr}_{A^{\prime\prime}B^{\prime\prime}%
}[S_{A^{\prime}A^{\prime\prime}B^{\prime}B^{\prime\prime}}^{2}]\right)
(|\Gamma\rangle_{A^{\prime}A^{\prime}}\otimes|\Gamma\rangle_{B^{\prime
}B^{\prime}})]\right\Vert _{\infty}\\
&  \leq\left\Vert \operatorname{Tr}_{A^{\prime\prime}B^{\prime\prime}%
}[S_{A^{\prime}A^{\prime\prime}B^{\prime}B^{\prime\prime}}^{2}]\right\Vert
_{\infty}\cdot\nonumber\\
&  \qquad\left\Vert \lbrack\left(  \langle\Gamma|_{A^{\prime}A^{\prime}%
}\otimes\langle\Gamma|_{B^{\prime}B^{\prime}}\right)  \left(  S_{AA^{\prime
}BB^{\prime}}^{1}\otimes I_{A^{\prime}B^{\prime}}\right)  (|\Gamma
\rangle_{A^{\prime}A^{\prime}}\otimes|\Gamma\rangle_{B^{\prime}B^{\prime}%
})]\right\Vert _{\infty}\\
&  =\left\Vert \operatorname{Tr}_{A^{\prime\prime}B^{\prime\prime}%
}[S_{A^{\prime}A^{\prime\prime}B^{\prime}B^{\prime\prime}}^{2}]\right\Vert
_{\infty}\left\Vert \operatorname{Tr}_{A^{\prime}B^{\prime}}[S_{AA^{\prime
}BB^{\prime}}^{1}]\right\Vert _{\infty}.
\end{align}
Since $S_{AA^{\prime\prime}BB^{\prime\prime}}^{3}$ and $V_{AA^{\prime\prime
}BB^{\prime\prime}}^{3}$ are particular choices that satisfy the constraints
in \eqref{eq-CF:concat-constraint-1}--\eqref{eq-CF:concat-constraint-3}, we
conclude that%
\begin{equation}
\beta(\mathcal{M}_{AB\rightarrow A^{\prime\prime}B^{\prime\prime}}^{3}%
)\leq\left\Vert \operatorname{Tr}_{A^{\prime\prime}B^{\prime\prime}%
}[S_{A^{\prime}A^{\prime\prime}B^{\prime}B^{\prime\prime}}^{2}]\right\Vert
_{\infty}\left\Vert \operatorname{Tr}_{A^{\prime}B^{\prime}}[S_{AA^{\prime
}BB^{\prime}}^{1}]\right\Vert _{\infty}.
\end{equation}
Since $S_{AA^{\prime}BB^{\prime}}^{1}$ and $V_{AA^{\prime}BB^{\prime}}^{1}$
are arbitrary Hermitian operators satisfying the constraints in
\eqref{eq-CF:map-1-constraint-1}--\eqref{eq-CF:map-1-constraint-3} and
$S_{A^{\prime}A^{\prime\prime}B^{\prime}B^{\prime\prime}}^{2}$ and
$V_{A^{\prime}A^{\prime\prime}B^{\prime}B^{\prime\prime}}^{2}$ are arbitrary
Hermitian operators satisfying the constraints in
\eqref{eq-CF:map-2-constraint-1}--\eqref{eq-CF:map-2-constraint-3}, we
conclude \eqref{eq-CF:subadd-serial-comp-maps}.
\end{Proof}

\begin{corollary*}{Data Processing under Local Channels}{cor:DP-LC-beta-p-bipartite-ch}Let
$\mathcal{M}_{AB\rightarrow A^{\prime}B^{\prime}}$ be a completely positive
bipartite map. Let $\mathcal{K}_{\hat{A}\rightarrow A}$, $\mathcal{L}_{\hat
{B}\rightarrow B}$, $\mathcal{N}_{A^{\prime}\rightarrow A^{\prime\prime}}$,
and $\mathcal{P}_{B^{\prime}\rightarrow B^{\prime\prime}}$ be local quantum
channels, and define the bipartite completely positive map $\mathcal{F}%
_{\hat{A}\hat{B}\rightarrow A^{\prime\prime}B^{\prime\prime}}$ as follows:%
\begin{equation}
\mathcal{F}_{\hat{A}\hat{B}\rightarrow A^{\prime\prime}B^{\prime\prime}%
}\coloneqq(\mathcal{N}_{A^{\prime}\rightarrow A^{\prime\prime}}\otimes
\mathcal{P}_{B^{\prime}\rightarrow B^{\prime\prime}})\mathcal{M}%
_{AB\rightarrow A^{\prime}B^{\prime}}(\mathcal{K}_{\hat{A}\rightarrow
A}\otimes\mathcal{L}_{\hat{B}\rightarrow B}).
\end{equation}
Then%
\begin{equation}
C_{\beta}(\mathcal{F}_{\hat{A}\hat{B}\rightarrow A^{\prime\prime}%
B^{\prime\prime}})\leq C_{\beta}(\mathcal{M}_{AB\rightarrow A^{\prime
}B^{\prime}}).
\end{equation}

\end{corollary*}

\begin{Proof}
Apply Propositions~\ref{prop-CF:zero-local-chs}\ and
\ref{prop-CF:subadditivity-bipartite-CP-maps}\ to find that%
\begin{align}
&  C_{\beta}(\mathcal{F}_{\hat{A}\hat{B}\rightarrow A^{\prime\prime}%
B^{\prime\prime}})\nonumber\\
&  \leq C_{\beta}(\mathcal{N}_{A^{\prime}\rightarrow A^{\prime\prime}}%
\otimes\mathcal{P}_{B^{\prime}\rightarrow B^{\prime\prime}})+C_{\beta
}(\mathcal{M}_{AB\rightarrow A^{\prime}B^{\prime}})+C_{\beta}(\mathcal{K}%
_{\hat{A}\rightarrow A}\otimes\mathcal{L}_{\hat{B}\rightarrow B})\\
&  =C_{\beta}(\mathcal{M}_{AB\rightarrow A^{\prime}B^{\prime}}).
\end{align}
This concludes the proof.
\end{Proof}

\begin{corollary*}
{Invariance under Local Unitary Channels}{cor:I-LUC-beta-p-bipartite-ch}%
Let $\mathcal{M}_{AB\rightarrow A^{\prime}B^{\prime}}$ be a completely
positive bipartite map. Let $\mathcal{U}_{A}$, $\mathcal{V}_{B}$,
$\mathcal{W}_{A^{\prime}}$, and $\mathcal{Y}_{B^{\prime}}$ be local unitary
channels, and define the bipartite completely positive map $\mathcal{F}%
_{\hat{A}\hat{B}\rightarrow A^{\prime\prime}B^{\prime\prime}}$ as follows:%
\begin{equation}
\mathcal{F}_{AB\rightarrow A^{\prime}B^{\prime}}\coloneqq(\mathcal{W}%
_{A^{\prime}}\otimes\mathcal{Y}_{B^{\prime}})\mathcal{M}_{AB\rightarrow
A^{\prime}B^{\prime}}(\mathcal{U}_{A}\otimes\mathcal{V}_{B}).
\end{equation}
Then%
\begin{equation}
C_{\beta}(\mathcal{F}_{AB\rightarrow A^{\prime}B^{\prime}})=C_{\beta
}(\mathcal{M}_{AB\rightarrow A^{\prime}B^{\prime}}).
\end{equation}

\end{corollary*}

\begin{Proof}
Apply Corollary~\ref{cor:DP-LC-beta-p-bipartite-ch} twice to conclude that
$C_{\beta}(\mathcal{M}_{AB\rightarrow A^{\prime}B^{\prime}})\geq C_{\beta
}(\mathcal{F}_{AB\rightarrow A^{\prime}B^{\prime}})$ and $C_{\beta
}(\mathcal{F}_{AB\rightarrow A^{\prime}B^{\prime}})\geq C_{\beta}%
(\mathcal{M}_{AB\rightarrow A^{\prime}B^{\prime}})$.
\end{Proof}

\begin{proposition*}
{Convexity}{prop-CF:convexity-beta}The measure $\beta$ is convex, in the
following sense:%
\begin{equation}
\beta(\mathcal{M}_{AB\rightarrow A^{\prime}B^{\prime}}^{\lambda})\leq
\lambda\beta(\mathcal{M}_{AB\rightarrow A^{\prime}B^{\prime}}^{1})+\left(
1-\lambda\right)  \beta(\mathcal{M}_{AB\rightarrow A^{\prime}B^{\prime}}^{0}),
\label{eq-CF:convexity-beta}%
\end{equation}
where $\mathcal{M}_{AB\rightarrow A^{\prime}B^{\prime}}^{0}$ and
$\mathcal{M}_{AB\rightarrow A^{\prime}B^{\prime}}^{1}$ are completely positive
bipartite maps, $\lambda\in\left[  0,1\right]  $, and%
\begin{equation}
\mathcal{M}_{AB\rightarrow A^{\prime}B^{\prime}}^{\lambda}\coloneqq \lambda
\mathcal{M}_{AB\rightarrow A^{\prime}B^{\prime}}^{1}+\left(  1-\lambda\right)
\mathcal{M}_{AB\rightarrow A^{\prime}B^{\prime}}^{0}.
\end{equation}

\end{proposition*}

\begin{Proof}
Let $S_{AA^{\prime}BB^{\prime}}^{x}$ and $V_{AA^{\prime}BB^{\prime}}^{x}$
satisfy the constraints in \eqref{eq-CF:basic-measure-NS-PPT-bi}\ for
$\mathcal{M}_{AB\rightarrow A^{\prime}B^{\prime}}^{x}$ for $x\in\left\{
0,1\right\}  $. Then%
\begin{align}
S_{AA^{\prime}BB^{\prime}}^{\lambda}  &  \coloneqq \lambda S_{AA^{\prime}BB^{\prime}%
}^{1}+\left(  1-\lambda\right)  S_{AA^{\prime}BB^{\prime}}^{0},\\
V_{AA^{\prime}BB^{\prime}}^{\lambda}  &  \coloneqq \lambda V_{AA^{\prime}BB^{\prime}%
}^{1}+\left(  1-\lambda\right)  V_{AA^{\prime}BB^{\prime}}^{0},
\end{align}
satisfy the constraints in \eqref{eq-CF:basic-measure-NS-PPT-bi}\ for
$\mathcal{M}_{AB\rightarrow A^{\prime}B^{\prime}}^{\lambda}$.\ Then it follows
that%
\begin{align}
\beta(\mathcal{M}_{AB\rightarrow A^{\prime}B^{\prime}}^{\lambda})  &
\leq\left\Vert \operatorname{Tr}_{A^{\prime}B^{\prime}}[S_{AA^{\prime
}BB^{\prime}}^{\lambda}]\right\Vert _{\infty}\\
&  \leq\lambda\left\Vert \operatorname{Tr}_{A^{\prime}B^{\prime}%
}[S_{AA^{\prime}BB^{\prime}}^{1}]\right\Vert _{\infty}+\left(  1-\lambda
\right)  \left\Vert \operatorname{Tr}_{A^{\prime}B^{\prime}}[S_{AA^{\prime
}BB^{\prime}}^{0}]\right\Vert _{\infty},
\end{align}
where the second inequality follows from convexity of the $\infty$-norm. Since
the inequality holds for all $S_{AA^{\prime}BB^{\prime}}^{x}$ and
$V_{AA^{\prime}BB^{\prime}}^{x}$ satisfying the constraints in
\eqref{eq-CF:basic-measure-NS-PPT-bi}\ for $\mathcal{M}_{AB\rightarrow
A^{\prime}B^{\prime}}^{x}$ for $x\in\left\{  0,1\right\}  $, we conclude \eqref{eq-CF:convexity-beta}.
\end{Proof}

\subsubsection{Related Measures}

We now define variations of the bipartite channel measure from
\eqref{eq-CF:basic-measure-NS-PPT-bi}. We employ generalized divergences to do
so, and in doing so, we arrive at a large number of variations of the basic
bipartite channel measure.

Using the generalized channel divergence from Definition~\ref{def-gen_channel_div}, we define the following:

\begin{definition}{$\Upsilon$-Measure of Classical Communication for Bipartite Channels}{def-CF:beta-p2p}
For a bipartite channel $\mathcal{N}_{AB\rightarrow A^{\prime}B^{\prime}}$, we
define the following measure of forward classical communication:%
\begin{equation}
\boldsymbol{\Upsilon}(\mathcal{N}_{AB\rightarrow A^{\prime}B^{\prime}%
})\coloneqq\inf_{\mathcal{M}_{AB\rightarrow A^{\prime}B^{\prime}}%
:\beta(\mathcal{M}_{AB\rightarrow A^{\prime}B^{\prime}})\leq1}\boldsymbol{D}%
(\mathcal{N}_{AB\rightarrow A^{\prime}B^{\prime}}\Vert\mathcal{M}%
_{AB\rightarrow A^{\prime}B^{\prime}}),
\label{eq-CF:gen-div-ch-ups-meas-bi-map}%
\end{equation}
where the optimization is with respect to completely positive bipartite maps
$\mathcal{M}_{AB\rightarrow A^{\prime}B^{\prime}}$.
\end{definition}

Using the quantum relative entropy, the sandwiched R\'{e}nyi relative entropy,
the Belavkin--Staszewski relative entropy, and the geometric R\'{e}nyi
relative entropy, we then obtain the following respective channel measures:
$\Upsilon(\mathcal{N}_{AB\rightarrow A^{\prime}B^{\prime}})$, $\widetilde
{\Upsilon}_{\alpha}(\mathcal{N}_{AB\rightarrow A^{\prime}B^{\prime}})$,
$\widehat{\Upsilon}(\mathcal{N}_{AB\rightarrow A^{\prime}B^{\prime}})$, and
$\widehat{\Upsilon}_{\alpha}(\mathcal{N}_{AB\rightarrow A^{\prime}B^{\prime}%
})$, defined by substituting $\boldsymbol{D}$ with $D$, $\widetilde{D}%
_{\alpha}$, $\widehat{D}$, and $\widehat{D}_{\alpha}$.

We now establish some properties of $\boldsymbol{\Upsilon}(\mathcal{N}%
_{AB\rightarrow A^{\prime}B^{\prime}})$, analogous to those established
earlier for $C_{\beta}(\mathcal{N}_{AB\rightarrow A^{\prime}B^{\prime}})$ in
Section~\ref{sec-CF:props-basic-measure}. We assume that the underlying
generalized divergence satisfies the minimal assumptions in
\eqref{eq-QEI:min-prop-gen-div} and \eqref{eq-QEI:min-prop-gen-div-2}; that
is, $\boldsymbol{D}(1\Vert c)\geq0$ for $c\in(0,1]$ and $\boldsymbol{D}%
(\rho\Vert\rho)=0$ for every state~$\rho$.

\begin{proposition*}
{Non-Negativity}{prop-CF:non-neg-gen-div-ups}Let $\mathcal{N}%
_{AB\rightarrow A^{\prime}B^{\prime}}$ be a bipartite channel. Then
\begin{equation}
\boldsymbol{\Upsilon}(\mathcal{N}_{AB\rightarrow A^{\prime}B^{\prime}})\geq0.
\end{equation}

\end{proposition*}

\begin{Proof}
We prove the first inequality and the proof of the second inequality is
similar. Consider that%
\begin{align}
&  \boldsymbol{\Upsilon}(\mathcal{N}_{AB\rightarrow A^{\prime}B^{\prime}%
})\nonumber\\
&  =\inf_{\substack{\mathcal{M}_{AB\rightarrow A^{\prime}B^{\prime}}%
:\\\beta(\mathcal{M}_{AB\rightarrow A^{\prime}B^{\prime}})\leq1}%
}\boldsymbol{D}(\mathcal{N}_{AB\rightarrow A^{\prime}B^{\prime}}%
\Vert\mathcal{M}_{AB\rightarrow A^{\prime}B^{\prime}})\nonumber\\
&  \geq\inf_{\substack{\mathcal{M}_{AB\rightarrow A^{\prime}B^{\prime}%
}:\\\beta(\mathcal{M}_{AB\rightarrow A^{\prime}B^{\prime}})\leq1}%
}\boldsymbol{D}(\mathcal{N}_{AB\rightarrow A^{\prime}B^{\prime}}(\Phi
_{RA}\otimes\Phi_{BS})\Vert\mathcal{M}_{AB\rightarrow A^{\prime}B^{\prime}%
}(\Phi_{RA}\otimes\Phi_{BS}))\nonumber\\
&  \geq\inf_{\substack{\mathcal{M}_{AB\rightarrow A^{\prime}B^{\prime}%
}:\\\beta(\mathcal{M}_{AB\rightarrow A^{\prime}B^{\prime}})\leq1}%
}\boldsymbol{D}(\operatorname{Tr}[\mathcal{N}_{AB\rightarrow A^{\prime
}B^{\prime}}(\Phi_{RA}\otimes\Phi_{BS})]\Vert\operatorname{Tr}[\mathcal{M}%
_{AB\rightarrow A^{\prime}B^{\prime}}(\Phi_{RA}\otimes\Phi_{BS})])\nonumber\\
&  =\inf_{\mathcal{M}_{AB\rightarrow A^{\prime}B^{\prime}}:\beta
(\mathcal{M}_{AB\rightarrow A^{\prime}B^{\prime}})\leq1}\boldsymbol{D}%
(1\Vert\operatorname{Tr}[\mathcal{M}_{AB\rightarrow A^{\prime}B^{\prime}}%
(\Phi_{RA}\otimes\Phi_{BS})])
\end{align}
The first inequality follows because $\boldsymbol{D}(\mathcal{N}%
_{AB\rightarrow A^{\prime}B^{\prime}}\Vert\mathcal{M}_{AB\rightarrow
A^{\prime}B^{\prime}})$ involves an optimization over all possible input
states, and we have chosen the product of maximally entangled states. The
second inequality follows from the data-processing inequality for the
generalized divergence. Thus, the inequality follows if we can show that%
\begin{equation}
\operatorname{Tr}[\mathcal{M}_{AB\rightarrow A^{\prime}B^{\prime}}(\Phi
_{RA}\otimes\Phi_{BS})]\leq1. \label{eq-CF:trace-M-less-one}%
\end{equation}
Let $\lambda$, $S_{AA^{\prime}BB^{\prime}}$, and $V_{AA^{\prime}BB^{\prime}}$
be arbitrary Hermitian operators satisfying the constraints in
\eqref{eq-CF:basic-measure-NS-PPT-bi-rewrite} for $\mathcal{M}_{AB\rightarrow
A^{\prime}B^{\prime}}$. Then, we find that%
\begin{align}
\lambda d_{A}d_{B}  &  =\lambda\operatorname{Tr}_{AB}[I_{AB}]\\
&  \geq\operatorname{Tr}_{AA^{\prime}BB^{\prime}}[S_{AA^{\prime}BB^{\prime}%
}]\\
&  \geq\operatorname{Tr}_{AA^{\prime}BB^{\prime}}[V_{AA^{\prime}BB^{\prime}%
}]\\
&  =\operatorname{Tr}_{AA^{\prime}BB^{\prime}}[T_{BB^{\prime}}(V_{AA^{\prime
}BB^{\prime}})]\\
&  \geq\operatorname{Tr}_{AA^{\prime}BB^{\prime}}[T_{BB^{\prime}}%
(\Gamma_{AA^{\prime}BB^{\prime}}^{\mathcal{M}})]\\
&  =\operatorname{Tr}_{AA^{\prime}BB^{\prime}}[\Gamma_{AA^{\prime}BB^{\prime}%
}^{\mathcal{M}}]\\
&  =\operatorname{Tr}[\Gamma_{AA^{\prime}BB^{\prime}}^{\mathcal{M}}],
\end{align}
which is equivalent to%
\begin{equation}
\lambda\geq\operatorname{Tr}[\mathcal{M}_{AB\rightarrow A^{\prime}B^{\prime}%
}(\Phi_{RA}\otimes\Phi_{BS})].
\end{equation}
Taking an infimum over $\lambda$, $S_{AA^{\prime}BB^{\prime}}$, and
$V_{AA^{\prime}BB^{\prime}}$ satisfying the constraints in
\eqref{eq-CF:basic-measure-NS-PPT-bi-rewrite} for $\mathcal{M}_{AB\rightarrow
A^{\prime}B^{\prime}}$ and applying the assumption $\beta(\mathcal{M}%
_{AB\rightarrow A^{\prime}B^{\prime}})\leq1$, we conclude \eqref{eq-CF:trace-M-less-one}.
\end{Proof}

\begin{proposition*}
{Stability}{prop-CF:stability-gen-div-ups}Let $\mathcal{N}%
_{AB\rightarrow A^{\prime}B^{\prime}}$ be a bipartite channel. Then
\begin{equation}
\boldsymbol{\Upsilon}(\mathcal{N}_{AB\rightarrow A^{\prime}B^{\prime}%
})=\boldsymbol{\Upsilon}(\operatorname{id}_{\bar{A}\rightarrow\tilde{A}}%
\otimes\mathcal{N}_{AB\rightarrow A^{\prime}B^{\prime}}\otimes
\operatorname{id}_{\bar{B}\rightarrow\tilde{B}}).
\label{eq-CF:stability-gen-div-ups}%
\end{equation}

\end{proposition*}

\begin{Proof}
The definition of the generalized channel divergence in
Definition~\ref{def-gen_channel_div} implies that it is stable, in the sense that%
\begin{multline}
\boldsymbol{D}(\mathcal{N}_{AB\rightarrow A^{\prime}B^{\prime}}\Vert
\mathcal{M}_{AB\rightarrow A^{\prime}B^{\prime}})=\\
\boldsymbol{D}(\operatorname{id}_{\bar{A}\rightarrow\tilde{A}}\otimes
\mathcal{N}_{AB\rightarrow A^{\prime}B^{\prime}}\otimes\operatorname{id}%
_{\bar{B}\rightarrow\tilde{B}}\Vert\operatorname{id}_{\bar{A}\rightarrow
\tilde{A}}\otimes\mathcal{M}_{AB\rightarrow A^{\prime}B^{\prime}}%
\otimes\operatorname{id}_{\bar{B}\rightarrow\tilde{B}}),
\end{multline}
for every channel $\mathcal{N}_{AB\rightarrow A^{\prime}B^{\prime}}$ and
completely positive map $\mathcal{M}_{AB\rightarrow A^{\prime}B^{\prime}}$.
Combining with Proposition~\ref{prop-CF:stability}\ and the definition in
\eqref{eq-CF:gen-div-ch-ups-meas-bi-map}, we conclude \eqref{eq-CF:stability-gen-div-ups}.
\end{Proof}

\begin{proposition*}
{Zero on Classical Feedback Channels}{prop-CF:zero-feedback-renyi-ups}%
Let $\overline{\Delta}_{B\rightarrow A^{\prime}}$ be a classical feedback
channel:%
\begin{equation}
\overline{\Delta}_{B\rightarrow A^{\prime}}(\cdot)\coloneqq\sum_{i=0}%
^{d-1}|i\rangle_{A^{\prime}}\langle i|_{B}(\cdot)|i\rangle_{B}\langle
i|_{A^{\prime}},
\end{equation}
where system $A^{\prime}$ is isomorphic to $ B$ and $d=d_{A^{\prime}}=d_{B}$. Then%
\begin{equation}
\boldsymbol{\Upsilon}(\overline{\Delta}_{B\rightarrow A^{\prime}})=0.
\end{equation}

\end{proposition*}

\begin{Proof}
This follows from Proposition~\ref{prop-CF:zero-feedback}. Since
$\beta(\overline{\Delta}_{B\rightarrow A^{\prime}})=1$, we can pick
$\mathcal{M}_{B\rightarrow A^{\prime}}=\overline{\Delta}_{B\rightarrow
A^{\prime}}$, and then%
\begin{equation}
\boldsymbol{D}(\overline{\Delta}_{B\rightarrow A^{\prime}}\Vert\mathcal{M}%
_{B\rightarrow A^{\prime}})=\boldsymbol{D}(\overline{\Delta}_{B\rightarrow
A^{\prime}}\Vert\overline{\Delta}_{B\rightarrow A^{\prime}})=0.
\end{equation}
So this establishes that $\boldsymbol{\Upsilon}(\overline{\Delta}_{B\rightarrow
A^{\prime}})\leq0$, and the other inequality $\boldsymbol{\Upsilon}(\overline
{\Delta}_{B\rightarrow A^{\prime}})\geq0$ follows from
Proposition~\ref{prop-CF:non-neg-gen-div-ups}.
\end{Proof}

\begin{proposition*}
{Zero on Tensor Products of Local Channels}
{prop-CF:zero-local-chs-gen-div-ups}Let $\mathcal{E}_{A\rightarrow
A^{\prime}}$ and $\mathcal{F}_{B\rightarrow B^{\prime}}$ be quantum channels.
Then%
\begin{equation}
\boldsymbol{\Upsilon}(\mathcal{E}_{A\rightarrow A^{\prime}}\otimes\mathcal{F}%
_{B\rightarrow B^{\prime}})=0.
\end{equation}

\end{proposition*}

\begin{Proof}
Same argument as given for Proposition~\ref{prop-CF:zero-feedback-renyi-ups},
but use Proposition~\ref{prop-CF:zero-local-chs}\ instead.
\end{Proof}

\bigskip

We now establish some properties that are more specific to the
Belavkin--Staszewski and geometric R\'{e}nyi relative entropies (however the
first actually holds also for the quantum relative entropy and other quantum
R\'{e}nyi relative entropies).

\begin{proposition}
{prop-CF:ups-to-cbeta-bipartite}Let $\mathcal{N}_{AB\rightarrow
A^{\prime}B^{\prime}}$ be a bipartite channel. Then for all $\alpha\in(1,2]$,%
\begin{equation}
\widehat{\Upsilon}(\mathcal{N}_{AB\rightarrow A^{\prime}B^{\prime}}%
)\leq\widehat{\Upsilon}_{\alpha}(\mathcal{N}_{AB\rightarrow A^{\prime
}B^{\prime}})\leq C_{\beta}(\mathcal{N}_{AB\rightarrow A^{\prime}B^{\prime}}).
\end{equation}

\end{proposition}

\begin{Proof}
Pick $\mathcal{M}_{AB\rightarrow A^{\prime}B^{\prime}}=\frac{1}{\beta
(\mathcal{N}_{AB\rightarrow A^{\prime}B^{\prime}})}\mathcal{N}_{AB\rightarrow
A^{\prime}B^{\prime}}$ in the definition of $\widehat{\Upsilon}(\mathcal{N}%
_{AB\rightarrow A^{\prime}B^{\prime}})$ and $\widehat{\Upsilon}_{\alpha
}(\mathcal{N}_{AB\rightarrow A^{\prime}B^{\prime}})$ and use the fact that,
for $c>0$, $\widehat{D}(\rho\Vert c\sigma)=\widehat{D}(\rho\Vert\sigma
)-\log_{2}c$ and $\widehat{D}_{\alpha}(\rho\Vert c\sigma)=\widehat{D}_{\alpha
}(\rho\Vert\sigma)-\log_{2}c$ for all $\alpha\in(1,2]$. We also require
the monotonicity in $\alpha$ property from Proposition~\ref{prop:geometric-renyi-props}.
\end{Proof}

\begin{proposition*}
{Subadditivity}{prop-CF:subadd-geo-ups}For bipartite channels
$\mathcal{N}_{AB\rightarrow A^{\prime}B^{\prime}}^{1}$ and $\mathcal{N}%
_{A^{\prime}B^{\prime}\rightarrow A^{\prime\prime}B^{\prime\prime}}^{2}$, the
following inequality holds for all $\alpha\in(0,1)\cup(1,2]$:%
\begin{equation}
\widehat{\Upsilon}_{\alpha}(\mathcal{N}_{A^{\prime}B^{\prime}\rightarrow
A^{\prime\prime}B^{\prime\prime}}^{2}\circ\mathcal{N}_{AB\rightarrow
A^{\prime}B^{\prime}}^{1})\leq\widehat{\Upsilon}_{\alpha}(\mathcal{N}%
_{A^{\prime}B^{\prime}\rightarrow A^{\prime\prime}B^{\prime\prime}}%
^{2})+\widehat{\Upsilon}_{\alpha}(\mathcal{N}_{AB\rightarrow A^{\prime
}B^{\prime}}^{1}).
\end{equation}

\end{proposition*}

\begin{Proof}
This inequality is a direct consequence of the subadditivity inequality in
[REF - GEOMETRIC\ CH\ RENYI\ SUBADD], and the fact that if $\mathcal{M}^{1}$
and $\mathcal{M}^{2}$ are completely positive bipartite maps satisfying
$\beta(\mathcal{M}^{1}),\beta(\mathcal{M}^{2})\leq1$, then $\beta
(\mathcal{M}^{2}\circ\mathcal{M}^{1})\leq1$ (see
Proposition~\ref{prop-CF:subadditivity-bipartite-CP-maps}).
\end{Proof}

\subsubsection{Measure of Classical Communication for a Point-to-Point
Channel}

Let $\mathcal{M}_{A\rightarrow B^{\prime}}$ be a point-to-point completely
positive map, which is a special case of a completely positive bipartite map
with the Bob input $B$ trivial and the Alice output $A^{\prime}$ trivial.\ We
first show that $\beta$ in \eqref{eq-CF:basic-measure-NS-PPT-bi}\ reduces to
the measure from \eqref{eq-cc_beta_SDP}.

\begin{proposition}{def-CF:beta-reduce-p2p}
Let $\mathcal{M}_{A\rightarrow B^{\prime}}$ be a point-to-point completely
positive map. Then%
\begin{equation}
\beta(\mathcal{M}_{A\rightarrow B^{\prime}})\coloneqq \inf_{S_{B^{\prime}%
},V_{AB^{\prime}}\in\operatorname{Herm}}\left\{
\begin{array}
[c]{c}%
\operatorname{Tr}[S_{B^{\prime}}]:\\
T_{BB^{\prime}}(V_{AB^{\prime}}\pm\Gamma_{AB^{\prime}}^{\mathcal{M}})\geq0,\\
I_{A}\otimes S_{B^{\prime}}\pm V_{AB^{\prime}}\geq0
\end{array}
\right\}  . \label{eq-CF:beta-p2p-exp}%
\end{equation}

\end{proposition}

\begin{Proof}
In this case, the systems $A^{\prime}$ and $B$ are trivial. So then the
definition in \eqref{eq-CF:basic-measure-NS-PPT-bi} reduces to%
\begin{equation}
\beta(\mathcal{M}_{A\rightarrow B^{\prime}})=\inf_{S_{AB^{\prime}%
},V_{AB^{\prime}}\in\operatorname{Herm}}\left\{
\begin{array}
[c]{c}%
\left\Vert \operatorname{Tr}_{B^{\prime}}[S_{AB^{\prime}}]\right\Vert
_{\infty}:\\
T_{BB^{\prime}}(V_{AB^{\prime}}\pm\Gamma_{AB^{\prime}}^{\mathcal{M}})\geq0,\\
S_{AB^{\prime}}\pm V_{AB^{\prime}}\geq0,\\
S_{AB^{\prime}}=\pi_{A}\otimes\operatorname{Tr}_{A}[S_{AB^{\prime}}]
\end{array}
\right\}  .
\end{equation}
The last constraint implies that the optimization simplifies to%
\begin{align}
\beta(\mathcal{M}_{A\rightarrow B^{\prime}})  &  =\inf_{S_{AB^{\prime}%
},V_{AB^{\prime}}\in\operatorname{Herm}}\left\{
\begin{array}
[c]{c}%
\left\Vert \operatorname{Tr}_{B^{\prime}}[\pi_{A}\otimes\operatorname{Tr}%
_{A}[S_{AB^{\prime}}]]\right\Vert _{\infty}:\\
T_{BB^{\prime}}(V_{AB^{\prime}}\pm\Gamma_{AB^{\prime}}^{\mathcal{M}})\geq0,\\
\pi_{A}\otimes\operatorname{Tr}_{A}[S_{AB^{\prime}}]\pm V_{AB^{\prime}}\geq0
\end{array}
\right\} \\
&  =\inf_{S_{B^{\prime}}^{\prime},V_{AB^{\prime}}\in\operatorname{Herm}%
}\left\{
\begin{array}
[c]{c}%
\left\Vert \operatorname{Tr}_{B^{\prime}}[\pi_{A}\otimes S_{B^{\prime}%
}^{\prime}]\right\Vert _{\infty}:\\
T_{BB^{\prime}}(V_{AB^{\prime}}\pm\Gamma_{AB^{\prime}}^{\mathcal{M}})\geq0,\\
\pi_{A}\otimes S_{B^{\prime}}^{\prime}\pm V_{AB^{\prime}}\geq0
\end{array}
\right\} \\
&  =\inf_{S_{B^{\prime}}^{\prime},V_{AB^{\prime}}\in\operatorname{Herm}%
}\left\{
\begin{array}
[c]{c}%
\operatorname{Tr}[S_{B^{\prime}}^{\prime}]\left\Vert \pi_{A}\right\Vert
_{\infty}:\\
T_{BB^{\prime}}(V_{AB^{\prime}}\pm\Gamma_{AB^{\prime}}^{\mathcal{M}})\geq0,\\
\pi_{A}\otimes S_{B^{\prime}}^{\prime}\pm V_{AB^{\prime}}\geq0
\end{array}
\right\} \\
&  =\inf_{S_{B^{\prime}}^{\prime},V_{AB^{\prime}}\in\operatorname{Herm}%
}\left\{
\begin{array}
[c]{c}%
\frac{1}{d_{A}}\operatorname{Tr}[S_{B^{\prime}}^{\prime}]:\\
T_{BB^{\prime}}(V_{AB^{\prime}}\pm\Gamma_{AB^{\prime}}^{\mathcal{M}})\geq0,\\
\pi_{A}\otimes S_{B^{\prime}}^{\prime}\pm V_{AB^{\prime}}\geq0
\end{array}
\right\} \\
&  =\inf_{S_{B^{\prime}},V_{AB^{\prime}}\in\operatorname{Herm}}\left\{
\begin{array}
[c]{c}%
\operatorname{Tr}[S_{B^{\prime}}]:\\
T_{BB^{\prime}}(V_{AB^{\prime}}\pm\Gamma_{AB^{\prime}}^{\mathcal{M}})\geq0,\\
\pi_{A}\otimes S_{B^{\prime}}\pm V_{AB^{\prime}}\geq0
\end{array}
\right\}  .
\end{align}
This concludes the proof.
\end{Proof}

\bigskip

More generally, consider that the definition in
\eqref{eq-CF:gen-div-ch-ups-meas-bi-map}\ becomes as follows for a
point-to-point channel $\mathcal{N}_{A\rightarrow B^{\prime}}$:%
\begin{equation}
\boldsymbol{\Upsilon}(\mathcal{N}_{A\rightarrow B^{\prime}})\coloneqq\inf
_{\mathcal{M}_{A\rightarrow B^{\prime}}:\beta(\mathcal{M}_{A\rightarrow
B^{\prime}})\leq1}\boldsymbol{D}(\mathcal{N}_{A\rightarrow B^{\prime}}%
\Vert\mathcal{M}_{A\rightarrow B^{\prime}}), \label{eq-CF:p2p-gen-div-ups}%
\end{equation}
which leads to the quantities $\widehat{\Upsilon}(\mathcal{N}_{A\rightarrow
B^{\prime}})$ and $\widehat{\Upsilon}_{\alpha}(\mathcal{N}_{A\rightarrow
B^{\prime}})$, for which we have the following bounds for $\alpha\in(1,2]$:%
\begin{equation}
\widehat{\Upsilon}(\mathcal{N}_{A\rightarrow B^{\prime}})\leq\widehat
{\Upsilon}_{\alpha}(\mathcal{N}_{A\rightarrow B^{\prime}})\leq C_{\beta
}(\mathcal{N}_{A\rightarrow B^{\prime}}).
\label{eq-CF:ups-p2p-relations-bs-geo-c-beta}%
\end{equation}

The next proposition is critical for establishing our upper bound proofs in
Section~\ref{sec-CF:geo-ups-proof-details}. It states that if one share of a maximally
classically correlated state passes through a completely positive\ map
$\mathcal{M}_{A\rightarrow B^{\prime}}$\ for which $\beta(\mathcal{M}%
_{A\rightarrow B^{\prime}})\leq1$, then the resulting operator has a very
small chance of passing the comparator test, as defined in
\eqref{eq-CF:comparator-test-def}. (Recall that we previously used the
comparator test in \eqref{eq-eacc_comparator_test} and \eqref{eq-comparator_test}.)

\begin{proposition*}
{Bound for Comparator Test Success Probability}{prop-CF:comp-test}Let%
\begin{equation}
\overline{\Phi}_{\hat{A}A}\coloneqq\frac{1}{d}\sum_{i=0}^{d-1}|i\rangle
\!\langle i|_{\hat{A}}\otimes|i\rangle\!\langle i|_{A}%
\end{equation}
denote the maximally classically correlated state, and let $\mathcal{M}%
_{A\rightarrow B^{\prime}}$\ be a completely positive\ map $\mathcal{M}%
_{A\rightarrow B^{\prime}}$\ for which $\beta(\mathcal{M}_{A\rightarrow
B^{\prime}})\leq1$. Then%
\begin{equation}
\operatorname{Tr}[\Pi_{\hat{A}B^{\prime}}\mathcal{M}_{A\rightarrow B^{\prime}%
}(\overline{\Phi}_{\hat{A}A})]\leq\frac{1}{d},
\end{equation}
where $\Pi_{\hat{A}B^{\prime}}$ is the comparator test:%
\begin{equation}
\Pi_{\hat{A}B^{\prime}}\coloneqq\sum_{i=0}^{d-1}|i\rangle\!\langle i|_{\hat
{A}}\otimes|i\rangle\!\langle i|_{B^{\prime}},
\label{eq-CF:comparator-test-def}%
\end{equation}
and the following systems are isomorphic:$\ \hat{A}$, $A$, and $B^{\prime}$.
\end{proposition*}

\begin{Proof}
Recall the expression for $\beta(\mathcal{M}_{A\rightarrow B^{\prime}})$ in
\eqref{eq-CF:beta-p2p-exp}. Let $S_{B^{\prime}}$ and $V_{AB^{\prime}}$ be
arbitrary Hermitian operators satisfying the constraints for $\beta
(\mathcal{M}_{A\rightarrow B^{\prime}})$. An application of
\eqref{eq-QM:post-selected-TP-Choi-op}\ implies that%
\begin{equation}
\mathcal{M}_{A\rightarrow B^{\prime}}(\overline{\Phi}_{\hat{A}A}%
)=\langle\Gamma|_{A\tilde{A}}\overline{\Phi}_{\hat{A}A}\otimes\Gamma
_{\tilde{A}B^{\prime}}^{\mathcal{M}}|\Gamma\rangle_{A\tilde{A}},
\end{equation}
where $\tilde{A}\simeq A$. This means that%
\begin{align}
\operatorname{Tr}[\Pi_{\hat{A}B^{\prime}}\mathcal{M}_{A\rightarrow B^{\prime}%
}(\overline{\Phi}_{\hat{A}A})]  &  =\operatorname{Tr}[\Pi_{\hat{A}B^{\prime}%
}\langle\Gamma|_{A\tilde{A}}\overline{\Phi}_{\hat{A}A}\otimes\Gamma_{\tilde
{A}B^{\prime}}^{\mathcal{M}}|\Gamma\rangle_{A\tilde{A}}]\\
&  =\operatorname{Tr}[T_{B^{\prime}}(\Pi_{\hat{A}B^{\prime}})\langle
\Gamma|_{A\tilde{A}}\overline{\Phi}_{\hat{A}A}\otimes\Gamma_{\tilde
{A}B^{\prime}}^{\mathcal{M}}|\Gamma\rangle_{A\tilde{A}}]\\
&  =\operatorname{Tr}[\Pi_{\hat{A}B^{\prime}}\langle\Gamma|_{A\tilde{A}%
}\overline{\Phi}_{\hat{A}A}\otimes T_{B^{\prime}}(\Gamma_{\tilde{A}B^{\prime}%
}^{\mathcal{M}})|\Gamma\rangle_{A\tilde{A}}]\\
&  \leq\operatorname{Tr}[\Pi_{\hat{A}B^{\prime}}\langle\Gamma|_{A\tilde{A}%
}\overline{\Phi}_{\hat{A}A}\otimes T_{B^{\prime}}(V_{\tilde{A}B^{\prime}%
})|\Gamma\rangle_{A\tilde{A}}]\\
&  =\operatorname{Tr}[T_{B^{\prime}}(\Pi_{\hat{A}B^{\prime}})\langle
\Gamma|_{A\tilde{A}}\overline{\Phi}_{\hat{A}A}\otimes V_{\tilde{A}B^{\prime}%
}|\Gamma\rangle_{A\tilde{A}}]\\
&  =\operatorname{Tr}[\Pi_{\hat{A}B^{\prime}}\langle\Gamma|_{A\tilde{A}%
}\overline{\Phi}_{\hat{A}A}\otimes V_{\tilde{A}B^{\prime}}|\Gamma
\rangle_{A\tilde{A}}]\\
&  \leq\operatorname{Tr}[\Pi_{\hat{A}B^{\prime}}\langle\Gamma|_{A\tilde{A}%
}\overline{\Phi}_{\hat{A}A}\otimes I_{\tilde{A}}\otimes S_{B^{\prime}}%
|\Gamma\rangle_{A\tilde{A}}]\\
&  =\operatorname{Tr}[\Pi_{\hat{A}B^{\prime}}\langle\Gamma|_{A\tilde{A}%
}\overline{\Phi}_{\hat{A}A}\otimes I_{\tilde{A}}|\Gamma\rangle_{A\tilde{A}%
}\otimes S_{B^{\prime}}]\\
&  =\operatorname{Tr}[\Pi_{\hat{A}B^{\prime}}\operatorname{Tr}_{A}%
[\overline{\Phi}_{\hat{A}A}]\otimes S_{B^{\prime}}]\\
&  =\frac{1}{d}\operatorname{Tr}[\Pi_{\hat{A}B^{\prime}}I_{\hat{A}}\otimes
S_{B^{\prime}}]\\
&  =\frac{1}{d}\operatorname{Tr}[S_{B^{\prime}}].
\end{align}
Since this holds for all $S_{B^{\prime}}$ and $V_{AB^{\prime}}$ satisfying the
constraints for $\beta(\mathcal{M}_{A\rightarrow B^{\prime}})$, we conclude
that%
\begin{equation}
\operatorname{Tr}[\Pi_{\hat{A}B^{\prime}}\mathcal{M}_{A\rightarrow B^{\prime}%
}(\overline{\Phi}_{\hat{A}A})]\leq\frac{1}{d}.
\end{equation}
This concludes the proof.
\end{Proof}

\bigskip

We finally state another proposition that plays an essential role in our upper
bound proofs in Section~\ref{sec-CF:geo-ups-proof-details}.

\begin{proposition}
{prop-CF:err-bound}Suppose that $\mathcal{N}_{A\rightarrow B}$ is a
channel with $A$ isomorphic to $B$ that satisfies%
\begin{equation}
\frac{1}{2}\left\Vert \mathcal{N}_{A\rightarrow B}(\overline{\Phi}%
_{RA})-\overline{\Phi}_{RB}\right\Vert _{1}\leq\varepsilon,
\end{equation}
for $\varepsilon\in\lbrack0,1)$ and where $\overline{\Phi}_{RB}\coloneqq\frac
{1}{d}\sum_{i}|i\rangle\!\langle i|_{R}\otimes|i\rangle\!\langle i|_{B}$ and
$d=d_{R}=d_{A}=d_{B}$. Then%
\begin{equation}
\log_{2}d\leq\inf_{\mathcal{M}_{A\rightarrow B}:\beta(\mathcal{M}%
_{A\rightarrow B})\leq1}D_{H}^{\varepsilon}(\mathcal{N}_{A\rightarrow
B}(\overline{\Phi}_{RA})\Vert\mathcal{M}_{A\rightarrow B}(\overline{\Phi}%
_{RA})), \label{eq-CF:err-bound-HTRE}%
\end{equation}
and for all $\alpha\in(1,2]$,%
\begin{multline}
\log_{2}d\leq\inf_{\mathcal{M}_{A\rightarrow B}:\beta(\mathcal{M}%
_{A\rightarrow B})\leq1}\widehat{D}_{\alpha}(\mathcal{N}_{A\rightarrow
B}(\overline{\Phi}_{RA})\Vert\mathcal{M}_{A\rightarrow B}(\overline{\Phi}%
_{RA}))\label{eq-CF:err-bound}\\
+\frac{\alpha}{\alpha-1}\log_{2}\!\left(  \frac{1}{1-\varepsilon}\right)  .
\end{multline}

\end{proposition}

\begin{Proof}
We begin by proving \eqref{eq-CF:err-bound-HTRE}. The condition%
\begin{equation}
\frac{1}{2}\left\Vert \mathcal{N}_{A\rightarrow B}(\overline{\Phi}%
_{RA})-\overline{\Phi}_{RB}\right\Vert _{1}\leq\varepsilon
\end{equation}
implies that%
\begin{equation}
\operatorname{Tr}[\Pi_{RB}\mathcal{N}_{A\rightarrow B}(\overline{\Phi}%
_{RA})]\geq1-\varepsilon,
\end{equation}
where $\Pi_{RB}\coloneqq\sum_{i}|i\rangle\!\langle i|_{R}\otimes
|i\rangle\!\langle i|_{B}$ is the comparator test. Indeed, applying a
completely dephasing channel $\overline{\Delta}_{B}(\cdot)\coloneqq\sum
_{i}|i\rangle\!\langle i|(\cdot)|i\rangle\!\langle i|$\ to the output of the
channel $\mathcal{N}_{A\rightarrow B}$ and applying the data-processing
inequality for trace distance, we conclude that%
\begin{align}
\varepsilon &  \geq\frac{1}{2}\left\Vert \mathcal{N}_{A\rightarrow
B}(\overline{\Phi}_{RA})-\overline{\Phi}_{RB}\right\Vert _{1}\\
&  \geq\frac{1}{2}\left\Vert (\overline{\Delta}_{B}\circ\mathcal{N}%
_{A\rightarrow B})(\overline{\Phi}_{RA})-\overline{\Delta}_{B}(\overline{\Phi
}_{RB})\right\Vert _{1}\\
&  =\frac{1}{2}\left\Vert (\overline{\Delta}_{B}\circ\mathcal{N}_{A\rightarrow
B})(\overline{\Phi}_{RA})-\overline{\Phi}_{RB}\right\Vert _{1}.
\end{align}
Let $\omega_{RB}\coloneqq(\overline{\Delta}_{B}\circ\mathcal{N}_{A\rightarrow
B})(\overline{\Phi}_{RA})$ and observe that it can be written as%
\begin{equation}
\omega_{RB}=\frac{1}{d}\sum_{i,j}p(j|i)|i\rangle\!\langle i|_{R}%
\otimes|j\rangle\!\langle j|_{B}%
\end{equation}
for some conditional probability distribution $p(j|i)$. Then%
\begin{align}
&  \frac{1}{2}\left\Vert (\overline{\Delta}_{B}\circ\mathcal{N}_{A\rightarrow
B})(\overline{\Phi}_{RA})-\overline{\Phi}_{RB}\right\Vert _{1}\nonumber\\
&  =\frac{1}{2}\left\Vert \frac{1}{d}\sum_{i,j}p(j|i)|i\rangle\!\langle
i|_{R}\otimes|j\rangle\!\langle j|_{B}-\frac{1}{d}\sum_{i,j}\delta
_{i,j}|i\rangle\!\langle i|_{R}\otimes|j\rangle\!\langle j|_{B}\right\Vert
_{1}\\
&  =\frac{1}{2}\frac{1}{d}\sum_{i}\left\Vert \sum_{j}(p(j|i)-\delta
_{i,j})|j\rangle\!\langle j|_{B}\right\Vert _{1}\\
&  =\frac{1}{2}\frac{1}{d}\sum_{i}\left[  \left(  1-p(i|i)\right)
+\sum_{j\neq i}p(j|i)\right] \\
&  =\frac{1}{d}\sum_{i}\left(  1-p(i|i)\right) \\
&  =1-\sum_{i}\frac{1}{d}p(i|i).
\end{align}
This implies that%
\begin{equation}
\sum_{i}\frac{1}{d}p(i|i)\geq1-\varepsilon.
\end{equation}
Now consider that%
\begin{align}
\operatorname{Tr}[\Pi_{RB}\mathcal{N}_{A\rightarrow B}(\overline{\Phi}_{RA})]
&  =\operatorname{Tr}[\overline{\Delta}_{B}(\Pi_{RB})\mathcal{N}_{A\rightarrow
B}(\overline{\Phi}_{RA})]\\
&  =\operatorname{Tr}[\Pi_{RB}(\overline{\Delta}_{B}\circ\mathcal{N}%
_{A\rightarrow B})(\overline{\Phi}_{RA})]\\
&  =\operatorname{Tr}[\Pi_{RB}\omega_{RB}]\\
&  =\sum_{i}\frac{1}{d}p(i|i).
\end{align}
So we conclude that%
\begin{equation}
\operatorname{Tr}[\Pi_{RB}\mathcal{N}_{A\rightarrow B}(\overline{\Phi}%
_{RA})]\geq1-\varepsilon.
\end{equation}
Applying the definition of the hypothesis testing relative entropy from
Definition~\ref{def-hypo_testing_rel_ent}, we conclude that%
\begin{align}
&  \inf_{\mathcal{M}_{A\rightarrow B}:\beta(\mathcal{M}_{A\rightarrow B}%
)\leq1}D_{H}^{\varepsilon}(\mathcal{N}_{A\rightarrow B}(\overline{\Phi}%
_{RA})\Vert\mathcal{M}_{A\rightarrow B}(\overline{\Phi}_{RA}))\nonumber\\
&  =\inf_{\substack{\mathcal{M}_{A\rightarrow B}:\\\beta(\mathcal{M}%
_{A\rightarrow B})\leq1}}\left[  -\log_{2}\inf_{\Lambda_{RB}\geq0}\left\{
\begin{array}
[c]{c}%
\operatorname{Tr}[\Lambda_{RB}\mathcal{M}_{A\rightarrow B}(\overline{\Phi
}_{RA})]:\\
\operatorname{Tr}[\Lambda_{RB}\mathcal{N}_{A\rightarrow B}(\overline{\Phi
}_{RA})]\geq1-\varepsilon,\ \Lambda_{RB}\leq I_{RB}%
\end{array}
\right\}  \right] \\
&  =-\log_{2}\sup_{\substack{\mathcal{M}_{A\rightarrow B}:\\\beta
(\mathcal{M}_{A\rightarrow B})\leq1}}\inf_{\Lambda_{RB}\geq0}\left\{
\begin{array}
[c]{c}%
\operatorname{Tr}[\Lambda_{RB}\mathcal{M}_{A\rightarrow B}(\overline{\Phi
}_{RA})]:\\
\operatorname{Tr}[\Lambda_{RB}\mathcal{N}_{A\rightarrow B}(\overline{\Phi
}_{RA})]\geq1-\varepsilon,\ \Lambda_{RB}\leq I_{RB}%
\end{array}
\right\}  .
\end{align}
Now consider that%
\begin{align}
&  \sup_{\substack{\mathcal{M}_{A\rightarrow B}:\\\beta(\mathcal{M}%
_{A\rightarrow B})\leq1}}\inf_{\Lambda_{RB}\geq0}\left\{
\begin{array}
[c]{c}%
\operatorname{Tr}[\Lambda_{RB}\mathcal{M}_{A\rightarrow B}(\overline{\Phi
}_{RA})]:\\
\operatorname{Tr}[\Lambda_{RB}\mathcal{N}_{A\rightarrow B}(\overline{\Phi
}_{RA})]\geq1-\varepsilon,\ \Lambda_{RB}\leq I_{RB}%
\end{array}
\right\} \nonumber\\
&  \leq\sup_{\mathcal{M}_{A\rightarrow B}:\beta(\mathcal{M}_{A\rightarrow
B})\leq1}\operatorname{Tr}[\Pi_{RB}\mathcal{M}_{A\rightarrow B}(\overline
{\Phi}_{RA})]\\
&  \leq\frac{1}{d},
\end{align}
where the last inequality follows from Proposition~\ref{prop-CF:comp-test}.
Then applying a negative logarithm gives \eqref{eq-CF:err-bound-HTRE}.

The inequality in \eqref{eq-CF:err-bound}\ follows as direct application of
the following relationship between hypothesis testing relative entropy and the
geometric R\'{e}nyi relative entropy:%
\begin{equation}
D_{H}^{\varepsilon}(\rho\Vert\sigma)\leq\widehat{D}_{\alpha}(\rho\Vert
\sigma)+\frac{\alpha}{\alpha-1}\log_{2}\!\left(  \frac{1}{1-\varepsilon
}\right)  , \label{eq-CF:HTRE-to-geo-Reny}%
\end{equation}
as well as the previous proposition. The proof of
\eqref{eq-CF:HTRE-to-geo-Reny} follows the same proof given for Proposition~\ref{prop:sandwich-to-htre}.
\end{Proof}

\subsubsection{Proof of Geometric $\Upsilon$-Information Upper Bound}

\label{sec-CF:geo-ups-proof-details}

We now have everything that we need to establish that the geometric $\Upsilon
$-information is an upper bound on the number of bits that can be transmitted
by means of a quantum channel assisted by a classical feedback channel. By
examining the protocol in Section~\ref{sec-CF:FB-assisted-protocol}, consider
that the final state $\omega_{M\hat{M}}^{p}$ of the protocol can be written as
follows:%
\begin{equation}
\omega_{M\hat{M}}^{p}=\mathcal{P}_{M^{\prime}\rightarrow\hat{M}}%
(\overline{\Phi}_{MM^{\prime}}^{p}), \label{eq-CF:omega-final-state-geo-ups}%
\end{equation}
where%
\begin{multline}
\mathcal{P}_{M^{\prime}\rightarrow\hat{M}}\coloneqq \mathcal{D}^{n}\circ
\mathcal{N}\circ\mathcal{E}^{n-1}\circ\overline{\Delta}\circ\mathcal{D}%
^{n-1}\circ\mathcal{N}\circ\mathcal{E}^{n-2}\circ\overline{\Delta}%
\circ\mathcal{D}^{n-2}\circ\label{eq-CF:whole-protocol-ups-bound}\\
\cdots\circ\mathcal{D}^{2}\circ\mathcal{N}\circ\mathcal{E}^{1}\circ
\overline{\Delta}\circ\mathcal{D}^{1}\circ\mathcal{N}\circ\mathcal{E}^{0}%
\circ\mathcal{A},
\end{multline}
and $\mathcal{A}$ is an appending channel that appends the state
$\overline{\Delta}_{F_{0}}(\Psi_{F_{0}B_{0}^{\prime}})$ to the input state
$\overline{\Phi}_{MM^{\prime}}^{p}$. In
\eqref{eq-CF:whole-protocol-ups-bound}, we have omitted all system labels for simplicity.

We now state the main result of this section:

\begin{theorem}{thm-CF:geo-ups-alpha-bound-CF-cap}
Fix $n\in\mathbb{N}$, $\varepsilon
\in\lbrack0,1)$, and $\alpha\in(1,2]$, and let $\mathcal{N}_{A\rightarrow B}$
be a quantum channel. For all $(n,\left\vert \mathcal{M}\right\vert
,\varepsilon)$ classical-feedback-assisted classical communication protocols
over the channel $\mathcal{N}_{A\rightarrow B}$, the following bound holds%
\begin{equation}
\frac{\log_{2}\left\vert \mathcal{M}\right\vert }{n}\leq\widehat{\Upsilon
}_{\alpha}(\mathcal{N}_{A\rightarrow B})+\frac{\alpha}{n\left(  \alpha
-1\right)  }\log_{2}\!\left(  \frac{1}{1-\varepsilon}\right)  ,
\label{eq-CF:geo-ups-info-up-bound}%
\end{equation}
where $\widehat{\Upsilon}_{\alpha}(\mathcal{N}_{A\rightarrow B})$ is the
geometric $\Upsilon$-information of $\mathcal{N}_{A\rightarrow B}$, as defined
in \eqref{eq-CF:p2p-gen-div-ups}, with $\boldsymbol{D}$ set to $\widehat
{D}_{\alpha}$.
\end{theorem}

\begin{Proof}
Consider an arbitrary $(n,\left\vert \mathcal{M}\right\vert ,\varepsilon)$
protocol of the form described in Section~\ref{sec-CF:FB-assisted-protocol},
with final state as given in \eqref{eq-CF:omega-final-state-geo-ups}. Let the
distribution $p$ over the messages be the uniform distribution. Since the
condition $p_{\text{err}}^{\ast}(\mathcal{C})\leq\varepsilon$ holds, with
$p_{\text{err}}^{\ast}$ defined in \eqref{eq-CF:maximal-error}, we can apply
\eqref{eq-CF:err-bound} of\ Proposition~\ref{prop-CF:err-bound} to conclude
that%
\begin{align}
\log_{2}\left\vert \mathcal{M}\right\vert  &  \leq\inf_{\mathcal{M}%
_{M^{\prime}\rightarrow\hat{M}}:\beta(\mathcal{M}_{M^{\prime}\rightarrow
\hat{M}})\leq1}\widehat{D}_{\alpha}(\mathcal{P}_{M^{\prime}\rightarrow\hat{M}%
}(\overline{\Phi}_{MM^{\prime}})\Vert\mathcal{M}_{M^{\prime}\rightarrow\hat
{M}}(\overline{\Phi}_{MM^{\prime}}))\nonumber\\
&  \qquad+\frac{\alpha}{\alpha-1}\log_{2}\!\left(  \frac{1}{1-\varepsilon
}\right) \\
&  \leq\widehat{\Upsilon}_{\alpha}(\mathcal{P}_{M^{\prime}\rightarrow\hat{M}%
})+\frac{\alpha}{\alpha-1}\log_{2}\!\left(  \frac{1}{1-\varepsilon}\right)  ,
\end{align}
where the second inequality follows from the definition in with
$\boldsymbol{D}$ set to $\widehat{D}_{\alpha}$.
Eq.~\eqref{eq-CF:whole-protocol-ups-bound} indicates that the whole protocol
is a serial composition of bipartite channels. Then we find that%
\begin{align}
&  \widehat{\Upsilon}_{\alpha}(\mathcal{P}_{M^{\prime}\rightarrow\hat{M}%
})\nonumber\\
&  =\widehat{\Upsilon}_{\alpha}(\mathcal{D}^{n}\circ\mathcal{N}\circ
\mathcal{E}^{n-1}\circ\overline{\Delta}\circ\mathcal{D}^{n-1}\circ
\mathcal{N}\circ\mathcal{E}^{n-2}\circ\overline{\Delta}\circ\mathcal{D}%
^{n-2}\circ\\
&  \qquad\qquad\cdots\circ\mathcal{D}^{2}\circ\mathcal{N}\circ\mathcal{E}%
^{1}\circ\overline{\Delta}\circ\mathcal{D}^{1}\circ\mathcal{N}\circ
\mathcal{E}^{0}\circ\mathcal{A})\\
&  \leq n\widehat{\Upsilon}_{\alpha}(\mathcal{N})+n\widehat{\Upsilon}_{\alpha
}(\overline{\Delta})+\sum_{i=1}^{n}\widehat{\Upsilon}_{\alpha}(\mathcal{D}%
^{i})+\sum_{i=0}^{n-1}\widehat{\Upsilon}_{\alpha}(\mathcal{E}^{i}%
)+\widehat{\Upsilon}_{\alpha}(\mathcal{A})\\
&  =n\widehat{\Upsilon}_{\alpha}(\mathcal{N}).
\end{align}
The inequality follows from Proposition~\ref{prop-CF:subadd-geo-ups}. The last
equality follows from Propositions~\ref{prop-CF:stability-gen-div-ups},
\ref{prop-CF:zero-local-chs-gen-div-ups}, and
\ref{prop-CF:zero-feedback-renyi-ups} because each encoding channel
$\mathcal{E}^{i}$ and decoding channel $\mathcal{D}^{i}$\ is a local channel
and $\overline{\Delta}$ is a classical feedback channel. We also implicitly
used the stability property in Proposition~\ref{prop-CF:stability-gen-div-ups}%
. Putting everything together, we conclude that%
\begin{equation}
\log_{2}\left\vert \mathcal{M}\right\vert \leq n\widehat{\Upsilon}_{\alpha
}(\mathcal{N})+\frac{\alpha}{\alpha-1}\log_{2}\!\left(  \frac{1}%
{1-\varepsilon}\right)  ,
\end{equation}
which is equivalent to the desired bound in \eqref{eq-CF:geo-ups-info-up-bound}.
\end{Proof}

\section{Classical Capacity of a Quantum Channel Assisted by Classical
Feedback}

\label{sec-CF:capacity-defs-and-thms}In this section, we analyze the
asymptotic case of feedback-assisted communication, in which we allow for an
arbitrary large number of rounds of feedback. The definitions in this section
are similar to those in previous chapters, and so we keep this section brief.

\begin{definition}
{Achievable Rate for Classical-Feedback-Assisted Classical Communication}
{def-cfacc-ach_rate} Given a quantum channel $\mathcal{N}$, a rate
$R\in\mathbb{R}^{+}$ is called an achievable rate for
classical-feedback-assisted classical communication over $\mathcal{N}$ if for
all $\varepsilon\in(0,1]$, all $\delta>0$, and all sufficiently large $n$,
there exists an $(n,2^{n(R-\delta)},\varepsilon)$ classical-feedback-assisted
classical communication protocol.
\end{definition}

\begin{definition}
{Classical-Feedback-Assisted Classical Capacity of a Quantum Channel}
{def-cfacc-cap} The classical-feedback-assisted classical capacity of a
quantum channel $\mathcal{N}$, denoted by $C_{\operatorname{CFB}}%
(\mathcal{N})$, is defined as the supremum of all achievable rates, i.e.,
\begin{equation}
C_{\operatorname{CFB}}(\mathcal{N})\coloneqq\sup\{R:R\text{ is an achievable
rate for }\mathcal{N}\}.
\end{equation}

\end{definition}

\begin{definition}
{Strong Converse Rate for Classical-Feedback-Assisted Classical Communication}
{def-cfacc-strong_conv_rate} Given a quantum channel $\mathcal{N}$, a
rate $R\in\mathbb{R}^{+}$ is called a strong converse rate for
classical-feedback-assisted classical communication over $\mathcal{N}$ if for
all $\varepsilon\in\lbrack0,1)$, all $\delta>0$, and all sufficiently large
$n$, there does not exist an $(n,2^{n(R+\delta)},\varepsilon)$
classical-feedback-assisted classical communication protocol.
\end{definition}

\begin{definition}
{Strong Converse Classical-Feedback-Assisted Classical Capacity of a Quantum
Channel}{def-cfacc-strong_conv_cap} The strong converse
classical-feedback-assisted classical capacity of a quantum channel
$\mathcal{N}$, denoted by $\widetilde{C}_{\operatorname{CFB}}(\mathcal{N})$,
is defined as the infimum of all strong converse rates, i.e.,
\begin{equation}
\widetilde{C}_{\operatorname{CFB}}(\mathcal{N})\coloneqq\inf\{R:R\text{ is a
strong converse rate for }\mathcal{N}\}.
\end{equation}

\end{definition}

We conclude several theorems, based on the bounds given in Section~\ref{}:

\begin{theorem*}
{Classical-Feedback-Assisted Classical Capacity of Entanglement-Breaking
Channels}{thm-cfacc-cap-ent-break} For an entanglement-breaking channel
$\mathcal{N}$, its classical-feedback-assisted classical capacity
$C_{\operatorname{CFB}}(\mathcal{N})$ and its strong converse
quantum-feedback-assisted classical capacity are both equal to its Holevo
information $\chi(\mathcal{N})$, i.e.,
\begin{equation}
C_{\operatorname{CFB}}(\mathcal{N})=\widetilde{C}_{\operatorname{CFB}%
}(\mathcal{N})=\chi(\mathcal{N}),
\end{equation}
where $\chi(\mathcal{N})$ is defined in \eqref{eq-Hol_inf_chan}.
\end{theorem*}

\begin{Proof}
The lower bound $\chi(\mathcal{N})\leq C_{\operatorname{CFB}}(\mathcal{N})$
follows from Theorem~\ref{thm-classical_capacity} (i.e., not making use of the classical feedback
channel at all). The upper bound $\widetilde{C}_{\operatorname{CFB}%
}(\mathcal{N})\leq\chi(\mathcal{N})$ follows from
\eqref{eq-CF:EB-bnd-renyi-n-shot}\ of\ Theorem~\ref{thm-CF:ent-break-upp-bnd},
and by reasoning similar to that given in the proof of Theorem~\ref{thm-str_conv_additive}.
\end{Proof}

\begin{theorem*}
{Average Entropy Weak Converse Bound for Classical Capacity Assisted by
Classical Feedback}{thm-CF:prob-mix-entropy-bnd}Let $\mathcal{N}%
_{A\rightarrow B}$ be a quantum channel that can be written as the following
probabilistic mixture of quantum channels:%
\begin{equation}
\mathcal{N}_{A\rightarrow B}=\sum_{x}p_{X}(x)\mathcal{N}_{A\rightarrow B}^{x},
\end{equation}
where $p_{X}$ is a probability distribution and $\{\mathcal{N}_{A\rightarrow
B}^{x}\}_{x}$ is a set of quantum channels. The following upper bound holds
for the classical capacity of a quantum channel assisted by classical
feedback:%
\begin{equation}
C_{\operatorname{CFB}}(\mathcal{N}_{A\rightarrow B})\leq\sup_{\rho_{A}}%
\sum_{x}p_{X}(x)H(\mathcal{N}_{A\rightarrow B}^{x}(\rho_{A})).
\end{equation}

\end{theorem*}

\begin{Proof}
This is a direct consequence of Theorem~\ref{cor-CF:avg-ent-upper-bnd}\ and
reasoning similar to that given for the proof around \eqref{eq-CC:weak-converse-pf-last-step}.
\end{Proof}

\begin{theorem*}
{Geometric $\Upsilon$-Information Strong Converse Bound for Classical Capacity
Assisted by Classical Feedback}{thm-CF:BS-ent-bnd-cfacc}The following
upper bound holds for the strong converse classical capacity of a quantum
channel $\mathcal{N}_{A\rightarrow B}$ assisted by classical feedback:%
\begin{equation}
\widetilde{C}_{\operatorname{CFB}}(\mathcal{N}_{A\rightarrow B})\leq
\widehat{\Upsilon}(\mathcal{N}_{A\rightarrow B}),
\end{equation}
where $\widehat{\Upsilon}(\mathcal{N}_{A\rightarrow B})$ is the $\Upsilon
$-information defined from the Belavkin--Staszewski relative entropy (see
\eqref{eq-CF:p2p-gen-div-ups} and Definition~\ref{def:belavkin-sta-rel-ent}).
\end{theorem*}

\begin{Proof}
This is a direct consequence of the upper bound in
Theorem~\ref{thm-CF:geo-ups-alpha-bound-CF-cap}\ and reasoning similar to that
given in the proof of Theorem~. We also require the fact that the geometric
R\'enyi relative entropy converges to the Belavkin--Staszewski relative entropy
in the limit as $\alpha\rightarrow1$ (see Proposition~\ref{prop:BS-rel-ent-to-geometric}).
\end{Proof}

\section{Examples}

In this section, we briefly provide some examples of channels for which we
evaluate the capacity upper bounds in
Section~\ref{sec-CF:capacity-defs-and-thms}. We begin with the quantum erasure
channel (see Section~\ref{sec:QM-over:qudit-erasure}). Recall that a quantum erasure channel acts as
follows on an input density operator $\rho$:%
\begin{equation}
\mathcal{E}_{p}(\rho)\coloneqq \left(  1-p\right)  \rho+p|e\rangle\!\langle e|,
\end{equation}
where $p\in\left[  0,1\right]  $ is the erasure probability and $|e\rangle\!
\langle e|$ is an erasure state orthogonal to every possible input. Let $d$ be
the dimension of the input to the channel. By inspection, we see that the
erasure channel is a probabilistic mixture of an identity channel and a
channel that traces out the input and replaces with the erasure state. Thus,
we apply Theorem~\ref{thm-CF:prob-mix-entropy-bnd}\ to conclude that%
\begin{align}
C_{\operatorname{CFB}}(\mathcal{E}_{p})  &  \leq\sup_{\rho_{A}}\left[  \left(
1-p\right)  H(\operatorname{id}(\rho_{A}))+pH(|e\rangle\!\langle e|)\right] \\
&  =\left(  1-p\right)  \sup_{\rho_{A}}H(\operatorname{id}(\rho_{A}))\\
&  =\left(  1-p\right)  \log_{2}d.
\end{align}
Since this upper bound is an achievable for classical communication over the
erasure channel without feedback (see Theorem~\ref{thm-cc_erasure_chan_cap}), we then conclude that%
\begin{equation}
C(\mathcal{E}_{p})=C_{\operatorname{CFB}}(\mathcal{E}_{p})=\left(  1-p\right)
\log_{2}d.
\end{equation}
That is, classical feedback does not increase the classical capacity of the
erasure channel.

\begin{figure}[ptb]
\begin{center}
\includegraphics[
width=4in
]{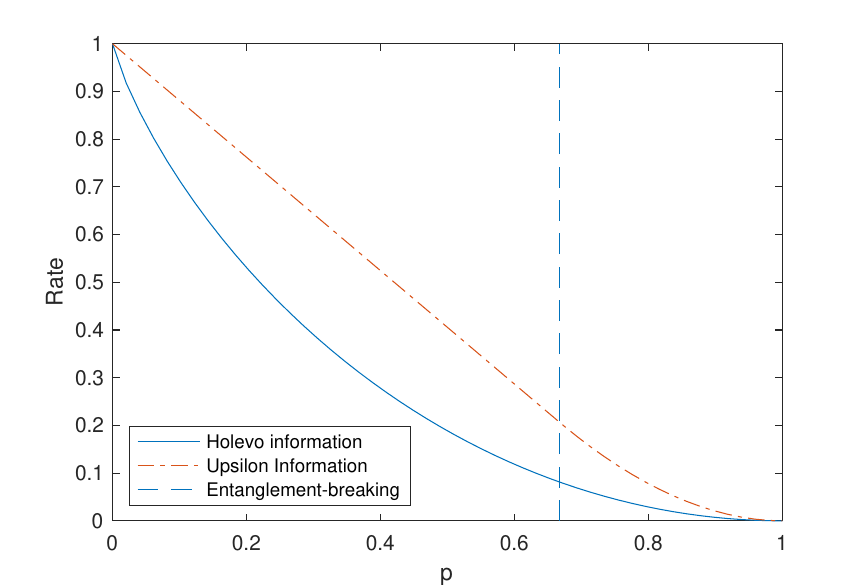}
\end{center}
\caption{Lower and upper bounds on the classical-feedback-assisted classical
capacity of the qubit depolarizing channel in \eqref{eq-CF:dep-ch-example},
with $d=2$. The dashed vertical line indicates that the qubit depolarizing
channel is entanglement breaking for $p\geq2/3$, so that the Holevo
information is equal to the feedback-assisted capacity for these values,
according to Theorem~\ref{thm-cfacc-cap-ent-break}.}%
\label{fig-CF:depola-plot}%
\end{figure}

Finally, we evaluate the bound in Theorem~\ref{thm-CF:BS-ent-bnd-cfacc} for
the qubit depolarizing channel. Recall from Section~\ref{} that it is defined
as
\begin{align}
\mathcal{D}^{p}(X)  &  \coloneqq \left(  1-p\right)  X+p\operatorname{Tr}%
[X]\pi,\label{eq-CF:dep-ch-example}\\
\pi &  \coloneqq I/d.
\end{align}
It was already established in Section~\ref{} that $\Upsilon(\mathcal{D}^{p})$
is an upper bound on its (unassisted) classical capacity, and we discussed in
Section~\ref{} how the Holevo information is equal to its classical capacity.
What we find now is that $\Upsilon(\mathcal{D}^{p})$ is an upper bound on its
classical capacity assisted by a classical feedback channel.
Figure~\ref{fig-CF:depola-plot}\ plots this upper bound and also plots the
Holevo information lower bound when $d=2$. The latter is given by
$1-h_{2}(p/2)$, where $h_{2}$ is the binary entropy function. Note that the
depolarizing channel is entanglement breaking for $p\geq\frac{d}{d+1}$. As
such, the bounds from Theorem~\ref{thm-cfacc-cap-ent-break} apply, so that,
for $p\geq\frac{d}{d+1}$, the Holevo information $1-h_{2}(p/2)$ is equal to
the classical capacity assisted by classical feedback. 

\section{Summary}

In this chapter, we developed the general theory of classical communication
over a quantum channel assisted by classical feedback from receiver to sender.
Our main focus was on establishing upper bounds on this capacity. The main
findings of this chapter are as follows:

\begin{enumerate}
\item We first proved that classical feedback does not enhance the classical
capacity of an entanglement-breaking channel.

\item Next, we established that the average output entropy of a channel is a
weak converse upper bound on the feedback-assisted capacity. The method for
establishing this average entropy bound involves identifying an information
measure that has two key properties: 1) it does not increase under a one-way
local operations and classical communication channel from the receiver to the
sender and 2) a quantum channel from sender to receiver cannot increase the
information measure by more than the maximum average output entropy of the
channel. This information measure can be understood as the sum of two terms,
with one corresponding to classical correlation and the other to entanglement.

\item We finally established a general strong converse upper bound on the
feedback-assisted capacity, in terms of the geometric $\Upsilon$-information
of a quantum channel. The main method for doing was to devise an information
measure for bipartite channels that is equal to zero for classical feedback
channels and products of local channels.
\end{enumerate}

\section{Bibliographic Notes}

\label{sec-CF:bib-notes}The classical capacity of a quantum channel assisted
by a classical feedback channel was first studied by \citet{BN05}, who proved
that classical feedback does not increase the capacity of
entanglement-breaking channels. This result was strengthened to a strong
converse statement by \citet{DW15}. \citet{SS09} provided an example of a
channel for which classical feedback can signficantly enhance the classical
capacity. \citet{BDSS06} related the feedback-assisted capacity to other
capacities in quantum Shannon theory, and \citet{GMW18} related it to other
notions of feedback-assisted capacity. \citet{DQSW19} established the entropy
upper bound on the feedback-assisted capacity, and \citet{DKQSWW20} established
the geometric $\Upsilon$-information upper bound on the strong converse
feedback-assisted capacity.

\chapter{LOCC-Assisted Quantum Communication}\label{chap-LOCC-QC}

	This chapter develops an important variation of quantum communication, in which we allow the sender and receiver the free use of classical communication. That is, in between every use of a given quantum communication channel $\mathcal{N}_{A\rightarrow B}$, the sender and receiver are allowed to perform local operations and classical communication (LOCC). For this reason, the capacity considered in this chapter is called the LOCC-assisted quantum capacity.

	The practical motivation for this kind of feedback-assisted quantum capacity comes from the fact that, these days, classical communication is rather cheap and plentiful. Thus, from a resource-theoretic perspective, it can be sensible to simply allow classical communication as a free resource (similar to how we did for entanglement-assisted communication in Chapter~\ref{chap-EA_capacity}). Then our goal is to place informative bounds on the rate at which quantum information can be communicated from the sender to the receiver in this setting. Furthermore, these bounds are relevant for understanding and placing limitations on the speed at which distributed quantum computation can be carried out.

	In order to establish upper bounds on LOCC-assisted quantum capacity, we revisit the concept of amortization introduced in Section~\ref{sec-QFBA:amortized-persp}. However, in this context, we proceed somewhat differently, instead employing entanglement measures to quantify how much entanglement can be generated by multiple uses of a quantum channel. Then we define the amortized entanglement of a quantum channel as the largest difference between the output and input entanglement of the channel. In particular, two entanglement measures on which we focus are the squashed entanglement and the Rains relative entropy, as well as a variant of the latter called max-Rains relative entropy. One key property of the squashed entanglement and the max-Rains relative entropy is that these entanglement measures do not increase under amortization, similar to how we previously observed that the mutual information of a channel does not increase under amortization. This key property leads to the conclusion that these entanglement measures can serve as upper bounds on the LOCC-assisted quantum capacity of a quantum channel, and arriving at this conclusion is one of the main goals of this chapter. At the end of the chapter, we demonstrate the utility of the squashed entanglement and Rains family of bounds by evaluating them for several example quantum channels of interest.

\subsubsection*{Combining Entanglement Distillation and Teleportation to Obtain a Quantum Communication Protocol}
	
\begin{figure}
		\centering
		\includegraphics[scale=0.8]{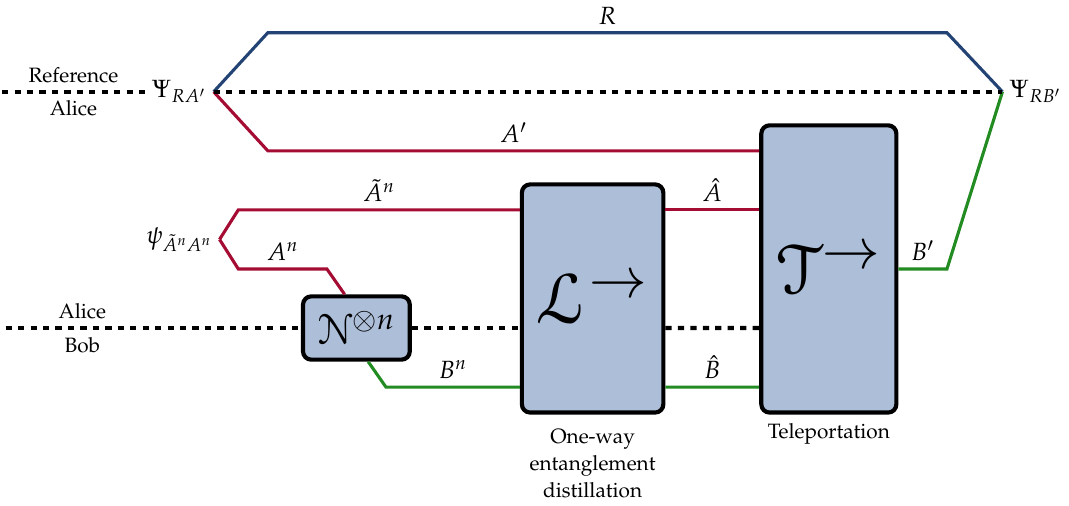}
		\caption{Sketch of a quantum communication protocol over $n$ uses of the channel $\mathcal{N}$, which makes use of entanglement distillation and teleportation. The arrow ``$\rightarrow$'' indicates that the channel contains classical communication from Alice to Bob only. We show in [] that, in the asymptotic limit, this strategy achieves the quantum capacity of $\mathcal{N}$.}\label{fig-ent_distill_plus_teleportation}
	\end{figure}

	We can use entanglement distillation along with teleportation in the asymptotic setting to obtain a lower bound on the number of transmitted qubits in a quantum communication protocol. The strategy is as follows; see Figure~\ref{fig-ent_distill_plus_teleportation}.
	\begin{enumerate}
		\item Alice prepares several copies, say $n$, of a pure state $\psi_{\tilde{A}A}$, with the dimension of $\tilde{A}$ equal to the dimension of $A$, and sends each of the $A$ systems through the channel $\mathcal{N}_{A\to B}$ to Bob.
		\item Alice and Bob now share $n$ copies of the state $\omega_{\tilde{A}B}=\mathcal{N}_{A\to B}(\psi_{\tilde{A}A})$. They perform a one-way entanglement distillation protocol to convert these mixed entangled states to a perfect, maximally entangled state $\Phi_{\hat{A}\hat{B}}$ of Schmidt rank $d\geq 2$. Roughly speaking, as shown in Chapter~\ref{chap-ent_distill}, a Schmidt rank of $2^{nI(\tilde{A}\rangle B)_{\omega}}$ is achievable for $n$ copies of $\omega_{\tilde{A}B}$, as $n\to\infty$.
		\item Using the distilled maximally entangled state, along with $2\log_2 d$ bits of classical communication, Alice and Bob perform the quantum teleportation protocol to transmit the $A'$ part of an arbitrary pure state $\Psi_{RA'}$ from Alice to Bob, with $d_{A'}=d$.
	\end{enumerate}
	Since there are $n$ uses of the channel in this strategy, we see that as $n\to\infty$, the \textit{rate} of this strategy (the number of qubits per channel use) is $\frac{\log_2 d}{n}=I(A\rangle B)_{\omega}$. By optimizing over all initial pure states $\psi_{\tilde{A}A}$ prepared by Alice, we find that, in the asymptotic setting, the rate $\sup_{\psi_{\tilde{A}A}}I(\tilde{A}\rangle B)_{\omega}=I^c(\mathcal{N})$ is achievable, where we recognize the coherent information $I^c(\mathcal{N})$ of the channel $\mathcal{N}$, which we define in \eqref{eq-coh_inf_chan}. Note that the strategy we have outlined here also involves classical communication between Alice and Bob during the entanglement distillation step and during the teleportation step. The results of Section~\ref{subsec-qcomm_one_shot_lower_bound} and Section~\ref{sec-qcomm_ach} show that this strategy is nonetheless a valid quantum communication protocol, in the sense that the rate $I^c(\mathcal{N})$ can be achieved by a strategy that does not make use of forward classical communication. 
	
	Now, Alice can do better than prepare the state $\psi_{\tilde{A}A}^{\otimes n}$ and send each $A$ system through the channel: she can prepare a state $\psi_{\tilde{A}_1\dotsb \tilde{A}_nA_1\dotsb A_n}\equiv\psi_{\tilde{A}^nA^n}$ such that the systems $A_1,\dotsc,A_n$ being sent through $\mathcal{N}$ are \textit{entangled}. One can then achieve a rate of $\frac{1}{n}\sup_{\psi_{\tilde{A}^n A^n}}I(\tilde{A}^n\rangle B^n)_{\omega}=\frac{1}{n}I^c(\mathcal{N}^{\otimes n})$. Since we are free to use the channel $\mathcal{N}$ as many times as we want, we can optimize over $n$ to obtain the communication rate
	\begin{equation}
		\sup_{n\in\mathbb{N}}\frac{1}{n}I^c(\mathcal{N}^{\otimes n})\eqqcolon I_{\text{reg}}^c(\mathcal{N}),
	\end{equation}
	and it is this \textit{regularized} coherent information of $\mathcal{N}$ that is optimal for quantum communication over the channel $\mathcal{N}$. In other words, $Q(\mathcal{N})=I_{\text{reg}}^c(\mathcal{N})$, and we prove this in Section~\ref{sec-qcomm_weak_conv}.

\section{$n$-Shot LOCC-Assisted Quantum Communication Protocol}\label{sec:LAQC-LAQC-protocols}

	This section discusses the most general form for an $n$-shot LOCC-assisted quantum communication protocol.
	
	Before starting, we should clarify that the goal of such a protocol is to produce an approximate maximally entangled state, with the Schmidt rank of the ideal target state as large as possible. Due to the assumption of free classical communication, as well as the quantum teleportation protocol discussed in Section~\ref{sec-teleportation}, generating an approximate
maximally entangled state is equivalent to generating an approximate identity quantum channel, such that the dimension of the ideal target identity channel is equal to the Schmidt rank of the target maximally entangled state. To make this statement more quantitative, suppose that $\omega_{A'B'}$ is an approximate maximally entangled state; i.e., suppose that it is $\varepsilon$-close in normalized trace distance to a maximally entangled state $\Phi_{A'B'}$ of Schmidt rank $d$:
\begin{equation}
\frac{1}{2}\norm{\omega_{A'B'} - \Phi_{A'B'}}_1 \leq \varepsilon.
\end{equation}
Let $\mathcal{T}_{AA'B'\to B}$ denote the one-way LOCC channel corresponding to quantum teleportation (as discussed around \eqref{eq-teleportation_channel}). Then by applying \eqref{eq-teleportation_sim_id}, it follows that teleportation over the ideal resource state $\Phi_{A'B'}$ realizes an identity channel $\id_{A\to B}$ of dimension $d$:
\begin{equation}
		\mathcal{T}_{AA'B'\to B}((\cdot)\otimes \Phi_{A'B'})=\id_{A\to B}(\cdot).
\end{equation}
Let $\mathcal{T}^{\omega}_{A\to B}$ denote the channel realized by teleportation over the unideal state~$\omega_{A'B'}$:
\begin{equation}
		\mathcal{T}^{\omega}_{A\to B}(\cdot) \coloneqq \mathcal{T}_{AA'B'\to B}((\cdot)\otimes \omega_{A'B'}).
\end{equation}
We would then like to determine the deviation of the ideal channel from $\mathcal{T}^{\omega}_{A\to B}$, and to do so, we can employ the normalized diamond distance. Then consider that, from the data-processing inequality for trace distance, 
\begin{align}
& \norm{\mathcal{T}^{\omega}_{A\to B} - \id_{A\to B}}_{\diamond} \notag \\
& =
\sup_{\psi_{RA}}\norm{\mathcal{T}^{\omega}_{A\to B}(\psi_{RA}) - \id_{A\to B}(\psi_{RA})}_{1} \\
& =  \sup_{\psi_{RA}}\norm{\mathcal{T}_{AA'B'\to B}(\psi_{RA} \otimes \omega_{A'B'}) - \mathcal{T}_{AA'B'\to B}(\psi_{RA}\otimes \Phi_{A'B'})}_{1} \\
& \leq \sup_{\psi_{RA}}\norm{\psi_{RA} \otimes \omega_{A'B'} - \psi_{RA}\otimes \Phi_{A'B'}}_{1} \\
& = \norm{ \omega_{A'B'} -  \Phi_{A'B'}}_{1} \leq 2 \varepsilon,
\end{align}
so that we arrive at the desired statement mentioned above:
\begin{equation}
\frac{1}{2}\norm{\omega_{A'B'} - \Phi_{A'B'}}_1 \leq \varepsilon \qquad \Rightarrow \qquad 
\frac{1}{2}\norm{\mathcal{T}^{\omega}_{A\to B} - \id_{A\to B}}_{\diamond} \leq \varepsilon .
\end{equation}

Thus, for the above reason, we focus exclusively on LOCC-assisted protocols whose aim is to generate an approximate maximally entangled state. In what follows, all bipartite cuts for separable states or LOCC\ channels should be understood as being between Alice's and Bob's systems.

	A protocol for LOCC-assisted quantum communication is depicted in Figure~[REF], and it is defined by the following elements:%
	\begin{equation}
		(\rho_{A_{1}'A_{1}B_{1}'}^{(1)},\{\mathcal{L}_{A_{i-1}%
'B_{i-1}B_{i-1}'\rightarrow A_{i}'A_{i}B_{i}'%
}^{(i)}\}_{i=2}^{n},\mathcal{L}_{A_{n}'B_{n}B_{n}'\rightarrow
M_{A}M_{B}}^{(n+1)}),
	\end{equation}
	where $\rho_{A_{1}'A_{1}B_{1}'}^{(1)}$ is a separable state,
$\mathcal{L}_{A_{i-1}'B_{i-1}B_{i-1}'\rightarrow A_{i}%
'A_{i}B_{i}'}^{(i)}$ is an LOCC\ channel for $i\in\left\{
2,\dotsc,n\right\}  $, and $\mathcal{L}_{A_{n}'B_{n}B_{n}^{\prime
}\rightarrow M_{A}M_{B}}^{(n+1)}$ is a final LOCC\ channel that generates the approximate maximally entangled state in systems $M_{A}$ and $M_{B}$. Let $\mathcal{C}$ denote all of these elements, which together constitute the LOCC-assisted quantum communication code. All systems with primed labels should be understood as local quantum memory or scratch registers that Alice or Bob can employ in this information-processing task. They are also assumed to be finite-dimensional, but could be arbitrarily large. The unprimed systems are the ones that are either input to or output from the quantum communication channel $\mathcal{N}_{A\rightarrow B}$.

	The LOCC-assisted quantum communication protocol begins with Alice and Bob performing an LOCC channel $\mathcal{L}_{\emptyset\rightarrow A_{1}^{\prime}A_{1}B_{1}'}^{(1)}$, which leads to the separable state $\rho_{A_{1}'A_{1}B_{1}'}^{(1)}$ mentioned above, where $A_{1}'$ and $B_{1}'$ are systems that are finite-dimensional
but arbitrarily large. The system $A_{1}$ is such that it can be fed into the first channel use. Alice sends system $A_{1}$ through the first channel use, leading to a state%
	\begin{equation}
		\omega_{A_{1}'B_{1}B_{1}'}^{(1)}\coloneqq\mathcal{N}%
_{A_{1}\rightarrow B_{1}}(\rho_{A_{1}'A_{1}B_{1}'}^{(1)}).
	\end{equation}
	Alice and Bob then perform the LOCC channel $\mathcal{L}_{A_{1}'B_{1}B_{1}'\rightarrow A_{2}'A_{2}B_{2}'}^{(2)}$, which leads to the state%
	\begin{equation}
		\rho_{A_{2}'A_{2}B_{2}'}^{(2)}\coloneqq\mathcal{L}%
_{A_{1}'B_{1}B_{1}'\rightarrow A_{2}'A_{2}%
B_{2}'}^{(2)}(\omega_{A_{1}'B_{1}B_{1}'}^{(1)}).
	\end{equation}
	Alice sends system $A_{2}$ through the second channel use $\mathcal{N}_{A_{2}\rightarrow B_{2}}$, leading to the state%
	\begin{equation}
		\omega_{A_{2}'B_{2}B_{2}'}^{(2)}\coloneqq\mathcal{N}%
_{A_{2}\rightarrow B_{2}}(\rho_{A_{2}'A_{2}B_{2}'}^{(2)}).
	\end{equation}
	This process iterates:\ the protocol uses the channel $n$ times. In general, we have the following states for all $i\in\{2,\dotsc,n\}$:%
	\begin{align}
		\rho_{A_{i}'A_{i}B_{i}'}^{(i)}  &  \coloneqq\mathcal{L}%
_{A_{i-1}'B_{i-1}B_{i-1}'\rightarrow A_{i}'A_{i}B_{i}'}^{(i)}(\omega_{A_{i-1}'B_{i-1}B_{i-1}'}^{(i-1)}),\\
		\omega_{A_{i}'B_{i}B_{i}'}^{(i)}  &  \coloneqq\mathcal{N}_{A_{i}\rightarrow B_{i}}(\rho_{A_{i}'A_{i}B_{i}'}^{(i)}),
	\end{align}
	where $\mathcal{L}_{A_{i-1}'B_{i-1}B_{i-1}'\rightarrow
A_{i}'A_{i}B_{i}'}^{(i)}$ is an LOCC channel. The final step
of the protocol consists of an LOCC channel $\mathcal{L}_{A_{n}'%
B_{n}B_{n}'\rightarrow M_{A}M_{B}}^{(n+1)}$, which generates the
systems $M_{A}$ and $M_{B}$ for Alice and Bob, respectively. The protocol's final state is as follows:%
	\begin{equation}
		\omega_{M_{A}M_{B}}\coloneqq\mathcal{L}_{A_{n}'B_{n}B_{n}^{\prime}\rightarrow M_{A}M_{B}}^{(n+1)}(\omega_{A_{n}'B_{n}B_{n}'}^{(n)}).
	\end{equation}

	The goal of the protocol is for the final state $\omega_{M_{A}M_{B}}$ to be close to a maximally entangled state, and we define the quantum error probability of the code as follows:%
	\begin{align}
		q_{\text{err}}(\mathcal{C})  &  \coloneqq1-F(\omega_{M_{A}M_{B}},\Phi_{M_{A}M_{B}})\label{eq:LAQC-error-criterion}\\
		&  =1-\langle\Phi|_{M_{A}M_{B}}\omega_{M_{A}M_{B}}|\Phi\rangle_{M_{A}M_{B}},
	\end{align}
	where $F$ denotes the quantum fidelity (Definition~\ref{def-fidelity}) and the maximally entangled state $\Phi_{M_{A}M_{B}}=|\Phi\rangle\!\langle\Phi|_{M_{A}M_{B}}$ is defined from%
	\begin{equation}
		|\Phi\rangle_{M_{A}M_{B}}\coloneqq\frac{1}{\sqrt{M}}\sum_{m=1}^{M}|m\rangle_{M_{A}}\otimes|m\rangle_{M_{B}},
	\end{equation}
	such that it has Schmidt rank $M$. Intuitively, the quantum error probability $q_{\text{err}}(\mathcal{C})$ is equal to the probability that one obtains the outcome ``not maximally entangled state $\Phi_{M_{A}M_{B}}$'' when performing the test or measurement $\{\Phi_{M_{A}M_{B}},I_{M_{A}M_{B}}-\Phi_{M_{A}M_{B}}\}$ on the final state $\omega_{M_{A}M_{B}}$ of the protocol.

	\begin{definition}{$(n,M,\varepsilon)$ LOCC-Assisted Quantum Communication Protocol}{def:LAQC-LOCC-assist-code}
		Let $\mathcal{C} \coloneqq (\rho_{A_{1}'A_{1}B_{1}^{\prime}}^{(1)},\{\mathcal{L}_{A_{i-1}'B_{i-1}B_{i-1}'\rightarrow A_{i}'A_{i}B_{i}'}^{(i)}\}_{i=2}^{n},\mathcal{L}_{A_{n}'B_{n}B_{n}'\rightarrow M_{A}M_{B}}^{(n+1)})$ be the elements of an $n$-round LOCC-assisted quantum communication protocol over the channel $\mathcal{N}_{A\rightarrow B}$. The protocol is called an $(n,M,\varepsilon)$ protocol, with $\varepsilon\in\left[  0,1\right]$, if $q_{\text{err}}(\mathcal{C})\leq\varepsilon$.
	\end{definition}

\subsection{Lower Bound on the Number of Transmitted Qubits}

[IN PROGRESS]

one-shot lower bound in terms of coherent information of a state. This achieves coherent information of a channel, as well as reverse coherent information of a channel.

\subsection{Amortized Entanglement as a General Upper Bound for LOCC-Assisted Quantum Communication Protocols}

	It is an interesting question to determine whether the inequality opposite to that in Lemma~\ref{lem:LAQC-amort-to-unamort}\ holds, and as the following proposition demonstrates, this question is intimately related to finding useful upper bounds on the rate at which maximal entanglement can be distilled by employing an LOCC-assisted quantum communication protocol. The following proposition represents our fundamental starting point when analyzing limitations on LOCC-assisted quantum communication protocols.

	\begin{proposition}{prop:LAQC-meta-converse-amortized}
		Let $\mathcal{N}_{A\rightarrow B}$ be a quantum channel, let $\varepsilon\in\left[  0,1\right]  $, and let $E$ be an entanglement measure that is equal to zero for all separable states. For an $(n,M,\varepsilon)$ LOCC-assisted quantum communication protocol with final state $\omega_{M_{A}M_{B}}$, the following bound holds%
		\begin{equation}\label{eq:LAQC-amortized-bnd}%
			E(M_{A};M_{B})_{\omega}\leq n\cdot E^{\mathcal{A}}(\mathcal{N}).
		\end{equation}
	\end{proposition}

	\begin{Proof}
		Consider an LOCC-assisted quantum communication protocol as presented in Section~\ref{sec:LAQC-LAQC-protocols}. From the monotonicity of the entanglement measure~$E$ with respect to LOCC channels, we find that%
		\begin{align}
			&  \!\!\!\!E(M_{A};M_{B})_{\omega}\nonumber\\
			&  \leq E(A_{n}';B_{n}B_{n}')_{\omega^{(n)}}\\
			&  =E(A_{n}';B_{n}B_{n}')_{\omega^{(n)}}-E(A_{1}'A_{1};B_{1}')_{\rho^{(1)}}\\
			&  =E(A_{n}';B_{n}B_{n}')_{\omega^{(n)}}+\left[  \sum_{i=2}^{n}E(A_{i}'A_{i};B_{i}')_{\rho^{(i)}}-E(A_{i}^{\prime}A_{i};B_{i}')_{\rho^{(i)}}\right] \nonumber\\
			&  \qquad-E(A_{1}'A_{1};B_{1}')_{\rho^{(1)}}\\
			&  \leq\sum_{i=1}^{n}\left[  E(A_{i}';B_{i}B_{i}')_{\omega^{(i)}}-E(A_{i}'A_{i};B_{i}')_{\rho^{(i)}}\right] \\
			&  \leq n\cdot E^{\mathcal{A}}(\mathcal{N}). \label{eq:LAQC-last-bound-amortized-proof}%
		\end{align}
		The first equality follows because the state $\rho_{A_{1}'A_{1}B_{1}'}^{(1)}$ is a separable state, and by assumption, the entanglement measure $E$ vanishes for all such states. The second equality follows by adding and subtracting equal terms. The second inequality follows because $E(A_{i}'A_{i};B_{i}')_{\rho^{(i)}}\leq E(A_{i-1}^{\prime};B_{i-1}B_{i-1}')_{\omega^{(i-1)}}$ for all $i\in\{2,\dotsc,n\}$, due to monotonicity of the entanglement measure $E$ with respect to LOCC channels. The final inequality follows from the definition of amortized entanglement and the fact that the states $\omega^{(i)}_{A_{i}'B_{i}B_{i}'}$
and $\rho^{(i)}_{A_{i}'A_{i}B_{i}'}$ are particular states to
consider in its optimization.
\end{Proof}

	The inequality in \eqref{eq:LAQC-amortized-bnd} states that the
entanglement of the final output state $\omega_{M_{A}M_{B}}$, as quantified by $E$, cannot exceed $n$ times the amortized entanglement of the channel $\mathcal{N}_{A\rightarrow B}$. Intuitively, the only resource allowed in the protocol, which has the potential to generate entanglement, is the quantum communication channel $\mathcal{N}_{A\rightarrow B}$. All of the LOCC\ channels allowed for free have no ability to generate entanglement on their own. Thus, the entanglement of the final state should not exceed the largest possible amount of entanglement that could ever be generated with $n$ calls to the channel, and this largest entanglement is exactly the amortized
entanglement of the channel.

	Observe that the bound in Proposition~\ref{prop:LAQC-meta-converse-amortized} depends on the final state $\omega_{M_{A}M_{B}}$, and thus it is not a universal bound, depending only on the parameters $n$, $M$, and $\varepsilon$, because this state in turn depends on the entire protocol. Similar to the upper bounds established in previous chapters, it is desirable to refine this bound such that it depends only on $n$, $M$, and $\varepsilon$, which are the parameters characterizing any generic LOCC-assisted quantum communication protocol. In the forthcoming sections, we consider particular entanglement measures, such as squashed entanglement and Rains relative entropy, which allow us to relate the parameters $M$ and $\varepsilon$ to the final state $\omega_{M_{A}M_{B}}$.

	To end this section, we note here that the bound in Proposition~\ref{prop:LAQC-meta-converse-amortized}\ simplifies for teleportation-simulable channels and for entanglement measures that are subadditive with respect to states and equal to zero for all separable states. This conclusion is a consequence of Propositions~\ref{prop:LAQC-tp-upper-bound} and \ref{prop:LAQC-meta-converse-amortized}:

	\begin{corollary*}{Reduction by Teleportation}{cor:LAQC-reduction-by-tele}
		Let $E_{S}$ be an entanglement measure that is subadditive with respect to states and equal to zero for all separable states. Let $\mathcal{N}_{A\rightarrow B}$ be a channel that\ is teleportation-simulable with associated resource state $\theta_{RB'}$. Let $\varepsilon\in\left[  0,1\right]$. For an $(n,M,\varepsilon)$ LOCC-assisted quantum communication protocol with final state $\omega_{M_{A}M_{B}}$, the following bound holds%
		\begin{equation}
			E_{S}(M_{A};M_{B})_{\omega}\leq n\cdot E_{S}(R;B')_{\theta}.
		\end{equation}
	\end{corollary*}

\subsection{Squashed Entanglement Upper Bound on the Number of Transmitted Qubits}\label{sec:LAQC-sq-ent-bnd-non-as}

	We now establish the squashed entanglement upper bound on the number of qubits that a sender can transmit to a receiver by employing an LOCC-assisted quantum communication protocol:

	\begin{theorem*}{$n$-Shot Squashed Entanglement Upper Bound}
{thm:LAQC-sq-ent-bnd-LOCC-as-cap}
		Let $\mathcal{N}_{A\rightarrow B}$ be a quantum channel, and let $\varepsilon\in[0,1)$. For all $(n,M,\varepsilon)$ LOCC-assisted quantum communication protocols over the channel $\mathcal{N}_{A\rightarrow B}$, the following bound holds%
		\begin{equation}
			\log_{2}M\leq\frac{1}{1-\sqrt{\varepsilon}}\left[  n\cdot E_{\operatorname{sq}}(\mathcal{N})+g_{2}(\sqrt{\varepsilon})\right]  .
		\end{equation}
	\end{theorem*}

	\begin{Proof}
	Consider an arbitrary $(n,M,\varepsilon)$ LOCC-assisted quantum communication protocol over the channel $\mathcal{N}_{A\rightarrow B}$, as defined in Section~\ref{sec:LAQC-LAQC-protocols}.
		The squashed entanglement is an entanglement measure (monotone under LOCC\ as shown in Theorem~\ref{thm:LAQC-mono-LOCC-sq}) and it is equal to zero for separable states (Proposition~\ref{prop:LAQC-squashed-sep}). Thus, Proposition~\ref{prop:LAQC-meta-converse-amortized}\ applies, and we find that%
		\begin{equation}
			E_{\operatorname{sq}}(M_{A};M_{B})_{\omega}\leq n\cdot E_{\operatorname{sq}}^{\mathcal{A}}(\mathcal{N})=n\cdot E_{\operatorname{sq}}(\mathcal{N}),
		\end{equation}
		where the equality follows from Theorem~\ref{thm:LAQC-amort-collapse-squashed}. Applying Definition~\ref{def:LAQC-LOCC-assist-code} leads to
		\begin{equation}
			F(\Phi_{M_{A}M_{B}},\omega_{M_{A}M_{B}})\geq1-\varepsilon.
		\end{equation}
		As a consequence of Proposition~\ref{prop:LAQC-cont-sq-unif}, we find that%
		\begin{align}
			&  E_{\operatorname{sq}}(M_{A};M_{B})_{\omega}\nonumber\\
			&  \geq E_{\operatorname{sq}}(M_{A};M_{B})_{\Phi}-\left[  \sqrt{\varepsilon}\log_{2}\min\left\{  \left\vert M_{A}\right\vert ,\left\vert M_{B}\right\vert\right\}  +g_{2}(\sqrt{\varepsilon})\right] \\
			&  =\log_{2}M-\left[  \sqrt{\varepsilon}\log_{2}M+g_{2}(\sqrt{\varepsilon })\right] \\
			&  =(1-\sqrt{\varepsilon})\log_{2}M-g_{2}(\sqrt{\varepsilon}).
		\end{align}
		The first equality follows from Proposition~\ref{prop:LAQC-reduction-for-pure-squashed}.\ We can finally rearrange the established inequality $n\cdot E_{\operatorname{sq}}(\mathcal{N})\geq(1-\sqrt{\varepsilon})\log_{2}M-g_{2}(\sqrt{\varepsilon})$ to be in the form stated in the theorem.
	\end{Proof}

\section{\texorpdfstring{$n$}{n}-Shot PPT-Assisted Quantum Communication Protocol}\label{sec-LOCC-QC:PPT-assist-prot}

	Recalling the completely PPT-preserving channels of Definition~\ref{def-PPT_pres_chan} as forming a superset of LOCC\ channels, we can also consider quantum communication protocols assisted by completely PPT-preserving channels (abbreviated as C-PPT-P channels). Such a PPT-assisted protocol has exactly the same form as given in Section~\ref{sec:LAQC-LAQC-protocols}, but it is instead defined by the following elements:%
	\begin{equation}
		(\rho_{A_{1}'A_{1}B_{1}'}^{(1)},\{\mathcal{P}_{A_{i-1}'B_{i-1}B_{i-1}'\rightarrow A_{i}'A_{i}B_{i}'}^{(i)}\}_{i=2}^{n},\mathcal{P}_{A_{n}'B_{n}B_{n}'\rightarrow M_{A}M_{B}}^{(n+1)}),
	\end{equation}
	where $\rho_{A_{1}'A_{1}B_{1}'}^{(1)}$ is a PPT\ state, $\mathcal{P}_{A_{i-1}'B_{i-1}B_{i-1}'\rightarrow A_{i}'A_{i}B_{i}'}^{(i)}$ is a C-PPT-P channel for $i\in\left\{  2,\dotsc,n\right\}  $, and $\mathcal{P}_{A_{n}'B_{n}B_{n}'\rightarrow M_{A}M_{B}}^{(n+1)}$ is a final C-PPT-P\ channel that generates an approximate maximally entangled state in systems $M_{A}$ and $M_{B}$. Denoting the final state of the protocol again by $\omega_{M_{A}M_{B}}$, the criterion for such a protocol is again given by \eqref{eq:LAQC-error-criterion}. We then arrive at the following definition:

	\begin{definition}{$(n,M,\varepsilon)$ PPT-Assisted Quantum Communication Protocol}{def:LAQC-PPT-assist-code}
		Let $\mathcal{C} \coloneqq (\rho_{A_{1}'A_{1}B_{1}'}^{(1)},\{\mathcal{P}_{A_{i-1}'B_{i-1}B_{i-1}'\rightarrow A_{i}'A_{i}B_{i}'}^{(i)}\}_{i=2}^{n},\mathcal{P}_{A_{n}'B_{n}B_{n}'\rightarrow M_{A}M_{B}}^{(n+1)})$ be the elements of an $n$-round PPT-assisted quantum communication protocol over the channel $\mathcal{N}_{A\rightarrow B}$. The protocol is called an $(n,M,\varepsilon)$ protocol, with $\varepsilon\in\left[0,1\right]  $, if $q_{\text{err}}(\mathcal{C})\leq\varepsilon$.
	\end{definition}

	Since every LOCC\ channel is a C-PPT-P channel, we can make the following observation immediately:

	\begin{remark}\label{rem:LAQC-LOCC-prot-is-PPT}
		Every $(n,M,\varepsilon)$ LOCC-assisted quantum communication protocol is also an $(n,M,\varepsilon)$ PPT-assisted quantum communication protocol.
	\end{remark}

	As a consequence of the above observation, any converse bound or limitation that we establish for an arbitrary $(n,M,\varepsilon)$ PPT-assisted protocol is also a converse bound for an $(n,M,\varepsilon)$ LOCC-assisted protocol.

	By the exact same reasoning as in the proof of Proposition~\ref{prop:LAQC-meta-converse-amortized}, but replacing LOCC\ channels with C-PPT-P channels, we arrive at the following proposition:

	\begin{proposition}{prop:LAQC-meta-converse-amortized-PPT}
		Let $\mathcal{N}_{A\rightarrow B}$ be a quantum channel, let $\varepsilon\in\left[  0,1\right]  $, and let $E$ be an entanglement measure that is monotone under completely PPT-preserving channels and is equal to zero for all PPT\ states. For an $(n,M,\varepsilon)$ PPT-assisted quantum communication protocol with final state $\omega _{M_{A}M_{B}}$, the following bound holds%
		\begin{equation}
			E(M_{A};M_{B})_{\omega}\leq n\cdot E^{\mathcal{A}}(\mathcal{N}).
		\end{equation}
	\end{proposition}

	Recalling Definition~\ref{def-PPT_sim_chan}, a channel $\mathcal{N}_{A\rightarrow B}$ with input system $A$ and output system $B$ is defined to be PPT-simulable with associated resource state $\omega_{RB'}$ if the following equality holds for all input states $\rho_{A}$:%
	\begin{equation}
		\mathcal{N}_{A\rightarrow B}(\rho_{A})=\mathcal{P}_{ARB'\rightarrow B}(\rho_{A}\otimes\omega_{RB'}), \label{eq:LAQC-PPT-sim-chs}%
	\end{equation}
	where $\mathcal{P}_{ARB'\rightarrow B}$ is a completely PPT-preserving channel between the sender, who has systems $A$ and $R$, and the receiver, who has system $B'$.

	By the same reasoning used to arrive at Corollary~\ref{cor:LAQC-reduction-by-tele}, but replacing LOCC\ channels with completely PPT-preserving ones, we find the following:

	\begin{corollary}{cor:LAQC-reduction-for-PPT-sim}
		Let $E_{S}$ denote an entanglement measure that is monotone non-increasing with respect to completely PPT-preserving channels, subadditive with respect to states, and equal to zero for all PPT\ states. Let $\mathcal{N}_{A\rightarrow B}$ be a channel that\ is PPT-simulable with associated resource state $\theta_{RB'}$. Let $\varepsilon\in\left[  0,1\right]  $. For an $(n,M,\varepsilon)$ PPT-assisted quantum communication protocol with final state $\omega_{M_{A}M_{B}}$, the following bound holds%
		\begin{equation}
			E_{S}(M_{A};M_{B})_{\omega}\leq n\cdot E_{S}(R;B')_{\theta}.
		\end{equation}
	\end{corollary}

\subsection{R\'enyi--Rains Information Upper Bounds on the Number of Transmitted Qubits}\label{sec:LAQC-bnd-Rains}

	We now establish the max-Rains information upper bound on the number of qubits that a sender can transmit to a receiver by employing a PPT-assisted quantum communication protocol:

	\begin{theorem*}{$n$-Shot Max-Rains Upper Bound}{thm:LAQC-max-Rains-bnd-PPT-as-cap}
		Let $\mathcal{N}_{A\rightarrow B}$ be a quantum channel, and let $\varepsilon \in[0,1)$. For all $(n,M,\varepsilon)$ PPT-assisted quantum communication protocols over the channel $\mathcal{N}_{A\rightarrow B}$, the following bound holds%
		\begin{equation}
			\log_{2}M\leq n\cdot R_{\max}(\mathcal{N})+\log_{2}\!\left(  \frac{1}{1-\varepsilon}\right)  .
		\end{equation}
	\end{theorem*}

	\begin{Proof}
		Consider an arbitrary $(n,M,\varepsilon)$ LOCC-assisted quantum communication protocol over the channel $\mathcal{N}_{A\rightarrow B}$, as defined in Section~\ref{sec:LAQC-LAQC-protocols}.
		The max-Rains relative entropy is an entanglement measure (monotone under completely PPT-preserving channels\ as shown in Proposition~\ref{prop:E-meas:max-rains-strong-ppt-mono}), and it is equal to zero for PPT states. 
		Thus, Proposition~\ref{prop:LAQC-meta-converse-amortized-PPT}\ applies, and we find that%
		\begin{equation}\label{eq:LAQC-max-Rains-bnd-1}%
			R_{\max}(M_{A};M_{B})_{\omega}\leq n\cdot R_{\max}^{\mathcal{A}}(\mathcal{N})=n\cdot R_{\max}(\mathcal{N}), 
		\end{equation}
		where the equality follows from Theorem~\ref{prop:LAQC-amort-doesnt-help-max-Rains}. Applying Definition~\ref{def:LAQC-PPT-assist-code} leads to
		\begin{equation}
			F(\Phi_{M_{A}M_{B}},\omega_{M_{A}M_{B}})\geq1-\varepsilon.
		\end{equation}
		As a consequence of Propositions~\ref{prop:core-meta-converse-privacy_a} and
\ref{prop:sandwich-to-htre}, we find that%
		\begin{align}
			\log_{2}M  &  \leq R^{\varepsilon}(M_{A};M_{B})_{\omega}\\
			&  \leq R_{\max}(M_{A};M_{B})_{\omega}+\log_{2}\!\left(  \frac{1}{1-\varepsilon}\right)  . \label{eq:LAQC-max-Rains-bnd-2}%
		\end{align}
		Combining \eqref{eq:LAQC-max-Rains-bnd-1} and \eqref{eq:LAQC-max-Rains-bnd-2}, we conclude the proof.
	\end{Proof}

	At this point, it is worthwhile to compare the squashed-entanglement bound in Theorem~\ref{thm:LAQC-sq-ent-bnd-LOCC-as-cap}\ with the max-Rains information bound in Theorem~\ref{thm:LAQC-max-Rains-bnd-PPT-as-cap}. First, both bounds hold for all quantum channels, and so this is an advantage that they both possess. The squashed entanglement and max-Rains information are rather different quantities, and so the quantities on their own can vary based on the channel for which they are evaluated. The squashed-entanglement bound in Theorem~\ref{thm:LAQC-sq-ent-bnd-LOCC-as-cap}\ is a weak-converse bound, whereas the max-Rains information bound in Theorem~\ref{thm:LAQC-max-Rains-bnd-PPT-as-cap}\ is a strong-converse bound. The max-Rains information bound has the advantage that it is efficiently computable by a semi-definite program, whereas it is not known how to compute the squashed entanglement. However, one can apply Proposition~\ref{prop:LAQC-sq-chann-form}\ to see that the squashed entanglement bound gives a whole host of upper bounds related to the choice of a squashing channel, and one can potentially obtain tight bounds by making a clever choice of a squashing channel.

	For channels that are PPT-simulable with associated resource states, as recalled in \eqref{eq:LAQC-PPT-sim-chs}, we obtain upper bounds that can be even stronger:

	\begin{theorem*}{$n$-Shot R\'enyi--Rains Upper Bounds for PPT-Simulable Channels}{thm:LAQC-Renyi-Rains-bnd-PPT-sim}
		Let $\mathcal{N}_{A\rightarrow B}$ be a quantum channel that is PPT-simulable with associated resource state $\theta_{SB'}$, and let $\varepsilon\in\lbrack0,1)$. For all $(n,M,\varepsilon)$ PPT-assisted quantum communication protocols over the channel $\mathcal{N}_{A\rightarrow B}$, the following bounds hold for all $\alpha>1$:%
		\begin{align}
			\log_{2}M  &  \leq n\cdot\widetilde{R}_{\alpha}(S;B')_{\theta}+\frac{\alpha}{\alpha-1}\log_{2}\!\left(  \frac{1}{1-\varepsilon}\right)  ,\\
			\log_{2}M  &  \leq\frac{1}{1-\varepsilon}\left[  n\cdot R(S;B^{\prime})_{\theta}+h_{2}(\varepsilon)\right]  .
		\end{align}
	\end{theorem*}

	\begin{Proof}
	Consider an arbitrary $(n,M,\varepsilon)$ LOCC-assisted quantum communication protocol over the channel $\mathcal{N}_{A\rightarrow B}$, as defined in Section~\ref{sec:LAQC-LAQC-protocols}.
		The R\'{e}nyi--Rains relative entropy and Rains relative entropy are monotone non-increasing under completely PPT-preserving channels (Proposition~\ref{prop-gen_Rains_rel_ent_properties}), equal to zero for PPT states, 
		 and subadditive with respect to states (Proposition~\ref{prop-gen_Rains_rel_ent_properties}). As such, Corollary~\ref{cor:LAQC-reduction-for-PPT-sim}\ applies, and we find for $\alpha>1$ that%
		\begin{align}
			\widetilde{R}_{\alpha}(M_{A};M_{B})_{\omega} &  \leq n\cdot\widetilde {R}_{\alpha}(S;B')_{\theta},\label{eq:LAQC-PPT-sim-conv-1}\\
			R(M_{A};M_{B})_{\omega} &  \leq n\cdot R(S;B')_{\theta}.\label{eq:LAQC-PPT-sim-conv-2}%
		\end{align}
		Applying Definition~\ref{def:LAQC-PPT-assist-code} leads to
		\begin{equation}
			F(\Phi_{M_{A}M_{B}},\omega_{M_{A}M_{B}})\geq1-\varepsilon.
		\end{equation}
		As a consequence of Proposition~\ref{prop:core-meta-converse-privacy_a}, we have that%
		\begin{equation}
			\log_{2}M\leq R^{\varepsilon}(M_{A};M_{B})_{\omega}.
		\end{equation}
		Applying Propositions~\ref{prop-hypo_to_rel_ent} and \ref{prop:sandwich-to-htre}, we find that%
		\begin{align}
			\log_{2}M &  \leq\widetilde{R}_{\alpha}(M_{A};M_{B})_{\omega}+\frac{\alpha}{\alpha-1}\log_{2}\!\left(  \frac{1}{1-\varepsilon}\right),\label{eq:LAQC-PPT-sim-conv-3}\\
			\log_{2}M &  \leq\frac{1}{1-\varepsilon}\left[  R(M_{A};M_{B})_{\omega}+h_{2}(\varepsilon)\right]  .\label{eq:LAQC-PPT-sim-conv-4}
		\end{align}
		Putting together \eqref{eq:LAQC-PPT-sim-conv-1}, \eqref{eq:LAQC-PPT-sim-conv-2}, \eqref{eq:LAQC-PPT-sim-conv-3}, and \eqref{eq:LAQC-PPT-sim-conv-4} concludes the proof.
	\end{Proof}

\section{LOCC- and PPT-Assisted Quantum Capacities of Quantum Channels}

	In this section, we analyze the asymptotic capacities, and as before, the upper bounds for the asymptotic capacities are straightforward consequences of the non-asymptotic bounds given in Sections~\ref{sec:LAQC-sq-ent-bnd-non-as} and \ref{sec:LAQC-bnd-Rains}. The definitions of these capacities are similar to what we have given previously, and so we only state them here briefly.

	\begin{definition}{Achievable Rate for LOCC-Assisted Quantum Communication}{def-LOCC_ast_q_comm_ach_rate}
		Given a quantum channel $\mathcal{N}$, a rate $R\in\mathbb{R}^{+}$ is called an achievable rate for LOCC-assisted quantum communication over $\mathcal{N}$ if for all $\varepsilon\in(0,1]$, all $\delta>0$, and all sufficiently large $n$, there exists an $(n,2^{n(R-\delta)},\varepsilon)$ LOCC-assisted quantum communication protocol.
	\end{definition}

	\begin{definition}{LOCC-Assisted Quantum Capacity of a Quantum Channel}{def-LOCC_ast_q_com_cap}
		The LOCC-assisted quantum capacity of a quantum channel $\mathcal{N}$, denoted by $Q^{\leftrightarrow}(\mathcal{N})$, is defined as the supremum of all achievable rates, i.e.,%
		\begin{equation}
			Q^{\leftrightarrow}(\mathcal{N})\coloneqq\sup\{R:R\text{ is an achievable rate for }\mathcal{N}\}.
		\end{equation}
	\end{definition}

	\begin{definition}{Weak Converse Rate for LOCC-Assisted Quantum Communication}{def-LOCC_ast_q_comm_weak_conv_rate}
		Given a quantum channel $\mathcal{N}$, a rate $R\in\mathbb{R}^{+}$ is called a weak converse rate for LOCC-assisted quantum communication over $\mathcal{N}$ if every $R'>R$ is not an achievable rate for $\mathcal{N}$.
	\end{definition}

	\begin{definition}{Strong Converse Rate for LOCC-Assisted Quantum Communication}{def-LOCC_ast_q_comm_str_conv_rate}
		Given a quantum channel $\mathcal{N}$, a rate $R\in\mathbb{R}^{+}$ is called a strong converse rate for LOCC-assisted quantum communication over $\mathcal{N}$ if for all $\varepsilon\in[0,1)$, all $\delta>0$, and all sufficiently large $n$, there does not exist an $(n,2^{n(R+\delta)},\varepsilon)$ LOCC-assisted quantum communication protocol.
	\end{definition}

	\begin{definition}{Strong Converse LOCC-Assisted Quantum Capacity of a Quantum Channel}{def-LOCC_ast_q_comm_str_conv_cap}
		The strong converse LOCC-assisted quantum capacity of a quantum channel $\mathcal{N}$, denoted by $\widetilde{Q}^{\leftrightarrow}(\mathcal{N})$, is defined as the infimum of all strong converse rates, i.e.,%
		\begin{equation}
			\widetilde{Q}^{\leftrightarrow}(\mathcal{N})\coloneqq\inf\{R:R\text{ is a strong converse rate for }\mathcal{N}\}.
		\end{equation}
	\end{definition}

	We have the exact same definitions for PPT-assisted quantum communication, and we use the notation $Q_{\operatorname{PPT}}^{\leftrightarrow}$ to refer to the PPT-assisted quantum capacity and $\widetilde{Q}_{\operatorname{PPT}}^{\leftrightarrow}$ for the strong converse PPT-assisted quantum capacity.

	Recall that, by definition, the following bounds hold%
	\begin{align}
		Q^{\leftrightarrow}(\mathcal{N})  &  \leq\widetilde{Q}^{\leftrightarrow}(\mathcal{N})\leq\widetilde{Q}_{\operatorname{PPT}}^{\leftrightarrow}(\mathcal{N}),\\
		Q^{\leftrightarrow}(\mathcal{N})  &  \leq Q_{\operatorname{PPT}}^{\leftrightarrow}(\mathcal{N})\leq\widetilde{Q}_{\operatorname{PPT}}^{\leftrightarrow}(\mathcal{N}).
	\end{align}

	As a direct consequence of the bound in Theorem~\ref{thm:LAQC-sq-ent-bnd-LOCC-as-cap} and methods similar to those given in the proof of Theorem~\ref{thm-ea_classical_comm_weak_converse}, we find the following:

	\begin{theorem*}{Squashed-Entanglement Weak-Converse Bound}{prop-LOCC_q_comm_sq_ent_weak_conv}
		The squashed entanglement of a channel $\mathcal{N}$ is a weak converse rate for LOCC-assisted quantum communication:%
		\begin{equation}
			Q^{\leftrightarrow}(\mathcal{N})\leq E_{\operatorname{sq}}(\mathcal{N}).
		\end{equation}
	\end{theorem*}

As a direct consequence of the bound in Theorem~\ref{thm:LAQC-max-Rains-bnd-PPT-as-cap}\ and methods similar to those given in Section~\ref{sec-eacc_str_conv}, we find that

	\begin{theorem*}{Max-Rains Strong-Converse Bound}{thm-LOCC_q_comm_max_rains_str_conv}
		The max-Rains information of a channel $\mathcal{N}$ is a strong converse rate for PPT-assisted quantum communication:%
		\begin{equation}
			\widetilde{Q}_{\operatorname{PPT}}^{\leftrightarrow}(\mathcal{N})\leq R_{\max}(\mathcal{N}).
		\end{equation}
	\end{theorem*}

	As a direct consequence of the bound in Theorem~\ref{thm:LAQC-Renyi-Rains-bnd-PPT-sim}\ and methods similar to those given in Section~\ref{sec-eacc_str_conv}, we find that

	\begin{theorem*}{Rains Strong-Converse Bound for PPT-Simulable Channels}{thm-LOCC_q_comm_str_conv_PPT_sim}
		Let $\mathcal{N}$ be a quantum channel that is PPT-simulable with associated resource state $\theta_{SB'}$. Then the Rains information of $\mathcal{N}$ is a strong converse rate for PPT-assisted quantum communication:%
		\begin{equation}
			\widetilde{Q}_{\operatorname{PPT}}^{\leftrightarrow}(\mathcal{N})\leq R(\mathcal{N}).
		\end{equation}
	\end{theorem*}

\section{Examples}

[IN PROGRESS]

erasure channel - get Rains information as a strong converse rate - will match
lower bound in terms of reverse coherent information (Hayashi called this
pseudocoherent information)

covariant dephasing channels - get Rains information as a strong converse rate
and then coherent information matches this (will evaluate Rains information
bound in unassisted quantum capacity chapter)

depolarizing channel - evaluate Rains information

use squashed entanglement to give upper bound for amplitude damping channel

\section{Bibliographic Notes}

The concept of LOCC-assisted quantum communication over a quantum channel was presented in \citep[Section~V]{BDSW96}. The same Section~V of \citep{BDSW96} also showed how to use the notion of teleportation simulation of a quantum channel and entanglement measures in order to bound the LOCC-assisted quantum capacity from above by a resource state that can realize the channel by teleportation simulation. \citet{Mul12} presented a more detailed analysis of this bounding technique. Other papers that make use of the teleportation-simulation technique in this and other contexts include those by \citet{HHH99,GC99,Zhou2000,BB01,TBS02,GI02,WPG07,NFC09,CDP09,STM11,LM15,PLOB15,TSW16,WTB16,TSW17,KW17}.

A precise mathematical definition of an LOCC-assisted quantum communication protocol conducted over a quantum channel was presented in \citep[Definition~12]{Mul12} and \citep[Section~IV]{TGW14IEEE}. 

That the entanglement of the final state of an $n$-round LOCC-assisted quantum communication is bounded from above by $n$ times the channel's amortized entanglement (Proposition~\ref{prop:LAQC-meta-converse-amortized}) was anticipated by \citet{BHLS03} and proven by \citet{KW17}. Corollary~\ref{cor:LAQC-reduction-by-tele} was anticipated by \citet{BDSW96} and presented in more detail in \citep[Chapter~4]{Mul12}, while the form in which we have presented it here is closely related to the presentation by \citet{KW17}. 

The $n$-round PPT-assisted quantum communication protocols presented in Section~\ref{sec-LOCC-QC:PPT-assist-prot} were considered by \citet{KW17}, with PPT-assisted quantum communication over a single or parallel use of a quantum channel considered by \citet{LM15,WD16b,WFD17}. The bound in Proposition~\ref{prop:LAQC-meta-converse-amortized-PPT} was established by \citet{KW17}.

Theorem~\ref{thm:LAQC-sq-ent-bnd-LOCC-as-cap} is due to \citet{TGW14IEEE}.

\citet{WD16a} defined a semi-definite programming upper bound on distillable entanglement of a bipartite state, and \citet{WD16b} defined a semi-definite programming upper bound on the quantum capacity of a quantum channel. \citet{WFD17} observed that the quantity defined by \citet{WD16a} is equal to the max-Rains relative entropy, while also observing that the quantity defined by \citet{WD16b} is equal to the max-Rains information of a quantum channel.
  \citet{BW17} established the max-Rains information as an upper bound on the $n$-round non-asymptotic PPT-assisted quantum capacity (Theorem~\ref{thm:LAQC-max-Rains-bnd-PPT-as-cap}). The upper bounds in Theorem~\ref{thm:LAQC-Renyi-Rains-bnd-PPT-sim} are due to \citet{KW17}.

\chapter{Secret Key Agreement}\label{chap-SKA}

		This chapter continues with the theme of feedback-assisted communication. Here, we consider secret-key-agreement protocols, where the goal is for the sender and receiver of a quantum channel $\mathcal{N}_{A\rightarrow B}$ to establish secret key, by using the quantum channel $\mathcal{N}_{A\rightarrow B}$ along with the free use of public classical communication. That is, between every channel use in a secret-key-agreement protocol, the sender and receiver are allowed to perform local operations and public classical communication (LOPC). The notion of capacity developed in this chapter is known as secret-key-agreement capacity.

	The secret key distilled in such a secret-key-agreement protocol should be protected from an eavesdropper. The model we assume here is that the eavesdropper is quite powerful, having access to the full environment of every use of the quantum channel $\mathcal{N}_{A\rightarrow B}$, as well as a copy of all of the classical data exchanged by the sender and receiver when they conduct a round of LOPC. To understand this model in a physical context, suppose that the quantum channel connecting the sender and receiver is a fiber-optic cable, which we can model as a bosonic loss channel, and suppose that the sender employs quantum states of light to distill a secret key with the receiver. Then, in this eavesdropper model, we are assuming that all of the light that does not make it to the receiver is collected by the eavesdropper in a quantum memory. In this way, the secret key distilled by such a protocol is guaranteed to be secure against a quantum-enabled eavesdropper.

	The practical motivation for secret-key-agreement protocols is related to the motivation that we considered in the previous chapter on LOCC-assisted quantum communication. Classical communication is cheap and plentiful these days, and so from a resource-theoretic perspective, it seems sensible to allow it for free in a theoretical model of communication. Once a secret key has been established, it can be used in conjuction with the well known one-time pad protocol as a scheme for private communication of an arbitrary message that has the same size as the key. Thus, as a consequence of the one-time pad protocol, it follows that secret key agreement and private communication are equivalent information-processing tasks when public classical communication is available for free. Furthermore, the model of LOPC-assisted secret key agreement is essentially the same model considered in quantum key distribution, which is one of the most famous applications in quantum information science.  One of the main goals of this chapter is to place bounds on the rate at which secret-key-agreement is possible. Due to the strong connection between the model of secret-key-agreement and quantum key distribution, these bounds then place limitations on the rates at which it is possible to generate secret key in a quantum key distribution protocol.

	The main method for placing limitations on the rates of secret-key-agreement protocols is similar to the approach that we took in the last chapter. In fact, there are many parallels. We again use the concept of amortization and entanglement measures, such as squashed entanglement and relative entropy of entanglement (and several variants of the latter).

	However, the main difference between this chapter and the previous one is that the communication model is different. As discussed above, a secret-key-agreement protocol is a three-party protocol, consisting of the legitimate sender and receiver, as well as the eavesdropper. Thus, \textit{a priori}, it is not obvious how to connect entanglement measures, which are used in two-party protocols, to secret-key-agreement protocols. To overcome this problem, we exploit the purification principle to establish a powerful equivalence between tripartite secret-key-agreement protocols and bipartite private-state distillation protocols. After doing so, we can apply entanglement measures to bound the rate at which it is possible to distill bipartite private states in a bipartite private-state distillation protocol, and then by appealing to the aforementioned equivalence, we can bound the rate at which it is possible to distill secret key in a tripartite secret-key-agreement protocol.
	
	The main conclusion of this chapter is that entanglement measures such as squashed entanglement and relative entropy of entanglement (and the latter's variations) are upper bounds on the secret-key-agreement capacity of quantum channels. At the end of the chapter, we evaluate these bounds for various channels of interest in order to determine the fundamental limitations on secret key agreement for these channels.

\section{$n$-Shot Secret-Key-Agreement Protocol}\label{sec:SKA-n-shot-SKA-prot}

	We begin by discussing the most general form for a secret-key-agreement protocol conducted over a quantum channel. The most important point to clarify before starting is the communication model, in particular, to address the question of who has access to what. First, we suppose that there is a quantum channel $\mathcal{N}_{A\rightarrow B}$ connecting the legitimate sender Alice to the legitimate receiver Bob. Alice has exclusive access to the input system $A$, and Bob has exclusive access to the output system $B$. As we know from Chapter~\ref{chap-QM_channels}, every quantum channel has an isometric channel $\mathcal{U}_{A\rightarrow BE}$\ extending it, such that the original channel $\mathcal{N}_{A\rightarrow B}$ is recovered by tracing over the purifying or environment system $E$:%
	\begin{equation}
		\mathcal{N}_{A\rightarrow B}=\operatorname{Tr}_{E}\circ\mathcal{U}_{A\rightarrow BE}.
	\end{equation}
	Taking the same perspective as that in Chapter~\ref{chap-private_capacity}, with the idea that a powerful, fully quantum eavesdropper could have access to every system to which the legitimate parties do not have access, we suppose that the quantum eavesdropper has access to the environment system $E$. Furthermore, in a secret-key-agreement protocol, the legitimate parties are allowed to use a public, classical communication channel, in addition to the quantum channel $\mathcal{N}_{A\rightarrow B}$, in order to generate a secret key. Since this channel is public, we suppose that the eavesdropper has access to all of the classical data exchanged between the legitimate parties.

	In more detail, an $n$-shot protocol for secret key agreement consists of $n$~calls to the quantum channel $\mathcal{N}_{A\rightarrow B}$, interleaved by LOPC channels. Since all of the classical data exchanged between Alice and Bob is assumed to be public and available to the eavesdropper as well, we call these channels ``LOPC'' channels, which is an abbreviation of  ``local
operations and public communication.'' In fact, a protocol for secret key agreement has essentially the same structure as a protocol for LOCC-assisted quantum communication, as discussed in Section~\ref{sec:LAQC-LAQC-protocols}, with the exception that the systems at the end should hold a secret key instead of a maximally entangled state.

	A protocol for secret key agreement is depicted in Figure~[REF],
	and consists of the following elements:%
	\begin{equation}
		\Big(\rho_{A_{1}^{\prime}A_{1}B_{1}^{\prime}Y_{1}},\{\mathcal{L}_{A_{i-1}^{\prime}B_{i-1}B_{i-1}^{\prime}\rightarrow A_{i}^{\prime}A_{i}B_{i}^{\prime}Y_{i}}^{(i)}\}_{i=2}^{n},\mathcal{L}_{A_{n}^{\prime}B_{n}B_{n}^{\prime}\rightarrow K_{A}K_{B}Y_{n+1}}^{(n+1)}\Big)\,.
	\end{equation}
	All systems labeled by $A$ belong to Alice, those labeled by $B$ belong to Bob, and those labeled by $Y$ are classical systems belonging to Eve, representing a copy of the classical data exchanged by Alice and Bob in a round of LOPC. In the above, $\rho_{A_{1}^{\prime}A_{1}B_{1}^{\prime}Y_{1}}$ is a separable state, $\mathcal{L}_{A_{i-1}^{\prime}B_{i-1}B_{i-1}^{\prime}\rightarrow A_{i}^{\prime}A_{i}B_{i}^{\prime}Y_{i}}^{(i)}$ is an LOPC channel for $i\in\left\{  2,\ldots,n\right\}  $, and $\mathcal{L}_{A_{n}^{\prime}B_{n}B_{n}^{\prime}\rightarrow K_{A}K_{B}Y_{n+1}}^{(n+1)}$ is a final LOPC channel that generates the approximate secret key in systems $K_{A}$ and $K_{B}$. Let $\mathcal{C}$ denote all of these elements, which together constitute the secret-key-agreement protocol. As with LOCC-assisted
quantum communication, all systems with primed labels should be understood as local quantum memory or scratch registers that Alice and Bob can employ in this information-processing task. We also assume that they are finite-dimensional, yet arbitrarily large. The unprimed systems are the ones that are either input to or output from the quantum communication channel $\mathcal{N}_{A\rightarrow B}$.

	The secret-key-agreement protocol begins with Alice and Bob performing an LOPC channel $\mathcal{L}_{\emptyset\rightarrow A_{1}^{\prime}A_{1}B_{1}^{\prime}Y_{1}}^{(1)}$, which leads to the separable state $\rho_{A_{1}^{\prime}A_{1}B_{1}^{\prime}Y_{1}}^{(1)}$ mentioned above, where $A_{1}^{\prime}$ and $B_{1}^{\prime}$ are systems that are finite-dimensional yet arbitrarily large. In particular, the state $\rho_{A_{1}^{\prime}A_{1}B_{1}^{\prime}Y_{1}}^{(1)}$ has the following form:%
	\begin{equation}\label{eq:SKA-init-state-rho}
		\rho_{A_{1}^{\prime}A_{1}B_{1}^{\prime}Y_{1}}^{(1)}\coloneqq\sum_{y_{1}}p_{Y_{1}}(y_{1})\tau_{A_{1}^{\prime}A_{1}}^{y_{1}}\otimes\zeta_{B^{\prime}_{1}}^{y_{1}}\otimes|y_{1}\rangle\!\langle y_{1}|_{Y_{1}},
	\end{equation}
	where $Y_{1}$ is a classical random variable corresponding to the message exchanged between Alice and Bob, which is needed to establish this state. The classical system $Y_{1}$ belongs to the eavesdropper. Also, $\{\tau_{A_{1}^{\prime}A_{1}}^{y_{1}}\}_{y_{1}}$ and $\{\zeta_{B^{\prime}_{1}}^{y_{1}}\}_{y_{1}}$ are sets of quantum states and $p_{Y_{1}}$ is a probability distribution. Note that the reduced state for Alice and Bob is a generic separable state of the following form:%
	\begin{equation}
		\rho_{A_{1}^{\prime}A_{1}B_{1}^{\prime}}^{(1)}=\sum_{y_{1}}p_{Y_{1}}(y_{1})\tau_{A_{1}^{\prime}A_{1}}^{y_{1}}\otimes\zeta_{B^{\prime}_{1}}^{y_{1}}.
	\end{equation}
	The system $A_{1}$ of $\rho_{A_{1}^{\prime}A_{1}B_{1}^{\prime}Y_{1}}^{(1)}$ is such that it can be fed into the first channel use. Alice then sends system $A_{1}$ through the first channel use, leading to a state%
	\begin{equation}
		\omega_{A_{1}^{\prime}B_{1}B_{1}^{\prime}E_{1}Y_{1}}^{(1)}\coloneqq\mathcal{U}_{A_{1}\rightarrow B_{1}E_{1}}^{\mathcal{N}}(\rho_{A_{1}^{\prime}A_{1}B_{1}^{\prime}Y_{1}}^{(1)}).
	\end{equation}
	Note that we write the channel use as the isometric channel $\mathcal{U}_{A_{1}\rightarrow B_{1}E_{1}}^{\mathcal{N}}$ that extends $\mathcal{N}_{A_{1}\rightarrow B_{1}}$, since we would like to incorporate the eavesdropper's system~$E_{1}$ explicitly into the description of the protocol. Alice and Bob then perform the LOPC channel $\mathcal{L}_{A_{1}^{\prime}B_{1}B_{1}^{\prime}\rightarrow A_{2}^{\prime}A_{2}B_{2}^{\prime}Y_{2}}^{(2)}$,
which leads to the state%
	\begin{equation}
		\rho_{A_{2}^{\prime}A_{2}B_{2}^{\prime}E_{1}Y_{1}Y_{2}}^{(2)}\coloneqq\mathcal{L}_{A_{1}^{\prime}B_{1}B_{1}^{\prime}\rightarrow A_{2}^{\prime}A_{2}B_{2}^{\prime}Y_{2}}^{(2)}(\omega_{A_{1}^{\prime}B_{1}B_{1}^{\prime}E_{1}Y_{1}}^{(1)}).
	\end{equation}
	The LOPC\ channel $\mathcal{L}_{A_{1}^{\prime}B_{1}B_{1}^{\prime}\rightarrow A_{2}^{\prime}A_{2}B_{2}^{\prime}Y_{2}}^{(2)}$ can be written as%
	\begin{equation}
		\mathcal{L}_{A_{1}^{\prime}B_{1}B_{1}^{\prime}\rightarrow A_{2}^{\prime} A_{2}B_{2}^{\prime}Y_{2}}^{(2)}\coloneqq\sum_{y_{2}}\mathcal{E}_{A_{1}^{\prime}\rightarrow A_{2}^{\prime}A_{2}}^{y_{2}}\otimes\mathcal{F}_{B_{1}B_{1}^{\prime}\rightarrow B_{2}^{\prime}}^{y_{2}}\otimes|y_{2}\rangle\!\langle y_{2}|_{Y_{2}}.
	\end{equation}
	In the above, $\{\mathcal{E}_{A_{1}^{\prime}\rightarrow A_{2}^{\prime}A_{2}}^{y_{2}}\}_{y_{2}}$ and $\{\mathcal{F}_{B_{1}B_{1}^{\prime}\rightarrow B_{2}^{\prime}}^{y_{2}}\}_{y_{2}}$ are sets of completely positive maps such that the sum map $\sum_{y_{2}}\mathcal{E}_{A_{1}^{\prime}\rightarrow A_{2}^{\prime}A_{2}}^{y_{2}}\otimes\mathcal{F}_{B_{1}B_{1}^{\prime}\rightarrow B _{2}^{\prime}}^{y_{2}}$ is trace preserving. The classical system $Y_{2}$ represents the eavesdropper's copy of the classical data exchanged by Alice and Bob in this round of LOPC. Note that the reduced channel acting on Alice and Bob's systems is as follows:%
	\begin{equation}
		\mathcal{L}_{A_{1}^{\prime}B_{1}B_{1}^{\prime}\rightarrow A_{2}^{\prime}A_{2}B_{2}^{\prime}}^{(2)}=\sum_{y_{2}}\mathcal{E}_{A_{1}^{\prime}\rightarrow A_{2}^{\prime}A_{2}}^{y_{2}}\otimes\mathcal{F}_{B_{1}B_{1}^{\prime}\rightarrow B_{2}^{\prime}}^{y_{2}}.
	\end{equation}
	Alice sends system $A_{2}$ through the second channel use $\mathcal{U}_{A_{2}\rightarrow B_{2}E_{2}}^{\mathcal{N}}$, leading to the state%
	\begin{equation}
		\omega_{A_{2}^{\prime}B_{2}B_{2}^{\prime}E_{1}E_{2}Y_{1}Y_{2}}^{(2)}\coloneqq\mathcal{U}_{A_{2}\rightarrow B_{2}E_{2}}^{\mathcal{N}}(\rho_{A_{2}^{\prime}A_{2}B_{2}^{\prime}E_{1}Y_{1}Y_{2}}^{(2)}).
	\end{equation}
	This process iterates:\ the protocol uses the channel $n$ times. In general, we have the following states for all $i\in\{2,\ldots,n\}$:%
	\begin{align}
		\rho_{A_{i}^{\prime}A_{i}B_{i}^{\prime}E_{1}^{i-1}Y_{1}^{i}}^{(i)}  & \coloneqq\mathcal{L}_{A_{i-1}^{\prime}B_{i-1}B_{i-1}^{\prime}\rightarrow A_{i}^{\prime}A_{i}B_{i}^{\prime}Y_{i}}^{(i)}(\omega_{A_{i-1}^{\prime} B_{i-1}B_{i-1}^{\prime}E_{1}^{i-1}Y_{1}^{i-1}}^{(i-1)}),\label{eq:SKA-rho-states}\\
\omega_{A_{i}^{\prime}B_{i}B_{i}^{\prime}E_{1}^{i}Y_{1}^{i}}^{(i)}  & \coloneqq\mathcal{U}_{A_{i}\rightarrow B_{i}E_{i}}^{\mathcal{N}}(\rho_{A_{i}^{\prime}A_{i}B_{i}^{\prime}E_{1}^{i-1}Y_{1}^{i}}^{(i)}),\label{eq:SKA-omega-states}%
	\end{align}
	where $\mathcal{L}_{A_{i-1}^{\prime}B_{i-1}B_{i-1}^{\prime}\rightarrow A_{i}^{\prime}A_{i}B_{i}^{\prime}Y_{i}}^{(i)}$ is an LOPC channel that can be written as%
	\begin{equation}\label{eq:SKA-LOCC-channels}
		\mathcal{L}_{A_{i-1}^{\prime}B_{i-1}B_{i-1}^{\prime}\rightarrow A_{i}^{\prime}A_{i}B_{i}^{\prime}Y_{i}}^{(i)}\coloneqq\sum_{y_{i}}\mathcal{E}_{A_{i-1}^{\prime}\rightarrow A_{i}^{\prime}A_{i}}^{y_{i}}\otimes \mathcal{F}_{B_{i-1}B_{i-1}^{\prime}\rightarrow B_{i}^{\prime}}^{y_{i}}\otimes|y_{i}\rangle\!\langle y_{i}|_{Y_{i}}. %
	\end{equation}
	In the above, $\{\mathcal{E}_{A_{i-1}^{\prime}\rightarrow A_{i}^{\prime}A_{i}}^{y_{i}}\}_{y_{i}}$ and $\{\mathcal{F}_{B_{i-1}B_{i-1}^{\prime}\rightarrow B_{i}^{\prime}}^{y_{i}}\}_{y_{i}}$ are sets of completely positive maps such that the sum map $\sum_{y_{i}}\mathcal{E}_{A_{i-1}^{\prime}\rightarrow A_{i}^{\prime}A_{i}}^{y_{i}}\otimes\mathcal{F}_{B_{i-1}B_{i-1}^{\prime}\rightarrow B_{i}^{\prime}}^{y_{i}}$ is trace preserving. The classical system $Y_{i}$ represents the eavesdropper's copy of the classical data exchanged by Alice and Bob in this round of LOPC. Note that the reduced channel acting on Alice and Bob's systems is as follows:%
	\begin{equation}
		\mathcal{L}_{A_{i-1}^{\prime}B_{i-1}B_{i-1}^{\prime}\rightarrow A_{i}^{\prime}A_{i}B_{i}^{\prime}}^{(i)}=\sum_{y_{i}}\mathcal{E}_{A_{i-1}^{\prime}\rightarrow A_{i}^{\prime}A_{i}}^{y_{i}}\otimes\mathcal{F}_{B_{i-1}B_{i-1}^{\prime}\rightarrow B_{i}^{\prime}}^{y_{i}}.
	\end{equation}
	In \eqref{eq:SKA-rho-states}--\eqref{eq:SKA-omega-states}, we have employed the following shorthands: $E_{1}^{i}\equiv E_{1}\cdots E_{i}$ and $Y_{1}^{i}\equiv Y_{1}\cdots Y_{i}$. The final step of the protocol consists of an LOPC channel $\mathcal{L}_{A_{n}^{\prime}B_{n}B_{n}^{\prime}\rightarrow K_{A}K_{B}Y_{n+1}}^{(n+1)}$, which generates the key systems $K_{A}$ and $K_{B}$ for Alice and Bob, respectively. The protocol's final state is as
follows:%
	\begin{equation}
		\omega_{K_{A}K_{B}E_{1}^{n}Y_{1}^{n+1}}\coloneqq\mathcal{L}_{A_{n}^{\prime}B_{n}B_{n}^{\prime}\rightarrow K_{A}K_{B}Y_{n+1}}^{(n+1)}(\omega_{A_{n}^{\prime}B_{n}B_{n}^{\prime}E_{1}^{n}Y_{1}^{n}}^{(n)}).
		\label{eq-SKA:final-omega-state}
	\end{equation}
	Note that the final LOPC\ channel can be written as%
	\begin{equation}\label{eq:SKA-final-LOCC-ch}
		\mathcal{L}_{A_{n}^{\prime}B_{n}B_{n}^{\prime}\rightarrow K_{A}K_{B}Y_{n+1}}^{(n+1)}\coloneqq\sum_{y_{n+1}}\mathcal{E}_{A_{n}^{\prime}\rightarrow K_{A}}^{y_{n+1}}\otimes\mathcal{F}_{B_{n}B_{n}^{\prime}\rightarrow K_{B}}^{y_{n+1}}\otimes|y_{n+1}\rangle\!\langle y_{n+1}|_{Y_{n+1}},
	\end{equation}
	and the reduced final channel acting on Alice and Bob's systems is as follows:%
	\begin{equation}
		\mathcal{L}_{A_{n}^{\prime}B_{n}B_{n}^{\prime}\rightarrow K_{A}K_{B}}^{(n+1)}=\sum_{y_{n+1}}\mathcal{E}_{A_{n}^{\prime}\rightarrow K_{A}}^{y_{n+1}}\otimes\mathcal{F}_{B_{n}B_{n}^{\prime}\rightarrow K_{B}}^{y_{n+1}}.
	\end{equation}

	The goal of the protocol is for the final state $\omega_{K_{A}K_{B}E_{1}^{n}Y_{1}^{n+1}}$ to be nearly indistinguishable from a tripartite secret-key state, and we define the privacy error of the code to be as follows:%
	\begin{equation}\label{eq:SKA-error-criterion}%
		p_{\text{err}}(\mathcal{C})\coloneqq1-F(\omega_{K_{A}K_{B}E_{1}^{n}Y_{1}^{n+1}},\overline{\Phi}_{K_{A}K_{B}}\otimes\sigma_{E_{1}^{n}Y_{1}^{n+1}}),
	\end{equation}
	where $\sigma_{E_{1}^{n}Y_{1}^{n+1}}$ is some state of the eavesdropper's systems, $F$ denotes the quantum fidelity (Definition~\ref{def-fidelity}) and the maximally classically correlated state $\overline{\Phi}_{K_{A}K_{B}}$ is defined as%
	\begin{equation}
		\overline{\Phi}_{K_{A}K_{B}}\coloneqq\frac{1}{K}\sum_{k=1}^{K}|k\rangle\!\langle k|_{K_{A}}\otimes|k\rangle\!\langle k|_{K_{B}}.
	\end{equation}
	Intuitively, the privacy error $p_{\text{err}}(\mathcal{C})$ quantifies how distinguishable the final state $\omega_{K_{A}K_{B}E_{1}^{n}Y_{1}^{n+1}}$ is from an ideal tripartite secret-key state $\overline{\Phi}_{K_{A}K_{B}}\otimes\sigma_{E_{1}^{n}Y_{1}^{n+1}}$, in which the key values in $K_{A}$ and $K_{B}$ are perfectly correlated and uniformly random and in tensor product
with the eavesdropper's systems $E_{1}^{n}Y_{1}^{n+1}$. For an ideal tripartite secret-key state, it is difficult for an eavesdropper to guess the value of the key by observing the content of her quantum systems $E_{1}^{n}Y_{1}^{n+1}$. In fact, the chance for an eavesdropper to guess the key value of an ideal secret-key state is equal to $1/K$, which is no better than
random guessing.

	Due to the isometric invariance of the fidelity and the fact that all isometric extensions of a channel are related by an isometry acting on the environment system, the privacy error in \eqref{eq:SKA-error-criterion} is invariant under any choice of an isometric channel $\mathcal{U}_{A\rightarrow BE}^{\mathcal{N}}$ that extends the original channel $\mathcal{N}_{A\rightarrow B}$. Thus, the relevant performance parameters for a secret-key agreement protocol do not change with the particular isometric extension chosen. This is to be expected since the actual information that the eavesdropper gains in the protocol does not depend on the particular isometric extension chosen.

	\begin{definition}{$(n,K,\varepsilon)$ Secret-Key-Agreement Protocol}{def:SKA-LOPC-assist-code}
		Let $(\rho_{A_{1}^{\prime}A_{1}B_{1}^{\prime}Y_{1}}^{(1)},\{\mathcal{L}_{A_{i-1}^{\prime}B_{i-1}B_{i-1}^{\prime}\rightarrow A_{i}^{\prime}A_{i}B_{i}^{\prime}Y_{i}}^{(i)}\}_{i=2}^{n},\mathcal{L}_{A_{n}^{\prime}B_{n}B_{n}^{\prime}\rightarrow K_{A}K_{B}Y_{n+1}}^{(n+1)})$ be the elements of an $n$-round LOPC-assisted secret-key-agreement protocol over the channel $\mathcal{N}_{A\rightarrow B}$. The protocol is called an $(n,K,\varepsilon)$ protocol, with $\varepsilon \in\left[  0,1\right]  $, if the privacy error $p_{\operatorname{err}}(\mathcal{C})\leq\varepsilon$.
	\end{definition}
	
	\subsection{Equivalence between Secret Key Agreement and LOPC-Assisted Private Communication}
	
	 The goal of a secret-key-agreement protocol is for Alice and Bob to establish an approximation of an ideal secret key, the latter begin uniformly distributed, perfectly correlated, and independent of Eve's quantum systems. What is the use of this secret key? As it turns out, it can be used for private communication by means of the one-time pad protocol. This in turn means that secret key agreement and private classical communication are equivalent information processing tasks when public classical communication is available for free, and the goal of this section is to clarify this point.
	 
	 An LOPC-assisted private communication protocol  uses a quantum channel $n$ times along with public classical communication to transmit an arbitrary message of size $K$ privately from Alice to Bob in such a way that the fidelity of the actual state at the end of the protocol and the ideal state is no smaller than $1 - \varepsilon$. In more detail, let $\overline{\Phi}^p_{M_{A}M_{B}}$ denote the following state in which there is an arbitrary distribution $p$ over the message:
	\begin{equation}
	\overline{\Phi}^p_{M_{A}M_{B}} \coloneqq \sum_{m=1}^{K}p(m) |m\rangle\!\langle m|_{M_{A}}\otimes|m\rangle\!\langle m|_{M_{B}}.
	\end{equation}
	Let $\omega^p_{M_{A}M_{B}E_{1}^{n}Y_{1}^{n+1}}$ denote the final state of the protocol, which is defined in the same way as \eqref{eq-SKA:final-omega-state}, with the exception that the message distribution $p$ is no longer uniform.
	Then an $(n,K,\varepsilon)$ LOPC-assisted private communication protocol is defined similarly to an $(n,K,\varepsilon)$ secret-key-agreement protocol as given above, except that the following inequality holds
		\begin{equation}
\max_{p:\mathcal{M} \to [0,1]}
1-F(\omega^p_{M_{A}M_{B}E_{1}^{n}Y_{1}^{n+1}},\overline{\Phi}^p_{M_{A}M_{B}}\otimes\sigma_{E_{1}^{n}Y_{1}^{n+1}}) \leq \varepsilon.
\label{eq-SKA:infidelity-SKA-to-priv-comm}
	\end{equation}
	where the maximization is over all message distributions $p$ and $\sigma_{E_{1}^{n}Y_{1}^{n+1}}$ is some fixed state of the eavesdropper's systems that is independent of the message transmitted.
	
	By the use of the one-time pad protocol, it follows that an $(n,K,\varepsilon)$ secret-key-agreement protocol leads to an $(n,K,\varepsilon)$ LOPC-assisted private communication protocol. To see how the one-time pad protocol works in conjunction with secret key agreement, suppose that Alice and Bob have completed an $(n,K,\varepsilon)$ secret-key-agreement protocol as described in the previous section, with the final state $\omega_{K_{A}K_{B}E_{1}^{n}Y_{1}^{n+1}}$ satisfying $p_{\text{err}}(\mathcal{C})\leq \varepsilon$. Alice then brings in her local message registers $M_A$ and $M_{A'}$, so that the overall quantum state is
	\begin{equation}
	\overline{\Phi}^p_{M_{A}M_{A'}} \otimes \omega_{K_{A}K_{B}E_{1}^{n}Y_{1}^{n+1}}
	\end{equation}
	The one-time pad protocol is an LOPC protocol in which Alice then performs the following classical computation, represented as a quantum channel, on her message register $M_{A'}$ and her key register $K_A$:
	\begin{equation}
	\sum_{k,m} \ket{m \oplus k}_{C_A} \bra{m}_{M_{A'}} \bra{k}_{K_A} (\cdot)  \ket{m}_{M_{A'}} \ket{k}_{K_A} \bra{m \oplus k}_{C_A},
	\end{equation}
	where the addition $\oplus$ is modulo $K$.
	She then transmits the classical register $C_A$ over a public classical channel to Bob. Eve can make a copy $C_{A'}$ of this classical register containing the value $m\oplus k$, but since Bob's key register $K_B$ is not available to her, the register $C_{A'}$ is nearly independent of Alice's message register $M_A$ (depending on how small $\varepsilon$ is). Bob then performs the following classical computation, represented as a quantum channel, on his received register $C_A$ and his key register $K_B$:
		\begin{equation}
	\sum_{c,k} \ket{c \ominus k}_{M_B} \bra{c}_{C_A} \bra{k}_{K_B} (\cdot)  \ket{c}_{C_A} \ket{k}_{K_B} \bra{c \ominus k}_{M_B},
	\end{equation}
	where the subtraction $\ominus$ is modulo $K$. Let $\omega^p_{M_{A}M_{B}E_{1}^{n}Y_{1}^{n+1}C_{A'}}$ denote the final state of the protocol. By applying the data-processing inequality to \eqref{eq-SKA:infidelity-SKA-to-priv-comm}, as well as the fact mentioned above that $C_{A'}$ is independent of $M_A$ and $M_B$ in the ideal case, the following inequality holds
	\begin{equation}
\max_{p:\mathcal{M}\to [0,1]}1-F(\omega^p_{M_{A}M_{B}E_{1}^{n}Y_{1}^{n+1}C_{A'}},\overline{\Phi}^p_{M_{A}M_{B}}\otimes\sigma_{E_{1}^{n}Y_{1}^{n+1}C_{A'}}) \leq \varepsilon,
	\end{equation}
	where $\sigma_{E_{1}^{n}Y_{1}^{n+1}}C_{A'}$ is a fixed state of the eavesdropper's systems.
	Thus, an $(n,K,\varepsilon)$ secret-key-agreement protocol leads to an $(n,K,\varepsilon)$ LOPC-assisted private communication protocol, as claimed.
	
	The other implication is trivial: while employing an $(n,K,\varepsilon)$ LOPC-assisted private communication protocol, Alice can choose the distribution of the message to be uniform, and then an $(n,K,\varepsilon)$ LOPC-assisted private communication protocol leads to an $(n,K,\varepsilon)$ secret-key-agreement protocol. Thus, it follows that LOPC-assisted private communication and secret key agreement are equivalent whenever public classical communication is available for free.

\section{Equivalence between Secret Key Agreement and LOCC-Assisted
Private-State Distillation}

	There is a deep and powerful equivalence between a secret-key-agreement protocol as described above and a protocol that uses LOCC assistance to distill a bipartite private state (recall Definition~\ref{def:private-state-bi}). This equivalence is helpful in the analysis of secret-key-agreement protocols, in the sense that one can use tools from entanglement theory in order to establish bounds on the rate at which secret key agreement is possible.

	The main idea behind this equivalence is to apply the purification principle to a secret-key-agreement protocol and then examine the consequences. That is, we can purify each step of the secret-key-agreement protocol discussed in the previous section, and then we can examine various reduced states at each step. In what follows, we detail such a purified protocol.

\subsection{The Purified Protocol}\label{sec:SKA-purified-protocol}

	To begin with, recall that the initial state $\rho_{A_{1}^{\prime}A_{1}B_{1}^{\prime}}^{(1)}$ of a secret-key-agreement protocol is a separable state of the form in \eqref{eq:SKA-init-state-rho}. The state $\rho_{A_{1}^{\prime}A_{1}B_{1}^{\prime}}^{(1)}$ can be purified as follows%
	\begin{equation}
		|\rho^{(1)}\rangle_{A_{1}^{\prime}A_{1}S_{A_{1}}B_{1}^{\prime}S_{B_{1}}Y_{1}}\coloneqq\sum_{y_{1}}\sqrt{p_{Y_{1}}(y_{1})}|\tau^{y_{1}}\rangle_{A_{1}^{\prime}A_{1}S_{A_{1}}}\otimes|\zeta^{y_{1}}\rangle_{B^{\prime}_{1}S_{B_{1}}}\otimes|y_{1}\rangle_{Y_{1}},
	\end{equation}
	where the systems $S_{A_{1}}$ and $S_{B_{1}}$ are known as local ``shield'' systems. In principle, the shield systems $S_{A_{1}}$ and $S_{B_{1}}$ could be held by Alice and Bob, respectively, and the states $|\tau^{y_{1}}\rangle_{A_{1}^{\prime}A_{1}S_{A_{1}}}$ and $|\zeta^{y_{1}}\rangle_{B^{\prime}_{1}S_{B_{1}}}$ purify $\tau_{A_{1}^{\prime}A_{1}}^{y_{1}}$ and $\zeta_{B^{\prime}_{1}}^{y_{1}}$ in \eqref{eq:SKA-init-state-rho}, respectively. We assume without loss of generality that the shield systems contain a coherent classical copy of the classical random variable~$Y_{1}$, such that tracing over systems $S_{A_{1}}$ and $S_{B_{1}}$ recovers the original state in \eqref{eq:SKA-init-state-rho}. As before, Eve possesses system $Y_{1}$, which contains a coherent classical copy of the classical data exchanged.

	Each LOPC channel $\mathcal{L}_{A_{i-1}^{\prime}B_{i-1}B_{i-1}^{\prime}\rightarrow A_{i}^{\prime}A_{i}B_{i}^{\prime}}^{(i)}$ for $i\in \{2,\ldots,n\}$ is of the form in \eqref{eq:SKA-LOCC-channels} and can be purified to an isometry in the following way:%
	\begin{multline}
		U_{A_{i-1}^{\prime}B_{i-1}B_{i-1}^{\prime}\rightarrow A_{i}^{\prime}A_{i}S_{A_{i}}B_{i}^{\prime}S_{B_{i}}Y_{i}}^{\mathcal{L}^{(i)}}\coloneqq\label{eq:SKA-LOCC-as-separable-iso-ext}\\
		\sum_{y_{i}}U_{A_{i-1}^{\prime}\rightarrow A_{i}^{\prime}A_{i}S_{A_{i}}}^{\mathcal{E}^{y_{i}}}\otimes U_{B_{i-1}B_{i-1}^{\prime}\rightarrow B_{i}^{\prime}S_{B_{i}}}^{\mathcal{F}^{y_{i}}}\otimes|y_{i}\rangle_{Y_{i}},
	\end{multline}
	where $\{U_{A_{i-1}^{\prime}\rightarrow A_{i}^{\prime}A_{i}S_{A_{i}}}^{\mathcal{E}^{y_{i}}}\}_{y_{i}}$ and $\{U_{B_{i-1}B_{i-1}^{\prime}\rightarrow B_{i}^{\prime}S_{B_{i}}}^{\mathcal{F}^{y_{i}}}\}_{y_{i}}$ are collections of linear operators, each of which is a contraction, that is,%
	\begin{equation}
		\Vert U_{A_{i-1}^{\prime}\rightarrow A_{i}^{\prime}A_{i}S_{A_{i}}}^{\mathcal{E}^{y_{i}}}\Vert_{\infty},\ \Vert U_{B_{i-1}B_{i-1}^{\prime}\rightarrow B_{i}^{\prime}S_{B_{i}}}^{\mathcal{F}^{y_{i}}}\Vert_{\infty}\leq 1,
	\end{equation}
	such that the linear operator in \eqref{eq:SKA-LOCC-as-separable-iso-ext} is an isometry.

	It is important to note here that the isometry $U_{A_{i-1}^{\prime}B_{i-1}B_{i-1}^{\prime}\rightarrow A_{i}^{\prime}A_{i}S_{A_{i}}B_{i}^{\prime}S_{B_{i}}Y_{i}}^{\mathcal{L}^{(i)}}$ results from purifying each step of an LOPC\ channel. That is, an LOPC\ channel is implemented as a sequence of one-way LOPC channels, which each consist of a generalized measurement by one party, classical communication of the measurement outcome to the other, and a channel by the other party, conditioned on the outcome of the measurement. So when purifying the LOPC\ channel, we purify each of these steps, and the resulting purified channel is what is represented in \eqref{eq:SKA-LOCC-as-separable-iso-ext}.

	The systems $S_{A_{i}}$ and $S_{B_{i}}$ in \eqref{eq:SKA-LOCC-as-separable-iso-ext} are shield systems belonging to Alice and Bob, respectively, and we assume without loss of generality that they contain a coherent classical copy of the classical random variable $Y_{i}$, such that tracing over the systems $S_{A_{i}}$ and $S_{B_{i}}$ recovers the original LOPC channel in \eqref{eq:SKA-LOCC-channels}. As before, $Y_{i}$ is a system held by Eve, containing a coherent classical copy of the classical data exchanged in this round.

	Thus, a purification of the state $\rho_{A_{i}^{\prime}A_{i}B_{i}^{\prime}}^{(i)}$ after each LOPC channel is as follows:
	\begin{multline}
		|\rho^{(i)}\rangle_{A_{i}^{\prime}A_{i}S_{A_{1}^{i}}B_{i}^{\prime}S_{B_{1}^{i}}E_{1}^{i-1}Y_{1}^{i}}\coloneqq\\
		U_{A_{i-1}^{\prime}B_{i-1}B_{i-1}^{\prime}\rightarrow A_{i}^{\prime}A_{i}S_{A_{i}}B_{i}^{\prime}S_{B_{i}}Y_{i}}^{\mathcal{L}^{(i)}}|\omega^{(i-1)}\rangle_{A_{i-1}^{\prime}B_{i-1}B_{i-1}^{\prime}S_{A_{1}^{i-1}}S_{B_{1}^{i-1}}E_{1}^{i-1}Y_{1}^{i-1}},
	\end{multline}
	where we have employed the shorthands $S_{A_{1}^{i}}\equiv S_{A_{1}}\cdots S_{A_{i}}$ and $S_{B_{1}^{i}}\equiv S_{B_{1}}\cdots S_{B_{i}}$, with a similar shorthand for $E_{1}^{i-1}$ and $Y_{1}^{i}$ as before. A purification of the state $\omega_{A_{i}^{\prime}B_{i}B_{i}^{\prime}}^{(i)}$ after each use of the channel $\mathcal{N}_{A\rightarrow B}$ is
	\begin{equation}
		|\omega^{(i)}\rangle_{A_{i}^{\prime}B_{i}S_{A_{1}^{i}}B_{i}^{\prime}S_{B_{1}^{i}}E_{1}^{i}Y_{1}^{i}}\coloneqq U_{A_{i}\rightarrow B_{i}E_{i}}^{\mathcal{N}}|\rho^{(i)}\rangle_{A_{i}^{\prime}A_{i}S_{A_{1}^{i}}B_{i}^{\prime}S_{B_{1}^{i}}E_{1}^{i-1}Y_{1}^{i}},
	\end{equation}
	where $U_{A_{i}\rightarrow B_{i}E_{i}}^{\mathcal{N}}$ is an isometric extension of the $i$th channel use $\mathcal{N}_{A_{i}\rightarrow B_{i}}$.

	The final LOPC channel takes the form in \eqref{eq:SKA-final-LOCC-ch}, and it can be purified to an isometry similarly as
	\begin{multline}
		U_{A_{n}^{\prime}B_{n}B_{n}^{\prime}\rightarrow K_{A}S_{A_{n+1}}K_{B}S_{B_{n+1}}Y_{n+1}}^{\mathcal{L}^{(n+1)}}\coloneqq\\
		\sum_{y_{n+1}}U_{A_{n}^{\prime}\rightarrow K_{A}S_{A_{n+1}}}^{\mathcal{E}^{y_{n+1}}}\otimes U_{B_{n}B_{n}^{\prime}\rightarrow K_{B}S_{B_{n+1}}}^{\mathcal{F}^{y_{n+1}}}\otimes|y_{n+1}\rangle_{Y_{n+1}}.
	\end{multline}
	The systems $S_{A_{n+1}}$ and $S_{B_{n+1}}$ are again shield systems belonging to Alice and Bob, respectively, and we assume again that they contain a coherent classical copy of the classical random variable $Y_{n+1}$, such that tracing over $S_{A_{n+1}}$ and $S_{B_{n+1}}$ recovers the original LOPC channel in \eqref{eq:SKA-final-LOCC-ch}. As before, $Y_{n+1}$ is a system held by Eve, containing a coherent classical copy of the classical data exchanged in this round.

	The final state at the end of the purified protocol is a pure state $|\omega\rangle_{K_{A}S_{A}K_{B}S_{B}E^{n}Y^{n+1}}$, given by
	\begin{multline}
		|\omega\rangle_{K_{A}S_{A}K_{B}S_{B}E^{n}Y^{n+1}}\coloneqq\\
		U_{A_{n}^{\prime}B_{n}B_{n}^{\prime}\rightarrow K_{A}S_{A_{n+1}}K_{B}S_{B_{n+1}}Y_{n+1}}^{\mathcal{L}^{(n+1)}}|\omega^{(n)}\rangle_{A_{n}^{\prime}B_{n}S_{A_{1}^{n}}B_{n}^{\prime}S_{B_{1}^{n}}E_{1}^{n}Y_{1}^{n}}.
	\end{multline}
	Alice is in possession of the key system $K_{A}$ and the shield systems $S_{A}\equiv S_{A_{1}}\cdots S_{A_{n+1}}$, Bob possesses the key system $K_{B}$ and the shield systems $S_{B}\equiv S_{B_{1}}\cdots S_{B_{n+1}}$, and Eve holds the environment systems $E^{n}\equiv E_{1}\cdots E_{n}$. Additionally, Eve has coherent copies $Y^{n+1}\equiv Y_{1}\cdots Y_{n+1}$ of all the classical data exchanged.

\subsection{LOCC-Assisted Bipartite Private-State Distillation}\label{sec:SKA-bipartite-PS-dist}

	As a consequence of the purification principle, on the one hand, if we trace over the shield systems at every step, then we simply recover the original tripartite secret-key-agreement protocol detailed in Section~\ref{sec:SKA-n-shot-SKA-prot}. On the other hand, suppose that we instead trace over all of Eve's systems at each step. Due to the fact that each state of the $Y$ systems is a coherent classical copy, the resulting reduced states consist of a classical mixture of various states of Alice and Bob's systems, as would arise in an LOCC-assisted quantum communication protocol.

	It is worthwhile to examine how each step changes after tracing over Eve's systems of the purified protocol. For the first step, the reduced state of Alice and Bob's systems is a separable state of the following form:%
	\begin{equation}\label{eq:SKA-bi-priv-state-distill-1}%
		\rho_{A_{1}^{\prime}A_{1}S_{A_{1}}B_{1}^{\prime}S_{B_{1}}}^{(1)}=\sum_{y_{1}}p_{Y_{1}}(y_{1})\tau_{A_{1}^{\prime}A_{1}S_{A_{1}}}^{y_{1}}\otimes\zeta_{B^{\prime}_{1}S_{B_{1}}}^{y_{1}},
	\end{equation}
	where $\tau_{A_{1}^{\prime}A_{1}S_{A_{1}}}^{y_{1}}=|\tau^{y_{1}}\rangle \!\langle\tau^{y_{1}}|_{A_{1}^{\prime}A_{1}S_{A_{1}}}$ and $\zeta_{B^{\prime}_{1}S_{B_{1}}}^{y_{1}}=|\zeta^{y_{1}}\rangle\!\langle\zeta^{y_{1}}|_{B^{\prime}_{1}S_{B_{1}}}$. Tracing over Eve's system $Y_{i}$ of each isometry $U_{A_{i-1}^{\prime}B_{i-1}B_{i-1}^{\prime}\rightarrow A_{i}^{\prime}A_{i}S_{A_{i}}B_{i}^{\prime}S_{B_{i}}Y_{i}}^{\mathcal{L}^{(i)}}$ leads to the following LOCC\ channel:%
	\begin{equation}\label{eq:SKA-bi-priv-state-distill-2}
		\mathcal{L}_{A_{i-1}^{\prime}B_{i-1}B_{i-1}^{\prime}\rightarrow A_{i}^{\prime}A_{i}S_{A_{i}}B_{i}^{\prime}S_{B_{i}}}^{(i)}=\sum_{y_{i}}\mathcal{U}_{A_{i-1}^{\prime}\rightarrow A_{i}^{\prime}A_{i}S_{A_{i}}}^{\mathcal{E}^{y_{i}}}\otimes\mathcal{U}_{B_{i-1}B_{i-1}^{\prime}\rightarrow B_{i}^{\prime}S_{B_{i}}}^{\mathcal{F}^{y_{i}}}, %
	\end{equation}
	where%
	\begin{align}
		\mathcal{U}_{A_{i-1}^{\prime}\rightarrow A_{i}^{\prime}A_{i}S_{A_{i}}}^{\mathcal{E}^{y_{i}}}(\cdot)  &  \coloneqq U_{A_{i-1}^{\prime}\rightarrow A_{i}^{\prime}A_{i}S_{A_{i}}}^{\mathcal{E}^{y_{i}}}(\cdot)[U_{A_{i-1}^{\prime}\rightarrow A_{i}^{\prime}A_{i}S_{A_{i}}}^{\mathcal{E}^{y_{i}}}]^{\dagger},\label{eq:SKA-reduced-maps-purified-1}\\
		\mathcal{U}_{B_{i-1}B_{i-1}^{\prime}\rightarrow B_{i}^{\prime}S_{B_{i}}}^{\mathcal{F}^{y_{i}}}(\cdot)  &  \coloneqq U_{B_{i-1}B_{i-1}^{\prime}\rightarrow B_{i}^{\prime}S_{B_{i}}}^{\mathcal{F}^{y_{i}}}(\cdot)[U_{B_{i-1}B_{i-1}^{\prime}\rightarrow B_{i}^{\prime}S_{B_{i}}}^{\mathcal{F}^{y_{i}}}]^{\dag}. \label{eq:SKA-reduced-maps-purified-2}%
	\end{align}
	Tracing over Eve's system $Y_{n+1}$ of the final isometry leads to the following LOCC channel:
	\begin{equation}\label{eq:SKA-bi-priv-state-distill-3}%
		\mathcal{L}_{A_{n}^{\prime}B_{n}B_{n}^{\prime}\rightarrow K_{A}S_{A_{n+1}}K_{B}S_{B_{n+1}}}^{(n+1)}=\sum_{y_{n+1}}\mathcal{U}_{A_{n}^{\prime}\rightarrow K_{A}S_{A_{n+1}}}^{\mathcal{E}^{y_{n+1}}}\otimes\mathcal{U}_{B_{n}B_{n}^{\prime}\rightarrow K_{B}S_{B_{n+1}}}^{\mathcal{F}^{y_{n+1}}},
	\end{equation}
	with a similar convention as in \eqref{eq:SKA-reduced-maps-purified-1}--\eqref{eq:SKA-reduced-maps-purified-2} for the maps $\mathcal{U}_{A_{n}^{\prime}\rightarrow K_{A}S_{A_{n+1}}}^{\mathcal{E}^{y_{n+1}}}$ and $\mathcal{U}_{B_{n}B_{n}^{\prime}\rightarrow K_{B}S_{B_{n+1}}}^{\mathcal{F}^{y_{n+1}}}$.

	The states at every step of the protocol are then given by the following for all $i\in\{2,\ldots,n\}$:%
	\begin{align}
		\rho_{A_{i}^{\prime}A_{i}S_{A_{1}^{i}}B_{i}^{\prime}S_{B_{1}^{i}}}^{(i)}  & \coloneqq\mathcal{L}_{A_{i-1}^{\prime}B_{i-1}B_{i-1}^{\prime}\rightarrow A_{i}^{\prime}A_{i}S_{A_{i}}B_{i}^{\prime}S_{B_{i}}}^{(i)}(\omega_{A_{i-1}^{\prime}S_{A_{1}^{i-1}}B_{i-1}B_{i-1}^{\prime}S_{B_{1}^{i-1}}}^{(i-1)}),\\
		\omega_{A_{i}^{\prime}S_{A_{1}^{i}}B_{i}B_{i}^{\prime}S_{B_{1}^{i}}}^{(i)}  & \coloneqq\mathcal{N}_{A_{i}\rightarrow B_{i}}(\rho_{A_{i}^{\prime}A_{i}S_{A_{1}^{i}}B_{i}^{\prime}S_{B_{1}^{i}}}^{(i)}),
	\end{align}
	and the final state of the protocol is given by%
	\begin{equation}
		\omega_{K_{A}S_{A}K_{B}S_{B}}\coloneqq\mathcal{L}_{A_{n}^{\prime}B_{n}B_{n}^{\prime}\rightarrow K_{A}S_{A_{n+1}}K_{B}S_{B_{n+1}}}^{(n+1)}(\omega_{A_{n}^{\prime}B_{n}S_{A_{1}^{n}}B_{n}^{\prime}S_{B_{1}^{n}}}^{(n)}).
		\label{eq-SKA:final-state-equiv-bipa-priv-st-prot}
	\end{equation}

	Finally, by employing \eqref{eq:SKA-error-criterion} and Proposition~\ref{prop-approx_key_state}, the following condition holds%
	\begin{equation}
		p_{\text{err}}(\mathcal{C}) = 1- F(\omega_{K_{A}S_{A}K_{B}S_{B}},\gamma _{K_{A}S_{A}K_{B}S_{B}}),
	\end{equation}
	where $\gamma_{K_{A}S_{A}K_{B}S_{B}}$ is a bipartite private state of the form in Theorem~\ref{thm-private_state}. Thus, applying Definition~\ref{def:SKA-LOPC-assist-code}, it follows that
	\begin{equation}\label{eq:SKA-error-crit-bipartite-PS-dist}%
		F(\omega_{K_{A}S_{A}K_{B}S_{B}},\gamma_{K_{A}S_{A}K_{B}S_{B}}) \geq1-\varepsilon. 
	\end{equation}

	We can now make a critical observation. By tracing over Eve's systems $E^{n}$ and $Y^{n+1}$ at every step of the purified protocol as we did above, it is clear that the resulting protocol is an LOCC-assisted protocol that distills an approximate bipartite private state on the systems $K_{A}S_{A}K_{B}S_{B}$, with performance parameter given by \eqref{eq:SKA-error-crit-bipartite-PS-dist}. Indeed, the initial state in \eqref{eq:SKA-bi-priv-state-distill-1} is a separable state, and the channels in \eqref{eq:SKA-bi-priv-state-distill-2} and \eqref{eq:SKA-bi-priv-state-distill-3} are LOCC channels, and so the protocol has the same form as an LOCC-assisted quantum communication protocol, as we studied in the previous chapter. However, the goal of this LOCC-assisted bipartite private-state distillation is not as stringent as it was in the previous chapter, for LOCC-assisted quantum communication. Namely, it is only necessary to distill an approximate bipartite private state and not necessarily an approximate maximally entangled state; but keep in mind that a maximally entangled state is a particular kind of bipartite private state. Thus, starting with a tripartite secret-key-agreement protocol, we can apply the purification principle, then trace over the systems of the eavesdropper, and the result is a bipartite private-state distillation protocol assisted by~LOCC.

	Alternatively, this reasoning can go in the opposite direction. Suppose instead that we had started with an LOCC-assisted bipartite private-state distillation protocol of the above form. Then we could purify it as we did in the previous subsection, and after doing so, we could trace over Alice and Bob's shield systems. If the protocol satisfies the condition in \eqref{eq:SKA-error-crit-bipartite-PS-dist}, then the resulting protocol would be an $(n,K,\varepsilon)$ tripartite secret-key-agreement protocol, which follows as a consequence of the equivalence between approximate bipartite private states and tripartite secret-key states, as given in Proposition~\ref{prop-approx_key_state}.

	As a consequence of this equivalence between tripartite secret-key-agreement protocols and bipartite private-state distillation protocols, we can employ the tools of entanglement theory in order to analyze bipartite private-state distillation protocols, in a way similar to how we did in the last chapter. For example, if our goal is to determine upper bounds on the rate at which it is possible to generate secret key in a secret-key-agreement protocol, then we can employ an entanglement measure to analyze the equivalent bipartite private-state distillation protocol in order to accomplish the goal. In fact, this is exactly what we accomplish in this chapter, demonstrating that the squashed entanglement of a channel serves as an upper bound on secret-key rates, and that variations of the relative entropy of entanglement, similar in spirit to the Rains relative entropy, serve as upper bounds on secret-key rates as well.

\subsubsection{Unboundedness of Shield Systems in a Bipartite Private-State Distillation Protocol}

	One observation that we make here is that the shield systems in a bipartite private-state distillation protocol are finite-dimensional, yet arbitrarily large. That is, there is no bound that we can establish on their dimension for a generic private-state distillation protocol, and this unboundedness is a consequence of the fact that the shield systems result from purifying the local memory or scratch registers of Alice and Bob, which in turn have no bound on their dimension. This unboundedness poses a challenge when trying to establish upper bounds on the rate at which secret key agreement, or equivalently, bipartite private-state distillation is possible. However, there are methods for handling this unboundedness that we detail later.

\subsection{Relation between Secret Key Agreement and LOCC-Assisted Quantum Communication}

	Due to the fact that a maximally entangled state is a particular kind of bipartite private state and due to the equivalence between secret key agreement and LOCC-assisted bipartite private-state distillation, we arrive at the following conclusion, which relates LOCC-assisted quantum communication to secret key agreement:

	\begin{proposition}{prop-LOCC_q_comm_to_SKA}
		Let $\mathcal{N}_{A\rightarrow B}$ be a quantum channel, let $n,K\in\mathbb{N}$, and let $\varepsilon\in\left[  0,1\right]  $. Then an $(n,K,\varepsilon)$ LOCC-assisted quantum communication protocol is also an $(n,K,\varepsilon)$ protocol for secret key agreement.
	\end{proposition}

	This statement is a rather simple observation, but it has consequences for capacities related to these tasks. That is, from this statement, we can conclude that the LOCC-assisted quantum capacity of a given quantum channel is bounded from above by its secret-key-agreement capacity. Thus, any upper bound established on the secret-key-agreement capacity of a quantum channel is also an upper bound on its LOCC-assisted quantum capacity. Furthermore, if a given quantity is a lower bound on the LOCC-assisted quantum capacity of a quantum channel, then it is also a lower bound on its secret-key-agreement capacity.

\subsection{$n$-Shot Secret-Key-Agreement Protocol Assisted by Public Separable Channels}

	\label{sec-SKA:SKA-assisted-pub-sep-ch}
	
	In Section~\ref{sec-LOCC-QC:PPT-assist-prot}, we generalized the notion of an LOCC-assisted quantum communication protocol to one that is assisted by PPT-preserving channels. We note here that we can consider a similar kind of generalization for secret-key-agreement protocols.

	Recall from Section~\ref{subsec-LOCC_channels} that any LOCC channel $\mathcal{L}_{AB}$\ can be written as a separable channel of the following form:%
	\begin{equation}
		\mathcal{L}_{AB\rightarrow A^{\prime}B^{\prime}}=\sum_{y}\mathcal{E}_{A\rightarrow A^{\prime}}^{y}\otimes\mathcal{F}_{B\rightarrow B^{\prime}}^{y},
	\end{equation}
	where $\{\mathcal{E}_{A\rightarrow A^{\prime}}^{y}\}_{y}$ and$\ \{\mathcal{F}_{B\rightarrow B^{\prime}}^{y}\}_{y}$ are sets of completely positive maps such that the sum map $\sum_{y}\mathcal{E}_{A}^{y}\otimes\mathcal{F}_{B}^{y}$ is trace preserving. However, the converse statement is not true. That is, it is not possible in general to implement an arbitrary separable channel of the form above as an LOCC\ channel.

	Thus, we can allow for a slight generalization of a secret-key-agreement protocol to one that is assisted by public separable channels. Indeed, we define a public separable channel to be the following generalization of an LOPC\ channel:
	\begin{equation}
		\mathcal{L}_{AB\rightarrow A^{\prime}B^{\prime}Y}=\sum_{y}\mathcal{E}_{A\rightarrow A^{\prime}}^{y}\otimes\mathcal{F}_{B\rightarrow B^{\prime}}^{y}\otimes|y\rangle\!\langle y|_{Y},
	\end{equation}
	where Alice and Bob have access to the $A$ and $B$ systems, respectively, and the eavesdropper has access to the system $Y$. The only requirement for a public separable channel is that $\{\mathcal{E}_{A\rightarrow A^{\prime}}^{y}\}_{y}$ and $\{\mathcal{F}_{B\rightarrow B^{\prime}}^{y}\}_{y}$ are sets of completely positive maps such that the sum map $\sum_{y}\mathcal{E}_{A}^{y}\otimes\mathcal{F}_{B}^{y}$ is trace preserving. Similar to the distinction between LOCC and separable channels, it is not possible in general to implement a public separable channel via local operations and public classical communication.

The main point that we make in this section is that we can generalize a secret-key-agreement protocol to be assisted by public separable channels rather than just LOPC\ channels. For fixed privacy error, the resulting protocol achieves a rate of communication that is either the same or higher than that achieved by an LOPC-assisted protocol, due to the fact that every LOPC channel is a public separable channel. Such a protocol is defined in the same way as we did in Section~\ref{sec:SKA-n-shot-SKA-prot}, and then we arrive at the following definition:

	\begin{definition}{$(n,K,\varepsilon)$ Secret-Key-Agreement Protocol\ Assisted by Public Separable Channels}{def:SKA-pub-sep-assist-code}
		Let $\mathcal{C}\coloneqq (\rho_{A_{1}^{\prime}A_{1}B_{1}^{\prime}Y_{1}}^{(1)},\{\mathcal{L}_{A_{i-1}^{\prime}B_{i-1}B_{i-1}^{\prime}\rightarrow A_{i}^{\prime}A_{i}B_{i}^{\prime}Y_{i}}^{(i)}\}_{i=2}^{n},\mathcal{L}_{A_{n}^{\prime}B_{n}B_{n}^{\prime}\rightarrow K_{A}K_{B}Y_{n+1}}^{(n+1)})$ be the elements of an $n$-round public-separable-assisted secret-key-agreement protocol over the channel $\mathcal{N}_{A\rightarrow B}$. The protocol is called an $(n,K,\varepsilon)$ protocol, with $\varepsilon\in\left[  0,1\right]$, if the privacy error $p_{\operatorname{err}}(\mathcal{C})\leq\varepsilon$.
	\end{definition}

	Furthermore, the equivalence between secret key agreement and bipartite private-state distillation, as outlined in Sections~\ref{sec:SKA-purified-protocol} and \ref{sec:SKA-bipartite-PS-dist}, still holds under this generalization (one can check that all of the steps given in Sections~\ref{sec:SKA-purified-protocol} and \ref{sec:SKA-bipartite-PS-dist}\ still hold). However, the correspondence changes as follows:\ to any tripartite $(n,K,\varepsilon)$ secret-key-agreement protocol assisted by public separable channels, there exists an $(n,K,\varepsilon)$ bipartite private-state distillation protocol assisted by separable channels and vice versa. Thus, we can again employ the tools of entanglement theory to analyze secret-key-agreement protocols assisted by public separable channels.

\subsection{Amortized Entanglement Bound for Secret-Key-Agreement Protocols}

	Due to the equivalence between tripartite secret-key-agreement protocols and bipartite private-state distillation protocols, we can use the tools of entanglement theory to establish upper bounds on the rate at which secret-key-agreement is possible. Namely, we can apply the idea of amortized entanglement from the previous chapter in order to establish a generic upper bound in terms of an amortized entanglement measure. In fact, by the same steps used to arrive at Proposition~\ref{prop:LAQC-meta-converse-amortized},
we find the following bound:

	\begin{proposition}{prop:SKA-amortized-bound}
		Let $\mathcal{N}_{A\rightarrow B}$ be a quantum channel, let $\varepsilon\in\left[  0,1\right]  $, and let $E$ be an entanglement measure that is equal to zero for all separable states. For an $(n,K,\varepsilon)$ secret-key-agreement protocol, the following bound holds%
		\begin{equation}\label{eq:SKA-amortized-bnd-SKA}%
			E(K_{A}S_{A};K_{B}S_{B})_{\omega}\leq n\cdot E^{\mathcal{A}}(\mathcal{N}),
		\end{equation}
		where $\omega_{K_{A}S_{A}K_{B}S_{B}}$ is the final state resulting from the equivalent $(n,K,\varepsilon)$ bipartite private-state distillation protocol (see \eqref{eq-SKA:final-state-equiv-bipa-priv-st-prot}) and $E^{\mathcal{A}}(\mathcal{N})$ is the amortized entanglement of the channel $\mathcal{N}_{A\rightarrow B}$, as given in Definition~\ref{def:LAQC-amortized-ent}.
	\end{proposition}

	Just as the bound from Proposition~\ref{prop:LAQC-meta-converse-amortized} depends on the final state of the LOCC-assisted quantum communication protocol, the same is true for the bound in \eqref{eq:SKA-amortized-bnd-SKA}. The bound is thus not a universal bound (a universal bound would depend only on the protocol parameters $n$, $K$, and $\varepsilon$). Thus, one of the main goals of the forthcoming sections is to employ particular entanglement measures in order to arrive at universal bounds for secret-key-agreement protocols.

	We should also observe that the quantity $E(K_{A}S_{A};K_{B}S_{B})_{\omega}$ in the bound in \eqref{eq:SKA-amortized-bnd-SKA} can be understood as quantifying amount of entanglement between the systems $K_{A}S_{A}$ and $K_{B}S_{B}$. As such, the shield systems $S_{A}$ and $S_{B}$ are involved, and they can have arbitrarily large dimension. Thus, one must account for this in the analysis of the entanglement $E(K_{A}S_{A};K_{B}S_{B})_{\omega}$. For example, in the previous chapter, we analyzed the analogous quantity $E(M_{A};M_{B})_{\omega}$ by employing squashed entanglement. In particular, since the state $\omega_{M_{A}M_{B}}$ there was an approximate maximally entangled state, we applied the asymptotic continuity of squashed entanglement from Proposition~\ref{prop:LAQC-cont-sq-unif} in order to evaluate $E(M_{A};M_{B})_{\omega}$. When we did so, the dimension of the maximally entangled state appeared in the continuity bound, and this was acceptable there because the dimension of the maximally entangled state is directly related to the rate of entanglement distillation. However, it is not clear whether we can take such an approach, via asymptotic continuity of squashed entanglement, when analyzing bipartite private-state
distillation, due to the fact that the shield systems do not necessarily have a bounded dimension. As such, we employ another method to analyze such protocols.

	Just as the bound in Proposition~\ref{prop:LAQC-meta-converse-amortized} simplifies for teleportation-simulable channels and particular entanglement measures, the same is true for the bound given in Proposition~\ref{prop:SKA-amortized-bound}, by employing the same reasoning:

	\begin{corollary*}{Reduction by Teleportation}{cor-SKA:reduc-by-TP}
		Let $E_{S}$ denote an entanglement measure that is
subadditive with respect to states (Definition~\ref{def-ent_meas_subadditive}) and equal to zero for all separable states. Let $\mathcal{N}_{A\rightarrow B}$ be a channel that is LOCC-simulable with associated resource state $\theta_{RB^{\prime}}$ (Definition~\ref{def-LOCC_sim_chan}). Let $\varepsilon\in\left[0,1\right]$. For an $(n,K,\varepsilon)$ secret-key-agreement protocol, the following bound holds%
		\begin{equation}
			E_S(K_{A}S_{A};K_{B}S_{B})_{\omega}\leq n\cdot E_{S}(R;B^{\prime})_{\theta},
		\end{equation}
		where $\omega_{K_{A}S_{A}K_{B}S_{B}}$ is the final state resulting from the equivalent $(n,K,\varepsilon)$ bipartite private-state distillation protocol.
	\end{corollary*}

	Just as we found bounds that apply to PPT-assisted quantum communication in terms of entanglement measures that are monotone with respect to PPT-preserving channels, we can also find bounds that apply to secret-key-agreement protocols that are assisted by public separable channels:

	\begin{proposition}{prop-SKA_ent_meas_to_amort_ent}
		Let $\mathcal{N}_{A\rightarrow B}$ be a quantum channel, let $\varepsilon \in\left[  0,1\right]  $, and let $E$ be an entanglement measure that is monotone non-increasing with respect to separable channels and equal to zero for all separable states. For an $(n,K,\varepsilon)$ secret-key-agreement protocol assisted by public separable channels, the following bound holds%
	\begin{equation}
		E(K_{A}S_{A};K_{B}S_{B})_{\omega}\leq n\cdot E^{\mathcal{A}}(\mathcal{N}),
	\end{equation}
	where $\omega_{K_{A}S_{A}K_{B}S_{B}}$ is the final state resulting from the equivalent $(n,K,\varepsilon)$ bipartite private-state distillation protocol assisted by separable channels and $E^{\mathcal{A}}(\mathcal{N})$ is the amortized entanglement of the channel $\mathcal{N}_{A\rightarrow B}$, as given in Definition~\ref{def:LAQC-amortized-ent}.
	\end{proposition}

	Finally, this bound again simplifies for channels that are simulable by the action of a separable channel on a resource state $\theta_{RB^{\prime}}$ (separable-simulable channels):

	\begin{corollary}{cor:SKA-SEP-simulable-bnd-SK}
		Let $E_{S}$ denote an entanglement measure
that is that is monotone non-increasing with respect to separable channels, subadditive with respect to states (Definition~\ref{def-ent_meas_subadditive}), and equal to zero for separable states. Let $\mathcal{N}_{A\rightarrow B}$ be a channel that is separable-simulable with associated resource state $\theta_{RB^{\prime}}$ (Definition~\ref{def-SEP_sim_chan}). Let $\varepsilon\in\left[  0,1\right]  $. For an $(n,K,\varepsilon)$ secret-key-agreement protocol assisted by public separable channels, the following bound holds%
		\begin{equation}
			E_S(K_{A}S_{A};K_{B}S_{B})_{\omega}\leq n\cdot E_{S}(R;B^{\prime})_{\theta},
		\end{equation}
		where $\omega_{K_{A}S_{A}K_{B}S_{B}}$ is the final state resulting from the equivalent $(n,K,\varepsilon)$ bipartite private-state distillation protocol assisted by separable channels.
	\end{corollary}

\section{Squashed Entanglement Upper Bound on the Number of Transmitted Private Bits}

	We now employ the squashed entanglement in order to bound the number of private bits that an $n$-shot secret-key-agreement protocol can generate. We have already shown in Section~\ref{sec-LAQC:sq-ent-and-props} that the squashed entanglement satisfies all of the requirements needed to apply it in Proposition~\ref{prop:SKA-amortized-bound}. Namely, it is equal to zero for separable states, it is an entanglement measure (non-increasing under the action of an LOCC channel), and the squashed entanglement of a channel does not increase under amortization (Theorem~\ref{thm:LAQC-amort-collapse-squashed}). Putting all of these items together, we can already conclude the following bound for an $(n,K,\varepsilon)$ secret-key-agreement protocol:%
	\begin{equation}
		E_{\operatorname{sq}}(K_{A}S_{A};K_{B}S_{B})_{\omega}\leq n\cdot E_{\operatorname{sq}}(\mathcal{N}),
	\end{equation}
	where $E_{\operatorname{sq}}(\mathcal{N})$ is the squashed entanglement of a channel (Definition~\ref{def:LAQC-ent-channel}) and $\omega_{K_{A}S_{A}K_{B}S_{B}}$ is the final state resulting from the equivalent $(n,K,\varepsilon)$ bipartite private-state distillation protocol. Thus, what remains is to evaluate the squashed entanglement of an approximate bipartite private state, and this is the main technical problem that we consider in this section before concluding that squashed entanglement is an upper bound on secret-key rates.

\subsection{Squashed Entanglement and Approximate Private States}

This subsection establishes Proposition~\ref{thm:SKA-bipartite-bound}, which is an upper bound on the logarithm of the dimension $K$ of a key system of an $\varepsilon$-approximate private state, as given in Definition~\ref{def:approx-priv}, in terms of its squashed entanglement, plus another term depending only on $\varepsilon$ and $\log_{2}K$.

	In what follows, we suppose that $\gamma_{AA^{\prime}BB^{\prime}}$ is a private state with key systems $AB$ and shield systems $A^{\prime}B^{\prime}$. Recall from Theorem~\ref{thm-private_state} that a private state of $\log_{2}K$ private bits can be written in the following form:
	\begin{equation}\label{eq:SKA-private-1}%
		\gamma_{ABA^{\prime}B^{\prime}}=U_{ABA^{\prime}B^{\prime}}\left(  \Phi_{AB}\otimes\sigma_{A^{\prime}B^{\prime}}\right)  U_{ABA^{\prime}B^{\prime}}^{\dagger}, 
	\end{equation}
	where $\Phi_{AB}$ is a maximally entangled state of Schmidt rank $K$
	\begin{equation}\label{eq:SKA-max-ent-state}%
		\Phi_{AB}\coloneqq\frac{1}{K}\sum_{i,j}|i\rangle\!\langle j|_{A}\otimes|i\rangle\!\langle j|_{B}, 
	\end{equation}
	and%
	\begin{equation}
		U_{ABA^{\prime}B^{\prime}}=\sum_{i,j}|i\rangle\!\langle i|_{A}\otimes|j\rangle\!\langle j|_{B}\otimes U_{A^{\prime}B^{\prime}}^{ij}%
	\end{equation}
	is a controlled unitary known as a \textquotedblleft twisting
unitary,\textquotedblright\ with each $U_{A^{\prime}B^{\prime}}^{ij}$ a unitary operator.\ Due to the fact that the maximally entangled state $\Phi_{AB}$ is unextendible, any extension $\gamma_{AA^{\prime}BB^{\prime}E}$ of a private state $\gamma_{AA^{\prime}BB^{\prime}}$ necessarily has the following form:%
	\begin{equation}\label{eq:SKA-ext-private-state}
		\gamma_{AA^{\prime}BB^{\prime}E}=U_{AA^{\prime}BB^{\prime}}\left(  \Phi_{AB}\otimes\sigma_{A^{\prime}B^{\prime}E}\right)  U_{AA^{\prime}BB^{\prime}}^{\dag},
	\end{equation}
	where $\sigma_{A^{\prime}B^{\prime}E}$ is an extension of $\sigma_{A^{\prime}B^{\prime}}$.

	We start with the following lemma, which applies to any extension of a bipartite private state:

	\begin{Lemma}{lem:SKA-log-K-to-info-measures}
		Let $\gamma_{AA^{\prime}BB^{\prime}}$ be a bipartite private state, and let $\gamma_{AA^{\prime}BB^{\prime}E}$ be an extension of it, as given above. Then the following identity holds for any such extension:%
		\begin{equation}\label{eq:SKA-logK-to-info-measures}%
			2\log_{2}K=I(A;BB^{\prime}|E)_{\gamma}+I(A^{\prime};B|AB^{\prime}E)_{\gamma}.
		\end{equation}
	\end{Lemma}

	\begin{Proof}
		First consider that the following identity holds as a consequence of two applications of the chain rule for conditional quantum mutual information:%
		\begin{align}
			I(AA^{\prime};BB^{\prime}|E)_{\gamma}  &  =I(A;BB^{\prime}|E)_{\gamma}+I(A^{\prime};BB^{\prime}|AE)_{\gamma}\nonumber\\
			&=I(A;BB^{\prime}|E)_{\gamma}+I(A^{\prime};B^{\prime}|AE)_{\gamma}+I(A^{\prime};B|B^{\prime}AE)_{\gamma}. \label{eq:SKA-chain-rule-CMI}%
		\end{align}
		Combined with the following identity, which holds for an extension $\gamma_{AA^{\prime}BB^{\prime}E}$ of a private state $\gamma_{AA^{\prime}BB^{\prime}}$,%
		\begin{equation}\label{eq:SKA-christandl-thesis}%
			I(AA^{\prime};BB^{\prime}|E)_{\gamma}=2\log_{2}K+I(A^{\prime};B^{\prime
}|AE)_{\gamma}, 
		\end{equation}
		we recover the statement in \eqref{eq:SKA-logK-to-info-measures}. So it remains to prove \eqref{eq:SKA-christandl-thesis}. By definition, we have that%
		\begin{equation}\label{eq:SKA-def-CMI-for-PS}%
			I(AA^{\prime};BB^{\prime}|E)_{\gamma}=H(AA^{\prime}E)_{\gamma}+H(BB^{\prime}E)_{\gamma}-H(E)_{\gamma}-H(AA^{\prime}BB^{\prime}E)_{\gamma}.
		\end{equation}
		By applying \eqref{eq:SKA-max-ent-state}--\eqref{eq:SKA-ext-private-state}, we can write $\gamma_{AA^{\prime}BB^{\prime}E}$ as follows:%
		\begin{equation}
			\gamma_{AA^{\prime}BB^{\prime}E}=\frac{1}{K}\sum_{i,j}|i\rangle\!\langle j|_{A}\otimes|i\rangle\!\langle j|_{B}\otimes U_{A^{\prime}B^{\prime}}^{ii}\sigma_{A^{\prime}B^{\prime}E}(U_{A^{\prime}B^{\prime}}^{jj})^{\dagger}.
		\end{equation}
		Tracing over system $B$ leads to the following state:%
		\begin{equation}\label{eq:SKA-trace-out-B}%
			\gamma_{AA^{\prime}B^{\prime}E}=\frac{1}{K}\sum_{i}|i\rangle\!\langle i|_{A}\otimes\gamma_{A^{\prime}B^{\prime}E}^{i}, 
		\end{equation}
		where%
		\begin{equation}
			\gamma_{A^{\prime}B^{\prime}E}^{i}\coloneqq U_{A^{\prime}B^{\prime}}^{ii}\sigma_{A^{\prime}B^{\prime}E}(U_{A^{\prime}B^{\prime}}^{ii})^{\dag}.
		\end{equation}
		Similarly, tracing over system $A$ of $\gamma_{AA^{\prime}BB^{\prime}E}$ leads to%
		\begin{equation}\label{eq:SKA-trace-out-A}%
			\gamma_{BA^{\prime}B^{\prime}E}=\frac{1}{K}\sum_{i}|i\rangle\!\langle i|_{B}\otimes\gamma_{A^{\prime}B^{\prime}E}^{i}. 
		\end{equation}
		So these and the chain rule for conditional entropy imply that%
		\begin{equation}\label{eq:SKA-approx-priv-squashed-step-1}%
			H(AA^{\prime}E)_{\gamma}=H(A)_{\gamma}+H(A^{\prime}E|A)_{\gamma}=\log_{2}K+H(A^{\prime}E|A)_{\gamma}. 
		\end{equation}
		Similarly, we have that%
		\begin{equation}\label{eq:SKA-approx-priv-squashed-step-2}%
			H(BB^{\prime}E)_{\gamma}=\log_{2}K+H(B^{\prime}E|B)_{\gamma}=\log_{2}K+H(B^{\prime}E|A)_{\gamma}, 
		\end{equation}
		where we have used the symmetries in \eqref{eq:SKA-trace-out-B}--\eqref{eq:SKA-trace-out-A}. Since $\gamma_{E}=\gamma_{E}^{i}$ for all $i$ (this is a consequence of $\gamma_{ABA^{\prime}B^{\prime}}$ being an ideal private state), we find that%
		\begin{equation}\label{eq:SKA-approx-priv-squashed-step-3}%
			H(E)_{\gamma}=\frac{1}{K}\sum_{i}H(E)_{\gamma^{i}}=H(E|A)_{\gamma}.
		\end{equation}
		Finally, we have that%
		\begin{align}
			H(AA^{\prime}BB^{\prime}E)_{\gamma}  &  =H(ABA^{\prime}B^{\prime}E)_{\Phi\otimes\sigma}\\
			&  =H(AB)_{\Phi}+H(A^{\prime}B^{\prime}E)_{\sigma}\\
			&  =\frac{1}{K}\sum_{i}H(A^{\prime}B^{\prime}E)_{\gamma^{i}}\\
			&  =H(A^{\prime}B^{\prime}E|A)_{\gamma}. \label{eq:SKA-approx-priv-squashed-step-4}%
		\end{align}
		The first equality follows from unitary invariance of quantum entropy. The second equality follows because the entropy is additive for tensor-product states. The third equality follows because $H(AB)_{\Phi} = 0$ since $\Phi_{AB}$ is a pure state, and $\sigma_{A^{\prime}B^{\prime}E}$ is related to $\gamma^{i}_{A^{\prime}B^{\prime}E}$ by the unitary $U^{ii}_{A^{\prime
}B^{\prime}}$. The final equality follows by applying \eqref{eq:SKA-trace-out-B}, and the fact that conditional entropy is a convex combination of entropies for a classical-quantum state where the conditioning system is classical.

		Combining \eqref{eq:SKA-def-CMI-for-PS}, \eqref{eq:SKA-approx-priv-squashed-step-1}, \eqref{eq:SKA-approx-priv-squashed-step-2}, \eqref{eq:SKA-approx-priv-squashed-step-3}, \eqref{eq:SKA-approx-priv-squashed-step-4}, and the fact that
		\begin{equation}
			I(A^{\prime};B^{\prime}|AE)_{\gamma} = H(A^{\prime}E|A)_{\gamma} + H(B^{\prime}E|A)_{\gamma} - H(E|A)_{\gamma} -H(A^{\prime}B^{\prime}E|A)_{\gamma},
		\end{equation}
		we recover \eqref{eq:SKA-christandl-thesis}.
	\end{Proof}

	We can now establish the squashed entanglement bound for an approximate bipartite private state:

	\begin{proposition}{thm:SKA-bipartite-bound}
		Let $\gamma_{AA^{\prime}BB^{\prime}}$ be a private state, with key systems $AB$ and shield systems $A'B'$, and let $\omega_{AA^{\prime}BB^{\prime}}$ be an $\varepsilon$-approximate private state, in the sense that%
		\begin{equation}
			F(\gamma_{AA^{\prime}BB^{\prime}},\omega_{AA^{\prime}BB^{\prime}})\geq1-\varepsilon
		\end{equation}
		for $\varepsilon\in\left[  0,1\right]  $. Suppose that $|A|=|B|=K$. Then%
		\begin{equation}
			(1 - 2 \sqrt{\varepsilon}) \log_{2}K\leq E_{\operatorname{sq}}(AA^{\prime};BB^{\prime})_{\omega}+
			2g_{2}(\sqrt{\varepsilon}),
		\end{equation}
		where%
		\begin{equation}
			g_{2}(\delta)  \coloneqq\left(  \delta+1\right)  \log_{2}(\delta+1)-\delta\log_{2}\delta.
		\end{equation}
	\end{proposition}

	\begin{Proof}
		By applying Uhlmann's theorem for fidelity (Theorem~\ref{thm-Uhlmann_fidelity}) and the inequalities relating trace distance and fidelity from Theorem~\ref{thm-Fuchs_van_de_graaf}, for a given extension $\omega_{AA^{\prime}BB^{\prime}E}$ of $\omega_{AA^{\prime}BB^{\prime}}$, there exists an extension $\gamma_{AA^{\prime}BB^{\prime}E}$ of $\gamma_{AA^{\prime}BB^{\prime}}$ such that%
		\begin{equation}
			\frac{1}{2}\left\Vert \gamma_{AA^{\prime}BB^{\prime}E}-\omega_{AA^{\prime}BB^{\prime}E}\right\Vert_{1}\leq\sqrt{\varepsilon}.
		\end{equation}
		Defining $f_1(\delta,K)\coloneqq 2 \delta \log_2 K + 2 g_2(\delta)$,
		we then find that%
		\begin{align}
			2\log_{2}K  &  =I(A;BB^{\prime}|E)_{\gamma}+I(A^{\prime};B|AB^{\prime}E)_{\gamma}\\
			&  \leq I(A;BB^{\prime}|E)_{\omega}+I(A^{\prime};B|AB^{\prime}E)_{\omega}+2f_{1}(\sqrt{\varepsilon},K)\\
			&  \leq I(A;BB^{\prime}|E)_{\omega}+I(A^{\prime};B|AB^{\prime}E)_{\omega}\nonumber\\
			&  \qquad\qquad+I(A^{\prime};B^{\prime}|AE)_{\omega}+2f_{1}(\sqrt{\varepsilon},K)\\
			&  =I(AA^{\prime};BB^{\prime}|E)_{\omega}+2f_{1}(\sqrt{\varepsilon},K).
		\end{align}
		The first equality follows from Lemma~\ref{lem:SKA-log-K-to-info-measures}. The first inequality follows from two applications of Proposition~\ref{lem:LAQC-uniform-cont-CMI} (asymptotic continuity of conditional mutual information). The second inequality follows because $I(A^{\prime};B^{\prime}|AE)_{\omega}\geq0$ (this is strong
subadditivity from Theorem~\ref{thm-SSA}). The last equality is a consequence of the chain rule for conditional mutual information, as used in \eqref{eq:SKA-chain-rule-CMI}. Since the inequality%
		\begin{equation}
			2\log_{2}K\leq I(AA^{\prime};BB^{\prime}|E)_{\omega}+2f_{1}(\sqrt{\varepsilon},K)
		\end{equation}
		holds for any extension of $\omega_{AA^{\prime}BB^{\prime}}$, the statement of the proposition follows.
	\end{Proof}

\subsection{Squashed Entanglement Upper Bound}\label{sec:SKA-sq-ent-bnd-non-as}

	We now establish the squashed entanglement upper bound on the number of private bits that a sender can transmit to a receiver by employing a secret-key-agreement protocol. The proof is similar to that of Theorem~\ref{thm:LAQC-sq-ent-bnd-LOCC-as-cap}, but it instead invokes Proposition~\ref{thm:SKA-bipartite-bound}.

	\begin{theorem*}{$n$-Shot Squashed Entanglement Upper Bound}{thm:SKA-sq-ent-bnd-LOCC-as-cap}
		Let $\mathcal{N}_{A\rightarrow B}$ be a quantum channel, and let $\varepsilon\in\lbrack0,1/4)$. For all $(n,K,\varepsilon)$ secret-key-agreement  protocols over the channel $\mathcal{N}_{A\rightarrow B}$, the following bound holds%
		\begin{equation}
			\log_{2}M\leq\frac{1}{1-2\sqrt{\varepsilon}}\left[  n\cdot E_{\operatorname{sq}}(\mathcal{N})+2g_{2}(\sqrt{\varepsilon})\right]  .
		\end{equation}
	\end{theorem*}

	\begin{Proof}
		Given an arbitrary $(n,K,\varepsilon)$ secret-key-agreement protocol as outlined in Section~\ref{sec:SKA-n-shot-SKA-prot}, we consider its equivalent $(n,K,\varepsilon)$ LOCC-assisted bipartite private-state distillation protocol, as outlined in Section~\ref{sec:SKA-bipartite-PS-dist}. The squashed entanglement is an entanglement measure (monotone under LOCC\ as shown in Theorem~\ref{thm:LAQC-mono-LOCC-sq}) and it is equal to zero for separable states (Proposition~\ref{prop:LAQC-squashed-sep}). Thus, Proposition~\ref{prop:SKA-amortized-bound}\ applies, and we find that%
		\begin{equation}\label{eq:SKA-sq-ent-bnd-apply-1}
			E_{\operatorname{sq}}(K_{A}S_{A};K_{B}S_{B})_{\omega}\leq n\cdot E_{\operatorname{sq}}^{\mathcal{A}}(\mathcal{N})=n\cdot E_{\operatorname{sq}}(\mathcal{N}),
		\end{equation}
		where the equality follows from Theorem~\ref{thm:LAQC-amort-collapse-squashed}. Applying Definition~\ref{def:SKA-LOPC-assist-code} and \eqref{eq:SKA-error-crit-bipartite-PS-dist} leads to%
		\begin{equation}
			F(\gamma_{K_{A}S_{A}K_{B}S_{B}},\omega_{K_{A}S_{A}K_{B}S_{B}})\geq 1-\varepsilon,
		\end{equation}
		where $\gamma_{K_{A}S_{A}K_{B}S_{B}}$ is an exact private state of $\log_{2}K$ private bits. As a consequence of Proposition~\ref{thm:SKA-bipartite-bound}, we find that%
		\begin{equation}\label{eq:SKA-sq-ent-bnd-apply-2}%
			E_{\operatorname{sq}}(K_{A}S_{A};K_{B}S_{B})_{\omega}\geq(1-2\sqrt{\varepsilon})\log_{2}K-2g_{2}(\sqrt{\varepsilon}%
).
		\end{equation}
		Putting together \eqref{eq:SKA-sq-ent-bnd-apply-1} and
\eqref{eq:SKA-sq-ent-bnd-apply-2}, we arrive at the statement of the theorem.
	\end{Proof}


\section{Relative Entropy of Entanglement Upper Bounds on the Number of Transmitted Private Bits}\label{sec:SKA-bnd-REE}

	We now establish the max-relative entropy of
entanglement bound on the number of private bits that a sender can transmit to a receiver by employing a secret-key-agreement protocol assisted by public separable channels:

	\begin{theorem*}{$n$-Shot Max-Relative Entropy of Entanglement Upper Bound}{thm:LAQC-max-REE-bnd-SEP-as-cap}
		Let $\mathcal{N}_{A\rightarrow B}$ be a quantum channel, and let $\varepsilon\in\lbrack0,1)$. For all $(n,K,\varepsilon)$ secret-key-agreement protocols assisted by public separable channels, over the channel $\mathcal{N}_{A\rightarrow B}$, the following bound holds%
		\begin{equation}
			\log_{2}K\leq n\cdot E_{\max}(\mathcal{N})+\log_{2}\!\left(  \frac {1}{1-\varepsilon}\right)  .
		\end{equation}
	\end{theorem*}

	\begin{Proof}
		Given an arbitrary $(n,K,\varepsilon)$ secret-key-agreement protocol assisted by public separable channels as outlined in Section~\ref{sec-SKA:SKA-assisted-pub-sep-ch}, we consider its equivalent $(n,K,\varepsilon)$ separable-assisted bipartite private-state distillation protocol. The max-relative entropy of entanglement is an entanglement measure (monotone under separable channels channels\ as shown in Proposition~\ref{prop-gen_div_ent_properties}) and it is equal to zero for separable states. 
		Thus, Proposition~\ref{prop:SKA-amortized-bound}\ applies, and we find that%
		\begin{equation}\label{eq:SKA-max-REE-bnd-1}%
			E_{\max}(K_{A}S_{A};K_{B}S_{B})_{\omega}\leq n\cdot E_{\max}^{\mathcal{A}}(\mathcal{N})=n\cdot E_{\max}(\mathcal{N}),
		\end{equation}
		where the equality follows from Theorem~\ref{prop:SKA-amort-doesnt-help-e-max}. Applying Definition~\ref{def:SKA-LOPC-assist-code} and \eqref{eq:SKA-error-crit-bipartite-PS-dist} leads to%
		\begin{equation}
			F(\gamma_{K_{A}S_{A}K_{B}S_{B}},\omega_{K_{A}S_{A}K_{B}S_{B}})\geq 1-\varepsilon,
		\end{equation}
		where $\gamma_{K_{A}S_{A}K_{B}S_{B}}$ is an exact private state of $\log_{2}K$ private bits. As a consequence of Propositions~\ref{prop:core-meta-converse-privacy} and \ref{prop:sandwich-to-htre}, we find that%
		\begin{align}
			\log_{2}K &  \leq E_{R}^{\varepsilon}(S_{A}K_{A};S_{B}K_{B})_{\omega}\\
			&  \leq E_{\max}(S_{A}K_{A};S_{B}K_{B})_{\omega}+\log_{2}\!\left(  \frac{1}{1-\varepsilon}\right)  .\label{eq:SKA-max-REE-bnd-2}%
		\end{align}
		Combining \eqref{eq:SKA-max-REE-bnd-1} and \eqref{eq:SKA-max-REE-bnd-2}, we conclude the proof.
	\end{Proof}

	For channels that are separable-simulable with associated resource states, as given in Definition~\ref{def-SEP_sim_chan}, we obtain upper bounds that can be even stronger:

	\begin{theorem*}{$n$-Shot R\'{e}nyi--REE Upper Bounds for Separable-Simulable Channels}{thm:SKA-Renyi-REE-bnd-SEP-sim}
		Let $\mathcal{N}_{A\rightarrow B}$ be a quantum channel that is separable-simulable with associated resource state $\theta_{SB^{\prime}}$, and let $\varepsilon\in\lbrack0,1)$. For all $(n,K,\varepsilon)$ secret-key-agreement protocols assisted by public separable channels, over the channel $\mathcal{N}_{A\rightarrow B}$, the following bounds hold for all $\alpha>1$:%
		\begin{align}
			\log_{2}K &  \leq n\cdot\widetilde{E}_{\alpha}(S;B^{\prime})_{\theta}+\frac{\alpha}{\alpha-1}\log_{2}\!\left(  \frac{1}{1-\varepsilon}\right)  ,\\
			\log_{2}K &  \leq\frac{1}{1-\varepsilon}\left[  n\cdot E_{R}(S;B^{\prime})_{\theta}+h_{2}(\varepsilon)\right]  .
		\end{align}
	\end{theorem*}

	\begin{Proof}
	Given an arbitrary $(n,K,\varepsilon)$ secret-key-agreement protocol assisted by public separable channels as outlined in Section~\ref{sec-SKA:SKA-assisted-pub-sep-ch}, we consider its equivalent $(n,K,\varepsilon)$ separable-assisted bipartite private-state distillation protocol. 
		The R\'{e}nyi relative entropy of entanglement and relative entropy of entanglement are monotone non-increasing under separable channels (Proposition~\ref{prop-gen_div_ent_properties}), equal to zero for separable states, 
		and subadditive with respect to states (Proposition~\ref{prop-gen_div_ent_properties}). As such, Corollary~\ref{cor:SKA-SEP-simulable-bnd-SK}\ applies, and we find for $\alpha>1$ that%
		\begin{align}
			\widetilde{E}_{\alpha}(S_{A}K_{A};S_{B}K_{B})_{\omega} &  \leq n\cdot \widetilde{E}_{\alpha}(S;B^{\prime})_{\theta},\label{eq:SKA-SEP-sim-conv-1}\\
			E_{R}(S_{A}K_{A};S_{B}K_{B})_{\omega} &  \leq n\cdot E_{R}(S;B^{\prime})_{\theta}.\label{eq:SKA-SEP-sim-conv-2}%
		\end{align}
		Applying Definition~\ref{def:SKA-LOPC-assist-code} and
\eqref{eq:SKA-error-crit-bipartite-PS-dist} leads to%
		\begin{equation}
			F(\gamma_{K_{A}S_{A}K_{B}S_{B}},\omega_{K_{A}S_{A}K_{B}S_{B}})\geq 1-\varepsilon,
		\end{equation}
		where $\gamma_{K_{A}S_{A}K_{B}S_{B}}$ is an exact private state of $\log_{2}K$ private bits. As a consequence of Proposition~\ref{prop:core-meta-converse-privacy}, we have that
		\begin{equation}
			\log_{2}K\leq E_R^{\varepsilon}(S_{A}K_{A};S_{B}K_{B})_{\omega}.
		\end{equation}
		Applying Propositions~\ref{prop-hypo_to_rel_ent} and \ref{prop:sandwich-to-htre}, we find that%
		\begin{align}
			\log_{2}K &  \leq\widetilde{E}_{\alpha}(S_{A}K_{A};S_{B}K_{B})_{\omega}+\frac{\alpha}{\alpha-1}\log_{2}\!\left(  \frac{1}{1-\varepsilon}\right),\label{eq:SKA-SEP-sim-conv-3}\\
			\log_{2}K &  \leq\frac{1}{1-\varepsilon}\left[  E_{R}(S_{A}K_{A};S_{B}K_{B})_{\omega}+h_{2}(\varepsilon)\right].\label{eq:SKA-SEP-sim-conv-4}%
		\end{align}
		Putting together \eqref{eq:SKA-SEP-sim-conv-1}, \eqref{eq:SKA-SEP-sim-conv-2}, \eqref{eq:SKA-SEP-sim-conv-3}, and \eqref{eq:SKA-SEP-sim-conv-4} concludes the proof.
	\end{Proof}

\section{Secret-Key-Agreement Capacities of Quantum Channels}

	In this section, we analyze the asymptotic capacities, and as before, the upper bounds for the asymptotic capacities are straightforward consequences of the non-asymptotic bounds given in Sections~\ref{sec:SKA-sq-ent-bnd-non-as} and \ref{sec:SKA-bnd-REE}. The definitions of these capacities are similar to what we have given previously, and so we only state them here briefly.

	\begin{definition}{Achievable Rate for Secret Key Agreement}{def-SKA_ach_rate}
		Given a quantum channel $\mathcal{N}$, a rate $R\in\mathbb{R}^{+}$ is called an achievable rate for secret key agreement over $\mathcal{N}$ if for all $\varepsilon\in(0,1]$, all $\delta>0$, and all sufficiently large~$n$, there exists an $(n,2^{n(R-\delta)},\varepsilon)$ secret-key-agreement protocol.
	\end{definition}

	\begin{definition}{Secret-Key-Agreement Capacity of a Quantum Channel}{def-SKA_cap}
		The secret-key-agreement capacity of a quantum channel $\mathcal{N}$, denoted by $P^{\leftrightarrow}(\mathcal{N})$, is defined as the supremum of all achievable rates, i.e.,%
		\begin{equation}
			P^{\leftrightarrow}(\mathcal{N})\coloneqq\sup\{R:R\text{ is an achievable rate for }\mathcal{N}\}.
		\end{equation}
	\end{definition}

	\begin{definition}{Weak Converse Rate for Secret Key Agreement}{def-SKA_weak_converse}
		Given a quantum channel $\mathcal{N}$, a rate $R\in\mathbb{R}^{+}$ is called a weak converse rate for secret key agreement over $\mathcal{N}$ if every $R^{\prime}>R$ is not an achievable rate for~$\mathcal{N}$.
	\end{definition}

	\begin{definition}{Strong Converse Rate for Secret Key Agreement}{def-SKA_strong_converse}
		Given a quantum channel $\mathcal{N}$, a rate $R\in\mathbb{R}^{+}$ is called a strong converse rate for secret key agreement over $\mathcal{N}$ if for all $\varepsilon\in\lbrack0,1)$, all $\delta>0$, and all sufficiently large $n$, there does not exist an $(n,2^{n(R+\delta)},\varepsilon)$ secret-key-agreement protocol.
	\end{definition}

	\begin{definition}{Strong Converse Secret-Key-Agreement Capacity of a Quantum Channel}{def-SKA_str_conv_cap}
		The strong converse secret-key-agreement capacity of a quantum channel $\mathcal{N}$, denoted by $\widetilde{P}^{\leftrightarrow}(\mathcal{N})$, is defined as the infimum of all strong converse rates, i.e.,%
		\begin{equation}
			\widetilde{P}^{\leftrightarrow}(\mathcal{N})\coloneqq\inf\{R:R\text{ is a strong converse rate for }\mathcal{N}\}.
		\end{equation}
	\end{definition}

	We have the exact same definitions for secret key agreement assisted by public separable channels, and we use the notation $P_{\operatorname{SEP}}^{\leftrightarrow}$ to refer to the public-separable-assisted secret-key-agreement capacity and $\widetilde{P}_{\operatorname{SEP}}^{\leftrightarrow}$ for the strong converse secret-key-agreement capacity assisted by public separable channels.

	Recall that, by definition, the following bounds hold%
	\begin{align}
		P^{\leftrightarrow}(\mathcal{N}) &  \leq\widetilde{P}^{\leftrightarrow}(\mathcal{N})\leq\widetilde{P}_{\operatorname{SEP}}^{\leftrightarrow}(\mathcal{N}),\\
		P^{\leftrightarrow}(\mathcal{N}) &  \leq P_{\operatorname{SEP}}^{\leftrightarrow}(\mathcal{N})\leq\widetilde{P}_{\operatorname{SEP}}^{\leftrightarrow}(\mathcal{N}).
	\end{align}

	As a direct consequence of the bound in Theorem~\ref{thm:SKA-sq-ent-bnd-LOCC-as-cap} and methods similar to those given in the proof of Theorem~\ref{thm-ea_classical_comm_weak_converse}, we find the following:

	\begin{theorem*}{Squashed-Entanglement Weak-Converse Bound}{thm-squash_ent_weak_conv}
		The squashed entanglement of a channel $\mathcal{N}$ is a weak converse rate for secret key agreement:%
		\begin{equation}
			P^{\leftrightarrow}(\mathcal{N})\leq E_{\operatorname{sq}}(\mathcal{N}).
		\end{equation}
	\end{theorem*}

	As a direct consequence of the bound in Theorem~\ref{thm:LAQC-max-REE-bnd-SEP-as-cap}\ and methods similar to those given in Section~\ref{sec-eacc_str_conv}, we find that

	\begin{theorem*}{Max-Relative Entropy of Entanglement Strong-Converse Bound}{thm-max_rel_ent_entanglement_strong_conv}
		The max-relative entropy of entanglement of a channel $\mathcal{N}$ is a strong converse rate for secret key agreement assisted by public separable channels:%
		\begin{equation}
			\widetilde{P}_{\operatorname{SEP}}^{\leftrightarrow}(\mathcal{N})\leq E_{\max}(\mathcal{N}).
		\end{equation}
	\end{theorem*}

	As a direct consequence of the bound in Theorem~\ref{thm:SKA-Renyi-REE-bnd-SEP-sim}\ and methods similar to those given in Section~\ref{sec-eacc_str_conv}, we find that

	\begin{theorem*}{Relative Entropy of Entanglement Strong-Converse Bound for Separable-Simulable Channels}{thm-rel_ent_entanglement_str_conv_sep_sim}
		Let $\mathcal{N}$ be a quantum channel that is separable-simulable with associated resource state $\theta_{SB^{\prime}}$. Then the relative entropy of entanglement of $\theta_{SB^{\prime}}$ is a strong converse rate for secret key agreement assisted by public separable channels:%
		\begin{equation}
			\widetilde{P}_{\operatorname{SEP}}^{\leftrightarrow}(\mathcal{N})\leq E_{R}(S;B')_{\theta}.
		\end{equation}
	\end{theorem*}

\section{Examples}

[IN PROGRESS]

\section{Bibliographic Notes}

Quantum key distribution is one of the first examples of a secret-key-agreement protocol conducted over a quantum channel \citep{BB84,E91}. Secret key agreement was considered in classical information theory by \citet{M93,AC93}. Secret key distillation from a bipartite quantum state was studied by a number of researchers, including \citet{DW05,HHHO05,C06,HHHLO08,HHHO09,CEHHOR07,CSW12}, well before secret key agreement over quantum channels was considered. One of the seminal insights in this domain, that a tripartite secret-key-distillation protocol is equivalent to a bipartite private-state distillation protocol, was made by \citet{HHHO05,HHHO09}. These authors also established the relative entropy of entanglement \citep{VP98} as an upper bound on the rate of a secret-key-distillation protocol. In earlier work, \citet{CLL04} observed that a separable state has no distillable secret key.

Secret key agreement over a quantum channel was formally defined by \citet{TGW14IEEE}, who also observed that the aforementioned insight extends to this setting, with more details being given by \citet{KW17}. The issue of unbounded shield systems resulting from secret-key-distillation or secret-key-agreement protocols was somewhat implicit in \citep{HHHO05,HHHO09} and discussed in more detail by \citet{CSW12,Wilde2016a,WTB16}. The relation between LOCC-assisted quantum communication and secret key agreement was discussed by \citet{WTB16}. The notion of secret key agreement assisted by public separable channels is original to this book (including the observation that various previously known bounds apply to these more general protocols).

The use of teleportation simulation and relative entropy of entanglement as a method for bounding rates of secret-key-agreement protocols was presented by \citet{PLOB15}. 

The amortized entanglement bound for secret-key-agreement protocols (Proposition~\ref{prop:SKA-amortized-bound}) was contributed by \citet{KW17}. Corollary~\ref{cor-SKA:reduc-by-TP}, as presented here, is due to \citet{KW17}.

The squashed entanglement upper bound on the rate of a secret-key-agreement protocol was established by \citet{TGW14IEEE,Wilde2016a}. Lemma~\ref{lem:SKA-log-K-to-info-measures} and Proposition~\ref{thm:SKA-bipartite-bound} are due to \citet{Wilde2016a}.

The use of sandwiched relative entropy of entanglement for bounding key rates was contributed by \citet{WTB16}. The max-relative entropy of entanglement of a state was introduced by \citet{Datta2009b,Dat09}, and the generalization to channels by \citet{CM17}. Lemmas~\ref{lem:SKA-alt-emax} and \ref{lem:SKA-alt-e-max-channel} were established by \citet{BW17}. Proposition~\ref{prop:SKA-amort-doesnt-help-e-max} was proven by \citet{CM17}. The proof that we follow here was given by \citet{BW17}. The interpretation of Proposition~\ref{prop:SKA-amort-doesnt-help-e-max} in terms of ``amortization collapse'' was given by \citet{BW17}. Theorem~\ref{thm:LAQC-max-REE-bnd-SEP-as-cap} was proven by \citet{CM17}. The precise statement of Theorem~\ref{thm:SKA-Renyi-REE-bnd-SEP-sim} is due to \citet{WTB16,KW17} (although it was stated for LOCC-simulable channels in these papers).

\end{mainmatter}

\begin{backmatter}

\chapter{Summary}
\fancyhead[L]{\nouppercase \leftmark}

[IN PROGRESS]

\end{backmatter}

\begin{appendices}
\cleardoublepage\phantomsection


\addtocontents{toc}{\protect\setcounter{tocdepth}{0}}
\addtocontents{ptc}{\protect\setcounter{tocdepth}{1}}

\pagestyle{appendix}

\chapter{Analyzing General Communication Scenarios}\label{chap-str_conv}

[IN PROGRESS]

\end{appendices}

{\pagestyle{plain}
\renewcommand{\UrlFont}{\ttfamily\footnotesize}
\footnotesize
\bibliography{sc}
\bibliographystyle{plainnat}
\cleardoublepage\phantomsection}

\printindex

\end{document}